# Statistics for Machine Learning with Mathematica® Applications

**Mohamed M. Hammad**

**2023**

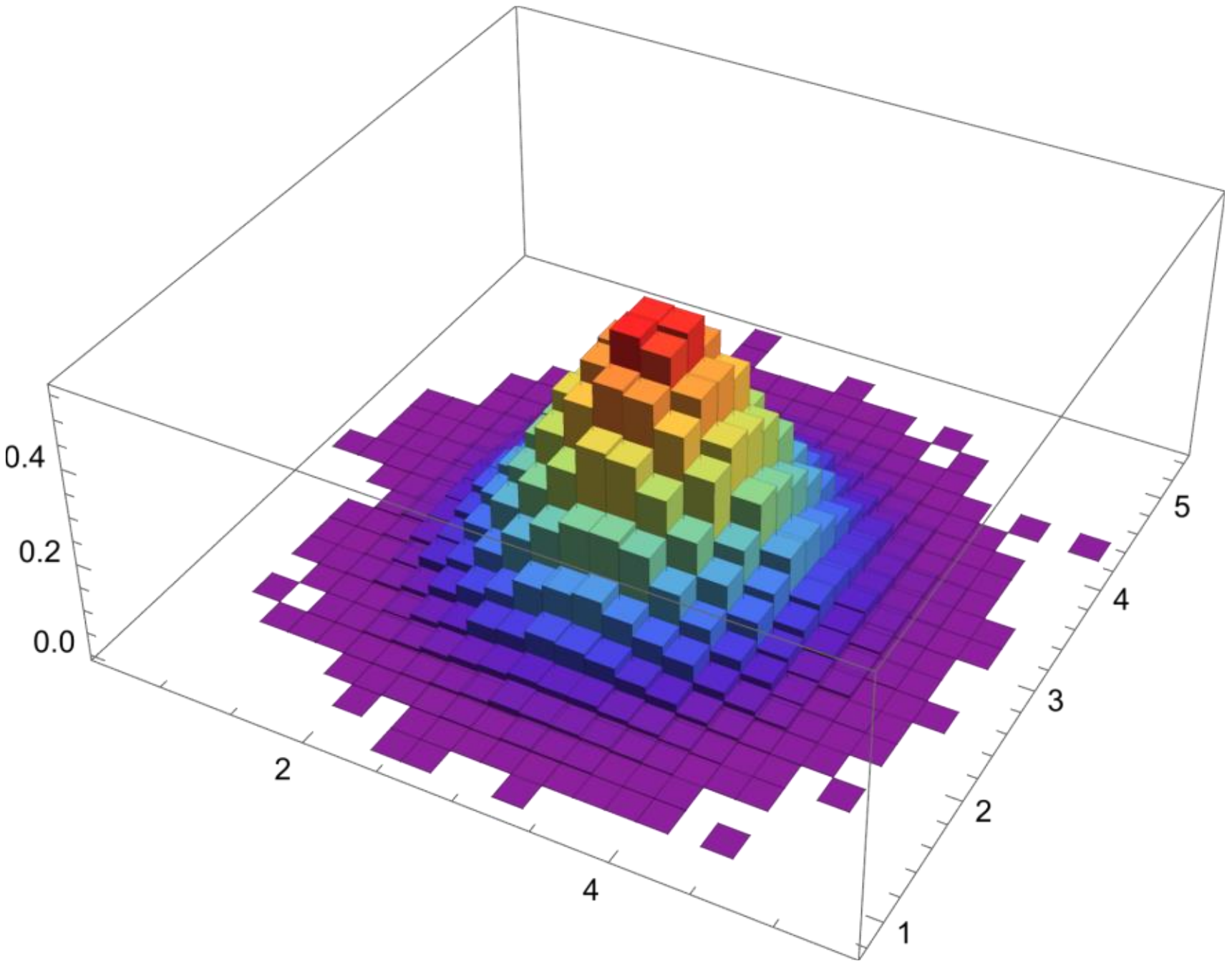

# Statistics for Machine Learning with Mathematica® Applications

**M. M. Hammad**

Department of Mathematics

Faculty of Science

Damanhour University, Egypt

https://orcid.org/0000-0003-0306-9719

m_hammad@sci.dmu.edu.eg

To

my

mother

# Preface

Statistics provides us with the means to make sense of data, uncover patterns, and draw meaningful conclusions from seemingly complex information. It has applications in a wide range of disciplines, including science, engineering, finance, social sciences, and many others. Meanwhile, Mathematica, a powerful computational software system, has been a fundamental tool for solving mathematical problems, performing symbolic calculations, and visualizing mathematical concepts. Bringing these two realms together opens up new possibilities for both novices and seasoned practitioners. By harnessing the capabilities of Mathematica, statisticians can handle large datasets, conduct sophisticated analyses, and communicate their findings more effectively.

In writing this book, our primary objective was to create a resource that would facilitate a comprehensive understanding of the fundamental concepts of statistics, essential for mastering the intricacies of machine learning, data science, and artificial intelligence. A central focus of this book is to introduce the reader to Mathematica, leveraging it as a powerful computational tool to bolster their statistical prowess. Not only does the book cater to beginners in the field, but it also aims to serve as a reference for seasoned data scientists, machine-learning practitioners, bio-statisticians, finance professionals, or engineers, who either possess prior knowledge of statistics or seek to fill gaps in their understanding.

There are several reasons why one might choose Mathematica for statistics:

- Comprehensive functionality: Mathematica provides a wide range of built-in statistical functions and capabilities. It can handle various statistical calculations, probability distributions, hypothesis testing, regression analysis, and more. This extensive functionality makes it a versatile tool for statisticians and data analysts.
- Symbolic computation: Mathematica is known for its strong symbolic computation capabilities. This allows users to work with mathematical expressions and perform symbolic manipulations, which can be particularly useful for theoretical statistical work and advanced modeling. This unique feature empowers users to perform analytical derivations, evaluate complex integrals, and solve differential equations involving statistical models. Such integration not only streamlines the workflow but also enhances the precision and rigor of statistical analyses.
- Data visualization: The software includes powerful data visualization tools that enable users to create informative plots and graphs to understand and present statistical results effectively. Utilizing built-in functions and interactive tools, Mathematica enables users to create a wide array of visually appealing charts, graphs, and plots. From simple histograms and scatter plots to 3D visualizations, users can efficiently present complex data patterns and relationships.
- Interactivity and dynamic interface: Mathematica has a dynamic and interactive interface that allows users to manipulate variables and parameters in real-time. This can be beneficial for exploring statistical concepts and conducting sensitivity analyses.
- Documentation and support: Mathematica is well-documented, and Wolfram Research, the company behind Mathematica, provides extensive resources and support to users. There are numerous tutorials, examples, and community forums to help users learn and troubleshoot any issues.
- Integration with other areas: Mathematica is not limited to statistics; it is a general-purpose computational platform. This means it can seamlessly integrate statistical analysis with other mathematical, engineering, or scientific computations within the same environment.
- Ease of use: For those familiar with the Mathematica language, performing statistical analyses can be straightforward and efficient. The language is designed to be expressive and readable, allowing users to perform complex tasks with concise code. With interactive notebooks, users can document and share their statistical workflows, promoting collaboration and reproducibility.

Whether you are a student seeking a thorough grounding in statistics or a professional researcher aiming to streamline your data analysis workflow, the book offers a comprehensive and practical guide to leveraging the combined power of statistics and Mathematica for data-driven decision-making and scientific discovery. By the end of the book, readers will have gained a deep understanding of statistical concepts and the proficiency to apply them effectively using Mathematica.

Most books on statistics tend to either be overly theoretical or present computational algorithms without enough mathematical background. The present work adopts a strategy that lies somewhere in the middle of these two directions. The computational statistics books present statistics methods to readers to carry out calculations manually or develop these algorithms on their own. This is obviously unrealistic for a typical introductory statistics course, which discusses a wide range of statistics strategies. By bridging the gap between theoretical principles and practical implementation, this book empowers learners to unlock the full potential of statistics in their pursuit of excellence in various domains.

Students studying engineering, operations research, data science, and mathematics at the undergraduate and graduate levels might find this book helpful. We have provided manually solved examples and examples solved using codes. We tried to provide proofs as simply as possible so that any reader with a background in calculus could easily follow them. In fact, the only prerequisites for engaging with the material are some basic programming experience and a fundamental understanding of mathematical notation. This ensures that learners from various backgrounds can dive into the world of statistics with confidence, using code as their guiding light to master the subject matter effectively.

The book takes a refreshingly code-centric approach. Nearly all the concepts we introduce are accompanied by illustrative code examples, lending practicality and tangibility to the learning experience. Even the figures in the book are generated using these code examples, emphasizing the code-first methodology. In order to ensure accessibility and ease of understanding, we have deliberately crafted the code examples in a simple format, prioritizing readability over efficiency and generality. In line with our instructional philosophy, each code example serves a dual purpose: not only does it demonstrate a specific statistical point but also simultaneously introduces and reinforces Mathematica programming concepts. Readers will learn how to leverage of Mathematica to perform complex statistical calculations, simulate data, and create visual representations of their findings.

The text covers topics like descriptive statistics, probability distributions, multivariate statistics, sampling theory, estimation theory, decision theory, hypothesis testing, and various inferential techniques, showcasing how Mathematica can handle sophisticated statistical models with ease. As readers progress through the book, they will find themselves empowered to explore a wide range of statistical techniques, from classical approaches to modern methodologies.

We offer our sincerest thanks to Professor Mohamed Abdalla Darwish the head of Department of Mathematics, Faculty of Science, Damanhour University, for his supporting. We express our deep gratitude to Professor Amr R. El Dhaba for his valuable discussions and for supporting this work. Also, we are grateful to many colleagues and friends for their invaluable feedback, thoughtful comments, constructive suggestions. These include Professor Hamed Awad, Dr. Fatma El-Safty, Dr. Hamdy El Shamy, Dr. Mohamed Elhaddad, Mohamed Yahia, Ayman A. Abdelaziz, Eman Farag, Hassan M. Shetawy, Walaa Mansour, Moaz El-Essawey, Aziza Salah, and Eman R. Hendawy.

We hope this book will give data scientists the tools they need to succeed in a data-driven world.

*knowledge itself is power - Sir Francis Bacon 1597*

Egypt 2023 M. M. Hammad

# Abstract

In recent years, the field of statistics has experienced a surge in interest and application, largely due to significant advances in computer technology. This progress has led to remarkable developments in statistics methods and algorithms, enabling their widespread adoption across various disciplines. Key areas benefiting from these advancements include machine learning, economics, finance, geophysics, molecular modeling, computational systems biology, operations research, and engineering. For example, in machine learning, statistics forms the foundation for algorithms used in regression, classification, clustering, and deep learning to analyze vast datasets and make predictions.

Mathematica, among other tools, has played a significant role in enabling the integration of statistics and computer technology, facilitating deeper exploration of data-driven insights and groundbreaking discoveries across diverse domains. With a rich library of functions, Mathematica allows users to calculate measures of central tendency, dispersion, and correlation, as well as perform hypothesis testing and estimation. Moreover, it supports probability distributions, making simulations and probabilistic modeling tasks more accessible.

This monograph presents the main theorems in mathematical statistics, ranging from basic descriptive statistics to sophisticated inferential techniques. In addition, we have created, more than 200 manipulates cover different scenarios in statistics, more than 500 light Mathematica codes (examples) and 25 programs (procedures) that follow the principles of testing hypotheses and estimation theory. The code will run as-is with no code from prior algorithms or third parties required beyond the installation of Mathematica.

## BRIEF CONTENTS

## CONTENTS

# CHAPTER 1

# MATHEMATICA LAB: INTRODUCTION TO MATHEMTICA

Mathematica is a powerful computational software program widely used in various fields of science, engineering, and mathematics. It provides a comprehensive environment for performing symbolic and numeric computations, creating visualizations, and solving complex problems. In this introductory chapter, we have covered the basic concepts and functions of Mathematica, including the following topics.

- Basic Concepts:
  Mathematica is built on a foundation of mathematical and computational concepts. It allows you to perform arithmetic operations, manipulate mathematical expressions, and solve equations. The software provides a vast collection of built-in functions for numerical calculations, algebraic manipulations, calculus, linear algebra, and more. These functions serve as fundamental tools for performing various tasks in Mathematica.
- Variables and Functions:
  In Mathematica, variables are used to store values that can be accessed and manipulated throughout a computation. You can assign values to variables using the assignment operator (`:=`) or the equal sign (`=`). Functions, on the other hand, are defined using the syntax: functionName[arguments]:= functionBody. Functions encapsulate a series of instructions that can be reused and called with different arguments.
- Lists:
  Lists are an essential data structure in Mathematica. They allow you to store collections of values, such as numbers, strings, or even other lists. Lists can be created using curly braces (`{}`) and elements are separated by commas. Mathematica provides a rich set of built-in functions for manipulating and operating on lists, including appending, deleting, sorting, and extracting elements.
- 2D and 3D Graphing:
  Mathematica offers powerful graphing capabilities for visualizing mathematical functions and data in two or three dimensions. You can plot functions using the `Plot` and `ParametricPlot` functions for 2D graphs, and `Plot3D` and `ParametricPlot3D` functions for 3D graphs. These functions allow you to customize various aspects of the plots, such as axes labels, plot ranges, colors, and styles.
- Control Structures:
  Control structures in Mathematica enable you to control the flow of execution in your programs. They include conditionals (`If`, `Switch`), loops (`For`, `While`), and functional programming constructs (`Map`, `Fold`, `Nest`). These control structures allow you to make decisions, iterate over lists or ranges of values, and perform operations on collections of data.
- Modules, Blocks, and Local Variables:
  Mathematica provides mechanisms for creating modular and reusable code through modules and blocks. Modules allow you to encapsulate a group of variables and functions, providing a local scope for their usage. Blocks are similar but are primarily used for scoping variables and expressions dynamically. Local variables defined within modules or blocks are not visible outside their scope, ensuring better code organization and reducing potential conflicts.

**Chapter 1 Outline**

# UNIT 1.1

# BASIC CONCEPTS

Mathematica is a computer algebra system that performs numeric, symbolic, and graphical computations. Although Mathematica can be used as a programming language, its high-level structure is more appropriate for performing sophisticated operations through the use of built-in functions. For example, Mathematica can find limits, derivatives, integrals, and determinants, as well as plot the graph of functions and perform symbolic computations. The number of built-in functions in Mathematica is enormous. Our goals in this introductory chapter are modest. Namely, we introduce a small subset of Mathematica commands necessary to explore Mathematica discussed in this book.

### Notebooks

A notebook is a document that allows us to interact with Mathematica. Each notebook is divided up into a sequence of individual units called cells, each containing a specific type of information such as text, graphics, input, or output. Text cells contain information to be read by the user but contain no executable Mathematica commands. The following cell, displaying `In[1]` $2^{20}$, is an example of an input cell containing executable Mathematica commands. Mathematica computes the value of $2^{20}$ and the results of the calculation are displayed as `Out[1]:1048576` in an output cell. When we create a new cell, the default cell type is an input cell. Suppose instead, we want to create a text cell. To do this, use the mouse to click on an area where we want to create a new cell and a horizontal line will appear. Then from the Format menu, select Style and then Text. A new text cell will then be created as soon as we begin typing. We can experiment with creating other types of cells by selecting a cell style of our choice, after first choosing Format and Style from the menu.

### Palettes

A palette is similar to a set of calculator buttons, providing shortcuts to entering commands and symbols into a notebook. The name of a useful palette is “Basic Math Assistant Input” and it can be found by selecting the Palettes menu and then Basic Math Assistant. After opening Basic Math Assistant, drag it to the right side of the screen and resize the notebook, if necessary, so that both the notebook and palette are visible in non-overlapping windows. To demonstrate the usefulness of palettes, suppose we wish to calculate $\sqrt{804609}$. The Mathematica command for computing the square root of n is `Sqrt[n]`. The following input cell was created by typing in the information exclusively from the keyboard.

```
Input     Sqrt[804609]
Output    897
```

A quicker and more natural way of entering $\sqrt{804609}$ can be accomplished by clicking on the square root button √□ in the palette and then entering 804609.

```
Input     √(804609)
Output    897
```

### Packages

Note that, many of Mathematica functions are available at startup, but additional specialized functions are available from add-in packages. You can load a built-in or installed package in two ways, with the `Needs[ ]` function or with the symbols `<<`. The package name has quotation marks if you use the `Needs[ ]` function, but does not has a mark with `<<`. Package names are always indicated with a backward apostrophe at the end of the name, `` ` ``.

```
Input      Needs["PackageName`"]
Input      <<PackageName`
```

## Help with Mathematica

There are four important tricks to keep in mind to help with Mathematica:

1- If you want to know something about a Mathematica function or procedure, just type ? followed by a Mathematica command name, and then enter the cell to get information on that command.

```
Input    ?FactorInteger
```

After you press Enter, Mathematica responds:

```
Output   FactorInteger[n] gives a list of the prime factors of the integer n, together
         with their exponents.
```

2- Mathematica also can finish typing a command for you if you provide the first few letters. Here is how it works: After typing a few letters choose Complete Selection from the Edit menu. If more than one completion is possible, you will be presented with a pop-up menu containing all of the options. Just click on the appropriate choice.

3- If you know the name of a command but have forgotten the syntax for its arguments, type the command name in an input cell, then choose Make Template from the Edit menu. Mathematica will paste a template into the input cell showing the syntax for the simplest form of the command. For example, if you typed `Plot`, and then choose Make Template, the input cell would look like this:

```
Input      Plot[f,{x,xmin, xmax}]
```

4- The Wolfram Documentation is the most useful feature imaginable; learn to use it and use it often. Go to the Help menu and choose Wolfram Documentation. A window will appear displaying the documentation home page.

## Document the Code

When you write programs in the Wolfram Language, there are various ways to document your code. As always, by far the best thing is to write clear code and to name the objects you define as explicitly as possible. Sometimes, however, you may want to add some "commentary text" to your code, to make it easier to understand. You can add such text at any point in your code simply by enclosing it in matching `(*  *)`. Notice that in the Wolfram Language, "comments" enclosed in `(*  *)` can be nested in any way.

| | |
|---|---|
| `(* text *)` | a comment that can be inserted anywhere in Wolfram Language code. |

***Mathematica Examples 1.1***

```
Input    If[a>b,(*then*)p,(*else*)q]
Output   If[a>b,p,q]
```

## Arithmetic Operations

Mathematica can be thought of as a sophisticated calculator, able to perform exact as well as approximate arithmetic computations. You can always control grouping the arithmetic computations by explicitly using parentheses. The following list summarizes the Mathematica symbols used for addition, subtraction, multiplication, division, and powers.

| | |
|---|---|
| `x+y+z` | gives the sum of three numbers. |
| `x*y*z, x×y×z, or x y z` | represents a product of terms. |
| `x-y` | is equivalent to $x + (-1 * y)$. |
| `x^y` | gives x to the power y. |
| `x/y` | is equivalent to x y^ − 1. |

***Mathematica Examples 1.2***

```
Input    2.3+5.63
```

```
Output   7.93

Input    2.4/8.9^2
Output   0.0302992

Input    2*3*4
Output   24

Input    (3+4)^2-2(3+1)
Output   41
```

The precedence of common operators is generally defined so that "higher-level" operations are performed first. For simple expressions, operations are typically ordered from highest to lowest in order: 1. Parenthesization, 2. Factorial, 3. Exponentiation, 4. Multiplication and division, 5. Addition and subtraction. Consider the expression `3×7+2^2`. This expression has a value `(3×7)+(2^2)=25`.

Mathematica has several built-in constants. The three most commonly used constants are $\pi$, $e$, and $i$. You can find each of these constants on the Basic Math Assistant palette. Some built-in constants are listed below.

| | |
|---|---|
| `I` | ($i = \sqrt{(-1)}$). |
| `E` | (2.71828). |
| `Pi` | ($\pi = 3.14159$). |

### Relational and Logical Operators

Relational and logical operators are instrumental in program flow control. They are used in Mathematica to test various conditions involving variables and expressions. The relational operators are listed below.

| | |
|---|---|
| `lhs==rhs` | returns True if lhs and rhs are identical. |
| `lhs!=rhs or lhs≠rhs` | returns False if lhs and rhs are identical. |
| `x>y` | yields True if x is determined to be greater than y. |
| `x>=y or x≥y` | yields True if x is determined to be greater than or equal to y. |
| `x<y` | yields True if x is determined to be less than y. |
| `x<=y or x≤y` | yields True if x is determined to be less than or equal to y. |

Logical operators are used to negate or combine relational expressions. The standard logical operators are listed below.

| | |
|---|---|
| $e_1$`&&`$e_2$`&&...` | is the logical AND function. It evaluates its arguments in order, giving False immediately if any of them are False, and True if they are all True. |
| $e_1$`\|\|`$e_2$`\|\|...` | is the logical OR function. It evaluates its arguments in order, giving True immediately if any of them are True, and False if they are all False. |
| `!expr` | is the logical NOT function. It gives False if expr is True, and True if it is False. |

***Mathematica Examples 1.3***

```
Input    10<7
Output   False

Input    Pi^E<E^Pi
Output   True

Input    2+2==4
Output   True

Input    (*Represent an equation:*)
         x^2==1+x
```

```
Output   x^2==1+x

Input    (* Returns True if elements are guaranteed unequal, and otherwise stays unevaluated:
         *)
         a!=b
Output   a!=b

Input    1!=2
Output   True

Input    1>2||Pi>3
Output   True

Input    2>1&&Pi>3
Output   True

Input    (3<5)||(4<5)
Output   True

Input    (3<5)&&!(4>5)
Output   True
```

### Elementary Functions

In the following, we discuss some of the more commonly used functions Mathematica offers. The Wolfram Language has nearly 6000 built-in functions. All have names in which each word starts with a capital letter. Remember that the argument of a function must be contained within square brackets, [ ]. Arguments to functions are always separated by commas.

### Common Functions

| | |
|---|---|
| Log[z] | gives the natural logarithm of z (logarithm to base *e*). |
| Log[b,z] | gives the logarithm to base b. |
| Exp[z] | gives the exponential of z. |
| Sqrt[z] or √z | gives the square root of z. |
| N[expr] | gives the numerical value of expr. |
| Abs[z] | gives the absolute value of the real or complex number z. |
| Floor[x] | gives the greatest integer less than or equal to x. |

### Trigonometric Functions

| | |
|---|---|
| Sin[z] | gives the sine of z. |
| Cos[z] | gives the cosine of z. |
| Tan[z] | gives the tangent of z. |

### Hyperbolic Functions

| | |
|---|---|
| Sinh[z] | gives the hyperbolic sine of z. |
| Cosh[z] | gives the hyperbolic cosine of z. |
| Tanh[z] | gives the hyperbolic tangent of z. |

### Numerical Functions

| | |
|---|---|
| IntegerPart[x] | integer part of x. |
| `FractionalPart[x]` | fractional part of x. |
| `Round[x]` | integer x closest to x. |
| `Max[x₁,x₂,...]` | the maximum of $x_1, x_2, \ldots$. |
| `Min[x₁,x₂,...]` | the minimum of $x_1, x_2, \ldots$. |
| `Re[z]` | the real part Re z. |
| `Im[z]` | the imaginary part Im z. |
| `Conjugate[z]` | the complex conjugate $z^*$. |

### Combinatorial Functions

| | |
|---|---|
| `n!` | factorial $n(n-1)(n-2)\ldots \times 2 \times 1$. |
| `n!!` | double factorial $n(n-2)(n-4)\ldots \times 3 \times 1$. |
| `Binomial[n,m]` | binomial coefficient $\binom{n}{m} = \frac{(n!)}{(m!(n-m)!)}$. |
| `Multinomial[n₁,n₂,...]` | multinomial coefficient $(n_1 + n_2 + \ldots)/(n_1!\, n_2! \ldots)$. |

***Mathematica Examples 1.4***

```
Input    Log[10,1000]
Output   3

Input    Exp[I Pi/5]
Output   E^((I \[Pi])/5)

Input    Sin[Pi/3]
Output   Sqrt[3]/2

Input    Sinh[1.4]
Output   1.9043

Input    N[1/7]
Output   0.142857

Input    Floor[2.4]
Output   2

Input    30!
Output   265252859812191058636308480000000

Input    Binomial[n,2]
Output   1/2 (-1+n) n

Input    Multinomial[6,5]
Output   462
```

### Sum and Product Functions

Sums and products are of fundamental importance in mathematics, and Mathematica makes their computation simple. Unlike other computer languages, initialization is automatic and the syntax is easy to apply, particularly if the Basic Math Assistant Input palette is used. Any symbol may be used as the index of summation. Negative increments are permitted wherever an increment is used.

| | |
|---|---|
| `Sum[f,{i,imax}]` | evaluates the $\sum_{i=1}^{i_{max}} f$. |
| `Sum[f,{i,imin,imax}]` | starts with $i = i_{min}$. |
| `Sum[f,{i,imin,imax,di}]` | uses steps di. |
| `Sum[f,{i,{i₁,i₂,…}}]` | uses successive values $i_1, i_2, \ldots$. |
| `Sum[f,{i,imin,imax},{j,jmin,jmax},…]` | evaluates the multiple sum $\sum_{i=1}^{i_{max}} \sum_{j=1}^{j_{max}} f$. |
| `Product[f,{i,imax}]` | evaluates the $\prod_{i=1}^{i_{max}} f$. |
| `Product[f,{i,imin,imax}]` | starts with $i = i_{min}$. |
| `Product[f,{i,imin,imax,di}]` | uses steps di. |
| `Product[f,{i,{i₁,i₂,…}}]` | uses successive values $i_1, i_2, \ldots$. |
| `Product[f,{i,imin,imax},{j,jmin,jmax},…]` | evaluates the multiple sum $\prod_{i=1}^{i_{max}} \prod_{j=1}^{j_{max}} f$. |

***Mathematica Examples 1.5***

```
Input   (* Numeric sum: *)
        Sum[i^2,{i,10}]
Output  385

Input   (* Symbolic sum: *)
        Sum[i^2,{i,1,n}]
Output  1/6 n (1+n) (1+2 n)

Input   Sum[1/i^6,{i,1,Infinity}]
Output  π^6/945

Input   (* Multiple sum with summation over j performed first: *)
        Sum[1/(j^2 (i+1)^2),{i,1,Infinity},{j,1,i}]
Output  π^4/120

Input   Product[i^2,{i,1,6}]
Output  518400

Input   Product[i^2,{i,1,n}]
Output  (n!)^2

Input   Product[2^(j+i),{i,1,p},{j,1,i}]
Output  2^(1/2 p (1 + p)^2)
```

### Limit and Series Functions

| | |
|---|---|
| `Limit[expr,x->x0]` | finds the limiting value of expr when x approaches $x_0$. |
| `Series[f,{x,x0,n}]` | generates a power series expansion for f about the point $x=x_0$ to order $(x-x_0)^n$. |

***Mathematica Examples 1.6***

```
Input   Limit[(Sin[x])/x,x->0]
Output  1

Input   Limit[(1+x/n)^n,n->Infinity]
Output  E^x

Input   (* Power series for the exponential function around x=0: *)
        Series[Exp[x],{x,0,10}]
Output  1+x+x^2/2+x^3/6+x^4/24+x^5/120+x^6/720+x^7/5040+x^8/40320+x^9/362880+x^10/3628800+O[x]^11

Input   (* Power series of an arbitrary function around x=a: *)
        Series[f[x],{x,a,3}]
Output  f[a]+f'[a] (x-a)+1/2 f''[a] (x-a)^2+1/6 f^(3)[a] (x-a)^3+O[x-a]^4

Input   Series[x^x,{x,0,4}]
Output  1 + Log[x] x + 1/2 Log[x]^2 x^2 + 1/6 Log[x]^3 x^3 + 1/24 Log[x]^4 x^4 + O[x]^5
```

### Differentiation Function

| | |
|---|---|
| `D[f,x]` | gives the partial derivative $\frac{\partial}{\partial x} f$. |
| `D[f,x,y,...]` | gives the derivative $\frac{\partial}{\partial x}\frac{\partial}{\partial y} \ldots f$. |
| `D[f,{x,n}]` | gives the multiple derivative $\frac{\partial^n}{\partial x^n} f$. |

***Mathematica Examples 1.7***

```
Input   (* Derivative with respect to x: *)
        D[x^n,x]
Output  n x^(-1 + n)

Input   (* Fourth derivative with respect to x: *)
        D[Sin[x]^10,{x,4}]
Output  5040 Cos[x]^4 Sin[x]^6 - 4680 Cos[x]^2 Sin[x]^8 + 280 Sin[x]^10

Input   (* Derivative with respect to x and y: *)
        D[(Sin[x y])/((x^2+y^2)),x,y]
Output  -((2 x^2 Cos[x y])/(x^2 + y^2)^2) - (2 y^2 Cos[x y])/(x^2 + y^2)^2 + Cos[x
        y]/(x^2 + y^2) + (8 x y Sin[x y])/(x^2 + y^2)^3 - (x y Sin[x y])/(x^2 + y^2)

Input   (* Derivative involving a symbolic function f: *)
        D[x f[x] f'[x],x]
Output  f[x] f'[x]+x f'[x]^2+x f[x] f''[x]

Input   D[Sin[x] Cos[x+y],x,y]
Output  -Cos[x+y] Sin[x]-Cos[x] Sin[x+y]

Input   D[ArcCoth[x],{x,2}]
Output  (2 x)/(1 - x^2)^2
```

### Integration Functions

| | |
|---|---|
| `Integrate[f,x]` | gives the indefinite integral $\int f\, d x$. |
| `Integrate[f,{x,xmin,xmax}]` | gives the definite integral $\int_{x_{min}}^{x_{max}} f\, d x$. |
| `Integrate[f,{x,xmin,xmax},{y,ymin,ymax},…]` | gives the multiple integral $\int_{x_{min}}^{x_{max}} d x \int_{y_{min}}^{y_{max}} dy \ldots f$ . |

***Mathematica Examples 1.8***

```
Input   (* Compute an indefinite integral: *)
        Integrate[1/((x^3+1)),x]
Output  ArcTan[(-1 + 2 x)/Sqrt[3]]/Sqrt[3] + 1/3 Log[1 + x] - 1/6 Log[1 - x + x^2]

Input   \[Integral]Sqrt[x+Sqrt[x]]\[DifferentialD]x
Output  1/12 Sqrt[Sqrt[x] + x] (-3 + 2 Sqrt[x] + 8 x) + 1/4 ArcTanh[Sqrt[Sqrt[x] +
        x]/Sqrt[x]]

Input   Integrate[1/((x^4+x^2+1)),{x,0,Infinity}]
Output  \[Pi]/(2 Sqrt[3])

Input   (* Compute an definite integral: *)
        Integrate[x/(Sqrt[1-x]),{x,0,1}]
Output  4/3

Input   Integrate[1/(((2+x^2) Sqrt[4+3 x^2])),{x,-Infinity,Infinity}]
Output  ArcCosh[Sqrt[3/2]]

Input   Integrate[x^2+y^2,{x,0,1},{y,0,x}]
Output  1/3
```

### Algebraic Operations

Mathematica has many functions for transforming algebraic expressions. The following list summarizes them.

| | |
|---|---|
| `Simplify[expr]` | performs a sequence of algebraic and other transformations on expr and returns the simplest form it finds. |
| `Expand[expr]` | expands out products and positive integer powers in expr. |
| `Factor[expr]` | factors a polynomial over the integers. |
| `Together[expr]` | puts terms in a sum over a common denominator, and cancels factors in the result. |
| `ExpandAll[expr]` | expands out all products and integer powers in any part of expr. |
| `FunctionExpand[expr]` | tries to expand out special and certain other functions in expr when possible reducing compound arguments to simpler ones. |
| `Reduce[expr,vars]` | reduces the statement expr by solving equations or inequalities for vars and eliminating quantifiers. |

***Mathematica Examples 1.9***

```
Input    Simplify[Sin[x]^2+Cos[x]^2]
Output   1

Input    Expand[(1+x)^10]
Output   1 + 10 x + 45 x^2 + 120 x^3 + 210 x^4 + 252 x^5 + 210 x^6 + 120 x^7 + 45 x^8
         + 10 x^9 + x^10

Input    Factor[x^10-1]
Output   (-1 + x) (1 + x) (1 - x + x^2 - x^3 + x^4) (1 + x + x^2 + x^3 + x^4)

Input    Together[x^2/(x^2-1)+x/(x^2-1)]
Output   x/(-1+x)

Input    (*Expand polynomials anywhere inside an expression:*)
         ExpandAll[1/(1+x)^3+Sin[(1+x)^3]]
Output   1/(1 + 3 x + 3 x^2 + x^3) + Sin[1 + 3 x + 3 x^2 + x^3]

Input    FunctionExpand[Sin[24 Degree]]
Output   -(1/8) Sqrt[3] (-1 - Sqrt[5]) - 1/4 Sqrt[1/2 (5 - Sqrt[5])]
```

### Solving Equations

Solutions of general algebraic equations may be found using the `Solve` command. `Solve` always tries to give you explicit formulas for the solutions to equations. However, it is a basic mathematical result that, for sufficiently complicated equations, explicit algebraic formulas in terms of radicals cannot be given. If you have an algebraic equation in one variable, and the highest power of the variable is at most four, then the Wolfram Language can always give you formulas for the solutions. However, if the highest power is five or more, it may be mathematically impossible to give explicit algebraic formulas for all the solutions.

You can also use the Wolfram Language to solve sets of simultaneous equations. You simply give the list of equations and specify the list of variables to solve for. Not all algebraic equations are solvable by Mathematica, even if theoretical solutions exist. If Mathematica is unable to solve an equation, it will represent the solution in a symbolic form. For the most part, such solutions are useless, and a numerical approximation is more appropriate. Numerical approximations are obtained with the command `NSolve`.

| | |
|---|---|
| `Solve[lhs==rhs,x]` | solve an equation for x. |
| `Solve[{lhs1==rhs1,lhs2==rhs2,…},{x,y,…}]` | solve a set of simultaneous equations for x, y, .... |
| `Eliminate[{lhs1==rhs1,lhs2==rhs2,…},{x,…}]` | eliminate x, ... in a set of simultaneous equations. |
| `Reduce[{lhs1==rhs1,lhs2==rhs2,…},{x,y,…}]` | give a set of simplified equations, including all possible solutions. |
| `NSolve[expr,vars]` | attempts to find numerical approximations to the solutions of the system expr of equations or inequalities for the variables vars. |
| `FindRoot[f,{x,x0}]` | searches for a numerical root of f, starting from the point $x = x_0$. |

***Mathematica Examples 1.10***

```
Input    Solve[x^2+a x+1==0,x]
Output   {{x -> 1/2 (-a - Sqrt[-4 + a^2])}, {x -> 1/2 (-a + Sqrt[-4 + a^2])}}

Input    Solve[a x+y==7&&b x-y==1,{x,y}]
Output   {{x->8/(a+b),y->-((a-7 b)/(a+b))}}

Input    (* Eliminate the variable y between two equations: *)
         Eliminate[{x==2+y,y==z},y]
Output   2+z==x

Input    NSolve[x^5-2 x+3==0,x,Reals]
Output   {{x->-1.42361}}

Input    Reduce[x^2-y^3==1,{x,y}]
Output   y == (-1 + x^2)^(1/3) || y == -(-1)^(1/3) (-1 + x^2)^(1/3) || y == (-1)^(2/3)
         (-1 + x^2)^(1/3)

Input    FindRoot[Sin[x]+Exp[x],{x,0}]
Output   {x->-0.588533}
```

## Some Notes

1- In doing calculations, you will often need to use previous results that you have got. In the Wolfram Language, `%` always stands for your last result.

| | |
|---|---|
| `%` | the last result generated. |
| `%%` | the next-to-last result. |
| `% n` | the result on output line Out[n]. |
| `Out[n]` | is a global object that is assigned to be the value produced on the $n^{th}$ output line. |

***Mathematica Examples 1.11***

```
Input    77^2
Output   5929

Input    %+1
Output   5930

Input    3 %+%^2+%%
Output   35188619

Input    % 2+% 3
Output   175943095
```

2- Although Mathematica is a powerful calculating tool, it has its limits. Sometimes it will happen that the calculations you tell Mathematica to do are too complicated or may be the output produced is too long. In these cases, Mathematica could be calculating for too long to get an output so you might want to stop these calculations. To abort a calculation: go to "Kernel" and select "Abort evaluation". It can take long to abort a calculation. If the computer does not respond an alternative is to close down the Kernel. By doing this you do not lose the data displayed in your notebooks, but you do lose all the results obtained so far from the Kernel, so in case you are running a series of calculations, you would have to start again. To close down the Kernel: go to "Kernel" and select "Quit Kernel" and then "Local". Closing down the Kernel is not a practice that is done only when you want to stop a calculation. Sometimes, when you have been using Mathematica for a long time you forget about the definitions and calculations that you have done before (you might have defined values for variables or functions, for example). Those definitions can clash with the calculations you are doing, so you might want to close down the Kernel and start your new calculations from scratch. In general, it is a good idea to close down the Kernel after you have finished with a series of calculations so that when you move to a different problem your new calculations do not interact with the previous ones.

# UNIT 1.2

# VARIABLES AND FUNCTIONS

When you perform long calculations, it is often convenient to give names to your intermediate results. Just as in standard mathematics, or other computer languages, you can do this by introducing named variables. It is very important to realize that the values you assign to variables are permanent. Once you have assigned a value to a particular variable, the value will be kept until you explicitly remove it. The value will, of course, disappear if you start a whole new Wolfram Language session.

| | |
|---|---|
| `x=value` | assign a value to the variable x. |
| `x=y=value` | assign a value to both x and y. |
| `x=.or Clear[x]` | remove any value assigned to x. |
| `{x,y}={value1,value2}` | assign different values to x and y. |
| `{x,y}={y,x}` | interchange the values of x and y. |

***Mathematica Examples 1.12***

```
Input     x=5
Output    5

Input     x^2
Output    25

Input     x=7+4
Output    11
```

In Mathematica, one can substitute an expression with another using rules. In particular one can substitute a variable with a value without assigning the value to the variable.

| | |
|---|---|
| `lhs:=rhs` | assigns rhs to be the delayed value of lhs. rhs is maintained in an unevaluated form. When lhs appears, it is replaced by rhs, evaluated afresh each time. |
| `expr/.rules` | applies a rule or list of rules in an attempt to transform each subpart of an expression expr. |
| `lhs->rhs or lhs->rhs` | represents a rule that transforms lhs to rhs. |

***Mathematica Examples 1.13***

```
Input     x=5
Input     y:=x+2
Input     y
Output    5
Output    7

Input     x=10
Output    10

Input     y
Output    12
```

***Mathematica Examples 1.14***

```
Input     x+y/. x->2
Output    2+y
```

```
Input    x+y/. {x->a,y->b}
Output   a+b

Input    x^2+y/.x->y/.y->x
Output   x + x^2

Input    x+2 y/.{x->y,y->a}
Output   2 a+y
```

The last example reveals that Mathematica goes through the expression only once and replaces the rules. If we need Mathematica to go through the expression again and replace any expression which is possible until no substitution is possible, one uses `//.` . In fact `/.` and `//.` are shorthand for `Replace` and `ReplaceRepeated`, respectively.

***Mathematica Examples 1.15***

```
Input    {x,x^2,a,b}/. x->3
Output   {3,9,a,b}

Input    x+2 y//.{x->y,y->a}
Output   3 a

Input    x+2 y//.{x->b,y->a,b->c}
Output   2 a+c

Input    x+2 y/. {x->b,y->a,b->c}
Output   2 a+b

Input    Sin[x]/. Sin->Cos
Output   Cos[x]
```

There are many functions that are built into the Wolfram Language. Here we discuss how you can add your own simple functions to the Wolfram Language. As a first example, consider adding a function called f which squares its argument. The Wolfram Language command to define this function is `f[x]:=x^2`. The names like f that you use for functions in the Wolfram Language are just symbols. Because of this, you should make sure to avoid using names that begin with capital letters, to prevent confusion with built-in Wolfram Language functions. You should also make sure that you have not used the names for anything else earlier in your session.

| | |
|---|---|
| `f[x]=value` | definition for a specific expression x. |
| `f[x_]=value` | definition for any expression, referred to as x. |
| `Clear[f]` | clear all definitions for f. |
| `Function[x,body]` | is a pure function with a single formal parameter x. |
| `Function[{x_1,x_2,…},body]` | is a pure function with a list of formal parameters. |
| `Map[f,expr] or f/@expr` | applies f to each element on the first level in expr. |
| `Map[f,expr,levelspec]` | applies f to parts of expr specified by levelspec. |

The character `_` (referred to as "blank") on the left-hand side is very important.

***Mathematica Examples 1.16***

```
Input    f[x_]:=x^2
         f[a+1]
Output   (1+a)^2

Input    f[4]
Output   16

Input    f[3 x+x^2]
Output   (3 x+x^2)^2
```

```
Input     Expand[f[(x+1+y)]]
Output    1+2 x+x^2+2 y+2 x y+y^2

Input     Function[u,3+u][x]
Output    3+x

Input     Function[{u,v},u^2+v^4][x,y]
Output    x^2+y^4

Input     (* Evaluate f on each element of a list: *)
          Map[f,{a,b,c,d,e}]
Output    {f[a],f[b],f[c],f[d],f[e]}

Input     f/@{a,b,c,d,e}
Output    {f[a],f[b],f[c],f[d],f[e]}

Input     (* Map at top level: *)
          Map[f,{{a,b},{c,d,e}}]
Output    {f[{a,b}],f[{c,d,e}]}

Input     (* Map at level 2: *)
          Map[f,{{a,b},{c,d,e}},{2}]
Output    {{f[a],f[b]},{f[c],f[d],f[e]}}

Input     (*Map at levels 1 and 2:*)
          Map[f,{{a,b},{c,d,e}},2]
Output    {f[{f[a],f[b]}],f[{f[c],f[d],f[e]}]}
```

One can define functions of several variables. Here is a simple example defining $f(x,y) = \sqrt{x^2 + y^2}$.

***Mathematica Examples 1.17***

```
Input     f[x_,y_]:=Sqrt[x^2+y^2]
Input     f[3,4]

Output    5
```

## Some Notes

1- There are four kinds of bracketing used in the Wolfram Language. Each kind of bracketing has a very different meaning.

| | |
|---|---|
| `(term)` | parentheses for grouping. |
| `f[x]` | square brackets for functions. |
| `{a,b,c}` | curly braces for lists. |
| `v[[i]]` | double brackets for indexing (Part[v,i]). |

2- Compound expression

| | |
|---|---|
| `expr₁;expr₂;expr₃` | do several operations and give the result of the last one. |
| `expr₁;expr₂;` | do the operations but print no output. |
| `expr;` | do an operation but display no output. |

***Mathematica Examples 1.18***

```
Input     x=4;y=6;z=y+6
Output    12

Input     a=2;b=3;a+b
Output    5
```

3- Particularly when you write procedural programs in the Wolfram Language, you will often need to modify the value of a particular variable repeatedly. You can always do this by constructing the new value and explicitly performing an assignment such as `x=value`. The Wolfram Language, however, provides special notations for incrementing the values of variables, and for some other common cases.

| | |
|---|---|
| `i++` | increment the value of i, by 1 returning the old value of i. |
| `i--` | decrement the value of i, by 1 returning the old value of i. |
| `++i` | pre-increment i, returning the new value of i. |
| `--i` | pre-decrement i, returning the new value of i. |
| `i+=di` | add di to the value of i and returns the new value of i. |
| `i-=di` | subtract di from i and returns the new value of i. |
| `x*=c` | multiply x by c. |
| `x/=c` | divide x by c. |

***Mathematica Examples 1.19***

```
Input    k=1;k++
Output   1

Input    k
Output   2

Input    k=x
Output   x

Input    k++
Output   x

Input    k
Output   1+x

Input    k=1;++k
Output   2

Input    k
Output   2

Input    k=1;k--
Output   1

Input    k
Output   0

Input    k=1;k-=5
Output   -4
Input    k
Output   -4
```

4- Primarily there are three equalities in Mathematica, `=`, `:=`, `==`. There is a fundamental differences between `=` and `:=` explained in the following examples:

***Mathematica Examples 1.20***

```
Input    x=5;y=x+2;
Input    y
Output   7

Input    x=10
Output   10
```

```
Input    x=5;y:=x+2;
Input    y
Output   7

Input    x=10
Output   10
```

```
Input     y                  Input     y
Output    7                  Output    12

Input     x=15               Input     x=15
Output    15                 Output    15

Input     y                  Input     y
Output    7                  Output    17
```

It is clear that when we defined `y=x+2` then y takes the value of x+2 and this will be assigned to y. No matter if x changes its value, the value of y remains the same. In other words, y is independent of x. But in `y:=x+2`, y is dependent on x, and when x changes, the value of y changes too. Namely using `:=` then y is a function with variable x. Finally, the equality `==` is used to compare:

***Mathematica Examples 1.21***

```
Input     5==5
Output    True

Input     3==5
Output    False
```

# UNIT 1.3

# LISTS

Lists are extremely important objects. In doing calculations, it is often convenient to collect together several objects and treat them as a single entity. Lists give you a way to make collections of objects. Lists are sequences of Mathematica objects separated by commas and enclosed by curly brackets. A list such as `{3,5,1}` is a collection of three objects. But in many ways, you can treat the whole list as a single object. You can, for example, do arithmetic on the whole list at once, or assign the whole list to be the value of a variable.

Defining your own lists is easy. You can, for example, type them in full, like this:

```
Input     oddList = {81, 3, 5, 7, 9, 11, 13, 15, 17}
Output    {81, 3, 5, 7, 9, 11, 13, 15, 17}
```

Alternatively, if (as here) the list elements correspond to a rule of some kind, the command `Table` can be used, like this:

```
Input     oddList = Table[2 n + 1, {n, 0, 8}]
Output    {1, 3, 5, 7, 9, 11, 13, 15, 17}
```

The functions for obtaining elements of lists are

| | |
|---|---|
| `First[list]` | the first element. |
| `Last[list]` | the last element. |
| `Part[list,n] or list[[n]]` | the nth element. |
| `Part[list,-n] or list[[-n]]` | the nth element from the end. |
| `Part[list,{n1,n2,...}] or list[[{n1,n2,...}]]` | the list of the n1th, n2th, ... elements. |
| `Take[list,n]` | the list of the first n elements. |
| `Take[list,-n]` | the list of the last n elements. |
| `Take[list,{m,n}]` | the list of the mth through nth elements. |
| `Rest[list]` | list without the first element. |
| `Most[list]` | list without the last element. |
| `Drop[list,n]` | list without the first n elements. |
| `Drop[list,-n]` | list without the last n elements. |
| `Drop[list,{m,n}]` | list without the mth through nth elements. |

***Mathematica Examples 1.22***

```
Input     First[{a,b,c}]
Output    a

Input     First[{{a,b},{c,d}}]
Output    {a,b}

Input     Last[{a,b,c}]
Output    c

Input     {a,b,c,d,e,f}[[3]]
Output    c

Input     {{a,b,c},{d,e,f},{g,h,i}}[[2,3]]
Output    f

Input     Take[{a,b,c,d,e,f},4]
Output    {a,b,c,d}
```

```
Input    Rest[{a,b,c,d}]
Output   {b,c,d}

Input    Most[{a,b,c,d}]
Output   {a,b,c}

Input    Drop[{a,b,c,d,e,f},2]
Output   {c,d,e,f}
```

Some functions for inserting, deleting, and replacing list and sublist elements are

| | |
|---|---|
| `Prepend[list,elem]` | insert elem at the beginning of list. |
| `Append[list,elem]` | insert elem at the end of list. |
| `Insert[list,elem,i]` | insert elem at position i in list. |
| `Insert[list,elem,{i,j,...}]` | insert elem at position {i, j, ...} in list. |
| `Insert[list,elem,{{i1,j1,...},{i2,...},...}]` | insert elem at positions {i1, j1, ...},{i2,...}, ... in list. |
| `Delete[list,i]` | delete the element at position i in list. |
| `Delete[list,{i,j,...}]` | delete the element at position {i, j, ...} in list. |
| `Delete[list,{{i1,j1,...},{i2,...},...}]` | delete elements at positions {i1, j1, ...},{i2,... }, ... in list. |
| `ReplacePart[list,elem,i]` | replace the element at position i in list with elem. |
| `ReplacePart[list,elem,{i,j,...}]` | replace the element at position {i, j, ...} with elem. |
| `ReplacePart[list,elem,{{i1,j1,...},{i2,...},...}]` | replace elements at positions {i1, j1, ...},{i2,...}, ... with elem. |

***Mathematica Examples 1.23***

```
Input    Prepend[{a,b,c,d},x]
Output   {x,a,b,c,d}

Input    Append[{a,b,c,d},x]
Output   {a,b,c,d,x}

Input    Insert[{a,b,c,d,e},x,3]
Output   {a,b,x,c,d,e}

Input    Insert[{a,b,c,d,e},x,-2]
Output   {a,b,c,d,x,e}

Input    Delete[{a,b,c,d},3]
Output   {a,b,d}

Input    Delete[{a,b,c,d},{{1},{3}}]
Output   {b,d}

Input    ReplacePart[{a,b,c,d,e},3->xxx]
Output   {a,b,xxx,d,e}

Input    ReplacePart[{a,b,c,d,e},{2->xx,5->yy}]
Output   {a,xx,c,d,yy}
```

Some functions for rearranging lists are

| | |
|---|---|
| `Sort[list]` | sort the elements of list into canonical order. |
| `Union[list]` | give a sorted version of list, in which all duplicated elements have been dropped. |
| `Reverse[list]` | reverse the order of the elements in list. |
| `RotateLeft[list]` | cycle the elements in list one position to the left. |
| `RotateLeft[list,n]` | cycle the elements in list n positions to the left. |
| `RotateRight[list]` | cycle the elements in list one position to the right. |
| `RotateRight[list,n]` | cycle the elements in list n positions to the right. |
| `Permutations[list]` | generate a list of all possible permutations of the elements in list. |
| `Partition[list,n]` | partition list into nonoverlapping sublists of length n. |
| `Partition[list,n,d]` | generate sublists with offset d. |
| `Split[list]` | split list into sublists consisting of runs of identical elements. |
| `Transpose[list]` | transpose the first two levels in list. |
| `Flatten[list]` | flatten out nested lists. |
| `Flatten[list,n]` | flatten out the top n levels. |
| `FlattenAt[list,i]` | flatten out a sublist that appears as the ith element of list. |
| `FlattenAt[list,{i,j,...}]` | flatten out the element of list at position {i,j, ...}. |
| `FlattenAt[list,{{i1,j1,...},{i2,j2,...}}]` | flatten out elements of list at several positions. |
| `Join[list1,list2,...]` | concatenate lists together. |
| `Union[list1,list2,...]` | give a sorted list of all the distinct elements that appear in any of the listi. |

***Mathematica Examples 1.24***

```
Input    Sort[{d,b,c,a}]
Output   {a,b,c,d}

Input    Sort[{4,1,3,2,2},Greater]
Output   {4,3,2,2,1}

Input    Union[{1,2,1,3,6,2,2}]
Output   {1,2,3,6}

Input    Union[{a,b,a,c},{d,a,e,b},{c,a}]
Output   {a,b,c,d,e}

Input    Reverse[{a,b,c,d}]
Output   {d,c,b,a}

Input    RotateLeft[{a,b,c,d,e},2]
Output   {c,d,e,a,b}

Input    RotateRight[{a,b,c,d,e},2]
Output   {d,e,a,b,c}

Input    Permutations[{a,b,c}]
Output   {{a,b,c},{a,c,b},{b,a,c},{b,c,a},{c,a,b},{c,b,a}}

Input    Partition[{a,b,c,d,e,f},2]
Output   {{a,b},{c,d},{e,f}}

Input    Flatten[{{a,b},{c,{d},e},{f,{g,h}}}]
Output   {a,b,c,d,e,f,g,h}

Input    Transpose[{{a,b,c},{x,y,z}}]
Output   {{a,x},{b,y},{c,z}}

Input    Join[{a,b,c},{x,y},{u,v,w}]
Output   {a,b,c,x,y,u,v,w}
```

Vectors and matrices in the Wolfram Language are simply represented by lists and by lists of lists, respectively. Functions for generating lists are `Range[ ]`, `Table[ ]`, and `Array[ ]`.

### Vectors

Mathematica has many functions for generating vectors. The following list summarizes them.

| | |
|---|---|
| `{e₁,e₂,...}` | is a list of elements. |
| `Range[n]` | create the list $\{1,2,3,\dots,n\}$. |
| `Range[n₁,n₂]` | create the list $\{n_1, n_1+1,\dots,n_2\}$. |
| `Range[n₁,n₂,dn]` | create the list $\{n_1, n_1+dn,\dots,n_2\}$. |
| `Table[f,{i,n}]` | build a length-n vector by evaluating f with $i = 1,2,\dots,n$. |
| `Length[list]` | give the number of elements in list. |
| `List[[i]] or Part[list,i]` | give the $i^{th}$ element in the vector list. |

**Mathematica Examples 1.25**

```
Input    List[a,b,c,d]
Output   {a,b,c,d}

Input    v={x,y}
Output   {x,y}

Input    Range[4]
Output   {1,2,3,4}

Input    Range[x,x+4]
Output   {x,1+x,2+x,3+x,4+x}

Input    Table[i^2,{i,10}]
Output   {1,4,9,16,25,36,49,64,81,100}

Input    Length[{a,b,c,d}]
Output   4

Input    {5,8,6,9}[[2]]
Output   8
```

| | |
|---|---|
| `c v` | multiply a vector v by a scalar. |
| `a.b` | dot product of two vectors a. b. |
| `Cross[a,b]` | cross product of two vectors (also input as $a \times b$). |
| `Norm[v]` | Euclidean norm of a vector v. |
| `Normalize[v]` | gives the normalized form of a vector v. |
| `Orthogonalize [{v₁,v₂,…}]` | gives an orthonormal basis found by orthogonalizing the vectors $v_i$. |

**Mathematica Examples 1.26**

```
Input    {a,b,c}.{x,y,z}
Output   a x+b y+c z

Input    Cross[{a,b,c},{x,y,z}]
Output   {-c y+b z,c x-a z,-b x+a y}

Input    Norm[{x,y,z}]
Output   Sqrt[Abs[x]^2 + Abs[y]^2 + Abs[z]^2]

Input    Normalize[{1,5,1}]
Output   {1/(3 Sqrt[3]), 5/(3 Sqrt[3]), 1/(3 Sqrt[3])}
```

```
Input    Orthogonalize[{{1,0,1},{1,1,1}}]
Output   {1/Sqrt[2], 0, 1/Sqrt[2]}, {0, 1, 0}}
```

### Matrix

Mathematica has many functions for generating matrices. The following list summarizes them.

| | |
|---|---|
| `{{a,b},{c,d}}` | matrix $\begin{pmatrix} a & b \\ c & d \end{pmatrix}$. |
| `Table[f,{i,m},{j,n}]` | build an m × n matrix by evaluating f with i ranging from 1 to m and j ranging. |
| `List[[i,j]] or Part[list,i,j]` | give the i, j th element in the matrix list. |
| `DiagonalMatrix[list]` | generate a square matrix with the elements in list on the main. |
| `Dimensions[list]` | give the dimensions of a matrix represented by list. |
| `Column[list]` | display the elements of list in a column. |
| `c m` | multiply a matrix m by a scalar. |
| `a.b` | dot product of two matrices a. b. |
| `Inverse[m]` | matrix inverse m. |
| `MatrixPower[m,n]` | gives the $n^{th}$ power of a matrix m. |
| `Det[m]` | Determinant m. |
| `Tr[m]` | Trace m. |
| `Transpose[m]` | Transpose m. |

***Mathematica Examples 1.27***

```
Input    m={{a,b},{c,d}}
Output   {{a,b},{c,d}}

Input    m[[1]]
Output   {a,b}

Input    m[[1,2]]
Output   b

Input    v={x,y}
Output   {x,y}

Input    m.v
Output   {a x+b y,c x+d y}

Input    m.m
Output   {{a^2+b c,a b+b d},{a c+c d,b c+d^2}}

Input    s=Table[i+j,{i,3},{j,3}]
Output   {{2,3,4},{3,4,5},{4,5,6}}

Input    DiagonalMatrix[{a,b,c}]
Output   {{a,0,0},{0,b,0},{0,0,c}}

Input    Det[m]
Output   -b c+a d

Input    Transpose[m]
Output   {{a,c},{b,d}}

Input    h=Table[1/(i+j-1),{i,3},{j,3}]
Output   {{1,1/2,1/3},{1/2,1/3,1/4},{1/3,1/4,1/5}}
```

```
Input     Inverse[h]
Output    {{9,-36,30},{-36,192,-180},{30,-180,180}}
```

## Array

| | |
|---|---|
| `Array[f,n]` | generates a list of length n, with elements f[i]. |
| `Array[f,n,r]` | generates a list using the index origin r. |
| `Array[f,n,{a,b}]` | generates a list using n values from a to b. |
| `Array[f,{n1,n2,...}]` | generates an $n_1 \times n_2 \times$ ... array of nested lists, with elements $f[i_1, i_2, \ldots]$. |

***Mathematica Examples 1.28***

```
Input     Array[f,10]
Output    {f[1],f[2],f[3],f[4],f[5],f[6],f[7],f[8],f[9],f[10]}

Input     Array[f,{3,2}]
Output    {{f[1,1],f[1,2]},{f[2,1],f[2,2]},{f[3,1],f[3,2]}}
```

## Layout & Tables

| | |
|---|---|
| `Print[expr]` | prints expr as output. |
| `MatrixForm[list]` | prints with the elements of list arranged in a regular array. |
| `TableForm[list]` | prints with the elements of list arranged in an array of rectangular cells. |
| `Grid[{{expr11,expr12,...},{expr21,expr22,...},...}]` | is an object that formats with the $expr_{ij}$ arranged in a two-dimensional grid. |
| `Row[expr1,expr2,...]` | is an object that formats with the $expr_i$ arranged in a row, potentially extending over several lines. |
| `Row[list,s]` | inserts s as a separator between successive elements. |
| `Column[expr1,expr2,...]` | is an object that formats with the $expr_i$ arranged in a column, with $expr_1$ above $expr_2$, etc. |
| `Multicolumn[list,cols]` | is an object that formats with the elements of list arranged in a grid with the indicated number of columns. |
| `Multicolumn[list,{rows,Automatic}]` | formats as a grid with the indicated number of rows. |

***Mathematica Examples 1.29***

```
Input     MatrixForm[{{1,2},{3,4}}]
Output    (1  2)
          (3  4)

Input     MatrixForm[Table[1/(i+j),{i,4},{j,4}]]
Output    (1/2  1/3  1/4  1/5)
          (1/3  1/4  1/5  1/6)
          (1/4  1/5  1/6  1/7)
          (1/5  1/6  1/7  1/8)

Input     TableForm[Table[1/(i+j),{i,4},{j,4}]]
Output     1/2  1/3  1/4  1/5
           1/3  1/4  1/5  1/6
           1/4  1/5  1/6  1/7
           1/5  1/6  1/7  1/8

Input     Grid[{{a,b,c},{x,y,z}}]
Output     a  b  c
           x  y  z
```

Input Grid[{{a,b,c},{x,y^2,z^3}},Frame->All]

Output

| | | |
|---|---|---|
| a | b | c |
| x | y^2 | z^3 |

```
Input   Row[{aaa,b,cccc}]
Output  aaabcccc

Input   Row[{aaa,b,cccc},"----"]
Output  aaa----b----cccc

Input   Column[{1,12,123,1234}]
Output  1
        12
        123
        1234

Input   Column[{1,22,333,4444},Frame->True]
Output  1
        22
        333
        4444

Input   Multicolumn[Range[50],{6,Automatic}]
Output  1  7   13  19  25  31  37  43  49
        2  8   14  20  26  32  38  44  50
        3  9   15  21  27  33  39  45
        4  10  16  22  28  34  40  46
        5  11  17  23  29  35  41  47
        6  12  18  24  30  36  42  48
```

# UNIT 1.4

# 2D AND 3D GRAPHING

The graph of a function offers tremendous insight into the behavior of the function and can be of great value in the solution of problems in mathematics. One of the outstanding features of Mathematica is its graphing capabilities. Mathematica contains functions for 2D and 3D graphing of functions, lists, and arrays of data.

### Basic Plotting

| | |
|---|---|
| `Plot[f,{x,xmin,xmax}]` | plot f as a function of x from $x_{min}$ to $x_{max}$. |
| `Plot[{f1,f2,...},{x,xmin,xmax}]` | plot several functions together. |

When the Wolfram Language plots a graph for you, it has to make many choices. It has to work out what the scales should be, where the function should be sampled, how the axes should be drawn, and so on. Most of the time, the Wolfram Language will probably make pretty good choices. However, if you want to get the very best possible pictures for your particular purposes, you may have to help the Wolfram Language in making some of its choices.

There is a general mechanism for specifying "options" in Wolfram Language functions. Each option has a definite name. As the last argument to a function like `Plot`, you can include a sequence of rules of the form `name->value`, to specify the values for various options. Any option for which you do not give an explicit rule is taken to have its "default" value.

Some options for `Plot` function are

| | |
|---|---|
| `AspectRatio` | the height-to-width ratio for the plot; Automatic sets it from the absolute x and y coordinates |
| `Axes` | whether to include axes |
| `AxesLabel` | labels to be put on the axes; ylabel specifies a label for the y axis, {xlabel,ylabel} for both axes |
| `AxesOrigin` | the point at which axes cross |
| `BaseStyle` | the default style to use for the plot |
| `FormatType` | the default format type to use for text in the plot |
| `Frame` | whether to draw a frame around the plot |
| `FrameLabel` | labels to be put around the frame; give a list in clockwise order starting with the lower x axis |
| `FrameTicks` | what tick marks to draw if there is a frame; None gives no tick marks |
| `GridLines` | what grid lines to include; Automatic includes a grid line for every major tick mark |
| `PlotLabel` | an expression to be printed as a label for the plot |
| `PlotRange` | the range of coordinates to include in the plot; All includes all points |
| `Ticks` | what tick marks to draw if there are axes; None gives no tick marks |
| `PlotStyle` | a list of lists of graphics primitives to use for each curve (see "Graphics Directives and Options") |
| `ClippingStyle` | what to draw when curves are clipped |
| `Filling` | filling to insert under each curve |
| `FillingStyle` | style to use for filling |
| `PlotPoints` | the initial number of points at which to sample the function |
| `MaxRecursion` | the maximum number of recursive subdivisions allowed |

***Mathematica Examples 1.30***

Input
```
Plot[
  Sin[x],
  {x,0,2 Pi},
  ImageSize->200
```

```
        ]
Output
```

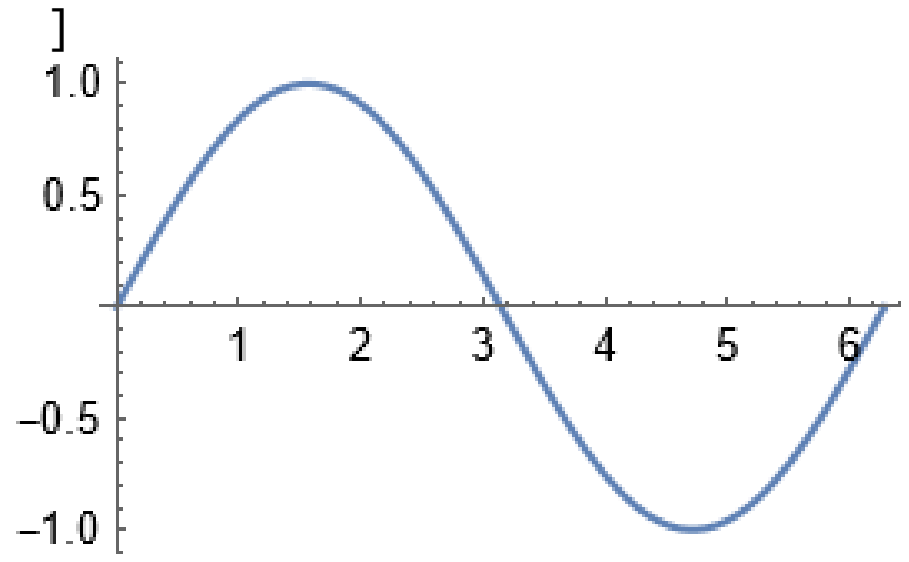

```
Input   Plot[
         {Sin[x],Sin[2 x],Sin[3 x]},
         {x,0,2 Pi},
         ImageSize->200
         ]
Output
```

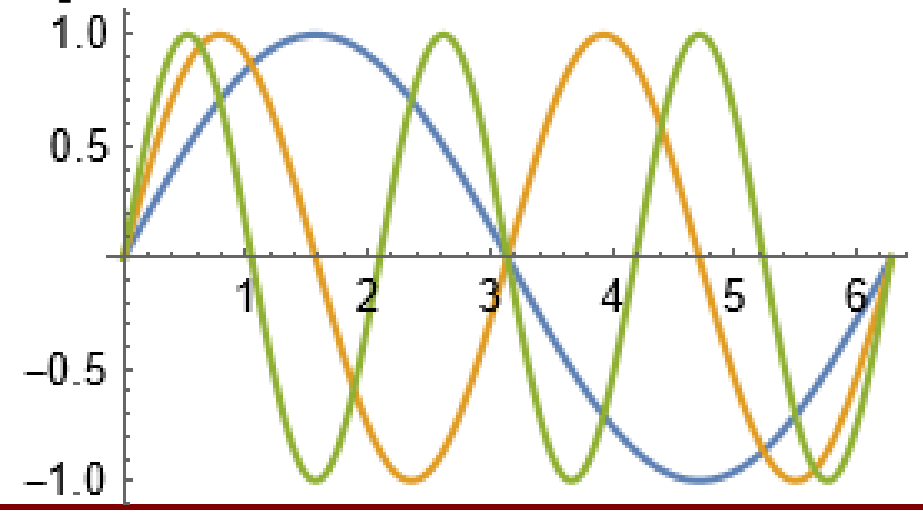

### 3D Plot

| | |
|---|---|
| `Plot3D[f,{x,xmin,xmax},{y,ymin,ymax}]` | make a three-dimensional plot of f as a function of the variables x and y. |

Some options for `Plot3D` function are

| | |
|---|---|
| `Axes` | whether to include axes |
| `AxesLabel` | labels to be put on the axes: zlabel specifies a label for the z axis, {xlabel,ylabel,zlabel} for all axes |
| `BaseStyle` | the default style to use for the plot |
| `Boxed` | whether to draw a three-dimensional box around the surface |
| `FaceGrids` | how to draw grids on faces of the bounding box; All draws a grid on every face |
| `LabelStyle` | style specification for labels |
| `Lighting` | simulated light sources to use |
| `Mesh` | whether an xy mesh should be drawn on the surface |
| `PlotRange` | the range of z or other values to include |
| `SphericalRegion` | whether to make the circumscribing sphere fit in the final display area |
| `ViewAngle` | angle of the field of view |
| `ViewCenter` | point to display at the center |
| `ViewPoint` | the point in space from which to look at the surface |
| `ViewVector` | position and direction of a simulated camera |
| `ViewVertical` | direction to make vertical |
| `BoundaryStyle` | how to draw boundary lines for surfaces |
| `ClippingStyle` | how to draw clipped parts of surfaces |
| `ColorFunction` | how to determine the color of the surfaces |
| `Filling` | filling under each surface |
| `FillingStyle` | style to use for filling |
| `PlotPoints` | the number of points in each direction at which to sample the function; {nx,ny} specifies different numbers in the x and y directions |
| `PlotStyle` | graphics directives for the style of each surface |

***Mathematica Examples 1.31***

Input
```
Plot3D[
 Sin[x+y^2],
 {x,-3,3},
 {y,-2,2},
 ImageSize->200
 ]
```
Output

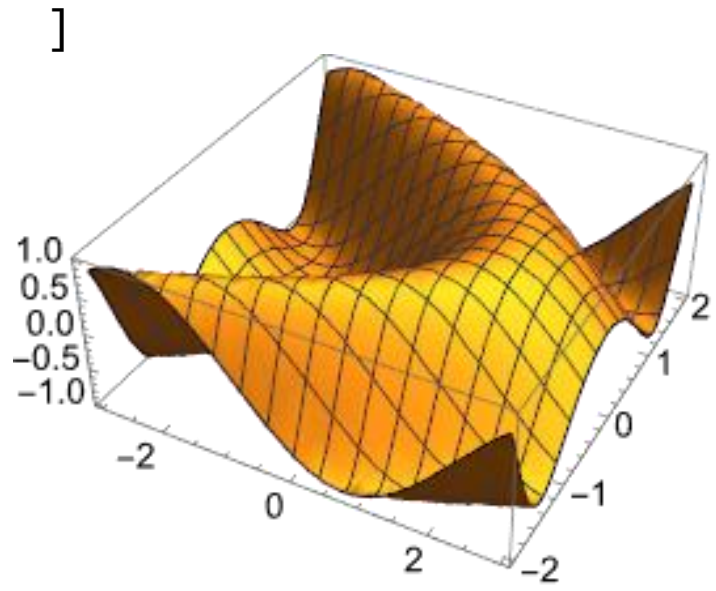

## Plotting Lists of Data

| | |
|---|---|
| `ListPlot[{y₁,y₂,...}]` | plot $y_1, y_2, \ldots$ at x values 1, 2, .... |
| `ListPlot[{{x₁,y₁},{x₂,y₂},...}]` | plot points $(x_1, y_1), \ldots$. |
| `ListLinePlot[list]` | join the points with lines. |
| `ListPlot3D[array]` | generates a three-dimensional plot of a surface representing an array of height values. |
| `ListPlot3D[{{x₁,y₁,z₁},{x₂,y₂,z₂},...}]` | generates a plot of the surface with heights $z_i$ at positions $x_i$, $y_i$. |
| `ListPlot3D[{data1,data2,...}]` | plots the surfaces corresponding to each of the $data_i$. |
| `ListPointPlot3D[array]` | generates a 3D scatter plot of points with a 2D array of height values. |
| `ListPointPlot3D[{{x₁,y₁,z₁},{x₂,y₂,z₂},...}]` | generates a 3D scatter plot of points with coordinates $x_i, y_i, z_i$ |
| `ListPointPlot3D[{data1,data2,...}]` | plots several collections of points, by default in different colors. |
| `DensityPlot[f,{x,xmin,xmax]},{y,ymin,ymax}]` | makes a density plot of f as a function of x and y. |
| `ContourPlot[f,{x,xmin,xmax},{y,ymin,ymax}]` | generates a contour plot of f as a function of x and y. |

***Mathematica Examples 1.32***

Input
```
ListPlot[
 Prime[Range[25]],
 ImageSize->200
 ]
```
Output

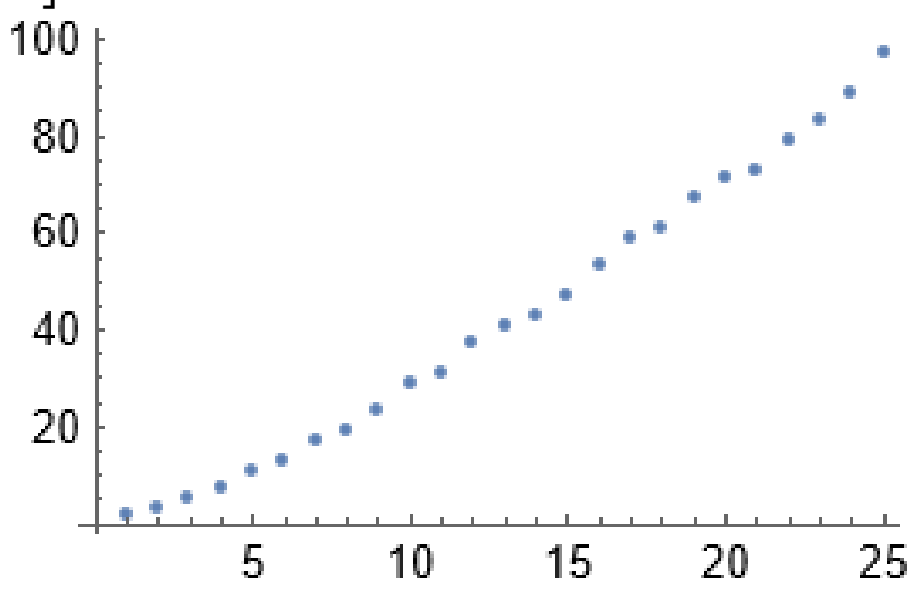

Input
```
ListPlot[
 Table[
  {Sin[n],Sin[2 n]},
  {n,50}
  ],
 ImageSize->200
```

```
]
```

Output

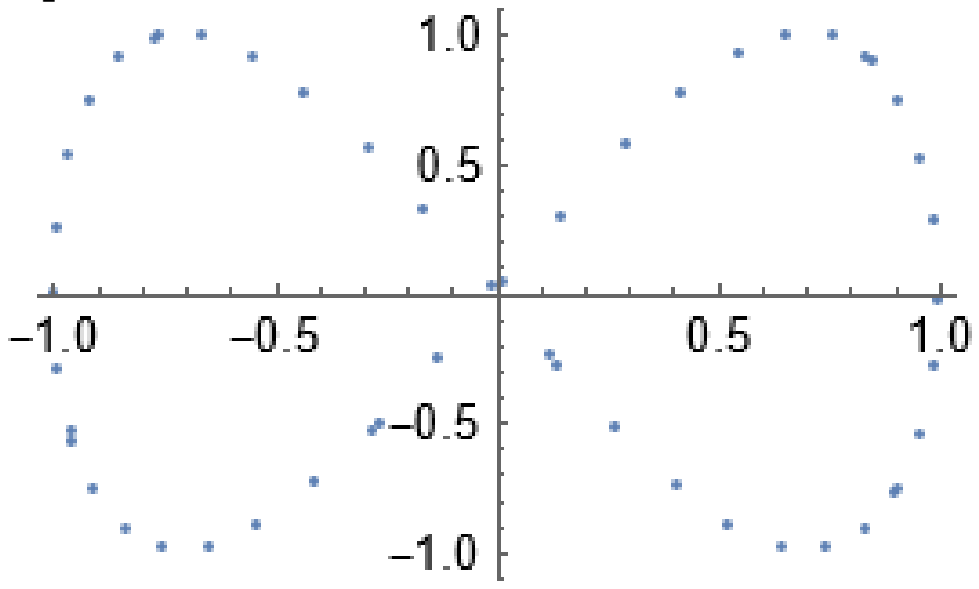

Input

```
ListLinePlot[
 {1,1,2,3,5,8},
 ImageSize->200
 ]
```

Output

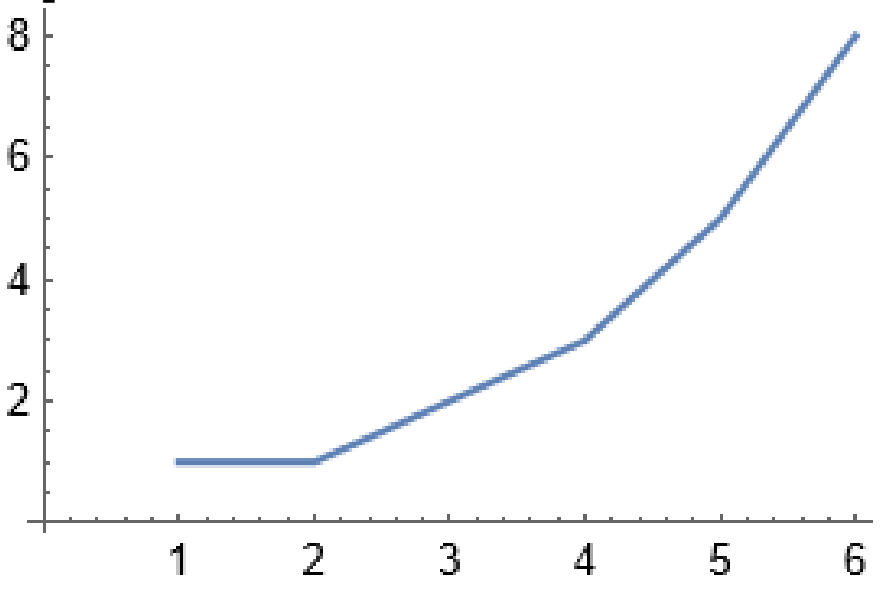

Input

```
ListPlot3D[
 {{1,1,1,1},{1,2,1,2},{1,1,3,1},{1,2,1,4}},
 Mesh->All,
 ImageSize->200
 ]
```

Output

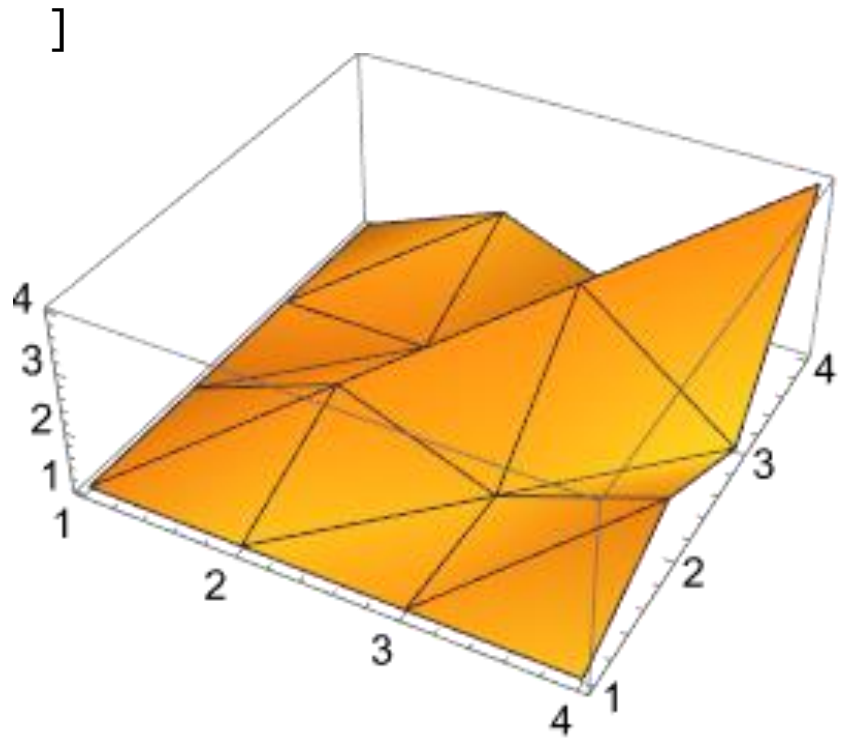

Input

```
data=Table[
    Sin[j^2+i],
    {i,0,Pi,Pi/5},
    {j,0,Pi,Pi/5}
    ];

ListPlot3D[
 data,
 Mesh->None,
 InterpolationOrder->3,
 ColorFunction->"SouthwestColors",
 ImageSize->200
 ]
```

Output

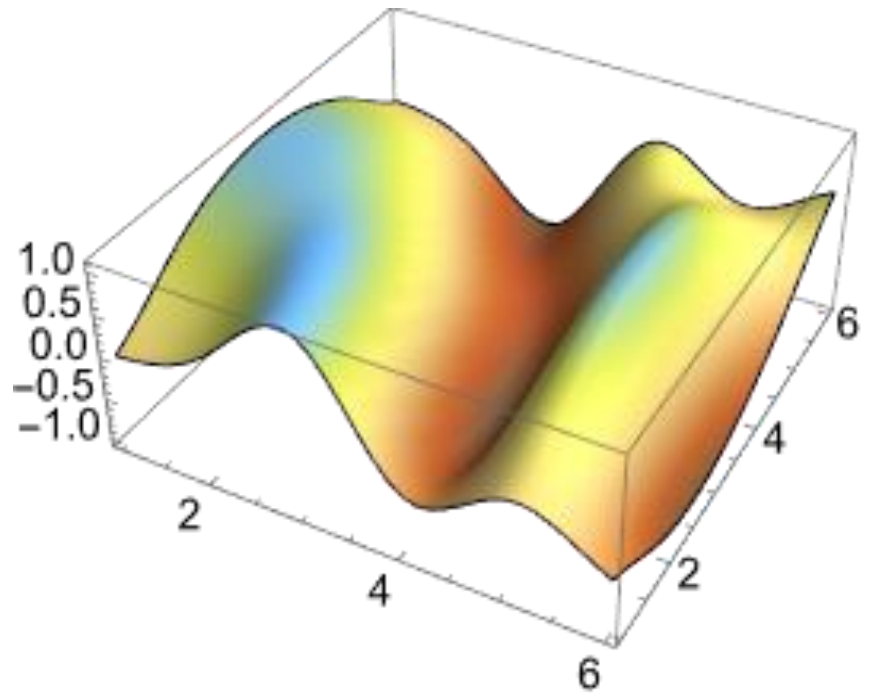

Input
```
ListPointPlot3D[
 Table[
  Sin[j^2+i],
  {i,0,3,0.1},
  {j,0,3,0.1}
  ],
 ImageSize->200
 ]
```
Output

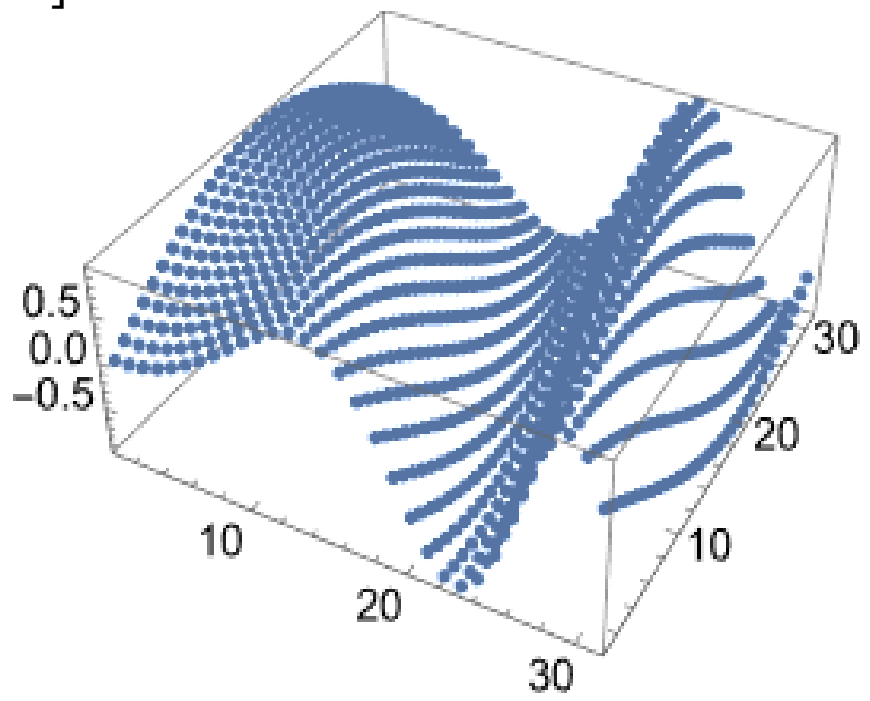

Input
```
ListPointPlot3D[
 {
  Table[
   Sin[j^2+i],
   {i,0,3,0.1},
   {j,0,3,0.1}
   ],
  Table[
   Sin[j^2+i]+3,
   {i,0,3,0.1},
   {j,0,3,0.1}
   ]
  },
 ImageSize->200
 ]
```
Output

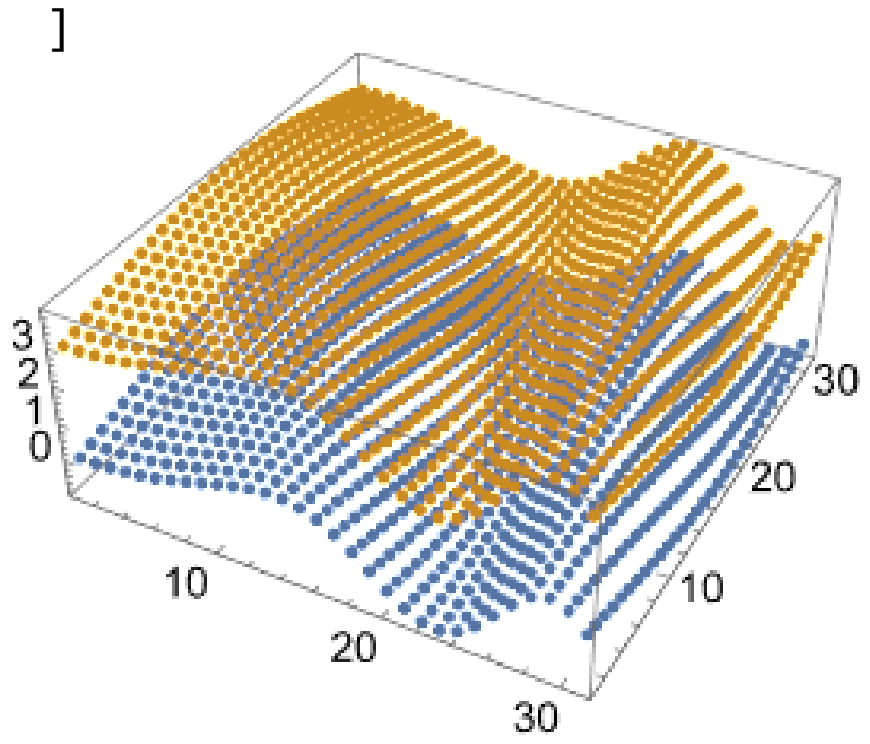

```
Input    DensityPlot[
          Sin[x] Sin[y],
          {x,-4,4},
          {y,-3,3},
          ColorFunction->"SunsetColors",
          PlotLegends->Automatic,
          ImageSize->200
          ]
```

Output

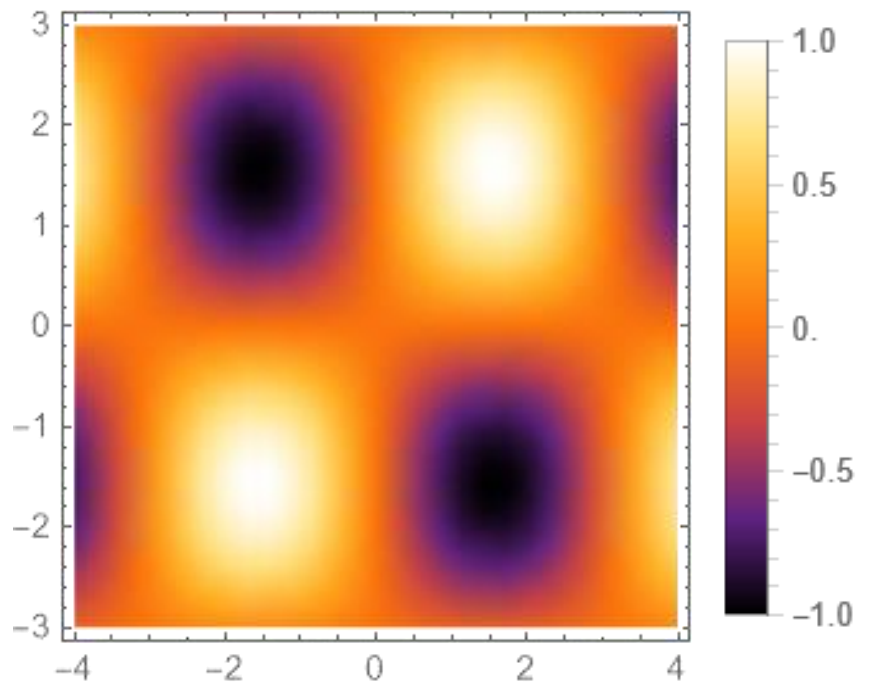

```
Input    ContourPlot[
          Cos[x]+Cos[y],
          {x,0,4 Pi},
          {y,0,4 Pi},
          PlotLegends->Automatic,
          ImageSize->200
          ]
```

Output

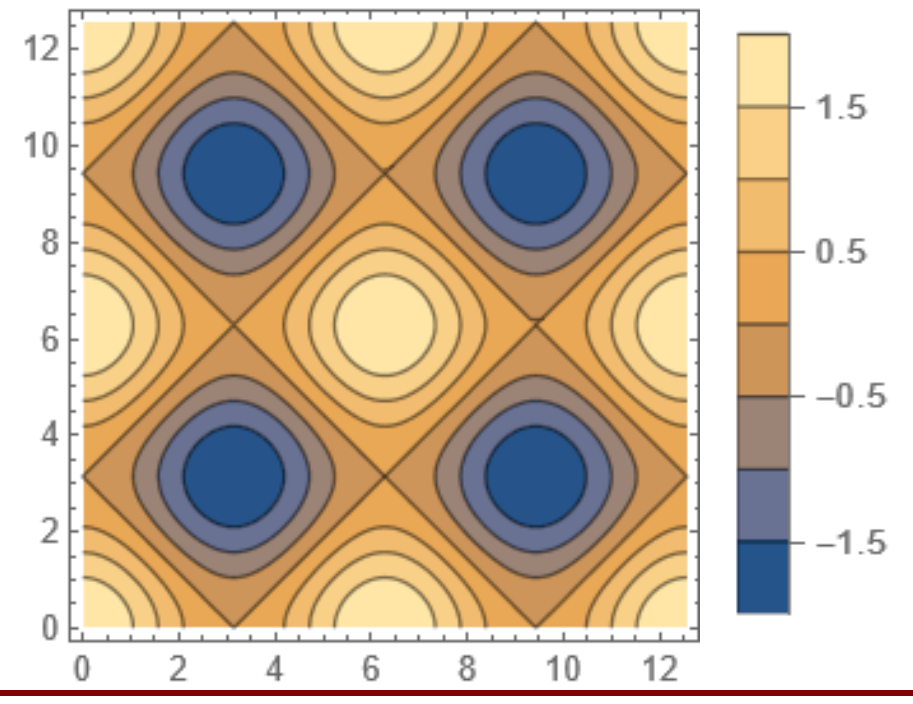

### Combining Plots

| | |
|---|---|
| `Show[plot₁,plot₂,...]` | combine several plots. |
| `GraphicsGrid[{plot₁,plot₂,...},...}]` | draw an array of plots. |
| `GraphicsRow[{plot₁,plot₂,...}]` | draw several plots side by side. |
| `GraphicsColumn[{plot₁,plot₂,...}]` | draw a column of plots. |

***Mathematica Examples 1.33***

```
Input    Show[
          Plot[
           x^2,
           {x,0,3.5},
           ImageSize->200
           ],
          ListPlot[
           {1,4,9},
           ImageSize->200]
```

```
]
```

Output 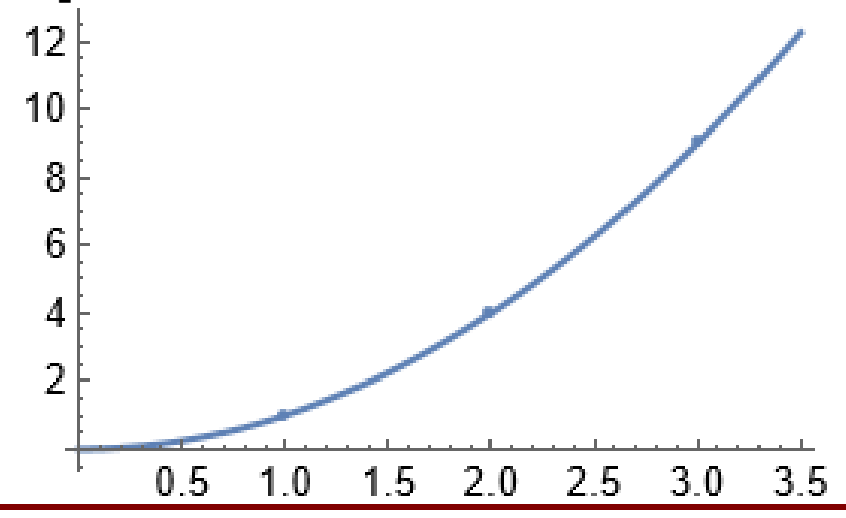

### Vector Field Plots

| | |
|---|---|
| `VectorPlot[{vx,vy },{x,xmin,xmax},{y,ymin,ymax }]` | generates a vector plot of the vector field $\{v_x, v_y\}$ as a function of x and y. |
| `VectorPlot3D[{vx,vy,vz},{x,xmin,xmax},{y,ymin,ymax },{z,zmin,zmax}]` | generates a 3D vector plot of the vector field $\{v_x, v_y, v_z\}$ as a function of x, y, and z. |
| `VectorDensityPlot[{{vx,vy},s},{x,xmin,xmax},{y,ymin,ymax}]` | generates a vector plot of the vector field $\{v_x, v_y\}$ as a function of x and y, superimposed on a density plot of the scalar field s. |

***Mathematica Examples 1.34***

Input
```
VectorPlot[
 {x,-y},
 {x,-3,3},
 {y,-3,3},
 ImageSize->200
 ]
```

Output 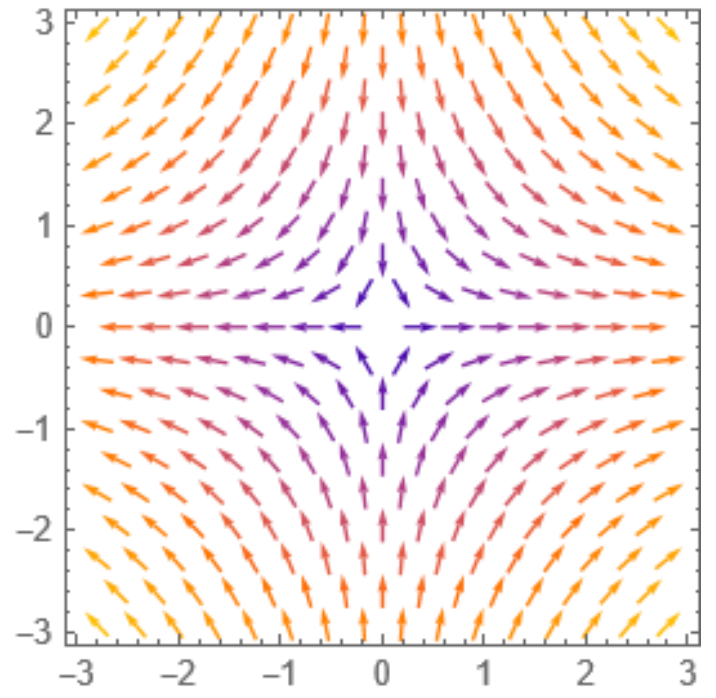

Input
```
VectorDensityPlot[
 {x,-y},
 {x,-3,3},
 {y,-3,3},
 ImageSize->200
 ]
```

Output 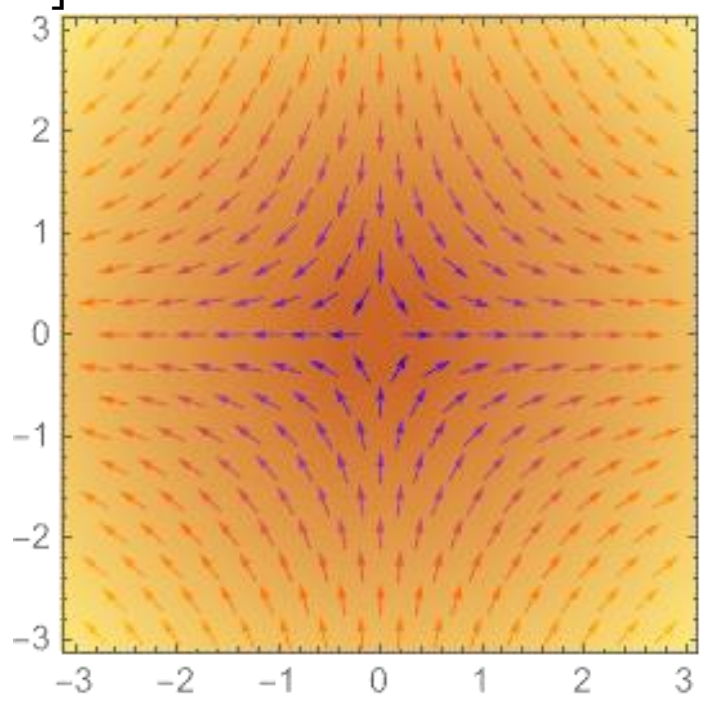

Input
```
VectorPlot3D[
 {x,y,z},
 {x,-1,1},
 {y,-1,1},
 {z,-1,1},
 ImageSize->200
 ]
```
Output

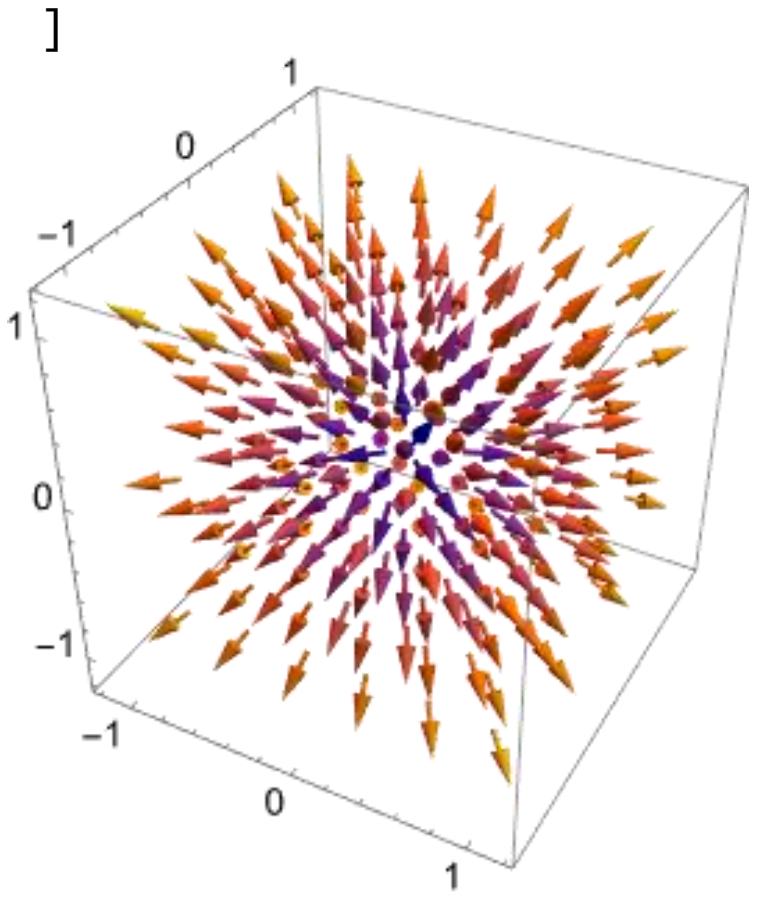

### Manipulate

The single command `Manipulate` lets you create an astonishing range of interactive applications with just a few lines of input. The output you get from evaluating a `Manipulate` command is an interactive object containing one or more controls (sliders, etc.) that you can use to vary the value of one or more parameters. The output is very much like a small applet or widget: it is not just a static result, it is a running program you can interact with.

| | |
|---|---|
| `Manipulate[expr,{u,umin,umax}]` | generates a version of expr with controls added to allow interactive manipulation of the value of u. |
| `Manipulate[expr,{u,umin,umax,du}]` | allows the value of u to vary between $u_{min}$ and $u_{max}$ in steps du. |
| `Manipulate[expr,{{u,uinit},umin,umax,...}]` | takes the initial value of u to be $u_{init}$. |

***Mathematica Examples 1.35***

Input
```
Manipulate[
 Plot[
  Sin[x (1+a x)],
  {x,0,6},
  ImageSize->200
  ],
 {a,0,2}
 ]
```
Output

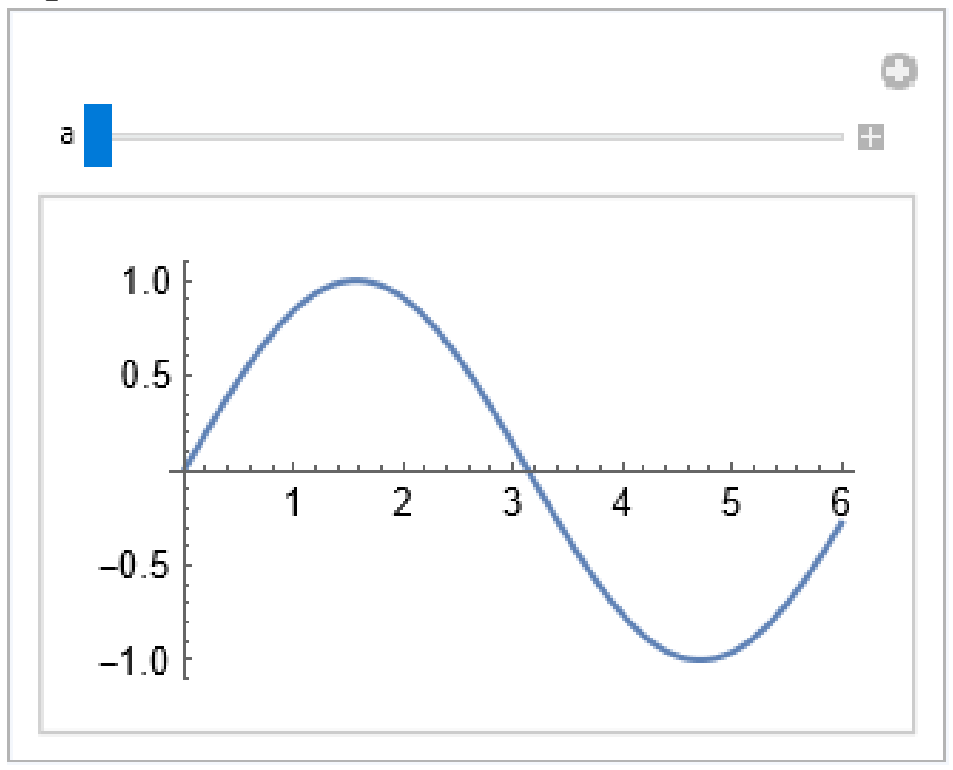

Input
```
Manipulate[
 Plot[
  Sin[a x+b],
  {x,0,6},
  ImageSize->200
  ],
 {a,1,4},
 {b,0,10}
 ]
```
Output

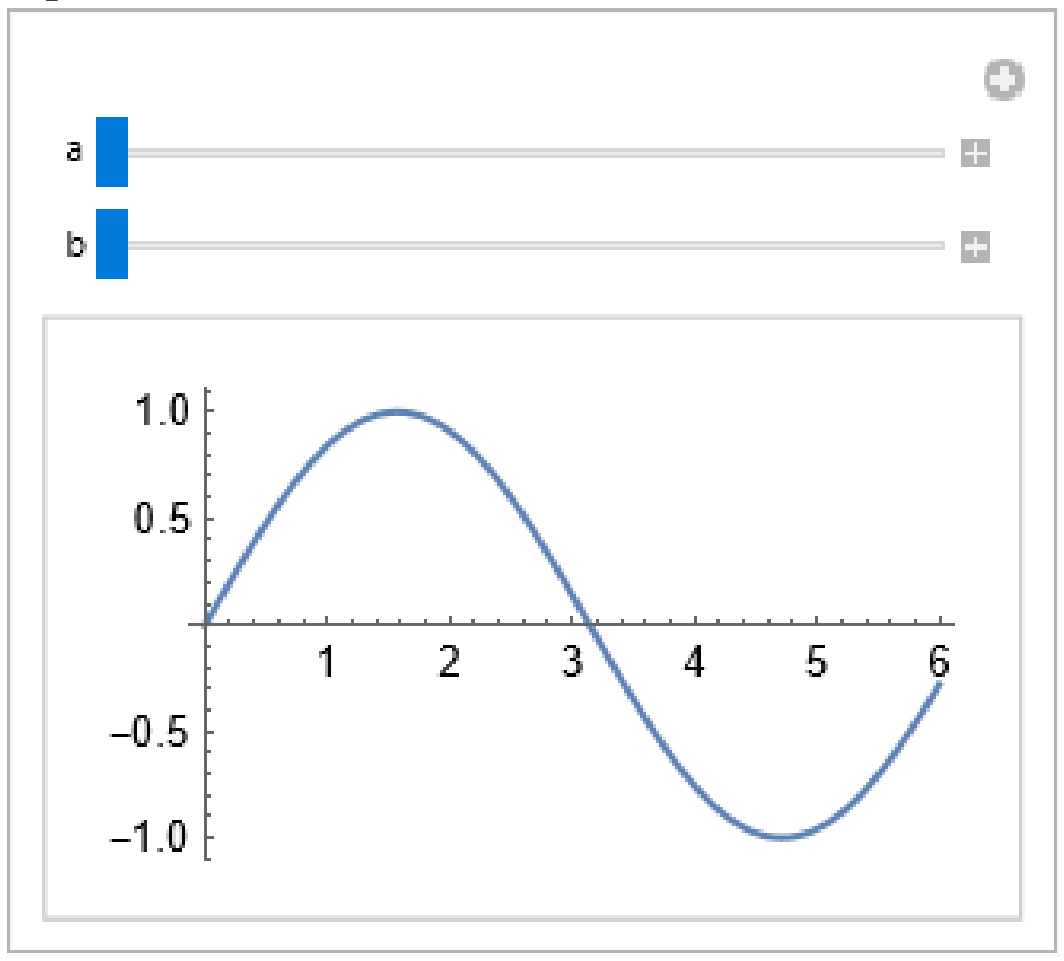

Input
```
Manipulate[
 ContourPlot3D[
  x^2+y^2+a z^3==1,
  {x,-2,2},
  {y,-2,2},
  {z,-2,2},
  Mesh->None,
  ImageSize->200
  ],
 {a,-2,2}
 ]
```
Output

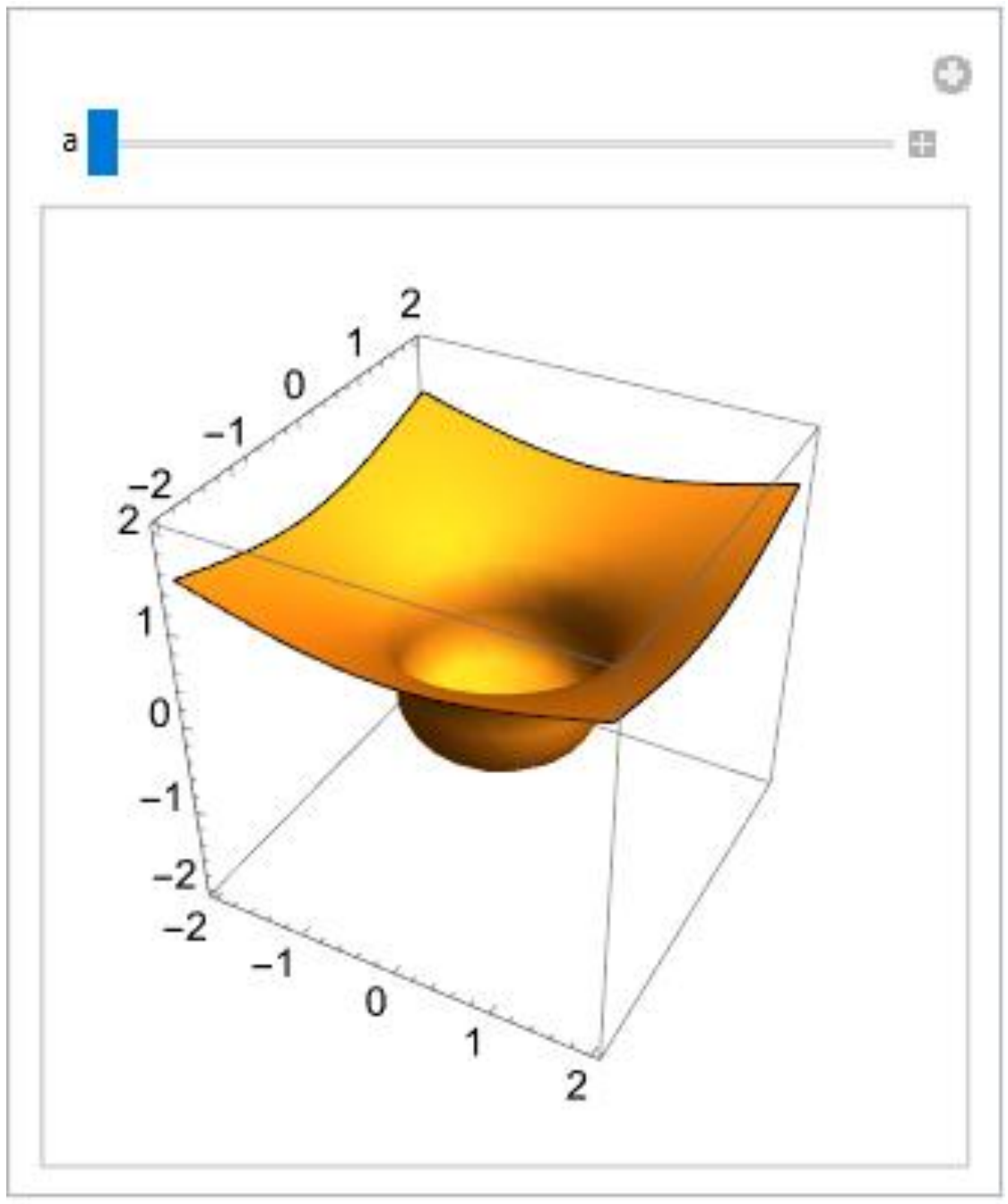

Input
```
Manipulate[
 Plot[
  If[t,Sin[x],Cos[x]],
  {x,0,10},
  ImageSize->200
  ],
 {t,{True,False}}
 ]
```
Output

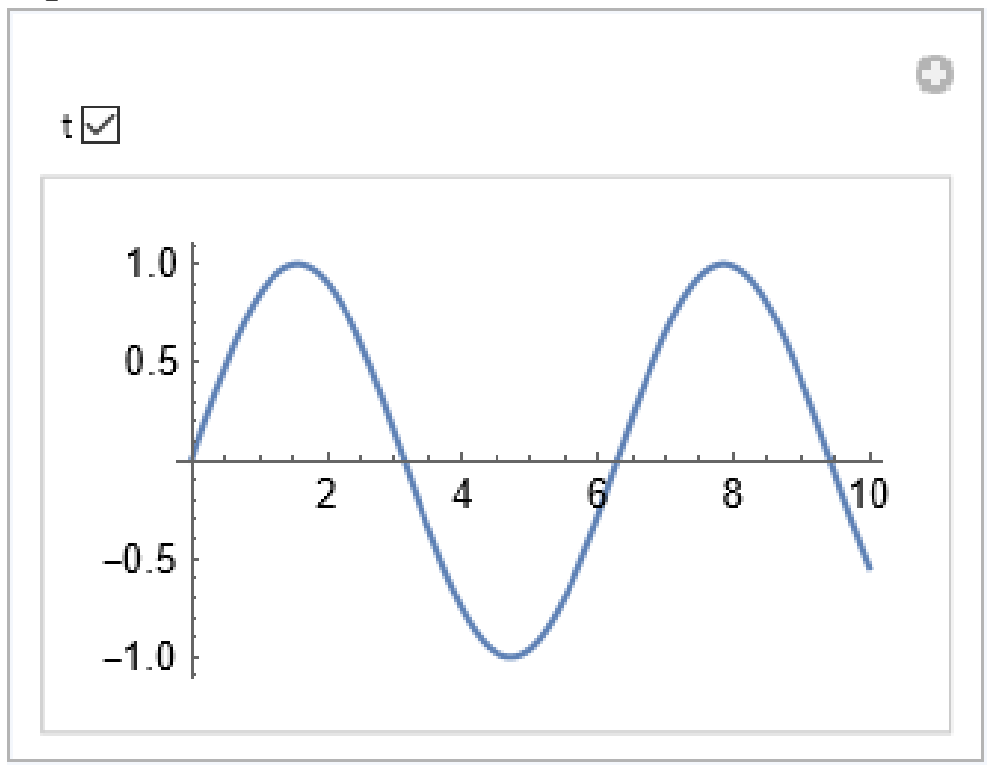

Input
```
Manipulate[
 Plot[
  f[x],
  {x,0,2 Pi},
  ImageSize->200
  ],
 {f,{Sin->"sine",Cos->"cosine",Tan->"tangent"}}
 ]
```
Output

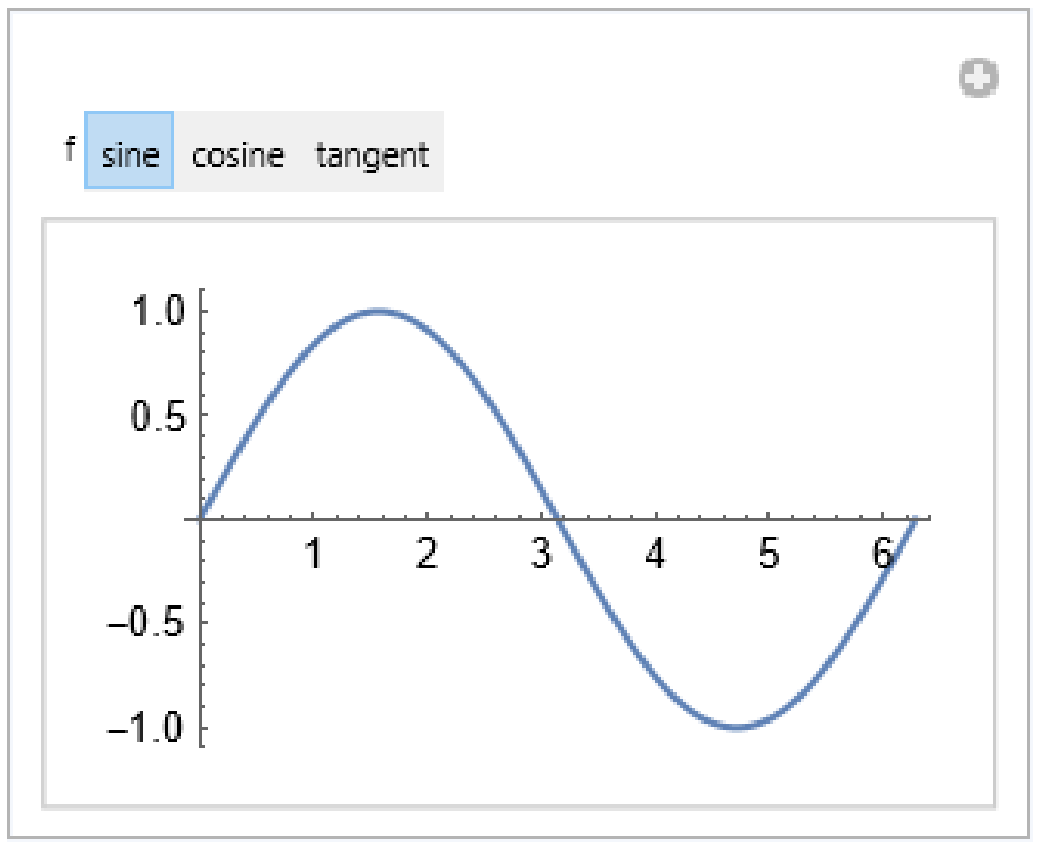

Input
```
Manipulate[
 ParametricPlot[
  {a1 Sin[n1 (x+p1)],a2 Cos[n2 (x+p2)]},
  {x,0,20 Pi},
  PlotRange->1,
  PerformanceGoal->"Quality",
  ImageSize->200
  ],
 Style["Vertical",Bold,Medium],
 {{n1,1,"Frequency"},1,4},
 {{a1,1,"Amplitude"},0,1},
 {{p1,0,"Phase"},0,2 Pi},
 Delimiter,
 Style["Horizontal",Bold,Medium],
```

```
         {{n2,5/4,"Frequency"},1,4},
         {{a2,1,"Amplitude"},0,1},
         {{p2,0,"Phase"},0,2 Pi},
         ControlPlacement->Left
         ]
```

Output

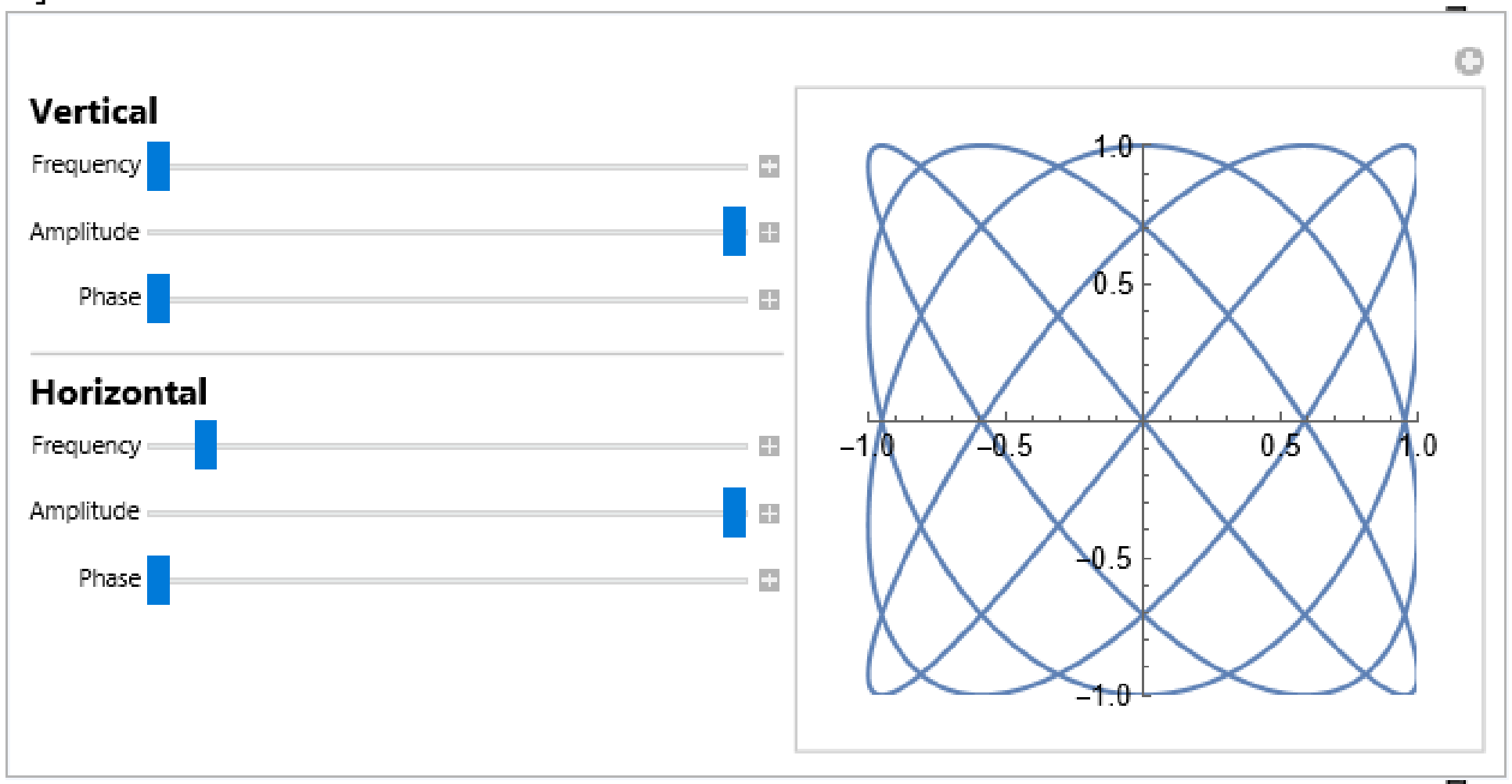

Input

```
Manipulate[
 Plot3D[
  Sin[x y+a],
  {x,0,3},
  {y,0,3},
  ImageSize->200
  ],
 {a,0,1}
 ]
```

Output

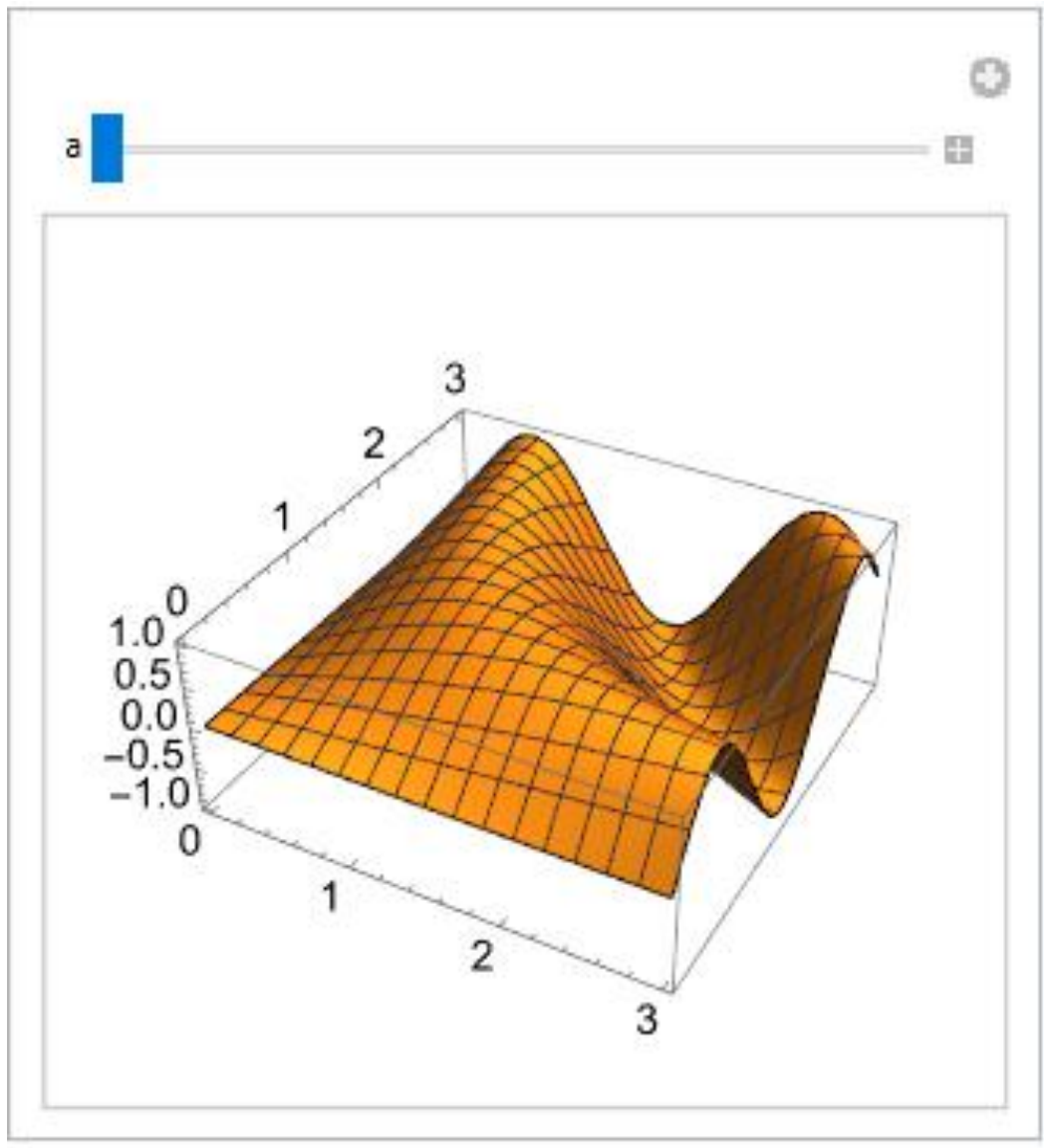

# UNIT 1.5

# CONTROL STRUCTURE

Most programming languages use control structures to control the flow of a program. The control structures include decision-making and loops. Decision-making is done by applying different conditions in the program. If the conditions are true, the statements following the condition are executed. The values in a condition are compared by using the comparison operators. The loops are used to run a set of statements several times until a condition is met. If the condition is true, the loop is executed. If the condition becomes false, the loop is terminated, and the control passes to the next statement that follows the loop block.

### Conditional Statements

Programmers often need to check the status of a computed intermediate result to branch the program to such or another block of instructions to pursue the computation. Several examples of the branching condition structures are next.

| | |
|---|---|
| `lhs:=rhs/;test` | is a definition to be used only if test yields True. |
| `If[test,then,else]` | evaluate then if test is True, and else if it is False. |
| `Which[test1,value1,test2,...]` | evaluate the $test_1$ in turn, giving the value associated with the first one that is True. |
| `Switch[expr,form1,value1,form2,...]` | compare expr with each of the $form_i$, giving the value associated with the first form it matches. |
| `Switch[expr,form1,value1,form2,...,_,def]` | use def as a default value. |
| `Piecewise[{{value1,test1 },{value1,test1 },...}]` | represents a piecewise function with values $value_i$ in the regions defined by the conditions $test_i$. |
| `Piecewise[{{value1,test1},...},def]` | give the value corresponding to the first $test_1$ which yields True. |

Note that,

1- `If[condition, t, f]` is left unevaluated if the condition evaluates to neither `True` nor `False`.
2- `If[condition, t]` gives `Null` if the condition evaluates to `False`.

***Mathematica Examples 1.36***

```
Input     (*If can be used as a statement:*)
          x=-2;
          If[
             x<0,
             y=-x,
             y=x
             ];
          y
Output    2

Input     (*If can also be used as an expression returning a value:*)x=-2;
          y=If[
             x<0,-x,x
             ]
Output    2

Input     If[
           7>8,x,y
```

```
          ]
Output    y

Input     x=2;
          If[
           x==0,Print["x is 0"],Print["x is different from 0"]
           ]
Output    x is different from 0

Input     x=3;
          y=0;
          If[
            x>1,y=Sqrt[x],y=x^2
            ];
          Print[y]

          m:=If[
              x>5,1,0
              ];
          Print[m]
Output    Sqrt[3]
Output    0

Input     a=2;
          Which[
           a==1,x,
           a==2,b
           ]
Output    b

Input     (* Use True for an else clause that always matches: *)
          sign[x_]:=Which[
            x<0,-1,
            x>0,1,
            True,
            Indeterminate
            ]
          {sign[-2],sign[0],sign[3]}
Output    {-1,Indeterminate,1}

Input     (* Use PiecewiseExpand to convert Which to Piecewise: *)
          PiecewiseExpand[
           Which[
            c1,a1,
            c2,a2,
            True,a3
            ]
           ]
```

Output $\begin{cases} a1 & c1 \\ a2 & !c1\&\&c2 \\ a3 & \text{True} \end{cases}$

```
Input     expr=3;
          Switch[
           expr,
           1,Print["expr is 1"],
           2,Print["expr is 2"],
           3,Print["expr is 3"],
           _,Print["expr has some other value"]
            ]
           ]
```

```
Output   expr is 3

Input    k=2;
         n=0;
         Switch[
            k,
            1,n=k+10,
            2,n=k^2+3,
            _,n=-1
            ];

         Print[n]
         k=5;
         n:=Switch[
              k,
              1,k+10,
              2,k^2+3,
              _,-1
              ];
         Print[n]
Output   7
Output   -1

Input    (* Find the derivative of a piecewise function: *)
         D[
          Piecewise[
            {
             {x^2,x<0},
             {x,x>0}
             }
            ],
          x
          ]
```

Output
$$\begin{cases} 2x & x<0 \\ 1 & x>0 \\ \text{Indeterminate} & \text{True} \end{cases}$$

```
Input    (* Define a piecewise function: *)
         pw=Piecewise[
            {
             {Sin[x]/x,x<0},
             {1,x==0}
             },
            -x^2/100+1
            ]
         (* Evaluate it at specific points: *)
         pw/. {{x->-5},{x->0},{x->5}}
```

Output
$$\begin{cases} \text{Sin}[x]/x & x<0 \\ 1 & x==0 \\ 1-\frac{x^2}{100} & \text{True} \end{cases}$$

```
Output   {Sin[5]/5,1,3/4}

Input    Piecewise[
          {
           {x^2,x<0},
           {x,x>0}
           }
          ]
```

Output
$$\begin{cases} x\text{^}2 & x<0 \\ x & x>0 \\ 0 & \text{True} \end{cases}$$

Input
```
Piecewise[
 {
  {Sin[x]/x,x<0},
  {1,x==0}
  },
 -x^2/100+1
 ]
```

Output
$$\begin{cases} \text{Sin}[x]/x & x<0 \\ 1 & x==0 \\ 1-\frac{x\text{^}2}{100} & \text{True} \end{cases}$$

Input
```
(*Remove unreachable cases:*)
Piecewise[
 {
  {e1,d1},
  {e2,d2},
  {e3,True},
  {e4,d4},
  {e5,d5}
  },
 e
 ]
```

Output
$$\begin{cases} e1 & d1 \\ e2 & d2 \\ e3 & \text{True} \end{cases}$$

Input
```
Piecewise[
 {
  {e1,d1},
  {e2,d2},
  {e3,d2&&d3},
  {e4,d4}
  },
 e
 ]
```

Output
$$\begin{cases} e1 & d1 \\ e2 & d2 \\ e4 & d4 \\ e & \text{True} \end{cases}$$

Input
```
x=4;
If[
 x>0,
 y=Sqrt[x],
 y=0
 ]
```

Output 2

Input
```
(* PiecewiseExpand converts nested piecewise functions into a single piecewise
function: *)
pw=Piecewise[
   {
    {Piecewise[{{1,x>=0}},2],Piecewise[{{x,x<=1}},x/2]^2>=1/2}
    },
   3
```

```
  ]
PiecewiseExpand[pw]
```

Output
$$\begin{cases} \begin{cases} 1 & x >= 0 \\ 2 & \text{True} \end{cases} & \left(\begin{cases} x & x <= 1 \\ \frac{x}{2} & \text{True} \end{cases}\right)^2 \geq \frac{1}{2} \\ 3 & \text{True} \end{cases}$$

Output
$$\begin{cases} 1 & x >= \sqrt{2} || \frac{1}{\sqrt{2}} <= x <= 1 \\ 2 & x >= \sqrt{2} || \frac{1}{\sqrt{2}} <= x <= 1 || x <= -\frac{1}{\sqrt{2}} \\ 3 & \text{True} \end{cases}$$

Input
```
(* Min, Max, UnitStep, and Clip are piecewise functions of real arguments: *)
  PiecewiseExpand/@{
   Min[x,y],
   Max[x,y,z],
   UnitStep[x],
   Clip[x,{a,b}]
  }
```

Output
$$\left\{\begin{cases} x & x - y <= 0 \\ y & \text{True} \end{cases}, \begin{cases} x & x - y >= 0 \&\& x - z >= 0 \\ y & x - y < 0 \&\& y - z >= 0 \\ z & \text{True} \end{cases}, \begin{cases} 1 & x >= 0 \\ 0 & \text{True} \end{cases}, \begin{cases} a & a - x > 0 \\ b & b - x < 0 \&\& a - x <= 0 \\ x & \text{True} \end{cases}\right\}$$

Input
```
(* Abs, Sign, and Arg are piecewise functions when their arguments are assumed to
be real: *)
Assuming[
 Element[x,Reals],
 PiecewiseExpand/@{Abs[x],Sign[x],Arg[x]}
 ]
```

Output
$$\left\{\begin{cases} -x & x < 0 \\ x & \text{True} \end{cases}, \begin{cases} -1 & x < 0 \\ 1 & x > 0 \\ 0 & \text{True} \end{cases}, \begin{cases} \pi & x < 0 \\ 0 & \text{True} \end{cases}\right\}$$

Input
```
(* KroneckerDelta and DiscreteDelta are piecewise functions of complex arguments:
*)
PiecewiseExpand/@{KroneckerDelta[x,y],DiscreteDelta[x,y]}
```

Output
$$\left\{\begin{cases} 1 & x - y == 0 \\ 0 & \text{True} \end{cases}, \begin{cases} 1 & x == 0 \&\& y == 0 \\ 0 & \text{True} \end{cases}\right\}$$

Input
```
(* Derivatives are computed piece-by-piece, unless the function is univariate in a
real variable: *)
D[
 Piecewise[
  {
   {(x^2-1)/(x-1),x!=1}
   },
  2],
 x]
```

Output
$$\begin{cases} \frac{2\,x}{-1 + x} - \frac{-1 + x\text{^}2}{(-1 + x)\text{^}2} & x \neq 1 \\ 0 & \text{True} \end{cases}$$

Input
```
x=-4;
y=Which[
  x>0,1/x,
  x<-3,x^2,
  True,0
  ]
```

Output 16

```
Input     a=-4;
          b=4;
          y=Switch[
             a^2,
             a b,1.0/a,
             b^2,1.0/b,
             _,01
             ]
Output    0.25
```

### Looping Statements

Mathematica has several looping functions, the most common of which is `Do[ ]`.

| | |
|---|---|
| `Do [expr,n]` | evaluates expr n times. |
| `Do [expr,{i,imax}]` | evaluates expr with the variable i successively taking on the values 1 through $i_{max}$ (in steps of 1). |
| `Do [expr,{i,imin,imax}]` | starts with $i = i_{min}$ |
| `Do [expr,{i,imin,imax},di}]` | uses steps di. |
| `Do [expr,{i,{i₁,i₂,…}}]` | uses the successive values $i_1, i_2$, .... |
| `Do [expr,{i,imin,imax},{j,jmin,jmax},…]` | evaluates expr looping over different values of j etc. for each i. |

***Mathematica Examples 1.37***

```
Input     Do[
           Print[5^k],
           {k,3,7}
           ]
Output    125
           625
           3125
           15625
           78125

Input     t=x;
          Do[
           Print[t=1/(1+k t)],
           {k,2,6,2}
           ]
Output    1/(1+2 x)
           1/(1+4/(1+2 x))
           1/(1+6/(1+4/(1+2 x)))

Input     Do[
           Print[{i,j}],
           {i,4},
           {j,i}
           ]
Output    {1,1}
           {2,1}
           {2,2}
           {3,1}
           {3,2}
           {3,3}
           {4,1}
           {4,2}
           {4,3}
           {4,4}
```

```
Input     sum=0;
          Do[
            Print[sum=sum+i],
            {i,1,4}
            ];
          sum
Output     1
           3
           6
           10
           10

Input     fact=1;
          Do[
            Print[fact=fact*i],
            {i,1,4}
            ];
          fact
Output     1
           2
           6
           24
           24

Input     Do[
           Do[
            Do[
             Print["i= ",i," j= ",j," k= ",k],
             {i,1,2}
             ],
            {j,1,2}
            ],
           {k,1,2}
           ]
Output     i=  1   j=  1   k=  1
           i=  2   j=  1   k=  1
           i=  1   j=  2   k=  1
           i=  2   j=  2   k=  1
           i=  1   j=  1   k=  2
           i=  2   j=  1   k=  2
           i=  1   j=  2   k=  2
           i=  2   j=  2   k=  2

Input     sum=0;
          Do[
            Print[i,",",sum=sum+i^2],
            {i,1,6,2}
            ];
          sum
Output     1 , 1
           3 , 10
           5 , 35
Output     35

Input     Do[
           Do[
            If[Sqrt[i^2+j^2]\[Element]Integers,Print[i," ",j]],
            {j,i,10}
            ],
           {i,1,10}
```

```
        ]
Output   3    4
         6    8

Input   Do[
         Print[k!],
         {k,3}
         ]
Output   1
         2
         6

Input   Do[
         Print[k," ",k^2," ",k^3],
         {k,3}
         ]
Output   1    1    1
         2    4    8
         3    9    27

Input   Do[
         Print[k," squared is ",k^2],
         {k,5}
         ]
Output   1  squared is  1
         2  squared is  4
         3  squared is  9
         4  squared is  16
         5  squared is  25

Input   Print["k k^2"]
        Print["-----"]
        Do[
         Print[k," ",k^2],
         {k,5}
         ]
Output   k  k^2
         -----
         1    1
         2    4
         3    9
         4    16
         5    25

Input   Do[
         Print[k],
         {k,1.6,5.7,1.2}
         ]
Output   1.6
         2.8
         4.
         5.2

Input   Do[
         Print[k],
         {k,3(a+b),8(a+b),2(a+b)}
         ]
Output   3 (a+b)
         5 (a+b)
         7 (a+b)
```

```
Input     (* The step can be negative: *)
          Do[
           Print[i],
           {i,10,8,-1}
           ]
Output     10
           9
           8

Input     (* The values can be symbolic: *)
          Do[
           Print[n],
           {n,x,x+3 y,y}
           ]

Output     x
           x+y
           x+2 y
           x+3 y

Input     (* Loop over i and j,with j running up to i-1: *)
          Do[
           Print[{i,j}],
           {i,4},
           {j,i-1}
           ]
Output     {2,1}
           {3,1}
           {3,2}
           {4,1}
           {4,2}
           {4,3}

Input     (*The body can be a procedure:*)
          t=67;
          Do[
           Print[t];t=Floor[t/2],
           {3}
           ]
Output     67
           33
           16
```

| | |
|---|---|
| `Nest[f,expr,n]` | apply f to expr n times. |
| `FixedPoint[f,expr]` | start with expr, and apply f repeatedly until the result no longer changes. |
| `NestList[f,expr,n]` | gives a list of the results of applying f to expr 0 through n times. |
| `While[test,body]` | evaluate body repetitively, so long as test is. |
| `For[start,test,incr,body]` | executes start, then repeatedly evaluates body and incr until test fails to give True. |
| `Break[]` | exits the nearest enclosing Do, For, or While. |

***Mathematica Examples 1.38***

```
Input     Nest[
           f,x,3
           ]
Output    f[f[f[x]]]

Input     Nest[
```

```
         Function[t,1/(1+t)],x,3
         ]
Output   1/(1+1/(1+1/(1+x)))

Input    NestList[
          f,x,4
          ]
Output   {x,f[x],f[f[x]],f[f[f[x]]],f[f[f[f[x]]]]}

Input    NestList[
          Cos,1.0,10
          ]
Output   {1.,0.540302,0.857553,0.65429,0.79348,0.701369,0.76396,0.722102,0.750418,0.731404,0
         .744237}

Input    FixedPoint[
          Function[t,Print[t];Floor[t/2]],67
          ]
Output    67
          33
          16
          8
          4
          2
          1
          0
Output    0

Input    n=17;
         While[
          n=Floor[n/2];n!=0,
          Print[n]
          ]
Output    8
          4
          2
          1

Input    n=1;
         While[
          n<4,Print[n];
          n=n+1
          ]
Output    1
          2
          3

Input    Do[
          Print[i];
          If[i>2,Break[]],
          {i,10}
          ]
Output    1
          2
          3

Input    For[
          i=1;t=x,
          i^2<10,
          i=i+1,
```

```
          t=t^2+i;
          Print[t]
          ]
Output    1+x^2
          2+(1+x^2)^2
          3+(2+(1+x^2)^2)^2

Input    For[
          sum=0.0;x=1.0,
          (1/x)>0.15,
          x=x+1,
          sum=sum+1/x;
          Print[sum]
          ]
Output    1.
          1.5
          1.83333
          2.08333
          2.28333
          2.45
```

# UNIT 1.6

# MODULES, BLOCKS, AND LOCAL VARIABLES

Global Variables are those variables declared in Main Program and can be used by Subprograms. Local Variables are those variables declared in Subprograms. The Wolfram Language normally assumes that all your variables are global. This means that every time you use a name like x, the Wolfram Language normally assumes that you are referring to the same object. Particularly when you write subprograms, however, you may not want all your variables to be global. You may, for example, want to use the name x to refer to two quite different variables in two different subprograms. In this case, you need the x in each subprogram to be treated as a local variable. You can set up local variables in the Wolfram Language using modules. Within each module, you can give a list of variables that are to be treated as local to the module.

| | |
|---|---|
| `Module[{x,y,...},body]` | a module with local variables x, y, .... |
| `Module[{x=x0,y=y0,…},body]` | a module with initial values for local variables. |

***Mathematica Examples 1.39***

```
Input     k=25
Output    25

Input     Module[
           {k},
           Do[
            Print[k," ",2^k],
            {k,3}
            ]
           ]
Output     1    2
           2    4
           3    8

Input     k
Output    25
```

Thus, we can create programs as a series of modules, each performing a specific task. For subtasks, we can embed modules within other modules to form a hierarchy of operations. The most common method for setting up modules is through function definitions,

***Mathematica Examples 1.40***

```
Input     k=25;
          integerPowers[x_Integer]:=Module[
            {k},
            Do[
             Print[k," ",x^k],
             {k,3}
             ]
            ]
          integerPowers[k]
Output     1    25
           2    625
           3    15625
```

```
Input    k
Output   25

Input    t=17
Output   17

Input    Module[
          {t},
          t=8;Print[t]
          ]
Output   8

Input    t
Output   17

Input    g[u_]:=Module[
           {t=u},
           t+=t/(1+u)
           ]
Input    g[a]
Output   a + a/(1+a)

Input    h[x_]:=Module[
           {t},
           t^2-1/;(t=x-4)>1
           ]
Input    h[10]
Output   35
```

The format of `Module` is `Module[{var1, var2, ...}, body]`, where `var1`, `var2`, ... are the variables we localize, and body is the body of the function. The value returned by `Module` is the value returned by the last operator in the body (unless an explicit `Return[]` statement is used within the body of `Module`. In this case, the argument of `Return[arg]` is returned). In particular, if one places the semicolon after this last operator, nothing (`Null`) is returned. As a variant, it is acceptable to initialize the local variables in the place of the declaration, with some global values: `Module[{var1 = value1, var2, ...}, body]`. However, one local variable (say, the one "just initialized" cannot be used in the initialization of another local variable inside the declaration list. The following would be a mistake: `Module[{var1 = value1, var2 = var1, ...}, body]`. Moreover, this will not result in an error, but just the global value for the symbol `var1` would be used in this example for the `var2` initialization (this is even more dangerous since no error message is generated and thus we don't see the problem.) In this case, it would be better to do initialization in steps: `Module[{var1=value1,var2,...}, var2=var1;body]`, that is, include the initialization of part of the variables in the body of `Module`. One can use `Return[value]` statement to return a value from anywhere within the `Module`. In this case, the rest of the code (if any) inside `Module` is slipped, and the result value is returned.

To show how this is done, the following code is an example of a module which will simulate a single gambler playing the game until the goal is achieved or the money is gone.

***Mathematica Examples 1.41***

```
Input    GamblersRuin[a_,c_,p_]:=Module[
           {ranval,var1,var2,var3},
           var1=a;
           var2=c;
           var3=p;
           While[
            0<var1<var2,
            ranval=Random[];
            If[
             ranval<var3,
             var1=var1+1,
```

```
          var1=var1-1
          ]
         ];
        Return[
         var1==var2
         ]
        ]
```

There are several things to notice in this example. First, this is the same thing we have done in the past to define a function. That is, we have a function name `GamblersRuin` with three input variables, a, c, and p. The operator `:=` is used to start the definition. Secondly, the function involves the Mathematica command `Module`. This just tells Mathematica to perform all the commands in the module (like a subroutine in Fortran or a method in C++). There are some special features we need to understand in the `Module` command. The `Module` command has two arguments. The first argument is a list of all the local variables that will only be used inside the module. In the above example, the local variable list is `{ranval,var1,var2,var3}`. These variables are only used in the module and are cleared once the module has been executed. The second argument is all the commands that will be executed each time the module is called. There are some assignment commands at the beginning that are used to make things cleaner. The module uses temporary variables so that the values of the input variables are not overwritten when the module executes.

The last command is added to our list of input lines to return a result from the work done by the `Module`. Without this we would never get any results from out calculation. Any recognized variable type or structure within Mathematica can be returned by a `Module`. In this example, the returned value is the result of testing two variables in the code for equality. The code fragment `var1==var2` tests to determine if the variables, `var1` and `var2`, are equal. If the two variables are equal, then the line outputs `True` and if they are not equal, the line outputs `False`.

Again, all but the last command must be ended by a semicolon. This is to make sure that the commands are separated in the execution. Commands separated by blank spaces will be considered as terms to be multiplied together. Leaving out the semicolon will give rise to lots of error messages, wrong results or both.

Modules in Mathematica allow one to treat the names as local. When one uses `Block` then the names are global, but the values are local.

| | |
|---|---|
| `Block[{x,y,...},body]` | evaluate expr using local values for x,y, .... |
| `Block[{x=x0,y=y0,...},body]` | assign initial values to x,y; and evaluate ... as above. |

`Block[]` is automatically used to localize values of iterators in iteration constructs such as `Do`, `Sum`, and `Table`. `Block[]` may be used to pack several expressions into one unity.

***Mathematica Examples 1.42***

```
Input     Clear[x,t,a]
Input     x^2+3
Output    3+x^2

Input     Block[{x=a+1},%]
Output    3+(1+a)^2

Input     x
Output    x

Input     t=17
Output    17

Input     Module[
           {t},
           Print[t]
           ]
```

```
Output   t$6220

Input    t
Output   17

Input    Block[
          {t},
          Print[t]
          ]
Output   t
```

# CHAPTER 2

# DESCRIPTIVE STATISTICS: FREQUENCY DISTRIBUTIONS AND HISTOGRAMS

Statistics is the science of data. An important aspect of dealing with data is organizing and summarizing the data in ways that facilitate its interpretation and subsequent analysis. This aspect of statistics is called descriptive statistics. Descriptive statistics are distinguished from inferential statistics (or inductive statistics) by its aim to summarize a sample, rather than use the data to learn about the population that the sample of data is thought to represent. This generally means that descriptive statistics, unlike inferential statistics, is not developed based on probability theory.

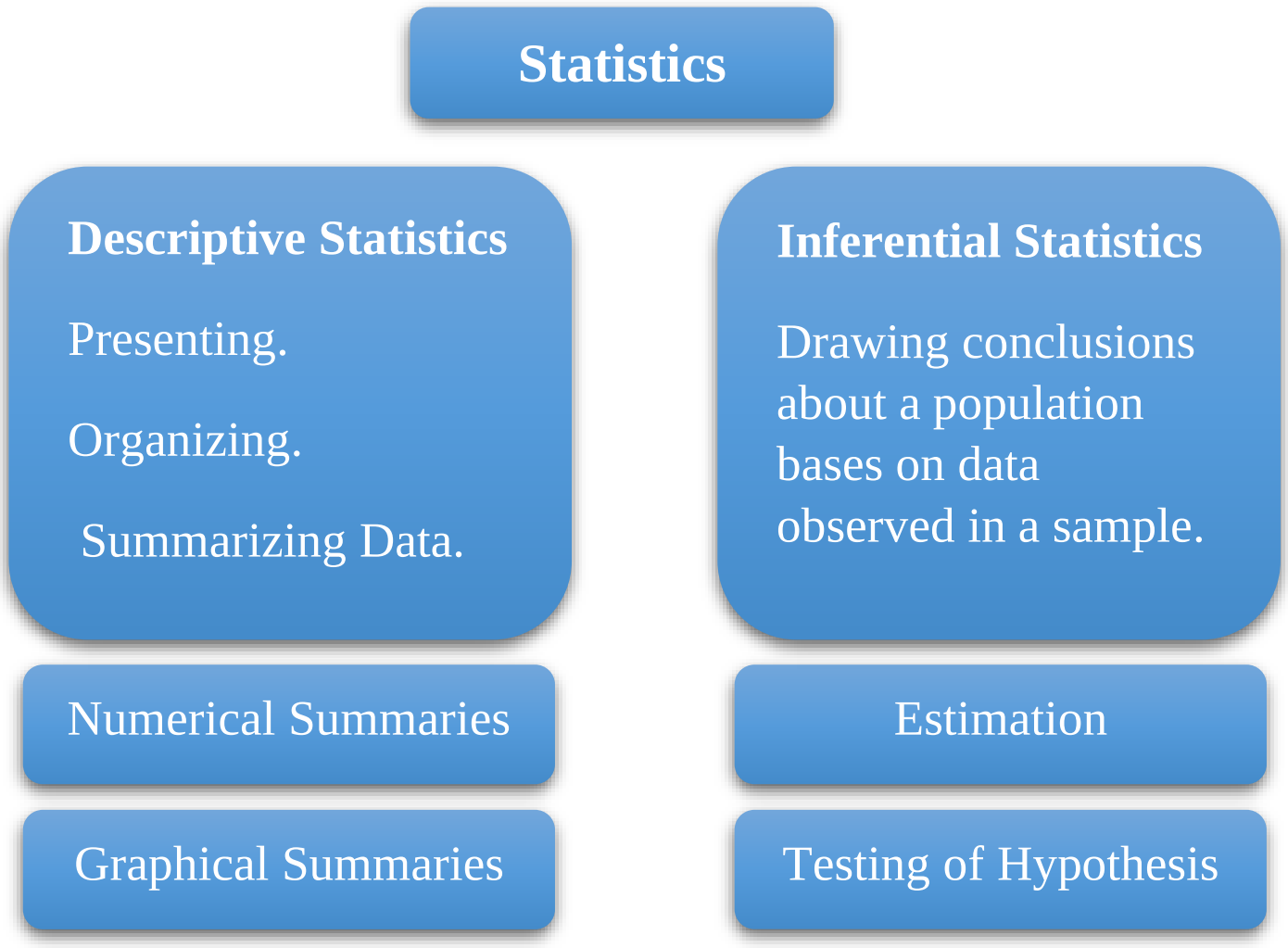

There are two general types of methods within descriptive statistics: graphical and numerical summaries. In this chapter, we will discuss the first of these types—representing a data set using visual techniques. Over the years it has been found that tables and graphs are particularly useful ways of presenting data, often revealing important features such as the range, the degree of concentration, and the symmetry of the data. Any good statistical analysis of data should always begin with plotting the data. In this chapter, we present some common graphical and tabular ways for presenting data. Many visual techniques may already be familiar to you: frequency distribution tables, histograms, polygons, pie charts, bar graphs, scatterplots, and the like. Here we focus on a selected few of these techniques that are most useful and relevant to probability and inferential statistics.

## 2.1 Frequency Distributions

**Definition (Population):** A population is the set of all measurements of interest to the investigator.

**Definition (Sample):** A sample is a subset of measurements selected from the population of interest (Figure 2.1).

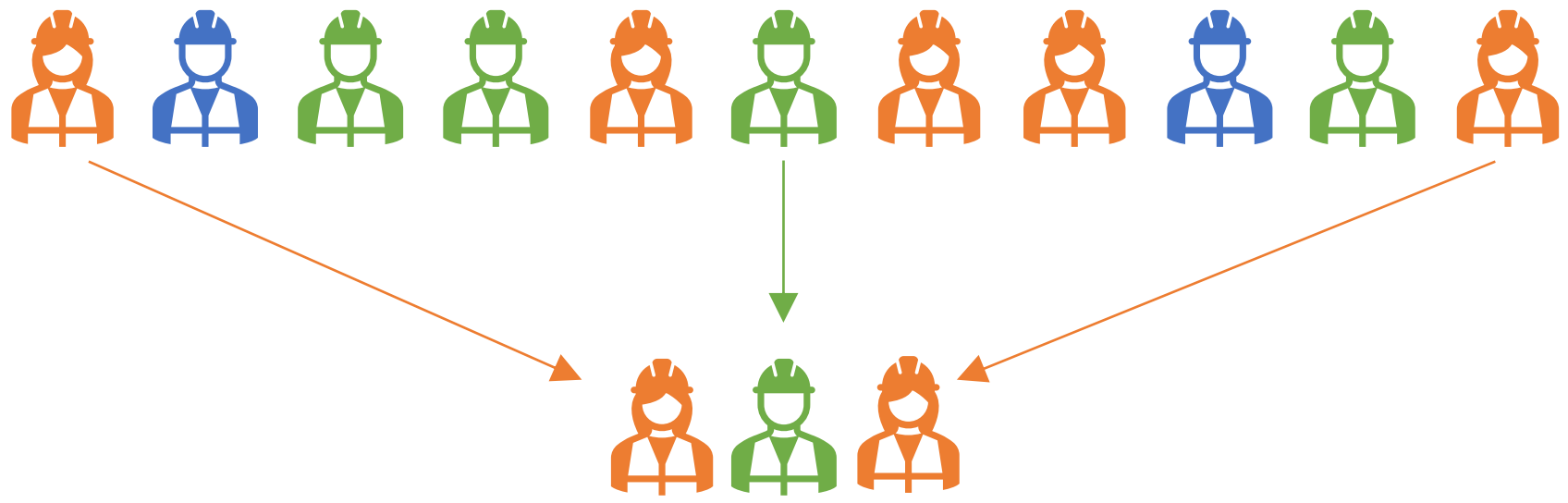

**Figure 2.1.** Difference between population and sample.

**Definition (Qualitative Variables):** Qualitative variables produce data that can be separated into categories. Hence, they are called categorical variables, and produce categorical data.

**Definition (Quantitative Variables):** Quantitative variables, often represented by the letter $x$, produce numerical data (see Table 2.1).

**Table 2.1.** Difference between quantitative and qualitative data.

| Quantitative Data | Qualitative Data |
|---|---|
| ► Countable or measurable, relating to numbers. | ► Descriptive, relating to words and language. |
| ► Tells of how many, how much, or how often. | ► Describe certain attributes, help us to understand the why or how behind certain behaviors. |
| ► Gathered by measuring and counting things. | ► Gathered through observations and interviews. |
| ► Analyzed using statistical analysis. | ► Analyzed by grouping the data into meaningful themes or categories. |

**Definitions**

**Univariate Data:**
Univariate data results when a single variable is measured.

**Bivariate Data:**
Bivariate data results when two variables are measured.

**Multivariate Data:**
Multivariate data results when more than two variables are measured.

**Discrete Variable:**
A discrete variable can assume only a finite or countable number of values.

**Continuous Variable:**
A continuous variable can assume the infinitely many values corresponding to the points on a line interval.

Once we obtain the sample data values, one way to become acquainted with them is through data visualization techniques such as displaying them in tables or graphically. These visual displays may reveal the patterns of behavior of the variables being studied. The statistical table is a list of the categories along with a measure of how often each value occurred. You can measure "how often" in three different ways:

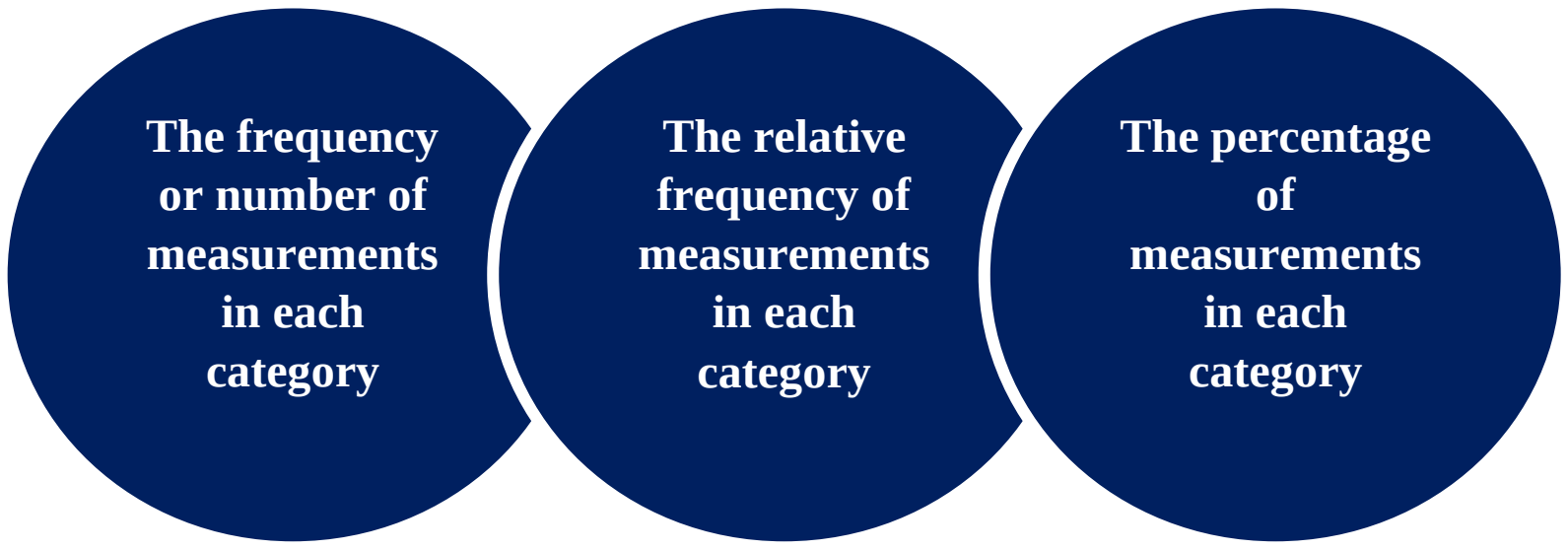

Let us consider the following example.

***Example 2.1***

In the following table the weights of 40 male students at Alex University are recorded to the nearest pound. Construct a frequency distribution.

| | | | | | | | |
|---|---|---|---|---|---|---|---|
| 138 | 164 | 150 | 132 | 144 | 125 | 149 | 157 |
| 146 | 158 | 140 | 147 | 136 | 148 | 152 | 144 |
| 168 | 126 | 138 | 176 | 163 | 119 | 154 | 165 |
| 146 | 173 | 142 | 147 | 135 | 153 | 140 | 135 |
| 161 | 145 | 135 | 142 | 150 | 156 | 145 | 128 |

***Solution***

The largest weight is 176 lb and the smallest weight is 119 lb, so that the range is $176 - 119 = 57$ lb. If five class intervals are used, the class-interval size is $57/5 = 11$ approximately; if 20 class intervals are used, the class-interval size is $57/20 = 3$ approximately.

One convenient choice for the class-interval size is 5 lb. Also, it is convenient to choose the class marks (midpoint) as 120, 125, 130, 135, . . . lb. Thus, the class intervals can be taken as 118– 122, 123– 127, 128– 132, … With this choice the class boundaries are 117.5, 122.5, 127.5, … , which do not coincide with the observed data. The required frequency distribution is shown in Table 2.2.

**Table 2.2.** Frequency distribution includes class intervals, frequency, relative frequency and percentage frequency.

| Inclusive Class Interval | Exclusive Class Interval | Frequency (f) | Relative Frequency | Percentage Frequency (rf) |
|---|---|---|---|---|
| 118 – 122 | 117.5 – 122.5 | 1 | 0.025 | 2.50 |
| 123 – 127 | 122.5 – 127.5 | 2 | 0.050 | 5.00 |
| 128 – 132 | 127.5 – 132.5 | 2 | 0.050 | 5.00 |
| 133 – 137 | 132.5 – 137.5 | 4 | 0.100 | 10.0 |
| 138 – 142 | 137.5 – 142.5 | 6 | 0.150 | 15.0 |
| 143 – 147 | 142.5 – 147.5 | 8 | 0.200 | 20.0 |
| 148 – 152 | 147.5 – 152.5 | 5 | 0.125 | 12.5 |
| 153 – 157 | 152.5 – 157.5 | 4 | 0.100 | 10.0 |
| 158 – 162 | 157.5 – 162.5 | 2 | 0.050 | 5.00 |
| 163 – 167 | 162.5 – 167.5 | 3 | 0.075 | 7.50 |
| 168 – 172 | 167.5 – 172.5 | 1 | 0.025 | 2.50 |
| 173 – 177 | 172.5 – 177.5 | 2 | 0.050 | 5.00 |
| Total | | 40 | 1 | 100 |

**Definitions**

**Raw Data:**
Raw data are collected data that have not been organized numerically.

**Array:**
An array is an arrangement of raw numerical data in ascending or descending order of magnitude.

**Range of Data:**
The difference between the largest and smallest numbers in an array is called the range of the data.

**Frequency of a Value:**
The frequency of a particular value is the number of times the value occurs.

**Class Frequency:**
When summarizing large masses of raw data, it is often useful to distribute the data into classes, or categories. The class frequency is the number of individuals belonging to each class.

**Frequency Distribution:**
A tabular arrangement of data by classes together with the corresponding class frequencies is called a frequency distribution, or frequency table.

**Lower and Upper Limits of a Class:**
The lower limit for a class is the smallest value in that class. On the other hand, the upper limit is the greatest value in that class.

**Class or Class Interval:**
The class interval is the difference between the upper-class limit and the lower-class limit. The class interval formula is given as follows:

Class interval or class: Upper Limit - Lower Limit.

**Exclusive Class Interval:**
In an exclusive class interval, the upper limit of one class is the same as the lower limit of the succeeding class. This method of sorting data ensures continuity between two consecutive classes. For example, $[140-144)$, $[144-148)$ are examples of exclusive class intervals.

In other words, we will adopt the left-end inclusion convention, which stipulates that a class interval contains its left-end but not its right-end boundary point. Thus, for instance, the class interval $20–30$ contains all values that are both greater than or equal to 20 and less than 30.

**Inclusive Class Interval:**
An inclusive class interval is created using the inclusive method of sorting data into a frequency distribution table. In such a method, the lower limit of a class does not get repeated in the upper limit of the preceding class.
For example, $[5-9]$, $[10-14]$, $[15-19]$ are examples of inclusive class intervals.

**Gap between Classes:**
The size of the gap between classes is the difference between the upper-class limit of one class and the lower-class limit of the next class.

When a frequency distribution is analyzed the inclusive class interval has to be converted to an exclusive class interval. This can be done by subtracting the correction factor (=gap between classes/2), usually = 0.5, from the lower class limit and adding correction factor to the upper class limit. For example, $[5-9]$, $[10-14]$, $[15-19]$ are inclusive class intervals. In exclusive class interval, we have $[4.5-9.5)$, $[9.5-14.5)$, $[14.5-19.5)$.

**Class Boundaries or True Class Limits:**
Class boundaries are endpoints used to separate the data into classes or groups. The boundary with the lower value is called the lower boundary while the one with a higher value is called the upper boundary.

In practice, the class boundaries are obtained by adding the upper limit of one class interval to the lower limit of the next-higher class interval and dividing by 2.

**Size, or Width, of a Class Interval:**
The size, or width, of a class interval is the difference between the lower- and upper-class boundaries and is also referred to as the class width, class size, or class length.

**Class Mark:**
The class mark is the midpoint of the class interval and is obtained by adding the lower- and upper-class limits and dividing by 2. The class mark is also called the class midpoint.

**Relative Frequency of a Class:**
The relative frequency of a class is the frequency of the class divided by the total frequency of all classes and is generally expressed as a percentage.

**Relative-Frequency Distribution:**
A tabular arrangement of data by classes together with the corresponding relative class frequencies is called a relative-frequency distribution, percentage distribution, or relative-frequency table.

**Remark:**

The number of class intervals chosen should be a trade-off between

(1) choosing too few classes at a cost of losing too much information about the actual data values in a class and
(2) choosing too many classes, which will result in the frequencies of each class being too small for a pattern to be discernible.

Although 5 to 10 class intervals are typical, the appropriate number is a subjective choice, and of course, you can try different numbers of class intervals to see which of the resulting charts appears to be most revealing about the data. It is common, although not essential, to choose class intervals of equal length.

**General Rules for Forming Frequency Distributions**

1. Determine the largest and smallest numbers in the raw data and thus find the range (the difference between the largest and smallest numbers).
2. Divide the range into a convenient number of class intervals having the same size.
3. Determine the number of observations falling into each class interval; that is, find the class frequencies.

## 2.2 Histogram and Frequency Polygons

Karl Pearson coined the term "histogram" to refer to an approximate representation of the distribution of numerical data. To create a histogram, the first step is to divide the range of values into intervals or bins, and then tally how many values fall within each interval. These bins are often non-overlapping and consecutive, and they may have either equal or unequal sizes. The resulting histogram can provide an estimate of the distribution of the data. We can say that histograms and frequency polygons are two graphic representations of frequency distributions.

1. A histogram consists of a set of rectangles having

- bases on a horizontal axis (the $X$ axis), with centers at the class marks and lengths equal to the class interval sizes,
- areas proportional to the class frequencies, see Figure 2.2.

2. A frequency polygon is a line graph of the class frequencies plotted against class marks. It can be obtained by connecting the midpoints of the tops of the rectangles in the histogram, see Figure 2.2.

The data used to construct a histogram are generated via a function $m_i$ that counts the number of observations that fall into each of the disjoint categories (bins). Thus, if we let $n$ be the total number of observations and $k$ be the total number of bins, the histogram data $m_i$ meet the following conditions:

$$n = \sum_{i=1}^{k} m_i. \tag{2.1}$$

There is no best number of bins, and different bin sizes can reveal different features of the data. The ideal number of bins may be determined or estimated by formula:

$$\text{number of bins} = 1 + 3.3 \log n\,, \tag{2.2}$$

or by the square-root choice formula:

$$\text{number of bins} = \sqrt{n}. \tag{2.3}$$

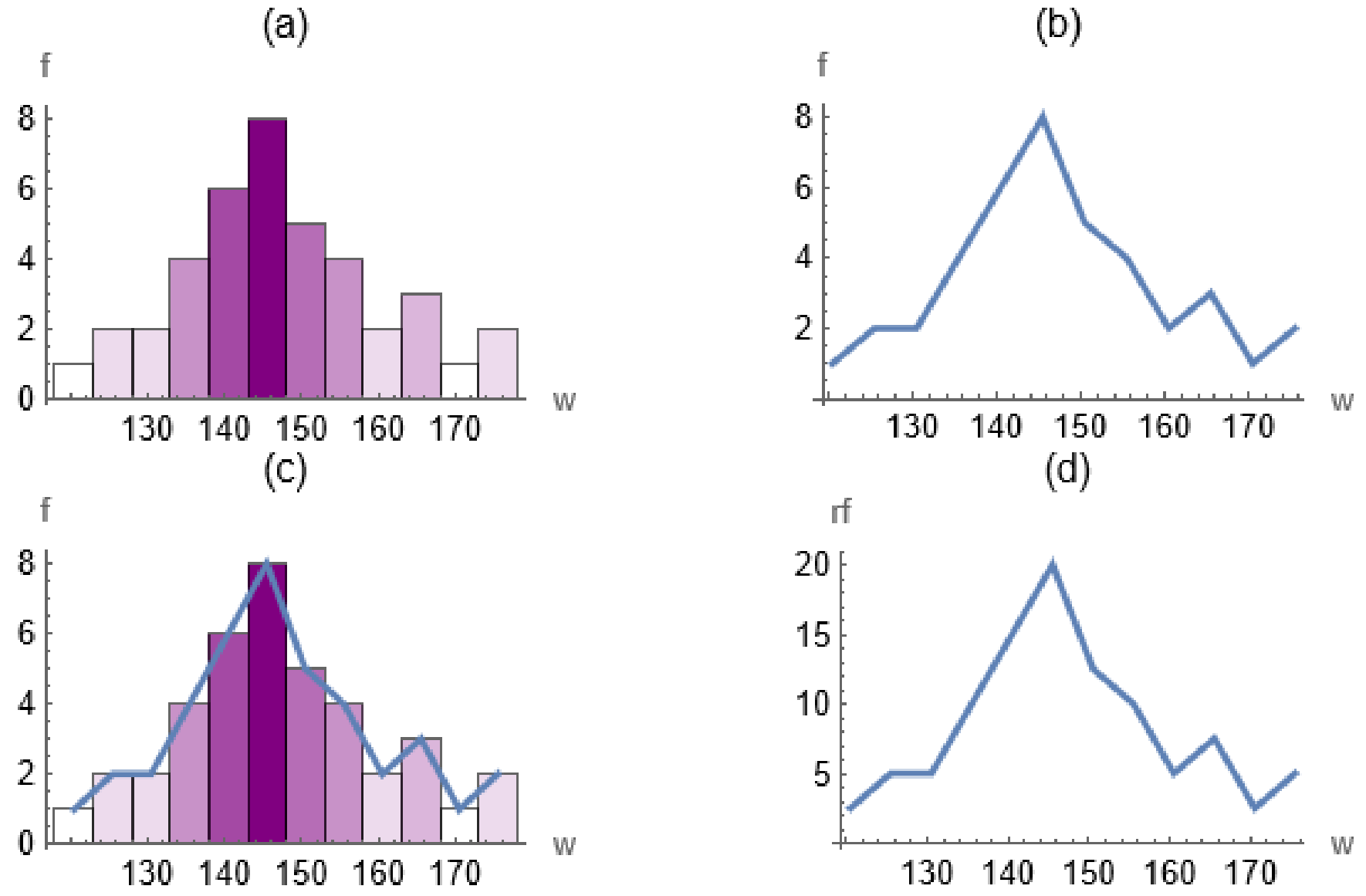

**Figure 2.2.** (a), (b), and (c) are the histogram, frequency polygon and the overlap of histogram and frequency polygon, respectively. However, (d) is the percentage polygon.

**Remarks:**

- Graphic representation of relative-frequency distributions can be obtained from the histogram or frequency polygon simply by changing the vertical scale from frequency to relative frequency, keeping exactly the same diagram. The resulting graphs are called relative-frequency histograms (or percentage histograms) and relative-frequency polygons (or percentage polygons), respectively. (See, for example, Figure 2.2, which uses data from Table 2.2)
- From the histogram we should be able to identify the center (i.e., the location) of the data, spread of the data, skewness of the data, presence of outliers, presence of multiple modes in the data, and whether the data can be capped with a bell-shaped curve. These properties provide indications of the proper distributional model for the data.
- A sample histogram provides valuable information about the population histogram, the graph that describes the distribution of the entire population. Remember, though, that different samples from the same population will produce different histograms, even if you use the same class boundaries. However, you can expect that the sample and population histograms will be similar. As you add more data to the sample, the two histograms become more alike.

## 2.3 Cumulative-Frequency Distributions and Ogives

### Cumulative-Frequency Distributions and Ogives

- The total frequency of all values less than the upper-class boundary of a given class interval is called the cumulative frequency up to and including that class interval.
- A table presenting such cumulative frequencies is called a cumulative distribution.
- A graph showing the cumulative frequency less than any upper-class boundary plotted against the upper-class boundary is called a cumulative-frequency polygon or ogive.
- For some purposes, it is desirable to consider a cumulative-frequency distribution of all values greater than or equal to the lower-class boundary of each class interval.

### Relative Cumulative-Frequency Distributions and Percentage Ogives

- The relative cumulative frequency, or percentage cumulative frequency, is the cumulative frequency divided by the total frequency.
- If the relative cumulative frequencies are used in cumulative-frequency table, in place of cumulative frequencies, the results are called relative cumulative-frequency distributions (or percentage cumulative distributions) and relative cumulative-frequency polygons (or percentage ogives), respectively. (See, for example, Figure 2.3)

***Example 2.2***

Construct (a) a cumulative-frequency distribution, (b) a percentage cumulative distribution, from the frequency distribution in Table 2.2.

***Solution***

The cumulative-frequency distribution and percentage cumulative distribution are shown in Table 2.3. Note that each entry in column 2 is obtained by adding successive entries from column 3 of Table 2.2, thus, $3 = 1 + 2$, $5 = 1 + 2 + 2$, etc. Each entry in column 3 is obtained from the previous column by dividing by $40$, the total frequency, and expressing the result as a percentage. Thus, $1/40 = 2.5\%$. (See Figure 2.3)

**Table 2.3.** Cumulative-frequency distribution and percentage cumulative distribution.

| Weight (w) | Cumulative Frequency (cf) | Percentage Cumulative Distribution (rcf) |
|---|---|---|
| Less than 118 | 0 | 0.0 |
| Less than 122 | 1 | 2.5 |
| Less than 127 | 3 | 7.5 |
| Less than 132 | 5 | 12.5 |
| Less than 137 | 9 | 22.5 |
| Less than 142 | 15 | 37.5 |
| Less than 147 | 23 | 57.5 |
| Less than 152 | 28 | 70.0 |
| Less than 157 | 32 | 80.0 |
| Less than 162 | 34 | 85.0 |
| Less than 167 | 37 | 92.5 |
| Less than 172 | 38 | 95.0 |
| Less than 177 | 40 | 100.0 |

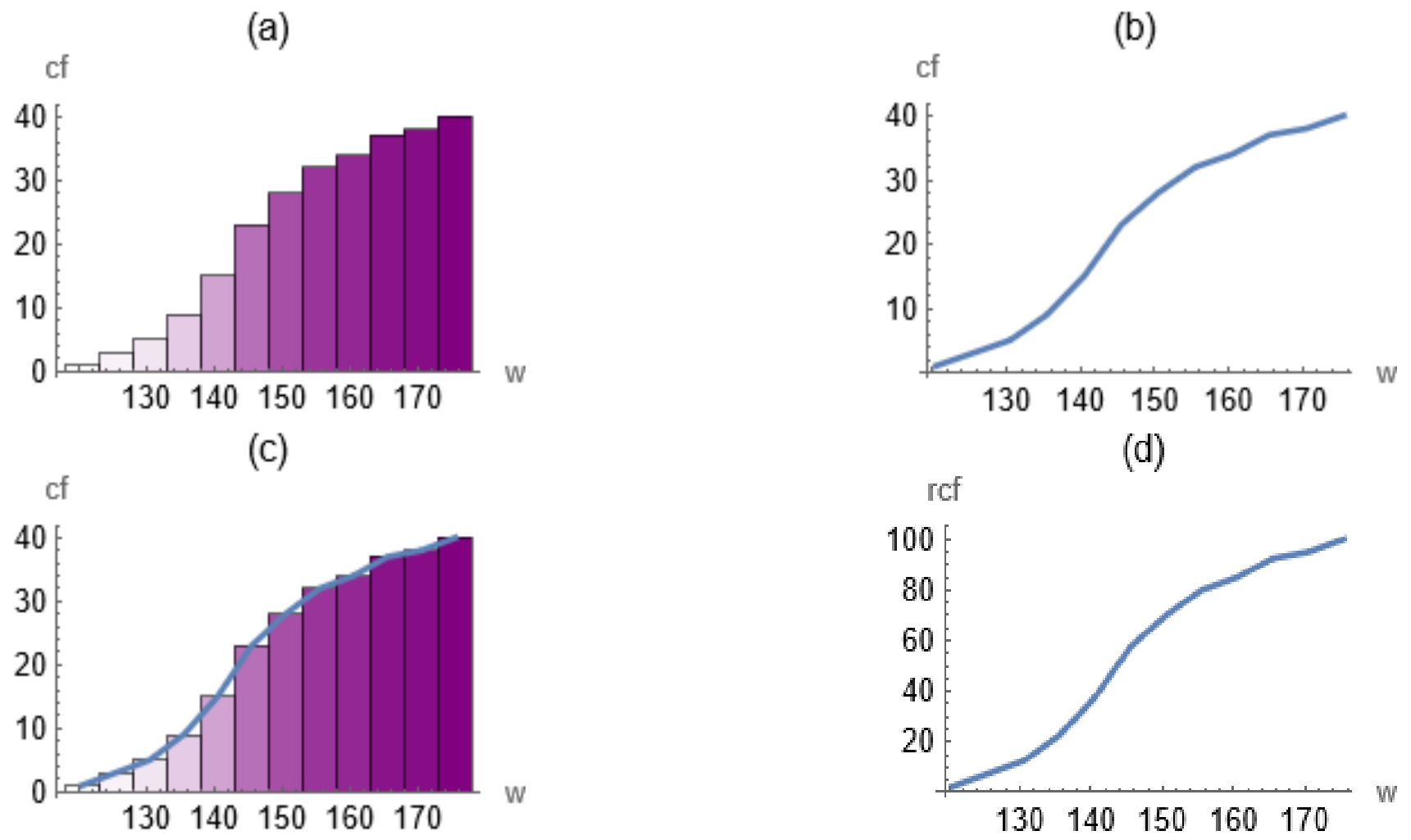

**Figure 2.3.** (a), (b), and (c) are the cumulative distribution, ogive and the overlap of cumulative distribution and ogive, respectively. However, (d) is the percentage ogives.

### Frequency Curves and Smoothed Ogives

- Collected data can usually be considered as belonging to a sample drawn from a large population. Since so many observations are available in the population, it is theoretically possible (for continuous data) to choose class intervals very small and still have sizable numbers of observations falling within each class. Thus, one would expect the frequency polygon or relative-frequency polygon for a large population to have so many small, broken line segments that they closely approximate curves, which we call frequency curves or relative-frequency curves, respectively (see Figure 2.4).
- It is reasonable to expect that such theoretical curves can be approximated by smoothing the frequency polygons or relative-frequency polygons of the sample, the approximation improving as the sample size is increased. For this reason, a frequency curve is sometimes called a smoothed frequency polygon.
- The advantage of using smooth curves to identify distribution shapes is that we need not worry about minor differences in shape. Instead, we can concentrate on overall patterns, which, in turn, allows us to classify most distributions by designating relatively few shapes.
- In a similar manner, smoothed ogives are obtained by smoothing the cumulative-frequency polygons, or ogives. It is usually easier to smooth an ogive than a frequency polygon (see Figure 2.4).

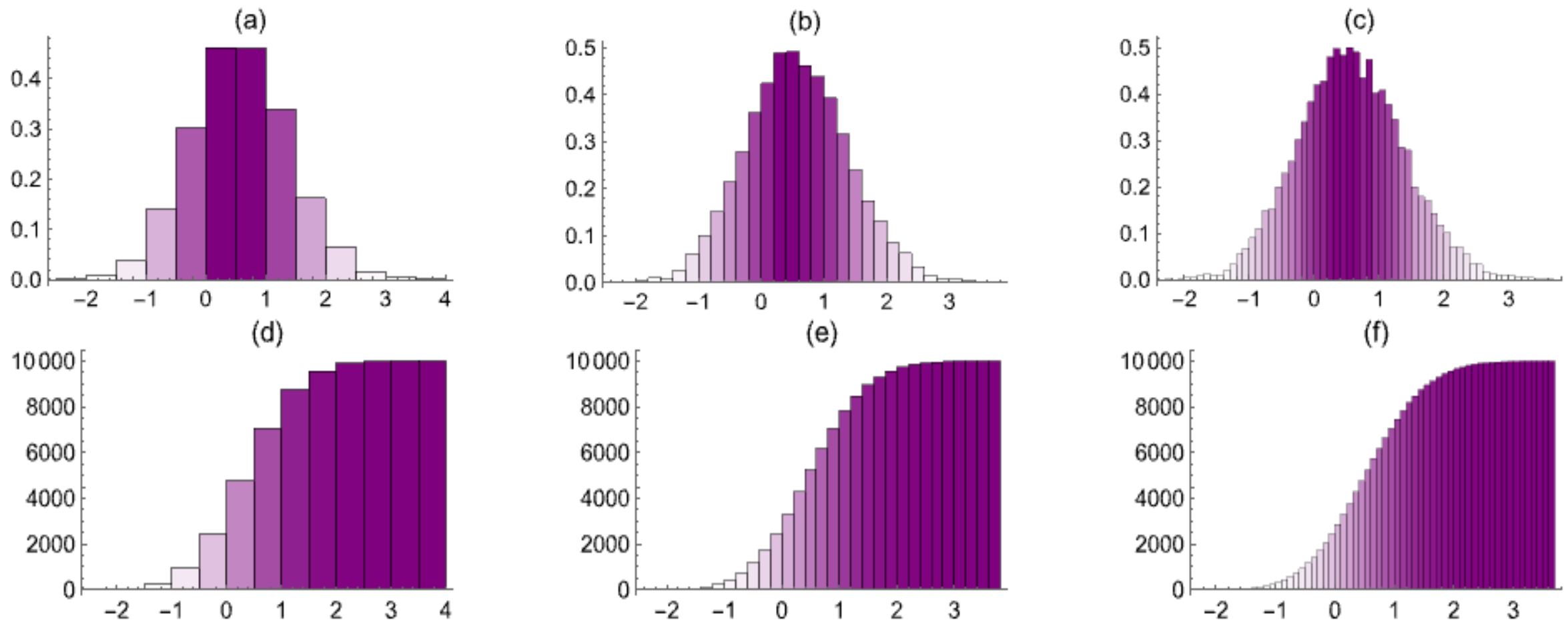

**Figure 2.4.** An ordinary and a cumulative histogram of the same data.

## 2.4 Types of Frequency Curves

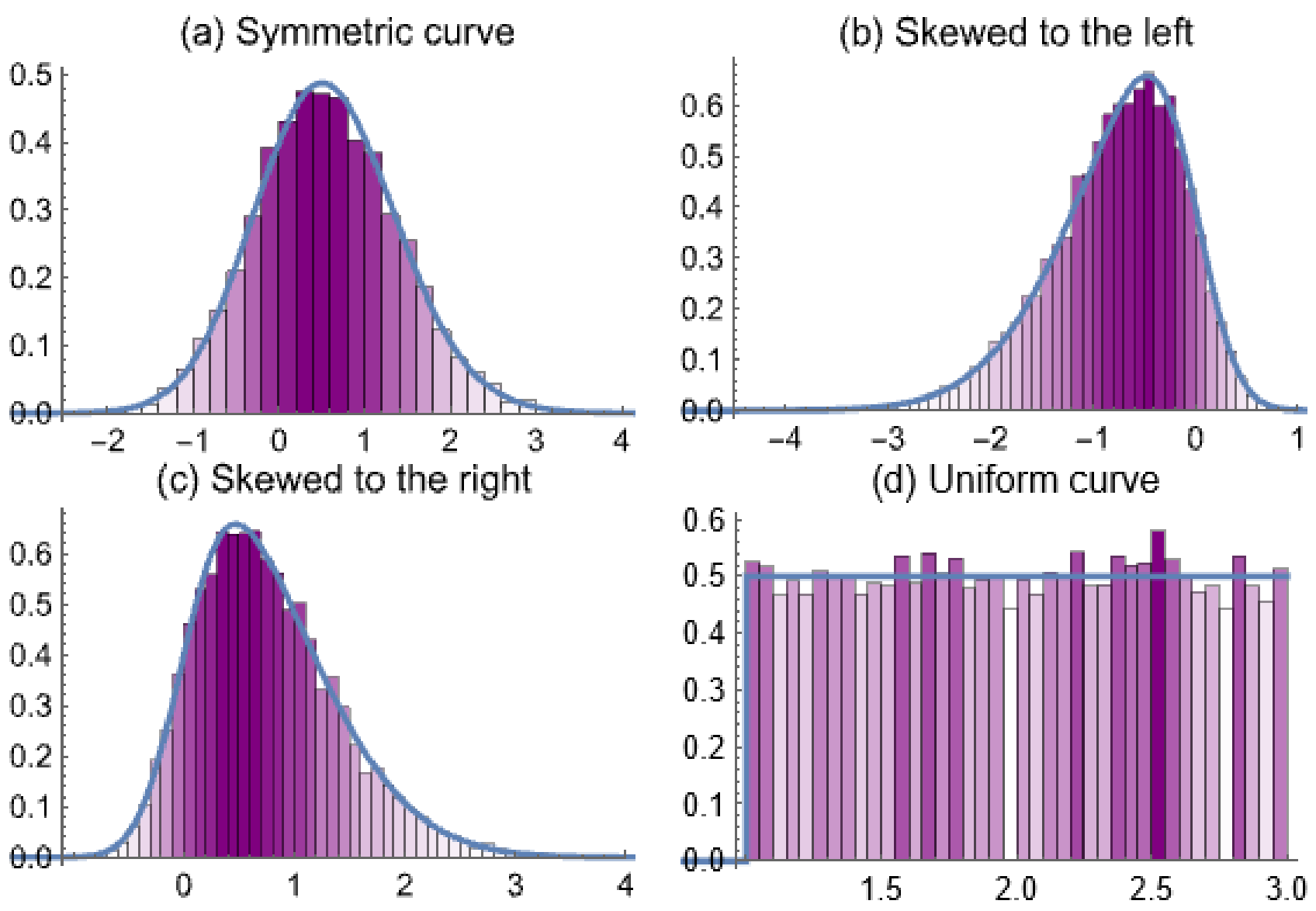

**Figure 2.5.** Four common distributions.

Frequency curves arising in practice take on certain characteristic shapes, (as shown in Figure 2.5).

1. Symmetrical or bell-shaped curves: These curves have a central peak and are perfectly symmetrical on either side of it. They are commonly observed in natural phenomena.
2. Skewed to the left: These curves have a long tail on the left side and a shorter tail on the right side. This is also known as a negative skew and is often observed in situations where there are a few extreme values that drag the mean towards the left.
3. Skewed to the right: These curves have a long tail on the right side and a shorter tail on the left side. This is also known as a positive skew and is often observed in situations where there are a few extreme values that drag the mean towards the right.
4. Uniformly distributed: These curves have a constant frequency across the entire range. They are often observed in situations where all outcomes are equally likely.
5. J-shaped: These curves have a low frequency at the beginning, followed by a sharp increase and then a gradual decrease. They are often observed in situations where there is a low probability of occurrence initially, followed by a higher probability at a certain point.
6. A U-shaped frequency curve: These curves have a high frequency at the beginning and end, with a dip or a trough in the middle.
7. Bimodal frequency curve: These curves have two distinct peaks, indicating two different groups or populations within the data.
8. Multimodal frequency curve: These curves have more than two distinct peaks, indicating multiple groups or populations within the data.

It is not necessary for a distribution to precisely exhibit one of these specific shapes in order to take the name: it need only approximate the shape, especially if the data set is small.

# CHAPTER 3

# MATHEMATICA LAB: DESCRIPTIVE STATISTICS PART 1

Descriptive statistics is an important part of data analysis. Mathematica provides powerful tools for analyzing and visualizing data, including built-in functions for random sampling, order and count statistics, frequency distributions, and distribution shapes. By using these functions, you can gain insights into your data, identify patterns and outliers, and make informed decisions based on statistical analysis.

- Random sampling is a method of selecting a subset of individuals or data points from a larger population in a way that each member of the population has an equal chance of being selected. In Mathematica, you can use built-in functions such as `RandomSample` and `RandomChoice` to generate random samples from a dataset or population. These functions can be useful for testing hypotheses, simulating experiments, and exploring datasets.
- Order and count statistics refer to the values in a dataset that are ranked according to their magnitude or position. In Mathematica, you can use the built-in functions `Min`, `Max`, `Gather`, `Sort`, `Ordering`, `DeleteDuplicates`, `TakeSmallest` and `TakeSmallest` to order and grouping data.
- Also, you can use the built-in function `Histogram` to generate a histogram of the frequency distribution of a dataset.
- Moreover, you can use the built-in functions `PDF`, and `CDF`, to visualize the shape of a distribution.

Therefore, we divided this chapter into four units to cover the following topics, random sampling, order and count statistics, frequency distributions, and distribution shapes. In the following table, we list the built-in functions that are used in this chapter.

| Random Sampling | Order and Grouping Data | Frequency Distributions | Distributions Shapes |
|---|---|---|---|
| RandomInteger | Gather | Tally | Histogram |
| RandomReal | GatherBy | Count | SmoothHistogram |
| RandomPoint | Sort | Counts | Histogram3D |
| RandomChoice | SortBy | Accumulate | SmoothHistogram3D |
| RandomSample | Ordering | BinLists | DensityHistogram |
| RandomVariate | TakeLargest | BinCounts | SmoothDensityHistogram |
| SeedRandom | TakeSmallest | HistogramList | |
| | DeleteDuplicates | | |

**Chapter 3 Outline**

# UNIT 3.1

# RANDOM SAMPLING

Mathematica provides a wide range of functions for generating different types of random data, such as integers, real numbers, vectors, matrices, and graphs which can be useful for a several applications, such as simulations, modeling, and data analysis.

1. When generating random data, it is important to specify the range and distribution of the data. For example, you can use the UniformDistribution function to generate data uniformly distributed between a specified minimum and maximum value, or the NormalDistribution function to generate data following a normal distribution with a specified mean and standard deviation.
2. The random data generated by Mathematica's built-in functions is pseudo-random, meaning that the sequence of numbers generated is deterministic and depends on the seed value. It is important to set a specific seed value using the SeedRandom function if you want to reproduce the same sequence of random numbers.
3. One of the key advantages of Mathematica's random data generation functions is their integration with other mathematical and statistical functions in the program. This allows users to easily incorporate generated random data sets into larger simulations and models, and to analyze the results using a wide range of tools.

| | |
|---|---|
| `RandomInteger[{imin,imax}]` | gives a pseudorandom integer in the range {imin,imax}. |
| `RandomInteger[imax]` | gives a pseudorandom integer in the range {0,...,imax}. |
| `RandomInteger[]` | pseudorandomly gives 0 or 1. |
| `RandomInteger[range,n]` | gives a list of n pseudorandom integers. |
| `RandomInteger[range,{n1,n2,…}]` | gives an n1×n2×… array of pseudorandom integers. |

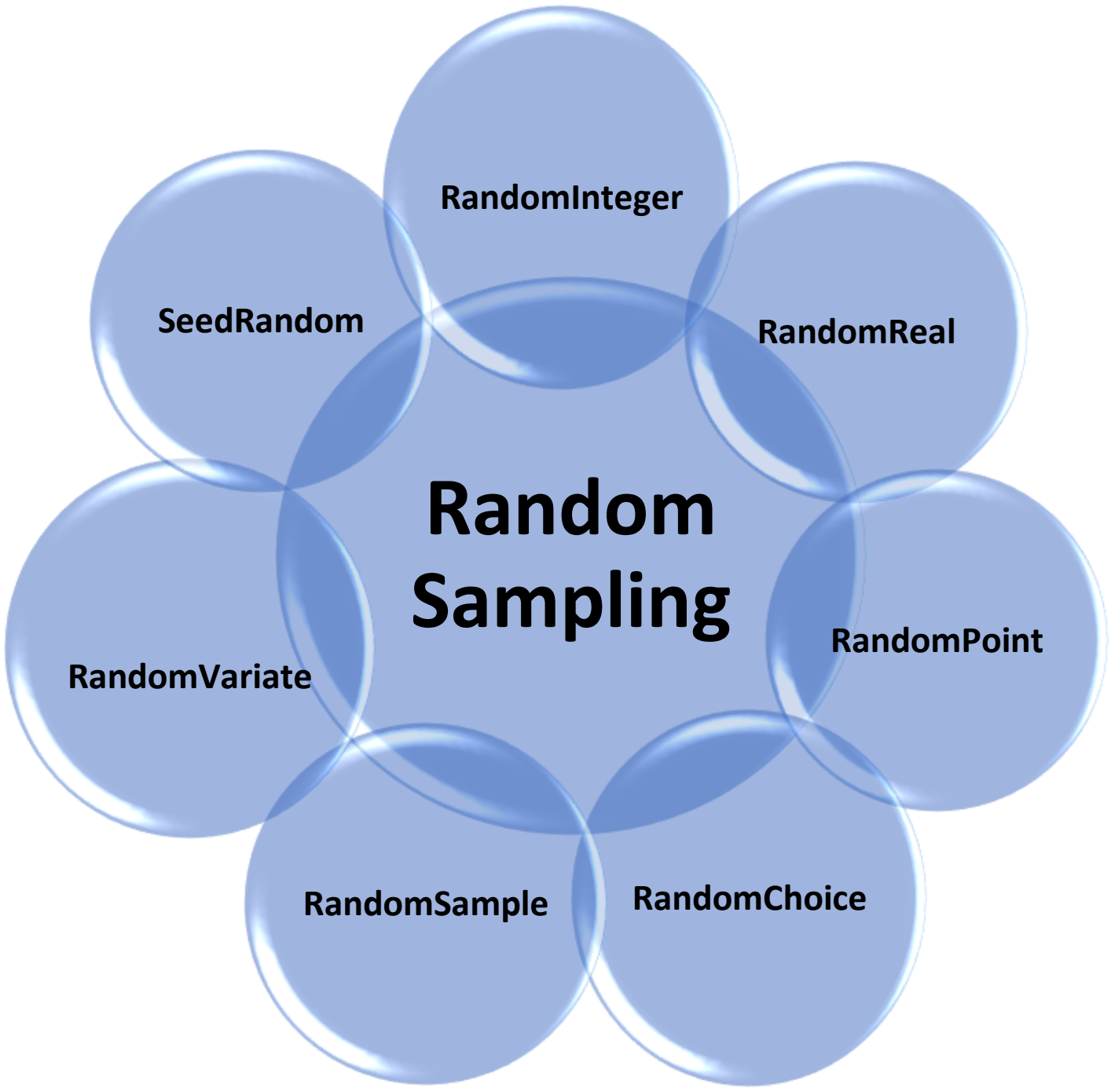

***Mathematica code 3.1*** RandomInteger

```
Input     (* A random integer in the range 1 through 5: *)
          RandomInteger[{1,5}]
Output    4

Input     (* Five random integers in the range 0 through 10: *)
          RandomInteger[10,5]
Output    {2,4,0,1,5}

Input     (* A 3*5 random array of 0s and 1s: *)
          RandomInteger[1,{3,5}]
Output    {{1,0,1,0,1},{1,1,1,1,0},{1,1,1,1,1}}

Input     (* This code defines a function that generates a random integer number between x
          and y: *)
          h[x_,y_]:=RandomInteger[{x,y}]
          h[5,10]
Output    7

Input     (* This code defines a function that generates an array of random integer numbers
          between -5 and 5 with dimensions x by y: *)
          j[x_,y_]:=RandomInteger[{-5,5},{x,y}]
          j[3,2]
Output    {{-1,2},{5,2},{5,2}}

Input     (* Simulate rolling two six-sided dice and adding the results: *)
          Total[
           RandomInteger[{1,6},2]
           ]
Output    9

Input     (* Simulate flipping a coin 10 times and counting the number of heads: *)
          Count[
           RandomInteger[{0(*tail*),1(*head*)},10],
           1
           ]
Output    4

Input     (* Simulate rolling a six-sided die until a 6 is rolled, and count the number of
          rolls: *)
          n=0;
          While[
             RandomInteger[{1,6}]!=6,
             n=n+1
             ];
          n
Output    2

Input     (* This code generates random data for a scatter plot by using the RandomInteger
          function with the interval {1,20}. The data variable is a list of n pairs of random
          values generated using the Table function. The data is plotted using the ListPlot
          function: *)

          n=100;
          data=Table[
              {RandomInteger[{1,20}],RandomInteger[{1,20}]},
              {n}
              ];

          ListPlot[
```

```
          data,
          PlotRange->{{0,21},{0,21}},
          PlotStyle->Directive[Purple,PointSize[Medium]],
          ImageSize->170
          ]
```

Output

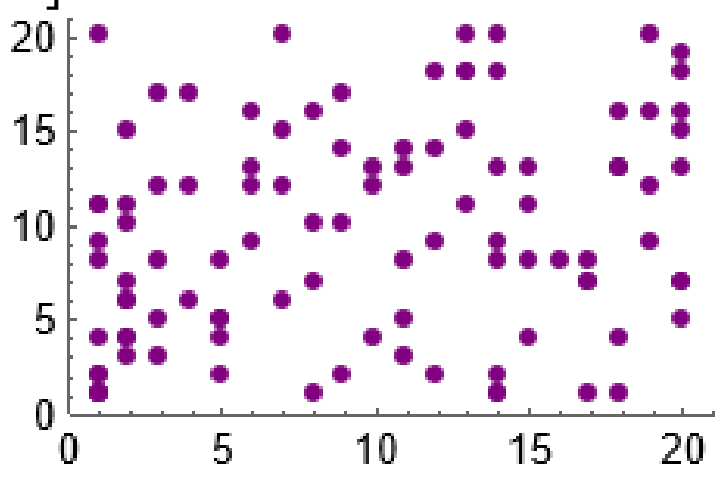

Input

```
(* This code generates random data for a list plot by using the RandomInteger
function with the interval {1,10}. The data variable is a list of n random values
generated using the Table function. The data is plotted using the ListPlot function
with the Joined option set to True: *)

n=100;
data=Table[
    RandomInteger[{1,10}],
    {n}
    ];

ListPlot[
 data,
 Joined->True,
 PlotRange->All,
 PlotStyle->Purple,
 ImageSize->170
 ]
```

Output

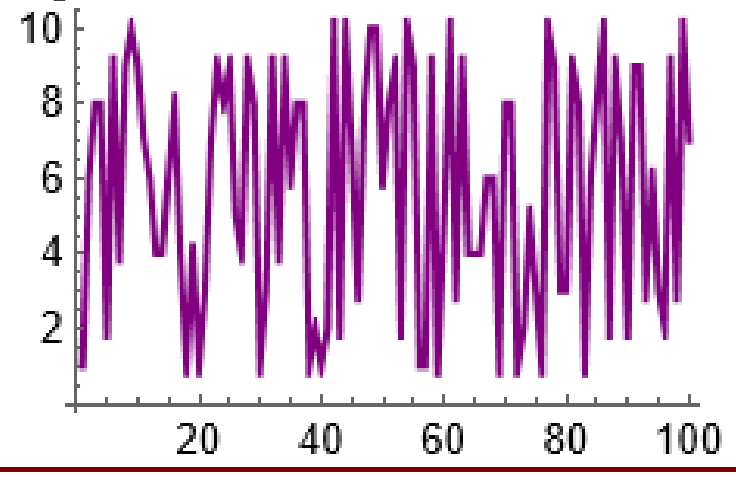

| | |
|---|---|
| `RandomReal[]` | gives a pseudorandom real number in the range 0 to 1. |
| `RandomReal[{xmin,xmax}]` | gives a pseudorandom real number in the range xmin to xmax. |
| `RandomReal[xmax]` | gives a pseudorandom real number in the range 0 to xmax. |
| `RandomReal[range,n]` | gives a list of n pseudorandom reals. |
| `RandomReal[range,{n1,n2,…}]` | gives an n1×n2×… array of pseudorandom reals. |

***Mathematica code 3.2*** RandomReal

```
Input     (* A random real number in the range 0 to 1: *)
          RandomReal[]
Output    0.399182

Input     (* A random real number in the range -5 to 5: *)
          RandomReal[{-5,5}]
Output    -1.70701

Input     (* Four random reals in the range 5 to 9: *)
```

```
        RandomReal[{5,9},4]
Output  {7.93855,6.0295,7.49032,8.26025}

Input   (* A 3*3 array of random reals in the range -2 to 2: *)
        RandomReal[{-2,2},{3,3}]
Output  {{1.88565,-1.90414,-0.938626},{-0.042738,-0.842162,-0.243077},{-1.68051,-
        1.51058,1.79921}}

Input   (* An array of 10 random real numbers with a normal distribution: *)
        RandomReal[NormalDistribution[],{10}]
Output  {0.209014,-1.21077,-0.442329,-0.569644,-0.402114,0.758743,0.386798,0.472529,-
        0.00123986,0.482507}

Input   (* This code defines a function that generates a random real number between x and
        y: *)
        h[x_,y_]:=RandomReal[{x,y}]
        h[5,10]
Output  9.09537

Input   (* This code defines a function that generates an array of random real numbers
        between -1 and 1 with dimensions x by y: *)
        j[x_,y_]:=RandomReal[{-1,1},{x,y}]
        j[3,2]
Output  {{-0.510808,-0.921107},{0.225556,0.083026},{0.52623,-0.732462}}

Input   (* This code defines a function that generates an array of n random real numbers
        between 0 and 1: *)
        k[n_]:=Table[RandomReal[],{n}]
        k[6]
Output  {0.977255,0.616266,0.305921,0.874535,0.399922,0.202713}

Input   (* Simulating a single coin toss: *)
        If[RandomReal[]<0.5,1,0]

        (* Simulating multiple coin tosses: *)
        Table[
         If[RandomReal[]<0.5,1,0],
         {10}
         ]

        (* Simulating a large number of coin tosses and plotting the results: *)
        tosses=Table[
            If[RandomReal[]<0.5,1,0],
            {100}
            ];

        ListPlot[
         tosses,
         PlotStyle->{Purple,PointSize[Small]},
         Filling->Axis,
         PlotRange->All,
         ImageSize->170
         ]
Output  1
Output  {0,0,0,1,1,1,0,1,1,1}
```

Output

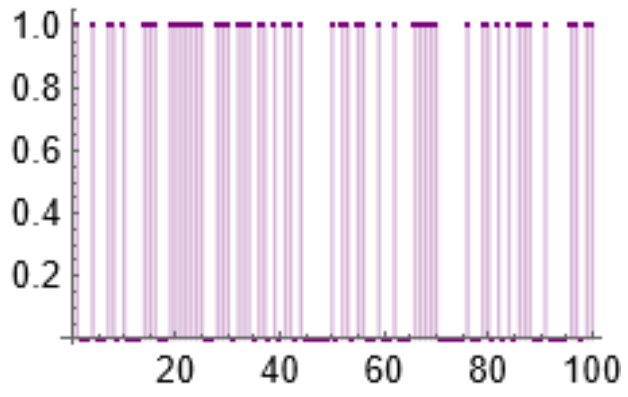

Input

```
(* This example uses Manipulate and ListPlot functions to generate a scatter plot of random points within a square with sides of length 2. The RandomReal function generates a matrix of n rows and 2 columns, each containing a random number between -1 and 1. The slider controls the number of points in the plot: *)

Manipulate[
 ListPlot[
  RandomReal[{-1,1},{n,2}],
  PlotStyle->Directive[Purple,PointSize[Medium],Opacity[0.6]],
  PlotRange->{{-1,1},{-1,1}},
  ImageSize->170
  ],
 {{n,20,"Number of points"},1,100,1}
 ]
```

Output

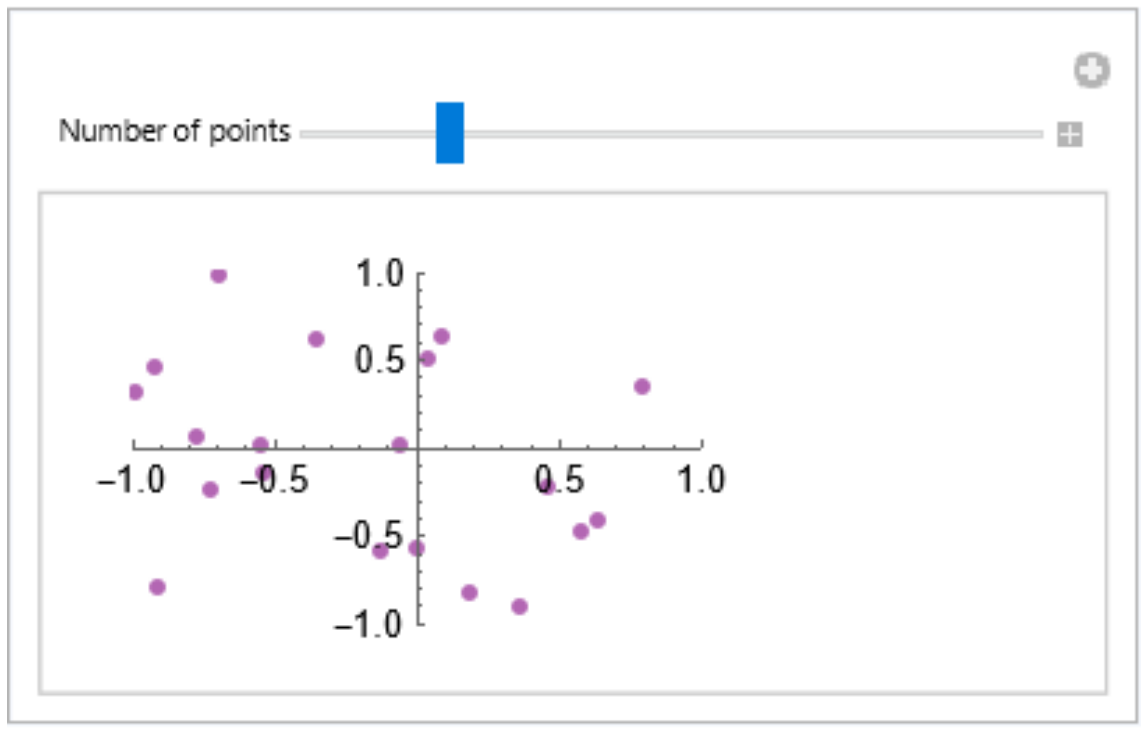

Input

```
(* This code generates random parameters for a linear model using the RandomReal function. The model function represents a linear model of the form y=a*x+b+epsilon, where epsilon is a random noise term generated using the RandomReal function with the interval {-0.1, 0.1}. The data variable stores the x and y values of the model for values of x between 0 and 10 with a step size of 0.1. The data points are plotted using the ListPlot function with the PlotRange option set to All: *)

model[a_,b_,x_]:=a x+b+RandomReal[{-0.1,0.1}];
data=Table[
    {x,model[2,1,x]},
    {x,0,10,0.1}
    ];

ListPlot[
 data,
 PlotRange->All,
 PlotStyle->Purple,
 ImageSize->170
 ]
```

Output

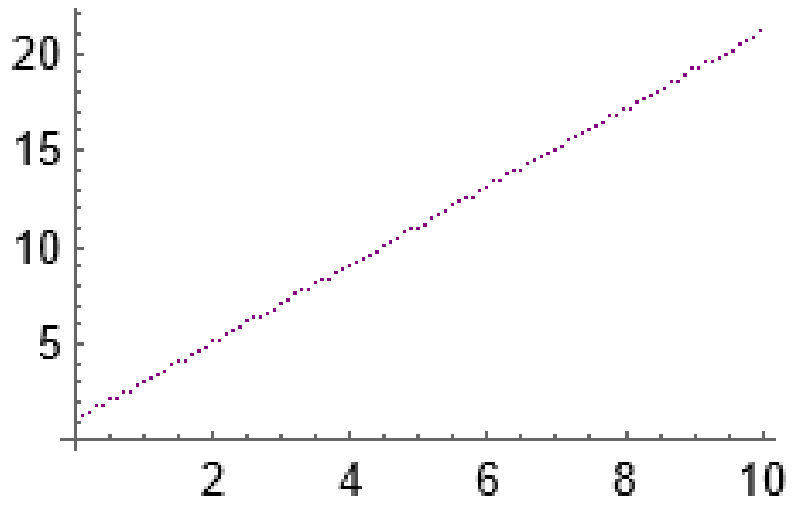

Input

```
(* This code generates random data points for a smooth curve by adding random noise to the function Sin[x]. The RandomReal function with the interval {-0.1, 0.1} is used to generate random noise. The data variable is a list of x and y values generated using the Table function with a step size of 10/n. The data points are plotted using the ListPlot function with the PlotRange option set to All: *)

n=1000;
data=Table[
    {x,Sin[x]+RandomReal[{-0.1,0.1}]},
    {x,0,10,10/n}
    ];

ListPlot[
 data,
 PlotRange->All,
 PlotStyle->Purple,
 ImageSize->170
 ]
```

Output

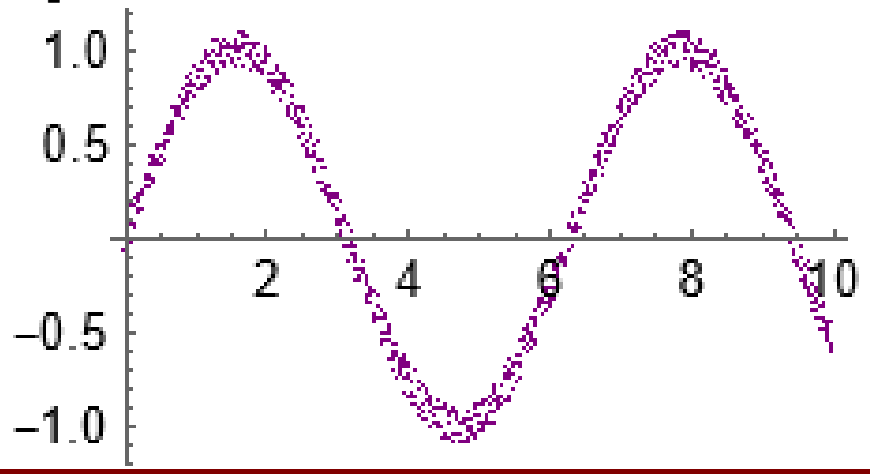

| | |
|---|---|
| RandomPoint[reg] | gives a pseudorandom point uniformly distributed in the region reg. |
| RandomPoint[reg,n] | gives a list of n pseudorandom points uniformly distributed in the region reg. |
| RandomPoint[reg,{n1,n2,…}] | gives an n1× n2×… array of pseudorandom points. |
| RandomPoint[reg,…,{{xmin,xmax},…}] | restricts to the bounds [xmin,xmax] ×…. |

***Mathematica code 3.3*** RandomPoint

Input

```
(* Generate a list of points in a unit disk: *)
pts=RandomPoint[Disk[],500];
Graphics[
 {PointSize[0.02],Purple,Opacity[0.3],Point[pts]},
 ImageSize->150
 ]
```

Output

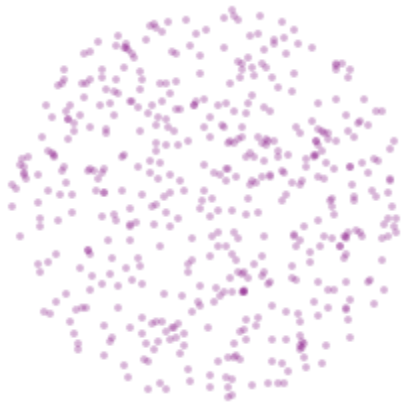

Input

```
(* Generate multiple lists of points on a unit circle: *)

pl=RandomPoint[Circle[],{4,50}];
Table[
 Graphics[
  {PointSize[0.02],Purple,Opacity[0.5],Point[pts]},
  ImageSize->100
  ],
 {pts,pl}
 ]
```

Output

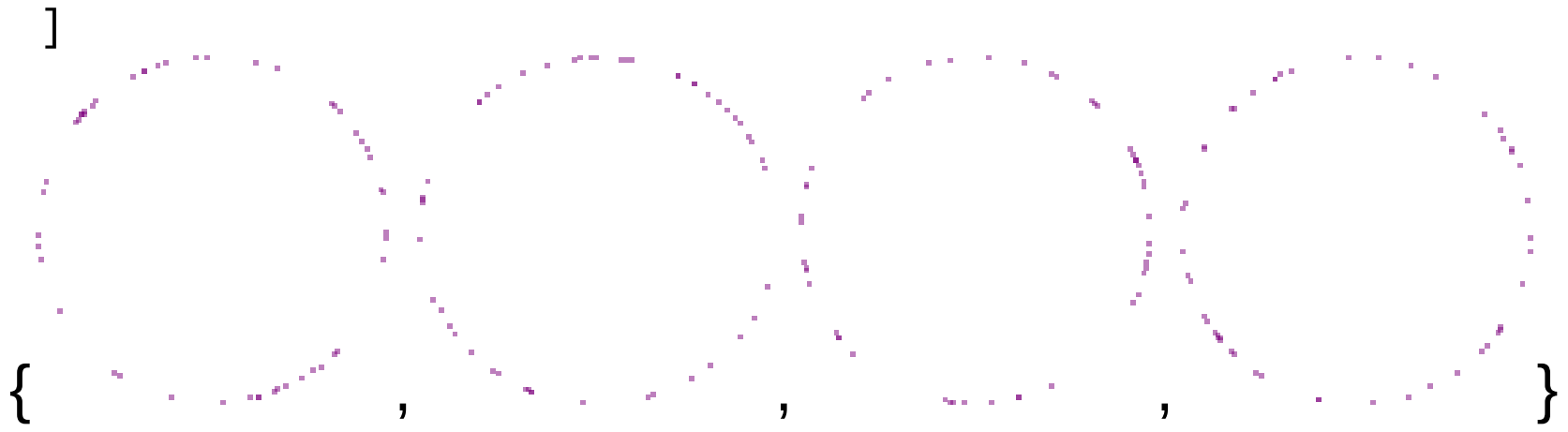

Input

```
(* This code generates a Manipulate with a Graphics object showing a number of randomly generated points within a triangle. The RandomPoint function generates n random points within a triangle with vertices at (0, 0),(1, 0),and (1/2, h),where h is the height of the triangle. The Manipulate allows you to adjust the number of points shown and the height of the triangle: *)

Manipulate[
 Graphics[

{PointSize[0.02],Purple,Opacity[0.4],Point/@RandomPoint[Triangle[{{0,0},{1,0},{1/2,h}}],n]}},
  Frame->True,
  PlotRange->{{-0.1,1.1},{-0.1,1.1}},
  ImageSize->300
  ],
 {{n,400,"Number of Points"},10,600,10},
 {{h,0.7,"Height of Triangle"},0.1,1,0.1}
 ]
```

Output

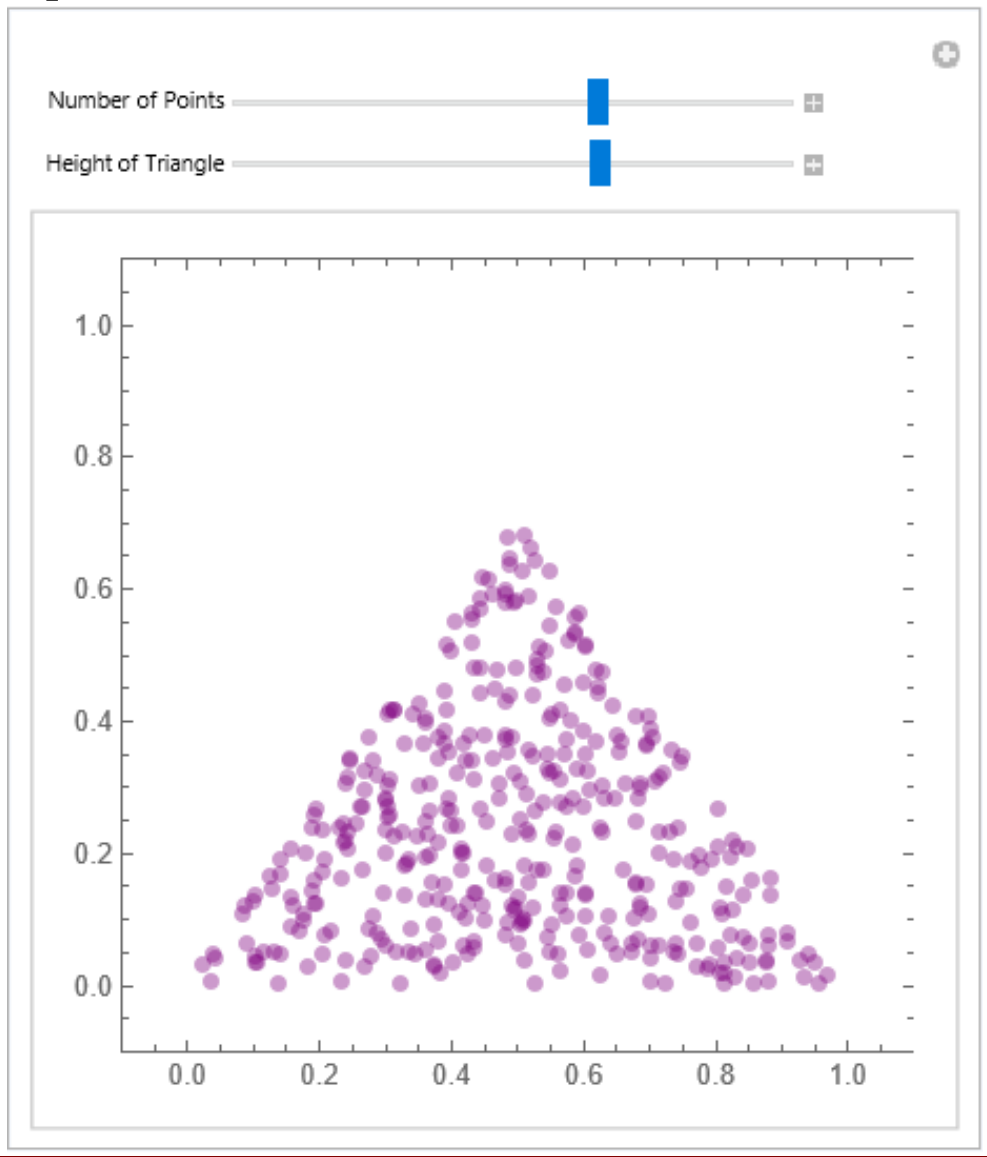

| | |
|---|---|
| `RandomChoice[{e1,e2,…}]` | gives a pseudorandom choice of one of the ei. |
| `RandomChoice[list,n]` | gives a list of n pseudorandom choices. |
| `RandomChoice[list,{n1,n2,…}]` | gives an n1×n2×… array of pseudorandom choices. |
| `RandomChoice[{w1,w2,…} → {e1,e2,…}]` | gives a pseudorandom choice weighted by the wi. |
| `RandomChoice[wlist → elist,n]` | gives a list of n weighted choices. |
| `RandomChoice[wlist → elist,{n1,n2,…}]` | gives an n1×n2×… array of weighted choices. |

***Mathematica code 3.4*** RandomChoice

```
Input     (* Choose among A, B, and C at random: *)
          RandomChoice[{A,B,C}]
Output    B

Input     (* A list of 15 random choices: *)
          RandomChoice[{A,B,C},15]
Output    {B,A,B,A,B,C,A,C,B,A,B,B,B,C,B}

Input     (* A 4*4 array of random choices: *)
          RandomChoice[{A,B,C},{4,4}]
Output    {{B,A,A,C},{B,C,B,C},{B,A,B,C},{B,B,B,A}}

Input     (* Choices weighted with probabilities: *)
          RandomChoice[{0.2,0.2,0.6}->{A,B,C},10]
          RandomChoice[{0.5,0.25,0.25}->{1,2,3},10]
Output    {B,C,C,B,C,C,C,C,B,A}
Output    {1,1,2,1,3,1,1,2,1,3}

Input     (* This code generates two random choices from nested lists: *)
          RandomChoice[{{1,2},{3},{4,5}},2]
Output    {{1,2},{3}}

Input     (* This code simulates flipping a coin 10 times and stores the results in the list
          results: *)
          results=RandomChoice[{"Heads","Tails"},10]
Output    {Tails,Tails,Heads,Heads,Tails,Heads,Heads,Heads,Heads,Tails}

Input     (* This code simulates flipping a coin 100 times and counts the number of times it
          comes up "Heads" and "Tails", respectively: *)
          results=RandomChoice[{"Heads","Tails"},100];
          Count[results,"Heads"]
          Count[results,"Tails"]
Output    44
Output    56

Input     (* This code simulates flipping a weighted coin 100 times, where the probability
          of "Heads" is 0.3 and the probability of "Tails" is 0.7: *)

          results=RandomChoice[{0.3,0.7}->{"Heads","Tails"},100];
          Count[results,"Heads"]
          Count[results,"Tails"]
Output    21
Output    79

Input     (* This code simulates rolling a six-sided die 100 times and counts the number of
          times the number 3 appears: *)
          results=RandomChoice[Range[6],100];
          Count[results,3]
Output    22

Input     (* This code simulates a coin toss experiment with 1000 flips, where each flip is
          randomly chosen to be either 0 (representing tails) or 1 (representing heads) using
          the RandomChoice function. The function is called with the list of coin choices as
```

```
the first argument, and the number of flips as the second argument. The resulting
list of flips is then accumulated using the Accumulate function to compute the
fraction of heads after each flip. The resulting fractions are plotted as a function
of the number of flips using the ListPlot function: *)

numFlips=1000;
coinChoices={0,1};
flipList=RandomChoice[coinChoices,numFlips];
headsFraction=Accumulate[flipList]/Range[1,numFlips];

ListPlot[
 headsFraction,
 PlotRange->All,
 PlotStyle->Purple,
 AxesLabel->{"Number of Flips","Fraction of Heads"},
 ImageSize->300
 ]
```

Output

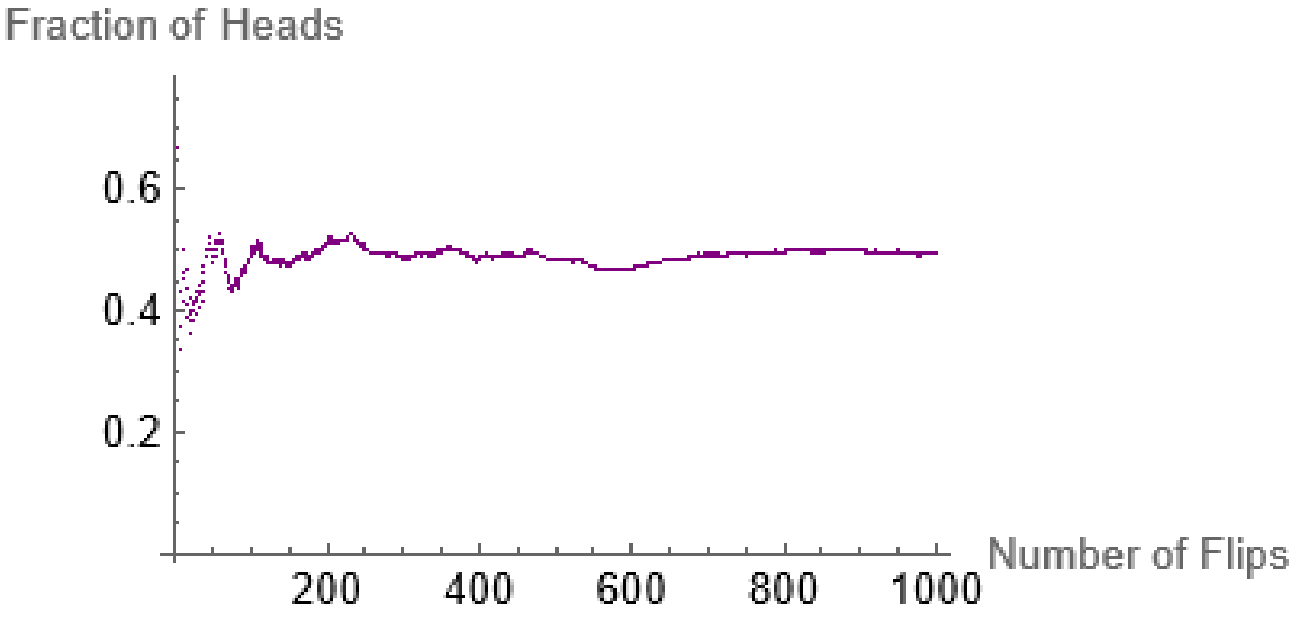

Input

```
(* This code simulates a dice rolling experiment with 1000 rolls, where each roll
is randomly chosen to be a number from 1 to 6 using the RandomChoice function. The
function is called with the list of dice choices as the first argument, and the
number of rolls as the second argument. The resulting list of rolls is then tabulated
using the Table function and the Count function to count the number of times each
face appears. The resulting counts are then plotted as a bar chart using the
BarChart function, with the x-axis labeled with the faces of the die and the y-axis
labeled with the number of rolls: *)

numRolls=1000;
diceChoices={1,2,3,4,5,6};
rollList=RandomChoice[diceChoices,numRolls];
rollCounts=Table[
    Count[rollList,i],
    {i,1,6}
    ];

BarChart[
 rollCounts,
 ChartLabels->Range[1,6],
 ColorFunction->Function[{height},Opacity[height]],
 ChartStyle->Purple,
 AxesLabel->{"Die Face","Number of Rolls"},
 ImageSize->250
 ]
```

Output

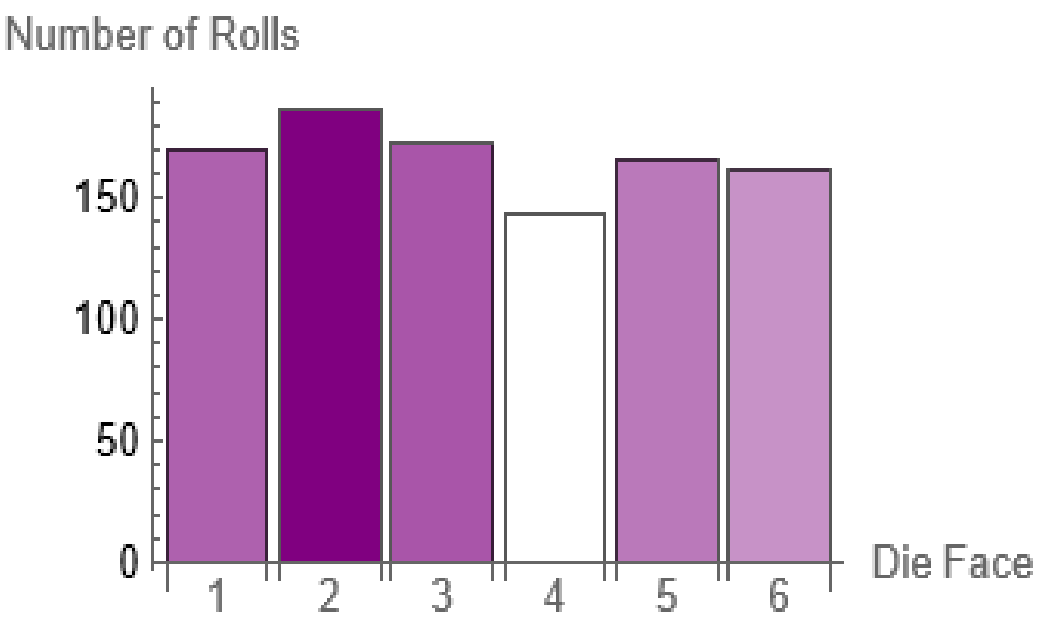

Input

```
(* This code uses the RandomChoice function to generate 10000 rolls of two six-
sided dice, sums the values of the dice, and stores the results in the variable
rolls. It then creates a table counts of the number of times each possible sum
occurs, from 2 to 12. Finally, it displays the frequency of each sum using a
BarChart with labeled bars: *)

rolls=Table[
    RandomChoice[Range[6]]+RandomChoice[Range[6]],
    {10000}
    ];

counts=Table[
    Count[rolls,i],
    {i,2,12}
    ];

BarChart[
 counts,
 ColorFunction->Function[{height},Opacity[height]],
 ChartStyle->Purple,
 ChartLabels->Range[2,12],
 ImageSize->220
 ]
```

Output

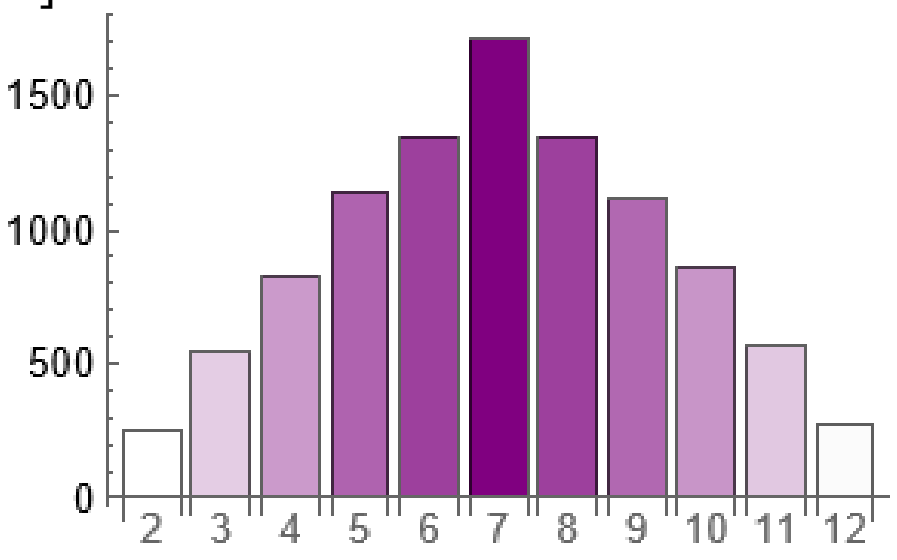

Input

```
(* This code uses the RandomChoice function to generate 10 random polynomial
functions of degree 5 with coefficients -1 or 1 and stores them in the variable
functions. It then displays the functions using a Plot over the interval [-1, 1] :
*)

functions=Table[
    Sum[
     RandomChoice[{-1,1}] x^i,
     {i,0,5}
     ],
    {10}
    ];

Plot[
 functions,
```

```
          {x,-1,1},
          ImageSize->200
          ]
Output
```

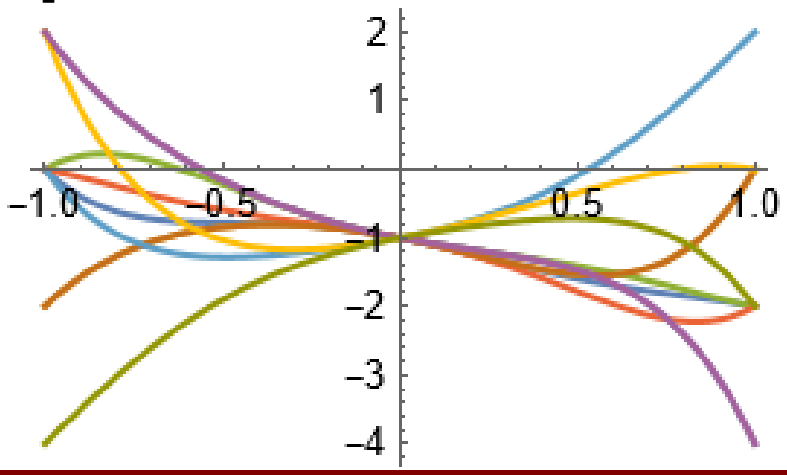

| | |
|---|---|
| `RandomSample[{e1,e2,…},n]` | gives a pseudorandom sample of n of the ei. |
| `RandomSample[{w1,w2,…} → {e1,e2,…},n]` | gives a pseudorandom sample of n of the ei chosen using weights wi. |
| `RandomSample[{e1,e2,…}]` | gives a pseudorandom permutation of the ei. |

***Mathematica code 3.5*** RandomSample

```
Input     (* Find a sample in which no elements ever occur more than once: *)
          RandomSample[Range[20],10]
Output    {11,15,17,20,13,6,2,19,9,16}

Input     (* Generate a random permutation: *)
          RandomSample[Range[5]]
Output    {4,5,3,1,2}

Input     (* Sample over all numbers between-20 and 10: *)
          RandomSample[-20;;10,5]
Output    {6,-4,-13,-10,-3}

Input     (* Sample over even numbers between -5 and 12: *)
          RandomSample[-10;;10;;2,5]
Output    {2,8,10,-4,-10}

Input     (* Guarantee that a set of random integers over a big range has no repetitions: *)
          RandomSample[1;;2^20,15]
Output    {885623,1040581,215821,994714,626881,428795,604106,536316,466021,445442,902056,13
          8851,389943,192821,21237}

Input     (* A function that generates a random sample of 3 elements from the range[x]: *)
          f[x_]:=RandomSample[Range[x],3]
          f[5]
Output    {4,1,5}

Input     (* A function that generates a random sample of 5 elements from the range x to y:
          *)
          g[x_,y_]:=RandomSample[Range[x,y],5]
          g[1,10]
Output    {2,7,9,1,8}

Input     (* This code plots the sine function f[x] between 0 and 2 Pi, and then superimposes
          10 Purple points on the plot. These points are randomly sampled from a list of
          equidistant points on the function. The RandomSample function is called with the
          list of points as the first argument, and 10 as the second argument to select 10
          random points from the list: *)

          f[x_]:=Sin[x];
          Plot[
           f[x],
```

```
 {x,0,2 Pi},
 PlotStyle->{Red,Opacity[0.4]},
 Epilog->{
   PointSize[0.02],
   Purple,
   Point[
    RandomSample[
     Table[{x,f[x]},{x,0,2 Pi,0.1}],
     10
     ]
    ]
   },
 ImageSize->220
 ]
```

Output

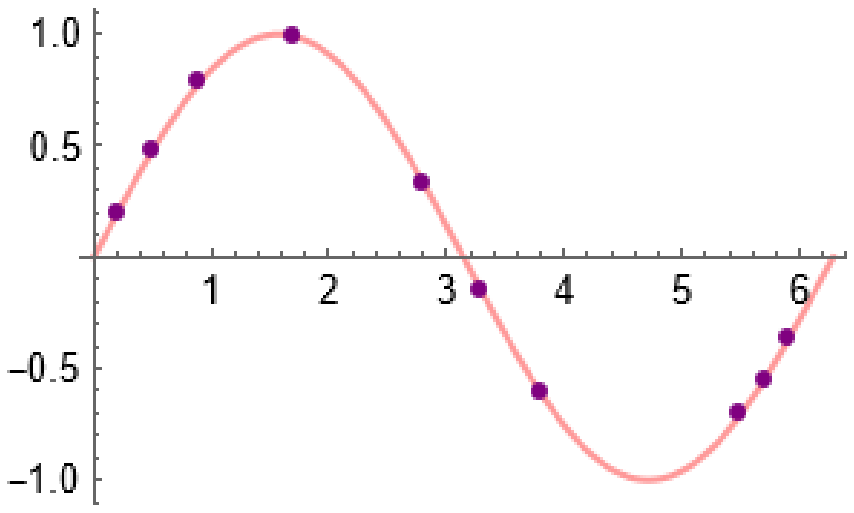

Input

```
(* This code creates a list of six colors, and then selects a random sample of three colors using the RandomSample function. The function is called with the list of colors as the first argument, and 3 as the second argument to select 3 random colors from the list. The selected colors are then used to draw three equally spaced disks in a row using the Graphics function: *)

colors={Red,Green,Blue,Yellow,Purple,Orange};
sampledColors=RandomSample[colors,3];
Graphics[
 {
  EdgeForm[Black],
  sampledColors[[1]],
  Disk[{0,0},1/5],
  sampledColors[[2]],
  Disk[{1/3,0},1/5],
  sampledColors[[3]],
  Disk[{2/3,0},1/5]
  }
 ]
```

Output

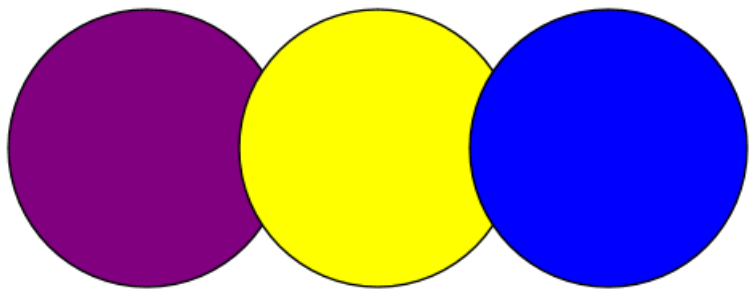

Input

```
(* This code uses the RandomSample function to randomly sample n points in the unit square, where each point has coordinates chosen uniformly at random from [0, 1]. The Manipulate allows the user to control the number of points n, and then displays the resulting scatter plot using ListPlot: *)

Manipulate[
 ListPlot[
  RandomSample[
   Table[{RandomReal[],RandomReal[]},{i,n}],
   n
```

```
      ],
     PlotStyle->{Purple,PointSize[0.02],Opacity[0.4]},
     ImageSize->220
     ],
    {{n,200},1,1000,1}
    ]
```

Output

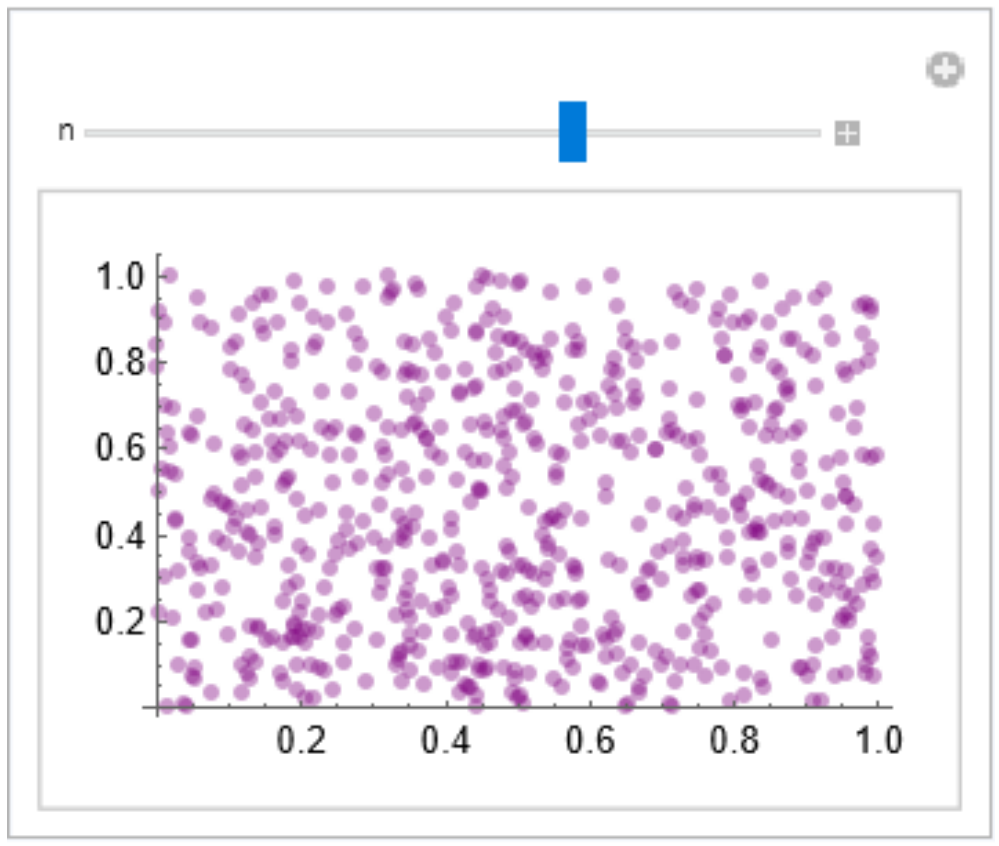

| | |
|---|---|
| `RandomVariate[dist]` | gives a pseudorandom variate from the symbolic distribution dist. |
| `RandomVariate[dist,n]` | gives a list of n pseudorandom variates from the symbolic distribution dist. |
| `RandomVariate[dist,{n1,n2,…}]` | gives an n1× n2×… array of pseudorandom variates from the symbolic distribution dist. |

***Mathematica code 3.6*** RandomVariate

```
Input     (* Simulate a continuous probability distribution: *)
          RandomVariate[NormalDistribution[]]
Output    0.10328

Input     (* Use RandomVariate to generate arrays of different sizes and dimensions: *)
          dist=NormalDistribution[3,5];

          (* A vector: *)
          RandomVariate[dist,{2}]

          (* A matrix: *)
          RandomVariate[dist,{2,3}]

          (* A rank-3 tensor: *)
          RandomVariate[dist,{2,3,4}]

Output    {-0.149637,4.25243}
Output    {{-4.01602,-0.727909,1.02559},{3.75503,-0.072331,16.9712}}
Output    {{{0.0557324,5.85955,14.4352,-4.05258},{3.33552,7.58878,-5.24877,-1.47722},
          {1.62014,-1.80374,1.58631,5.12447}}, {{0.0360298,4.5599,6.24072,13.0314},
          {0.257872,8.69796,5.67066,1.58548},{-1.4172,5.65143,6.08237,-4.98734}}}

Input     (* Simulate a discrete probability distribution: *)
          RandomVariate[PoissonDistribution[3],10]
Output    {6,3,0,2,0,4,8,1,2,4}

Input     (* Simulate a multivariate continuous distribution: *)
          RandomVariate[BinormalDistribution[1/2],5]
Output    {{0.524443,0.29608},{0.108225,0.503184},{-0.933878,1.07482}, {0.762521,0.867619},
          {1.6852,1.67307}}
```

Input

```
(* The code uses the Table function to generate a list of n random variables drawn
from a normal distribution with mean mu and standard deviation sigma using the
RandomVariate function. This list is then passed to ListPlot to create the plot.
The Manipulate function allows for interactive exploration of the plot by adjusting
the values of n, mu, and sigma. This code can be used as a starting point for more
advanced statistical simulations and explorations. For example, one could modify
the code to simulate data from a different distribution, or to add additional
controls for exploring different aspects of the distribution (such as skewness or
kurtosis): *)

Manipulate[
 ListPlot[
  Table[
   RandomVariate[NormalDistribution[μ,σ]],
   {n}
   ],
  Filling->Axis,
  PlotStyle->{Directive[Purple,Opacity[0.8]]},
  PlotRange->{{0,100},{-10,10}},
  ImageSize->300
  ],
 {{n,50},1,100,1},
 {{μ,0},-5,5,0.1},
 {{σ,1},0.1,5,0.1}
 ]
```

Output

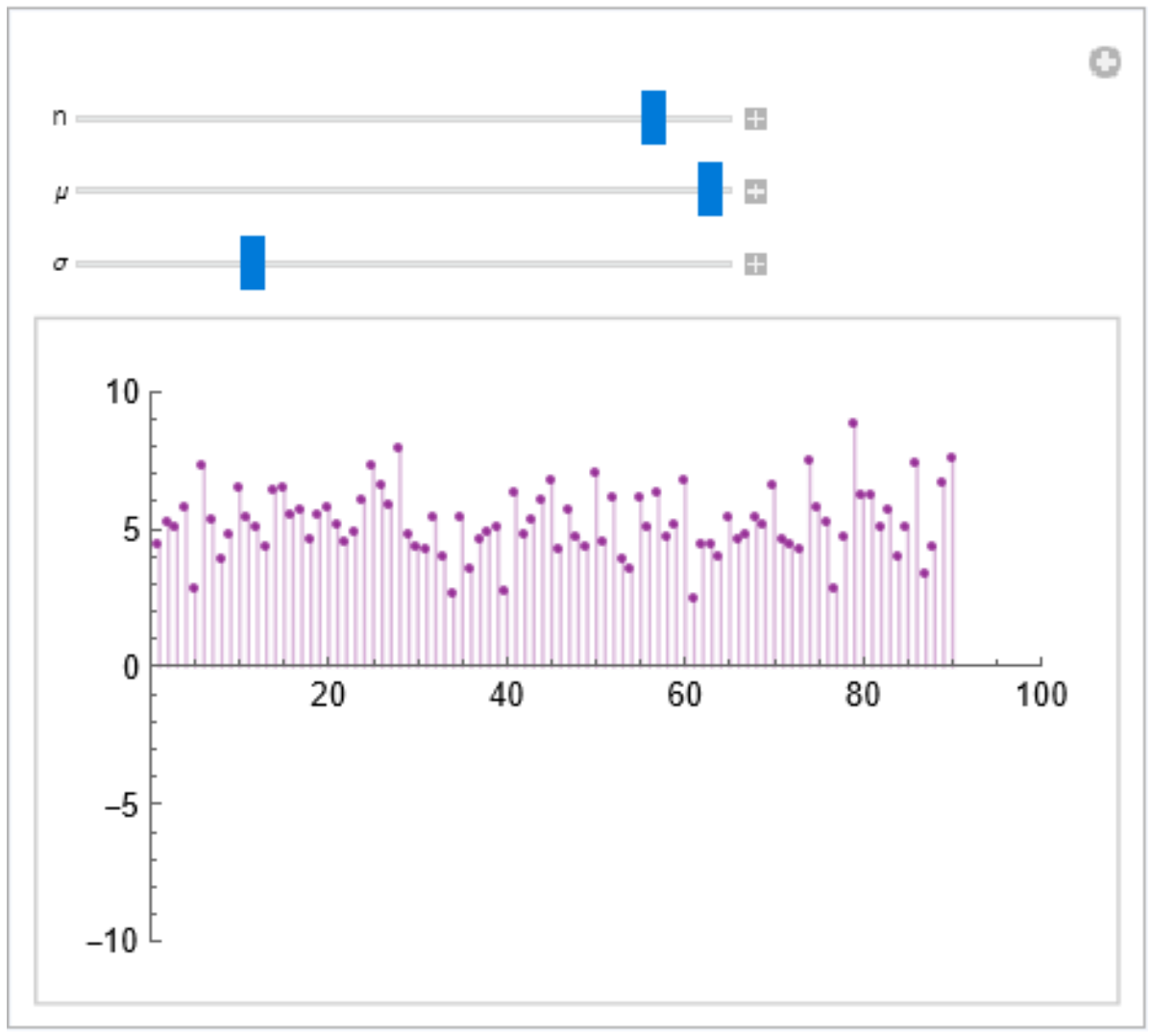

Input

```
(* This code generates a dataset of size n with a normal distribution with mean mu
and standard deviation sigma, and then plots the probability density function (PDF)
of the distribution with the generated data superimposed as points on the plot.
The user can adjust the mean, standard deviation, and sample size using sliders:
*)

Manipulate[
 data=RandomVariate[NormalDistribution[mu,sigma],n];
 Plot[
  PDF[NormalDistribution[mu,sigma],x],
  {x,-10,10},
  PlotRange->All,
  PlotStyle->{Directive[Purple,Opacity[0.8]]},
  Epilog->{
    Directive[Blue,Opacity[0.4]],
    PointSize[0.02],
    Point[Transpose[{data,ConstantArray[0,n]}]]
```

```
   },
  ImageSize->300
  ],
 {{mu,0,"Mean"},-5,5,Appearance->"Labeled"},
 {{sigma,1,"Standard Deviation"},0.1,5,Appearance->"Labeled"},
 {{n,30,"Sample Size"},10,100,10,Appearance->"Labeled"}
 ]
```

Output

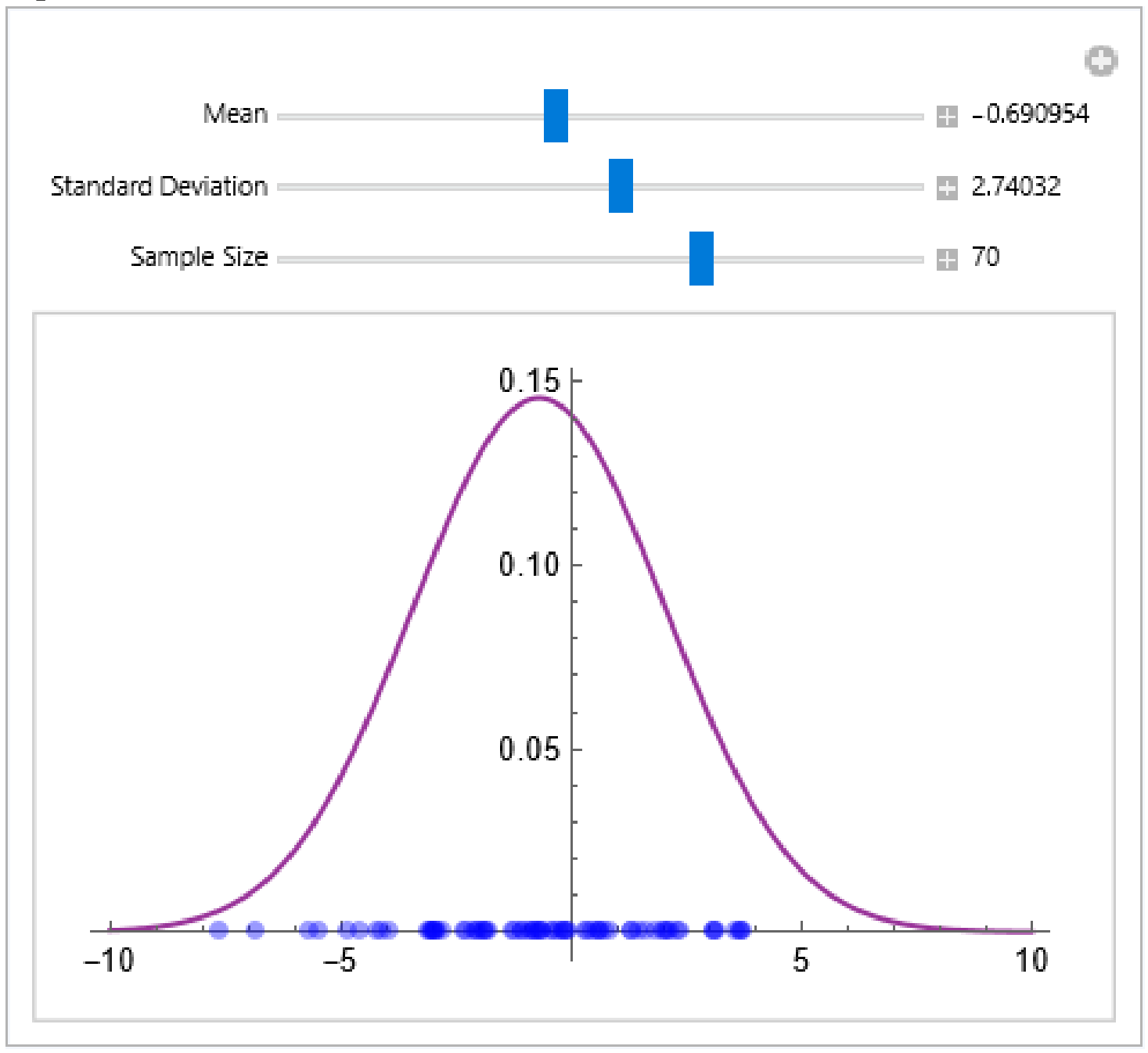

| | |
|---|---|
| SeedRandom[s] | resets the pseudorandom generator, using s as a seed. |
| SeedRandom[] | resets the generator, using as a seed the time of day and certain attributes of the current Wolfram System session. |

***Mathematica code 3.7*** RandomVariate

Input
```
(* This code sets the random number generator to a fixed seed of 1234 using the SeedRandom function. It then generates a list of 10 random real numbers between 0 and 1 using the RandomReal function. Because the seed is fixed, this list of random numbers will always be the same: *)
SeedRandom[1234];
srl=RandomReal[1,10]
```
Output
```
{0.876608,0.521964,0.0862234,0.377913,0.0116446,0.927266,0.543757,0.479332,0.245349,0.759896}
```

Input
```
(* The seed can be a string: *)
SeedRandom["password"];
RandomReal[]
```
Output
```
0.109633
```

Input
```
(* This code sets the random number generator to a fixed seed of 1234 using the SeedRandom function. It then defines a random function f[x] that generates a random real number between 0 and 1 and multiplies it by the sine of x. Finally, it plots the function f[x] over the range [0, 10]. Because the seed is fixed, the random function will always generate the same values, and the plot will be the same every time the code is run: *)

SeedRandom[1234];
```

```
f[x_]:=RandomReal[{0,1}] Sin[x];
Plot[
 f[x],
 {x,0,10},
 PlotStyle->{Directive[Purple,Opacity[0.5]]},
 ImageSize->300
 ]
```

Output

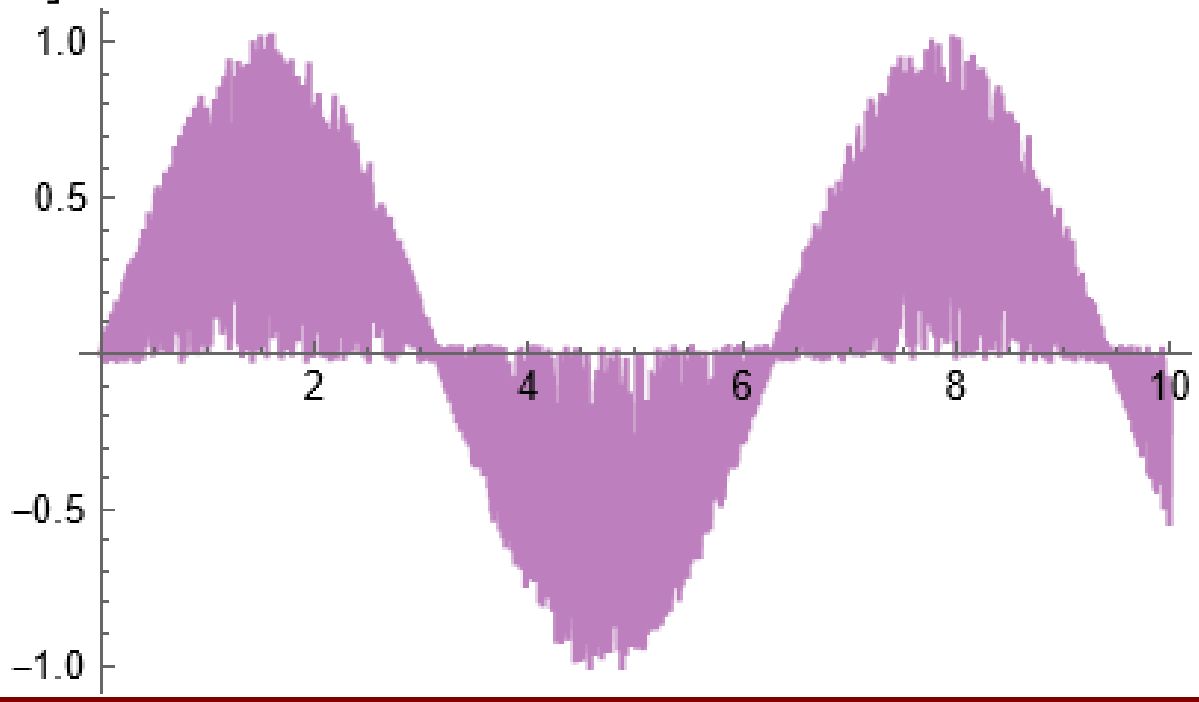

# UNIT 3.2

# ORDER AND GROUPING DATA

Order and grouping data functions are particularly useful when dealing with large datasets that have a lot of variability. By extracting key statistics from the dataset using these functions, you can get a clearer picture of the data and identify patterns and outliers that may not be immediately apparent. `Mathematica` provides a wide range of order and grouping data functions that can be used to analyze and organize data sets. Some of the most commonly used functions include `Min`, `Max`, `Gather`, `Sort`, `Ordering`, `TakeSmallest` and `TakeSmallest`.

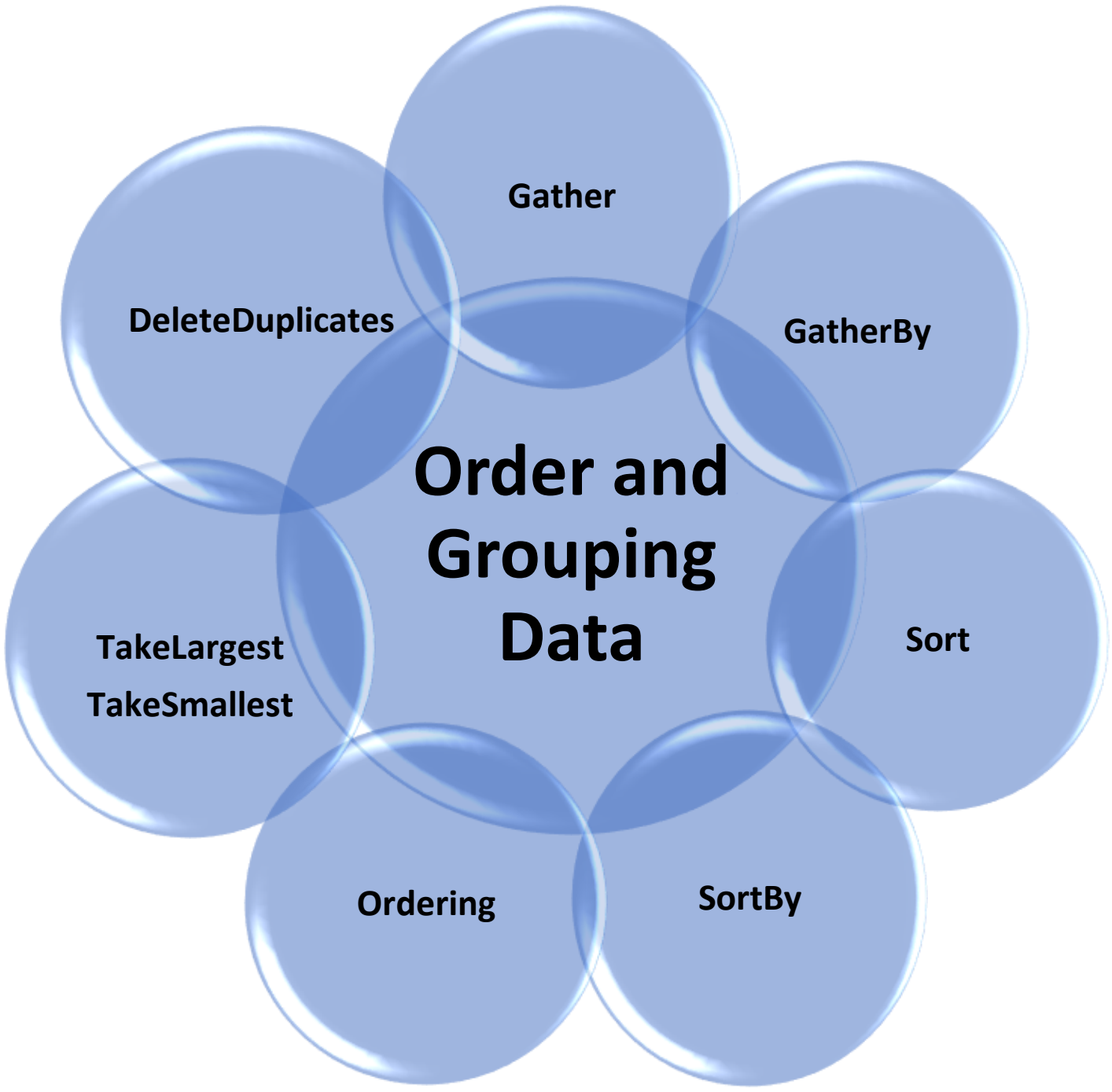

| | |
|---|---|
| `Gather[list]` | gathers the elements of list into sublists of identical elements. |
| `GatherBy[list,f]` | gathers into sublists each set of elements in list that gives the same value when f is applied. |
| `DeleteDuplicates[list]` | deletes all duplicates from list. |
| `TakeLargest[list,n]` | gives the n numerically largest elements in list, sorted in descending order. |
| `TakeLargest[list → prop,n]` | gives the property prop for the n largest elements in list. |
| `TakeLargest[n]` | represents an operator form of TakeLargest that can be applied to an expression. |
| `TakeSmallest[list,n]` | gives the n numerically smallest elements in list, sorted in ascending order. |
| `TakeSmallest[list → prop,n]` | gives the property prop for the n smallest elements in list. |
| `TakeSmallest[n]` | represents an operator form of TakeSmallest that can be applied to an expression. |
| `Ordering[list]` | gives the positions in list at which each successive element of Sort[list] appears. |
| `Ordering[list,n]` | gives the positions in list at which the first n elements of Sort[list] appear. |
| `Ordering[list,-n]` | gives the positions of the last n elements of Sort[list]. |
| `Ordering[list,n,p]` | gives positions in list of elements of Sort[list,p]. |
| `Sort[list]` | sorts the elements of list into canonical order. |
| `Sort[list,p]` | sorts using the ordering function p. |
| `SortBy[list,f]` | sorts the elements of list in the order defined by applying f to each of them. |
| `SortBy[list,{f1,f2,…}]` | breaks ties by successively using the values obtained from the fi. |
| `SortBy[list,f,p]` | sorts the elements of list using the function p to compare the results of applying f to each element. |
| `SortBy[f]` | represents an operator form of SortBy that can be applied to an expression. |

***Mathematica code 3.8*** Order and Grouping Data

```
Input     (* Gather elements into sublists of identical elements: *)
          Gather[{a,b,a,d,b}]
Output    {{a,a},{b,b},{d}}

Input     (* Gather data in odd and even lists: *)
          GatherBy[{1,2,3,4,5,6,7,8,9,10},OddQ]
Output    {{1,3,5,7,9},{2,4,6,8,10}}

Input     (* Gather by the first part: *)
          GatherBy[{{a,1},{b,1},{a,2},{d,1},{b,3}},First]
Output    {{{a,1},{a,2}},{{b,1},{b,3}},{{d,1}}}

Input     (* Delete duplicated elements: *)
          DeleteDuplicates[{1,7,8,4,3,4,1,9,9,2}]
Output    {1,7,8,4,3,9,2}

Input     (* Take the two smallest numbers in a list: *)
          TakeSmallest[{1,3,5,4,7,8,9},2]
Output    {1,3}

Input     (* Take the two largest numbers in a list: *)
          TakeLargest[{1,3,5,4,7,8,9},2]
Output    {9,8}

Input     (* Sort a list: *)
          Sort[{d,b,c,a}]
Output    {a,b,c,d}

Input     (* Sort a list of lists by the last element of each sublist: *)
          SortBy[{{1,2,3},{2,3,1},{3,1,2},{2,2}},Last]
Output    {{2,3,1},{2,2},{3,1,2},{1,2,3}}

Input     (* Sort by the total of each sublist: *)
          SortBy[{{1,2,3},{2,3,1},{3,1,2},{2,2}},Total]
Output    {{2,2},{1,2,3},{2,3,1},{3,1,2}}

Input     (* Find the ordering that sorts a list: *)
          Ordering[{c,a,b}]
          (* Apply the ordering: *)
          {c,a,b}[[{2,3,1}]]
Output    {2,3,1}
Output    {a,b,c}

Input     (* Find the positions of the 4 smallest elements in a list: *)
          Ordering[{2,6,1,9,1,2,3},4]
Output    {3,5,1,6}

Input     (* Cumulative sums: *)
          Accumulate[{a,b,c,d}]
```

```
Output    {a,a+b,a+b+c,a+b+c+d}

Input     (* Accumulate within columns: *)
          Accumulate[{{a,b},{c,d},{e,f}}]
Output    {{a,b},{a+c,b+d},{a+c+e,b+d+f}}

Input     (* Triangular numbers: *)
          Accumulate[Range[10]]
Output    {1,3,6,10,15,21,28,36,45,55}

Input     (* Random walk: *)
          ListLinePlot[
           Accumulate[RandomReal[{-1,1},100]],
           PlotStyle->Directive[Purple,Thickness[0.005]],
           ImageSize->250
           ]
Output
```

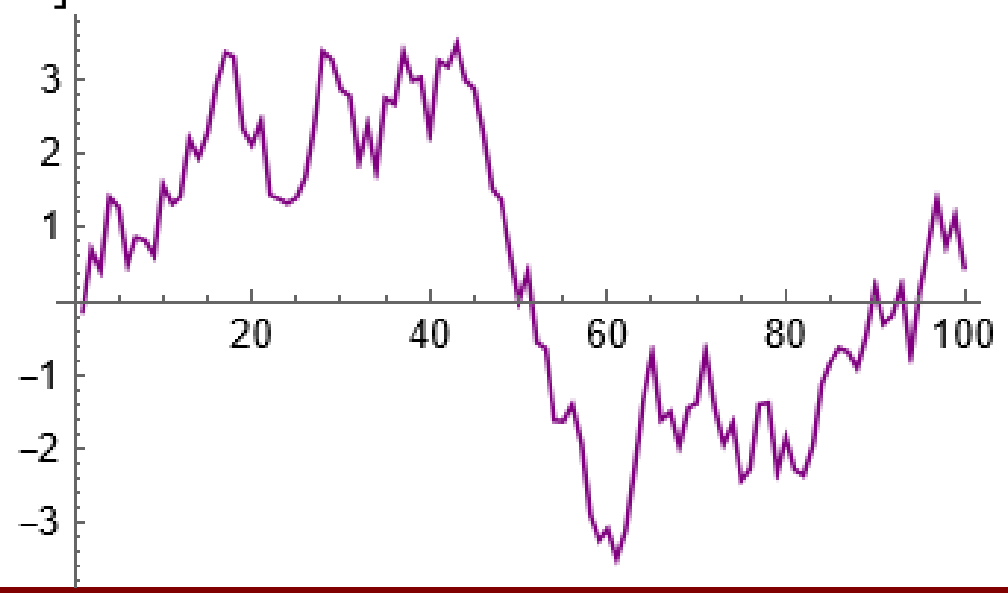

# UNIT 3.3

# COUNT STATISTICS AND FREQUENCY DISTRIBUTIONS

Count statistics refer to the analysis of data that consists of counts or frequencies of events or occurrences. In `Mathematica`, count statistics can be analyzed using a variety of built-in functions, including `Count`, `Tally`, and `BinCounts`. These functions can be used to generate frequency tables and visualize count data.

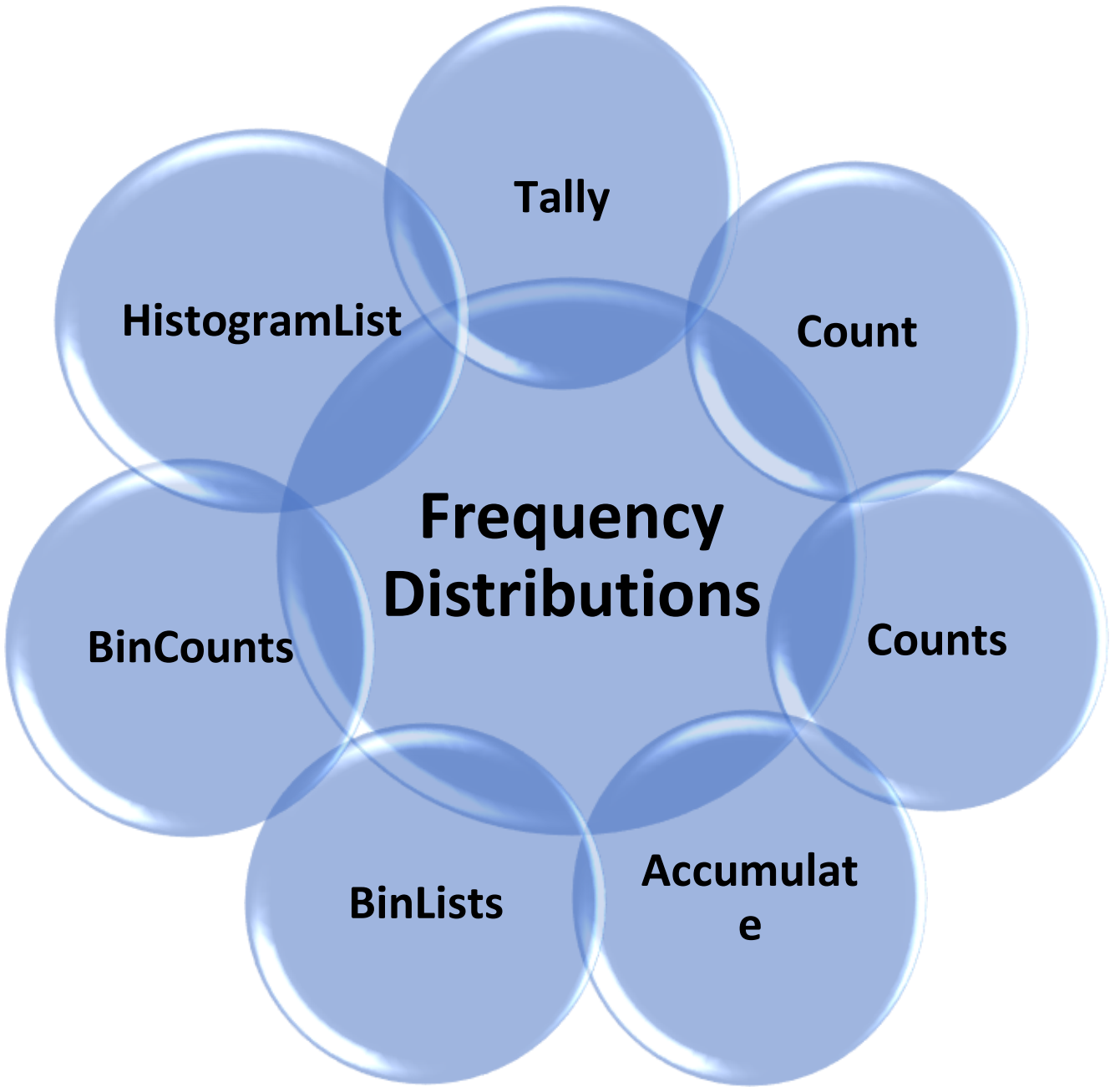

| | |
|---|---|
| `Tally[list]` | tallies the elements in list, listing all distinct elements together with their multiplicities. |
| `BinLists[{x1,x2,…}]` | gives lists of the elements xi whose values lie in successive integer bins. |
| `BinLists[{x1,x2,…},dx]` | gives lists of the elements xi whose values lie in successive bins of width dx. |
| `BinLists[{x1,x2,…},{xmin,xmax,dx}]` | gives lists of the xi that lie in successive bins of width dx from xmin to xmax. |
| `BinLists[{x1,x2,…},{{b1,b2,…}}]` | gives lists of the xi that lie in the intervals [b1,b2), [b2,b3), …. |
| `BinLists[{{x1,y1,…},{x2,y2,…},…},xbins,ybins,…]` | gives an array of lists where the first index corresponds to x bins, the second to y, and so on. |
| `BinCounts[{x1,x2,…}]` | counts the number of elements xi whose values lie in successive integer bins. |
| `BinCounts[{x1,x2,…},dx]` | counts the number of elements xi whose values lie in successive bins of width dx. |
| `BinCounts[{x1,x2,…},{xmin,xmax,dx}]` | counts the number of xi in successive bins of width dx from xmin to xmax. |
| `BinCounts[{x1,x2,…},{{b1,b2,…}}]` | counts the number of xi in the intervals [b1,b2), [b2,b3), …. |
| `BinCounts[{{x1,y1,…},{x2,y2,…},…},xbins,ybins,…]` | gives an array of counts where the first index corresponds to x bins, the second to y, and so on. |

***Mathematica code 3.9*** Tally

```
Input     (* Obtain tallies for a list of symbols: *)
          Tally[{a,a,b,a,c,b,a}]
Output    {{a,4},{b,2},{c,1}}

Input     (* Results are returned in order of first occurrence in the list: *)
          Tally[{b,a,b,a,c,b,a}]
Output    {{b,3},{a,3},{c,1}}

Input     (* Count how many times each element appears in a list: *)
          Tally[{1,2,3,2,1}]
Output    {{1,2},{2,2},{3,1}}

Input     (* Count how many times each sublist appears in a list of sublists: *)
          Tally[{{1,2},{3,4},{1,2},{5,6}}]
Output    {{{1,2},2},{{3,4},1},{{5,6},1}}

Input     (* Count how many times each character appears in a string: *)
          Tally[
           Characters[
            "the quick brown fox jumps over the lazy dog"
            ]
           ]
Output    {{t,2},{h,2},{e,3},{ ,8}, {q,1}, {u,2}, {i,1}, {c,1}, {k,1}, {b,1}, {r,2}, {o,4},
          {w,1},{n,1},{f,1},{x,1},{j,1},{m,1},{p,1},{s,1},{v,1},{l,1},{a,1},{z,1},{y,1},{d,
          1},{g,1}}

Input     (* This code creates a Manipulate slider for the number of random integers (n) to
          generate and counts the number of occurrences of each integer using Tally. The
          resulting counts are plotted using ListPlot, with the integer values on the x-axis
          and their corresponding counts on the y-axis: *)
          Manipulate[
           data=RandomInteger[{1,10},n];
           count=Tally[data];
           ListPlot[
            count,
            Filling->Axis,
            PlotStyle->{Directive[Purple,Opacity[0.8],PointSize[Medium]]},
            PlotRange->{{0,11},{0,200}},
            ImageSize->300
            ],
           {{n,100},10,1000,10}
           ]
Output
```

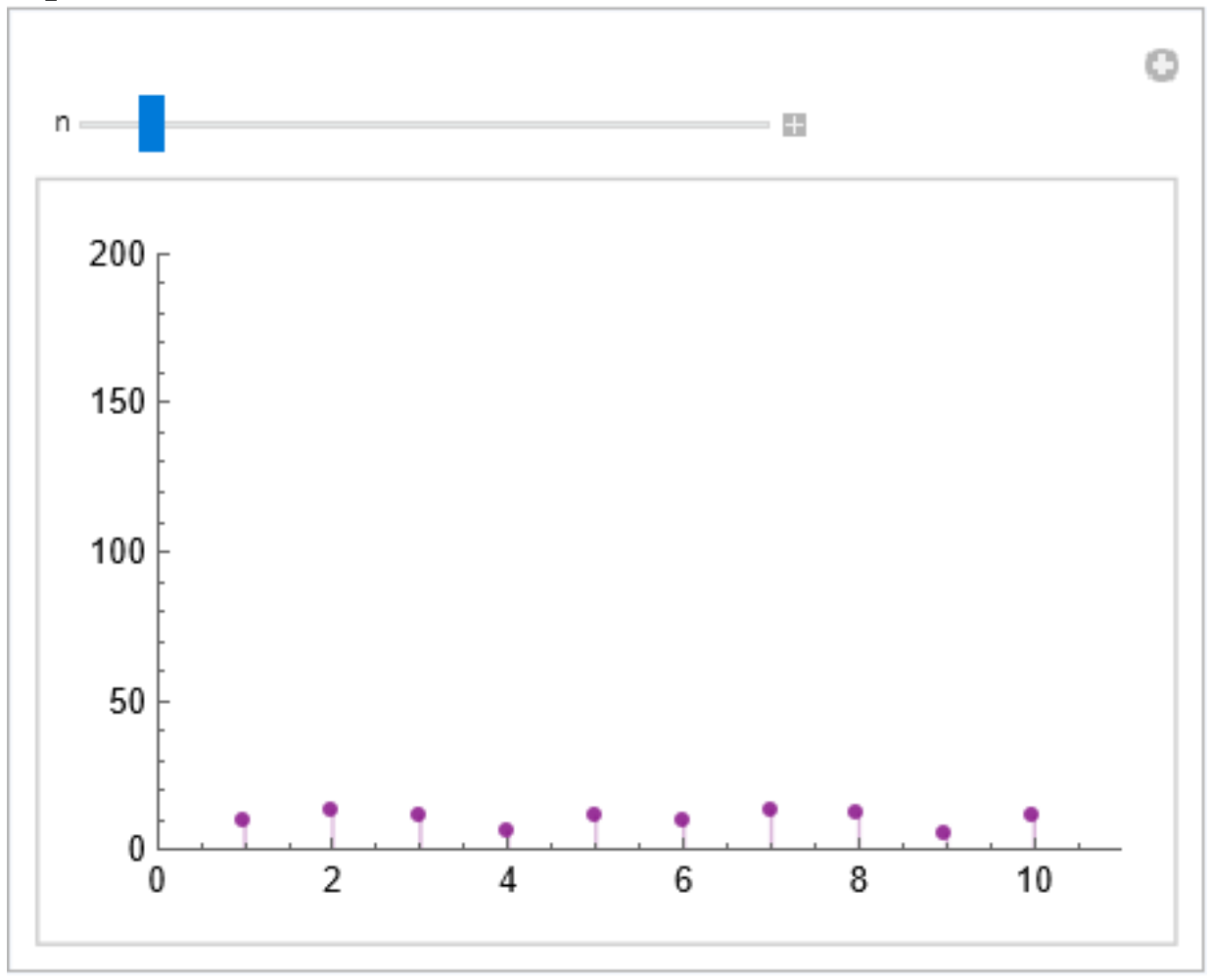

Input

```
(* This code creates a Manipulate slider for the number of random numbers (n) to generate from a normal distribution with mean (μ) and standard deviation (σ). The numbers are rounded to the nearest 0.1 and counted using Tally. The resulting counts are plotted using ListPlot, with the rounded values on the x-axis and their corresponding counts on the y-axis: *)

Manipulate[
 data=RandomVariate[NormalDistribution[μ,σ],n];
 count=Tally[Round[data,0.1]];
 ListPlot[
  count,
  Filling->Axis,
  PlotStyle->{Directive[Purple,Opacity[0.8],PointSize[Small]]},
  PlotRange->{{-15,15},{0,500}},
  ImageSize->350,
  Frame->{True,True,False,False},
  FrameTicks->{{Automatic,None},{None,None}},
  FrameLabel->{"Value","Count"}],
 {{n,5000},100,10000,100},
 {{μ,0},-5,5,0.1},
 {{σ,2},0.1,5,0.1}
 ]
```

Output

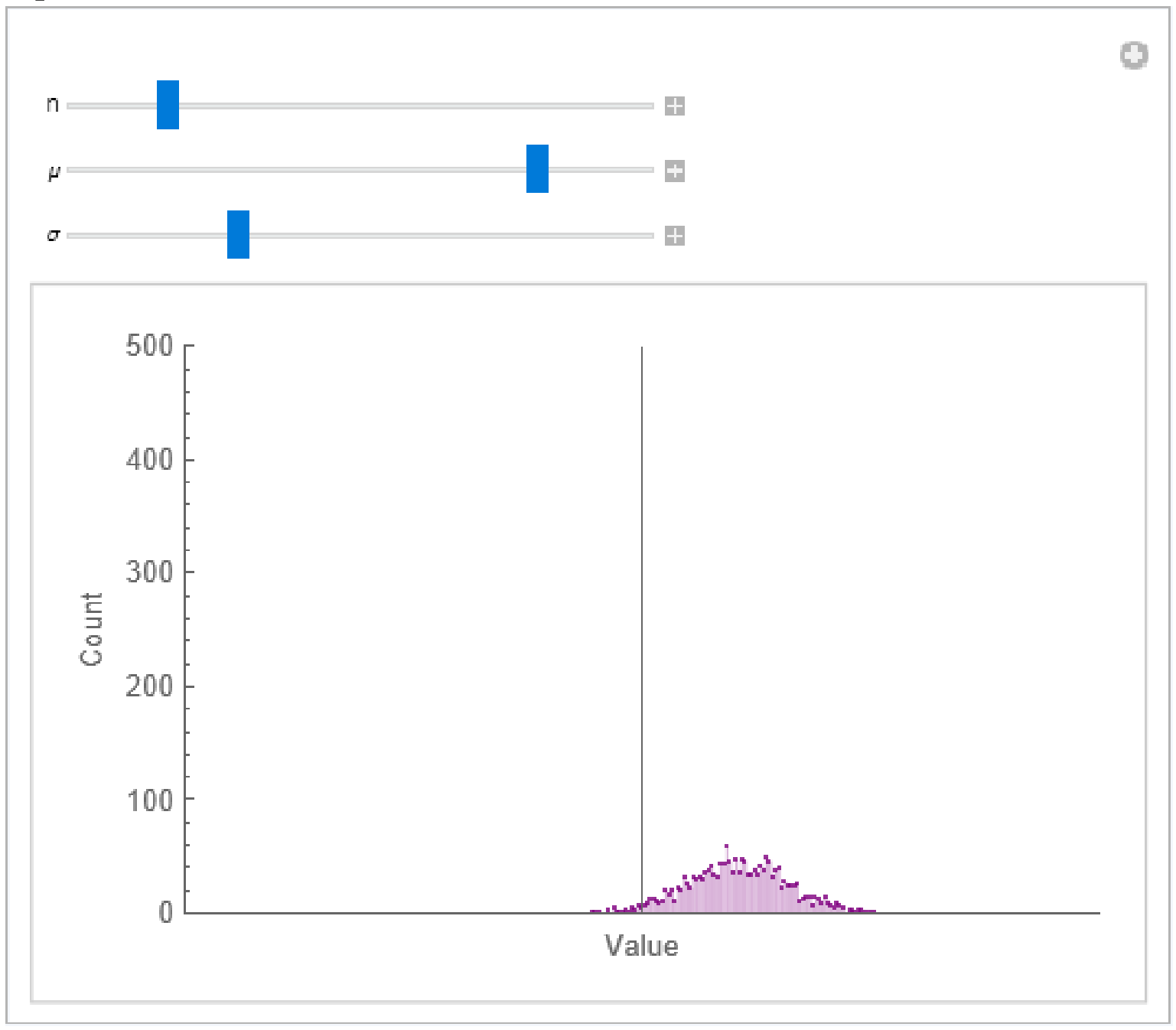

Input

```
(* The code is generating a list of integers data and then using the Tally function to count the number of occurrences of each unique element in the list. The resulting list tally will have pairs of elements, where the first element corresponds to a unique value in data and the second element corresponds to the count of how many times that value appears. This code is useful for generating a formatted frequency table for a small dataset, such as when manually inspecting the frequency distribution of a sample. It is similar to using the Counts function, but with the difference that Tally returns a list of pairs instead of an association object. Additionally, Tally can be useful when the frequency counts need to be sorted according to some criterion, such as when displaying the most common values first: *)

data={1,2,3,3,3,4,4,5,5};
```

```
            tally=Tally[data];
            TableForm[
             tally,
             TableHeadings->{None,{"Value","Frequency"}}
             ]
Output      {
             {Value, Frequency},
             {1, 1},
             {2, 1},
             {3, 3},
             {4, 2},
             {5, 2}
            }
```

***Mathematica code 3.10*** Count and Counts

```
Input       (* Count how many times b occurs: *)
            Count[{a,b,a,a,b,c,b},b]
Output      3

Input       (* Count works with patterns: *)
            Count[{a,2,a,a,1,c,b,3,3},_Integer]
Output      4

Input       (* Count the number of elements not matching b: *)
            Count[{a,b,a,a,b,c,b,a,a},Except[b]]
Output      6

Input       (* The Count function is used to count the number of elements in the list myList
            that are greater than 5. The /; pattern is used to select only those elements that
            satisfy the condition of being greater than 5: *)

            myList={3,7,2,5,1,9,4};
            Count[myList,x_/;x>5]
Output      2

Input       (* The Count function is used to count the number of pairs of adjacent elements in
            the list myList that satisfy the condition of having the first element less than
            the second element. The Partition function is used to create a list of all such
            pairs, and the {x_,y_}/;x<y pattern is used to select only those pairs that satisfy
            the given condition: *)

            myList={3,5,2,8,6,7,10,9};
            Partition[myList,2]
            Count[Partition[myList,2],{x_,y_}/;x<y]
Output      {{3,5},{2,8},{6,7},{10,9}}
Output      3

Input       (* The Count function is used to count the number of occurrences of the value 5 in
            the matrix myMatrix. The Flatten function is used to convert myMatrix to a list,
            so that Count can be applied to the individual elements: *)

            myMatrix={{1,2,3},{4,5,6},{7,8,9}};
            Count[Flatten[myMatrix],5]
Output      1

Input       (*  The built-in Counts function can be useful for quickly analyzing the frequency
            distribution of categorical data, such as counting the number of times each word
            appears in a text document, or the number of occurrences of each category in a
            dataset. It is also a simple and efficient way to generate a frequency table, as
```

```
        the function takes care of both grouping the data by unique elements and counting
        their frequencies in one step. In this code, we count the number of occurrences of
        each element in the list {a, b, c, a}: *)

        Counts[{a,b,c,a}]
Output  <|a->2,b->1,c->1|>

Input   (* This code is generating a list of 100 random integers between 0 and 1. Then,
        the Counts function is applied to this list to count the number of occurrences of
        each unique element. Since there are only two possible values in the list (0 and
        1), the resulting association object will have keys of 0 and 1, each with a
        corresponding count of how many times it appears in the list. This code is useful
        for generating a frequency table for binary data, such as the results of a coin
        toss or the outcomes of a binary classification problem. It can also be used to
        simulate random binary data and analyze its frequency distribution: *)

        Counts[RandomInteger[{0,1},100]]
Output  <|0->51,1->49|>

Input   (* This code is useful for generating a formatted frequency table for a small
        dataset, such as when manually inspecting the frequency distribution of a sample.
        It can also be used to quickly visualize the frequency distribution of discrete
        data, by displaying the frequency counts in a table format: *)

        data={1,2,3,3,3,4,4,5,5};
        counts=Counts[data]
        TableForm[
         Transpose[{Keys[counts],Values[counts]}],
         TableHeadings->{None,{"Value","Frequency"}}
         ]
Output  <|1->1,2->1,3->3,4->2,5->2|>
Output  {
         {Value, Frequency},
         {1, 1},
         {2, 1},
         {3, 3},
         {4, 2},
         {5, 2}
        }
```

***Mathematica code 3.11*** BinLists and BinCounts

```
Input   (* Make lists of elements in bins of width 1 from 0 to 12: *)
        BinLists[{1,3,2,1,4,5,6,2,10,2},{0,12,1}]

        (* Count the number of elements in bins of width 1 from 0 to 12: *)
        BinCounts[{1,3,2,1,4,5,6,2,10,2},{0,12,1}]
Output  {{},{1,1},{2,2,2},{3},{4},{5},{6},{},{},{},{10},{}}
Output  {0,2,3,1,1,1,1,0,0,0,1,0}

Input   (* List elements in a sequence of ranges: *)
        BinLists[{1,3,2,1,4,5,6,2,3,5,9,1},{{-Infinity,1,2,4,6,Infinity}}]

        (* Count the number of elements in a sequence of ranges: *)
        BinCounts[{1,3,2,1,4,5,6,2,3,5,9,1},{{-Infinity,1,2,4,6,Infinity}}]
Output  {{},{1,1,1},{3,2,2,3},{4,5,5},{6,9}}
Output  {0,3,4,3,2}

Input   (* List elements in bins of a specified width: *)
        BinLists[{1,3,2,1,4,5,6,2},3]
```

```
        (* Count the number of elements in bins of a specified width: *)
        BinCounts[{1,3,2,1,4,5,6,2},3]
Output  {{1,2,1,2},{3,4,5},{6}}
Output  {4,3,1}

Input   (* The length of BinLists is equivalent to the results from BinCounts: *)
        data=RandomReal[{-3,3},1000];
        listoflengths1=Map[Length,BinLists[data]]
        listoflengths2=BinCounts[data]

        Total[listoflengths1]
        Total[listoflengths2]
Output  {181,170,175,152,155,167}
Output  {181,170,175,152,155,167}
Output  1000
Output  1000

Input   (* we first generate a list of 20 random real numbers between -10 and 10 using
        RandomReal. Then, we define the bin boundaries as {-10,-5,0,5,10}. We use BinCounts
        to count the number of values that fall within each bin, and BinLists to create a
        list of sublists, where each sublist contains the values that fall within a
        particular bin: *)

        data=RandomReal[{-10,10},20];
        bins={-10,-5,0,5,10};
        binCounts=BinCounts[data,{bins}]
        binLists=BinLists[data,{bins}]
Output  {7,6,3,4}
Output  {{-7.05881,-9.47133,-5.01999,-6.65485,-5.38416,-9.14236,-6.89645},{-0.642216,-
        0.034109,-1.71578,-0.3198,-0.627258,-2.56717}, {2.80949,2.99958,3.2704},
        {5.72046,7.88653,9.00539,9.60528}}

Input   (* This code generates a list of random real numbers between 0 and 10 and then
        creating two different plots based on the data. The first plot shows the binned
        data using the BinLists function, which splits the data into bins defined by the
        bins list. The second plot shows the count data using the BinCounts function,
        which counts the number of data points in each bin defined by bins. Note that,
        RandomReal chooses reals with a uniform probability distribution: *)

        data=RandomReal[10,100];
        bins=Range[0,10,1];
        binnedData=BinLists[data,{bins}]
        countData=BinCounts[data,{bins}]
        ListPlot[
         binnedData,
         Joined->True,
         Mesh->All,
         ImageSize->170
         ]

        ListPlot[
         countData,
         Filling->Axis,
         PlotStyle->{Directive[Purple,Opacity[0.8],PointSize[Large]]},
         ImageSize->170
         ]
Output  {{0.974741,0.237327,0.281386,0.389847,0.66638,0.944951,0.942486,0.594528},{1.466
        93,1.5268,1.60659,1.98598,1.05362,1.40106},{2.39582,2.1156,2.32664,2.29535,2.158
        34,2.73596,2.44445,2.912,2.57553,2.95727,2.12335},{3.48122,3.53543,3.75044,3.093
        48,3.16114,3.97694,3.34258,3.38654,3.7172,3.2494},{4.10362,4.05316,4.45516,4.881
```

```
38,4.89215,4.78291,4.19745,4.23749,4.29946,4.81448},{5.20934,5.15297,5.2664,5.81
182,5.64929,5.37305,5.35408,5.50074,5.64166,5.333,5.58939},{6.76023,6.6007,6.224
42,6.66529,6.65225,6.01426,6.85846,6.63136,6.75016},{7.32637,7.69574,7.72214,7.7
6053,7.182,7.50633,7.4313,7.69111,7.55321,7.54387},{8.64312,8.36571,8.82821,8.38
553,8.16058,8.36009,8.42748,8.50908,8.8086,8.58884,8.73584},{9.2052,9.23458,9.82
959,9.63361,9.16219,9.30296,9.0229,9.11091,9.04083,9.30531,9.20265,9.38045,9.711
14,9.08949}}
```

Output {8,6,11,10,10,11,9,10,11,14}

Output

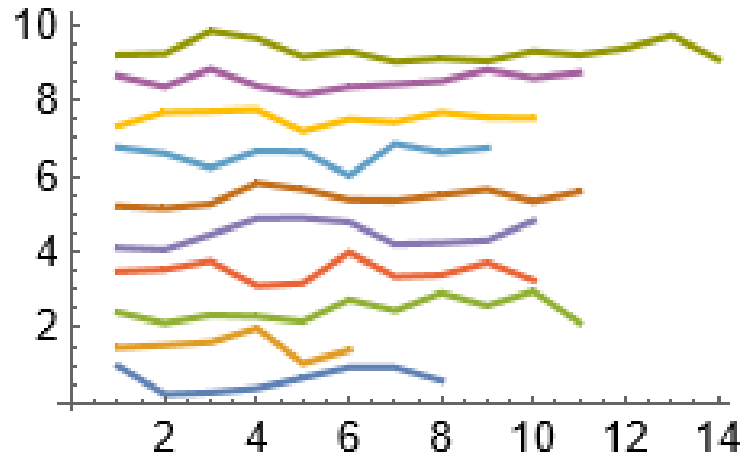

Output

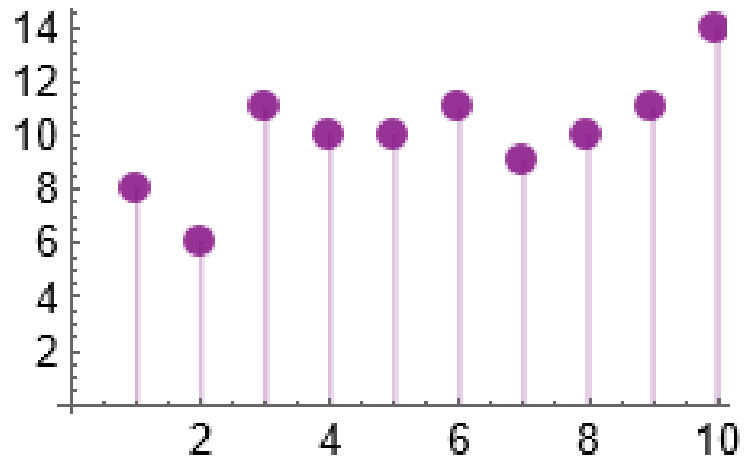

Input

```
(* This code generates a list of 1000 random numbers sampled from a normal distribution with mean 0 and standard deviation 1 using the RandomVariate function. Then it defines the bin specification as a list of cut-off points. The BinCounts function is then used to count the number of values that fall into each bin, based on the binSpec list, and stores the counts in the binCounts list. Finally, ListPlot is used to create a plot of the bin counts: *)

data=RandomVariate[NormalDistribution[0,1],1000];
binSpec={-3,-2,-1,0,1,2,3};
binCounts=BinCounts[data,{binSpec}]
ListPlot[
 binCounts,
 Filling->Axis,
 PlotStyle->{Directive[Purple,Opacity[0.8],PointSize[Large]]},
 ImageSize->170
 ]
```

Output {23,123,349,344,142,18}

Output

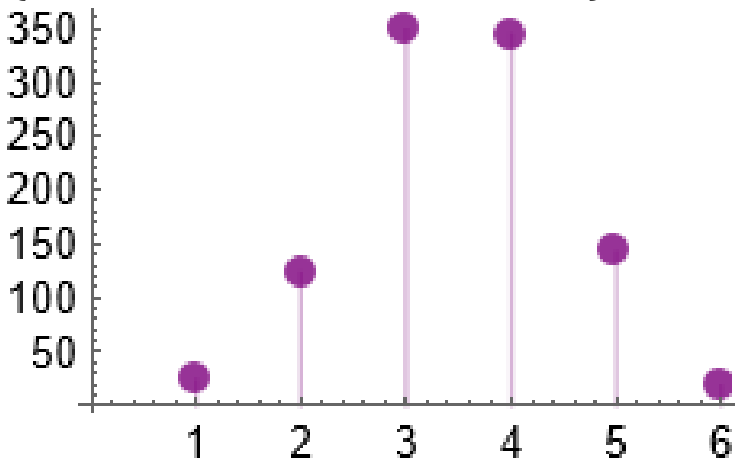

Input

```
(* This code creates a Manipulate slider for the number of random numbers (n) to generate from a normal distribution with mean (μ) and standard deviation (σ). The generated numbers are then binned using BinCounts with a minimum value (min), maximum value (max),and bin width (Δ) also specified by sliders. The resulting bin counts are plotted using ListPlot, with the bin index on the x-axis and the count on the y-axis: *)

Manipulate[
 data=RandomVariate[NormalDistribution[μ,σ],n];
 counts=BinCounts[data,{min,max,Δ}];
```

```
ListPlot[
 counts,
 Filling->Axis,
 PlotStyle->{Directive[Purple,Opacity[0.8],PointSize[Small]]},
 PlotRange->All,
 ImageSize->300,
 Frame->{True,True,False,False},
 FrameTicks->{{Automatic,None},{None,None}},
 FrameLabel->{"Bin","Count"}],
{{n,1000},100,10000,100},
{{μ,0},-5,5,0.1},
{{σ,1},0.1,5,0.1},
{{min,-5},-5,5,0.1},
{{max,5},-5,5,0.1},
{{Δ,0.1},0.1,1,0.01}
]
```

Output

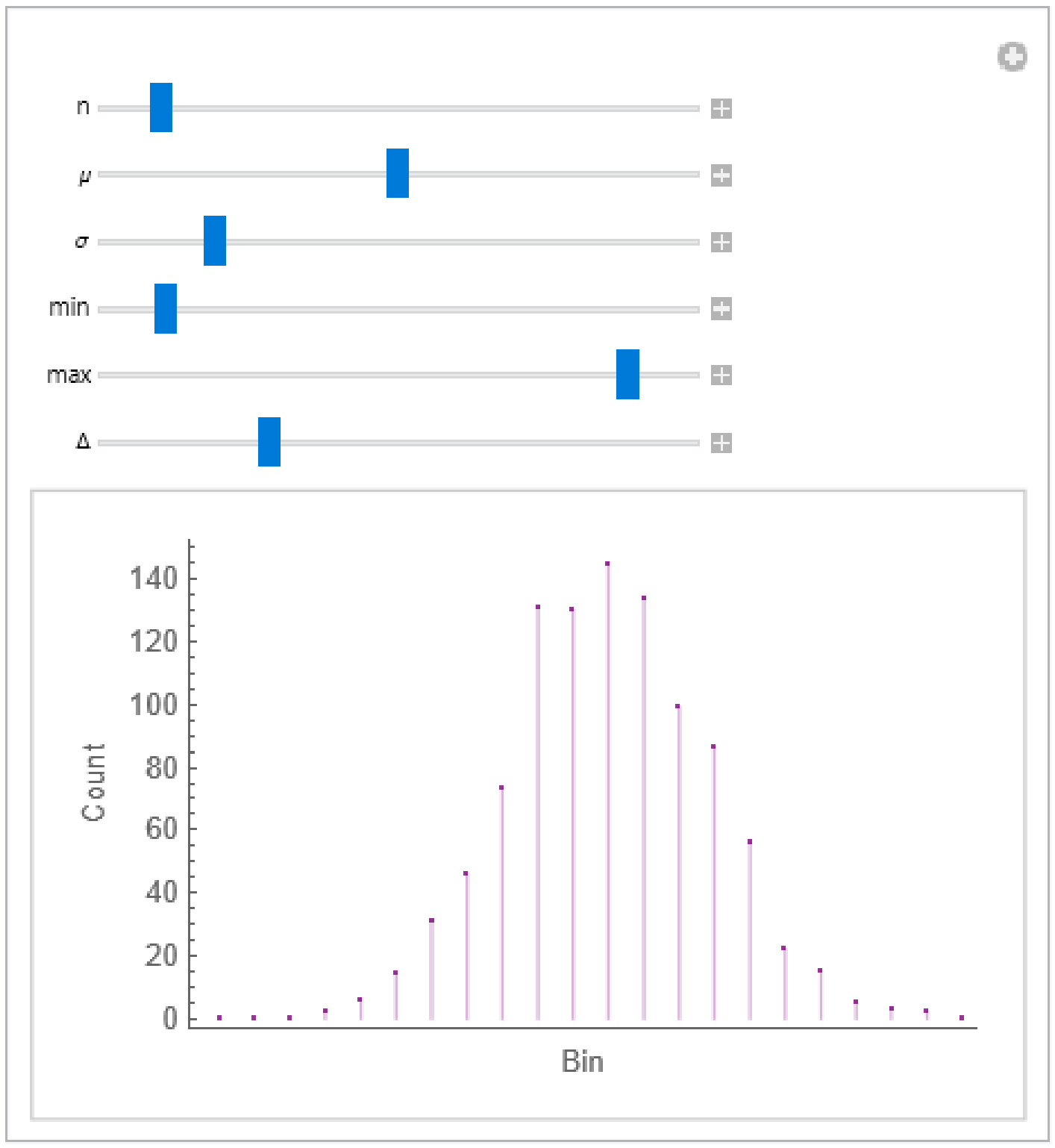

Input

```
(* This code generates a list of data, creates a frequency table with bins of width 1 ranging from 1 to 6 using BinCounts, calculates the relative frequencies of each bin, and displays the table with appropriate labels using TableForm and TableHeadings: *)

data={1,2,3,3,3,4,4,5};
binCounts=BinCounts[data,{1,6,1}]
relativeFrequencies=Round[binCounts/Total[binCounts],0.001]

TableForm[
 Transpose[{{"1-2","2-3","3-4","4-5","5-6"},binCounts,relativeFrequencies}],
 TableHeadings->{None,{"Value","Frequency","Relative Frequency"}}
 ]
```

Output {1,1,3,2,1}

Output {0.125,0.125,0.375,0.25,0.125}

Output

```
{
 {Value, Frequency, Relative Frequency},
```

```
 {1-2, 1, 0.125},
 {2-3, 1, 0.125},
 {3-4, 3, 0.375},
 {4-5, 2, 0.25},
 {5-6, 1, 0.125}
}
```

Input

```
(* Generating a frequency table, frequency polygon, and ogive for a list of data
with specified bin widths using BinCounts, TableForm, and ListLinePlot: *)

data={3,5,6,8,9,10,12,13,14,15,17,20};
binWidth=3;
binCounts=BinCounts[data,{Min[data],Max[data]+binWidth,binWidth}];
frequencies=binCounts/Total[binCounts];
ranges=Partition[Range[Min[data],Max[data]+binWidth,binWidth],2,1];
intevals=Table[{Row[{ToString[ranges[[i]][[1]]],"-",
ToString[ranges[[i]][[2]]]}]},{i,1,Length[ranges]}];

TableForm[
Transpose[{intevals,binCounts,Accumulate[binCounts],frequencies,Accumulate[frequ
encies]}],
 TableHeadings->{None,{"Bin Range","Frequency","Cumulative Frequency","Relative
Frequency","Accumulate Relative frequencies"}}
 ]
ListLinePlot[
 Transpose[{Mean/@ranges,Accumulate[binCounts]}],
 Frame->True,
 FrameLabel->{"Data","Cumulative Frequency"},
 PlotRange->{{Min[data],Max[data]},{0,Length[data]}},
 PlotStyle->Directive[Purple,Thickness[0.005]],
 ImageSize->250
 ]
ListLinePlot[
 Transpose[{Mean/@ranges,Accumulate[frequencies]}],
 Frame->True,
 FrameLabel->{"Data","Cumulative Relative Frequency"},
 PlotRange->{{Min[data],Max[data]},{0,1}},
 PlotStyle->Directive[Purple,Thickness[0.005]],
 ImageSize->250
 ]
```

Output

```
{
 {Bin Range, Frequency, Cumulative Frequency, Relative Frequency, Accumulate
Relative frequencies},
 {{{3 - 6}}, 2, 2, 1/6, 1/6},
 {{{6 - 9}}, 2, 4, 1/6, 1/3},
 {{{9 - 12}}, 2, 6, 1/6, 1/2},
 {{{12 - 15}}, 3, 9, 1/4, 3/4},
 {{{15 - 18}}, 2, 11, 1/6, 11/12},
 {{{18 - 21}}, 1, 12, 1/12, 1}
}
```

Output

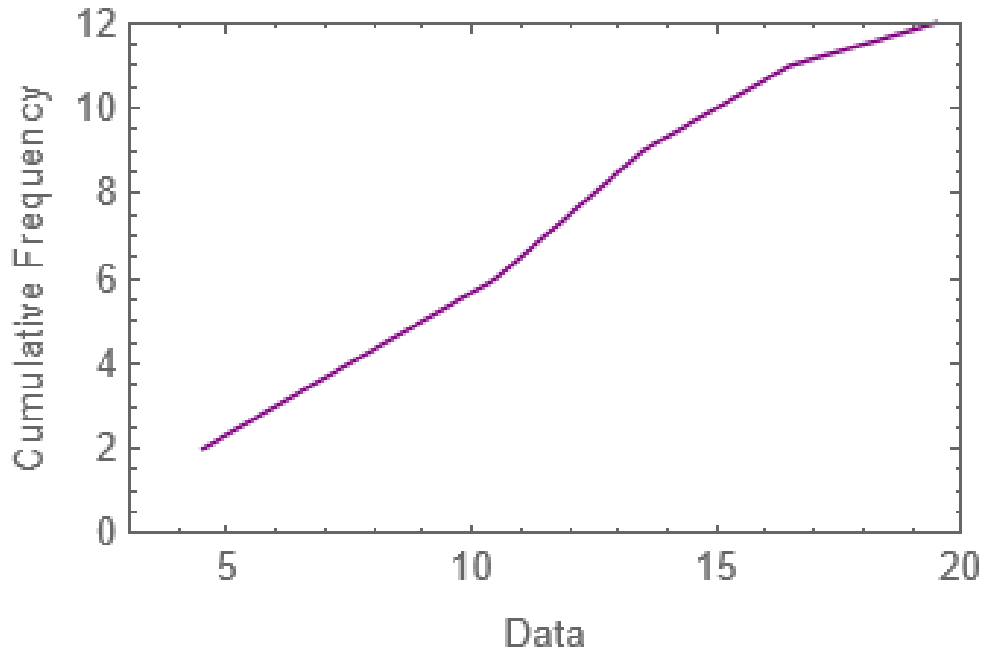

Output

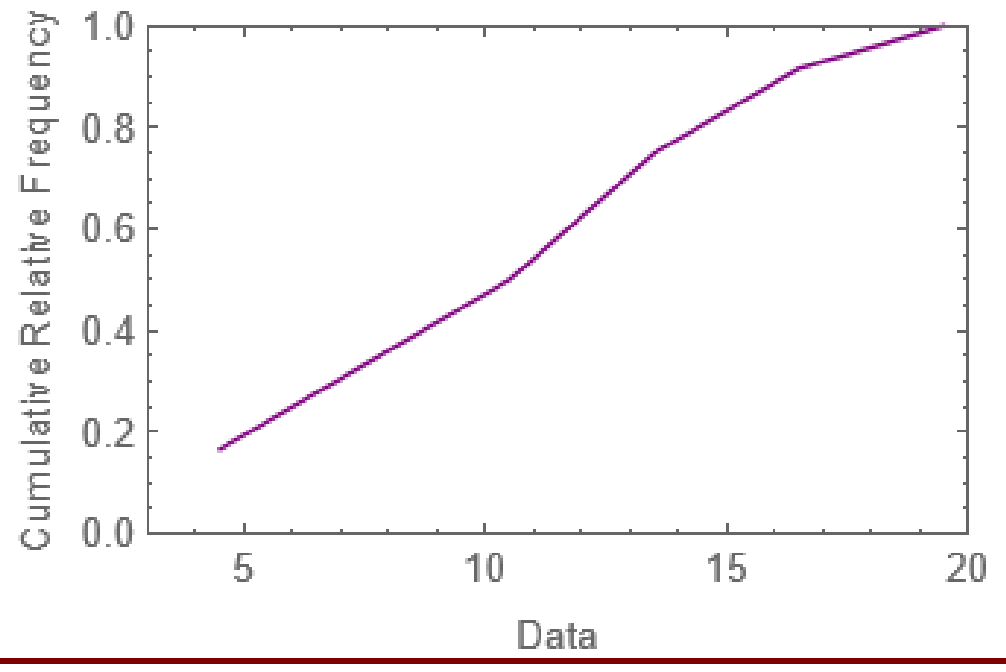

| | |
|---|---|
| `HistogramList[{{x1,y1,…},{x2,y2,…},…}]` | gives a list of bins and histogram heights of the values {xi,yi,…}. |
| `HistogramList[…,bspec]` | gives a list of bins and histogram heights with bins specified by bspec. |
| `HistogramList[…,bspec,hspec]` | gives a list of bins and histogram heights with bin heights computed according to the specification hspec. |

**Remarks:**

- `HistogramList` produces a list of bin delimiters `{b1,b2,…}` for each dimension and a depth- array of values for each bin.
- For 2D data, the output has the form `{{{bx1,bx2,…},{by1,…}},{{v1,1,v1,2,…},{v2,1,…},…}}` where `vi,j` is the value corresponding to the bin `[bxi,bxi+1)×[byj,byj+1)`.
- The following bin specifications `bpsec` can be given:

| | |
|---|---|
| `n` | use n bins |
| `{w}` | use bins of width w |
| `{min,max,w}` | use bins of width w from min to max |
| `{{b1,b2,…}}` | use bins [b1,b2),[b2,b3),… |
| `Automatic` | determine bin widths automatically |
| `"name"` | use a named binning method |
| `{"Log",bspec}` | apply binning bspec on log-transformed data |
| `fw` | apply fw to get an explicit bin specification {b1,b2,…} |
| `{xspec,yspec,…}` | give different x, y, etc. specifications |

- Possible named binning methods include:

| | |
|---|---|
| `"Sturges"` | compute the number of bins based on the length of data |
| `"Scott"` | asymptotically minimize the mean square error |
| `"FreedmanDiaconis"` | twice the interquartile range divided by the cube root of sample size |
| `"Knuth"` | balance likelihood and prior probability of a piecewise uniform model |
| `"Wand"` | one-level recursive approximate Wand binning |

- Different forms of histogram data can be obtained by giving different bin height specifications `hspec` in `HistogramList[data,bspec,hspec]`. The following forms can be used:

| | |
|---|---|
| `"Count"` | the number of values lying in each bin |
| `"CumulativeCount"` | cumulative counts |
| `"SurvivalCount"` | survival counts |
| `"Probability"` | fraction of values lying in each bin |
| `"PDF"` | probability density function |
| `"CDF"` | cumulative distribution function |
| `"SF"` | survival function |
| `"HF"` | hazard function |
| `"CHF"` | cumulative hazard function |
| `fh` | heights obtained by applying fh to bins and counts |

***Mathematica code 3.12*** HistogramList

```
Input       (* Generate a list of bin delimiters and counts for a dataset: *)
            datasample=RandomVariate[NormalDistribution[],50];

            {bins,counts}=HistogramList[datasample,{1}(* width bins: *)];

            bins
            partitionbins=Partition[bins,2,1]
            counts

            TableForm[
             Table[
              {Row[partitionbins[[i]],","],counts[[i]]},
              {i,1,Length[counts]}
              ],
             TableAlignments->Center,
             TableHeadings->{None,{"Bin Interval","Counts"}}
             ]
Output      {-3,-2,-1,0,1,2,3}
Output      {{-3,-2},{-2,-1},{-1,0},{0,1},{1,2},{2,3}}
Output      {1,4,24,14,4,3}
Output      {
             {Bin Interval, Counts},
             {-3,-2, 1},
             {-2,-1, 4},
             {-1,0, 24},
             {0,1, 14},
             {1,2, 4},
             {2,3, 3}
            }

Input       (* Use different height functions: *)
            sampledata=RandomVariate[NormalDistribution[],100];

            {bins1,counts1}=HistogramList[sampledata,{1},"Count"]
            {bins2,counts2}=HistogramList[sampledata,{1},"PDF"]
            {bins3,counts3}=HistogramList[sampledata,{1},"CDF"]

            partitionbins=Partition[bins1,2,1]

            TableForm[
             Table[
              {Row[partitionbins[[i]],","],counts1[[i]],counts2[[i]],counts3[[i]]},
              {i,1,Length[counts1]}
              ],
             TableAlignments->Center,
             TableHeadings->{None,{"Bin Interval","Count","PDF","CDF"}}
             ]

Output      {{-3,-2,-1,0,1,2},{3,11,37,33,16}}
Output      {{-3,-2,-1,0,1,2},{3/100,11/100,37/100,33/100,4/25}}
Output      {{-3,-2,-1,0,1,2},{0.03,0.14,0.51,0.84,1.}}
Output      {{-3,-2},{-2,-1},{-1,0},{0,1},{1,2}}
Output      {
             {Bin Interval, Count, PDF, CDF},
             {-3,-2, 3, 3/100, 0.03},
             {-2,-1, 11, 11/100, 0.14},
             {-1,0, 37, 37/100, 0.51},
             {0,1, 33, 33/100, 0.84},
             {1,2, 16, 4/25, 1.}
```

```
              }

Input         (* Specify the number of bins to use: *)
              data=RandomVariate[NormalDistribution[0,1],50];
              HistogramList[data,5]

              (* Specify the bin width: *)
              HistogramList[data,{.5}]

              (* Specify limits and bin size: *)
              HistogramList[data,{-4,4,2}]

              (* List specific bin delimiters: *)
              HistogramList[data,{{-3,-1,0,1,3}}]
Output        {{-2,-1,0,1,2,3},{4,20,16,9,1}}
Output        {{-2.,-1.5,-1.,-0.5,0.,0.5,1.,1.5,2.,2.5},{1,3,8,12,8,8,9,0,1}}
Output        {{-4,-2,0,2,4},{0,24,25,1}}
Output        {{-3,-1,0,1,3},{4,20,16,10}}

Input         (* This code generates a table that shows how different named binning methods
              affect the output of the HistogramList function when applied to a random data set
              generated from a normal distribution: *)

              data=RandomVariate[NormalDistribution[0,1],50];

              Grid[
               Table[
                {w,First[HistogramList[data,w]]},
                {w,{"Sturges","Scott","FreedmanDiaconis","Wand","Knuth"}}
                ],
               Frame->All,
               Alignment->{{Right,Left}}
               ]
Output        {
               {Sturges, {-(5/2),-2,-(3/2),-1,-(1/2),0,1/2,1,3/2,2,5/2}},
               {Scott, {-3,-2,-1,0,1,2,3}},
               {FreedmanDiaconis, {-(5/2),-2,-(3/2),-1,-(1/2),0,1/2,1,3/2,2,5/2}},
               {Wand, {-3,-2,-1,0,1,2,3}},
               {Knuth, {-(5/2),-2,-(3/2),-1,-(1/2),0,1/2,1,3/2,2,5/2}}
              }

Input         (* This code generates a table that shows how different height specifications
              affect the output of the HistogramList function when applied to a random data set
              generated from a normal distribution: *)

              data=RandomVariate[NormalDistribution[0,1],50];

              Grid[
               Table[
                {h,Last[HistogramList[data,{1},h]]},

              {h,{"Count","CumulativeCount","Probability","CumulativeProbability","PDF","CDF"}
              }
                ],
               Frame->All,
               Alignment->{{Right,Left}}
               ]
Output        {
               {Count, {10,17,16,6,1}},
               {CumulativeCount, {10.,27.,43.,49.,50.}},
```

```
 {Probability, {0.2,0.34,0.32,0.12,0.02}},
 {CumulativeProbability, {0.2,0.54,0.86,0.98,1.}},
 {PDF, {1/5,17/50,8/25,3/25,1/50}},
 {CDF, {0.2,0.54,0.86,0.98,1.}}
}
```

Input

```
(* This code generates a frequency table using the default height function, which is "Count". The HistogramList function returns the bin boundaries in bins and the counts of data points in each bin in counts. The Transpose function is used to create a table with bin boundaries and counts as columns: *)

data={1,2,3,3,4,4,4,5,5,6};
{bins,counts}=HistogramList[data];
freqTable=TableForm[
  Transpose[{Most[bins],Rest[bins],counts}],
  TableHeadings->{None,{"Left limit of Bin","Right limit of Bin","Count"}}
  ]

Histogram[
 data,
 ColorFunction->Function[Opacity[0.5]],
 ChartStyle->Purple,
 ImageSize->170
 ]
```

Output

```
{
 {Left limit of Bin, Right limit of Bin, Count},
 {0.5, 1.5, 1},
 {1.5, 2.5, 1},
 {2.5, 3.5, 2},
 {3.5, 4.5, 3},
 {4.5, 5.5, 2},
 {5.5, 6.5, 1}
}
```

Output

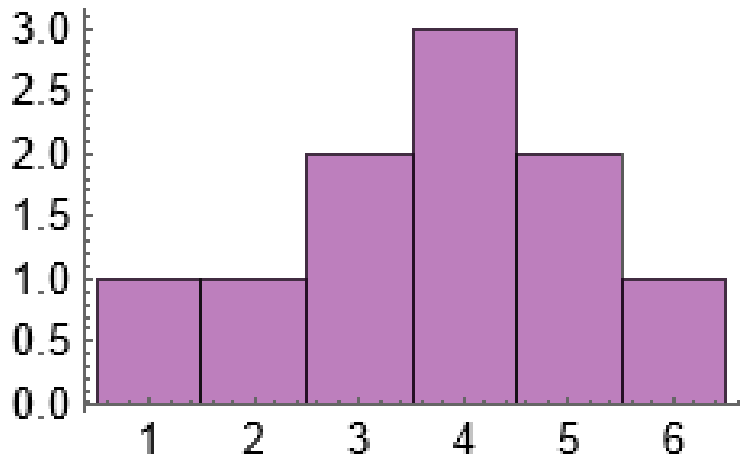

Input

```
(* This code generates a frequency table using the default height function, which is "Count". The HistogramList function returns the bin boundaries in bins and the counts of data points in each bin in counts. Most[bins] gives the lower bin boundaries, Rest[bins] gives the upper bin boundaries, and counts gives the counts of data points in each bin. The Transpose function is used to create a table with three columns: lower bin boundaries, upper bin boundaries, and counts: *)

data={1,2,3,3,4,4,4,5,5,6};
{bins,counts}=HistogramList[data,{0,7,1},"Count"];
freqTable=TableForm[
  Transpose[{Most[bins],Rest[bins],counts}],
  TableHeadings->{None,{"Left limit of Bin","Right limit of Bin","Count"}}
  ]

Histogram[
 data,
 {0,7,1},
```

```
             "Count",
             ColorFunction->Function[Opacity[0.5]],
             ChartStyle->Purple,
             ImageSize->170
             ]
Output      {
             {Left limit of Bin, Right limit of Bin, Count},
             {0, 1, 0},
             {1, 2, 1},
             {2, 3, 1},
             {3, 4, 2},
             {4, 5, 3},
             {5, 6, 2},
             {6, 7, 1}
            }
Output
```

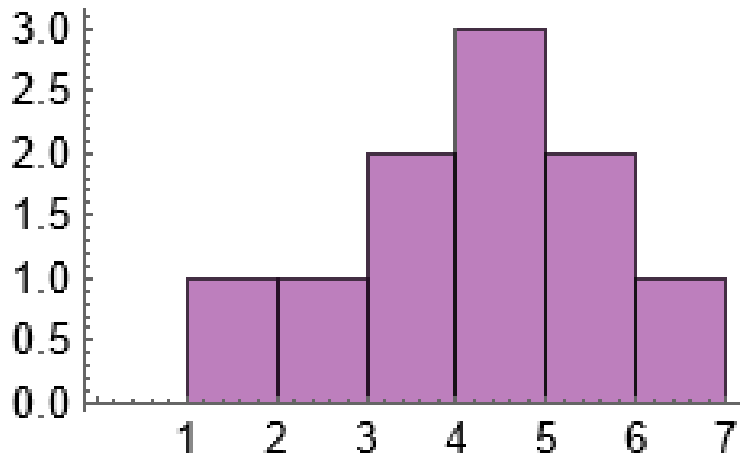

```
Input       (* This code generates a frequency table using "Probability" as the height
            function. The HistogramList function returns the bin boundaries in bins and the
            normalized frequencies in freq. Most[bins] gives the lower bin boundaries,
            Rest[bins] gives the upper bin boundaries, and freq gives the normalized
            frequencies in each bin. The Transpose function is used to create a table with
            three columns: lower bin boundaries, upper bin boundaries, and normalized
            frequencies: *)

            data={1,2,3,3,4,4,4,5,5,6};
            {bins,freq}=HistogramList[data,3,"Probability"];
            freqTable=TableForm[
              Transpose[{Most[bins],Rest[bins],freq}],
              TableHeadings->{None,{"Left limit of Bin","Right limit of Bin","Probability"}}
              ]

            Histogram[
             data,
             3,
             "Probability",
             ColorFunction->Function[Opacity[0.5]],
             ChartStyle->Purple,
             ImageSize->170
             ]
Output      {
             {Left limit of Bin, Right limit of Bin, Probability},
             {0, 2, 0.1},
             {2, 4, 0.3},
             {4, 6, 0.5},
             {6, 8, 0.1}
            }
```

Output

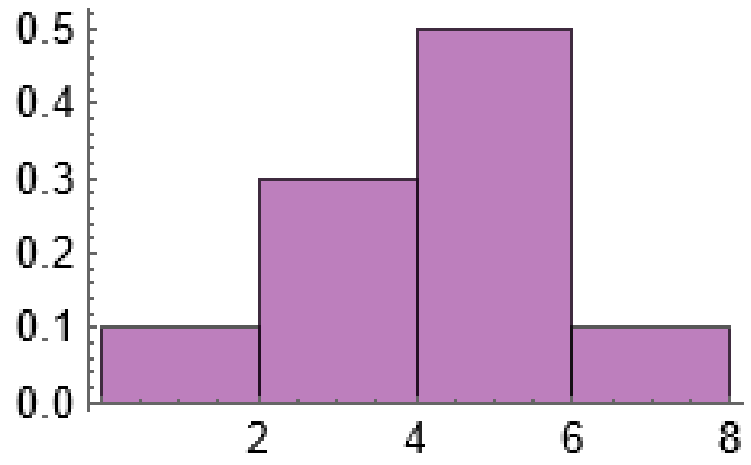

# UNIT 3.4

# DISTRIBUTION SHAPES

Some of the most commonly used `Mathematica` functions include `Histogram`, and `Histogram3D`. By using these functions, you can gain a deeper understanding of the underlying patterns in your data and make informed decisions based on your findings.

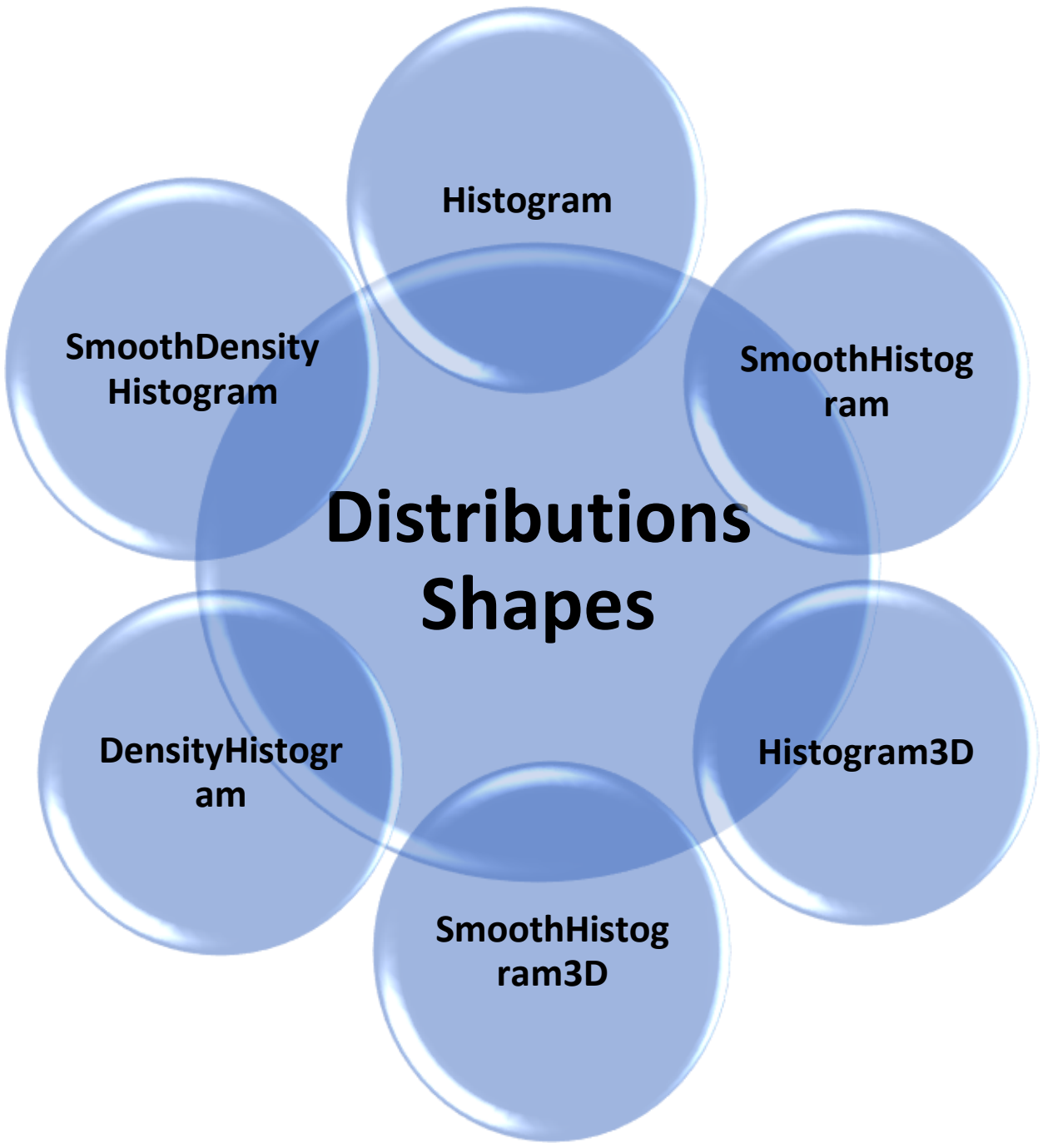

| | |
|---|---|
| `Histogram[{x1,x2,…}]` | plots a histogram of the values xi. |
| `Histogram[{x1,x2,…},bspec]` | plots a histogram with bin width specification bspec. |
| `Histogram[{x1,x2,…},bspec,hspec]` | plots a histogram with bin heights computed according to the specification hspec. |
| `Histogram[{data1,data2,…},…]` | plots histograms for multiple datasets datai. |

**Remarks:**

- The following bin specifications `bpsec` can be given:

| | |
|---|---|
| `n` | use n bins |
| `{dx}` | use bins of width dx |
| `{xmin,xmax,dx}` | use bins of width dx from xmin to xmax |
| `{{b1,b2,…}}` | use bins [b1,b2),[b2,b3),… |
| `Automatic` | determine bin widths automatically |
| `"name"` | use a named binning method |
| `{"Log",bspec}` | apply binning bspec on log-transformed data |
| `fb` | apply fb to get an explicit bin specification {b1,b2,…} |

- Possible named binning methods include:

| | |
|---|---|
| "Sturges" | compute the number of bins based on the length of data |
| "Scott" | asymptotically minimize the mean square error |
| "FreedmanDiaconis" | twice the interquartile range divided by the cube root of sample size |
| "Knuth" | balance likelihood and prior probability of a piecewise uniform model |
| "Wand" | one-level recursive approximate Wand binning |

- Different forms of histogram can be obtained by giving different bin height specifications `hspec` in `Histogram[data,bspec,hspec]`. The following forms can be used:

| | |
|---|---|
| "Count" | the number of values lying in each bin |
| "CumulativeCount" | cumulative counts |
| "SurvivalCount" | survival counts |
| "Probability" | fraction of values lying in each bin |
| "Intensity" | count divided by bin width |
| "PDF" | probability density function |
| "CDF" | cumulative distribution function |
| "SF" | survival function |
| "HF" | hazard function |
| "CHF" | cumulative hazard function |
| {"Log",hspec} | log-transformed height specification |
| fh | heights obtained by applying fh to bins and counts |

***Mathematica code 3.13*** Histogram

Input

```
(* This code generates histogram of 200 random samples drawn from a normal distribution with mean 0 and standard deviation 1: *)

Histogram[
 RandomVariate[
  NormalDistribution[0,1],
  200
  ],
 ColorFunction->Function[{height},Opacity[height]],
 ChartStyle->Purple,
 ImageSize->170
 ]
```

Output

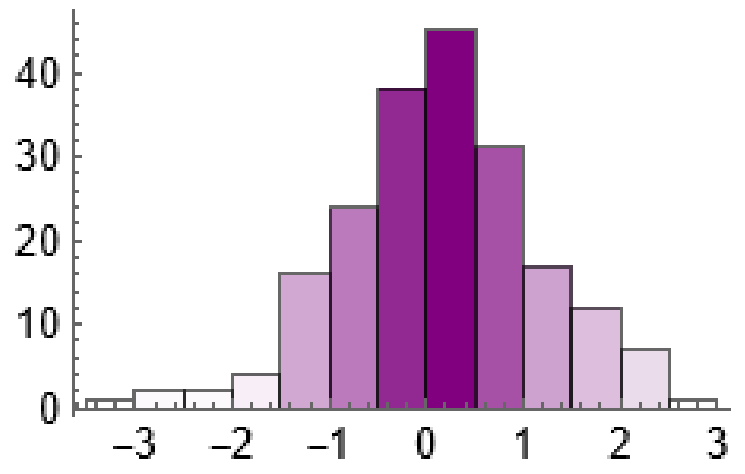

Input

```
(* Specify the number of bins to use: *)

sampledata=RandomVariate[
    NormalDistribution[0,1],
    300
    ];

Histogram[
```

```
 sampledata,
 6,
 ColorFunction->Function[{height},Opacity[height]],
 ChartStyle->Purple,
 ImageSize->170
 ]
```

Output

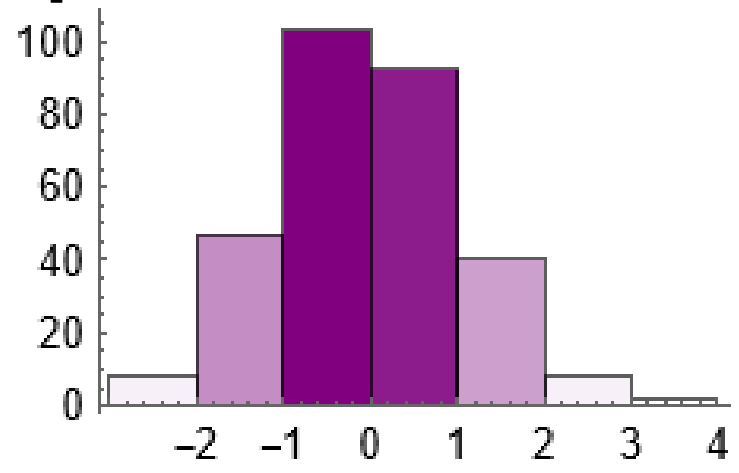

Input

```
(* Specify the bin width: *)

Histogram[
 sampledata,
 {.4},
 ColorFunction->Function[{height},Opacity[height]],
 ChartStyle->Purple,
 ImageSize->170
 ]
```

Output

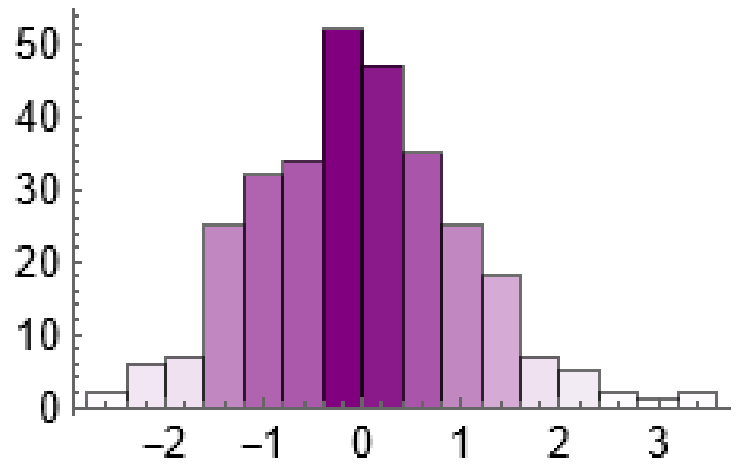

Input

```
(* This code generates four histograms of 1000 random samples drawn from a normal
distribution with mean 0 and standard deviation 1, using different binning methods.
The different methods used are: "Sturges", "Scott", "FreedmanDiaconis", and "Wand".
This allows for a comparison of different ways of visualizing the same data, which
can help in understanding the data and selecting the most appropriate method for a
particular analysis: *)

sampledata=RandomVariate[
   NormalDistribution[0,1],
   1000
   ];

Table[
 Histogram[
  sampledata,
  binmethod,
  PlotLabel->binmethod,
  ColorFunction->Function[{height},Opacity[height]],
  ChartStyle->Purple,
  ImageSize->170
  ],
 {binmethod,{"Sturges","Scott","FreedmanDiaconis","Wand"}}
 ]
```

Output

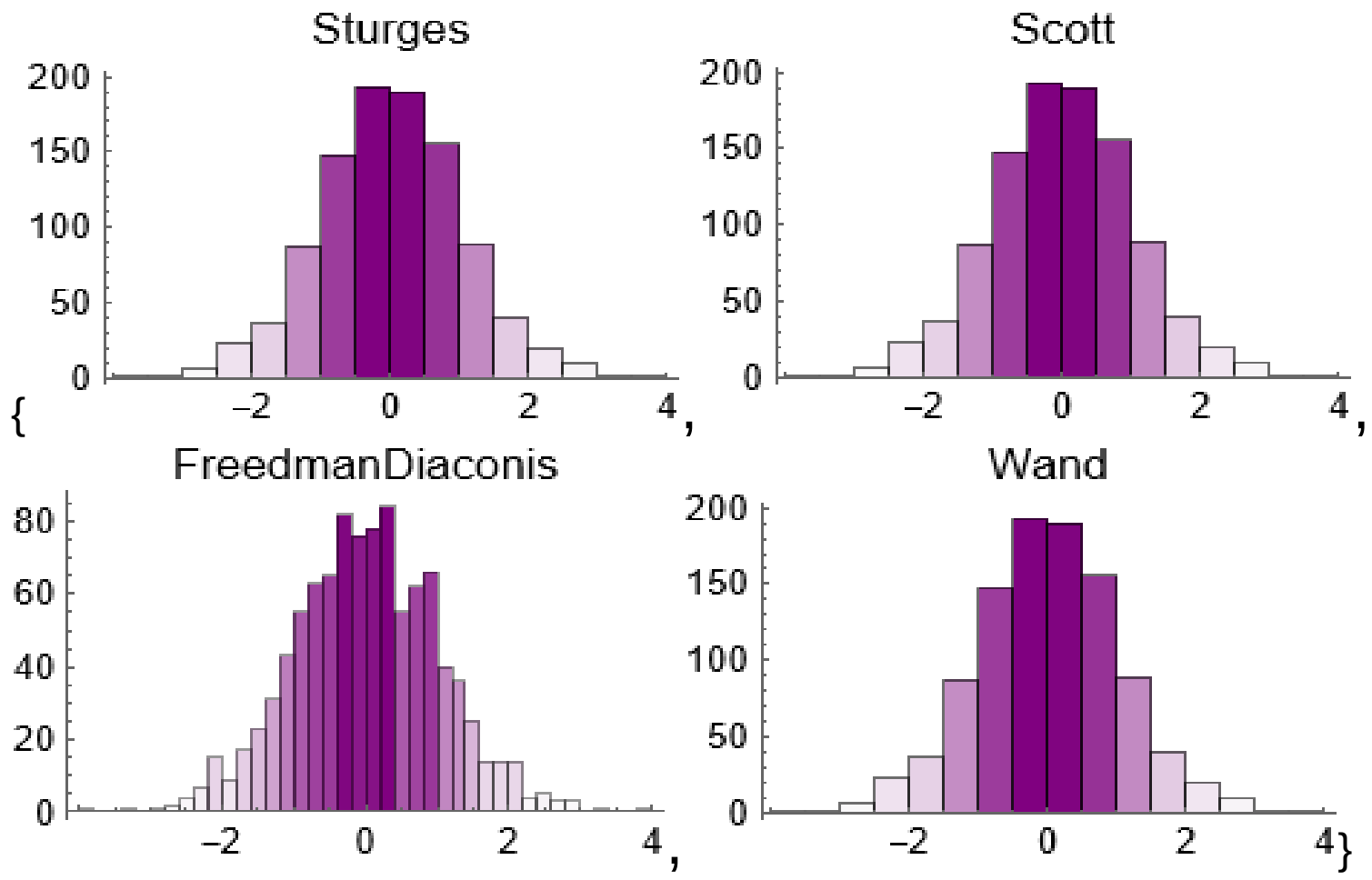

Input

```
(* This code generates six histograms of 500 random samples drawn from a normal distribution with mean 0 and standard deviation 1, using different methods to set the vertical scale of the histograms. The different methods used are: "Count", "Probability", "PDF", "CumulativeCount", "CDF", and "SF". This allows for a comparison of different ways of visualizing the same data, which can help in understanding the data and selecting the most appropriate method for a particular analysis: *)

sampledata=RandomVariate[
   NormalDistribution[0,1],
   500
   ];

Table[
 Histogram[
  sampledata,
  Automatic,
  heightmethod,
  PlotLabel->heightmethod,
  ColorFunction->Function[{height},Opacity[height]],
  ChartStyle->Purple,
  ImageSize->170
  ],
 {heightmethod,{"Count","Probability","PDF","CumulativeCount","CDF","SF"}}
 ]
```

Output

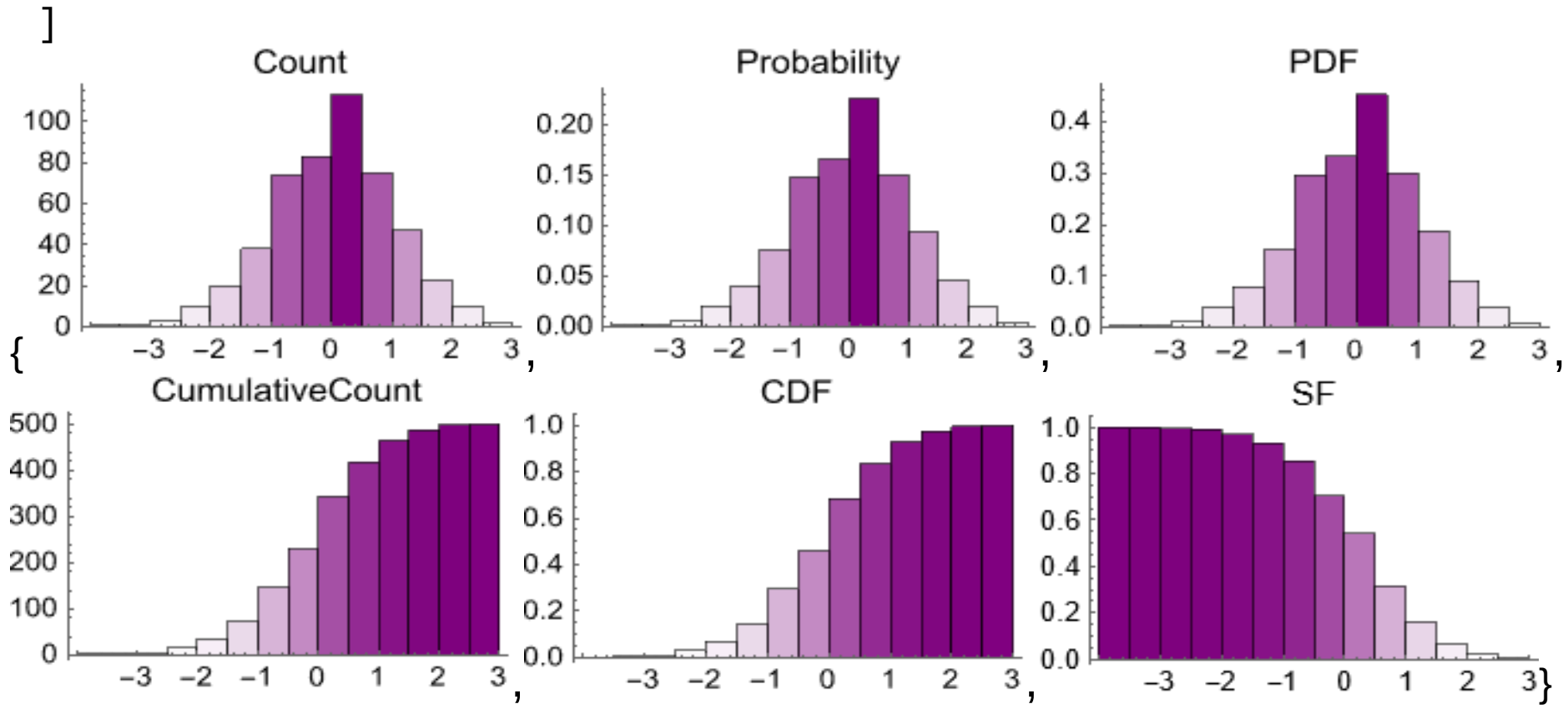

Input

```
(* Multiple datasets: *)

sampledata1=RandomVariate[
```

```
    NormalDistribution[0,1],
    1000
    ];
sampledata2=RandomVariate[
    NormalDistribution[4,3/4],
    1000
    ];

Histogram[
 {sampledata1,sampledata2},
 ColorFunction->Function[{height},Opacity[height]],
 ImageSize->170
 ]
```

Output

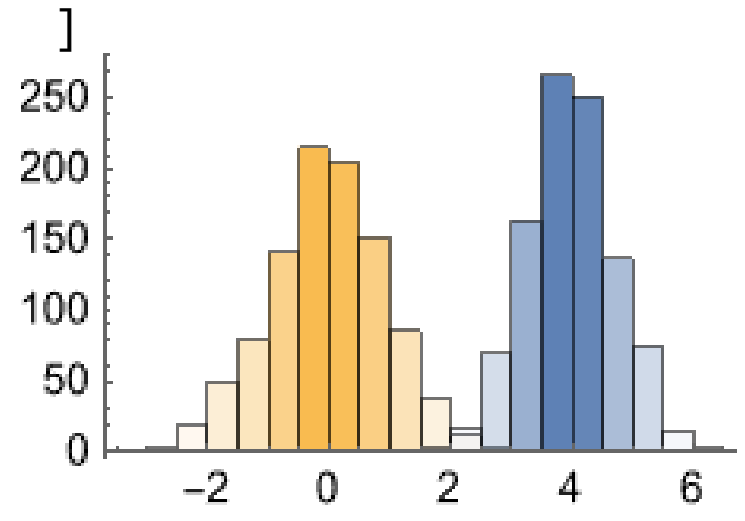

Input

```
(* Use different layouts to display multiple datasets: *)

sampledata1=RandomVariate[
    NormalDistribution[0,1],
    1000
    ];

sampledata2=RandomVariate[
    NormalDistribution[0,1],
    1000
    ];

Table[
 Histogram[
  {sampledata1,sampledata2},
  PlotLabel->layoutsmethod,
  ChartLayout->layoutsmethod,
  ColorFunction->Function[{height},Opacity[height]],
  ImageSize->170
  ],
 {layoutsmethod,{"Overlapped","Stacked"}}
 ]
```

Output

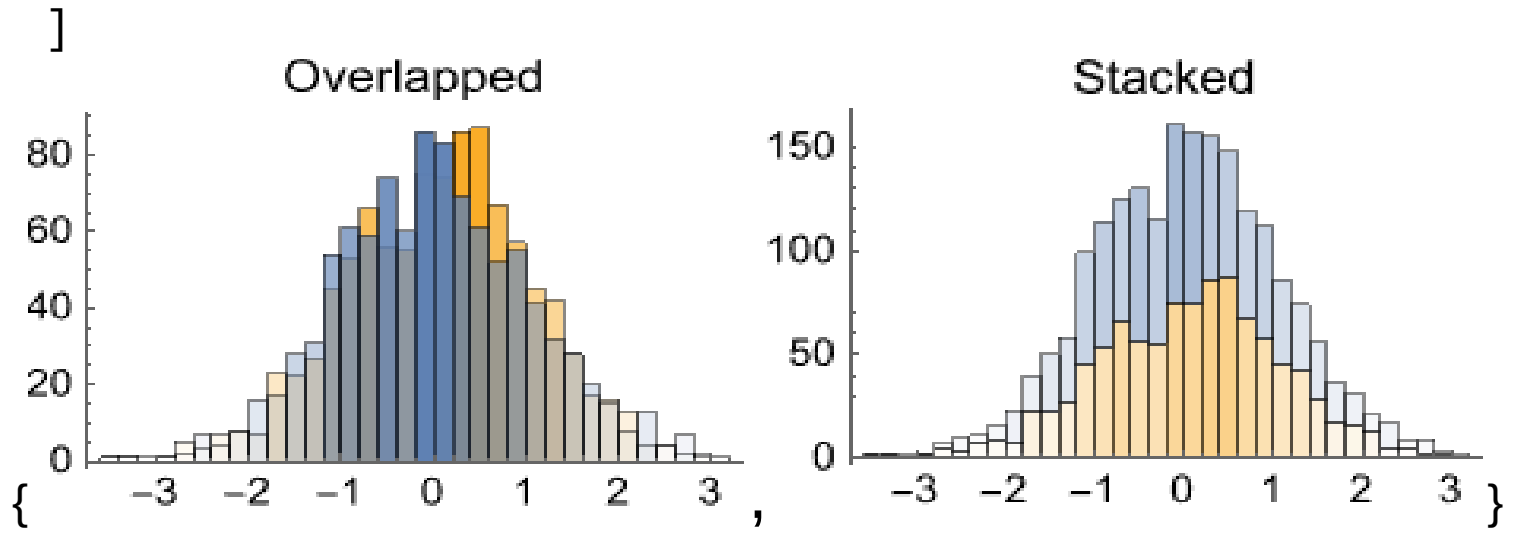

Input

```
(* Provide value labels for bars by using LabelingFunction: *)

Histogram[
 RandomVariate[
  NormalDistribution[0,1],
  400
```

```
  ],
 LabelingFunction->Above,
 ColorFunction->Function[{height},Opacity[height]],
 ChartStyle->Purple,
 ImageSize->170
 ]
```

Output

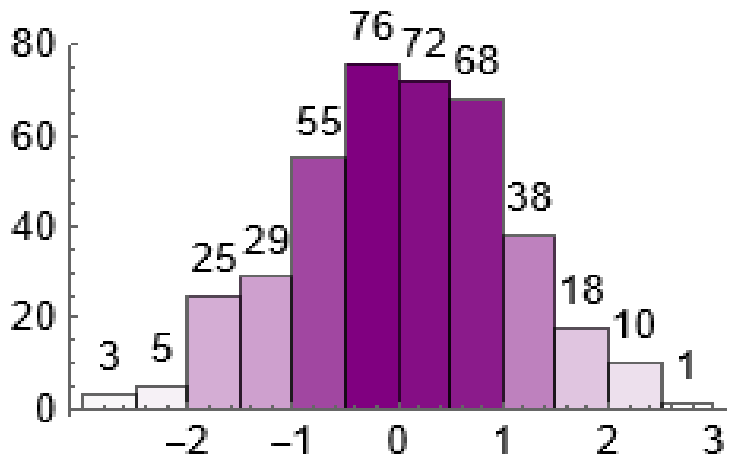

Input

```
(* This code combines a histogram of 2000 random samples drawn from a normal
distribution with mean 1 and standard deviation 1, along with a plot of the
probability density function (PDF) of the same normal distribution: *)

Show[
 Histogram[
  RandomVariate[
   NormalDistribution[1,1],
   2000
   ],
  Automatic,
  "PDF",
  ColorFunction->Function[{height},Opacity[height]],
  ChartStyle->Purple,
  ImageSize->170
  ],
 Plot[
  PDF[
   NormalDistribution[1,1],
   x
   ],
  {x,-5,5},
  PlotStyle->RGBColor[0.88,0.61,0.14]
  ]
 ]
```

Output

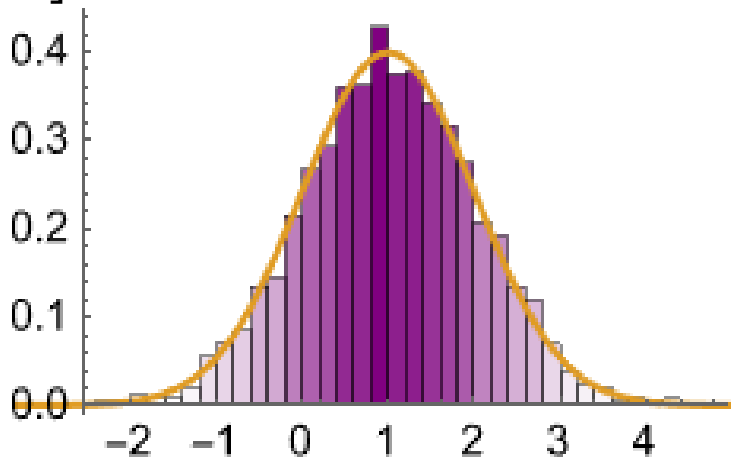

Input

```
(* This code generates a histogram of 1000 random samples drawn from a normal
distribution with mean 0 and standard deviation 1. The resulting histogram has no
outline or borders around the bars and is colored in purple. Without the borders
around the bars, the shape of the distribution may be more apparent. This can be
especially useful when comparing multiple histograms with different data sets or
parameters: *)

Histogram[
 RandomVariate[
  NormalDistribution[0,1],
  1000
```

```
  ],
ChartBaseStyle->EdgeForm[None],
ChartStyle->Purple,
ImageSize->170
]
```

Output

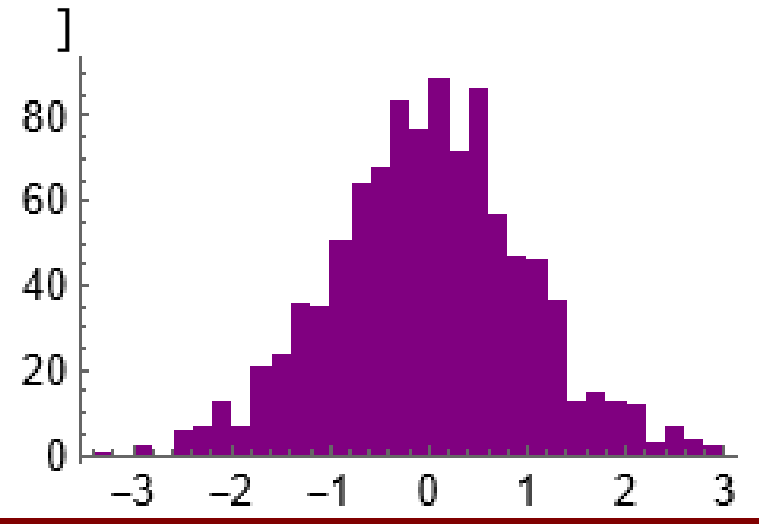

| | |
|---|---|
| `Histogram3D[{{x1,y1},{x2,y2},…}]` | plots a 3D histogram of the values {xi,yi}. |
| `Histogram3D[{{x1,y1},{x2,y2},…},bspec]` | plots a 3D histogram with bins specified by bspec. |
| `Histogram3D[{{x1,y1},{x2,y2},…},bspec, hspec]` | plots a 3D histogram with bin heights computed according to the specification hspec. |
| `Histogram3D[{data1,data2,…}]` | plots 3D histograms for multiple datasets datai. |

**Remarks:**

- The following bin specifications `bpsec` can be given:

| | |
|---|---|
| `n` | use n bins |
| `{w}` | use bins of width w |
| `{min,max,w}` | use bins of width w from min to max |
| `{{b1,b2,…}}` | use bins [b1,b2),[b2,b3),… |
| `Automatic` | determine bin widths automatically |
| `"name"` | use a named binning method |
| `{"Log",bspec}` | apply binning bspec on log-transformed data |
| `fb` | apply fb to get an explicit bin specification {b1,b2,…} |
| `{xspec,yspec,…}` | give different x, y, etc. specifications |

- Possible named binning methods include:

| | |
|---|---|
| `"Sturges"` | compute the number of bins based on the length of data |
| `"Scott"` | asymptotically minimize the mean square error |
| `"FreedmanDiaconis"` | twice the interquartile range divided by the cube root of sample size |
| `"Knuth"` | balance likelihood and prior probability of a piecewise uniform model |
| `"Wand"` | one-level recursive approximate Wand binning |

- Different forms of 3D histograms can be obtained by giving different bin height specifications `hspec` in `Histogram3D[data,bspec,hspec]`. The following forms can be used:

| | |
|---|---|
| `"Count"` | the number of values lying in each bin |
| `"CumulativeCount"` | cumulative counts |
| `"SurvivalCount"` | survival counts |
| `"Probability"` | fraction of values lying in each bin |
| `"Intensity"` | count divided by bin area |
| `"PDF"` | probability density function |
| `"CDF"` | cumulative distribution function |
| `"SF"` | survival function |
| `"HF"` | hazard function |
| `"CHF"` | cumulative hazard function |
| `{"Log",hspec}` | log-transformed height specification |
| `fh` | heights obtained by applying fh to bins and counts |

***Mathematica code 3.14*** Histogram3D

Input
```
(* This code generates a 3D histogram for a random sample of size 400 from a bivariate normal distribution with mean 1 and standard deviation 2 in each dimension. The code provides a simple way to visualize the distribution of a bivariate normal sample using a 3D histogram: *)

Histogram3D[
 RandomVariate[
  NormalDistribution[1,2],
  {400,2}
  ],
 ColorFunction->Function[{height},Opacity[0.9]],
 ChartStyle->RGBColor[0.6,0.30,0.60],
 ImageSize->170
 ]
```

Output

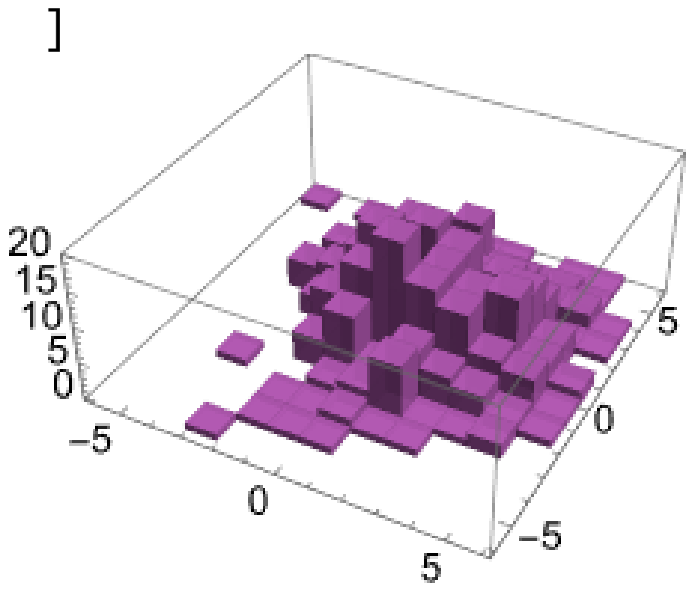

Input
```
(* Multiple datasets: *)

sampledata1=RandomVariate[
   NormalDistribution[0,1/2],
   {750,2}
   ];

sampledata2=RandomVariate[
   NormalDistribution[4,1],
   {750,2}
   ];

Histogram3D[
 {sampledata1,sampledata2},
 ColorFunction->Function[{height},Opacity[0.6]],
 ImageSize->170
 ]
```

Output

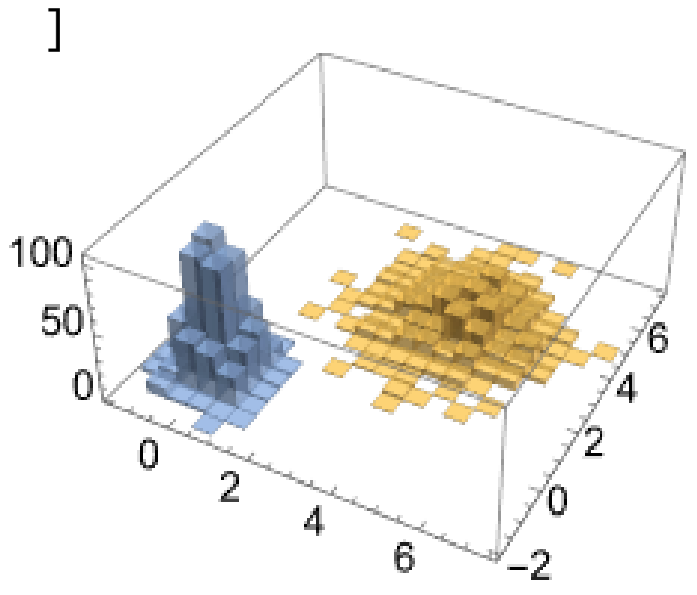

Input
```
(* This code generates a 3D histogram for a random sample of size 400 from a bivariate normal distribution with mean 0 and standard deviation 1 in each dimension. The binning specification {3,5} is used to divide the data into 3 equally spaced bins in the first dimension and 5 equally spaced bins in the second dimension: *)

sampledata=RandomVariate[
   NormalDistribution[0,1],
```

```
    {400,2}
    ];
Histogram3D[
 sampledata,
 {3,5},
 ColorFunction->Function[{height},Opacity[0.9]],
 ChartStyle->RGBColor[0.6,0.30,0.60],
 ImageSize->170
 ]
```

Output

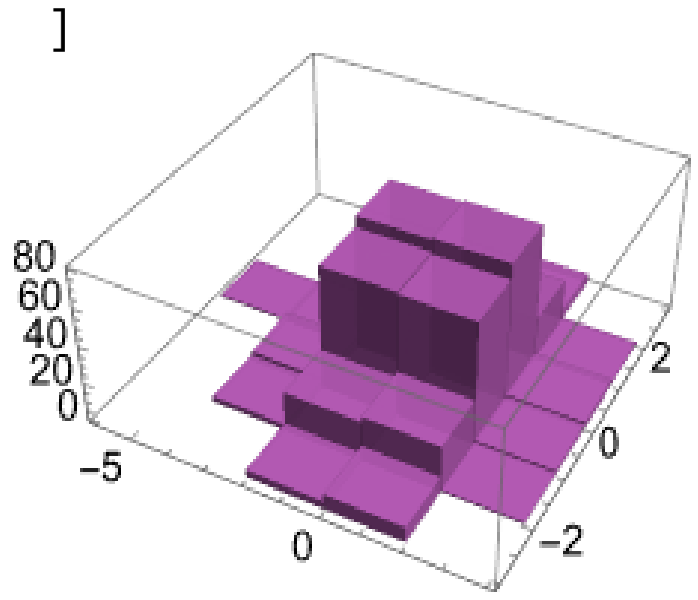

Input

```
(* Specify a different bin width to use in x and y: *)

sampledata=RandomVariate[
    NormalDistribution[0,1],
    {400,2}
    ];

Histogram3D[
 sampledata,
 {{.5},{2}},
 ColorFunction->Function[{height},Opacity[0.9]],
 ChartStyle->RGBColor[0.6,0.30,0.60],
 ImageSize->170
 ]
```

Output

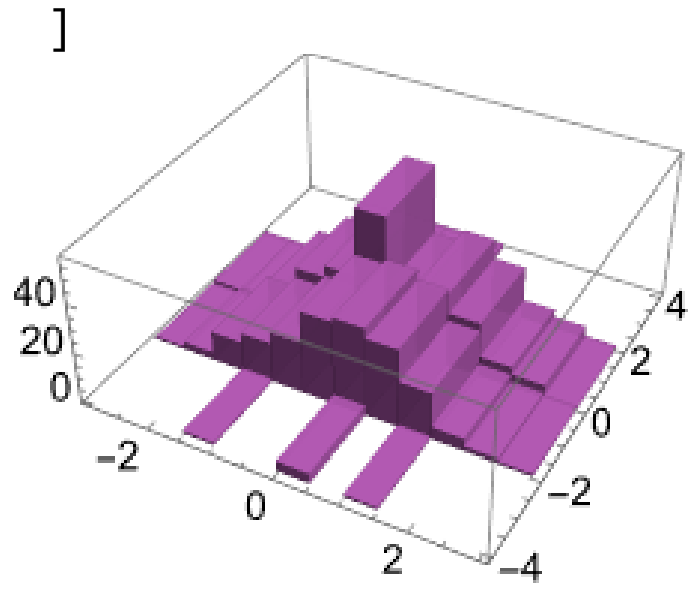

Input

```
(* This code generates a 3D histogram of a random sample of size 500 from a bivariate
normal distribution with mean 0 and standard deviation 1 in each dimension. The
code is designed to demonstrate how different binning methods can affect the
appearance of a histogram of bivariate data. By visualizing the data in this way,
it can be easier to identify patterns and relationships within the data: *)

sampledata=RandomVariate[
    NormalDistribution[0,1],
    {500,2}
    ];

Table[
 Histogram3D[
  sampledata,
  methods,
  PlotLabel->methods,
  ColorFunction->Function[{height},Opacity[0.9]],
```

```
   ChartStyle->RGBColor[0.6,0.30,0.60],
   ImageSize->170
   ],
 {methods,{"Sturges","Scott","FreedmanDiaconis","Wand"}}
 ]
```

Output

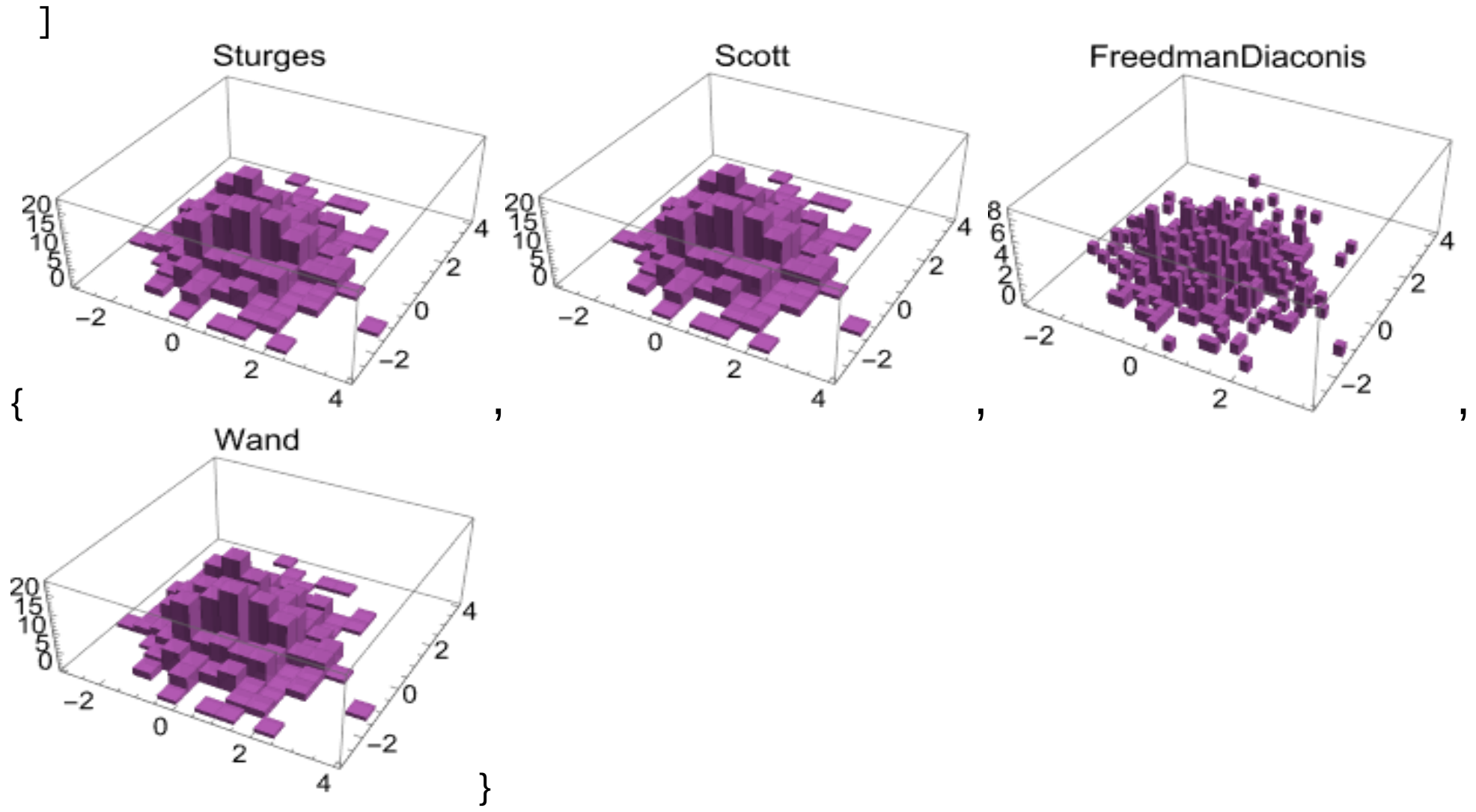

Input

```
(* This code generates a sequence of 3D histograms for a random sample of size 300
from a bivariate normal distribution with mean 0 and standard deviation 1 in each
dimension. The use of different height functions allows for a more comprehensive
understanding of the sample's distribution and provides a range of useful
visualization tools for data analysis: *)

sampledata=RandomVariate[
   NormalDistribution[0,1],
   {300,2}
   ];

Table[
 Histogram3D[
   sampledata,
   Automatic,
   height,
   PlotLabel->height,
   ColorFunction->Function[{height},Opacity[0.9]],
   ChartStyle->RGBColor[0.6,0.30,0.60],
   ImageSize->170
   ],
 {height,{"Count","Probability","PDF","CDF","SF","HF"}}
 ]
```

Output

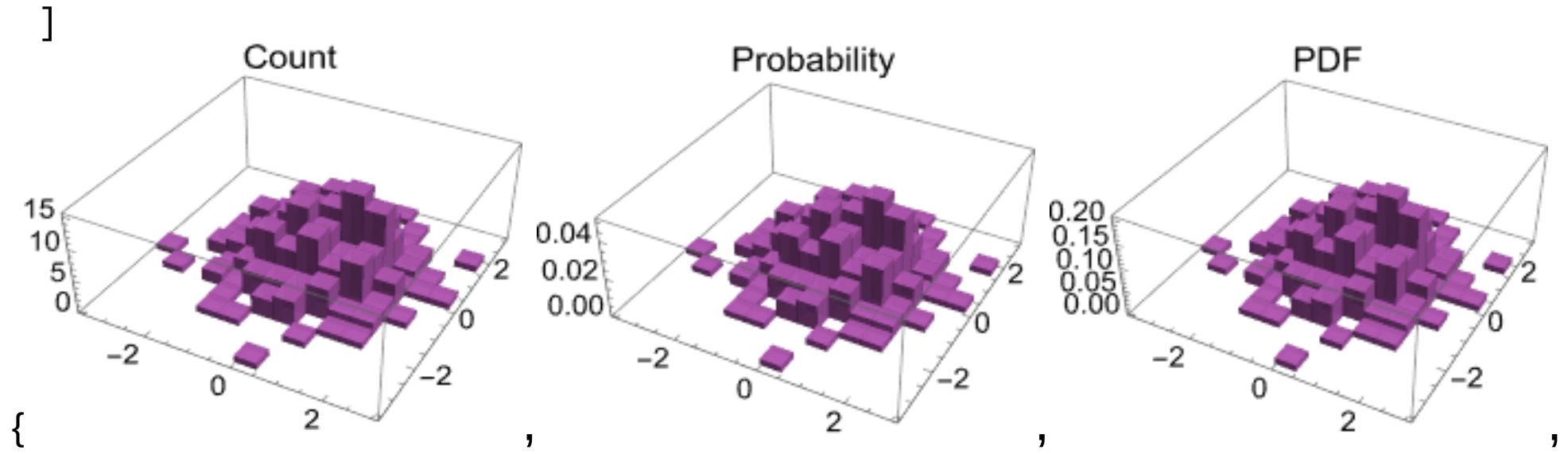

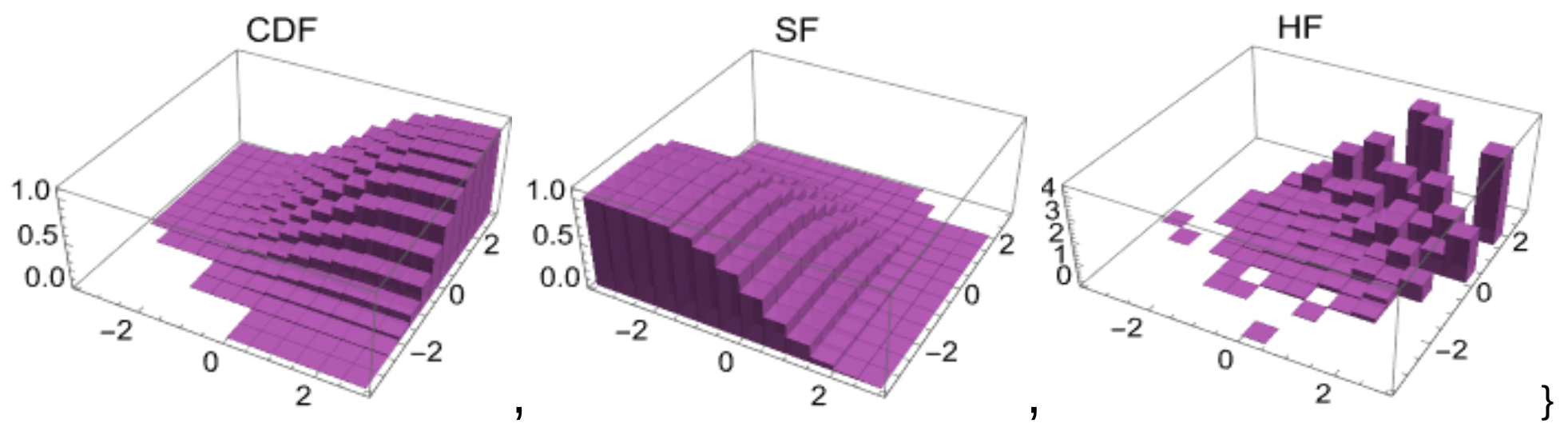

Input
```
(* This code generates three 3D histograms to visualize the distributions of three different random samples. The three histograms provide a way to compare the distributions of three different random samples in a 3D space. The customization options allow for different visual styles to be applied to the plot to suit different needs and preferences. The code demonstrates how to use the Style function to customize the color of a single dataset or all datasets in a multi-dataset plot: *)
sampledata1=RandomVariate[
    NormalDistribution[0,1],
    {400,2}
    ];

sampledata2=RandomVariate[
    NormalDistribution[3,1/2],
    {400,2}
    ];

sampledata3=RandomVariate[
    NormalDistribution[5,1/3],
    {400,2}
    ];

{
 Histogram3D[
  {sampledata1,sampledata2,sampledata3},
  ImageSize->170
  ],

 Histogram3D[
  {sampledata1,Style[sampledata2,RGBColor[0.6,0.30,0.60]],sampledata3},
  ImageSize->170
  ],

 Histogram3D[
  Style[
   {sampledata1,sampledata2,sampledata3},
   RGBColor[0.6,0.30,0.60]
   ],
  ImageSize->170
  ]
 }
```
Output

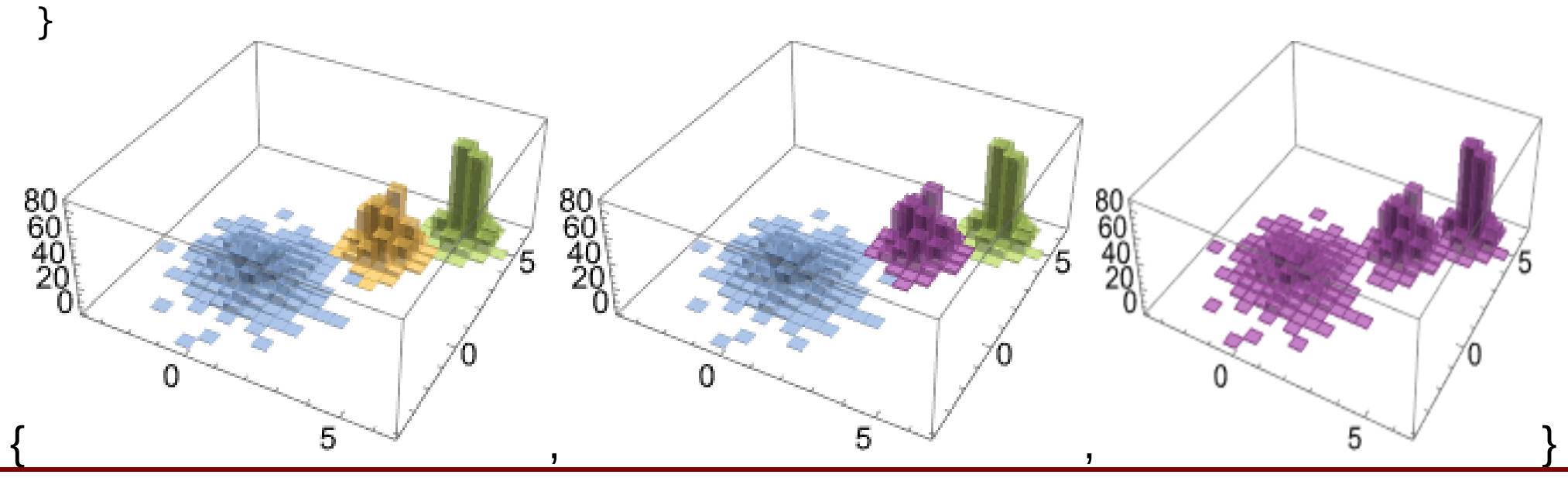

# CHAPTER 4

# DESCRIPTIVE STATISTICS: MEASURES OF CENTRAL TENDENCY

Descriptive statistics are a set of statistical techniques that are used to summarize and describe the main features of a dataset. Descriptive statistics break down into several types (measures of central tendency, measures of dispersion, and measures of symmetry).

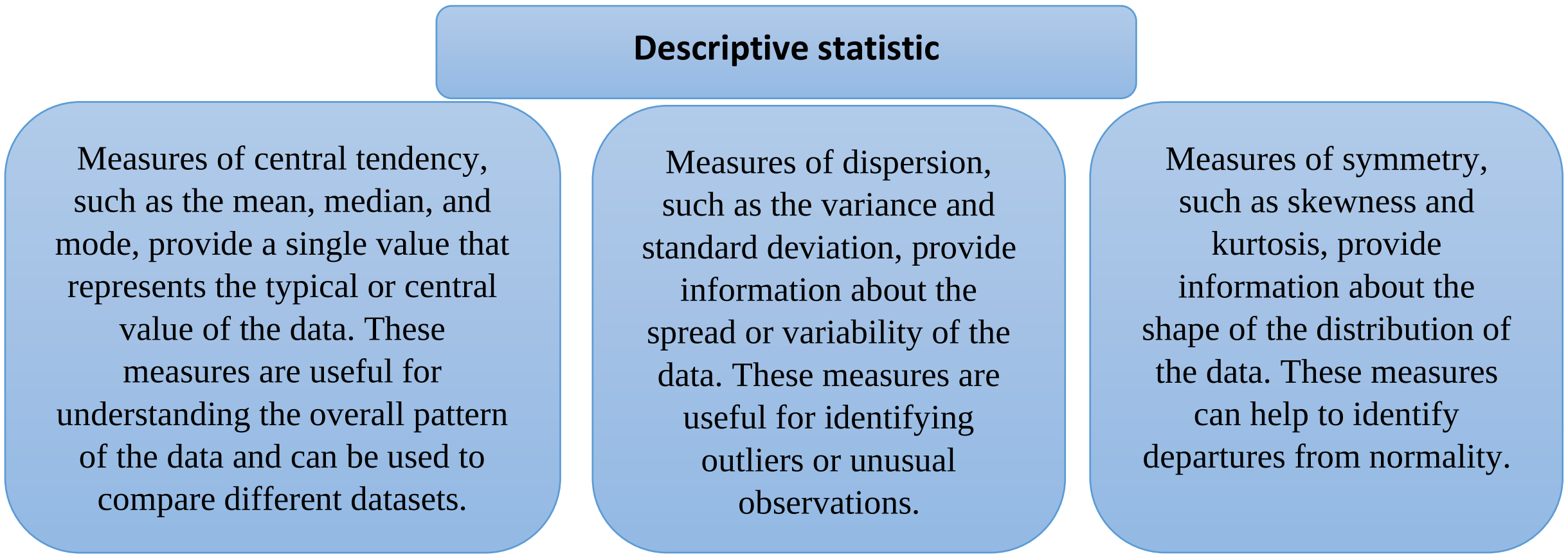

In this chapter, we will discuss the first of these types— measures of central tendency.

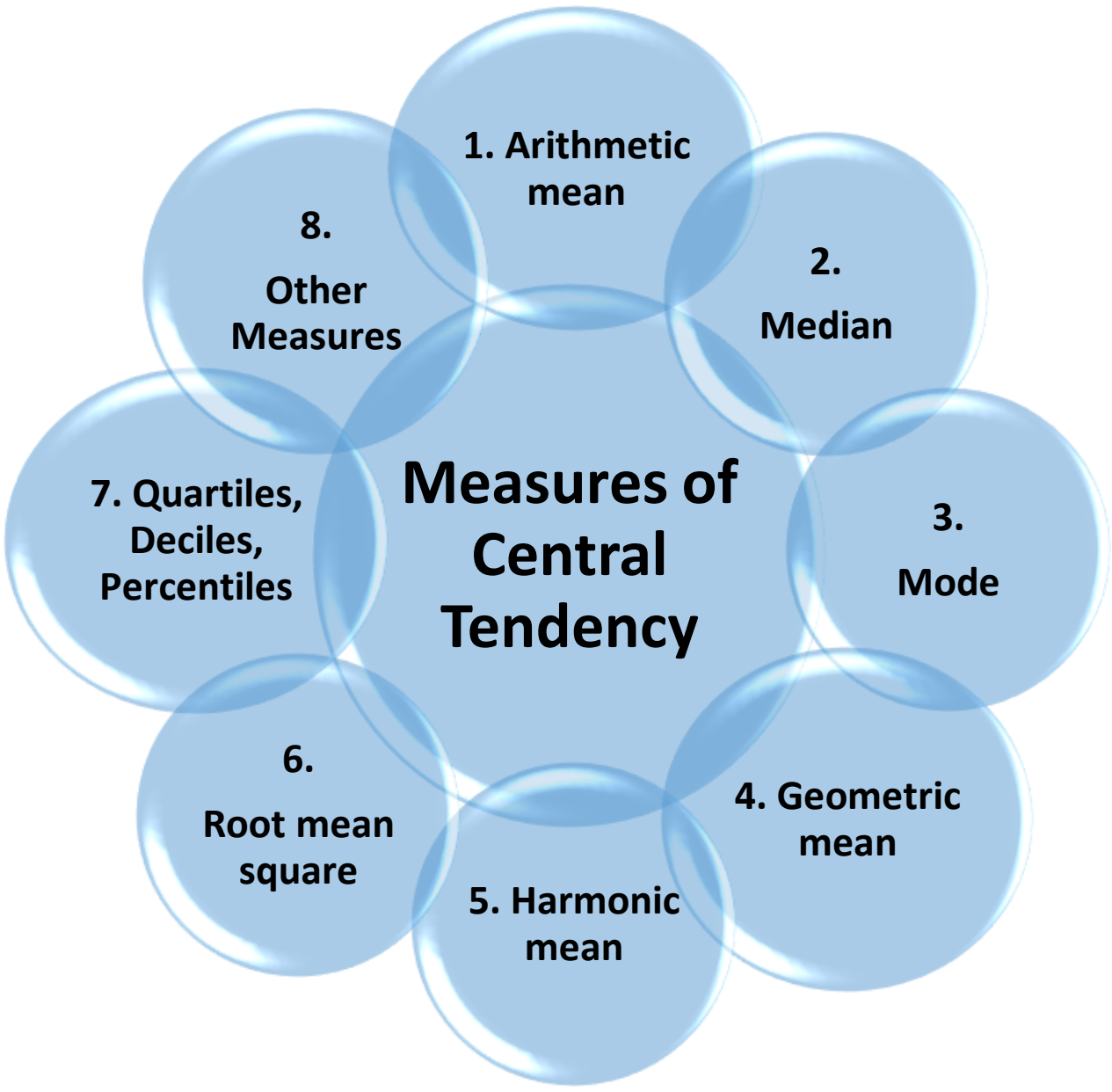

## 4.1 Mean

An average is a value that is typical, or representative, of a set of data. Since such typical values tend to lie centrally within a set of data arranged according to magnitude, averages are also called measures of central tendency. Several types of measures of central tendency can be defined, such as the arithmetic mean, median, mode, geometric mean, harmonic mean, root mean square, trimmed mean, winsorized mean, quartiles, deciles, and percentiles. Each has advantages and disadvantages, depending on the data and the intended purpose.

**Definition (The Arithmetic Mean):** The arithmetic mean, or briefly the mean, of a set of $N$ numbers $v_1, v_2, v_3, \ldots, v_N$ is denoted by $\bar{v}$ and is defined as

$$\bar{v} = \frac{v_1 + v_2 + \cdots + v_N}{N} = \frac{\sum_{j=1}^{N} v_j}{N}. \tag{4.1}$$

For instance, the arithmetic mean of the numbers in the set

$$h = \{133, 136, 149, 133, 123, 121, 140, 139, 117, 117, 136, 108, 126, 104, 116, 147, 140, 148, 150, 122, 135, 146, 133, 144, 117, 124, 135, 117, 120, 121, 110, 124, 103, 137, 101, 119, 104, 113, 139, 133\}, \tag{4.2}$$

is $\bar{v} = 127$. The mean is useful because it shows where the "center of gravity" exists for an observed set of values (see Figure 4.1).

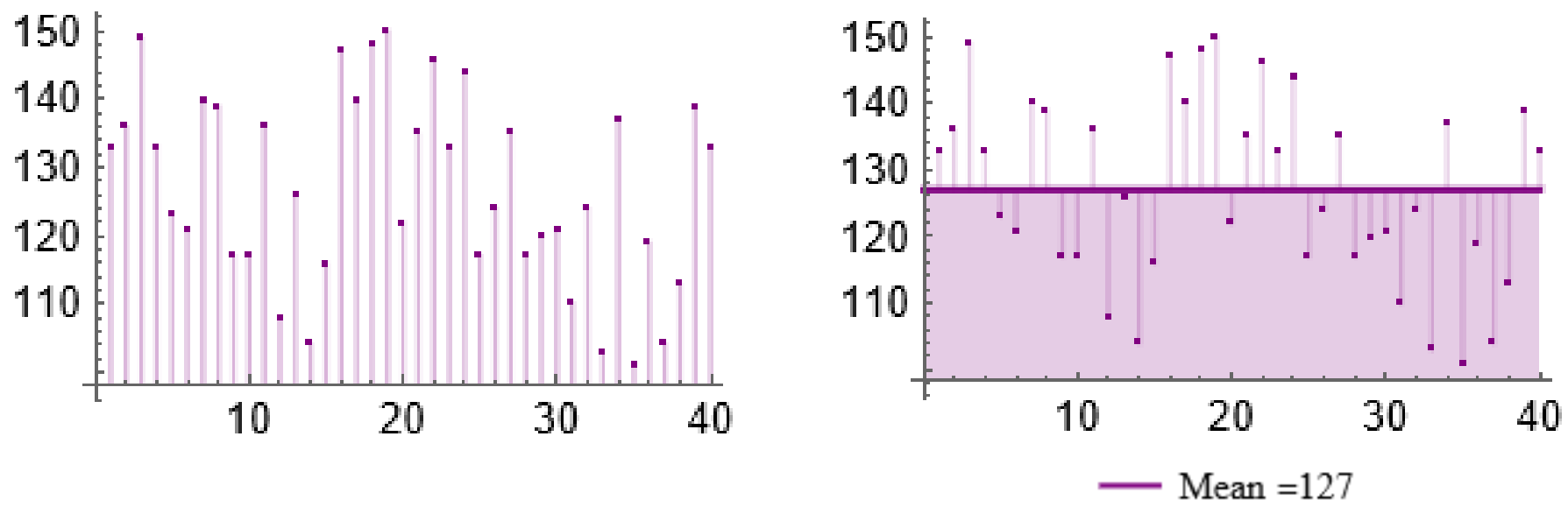

**Figure 4.1** The left plot displays the values in the set $h$ (4.2) as a list plot with the area under the points filled in to the axis. The right plot displays the same set of data, $h$, and a horizontal line at the mean value $\bar{v} = 127$. In the right plot, the data points display as a list plot filled up to the mean value to explain how the values in $h$ are distributed around the mean.

If the numbers $v_1, v_2, v_3, \ldots, v_K$ occur with frequencies $f_1, f_2, f_3, \ldots, f_K$, (grouped data), the arithmetic mean is

$$\begin{aligned}\bar{v} &= \frac{f_1 v_1 + f_2 v_2 + \cdots + f_K v_K}{f_1 + f_2 + \cdots + f_K} \\ &= \frac{\sum_{j=1}^{K} f_j v_j}{\sum_{j=1}^{K} f_j} \\ &= \frac{\sum_{j=1}^{K} f_j v_j}{N},\end{aligned} \tag{4.3}$$

where $\sum_{j=1}^{K} f_j = N$ is the total frequency. For instance, if 5, 8, 6, and 2 occur with frequencies 3, 2, 4, and 1, respectively, the arithmetic mean is $\bar{v} = \frac{(5)(3)+(8)(2)+(6)(4)+(2)(1)}{3+2+4+1} = 5.7$.

Sometimes we associate with the numbers $v_1, v_2, v_3, \ldots, v_K$ certain weighting factors (or weights) $w_1, w_2, \ldots, w_K$, depending on the significance or importance attached to the numbers. This can be helpful when we want some values to contribute to the mean more than others. A common example of this is weighting academic exams to give a final grade. If you have three exams and a final exam, and we give each of the three exams 20% weight and the final exam 40% weight of the final grade. The mean for a set of weighted numbers is given by,

$$\bar{v} = \frac{w_1 v_1 + w_2 v_2 + \cdots + w_K v_K}{w_1 + w_2 + \cdots + w_K} = \frac{\sum_{j=1}^{K} w_j v_j}{\sum_{j=1}^{K} w_j}, \tag{4.4}$$

and is called the weighted arithmetic mean. For instance, if a final examination in a course is weighted 3 times as much as a quiz and a student has a final examination grade of 85 and quiz grades of 70 and 90, the mean grade is $\bar{v} = \frac{(70)(1)+(90)(1)+(85)(3)}{1+1+3} = 83$.

If $f_1$ numbers have mean $m_1$, $f_2$ numbers have mean $m_2, \dots, f_K$ numbers have mean $m_K$, then the mean of all the numbers is

$$\bar{v} = \frac{f_1 m_1 + f_2 m_2 + \cdots + f_K m_K}{f_1 + f_2 + \cdots + f_K}, \tag{4.5}$$

that is a weighted arithmetic mean of all the means.

**Theorem 4.1:** The sum of the deviations of $v_1, v_2, v_3, \dots, v_N$ from their mean $\bar{v}$ is equal to zero.

**Proof:**

Let $d_j = v_j - \bar{v}$, $j = 1, \dots, N$, be the deviations of $v_1, v_2, v_3, \dots, v_N$ from their mean $\bar{v}$. Then

$$\begin{aligned}\text{Sum of the deviations} &= \sum_{j=1}^{N} d_j \\ &= \sum_{j=1}^{N} (v_j - \bar{v}) \\ &= \sum_{j=1}^{N} v_j - N\bar{v} \\ &= \sum_{j=1}^{N} v_j - N\left(\frac{\sum_{j=1}^{N} v_j}{N}\right) \\ &= \sum_{j=1}^{N} v_j - \sum_{j=1}^{N} v_j \\ &= 0.\end{aligned}$$

■

**Theorem 4.2:** If $z_j = x_j + y_j$, $j = 1, \dots, N$, then $\bar{z} = \bar{x} + \bar{y}$.

**Proof:**

By definition,

$$\bar{x} = \frac{\sum_{j=1}^{N} x_j}{N}, \qquad \bar{y} = \frac{\sum_{j=1}^{N} y_j}{N}, \qquad \bar{z} = \frac{\sum_{j=1}^{N} z_j}{N}.$$

Hence,

$$\bar{z} = \frac{\sum_{j=1}^{N} z_j}{N}$$

$$
\begin{aligned}
&= \frac{\sum_{j=1}^{N} x_j + y_j}{N} \\
&= \frac{\sum_{j=1}^{N} x_j + \sum_{j=1}^{N} y_j}{N} \\
&= \frac{\sum_{j=1}^{N} x_j}{N} + \frac{\sum_{j=1}^{N} y_j}{N} \\
&= \bar{x} + \bar{y}.
\end{aligned}
$$

■

**Theorem 4.3:** If $N$ numbers $v_1, v_2, v_3, \dots, v_N$ have deviations from any number $A$ given by $d_j = v_j - A$, $j = 1, \dots, N$; respectively, then

$$\bar{v} = A + \frac{\sum_{j=1}^{N} d_j}{N}. \tag{4.6}$$

If $K$ numbers $v_1, v_2, v_3, \dots, v_K$ have respective frequencies $f_1, f_2, f_3, \dots, f_K$ and $d_j = v_j - A$, $j = 1, \dots, K$; respectively, then

$$\bar{v} = A + \frac{\sum_{j=1}^{K} f_j d_j}{N}, \tag{4.7}$$

where $\sum_{j=1}^{K} f_j = N$.

**Proof:**

(a) Since $d_j = v_j - A$, and $v_j = d_j + A$, we have

$$
\begin{aligned}
\bar{v} &= \frac{\sum_{j=1}^{N} v_j}{N} \\
&= \frac{\sum_{j=1}^{N} (d_j + A)}{N} \\
&= \frac{NA + \sum_{j=1}^{N} d_j}{N} \\
&= A + \frac{\sum_{j=1}^{N} d_j}{N}.
\end{aligned}
$$

(b)

$$
\begin{aligned}
\bar{v} &= \frac{\sum_{j=1}^{K} f_j v_j}{\sum_{j=1}^{K} f_j} \\
&= \frac{\sum_{j=1}^{K} f_j v_j}{N} \\
&= \frac{\sum_{j=1}^{K} f_j (d_j + A)}{N} \\
&= \frac{\sum_{j=1}^{K} f_j d_j + \sum_{j=1}^{K} f_j A}{N} \\
&= \frac{\sum_{j=1}^{K} f_j d_j + A \sum_{j=1}^{K} f_j}{N} \\
&= \frac{\sum_{j=1}^{K} f_j d_j + AN}{N} \\
&= A + \frac{\sum_{j=1}^{K} f_j d_j}{N}.
\end{aligned}
$$

■

***Example 4.1***

Find the arithmetic mean of the numbers 5, 8, 11, 9, 12, 6, 14, and 10, choosing as the ‘‘guessed mean’’ $A$ the value 9.

***Solution***

The deviations of the given numbers from 9 are −4, −1, 2, 0, 3, −3, 5, and 1, and the sum of the deviations is $-4-1+2+0+3-3+5+1=3$. Thus

$$\bar{v}=A+\frac{\sum_{j=1}^{N}d_j}{N}=9+\frac{3}{8}=9.375.$$

Figure 4.2 shows the position of the mean for the symmetric and skewed to the right frequency curves.

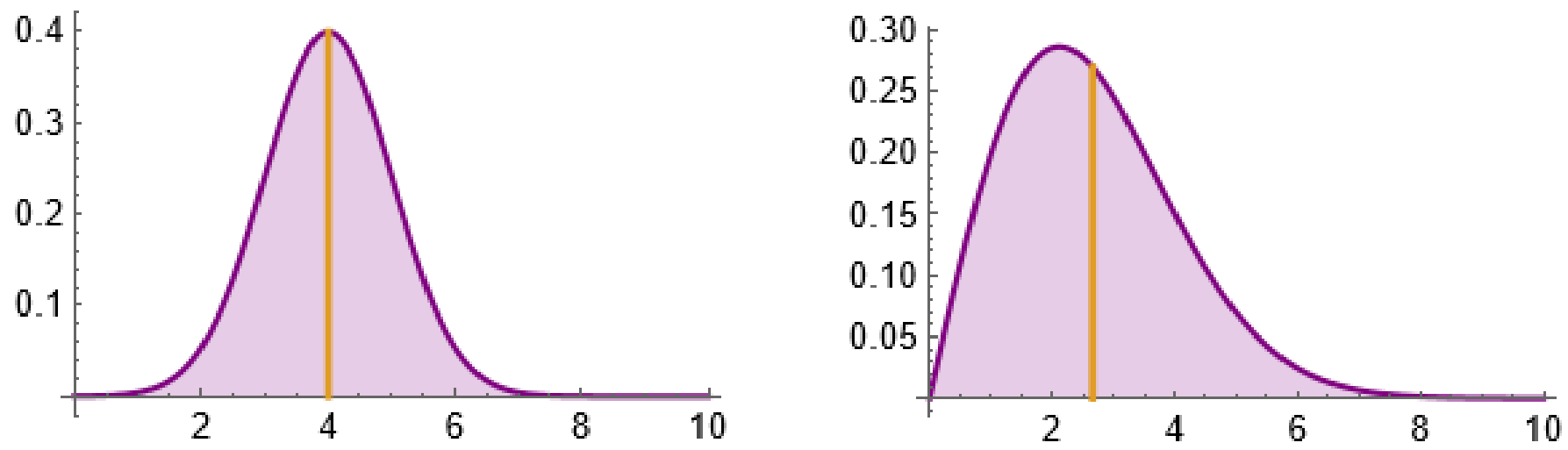

**Figure 4.2** The position of the mean for symmetric and right-skewed frequency curve.

When data are presented in a frequency distribution, all values falling within a given class interval are considered to be coincident with the class mark of the interval. Formulas (4.3) and (4.7) are valid for such grouped data if we interpret $v_j$ as the class mark, $f_j$ as its corresponding class frequency, $A$ as any guessed or assumed class mark, and $d_j=v_j-A$ as the deviations of $v_j$ from $A$. Computations using formulas (4.3) and (4.7) are sometimes called the long and short methods, respectively.

**Theorem 4.4:** Let $d_j=v_j-A$ denote the deviations of any class mark $v_j$ in a frequency distribution from a given class mark $A$. If all class intervals have equal size $c$, then
(a) the deviations are all multiples of $c$ (i.e., $d_j=cu_j$, where $u_j=0,\pm1,\pm2,\pm3,\ldots$) and
(b) the arithmetic mean can be computed from the formula

$$\bar{v}=A+\left(\frac{\sum_{j=1}^{K}f_ju_j}{N}\right)c. \tag{4.8}$$

**Proof:**

(a) If $v_1,v_2,v_3,\ldots$ are successive class marks, their common difference will for this case be equal to $c$, so that $v_2=v_1+c$, $v_3=v_2+c=v_1+2c$ and in general $v_j=v_1+(j-1)c$. Then any two class marks $v_p$ and $v_q$ will differ by

$$\begin{aligned}v_p-v_q&=[v_1+(p-1)c]-[v_1+(q-1)c]\\&=(p-q)c,\end{aligned}$$

which is a multiple of $c$.

(b) By part (a), the deviations of all the class marks from any given one are multiples of $c$ (i.e., $d_j=cu_j$). Then, we have

$$\bar{v}=A+\left(\frac{\sum_{j=1}^{K}f_jd_j}{N}\right)=A+\left(\frac{\sum_{j=1}^{K}f_jcu_j}{N}\right)=A+\left(\frac{\sum_{j=1}^{K}f_ju_j}{N}\right)c.$$

■

## 4.2 Median

**Definition (The Median):** The median of a set of numbers arranged in order of magnitude (i.e., in an array) is either the middle value or the arithmetic mean of the two middle values. Sometimes, the median denotes by $\tilde{v}$.

Loosely speaking, order the values of a data set of size $n$ from smallest to largest. If $n$ is odd, the sample median is the value in position $(n+1)/2$; if $n$ is even, it is the average of the values in positions $n/2$ and $n/2+1$ (see Figure 4.3). For instance, the set of numbers 2, 5, 6, 9, and 11 has a median 6, (rank the $n=5$ measurements from smallest to largest: 2, 5, 6, 9, 11), however, the set of numbers 2, 9, 11, 5, 6 and 27 has a median $\frac{1}{2}(6+9)=7.5$, (rank the measurements from smallest to largest: 2, 5, 6, 9, 11, 27) .

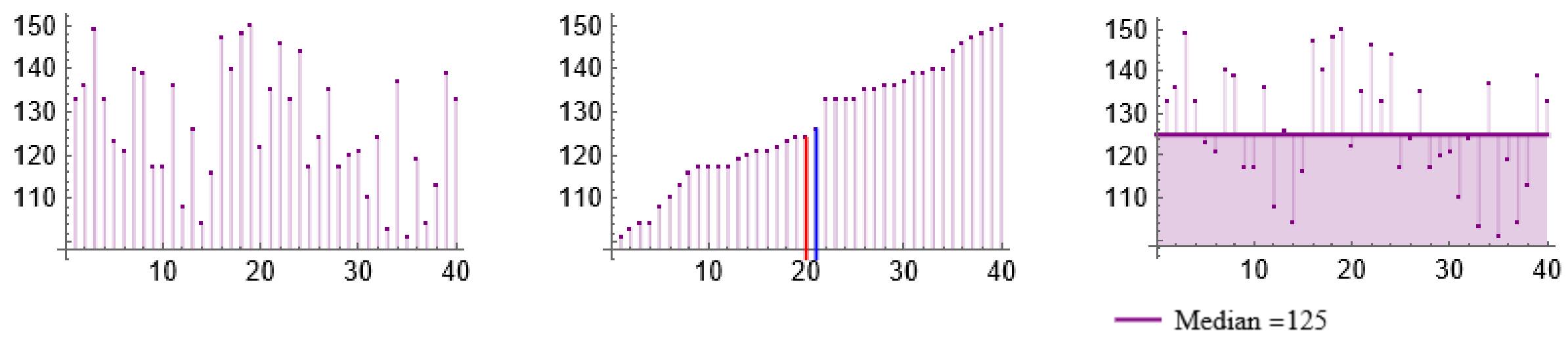

**Figure 4.3** The left plot displays the values in the set $h$ (4.2) as a list plot with the area under the points filled in to the axis. The medial plot displays the same data set, $h$, afer sorting the data points from smallest to largest. The red and blue lines represent positions $n/2=20$ and $n/2+1=21$, respectively. The right plot displays the same set of data, $h$, and a horizontal line at the median value $\tilde{v}=125$. In the right plot, the data points display as a list plot filled up to the median value to explain how the values in $h$ are distributed around the median.

For grouped data, the median is given by

$$\text{Median} = L_1 + \left(\frac{\frac{N}{2} - \left(\sum_{j=1}^{K} f_j\right)_1}{f_{\text{median}}}\right) c, \tag{4.9}$$

where $L_1$= lower class boundary of the median class (i.e., the class containing the median)

$N$ =number of items in the data (i.e., total frequency)

$\left(\sum_{j=1}^{K} f_j\right)_1$=sum of frequencies of all classes lower than the median class, (cumulative frequency of the preceding class).

$f_{\text{median}}$=frequency of the median class.

$c$ =size of the median class interval.

***Example 4.2***

Find the median weight of the 40 male students at Alex university by using the frequency distribution in Table 4.1.

**Table 4.1.** Frequency distribution.

| Class Interval | Frequency (f) |
|---|---|
| 118 – 126 | 3 |
| 127 – 135 | 5 |
| 136 – 144 | 9 |
| 145 – 153 | 12 |
| 154 – 162 | 5 |
| 163 – 171 | 4 |
| 172 – 180 | 2 |
| Total | 40 |

***Solution***

**First method**

The median is that weight for which half the total frequency $(40/2 = 20)$ lies above it and half lies below it. Now the sum of the first three class frequencies is $3 + 5 + 9 = 17$. Thus, to give the desired 20, we require three more of the 12 cases in the fourth class. Since the fourth-class interval, 145– 153, actually corresponds to weights 144.5 to 153.5, the median must lie $3/12$ of the way between 144.5 and 153.5; that is, the median is

$$144.5 + \frac{3}{12}(153.5 - 144.5) = 146.8 \text{ lb.}$$

**Second method**

Since the sum of the first three and first four class frequencies are $3 + 5 + 9 = 17$ and $3 + 5 + 9 + 12 = 29$, respectively, it is clear that the median lies in the fourth class, which is, therefore, the median class. Then $L_1 = 144.5$, $N = 40$, $\left(\sum_{j=1}^{K} f_j\right)_1 = 17$, $f_{\text{median}} = 12$, $c = 9$ and thus

$$\begin{aligned}\text{Median} &= L_1 + \left(\frac{\frac{N}{2} - \left(\sum_{j=1}^{K} f_j\right)_1}{f_{\text{median}}}\right)c \\ &= 144.5 + \left(\frac{\frac{40}{2} - 17}{12}\right)(9) \\ &= 146.8 \text{ lb.}\end{aligned}$$

## Remarks:

- The mean and median are both useful statistics for describing the central tendency of a data set. The mean makes use of all the data values and is affected by extreme values that are much larger or smaller than the others; the median makes use of only one or two of the middle values and is thus not affected by extreme values. Which of them is more useful depends on what one is trying to learn from the data.
- When your median is very different from your mean, that means you have a skewed dataset with outliers.
- The relationship between the median and mean can provide valuable information about the shape of a frequency distribution. Specifically, the positioning of the median and mean can provide insight into the skewness of the distribution. In a symmetric frequency curve, where the data is evenly distributed around the center, the median and mean will be located at the same point. This is because the median represents the center of the data, and the mean is calculated as the sum of all the data points divided by the number of data points, resulting in a value that is also at the center of the data. In contrast, for a right-skewed frequency distribution, the mean will be greater than the median. This occurs because the long tail of the distribution is pulling the mean in that direction, while the median is less affected by extreme values. In this case, the median is a better measure of central tendency than the mean, as it is less influenced by outliers. Figure 4.4 shows the relative positions of the mean and median for the symmetric and skewed to the right frequency curves. For symmetrical curves, the mean and median coincide.

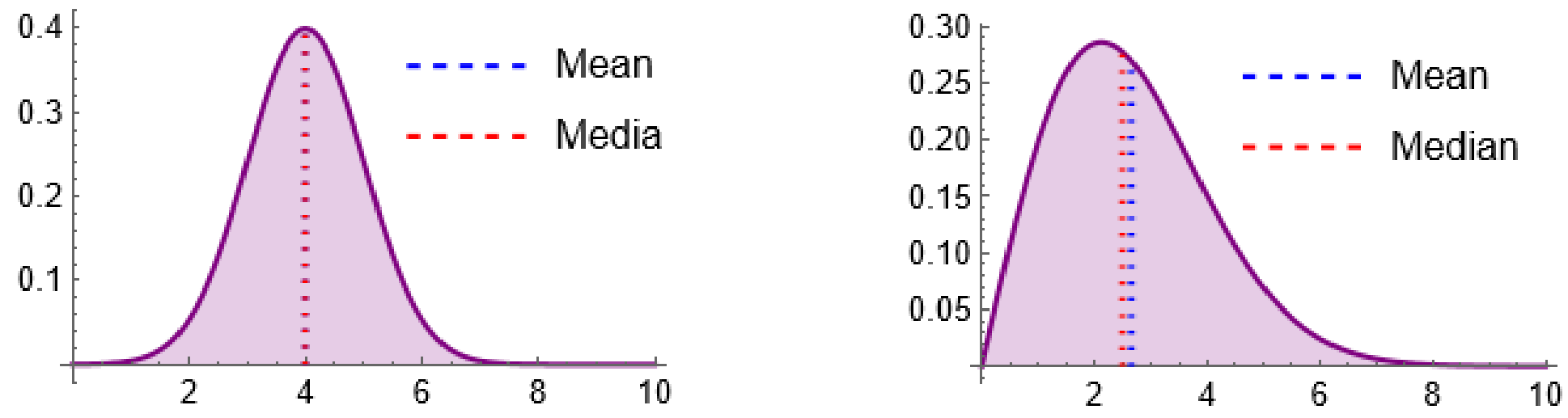

**Figure 4.4** Relative positions of median and mean for symmetric and right-skewed frequency curve.

## 4.3 Mode

**Definition (The Mode):** The mode of a set of numbers is that value which occurs with the greatest frequency. For instance, the set of numbers 2, 4, 7, 8, 8, 8, 10, 10, 11, 12 and 18 has mode 8, however, the set of numbers 2, 3, 8, 11, 12, 14, and 16 has no mode. Set 1, 2, 3, 3, 3, 5, 5, 8, 8, 8, and 9 has two modes, 3 and 8.

**Remarks:**

- The mode may not exist, and even if it does exist it may not be unique. When no value occurs more than once, there is no mode.
- If no single value occurs most frequently, then all the values that occur at the highest frequency are called modal values. When two values occur with an equal amount of frequency, then the dataset is considered bimodal.
- A distribution having only one mode is called unimodal.
- In the case of grouped data where a frequency curve has been constructed to fit the data, the mode will be the value (or values) of $v$ corresponding to the maximum point (or points) on the curve. This value of $v$ is sometimes denoted by $\hat{v}$.
- The mode is not unduly affected by extreme values, that is, values that are extremely high or extremely low. For example, if we are given the following set of observations: 1,1,1,1,1,2,2,100. The mean of the given set of data values is 13.625 which is clearly not representative of the above data values. However, the mode which is equal to 1 is clearly representative of a typical value from the above data set. This is one advantage of the mode compared to the mean.
- From a frequency distribution or histogram, the mode can be obtained from the formula

$$\text{Mode} = L + \frac{f_1 - f_0}{(f_1 - f_0) + (f_1 - f_2)} h, \tag{4.10}$$

where,

$L$= the lower limit of the modal class (i.e., the class containing the mode)
$f_1$= the frequency of the modal class
$f_0$= the frequency of the class preceding the modal class
$f_2$= the frequency of the class succeeding the modal class
$h$ = size of the modal class interval

**Definition (The Empirical Relation):** For unimodal frequency curves that are moderately skewed (asymmetrical), we have the empirical relation,

$$\text{Mean} - \text{mode} = 3(\text{mean} - \text{median}). \tag{4.11}$$

***Example 4.3***

Following is the distribution of height (in cm) of 50 students.

| Height ($x$) | 125 − 130 | 130 − 135 | 135 − 140 | 140 − 145 | 145 − 150 |
|---|---|---|---|---|---|
| Number of students ($f$) | 7 | 14 | 10 | 10 | 9 |

Find the modal height of the students.

***Solution***

Here the maximum frequency is 14, therefore, the modal class is 130 − 135. Thus, we have $L = 130$, $h = 5$, $f_1 = 14$, $f_0 = 7$, and $f_2 = 10$.

$$\text{Mode} = L + \left(\frac{\Delta_1}{(f_1 - f_0) + (f_1 - f_2)}\right) h$$

$$= 130 + \left(\frac{14 - 7}{(14 - 7) + (14 - 10)}\right)(5) = 133.18.$$

## 4.4 Other Measures of Central Tendency

### Geometric Mean

The geometric mean is a mean or average that uses the product of the values of a finite set of real numbers to indicate a central tendency (as opposed to the arithmetic mean, which uses the sum of the values).

**Definition (The Geometric Mean $G$):** The geometric mean $G$ of a set of $N$ positive numbers $v_1, v_2, v_3, \dots, v_N$ is the $N$th root of the product of the numbers:

$$G = \sqrt[N]{v_1 v_2 v_3 \dots v_N}. \tag{4.12}$$

For instance, the geometric mean of the numbers 3, 9, and 27 is $G = \sqrt[3]{(3)(9)(27)} = 9$.

**Theorem 4.5:** The geometric mean $G$ of a set of $N$ positive numbers $v_1, v_2, v_3, \dots, v_N$ can also be expressed as the exponential of the arithmetic mean of logarithms

$$G = e^{\frac{1}{N}\sum_{i=}^{N} \ln v_i}. \tag{4.13}$$

**Proof:**

$$\begin{aligned} G &= \sqrt[N]{v_1 v_2 v_3 \dots v_N} \\ &= (v_1 v_2 v_3 \dots v_N)^{\frac{1}{N}} \\ &= e^{\ln (v_1 v_2 v_3 \dots v_N)^{\frac{1}{N}}} \\ &= e^{\frac{1}{N}\ln (v_1 v_2 v_3 \dots v_N)} \\ &= e^{\frac{1}{N}(\ln v_1 + \ln v_2 + \dots + \ln v_N)} \\ &= e^{\frac{1}{N}\sum_{i=}^{N} \ln v_i}, \end{aligned}$$

i.e.,

$$\text{Geometric mean} = e^{\text{arithmetic mean of logarithms}}.$$

■

**Theorem 4.6:** Let the numbers $v_1, v_2, v_3, \dots, v_K$ occur with frequencies $f_1, f_2, f_3, \dots, f_K$, where $\sum_{j=1}^{K} f_j = N$ is the total frequency, then the weighted geometric mean is

$$G = \sqrt[N]{v_1^{f_1} v_2^{f_2} \dots v_K^{f_K}}, \tag{4.14}$$

and

$$\log G = \frac{1}{N}\sum_{j=1}^{K} f_j \log v_j. \tag{4.15}$$

**Proof:**

Part 1.

$$G = \sqrt[N]{(v_1 v_1 \dots v_1)(v_2 v_2 \dots v_2) \dots (v_K \dots v_K)} = \sqrt[N]{v_1^{f_1} v_2^{f_2} \dots v_K^{f_K}}.$$

Part 2.

$$\begin{aligned} \log G &= \log\left(v_1^{f_1} v_2^{f_2} \dots v_K^{f_K}\right)^{\frac{1}{N}} \\ &= \frac{1}{N}\log\left(v_1^{f_1} v_2^{f_2} \dots v_K^{f_K}\right) \end{aligned}$$

$$= \frac{1}{N}(f_1 \log v_1 + f_2 \log v_2 + \cdots + f_K \log v_K)$$
$$= \frac{1}{N}\sum_{j=1}^{K} f_j \log v_j.$$

■

**Remark:**

- One limitation of the geometric mean is that it cannot be used with negative numbers, as the $n$th root of a negative number is not defined.
- The geometric mean is useful in data normalization, where it is used to scale data to a common baseline. For example, if you have a set of data with values that span several orders of magnitude, you can use the geometric mean to calculate a "typical" value for the data set, which can then be used to scale the data to a more manageable range. One typically divides each data point by the geometric mean of the entire dataset. This has the effect of scaling the data so that it is centered around 1.
- The geometric mean is useful also in situations where the data exhibits exponential growth or decay, such as in population growth or radioactive decay.

## Harmonic Mean

**Definition (The Harmonic Mean *H*):** The harmonic mean $H$ of a set of $N$ numbers $v_1, v_2, v_3, \ldots, v_N$ is the reciprocal of the arithmetic mean of the reciprocals of the numbers:

$$H = \frac{1}{\frac{1}{N}\sum_{j=1}^{N}\frac{1}{v_j}}. \tag{4.16}$$

Hence, the harmonic mean is calculated by dividing the number of observations by the sum of their reciprocals.

$$H = \frac{N}{\sum_{j=1}^{N}\frac{1}{v_j}}. \tag{4.17}$$

For instance, the harmonic mean of the numbers 2, 4, and 8 is $H = \frac{3}{\frac{1}{2}+\frac{1}{4}+\frac{1}{8}} = 3.43$.

**Remarks:**

- The harmonic mean is useful when dealing with rates or ratios, where the data has an inverse relationship. For example, when calculating the average speed of a journey where the distance and time taken are known, the harmonic mean can be used to give a more accurate measure than the arithmetic mean. If a vehicle travels a certain distance $d$ outbound at a speed $x$ (e.g. 60 km/h) and returns the same distance at a speed $y$ (e.g. 20 km/h), then its average speed is the harmonic mean of $x$ and $y$ (30 km/h), not the arithmetic mean (40 km/h),

$$\text{Average speed for the entire journey} = \frac{\text{Total distance traveled}}{\text{Sum of time for each segment}} = \frac{2d}{\frac{d}{x}+\frac{d}{y}} = \frac{2}{\frac{1}{x}+\frac{1}{y}}.$$

- The geometric mean of a set of positive numbers $v_1, v_2, v_3, \ldots, v_N$ is less than or equal to their arithmetic mean but is greater than or equal to their harmonic mean. In symbols,

$$H \leq G \leq \bar{v}. \tag{4.18}$$

The equality signs hold only if all the numbers $v_1, v_2, v_3, \ldots, v_N$ are identical. For example, the set 2, 4, 8 has an arithmetic mean 4.67, geometric mean 4, and harmonic mean 3.43.

### Root Mean Square

**Definition (The Root Mean Square (RMS) or Quadratic Mean):** The RMS is calculated by taking the square root of the average of the squares of a set of values. The RMS of a set of numbers $v_1, v_2, v_3, \ldots, v_N$ is defined by

$$\text{RMS} = \sqrt{\overline{v^2}} = \sqrt{\frac{\sum_{j=1}^{N} v_j^2}{N}}. \tag{4.19}$$

For instance, the RMS of the set 1, 3, 4, 5, and 7 is RMS $= \sqrt{\frac{1^2+3^2+4^2+5^2+7^2}{5}} = 4.47$.

**Remarks:**

- The RMS takes into account the magnitude of both positive and negative values. This property makes the RMS particularly useful in situations where both positive and negative values are present. This is because when you square a negative number, the result is positive. So, when you calculate the average of the squares of a set of values and then take the square root of that average, the negative values contribute just as much to the result as the positive values. This means that the RMS provides a more complete picture of the overall magnitude of a set of values.
- The RMS can be used to calculate the standard deviation of a set of values. Specifically, the standard deviation is equal to the RMS of the deviations from the mean.

### Truncated Mean or Trimmed Mean

**Definition (The Trimmed Mean):** Trimmed mean is a statistical measure that is calculated by first removing a certain percentage of the largest and smallest values from a dataset and then taking the arithmetic mean of the remaining values. The percentage of values that are trimmed is typically between 5% and 25%.

**Remarks:**

- One advantage of using trimmed mean is that it is less affected by outliers than the standard arithmetic mean. This is because the extreme values are removed before calculating the mean, which reduces the impact of outliers on the final result.
- The truncated mean uses more information from the data set than the median, but unless the underlying data set is symmetric, the truncated mean of a sample is unlikely to produce an unbiased estimator for either the mean or the median.

Here are the steps to calculate the trimmed mean:

- Decide on the total percentage of values to be trimmed. This is typically denoted as '$p$' and is usually between 5% and 25%. For example, if you decide to trim a total 10% of values, $p$ would be 0.1.
- Sort the dataset in ascending or descending order, depending on your preference.
- Calculate the number of values to be trimmed on each end of the dataset. This can be done using the formula:

  $$\text{number of values to be trimmed on each end} = N \times (p/2)$$

  where $N$ is the total number of values in the dataset. Round this number to the nearest integer.

- Remove the calculated number of values from each end of the dataset. For example, if the dataset contains 20 values, and you decided to trim a total 10%, (i.e., 5% from each side) then you would remove 1 value from each end, leaving 18 values in the trimmed dataset.
- Calculate the arithmetic mean of the remaining values in the trimmed dataset. This is your trimmed mean.

***Example 4.4***

Find the totally 40% trimmed mean for the dataset $\{1, 2, 3, 4, 5, 6, 7, 8, 9, 10\}$.

***Solution***

- $p = 0.4$. The number of values to be trimmed on each end is:

  number of values to be trimmed on each end = $N \times (p/2) = 10 \times 0.2 = 2$.
- Remove 2 values from each end of the dataset, leaving: $3, 4, 5, 6, 7, 8$.
- The trimmed mean is then calculated as the arithmetic mean of these values:

  trimmed mean = $(3 + 4 + 5 + 6 + 7 + 8)/6 = 5.5$.
- So, the trimmed mean of this dataset, with 40% of values trimmed, is 5.5.

### Winsorized Mean

**Definition (The Winsorized Mean):** The Winsorized mean is a statistical method used to reduce the influence of extreme values (outliers) on the calculation of the mean. Instead of completely removing the outliers, the Winsorized mean replaces them with the nearest non-outlying value.

**Remarks:**

- Winsorized mean has several advantages over other methods, such as the mean or median. It is less sensitive to outliers and can provide a more accurate representation of the data.
- The winsorized mean includes modifying data points, while the trimmed mean involves removing data points. It is common for the winsorized mean and trimmed mean to be close or sometimes equal in value to each other.

Here are the steps to calculate Winsorized mean:

- Sort the dataset in ascending order.
- Decide on the percentage of values to be trimmed from each tail of the dataset. For example, if you decide to trim 10% from each tail, then the top and bottom 10% of values will be replaced.
- Calculate the number of values to be trimmed from each tail of the dataset. This can be done by multiplying the percentage to be trimmed by the total number of observations in the dataset and then rounding to the nearest integer. For example, if you have a dataset of 100 observations and you decide to trim 10% from each tail, you will need to trim 10 values from each tail.
- Replace extreme values in the dataset. For example, if you are trimming 10% from each tail of the dataset, you would replace the top 10% of values with the value at the 90th percentile and replace the bottom 10% of values with the value at the 10th percentile.
- Calculate the arithmetic mean of the trimmed data. This is the Winsorized Mean.

***Example 4.5***

Find the total 20% winsorized mean for the dataset

$$\{2, 4, 7, 8, 11, 14, 18, 23, 23, 27, 35, 40, 49, 50, 55, 60, 61, 61, 62, 75\}.$$

***Solution***

- A total 20% winsorized mean takes the top 10% and bottom 10% and replaces them with their next closest value.

- The two smallest and two largest data points—20% of the 20 data points—will be replaced with their next closest value.
- Thus, the new data set is as follows: 7, 7, 7, 8, 11, 14, 18, 23, 23, 27, 35, 40, 49, 50, 55, 60, 61, 61, 61, 61. The winsorized mean is 33.9.

### Quartiles, Deciles, and percentiles

- If a set of data is arranged in order of magnitude, the middle value (or arithmetic mean of the two middle values) that divides the set into two equal parts is the median.
- By extending this idea, quartiles, deciles, and percentiles divide a dataset into equal parts based on their rank or order.
- Quartiles divide a dataset into four equal parts, where the first quartile ($Q_1$) represents the 25th percentile, the second quartile ($Q_2$) represents the 50th percentile (the median), and the third quartile ($Q_3$) represents the 75th percentile. Quartiles are useful in describing the spread of a dataset and identifying outliers (see Figure 4.5).
- Deciles divide a dataset into ten equal parts, where the first decile represents the 10th percentile, the second decile represents the 20th percentile, and so on. Deciles are denoted by $D_1, D_2, \dots, D_9$ (see Figure 4.6).
- Percentiles divide a dataset into 100 equal parts, where the first percentile represents the smallest value, and the 100th percentile represents the largest value. For example, the 75th percentile represents the value below which 75% of the data falls. Percentiles are denoted by $P_1, P_2, \dots, P_{99}$.
- The fifth decile and the 50th percentile correspond to the median.
- The 25th and 75th percentiles correspond to the first and third quartiles, respectively.
- Collectively, quartiles, deciles, percentiles, and other values obtained by equal subdivisions of the data are called quantiles.

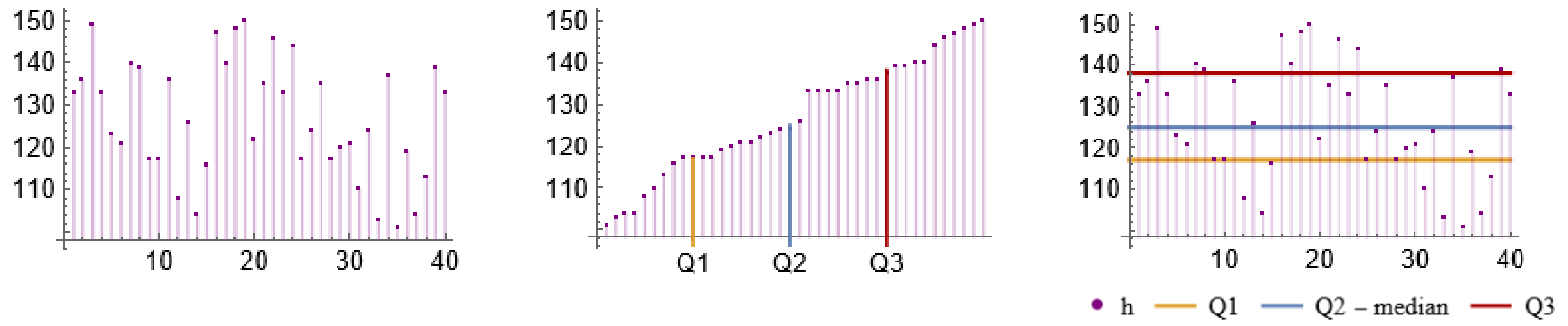

**Figure 4.5** The left plot displays the values in the set $h$ (4.2) as a list plot with the area under the points filled in to the axis. The medial plot displays the same data set, $h$, after sorting the data points from smallest to largest. The $Q1$, $Q2$, and $Q3$ lines represent positions 25%, 50%, and 75% of data, respectively. The right plot displays the same set of data, $h$, and horizontal lines at the quartiles $Q1 = 117$, $Q2 = 125$, and $Q3 = 138$.

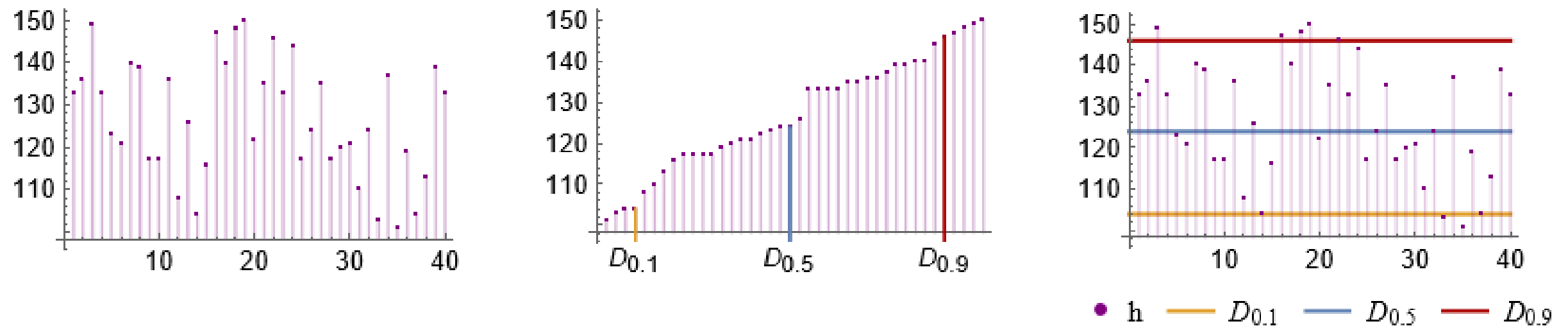

**Figure 4.6** The left plot displays the values in the set $h$ (4.2) as a list plot with the area under the points filled in to the axis. The medial plot displays the same data set, $h$, after sorting the data points from smallest to largest. The $D_{0.1}$, $D_{0.5}$ and $D_{0.9}$ lines represent positions 10%, 50%, and 90% of the data, respectively. The right plot displays the same set of data, $h$, and horizontal lines at the deciles $D_{0.1} = 104$, $D_{0.5} = 124$ and $D_{0.9} = 146$.

To calculate quartiles, there is no universal agreement on selecting quartile values.

Method 1

- Use the median to divide the ordered data set into two halves.
  1- If there is an odd number of data points in the original ordered data set, do not include the median (the central value in the ordered list) in either half.
  2- If there is an even number of data points in the original ordered data set, split this data set exactly in half.
- The lower quartile value is the median of the lower half of the data. The upper quartile value is the median of the upper half of the data.

Method 2

- Use the median to divide the ordered data set into two halves.
  1- If there are an odd number of data points in the original ordered data set, include the median (the central value in the ordered list) in both halves.
  2- If there are an even number of data points in the original ordered data set, split this data set exactly in half.
- The lower quartile value is the median of the lower half of the data. The upper quartile value is the median of the upper half of the data.

Here are the steps to calculate quartiles using the Tukey method:

- Sort the data in ascending order.
- Calculate the median of the entire dataset, which represents the 50th percentile. This is $Q_2$, the second quartile.
- Divide the data into two halves: the lower half (below $Q_2$) and the upper half (above $Q_2$).
- Calculate the median of the lower half, which is $Q_1$, the first quartile.
- Calculate the median of the upper half, which is $Q_3$, the third quartile.

Here is an example calculation of quartiles using the Tukey method:

***Example 4.6***

Find the quartiles for the dataset $\{2, 5, 1, 8, 4, 7, 3, 6\}$.

***Solution***

- Sort the data in ascending order: $\{1, 2, 3, 4, 5, 6, 7, 8\}$
- Calculate the median of the entire dataset: $Q_2 = 4.5$
- Divide the data into two halves: $\{1, 2, 3, 4\}$ (lower half) and $\{5, 6, 7, 8\}$ (upper half)
- Calculate the median of the lower half: $Q_1 = 2.5$
- Calculate the median of the upper half: $Q_3 = 6.5$
- Therefore, the quartiles for this dataset are $Q_1 = 2.5$, $Q_2 = 4.5$, and $Q_3 = 6.5$.

To calculate deciles, you can follow the following steps:

- Sort the data in ascending order from smallest to largest.
- Count the number of data points, denoted as "$n$".
- The decile formulas are $D(x) = \frac{(n+1)x}{10}$, where $x$ is the value of the decile that needs to be calculated and ranges from 1 to 9 and $n$ is the total number of observations in that data set.

***Example 4.7***

Find the first two deciles for the following dataset,

$$\{24, 32, 27, 32, 23, 62, 45, 80, 59, 63, 36, 54, 57, 36, 72, 55, 51, 32, 56, 33, 42, 55, 30\}.$$

***Solution***

The steps required are as follows:

1. Arrange the data in increasing order. This gives 23, 24, 27, 30, 32, 32, 32, 33, 36, 36, 42, 45, 51, 54, 55, 55, 56, 57, 59, 62, 63, 72, 80.
2. Identify the total number of points. Here, $n = 23$
3. Apply the decile formula to calculate the position of the required data point. $D(1) = (n + 1)/10 = 2.4$. This implies the value of the 2.4th data point has to be determined. This will lie between the scores in the 2nd and 3rd positions. In other words, the 2.4th data is 0.4 of the way between the scores 24 and 27
4. The value of the decile can be determined as
   a. [lower score + (distance)*(higher score - lower score)].
   b. This is given as $24 + 0.4 * (27 - 24) = 25.2$.
5. Apply steps 3 and 4 to determine the rest of the deciles. $D(2) = 2(n + 1)/10 = 4.8$th data between digit number 4 and 5. Thus, $30 + 0.8 * (32 - 30) = 31.6$.

## 4.5 Five-Number Summary and the Box Plot

An outlier may result from transposing digits when recording a measurement, from incorrectly reading an instrument dial, from a broken piece of equipment, or from other problems. Even when there are no recording errors, a data set may contain one or more measurements that, for one reason or another, are very different from the others in the set. These outliers can cause a distortion in commonly used numerical measures such as mean and standard deviation. In fact, outliers may themselves contain important information not shared with the other measurements in the set. Therefore, isolating outliers, if they are present, is an important first step in analyzing a data set. The box plot and five-number summary are designed exactly for this purpose.

The median and the upper and lower quartiles divide the data into four sets, each containing an equal number of measurements. If we add the largest number (Max) and the smallest number (Min) in the data set to this group, we will have a set of numbers that give insight into the spread, central tendency, and outliers in the data.

The five-number summary consists of the following numerical measures:

(Min, $Q_1$, Median, $Q_3$, Max).

- The median ($Q_2$) represents the center of the dataset. It divides the data into two equal halves, with 50% of the observations below and 50% above it. It is robust to outliers.
- The range, calculated as the difference between the maximum and minimum values, gives an idea of the overall spread of the data.
- The interquartile range (IQR), calculated as the difference between the third and first quartiles ($Q_3 - Q_1$), represents the spread of the middle 50% of the data and is resistant to outliers.

The box plot, also known as the box-and-whisker plot, is a graphical representation of the five-number summary. A box plot is a useful tool for identifying outliers, comparing distributions, and identifying trends in data.

- The median is represented by the vertical line in the box plot.
- The first quartile ($Q_1$) is represented by the left end of the box.
- The third quartile ($Q_3$) is represented by the right end of the box.
- The interquartile range (IQR) is the distance between $Q_1$ and $Q_3$. It is a measure of the spread of data.
- The whiskers extend from the box to the minimum and maximum values.

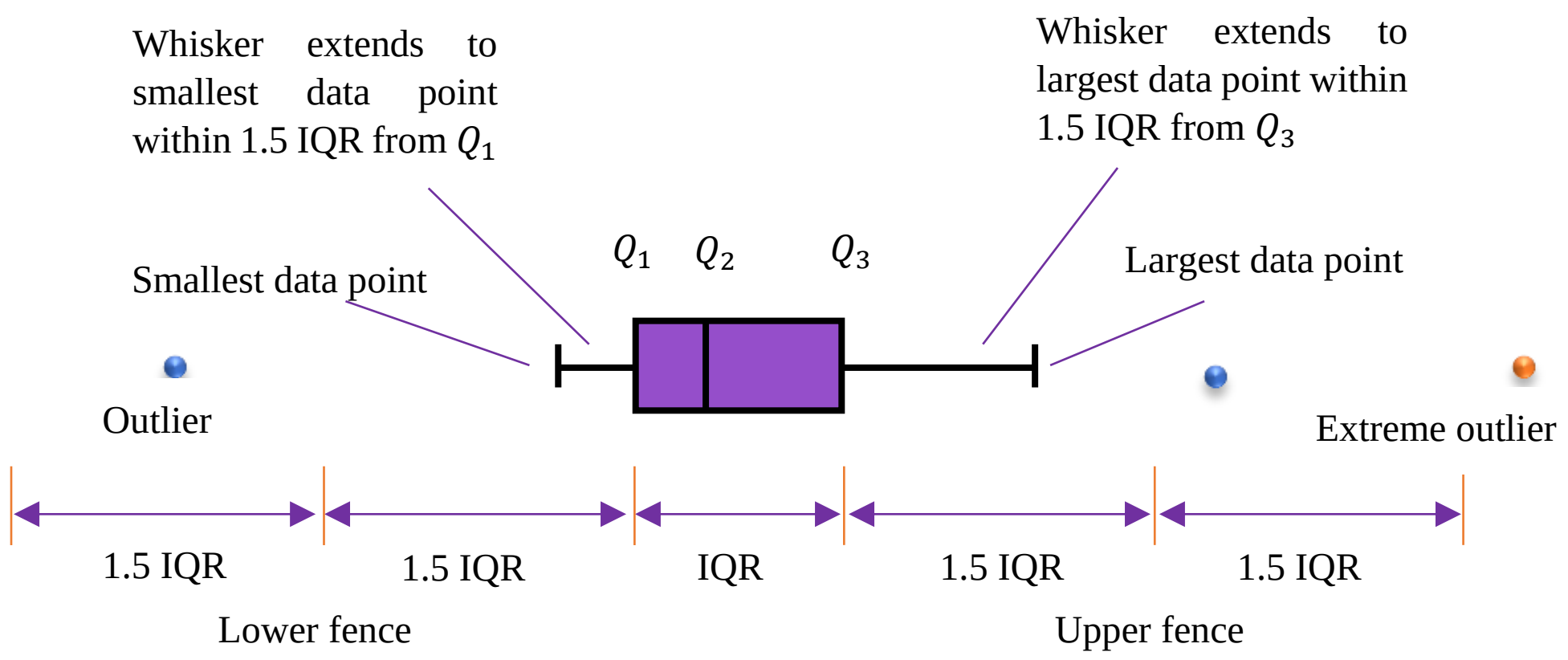

**Figure 4.7.** Description of a box plot.

### Detecting Outliers

Outliers are observations that are beyond the:

- Lower fence: $Q_1 - 1.5$ (IQR)
- Upper fence: $Q_3 + 1.5$ (IQR)

The upper and lower fences are shown in Figure 4.7, but they are not usually drawn on the box plot. Any measurement beyond the upper or lower fence is an outlier; the rest of the measurements, inside the fences, are not unusual. Finally, the box plot marks the range of the data set using "whiskers" to connect the smallest and largest measurements (excluding outliers) to the box.

Here are some examples of how box plots can be used:

- One of the significant advantages of the box plot is its ability to identify potential outliers in the dataset. Outliers are data points that lie significantly above or below the whiskers of the plot. By indicating these extreme values, the box plot highlights the presence of unusual or influential observations.
- Box plots are highly effective for comparing distributions across different groups or categories. By placing multiple box plots side by side, it becomes easier to identify differences in central tendency, spread, and variability between the groups. This comparative aspect is particularly useful in exploratory data analysis and hypothesis testing.
- The box plot can indicate the skewness and symmetry of the data distribution. If the median is roughly centered in the box, and the whiskers are approximately equal in length, the distribution is likely symmetrical. Skewed distributions may have unequal whisker lengths and a median closer to one quartile than the other.
- The box plot enables a quick assessment of the range and variability of the data. The length of the whiskers gives an indication of the spread, while the width of the box represents IQR.
- Box plots can be used to identify trends in data over time. By looking at how the median, IQR, and whiskers change over time, you can get a sense of whether the data is increasing, decreasing, or staying the same.

# CHAPTER 5

# MATHEMATICA LAB: DESCRIPTIVE STATISTICS PART 2

Mathematica offers various functions to compute central tendency measures. These functions provide quick and accurate calculations for determining the typical or central value of a dataset and for visualizing location statistics.

- The `Mean` function in Mathematica calculates the arithmetic mean of a list of numbers. It is a commonly used measure of central tendency and provides the average value of the dataset.
- The `Median` function computes the middle value of a sorted dataset. It is useful for finding a representative value that is not influenced by extreme values or outliers.
- The `Commonest` function determines the mode(s) of a dataset, which representsthe most frequently occurring value(s). This can be useful when dealing with categorical or discrete data.
- Mathematica provides functions like `Quartiles` and `Quantile` to calculate specific quantiles of a dataset. These functions allow you to find values that divide the dataset into equal proportions, such as the first quartile (25th percentile) or the median (50th percentile).
- The `TrimmedMean` function in Mathematica calculates the mean of a dataset after excluding a specified percentage of extreme values from both ends. It can be useful in situations where outliers may significantly affect the overall mean.
- The `WinsorizedMean` function provides a robust measure of central tendency by reducing the impact of outliers or extreme values on the calculated mean. It achieves this by replacing extreme values with values from a specified percentile.
- The `SpatialMedian` function calculates the spatial median, also known as the geometric median, which is a robust measure of central tendency in spatial data analysis. It provides a robust estimate of the "center" of a spatial dataset, minimizing the sum of distances to all other points.
- Both the `HarmonicMean` and `GeometricMean` functions provide alternative measures of central tendency that are suitable for specific types of data. While the `HarmonicMean` is useful for rates and ratios, the `GeometricMean` is applicable for multiplicative relationships and positive values.
- Mathematica offers built-in functions for visualizing location statistics. For example, you can create box plots using `BoxWhiskerChart` to display the median, quartiles, and potential outliers in a dataset.

Therefore, we divided this chapter into two units to cover the following topics, location statistics and visualizing location statistics.

In the following table, we list the built-in functions that are used in this chapter.

| Location Statistics | | Box Whisker Chart |
|---|---|---|
| `Mean` | `SpatialMedian` | `BoxWhiskerChart` |
| `Median` | `HarmonicMean` | |
| `Commonest` | `GeometricMean` | |
| `TrimmedMean` | `RootMeanSquare` | |
| `WinsorizedMean` | `Quartiles` | |
| | `Quantile` | |

**Chapter 5 Outline**

# UNIT 5.1

# LOCATION STATISTICS

The Wolfram Language's descriptive statistics functions operate both on explicit data and on symbolic representations of statistical distributions, making it a valuable tool for data analysis and exploratory data science tasks.

| | |
|---|---|
| `Mean[list]` | gives the statistical mean of the elements in list. |
| `Mean[dist]` | gives the mean of the distribution dist. |
| `Median[list]` | gives the median of the elements in list. |
| `Median[dist]` | gives the median of the distribution dist. |
| `Commonest[list]` | gives a list of the elements that are the most common in list. |
| `Commonest[list,n]` | gives a list of the n most common elements in list. |

***Mathematica Examples 5.1*** Mean

```
Input     (* Mean of numeric values: *)
          Mean[{1,2,3,4,5}]
Output    3

Input     (* Mean of symbolic values: *)
          Mean[{a,b,c,d}]
Output    1/4 (a+b+c+d)

Input     (* Means of elements in each column: *)
          Mean[{{a,x},{b,y},{c,z},{d,w}}]
Output    {1/4 (a+b+c+d),1/4 (w+x+y+z)}

Input     (* Find the mean of WeightedData: *)
          sampledata={8,3,5,4,9,0,4,2,2,3};
          weights={0.15,0.09,0.12,0.10,0.16,0.,0.11,0.08,0.08,0.09};
          Mean[
           WeightedData[sampledata,weights]
           ]
Output    5.04082

Input     (* Find the mean for univariate distributions: *)
          Mean[
           BinomialDistribution[n,p]
           ]
Output    n p

Input     (* Mean is Total divided by Length: *)
          Mean[{x,y,z,w}]
          Total[{x,y,z,w}]/Length[{x,y,z,w}]
Output    1/4 (w+x+y+z)
Output    1/4 (w+x+y+z)

Input     (* The code analyzes a dataset of heights by calculating the mean and generating two
          plots. The first plot displays the data set, while the second plot displays the data
          set along with horizontal line for the mean: *)

          h={133,136,149,133,123,121,140,139,117,117,136,108,126,104,116,147,140,148,150,122,
          135,146,133,144,117,124,135,117,120,121,110,124,103,137,101,119,104,113,139,133};
```

```
m=N[Mean[h]]
n=Length[h];

ListPlot[
 h,
 Filling->Axis,
 PlotStyle->Purple,
 ImageSize->170
 ]

ListPlot[
 {h,{{0,m},{n,m}}},
 Joined->{False,True},
 Filling->{1->m,2->Axis},
 PlotStyle->Purple,
 ImageSize->170,
 PlotLegends->{None,"Mean =127"}
 ]
```

Output 127.

Output

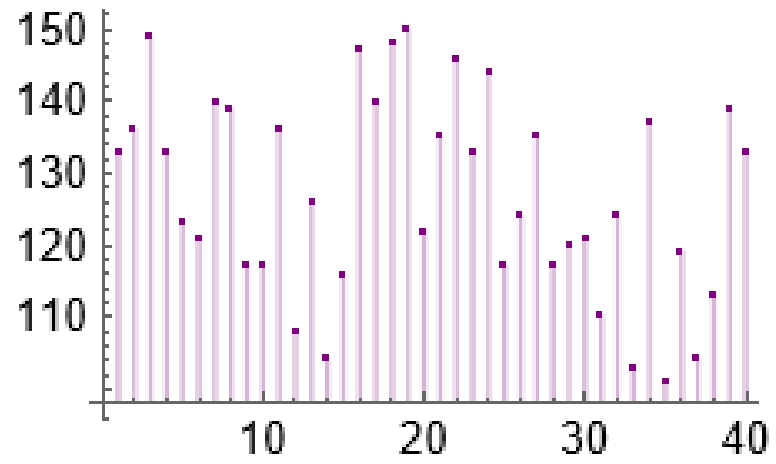

Output

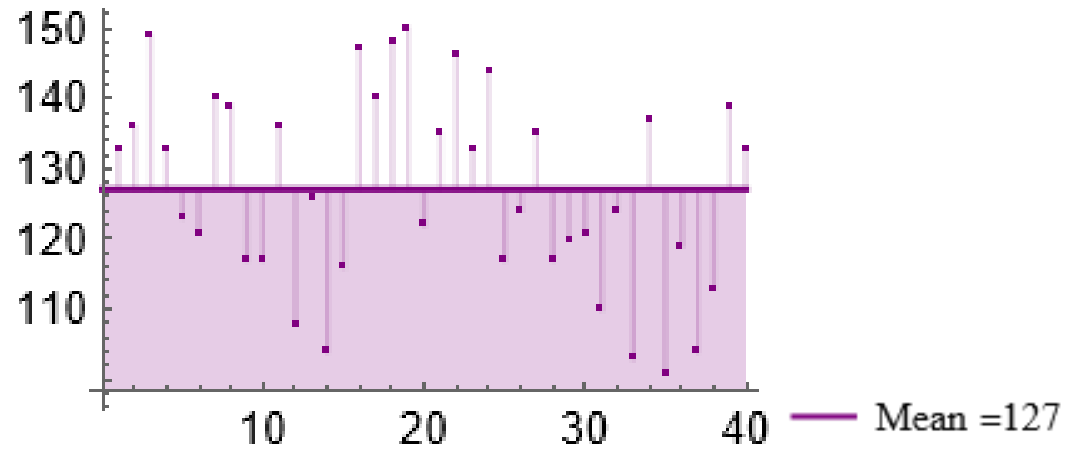

Input

```
(* In this code, we generate a random sample of size n from a standard normal
distribution. We then calculate the deviations from the mean for each data point by
subtracting the sample mean from each data point, and store them in the deviations
variable. Finally, we calculate the mean of the deviations. We use the Histogram
function to create a histogram of the deviations from the mean. The plot shows that
the deviations from the mean are centered around 0, and that the distribution is
relatively symmetric and unimodal. This explains that the mean of deviations from
the Mean is zero: *)

n=1000;   (* sample size: *)

(* Define a distribution: *)
dist=NormalDistribution[0,1];

(* Generate a random sample from the distribution: *)
sample=RandomVariate[dist,n];

(* Calculate the deviations from the mean: *)
deviations=sample-Mean[sample];
sum=Total[deviations]

(* Calculate the mean of the deviations: *)
meanDeviations=Mean[deviations]
```

```
(* Create a histogram of the deviations from the mean: *)
Histogram[
 deviations,
 60,
 "PDF",
 PlotLabel->StringForm["Distribution of Deviations from the Mean\nMean of Deviations = ``",meanDeviations],
 ColorFunction->Function[Opacity[0.7]],
 ChartStyle->Purple,
 ImageSize->300
 ]

ListPlot[
 deviations,
 PlotLabel->StringForm["Deviations from the Mean\nMean of Deviations = ``",meanDeviations],
 ColorFunction->Function[Opacity[0.7]],
 Filling->Axis,
 PlotStyle->Purple,
 ImageSize->300
 ]
```

Output $-1.95399*10^{-14}$

Output $-1.95399*10^{-17}$

Output

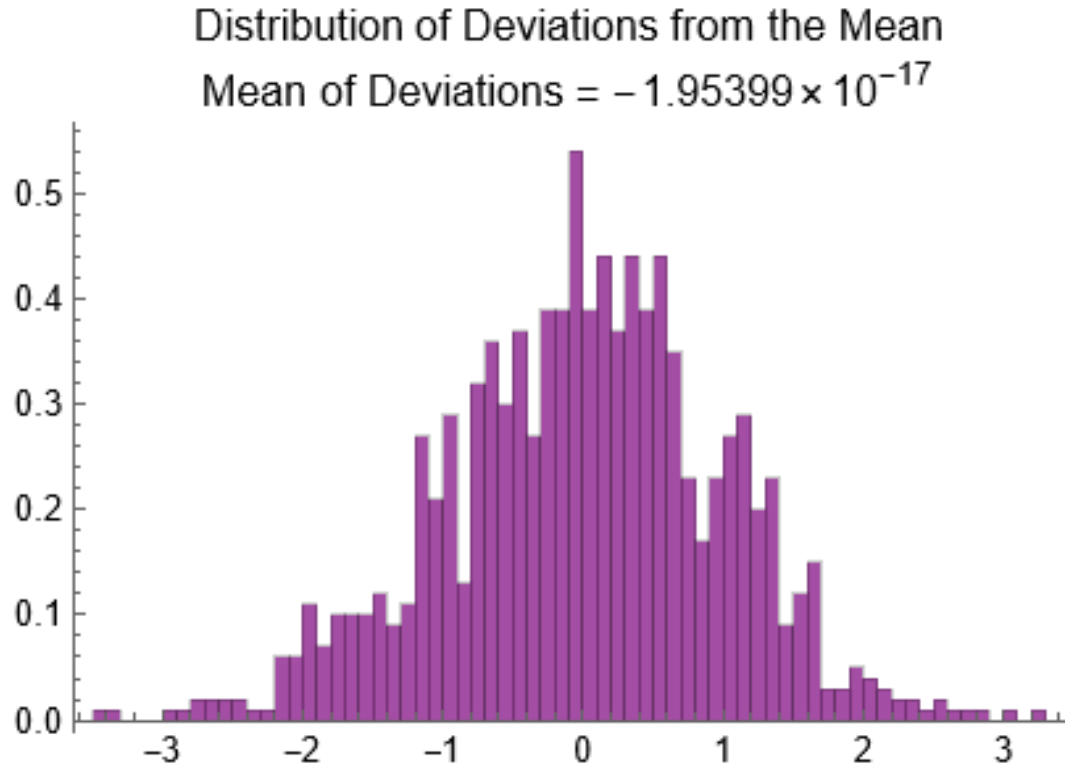

Output

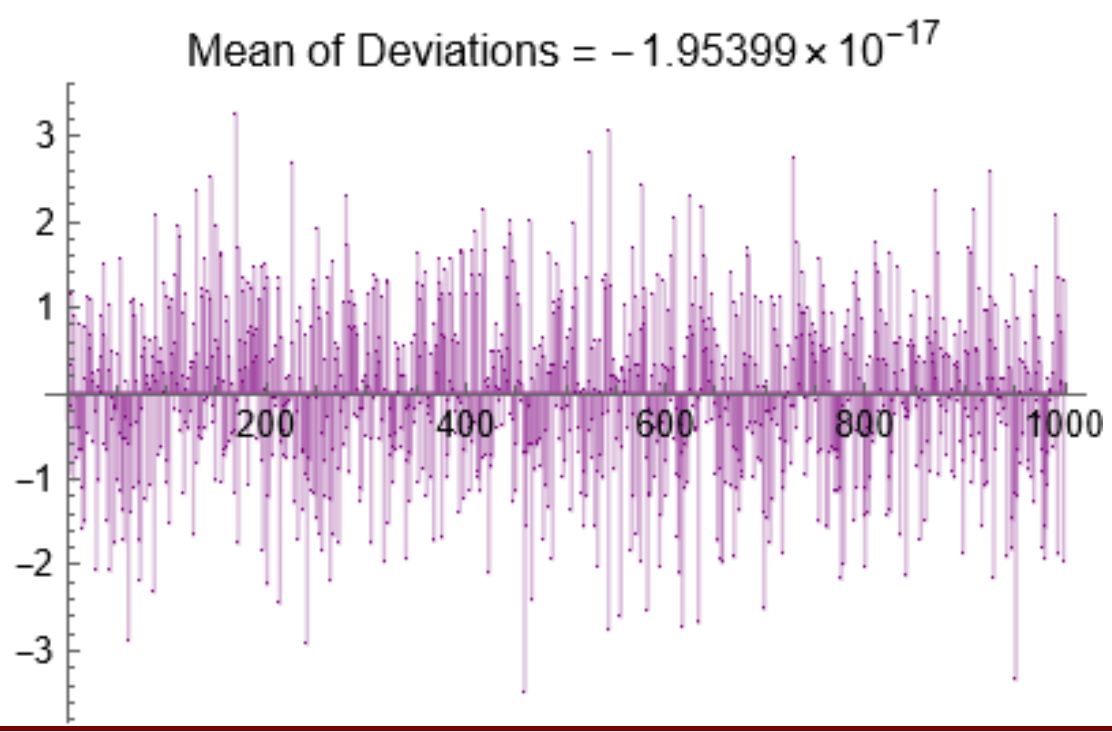

***Mathematica Examples 5.2*** Median

Input
```
(* Find the middle value in the list: *)
Median[{1,2,3,4,5,6,7}]
```
Output 4

Input
```
(* Average the two middle values: *)
Median[{1,2,3,4,5,6,7,8}]
```
Output 9/2

```
Input    (* Median of a parametric distribution: *)
         Median[
          ExponentialDistribution[λ]
          ]
Output   Log[2]/λ

Input    (* Median for a matrix gives column-wise medians: *)
         Median[{{1,2},{3,4},{5,6}}]
Output   {3,4}

Input    (* Median for Weighted Data: *)
         data={7,8,2,4,5,1,3,2,2,1};
         w={0.15,0.09,0.12,0.10,0.16,0.,0.11,0.08,0.08,0.09};
         Median[
          WeightedData[data,w]
          ]
Output   4

Input    (* The code reads in a dataset h of heights and calculates its median value. It then
         produces three plots: a scatter plot of the dataset h, a scatter plot of the sorted
         dataset s, and a plot that shows both h and a horizontal line at the median value.
         These plots are used for visualizing the distribution of the dataset and highlighting
         the location of the median value: *)

         h={133,136,149,133,123,121,140,139,117,117,136,108,126,104,116,147,140,148,150,122,
         135,146,133,144,117,124,135,117,120,121,110,124,103,137,101,119,104,113,139,133};

         m=N[Median[h]]
         n=Length[h];
         s=Sort[h];

         ListPlot[
          h,
          Filling->Axis,
          PlotStyle->Purple,
          ImageSize->170
          ]

         ListPlot[
          s,
          Filling->Axis,
          PlotStyle->Purple,
          Epilog->{Red,Line[{{20,0},{20,124}}],Blue,Line[{{21,0},{21,126}}]},
          ImageSize->170
          ]

         ListPlot[
          {h,{{0,m},{n,m}}},
          Joined->{False,True},
          Filling->{1->m,2->Axis},
          PlotStyle->Purple,
          ImageSize->170,
          PlotLegends->{None,"Median =125"}
          ]
Output   125.
```

Output 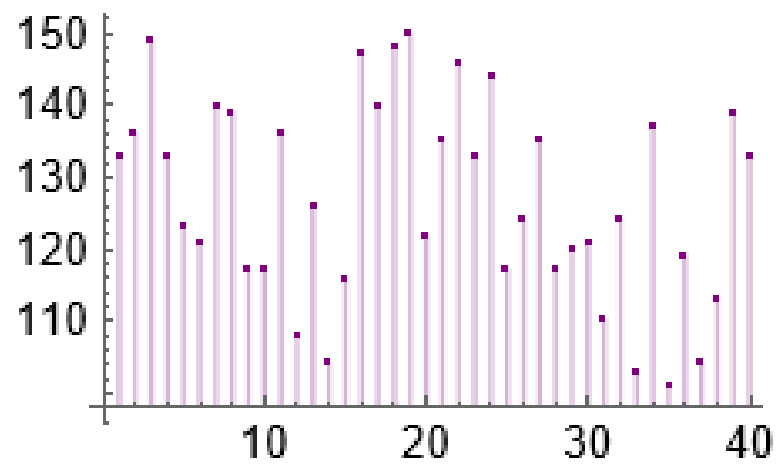

Output 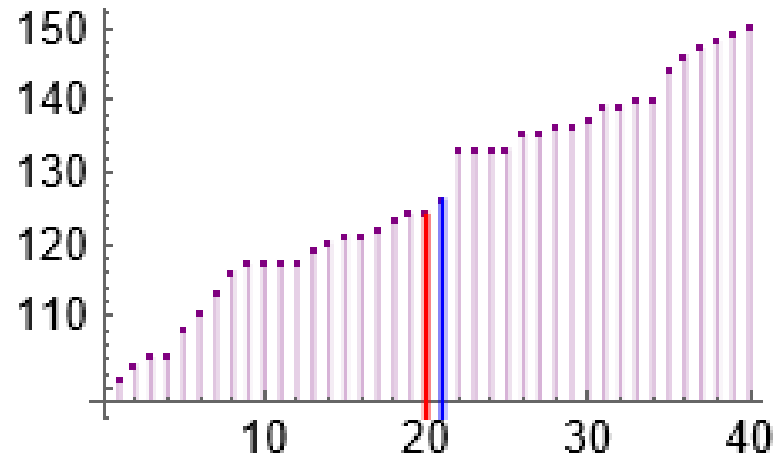

Output 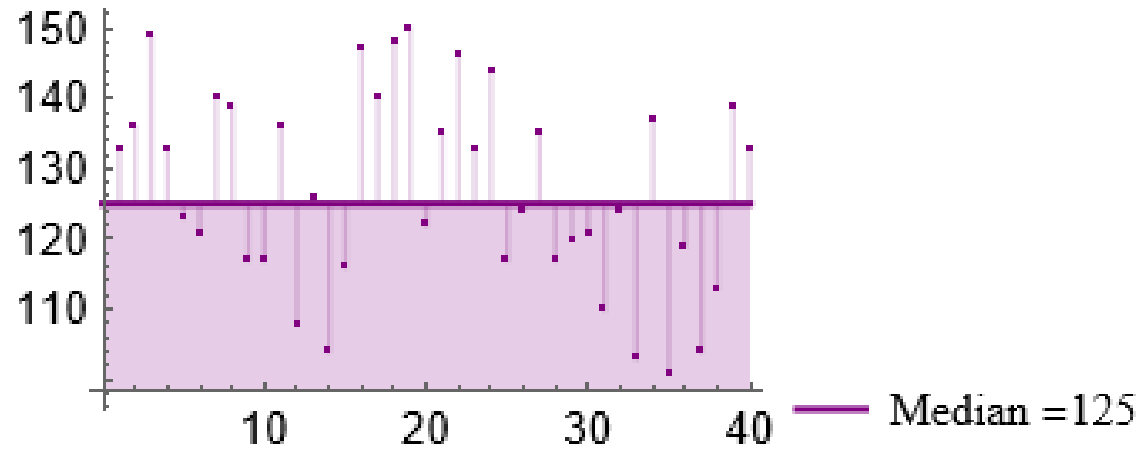

Input

```
(* The code is used to compute the median deviation of a given dataset using two different approaches. The first approach uses the built-in function MedianDeviation to directly compute the median absolute deviation of the dataset, while the second approach involves manually computing the median of the absolute deviations from the median of the dataset using the Median function: *)

data=RandomReal[10,5];
MedianDeviation[data]
Median[
 Abs[
  data-Median[data]
  ]
 ]
```

Output 0.363813

Output 0.363813

Input

```
(* The code explain that for nearly symmetric samples, Mean and Median are nearly the same. The code is designed to generate n random samples from a nearly symmetric distribution, calculate the differences between the mean and median for each sample, calculate the mean of the differences, and finally create a histogram and ListPlot of the differences with a label displaying the mean difference. The plot shows that the differences are centered around 0, and that the distribution is relatively symmetric and unimodal. This explain that for nearly symmetric samples, Mean and Median are nearly the same: *)

n=1000;   (* number of samples: *)
m=1000;   (* sample size: *)

(* Define a nearly symmetric distribution: *)
dist=NormalDistribution[0,1];

(* Generate n random samples from the distribution: *)
samples=RandomVariate[dist,{n,m}];

(* Calculate the differences between the mean and median for each sample: *)
```

```
diffs=Map[Mean,samples]-Map[Median,samples];

(* Calculate the mean of the differences: *)
meanDiff=Mean[diffs];

(* Create a histogram of the differences: *)
Histogram[
 diffs,
 60,
 "PDF",
 PlotLabel->StringForm["Distribution of Differences (Mean - Median)\nMean = ``
",meanDiff],
 ColorFunction->Function[Opacity[0.7]],
 ChartStyle->Purple,
 ImageSize->250
 ]
ListPlot[
 diffs,
 PlotLabel->StringForm["Differences (Mean - Median)"],
 ColorFunction->Function[Opacity[0.7]],
 Filling->Axis,
 PlotStyle->Purple,
 ImageSize->250
 ]
```

Output

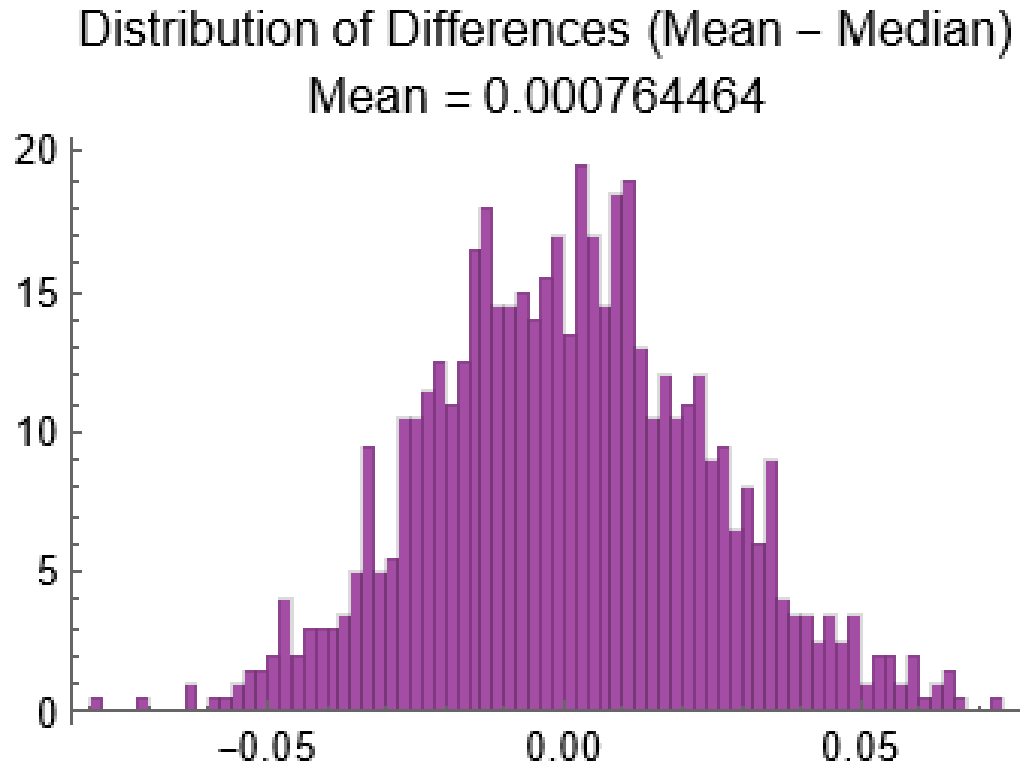

Output

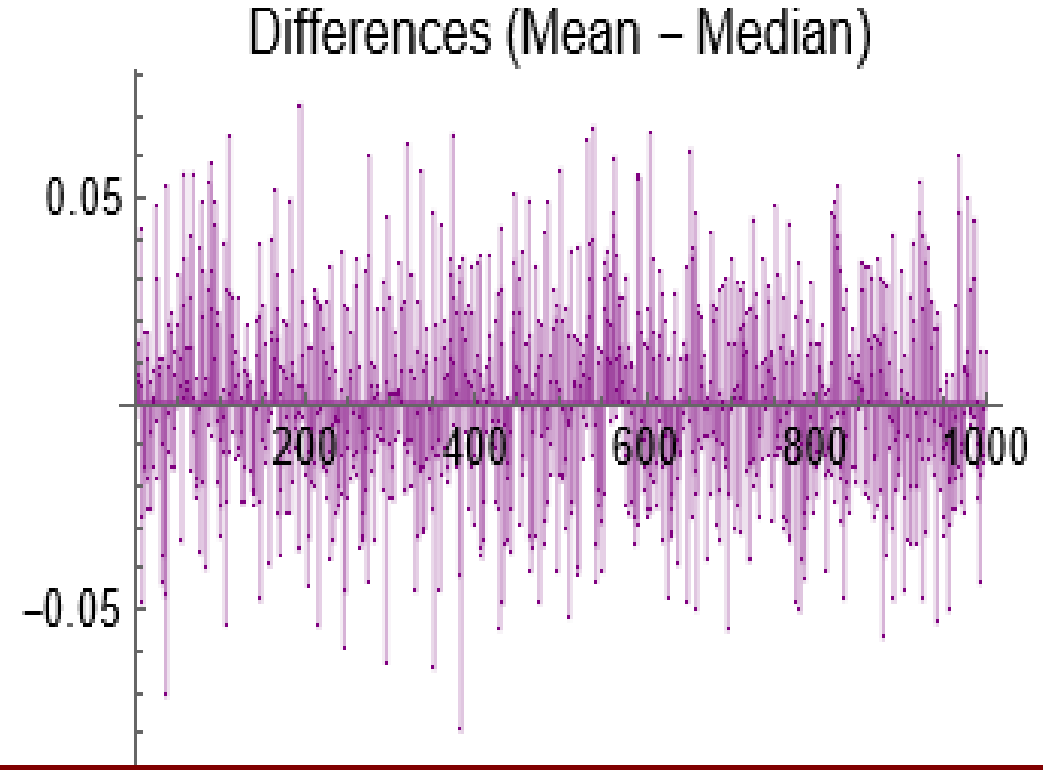

***Mathematica Examples 5.3*** Commonest

```
Input   (* Obtain the elements with the highest frequency: *)
        Commonest[{b,a,c,2,a,b,1,2,1,2,a}]
Output  {a,2}

Input   (* Obtain the 4 most common elements: *)
        Commonest[{b,a,c,a,a,b,1,1,1,2},4]
Output  {b,a,c,1}
```

```
Input   (* The code reads in a dataset h of heights and calculates Commonest elements. It
        then produces two plots: a scatter plot of the dataset h, and a plot that shows both
        h and three horizontal lines at the commonest elements and mean. These plots are used
        for visualizing the distribution of the dataset and highlighting the location of the
        commonest elements and mean: *)

        h={133,136,149,133,123,121,140,139,117,117,136,108,126,104,116,147,140,148,150,122,
        135,146,133,144,117,124,135,117,120,121,110,124,103,137,101,119,104,113,139,133};

        Tally[h]
        c=N[Commonest[h]]
        m=Mean[h]
        n=Length[h];

        ListPlot[
         h,
         Filling->Axis,
         PlotStyle->Purple,
         ImageSize->170
         ]

        ListPlot[
         {h,{{0,c[[1]]},{n,c[[1]]}},{{0,c[[2]]},{n,c[[2]]}},{{0,m},{n,m}}},
         Joined->{False,True,True,True},
         Filling->{1->m,2->{3}},
         PlotStyle->{Purple,Purple,Purple,Red},
         ImageSize->170,
         PlotLegends->{"h","Commonest=133","Commonest=117","Mean =127"}
         ]
Output  {{133,4},{136,2},{149,1},{123,1},{121,2},{140,2},{139,2},{117,4},{108,1},{126,1},{1
        04,2},{116,1},{147,1},{148,1},{150,1},{122,1},{135,2},{146,1},{144,1},{124,2},{120,
        1},{110,1},{103,1},{137,1},{101,1},{119,1},{113,1}}
Output  {133.,117.}
Output  127
Output
```

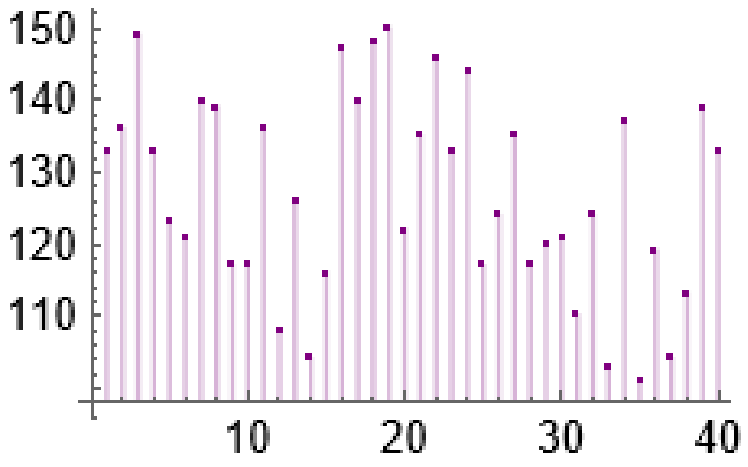

Output

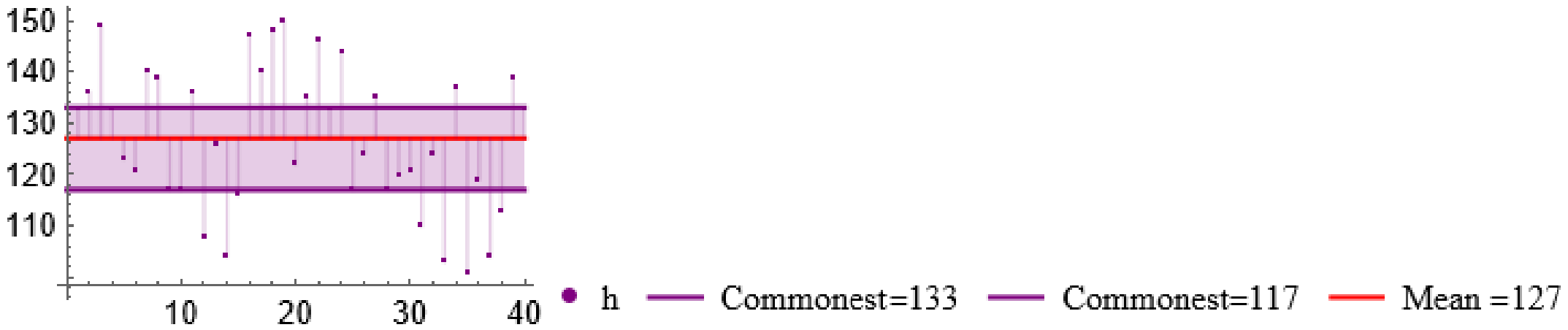

| | |
|---|---|
| `TrimmedMean[list,f]` | gives the mean of the elements in list after dropping a fraction f of the smallest and largest elements. |
| `TrimmedMean[list,{f1,f2}]` | gives the mean when a fraction f1 of the smallest elements and a fraction f2 of the largest elements are removed. |
| `TrimmedMean[list]` | gives the 5% trimmed mean TrimmedMean[list,0.05]. |
| `TrimmedMean[dist,…]` | gives the trimmed mean of a univariate distribution dist. |

***Mathematica Examples 5.4*** TrimmedMean

```
Input    (* A 0% TrimmedMean is equivalent to Mean: *)
         TrimmedMean[Range[10],0]
         Mean[Range[10]]
Output   11/2
Output   11/2

Input    (* The code demonstrates three different ways to calculate the trimmed mean by
         removing extreme values from either or both ends of the data set. *)
         (* Trimmed mean after removing extreme values for two sides: *)
         TrimmedMean[{-10,1,1,1,1,20},0.2]

         (* Trimmed mean after removing the smallest extreme values: *)
         TrimmedMean[{-10,1,1,1,1,20},{0.2,0}]

         (* Trimmed mean after removing the largest extreme values: *)
         TrimmedMean[{-10,1,1,1,1,20},{0,0.2}]
Output   1
Output   24/5
Output   -(6/5)

Input    (* Trimmed mean of a symbolic distribution: *)
         TrimmedMean[
          ExponentialDistribution[λ]
          ]
Output   (18-19 Log[19]+18 Log[20])/(18 λ)

Input    (* The code demonstrates the importance of using a robust estimate of location, such
         as the trimmed mean, when outliers are present in a data set. The code also highlights
         the potential problems with using the arithmetic mean in the presence of outliers,
         as it can be heavily influenced by extreme values and lead to inaccurate or misleading
         results: *)

         (* Obtain a robust estimate of location when outliers are present: *)
         N[
          TrimmedMean[{1,5,2,6,10,10^6,5,4,-2000,5},.2]
          ]

         (* Extreme values have a large influence on the Mean: *)
         N[
          Mean[{1,5,2,6,10,10^6,5,4,-2000,5}]
          ]
Output   4.5
Output   99803.8

Input    (* The code is calculating the trimmed mean of a data set using two different methods.
         The first method involves manually trimming a percentage of the extreme values from
         the data set using built-in Mathematica functions, while the second method uses the
         built-in "TrimmedMean" function. The manual trimming approach may be useful in cases
         where more control over the trimming process is required. The code first defines a
         data set, "data", containing a series of integer values. It then sets the trimming
         percentage, "p", to 0.2, indicating that 20% of the extreme values from each end of
         the data set should be removed. The code then calculates the length of the data set,
         "n", and the number of values to trim,"numToTrim", using the rounding function
         "Round". The data set is then sorted in ascending order using the "Sort" function,
         and the extreme values are trimmed using the "Take" function to create a new data
         set, "trimmedData". Finally, the trimmed mean is calculated using the "Mean" function
         on the "trimmedData" data set: *)

         data={12,15,19,20,22,23,25,26,27,28,31,35,39,42,45,46,47,50,55,62};
```

```
        p=0.2; (* percentage of data to trim from each ends: *)
        n=Length[data];
        numToTrim=Round[p*n];
        sorteddata=Sort[data]
        trimmedData=Take[sorteddata,{numToTrim+1,n-numToTrim}]
        trimmedMean=Mean[trimmedData]
        TrimmedMean[data,0.2]
Output  {12,15,19,20,22,23,25,26,27,28,31,35,39,42,45,46,47,50,55,62}
Output  {22,23,25,26,27,28,31,35,39,42,45,46}
Output  389/12
Output  389/12

Input   (* In this code, the trimmedMean function takes two arguments: data, which is a list
        of numeric data, and p, which is an integer specifying the percentage of trimming to
        apply. The function first sorts the data and then takes the middle (100-p) percent
        of the sorted data to compute the mean. We generate a sample of 100 normally
        distributed random numbers and compute the trimmed mean for 10%, and 20% trimming:
        *)

        (* Define a function to compute the trimmed mean: *)
        trimmedMean[data_,p_]:=Module[
          {n=Length[data]},
          Mean[
           Take[
            Sort[data],
            {Ceiling[n*p/200]+1,Floor[n*(100-(p/2))/100]}
            ]
           ]
          ]
        (* Generate some sample data: *)
        data=RandomVariate[NormalDistribution[0,1],100];

        (* Compute the trimmed mean for various percentages of trimming: *)
        trimmedMean[data,10] (* 10% trimming: *)
        trimmedMean[data,20] (* 20% trimming: *)

        TrimmedMean[data,0.05]
        TrimmedMean[data,0.1]
Output  -0.0977251
Output  -0.0902409
Output  -0.0977251
Output  -0.0902409

Input   (* The code analyzes a dataset of heights by calculating the mean and trimmed mean
        and generating two plots. The first plot displays the data set, while the second plot
        displays the data set along with horizontal lines for the mean and trimmed mean. This
        allows for a clear visual comparison of the two measures and demonstrates how the
        trimmed mean is less sensitive to outliers than the mean: *)

        h={133,136,149,133,123,121,140,139,117,117,136,108,126,104,116,147,140,148,150,122,
        135,146,133,144,117,124,135,117,120,121,110,124,103,137,101,119,104,113,139,133,600
        };

        n=Length[h];
        m=N[Mean[h]]
        tmean=N[TrimmedMean[h,0.49]]

        ListPlot[
         h,
         Filling->Axis,
         PlotStyle->Purple,
```

```
 ImageSize->170
 ]

ListPlot[
 {h,{{0,tmean},{n,tmean}},{{0,m},{n,m}}},
 Joined->{False,True,True},
 Filling->{1->tmean,2->Axis},
 PlotStyle->{Purple,Purple,Red},
 ImageSize->170,
 PlotLegends->{"h","Trimmed Mean .49","Mean"}
 ]
```

Output 138.537

Output 126.

Output

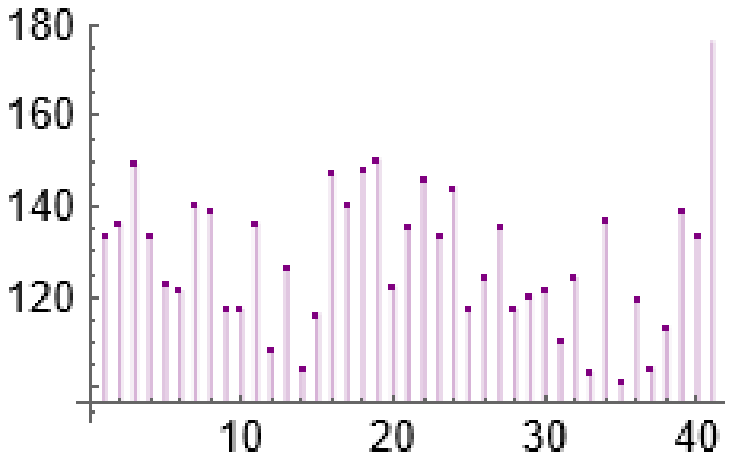

Output

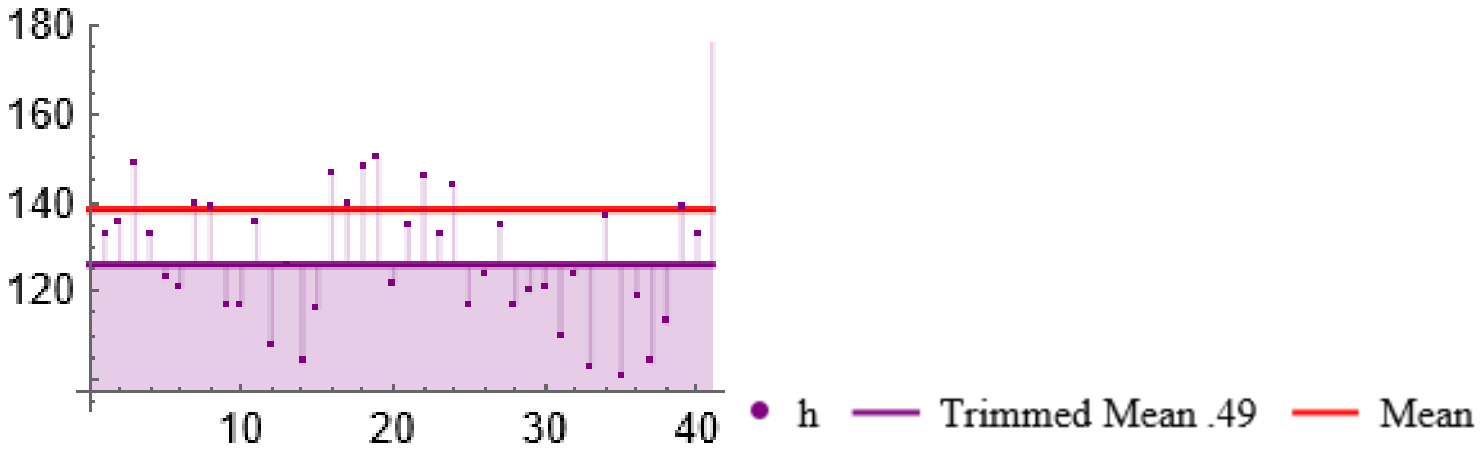

Input

```
(* This code creates a Manipulate function that allows the user to explore the effects of changing the parameters of a gamma distribution (α and β) on the mean and trimmed mean of a sample of random variates drawn from the distribution. The Manipulate function creates two plots: a histogram of the data and a list plot of the data points. The mean and trimmed mean of the data are plotted as red and blue lines, respectively, in both plots. This code is a tool for exploring the differences between the mean and trimmed mean of a sample of random variates from a gamma distribution. The Manipulate function allows the user to easily adjust the parameters of the gamma distribution and the sample size and trimming percentage, making it a flexible tool for exploring the effects of different factors on the mean and trimmed mean: *)

Manipulate[
 data=RandomVariate[GammaDistribution[α,β],n];
 tm=TrimmedMean[data,p];
 m=Mean[data];
 Row[
  Show[
   Histogram[
    data,
    {0.3},
    "PDF",
    PlotRange->{{0,20},{0,0.3}},
    PlotLabel->"Histogram of Data",
    ImageSize->250,
    ColorFunction->Function[Opacity[0.5]],
    ChartStyle->Purple
    ],
   Graphics[
    {Red,Thick,Line[{{tm,0},{tm,1}}],Blue,Thick,Line[{{m,0},{m,1}}]}
    ]
```

```
    ],
   Show[
    ListPlot[
     data,
     PlotStyle->Directive[PointSize[0.02],Opacity[0.5],Purple],
     PlotLabel->"List Plot of Data",
     ImageSize->250
     ],
    Graphics[
     {Red,Thick,Line[{{1,tm},{n,tm}}],Blue,Thick,Line[{{1,m},{n,m}}]}
     ]
    ]
   ],
  {{α,2},2,5,Appearance->"Labeled"},
  {{β,2},1,3,Appearance->"Labeled"},
  {{n,500},100,1000,10,Appearance->"Labeled"},
  {{p,0.1},0,0.49,Appearance->"Labeled"}
  ]
```

Output

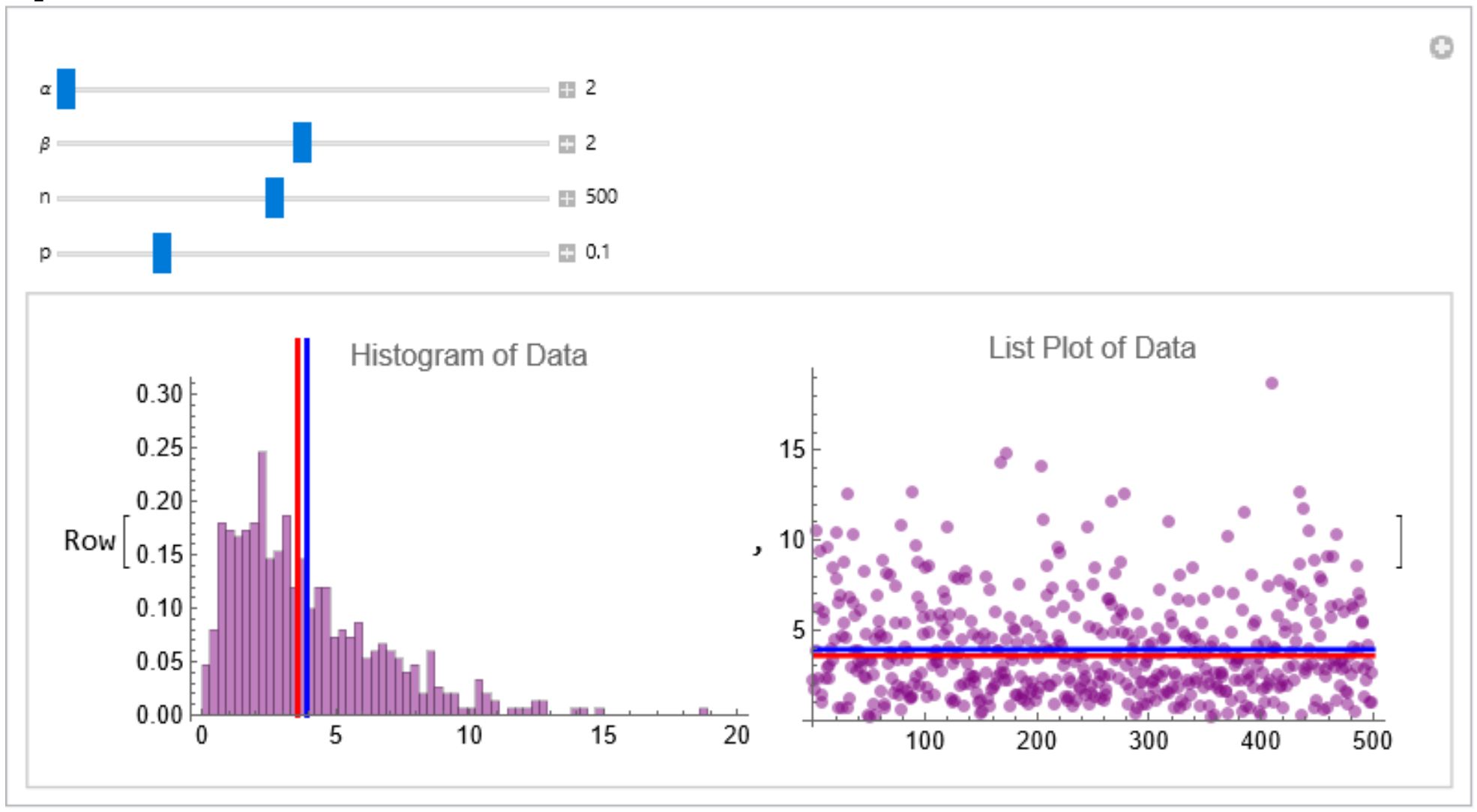

Input

```
(* Suppose we have a square with vertices at (xmin, ymin),(xmin, ymax),(xmax, ymin),and (xmax, ymax). We can generate a random sample of points within this square using the following code. The variable points now contain a list of 1000 points that lie within the square. The trimmed mean of the x and y coordinates of these points is then used to find the center of the square, which is plotted as a blue point in the final plot: *)

(* Generate random sample of x and y coordinates: *)
n=1000; (* Number of points in the sample: *)
xmin=0;xmax=5; (* Range of x values: *)
ymin=0;ymax=10; (* Range of y values: *)

points=Table[
    {RandomReal[{xmin,xmax}],RandomReal[{ymin,ymax}]},
    {i,1,n}
    ];

(* Find the trimmed mean of the x and y coordinates: *)
alpha=0.2; (* Trimming parameter: *)
xTrimmed=TrimmedMean[points[[All,1]],alpha];
yTrimmed=TrimmedMean[points[[All,2]],alpha];
center={xTrimmed,yTrimmed};
```

```
(* Plot the points and the center: *)
ListPlot[
 points,
 AspectRatio->1,
 Frame->True,
 Axes->False,
 PlotStyle->Directive[Purple,PointSize[0.015],Opacity[0.5]],
 PlotRange->{{xmin,xmax},{ymin,ymax}},
 Epilog->{Blue,Opacity[0.9],PointSize[0.04],Point[center]},
 ImageSize->200
 ]
```

Output

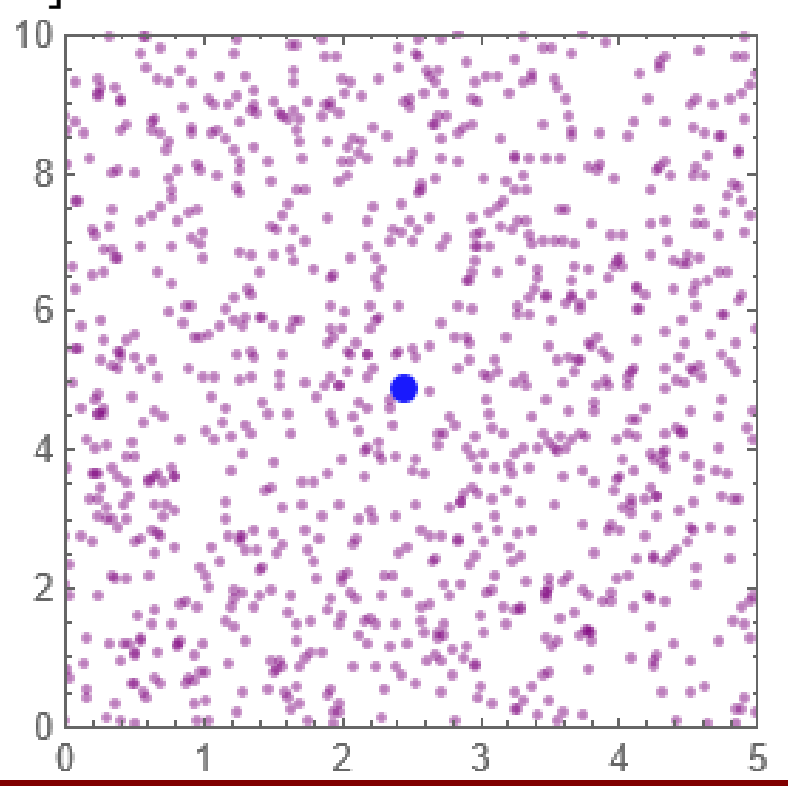

| | |
|---|---|
| `WinsorizedMean[list,f]` | gives the mean of the elements in list after replacing the fraction f of the smallest and largest elements by the remaining extreme values. |
| `WinsorizedMean[list,{f1,f2}]` | gives the mean when the fraction f1 of the smallest elements and the fraction f2 of the largest elements are replaced by the remaining extreme values. |
| `WinsorizedMean[list]` | gives the 5% winsorized mean WinsorizedMean[list,0.05]. |
| `WinsorizedMean[dist,…]` | gives the winsorized mean of a univariate distribution dist. |

***Mathematica Examples 5.5*** WinsorizedMean

Input
```
(* The code demonstrates three different ways to calculate the Winsorized mean by removing extreme values from either or both ends of the data set: *)

(* Winsorized mean after removing extreme values: *)
WinsorizedMean[{-10,1,1,1,1,20},0.2]

(* Winsorized mean after removing the smallest extreme values: *)
WinsorizedMean[{-10,1,1,1,1,20},{0.2,0}]

(* Winsorized mean after removing the largest smallest extreme values: *)
WinsorizedMean[{-10,1,1,1,1,20},{0,0.2}]
```
Output
```
1
```
Output
```
25/6
```
Output
```
-(5/6)
```

Input
```
(* Winsorized mean of a matrix gives column-wise means: *)
WinsorizedMean[
 RandomReal[1,{50,2}],
 .1
 ]
```
Output
```
{0.479019,0.474528}
```

Input
```
(* The code demonstrates the importance of using a robust estimate of location, such as the Winsorized mean, when outliers are present in a data set. The code also
```

```
         highlights the potential problems with using the arithmetic mean in the presence of
         outliers, as it can be heavily influenced by extreme values and lead to inaccurate
         or misleading results: *)

         (* Obtain a robust estimate of the location when outliers are present: *)
         N[
          WinsorizedMean[
           {1,5,2,6,10,10^6,5,4,-2000,5},
           .1
           ]
          ]
         (* Extreme values have a large influence on the Mean: *)
         N[
          Mean[
           {1,5,2,6,10,10^6,5,4,-2000,5}
           ]
          ]
Output   4.9
Output   99803.8

Input    (* The code is calculating the winsorized mean of a data set using two different
         methods. The first method involves manually trimming a percentage of the extreme
         values from the data set using built-in Mathematica functions, while the second
         method uses the built-in "WinsorizedMean" function. The manual trimming approach may
         be useful in cases where more control over the trimming process is required. In this
         code, we first define the dataset using a list called data. Then, we define the
         trimming percentage using a variable called trim. The number of values to be trimmed
         from both ends of the distribution is calculated using the Ceiling function, and
         stored in a variable called n. We then sort the data using the Sort function and
         replace the outliers with the next largest or smallest non-outlying value using a
         combination of Table and Take functions. The resulting modified dataset is stored in
         a variable called winsorizedData. Finally, we calculate the arithmetic mean of the
         modified dataset using the Mean function, and store the result in a variable called
         winsorizedMean: *)

         (* Define the dataset: *)
         data={2,5,8,10,11,13,14,16,18,20};

         (* Define the trimming percentage: *)
         trim=0.2;

         (* Calculate the number of values to be trimmed from both ends of the distribution:
         *)
         n=Ceiling[trim*Length[data]];

         (* Sort the data: *)
         sortedData=Sort[data]

         (* Replace the outliers with the next largest or smallest non-outlying value: *)
         winsorizedData=Join[
           Table[sortedData[[n+1]],{n}],
           Take[sortedData,{n+1,-n-1}],
           Table[sortedData[[-n-1]],{n}]
           ]

         (* Calculate the Winsorized Mean: *)
         winsorizedMean=Mean[winsorizedData]
         WinsorizedMean[data,0.2]
Output   {2,5,8,10,11,13,14,16,18,20}
Output   {8,8,8,10,11,13,14,16,16,16}
```

```
Output   12
Output   12

Input    (* In this code, the winsorizedmean function takes two arguments: data, which is a
         list of numeric data, and p, which is an integer specifying the percentage of trimming
         to apply. The function first sorts the data and then replaces the lowest and highest
         m values with the corresponding minimum and maximum values. Finally, it computes the
         mean of the modified data to get the Winsorized Mean: *)

         winsorizedmean[data_,p_]:=Module[
           {sortedData,n,m},
           sortedData=Sort[data];
           n=Length[data];
           m=Floor[p*n];
           Mean[
            Join[
             ConstantArray[sortedData[[m+1]],m],
             Drop[Drop[sortedData,m],-m],
             ConstantArray[sortedData[[-m-1]],m]
             ]
            ]
           ]

         (* Define the dataset: *)
         data={2,5,8,10,11,13,14,16,18,20};
         winsorizedmean[data,0.2]
         WinsorizedMean[data,0.2]
Output   12
Output   12

Input    (* The code analyzes a dataset of heights by calculating the mean and Winsorized mean
         and generating two plots. The first plot displays the data set, while the second plot
         displays the data set along with horizontal lines for the mean and Winsorized mean.
         This allows for a clear visual comparison of the two measures and demonstrates how
         the Winsorized mean is less sensitive to outliers than the mean: *)

         h={133,136,149,133,123,121,140,139,117,117,136,108,126,104,116,147,140,148,150,122,
         135,146,133,144,117,124,135,117,120,121,110,124,103,137,101,119,104,113,139,133,600
         };

         n=Length[h];
         m=N[Mean[h]]
         wmean=N[WinsorizedMean[h,0.49]]

         ListPlot[
          h,
          Filling->Axis,
          PlotStyle->Purple,
          ImageSize->170
          ]

         ListPlot[
          {h,{{0,wmean},{n,wmean}},{{0,m},{n,m}}},
          Joined->{False,True,True},
          Filling->{1->wmean,2->Axis},
          PlotStyle->{Purple,Purple,Red},
          ImageSize->170,
          PlotLegends->{"h","Winsorized Mean .49","mean"}
          ]
Output   138.537
Output   126.
```

Output
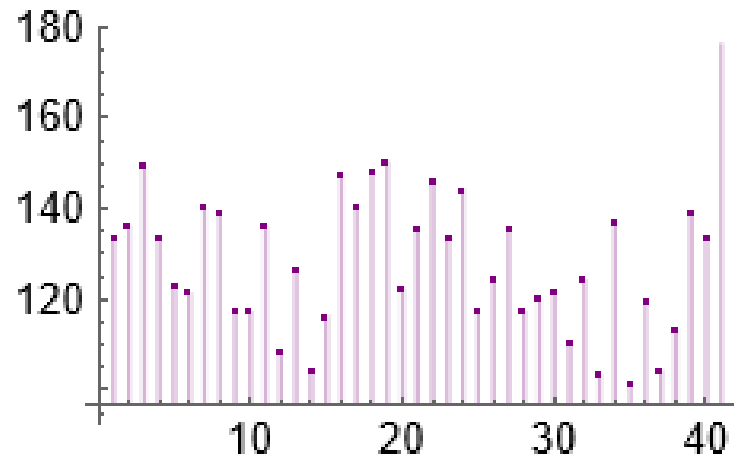

Output
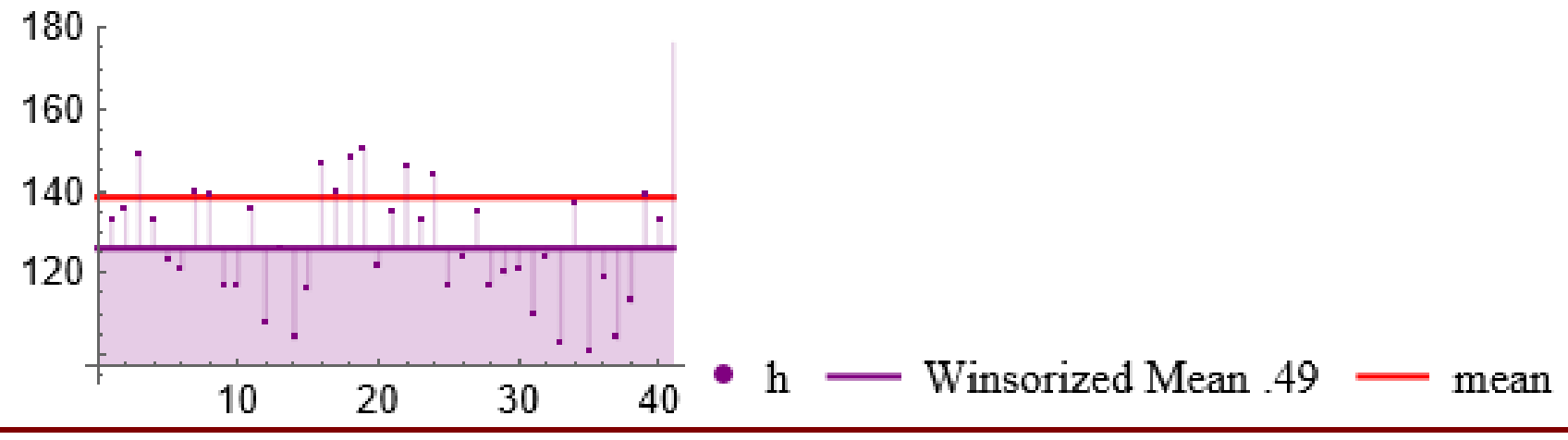

| | |
|---|---|
| SpatialMedian[{x1,x2,…}] | gives the spatial median of the elements xi. |
| SpatialMedian[data] | gives the spatial median for several different forms of data. |

***Mathematica Examples 5.6*** SpatialMedian

```
Input   (* Find the spatial median of a list of vectors: *)
        SpatialMedian[{{1.,3.,5.},{7.,1.,2.},{9.,3.,1.},{4.,5.,6.}}]
Output  {5.65837,2.74486,3.251}

Input   (* Spatial median works with WeightedData: *)
        wd=WeightedData[{{1.,3.},{-4.,2.},{3.,1.},{5.,6.}},{1.,3.,4.,5.}];
        SpatialMedian[wd]
Output  {2.56322,2.7113}

Input   (* For univariate data, Median coincides with SpatialMedian: *)
        Median[{1,2,3,4,4,3,1,1,5,6}]
        SpatialMedian[{1,2,3,4,4,3,1,1,5,6}]
Output  3
Output  3

Input   (* SpatialMedian under ManhattanDistance for multivariate data is the same as Median:
        *)
        data={{1.5,3.},{2.6,4.},{9.2,5.},{7.3,-5.}};
        SpatialMedian[data,DistanceFunction->ManhattanDistance]
        Median[data]
Output  {4.95,3.5}
Output  {4.95,3.5}

Input   (* Obtain a robust estimate of a multivariate location when outliers are present: *)
        data={{3.,-5.},{2.,-5.},{0.,2.},{-4.,-3.},{10^8,-1.},{8.,-20000.}};
        SpatialMedian[data]

        (* Extreme values have a large influence on the Mean: *)
        Mean[data]
Output  {2.,-5.}
Output  {1.66667*10^7,-3335.33}

Input   (* The code generates a random set of 500 real numbers between-10 and 10 and stores
        them in the variable "data". The "ArgMin" function is then used to find the value of
        "p" that minimizes the sum of the square root of the distance between each element
        in "data" and "p". This effectively finds the value of "p" that is closest to the
        median of "data". The result is stored in the variable "result". Finally, the code
```

```
        calls the "SpatialMedian" function on "data", which returns the spatial median of
        the dataset. The spatial median is a generalization of the median to higher-
        dimensional spaces: *)

        data=RandomReal[{-10,10},500];
        result=ArgMin[
          Sum[
           Sqrt[(data[[i]]-p)^2],
           {i,1,Length[data]}
           ],
          p
          ]
        SpatialMedian[data]
Output  -0.563169
Output  -0.561731

Input   (* The code generates a random set of 500 two-dimensional points with coordinates
        between-10 and 10 and stores them in the variable "data". The "ArgMin" function is
        then used to find the point (x, y) that minimizes the sum of the Euclidean distances
        between each point in "data" and the point (x, y). This effectively finds the point
        that is closest to the geometric median of "data". The result is stored in the
        variable "result". Finally, the code calls the "SpatialMedian" function on "data",
        which returns the spatial median of the dataset: *)

        data=RandomReal[{-10,10},{500,2}];
        result=ArgMin[
          Total[Sqrt[Total[(Transpose[data]-{x,y})^2]]],
          {x,y}
          ]
        SpatialMedian[data]
Output  {-0.0300229,-0.0964838}
Output  {-0.0300227,-0.0964845}

Input   (* Sample points from a convex polygon: *)
        poly1=Polygon[{{-5,0},{-1,3/2},{1,3/2},{2,0},{1,-3/2},{-1,-3/2}}];
        poly2=Polygon[{{-2,0},{-1,3/2},{1,3/2},{2,0},{1,-3/2},{-1,-3/2}}];

        pts1=RandomPoint[poly1,1500];
        pts2=RandomPoint[poly2,1500];

        Graphics[
         {Lighter[Purple,.9],poly1,Purple,PointSize[Small],Point[pts1]}
         ]

        Graphics[
         {Lighter[Purple,.9],poly2,Purple,PointSize[Small],Point[pts2]}
         ]

        (* Estimate the center of the polygon by computing the spatial median of random
        points: *)
        smed1=SpatialMedian[pts1]
        m1=Mean[pts1]
        Graphics[

        {Lighter[Purple,.9],poly1,Purple,PointSize[Small],Point[pts1],PointSize[0.03],Blue,
        Point[smed1],PointSize[0.03],Red,Point[m1]}
         ]

        smed2=SpatialMedian[pts2]
        m2=Mean[pts2]
```

```
        Graphics[

        {Lighter[Purple,.9],poly2,Purple,PointSize[Small],Point[pts2],PointSize[0.03],Blue,
        Point[smed2],PointSize[0.03],Red,Point[m2]}
         ]
Output
```

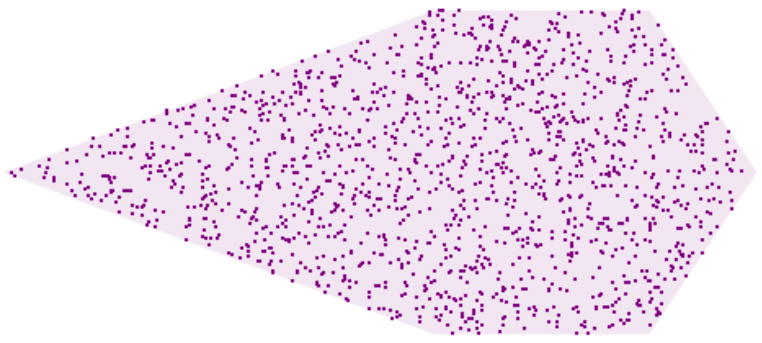

```
Output
```

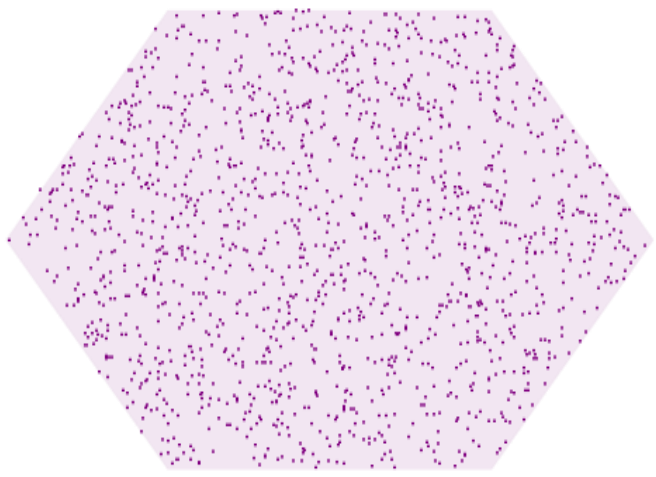

```
Output  {-0.823396,0.0101705}
Output  {-0.909217,-0.00776065}
Output
```

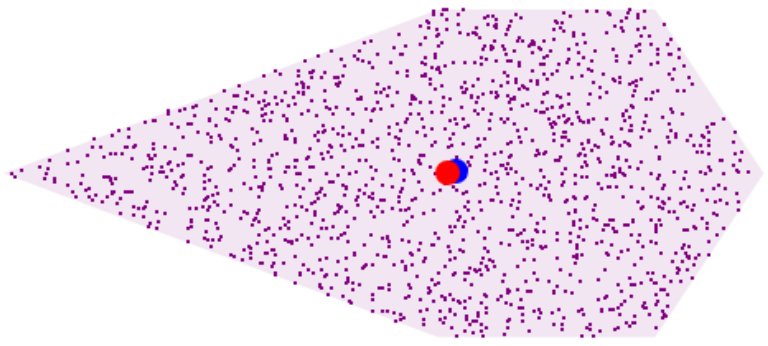

```
Output  {-0.0258644,0.00723824}
Output  {-0.0164048,0.0124507}
Output
```

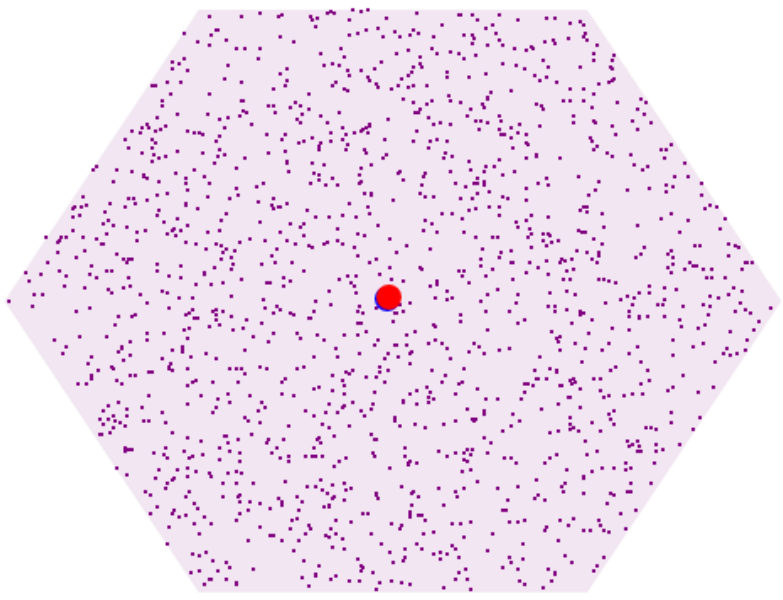

| | |
|---|---|
| HarmonicMean[list] | gives the harmonic mean of the values in list. |
| GeometricMean[list] | gives the geometric mean of the values in list. |

***Mathematica Examples 5.7*** GeometricMean

```
Input   (* Geometric mean of a list: *)
        GeometricMean[{a,b,c,d,e,f}]
Output  (a b c d e f)^(1/6)

Input   (* Geometric mean of columns of a matrix: *)
        GeometricMean[{{1,2},{5,10},{2,1},{4,3},{12,15}}]
Output  {2 15^(1/5),30^(2/5)}

Input   (* Find the geometric mean of WeightedData: *)
        data={8,3,5,4,9,1,4,2,2,3};
        w={0.15,0.09,0.12,0.10,0.16,0.3,0.11,0.08,0.08,0.09};
        GeometricMean[
```

```
         WeightedData[data,w]
         ]
Output   3.11989

Input    (* Mean is logarithmically related to GeometricMean for positive values: *)
         Log[GeometricMean[{a,b,c,d,e}]]//PowerExpand
         Mean[Log[{a,b,c,d,e}]]
Output   1/5 (Log[a]+Log[b]+Log[c]+Log[d]+Log[e])
Output   1/5 (Log[a]+Log[b]+Log[c]+Log[d]+Log[e])
```

***Mathematica Examples 5.8*** HarmonicMean

```
Input    (* Harmonic mean of symbolic values: *)
         HarmonicMean[{a,b,c,d,e,f}]
Output   6/(1/a+1/b+1/c+1/d+1/e+1/f)

Input    (* Harmonic mean of columns of a matrix: *)
         HarmonicMean[{{3,2},{2,2},{5,10},{5,1},{4,8}}]
Output   {300/89,200/89}

Input    (* Find the harmonic mean of WeightedData: *)
         sampledata={8,3,5,4,9,1,4,2,2,3};
         w={0.15,0.09,0.12,0.10,0.16,0.3,0.11,0.08,0.08,0.09};
         HarmonicMean[
          WeightedData[sampledata,w]
          ]
Output   2.31453

Input    (* The code generates random data, computes the harmonic mean, and verifies that the
         harmonic mean is the inverse of the mean of the inverse of the data: *)
         (* Generate some random data *)
         data=RandomReal[{1,10},100];

         (* Compute the harmonic mean: *)
         harmean1=HarmonicMean[data]
         harmean2=1/Mean[1/data]

         (* Verify that the harmonic mean is the inverse of the mean of the inverse: *)
         harmean1==harmean2
Output   3.786
Output   3.786
Output   True

Input    (* HarmonicMean is logarithmically related to GeometricMean for positive values: *)
         1/Log[GeometricMean[{a,b,c}]]//PowerExpand
         HarmonicMean[1/Log[{a,b,c}]]

         (* For numerical data*)
         data=RandomReal[{1,10},100];
         1/Log[GeometricMean[data]]
         HarmonicMean[1/Log[data]]
Output   3/(Log[a]+Log[b]+Log[c])
Output   3/(Log[a]+Log[b]+Log[c])
Output   0.653438
Output   0.653438

Input    (* The code will generate a list of random numbers, calculate their Harmonic Mean,
         Geometric Mean, and Arithmetic Mean, and then plot the means against the number of
         elements in the list. For positive data, we can see that Harmonic Mean <= Geometric
         Mean <= Arithmetic Mean: *)
```

```
SeedRandom[1234];
n=100; (* Number of elements in the list: *)
d=RandomReal[{1,10},n]; (* Generate random list: *)
hm=HarmonicMean[d]; (* Calculate harmonic mean: *)
gm=GeometricMean[d]; (* Calculate geometric mean: *)
am=Mean[d]; (* Calculate arithmetic mean: *)
means={hm,gm,am} (* Combine means into a list: *)
labels={"data points","Harmonic Mean","Geometric Mean","Arithmetic Mean"}; (*
Labels for the plot: *)

(* Create the plot: *)
plot=ListPlot[
  {d,{{0,hm},{n,hm}},{{0,gm},{n,gm}},{{0,am},{n,am}}},
  Joined->{False,True,True,True},
  Filling->{1->Axis},
  Frame->True,
  PlotRange->All,
  PlotStyle->{Purple,Red,Green,Blue},
  PlotLegends->labels,
  ImageSize->250
  ]
```

Output {4.05091,4.96548,5.76365}

Output

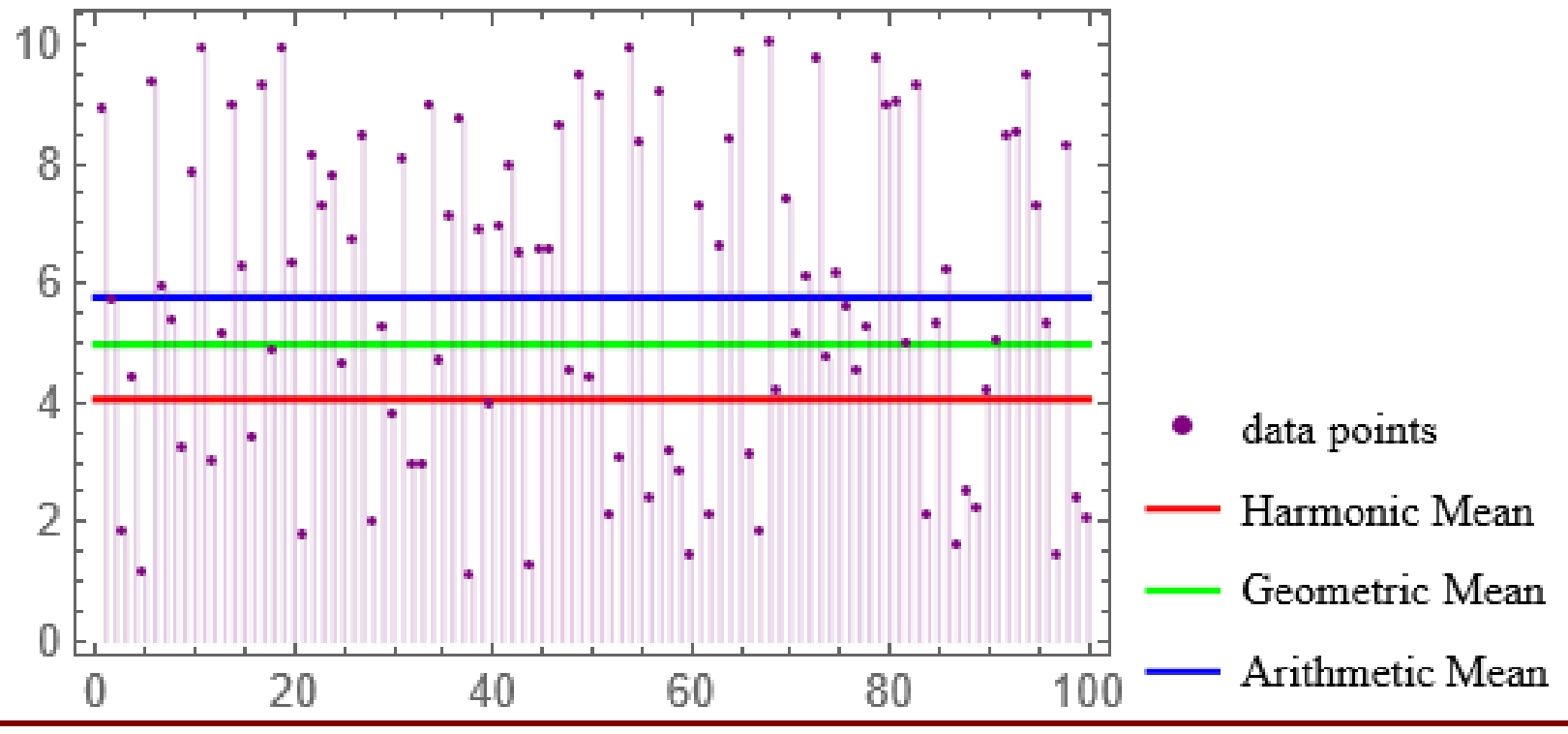

| | |
|---|---|
| `RootMeanSquare[list]` | gives the root mean square of values in list. |
| `RootMeanSquare[dist]` | gives the root mean square of the distribution dist. |

***Mathematica Examples 5.9*** RootMeanSquare

Input
```
(* RootMeanSquare of a list: *)
RootMeanSquare[{a,b,c,d,e}]
```
Output $\frac{\sqrt{a^2+b^2+c^2+d^2+e^2}}{\sqrt{5}}$

Input
```
(* RootMeanSquare of columns of a matrix: *)
RootMeanSquare[{{3,4},{1,2},{5,10},{5,2},{4,8}}]
```
Output $\{2\sqrt{\frac{19}{5}},2\sqrt{\frac{47}{5}}\}$

Input
```
(* Exact input yields exact output: *)
RootMeanSquare[{1,2,3,4,5,6,7,8,9,10}]
```
Output $\sqrt{\frac{77}{2}}$

Input
```
(* RootMeanSquare for WeightedData: *)
```

```
        data={8,3,5,4,9,1,4,2,2,3};
        w={0.15,0.09,0.12,0.10,0.16,0.3,0.11,0.08,0.08,0.09};
        RootMeanSquare[
         WeightedData[data,w]
         ]
Output  4.95921

Input   (* The code analyzes a dataset of heights by calculating the Root Mean Square and
        generating two plots. The first plot displays the data set, while the second plot
        displays the data set along with horizontal line for the Root Mean Square: *)

        h={133,136,149,133,123,121,140,139,117,117,136,108,126,104,116,147,140,148,150,122,
        135,146,133,144,117,124,135,117,120,121,110,124,103,137,101,119,104,113,139,133};

        rms=N[RootMeanSquare[h]]
        n=Length[h];

        ListPlot[
         h,
         Filling->Axis,
         PlotStyle->Purple,
         ImageSize->170
         ]

        ListPlot[
         {h,{{0,rms},{n,rms}}},
         Joined->{False,True},
         Filling->{1->rms,2->Axis},
         PlotStyle->Purple,
         ImageSize->170,
         PlotLegends->{None,"RMS =127.744"}
         ]
Output  127.744
Output
```

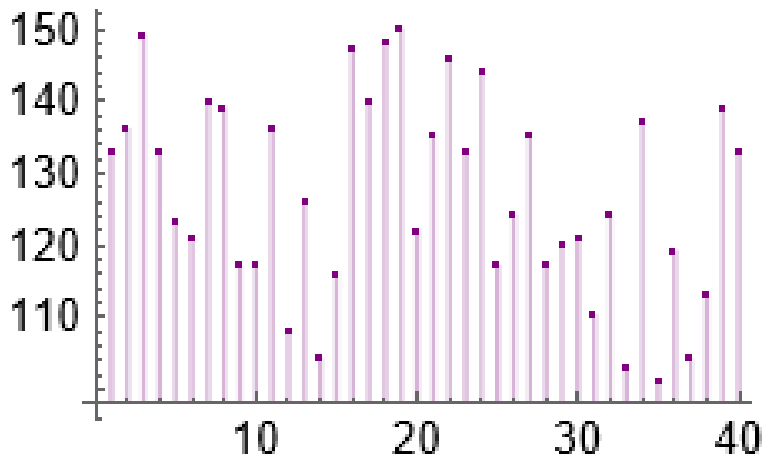

```
Output
```

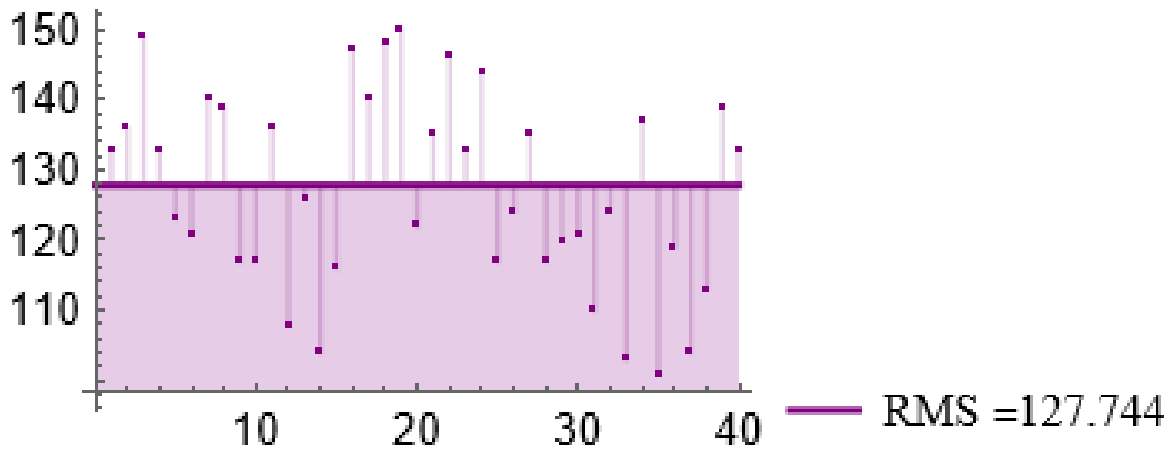

```
Input   (* RootMeanSquare is the square root of the Mean of the data squared: *)
        data=RandomReal[10,10];
        Sqrt[Mean[data^2]]
        RootMeanSquare[data]
Output  6.45995
Output  6.45995

Input   (* RootMeanSquare is equivalent to a scaled Norm: *)
        data=RandomReal[10,10];
        Norm[data]/Sqrt[Length[data]]
        RootMeanSquare[data]
```

```
Output   6.18667
Output   6.18667

Input    (* RootMeanSquare is a scaled EuclideanDistance from the Mean: *)
         data=RandomReal[10,10];
         mean=Mean[data]
         RootMeanSquare[data-mean]
         EuclideanDistance[data,mean]/Sqrt[Length[data]]
Output   4.50879
Output   2.51917
Output   2.51917
```

| | |
|---|---|
| `Quartiles[list]` | gives a list of the 1/4, 1/2 and 3/4 quantiles of the elements in list. |
| `Quartiles[dist]` | gives a list of the 1/4, 1/2 and 3/4 quantiles of the distribution dist. |
| `Quantile[list,q]` | gives the q ^(th) quantile of list. |
| `Quantile[list,{q1,q2,…}]` | gives a list of quantiles q1, q2, …. |
| `Quantile[list,q,{{a,b},{c,d}}]` | uses the quantile definition specified by parameters a, b, c, d. |
| `Quantile[dist,q]` | gives a quantile of the distribution dist. |

***Mathematica Examples 5.10*** Quartiles

```
Input    (* Quartiles for a list of exact numbers: *)
         Quartiles[{1,3,4,2,5,6}]
Output   {2,7/2,5}

Input    (* Quartiles for a matrix gives columnwise quartiles: *)
         data={{1,6,2},{4,11,7},{5,7,8},{10,12,13},{20,21,22}};

         Quartiles[data]
         Quartiles[{1,4,5,10,20}]
         Quartiles[{6,11,7,12,21}]
         Quartiles[{2,7,8,13,22}]
Output   {{13/4,5,25/2},{27/4,11,57/4},{23/4,8,61/4}}
Output   {13/4,5,25/2}
Output   {27/4,11,57/4}
Output   {23/4,8,61/4}

Input    (* Find the quartiles of WeightedData: *)
         data={8,3,5,4,9,0,4,2,2,3};
         w={0.15,0.09,0.12,0.10,0.16,0.,0.11,0.08,0.08,0.09};
         Quartiles[
          WeightedData[data,w]
          ]
Output   {3,4,8}

Input    (* The second quartile is the Median: *)
         data=RandomReal[15,10];
         Quartiles[data][[2]]
         Median[data]
Output   12.6404
Output   12.6404

Input    (* This code is designed to calculate the quartiles of a given dataset. The code
         sorts the dataset, calculates the median (Q2), and then divides the dataset into the
         lower and upper halves to calculate the first (Q1) and third (Q3) quartiles: *)
         data={2,4,5,7,9,10,12,15,18,22};
```

```
        (* Step 1 sorts the given data in ascending order, which is a necessary step before
        calculating the quartiles: *)
        sortedData=Sort[data];

        (* Step 2 calculates the second quartile (Q2), i.e., the median. If the dataset has
        an even number of elements, Q2 is defined as the mean of the two middle values. If
        the dataset has an odd number of elements, Q2 is defined as the middle value: *)
        n=Length[sortedData];
        Q2=If[
            EvenQ[n],
            Mean[
             Take[sortedData,{n/2,n/2+1}]
             ],
            sortedData[[Ceiling[n/2]]]
            ];

        (* Step 3 calculates the first quartile (Q1). It considers the lower half of the
        sorted data and follows the same logic as in Step 2: *)
        lowerHalf=Take[sortedData,Floor[n/2]];
        Q1=If[
            EvenQ[Length[lowerHalf]],
            Mean[
             Take[lowerHalf,{Length[lowerHalf]/2,Length[lowerHalf]/2+1}]
             ],
            lowerHalf[[Ceiling[Length[lowerHalf]/2]]]
            ];

        (* Step 4 calculates the third quartile (Q3). It considers the upper half of the
        sorted data and follows the same logic as in Step 2: *)
        upperHalf=Take[sortedData,-Ceiling[n/2]];
        Q3=If[
            EvenQ[Length[upperHalf]],
            Mean[
             Take[upperHalf,{Length[upperHalf]/2,Length[upperHalf]/2+1}]
             ],
            upperHalf[[Ceiling[Length[upperHalf]/2]]]
            ];

        (* Finally, the quartiles Q1, Q2, and Q3 are outputted as a list, and the Quartiles
        function is called and returns the same output: *)
        {Q1,Q2,Q3}
        Quartiles[data]
Output  {5,19/2,15}
Output  {5,19/2,15}

Input   (* The code reads in a dataset h of heights and calculates its quartiles. It then
        produces three plots: a scatter plot of the dataset h, a scatter plot of the sorted
        dataset s, and a plot that shows both h and three horizontal lines at the quartiles.
        These plots are used for visualizing the distribution of the dataset and highlighting
        the location of the quartiles: *)

        h={133,136,149,133,123,121,140,139,117,117,136,108,126,104,116,147,140,148,150,122,
        135,146,133,144,117,124,135,117,120,121,110,124,103,137,101,119,104,113,139,133};

        qrs=N[Quartiles[h]]
        n=Length[h]
        s=Sort[h];

        ListPlot[
         h,
```

```
 Filling->Axis,
 PlotStyle->Purple,
 ImageSize->170
 ]

ListPlot[
 s,
 Filling->Axis,
 Epilog-
>{RGBColor[0.88,0.61,0.14],Line[{{10,0},{10,qrs[[1]]}}],RGBColor[0.37,0.5,0.7],Line
[{{20,0},{20,qrs[[2]]}}],Darker[Red],Line[{{30,0},{30,qrs[[3]]}}]},
 Ticks->{{{10,"Q1",{0,.01}},{20,"Q2",{0,.01}},{30,"Q3",{0,.01}}},True},
 PlotStyle->Purple,
 ImageSize->170
 ]

ListPlot[

{h,{{0,qrs[[1]]},{n,qrs[[1]]}},{{0,qrs[[2]]},{n,qrs[[2]]}},{{0,qrs[[3]]},{n,qrs[[3]
]}}},
 Joined->{False,True,True,True},
 Filling->{1->Axis},
 PlotStyle->{Purple,RGBColor[0.88,0.61,0.14],RGBColor[0.37,0.5,0.7],Darker[Red]},
 PlotLegends->{"h","Q1","Q2 - median","Q3"},
 AxesLabel->Automatic,
 ImageSize->170
 ]
```

Output {117.,125.,138.}

Output 40

Output

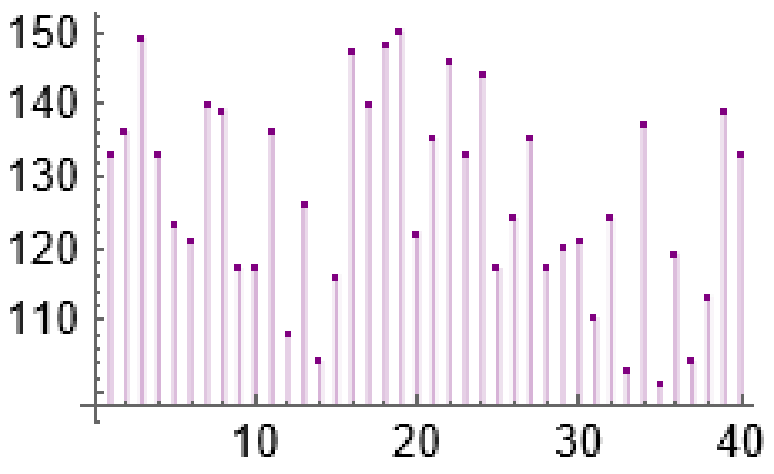

Output

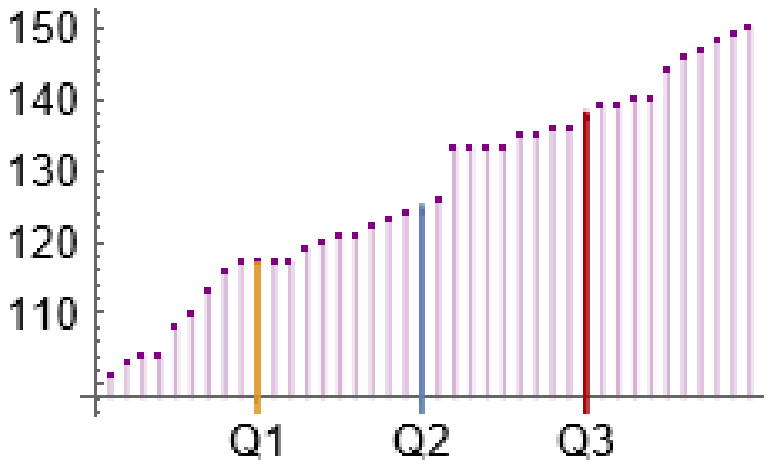

Output

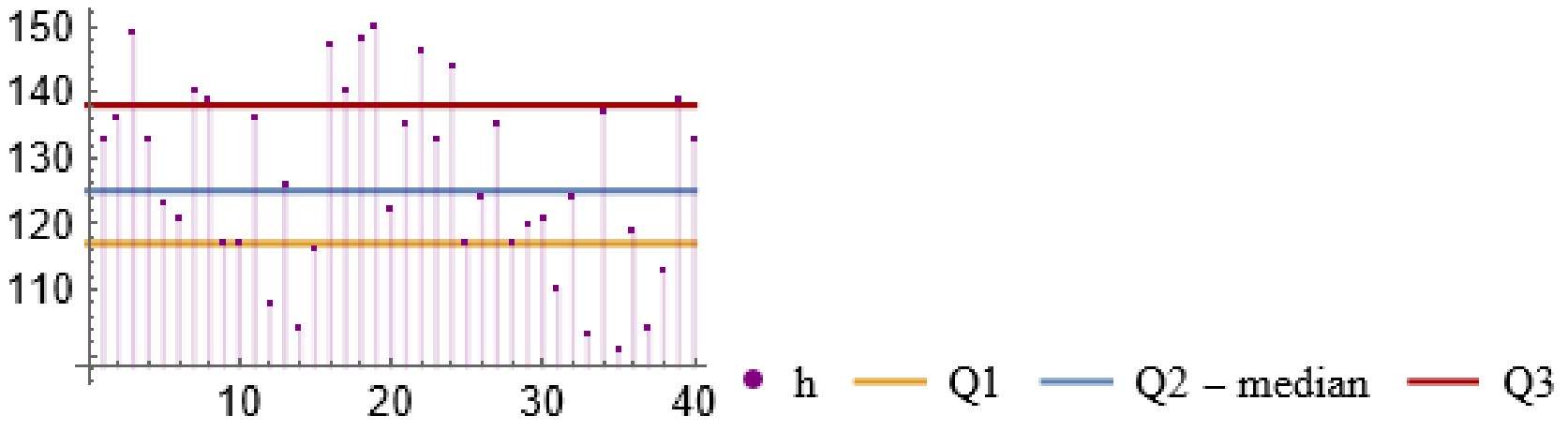

Input

```
(* This code generates a plot of the probability density function (PDF) of a chi-
squared distribution with five degrees of freedom, and displays the quartiles of the
distribution as vertical lines on the plot: *)

(* Defines the chi-squared distribution with five degrees of freedom and assigns it
to the variable d: *)
```

```
        d=ChiSquareDistribution[5];

        (* Uses the Quartiles function to compute the first, second, and third quartiles of
        the distribution, and assigns them to the variables Q1, Q2, and Q3, respectively: *)
        {Q1,Q2,Q3}=N[Quartiles[d]]

        (* Uses the PDF function to compute the PDF of the chi-squared distribution at the
        quartile values, and assigns them to the variables v1, v2, and v3, respectively: *)
        {v1,v2,v3}=PDF[d,{Q1,Q2,Q3}]

        (* Plots the chi-squared distribution and displays the quartiles of the distribution
        as vertical lines on the plot: *)
        Plot[
         PDF[d,x],
         {x,0,14},
         Filling->Axis,
         Epilog->{
            Directive[Blue],Line[{{Q1,0},{Q1,v1}}],
            Directive[Green],Line[{{Q2,0},{Q2,v2}}],
            Directive[Red],Line[{{Q3,0},{Q3,v3}}]
            },
         MeshFunctions->{#1&},
         Mesh->{{Q1,Q2,Q3}},
         MeshStyle->Black,
         MeshShading->ColorData[97,"ColorList"],
         PlotStyle->Purple,
         Ticks->{{{Q1,"Q1",{0,.01}},{Q2,"Q2",{0,.01}},{Q3,"Q3",{0,.01}}}},
         ImageSize->170
         ]
Output  {2.6746,4.35146,6.62568}
Output  {0.15272,0.137036,0.0825822}
Output
```

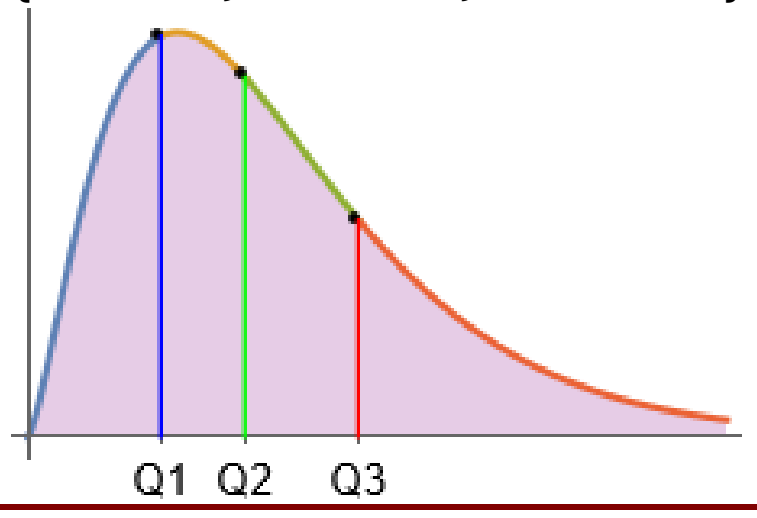

**Mathematica Examples 5.11** Quantile

```
Input   (* Find the halfway value (median) of a list: *)
        Quantile[{1,2,3,4,5,6,7,8,9},1/2]
Output  5

Input   (* Find the quarter-way value (lower quartile) of a list: *)
        Quantile[{1,2,3,4,5,6,7,8,9},1/4]
Output  3

Input   (* Lower and upper quartiles: *)
        Quantile[{1,2,3,4,5,6,7,8,9},{1/4,3/4}]
Output  {3,7}

Input   (* The Quantile function is used to calculate the values of the dataset that
        correspond to the 10th, 20th, 80th, and 90th percentiles: *)
        Quantile[{1,2,3,4,5,6,7,8,9,10},{0.1,0.2,0.8,0.9}]
Output  {1,2,8,9}

Input   (* The q^(th) quantile for a normal distribution: *)
```

```
       Quantile[NormalDistribution[μ,σ],q]
Output μ- √2 σ InverseErfc[2 q]   if 0<=q<=1

Input  (* Find quantiles of elements in each column: *)
       Quantile[{{1,5 },{2,4},{3,3},{4,2},{5,1}},1/4]
       Quantile[{1,2,3,4,5 },1/4]
       Quantile[{5,4,3,2,1 },1/4]
Output {2,2}
Output 2
Output 2

Input  (* Find quantiles for WeightedData: *)
       data={8,3,5,4,9,0,4,2,2,3};
       w={0.15,0.09,0.12,0.10,0.16,0.,0.11,0.08,0.08,0.09};
       Quantile[WeightedData[data,w],0.4]
Output 4

Input  (* The code reads in a dataset h of heights and calculates its quantile. It then
       produces three plots: a scatter plot of the dataset h, a scatter plot of the sorted
       dataset s, and a plot that shows both h and three horizontal lines at the values of
       the 10th, 50th (median), and 90th percentiles of the data: *)

       h={133,136,149,133,123,121,140,139,117,117,136,108,126,104,116,147,140,148,150,122,
       135,146,133,144,117,124,135,117,120,121,110,124,103,137,101,119,104,113,139,133};

       qs=Quantile[h,{0.1,0.5,0.9}]
       n=Length[h];
       s=Sort[h];

       ListPlot[
        h,
        Filling->Axis,
        PlotStyle->Purple,
        ImageSize->170
        ]

       ListPlot[
        s,
        Filling->Axis,
        Epilog-
       >{RGBColor[0.88,0.61,0.14],Line[{{4,0},{4,qs[[1]]}}],RGBColor[0.37,0.5,0.7],Line[{{
       20,0},{20,qs[[2]]}}],Darker[Red],Line[{{36,0},{36,qs[[3]]}}]},
        Ticks->{{{4,"D0.1",{0,.01}},{20,"D0.5",{0,.01}},{36,"D0.9",{0,.01}}},True},
        PlotStyle->Purple,
        ImageSize->170
        ]

       ListPlot[
        {h,{{0,qs[[1]]},{n,qs[[1]]}},{{0,qs[[2]]},{n,qs[[2]]}},{{0,qs[[3]]},{n,qs[[3]]}}},
        Joined->{False,True,True,True},
        Filling->{1->Axis},
        PlotStyle->{Purple,RGBColor[0.88,0.61,0.14],RGBColor[0.37,0.5,0.7],Darker[Red]},
        PlotLegends->{"h","D0.1","D0.5","D0.9"},
        AxesLabel->Automatic,
        ImageSize->170
        ]
Output {104,124,146}
```

Output

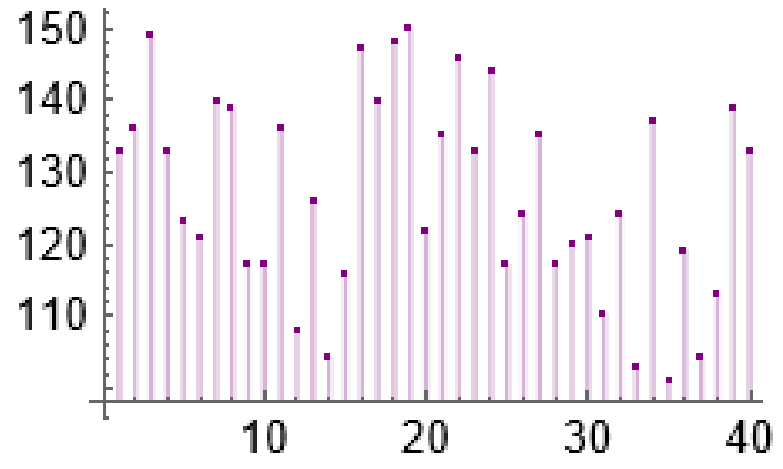

Output

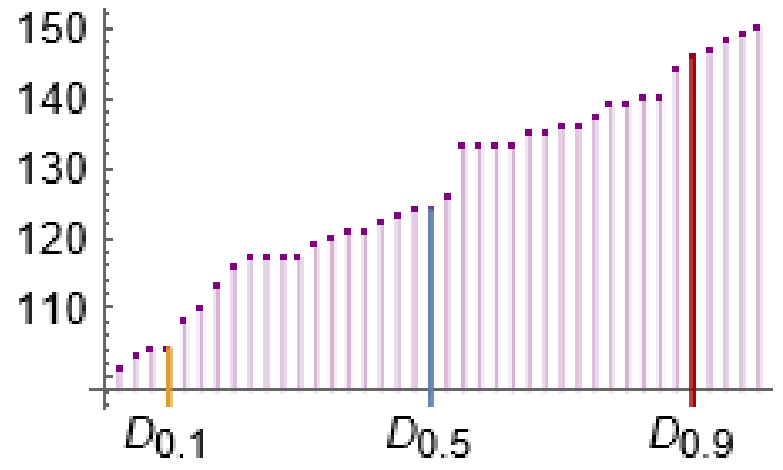

Output

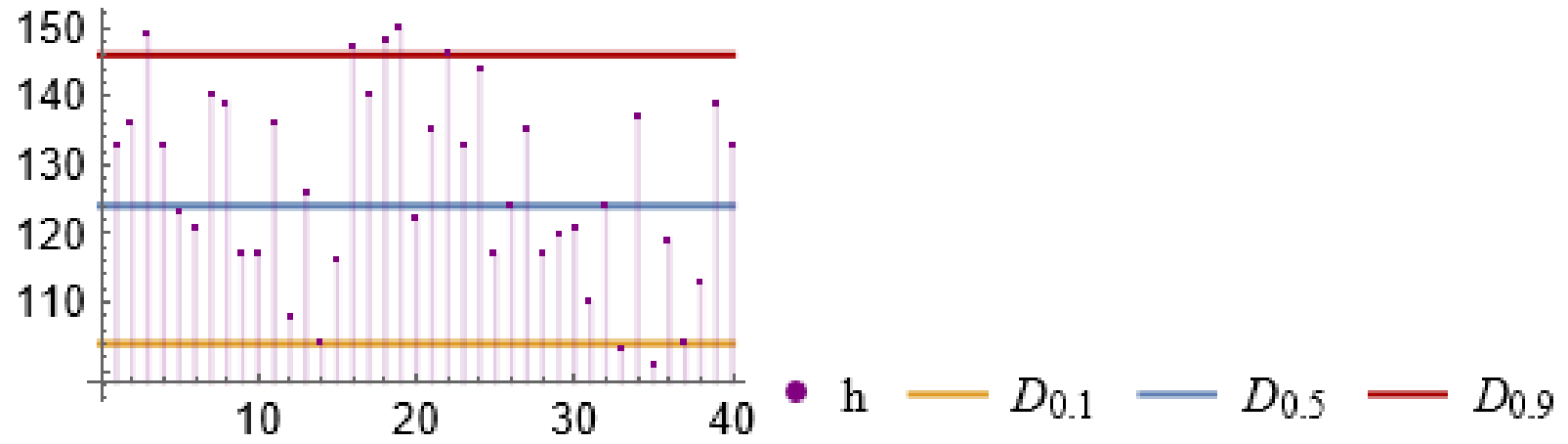

# UNIT 5.2

# BOX WHISKER CHART

The Mathematica function `BoxWhiskerChart` is a powerful tool for visualizing and analyzing data distributions. Here are some features of this function:

- **Clear representation of data:** `BoxWhiskerChart` provides a clear and concise representation of the distribution of a dataset. It displays important statistical measures such as the median, quartiles, and outliers, making it easy to understand the central tendency and spread of the data.
- **Customizable appearance:** The function offers a wide range of options to customize the appearance of the box-and-whisker plot. You can adjust the colors, styles, and sizes of the boxes, whiskers, outliers, and other elements to suit your preferences or match your presentation or publication style.
- **Comparative analysis:** `BoxWhiskerChart` allows for easy comparison of multiple datasets. You can plot several box-and-whisker diagrams side by side or in a stacked manner, making it straightforward to identify differences or similarities in distributions.
- **Interaction and exploration:** The resulting chart is interactive, meaning you can hover over different elements to obtain more detailed information about specific data points or summary statistics. This interactivity enhances the exploratory data analysis process and allows for a deeper understanding of the underlying distribution.

| | |
|---|---|
| `BoxWhiskerChart[{x1,x2,…}]` | makes a box-and-whisker chart for the values xi. |
| `BoxWhiskerChart[{x1,x2,…},bwspec]` | makes a chart with box-and-whisker symbol specification bwspec. |
| `BoxWhiskerChart[{data1,data2,…},…]` | makes a chart with box-and-whisker symbol for each datai. |
| `BoxWhiskerChart[{{data1,data2,…},…},…]` | makes a box-and-whisker chart from multiple groups of datasets {data1,data2,…}. |

`BoxWhiskerChart` draws a box-and-whisker summary of the distribution of values in each `datai`. See the following figure

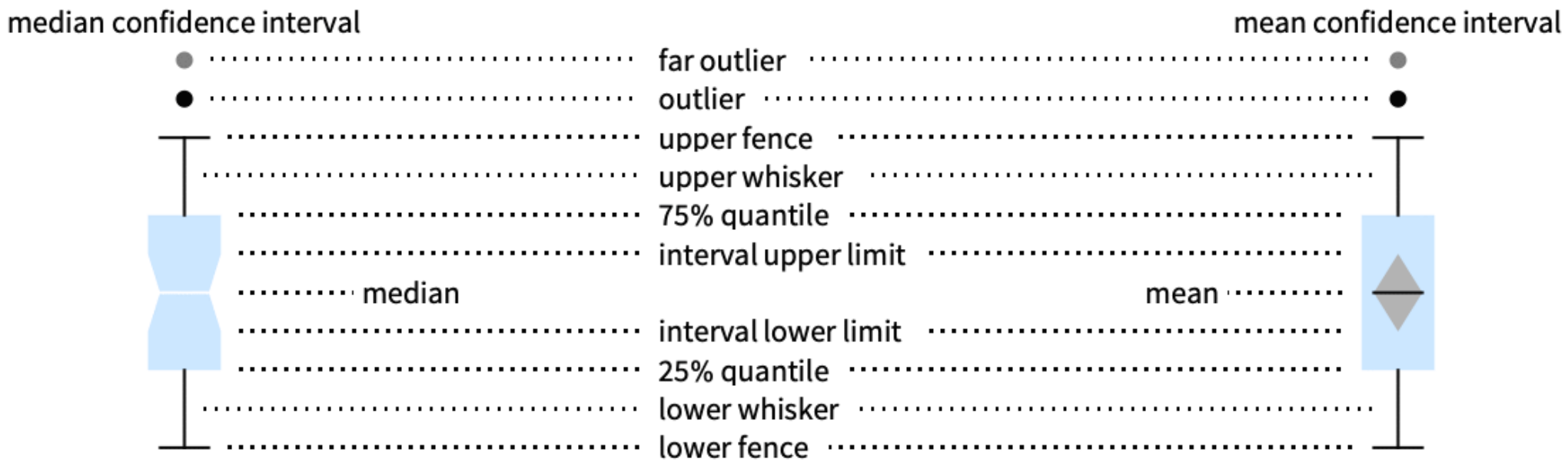

The following box-and-whisker specifications `bwspec` can be given:

| | |
|---|---|
| `"Notched"` | median confidence interval notch |
| `"Outliers"` | outlier markers |
| `"Median"` | median marker |
| `"Basic"` | box-and-whisker only |
| `"Mean"` | mean marker |
| `"Diamond"` | mean confidence interval diamond |
| `{{elem1,val11,…},…}` | box-and-whisker element specification |
| `{"name",{elem1,val11,…},…}` | named bwspec with element modifica |

***Mathematica Examples 5.12*** BoxWhiskerChart

Input

```
(* The code uses the RandomVariate function to generate a data vector with 200 values. It samples from a normal distribution with a mean of 1 and a standard deviation of 2. then, the code generates a basic box-and-whisker chart using randomly generated data: *)

BoxWhiskerChart[
 RandomVariate[
  NormalDistribution[1,2],
  200
  ],
 ChartStyle->Purple,
 ImageSize->170
 ]
```

Output

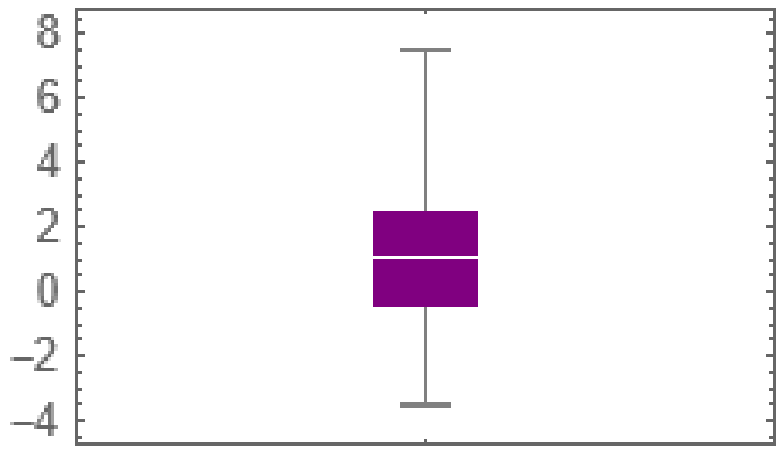

Input

```
(* The code generates a box-and-whisker chart for multiple datasets created using a loop. The code uses a Table construct to generate four sets of data points. For each iteration of the loop, it generates 100 data points by sampling from a normal distribution. The mean of the normal distribution varies based on the values in the list {0,2,4,6}. The standard deviation is fixed at 2. This approach allows you to create multiple datasets with different means for comparison or analysis purposes. The BoxWhiskerChart function is used to create a chart from the generated data. The data variable contains the four sets of data points generated in the loop: *)

data=Table[
    RandomVariate[
     NormalDistribution[x,2],
     100
     ],
    {x,{0,2,4,6}}
    ];

BoxWhiskerChart[
 data,
 ChartStyle->Purple,
 ImageSize->170
 ]
```

Output

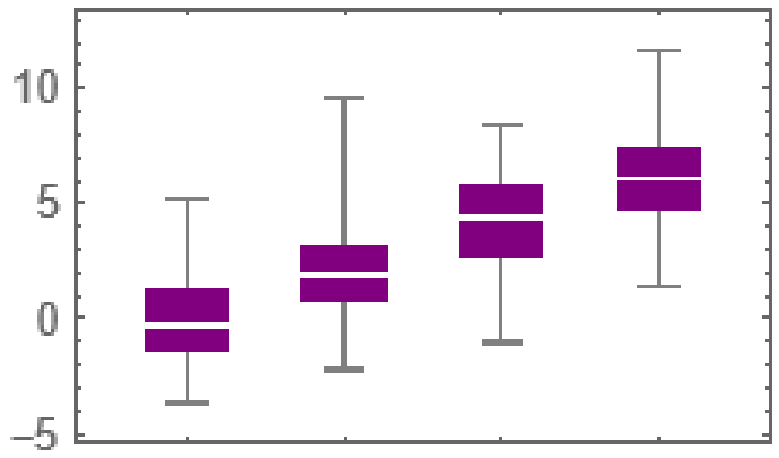

Input

```
(* The code demonstrates customization options for a box-and-whisker chart. It generates a list of seven datasets, each containing 200 data points randomly sampled from normal distributions. The mean of each dataset is randomly chosen from the integers 0 to 6, while the standard deviation is fixed at 1. The first BoxWhiskerChart call creates a notched box-and-whisker chart, displaying confidence intervals around
```

```
the medians of each dataset. The second BoxWhiskerChart call explicitly shows
outliers, highlighting extreme or unusual data points: *)

data=Table[
    RandomVariate[
     NormalDistribution[RandomInteger[6],1],
     200
     ],
    {7}
    ];

BoxWhiskerChart[
 data,
 "Notched",
 ChartStyle->Purple,
 ImageSize->200
 ]
(* Show outliers: *)
BoxWhiskerChart[
 data,
 "Outliers",
 ChartStyle->Purple,
 ImageSize->200
 ]
```

Output

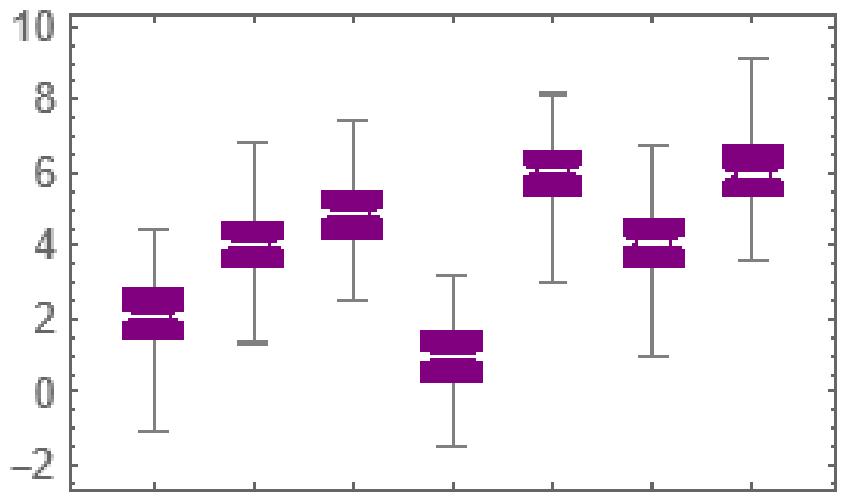

Output

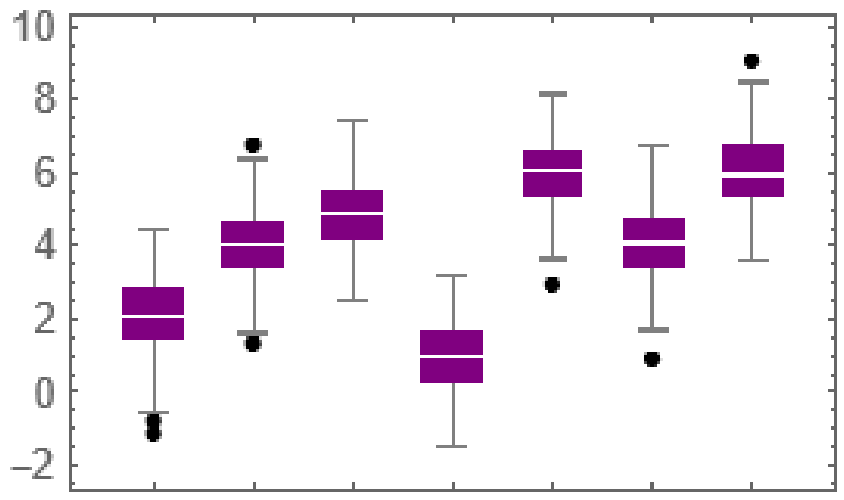

Input

```
(* The code utilizes named presets to create multiple box-and-whisker charts with
different styles. Inside the Table loop, each iteration generates a chart with a
specific preset style. The available preset styles used in the code include "Basic",
"Outliers", "Notched", "Median", "Mean", and "Diamond". These presets provide
different visual representations and statistical measures for the box-and-whisker
charts: *)

data=Table[
    RandomReal[
     BetaDistribution[x,2],
     300
     ],
    {x,1,5,0.7}
    ];

Table[
 BoxWhiskerChart[
  data,
```

```
   j,
   PlotLabel->Text[j],
   ChartStyle->Purple,
   ImageSize->170
   ],
  {j,{"Basic","Outliers","Notched","Median","Mean","Diamond"}}
  ]
```

Output

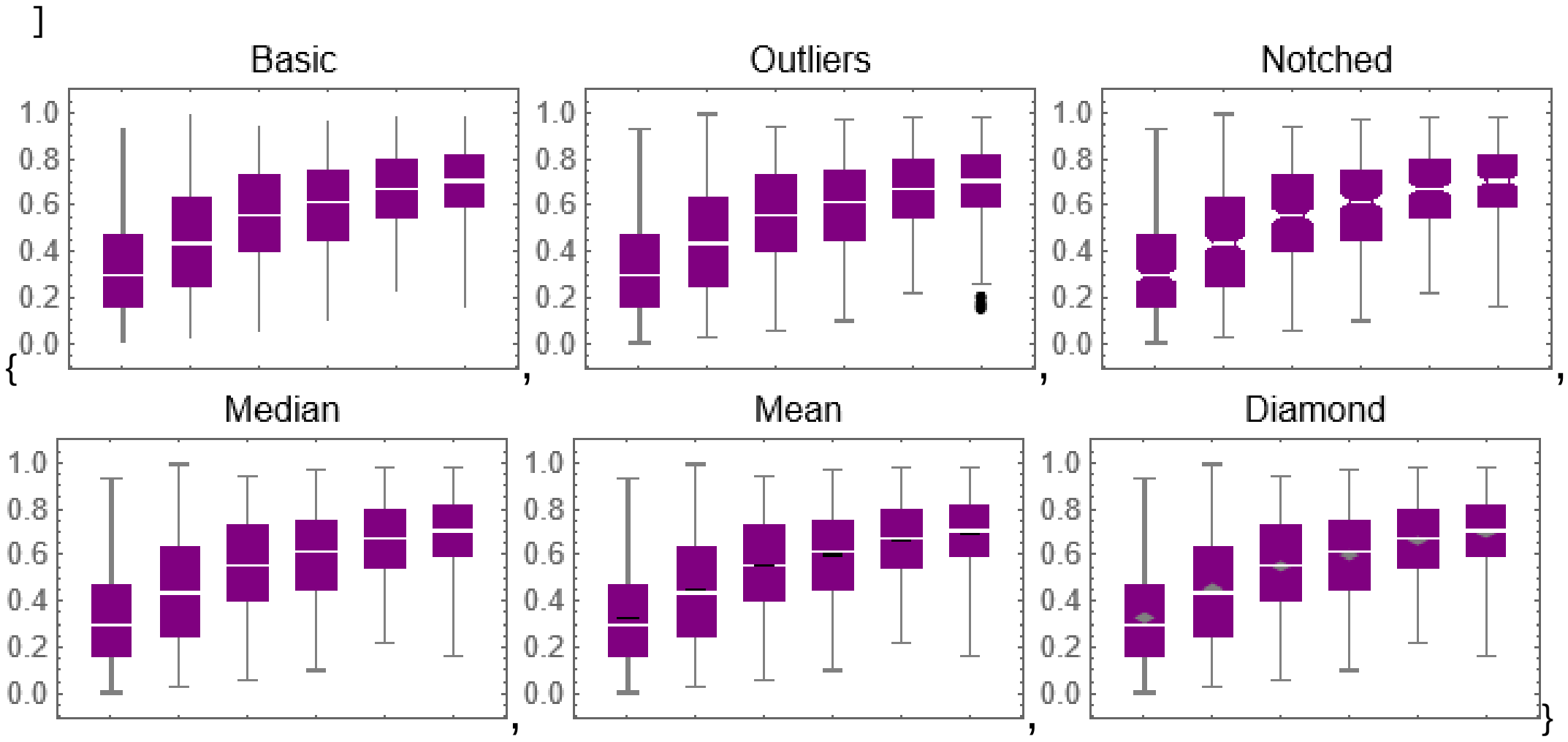

Input

```
(* The code generates a list of datasets using a Table construct. Each dataset
contains 200 data points generated from a Pareto distribution. The shape parameter
of the Pareto distribution varies from 4 to 8 in increments of 1, resulting in
multiple datasets with different shapes. The first Table loop generates box-and-
whisker charts with outliers represented by different shapes. Each iteration uses a
different shape symbol, including a bullet (•), a square (\[Square]), and a right
triangle (\[RightTriangle]). This provides visual differentiation for the outliers
in the charts. The second Table loop demonstrates how to style the outliers in the
box-and-whisker charts with different colors. Each iteration assigns a different
color (red, green, blue) to the outliers: *)

data=Table[
    RandomReal[
     ParetoDistribution[2,x],
     200
     ],
    {x,4,8,1}
    ];

Table[
 BoxWhiskerChart[
  data,
  {{"Outliers",shapes}},
  ChartStyle->Purple,
  ImageSize->200
  ],
 {shapes,{"•","\[Square]","\[RightTriangle]"}}
 ]
(* Style the outliers: *)
Table[
 BoxWhiskerChart[
  data,
  {{"Outliers","\[RightTriangle]",colors}},
  ChartStyle->Purple,
  ImageSize->200
  ],
```

```
 {colors,{Red,Green,Blue}}
 ]
```

Output

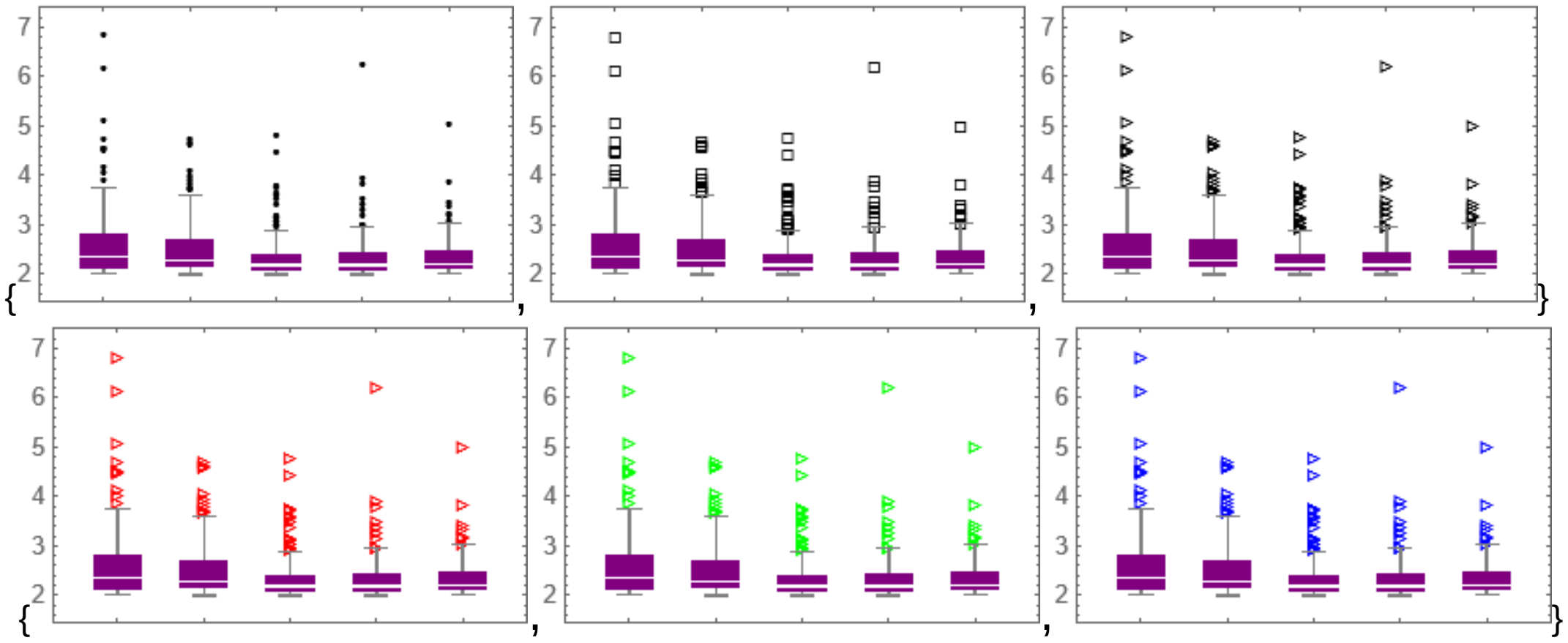

Input

```
(* By adjusting the width, style, shape, or removing the median marker, you can customize the appearance of the box-and-whisker chart to suit your preferences and highlight specific aspects of the data. The code generates a list of datasets using a Table construct. Each dataset contains 200 data points generated from a beta distribution. The shape parameter of the beta distribution varies from 1 to 4, resulting in multiple datasets with different shapes. The first Table loop generates box-and-whisker charts with the median marker having different widths. The second Table loop demonstrates how to style the median marker using different line styles. The third Table loop shows how to use different shape symbols for the median marker: *)

data=Table[
    RandomReal[
     BetaDistribution[α,2],
     200
     ],
    {α,1,4,6}
    ];

Table[
 BoxWhiskerChart[
  data,
  {{"MedianMarker",width,Green}},
  ChartStyle->Purple,
  ImageSize->170
  ],
 {width,{0.1,0.5,2}}
 ]

(* Style the median marker: *)
Table[
 BoxWhiskerChart[
  data,
  {{"MedianMarker",0.8,styles}},
  ChartStyle->Purple,
  ImageSize->170
  ],
 {styles,{Green,Directive[Dotted,Green],Directive[Thickness[0.02],Green]}}
 ]

(* Use a different shape of median marker: *)
Table[
```

```
  BoxWhiskerChart[
   data,
   {{"MedianMarker",shapes,Green}},
   ChartStyle->Purple,
   ImageSize->170
   ],
  {shapes,{"◦","\[CircleTimes]","*"}}
  ]

(* Do not show the median marker: *)
BoxWhiskerChart[
 data,
 {{"MedianMarker",None}},
 ChartStyle->Purple,
 ImageSize->170
 ]
```

Output

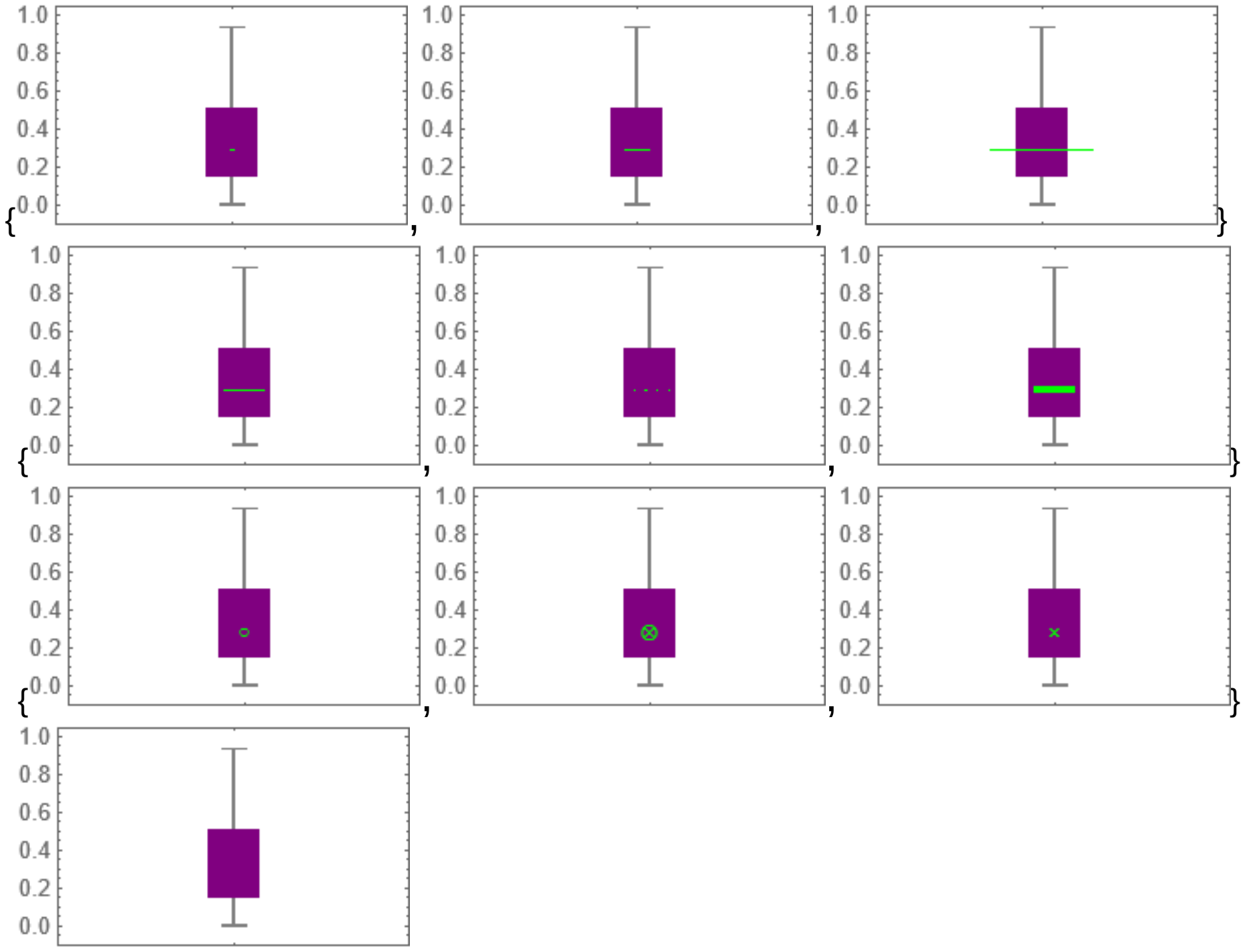

Input

```
(* The code generates a series of BoxWhiskerCharts to visualize the distribution of a dataset called 'data' using the SkewNormalDistribution. The dataset has dimensions 8 rows by 200 columns. The code uses the BoxWhiskerChart function to create a box and whisker plot for each column in the 'data' dataset. Each chart represents statistical measures such as the minimum, mean, median, and maximum values with different line styles and colors: *)

data=RandomVariate[
    SkewNormalDistribution[0,2,4],
    {8,200}
    ];

Table[
 BoxWhiskerChart[
  data,
  {
   {"Whiskers",Directive[Thick,s[[2]],Opacity[0.8]]},
   {"Fences",Directive[Thick,s[[2]],Opacity[0.8]]}
```

```
   },
  Joined->s[[1]],
  PlotLabel->Style[Row[{"Joined - ",s[[1]]}]],
  ChartStyle->"Pastel",
  ImageSize->200
  ],
 {s,{{"Min",Red},{"Mean",Green},{"Median",Blue},{"Max",Orange}}}
 ]
```

Output

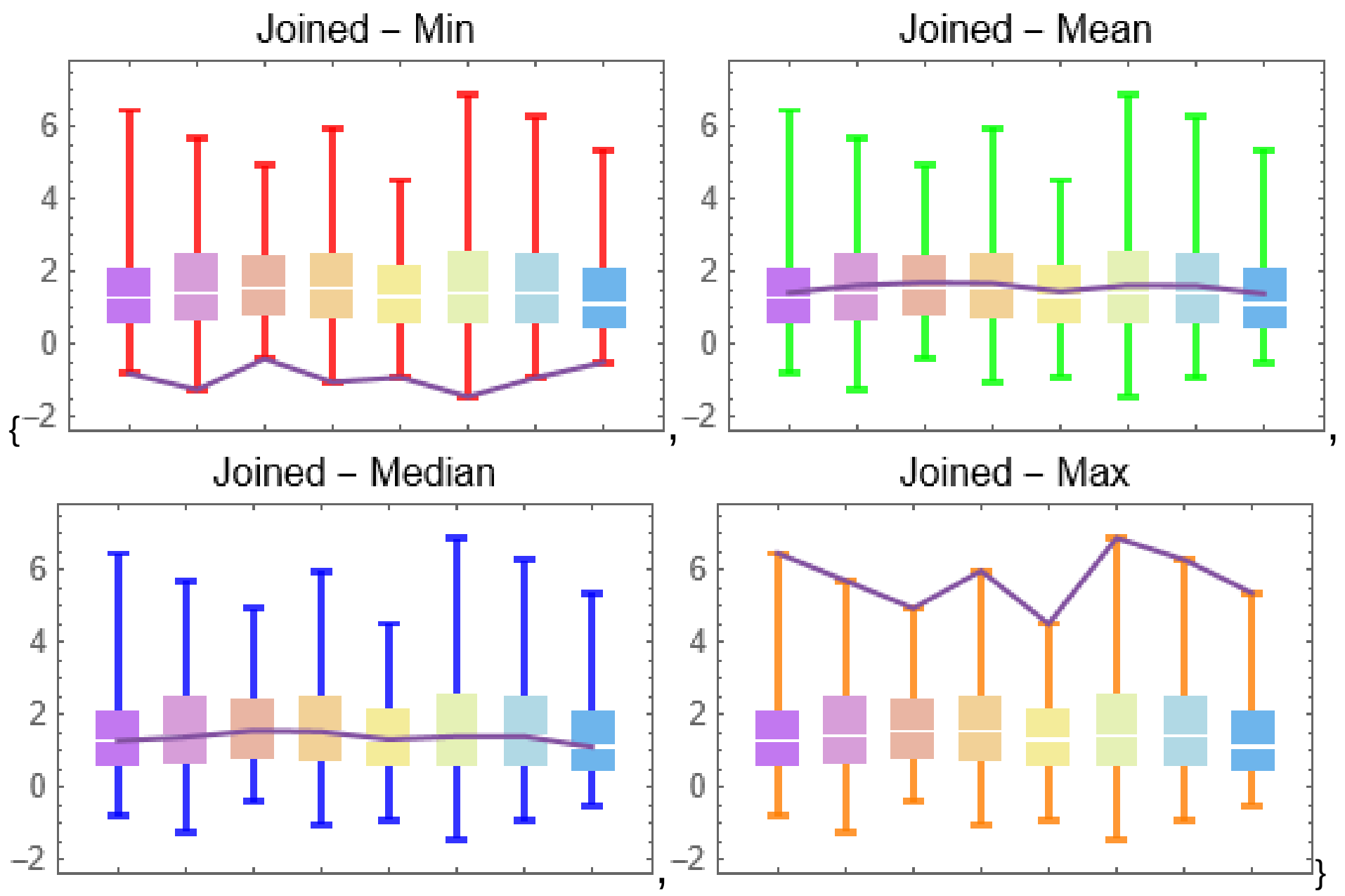

Input

```
(* The code generates three sets of random data (data1, data2, data3) by sampling
from a normal distribution with different means (μ) using RandomVariate. Each set
consists of 40 data points. The BoxWhiskerChart function is used to create a box-
and-whisker plot. Additionally, points are overlaid on the chart to show the
individual data points of each dataset. The Epilog option allows additional elements
to be added to the chart. Three sets of points are created using Map and Transpose,
corresponding to the three datasets. Each point is randomly displaced in the x-axis
within a small range, and the y-coordinate is the actual data value: *)

{data1,data2,data3}=Table[
   RandomVariate[NormalDistribution[μ,1],40],
   {μ,{0,4,2}}
   ];
BoxWhiskerChart[
 {data1,data2,data3},
 ChartStyle-
>{Directive[Blue,Opacity[0.4]],Directive[Red,Opacity[0.4]],Directive[White,EdgeForm
[Thickness[0.001]]]},
 ImageSize->350,
 Epilog->{
   Directive[Purple,Opacity[0.4]],
   PointSize[0.016],
   Map[
    Point,{
     Transpose[{ConstantArray[1,Length[data1]]+RandomReal[{-
0.06,0.06},Length[data1]],data1}],
     Transpose[{ConstantArray[2,Length[data2]]+RandomReal[{-
0.06,0.06},Length[data2]],data2}],
```

```
      Transpose[{ConstantArray[3,Length[data3]]+RandomReal[{-
0.06,0.06},Length[data3]],data3}]}
     ]
    }
  ]
```

Output

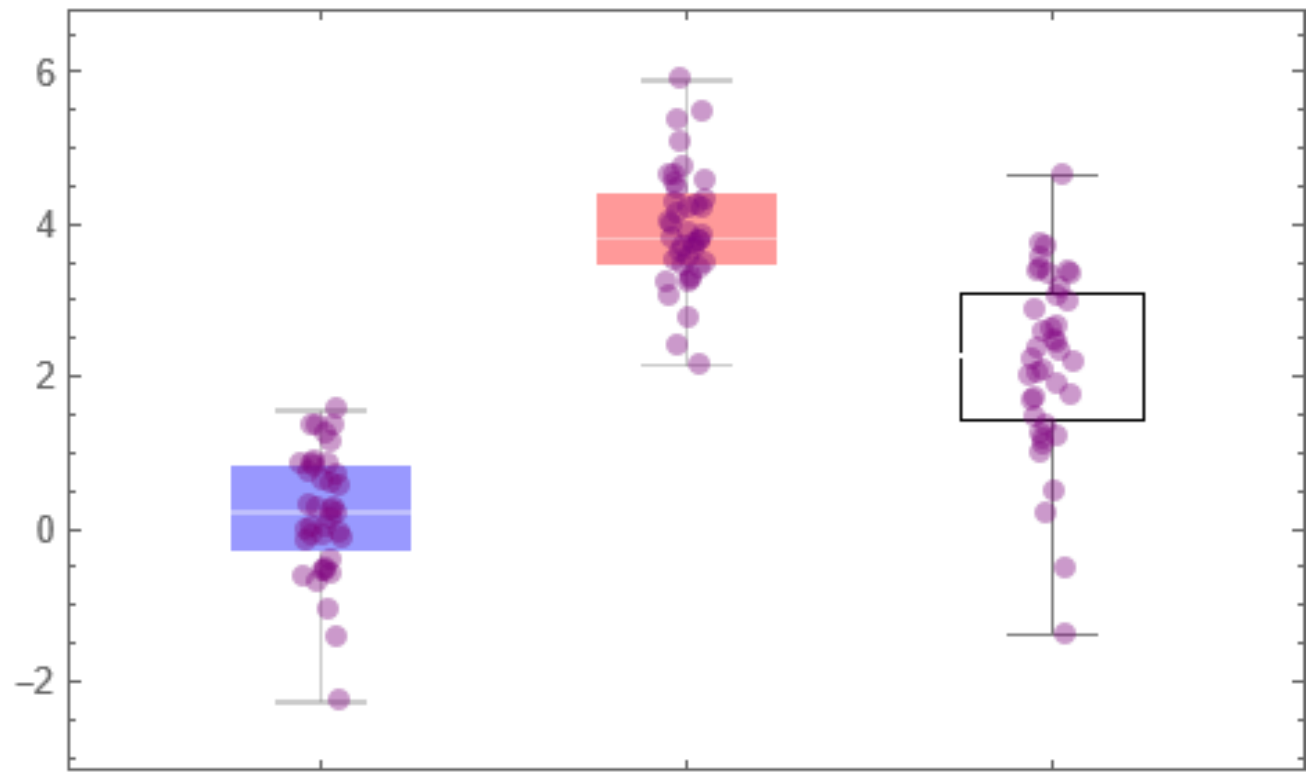

Input

```
(* The code generates three sets of random data (data1, data2, data3) by sampling
from a normal distribution with different means (μ) using RandomVariate. Each set
now consists of 500 data points, which is an increase from the previous 40 data
points. The BoxWhiskerChart function is used to create a box-and-whisker plot, similar
to the previous code. The Epilog option is used to add additional elements to the
chart. In this code, points are overlaid on the chart to represent the individual
data points of each dataset. The points are created using Map and Transpose, similar
to the previous code. However, the random displacement in the x-axis is now increased
to a range of-0.25 to 0.25, which is larger than before. The points are also styled
with a different color, increased opacity, and reduced size compared to the previous
code: *)

{data1,data2,data3}=Table[
   RandomVariate[NormalDistribution[μ,1],500],
   {μ,{0,4,2}}
   ];
BoxWhiskerChart[
 {data1,data2,data3},
 ChartStyle-
>{Directive[Blue,Opacity[0.4]],Directive[Red,Opacity[0.4]],Directive[White,EdgeForm
[Thickness[0.001]]]},
 ImageSize->350,
 Epilog->{
   Directive[Purple,Opacity[0.5]],
   PointSize[0.0045],
   Map[
    Point,{
     Transpose[{ConstantArray[1,Length[data1]]+RandomReal[{-
0.25,0.25},Length[data1]],data1}],
     Transpose[{ConstantArray[2,Length[data2]]+RandomReal[{-
0.25,0.25},Length[data2]],data2}],
     Transpose[{ConstantArray[3,Length[data3]]+RandomReal[{-
0.25,0.25},Length[data3]],data3}]}
     ]
    }
  ]
```

Output

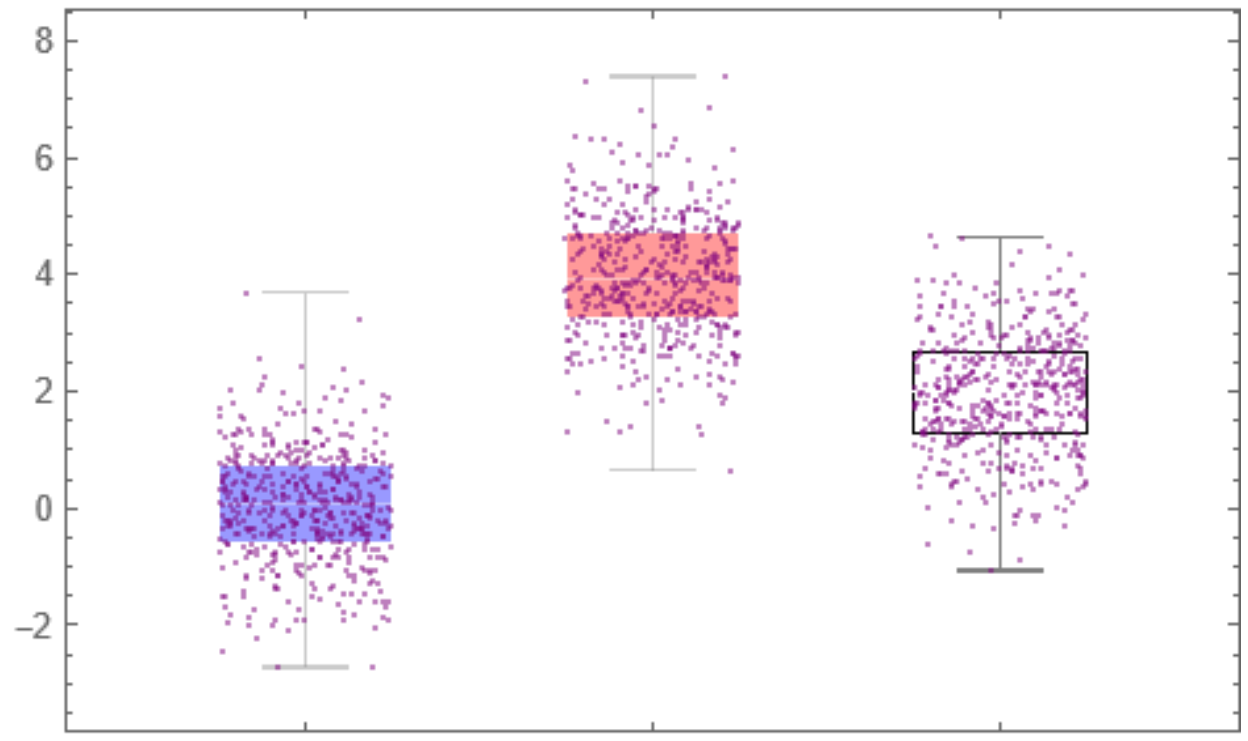

Input

```
(* Similar to the previous code but, in this code, the Epilog contains points overlaid on the chart. Three sets of points are created using Map and Transpose, corresponding to the three datasets. Each point is located at a fixed x-coordinate based on the dataset it belongs to, and the y-coordinate is the actual data value: *)

{data1,data2,data3}=Table[
    RandomVariate[NormalDistribution[μ,1],40],
    {μ,{0,4,2}}
    ];
BoxWhiskerChart[
 {data1,data2,data3},
 ChartStyle-
>{Directive[Blue,Opacity[0.4]],Directive[Red,Opacity[0.4]],Directive[White,EdgeForm
[Thickness[0.001]]]},
 ImageSize->350,
 Epilog->{
    Directive[Purple,Opacity[0.4]],
    PointSize[0.016],
    Map[
     Point,{
      Transpose[{ConstantArray[1,Length[data1]],data1}],
      Transpose[{ConstantArray[2,Length[data2]],data2}],
      Transpose[{ConstantArray[3,Length[data3]],data3}]}
     ]
    }
 ]
```

Output

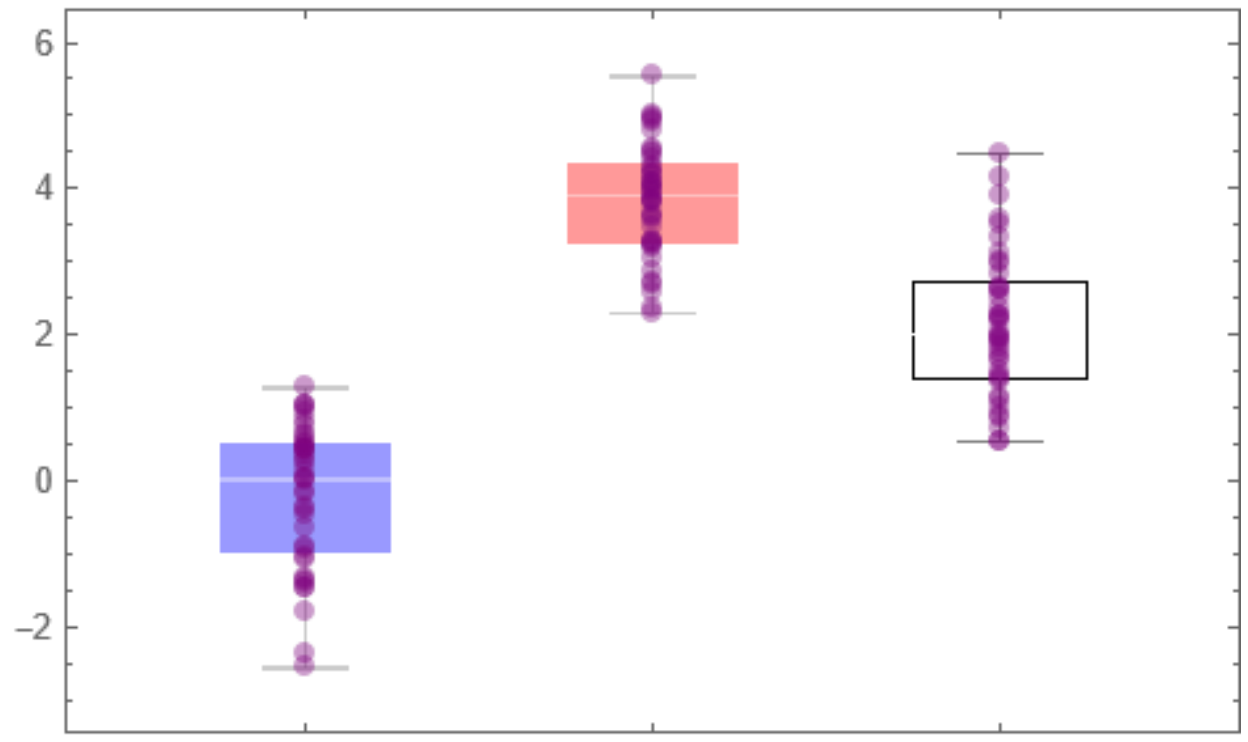

Input

```
(* Similar to the previous code but, in this code, the chart's appearance, including colors, opacity, and point style, is customized to enhance visual clarity: *)

{data1,data2,data3}=Table[
    RandomVariate[NormalDistribution[μ,1],40],
    {μ,{0,4,2}}
```

```
  ];

Show[
 BoxWhiskerChart[
  {data1,data2,data3},
  {
   {"Whiskers",Directive[Thick,Opacity[0.8]]},
   {"Fences",Directive[Thick,Opacity[0.8]]}
   },
  ChartStyle-
>{Directive[Blue,Opacity[0.4]],Directive[Red,Opacity[0.4]],Directive[White,EdgeForm
[Thickness[0.001]]]},
  ImageSize->350
  ],
 ListPlot[
  {
   Transpose[{ConstantArray[1,Length[data1]],data1}],
   Transpose[{ConstantArray[2,Length[data2]],data2}],
   Transpose[{ConstantArray[3,Length[data3]],data3}]
   },
  ImageSize->350,
  PlotMarkers->{"-",PointSize[0.06]}
  ]
 ]
```

Output

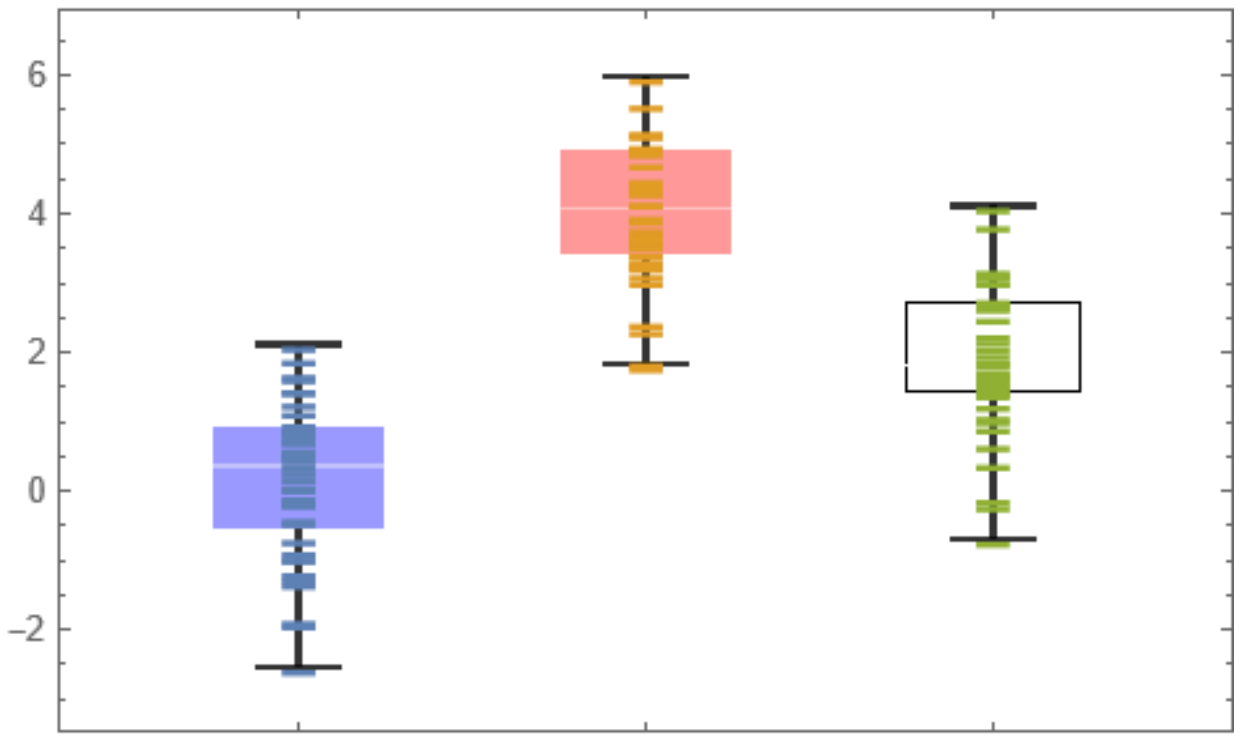

Input

```
(* The code generates multiple density plots using the DistributionChart function. It generates a 2D array of random data called data. It consists of 10 rows and 100 columns, where each element is sampled from a standard normal distribution. The code then uses a combination of Table and DistributionChart to generate density plots for each element in the list of chart element functions. The Joined option is set to "Mean" to connect the mean values of the distributions with a line. The ChartElementFunction option is used to specify different chart element functions, such as "PointDensity" and "LineDensity". *)

data=RandomVariate[NormalDistribution[],{10,100}];

Table[
 DistributionChart[
  data,
  Joined->"Mean",
  ChartElementFunction->s,
  ChartStyle->"Pastel",
  ImageSize->300
  ],
 {s,{"PointDensity","LineDensity"}}
 ]
```

Output

{ 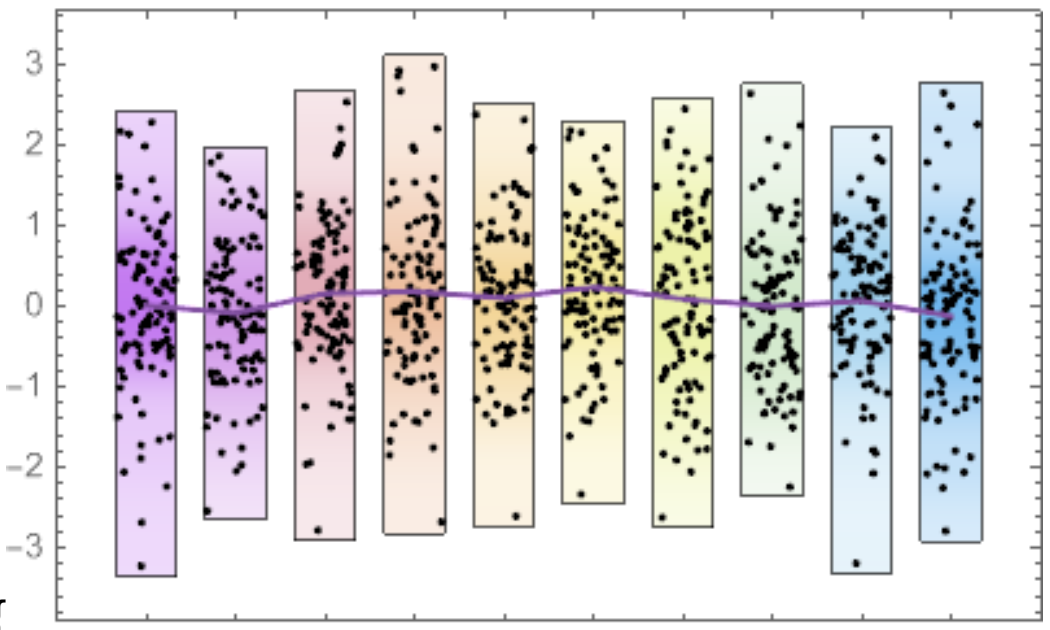

, 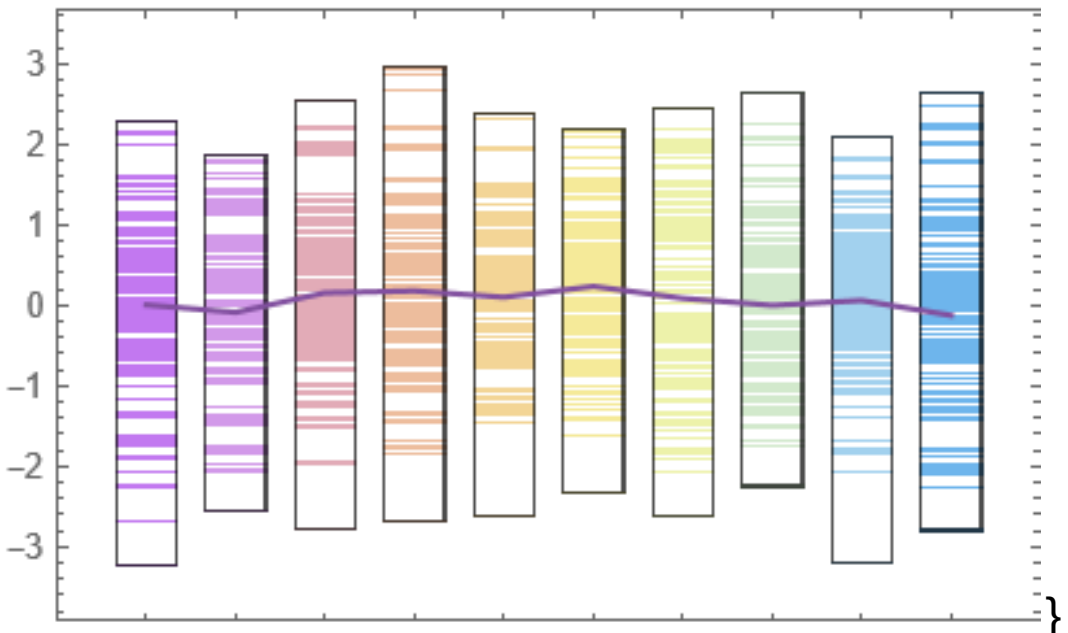

}

# CHAPTER 6

# DESCRIPTIVE STATISTICS: MEASURES OF DISPERSION AND SYMMETRY

Data sets may have the same center but look different because of the way the numbers spread out from the center. The degree to which numerical data tends to spread about an average value is called the dispersion, or variation, of the data. Various measures of this variation are available, the most common being the range, mean deviation, semi-interquartile range, 10–90 percentile range, and standard deviation. On the other hand, measures of symmetry assess the degree to which the dataset is symmetrical or skewed.  The most common measure of symmetry is skewness and kurtosis. In this chapter, we will discuss these two measures.

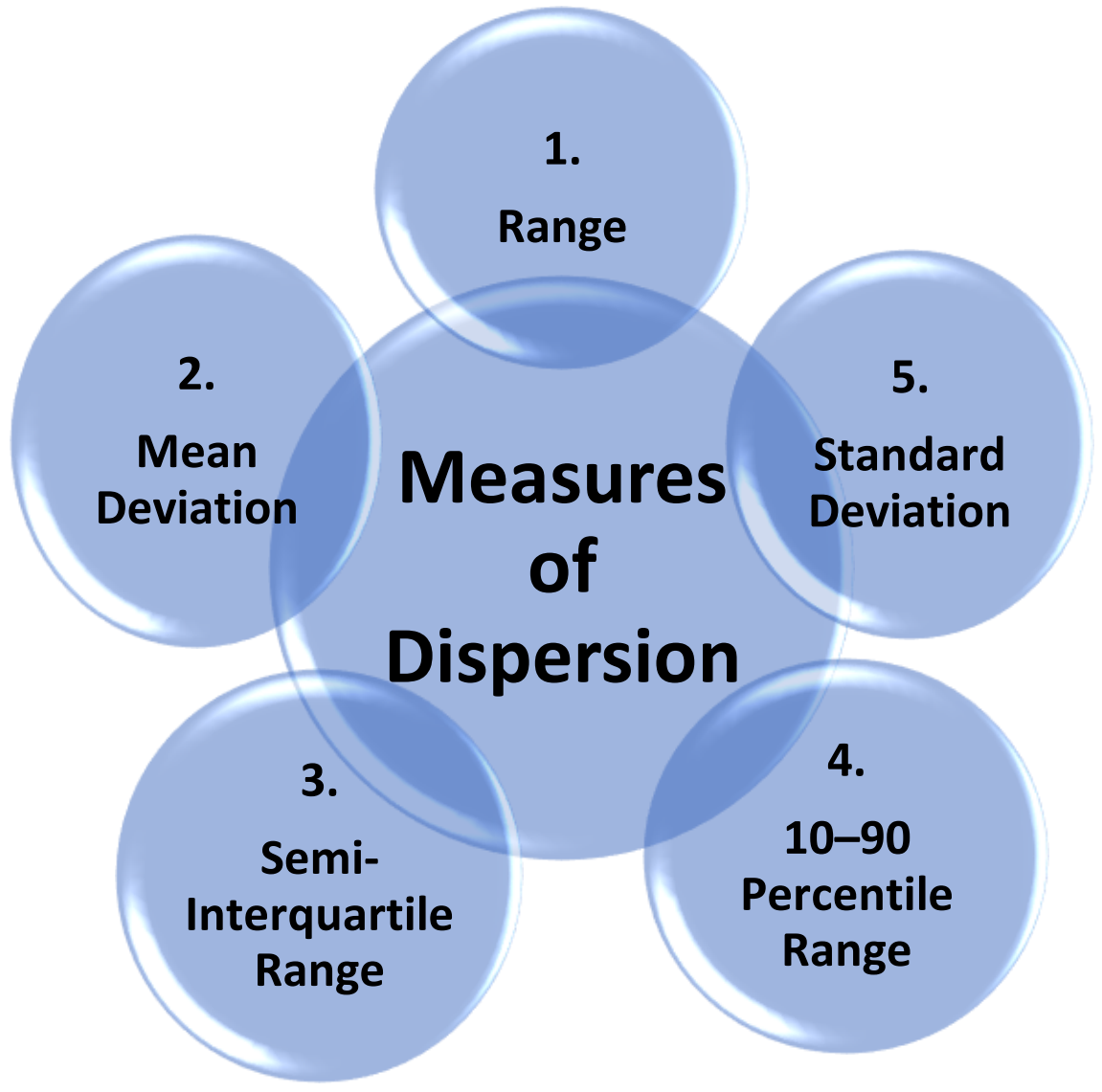

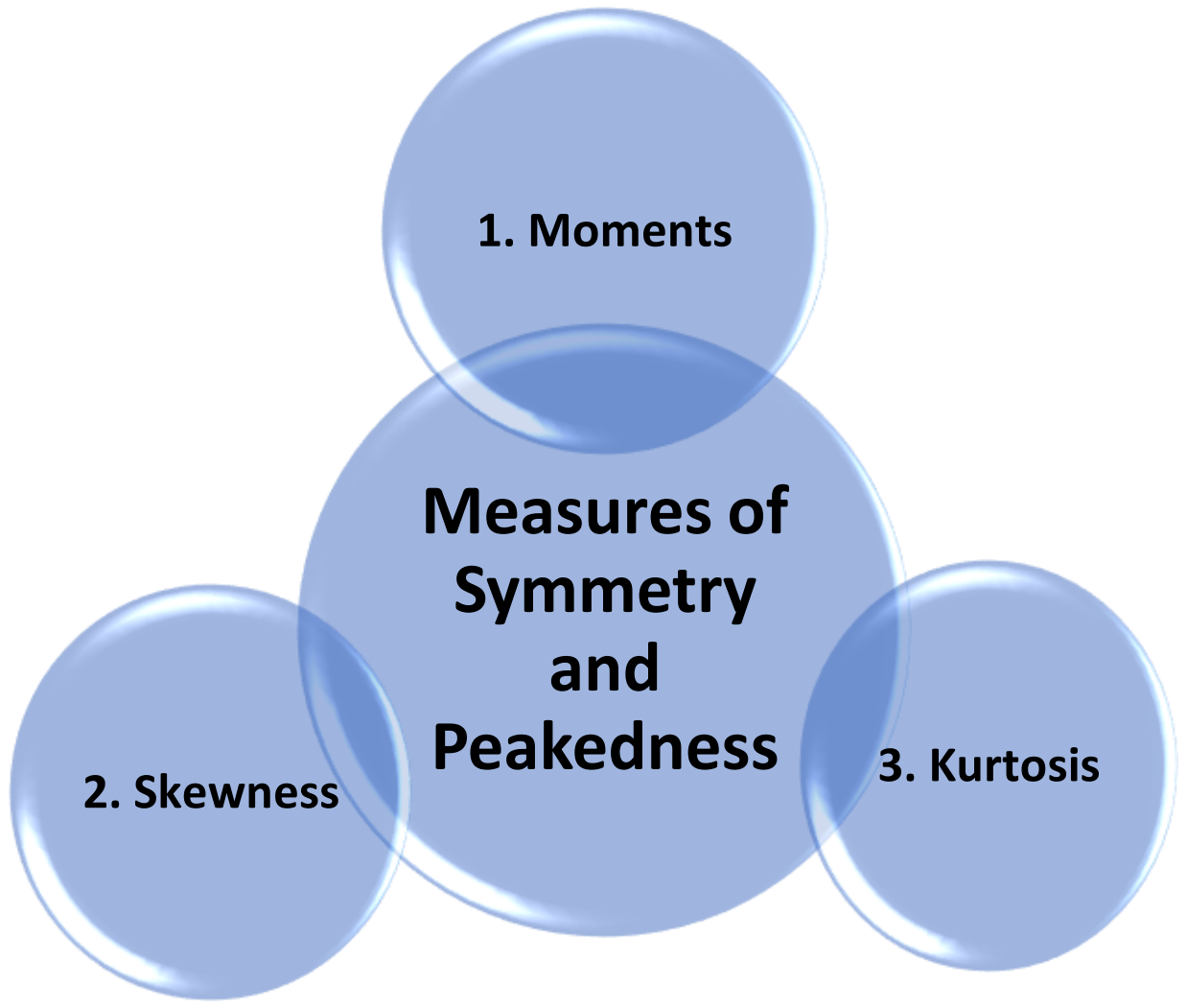

## 6.1 The Standard Deviation and Other Measures of Dispersion

### Range

**Definition (The Range):** The range of a set of numbers is the difference between the largest and smallest numbers in the set.

For instance, the range of the set $1, 3, 3, 5, 5, 5, 8, 10, 14$ is $14 - 1 = 13$.

**Remarks:**

1- The range is a straightforward way to measure the variability of a dataset. It tells us how much the data is spread out from the minimum value to the maximum value. A larger range indicates greater variability or diversity in the dataset, while a smaller range indicates less variability dataset.
2- One limitation of the range is that it only considers the two extreme values in the dataset - the maximum and minimum values. It doesn't consider the distribution of the remaining values, which can be important in understanding the overall shape of the data.
3- Hence, the range may not accurately reflect the spread of the central portion of the data.
4- It is important to note that the range is sensitive to outliers, which are extreme values that are far from the rest of the data. Outliers can distort the range and make it less informative, so it is often useful to complement the range with other measures of dispersion, such as the interquartile range or the standard deviation.

### 10–90 Percentile Range

**Definition (The 10–90 Percentile Range):** The 10–90 percentile range of a set of data is defined by

$$10 - 90 \text{ percentile range} = P_{90} - P_{10}, \quad (6.1)$$

where $P_{10}$ and $P_{90}$ are the 10th and 90th percentiles for the data.

Let us consider the sets (a) $12, 6, 7, 3, 15, 10, 18, 5$ and (b) $9, 3, 8, 8, 9, 8, 9, 18$. In both cases, range=(largest number -smallest number)$= 18 - 3 = 15$. However, as seen from the arrays of sets (a) and (b),

(a) $3, 5, 6, 7, 10, 12, 15, 18$ (b) $3, 8, 8, 8, 9, 9, 9, 18$,

there is much more variation, or dispersion, in (a) than in (b). In fact, (b) consists mainly of $8$'s and $9$'s. Since the range indicates no difference between the sets, it is not a very good measure of dispersion in this case (where extreme values are present, the range is generally a poor measure of dispersion). An improvement is achieved by throwing out the extreme cases, 3 and 18. Then for set (a) the range is $15 - 5 = 10$, while for set (b) the range is $9 - 8 = 1$, clearly showing that (a) has greater dispersion than (b). However, this is not the way the range is defined. The $10 - 90$ percentile range were designed to improve on the range by eliminating extreme cases.

### Semi-Interquartile Range

**Definition (The Semi-Interquartile Range):** The semi-interquartile range, or quartile deviation, of a set of data is denoted by $Q$ and is defined by

$$Q = \frac{Q_3 - Q_1}{2}, \quad (6.2)$$

where $Q_1$ and $Q_3$ are the first and third quartiles for the data.

For instance, let us consider the semi-interquartile range of the set $6, 47, 49, 15, 43, 41, 7, 39, 43, 41, 36$ (or $6, 7, 15, 36, 39, 41, 41, 43, 43, 47, 49$). The rank of the median is 6, which means there are five points on each side. Then we need to split the lower half of the data in two again to find the lower quartile. The lower quartile will be the point of rank $(5 + 1)/2 = 3$. The result is $Q_1 = 15$. The second half must also be split in two to find the value of the upper quartile. The rank of the upper quartile will be $6 + 3 = 9$. So $Q_3 = 43$. The interquartile range will be $Q_3 - Q_1 = 28$. The semi-interquartile range is $14$ and the range is $49 - 6 = 43$.

**Remarks:**

- In some cases, it may be more appropriate to use the full interquartile range (i.e., the difference between the first and third quartiles) instead of the semi-interquartile range. This is particularly true when the data is symmetric or when there are no extreme values.
- It is important to note that, the semi-interquartile range is a measure of the spread of the middle $50\%$ of a dataset. The 10-90 percentile range, on the other hand, is a measure of the spread of the middle $80\%$ of a dataset. Hence, the semi-interquartile range is more resistant to outliers than the $10 - 90$ percentile range, but it is less precise. The $10 - 90$ percentile range is more precise than the semi-interquartile range, but it is less resistant to outliers.

### Mean Absolute Deviation

**Definition (The Mean Absolute Deviation):** The mean absolute deviation, or average deviation, of a set of $N$ numbers $v_1, v_2, v_3, \ldots, v_N$ is abbreviated (MD) and is defined by

$$\text{Mean deviation (MD)} = \frac{\sum_{j=1}^{N} |v_j - \bar{v}|}{N}, \tag{6.3}$$

where $\bar{v}$ is the arithmetic mean of the numbers and $|v_j - \bar{v}|$ is the absolute value of the deviation of $v_j$ from $\bar{v}$. Simply, it is the average distance between each data point and the mean of the dataset (see Figure 6.1).

For instance, the mean deviation of the set 2, 3, 6, 8, 11 is $\frac{|2-6|+|3-6|+|6-6|+|8-6|+|11-6|}{5} = 2.8$, where we use the mean $\bar{v} = \frac{2+3+6+8+11}{5} = 6$.

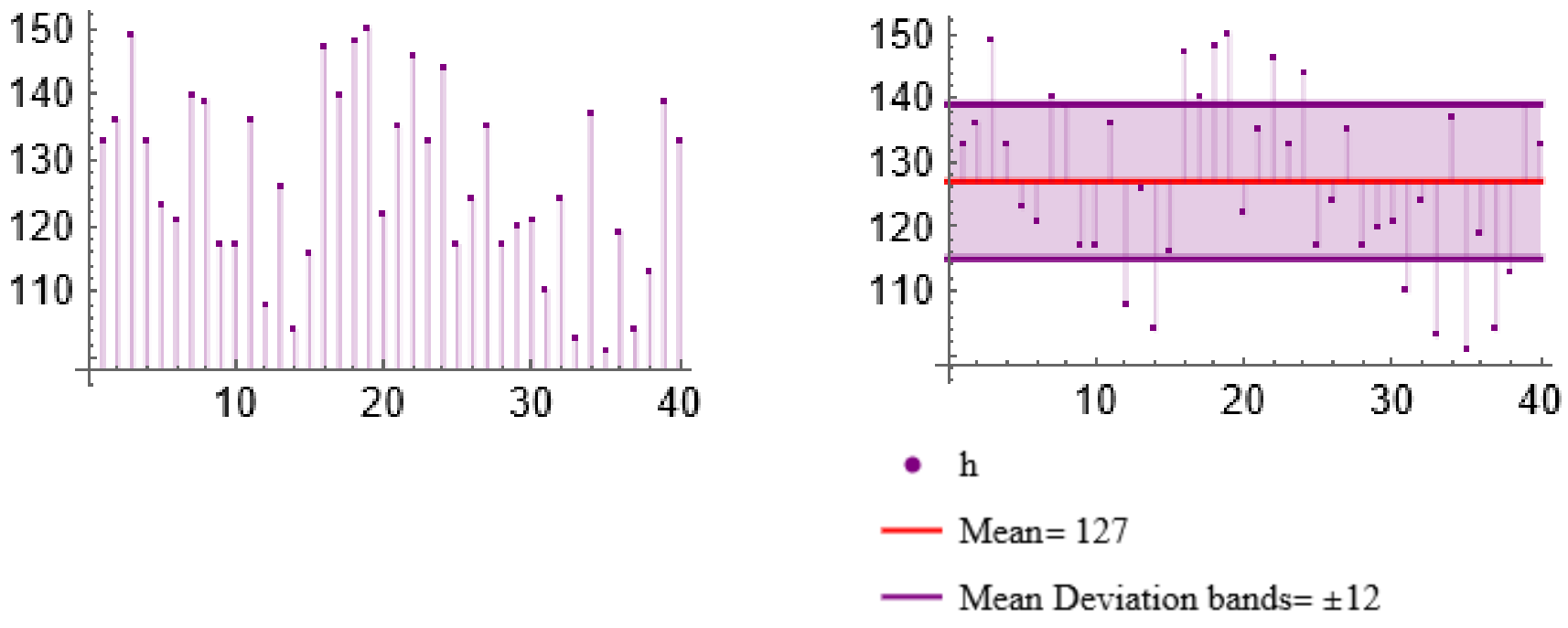

**Figure 6.1** The left plot displays the values in the set $h$ (4.2) as a list plot with the area under the points filled in to the axis. The right plot displays the same set of data, $h$, and horizontal lines at the mean value $\bar{v} = 127$ and mean absolute deviation band. In the right plot, the data points display as a list plot filled up to the mean value to explain how the values in $h$ are distributed around the mean.

If $v_1, v_2, v_3, \ldots, v_K$ occur with frequencies $f_1, f_2, f_3, \ldots, f_K$, respectively, the mean deviation can be written as

$$\text{Mean deviation (MD)} = \frac{\sum_{j=1}^{K} f_j |v_j - \bar{v}|}{N}, \tag{6.4}$$

where $\sum_{j=1}^{K} f_j = N$. This form is useful for grouped data, where the $v_j$'s represent class marks and the $f_j$'s are the corresponding class frequencies. Occasionally the mean deviation is defined in terms of absolute deviations from the median or other average instead of from the mean. An interesting property of the sum $\sum_{j=1}^{N} |v_j - a|$ is that it is a minimum when $a$ is the median (i.e., the mean deviation about the median is a minimum).

***Example 6.1***

Find the mean deviation for the data in the frequency distribution (Table 6.1), where $\bar{x} = 10.80$.

**Table 6.1.** Frequency distribution.

| Class Interval | Frequency $(f)$ | Class mark | $\lvert x - \bar{x} \rvert$ | $f * \lvert x - \bar{x} \rvert$ |
|---|---|---|---|---|
| $0 - 4$ | 7 | 2 | $2 - 10.80 = 8.80$ | 61.6 |
| $4 - 8$ | 4 | 6 | $6 - 10.80 = 4.80$ | 19.2 |
| $8 - 12$ | 19 | 10 | $10 - 10.80 = 0.80$ | 15.2 |
| $12 - 16$ | 12 | 14 | $14 - 10.80 = 3.2$ | 38.4 |
| $16 - 20$ | 8 | 18 | $18 - 10.80 = 7.2$ | 57.6 |
| Total | 50 | | | 192 |

***Solution***

$$\text{Mean deviation (MD)} = \frac{\sum_{j=1}^{K} f_j \lvert v_j - \bar{v} \rvert}{N} = \frac{192}{50} = 3.84.$$

## Standard Deviation

In contrast to the range, the standard deviation considers all the observations. Roughly speaking, the standard deviation measures variation by indicating how far, on average, the observations are from the mean. For a data set with a large amount of variation, the observations will, on average, be far from the mean; so the standard deviation will be large. For a data set with a small amount of variation, the observations will, on average, be close to the mean; so the standard deviation will be small.

**Definition (The Standard Deviation of a Population):**

(a) The standard deviation of a set of $N$ numbers $v_1, v_2, v_3, \dots, v_N$ is denoted by $S$ and is defined by

$$S = \sqrt{\frac{\sum_{j=1}^{N} (v_j - \bar{v})^2}{N}}. \qquad (6.5.1)$$

Thus $S$ is sometimes called, the root-mean-square deviation (see Figure 6.2).

(b) If $v_1, v_2, v_3, \dots, v_K$ occur with frequencies $f_1, f_2, f_3, \dots, f_K$, respectively, the standard deviation can be written

$$S = \sqrt{\frac{\sum_{j=1}^{K} f_j (v_j - \bar{v})^2}{N}}, \qquad (6.6.1)$$

where $\sum_{j=1}^{K} f_j = N$. This form is useful for grouped data.

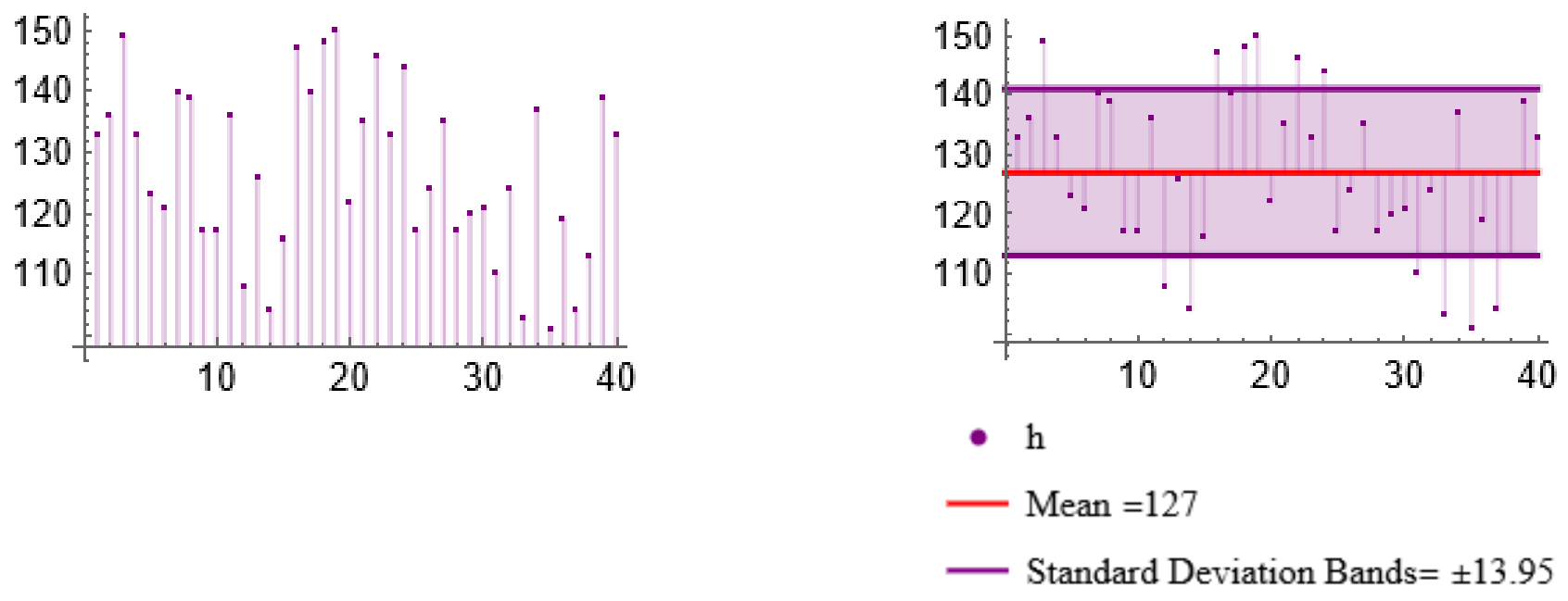

**Figure 6.2** The left plot displays the values in the set $h$ (4.2) as a list plot with the area under the points filled in to the axis. The right plot displays the same set of data, $h$, and horizontal lines at the mean value $\bar{v} = 127$ and standrad deviation band. In the right plot, the data points display as a list plot filled up to the mean value to explain how the values in $h$ are distributed around the mean.

Sometimes the standard deviation of a sample's data is defined with $N-1$ replacing $N$ in the denominators of the expressions in (6.5) and (6.6) because the resulting value represents a better estimate of the standard deviation of a population from which the sample is taken (we will discuss this point in detail in Chapter 18). For large values of $N$ (certainly $N > 30$), there is practically no difference between the two definitions.

**Definition (The Standard Deviation for Sample):**
(a) The standard deviation of a set of $N$ numbers $v_1, v_2, v_3, \dots, v_N$ is denoted by $S$ and is defined by

$$S = \sqrt{\frac{\sum_{j=1}^{N}(v_j - \bar{v})^2}{N-1}}. \tag{6.5.2}$$

Thus $S$ is sometimes called, the root-mean-square deviation.
(b) If $v_1, v_2, v_3, \dots, v_K$ occur with frequencies $f_1, f_2, f_3, \dots, f_K$, respectively, the standard deviation can be written

$$S = \sqrt{\frac{\sum_{j=1}^{K} f_j(v_j - \bar{v})^2}{N-1}}, \tag{6.6.2}$$

where $\sum_{j=1}^{K} f_j = N$. This form is useful for grouped data.

***Example 6.2***

Find the standard deviation for the data in the frequency distribution (Table 6.2), where $\bar{x} = 10.8$.

**Table 6.2.** Frequency distribution.

| Class Interval | Frequency ($f$) | Class mark | $x - \bar{x}$ | $(x-\bar{x})^2$ | $f * (x-\bar{x})^2$ |
|---|---|---|---|---|---|
| $0-4$ | 7 | 2 | $2 - 10.80 = -8.80$ | 77.44 | 542.08 |
| $4-8$ | 4 | 6 | $6 - 10.80 = -4.80$ | 23.04 | 92.16 |
| $8-12$ | 19 | 10 | $10 - 10.80 = -0.80$ | 0.64 | 12.16 |
| $12-16$ | 12 | 14 | $14 - 10.80 = 3.2$ | 10.24 | 122.88 |
| $16-20$ | 8 | 18 | $18 - 10.80 = 7.2$ | 51.84 | 414.72 |
| Total | 50 | | | | 1184 |

***Solution***

$$S = \sqrt{\frac{\sum_{j=1}^{K} f_j\left(v_j - \bar{v}\right)^2}{N-1}}$$

$$= \sqrt{\frac{1184}{49}} = \sqrt{24.16} = 4.92.$$

## Variance

Variance is a measure of the spread of data points around the mean. Variance is a useful tool for understanding and comparing data sets. For example, if you have two data sets with the same mean, but different variances, you can tell that the data points in one data set are more spread out than the data points in the other data set.

**Definition (The Variance):** The variance of a set of data is defined as the square of the standard deviation and is thus given by $S^2$ in (6.5) and (6.6).

For instance, the variances of the data set 3, 4, 6, 7, 10 is 6 (the mean is $(3+4+6+7+10)/5 = 6$ and $S^2 = ((-3)^2 + (-2)^2 + (0)^2 + (1)^2 + (4)^2))/5 = 6$).

When it is necessary to distinguish the standard deviation of a population from the standard deviation of a sample drawn from this population, we often use the symbol $S$ for the latter and $\sigma$ for the former. Thus $S^2$ and $\sigma^2$ would represent the sample variance and population variance, respectively.

**Theorem 6.1:**
(a) The standard deviation of a set of $N$ numbers $v_1, v_2, v_3, \ldots, v_N$ is defined by

$$S = \sqrt{\frac{\sum_{j=1}^{N} v_j^2}{N} - \left(\frac{\sum_{j=1}^{N} v_j}{N}\right)^2} = \sqrt{\overline{v^2} - \bar{v}^2}, \tag{6.7}$$

where $\overline{v^2}$ denotes the mean of the squares of the various values of $v$, while $\bar{v}^2$ denotes the square of the mean of the various values of $v$.
(b) If $v_1, v_2, v_3, \ldots, v_K$ occur with frequencies $f_1, f_2, f_3, \ldots, f_K$, respectively, the standard deviation can be written

$$S = \sqrt{\frac{\sum_{j=1}^{K} f_j v_j^2}{N} - \left(\frac{\sum_{j=1}^{K} f_j v_j}{N}\right)^2} = \sqrt{\overline{v^2} - \bar{v}^2}. \tag{6.8}$$

**Proof:**

(a) By definition, $S = \sqrt{\frac{\sum_{j=1}^{N}(v_j - \bar{v})^2}{N}}$. Then,

$$\begin{aligned}
S^2 &= \frac{\sum_{j=1}^{N}\left(v_j - \bar{v}\right)^2}{N} \\
&= \frac{\sum_{j=1}^{N}\left(v_j^2 - 2v_j\bar{v} + \bar{v}^2\right)}{N} \\
&= \frac{\sum_{j=1}^{N} v_j^2 - 2\bar{v}\sum_{j=1}^{N} v_j + N\bar{v}^2}{N} \\
&= \frac{\sum_{j=1}^{N} v_j^2}{N} - 2\bar{v}\frac{\sum_{j=1}^{N} v_j}{N} + \bar{v}^2 \\
&= \frac{\sum_{j=1}^{N} v_j^2}{N} - 2\bar{v}^2 + \bar{v}^2 \\
&= \frac{\sum_{j=1}^{N} v_j^2}{N} - \bar{v}^2 \\
&= \overline{v^2} - \bar{v}^2.
\end{aligned}$$

(b) Using $S = \sqrt{\frac{\sum_{j=1}^{K} f_j(v_j - \bar{v})^2}{N}}$, we have

$$\begin{aligned}
S^2 &= \frac{\sum_{j=1}^{K} f_j\left(v_j - \bar{v}\right)^2}{N} \\
&= \frac{\sum_{j=1}^{K} f_j\left(v_j^2 - 2v_j\bar{v} + \bar{v}^2\right)}{N} \\
&= \frac{\sum_{j=1}^{K}\left(f_j v_j^2 - 2f_j v_j\bar{v} + f_j\bar{v}^2\right)}{N} \\
&= \frac{\sum_{j=1}^{K} f_j v_j^2 - 2\sum_{j=1}^{K} f_j v_j\bar{v} + \sum_{j=1}^{K} f_j\bar{v}^2}{N} \\
&= \frac{\sum_{j=1}^{K} f_j v_j^2 - 2\bar{v}\sum_{j=1}^{K} f_j v_j + \bar{v}^2\sum_{j=1}^{K} f_j}{N} \\
&= \frac{\sum_{j=1}^{K} f_j v_j^2}{N} - 2\bar{v}\frac{\sum_{j=1}^{K} f_j v_j}{N} + \bar{v}^2\frac{\sum_{j=1}^{K} f_j}{N} \\
&= \frac{\sum_{j=1}^{K} f_j v_j^2}{N} - 2\bar{v}^2 + \bar{v}^2 \\
&= \frac{\sum_{j=1}^{K} f_j v_j^2}{N} - \bar{v}^2
\end{aligned}$$

$$= \frac{\sum_{j=1}^{K} f_j v_j^2}{N} - \left(\frac{\sum_{j=1}^{K} f_j v_j}{N}\right)^2,$$

where $\sum_{j=1}^{K} f_j = N$.

■

**Theorem 6.2:**
If $d_j = v_j - A$ are the deviations of $v_j$ from some arbitrary constant $A$, results (6.5) and (6.6) become, respectively,

$$S = \sqrt{\frac{\sum_{j=1}^{N} d_j^2}{N} - \left(\frac{\sum_{j=1}^{N} d_j}{N}\right)^2} = \sqrt{\overline{d^2} - \bar{d}^2}, \tag{6.9}$$

$$S = \sqrt{\frac{\sum_{j=1}^{K} f_j d_j^2}{N} - \left(\frac{\sum_{j=1}^{K} f_j d_j}{N}\right)^2} = \sqrt{\overline{d^2} - \bar{d}^2}. \tag{6.10}$$

**Proof:**

Since $d_j = v_j - A$, $v_j = d_j + A$, and $\bar{v} = A + \bar{d}$, we have

$$v_j - \bar{v} = (d_j + A) - (A + \bar{d}) = d_j - \bar{d},$$

so that

$$S = \sqrt{\frac{\sum_{j=1}^{K} f_j (v_j - \bar{v})^2}{N}}$$

$$= \sqrt{\frac{\sum_{j=1}^{K} f_j (d_j - \bar{d})^2}{N}}$$

$$= \sqrt{\frac{\sum_{j=1}^{K} f_j (d_j^2 - 2d_j\bar{d} + \bar{d}^2)}{N}}$$

$$= \sqrt{\frac{\sum_{j=1}^{K} (f_j d_j^2 - 2f_j d_j \bar{d} + f_j \bar{d}^2)}{N}}$$

$$= \sqrt{\frac{\sum_{j=1}^{K} f_j d_j^2 - 2\sum_{j=1}^{K} f_j d_j \bar{d} + \sum_{j=1}^{K} f_j \bar{d}^2}{N}}$$

$$= \sqrt{\frac{\sum_{j=1}^{K} f_j d_j^2}{N} - 2\bar{d}\frac{\sum_{j=1}^{K} f_j d_j}{N} + \bar{d}^2 \frac{\sum_{j=1}^{K} f_j}{N}}$$

$$= \sqrt{\frac{\sum_{j=1}^{K} f_j d_j^2}{N} - 2\bar{d}^2 + \bar{d}^2}$$

$$= \sqrt{\frac{\sum_{j=1}^{K} f_j d_j^2}{N} - \bar{d}^2}$$

$$= \sqrt{\frac{\sum_{j=1}^{K} f_j d_j^2}{N} - \left(\frac{\sum_{j=1}^{K} f_j d_j}{N}\right)^2} = \sqrt{\overline{d^2} - \bar{d}^2},$$

where $\sum_{j=1}^{K} f_j = N$, and $\bar{d} = \frac{\sum_{j=1}^{K} f_j d_j}{N}$.

■

**Theorem 6.3:**
When data are grouped into a frequency distribution whose class intervals have equal size $c$, we have $d_j = cu_j$ or $v_j = A + cu_j$ and result (6.10) becomes,

$$S = c\sqrt{\frac{\sum_{j=1}^{K} f_j u_j^2}{N} - \left(\frac{\sum_{j=1}^{K} f_j u_j}{N}\right)^2} = c\sqrt{\overline{u^2} - \bar{u}^2}. \tag{6.11}$$

**Proof:**

Since $d_j = v_j - A = cu_j$. Thus, since $c$ is a constant,

$$\begin{aligned} S &= \sqrt{\frac{\sum_{j=1}^{K} f_j d_j^2}{N} - \left(\frac{\sum_{j=1}^{K} f_j d_j}{N}\right)^2} \\ &= \sqrt{\frac{\sum_{j=1}^{K} f_j (cu_j)^2}{N} - \left(\frac{\sum_{j=1}^{K} f_j (cu_j)}{N}\right)^2} \\ &= \sqrt{\frac{c^2 \sum_{j=1}^{K} f_j u_j^2}{N} - \left(\frac{c \sum_{j=1}^{K} f_j u_j}{N}\right)^2} \\ &= \sqrt{c^2 \frac{\sum_{j=1}^{K} f_j u_j^2}{N} - c^2 \left(\frac{\sum_{j=1}^{K} f_j u_j}{N}\right)^2} \\ &= c\sqrt{\frac{\sum_{j=1}^{K} f_j u_j^2}{N} - \left(\frac{\sum_{j=1}^{K} f_j u_j}{N}\right)^2} \\ &= c\sqrt{\overline{u^2} - \bar{u}^2}. \end{aligned}$$

■

**Remarks:**

1. The standard deviation can be defined as

$$S = \sqrt{\frac{\sum_{j=1}^{N} (v_j - a)^2}{N}}, \tag{6.12}$$

where $a$ is an average besides the arithmetic mean.

2. Suppose that two sets consisting of $N_1$ and $N_2$ numbers have variances given by $S_1^2$ and $S_2^2$, respectively, and have the same mean. Then the combined variance or pooled variance of both sets is given by

$$S^2 = \frac{N_1 S_1^2 + N_2 S_2^2}{N_1 + N_2}. \tag{6.13}$$

***Example 6.3***

Given the sets $\{2, 5, 8, 11, 14\}$, and $\{2, 8, 14\}$, find (a) the mean of each set, (b) the variance of each set, (c) the mean of the combined (or pooled) sets, and (d) the variance of the combined sets.

***Solution***

(a) Mean of first set $= \frac{1}{5}(2 + 5 + 8 + 11 + 14) = 8$. Mean of second set $= \frac{1}{3}(2 + 8 + 14) = 8$.

(b) Variance of first set $= S_1^2 = \frac{1}{5}((2-8)^2 + (5-8)^2 + (8-8)^2 + (11-8)^2 + (14-8)^2) = 18$. Variance of second set $= S_2^2 = \frac{1}{5}((2-8)^2 + (8-8)^2 + (14-8)^2) = 24$.

(c) The mean of the combined sets is

$$\frac{2+5+8+11+14+2+8+14}{5+3} = 8.$$

(d) The variance of the combined sets is

$$S^2 = \frac{(2-8)^2+(5-8)^2+(8-8)^2+(11-8)^2+(14-8)^2+(2-8)^2+(8-8)^2+(14-8)^2}{5+3} = 20.25$$

**Another method**

$$S^2 = \frac{N_1 s_1^2 + N_2 s_2^2}{N_1 + N_2} = \frac{(5)(18)+(3)(24)}{5+3} = 20.25.$$

The actual variation, or dispersion, as determined from the standard deviation or other measure of dispersion is called the absolute dispersion.

**Definition (The Relative Dispersion):** The relative dispersion is defined by

$$\text{Relative dispersion} = \frac{\text{absolute dispersion}}{\text{average}}. \qquad (6.14)$$

**Definition (The Coefficient of Variation):** If the absolute dispersion is the standard deviation $S$ and if the average is the mean $\bar{v}$, then the relative dispersion is called the coefficient of variation; it is denoted by CV and is given by

$$\text{Coefficient of variation (CV)} = \frac{S}{\bar{v}}. \qquad (6.15)$$

**Remarks:**

1- The actual value of the CV is independent of the unit in which the measurement has been taken, so it is a dimensionless number.
2- Relative dispersion or coefficient of variation is a useful measure because it allows us to compare the variability of datasets that have different units of measurement or different scales. For example, if we are comparing the variability of salaries in two different countries, we may find that the standard deviation of salaries is much higher in one country than in the other. However, this may not necessarily mean that the salaries in one country are more variable than the other, as the mean salary may also be much higher in that country. By calculating the relative dispersion, we can compare the variability of the two datasets in a way that takes into account their different scales.
3- Hence, for comparison between data sets with different units or widely different means, one should use the coefficient of variation instead of the standard deviation.
4- A disadvantage of the coefficient of variation is that it fails to be useful when $\bar{v}$ is close to zero.

For example, in the following, we will take the values given as randomly chosen from a larger population of values.
1- The data set $[100, 100, 100]$ has constant values. Its standard deviation is $0$ and average is $100$, giving the coefficient of variation as $0/100 = 0$.
2- The data set $[90, 100, 110]$ has more variability. Its standard deviation is $10$ and its average is $100$, giving the coefficient of variation as $10/100 = 0.1$.
3- The data set $[1, 5, 6, 8, 10, 40, 65, 88]$ has still more variability. Its standard deviation is $32.9$ and its average is $27.9$, giving a coefficient of variation of $32.9/27.9 = 1.18$.

**Definition (The Standardized Variable):** The variable that measures the deviation from the mean in units of the standard deviation is called a standardized variable, is a dimensionless quantity and is given by

$$z = \frac{v - \bar{v}}{S}. \qquad (6.16)$$

**Definition (The Standard Scores):** If the deviations from the mean are given in units of the standard deviation, they are said to be expressed in standard units, or standard scores.

*Example 6.4*

A student received a grade of 84 on a final examination in mathematics for which the mean grade was 76 and the standard deviation was 10. On the final examination in physics, for which the mean grade was 82 and the standard deviation was 16, she received a grade of 90. In which subject was her relative standing higher?

***Solution***

The standardized variable $z = \frac{v-\bar{v}}{S}$ measures the deviation of $v$ from the mean $\bar{v}$ in terms of standard deviation $S$. For mathematics, $z = 0.8$; for physics, $z = 0.5$. Thus, the student had a grade 0.8 of a standard deviation above the mean in mathematics, but only 0.5 of a standard deviation above the mean in physics. Thus, her relative standing was higher in mathematics.

### Trimmed and Winsorized Variance

One limitation of variance is that it can be affected by outliers or extreme values in the dataset. In such cases, alternative measures such as trimmed variance or winsorized variance may be more appropriate.

**Definition (The Trimmed Variance):** In trimmed variance, a certain percentage of the data points at the very top and very bottom of the distribution are trimmed or removed. For example, a 10% trimmed variance would remove the top 10% and bottom 10% of the data points.

**Definition (The Winsorized Variance):** In Winsorized variance, a certain percentage of the data points at the very top and very bottom of the distribution are replaced by the value of the nearest data point that is not trimmed. For example, a 10% Winsorized variance would replace the top 10% and bottom 10% of the data points with the value of the 10th and 90th percentiles, respectively.

## 6.2 Moments, Skewness, and Kurtosis

### Moment

**Definition (The moment):** If $v_1, v_2, v_3, \dots, v_N$ are the $N$ values assumed by the variable $v$, we define the quantity,

$$\overline{v^r} = \frac{v_1^r + v_2^r + \cdots + v_N^r}{N} = \frac{\sum_{j=1}^{N} v_j^r}{N}, \quad (6.17)$$

called the $r$th moment. The first moment with $r = 1$ is the arithmetic mean $\bar{v}$.

The $r$th moment about the mean $\bar{v}$ is defined as

$$m_r = \frac{\sum_{j=1}^{N} (v_j - \bar{v})^r}{N}. \quad (6.18)$$

If $r = 1$, then $m_1 = 0$. If $r = 2$, then $m_2 = S^2$, the variance.

The $r$th moment about any origin $A$ is defined as

$$m_r' = \frac{\sum_{j=1}^{N} (v_j - A)^r}{N}. \quad (6.19)$$

If $A = 0$, (6.19) reduces to equation (6.17). If $r = 1$, $m_1' = \frac{\sum_{j=1}^{N} v_j - A}{N} = \frac{\sum_{j=1}^{N} v_j}{N} - \frac{AN}{N} = \bar{v} - A$.

If $v_1, v_2, v_3, \dots, v_K$ occur with frequencies $f_1, f_2, f_3, \dots, f_K$, respectively, the above moments are given by

$$\overline{v^r} = \frac{f_1 v_1^r + f_2 v_2^r + \cdots + f_K v_K^r}{N} = \frac{\sum_{j=1}^{K} f_j v_j^r}{N}, \tag{6.20.1}$$

$$m_r = \frac{\sum_{j=1}^{K} f_j (v_j - \bar{v})^r}{N}, \tag{6.20.2}$$

$$m'_r = \frac{\sum_{j=1}^{K} f_j (v_j - A)^r}{N} = \frac{\sum_{j=1}^{K} f_j d_j^r}{N}, \tag{6.20.3}$$

where $\sum_{j=1}^{K} f_j = N$ and $d_j = v_j - A$.

***Example 6.5***

1. Find the first four moments of the set 1, 3, 5, 6, 11, 14.
2. Find the first four moments about the mean for the set 1, 3, 5, 6, 11, 14.
3. Find the first four moments about the origin 6 for the set 1, 3, 5, 6, 11, 14.

***Solution***

1. The first moment, or arithmetic mean, is

$$\bar{v} = \frac{\sum_i v_i}{N} = \frac{1+3+5+6+11+14}{6} = 6.66667.$$

The second moment is

$$\overline{v^2} = \frac{\sum_i v_i^2}{N} = \frac{1^2+3^2+5^2+6^2+11^2+14^2}{6} = 64.6667.$$

The third moment is

$$\overline{v^3} = \frac{\sum_i v_i^3}{N} = \frac{1^3+3^3+5^3+6^3+11^3+14^3}{6} = 740.667.$$

The fourth moment is

$$\overline{v^4} = \frac{\sum_i v_i^4}{N} = \frac{1^4+3^4+5^4+6^4+11^4+14^4}{6} = 9176.67.$$

2.

$$m_1 = \frac{\sum_i v_i - \bar{v}}{N} = \frac{(1-6.7)+(3-6.7)+(5-6.7)+(6-6.7)+(11-6.7)+(14-6.7)}{6} = 0$$

$$m_2 = \frac{\sum_i (v_i - \bar{v})^2}{N} = \frac{(1-6.7)^2+(3-6.7)^2+(5-6.7)^2+(6-6.7)^2+(11-6.7)^2+(14-6.7)^2}{6} = 20.22$$

$$m_3 = \frac{\sum_i (v_i - \bar{v})^3}{N} = \frac{(1-6.7)^3+(3-6.7)^3+(5-6.7)^3+(6-6.7)^3+(11-6.7)^3+(14-6.7)^3}{6} = 39.92$$

$$m_4 = \frac{\sum_i (v_i - \bar{v})^4}{N} = \frac{(1-6.7)^4+(3-6.7)^4+(5-6.7)^4+(6-6.7)^4+(11-6.7)^4+(14-6.7)^4}{6} = 744.1$$

3.

$$m'_1 = \frac{\sum_i v_i - 6}{N} = \frac{(1-6)+(3-6)+(5-6)+(6-6)+(11-6)+(14-6)}{6}.$$

$$m'_2 = \frac{\sum_i (v_i - 6)^2}{N} = \frac{(1-6)^2+(3-6)^2+(5-6)^2+(6-6)^2+(11-6)^2+(14-6)^2}{6}.$$

$$m'_3 = \frac{\sum_i (v_i - 6)^3}{N} = \frac{(1-6)^3+(3-6)^3+(5-6)^3+(6-6)^3+(11-6)^3+(14-6)^3}{6}.$$

$$m'_4 = \frac{\sum_i (v_i - 6)^4}{N} = \frac{(1-6)^4+(3-6)^4+(5-6)^4+(6-6)^4+(11-6)^4+(14-6)^4}{6}.$$

**Theorem 6.4:**

The following relations exist between moments about the mean $m_r$ and moments about an arbitrary origin $m'_r$:

$$m_2 = m'_2 - m'^2_1, \tag{6.21.1}$$

$$m_3 = m'_3 - 3m'_1 m'_2 + 2m'^3_1, \tag{6.21.2}$$

$$m_4 = m'_4 - 4m'_1 m'_3 + 6m'^2_1 m'_2 - 3m'^4_1. \tag{6.21.3}$$

**Proof:**

By definition,

$$\begin{aligned} m_r &= \frac{1}{N}\sum_{j=1}^{K} f_j\left(v_j - \bar{v}\right)^r \\ &= \frac{1}{N}\sum_{j=1}^{K} f_j\left(v_j - A + A - \bar{v}\right)^r \\ &= \frac{1}{N}\sum_{j=1}^{K} f_j(d_j + A - \bar{v})^r, \end{aligned}$$

where $d_j = v_j - A$. Since,

$$\begin{aligned} m_1' &= \frac{\sum_{j=1}^{K} f_j\left(v_j - A\right)}{N} \\ &= \frac{\sum_{j=1}^{K}\left(f_j v_j - f_j A\right)}{N} \\ &= \frac{\sum_{j=1}^{K} f_j v_j - \sum_{j=1}^{K} f_j A}{N} = \bar{v} - A. \end{aligned}$$

We have

$$\begin{aligned} m_r &= \frac{1}{N}\sum_{j=1}^{K} f_j\left(d_j + A - \bar{v}\right)^r \\ &= \frac{1}{N}\sum_{j=1}^{K} f_j\left(d_j - m_1'\right)^r \\ &= \frac{1}{N}\sum_{j=1}^{K} f_j\left(d_j^r - {}^rC_1\ d_j^{r-1}m_1' + {}^rC_2\ d_j^{r-2}{m_1'}^2 - {}^rC_3\ d_j^{r-3}{m_1'}^3 + \cdots + (-1)^r {m_1'}^r\right) \\ &= \sum_{j=1}^{K}\left(\frac{1}{N} f_j d_j^r - {}^rC_1\ \frac{1}{N} f_j d_j^{r-1}m_1' + {}^rC_2\ \frac{1}{N} f_j d_j^{r-2}{m_1'}^2 - {}^rC_3\ \frac{1}{N} f_j d_j^{r-3}{m_1'}^3 + \cdots + (-1)^r \frac{1}{N} f_j {m_1'}^r\right) \\ &= m_r' - {}^rC_1\ m_{r-1}'m_1' + {}^rC_2\ m_{r-2}'{m_1'}^2 - \cdots + (-1)^r {m_1'}^r. \end{aligned}$$

In particular, on putting $r = 2$, 3 and 4, we get (6.21).

■

The coding method can also be used to provide a short method for computing moments. This method uses the fact that $v_j = A + cu_j$, (The class intervals have equal size $c$, $d_j = cu_j$, where $u_j = 0, \pm 1, \pm 2, \pm 3, \ldots$), so that from equation (6.20.3) we have

$$\begin{aligned} m_r' &= \frac{\sum_{j=1}^{K} f_j\left(v_j - A\right)^r}{N} \\ &= \frac{\sum_{j=1}^{K} f_j\left(cu_j\right)^r}{N} \\ &= c^r \frac{\sum_{j=1}^{K} f_j u_j^r}{N}, \end{aligned} \tag{6.22}$$

which can be used to find $m_r$ by applying equations (6.21).

To avoid particular units, we can define the dimensionless moments about the mean as

$$a_r = \frac{m_r}{S^r} = \frac{m_r}{(\sqrt{m_2})^r} = \frac{m_r}{\sqrt{m_2^r}}, \tag{6.23}$$

where $S = \sqrt{m_2}$ is the standard deviation. Since $m_1 = 0$ and $m_2 = S^2$, we have $a_1 = 0$ and $a_2 = 1$.

***Example 6.6***

Calculate the first four moments distribution about the mean.

| $v_j$ | 0 | 1 | 2 | 3 | 4 | 5 | 6 | 7 | 8 |
|---|---|---|---|---|---|---|---|---|---|
| $f_j$ | 1 | 8 | 28 | 56 | 70 | 56 | 28 | 8 | 1 |

***Solution***

**Table 6.3.** Frequency distribution.

| $v_j$ | $f_j$ | $u_j = v_j - 4$ | $f_j u_j$ | $f_j u_j^2$ | $f_j u_j^3$ | $f_j u_j^4$ |
|---|---|---|---|---|---|---|
| 0 | 1 | −4 | −4 | 16 | −64 | 256 |
| 1 | 8 | −3 | −24 | 72 | −216 | 648 |
| 2 | 28 | −2 | −56 | 112 | −224 | 448 |
| 3 | 56 | −1 | −56 | 56 | −56 | 56 |
| 4 | 70 | 0 | 0 | 0 | 0 | 0 |
| 5 | 56 | 1 | 56 | 56 | 56 | 56 |
| 6 | 28 | 2 | 56 | 112 | 224 | 448 |
| 7 | 8 | 3 | 24 | 72 | 216 | 648 |
| 8 | 1 | 4 | 4 | 16 | 64 | 256 |
| Total | 256 | 0 | 0 | 512 | 0 | 2816 |

Using Table 6.3, moments about the point $x = 4$ are

$$m_1' = \frac{\sum_j f_j u_j}{N} = 0,$$
$$m_2' = \frac{\sum_j f_j u_j^2}{N} = \frac{512}{256} = 2,$$
$$m_3' = \frac{\sum_j f_j u_j^3}{N} = 0,$$
$$m_4' = \frac{\sum_j f_j u_j^4}{N} = \frac{2816}{256} = 11.$$

Moments about mean are:

$$m_1 = 0,$$
$$m_2 = m_2' - m_1'^2 = 2,$$
$$m_3 = m_3' - 3m_1'm_2' + 2m_1'^3 = 0,$$
$$m_4 = m_4' - 4m_1'm_3' + 6m_1'^2 m_2' - 3m_1'^4 = 11.$$

## Skewness

**Definition (The Skewness):** Skewness is the degree of asymmetry, or departure from symmetry, of a distribution.

Remember, if the frequency curve (smoothed frequency polygon) of a distribution has a longer tail to the right of the central maximum than to the left, the distribution is said to be skewed to the right, or to have positive skewness. If the reverse is true, it is said to be skewed to the left, or to have negative skewness.

For skewed distributions, the mean tends to lie on the same side of the mode as the longer tail. Thus, a measure of the asymmetry is supplied by the difference: mean–mode. This can be made dimensionless if we divide it by a measure of dispersion, such as the standard deviation, leading to the definition

$$S_K = \text{Skewness} = \frac{\text{mean} - \text{mode}}{\text{standard deviation}} = \frac{\bar{v} - \text{mode}}{S}. \qquad (6.24)$$

To avoid using the mode, we can define

$$S_K = \text{Skewness} = \frac{3(\text{mean} - \text{median})}{\text{standard deviation}} = \frac{3(\bar{v} - \text{median})}{S}. \qquad (6.25)$$

Equations (6.24) and (6.25) are called, respectively, Pearson's first and second coefficients of skewness. The limits for Karl Pearson's coefficient of skewness are $\pm 3$.

Other measures of skewness, (Bowley's Coefficient of Skewness), defined in terms of quartiles and percentiles, are as follows:

$$\text{Quartile coefficient of skewness} = S_K = \frac{(Q_3 - Q_2) - (Q_2 - Q_1)}{(Q_3 - Q_1)} = \frac{Q_3 - 2Q_2 + Q_1}{Q_3 - Q_1}, \quad (6.26)$$

$$10 - 90 \text{ percentile coefficient of skewness} = S_K = \frac{(P_{90} - P_{50}) - (P_{50} - P_{10})}{(P_{90} - P_{10})} = \frac{P_{90} - 2P_{50} + P_{10}}{P_{90} - P_{10}}. \quad (6.27)$$

From (6.27) we observe that $S_K = 0$, if $Q_3 - Q_2 = Q_2 - Q_1$. This implies that for a symmetrical distribution ($S_K = 0$), median is equidistant from the upper and lower quartiles. Moreover, skewness is positive if:

$$Q_3 - Q_2 > Q_2 - Q_1 \Rightarrow Q_3 + Q_1 > 2Q_2, \quad (6.28)$$

and skewness is negative if

$$Q_3 - Q_2 < Q_2 - Q_1 \Rightarrow Q_3 + Q_1 < 2Q_2. \quad (6.29)$$

We know that for two real positive numbers $a$ and $b$ (i.e., $a > 0$ and $b > 0$), the moduls value of the difference $(a - b)$ is always less than or equal to the modules value of the sum $(a + b)$, i.e.,

$$|a - b| \leq |a + b| \Rightarrow \left|\frac{a - b}{a + b}\right| \leq 1. \quad (6.30)$$

We also know that $(Q_3 - Q_2)$ and $(Q_2 - Q_1)$ are both non-negative. Thus, taking $a = Q_3 - Q_2$ and $b = Q_2 - Q_1$, in (6.30), we get

$$\left|\frac{(Q_3 - Q_2) - (Q_2 - Q_2)}{(Q_3 - Q_2)}\right| \leq 1 \Rightarrow |\, S_K(\text{Bowley})| \leq 1$$
$$\Rightarrow -1 \leq S_K(\text{Bowley}) \leq 1. \quad (6.31)$$

Thus, Bowley's coefficient of skewness ranges from $-1$ to 1.

An important measure of skewness uses the third moment about the mean expressed in dimensionless form and is given by

$$\text{Moment coefficient of skewness} = a_3 = \frac{m_3}{S^3} = \frac{m_3}{(\sqrt{m_2})^3} = \frac{m_3}{\sqrt{m_2^3}}. \quad (6.32)$$

Another measure of skewness is sometimes given by $b_1 = a_3^2$. For perfectly symmetrical curves, such as the normal curve, $a_3$ and $b_1$ are zero.

***Example 6.7***

Table 6.4 shows the frequency distribution of the weekly wages of 65 employees at the Orange Company.

**Table 6.4.** Frequency distribution.

| Wages | Number of employees |
|---|---|
| \$250.00 – \$259.99 | 8 |
| \$260.00 – \$269.99 | 10 |
| \$270.00 – \$279.99 | 16 |
| \$280.00 – \$289.99 | 14 |
| \$290.00 – \$299.99 | 10 |
| \$200.00 – \$209.99 | 5 |
| \$210.00 – \$219.99 | 2 |

1. Find Pearson's (a) first and (b) second coefficients of skewness for the wage distribution of the 65 employees at the Orange Company.
2. Find the (a) quartile and (b) percentile coefficients of skewness for the above frequency distribution.

***Solution***

1.

Mean = \$279.76, median = \$279.06, mode = \$277.50, and standard deviation $S$ = \$15.60. Thus:

First coefficient of skewness $= \frac{\text{mean}-\text{mode}}{S} = \frac{\$279.76-\$277.50}{\$15.60} = 0.1448.$

Second coefficient of skewness $= \frac{3(\text{mean}-\text{median})}{S} = \frac{3(\$279.76-\$279.06)}{\$15.60} = 0.1346.$

2.
$Q_1 = \$268.25, Q_2 = P_{50} = \$279.06, Q_3 = \$290.75, P_{10} = D_1 = \$258.12$, and $P_{90} = D_9 = \$301.00$: Thus:

Quartile coefficient of skewness $= \frac{Q_3-2Q_2+Q_1}{Q_3-Q_1} = \frac{\$290.75-2(\$279.06)+\$268.25}{\$290.75-\$268.25} = 0.0391.$

Percentile coefficient of skewness $= \frac{P_{90}-2P_{50}+P_{10}}{P_{90}-P_{10}} = \frac{\$301.00-2(\$279.06)+\$258.12}{\$301.00-\$258.12} = 0.0233.$

## Kurtosis

If we know the measures of central tendency, dispersion and skewness, we still cannot form a complete idea about the distribution. In addition to these measures, we should know one more measure which Prof. Karl Pearson calls the "Convexity of curve or Kurtosis". Kurtosis enables us to have an idea about the flatness or peakedness of the curve,

**Definition (The Kurtosis):** Kurtosis is the degree of peakedness of a distribution, usually taken relative to a normal distribution.

Kurtosis is based on the size of a distribution's tails. Positive kurtosis indicates too few observations in the tails, whereas negative kurtosis indicates too many observations in the tail of the distribution.

A distribution having a relatively high peak is called leptokurtic, while one which is flat-topped is called platykurtic. A normal distribution, which is not very peaked or very flat-topped, is called mesokurtic. One measure of kurtosis uses the fourth moment about the mean expressed in dimensionless form and is given by

$$\text{Moment coefficient of kurtosis} = a_4 = \frac{m_4}{S^4} = \frac{m_4}{m_2^2}, \tag{6.33}$$

which is often denoted by $b_2$. For the normal distribution, $b_2 = a_4 = 3$. For this reason, the kurtosis is sometimes defined by $b_2 - 3$, which is positive for a leptokurtic distribution, negative for a platykurtic distribution, and zero for the normal distribution.

Another measure of kurtosis is based on both quartiles and percentiles and is given by

$$\kappa = \frac{Q}{P_{90} - P_{10}}, \tag{6.34}$$

where $Q = (Q_3 - Q_1)/2$ is the semi-interquartile range. We refer to $\kappa$ (the lowercase Greek letter kappa) as the percentile coefficient of kurtosis; for the normal distribution, $\kappa$ has the value 0.263.

***Example 6.8***

Find fourth moment of kurtosis for the data set: 26, 12, 16, 56, 112, 24.

***Solution***

$$\text{Mean} = \frac{26 + 12 + 16 + 56 + 112 + 24}{6} = 41.$$

$$m_2 = \frac{[(26-41)^2 + (12-41)^2 + (16-41)^2 + (56-41)^2 + (112-41)^2 + (24-41)^2]}{6} = 1207.67.$$

$$m_4 = \frac{[(26-41)^4 + (12-41)^4 + (16-41)^4 + (56-41)^4 + (112-41)^4 + (24-41)^4]}{6} = 4449059.67.$$

$$a_4 = \frac{m_4}{m_2^2} = \frac{4449059.667}{(1207.667)^2} = 3.05.$$

### Notations:

When it is necessary to distinguish a sample's moments, measures of skewness, and measures of kurtosis from those corresponding to a population of which the sample is a part, it is often the custom to use Latin symbols for the former and Greek symbols for the latter. Thus, if the sample's moments are denoted by $m_r$ and $m_r'$, the corresponding Greek symbols would be $\mu_r$ and $\mu_r'$. Subscripts are always denoted by Latin symbols. Similarly, if the sample's measures of skewness and kurtosis are denoted by $a_3$ and $a_4$, respectively, the population's skewness and kurtosis would be $\alpha_3$ and $\alpha_4$. We already know from that the standard deviation of a sample and of a population are denoted by $S$ and $\sigma$, respectively.

# CHAPTER 7

# MATHEMATICA LAB: DESCRIPTIVE STATISTICS PART 3

Mathematica offers various functions to compute dispersion and shape measures.

- The `InterquartileRange` function calculates the range between the upper quartile and the lower quartile in a dataset. It is useful for identifying the spread or dispersion of the middle 50% of the data.
- The `QuartileDeviation` function calculates the semi-interquartile range, which is half of the Interquartile Range. It provides a measure of dispersion around the median and is less affected by extreme values.
- The `MeanDeviation` function computes the average absolute deviation of each data point from the mean. It gives an indication of the average distance between individual data points and the mean. `MeanDeviation` is less influenced by extreme values and provides a robust measure of dispersion.
- The `StandardDeviation` function calculates the standard deviation, which is a widely used measure of dispersion. It quantifies the amount of variation or spread in a dataset by measuring the average distance between each data point and the mean. A higher standard deviation indicates greater variability.
- The `Variance` function computes the average squared deviation of each data point from the mean. It provides a measure of the overall variability in a dataset.
- The `TrimmedVariance` function calculates the variance after trimming a certain percentage of extreme values from both ends of the dataset. Trimming reduces the impact of outliers and extreme values on the variance calculation, providing a more robust measure of dispersion.
- The `WinsorizedVariance` function is similar to `TrimmedVariance`, but instead of removing extreme values, it replaces them with values closer to the mean.
- The `Moment` function computes the nth moment of a dataset.
- The `CentralMoment` function calculates the nth central moment of a dataset. It measures the dispersion of data around the mean.
- `FactorialMoment` function computes the nth factorial moment of a dataset.
- The `Skewness` function measures the asymmetry of a dataset's distribution. It indicates whether the dataset is skewed to the left (negative skewness) or to the right (positive skewness) relative to the mean.
- The `QuartileSkewness` function is a measure of skewness based on quartiles.
- The `Kurtosis` function measures the peakedness or flatness of a dataset's distribution. It provides insights into the tail behavior and presence of outliers.

Therefore, we divided this chapter into two units to cover the following topics, dispersion statistics and shape statistics. In the following table, we list the built-in functions that are used in this chapter.

| Dispersion Statistics | | Shape Statistics | |
|---|---|---|---|
| `InterquartileRange` | `Variance` | `Moment` | `QuartileSkewness` |
| `QuartileDeviation` | `TrimmedVariance` | `CentralMoment` | `Kurtosis` |
| `MeanDeviation` | `WinsorizedVariance` | `FactorialMoment` | |
| `StandardDeviation` | | `Skewness` | |

**Chapter 7 Outline**

# UNIT 7.1

# DISPERSION STATISTICS

Dispersion statistics summarize the scatter or spread of the data. Most of these functions describe deviation from a particular location. For instance, variance is a measure of deviation from the mean. Mathematica provides a set of functions and tools for calculating, visualizing, and analyzing dispersion statistics, allowing users to gain deeper insights into the variability and distribution of their data. Let us go through them in detail.

| | |
|---|---|
| `InterquartileRange[list]` | gives the difference between the upper and lower quartiles for the elements in list. |
| `InterquartileRange[dist]` | gives the difference between the upper and lower quartiles for the distribution dist. |
| `QuartileDeviation[list]` | gives the quartile deviation or semi-interquartile range of the elements in list. |
| `QuartileDeviation[dist]` | gives the quartile deviation or semi-interquartile range of the distribution dist. |

***Mathematica Examples 7.1*** InterquartileRange

```
Input    (* This code defines a function IQR that takes a list of data as input and uses the
         built-in Quartiles function to calculate the first and third quartiles, which are
         used to calculate the interquartile range. *)

         data={1,2,3,4,5,6,7,8,9};
         IQR[data_]:=Quartiles[data][[3]]-Quartiles[data][[1]]
         IQR[data]
         InterquartileRange[data]
Output   9/2
Output   9/2

Input    (* InterquartileRange for a matrix gives columnwise ranges: *)
         InterquartileRange[{{1,5},{3,5},{1,8},{5,6},{7,8},{2,4}}]
         InterquartileRange[{1,3,1,5,7,2}]
         InterquartileRange[{5,5,8,6,8,4}]
Output   {4,3}
Output   4
Output   3

Input    (* Find the interquartile range for WeightedData: *)
         data={8,3,5,4,9,0,4,2,2,3};
         w={0.15,0.09,0.12,0.10,0.16,0.,0.11,0.08,0.08,0.09};
         InterquartileRange[
          WeightedData[data,w]
          ]
Output   5

Input    (* Interquartile range of a parametric distribution: *)
         InterquartileRange[
          NormalDistribution[μ,σ]
          ]
Output   2 √2 σ InverseErfc[1/2]

Input    (* The code generates three plots, each representing a different normal distribution
         with varying standard deviations. The InterquartileRange (IQR) value for each
```

```
distribution is displayed as a label on the plot. The IQR indicates the spread of
values within the middle 50% of the data: *)

dists={
    NormalDistribution[0,1],
    NormalDistribution[0,2],
    NormalDistribution[0,4]
    };

Table[
 Plot[
  PDF[d,x],
  {x,-10,10},
  PlotStyle->Directive[Purple,Opacity[0.7]] ,
  Filling->Axis,
  Ticks->{Automatic,None},
  PlotRange->{Automatic,{0,0.4}},
  PlotLabel->N[InterquartileRange[d]],
  ImageSize->170
  ],
 {d,dists}
 ]
```

Output

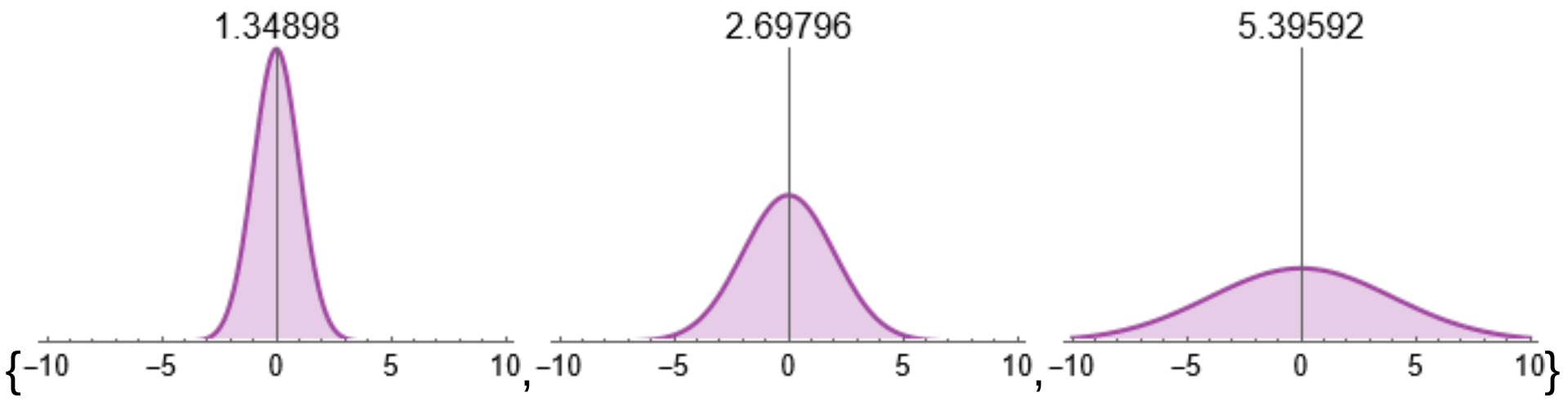

Input

```
(* The code reads in a dataset h of heights and calculates its interquartile range,
median, and length of the data. It then produces two plots, a scatter plot of data
h and a plot that shows both h and three horizontal lines representing the median
and interquartile range: *)

h={133,136,149,133,123,121,140,139,117,117,136,108,126,104,116,147,140,148,150,122,
135,146,133,144,117,124,135,117,120,121,110,124,103,137,101,119,104,113,139,133};

iqr=N[InterquartileRange[h]]
m=N[Median[h]]
n=Length[h];

ListPlot[
 h,
 Filling->Axis,
 PlotStyle->Purple,
 ImageSize->170
 ]

ListPlot[
 {h,{{0,m},{n,m}},{{0,m-iqr},{n,m-iqr}},{{0,m+iqr},{n,m+iqr}}},
 Joined->{False,True,True,True},
 Filling->{1->m,3->{4}},
 PlotStyle->{Purple,Automatic,Automatic,Automatic},
 PlotLegends->{"h","Median","l_Interquartile Range","u_Interquartile Range"},
 AxesLabel->Automatic,
 ImageSize->170
 ]
```

Output 21.

Output 125.

Output

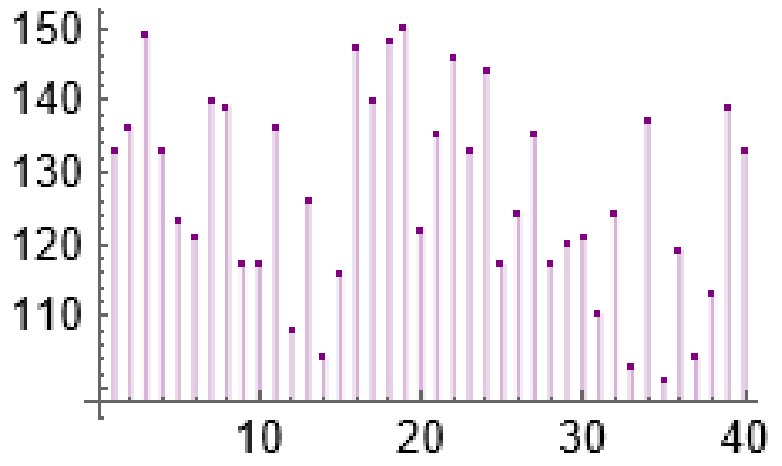

Output

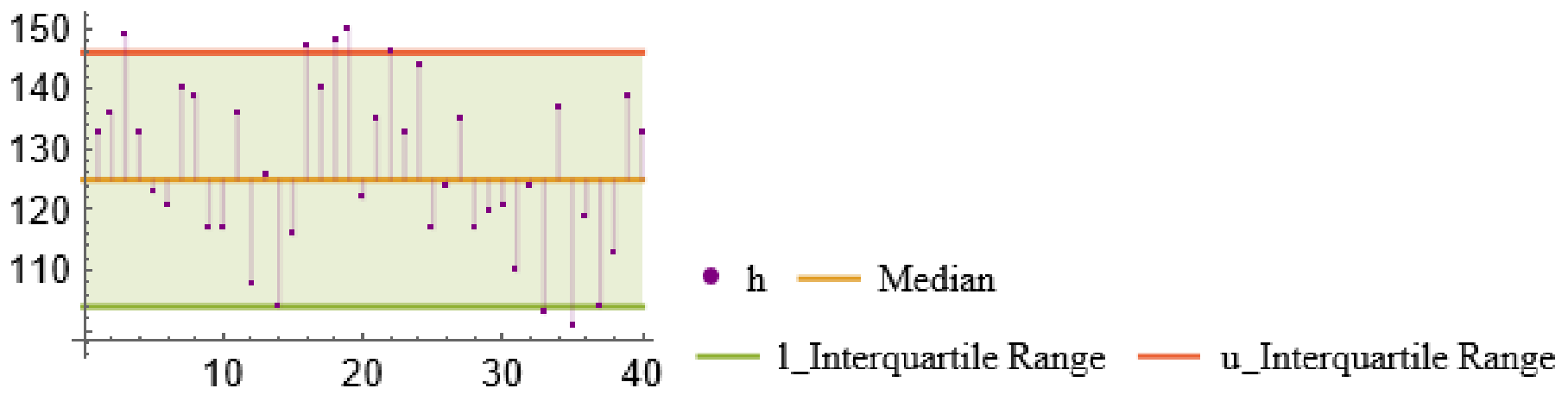

Input

```
(* BoxWhiskerChart shows the interquartile range for data. The code generates a list of 100 random numerical data points sampled from a normal distribution with mean 0 and standard deviation 1. The code creates BoxWhiskerChart and uses Show function to overlay additional graphics on the box-and-whisker plot. It adds lines representing the upper and lower bounds of the interquartile range (IQR), and a line connecting Q1 and Q3. The lines are drawn using the Line function, and different colors (blue, red, green) are assigned to differentiate them: *)

(* Generate random numerical data *)
data=RandomVariate[
    NormalDistribution[0,1],
    {100}
    ];

(* Calculate the quartiles and interquartile range *)
quartiles=Quartiles[data];
iqr=InterquartileRange[data];

(* Create BoxWhiskerChart *)
boxWhisker=BoxWhiskerChart[
    data,
    "Outliers",
    ChartStyle->{Purple},
    ImageSize->250
    ];

(* Display the interquartile range as a line on the chart *)
boxWhiskerWithIQR=Show[
   boxWhisker,
   Graphics[
    {
     Thick,
     Blue,
     Line[{{1.25,quartiles[[3]]},{1.25,quartiles[[3]]+1.5*iqr}}],
     Inset["Q3+1.5*IQR",{1.75,2}],
     Red,
     Line[{{1.25,quartiles[[1]]},{1.25,quartiles[[1]]-1.5*iqr}}],
     Inset["Q1-1.5*IQR",{1.75,-2}],
     Green,
     Line[{{1.25,quartiles[[1]]},{1.25,quartiles[[3]]}}],
```

```
          Inset["IQR",{1.75,0}]
          }
        ]
      ]
```

Output

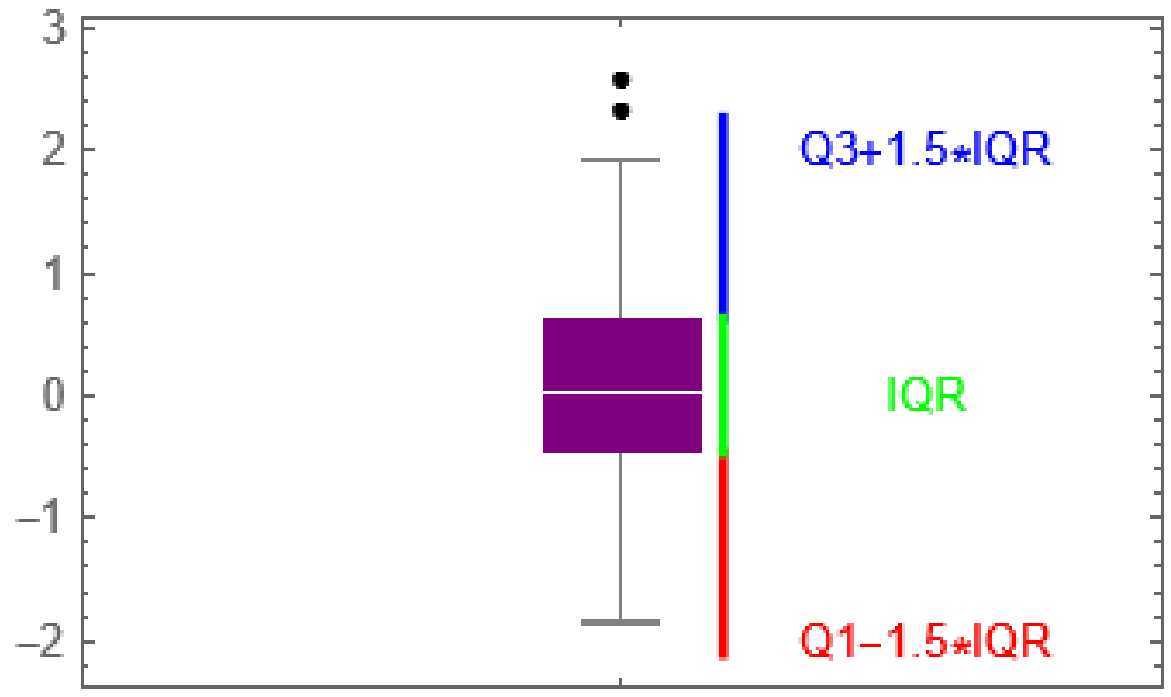

**Mathematica Examples 7.2** QuartileDeviation

```
Input   (* QuartileDeviation is half the difference between the first and third quartiles:
        *)
        data=RandomReal[10,20];
        qd=(Quartiles[data][[3]]-Quartiles[data][[1]])/2
        QuartileDeviation[data]
Output  1.84785
Output  1.84785

Input   (* Find the quartile deviation for WeightedData: *)
        data={8,3,5,4,9,0,4,2,2,3};
        w={0.15,0.09,0.12,0.10,0.16,0.,0.11,0.08,0.08,0.09};
        QuartileDeviation[
         WeightedData[data,w]
         ]
Output  5/2

Input   (* Obtain a robust estimate of dispersion when extreme values are present: *)
        N[QuartileDeviation[{5,3,10^5,20,15,6}]]

        (* Measures based on the Mean are heavily influenced by extreme values: *)
        N[StandardDeviation[{5,3,10^5,20,15,6}]]
        N[MeanDeviation[{5,3,10^5,20,15,6}]]
Output  7.5
Output  40820.8
Output  27775.1

Input   (* The code begins by calculating the quartile deviation, and mean of the dataset h.
        It then proceeds to create two plots: The first plot is a line plot of the dataset
        h with filled areas below the points. The second plot displays a dataset and lines.
        It includes the line plot of h, horizontal lines representing the mean, and upper
        and lower quartile deviation bands around the mean: *)

        h={133,136,149,133,123,121,140,139,117,117,136,108,126,104,116,147,140,148,150,122,
        135,146,133,144,117,124,135,117,120,121,110,124,103,137,101,119,104,113,139,133};

        qd=N[QuartileDeviation[h]]
        m=N[Median[h]]
        n=Length[h];

        ListPlot[
         h,
```

```
         Filling->Axis,
         PlotStyle->Purple,
         ImageSize->170
         ]
        (*Plot the quartile deviation respective of the median: *)
        ListPlot[
         {h,{{0,m},{n,m}},{{0,m-qd},{n,m-qd}},{{0,m+qd},{n,m+qd}}},
         Joined->{False,True,True,True},
         Filling->{1->m,3->{4}},
         PlotStyle->{Purple,Automatic,Automatic,Automatic},
         PlotLegends->{"h","median","l_quartile deviation band","u_quartile deviation
        band"},
         AxesLabel->Automatic,
         ImageSize->170
         ]
Output  10.5
Output  125.
Output
```

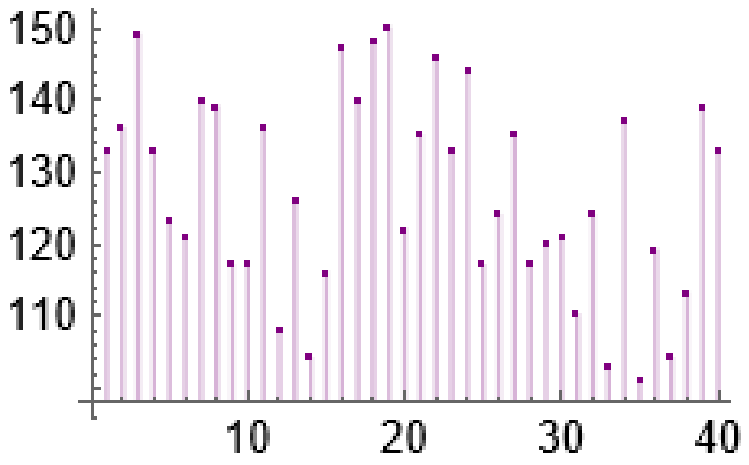

```
Output
```

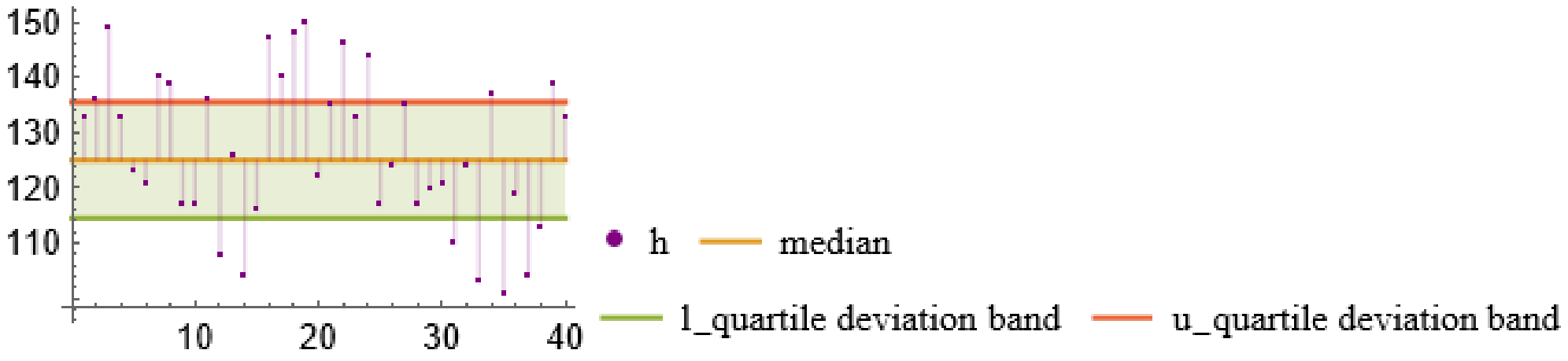

| | |
|---|---|
| `MeanDeviation[list]` | gives the mean absolute deviation from the mean of the elements in list. |
| `StandardDeviation[list]` | gives the sample standard deviation of the elements in list. |
| `StandardDeviation[dist]` | gives the standard deviation of the distribution dist. |
| `Variance[list]` | gives the sample variance of the elements in list. |
| `Variance[dist]` | gives the variance of the distribution dist. |

***Mathematica Examples 7.3*** MeanDeviation

```
Input   (* MeanDeviation of a list: *)
        Simplify[
         MeanDeviation[
          {a,b,c,d}
          ]
         ]
Output  1/16 (Abs[a+b+c-3 d]+Abs[3 a-b-c-d]+Abs[a+b-3 c+d]+Abs[a-3 b+c+d])

Input   (* MeanDeviation of columns of a matrix: *)
        MeanDeviation[{{2,1},{1,2},{4,8},{5,3},{2,15}}]
        MeanDeviation[{2,1,4,5,2}]
        MeanDeviation[{1,2,8,3,15}]
Output  {34/25,114/25}
Output  34/25
Output  114/25

Input   (* Find the mean deviation of WeightedData: *)
```

```
         Simplify[
          MeanDeviation[
           WeightedData[{1,2,3},{0.6,0.2,0.2}]
            ]
          ]
Output   0.72

Input    (* MeanDeviation is the Mean of absolute deviations from the Mean: *)
         data=RandomReal[10,5];
         mean=Mean[data];
         meanDeviation=Mean[Abs[data-mean]]
         MeanDeviation[data]
Output   1.1376
Output   1.1376

Input    (* MeanDeviation is equivalent to the 1-norm of the deviations divided by the Length.
         Mean deviation measures the average absolute deviation of data points from the mean,
         while the 1-norm of the deviations represents the sum of the absolute values of the
         differences. By dividing the 1-norm of the deviations by the length of the dataset,
         we obtain the mean deviation. This relationship highlights the connection between
         the statistical concept of mean deviation and the mathematical concept of the 1-
         norm, providing a concise and computationally efficient way to calculate the mean
         deviation: *)

         data=RandomReal[10,5];
         mean=Mean[data];
         meanDeviation=Norm[data-mean,1]/Length[data]
         MeanDeviation[data]
Output   3.09703
Output   3.09703

Input    (* MeanDeviation as a scaled ManhattanDistance from the Mean. Mean deviation
         represents the average absolute deviation of data points from the mean, while
         Manhattan distance measures the total absolute difference between corresponding
         elements of two vectors. By scaling the Manhattan distance with a factor of 1 divided
         by the length of the dataset, we obtain the mean deviation. This relationship
         highlights the connection between the statistical concept of mean deviation and the
         geometric concept of Manhattan distance: *)

         data={1,2,3,4,5};
         mean=Mean[data];
         meanDeviation=(1/Length[data])*ManhattanDistance[data,Table[mean,Length[data]]]
         MeanDeviation[data]
Output   6/5
Output   6/5

Input    (* The code provides a visual representation of the dataset h and highlights the mean
         and mean deviation. The data is then visualized using two ListPlot functions. The
         first ListPlot displays the dataset h with the Filling option set to Axis. This plot
         represents the distribution of data points along the vertical axis, providing a
         visual overview of the dataset. The second ListPlot showcases the dataset h along
         with additional lines representing the mean value (m) and the mean deviation bands.
         The mean deviation bands are illustrated by two lines parallel to the mean line, with
         a distance of mean deviation (md) from the mean: *)

         h={133,136,149,133,123,121,140,139,117,117,136,108,126,104,116,147,140,148,150,122,
         135,146,133,144,117,124,135,117,120,121,110,124,103,137,101,119,104,113,139,133};

         md=N[MeanDeviation[h]]
         m=N[Mean[h]]
         n=Length[h];
```

```
ListPlot[
 h,
 Filling->Axis,
 PlotStyle->Purple,
 ImageSize->170
 ]

ListPlot[
 {h,{{0,m},{n,m}},{{0,m-md},{n,m-md}},{{0,m+md},{n,m+md}}},
 Joined->{False,True,True,True},
 Filling->{1->m,3->{4}},
 PlotStyle->{Purple,Automatic,Automatic,Automatic},
 PlotLegends->{"h","mean","l_mean deviation band","u_mean deviation band"},
 AxesLabel->Automatic,
 ImageSize->170
 ]
```

Output 12.

Output 127.

Output

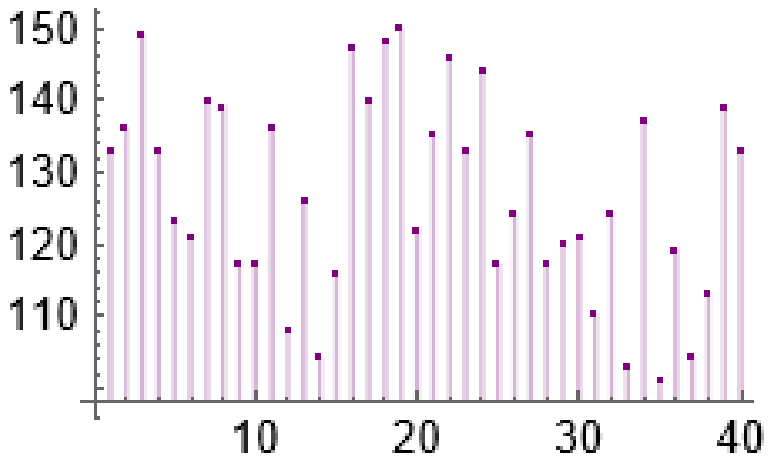

Output

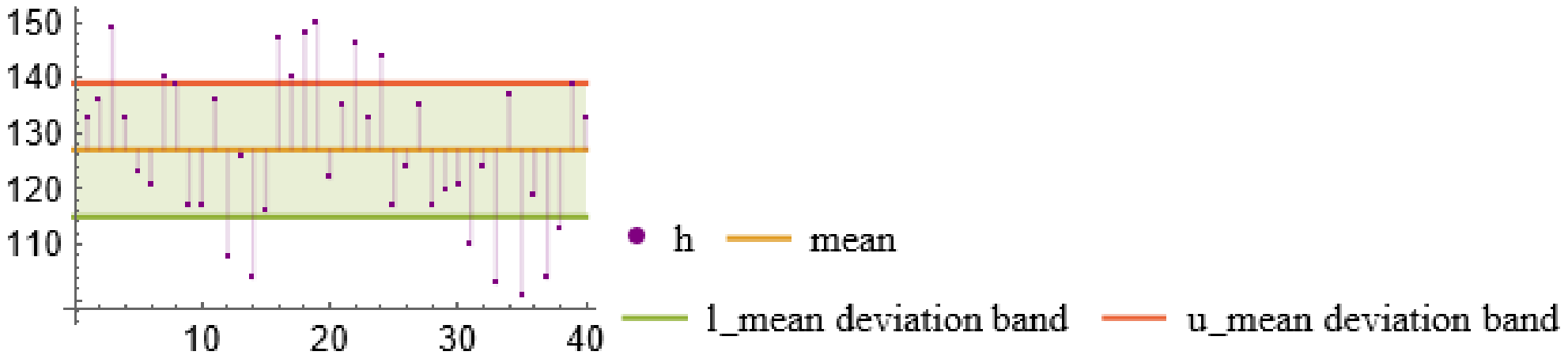

***Mathematica Examples 7.4*** StandardDeviation

Input
```
(* To calculate the standard deviation of a list of numbers: *)
list={1,2,3,4,5};
StandardDeviation[list]
```
Output $\sqrt{5/2}$

Input
```
(* The square of StandardDeviation is Variance: *)
Variance[{1,2,3,4,5,6}]
StandardDeviation[{1,2,3,4,5,6}]
```
Output 7/2

Output $\sqrt{7/2}$

Input
```
(* To calculate the standard deviation of a matrix: *)
matrix={{1,2,3},{4,5,6},{7,8,9}};
StandardDeviation[matrix]
StandardDeviation[{1,4,7}]
StandardDeviation[{2,5,8}]
StandardDeviation[{3,6,9}]
```
Output {3,3,3}

Output 3

Output 3

Output 3

```
Input   (* Find the standard deviation of WeightedData: *)
        data={8,3,5,4,9,0,4,2,2,3};
        w={0.15,0.09,0.12,0.10,0.16,0.,0.11,0.08,0.08,0.09};
        StandardDeviation[
         WeightedData[data,w]
         ]
Output  2.69554

Input   (* The code implements a function standardDeviation that calculates the sample
        standard deviation of a given dataset. It utilizes a Module to encapsulate local
        variables and avoids unintended side effects. The code calculates the mean of the
        dataset and determines the deviations of each data point from the mean. It squares
        the deviations, sums them up, and divides by (n-1) to obtain the sample variance.
        Finally, the square root of the sample variance is taken to compute the standard
        deviation. The code also includes a test case using a specific dataset and compares
        the result with the built-in StandardDeviation function: *)

        standardDeviation[data_]:=Module[
          {n,mean,deviations},
          n=Length[data];
          mean=Total[data]/n;
          deviations=data-mean;
          Sqrt[Total[deviations^2]/(n-1)]
          ]

        data={1,2,3,4,5};
        result=standardDeviation[data]
        StandardDeviation[data]
```

Output $\sqrt{5/2}$

Output $\sqrt{5/2}$

```
Input   (* The code explores multiple approaches to calculate the standard deviation of a
        given dataset. The code covers definitions based on norms, means, root mean square,
        and Euclidean distance. Each approach is mathematically valid and provides a way to
        compute the standard deviation. The code then creates a list containing the results
        of each definition, allowing for a comparison between them: *)

        data=RandomReal[10,20];
        StandardDeviation[data]

        (* StandardDeviation is a scaled Norm of deviations from the Mean: *)
        def1=Norm[data-Mean[data]]/Sqrt[Length[data]-1];

        (* StandardDeviation is the square root of a scaled Mean of squared deviations: *)
        def2=Sqrt[Mean[(data-Mean[data])^2] Length[data]/(Length[data]-1)];

        (* StandardDeviation is a scaled RootMeanSquare of the deviations: *)
        def3=RootMeanSquare[data-Mean[data]] Sqrt[Length[data]/(Length[data]-1)];

        (* StandardDeviation as a scaled EuclideanDistance from the Mean: *)
        mean=Mean[data];
        len=Length[data];
        def4=EuclideanDistance[data,Table[mean,{len}]]/Sqrt[(len-1)];
        {def1,def2,def3,def4}
Output  2.62297
Output  {2.62297,2.62297,2.62297,2.62297}
```

Input
```
(* The code generates a list of 1000 random numbers from a standard normal distribution. It then calculates the standard deviation of the data and creates a histogram plot with the probability density function (PDF). The plot includes red points representing the mean and ± standard deviation of the data: *)

data=RandomVariate[
   NormalDistribution[0,1],
   1000
   ];

StandardDeviation[data]

Histogram[
 data,
 Automatic,
 "PDF",
 Epilog->{
   Red,
   PointSize[0.03],
   Point[{{StandardDeviation[data],0},{Mean[data],0},{-StandardDeviation[data],0}}]
   },
 ColorFunction->Function[{height},Opacity[height]],
 ChartStyle->Purple,
 ImageSize->200
 ]
```
Output 0.96111

Output

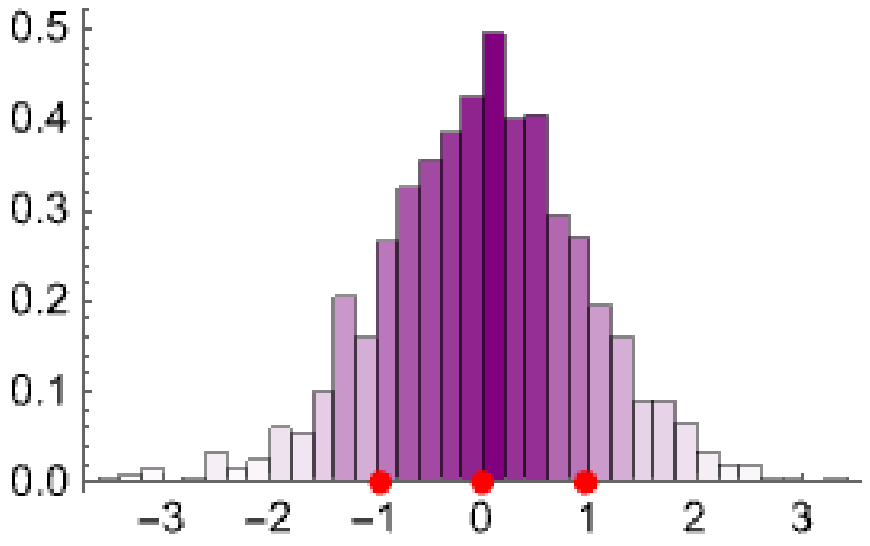

Input
```
(* The code begins by calculating the standard deviation, and mean of the dataset h. It then proceeds to create two plots: The first plot is a line plot of the dataset h with filled areas below the points, visualizing the distribution of the data. The second plot displays a dataset and lines. It includes the line plot of h, horizontal lines representing the mean, and upper and lower standard deviation bands around the mean: *)

h={133,136,149,133,123,121,140,139,117,117,136,108,126,104,116,147,140,148,150,122,135,146,133,144,117,124,135,117,120,121,110,124,103,137,101,119,104,113,139,133};

sd=N[StandardDeviation[h]]
m=N[Mean[h]]
n=Length[h];

ListPlot[
 h,
 Filling->Axis,
 PlotStyle->Purple,
 ImageSize->170
 ]

ListPlot[
 {h,{{0,m},{n,m}},{{0,m-sd},{n,m-sd}},{{0,m+sd},{n,m+sd}}},
```

```
          Joined->{False,True,True,True},
          Filling->{1->m,3->{4}},
          PlotStyle->{Purple,Automatic,Automatic,Automatic},
          PlotLegends->{"h","mean","l_standard deviation band","u_standard deviation band"},
          AxesLabel->Automatic,
          ImageSize->170
          ]
Output    13.9468
Output    127.
Output
```

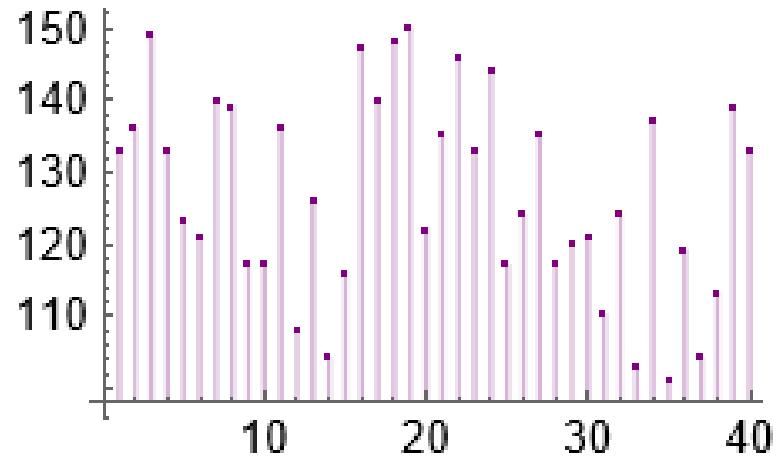

Output

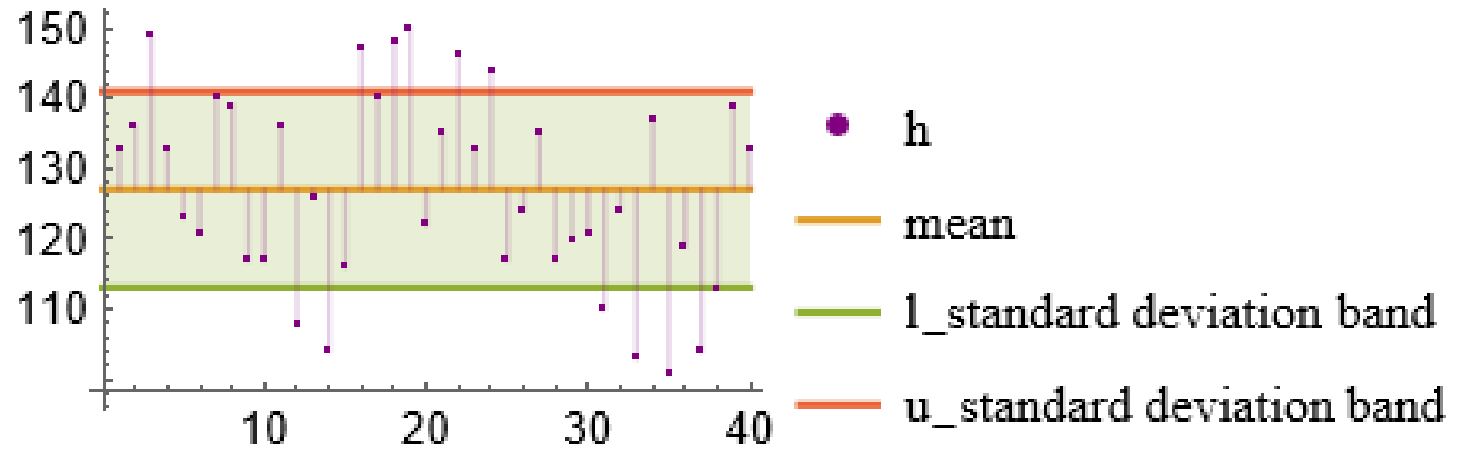

```
Input     (* This code generates a random sample of 1000 points from a standard normal
          distribution, computes various statistical measures such as the quartile deviation,
          standard deviation, mean deviation and mean of the sample, and plots a histogram of
          the sample with lines representing these statistical measures: *)

          data=RandomVariate[
             NormalDistribution[0,1],
             1000
             ];

          Quartiles[data];
          InterquartileRange[data];
          qd=QuartileDeviation[data];
          sd=N[StandardDeviation[data]];
          md=N[MeanDeviation[data]];
          m=N[Mean[data]];

          Print[
           "Mean= " ,m ", ",
           "QuartileDeviation= ",qd,", ",
           "MeanDeviation= " ,md,", ",
            "StandardDeviation= ", sd,", "
           ]

          Histogram[
           data,
           Automatic,
           "PDF",
           Epilog->{
             Directive[Thickness[0.008],Dashed],
             Red,Line[{{qd,0},{qd,0.5}}],
             Blue, Line[{{sd,0},{sd,0.5}}],
             Green,Line[{{md,0},{md,0.5}}],
             Black,Line[{{m,0},{m,0.5}}]
```

```
   },
 ChartStyle->Directive[Purple,Opacity[0.5]],
 ImageSize->220
 ]
```

Output Mean= -0.0471451 , QuartileDeviation= 0.664779 , MeanDeviation= 0.792928 , StandardDeviation= 1.00631 .

Output

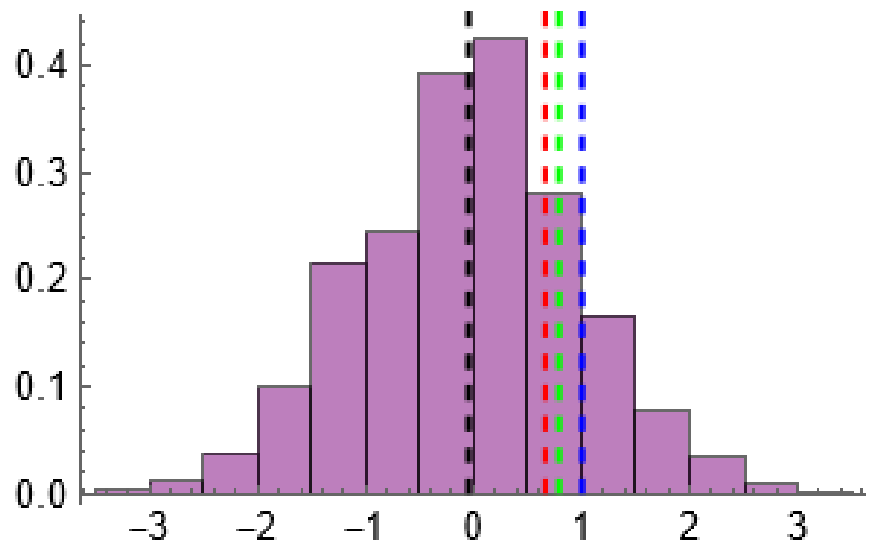

Input

```
(* This code generates a 2D dataset from a standard normal distribution and calculates various statistical measures such as quartile deviation, standard deviation, and mean for each component of the dataset. It then displays the results and creates a scatter plot with highlighted points representing the statistical measures: *)

data=RandomVariate[
    NormalDistribution[0,1],
    {2000,2}
    ];
qdx=QuartileDeviation[data[[All,1]]];
qdy=QuartileDeviation[data[[All,2]]];

sdx=StandardDeviation[data[[All,1]]];
sdy=StandardDeviation[data[[All,2]]];

mx=Mean[data[[All,1]]];
my=Mean[data[[All,2]]];
Print[
 "Mean= {" ,mx,",",my,"}, ",
 "QuartileDeviation= {",qdx,",",qdy,"}, ",
  "StandardDeviation= {", sdx,",",sdy,"} "

 ]
ListPlot[
 data,
 Epilog->{
    Red, PointSize[0.02],Point[{qdx,qdy}],
    Blue, PointSize[0.02],Point[{sdx,sdy}],
    Green, PointSize[0.02],Point[{mx,my}]
    },
 ImageSize->220
 ]
```

Output Mean= { -0.0169259 , -0.00549935 }, QuartileDeviation= { 0.676502 , 0.655909 }, StandardDeviation= { 0.998782 , 1.0005 }

Output

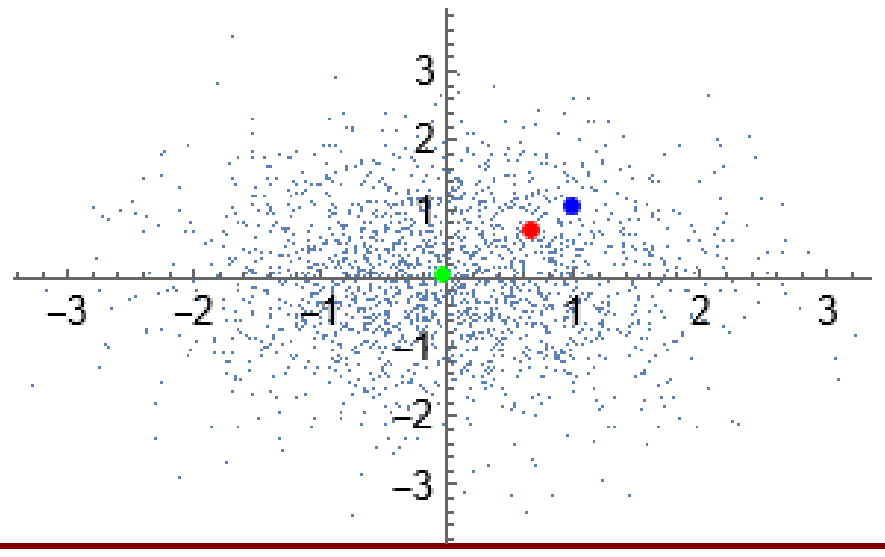

***Mathematica Examples 7.5*** Variance

```
Input    (* Variance of a list of numbers: *)
         Variance[{3.2,1.21,3.4,2,4.66,1.5,5.61,7.22}]
Output   4.45003

Input    (* Variance of elements in each column: *)
         Variance[{{4,1},{5.2,7},{5.3,8},{5.4,9}}]
         Variance[{4,5.2,5.3,5.4}]
         Variance[{1,7,8,9}]
Output   {0.429167,155/12}
Output   0.429167
Output   155/12

Input    (* Find the variance of WeightedData: *)
         data={8,3,5,4,9,0,4,2,2,3};
         w={0.15,0.09,0.12,0.10,0.16,0.,0.11,0.08,0.08,0.09};
         Variance[WeightedData[data,w]]
Output   7.26594

Input    (* The code explores multiple approaches to calculate the variance of a given dataset.
         To calculate the variance of a set of numbers, you can follow these steps. step 1
         Find the mean of the set of numbers by adding all the numbers and dividing the sum
         by the total number of values. step 2 For each number in the set, subtract the mean
         and then square the result. step 3 Add up all the squared differences from step 2.
         step 4 Divide the sum from step 3 by the total number of values in the set minus one.
         This is the variance: *)

         data={1,2,3,4,5}; (* Example list of data *)
         n=Length[data];(* Length of data *)
         mean=Mean[data]; (* Step 1: calculate the mean *)
         differences=(#-mean)^2&/@data; (* Step 2: subtract the mean and square *)
         sum=Total[differences]; (* Step 3: add up the squared differences *)
         variance1=sum/(Length[data]-1) (* Step 4: divide by n-1 to get the variance *);

         (*or *)
         variance2=Total[(data-Mean[data])^2]/(n-1);

         (*or *)
         variance3=Mean[(data-Mean[data])^2](n/(n-1));

         (*or *)
         Variance[data]
         {variance1,variance2,variance3}
Output   5/2
Output   {5/2,5/2,5/2}

Input    (* This code defines a function called variance that takes a list of data as its
         input. Inside the Module function, the mean of the data is calculated using the
         built-in Mean function, and the deviations from the mean are calculated by subtracting
```

```
        the mean from the data. The variance is then calculated using, the sum of squared
        deviations divided by n-1: *)

        variance[data_]:=Module[
          {mean,deviations},
          mean=Mean[data];
          deviations=data-mean;
          (1/(Length[data]-1)) Total[deviations^2]
          ]
        data={1,2,3,4,5};
        variance[data]
Output  5/2

Input   (* Variance is a scaled squared Norm of deviations from the Mean: *)

        data={1,2,3,4,5}; (* example list of data *)
        mean=Mean[data]; (* calculate the mean *)
        deviations=data-mean; (* calculate the deviations from the mean *)
        variance=Variance[data]; (* calculate the variance *)
        scaledNorm=(1/(Length[data]-1)) Norm[deviations]^2; (* calculate the scaled
        squared norm *)
        variance==scaledNorm (* compare the variance to the scaled squared norm *)
Output  True

Input   (* The square root of Variance is a scaled RootMeanSquare of the deviations: *)

        data={1,2,3,4,5}; (* example list of data *)
        mean=Mean[data]; (* calculate the mean *)
        deviations=data-mean; (* calculate the deviations from the mean *)
        variance=Variance[data]; (* calculate the variance *)
        scaledRMS=Sqrt[ (Length[data]/(Length[data]-1))Mean[deviations^2]]; (* calculate
        the scaled root mean square *)
        Sqrt[variance]==scaledRMS (* compare the square root of variance to the scaled root
        mean square *)
Output  True

Input   (* Variance is a scaled SquaredEuclideanDistance from the Mean: *)
        data=RandomReal[10,5];
        mean=Mean[data]
        len=Length[data]
        SquaredEuclideanDistance[data,Table[mean,{len}]]/(len-1)

        Variance[data]
Output  3.75454
Output  5
Output  10.5972
Output  10.5972

Input   (* In the code, the variance of a normal distribution is calculated using the Variance
        function, which provides a measure of the spread or variation of the distribution.
        By varying the value of σ (standard deviation) in the loop and generating plots for
        each value, the code visually illustrates how changing the standard deviation affects
        the shape and spread of the normal distribution. The variance itself represents the
        average squared deviation of data points from the mean. As the standard deviation
        increases (larger σ values), the variance will also increase, indicating a greater
        dispersion or variation of the data: *)

        d=NormalDistribution[0,σ];
        Variance[d]
        Table[
         Plot[
```

```
      PDF[d,x],
      {x,-4,4},
      PlotRange->{0,0.4},
      Filling->Axis,
      PlotStyle->Purple,
      Ticks->{Automatic,None},
      PlotLabel->Row[{"\!\(\*SuperscriptBox[\(σ\), \(2\)]\) = ",σ^2}],
      ImageSize->170
      ],
     {σ,{1,1.5,2}}
     ]
```

Output σ^2

Output

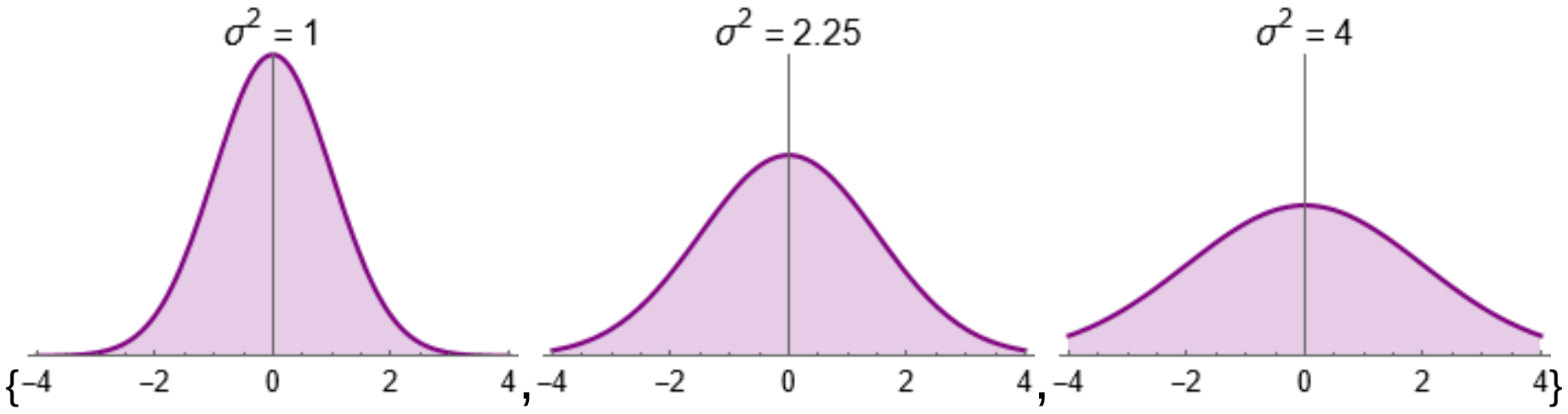

Input

```
(* This code creates a Manipulate function with slider and a dropdown menu. The n slider allows the user to adjust the sample size, while the dist dropdown menu allows the user to choose the distribution that the sample is drawn from. The code then uses RandomVariate to generate a sample of size n from the selected distribution, calculates the variance using Variance, and displays both a ListPlot and a Histogram of the sample data along with the calculated variance. *)

Manipulate[
 Module[
  {data,var},
  data=RandomVariate[dist,n];
  var=Variance[data];

  Grid[
   {
    {
     ListPlot[
      data,
      PlotRange->All,
      ImageSize->250,
      Filling->Axis,
      PlotStyle->Purple
      ],
     Histogram[
      data,
      "FreedmanDiaconis",
      "PDF",
      Frame->True,
      FrameLabel->{"Data","PDF"},
      ImageSize->250,
      ColorFunction->Function[Opacity[0.7]],
      ChartStyle->Purple
      ]
     },
    {Null,Text["Variance: "<>ToString[var]]}
    }
   ]
  ],
```

```
 {{n,300,"Sample size"},10,1000,10,Appearance-
>"Labeled"},{{dist,NormalDistribution[0,1],"Distribution"},{NormalDistribution[0,1]
,StudentTDistribution[3],ExponentialDistribution[1],UniformDistribution[{-1,1}]}},
 Alignment->Center
 ]
```

Output

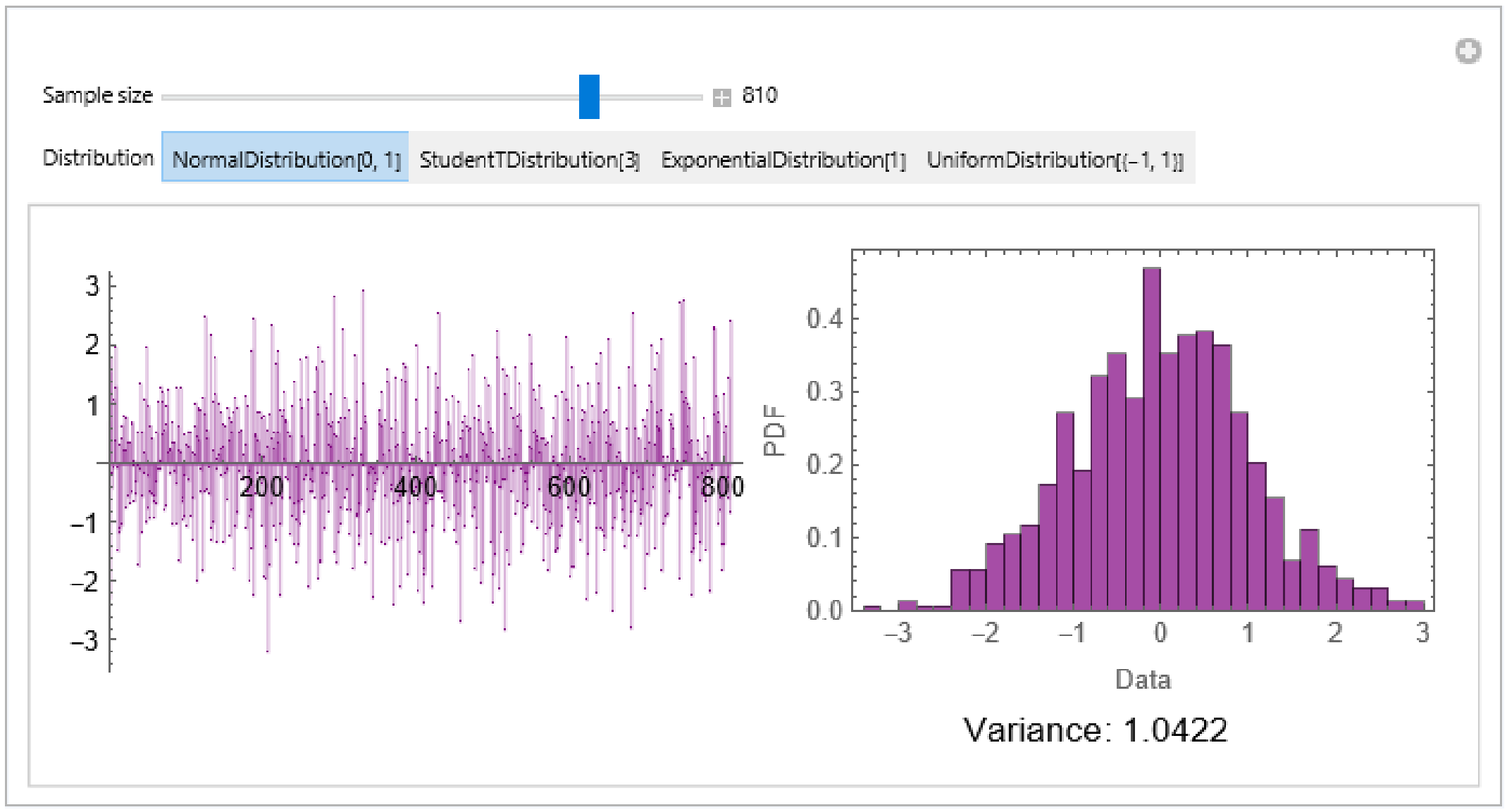

Input

```
(* This code first generates a sample distribution of variance for 1000 samples of
size 5, using a Table function and the built-in Variance function. It then plots a
histogram of the variances using Histogram. Finally, it plots a list of the variances
using ListPlot, with the sample number on the x-axis and the variance on the y-axis
and labeled axes. *)

(*Generate a sample distribution of variance for 1000 samples of size 5*)
variances=Table[
    Variance[
     RandomReal[10,5]
     ],
    {i,1,500}
    ];

(*Plot a histogram of the variances*)
Histogram[
 variances,
 "FreedmanDiaconis",
 "PDF",
 Frame->True,
 ColorFunction->Function[Opacity[0.7]],
 ChartStyle->Purple,
 FrameLabel->{"Variance","PDF"},
 ImageSize->250
 ]

(*Plot a list of the variances*)
ListPlot[
 variances,
 ColorFunction->Function[Opacity[0.7]],
 Filling->Axis,
 PlotStyle->Purple,
 PlotRange->All,
 Frame->True,
```

```
    FrameLabel->{"Sample","Variance"},
    ImageSize->250
    ]
```

Output

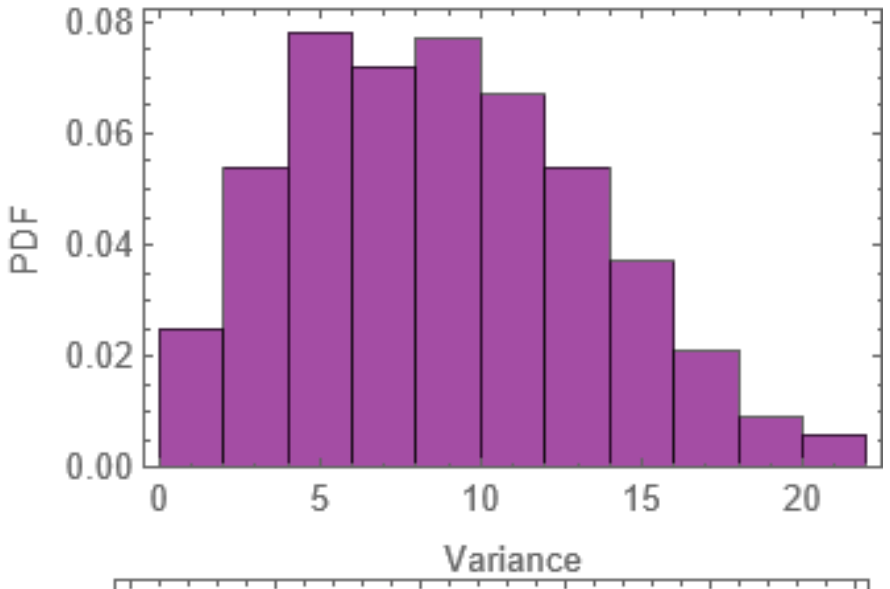

Output

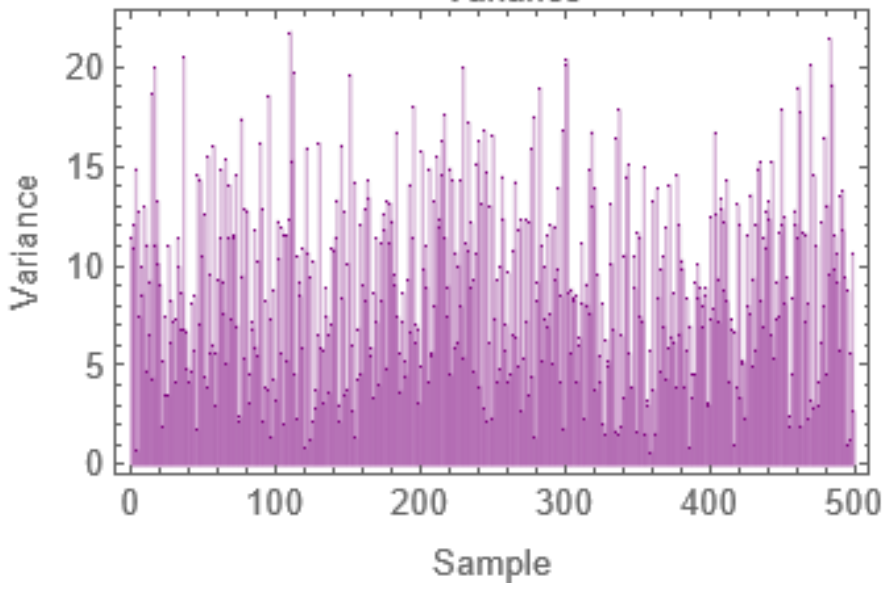

Input

```
(* Calculate the mean, mean deviation, standard deviation, and variance for grouped
data: *)
(* Define the grouped data *)
data={
    {0,4,8,12,16,20}(* Class boundaries *),
    {7,4,19,12,8} (* Frequencies *)
    };

(* Calculate the class marks *)
classMarks=Mean/@Partition[data[[1]],2,1];

(* Calculate the mean *)
mean=N[
    Total[classMarks*data[[2]]]/Total[data[[2]]]
    ];

(* Calculate the mean deviation *)
meanDeviation=N[
    Total[Abs[classMarks-mean]*data[[2]]]/Total[data[[2]]]
    ];

(* Calculate the standard deviation *)
standardDeviation=N[
    Sqrt[Total[((classMarks-mean)^2)*data[[2]]]/(Total[data[[2]]]-1)]
    ];

(* Calculate the variance *)
variance=N[standardDeviation^2];

(* Print the results *)
Print["Mean:",mean];
Print["Mean Deviation:",meanDeviation];
Print["Standard Deviation:",standardDeviation];
Print["Variance:",variance];
```

Output

```
Mean: 10.8
Mean Deviation: 3.84
```

```
Standard Deviation: 4.91561
Variance: 24.1633
```

| | |
|---|---|
| `TrimmedVariance[list,f]` | gives the variance of the elements in list after dropping a fraction f of the smallest and largest elements. |
| `TrimmedVariance[list,{f1,f2}]` | gives the variance when a fraction f1 of the smallest elements and a fraction f2 of the largest elements are removed. |
| `TrimmedVariance[list]` | gives the 5% trimmed variance TrimmedVariance[list,0.05]. |
| `TrimmedVariance[dist,…]` | gives the trimmed variance of a univariate distribution dist. |
| `WinsorizedVariance[list,f]` | gives the variance of the elements in list after replacing the fraction f of the smallest and largest elements by the remaining extreme values. |
| `WinsorizedVariance[list,{f1,f2}]` | gives the variance when the fraction f1 of the smallest elements and the fraction f2 of the largest elements are replaced by the remaining extreme values. |
| `WinsorizedVariance[list]` | gives the 5% winsorized variance WinsorizedVariance[list,0.05]. |
| `WinsorizedVariance[dist,…]` | gives the winsorized variance of a univariate distribution dist. |

***Mathematica Examples 7.6*** TrimmedVariance

```
Input     (* Trimmed variance after removing extreme values: *)
          TrimmedVariance[{-10,1,1,1,1,20},0.2]
Output    0

Input     (* Trimmed variance after removing the smallest extreme values: *)
          TrimmedVariance[{-10,1,1,1,1,20},{0.2,0}]
Output    361/5

Input     (* Trimmed variance of a symbolic distribution: *)
          TrimmedVariance[ExponentialDistribution[λ]]
Output    (324-19 Log[19]^2)/(324 λ^2)

Input     (* Obtain a robust estimate of dispersion when outliers are present: *)
          N[
           TrimmedVariance[{1,5,2,6,10,10^6,5,4,-2000,5},.1]
           ]

          (* Extreme values have a large influence on the Variance: *)
          N[
           Variance[{1,5,2,6,10,10^5,5,4,-200,5}]
           ]
Output    7.35714
Output    1.00036*10^9

Input     (* A 0% TrimmedVariance is equivalent to Variance: *)
          TrimmedVariance[Range[10],0]
          Variance[Range[10]]
Output    55/6
Output    55/6

Input     (* The code is calculating the trimmed variance of a data set using two different
          methods. The first method defines a function called trimmedvariance that calculates
          the trimmed variance of a dataset. The trimmed variance is obtained by excluding a
          certain percentage of extreme values from the dataset and then calculating the
          variance of the remaining values while the second method uses the built-in
          "TrimmedVariance" function: *)

          trimmedvariance[data_,p_]:=Module[
            {n,m,sortedData,trimLength},
            n=Length[data];
```

```
          m=Floor[n*p];
          sortedData=Sort[data];
          trimLength=n-2*m;
          If[
           trimLength<=0,
           Message[TrimmedVariance::trimtoolarge,p];
           Return[$Failed]
           ];
          newsorteddata=Take[sortedData,{m+1,-m-1}];
          Variance[newsorteddata]
          ]

        data={1,2,3,4,5,6,7,8,9,10};
        trimPercentage=0.2; (*20%*)
        trimmedvariance[data,trimPercentage]
        TrimmedVariance[data,trimPercentage]
Output  7/2
Output  7/2
```

**Mathematica Examples 7.7** WinsorizedVariance

```
Input   (* Winsorized variance after removing extreme values: *)
        WinsorizedVariance[{-100,1,1,1,1,200},0.2]
Output  0

Input   (* Winsorized variance after removing the smallest extreme values: *)
        WinsorizedVariance[{-100,1,1,1,1,200},{0.2,0}]
Output  39601/6

Input   (* Obtain a robust estimate of location when outliers are present: *)
        N[
         WinsorizedVariance[{1,5,2,6,10,10^6,5,4,-2000,5},.1]
         ]

        (* Extreme values have a large influence on the variance: *)
        N[
         Variance[{1,5,2,6,10,10^6,5,4,-2000,5}]
         ]
Output  10.3222
Output  1.00044*10^11

Input   (* A 0% WinsorizedVariance is equivalent to Variance: *)
        WinsorizedVariance[Range[10],0]
        Variance[Range[10]]
Output  55/6
Output  55/6

Input   (* The code begins by calculating the variance, Winsorized variance, trimmed variance,
        mean, Winsorized mean, trimmed mean, Sqrt of variance, Sqrt of Winsorized variance,
        Sqrt of trimmed variance, and length of the dataset h. It then proceeds to create
        two plots: The first plot is a line plot of the dataset h with filled areas below
        the points, visualizing the distribution of the data. The second plot displays a
        dataset and lines. It includes the line plot of h, a horizontal line representing
        the Winsorized mean, and upper and lower bands around the Winsorized mean, which
        represent square-root-winsorized-variance above and below the Winsorized mean: *)

        h={133,136,149,133,123,121,140,139,117,117,136,108,126,104,116,147,140,148,150,122,
        135,146,133,144,117,124,135,117,120,121,110,124,103,137,101,119,104,113,139,133,500
        };
```

```
v=N[Variance[h]];
wv=N[WinsorizedVariance[h]];
tv=N[TrimmedVariance[h]];
m=N[Mean[h]];
wm=N[WinsorizedMean[h]];
tm=N[TrimmedMean[h]];
sv=N[Sqrt[v]];
swv=N[Sqrt[wv]];
stv=N[Sqrt[tv]];

n=Length[h];

{v,wv,tv}
{m,wm,tm}
{sv,swv,stv}

ListPlot[
 h,
 Filling->Axis,
 PlotStyle->Purple,
 ImageSize->170
 ]

ListPlot[
 {
  h,
  {{0,wm},{n,wm}},
  {{0,wm-swv},{n,wm-swv}},
  {{0,wm+swv},{n,wm+swv}}
  },
 Joined->{False,True,True,True},
 Filling->{1->wm,3->{4}},
 PlotStyle->{Purple,Automatic,Automatic,Automatic},
 PlotLegends->{"h","winsorized mean","l_square-root-winsorized-variance","u_square-
root-winsorized-variance"},
 AxesLabel->Automatic,
 ImageSize->250
 ]
```

Output {3583.04,195.394,160.703}
Output {136.098,127.61,127.73}
Output {59.8585,13.9783,12.6769}
Output

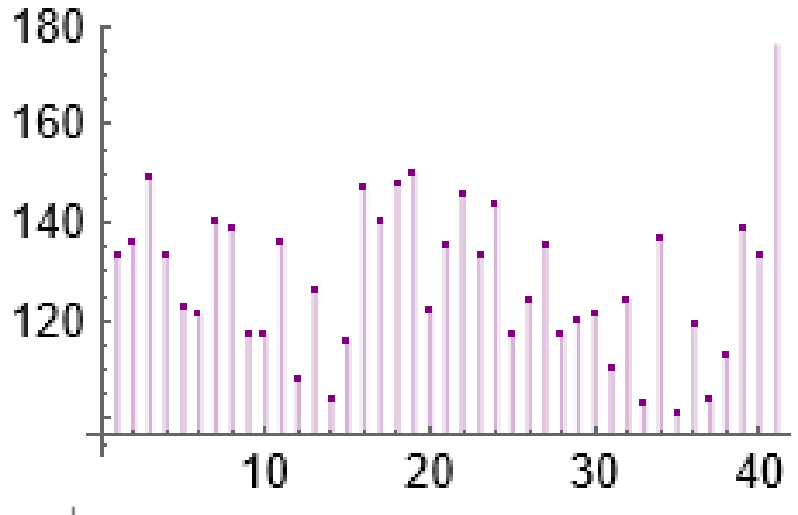

Output

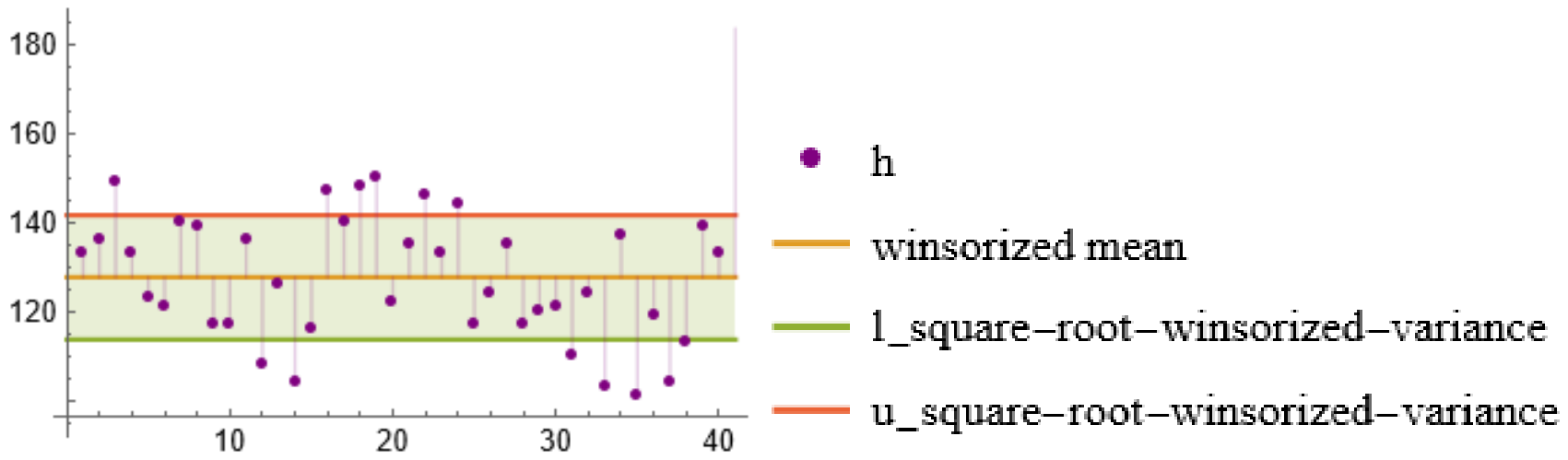

# UNIT 7.2

# SHAPE STATISTICS

A variety of moments are used to summarize a distribution or data. Mean is used to indicate a center location, variance and standard deviation are used to indicate dispersion, etc. The Wolfram Language fully supports moments of any order, univariate or multivariate, for symbolic distributions and data. Moreover, you can get some information about the shape of a distribution using shape statistics functions. Skewness describes the amount of asymmetry. Kurtosis measures the concentration of data around the peak and in the tails versus the concentration in the flanks. let us start by considering moment functions.

| | |
|---|---|
| `Moment[list,r]` | gives the r^(th) sample moment of the elements in list. |
| `Moment[dist,r]` | gives the r^(th) moment of the distribution dist. |
| `Moment[…,{r1,r2,…}]` | gives the {r1,r2,…}^(th) multivariate moment. |
| `Moment[r]` | represents the r^(th) formal moment. |
| `CentralMoment[list,r]` | gives the r^(th) central moment of the elements in list with respect to their mean. |
| `CentralMoment[dist,r]` | gives the r^(th) central moment of the distribution dist. |
| `CentralMoment[r]` | represents the r^(th) formal central moment. |
| `FactorialMoment[list,r]` | gives the r^(th) moment of the elements in the list. |
| `FactorialMoment[dist,r]` | gives the r^(th) moment of the distribution dist. |
| `FactorialMoment[r]` | represents the r^(th) factorial moment. |

***Mathematica Examples 7.8*** Moment

```
Input    (* First Moment: The first moment of a set of data is the mean: *)
         list={1,2,3,4,5};
         moment1=Mean[list]
         Moment[list,1]
Output   3
Output   3

Input    (* Use symbolic data: *)
         Moment[{x,y,z},2]
Output   1/3 (x²+y²+z²)

Input    (* Find moments of WeightedData: *)
         data={8,3,5,4,9,0,4,2,2,3};
         w={0.15,0.09,0.12,0.10,0.16,0.,0.11,0.08,0.08,0.09};
         Moment[
          WeightedData[data,w],
          2
          ]
Output   31.8163

Input    (* Find the moments for univariate distributions: *)
         Moment[
          BinomialDistribution[n,p],
          1
          ]
Output   n p
```

```
Input   (* Moment of order one is the Mean for univariate distributions: *)
        Moment[NormalDistribution[μ,σ],1]==Mean[NormalDistribution[μ,σ]]
        Moment[{x,y,z,t,v},1]==Mean[{x,y,z,t,v}]
Output  True
Output  True

Input   (* The code provides two alternative functions for calculating moments of a dataset.
        The first function (moment1) computes the moment by taking the mean of the data
        raised to the specified power, while the second function (moment2) calculates it by
        averaging the data points raised to the power of k: *)

        moment1[data_,k_]:=Mean[data^k]

        (* or *)
        moment2[data,k_]:=Total[data^k]/Length[data]
        data={1,2,3,4,5};

        moment1[data,3]
        moment2[data,3]
        Moment[data,3]
Output  45
Output  45
Output  45

Input   (* The code creates an interactive interface using the Manipulate function. The
        Manipulate function allows the user to select different distributions from a dropdown
        menu and adjust the maximum order of moments to display using sliders. The code
        calculates the moments based on the selected distribution and maximum order, and
        displays the PDF plot and moments table within the Manipulate output: *)

        Manipulate[
         moments=Table[
           Moment[distribution,k],
           {k,0,maxOrder}
           ];
         Grid[
          {
           {
            Plot[
             PDF[distribution,x],
             {x,-3,3},
             PlotRange->All,
             PlotLabel->"Probability Density Function (PDF)",
             AxesLabel->{"x","PDF"}]
            },
           {
            TableForm[
             Transpose[{Range[0,maxOrder],moments}],
             TableHeadings->{None,{"Order","Moment"}}
             ]
            }
           }
          ],
         {{distribution,NormalDistribution[0,1]},{NormalDistribution[0,1]-
        >"Normal",GammaDistribution[2,1]->"Gamma",UniformDistribution[{-1,1}]->"Uniform"}},
         {{maxOrder,4},0,10,1}
         ]
```

Output

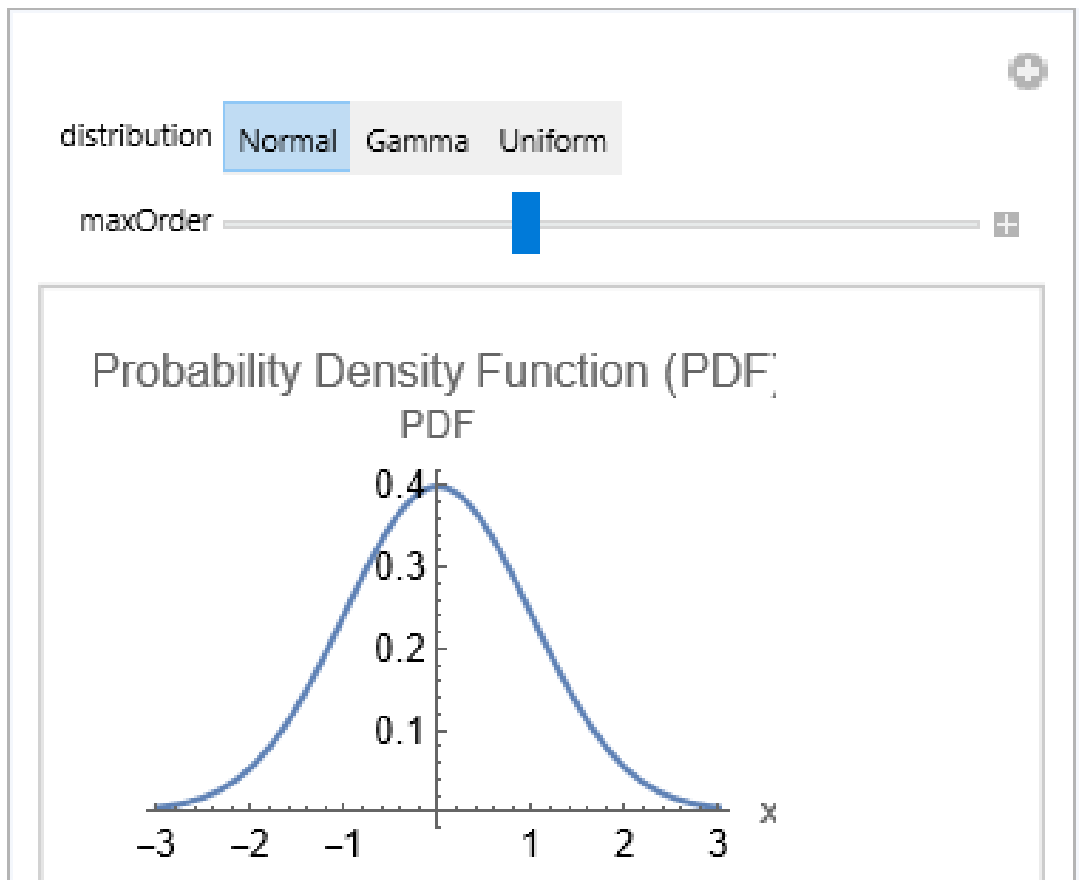

| Order | Moment |
|---|---|
| 0 | 1 |
| 1 | 0 |
| 2 | 1 |
| 3 | 0 |
| 4 | 3 |

***Mathematica Examples 7.9*** CentralMoment

```
Input    (* Use symbolic data: *)
         CentralMoment[{x,y,z},2]
Output   1/3 ((x+1/3 (-x-y-z))^2+(y+1/3 (-x-y-z))^2+(1/3 (-x-y-z)+z)^2)

Input    (* Exact input yields exact output: *)
         CentralMoment[{1,2,3,4,5,6},4]
Output   707/48

Input    (* Find central moments of WeightedData: *)
         CentralMoment[
          WeightedData[{1,2,3},{x,y,z}],
          2
          ]
Output   (x y+4 x z+y z)/(x+y+z)^2

Input    (* The second central moment is a scaled Variance: *)
         data=RandomReal[10,20];
         Variance[data] (Length[data]-1)/Length[data]
         CentralMoment[data,2]
Output   8.03318
Output   8.03318

Input    (*RootMeanSquare of deviations is the square root of a CentralMoment:*)
         data=RandomReal[10,10];
         Sqrt[
          CentralMoment[data,2]
          ]

         RootMeanSquare[data-Mean[data]]
Output   2.25406
Output   2.25406

Input    (* The n^(th) CentralMoment is the Mean of deviations raised to the n^(th) power: *)
         CentralMoment[{a,b,c},n]
         Mean[({a,b,c}-Mean[{a,b,c}])^n]
```

```
Output   1/3 ((a+1/3 (-a-b-c))^n+(b+1/3 (-a-b-c))^n+(1/3 (-a-b-c)+c)^n)
Output   1/3 ((a+1/3 (-a-b-c))^n+(b+1/3 (-a-b-c))^n+(1/3 (-a-b-c)+c)^n)

Input    (* StandardDeviation is the square root of a scaled CentralMoment: *)
         data=RandomReal[10,5];
         StandardDeviation[data]
         Sqrt[CentralMoment[data,2] Length[data]/(Length[data]-1)]
Output   2.41201
Output   2.41201

Input    (* The code defines a function called computeCentralMoment using the Module function.
         This function takes two parameters: data_, representing the dataset, and n_,
         representing the order of the central moment. The code uses a summation formula to
         calculate the central moment manually. It iterates over each element of the dataset,
         subtracts the mean from the element, raises the result to the 4th power, and
         accumulates the values in the centralMoment variable. Finally, it divides the sum by
         the length of the dataset to obtain the average.  Also it provides an example usage
         by assigning a dataset (data) and an order of the central moment (n). Additionally,
         it calculates the central moment using the built-in CentralMoment function: *)

         computeCentralMoment[data_,n_]:=Module[
           {mean,centralMoment,length},
           mean=Mean[data];
           length=Length[data];
           centralMoment=Sum[(data[[i]]-mean)^n,{i,length}]/length;
           centralMoment
           ]
         data={1,2,3,4,5,6,7,8};
         n=4;
         centralMoment=computeCentralMoment[data,n]
         CentralMoment[data,n]
Output   777/16
Output   777/16

Input    (* In this code, the Manipulate function creates an interactive interface that allows
         you to explore the effects of changing the distributions and the order of the central
         moment. The code includes a histogram plot of the sample data using the Histogram
         function. The plot is displayed as a probability density function, and the calculated
         central moment is shown as a text label. The Manipulate parameters include the choice
         of distribution, sample size, and order of the central moment. The user can select
         different distributions, adjust the sample size using a slider, and modify the order
         of the central moment: *)

         Manipulate[
          Module[
           {data,mean,centralMoment},
           (*Generate sample data from a distribution*)
           data=RandomVariate[Distribution,sampleSize];
           (*Calculate central moment*)
           centralMoment=CentralMoment[data,n];
           (*Plotting the histogram*)
           Histogram[
             data,
             Automatic,
             "PDF",
             PlotRange->All,
             PlotLabel->{Text["Central Moment ("<>ToString[n]<>"):
         "<>ToString[centralMoment]]},
             ImageSize->300,
             ColorFunction->Function[Opacity[0.5]],
             ChartStyle->Purple
```

```
    ]
   ],
  (*Manipulate parameters*)
  {{Distribution,NormalDistribution[0,1],"Distribution:"},{NormalDistribution[0,1]->"Normal",GammaDistribution[2,1]->"Gamma",UniformDistribution[{-1,1}]->"Uniform"}},{{sampleSize,1000,"Sample Size:"},{100,500,1000,5000}},
  {{n,2,"Order of Central Moment:"},{1,2,3,4}}
  ]
```

Output

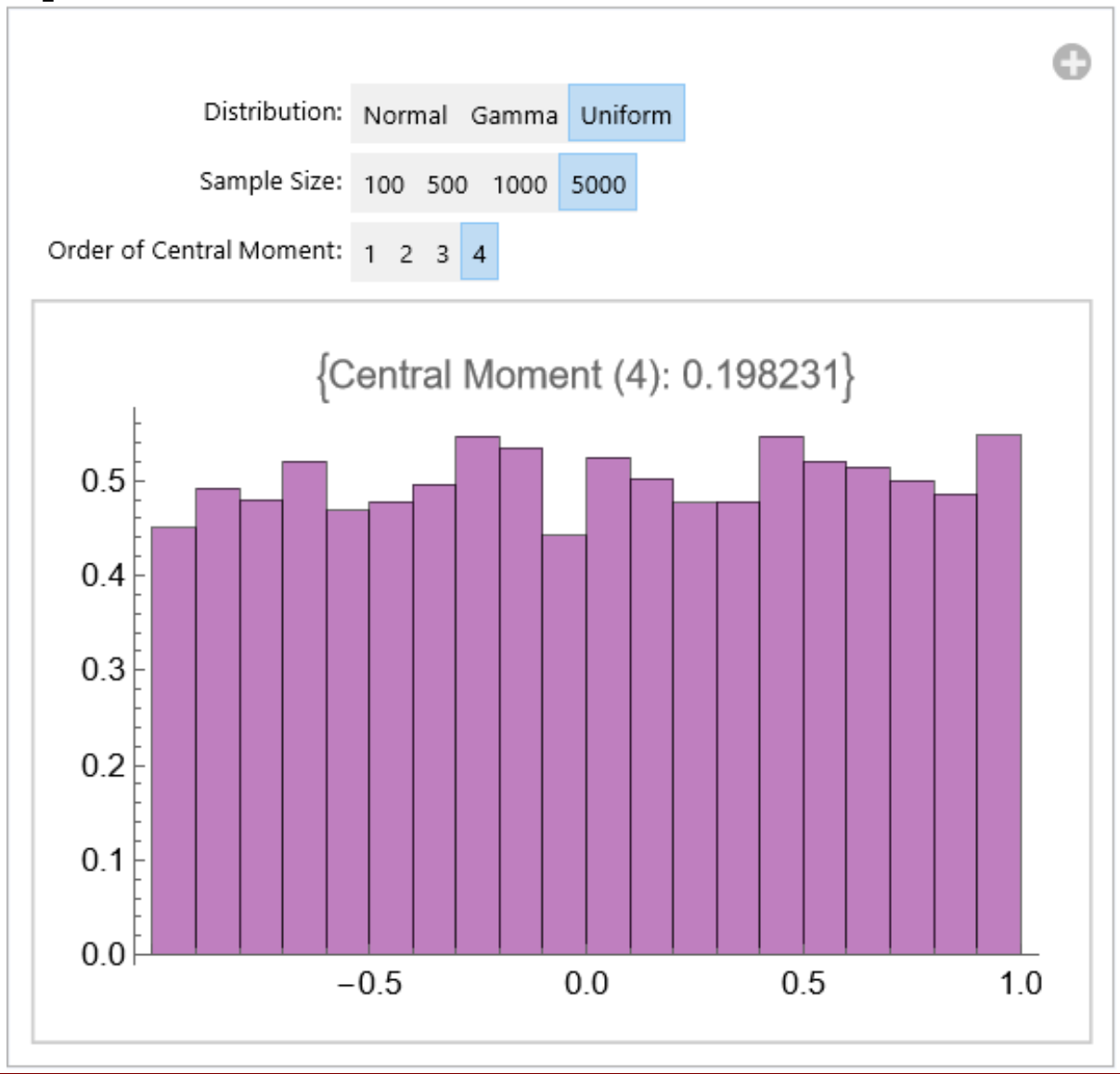

***Mathematica Examples 7.10*** FactorialMoment

```
Input   (* Use symbolic data: *)
        FactorialMoment[{x,y,z},3]
Output  1/3 ((-2+x) (-1+x) x+(-2+y) (-1+y) y+(-2+z) (-1+z) z)

Input   (* Exact input yields exact output: *)
        FactorialMoment[{1,2,3,4,5},1]
Output  3

Input   (* Find factorial moments of WeightedData: *)
        FactorialMoment[
         WeightedData[{1,2,3},{x,y,z}],
         1
         ]
Output  (x+2 y+3 z)/(x+y+z)
Input   (* First factorial moment is equivalent to Mean: *)
        FactorialMoment[{x,y,z,u,v},1]
        Mean[{x,y,z,u,v}]
Output  1/5 (u+v+x+y+z)
Output  1/5 (u+v+x+y+z)
```

| | |
|---|---|
| `Skewness[list]` | gives the coefficient of skewness for the elements in list. |
| `Skewness[dist]` | gives the coefficient of skewness for the distribution dist. |
| `QuartileSkewness[list]` | gives the coefficient of quartile skewness for the elements in list. |
| `QuartileSkewness[dist]` | gives the coefficient of quartile skewness for the distribution dist. |
| `Kurtosis[list]` | gives the coefficient of kurtosis for the elements in list. |
| `Kurtosis[dist]` | gives the coefficient of kurtosis for the distribution dist. |

**Mathematica Examples 7.11** Skewness

```
Input   (* Skewness for a list of values: *)
        Skewness[{a,b,c}]
Output  (√3 ((a+1/3 (-a-b-c))^3+(b+1/3 (-a-b-c))^3+(1/3 (-a-b-c)+c)^3))/((a+1/3 (-a-b-
        c))^2+(b+1/3 (-a-b-c))^2+(1/3 (-a-b-c)+c)^2)^3/2

Input   (* Skewness for a matrix gives column-wise skewness: *)
        Skewness[N[{{1,3,4},{4,6,1},{12,1,6}}]]
        Skewness[N[{1,4,12}]]
        Skewness[N[{3,6,1}]]
        Skewness[N[{4,1,6}]]
Output  {0.492208,0.239063,-0.239063}
Output  0.492208
Output  0.239063
Output  -0.239063

Input   (* Find the skewness of WeightedData: *)
        data={8,3,5,4,9,0,4,2,2,3};
        w={0.15,0.09,0.12,0.10,0.16,0.,0.11,0.08,0.08,0.09};
        Skewness[
         WeightedData[data,w]
         ]
Output  0.475178

Input   (* Skewness is a ratio of powers of third and second central moments: *)
        data=RandomReal[15,5];
        Skewness[data]
        CentralMoment[data,3]/(CentralMoment[data,2]^(3/2))
Output  0.158081
Output  0.158081

Input   (* Computing the skewness using the formula: *)
        data={1,2,3,4,5};
        n=Length[data];
        mean=Mean[data];
        sd=StandardDeviation[data];
        skewness=(1/sd^3)*Sum[
            (1/n)(data[[i]]-mean)^3,
            {i,1,n}
            ]
        Skewness[data]
Output  0
Output  0

Input   (* In this code, we define a function called skewness that takes a list of data
        points as an argument. Within the Module, we create local variables n, mean, stdDev,
        and skew to store intermediate results. Next, we calculate the skewness using the
        formula:(1/n) Sum[((data[[i]]-mean)/stdDev)^3,{i,1,n}]. This formula calculates the
        third standardized moment, which represents the skewness of the data: *)

        skewness[data_]:=Module[
          {n,mean,stdDev,skew},
          n=Length[data];
          mean=Mean[data];
          stdDev=StandardDeviation[data];
          skew=(1/n) Sum[((data[[i]]-mean)/stdDev)^3,{i,1,n}];
          skew
          ]
```

```
        data={1,2,3,4,5};
        result=skewness[data];
        Print["Skewness: ",result]
        Print["Skewness: ",Skewness[data]]
Output  Skewness:  0
Output  Skewness:  0

Input   (* The code creates three skew normal distributions with different shape parameter α
        and visualizes their probability density functions (PDFs) on a plot. It computes the
        skewness values for each distribution and displays them in the plot legend. The code
        demonstrates how the shape parameter α affects the shape of the distributions: *)

        dist1=SkewNormalDistribution[0,1,-3];   (* α=-3*)
        dist2=SkewNormalDistribution[0,1,0];    (* α=0*)
        dist3=SkewNormalDistribution[0,1,3];    (* α=3*)

        skewness1=N[Skewness[dist1]];
        skewness2=N[Skewness[dist2]];
        skewness3=N[Skewness[dist3]];

        Plot[
         {PDF[dist1,x],PDF[dist2,x],PDF[dist3,x]},
         {x,-5,5},
         PlotStyle->{Purple,Darker[Red],Darker[Blue]},
         Filling->Axis,
         PlotLegends->{
           "Skewness = "<>ToString[skewness1],
           "Skewness = "<>ToString[skewness2],
           "Skewness = "<>ToString[skewness3]
           },
         Frame->True,
         FrameLabel->{"x","PDF"},
         PlotRange->All,
         ImageSize->250
         ]
Output
```

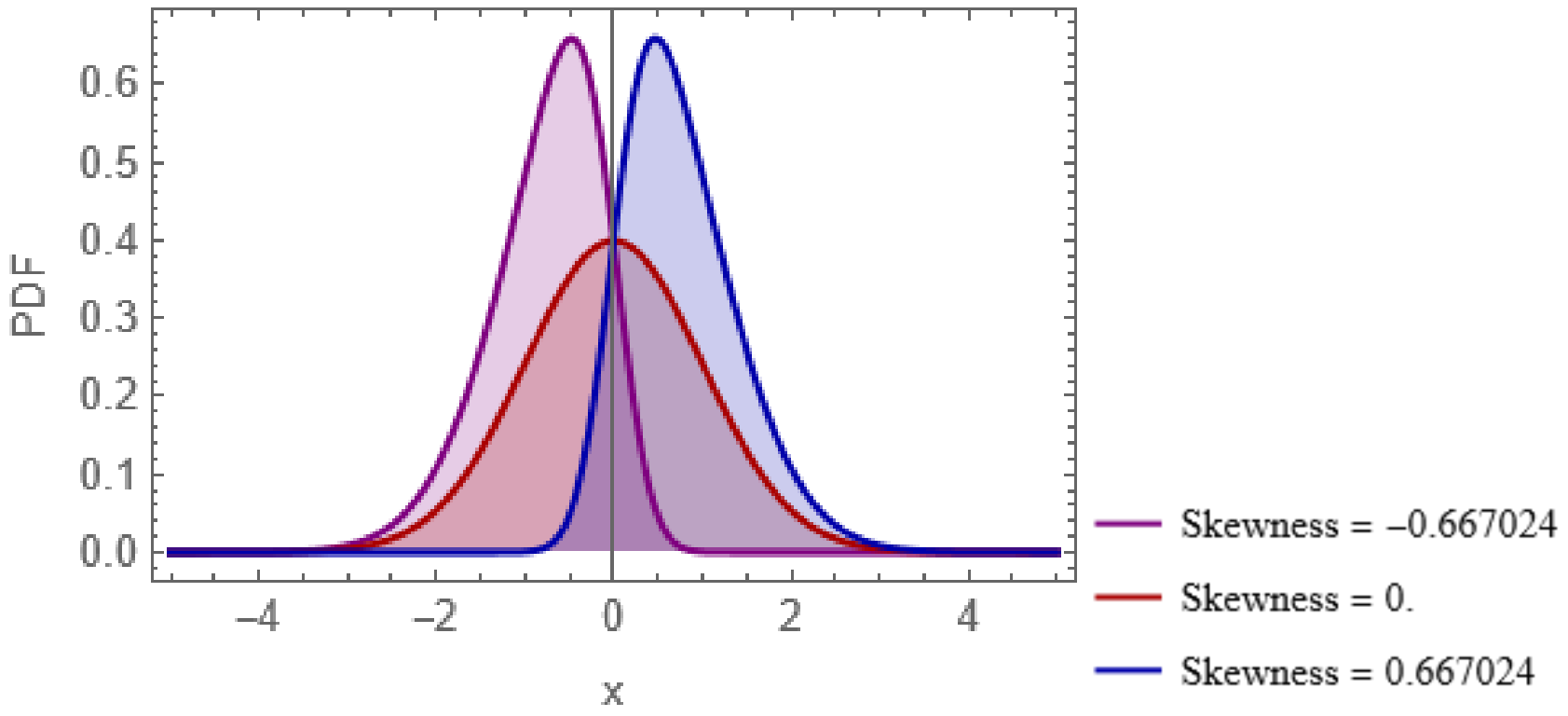

```
Input   (* The code defines a function dist[x_,α_] that represents a skewed normal
        distribution with mean 0, standard deviation 1, and the shape parameter α. The code
        then uses the Manipulate function to create an interactive plot. The plot displays
        the probability density function (PDF) of the skewed normal distribution for different
        values of α. Inside the Manipulate function, the PlotLabel option is set to display
        the skewness value of the distribution corresponding to the current value of α. The
        skewness is calculated using the Skewness function, which takes the
        SkewNormalDistribution with parameters 0, 1, and α as an argument. By including the
        skewness value in the plot label, the code helps to visually explain the skewness of
        the dist distribution. As you adjust the α value using the Manipulate slider, the
```

```
plot label will update in real-time to reflect the current skewness of the
distribution: *)

dist[x_,α_]:=SkewNormalDistribution[0,1,α]

Manipulate[
 Plot[
  PDF[dist[x,α],x],
  {x,-5,5},
  PlotRange->All,
  Filling->Axis,
  PlotStyle->Purple,
  Frame->True,
  FrameLabel->{"x","PDF"},
  PlotLabel->Row[{"Skewness: ",Skewness[SkewNormalDistribution[0,1,α]]}],
  ImageSize->300
  ],
 {{α,0,"α"},-5,5,0.1,Appearance->"Labeled"}
 ]
```

Output

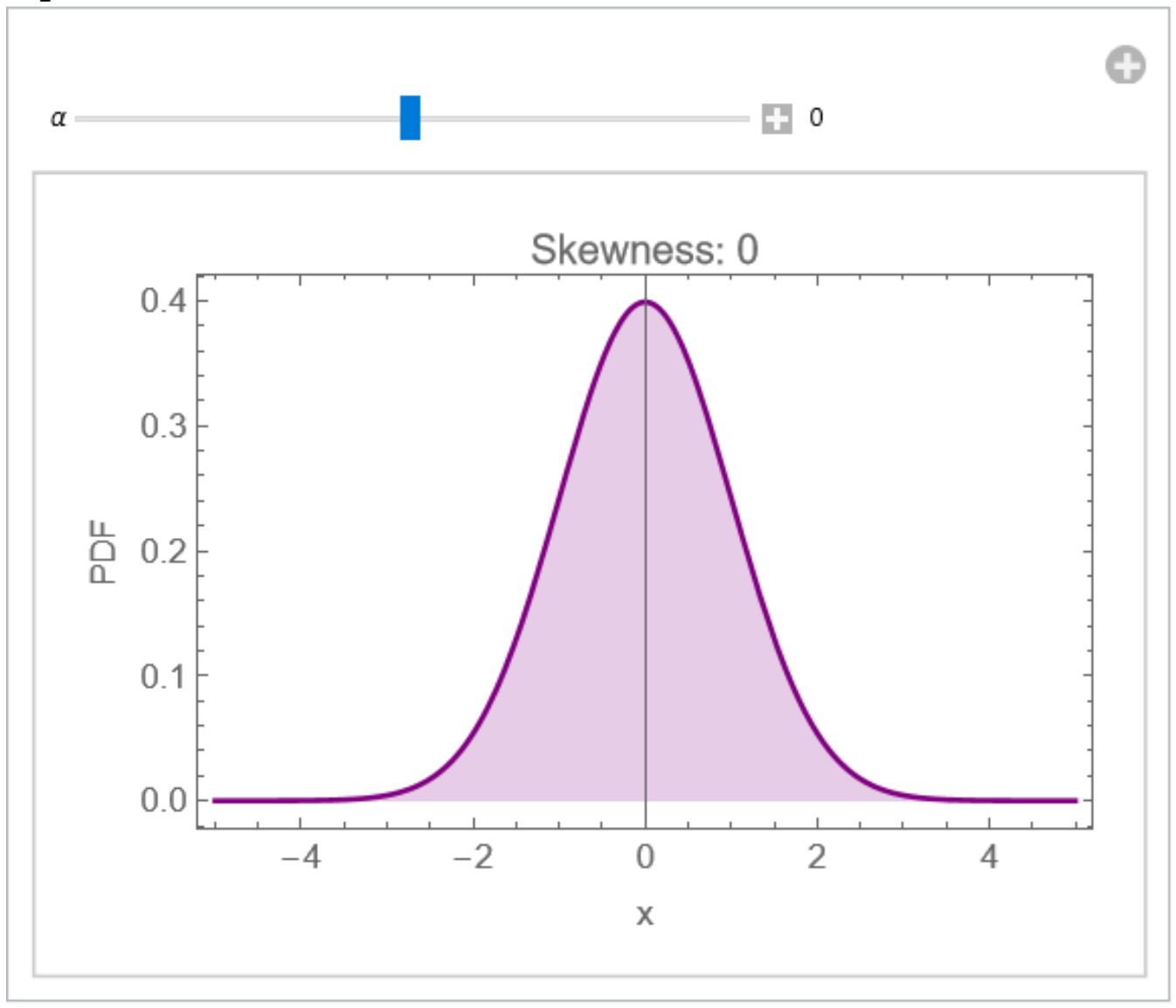

Input

```
(* The code generates a sample dataset from a skewed normal distribution and performs
various visualizations and calculations on the data. The code begins by generating
1000 random numbers from a skewed normal distribution and storing them in the variable
data. It then calculates the skewness of the data using the Skewness function and
prints the result. Next, the code creates a histogram of the data using the Histogram
function, displaying the probability density function (PDF). It also creates two
scatter plots: listplot1 shows the original data points, and listplot2 displays the
sorted data with respect to their ranks. In the case of a skewed distribution, the
plot from listplot2 can help identify if the points are more concentrated towards
one end, suggesting a skew in that direction. If the points are predominantly
clustered towards the lower ranks (left side of the plot),it indicates a negative
skew, while clustering towards the higher ranks (right side of the plot) indicates a
positive skew: *)

(*Generate sample data*)
data=RandomVariate[
    SkewNormalDistribution[0,1,2],
    1000
    ];
```

```
(* Calculate skewness of the data *)
skewness=Skewness[data];
Print["Skewness: ",skewness]

(* Create histogram *)
histogram=Histogram[
    data,
    Automatic,
    "PDF",
    ColorFunction->Function[Opacity[0.7]],
    ChartStyle->Purple,
    Frame->True,
    FrameLabel->{"x","PDF"},
    ImageSize->250
    ]  ;

(* Create ListPlot1 for data *)
listplot1=ListPlot[
    data,
    PlotRange->All,
    ColorFunction->Function[Opacity[0.7]],
    PlotStyle->Directive[Purple,PointSize[0.01]],
    ImageSize->250
    ]  ;
(* Create ListPlot2 for sorted data *)
listplot2=ListPlot[
    {Transpose[{Range[Length[data]],Sort[data]}]},
    PlotRange->All,
    PlotStyle->Directive[PointSize[0.002],Purple],
    Frame->True,
    FrameLabel->{"Rank","Value"},
    ImageSize->250
    ];

Column[
 {histogram,listplot1,listplot2}
 ]
```

Output Skewness: 0.443087

Output

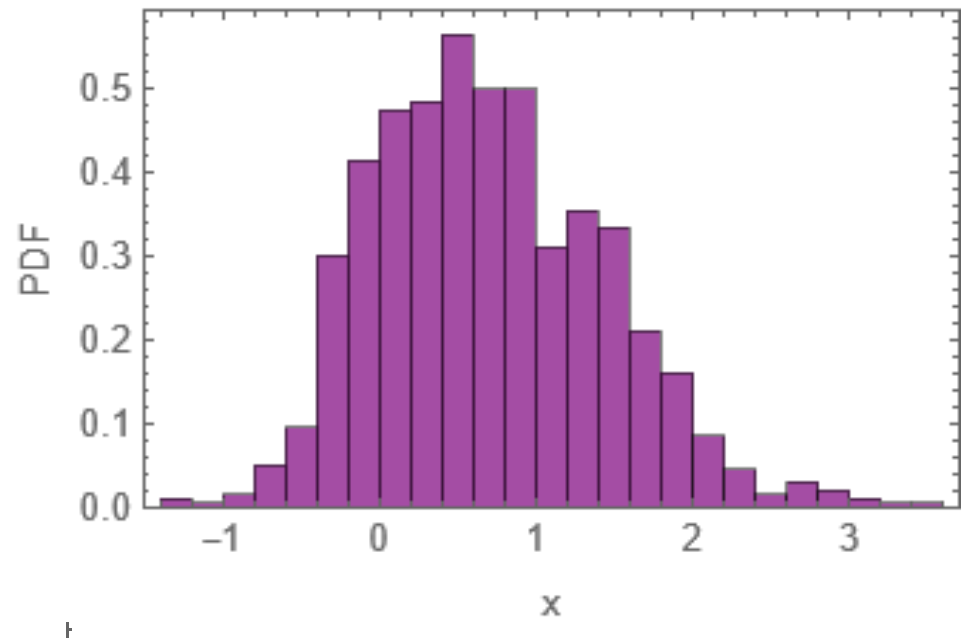

Output

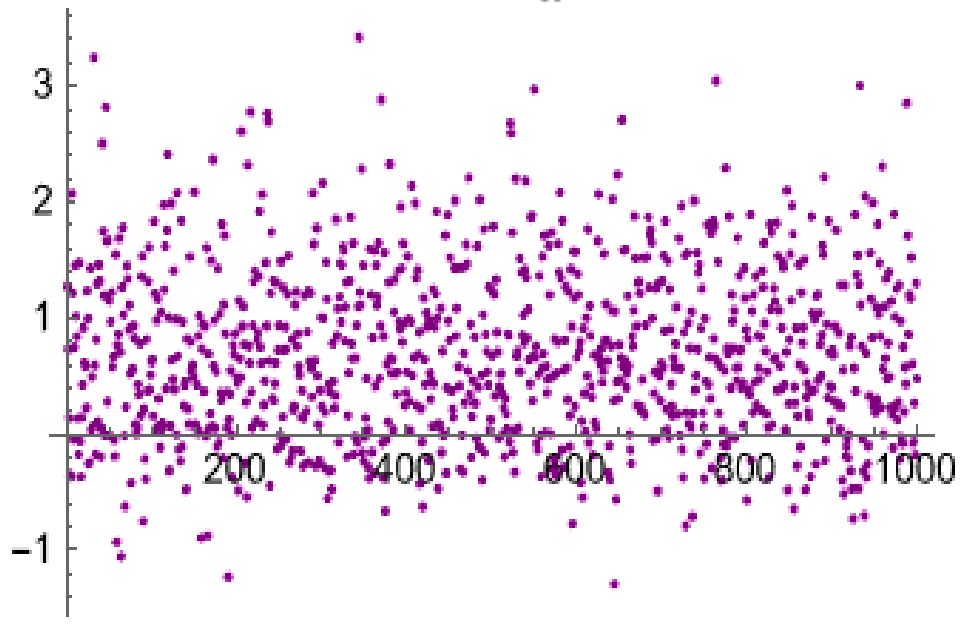

Output

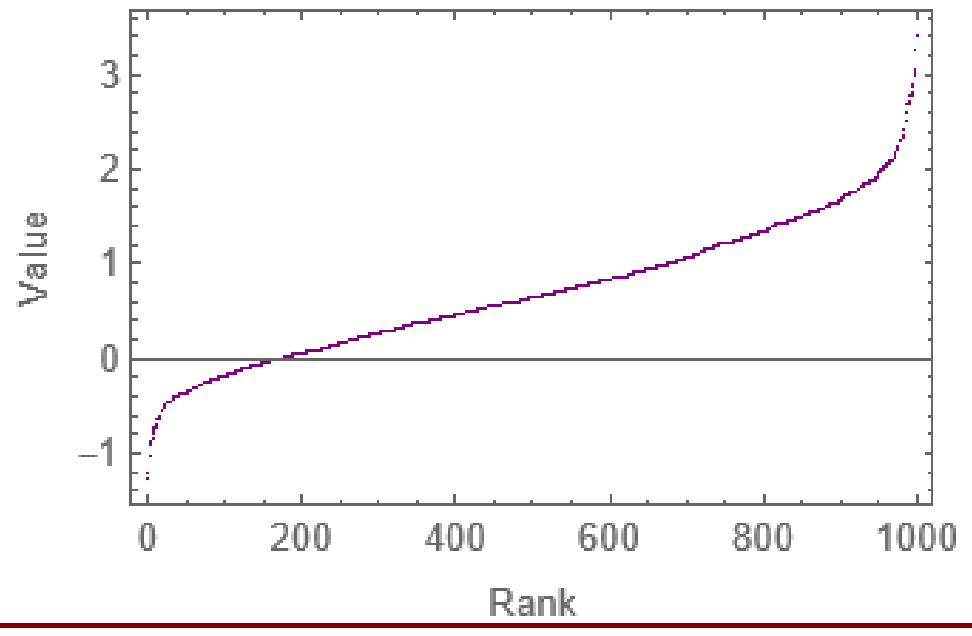

***Mathematica Examples 7.12*** QuartileSkewness

```
Input   (* Quartile skewness for a list of exact numbers: *)
        list={1,2,3,4,5,6,7,8,9,10};
        QuartileSkewness[list]
Output  0

Input   (* QuartileSkewness for a matrix gives column-wise ranges: *)
        QuartileSkewness[{{1,2},{3,5},{1,8},{5,6},{7,8},{2,4}}]
        QuartileSkewness[{1,3,1,5,7,2}]
        QuartileSkewness[{2,5,8,6,8,4}]
Output  {1/4,1/4}
Output  1/4
Output  1/4

Input   (* Find the quartile skewness for WeightedData: *)
        data={8,3,5,4,9,0,4,2,2,3};
        w={0.15,0.09,0.12,0.10,0.16,0.,0.11,0.08,0.08,0.09};
        QuartileSkewness[
         WeightedData[data,w]
         ]
Output  3/5

Input   (* Obtain a robust estimate of asymmetry when extreme values are present: *)
        N[
         QuartileSkewness[{5,3,10^5,2,1,400,8}]
         ]

        (* Measures based on the Mean are heavily influenced by extreme values: *)
        N[
         Skewness[{5,3,10^5,2,1,400,8}]
         ]
Output  0.981651
Output  2.04118

Input   (* QuartileSkewness is a function of quartiles: *)

        data=RandomReal[10,20];
        QuartileSkewness[data]
        {q1,q2,q3}=Quartiles[data]
        (q1-2 q2+q3)/(q3-q1)
Output  0.00634515
Output  {1.64113,5.17119,8.74634}
Output  0.00634515

Input   (* This code provides a visual representation of a sample dataset from a normal
        distribution through the histogram and includes the quartile skewness value to provide
        insights into its asymmetry. The quartile skewness measure highlights the skewness
```

```
characteristics based on the positions of the quartiles, enhancing the understanding
of the dataset: *)

dist=NormalDistribution[0,1];
qs=QuartileSkewness[dist];
Histogram[
 RandomVariate[dist,1000],
 Automatic,
 "PDF",
 Epilog->{Text["QuartileSkewness = "<>ToString[qs],{1.5,0.4}]},
 ColorFunction->Function[{height},Opacity[height]],
 ChartStyle->Purple,
 ImageSize->250
 ]
```

Output

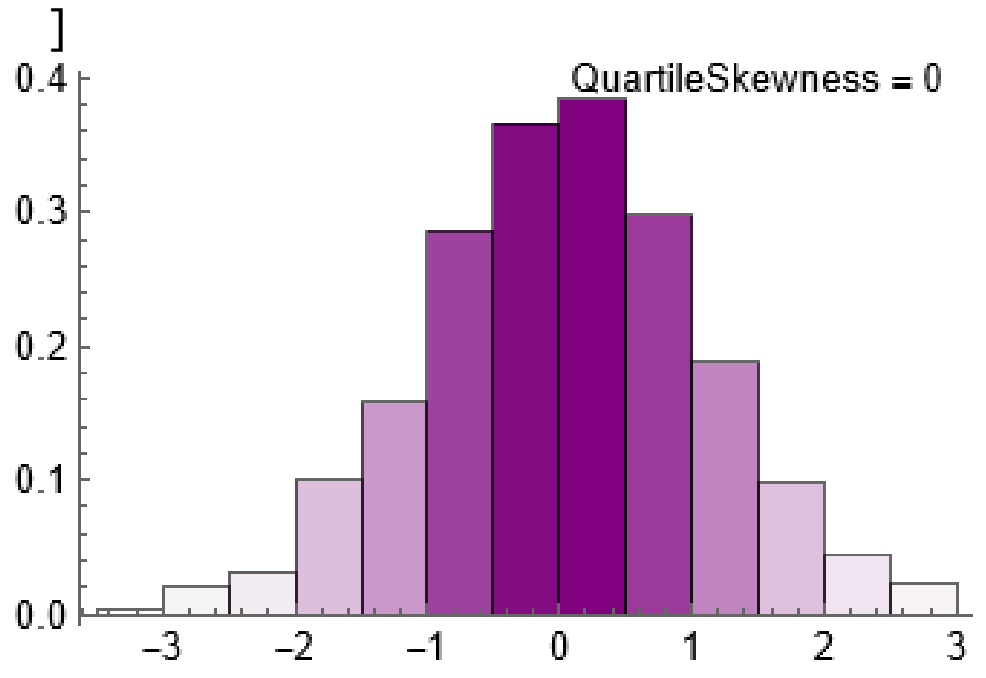

Input

```
(* The code begins by defining a dataset h consisting of a series of numerical values.
It then calculates the quartile skewness of h using the QuartileSkewness function,
storing the result in the variable qs after converting it to a numerical value using
N. Two visualizations are created based on the dataset. The list plot shows the trend
of the values, while the histogram illustrates the distribution and frequency of the
data points. Additionally, the quartile skewness calculation offers a numerical
measure of the skewness of the dataset, allowing for a quantitative assessment of
its asymmetry: *)

h={133,136,149,133,123,121,140,139,117,117,136,108,126,104,116,147,140,148,150,122,
135,146,133,144,117,124,135,117,120,121,110,124,103,137,101,119,104,113,139,133};

qs=N[QuartileSkewness[h]]

ListPlot[
 h,
 Filling->Axis,
 PlotStyle->Purple,
 ImageSize->170
 ]

Histogram[
 h,
 Automatic,
 ColorFunction->Function[{height},Opacity[height]],
 ChartStyle->Purple,
 LabelingFunction->Above,
 ImageSize->170
 ]
```

Output 0.238095

Output 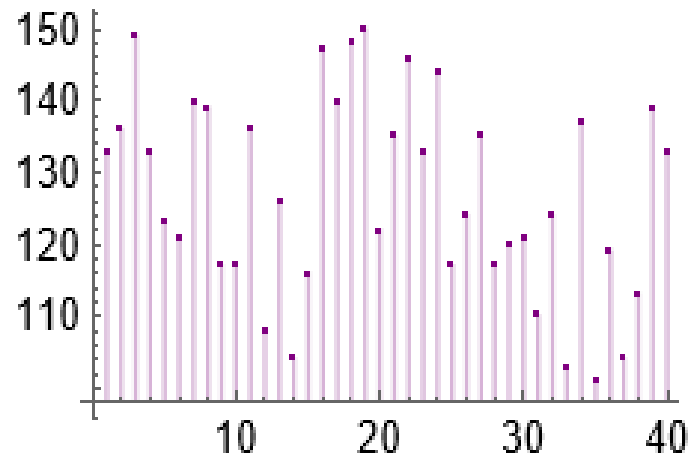

Output 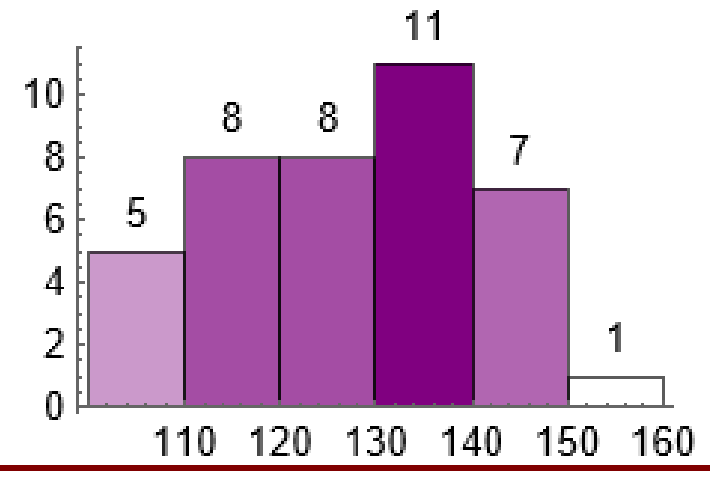

***Mathematica Examples 7.13*** Kurtosis

```
Input    (* Kurtosis for a list of values: *)
         Kurtosis[{a,b,c}]
Output   (3 ((a+1/3 (-a-b-c))^4+(b+1/3 (-a-b-c))^4+(1/3 (-a-b-c)+c)^4))/((a+1/3 (-a-b-
         c))^2+(b+1/3 (-a-b-c))^2+(1/3 (-a-b-c)+c)^2)^2

Input    (* Kurtosis for a matrix gives column-wise kurtosis: *)
         Kurtosis[{{1,3,4},{4,6,1},{12,1,6}}]//Together
         Kurtosis[{1,4,12}]
         Kurtosis[{3,6,1}]
         Kurtosis[{4,1,6}]
Output   {3/2,3/2,3/2}
Output   3/2
Output   3/2
Output   3/2

Input    (* Find the kurtosis of WeightedData: *)
         data={8,3,5,4,9,0,4,2,2,3};
         w={0.15,0.09,0.12,0.10,0.16,0.,0.11,0.08,0.08,0.09};
         Kurtosis[
          WeightedData[data,w]
          ]
Output   1.68723

Input    (* Kurtosis is a ratio of powers of fourth and second central moments: *)
         data=RandomReal[15,5];
         Kurtosis[data]
         CentralMoment[data,4]/CentralMoment[data,2]^2
Output   2.03
Output   2.03

Input    (* The code defines a function named kurtosis that calculates the kurtosis of a data
         set. It computes the mean, variance, third moment, and fourth moment of the data
         using built-in functions. The kurtosis is then calculated using a formula that
         normalizes the fourth central moment by dividing it by the square of the sample
         variance multiplied by (n-1)/n, where n is the length of the data set: *)

         kurtosis[data_]:=Module[
           {mean,variance,thirdMoment,fourthMoment},
           mean=Mean[data];
           variance=Variance[data];
           thirdMoment=CentralMoment[data,3];
```

```
    fourthMoment=CentralMoment[data,4];
    CentralMoment[data,4]/(Variance[data](Length[data]-1)/Length[data])^2
    ]

  data={1,2,3,4,5,6,7,8,9,10};   (*Replace with your own data*)
  kurtosis[data]
  Kurtosis[data]
```

Output 293/165

Output 293/165

Input (* The code generates a histogram of 1000 random variates sampled from a standard normal distribution. It calculates the kurtosis of the distribution and displays it as an annotation on the plot. Kurtosis measures the "tailedness" or the shape of the distribution: *)

```
dist=NormalDistribution[0,1];
ku=Kurtosis[dist]
Histogram[
 RandomVariate[dist,1000],
 Automatic,
 "PDF",
 Epilog->{Text["Kurtosis = "<>ToString[ku],{1.5,0.35}]},
 ColorFunction->Function[{height},Opacity[height]],
 ChartStyle->Purple,
 ImageSize->220
 ]
```

Output 3

Output

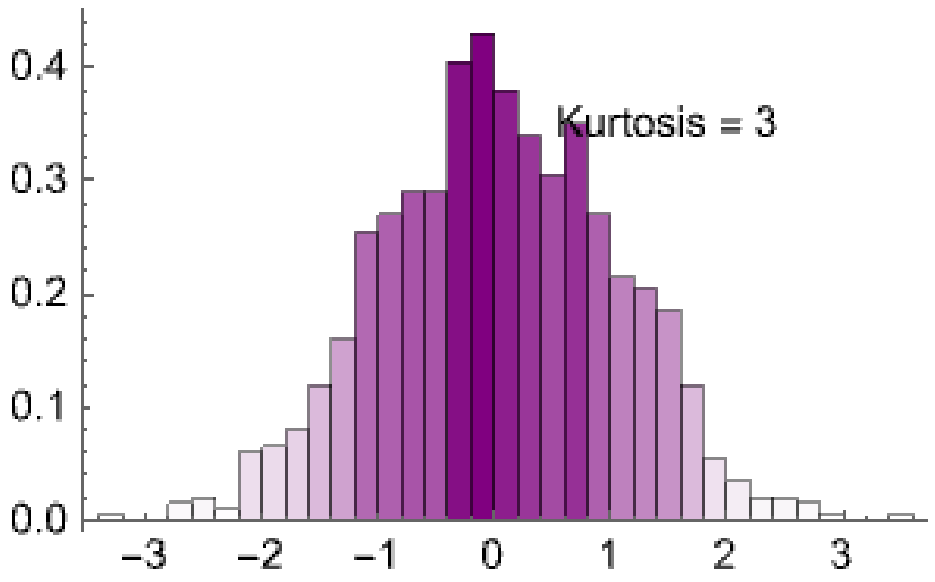

Input (* The code generates a plot of the probability density functions (PDFs) for three different distributions: normal distribution, Student's t-distribution, and Laplace distribution. It also displays the kurtosis values for each distribution in a table placed in the top-right corner of the plot: *)

```
dist1=NormalDistribution[0,1];
dist2=StudentTDistribution[8];
dist3=LaplaceDistribution[0,Sqrt[2]];

pdf1=PDF[dist1,x];
pdf2=PDF[dist2,x];
pdf3=PDF[dist3,x];

kurtosis1=Kurtosis[dist1];
kurtosis2=Kurtosis[dist2];
kurtosis3=Kurtosis[dist3];

Plot[
 {pdf1,pdf2,pdf3},
 {x,-4,4},
```

```
 PlotLegends->{"Normal Distribution","Student's t-Distribution","Laplace
Distribution"},
 PlotRange->All,
 Frame->True,
 FrameLabel->{"x","PDF"},
 ImageSize->300,
 Epilog->Inset[
   Grid[
    {
     {"Kurtosis","Distribution"},
     {kurtosis1,"Normal Distribution"},
     {kurtosis2,"Student's t-Distribution"},
     {kurtosis3,"Laplace Distribution"}
     }
    ],
   {Right,Top},{Right,Top}
   ]
 ]
```

Output

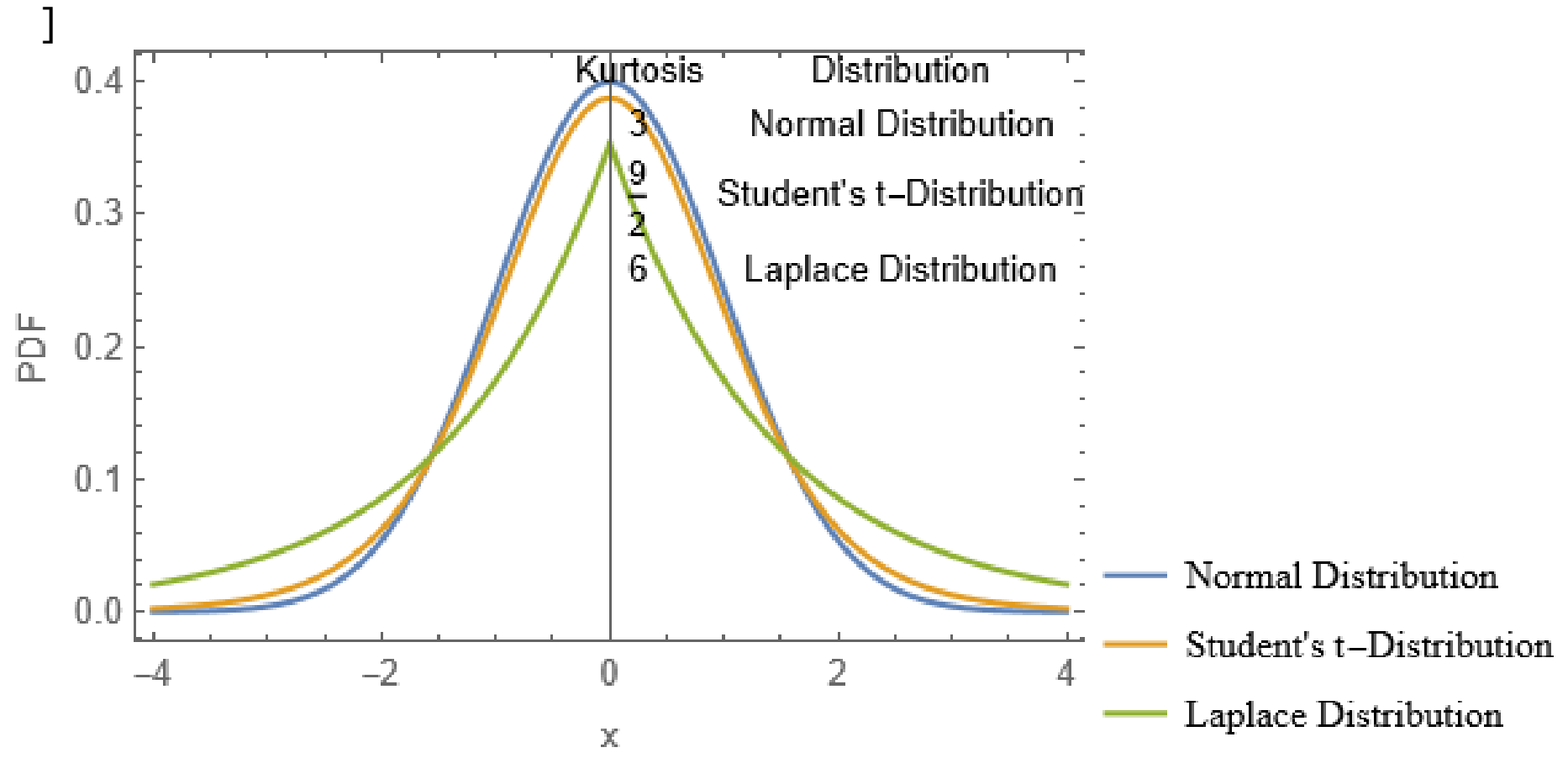

Input

```
(* The code performs weighted descriptive statistics on a dataset. It calculates
several statistics including the mean, median, variance, standard deviation,
kurtosis, skewness, and interquartile range. The dataset is accompanied by
corresponding weights, which are taken into account during the calculations. The
results are displayed in a tabular format using the Grid function: *)

(* Create weighted univariate data: *)
data={8,3,5,4,9,0,4,2,2,3};
w={0.15,0.09,0.12,0.10,0.16,0.,0.11,0.08,0.08,0.09};
a=WeightedData[data,w];
(* Some weighted descriptive statistics: *)
Grid[
 Table[
  {i,i[a]},

{i,{Mean,Median,Variance,StandardDeviation,Kurtosis,Skewness,InterquartileRange}}
  ]
 ]
```

Output

```
{
 {Mean, 5.04082},
 {Median, 4},
 {Variance, 7.26594},
 {StandardDeviation, 2.69554},
 {Kurtosis, 1.68723},
 {Skewness, 0.475178},
 {InterquartileRange, 5}
}
```

# CHAPTER 8

# FUNDAMENTAL PRINCIPLES OF PROBABILITY

Probability theory is a branch of mathematics that deals with the study of uncertainty and randomness. It provides a framework for analyzing and quantifying the likelihood of events occurring in various scenarios. In this chapter, we consider three principal concepts in probability namely, interpretations of probability, counting techniques, and conditional probability.

- In probability theory, there are different interpretations of what probability represents. These interpretations reflect different philosophical viewpoints and have implications for how probabilities are understood and used. Some common interpretations include:
  - Classical Interpretation: The classical interpretation assumes that all outcomes of an experiment are equally likely, and probability is calculated as the ratio of favorable outcomes to the total number of equally likely outcomes.
  - Frequency Interpretation: The frequency interpretation views probability as the long-term relative frequency of an event occurring in a repeated experiment or observation.
- Counting techniques form another fundamental concept of probability theory. These techniques provide methods to determine the number of possible outcomes in a given situation. The two fundamental techniques are permutations and combinations:
  - Permutations deal with the arrangement of objects in a specific order. It involves counting the number of ways to select and arrange a subset of objects from a larger set.
  - Combinations focus on selecting objects without considering their order.
- Conditional probability is a crucial concept in probability theory. Conditional probability is essential in understanding dependent events, where the occurrence of one event affects the likelihood of the other. It helps in modeling real-world scenarios.

Probability theory also forms the foundation for more advanced concepts, such as random variables, probability distributions, and statistical inference. These concepts enable us to model and analyze complex systems, estimate parameters, and make predictions based on observed data. In the following chapters, we represent these concepts in detail.

## 8.1 Elementary Probability Theory

Statisticians use the word experiment to describe any process that generates a set of data. A simple example of a statistical experiment is the tossing of a coin. In this experiment, there are only two possible outcomes, heads or tails.

**Definition (Random Experiment):** An experiment that can result in different outcomes, even though it is repeated in the same manner every time, is called a random experiment.

Assume that the experiment can be repeated any number of times under identical conditions. Each repetition is called a trial. A (random) experiment satisfies the following three conditions:

1. The set of all possible outcomes is known in advance in each trial;
2. In any particular trial, it is not known which particular outcome will happen; and
3. The experiment can be repeated under identical conditions.

**Definition (Sample Space):** The set of all possible outcomes of a random experiment is called the sample space of the experiment. The sample space is denoted as $S$.

**Definition (Discrete Sample Space):** A sample space is discrete if it consists of a finite or countable infinite set of outcomes.
**Definition (Continuous Sample Space):** A sample space is continuous if it contains an interval (either finite or infinite) of real numbers.

***Example 8.1***

Consider the experiment of tossing a die. If we are interested in the number that shows on the top face, the sample space is

$$S_1 = \{1, 2, 3, 4, 5, 6\}.$$

If we are interested only in whether the number is even or odd, the sample space is simply

$$S_2 = \{\text{even}, \text{odd}\}.$$

Note that, Example 8.1 illustrates the fact that more than one sample space can be used to describe the outcomes of an experiment. In some experiments, it is helpful to list the elements of the sample space systematically utilizing a tree diagram.

***Example 8.2***

An experiment consists of flipping a coin and then flipping it a second time if a head occurs. If a tail occurs on the first flip, then a die is tossed once. To list the elements of the sample space providing the most information, we construct the following tree diagram

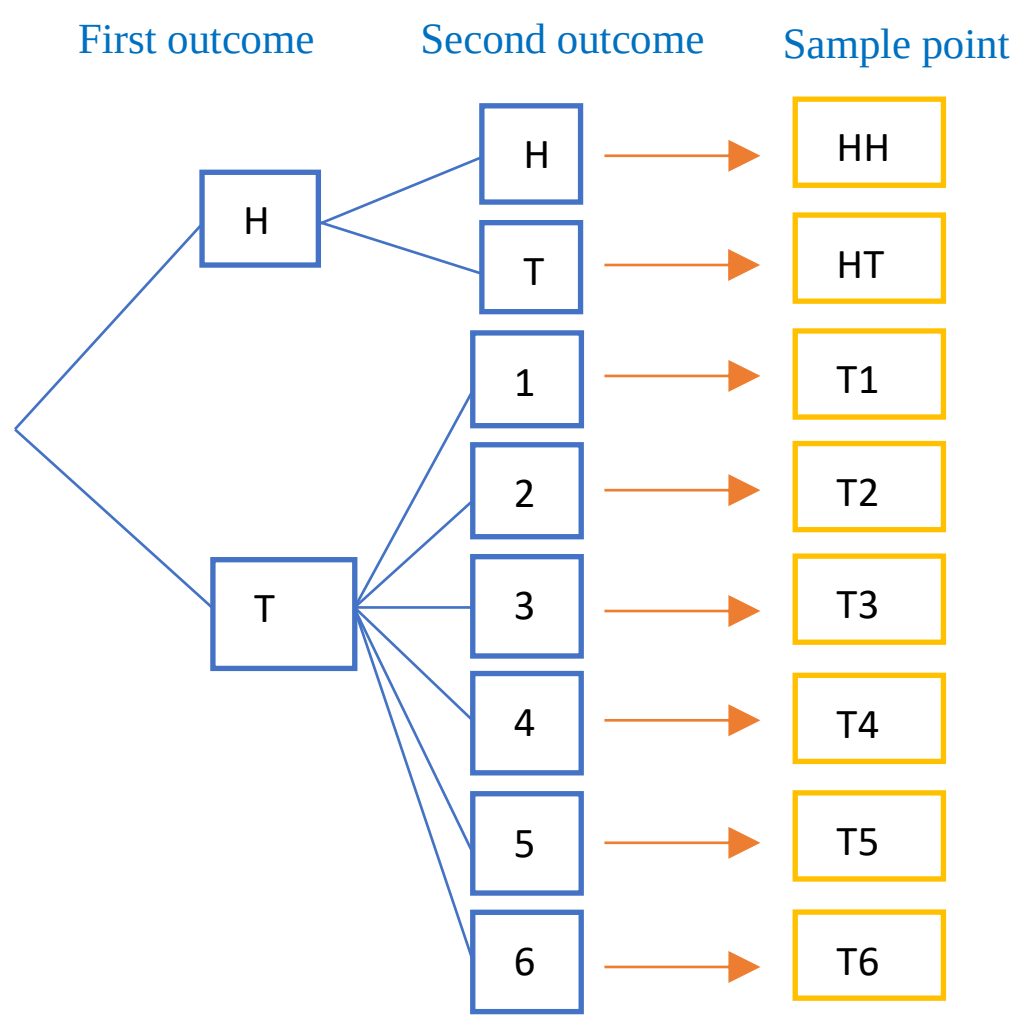

An outcome in $S$ is called a sample point or element. For any given experiment, we may be interested in the occurrence of certain events rather than in the occurrence of a specific element in the sample space. For instance, we may be interested in the event $A$ that the outcome when a die is tossed is divisible by 3. This will occur if the outcome is an element of the subset $A = \{3,6\}$ of the sample space $S_1$ in Example 8.1.

**Definition (Event):** An event is a subset of the sample space of a random experiment.

***Example 8.3***

Define the events $A$ and $B$ for the die-tossing experiment:
$A$: Observe an odd number
$B$: Observe a number less than 4

***Solution***

Since event $A$ occurs if the upper face is 1, 3, or 5, it is a collection of three simple events (sample points) and we write $A = \{1,3,5\}$. Similarly, the event $B$ occurs if the upper face is 1, 2, or 3 and is defined as a collection of three simple events: $B = \{1, 2,3\}$.

Some of the basic set operations are summarized here in terms of events:

- The union of two events is the event that consists of all outcomes that are contained in either of the two events. We denote the union as $E_1 \cup E_2$. Note that $E_1 \cup E_2 = E_2 \cup E_1$.
- The intersection of two events is the event that consists of all outcomes that are contained in both of the two events. We denote the intersection as $E_1 \cap E_2$. Note that $E_1 \cap E_2 = E_2 \cap E_1$.
- The distributive law for set operations implies that
$$(A \cup B) \cap C = (A \cap C) \cup (B \cap C) \text{ and } (A \cap B) \cup C = (A \cup C) \cap (B \cup C). \quad (8.1)$$
- The complement of an event in a sample space is the set of outcomes in the sample space that are not in the event. We denote the complement of the event $E$ as $E'$. The notation $E^C$ is also used in other literature to denote the complement.
- The definition of the complement of an event implies that
$$(E')' = E. \quad (8.2)$$
- DeMorgan's laws imply that
$$(A \cup B)' = A' \cap B' \text{ and } (A \cap B)' = A' \cup B'. \quad (8.3)$$

**Definition (Mutually Exclusive):** Two events are mutually exclusive if, when one event occurs, the other cannot, and vice versa. Hence, two events, denoted as $E_1$ and $E_2$, such that
$$E_1 \cap E_2 = \emptyset, \quad (8.4)$$
are said to be mutually exclusive.

***Example 8.4***

In the die-tossing experiment, Example 8.3, events $A$ and $B$ are not mutually exclusive, because they have two outcomes in common—observing a 1 or a 3. Both events $A$ and $B$ will occur if either 1 or 3 is observed when the experiment is performed. In contrast, the six simple events 1, 2, ..., and 6 form a set of all mutually exclusive outcomes of the experiment. When the experiment is performed once, one and only one of these simple events can occur.

We can use Venn diagrams to represent a sample space and events in a sample space, see for example, Figure 8.1.

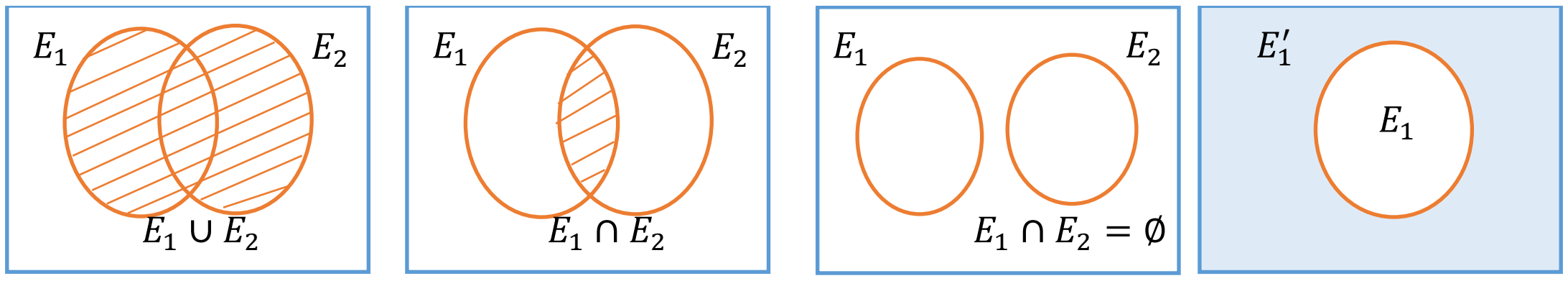

**Figure 8.1.** Venn diagrams.

## 8.2 Counting Techniques

In a sample space with a large number of outcomes, determining the number of outcomes associated with the events through direct enumeration could be tedious. In this section, we consider three counting techniques (multiplication principle, permutations and combinations) and use them in probability computations.

Counting Techniques

The multiplication principle allow us to determine the total number of outcomes in a given scenario. Whether it is arranging objects in a specific order or selecting items from a set, this technique help us systematically count the possibilities.

Permutations refer to the arrangements or orderings of a set of objects. They are used when the order of elements matters.

Combinations, on the other hand, are concerned with the selection of objects without considering their order.

**Definition (Multiplication Principle):**
Assume an operation can be described as a sequence of $k$ experiments $A_1$, $A_2$, ., $A_k$ contain, respectively, $n_1$, $n_2$, ., $n_k$ outcomes, such that for each possible outcome of $A_1$ there are $n_2$ possible outcomes for $A_2$, and so on, then there are a total of

$$n_1 \times n_2 \times \cdots \times n_k, \tag{8.5}$$

possible outcomes for the composite experiment $A_1$, $A_2$, ., $A_k$.

The multiplication principle is particularly useful when dealing with situations where events occur one after another in a sequence. By multiplying the number of possibilities at each step, we can determine the total number of outcomes for the entire sequence. The multiplication principle is often visualized using tree diagrams.

***Example 8.5***

How many sample points are there in the sample space when a pair of dice is thrown once?

***Solution***

(1,1), (2,1), (3,1), (4,1), (5,1), (6,1),
(1,2), (2,2), (3,2), (4,2), (5,2), (6,2),
(1,3), (2,3), (3,3), (4,3), (5,3), (6,3),
(1,4), (2,4), (3,4), (4,4), (5,4), (6,4),
(1,5), (2,5), (3,5), (4,5), (5,5), (6,5),
(1,6), (2,6), (3,6), (4,6), (5,6), (6,6),

The first die can land face-up in any one of $n_1 = 6$ ways. For each of these 6 ways, the second die can also land face-up in $n_2 = 6$ ways. Therefore, the pair of dice can land in $n_1 \times n_2 = 6 \times 6 = 36$ possible ways.

When a random sample of size $k$ is taken from a total of $n$ objects, the total number of ways in which the random sample of size $k$ can be selected depends on the particular sampling method we employ. Here we will consider four sampling methods:

- Sampling with replacement and the objects are ordered,
- Sampling without replacement and the objects are ordered,
- Sampling without replacement and the objects are not ordered, and
- Sampling with replacement and the objects are not ordered.

### Sampling with Replacement and the Objects Are Ordered

- Sampling with replacement refers to a sampling method where an object or element is selected from a set, and after selection, it is returned to the set before the next selection is made. This means that the same object can be chosen more than once in the sampling process.
- When the objects are ordered, it means that the arrangement or sequence of the selected objects is considered significant. The order in which the objects are selected or arranged affects the outcome.
- When a random sample of size $k$ is taken with replacement from a total of $n$ objects and the objects being ordered, then there are $n^k$ possible ways of selecting $k$-tuples.

| 1 | 2 | … | $k$ |
|---|---|---|---|
| Choice 1 | Choice 1 | … | Choice 1 |
| … | … | … | … |
| Choice $n$ | Choice $n$ | … | Choice $n$ |

For example,

(1) if a die is rolled four times, then the sample space will consist of $6^4$ 4-tuples.

(2) If an urn contains nine balls numbered 1 to 9, and a random sample with replacement of size $k = 6$ is taken, then the sample space $S$ will consist of $9^6$ 6-tuples.

### Sampling without Replacement and the Objects Are Ordered

Sampling without replacement refers to a sampling method where each object or element is selected from a set, and once selected, it is not returned to the set before the next selection is made. This means that each object can only be chosen once in the sampling process.

Consider a set of elements, such as $S = \{a, b, c\}$. A permutation of the elements is an ordered sequence of the elements. For example, $abc$, $acb$, $bac$, $bca$, $cab$, and $cba$ are all of the permutations of the elements of $S$. There are $n_1 = 3$ choices for the first position. No matter which letter is chosen, there are always $n_2 = 2$ choices for the second position. No matter which two letters are chosen for the first two positions, there is only $n_3 = 1$ choice for the last position, giving a total of $n_1 n_2 n_3 = (3)(2)(1) = 6$ permutations.

| 1 | 2 | ... | $n$ |
|---|---|---|---|
| Choice 1 | Choice 1 | ... | Choice 1 |
| ... | ... | ... | |
| | | ... | |
| | Choice $n - 1$ | | |
| Choice $n$ | | | |

**Definition (Number of Permutations):** The number of permutations of $n$ different elements is $n!$ where

$$n! = n \times (n-1) \times (n-2) \times \cdots \times 2 \times 1. \tag{8.6}$$

In some situations, we are interested in the number of arrangements of only some of the elements of a set. The following result also follows from the multiplication rule and the previous discussion.

| 1 | 2 | … | $r$ |
|---|---|---|---|
| Choice 1 | Choice 1 | … | Choice 1 |
| … | … | … | … |
| | | | Choice $n-(r-1)$ |
| | | … | |
| | Choice $n-1$ | | |
| Choice $n$ | | | |

**Definition (Permutations of Subsets):** The number of permutations of subsets of $r$ elements selected from a set of $n$ different elements is

$$P_r^n = n \times (n-1) \times (n-2) \times \cdots \times (n-r+1) = \frac{n!}{(n-r)!}. \tag{8.7}$$

***Example 8.6***

The number of permutations of the four letters $a$, $b$, $c$, and $d$ will be $4! = 24$. We have

$$\{abcd, abdc, acbd, acdb, adbc, adcb, bacd, badc, bdac, bdca, bcad, bcda,$$
$$cabd, cadb, cbad, cbda, cdab, cdba, dabc, dacb, dbac, dbca, dcab, dcba\}$$

Now consider the number of permutations that are possible by taking two letters at a time from four. These would be

$$\{ab, ac, ad, ba, bc, bd, ca, cb, cd, da, db, dc\}.$$

We have two positions to fill, with $n_1 = 4$ choices for the first and then $n_2 = 3$ choices for the second, for a total of $n_1 n_2 = (4)(3) = 12$ permutations. In general, $n$ distinct objects taken $r$ at a time can be arranged in $n(n-1)(n-2)\cdots(n-r+1)$ ways$= 4!/2! = (4)(3) = 12$.

### Permutations of Similar Objects

When dealing with permutations of similar objects, we often encounter repetitions of elements. This is because there are multiple objects with the same characteristics and arranging them in different orders results in equivalent outcomes.

If the objects are all distinct, then we have seen that the number of permutations without repetition is $n!$. For each of these permutations, we can permute the $n_1$ identical objects of type 1 in $n_1!$ possible ways; since these objects are considered identical, the arrangement is unchanged. Similarly, we can take any of the $n_2!$ permutations of the identical objects of type 2 and obtain the same arrangement. Continuing this argument, we account for these repeated arrangements by dividing by the number of repetitions. This gives the following result for the total number of permutations:

**Definition (Permutations of Similar Objects):** The number of distinct permutations of $n = n_1 + n_2 + \cdots + n_r$ objects of which $n_1$ are of one type, $n_2$ are of a second type, …, and $n_r$ are of an $r$th type is

$$\frac{n!}{n_1!\, n_2!\, n_3! \ldots n_r!}. \tag{8.8}$$

***Example 8.7***

With 4 different letters $a$, $b$, $c$, and $d$, we have 24 distinct permutations. If we let $a = b = x$ and $c = d = y$, we can list only the following distinct permutations:

$$\{abcd, abdc, acbd, acdb, adbc, adcb, bacd, badc, bdac, bdca, bcad, bcda, cabd, cadb, cbad, cbda, cdab, cdba, dabc, dacb, dbac, dbca, dcab, dcba\}$$

$$\Downarrow$$

$$\{xxyy, xxyy, xyxy, xyyx, xyxy, xyyx, xxyy, xxyy, xyxy, xyyx, xyxy, xyyx, yxxy, yxyx, yxxy, yxyx, yyxx, yyxx, yxxy, yxyx, yxxy, yxyx, yyxx, yyxx\}$$

$$\Downarrow$$

$$\{xxyy, xyxy, yxxy, yyxx, xyyx, yxyx\}.$$

Thus, we have $4!/(2!)(2!) = 6$ distinct permutations.

### Sampling without Replacement and the Objects Are Not Ordered

When the objects are not ordered, it means that the arrangement or sequence of the selected objects is not considered significant. The order in which the objects are selected or arranged does not affect the outcome.

Note that in a permutation, the order in which each object is selected becomes important. When the order of arrangement is not important for example, if we do not distinguish between $AB$ and $BA$, the arrangement is called a combination. We give the following result for the number of combinations.

**Definition (Combinations):** The number of ways in which $r$ objects can be selected (without replacement) from a collection of $n$ distinct objects is, (the number of combinations is denoted as $\binom{n}{r}$ or $C_r^n$)

$$C_r^n = \binom{n}{r} = \frac{n!}{r!\,(n-r)!} = \frac{n(n-1)(n-2)\ldots(n-r+1)}{r!}. \tag{8.9}$$

***Example 8.8***

How many different methods can the admissions committee of the Math department select three Egyptian graduate students from 15 Egyptian applicants and four foreign graduate students from 25 foreign applicants?

***Solution***

The three Egyptian students can be chosen in $\binom{15}{3}$ ways, and the four foreign students can be chosen in $\binom{25}{4}$ ways. Hence, by the multiplication principle, the whole selection can be made in $\binom{25}{4}\binom{15}{3}$ ways.

### Sampling with Replacement and the Objects Are Not Ordered

In obtaining an unordered sample of size $r$, with replacement, from a total of $n$ objects, $(r-1)$ replacements will be made before sampling ceases. Hence, the number of possible samples can be obtained by using the formula,

$$\binom{n+r-1}{r} = \frac{(n+r-1)!}{r!\,(n-1)!}. \tag{8.10}$$

***Example 8.9***

Suppose that we want to sample from the set $A = \{1,2,3\}$, $r = 2$ times such that repetition is allowed and order does not matter.

There are 6 different ways of doing this

- 1,1;
- 1,2;
- 1,3;
- 2,2;
- 2,3;
- 3,3;

How can we get the number 6 without actually listing all the possibilities? One way to think about this is to note that any of the pairs in the above list can be represented by the number of 1 's, 2 's and 3 's it contains. That is, if $x_1$ is the number of ones, $x_2$ is the number of twos, and $x_3$ is the number of threes, we can equivalently represent each pair by a vector $(x_1, x_2, x_3)$ , i.e.,

- 1,1 $\rightarrow (x_1, x_2, x_3) = (2,0,0)$;
- 1,2 $\rightarrow (x_1, x_2, x_3) = (1,1,0)$;
- 1,3 $\rightarrow (x_1, x_2, x_3) = (1,0,1)$;
- 2,2 $\rightarrow (x_1, x_2, x_3) = (0,2,0)$;
- 2,3 $\rightarrow (x_1, x_2, x_3) = (0,1,1)$;
- 3,3 $\rightarrow (x_1, x_2, x_3) = (0,0,2)$.

Note that here $x_i \geq 0$ are integers and $x_1 + x_2 + x_3 = 2$. Thus, we can claim that the number of ways we can sample two elements from the set $A = \{1,2,3\}$ such that ordering does not matter and repetition is allowed is the same as solutions to the following equation $x_1 + x_2 + x_3 = 2$, where $x_i \in \{0,1,2\}$.

**Theorem 8.1:** The total number of distinct $r$ samples from an $n$-element set such that repetition is allowed and order does not matter is the same as the number of distinct solutions to the equation

$$x_1 + x_2 + \ldots + x_n = r, \quad \text{where } x_i \in \{0,1,2,3,\ldots\}. \tag{8.11}$$

and is equal to $\binom{n+r-1}{r} = \frac{(n+r-1)!}{r!(n-1)!}$.

## 8.3 Interpretations and Axioms of Probability

Probability is used to quantify the likelihood, or chance, that an outcome of a random experiment will occur. The likelihood of an outcome is quantified by assigning a number from the interval $[0,1]$ to the outcome (or a percentage from 0 to 100%).

### Classical Interpretation

The classical probability rule is applied to compute the probabilities of events for an experiment for which all outcomes are equally likely. For example, head and tail are two equally likely outcomes when a fair coin is tossed once. Each of these two outcomes has the same chance of occurrence.

According to the classical probability rule, to find the probability of a simple event, we divide 1.0 by the total number of outcomes for the experiment. On the other hand, to find the probability of a compound event $E$, we divide the number of outcomes favorable to event $E$ by the total number of outcomes for the experiment.

**Definition (Equally Likely Outcomes):** Whenever a sample space consists of $N$ possible outcomes that are equally likely, the probability of each outcome is $1/N$.

**Definition (Probability of an Event):** For a discrete sample space, the probability of an event $E$, denoted as $P(E)$, equals the sum of the probabilities of the outcomes in $E$.

### Relative Frequency Interpretation

The relative frequency of an event is used as an approximation for the probability of that event. Because relative frequencies are determined by performing an experiment, the probabilities calculated using relative frequencies may change when an experiment is repeated. The probability of an outcome is interpreted as the limiting value of the proportion of times the outcome occurs in $n$ repetitions of the random experiment as $n$ increases beyond all bounds.

If an experiment is repeated $n$ times and an event $A$ is observed $f$ times where $f$ is the frequency, then, according to the relative frequency concept of probability:

$$P(A) = \frac{f}{n} = \frac{\text{Frequency of A}}{\text{Sample size}}. \quad (8.12)$$

### Axioms of Probability

Axioms are the foundational principles upon which probability theory is constructed. The axiomatic approach defines the properties that probabilities must satisfy to be considered valid measures of uncertainty. The Kolmogorov axioms are introduced by Russian mathematician Andrey Kolmogorov in 1933.

**Definition (Kolmogorov Axioms- Axioms of Probability):** Probability is a number that is assigned to each member of a collection of events from a random experiment that satisfies the following properties:
(1) $P(S) = 1$ where $S$ is the sample space
(2) $0 \leq P(E) \leq 1$ for any event $E$
(3) For two events $E_1$ and $E_2$ with $E_1 \cap E_2 = \emptyset$

$$P(E_1 \cup E_2) = P(E_1) + P(E_2). \quad (8.13)$$

**Remark:**

$$P(\emptyset) = 0, \quad (8.14)$$

and for any event $E$,

$$P(E') = 1 - P(E). \quad (8.15)$$

***Example 8.10***

A coin is tossed twice. What is the probability that at least 1 head occurs?

***Solution***

The sample space for this experiment is

$$S = \{HH, HT, TH, TT\}.$$

If the coin is balanced, each of these outcomes is equally likely to occur. Therefore, we assign a probability of $\omega$ to each sample point. Then $4\omega = 1$, or $\omega = 1/4$. If $A$ represents the event of at least 1 head occurring, then

$$A = \{HH, HT, TH\} \text{ and } P(A) = \frac{1}{4} + \frac{1}{4} + \frac{1}{4} = \frac{3}{4}.$$

Hence, if an experiment can result in any one of $N$ different equally likely outcomes, and if exactly $n$ of these outcomes correspond to event $A$, then the probability of event $A$ is

$$P(A) = \frac{n}{N}. \quad (8.16)$$

### Unions of Events and Addition Rules

Joint events are generated by applying basic set operations to individual events. Unions of events, such as $A \cup B$; intersections of events, such as $A \cap B$; and complements of events, such as $A'$— are common of interest. The probability of a joint event can often be determined from the probabilities of the individual events that it comprises. Basic set operations are also sometimes helpful in determining the probability of a joint event.

**Theorem 8.2 (Probability of a Union):**

$$P(A \cup B) = P(A) + P(B) - P(A \cap B). \quad (8.17)$$

If $A$ and $B$ are mutually exclusive events, ($A \cap B = \emptyset$ and $P(A \cap B) = 0$), then

$$P(A \cup B) = P(A) + P(B). \quad (8.18)$$

For three events, we have

$$P(A \cup B \cup C) = P(A) + P(B) + P(C) - P(A \cap B) - P(A \cap C) - P(B \cap C) + P(A \cap B \cap C). \quad (8.19)$$

Moreover, a collection of events, $E_1, E_2, \ldots, E_k$, is said to be mutually exclusive if for all pairs,

$$E_i \cap E_j = \emptyset. \quad (8.20)$$

For a collection of mutually exclusive events,

$$P(E_1 \cup E_2 \cup \ldots \cup E_k) = P(E_1) + P(E_2) + \cdots + P(E_k). \quad (8.21)$$

**Proof:**

The $P(A \cup B)$ is the sum of the probabilities of the sample points in $A \cup B$. Now $P(A) + P(B)$ is the sum of all the probabilities in $A$ plus the sum of all the probabilities in $B$. Therefore, we have added the probabilities in $(A \cap B)$ twice. Since these probabilities add up to $P(A \cap B)$, we must subtract this probability once to obtain the sum of the probabilities in $A \cup B$.

■

***Example 8.11***

What is the probability of getting a total of 7 or 11 when a pair of fair dice is tossed?

***Solution***

Let $A$ be the event that 7 occurs and $B$ the event that 11 comes up. Now, a total of 7 occurs for 6 of the 36 sample points, $\{(1,6), (2,5), (3,4), (4,3), (5,2), (6,1)\}$, and a total of 11 occurs for only 2 of the sample points, $\{(5,6), (6,5)\}$. Since all sample points are equally likely, we have $P(A) = 6/36 = 1/6$ and $P(B) = 2/36 = 1/18$. The events $A$ and $B$ are mutually exclusive since a total of 7 and 11 cannot both occur on the same toss. Therefore,

$$P(A \cup B) = P(A) + P(B) = \frac{1}{6} + \frac{1}{18} = \frac{2}{9}.$$

## 8.4 Conditional Probability

One very important concept in probability theory is conditional probability. In some applications, the practitioner is interested in the probability structure under certain restrictions. The probability of an event $B$ under the knowledge that the outcome will be in event $A$ is denoted as $P(B|A)$ and this is called the conditional probability of $B$ given $A$.

**Definition (Conditional Probability):** The conditional probability of an event $B$ given that an event $A$ has occurred, $P(A) > 0$, is

$$P(B|A) = \frac{P(A \cap B)}{P(A)}. \tag{8.22}$$

This definition can be understood in a special case in which all outcomes of a random experiment are equally likely. If there are $N$ total outcomes,

$$P(A) = (\text{number of outcomes in } A)/N. \tag{8.23}$$

Also,

$$P(A \cap B) = (\text{number of outcomes in } A \cap B)/N. \tag{8.24}$$

Consequently,

$$\frac{P(A \cap B)}{P(A)} = \frac{\text{number of outcomes in } A \cap B}{\text{number of outcomes in } A}. \tag{8.25}$$

Therefore, $P(B|A)$ can be interpreted as the relative frequency of event $B$ among the trials that produce an outcome in event $A$.

***Example 8.12***

What is the probability that the total of two dice will be greater than 8 given that the first die is a 6?

***Solution***

This can be computed by considering only outcomes for which the first die is a 6. Then, determine the proportion of these outcomes that total more than 8. There are 6 outcomes for which the first die is a 6, $\{(6,1), (6,2), (6,3), (6,4), (6,5), (6,6)\}$, and of these, there are four that total more than 8 $\{(6,3), (6,4), (6,5), (6,6)\}$. The probability of a total greater than 8 given that the first die is 6 is therefore $\frac{4}{6} = \frac{2}{3}$.

More formally, this probability can be written as:

$$P(\text{total} > 8|\text{Die } 1 = 6) = 2/3.$$

**Definition (Multiplication Rule):**

$$P(A \cap B) = P(B|A)P(A) = P(A|B)P(B). \quad (8.26)$$

Thus, the probability that both $A$ and $B$ occur is equal to the probability that $A$ occurs multiplied by the conditional probability that $B$ occurs, given that $A$ occurs.

***Example 8.13***

Suppose we have a bag containing 5 balls of which 3 are green and 2 are red. Suppose we draw two balls in succession from this bag without replacement.

- what is the probability of drawing 2 green balls?
- what is the probability of drawing a red ball followed by a green ball?

***Solution***

Let's define the following events: event $G_1$ - selecting a green ball for the first draw, event $G_2$- selecting a green ball for the second draw, and event $R_1$- selecting a red ball for the first draw. The probability of drawing 2 green balls is:

$$P(G_1 \cap G_2) = P(G_2|G_1)P(G_1) = \left(\frac{2}{4}\right)\left(\frac{3}{5}\right) = \frac{3}{10}.$$

The probability of drawing a red ball followed by a green ball is:

$$P(R_1 \cap G_2) = P(G_2|R_1)P(R_1) = \left(\frac{3}{4}\right)\left(\frac{2}{5}\right) = \frac{3}{10}.$$

For any event $B$, we can write $B$ as the union of the part of $B$ in $A$ and the part of $B$ in $A'$. That is,

$$B = (A \cap B) \cup (A' \cap B). \quad (8.27)$$

This result is shown in the Venn diagram in Figure 8.2. Because $A$ and $A'$ are mutually exclusive, $A \cap B$ and $A' \cap B$ are mutually exclusive.

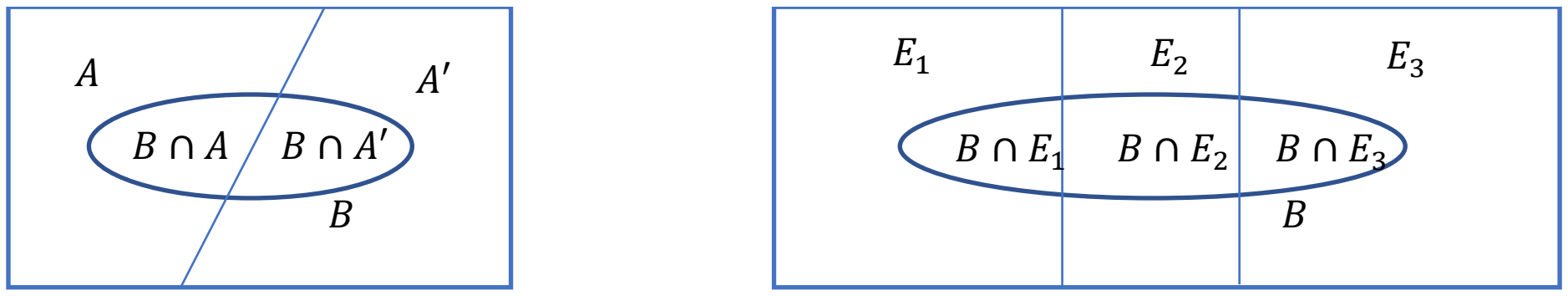

**Figure 8.2.** (Left figure) Partitioning an event into two mutually exclusive subsets. (Right figure) Partitioning an event into several mutually exclusive subsets.

Therefore, the following total probability rule is obtained.

**Definition (Total Probability Rule):** For any events A and B,

$$\begin{aligned} P(B) &= P(B \cap A) + P(B \cap A') \\ &= P(B|A)P(A) + P(B|A')P(A'). \end{aligned} \quad (8.28)$$

Moreover, assume $E_1, E_2, \dots, E_k$ are $k$ mutually exclusive and exhaustive sets. Then

$$\begin{aligned} P(B) &= P(B \cap E_1) + P(B \cap E_2) + \cdots + P(B \cap E_k) \\ &= P(B|E_1)P(E_1) + P(B|E_2)P(E_2) + \cdots + P(B|E_k)P(E_k). \end{aligned} \quad (8.29)$$

## Independence

Two events are said to be independent if the occurrence (or non-occurrence) of one event does not affect the probability that the other event will occur. In this case, the conditional probability of $P(B|A)$ might equal $P(B)$, i.e., the knowledge that the outcome of the experiment is in event $A$ does not affect the probability that the outcome is in event $B$. So that, we obtain

$$P(A \cap B) = P(B|A)P(A) = P(B)P(A), \quad (8.30)$$

and

$$P(A|B) = \frac{P(A \cap B)}{P(B)} = \frac{P(A)P(B)}{P(B)} = P(A). \quad (8.31)$$

These conclusions lead to an important definition.

**Definition (Independence, two events):** Two events are independent if any one of the following equivalent statements is true:
(1) $P(A|B) = P(A)$,
(2) $P(B|A) = P(B)$,
(3) $P(A \cap B) = P(A)P(B)$.

***Example 8.14***

Suppose we toss a coin twice and we define the following events:
event $A$- obtaining heads for the first toss, and event $B$- obtaining heads for the second toss.

***Solution***

Because the outcome of the first coin toss does not affect the outcome of the next coin toss, these events are independent.

***Example 8.15***

Suppose we roll two fair dice. What is the probability that the outcome of both the dice is even?

***Solution***

Let events $A$ and $B$ represent the outcome of the first and second dice respectively. Events $A$ and $B$ are independent because the outcome of the first dice does not affect the outcome of the second dice. Therefore, the probability that the outcome of both dice is even given by:

$$P(A \text{ is even} \cap B \text{ is even}) = P(A \text{ is even})P(B \text{ is even}) = \left(\frac{3}{6}\right)\left(\frac{3}{6}\right) = \frac{1}{4}.$$

It is simple to show that independence implies related results such as

$$P(A' \cap B') = P(A')P(B'). \quad (8.32)$$

**Remark:**

A mutually exclusive relationship between two events is based only on the outcomes that compose the events. However, an independence relationship depends on the probability model used for the random experiment. Often, independence is assumed to be part of the random experiment that describes the physical system under study. The concepts of mutually independent events and mutually exclusive events are separate and distinct. The following table contrasts the results for the two cases (provided that the probability of the conditioning event is not zero).

| | If statistically independent | If mutually exclusive |
|---|---|---|
| $P(A\|B) =$ | $P(A)$ | 0 |
| $P(B\|A) =$ | $P(B)$ | 0 |
| $P(A \cap B) =$ | $P(A)P(B)$ | 0 |

**Definition (Independence, multiple events):** The events $E_1, E_2, \dots, E_n$ are independent if and only if

$$P\left(E_{i_1} \cap E_{i_2} \cap \dots \cap E_{i_k}\right) = P\left(E_{i_1}\right) \times P\left(E_{i_2}\right) \times \dots \times P\left(E_{i_k}\right). \quad (8.33)$$

**Definition (Pairwise Independent, multiple events):** The events $E_1, E_2, \dots, E_n$ are pairwise independent if and only if

$$P\left(E_{i_j} \cap E_{i_k}\right) = P\left(E_{i_j}\right) \times P\left(E_{i_k}\right) \text{ for each } j \neq k. \quad (8.34)$$

### Bayes Theorem

From the definition of conditional probability,

$$P(A \cap B) = P(A|B)P(B) = P(B \cap A) = P(B|A)P(A). \tag{8.35}$$

Now, considering the second and last terms in the preceding expression, we can write

$$P(A|B) = \frac{P(B|A)P(A)}{P(B)} \text{ for } P(B) > 0. \tag{8.36}$$

This is a useful result that enables us to solve for $P(A|B)$ in terms of $P(B|A)$.

**Theorem 8.3 (Bayes Theorem):** If $E_1, E_2, \ldots, E_k$ are $k$ mutually exclusive and exhaustive events and $B$ is any event,

$$P(E_1|B) = \frac{P(B|E_1)P(E_1)}{P(B|E_1)P(E_1) + P(B|E_2)P(E_2) + \cdots + P(B|E_k)P(E_k)}, \tag{8.37}$$

for $P(B) > 0$

***Example 8.16***

Suppose we have the following two bags: bag $x$, which contains 1 red and 3 green balls. bag $y$, which contains 2 red and 1 green balls. We pick one bag at random and then draw a single ball. Given that the drawn ball is red, compute the probability that it came from bag $y$.

***Solution***

We define the following events: event $x$- bag $x$ is chosen, event $y$- bag $y$ is chosen, and event red - a red ball is drawn.

The probability we are after is, therefore, $P(y|\text{red})$. Using Bayes theorem, we have that:

$$P(y|\text{red}) = \frac{P(\text{red}|y)P(y)}{P(\text{red})}.$$

We are not provided with $P(\text{red})$, so we must use the second form of Bayes theorem in which we express the denominator $P(\text{red})$ using the law of total probability:

$$P(y|\text{red}) = \frac{P(\text{red}|y)P(y)}{P(\text{red}|x)P(x) + P(\text{red}|y)P(y)} = \frac{\left(\frac{2}{3}\right)\left(\frac{1}{2}\right)}{\left(\frac{1}{4}\right)\left(\frac{1}{2}\right) + \left(\frac{2}{3}\right)\left(\frac{1}{2}\right)} = \frac{\frac{1}{3}}{\frac{11}{24}} = 0.73.$$

***Example 8.17***

Suppose a statistics class contains 70% male and 30% female students. It is known that in a test, 5% of males and 10% of females got an "A" grade. If one student from this class is randomly selected and observed to have an "A" grade, what is the probability that this is a male student?

***Solution***

Let $A_1$ denote that the selected student is a male, and $A_2$ denote that the selected student is a female. Here the sample space $S = A_1 \cup A_2$. Let $D$ denote that the selected student has an "A" grade. We are given $P(A_1) = 0.7$, $P(A_2) = 0.3$, $P(D|A_1) = 0.05$, and $P(D|A_2) = 0.10$. Then by the total probability rule,

$$\begin{aligned} P(D) &= P(D|A_1)P(A_1) + P(D|A_2)P(A_2) \\ &= 0.035 + 0.030 \\ &= 0.065. \end{aligned}$$

Now by Bayes' rule,

$$\begin{aligned} P(A_1|D) &= \frac{P(D|A_1)P(A_1)}{P(D|A_1)P(A_1) + P(D|A_2)P(A_2)} \\ &= \frac{(0.7)(0.05)}{0.065} \\ &= \frac{7}{13} \\ &= 0.538. \end{aligned}$$

# CHAPTER 9

# MATHEMATICA LAB: PRINCIPLES OF PROBABILITY

The Wolfram Language provides a comprehensive set of functions for working with sets and combinatorial objects. For example

- The union of two sets A and B can be computed using the `Union` function in Wolfram Language. It returns a new set that contains all the unique elements from both sets.
- The intersection of two sets A and B can be calculated using the `Intersection` function. It returns a new set that contains the common elements present in both sets.
- The `Complement` function finds the complement with respect to a universal set.
- `Factorial`, `Binomial` and `Multinomial` functions calculate the factorial of a number, the binomial coefficient and the multinomial coefficient, respectively.

Moreover, there are various built-in functions and techniques to handle different counting scenarios.

- When sampling with replacement and the objects are ordered, you can use the `Tuples` function. It generates all possible combinations with repetition of a given set of elements.
- When sampling without replacement and the objects are ordered, you can use the `Permutations` function. It generates all possible permutations of a given list of elements.
- When sampling without replacement and the objects are not ordered, you can use the `Subsets` function. It generates all possible subsets of a given list of elements.
- When sampling with replacement and the objects are not ordered, you can use the `Tuples` function followed by `DeleteDuplicates`. The `Tuples` function generates all possible combinations with repetition, and then `DeleteDuplicates` removes any duplicate samples.

Therefore, we divided this chapter into three units to cover the following topics, operations on sets, combinatorial functions and different counting scenarios and probability.

In the following table, we list the built-in functions that are used in this chapter.

| Operations on Sets | | | | Combinatorial Functions |
|---|---|---|---|---|
| `Tuples` | `Union` | `IntersectingQ` | `ContainsNone` | `Factorial (!)` |
| `Subsets` | `Intersection` | `DisjointQ` | `ContainsAny` | `Binomial` |
| `Subsequences` | `Complement` | `SubsetQ` | `ContainsOnly` | `Multinomial` |
| `Permutations` | | `ContainsAll` | `ContainsExactly` | |

**Chapter 9 Outline**

# UNIT 9.1

# OPERATIONS ON SETS

In the Wolfram Language, sets are represented by sorted lists.

| | |
|---|---|
| `Tuples[list,n]` | generates a list of all possible n-tuples of elements from list. |
| `Tuples[{list1,list2,…}]` | generates a list of all possible tuples whose i^(th) element is from listi. |

***Mathematica Examples 9.1*** Tuples

```
Input   (* The elements of list are treated as distinct, so that Tuples[list,n] for a list
        of length k gives output of length k^n: *)
        (* Generate all possible 2-tuples (pairs) from a list of elements: *)
        elements={x,y,z};
        pairs=Tuples[elements,2]
        Length[pairs]
Output  {{x,x},{x,y},{x,z},{y,x},{y,y},{y,z},{z,x},{z,y},{z,z}}
Output  9

Input   (* Generate all possible 3-tuples (triplets) from a range of numbers: *)
        triplets=Tuples[Range[1,3],3]
        Length[triplets]
Output  {{1,1,1},{1,1,2},{1,1,3},{1,2,1},{1,2,2},{1,2,3},{1,3,1},{1,3,2},{1,3,3},{2,1,1},
        {2,1,2},{2,1,3},{2,2,1},{2,2,2},{2,2,3},{2,3,1},{2,3,2},{2,3,3},{3,1,1},{3,1,2},
        {3,1,3},{3,2,1},{3,2,2},{3,2,3},{3,3,1},{3,3,2},{3,3,3}}
Output  27
```

| | |
|---|---|
| `Subsets[list]` | gives a list of all possible subsets of list. |
| `Subsets[list,n]` | gives all subsets containing at most n elements. |
| `Subsets[list,{n}]` | gives all subsets containing exactly n elements. |

***Mathematica Examples 9.2*** Subsets

```
Input   (* Generate all subsets of a list: *)
        Subsets[{x,y,z}]
Output  {{},{x},{y},{z},{x,y},{x,z},{y,z},{x,y,z}}

Input   (* All possible subsets containing up to 2 elements: *)
        Subsets[{x,y,z,u,v},2]
Output  {{},{x},{y},{z},{u},{v},{x,y},{x,z},{x,u},{x,v},{y,z},{y,u},{y,v},{z,u},{z,v},{u,v}
        }

Input   (* Generate subsets of a certain length: *)
        Subsets[{x,y,z,u,v},{2}]
Output  {{x,y},{x,z},{x,u},{x,v},{y,z},{y,u},{y,v},{z,u},{z,v},{u,v}}

Input   (* The first 5 subsets containing 3 elements: *)
        Subsets[{x,y,z,u,v},{3},5]
Output  {{x,y,z},{x,y,u},{x,y,v},{x,z,u},{x,z,v}}

Input   (* Find all ways to pick 3 elements from 4: *)
        Subsets[{1,2,3,4},{3}]
        Binomial[4,3]
Output  {{1,2,3},{1,2,4},{1,3,4},{2,3,4}}
```

```
Output   4

Input    (* Different occurrences of the same element are treated as distinct: *)
         Subsets[{a,b,b,b}]
Output   {{},{a},{b},{b},{b},{a,b},{a,b},{a,b},{b,b},{b,b},{b,b},{a,b,b},{a,b,b},{a,b,b},
         {b,b,b},{a,b,b,b}}

Input    (* Generate subsets satisfying a specific criterion: *)
         list={1,2,3,4};
         Subsets[list]
         subsets=Select[Subsets[list],Total[#]>5&]
Output   {{},{1},{2},{3},{4},{1,2},{1,3},{1,4},{2,3},{2,4},{3,4},{1,2,3},{1,2,4},{1,3,4},
         {2,3,4},{1,2,3,4}}

Output   {{2,4},{3,4},{1,2,3},{1,2,4},{1,3,4},{2,3,4},{1,2,3,4}}

Input    (* In this example, the "Select" function is used to filter the subsets based on the
         condition that the total of the elements in the subset is equal to 5: *)
         list={1,2,3,4};
         total=5;
         Subsets[list]
         subsets=Select[Subsets[list],Total[#]==total&]
Output   {{},{1},{2},{3},{4},{1,2},{1,3},{1,4},{2,3},{2,4},{3,4},{1,2,3},{1,2,4},{1,3,4},{2,
         3,4},{1,2,3,4}}
Output   {{1,4},{2,3}}

Input    (* In this example, the "Select" function is used to filter the subsets based on the
         condition that the length of the subset is greater than or equal to 2: *)
         list={1,2,3,4};
         minLength=2;
         Subsets[list]
         subsets=Select[Subsets[list],Length[#]>=minLength&]
Output   {{},{1},{2},{3},{4},{1,2},{1,3},{1,4},{2,3},{2,4},{3,4},{1,2,3},{1,2,4},{1,3,4},
         {2,3,4},{1,2,3,4}}
Output   {{1,2},{1,3},{1,4},{2,3},{2,4},{3,4},{1,2,3},{1,2,4},{1,3,4},{2,3,4},{1,2,3,4}}

Input    (* Tuples gives all possible combinations and reorderings of elements: *)
         Tuples[{a,b,c},2]
         Subsets[{a,b,c},{2}]
Output   {{a,a},{a,b},{a,c},{b,a},{b,b},{b,c},{c,a},{c,b},{c,c}}
Output   {{a,b},{a,c},{b,c}}
```

| | |
|---|---|
| `Subsequences[list]` | gives the list of all possible subsequences of list. |
| `Subsequences[list,n]` | gives all subsequences containing at most n elements. |
| `Subsequences[list,{n}]` | gives all subsequences containing exactly n elements. |

***Mathematica Examples 9.3*** Subsequences

```
Input    (* Generate all subsequences of a list: *)
         Subsequences[{x,y,z}]
Output   {{},{x},{y},{z},{x,y},{y,z},{x,y,z}}

Input    (* Note the difference between Subsequences and Subsets functions. Subsets[list]
         function generates all possible subsets of a given list. A subset is a collection of
         elements that can be selected from the original list, regardless of their order.
         While Subsequences[list] function generates all subsequences that maintain the order
         of the original list: *)

         Subsets[{1,2,3,4,5}]
         Subsequences[{1,2,3,4,5}]
```

```
Output  {{},{1},{2},{3},{4},{5},{1,2},{1,3},{1,4},{1,5},{2,3},{2,4},{2,5},{3,4},{3,5},{4,5}
        ,{1,2,3},{1,2,4},{1,2,5},{1,3,4},{1,3,5},{1,4,5},{2,3,4},{2,3,5},{2,4,5},{3,4,5},{1
        ,2,3,4},{1,2,3,5},{1,2,4,5},{1,3,4,5},{2,3,4,5},{1,2,3,4,5}}

Output  {{},{1},{2},{3},{4},{5},{1,2},{2,3},{3,4},{4,5},{1,2,3},{2,3,4},{3,4,5},{1,2,3,4},{
        2,3,4,5},{1,2,3,4,5}}

Input   (* All possible subsequences containing up to 2 elements: *)
        Subsequences[{x,y,z,w},2]
Output  {{},{x},{y},{z},{w},{x,y},{y,z},{z,w}}

Input   (* Subsequences containing exactly 2 elements: *)
        Subsequences[{x,y,z,w},{2}]
Output  {{x,y},{y,z},{z,w}}

Input   (* The first 2 subsequences containing 3 elements: *)
        Subsequences[{x,y,z,w,u,v},{3},2]
Output  {{x,y,z},{y,z,w}}

Input   (* All subsequences with even length: *)
        Subsequences[{x,y,z,w,u},{0,5,2}]
Output  {{},{x,y},{y,z},{z,w},{w,u},{x,y,z,w},{y,z,w,u}}

Input   (* Find all subsequences that sum of the elements in the subset equal to a specific
        value: *)
        list={1,2,3,4,5};
        targetSum=7;
        subsequences=Subsequences[list]
        desiredSubsequences=Select[subsequences,Total[#]==targetSum&]
Output  {{},{1},{2},{3},{4},{5},{1,2},{2,3},{3,4},{4,5},{1,2,3},{2,3,4},{3,4,5},{1,2,3,4},{
        2,3,4,5},{1,2,3,4,5}}

Output  {{3,4}}
```

| | |
|---|---|
| `Permutations[list]` | generates a list of all possible permutations of the elements in list. |
| `Permutations[list,n]` | gives all permutations containing at most n elements. |
| `Permutations[list,{n}]` | gives all permutations containing exactly n elements. |

***Mathematica Examples 9.4*** Permutations

```
Input   (* Compute all permutations of a list: *)
        list={1,2,3};
        perms=Permutations[list]
        Length[perms]
Output  {{1,2,3},{1,3,2},{2,1,3},{2,3,1},{3,1,2},{3,2,1}}
Output  6

Input   (* Find all permutations of a string: *)
        string="abcd";
        perms=Permutations[Characters[string]]
        Length[perms]
Output  {{a,b,c,d},{a,b,d,c},{a,c,b,d},{a,c,d,b},{a,d,b,c},{a,d,c,b},{b,a,c,d},{b,a,d,c},{b
        ,c,a,d},{b,c,d,a},{b,d,a,c},{b,d,c,a},{c,a,b,d},{c,a,d,b},{c,b,a,d},{c,b,d,a},{c,d,
        a,b},{c,d,b,a},{d,a,b,c},{d,a,c,b},{d,b,a,c},{d,b,c,a},{d,c,a,b},{d,c,b,a}}
Output  24

Input   (* Find the permutations of a list with a specific length: *)
        list={1,2,3};
        length=2;
        perms=Permutations[list,{length}]
Output  {{1,2},{1,3},{2,1},{2,3},{3,1},{3,2}}
```

```
Input    (* Generate and display permutations of a list, one at a time: *)
         list={1,2,3};
         perms=Permutations[list];
         Do[
           Print[perms[[i]]],
           {i,Length[perms]}
           ];
Output    {1,2,3}
          {1,3,2}
          {2,1,3}
          {2,3,1}
          {3,1,2}
          {3,2,1}

Input    (* Find all permutations of a list and select those that satisfy a specific criterion:
         *)
         list={1,2,3,4};
         perms=Permutations[list];
         length=Length[perms]
         filteredPerms=Select[perms,#[[1]]>#[[4]]&]
Output   24
Output   {{2,3,4,1},{2,4,3,1},{3,1,4,2},{3,2,4,1},{3,4,1,2},{3,4,2,1},{4,1,2,3},{4,1,3,2},
         {4,2,1,3},{4,2,3,1},{4,3,1,2},{4,3,2,1}}

Input    (* Generate permutations of a string and check if a specific permutation exists: *)
         string="abcd";
         perms=Permutations[Characters[string]];
         desiredPerm={"b","d","a","c"};
         containsPerm=MemberQ[perms,desiredPerm];
         Print["Does the permutation exist? ",containsPerm];
Output   Does the permutation exist?  True
```

| | |
|---|---|
| `Union[list1,list2,…]` | gives a sorted list of all the distinct elements that appear in any of the listi. |
| `Union[list]` | gives a sorted version of a list, in which all duplicated elements have been dropped. |
| `Intersection[list1,list2,…]` | gives a sorted list of the elements common to all the listi. |
| `Complement[eall,e1,e2,…]` | gives the elements in eall that are not in any of the ei. |

***Mathematica Examples 9.5*** Union

```
Input    (* Give a sorted list of distinct elements: *)
         Union[{1,2,1,3,6,2,2}]
Output   {1,2,3,6}

Input    (* Give a sorted list of distinct elements from all the lists: *)
         Union[{a,b,a,c},{d,a,e,b},{c,a}]
Output   {a,b,c,d,e}

Input    (* Union of multiple lists: *)
         list1={1,2,3};
         list2={3,4,5};
         list3={5,6,7};
         unionList=Union[list1,list2,list3]
Output   {1,2,3,4,5,6,7}

Input    (* Union of a list and an array: *)
         list1={{1,2,3,3}};
         list2={1,2,3,3};
         array={{3,4,5},{6,7,8}};
         unionList1=Union[list1,array]
         unionList2=Union[list2,array]
```

```
Output    {{3,4,5},{6,7,8},{1,2,3,3}}
Output    {1,2,3,{3,4,5},{6,7,8}}

Input     (* Union of two strings: *)
          string1="Hello";
          string2="World";
          Characters[string1]
          Characters[string2]
          unionString=Union[Characters[string1],Characters[string2]]
Output    {H,e,l,l,o}
Output    {W,o,r,l,d}
Output    {d,e,H,l,o,r,W}

Input     expr1={x^2+y^2};
          expr2={x^2-y^2};
          unionExpr=Union[expr1,expr2]
Output    {x^2-y^2,x^2+y^2}
```

**Mathematica Examples 9.6** Intersection

```
Input     (* Finding the common elements between two lists: *)
          list1={1,2,3,4,5};
          list2={4,5,6,7,8};
          commonElements=Intersection[list1,list2]
Output    {4,5}

Input     (* Finding the common elements among multiple lists: *)
          list1={1,2,3,4,5};
          list2={4,5,6,7,8};
          list3={3,4,5,9,10};
          commonElements=Intersection[list1,list2,list3]
Output    {4,5}

Input     (* Finding the common elements between two arrays: *)
          array1={{1,2,3},{4,5,6},{7,8,9}};
          array2={{4,5,6},{7,8,9},{10,11,12}};
          commonElements=Intersection[array1,array2]
Output    {{4,5,6},{7,8,9}}

Input     (* Finding common elements between two strings: *)
          string1="Hello";
          string2="World";
          Characters[string1]
          Characters[string2]
          commonCharacters=Intersection[Characters[string1],Characters[string2]]
Output    {H,e,l,l,o}
Output    {W,o,r,l,d}
Output    {l,o}
```

**Mathematica Examples 9.7** Complement

```
Input     (* Find which elements in the first list are not in any of the subsequent lists: *)
          Complement[{a,b,c,d,e},{a,c},{d}]
Output    {b,e}

Input     (* Finding the elements unique to the first list: *)
          list1={1,2,3,4,5};
          list2={3,4,5,6,7};
          complementList=Complement[list1,list2]
Output    {1,2}
```

```
Input   (* Finding the elements unique to both lists: *)
        list1={1,2,3,4,5};
        list2={3,4,5,6,7};
        Union[Complement[list1,list2] , Complement[list2,list1]]
Output  {1,2,6,7}
```

| | |
|---|---|
| `IntersectingQ[list1,list2]` | yields True if list1 and list2 have at least one element in common, and False otherwise. |
| `DisjointQ[list1,list2]` | yields True if list1 and list2 do not share any common elements, and False otherwise. |
| `SubsetQ[list1,list2]` | yields True if list2 is a subset of list1, and False otherwise. |
| `ContainsAll[e1,e2]` | yields True if e1 contains all of the elements of e2. |
| `ContainsNone[e1,e2]` | yields True if e1 contains none of the elements in e2. |
| `ContainsAny[e1,e2]` | yields True if e1 contains any of the elements of e2. |
| `ContainsOnly[e1,e2]` | yields True if e1 contains only elements that appear in e2. |
| `ContainsExactly[e1,e2]` | yields True if e1 contains exactly the same elements as e2. |

***Mathematica Examples 9.8*** IntersectingQ, …, and ContainsExactly

```
Input   (* Check if two lists have any common elements: *)
        list1={1,2,3,4};
        list2={3,4,5,6};
        list3={7,8,9,10};

        result1=IntersectingQ[list1,list2]
        result2=IntersectingQ[list1,list3]
Output  True
Output  False

Input   (* Checking if a list intersects with any sublist in a list of lists: *)
        listOfLists={{1,2},{3,4,5},{6,7},{8,9}};
        subList2={1,2};
        subList3={{1,2}};

        IntersectingQ[listOfLists,subList2]
        IntersectingQ[listOfLists,subList3]
Output  False
Output  True

Input   (* Checking if two sets of integers are disjoint: *)
        set1={1,2,3};
        set2={4,5,6};
        set3={1,8,9};

        DisjointQ[set1,set2]
        DisjointQ[set1,set3]
Output  True
Output  False

Input   (* Checking if two lists of strings are disjoint: *)
        list1={"apple","banana","orange"};
        list2={"car","bus","train"};

        DisjointQ[list1,list2]
Output  True

Input   (* Checking if two sets of real numbers are disjoint: *)
        set3={1.5,2.7,3.1};
        set4={2.6,5.8,3.1};

        DisjointQ[set3,set4]
Output  False
```

```
Input    (* Checking if two lists of expressions are disjoint: *)
         list3={x^2,x^3,x^4};
         list4={Sin[x],Cos[x],Exp[x]};

         DisjointQ[list3,list4]
Output   True

Input    (* Test if a set is a subset of another set: *)
         SubsetQ[{1,2,3},{3,1}]
Output   True

Input    (* Check if a list is a subset of another list: *)
         list1={1,2,3,4};
         list2={2,4};

         SubsetQ[list2,list1]
         SubsetQ[list1,list2]
Output   False
Output   True

Input    (* Check if a set of lists is a subset of another set of lists: *)
         set1={{1,2},{3,4},{5,6}};
         set2={{1,2},{3,4}};

         SubsetQ[set2,set1]
         SubsetQ[set1,set2]
Output   False
Output   True

Input    (* The first list contains all elements of the second list: *)
         ContainsAll[{b,a,b,c},{a,b}]

         (* The first list does not contain all elements of the second list: *)
         ContainsAll[{b,a,b,c},{a,b,d}]
Output   True
Output   False

Input    (* Check if a list contains all the elements of multiple other lists: *)
         list4={1,2,3,4,5};
         list5={2,4};
         list6={1,3};
         ContainsAll[list4,{list5,list6}]

         list7={1,2,3,4,5};
         list8={2,4};
         list9={1,3};
         ContainsAll[list7,Flatten[{list8,list9}]]
Output   False
Output   True

Input    (* The first list contains none of the elements of the second list: *)
         ContainsNone[{c,d,c},{a,b}]

         (* The first list contains elements of the second list: *)
         ContainsNone[{c,d,c,a},{a,b}]
Output   True
Output   False

Input    (* The first list contains some of the elements of the second list: *)
         ContainsAny[{b,a,b},{a,b,c}]
```

```
        (* The first list does not contain any of the elements of the second list: *)
        ContainsAny[{d,f,e},{a,b,c}]
Output  True
Output  False

Input   (* The first list contains only elements in the second list: *)
        ContainsOnly[{2,1,1},{1,2,3}]

        (* The first list contains elements not present in the second list: *)
        ContainsOnly[{2,1,3},{1,2,4}]
Output  True
Output  False

Input   (* Both lists contain exactly the same elements: *)
        list1={1,2,3,4};
        list2={2,1,4,3};
        list3={3,2,4,1};
        list4={1,2,2,3,4};

        (* The lists contain different elements: *)
        list5={3,2};
        list6={1,2,3,4,5};

        ContainsExactly[list1,list2]
        ContainsExactly[list1,list3]
        ContainsExactly[list1,list4]
        ContainsExactly[list1,list5]
        ContainsExactly[list1,list6]
Output  True
Output  True
Output  True
Output  False
Output  False
```

# UNIT 9.2

# COMBINATORIAL FUNCTIONS

Mathematica provides a wide range of built-in functions for working with combinatorial objects.

| | |
|---|---|
| `n!` | gives the factorial of n. |
| `Binomial[n,m]` | gives the binomial coefficient $\binom{n}{m}$. |
| `Multinomial[n1,n2,…]` | gives the multinomial coefficient (n1+n2+…)!/(n1!n2!...). |

***Mathematica Examples 9.9*** n!

```
Input    (* Evaluate numerically: *)
         4!
         (* Values at zero: *)
         0!
Output   24
Output   1

Input    (* Factorial threads elementwise over lists: *)
         {2,3,5,7,11}!
Output   {2,6,120,5040,39916800}

Input    (* Use FullSimplify to simplify expressions involving Factorial: *)
         FullSimplify[(n+3)!/n!]
Output   (1+n) (2+n) (3+n)

Input    (* Calculate the factorial of a matrix element-wise:*)
         matrix={{1,3},{4,5}};
         factorials=Factorial[matrix]
Output   {{1,6},{24,120}}

Input    (* Plot the factorial function for a range of numbers: *)
         DiscretePlot[
          Factorial[n],
          {n,0,5},
          PlotRange->All,
          PlotMarkers->Automatic,
          PlotStyle->Purple,
          ImageSize->200
          ]
```

Output

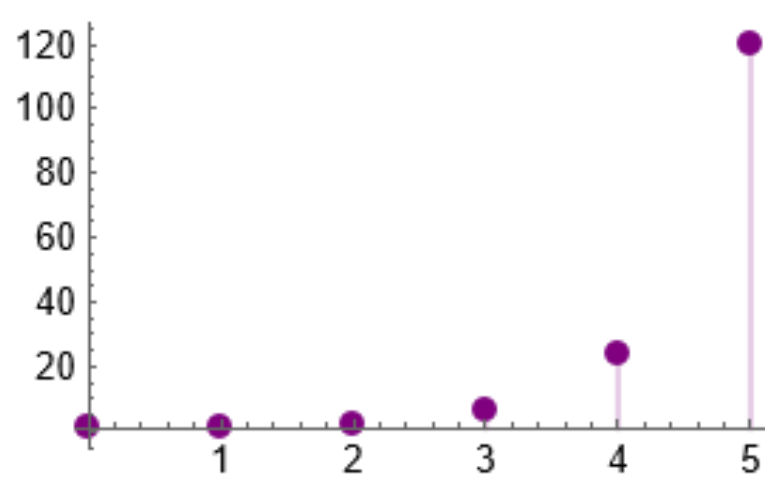

```
Input    (* The code calculates the factorial of a given number n using a For loop. The
         factorial function is defined with a single parameter n. Inside the function, a local
         variable result is initialized to 1. The For loop is used to iterate over the values
```

```
from 2 to n, incrementing i by 1 in each iteration. During each iteration, the result
variable is multiplied by the current value of i. Finally, the calculated factorial
value is stored in the variable result. In the code, n is assigned a value of 5, and
the factorial function is called with this value. The resulting factorial value is
then stored in the variable result: *)

factorial[n_]:=Module[
  {result=1},
  For[
   i=2,
   i<=n,
   i++,
   result=result*i;
   ];
  result]

n=5;
result=factorial[n]
```

Output 120

Input

```
(* The code calculates the factorial of a given number n using a recursive method.
The factorial function is defined with a single parameter n. Inside the function, an
If statement is used to check if n is equal to 0. If it is, the function returns 1,
indicating the base case of the factorial. Otherwise, the function recursively calls
itself with the argument n-1 and multiplies it by n, effectively calculating the
factorial by multiplying n with the factorial of n-1. In the code, n is assigned a
value of 5, and the factorial function is called with this value. The resulting
factorial value is then stored in the variable result: *)

factorial[n_]:=If[n==0,1,n*factorial[n-1]]

n=5;
result=factorial[n]
```

Output 120

Input

```
(* The code calculates the factorial of a given number n using an iterative method.
The factorial function is defined with a single parameter n. Inside the function, a
local variable result is initialized to 1. The Do loop is used to iterate over the
values from 2 to n. In each iteration, the result variable is multiplied by the
current value of i, updating the factorial value. Finally, the calculated factorial
value is stored in the variable result. In the code, n is assigned a value of 5, and
the factorial function is called with this value. The resulting factorial value is
then stored in the variable result: *)

factorial[n_]:=Module[
  {result=1},
  Do[
   result=result*i,
   {i,2,n}
   ];
  result]

n=5;
result=factorial[n]
```

Output 120

Input

```
(* The code calculates the factorial of a given number n using a functional approach
with the Product function. The factorialProduct function is defined with a single
parameter n. Inside the function, the Product function is used, which takes the
variable i and iterates over the values from 1 to n. In each iteration, the i value
is multiplied, effectively calculating the factorial by taking the product of all
```

```
        the values from 1 to n. In the code, the factorialProduct function is called with a
        value of 5, which calculates the factorial of 5 using the Product function. The
        resulting factorial value is returned as the output: *)

        factorialProduct[n_]:=Product[i,{i,1,n}]
        factorialProduct[5]
Output  120

Input   (* Plot over a subset of the reals: *)
        Plot[
         Factorial[x],
         {x,-5,5},
         PlotStyle->Purple,
         ImageSize->200
         ]
```

Output

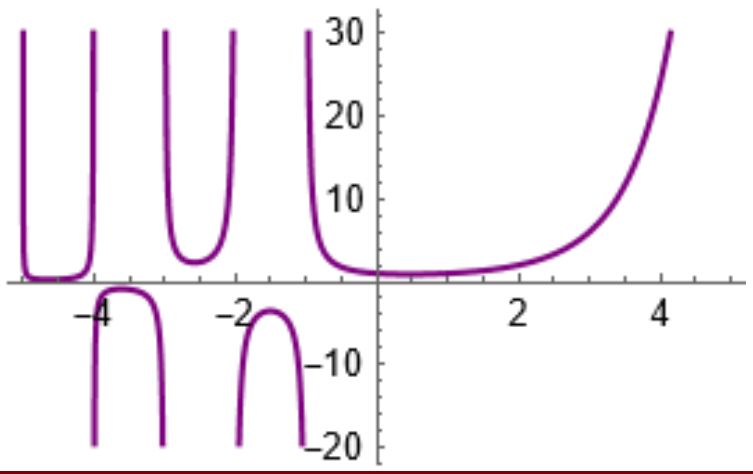

**Mathematica Examples 9.10** Binomial[n,m]

```
Input   (* The code calculates the binomial using the built-in Binomial function and the
        factorial formula: *)

        n=5;  (* Number of trials *)
        k=2;  (* Number of successes *)
        Binomial[n,k]

        (* Alternatively, if you want to implement your own function to calculate the binomial
        coefficient, you can use the formula: *)
        binomial[n,k]=n!/(k!*(n-k)!)
Output  10
Output  10

Input   (* In this code, we set n to 10, representing the number of trials. We then generate
        a table of {m,Binomial[n,m]} pairs using the Table function, where m ranges from 0
        to n. This table stores the values of the binomial coefficient for different values
        of m. The ListPlot function is used to create a plot of these data points: *)

        n=10;  (* number of trials *)
        data=Table[
            {m,Binomial[n,m]},
            {m,0,n}
            ];
        ListPlot[
         data,
         PlotStyle->Directive[PointSize[Large],Purple],
         Frame->True,
         FrameLabel->{"m","Binomial[n, m]"},
         PlotLabel->"Binomial Coefficient",
         ImageSize->220
         ]
```

Output

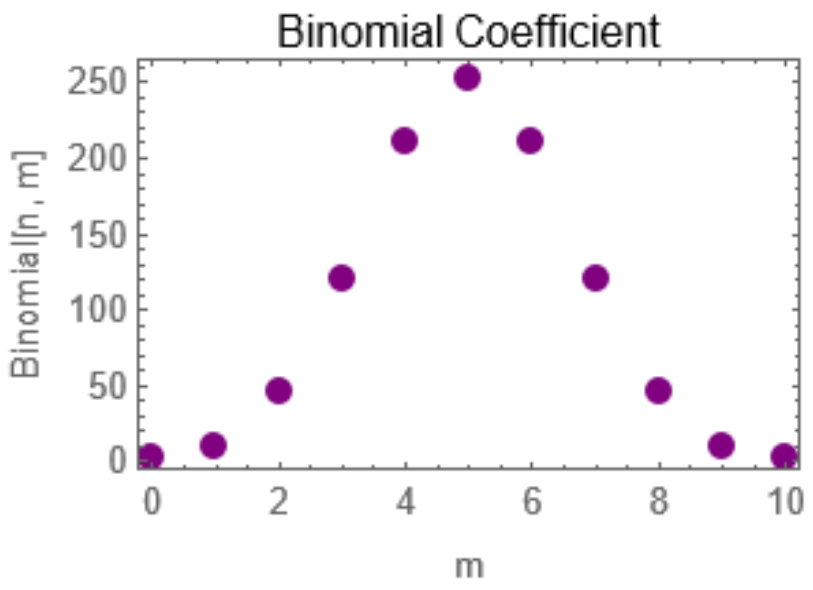

Input
```
(* Pascal's triangle is a triangular array of the binomial coefficients that arises in probability theory, combinatorics, and algebra. In row 0 (the topmost row), there is a unique nonzero entry 1. Each entry of each subsequent row is constructed by adding the number above and to the left with the number above and to the right, treating blank entries as 0. In this code, we set n to 7, which determines the number of rows in Pascal's triangle. We use nested Table functions to generate a square matrix triangle representing Pascal's triangle. Each element in the matrix is calculated using the Binomial function with row i and column j. The ListPlot function is then used to visualize the constructed Pascal's triangle. We use Flatten[triangle,1] to convert the matrix into a 1D list of values, which will be plotted: *)

n=7;  (* Number of rows in Pascal's triangle *)
Column[
 Table[
  Binomial[i,j],
  {i,0,n},
  {j,0,i}
  ],
 Center
 ]

triangle=Table[
    Binomial[i,j],
    {i,0,n},
    {j,0,i}
    ];

ListPlot[
 Flatten[triangle,1],
 Joined->True,
 Mesh->True,
 PlotStyle->Directive[PointSize[Medium],Purple,Opacity[0.8]],
 Frame->True,
 FrameLabel->{"Row","Value"},
 PlotLabel->"Pascal's Triangle",
 ImageSize->220
 ]
```

Output
```
{
{{1}},
{{1,1}},
{{1,2,1}},
{{1,3,3,1}},
{{1,4,6,4,1}},
{{1,5,10,10,5,1}},
{{1,6,15,20,15,6,1}},
{{1,7,21,35,35,21,7,1}}
}
```

Output

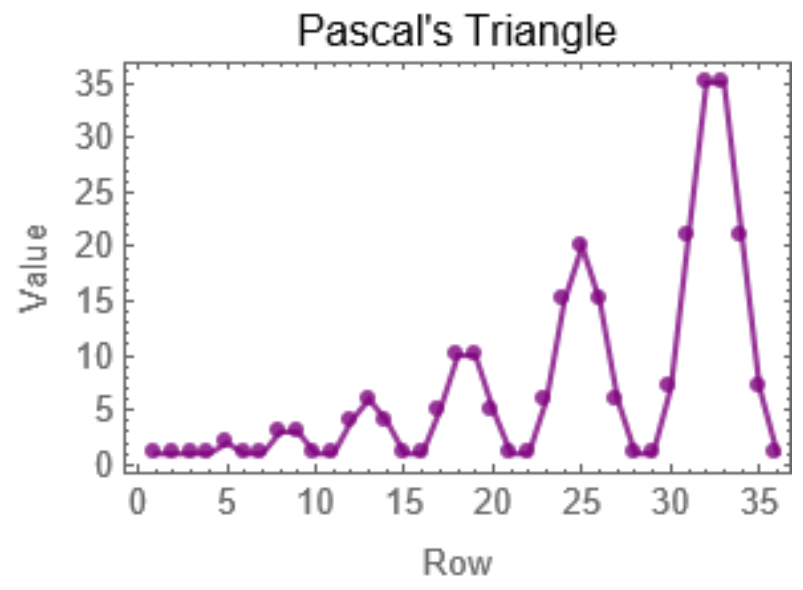

```
Input    (* This code calculates the number of ways to choose 3 elements without replacement
         from a set of 5 elements using the Binomial function. We use the Subsets function to
         generate all possible combinations of "k" elements from the set of numbers ranging
         from 1 to "n" and then calculate the length of the resulting list: *)

         n=5; (* Total number of elements *)
         k=3; (* Number of elements to choose *)

         numWaysBinomial=Binomial[n,k]

         (* Direct enumeration *)
         Subsets[Range[n],{k}]
         numWaysEnumeration=Length[Subsets[Range[n],{k}]]
         numWaysBinomial==numWaysEnumeration

Output   10
Output   {{1,2,3},{1,2,4},{1,2,5},{1,3,4},{1,3,5},{1,4,5},{2,3,4},{2,3,5},{2,4,5},{3,4,5}}
Output   10
Output   True

Input    (* Number of ways to choose m elements with replacement from a set of n elements:
         *)
         n=5;  (* Number of elements in the set *)
         k=2;  (* Number of elements to choose *)

         (*Using the binomial function*)
         numWaysBinomial=Binomial[n+k-1,k]

         (* Direct enumeration *)
         numWaysEnumeration1=Tuples[Range[n],{k}](* with Duplicates *)
         Length[numWaysEnumeration1]

         numWaysEnumeration2=DeleteDuplicates[
           Map[Sort,Tuples[Range[n],{k}]]
           ](* without Duplicates*)
         Length[numWaysEnumeration2]
Output   15
Output   {{1,1},{1,2},{1,3},{1,4},{1,5},{2,1},{2,2},{2,3},{2,4},{2,5},{3,1},{3,2},{3,3},{3,4
         },{3,5},{4,1},{4,2},{4,3},{4,4},{4,5},{5,1},{5,2},{5,3},{5,4},{5,5}}

Output   25
Output   {{1,1},{1,2},{1,3},{1,4},{1,5},{2,2},{2,3},{2,4},{2,5},{3,3},{3,4},{3,5},{4,4},{4,5
         },{5,5}}

Output   15

Input    (* The code calculates the number of ways to arrange 5 indistinguishable objects of
         one kind and 2 indistinguishable objects of another kind (represented by the numbers
         0 and 1) using the binomial coefficient and then checks the result by generating all
         possible permutations and counting them through direct enumeration: *)
```

```
        n=5;
        m=2;
        arrangements[m_,n_]:=Binomial[m+n,m]
        arrangements[5,2]

        (* Direct enumeration *)
        Permutations[Flatten[{Table[0,m],Table[1,n]}]]
        directEnumeration[m_,n_]:=Length[Permutations[Flatten[{Table[0,m],Table[1,n]}]]]
        directEnumeration[5,2]
Output  21
Output  {{0,0,1,1,1,1,1},{0,1,0,1,1,1,1},{0,1,1,0,1,1,1},{0,1,1,1,0,1,1},{0,1,1,1,1,0,1},{0
        ,1,1,1,1,1,0},{1,0,0,1,1,1,1},{1,0,1,0,1,1,1},{1,0,1,1,0,1,1},{1,0,1,1,1,0,1},{1,0,
        1,1,1,1,0},{1,1,0,0,1,1,1},{1,1,0,1,0,1,1},{1,1,0,1,1,0,1},{1,1,0,1,1,1,0},{1,1,1,0
        ,0,1,1},{1,1,1,0,1,0,1},{1,1,1,0,1,1,0},{1,1,1,1,0,0,1},{1,1,1,1,0,1,0},{1,1,1,1,1,
        0,0}}
Output  21

Input   (* The code illustrates the binomial theorem, which states that for any non-negative
        integer n and real numbers a and b, the expansion of the binomial (a+b)^n can be
        expressed as the sum of terms of the form C(n,k)*a^(n-k)*b^k, where C(n,k) represents
        the binomial coefficient. The code begins by defining the variables a and b, which
        are set to x and y, respectively. The exponent n is also defined as 5. Next, the code
        calculates the binomial coefficients using the Binomial function and stores them in
        the coefficients list. After that, the code generates the terms of the expansion by
        multiplying each coefficient with the corresponding powers of a and b, as specified
        by the binomial theorem. The expansion is then printed by summing up the terms using
        the Sum function and displaying the result. A dditionally, the code demonstrates the
        expansion using the built-in Expand function, which directly computes (a+b)^n and
        prints the expanded form. Lastly, a function named binomialTheorem is defined. This
        function takes an integer n as input and generates a bar chart to visualize the
        binomial coefficients. The coefficients are computed and stored in the coefficients
        list using the same approach as before. To observe different cases, the function
        binomialTheorem is called with a specific input value of n, which can be modified to
        visualize different scenarios: *)

        (* Define the variables *)
        a=x;
        b=y;

        (* Define the exponent *)
        n=5;

        (* Calculate the binomial coefficients *)
        coefficients=Table[
          Binomial[n,k],
          {k,0,n}
          ]

        (* Generate the terms of the expansion *)
        terms=Table[
          coefficients[[k+1]]*a^(n-k)*b^k,
          {k,0,n}
          ]

        (* Print the expansion *)
        Sum[terms[[k+1]],{k,0,n}]

        (* Print the expansion using Expand function *)
        Expand[(a+b)^n]
```

```
          binomialTheorem[n_]:=Module[
            {coefficients},

            coefficients=Table[
              Binomial[n,k],
              {k,0,n}
              ];

            BarChart[
             coefficients,
             ChartLabels->Range[0,n],
             ChartStyle->"Rainbow",
             Frame->True,
             FrameLabel->{None,"Coefficient"},
             ImageSize->250
             ]
            ]

          binomialTheorem[n]   (* Change the input value to visualize different cases *)
Output    {1,5,10,10,5,1}
Output    {x^5,5 x^4 y,10 x^3 y^2,10 x^2 y^3,5 x y^4,y^5}
Output    x^5+5 x^4 y+10 x^3 y^2+10 x^2 y^3+5 x y^4+y^5
Output    x^5+5 x^4 y+10 x^3 y^2+10 x^2 y^3+5 x y^4+y^5
Output
```

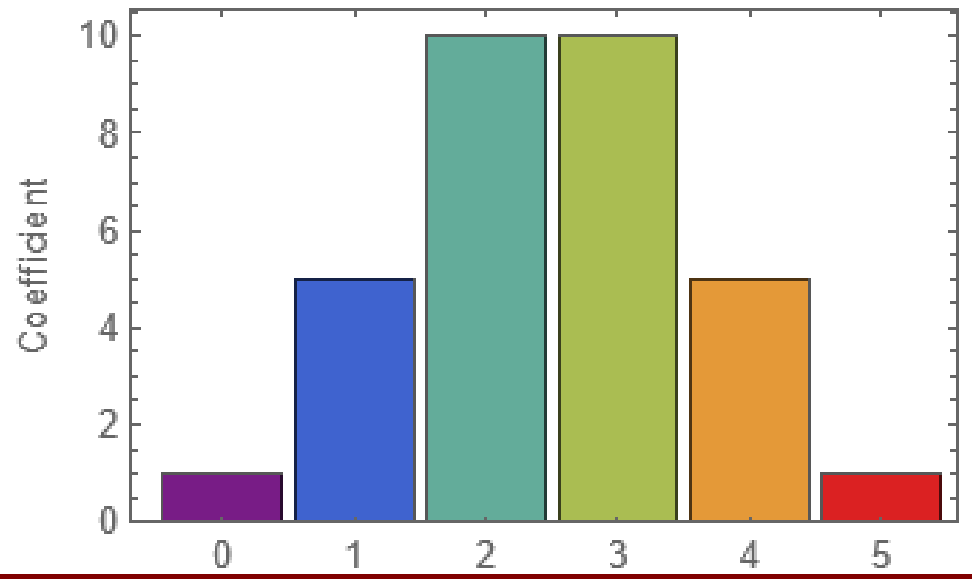

***Mathematica Examples 9.11*** Multinomial

```
Input     (* Evaluate numerically: *)
          Multinomial[1,2,1]
          multinomialcoefficient=(1+2+1)!/(1!2!1!)

          (* The 1,2,1 multinomial coefficient appears as the coefficient of x y^2 z in the
          expansion of (x+y+z)^4 : *)
          ex=Expand[(x+y+z)^4]
          Coefficient[ex,x y^2 z]
Output    12
Output    12

Output    X^4+4 x^3 y+6 x^2 y^2+4 x y^3+y^4+4 x^3 z+12 x^2 y z+12 x y^2 z+4 y^3 z+6 x^2
          z^2+12 x y z^2+6 y^2 z^2+4 x z^3+4 y z^3+z^4

Output    12

Input     (* Illustrate the multinomial theorem: *)
          Expand[(x+y+z)^4]

          sum=Sum[
            KroneckerDelta[4-(n1+n2+n3)] Multinomial[n1,n2,n3] x^n1 y^n2 z^n3,
            {n1,0,4},
            {n2,0,4},
            {n3,0,4}
            ]
```

```
        Factor[sum]

Output  X^4+4 x^3 y+6 x^2 y^2+4 x y^3+y^4+4 x^3 z+12 x^2 y z+12 x y^2 z+4 y^3 z+6 x^2
        z^2+12 x y z^2+6 y^2 z^2+4 x z^3+4 y z^3+z^4

Output  X^4+4 x^3 y+6 x^2 y^2+4 x y^3+y^4+4 x^3 z+12 x^2 y z+12 x y^2 z+4 y^3 z+6 x^2
        z^2+12 x y z^2+6 y^2 z^2+4 x z^3+4 y z^3+z^4

Output  (x+y+z)^4

Input   (* In this code, the function multinomial takes a list n as input, where each element
        represents the number of occurrences of a particular item. For example, if n={a,b,c},
        it means we have a occurrences of the first item, b occurrences of the second item,
        and c occurrences of the third item. Inside the Module, we define two local variables:
        numerator and denominator. The numerator is calculated by taking the factorial of
        the total number of occurrences (Total[n]), while the denominator is obtained by
        taking the product of the factorials of each element in the list n. Finally, the
        numerator is divided by the denominator to obtain the multinomial coefficient. You
        can call this function with a list of numbers to calculate the multinomial
        coefficient. Finally, the code calculates the multinomial using the built-in
        multinomial function: *)

        multinomial[n_]:=Module[
          {numerator,denominator},
          numerator=Factorial[Total[n]];
          denominator=Apply[Times,Factorial[n]];
          numerator/denominator
          ]
        multinomial[{2,3,4}]
        multinomialcoefficient=(2+3+4)!/(2!3!4!)
        Multinomial[2,3,4]
Output  1260
Output  1260
Output  1260

Input   (* With two arguments, Multinomial gives binomial coefficients: *)
        Binomial[10,3]
        Multinomial[7,3]
Output  120
Output  120
```

# UNIT 9.3

# DIFFERENT COUNTING SCENARIOS AND PROBABILITY

***Mathematica Examples 9.12*** Sampling with replacement and the objects being ordered

```
Input    (* Sampling with replacement and the objects being ordered refers to a sampling
         technique where elements are randomly selected from a set, and each element can be
         selected multiple times (with replacement), and the order of the selected elements
         matters. In this code, n represents the number of elements in the set, and k represents
         the number of elements to be selected. The function Tuples generates all possible
         combinations of length k from the range Range[n]. This function simulates the sampling
         process with replacement and ordered objects. Also, this code calculates the total
         number of possible combinations when sampling with replacement and ordered objects.
         It uses the formula n^k. The result is stored in the variable numCombinations: *)

         n=5;  (* Number of elements in the set. *)
         k=2;  (* Number of elements to be selected. *)

         samples=Tuples[Range[n],k]
         numCombinations=n^k

Output   {{1,1},{1,2},{1,3},{1,4},{1,5},{2,1},{2,2},{2,3},{2,4},{2,5},{3,1},{3,2},{3,3},
         {3,4},{3,5},{4,1},{4,2},{4,3},{4,4},{4,5},{5,1},{5,2},{5,3},{5,4},{5,5}}
Output   25

Input    (* In this code, n represents the number of elements in the set, and k represents
         the number of elements to be selected. To calculate the number of possible ways when
         sampling with replacement and ordered objects, we use the formula n^k. This is because
         for each position in the selection, we have n choices (as replacement is allowed),
         and since the objects are ordered, we consider all possible arrangements. The result
         is stored in the variable numWays. To calculate the probability of an event, assuming
         each event is equally likely, we use the formula 1/n^k. This is because there are
         n^k total possible outcomes, and each event has exactly one outcome. Since the events
         are mutually exclusive and equally likely, the probability of each event is 1/n^k.
         The result is stored in the variable probEvent: *)

         n=5;  (* Number of elements in the set. *)
         k=2;  (* Number of elements to be selected. *)

         numWays=n^k   (* Number of possible ways. *)
         probEvent=1/n^k   (* Probability of an event. *)

Output   25
Output   1/25

Input    (* In this code, the samplingWithReplacementOrdered function takes two parameters as
         explained before n and k. Inside the function, the Module function is used to define
         local variables numWays and probEvent. The variable numWays is assigned the value
         n^k, which represents the number of possible ways when sampling with replacement and
         ordered objects. The variable probEvent is assigned the value 1/n^k, which represents
         the probability of an event. The function returns a list containing numWays and
         probEvent: *)

         samplingWithReplacementOrdered[n_,k_]:=Module[
           {numWays,probEvent},
           numWays=n^k;(* Number of possible ways. *)
```

```
    probEvent=1/n^k;(* Probability of an event. *)
    {numWays,probEvent}
    ]

(* Example usage: *)
{n,k}={5,3};  (* Number of elements in the set and number of elements to be
selected. *)
{numWays,probEvent}=samplingWithReplacementOrdered[n,k]
```

Output {125,1/125}

Input

```
(* In this code, the samplingWithReplacementOrdered function takes two parameters as
explained befor n and k. Inside the function, the Module function is used to define
local variables sampleSpace, numWays, and probEvent. The sampleSpace variable is
generated using Tuples to represent all possible combinations. The numWays variable
is assigned the length of sampleSpace, representing the number of possible ways. The
probEvent variable is assigned 1/numWays, representing the probability of an event.
The function returns a list containing the sampleSpace, numWays, and probEvent. In
the example usage, you can provide the values of n and k in the
samplingWithReplacementOrdered function call. The result will be assigned to
sampleSpace, numWays, and probEvent, which you can then use for further analysis. To
visualize the sample space and probabilities, the code includes a histogram of the
sums of the elements in the sample space, using Histogram. This histogram displays
the probabilities of each sum occurring. Additionally, a ListPlot is used to show
the probabilities for each element sum using Tally and normalized by numWays: *)

samplingWithReplacementOrdered[n_,k_]:=Module[
    {sampleSpace,numWays,probEvent},
    sampleSpace=Tuples[Range[n],k];(* Generate sample space. *)
    numWays=Length[sampleSpace];(* Number of possible ways. *)
    probEvent=1/numWays;(* Probability of an event. *)
    {sampleSpace,numWays,probEvent}
    ]

(* Example usage: *)
{n,k}={6,2};  (* Number of elements in the set and number of elements to be
selected. *)
{sampleSpace,numWays,probEvent}=samplingWithReplacementOrdered[n,k];

sampleSpace
numWays
probEvent
Total/@sampleSpace

(* Histogram of sample space: *)
Histogram[
 Total/@sampleSpace,
 Automatic,
 "Probability",
 ImageSize->250,
 ColorFunction->Function[Opacity[0.7]],
 ChartStyle->Purple
 ]

Tally[Total/@sampleSpace]
N[Tally[Total/@sampleSpace]/numWays]

(* ListPlot of probabilities for each element sum: *)
ListPlot[
 Tally[Total/@sampleSpace]/numWays,
```

```
PlotMarkers->Automatic,
PlotRange->All,
PlotStyle->Directive[PointSize[0.009],Purple,Opacity[0.7]],
Filling->Axis,
ImageSize->250
]
```

Output {{1,1},{1,2},{1,3},{1,4},{1,5},{1,6},{2,1},{2,2},{2,3},{2,4},{2,5},{2,6},{3,1},{3,2},{3,3},{3,4},{3,5},{3,6},{4,1},{4,2},{4,3},{4,4},{4,5},{4,6},{5,1},{5,2},{5,3},{5,4},{5,5},{5,6},{6,1},{6,2},{6,3},{6,4},{6,5},{6,6}}

Output 36

Output 1/36

Output {2,3,4,5,6,7,3,4,5,6,7,8,4,5,6,7,8,9,5,6,7,8,9,10,6,7,8,9,10,11,7,8,9,10,11,12}

Output

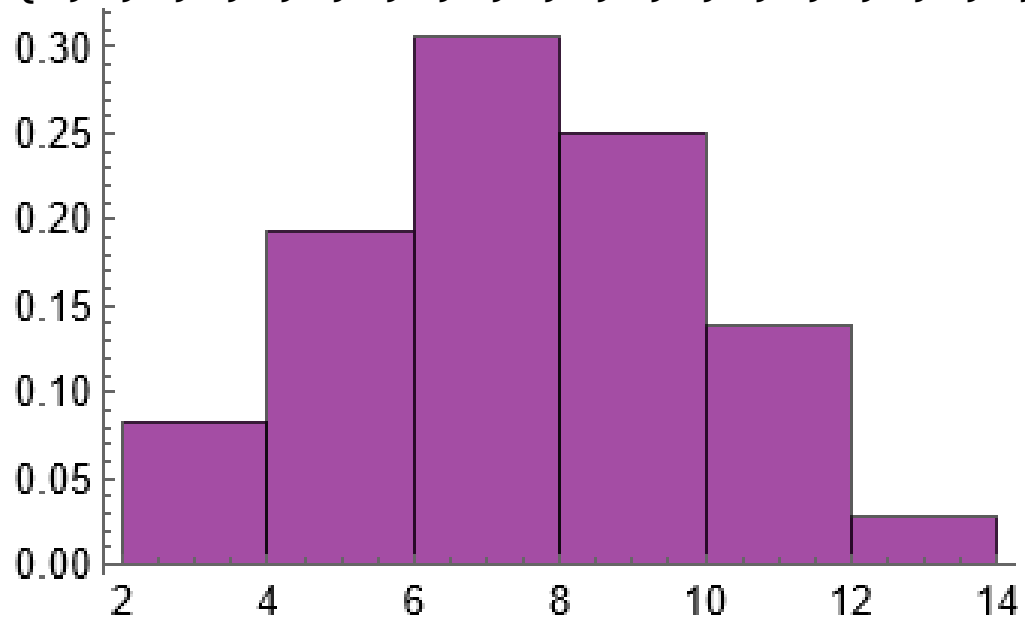

Output {{2,1},{3,2},{4,3},{5,4},{6,5},{7,6},{8,5},{9,4},{10,3},{11,2},{12,1}}

Output {{0.0555556,0.0277778},{0.0833333,0.0555556},{0.111111,0.0833333},{0.138889,0.111111},{0.166667,0.138889},{0.194444,0.166667},{0.222222,0.138889},{0.25,0.111111},{0.277778,0.0833333},{0.305556,0.0555556},{0.333333,0.0277778}}

Output

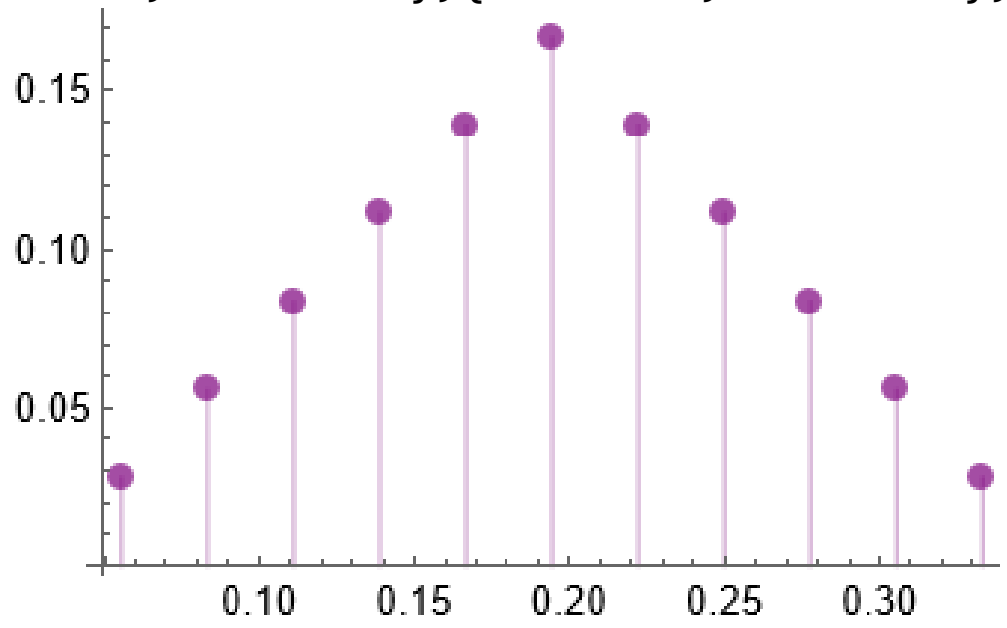

Input

```
(* In this code, after generating the samples using the Tuples function as explained before, the ListPointPlot3D function is used to create a 3D plot. Each point in the plot represents a selection, where the x-coordinate corresponds to the object chosen in the first selection, the y-coordinate corresponds to the object chosen in the second selection, and the z-coordinate corresponds to the object chosen in the third selection. Running this code will display a 3D plot where each point represents a specific combination of selections. The plot allows you to visualize the distribution of the selections in the three-dimensional space, giving you an intuitive representation of the sampling process. Note: The resulting plot may become crowded if the number of objects or selections is large, making it difficult to visualize each point clearly. In such cases, you can consider sampling a subset of the results or using alternative visualization techniques: *)

nObjects=5;  (* Total number of objects. *)
nSelections=3;  (* Number of selections to be made. *)

samples=Tuples[Range[nObjects],nSelections];

(* Generate random colors for each point. *)
colors=RandomColor[
    Length[samples]
```

```
   ];

(* Generate 3D plot with colored points. *)
ListPointPlot3D[
 Style[#,colors[[Position[samples,#][[1,1]]]]]&/@samples,
 PlotStyle->PointSize[0.03],
 BoxRatios->{1,1,1},
 AxesLabel->{"Selection 1","Selection 2","Selection 3"}
 ]
```

Output

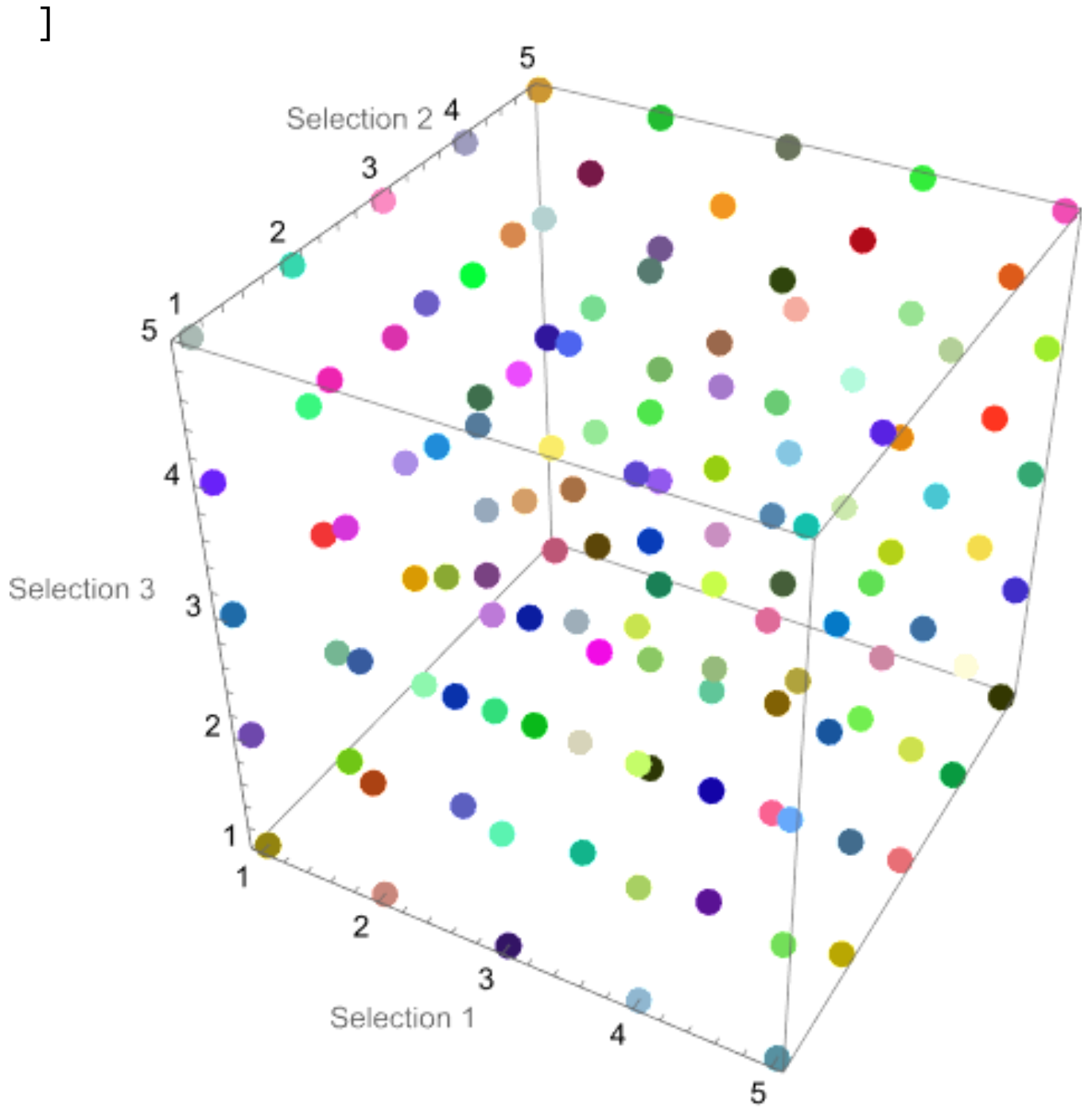

Input

```
(* In this updated code, the Manipulate function wraps around the code. Inside the
Manipulate, there are two control variables nObjects and nSelections that represent
the number of objects and selections, respectively. The control variables are defined
using the {{variable, initialValue, label}, min, max, step} syntax. Whenever you
adjust the sliders for the number of objects, the samples and colors are updated
accordingly, and the plot is automatically refreshed to reflect the changes. This
allows you to interactively explore different combinations of objects and selections:
*)

Manipulate[
 samples=Tuples[Range[nObjects],3];
 colors=RandomColor[
   Length[samples]
   ];
 ListPointPlot3D[
  Style[#,colors[[Position[samples,#][[1,1]]]]]&/@samples,
  PlotStyle->PointSize[0.03],
  BoxRatios->{1,1,1},
  AxesLabel->{"Selection 1","Selection 2","Selection 3"}
  ],
 {{nObjects,5,"Number of Objects"},3,10,1}
 ]
```

Output

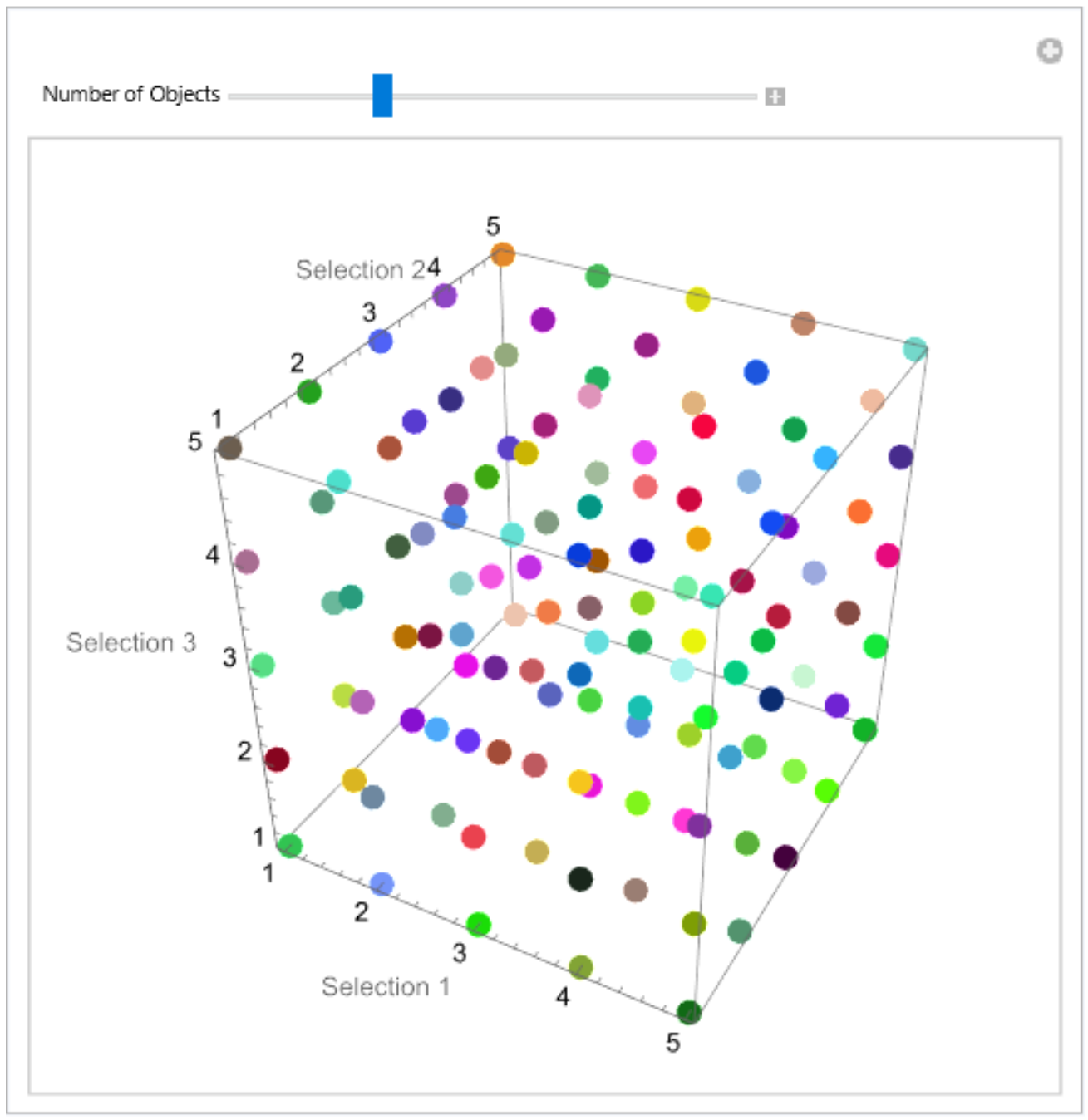

Input
```
(* Set of Events: *)
(* Event 1: The sum of the selected elements is greater than 10. In this code, the Select function is used to filter the sums of all possible combinations and select only those that are greater than 10. The result is stored in the variable event. The Tally function counts the occurrences of each sum, resulting in a list of pairs where the first element is a sum and the second element is its frequency: *)

n=6;  (* Number of elements in the set. *)
k=2;  (* Number of elements to be selected. *)
values={1,2,3,4,5,6};  (* Values of each element. *)

pairs=Tuples[values,k]
sumofeachpairs=Total/@pairs

event=Select[sumofeachpairs,#>10&]
Tally[event]
```

Output
```
{{1,1},{1,2},{1,3},{1,4},{1,5},{1,6},{2,1},{2,2},{2,3},{2,4},{2,5},{2,6},{3,1},{3,2},{3,3},{3,4},{3,5},{3,6},{4,1},{4,2},{4,3},{4,4},{4,5},{4,6},{5,1},{5,2},{5,3},{5,4},{5,5},{5,6},{6,1},{6,2},{6,3},{6,4},{6,5},{6,6}}
```
Output
```
{2,3,4,5,6,7,3,4,5,6,7,8,4,5,6,7,8,9,5,6,7,8,9,10,6,7,8,9,10,11,7,8,9,10,11,12}
```
Output
```
{11,11,12}
```
Output
```
{{11,2},{12,1}}
```

Input
```
(* Event 2: The first selected element is even. This code generates all possible combinations and selects only those where the first element is even, using the EvenQ function. The result is stored in the variable event: *)

n=6;  (* Number of elements in the set. *)
k=3;  (* Number of elements to be selected. *)
values={1,2,3,4,5,6};  (* Values of each element. *)

event=Select[Tuples[values,k],EvenQ[First[#]]&]
```

Output

```
{{2,1,1},{2,1,2},{2,1,3},{2,1,4},{2,1,5},{2,1,6},{2,2,1},{2,2,2},{2,2,3},{2,2,4},{2
,2,5},{2,2,6},{2,3,1},{2,3,2},{2,3,3},{2,3,4},{2,3,5},{2,3,6},{2,4,1},{2,4,2},{2,4,
3},{2,4,4},{2,4,5},{2,4,6},{2,5,1},{2,5,2},{2,5,3},{2,5,4},{2,5,5},{2,5,6},{2,6,1},
{2,6,2},{2,6,3},{2,6,4},{2,6,5},{2,6,6},{4,1,1},{4,1,2},{4,1,3},{4,1,4},{4,1,5},{4,
1,6},{4,2,1},{4,2,2},{4,2,3},{4,2,4},{4,2,5},{4,2,6},{4,3,1},{4,3,2},{4,3,3},{4,3,4
},{4,3,5},{4,3,6},{4,4,1},{4,4,2},{4,4,3},{4,4,4},{4,4,5},{4,4,6},{4,5,1},{4,5,2},{
4,5,3},{4,5,4},{4,5,5},{4,5,6},{4,6,1},{4,6,2},{4,6,3},{4,6,4},{4,6,5},{4,6,6},{6,1
,1},{6,1,2},{6,1,3},{6,1,4},{6,1,5},{6,1,6},{6,2,1},{6,2,2},{6,2,3},{6,2,4},{6,2,5}
,{6,2,6},{6,3,1},{6,3,2},{6,3,3},{6,3,4},{6,3,5},{6,3,6},{6,4,1},{6,4,2},{6,4,3},{6
,4,4},{6,4,5},{6,4,6},{6,5,1},{6,5,2},{6,5,3},{6,5,4},{6,5,5},{6,5,6},{6,6,1},{6,6,
2},{6,6,3},{6,6,4},{6,6,5},{6,6,6}}
```

Input

```
(* Event 3: The selected elements are in ascending order. In this code, the OrderedQ
function is used to check if each combination is in ascending order. Only those
combinations that satisfy this condition are selected and stored in the variable
event: *)

n=6;  (* Number of elements in the set. *)
k=3;  (* Number of elements to be selected. *)
values={1,2,3,4,5,6};  (* Values of each element. *)

event=Select[Tuples[values,k],OrderedQ[#]&]
```

Output

```
{{1,1,1},{1,1,2},{1,1,3},{1,1,4},{1,1,5},{1,1,6},{1,2,2},{1,2,3},{1,2,4},{1,2,5},{1
,2,6},{1,3,3},{1,3,4},{1,3,5},{1,3,6},{1,4,4},{1,4,5},{1,4,6},{1,5,5},{1,5,6},{1,6,
6},{2,2,2},{2,2,3},{2,2,4},{2,2,5},{2,2,6},{2,3,3},{2,3,4},{2,3,5},{2,3,6},{2,4,4},
{2,4,5},{2,4,6},{2,5,5},{2,5,6},{2,6,6},{3,3,3},{3,3,4},{3,3,5},{3,3,6},{3,4,4},{3,
4,5},{3,4,6},{3,5,5},{3,5,6},{3,6,6},{4,4,4},{4,4,5},{4,4,6},{4,5,5},{4,5,6},{4,6,6
},{5,5,5},{5,5,6},{5,6,6},{6,6,6}}
```

Input

```
(* Event 4: The maximum selected element is equal to 6. In this code, the Select
function is used to filter the combinations and select only those where the maximum
element is equal to 6. The result is stored in the variable event: *)

n=6;  (* Number of elements in the set. *)
k=3;  (* Number of elements to be selected. *)
values={1,2,3,4,5,6};  (* Values of each element. *)

event=Select[Tuples[values,k],Max[#]==6&]
```

Output

```
{{1,1,6},{1,2,6},{1,3,6},{1,4,6},{1,5,6},{1,6,1},{1,6,2},{1,6,3},{1,6,4},{1,6,5},{1
,6,6},{2,1,6},{2,2,6},{2,3,6},{2,4,6},{2,5,6},{2,6,1},{2,6,2},{2,6,3},{2,6,4},{2,6,
5},{2,6,6},{3,1,6},{3,2,6},{3,3,6},{3,4,6},{3,5,6},{3,6,1},{3,6,2},{3,6,3},{3,6,4},
{3,6,5},{3,6,6},{4,1,6},{4,2,6},{4,3,6},{4,4,6},{4,5,6},{4,6,1},{4,6,2},{4,6,3},{4,
6,4},{4,6,5},{4,6,6},{5,1,6},{5,2,6},{5,3,6},{5,4,6},{5,5,6},{5,6,1},{5,6,2},{5,6,3
},{5,6,4},{5,6,5},{5,6,6},{6,1,1},{6,1,2},{6,1,3},{6,1,4},{6,1,5},{6,1,6},{6,2,1},{
6,2,2},{6,2,3},{6,2,4},{6,2,5},{6,2,6},{6,3,1},{6,3,2},{6,3,3},{6,3,4},{6,3,5},{6,3
,6},{6,4,1},{6,4,2},{6,4,3},{6,4,4},{6,4,5},{6,4,6},{6,5,1},{6,5,2},{6,5,3},{6,5,4}
,{6,5,5},{6,5,6},{6,6,1},{6,6,2},{6,6,3},{6,6,4},{6,6,5},{6,6,6}}
```

Input

```
(* Event 5: The selected elements form a palindrome. This code generates all possible
combinations and selects only those combinations where the elements form a palindrome,
using the Reverse function. The result is stored in the variable event: *)

n=6;  (*Number of elements in the set*)
k=3;  (*Number of elements to be selected*)
values={1,2,3,4,5,6};  (*Values of each element*)

event=Select[Tuples[values,k],#==Reverse[#]&]
```

Output {{1,1,1},{1,2,1},{1,3,1},{1,4,1},{1,5,1},{1,6,1},{2,1,2},{2,2,2},{2,3,2},{2,4,2},{2,5,2},{2,6,2},{3,1,3},{3,2,3},{3,3,3},{3,4,3},{3,5,3},{3,6,3},{4,1,4},{4,2,4},{4,3,4},{4,4,4},{4,5,4},{4,6,4},{5,1,5},{5,2,5},{5,3,5},{5,4,5},{5,5,5},{5,6,5},{6,1,6},{6,2,6},{6,3,6},{6,4,6},{6,5,6},{6,6,6}}

Input

```
(* Event 6: The selected elements contain duplicates. In this code, the Union function is used to remove duplicates from each combination. The Length function is then used to check if the resulting set has fewer elements than k, indicating the presence of duplicates. Only those combinations that satisfy this condition are selected and stored in the variable event: *)

n=6;  (* Number of elements in the set. *)
k=3;  (* Number of elements to be selected. *)
values={1,2,3,4,5,6};  (* Values of each element. *)

event=Select[Tuples[values,k],Length[Union[#]]<k&]
```

Output {{1,1,1},{1,1,2},{1,1,3},{1,1,4},{1,1,5},{1,1,6},{1,2,1},{1,2,2},{1,3,1},{1,3,3},{1,4,1},{1,4,4},{1,5,1},{1,5,5},{1,6,1},{1,6,6},{2,1,1},{2,1,2},{2,2,1},{2,2,2},{2,2,3},{2,2,4},{2,2,5},{2,2,6},{2,3,2},{2,3,3},{2,4,2},{2,4,4},{2,5,2},{2,5,5},{2,6,2},{2,6,6},{3,1,1},{3,1,3},{3,2,2},{3,2,3},{3,3,1},{3,3,2},{3,3,3},{3,3,4},{3,3,5},{3,3,6},{3,4,3},{3,4,4},{3,5,3},{3,5,5},{3,6,3},{3,6,6},{4,1,1},{4,1,4},{4,2,2},{4,2,4},{4,3,3},{4,3,4},{4,4,1},{4,4,2},{4,4,3},{4,4,4},{4,4,5},{4,4,6},{4,5,4},{4,5,5},{4,6,4},{4,6,6},{5,1,1},{5,1,5},{5,2,2},{5,2,5},{5,3,3},{5,3,5},{5,4,4},{5,4,5},{5,5,1},{5,5,2},{5,5,3},{5,5,4},{5,5,5},{5,5,6},{5,6,5},{5,6,6},{6,1,1},{6,1,6},{6,2,2},{6,2,6},{6,3,3},{6,3,6},{6,4,4},{6,4,6},{6,5,5},{6,5,6},{6,6,1},{6,6,2},{6,6,3},{6,6,4},{6,6,5},{6,6,6}}

Input

```
(* Event 7: The selected elements are all odd. In this code, the AllTrue function is used to check if all elements in each combination are odd using the OddQ function. Only those combinations that satisfy this condition are selected and stored in the variable event: *)

n=6;  (* Number of elements in the set. *)
k=3;  (* Number of elements to be selected. *)
values={1,2,3,4,5,6};  (* Values of each element. *)

event=Select[Tuples[values,k],AllTrue[#,OddQ]&]
```

Output {{1,1,1},{1,1,3},{1,1,5},{1,3,1},{1,3,3},{1,3,5},{1,5,1},{1,5,3},{1,5,5},{3,1,1},{3,1,3},{3,1,5},{3,3,1},{3,3,3},{3,3,5},{3,5,1},{3,5,3},{3,5,5},{5,1,1},{5,1,3},{5,1,5},{5,3,1},{5,3,3},{5,3,5},{5,5,1},{5,5,3},{5,5,5}}

Input

```
(* Event 8: The sum of the selected elements is divisible by 3. This code calculates the sums of all possible combinations and selects only those sums that are divisible by 3 using the Divisible function: *)

n=6;  (* Number of elements in the set. *)
k=3;  (* Number of elements to be selected. *)
values={1,2,3,4,5,6};  (* Values of each element. *)

event=Select[Total/@Tuples[values,k],Divisible[#,3]&]
```

Output {3,6,6,9,6,9,6,9,9,12,9,12,6,9,6,9,6,9,9,12,9,12,9,12,6,9,6,9,9,12,9,12,9,12,12,15,6,9,9,12,9,12,9,12,12,15,12,15,9,12,9,12,9,12,12,15,12,15,12,15,9,12,9,12,12,15,12,15,12,15,15,18}

Input

```
(* Event 9: The selected elements contain a specific element. In this code, the MemberQ function is used to check if the specific element is present in each
```

```
        combination. Only those combinations that contain the specific element are selected
        and stored in the variable event: *)

        n=6;  (* Number of elements in the set. *)
        k=3;  (* Number of elements to be selected. *)
        values={1,2,3,4,5,6};  (* Values of each element. *)
        specificElement=4;  (* Specific element to check for. *)

        event=Select[Tuples[values,k],MemberQ[#,specificElement]&]

Output  {{1,1,4},{1,2,4},{1,3,4},{1,4,1},{1,4,2},{1,4,3},{1,4,4},{1,4,5},{1,4,6},{1,5,4},{1
        ,6,4},{2,1,4},{2,2,4},{2,3,4},{2,4,1},{2,4,2},{2,4,3},{2,4,4},{2,4,5},{2,4,6},{2,5,
        4},{2,6,4},{3,1,4},{3,2,4},{3,3,4},{3,4,1},{3,4,2},{3,4,3},{3,4,4},{3,4,5},{3,4,6},
        {3,5,4},{3,6,4},{4,1,1},{4,1,2},{4,1,3},{4,1,4},{4,1,5},{4,1,6},{4,2,1},{4,2,2},{4,
        2,3},{4,2,4},{4,2,5},{4,2,6},{4,3,1},{4,3,2},{4,3,3},{4,3,4},{4,3,5},{4,3,6},{4,4,1
        },{4,4,2},{4,4,3},{4,4,4},{4,4,5},{4,4,6},{4,5,1},{4,5,2},{4,5,3},{4,5,4},{4,5,5},{
        4,5,6},{4,6,1},{4,6,2},{4,6,3},{4,6,4},{4,6,5},{4,6,6},{5,1,4},{5,2,4},{5,3,4},{5,4
        ,1},{5,4,2},{5,4,3},{5,4,4},{5,4,5},{5,4,6},{5,5,4},{5,6,4},{6,1,4},{6,2,4},{6,3,4}
        ,{6,4,1},{6,4,2},{6,4,3},{6,4,4},{6,4,5},{6,4,6},{6,5,4},{6,6,4}}

Input   (* This code generates a specified number of random samples of combinations when
        sampling with replacement and ordered objects. The RandomChoice function is used to
        randomly select k elements from the range Range[n], and the function is called
        numSamples times. The result is stored in the variable samples: *)

        n=5;  (* Number of elements in the set. *)
        k=3;  (* Number of elements to be selected. *)
        numSamples=5;  (* Number of random samples to generate. *)

        samples=RandomChoice[Range[n],{numSamples,k}]

Output  {{1,3,1},{1,4,3},{1,4,5},{5,2,4},{4,5,2}}
```

***Mathematica Examples 9.13*** Sample Without Replacement and the objects being ordered

```
Input   (* Sampling without replacement and ordered: *)

        (* Define the values: *)
        values=Range[5];
        (* Define the sample size: *)
        sampleSize=2;
        (* Perform sampling without replacement and ordered: *)
        samples=Permutations[values,{sampleSize}]
        Length[samples]

Output  {{1,2},{1,3},{1,4},{1,5},{2,1},{2,3},{2,4},{2,5},{3,1},{3,2},{3,4},{3,5},{4,1},
        {4,2},{4,3},{4,5},{5,1},{5,2},{5,3},{5,4}}
Output  20

Input   (* The formula to count permutations in the case of sampling without replacement and
        ordered is given by the expression: n!/(n-k)!, where "n" represents the total number
        of items available and "k" represents the number of items to select: *)

        n=5;  (* Total number of people. *)
        k=2;   (* Number of committee members. *)

        numPermutations=Factorial[n]/Factorial[n-k]

Output  20

Input   (* Examples: Sampling without replacement and ordered:*)
```

```
(* Example 1: Choosing a Committee. Suppose you have a group of 10 people, and you want to form a committee of 4 members. Let us calculate the number of possible committees: *)

n=10;  (* Total number of people. *)
k=4;   (* Number of committee members. *)
numPermutations=Factorial[n]/Factorial[n-k]


(* Example 2: Arranging Letters in a Word. Consider the word "MATH" How many different arrangements of letters can be formed using all the letters: *)

word="MATH";
numPermutations=Factorial[StringLength[word]]

(* Example 3: Password Combinations. Suppose you need to create a password using exactly 6 lowercase letters (a-z) without repetition. Let us calculate the number of possible password combinations: *)

n=26;  (* Total number of lowercase letters. *)
k=6;   (* Number of letters in the password. *)
numPermutations=Factorial[n]/Factorial[n-k]

(* Example 4: Assigning Seats. Suppose you have 8 people attending a dinner party, and there are 8 seats arranged in a row. How many ways can the people be assigned to the seats: *)

n=8;  (* Total number of people. *)
k=8;  (* Number of seats. *)
numPermutations=Factorial[n]

(* Example 5: Creating License Plates. In a certain country, license plates consist of 3 letters followed by 3 digits (e.g., ABC123). How many unique license plates can be created?: *)

numLetters=26;  (* Total number of letters (A-Z). *)
numDigits=10;   (* Total number of digits (0-9). *)
numPositions=6; (* Total number of positions on the license plate. *)
numPermutations=Factorial[numLetters+numDigits]/Factorial[numPositions]

Output  5040
Output  24
Output  165765600
Output  40320
Output  516657398319307246483332566876160000000

Input   (* Sample without replacement and ordered: *)

        (* Define the number of objects and the sample size: *)
        n=5; (* Number of objects. *)
        k=3; (* Sample size. *)
        values=Range[1,n]; (* Values of each element. *)

        (* Generate all possible permutations of the objects: *)
        permutations=Permutations[values];

        (* Select a random permutation from the list: *)
        randomPermutation=RandomChoice[permutations]

        (* Take the first 4 elements from the random permutations: *)
```

```
        sample=Take[permutations,4]

Output  {2,4,5,1,3}
Output  {{1,2,3,4,5},{1,2,3,5,4},{1,2,4,3,5},{1,2,4,5,3}}

Input   (* This function takes two arguments: n represents the total number of objects, and
        k represents the number of objects being sampled. It returns a list with two elements.
        The first element is the number of elements in the sample space, and the second
        element is a list of all the elements in the sample space: *)

        sampleSpaceOrderNoReplacement[n_,k_]:=Module[
          {sampleSpace,numElements},
          sampleSpace=Permutations[Range[n],{k}];
          numElements=Length[sampleSpace];
          {numElements,sampleSpace}
          ]

        (* Selecting 2 marbles from a bag containing 5 red marbles and 3 blue marbles without
        replacement and with ordering: *)
        sampleSpaceOrderNoReplacement[8,2]

Output  {56,{{1,2},{1,3},{1,4},{1,5},{1,6},{1,7},{1,8},{2,1},{2,3},{2,4},{2,5},{2,6},{2,7},
        {2,8},{3,1},{3,2},{3,4},{3,5},{3,6},{3,7},{3,8},{4,1},{4,2},{4,3},{4,5},{4,6},{4,7}
        ,{4,8},{5,1},{5,2},{5,3},{5,4},{5,6},{5,7},{5,8},{6,1},{6,2},{6,3},{6,4},{6,5},{6,7
        },{6,8},{7,1},{7,2},{7,3},{7,4},{7,5},{7,6},{7,8},{8,1},{8,2},{8,3},{8,4},{8,5},{8,
        6},{8,7}}}

Input   (* In this example, we start by defining the values as a range from 1 to 6. The
        sample size is set to 3,and we generate all possible ordered samples without
        replacement using Permutations. Next, we define the condition for the event using
        AllTrue and the EvenQ function. This condition ensures that all elements in a sample
        are even numbers. We filter the generated samples using Select and apply the condition
        to keep only the samples that satisfy the condition. The total number of samples is
        calculated using Length, and the number of favorable outcomes is obtained by
        calculating the length of the filtered samples. Finally, we calculate the probability
        of the event by dividing the number of favorable outcomes by the total number of
        samples: *)

        (* Total number of objects: *)
        n=6;

        (* Number of objects selected in each sample: *)
        sampleSize=3;

        (* Values of each element: *)
        values=Range[1,n];

        (* Generate all possible ordered samples without replacement: *)
        samples=Permutations[values,{sampleSize}];

        (* Define the condition for the event: *)
        condition=AllTrue[#,EvenQ]&;

        (* Filter the samples based on the condition: *)
        filteredSamples=Select[samples,condition];

        (* Calculate the total number of samples: *)
        totalSamples=Length[samples];

        (* Calculate the number of favorable outcomes: *)
```

```
favorableOutcomes=Length[filteredSamples];

(* Calculate the probability of the event: *)
probability=favorableOutcomes/totalSamples;

(* Display the total number of samples and favorable outcomes: *)
Print["Total number of samples: ",totalSamples];
Print["Favorable outcomes: ",favorableOutcomes];

(* Display the probability of the event: *)
Print["Probability of the event: ",probability];
```

Output

```
Total number of samples:  120
Favorable outcomes:  6
Probability of the event:  1/20
```

Input

```
(* In this code, the sampleWithoutReplacementOrdered function takes three arguments:
n (total number of objects), r (number of objects selected in each sample), and event
(the event we are interested in). The function calculates the sample space by
generating all permutations of length r from the range of 1 to n. It then filters
the sample space to find the favorable outcomes that satisfy the given event. The
count of favorable outcomes and the probability of the event are calculated using
the lengths of the corresponding lists: *)

sampleWithoutReplacementOrdered[n_,r_,event_]:=Module[
  {sampleSpace,favorableOutcomes,probability,count},
  sampleSpace=Permutations[Range[n],{r}];
  favorableOutcomes=Select[sampleSpace,event];
  count=Length[favorableOutcomes];
  probability=count/Length[sampleSpace];
  {probability,count}
  ]

(* Example usage: *)
n=5; (* Total number of objects. *)
r=3 ;(* Number of objects selected in each sample. *)
event[x_]:=x[[1]]<x[[2]](* Example event: First element is less than the second
element. *)

sampleSpace=Permutations[Range[n],{r}]
favorableOutcomes=Select[sampleSpace,event]

{probability,count}=sampleWithoutReplacementOrdered[n,r,event];
Print["Probability of the event:",probability]
Print["Count of the event:",count]
```

Output

```
{{1,2,3},{1,2,4},{1,2,5},{1,3,2},{1,3,4},{1,3,5},{1,4,2},{1,4,3},{1,4,5},{1,5,2},{1
,5,3},{1,5,4},{2,1,3},{2,1,4},{2,1,5},{2,3,1},{2,3,4},{2,3,5},{2,4,1},{2,4,3},{2,4,
5},{2,5,1},{2,5,3},{2,5,4},{3,1,2},{3,1,4},{3,1,5},{3,2,1},{3,2,4},{3,2,5},{3,4,1},
{3,4,2},{3,4,5},{3,5,1},{3,5,2},{3,5,4},{4,1,2},{4,1,3},{4,1,5},{4,2,1},{4,2,3},{4,
2,5},{4,3,1},{4,3,2},{4,3,5},{4,5,1},{4,5,2},{4,5,3},{5,1,2},{5,1,3},{5,1,4},{5,2,1
},{5,2,3},{5,2,4},{5,3,1},{5,3,2},{5,3,4},{5,4,1},{5,4,2},{5,4,3}}
```

Output

```
{{1,2,3},{1,2,4},{1,2,5},{1,3,2},{1,3,4},{1,3,5},{1,4,2},{1,4,3},{1,4,5},{1,5,2},{1
,5,3},{1,5,4},{2,3,1},{2,3,4},{2,3,5},{2,4,1},{2,4,3},{2,4,5},{2,5,1},{2,5,3},{2,5,
4},{3,4,1},{3,4,2},{3,4,5},{3,5,1},{3,5,2},{3,5,4},{4,5,1},{4,5,2},{4,5,3}}
```

Output

```
Probability of the event: 1/2
Count of the event: 30
```

Input

```
(* In this code, the event is defined as the sum of the elements in the sample being
less than or equal to 20. You can modify the event function according to your specific
```

```
requirements, such as checking for a different sum condition or any other criterion
based on the elements of the sample: *)

sampleWithoutReplacementOrdered[n_,r_,event_]:=Module[
  {sampleSpace,favorableOutcomes,probability,count},
  sampleSpace=Permutations[Range[n],{r}];
  favorableOutcomes=Select[sampleSpace,event];
  count=Length[favorableOutcomes];
  probability=count/Length[sampleSpace];
  {probability,count}]

(* Example usage: *)
n=10; (* Total number of objects. *)
r=3; (* Number of objects selected in each sample. *)
event[x_]:=Total[x]<=20; (* Example event:the sum of the elements in the sample is
less than or equal to 20. *)

{probability,count}=sampleWithoutReplacementOrdered[n,r,event];
Print["Probability of the event:",probability]
Print["Count of the event:",count]
```

Output

```
Probability of the event: 97/120
Count of the event: 582
```

Input

```
(* The method to calculate the number of elements in the sample space for the case
of "Permutations of Similar Objects"*)(* The formula to calculate the number of
permutations of similar objects is: n!/(n1!*n2!*...*nk!), Where: n is the total
number of objects in the set. n1,n2,..., nk are the counts of each type of similar
object: *)

(* In this code, objectCounts represents the number of each type of similar objects.
The Total function calculates the total number of objects by summing up the counts.
The Factorial function computes the factorial of the total number of objects and each
count of similar objects. The Apply[Times,...] multiplies all the factorials
together. The result is the number of elements in the sample space, considering the
permutations of similar objects as distinct: *)

objectCounts={2,1,1}; (* Number of each type of similar objects. *)
numElements=Factorial[Total[objectCounts]]/Apply[Times,Factorial[objectCounts]];
Print["Number of Elements: ",numElements]

(* The Multinomial function computes the multinomial coefficient, which gives the
number of ways to arrange the objects when they are considered indistinguishable
within each group but distinguishable between groups: *)

objectCounts={2,1,1}; (* Number of each type of similar objects. *)
numElements=Apply[Multinomial,objectCounts];
Print["Number of Elements: ",numElements]
```

Output

```
Number of Elements:  12
Number of Elements:  12
```

Input

```
(* In this code, we start by defining the set of objects as a list objects with
repeated elements (a,a,b,c). Then, we use the Permutations function to generate all
possible permutations of the objects. The resulting permutations are stored in the
permutations variable. Finally, we use the TableForm function to display the elements
of the sample space in a tabular format. The TableHeadings option is used to label
the columns with the original objects: *)

(* Define the set of objects: *)
```

```
objects={a,a,b,c};

(* Generate all permutations of the objects: *)
permutations=Permutations[objects];(* In Permutations function: Repeated elements are treated as identical. *)

(* Display the elements of the sample space: *)
TableForm[permutations,TableHeadings->{None,objects}]
```

Output
```
{
 {a, a, b, c},
 {a, a, b, c},
 {a, a, c, b},
 {a, b, a, c},
 {a, b, c, a},
 {a, c, a, b},
 {a, c, b, a},
 {b, a, a, c},
 {b, a, c, a},
 {b, c, a, a},
 {c, a, a, b},
 {c, a, b, a},
 {c, b, a, a}
}
```

***Mathematica Examples 9.14*** Sampling without Replacement and the Objects Are Not Ordered

Input
```
(* Sampling without Replacement and the Objects Are Not Ordered: *)
(* In this code, the sampleSpaceSize function calculates the number of elements in the sample space using the binomial coefficient Binomial[n,k]. The generateSampleSpace function generates all possible subsets of length k from the given list of elements using the Subsets function: *)

sampleSpaceSize[n_,k_]:=Binomial[n,k]
generateSampleSpace[n_,k_,elements_]:=Subsets[elements,{k}]

(* Example usage: *)
n=5; (* Total number of objects. *)
k=3; (* Number of objects to sample. *)
elements={a,b,c,d,e}; (*List of objects. *)
(* Calculate the size of the sample space. *)
size=sampleSpaceSize[n,k]
(* Generate the sample space: *)
sampleSpace=generateSampleSpace[n,k,elements]
```

Output
```
10
```
Output
```
{{a,b,c},{a,b,d},{a,b,e},{a,c,d},{a,c,e},{a,d,e},{b,c,d},{b,c,e},{b,d,e},{c,d,e}}
```

Input
```
sampleSpaceSize[n_,k_]:=Binomial[n,k]
generateSampleSpace[n_,k_,elements_]:=Module[
  {sampleSpace},
  sampleSpace=Subsets[elements,{k}];
  sampleSpace
  ]

(* Example usage: *)
n=5; (* Total number of objects. *)
k=3; (* Number of objects to sample. *)
elements={a,b,c,d,e}; (* List of objects. *)
(* Calculate the size of the sample space: *)
```

```
        size=sampleSpaceSize[n,k]
        (* Generate the sample space: *)
        sampleSpace=generateSampleSpace[n,k,elements]

Output  10
Output  {{a,b,c},{a,b,d},{a,b,e},{a,c,d},{a,c,e},{a,d,e},{b,c,d},{b,c,e},{b,d,e},{c,d,e}}

Input   (* More Examples: *)

        (* In this code, the sampleSpace function takes two arguments n and r, which represent
        the total number of objects and the number of objects to be selected, respectively.
        It generates all possible combinations of size r from the list of objects and returns
        the sample space: *)

        (* Module function to generate sample space: *)
        sampleSpace[n_,r_]:=Module[
          {objects,combinations},
          objects=Range[n];
          (* Create a list of objects. *)
          combinations=Subsets[objects,{r}];
          (* Generate all combinations of size r. *)
          combinations
          ]

        (* Example 1: Selecting Students for a Project. Suppose you have a group of 7
        students,and you need to select a team of 4 students for a project. You want to
        select the team members without replacement, and the order of their selection does
        not matter: *)

        n=7; (* Total number of students. *)
        r=3; (* Number of students to be selected. *)

        elements=sampleSpace[n,r];
        numberOfElements=Binomial[n,r];

        Print["Sample Space: ",elements];
        Print["Number of Elements: ",numberOfElements];


        (* Example 2: Picking Marbles from a Bag. Suppose you have a bag containing 6 marbles
        of different colors (red, blue, green, yellow, and orange). You want to select 3
        marbles from the bag without replacement, and the order of the marbles does not
        matter: *)

        n=6; (* Total number of marbles in the bag. *)
        r=2; (* Number of marbles to be selected. *)

        elements=sampleSpace[n,r];
        numberOfElements=Binomial[n,r];

        Print["Sample Space: ",elements];
        Print["Number of Elements: ",numberOfElements];

Output  Sample Space:
        {{1,2,3},{1,2,4},{1,2,5},{1,2,6},{1,2,7},{1,3,4},{1,3,5},{1,3,6},{1,3,7},{1,4,5},{1
        ,4,6},{1,4,7},{1,5,6},{1,5,7},{1,6,7},{2,3,4},{2,3,5},{2,3,6},{2,3,7},{2,4,5},{2,4,
        6},{2,4,7},{2,5,6},{2,5,7},{2,6,7},{3,4,5},{3,4,6},{3,4,7},{3,5,6},{3,5,7},{3,6,7},
        {4,5,6},{4,5,7},{4,6,7},{5,6,7}}
Output  Number of Elements:  35
```

```
Output   Sample Space:
         {{1,2},{1,3},{1,4},{1,5},{1,6},{2,3},{2,4},{2,5},{2,6},{3,4},{3,5},{3,6},{4,5},{4,6
         },{5,6}}
Output   Number of Elements:  15

Input    (* In this code, the sampleSpaceSize function remains the same as before, which
         calculates the number of elements in the sample space using the Binomial function.
         The generateSampleSpace function also remains the same as before, which generates
         the sample space using the Subsets function. The probability function takes two
         arguments: event and sampleSpace. It calculates the probability of an event by
         dividing the length of the event by the length of the sample space. The event variable
         represents the event of interest The event uses Select to filter the sample space
         based on a condition: *)

         (* Function to calculate the number of elements in sample space: *)
         sampleSpaceSize[n_,k_]:=Binomial[n,k]

         (* Function to generate sample space: *)
         generateSampleSpace[n_,k_]:=Module[
           {objects,combinations},
           objects=Range[n];
           combinations=Subsets[objects,{k}];
           combinations]

         (* Function to calculate the probability of an event: *)
         probability[event_,sampleSpace_]:=Length[event]/Length[sampleSpace]

         (* Example usage: *)
         n=5;  (* Total number of objects. *)
         k=3;  (* Number of objects to be selected. *)

         size=sampleSpaceSize[n,k];  (* Calculate the number of elements in sample space.
         *)
         sampleSpace=generateSampleSpace[n,k];  (* Generate the sample space. *)

         Print["Number of Elements: ",size]; (* Output the number of elements in the sample
         space. *)
         Print["Sample Space: ",sampleSpace ]; (* Output the generated sample space. *)

         (* Example event 1: Sum of selected objects is even: *)
         event=Select[sampleSpace,EvenQ[Total[#]]&]
         prob=probability[event,sampleSpace] (* Calculate the probability of the event. *)

         (* Example event: The first selected object is even: *)
         event=Select[sampleSpace,EvenQ[First[#]]&]
         prob=probability[event,sampleSpace]  (* Calculate the probability of the event. *)

         (* Example event: The sum of the selected objects is greater than 10: *)
         event=Select[sampleSpace,Total[#]>10&]
         prob=probability[event,sampleSpace] (* Calculate the probability of the event. *)

Output   Number of Elements:  10
Output   Sample Space:
         {{1,2,3},{1,2,4},{1,2,5},{1,3,4},{1,3,5},{1,4,5},{2,3,4},{2,3,5},{2,4,5},{3,4,5}}
          {{1,2,3},{1,2,5},{1,3,4},{1,4,5},{2,3,5},{3,4,5}}
Output   3/5
Output   {{2,3,4},{2,3,5},{2,4,5}}
Output   3/10
Output   {{2,4,5},{3,4,5}}
Output   1/5
```

Input

```
(* The code focused on sampling without replacement and unordered objects. The code includes functions to calculate the size of the sample space, generate the sample space itself, and compute the probability of events. Additionally, the code demonstrates visualizations using a histogram and a list plot. The sampleSpaceSize function uses the Binomial function to calculate the number of elements in the sample space based on the total number of objects and the desired number of objects to be selected. The generateSampleSpace function utilizes the Subsets function to generate all possible combinations of objects from the given sample space. To demonstrate the sampling process, we randomly generate a specified number of samples from the sample space using the RandomChoice function. The frequencies of the samples are then calculated using the Tally function, and two visualizations are created. The histogram represents the frequency distribution of the samples, while the list plot displays the frequencies in a point-based representation: *)

(* Function to calculate the number of elements in sample space: *)
sampleSpaceSize[n_,k_]:=Binomial[n,k]

(* Function to generate sample space: *)
generateSampleSpace[n_,k_]:=Module[
  {objects,combinations},
  objects=Range[n];
  combinations=Subsets[objects,{k}];
  combinations]

(* Example usage. *)
n=10;  (* Total number of objects. *)
k=4;  (* Number of objects to be selected. *)

size=sampleSpaceSize[n,k];  (* Calculate the number of elements in sample space.
*)
sampleSpace=generateSampleSpace[n,k];  (* Generate the sample space. *)

(* Generate random samples from the sample space: *)
numSamples=1000;  (* Number of samples to generate. *)
samples=RandomChoice[sampleSpace,numSamples];

(* Calculate the frequencies of the samples: *)
frequencies=Tally[samples];

(* Create a histogram of the frequencies:*)
Histogram[
 frequencies[[All,2]],
 Automatic,
 "Probability",
 Frame->True,
 FrameLabel->{"Frequency","Probability"},
 PlotLabel->"Sample - Probability",
 ImageSize->250,
 ColorFunction->Function[Opacity[0.7]],
 ChartStyle->Purple
 ]

(* Create a list plot of the frequencies: *)
ListPlot[
 frequencies[[All,2]],
 PlotStyle->Directive[PointSize[0.009],Purple,Opacity[0.7]],
 Filling->Axis,
 Frame->True,
 FrameLabel->{"Sample","Frequency"},
```

```
PlotLabel->"Sample - Frequencies",
ImageSize->250
]
```

Output

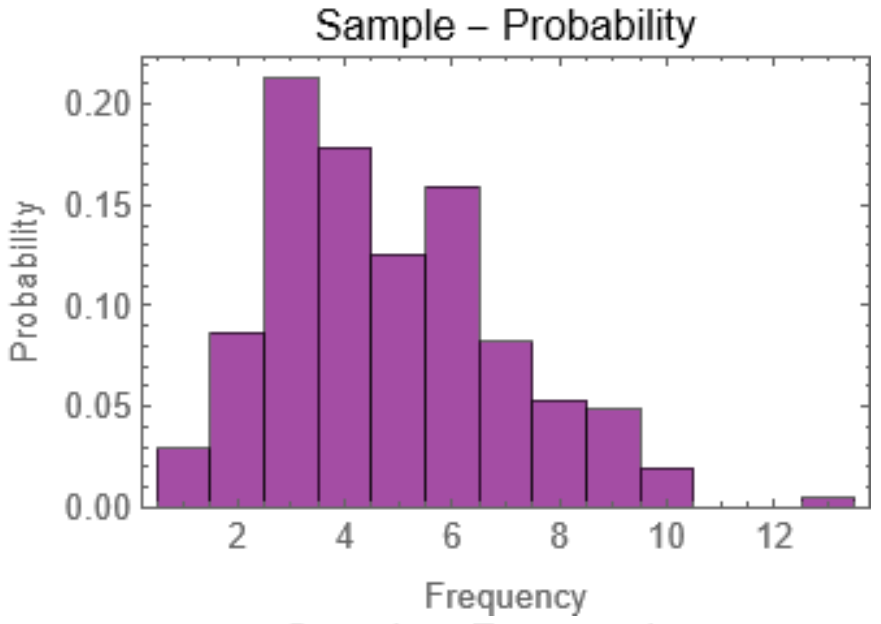

Output

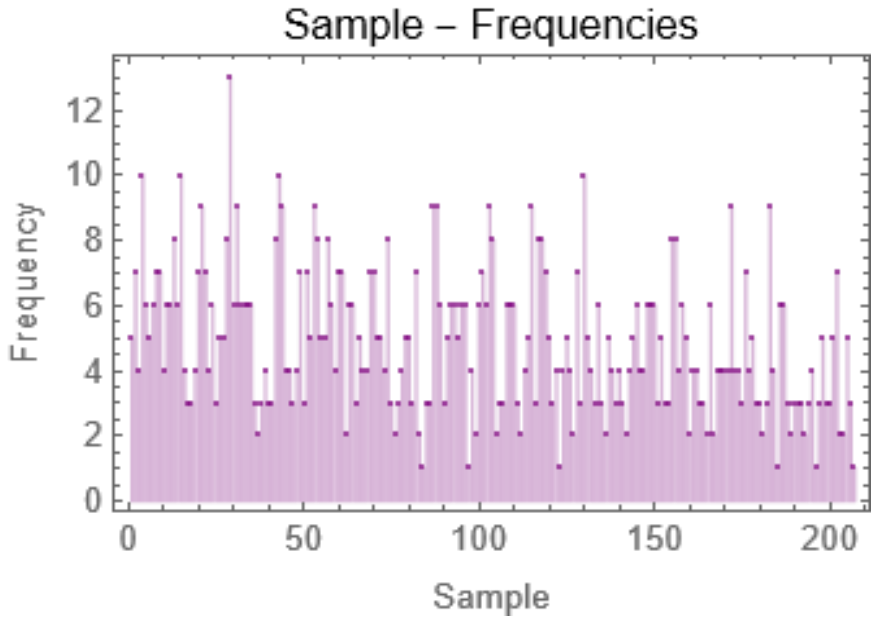

Input

```
(* The updated code includes a Manipulate function to provide an interactive experience for exploring sampling without replacement and unordered objects. The code allows you to adjust parameters such as the total number of objects, the number of objects to be selected, the number of samples to generate, and the condition for the event. Inside the Manipulate module, the code calculates the size of the sample space, generates the sample space itself, generates random samples, calculates frequencies, and determines the probability of the event based on the specified condition. The code also creates a histogram and a list plot to visualize the frequency distribution of the samples. Additionally, it displays the size of the sample space and the probability of the event. By interacting with the sliders or input fields in the Manipulate interface, you can dynamically explore different scenarios, observe the corresponding plots, and view the calculated probabilities and sample space sizes: *)

(* Function to calculate the number of elements in sample space: *)
sampleSpaceSize[n_,k_]:=Binomial[n,k]

(* Function to generate sample space: *)
generateSampleSpace[n_,k_]:=Module[
  {objects,combinations},
  objects=Range[n];
  combinations=Subsets[objects,{k}];
  combinations
  ]

(* Function to calculate the probability of an event: *)
probability[event_,sampleSpace_]:=Length[event]/Length[sampleSpace]

Manipulate[
 Module[
  {size,sampleSpace,event,prob,frequencies},

  (* Calculate the size of the sample space: *)
  size=sampleSpaceSize[n,k];

  (* Generate the sample space: *)
```

```
    sampleSpace=generateSampleSpace[n,k];

    (* Generate random samples from the sample space: *)
    samples=RandomChoice[sampleSpace,numSamples];

    (* Calculate the frequencies of the samples: *)
    frequencies=Tally[samples];

    (* Calculate the probability of the event: *)
    event=Select[samples,condition];
    prob=probability[event,sampleSpace];

    (* Output the size of the sample space, the probability of the event, Histogram
  and ListPlot: *)
    Column[
     {
      Row[
       {"Sample Space Size: ",size}
       ],
      Row[
       {"Probability of Event: ",prob}
       ],
      (* Create a histogram of the frequencies: *)
      Histogram[
       frequencies[[All,2]],
       Automatic,
       "Probability",
       Frame->True,
       FrameLabel->{"Frequency","Probability"},
       PlotLabel->"Sample - Frequencies",
       ImageSize->250,
       ColorFunction->Function[Opacity[0.7]],
       ChartStyle->Purple
       ],
      (* Create a list plot of the frequencies: *)
      ListPlot[
       frequencies[[All,2]],
       PlotStyle->Directive[PointSize[0.009],Purple,Opacity[0.7]],
       Filling->Axis,
       Frame->True,
       FrameLabel->{"Sample","Frequency"},
       PlotLabel->"Sample - Frequencies",
       ImageSize->250
       ]
      }
     ]
    ],
   (* Manipulate parameters: *)
   {{n,10,"Total Objects"},5,15,1},
   {{k,4,"Selected Objects"},2,n,1},
   {{numSamples,1000,"Number of Samples"},100,5000,100},
   {{condition,EvenQ[First[#]]&,"Event
  Condition"},{EvenQ[First[#]]&,OddQ[Total[#]]&,EvenQ[Total[#]]&}}
   ]
```

Output

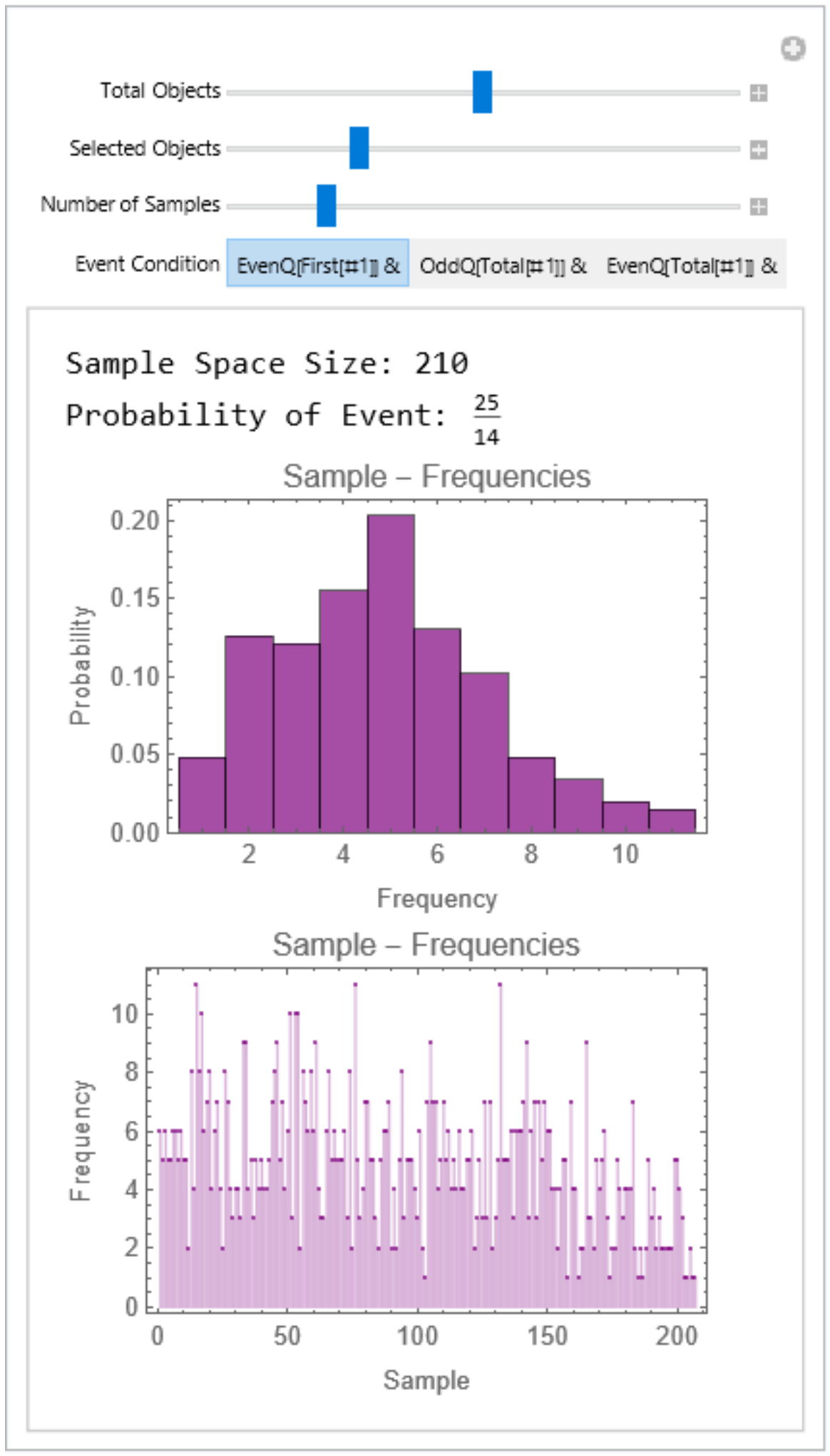

**Mathematica Examples 9.15** Sampling with replacement and the objects are not ordered

Input

```
(* Sampling with replacement and the objects are not ordered: *)
(* In this code, the sampleSpaceSize function calculates the number of elements in the sample space using the binomial coefficient formula (n+k\[Minus]1 choose k). The generateSampleSpace function generates the sample space by using the Tuples function to generate all possible combinations of elements with repetition, and then removes duplicate and equivalent combinations using DeleteDuplicates and sorting: *)

sampleSpaceSize[n_,k_]:=Binomial[n+k-1,k]

generateSampleSpace[n_,k_]:=Module[
   {elements,combinations},
   elements=Range[n];
   combinations=Tuples[elements,k];
   DeleteDuplicates[Sort/@combinations]
   ]

n=3; (* Number of distinct objects. *)
k=2; (* Number of selections. *)

size=sampleSpaceSize[n,k];
sampleSpace=generateSampleSpace[n,k];
```

```
        Print["Number of elements in the sample space: ",size];
        Print["Sample space: ",sampleSpace];

Output  Number of elements in the sample space:  6
Output  Sample space:  {{1,1},{1,2},{1,3},{2,2},{2,3},{3,3}}

Input   (* In this code, the sampleSpaceSize function calculates the number of elements in
        the sample space using the binomial coefficient formula (n+k\[Minus]1 choose k). The
        generateSampleSpace function generates the sample space by using the Tuples function
        to generate all possible combinations of elements with repetition, and then removes
        duplicate and equivalent combinations using DeleteDuplicates and sorting. The
        calculateProbability function calculates the probability of an event by dividing the
        count of distinct elements in the event by the total number of elements in the sample
        space: *)

        sampleSpaceSize[n_,k_]:=Binomial[n+k-1,k]

        generateSampleSpace[n_,k_]:=Module[{elements,combinations},elements=Range[n];
          combinations=Tuples[elements,k];
          DeleteDuplicates[Sort/@combinations]]

        calculateProbability[event_,sampleSpace_]:=Length[event]/Length[sampleSpace]

        n=3; (* Number of distinct objects. *)
        k=2; (* Number of selections. *)

        size=sampleSpaceSize[n,k];
        sampleSpace=generateSampleSpace[n,k];

        Print["Number of elements in the sample space: ",size];
        Print["Sample space: ",sampleSpace];

        condition1=#[[1]]<#[[2]]&; (* Example condition 1: The first element is less than
        the second element. *)
        event1=Select[sampleSpace,condition1];
        probability1=calculateProbability[event1,sampleSpace];
        Print["Event 1 is : ",event1];
        Print["Probability of event 1: ",probability1];


        condition2=#[[1]]==#[[2]]&; (* Example condition 2: The first element is equal to
        the second element. *)
        event2=Select[sampleSpace,condition2];
        probability2=calculateProbability[event2,sampleSpace];
        Print["Event 2 is : ",event2];
        Print["Probability OF event 2: ",probability2];

        condition3=#[[1]]+#[[2]]==6&; (* Example condition 3: The sum of the first two
        elements is equal to 6. *)
        event3=Select[sampleSpace,condition3]; (* Example conditional event. *)
        probability3=calculateProbability[event3,sampleSpace];
        Print["Event 3 is : ",event3];
        Print["Probability OF event 2: ",probability3];

Output   Number of elements in the sample space:  6
         Sample space:  {{1,1},{1,2},{1,3},{2,2},{2,3},{3,3}}

         Event 1 is :  {{1,2},{1,3},{2,3}}
         Probability of event 1:  1/2
```

```
Event 2 is :  {{1,1},{2,2},{3,3}}
Probability OF event 2:  1/2

Event 3 is :  {{3,3}}
Probability OF event 2:  1/6
```

Input

```
(* The code calculates the number of elements in the sample space and generates a random sample using the "Sampling with replacement and the objects are not ordered" approach. It then calculates the frequency distribution of the random sample using Tally and visualizes the distribution using a histogram and ListPlot. The sampleSpaceSize function calculates the number of elements in the sample space using the binomial coefficient formula (n+k\[Minus]1 choose k). The generateSampleSpace function generates the sample space by using Tuples to create all possible combinations of elements with repetition. Duplicate and equivalent combinations are removed using DeleteDuplicates and sorting. The values of n and k represent the number of distinct objects and the number of selections, respectively. A random sample is generated from the sample space using RandomChoice. The frequency of each element in the random sample is calculated using Tally. The frequency distribution is visualized using a histogram, where the x-axis represents the elements and the y-axis represents the frequency. The frequency distribution is also visualized using a ListPlot,where the x-axis represents the elements and the y-axis represents the frequency: *)

sampleSpaceSize[n_,k_]:=Binomial[n+k-1,k]

generateSampleSpace[n_,k_]:=Module[
   {elements,combinations},elements=Range[n];
  combinations=Tuples[elements,k];
  DeleteDuplicates[Sort/@combinations]
  ]

n=10; (* Number of distinct objects. *)
k=3; (* Number of selections. *)

size=sampleSpaceSize[n,k];
sampleSpace=generateSampleSpace[n,k];

Print["Number of elements in the sample space: ",size];

(* Random sampling from the sample space: *)
randomSample=RandomChoice[sampleSpace,1000];

(* Calculate frequency using Tally:*)
frequency=Tally[randomSample];

(* Histogram of the frequency: *)
Histogram[
 frequency[[All,2]],
 Automatic,
 "Count",
 Frame->True,
 FrameLabel->{"Frequency","Frequency"},
 PlotLabel->"Sample - Frequency",
 ImageSize->250,
 ColorFunction->Function[Opacity[0.7]],
 ChartStyle->Purple
 ]

(* ListPlot of the frequency:*)
```

```
ListPlot[
 frequency[[All,2]],
 PlotStyle->Directive[PointSize[0.009],Purple,Opacity[0.7]],
 Filling->Axis,
 Frame->True,
 FrameLabel->{"Sample","Frequency"},
 PlotLabel->"Sample - Frequencies",
 ImageSize->250
 ]
```

Output Number of elements in the sample space: 220

Output

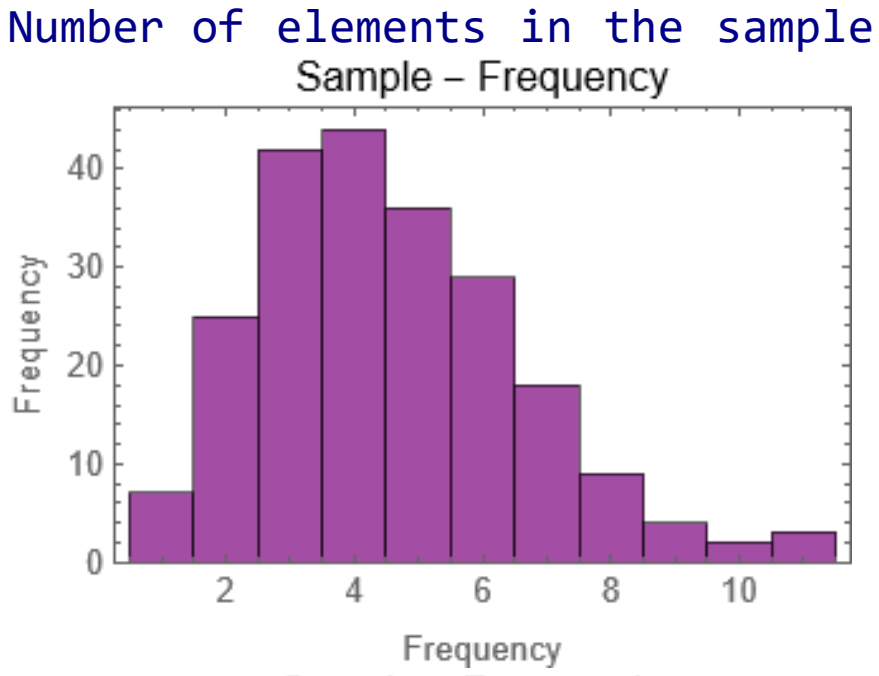

Output

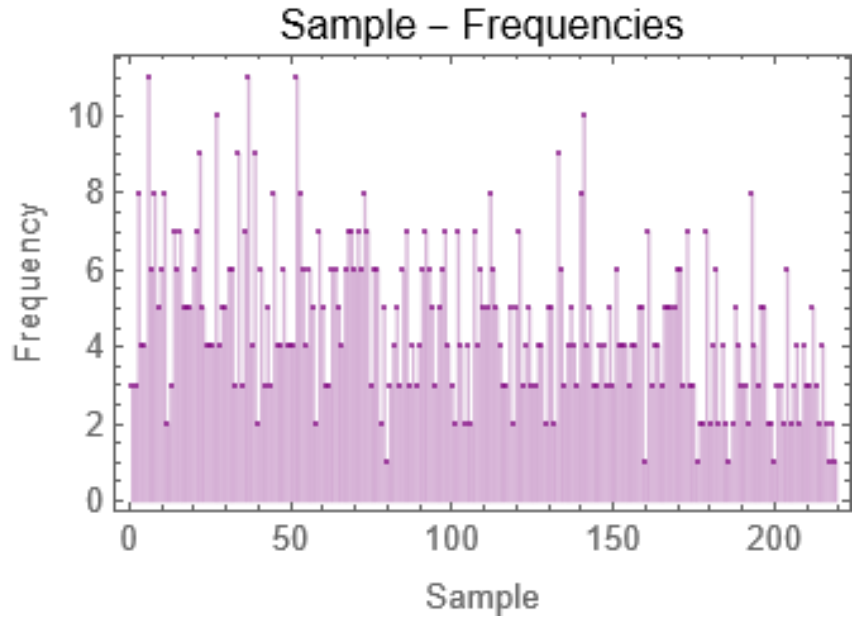

Input

```
(* In this code, the Manipulate function is used to create a dynamic interface. The
values of n and k are initialized to 3 and 2, respectively, and can be adjusted using
sliders. Whenever the values of n and k are changed, the sample space is regenerated,
a new random sample is generated, and the frequency distribution is recalculated: *)

sampleSpaceSize[n_,k_]:=Binomial[n+k-1,k]

generateSampleSpace[n_,k_]:=Module[
   {elements,combinations},
   elements=Range[n];
   combinations=Tuples[elements,k];
   DeleteDuplicates[Sort/@combinations]
   ]

Manipulate[size=sampleSpaceSize[n,k];
 sampleSpace=generateSampleSpace[n,k];
 randomSample=RandomChoice[sampleSpace,1000];
 frequency=Tally[randomSample];
 Column[
   {
    Histogram[
     frequency[[All,2]],
     Automatic,
     "Count",
     Frame->True,
     FrameLabel->{"Frequency","Frequency"},
     PlotLabel->"Sample - Frequency",
     ImageSize->250,
```

```
            ColorFunction->Function[Opacity[0.7]],
            ChartStyle->Purple
            ],
           ListPlot[
            frequency[[All,2]],
            PlotStyle->Directive[PointSize[0.009],Purple,Opacity[0.7]],
            Filling->Axis,
            Frame->True,
            FrameLabel->{"Sample","Frequency"},
            PlotLabel->"Sample - Frequencies",
            ImageSize->250
            ],
           Row[{"Number of elements in the sample space: ",size}],
           Row[{"Sample space: ",sampleSpace}]
           }
          ],
         {{n,3},1,10,1},
         {{k,2},1,10,1}
         ]
```

Output

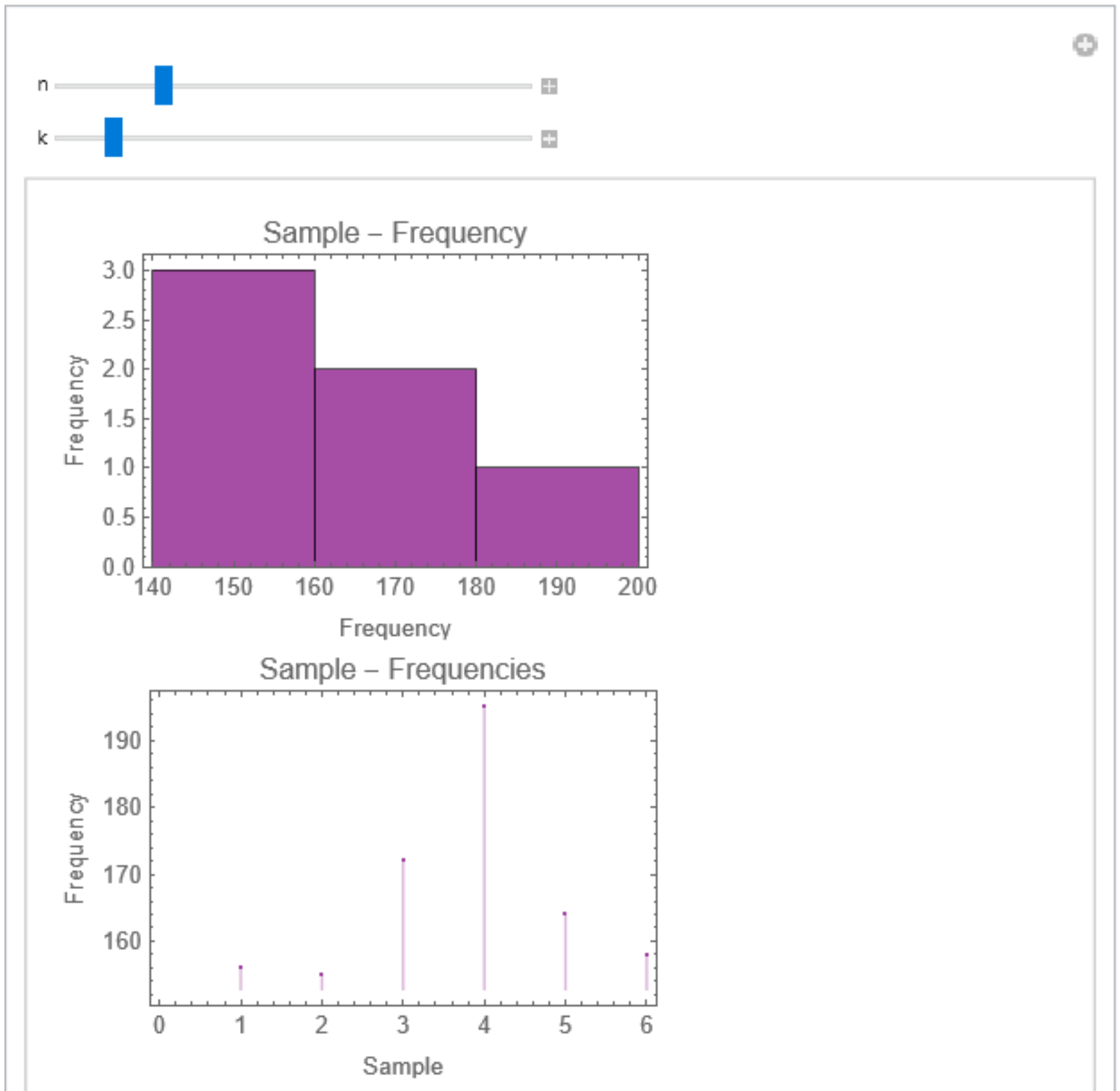

Number of elements in the sample space: 6

Sample space: {{1, 1}, {1, 2}, {1, 3}, {2, 2}, {2, 3}, {3, 3}}

***Mathematica Examples 9.16*** Conditional Probability

Input

```
(* Conditional Probability: *)
(* This code begins by defining the sample space, which represents the possible
outcomes. In this example, it is a fair six-sided die, so the sample space is defined
as the list {1,2,3,4,5,6}. Next, the events A and B are defined as lists that satisfy
```

```
        certain conditions. The probabilities of events A and B are calculated by dividing
        the length of each event by the length of the sample space. The intersection of
        events A and B is then computed using the Intersection function. Finally, the
        conditional probability P(A|B) is calculated by dividing the length of the
        intersection of events A and B by the length of event B: *)

        (*Define the sample space and events: *)
        sampleSpace={1,2,3,4,5,6}; (* Sample space for a fair six-sided die. *)
        eventA={2,4,6};
        eventB={3,4,5};

        (* Calculate the probability of event A: *)
        probA=Length[eventA]/Length[sampleSpace];

        (* Calculate the probability of event B: *)
        probB=Length[eventB]/Length[sampleSpace];

        (* Calculate the intersection of events A and B: *)
        intersectionAB=Intersection[eventA,eventB];

        (* Calculate the conditional probability P(A|B): *)
        conditionalProb=Length[intersectionAB]/Length[eventB];

        (* Output the results: *)
        Print["The probability of event A is ",N[probA]]
        Print["The probability of event B is ",N[probB]]
        Print["The conditional probability P(A|B) is ",N[conditionalProb]]

Output   The probability of event A is  0.5
         The probability of event B is  0.5
         The conditional probability P(A|B) is  0.333333

Input   (* In this code, we defined a new function called conditionalProbability that
        encapsulates the calculations for conditional probability within a Module block. The
        function takes three arguments: sampleSpace, eventA, and eventB. Inside the Module
        block, the probabilities of events A and B, the intersection of events A and B, and
        the conditional probability are computed, just as in the previous code. Finally,
        outside the Module block, we call the conditionalProbability function with the
        appropriate arguments: *)

        conditionalProbability[sampleSpace_,eventA_,eventB_]:=Module[
          {probA,probB,intersectionAB,conditionalProb},

          (* Calculate the probability of event A: *)
          probA=Length[eventA]/Length[sampleSpace];

          (* Calculate the probability of event B: *)
          probB=Length[eventB]/Length[sampleSpace];

          (* Calculate the intersection of events A and B: *)
          intersectionAB=Intersection[eventA,eventB];

          (* Calculate the conditional probability P(A|B): *)
          conditionalProb=Length[intersectionAB]/Length[eventB];

          (* Output the results*)
          Print["The probability of event A is ",N[probA]];
          Print["The probability of event B is ",N[probB]];
          Print["The conditional probability P(A|B) is ",N[conditionalProb]];]
```

```
(*Define the sample space and events: *)
sampleSpace={1,2,3,4,5,6}; (* Sample space for a fair six-sided die. *)
eventA={2,4,6};
eventB={3,4,5};

(* Call the conditionalProbability function: *)
conditionalProbability[sampleSpace,eventA,eventB];
```

Output

```
The probability of event A is  0.5
The probability of event B is  0.5
The conditional probability P(A|B) is  0.333333
```

Input

```
(* In this updated code, the sample space is defined as all possible outcomes of
rolling two fair six-sided dice using the Tuples function. The conditions for event
A and event B are also updated. In this example, event A is defined as the sum of
the dice being equal to 7 (Total[#]==7&),and event B is defined as at least one of
the dice shows a 3 (MemberQ[#,3]). By calling the conditionalProbability function
with the updated sample space and conditions, the code will simulate rolling two
dice, calculate the probabilities, and print the results including the conditional
probability: *)

conditionalProbability[sampleSpace_,conditionA_,conditionB_]:=Module[
  {probA,probB,intersectionAB,conditionalProb,eventA,eventB},

  (* Define event A based on the condition: *)
  eventA=Select[sampleSpace,conditionA];

  (* Define event B based on the condition: *)
  eventB=Select[sampleSpace,conditionB];

  (* Calculate the probability of event A: *)
  probA=Length[eventA]/Length[sampleSpace];

  (* Calculate the probability of event B: *)
  probB=Length[eventB]/Length[sampleSpace];

  (* Calculate the intersection of events A and B: *)
  intersectionAB=Intersection[eventA,eventB];

  (* Calculate the conditional probability P(A|B): *)
  conditionalProb=Length[intersectionAB]/Length[eventB];

  (* Output the results*)
  Print["The probability of event A is ",N[probA]];
  Print["The probability of event B is ",N[probB]];
  Print["The conditional probability P(A|B) is ",N[conditionalProb]];
  ]

(* Define the sample space: *)
sampleSpace=Tuples[Range[6],2] (* Sample space for rolling two dice. *)

(* Define the conditions for event A and event B:*)
conditionA=Total[#]==7& ;(* Event A:The sum of the dice is 7. *)
conditionB=MemberQ[#,3]&; (* Event B:At least one of the dice shows a 3. *)

(* Call the conditionalProbability function: *)
conditionalProbability[sampleSpace,conditionA,conditionB];
```

```
Output   {{1,1},{1,2},{1,3},{1,4},{1,5},{1,6},{2,1},{2,2},{2,3},{2,4},{2,5},{2,6},{3,1},
         {3,2},{3,3},{3,4},{3,5},{3,6},{4,1},{4,2},{4,3},{4,4},{4,5},{4,6},{5,1},{5,2},
         {5,3},{5,4},{5,5},{5,6},{6,1},{6,2},{6,3},{6,4},{6,5},{6,6}}
Output    The probability of event A is  0.166667
          The probability of event B is  0.305556
          The conditional probability P(A|B) is  0.181818

Input    (* In this updated code, the sample space is defined as all possible outcomes of
         flipping three coins using the Tuples function with the elements {0,1} representing
         tails and heads, respectively. The conditions for event A and event B are defined as
         follows: Event A: All three coins show heads. This is represented by
         Count[#,1]==3&,which checks if the number of heads (1s) in a given outcome is equal
         to 3.
         Event B: At least one coin shows heads. This is represented by MemberQ[#,1]&,which
         checks if the value 1 (heads) is present in the outcome. By calling the
         conditionalProbability function with the updated sample space and conditions, the
         code will simulate flipping three coins, calculate the probabilities, and print the
         results including the conditional probability based on the new conditions for event
         A and event B: *)

         conditionalProbability[sampleSpace_,conditionA_,conditionB_]:=Module[
           {probA,probB,intersectionAB,conditionalProb,eventA,eventB},

           (* Define event A based on the condition: *)
           eventA=Select[sampleSpace,conditionA];

           (* Define event B based on the condition: *)
           eventB=Select[sampleSpace,conditionB];

           (* Calculate the probability of event A: *)
           probA=Length[eventA]/Length[sampleSpace];

           (* Calculate the probability of event B: *)
           probB=Length[eventB]/Length[sampleSpace];

           (* Calculate the intersection of events A and B: *)
           intersectionAB=Intersection[eventA,eventB];

           (* Calculate the conditional probability P(A|B): *)
           conditionalProb=Length[intersectionAB]/Length[eventB];

           (* Output the results*)
           Print["The probability of event A is ",N[probA]];
           Print["The probability of event B is ",N[probB]];
           Print["The conditional probability P(A|B) is ",N[conditionalProb]];
           ]

         (* Define the sample space:*)
         sampleSpace=Tuples[{0,1},3] (* Sample space for flipping three coins. *)

         (* Define the conditions for event A and event B: *)
         conditionA=Count[#,1]==3&; (* Event A:All three coins show heads.*)
         conditionB=MemberQ[#,1]&; (* Event B:At least one coin shows heads. *)

         (* Call the conditionalProbability function: *)
         conditionalProbability[sampleSpace,conditionA,conditionB];

Output   {{0,0,0},{0,0,1},{0,1,0},{0,1,1},{1,0,0},{1,0,1},{1,1,0},{1,1,1}}
Output    The probability of event A is  0.125
          The probability of event B is  0.875
```

```
The conditional probability P(A|B) is  0.142857
```

***Mathematica Examples 9.17*** Independence

Input

```
(* Independence: *)
(* In this code, we define two events, eventA and eventB, as lists of outcomes. We then calculate the probabilities of each event by dividing the length of the event by the total number of possible outcomes (assuming a six-sided die). Next, we calculate the joint probability of events A and B by finding the length of their intersection and dividing it by the total number of possible outcomes. Finally, we check for independence by comparing the joint probability with the product of the individual probabilities. If they are equal, the events are considered independent: *)

(* Define the events: *)
eventA={1,2,3}; (* Event A. *)
eventB={2,4,6}; (* Event B. *)

(* Calculate the probabilities: *)
probA=Length[eventA]/6; (* Probability of event A. *)
probB=Length[eventB]/6; (* Probability of event B. *)

(* Calculate the joint probability: *)
jointProb=Length[Intersection[eventA,eventB]]/6; (* Joint probability of events A and B. *)

(* Check for independence: *)
independence=jointProb==probA*probB;

(*Print the results*)
Print["Probability of event A: ",probA];
Print["Probability of event B: ",probB];
Print["Joint probability of events A and B: ",jointProb];
Print["Are events A and B independent? ",independence];
```

Output

```
Probability of event A:  1/2
Probability of event B:  1/2
Joint probability of events A and B:  1/6
Are events A and B independent?  False
```

Input

```
(* In this code, the calculateIndependence function takes three arguments: eventA, eventB, and sampleSpace. It calculates the probabilities of event A, event B, and the intersection of events A and B based on the provided sample space. It then compares the calculated probability of the intersection with the product of the probabilities of the individual events to check for independence. The sample space is defined using the Tuples function with the elements {0,1} representing possible outcomes of the two events. The events A and B are defined as functions using the & notation. In this example, event A is defined as the first element being 1, and event B is defined as the second element being 1. You can modify these conditions according to your specific scenario. By calling the calculateIndependence function with the defined events and sample space, the code will check if the events are independent and print the corresponding message: *)

calculateIndependence[eventA_,eventB_,sampleSpace_]:=Module[
  {probA,probB,probAB},

  (* Calculate the probability of event A: *)
  probA=Length[eventA]/Length[sampleSpace];

  (* Calculate the probability of event B: *)
```

```
    probB=Length[eventB]/Length[sampleSpace];

    (* Calculate the probability of the intersection of events A and B: *)
    probAB=Length[eventA&&eventB]/Length[sampleSpace];

    (* Check if the events are independent: *)
    If[Abs[probAB-probA*probB]<10^(-6),
     Print["The events are independent:","probAB= ",probAB,", " ,"probA= ",probA,",
" ,"probB= ",probB],
     Print["The events are not independent:","probAB= ",probAB,", " ,"probA=
",probA,", " ,"probB= ",probB]
     ];
    ]

(* Define the sample space: *)
sampleSpace=Tuples[{0,1},2] (* Sample space for two events. *)

(* Define the events A and B: *)
eventA=#[[1]]==1&; (* Event A:The first element is 1. *)
eventB=#[[2]]==1&; (* Event B:The second element is 1. *)

(* Call the calculateIndependence function: *)
calculateIndependence[eventA,eventB,sampleSpace];

Output  {{0,0},{0,1},{1,0},{1,1}}
Output  The events are not independent: probAB=  1/2 ,  probA=  1/4 ,  probB=  1/4
```

# CHAPTER 10

# DISCRETE RANDOM VARIABLES AND DISTRIBUTIONS

In this chapter, we delve into the world of discrete random variables (RVs)and their associated probability distributions. Discrete RVs play a crucial role in probability theory and statistics, providing a framework to understand and analyze phenomena that can be counted or measured in discrete units. From simple coin flips to the occurrence of rare events, discrete RVs enable us to model and make predictions about a wide range of real-world scenarios.

- We begin this chapter by introducing the concept of discrete RVs and their probability mass functions (PMFs). A discrete RV represents the outcomes of an experiment or event that can take on a finite or countably infinite set of values. The PMF assigns probabilities to each possible value that the RV can assume, forming the building blocks of its probability distribution.
- Next, we will discuss the cumulative distribution function (CDF), which provides a comprehensive view of the probability distribution. The CDF characterizes the probability that a discrete RV takes on a value less than or equal to a specified threshold. By understanding the CDF, we gain valuable insights into the overall behavior and properties of the RV.
- We then shift our focus to moment-generating functions (MGFs), which offer a powerful tool for studying the properties of discrete RVs. MGFs encode the moments of a RV, including its mean, variance, and higher-order moments, providing a concise representation of its distribution. Through MGFs, we can derive important statistical measures.
- The subsequent sections of this chapter will discuss specific discrete probability distributions that are commonly encountered in various fields. We will explore the following distributions:
    - Bernoulli distribution: A simple yet fundamental distribution that models a binary outcome (success or failure) with a single parameter representing the probability of success.
    - Binomial distribution: A distribution that describes the number of successes in a fixed number of independent Bernoulli trials. It is characterized by two parameters: the number of trials and the probability of success.
    - Geometric distribution: A distribution that models the number of trials required to achieve the first success in a sequence of independent Bernoulli trials with a constant probability of success.
    - Negative binomial distribution: A distribution that describes the number of trials required to achieve a fixed number of successes in a sequence of independent Bernoulli trials with a constant probability of success.
    - Poisson distribution: A distribution that models the number of events occurring within a fixed interval of time or space. It is commonly used to represent rare events with a known average rate.
    - Hypergeometric distribution: A distribution that models the probability of obtaining a specified number of successes in a fixed number of draws from a finite population without replacement.
    - Discrete uniform distribution: A distribution where all possible outcomes have an equal probability of occurring. It is often used when the outcome of an experiment is equally likely to be any value within a given range.

Throughout this chapter, we present theoretical concepts, provide mathematical derivations, and illustrate the practical relevance of each distribution with real-world examples. By mastering these key discrete RVs and distributions, readers will develop a solid foundation for probabilistic reasoning and statistical analysis, empowering them to solve complex problems and make informed decisions in a wide range of fields.

## 10.1 Discrete RVs

### RV

The concept of a RV allows us to pass from the experimental outcomes to a numerical function of the outcomes, often simplifying the sample space. In simpler terms, a RV is like a function that maps the outcomes of a random event to numerical values. A RV is denoted by an uppercase letter such as $X$. After an experiment is conducted, the measured value of the RV is denoted by a lowercase letter such as $x$. Just like how we assign probabilities to events, we can do the same to RVs. Consider the following examples.

***Example 10.1***

In tossing dice, we are often interested in the sum of the two dice and are not really concerned about the values of the individual dice. That is, we may be interested in knowing that the sum is 7 and not be concerned over whether the actual outcome was (1,6) or (2,5) or (3,4) or (4,3) or (5,2) or (6,1). Letting $X$ denote the RV that is defined as the sum of two fair dice, then:

$$P(X=2)=P((1,1))=1/36,$$
$$P(X=3)=P((1,2),(2,1))=2/36,$$
$$P(X=4)=P((1,3),(2,2),(3,1))=3/36,$$
$$P(X=5)=P((1,4),(2,3),(3,2),(4,1))=4/36,$$
$$P(X=6)=P((1,5),(2,4),(3,3),(4,2),(5,1))=5/36,$$
$$P(X=7)=P((1,6),(2,5),(3,4),(4,3),(5,2),(6,1))=6/36,$$
$$P(X=8)=P((2,6),(3,5),(4,4),(5,3),(6,2))=5/36,$$
$$P(X=9)=P((3,6),(4,5),(5,4),(6,3))=4/36,$$
$$P(X=10)=P((4,6),(5,5),(6,4))=3/36,$$
$$P(X=11)=P((5,6),(6,5))=2/36,$$
$$P(X=12)=P((6,6))=1/36.$$

In other words, the RV $X$ can take on any integral value between 2 and 12. Moreover, we must have

$$1=P(S)=P\left(\bigcup_{x=2}^{12}(X=x)\right)=\sum_{x=2}^{12}P(X=x).$$

***Example 10.2***

For our dice-rolling experiment, let us define the events of success ($S$) and failure ($F$) as follows: success - the outcome of rolling a 3, and failure - the outcome of not rolling a 3. Therefore, the sample space $G$ is:

$$G=\{FF,SF,FS,SS\}.$$

Here, $SF$, for instance, represents a success followed by a failure. We defined the RV $X$ as the number of times we roll a 3, which is equivalent to the number of times we observe a success $S$. The RV $X$ maps each sample point to a specific value $x$ like so:

$$X(FF)=0, X(SF)=1, X(FS)=1 \text{ and } X(FF)=2.$$

In other words, the RV $X$ can take on any integral value between 0 and 2.

$$P(X=0)=P((1,1),(1,2),(2,1),(2,2),(1,4),$$
$$(4,1),(1,5),(2,4),(4,2),(5,1),$$
$$(1,6),(2,5),(5,2),(6,1),(2,6),$$
$$(4,4),(6,2),(4,5),(5,4),(4,6),$$
$$(5,5),(6,4),(5,6),(6,5),(6,6))=25/36,$$

$$P(X=1)=P((3,1),(3,2),(3,4),(3,5),(3,6),$$
$$(1,3),(2,3),(4,3),(5,3),(6,3))=10/36,$$
$$P(X=2)=P((3,3))=1/36.$$

Moreover, we must have $P(S)=\sum_{x=0}^{2}P(X=x)=1$.

***Example 10.3***

Another RV of possible interest in the dice-rolling experiment is the value of the first die. Letting $Y$ denote this RV, then $Y$ is equally likely to take on any of the values 1 through 6. That is,

$$P(Y = y) = 1/6, \qquad y = 1,2,3,4,5,6.$$

***Example 10.4***

Suppose that an individual purchases two electronic components, each of which may be either defective or acceptable. In addition, suppose that the four possible results — $(d, d)$, $(d, a)$, $(a, d)$, $(a, a)$ — have respective probabilities 0.09, 0.21, 0.21, 0.49 [where $(d, d)$ means that both components are defective, $(d, a)$ that the first component is defective and the second acceptable, and so on]. If we let $X$ denote the number of acceptable components obtained in the purchase, then $X$ is a RV taking on one of the values 0, 1, 2 with respective probabilities,

$$P(X = 0) = 0.09,$$
$$P(X = 1) = 0.42,$$
$$P(X = 2) = 0.49.$$

***Example 10.5***

Suppose we draw two balls from a bag containing many red and green balls. We are interested in the number of green balls we draw. How should we define the RV in this case?

***Solution***

We should define RV $X$ as the number of green balls we draw. The sample space $G$ in this case is:

$$G = \{RR, RG, GR, GG\}.$$

Here, $R$ represents the event of drawing a red ball, and $G$ represents the event of drawing a green ball. The RV $X$ maps each of the sample points to a real value $x$, which is the number of green balls we draw for each sample point:

$$X(RR) = 0, X(RG) = 1, X(GR) = 1 \text{ and } X(GG) = 2.$$

***Example 10.6***

Suppose we draw two balls with-replacement from a bag containing 2 red and 3 green balls. What is the probability of drawing:

(a) no red balls? (b) one red ball? (c) two red balls?

***Solution***

Let us define RV $X$ as the number of red balls we draw. The probabilities of interest are:

$P(X = 0)$ - the probability of drawing no red balls.
$P(X = 1)$ - the probability of drawing one red ball.
$P(X = 2)$ - the probability of drawing two red balls.

The RV $X$ maps each of the sample points to a real value $x$, which is the number of red balls we draw for each sample point:

$$X(RR) = 2, X(RG) = 1, X(GR) = 1 \text{ and } X(GG) = 0.$$

We can compute these probabilities by referring to the sample space:

$$P(X = 0) = P(GG) = \left(\frac{3}{5}\right)\left(\frac{3}{5}\right) = \frac{9}{25},$$

$$P(X = 2) = P(RR) = \left(\frac{2}{5}\right)\left(\frac{2}{5}\right) = \frac{4}{25},$$

$$P(X = 1) = P(RG) + P(GR) = \left(\frac{2}{5}\right)\left(\frac{3}{5}\right) + \left(\frac{3}{5}\right)\left(\frac{2}{5}\right) = \frac{12}{25}.$$

Notice the following:

$$P(X = 0) + P(X = 1) + P(X = 2) = P(GG) + P(RR) + P(RG) + P(GR) = 1.$$

**Definition (RV):** A RV is a function that assigns a real number to each outcome in the sample space, $S$, of a random experiment,

$$X(\omega) = x, \tag{10.1}$$

with each outcome $\omega$ in $S$. (See Figure 10.1)

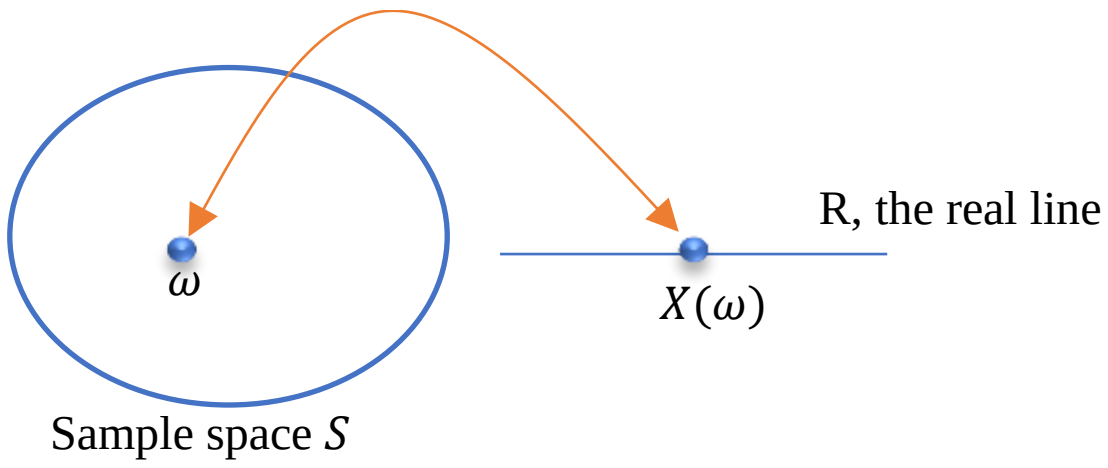

**Figure 10.1.** RV as a function.

In the above examples, the RVs of interest took on a finite number of possible values. RVs whose set of possible values can be written either as a finite sequence $x_1, \ldots, x_n$, or as an infinite sequence $x_1, \ldots$ are said to be discrete.

**Definition (Discrete RVs):** A discrete RV is a RV with a countable finite (or countably infinite) range. Or a discrete RV is a RV whose possible values can be listed.

**Definition (Continuous RVs):** A continuous RV is a RV with an interval (either finite or infinite) of real numbers for its range.

Here are a few examples of discrete RVs:

1. The number of cars sold at a dealership during a given month.
2. The number of houses in a certain block.
3. The number of fish caught on a fishing trip.
4. The number of complaints received at the office of an airline on a given day.
5. The number of customers who visit a bank during any given hour.
6. The number of heads obtained in three tosses of a coin.
7. The number of emails received in a day.
8. The number of defective items in a batch of products.
9. The number of children in a family.
10. The number of successful attempts in a series of trials.
11. The number of books borrowed by a student from the library.
12. The number of accidents that occur in a given time period, such as the number of car accidents in a month.
13. The number of phone calls received by a call center in a day.
14. The number of students present in a classroom on a given day.
15. The number of flights delayed at an airport in a given hour or day.
16. The number of rainy days in a particular month.

## 10.2 Probability Distributions and PMFs

RVs are so important in random experiments that sometimes we essentially ignore the original sample space of the experiment and focus on the probability distribution of the RV. The probability distribution of a RV is a description of the probabilities associated with the possible values of the RV. For a discrete RV, the distribution is often specified by just a list of the possible values along with the probability of each. For other cases, probabilities are expressed in terms of a formula.

**Definition (PMF):** For a discrete RV $X$ with possible values $x_1, x_2, \dots, x_n$, a PMF is a function such that:
(1) $f(x_i) \geq 0$.
(2) $\sum_{i=1}^{n} f(x_i) = 1$.
(3) $f(x_i) = P(X = x_i)$.
PMFs are also commonly referred to as discrete probability distribution.

In general, for any discrete RV with possible values $x_1$, $x_2$, ..., the events $(X = x_1)$, $(X = x_2)$, ... are mutually exclusive. Therefore, $P(X \leq x) = \sum_{x_i \leq x} f(x_i)$. This leads to,

**Definition (CDF):** The CDF of a discrete RV $X = x$, denoted as $F(x)$, is:

$$F(x) = P(X \leq x) = \sum_{x_i \leq x} f(x_i). \tag{10.2}$$

For a discrete RV $X = x$, $F(x)$ satisfies the following properties.
(1) $F(x) = P(X \leq x) = \sum_{x_i \leq x} f(x_i)$.
(2) $0 \leq F(x) \leq 1$.
(3) If $x \leq y$, then $F(x) \leq F(y)$.
That is, $F(x)$ is the probability that the RV $X$ takes on a value that is less than or equal to $x$.

**Remarks:**

- Note that this is analogous to a relative-frequency distribution with probabilities replacing the relative frequencies. Thus, we can think of probability distributions as theoretical or ideal limiting forms of relative-frequency distributions when the number of observations made is very large.
- Notation: We will use the notation $X \sim F$ to signify that $F$ is the CDF of $X$.
- All probability questions about $X$ can be answered in terms of its CDF $F$. For example, suppose we wanted to compute $P(a < X \leq b)$. This can be accomplished by first noting that the event $(X \leq b)$ can be expressed as the union of the two mutually exclusive events $(X \leq a)$ and $(a < X \leq b)$. Therefore, we obtain that

$$P(X \leq b) = P(X \leq a) + P(a < X \leq b), \tag{10.3}$$

or

$$\begin{aligned} P(a < X \leq b) &= P(X \leq b) - P(X \leq a) \\ &= F(b) - F(a). \end{aligned} \tag{10.4}$$

***Example 10.7***

Suppose the RV $X$ has CDF

$$F(x) = \begin{cases} 0 & x \leq 0 \\ 1 - exp\{-x^2\} & x > 0 \end{cases}.$$

What is the probability that $X$ exceeds 1?

***Solution***

The desired probability is computed as follows:

$$\begin{aligned} P(X > 1) &= 1 - P(X \leq 1) \\ &= 1 - F(1) \\ &= e^{-1} \\ &= 0.368. \end{aligned}$$

It is often helpful to look at a probability distribution in graphic form. One might plot the points $(x, f(x))$. By joining the points to the $x$ axis either with a dashed or with a solid line, we obtain a PMF plot. The graph of the probability distribution makes it easy to see what values of $X$ are most likely to occur, and it also indicates a symmetry of the probability distribution. Instead of plotting the points $(x, f(x))$, we more frequently construct a probability histogram. To understand that well, let us consider the following examples.

***Example 10.8***

When a balanced coin is tossed three times, eight equally likely outcomes are possible. For instance, HHT means that the first two tosses are heads and the third is tails. Let $X$ denote the total number of heads obtained in the three tosses. Then $X$ is a discrete RV whose possible values are 0, 1, 2, and 3, as shown in Table 10.1.
a. Use random-variable notation to represent the event that exactly two heads are tossed.
b. Determine $P(X = 2)$.
c. Find the probability distribution of $X$.
d. Use random-variable notation to represent the event that at most two heads are tossed.
e. Find $P(X \leq 2)$.

***Solution***

Possible outcomes

HHH, HTH, THH, TTH, HHT, HTT, THT, TTT.

a. The event that exactly two heads are tossed can be represented as $(X = 2)$.
b. There are three ways to get exactly two heads and that there are eight possible (equally likely) outcomes altogether. So,

$$P(X = 2) = 3/8 = 0.375.$$

c. The remaining probabilities for $X$ are computed as in part (b) and are shown in Table 10.1.

**Table 10.1**

| $X$ | 0 | 1 | 2 | 3 |
|---|---|---|---|---|
| $P(X)$ | 0.125 | 0.375 | 0.375 | 0.125 |

d. The event that at most two heads are tossed can be represented as $(X \leq 2)$, read as "$X$ is less than or equal to two."
e. The event that at most two heads are tossed can be expressed as

$$(X \leq 2) = ((X = 0) \text{ or } (X = 1) \text{ or } (X = 2)).$$

Because the three events on the right are mutually exclusive, we have,

$$\begin{aligned} P(X \leq 2) &= P(X = 0) + P(X = 1) + P(X = 2) \\ &= 0.125 + 0.375 + 0.375 \\ &= 0.875. \end{aligned}$$

***Example 10.9***

If $X$ is the RV we associated with rolling a fair six-sided die, then we can easily write down the CDF of $X$. The probability distribution and the CDFs are as shown in Table 10.2.

**Table 10.2.**

| $X$ | 1 | 2 | 3 | 4 | 5 | 6 |
|---|---|---|---|---|---|---|
| $P(X)$ | 1/6 | 1/6 | 1/6 | 1/6 | 1/6 | 1/6 |
| $F(x)$ | 1/6 | 2/6 | 3/6 | 4/6 | 5/6 | 6/6 |

Notice that $P(X \leq x) = 0$ for any $x < 1$ since $X$ cannot take values less than 1. Also, notice that $P(X \leq x) = 1$ for any $x > 6$. Finally, note that the probabilities $P(X \leq x)$ are constant on any interval of the form $[x, x + 1)$ as required. Figure 10.2 represents the probability histogram, PMF and CDF.

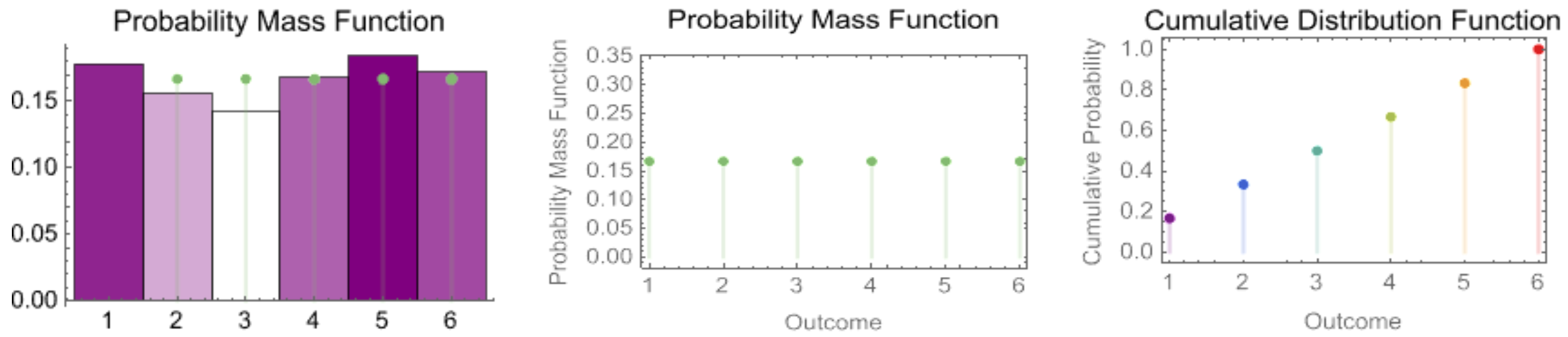

**Figure 10.2.** The probability histogram, PMF and CDF for 500 samples from a fair six-sided die.

***Example 10.10***

Let a pair of fair dice be tossed and let $X$ denote the sum of the points obtained. Then the probability distribution and the CDFs are as shown in Table 10.3. For example, the probability of getting sum 5 is $4/36 = 1/9$; thus in 900 tosses of the dice we would expect 100 tosses to give the sum 5.

**Table 10.3.**

| $X$ | 2 | 3 | 4 | 5 | 6 | 7 | 8 | 9 | 10 | 11 | 12 |
|---|---|---|---|---|---|---|---|---|---|---|---|
| $P(X)$ | $1/36$ | $2/36$ | $3/36$ | $4/36$ | $5/36$ | $6/36$ | $5/36$ | $4/36$ | $3/36$ | $2/36$ | 1/36 |
| $F(x)$ | $1/36$ | $3/36$ | $6/36$ | $10/36$ | $15/36$ | $21/36$ | $26/36$ | $30/36$ | $33/36$ | $35/36$ | 36/36 |

Notice that the CDF is constant over any half-closed integer interval from 2 to 12. For example, $F(X) = 3/36$ for all $X$ in the interval $[3,4)$. Figure 10.3 represents the probability histogram, PMF and CDF.

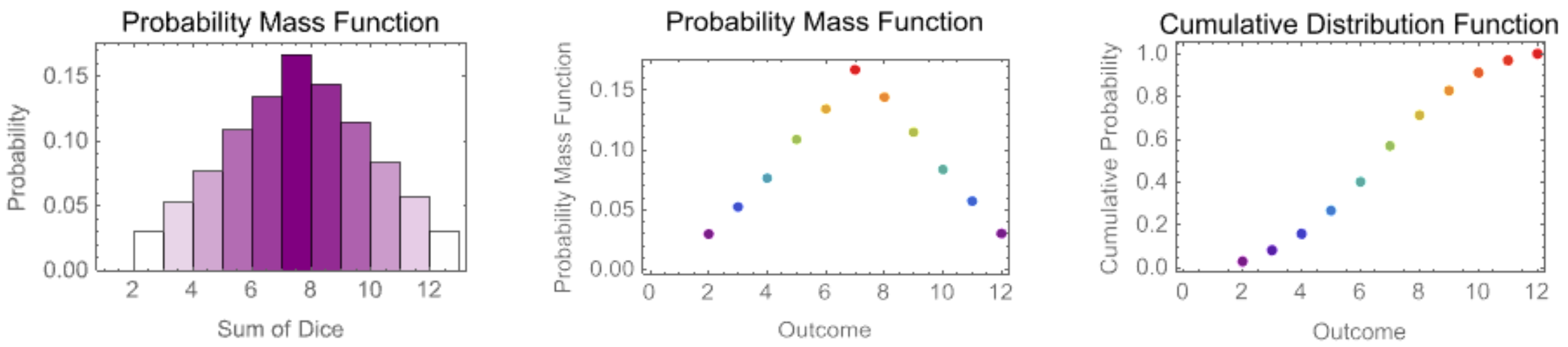

**Figure 10.3.** The probability histogram, PMF and CDF for 10000 samples from a pair of fair dice.

## 10.3 Mean, Variance, and Standard Deviation

The concept of expectation is easily extended. If $X$ denotes a discrete RV that can assume the values $x_1, x_2, \ldots, x_k$ with respective probabilities $P_1, P_2, \ldots, P_k$, where $P_1 + P_2 + \ldots + P_k = 1$, the mathematical expectation of $X$ (or simply the expectation of $X$), denoted by $E(X)$, is defined as

$$E(X) = P_1 x_1 + P_2 x_2 + \ldots + P_k x_k = \sum_{j=1}^{k} P_j x_j. \tag{10.5}$$

If the probabilities $P_j$ in this expectation are replaced with the relative frequencies $f_j/N$, where $N = \sum f_j$, the expectation reduces to $\sum_{j=1}^{k} f_j x_j / N$, which is the arithmetic mean $\bar{X}$ of a sample of size $N$ in which $x_1, x_2, \ldots, x_k$ appear with these relative frequencies. As $N$ gets larger and larger, the relative frequencies $f_j/N$ approach the probabilities $P_j$. Thus, we are led to the interpretation that $E(X)$ represents the mean of the population from which the sample is drawn.

**Definition (Mean):** The mean or expected value of the discrete RV $X$, denoted as $\mu$ or $E(X)$, is

$$\mu = E(X) = \sum_x x f(x). \tag{10.6}$$

**Definition (Expected Value):** The expected value of a function $h(X)$ of a discrete RV $X$ with PMF $f(x)$ is defined as

$$E[h(X)] = \sum_x h(x) f(x). \tag{10.7}$$

**Theorem 10.1:** If $a$ and $b$ are constants, then,

$$E[aX + b] = aE[X] + b. \tag{10.8}$$

**Proof**

$$
\begin{aligned}
E[aX+b] &= \sum_x (ax+b)P(x) \\
&= a\sum_x xP(x) + b\sum_x P(x) \\
&= aE[X] + b.
\end{aligned}
$$

■

**Remarks:**

- If we take $a = 0$ in Theorem 10.1, we see that,
  $$E[b] = b. \tag{10.9}$$
  That is, the expected value of a constant is just its value.
- If we take $b = 0$, then we obtain,
  $$E[aX] = aE[X]. \tag{10.10}$$
  The expected value of a constant multiplied by a RV is just the constant times the expected value of the RV.
- The quantity $E[X^n]$, $n \geq 1$, is called the $n$th moment of $X$. By (10.7), we note that
  $$E[X^n] = \sum_x x^n P(x). \tag{10.11}$$

**Definition (Variance):** If $X$ is a RV with mean $\mu$, then the variance of $X$, denoted as $\sigma^2$ or $V(X) \equiv \mathrm{Var}(X)$ is defined by

$$
\begin{aligned}
\sigma^2 = V(X) &= \mathrm{Var}(X) \\
&= E[(X-\mu)^2] \\
&= \sum_x (x-\mu)^2 f(x).
\end{aligned} \tag{10.12}
$$

**Definition (Standard Deviation):** The standard deviation of $X$ is

$$\sigma = \sqrt{\sigma^2}. \tag{10.13}$$

An alternative formula for $\mathrm{Var}(X)$ can be derived as follows:

$$
\begin{aligned}
\mathrm{Var}(X) &= E[(X-\mu)^2] \\
&= E[X^2 - 2\mu X + \mu^2] \\
&= E[X^2] - E[2\mu X] + E[\mu^2] \\
&= E[X^2] - 2\mu E[X] + \mu^2 \\
&= E[X^2] - \mu^2.
\end{aligned}
$$

**Theorem 10.2:** The variance of $X$ is equal to the expected value of the square of $X$ minus the square of the expected value of $X$.

$$\mathrm{Var}(X) = E[X^2] - (E[X])^2. \tag{10.14}$$

***Example 10.11***

Compute $E[X]$ and $\mathrm{Var}(X)$ when $X$ represents the outcome when we roll a fair die.

***Solution***

Since $P(X = x) = 1/6$, $x = 1,2,3,4,5,6$, we obtain

$$E[X] = 1\left(\frac{1}{6}\right) + 2\left(\frac{1}{6}\right) + 3\left(\frac{1}{6}\right) + 4\left(\frac{1}{6}\right) + 5\left(\frac{1}{6}\right) + 6\left(\frac{1}{6}\right) = 7/2.$$

$$E[X^2] = \sum_{x=1}^{6} x^2 P(X = x)$$
$$= 1^2\left(\frac{1}{6}\right) + 2^2\left(\frac{1}{6}\right) + 3^2\left(\frac{1}{6}\right) + 4^2\left(\frac{1}{6}\right) + 5^2\left(\frac{1}{6}\right) + 6^2\left(\frac{1}{6}\right) = 91/6.$$

Hence, we obtain from (10.14) that

$$\text{Var}(X) = E[X^2] - (E[X])^2 = \frac{91}{6} - \left(\frac{7}{2}\right)^2 = \frac{35}{12}.$$

**Theorem 10.3:** For any constants $a$ and $b$,

$$\text{Var}(aX + b) = a^2\text{Var}(X). \tag{10.15}$$

**Proof:**

Let $\mu = E[X]$ and recall that $E[aX + b] = a\mu + b$. Thus, by the definition of variance, we have

$$\begin{aligned}\text{Var}(aX + b) &= E[(aX + b - E[aX + b])^2] \\ &= E[(aX + b - a\mu - b)^2] \\ &= E[(aX - a\mu)^2] \\ &= E[a^2(X - \mu)^2] \\ &= a^2E[(X - \mu)^2] \\ &= a^2\text{Var}(X).\end{aligned}$$

■

**Remarks:**

- Specifying particular values for $a$ and $b$ in (10.15) leads to some interesting results. For instance, by setting $a = 0$ in (10.15) we obtain that
$$\text{Var}(b) = 0. \tag{10.16}$$
That is, the variance of a constant is 0.
- Similarly, by setting $a = 1$ we obtain,
$$\text{Var}(X + b) = Var(X). \tag{10.17}$$
That is, the variance of a constant plus a RV is equal to the variance of the RV.
- Finally, setting $b = 0$ yields
$$\text{Var}(aX) = a^2\text{Var}(X). \tag{10.18}$$

## 10.4 MGFs

**Definition (MGF):** The MGF $M_X(t)$ of the RV $X$ is defined for all values $t$ by

$$\begin{aligned}M_X(t) &= E[e^{tX}] \\ &= \sum_x e^{tx}P(x).\end{aligned} \tag{10.19}$$

We call $M_X(t)$ the MGF because all of the moments of $X$ can be obtained by successively differentiating $M_X(t)$. Recall that the Maclaurin series of the function $e^{tx}$ is

$$e^{tx} = 1 + tx + \frac{(tx)^2}{2!} + \frac{(tx)^3}{3!} + \cdots + \frac{(tx)^n}{n!} + \cdots. \tag{10.20}$$

By using the fact that the expected value of the sum equals the sum of the expected values, the MGF can be written as

$$\begin{aligned} M_X(t) &= E[e^{tX}] \\ &= E\left[1 + tX + \frac{(tX)^2}{2!} + \frac{(tX)^3}{3!} + \cdots + \frac{(tX)^n}{n!} + \cdots\right] \\ &= 1 + E[tX] + E\left[\frac{(tX)^2}{2!}\right] + E\left[\frac{(tX)^3}{3!}\right] + \cdots + E\left[\frac{(tX)^n}{n!}\right] + \cdots \\ &= 1 + tE[X] + \frac{t^2}{2!}E[X^2] + \frac{t^3}{3!}E[X^3] + \cdots + \frac{t^n}{n!}E[X^n] + \cdots. \end{aligned} \tag{10.21}$$

Note that $M_X(0) = 1$ for all the distributions. Taking the derivative of $M_X(t)$ with respect to $t$, we obtain

$$\begin{aligned} \frac{d}{dt}M_X(t) &= M_X'(t) \\ &= E[X] + tE[X^2] + \frac{t^2}{2!}E[X^3] + \frac{t^3}{3!}E[X^3] + \cdots + \frac{t^{n-1}}{(n-1)!}E[X^n] + \cdots. \end{aligned} \tag{10.22}$$

Evaluating this derivative at $t = 0$, all terms except $E[X]$ become zero. We have

$$M_X'(0) = E[X]. \tag{10.23}$$

Or,

$$\begin{aligned} M_X'(t) &= \frac{d}{dt}E[e^{tX}] \\ &= E\left[\frac{d}{dt}e^{tX}\right] \\ &= E[Xe^{tX}]. \end{aligned} \tag{10.24}$$

Hence, $M_X'(0) = E[X]$. Similarly, taking the second derivative of $M_X(t)$, we obtain

$$M_X''(0) = E[X^2], \tag{10.25}$$

where,

$$\begin{aligned} M_X''(t) &= \frac{d}{dt}M_X'(t) \\ &= \frac{d}{dt}E[Xe^{tX}] \\ &= E\left[\frac{d}{dt}(Xe^{tX})\right] \\ &= E[X^2e^{tX}], \end{aligned} \tag{10.26}$$

and so $M_X''(0) = E[X^2]$. Continuing in this manner, from the $n$th derivative $M_X^{(n)}(t)$ with respect to $t$, we obtain all the moments to be

$$M_X^{(n)}(0) = E[X^n], \qquad n = 1,2,3,\ldots. \tag{10.27}$$

We summarize these calculations in the following theorem.

**Theorem 10.4:** If $M_X(t)$ exists, then for any positive integer $k$,

$$\left.\frac{d^k}{dt^k}M_X(t)\right|_{t=0} = M_X^{(k)}(0) = E\left[X^{(k)}\right]. \tag{10.28}$$

The usefulness of the foregoing theorem lies in the fact that, if the MGF can be found, the often difficult process of summation involved in calculating different moments can be replaced by the much easier process of differentiation.

## 10.5 Discrete Distributions

A list of some commonly encountered discrete probability distributions:

1. Bernoulli distribution
2. Binomial distribution
3. Geometric distribution
4. Negative binomial distribution
5. Poisson distribution
6. Hypergeometric distribution
7. Discrete uniform distribution
8. Multinomial distribution
9. Zipf distribution
10. Rademacher distribution
11. Logarithmic distribution
12. Conway-Maxwell-Poisson distribution
13. Skellam distribution
14. Pólya distribution
15. Zeta distribution
16. Negative hypergeometric distribution
17. Wallenius noncentral hypergeometric distribution
18. Yule-Simon distribution
19. Zero-truncated Poisson distribution

Each distribution has its own specific characteristics and applications, making them useful for modeling and analyzing various types of discrete RVs. The following seven distributions form the core set of commonly used discrete probability distributions. They are frequently encountered in various fields, including statistics, probability theory, engineering, finance, and social sciences. Understanding these distributions and their properties is crucial for many statistical analyses and modeling applications. In this section, we discuss these distributions in some detail.

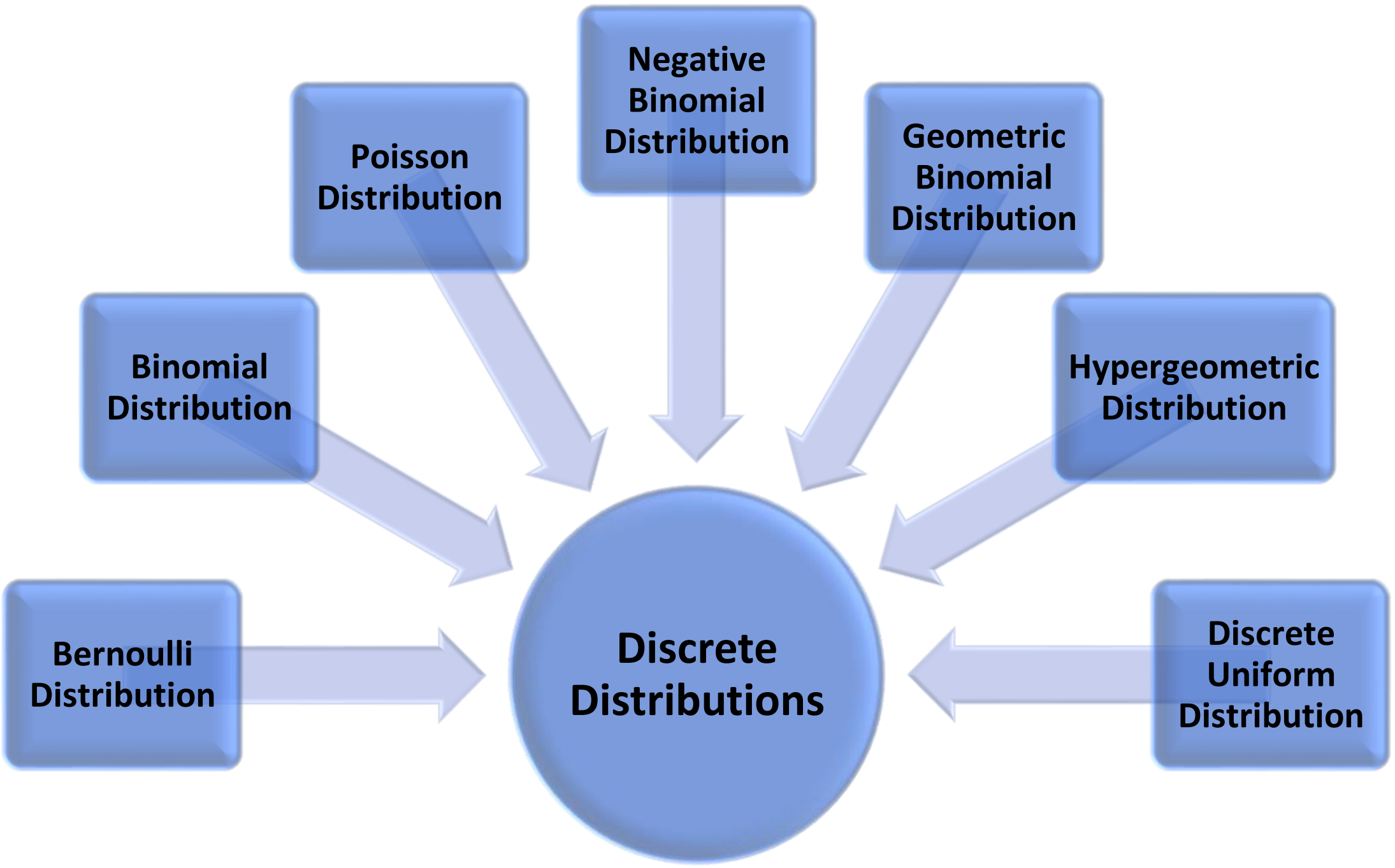

### 10.5.1 Bernoulli Distribution

Random experiments having exactly two mutually exclusive outcomes are called dichotomous experiments or Bernoulli trials. For instance, when a coin is tossed, the two possible outcomes are head $(H)$ and tail $(T)$. Even if an experiment has more than two mutually exclusive outcomes, we can consider it as a dichotomous experiment. For example, when a die is thrown, the set of all mutually exclusive outcomes is $\Omega = \{1, 2, 3, 4, 5, 6\}$. If we consider getting faces 3 or 5 as an event, say $A$, and its complement in $\Omega$ as the other event, the experiment reduces to a dichotomous experiment. Here, $A = \{3,5\}$ and $A' = \{1,2,4,6\}$.

In fact, all non-trivial experiments can be viewed as Bernoulli trials or dichotomous experiments. The event in which we are interested is labeled as success $(S)$ having probability $P(S) = p$, $0 < p < 1$ and its complementary event as failure $(F)$ with probability $P(F) = 1 - p = q$. If for such an experiment, a RV $X$ is defined such that it takes value 1 when success occurs and 0 when failure occurs, then $X$ follows a Bernoulli distribution. Hence, Bernoulli distribution, is the discrete probability distribution of a RV which takes only two values 1 and 0 with respective probabilities $p$ and $1 - p$, see Figure 10.4. This distribution is named after Swiss mathematician James Bernoulli (1654-1705). In above example, if the die is fair and we are interested in the occurrence of event $A$, then $p = 2/6$.

**Definition (Bernoulli Distribution):** A RV $X$ is said to follow Bernoulli distribution with parameter $p$ if its PMF is given by,

$$f_X(x) = \begin{cases} p^x q^{1-x}\,; & x = 0, 1 \\ 0\,; & \text{otherwise}' \end{cases} \tag{10.29}$$

where $0 < p < 1$ and $p + q = 1$.

**Theorem 10.5:** If $X$ is a Bernoulli RV with parameter $p$,

$$\mu = E(X) = p \text{ and } \sigma^2 = V(X) = p(1-p). \tag{10.30}$$

**Proof:**

We know that for $X$, $P(X = 1) = p$, and $P(X = 0) = q$, so

$$\begin{aligned} E[X] &= (1)P(X = 1) + (0)P(X = 0) \\ &= (1)p + (0)q \\ &= p. \end{aligned}$$

Thus, the mean or expected value of a Bernoulli distribution is given by $E[X] = p$.

The variance can be written as follows: $\text{Var}[X] = E[X^2] - (E[X])^2$. Using the properties of $E[X^2]$, we get,

$$\begin{aligned} E[X^2] &= \sum X^2 P(X = x) \\ &= 1^2(p) + 0^2(q) \\ &= p. \end{aligned}$$

Substituting this value, we have

$$\begin{aligned} \text{Var}[X] &= p - p^2 \\ &= p(1-p) \\ &= pq. \end{aligned}$$

Hence, the variance of a Bernoulli distribution is $\text{Var}[X] = p(1-p) = pq$.

■

**Remarks:**

- It is commonly used to model binary outcomes, such as success/failure, yes/no, or presence/absence.
- The distribution is characterized by a single parameter, usually denoted as $p$, which represents the probability of success. The probability of failure is then given by $1-p$.
- The Bernoulli distribution serves as the building block for more complex distributions, such as the binomial distribution, which models the number of successes in a fixed number of Bernoulli trials. Hence, Bernoulli distribution is a special case of the Binomial distribution when the number of trials = 1.

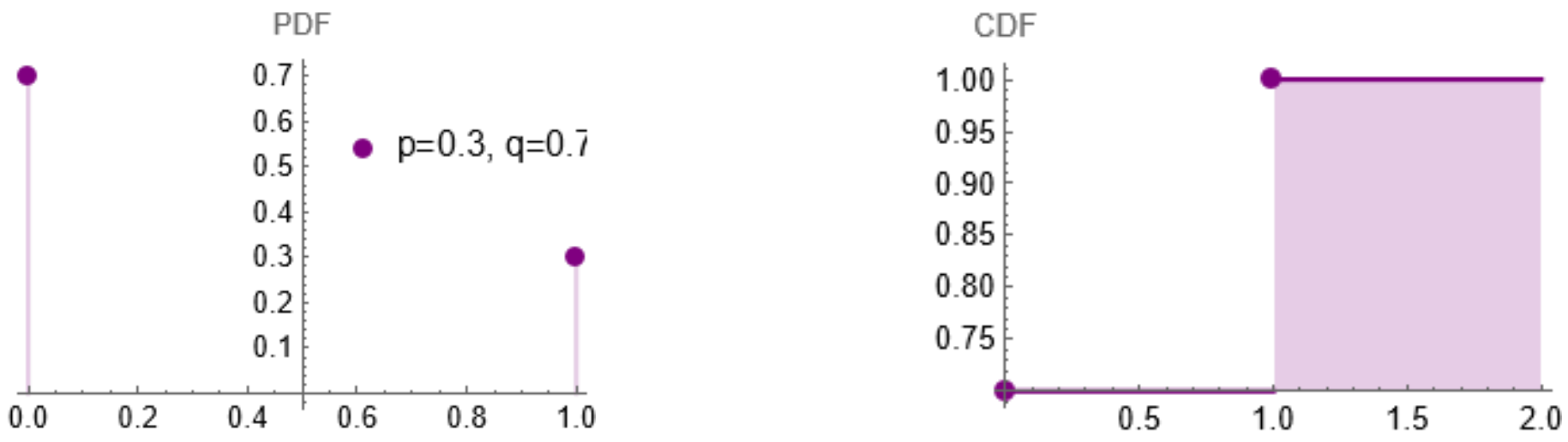

**Figure 10.4.** PMF (left) and CDF (right) of Bernoulli distribution. The graph shows that the probability of success is $p$ when $X=1$ and the probability of failure of $X$ is $(1-p)$ or $q$ if $X=0$.

### 10.5.2 Binomial Distribution

A trial with only two possible outcomes is used so frequently as a building block of a random experiment that it is called a Bernoulli trial. It is usually assumed that the trials that constitute the random experiment are independent. This implies that the outcome from one trial has no effect on the outcome to be obtained from any other trial. Furthermore, it is often reasonable to assume that the probability of a success in each trial is constant. The binomial distribution with parameters $n$ and $p$ is the discrete probability distribution of the number of successes in a sequence of $n$ independent experiments, each asking a yes–no question, and each with its own Boolean-valued outcome: success (with probability $p$) or failure (with probability $q=1-p$).

**Definition (Binomial Distribution):** The binomial distribution has certain conditions that need to be met for its application. A random experiment consists of $n$ Bernoulli trials such that
(1) The trials are independent.
(2) Each trial results in only two possible outcomes, labeled as "success" and "failure."
(3) The probability of a success in each trial, denoted as $p$, remains constant.

If $p$ is the probability that an event will happen in any single trial (called the probability of a success) and $q=1-p$ is the probability that it will fail to happen in any single trial (called the probability of a failure), then the probability that the event will happen exactly $x$ times in $n$ trials (i.e., $x$ successes and $n-x$ failures will occur) is given by (the PMF of $X$ is)

$$f_X(x)=\binom{n}{x}p^x(1-p)^{n-x},\ x=0,1,\dots,n. \quad (10.31)$$

**Remarks:**

- In $n$ trials if we are getting $x$ successes, then there will be $n-x$ failures. Since the trials are independent and $p$ is same in all trials, probability of getting $x$ successes is $p\times p\times\ldots\times p$ ($x$ times) $=p^x$ and probability of getting $n-x$ failures is $q\times q\times\ldots\times q$ ($n-x$ times) $=q^{n-x}$. Hence, the probability of getting $x$ successes and $n-x$ failures is $p^xq^{n-x}$. The number of ways in which $x$ successes can occur in $n$ trials is $n!/x!\,(n-x)!=\binom{n}{x}$. Hence, the probability of getting $x$ successes in $n$ trials in any order is given by, $\binom{n}{x}p^xq^{n-x}$. This probability distribution of the RV $X$ is denoted by $X\sim B(n,p)$, see Figure 10.5.

- This discrete probability distribution is often called the binomial distribution since for $x = 0, 1, 2, \ldots, n$ it corresponds to successive terms of the binomial formula, or binomial expansion,

$$(q + p)^n = q^n + \binom{n}{1} q^{n-1} p^1 + \binom{n}{2} q^{n-2} p^2 + \cdots + p^n.$$

For example

$$(q + p)^4 = q^4 + \binom{4}{1} q^3 p^1 + \binom{4}{2} q^2 p^2 + \binom{4}{3} q^1 p^3 + p^4 = q^4 + 4q^3 p^1 + 6q^2 p^2 + 4q^1 p^3 + p^4.$$

***Example 10.12***

The probability of getting exactly 2 heads in 6 tosses of a fair coin is

$$f(x) = \binom{n}{x} p^x (1 - p)^{n-x} = \binom{6}{2} \left(\frac{1}{2}\right)^2 \left(\frac{1}{2}\right)^{6-2} = \frac{6!}{2!\,4!} \left(\frac{1}{2}\right)^6 = \frac{15}{64}.$$

Using $n = 6$, $x = 2$, and $p = q = \frac{1}{2}$.

**Remark:**

- If $n = 1$, the binomial RV reduces to Bernoulli RV, denoted by $B(1, p)$.
- If $X \sim B(n, p)$, then $\sum_{x=0}^{n} f_X(x) = \sum_{x=0}^{n} \binom{n}{x} p^x q^{n-x} = (q + p)^n = 1$.
- Let $X \sim B(n, p)$. Then $X$ gives number of success in $n$ independent trials with probability $p$ for success in each trial. Note that $n - X$ gives number of failures in $n$ independent trials with probability $1 - p = q$ for failure in each trial. Therefore, $n - X \sim B(n, q)$.

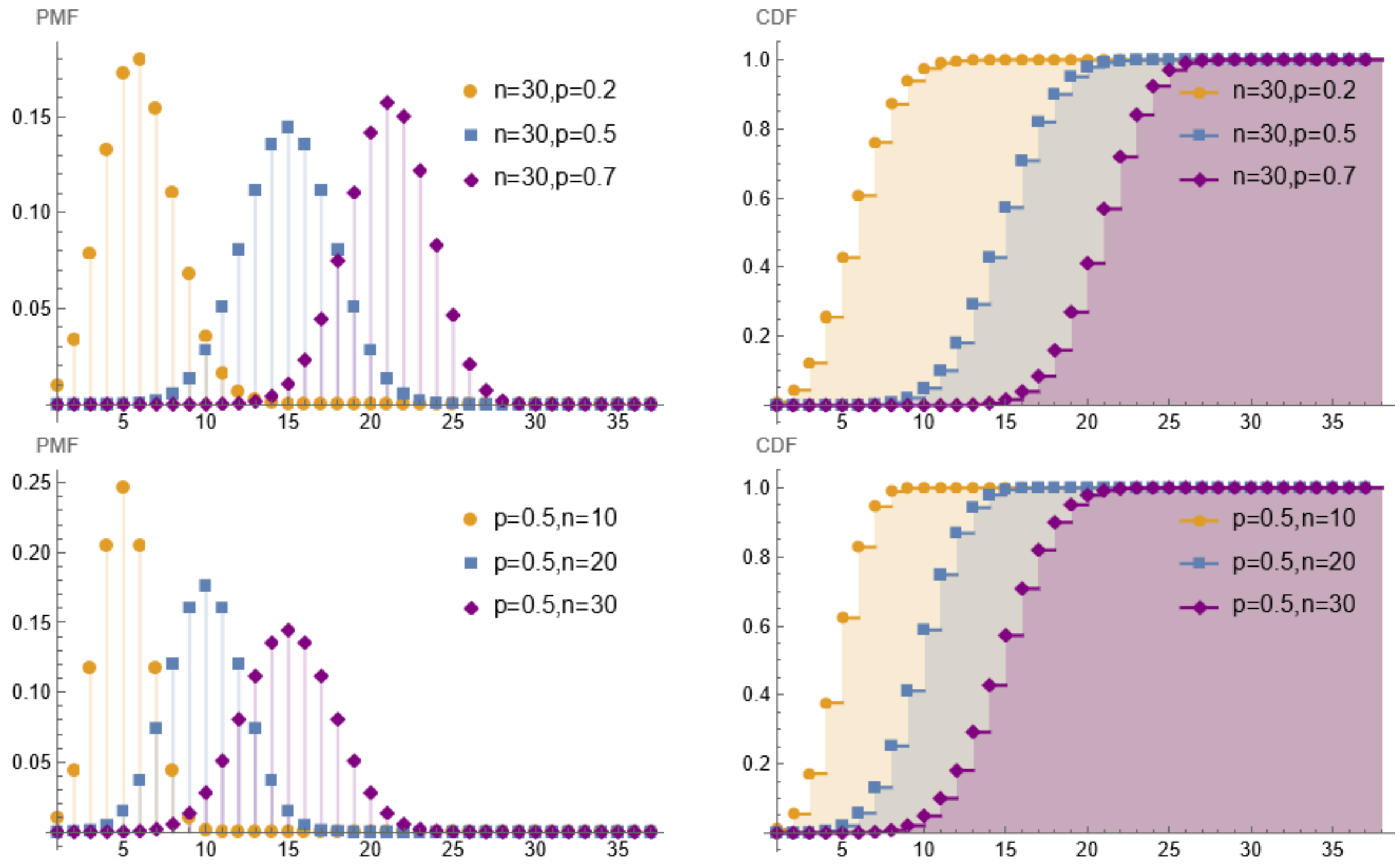

**Figure 10.5.** PMFs (left) and CDFs (right) of the binomial distribution. The curve of the binomial-distribution depends on the two parameters $p$ and $n$.

**Applications:**

1- The binomial distribution can be used to model the number of defective items in a production batch. In a production process, items are often inspected or tested to determine whether they meet certain quality standards. The binomial distribution can be applied in this context by considering each item produced as a trial, and the outcome of each trial is whether the item is defective or not. The binomial distribution assumes the following conditions are met:

- Fixed number of trials: The production batch consists of a fixed number of items, denoted by '$n$'.
- Two possible outcomes: Each item can either be defective or non-defective, representing the two possible outcomes.
- Independent trials: The production process assumes that the items are produced independently of each other. The probability of an item being defective does not depend on the outcomes of other items.
- Constant probability of defect: The probability of an item being defective, denoted by '$p$', remains the same for each item produced in the batch.
- By using the binomial distribution, we can calculate the probabilities associated with different numbers of defective items in the batch. The PMF of the binomial distribution allows us to determine the likelihood of observing a specific number of defective items in the batch, given the fixed number of items and the probability of a defect.

2- In clinical trials and drug testing, binomial distribution is often used to analyze and interpret the results. It provides a framework for understanding the probability of success or response to treatment among a group of participants. Here is a breakdown of why the binomial distribution is applicable in this context:

- Binary outcome: In clinical trials, the primary outcome of interest is typically a binary outcome, such as whether a patient experiences a positive response to a treatment or not.
- Fixed number of trials: In a clinical trial, there is a fixed number of participants who undergo the treatment being tested. Each participant can be considered a trial in the context of the binomial distribution. For example, if there are 100 participants, the binomial distribution allows us to analyze the number of participants who respond positively to the treatment out of the 100.
- Independent trials: The binomial distribution assumes that each participant's response to the treatment is independent of others. That means the response of one participant does not influence the response of another participant.
- Constant probability of success: In the context of drug testing, the probability of success refers to the likelihood of a positive response to the treatment. The binomial distribution assumes that this probability remains constant for each participant in the trial. This assumption allows us to model the probability of observing a certain number of successes (positive responses) among the fixed number of trials (participants).

***Example 10.13***

Suppose that 20% of the population is left-handed. Find the probability that in group of 60 individuals there will be,
(a) at most 10 left-handers,
(b) at least 10 left-handers,
(c) between 7 and 9 left-handers inclusive, and
(d) exactly 13 left-handers. Use Mathematica to find the solutions.

***Solution***

The probability of left-handed is $p = 20/100$, $n = 60$.

(a)

```
Probability[x<=10,x\[Distributed]BinomialDistribution[60,0.20]]
0.323403
```

(b)

```
Probability[x>=10,x\[Distributed]BinomialDistribution[60,0.20]]
0.786785
```

(c)
```
Probability[7<=x<=9,x\[Distributed]BinomialDistribution[60,0.20]]
 0.182377
```
(d)
```
 Probability[x==13,x\[Distributed]BinomialDistribution[60,0.20]]
 0.11799
```

***Example 10.14***

Find the probability that in four tosses of a fair die a 3 appears,
(a) at no time,
(b) once,
(c) twice,
(d) three times,
(e) four times,
(f ) five times. Use Mathematica to find the solutions.

***Solution***

The probability of 3 in a single toss $p = 1/6, n = 4$.
(a)
```
 Probability[x==0,x\[Distributed]BinomialDistribution[4,1/6]]
 625/1296
```
(b)
```
 Probability[x==1,x\[Distributed]BinomialDistribution[4,1/6]]
 125/324
```
(c)
```
 Probability[x==2,x\[Distributed]BinomialDistribution[4,1/6]]
 25/216
```
(d)
```
 Probability[x==3,x\[Distributed]BinomialDistribution[4,1/6]]
 5/324
```
(e)
```
 Probability[x==4,x\[Distributed]BinomialDistribution[4,1/6]]
 1/1296
```
(f )
```
 Probability[x==5,x\[Distributed]BinomialDistribution[4,1/6]]
 0
```

***Example 10.15***

Find the probability that in a family of 5 children there will be,
(a) at least 1 boy
(b) at least 1 boy and 1 girl. Assume that the probability of a male birth is 1/2.
(c) Out of 4000 families with 5 children each, how many would you expect to have at least 1 boy.

***Solution***

(a)
```
 Probability[x>=1,x\[Distributed]BinomialDistribution[5,1/2]]
 31/32
```
(b)
```
 Probability[(x>=1)&&(0<=x<=4),x\[Distributed]BinomialDistribution[5,1/2]]
 Probability[(1<=x<=4),x\[Distributed]BinomialDistribution[5,1/2]]
 15/16
 15/16
```
(c) Expected number of families with at least 1 boy = (total number of families)( probability at least 1 boy).
```
 4000*15/16
 3750
```

***Example 10.16***

Find the probability that in ten tosses of a coin a head appears.
(d) Obtain a table of probabilities for the RV $X$, the number of head appears, $X = 0$ to $X = 10$.
(e) Plot the probabilities for each value of $X$.

***Solution***

```
probabilitydistribution=Table[
   {
    i,
    N[
     Probability[x==i,x\[Distributed]BinomialDistribution[10,1/2]]
     ]
    },
   {i,0,10}
   ]
{{0,0.000976563},{1,0.00976563},{2,0.0439453},{3,0.117188},{4,0.205078},{5,0.246094},{6,0.20
5078},{7,0.117188},{8,0.0439453},{9,0.00976563},{10,0.000976563}}
 ListPlot[
 probabilitydistribution,
 PlotStyle->Purple,
 Filling->Axis,
 Mesh->All,
 ImageSize->200,
 AxesLabel->{"X","probability"}
 ]
```

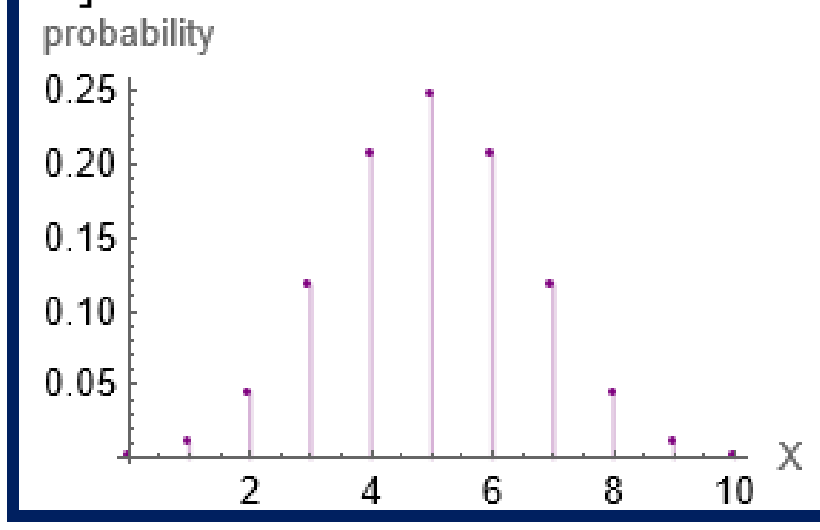

**Theorem 10.6:** If $X$ is a binomial RV with parameters $p$ and $n$,

$$\mu = E[(X)] = np. \tag{10.32}$$

**Proof:**

$$\begin{aligned}
\mu_1' &= E[X] \\
&= \sum_{x=0}^{n} x \binom{n}{x} p^x q^{n-x} \\
&= \sum_{x=1}^{n} x \frac{n!}{x!\,(n-x)!} p^x q^{n-x} \\
&= \sum_{x=1}^{n} \frac{n!}{(x-1)!\,(n-x)!} p^x q^{n-x} \\
&= np \sum_{x=1}^{n} \frac{(n-1)!}{(x-1)!\,(n-x)!} p^{x-1} q^{n-x} \\
&= np \sum_{x=1}^{n} \binom{n-1}{x-1} p^{x-1} q^{n-x} \\
&= np(q+p)^{n-1} \\
&= np.
\end{aligned}$$

■

**Theorem 10.7:** If $X$ is a binomial RV with parameters $p$ and $n$,

$$V(X) = nqp. \tag{10.33}$$

**Proof:**

$$V(X) = \mu_2' - (\mu_1')^2 = E[X^2] - [E(X)]^2,$$

and

$$\begin{aligned}
\mu_2' &= E[X^2] \\
&= \sum_{x=0}^{n} [x(x-1) + x] \binom{n}{x} p^x q^{n-x} \\
&= \sum_{x=0}^{n} x(x-1) \binom{n}{x} p^x q^{n-x} + \sum_{x=0}^{n} x \binom{n}{x} p^x q^{n-x} \\
&= \sum_{x=2}^{n} x(x-1) \frac{n!}{x!\,(n-x)!} p^x q^{n-x} + E(X) \\
&= n(n-1)p^2 \sum_{x=2}^{n} \frac{(n-2)!}{(x-2)!\,(n-x)!} p^{x-2} q^{n-x} + np \\
&= n(n-1)p^2 \sum_{x=2}^{n} \binom{n-2}{x-2} p^{x-2} q^{n-x} + np \\
&= n(n-1)p^2 + np.
\end{aligned}$$

Therefore,

$$\begin{aligned}
V(X) &= n(n-1)p^2 + np - (np)^2 \\
&= n^2p^2 - np^2 + np - n^2p^2 \\
&= np - np^2 \\
&= np(1-p) \\
&= nqp.
\end{aligned}$$

■

**Theorem 10.8 (Third and Fourth Raw Moment):** If $X$ is a binomial RV with parameters $p$ and $n$,

$$\mu_3 = npq(q-p), \tag{10.34}$$

$$\mu_4 = 3n^2p^2q^2 + npq(1-6pq). \tag{10.35}$$

**Proof:**

The third raw moment,

$$\begin{aligned}
\mu_3' &= E[X^3] \\
&= \sum_{x=0}^{n} x^3 \binom{n}{x} p^x q^{n-x} \\
&= \sum_{x=0}^{n} [x(x-1)(x-2) + 3x(x-1) + x] \binom{n}{x} p^x q^{n-x} \\
&= n(n-1)(n-2)p^3 + 3n(n-1)p^2 + np.
\end{aligned}$$

Therefore,

$$\begin{aligned}
\mu_3 &= \mu_3' - 3\mu_2'\mu_1' + 2(\mu_1')^3 \\
&= npq(1-2p) \\
&= npq(q-p).
\end{aligned}$$

The fourth raw moment,

$$\begin{aligned}\mu_4' &= E[X^4]\\ &= \sum_{x=0}^{n} x^4 \binom{n}{x} p^x q^{n-x}\\ &= \sum_{x=0}^{n} [x(x-1)(x-2)(x-3) + 6x(x-1)(x-2) + 7x(x-1) + x] \binom{n}{x} p^x q^{n-x}\\ &= n(n-1)(n-2)(n-3)p^4 + 6n(n-1)(n-2)p^3 + 7n(n-1)p^2 + np.\end{aligned}$$

Therefore,

$$\begin{aligned}\mu_4 &= \mu_4' - 4\mu_3'\mu_1' + 6\mu_2'(\mu_1')^2 - 3(\mu_1')^4\\ &= 3n^2p^2q^2 + npq(1-6pq).\end{aligned}$$

■

**Theorem 10.9 (Skewness):** If $X$ is a binomial RV with parameters $p$ and $n$,

$$\gamma_1 = \frac{q-p}{\sqrt{npq}}. \tag{10.36}$$

**Proof:**

$$\begin{aligned}\beta_1 &= \frac{\mu_3^2}{\mu_2^3}\\ &= \frac{[npq(q-p)]^2}{(npq)^3}\\ &= \frac{(q-p)^2}{npq}.\end{aligned}$$

Therefore,

$$\gamma_1 = \sqrt{\beta_1} = \frac{q-p}{\sqrt{npq}}.$$

Hence, binomial distribution is:

1. Positively skewed if $\gamma_1 > 0$; i.e., $q > p$.

2. Symmetric if $\gamma_1 = 0$, i.e.; $q = p$.

3. Negatively skewed if $\gamma_1 < 0$; i.e., $q < p$.

■

**Theorem 10.10 (Kurtosis):** If $X$ is a binomial RV with parameters $p$ and $n$,

$$\gamma_2 = \frac{1-6pq}{npq}. \tag{10.37}$$

**Proof:**

$$\begin{aligned}\beta_2 &= \frac{\mu_4}{\mu_2^2}\\ &= \frac{3n^2p^2q^2 + npq(1-6pq)}{(npq)^2}\\ &= 3 + \frac{1-6pq}{npq}.\end{aligned}$$

Therefore,

$$\gamma_2 = \beta_2 - 3 = \frac{1-6pq}{npq}.$$

Hence, binomial distribution is:

1. Leptokurtic if $\beta_2 > 3$; i.e., $\gamma_2 > 0$; i.e., $pq < 1/6$.

2. Mesokurtic if $\beta_2 = 3$; i.e., $\gamma_2 = 0$; i.e., $pq = 1/6$.

3. Platykurtic if $\beta_2 < 3$; i.e., $\gamma_2 < 0$; i.e., $pq > 1/6$.

■

**Theorem 10.11:** If $X$ is a binomial RV with parameters $p$ and $n$,

$$M_X(t) = (q + pe^t)^n. \tag{10.38}$$

**Proof:**

$$\begin{aligned} M_X(t) &= E[e^{tX}] \\ &= \sum_{x=0}^{n} e^{tx} \binom{n}{x} p^x q^{n-x} \\ &= \sum_{x=0}^{n} \binom{n}{x} (pe^t)^x q^{n-x} \\ &= (q + pe^t)^n. \end{aligned}$$

■

**Theorem 10.12:** If $X \sim B(n,p)$, $Y \sim B(m,p)$ and $X$ and $Y$ are independent, then

$$X + Y \sim B(n+m, p). \tag{10.39}$$

**Proof:**

$X \sim B(n,p)$ and $Y \sim B(m,p)$ implies $M_X(t) = (q+pe^t)^n$ and $M_Y(t) = (q+pe^t)^m$ respectively. Since $X$ and $Y$ are independent,

$$\begin{aligned} M_{X+Y}(t) &= M_X(t) \times M_Y(t) \\ &= (q+pe^t)^n (q+pe^t)^m \\ &= (q+pe^t)^{n+m}, \end{aligned}$$

which is the MGF of $B(n+m, p)$.

■

**Theorem 10.13:** When $X \sim B(n,p)$,

$$\mu_{r+1} = pq\left(nr\mu_{r-1} + \frac{d\mu_r}{dp}\right). \tag{10.40}$$

**Proof:**

$$\mu_r = E[X - E(X)]^r = E[X - np]^r = \sum_{x=0}^{n} (x-np)^r \binom{n}{x} p^x q^{n-x}.$$

Therefore,

$$\frac{d}{dp}\mu_r = \frac{d}{dp}\left[\sum_{x=0}^{n} (x-np)^r \binom{n}{x} p^x q^{n-x}\right]$$

$$= \sum_{x=0}^{n} \frac{d}{dp}\left((x-np)^r \binom{n}{x} p^x q^{n-x}\right)$$
$$= \sum_{x=0}^{n} \Big[ r(x-np)^{r-1}(-n) \binom{n}{x} p^x q^{n-x} + (x-np)^r \binom{n}{x} x p^{x-1}(1-p)^{n-x}$$
$$+ (x-np)^r \binom{n}{x} p^x (n-x)(1-p)^{n-x-1}(-1)\Big]$$
$$= -nr \sum_{x=0}^{n} (x-np)^{r-1} \binom{n}{x} p^x q^{n-x} + \sum_{x=0}^{n} (x-np)^r \binom{n}{x} p^x q^{n-x} \left(\frac{x}{p}\right) + \sum_{x=0}^{n} (x-np)^r \binom{n}{x} p^x q^{n-x} \left(-\frac{n-x}{1-p}\right)$$
$$= -nr\mu_{r-1} + \sum_{x=0}^{n} (x-np)^r \binom{n}{x} p^x q^{n-x} \left(\frac{x}{p} - \frac{n-x}{1-p}\right)$$
$$= -nr\mu_{r-1} + \sum_{x=0}^{n} (x-np)^r \binom{n}{x} p^x q^{n-x} \left(\frac{x-np}{pq}\right)$$
$$= -nr\mu_{r-1} + \frac{1}{pq} \sum_{x=0}^{n} (x-np)^{r+1} \binom{n}{x} p^x q^{n-x}$$
$$= -nr\mu_{r-1} + \frac{1}{pq}\mu_{r+1}.$$

Therefore,

$$\mu_{r+1} = pq\left(nr\mu_{r-1} + \frac{d\mu_r}{dp}\right).$$

■

**Theorem 10.14:** When $X \sim B(n,p)$,

$$\frac{f_X(x+1)}{f_X(x)} = \frac{n-x}{x+1}\frac{p}{q}. \tag{10.41}$$

**Proof:**

The PMF is

$$f_X(x) = \binom{n}{x} p^x q^{n-x},$$

hence, we have

$$f_X(x+1) = \binom{n}{x+1} p^{x+1} q^{n-x-1},$$

and

$$\frac{f_X(x+1)}{f_X(x)} = \frac{\binom{n}{x+1} p^{x+1} q^{n-x-1}}{\binom{n}{x} p^x q^{n-x}}$$
$$= \frac{\frac{n!}{(x+1)!\,(n-x-1)!} p^{x+1} q^{n-x-1}}{\frac{n!}{x!\,(n-x)!} p^x q^{n-x}}$$
$$= \frac{n-x}{x+1}\frac{p}{q}.$$

■

### 10.5.3 Poisson Distribution

The Poisson distribution is named after the French mathematician Siméon Denis Poisson, who introduced it in the early 19th century. He approached the distribution by considering the limit of a binomial distribution in which $n$ tends to infinity, $p$ tends to zero and $np$ remains finite and equal to $\lambda$. The Poisson distribution is widely used to model the occurrence of rare events in a given time period or space interval.

**Definition (Poisson Distribution):** The Poisson distribution is a discrete probability distribution that describes the number of events that occur in a fixed interval of time or space, given a known average rate of occurrence and the events happening independently of each other.

**Definition (Poisson Distribution):** The RV $X$ that equals the number of events in a Poisson process (a process in which events occur randomly and independently at a constant average rate) is a Poisson RV with parameter $0 < \lambda$, and the PMF of $X$ is given by

$$f_X(x) = \begin{cases} \dfrac{e^{-\lambda}\lambda^x}{x!} ; & x = 0, 1,2, \dots \\ 0 ; & \text{otherwise}, \end{cases} \tag{10.42}$$

where $\lambda > 0$ is the parameter of the Poisson distribution, see Figure 10.6. We can write $X \sim P(\lambda)$.

#### The assumptions of Poisson distribution

- Independent events:
  The events or occurrences being counted must be independent of each other. This means that the occurrence of one event does not affect the occurrence of another event.
- Fixed time or space interval:
  The Poisson distribution models the number of events that occur in a fixed interval of time or space. This interval must be well-defined and consistent.
- Constant average rate:
  The events must occur at a constant average rate throughout the interval. This average rate is denoted by λ. It represents the expected number of events occurring in the given interval.
- Infinitesimally small sub-intervals:
  The probability of more than one event occurring in an infinitesimally small sub-interval must be negligible. In other words, the events should be relatively rare or low in frequency within the chosen interval.

#### Some applications of the Poisson distribution:

- Queuing theory:
  The Poisson distribution is commonly used to model the arrival rate of customers in queuing systems, such as waiting lines at banks, call centers, or supermarkets. It helps analyze the waiting times, service rates, and the probability of queue lengths exceeding certain thresholds.
- Network traffic analysis:
  In computer networks and telecommunications, the Poisson distribution is often used to model the arrival rate of packets or messages. It helps in studying network congestion, estimating bandwidth requirements, and designing efficient routing algorithms.
- Epidemiology:
  The Poisson distribution is employed in epidemiological studies to analyze the occurrence of disease outbreaks or the number of cases in a specific population over time. It aids in understanding the spread of infectious diseases, estimating disease rates, and assessing the effectiveness of interventions.
- Environmental studies:
  The Poisson distribution is utilized in environmental research to analyze the frequency of natural events like earthquakes, floods, or forest fires. It helps in assessing the risk and severity of such events, planning for disaster management, and evaluating the impact of environmental factors.

- Manufacturing and quality control:
  The Poisson distribution is applied in quality control processes to model the occurrence of defects in a production line or the number of errors in a sample. It aids in setting quality standards, monitoring process performance, and making decisions regarding product acceptance or rejection.
- Sports analytics:
  The Poisson distribution finds application in sports analytics, particularly in modeling the number of goals, points, or scores in various sports. It aids in predicting match outcomes, and assessing player performance.

### How the Poisson distribution is applied in modeling, for example, the arrival rates in queuing systems:

- The Poisson distribution assumes that customers arrive at a queuing system at a constant average rate, such as $\lambda$ customers per unit of time (e.g., per hour, per day). This rate is often estimated based on historical data or assumptions about the system.
- The Poisson distribution assumes that customer arrivals occur independently of each other. In other words, the arrival of one customer does not affect the arrival of another.
- Using the Poisson distribution's PMF, the probability of a specific number of customer arrivals within a given time period can be calculated.
- By considering the arrival rate and using the Poisson distribution, queuing theory can estimate the expected number of customers in the system or the average queue length. Queuing theory enables the calculation of key performance measures, such as the utilization rate of the service facility, the probability of the system being idle, or the average service rate required to meet a target level of customer service. These metrics guide decision-making for system design, capacity planning, and service level agreements.

### More examples of Poisson distribution modeling

- Consider the number of telemarketing phone calls received by a household during a given day. In this example, the receiving of a telemarketing phone call by a household is called an occurrence, the interval is one day (an interval of time), and the occurrences are random (that is, there is no specified time for such a phone call to come in) and discrete. The total number of telemarketing phone calls received by a household during a given day may be 0, 1, 2, 3, 4, and so forth. The independence of occurrences in this example means that the telemarketing phone calls are received individually and none of two (or more) of these phone calls are related.

In contrast, consider the arrival of patients at a physician’s office. These arrivals are nonrandom if the patients have to make appointments to see the doctor. The arrival of commercial airplanes at an airport is nonrandom because all planes are scheduled to arrive at certain times, and airport authorities know the exact number of arrivals for any period (although this number may change slightly because of late or early arrivals and cancellations). The Poisson probability distribution cannot be applied to these examples.

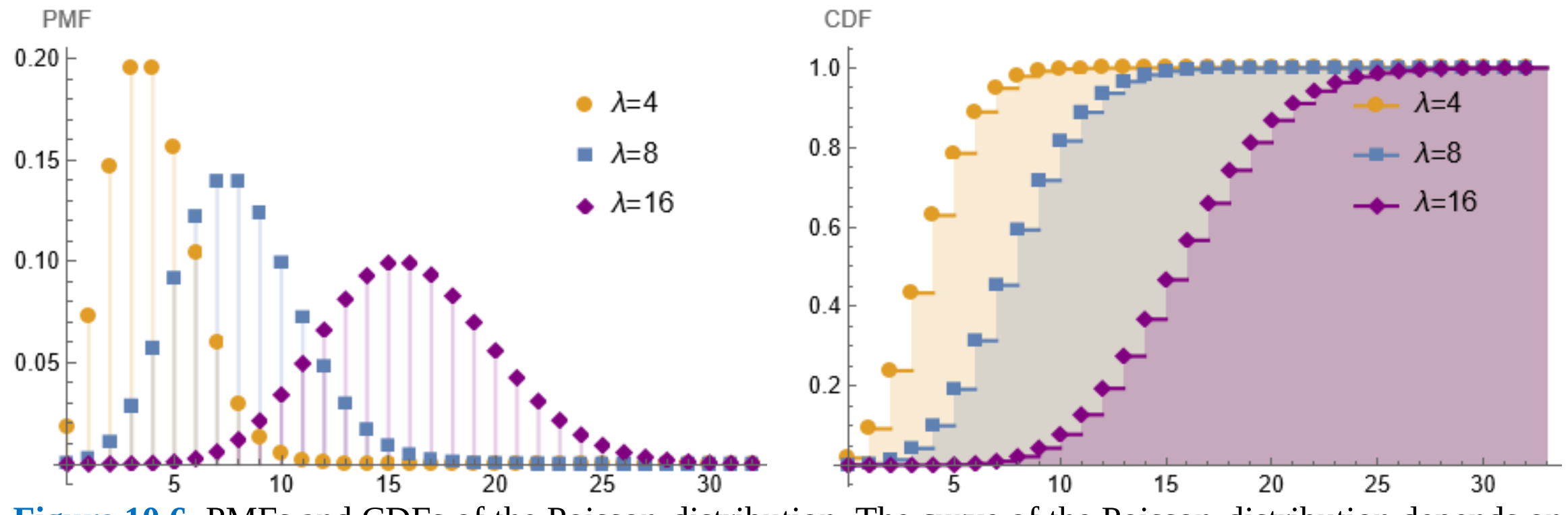

**Figure 10.6.** PMFs and CDFs of the Poisson-distribution. The curve of the Poisson-distribution depends on the parameters $\lambda$.

***Example 10.17***

In the World Cup, an average of 2.5 goals are scored each game. Modeling this situation with a Poisson distribution, what is the probability that $k$ goals are scored in a game?

***Solution***

In this instance, $\lambda = 2.5$. The formula (10.42) applies directly:

$$P(x=0) = \frac{e^{-2.5}(2.5)^0}{0!} = 0.082,$$
$$P(x=1) = \frac{e^{-2.5}(2.5)^1}{1!} = 0.205,$$
$$P(x=2) = \frac{e^{-2.5}(2.5)^2}{2!} = 0.257.$$

***Example 10.18***

The number of admissions per day at an emergency room has a Poisson distribution and the mean is 5.
(a) Find the probability of at most 2 admissions per day
(b) The probability of at least 6 admissions per day.

***Solution***

(a)
```
N[Probability[x<=2,x\[Distributed]PoissonDistribution[5]]]
0.124652
```

(b)
```
N[Probability[x>=6,x\[Distributed]PoissonDistribution[5]]]
0.384039
```

***Example 10.19***

The average number of traffic accidents on a certain section of highway is three per week. Assume that the number of accidents follows a Poisson distribution with $\lambda = 3$.
(a) Find the probability of no accidents on this section of highway during a 1-week period.
(b) Find the probability of at most four accidents on this section of highway during a 2-week period.

***Solution***

(a)
```
N[Probability[x==0,x\[Distributed]PoissonDistribution[3]]]
0.0497871
```

(b)
```
N[Probability[x<=4,x\[Distributed]PoissonDistribution[6]]]
0.285057
```

***Example 10.20***

Emergency Room Traffic Desert Samaritan Hospital keeps records of emergency room (ER) traffic. Those records indicate that the number of patients arriving between 6:00 P.M. and 7:00 P.M. has a Poisson distribution with parameter $\lambda = 5.8$. Determine the probability that, on a given day, the number of patients who arrive at the emergency room between 6:00 P.M. and 7:00 P.M. will be
(a) exactly 3.
(b) at most 3.
(c) between 5 and 9, inclusive.
(d) Obtain a table of probabilities for the RV $X$, the number of patients arriving between 6:00 P.M. and 7:00 P.M, $X = 0$ to $X = 20$.
(e) Plot the probabilities for each value of $X$.

***Solution***

(a)
```
N[Probability[x==3,x\[Distributed]PoissonDistribution[5.8]]]
0.0951816
```
(b)
```
N[Probability[x<=3,x\[Distributed]PoissonDistribution[5.8]]]
0.0319518
```
(c)
```
N[Probability[5<=x<=9,x\[Distributed]PoissonDistribution[5.8]]]
0.821296
```
(d)
```
probabilitydistribution=Table[
 {
  i,
  N[
   Probability[x==i,x\[Distributed]PoissonDistribution[5.8]]
   ]
  },
 {i,0,20}
 ]

{{0,0.00302755},{1,0.0175598},{2,0.0509235},{3,0.098452},{4,0.142755},{5,0.165596},{6,0.160076},{7,0.132635},{8,0.0961602},{9,0.0619699},{10,0.0359426},{11,0.0189515},{12,0.0091599},{13,0.00408673},{14,0.00169307},{15,0.000654655},{16,0.000237312},{17,0.0000809654},{18,0.0000260888},{19,7.96396*10-6},{20,2.30955*10-6}}
```

(e)
```
ListPlot[
probabilitydistribution,
PlotStyle->Purple,
Filling->Axis,
Mesh->All,
ImageSize->200,
AxesLabel->{"X","probability"}
]
```
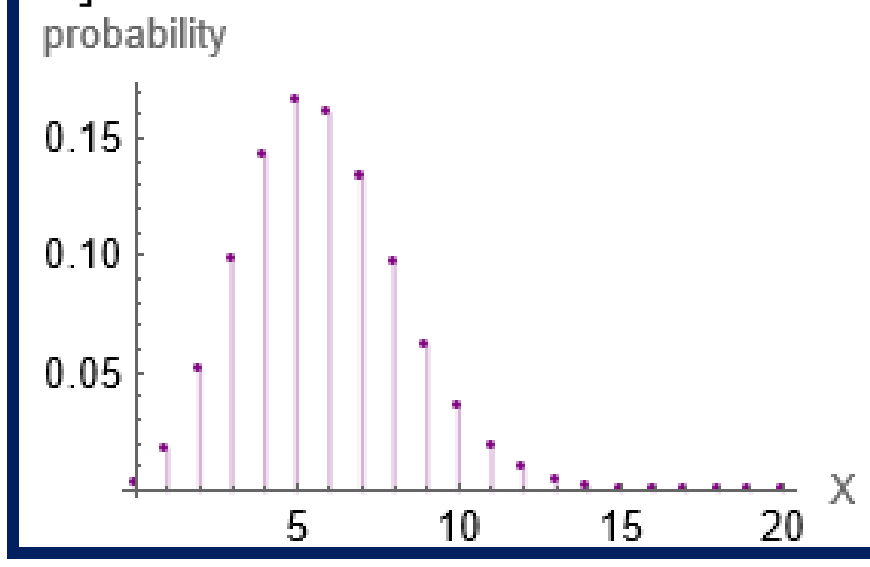

There are connections between the Poisson distribution and several other probability distributions.

**Relationship with the Binomial Distribution:**

The Poisson distribution can be derived as an approximation of the binomial distribution when the number of trials $n$ is large and the probability of success $(p)$ is small, while the product $\lambda = np$ remains moderate or large. In this scenario, the Poisson distribution with parameter $\lambda$ provides a good approximation to the binomial distribution. This connection is known as the Poisson approximation to the binomial distribution.

**Relationship with the Exponential Distribution:**

The inter-arrival times in a Poisson process follow an exponential distribution. This means that if the events in a system follow a Poisson process, the time between consecutive events (inter-arrival time) will be exponentially distributed. Similarly, if the inter-arrival times are exponentially distributed, it implies that the events are occurring according to a Poisson process.

### Poisson distribution as a limiting form of binomial distribution

**Theorem 10.15:** The Poisson distribution is obtained as an approximation to the binomial distribution under the conditions:
i) $n$ is very large ($n \to \infty$),
ii) $p$ is very small ($p \to 0$),
iii) $np = \lambda$, a finite quantity.

**Proof:**

For binomial distribution,

$$f(x) = \binom{n}{x} p^x q^{n-x}; \qquad x = 0,1,2,\ldots,n,$$

where $0 < p < 1$ and $p + q = 1$. Now,

$$\begin{aligned} f(x) &= \binom{n}{x} p^x q^{n-x} \\ &= \frac{n!}{x!\,(n-x)!} p^x q^{n-x} \\ &= \frac{n(n-1)(n-2)\ldots(n-x+1)}{x!} p^x q^{n-x} \\ &= \frac{n^x \left(1-\frac{1}{n}\right)\left(1-\frac{2}{n}\right)\ldots\left(1-\frac{x-1}{n}\right)}{x!\,(1-p)^x} p^x q^n \\ &= \frac{\left(1-\frac{1}{n}\right)\left(1-\frac{2}{n}\right)\ldots\left(1-\frac{x-1}{n}\right)}{q^x} \frac{(np)^x q^n}{x!}. \end{aligned}$$

Now,

$$\lim_{n\to\infty} \left(1-\frac{1}{n}\right)\left(1-\frac{2}{n}\right)\ldots\left(1-\frac{x-1}{n}\right) = 1.$$

Also,

$$np = \lambda \Rightarrow p = \frac{\lambda}{n}.$$

Therefore,

$$\lim_{n\to\infty} q^x = \lim_{n\to\infty} (1-p)^x = \lim_{n\to\infty} \left(1-\frac{\lambda}{n}\right)^x = 1, \qquad \lim_{n\to\infty} q^n = \lim_{n\to\infty} (1-p)^n = \lim_{n\to\infty} \left(1-\frac{\lambda}{n}\right)^n = e^{-\lambda}.$$

Using these limits, we get

$$f(x) = \frac{e^{-\lambda}\lambda^x}{x!}; \; x = 0,1,2,\ldots.$$

which is the PMF of a Poisson distribution.

■

***Example 10.21***

Ten percent of the tools produced in a certain manufacturing process turn out to be defective. Find the probability that in a sample of 10 tools chosen at random exactly 2 will be defective by using

(a) the binomial distribution and
(b) the Poisson approximation to the binomial distribution.
$\lambda = np = 10(0.1) = 1$.

***Solution***

```
Probability[x==2,x\[Distributed]BinomialDistribution[10,0.1]]
0.19371
N[Probability[x==2,x\[Distributed]PoissonDistribution[1]]]
0.18394
```

In general, the approximation is good if $p = 0.1$ and $np \leq 5$.

**Theorem 10.16:** If $X$ is a Poisson RV with parameter $\lambda$,

$$\mu = E[X] = \lambda. \tag{10.43}$$

**Proof:**

$$\begin{aligned}\mu_1' &= E[X]\\ &= \sum_{x=0}^{\infty} x\frac{e^{-\lambda}\lambda^x}{x!}\\ &= \lambda e^{-\lambda}\sum_{x=1}^{\infty}\frac{\lambda^{x-1}}{(x-1)!}\\ &= \lambda e^{-\lambda}e^{\lambda}\\ &= \lambda.\end{aligned}$$

■

**Theorem 10.17:** If $X$ is a Poisson RV with parameter $\lambda$,

$$V(X) = \lambda. \tag{10.44}$$

**Proof:**

$$V(X) = \mu_2' - (\mu_1')^2 = E[X^2] - (E[X])^2,$$

and

$$\begin{aligned}\mu_2' &= E[X^2]\\ &= \sum_{x=0}^{\infty}[x(x-1)+x]\frac{e^{-\lambda}\lambda^x}{x!}\\ &= \sum_{x=2}^{\infty}x(x-1)\frac{e^{-\lambda}\lambda^x}{x!} + \sum_{x=1}^{\infty}x\frac{e^{-\lambda}\lambda^x}{x!}\\ &= \lambda^2 e^{-\lambda}\sum_{x=2}^{\infty}\frac{\lambda^{x-2}}{(x-2)!} + E(X)\\ &= \lambda^2 e^{-\lambda}e^{\lambda} + \lambda = \lambda^2 + \lambda.\end{aligned}$$

Therefore,

$$V(X) = \lambda^2 + \lambda - \lambda^2 = \lambda,$$

and $\sigma = \sqrt{\lambda}$.

■

**Theorem 10.18:** If $X$ is a Poisson RV with parameter $\lambda$, in a similarly way, we can obtain $\mu_3$ and $\mu_4$

$$\mu_3 = \lambda. \tag{10.45}$$

$$\mu_4 = 3\lambda^2 + \lambda. \tag{10.46}$$

**Theorem 10.19 (Skewness and Kurtosis):** If $X$ is a Poisson RV with parameter $\lambda$,

$$\gamma_1 = \frac{1}{\sqrt{\lambda}}, \qquad \gamma_2 = \frac{1}{\lambda}. \tag{10.47}$$

**Proof:**

$$\beta_1 = \frac{\mu_3^2}{\mu_2^3} = \frac{\lambda^2}{\lambda^3} = \frac{1}{\lambda}.$$

Therefore,

$$\gamma_1 = \sqrt{\beta_1} = \frac{1}{\sqrt{\lambda}},$$

and

$$\beta_2 = \frac{\mu_4}{\mu_2^2} = \frac{3\lambda^2 + \lambda}{\lambda^2} = 3 + \frac{1}{\lambda}.$$

Therefore,

$$\gamma_2 = \beta_2 - 3 = \frac{1}{\lambda}.$$

■

**Theorem 10.20:** If $X$ is a Poisson RV with parameter $\lambda$,

$$M_X(t) = e^{\lambda(e^t - 1)}. \tag{10.48}$$

**Proof:**

$$\begin{aligned} M_X(t) &= E[e^{tX}] \\ &= \sum_{x=0}^{\infty} e^{tx} \frac{e^{-\lambda}\lambda^x}{x!} \\ &= e^{-\lambda} \sum_{x=0}^{\infty} \frac{(e^t\lambda)^x}{x!} \\ &= e^{-\lambda} e^{\lambda e^t} \\ &= e^{\lambda(e^t-1)}. \end{aligned}$$

■

**Theorem 10.21:** If $X \sim P(\lambda_1)$, $Y \sim P(\lambda_2)$ and $X$ and $Y$ are independent, then

$$X + Y \sim P(\lambda_1 + \lambda_2). \tag{10.49}$$

**Proof:**

$X \sim P(\lambda_1)$ and $Y \sim P(\lambda_2)$ implies $M_X(t) = e^{\lambda_1(e^t-1)}$ and $M_Y(t) = e^{\lambda_2(e^t-1)}$ respectively. Since $X$ and $Y$ are independent,

$$\begin{aligned} M_{X+Y}(t) &= M_X(t) \times M_Y(t) \\ &= e^{\lambda_1(e^t-1)} e^{\lambda_2(e^t-1)} \\ &= e^{(\lambda_1+\lambda_2)(e^t-1)}, \end{aligned}$$

which is the MGF of $P(\lambda_1 + \lambda_2)$.

■

**Theorem 10.22.** When $X \sim P(\lambda)$,

$$\mu_{r+1} = \lambda\left(r\mu_{r-1} + \frac{d\mu_r}{d\lambda}\right). \tag{10.50}$$

**Proof:**

$$\mu_r = E\big[X - E[X]\big]^r = E[X-\lambda]^r = \sum_{x=0}^{\infty}(x-\lambda)^r \frac{e^{-\lambda}\lambda^x}{x!}.$$

Therefore,

$$\begin{aligned}
\frac{d}{d\lambda}\mu_r &= \frac{d}{d\lambda}\left[\sum_{x=0}^{\infty}(x-\lambda)^r \frac{e^{-\lambda}\lambda^x}{x!}\right] \\
&= \sum_{x=0}^{\infty}\frac{d}{d\lambda}\left((x-\lambda)^r \frac{e^{-\lambda}\lambda^x}{x!}\right) \\
&= \sum_{x=0}^{\infty}\frac{1}{x!}\left[r(x-\lambda)^{r-1}(-1)e^{-\lambda}\lambda^x + (x-\lambda)^r\left(-e^{-\lambda}\right)\lambda^x + (x-\lambda)^r e^{-\lambda}x\lambda^{x-1}\right] \\
&= -r\sum_{x=0}^{\infty}(x-\lambda)^{r-1}\frac{e^{-\lambda}\lambda^x}{x!} - \sum_{x=0}^{\infty}(x-\lambda)^r\frac{e^{-\lambda}\lambda^x}{x!} + \sum_{x=0}^{\infty}\frac{x}{\lambda}(x-\lambda)^r\frac{e^{-\lambda}\lambda^x}{x!} \\
&= -r\mu_{r-1} + \sum_{x=0}^{\infty}(x-\lambda)^r\frac{e^{-\lambda}\lambda^x}{x!}\left(-1+\frac{x}{\lambda}\right) \\
&= -r\mu_{r-1} + \sum_{x=0}^{\infty}(x-\lambda)^r\frac{e^{-\lambda}\lambda^x}{x!}\left(\frac{x-\lambda}{\lambda}\right) \\
&= -r\mu_{r-1} + \frac{1}{\lambda}\sum_{x=0}^{\infty}(x-\lambda)^{r+1}\frac{e^{-\lambda}\lambda^x}{x!} \\
&= -r\mu_{r-1} + \frac{1}{\lambda}\mu_{r+1}.
\end{aligned}$$

Therefore,

$$\mu_{r+1} = \lambda\left(r\mu_{r-1} + \frac{d\mu_r}{d\lambda}\right).$$

■

### 10.5.4 Negative Binomial Distribution

The negative binomial distribution is a discrete probability distribution that models the number of failures in a sequence of independent and identically distributed Bernoulli trials before a specified number of successes (denoted as $r$) occurs. The distribution is characterized by two parameters: the probability of success in each trial (denoted as $p$) and the number of successes that must occur before the experiment is stopped (denoted as $r$). For example, we can define rolling a 6 on a dice as a success, and rolling any other number as a failure, and ask how many failure rolls will occur before we see the third success ($r = 3$). In such a case, the probability distribution of the number of failures that appear will be a negative binomial distribution.

**Definition (Negative Binomial Distribution):** A RV $X$ is said to follow negative binomial distribution with parameters $r$ and $p$ if its PMF is given by,

$$f_X(x) = \begin{cases} \binom{x+r-1}{x} p^r q^x; & x = 0, 1,2, \ldots \\ 0\,; & \text{otherwise}, \end{cases} \tag{10.51}$$

where $r$ is the number of successes, $x$ is the number of failures, $0 < p < 1$ and $p + q = 1$ and we write $X \sim \mathrm{NB}(r, p)$. (See Figure10.7)

The negative binomial distribution is an extension of the binomial distribution. While the binomial distribution models the number of successes in a fixed number of independent Bernoulli trials, the negative binomial distribution models the number of trials required to achieve a fixed number of successes. Now suppose that a Bernoulli trial is repeated $n$ times keeping the probability, $p$, of success same through out the trials and the trials independent. If we are interested in finding the probability distribution of $X$, the number of successes in $n$ trials, then $X \sim B(n,p)$. Here, the number of trials '$n$' is fixed. Instead, suppose we are interested in finding the probability of number of trials required to get $r$ successes. Then we have the negative binomial distribution. Here, the number of successes is fixed, not the number of trials '$n$'. Hence the name negative binomial.

Let the RV $Y$ be the number of trials required to get $r$ successes. Then $y = r, r+1, \dots$ . Suppose $Y$ takes the value $y$. i.e., $y$ trials are required to get $r$ success. Then, out of these $y$ trials $r$ are successes including the $y$th one. Hence, there will be $y - r$ failures preceding the $r$th success. Let $X$ be the number of failures preceding $r$th success. Clearly, $X$ takes values $0,1,2, \dots$ .

Note that both the RVs $X$ and $Y$ follow negative binomial distribution where $X$ assumes values $0,1,2, \dots$ and $Y$ assumes values $1,2,3, \dots$ .

$$\begin{aligned} P(\mathrm{X} = \mathrm{x}) &= P\left(x \text{ failures preceding the } r^{\text{th}} \text{ success}\right) \\ &= P\left(\text{Getting } r-1 \text{ successes in } x+r-1 \text{ trials and a success in } (x+r)^{\text{th}} \text{ trial}\right) \\ &= \binom{x+r-1}{r-1} p^{r-1} q^{(x+r-1)-(r-1)} p \\ &= \binom{x+r-1}{x} p^r q^x; x = 0,1,2, \dots . \end{aligned}$$

### The assumptions of negative binomial distribution

- Independent trials:
  The trials or events being considered must be independent of each other. The outcome of one trial should not influence the outcome of subsequent trials.
- Fixed probability of success:
  The probability of success ($p$) remains constant across all trials.
- Counting discrete events:
  The negative binomial distribution is appropriate for situations where the variable of interest is a count of discrete events.
- No upper limit on trials:
  There is no upper limit on the number of trials. The negative binomial distribution allows for an indefinite number of trials until a specified number of successes is achieved.

### Some applications of the negative binomial distribution:

- Quality control:
  In manufacturing or production processes, the negative binomial distribution can be used to model the number of defective items found before a certain number of acceptable items are produced. This helps in assessing the quality of the process and determining the appropriate inspection levels.
- Sports analytics:
  The negative binomial distribution can be applied in sports analytics to model the number of attempts or games needed for a player to achieve a certain number of goals, points, or victories. This assists in evaluating player performance, predicting future outcomes, and making strategic decisions.
- Customer relationship management:
  In marketing and customer relationship management, the negative binomial distribution can be used to model the number of contacts or interactions required to achieve a certain number of successful sales or conversions.

This aids in analyzing customer behavior, optimizing marketing campaigns, and predicting customer acquisition rates.

- Website analytics:
  In web analytics, the negative binomial distribution can be employed to model the number of page views or clicks before a certain number of conversions or desired actions are completed by users. This helps in measuring website effectiveness, evaluating marketing campaigns, and optimizing user experiences.
- Social sciences:
  The negative binomial distribution can be used in social sciences, such as sociology or criminology, to model the number of events (such as criminal offenses or social interactions) before a specific number of desired outcomes or patterns are observed. This assists in understanding social phenomena, predicting behavior, and evaluating intervention programs.

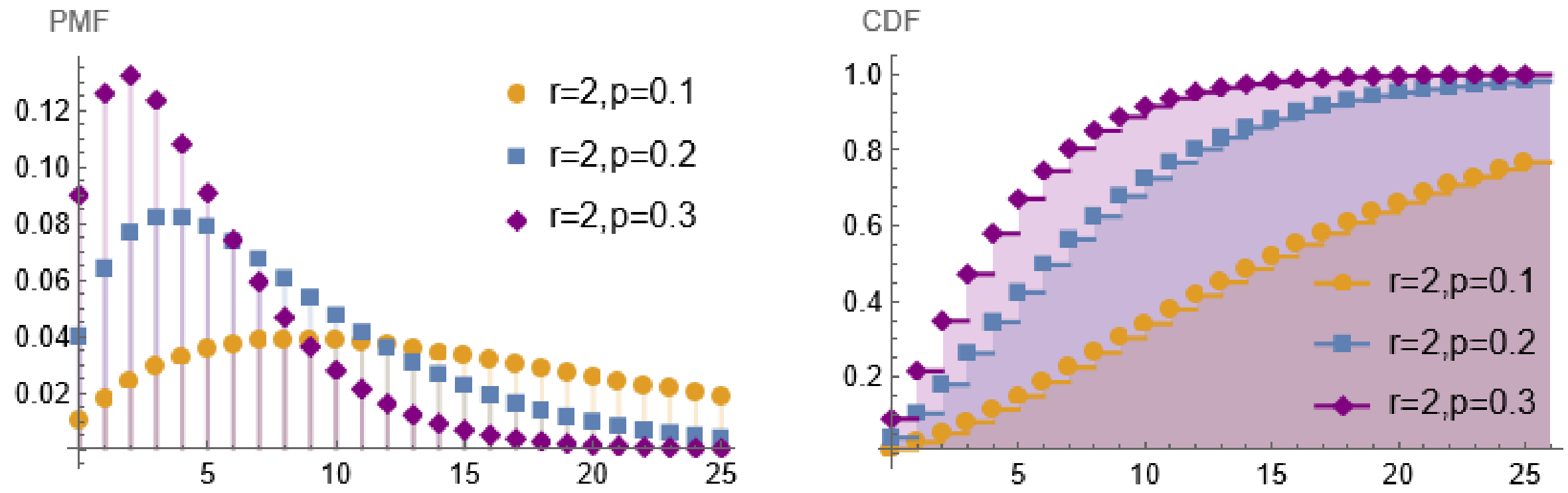

**Figure 10.7.** PMFs (left) and CDFs (right) of the negative binomial distribution. The curve of the negative binomial distribution depends on the parameters $r$ and $p$.

**Theorem 10.23:** If $X$ is a negative binomial RV with parameters $p$ and $r$, the mean of a RV following NB$(r,p)$ is given by

$$E[X] = \frac{rq}{p}, \tag{10.52}$$

and the variance is given by

$$V(X) = \frac{rq}{p^2}. \tag{10.53}$$

***Example 10.22***

In an NBA championship series, the team that wins three games out of seven is the winner. Suppose that teams $A$ and $B$ face each other in the championship games and that team $A$ has probability 0.45 of winning a game over team $B$.

(a) What is the probability that team $A$ will win the series in 6 games?

(b) What is the probability that team $A$ will win the series?

***Solution***

(a)

```
N[Probability[x==3,x\[Distributed]NegativeBinomialDistribution[3,0.45]]]
0.151609
```

(b)

```
N[Probability[0<=x<=4,x\[Distributed]NegativeBinomialDistribution[3,0.45]]]
0.68356
```

***Example 10.23***

Obtain a table of probabilities for the RV $X \sim NB(r,p)$, $r=5$, $p=0.4$, and $X=0$ to $X=20$. Plot the probabilities for each value of $X$.

***Solution***

```
probabilitydistribution=Table[
  {
   i,
   N[
    Probability[x==i,x\[Distributed]NegativeBinomialDistribution[5,0.4]]
    ]
   },
  {i,0,20}
  ]

{{0,0.01024},{1,0.03072},{2,0.055296},{3,0.0774144},{4,0.0928973},{5,0.100329},{6,0.100329},
{7,0.094596},{8,0.0851364},{9,0.0737849},{10,0.0619793},{11,0.0507103},{12,0.0405683},{13,0.
0318305},{14,0.0245549},{15,0.0186618},{16,0.0139963},{17,0.0103737},{18,0.00760741},{19,0.0
0552538},{20,0.00397827}}
 ListPlot[
 probabilitydistribution,
 PlotStyle->Purple,
 Filling->Axis,
 Mesh->All,
 ImageSize->200,
 AxesLabel->{"X","probability"}
 ]
```

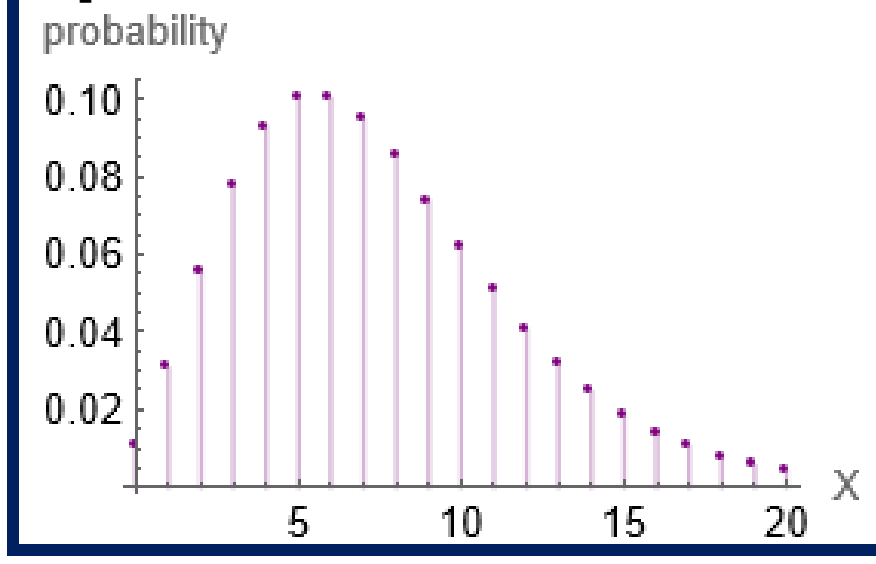

### 10.5.5 Geometric Distribution

Consider a sequence of trials, where each trial has only two possible outcomes (designated failure and success). The probability of success is assumed to be the same for each trial. In such a sequence of trials, the geometric distribution is useful to model the number of failures before the first success since the experiment can have an indefinite number of trials until success, unlike the binomial distribution which has a set number of trials. The distribution gives the probability that there are zero failures before the first success, one failure before the first success, two failures before the first success, and so on. For example, if you are flipping a fair coin until you get heads (success), the number of tails you get before the first heads follows a geometric distribution.

Negative binomial RV gives number of failures preceding the $r$th success. When $r=1$ it reduces to geometric RV and its distribution is called geometric distribution. Hence, it gives the probability distribution of number of failures preceding the 1st success and takes values $0,1,2,\ldots$ .

**Definition (Geometric Distribution):** In a series of Bernoulli trials (independent trials with constant probability $p$ of a success), the RV $X$ that equals the number of failures before the first success is a geometric RV with parameter $0<p<1$. Its PMF is given by,

$$f_X(x)=\begin{cases} pq^x; & x=0,1,2,\ldots \\ 0; & \text{otherwise,} \end{cases} \tag{10.54}$$

where $0<p<1$ and $p+q=1$. Here we write $X \sim \text{GEO}_0(p)$. (See Figure 10.8).

**Remarks:**

- Since the probabilities for $x = 0,1,2,\ldots$ are the terms of geometric progression series, the distribution has the name geometric distribution. Sometimes it is called Furry distribution.
- Some authors define geometric distribution as the number of trials required to obtain the first success. In this case the RV $X$ takes values $x = 1,2,3,\ldots$ and the PMF is given by

$$f_X(x) = \begin{cases} pq^{x-1}; & x = 1,2,\ldots \\ 0; & \text{otherwise,} \end{cases} \tag{10.55}$$

where $0 < p < 1$ and $p + q = 1$. Here we write $X \sim \text{GEO}_1(p)$.

***Example 2.24***

Letting $S$ denote the outcome of "success" and $F$ denote the outcome of "failure", we can summarize the possible outcomes of a geometric experiment and their likelihoods (the PMF) in Table 10.4. Here, we write $p$ for the probability of success and $q$ for the probability of failure.

**Table 10.4.**

| Experimental Outcome | Value of the RV, $X = x$ | Probability |
|---|---|---|
| $S$ | $x = 0$ | $p$ |
| $FS$ | $x = 1$ | $q.p$ |
| $FFS$ | $x = 2$ | $q^2.p$ |
| $FFFS$ | $x = 3$ | $q^3.p$ |
| $FFFFS$ | $x = 4$ | $q^4.p$ |
| ... | ... | ... |

**The geometric distribution makes several key assumptions:**

- Independent trials:
  Geometric distribution assumes that each trial or attempt is independent of the others. The outcome of one trial does not affect the outcome of subsequent trials.
- Binary outcomes:
  Geometric distribution assumes that each trial has two possible outcomes, often referred to as success and failure. These outcomes are mutually exclusive, meaning that only one of them can occur in each trial.
- Constant probability of success:
  Geometric distribution assumes that the probability of success remains constant across all trials.
- Discrete outcomes:
  Geometric distribution deals with discrete outcomes, specifically the count of trials until the first success.
- Infinite range (theoretically):
  Geometric distribution theoretically has an infinite range, as it is possible for an infinite number of trials to be required before the first success occurs. However, in practice, the distribution is often truncated to a finite range for practicality and ease of analysis.

**Geometric distribution has several applications in various fields.**

- Sports analytics:
  Geometric distribution finds applications in sports analytics, particularly in scenarios where the focus is on the number of attempts or trials needed to achieve a particular outcome. For example, it can be used to analyze the number of shots needed to score a goal in soccer or the number of attempts needed to make a successful basketball free throw.
- Marketing and sales forecasting:

Geometric distribution can be used to model customer behavior, such as the number of marketing interactions or sales calls needed to secure the first purchase from a potential customer. It aids in forecasting conversion rates and optimizing sales strategies.

- Search algorithms:
  In algorithms and search theory, the geometric distribution can be used to model the number of trials (e.g., iterations or comparisons) needed until finding the desired solution or item in a search process.

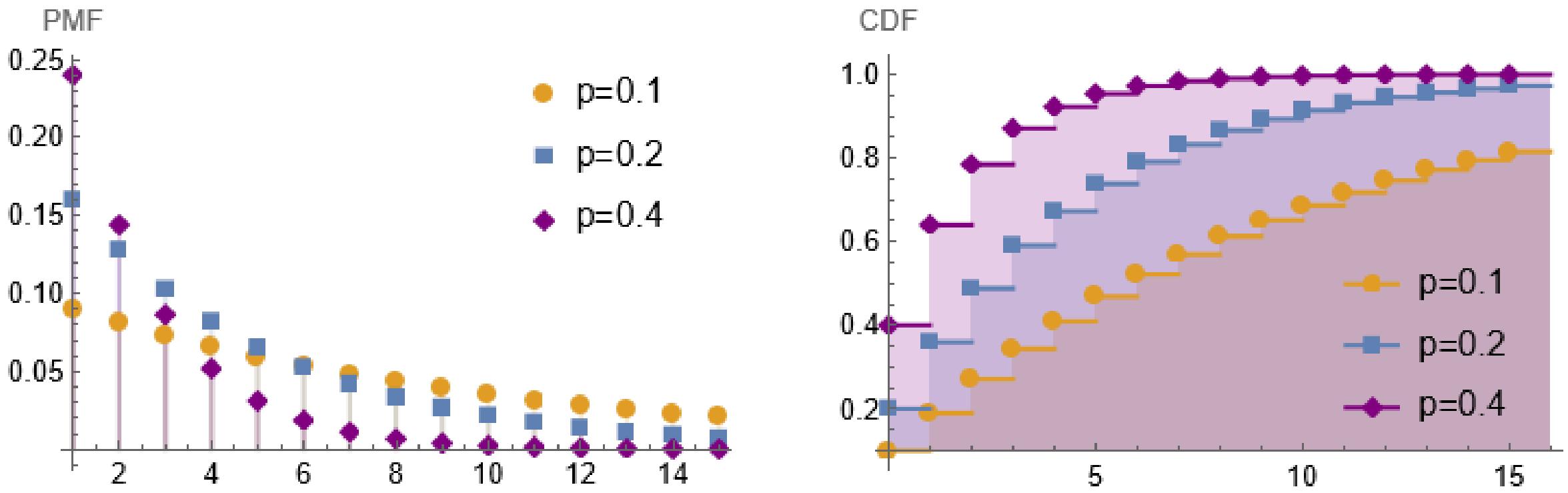

**Figure 10.8.** PMFs (left) and CDFs (right) of the geometric distribution. The curve of the geometric distribution depends on the parameters $p$.

**Theorem 10.24:** If $X$ is a geometric RV with parameter $p$,

$$\mu = E[X] = \frac{q}{p}, \tag{10.56}$$

$$\sigma^2 = V(X) = \frac{q}{p^2}, \tag{10.57}$$

$$M_X(t) = \frac{p}{1 - qe^t}. \tag{10.58}$$

**Proof:**

Mean

$$\begin{aligned}\mu_1' &= E[X] \\ &= \sum_{x=0}^{\infty} xpq^x \\ &= p[q + 2q^2 + 3q^3 + \cdots] \\ &= pq[1 + 2q + 3q^2 + \cdots] \\ &= pq(1-q)^{-2} \\ &= \frac{pq}{p^2} \\ &= \frac{q}{p}.\end{aligned}$$

Variance

$$V(X) = \mu_2' - (\mu_1')^2 = E(X^2) - [E(X)]^2,$$

and

$$\begin{aligned}\mu_2' &= E(X^2) \\ &= \sum_{x=0}^{\infty} [x(x-1) + x]pq^x\end{aligned}$$

$$
\begin{aligned}
&= \sum_{x=2}^{\infty} x(x-1)pq^x + \sum_{x=1}^{\infty} xpq^x \\
&= p\big((2)(1)q^2 + (3)(2)q^3 + (4)(3)q^4 + \cdots\big) + E(X) \\
&= 2pq^2(1-q)^{-3} + \frac{q}{p} \\
&= 2q^2p^{-2} + \frac{q}{p}.
\end{aligned}
$$

Therefore,

$$
\begin{aligned}
V(X) &= \frac{2q^2}{p^2} + \frac{q}{p} - \left(\frac{q}{p}\right)^2 \\
&= \frac{q^2}{p^2} + \frac{q}{p} \\
&= \frac{q^2 + pq}{p^2} \\
&= \frac{q(q+p)}{p^2} \\
&= \frac{q}{p^2}.
\end{aligned}
$$

MGF,

$$
\begin{aligned}
M_X(t) &= E(e^{tX}) \\
&= \sum_{x=0}^{\infty} e^{tX} pq^x \\
&= p\sum_{x=0}^{\infty} (qe^t)^x \\
&= p(1 + qe^t + (qe^t)^2 + \cdots) \\
&= p(1 - qe^t)^{-1} \\
&= \frac{p}{1 - qe^t}.
\end{aligned}
$$

■

***Example 10.25***

For a certain manufacturing process, it is known that, on average, 2 in every 100 items is defective. What is the probability that the seventh item inspected is the first defective item found?

***Solution***

```
Probability[x==6,x\[Distributed]GeometricDistribution[0.02]]
0.0177168
```

***Example 10.26***

Obtain a table of probabilities for the RV $X \sim \mathrm{GEO}_0(p)$, $p = 0.2$, and $X = 0$ to $X = 20$. Plot the probabilities for each value of $X$.

***Solution***

```
probabilitydistribution=Table[
  {
   i,
   N[
    Probability[x==i,x\[Distributed]GeometricDistribution[0.2]]
```

```
    ]
   },
  {i,0,20}
  ]

{{0,0.2},{1,0.16},{2,0.128},{3,0.1024},{4,0.08192},{5,0.065536},{6,0.0524288},{7,0.041943},{8,0.0335544},{9,0.0268435},{10,0.0214748},{11,0.0171799},{12,0.0137439},{13,0.0109951},{14,0.00879609},{15,0.00703687},{16,0.0056295},{17,0.0045036},{18,0.00360288},{19,0.0028823},{20,0.00230584}}
 ListPlot[
 probabilitydistribution,
 PlotStyle->Purple,
 Filling->Axis,
 Mesh->All,
 ImageSize->200,
 AxesLabel->{"X","probability"}
 ]
```

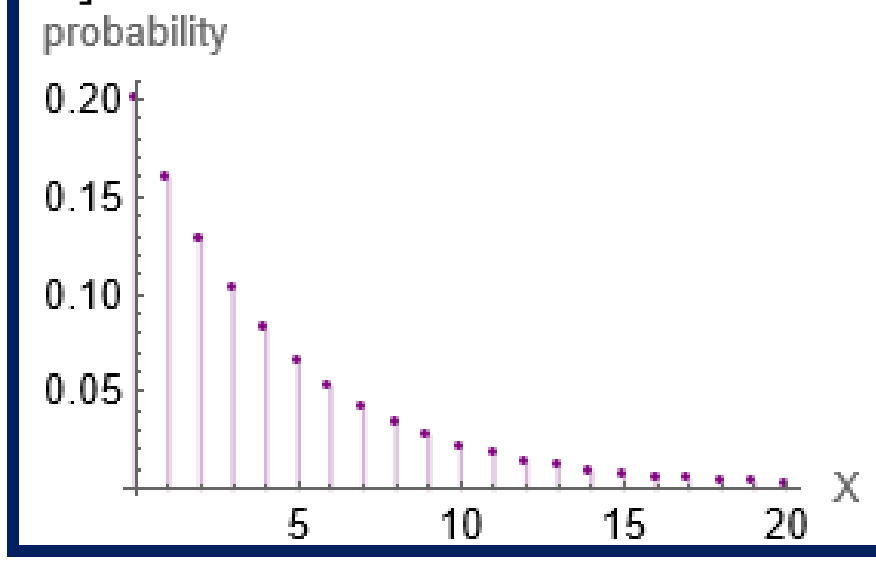

### 10.5.6 Hypergeometric Distribution

We learned that one of the conditions required to apply the binomial distribution is that the trials are independent, so that the probabilities of the two outcomes or events (success and failure) remain constant. If the trials are not independent, we cannot apply the binomial probability distribution to find the probability of $x$ successes in $n$ trials. In such cases we replace the binomial probability distribution by the hypergeometric probability distribution. Such a case occurs when a sample is drawn without replacement from a finite population. When sampling is done without replacement, it means that each item selected from the population is not put back or replaced before the next selection. As a result, the composition of the population changes with each selection, and the probability of drawing subsequent items is affected by previous selections. In the context of hypergeometric distribution, sampling without replacement is a fundamental assumption. The distribution calculates the probability of obtaining a specific number of successes (items of interest) in a fixed-sized sample, drawn without replacement from a finite population.

For example, let us consider a bag of colored balls: 10 red balls, and 5 blue balls. If we randomly draw 4 balls from the bag without replacement, the probability of obtaining a specific number of red balls (successes) can be calculated using the hypergeometric distribution. The hypergeometric distribution takes into account the fact that the probability of selecting a red ball changes with each draw. As we remove red balls from the bag, the probability of selecting another red ball decreases. This is different from the binomial distribution, which assumes sampling with replacement and treats each draw as independent.

**Definition (Hypergeometric Distribution):** A set of $N$ objects contains $K$ objects classified as successes and $N-K$ objects classified as failures. A sample of size $n$ objects is selected randomly (without replacement) from the $N$ objects where $K \leq N$ and $n \leq N$. The RV $X$ that equals the number of successes in the sample is a hypergeometric RV (see Figure 10.9). The PMF is

$$f_X(x) = \frac{\binom{K}{x}\binom{N-K}{n-x}}{\binom{N}{n}}, x = \max\{0, n+K-N\} \text{ to } \min\{K, n\}. \tag{10.59}$$

### The assumptions of hypergeometric distribution

- Finite population:
  The hypergeometric distribution assumes that the population size ($N$) is fixed and known. This means that there is a specific number of items or individuals in the population from which the sample is drawn.
- Sampling without replacement:
  The hypergeometric distribution assumes that sampling is done without replacement. This means that each item selected from the population is not returned or replaced before the next selection.
- Two mutually exclusive outcomes:
  The hypergeometric distribution deals with two mutually exclusive outcomes, typically referred to as successes and failures. The distribution calculates the probability of obtaining a specific number of successes in the sample. For example, in a bag of red and blue balls, success may refer to selecting a red ball, and failure may refer to selecting a blue ball.
- Fixed number of successes in the population:
  The hypergeometric distribution assumes that the number of successes in the population ($K$) is known.
- Fixed sample size:
  The hypergeometric distribution assumes a fixed sample size ($n$). This means that the number of items or individuals selected from the population remains constant throughout the sampling process. The distribution calculates the probability of obtaining a specific number of successes within this fixed sample size.

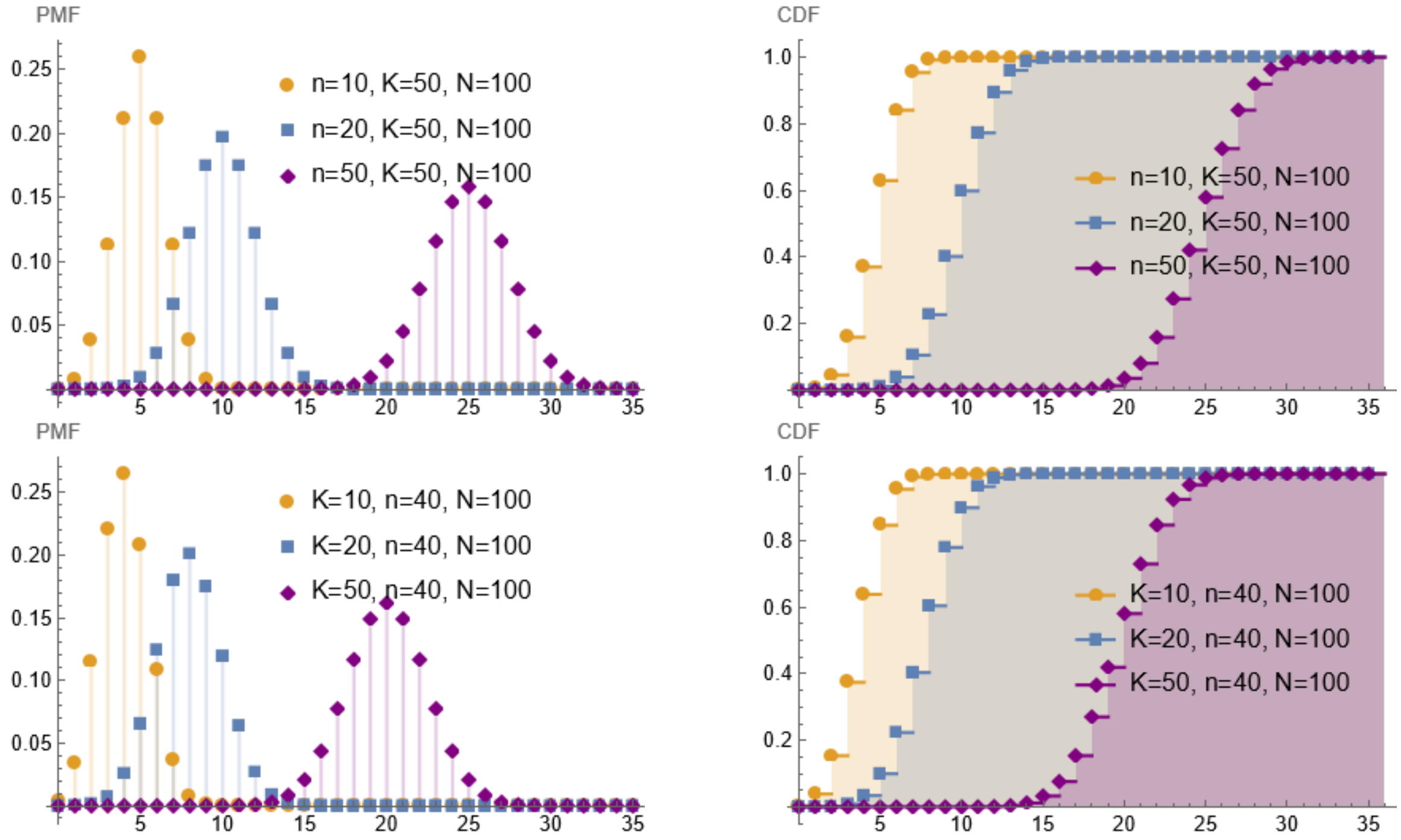

**Figure 10.9.** PMFs (left) and CDFs (right) of the hypergeometric distribution. The curve of the hypergeometric distribution depends on the parameters $n$, $K$ and $N$.

**Definition:** If $X$ is a hypergeometric RV with parameters $N$, $K$, and $n$, then

$$\mu = E[X] = np \text{ and } \sigma^2 = V(X) = np(1-p)\left(\frac{N-n}{N-1}\right), \tag{10.60}$$

where $p = K/N$.

For a hypergeometric RV, $E[X]$ is similar to the mean of a binomial RV. Also, $V(X)$ differs from the result for a binomial RV only by the following term.

**Definition (Finite Population Correction Factor):** The term in the variance of a hypergeometric RV

$$\frac{N-n}{N-1}, \tag{10.61}$$

is called the finite population correction factor.

***Example 10.27***

Dawn Corporation has 15 employees who hold managerial positions. Of them, 6 are females and 9 are males. The company is planning to send 4 of these 15 managers to a conference. If 4 managers are randomly selected out of 15,
(a) find the probability that all 4 of them are females
(b) find the probability that at most 1 of them is a female

***Solution***

Let the selection of a female be called a success and the selection of a male be called a failure.
(a)
From the given information,
$N$ = total number of managers in the population = 15
$K$ = number of successes (females) in the population = 6
$n$ = number of selections (sample size) = 4
$x$ = number of successes (females) in four selections = 4

```
N[Probability[x==4,x\[Distributed]HypergeometricDistribution[4,6,15]]]
0.010989
```

(b)
The probability that at most 1 of them is a female.

$N$ = total number of managers in the population = 15
$K$ = number of successes (females) in the population = 6
$n$ = number of selections (sample size) = 4
$x$ = number of successes (females) in four selections (at most 1)

```
N[Probability[x<=1,x\[Distributed]HypergeometricDistribution[4,6,15]]]
0.461538
```

***Example 10.28***

Obtain a table of probabilities for the RV $X$ ~Hypergeometric Distribution ($N = 30, K = 12, n = 10$). Plot the probabilities for each value of $X$.

***Solution***

```
probabilitydistribution=Table[
  {
   i,
   N[
    Probability[x==i,x\[Distributed]HypergeometricDistribution[10,12,30]]
    ]
   },
  {i,1,15}
  ]

{{1,0.0194189},{2,0.0961234},{3,0.233026},{4,0.305847},{5,0.225856},{6,0.0941068},{7,0.02151
01},{8,0.00252072},{9,0.000131802},{10,2.1967*10-6},{11,0.},{12,0.},{13,0.},{14,0.},{15,0.}}
 ListPlot[
 probabilitydistribution,
 PlotStyle->Purple,
```

```
 Filling->Axis,
 Mesh->All,
 ImageSize->200,
 AxesLabel->{"X","probability"}
 ]
```

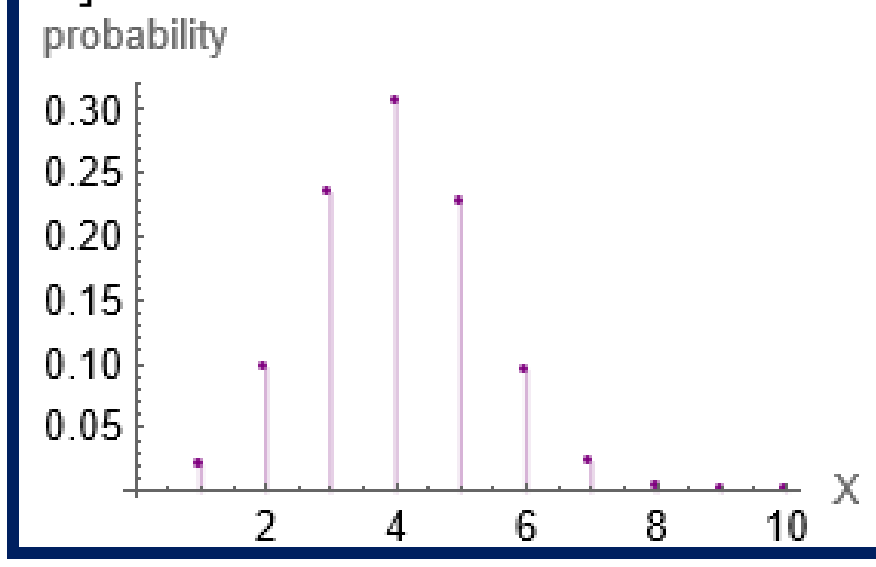

### 10.5.7 Discrete Uniform Distribution

The discrete uniform distribution is a symmetric probability distribution wherein a finite number of values are equally likely to be observed; every one of $n$ values has equal probability $1/n$. A simple example of the discrete uniform distribution is throwing a fair dice. The possible values are 1, 2, 3, 4, 5, 6, and each time the die is thrown the probability of a given score is $1/6$. If two dice are thrown and their values added, the resulting distribution is no longer uniform because not all sums have equal probability.

#### Parameters of the discrete uniform distribution

- The parameters of the discrete uniform distribution are typically denoted as follows: $a$ represents the lowest value that the RV can take. $b$ represents the highest value that the RV can take. Simply, we have consecutive integers $\{a, a+1, a+2, \ldots, b\}$. The PMF of the discrete uniform distribution assigns an equal probability to each value within the range. The PMF is defined as:

$$P(X = x) = \frac{1}{b - a + 1}, \qquad \text{for} \quad a \leq x \leq b,$$

  where $X$ is the RV, $x$ is a specific value within the range, and $P(X = x)$ represents the probability of $X$ taking the value $x$.

- In some cases, the discrete uniform distribution is defined using the parameter $n$, which represents the total number of possible outcomes or elements in the range. It is important to note that when using $n$ as the parameter, the range of outcomes is implicitly defined as $\{1, 2, \ldots, n\}$, and each outcome has an equal probability of $1/n$. The minimum value $a$ can be defined as 1, representing the lowest value in the range. The maximum value $b$ can be defined as $n$, representing the highest value in the range. Using these parameters, the PMF of the discrete uniform distribution becomes:

$$P(X = x) = \frac{1}{n}, \qquad \text{for} \quad 1 \leq x \leq n.$$

**Definition (Discrete Uniform Distribution):** A RV $X$ has a discrete uniform distribution if each of the $n$ values in its range, $\{x_1, x_2, \ldots, x_n\}$, has equal probability (see Figure 10.10). Its PMF is of the form,

$$f_X(x) = \begin{cases} \frac{1}{n}; & x = 1, 2, \ldots n \\ 0; & \text{otherwise}, \end{cases} \tag{10.62}$$

or

$$f_X(x) = \begin{cases} \frac{1}{b - a + 1}; & x = a, a+1, a+2, \ldots b, \\ 0; & \text{otherwise}. \end{cases} \tag{10.63}$$

### The assumptions of discrete uniform distribution

- Finite range:
  The distribution assumes a finite range of values, typically denoted by $\{x_1, x_2, \ldots, x_n\}$. These values are equally likely to occur, and there is no preference or bias towards any specific value within the range.
- Equal probability:
  Each value in the range has an equal probability of occurring. This means that the PMF assigns the same probability to each outcome.
- Independence:
  The occurrences of the values within the range are assumed to be independent of each other. In other words, the outcome of one observation does not affect the probabilities or outcomes of subsequent observations.
- Exhaustive and mutually exclusive:
  The values in the range are assumed to be exhaustive and mutually exclusive. This means that one and only one of the possible values can occur at any given time.
- Discreteness:
  The distribution deals with discrete values rather than continuous values. The RV can only take on specific values within the range, and there are no intermediate values.

### Discrete uniform distribution has several applications in various fields.

- Random number generation:
  The discrete uniform distribution is often used as a basis for generating random numbers. Random number generators aim to produce sequences of numbers that appear uniformly distributed over a specified range. The discrete uniform distribution ensures that each number within the range has an equal chance of being selected, making it useful for simulations.
- Monte Carlo simulations:
  The discrete uniform distribution is frequently employed in Monte Carlo simulations, which use random sampling to model and analyze complex systems. Monte Carlo simulations are widely used in physics, engineering, finance, and other fields. The uniform distribution is often utilized to generate RVs within a specific range to simulate uncertain events or variables in the system being studied.

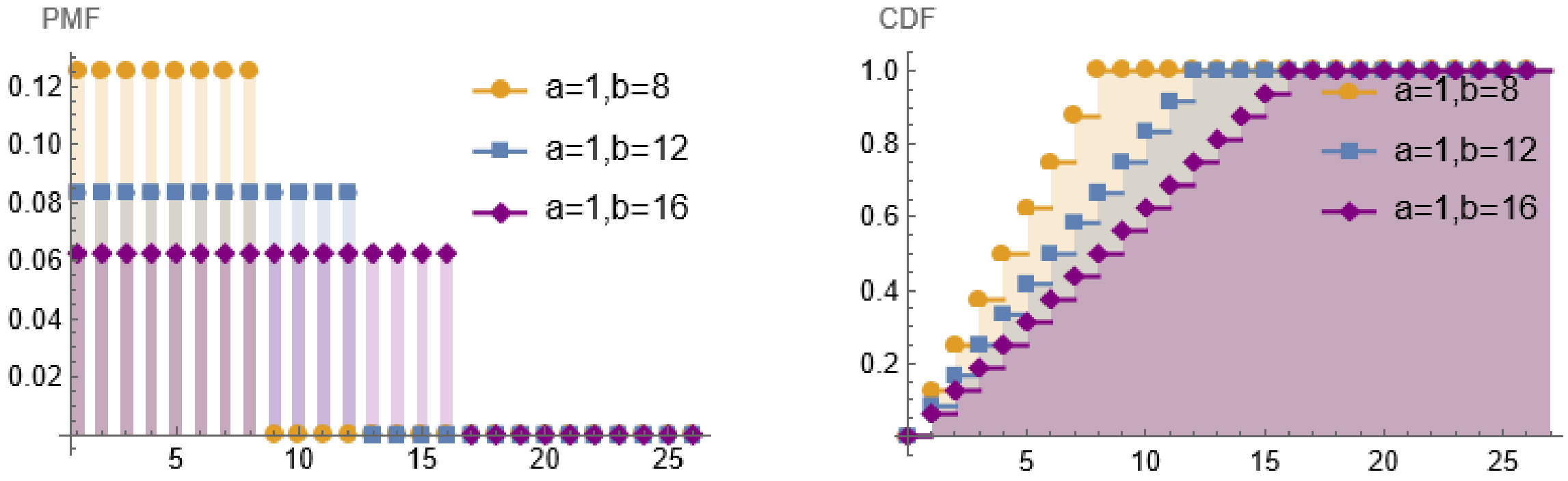

**Figure 10.10.** PMFs (left) and CDFs (right) of the discrete uniform distribution. The curve of the discrete uniform distribution depends on the parameters $a$ and $b$.

**Theorem 10.25:** Suppose that $X$ is a discrete uniform RV on the consecutive integers $\{a, a+1, a+2, \ldots, b\}$, for $a \leq b$. The mean of $X$ is

$$\mu = E[X] = \frac{b+a}{2}. \tag{10.64}$$

The variance of $X$ is

$$\sigma^2 = \frac{(b-a+1)^2 - 1}{12}. \tag{10.65}$$

Suppose that $X$ is a discrete uniform RV on the consecutive integers $\{1,2,\ldots,n\}$ i.e., in this case $a = 1$ and $b = n$. The mean of $X$ is

$$E[X] = \frac{n+1}{2}. \tag{10.66}$$

The variance of $X$ is

$$\mathrm{Var}(X) = \frac{n^2-1}{12}. \tag{10.67}$$

The MGF is

$$M_X(t) = \frac{e^t}{n}\frac{e^{nt}-1}{e^t-1}. \tag{10.68}$$

**Proof:**

Mean,

$$\begin{aligned}
\mu_1' &= E[X] \\
&= \sum_{x=1}^{n} x\frac{1}{n} \\
&= \frac{1}{n}\sum_{x=1}^{n} x \\
&= \frac{1}{n}[1+2+3+\cdots+n] \\
&= \frac{n(n+1)}{2n} \\
&= \frac{(n+1)}{2}.
\end{aligned}$$

Variance

$$V(X) = \mu_2' - (\mu_1')^2 = E[X^2] - (E[X])^2,$$

and

$$\begin{aligned}
\mu_2' &= E[X^2] \\
&= \sum_{x=1}^{n} x^2\frac{1}{n} \\
&= \frac{1}{n}\sum_{x=1}^{n} x^2 \\
&= \frac{1}{n}(1^2+2^2+\cdots+n^2) \\
&= \frac{n(n+1)(2n+1)}{6n} \\
&= \frac{(n+1)(2n+1)}{6}.
\end{aligned}$$

Therefore,

$$\begin{aligned}
V(X) &= \frac{(n+1)(2n+1)}{6} - \left(\frac{n+1}{2}\right)^2 \\
&= \frac{n+1}{2}\left[\frac{4n+2-3n-3}{6}\right] \\
&= \left(\frac{n+1}{2}\right)\left(\frac{n-1}{6}\right)
\end{aligned}$$

$$= \frac{n^2 - 1}{12}.$$

MGF,

$$\begin{aligned} M_X(t) &= E(e^{tX}) \\ &= \sum_{x=1}^{n} e^{tX} \frac{1}{n} \\ &= \frac{1}{n} \sum_{x=1}^{n} e^{tX} \\ &= \frac{1}{n}(e^t + e^{2t} + e^{3t} + \cdots + e^{nt}) \\ &= \frac{1}{n} e^t \left(1 + e^t + e^{2t} + \cdots + e^{(n-1)t}\right) \\ &= \frac{e^t}{n} \frac{e^{nt} - 1}{e^t - 1}. \end{aligned}$$

■

***Example 10.29***

Let us consider an example where we roll a fair six-sided die. In this case, the possible outcomes are the numbers 1, 2, 3, 4, 5, and 6, and each outcome has a probability of $1/6$ since the die is fair.
(a) What is the probability of rolling a 3?
(b) What is the probability of rolling an even number?
(c) What is the probability of rolling a number greater than 6?

***Solution***

(a)

Since there is only one 3 on the die, the probability of rolling a 3 is $1/6$.

```
Probability[x==3,x\[Distributed]DiscreteUniformDistribution[{1,6}]]
1/6
```

(b)

There are three even numbers on the die (2, 4, and 6), so the probability of rolling an even number is $3/6$ or $1/2$.

```
Probability[x==2||x==4||x==6,x\[Distributed]DiscreteUniformDistribution[{1,6}]]
1/2
```

(c)

Since the die only has numbers from 1 to 6, the probability of rolling a number greater than 6 is 0.

```
Probability[x>6,x\[Distributed]DiscreteUniformDistribution[{1,6}]]
0
```

***Example 10.30***

Suppose we have a random number generator that generates integers from 5 to 20, inclusive. Each number between 5 and 20 has an equal probability of being generated, which is $\frac{1}{20-5+1} = \frac{1}{16} = 0.0625$.

***Solution***

```
probabilitydistribution=Table[
  {
   i,
   N[
    Probability[x==i,x\[Distributed]DiscreteUniformDistribution[{5,20}]]
    ]
   },
  {i,1,25}
  ]
```

```
{{1,0.},{2,0.},{3,0.},{4,0.},{5,0.0625},{6,0.0625},{7,0.0625},{8,0.0625},{9,0.0625},{10,0.06
25},{11,0.0625},{12,0.0625},{13,0.0625},{14,0.0625},{15,0.0625},{16,0.0625},{17,0.0625},{18,
0.0625},{19,0.0625},{20,0.0625},{21,0.},{22,0.},{23,0.},{24,0.},{25,0.}}
 ListPlot[
 probabilitydistribution,
 PlotStyle->Purple,
 Filling->Axis,
 Mesh->All,
 ImageSize->200,
 AxesLabel->{"X","probability"}
 ]
```

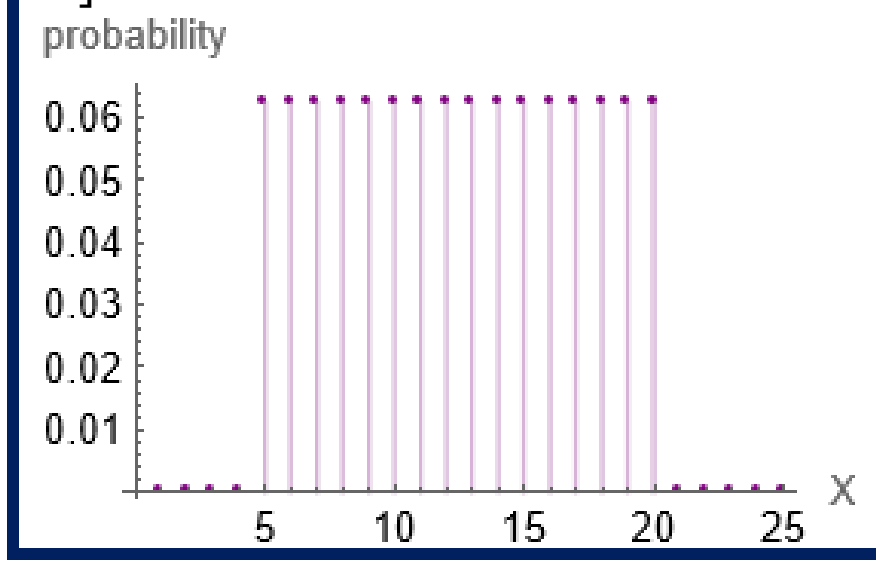

# CHAPTER 11

# MATHEMATICA LAB: DISCRETE RANDOM VARIABLES AND PROBABILITY DISTRIBUTIONS

In this chapter, we delve into the world of discrete random variables, which play a crucial role in modeling various phenomena with countable outcomes. We explore their PMFs, CDFs, and MGFs, while leveraging the power of Mathematica to perform computations and gain insights.

- Mathematica provides a range of built-in functions designed to facilitate the analysis of discrete random variables. These functions include `Probability`, `NProbability`, `PDF`, `CDF`, `Expectation`, `NExpectation`, `MomentGeneratingFunction` and `CentralMomentGeneratingFunction`. The `Probability` function allows users to estimate the likelihood of specific events occurring for discrete random variables. `NProbability` enables numerical estimation of probabilities. `PDF` and `CDF` assist in determining the probability density and cumulative probability of discrete random variables, respectively. `Expectation` and `NExpectation` functions calculate the expected value or mean of a discrete random variable, providing insights into its central tendency. `MomentGeneratingFunction` and `CentralMomentGeneratingFunction` allow for the calculation of higher-order moments and central moments, offering a more comprehensive understanding of the distribution of the random variable. By utilizing these functions, mathematicians and data scientists can efficiently analyze and extract valuable insights from discrete random variables.
- Moreover, Mathematica offers a comprehensive set of built-in functions to handle various probability distributions effortlessly. In this chapter, we also explore six essential probability distributions: Binomial distribution, Poisson distribution, negative binomial distribution, geometric distribution, hypergeometric distribution, and discrete uniform distribution.

Therefore, we divided this chapter into seven units to cover the above topics. In the subsequent units, we delve deeper into each distribution, exploring their properties, statistical measures, and applications.

In the following table, we list the built-in functions that are used in this chapter.

| | | |
|---|---|---|
| `Distributed` | `Expectation` | `BinomialDistribution` |
| `Conditioned` | `NExpectation` | `PoissonDistribution` |
| `Probability` | `MomentGeneratingFunction` | `NegativeBinomialDistribution` |
| `NProbability` | `CentralMomentGeneratingFunction` | `GeometricDistribution` |
| `PDF` | `EstimatedDistribution` | `HypergeometricDistribution` |
| `CDF` | | `DiscreteUniformDistribution` |

**Chapter 11 Outline**

# UNIT 11.1

# DISCRETE RANDOM VARIABLES

**Mathematica Examples 11.1** Discrete Random Variable (Manual study`)

Input

```
(* In this code, we start by defining the probability mass function (PMF) as a list of probabilities for each outcome of the random variable X. Then, we define the random variable X as a list of corresponding values. Next, we compute the expected value by summing the product of each outcome with its corresponding probability. The variance is computed similarly, but with the squared difference between each outcome and the mean. The cumulative distribution function (CDF) is obtained by accumulating the probabilities of the pmf. Finally, we define a function pXleqk[k] to calculate the probability that X is less than or equal to a given value k. We demonstrate this by calculating P(X<=3) and display the results using the Print function: *)

(* Define the PMF: *)
pmf={1/6,1/6,1/6,1/6,1/6,1/6};

(* Define the random variable: *)
X={1,2,3,4,5,6};

(* Compute the expected value: *)
mean=Sum[
    pmf[[i]]*X[[i]],
    {i,Length[X]}
    ];

(* Compute the variance: *)
variance=Sum[
    pmf[[i]]*(X[[i]]-mean)^2,
    {i,Length[X]}
    ];

(* Compute the CDF: *)
cdf=Accumulate[pmf];

(* Compute the probability of X<=k: *)
pXleqk[k_]:=cdf[[k]];

(* Display results: *)
Print["Probability Mass Function (PMF): ",pmf];
Print["Random Variable (X): ",X];
Print["Expected Value: ",mean];
Print["Variance: ",variance];
Print["Cumulative Distribution Function (CDF): ",cdf];
Print["P(X <= k): ",pXleqk[3]];
```

```
Output   Probability Mass Function (PMF):   {1/6,1/6,1/6,1/6,1/6,1/6}
Output   Random Variable (X):   {1,2,3,4,5,6}
Output   Expected Value:   7/2
Output   Variance:   35/12
Output   Cumulative Distribution Function (CDF):   {1/6,1/3,1/2,2/3,5/6,1}
Output   P(X <= k):   1/2
```

Input

```
(* In this code, we define a discrete random variable x with a sample space of {1,2,3,4,5} and the corresponding PMF p with values {0.1,0.2,0.3,0.2,0.2}. We then calculate the mean and variance of the random variable using the formulas mean=Sum[x[[i]]*p[[i]],{i,Length[x]}] and variance = Sum[(x[[i]] - mean)^2 *p[[i]] ,{i,Length[x]}]. Next, we calculate the CDF using cdf=Accumulate[p], which gives the cumulative probabilities. Finally, we plot the PMF and the CDF using ListPlot. The Transpose[{x,p}] command combines the x and p lists into a single list for plotting: *)

(* Define a discrete random variable: *)
x={1,2,3,4,5};  (* Sample space. *)
p={0.1,0.2,0.3,0.2,0.2};  (* PMF values. *)

(* Calculate the mean: *)
mean=Sum[
  x[[i]]*p[[i]],
  {i,Length[x]}
  ]

(* Calculate the variance: *)
variance=Sum[
  (x[[i]]-mean)^2*p[[i]],
  {i,Length[x]}]

(* Calculate the CDF: *)
cdf=Accumulate[p]

(* Plot the PMF: *)
ListPlot[
 Transpose[{x,p}],
 PlotRange->All,
 PlotStyle->Directive[PointSize[0.03],Purple],
 AxesLabel->{"x","P(X = x)"},
 PlotLabel->"Probability Mass Function",
 ImageSize->250
 ]
(* Plot the CDF: *)
ListPlot[
 Transpose[{x,cdf}],
 PlotRange->All,
 PlotStyle->Directive[PointSize[0.03],Purple],
 AxesLabel->{"x","P(X <= x)"},
 PlotLabel->"Cumulative Distribution Function",
 ImageSize->250
 ]
```

Output 3.2

Output 1.56

Output {0.1,0.3,0.6,0.8,1.}

Output

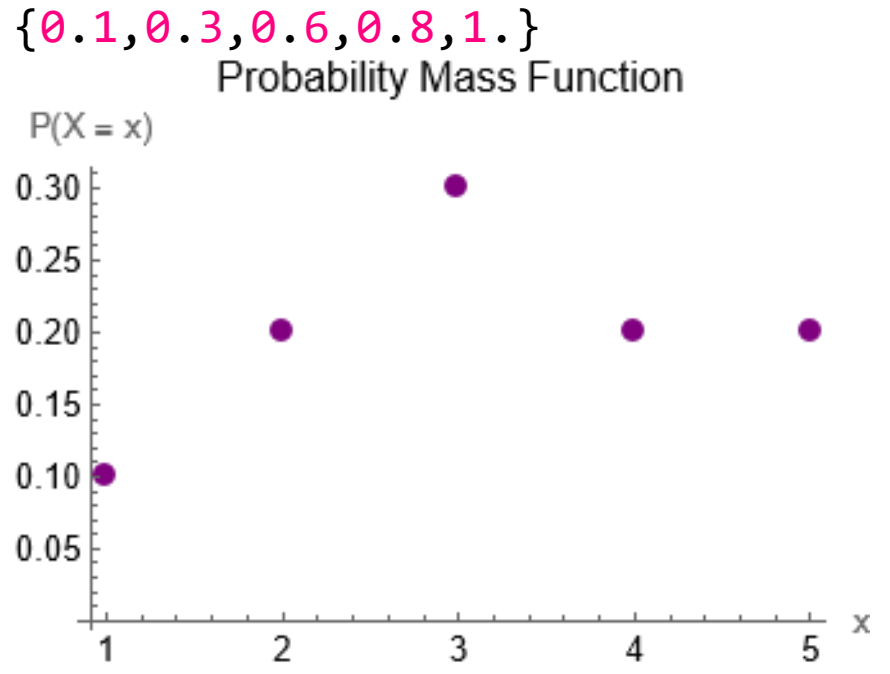

Output

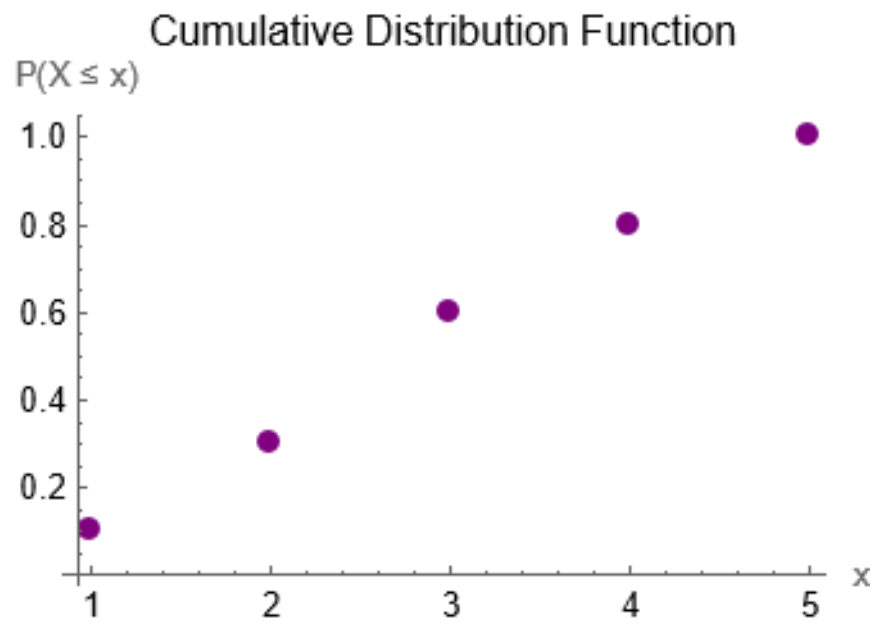

Input

```
(* This code demonstrates how to define a discrete random variable using the Module function and calculate its PMF, CDF, expected value, variance, and standard deviation. In this code, the Module function is used to encapsulate the variables and calculations. The x variable represents the possible values of the random variable, and the probabilities variable represents the probabilities associated with each value. The PMF is calculated by creating a list pmf that contains the probabilities corresponding to each value. The CDF is computed using the Accumulate function on the PMF. The expected value is calculated by multiplying each value of x with its corresponding probability and summing them up. The variance is computed by summing the squared deviations of the random variable from its expected value, and the standard deviation is obtained by taking the square root of the variance: *)

(* Define the discrete random variable and its probabilities: *)
Module[
 {x,probabilities},
 x={1,2,3,4};(* Possible values of the random variable. *)
 probabilities={0.2,0.3,0.4,0.1};(* Probabilities of each value. *)
 (* Calculate the PMF: *)
 pmf=Table[probabilities[[i]],{i,1,Length[x]}];
 (* Calculate the CDF: *)
 cdf=Accumulate[pmf];
 (* Calculate the expected value: *)
 expectedValue=Total[x*pmf];
 (* Calculate the variance: *)
 variance=Total[((x-expectedValue)^2)*pmf];
 (* Calculate the standard deviation: *)
 standardDeviation=Sqrt[variance];
 (* Print the results: *)
 Print["Discrete Random Variable: ",x];
 Print["PMF: ",pmf];
 Print["CDF: ",cdf];
 Print["Expected Value: ",expectedValue];
 Print["Variance: ",variance];
 Print["Standard Deviation: ",standardDeviation];]
```

Output Discrete Random Variable: {1,2,3,4}
Output PMF: {0.2,0.3,0.4,0.1}
Output CDF: {0.2,0.5,0.9,1.}
Output Expected Value: 2.4
Output Variance: 0.84
Output Standard Deviation: 0.916515

Input

```
(* In this example, the DiscreteRandomVariable function takes a list of data as input and returns a list containing the PMF, CDF, expected value, variance, and standard deviation of the random variable defined by the data. The PMF and CDF are defined using local functions pmf and cdf, respectively, within the Module function. The expected value, variance, and standard deviation are calculated using summations over the data: *)

DiscreteRandomVariable[data_]:=Module[
```

```
    {pmf,cdf,expectedValue,variance,standardDeviation},
    (* Calculate the PMF: *)
    pmf[x_]:=Count[data,x]/Length[data];
    (* Calculate the CDF: *)
    cdf[x_]:=Sum[pmf[value],{value,Min[data],x}];
    (* Calculate the expected value: *)
    expectedValue=Sum[x*pmf[x],{x,Min[data],Max[data]}];
    (* Calculate the variance: *)
    variance=Sum[(x-expectedValue)^2*pmf[x],{x,Min[data],Max[data]}];
    (* Calculate the standard deviation: *)
    standardDeviation=Sqrt[variance];
    (* Return a list containing the PMF,CDF,expected value,variance,and standard
deviation: *)
    {pmf,cdf,expectedValue,variance,standardDeviation}
    ]

(* Example usage: *)
data={1,2,2,3,4,4,4,5};
{pmf,cdf,mean,var,sd}=DiscreteRandomVariable[data];

(* Output the results: *)
pmf/@Range[1,5]
cdf/@Range[1,5]
mean
var
sd
Output  {1/8,1/4,1/8,3/8,1/8}
Output  {1/8,3/8,1/2,7/8,1}
Output  25/8
Output  103/64
Output  √103/8
```

| | |
|---|---|
| `Distributed[x,dist]` | or x \[Distributed] dist asserts that the random variable x is distributed according to the probability distribution dist. |
| `Distributed[{x1,x2,…},dist]` | or {x1,x2,…} \[Distributed] dist asserts that the random vector {x1,x2,…} is distributed according to the multivariate probability distribution dist. |
| `Conditioned[expr,cond]` | or expr \[Conditioned] cond represents expr conditioned by the predicate cond. |
| `Probability[pred,x\[Distributed]dist]` | gives the probability for an event that satisfies the predicate pred under the assumption that x follows the probability distribution dist. |
| `Probability[pred,x\[Distributed]data]` | gives the probability for an event that satisfies the predicate pred under the assumption that x follows the probability distribution given by data. |
| `Probability[pred,{x1,x2,…}\[Distributed]dist]` | gives the probability that an event satisfies pred under the assumption that {x1,x2,…} follows the multivariate distribution dist. |
| `Probability[pred,{x1\[Distributed]dist1,x2\[Distributed]dist2,…}]` | gives the probability that an event satisfies pred under the assumption that x1, x2, … are independent and follow the distributions dist1, dist2, …. |
| `Probability[pred1\[Conditioned]pred2,…]` | gives the conditional probability of pred1 given pred2. |
| `NProbability[pred,x\[Distributed]dist]` | gives the numerical probability for an event that satisfies the predicate pred under the assumption that x follows the probability distribution dist. |
| `NProbability[pred,{x1,x2,…}\[Distributed]dist]` | gives the numerical probability that an event satisfies pred under the assumption that {x1,x2,…} follows the multivariate distribution dist. |
| `NProbability[pred,{x1\[Distributed]dist1,x2\[Distributed]dist2,…}]` | gives the numerical probability that an event satisfies pred under the assumption that x1, x2, … are independent and follow the distributions dist1, dist2, …. |
| `NProbability[pred1\[Conditioned]pred2,…]` | gives the numerical conditional probability of pred1 given pred2. |

***Mathematica Examples 11.2*** Distributed, Conditioned, Probability and NProbability

```
Input   (* For a distribution specified by a list, Probability computes relative frequencies:
        *)
        data={1,2,3,4,5,6,7,8,9,10,11};
        Probability[3<=x<=7,x\[Distributed]data]
        Length[Table[data[[i]],{i,3,7}]]/Length[data]
Output  5/11
Output  5/11

Input   dist={1,2,3,4,5,6};

        (*  Probability of x is less than 6 and greater than 1: *)
        p1=Probability[x<6&&x>1,x\[Distributed]dist];

        (*  Probability of x is less than 5 and greater than 1: *)
        p2=Probability[x<5&&x>1,x\[Distributed]dist];

        (*  Probability of x is less than 5 or x is less than 3: *)
        p3=Probability[x<5||x<3,x\[Distributed]dist];

        (*  Probability of x is less than 4 or x is less than 3: *)
        p4=Probability[x<4||x<3,x\[Distributed]dist];

        (*  Probability of (x is less than 4 or x is less than 3) and x is greater than 1:
        *)
        p5=Probability[(x<4||x<3)&&x>1,x\[Distributed]dist];

        (*  Probability of (x is less than or equal to 4 or x is less than 3) and x is greater
        than 1: *)
        p6=Probability[(x<=4||x<3)&&x>1,x\[Distributed]dist];

        (* Probability of x being equal to 3: *)
        p7=Probability[x==3,x\[Distributed]dist];

        (* Probability of x being an even number: *)
        p8=Probability[EvenQ[x],x\[Distributed]dist];

        (* Probability of x being divisible by 3: *)
        p9=Probability[Mod[x,3]==0,x\[Distributed]dist];

        (* Probability of x being less than or equal to 2 or greater than or equal to 5: *)
        p10=Probability[x<=2||x>=5,x\[Distributed]dist];

        (* Probability of x being an odd number and less than or equal to 4: *)
        p11=Probability[OddQ[x]&&x<=4,x\[Distributed]dist];

        (* Probability of x being a prime number: *)
```

```
        p12=Probability[PrimeQ[x],x\[Distributed]dist];

        (* Probability of x being between 2 and 4 (inclusive): *)
        p13=Probability[2<=x<=4,x\[Distributed]dist];

        (* Probability of x being an even number or a multiple of 3: *)
        p14=Probability[EvenQ[x]||Mod[x,3]==0,x\[Distributed]dist];

        (* Probability of x being a multiple of 4: *)
        p15=Probability[Mod[x,4]==0,x\[Distributed]dist];

        (* Probability of x being an even number and less than 4: *)
        p16=Probability[EvenQ[x]&&x<4,x\[Distributed]dist];

        (* Probability of x being a perfect square: *)
        p17=Probability[Sqrt[x]\[Element]Integers,x\[Distributed]dist];

        (* Probability of the square of x being greater than 10: *)
        p18=Probability[x^2>10,x\[Distributed]dist];

        (* Probability of the logarithm of x being greater than 2: *)
        p19=Probability[Log[x]>2,x\[Distributed]dist];

        (* Probability of the reciprocal of x being less than 0.2: *)
        p20=Probability[1/x<0.2,x\[Distributed]dist];

        {p1,p2,p3,p4,p5,p6,p7,p8,p9,p10,p11,p12,p13,p14,p15,p16,p17,p18,p19,p20}
Output  {2/3,1/2,2/3,1/2,1/3,1/2,1/6,0,1/3,2/3,0,0,1/2,1/3,1/6,0,1/3,1/2,0,1/6}

Input   (* The probability of an impossible event is 0: *)
        Probability[x==7,x\[Distributed]{1,2,3,4,5,6}]
Output  0

Input   (* The probability of a certain event is 1: *)
        Probability[1<=x<=6,x\[Distributed]{1,2,3,4,5,6}]
Output  1

Input   (* The probability of an arbitrary event must lie between 0 and 1: *)
        Probability[1<x<4,x\[Distributed]{1,2,3,4,5,6}]
Output  1/3

Input   (* The probability of a discrete univariate distribution is given by the PDF: *)
        PDF[PoissonDistribution[m],5]
        Probability[x==5,x\[Distributed]PoissonDistribution[m]]
Output  1/120 E^-m m^5
Output  1/120 E^-m m^5

Input   (* Compute the probability of a simple event: *)
        Probability[x<=5,x\[Distributed]PoissonDistribution[m]]

        Simplify[
         Total[
          Table[
           Probability[x==i,x\[Distributed]PoissonDistribution[m]],
           {i,0,5}
           ]
          ]
         ]
Output  1/120 E^-m (120+120 m+60 m^2+20 m^3+5 m^4+m^5)
Output  1/120 E^-m (120+120 m+60 m^2+20 m^3+5 m^4+m^5)
```

```
Input   (* Compute the probability of nonlinear and logical combination of inequalities: *)
        Probability[x^2>3\[Or]Abs[x]<1,x\[Distributed]{1,2,3}]
Output  2/3

Input   (* With no Assumptions, conditions are generated: *)
        Probability[x<a,x\[Distributed]DiscreteUniformDistribution[{2,8}]]
```

Output

$$\begin{cases} 1 & \text{Ceiling}[a] \geq 9 \\ \frac{1}{7}(-2+\text{Ceiling}[a]) & 3 \leq \text{Ceiling}[a] < 9 \\ 0 & \text{True} \end{cases}$$

```
Input   (* With Assumptions, a result valid under the given assumptions is returned: *)
        Probability[x<a,x\[Distributed]DiscreteUniformDistribution[{2,8}],Assumptions-
        >5<a<7]
        Assuming[5<a<7,Probability[x<a,x\[Distributed]DiscreteUniformDistribution[{2,8}]]]
```

Output

$$\begin{cases} 4/7 & a \leq 6 \\ 5/7 & \text{True} \end{cases}$$

Output

$$\begin{cases} 4/7 & a \leq 6 \\ 5/7 & \text{True} \end{cases}$$

```
Input   (* Use NProbability to find the numerical value for the probability of an event: *)
        dist=PoissonDistribution[1];
        {N[Probability[x<=2,x\[Distributed]dist]],NProbability[x<=2,x\[Distributed]dist]}
Output  {0.919699,0.919699}

Input   (* The code demonstrates the principle that the probability of the union of disjoint
        events is the sum of the individual probabilities, and when the events are not
        disjoint, the intersection probability needs to be subtracted to calculate the correct
        union probability: *)

        (* Define the individual events: *)
        event1=x<5;
        event2=x>10;
        event3=8<x<15;

        (* Define the probability distributions for the events: *)
        dist=DiscreteUniformDistribution[{0,20}];

        (* Calculate the individual probabilities: *)
        prob1=Probability[event1,x\[Distributed]dist];
        prob2=Probability[event2,x\[Distributed]dist];
        prob3=Probability[event3,x\[Distributed]dist];
        prob4=Probability[event2\[And]event3,x\[Distributed]dist];

        (* Calculate the union probability: *)
        unionProbability1=Probability[event1||event2,x\[Distributed]dist];

        (* The probability of the union of disjoint events is the sum of the individual
        probabilities: *)
        unionProbability2=prob1+prob2;

        (* Calculate the union probability: *)
        unionProbability3=Probability[event2||event3,x\[Distributed]dist];

        (* For non-disjoint events, one needs to subtract the probability of an intersection
        event: *)
        unionProbability4=prob2+prob3-prob4;
```

```
        (*Display the results*)
        Print["Probability of event 1: ",prob1];
        Print["Probability of event 2: ",prob2];
        Print["Probability of event 3: ",prob3];
        Print["Probability of event 4 (intersection event): ",prob4];

        Print["Probability of the union of disjoint events (event1 and event2):
        ",unionProbability1," , ",unionProbability2];
        Print["Probability of the union of non-disjoint events (event2 and event3):
        ",unionProbability3," , ",unionProbability4];
Output  Probability of event 1:  5/21
Output  Probability of event 2:  10/21
Output  Probability of event 3:  2/7
Output  Probability of event 4 (intersection event):  4/21
Output  Probability of the union of disjoint events (event1 and event2):  5/7, 5/7
Output  Probability of the union of non-disjoint events (event2 and event3):  4/7, 4/7

Input   (* Define the probability distribution for random variable X: *)
        d={1,2,3,4,5,6,7,8,9};

        (* Define the events  a and b: *)
        a=x>=2;
        b=x<4;

        (* Calculate the conditional probability of A given B: *)
        pAgivenB1=Probability[a\[Conditioned]b,x\[Distributed]d];
        pAgivenB2=Probability[a\[And]b,x\[Distributed]d]/Probability[b,x\[Distributed]d];
        pAgivenB1==pAgivenB2

        (* Calculate the conditional probability of B given A:*)
        pBgivenA3=Probability[b\[Conditioned]a,x\[Distributed]d];
        pBgivenA4=Probability[a\[And]b,x\[Distributed]d]/Probability[a,x\[Distributed]d];
        pBgivenA3==pBgivenA4

        (*Display the results*)
        Print["Conditional probability P(A|B) = ",pAgivenB1,", Conditional probability
        P(B|A) = ",pBgivenA3];
Output  True
Output  True
Output  Conditional probability P(A|B) =  2/3 , Conditional probability P(B|A) =  1/4

Input   (* The code generates a list of 1000 random values from a discrete uniform
        distribution between 1 and 6. It defines two events, A and B, where event A represents
        values greater than 3 and less than 5,and event B represents values greater than 2
        and less than 5. The code then estimates the conditional probability P(A|B) and
        P(B|A) using the Probability function with the sample data. It also calculates the
        frequency of each value in the sample data using Tally. The results are displayed,
        showing the frequency of each value in the sample data, the estimated conditional
        probability P(A|B), and the estimated conditional probability P(B|A).*)

        (* Generate a list of 1000 random values from a discrete uniform distribution: *)
        sampleData=RandomVariate[
           DiscreteUniformDistribution[{1,6}],
           1000
           ];

        (* Define events A and B: *)
        A[x_]:=5>x>3; (* Event A:5>x>3.*)
        B[x_]:=5>x>2; (* Event B:5>x>2.*)
```

```
        (* Estimate the conditional probability P(A|B): *)
        conditionalProbab=Probability[A[x]\[Conditioned]B[x],x\[Distributed]sampleData];

        (* Estimate the conditional probability P(B|A): *)
        conditionalProbba=Probability[B[x]\[Conditioned]A[x],x\[Distributed]sampleData];

        (* Calculate frequency of each value in the sample data: *)
        freq=Tally[sampleData];

        (* Display the results: *)
        Print["freq = ",freq] (* Display the frequency of each value in the sample data. *)
        Print["Estimated conditional probability P(A|B) = ",conditionalProbab] (* Display
        the estimated conditional probability P(A|B). *)
        Print["Estimated conditional probability P(B|A) = ",conditionalProbba] (* Display
        the estimated conditional probability P(B|A). *)
Output  freq =  {{2,168},{5,146},{3,166},{6,162},{1,177},{4,181}}
Output  Estimated conditional probability P(A|B) =  181/347
Output  Estimated conditional probability P(B|A) =  1

Input   (* In this example, the sample data represents the grades of 100 students, randomly
        chosen from the set {"A","B","C","D","F"}. The events A and B are defined based on
        the grades, where event A represents getting an A or B grade, and event B represents
        getting a B or C grade. The conditional probabilities P(A|B) and P(B|A) are estimated
        using the Probability function with the sample data. The frequency of each grade
        value in the sample data is calculated using Tally, and the results are displayed
        using Print: *)

        (*Generate sample data*)
        sampleData=RandomChoice[{"A","B","C","D","F"},100];

        (*Define events A and B*)
        A[x_]:=x=="A"||x=="B"; (* Event A: Getting an A or B grade. *)
        B[x_]:=x=="B"||x=="C"; (* Event B: Getting a B or C grade. *)

        (* Estimate the conditional probability P(A|B): *)
        conditionalProbab=Probability[A[x]\[Conditioned]B[x],x\[Distributed]sampleData];

        (* Estimate the conditional probability P(B|A): *)
        conditionalProbba=Probability[B[x]\[Conditioned]A[x],x\[Distributed]sampleData];

        (* Calculate frequency of each value in the sample data: *)
        freq=Tally[sampleData];

        (* Display the results: *)
        Print["freq = ",freq] (* Display the frequency of each value in the sample data. *)
        Print["Estimated conditional probability P(A|B) = ",conditionalProbab] (* Display
        the estimated conditional probability P(A|B).*)
        Print["Estimated conditional probability P(B|A) = ",conditionalProbba] (* Display
        the estimated conditional probability P(B|A). *)
Output  freq =  {{B,21},{A,20},{F,18},{C,21},{D,20}}
Output  Estimated conditional probability P(A|B) =  1/2
Output  Estimated conditional probability P(B|A) =  21/41

Input   (* A conditional probability is a ratio of two probabilities:*)
        expr1=x^2<30;
        expr2=x>1;
        Probability[expr1\[Conditioned]expr2,x\[Distributed]PoissonDistribution[2]]
        Probability[expr1&&expr2,x\[Distributed]PoissonDistribution[2]]/Probability[expr2,x
        \[Distributed]PoissonDistribution[2]]
Output  64/(15 (-3+E^2))
```

```
Output   64/(15 (-3+E^2))

Input    (* The conditional probability is 0 if the events are mutually exclusive:*)
         expr1=3<x<5;
         expr2=x>7;
         Probability[expr1\[Conditioned]expr2,x\[Distributed]PoissonDistribution[2]]
         Probability[expr2\[Conditioned]expr1,x\[Distributed]PoissonDistribution[2]]
Output   0
Output   0

Input    (* Discrete univariate distribution:*)
         p=Probability[x^2>12\[Conditioned]x>2,x\[Distributed]PoissonDistribution[m]]

         Plot[
          p,
          {m,0,10},
          PlotStyle->Purple,
          ImageSize->200
          ]
Output   (6-6 E^m+6 m+3 m^2+m^3)/(6-6 E^m+6 m+3 m^2)
Output
```

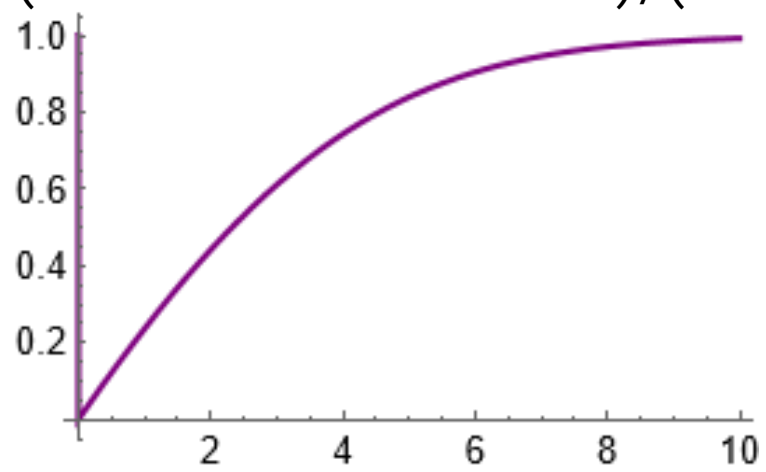

| | |
|---|---|
| `PDF[dist,x]` | gives the probability density function for the distribution dist evaluated at x. |
| `PDF[dist,{x1,x2,…}]` | gives the multivariate probability density function for a distribution dist evaluated at {x1,x2,…}. |
| `PDF[dist]` | gives the PDF as a pure function. |

***Mathematica Examples 11.3*** PDF

```
Input    (* The first line defines the Negative Binomial distribution with parameters 20 and
         1/3, the second line calculates the PDF for a specific value (5), and the third line
         defines a symbolic expression for the PDF as a function of the variable x: *)

         dist=NegativeBinomialDistribution[20,1/3]
         PDF[NegativeBinomialDistribution[20,1/3],5]
         PDF[NegativeBinomialDistribution[20,1/3],x]
Output   NegativeBinomialDistribution[20,1/3]
Output
```

$$\begin{cases} 2^{x}3^{-20-x}\text{Binomial}[19+x,19] & x>=0 \\ 0 & \text{True} \end{cases}$$

```
Output   453376/282429536481

Input    (* This code creates a Binomial distribution and plots its PDF over the range of 0
         to 20: *)

         pdf1=PDF[BinomialDistribution[n,p],x]
         dist=BinomialDistribution[20,1/2](* Create a Binomial distribution with n=20 and
         p=1/2. *)
         pdf=PDF[BinomialDistribution[20,1/2],x] (* Generate PDF of a Binomial distribution
         with n=20 and p=1/2. *)

         DiscretePlot[
          pdf,
          {x,0,20},
          ExtentSize->0.5,
```

```
 PlotStyle->Purple,
 ImageSize->200
 ]

DiscretePlot[
 pdf,
 {x,0,20},
 PlotStyle->Purple,
 ImageSize->200
 ]
```

Output $\begin{cases} (1-p)^{n-x}p^x \text{Binomial}[n,x] & 0 <= x <= n \\ 0 & \text{True} \end{cases}$

Output `BinomialDistribution[20,1/2]`

Output $\begin{cases} \frac{\text{Binomial}[20,x]}{1048576} & 0 <= x <= 20 \\ 0 & \text{True} \end{cases}$

Output

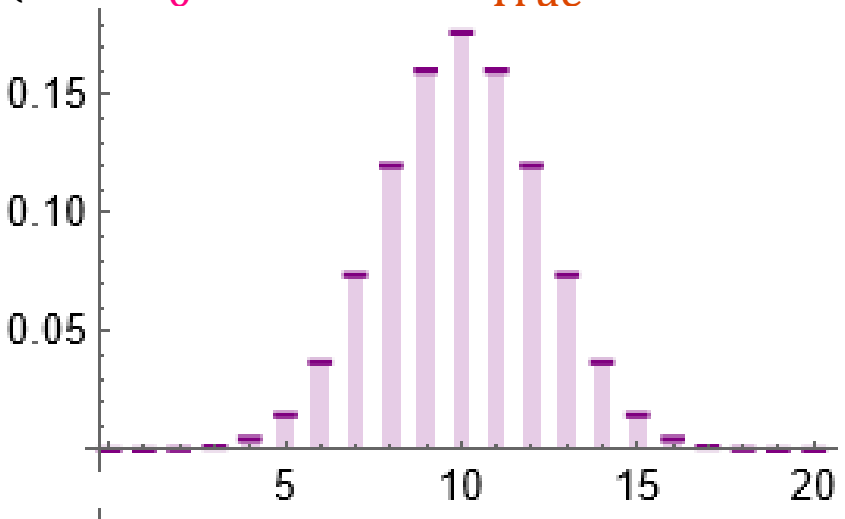

Output

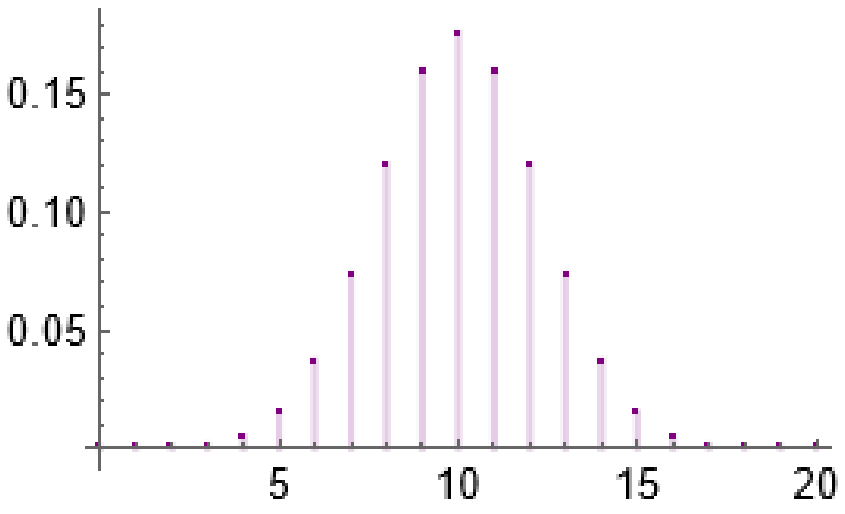

Input

```
(* The resulting 3D plot visualizes the PMF of the Multivariate Poisson distribution
with specific parameter values n=1, a=2,and b=3. The plot allows you to observe the
PMF values for different combinations of x and y within the specified ranges. Each
cell in the plot represents the PMF value for a specific x and y combination: *)

pdf1=PDF[MultivariatePoissonDistribution[n,{a,b}],{x,y}]
pdf=PDF[MultivariatePoissonDistribution[1,{2,3}],{x,y}]

DiscretePlot3D[
 pdf,
 {x,0,10},
 {y,0,10},
 ExtentSize->0.6,
 PlotStyle->Purple,
 ImageSize->200
 ]

DiscretePlot3D[
 pdf,
 {x,0,10},
 {y,0,10},
 PlotRange->All,
 PlotStyle->Purple,
 ImageSize->200
 ]
```

Output $\begin{cases} \frac{b^{-x+y}e^{-a-b-n}(-n)^x \text{HypergeometricU}[-x, 1-x+y, -((a\,b)/n)]}{x!\ y!} & x >= 0 \&\& y >= 0 \\ 0 & \text{True} \end{cases}$

Output $\begin{cases} \frac{(-1)^x 3^{-x+y} \text{HypergeometricU}[-x, 1-x+y, -6]}{e^6\ x!\ y!} & x >= 0 \&\& y >= 0 \\ 0 & \text{True} \end{cases}$

Output

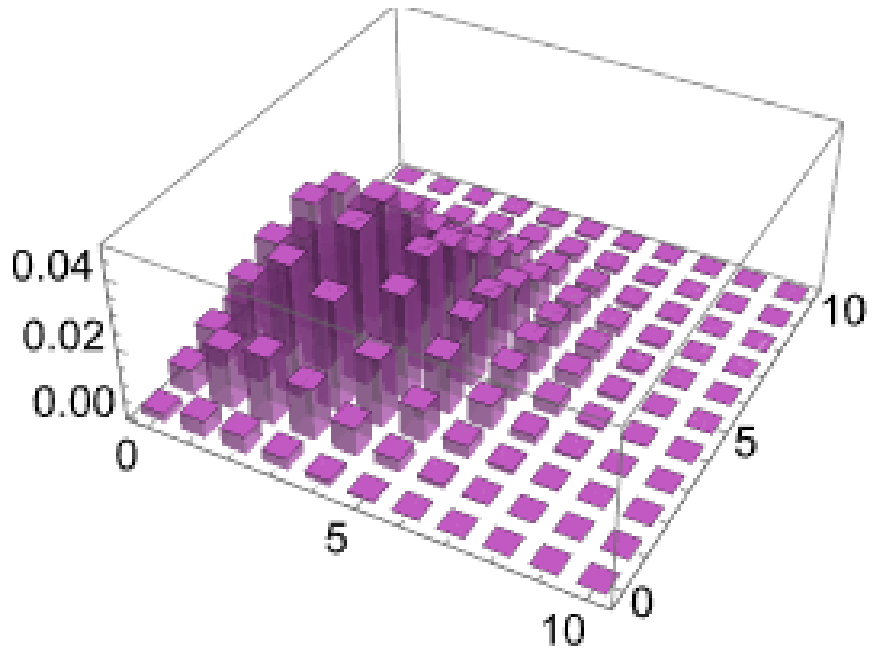

Output

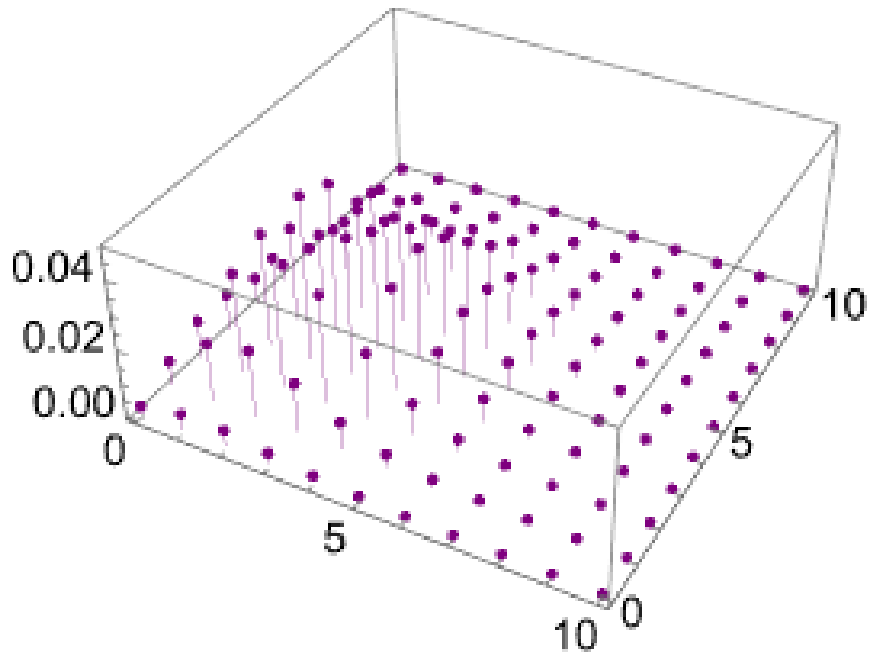

| | |
|---|---|
| Expectation[ expr, x\[Distributed]dist] | gives the expectation of expr under the assumption that x follows the probability distribution dist. |
| Expectation[ expr, x\[Distributed]data] | gives the expectation of expr under the assumption that x follows the probability distribution given by data. |
| Expectation[ expr,{x1,x2,…}\[Distributed]dist] | gives the expectation of expr under the assumption that {x1,x2,…} follows the multivariate distribution dist. |
| Expectation[ expr,{x1\[Distributed]dist1, x2\[Distributed]dist2,…}] | gives the expectation of expr under the assumption that x1, x2, … are independent and follow the distributions dist1, dist2, …. |
| Expectation[ expr\[Conditioned]pred,…] | gives the conditional expectation of expr given pred. |
| NExpectation[ expr,x\[Distributed] dist] | gives the numerical expectation of expr under the assumption that x follows the probability distribution dist. |
| NExpectation[expr, {x1,x2,…}\[Distributed]dist] | gives the numerical expectation of expr under the assumption that {x1,x2,…} follows the multivariate distribution dist. |
| NExpectation[expr, {x1\[Distributed]dist1, x2\[Distributed]dist2,…}] | gives the numerical expectation of expr under the assumption that x1, x2, … are independent and follow the distributions dist1, dist2, …. |
| `NExpectation[expr \[Conditioned]pred,…]` | gives the numerical conditional expectation of expr given pred. |

***Mathematica Examples 11.4*** Expectation and NExpectation

Input
```
(* The code showcases the steps involved in determining the expectation. It begins by defining the discrete distribution, specifying its parameters, and assigning it to a variable. The function g(x) is defined to represent the desired function for which the expectation is being calculated. The expectation is then computed using the Expectation function, which takes the function g(x) and the random variable x distributed according to the defined distribution. This function provides a direct method to calculate the expectation. Additionally, the code includes an explaining the concept of the expectation for a discrete distribution as the PDF-weighted sum: *)

d=PoissonDistribution[μ];
g[x_]=x^2;

directExpectation=Expectation[g[x],x\[Distributed]d];
manualExpectation=Sum[g[x] PDF[d, x], {x, 0, \[Infinity]}];

{directExpectation,manualExpectation}
```
Output
```
{μ+μ^2,μ+μ^2}
```

Input
```
(* Compute the expectation of an arbitrary expression: *)
Expectation[x^2+7 x+8,x\[Distributed]PoissonDistribution[μ]]
```
Output
```
8+8 μ+μ^2
```

Input
```
(* For a distribution specified by a list, Expectation is equivalent to using Mean. When the distribution is specified by a list of values, the probabilities associated with each value can be assumed to be uniform or equally likely. In this case, both Expectation and Mean will give you the same result because they essentially perform the same calculation: *)

data={1,2,3,4,5,6,7,8,9,10};
e=Expectation[x^2+3,x\[Distributed]data];

m=Mean[
    Table[
     x^2+3,
     {x,data}
     ]
    ];
{e,m}
```
Output
```
{83/2,83/2}
```

Input
```
(* The code demonstrates the equivalence between Expectation and Mean in finding the expected value of an expression in a distribution specified by a list. The code begins by generating a list r of random variates drawn from a Poisson distribution with a mean of 0.5. Then, it calculates the expectation of the expression x^2+3 x+11 by using the Expectation function and specifying x as distributed according to the list r. To compare this with the mean of the expression, the code explicitly constructs a table where each element is obtained by evaluating the expression for each value in the list r. The Mean function is then used to calculate the arithmetic average of these values: *)

r=RandomVariate[
    PoissonDistribution[0.5],
    1000
    ];
```

```
        e=Expectation[x^2+3 x+11,x\[Distributed]r];

        m=Mean[
            Table[
             x^2+3 x+11,
             {x,r}
             ]
            ];
        {e,m}
Output  {3327/250,3327/250}

Input   (* Use NExpectation to find the numerical value of an expectation: *)
        dist=PoissonDistribution[1];

        N[Expectation[E^(-x^2),x\[Distributed]dist]]
        NExpectation[E^(-x^2),x\[Distributed]dist]
Output  0.506591
Output  0.506591

Input   (* The code computes the conditional expectation of the expression x^2+1, considering
        a specific condition x>1/2. It utilizes the Expectation function and specifies the
        variable x to follow a Poisson distribution with a mean of 0.5. By using the
        conditioning operator \ [Conditioned], the code narrows down the calculation to
        values of x that satisfy the given condition. The result is the average value of
        x^2+1 under the specified condition: *)

        Expectation[x^2+1\[Conditioned]x>1/2,x\[Distributed]PoissonDistribution[0.5]]
Output  2.90612

Input   (* Find the conditional expectation with general nonzero probability conditioning:
        *)
        e=Expectation[x^2+12\[Conditioned]x>2,x\[Distributed]PoissonDistribution[λ]]

        Plot[
         e,
         {λ,0,10},
         PlotStyle->Purple,
         ImageSize->200
         ]

Output  -((2 (-12-13 λ-8 λ^2+E^λ (12+λ+λ^2)))/(2-2 E^λ+2 λ+λ^2))
Output
```

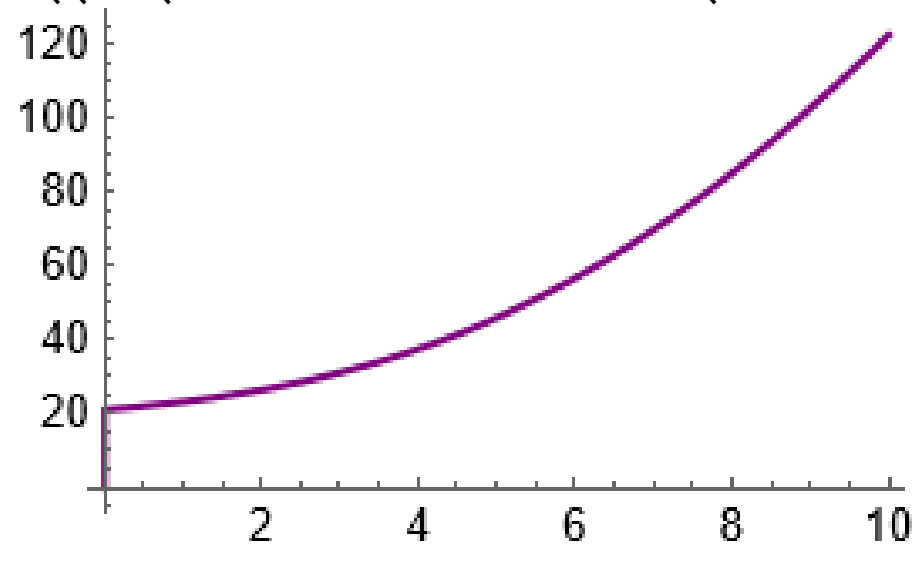

```
Input   (* A conditional expectation is defined by a ratio of expectation and probability:
        *)
        d=ExponentialDistribution[λ];
        expr=x^2;
        cond=x>1;
        Expectation[expr\[Conditioned]cond,x\[Distributed]d]==Expectation[Piecewise[{{expr,
        cond}}],x\[Distributed]d]/Probability[cond,x\[Distributed]d]
Output  True
```

```
Input    (* Suppose we have a random variable X that represents the outcome of rolling a fair
         six-sided die, and event A represents the event of rolling an even number. The
         possible values of X are {1,2,3,4,5,6},and the corresponding conditional
         probabilities P(X=x|A)are {0,1/3,0,1/3,0,1/3}. The conditional probability P(X=x|A)
         can be calculated using the following formula: P(X=x|A)=P(X=x ∩ A)/P(A). To calculate
         E[X|A], we apply the formula :E[X|A]=(1*0)+(2*1/3)+(3*0)+(4*1/3)+(5*0)+(6*1/3)
         =0+2/3+0+4/3+0+2=4. So, in this example, the conditional expectation E[X|A] is equal
         to 4, which represents the average value of X when the event A (rolling an even
         number) has occurred: *)

         (* Define the possible values of X: *)
         valuesOfX={1,2,3,4,5,6};

         (* Define the conditional probabilities P(X=x|A): *)
         conditionalProbs={0,1/3,0,1/3,0,1/3};

         (* Calculate the conditional expectation E[X|A]: *)
         conditionalExpectation=Sum[
             x*conditionalProbs[[x]],
             {x,valuesOfX}
             ];

         (* Print the result: *)
         Print["The conditional expectation E[X|A] is ",conditionalExpectation]

         Expectation[x\[Conditioned](x==2||x==4||x==6),x\[Distributed]valuesOfX]
         Expectation[Piecewise[{{x,(x==2||x==4||x==6)}}],x\[Distributed]valuesOfX]/Probabili
         ty[(x==2||x==4||x==6),x\[Distributed]valuesOfX]
Output   The conditional expectation E[X|A] is  4
Output   4
Output   4

Input    (* The code demonstrates the calculation of statistical properties such as mean,
         variance, moment, and central moment using the Expectation function in Mathematica.
         It defines custom functions that utilize Expectation to compute these properties
         based on a given probability distribution. The code includes comparisons between the
         results obtained from the custom functions and the built-in Mathematica functions
         (Mean, Variance, Moment, CentralMoment) for the Poisson distribution: *)

         (* Define a probability distribution: *)
         dist=PoissonDistribution[λ];

         (* Define function to calculate mean: *)
         mean[dist_]:=Module[
             {x},
             Expectation[x,x\[Distributed]dist]
             ];

         {Mean[dist],mean[dist]}

         (* Define function to calculate variance: *)
         variance[dist_]:=Module[
             {x,m},
             m=mean[dist];
             Expectation[(x-m)^2,x\[Distributed]dist]
             ];

         {Variance[dist],variance[dist]}

         (* Define function to calculate moment: *)
```

```
        moment[dist_,n_]:=Module[
           {x},
           Expectation[x^n,x\[Distributed]dist]
           ];

        {Moment[dist,5],moment[dist,5]}

        (* Define function to calculate central moment: *)
        centralMoment[dist_,n_]:=Module[
           {x,m},
           m=mean[dist];
           Expectation[(x-m)^n,x\[Distributed]dist]
           ];

        {CentralMoment[dist,5],centralMoment[dist,5]}
Output  {λ,λ}
Output  {λ,λ}
Output  {λ+15 λ^2+25 λ^3+10 λ^4+λ^5,λ+15 λ^2+25 λ^3+10 λ^4+λ^5}
Output  {λ+10 λ^2,λ+10 λ^2}
```

| | |
|---|---|
| `CDF[dist,x]` | gives the cumulative distribution function for the distribution dist evaluated at x. |
| `CDF[dist,{x1,x2,…}]` | gives the multivariate cumulative distribution function for the distribution dist evaluated at {x1,x2,…}. |
| `CDF[dist]` | gives the CDF as a pure function. |

***Mathematica Examples 11.5*** CDF

```
Input   (* In this code, we define the distribution using the BinomialDistribution function
        with parameters 10 and 0.5. Next, we define a function manualCDF[n] that calculates
        the CDF manually by summing up the probability density function (PDF) values from 0
        to n. We then generate a list of values nValues from 0 to 10. Using the manually
        calculated CDF function and the built-in CDF function, we calculate the CDF values
        for each n in nValues. Finally, we display the results in a table format using
        TableForm: *)

        (* Define the distribution: *)
        dist=BinomialDistribution[10,0.5];

        (* Calculate the CDF manually: *)
        manualCDF[n_]:=Sum[
          PDF[dist,k],
          {k,0,n}
          ]

        (* Compare with the built-in CDF: *)
        nValues=Range[0,10];
        manualCDFValues=manualCDF/@nValues;
        builtInCDFValues=CDF[dist,nValues];

        (* Print the results: *)
        TableForm[
         Transpose[
          {nValues,manualCDFValues,builtInCDFValues}
          ],
         TableHeadings->{None,{"n","Manual CDF","Built-in CDF"}}
         ]
Output  {
         {n, Manual CDF, Built-in CDF},
         {0, 0.000976563, 0.000976563},
         {1, 0.0107422, 0.0107422},
         {2, 0.0546875, 0.0546875},
```

```
 {3, 0.171875, 0.171875},
 {4, 0.376953, 0.376953},
 {5, 0.623047, 0.623047},
 {6, 0.828125, 0.828125},
 {7, 0.945313, 0.945312},
 {8, 0.989258, 0.989258},
 {9, 0.999023, 0.999023},
 {10, 1., 1}
}
```

Input

```
(* The CDF is the sum of the PDF for discrete distributionsF(x)=\!\(
\*UnderoverscriptBox[\(\[Sum]\), \(ξ = \(-∞\)\), \(x\)]\(f\((ξ)\)\)\): *)
PDF[GeometricDistribution[p],m]

FullSimplify[
 Sum[
  PDF[GeometricDistribution[p],m],
  {m,-∞,Floor[n]}
  ]
 ]

CDF[GeometricDistribution[p],n]
```

Output $\begin{cases} (1-p)^m p & m >= 0 \\ 0 & \text{True} \end{cases}$

Output $\begin{cases} 1-(1-p)^{1+\text{Floor}[n]} & \text{Floor}[n] >= 0 \\ 0 & \text{True} \end{cases}$

Output $\begin{cases} 1-(1-p)^{1+\text{Floor}[n]} & n >= 0 \\ 0 & \text{True} \end{cases}$

Input

```
(* Define the distribution: *)
dist=BinomialDistribution[10,0.5];

(* Calculate the PMF: *)
pmf[k_]:=PDF[dist,k]

(* Calculate the CDF: *)
cdf[n_]:=Sum[
  pmf[k],
  {k,0,n}
  ]

(* Plot the PMF and CDF: *)
DiscretePlot[
 {pmf[k],cdf[k]},
 {k,0,10},
 PlotRange->All,
 PlotLegends->{"PMF","CDF"},
 ImageSize->200
 ]
```

Output

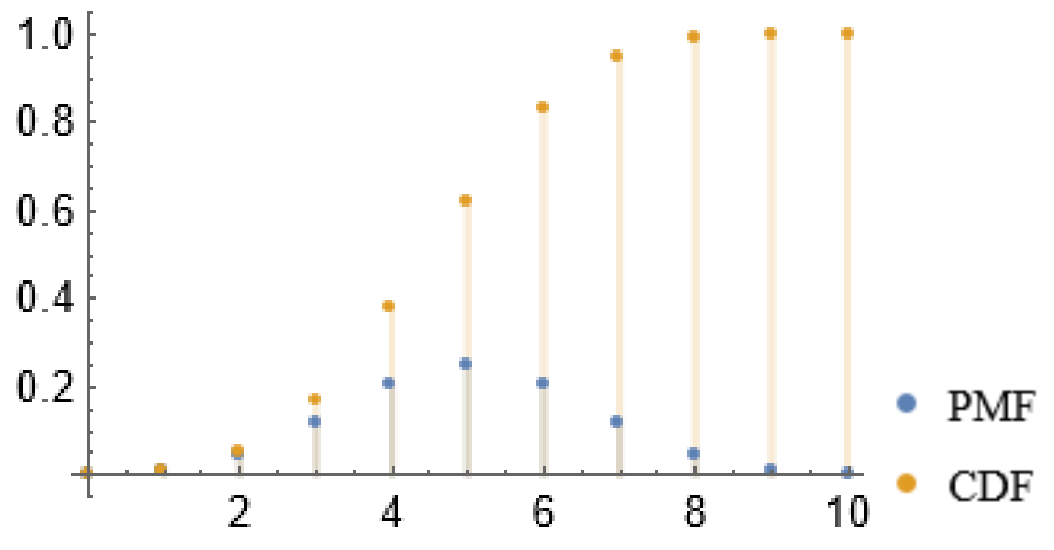

Input

```
(* The CDF of a univariate discrete distribution:*)
```

```
        CDF[PoissonDistribution[μ],k]

        DiscretePlot[
         CDF[PoissonDistribution[3],k],
         {k,0,10},
         PlotStyle->Purple,
         ImageSize->200,
         ExtentSize->Right,
         ExtentMarkers->{"Filled","Empty"}
         ]
Output  GammaRegularized[3,μ]
Output
```

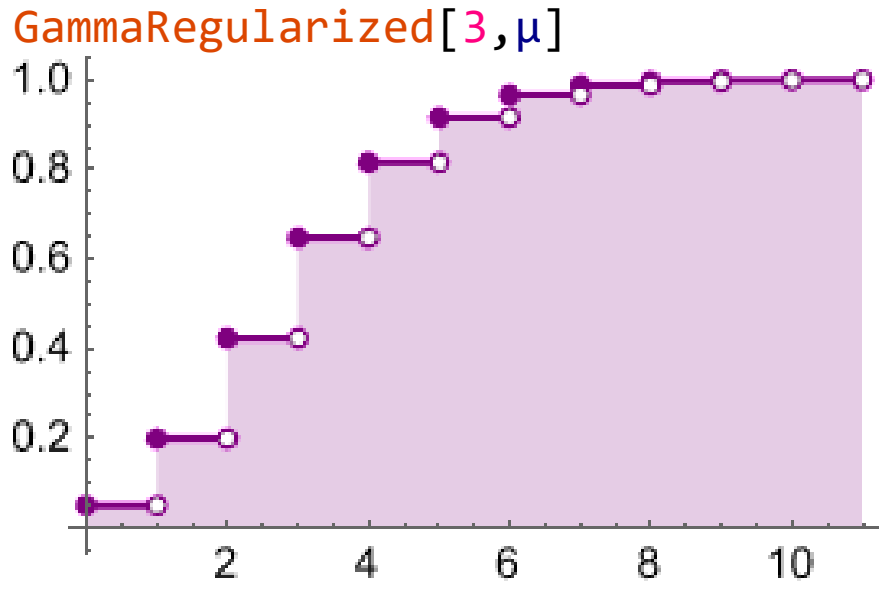

```
Input   (* The CDF for a multivariate Poisson distribution:*)
        DiscretePlot3D[
         CDF[MultivariatePoissonDistribution[5,{2,3}],{x,y}],
         {x,0,12},
         {y,0,12},
         ExtentSize->Right,
         PlotStyle->Lighter[Purple,0.1],
         ImageSize->200
         ]
Output
```

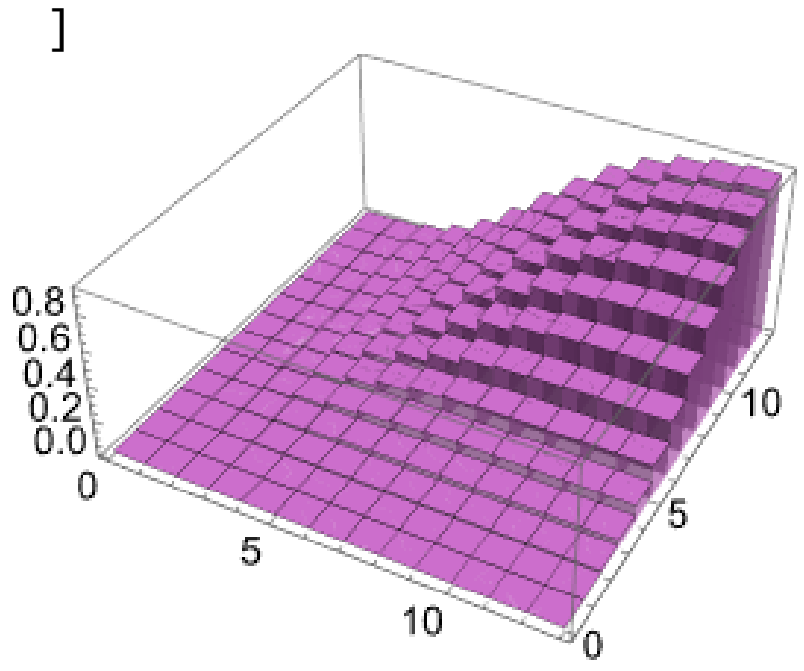

```
Input   (* Obtain exact numeric results:*)
        CDF[NegativeBinomialDistribution[20,1/3],5]
Output  23731/10460353203

Input   (* The probability of x<=a for a univariate distribution is given by its CDF:*)
        {
         Probability[x<=5,x\[Distributed]GeometricDistribution[1/3]],
         CDF[GeometricDistribution[1/3],5]
         }
Output  {665/729,665/729}

Input   (* A univariate CDF is 0 at -∞ and 1 at ∞:*)
        CDF[DiscreteUniformDistribution[{0,5}],-∞]
        CDF[DiscreteUniformDistribution[{0,5}],∞]
Output  0
Output  1
```

| | |
|---|---|
| `MomentGeneratingFunction[dist,t]` | gives the moment-generating function for the distribution dist as a function of the variable t. |
| `MomentGeneratingFunction[dist,{t1,t2,…}]` | gives the moment-generating function for the multivariate distribution dist as a function of the variables t1, t2, … . |
| `CentralMomentGeneratingFunction[dist,t]` | gives the central moment-generating function for the distribution dist as a function of the variable t. |
| `CentralMomentGeneratingFunction[dist,{t1,t2,…}]` | gives the central moment-generating function for the multivariate distribution dist as a function of the variables t1, t2, …. |

***Mathematica Examples 11.6*** MomentGeneratingFunction and CentralMomentGeneratingFunction

Input
```
(* The mgf for a univariate discrete distribution: *)
MomentGeneratingFunction[PoissonDistribution[μ],t]
```
Output $e^{(-1+e^t)\mu}$

Input
```
(* In this code, we define the mgf function which takes two arguments: t and dist.
The t represents the value at which we want to evaluate the MGF, and dist represents
the probability distribution of the random variable. The MGF is calculated using the
Expectation function in Mathematica. We use the Exp[t*x] term as the function inside
the Expectation, which represents the exponential function raised to the power of
t*x. This is the characteristic function of the random variable: *)

(* Define the Moment Generating Function: *)
mgf[t_,dist_]:=Expectation[Exp[t*x],x\[Distributed]dist]

(* Example usage: *)
dist=PoissonDistribution[μ] (* Poisson distribution with μ. *)
mgf[t,dist] (* Calculate the MGF at a given value of t. *)
MomentGeneratingFunction[dist,t]
```
Output `PoissonDistribution[μ]`
Output $e^{(-1+e^t)\mu}$
Output $e^{(-1+e^t)\mu}$

Input
```
(* In this code, we first define the probability distribution of interest. In the
example, we used a Poisson distribution. Next, we define the mgf function, which
computes the Moment Generating Function using the MomentGeneratingFunction function
in Mathematica. To find the first three moments,we differentiate the MGF with
respect to t and then evaluate the result at t=0. The first derivative gives the
first moment (mean),the second derivative gives the second moment (variance), and
the third derivative gives the third moment: *)

(* Define the probability distribution: *)
dist=PoissonDistribution[μ] ;(* Poisson distribution with μ.*)

(*Define the Moment Generating Function*)
mgf[t_,dist_]:=MomentGeneratingFunction[dist,t]

(*Find the first three moments*)
moment1=D[mgf[t,dist],{t,1}]/. t->0 (* First moment (mean). *);
moment2=D[mgf[t,dist],{t,2}]/. t->0 (* Second moment (variance). *);
moment3=D[mgf[t,dist],{t,3}]/. t->0 (* Third moment. *);

(* Display the results: *)
moment1
moment2
moment3
```

```
Output   μ
Output   μ (1+μ)
Output   μ^2+μ (1+μ)^2

Input    (* The  central moment-generating function ((cmgf) for a univariate discrete
         distribution:*)
         CentralMomentGeneratingFunction[PoissonDistribution[μ],t]
Output   e^((-1+e^t)μ-tμ)

Input    (*The cmgf is the moment-generating function times exp(-t μ):*)
         d=PoissonDistribution[μ];
         CentralMomentGeneratingFunction[d,t]
         MomentGeneratingFunction[d,t] Exp[-t Mean[d]]
Output   e^((-1+e^t)μ-tμ)
Output   e^((-1+e^t)μ-tμ)

Input    (*Generating functions including MomentGeneratingFunction are defined by an
         expectation:*)
         dist=PoissonDistribution[3];
         Simplify[
          {
           MomentGeneratingFunction[dist,t],
           Expectation[E^(t x),x\[Distributed]dist]
           }
          ]
         Simplify[
          {
           CentralMomentGeneratingFunction[dist,t],
           Expectation[E^(t (x-Mean[dist])),x\[Distributed]dist]
           }
          ]
Output   {e^(3(-1+e^t)), e^(3(-1+e^t))}
Output   {e^(3(-1+e^t-t)), e^(3(-1+e^t-t))}
```

| | |
|---|---|
| `EstimatedDistribution[data,dist]` | estimates the parametric distribution dist from data. |

***Mathematica Examples 11.7*** EstimatedDistribution

```
Input    (* The code generates a sample of 10000 values from a discrete Poisson distribution
         and estimates the distribution parameters (μ) from the sample using the
         EstimatedDistribution function. It then compares the density histogram of the sample
         with the PDF of the estimated distribution: *)
         sampledata=RandomVariate[
             PoissonDistribution[4],
             10^4
             ];
         (* Estimate the distribution parameters from sample data: *)
         ed=EstimatedDistribution[
           sampledata,
           PoissonDistribution[μ]
           ]
         (* Compare the density histogram of the sample with the PDF of the estimated
         distribution: *)
         Show[
          Histogram[
           sampledata,
           {1},
           "PDF",
           ColorFunction->Function[{height},Opacity[height]],
           ChartStyle->Purple,
```

```
        ImageSize->320
        ],
       DiscretePlot[
        PDF[ed,x],
        {x,0,Max[sampledata]},
        PlotStyle->PointSize[Medium],
        ColorFunction->"Rainbow"
        ]
       ]
Output   PoissonDistribution[4.0022]
Output
```

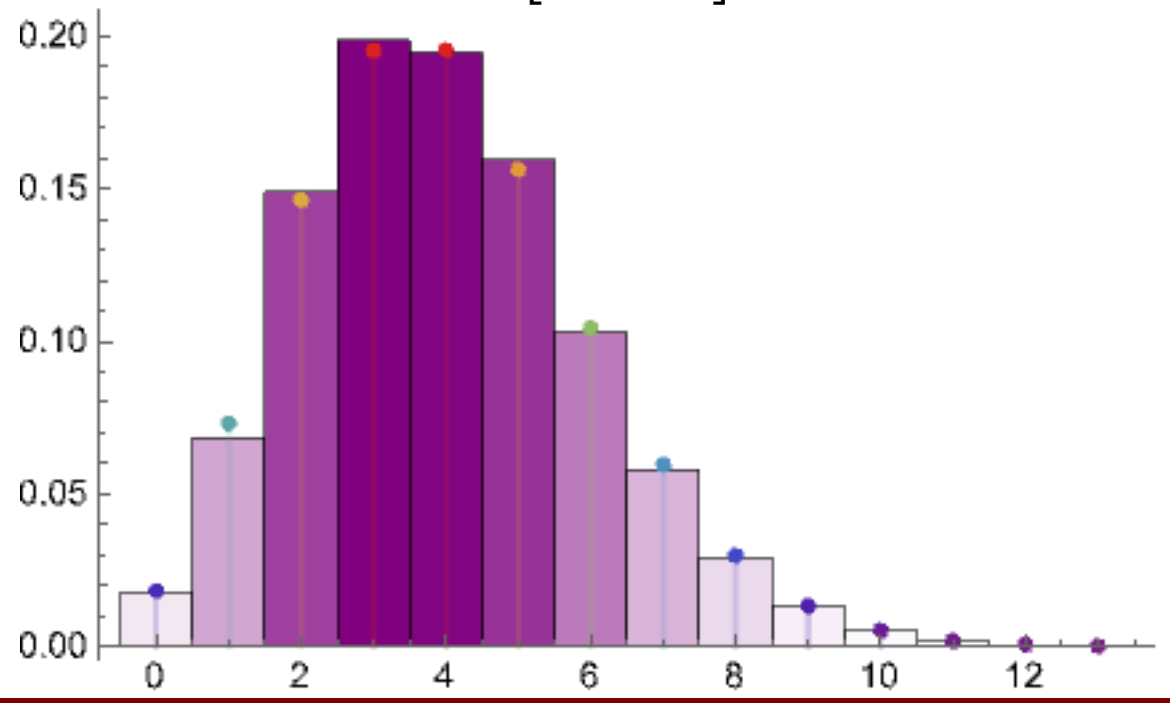

# UNIT 11.2

# BINOMIAL DISTRIBUTION

***Mathematica Examples 11.8***

Input
```
(* In this example, we use the binomial function with parameters n=10 (number of trials) and p=0.5 (probability of success). We calculate the probability mass function (PMF) using the Binomial function and plot it using ListPlot: *)

n=10;   (* Number of trials. *)
p=0.5;   (* Probability of success. *)

(* Define a discrete random variable with a binomial distribution: *)
x=Range[0,n];
pmf=Table[
    Binomial[n,k]*p^k*(1-p)^(n-k),
    {k,0,n}
    ];

(* Plot the PMF: *)
ListPlot[
 Transpose[{x,pmf}],
 Filling->Axis,
 PlotRange->All,
 PlotStyle->Purple,
 AxesLabel->{"x","P(X = x)"},
 PlotLabel->"Binomial Distribution PMF",
 ImageSize->200
 ]
```

Output

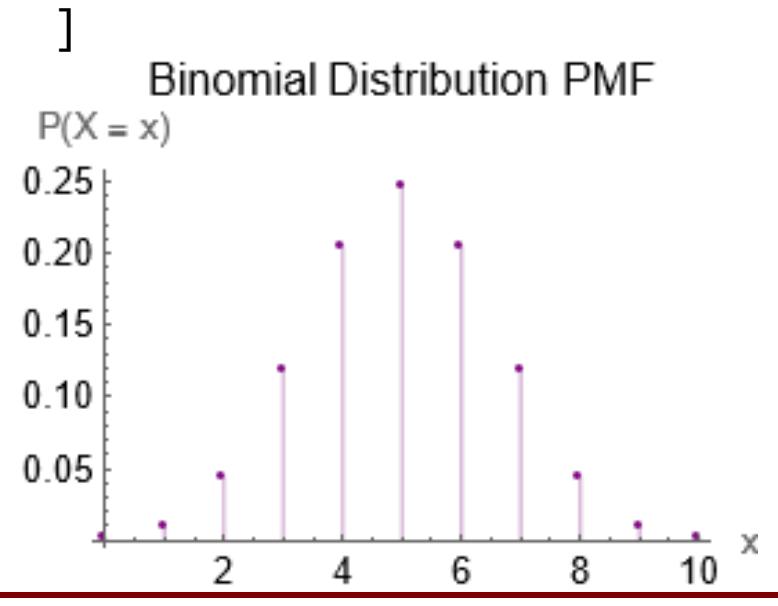

***Mathematica Examples 11.9***

Input
```
(* The code generates a discrete plot of the PMF for a binomial distribution with parameters n=30 and three different values of p (0.2, 0.5, and 0.7). The plot shows the values of the PMF for all possible values of k between 0 and 37: *)

DiscretePlot[
 Evaluate[
  Table[
   PDF[
    BinomialDistribution[30,p],
    k
    ],
   {p,{0.2,0.5,0.7}}
   ]
  ],
```

```
 {k,37},
 PlotRange->All,
 PlotMarkers->Automatic,
 PlotLegends->Placed[{"n=30,p=0.2","n=30,p=0.5","n=30,p=0.7"},{0.8,0.75}],
 PlotStyle->{RGBColor[0.88,0.61,0.14],RGBColor[0.37,0.5,0.7],Purple},
 ImageSize->320,
 AxesLabel->{None,"PMF"}
 ]
```

Output

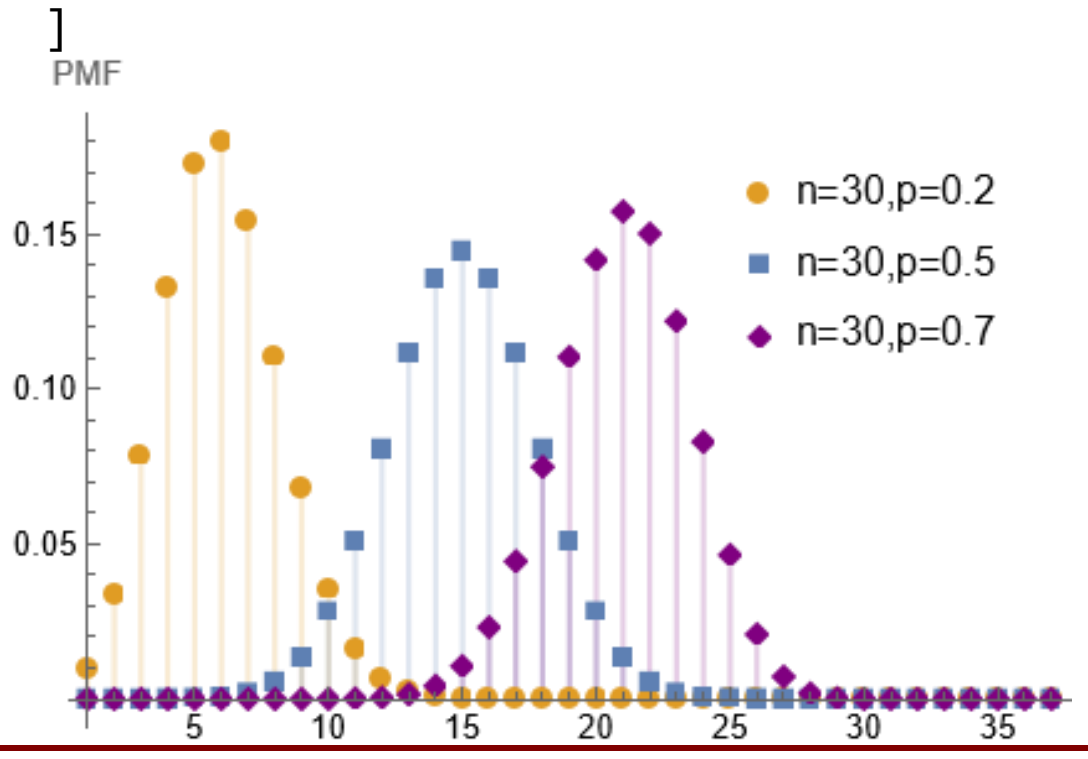

## *Mathematica Examples 11.10*

Input

```
(* The code generates a discrete plot of the cumulative distribution function (CDF)
for a binomial distribution with parameters n=30 and three different values of p
(0.2, 0.5, and 0.7). The plot shows the values of the CDF for all possible values of
k between 0 and 37: *)

DiscretePlot[
 Evaluate[
  Table[
   CDF[
    BinomialDistribution[30,p],
    k
    ],
   {p,{0.2,0.5,0.7}}
   ]
  ],
 {k,37},
 ExtentSize->Right,
 PlotRange->All,
 PlotMarkers->Automatic,
 PlotLegends->Placed[{"n=30,p=0.2","n=30,p=0.5","n=30,p=0.7"},{0.8,0.75}],
 PlotStyle->{RGBColor[0.88,0.61,0.14],RGBColor[0.37,0.5,0.7],Purple},
 ImageSize->320,
 AxesLabel->{None,"CDF"}
 ]
```

Output

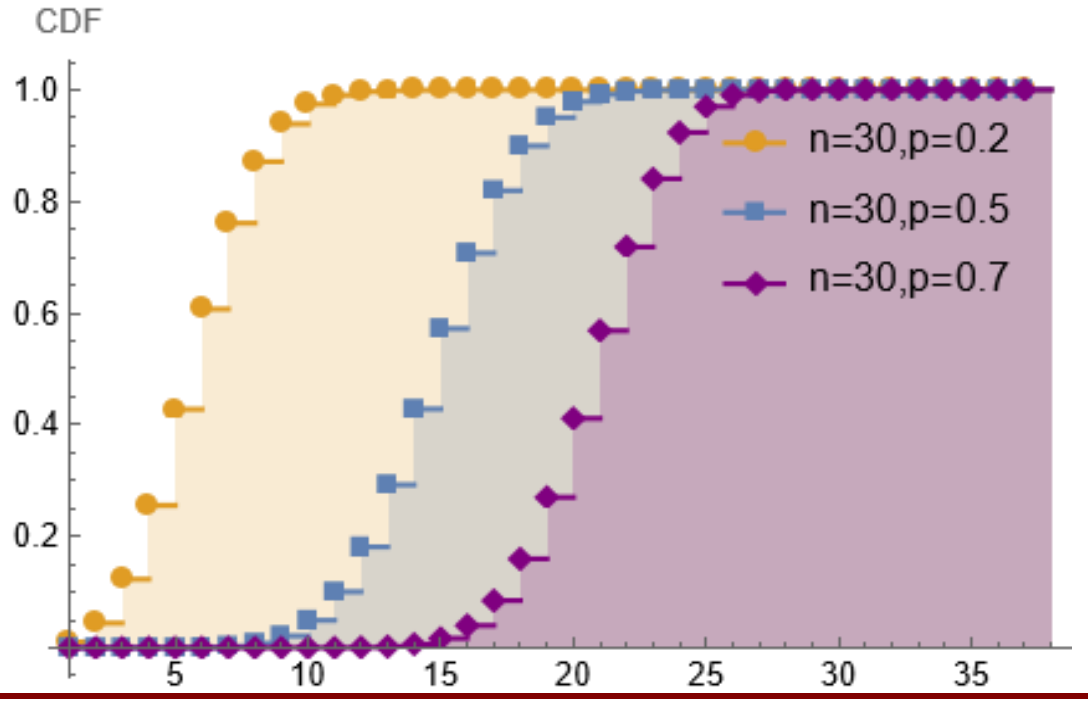

***Mathematica Examples 11.11***

Input

```
(* The code generates a discrete plot of the PMF for a binomial distribution with
three different values of n (10, 20, and 30) and p=0.5. The plot shows the values of
the PMF for all possible values of k between 0 and 37: *)

DiscretePlot[
 Evaluate[
  Table[
   PDF[
    BinomialDistribution[n,.5],
    k
    ],
   {n,{10,20,30}}
   ]
  ],
 {k,37},
 PlotRange->All,
 PlotMarkers->Automatic,
 PlotLegends->Placed[{"p=0.5,n=10","p=0.5,n=20","p=0.5,n=30"},{0.8,0.75}],
 PlotStyle->{RGBColor[0.88,0.61,0.14],RGBColor[0.37,0.5,0.7],Purple},
 ImageSize->320,
 AxesLabel->{None,"PMF"}
 ]
```

Output

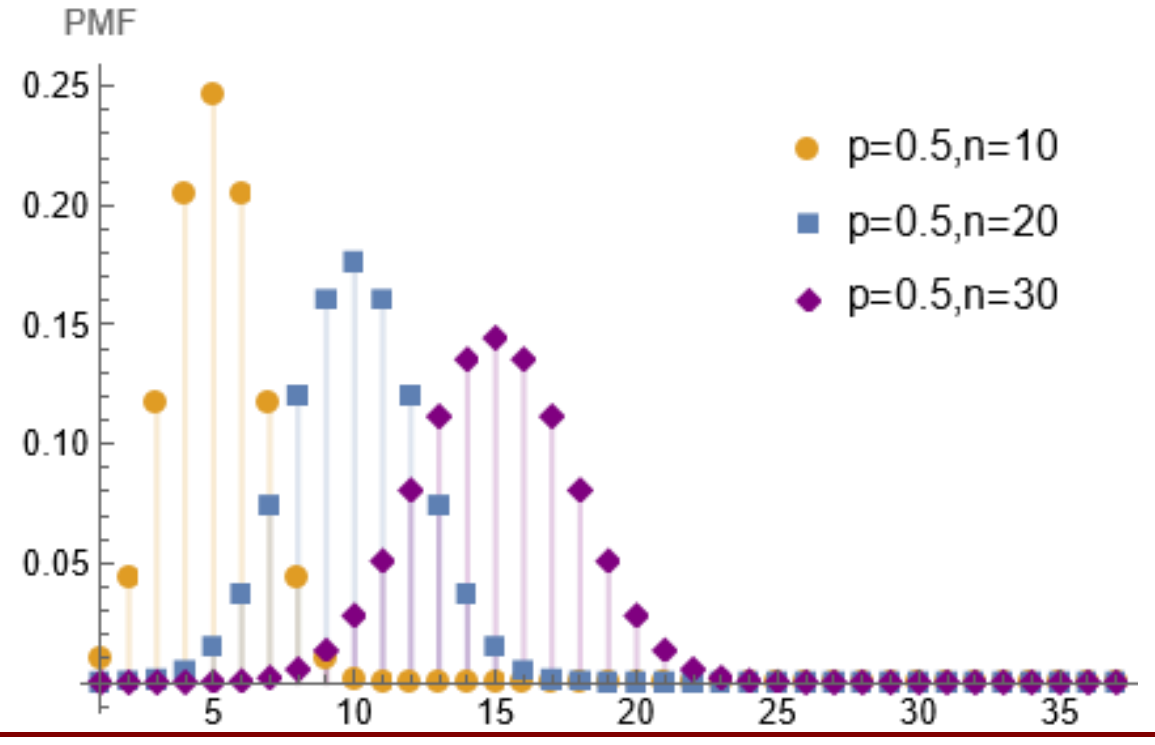

***Mathematica Examples 11.12***

Input

```
(* The code generates a discrete plot of the CDF for a binomial distribution with
three different values of n (10, 20, and 30) and p=0.5. The plot shows the values of
the CDF for all possible values of k between 0 and 37: *)

DiscretePlot[
 Evaluate[
  Table[
   CDF[
    BinomialDistribution[n,.5],
    k
    ],
   {n,{10,20,30}}
   ]
  ],
 {k,37},
 ExtentSize->Right,
 PlotRange->All,
 PlotMarkers->Automatic,
 PlotLegends->Placed[{"p=0.5,n=10","p=0.5,n=20","p=0.5,n=30"},{0.8,0.75}],
 PlotStyle->{RGBColor[0.88,0.61,0.14],RGBColor[0.37,0.5,0.7],Purple},
 ImageSize->320,
 AxesLabel->{None,"CDF"}
```

```
]
```

Output

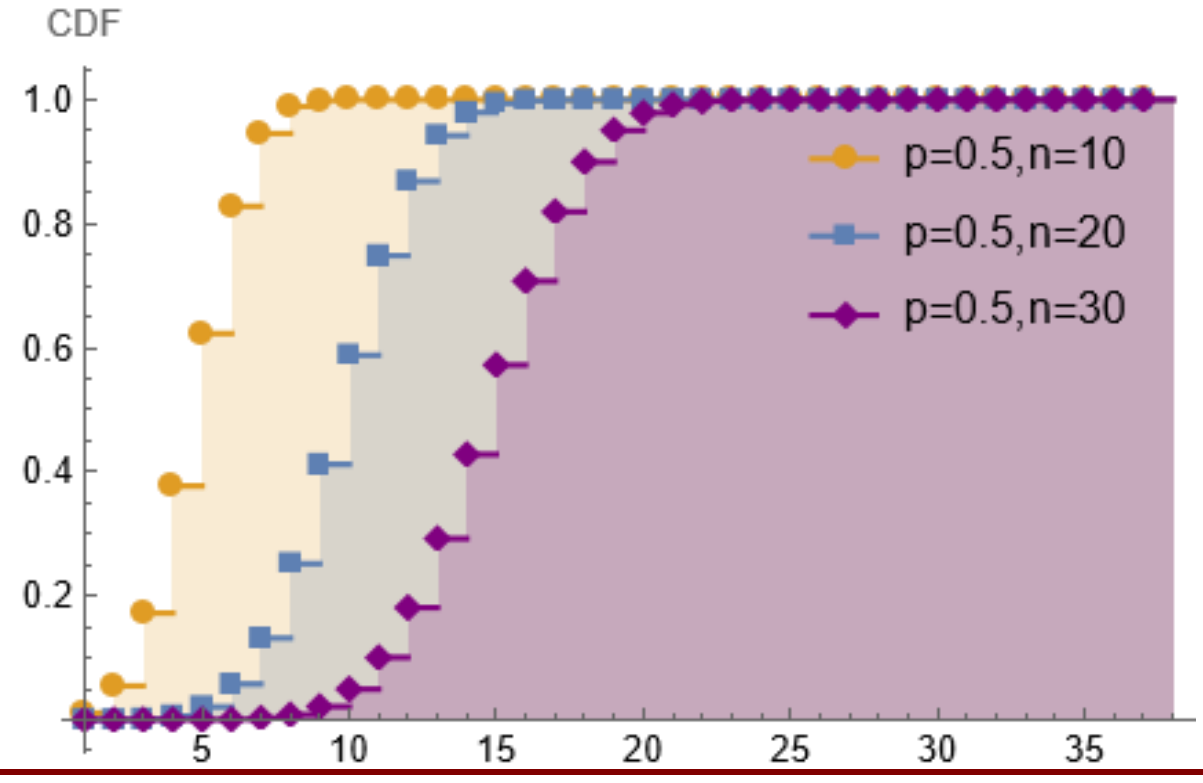

***Mathematica Examples 11.13***

Input

```
(* The code generates a histogram and a discrete plot of the PMF for a binomial
distribution with parameters n=10, p=0.4 and sample size 10000: *)

data=RandomVariate[
    BinomialDistribution[10,0.4],
    10^4
    ];

Show[
 Histogram[
  data,
  {1},
  "PDF",
  ColorFunction->Function[{height},Opacity[height]],
  ChartStyle->Purple,
  ImageSize->320,
  AxesLabel->{None,"PDF"}
  ],
 DiscretePlot[
  PDF[
   BinomialDistribution[10,0.4],
   x
   ],
  {x,0,10},
  PlotStyle->PointSize[Medium],
  ColorFunction->"Rainbow"
  ]
 ]
```

Output

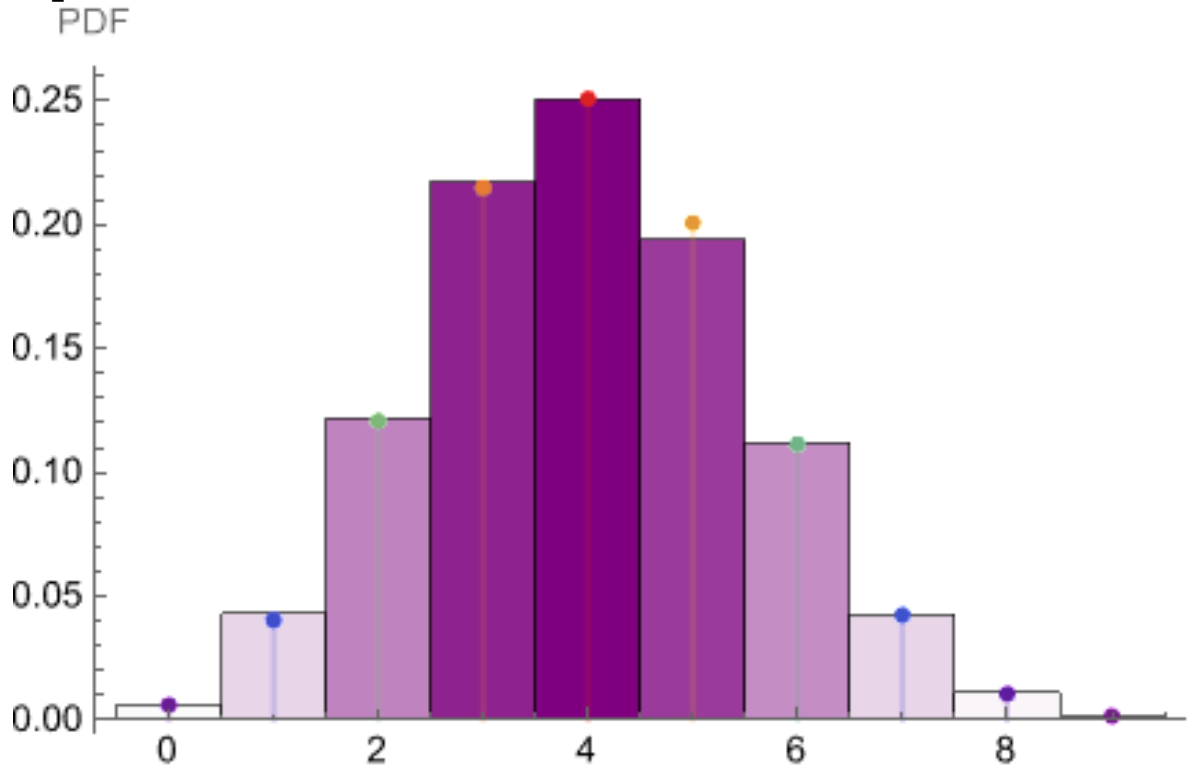

***Mathematica Examples 11.14***

Input

```
(* The code calculates and displays some descriptive statistics (mean, variance, standard deviation, kurtosis, and skewness) for a binomial distribution with parameters n and p: *)

Grid[
 Table[
  {
   statistics,
   FullSimplify[statistics[BinomialDistribution[n,p]]]
   },
  {statistics,{Mean,Variance,StandardDeviation,Kurtosis,Skewness}}
  ],
 ItemStyle->12,
 Alignment->{{Right,Left}},
 Frame->All,
 Spacings->{Automatic,0.8}
 ]
```

Output

| | |
|---|---|
| Mean | n p |
| Variance | -n (-1+p) p |
| StandardDeviation | Sqrt[-n (-1 + p) p] |
| Kurtosis | 3-6/n+1/(n p-n p^2) |
| Skewness | (1 - 2 p)/Sqrt[-n (-1 + p) p] |

***Mathematica Examples 11.15***

Input

```
(* The code calculates and displays some additional descriptive statistics (moments, central moments, and factorial moments) for a binomial distribution with parameters n and p: *)

Grid[
 Table[
  {
   statistics,
   FullSimplify[statistics[BinomialDistribution[n,p],1]],
   FullSimplify[statistics[BinomialDistribution[n,p],2]]
   },
  {statistics,{Moment,CentralMoment,FactorialMoment}}
  ],
 ItemStyle->12,
 Alignment->{{Right,Left}},
 Frame->All,
 Spacings->{Automatic,0.8}
 ]
```

Output

| | | |
|---|---|---|
| Moment | n p | n p (1+(-1+n) p) |
| CentralMoment | 0 | -n (-1+p) p |
| FactorialMoment | n p | (-1+n) n p^2 |

***Mathematica Examples 11.16***

Input

```
(* This code generates a random sample of size 10,000 from a binomial distribution with parameters n=40 and p=0.5, estimates the distribution parameters using the EstimatedDistribution function, and then compares the histogram of the sample with the estimated PDF of the binomial distribution using a histogram and a discrete plot of the PDF: *)

sampledata=RandomVariate[
   BinomialDistribution[40,.5],
   10^4
   ];
```

```
(* Estimate the distribution parameters from sample data: *)
ed=EstimatedDistribution[
   sampledata,
   BinomialDistribution[n,p]
   ]
(* Compare a density histogram of the sample with the PDF of the estimated
distribution: *)
Show[
 Histogram[
   sampledata,
   {1},
   "PDF",
   ColorFunction->Function[{height},Opacity[height]],
   ChartStyle->Purple,
   ImageSize->320
   ],
 DiscretePlot[
   PDF[ed,x],
   {x,0,40},
   PlotStyle->PointSize[Medium],
   ImageSize->320,
   ColorFunction->"Rainbow"
   ]
 ]
```

Output BinomialDistribution[41,0.4877]

Output

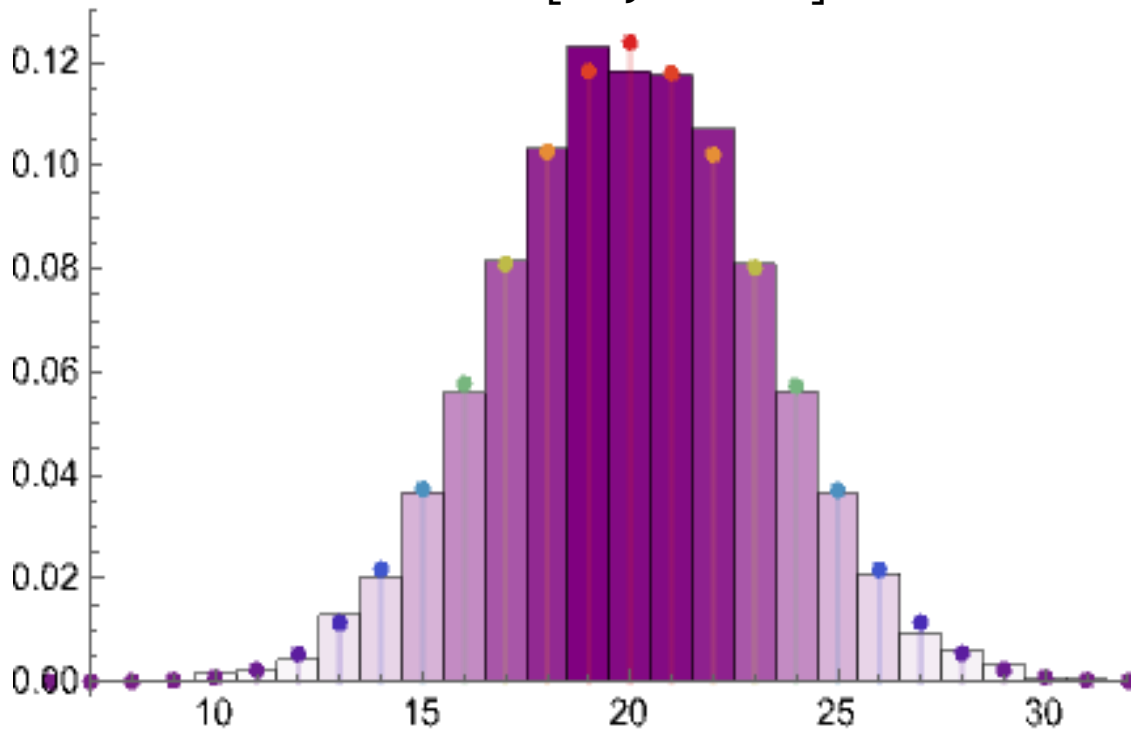

***Mathematica Examples 11.17***

Input

```
(* The code generates a dataset of 1000 observations from a binomial distribution
with parameters n=30 and p=0.3. Then, it computes the sample mean and quartiles of
the data, and plots a histogram of the data using the "PDF" option to display the
probability density function. Additionally, the code adds vertical lines to the plot
corresponding to the sample mean and quartiles: *)

data=RandomVariate[
    BinomialDistribution[30,0.3],
    1000
    ];
mean=Mean[data];
quartiles=Quantile[
    data,
    {0.25,0.5,0.75}
    ];
Histogram[
 data,
 Automatic,
 "PDF",
 Epilog->{
```

```
      Directive[Red,Thickness[0.006]],
      Line[{{mean,0},{mean,0.25}}],
      Directive[Green,Dashed],
      Line[{{quartiles[[1]],0},{quartiles[[1]],0.25}}],
      Line[{{quartiles[[2]],0},{quartiles[[2]],0.25}}],
      Line[{{quartiles[[3]],0},{quartiles[[3]],0.25}}]
      },
    ColorFunction->Function[{height},Opacity[height]],
    ImageSize->320,
    ChartStyle->Purple,
    PlotRange->{{0,20},{0,0.2}}
    ]
```

Output

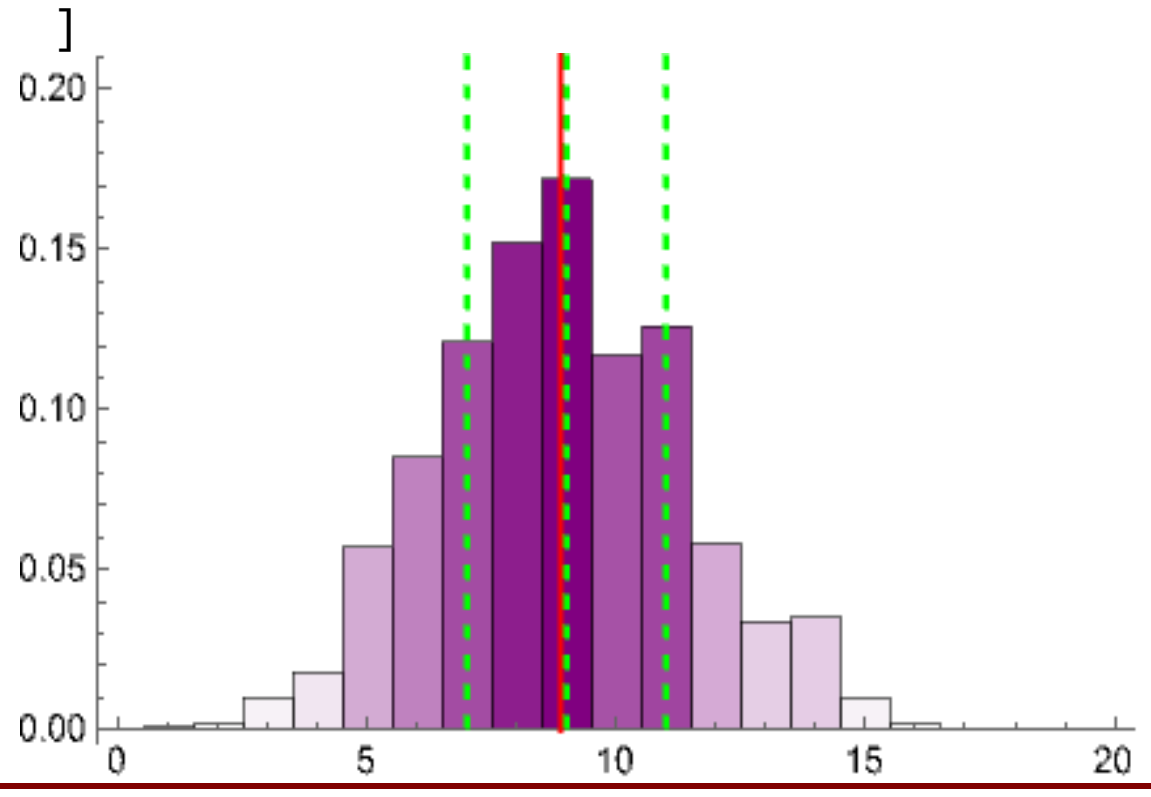

***Mathematica Examples 11.18***

Input

```
(* The code creates a dynamic histogram of data generated from a binomial distribution
using the Manipulate function. The Manipulate function creates interactive controls
for the user to adjust the values of n, p, and m, which are the parameters of the
binomial distribution and the sample size: *)

Manipulate[
 Module[
  {
   data=RandomVariate[
     BinomialDistribution[n,p],
     m
     ]
   },
  Show[
   Histogram[
    data,
    {1},
    "PDF",
    PlotRange->{{0,n},All},
    ColorFunction->Function[{height},Opacity[height]],
    ImageSize->320,
    ChartStyle->Purple
    ]
   ]
  ],
 {{n,10,"n"},1,100,1},
 {{p,0.5,"p"},0,1,0.01},
 {{m,100,"m"},10,1000,10}
 ]
```

Output

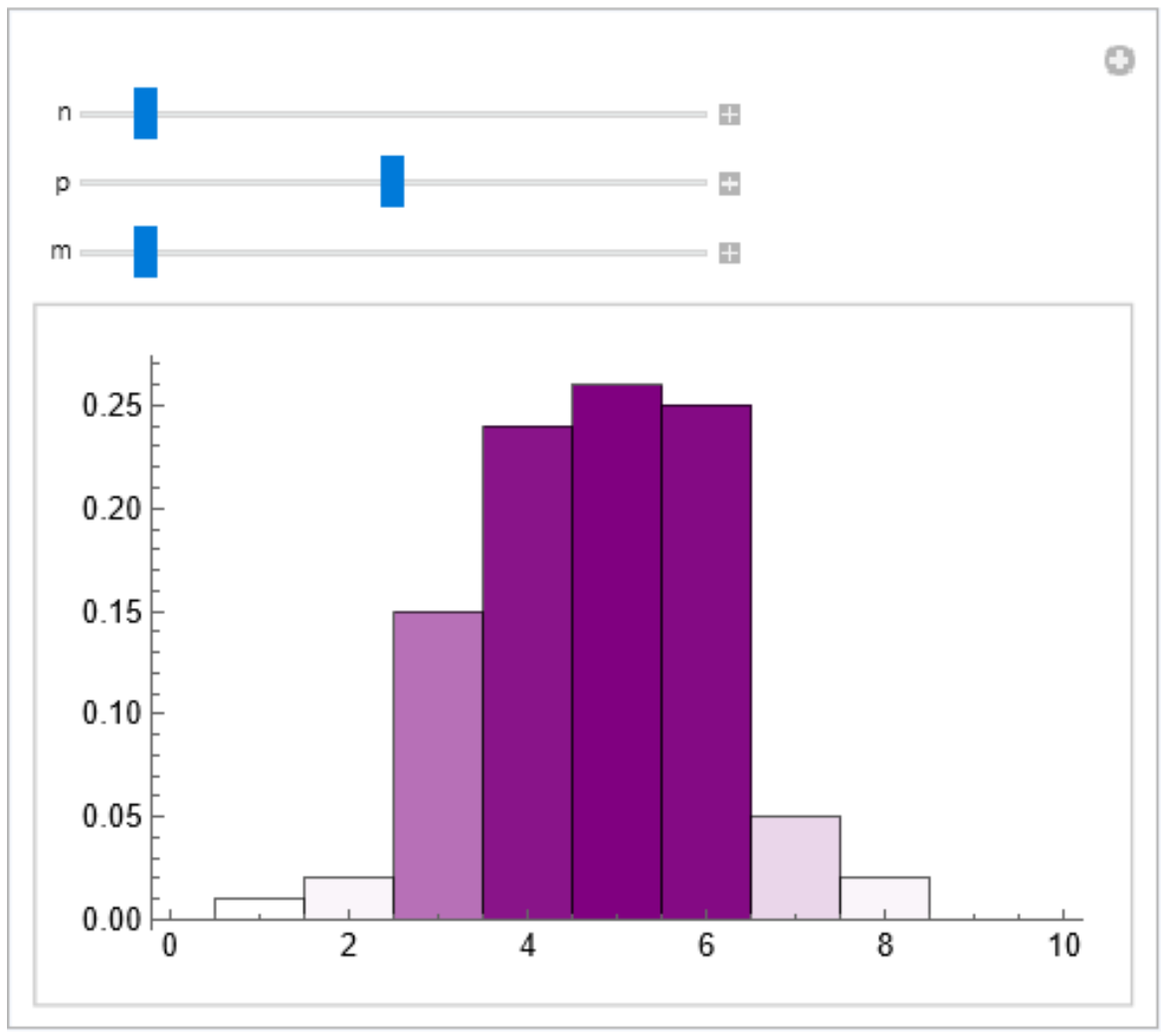

### *Mathematica Examples 11.19*

Input

```
(* The code creates a plot of the CDF of a binomial distribution using the Manipulate
function. The Manipulate function allows you to interactively change the values of
the parameters n and p, respectively: *)
Manipulate[
 Plot[
  CDF[
   BinomialDistribution[n,p],
   x
   ],
  {x,0,n},
  Filling->Axis,
  FillingStyle->LightPurple,
  PlotRange->{{0,n},{0,1}},
  Epilog->{Text[StringForm["n = `` & p = ``",n,p],{n/2,0.9}]},
  AxesLabel->{"x","CDF"},
  ImageSize->320,
  PlotStyle->Purple
  ],
 {{n,10},1,100,1},
 {{p,0.5},0,1,0.01}
 ]
```

Output

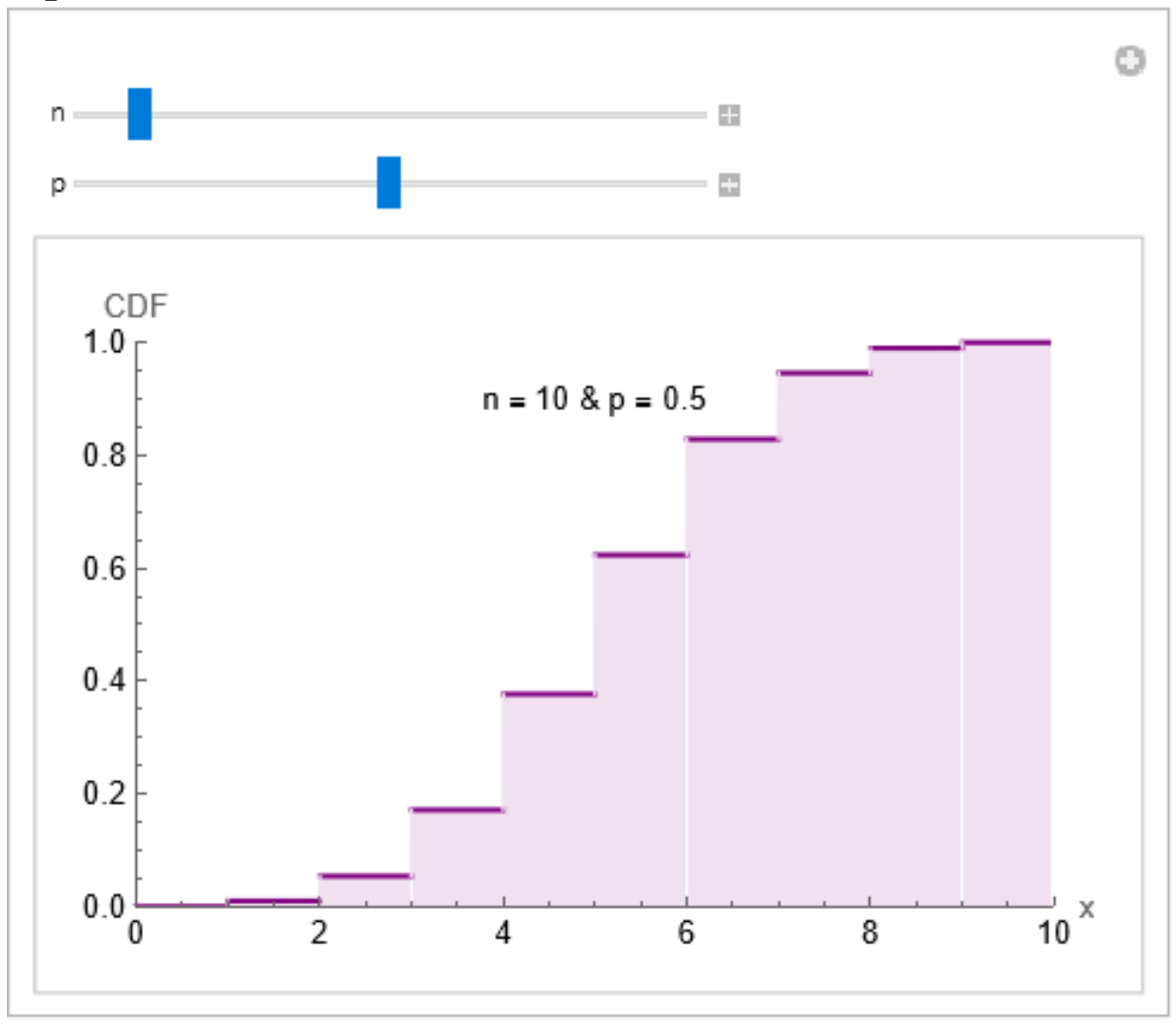

***Mathematica Examples 11.20***

Input

```
(* The code uses the Grid function to create a grid of two plots, one for the PDF
and one for the CDF, both of which are discrete plots. The DiscretePlot function is
used to create the plots, with the probability density or cumulative probability on
the y-axis and the number of successes (k) on the x-axis. The code uses slider
controls to adjust the values of n and p: *)

Manipulate[
 Grid[
  {
   {DiscretePlot[
     PDF[
      BinomialDistribution[n,p],
      k
      ],
     {k,0,n},
     PlotRange->{{0,n},{0,1}},
     PlotStyle->{Purple,PointSize[0.03]},
     PlotLabel->"PDF of Binomial Distribution",
     AxesLabel->{"k","PDF"}
     ],
    DiscretePlot[
     CDF[
      BinomialDistribution[n,p],
      k
      ],
     {k,0,n},
     PlotRange->{{0,n},{0,1}},
     PlotStyle->{Purple,PointSize[0.03]},
     PlotLabel->"CDF of Binomial Distribution",
     AxesLabel->{"k","CDF"}
     ]
    }
   },
  Spacings->{5,5}
  ],
 {{n,10,"n"},1,50,1},
 {{p,0.5,"p"},0,1,0.01}
 ]
```

Output

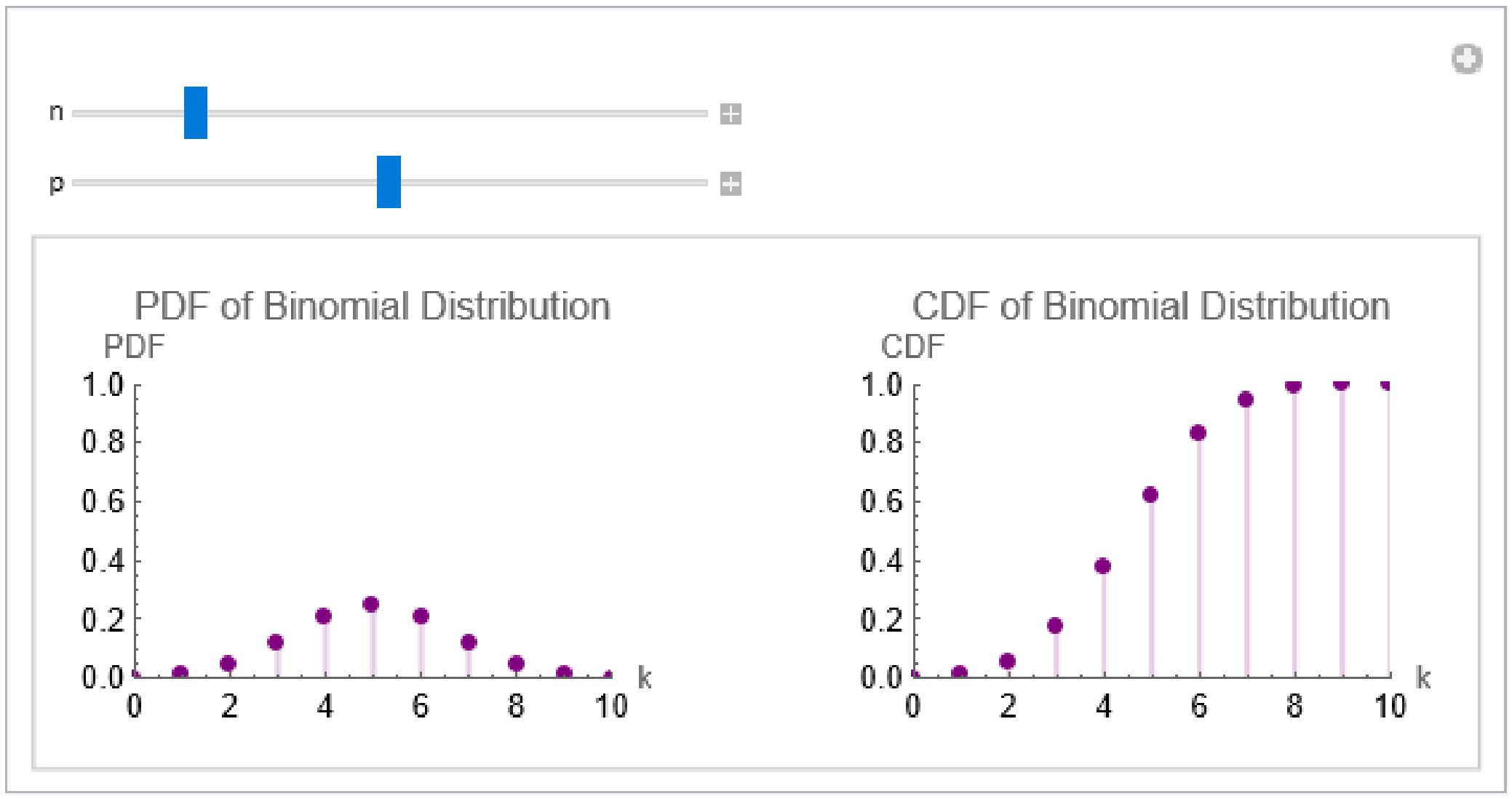

***Mathematica Examples 11.21***

```
Input    (* Suppose a basketball player has a free throw success rate of 80%. Simulate 15 free
         throws. Also, we can use the binomial distribution to calculate the probability of
         the player hits 2 out of 4 free throws in a game: *)

         RandomVariate[BernoulliDistribution[0.8],15]/. {0->"Miss",1->"Hit"}
         Probability[x==2,x\[Distributed]BinomialDistribution[4,0.8]]
Output   {Miss,Hit,Hit,Hit,Hit,Hit,Hit,Hit,Hit,Hit,Hit,Hit,Hit,Hit,Hit}
Output   0.1536
```

***Mathematica Examples 11.22***

```
Input    (* Suppose we flip a fair coin 100 times. We can use the binomial distribution to
         compute the probability that there are between 60 and 70 heads in 100 coin flips: *)

         h[n_]:=BinomialDistribution[n,0.5]
         N[
          Probability[
           60<=x<=70,
           x\[Distributed]h[100]
           ]
          ]
         (* Now, suppose that for an unfair coin the probability of heads is 0.58: *)
         uh[n_]:=BinomialDistribution[n,0.58];
         N[
          Probability[
           60<=x<=70,
           x\[Distributed]uh[100]
           ]
          ]
Output   0.0284279
Output   0.377568
```

***Mathematica Examples 11.23***

```
Input    (* The code computes the probability that there are between 60 and 80 heads in n coin
         flips of an unfair coin with probability of heads given by p and generates a table
         of probabilities for different values of n between 100 and 200 in steps of 25. The
         plot shows how the probability varies with the value of p, the probability of heads:
         *)

         (* Compute the probability that there are between 60 and 80 heads in n coin flips of
         an unfair coin with the probability of heads is p: *)

         prob1=Probability[
             60<=x<=70,
             x\[Distributed]BinomialDistribution[n,p]
             ];

         (* n coin flips takes the values 100, 125, 150, 175 and 200: *)
         newprob=Table[
             prob1,
             {n,100,200,25}
             ];

         (* Plot of an unfair coin with {p,0.1,0.9}: *)
         Plot[
          newprob,
          {p,0.1,0.9},
          PlotRange->All,
          Filling->Axis,
          ImageSize->320,
```

```
 PlotLegends->Placed[{"n=100","n=125","n=150","n=175","n=200"},{0.15,0.7}]
 ]
```

Output

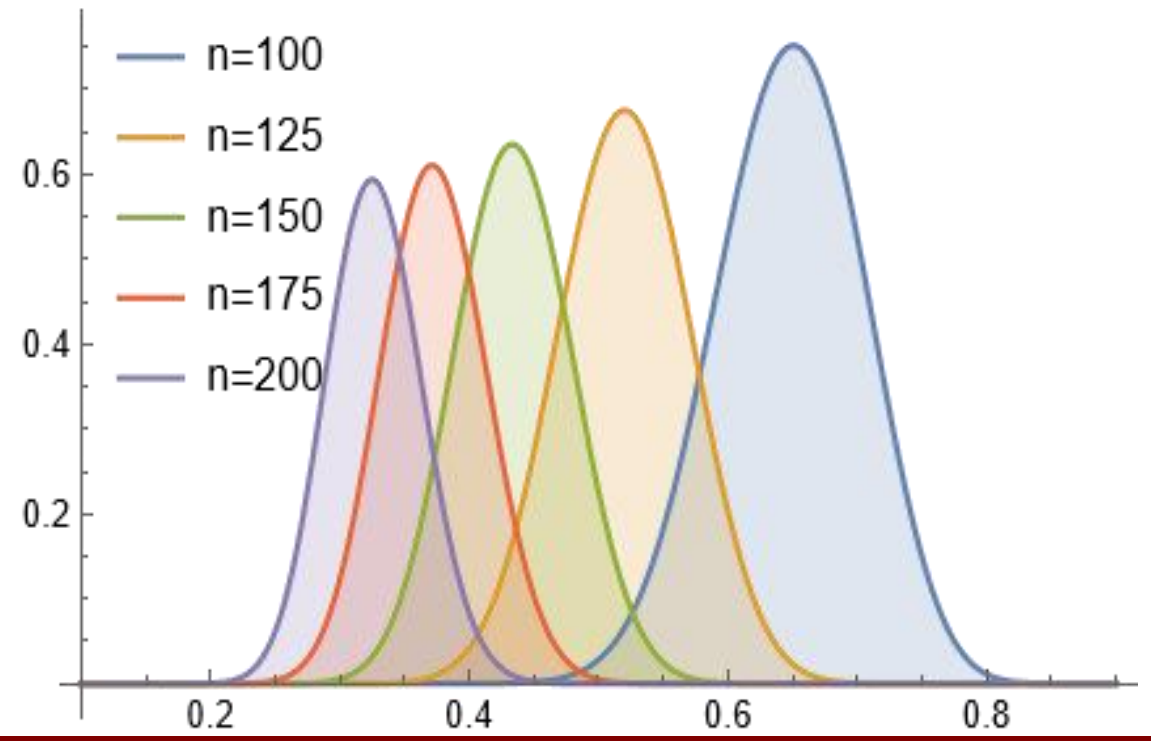

**Mathematica Examples 11.24**

Input

```
(* Two players, A and B, are playing a game where their chances of winning are in the ratio of 3 to 2. We want to determine the probability that A wins at least three games out of a total of five games played: *)

p=3/5;(* probability that A wins a game. *)
q=1-p;(* probability that A loses a game. *)
n=5;(* total number of games played. *)

(* The probability that out of n games played, A wins 'r' games is given by binomial distribution. *)
(* create binomial distribution with parameters n and p. *)
dist=BinomialDistribution[n,p];
N[
 Probability[
  r==3,
  r\[Distributed]dist
  ]
 ]
(* probability that A wins at least 3 games: *)
N[
 Probability[
  r>=3,
  r\[Distributed]dist
  ]
 ]

(* Or, by specifying the event r>=3 (i.e., A wins three, four, or five games): *)
f[i_]=N[
    Probability[
     r==i,
     r\[Distributed]dist
     ]
    ];

Sum[
 f[i],
 {i,3,5}
 ]
```

Output 0.3456

Output 0.68256

Output 0.68256

***Mathematica Examples 11.25***

Input

```
(* The number of heads in n flips with a fair coin can be modeled with BinomialDistribution: *)
heads[n_]:=BinomialDistribution[n,1/2]

(* Show the distribution of heads for 100 coin flips: *)
DiscretePlot[
 PDF[heads[100],k],
 {k,0,100},
 PlotRange->All,
 PlotStyle->{Purple,PointSize[0.005]},
 ImageSize->250
 ]
(* Compute the probability that there are between 60 and 80 heads in 100 coin flips: *)
NProbability[60<=x<=80,x\[Distributed]heads[100]]

(* Now, suppose that for an unfair coin the probability of heads is 0.6: *)
uheads[n_]:=BinomialDistribution[n,0.6];

(* The distribution and the corresponding probabilities have changed: *)
DiscretePlot[
 PDF[uheads[100],k],
 {k,0,100},
 PlotRange->All,
 PlotStyle->{Purple,PointSize[0.005]},
 ImageSize->250
 ]
Probability[60<=x<=80,x\[Distributed]uheads[100]]
```

Output

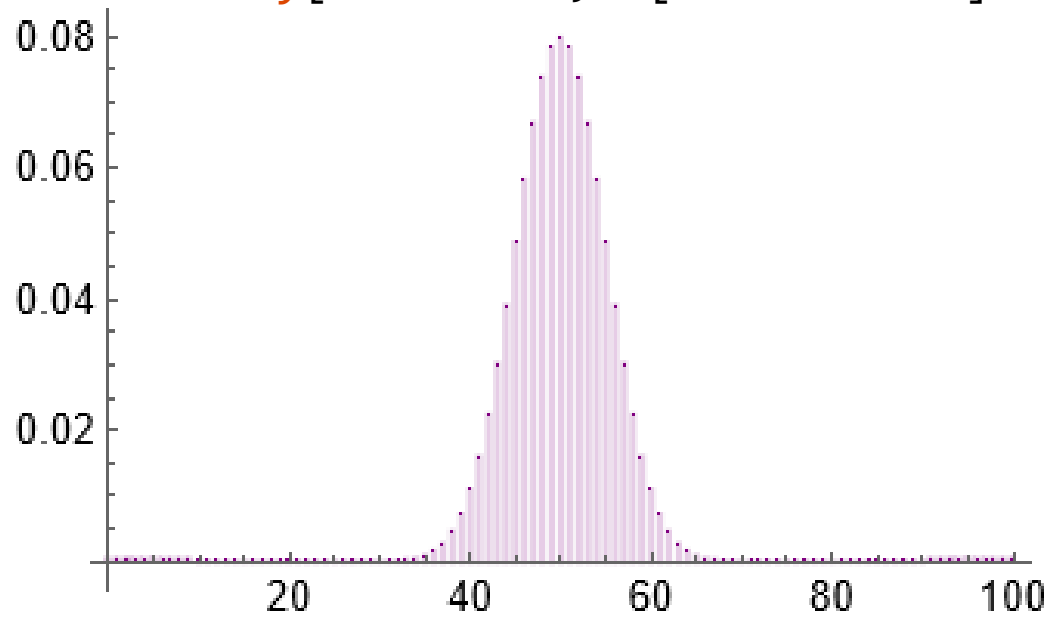

Output 0.028444

Output

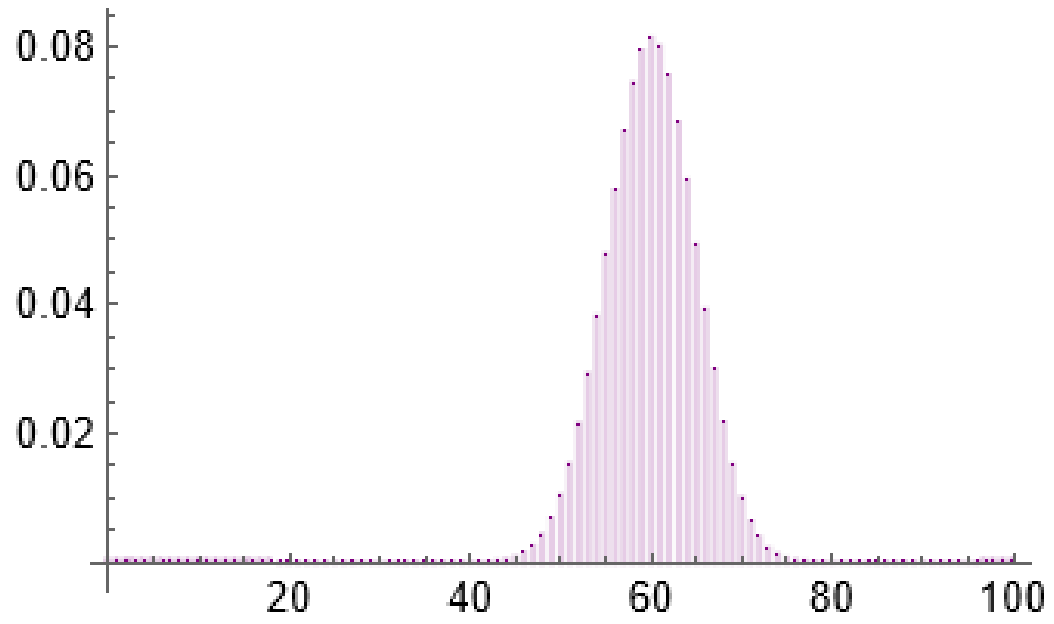

Output 0.543289

***Mathematica Examples 11.26***

Input

```
(* In this example, we assume a binomial distribution with parameters n=10 and p=0.5. We define the distribution using dist=BinomialDistribution[10,0.5]. We define the support of the distribution using support=Range[0,10]. We calculate the PMF for each support value using pmf=PDF[dist,#]&/@support. This applies the PDF function to each value in the support list and stores the results in the pmf list. We sum the PMF
```

```
         values using Total[pmf]. This sums all the elements in the pmf list, providing the
         sum over the support of the distribution. The result should be 1, indicating that
         the sum over the support of the distribution is equal to unity: *)

         dist=BinomialDistribution[10,0.5];   (* Example distribution. *)
         support=Range[0,10]   (* Support of the distribution. *)
         pmf=PDF[dist,#]&/@support   (* Calculate the PMF for each support value. *)
         Total[pmf]   (* Sum the PMF values. *)

Output   {0,1,2,3,4,5,6,7,8,9,10}
Output   {0.000976563,0.00976562,0.0439453,0.117188,0.205078,0.246094,0.205078,0.117188,0.04
         39453,0.00976562,0.000976563}
Output   1.
```

# UNIT 11.3

# POISSON DISTRIBUTION

**Mathematica Examples 11.27**

Input

```
(* In this example, we calculate the probability mass function (PMF) using the formula Exp[-lambda]*lambda^k/k! and plot it using ListPlot, lambda=4 (average number of events): *)

lambda=4;   (* Average number of events. *)

(* Define a discrete random variable and calculate the PMF: *)
x=Range[0,10];
pmf=Table[
    Exp[-lambda]*lambda^k/k!,
    {k,0,10}
    ];

(* Plot the PMF: *)
ListPlot[
 Transpose[{x,pmf}],
 Filling->Axis,
 PlotRange->All,
 PlotStyle->Purple,
 AxesLabel->{"x","P(X = x)"},
 PlotLabel->"Poisson Distribution PMF",
 ImageSize->200
 ]
```

Output

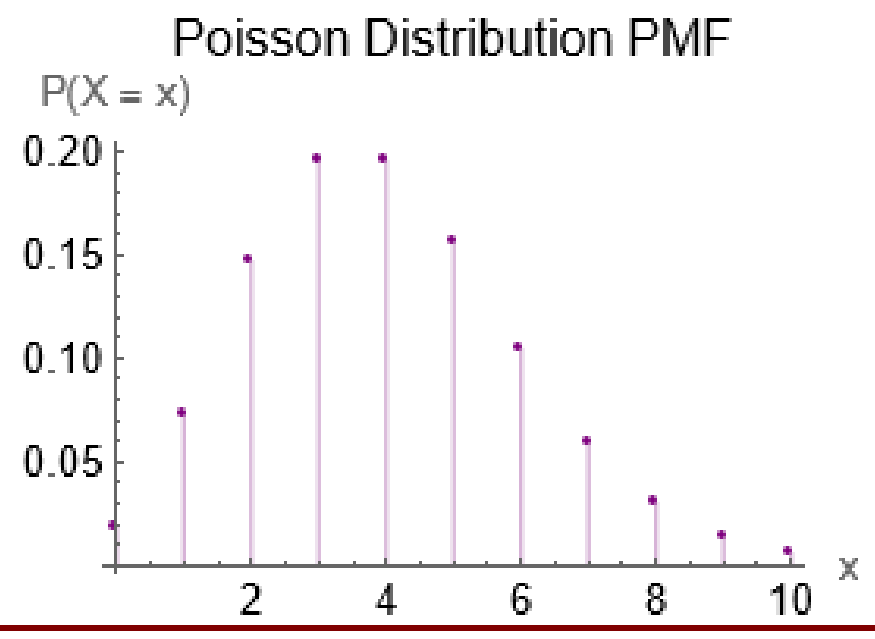

**Mathematica Examples 11.28**

Input

```
(* The code generates a discrete plot of the PMF for a Poisson distribution with three different values of parameter λ= (4, 8, and 16). The plot shows the values of the PMF for all possible values of i between 0 and 32: *)

DiscretePlot[
 Evaluate[
  Table[
   PDF[
    PoissonDistribution[λ],
    i
    ],
   {λ,{4,8,16}}
   ]
```

```
   ],
  {i,0,32},
  PlotRange->All,
  PlotMarkers->Automatic,
  PlotLegends->Placed[{"λ=4","λ=8","λ=16"},{0.8,0.75}],
  PlotStyle->{RGBColor[0.88,0.61,0.14],RGBColor[0.37,0.5,0.7],Purple},
  ImageSize->320,
  AxesLabel->{None,"PMF"}
  ]
```

Output

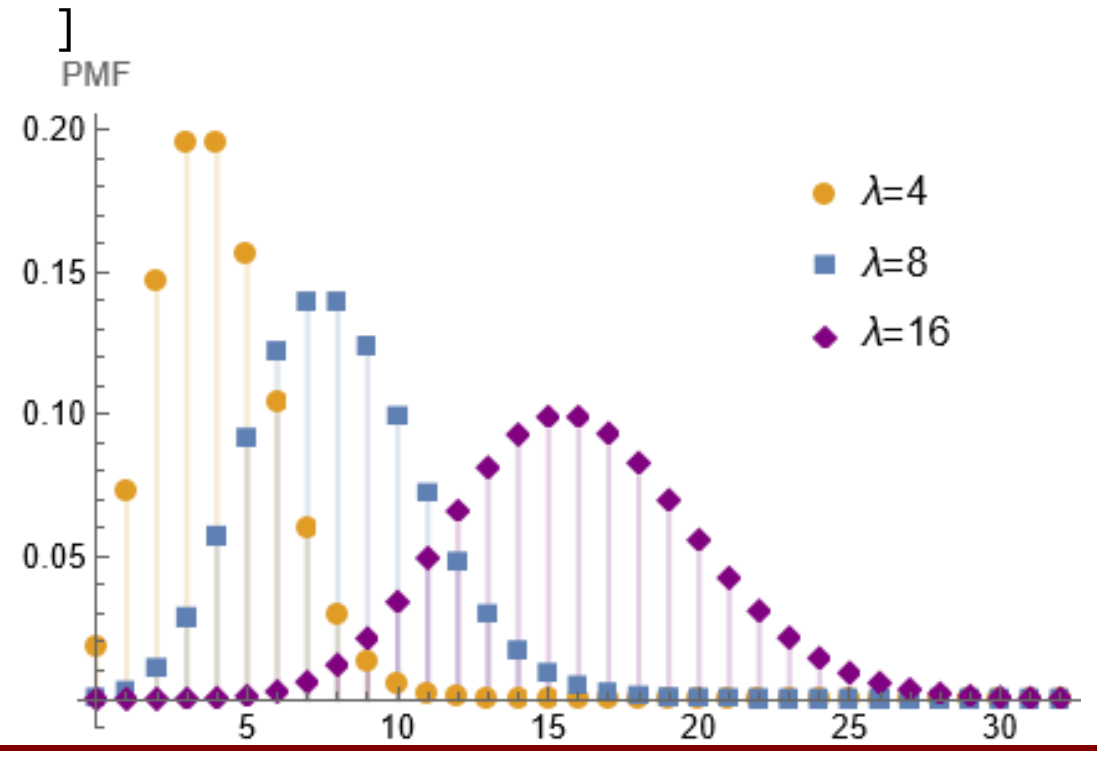

**Mathematica Examples 11.29**

Input

```
(* The code generates a discrete plot of the cumulative distribution function (CDF)
for a Poisson distribution with three different values of parameter λ= (4, 8, and
16). The plot shows the values of the CDF for all possible values of k between 0 and
32: *)

DiscretePlot[
 Evaluate[
  Table[
   CDF[
    PoissonDistribution[λ],
    i
    ],
   {λ,{4,8,16}}
   ]
  ],
 {i,0,32},
 ExtentSize->Right,
 PlotRange->All,
 PlotMarkers->Automatic,
 PlotLegends->Placed[{"λ=4","λ=8","λ=16"},{0.8,0.75}],
 PlotStyle->{RGBColor[0.88,0.61,0.14],RGBColor[0.37,0.5,0.7],Purple},
 ImageSize->320,
 AxesLabel->{None,"CDF"}
 ]
```

Output

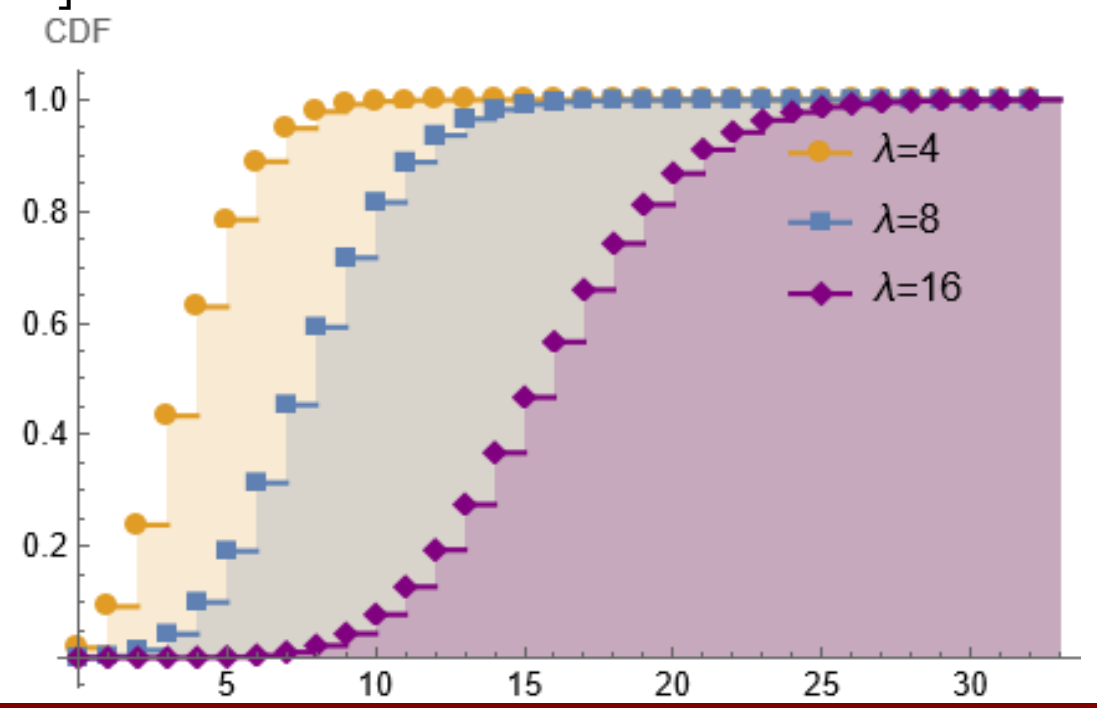

***Mathematica Examples 11.30***

Input
```
(* The code generates a histogram and a discrete plot of the PMF for a Poisson distribution with parameters λ=4 and sample size 10000: *)

sampledata=RandomVariate[
    PoissonDistribution[4],
    10^4
    ];

Show[
 Histogram[
  sampledata,
  {1},
  "PDF",
  ColorFunction->Function[{height},Opacity[height]],
  ChartStyle->Purple,
  ImageSize->320,
  AxesLabel->{None,"PDF"}
  ],
 DiscretePlot[
  PDF[
   PoissonDistribution[4],
   x
   ],
  {x,0,Max[data]},
  PlotStyle->PointSize[Medium],
  ColorFunction->"Rainbow"
  ]
 ]
```

Output

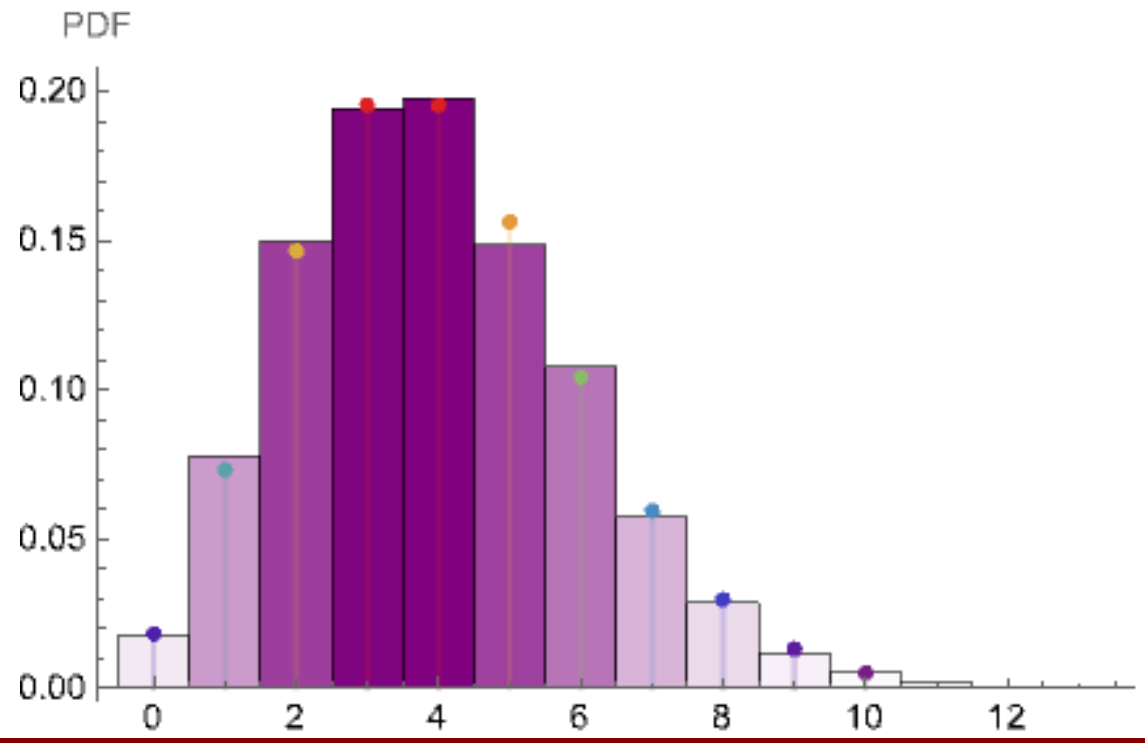

***Mathematica Examples 11.31***

Input
```
(* The code calculates and displays some descriptive statistics (mean, variance, standard deviation, kurtosis, and skewness) for a Poisson distribution with parameter λ: *)

Grid[
 Table[
  {
   statistics,
   FullSimplify[statistics[PoissonDistribution[λ]]],
   FullSimplify[statistics[PoissonDistribution[λ]]],
   FullSimplify[statistics[PoissonDistribution[λ]]]
   },
  {statistics,{Mean,Variance,StandardDeviation,Kurtosis,Skewness}}
  ],
 ItemStyle->12,
```

```
 Alignment->{{Right,Left}},
 Frame->All,
 Spacings->{Automatic,0.8}
 ]
```

Output

| | | | |
|---|---|---|---|
| Mean | λ | λ | λ |
| Variance | λ | λ | λ |
| StandardDeviation | Sqrt[λ] | Sqrt[λ] | Sqrt[λ] |
| Kurtosis | 3+1/λ | 3+1/λ | 3+1/λ |
| Skewness | 1/Sqrt[λ] | 1/Sqrt[λ] | 1/Sqrt[λ] |

***Mathematica Examples 11.32***

Input

```
(* The code calculates and displays some additional descriptive statistics (moments, central moments, and factorial moments) for a Poisson distribution with parameter λ: *)

Grid[
 Table[
  {
   statistics,
   FullSimplify[statistics[PoissonDistribution[λ],1]],
   FullSimplify[statistics[PoissonDistribution[λ],2]],
   FullSimplify[statistics[PoissonDistribution[λ],3]]
   },
  {statistics,{Moment,CentralMoment,FactorialMoment}}
  ],
 ItemStyle->12,
 Alignment->{{Right,Left}},
 Frame->All,
 Spacings->{Automatic,0.8}
 ]
```

Output

| | | | |
|---|---|---|---|
| Moment | λ | λ (1+λ) | λ (1+λ (3+λ)) |
| CentralMoment | 0 | λ | λ |
| FactorialMoment | λ | λ^2 | λ^3 |

***Mathematica Examples 11.33***

Input

```
(* The code generates a sample of 10000 values from a discrete Poisson distribution and estimates the distribution parameters (μ) from the sample using the EstimatedDistribution function. It then compares the density histogram of the sample with the PDF of the estimated distribution: *)

sampledata=RandomVariate[
    PoissonDistribution[4],
    10^4
    ];

(* Estimate the distribution parameters from sample data: *)
ed=EstimatedDistribution[
   sampledata,
   PoissonDistribution[μ]
   ]

(* Compare the density histogram of the sample with the PDF of the estimated distribution: *)
Show[
 Histogram[
  sampledata,
  {1},
```

```
          "PDF",
          ColorFunction->Function[{height},Opacity[height]],
          ChartStyle->Purple,
          ImageSize->320
          ],
         DiscretePlot[
          PDF[ed,x],
          {x,0,Max[sampledata]},
          PlotStyle->PointSize[Medium],
          ColorFunction->"Rainbow"
          ]
         ]
```

Output PoissonDistribution[3.9556]

Output

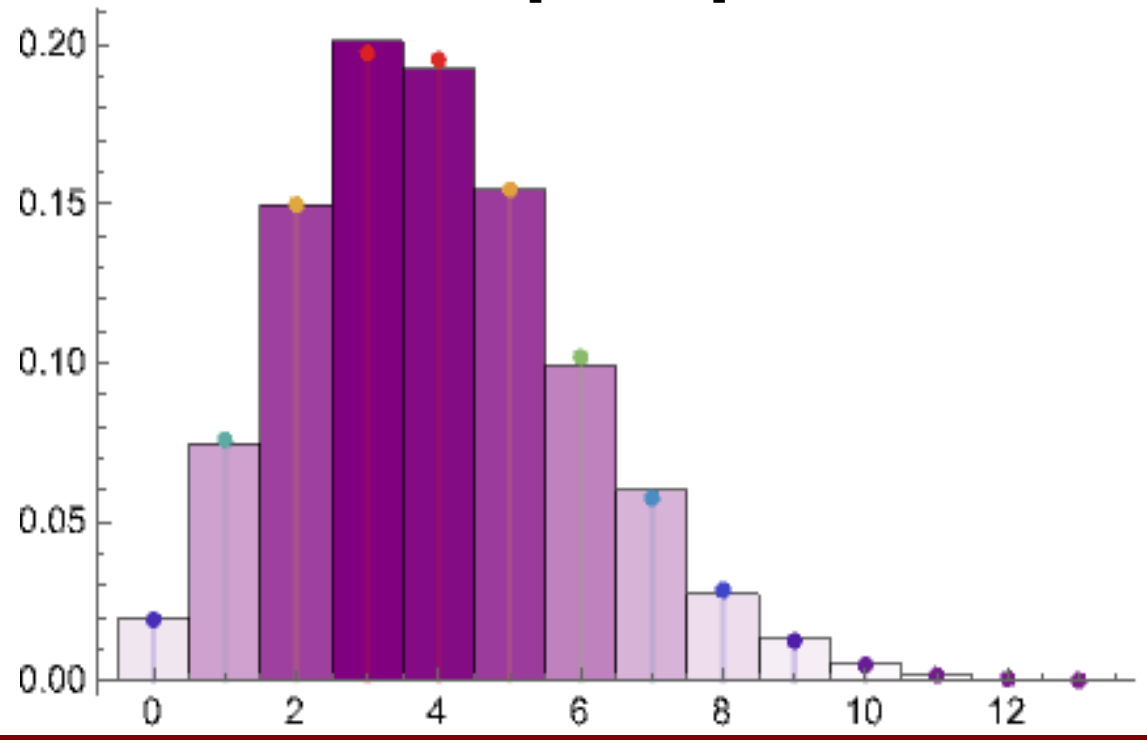

**Mathematica Examples 11.34**

Input

```
(* The code generates a histogram of 500 random samples drawn from a Poisson
distribution with a range of 0 to 10. The histogram is displayed with the "PDF"
option, which normalizes the bin heights to represent a probability density function.
The x-axis ranges from 0 to 10 with a bin size of 1. The mean and quartiles of the
data are also displayed as vertical lines in red and green, respectively: *)

dist=PoissonDistribution[5];
data=RandomVariate[dist,500];
Histogram[
 data,
 {0,10,1},
 "PDF",
 Epilog->{
   Directive[Red,Thickness[0.006]],
   Line[{{Mean[data],0},{Mean[data],0.15}}],
   Directive[Green,Dashed],
   Line[{{Quantile[data,0.25],0},{Quantile[data,0.25],0.15}}],
   Line[{{Quantile[data,0.5],0},{Quantile[data,0.5],0.15}}],
   Line[{{Quantile[data,0.75],0},{Quantile[data,0.75],0.15}}]
   },
 Frame->True,
 FrameLabel->{"x","PDF"},
 LabelStyle->Directive[Bold,Medium],
 ColorFunction->Function[{height},Opacity[height]],
 ChartStyle->Purple,
 ImageSize->320,
 AxesLabel->{None,"PDF"}
 ]
```

Output

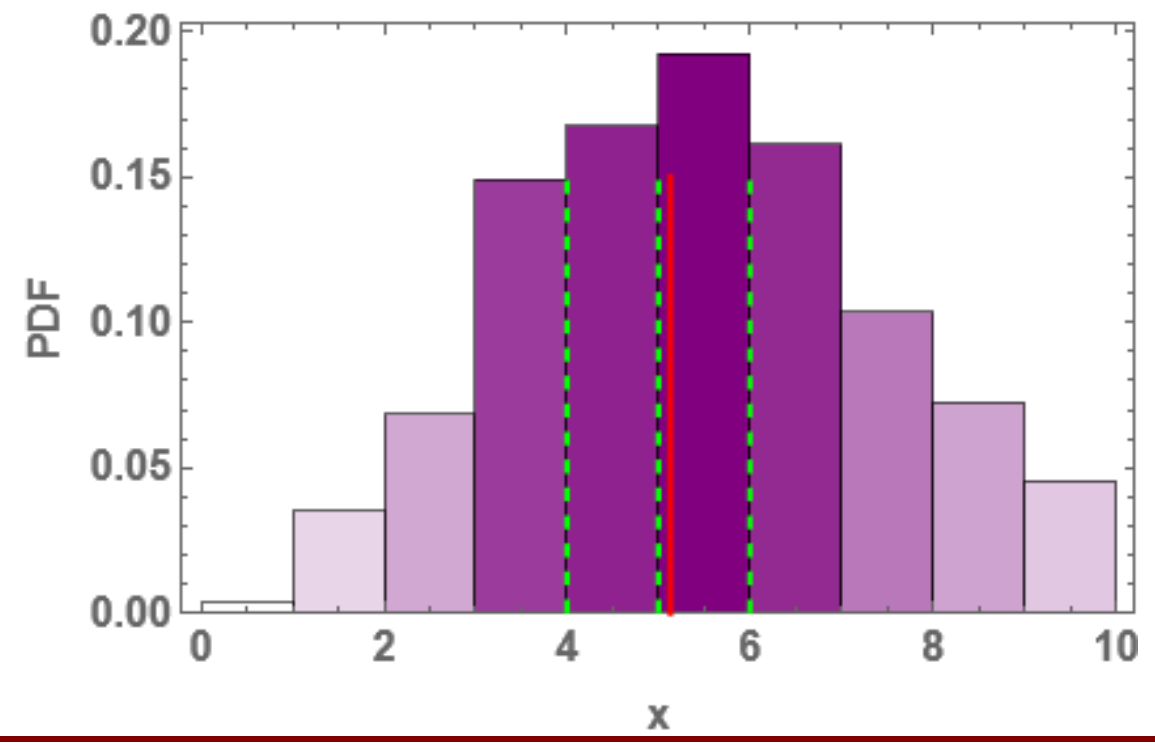

***Mathematica Examples 11.35***

Input

```
(* This code generates two sets of random data from Poisson distributions with
different means (λ1=3 and λ2=6) and creates a histogram to compare the distributions:
*)

lambda1=3;
lambda2=6;
n=1000;
data1=RandomVariate[
    PoissonDistribution[lambda1],
    n
    ];

data2=RandomVariate[
    PoissonDistribution[lambda2],
    n
    ];

Histogram[
 {data1,data2},
 LabelingFunction->Above,
 ChartLegends->{"Poisson with mean 3","Poisson with mean 6"},
 ChartStyle->{Directive[Opacity[0.5],Red],Directive[Opacity[0.6],Purple]},
 ImageSize->Medium,
 ImageSize->320
 ]
```

Output

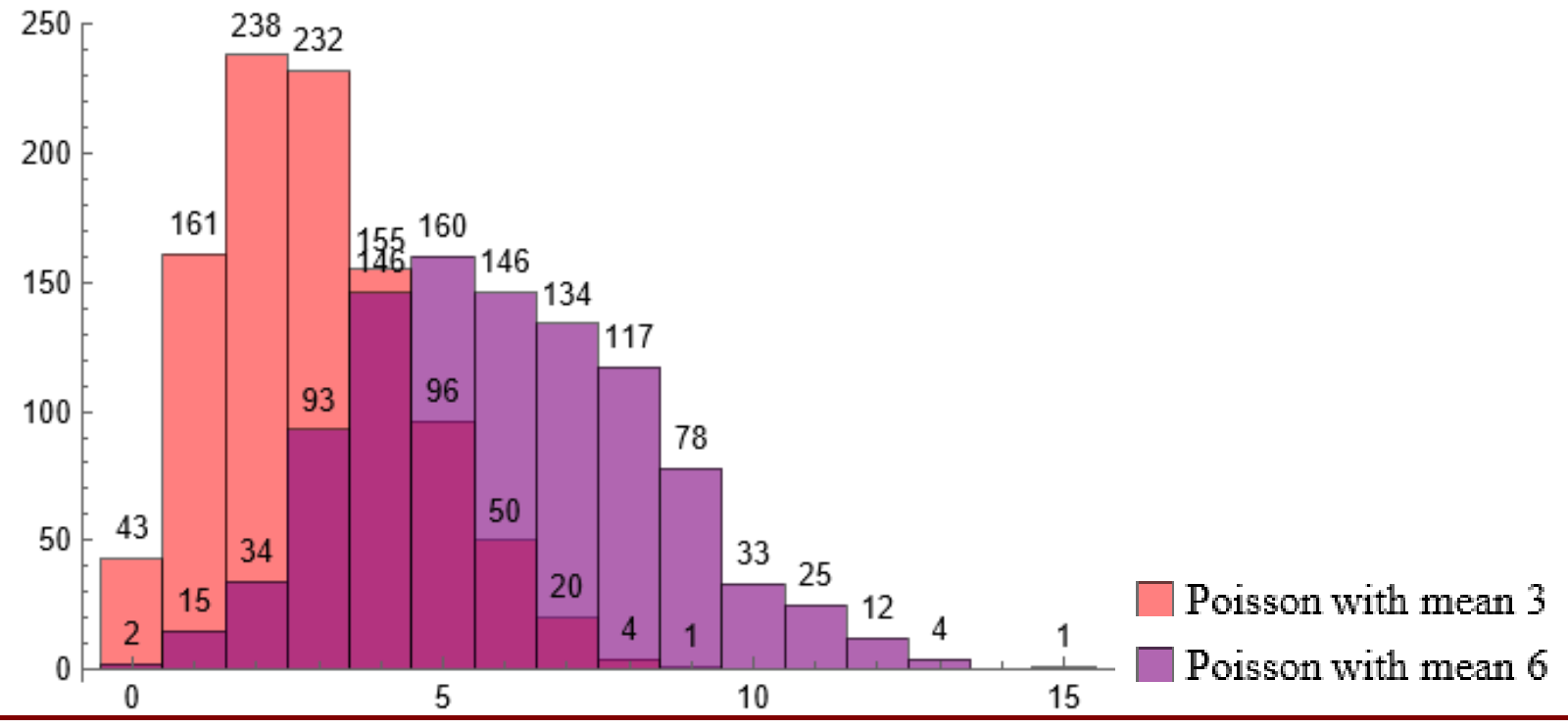

***Mathematica Examples 11.36***

Input

```
(* The code defines a Manipulate that creates a discrete Poisson distribution with
parameter lambda, generates n random variates from the distribution, and plots a
histogram of the sample. The Manipulate includes sliders to control the value lambda
and n: *)
```

```
Manipulate[
 data=RandomVariate[
   PoissonDistribution[lambda],
   n
   ];
 Histogram[
  data,
  ColorFunction->Function[{height},Opacity[height]],
  ChartStyle->Purple,
  ImageSize->320
  ],
 {{lambda,5},1,20},
 {{n,1000},100,5000,100}
 ]
```

Output

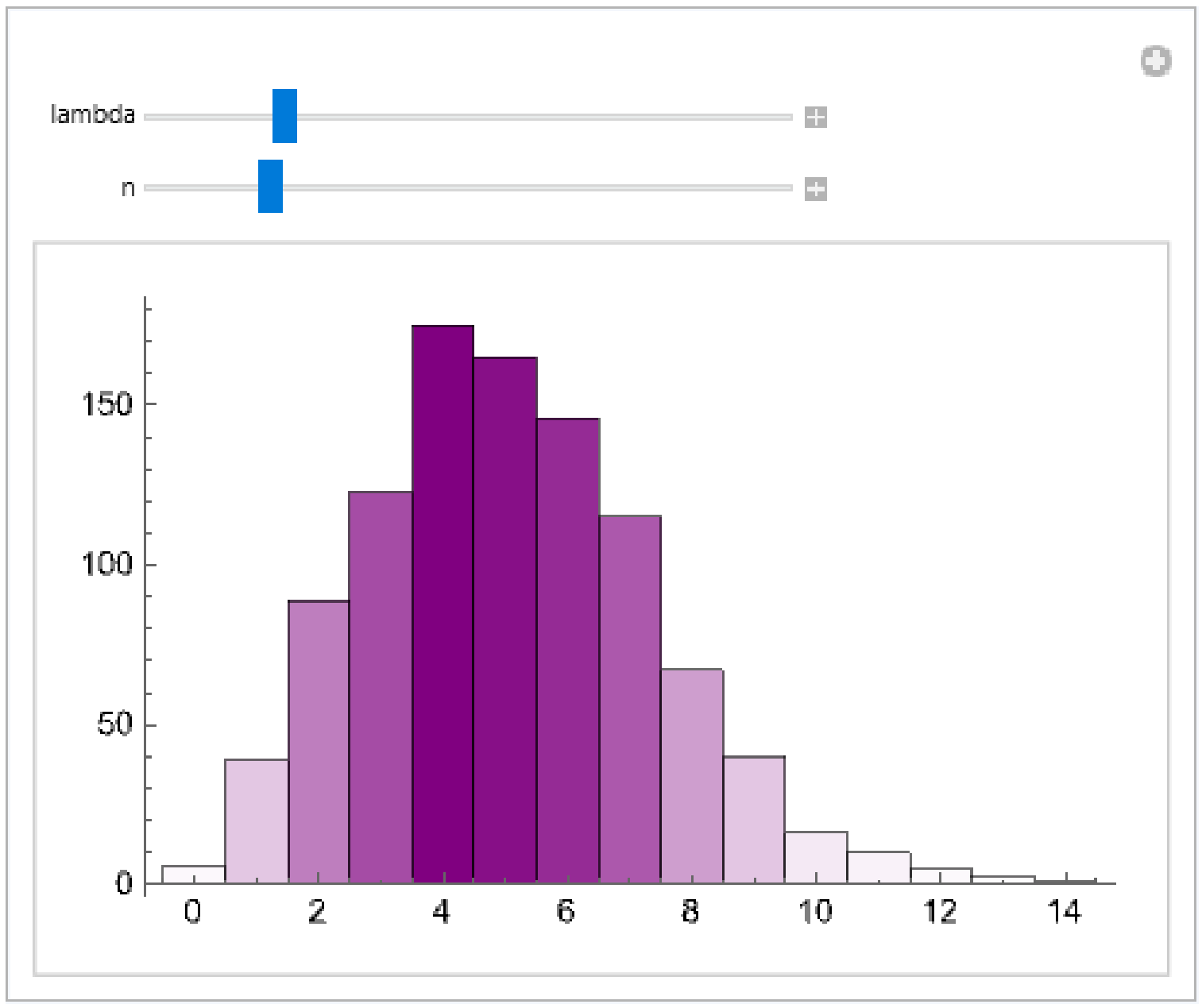

***Mathematica Examples 11.37***

Input

```
(* The code creates a plot of the CDF of a Poisson distribution using the Manipulate
function. The Manipulate function allows you to interactively change the values of
the parameter (λ) lambda: *)

Manipulate[
 Plot[
  CDF[
   PoissonDistribution[lambda],
   x
   ],
  {x,0,20},
  PlotRange->All,
  PlotLabel->StringForm[
    "Cumulative Distribution Function of
Poisson Distribution with λ = ``"
    ,lambda],
  Filling->Axis,
  FillingStyle->LightPurple,
  ImageSize->320,
  PlotStyle->Purple
```

```
    ],
   {{lambda,1},0.1,10,0.1}
   ]
```

Output

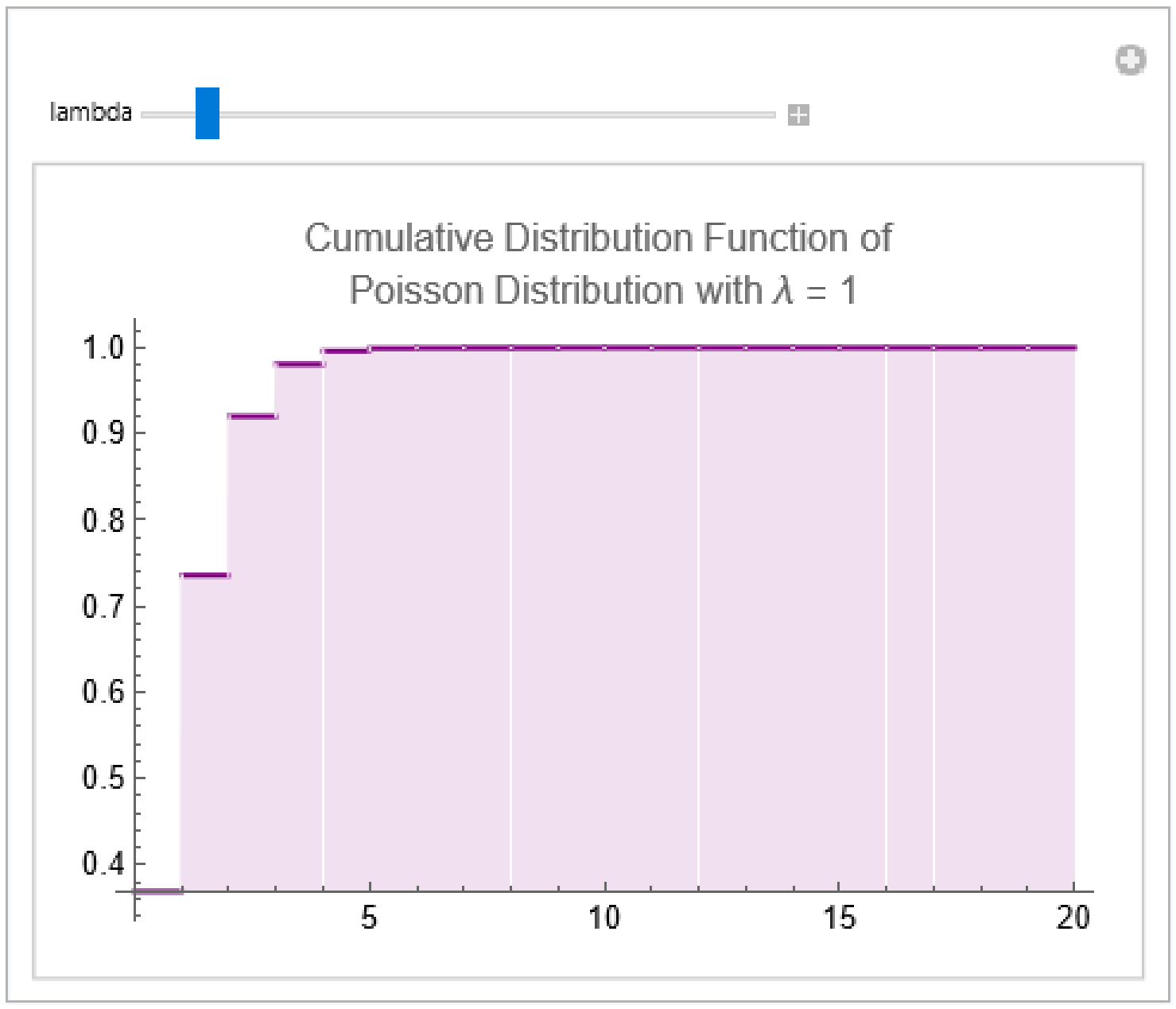

***Mathematica Examples 11.38***

Input

```
(* The code uses the Manipulate function to create an interactive plot that shows
the PMF and CDF of a Poisson distribution with parameter λ. The user can adjust the
value of λ using a slider: *)

Manipulate[
 Module[
  {dist},
  dist=PoissonDistribution[lambda];
  Grid[
   {
    {
     DiscretePlot[
      PDF[dist,x],
      {x,0,40},
      PlotRange->{{0,40},{0,0.4}},
      PlotStyle->{Purple,PointSize[0.02]},
      ExtentSize->Full,
      AxesLabel->{"x","PMF"},
      ImageSize->250,
      Filling->Axis,
      FillingStyle->Purple,
      PlotLabel->StringForm[" λ = ``",lambda]
      ],
     Plot[
      CDF[dist,x],
      {x,0,40},
      PlotStyle->{Purple,Thickness[0.007]},
      AxesLabel->{"k","CDF"},
      ImageSize->250,
      PlotLabel->StringForm[" λ = ``",lambda]
      ]
     }
```

```
   },
   Frame->All
   ]
  ],
 {{lambda,5,"lambda"},1,20,0.1}
 ]
```

Output

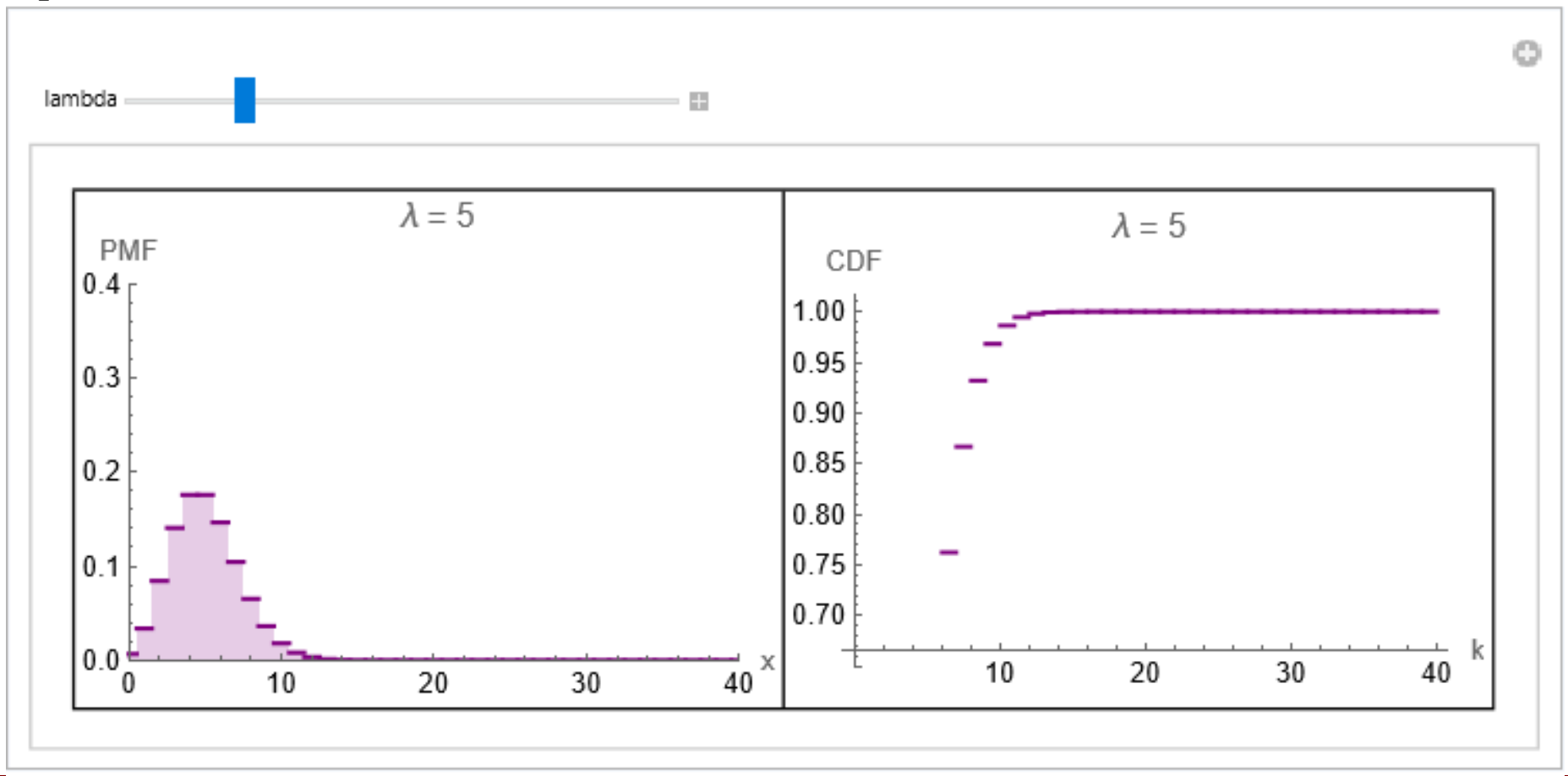

## *Mathematica Examples 11.39*

Input

```
(* The code generates 10,000 random samples from both the binomial and Poisson
distributions, and then plots their histograms using the Histogram function. The
resulting plot shows that the two histograms are nearly identical, indicating that
the two distributions converge as n becomes large and p becomes small: *)
n=1000;
p=0.01;
lambda=n*p;
binomSamples=RandomVariate[
    BinomialDistribution[n,p],
    10000
    ];
poissonSamples=RandomVariate[
    PoissonDistribution[lambda],
    10000
    ];
Histogram[
 {binomSamples,poissonSamples},
 Automatic,
 "PDF",
 ImageSize->320,
 ChartStyle->{Directive[Opacity[0.5],Blue],Directive[Opacity[0.4],Purple]}
 ]
```

Output

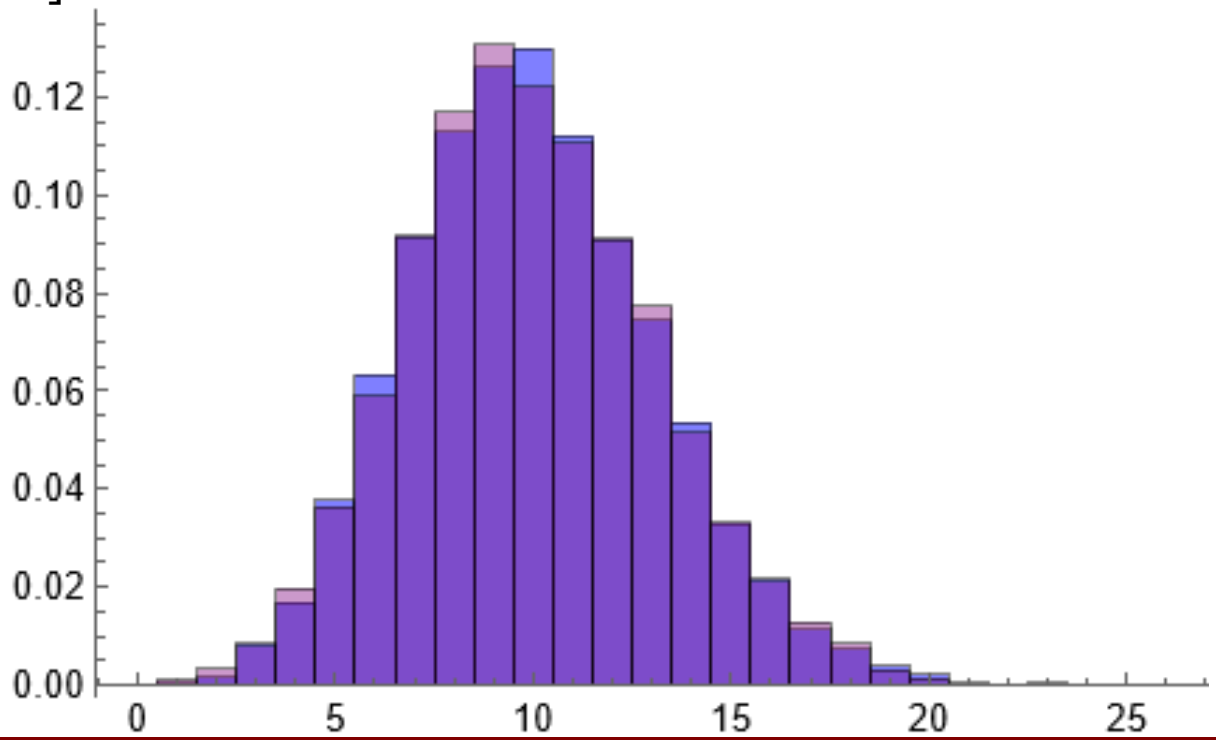

***Mathematica Examples 11.40***

Input

```
(* The code uses Manipulate to allow the user to adjust the number of trials n and
the success probability p of the binomial distribution. The code then generates 10000
samples from both the binomial and Poisson distributions, and plots the histograms
of these samples on the same graph, with different colors: *)

Manipulate[
 Module[
  {lambda=n*p},
  Show[
   Histogram[
    RandomVariate[
     BinomialDistribution[n,p],
     10000
     ],
    {0,n+1,1},
    "PDF",
    ChartStyle->Directive[Opacity[0.5],Blue],
    PlotRange->Full,
    PlotLabel->"Binomial and Poisson Distributions"
    ],
   Histogram[
    RandomVariate[
     PoissonDistribution[lambda],
     10000
     ],
    {0,n+1,1},
    "PDF",
    PlotRange->Full,
    ImageSize->320,
    ChartStyle->Directive[Opacity[0.5],Purple]
    ]
   ]
  ],
 {{n,10,"Number of trials"},1,200,10},
 {{p,0.5,"Success probability"},0.05,1,0.05}
 ]
```

Output

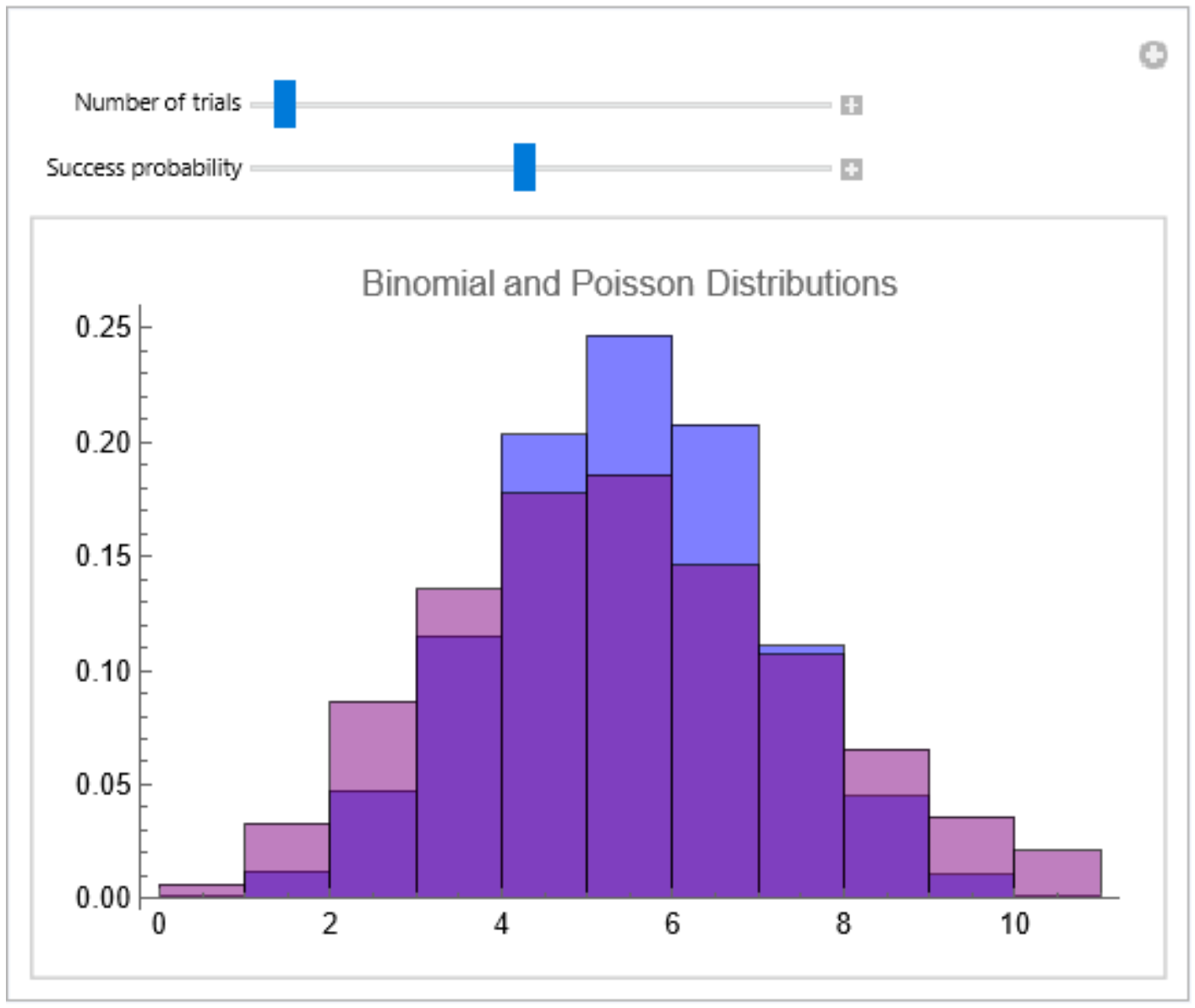

***Mathematica Examples 11.41***

Input
```
(* The code uses Manipulate to allow the user to adjust the rate parameter lambda of the Poisson distribution. The code then generates 10000 samples from both the binomial and Poisson distributions with fixed values of n=50 and p=0.2, and plots the histograms of these samples on the same graph, showing how the Poisson distribution becomes a better approximation of the binomial distribution as lambda increases: *)

Manipulate[
 Module[
  {n=100,p=0.2},
  Show[
   Histogram[
    RandomVariate[
     BinomialDistribution[n,p],
     10000
     ],
    {0,n+1,1},
    "PDF",
    PlotRange->Full,
    ImageSize->320,
    ChartStyle->Directive[Opacity[0.5],Blue],
    PlotLabel->"Binomial and Poisson Distributions"
    ],
   Histogram[
    RandomVariate[
     PoissonDistribution[lambda],
     10000
     ],
    {0,n+1,1},
    "PDF",
    ImageSize->320,
    PlotRange->Full,
    ChartStyle->Directive[Opacity[0.5],Purple]
    ]
   ]
  ],
 {{lambda,20,"lambda"},1,50,1}
 ]
```

Output

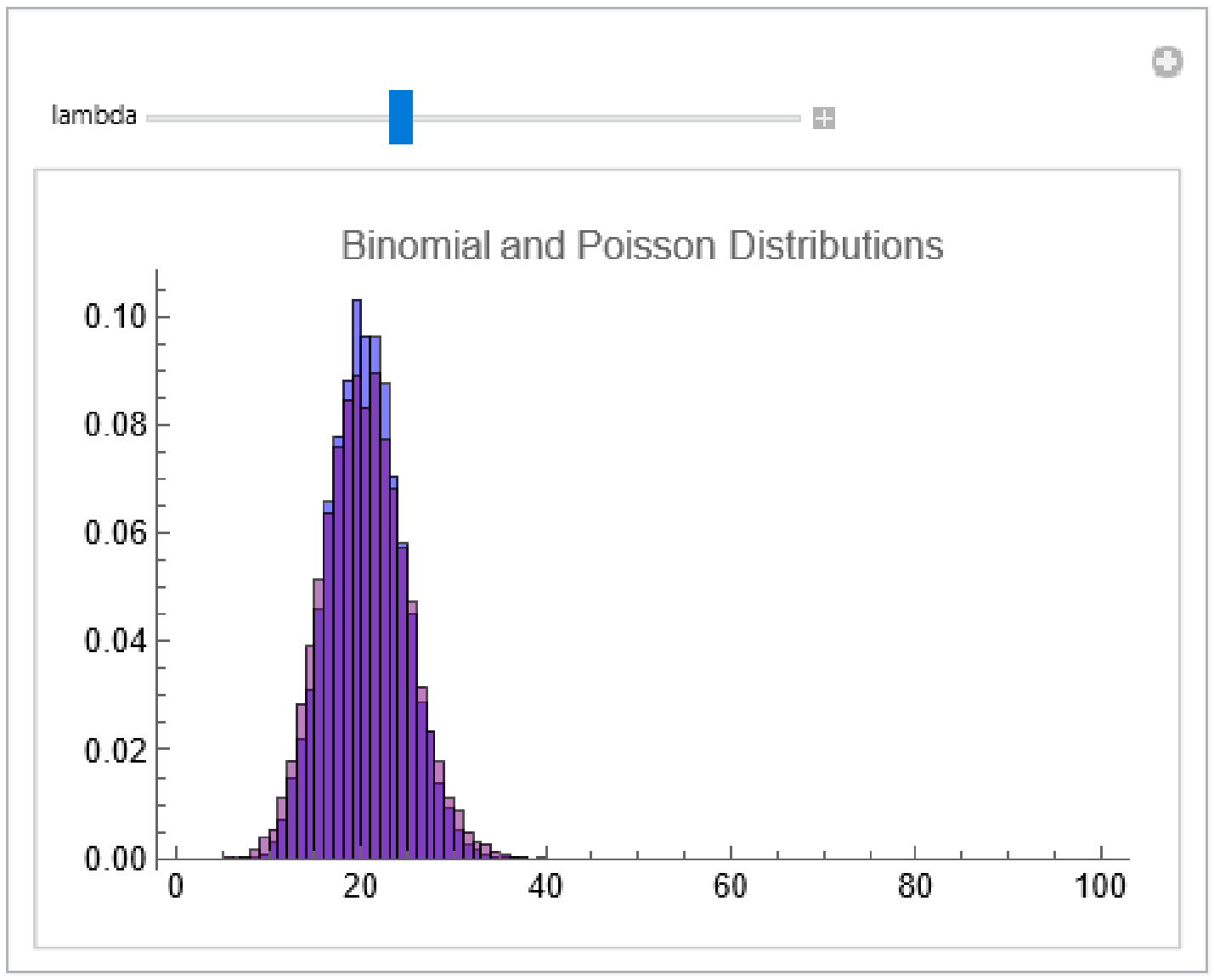

***Mathematica Examples 11.42***

Input
```
(* Prove that Poisson distribution is a limiting case for binomial distribution with p=μ/n: *)

bd=PDF[BinomialDistribution[n,μ/n],k]
lbd=Limit[bd,n->∞]
FullSimplify[lbd-PDF[PoissonDistribution[μ],k],k>=0]
```

Output
```
{
 {\[Piecewise], {
   {(μ/n)^k (1-μ/n)^(-k+n) Binomial[n,k], 0<=k<=n},
   {0, True}
  }}
}
 {
 {\[Piecewise], {
   {(E^-μ μ^k)/Gamma[1+k], k>=0},
   {0, True}
```

Output
```
0
```

***Mathematica Examples 11.43***

Input
```
(* Simulate the daily accidents: *)
(* The first part of the code generates a list of 20 random values from a Poisson distribution with λ=50, which simulates the number of accidents in a city on a daily basis: *)

sampledata=RandomVariate[
   PoissonDistribution[50],
   20
   ]

ListPlot[
 sampledata,
 Filling->Axis,
 PlotStyle->Purple,
 AxesLabel->Automatic,
 ImageSize->170
 ]

(* The second part of the code calculates the probability of having 40 or more accidents in a single day: *)
Probability[
 x>=40,
 x\[Distributed]PoissonDistribution[50.]
 ]

(* The third part of the code calculates the standard deviation of the Poisson distribution with λ=50: *)
StandardDeviation[
 PoissonDistribution[50.]
 ]
```

Output
```
{41,52,55,62,58,57,49,53,50,47,53,68,50,55,45,54,45,64,59,54}
```

Output 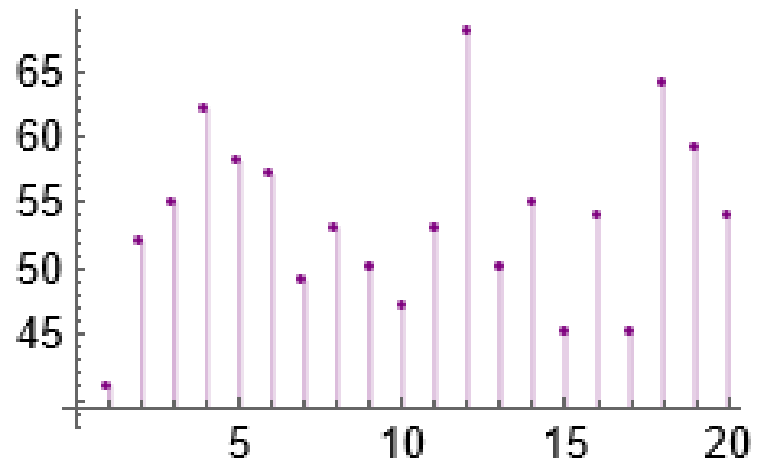

Output 0.93543

Output 7.07107

*Mathematica Examples 11.44*

Input
```
(* Simulate the raindrop count for each 5-second interval*)
(* The code simulates the count of raindrops falling into a bucket in a 5 -second interval based on a Poisson distribution with a mean of 50. It then plots the simulated data using ListPlot. The code then calculates the probability of exactly 20 raindrops falling into the bucket in a 5-second interval: *)

sampledata=RandomVariate[
  PoissonDistribution[50.],
  30
  ]

ListPlot[
 sampledata,
 Filling->Axis,
 PlotStyle->Purple,
 AxesLabel->Automatic,
 ImageSize->170
 ]

Probability[
 x==20,
 x\[Distributed]PoissonDistribution[50.]
 ]
```

Output {39,42,51,47,52,49,50,42,49,48,47,52,42,46,57,46,51,60,62,45,52,55,61,46,47,56,47,47,55,40}

Output 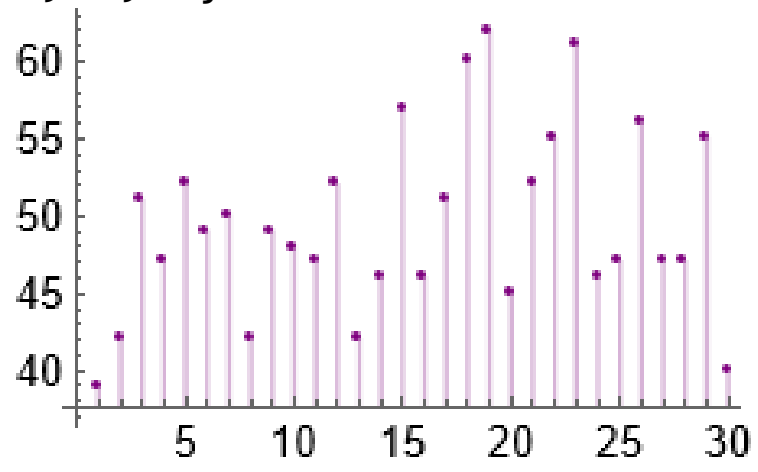

Output 7.56051*10^-7

*Mathematica Examples 11.45*

Input
```
(* In a book of 650 pages 200 typographical errors occur. Assuming Poisson law for the number of errors per page: *)
(* Tbe average number of typograpbical errors per page in the book is given by: *)
λ=200/650.0

(* The probability that a page will contain no error is: *)
Probability[x==0,x\[Distributed]PoissonDistribution[λ]]

(* The probabilities that a page will contain {0,1,2,...,10} errors are: *)
errorsofapage=Table[
```

```
    Probability[x==i,x\[Distributed]PoissonDistribution[λ]],
    {i,0,10}
    ]

  ListPlot[
   errorsofapage,
   Filling->Axis,
   PlotStyle->Purple,
   AxesLabel->Automatic,
   ImageSize->200
   ]

  (* The following code will simulate the above results by generating a large number
  of trials (in this case, 1 million trials). For each trial, it simulates the number
  of errors on 1 page using the Poisson distribution with an average of λ. It then
  counts the number of trials where there are no errors in the 1 page. Finally, it
  calculates the estimated probability by dividing the count of no errors by the total
  number of trials: *)

  λ=200/650; (*Average number of errors per page*)
  numTrials=10^6; (*Number of trials for simulation*)
  countNoErrors=0; (*Counter for no errors in 1 page*)

  Do[
   errors=RandomVariate[PoissonDistribution[λ],1];(*Simulate errors for 1 page*)
   If[Total[errors]==0,countNoErrors++],(*Check if no errors*)
   {numTrials}
   ]
  countNoErrors
  estimatedProbability=N[countNoErrors/numTrials] (*Estimate the probability*)
```

```
Output  0.307692
Output  0.735141
Output  {0.735141,0.226197,0.0347996,0.00356919,0.000274553,0.0000168956,8.6644*10^-
        7,3.80853*10^-8,1.46482*10^-9,5.00792*10^-11,1.5409*10^-12}
Output
```

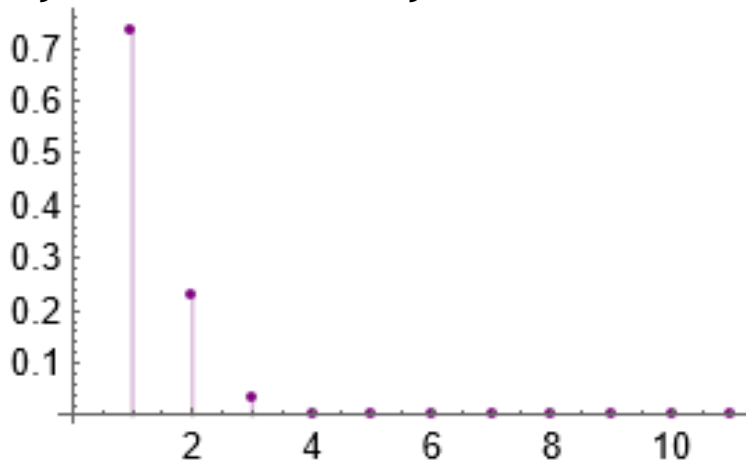

```
Output  735118
Output  0.735118
```

***Mathematica Examples 11.46***

```
Input   (* In a book of 650 pages 200 typographical errors occur. Assuming Poisson law for
        the number of errors per page: *)

        (* Tbe average number of typograpbical errors per 3 page in the book is given by: *)
        λ=(200/650.0);

        (* The required probability that a random sample of 3 pages will contain no error
        is: *)
        (Probability[x==0,x\[Distributed]PoissonDistribution[λ]])^3

        (* Also we can use the following code: *)
        N[
         Probability[
```

```
    x==0,
    x\[Distributed]PoissonDistribution[(200/650.0)*3]
    ]
   ]

  (* The probabilities that 3 pages will contain {0,1,2,...,10} errors are: *)
  errorsof3page=Table[
    Probability[x==i,x\[Distributed]PoissonDistribution[(200/650.0)*3]],
    {i,0,10}
    ]

  ListPlot[
   errorsof3page,
   Filling->Axis,
   PlotStyle->Purple,
   AxesLabel->Automatic,
   ImageSize->200
   ]

  (* The following code will simulate the above results by generating a large number
  of trials (in this case, 1 million trials). For each trial, it simulates the number
  of errors on 3 pages using the Poisson distribution with an average of λ. It then
  counts the number of trials where there are no errors in the 3 pages. Finally, it
  calculates the estimated probability by dividing the count of no errors in the 3
  pages by the total number of trials: *)

  λ=200/650; (*Average number of errors per page*)
  numTrials=10^6; (*Number of trials for simulation*)
  countNoErrors=0; (*Counter for no errors in 3 pages*)

  Do[
   errors=RandomVariate[PoissonDistribution[λ],3];(*Simulate errors for 3 pages*)
   If[Total[errors]==0,countNoErrors++],(*Check if no errors*)
   {numTrials}
   ]
  countNoErrors
  estimatedProbability=N[countNoErrors/numTrials] (*Estimate the probability*)

Output  0.397295
Output  0.397295
Output  {0.397295,0.366734,0.169262,0.0520805,0.0120186,0.00221881,0.000341356,0.000045014,
        5.19392*10^-6,5.3271*10^-7,4.91732*10^-8}
Output
```

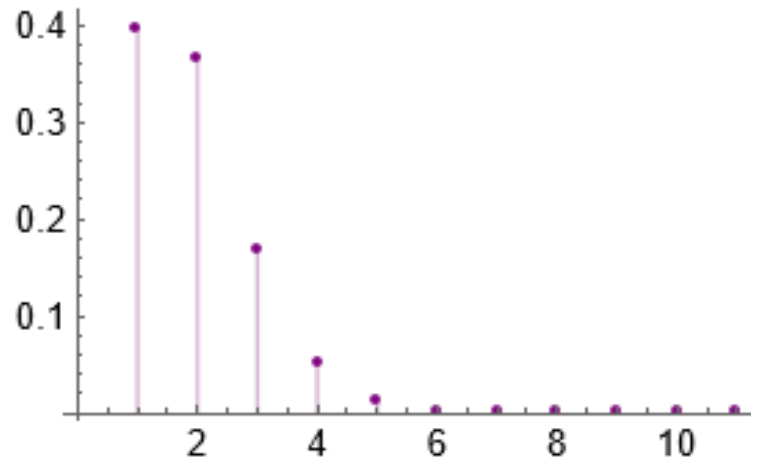

```
Output  396933
Output  0.396933
```

***Mathematica Examples 11.47***

```
Input   (* Assuming that the quantity of phone calls received by a phone exchange within a
        particular time frame is random, let us say that the number of calls received between
        10 A.M. and 11 A.M. is represented by X1, and it follows a Poisson distribution with
        a parameter of 2. In the same way, the number of calls received between 11 A.M. and
        12 noon, X2, follows a Poisson distribution with a parameter of 6. Given that X1 and
```

```
        X2 are independent of each other, what is the probability the total number of calls
        received between 10 A.M. and 12 noon is more than 5?:  *)

        (* We use the Poisson distribution with parameter 2+6=8 to model the total number of
        calls in 2 hours, and use the Probability function to find the probability that the
        total number of calls is less than or equal to 5. Then, we subtract this probability
        from 1 to get the probability that the total number of calls is more than 5: *)

        dist=PoissonDistribution[2+6];

        (* The probability that the total number of calls is less than 5 using Poisson
        distribution: *)
        N[
         Probability[
          X1+X2<=5,
          X1+X2\[Distributed]dist
          ]
         ]

        (* The probability that the total number of calls is greater than 5 using Poisson
        distribution: *)
        P=N[
          1-Probability[X1+X2<=5,X1+X2\[Distributed]dist]
          ]
        (* or *)
        P=N[
          Probability[(X1+X2)>5,(X1+X2)\[Distributed]dist]
          ]

Output  0.191236
Output  0.808764
Output  0.808764
```

# UNIT 11.4

# NEGATIVE BINOMIAL DISTRIBUTION

*Mathematica Examples 11.48*

Input

```
(* The code generates a discrete plot of the probability mass function (PMF) for a
negative binomial distribution with parameters r=2 and three different values of p
(0.1, 0.2, and 0.3). The plot shows the values of the PMF for all possible values of
k between 0 and 25: *)

DiscretePlot[
 Evaluate[
  Table[
   PDF[
    NegativeBinomialDistribution[2,p],
    k
    ],
   {p,{0.1,0.2,0.3}}
   ]
  ],
 {k,0,25},
 PlotRange->All,
 PlotMarkers->Automatic,
 PlotLegends->Placed[{"r=2,p=0.1","r=2,p=0.2","r=2,p=0.3"},{0.8,0.75}],
 PlotStyle->{RGBColor[0.88,0.61,0.14],RGBColor[0.37,0.5,0.7],Purple},
 ImageSize->250,
 AxesLabel->{None,"PMF"}
 ]
```

Output

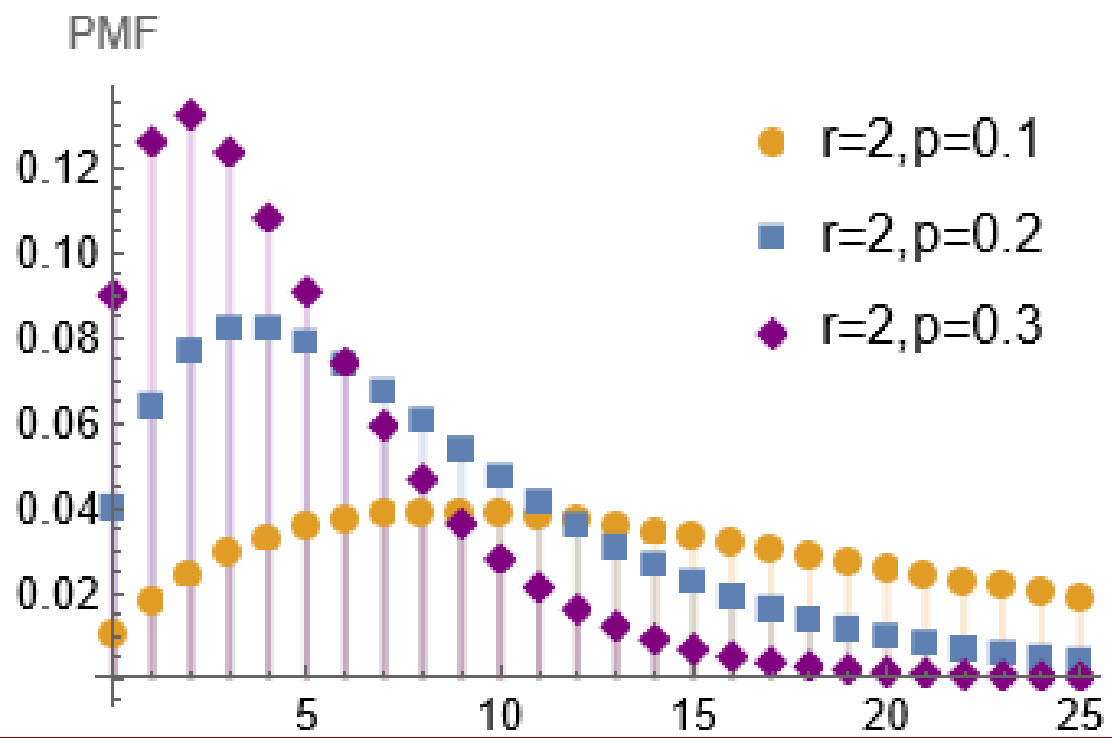

*Mathematica Examples 11.49*

Input

```
(* The code generates a discrete plot of the cumulative distribution function (CDF)
for a negative binomial distribution with parameters r=2 and three different values
of p (0.1, 0.2, and 0.3). The plot shows the values of the CDF for all possible
values of k between 0 and 25: *)

DiscretePlot[
 Evaluate[
  Table[
   CDF[
    NegativeBinomialDistribution[2,p],
```

```
      k
      ],
     {p,{0.1,0.2,0.3}}
     ]
    ],
   {k,0,25},
   ExtentSize->Right,
   PlotRange->All,
   PlotMarkers->Automatic,
   PlotLegends->Placed[{"r=2,p=0.1","r=2,p=0.2","r=2,p=0.3"},{0.8,0.3}],
   PlotStyle->{RGBColor[0.88,0.61,0.14],RGBColor[0.37,0.5,0.7],Purple},
   ImageSize->250,
   AxesLabel->{None,"CDF"}
   ]
```

Output

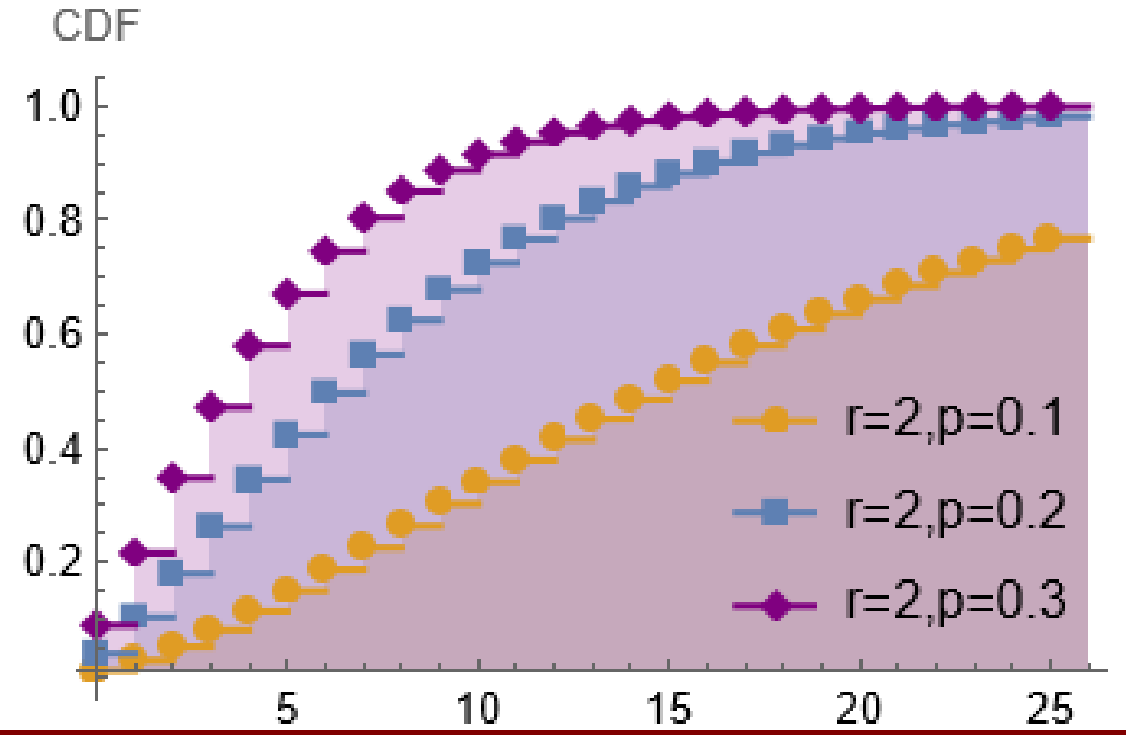

***Mathematica Examples 11.50***

Input

```
(* The code generates a histogram and a discrete plot of the PMF for a negative
binomial distribution with parameters r=10, p=0.3 and sample size 10000: *)

data=RandomVariate[NegativeBinomialDistribution[10,0.3],10^4];

Show[
 Histogram[
  data,
  {2},
  "PDF",
  ColorFunction->Function[{height},Opacity[height]],
  ChartStyle->Purple,
  ImageSize->320,
  AxesLabel->{None,"PDF"}
  ],
 DiscretePlot[
  PDF[
   NegativeBinomialDistribution[10,0.3],
   x
   ],
  {x,0,50},
  ImageSize->320,
  PlotStyle->PointSize[Medium],
  ColorFunction->"Rainbow"
  ]
 ]
```

Output

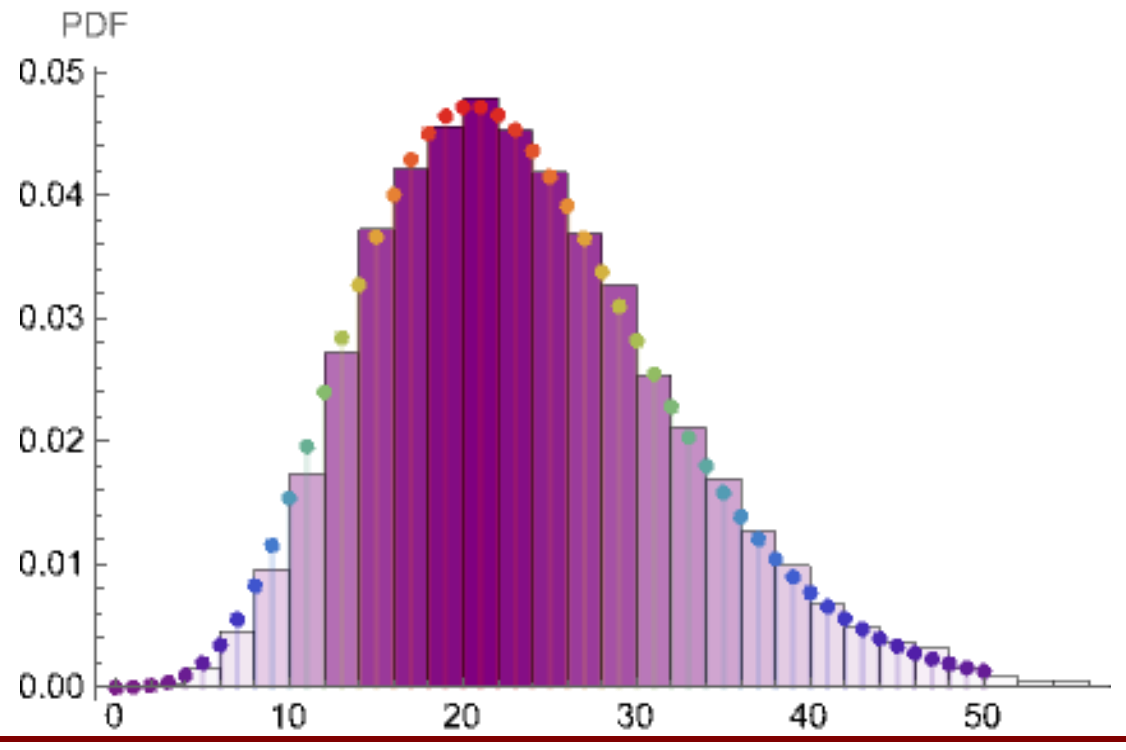

**Mathematica Examples 11.51**

Input

```
(* The code calculates and displays some descriptive statistics (mean, variance, standard deviation, kurtosis, and skewness) for a negative binomial distribution with parameters r and p: *)

Grid[
 Table[
  {
   statistics,
   FullSimplify[statistics[NegativeBinomialDistribution[r,p]]]
   },
  {statistics,{Mean,Variance,StandardDeviation,Kurtosis,Skewness}}
  ],
 ItemStyle->12,
 Alignment->{{Right,Left}},
 Frame->All,
 Spacings->{Automatic,0.8}
 ]
```

Output

| | |
|---|---|
| Mean | (-1+1/p) r |
| Variance | (r-p r)/p^2 |
| StandardDeviation | Sqrt[r - p r]/p |
| Kurtosis | 3+(6+(-6+p) p)/(r-p r) |
| Skewness | (2-p)/ Sqrt[r - p r] |

**Mathematica Examples 11.52**

Input

```
(* The code calculates and displays some additional descriptive statistics (moments, central moments, and factorial moments) for a negative binomial distribution with parameters r and p: *)

Grid[
 Table[
  {
   statistics,
   FullSimplify[statistics[NegativeBinomialDistribution[r,p],1]],
   FullSimplify[statistics[NegativeBinomialDistribution[r,p],2]]
   },
  {statistics,{Moment,CentralMoment,FactorialMoment}}
  ],
 ItemStyle->12,
 Alignment->{{Right,Left}},
 Frame->All,
 Spacings->{Automatic,0.8}
 ]
```

Output

| | | |
|---|---|---|
| Moment | (-1+1/p) r | ((-1+p) r (-1+(-1+p) r))/p^2 |
| CentralMoment | 0 | (r-p r)/p^2 |
| FactorialMoment | (-1+1/p) r | (-1+1/p)^2 r (1+r) |

***Mathematica Examples 11.53***

Input

```
(* This code generates a random sample of size 10,000 from a negative binomial distribution with parameters r=30 and p=0.5, estimates the distribution parameters using the EstimatedDistribution function, and then compares the histogram of the sample with the estimated PDF of the negative binomial distribution using a histogram and a discrete plot of the PDF: *)

sampledata=RandomVariate[
   NegativeBinomialDistribution[30,.5],
   10^4
   ];

(* Estimate the distribution parameters from sample data: *)
ed=EstimatedDistribution[
  sampledata,
  NegativeBinomialDistribution[r,p]
  ]

(* Compare a density histogram of the sample with the PDF of the estimated distribution: *)
Show[
 Histogram[
  sampledata,
  {1},
  "PDF",
  ColorFunction->Function[{height},Opacity[height]],
  ChartStyle->Purple,
  ImageSize->320
  ],
 DiscretePlot[
  PDF[ed,x],
  {x,0,80},
  PlotStyle->PointSize[Medium],
  ImageSize->320,
  ColorFunction->"Rainbow"
  ]
 ]
```

Output NegativeBinomialDistribution[31.4471,0.51108]

Output

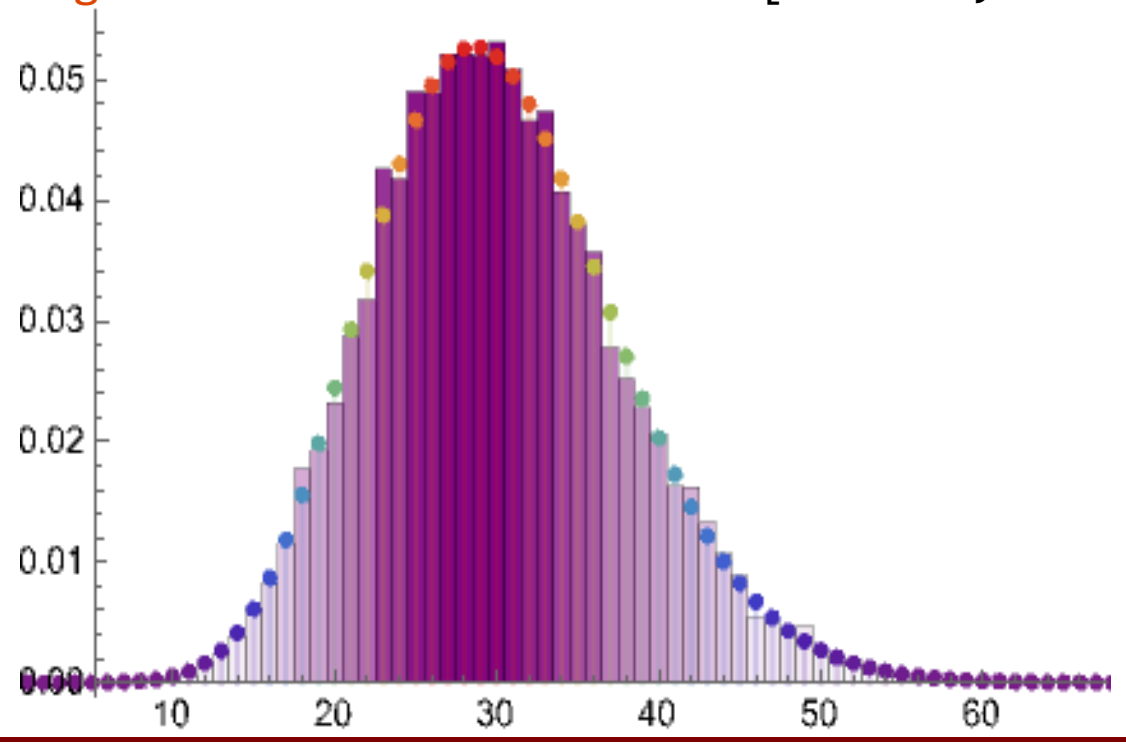

***Mathematica Examples 11.54***

Input

```
(* The code generates a dataset of 1000 observations from a negative binomial distribution with parameters r=30 and p=0.3. Then, it computes the sample mean and
```

```
quartiles of the data, and plots a histogram of the data using the "PDF" option to
display the probability density function. Additionally, the code adds vertical lines
to the plot corresponding to the sample mean and quartiles: *)

data=RandomVariate[
   NegativeBinomialDistribution[30,0.5],
   1000
   ];
mean=Mean[data];
quartiles=Quantile[
   data,
   {0.25,0.5,0.75}
   ];
Histogram[
 data,
 Automatic,
 "PDF",
 Epilog->{
   Directive[Red,Thickness[0.006]],
   Line[{{mean,0},{mean,0.25}}],
   Directive[Green,Dashed],
   Line[{{quartiles[[1]],0},{quartiles[[1]],0.25}}],
   Line[{{quartiles[[2]],0},{quartiles[[2]],0.25}}],
   Line[{{quartiles[[3]],0},{quartiles[[3]],0.25}}]
   },
 ColorFunction->Function[{height},Opacity[height]],
 ImageSize->320,
 ChartStyle->Purple,
 PlotRange->{{0,60},{0,0.06}}
 ]
```

Output

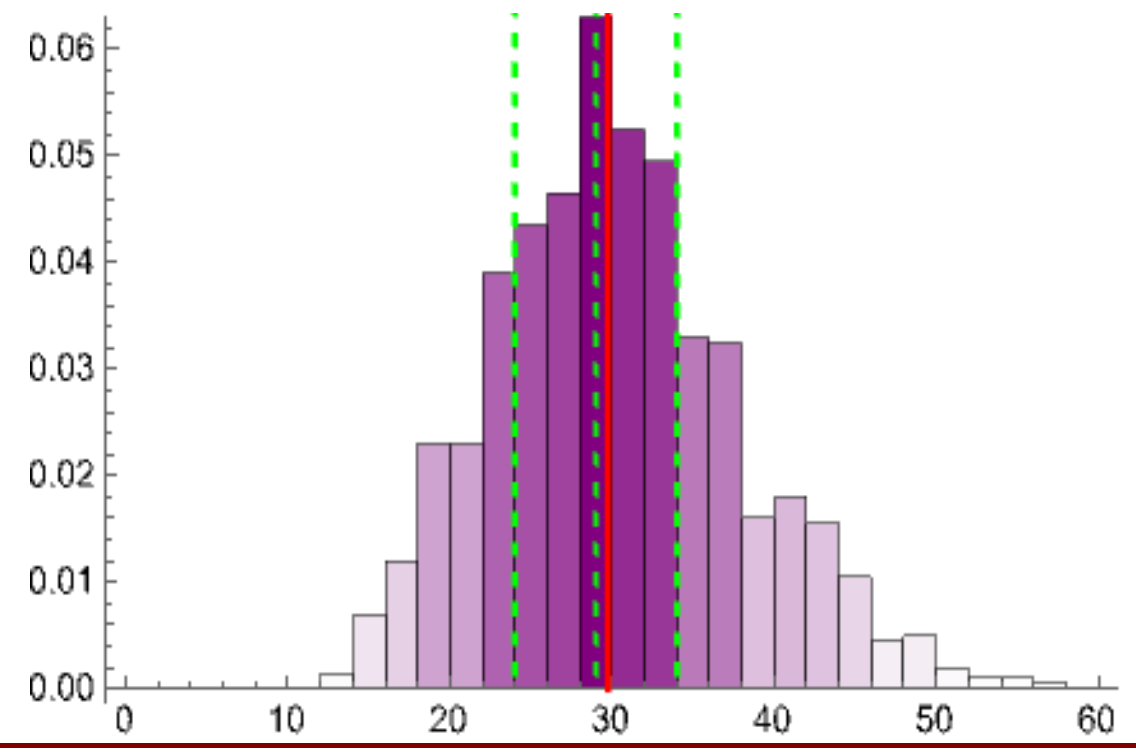

***Mathematica Examples 11.55***

Input

```
(* The code creates a dynamic histogram of data generated from a negative binomial
distribution using the Manipulate function. The Manipulate function creates
interactive controls for the user to adjust the values of r, p, and m, which are the
parameters of the negative binomial distribution and the sample size: *)

Manipulate[
 Module[
  {
   data=RandomVariate[
      NegativeBinomialDistribution[r,p],
      m
      ]
   },
  Show[
```

```
      Histogram[
       data,
       {1},
       "PDF",
       ColorFunction->Function[{height},Opacity[height]],
       ImageSize->320,
       ChartStyle->Purple
       ]
      ]
     ],
    {{r,10,"r"},1,50,1},
    {{p,0.5,"p"},0,1,0.01},
    {{m,500,"m"},100,10000,100}
    ]
```

Output

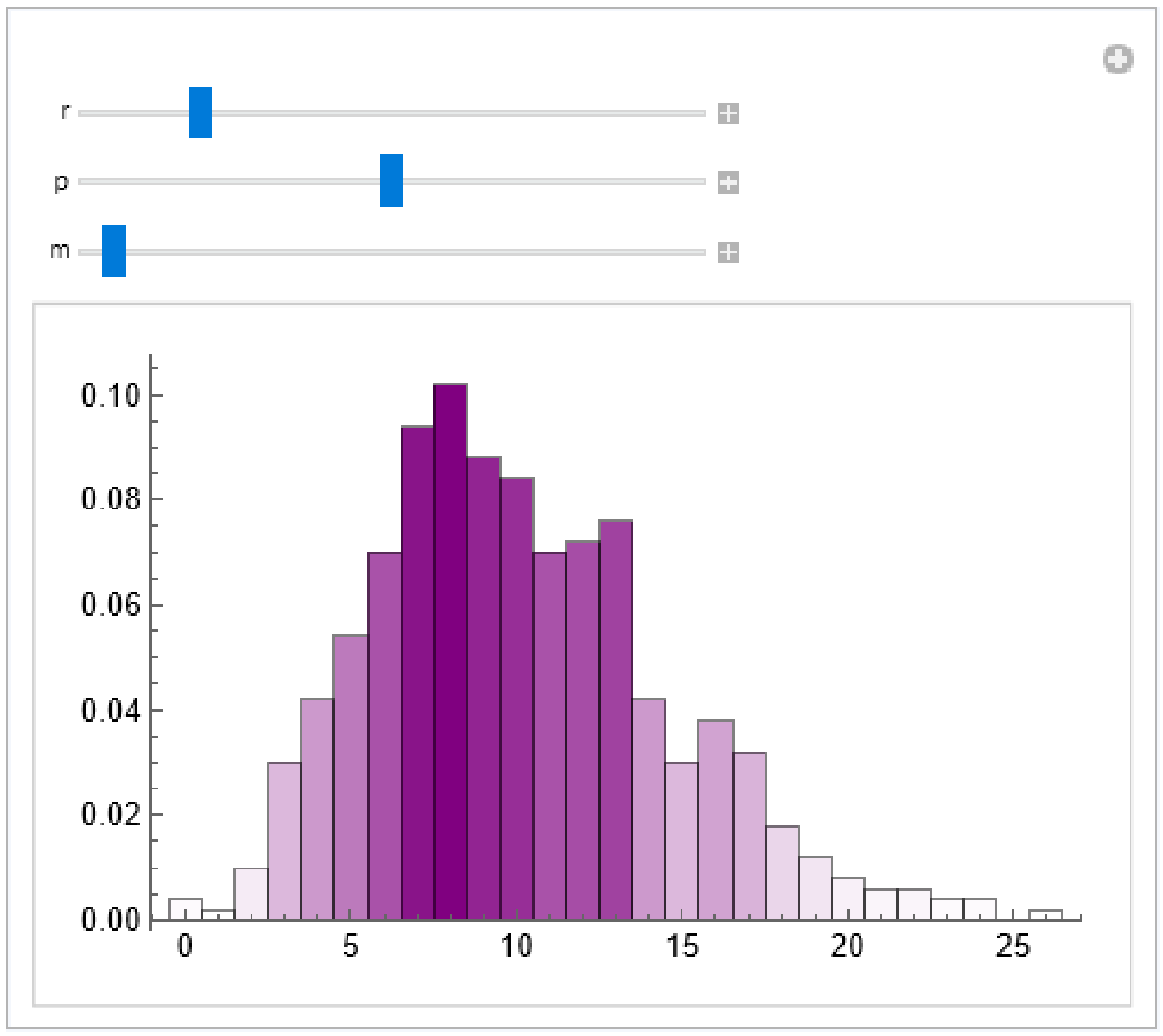

***Mathematica Examples 11.56***

Input

```
(* The code creates a plot of the CDF of a negative binomial distribution using the
Manipulate function. The Manipulate function allows you to interactively change the
values of the parameters r and p, respectively: *)
Manipulate[
 Plot[
  CDF[
   NegativeBinomialDistribution[r,p],
   x
   ],
  {x,0,r},
  Filling->Axis,
  FillingStyle->LightPurple,
  PlotRange->{{0,r},{0,1}},
  Epilog->{Text[StringForm["r = `` & p = ``",r,p],{r/2,0.9}]},
  AxesLabel->{"x","CDF"},
  ImageSize->320,
  PlotStyle->Purple
  ],
 {{r,10},1,100,1},
 {{p,0.5},0,1,0.01}
 ]
```

Output

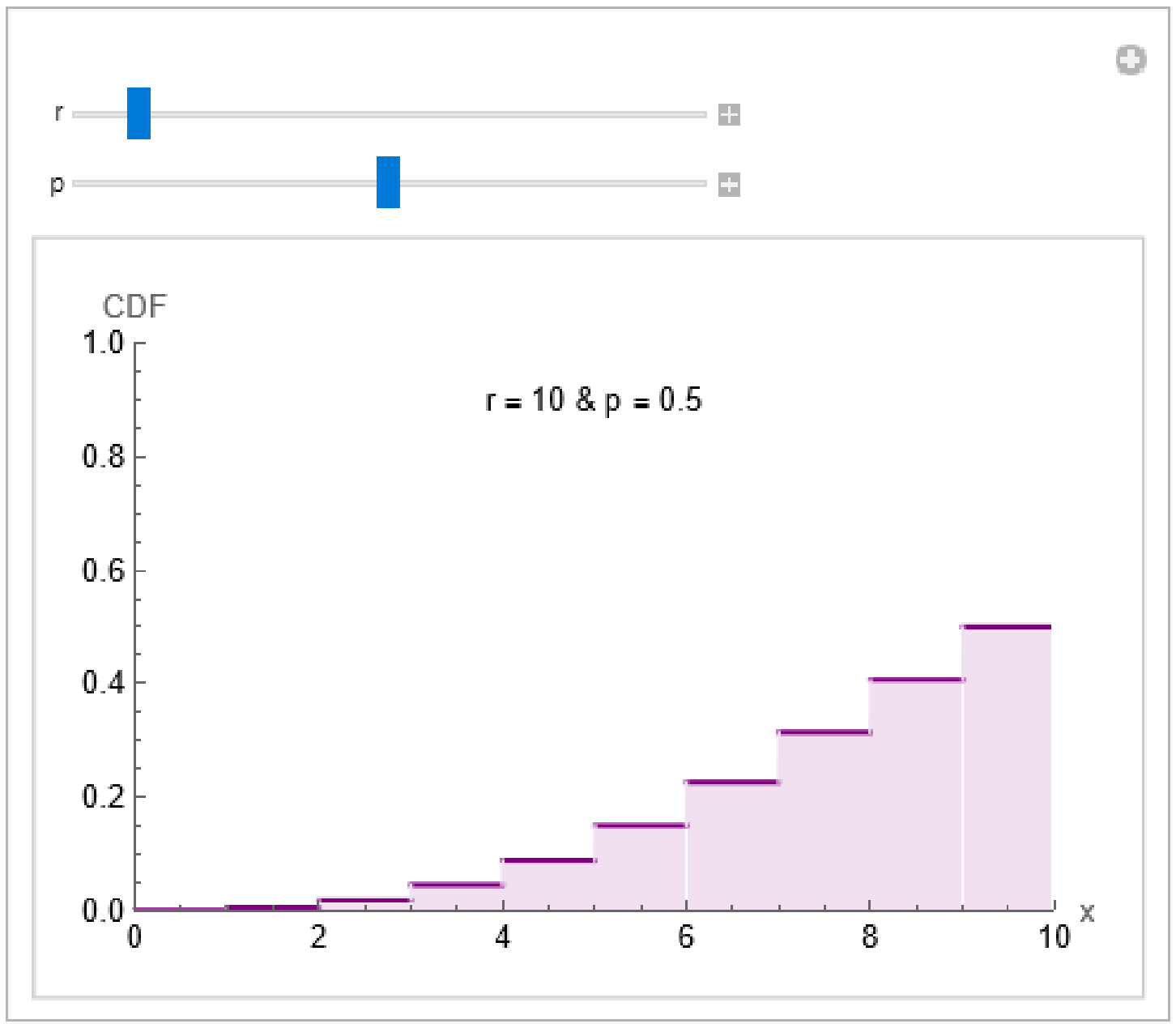

***Mathematica Examples 11.57***

Input

```
(* The code uses the Grid function to create a grid of two plots, one for the PDF
and one for the CDF, both of which are discrete plots. The DiscretePlot function is
used to create the plots, with the probability density or cumulative probability on
the y-axis and the number (k) on the x-axis. The code uses slider controls to adjust
the values of r and p: *)

Manipulate[
 Grid[
  {
   {DiscretePlot[
     PDF[
      NegativeBinomialDistribution[r,p],
      k
      ],
     {k,0,r},
     PlotRange->{{0,r},{0,1}},
     PlotStyle->{Purple,PointSize[0.03]},
     PlotLabel->"PDF of Negative Binomial Distribution",
     AxesLabel->{"k","PDF"},
     ImageSize->250

     ],
    DiscretePlot[
     CDF[
      NegativeBinomialDistribution[r,p],
      k
      ],
     {k,0,r},
     PlotRange->{{0,r},{0,1}},
     PlotStyle->{Purple,PointSize[0.03]},
     PlotLabel->"CDF of Negative Binomial Distribution",
     AxesLabel->{"k","CDF"},
     ImageSize->250
     ]
    }
```

```
     },
    Spacings->{5,5}
    ],
   {{r,20,"r"},20,50,1},
   {{p,0.5,"p"},0,1,0.01}
   ]
```

Output

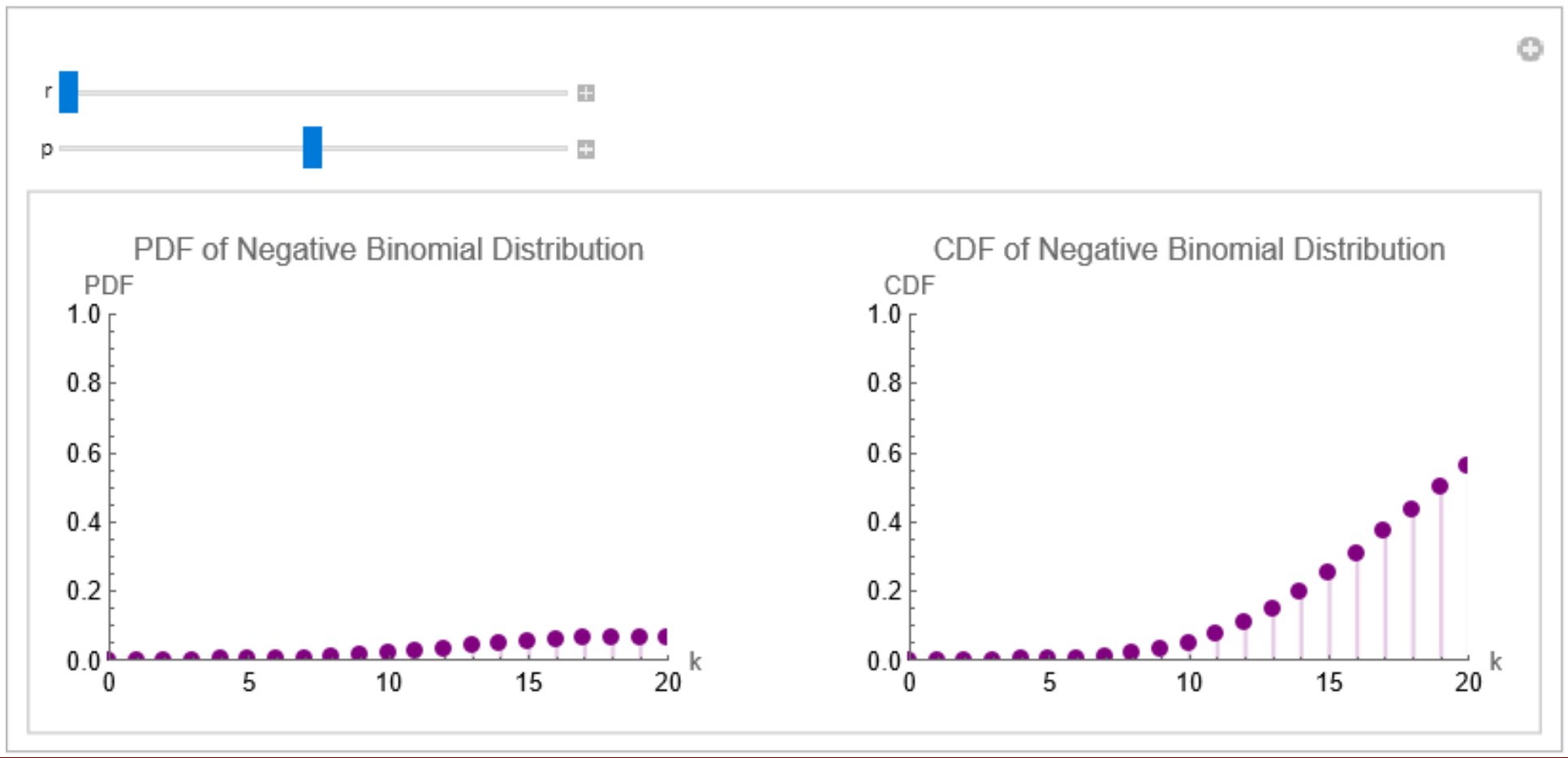

***Mathematica Examples 11.58***

Input

```
(* The number of tails before getting 3 heads with a fair coin: *)
heads3=NegativeBinomialDistribution[3,1/2];

(* Plot the distribution of tail counts: *)
DiscretePlot[
 PDF[heads3,k],
 {k,0,10},
 ImageSize->170,
 PlotStyle->Purple
 ]

(* Compute the probability of getting at least 7 tails before getting 3 heads: *)
N[
 Probability[
  tails>=7,
  tails\[Distributed]heads3
  ]
 ]
```

Output

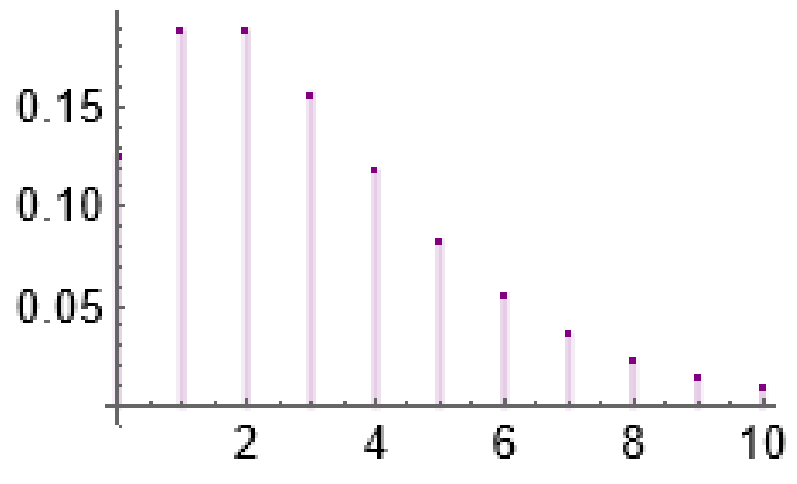

Output 0.0898438

# UNIT 11.5

# GEOMETRIC DISTRIBUTION

***Mathematica Examples 11.59***

Input

```
(* In this example, we calculate the probability mass function (PMF)of the geometric distribution with parameter p=0.3 (probability of success) using the formula p*(1-p)^k and plot it using ListPlot: *)

p=0.3;   (* Probability of success: *)

(* Define a discrete random variable with a geometric distribution: *)
x=Range[0,10];
pmf=Table[
    p*(1-p)^k,
    {k,0,10}
    ];

(* Plot the PMF: *)
ListPlot[
 Transpose[{x,pmf}],
 PlotRange->All,
 AxesLabel->{"x","P(X = x)"},
 PlotLabel->"Geometric Distribution PMF",
 Filling->Axis,
 ImageSize->200,
 PlotStyle->Purple
 ]
```

Output

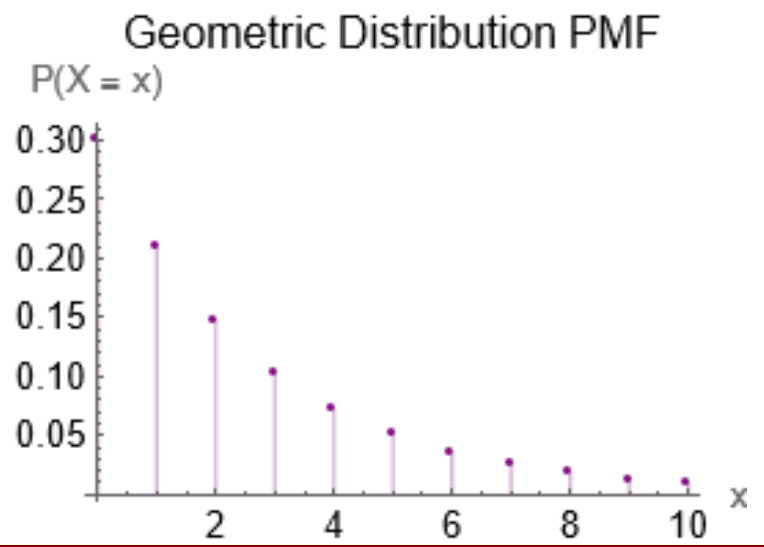

***Mathematica Examples 11.60***

Input

```
(* The code generates a discrete plot of the PMF for a geometric distribution with three different values of parameter p= (0.1, 0.2, and 0.4). The plot shows the values of the PMF for all possible values of k between 0 and 15: *)

DiscretePlot[
 Evaluate[
  Table[
   PDF[
    GeometricDistribution[p],
    k
    ],
   {p,{0.1,0.2,0.4}}
   ]
  ],
```

```
        {k,15},
        PlotRange->All,
        PlotMarkers->Automatic,
        PlotLegends->Placed[{"p=0.1","p=0.2","p=0.4"},{0.8,0.75}],
        PlotStyle->{RGBColor[0.88,0.61,0.14],RGBColor[0.37,0.5,0.7],Purple},
        ImageSize->250,
        AxesLabel->{None,"PMF"}
        ]
```

Output

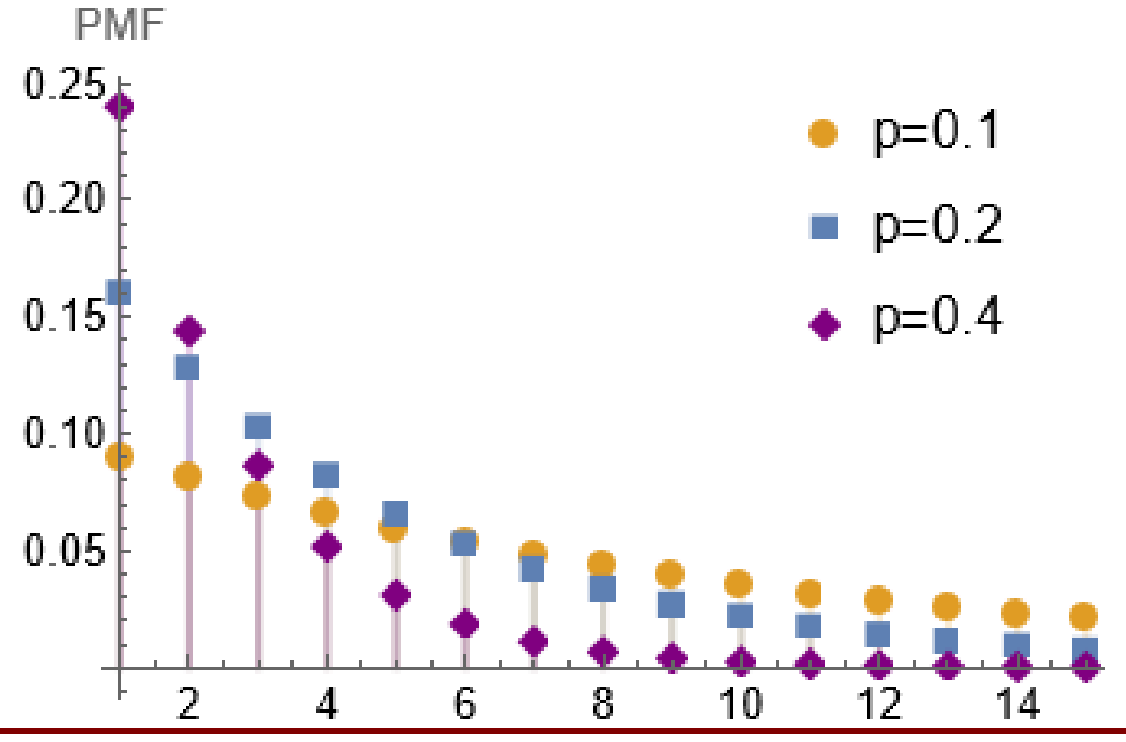

***Mathematica Examples 11.61***

Input

```
(* The code generates a discrete plot of the cumulative distribution function (CDF)
for a geometric distribution with three different values of parameter p= (0.1, 0.2,
and 0.4). The plot shows the values of the CDF for all possible values of k between
0 and 15: *)
DiscretePlot[
 Evaluate[
  Table[
   CDF[
    GeometricDistribution[p],
    k
    ],
   {p,{0.1,0.2,0.4}}
   ]
  ],
 {k,0,15},
 ExtentSize->Right,
 PlotRange->All,
 PlotMarkers->Automatic,
 PlotLegends->Placed[{"p=0.1","p=0.2","p=0.4"},{0.8,0.3}],
 PlotStyle->{RGBColor[0.88,0.61,0.14],RGBColor[0.37,0.5,0.7],Purple},
 ImageSize->250,
 AxesLabel->{None,"CDF"}
 ]
```

Output

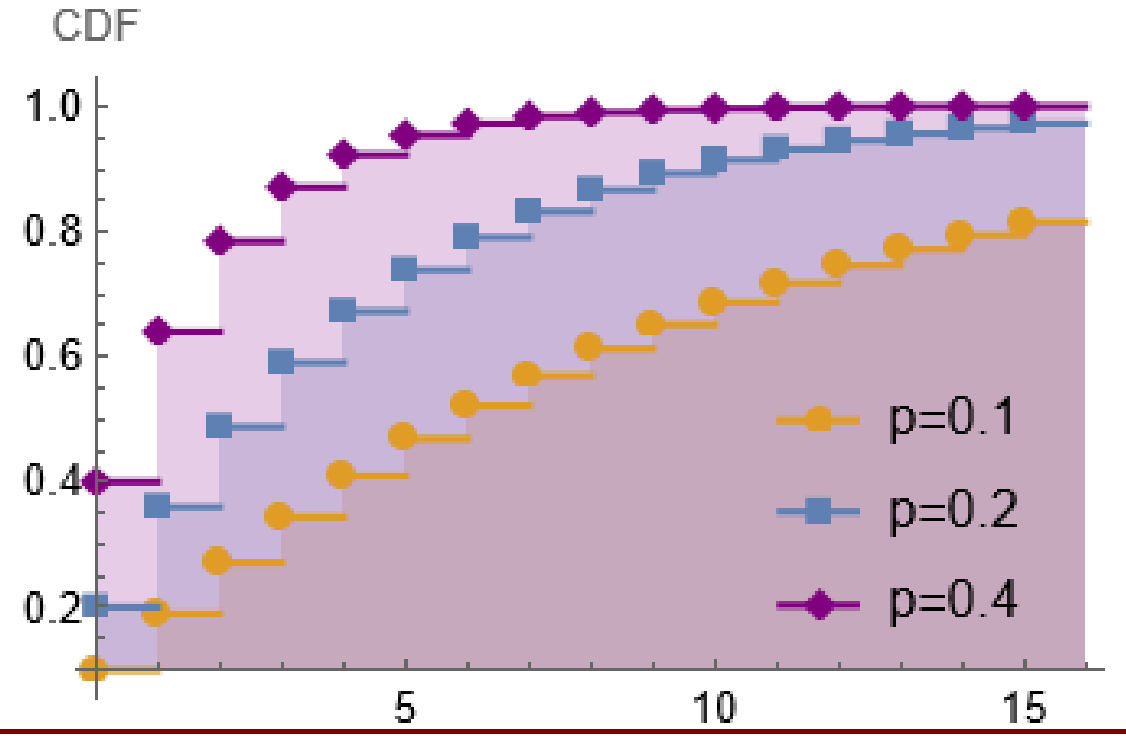

## Mathematica Examples 11.62

Input

```
(* The code generates a histogram and a discrete plot of the PMF for a Geometric Distribution with parameters p=0.1 and sample size 10000: *)

data=RandomVariate[d=GeometricDistribution[.1],10^4];
Show[
 Histogram[
  data,
  {2},
  "PDF",
  ColorFunction->Function[{height},Opacity[height]],
  ChartStyle->Purple,
  ImageSize->320,
  AxesLabel->{None,"PDF"}
  ],
 DiscretePlot[
  PDF[
   d,
   x
   ],
  {x,0,40},
  ImageSize->320,
  PlotStyle->PointSize[Medium],
  ColorFunction->"Rainbow"
  ]
 ]
```

Output

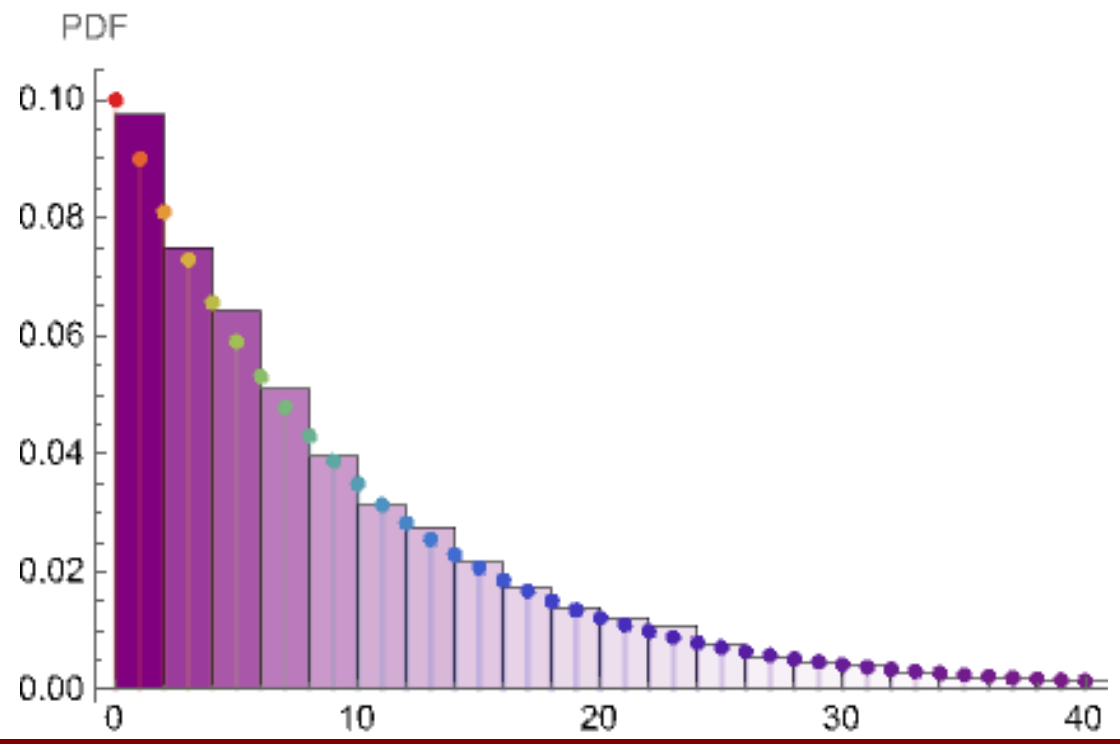

## Mathematica Examples 11.63

Input

```
(* The code generates a sample of 10000 values from a discrete geometric distribution and estimates the distribution parameters (p) from the sample using the EstimatedDistribution function. It then compares the density histogram of the sample with the PDF of the estimated distribution: *)

sampledata=RandomVariate[
    GeometricDistribution[.5],
    10^4
    ];

(* Estimate the distribution parameters from sample data: *)
ed=EstimatedDistribution[
  sampledata,
  GeometricDistribution[p]
  ]

(* Compare the density histogram of the sample with the PDF of the estimated distribution: *)
```

```
Show[
 Histogram[
  sampledata,
  {1},
  "PDF",
  ColorFunction->Function[{height},Opacity[height]],
  ChartStyle->Purple,
  ImageSize->320
  ],
 DiscretePlot[
  PDF[ed,x],
  {x,0,15},
  ImageSize->320,
  PlotStyle->PointSize[Medium],
  ColorFunction->"Rainbow"
  ]
 ]
```

Output GeometricDistribution[0.504745]

Output

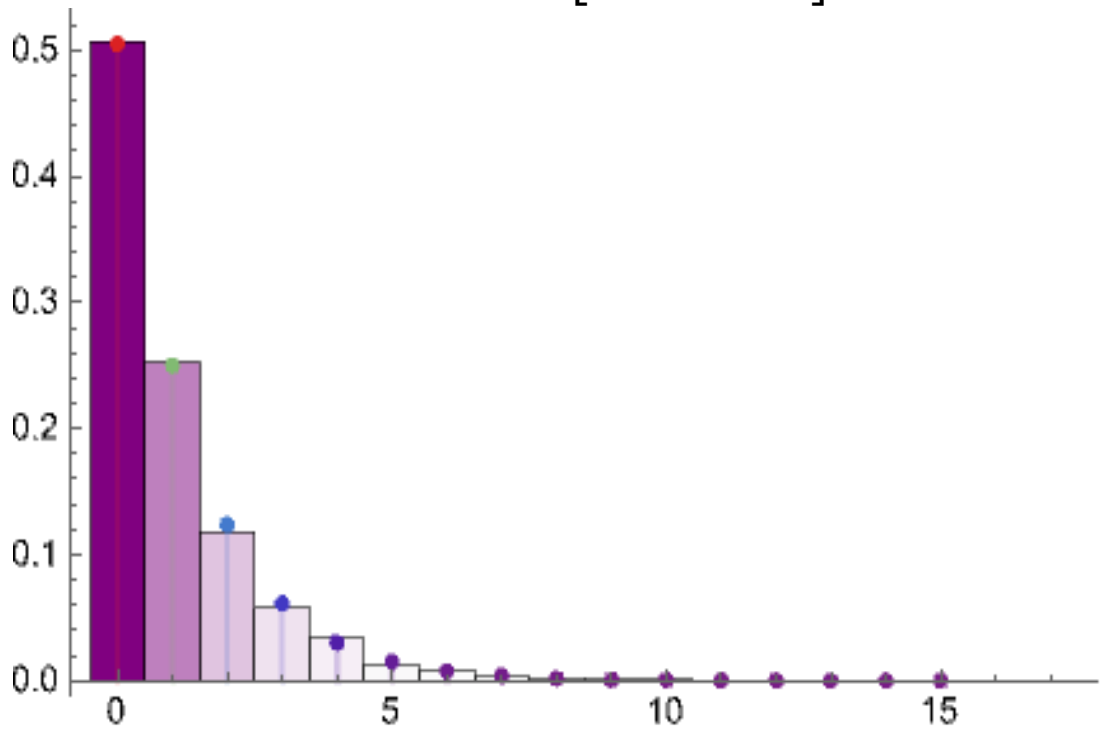

***Mathematica Examples 11.64***

Input

```
(* The code calculates and displays some descriptive statistics (mean, variance, standard deviation, kurtosis, and skewness) for a geometric distribution with parameter λ: *)

Grid[
 Table[
  {
   statistics,
   FullSimplify[statistics[GeometricDistribution[p]]]
   },
  {statistics,{Mean,Variance,StandardDeviation,Kurtosis,Skewness}}
  ],
 ItemStyle->12,
 Alignment->{{Right,Left}},
 Frame->All,
 Spacings->{Automatic,0.8}
 ]
```

Output

| | |
|---|---|
| Mean | -1+1/p |
| Variance | (1-p)/p^2 |
| StandardDeviation | Sqrt[1 - p]/p |
| Kurtosis | 8+1/(1-p)-p |
| Skewness | (2 - p)/Sqrt[1 - p] |

***Mathematica Examples 11.65***

Input

```
(* The code calculates and displays some additional descriptive statistics (moments, central moments, and factorial moments) for a geometric distribution with parameter λ: *)

Grid[
 Table[
  {
   statistics,
   FullSimplify[statistics[GeometricDistribution[p],1]],
   FullSimplify[statistics[GeometricDistribution[p],2]]
   },
  {statistics,{Moment,CentralMoment,FactorialMoment}}
  ],
 ItemStyle->12,
 Alignment->{{Right,Left}},
 Frame->All,
 Spacings->{Automatic,0.8}
 ]
```

Output

| | | |
|---|---|---|
| Moment | -1+1/p | ((-2+p) (-1+p))/p^2 |
| CentralMoment | 0 | (1-p)/p^2 |
| FactorialMoment | -1+1/p | 2 (-1+1/p)^2 |

***Mathematica Examples 11.66***

Input

```
(* The code generates a histogram of 10000 random samples drawn from a geometric distribution with a range of 0 to 50. The histogram is displayed with the "PDF" option, which normalizes the bin heights to represent a probability density function. The x-axis ranges from 0 to 50 with a bin size of 1. The mean and quartiles of the data are also displayed as vertical lines in red and green, respectively: *)

dist=GeometricDistribution[0.1];
data=RandomVariate[dist,10000];
Histogram[
 data,
 {0,50,1},
 "PDF",

 Epilog->{
   Directive[Red,Thickness[0.005]],
   Line[{{Mean[data],0},{Mean[data],0.1}}],
   Directive[Green,Dashed],
   Line[{{Quantile[data,0.25],0},{Quantile[data,0.25],0.1}}],
   Line[{{Quantile[data,0.5],0},{Quantile[data,0.5],0.1}}],
   Line[{{Quantile[data,0.75],0},{Quantile[data,0.75],0.1}}]
   },
 Frame->True,
 FrameLabel->{"x","PDF"},
 LabelStyle->Directive[Bold,Medium],
 ColorFunction->Function[{height},Opacity[height]],
 ChartStyle->Purple,
 ImageSize->320,
 AxesLabel->{None,"PDF"}
 ]
```

Output

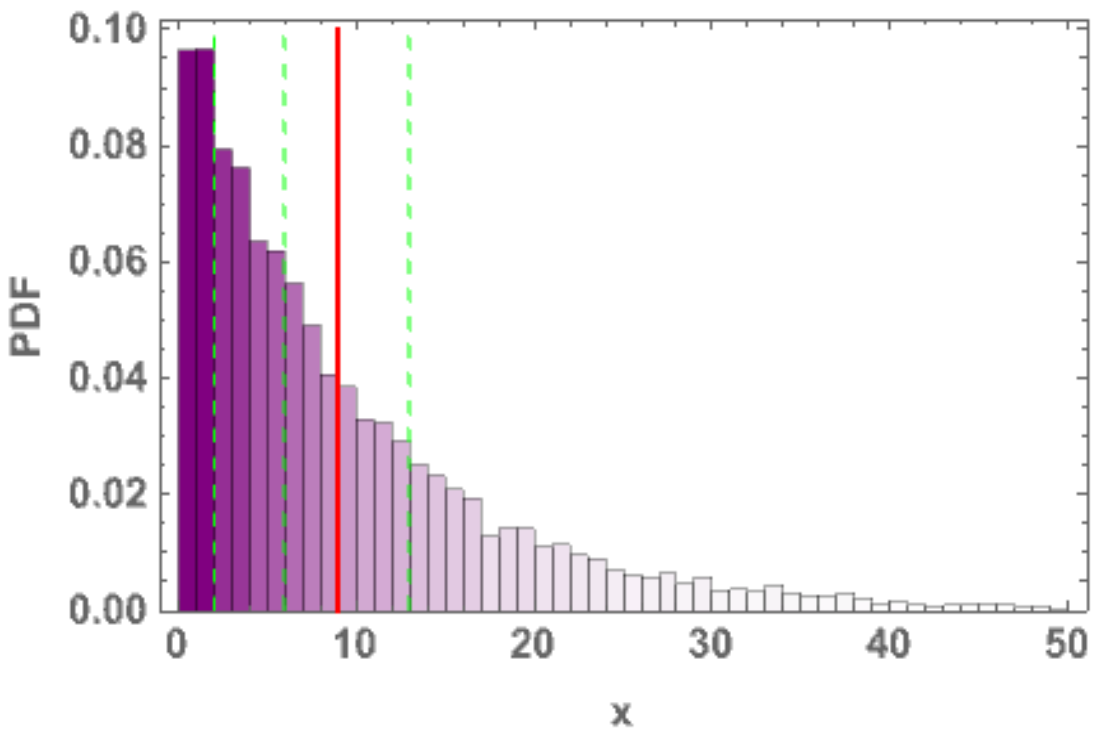

***Mathematica Examples 11.67***

Input

```
(* This code generates two sets of random data from geometric distributions with different means (p1=0.2 and p2=0.4) and creates a histogram to compare the distributions: *)

p1=0.2;
p2=0.4;
n=1000;
data1=RandomVariate[
   GeometricDistribution[p1],
   n
   ];

data2=RandomVariate[
   GeometricDistribution[p2],
   n
   ];

Histogram[
 {data1,data2},
 LabelingFunction->Above,
 ChartLegends->{"p1=0.2","p2=0.4"},
 ChartStyle->{Directive[Opacity[0.5],Red],Directive[Opacity[0.6],Purple]},
 ImageSize->Medium,
 ImageSize->320
 ]
```

Output

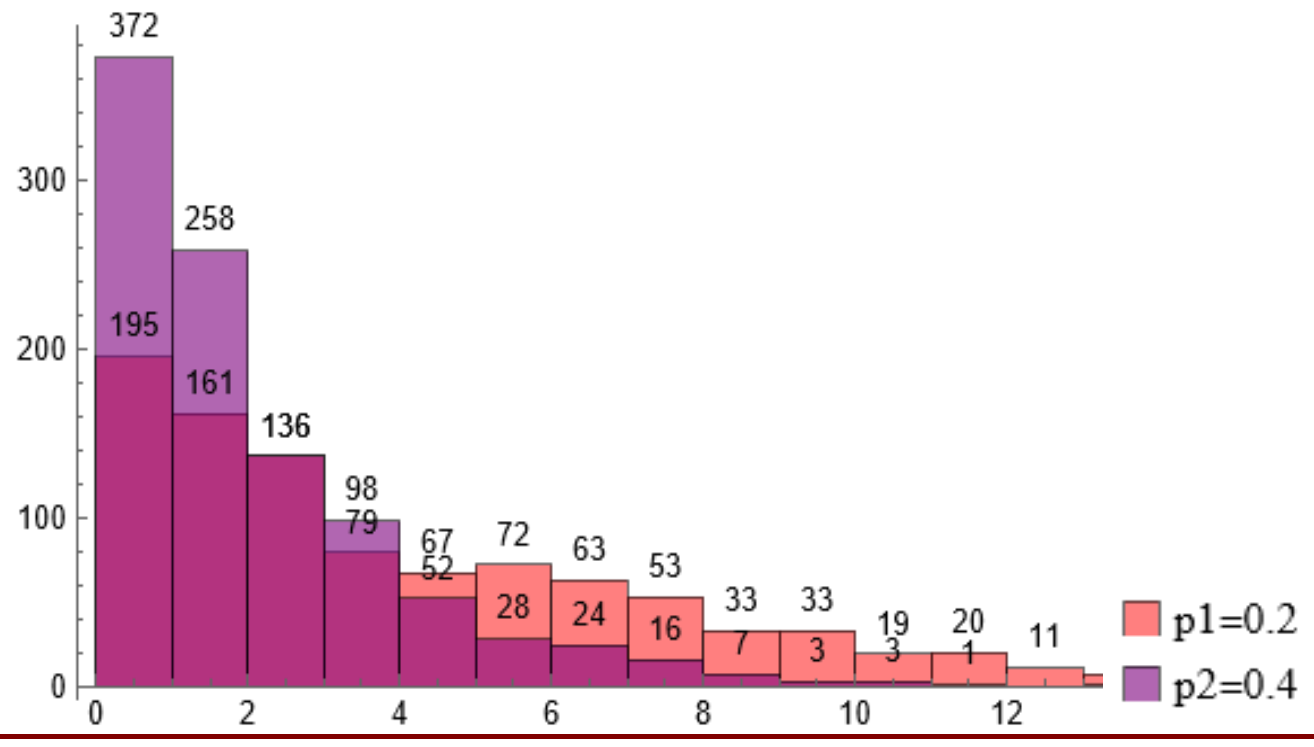

***Mathematica Examples 11.68***

Input

```
(* The code defines a Manipulate that creates a discrete geometric distribution with parameter p, generates n (Sample size) random variates from the distribution, and plots a histogram of the sample. The Manipulate includes sliders to control the value p and n (Sample size): *)
```

```
Manipulate[
 data=RandomVariate[
   GeometricDistribution[p],
   n
   ];
 Histogram[
  data,
  Automatic,
  "PDF",
  ColorFunction->Function[{height},Opacity[height]],
  ChartStyle->Purple,
  ImageSize->320
  ],
 {{p,0.3,"p"},0.1,0.9,0.05},
 {{n,1000,"Sample size"},100,5000,100}
 ]
```

Output

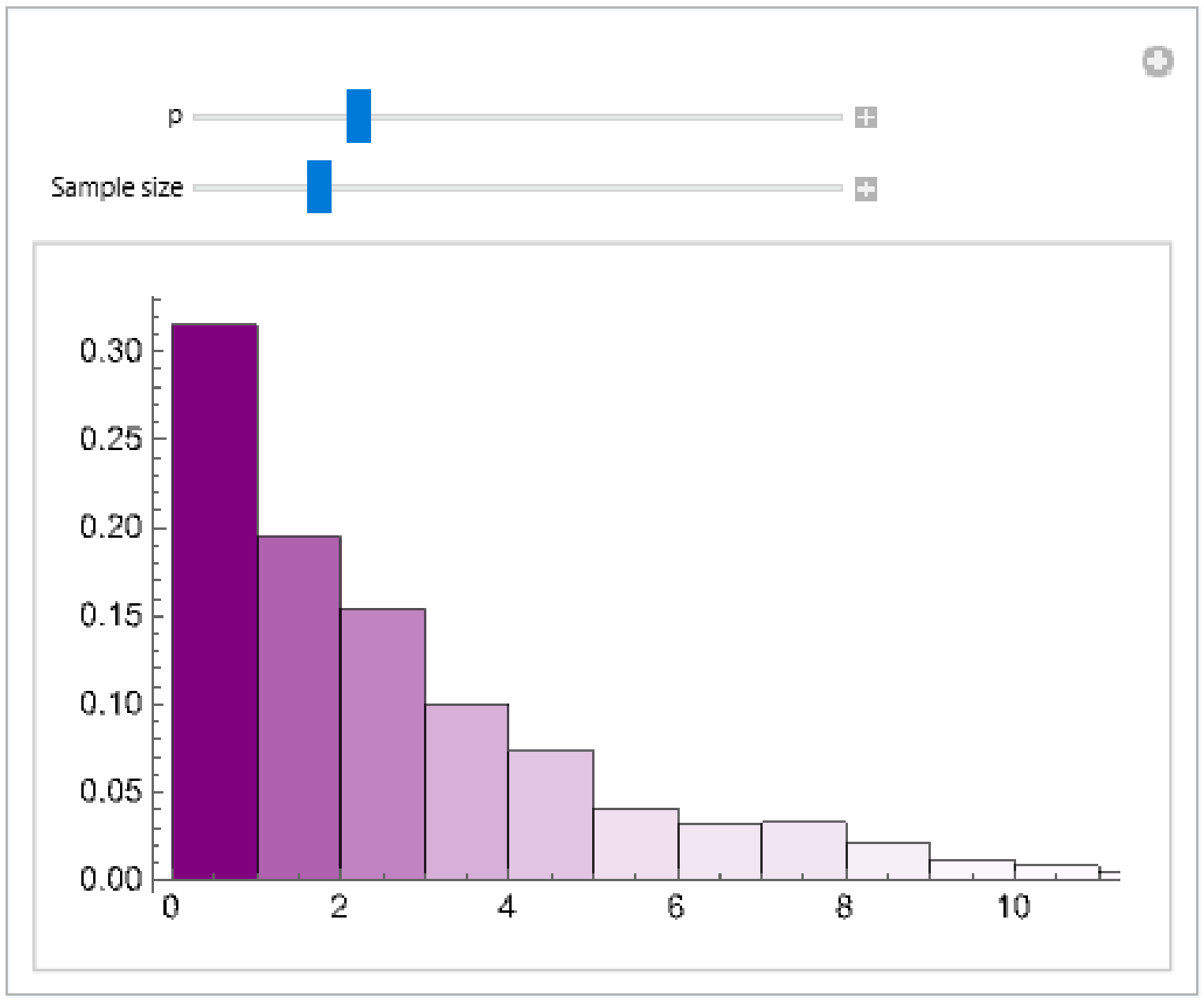

***Mathematica Examples 11.69***

Input

```
(* The code creates a plot of the CDF of a geometric distribution using the Manipulate
function. The Manipulate function allows you to interactively change the values of
the parameter p: *)

Manipulate[
 Plot[
  CDF[
   GeometricDistribution[p],
   x
   ],
  {x,0,20},
  PlotRange->All,
  PlotLabel->StringForm[
    "Cumulative Distribution Function
of Geometric Distribution with p = ``"
    ,p
    ],
  Filling->Axis,
```

```
    FillingStyle->LightPurple,
    ImageSize->320,
    PlotStyle->Purple
    ],
   {{p,0.5},0.1,0.9,0.01}
   ]
```

Output

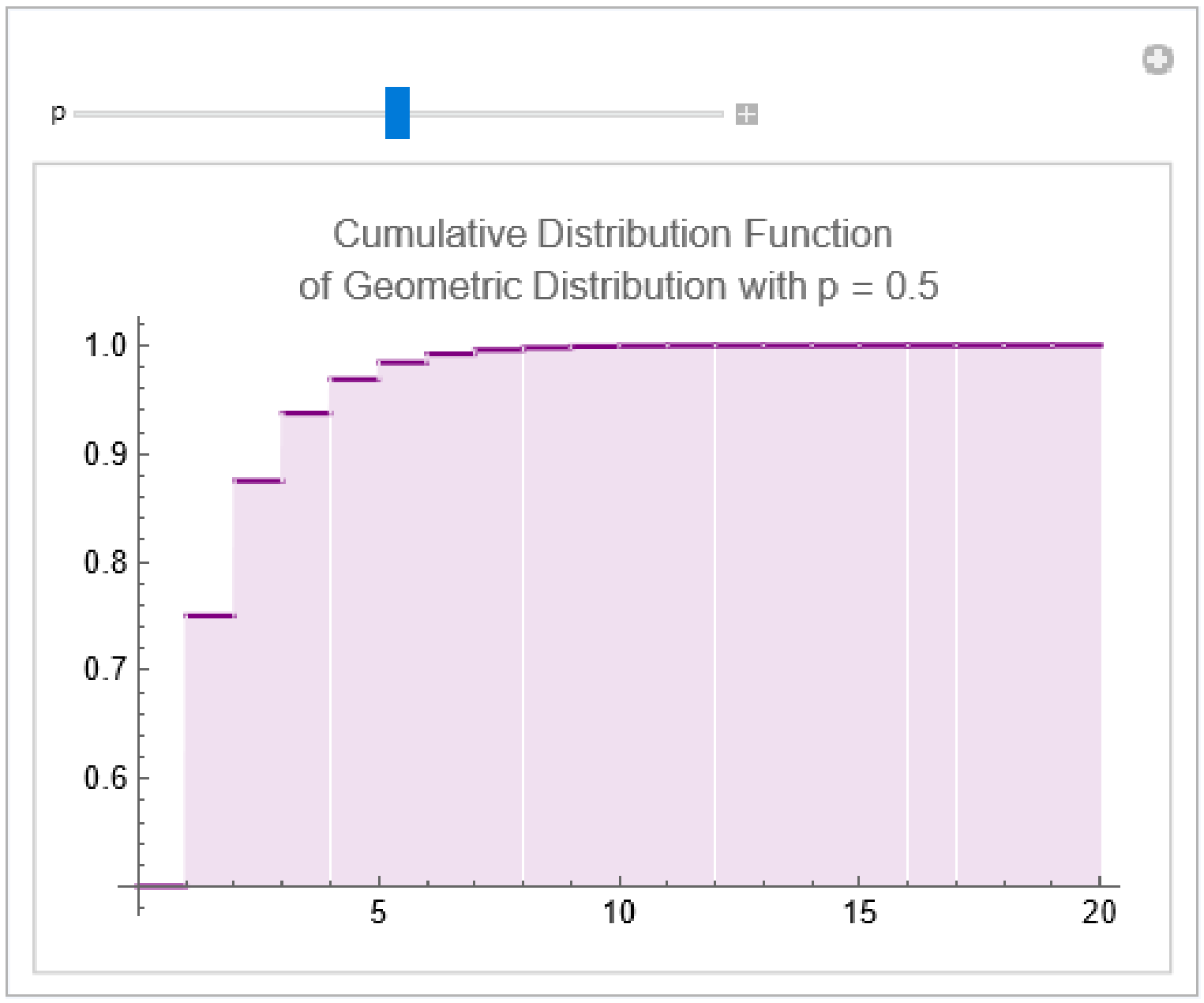

## *Mathematica Examples 11.70*

Input

```
(* The code uses the Manipulate function to create an interactive plot that shows
the PMF and CDF of a geometric distribution with parameter p. The user can adjust
the value of p using a slider: *)

Manipulate[
 Module[
  {dist},
  dist=GeometricDistribution[p];
  Grid[
   {
    {
     DiscretePlot[
      PDF[dist,x],
      {x,0,40},
      PlotRange->{{0,40},{0,0.4}},
      PlotStyle->{Purple,PointSize[0.02]},
      ExtentSize->Full,
      AxesLabel->{"x","PMF"},
      ImageSize->250,
      Filling->Axis,
      FillingStyle->Purple,
      PlotLabel->StringForm[" p = ``",p]
      ],
     Plot[
      CDF[dist,x],
      {x,0,40},
      PlotStyle->{Purple,Thickness[0.007]},
      AxesLabel->{"k","CDF"},
      ImageSize->250,
```

```
          PlotLabel->StringForm[" p = ``",p]
          ]
         }
        },
       Frame->All
       ]
      ],
     {{p,0.1,"p"},0.1,0.9,0.01}
     ]
```

Output

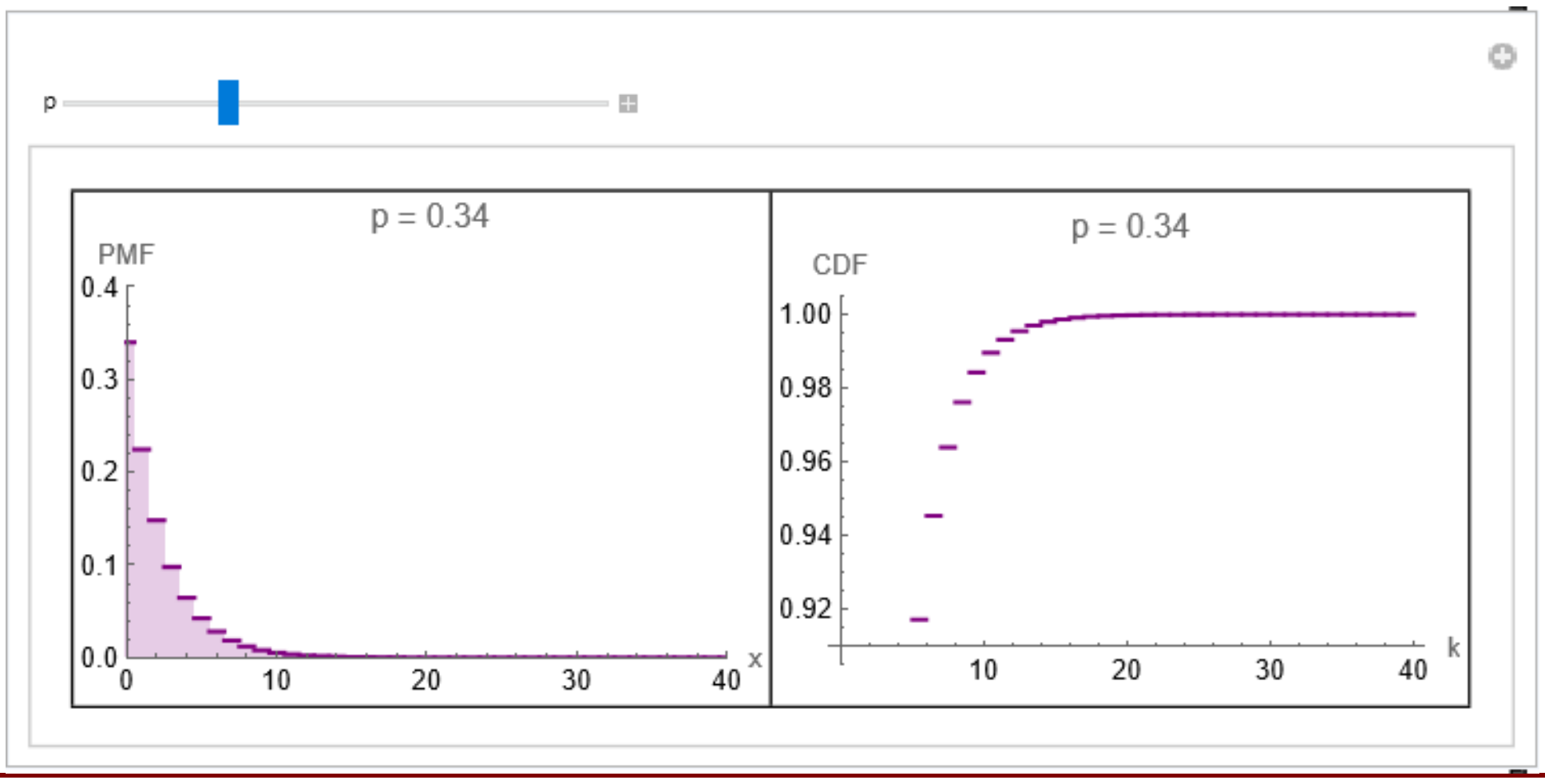

**Mathematica Examples 11.71**

Input

```
(* A coin-tossing experiment consists of tossing a fair coin repeatedly until a tail
results. Simulate the process: *)
sampledata=RandomVariate[
  GeometricDistribution[1/2],
  10
  ]

ListPlot[
 RandomVariate[
  GeometricDistribution[1/2],
  100
  ],
 Filling->Axis,
 ImageSize->200,
 PlotStyle->Purple
 ]
(* Compute the probability that at least 4 coin tosses will be necessary: *)
Probability[
 x>=4,
 x\[Distributed]GeometricDistribution[1/2]
 ]
```

Output {3,0,0,0,0,1,0,1,0,0}

Output

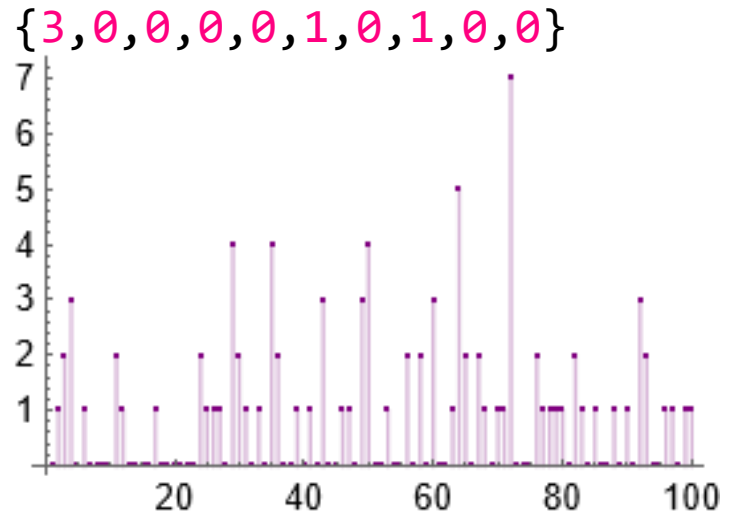

Output 1/16

***Mathematica Examples 11.72***

```
Input    (* The sum over the support of the distribution is unity: *)
         PDF[GeometricDistribution[p],k]

         Sum[
          PDF[GeometricDistribution[p],k],
          {k,0,Infinity}
          ]

Output   {
          {\[Piecewise], {
            {(1-p)^k p, k>=0},
            {0, True}
           }}
         }
Output   1
```

# UNIT 11.6

# HYPERGEOMETRIC DISTRIBUTION

***Mathematica Examples 11.73***

Input

```
(* The code generates a discrete plot of the probability mass function (PMF) for a hypergeometric distribution with K=50, N=100 and three different values of n= (10, 20 and 50). The plot shows the values of the PMF for all possible values of x between 0 and 35: *)

DiscretePlot[
 Evaluate[
  Table[
   PDF[
    HypergeometricDistribution[n,50,100],
    x
    ],
   {n,{10,20,50}}
   ]
  ],
 {x,0,35},
 PlotRange->All,
 PlotMarkers->Automatic,
 PlotLegends->Placed[{"n=10, K=50, N=100","n=20, K=50, N=100","n=50, K=50, N=100"},{0.6,0.75}],
 PlotStyle->{RGBColor[0.88,0.61,0.14],RGBColor[0.37,0.5,0.7],Purple},
 ImageSize->300,
 AxesLabel->{None,"PMF"}
 ]
```

Output

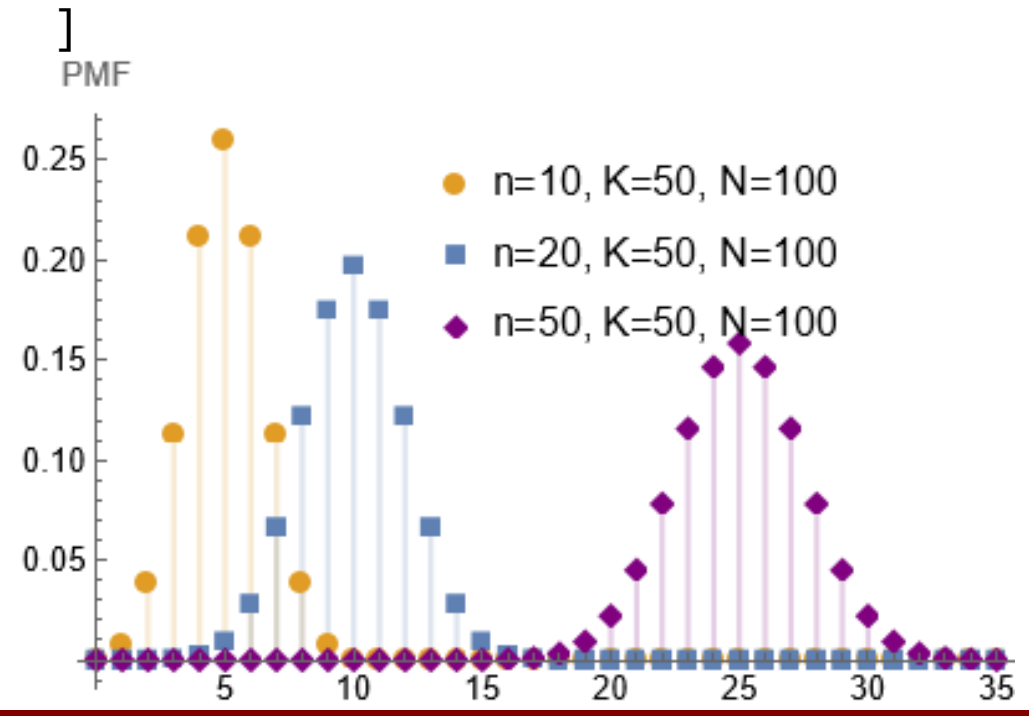

***Mathematica Examples 11.74***

Input

```
(* The code generates a discrete plot of the cumulative distribution function (CDF) for a hypergeometric distribution with K=50, N=100 and three different values of n= (10, 20 and 50). The plot shows the values of the CDF for all possible values of x between 0 and 35: *)
DiscretePlot[
 Evaluate[
  Table[
   CDF[
    HypergeometricDistribution[n,50,100],
    x
    ],
   {n,{10,20,50}}
```

```
      ]
     ],
    {x,0,35},
    ExtentSize->Right,
    PlotRange->All,
    PlotMarkers->Automatic,
    PlotLegends->Placed[{"n=10, K=50, N=100","n=20, K=50, N=100","n=50, K=50,
N=100"},{0.7,0.5}],
    PlotStyle->{RGBColor[0.88,0.61,0.14],RGBColor[0.37,0.5,0.7],Purple},
    ImageSize->300,
    AxesLabel->{None,"CDF"}
    ]
```

Output

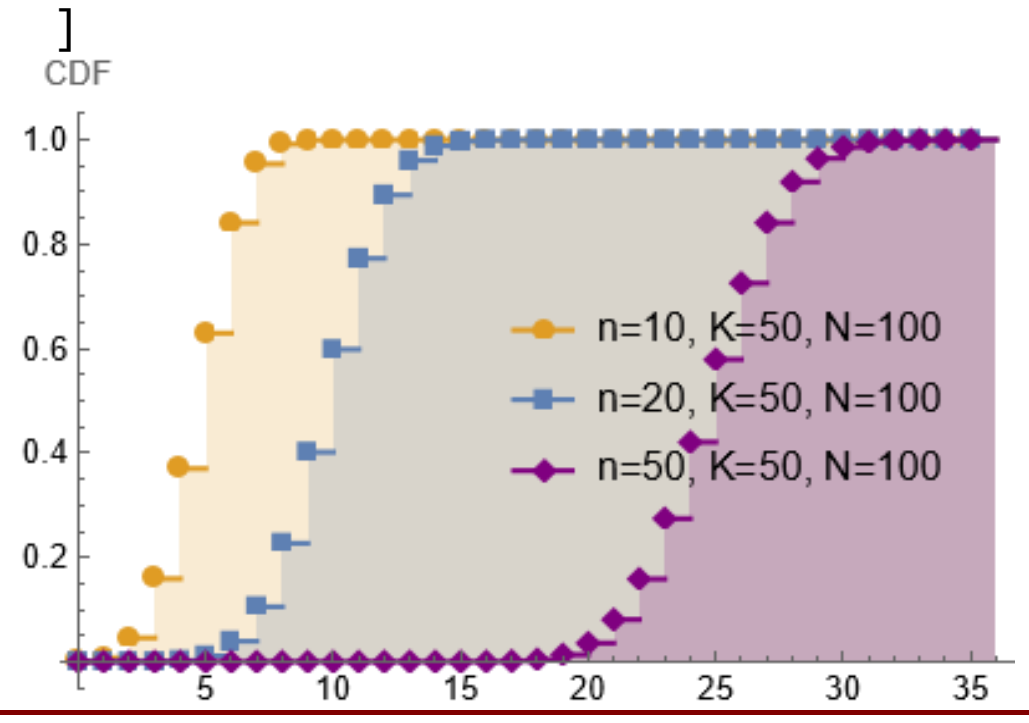

***Mathematica Examples 11.75***

Input

```
(* The code generates a discrete plot of the PMF for a hypergeometric distribution
with three different values of K= (10,20,50), N=100 and n=40. The plot shows the
values of the PMF for all possible values of x between 0 and 35: *)
DiscretePlot[
 Evaluate[
  Table[
   PDF[
    HypergeometricDistribution[40,K,100],
    x
    ],
   {K,{10,20,50}}
   ]
  ],
 {x,0,35},
 PlotRange->All,
 PlotMarkers->Automatic,
 PlotLegends->Placed[{"K=10, n=40, N=100","K=20, n=40, N=100","K=50, n=40,
N=100"},{0.6,0.75}],
 PlotStyle->{RGBColor[0.88,0.61,0.14],RGBColor[0.37,0.5,0.7],Purple},
 ImageSize->300,
 AxesLabel->{None,"PMF"}
 ]
```

Output

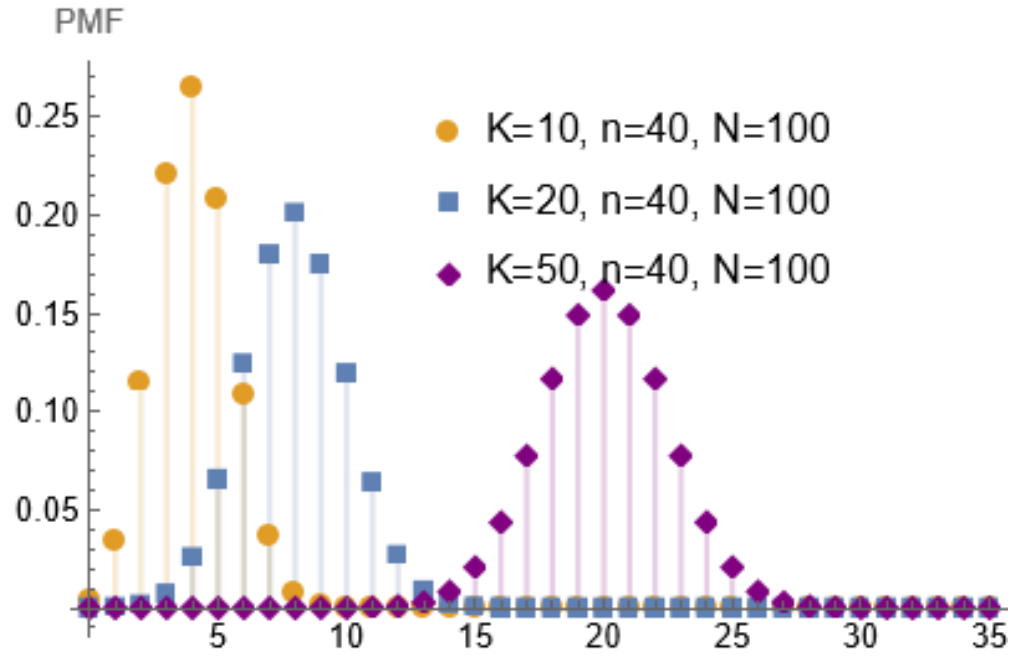

**Mathematica Examples 11.76**

Input

```
(* The code generates a discrete plot of the CDF for a hypergeometric distribution
with three different values of K= (10,20,50), N=100 and n=40. The plot shows the
values of the CDF for all possible values of x between 0 and 35: *)

DiscretePlot[
 Evaluate[
  Table[
   CDF[
    HypergeometricDistribution[40,K,100],
    x
    ],
   {K,{10,20,50}}
   ]
  ],
 {x,0,35},
 ExtentSize->Right,
 PlotRange->All,
 PlotMarkers->Automatic,
 PlotLegends->Placed[{"K=10, n=40, N=100","K=20, n=40, N=100","K=50, n=40,
N=100"},{0.7,0.5}],
 PlotStyle->{RGBColor[0.88,0.61,0.14],RGBColor[0.37,0.5,0.7],Purple},
 ImageSize->300,
 AxesLabel->{None,"CDF"}
 ]
```

Output

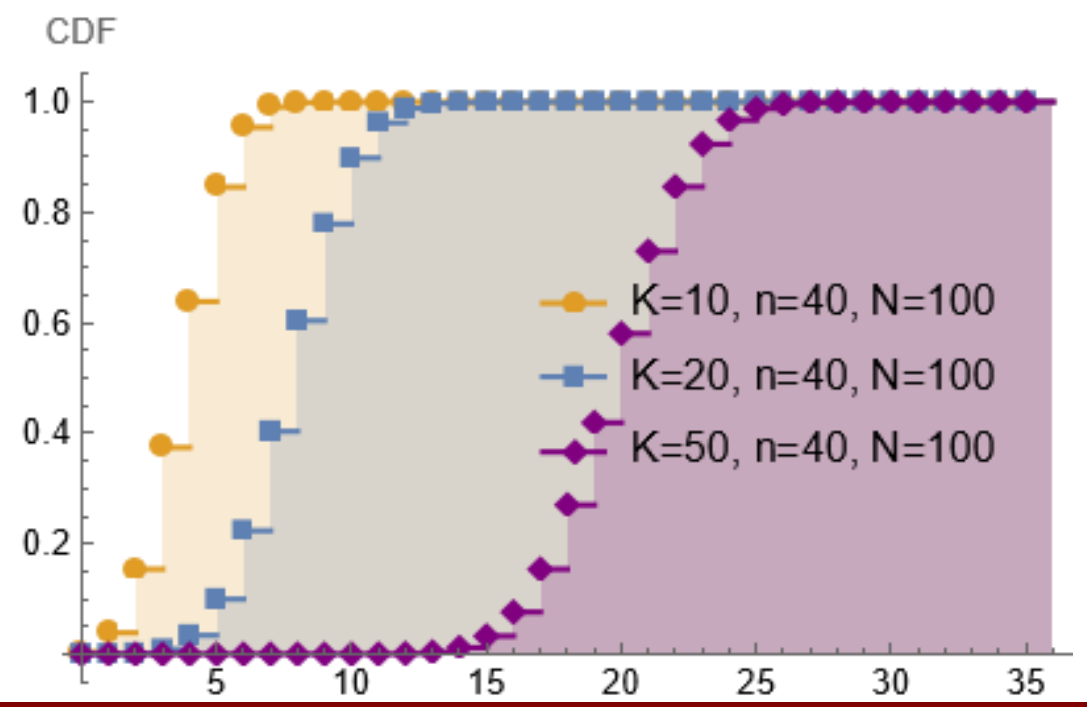

**Mathematica Examples 11.77**

Input

```
(* The code generates a discrete plot of the PMF for a hypergeometric distribution
with K=50, three different values of N= (60,80,100) and n=40. The plot shows the
values of the PMF for all possible values of x between 15 and 40: *)

DiscretePlot[
 Evaluate[
  Table[
   PDF[
    HypergeometricDistribution[40,50,N],
    x
    ],
   {N,{60,80,100}}
   ]
  ],
 {x,15,40},
 PlotRange->All,
 PlotMarkers->Automatic,
 PlotLegends->Placed[{"n=40, K=50, N=60","n=40, K=50, N=80","n=40, K=50,
N=100"},{0.25,0.8}],
 PlotStyle->{RGBColor[0.88,0.61,0.14],RGBColor[0.37,0.5,0.7],Purple},
```

```
ImageSize->320,
AxesLabel->{None,"PDF"}
]
```

Output

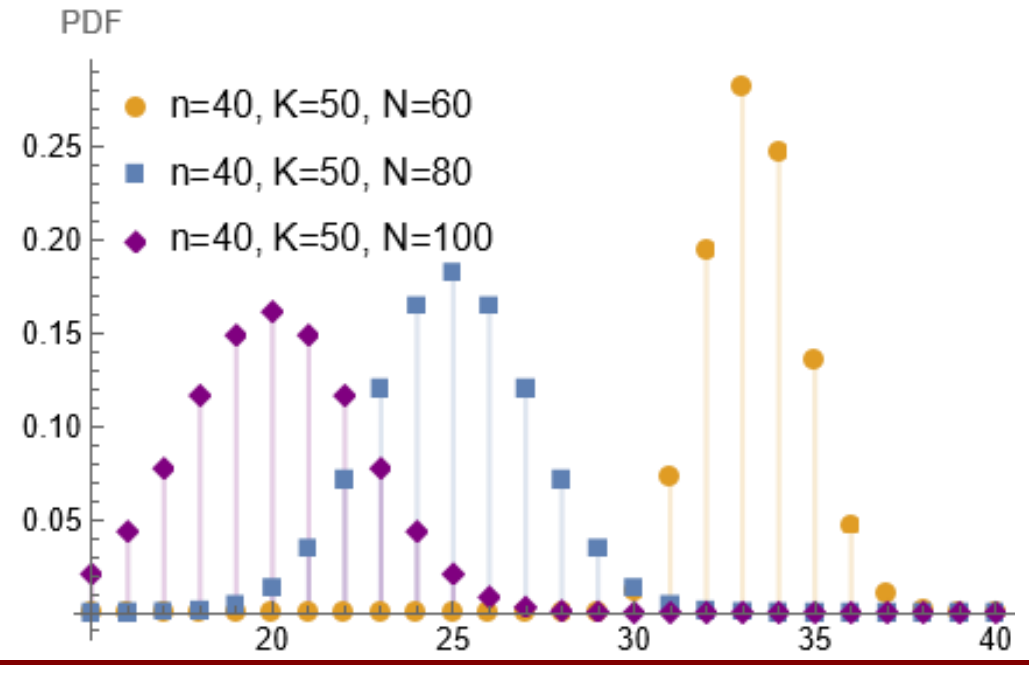

***Mathematica Examples 11.78***

Input

```
(* The code generates a discrete plot of the CDF for a hypergeometric distribution
with K=50, three different values of N= (60,80,100) and n=40. The plot shows the
values of the CDF for all possible values of k between 15 and 40: *)

DiscretePlot[
 Evaluate[
  Table[
   CDF[
    HypergeometricDistribution[40,50,N],
    x
    ],
   {N,{60,80,100}}
   ]
  ],
 {x,15,40},
 ExtentSize->Right,
 PlotRange->All,
 PlotMarkers->Automatic,
 PlotLegends->Placed[{"n=40, K=50, N=60","n=40, K=50, N=80","n=40, K=50,
N=100"},{0.7,0.5}],
 PlotStyle->{RGBColor[0.88,0.61,0.14],RGBColor[0.37,0.5,0.7],Purple},
 ImageSize->320,
 AxesLabel->{None,"CDF"}
 ]
```

Output

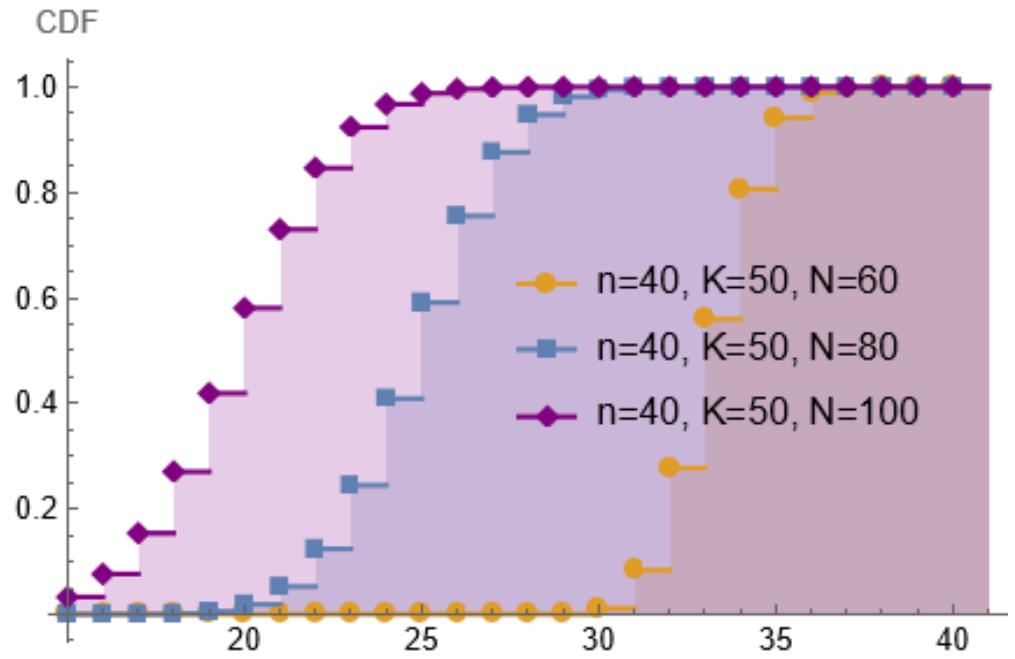

***Mathematica Examples 11.79***

Input

```
(* The code generates a histogram and a discrete plot of the PMF for a hypergeometric
distribution with parameters n=20, K=50, N=100 and sample size 10000: *)
```

```
data=RandomVariate[
   HypergeometricDistribution[20,50,100],
   10^4
   ];
Show[
 Histogram[
  data,
  {4.5,16.5,1},
  "PDF",
  ColorFunction->Function[{height},Opacity[height]],
  ChartStyle->Purple,
  ImageSize->320,
  AxesLabel->{None,"PDF"}
  ],

 DiscretePlot[
  PDF[
   HypergeometricDistribution[20,50,100],
   x
   ],
  {x,0,20},
  ImageSize->320,
  PlotStyle->PointSize[Medium],
  ColorFunction->"Rainbow"
  ]
 ]
```

Output

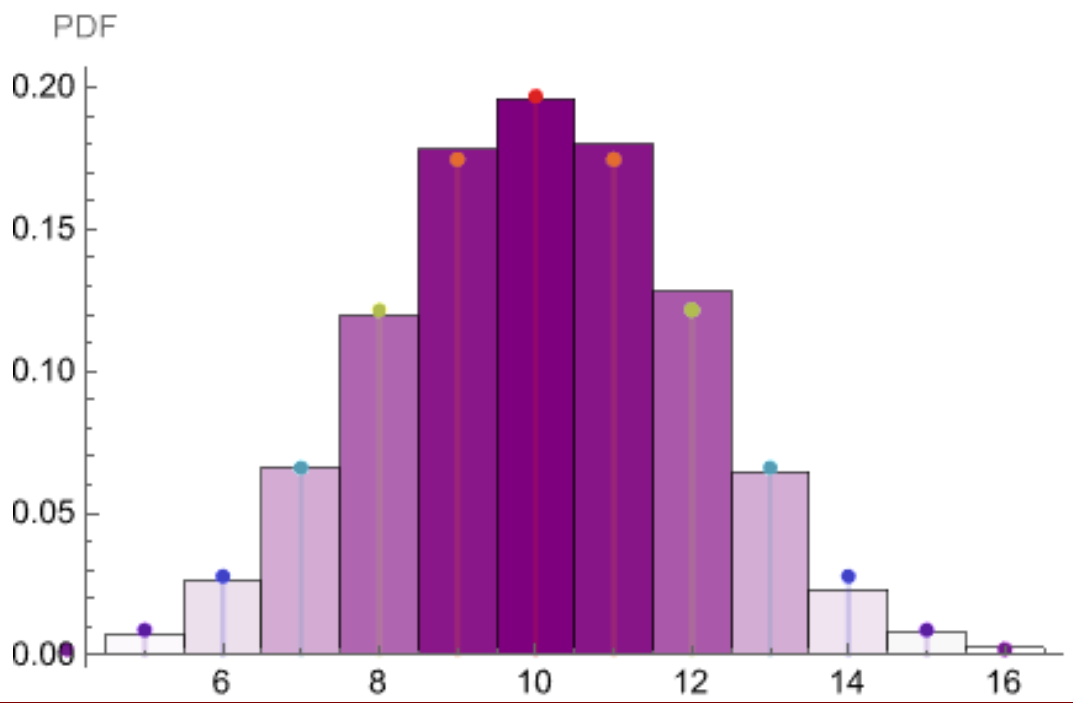

*Mathematica Examples 11.80*

Input

```
(* The code calculates and displays some descriptive statistics (mean, variance, standard deviation, kurtosis, and skewness) for a hypergeometric distribution with parameters n, K, and N: *)

Grid[
 Table[
  {
   statistics,
   FullSimplify[statistics[HypergeometricDistribution[n,K,N]]]
   },
  {statistics,{Mean,Variance,StandardDeviation,Kurtosis,Skewness}}
  ],
 ItemStyle->12,
 Alignment->{{Right,Left}},
 Frame->All,
 Spacings->{Automatic,0.8}
 ]
```

Output

| | |
|---|---|
| Mean | (K n)/N |
| Variance | (K n (K- N) (n - N))/((-1+ N) N ^2) |
| StandardDeviation | Sqrt[(K n (K- N) (n - N))/(-1 + N)]/N |
| Kurtosis | $\frac{(-1+N)N^2\left(6n(n-N)+N(1+N)+\frac{3K(K-N)(2N^2+n^2(6+N)-nN(6+N))}{N^2}\right)}{(Kn(K-N)(n-N)(-3+N)(-2+N))}$ |
| Skewness | (Sqrt[-1 + N](-2 K + N)(-2 n + N))/(Sqrt[K n (K - N)(n - N)](-2+N)) |

***Mathematica Examples 11.81***

Input

```
(* The code calculates and displays some additional descriptive statistics (moments, central moments, and factorial moments) for a hypergeometric distribution with parameters n, K, and N: *)

Grid[
 Table[
  {
   statistics,
   FullSimplify[statistics[HypergeometricDistribution[n,K,N],1]],
   FullSimplify[statistics[HypergeometricDistribution[n,K,N],2]]
   },
  {statistics,{Moment,CentralMoment,FactorialMoment}}
  ],
 ItemStyle->12,
 Alignment->{{Right,Left}},
 Frame->All,
 Spacings->{Automatic,0.8}
 ]
```

Output

| | | |
|---|---|---|
| Moment | $\frac{K n}{N}$ | $\begin{cases} \frac{K n}{N}, & N<2 \\ \frac{K n (K(-1+n)- n+N)}{(-1+N)N}, & \text{True} \end{cases}$ |
| CentralMoment | 0 | $\begin{cases} \frac{K n (-K n+N)}{N^2}, & N<2 \\ \frac{K n (K-N)(n-N)}{(-1+N) N^2}, & \text{True} \end{cases}$ |
| FactorialMoment | $\begin{cases} \frac{K n}{N}, & 1 <= N \\ 0, & \text{True} \end{cases}$ | $\begin{cases} \frac{(-1+K) K (-1+n) n}{(-1+N) N}, & N >= 2 \\ 0, & \text{True} \end{cases}$ |

***Mathematica Examples 11.82***

Input

```
(* This code generates a random sample of size 10,000 from a hypergeometric distribution with parameters n=10 and K=30, N=60, estimates the distribution parameters using the EstimatedDistribution function, and then compares the histogram of the sample with the estimated PDF of the hypergeometric distribution using a histogram and a discrete plot of the PDF: *)

sampledata=RandomVariate[
    HypergeometricDistribution[10,30,60],
    10000
    ];
(* Estimate the distribution parameters from sample data: *)
ed=EstimatedDistribution[
   sampledata,
   HypergeometricDistribution[10,K,N]
   ]

(* Compare a density histogram of the sample with the PDF of the estimated distribution: *)
Show[
```

```
  Histogram[
   sampledata,
   {1},
   "PDF",
   ColorFunction->Function[{height},Opacity[height]],
   ChartStyle->Purple,
   ImageSize->320
   ],
  DiscretePlot[
   PDF[ed,x],
   {x,0,40},
   PlotStyle->PointSize[Medium],
   ImageSize->320,
   ColorFunction->"Rainbow"
   ]
  ]
```

Output
```
NMaximize::cvmit: Failed to converge to the requested accuracy or precision within 100 iterations.
 HypergeometricDistribution[10,31,62]
```

Output

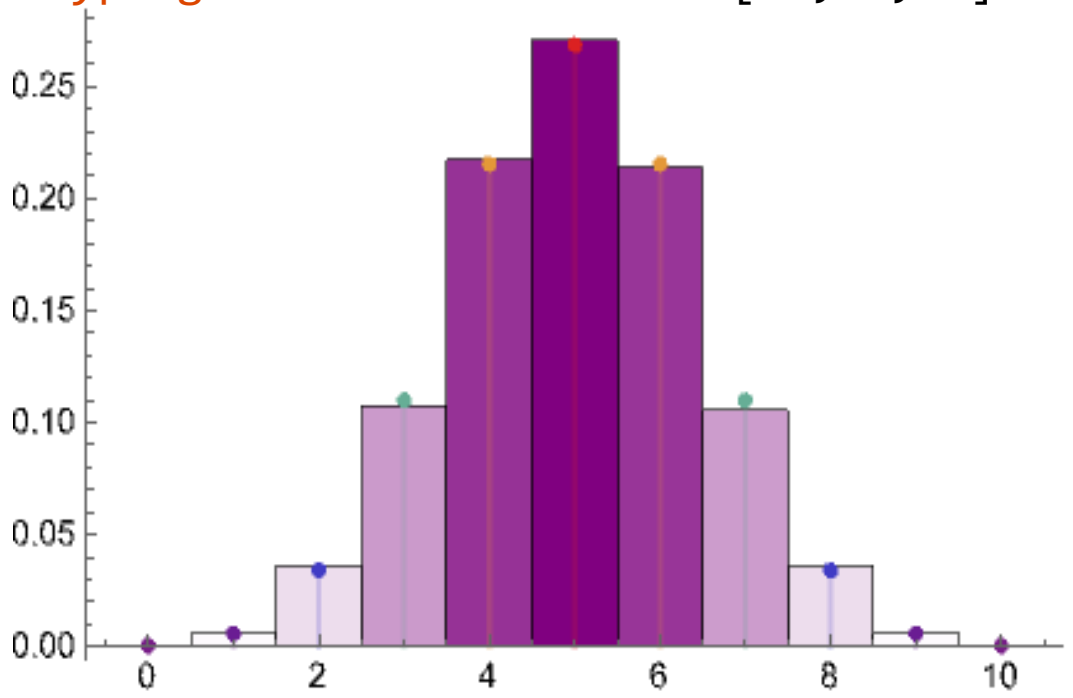

***Mathematica Examples 11.83***

Input
```
(* The code generates a dataset of 1000 observations from a hypergeometric distribution with parameters n=10, K=30 and N=60. Then, it computes the sample mean and quartiles of the data, and plots a histogram of the data using the "PDF" option to display the probability density function. Additionally, the code adds vertical lines to the plot corresponding to the sample mean and quartiles: *)

data=RandomVariate[
   HypergeometricDistribution[10,30,60],
   1000
   ];
mean=Mean[data];
quartiles=Quantile[
   data,
   {0.25,0.5,0.75}
   ];
Histogram[
 data,
 Automatic,
 "PDF",
 Epilog->{
   Directive[Red,Thickness[0.006]],
   Line[{{mean,0},{mean,0.3}}],
   Directive[Green,Dashed],
   Line[{{quartiles[[1]],0},{quartiles[[1]],0.3}}],
   Line[{{quartiles[[2]],0},{quartiles[[2]],0.3}}],
   Line[{{quartiles[[3]],0},{quartiles[[3]],0.3}}]
```

```
    },
   ColorFunction->Function[{height},Opacity[height]],
   ImageSize->320,
   ChartStyle->Purple,
   PlotRange->{{0,10},{0,0.3}}
   ]
```

Output

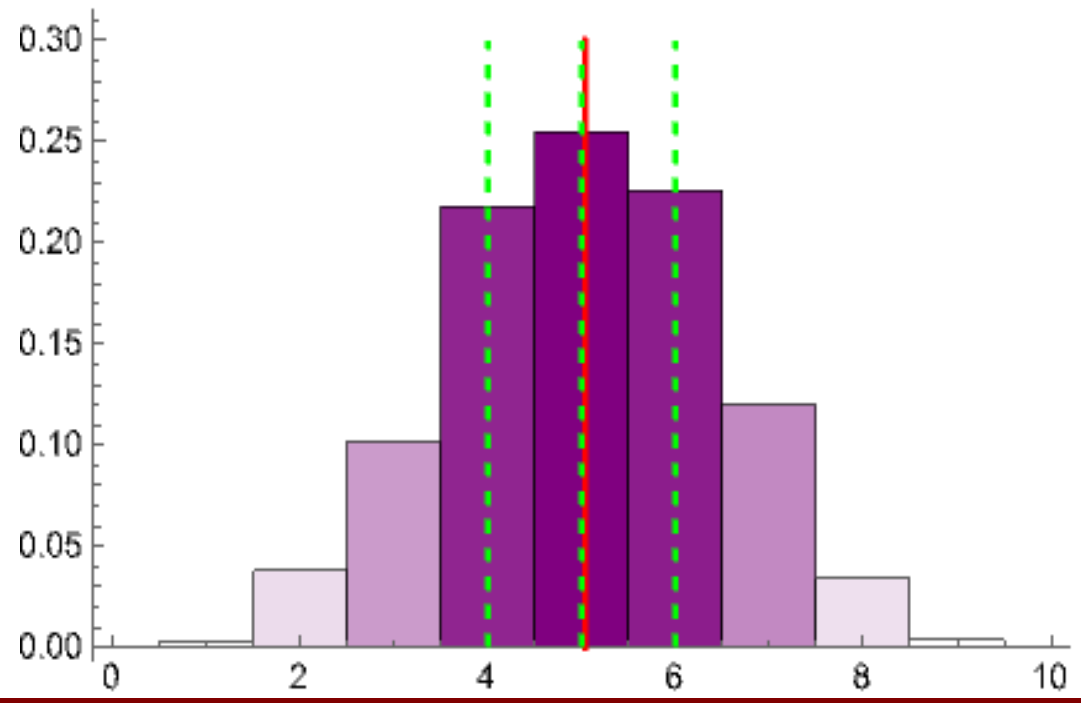

## Mathematica Examples 11.84

Input

```
(* The code creates a dynamic histogram of data generated from a hypergeometric distribution using the Manipulate function. The Manipulate function creates interactive controls for the user to adjust the values of n, K, N, and m, which are the parameters of the Hypergeometric Distribution and the sample size: *)

Manipulate[
 Module[
  {
   data=RandomVariate[
     HypergeometricDistribution[n,K,N],
     m
     ]
   },
  Show[
   Histogram[
    data,
    {1},
    "PDF",
    PlotRange->{{0,n},All},
    ColorFunction->Function[{height},Opacity[height]],
    ImageSize->320,
    ChartStyle->Purple
    ]
   ]
  ],
 {{n,5,"n"},5,10,1},
 {{K,10,"K"},10,20,1},
 {{N,20,"N"},20,60,1},
 {{m,100,"m"},10,1000,10}
 ]
```

Output

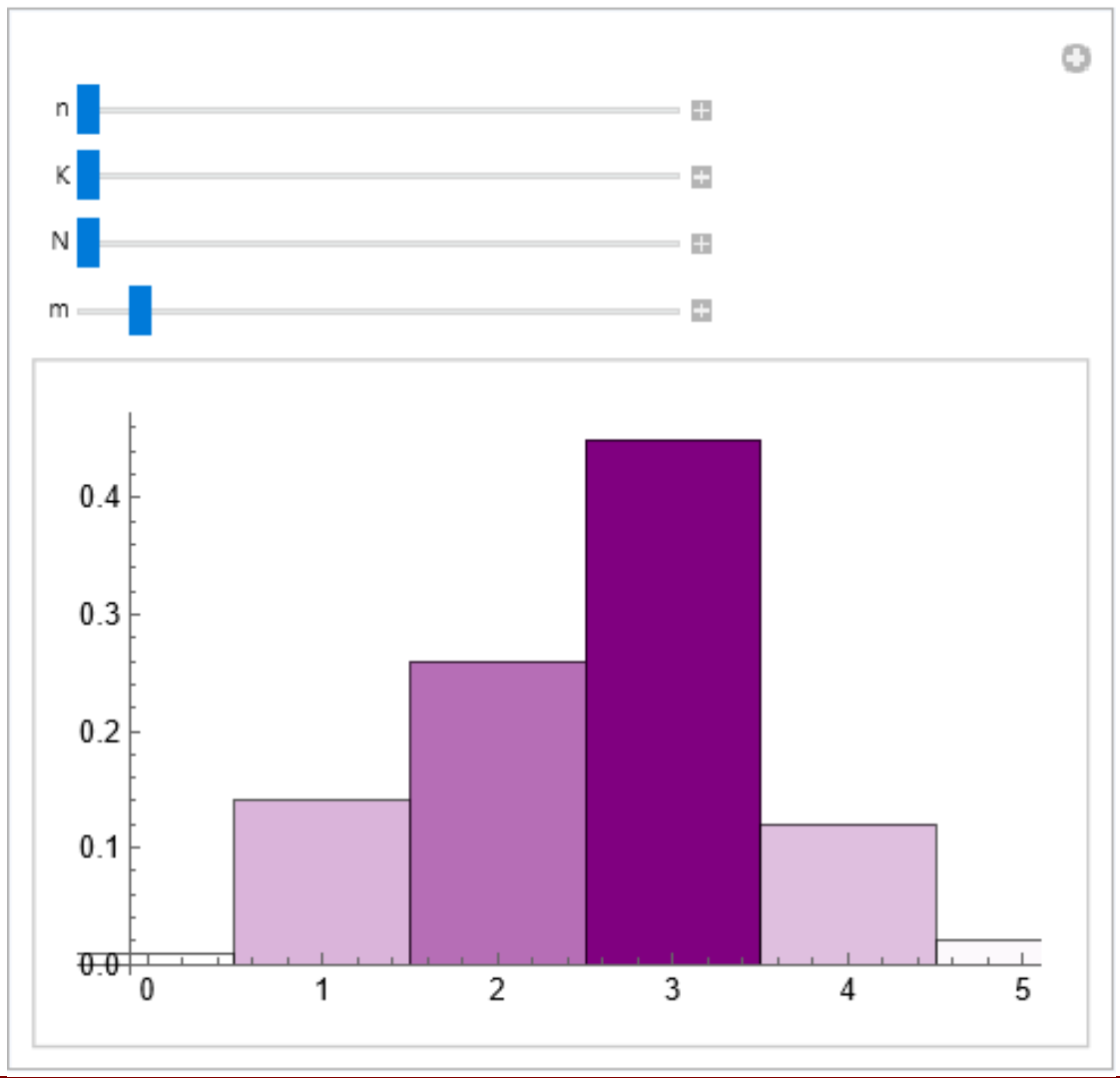

*Mathematica Examples 11.85*

Input

```
(* The code creates a plot of the CDF of a hypergeometric distribution using the
Manipulate function. The Manipulate function allows you to interactively change the
values of the parameters n, K and N, respectively: *)
Manipulate[
 Plot[
  CDF[
   HypergeometricDistribution[n,K,N],
   x
   ],
  {x,0,n},
  Filling->Axis,
  FillingStyle->LightPurple,
  PlotRange->{{0,n},{0,1}},
  Epilog->{Text[StringForm["n = `` & s = ``& t = ``",n,K,N],{n/2,0.9}]},
  AxesLabel->{"x","CDF"},
  ImageSize->320,
  PlotStyle->Purple],
 {{n,5},1,10,1},
 {{K,10},10,20,1},
 {{N,20},20,60,1}]
```

Output

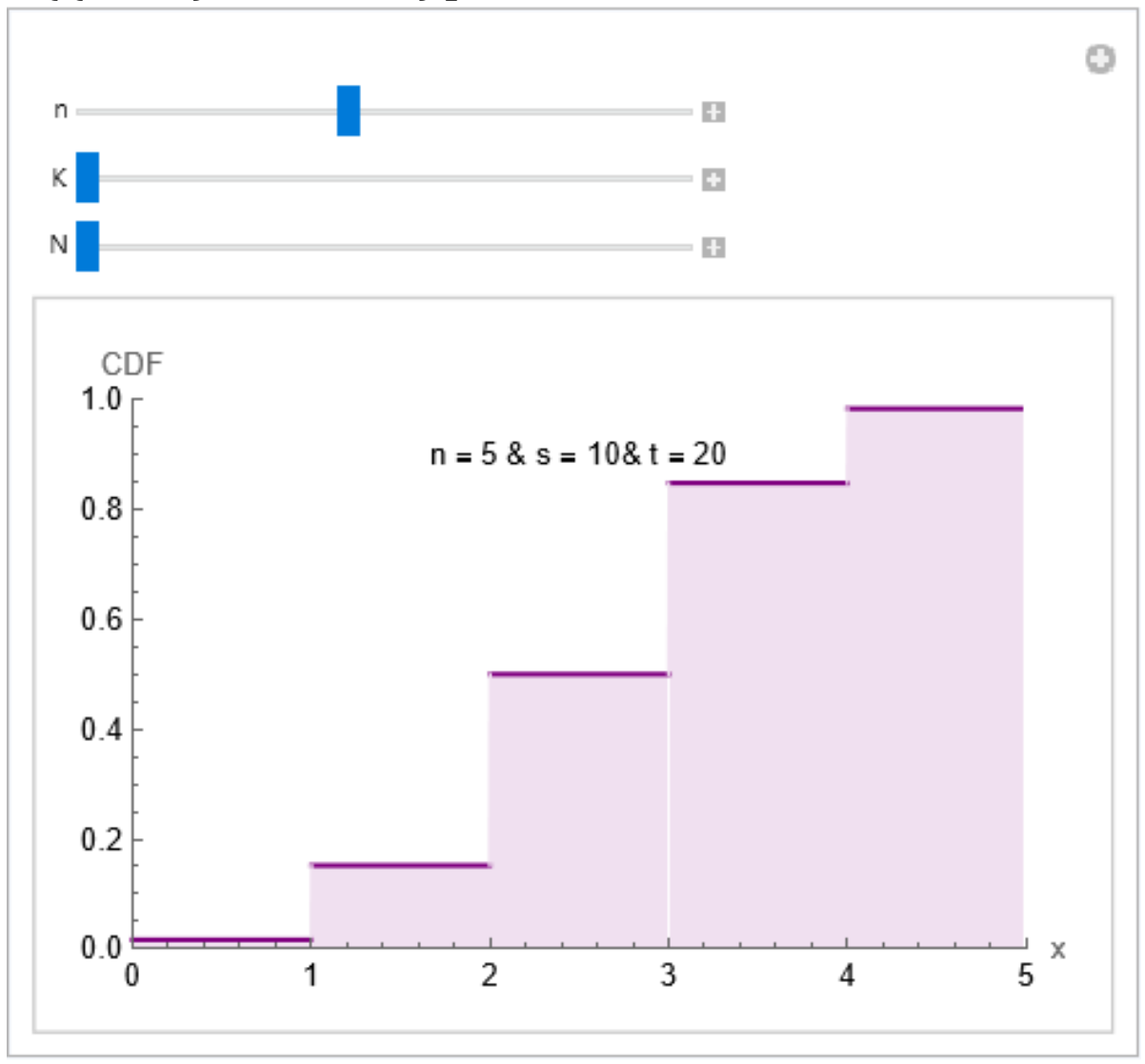

***Mathematica Examples 11.86***

Input

```
(* The code uses the Grid function to create a grid of two plots, one for the PDF
and one for the CDF, both of which are discrete plots. The code uses slider controls
to adjust the values of n, K and N: *)

Manipulate[
 Grid[
  {
   {DiscretePlot[
     PDF[
      HypergeometricDistribution[n,K,N],
      x
      ],
     {x,0,n},
     PlotRange->{{0, n},{0,1.5}},
     PlotStyle->{Purple,PointSize[0.03]},
     PlotLabel->"PDF of Hypergeometric Distribution",
     AxesLabel->{"x","PDF"},
     ImageSize->220
     ],
    DiscretePlot[
     CDF[
      HypergeometricDistribution[n,K,N],
      x
      ],
     {x,0,n},
     PlotRange->{{0,n},{0,1.5}},
     PlotStyle->{Purple,PointSize[0.03]},
     PlotLabel->"CDF of Hypergeometric Distribution",
     AxesLabel->{"x","CDF"},
     ImageSize->220
     ]
    }
   },
  Spacings->{5,5}
  ],
 {{n,5},1,10,1},
 {{K,10},10,20,1},
 {{N,20},20,60,1}
 ]
```

Output

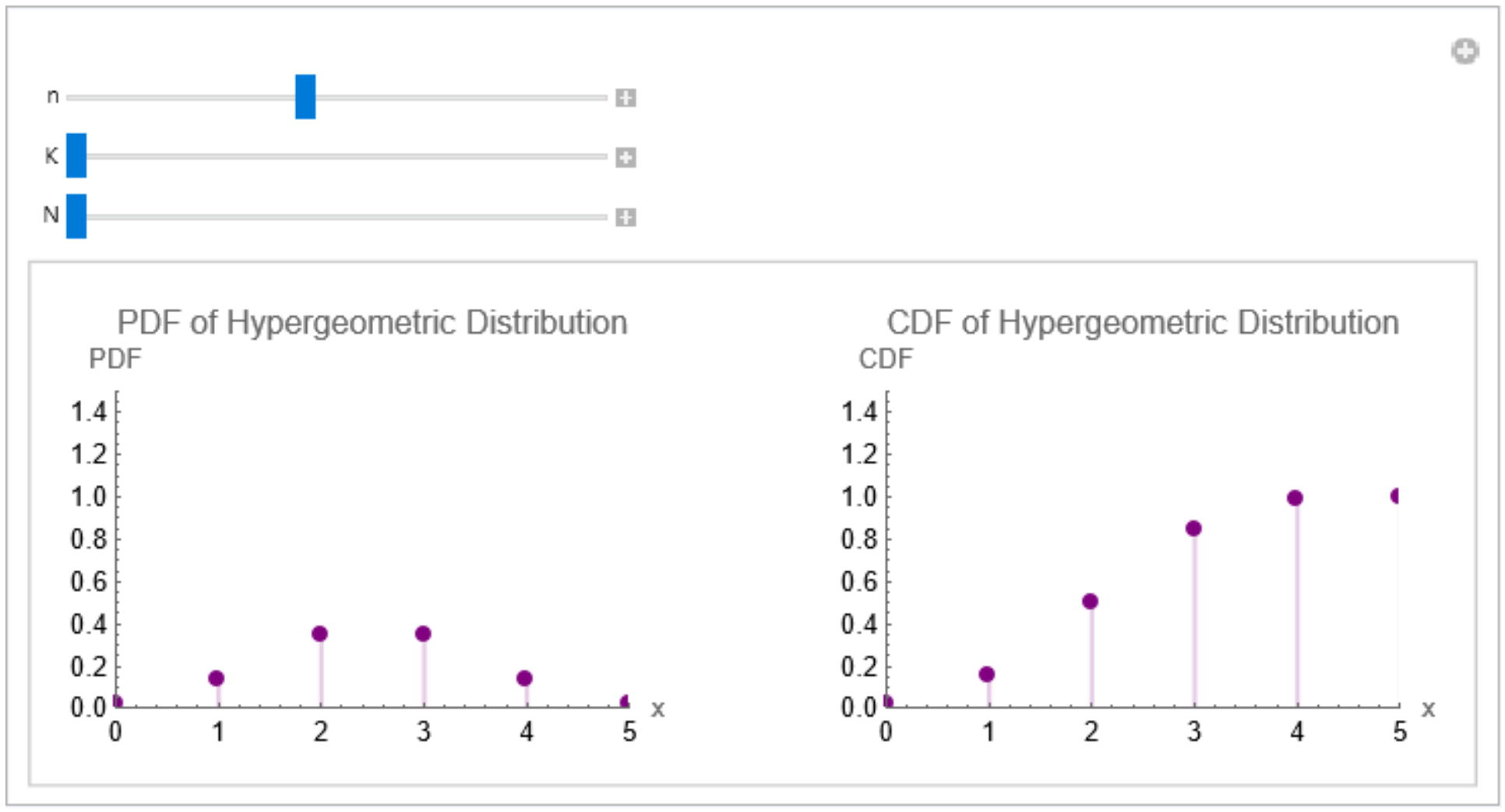

***Mathematica Examples 11.87***

Input

```
(* Suppose there are 5 defective items in a batch of 10 items, and 6 items are selected for testing. Simulate the process of testing when the number of defective items found is counted: *)

sampledata=RandomVariate[
  HypergeometricDistribution[6,5,10],
  20
  ]
ListPlot[
 sampledata,
 Filling->Axis,
 ImageSize->200,
 PlotStyle->Purple
 ]

(* Find the probability that there are 2 defective items in the sample: *)
NProbability[
 x==2,
 x\[Distributed]HypergeometricDistribution[6,5,10]
 ]
```

Output {2,2,4,2,3,3,2,2,3,4,3,2,4,2,4,4,2,3,3,5}

Output

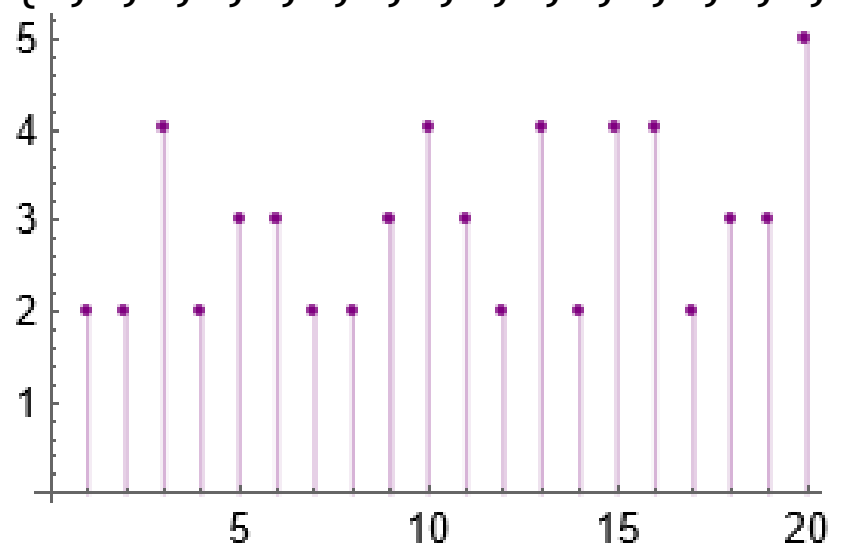

Output 0.238095

***Mathematica Examples 11.88***

Input

```
(* 30 US citizens and 40 non-US citizens pass through a security line at an airport. Ten are randomly selected for further screening. What is the probability that 3 or fewer of the selected passengers are US citizens: *)

(* Create the Hypergeometric distribution: *)
dist=HypergeometricDistribution[10,30,70];

(* Find the probability that 3 or fewer US citizens are selected: *)
p=NProbability[
  x<=3,
  x\[Distributed]dist
  ]
(*or*)
P=N[
  Sum[
   PDF[dist,i],
   {i,0,3}
   ]
  ]
```

Output 0.297948

Output 0.297948

***Mathematica Examples 11.89***

```
Input    (* Suppose an urn has 100 elements, of which 40 are special: *)
         urn[n_]=HypergeometricDistribution[n,40,100];

         (* The probability distribution that there are 20 special elements in a draw of 50
         elements: *)
         DiscretePlot[
          PDF[urn[50],k],
          {k,0,50},
          PlotRange->All,
          ImageSize->200,
          PlotStyle->Purple
          ]

         (* Compute the probability that there are more than 25 special elements in a draw of
         50 elements: *)
         Probability[X>25,X\[Distributed]urn[50]]

         (* Compute the expected number of special elements in a draw of 50 elements: *)
         Mean[urn[50]]

Output   43491718681268/3597278200209303
Output   20
```

# UNIT 11.7

# DISCRETE UNIFORM DISTRIBUTION

***Mathematica Examples 11.90***

Input

```
(* In this example, we define the random variable x as a range from 1 to n and set
the probability mass function (PMF) pmf as a uniform distribution. Each outcome has
an equal probability of 1/n. We then plot the PMF using ListPlot: *)

n=5;   (* Number of outcomes. *)

(* Define a discrete random variable with a uniform distribution: *)
x=Range[1,n];
pmf=Table[
    1/n,{n}
    ];

(* Plot the PMF: *)
ListPlot[
 Transpose[{x,pmf}],
 Filling->Axis,
 PlotRange->All,
 AxesLabel->{"x","P(X = x)"},
 PlotLabel->"Uniform Distribution PMF",
 PlotStyle->Purple,
 ImageSize->250
 ]
```

Output

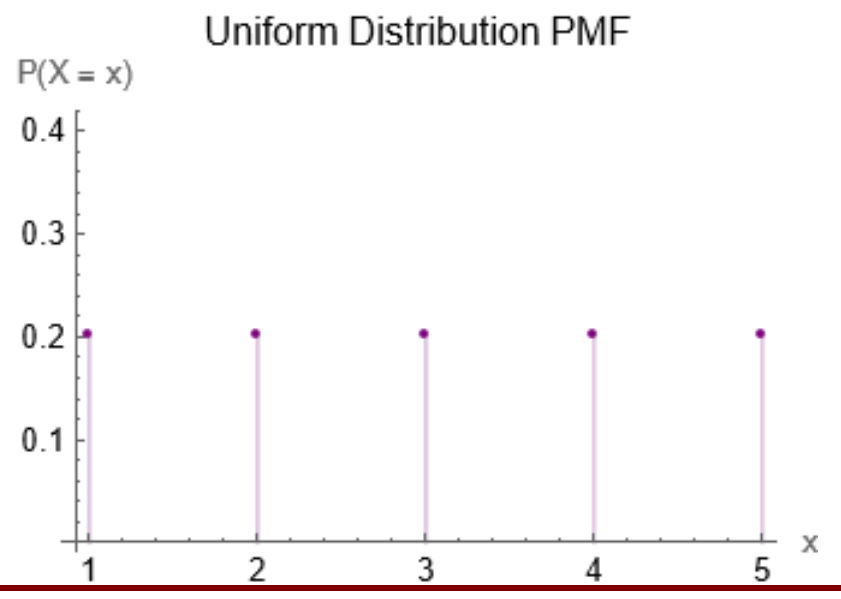

***Mathematica Examples 11.91***

Input

```
(* The code generates a discrete plot of the PMF for a discrete uniform distribution
with parameters a=1 and three different values of b (8, 12, and 16). The plot shows
the values of the PMF for all possible values of j between 0 and 26: *)

DiscretePlot[
 Evaluate[
  Table[
   PDF[
    DiscreteUniformDistribution[{1,b}],
    j
    ],
   {b,{8,12,16}}
   ]
  ],
```

```
{j,26},
ExtentSize->1/2,
PlotRange->All,
PlotMarkers->Automatic,
PlotLegends->Placed[{"a=1,b=8","a=1,b=12","a=1,b=16"},{0.8,0.75}],
PlotStyle->{RGBColor[0.88,0.61,0.14],RGBColor[0.37,0.5,0.7],Purple},
ImageSize->250,
AxesLabel->{None,"PMF"}
]
```

Output

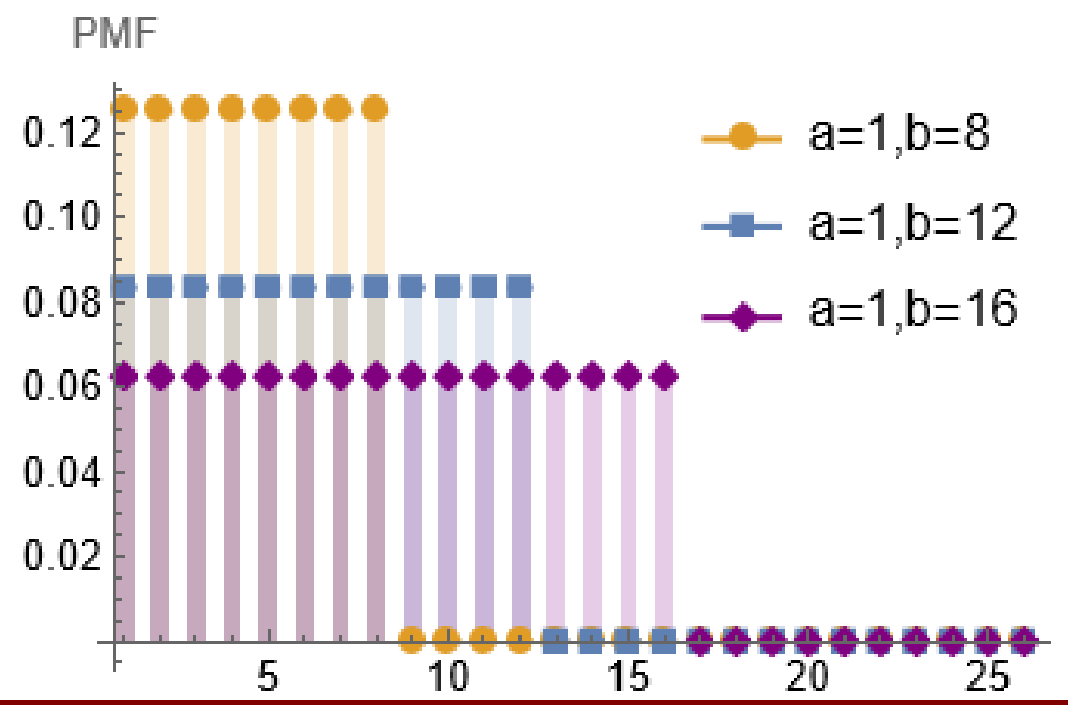

***Mathematica Examples 11.92***

Input

```
(* The code generates a discrete plot of the cumulative distribution function (CDF)
for a discrete uniform distribution with parameters a=1 and three different values
of b (8, 12, and 16). The plot shows the values of the CDF for all possible values
of k between 0 and 26: *)
DiscretePlot[
 Evaluate[
  Table[
   CDF[
    DiscreteUniformDistribution[{1,b}],
    k
    ],
   {b,{8,12,16}}
   ]
  ],
 {k,0,26},
 ExtentSize->Right,
 PlotRange->All,
 PlotMarkers->Automatic,
 PlotLegends->Placed[{"a=1,b=8","a=1,b=12","a=1,b=16"},{0.8,0.75}],
 PlotStyle->{RGBColor[0.88,0.61,0.14],RGBColor[0.37,0.5,0.7],Purple},
 ImageSize->250,
 AxesLabel->{None,"CDF"}
 ]
```

Output

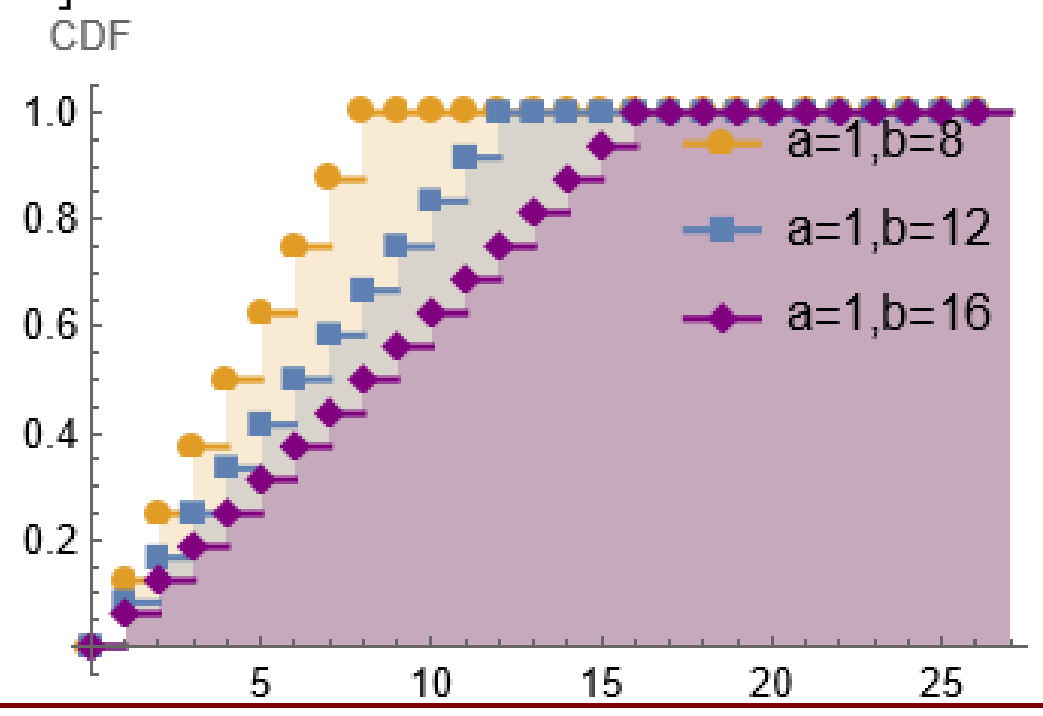

***Mathematica Examples 11.93***

Input
```
(* The code is generating a plot that shows the probability density function (PDF) of a discrete uniform distribution with a range of 20 to 40, superimposed on a histogram of a dataset 'data'. The histogram is generated with a bin size of 1 and the 'PDF' option, which normalizes the bin heights to represent a probability density function. The plot also includes a discrete plot of the PDF of the discrete uniform distribution for values of 'k' between 20 and 40: *)

data=RandomVariate[
   DiscreteUniformDistribution[{20,40}],
   10^4
   ];
Show[
 Histogram[
  data,
  {1},
  "PDF",
  ColorFunction->Function[{height},Opacity[height]],
  ChartStyle->Purple,
  ImageSize->320,
  AxesLabel->{None,"PDF"}
  ],

 DiscretePlot[
  PDF[
   DiscreteUniformDistribution[{20,40}],
   k
   ],
  {k,20,40},
  PlotStyle->PointSize[Medium],
  ColorFunction->"Rainbow"
  ]
 ]
```

Output

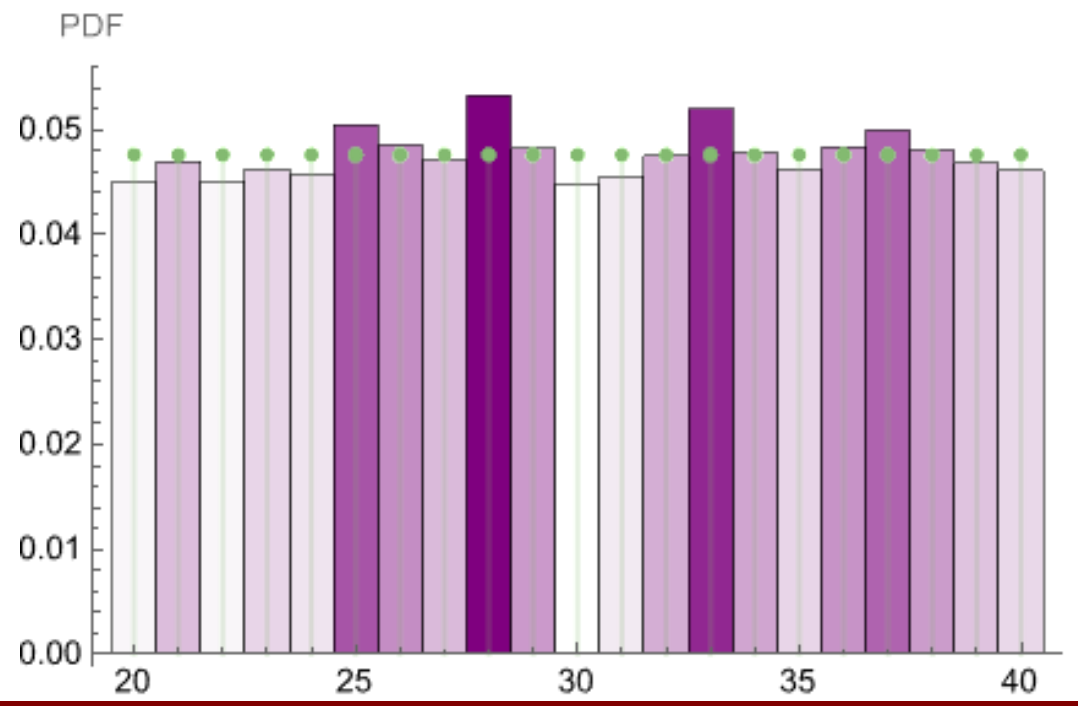

***Mathematica Examples 11.94***

Input
```
(* The code calculates and displays some descriptive statistics (mean, variance, standard deviation, kurtosis, and skewness) for a discrete uniform distribution with parameters a and b: *)

Grid[
 Table[
  {
   statistics,
   FullSimplify[statistics[DiscreteUniformDistribution[{a,b}]]]
   },
  {statistics,{Mean,Variance,StandardDeviation,Kurtosis,Skewness}}
  ],
```

```
 ItemStyle->12,
 Alignment->{{Right,Left}},
 Frame->All,
 Spacings->{Automatic,0.8}
 ]
```

Output

| | |
|---|---|
| Mean | (a + b)/2 |
| Variance | 1/12 (-1+(1- a + b)^2) |
| StandardDeviation | Sqrt[-1 + (1 - a + b)^2]/(2 Sqrt[3]) |
| Kurtosis | 3/5 (3- 4/(-1 + (1 - a + b)^2)) |
| Skewness | 0 |

***Mathematica Examples 11.95***

Input

```
(* The code calculates and displays some additional descriptive statistics (moments,
central moments, and factorial moments) for a discrete uniform distribution with
parameters a and b: *)

Grid[
 Table[
  {
   statistics,
   FullSimplify[statistics[DiscreteUniformDistribution[{a,b}],1]],
   FullSimplify[statistics[DiscreteUniformDistribution[{a,b}],2]]
   },
  {statistics,{Moment,CentralMoment,FactorialMoment}}
  ],
 ItemStyle->12,
 Alignment->{{Right,Left}},
 Frame->All,
 Spacings->{Automatic,0.8}
 ]
```

Output

| | | |
|---|---|---|
| Moment | (a+b)/2 | 1/6 (a (-1+2 a)+b+2 a b+2 b^2) |
| CentralMoment | 0 | 1/12 (-a+b) (2-a+b) |
| FactorialMoment | (a+b)/2 | 1/3 (a^2+a (-2+b)+(-1+b) b) |

***Mathematica Examples 11.96***

Input

```
(* The code generates a histogram of 500 random samples drawn from a discrete uniform
distribution with a range of -10 to 10. The histogram is displayed with the "PDF"
option, which normalizes the bin heights to represent a probability density function.
The x-axis ranges from -10.5 to 10.5 with a bin size of 1. The mean and quartiles of
the data are also displayed as vertical lines in red and green, respectively: *)

dist=DiscreteUniformDistribution[{-10,10}];
data=RandomVariate[dist,500];
Histogram[
 data,
 {-10.5,10.5,1},
 "PDF",
 Epilog->{
   Directive[Red,Thickness[0.006]],
   Line[{{Mean[data],0},{Mean[data],0.04}}],
   Directive[Green,Dashed],
   Line[{{Quantile[data,0.25],0},{Quantile[data,0.25],0.04}}],
   Line[{{Quantile[data,0.5],0},{Quantile[data,0.5],0.04}}],
   Line[{{Quantile[data,0.75],0},{Quantile[data,0.75],0.04}}]
   },
 Frame->True,
 FrameLabel->{"x","PDF"},
```

```
         LabelStyle->Directive[Bold,Medium],
         ColorFunction->Function[{height},Opacity[height]],
         ChartStyle->Purple,
         ImageSize->320,
         AxesLabel->{None,"PDF"}
         ]
```

Output

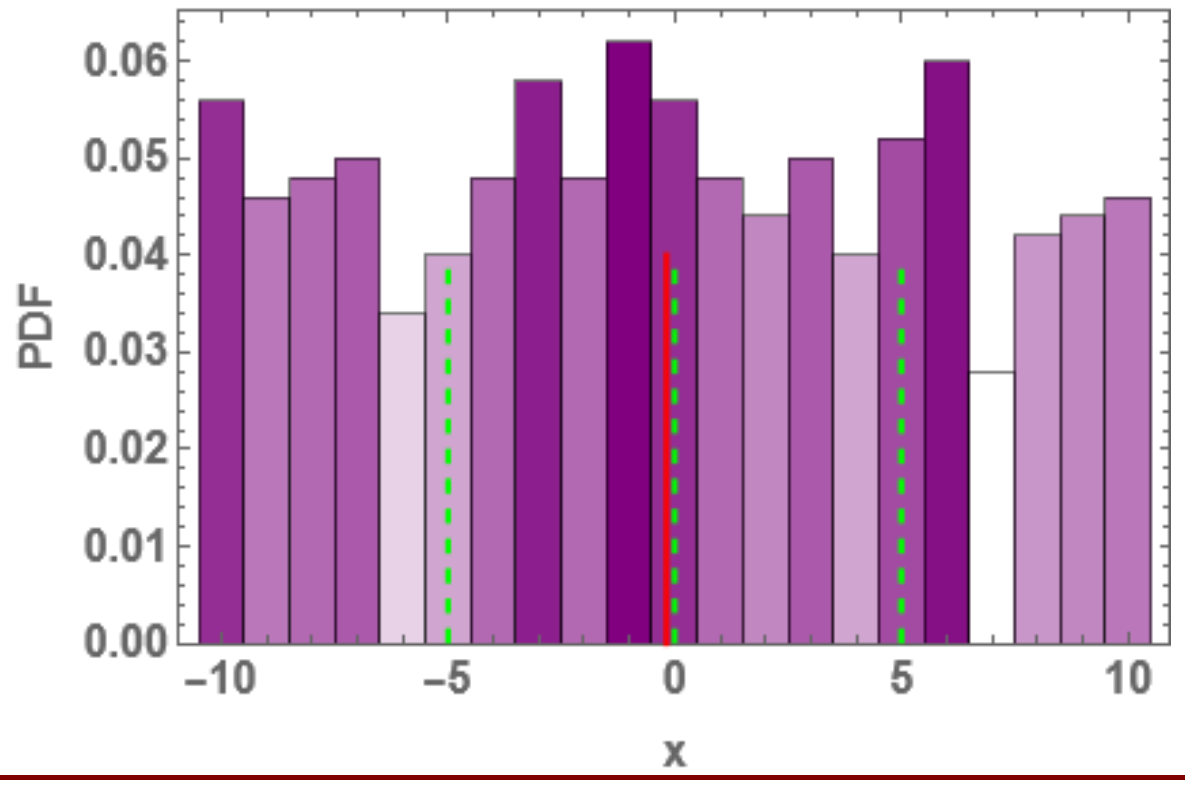

*Mathematica Examples 11.97*

Input

```
(* The code defines a Manipulate that creates a discrete uniform distribution with
parameters a and b, generates m random variates from the distribution, and plots a
histogram of the sample. The Manipulate includes sliders to control the values of a,
b, and m. The Epilog option of the Show function adds a text label to the plot that
displays the mean of the distribution. The mean is calculated as the average of a
and b, and is formatted using ToString and Style: *)

Manipulate[
 dist=DiscreteUniformDistribution[{a,b}];
 h=Histogram[
    RandomVariate[dist,m],
    {a,b,1},
    PlotRange->{{a-1,b+1},{0,Automatic}},
    ColorFunction->Function[{height},Opacity[height]],
    ChartStyle->Purple
    ];
 mean=Style[
    StringJoin[
     "Mean: ",ToString[N[(a+b)/2]
      ]
     ],
    14
    ];

 Show[
  h,
  Epilog->Text[mean,{a-0.5,2}],
  ImageSize->320
  ],
 {{a,-10,"Minimum Value a"},-10,10,1},
 {{b,11,"Maximum Value b"},-9,11,1},
 {{m,1000,"Sample Size"},10,1000,10},
 Initialization:>(dist=DiscreteUniformDistribution[{a,b}];)
 ]
```

Output

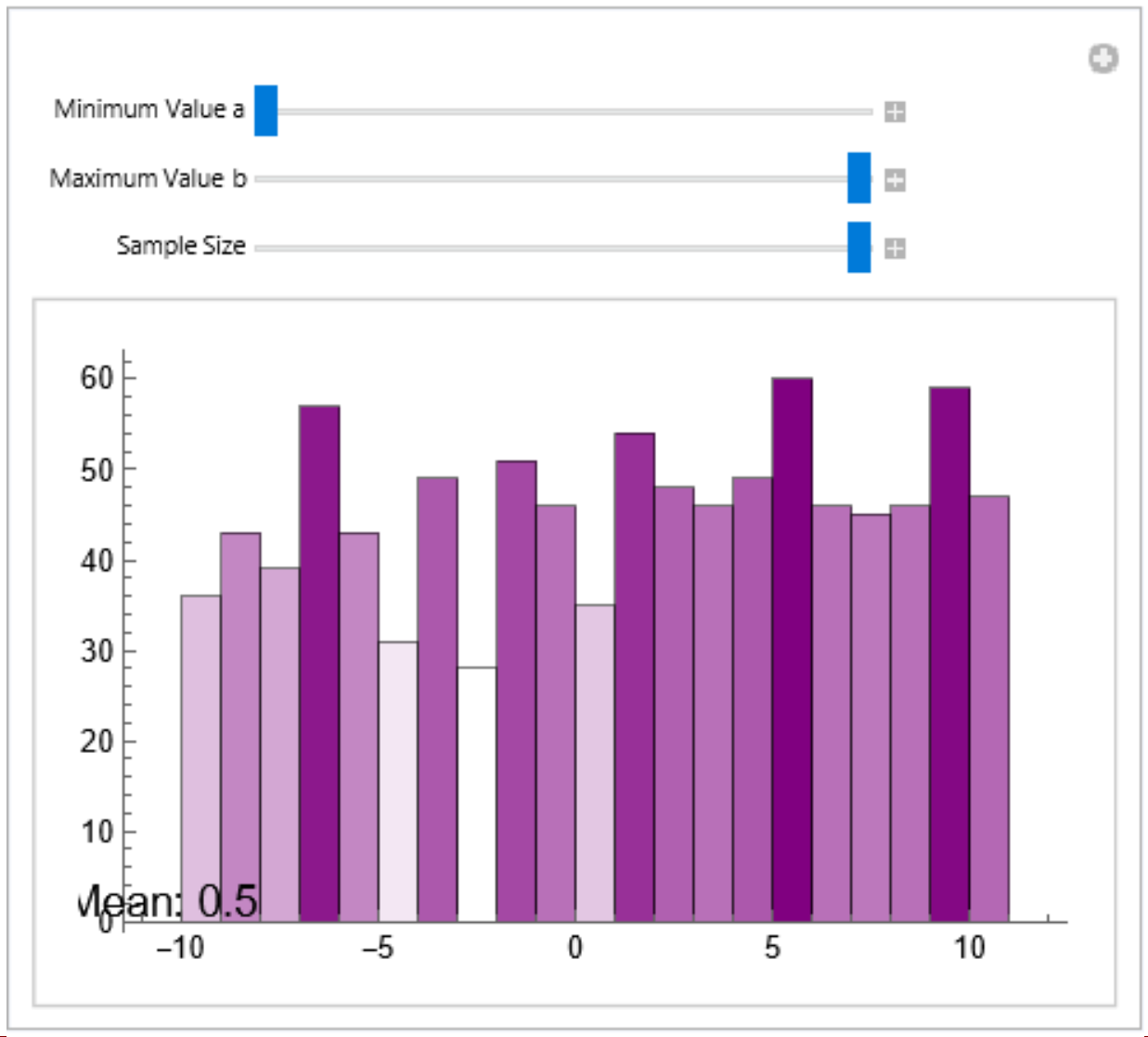

***Mathematica Examples 11.98***

Input

```
(* The code defines a Manipulate that creates a discrete uniform distribution with
minimum value 1 and maximum value n, and plots its CDF using the Plot function. The
Manipulate includes a slider to control the value of n. The plot displays the CDF of
the distribution, which is the probability that a random variate from the distribution
is less than or equal to a given value x: *)
Manipulate[
 dist=DiscreteUniformDistribution[{1,n}];
 Plot[
  CDF[dist,x],
  {x,0,n+1},
  PlotRange->{{0,n+1},{0,1}},
  Filling->Axis,
  FillingStyle->LightPurple,
  ImageSize->320,
  PlotStyle->Purple
  ],
 {{n,5,"Number of Elements"},2,30,1},
 Initialization:>(dist=DiscreteUniformDistribution[{1,n}];)
 ]
```

Output

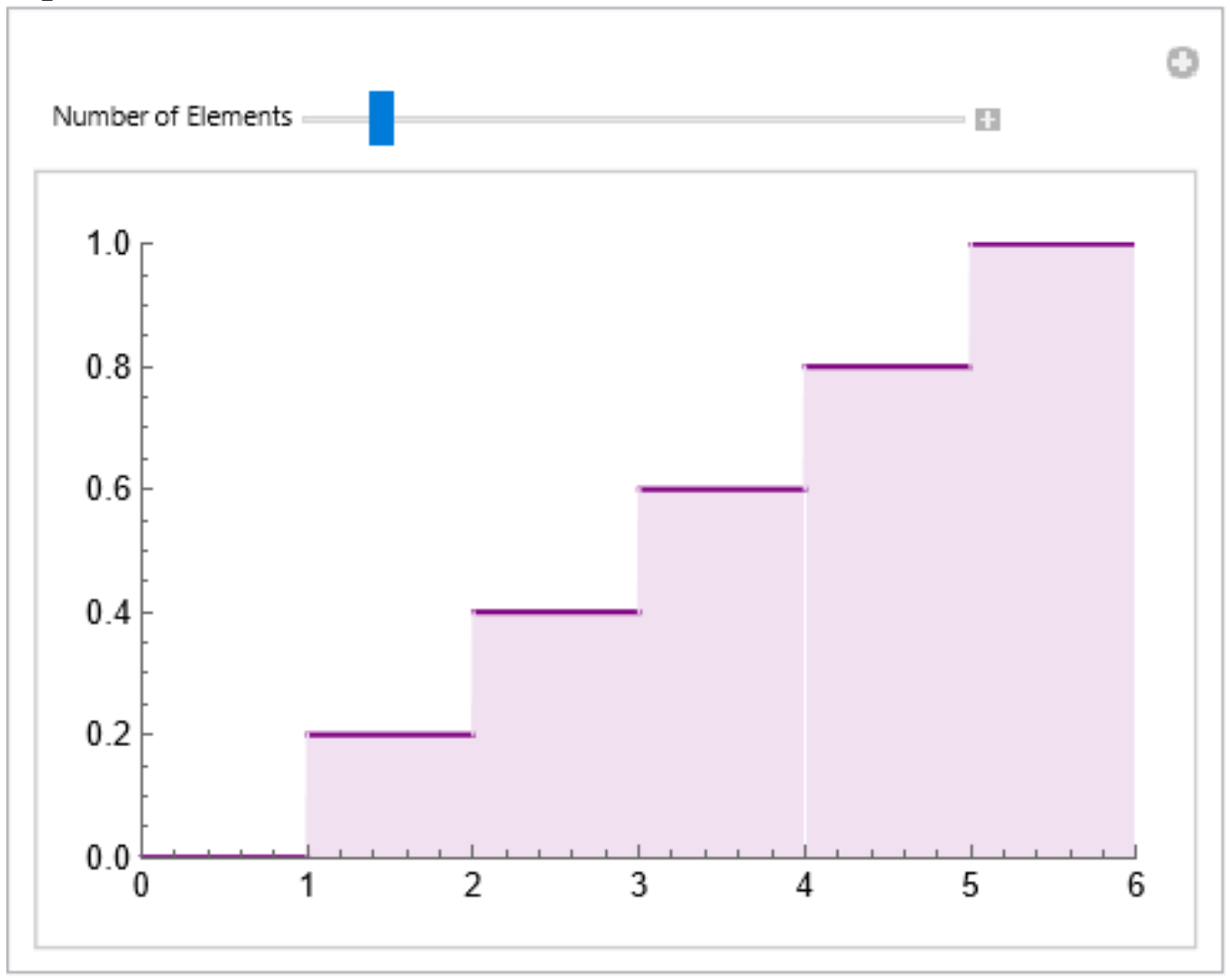

### *Mathematica Examples 11.99*

Input

```
(* The code creates a Manipulate with sliders for the lower and upper bounds of the
discrete uniform distribution. As you move the sliders, the code generates plots of
both the PDF and CDF for the new parameters: *)

Manipulate[
 {
  DiscretePlot[
   PDF[
    DiscreteUniformDistribution[
     {a,b}
     ],
    x
    ],
   {x,a,b},
   PlotRange->All
   ],
  DiscretePlot[
   CDF[
    DiscreteUniformDistribution[
     {a,b}
     ],
    x
    ],
   {x,a,b},
   PlotRange->All
   ]
  },
 {{a,1,"Lower Bound"},0,20,1},
 {{b,5,"Upper Bound"},1,30,1}
 ]
```

Output

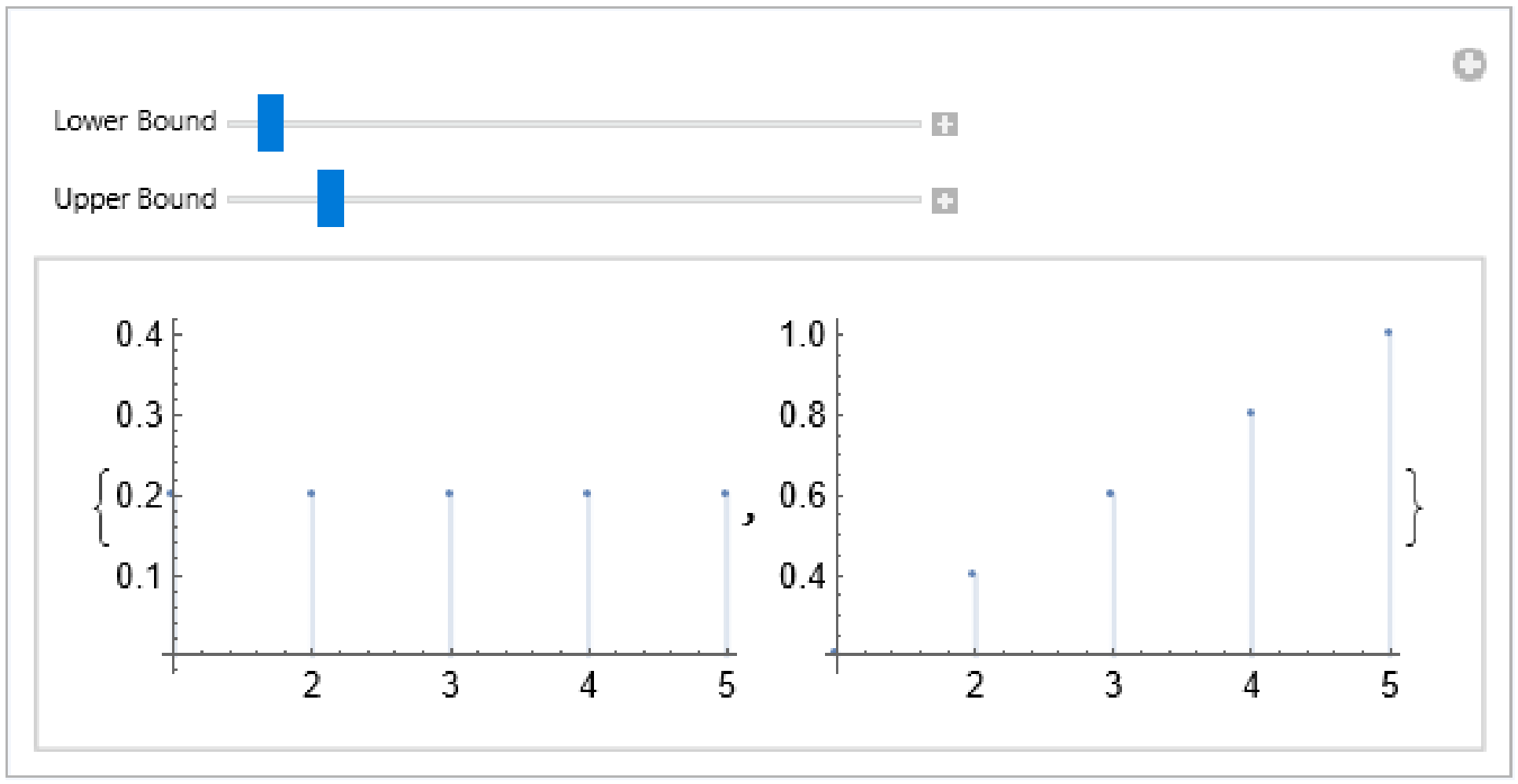

### *Mathematica Examples 11.100*

Input

```
(* The code generates a sample of 500 values from a discrete uniform distribution
with support {2, 3, 4, 5} and estimates the distribution parameters (a and b) from
the sample using the EstimatedDistribution function. It then compares the density
histogram of the sample with the PDF of the estimated distribution: *)

sampledata=RandomVariate[
   DiscreteUniformDistribution[{2,5}],
```

```
   500
   ];

(* Estimate the distribution parameters from sample data: *)
ed=EstimatedDistribution[
  sampledata,
  DiscreteUniformDistribution[
   {a,b}
   ]
  ]

(* Compare the density histogram of the sample with the PDF of the estimated distribution: *)
Show[
 Histogram[
  sampledata,
  {1},
  "PDF",
  ColorFunction->Function[{height},Opacity[height]],
  ChartStyle->Purple
  ],
 DiscretePlot[
  PDF[ed,x],
  {x,2,6},
  PlotStyle->PointSize[Medium],
  ColorFunction->"Rainbow",
  ImageSize->320
  ]
 ]
```

Output DiscreteUniformDistribution[{2,5}]

Output

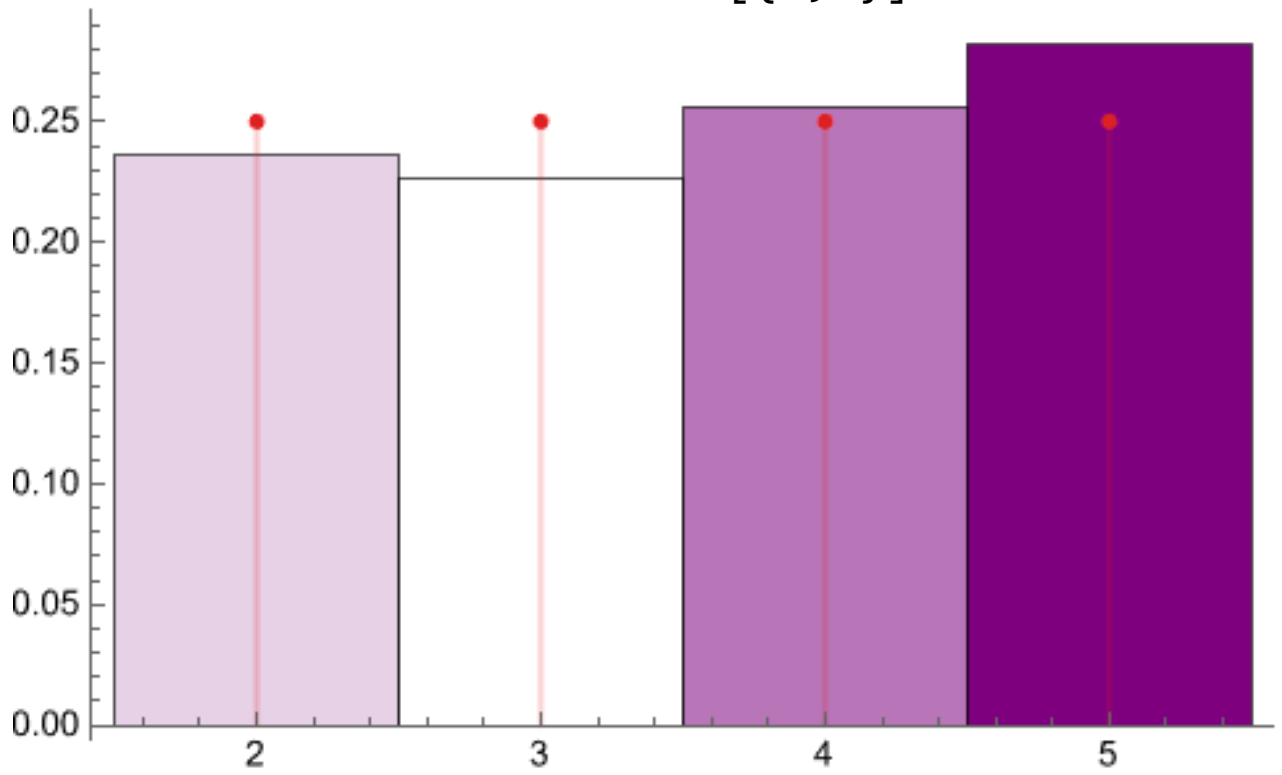

*Mathematica Examples 11.101*

Input

```
(* The code defines a discrete uniform distribution on the integers 1 through 6 and generates a random sample of size 1000 from the distribution. It then counts the number of times each possible outcome occurs in the sample: *)

dist=DiscreteUniformDistribution[{1,6}];
data=RandomVariate[dist,1000];
Count[data,#]&/@Range[6]
```

Output {174,187,162,158,152,167}

*Mathematica Examples 11.102*

Input

```
(* The code defines a discrete uniform distribution on the integers 1 through 6 and generates a random sample of size 1000 from the distribution. It then calculates the
```

```
        empirical and exact probabilities of each possible outcome and plots them on a bar
        chart: *)

        dist=DiscreteUniformDistribution[{1,6}];
        n=1000;

        data=Table[
            RandomVariate[dist],
            {n}
            ];

        empiricalprobs=N[
          Table[
            Count[data,i]/n,
            {i,6}
            ]
          ]

        exactprobs=N[
          Table[
            PDF[dist,i],
            {i,6}
            ]
          ]

        ListPlot[
         {empiricalprobs,exactprobs},
         PlotMarkers->{"◆","●"},
         PlotLegends->{"Empirical","Exact"},
         Frame->True,
         FrameLabel->{"Outcome","Probability"},
         LabelStyle->Directive[Bold,Medium],
         PlotStyle->{Purple,Blue},
         ImageSize->320
         ]
Output  {0.184,0.163,0.154,0.162,0.182,0.155}
Output  {0.166667,0.166667,0.166667,0.166667,0.166667,0.166667}
Output
```

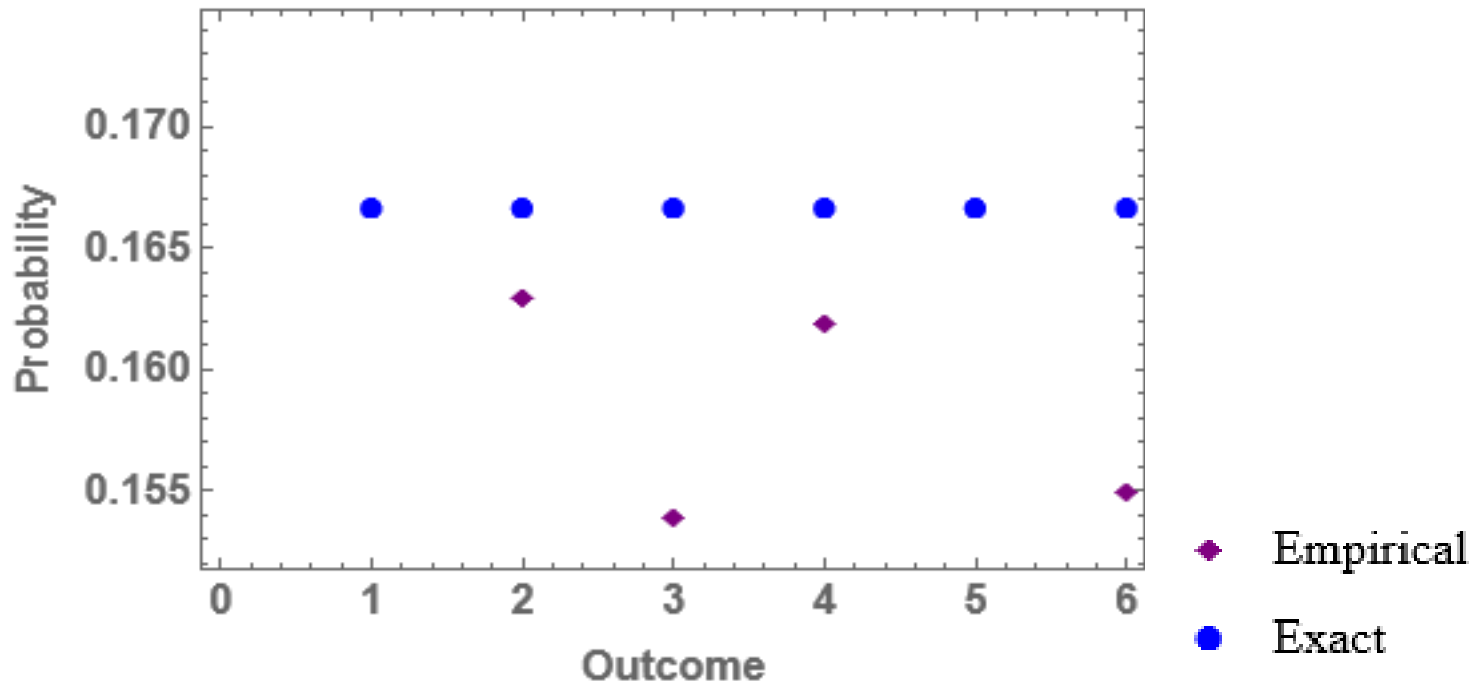

*Mathematica Examples 11.103*

```
Input   (* The code defines a discrete uniform distribution on the integers 1 through 6 and
        generates a random sample of size 100 from the distribution. It then plots the
        cumulative sum of the sample as a function of the trial number: *)
        dist=DiscreteUniformDistribution[{1,6}];
        n=100;

        data=Table[
          RandomVariate[dist],
          {n}
```

```
          ]
        Accumulate[data]
        ListPlot[
         Accumulate[data],
         Filling->Axis,
         Frame->True,
         FrameLabel->{"Trial","Cumulative Total"},
         LabelStyle->Directive[Bold,Medium],
         PlotStyle->Purple,
         ImageSize->320
         ]
Output  {4,3,6,6,5,6,6,2,4,3,5,5,4,5,3,6,1,2,6,6,4,5,4,4,4,5,1,5,2,3,5,4,5,3,1,4,4,1,2,6,6,
        3,1,6,3,1,3,5,3,3,5,3,2,1,6,2,6,6,6,6,3,4,2,6,4,6,2,2,3,4,6,1,1,1,4,1,1,6,3,6,5,2,6
        ,3,6,6,2,5,6,3,3,3,6,4,5,1,2,6,4,3}
Output  {4,7,13,19,24,30,36,38,42,45,50,55,59,64,67,73,74,76,82,88,92,97,101,105,109,114,11
        5,120,122,125,130,134,139,142,143,147,151,152,154,160,166,169,170,176,179,180,183,1
        88,191,194,199,202,204,205,211,213,219,225,231,237,240,244,246,252,256,262,264,266,
        269,273,279,280,281,282,286,287,288,294,297,303,308,310,316,319,325,331,333,338,344
        ,347,350,353,359,363,368,369,371,377,381,384}
Output
```

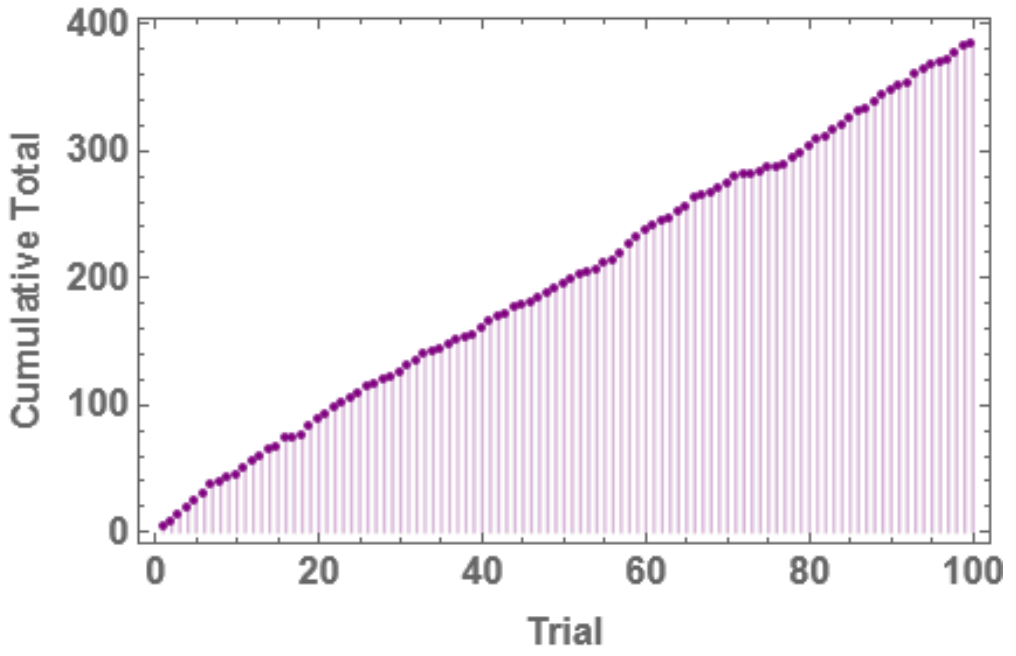

***Mathematica Examples 11.104***

```
Input   (* This code models a fair six-sided die using a discrete uniform distribution and
        then performs a few calculations to find the probability that the sum of four dice
        values is less than seven. It uses the Tuples function to generate all possible
        outcomes of four dice throws and selects only those outcomes where the sum of the
        four dice values is less than or equal to seven. The length of this list is divided
        by 6^4 to obtain the theoretical probability: *)

        (* Defines the distribution and assigns it to the variable "dice": *)
        dice=DiscreteUniformDistribution[{1,6}];

        (* Generates 15 throws of the die using the RandomVariate function: *)
        RandomVariate[dice,15]

        (* Calculates the probability that the sum of four dice values is less than seven:
        *)
        N[
         Probability[
          x1+x2+x3+x4<=7,

        {x1\[Distributed]dice,x2\[Distributed]dice,x3\[Distributed]dice,x4\[Distributed]dic
        e}
          ]
         ]

        (* Verify by generating random dice throws, in this case 10^5 times four dice throws:
        *)
        N[
```

```
            Count[
             RandomVariate[dice,{10^5,4}],_?(Total[#]<=7&)
             ]
            ]/10^5

          (* Verifies the result using an explicit enumeration of all possible dice outcomes:
          *)
          g=Select[
            Tuples[
             Range[6],
             4
             ],
            Total[#]<=7&
            ]

          N[Length[g]]/6^4

Output    {6,4,2,2,6,1,2,5,2,2,2,2,3,1,1}
Output    0.0270062
Output    0.02703
Output    {{1,1,1,1},{1,1,1,2},{1,1,1,3},{1,1,1,4},{1,1,2,1},{1,1,2,2},{1,1,2,3},{1,1,3,1},{1
          ,1,3,2},{1,1,4,1},{1,2,1,1},{1,2,1,2},{1,2,1,3},{1,2,2,1},{1,2,2,2},{1,2,3,1},{1,3,
          1,1},{1,3,1,2},{1,3,2,1},{1,4,1,1},{2,1,1,1},{2,1,1,2},{2,1,1,3},{2,1,2,1},{2,1,2,2
          },{2,1,3,1},{2,2,1,1},{2,2,1,2},{2,2,2,1},{2,3,1,1},{3,1,1,1},{3,1,1,2},{3,1,2,1},{
          3,2,1,1},{4,1,1,1}}
Output    0.0270062
```

# CHAPTER 12

# CONTINUOUS RANDOM VARIABLES AND DISTRIBUTIONS

In this chapter, we will delve into the world of continuous RVs and distributions. Unlike discrete RVs that take on only a finite or countable number of values, continuous RVs can assume any value within a specified range or interval. The study of continuous RVs is crucial in various fields, including statistics, probability theory, and applied mathematics, due to their widespread applicability in modeling real-world phenomena.

- We will begin by exploring the concept of probability density functions (PDFs). Continuous RVs are described by a PDF rather than a PMF as used for discrete RVs. The PDF gives the probability of a variable falling within a certain interval. Integration over the PDF within a specific interval gives the probability of the variable falling within that interval. PDF provides a means to quantify the likelihood of observing a value within a given interval. We will discuss the properties and interpretation of PDFs, as well as their role in calculating probabilities and expected values.
- Next, we will study CDFs of continuous RVs, which complement PDFs by providing a different perspective on the distribution of continuous RVs. CDFs give the probability of observing a value less than or equal to a specific point. We will examine the properties of CDFs and their connection to PDFs, enabling us to obtain valuable information about the behavior of continuous RVs.
- Another fundamental concept we will explore is the MGF of continuous RVs. The MGF allows us to derive moments of a RV, providing a comprehensive description of its distribution. We will investigate how MGFs can be used to determine moments, variance, and other statistical measures of continuous RVs.
- After establishing these foundational concepts, we will proceed to study specific continuous probability distributions.
    - We will begin with the uniform distribution, which represents a scenario where all outcomes within a given range are equally likely.
    - Next, we will examine the exponential distribution, which is widely used in modeling various real-world events, such as the time between occurrences of certain events.
    - Following that, we will study the gamma distribution, which is a versatile continuous probability distribution frequently encountered in fields such as reliability analysis, queuing theory, and insurance. We will explore the gamma distribution's shape, parameters, and applications, shedding light on its usefulness in modeling diverse phenomena.
    - Lastly, we will turn our attention to the normal distribution, also known as the Gaussian distribution. Normal distribution is one of the most important and widely used continuous probability distributions due to its remarkable properties and its natural occurrence in many real-world phenomena. We will investigate the characteristics of the normal distribution, including its bell-shaped curve, and the empirical rule.

Each of these distributions will be examined in detail, discussing their PDFs, CDFs, and moments. Real-world examples and applications will be provided to illustrate the practical relevance of these distributions in various fields of study. By studying continuous RVs and their associated distributions, we gain a deeper understanding of the probabilistic nature of continuous phenomena.

## 12.1 Continuous RVs and PDF

### Continuous RVs

The possible values that a continuous RV can assume are infinite and uncountable. For example, the variable that represents the time taken by a worker to commute from home to work is a continuous RV. Suppose 6 minutes is the minimum time and 120 minutes is the maximum time taken by all workers to commute from home to work. Let $X$ be a continuous RV that denotes the time taken to commute from home to work by a randomly selected worker. Then $X$ can assume any value in the interval 6 to 120 minutes. This interval contains an infinite number of values that are uncountable.

Examples of continuous RVs:

- Height and weight:
  The height of individuals can be considered a continuous RV since it can take on any value within a certain range. For example, a person's height can be measured in centimeters or inches and can have values like 150.5 cm, 175.2 cm, or any other value within the range of human heights. Also, the weight of objects is often considered a continuous RV. It can be measured in grams, kilograms, or pounds and can have any value within a certain range. For instance, the weight of a fruit can be 250 grams, 500 grams, or any other value within the weight distribution.
- Time:
  Time is often modeled as a continuous RV. For instance, the time it takes for a car to travel from one point to another, the duration of a phone call, or the time it takes to complete a task can be considered continuous RVs. These variables can have infinitely many possible values, such as 4.562 seconds, 12.987 milliseconds, or any other value in the range of possible times.
- Temperature:
  Temperature is another example of a continuous RV. It can be measured in degrees Celsius or Fahrenheit. For example, the temperature outside can be 27.3°C, 18.9°C, or any other value within the possible temperature range.
- Income:
  Income is often modeled as a continuous RV since it can take on any value within a certain range. For example, a person's annual income can be $50,000, $75,000, or any other value within the income distribution.
- Stock prices:
  The prices of stocks in financial markets are considered continuous RVs.
- Blood pressure:
  Blood pressure is a continuous RV that represents the force exerted by blood against the walls of blood vessels. It is typically measured in millimeters of mercury (mmHg). For instance, a person's blood pressure can be 120/80 mmHg, 140/90 mmHg or any other combination within the possible pressure range.
- Rainfall:
  The amount of rainfall in a particular area during a given time period can be modeled as a continuous RV. It can be measured in millimeters or inches. For instance, the rainfall in a region can be 10 millimeters, 50 millimeters, or any other value within the possible range of rainfall amounts.
- Arrival time:
  In queuing theory or transportation analysis, the time between arrivals of customers, vehicles, or events can be modeled as continuous RVs. For instance, the time between the arrivals of customers at a store, the time between buses arriving at a bus stop, or the inter-arrival time of requests to a server can all be represented by continuous RVs.
- Response time in customer service:
  The time it takes for a customer service representative to respond to a customer inquiry or request can be modeled as a continuous RV.

- Lifetime of electronic devices:
  The lifespan of electronic devices such as smartphones, laptops, or refrigerators can be treated as a continuous RV. It represents the duration until the device becomes inoperable or obsolete. For example, the lifetime of a smartphone can be 2 years, 4 years, or any other value within the range of possible lifetimes. The time until a light bulb burns out, or the lifespan of a battery can all be considered continuous RVs.
- Speed:
  Suppose we have a RV representing the speed of vehicles on a highway. The speed can take on any positive value within a certain range, such as 60 km/h, 65.5 km/h, or any other value in between. Speed is a continuous RV as it can vary continuously.

These examples illustrate how continuous RVs can represent quantities that can take on a wide range of values, potentially spanning an infinite number of possibilities.

### PDFs, CDFs and MGFs

Suppose you have a set of measurements on a continuous RV, and you create a relative frequency histogram to describe their distribution. For a small number of measurements, you could use a small number of classes; then as more and more measurements are collected, you can use more classes and reduce the class width. The outline of the histogram will change slightly, for the most part becoming less and less irregular, as shown in Figure 12.1.a As the number of measurements becomes very large and the class widths become very narrow, the relative frequency histogram appears more and more like the smooth curve shown in Figure 12.1.c. This smooth curve describes the probability distribution of the continuous RV.

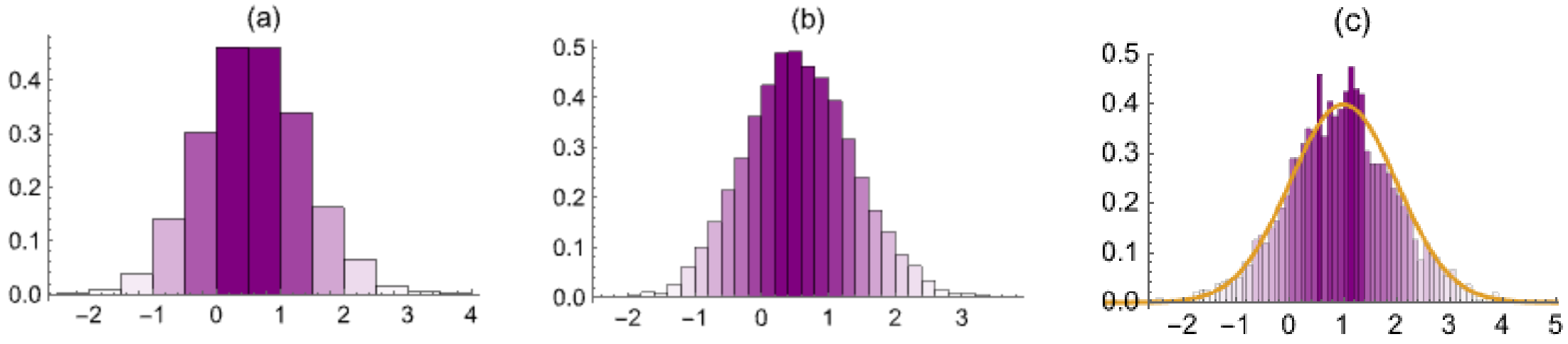

**Figure 12.1.** Relative frequency histograms for increasingly large sample sizes.

A PDF $f(x)$ can be used to describe the probability distribution of a continuous RV $X$. If an interval is likely to contain a value for $X$, its probability is large, and it corresponds to large values for $f(x)$. A histogram is an approximation to a PDF. For each interval of the histogram, the area of the bar equals the relative frequency (proportion) of the measurements in the interval. The relative frequency is an estimate of the probability that a measurement falls in the interval. Similarly, the area under $f(x)$ over any interval equals the true probability that a measurement falls in the interval.

Remember that, for discrete RVs,

- The sum of all the probabilities $P(x)$ equals 1 and
- The probability that $X$ falls into a certain interval is the sum of all the probabilities in that interval.

Continuous RVs have some parallel characteristics listed next.

- The area under a continuous probability distribution is equal to 1.
- The probability that $X$ will fall into a particular interval—say, from $a$ to $b$—is equal to the area under the curve between the two points $a$ and $b$. Hence, the probability that $X$ is between $a$ and $b$ is determined as the integral of $f(x)$ from $a$ to $b$.

**Definition (PDF):** For a continuous RV $X$, a PDF is a function such that

$$f(x) \geq 0, \tag{12.1.1}$$

$$\int_{-\infty}^{\infty} f(x)dx = 1, \tag{12.1.2}$$

$$P(a \leq X \leq b) = \int_a^b f(x)dx = \text{area under } f(x) \text{ from } a \text{ to } b \text{ for any } a \text{ and } b. \tag{12.1.3}$$

(See Figure 12.2.).

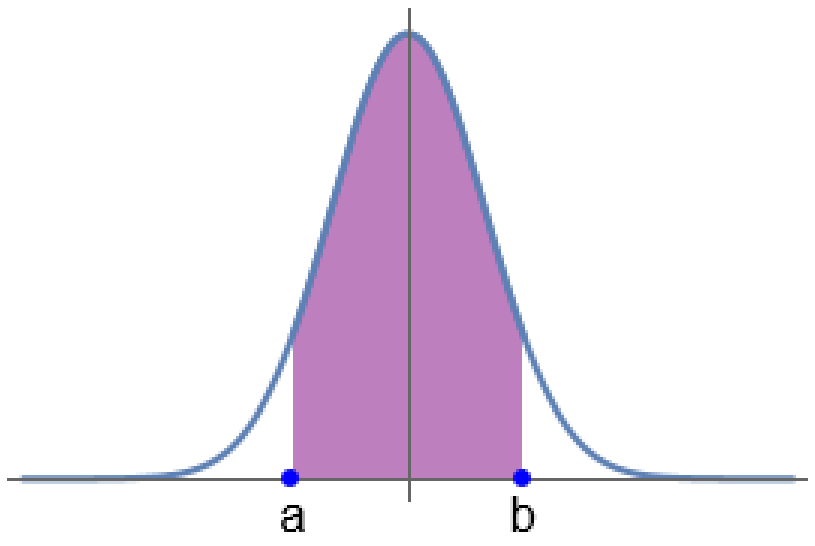

**Figure 12.2.** Probability as an area under a curve.

A PDF provides a simple description of the probabilities associated with a RV. As long as $f(x)$ is nonnegative and $\int_{-\infty}^{\infty} f(x)dx = 1$, $0 \leq P(a < X < b) \leq 1$ so that the probabilities are properly restricted. A PDF is zero for $x$ values that cannot occur.

There is an important difference between discrete and continuous RVs. Consider the probability that $X$ equals some particular value, say, $a$. Since there is no area above a single point, $X = a$, in the probability distribution for a continuous RV, our definition implies that the probability is always zero. For example, if $X$ is the height of a randomly selected female student from a university, then the probability that this student is exactly 66.8 inches tall is zero; that is, $P(X = 66.8) = 0$.

- For a continuous RV $X$ and any value $a$, $P(X = a) = \int_a^a f(x)dx = 0$.
- This implies that

$$P(x \geq a) = P(x > a) \quad \text{and} \quad P(x \leq a) = P(x < a).$$

- If $X$ is a continuous RV, for any $a$ and $b$,

$$P(a \leq X \leq b) = P(a < X \leq b) = P(a \leq X < b) = P(a < X < b).$$

- In other words, the probability that $X$ assumes a value in the interval $a$ to $b$ is the same whether or not the values $a$ and $b$ are included in the interval. For the example on the heights of female students, the probability that a randomly selected female student is between 65 and 68 inches tall is the same as the probability that this female is 65 to 68 inches tall.
- This is not true in general for discrete RVs.

**Definition (CDF):** The CDF of a continuous RV $X$ is

$$F(x) = P(X \leq x) = \int_{-\infty}^{x} f(u)du, \tag{12.2.1}$$

for $-\infty < x < \infty$. Given $F(x)$,

$$f(x) = \frac{dF(x)}{dx}, \tag{12.2.2}$$

as long as the derivative exists. That is, the density is the derivative of the CDF.
The importance of the CDF here, just as for discrete RV, is that probabilities of various intervals can be computed from a formula for $F(x)$. If we have the CDF $F(x)$, then

$$P(a \leq X \leq b) = F(b) - F(a). \tag{12.2.3}$$

***Example 12.1***

Suppose that $X$ is a continuous RV whose PDF is given by

$$f(x) = \begin{cases} C(4x - 2x^2), & 0 < x < 2 \\ 0, & \text{otherwise} \end{cases}.$$

(a) What is the value of $C$? (b) Find $P(X > 1)$. (c) Find $F(x)$.

***Solution***

(a) Since $f$ is a PDF, we must have that

$$\int_{-\infty}^{\infty} f(x)dx = 1,$$

implying that

$$\int_0^2 C(4x - 2x^2)dx = 1,$$

$$\left[C\left(4\frac{x^2}{2} - 2\frac{x^3}{3}\right)\right]_{x=0}^{x=2} = 1 \qquad \Rightarrow C\left(4\frac{4}{2} - 2\frac{8}{3}\right) - 0 = \frac{8}{3}C = 1.$$

Hence, $C = \frac{3}{8}$.

(b)

$$P(X > 1) = \int_1^{\infty} f(x)dx = \int_1^2 \frac{3}{8}(4x - 2x^2)dx = \frac{1}{2}.$$

(c)

$$F(x) = P(X \le x) = \int_{-\infty}^{x} \frac{3}{8}(4u - 2u^2)du = \int_0^x \frac{3}{8}(4u - 2u^2)du = \frac{3x^2}{4} - \frac{x^3}{4}.$$

The mean and variance can also be defined for a continuous RV. Integration replaces summation in the discrete definitions.

**Definition (Mean and Variance):** Suppose that $X$ is a continuous RV with PDF $f(x)$. The mean or expected value of $X$, denoted as $\mu$ or $E[X]$, is

$$\mu = E[X] = \int_{-\infty}^{\infty} xf(x)dx. \tag{12.3}$$

The variance of $X$, denoted as $V(X)$ or $\sigma^2$, is

$$\sigma^2 = V(X) = \int_{-\infty}^{\infty} (x - \mu)^2 f(x)dx = \int_{-\infty}^{\infty} x^2 f(x)dx - \mu^2. \tag{12.4}$$

The standard deviation of $X$ is $\sigma = \sqrt{\sigma^2}$.

**Definition (Expected Value):** The expected value of a function $h(X)$ of a continuous RV $X$ with PDF $f(x)$ is defined as

$$E[h(X)] = \int_{-\infty}^{\infty} h(x)f(x)dx. \tag{12.5}$$

Let us consider the example of the height of adult males in a given population. The height of an individual can be considered a continuous RV because it can take on any value within a certain range (e.g., from 150 cm to 200 cm), and there are infinitely many possible values between any two given heights.

- Let us say we have a population of 10,000 adult males, and we want to study their heights. We can represent the height of each individual as a continuous RV, denoted by the symbol "$X$." In this example, $X$ can take on any value between 150 cm and 200 cm, including decimal values.
- To analyze this continuous RV, we can look at its probability distribution, which describes how likely it is for $X$ to take on different values within the given range. In this case, the probability distribution of the height of adult males might resemble a bell-shaped curve, also known as a normal distribution.

- Using a normal distribution, we can calculate the probability of an individual having a height within a certain range. For example, we can calculate the probability of an adult male being between 170 cm and 180 cm tall. This calculation involves integrating the PDF of the normal distribution over the interval from 170 cm to 180 cm.
- Moreover, continuous RVs allow us to calculate various summary statistics. For instance, we can determine the mean height of the population, $\mu$, which represents the average height of adult males in the given population. We can also calculate the standard deviation, $\sigma$, which measures the variability or spread of heights in the population.

**Definition (MGF):** The MGF of the continuous RV $X$ with PDF $f(x)$ is defined as

$$M_X(t) = E[e^{tX}] = \int_{-\infty}^{\infty} e^{tx} f(x)dx. \tag{12.6}$$

***Example 12.2***

Let $X$ be a RV with PDF $f(x) = \frac{1}{\sqrt{2\pi}} e^{-\frac{x^2}{2}}, -\infty < x < \infty$. (We call such RV a standard normal RV.) Find the MGF of $X$.

***Solution***

By the definition of MGF, we have

$$\begin{aligned} M_X(t) &= E[e^{tX}] \\ &= \int_{-\infty}^{\infty} e^{tx} f(x)dx \\ &= \int_{-\infty}^{\infty} e^{tx} \frac{1}{\sqrt{2\pi}} e^{-\frac{x^2}{2}} dx \\ &= \frac{1}{\sqrt{2\pi}} \int_{-\infty}^{\infty} e^{\left(tx - \frac{x^2}{2}\right)} dx \\ &= \frac{1}{\sqrt{2\pi}} \int_{-\infty}^{\infty} e^{-\frac{1}{2}(x^2 - 2tx)} dx \\ &= \frac{1}{\sqrt{2\pi}} \int_{-\infty}^{\infty} e^{-\frac{1}{2}(x^2 - 2tx + t^2) + \frac{1}{2}t^2} dx \\ &= \frac{1}{\sqrt{2\pi}} \int_{-\infty}^{\infty} e^{-\frac{1}{2}(x-t)^2 + \frac{1}{2}t^2} dx \\ &= e^{\frac{1}{2}t^2} \frac{1}{\sqrt{2\pi}} \int_{-\infty}^{\infty} e^{-\frac{1}{2}(x-t)^2} dx = e^{\frac{1}{2}t^2}, \end{aligned}$$

where $\frac{1}{\sqrt{2\pi}} \int_{-\infty}^{\infty} e^{-\frac{1}{2}(x-t)^2} dx = 1$.

**Definition (Properties of the MGF):**

1. The MGF of $X$ is unique in the sense that, if two RVs $X$ and $Y$ have the same MGF ($M_X(t) = M_Y(t)$, for $t$ in an interval containing 0), then $X$ and $Y$ have the same distribution.
2. If $X$ and $Y$ are independent, then

$$M_{X+Y} = M_X(t) M_Y(t). \tag{12.7}$$

That is, the MGF of the sum of two independent RVs is the product of the MGFs of the individual RVs. The result can be extended to $n$ RVs.

3. Let $Y = aX + b$. Then

$$M_Y(t) = E\left[e^{(aX+b)t}\right] = E[e^{atX+bt}] = E[e^{bt} e^{atX}] = e^{bt} M_X(at). \tag{12.8}$$

## 12.2 Continuous Distributions

The following is a list of some commonly continuous probability distributions:

1. Uniform Distribution
2. Gamma Distribution
3. Exponential Distribution
4. Normal Distribution
5. Half Normal Distribution
6. Log Normal Distribution
7. Inverse Gaussian Distribution
8. Chi-Square Distribution
9. Inverse Chi-Square Distribution
10. F-Ratio Distribution
11. Student T Distribution
12. Noncentral Chi-Square Distribution
13. Noncentral Student T Distribution
14. Noncentral F-Ratio Distribution
15. Triangular Distribution
16. Beta Distribution
17. Cauchy Distribution
18. Chi Distribution
19. Gumbel Distribution
20. Inverse Gamma Distribution
21. Laplace Distribution
22. Levy Distribution
23. Logistic Distribution
24. Maxwell Distribution
25. Pareto Distribution
26. Rayleigh Distribution
27. Weibull Distribution

These are just a few examples of continuous random distributions. There are many other distributions with different properties and applications in statistics and probability theory. The following four distributions form the core set of commonly used continuous probability distributions. They are frequently encountered in various fields, including statistics, probability theory, engineering. In this section, we discuss these distributions in some detail.

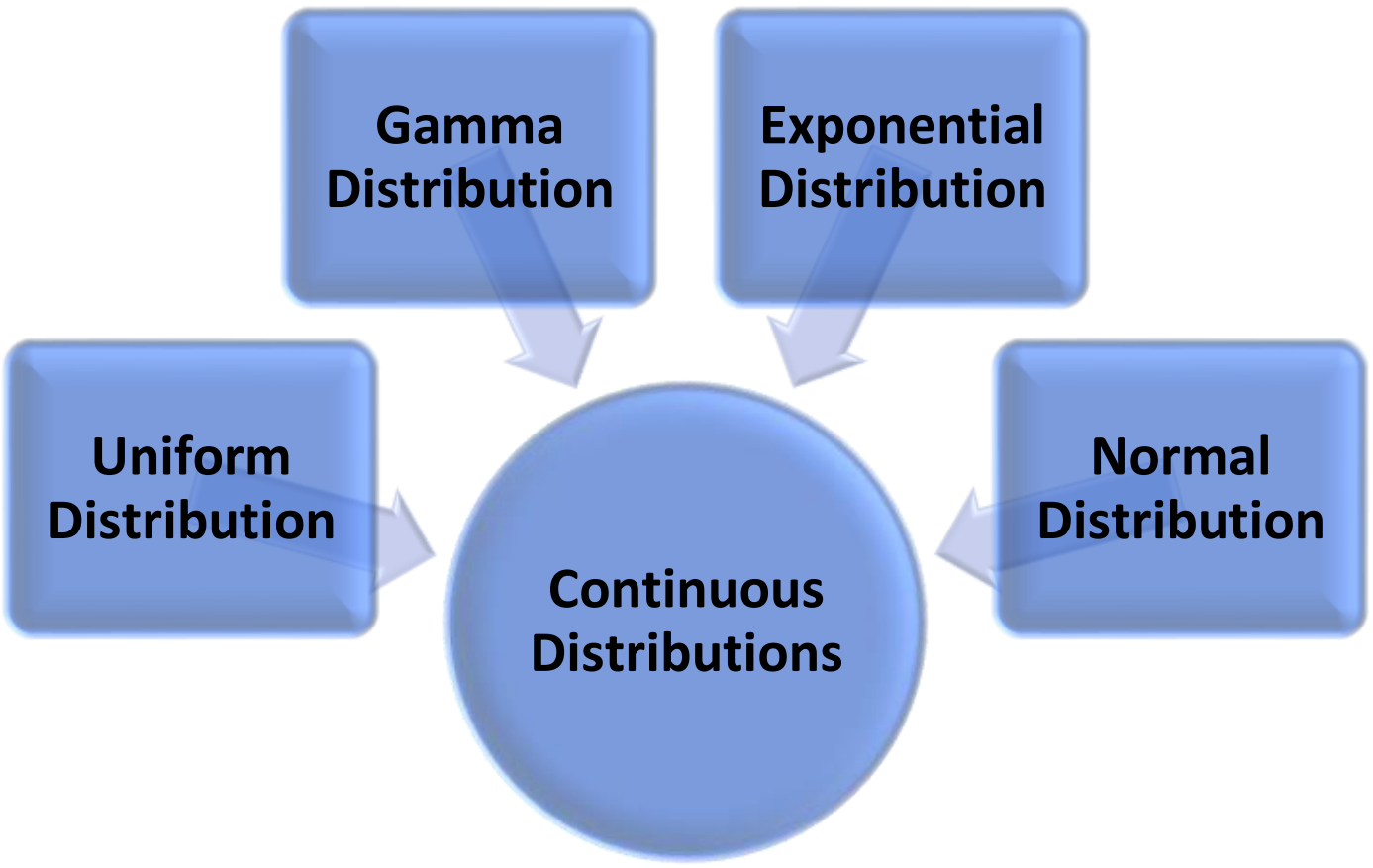

### 12.2.1 Continuous Uniform or Rectangular Distribution

One of the simplest continuous distributions in all of statistics is the continuous uniform distribution. This distribution is characterized by a density function that is "flat," and thus the probability is uniform in a closed interval, say $[a, b]$. It is a family of symmetric probability distributions. The distribution describes an experiment where there is an arbitrary number outcome that lies between certain bounds. The bounds are defined by the parameters, $a$ and $b$. The interval can be either closed (eg. $[a, b]$) or open (eg. $(a, b)$). Therefore, the distribution is often abbreviated $\mathrm{U}(a, b)$. The difference between the bounds defines the interval length; all intervals of the same length on the distribution's support are equally probable.

**Definition (PDF of Continuous Uniform Distribution):** A RV $X$ is said to have a continuous uniform distribution over an interval $[a, b]$ if its PDF is given by

$$f_X(x) = \begin{cases} \dfrac{1}{b-a}; & a \le x \le b, \\ 0; & \text{otherwise.} \end{cases} \tag{12.9}$$

**Definition (CDF of Uniform Continuous Distribution):** The CDF $F(x)$ of $\mathrm{U}(a, b)$ is given by

$$F_X(x) = \begin{cases} 0 \quad ; & x < a\,, \\ \dfrac{x-a}{b-a}; & a \le x \le b, \\ 1 \quad ; & x > b. \end{cases} \tag{12.10}$$

Since $F_X(x)$ is not continuous at $x = a$ and $x = b$, it is not differentiable at these points. Thus

$$\frac{d}{dx}F(x) = f(x) = \frac{1}{b-a} \neq 0,$$

exists everywhere except at the points $x = a$ and $x = b$ and consequently we get the PDF $f(x)$, (see Figure 12.3).

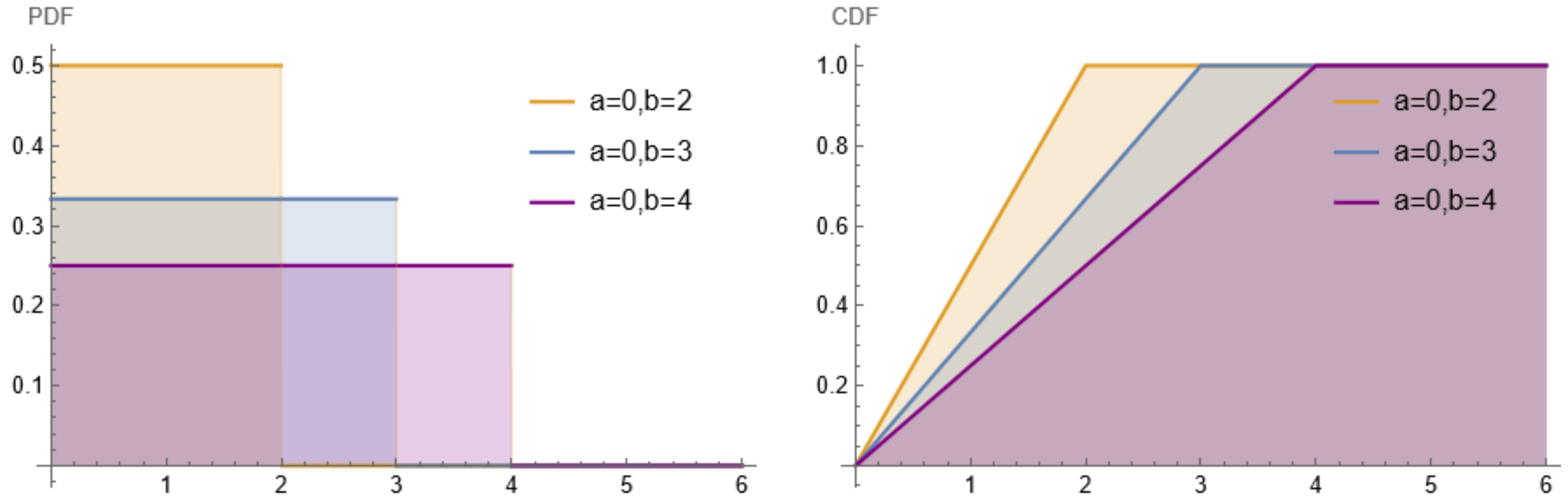

**Figure 12.3.** PMF (left) and CDF (right) of continuous uniform distribution. The curve of the continuous uniform distribution depends on the two parameters $a$ and $b$.

**Remarks:**

- The continuous uniform distribution requires specifying the interval within which the RV can take on values. This interval is usually defined by two parameters, a lower bound and an upper bound, which determine the range of possible outcomes. The width of the interval affects the spread and variability of the distribution.
- The continuous uniform distribution can be seen as the continuous counterpart of the discrete uniform distribution. While the discrete uniform distribution assigns equal probabilities to a finite number of discrete values, the continuous uniform distribution assigns equal probabilities to an infinite number of values within a continuous interval.
- The PDF of a continuous uniform distribution is constant within the interval and zero outside the interval. This reflects the uniformity of the distribution, as there are no peaks or valleys in the probability density.

- The distribution is known as rectangular distribution, since the curve $y = f(x)$ describes a rectangle over the $x$-axis and between the ordinates at $x = a$ and $x = b$.
- The CDF of a continuous uniform distribution is a linear function that increases uniformly from 0 to 1 as the value of the RV ranges from the lower bound to the upper bound. This property makes it easy to calculate probabilities and percentiles associated with specific values.

### The continuous uniform distribution has several applications in various fields:

- Random number generation:
  The continuous uniform distribution is often used to generate random numbers within a specified range. This is useful in simulations, computer graphics, cryptography, and various statistical applications.
- Monte Carlo simulations:
  Monte Carlo simulations involve using random numbers to estimate the outcome of complex systems or processes. The continuous uniform distribution is frequently used to generate random inputs for these simulations.

### How the continuous uniform distribution is applied in simulating various physical phenomena:

- Initial conditions:
  When simulating physical systems, it is often necessary to assign initial conditions to the variables involved. The continuous uniform distribution can be used to randomly generate initial values within the desired range. For example, in fluid dynamics simulations, the initial velocities or temperatures of particles can be assigned using the uniform distribution.
- Parameter variation:
  Simulations often involve varying parameters within certain ranges to study their effects on the system. The continuous uniform distribution can be used to randomly sample parameter values within their specified ranges. This allows for exploring a broad range of parameter configurations to understand the behavior of the system under different conditions.
- Monte Carlo simulations:
  Monte Carlo simulations involve repeated random sampling to estimate the behavior of a system or process. The continuous uniform distribution is often used to generate random inputs for these simulations. For example, in nuclear physics simulations, the position and direction of particles can be sampled using the uniform distribution to simulate the scattering or absorption processes.
- Sensor noise:
  In physical measurements and sensor simulations, noise is often introduced to account for uncertainties and measurement errors. The continuous uniform distribution can be employed to model sensor noise by adding random fluctuations to the measured values within a specified range. This helps in simulating realistic sensor data and assessing the performance of signal processing algorithms.
- Boundary conditions:
  Simulating physical systems often requires specifying boundary conditions. The continuous uniform distribution can be used to randomly assign boundary values within the desired range. For instance, in heat transfer simulations, the temperature at the boundary of an object can be assigned using the uniform distribution to simulate different thermal boundary conditions.
- Random perturbations:
  In dynamic systems, random perturbations are often used to model environmental or external influences. The continuous uniform distribution can be used to generate random perturbations within specified ranges. This allows for simulating the effects of random disturbances on the system's behavior over time.

**Theorem 12.1:** If $X$ is a continuous uniform RV over $a \le x \le b$,

$$E[X] = \frac{a+b}{2}, \tag{12.11}$$

$$V(X) = \frac{(b-a)^2}{12}, \tag{12.12}$$

$$M_X(t) = \frac{e^{bt} - e^{at}}{t(b-a)}. \tag{12.13}$$

**Proof:**

Mean,

$$\begin{aligned} E[X] &= \int_a^b x \frac{1}{b-a} dx \\ &= \frac{1}{b-a}\left[\frac{x^2}{2}\right]_a^b \\ &= \frac{1}{b-a}\frac{b^2-a^2}{2} = \frac{a+b}{2}. \end{aligned}$$

In other words, the expected value of a uniform $[a,b]$ RV is equal to the midpoint of the interval $[a,b]$.

Variance,

$$V(X) = E[X^2] - (E[X])^2,$$

$$\begin{aligned} E[X^2] &= \int_a^b x^2 \frac{1}{b-a} dx \\ &= \frac{1}{b-a}\left[\frac{x^3}{3}\right]_a^b \\ &= \frac{1}{b-a}\frac{b^3-a^3}{3} \\ &= \frac{1}{b-a}\frac{(b-a)(a^2+ab+b^2)}{3} \\ &= \frac{a^2+ab+b^2}{3}. \end{aligned}$$

Therefore,

$$V(X) = \frac{a^2+ab+b^2}{3} - \left(\frac{a+b}{2}\right)^2 = \frac{(b-a)^2}{12}.$$

MGF,

$$\begin{aligned} M_X(t) &= E[e^{tX}] \\ &= \int_a^b e^{tx} \frac{1}{b-a} dx \\ &= \frac{1}{b-a}\left[\frac{e^{tx}}{t}\right]_a^b \\ &= \frac{e^{bt}-e^{at}}{t(b-a)}, \qquad t \neq 0. \end{aligned}$$

■

***Example 12.3***

It is possible to consider the melting point, $X$, of a specific solid to be a continuous RV that is uniformly distributed between 120 and 145 C. Find the probability that such a solid will melt between 125 and 130 C.

***Solution***

The PDF is given by:

$$f_X(x) = \begin{cases} \frac{1}{25}; & 120 \leq x \leq 145, \\ 0; & \text{otherwise.} \end{cases}$$

Hence,

$$P(125 \leq X \leq 130) = \int_{125}^{130} \frac{1}{25} dx = \frac{1}{25}[x]_{125}^{130} = \frac{5}{25} = 0.2.$$

Thus, there is a 20% chance of this solid melting between 125 and 130 C.

```
NProbability[125<x<130,x \[Distributed] UniformDistribution[{120,145}]]
 0.2
```

***Example 12.4***

If $X$ is uniformly distributed over the interval [0,12], compute the probability that
(a) $3 < X < 8$,
(b) $1 < X < 7$,
(c) $X < 10$,
(d) $X > 5$.

***Solution***

The PDF is given by:

$$f_X(x) = \begin{cases} \frac{1}{12}; & 0 \leq x \leq 12, \\ 0; & \text{otherwise.} \end{cases}$$

```
NProbability[3<x<8,x \[Distributed] UniformDistribution[{0,12}]]
0.416667
NProbability[1<x<7,x \[Distributed] UniformDistribution[{0,12}]]
0.5
NProbability[x<10,x \[Distributed] UniformDistribution[{0,12}]]
0.833333
NProbability[5<x,x \[Distributed] UniformDistribution[{0,12}]]
0.583333
```

***Example 12.5***

If $X$ is uniformly distributed over the interval [0,10], compute the probability that
(a) $2 < X < 5$ and $X > 3$,
(b) $|X - 4| < 2$,

***Solution***

The PDF is given by:

$$f_X(x) = \begin{cases} \frac{1}{10}; & 0 \leq x \leq 10, \\ 0; & \text{otherwise.} \end{cases}$$

```
(*Define the continuous random variable*)
ux=UniformDistribution[{0,10}];
(*Calculate the probabilities of complex events*)
prob1=N[Probability[2<x<5&&x>3,x\[Distributed]ux]]
prob2=N[Probability[Abs[x-4]<2,x\[Distributed]ux]]
(*Plot the results*)
```

```
Plot[
 {
  Piecewise[{{0.1,2<x<5&&x>3}},0],
  Piecewise[{{0.1,Abs[x-4]<2}},0],
  Piecewise[{{0.1,0<x<10}},0]
  },
 {x,0,10},
 PlotStyle->{Directive[Blue,Thickness[0.025]],Directive[Red,Thickness[0.015]],Purple},
 Filling->Axis,
 PlotLegends->{"Event 1","Event 2","PDF"},
 ImageSize->250
 ]
 0.2
 0.4
```

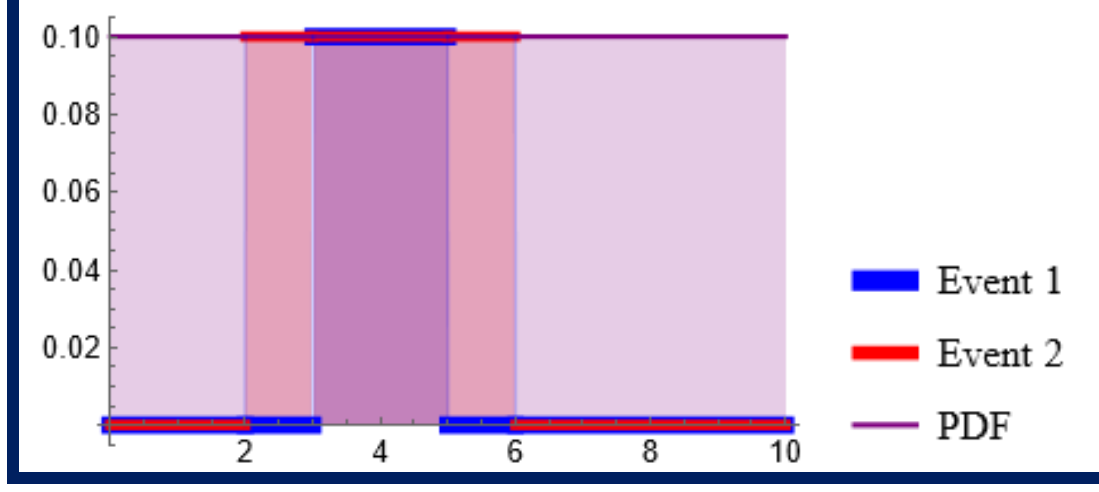

### 12.2.2 Exponential Distribution

The exponential distribution is a probability distribution that models the time it takes for an event to occur in a continuous and memoryless manner. In the exponential distribution, the RV represents the time until the next event occurs. In other words, it is used to model waiting times or inter-arrival times between events. In this context, an "event" refers to a specific occurrence or outcome of interest. For example, the arrival of a customer at a service counter, the failure of a machine, or the time until the next phone call in a call center can all be considered events. It is commonly used in various fields, including reliability engineering, queueing theory, and survival analysis. The Exponential Distribution is characterized by a single parameter, often denoted as $\lambda$, which represents the rate parameter. The rate parameter determines the average rate at which events occur. The higher the value of $\lambda$, the more frequent the events are. The exponential distribution may be viewed as a continuous counterpart of the geometric distribution, which describes the number of Bernoulli trials necessary for a discrete process to change state. In contrast, the exponential distribution describes the time for a continuous process to change state.

**Definition (Exponential Distribution):** A RV $X$ is said to have an exponential distribution with parameter $\lambda$, if its PDF is given by

$$f_X(x) = \begin{cases} \lambda e^{-\lambda x}; & x \geq 0, \\ 0; & \text{otherwise}, \end{cases} \tag{12.14}$$

where $\lambda > 0$. If a RV $X$ has this distribution, we write $X \sim \text{Exp}(\lambda)$. (See Figure 12.4)

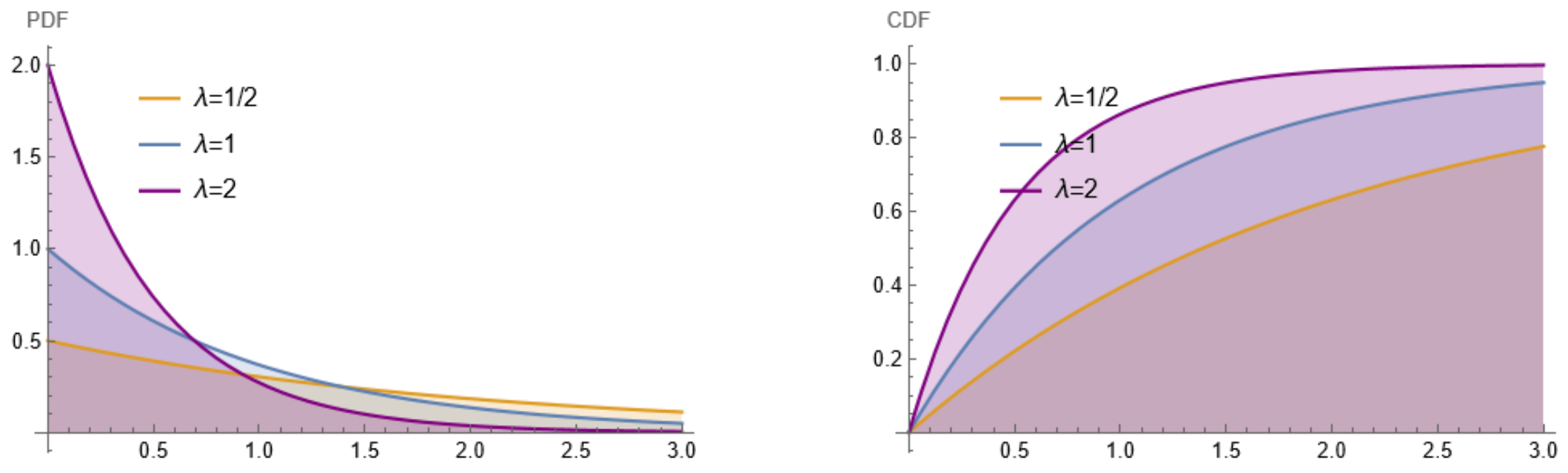

**Figure 12.4.** PMF (left) and CDF (right) of continuous exponential distribution. The curve of the continuous exponential distribution depends on the one parameter $\lambda$.

The PDF $f_X(x)$ assumes the value 0 for negative values of $x$, and then for positive values, it starts off at a value equal to $\lambda$. This is because if we put $x = 0$ in the PDF expression, you get $\lambda$ times $e^0$, which leaves us just with $\lambda$. So, it starts off with $\lambda$, and then it decays at the rate of $\lambda$. Notice that when $\lambda$ is small, the initial value of the PDF is small. But then the decay rate is also small, so that the PDF extends over a large range of $x$'s.

**Definition (CDF of Exponential Distribution):** The CDF is given by

$$F_X(x) = \int_0^x \lambda e^{-\lambda x} dx = \begin{cases} 1 - e^{-\lambda x}; & x \geq 0, \\ 0; & \text{otherwise.} \end{cases} \quad (12.15)$$

In the context of estimating the probability of the expected waiting time until the next event, let us consider an example. Suppose we are interested in estimating the waiting time until the next customer arrives at a store. We assume that customers arrive according to a Poisson process, where events occur continuously and independently at a constant average rate. Exponential distribution is used to model the time between consecutive customer arrivals. If we want to find the probability that the waiting time until the next event is less than or equal to a specific time $t$, we calculate the CDF of the exponential distribution. The CDF gives us the probability that the waiting time is less than or equal to $x$. Therefore, we can estimate the probability of the expected waiting time until the next event by plugging in the appropriate value of $x$ into the CDF equation.

***Example 12.6***

If the rate parameter $\lambda = 1/10$. What is the probability that the waiting time until the next customer arrival is less than or equal to 5 minutes.

***Solution***

We can calculate:

$$F(5) = 1 - e^{(-5/10)} \approx 0.393.$$

This means that there is approximately a 39.3% chance that the next customer will arrive within 5 minutes.

```
NProbability[x<=5,x \[Distributed] ExponentialDistribution[0.1]]
N[CDF[ ExponentialDistribution[0.1],5]]
 0.393469
 0.393469
```

***Example 12.7***

Suppose you are studying the time between earthquakes occurring in a particular region. Historical data suggests that earthquakes occur randomly and independently with an average interarrival time of 20 years. What is the probability that the waiting time until the next earthquake occurs between 10 to 25 years.

***Solution***

In this case, you might choose to model the time between earthquakes using an exponential distribution. Since the average interarrival time is 20 years, the rate parameter would be $\lambda = 1/20 = 0.05$.

To find the probability that the waiting time until the next earthquake occurs between 10 to 25 years, we need to calculate the cumulative probability within that range.

$$P(10 \leq X \leq 25) = F(25) - F(10) = 0.320026.$$

```
dist=ExponentialDistribution[1/20];
NProbability[10<=x<=25,x \[Distributed] dist]
N[CDF[ dist,25]-CDF[ dist,10]]
 0.320026
 0.320026
```

**Definition (Memoryless):** The exponential distribution is memoryless, which means that the time until the next event occurs does not depend on how much time has already elapsed.

To understand this property intuitively, consider an example. Let us say $X$ represents the time until a light bulb fails, following an exponential distribution with a mean lifetime of 100 hours. If the light bulb has already been working for 50 hours and has not failed yet, the memoryless property tells us that the probability of the light bulb failing within the next 20 hours is the same as the probability of a new light bulb failing within 20 hours, regardless of the time that has already passed. This is because the exponential distribution does not "remember" the past; the remaining lifetime of the light bulb is still governed by the same exponential distribution with a mean of 100 hours.

**Proof:**

Let $X$ be exponentially distributed with parameter $\lambda$. Suppose we know $X > s$. What is the probability that $X$ is also greater than some value $s + t$? We can demonstrate the memoryless property as follows:

$$\begin{aligned}
P(X > s + t | X > s) &= \frac{[P(X > s + t \text{ and } X > s)]}{P(X > s)} \\
&= \frac{[P(X > s + t)]}{P(X > s)} && [\text{since } X > s + t \text{ implies } X > s] \\
&= \frac{1 - P(X \leq s + t)}{1 - P(X \leq s)} && [\text{using the complementary probability}] \\
&= \frac{1 - \left[1 - e^{-\lambda(s+t)}\right]}{1 - [1 - e^{-\lambda s}]} && [\text{using CDF}] \\
&= \frac{e^{-\lambda(s+t)}}{e^{-\lambda s}} \\
&= e^{-\lambda t} \\
&= P(X > t).
\end{aligned}$$

Hence, we have shown that $P(X > s + t|X > s) = P(X > t)$. It turns out that the conditional probability does not depend on $s$. The probability of an exponential RV exceeding the value $s + t$ given $s$ is the same as the variable originally exceeding that value $t$, regardless of $s$, which confirms the memoryless property of the exponential distribution.

The exponential distribution is memoryless because the past has no bearing on its future behavior. Every instant is like the beginning of a new random period, which has the same distribution regardless of how much time has already elapsed.

■

The exponential distribution has some key properties:

- The exponential distribution is the only memoryless continuous RV.
- Lack of a maximum value:
  The exponential distribution has an unbounded support, meaning that the RV can take on any positive value. There is no maximum value for the time until the next event occurs.
- Exponential decay:
  The exponential distribution has a decaying exponential shape, with a higher probability density at shorter times and a lower probability density at longer times. This property reflects the decreasing likelihood of an event occurring as time goes on.

**The exponential distribution has several applications in various fields:**

- In real-world scenarios:

The assumption of a constant rate (or probability per unit time) is rarely satisfied. For example, the rate of incoming phone calls differs according to the time of day. But if we focus on a time interval during which the rate is roughly constant, such as from 2 to 4 p.m. during workdays, the exponential distribution can be used as a good approximate model for the time until the next phone call arrives. Similar caveats apply to the following examples which yield approximately exponentially distributed variables:

- The time until a radioactive particle decays, or the time between clicks of a Geiger counter
- The time it takes before your next telephone call
- The time until default (on payment to company debt holders) in reduced-form credit risk modeling

- Reliability engineering:
  The exponential distribution is commonly used to model the time to failure of components or systems in reliability engineering. It helps in estimating the reliability and mean time between failures of devices.
- Queueing theory:
  In queueing theory, exponential distribution is used to model the interarrival times of customers or entities in a queuing system. It helps analyze waiting times and queue lengths in various scenarios, such as telecommunications networks or service centers.
- Radioactive decay:
  The exponential distribution is used to model the decay of radioactive isotopes. The time it takes for a certain fraction of the radioactive material to decay follows an exponential distribution.
- Actuarial science:
  Actuaries use the exponential distribution to model the time until an event, such as death or insurance claim, occurs. It aids in determining life expectancy, pricing insurance policies, and assessing the financial risk associated with various events.

The exponential distribution is however not appropriate to model the overall lifetime of organisms or technical devices, because the "failure rates" here are not constant: more failures occur for very young and for very old systems.

**Theorem 12.2:** If $X$ is exponential RV, then

$$E[X] = \frac{1}{\lambda}, \tag{12.16}$$

$$V(X) = \frac{1}{\lambda^2}, \tag{12.17}$$

$$M_X(t) = \frac{\lambda}{\lambda - t}. \tag{12.18}$$

The mean of the exponential distribution is equal to $1/\lambda$, this means that as the rate parameter increases, the distribution becomes more concentrated around zero.

**Proof:**

Mean,

$$\begin{aligned} E[X] &= \int_0^\infty x\lambda e^{-\lambda x} dx \\ &= \lambda \int_0^\infty x e^{-\lambda x} dx \\ &= \lambda \frac{\Gamma(2)}{\lambda^2} \\ &= \frac{1}{\lambda}. \end{aligned}$$

Variance,

$$V(X) = E[X^2] - (E[X])^2$$

$$\begin{aligned} E[X^2] &= \int_0^\infty x^2 \lambda e^{-\lambda x} dx \\ &= \lambda \frac{\Gamma(3)}{\lambda^3} \\ &= \frac{2}{\lambda^2}. \end{aligned}$$

Therefore,

$$V(X) = \frac{2}{\lambda^2} - \left(\frac{1}{\lambda}\right)^2 = \frac{1}{\lambda^2}.$$

MGF,

$$\begin{aligned} M_X(t) &= E[e^{tX}] \\ &= \int_0^\infty e^{tx} \lambda e^{-\lambda x} dx \\ &= \left(1 - \frac{t}{\lambda}\right)^{-1} \\ &= \left(\frac{\lambda}{\lambda - t}\right). \end{aligned}$$

■

**Theorem 12.3:** If $X_1$, $X_2$, ..., $X_n$ are independent exponential RVs having respective parameters $\lambda_1$, $\lambda_1$, $\lambda_2$..., $\lambda_n$, then $\min(X_1, X_2, \dots, X_n)$ is exponential with parameter $\sum_{i=1}^{n} \lambda_i$.

**Proof:**

Since the smallest value of a set of numbers is greater than $x$ if and only if all values are greater than $x$, we have

$$\begin{aligned} P(\min(X_1, X_2, \dots, X_n) > x) &= P(X_1 > x, X_2 > x, \dots, X_n > x) \\ &= \prod_{i=1}^{n} P(X_i > x) \\ &= \prod_{i=1}^{n} e^{\lambda_i x} \\ &= e^{\sum_{i=1}^{n} \lambda_i x}. \end{aligned}$$

■

**Theorem 12.4:** $cX$ is exponential RV with parameter $\lambda/c$ when $X$ is exponential RV with parameter $\lambda$, and $c > 0$.

**Proof:**

This follows since

$$\begin{aligned} P(\mathrm{c}X \le x) &= P\left(X \le \frac{x}{\mathrm{c}}\right) \\ &= 1 - e^{-\frac{\lambda x}{c}}. \end{aligned}$$

■

### 12.2.3 Gamma Distribution

The Gamma distribution is a continuous probability distribution that is widely used in various fields, including statistics, engineering, and physics. It is a versatile distribution that allows us to model a wide range of real-world phenomena, such as waiting times, survival analysis, and the distribution of RVs that are always positive. The Gamma distribution was first introduced by mathematician Leonard Euler in the 18th century while studying the problem of waiting times. It was later developed and extensively studied by Karl Pearson and other statisticians in the early 20th century, solidifying its importance in statistical theory and practice. There are two equivalent parameterizations in common use:

- With a shape parameter $k$ and a scale parameter $\theta$.
- With a shape parameter $\alpha = k$ and an inverse scale parameter $\beta = 1/\theta$, called a rate parameter.

In each of these forms, both parameters are positive real numbers. These parameters allow us to control the shape, location, and spread of the distribution, making it a flexible tool for modeling various types of data. Moreover, the gamma distribution encompasses several other well-known distributions as special cases, such as the exponential distribution, chi-square distribution, and Erlang distribution.

The exponential distribution describes the time between events in a Poisson process, where events occur at a constant average rate. The exponential distribution is characterized by a single parameter, the rate parameter $\beta$, which represents the average rate of event occurrence. The gamma distribution is characterized by two parameters: $\alpha$ and $\beta$. The shape parameter $\alpha$ determines the number of sub-events or events that need to occur in sequence before a major event takes place. The parameter $\beta$ controls the rate at which these sub-events occur. If we consider a sequence of events where the waiting time for each event follows an exponential distribution with rate $\beta$, then the waiting time for the $n$-th event follows a gamma distribution with shape parameter $\alpha = n$. This means that as each sub-event occurs, the waiting time for the major event increases, and it follows a gamma distribution. The shape parameter $\alpha$ is equal to the number of sub-events or steps required to reach the major event. The construction of the gamma distribution using the exponential distribution allows it to model various phenomena where multiple sub-events, each following an exponential distribution, must happen in sequence for a major event to occur. Examples are the waiting time of cell-division events, the number of compensatory mutations for a given mutation, and the waiting time until a repair is necessary for a hydraulic system. In these cases, each sub-event contributes to the overall waiting time, and the gamma distribution can capture the variability and cumulative effect of these sub-events.

The construction of the gamma distribution using the exponential distribution where multiple sub-events, each following an exponential distribution, must happen in sequence for a major event to occur.

**Proof:**

Consider the distribution function $D(x)$ of waiting times until the $h$th Poisson event given a Poisson distribution with a rate of change $\lambda$,

$$\begin{aligned} D(x) &= P(X \leq x) \\ &= 1 - P(X > x) \\ &= 1 - P(0,1,2,\dots,(h-1) \text{ events in } [0,x]) \\ &= 1 - \sum_{k=0}^{h-1} \frac{(\lambda x)^k e^{-\lambda x}}{k!} \\ &= 1 - e^{-\lambda x} \sum_{k=0}^{h-1} \frac{(\lambda x)^k}{k!}. \end{aligned}$$

for $x$ in $[0,\infty)$. The corresponding probability function $P(x)$ of waiting times until the $h$th Poisson event is then obtained by differentiating $D(x)$,

$$\begin{aligned}
P(x) &= D'(x)\\
&= \frac{d}{dx}\left(1-\sum_{k=0}^{h-1}\frac{(\lambda x)^k e^{-\lambda x}}{k!}\right)\\
&= \frac{d}{dx}\left(1-e^{-\lambda x}-\sum_{k=1}^{h-1}\frac{(\lambda x)^k e^{-\lambda x}}{k!}\right)\\
&= -\frac{d}{dx}e^{-\lambda x}-\sum_{k=1}^{h-1}\frac{d}{dx}\frac{(\lambda x)^k e^{-\lambda x}}{k!}\\
&= \lambda e^{-\lambda x}-\sum_{k=1}^{h-1}\frac{1}{k!}\frac{d}{dx}\left[(\lambda x)^k e^{-\lambda x}\right]\\
&= \lambda e^{-\lambda x}-\sum_{k=1}^{h-1}\frac{1}{k!}\left[\left(\frac{d}{dx}(\lambda x)^k\right)e^{-\lambda x}+(\lambda x)^k\frac{d}{dx}e^{-\lambda x}\right]\\
&= \lambda e^{-\lambda x}-\sum_{k=1}^{h-1}\frac{1}{k!}\left[(k(\lambda x)^{k-1}\lambda)e^{-\lambda x}+(\lambda x)^k\left(-\lambda e^{-\lambda x}\right)\right]\\
&= \lambda e^{-\lambda x}-\lambda e^{-\lambda x}\sum_{k=1}^{h-1}\frac{1}{k!}\left[(k(\lambda x)^{k-1})-(\lambda x)^k\right]\\
&= \lambda e^{-\lambda x}\left(1-\sum_{k=1}^{h-1}\left[\left(\frac{k(\lambda x)^{k-1}}{k!}\right)-\frac{(\lambda x)^k}{k!}\right]\right)\\
&= \lambda e^{-\lambda x}\left(1-\sum_{k=1}^{h-1}\left[\left(\frac{(\lambda x)^{k-1}}{(k-1)!}\right)-\frac{(\lambda x)^k}{k!}\right]\right)\\
&= \lambda e^{-\lambda x}\left(1-\left[1-\frac{(\lambda x)^{h-1}}{(h-1)!}\right]\right)\\
&= \lambda e^{-\lambda x}\frac{(\lambda x)^{h-1}}{(h-1)!}\\
&= \frac{\lambda(\lambda x)^{h-1}}{(h-1)!}e^{-\lambda x}.
\end{aligned}$$

Now let $k=h$ (not necessarily an integer) and define $\theta$ is a reciprocal of the event rate $\theta=1/\lambda$, which is the mean wait time (the average time between event arrivals). Then the above equation can be written

$$\begin{aligned}
P(x) &= \frac{\lambda(\lambda x)^{k-1}}{(k-1)!}e^{-\frac{x}{\theta}}\\
&= \frac{\lambda\lambda^{k-1}x^{k-1}}{(k-1)!}e^{-\frac{x}{\theta}}\\
&= \frac{\lambda^k x^{k-1}}{\Gamma(k)}e^{-\frac{x}{\theta}}\\
&= \frac{\theta^{-k}}{\Gamma(k)}x^{k-1}e^{-\frac{x}{\theta}},
\end{aligned}$$

for $x$ in $[0,\infty)$, where $(k-1)!=\Gamma(k)$. This is the PDF for the gamma distribution.

■

**Definition (PDF of Gamma Distribution):** A RV $X$ is said to have a gamma distribution with parameters $k$ and $\theta$, if its PDF is given by

$$f_X(x) = f_X(x; k, \theta) = \begin{cases} \dfrac{\theta^{-k}}{\Gamma(k)} x^{k-1} e^{-\frac{x}{\theta}}; & x > 0, \\ 0; & \text{otherwise}, \end{cases} \quad (12.19.1)$$

where $k > 0$ and $\theta > 0$, and $\Gamma(k)$ is the gamma function. (See Figure 12.5).
A RV $X$ is said to have a gamma distribution with parameters $\alpha$ and $\beta$, if its PDF is given by

$$f_X(x) = f_X(x; \alpha, \beta) = \begin{cases} \dfrac{\beta^{\alpha}}{\Gamma(\alpha)} x^{\alpha-1} e^{-\beta x}; & x > 0, \\ 0; & \text{otherwise}, \end{cases} \quad (12.19.2)$$

where $\alpha > 0$ and $\beta > 0$, and $\Gamma(\alpha)$ is the gamma function.

**Definition (CDF of Gamma Distribution):** The CDF is the regularized gamma function:

$$F_X(x) = F_X(x; k, \theta) = \int_0^x f(u; k, \theta)du = \frac{\gamma(k, \frac{x}{\theta})}{\Gamma(\text{k})}, \quad (12.20.1)$$

$$F_X(x) = F_X(x; \alpha, \beta) = \int_0^x f(u; \alpha, \beta)du = \frac{\gamma(\alpha, \beta x)}{\Gamma(\alpha)}, \quad (12.20.2)$$

$\gamma(k, \frac{x}{\theta})$ or $\gamma(\alpha, \beta x)$ is the lower incomplete gamma function.

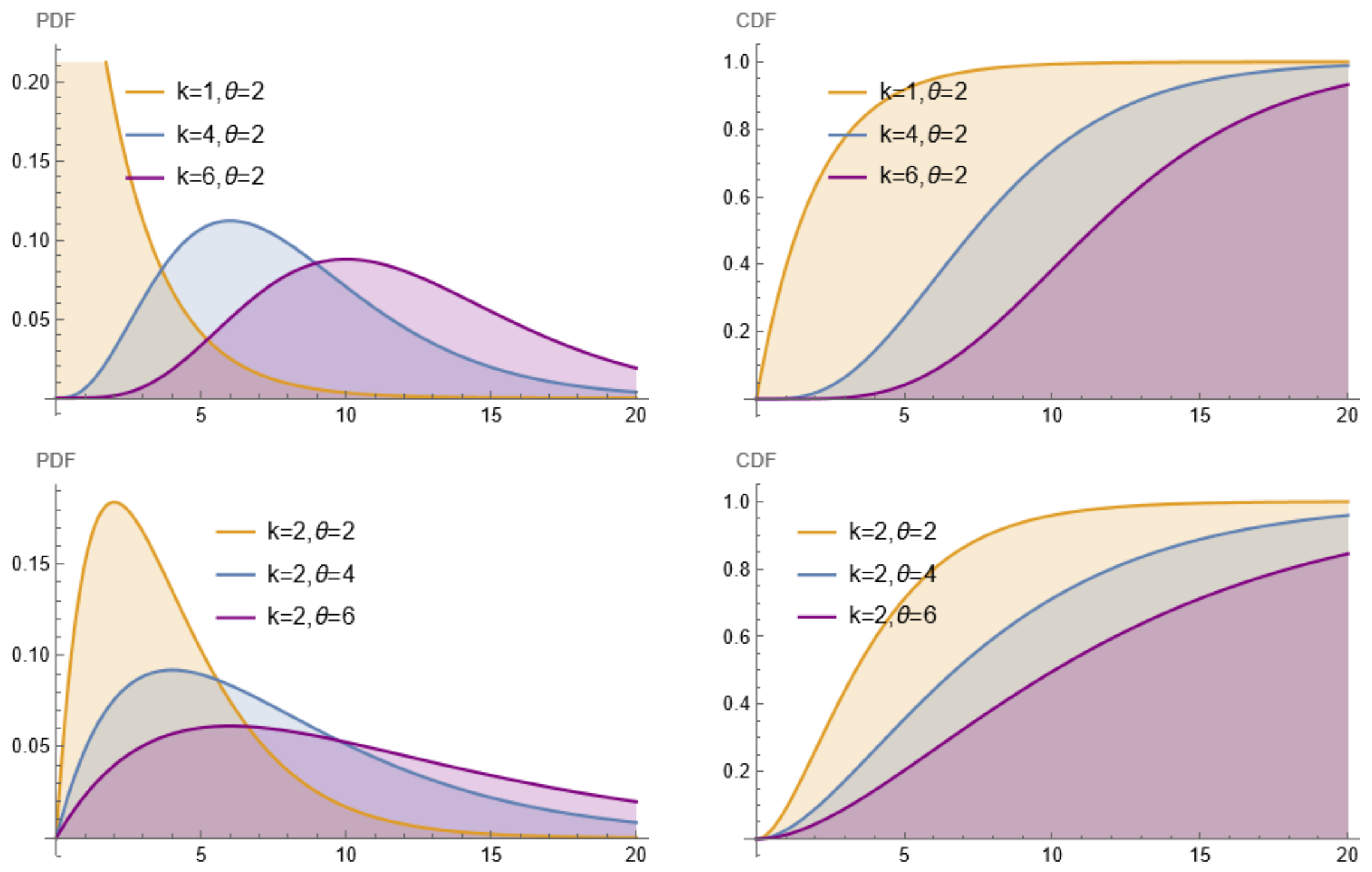

**Figure 12.5.** PMF (left) and CDF (right) of continuous gamma distribution. The curve of the continuous gamma distribution depends on the two parameters $k$ and $\theta$.

**Remark:**

The gamma distribution represents the sum of $k$ independent exponential RVs with the same rate parameter $\theta$. Therefore, when $k = 1$ in the gamma distribution, it reduces to the exponential distribution. Mathematically, when $k = 1$ (12.19.1), or $\alpha = 1$ (12.19.2), the PDF of the gamma distribution simplifies to: $f_X(x) = e^{-\frac{x}{\theta}}/\theta$ or $f_X(x) =$ $\beta e^{-\beta x}$, respectively. Comparing this with the PDF of the exponential distribution $f_X(x) = \lambda e^{-\lambda x}$, you can see that they are equivalent.

**Theorem 12.5:** If $X \sim \text{Gamma}(k, \theta)$, then,

$$M_X(t) = (1 - \theta t)^{-k} = \left(1 - \frac{t}{\beta}\right)^{-\alpha} \text{ for } t < \beta = \frac{1}{\theta}, \tag{12.21}$$

$$E[X] = k\theta = \frac{\alpha}{\beta}, \tag{12.22}$$

$$V(X) = k\theta^2 = \frac{\alpha}{\beta^2}. \tag{12.23}$$

**Proof:**

MGF,

$$\begin{aligned} M_X(t) &= E[e^{tX}] \\ &= \int_0^\infty e^{tx} \frac{\theta^{-k}}{\Gamma(k)} x^{k-1} e^{-\frac{x}{\theta}} dx \\ &= \int_0^\infty \frac{\theta^{-k}}{\Gamma(k)} x^{k-1} e^{tx - \frac{x}{\theta}} dx \\ &= \int_0^\infty \frac{\theta^{-k}}{\Gamma(k)} x^{k-1} e^{-(1-\theta t)\frac{x}{\theta}} dx. \end{aligned}$$

Let $x = \frac{y\theta}{1-\theta t}$, $y = (1 - \theta t)\frac{x}{\theta}$ and $dy = \frac{1-\theta t}{\theta} dx$, so

$$\begin{aligned} M_X(t) &= \int_0^\infty \frac{\theta^{-k}}{\Gamma(k)} \left(\frac{y\theta}{1-\theta t}\right)^{k-1} e^{-y} \frac{\theta}{1-\theta t} dy \\ &= \frac{\theta}{1-\theta t} \frac{\theta^{-k}}{\Gamma(k)} \left(\frac{\theta}{1-\theta t}\right)^{k-1} \int_0^\infty y^{k-1} e^{-y} dy \\ &= \frac{1}{(1-\theta t)^k} \frac{1}{\Gamma(k)} \int_0^\infty y^{k-1} e^{-y} dy \\ &= \frac{1}{(1-\theta t)^k}. \end{aligned}$$

Mean,

$$\begin{aligned} E[X] &= \frac{d}{dt} M_X(t) \bigg|_{t=0} \\ &= [-k(1-\theta t)^{-k-1}(-\theta)]|_{t=0} = k\theta. \end{aligned}$$

Variance,

$$\begin{aligned} E[X^2] &= \frac{d^2}{dt^2} M_X(t) \bigg|_{t=0} \\ &= \frac{d}{dt}[k\theta(1-\theta t)^{-k-1}] \bigg|_{t=0} \\ &= [(-k-1)k\theta(1-\theta t)^{-k-2}(-\theta)]|_{t=0} \\ &= [(k+1)k\theta^2]. \end{aligned}$$

Therefore,

$$V(X) = E[X^2] - (E[X])^2 = (k+1)k\theta^2 - (k\theta)^2 = k\theta^2.$$

■

**Theorem 12.6:** If $X_i$ has a Gamma$(k_i, \theta)$ distribution for $i = 1, 2, \ldots, n$ (i.e., all distributions have the same scale parameter $\theta$), then

$$\sum_{i=1}^{n} X_i \sim \text{Gamma}\left(\sum_{i=1}^{n} k_i, \theta\right), \tag{12.24}$$

provided all $X_i$ are independent.

**Theorem 12.7:** If $X \sim \text{Gamma}(k, \theta)$, then, for any $c > 0$,

$$cX \sim \text{Gamma}(k, c\theta), \tag{12.25}$$

by MGFs, or equivalently, if $X \sim \text{Gamma}(\alpha, \beta)$ (shape-rate parameterization)

$$cX \sim \text{Gamma}\left(\alpha, \frac{\beta}{c}\right). \tag{12.26}$$

***Example 12.8***

Suppose that telephone calls arriving at a particular switchboard follow a Poisson process with an average of 5 calls coming per minute. What is the probability that up to a minute will elapse by the time 2 calls have come in to the switchboard?

***Solution***

The Poisson process applies, with time until 2 Poisson events following a gamma distribution with $\theta = 1/5$ and $k = 2$. Denote by $X$ the time in minutes that transpires before 2 calls come. The required probability is given by

```
NProbability[0<=x<=1,x \[Distributed]GammaDistribution[2,1/5]]
 0.959572
```

### 12.2.4 Normal Distribution

Normal distribution plays a pivotal role in most of the statistical techniques used in applied statistics. The main reason for this is the central limit theorem, according to which normal distribution is found to be the approximation of most of the RVs. We may discuss it in detail later.

It was first introduced by a French mathematician, Abraham De-Moivre (1667-1754). He obtained it while working on certain problems in the games of chance. Later, two mathematical astronomers Pierre Laplace (1749-1827) and Karl Gauss (1777-1855) developed this distribution independently. They found that it can be used to model errors (the deviation of the observed value from the true value). Hence, this distribution is also known as Gaussian distribution and Laplace's distribution. But it is most commonly known as the normal distribution.

**Definition (Normal Distribution):** A RV $X$ is said to follow a normal distribution with parameters $\mu$ and $\sigma^2$ if its PDF is given by

$$f_X(x) = \frac{1}{\sigma\sqrt{2\pi}} e^{-\frac{(x-\mu)^2}{2\sigma^2}}, \qquad -\infty < x < \infty, \tag{12.27}$$

where $-\infty < \mu < \infty$ and $\sigma > 0$. In this case, we can write $X \sim N(\mu, \sigma^2)$ or $X \sim N(\mu, \sigma)$.

For a normal distribution with density $f$, mean $\mu$ and deviation $\sigma$, the CDF is given by,

$$F_X(x) = \frac{1}{2}\left(1 + \text{Erf}\left(\frac{x-\mu}{\sigma\sqrt{2}}\right)\right), \tag{12.28}$$

where, the error function Erf$(x)$ is

$$\text{Erf}(x) = \frac{2}{\sqrt{\pi}} \int_0^x e^{-t^2} dt. \tag{12.29}$$

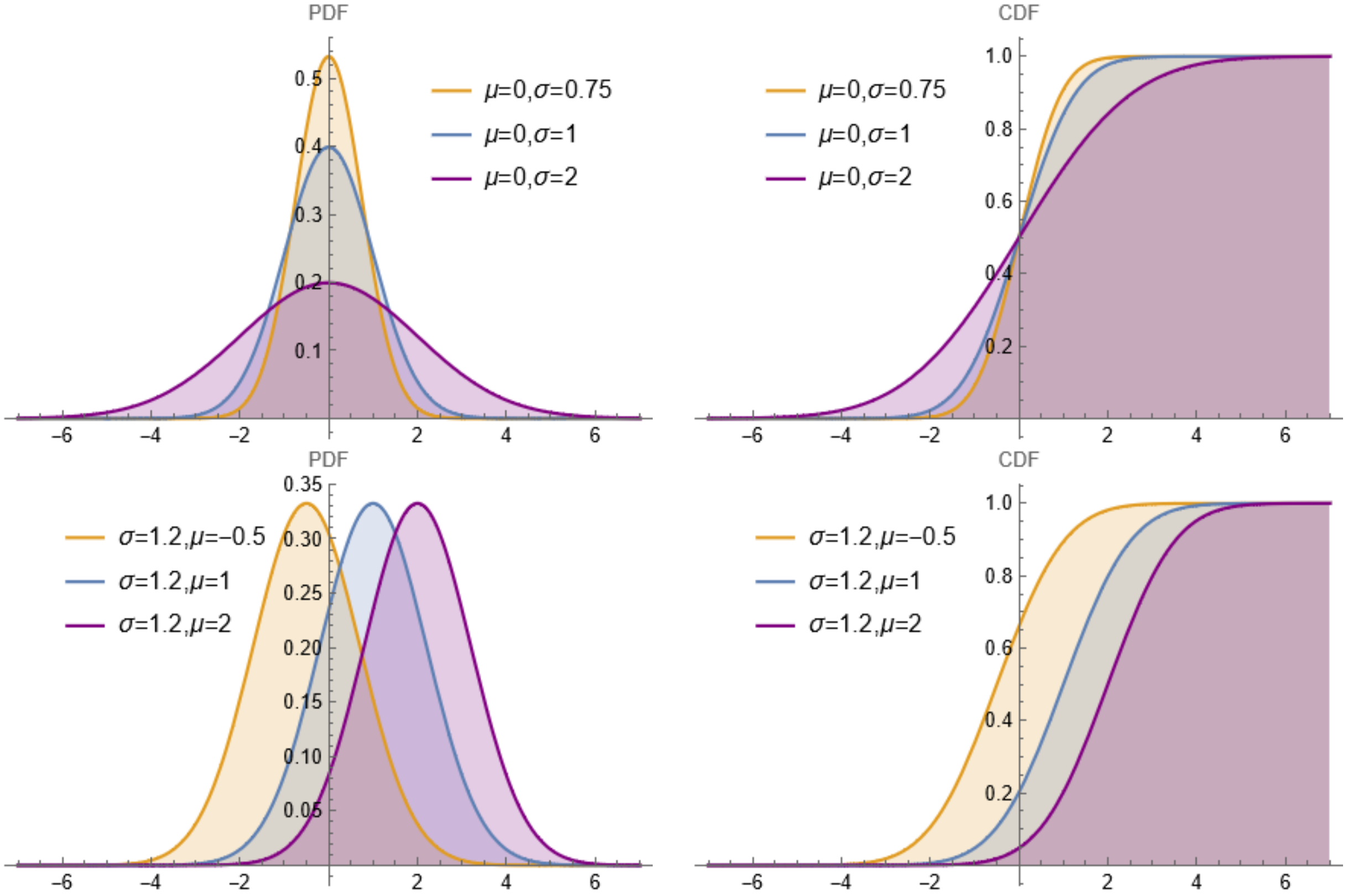

**Figure 12.6.** PMF (left) and CDF (right) of normal distribution. The curve of the normal distribution depends on the two parameters $\mu$ and $\sigma$.

Given the values of the mean, $\mu$, and the standard deviation, $\sigma$, we can find the area under a normal distribution curve for any interval. The value of $\mu$ determines the center of a normal distribution curve on the horizontal axis, and the value of $\sigma$ gives the spread of the normal distribution curve. The three normal distribution curves shown in Figure 12.6 (upper panel) have the same mean but different standard deviations. By contrast, the three normal distribution curves in Figure 12.6 (lower panel) have different means but the same standard deviation.

We list the following properties of the normal curve:

1. The normal curve is symmetrical about the ordinate at $x = \mu$, i.e., $f(\mu + c) = f(\mu - c)$ for any $c$.
2. 50% of the total area under a normal distribution curve lies on the left side of the mean, and 50% lies on the right side of the mean.
3. The mean, median, and mode are identical and occur at $x = \mu$.
4. The mode of the normal curve is at $x = \mu$, and is equal to $\frac{1}{\sigma\sqrt{2\pi}}$.
5. It is unimodal: its first derivative is positive for $x < \mu$ negative for $x > \mu$ and zero only at $x = \mu$.
6. The normal curve extends from $-\infty$ to $+\infty$.
7. In graphical form, normal distribution will appear as a bell curve.
8. For a normal distribution $\beta_1 = 0$ (i.e., symmetric) and $\beta_2 = 3$ (i.e., mesokurtic).
9. $x$-axis is an asymptote to the curve. That is, the curve touches the $x$-axis only at $\pm\infty$.
10. All odd-order central moments are zero. i.e.,
$$\mu_{2r+1} = 0, \qquad r = 0,1,2,\ldots$$
11. Even order central moments are given by
$$\mu_{2r} = 1.3.5\ldots(2r - 1)\sigma^{2r}, r = 0,1,2,\ldots$$

12. The points of inflection of the curve are $x = \mu \pm \sigma$
13. The lower and upper quartiles are equidistant from the median.
14. The area under the normal curve is distributed as:
    (a) 68.27% of the items lie between $\mu - \sigma$ and $\mu + \sigma$.
    $$\text{i.e.}, P(\mu - \sigma \leq X \leq \mu + \sigma) = 0.6827.$$
    (b) 95.45% of the items lie between $\mu - 2\sigma$ and $\mu + 2\sigma$.
    $$\text{i.e.}, P(\mu - 2\sigma \leq X \leq \mu + 2\sigma) = 0.9545.$$
    (c) 99.73% of the items lie between $\mu - 3\sigma$ and $\mu + 3\sigma$.
    $$\text{i.e.}, P(\mu - 3\sigma \leq X \leq \mu + 3\sigma) = 0.9973.$$
    (d) The total area under the curve and above the horizontal axis is equal to 1.

**The normal distribution has several applications in various fields:**

- Heights of adult males:
  The distribution of heights of adult males tends to follow a normal distribution. The mean and standard deviation can vary depending on the population being considered, but it is often modeled as a normal RV. Most people are of average height, with a small number of people being taller or shorter than average.
- Errors in measurements:
  In many scientific experiments and measurements, there is inherent uncertainty and variability. The errors associated with these measurements are often assumed to be normally distributed.
- Body temperatures:
  The distribution of body temperatures in a healthy adult population is often modeled as a normal RV with a mean around 98.6 degrees Fahrenheit (37 degrees Celsius) and a standard deviation of approximately 0.6 degrees Fahrenheit (0.3 degrees Celsius).
- Blood pressure:
  Systolic and diastolic blood pressures in a population are often assumed to follow normal distributions. The mean and standard deviation can vary depending on factors such as age and health conditions.
- Exam scores:
  In educational settings, the scores on exams are often assumed to be normally distributed. Most students score around the average, with a small number of students scoring much higher or lower. This assumption allows for the use of statistical methods to analyze performance and set grading criteria.
- Machine learning:
  The normal distribution serves as a fundamental assumption in many machine learning algorithms, such as linear regression, logistic regression, and Gaussian mixture models. It allows for probabilistic modeling and inference in various learning tasks.
- Sampling theory:
  The normal distribution plays a crucial role in sampling theory and the central limit theorem. The central limit theorem states that as the sample size increases, the distribution of the sample mean approaches a normal distribution. This is true even if the individual variables are not normally distributed.
- Normal RVs are also used in hypothesis testing, confidence intervals, and other statistical procedures.

**Theorem 12.8:** The mean and variance of $X \sim N(\mu, \sigma)$ are
$$E[X] = \mu, \tag{12.30}$$
$$V(X) = \sigma^2. \tag{12.31}$$

**Proof:**

Mean,

$$\mu_1' = E[X]$$
$$= \int_{-\infty}^{\infty} x \frac{1}{\sigma\sqrt{2\pi}} e^{-\frac{(x-\mu)^2}{2\sigma^2}} dx$$

$$= \int_{-\infty}^{\infty} (x-\mu+\mu)\frac{1}{\sigma\sqrt{2\pi}}e^{-\frac{(x-\mu)^2}{2\sigma^2}}dx$$
$$= \int_{-\infty}^{\infty} (x-\mu)\frac{1}{\sigma\sqrt{2\pi}}e^{-\frac{(x-\mu)^2}{2\sigma^2}}dx + \mu\int_{-\infty}^{\infty} \frac{1}{\sigma\sqrt{2\pi}}e^{-\frac{(x-\mu)^2}{2\sigma^2}}dx$$
$$= \int_{-\infty}^{\infty} \sigma z\frac{1}{\sqrt{2\pi}}e^{-\frac{z^2}{2}}dz + \mu \qquad (z = (x-\mu)/\sigma)$$
$$= \frac{\sigma}{\sqrt{2\pi}}\int_{-\infty}^{\infty} ze^{-\frac{z^2}{2}}dz + \mu$$
$$= \frac{\sigma}{\sqrt{2\pi}}\left[e^{-\frac{z^2}{2}}\right]_{-\infty}^{\infty} + \mu$$
$$= \frac{\sigma}{\sqrt{2\pi}} \times 0 + \mu$$
$$= \mu.$$

Variance,

$$V(X) = E\big(X - E(X)\big)^2$$
$$= E(X-\mu)^2$$
$$= \int_{-\infty}^{\infty} (x-\mu)^2\frac{1}{\sigma\sqrt{2\pi}}e^{-\frac{(x-\mu)^2}{2\sigma^2}}dx$$
$$= \frac{1}{\sigma\sqrt{2\pi}}\int_{-\infty}^{\infty} (\sigma z)^2 e^{-\frac{z^2}{2}}\sigma dz \qquad (z = (x-\mu)/\sigma)$$
$$= \frac{\sigma^2}{\sqrt{2\pi}}\int_{-\infty}^{\infty} z^2 e^{-\frac{z^2}{2}}dz \qquad \left(\text{With u = z and dv} = \text{z}e^{-\frac{z^2}{2}}\text{, the integration by parts formula}\right)$$
$$= \frac{\sigma^2}{\sqrt{2\pi}}\left[\left.-ze^{-\frac{z^2}{2}}\right|_{-\infty}^{\infty} + \int_{-\infty}^{\infty} e^{-\frac{z^2}{2}}dz\right] \qquad \left(\int \text{udv} = \text{uv} - \int \text{vdu}\right)$$
$$= \frac{\sigma^2}{\sqrt{2\pi}}\left[0 + \sqrt{2\pi}\right]$$
$$= \sigma^2.$$

■

**Theorem 12.9:** If $X$ is a normal RV with parameters $\mu$ and $\sigma$, then all odd-order central moments are zero,

$$\mu_{2r+1} = 0. \qquad (12.32)$$

**Proof:**

$$\mu_{2r+1} = E\big(X - E(X)\big)^{2r+1}$$
$$= E(X-\mu)^{2r+1}$$
$$= \int_{-\infty}^{\infty} (x-\mu)^{2r+1}\frac{1}{\sigma\sqrt{2\pi}}e^{-\frac{(x-\mu)^2}{2\sigma^2}}dx$$
$$= \frac{1}{\sigma\sqrt{2\pi}}\int_{-\infty}^{\infty} (\sigma z)^{2r+1}e^{-\frac{z^2}{2}}dx \qquad (z = (x-\mu)/\sigma)$$

$$= \frac{\sigma^{2r+1}}{\sqrt{2\pi}} \int_{-\infty}^{\infty} z^{2r+1} e^{-\frac{z^2}{2}} dx$$

$$= \frac{\sigma^{2r+1}}{\sqrt{2\pi}} \times 0. \quad \text{(being the integral of an odd function)} = 0$$

■

**Theorem 12.10:** If $X$ is a normal RV with parameters $\mu$ and $\sigma$, then the even order central moments are given by,

$$\mu_{2r} = (1)(3)(5) \dots (2r-1)\sigma^{2r}. \quad (12.33)$$

**Proof:**

$$\mu_{2r} = E\big(X - E(X)\big)^{2r}$$
$$= E(X-\mu)^{2r}$$
$$= \int_{-\infty}^{\infty} (x-\mu)^{2r} \frac{1}{\sigma\sqrt{2\pi}} e^{-\frac{(x-\mu)^2}{2\sigma^2}} dx$$
$$= \frac{1}{\sigma\sqrt{2\pi}} \int_{-\infty}^{\infty} (\sigma z)^{2r} e^{-\frac{z^2}{2}} \sigma dz \qquad (z = (x-\mu)/\sigma)$$
$$= \frac{\sigma^{2r}}{\sqrt{2\pi}} \int_{-\infty}^{\infty} z^{2r} e^{-\frac{z^2}{2}} dz$$
$$= \frac{2\sigma^{2r}}{\sqrt{2\pi}} \int_{0}^{\infty} z^{2r} e^{-\frac{z^2}{2}} dz \quad \text{(being the integral of an even function)}$$
$$= \frac{2\sigma^{2r}}{\sqrt{2\pi}} \int_{0}^{\infty} (2u)^r e^{-u} \frac{du}{\sqrt{2u}} \quad \left(\text{With } u = \frac{z^2}{2}\right)$$
$$= \frac{2^r\sigma^{2r}}{\sqrt{\pi}} \int_{0}^{\infty} u^{r-\frac{1}{2}} e^{-u} du$$
$$= \frac{2^r\sigma^{2r}}{\sqrt{\pi}} \frac{\Gamma\left(r+\frac{1}{2}\right)}{1^{r+\frac{1}{2}}}$$
$$= \frac{2^r\sigma^{2r}}{\sqrt{\pi}} \left(r-\frac{1}{2}\right)\left(r-\frac{3}{2}\right) \dots \frac{3}{2}\frac{1}{2}\Gamma\left(\frac{1}{2}\right)$$
$$= \frac{2^r\sigma^{2r}}{\sqrt{\pi}} \frac{(2r-1)(2r-3)\dots(3)(1)\sqrt{\pi}}{2^r} = (1)(3)(5)\dots(2r-1)\sigma^{2r}.$$

■

**Theorem 12.11 (Third and Fourth Raw Moment):** If $X$ is a normal RV with parameters $\mu$ and $\sigma$,

$$\mu_2 = \sigma^2 \quad \text{and } \mu_4 = 3\sigma^4. \quad (12.34)$$

**Proof:**

We have,

$$\mu_{2r} = (1)(3)(5) \dots (2r-1)\sigma^{2r},$$

$$\mu_{2r+2} = (1)(3)(5) \dots (2r-1)(2r+1)\sigma^{2r+2}.$$

Therefore,

$$\frac{\mu_{2r+2}}{\mu_{2r}} = (2r+1)\sigma^2.$$

Hence,

$$\mu_{2r+2} = (2r+1)\sigma^2\mu_{2r}.$$

With this recurrence formula and the information $\mu_0 = 1$, we can calculate $\mu_2$ and $\mu_4$ successively. Putting $r = 0$ we get $\mu_2 = \sigma^2$ and then substituting $r = 1$ we obtain $\mu_4 = 3\sigma^4$.

■

**Theorem 12.12:** The normal distribution is symmetric and mesokurtic.

$$\gamma_1 = 0, \gamma_2 = 0. \qquad (12.35)$$

**Proof:**

Skewness:

Since all the odd-order central moments are zero,

$$\beta_1 = \frac{\mu_3^2}{\mu_2^3} = 0.$$

Hence, $\gamma_1 = \sqrt{\beta_1} = 0$. That is, the normal distribution is symmetric.

Kurtosis:

$$\beta_2 = \frac{\mu_4}{\mu_2^2} = \frac{3\sigma^4}{\sigma^4} = 3.$$

Hence, $\gamma_2 = \beta_2 - 3 = 0$. That is, the distribution is mesokurtic.

■

**Theorem 12.13:** The MGF of normal distribution $N(\mu, \sigma)$ is given by,

$$M_X(t) = e^{\mu t + \frac{1}{2}t^2\sigma^2}. \qquad (12.36)$$

**Proof:**

$$\begin{aligned}
M_X(t) &= E[e^{tX}] \\
&= \int_{-\infty}^{\infty} e^{tx} \frac{1}{\sigma\sqrt{2\pi}} e^{-\frac{(x-\mu)^2}{2\sigma^2}} dx \qquad \left(z = \frac{x-\mu}{\sigma}\right) \\
&= \frac{1}{\sigma\sqrt{2\pi}} \int_{-\infty}^{\infty} e^{t(\mu+z\sigma)} e^{-\frac{z^2}{2}} dz \\
&= \frac{e^{\mu t}}{\sqrt{2\pi}} \int_{-\infty}^{\infty} e^{tz\sigma - \frac{z^2}{2}} dz \\
&= \frac{e^{\mu t}}{\sqrt{2\pi}} \int_{-\infty}^{\infty} e^{-\frac{1}{2}(z^2 - 2tz\sigma)} dz \\
&= \frac{e^{\mu t}}{\sqrt{2\pi}} \int_{-\infty}^{\infty} e^{-\frac{1}{2}(z^2 - 2tz\sigma + t^2\sigma^2) + \frac{1}{2}t^2\sigma^2} dz \\
&= \frac{e^{\mu t + \frac{1}{2}t^2\sigma^2}}{\sqrt{2\pi}} \int_{-\infty}^{\infty} e^{-\frac{1}{2}(z-t\sigma)^2} dz \qquad (u = z - t\sigma)
\end{aligned}$$

$$= \frac{e^{\mu t+\frac{1}{2}t^2\sigma^2}}{\sqrt{2\pi}} \int_{-\infty}^{\infty} e^{-\frac{u^2}{2}} du$$

$$= 2\frac{e^{\mu t+\frac{1}{2}t^2\sigma^2}}{\sqrt{2\pi}} \int_{0}^{\infty} e^{-\frac{u^2}{2}} du \quad \text{(being the integral of an even function)}$$

$$= 2\frac{e^{\mu t+\frac{1}{2}t^2\sigma^2}}{\sqrt{2\pi}} \int_{0}^{\infty} e^{-v} \frac{dv}{\sqrt{2v}} \quad \left(v = \frac{u^2}{2}\right)$$

$$= \frac{e^{\mu t+\frac{1}{2}t^2\sigma^2}}{\sqrt{\pi}} \int_{0}^{\infty} v^{\frac{1}{2}-1} e^{-v} dv$$

$$= \frac{e^{\mu t+\frac{1}{2}t^2\sigma^2}}{\sqrt{\pi}} \frac{\Gamma\left(\frac{1}{2}\right)}{1^{\frac{1}{2}}}$$

$$= \frac{e^{\mu t+\frac{1}{2}t^2\sigma^2}}{\sqrt{\pi}} \sqrt{\pi}$$

$$= e^{\mu t+\frac{1}{2}t^2\sigma^2}.$$

■

**Theorem 12.14 (Additive Property):** If $X_1 \sim N(\mu_1, \sigma_1^2)$, $X_2 \sim N(\mu_2, \sigma_2^2)$ and if $X_1$ and $X_2$ are independent, then,

$$X_1 + X_2 \sim N(\mu_1 + \mu_2, \sigma_1^2 + \sigma_2^2). \tag{12.37}$$

**Proof:**

Given $X_1 \sim N(\mu_1, \sigma_1^2)$, implies $M_{X_1}(t) = e^{\mu_1 t+\frac{1}{2}t^2\sigma_1^2}$ and $X_2 \sim N(\mu_2, \sigma_2^2)$, implies $M_{X_2}(t) = e^{\mu_2 t+\frac{1}{2}t^2\sigma_2^2}$. Since $X_1$ and $X_2$ are independent,

$$\begin{aligned} M_{X_1+X_2}(t) &= M_{X_1}(t) M_{X_2}(t) \\ &= e^{\mu_1 t+\frac{1}{2}t^2\sigma_1^2} e^{\mu_2 t+\frac{1}{2}t^2\sigma_2^2} \\ &= e^{(\mu_1+\mu_2)t+\frac{1}{2}t^2(\sigma_1^2+\sigma_2^2)}, \end{aligned}$$

which is the MGF of $N(\mu_1 + \mu_2, \sigma_1^2 + \sigma_2^2)$.

■

**Theorem 12.15 (Additive Property):** If $X_i$, $i = 1,2,\ldots,n$ are $n$ independent normal variates with mean $\mu_i$ and variance $\sigma_i^2$ respectively, then $Y = \sum_{i=1}^{n} X_i$ is normally distributed with mean $\sum_{i=1}^{n} \mu_i$ and variance $\sum_{i=1}^{n} \sigma_i^2$.

$$Y = \sum_{i=1}^{n} X_i \sim N\left(\sum_{i=1}^{n} \mu_i, \sum_{i=1}^{n} \sigma_i^2\right). \tag{12.38}$$

**Proof:**

Given $X_i \sim N(\mu_i, \sigma_i^2)$, implies $M_{X_i}(t) = e^{\mu_i t+\frac{1}{2}t^2\sigma_i^2}$. Now

$$\begin{aligned} M_Y(t) &= M_{\sum_{i=1}^{n} X_i}(t) \\ &= \prod_{i=1}^{n} M_{X_i}(t) \\ &= \prod_{i=1}^{n} e^{\mu_i t+\frac{1}{2}t^2\sigma_i^2} \end{aligned}$$

$$= e^{\sum_{i=1}^{n}\mu_i t + \frac{t^2}{2}\sum_{i=1}^{n}\sigma_i^2},$$

which is the MGF of normal variate with mean $\sum_{i=1}^{n}\mu_i$ and variance $\sum_{i=1}^{n}\sigma_i^2$.

■

**Theorem 12.16:** If $X_i$, $i = 1,2,\dots,n$ are $n$ independent normal variates with mean $\mu_i$ and variance $\sigma_i^2$ respectively, then their linear combination, $Y = \sum_{i=1}^{n} a_i X_i$, is normally distributed with mean $\sum_{i=1}^{n} a_i\mu_i$ and variance $\sum_{i=1}^{n} a_i^2\sigma_i^2$ where $a_i$'s are constants.

**Proof:**

Given $X_i \sim N(\mu_i, \sigma_i^2)$, implies $M_{X_i}(t) = e^{\mu_i t + \frac{1}{2}t^2\sigma_i^2}$. Now

$$\begin{aligned}
M_Y(t) &= M_{\sum_{i=1}^{n} a_i X_i}(t) \\
&= \prod_{i=1}^{n} M_{a_i X_i}(t) \\
&= \prod_{i=1}^{n} M_{X_i}(a_i t) \\
&= \prod_{i=1}^{n} e^{\mu_i a_i t + \frac{1}{2}t^2 a_i^2\sigma_i^2} \\
&= e^{\sum_{i=1}^{n}\mu_i a_i t + \frac{1}{2}t^2\sum_{i=1}^{n} a_i^2\sigma_i^2},
\end{aligned}$$

which is the MGF of normal variate with mean $\sum_{i=1}^{n} a_i\mu_i$ and variance $\sum_{i=1}^{n} a_i^2\sigma_i^2$.

■

**Theorem 12.17:** If $X$ is normal RV with mean $\mu$ and variance $\sigma^2$, then for any constants $a$ and $b$, $b \neq 0$, the RV $Y = a + bX$ is also a normal RV with parameters,

$$E[Y] = E[a + bX] = a + bE[X] = a + b\mu, \qquad (12.39)$$

and variance

$$V(Y) = V(a + bX) = b^2V(X) = b^2\sigma^2. \qquad (12.40)$$

**Proof:**

Let $F_Y(y)$ be the distribution function of $Y$. Then, for $b > 0$

$$\begin{aligned}
F_Y(y) &= P(Y \le y) \\
&= P(a + bX \le y) \\
&= P\left(X \le \frac{y-a}{b}\right) \\
&= F_X\left(\frac{y-a}{b}\right),
\end{aligned}$$

where $F_X$ is the distribution function of $X$. Similarly, if $b < 0$, then

$$\begin{aligned}
F_Y(y) &= P(a + bX \le y) \\
&= P\left(X \ge \frac{y-a}{b}\right) \\
&= 1 - F_X\left(\frac{y-a}{b}\right).
\end{aligned}$$

Differentiation yields that the density function of $Y$ is

$$f_Y(y) = \begin{cases} \frac{1}{b} f_X\left(\frac{y-a}{b}\right), & b > 0, \\ -\frac{1}{b} f_X\left(\frac{y-a}{b}\right), & b < 0, \end{cases}$$

which can be written as

$$\begin{aligned} f_Y(y) &= \frac{1}{|b|} f_X\left(\frac{y-a}{b}\right) \\ &= \frac{1}{\sqrt{2\pi}\sigma|b|} e^{-\frac{\left(\frac{y-a}{b}-\mu\right)^2}{2\sigma^2}} \\ &= \frac{1}{\sqrt{2\pi}\sigma|b|} e^{-\frac{(y-a-b\mu)^2}{2b^2\sigma^2}}, \end{aligned}$$

showing that $Y = a + bX$ is normal with mean $a + b\mu$ and variance $b^2\sigma^2$.

■

### 12.2.5 Standard Normal Distribution

From (12.39) and (12.40), if $b = \frac{1}{\sigma}$ and $a = -\frac{\mu}{\sigma}$, then

$$Y = a + bX = \frac{X-\mu}{\sigma} = Z,$$

is normal RV with mean $-\frac{\mu}{\sigma} + \frac{1}{\sigma}\mu = 0$ and variance $\left(\frac{1}{\sigma}\right)^2 \sigma^2 = 1$.

**Theorem 12.18 (Standardizing a Normal RV):** If $X$ is a normal RV with $E[X] = \mu$ and $V(X) = \sigma^2$, the RV

$$Z = \frac{X-\mu}{\sigma}, \tag{12.41}$$

is a normal RV with $E[Z] = 0$ and $V(Z) = 1$. $Z$ is called a standard normal RV. We write $Z \sim N(0,1)$.

Creating a new RV by this transformation is referred to as standardizing. The RV $Z$ represents the distance of $X$ from its mean in terms of standard deviations. It is the key step to calculating a probability for an arbitrary normal RV.

**Definition (Standard Normal Distribution):** A RV $Z$ is said to follow standard normal distribution if its PDF is given by

$$f_Z(z) = \frac{1}{\sqrt{2\pi}} e^{-\frac{z^2}{2}}, \qquad -\infty < z < \infty. \tag{12.42}$$

We can see that

$$E[Z] = E\left[\frac{X-\mu}{\sigma}\right] = 0, \tag{12.43}$$

$$V(Z) = V\left(\frac{X-\mu}{\sigma}\right) = 1, \tag{12.44}$$

$$M_Z(t) = e^{\frac{t^2}{2}}. \tag{12.45}$$

Standard normal distribution satisfies all the properties of normal distribution provided $\mu = 0$ and $\sigma = 1$. Some of them are the following.

1. The curve of $f(z)$ is symmetrical about the ordinate at $z = 0$.
2. The curve of $f(z)$ is maximum at $z = 0$ and the maximum ordinate is $\frac{1}{\sqrt{2\pi}}$.

3. The curve extends from $-\infty$ to $+\infty$.
4. Mean = Median = Mode =0.
5. In a standard normal distribution 68.27% of the items lies between $-1$ and $+1$, 95.45% of observations are lying between $-2$ and $+2$. and 99.73% of observations lies between $-3$ and $+3$.

Suppose $X \sim N(\mu, \sigma^2)$ and we are interested in finding the probability of the variate $X$ lying between two values, say, $a$ and $b$. To determine this, we first make the transformation $Z = \frac{X-\mu}{\sigma}$.

**Definition (Standardizing to Calculate a Probability):** Suppose that $X$ is a normal RV with mean $\mu$ and variance $\sigma^2$. Then,

$$P(X \le x) = P\left(\frac{X-\mu}{\sigma} \le \frac{x-\mu}{\sigma}\right) = P(Z \le z), \tag{12.46}$$

where $Z$ is a standard normal RV, and $z = \frac{x-\mu}{\sigma}$ is the $z$-value obtained by standardizing $X$.
Hence,

$$P(a < X < b) = P\left(\frac{a-\mu}{\sigma} < \frac{X-\mu}{\sigma} < \frac{b-\mu}{\sigma}\right) = P(z_1 < Z < z_2), \tag{12.47}$$

where $z_1 = \frac{a-\mu}{\sigma}$ and $z_2 = \frac{b-\mu}{\sigma}$. Therefore, $P(a < X < b)$ is the area under the standard normal curve between the abscissae $z_1$ and $z_2$.

**Definition ($z_\alpha$, $z$-Score):** The symbol $Z_\alpha$ is used to denote the $z$-score that has an area of $\alpha$ to its right under the standard normal curve.

***Example 12.9***

On a final examination in physics, the mean was 70 and the standard deviation was 14.
1- Determine the standard scores of students receiving the grades (a) 65, (b) 92, and (c) 74.
2- Find the grades corresponding to the standard scores (a) $-1$ and (b) 1.6.

***Solution***

1-
(a)

$$z = \frac{X-\mu}{\sigma} = \frac{65-70}{14} = -0.3571.$$

(b)

$$z = \frac{X-\mu}{\sigma} = \frac{92-70}{14} = 1.5714.$$

(c)

$$z = \frac{X-\mu}{\sigma} = \frac{74-70}{14} = 0.2857.$$

2-
(a)

$$X = \mu + z\sigma = 70 + (-1)14 = 56.$$

(b)

$$X = \mu + z\sigma = 70 + (1.6)14 = 92.4.$$

***Example 12.10***

$X \sim N(20,4)$. Find the probability that the value taken by $X$ is
(a) less than 24.
(b) greater than 24.
(c) between 18 and 22.

***Solution:***

(a) Given $X \sim N(20,4)$. Therefore, $\mu = 20$ and $\sigma = 2$.

$$P(X < 24) = P\left(\frac{X-\mu}{\sigma} < \frac{24-\mu}{\sigma}\right) = P\left(\frac{X-20}{2} < \frac{24-20}{2}\right) = P(Z < 2).$$

```
NProbability[x<24,x \[Distributed] NormalDistribution[20,2]]
NProbability[x<2,x \[Distributed] NormalDistribution[0,1]]
 0.97725
 0.97725
```

(b)

$$P(X > 24) = P\left(\frac{X-20}{2} > \frac{24-20}{2}\right) = P(Z > 2).$$

```
NProbability[x>24,x \[Distributed] NormalDistribution[20,2]]
NProbability[x>2,x \[Distributed] NormalDistribution[0,1]]
 0.0227501
 0.0227501
```

(c)

$$P(18 < X < 22) = P\left(\frac{18-20}{2} < \frac{X-20}{2} < \frac{22-20}{2}\right) = P(-1 < Z < 1).$$

```
NProbability[18<x<22,x \[Distributed] NormalDistribution[20,2]]
NProbability[-1<x<1,x \[Distributed] NormalDistribution[0,1]]
 0.682689
 0.682689
```

***Example 12.11***

Find the $z_\alpha$, $\alpha = \{0.025, 0.05, 0.10, 0.5, 0.90, 0.95, 0.975\}$.

***Solution***

**Part 1**

```
(* The symbol αleft denotes the areas under the standard normal curve to the left of the corresponding z-score: *)
αleft={0.025,0.05,0.10,0.5,0.90,0.95,0.975}

(* The variable αright represents the complement of the αleft, which denotes the areas under the standard normal curve to the right of the corresponding z-score: *)
αright=1-αleft

(* The Quantile function is used to calculate the z-scores corresponding to the specified αleft under the standard normal curve. This provides the critical values for confidence intervals or hypothesis testing: *)
N[Quantile[NormalDistribution[0,1],αleft]]

(* The InverseCDF function is an alternative method to compute the z-scores corresponding to the αleft under the standard normal curve. It yields the same critical values as the Quantile function: *)
InverseCDF[NormalDistribution[0,1],αleft]

 {0.025,0.05,0.1,0.5,0.9,0.95,0.975}
 {0.975,0.95,0.9,0.5,0.1,0.05,0.025}
 {-1.95996,-1.64485,-1.28155,0.,1.28155,1.64485,1.95996}
 {-1.95996,-1.64485,-1.28155,0.,1.28155,1.64485,1.95996}
```

**Part 2**

```
dist=NormalDistribution[0,1]; (*Define a normal distribution with mean 0 and standard
deviation 1*)

αleft={0.025,0.05,0.10,0.5,0.90,0.95,0.975};
αright=1-αleft;

q=N[Quantile[dist,αleft]]; (*Calculate quantiles based on the αleft*)

Plot[
 PDF[dist,x],(*Plot the probability density function (PDF) of the normal distribution*)
 {x,-3,3},
 (* The'Epilog' option adds white vertical lines at the quantiles corresponding to the
αleft, and blue text labels indicating the complement αleft: *)
 Epilog->{
   White,
   Thickness[0.008],
   Apply[
    Sequence,
    Table[
     Line[{{q[[i]],0},{q[[i]],PDF[dist,q[[i]]]}}],
     {i,1,7}
     ]
    ],
   Blue,
   Table[
    Text[αright[[i]],{q[[i]],0.05+PDF[dist,q[[i]]]}],
    {i,1,7}
    ]
   },
 PlotRange->{0,0.48},(*Set the y-axis plot range*)
 Filling->Axis,
 PlotStyle->Purple,
 ImageSize->250
 ]
```

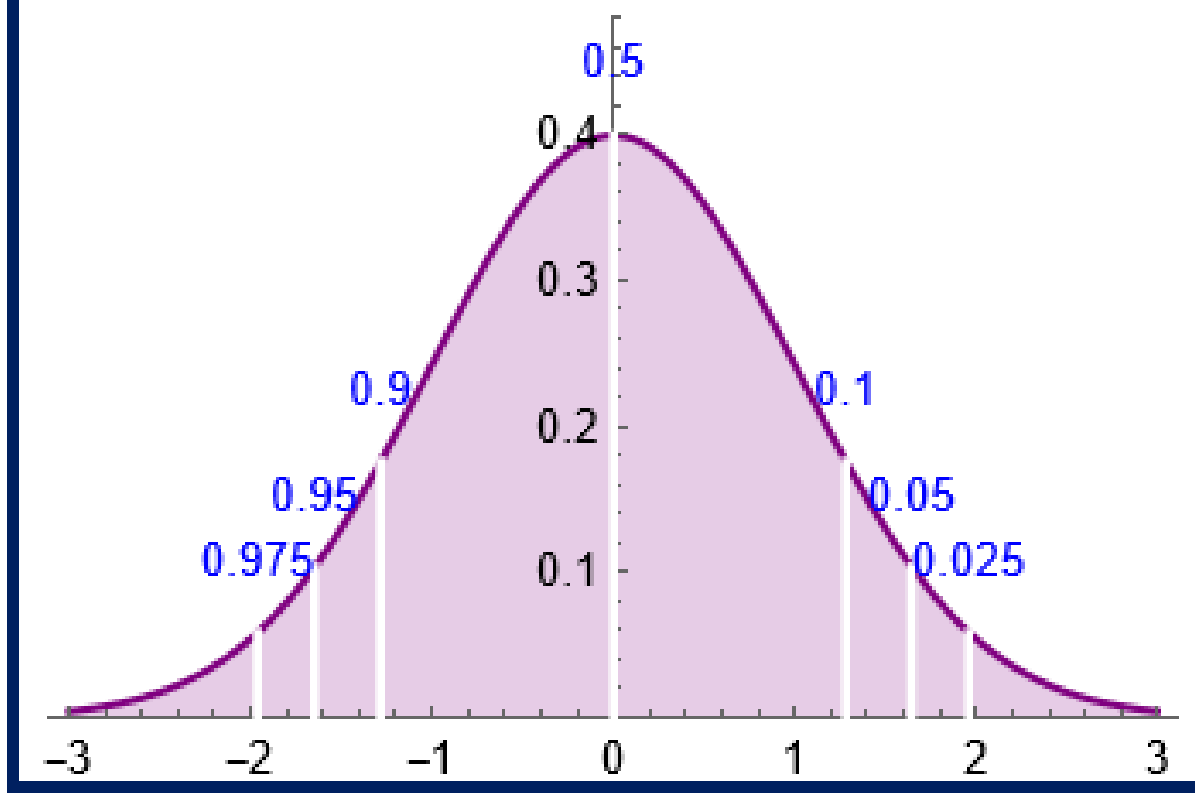

## 12.3 Normal Distribution as a Limiting Form of Binomial Distribution

The relation between the binomial and normal distributions can be understood through an approximation. When the number of trials $n$ in the binomial distribution is large and the probability of success $p$ is not extremely close to 0 or 1, the shape of the binomial distribution becomes more and more similar to the shape of the normal distribution.

More formally, if $X$ is a RV that follows a binomial distribution with parameters $n$ and $p$, then the mean of $X$ is given by $\mu = np$, and the standard deviation is given by $\sigma = \sqrt{npq}$. When $n$ is large, the binomial distribution can be approximated by a normal distribution with the same mean and standard deviation. That is, $X$ is approximately normally distributed with mean $\mu$ and standard deviation $\sigma$. (See Figure 12.7.)

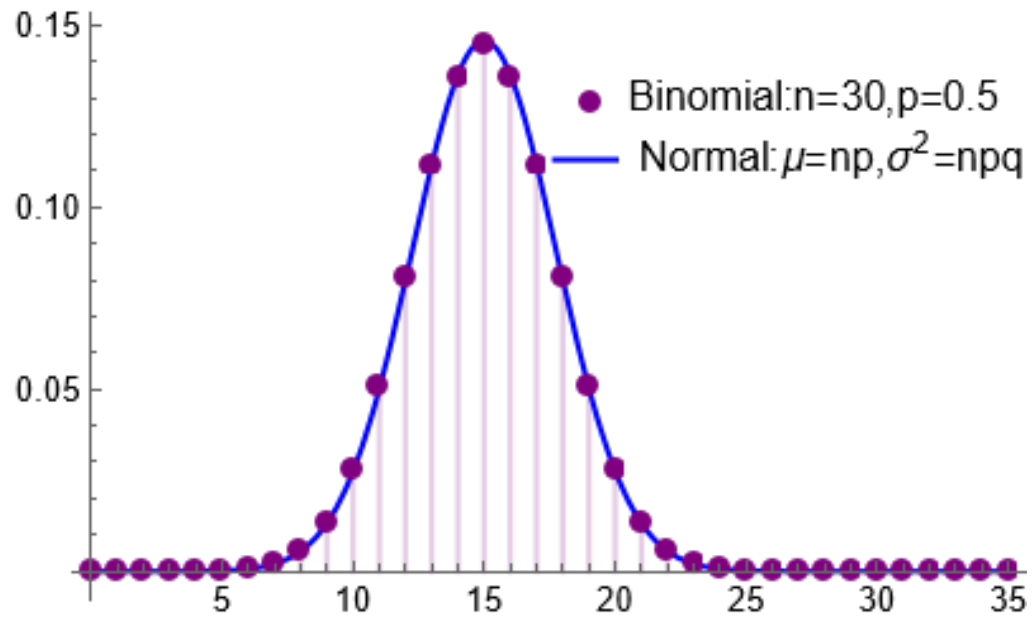

**Figure12.7.** Plot of binomial for $n = 30$ and $p = 0.5$ and normal curve with mean $\mu = np$ and standard deviation $\sigma = \sqrt{npq}$.

**Theorem 12.19:** Binomial distribution tends to normal distribution under the following conditions
1. $n$ is large ($n \to \infty$)
2. neither $p$ nor $q$ is very small
In practice the approximation is very good if both $np$ and $nq$ are greater than 5.

**Proof:**

Let $X \sim B(n, p)$. Then,

$$f_X(x) = \binom{n}{x} p^x q^{n-x};\ x = 0,1,2,\ldots,n,$$

where $0 < p < 1$ and $p + q = 1$. Therefore,

$$E[X] = np,$$

$$V(X) = npq,$$

and

$$M_X(t) = (q + pe^t)^n.$$

Define,

$$\begin{aligned} Z &= \frac{X - E[X]}{\sqrt{V(X)}} \\ &= \frac{X - np}{\sqrt{npq}} \\ &= \frac{X - \mu}{\sigma}. \end{aligned}$$

Now,

$$\begin{aligned} M_Z(t) &= M_{\frac{X-\mu}{\sigma}}(t) \\ &= e^{-\frac{\mu t}{\sigma}} M_X\left(\frac{t}{\sigma}\right) = e^{-\frac{\mu t}{\sigma}} \left(q + pe^{\frac{t}{\sigma}}\right)^n. \end{aligned}$$

Therefore,

$$\ln M_Z(t) = -\frac{\mu t}{\sigma} + n \ln\left(q + pe^{\frac{t}{\sigma}}\right)$$

$$= -\frac{\mu t}{\sigma} + n\ln\left(q + p\left(1 + \frac{\frac{t}{\sigma}}{1!} + \frac{\left(\frac{t}{\sigma}\right)^2}{2!} + \cdots\right)\right)$$

$$= -\frac{\mu t}{\sigma} + n\ln\left(q + p + p\left(\frac{\frac{t}{\sigma}}{1!} + \frac{\left(\frac{t}{\sigma}\right)^2}{2!} + \cdots\right)\right)$$

$$= -\frac{\mu t}{\sigma} + n\ln\left(q + p + p\left(\frac{t}{\sigma} + \frac{t^2}{2\sigma^2} + \cdots\right)\right)$$

$$= -\frac{\mu t}{\sigma} + n\ln\left(1 + p\left(\frac{t}{\sigma} + \frac{t^2}{2\sigma^2} + \cdots\right)\right)$$

$$= -\frac{\mu t}{\sigma} + n\left(p\left(\frac{t}{\sigma} + \frac{t^2}{2\sigma^2} + \cdots\right) - \frac{p^2}{2}\left(\frac{t}{\sigma} + \frac{t^2}{2\sigma^2} + \cdots\right)^2 + \cdots\right)$$

$$= -\frac{\mu t}{\sigma} + n\left(\frac{pt}{\sigma} + \frac{pt^2}{2\sigma^2} - \frac{p^2t^2}{2\sigma^2} + O\left(\frac{1}{n^{\frac{3}{2}}}\right)\right)$$

$$= -\frac{\mu t}{\sigma} + \frac{npt}{\sigma} + \frac{npt^2}{2\sigma^2}(1-p) + O\left(\frac{1}{n^{\frac{1}{2}}}\right)$$

$$= -\frac{\mu t}{\sigma} + \frac{npt}{\sigma} + \frac{npqt^2}{2\sigma^2} + O\left(\frac{1}{n^{\frac{1}{2}}}\right)$$

$$= -\frac{\mu t}{\sigma} + \frac{\mu t}{\sigma} + \frac{\sigma^2 t^2}{2\sigma^2} + O\left(\frac{1}{n^{\frac{1}{2}}}\right)$$

$$= \frac{t^2}{2} + O\left(1/n^{1/2}\right),$$

as $n \to \infty$

$$\ln M_Z(t) = \frac{t^2}{2} \quad \Rightarrow M_Z(t) = e^{\frac{t^2}{2}}.$$

This is the MGF of a standard normal variate. So $Z \to N(0,1)$ as $n \to \infty$.

■

**Definition (Continuity Correction):** Continuity correction is an adjustment done while approximating a discrete RV with a continuous RV, like approximating Binomial or Poisson RV with normal RV. Hence, while calculating the probability of a discrete RV using normal approximation, correction factor should be applied. This can be done by subtracting $-0.5$ from the lower limit and adding $0.5$ to the upper limit.

***Example 12.12***

Find the probability of getting between 4 and 7 heads inclusive in 12 tosses of a fair coin by using,
(a) the binomial distribution and
(b) the normal approximation to the binomial distribution.

***Solution:***

Note that even though the binomial distribution is discrete, it has the shape of the continuous normal distribution. When approximating the binomial probability at 4, 5, 6, and 7 heads by the area under the normal curve, find the normal curve area from $X = 3.5$ to $X = 7.5$. The 0.5 that you go on either side of $X = 4$ and $X = 7$ is the continuity correction. The following are the steps to follow when approximating the binomial with the normal. Choose the normal curve with mean $np = 12(0.5) = 6$ and standard deviation $\sqrt{npq} = \sqrt{12(0.5)(0.5)} = \sqrt{3}$. You are choosing the normal curve with the same center and variation as the binomial distribution. Then find the area under the curve from 3.5 to 7.5. This is the normal approximation to the binomial distribution.

```
(* The number of trials in the binomial distribution *)
n=12;
(* The probability of success in each trial *)
p=1/2;
(* Calculating the mean of the normal approximation distribution *)
μ=n*p
(* Calculating the standard deviation of the normal approximation distribution *)
σ=Sqrt[n*p*(1-p)]

p4=Probability[x==4,x\[Distributed]BinomialDistribution[n,p]] (* Probability of exactly 4 successes *)
p5=Probability[x==5,x\[Distributed]BinomialDistribution[n,p]] (* Probability of exactly 5 successes *)
p6=Probability[x==6,x\[Distributed]BinomialDistribution[n,p]](* Probability of exactly 6 successes *)
p7=Probability[x==7,x\[Distributed]BinomialDistribution[n,p]] (* Probability of exactly 7 successes *)

(* Total probability by summing individual probabilities *)
total=N[p4+p5+p6+p7]

(* Probability of getting a value between 4 and 7 (inclusive) in the binomial distribution *)
NProbability[4<=x<=7,x\[Distributed]BinomialDistribution[n,p]]

(* Probability of getting a value between 3.5 and 7.5 (inclusive) in a normal distribution *)
Probability[3.5<=x<=7.5,x\[Distributed]NormalDistribution[μ,σ]]

 6
 √3
 495/4096
 99/512
 231/1024
 99/512
 0.733154
 0.733154
 0.732305

(* The code plots the PMF of the binomial distribution with parameters n (number of trials) and p (probability of success). It also plots the PDF of the normal distribution with parameters μ=n*p (mean) and σ=Sqrt[n*p*(1-p)] (standard deviation): *)

n=12;(* The number of trials in the binomial distribution*)
p=0.5; (* The probability of success in each trial*)
μ=n*p;(* Calculating the mean of the normal approximation distribution *)
σ=Sqrt[n*p*(1-p)]; (* Calculating the standard deviation of the normal approximation distribution *)

Show[
 (* Plotting the PDF of the normal distribution: *)
 Plot[
  PDF[NormalDistribution[μ,σ],x],
  {x,0,12},
  PlotRange->All,
```

```
  PlotLegends->Placed[{"μ=np,σ2=npq"},{0.8,0.75}],
  PlotStyle->Blue,
  ImageSize->300
  ],
 (* Plotting the PMF of the binomial distribution: *)
 DiscretePlot[
  Evaluate[
   PDF[BinomialDistribution[n,p],k]
   ],
  {k,0,12},
  PlotRange->All,
  PlotMarkers->Automatic,
  PlotStyle->Purple,
  PlotLegends->Placed[{"n=12,p=0.5"},{0.8,0.85}],
  ImageSize->300,
  AxesLabel->{None,"PMF"}
  ]
 ]
```

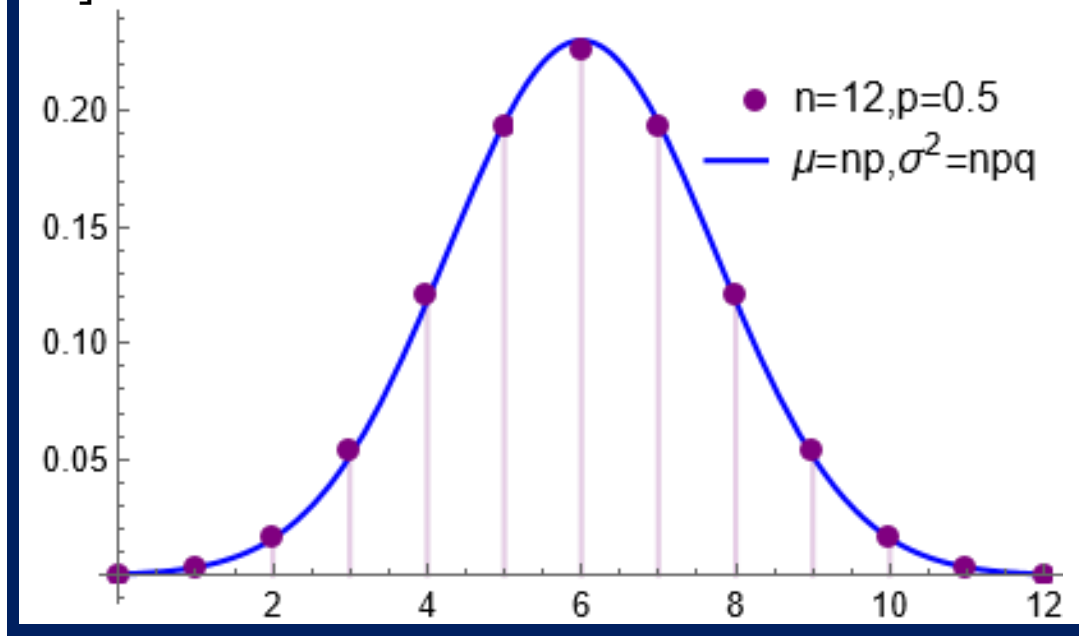

## 12.4 Normal Distribution as a Limiting Form of Poisson Distribution

The relation between the Poisson and normal distributions is another important concept in probability and statistics. It arises from the approximation of the Poisson distribution by the normal distribution under certain conditions.

When the average rate parameter $\lambda$ of the Poisson distribution is large, the Poisson distribution can be approximated by a normal distribution with the same mean and variance. That is, a Poisson RV $X$ with parameter $\lambda$ is approximately normally distributed with mean $\mu = \lambda$ and standard deviation $\sigma = \sqrt{\lambda}$.

**Definition (Normal Approximation to the Poisson Distribution):** If $X$ is a Poisson RV with $E[(X)] = \lambda$ and $V(X) = \lambda$,

$$Z = \frac{X - \lambda}{\sqrt{\lambda}}, \tag{12.48}$$

is approximately a standard normal RV. The same continuity correction used for the binomial distribution can also be applied. The approximation is good for $\lambda > 5$.

***Example 12.13***

Let $X$ is a Poisson RV with rate parameter $\lambda = 20$, find the probability of $X$ getting between 12 and 15 by using
(a) the Poisson distribution and
(b) the normal approximation to the Poisson distribution.

***Solution***

Note that even though the Poisson distribution is discrete, it has the shape of the continuous normal distribution. When approximating the Poisson probability at $X = 12, X = 13, X = 14$, and $X = 15$ by the area under the normal curve, find the normal curve area from $X = 11.5$ to $X = 15.5$. The 0.5 that you go on either side of $X = 12$ and $X = 15$ is the continuity correction. The following are the steps to follow when approximating the Poisson with the normal. Choose the normal curve with mean $\lambda = 20$ and standard deviation $\sqrt{\lambda} = \sqrt{20}$. You are choosing the normal curve with the same center and variation as the binomial distribution. Then find the area under the curve from 11.5 to 15.5. This is the normal approximation to the Poisson distribution.

```
λ=20; (* Poisson rate parameter*)
μ=λ;(* Calculating the mean of the normal approximation distribution *)
σ=Sqrt[λ];(* Calculating the standard deviation of the normal approximation distribution *)

p12=NProbability[x==12,x\[Distributed]PoissonDistribution[λ]] (* Probability of exactly 4
successes *)
p13=NProbability[x==13,x\[Distributed]PoissonDistribution[λ]] (* Probability of exactly 5
successes *)
p14=NProbability[x==14,x\[Distributed]PoissonDistribution[λ]](* Probability of exactly 6
successes *)
p15=NProbability[x==15,x\[Distributed]PoissonDistribution[λ]] (* Probability of exactly 7
successes *)
(* Total probability by summing individual probabilities *)
total=N[p12+p13+p14+p15]

(* Probability of getting a value between 4 and 7 (inclusive) in the binomial distribution *)
NProbability[12<=x<=15,x\[Distributed]PoissonDistribution[λ]]
(* Probability of getting a value between 3.5 and 7.5 (inclusive) in a normal distribution *)
Probability[11.5<=x<=15.5,x\[Distributed]NormalDistribution[μ,σ]]
 0.0176252
 0.0271156
 0.0387366
 0.0516489
 0.135126
 0.135126
 0.128479

(* The code plots the PMF of the Poisson distribution with parameter λ. It also plots the PDF
of the normal distribution with parameters μ=λ (mean) and σ=Sqrt[λ] (standard deviation). :*)
λ=20; (* Poisson rate parameter *)
μ=λ;(* Calculating the mean of the normal approximation distribution *)
σ=Sqrt[λ];(* Calculating the standard deviation of the normal approximation distribution *)

Show[
 (*Plotting the PDF of the normal distribution*)
 Plot[
  PDF[NormalDistribution[μ,σ],x],
  {x,0,40},
  PlotRange->All,
  PlotLegends->Placed[{"μ=λ,σ2=λ"},{0.8,0.75}],
  PlotStyle->Blue,
  ImageSize->300
  ],

 (*Plotting the PMF of the Poisson distribution*)
 DiscretePlot[
  Evaluate[
   PDF[PoissonDistribution[λ],k]
   ],
  {k,0,40},
  PlotRange->All,
  PlotMarkers->Automatic,
  PlotStyle->Purple,
  PlotLegends->Placed[{"λ=15"},{0.8,0.85}],
  ImageSize->300,
  AxesLabel->{None,"PMF"}
```

```
    ]
  ]
```

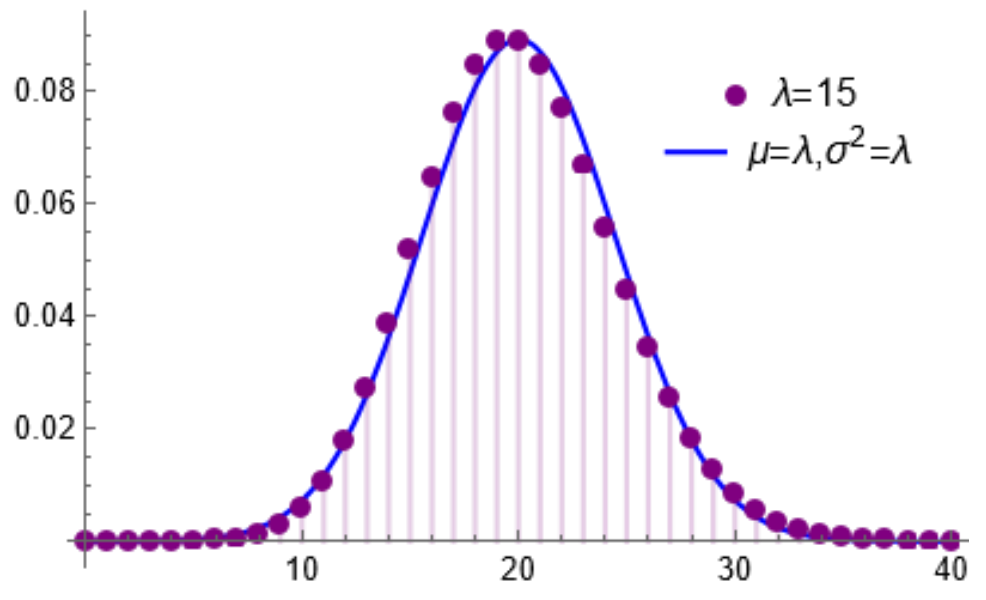

## 12.5 Probability Plots

If a simple random sample is taken from a population, the distribution of the observed values of a variable will approximate the distribution of the variable—and the larger the sample, the better the approximation tends to be. We can use this fact to help decide whether a variable is normally distributed. If a variable is normally distributed, then, for a large sample, a histogram of the observations should be roughly bell shaped; for a very large sample, even moderate departures from a bell shape cast doubt on the normality of the variable. However, for a relatively small sample, ascertaining a clear shape in a histogram and, in particular, whether it is bell shaped is often difficult. Thus, for relatively small samples, a more sensitive graphical technique than the ones we have presented so far is required for assessing normality. Probability plots (Quantile-Quantile plot (Q–Q plot) and Probability-Probability plot (P-P plot)) provide such techniques.

Q-Q plots and P-P plots are useful for assessing the normality assumption of a dataset, but they can also be used to compare data against other distributions, such as the exponential or uniform distribution. They provide visual evidence of whether the data conform to a particular distribution or exhibit departures from it.

### Q–Q plot

A Q–Q plot is a plot of the quantiles of two distributions against each other, or a plot based on estimates of the quantiles. A point $(x, y)$ on the plot corresponds to one of the quantiles of the second distribution ($y$-coordinate) plotted against the same quantile of the first distribution ($x$-coordinate). The pattern of points in the plot is used to compare the two distributions. If the two distributions being compared are similar, the points in the Q–Q plot will approximately lie on the straight line.

- A Q–Q plot is used to compare the shapes of distributions, providing a graphical view of how properties such as location, scale, and skewness are similar or different in the two distributions.
- Q–Q plots can be used to compare collections of data too.
- The main step in constructing a Q–Q plot is calculating or estimating the quantiles to be plotted. If one or both of the axes in a Q–Q plot is based on a theoretical distribution with a continuous CDF, all quantiles are uniquely defined and can be obtained by inverting the CDF.
- If a theoretical probability distribution with a discontinuous CDF is one of the two distributions being compared, some of the quantiles may not be defined, so an interpolated quantile may be plotted.
- If the Q–Q plot is based on data, there are multiple quantile estimators in use. Rules for forming Q–Q plots when quantiles must be estimated or interpolated are called plotting positions.

How a Q-Q plot is constructed:

- Sort the data:
  Start by sorting the dataset in ascending order. This sorted dataset will be used to calculate the quantiles.
- Calculate theoretical quantiles:
  Choose a theoretical distribution that you want to compare the dataset against. For example, if you want to check for normality, you would use the normal distribution. Calculate the theoretical quantiles corresponding to the sorted data points using the chosen distribution. There are various methods available to estimate the theoretical quantiles, depending on the distribution.
- Calculate empirical quantiles:
  Calculate the empirical quantiles for the sorted data points. The empirical quantiles represent the percentiles of the dataset. For example, the 25th percentile would be the value below which 25% of the data falls.
- Plot the Q-Q plot:
  Plot the empirical quantiles on the y-axis and the corresponding theoretical quantiles on the x-axis. Each data point represents a pair of quantiles (theoretical quantile, empirical quantile). Connect these points to visualize the Q-Q plot.

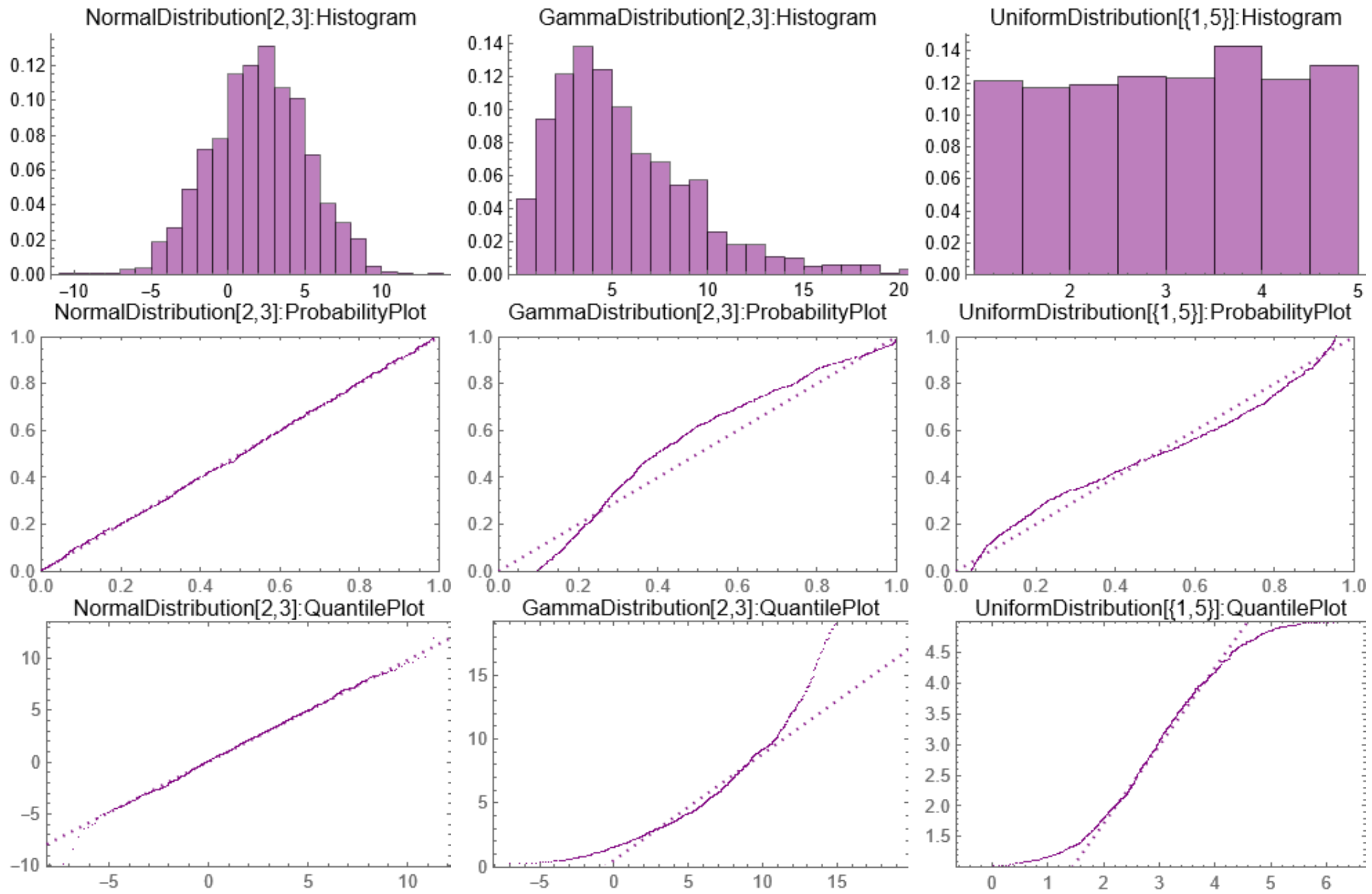

**Figure 12.8.** Histogram, ProbabilityPlot, and QuantilePlot of normal distribution (first column), gamma distribution (second column) and uniform distribution (third column). `ProbabilityPlot[list]` generates a plot of the CDF of list against the CDF of a normal distribution. `QuantilePlot[list]` generates a plot of quantiles of list against the quantiles of a normal distribution.

**P-P plot**

A P-P plot is another graphical tool used to assess the fit of a dataset to a specific distribution or to compare two datasets. The P-P plot compares the observed CDF values of the data against the expected CDFs under the assumed distribution. It helps visualize whether the data conform to a particular distribution or exhibit departures from it.

How a P-P plot is constructed:

- Sort the data in ascending order.
- Calculate the CDF values for each data point based on the rank (i.e., the position of the data point in the sorted list).
- Determine the corresponding expected CDFs for the assumed distribution.
- Plot the observed CDFs against the expected CDFs on a scatter plot.

If the observed CDFs fall approximately on a straight line, it suggests that the data follow the assumed distribution. Deviations from the straight line indicate deviations from the assumed distribution.

**Remarks:**

- As you can see, Figure 12.8, (first column), just because data come from a normal distribution does not imply that they will be perfectly linear. In general, the points are close to the line, but small patterns such as in the upper right or the gap in the lower left can occur without invalidating the normality assumption.
- In addition to the overall shape of the plot, specific patterns or deviations can be observed. Outliers, heavy tails, and multimodal distributions can be detected through departures from the straight line. S-shaped or J-shaped curves can indicate departures from normality. (See Figure 12.8.)

# CHAPTER 13

# MATHEMATICA LAB: CONTINUOUS RANDOM VARIABLES AND PROBABILITY DISTRIBUTIONS

In this chapter, we explore the world of continuous random variables, which are essential for modeling phenomena with uncountable outcomes. We study the continuous probability distributions, exploring their PDFs, CDFs, and MGFs. We will demonstrate how to define, manipulate, and analyze continuous random variables and probability distributions using Mathematica's syntax and functionality. By engaging in hands-on exercises and experimenting with different scenarios, readers will enhance their understanding of the concepts and develop proficiency in utilizing Mathematica for probability analysis.

- In Mathematica, the functions `Probability`, `NProbability`, `PDF`, `CDF`, `Expectation`, `MomentGeneratingFunction`, and `CentralMomentGeneratingFunction` are versatile tools that can be used with both discrete and continuous random variables. Users gain the advantage of a unified framework for analyzing probability distributions and can apply these functions to estimate probabilities, calculate expected values, and other characteristics of both discrete and continuous random variables.
- Additionally, Mathematica provides a comprehensive set of built-in functions to effortlessly handle various common continuous probability distributions. In this chapter, we study four fundamental probability distributions: Normal distribution, exponential distribution, uniform distribution, and gamma distribution. These distributions find applications in various fields, such as physics, finance, and engineering, and Mathematica equips users with the necessary tools to analyze, visualize, and extract meaningful information from them.
- Derived distributions in mathematics are modifications or transformations of existing distributions, achieved through functions of random variables or weighted mixtures. In this chapter, we also study in detail two important built-in Mathematica functions `TransformedDistribution` and `MixtureDistribution`.

In the following table, we list the built-in functions that are used in this chapter.

| | | |
|---|---|---|
| `Distributed` | `Expectation` | `UniformDistribution` |
| `Conditioned` | `NExpectation` | `ExponentialDistribution` |
| `Probability` | `MomentGeneratingFunction` | `GammaDistribution` |
| `NProbability` | `CentralMomentGeneratingFunction` | `Normal Distribution` |
| `PDF` | `EstimatedDistribution` | `TransformedDistribution` |
| `CDF` | `ProbabilityPlot` | `MixtureDistribution` |
| | `QuantilePlot` | |

Therefore, we divided this chapter into seven units to cover the above topics.

**Chapter 13 Outline**

Unit 13.1. Continuous Random Variables
Unit 13.2. Uniform Distribution
Unit 13.3. Exponential Distribution
Unit 13.4. Gamma Distribution
Unit 13.5. Normal Distribution
Unit 13.6. Probability Plots
Unit 13.7. Derived Statistical Distributions

# UNIT 13.1

# CONTINUOUS RANDOM VARIABLES

In this unit, our primary focus will be on the functions `Probability`, `CDF`, `InverseCDF`, and `Quantile`. We will examine how these functions are used in the practical applications, specifically in calculating probabilities and determining z-scores for continuous random variables.

***Mathematica Examples 13.1***

Input

```
(* In this code, we define the PDF using the pdf function, which represents a normal distribution with a mean of 0 and a standard deviation of 1. Then, the cdf function calculates the integral of the PDF from negative infinity to x, representing the CDF. Finally, we compute the CDF values using both our cdf function and the built-in CDF function for the given value 1.5. The output will demonstrate that the results from both approaches are the same: *)

(* Define a continuous distribution: *)
dist=NormalDistribution[0,1];

(* Define the PDF and CDF functions: *)
pdf[x_]:=PDF[dist,x]
cdf[x_]:=Integrate[pdf[t],{t,-Infinity,x}]

(* Choose a specific value to evaluate: *)
x=1.5;

(* Compute the CDF using the manual integration: *)
manualCDF=cdf[x]

(* Compute the CDF using the built-in CDF function: *)
builtInCDF=CDF[dist,x]

(* Compare the results: *)
difference=manualCDF-builtInCDF
```

Output 0.933193

Output 0.933193

Output -1.11022*10^-16

***Mathematica Examples 13.2***

Input

(* The CDF is the integral of the PDF for continuous distributions $F(x)=\int_{-\infty}^{x} f(\xi)d\xi$: *)

```
Simplify[
 Integrate[
  PDF[ExponentialDistribution[λ],x],
  {x,0,y},
  Assumptions->λ>0&&y\[Element]Reals
  ]]
CDF[ExponentialDistribution[λ],y]
```

Output $\begin{cases} 1-e^{-y\lambda} & y>0 \\ 0 & \text{True} \end{cases}$

Output $\begin{cases} 1-e^{-y\lambda} & y\geq 0 \\ 0 & \text{True} \end{cases}$

***Mathematica Examples 13.3***

Input
```
(* The PDF and CDF of a univariate continuous distribution: *)

PDF[NormalDistribution[μ,σ],x]
CDF[NormalDistribution[μ,σ],x]
MomentGeneratingFunction[NormalDistribution[μ,σ],x]
Mean[NormalDistribution[μ,σ]]
Variance[NormalDistribution[μ,σ]]
```

Output $\frac{e^{-\frac{(x-\mu)^2}{2\sigma^2}}}{\sqrt{2\pi}\sigma}$

Output $\frac{1}{2}\text{Erfc}[\frac{-x+\mu}{\sqrt{2}\,\sigma}]$

Output $e^{x\,\mu+\frac{x^2\sigma^2}{2}}$

Output μ

Output σ^2

***Mathematica Examples 13.4***

Input
```
(* The probability of x<=a for a univariate distribution is given by its CDF: *)
{
 Probability[x<=a,x\[Distributed]NormalDistribution[]],
 CDF[NormalDistribution[],a]
 }
```

Output $\{\frac{1}{2}\text{Erfc}\left[-\frac{a}{\sqrt{2}}\right],\frac{1}{2}\text{Erfc}\left[-\frac{a}{\sqrt{2}}\right]\}$

***Mathematica Examples 13.5***

Input
```
(* A univariate CDF is 0 at -∞ and 1 at ∞:*)
CDF[NormalDistribution[],-∞]
CDF[NormalDistribution[],∞]
```

Output 0

Output 1

***Mathematica Examples 13.6***

Input
```
(* Compute the probability of x<=3.5 for normal distribution with μ=3, σ=1: *)

p01=NProbability[x<=3.5,x\[Distributed]NormalDistribution[3,1]];
p02=N[CDF[NormalDistribution[3,1],3.5]];
p03=N[1-CDF[NormalDistribution[3,1],2.5]];(* From symmetry of normal distribution
about the mean at μ=3: *)
{p01,p02,p03}
```

Output {0.691462,0.691462,0.691462}

***Mathematica Examples 13.7***

Input
```
(* Compute the probability of x<=1 for normal distribution with μ=0, σ=2: *)

p04=NProbability[x<=1,x\[Distributed]NormalDistribution[0,2]];
p05=N[CDF[NormalDistribution[0,2],1]];
```

```
        p06=N[1-CDF[NormalDistribution[0,2],-1]];(* From symmetry of normal distribution
        about the mean at μ=0: *)
        {p04,p05,p06}

Output  {0.691462,0.691462,0.691462}
```

***Mathematica Examples 13.8***

```
Input   (* Compute the probability x>3.5 for normal distribution with μ=3, σ=1: *)

        p07=NProbability[x>3.5,x\[Distributed]NormalDistribution[3,1]];
        p08=1-CDF[NormalDistribution[3,1],3.5];
        p09=CDF[NormalDistribution[3,1],2.5];(* From symmetry of normal distribution about
        the mean at μ=3: *)
        {p07,p08,p09}

Output  {0.308538,0.308538,0.308538}
```

***Mathematica Examples 13.9***

```
Input   (* Compute the probability of x>1 for normal distribution with μ=0, σ=2: *)

        p010=NProbability[x>1,x\[Distributed]NormalDistribution[0,2]];
        p011=N[1-CDF[NormalDistribution[0,2],1]];
        p012=N[CDF[NormalDistribution[0,2],-1]];(* From symmetry of normal distribution
        about the mean at μ=0: *)
        {p010,p011,p012}

Output  {0.308538,0.308538,0.308538}
```

***Mathematica Examples 13.10***

```
Input   (* Compute the probability of |x|<3.5 for normal distribution with μ=3, σ=1: *)

        p013=NProbability[Abs[x]<3.5,x\[Distributed]NormalDistribution[3,1]];
        p014=NProbability[-3.5<x<3.5,x\[Distributed]NormalDistribution[3,1]];
        p015=CDF[NormalDistribution[3,1],3.5]-CDF[NormalDistribution[3,1],-3.5];
        (* Note that CDF[NormalDistribution[3,1],-3.5]=4.016000583859125`*10^-11=0 since
        the mean at μ=3: *)
        p016= CDF[NormalDistribution[3,1],3.5];
        p017=1-CDF[NormalDistribution[3,1],2.5]; (* From symmetry of normal distribution
        about the mean at μ=3: *)
        {p013,p014,p015,p016,p017}

Output  {0.691462,0.691462,0.691462,0.691462,0.691462}
```

***Mathematica Examples 13.11***

```
Input   (* Compute the probability of |x|<1 for normal distribution with μ=0, σ=2: *)

        p018=NProbability[Abs[x]<1,x\[Distributed]NormalDistribution[0,2]];
        p019=NProbability[-1<x<1,x\[Distributed]NormalDistribution[0,2]];
        p020=N[CDF[NormalDistribution[0,2],1]-CDF[NormalDistribution[0,2],-1]];
        p021=N[1- CDF[NormalDistribution[0,2],-1]-(1- CDF[NormalDistribution[0,2],1])];
        p022=N[1-2 CDF[NormalDistribution[0,2],-1]]; (* From symmetry of normal
        distribution about the mean at μ=0: *)
        p023=N[1-2(1- CDF[NormalDistribution[0,2],1])];(* From symmetry of normal
        distribution about the mean at μ=0: *)
        {p018,p019,p020,p021,p022, p023}

Output  {0.382925,0.382925,0.382925,0.382925,0.382925,0.382925}
```

***Mathematica Examples 13.12***

```
Input    (* Compute the probability of |x|>3.5 for normal distribution with μ=3, σ=1: *)

         p024=NProbability[Abs[x]>3.5,x\[Distributed]NormalDistribution[3,1]];
         p025=NProbability[x<-3.5||3.5<x,x\[Distributed]NormalDistribution[3,1]];
         p026=CDF[NormalDistribution[3,1],-3.5]+(1- CDF[NormalDistribution[3,1],3.5]);
         (* Note that CDF[NormalDistribution[3,1],-3.5]=4.016000583859125`*10^-11=0 since
         the mean at μ=3: *)
         p027=(1- CDF[NormalDistribution[3,1],3.5]);
         p028=CDF[NormalDistribution[3,1],2.5]; (* From symmetry of normal distribution
         about the mean at μ=3: *)
         {p024,p025,p026,p027,p028}

Output   {0.308538,0.308538,0.308538,0.308538,0.308538}
```

***Mathematica Examples 13.13***

```
Input    (* Compute the probability of |x|>1 for normal distribution with μ=0, σ=2: *)

         p029=NProbability[Abs[x]>1,x\[Distributed]NormalDistribution[0,2]];
         p030=NProbability[x<-1||1<x,x\[Distributed]NormalDistribution[0,2]];
         p031=N[CDF[NormalDistribution[0,2],-1]+(1- CDF[NormalDistribution[0,2],1])];
         p032=2 N[(1- CDF[NormalDistribution[0,2],1])]; (* From symmetry of normal
         distribution about the mean at μ=0: *)
         p033=2N[CDF[NormalDistribution[0,2],-1] ];(* From symmetry of normal distribution
         about the mean at μ=0: *)
         {p029,p030,p031,p032,p033}

Output   {0.617075,0.617075,0.617075,0.617075,0.617075}
```

***Mathematica Examples 13.14***

```
Input    (* Compute the z value of probability=area from left =0.05 for normal distribution
         with μ=0, σ=1: *)

         zscore=SolveValues[
            CDF[NormalDistribution[0,1],z]==0.05,
            z,
            Reals
            ]

Output   {-1.64485}
```

***Mathematica Examples 13.15***

```
Input    (* Compute the z value of probability=area from left = (0.025, 0.05, 0.10, 0.5, 0.90,
         0.95, 0.975) =area from right= (0.975, 0.95, 0.9, 0.5, 0.1, 0.05, 0.025) for normal
         distribution with μ=0, σ=1: *)

         leftarea={0.025,0.05,0.10,0.5,0.90,0.95,0.975}
         rightarea=1-leftarea
         zscores=Flatten[
            Table[
             SolveValues[
              CDF[NormalDistribution[0,1],z]==a,
              z,
              Reals
              ],
             {a,{0.025,0.05,0.10,0.5,0.90,0.95,0.975}}
             ]
            ]
```

```
Output    {0.025,0.05,0.1,0.5,0.9,0.95,0.975}
Output    {0.975,0.95,0.9,0.5,0.1,0.05,0.025}
Output    {-1.95996,-1.64485,-1.28155,0.,1.28155,1.64485,1.95996}
```

***Mathematica Examples 13.16***

```
Input     (* Compute the z value of probability=area from left = (0.025, 0.05, 0.10, 0.5, 0.90,
          0.95, 0.975) =area from right= (0.975, 0.95, 0.9, 0.5, 0.1, 0.05, 0.025) for normal
          distribution with μ=0, σ=1: *)

          leftarea={0.025,0.05,0.10,0.5,0.90,0.95,0.975}
          rightarea=1-leftarea
          z1=InverseCDF[NormalDistribution[0,1],leftarea]
          z2=N[Quantile[NormalDistribution[0,1],leftarea]]

Output    {0.025,0.05,0.1,0.5,0.9,0.95,0.975}
Output    {0.975,0.95,0.9,0.5,0.1,0.05,0.025}
Output    {-1.95996,-1.64485,-1.28155,0.,1.28155,1.64485,1.95996}
Output    {-1.95996,-1.64485,-1.28155,0.,1.28155,1.64485,1.95996}
```

***Mathematica Examples 13.17***

```
Input     (* Compute the z value of probability=area from medial =(0.90,0.95,0.975)=area from
          left =(0.05,0.025,0.0125) for normal distribution with μ=0, σ=1. Note that we have
          two values of z correspond to one area. For example, if the area from medial =0.90,
          there are two boundaries for this area, and the areas from left, that correspond to
          these boundaries, are 0.05 and 0.95: *)

          medialarea={0.90,0.95,0.975}
          leftarea={0.05,0.95,0.025,0.975,0.0125,0.9875}
          z1=InverseCDF[NormalDistribution[0,1],leftarea]
          z2=N[Quantile[NormalDistribution[0,1],leftarea]]

Output    {0.9,0.95,0.975}
Output    {0.05,0.95,0.025,0.975,0.0125,0.9875}
Output    {-1.64485,1.64485,-1.95996,1.95996,-2.2414,2.2414}
Output    {-1.64485,1.64485,-1.95996,1.95996,-2.2414,2.2414}
```

# UNIT 13.2

# UNIFORM DISTRIBUTION

***Mathematica Examples 13.18***

Input

```
(* The code generates a plot of the probability density function (PDF) for Uniform distribution with different values of max= (2, 3 and 4) and min=0. The plot shows the values of the PDF for all possible values of x between 0 and 6: *)

Plot[
 Evaluate[
  Table[
   PDF[
    UniformDistribution[{0,max}],
    x
    ],
   {max,{2,3,4}}
   ]
  ],
 {x,0,6},
 PlotRange->All,
 Filling->Axis,
 PlotLegends->Placed[{"a=0,b=2","a=0,b=3","a=0,b=4"},{0.8,0.75}],
 PlotStyle->{RGBColor[0.88,0.61,0.14],RGBColor[0.37,0.5,0.7],Purple},
 ImageSize->320,
 AxesLabel->{None,"PDF"}
 ]
```

Output

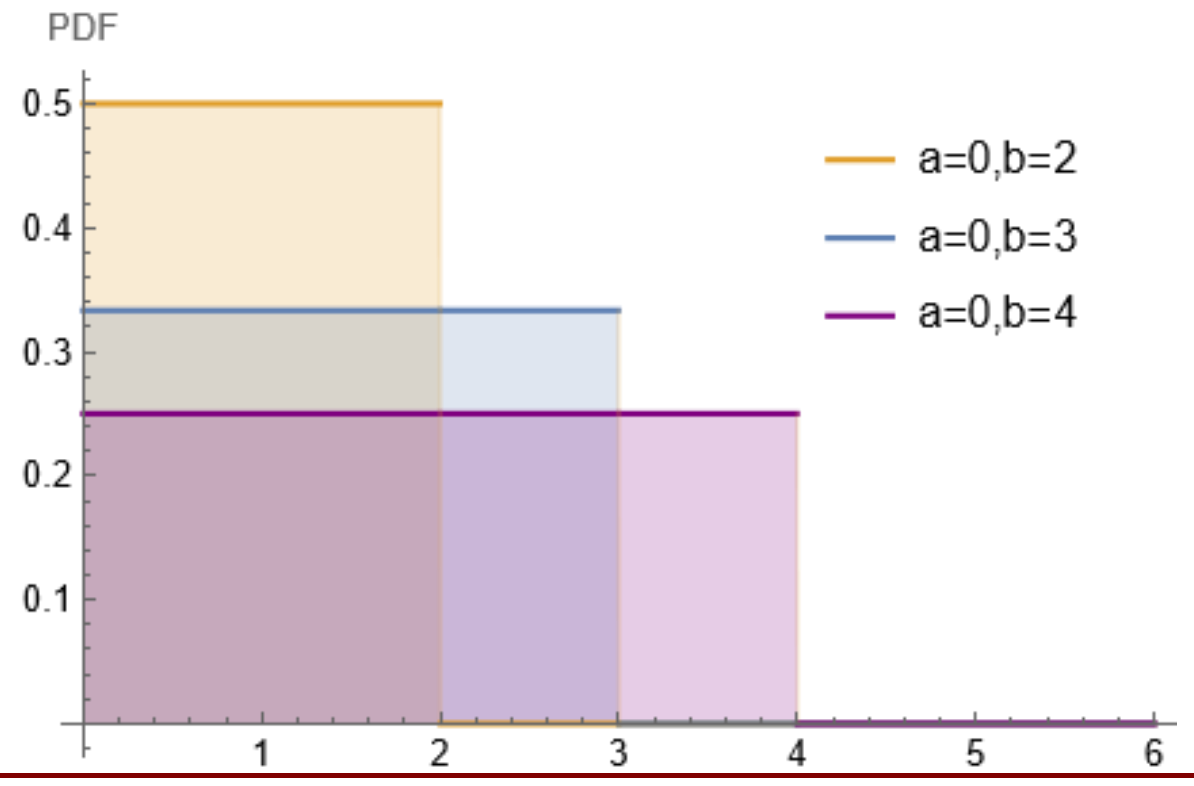

***Mathematica Examples 13.19***

Input

```
(* The code generates a plot of the CDF of Uniform distribution with different values of max= (2, 3 and 4) and min=0: *)

Plot[
 Evaluate[
  Table[
   CDF[
    UniformDistribution[{0,max}],
    x
    ],
   {max,{2,3,4}}
```

```
      ]
     ],
    {x,0,6},
    PlotRange->All,
    Filling->Axis,
    Exclusions->None,
    PlotLegends->Placed[{"a=0,b=2","a=0,b=3","a=0,b=4"},{0.8,0.75}],
    PlotStyle->{RGBColor[0.88,0.61,0.14],RGBColor[0.37,0.5,0.7],Purple},
    ImageSize->320,
    AxesLabel->{None,"CDF"}
    ]
```

Output

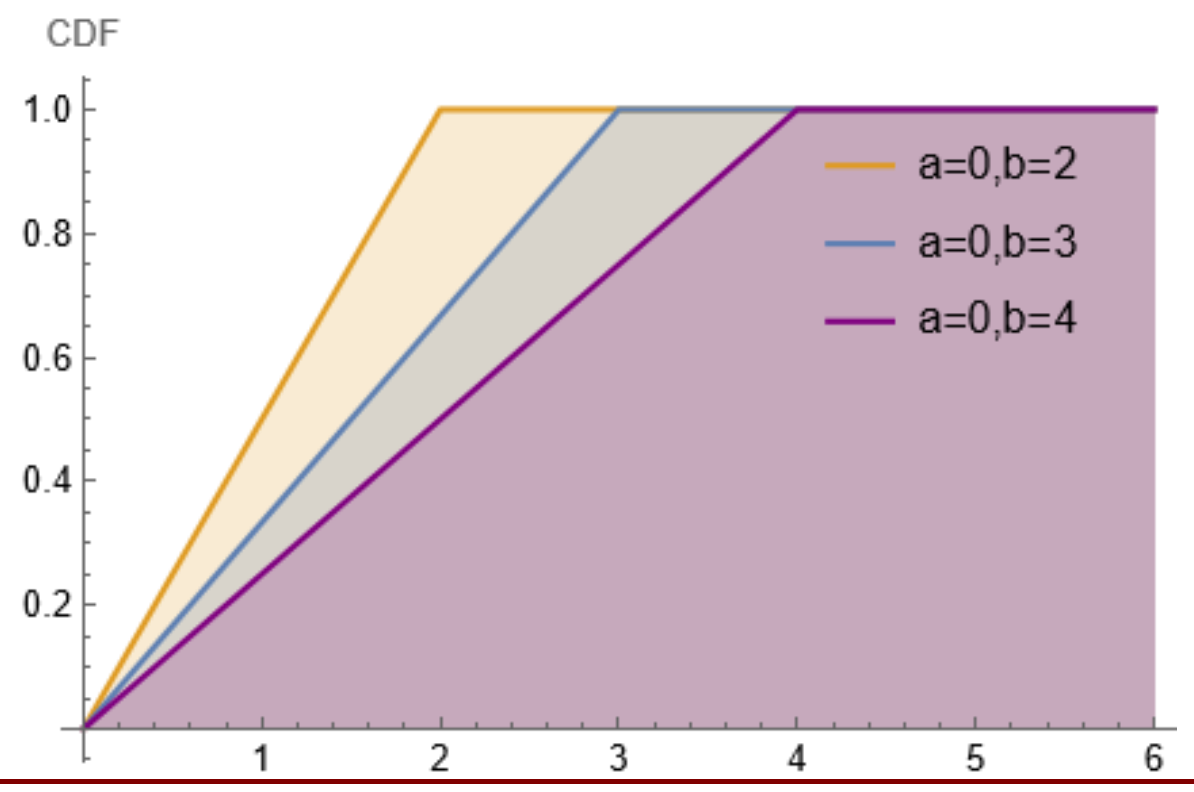

***Mathematica Examples 13.20***

Input

```
(* The code generates a histogram and a plot of the PDF for Uniform distribution with
parameters min=1 and max=3 and sample size 10000: *)

data=RandomVariate[
    UniformDistribution[
      {1,3}],
    10^4
    ];

Show[
 Histogram[
  data,
  20,
  "PDF",
  ColorFunction->Function[{height},Opacity[height]],
  ChartStyle->Purple,
  ImageSize->320,
  AxesLabel->{None,"PDF"}
  ],

 Plot[
  PDF[
   UniformDistribution[{1,3}],
   x
   ],
  {x,0,4},
  PlotStyle->RGBColor[0.88,0.61,0.14],
  PlotRange->{0,4}
  ]
 ]
```

Output

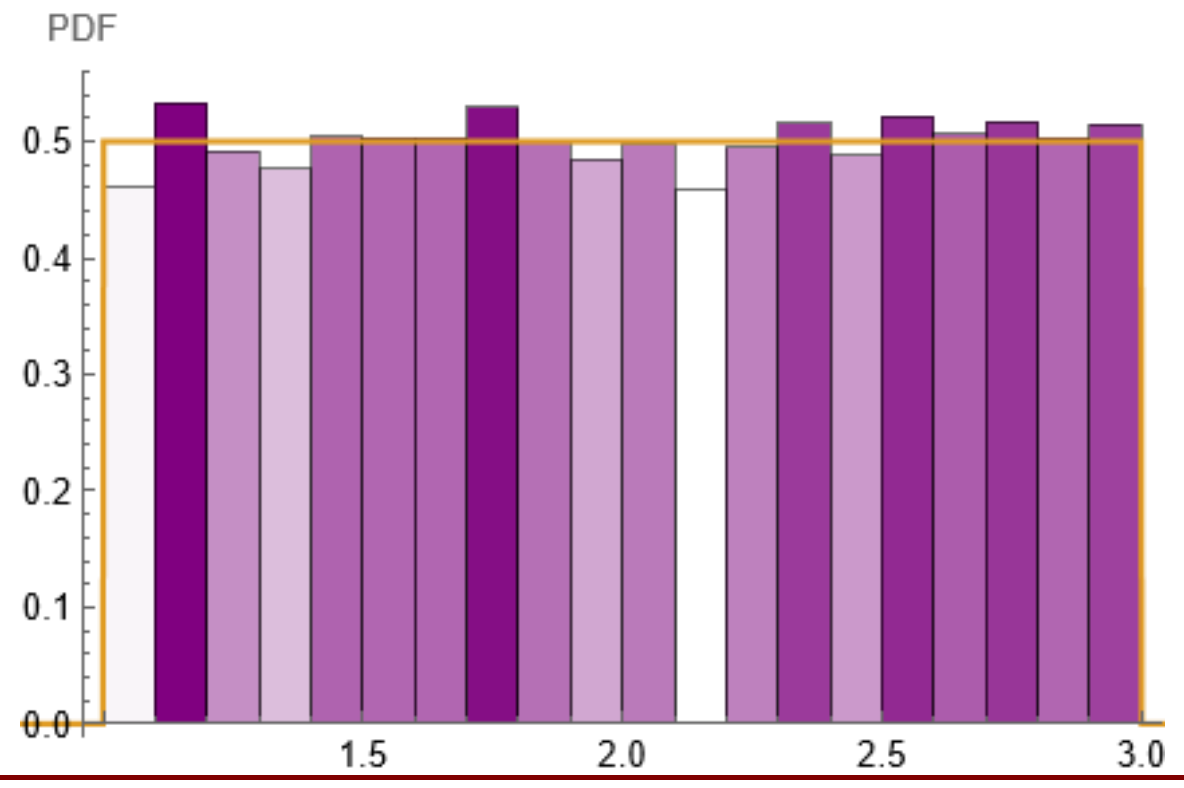

***Mathematica Examples 13.21***

Input

```
(* The code creates a dynamic histogram of data and a plot of the PDF generated from
Uniform distribution using the Manipulate function. The Manipulate function creates
interactive controls for the user to adjust the values of min and max, which are the
parameters of the Uniform distribution and the sample size: *)

Manipulate[
 Module[
  {
   data=RandomVariate[
     UniformDistribution[{min,max}],
     n
     ]
   },

  Show[
   Histogram[
    data,
    Automatic,
    "PDF",
    ColorFunction->Function[{height},Opacity[height]],
    ImageSize->320,
    ChartStyle->Purple
    ],
   Plot[
    PDF[
     UniformDistribution[{min,max}],
     x
     ],
    {x,0,7},
    ColorFunction->"Rainbow"
    ]
   ]
  ],
 {{min,1,"min"},1,4,0.1},
 {{max,4.5,"max"},4.5,7,0.1},
 {{n,300,"n"},100,1000,10}
 ]
```

Output

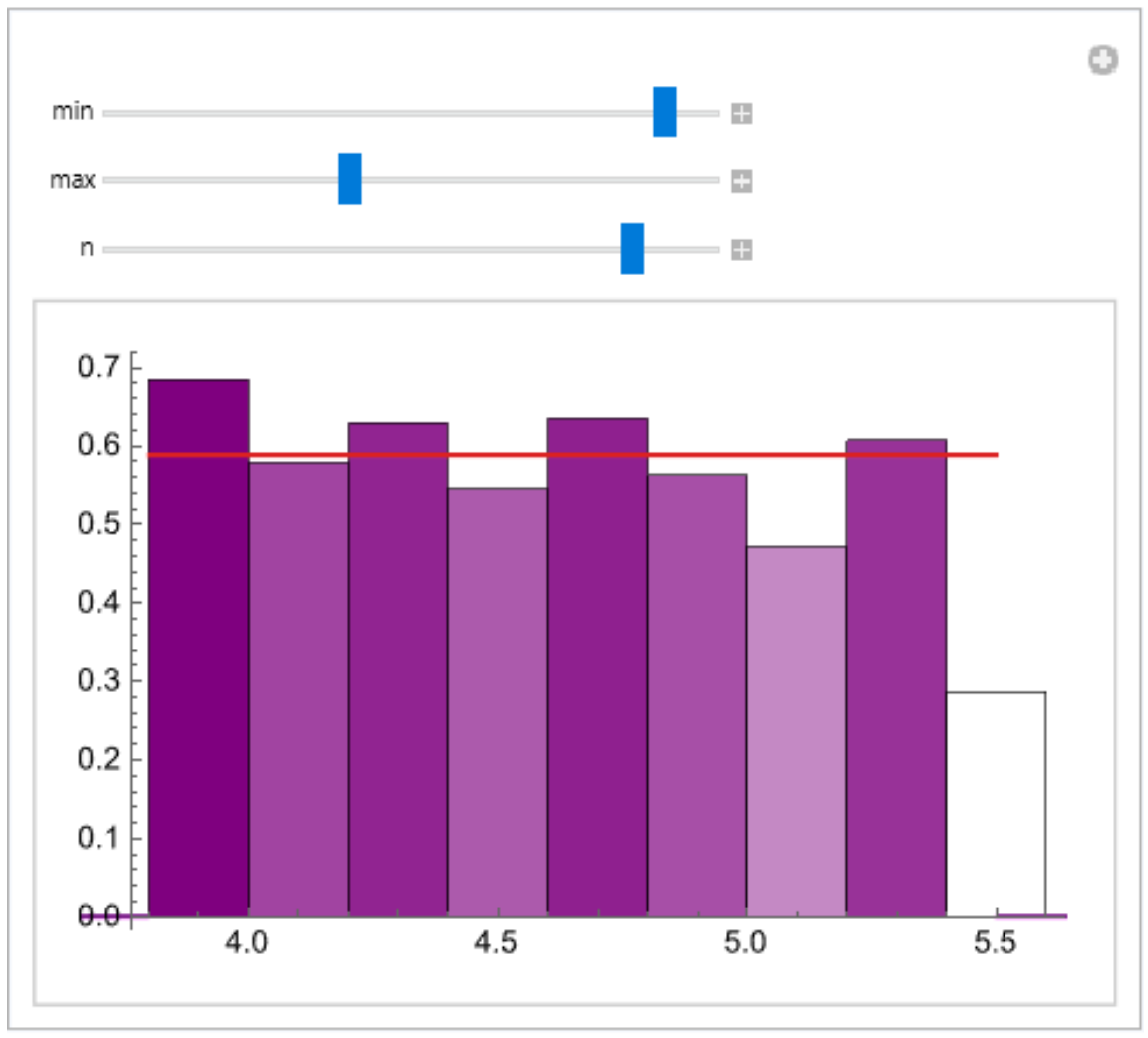

***Mathematica Examples 13.22***

Input

```
(* The code creates a plot of the CDF of Uniform distribution using the Manipulate
function. The Manipulate function allows you to interactively change the values of
the parameters min and max, respectively: *)
Manipulate[
 Plot[
  CDF[
   UniformDistribution[{min,max}],
   x
   ],
  {x,1,8},
  Filling->Axis,
  FillingStyle->LightPurple,
  PlotRange->All,
  PlotLabel->Row[{"min = ",min,", max = ",max}],
  AxesLabel->{"x","CDF"},
  ImageSize->320,
  PlotStyle->Purple
  ],
 {{min,1,"min"},1,4,0.1},
 {{max,4.5,"max"},4.5,7,0.1}
 ]
```

Output

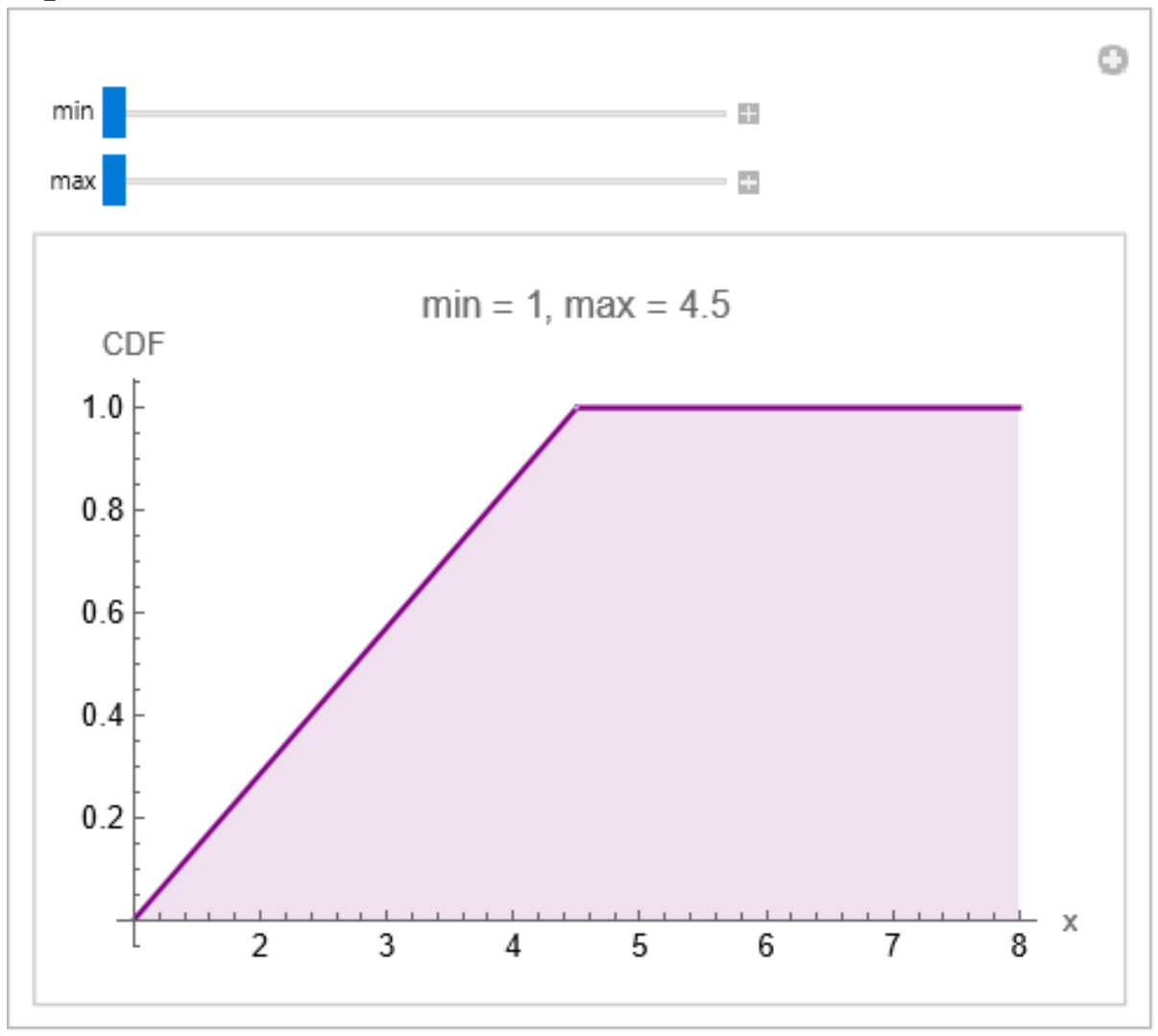

*Mathematica Examples 13.23*

Input

```
(* The code uses the Grid function to create a grid of two plots, one for the PDF and one for the CDF of Uniform distribution. The code uses slider controls to adjust the values of min and max: *)

Manipulate[
 Grid[
  {
   {Plot[
     PDF[
      UniformDistribution[{min,max}],
      x
      ],
     {x,0,8},
     PlotRange->All,
     PlotStyle->{Purple,PointSize[0.03]},
     PlotLabel->"PDF of Uniform distribution",
     AxesLabel->{"x","PDF"}
     ],
    Plot[
     CDF[
      UniformDistribution[{min,max}],
      x
      ],
     {x,0,8},
     PlotRange->All,
     PlotStyle->{Purple,PointSize[0.03]},
     PlotLabel->"CDF of Uniform distribution",
     AxesLabel->{"x","CDF"}
     ]
    }
   },
  Spacings->{5,5}
  ],
 {{min,1,"min"},1,4,0.1},
 {{max,4.5,"max"},4.5,7,0.1}
 ]
```

Output

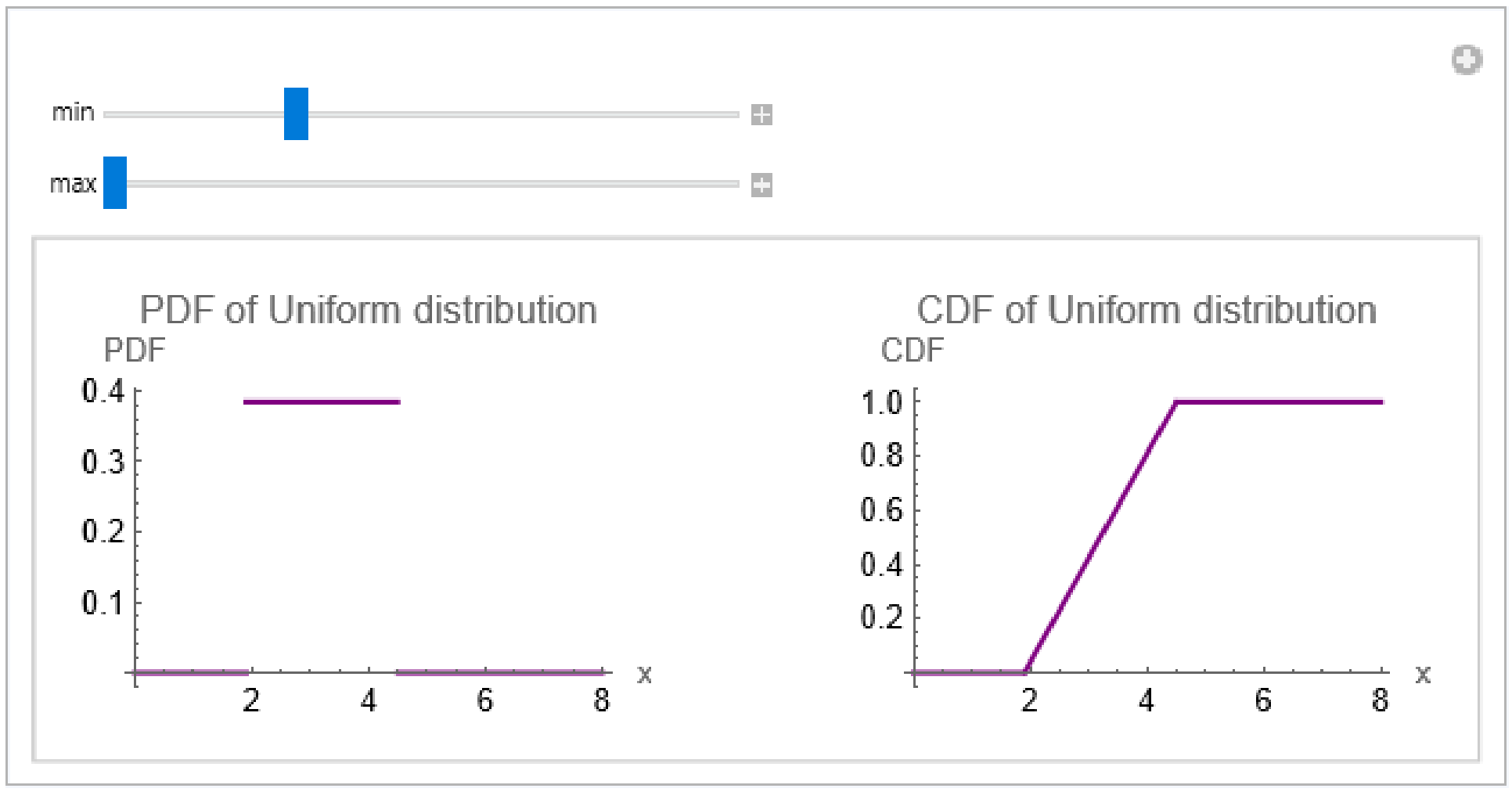

***Mathematica Examples 13.24***

Input

```
(* The code calculates and displays some descriptive statistics (mean, variance,
standard deviation, kurtosis and skewness) for Uniform distribution with parameters
a and b: *)

Grid[
 Table[
  {
   statistics,
   FullSimplify[statistics[UniformDistribution[{a,b}]]]
   },
  {statistics,{Mean,Variance,StandardDeviation,Kurtosis,Skewness}}
  ],
 ItemStyle->12,
 Alignment->{{Right,Left}},
 Frame->All,
 Spacings->{Automatic,0.8}
 ]
```

Output

| | |
|---|---|
| Mean | (a+b)/2 |
| Variance | 1/12 (a-b)^2 |
| StandardDeviation | (-a + b)/(2 Sqrt[3]) |
| Kurtosis | 9/5 |
| Skewness | 0 |

***Mathematica Examples 13.25***

Input

```
(* The code calculates and displays some additional descriptive statistics (moments,
central moments, and factorial moments) for Uniform distribution with parameters a
and b: *)

Grid[
 Table[
  {
   statistics,
   FullSimplify[statistics[UniformDistribution[{a,b}],1]],
   FullSimplify[statistics[UniformDistribution[{a,b}],2]]
   },
  {statistics,{Moment,CentralMoment,FactorialMoment}}
  ],
 ItemStyle->12,
 Alignment->{{Right,Left}},
 Frame->All,
 Spacings->{Automatic,0.8}
 ]
```

Output

| | | |
|---|---|---|
| Moment | (a+b)/2 | 1/3 (a^2+a b+b^2) |
| CentralMoment | 0 | 1/12 (a-b)^2 |
| FactorialMoment | (a+b)/2 | 1/6 (2 a^2+a (-3+2 b)+b (-3+2 b)) |

***Mathematica Examples 13.26***

Input

```
(* The code generates a dataset of 1000 observations from Uniform distribution with
parameters min=1 and max=7. Then, it computes the sample mean and quartiles of the
data, and plots a histogram of the data and plot of the PDF. Additionally, the code
adds vertical lines to the plot corresponding to the sample mean and quartiles: *)

data=RandomVariate[
   UniformDistribution[{1,7}],
   1000
   ];
```

```
mean=Mean[data];
quartiles=Quantile[
    data,
    {0.25,0.5,0.75}
    ];

Show[
 Histogram[
  data,
  Automatic,
  "PDF",
  Epilog->{
     Directive[Red,Thickness[0.006]],
     Line[{{mean,0},{mean,0.25}}],
     Directive[Green,Dashed],
     Line[{{quartiles[[1]],0},{quartiles[[1]],0.25}}],
     Line[{{quartiles[[2]],0},{quartiles[[2]],0.25}}],
     Line[{{quartiles[[3]],0},{quartiles[[3]],0.25}}]
     },
  ColorFunction->Function[{height},Opacity[height]],
  ImageSize->320,
  ChartStyle->Purple,
  PlotRange->{{0,8},{0,0.2}}
  ],
 Plot[
  PDF[UniformDistribution[{1,7}],x],
  {x,0,8},
  ImageSize->320,
  ColorFunction->"Rainbow"
  ]
 ]
```

Output

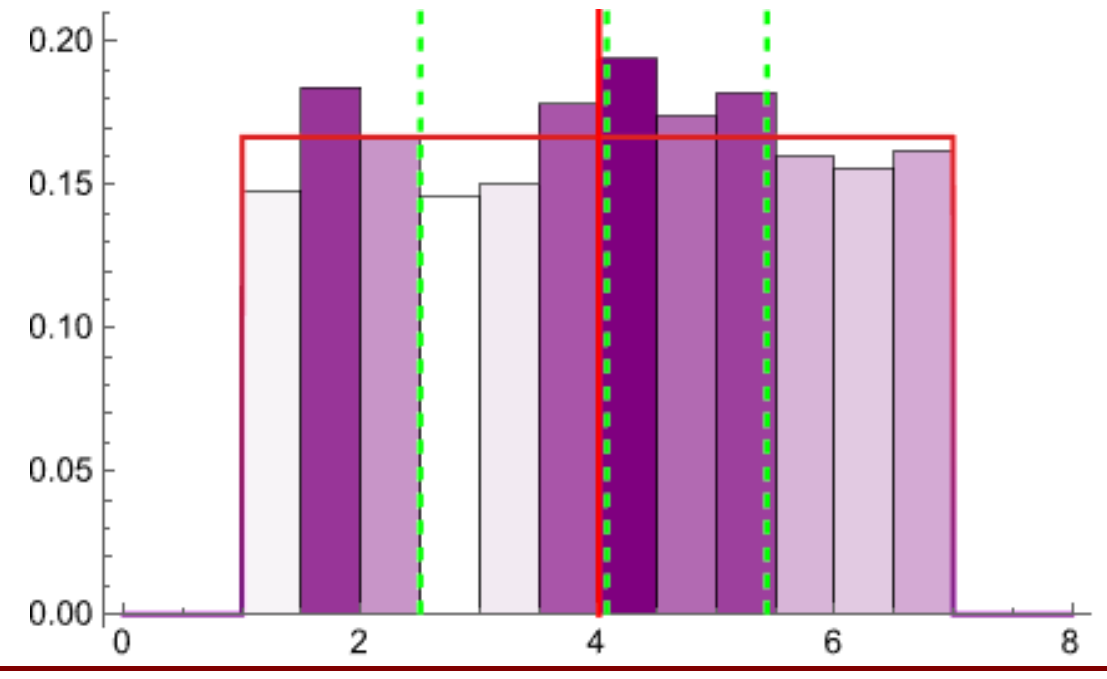

***Mathematica Examples 13.27***

Input

```
(* The code generates a random sample of size 10,000 from Uniform distribution with parameters min=1 and max=7, estimates the distribution parameters using the EstimatedDistribution function, and then compares the histogram of the sample with the estimated PDF of the normal distribution using a histogram and a plot of the PDF: *)

sampledata=RandomVariate[
    UniformDistribution[{1,7}],
    10^4
    ];
(* Estimate the distribution parameters from sample data: *)
ed=EstimatedDistribution[
   sampledata,
```

```
  UniformDistribution[{min,max}]
  ]
(* Compare a density histogram of the sample with the PDF of the estimated distribution: *)
Show[
 Histogram[
  sampledata,
  {1},
  "PDF",
  ColorFunction->Function[{height},Opacity[height]],
  ChartStyle->Purple,
  ImageSize->320
  ],
 Plot[
  PDF[ed,x],
  {x,0,8},
  ImageSize->320,
  ColorFunction->"Rainbow"
  ]
 ]
```

Output UniformDistribution[{1.00068,6.99972}]

Output

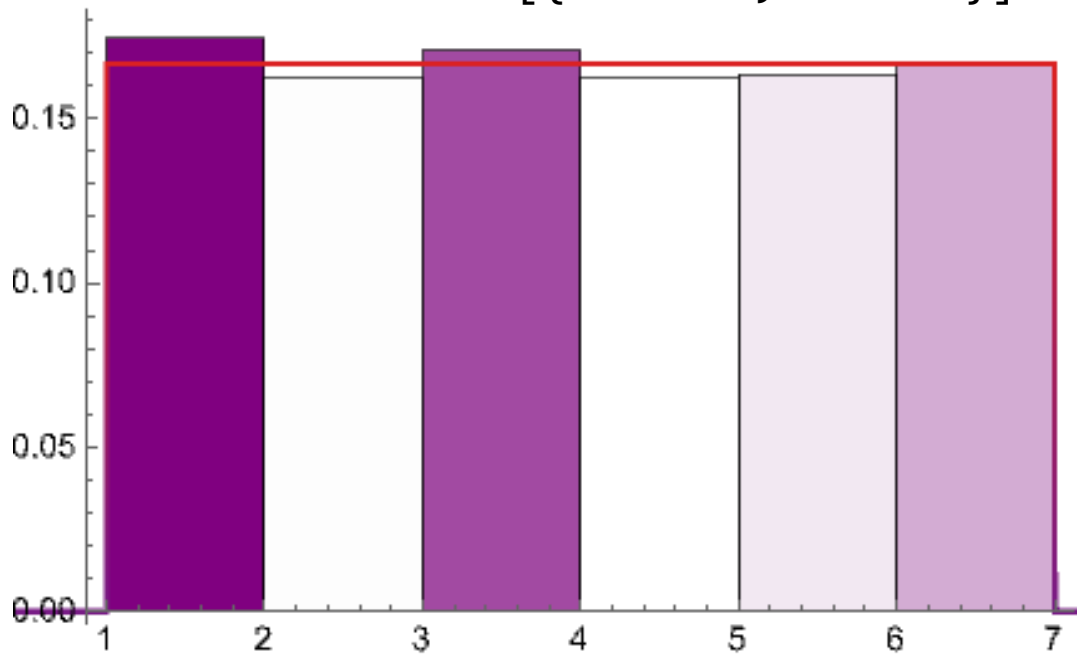

***Mathematica Examples 13.28***

Input

```
(* The code generates a 2D dataset with 1000 random points that follow Uniform distribution with min=0 and max=6. The dataset is then used to create a row of three plots. The first plot is a histogram of the X-axis values of the dataset. The second plot is a histogram of the Y-axis values of the dataset. It is similar to the first plot, but shows the distribution of the Y-axis values instead. The third plot is a scatter plot of the dataset, with the X-axis values on the horizontal axis and the Y-axis values on the vertical axis. Each point in the plot represents a pair of X and Y values from the dataset: *)

data=RandomVariate[
   UniformDistribution[{0,6}],
   {1000,2}
   ];
GraphicsRow[
 {
  Histogram[
   data[[All,1]],
   {0.1},
   PlotLabel->"X-axis",
   ColorFunction->Function[{height},Opacity[height]],
   ChartStyle->Purple
   ],
  Histogram[
   data[[All,2]],
   {0.1},
```

```
            PlotLabel->"Y-axis",
            ColorFunction->Function[{height},Opacity[height]],
            ChartStyle->Purple
            ],
           ListPlot[
            data,
            PlotStyle->{Purple,PointSize[0.015]},
            AspectRatio->1,
            Frame->True,
            Axes->False
            ]
           }
          ]
```

Output

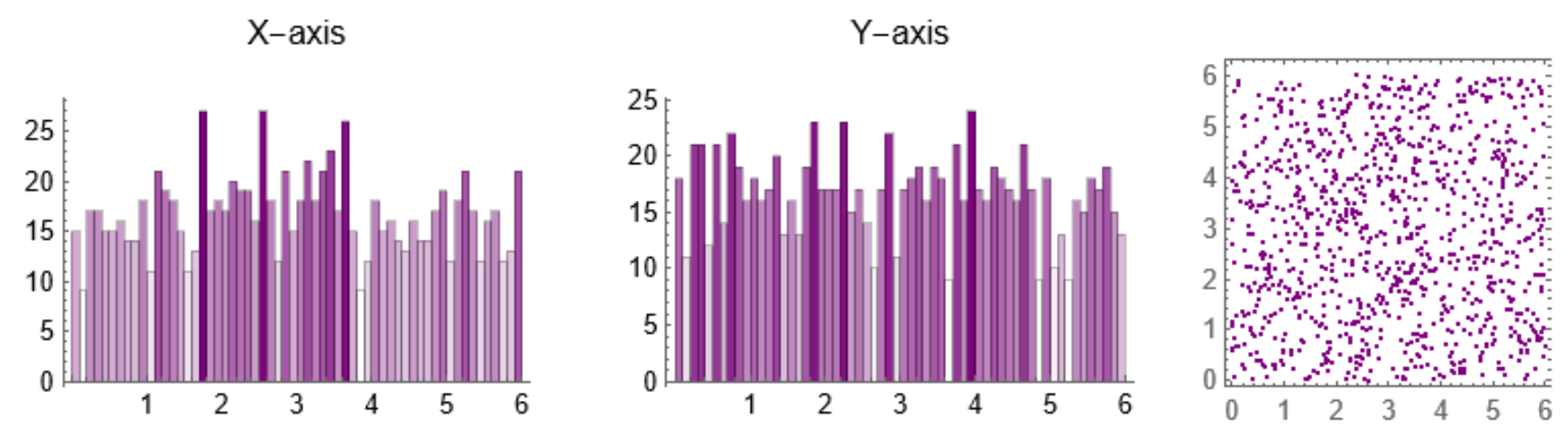

## Mathematica Examples 13.29

Input

```
(* The code generates a set of random data points with Uniform distribution in three dimensions, and then creates three histograms, one for each dimension, showing the distribution of the points along that axis. Additionally, it creates a 3D scatter plot of the data points: *)

data=RandomVariate[
    UniformDistribution[{0,6}],
    {1000,3}
    ];
GraphicsGrid[
 {
  {
   Histogram[
    data[[All,1]],
    Automatic,
    "PDF",
    PlotLabel->"X-axis",
    ColorFunction->Function[{height},Opacity[height]],
    ChartStyle->Purple
    ],
   Histogram[
    data[[All,2]],
    Automatic,
    "PDF",
    PlotLabel->"Y-axis",
    ColorFunction->Function[{height},Opacity[height]],
    ChartStyle->Purple
    ],
   Histogram[
    data[[All,3]],
    Automatic,
    "PDF",
    PlotLabel->"Z-axis",
    ColorFunction->Function[{height},Opacity[height]],
    ChartStyle->Purple
    ],
```

```
        ListPointPlot3D[
         data,
         BoxRatios->{1,1,1},
         PlotStyle->{Purple,PointSize[0.015]}
         ]
        }
      }
    ]
```

Output

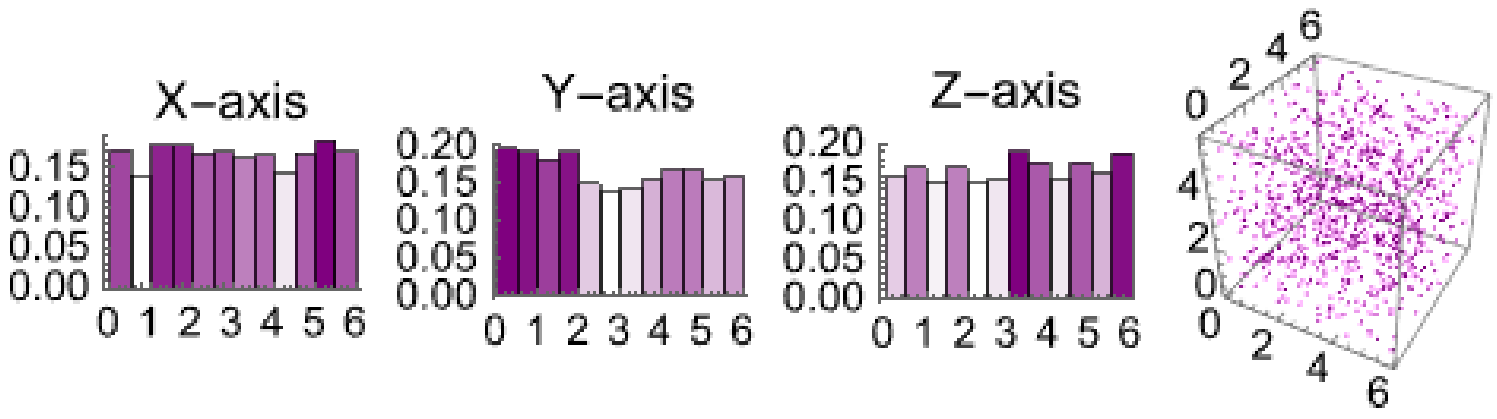

***Mathematica Examples 13.30***

Input

```
(* The code generates a 3D scatter plot of Uniform distributed points, where the x-
axis is red, y-axis is green, and z-axis is blue: *)

data=RandomVariate[
   UniformDistribution[{0,6}],
   {2000,3}
   ];
Graphics3D[
 {
  {PointSize[0.006],Purple,Point[data]},
  Thin,
  {Red,Opacity[0.4],Line[{{#,0,0},{#,0,-0.5}}]&/@data[[All,1]]},
  Thin,
  {Green,Opacity[0.4],Line[{{0,#,0},{0,#,-0.5}}]&/@data[[All,2]]},
  Thin,
  {Blue,Opacity[0.4],Line[{{0,0,#},{0,-0.5,#}}]&/@data[[All,3]]}
  },
 BoxRatios->{1,1,1},
 Axes->True,
 AxesLabel->{"X","Y","Z"},
 ImageSize->320
 ]
```

Output

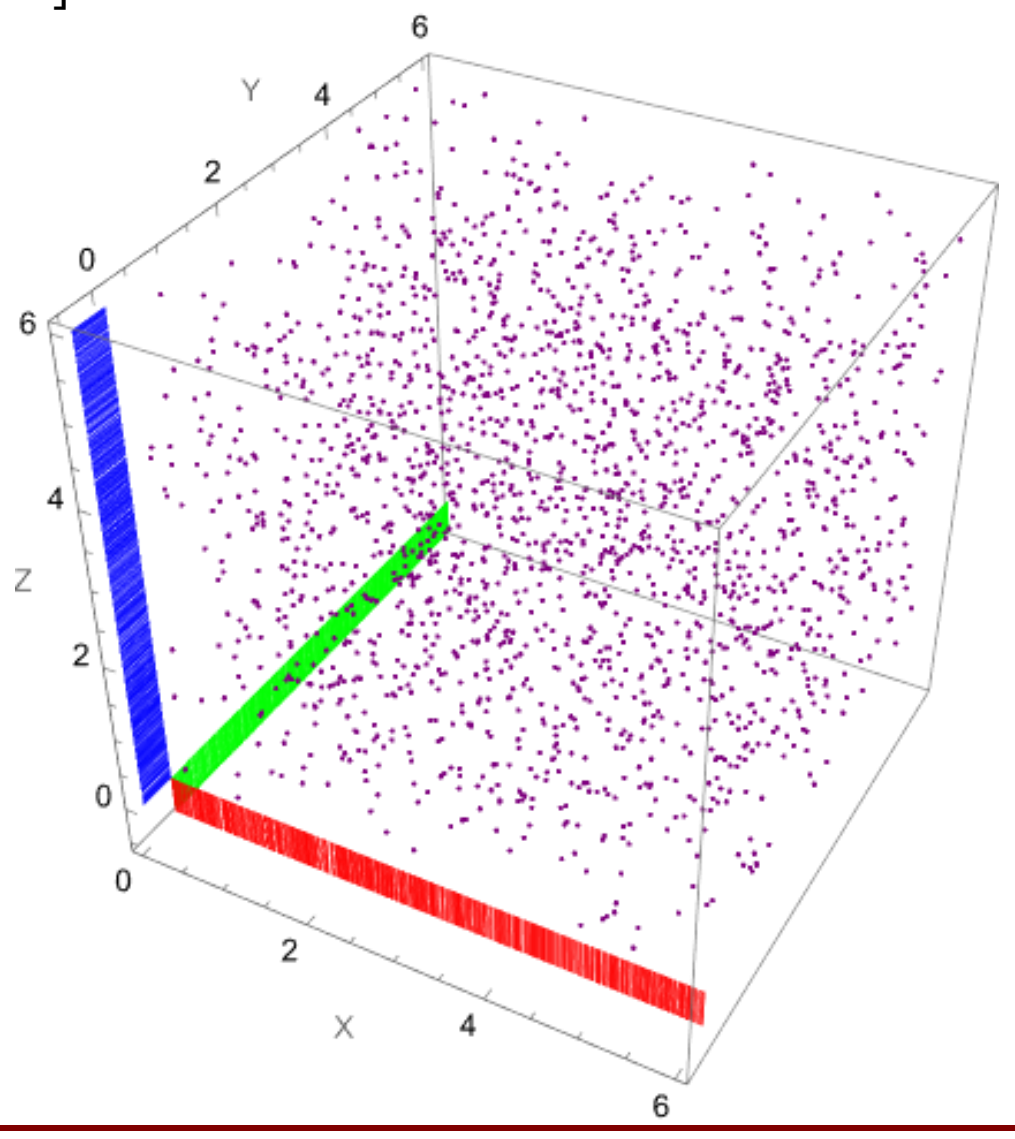

***Mathematica Examples 13.31***

Input

```
(* The code demonstrates a common technique in statistics and data analysis, which is the use of random sampling to estimate population parameters. The code generates random samples from Uniform distribution with min=0 and max=6, and then using these samples to estimate the parameters of another Uniform distribution with unknown min and max. This process is repeated 20 times, resulting in 20 different estimated distributions. The code also visualizes the resulting estimated distributions using the PDF function. The code plots the PDFs of these estimated distributions using the PDF function and the estimated parameters. The plot shows the PDFs in a range from-0 to 5. The code also generates a list plot of 2 sets of random samples from the Uniform distribution with λ=3. The plot shows the 100 random points generated from two random samples. The code generates also a histogram of the PDF for Uniform distribution of the two samples: *)

estim0distributions=Table[
  dist=UniformDistribution[{0,6}];

  sampledata=RandomVariate[
    dist,
    100
    ];

  ed=EstimatedDistribution[
    sampledata,
    UniformDistribution[{min,max}]
    ],
  {i,1,20}
  ]

pdf0ed=Table[
   PDF[estim0distributions[[i]],x],
   {i,1,20}
   ];

(* Visualizes the resulting estimated distributions *)
Plot[
 pdf0ed,
 {x,0,5},
 PlotRange->Full,
 ImageSize->320,
 PlotStyle->Directive[Purple,Opacity[0.3],Thickness[0.002]]
 ]

(* Visualizes 100 random points generated from two random samples *)
table=Table[
   dist=UniformDistribution[{0,6}];
   sampledata=RandomVariate[
     dist,
     100],
   {i,1,2}
   ];

ListPlot[
 table,
 ImageSize->320,
 PlotStyle->Directive[Opacity[0.5],Thickness[0.003]],
 Filling->Axis
 ]
```

```
Histogram[
 table,
 Automatic,
 LabelingFunction->Above,
 ChartLegends->{"Sample 1","Sample 2"},
 ChartStyle->{Directive[Opacity[0.2],Red],Directive[Opacity[0.2],Purple]},
 ImageSize->320
 ]
```

Output

```
{UniformDistribution[{0.00582607,5.94128}],UniformDistribution[{0.0373937,5.94007}]
,UniformDistribution[{0.0214787,5.90539}],UniformDistribution[{0.0693686,5.98272}],
UniformDistribution[{0.0824826,5.96349}],UniformDistribution[{0.0746822,5.83899}],U
niformDistribution[{0.15242,5.92519}],UniformDistribution[{0.113827,5.88749}],Unifo
rmDistribution[{0.0358717,5.86178}],UniformDistribution[{0.000896283,5.95311}],Unif
ormDistribution[{0.00413594,5.84059}],UniformDistribution[{0.0657736,5.97573}],Unif
ormDistribution[{0.0699305,5.94454}],UniformDistribution[{0.118469,5.94359}],Unifor
mDistribution[{0.173582,5.88439}],UniformDistribution[{0.000923629,5.93441}],Unifor
mDistribution[{0.0285627,5.96516}],UniformDistribution[{0.0358934,5.96964}],Uniform
Distribution[{0.0334921,5.98387}],UniformDistribution[{0.0855911,5.97137}]}
```

Output

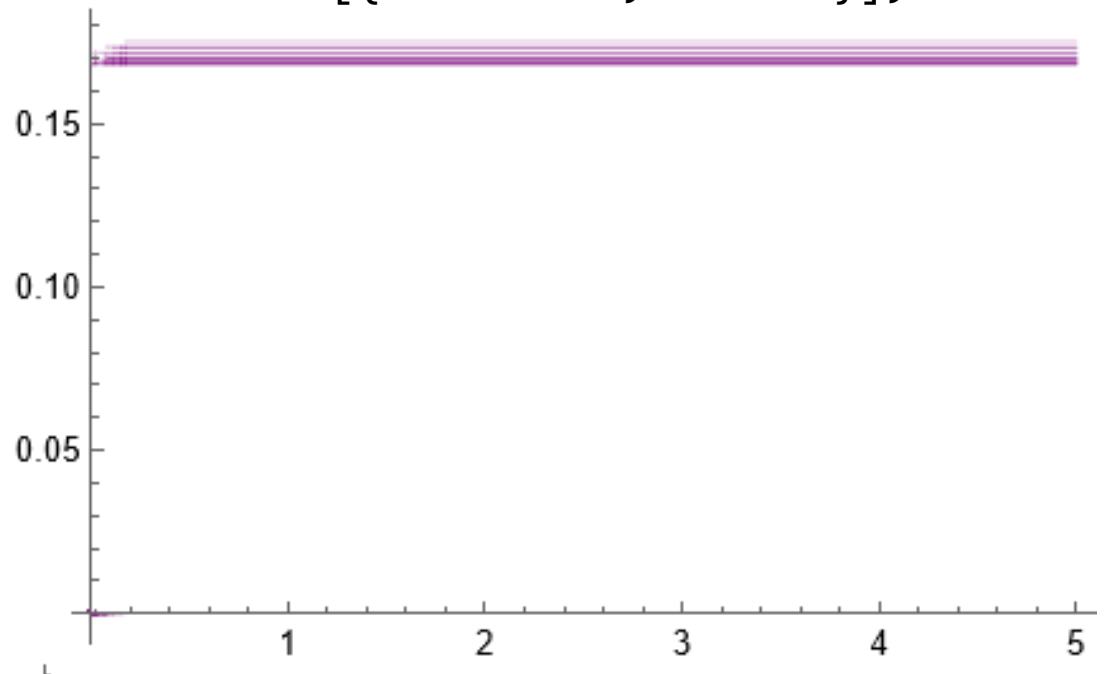

Output

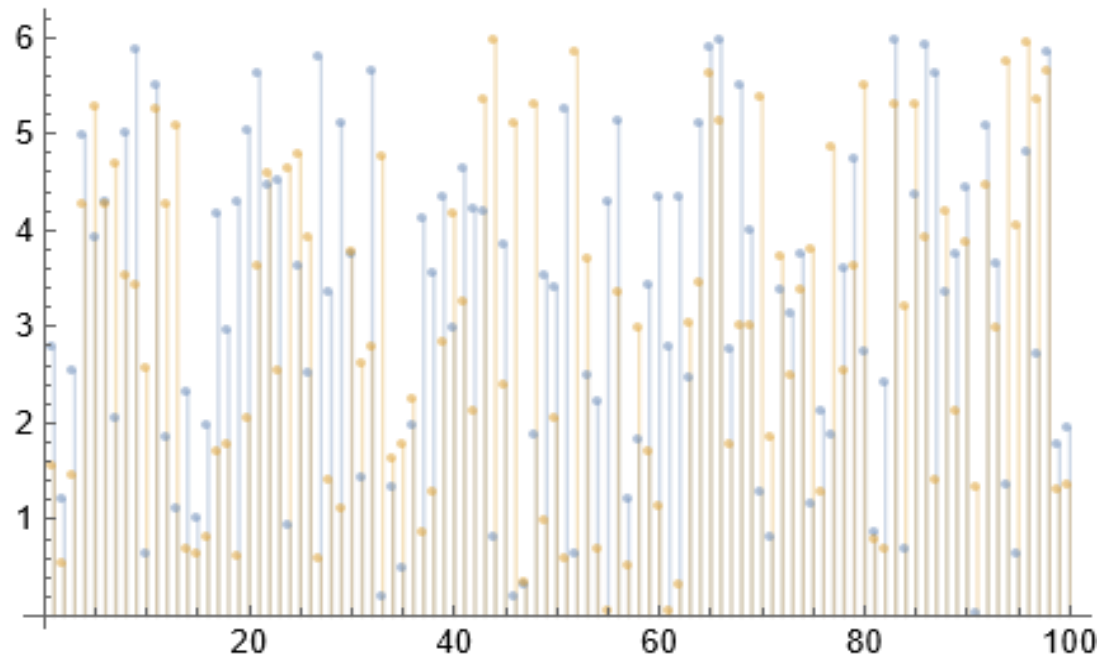

Output

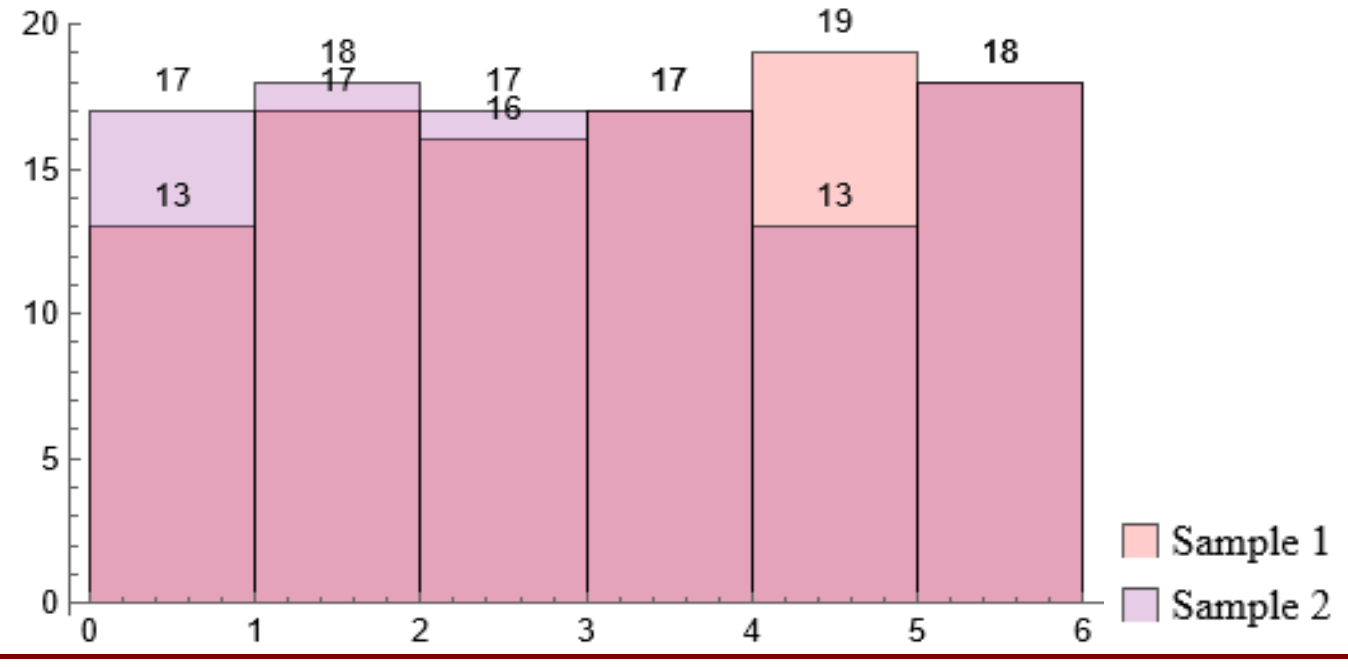

***Mathematica Examples 13.32***

Input

```
(* The code generates and compares the means of random samples drawn from Uniform
distribution with the given parameters a and b. The code uses the Manipulate function
to create a user interface with sliders to adjust the values of the a, b, number of
samples, and sample size. By varying the values of "Number of Samples" and "Sample
```

```
Size" sliders, the code allows the user to explore how changing these parameters
affects the means of the random samples. Specifically, increasing the number of
samples n tends to make the distribution of the means narrower and more concentrated
around the true mean of the underlying distribution. On the other hand, increasing
the sample size tends to reduce the variability in the means and make them more
precise estimators of the true mean: *)

Manipulate[
 Module[
  {means,dist},

  dist=UniformDistribution[{a,b}];

  means=Map[
    Mean,
    RandomVariate[dist,{n,samples}]
    ];
  m=N[Mean[means]];

  ListPlot[
   {means,{{0,m},{n,m}}},
   Joined->{False,True},
   Filling->Axis,
   PlotRange->{{1,50},{0,6}},
   PlotStyle->{Purple,Red},
   AxesLabel->{"Number of Samples","Sample Mean"},
   PlotLabel->Row[{"a = ",a,", b = ",b}],
   ImageSize->320
   ]
  ],
 {{a,2,"Shape"},0.1,4,0.1},
 {{b,4.5,"Scale"},4.5,6,0.1},
 {{n,50,"Number of Samples"},3,50,1},
 {{samples,50,"Sample Size"},1,100,1},
 TrackedSymbols:>{n,samples,a,b}
 ]
```

Output

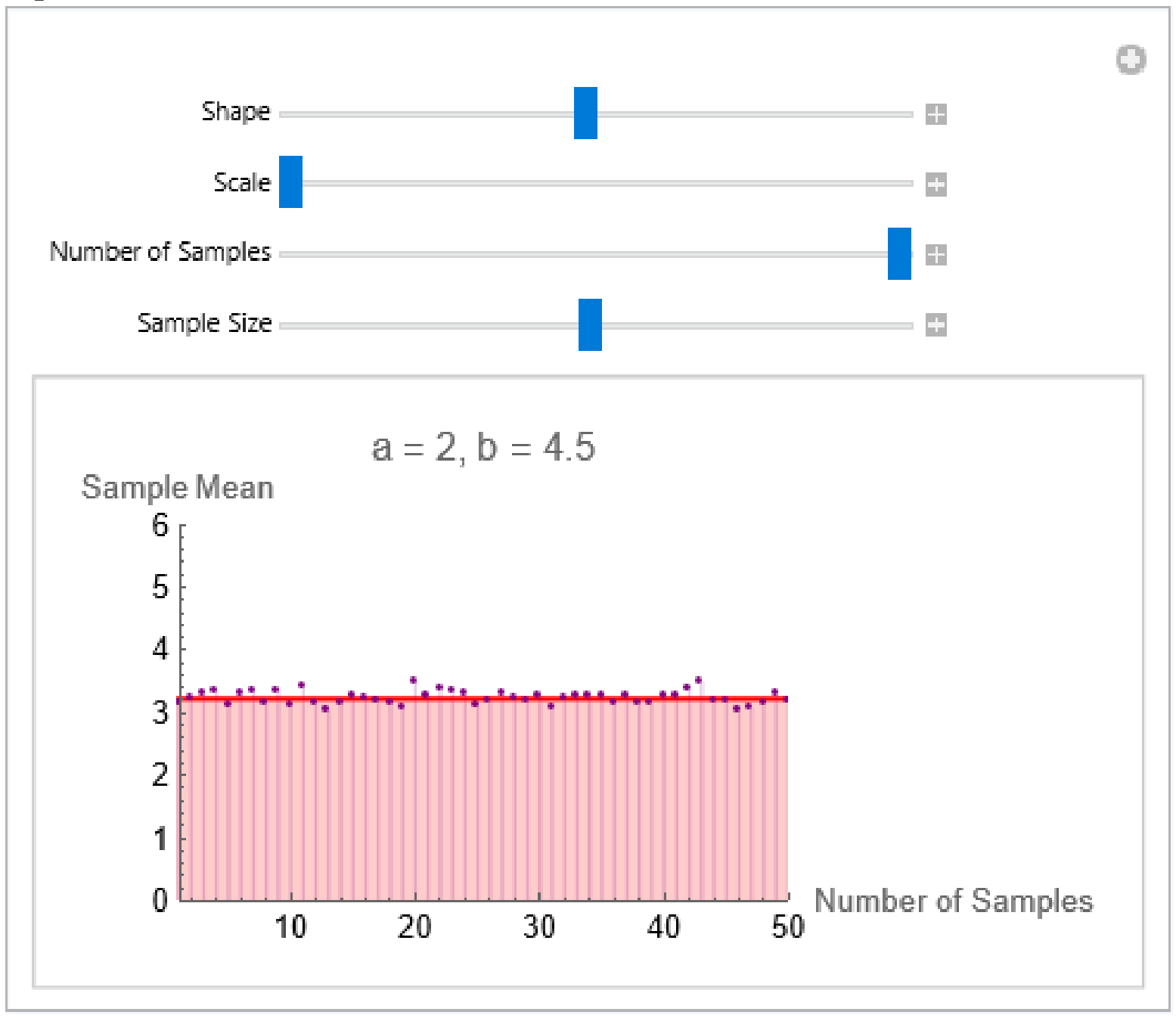

***Mathematica Examples 13.33***

Input

```
(* The code is designed to compare two Uniform distributions. It does this by
generating random samples from each distribution and displaying them in a histogram,
as well as plotting the PDFs of the two distributions. The code allows the user to
manipulate the parameters a, and b of both Uniform distributions through the sliders
for a1, b1, a2 and b2. By changing these parameters, the user can see how the
distributions change and how they compare to each other. The histograms display the
sample data for each distribution, with the first histogram showing the sample data
for the first Uniform distribution and the second histogram showing the sample data
for the second Uniform distribution. The histograms are overlaid on each other, with
the opacity of each histogram set to 0.2 to make it easier to see where the data
overlap. The PDFs of the two distributions are also plotted on the same graph, with
the first distribution shown in blue and the second distribution shown in red. The
legend indicates which color corresponds to which distribution. By looking at the
histograms and the PDFs, the user can compare the two Uniform distributions and see
how they differ in terms of shape, scale, and overlap of their sample data: *)

Manipulate[
 Module[
  {dist1,dist2,data1,data2},
  SeedRandom[seed];
  dist1=UniformDistribution[{a1,b1}];
  dist2=UniformDistribution[{a2,b2}];
  data1=RandomVariate[dist1,n];
  data2=RandomVariate[dist2,n];
  Column[
   {
    Show[
     ListPlot[
      data1,
      ImageSize->320,
      PlotStyle->Blue
      ],
     ListPlot[
      data2,
      ImageSize->320,
      PlotStyle->Red
      ]
     ],
    Show[
     Plot[
      {PDF[dist1,x],PDF[dist2,x]},
      {x,Min[{data1,data2}],Max[{data1,data2}]},
      PlotLegends->{"Distribution 1","Distribution 2"},
      PlotRange->All,
      PlotStyle->{Blue,Red},
      ImageSize->320
      ],
     Histogram[
      {data1,data2},
      Automatic,
      "PDF",
      ChartLegends->{"sample data1","sample data2"},
      ChartStyle->{Directive[Opacity[0.2],Red],Directive[Opacity[0.2],Purple]},
      ImageSize->320
      ]
     ]
    }
   ]
  ],
 {{a1,1},1,5,0.1},
```

```
{{b1,10},5.5,10,0.1},
{{a2,1},1,5,0.1},
{{b2,10},5.5,10,0.1},
{{n,500},{100,500,1000,2000}},
{{seed,1234},ControlType->None}
]
```

Output

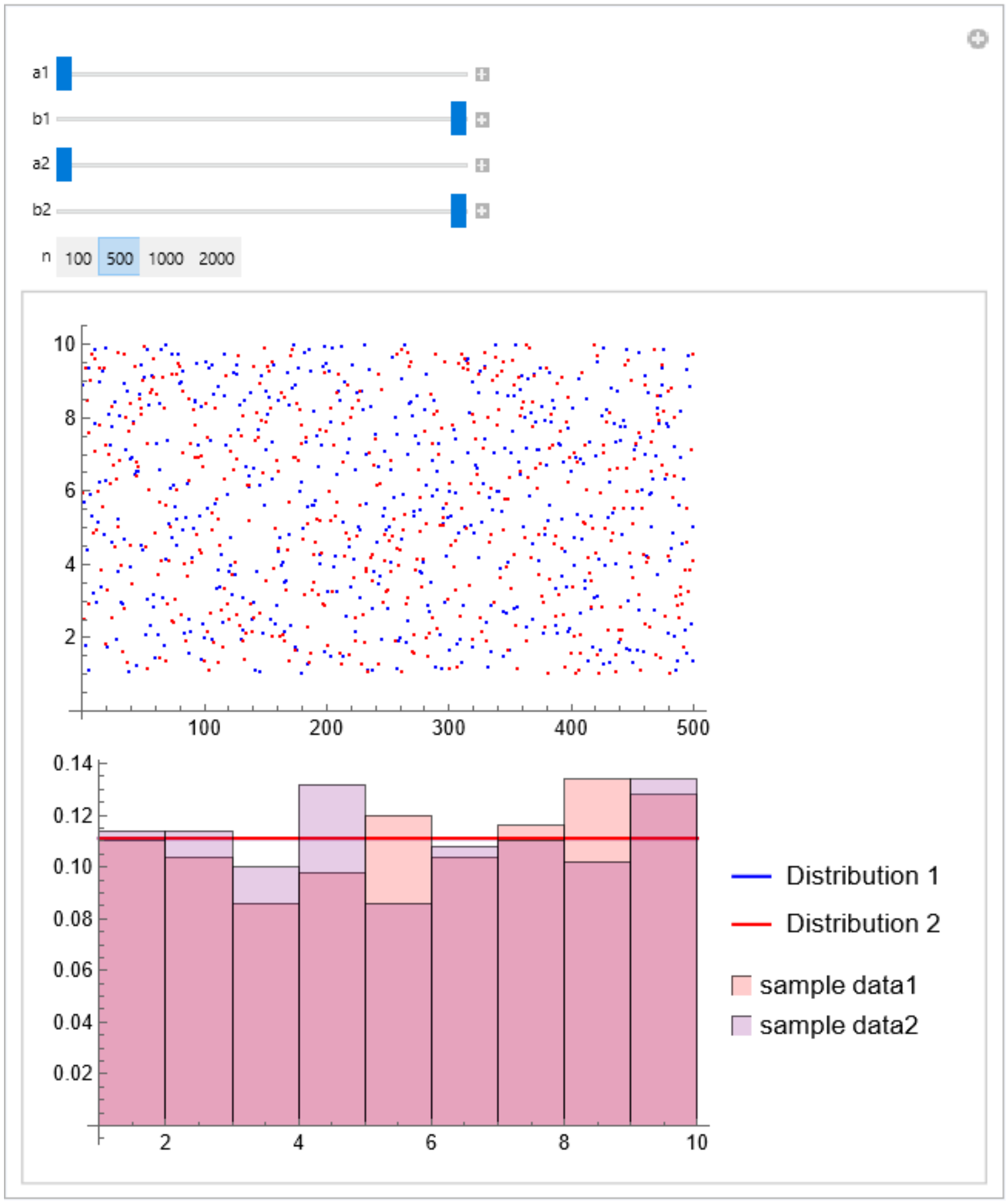

*Mathematica Examples 13.34*

Input

```
(* DiscreteUniformDistribution is the discrete analog of UniformDistribution: *)

PDF[
 UniformDistribution[{min,max}],
 x
 ]

PDF[
 DiscreteUniformDistribution[{min,max}],
 x
 ]
```

Output

$$\begin{cases} \frac{1}{\max - \min} & \min <= x <= \max \\ 0 & \text{True} \end{cases}$$

Output
$$\begin{cases} \frac{1}{1 + \max - \min} & \min <= x <= \max \\ 0 & \text{True} \end{cases}$$

***Mathematica Examples 13.35***

Input
```
(* Sidelength of a random rectangle is uniformly distributed between 10 and 20 meters.
Find the mean area of a rectangle whose height and width are independent and uniformly
distributed: *)

meanHeight=Mean[UniformDistribution[{10,20}]];
meanWidth=Mean[UniformDistribution[{10,20}]];
meanArea=meanHeight*meanWidth;
Print["The mean area of the rectangle is ",meanArea];

(* or *)
n=10^6;
heights=RandomReal[{10,20},n];
widths=RandomReal[{10,20},n];
meanArea=Mean[heights*widths];
Print["The mean area of the rectangle is ",meanArea];

(* or*)
meanArea=NExpectation[x*y,{x,y}\[Distributed]UniformDistribution[{{10,20},{10,20}}]
];
Print["The mean area of the rectangle is ",meanArea];
```

Output
```
The mean area of the rectangle is  225
The mean area of the rectangle is  224.995
The mean area of the rectangle is  225.
```

# UNIT 13.3

# EXPONENTIAL DISTRIBUTION

***Mathematica Examples 13.36***

Input

```
(* This code generates a plot of the probability density function (PDF) for a Exponential distribution with different values of the parameter λ= (1/2, 1, 2). The PDF is evaluated at various values of the random variable x between 0 and 3: *)

Plot[
 Evaluate[
  Table[
   PDF[
    ExponentialDistribution[λ],
    x
    ],
   {λ,{1/2,1,2}}
   ]
  ],
 {x,0,3},
 PlotRange->All,
 Filling->Axis,
 PlotLegends->Placed[{"λ=1/2","λ=1","λ=2"},{0.25,0.75}],
 PlotStyle->{RGBColor[0.88,0.61,0.14],RGBColor[0.37,0.5,0.7],Purple},
 ImageSize->320,
 AxesLabel->{None,"PDF"}
 ]
```

Output

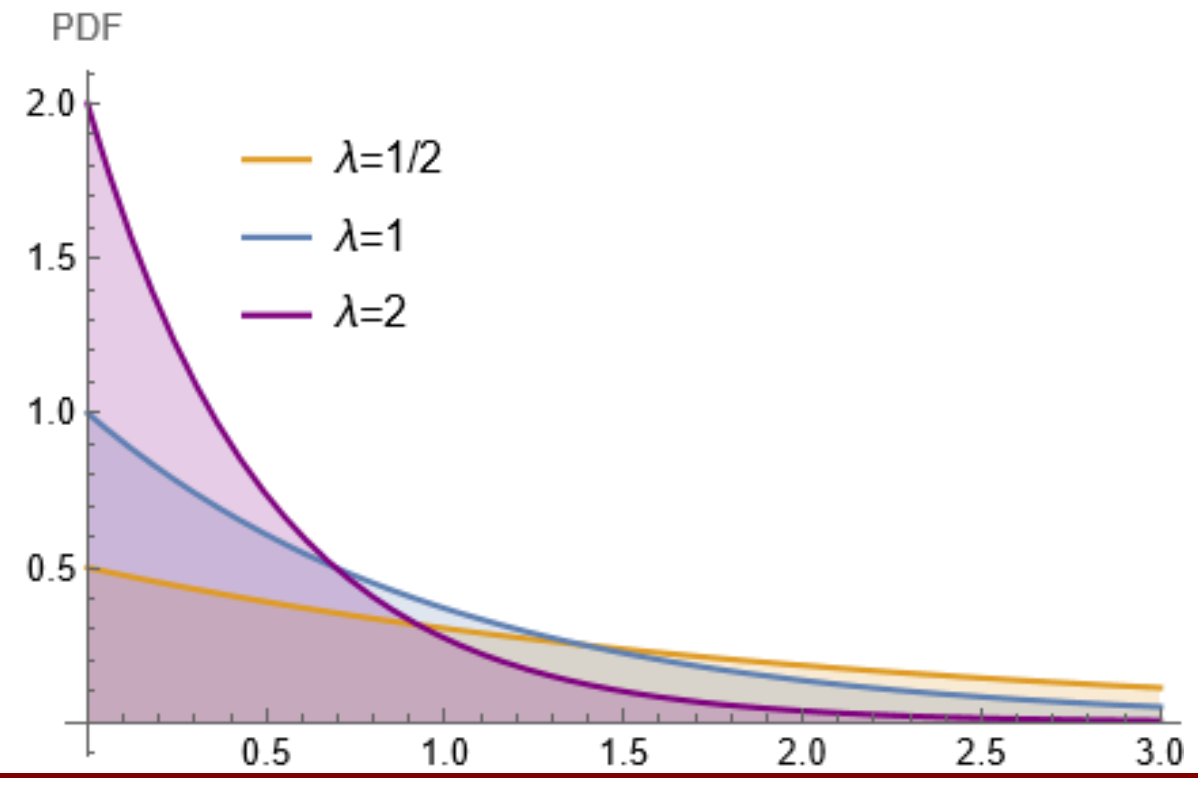

***Mathematica Examples 13.37***

Input

```
(* The code generates a plot of the CDF of the Exponential distribution with different values of the parameter λ= (1/2, 1, 2). The CDF is evaluated at various values of the random variable x between 0 and 3: *)

Plot[
 Evaluate[
  Table[
   CDF[
    ExponentialDistribution[λ],
    x
    ],
```

```
            {λ,{1/2,1,2}}
            ]
           ],
          {x,0,3},
          PlotRange->All,
          Filling->Axis,
          PlotLegends->Placed[{"λ=1/2","λ=1","λ=2"},{0.25,0.75}],
          PlotStyle->{RGBColor[0.88,0.61,0.14],RGBColor[0.37,0.5,0.7],Purple},
          ImageSize->320,
          AxesLabel->{None,"CDF"}
          ]
```

Output

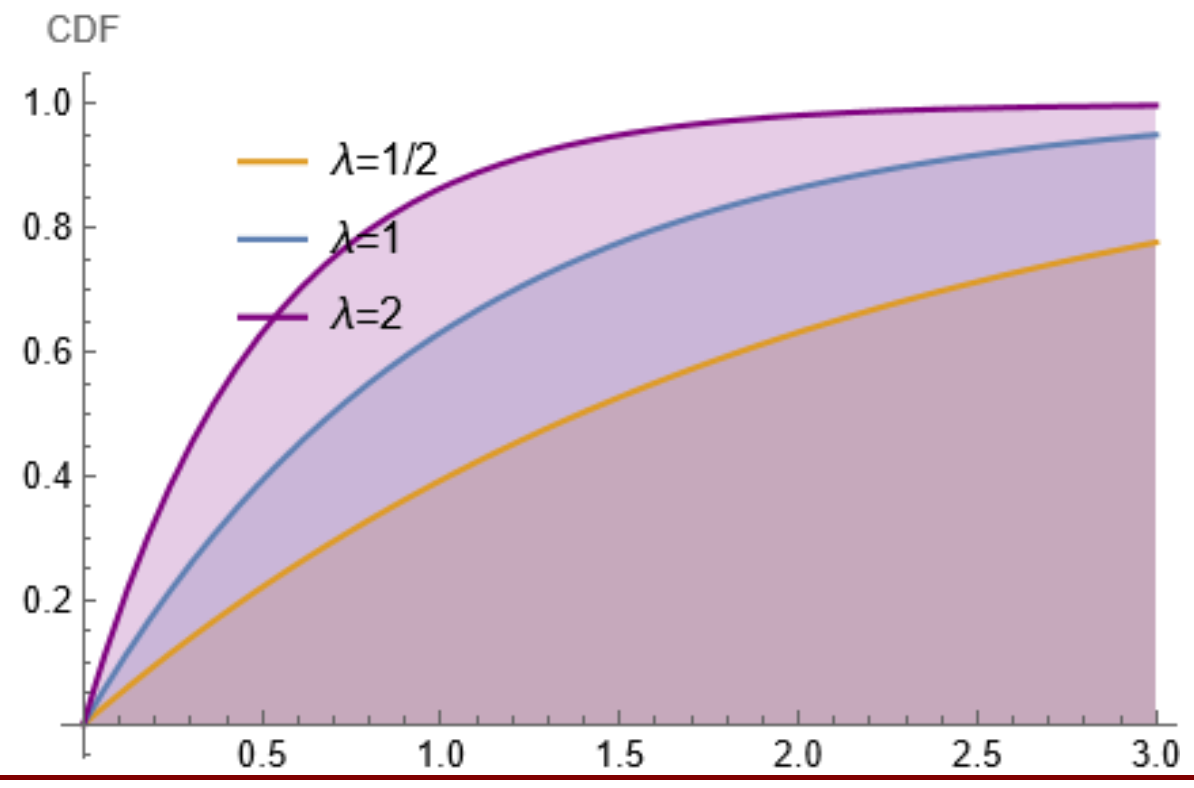

***Mathematica Examples 13.38***

Input

```
(* The code generates a histogram and a plot of the PDF for a Exponential distribution
with parameters λ=3.5 and sample size 10000: *)

data=RandomVariate[
   ExponentialDistribution[3.5],
   10^4
   ];
 Show[
 Histogram[
  data,
  {0,2,0.1},
  "PDF",
  ColorFunction->Function[{height},Opacity[height]],
  ChartStyle->Purple,
  ImageSize->320,
  AxesLabel->{None,"PDF"}
  ],

 Plot[
  PDF[
   ExponentialDistribution[3.5],
   x
   ],
  {x,0,2},
  PlotStyle->RGBColor[0.88,0.61,0.14],
  PlotRange->{0,4}
  ]
 ]
```

Output

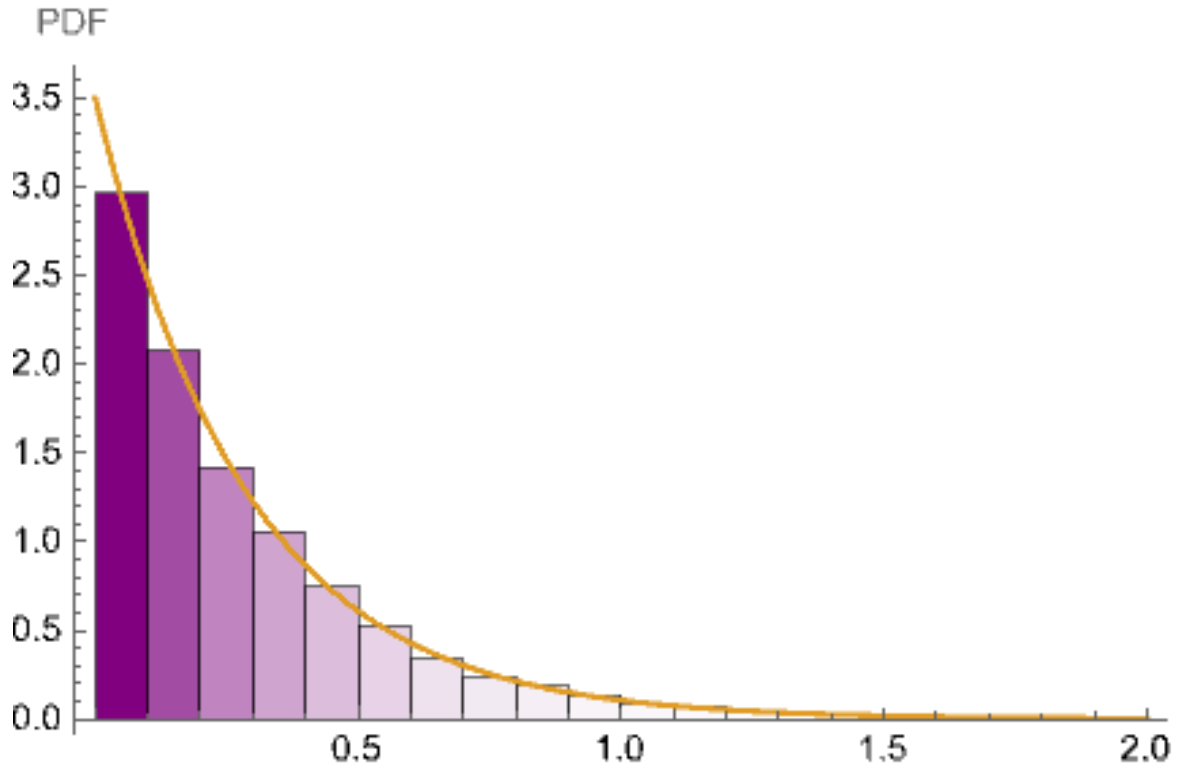

*Mathematica Examples 13.39*

Input

```
(* The code creates a dynamic histogram of data and a plot of the PDF generated from
a Exponential distribution using the Manipulate function. The Manipulate function
creates interactive controls for the user to adjust the values of λ and n, which are
the parameter of the Exponential distribution and the sample size: *)

Manipulate[
 Module[
  {
   data=RandomVariate[
     ExponentialDistribution[λ],
     n
     ]
   },

  Show[
   Histogram[
    data,
    Automatic,
    "PDF",
    PlotRange->All,
    ColorFunction->Function[{height},Opacity[height]],
    ImageSize->320,
    ChartStyle->Purple
    ],
   Plot[
    PDF[
     ExponentialDistribution[λ],
     x
     ],
    {x,0,10},
    PlotRange->All,
    ColorFunction->"Rainbow"
    ]
   ]
  ],
 {{λ,1,"λ"},1,6,0.1},
 {{n,300,"n"},100,1000,10}
 ]
```

Output

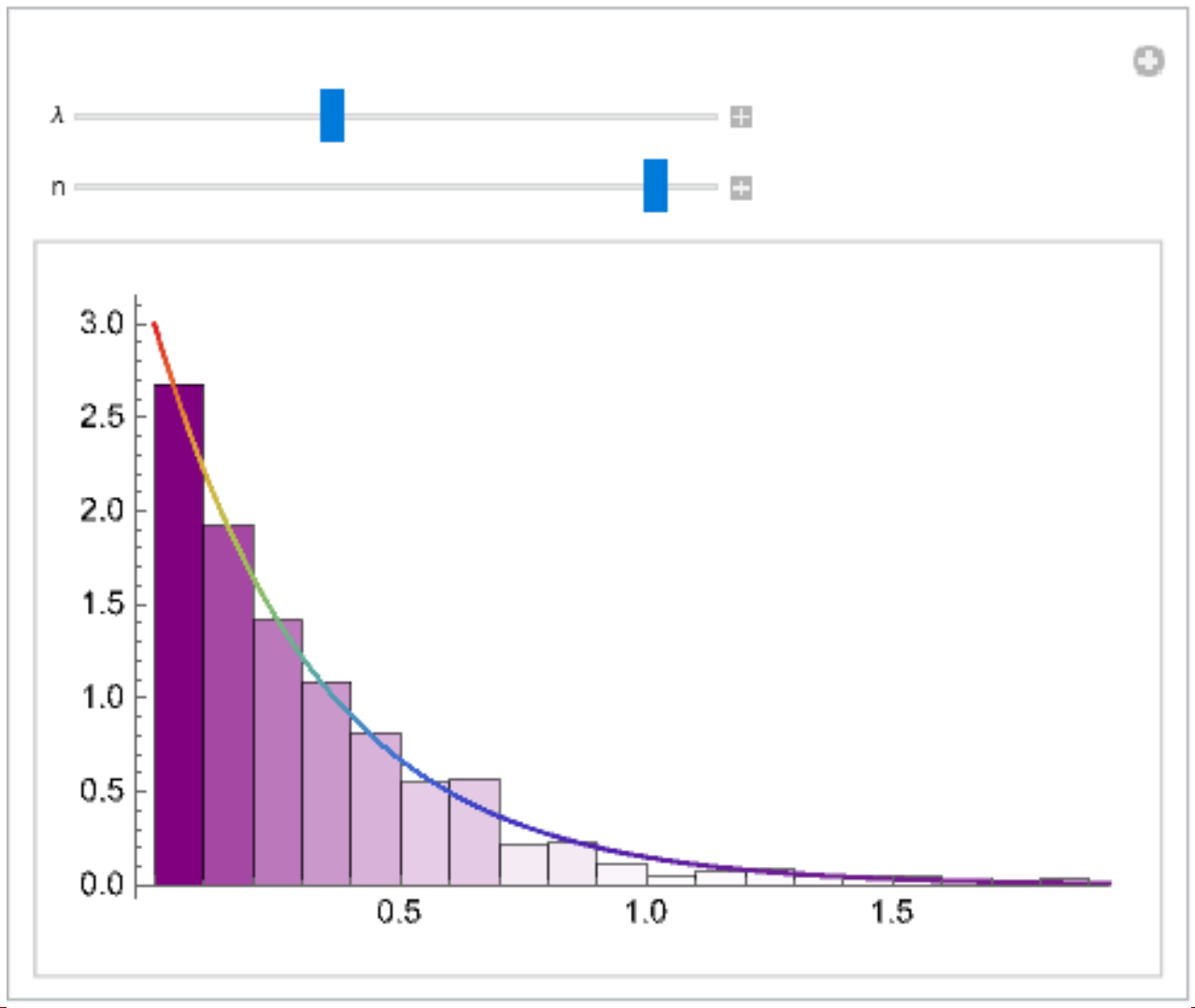

**Mathematica Examples 13.40**

Input

```
(* The code creates a plot of the CDF of a Exponential distribution using the Manipulate function. The Manipulate function allows you to interactively change the values of the parameters λ: *)
Manipulate[
 Plot[
  CDF[
   ExponentialDistribution[λ],
   x
   ],
  {x,0,7},
  Filling->Axis,
  FillingStyle->LightPurple,
  PlotRange->All,
  AxesLabel->{"x","CDF"},
  ImageSize->320,
  PlotStyle->Purple,
  PlotLabel->Row[{"λ = ",λ}]
  ],
 {{λ,1},1,6,0.1}
 ]
```

Output

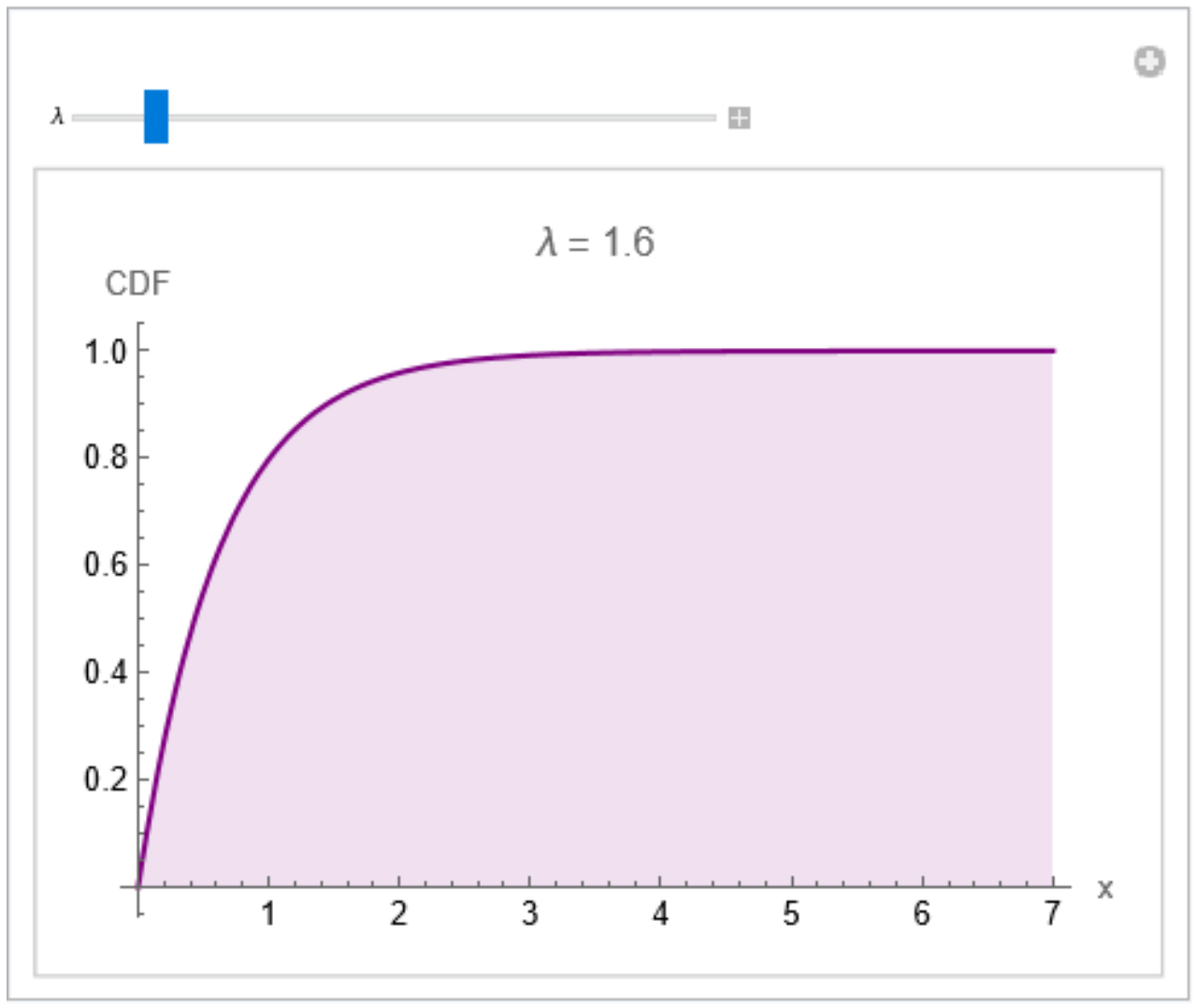

***Mathematica Examples 13.41***

Input

```
(* The code uses the Grid function to create a grid of two plots, one for the PDF and one for the CDF of Exponential distribution. The code uses slider controls to adjust the values of λ: *)

Manipulate[
 Grid[
  {
   {Plot[
     PDF[
      ExponentialDistribution[λ],
      x
      ],
     {x,0,7},
     PlotRange->All,
     PlotStyle->{Purple,PointSize[0.03]},
     PlotLabel->"PDF of Exponential distribution",
     AxesLabel->{"x","PDF"}
     ],
    Plot[
     CDF[
      ExponentialDistribution[λ],
      x
      ],
     {x,0,7},
     PlotRange->All,
     PlotStyle->{Purple,PointSize[0.03]},
     PlotLabel->"CDF of Exponential distribution",
     AxesLabel->{"x","CDF"}
     ]
    }
   },
  Spacings->{5,5}
  ],
 {{λ,1},1,6,0.1}
 ]
```

Output

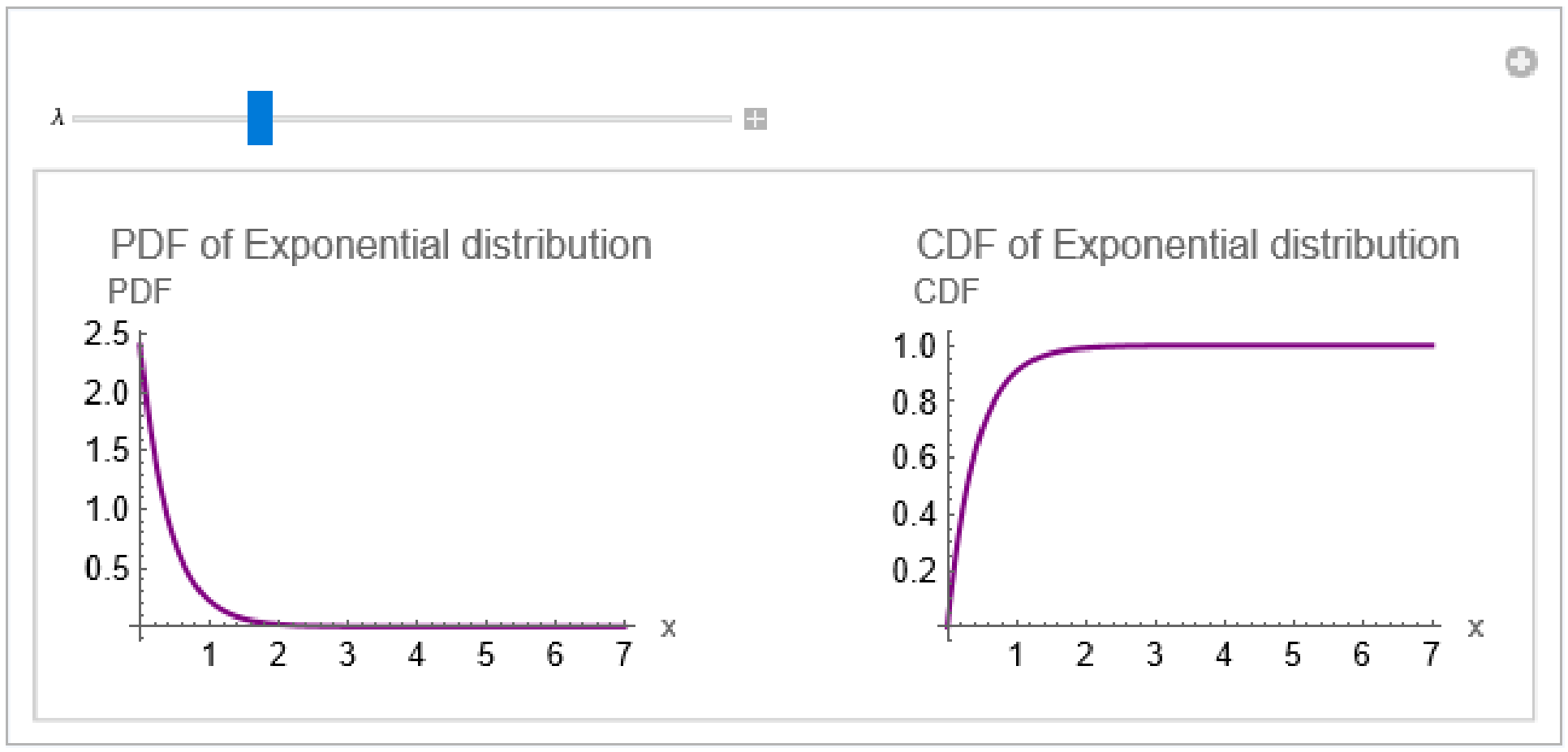

***Mathematica Examples 13.42***

Input

```
(* The code calculates and displays some descriptive statistics (mean, variance, standard deviation, kurtosis and skewness) for an Exponential distribution with parameter λ: *)
```

```
Grid[
 Table[
  {
   statistics,
   FullSimplify[statistics[ExponentialDistribution[λ]]]
   },
  {statistics,{Mean,Variance,StandardDeviation,Kurtosis,Skewness}}
  ],
 ItemStyle->12,
 Alignment->{{Right,Left}},
 Frame->All,
 Spacings->{Automatic,0.8}
 ]
```

Output

| | |
|---|---|
| Mean | 1/λ |
| Variance | 1/λ^2 |
| StandardDeviation | 1/λ |
| Kurtosis | 9 |
| Skewness | 2 |

***Mathematica Examples 13.43***

Input

```
(* The code calculates and displays some additional descriptive statistics (moments, central moments, and factorial moments) for an Exponential distribution with parameter λ: *)

Grid[
 Table[
  {
   statistics,
   FullSimplify[statistics[ExponentialDistribution[λ],1]],
   FullSimplify[statistics[ExponentialDistribution[λ],2]]
   },
  {statistics,{Moment,CentralMoment,FactorialMoment}}
  ],
 ItemStyle->12,
 Alignment->{{Right,Left}},
 Frame->All,
 Spacings->{Automatic,0.8}
 ]
```

Output

| | | |
|---|---|---|
| Moment | 1/λ | 2/λ^2 |
| CentralMoment | 0 | 1/λ^2 |
| FactorialMoment | 1/λ | (2-λ)/λ^2 |

***Mathematica Examples 13.44***

Input

```
(* The code generates a dataset of 10000 observations from an Exponential distribution with parameter λ=3. Then, it computes the sample mean and quartiles of the data, and plots a histogram of the data and plot of the PDF. Additionally, the code adds vertical lines to the plot corresponding to the sample mean and quartiles: *)

data=RandomVariate[
    ExponentialDistribution[3],
    10000
    ];

mean=Mean[data];
quartiles=Quantile[
    data,
```

```
      {0.25,0.5,0.75}
      ];

   Show[
    Histogram[
     data,
     Automatic,
     "PDF",
     Epilog->{
       Directive[Red,Thickness[0.006]],
       Line[{{mean,0},{mean,3}}],
       Directive[Green,Dashed],
       Line[{{quartiles[[1]],0},{quartiles[[1]],3}}],
       Line[{{quartiles[[2]],0},{quartiles[[2]],3}}],
       Line[{{quartiles[[3]],0},{quartiles[[3]],3}}]
       },
     ColorFunction->Function[{height},Opacity[height]],
     ImageSize->320,
     ChartStyle->Purple,
     PlotRange->{{0,3},{0,3}}
     ],
    Plot[
     PDF[
      ExponentialDistribution[3],
      x
      ],
     {x,0,3},
     PlotRange->{{0,3},{0,3}},
     ImageSize->320,
     ColorFunction->"Rainbow"
     ]
    ]
```

Output

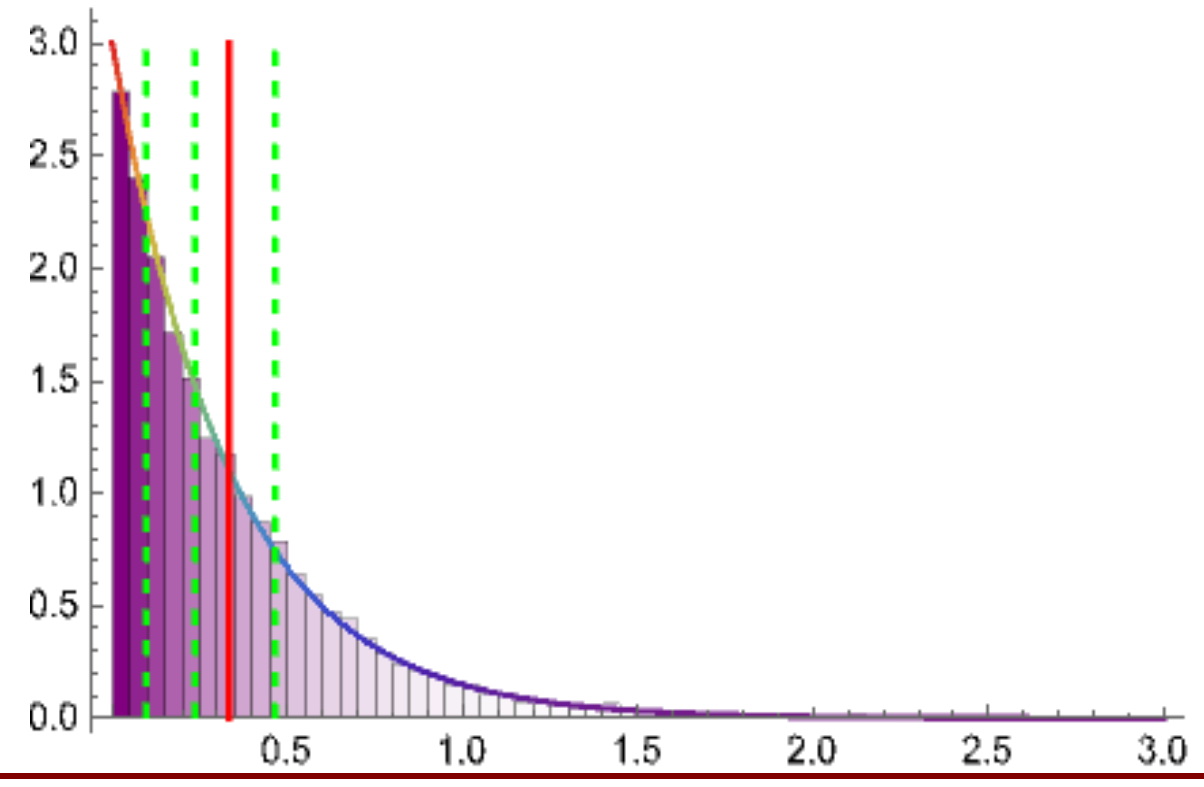

***Mathematica Examples 13.45***

Input

```
(* The code generates a random sample of size 10,000 from an Exponential distribution with parameter λ=3, estimates the distribution parameters using the EstimatedDistribution function, and then compares the histogram of the sample with the estimated PDF of the Exponential distribution using a histogram and a plot of the PDF: *)

sampledata=RandomVariate[
   ExponentialDistribution[3],
   10^4
   ];
(* Estimate the distribution parameters from sample data: *)
```

```
ed=EstimatedDistribution[
  sampledata,
  ExponentialDistribution[λ]
  ]
(* Compare a density histogram of the sample with the PDF of the estimated distribution: *)
Show[
 Histogram[
  sampledata,
  Automatic,
  "PDF",
  ColorFunction->Function[{height},Opacity[height]],
  ChartStyle->Purple,
  ImageSize->320
  ],
 Plot[
  PDF[ed,x],
  {x,0,2},
  ImageSize->320,
  ColorFunction->"Rainbow"
  ]
 ]
```

Output ExponentialDistribution[2.99728]
Output

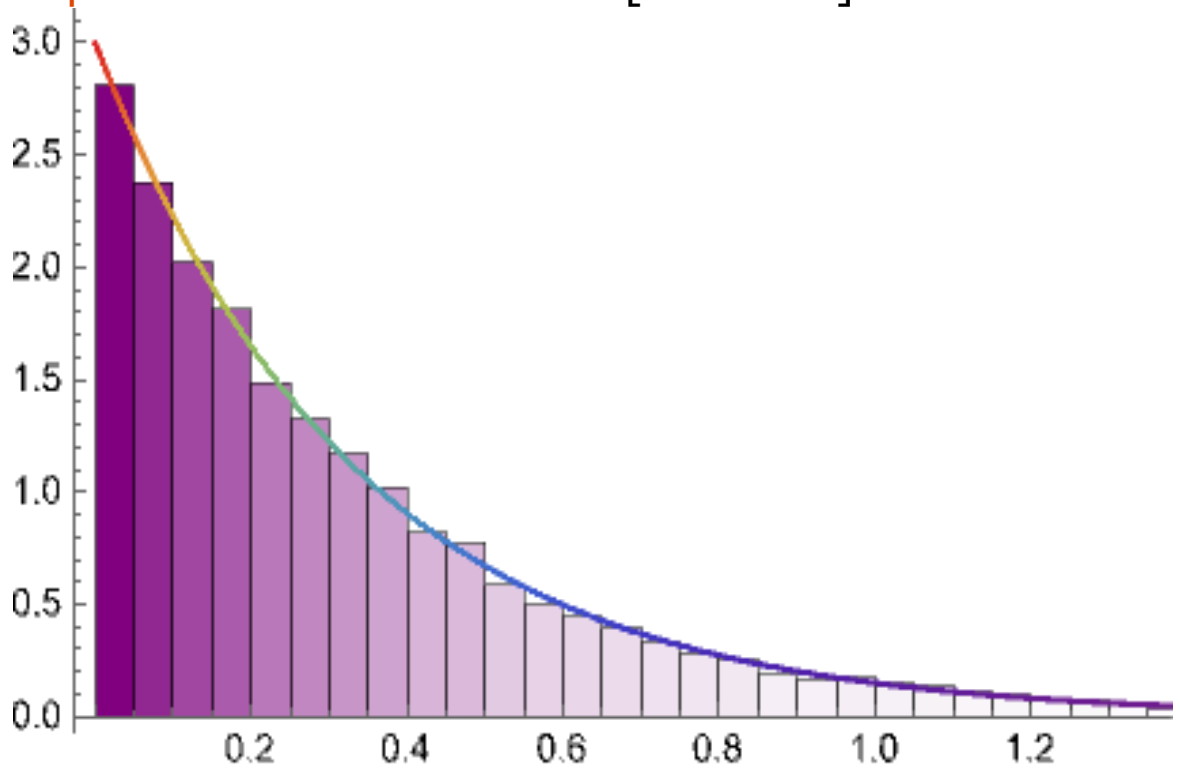

***Mathematica Examples 13.46***

Input

```
(* The code generates a 2D dataset with 1000 random points that follow an Exponential distribution with λ=3. The dataset is then used to create a row of three plots. The first plot is a histogram of the X-axis values of the dataset. The second plot is a histogram of the Y-axis values of the dataset. It is similar to the first plot, but shows the distribution of the Y-axis values instead. The third plot is a scatter plot of the dataset, with the X-axis values on the horizontal axis and the Y-axis values on the vertical axis. Each point in the plot represents a pair of X and Y values from the dataset: *)

data=RandomVariate[
   ExponentialDistribution[3],
   {1000,2}
   ];
GraphicsRow[
 {
  Histogram[
   data[[All,1]],
   {0.1},
   ImageSize->170,
   PlotLabel->"X-axis",
   ColorFunction->Function[{height},Opacity[height]],
```

```
          ChartStyle->Purple
          ],
         Histogram[
          data[[All,2]],
          {0.1},
          ImageSize->170,
          PlotLabel->"Y-axis",
          ColorFunction->Function[{height},Opacity[height]],
          ChartStyle->Purple
          ],
         ListPlot[
          data,
          ImageSize->170,
          PlotStyle->{Purple,PointSize[0.015]},
          AspectRatio->1,
          Frame->True,
          Axes->False
          ]
         }
        ]
```

Output

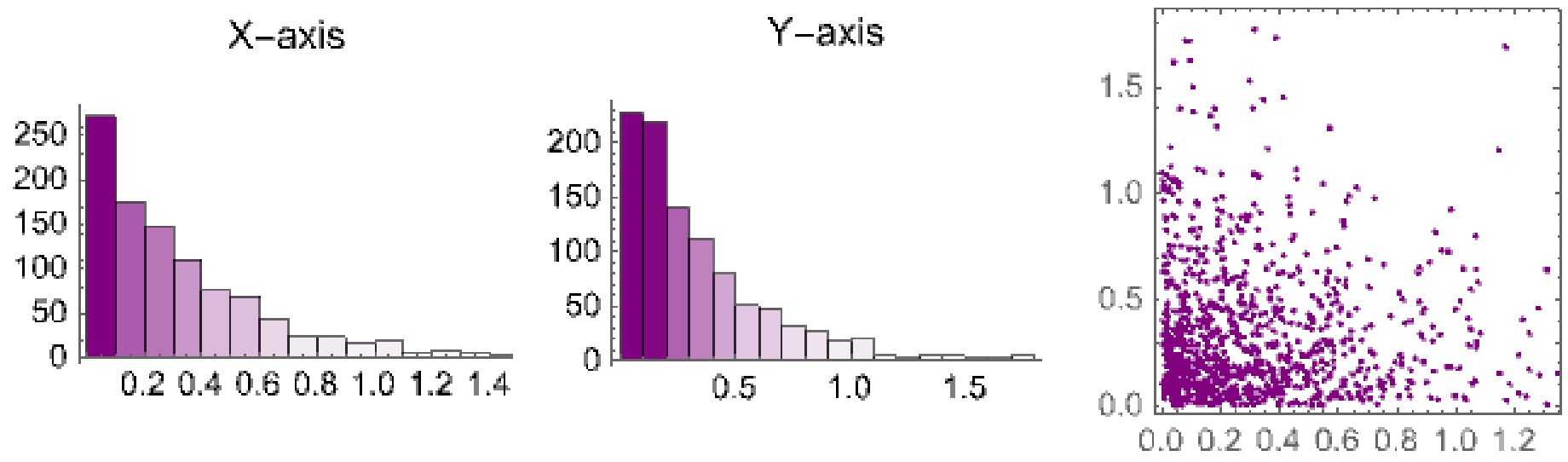

***Mathematica Examples 13.47***

Input

```
(* The code generates a set of random data points with an Exponential distribution
with λ=3 in three dimensions, and then creates three histograms, one for each
dimension, showing the distribution of the points along that axis. Additionally, it
creates a 3D scatter plot of the data points: *)

data=RandomVariate[
   ExponentialDistribution[3],
   {1000,3}
   ];

GraphicsGrid[
 {
  {
   Histogram[
    data[[All,1]],
    Automatic,
    "PDF",
    PlotLabel->"X-axis",
    ColorFunction->Function[{height},Opacity[height]],
    ChartStyle->Purple
    ],
   Histogram[
    data[[All,2]],
    Automatic,
    "PDF",
    PlotLabel->"Y-axis",
```

```
ColorFunction->Function[{height},Opacity[height]],
ChartStyle->Purple
],
Histogram[
data[[All,3]],
Automatic,
"PDF",
PlotLabel->"Z-axis",
ColorFunction->Function[{height},Opacity[height]],
ChartStyle->Purple
],
ListPointPlot3D[
data,
BoxRatios->{1,1,1},
PlotStyle->{Purple,PointSize[0.015]}
]
}
}
]
```

Output

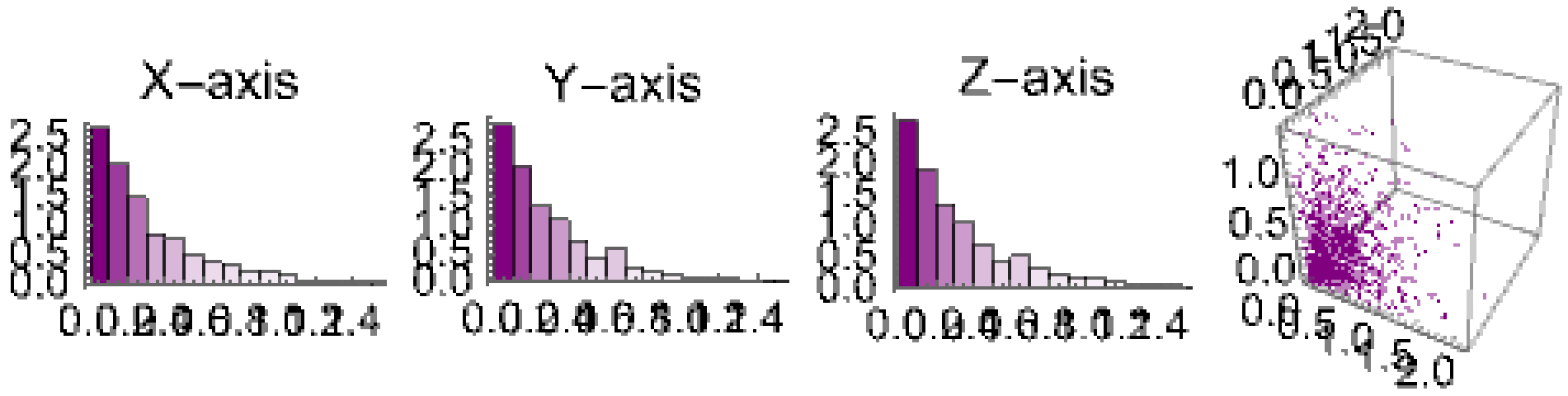

*Mathematica Examples 13.48*

Input

```
(* The code generates a 3D scatter plot of an Exponential distribution points with
λ=3, where the x-axis is red, y-axis is green, and z-axis is blue: *)

data=RandomVariate[
    ExponentialDistribution[3],
    {2000,3}
    ];
Graphics3D[
 {
  {PointSize[0.006],Purple,Opacity[0.6],Point[data]},
  Thin,
  {Red,Opacity[0.4],Line[{{#,0,0},{#,0,-0.3}}]&/@data[[All,1]]},
  Thin,
  {Green,Opacity[0.4],Line[{{0,#,0},{0,#,-0.3}}]&/@data[[All,2]]},
  Thin,
  {Blue,Opacity[0.4],Line[{{0,0,#},{0,-0.3,#}}]&/@data[[All,3]]}
  },
 BoxRatios->{1,1,1},
 Axes->True,
 AxesLabel->{"X","Y","Z"},
 ImageSize->320
 ]
```

Output

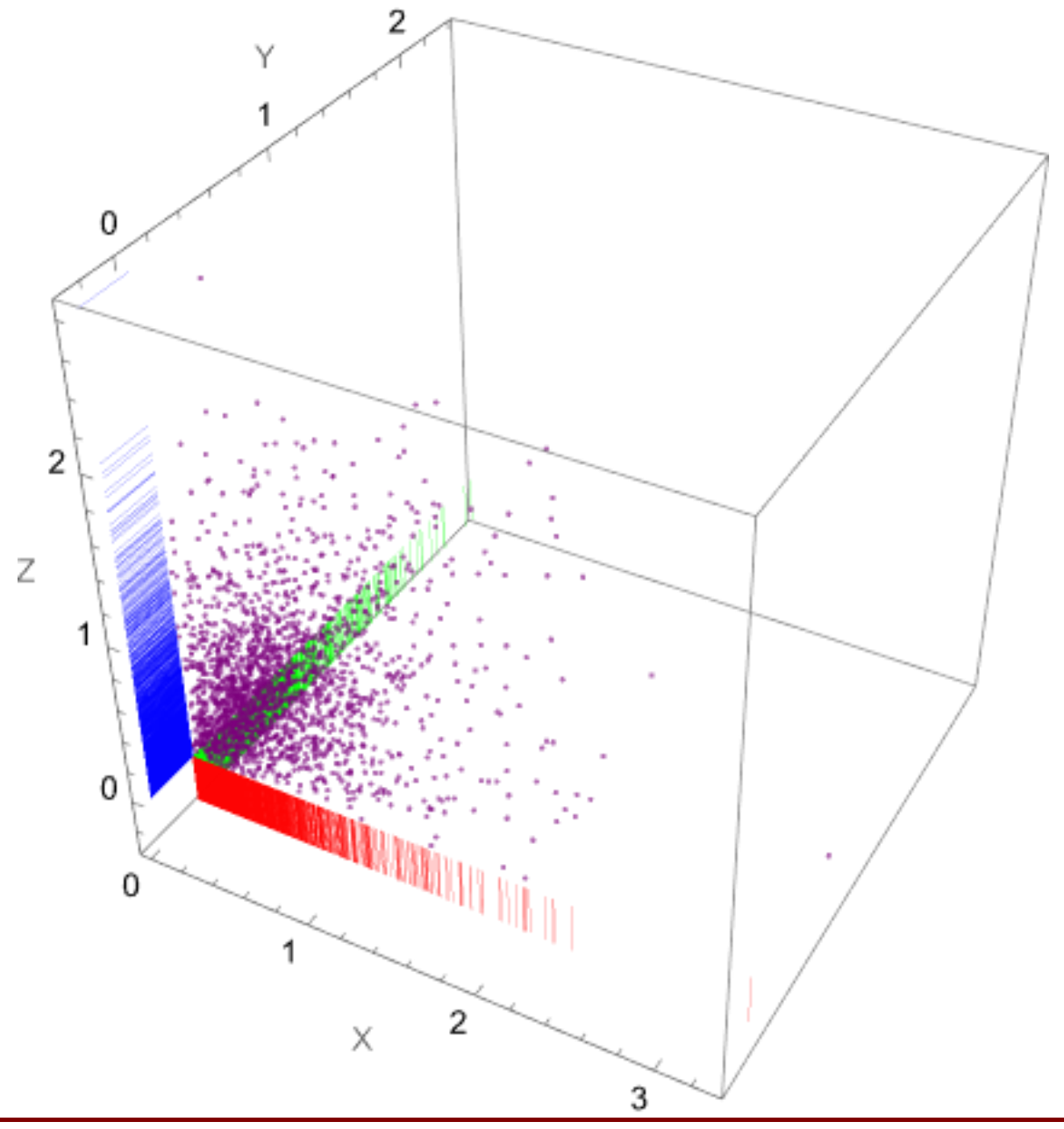

***Mathematica Examples 13.49***

Input

```
(* The code demonstrates a common technique in statistics and data analysis, which is the use of random sampling to estimate population parameters. The code generates random samples from Exponential distribution with λ=3, and then using these samples to estimate the parameters of another Exponential distribution with unknown λ. This process is repeated 20 times, resulting in 20 different estimated distributions. The code also visualizes the resulting estimated distributions using the PDF function. The code plots the PDFs of these estimated distributions using the PDF function and the estimated parameters. The plot shows the PDFs in a range from-0 to 1.5. The code also generates a list plot of 2 sets of random samples from the Exponential distribution with λ=3. The plot shows the 100 random points generated from two random samples. The code generates also a histogram of the PDF for Exponential distribution of the two samples: *)

estim0distributions=Table[
  dist=ExponentialDistribution[3];

  sampledata=RandomVariate[
    dist,
    100
    ];

  ed=EstimatedDistribution[
    sampledata,
    ExponentialDistribution[λ]
    ],
  {i,1,20}
  ]

pdf0ed=Table[
   PDF[estim0distributions[[i]],x],
   {i,1,20}
   ];
(* Visualizes the resulting estimated distributions *)
Plot[
 pdf0ed,
 {x,0,1.5},
```

```
 PlotRange->Full,
 ImageSize->400,
 PlotStyle->Directive[Purple,Opacity[0.3],Thickness[0.002]]
 ]
(* Visualizes 100 random points generated from two random samples *)
table =Table[
   dist=ExponentialDistribution[3];
   sampledata=RandomVariate[
     dist,
     100],
   {i,1,2}
   ];
ListPlot[
 table,
 ImageSize->320,
 Filling->Axis,
 PlotStyle->Directive[Opacity[0.5],Thickness[0.003]]
 ]
Histogram[
 table,
 Automatic,
 LabelingFunction->Above,
 ChartLegends->{"Sample 1","Sample 2"},
 ChartStyle->{Directive[Opacity[0.2],Red],Directive[Opacity[0.2],Purple]},
 ImageSize->320
 ]
```

Output {ExponentialDistribution[2.85543],ExponentialDistribution[2.59368],ExponentialDistribution[3.15706],ExponentialDistribution[3.96547],ExponentialDistribution[2.83275],ExponentialDistribution[2.82745],ExponentialDistribution[3.26445],ExponentialDistribution[3.34883],ExponentialDistribution[3.18225],ExponentialDistribution[2.84192],ExponentialDistribution[3.13402],ExponentialDistribution[2.52633],ExponentialDistribution[2.90694],ExponentialDistribution[3.55431],ExponentialDistribution[3.02981],ExponentialDistribution[3.03789],ExponentialDistribution[2.85686],ExponentialDistribution[2.6795],ExponentialDistribution[3.0948],ExponentialDistribution[3.35503]}

Output

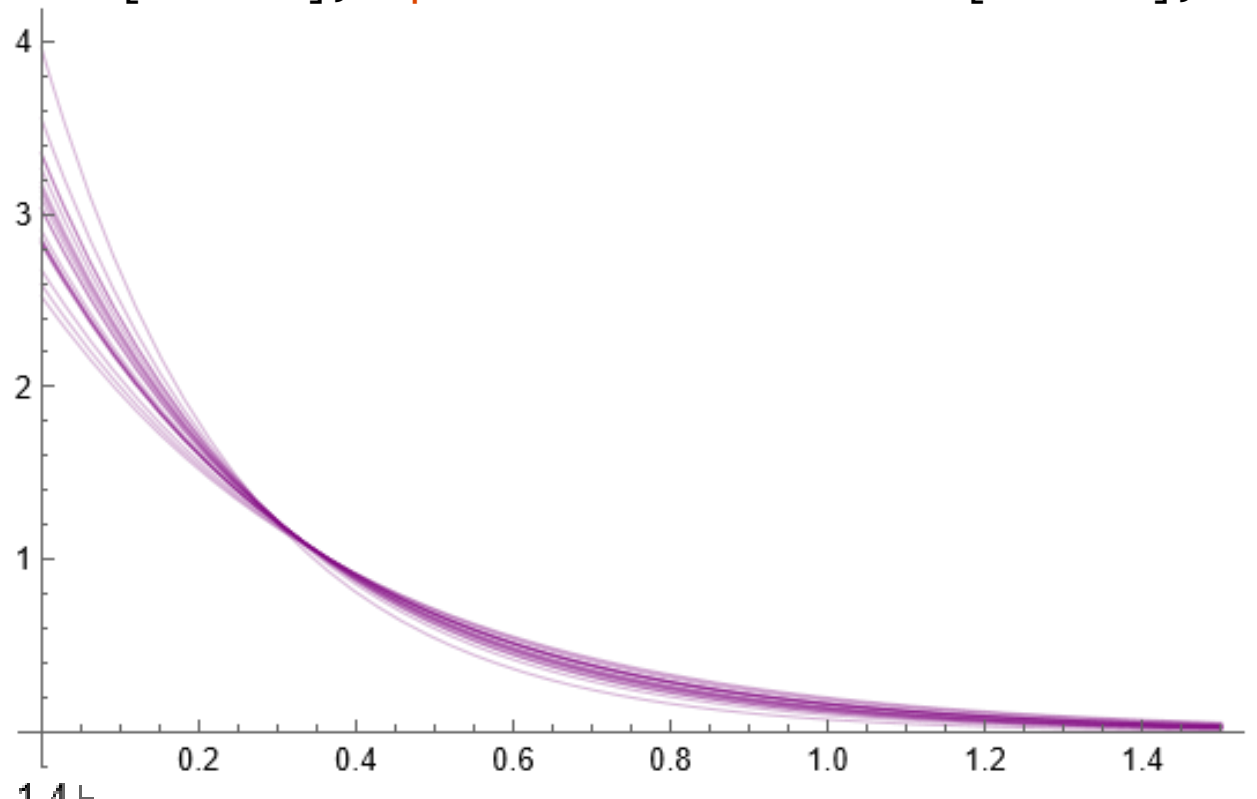

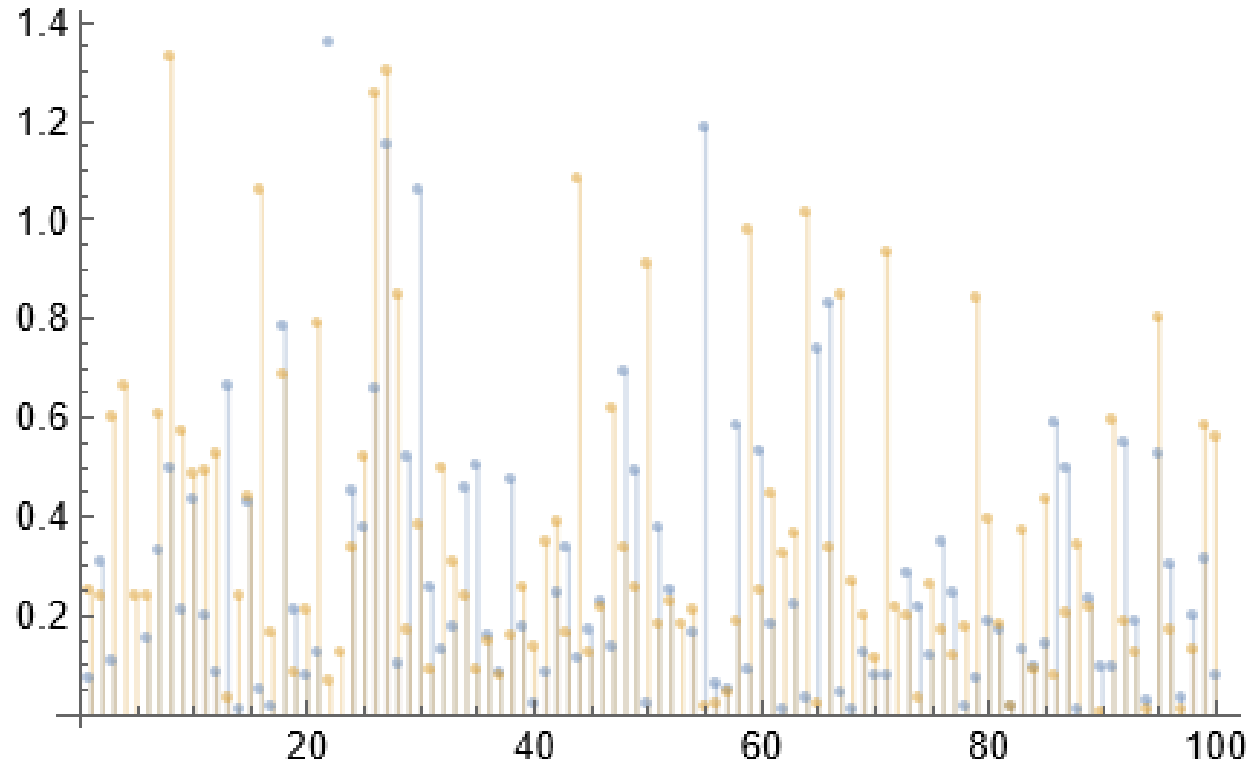

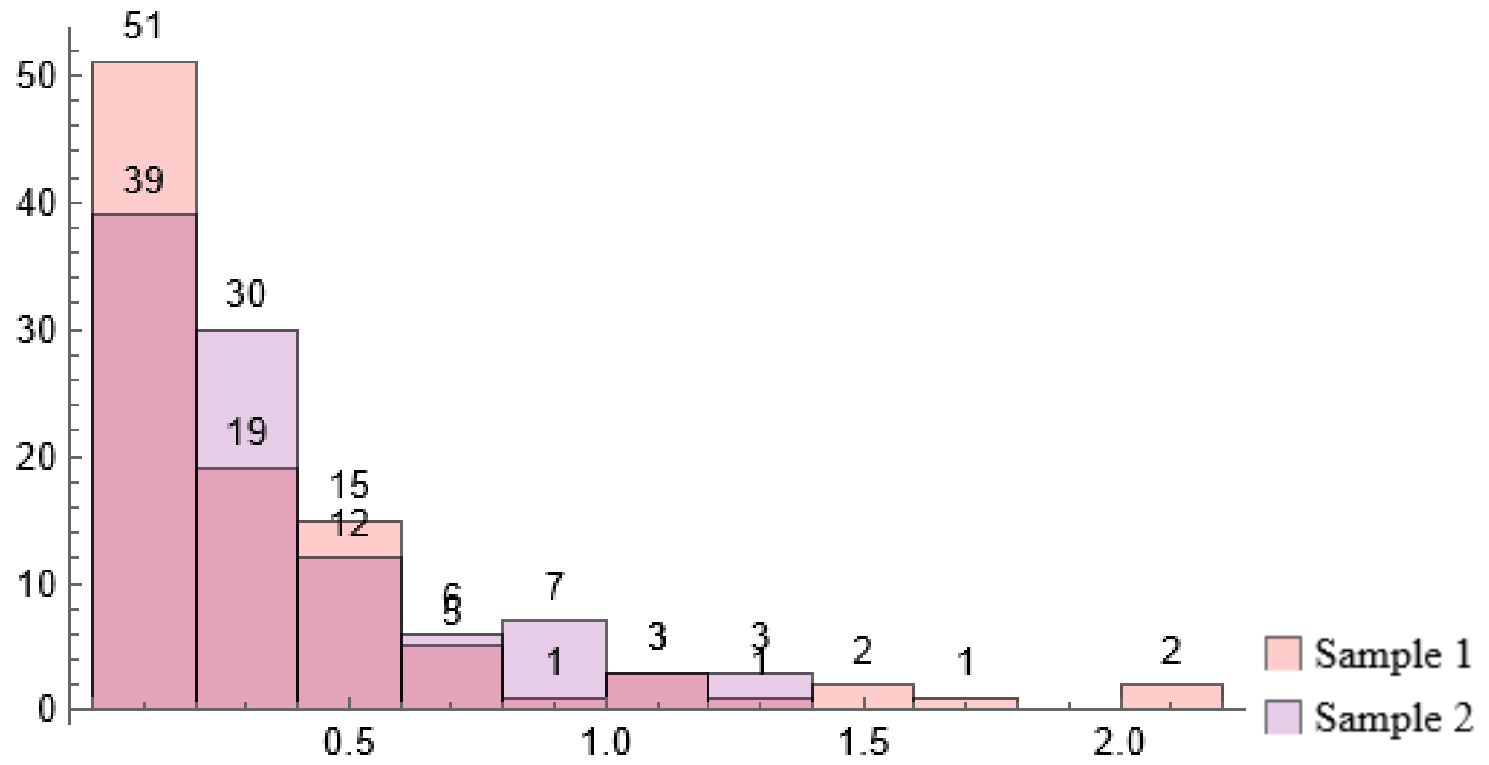

### *Mathematica Examples 13.50*

Input

```
(* The code generates and compares the means of random samples drawn from a Exponential distribution with the given parameter λ. The code uses the Manipulate function to create a user interface with sliders to adjust the values of the parameter λ, number of samples, and sample size. By varying the values of "Number of Samples" and "Sample Size" sliders, the code allows the user to explore how changing these parameters affects the means of the random samples. Specifically, increasing the number of samples n tends to make the distribution of the means narrower and more concentrated around the true mean of the underlying distribution. On the other hand, increasing the sample size tends to reduce the variability in the means and make them more precise estimators of the true mean: *)

Manipulate[
 Module[
  {means,dist},

  dist=ExponentialDistribution[λ];

  means=Map[
    Mean,
    RandomVariate[dist,{n,samples}]
    ];
  m=N[Mean[means]];

  ListPlot[
   {means,{{0,m},{n,m}}},
   Joined->{False,True},
   Filling->Axis,
   PlotRange->{{1,50},{0,10}},
   PlotStyle->{Purple,Red},
   AxesLabel->{"Number of Samples","Sample Mean"},
   PlotLabel->Row[{"λ = ",λ}],
   ImageSize->320
   ]
  ],
 {{λ,2,"λ"},0.6,4,0.1},
 {{n,50,"Number of Samples"},3,50,1},
 {{samples,50,"Sample Size"},1,100,1},
 TrackedSymbols:>{n,samples,λ}
 ]
```

Output

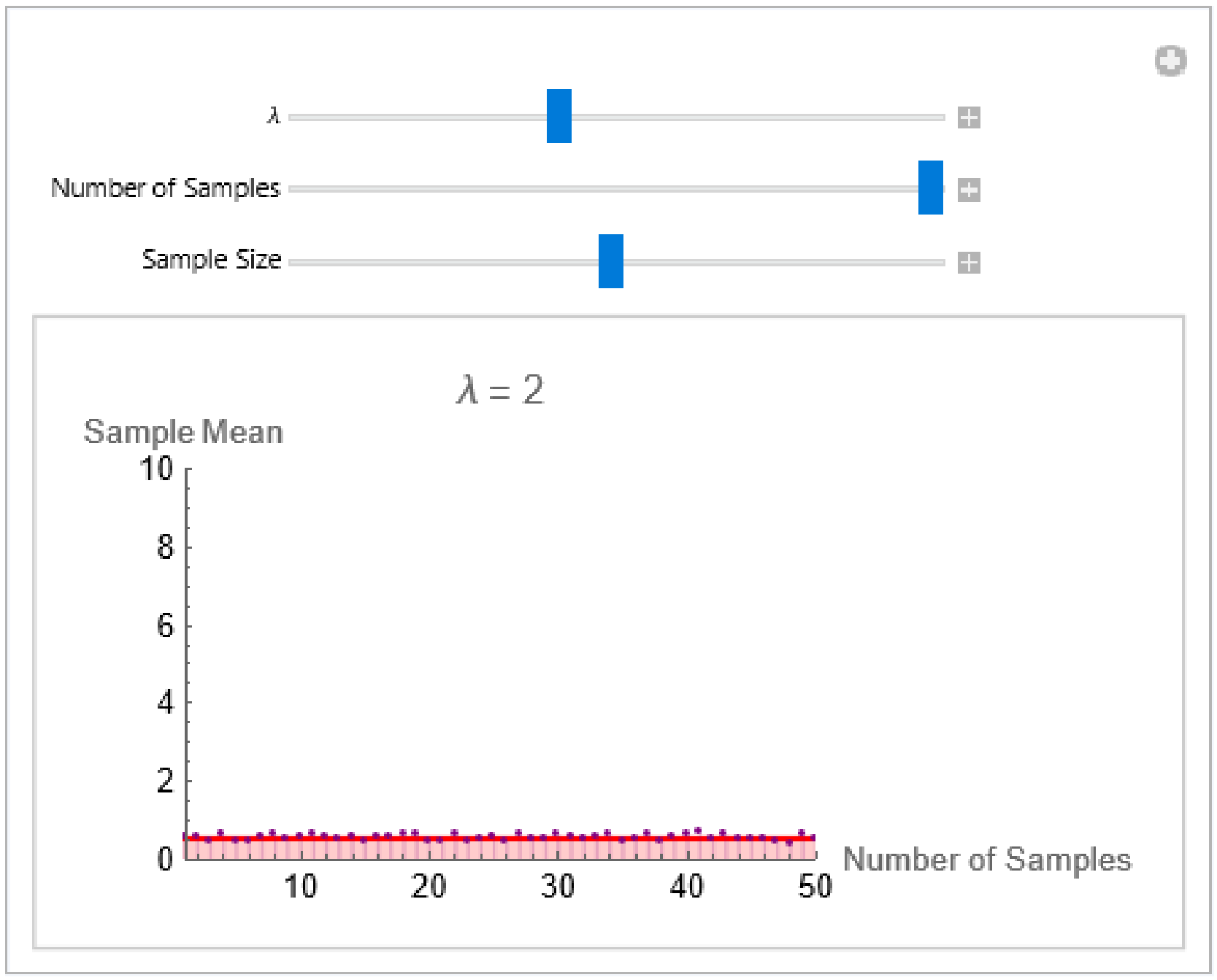

***Mathematica Examples 13.51***

Input

```
(* The code is designed to compare two Exponential distributions. It does this by generating random samples from each distribution and displaying them in a histogram, as well as plotting the PDFs of the two distributions. The code allows the user to manipulate the parameters λ1 and λ2 of both Exponential distributions through the sliders for λ1 and λ2. By changing these parameters, the user can see how the distributions change and how they compare to each other. The histograms display the sample data for each distribution, with the first histogram showing the sample data for the first Exponential distribution and the second histogram showing the sample data for the second Exponential distribution. The histograms are overlaid on each other, with the opacity of each histogram set to 0.2 to make it easier to see where the data overlap. The PDFs of the two distributions are also plotted on the same graph, with the first distribution shown in blue and the second distribution shown in red. The legend indicates which color corresponds to which distribution. By looking at the histograms and the PDFs, the user can compare the two Exponential distributions and see how they differ in terms of shape, scale, and overlap of their sample data: *)
Manipulate[
 Module[
  {dist1,dist2,data1,data2},
  SeedRandom[seed];
  dist1=ExponentialDistribution[λ1];
  dist2=ExponentialDistribution[λ2];
  data1=RandomVariate[dist1,n];
  data2=RandomVariate[dist2,n];
  Column[
   {
    Show[
     ListPlot[
      data1,
      ImageSize->320,
      PlotStyle->Blue
      ],
     ListPlot[
      data2,
      ImageSize->320,
```

```
              PlotStyle->Red
              ]
             ],
            Show[
             Plot[
              {PDF[dist1,x],PDF[dist2,x]},
              {x,Min[{data1,data2}],Max[{data1,data2}]},
              PlotLegends->{"Distribution 1","Distribution 2"},
              PlotRange->All,
              PlotStyle->{Blue,Red},
              ImageSize->320
              ],
             Histogram[
              {data1,data2},
              Automatic,
              "PDF",
              ChartLegends->{"sample data1","sample data2"},
              ChartStyle->{Directive[Opacity[0.2],Red],Directive[Opacity[0.2],Purple]},
              ImageSize->320
              ]
             ]
            }
           ]
          ],
         {{λ1,6},0.1,10,0.1},
         {{λ2,6},0.1,10,0.1},
         {{n,500},{100,500,1000,2000}},
         {{seed,1234},ControlType->None}
         ]
```

Output

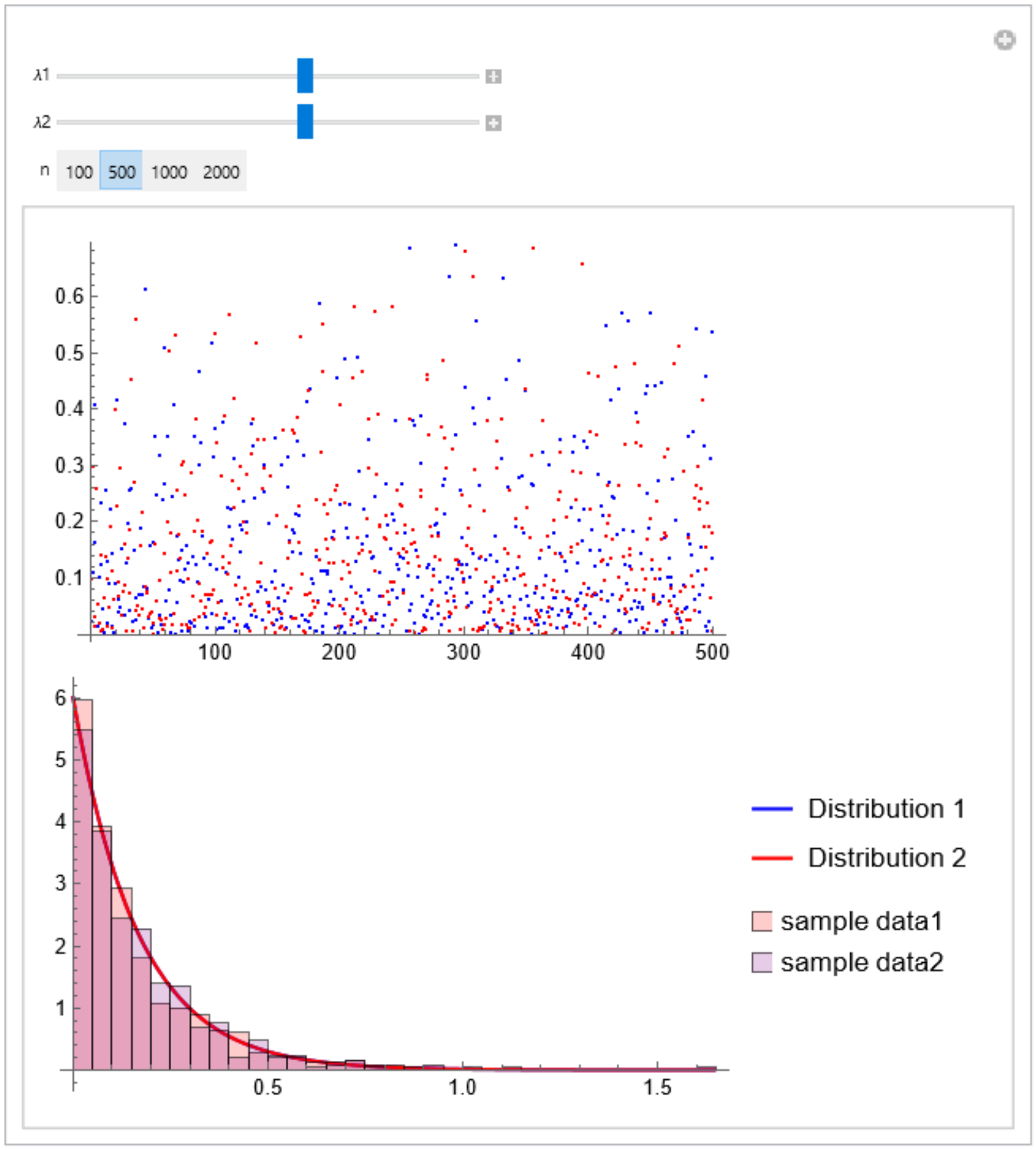

***Mathematica Examples 13.52***

```
Input    (* An exponential distribution with a rate parameter of 1/4000 per hour describes
         the lifespan of a battery. Calculate the probability that a random battery will
         operate for less than 3000 hours: *)

         λ=1/4000.;
         prob1=Probability[t<3000,t\[Distributed]ExponentialDistribution[λ]]
         (* or *)
         prob2=CDF[ExponentialDistribution[λ],3000]

Output   0.527633
Output   0.527633
```

# UNIT 13.4

# GAMMA DISTRIBUTION

***Mathematica Examples 13.53***

Input

```
(* This code generates a plot of the probability density function (PDF) for a Gamma distribution with different values of the parameter k= (1, 4, 6) and a fixed θ=2. The PDF is evaluated at various values of the random variable x between 0 and 20: *)

Plot[
 Evaluate[
  Table[
   PDF[
    GammaDistribution[k,2],
    x
    ],
   {k,{1,4,6}}
   ]
  ],
 {x,0,20},
 Filling->Axis,
 PlotLegends->Placed[{"k=1,θ=2","k=4,θ=2","k=6,θ=2"},{0.25,0.75}],
 PlotStyle->{RGBColor[0.88,0.61,0.14],RGBColor[0.37,0.5,0.7],Purple},
 ImageSize->320,
 AxesLabel->{None,"PDF"}
 ]
```

Output

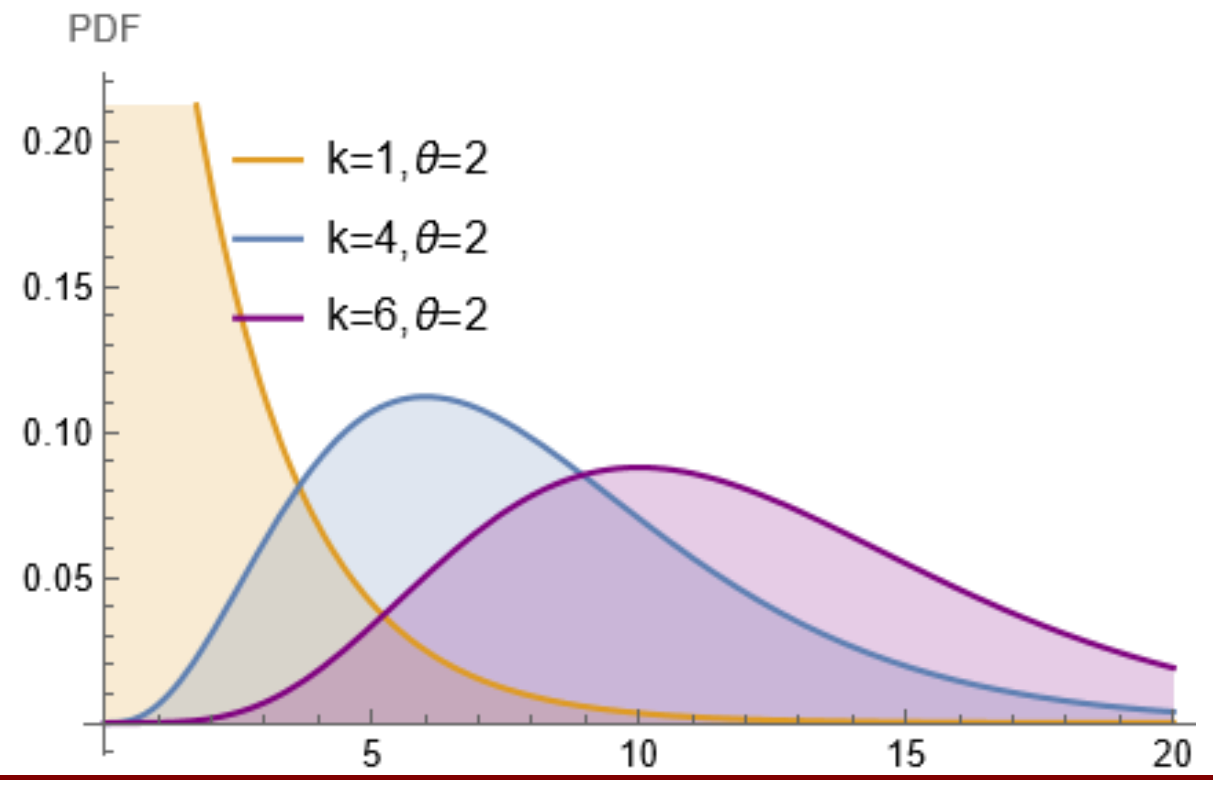

***Mathematica Examples 13.54***

Input

```
(* The code generates a plot of the cumulative distribution function (CDF) of the Gamma distribution with different values of the parameter k= (1, 4, 6) and a fixed θ=2. The CDF is evaluated at various values of the random variable x between 0 and 20: *)

Plot[
 Evaluate[
  Table[
   CDF[
    GammaDistribution[k,2],
    x
    ],
```

```
    {k,{1,4,6}}
    ]
   ],
  {x,0,20},
  Filling->Axis,
  PlotLegends->Placed[{"k=1,θ=2","k=4,θ=2","k=6,θ=2"},{0.25,0.75}],
  PlotStyle->{RGBColor[0.88,0.61,0.14],RGBColor[0.37,0.5,0.7],Purple},
  ImageSize->320,
  AxesLabel->{None,"CDF"}
  ]
```

Output

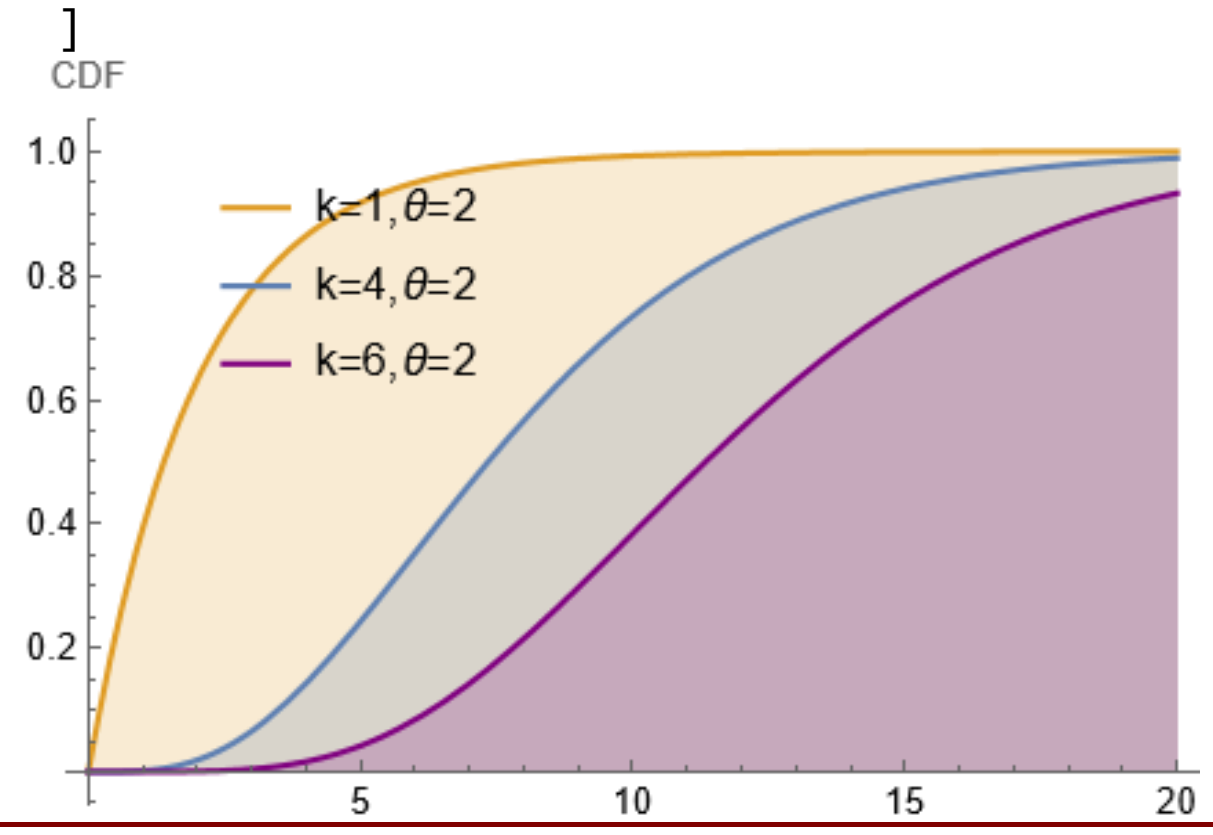

***Mathematica Examples 13.55***

Input

```
(* This code generates a plot of the PDF for a Gamma distribution with different
values of the parameter θ= (2, 4, 6) and a fixed k=2. The PDF is evaluated at various
values of the random variable x between 0 and 20: *)

Plot[
 Evaluate[
  Table[
   PDF[
    GammaDistribution[2,θ],
    x
    ],
   {θ,{2,4,6}}
   ]
  ],
 {x,0,20},
 Filling->Axis,
 PlotLegends->Placed[{"k=2,θ=2","k=2,θ=4","k=2,θ=6"},{0.4,0.75}],
 PlotStyle->{RGBColor[0.88,0.61,0.14],RGBColor[0.37,0.5,0.7],Purple},
 ImageSize->320,
 AxesLabel->{None,"PDF"}
 ]
```

Output

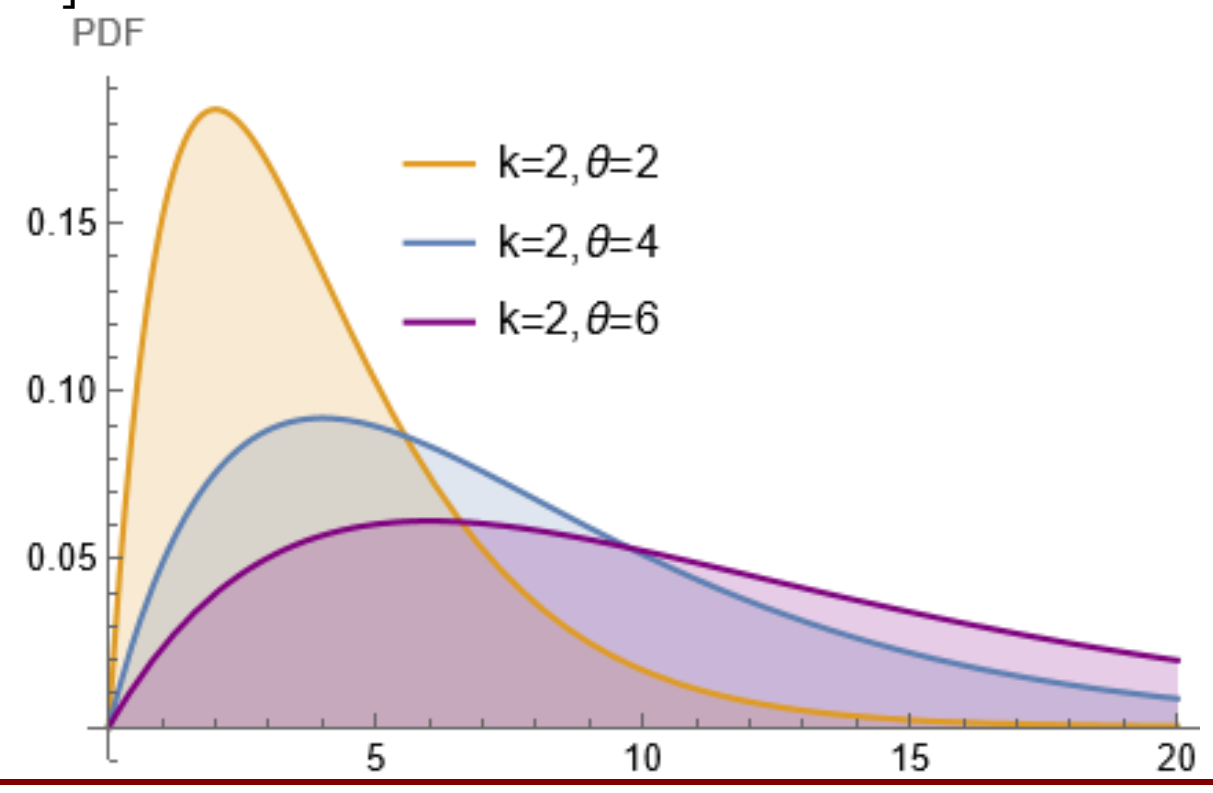

***Mathematica Examples 13.56***

Input

```
(* The code generates a plot of the CDF of the Gamma distribution with different values of the parameter θ= (2, 4, 6) and a fixed k=2. The CDF is evaluated at various values of the random variable x between 0 and 20: *)

Plot[
 Evaluate[
  Table[
   CDF[
    GammaDistribution[2,θ],
    x
    ],
   {θ,{2,4,6}}
   ]
  ],
 {x,0,20},
 Filling->Axis,
 PlotLegends->Placed[{"k=2,θ=2","k=2,θ=4","k=2,θ=6"},{0.2,0.75}],
 PlotStyle->{RGBColor[0.88,0.61,0.14],RGBColor[0.37,0.5,0.7],Purple},
 ImageSize->320,
 AxesLabel->{None,"CDF"}
 ]
```

Output

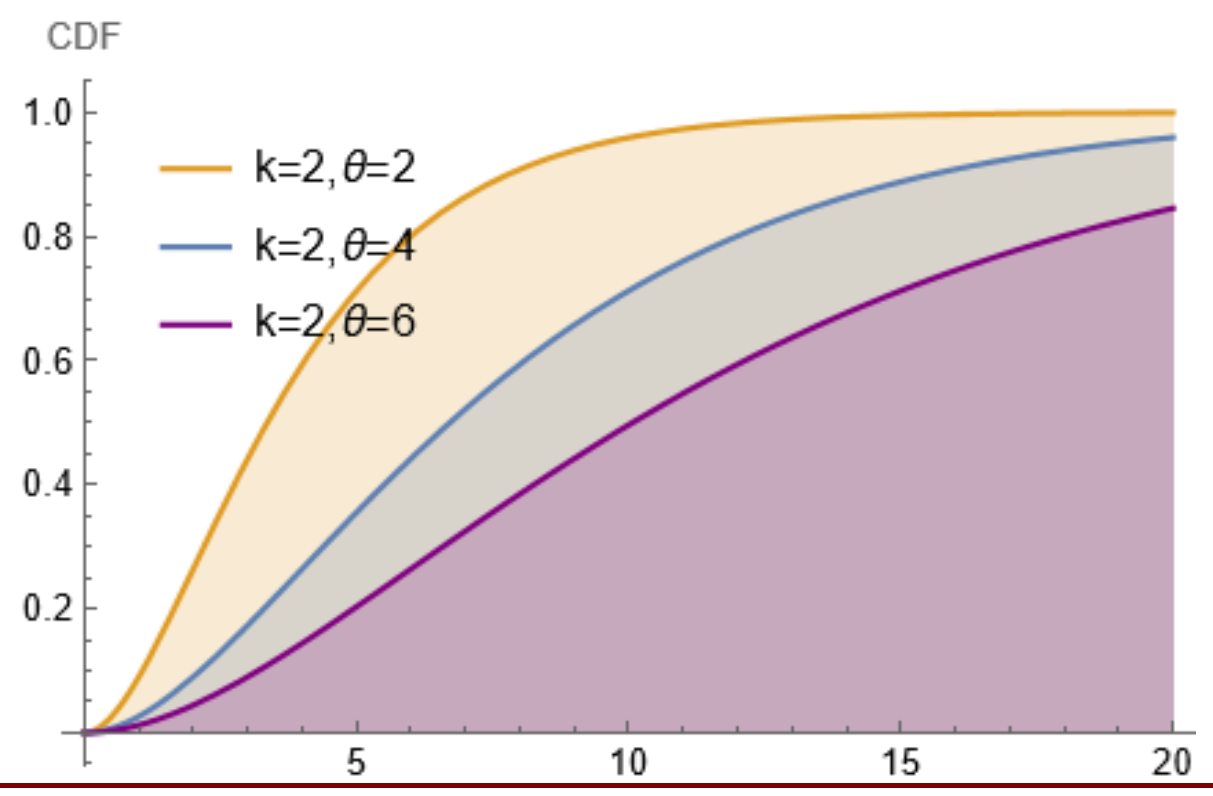

***Mathematica Examples 13.57***

Input

```
(* The code generates a histogram and a plot of the PDF for a Gamma distribution with parameters α=4 and β=2 and sample size 10000: *)

data=RandomVariate[
    GammaDistribution[4,2],
    10^4
    ];

Show[
 Histogram[
  data,
  15,
  "PDF",
  ColorFunction->Function[{height},Opacity[height]],
  ChartStyle->Purple,
  ImageSize->320,
  AxesLabel->{None,"PDF"}
  ],

 Plot[
```

```
          PDF[
           GammaDistribution[4,2],
           x
           ],
          {x,Min[data],Max[data]},
          PlotStyle->RGBColor[0.88,0.61,0.14]
          ]
         ]
```

Output

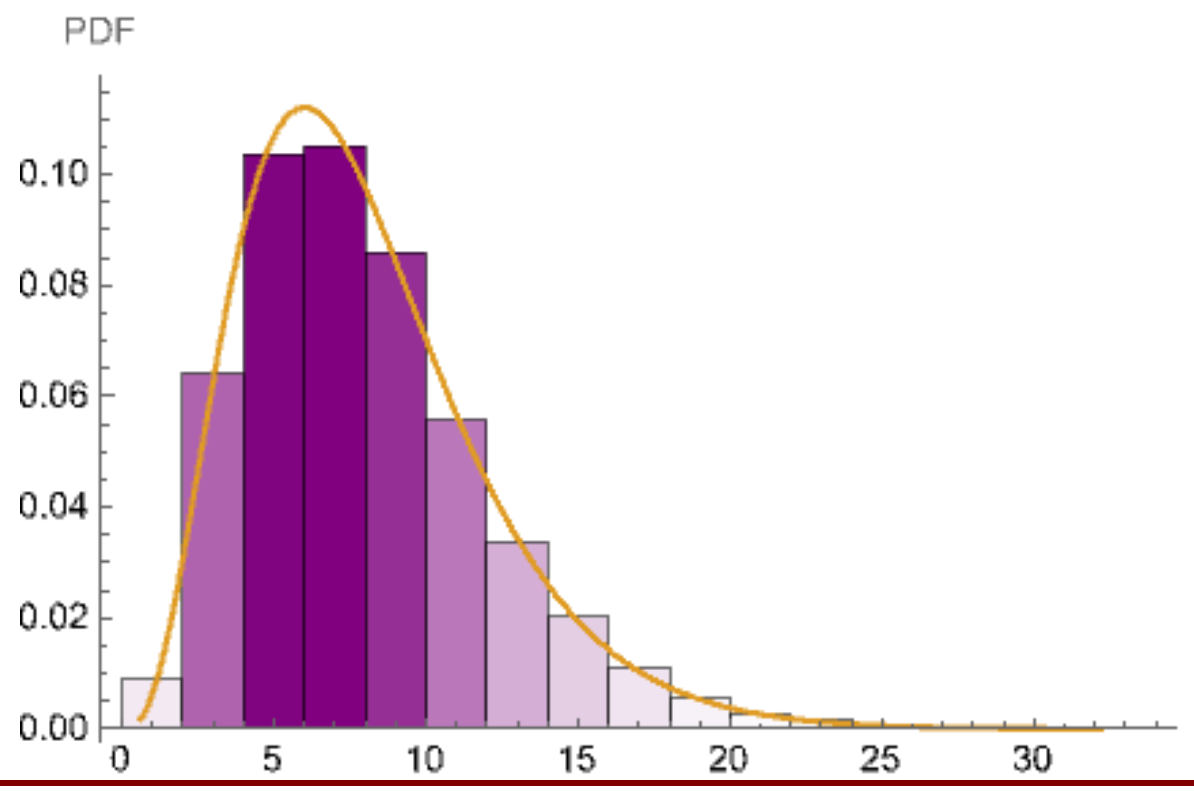

***Mathematica Examples 13.58***

Input

```
(* The code creates a dynamic histogram of data and a plot of the PDF generated from
a Gamma distribution using the Manipulate function. The Manipulate function creates
interactive controls for the user to adjust the values of α, β and n, which are the
parameters of the Gamma distribution and the sample size: *)

Manipulate[
 Module[
  {
   data=RandomVariate[
     GammaDistribution[α,β],
     n
     ]
   },

  Show[
   Histogram[
    data,
    {1},
    "PDF",
    PlotRange->{{0,11},All},
    ColorFunction->Function[{height},Opacity[height]],
    ImageSize->320,
    ChartStyle->Purple
    ],
   Plot[
    PDF[
     GammaDistribution[α,β],
     x
     ],
    {x,0,11},
    PlotRange->All,
    ColorFunction->"Rainbow"
    ]
   ]
  ],
```

```
{{α,0.1,"α"},0.1,4,0.1},
{{β,0.5,"β"},0.1,4,0.1},
{{n,300,"n"},100,1000,10}
]
```

Output

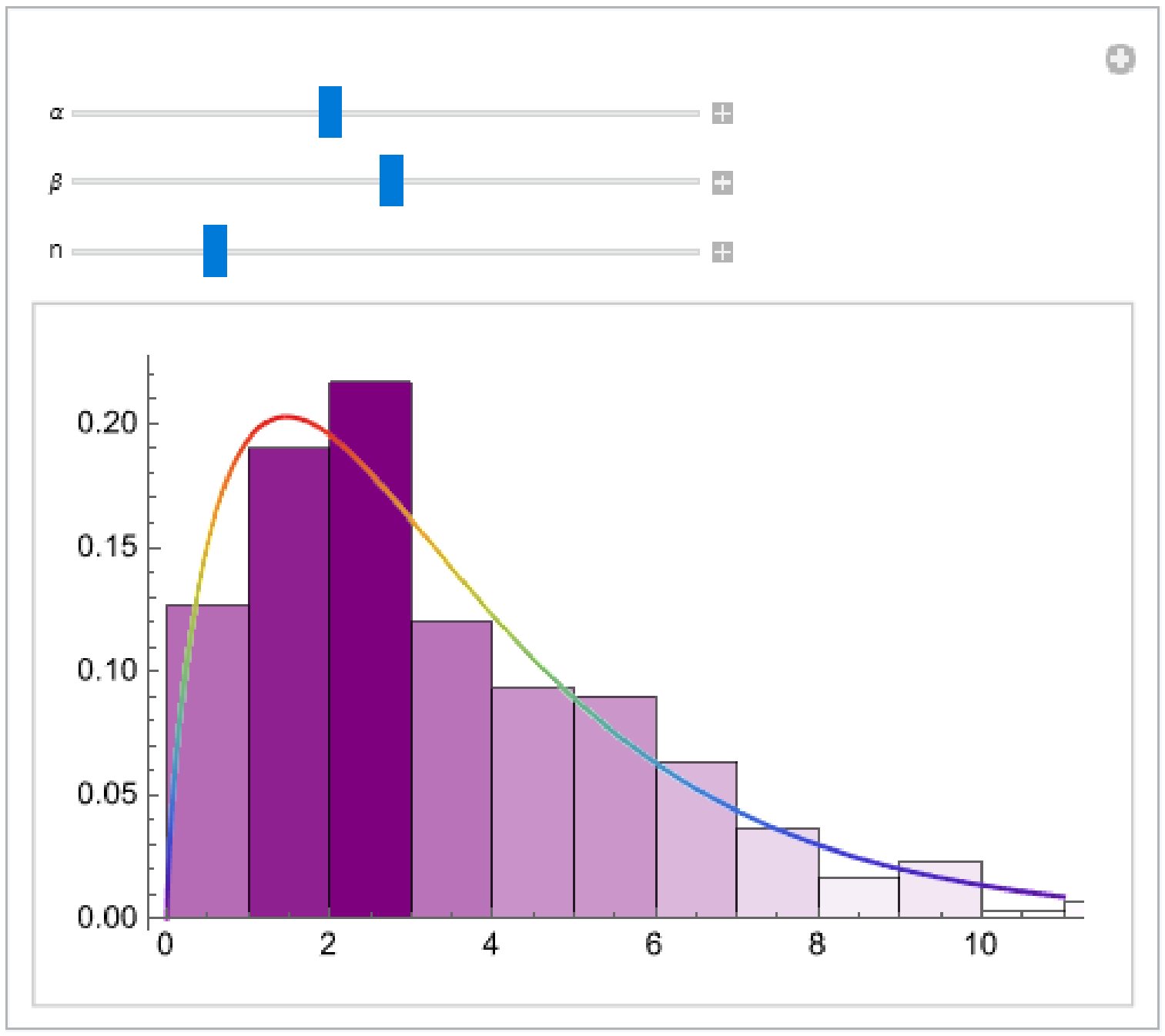

***Mathematica Examples 13.59***

Input

```
(* The code creates a plot of the CDF of a Gamma distribution using the Manipulate
function. The Manipulate function allows you to interactively change the values of
the parameters α and β, respectively: *)

Manipulate[
 Plot[
  CDF[
   GammaDistribution[α,β],
   x
   ],
  {x,0,10},
  Filling->Axis,
  FillingStyle->LightPurple,
  PlotRange->All,
  AxesLabel->{"x","CDF"},
  ImageSize->320,
  PlotStyle->Purple,
  PlotLabel->Row[{"α = ",α,", β = ",β}]
  ],
 {{α,0.1},0.1,4,0.1},
 {{β,0.5},0.1,4,0.1}
 ]
```

Output

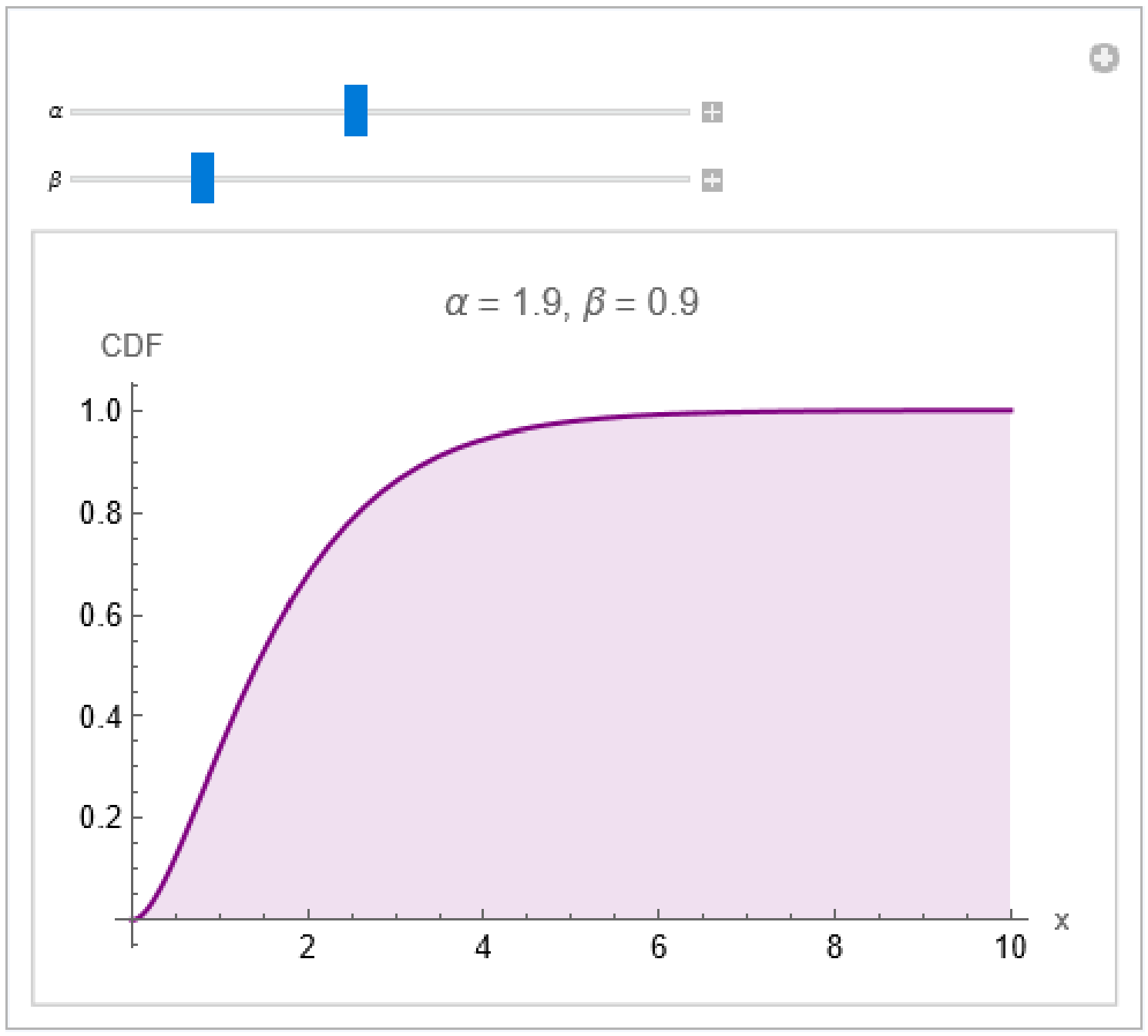

***Mathematica Examples 13.60***

Input

```
(* The code uses the Grid function to create a grid of two plots, one for the PDF
and one for the CDF of Gamma distribution. The code uses slider controls to adjust
the values of α and β: *)

Manipulate[
 Grid[
  {
   {Plot[
     PDF[
      GammaDistribution[α,β],
      x
      ],
     {x,0,10},
     PlotRange->All,
     PlotStyle->{Purple,PointSize[0.03]},
     PlotLabel->"PDF of Gamma distribution",
     AxesLabel->{"x","PDF"}
     ],
    Plot[
     CDF[
      GammaDistribution[α,β],
      x
      ],
     {x,0,10},
     PlotRange->All,
     PlotStyle->{Purple,PointSize[0.03]},
     PlotLabel->"CDF of Gamma distribution",
     AxesLabel->{"x","CDF"}
     ]
    }
   },
  Spacings->{5,5}
  ],
 {{α,1.7},0.1,4,0.1},
 {{β,2},0.1,4,0.1}
```

```
  ]
```

Output

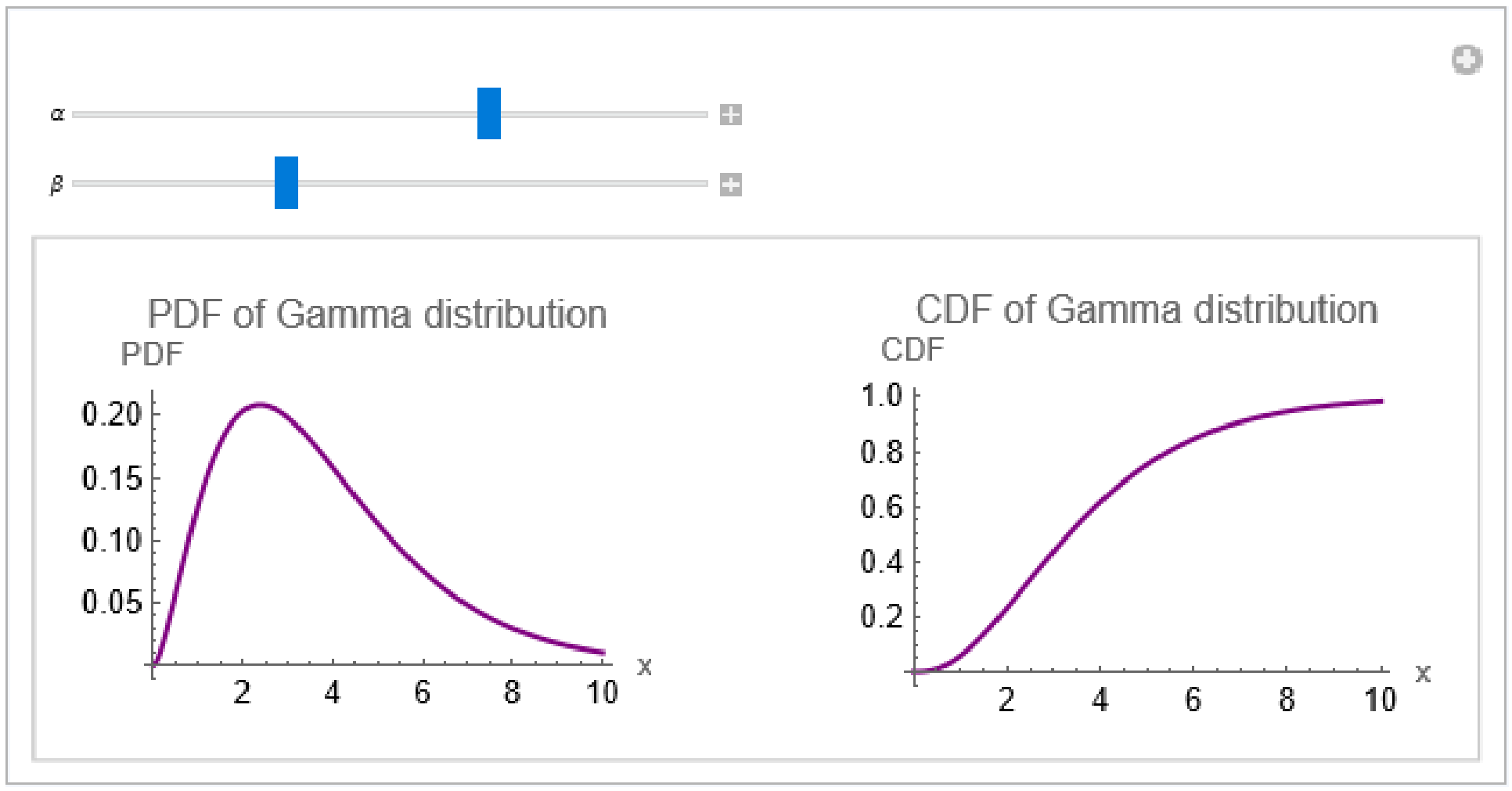

***Mathematica Examples 13.61***

Input

```
(* The code calculates and displays some descriptive statistics (mean, variance, standard deviation, kurtosis and skewness) for a Gamma distribution with parameters α and β: *)

Grid[
 Table[
  {
   statistics,
   FullSimplify[statistics[GammaDistribution[α,β]]]
   },
  {statistics,{Mean,Variance,StandardDeviation,Kurtosis,Skewness}}
  ],
 ItemStyle->12,
 Alignment->{{Right,Left}},
 Frame->All,
 Spacings->{Automatic,0.8}
 ]
```

Output

| | |
|---|---|
| Mean | α β |
| Variance | α β^2 |
| StandardDeviation | Sqrt[α] β |
| Kurtosis | 3+6/α |
| Skewness | 2/Sqrt[α] |

***Mathematica Examples 13.62***

Input

```
(* The code calculates and displays some additional descriptive statistics (moments, central moments, and factorial moments) for a Gamma distribution with parameters α and β: *)

Grid[
 Table[
  {
   statistics,
   FullSimplify[statistics[GammaDistribution[α,β],1]],
   FullSimplify[statistics[GammaDistribution[α,β],2]]
   },
  {statistics,{Moment,CentralMoment,FactorialMoment}}
```

```
  ],
 ItemStyle->12,
 Alignment->{{Right,Left}},
 Frame->All,
 Spacings->{Automatic,0.8}
 ]
```

Output

| | | |
|---|---|---|
| Moment | α β | α (1+α) β^2 |
| CentralMoment | 0 | α β^2 |
| FactorialMoment | α β | α β (-1+β+α β) |

***Mathematica Examples 13.63***

Input

```
(* The code generates a dataset of 1000 observations from a Gamma distribution with parameters α=1.7 and β=2. Then, it computes the sample mean and quartiles of the data, and plots a histogram of the data and plot of the PDF. Additionally, the code adds vertical lines to the plot corresponding to the sample mean and quartiles: *)

data=RandomVariate[
    GammaDistribution[1.7,2],
    10000
    ];

mean=Mean[data];
quartiles=Quantile[
    data,
    {0.25,0.5,0.75}
    ];

Show[
 Histogram[
  data,
  Automatic,
  "PDF",
  Epilog->{
    Directive[Red,Thickness[0.006]],
    Line[{{mean,0},{mean,0.25}}],
    Directive[Green,Dashed],
    Line[{{quartiles[[1]],0},{quartiles[[1]],0.25}}],
    Line[{{quartiles[[2]],0},{quartiles[[2]],0.25}}],
    Line[{{quartiles[[3]],0},{quartiles[[3]],0.25}}]
    },
  ColorFunction->Function[{height},Opacity[height]],
  ImageSize->320,
  ChartStyle->Purple,
  PlotRange->{{0,20},{0,0.25}}
  ],
 Plot[
  PDF[
   GammaDistribution[1.7,2],
   x
   ],
  {x,0,20},
  PlotRange->{{0,20},{0,0.25}},
  ImageSize->320,
  ColorFunction->"Rainbow"
  ]
 ]
```

Output

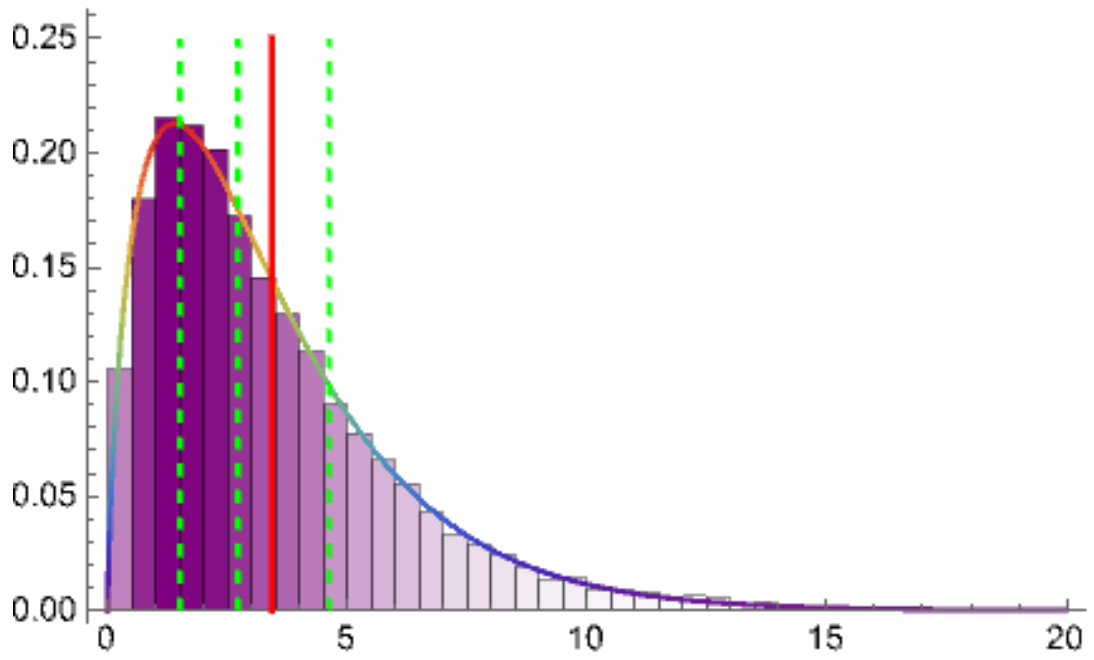

***Mathematica Examples 13.64***

Input

```
(* The code generates a random sample of size 10,000 from a Gamma distribution with parameters α=1.7 and β=2, estimates the distribution parameters using the EstimatedDistribution function, and then compares the histogram of the sample with the estimated PDF of the Gamma distribution using a histogram and a plot of the PDF: *)

sampledata=RandomVariate[
    GammaDistribution[1.7,2],
    10^4
    ];
(* Estimate the distribution parameters from sample data: *)
ed=EstimatedDistribution[
   sampledata,
   GammaDistribution[α,β]
   ]
(* Compare a density histogram of the sample with the PDF of the estimated distribution: *)
Show[
 Histogram[
   sampledata,
   {1},
   "PDF",
   ColorFunction->Function[{height},Opacity[height]],
   ChartStyle->Purple,
   ImageSize->320
   ],
 Plot[
   PDF[ed,x],
   {x,0,20},
   ImageSize->320,
   ColorFunction->"Rainbow"
   ]
 ]
```

Output

```
GammaDistribution[1.68648,2.00557]
```

Output

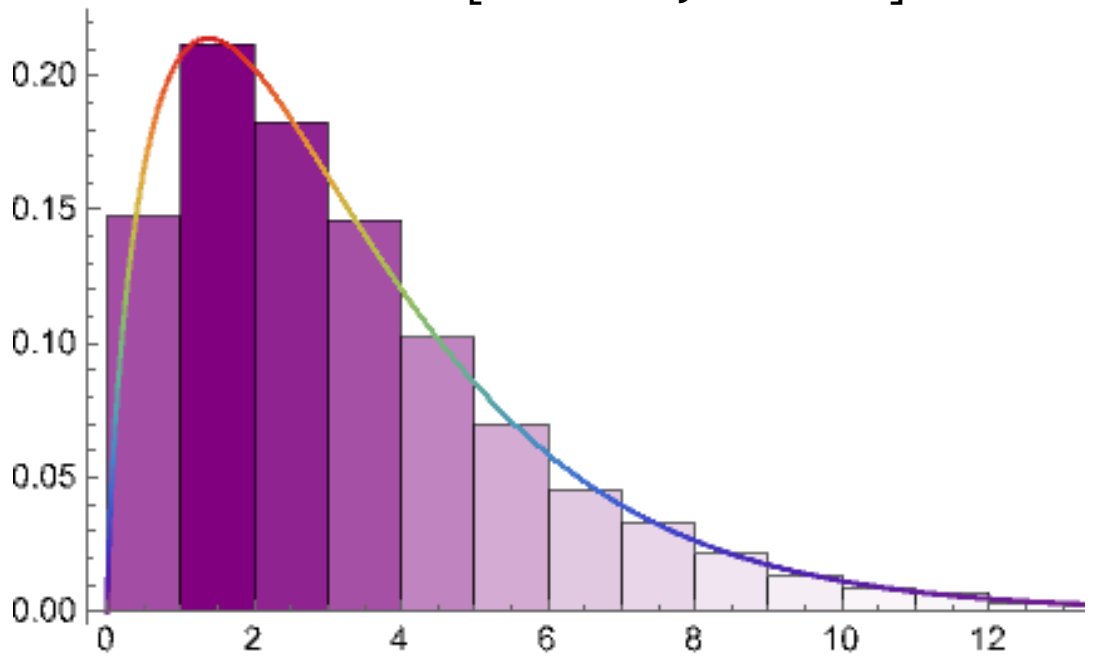

***Mathematica Examples 13.65***

Input

```
(* The code creates a Manipulate interface where the user can adjust the parameters of two Gamma distributions (dist1 and dist2) and see the resulting sum of these distributions (distSum) plotted on the same graph. The plot shows the PDFs of each distribution as well as the PDF of the sum of the distributions. The TransformedDistribution function is used to create the sum of two distributions by defining the distribution of the sum of two random variables, x and y, where x follows dist1 and y follows dist2. Hence, we prove that Gamma distribution is closed under addition: *)
Manipulate[
 dist1=GammaDistribution[α1,β1];
 dist2=GammaDistribution[α2,β2];

 distSum=TransformedDistribution[
   x+y,
   {Distributed[x,dist1],Distributed[y,dist2]}
   ];

 Plot[
  {
   PDF[dist1,x],
   PDF[dist2,x],
   PDF[distSum,x]
   },
  {x,0,20},
  ImageSize->400,
  PlotRange->All,
  AxesLabel->{"x","f(x)"},
  Filling->{1->{2},2->{3}},
  FillingStyle->{LightBlue,LightPurple},
  PlotLegends->{"Distribution 1","Distribution 2","Sum of Distributions"}
  ],
 {{α1,1.7,"α1"},0.1,5,0.1,Appearance->"Labeled"},
 {{β1,2,"β1"},1,3,0.1,Appearance->"Labeled"},
 {{α2,1.7,"α2"},0.1,5,0.1,Appearance->"Labeled"},
 {{β2,2,"β2"},1,3,0.1,Appearance->"Labeled"}]
```

Output

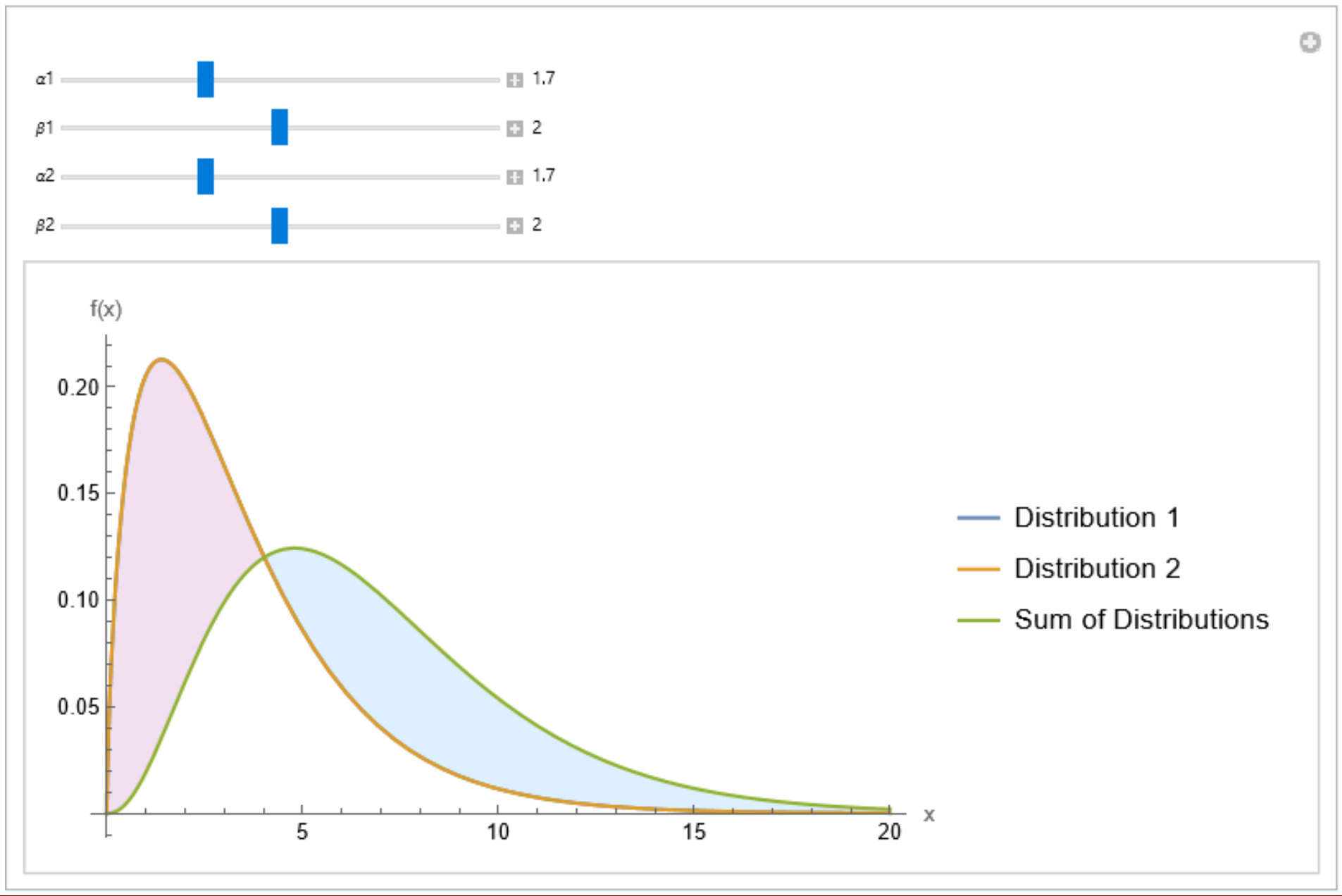

***Mathematica Examples 13.66***

Input

```
(* The code generates a 2D dataset with 1000 random points that follow a Gamma distribution with α=1.7 and β=2. The dataset is then used to create a row of three plots. The first plot is a histogram of the X-axis values of the dataset. The second plot is a histogram of the Y-axis values of the dataset. It is similar to the first plot, but shows the distribution of the Y-axis values instead. The third plot is a scatter plot of the dataset, with the X-axis values on the horizontal axis and the Y-axis values on the vertical axis. Each point in the plot represents a pair of X and Y values from the dataset: *)

data=RandomVariate[
    GammaDistribution[1.7,2],
    {1000,2}
    ];
GraphicsRow[
 {
  Histogram[
    data[[All,1]],
    {0.1},
    ImageSize->170,
    PlotLabel->"X-axis",
    ColorFunction->Function[{height},Opacity[height]],
    ChartStyle->Purple
    ],
  Histogram[
    data[[All,2]],
    {0.1},
    ImageSize->170,
    PlotLabel->"Y-axis",
    ColorFunction->Function[{height},Opacity[height]],
    ChartStyle->Purple
    ],
  ListPlot[
    data,
    ImageSize->170,
    PlotStyle->{Purple,PointSize[0.015]},
    AspectRatio->1,
    Frame->True,
    Axes->False
    ]
  }
 ]
```

Output

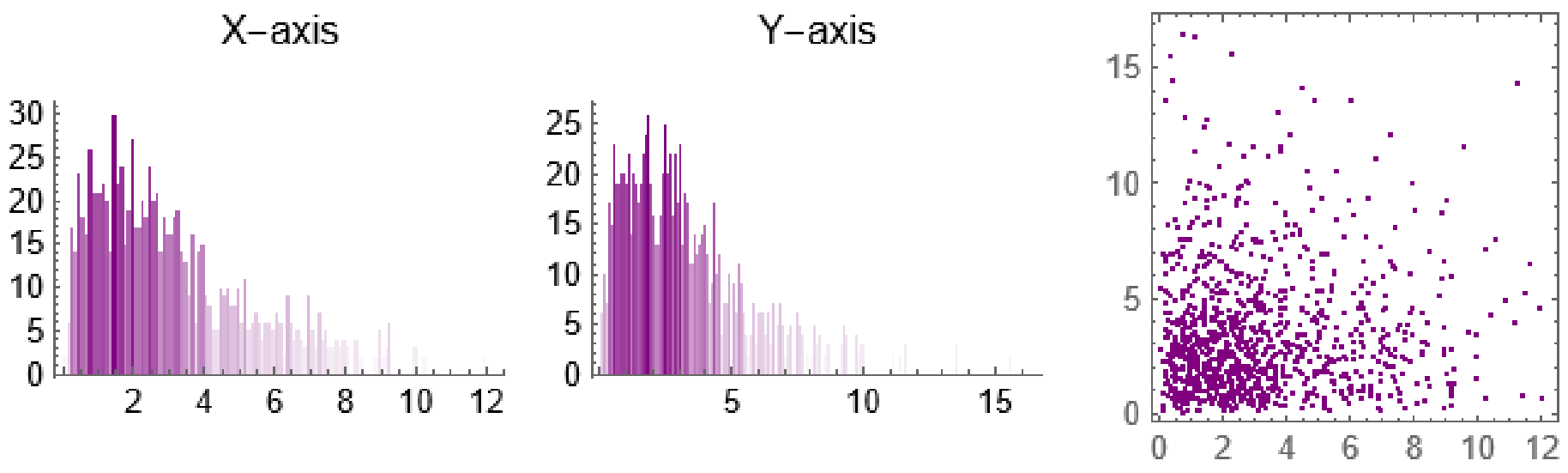

***Mathematica Examples 13.67***

Input

```
(* The code generates a set of random data points with a Gamma distribution with α=1.7 and β=2 in three dimensions, and then creates three histograms, one for each
```

```
dimension, showing the distribution of the points along that axis. Additionally, it
creates a 3D scatter plot of the data points: *)

data=RandomVariate[
    GammaDistribution[1.7,2],
    {1000,3}
    ];

GraphicsGrid[
 {
  {
   Histogram[
    data[[All,1]],
    Automatic,
    "PDF",
    PlotLabel->"X-axis",
    ColorFunction->Function[{height},Opacity[height]],
    ChartStyle->Purple
    ],
   Histogram[
    data[[All,2]],
    Automatic,
    "PDF",
    PlotLabel->"Y-axis",
    ColorFunction->Function[{height},Opacity[height]],
    ChartStyle->Purple
    ],
   Histogram[
    data[[All,3]],
    Automatic,
    "PDF",
    PlotLabel->"Z-axis",
    ColorFunction->Function[{height},Opacity[height]],
    ChartStyle->Purple
    ],
   ListPointPlot3D[
    data,
    BoxRatios->{1,1,1},
    PlotStyle->{Purple,PointSize[0.015]}
    ]
   }
  }
 ]
```

Output

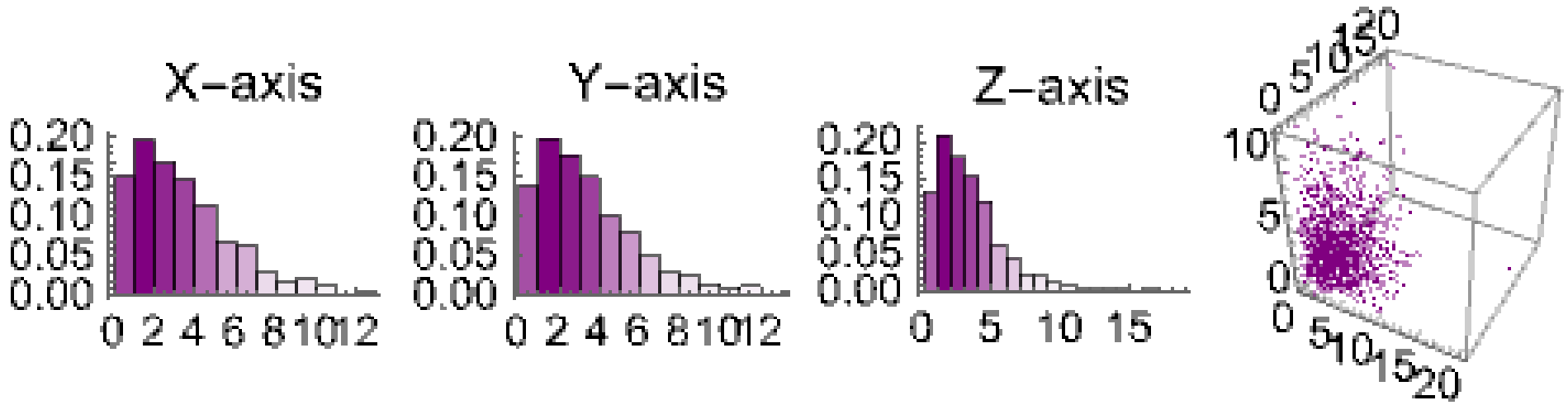

*Mathematica Examples 13.68*

Input

```
(* The code generates a 3D scatter plot of Gamma distribution points, where the x-
axis is red, y-axis is green, and z-axis is blue: *)

data=RandomVariate[
    GammaDistribution[1.7,2],
    {2000,3}
```

```
   ];
 Graphics3D[
  {
   {PointSize[0.006],Purple,Opacity[0.6],Point[data]},
   Thin,
   {Red,Opacity[0.4],Line[{{#,0,0},{#,0,-1}}]&/@data[[All,1]]},
   Thin,
   {Green,Opacity[0.4],Line[{{0,#,0},{0,#,-1}}]&/@data[[All,2]]},
   Thin,
   {Blue,Opacity[0.4],Line[{{0,0,#},{0,-1,#}}]&/@data[[All,3]]}
   },
  BoxRatios->{1,1,1},
  Axes->True,
  AxesLabel->{"X","Y","Z"},
  ImageSize->320
  ]
```

Output

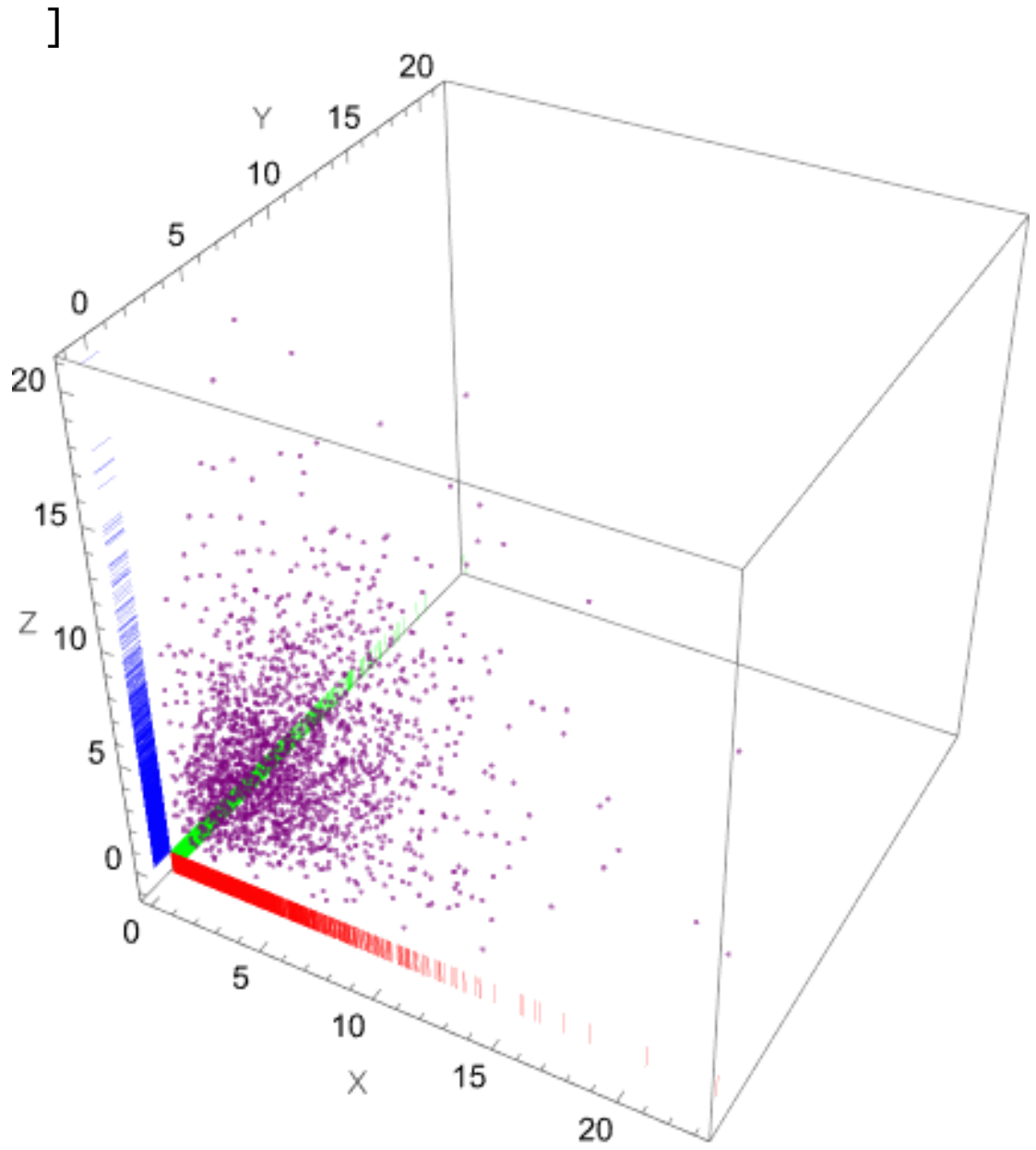

*Mathematica Examples 13.69*

Input

```
(* The code demonstrates a common technique in statistics and data analysis, which is the use of random sampling to estimate population parameters. The code generates random samples from a Gamma distribution with α=1.7 and β=2, and then using these samples to estimate the parameters of another Gamma distribution with unknown α and β. This process is repeated 20 times, resulting in 20 different estimated distributions. The code also visualizes the resulting estimated distributions using the PDF function. The code plots the PDFs of these estimated distributions using the PDF function and the estimated parameters. The plot shows the PDFs in a range from-0 to 15. The code also generates a list plot of 2 sets of random samples from the Gamma distribution with α=1.7 and β=2. The plot shows the 100 random points generated from two random samples. The code generates also a histogram of the PDF for Gamma distribution of the two samples: *)

estim0distributions=Table[
  dist=GammaDistribution[1.7,2];

  sampledata=RandomVariate[
    dist,
    100
    ];
```

```
    ed=EstimatedDistribution[
      sampledata,
      GammaDistribution[α,β]
      ],
    {i,1,20}
    ]

pdf0ed=Table[
   PDF[estim0distributions[[i]],x],
   {i,1,20}
   ];

(* Visualizes the resulting estimated distributions *)
Plot[
 pdf0ed,
 {x,0,15},
 PlotRange->Full,
 ImageSize->400,
 PlotStyle->Directive[Purple,Opacity[0.3],Thickness[0.002]]
 ]

(* Visualizes 100 random points generated from two random samples *)
table=Table[
   dist=GammaDistribution[1.7,2];
   sampledata=RandomVariate[
     dist,
     100],
   {i,1,2}
   ];

ListPlot[
 table,
 ImageSize->320,
 Filling->Axis,
 PlotStyle->Directive[Opacity[0.5],Thickness[0.003]]
 ]

Histogram[
 table,
 Automatic,
 LabelingFunction->Above,
 ChartLegends->{"Sample 1","Sample 2"},
 ChartStyle->{Directive[Opacity[0.2],Red],Directive[Opacity[0.2],Purple]},
 ImageSize->320
 ]
```

Output

```
{GammaDistribution[1.82845,1.90159],GammaDistribution[2.02431,1.66529],GammaDistrib
ution[1.69342,2.08188],GammaDistribution[1.77679,1.9663],GammaDistribution[1.56903,
2.1404],GammaDistribution[1.85814,1.89476],GammaDistribution[1.56202,1.98928],Gamma
Distribution[1.65214,2.14826],GammaDistribution[2.18432,1.41547],GammaDistribution[
1.65386,2.05039],GammaDistribution[1.7418,2.13234],GammaDistribution[1.62585,1.8542
6],GammaDistribution[1.77923,1.79503],GammaDistribution[1.52099,2.25253],GammaDistr
ibution[1.81334,1.79421],GammaDistribution[1.5542,2.13024],GammaDistribution[1.5141
9,2.23658],GammaDistribution[2.22074,1.43941],GammaDistribution[2.41646,1.39435],Ga
mmaDistribution[1.80294,1.80006]}
```

Output
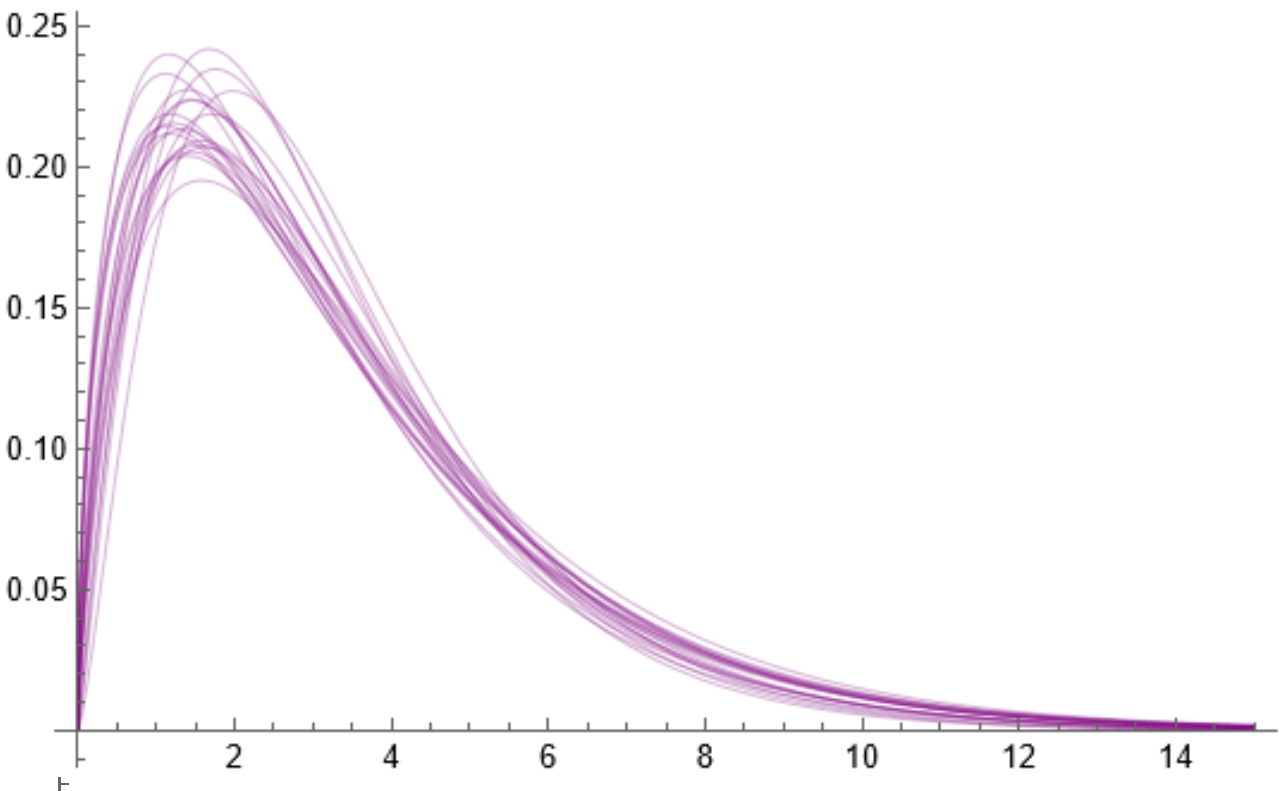

Output
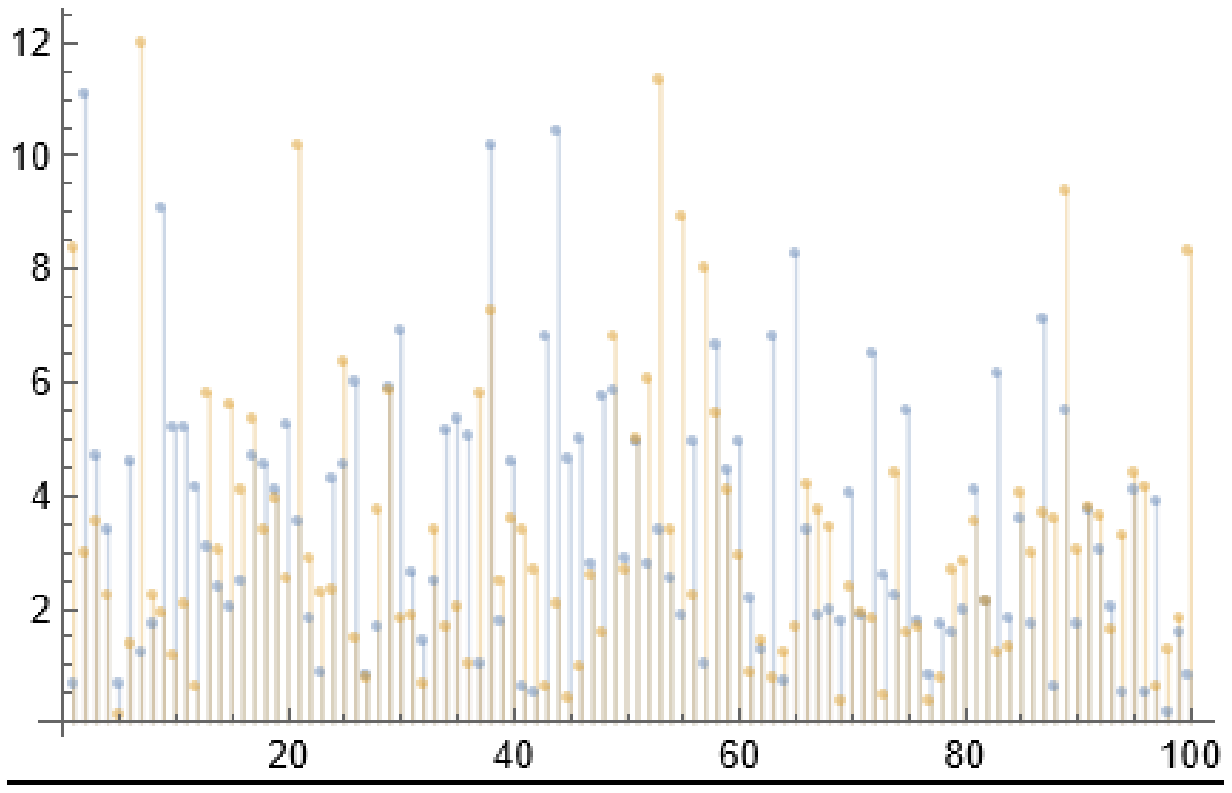

Output
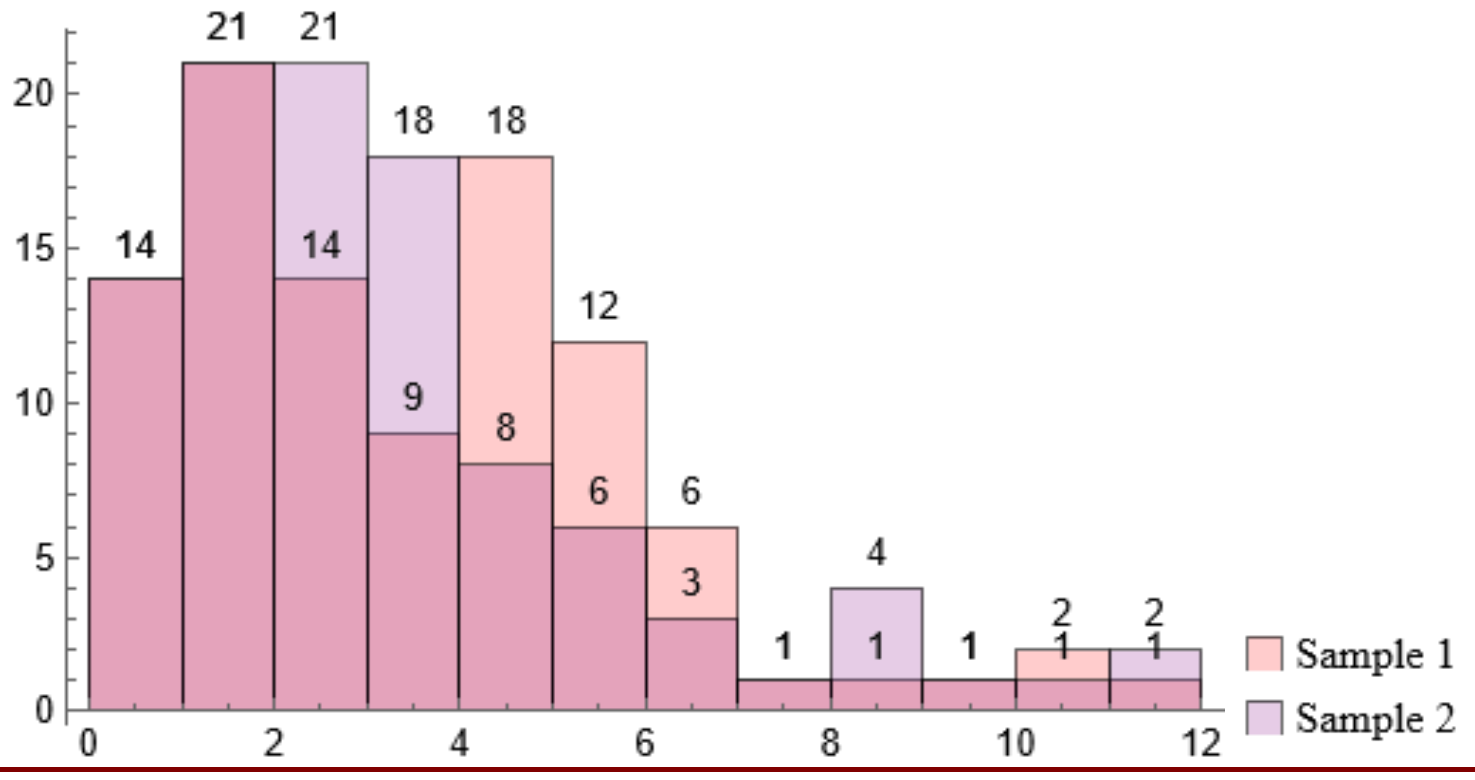

## *Mathematica Examples 13.70*

Input

```
(* The code generates and compares the means of random samples drawn from a gamma
distribution with the given shape and scale parameters. The code uses the Manipulate
function to create a user interface with sliders to adjust the values of the shape,
scale, number of samples, and sample size. By varying the values of "Number of
Samples" and "Sample Size" sliders, the code allows the user to explore how changing
these parameters affects the means of the random samples. Specifically, increasing
the number of samples n tends to make the distribution of the means narrower and more
concentrated around the true mean of the underlying distribution. On the other hand,
increasing the sample size tends to reduce the variability in the means and make them
more precise estimators of the true mean: *)
Manipulate[
 Module[
  {means,dist},

  dist=GammaDistribution[shape,scale];

  means=Map[
```

```
      Mean,
      RandomVariate[dist,{n,samples}]
      ];
    m=N[Mean[means]];

    ListPlot[
     {means,{{0,m},{n,m}}},
     Joined->{False,True},
     Filling->Axis,
     PlotRange->{{1,50},{0,20}},
     PlotStyle->{Purple,Red},
     AxesLabel->{"Number of Samples","Sample Mean"},
     PlotLabel->Row[{"Shape = ",shape,", Scale = ",scale}],
     ImageSize->320
     ]
    ],
   {{shape,2,"Shape"},0.1,4,0.1},
   {{scale,2,"Scale"},0.1,4,0.1},
   {{n,50,"Number of Samples"},3,50,1},
   {{samples,50,"Sample Size"},1,100,1},
   TrackedSymbols:>{n,samples,shape,scale}
   ]
```

Output

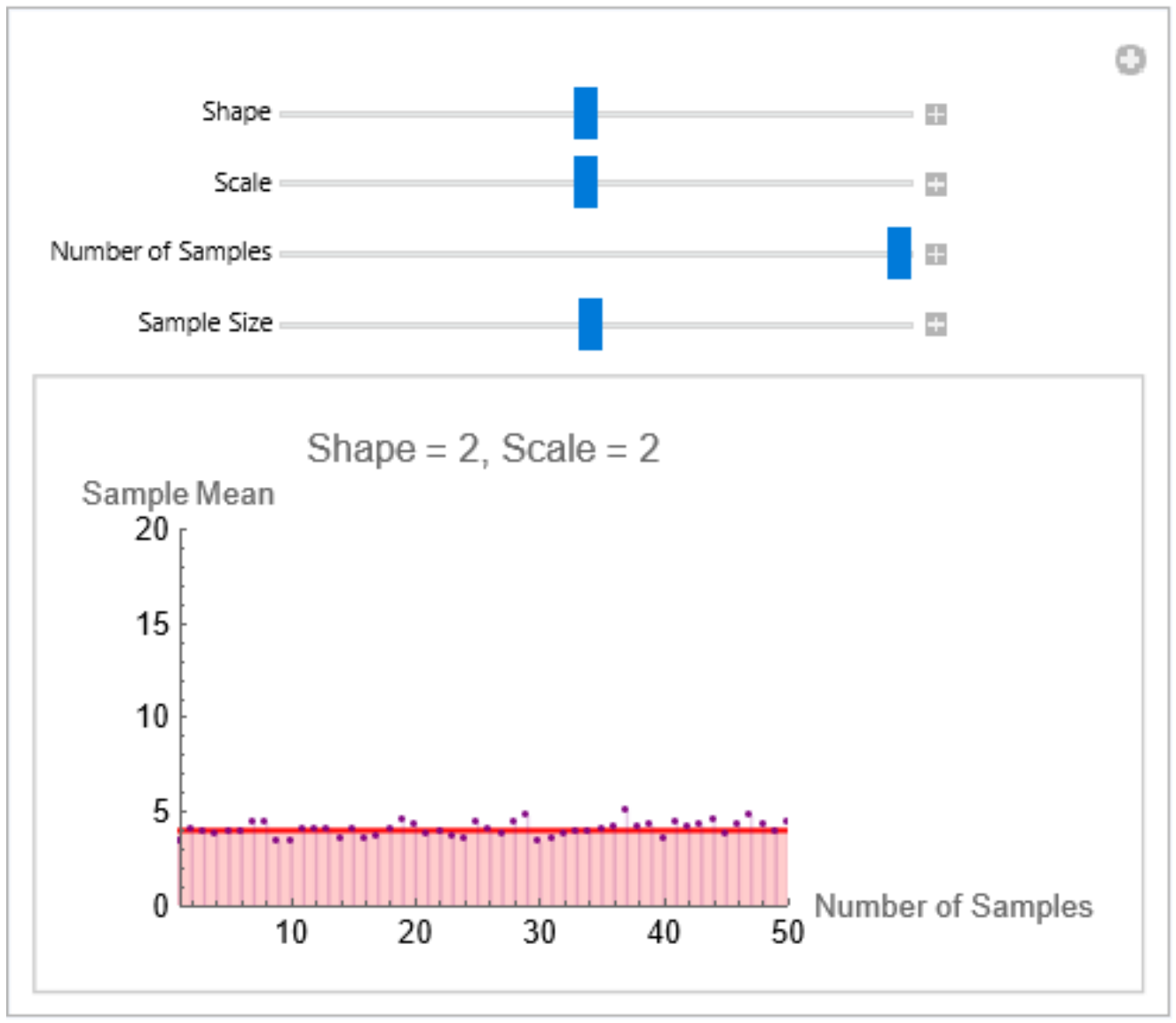

***Mathematica Examples 13.71***

Input

```
(* The code is designed to compare two gamma distributions. It does this by generating
random samples from each distribution and displaying them in a histogram, as well as
plotting the PDFs of the two distributions. The code allows the user to manipulate
the shape and scale parameters of both gamma distributions through the sliders for
α1, β1, α2 and β2. By changing these parameters, the user can see how the distributions
change and how they compare to each other. The histograms display the sample data
for each distribution, with the first histogram showing the sample data for the first
gamma distribution and the second histogram showing the sample data for the second
gamma distribution. The histograms are overlaid on each other, with the opacity of
each histogram set to 0.2 to make it easier to see where the data overlap. The PDFs
of the two distributions are also plotted on the same graph, with the first
distribution shown in blue and the second distribution shown in red. The legend
indicates which color corresponds to which distribution. By looking at the histograms
and the PDFs, the user can compare the two gamma distributions and see how they
differ in terms of shape, scale, and overlap of their sample data: *)
```

```
Manipulate[
 Module[
  {dist1,dist2,data1,data2},
  SeedRandom[seed];
  dist1=GammaDistribution[α1,β1];
  dist2=GammaDistribution[α2,β2];
  data1=RandomVariate[dist1,n];
  data2=RandomVariate[dist2,n];
  Column[
   {
    Show[
     ListPlot[
      data1,
      ImageSize->320,
      PlotStyle->Blue
      ],
     ListPlot[
      data2,
      ImageSize->320,
      PlotStyle->Red
      ]
     ],
    Show[
     Plot[
      {PDF[dist1,x],PDF[dist2,x]},
      {x,Min[{data1,data2}],Max[{data1,data2}]},
      PlotLegends->{"Distribution 1","Distribution 2"},
      PlotRange->All,
      PlotStyle->{Blue,Red},
      ImageSize->320
      ],
     Histogram[
      {data1,data2},
      Automatic,
      "PDF",
      ChartLegends->{"sample data1","sample data2"},
      ChartStyle->{Directive[Opacity[0.2],Red],Directive[Opacity[0.2],Purple]},
      ImageSize->320
      ]
     ]
    }
   ]
  ],
 {{α1,6},0.1,10,0.1},
 {{β1,2},0.1,10,0.1},
 {{α2,6},0.1,10,0.1},
 {{β2,2},0.1,10,0.1},
 {{n,500},{100,500,1000,2000}},
 {{seed,1234},ControlType->None}
 ]
```

Output

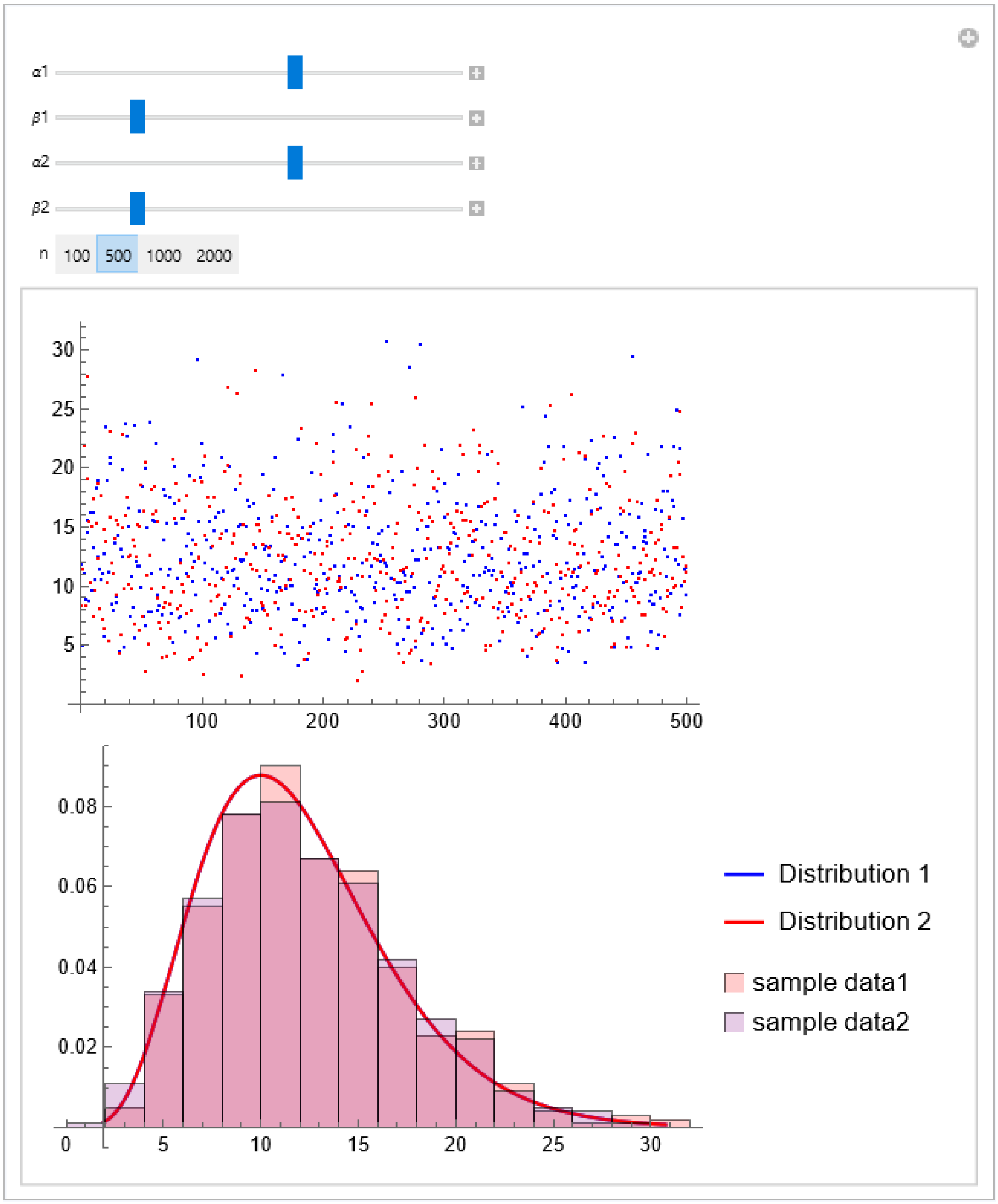

***Mathematica Examples 13.72***

Input

```
(* ChiSquareDistribution is a special case of gamma distribution: *)

PDF[
 ChiSquareDistribution[2 n],
 x
 ]

PDF[
 GammaDistribution[n,2],
 x
 ]
```

Output $\begin{cases} \frac{2^{-n} e^{-x/2} x^{-1+n}}{\text{Gamma}[n]} & x > 0 \\ 0 & \text{True} \end{cases}$

Output $\begin{cases} \frac{2^{-n} e^{-x/2} x^{-1+n}}{\text{Gamma}[n]} & x > 0 \\ 0 & \text{True} \end{cases}$

***Mathematica Examples 13.73***

Input

```
(* ChiDistribution is a special case of GammaDistribution: *)

PDF[
 GammaDistribution[1/2 v,Sqrt[2],2,0],
 x
 ]

PDF[
 ChiDistribution[v],
 x
 ]
```

Output $\begin{cases} \frac{2^{\frac{1}{2}+\frac{1-v}{2}} e^{-\frac{x^2}{2}} x^{-1+v}}{\text{Gamma}[v/2]} & x > 0 \\ 0 & \text{True} \end{cases}$

Output $\begin{cases} \frac{2^{\frac{1}{2}+\frac{1-v}{2}} e^{-\frac{x^2}{2}} x^{-1+v}}{\text{Gamma}[v/2]} & x > 0 \\ 0 & \text{True} \end{cases}$

***Mathematica Examples 13.74***

Input

```
(* ExponentialDistribution is a special case of gamma distribution: *)

PDF[
 ExponentialDistribution[1/λ],
 x
 ]

PDF[
 GammaDistribution[1,λ],
 x
 ]
```

Output $\begin{cases} \frac{e^{-\frac{x}{\lambda}}}{\lambda} & x >= 0 \\ 0 & \text{True} \end{cases}$

Output $\begin{cases} \frac{e^{-\frac{x}{\lambda}}}{\lambda} & x > 0 \\ 0 & \text{True} \end{cases}$

# UNIT 13.5

# NORMAL DISTRIBUTION

**Mathematica Examples 13.75**

Input

```
(* The code generates a plot of the probability density function (PDF) for a normal
distribution with different values of standard deviation σ= (0.75, 1 and 2) and a
fixed mean (μ=0). The plot shows the values of the PDF for all possible values of x
between -7 and 7. The resulting plot shows the bell-shaped curve of the normal
distribution with different shapes, reflecting the impact of varying the value of
the standard deviation: *)

Plot[
 Evaluate[
  Table[
   PDF[
    NormalDistribution[0,σ],
    x
    ],
   {σ,{.75,1,2}}
   ]
  ],
 {x,-7,7},
 PlotRange->All,
 Filling->Axis,
 PlotLegends->Placed[{"μ=0,σ=0.75","μ=0,σ=1","μ=0,σ=2"},{0.8,0.75}],
 PlotStyle->{RGBColor[0.88,0.61,0.14],RGBColor[0.37,0.5,0.7],Purple},
 ImageSize->320,
 AxesLabel->{None,"PDF"}
 ]
```

Output

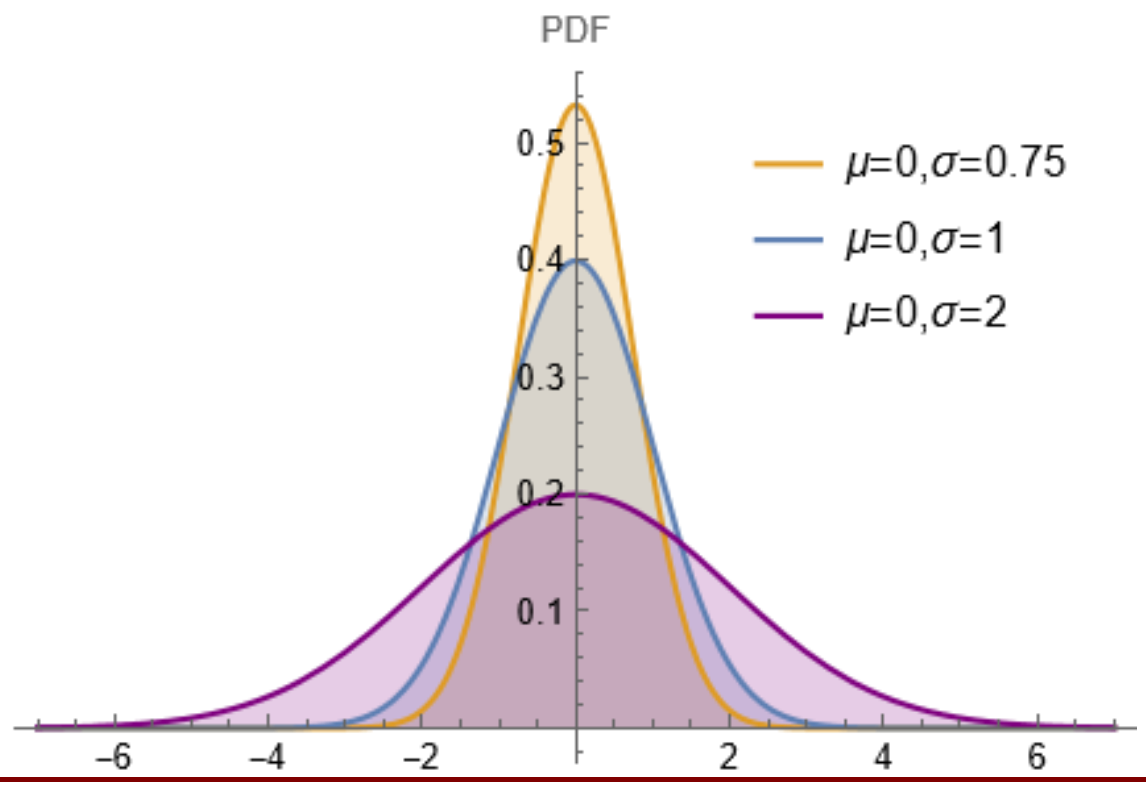

**Mathematica Examples 13.76**

Input

```
(* The code generates a plot of the cumulative distribution function (CDF) of the
normal distribution with different values of standard deviation σ= (0.75, 1 and 2)
and a fixed mean (μ=0). The resulting plot shows the S-shaped curve of the normal
distribution with different shapes, reflecting the impact of varying the value of
the standard deviation: *)

Plot[
 Evaluate[
```

```
    Table[
     CDF[
      NormalDistribution[0,σ],
      x
      ],
     {σ,{.75,1,2}}
     ]
    ],
   {x,-7,7},
   PlotRange->All,
   Filling->Axis,
   PlotLegends->Placed[{"μ=0,σ=0.75","μ=0,σ=1","μ=0,σ=2"},{0.25,0.75}],
   PlotStyle->{RGBColor[0.88,0.61,0.14],RGBColor[0.37,0.5,0.7],Purple},
   ImageSize->320,
   AxesLabel->{None,"CDF"}
   ]
```

Output

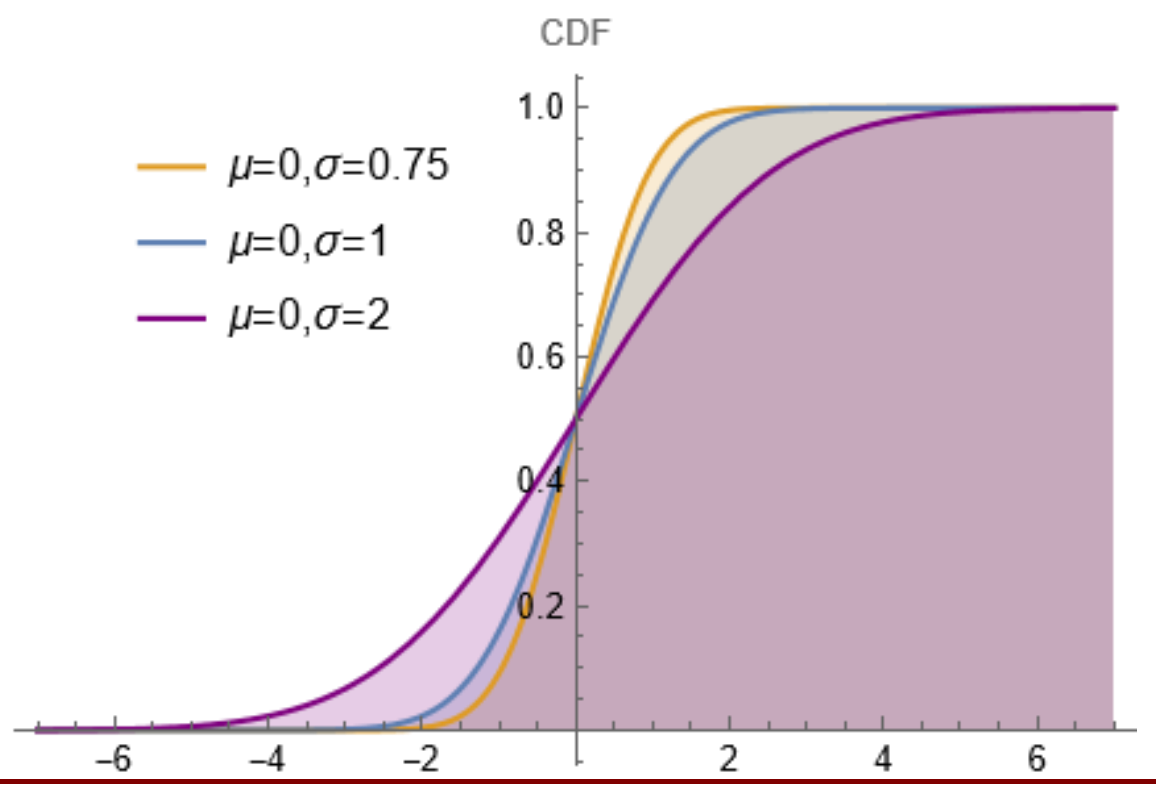

***Mathematica Examples 13.77***

Input

```
(* The code generates a plot of the PDF for a normal distribution with a fixed value of standard deviation σ=1.2 and different values of mean (μ=-0.5, 1, 2). The plot shows the values of the PDF for all possible values of x between -7 and 7. The resulting plot shows the bell-shaped curve of the normal distribution with different locations, reflecting the impact of varying the value of the mean: *)

Plot[
 Evaluate[
  Table[
   PDF[
    NormalDistribution[μ,1.2],
    x
    ],
   {μ,{-0.5,1,2}}
   ]
  ],
 {x,-7,7},
 PlotRange->All,
 Filling->Axis,
 PlotLegends->Placed[{"σ=1.2,μ=-0.5","σ=1.2,μ=1","σ=1.2,μ=2"},{0.25,0.75}],
 PlotStyle->{RGBColor[0.88,0.61,0.14],RGBColor[0.37,0.5,0.7],Purple},
 ImageSize->320,
 AxesLabel->{None,"PDF"}
 ]
```

Output

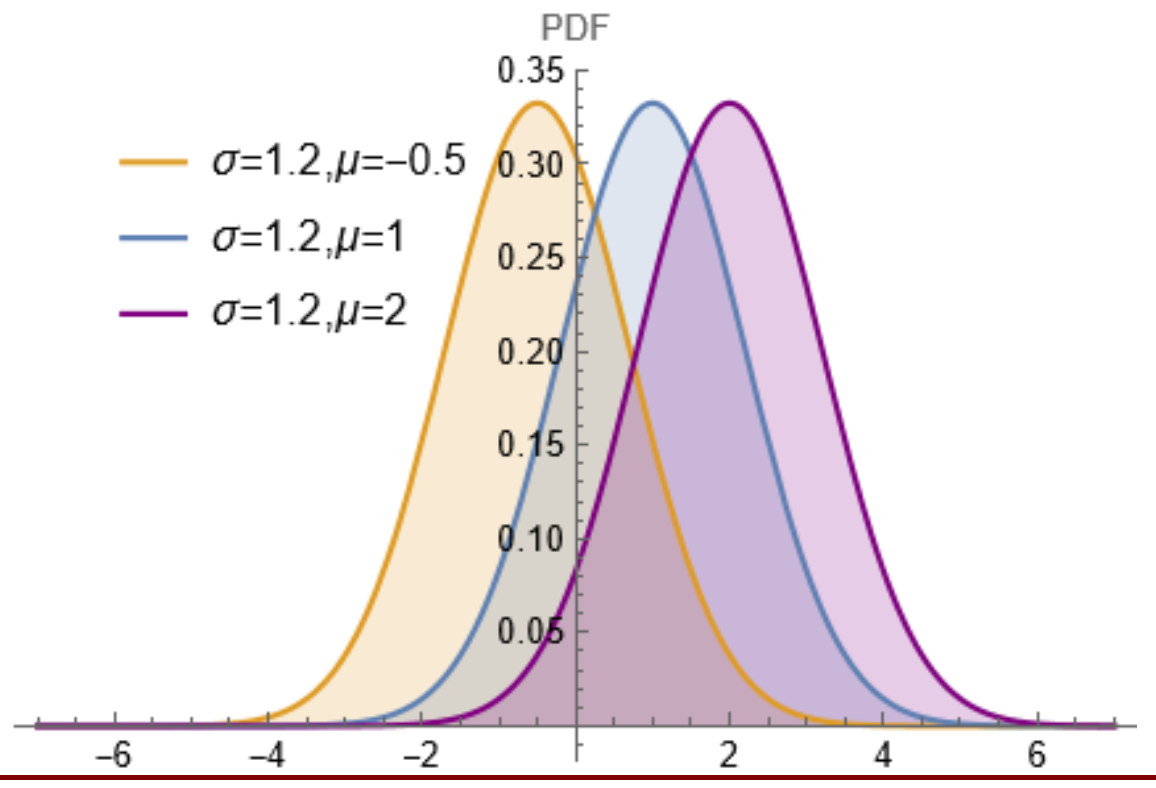

***Mathematica Examples 13.78***

Input

```
(* The given code generates a plot of the CDF of the normal distribution with a fixed
value of standard deviation σ=1.2 and different values of mean (μ=-0.5, 1, 2). The
resulting plot shows the S-shaped curve of the normal distribution with different
shapes, reflecting the impact of varying the value of the mean: *)

Plot[
 Evaluate[
  Table[
   CDF[
    NormalDistribution[μ,1.2],
    x
    ],
   {μ,{-0.5,1,2}}
   ]
  ],
 {x,-7,7},
 PlotRange->All,
 Filling->Axis,
 PlotLegends->Placed[{"σ=1.2,μ=-0.5","σ=1.2,μ=1","σ=1.2,μ=2"},{0.25,0.75}],
 PlotStyle->{RGBColor[0.88,0.61,0.14],RGBColor[0.37,0.5,0.7],Purple},
 ImageSize->320,
 AxesLabel->{None,"CDF"}
 ]
```

Output

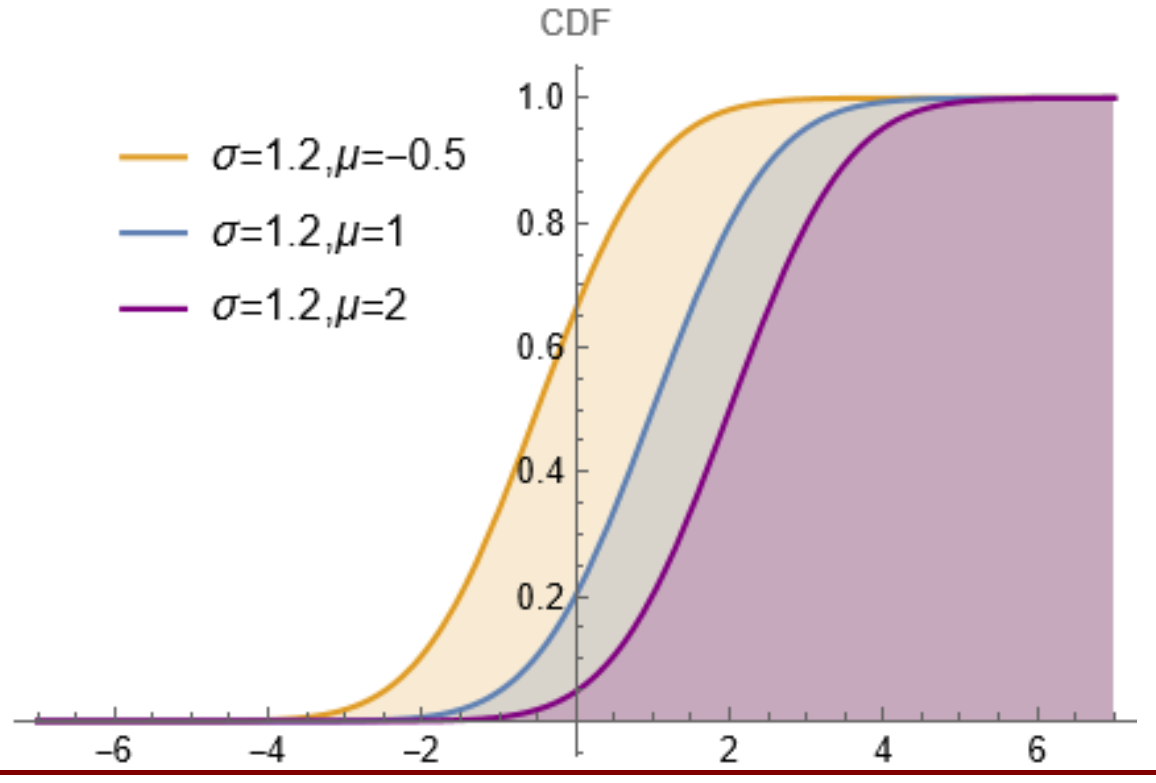

***Mathematica Examples 13.79***

Input

```
(* The code generates a histogram and a plot of the PDF for a Normal Distribution
with parameters μ=1 and σ=3 and sample size 10000: *)

data=RandomVariate[
   NormalDistribution[1,3],
```

```
    10^4
    ];
Show[
 Histogram[
  data,
  20,
  "PDF",
  ColorFunction->Function[{height},Opacity[height]],
  ChartStyle->Purple,
  ImageSize->320,
  AxesLabel->{None,"PDF"}
  ],
 Plot[
  PDF[
   NormalDistribution[1,3],
   x
   ],
  {x,-9,9},
  PlotStyle->RGBColor[0.88,0.61,0.14],
  PlotRange->{0,4}
  ]
 ]
```

Output

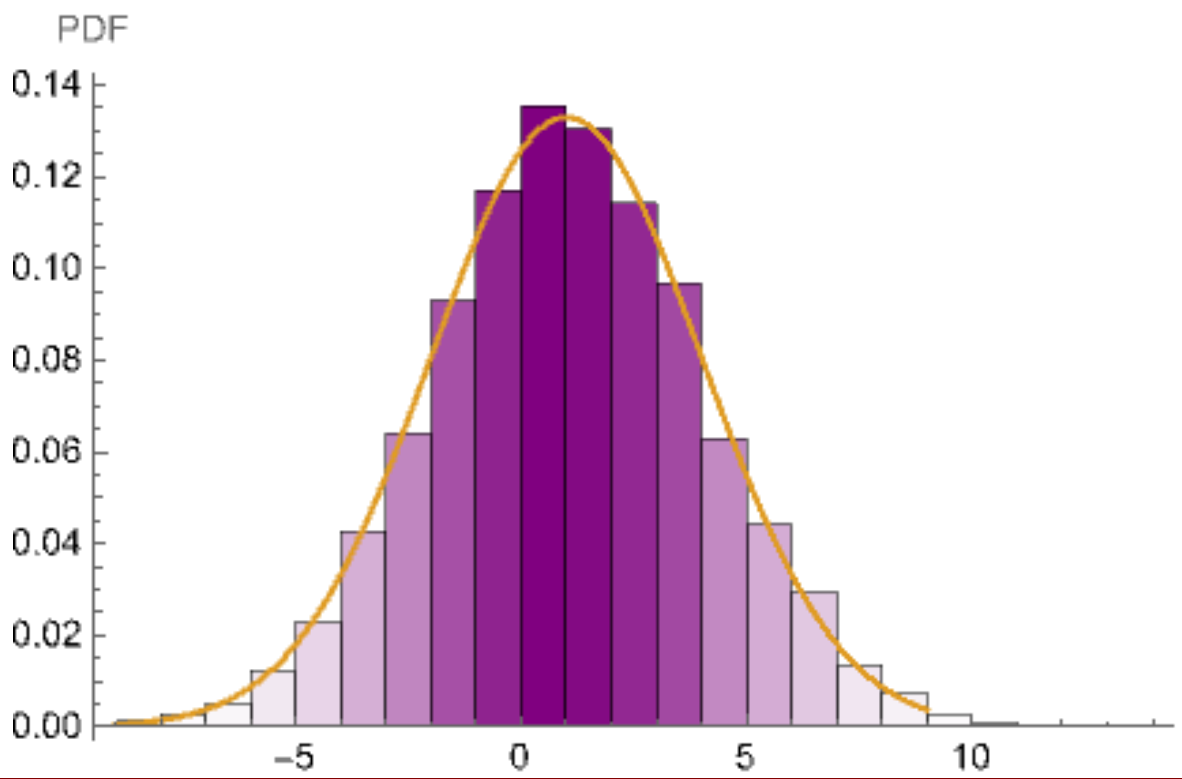

***Mathematica Examples 13.80***

Input

```
(* The code creates a dynamic histogram of data and a plot of the PDF generated from
a normal distribution using the Manipulate function. The Manipulate function creates
interactive controls for the user to adjust the values of μ, σ, and n, which are the
parameters of the normal distribution and the sample size: *)

Manipulate[
 Module[
  {
   data=RandomVariate[
     NormalDistribution[μ,σ],
     n
     ]
   },

  Show[
   Histogram[
    data,
    {1},
    "PDF",
    PlotRange->{{-11,11},All},
    ColorFunction->Function[{height},Opacity[height]],
```

```
      ImageSize->320,
      ChartStyle->Purple
      ],
     Plot[
      PDF[
       NormalDistribution[μ,σ],
       x
       ],
      {x,-11,11},
      PlotRange->All,
      ColorFunction->"Rainbow"
      ]
     ]
    ],
   {{μ,0,"μ"},-3,3,0.1},
   {{σ,1,"σ"},0.1,3,0.1},
   {{n,100,"n"},100,1000,10}
   ]
```

Output

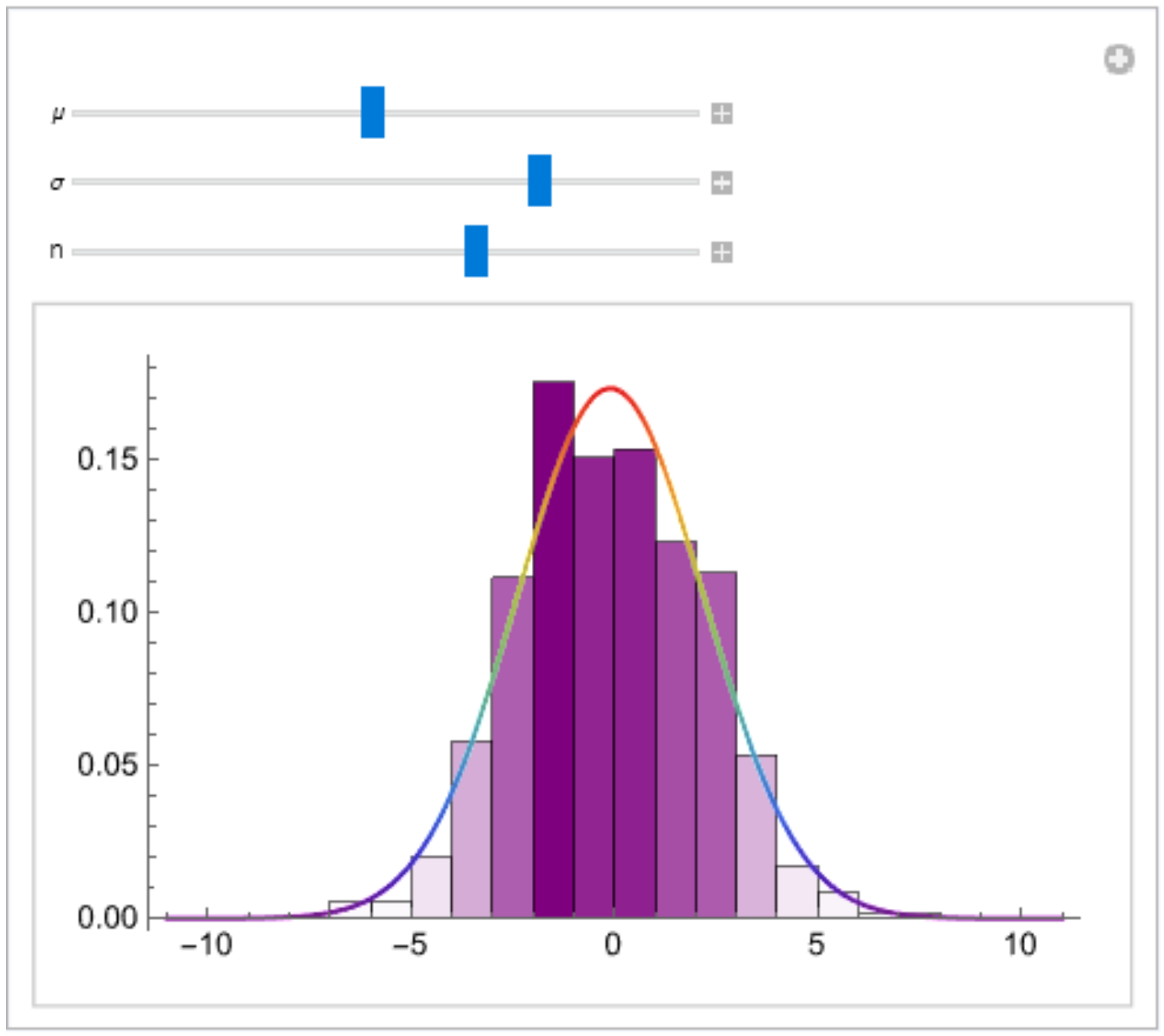

## *Mathematica Examples 13.81*

Input

```
(* The code creates a plot of the CDF of a normal distribution using the Manipulate
function. The Manipulate function allows you to interactively change the values of
the parameters n and p, respectively: *)

Manipulate[
 Plot[
  CDF[
   NormalDistribution[μ,σ],
   x
   ],
  {x,-10,10},
  Filling->Axis,
  FillingStyle->LightPurple,
  PlotRange->All,
  Epilog->{Text[StringForm["μ = `` & σ = ``",μ,σ],{μ/2,0.9}]},
  AxesLabel->{"x","CDF"},
  ImageSize->320,
```

```
   PlotStyle->Purple
   ],
  {{μ,0},-3,3,0.1},
  {{σ,0.5},0.1,3,0.1}
  ]
```

Output

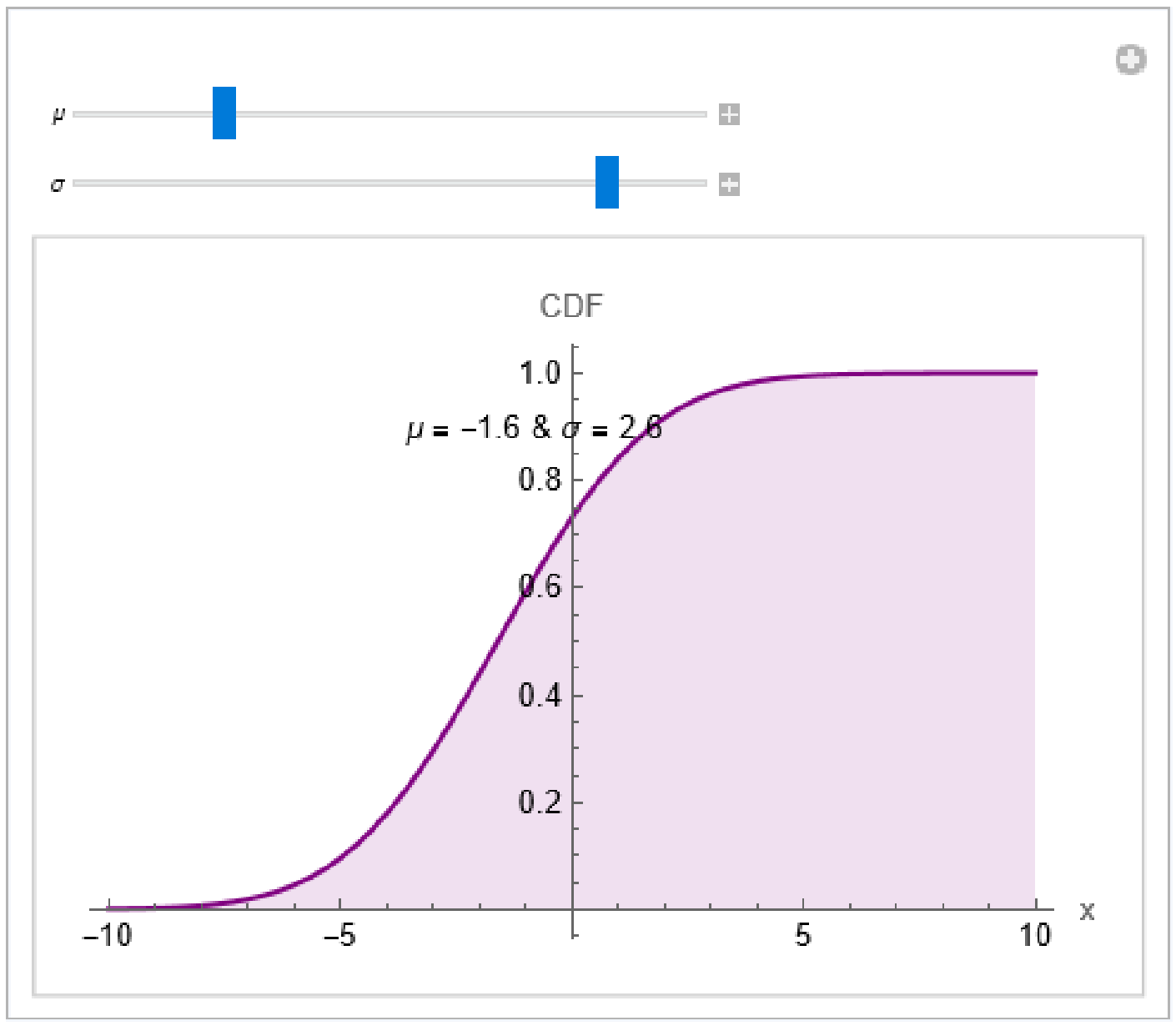

***Mathematica Examples 13.82***

Input

```
(* The code uses the Grid function to create a grid of two plots, one for the PDF
and one for the CDF of normal distribution. The code uses slider controls to adjust
the values of μ and σ: *)

Manipulate[
 Grid[
  {
   {Plot[
     PDF[
      NormalDistribution[μ,σ],
      x
      ],
     {x,-10,10},
     PlotRange->All,
     PlotStyle->{Purple,PointSize[0.03]},
     PlotLabel->"PDF of normal distribution",
     AxesLabel->{"x","PDF"}
     ],
    Plot[
     CDF[
      NormalDistribution[μ,σ],
      x
      ],
     {x,-10,10},
     PlotRange->All,
     PlotStyle->{Purple,PointSize[0.03]},
     PlotLabel->"CDF of normal distribution",
     AxesLabel->{"x","CDF"}
     ]
```

```
      }
     },
    Spacings->{5,5}
    ],
   {{μ,0},-3,3,0.1},
   {{σ,0.5},0.1,3,0.1}
   ]
```

Output

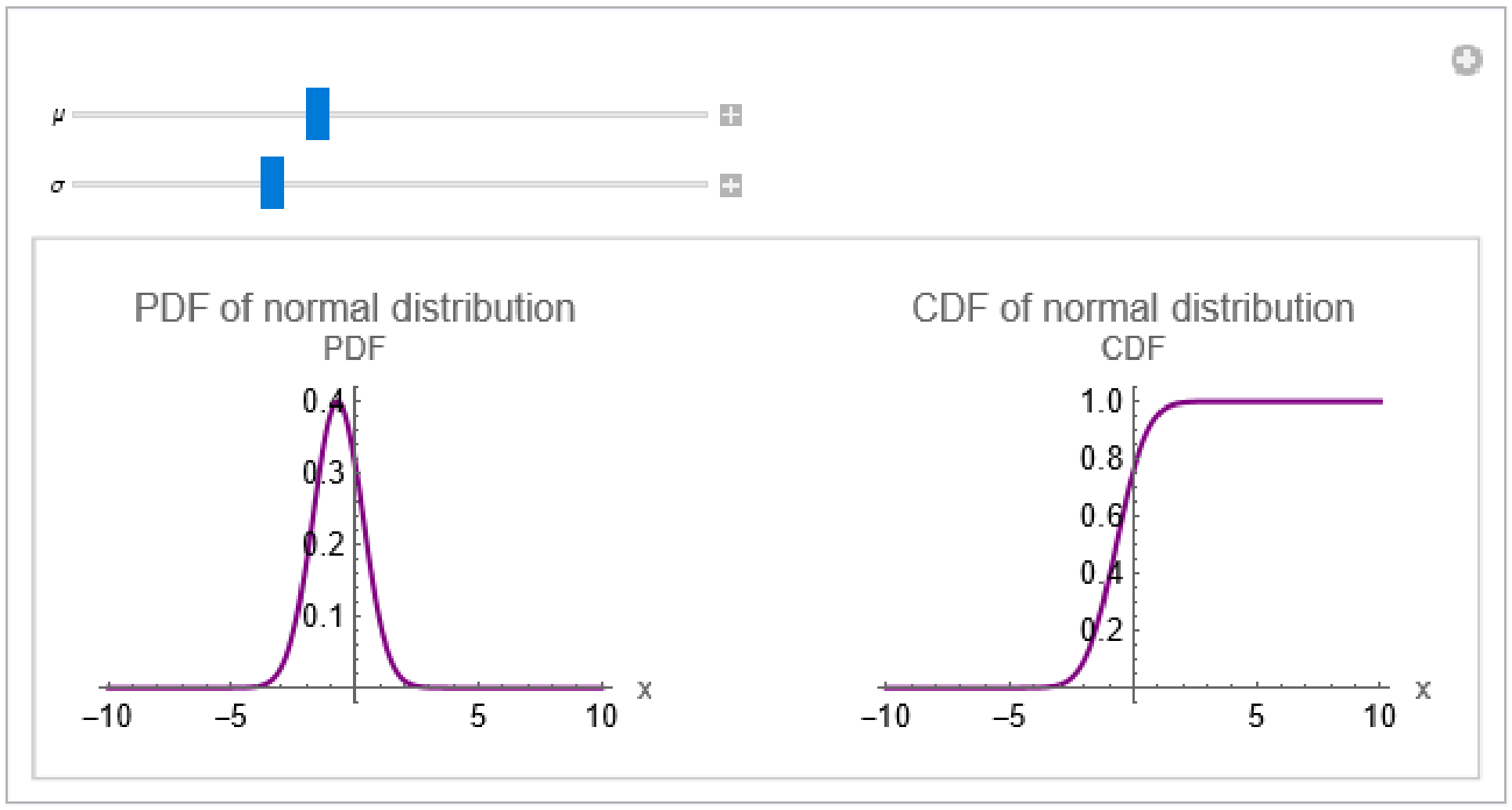

***Mathematica Examples 13.83***

Input

```
(* The code calculates and displays some descriptive statistics (mean, variance,
standard deviation, kurtosis and skewness) for a normal distribution with parameters
μ and σ: *)

Grid[
 Table[
  {
   statistics,
   FullSimplify[statistics[NormalDistribution[μ,σ]]]
   },
  {statistics,{Mean,Variance,StandardDeviation,Kurtosis,Skewness}}
  ],
 ItemStyle->12,
 Alignment->{{Right,Left}},
 Frame->All,
 Spacings->{Automatic,0.8}
 ]
```

Output

| | |
|---|---|
| Mean | μ |
| Variance | σ^2 |
| StandardDeviation | σ |
| Kurtosis | 3 |
| Skewness | 0 |

***Mathematica Examples 13.84***

Input

```
(* The code calculates and displays some additional descriptive statistics (moments,
central moments, and factorial moments) for a  Normal Distribution with parameters μ
and σ: *)

Grid[
 Table[
```

```
    {
     statistics,
     FullSimplify[statistics[NormalDistribution[μ,σ],1]],
     FullSimplify[statistics[NormalDistribution[μ,σ],2]]
     },
    {statistics,{Moment,CentralMoment,FactorialMoment}}
    ],
   ItemStyle->12,
   Alignment->{{Right,Left}},
   Frame->All,
   Spacings->{Automatic,0.8}
   ]
```

Output

| | | |
|---|---|---|
| Moment | μ | μ^2+σ^2 |
| CentralMoment | 0 | σ^2 |
| FactorialMoment | μ | (-1+μ) μ+σ^2 |

***Mathematica Examples 13.85***

Input

```
(* The code generates a dataset of 1000 observations from a normal distribution with
parameters μ=1 and σ=3. Then, it computes the sample mean and quartiles of the data,
and plots a histogram of the data and plot of the PDF. Additionally, the code adds
vertical lines to the plot corresponding to the sample mean and quartiles: *)

data=RandomVariate[
    NormalDistribution[1,3],
    1000
    ];

mean=Mean[data];
quartiles=Quantile[
    data,
    {0.25,0.5,0.75}
    ];

Show[
 Histogram[
  data,
  Automatic,
  "PDF",
  Epilog->{
    Directive[Red,Thickness[0.006]],
    Line[{{mean,0},{mean,0.25}}],
    Directive[Green,Dashed],
    Line[{{quartiles[[1]],0},{quartiles[[1]],0.25}}],
    Line[{{quartiles[[2]],0},{quartiles[[2]],0.25}}],
    Line[{{quartiles[[3]],0},{quartiles[[3]],0.25}}]
    },
  ColorFunction->Function[{height},Opacity[height]],
  ImageSize->320,
  ChartStyle->Purple,
  PlotRange->{{-10,10},{0,0.2}}
  ],
 Plot[
  PDF[NormalDistribution[1,3],x],
  {x,-10,10},
  ImageSize->320,
  ColorFunction->"Rainbow"
  ]
 ]
```

Output
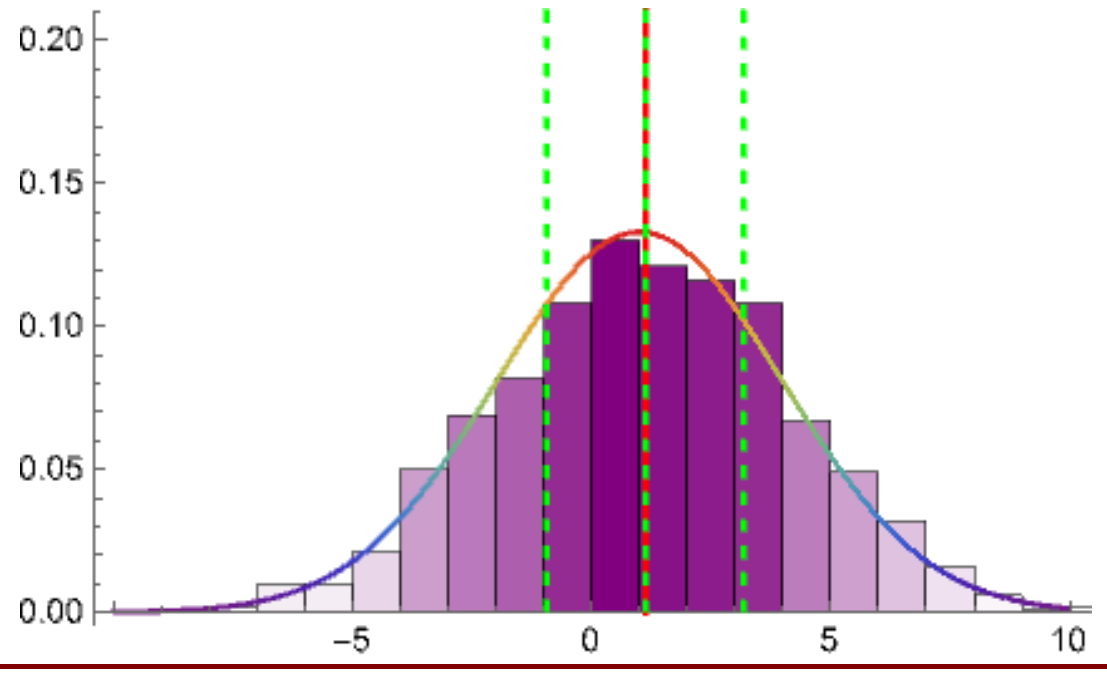

***Mathematica Examples 13.86***

Input
```
(* The code generates a random sample of size 10,000 from a normal distribution with parameters μ=1 and σ=3, estimates the distribution parameters using the EstimatedDistribution function, and then compares the histogram of the sample with the estimated PDF of the normal distribution using a histogram and a plot of the PDF: *)

sampledata=RandomVariate[
    NormalDistribution[1,3],
    10^4
    ];
(* Estimate the distribution parameters from sample data: *)
ed=EstimatedDistribution[
   sampledata,
   NormalDistribution[μ,σ]
   ]
(* Compare a density histogram of the sample with the PDF of the estimated distribution: *)
Show[
 Histogram[
   sampledata,
   {1},
   "PDF",
   ColorFunction->Function[{height},Opacity[height]],
   ChartStyle->Purple,
   ImageSize->320
   ],
 Plot[
   PDF[ed,x],
   {x,-10,10},
   ImageSize->320,
   ColorFunction->"Rainbow"
   ]
 ]
```

Output `NormalDistribution[1.00866,2.95354]`

Output
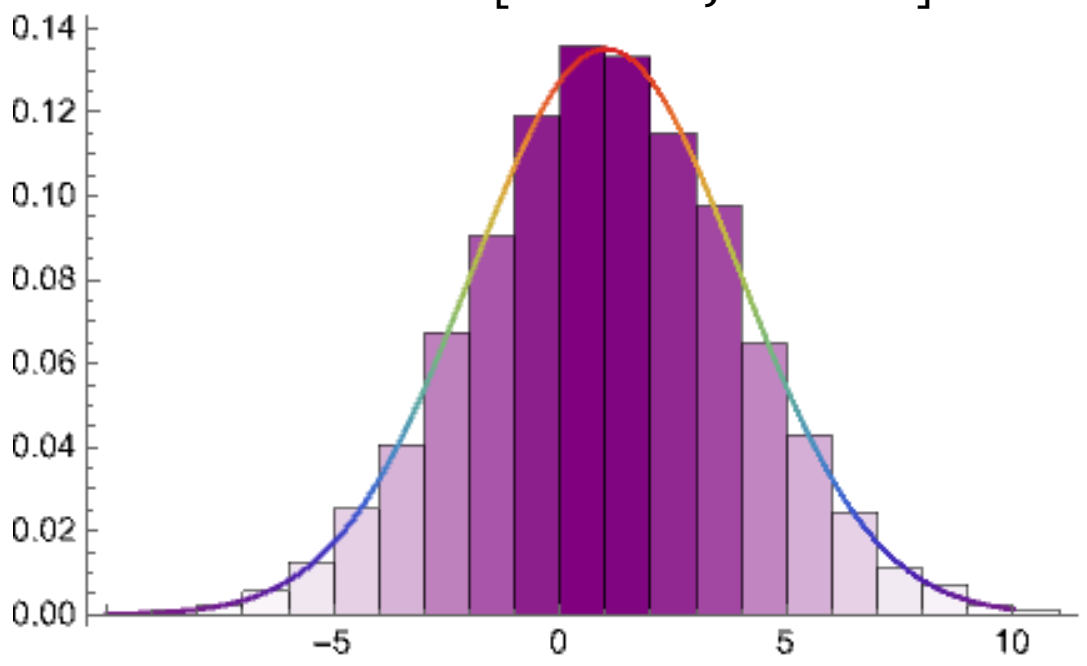

***Mathematica Examples 13.87***

Input

```
(* The code creates a Manipulate interface where the user can adjust the parameters of two normal distributions (dist1 and dist2) and see the resulting sum of these distributions (distSum) plotted on the same graph. The plot shows the PDFs of each distribution as well as the PDF of the sum of the distributions. The TransformedDistribution function is used to create the sum of two distributions by defining the distribution of the sum of two random variables, x and y, where x follows dist1 and y follows dist2. Hence, we prove that Normal distribution is closed under addition: *)

Manipulate[
 dist1=NormalDistribution[mean1,sd1];
 dist2=NormalDistribution[mean2,sd2];

 distSum=TransformedDistribution[
    x+y,
    {Distributed[x,dist1],Distributed[y,dist2]}
    ];

 Plot[
  {
    PDF[dist1,x],
    PDF[dist2,x],
    PDF[distSum,x]
    },
  {x,-5,5},
  PlotRange->All,
  AxesLabel->{"x","f(x)"},
  Filling->{1->{2},2->{3}},
  FillingStyle->{LightBlue,LightPurple},
  PlotLegends->{"Distribution 1","Distribution 2","Sum of Distributions"}
  ],
 {{mean1,0,"Mean 1"},-5,5,Appearance->"Labeled"},
 {{sd1,1,"Standard Deviation 1"},0.1,5,Appearance->"Labeled"},
 {{mean2,0,"Mean 2"},-5,5,Appearance->"Labeled"},
 {{sd2,1,"Standard Deviation 2"},0.1,5,Appearance->"Labeled"}
 ]
```

Output

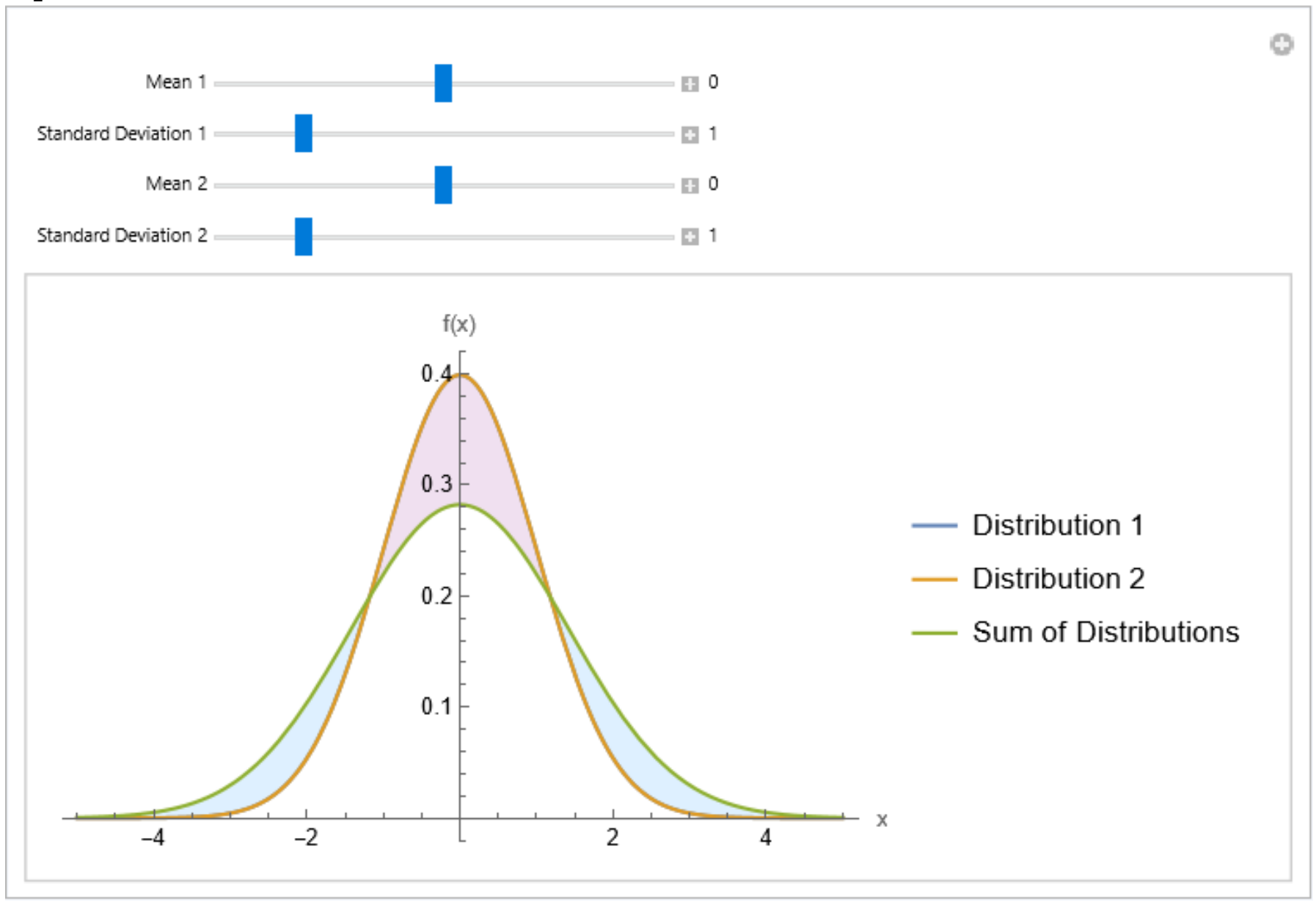

*Mathematica Examples 13.88*

Input

```
(* The code demonstrate that the normal distribution is symmetric about its mean.
The PDF function used in the code calculates the PDF of the normal distribution,
which is symmetric about its mean. The plot of the PDF clearly shows this symmetry,
with the peak of the distribution located at the mean and the same amount of
probability density on both sides of the mean. Additionally, the Epilog option in
the code adds vertical lines at μ-σ, μ, and μ+σ, indicating one standard deviation
away from the mean on both sides. Since the normal distribution is symmetric, these
lines are equidistant from the mean and therefore, the distance between μ and
μ\[PlusMinus]σ is the same. This property is true for all normal distributions, and
is one of the defining characteristics of the normal distribution: *)

Manipulate[
 Plot[
  PDF[
   NormalDistribution[μ,σ],
   x
   ],
  {x,-10,10},
  Filling->{1->Axis},
  FillingStyle->LightPurple,
  PlotStyle->Purple,
  PlotRange->{{-10,10},{0,0.45}},
  Epilog->{
    Directive[Red,Dashed],
    Line[{{μ-σ,0},{μ-σ,0.4}}],
    Line[{{μ+σ,0},{μ+σ,0.4}}],
    Line[{{μ,0},{μ,0.4}}]
    },
  PlotLabel->Row[{"μ = ",μ,", σ = ",σ,", Distance = ",σ}]
  ],
 {{μ,0},-5,5},
 {{σ,1},0.1,5},
 ControlPlacement->Top
 ]
```

Output

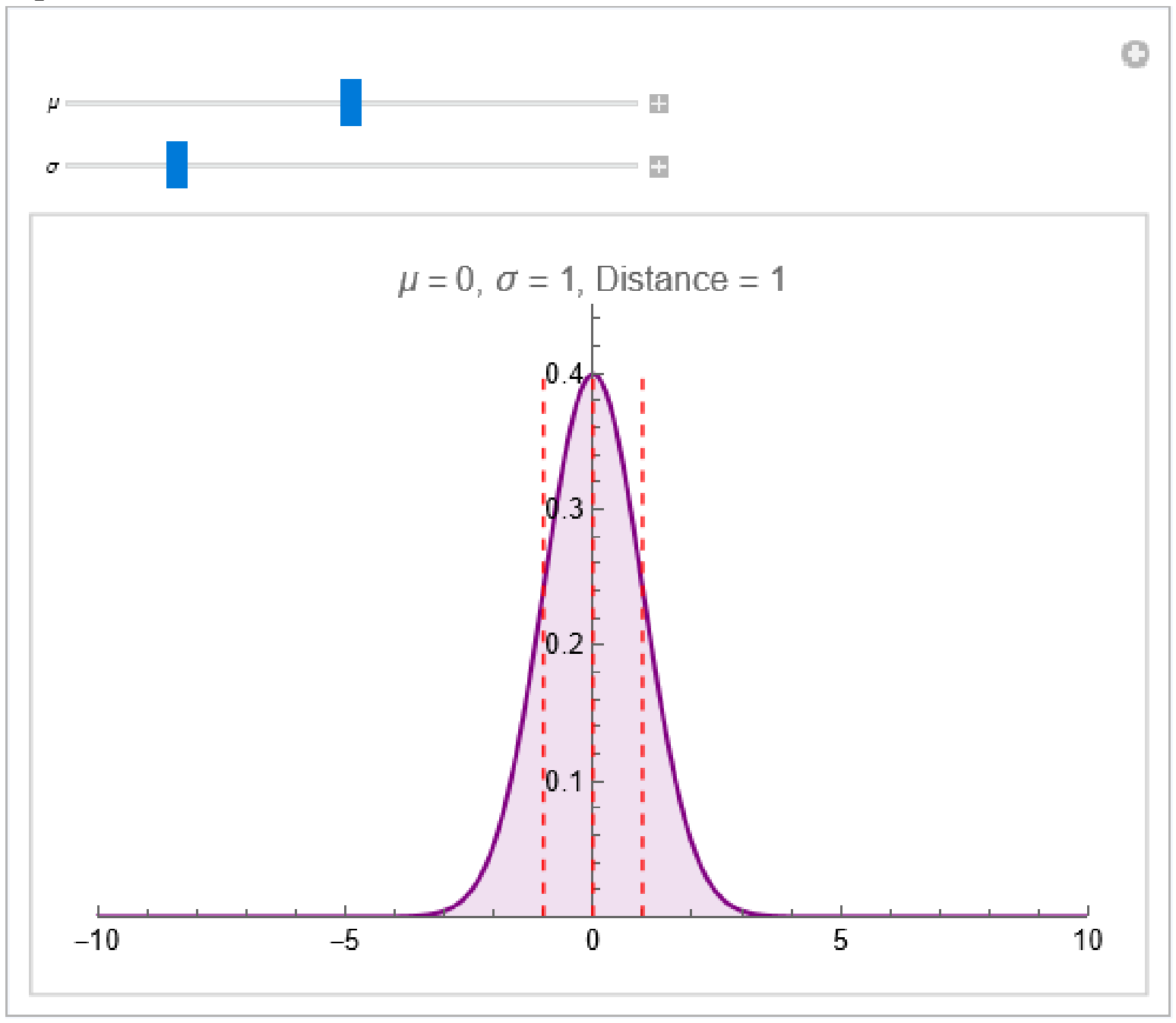

***Mathematica Examples 13.89***

Input
```
(* The code generates a 2D dataset with 1000 random points that follow a normal distribution with a mean of 0 and a standard deviation of 1. The dataset is then used to create a row of three plots. The first plot is a histogram of the X-axis values of the dataset. The second plot is a histogram of the Y-axis values of the dataset. It is similar to the first plot, but shows the distribution of the Y-axis values instead. The third plot is a scatter plot of the dataset, with the X-axis values on the horizontal axis and the Y-axis values on the vertical axis. Each point in the plot represents a pair of X and Y values from the dataset: *)

data=RandomVariate[
   NormalDistribution[0,1],
   {1000,2}
   ];
GraphicsRow[
 {
  Histogram[
   data[[All,1]],
   {0.1},
   PlotLabel->"X-axis",
   ColorFunction->Function[{height},Opacity[height]],
   ChartStyle->Purple
   ],
  Histogram[
   data[[All,2]],
   {0.1},
   PlotLabel->"Y-axis",
   ColorFunction->Function[{height},Opacity[height]],
   ChartStyle->Purple
   ],
  ListPlot[
   data,
   PlotStyle->{Purple,PointSize[0.015]},
   AspectRatio->1,
   Frame->True,
   Axes->False
   ]
  }
 ]
```

Output

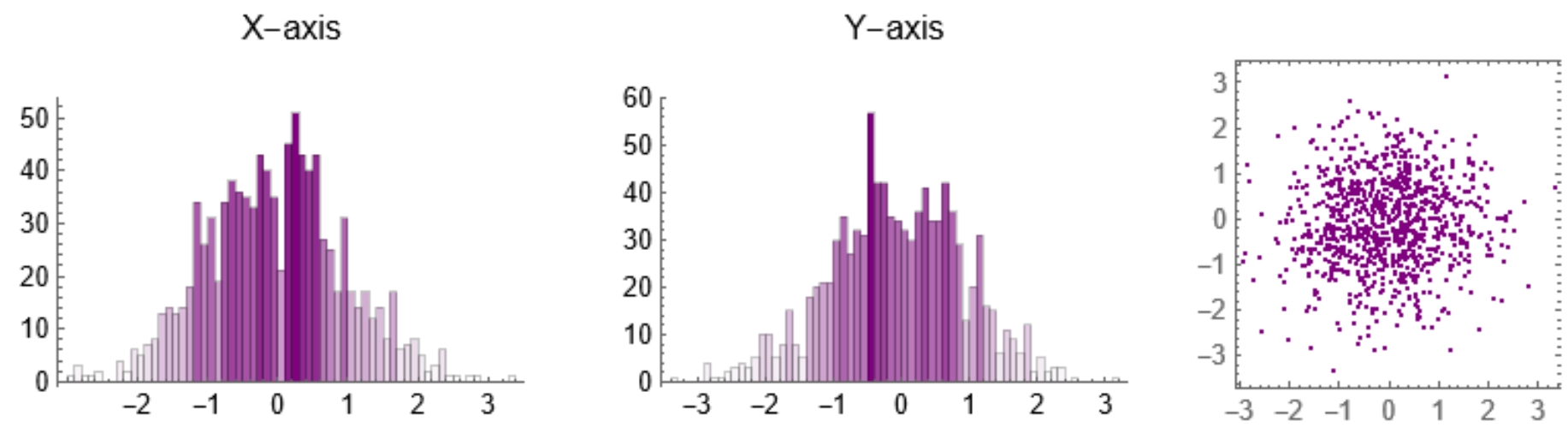

***Mathematica Examples 13.90***

Input
```
(* The code generates a set of random data points with a normal distribution in three dimensions, and then creates three histograms, one for each dimension, showing the distribution of the points along that axis. Additionally, it creates a 3D scatter plot of the data points: *)

data=RandomVariate[
   NormalDistribution[0,1],
   {1000,3}
```

```
        ];

    GraphicsGrid[
     {
      {
       Histogram[
        data[[All,1]],
        Automatic,
        "PDF",
        PlotLabel->"X-axis",
        ColorFunction->Function[{height},Opacity[height]],
        ChartStyle->Purple
        ],
       Histogram[
        data[[All,2]],
        Automatic,
        "PDF",
        PlotLabel->"Y-axis",
        ColorFunction->Function[{height},Opacity[height]],
        ChartStyle->Purple
        ],
       Histogram[
        data[[All,3]],
        Automatic,
        "PDF",
        PlotLabel->"Z-axis",
        ColorFunction->Function[{height},Opacity[height]],
        ChartStyle->Purple
        ],
       ListPointPlot3D[
        data,
        BoxRatios->{1,1,1},
        PlotStyle->{Purple,PointSize[0.015]}
        ]
       }
      }
     ]
```

Output

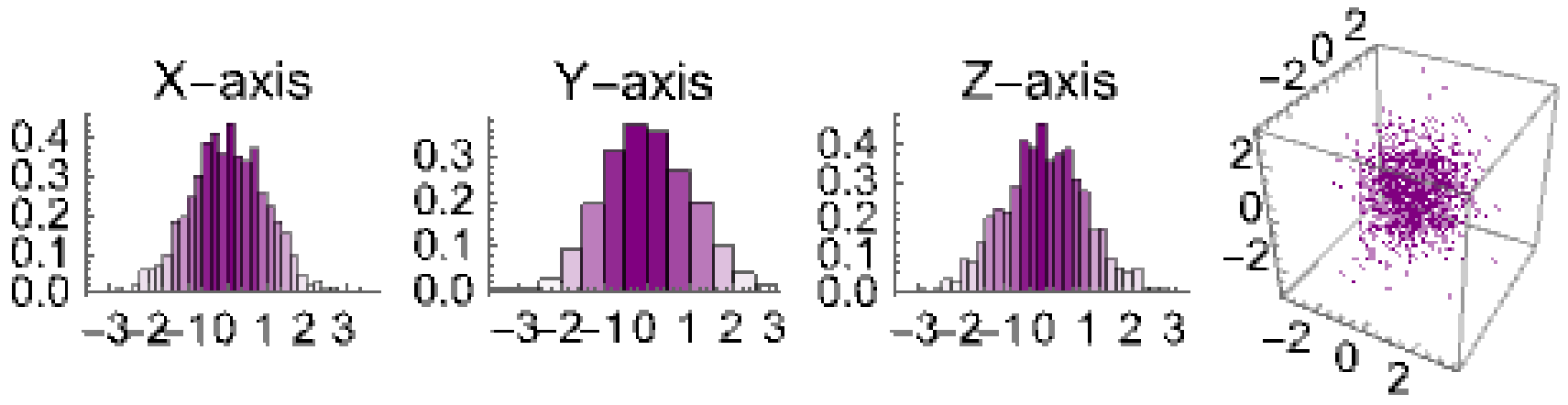

***Mathematica Examples 13.91***

Input

```
(* The code generates a 3D scatter plot of normally distributed points, where the x-
axis is red, y-axis is green, and z-axis is blue: *)

data=RandomVariate[
    NormalDistribution[0,1],
    {2000,3}
    ];
Graphics3D[
 {
  {PointSize[0.006],Purple,Point[data]},
  Thin,
  {Red,Opacity[0.4],Line[{{#,0,0},{#,0,-0.5}}]&/@data[[All,1]]},
```

```
     Thin,
     {Green,Opacity[0.4],Line[{{0,#,0},{0,#,-0.5}}]&/@data[[All,2]]},
     Thin,
     {Blue,Opacity[0.4],Line[{{0,0,#},{0,-0.5,#}}]&/@data[[All,3]]}
     },
    BoxRatios->{1,1,1},
    Axes->True,
    AxesLabel->{"X","Y","Z"},
    ImageSize->320
    ]
```

Output

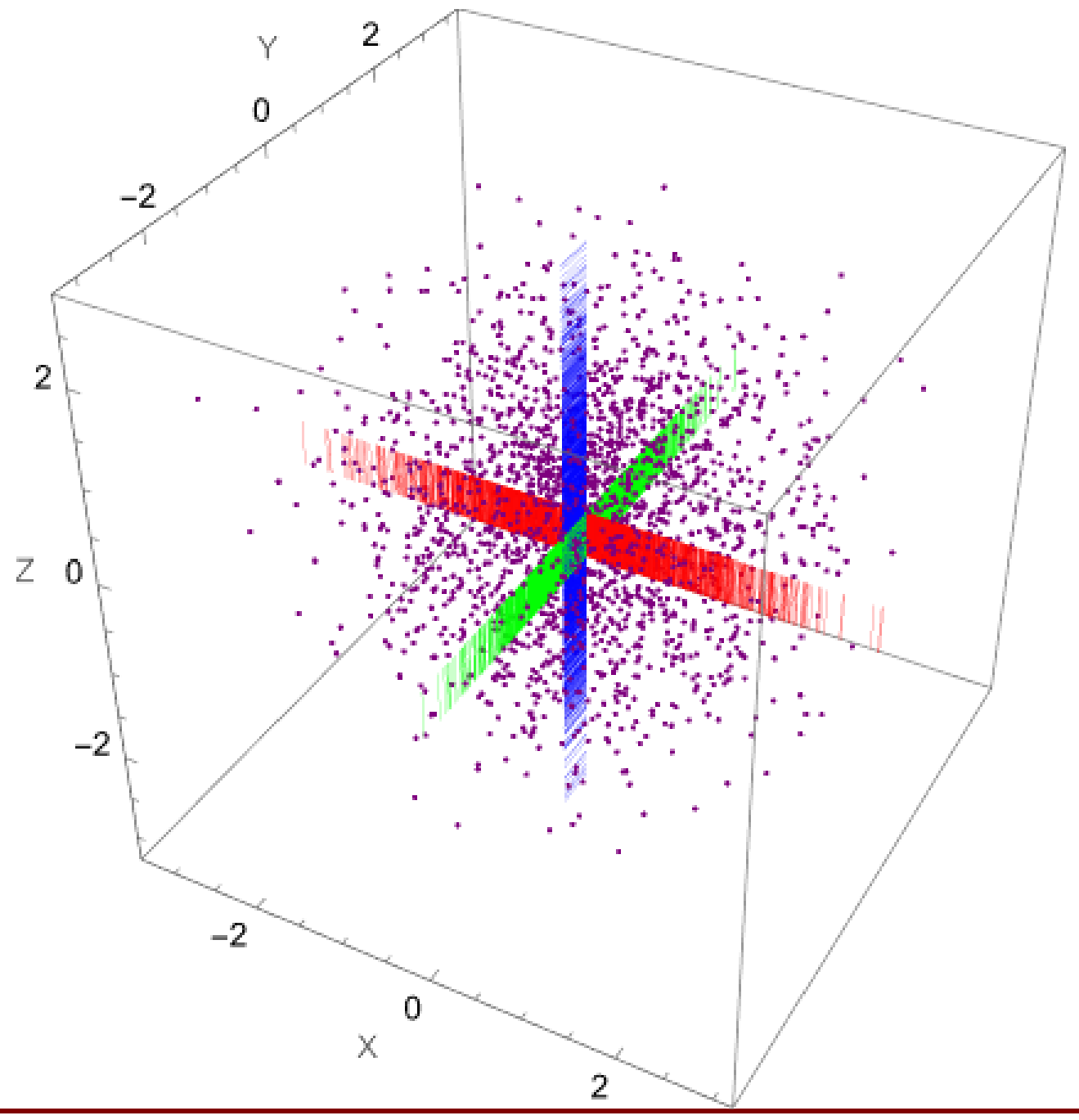

*Mathematica Examples 13.92*

Input

```
(* The code demonstrates a common technique in statistics and data analysis, which is the use of random sampling to estimate population parameters. The code generates random samples from a normal distribution with mean 0 and standard deviation 1, and then using these samples to estimate the parameters of another normal distribution with unknown mean and standard deviation. This process is repeated 20 times, resulting in 20 different estimated distributions. The code also visualizes the resulting estimated distributions using the PDF function. The code plots the PDFs of these estimated distributions using the PDF function and the estimated parameters. The plot shows the PDFs in a range from-3.5 to 3.5. The code also generates a list plot of 2 sets of random samples from the normal distribution with mean 0 and standard deviation 1. The plot shows the 100 random points generated from two random samples. The code generates also a histogram of the PDF for Normal distribution of the two samples: *)

estim0distributions=Table[
  dist=NormalDistribution[0,1];

  sampledata=RandomVariate[
    dist,
    100
    ];

  ed=EstimatedDistribution[
    sampledata,
```

```
          NormalDistribution[α,β]
          ],
        {i,1,20}
        ]

pdf0ed=Table[
    PDF[estim0distributions[[i]],x],
    {i,1,20}
    ];

(* Visualizes the resulting estimated distributions *)
Plot[
 pdf0ed,
 {x,-3.5,3.5},
 PlotRange->Full,
 ImageSize->400,
 PlotStyle->Directive[Purple,Opacity[0.3],Thickness[0.002]]
 ]

(* Visualizes 100 random points generated from two random samples *)
table=Table[
    dist=NormalDistribution[0,1];
    sampledata=RandomVariate[
      dist,
      100],
    {i,1,2}
    ];

ListPlot[
 table,
 ImageSize->320,
 Filling->Axis,
 PlotStyle->Directive[Opacity[0.5],Thickness[0.003]]
 ]

Histogram[
 table,
 Automatic,
 LabelingFunction->Above,
 ChartLegends->{"Sample 1","Sample 2"},
 ChartStyle->{Directive[Opacity[0.2],Red],Directive[Opacity[0.2],Purple]},
 ImageSize->320
 ]
```

Output

```
{NormalDistribution[0.196054,0.984213], NormalDistribution[0.0892481,0.965441],
NormalDistribution[0.0794514,0.963423], NormalDistribution[0.137788,1.03554],
NormalDistribution[-0.0493939,0.921769], NormalDistribution[0.124007,1.04963],
NormalDistribution[0.179146,0.909966], NormalDistribution[0.0916121,0.952755],
NormalDistribution[-0.130355,1.05336], NormalDistribution[-0.158586,0.971455],
NormalDistribution[-0.129099,1.02365], NormalDistribution[-0.0833337,0.947885],
NormalDistribution[0.0966752,1.00656], NormalDistribution[0.22479,0.848663],
NormalDistribution[0.150477,1.03951], NormalDistribution[0.0340835,1.05999],
NormalDistribution[-0.0886306,0.994703], NormalDistribution[0.00108821,0.986002],
NormalDistribution[0.0711757,1.0596], NormalDistribution[0.151034,0.977165]}
```

Output

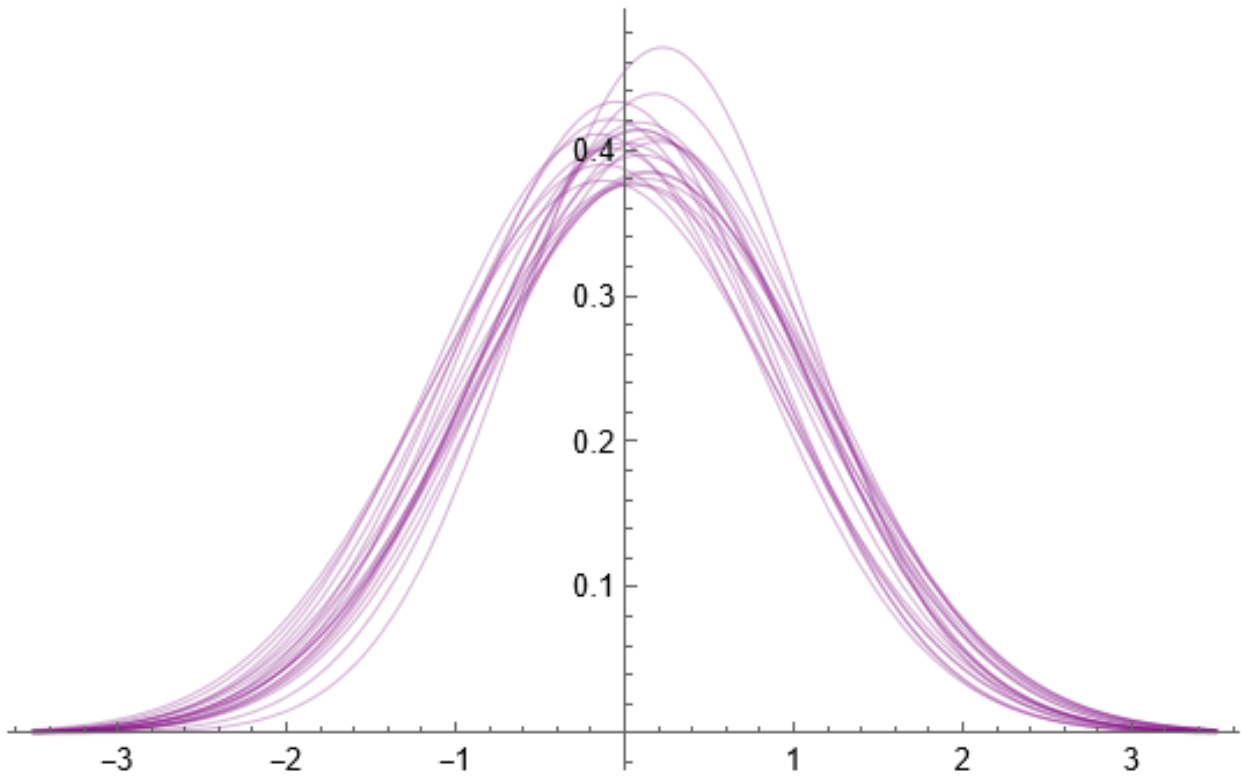

Output

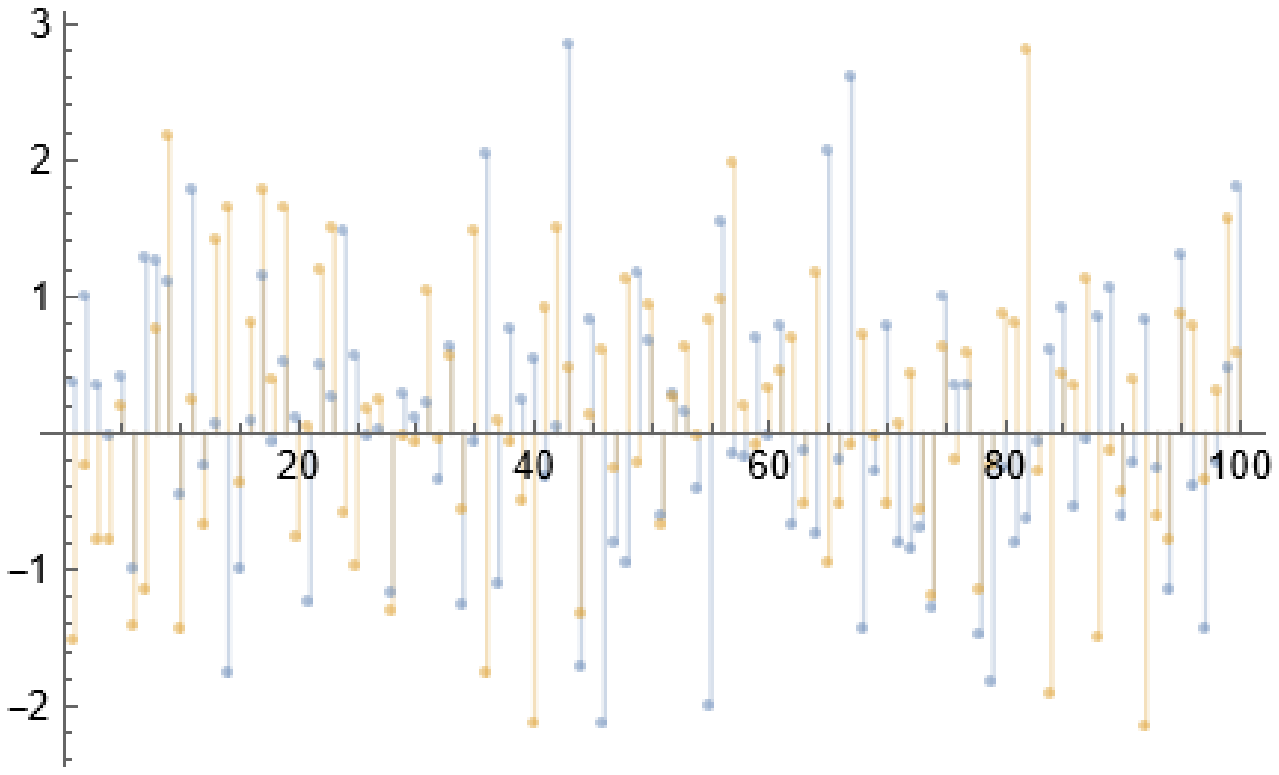

Output

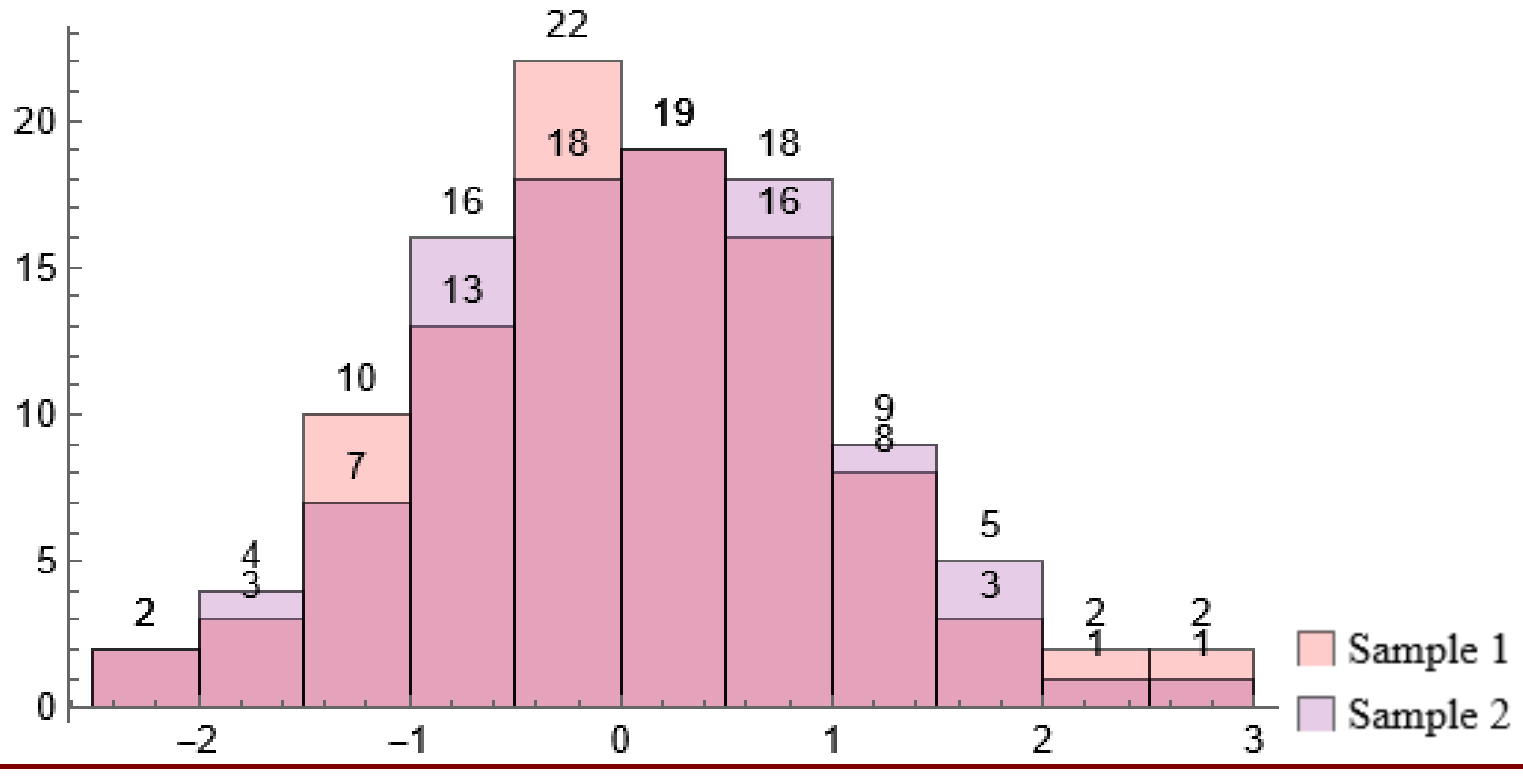

## *Mathematica Examples 13.93*

Input

```
(* The code generates and compares the means of random samples drawn from a Normal distribution with the given parameters μ and σ. The code uses the Manipulate function to create a user interface with sliders to adjust the values of the μ, σ, number of samples, and sample size. By varying the values of "Number of Samples" and "Sample Size" sliders, the code allows the user to explore how changing these parameters affects the means of the random samples. Specifically, increasing the number of samples n tends to make the distribution of the means narrower and more concentrated around the true mean of the underlying distribution. On the other hand, increasing the sample size tends to reduce the variability in the means and make them more precise estimators of the true mean: *)

Manipulate[
 Module[
  {means,dist},

  dist=NormalDistribution[μ,σ];
```

```
    means=Map[
      Mean,
      RandomVariate[dist,{n,samples}]
      ];
    m=N[Mean[means]];

    ListPlot[
     {means,{{0,m},{n,m}}},
     Joined->{False,True},
     Filling->Axis,
     PlotRange->{{1,50},{0,20}},
     PlotStyle->{Purple,Red},
     AxesLabel->{"Number of Samples","Sample Mean"},
     PlotLabel->Row[{"μ = ",μ,", σ = ",σ}],
     ImageSize->320
     ]
    ],
  {{μ,2,"Shape"},0.1,4,0.1},
  {{σ,2,"Scale"},0.1,4,0.1},
  {{n,50,"Number of Samples"},3,50,1},
  {{samples,50,"Sample Size"},1,100,1},
  TrackedSymbols:>{n,samples,μ,σ}
  ]
```

Output

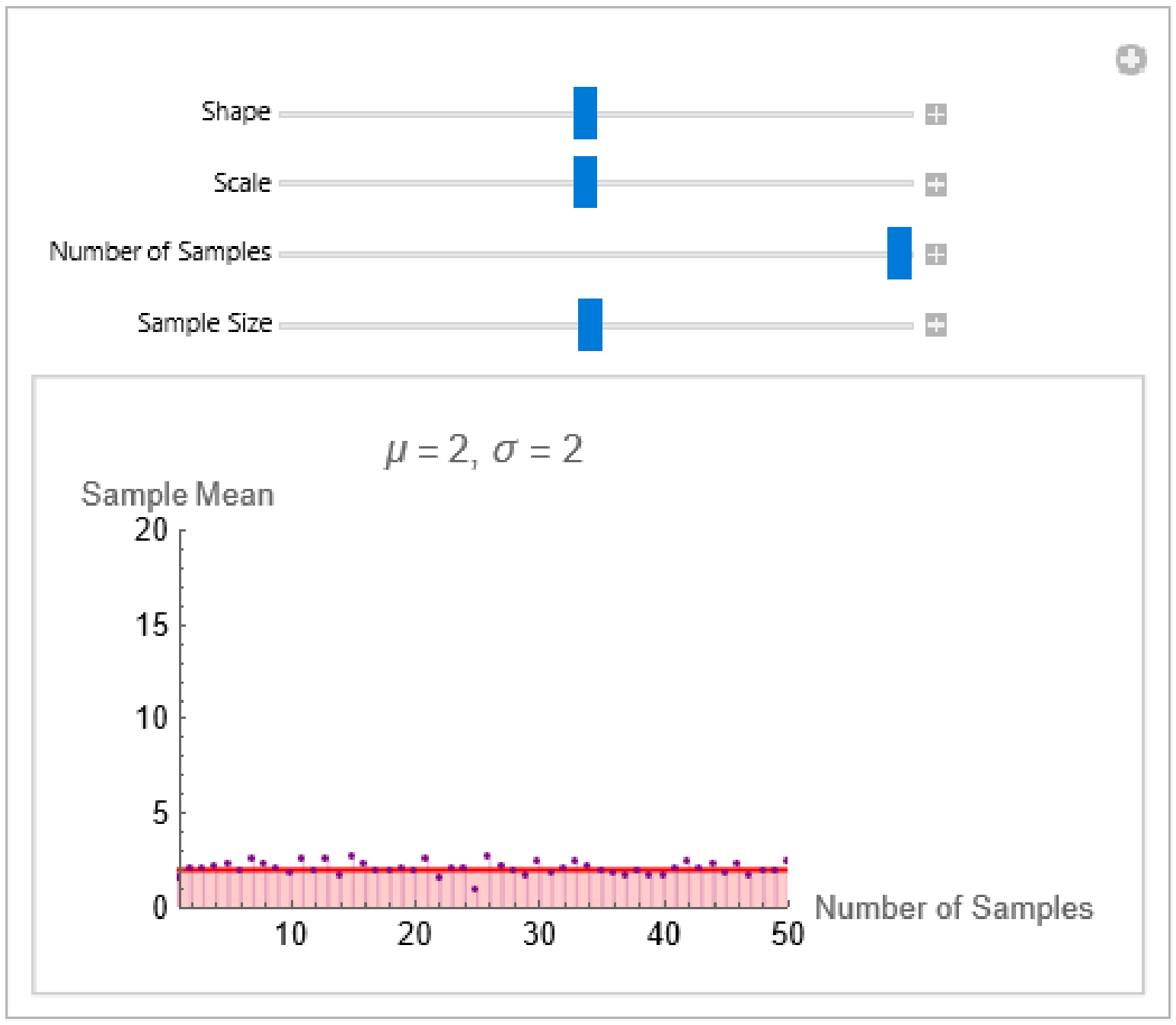

***Mathematica Examples 13.94***

Input (* The code is designed to compare two Normal distributions. It does this by generating random samples from each distribution and displaying them in a histogram, as well as plotting the PDFs of the two distributions. The code allows the user to manipulate the parameters μ and σ of both Normal distributions through the sliders for μ1, σ1, μ2 and σ2. By changing these parameters, the user can see how the distributions change and how they compare to each other. The histograms display the sample data for each distribution, with the first histogram showing the sample data for the first Normal distribution and the second histogram showing the sample data for the second Normal distribution. The histograms are overlaid on each other, with the opacity of each histogram set to 0.2 to make it easier to see where the data

```
overlap. The PDFs of the two distributions are also plotted on the same graph, with
the first distribution shown in blue and the second distribution shown in red. The
legend indicates which color corresponds to which distribution. By looking at the
histograms and the PDFs, the user can compare the two Normal distributions and see
how they differ in terms of shape, scale, and overlap of their sample data: *)

Manipulate[
 Module[
  {dist1,dist2,data1,data2},
  SeedRandom[seed];
  dist1=NormalDistribution[μ1,σ1];
  dist2=NormalDistribution[μ2,σ2];
  data1=RandomVariate[dist1,n];
  data2=RandomVariate[dist2,n];
  Column[
   {
    Show[
     ListPlot[
      data1,
      ImageSize->320,
      PlotStyle->Blue
      ],
     ListPlot[
      data2,
      ImageSize->320,
      PlotStyle->Red
      ]
     ],
    Show[
     Plot[
      {PDF[dist1,x],PDF[dist2,x]},
      {x,Min[{data1,data2}],Max[{data1,data2}]},
      PlotLegends->{"Distribution 1","Distribution 2"},
      PlotRange->All,
      PlotStyle->{Blue,Red},
      ImageSize->320
      ],
     Histogram[
      {data1,data2},
      Automatic,
      "PDF",
      ChartLegends->{"sample data1","sample data2"},
      ChartStyle->{Directive[Opacity[0.2],Red],Directive[Opacity[0.2],Purple]},
      ImageSize->320
      ]
     ]
    }
   ]
  ],
 {{μ1,6},0.1,10,0.1},
 {{σ1,2},0.1,10,0.1},
 {{μ2,6},0.1,10,0.1},
 {{σ2,2},0.1,10,0.1},
 {{n,500},{100,500,1000,2000}},
 {{seed,1234},ControlType->None}
 ]
```

Output

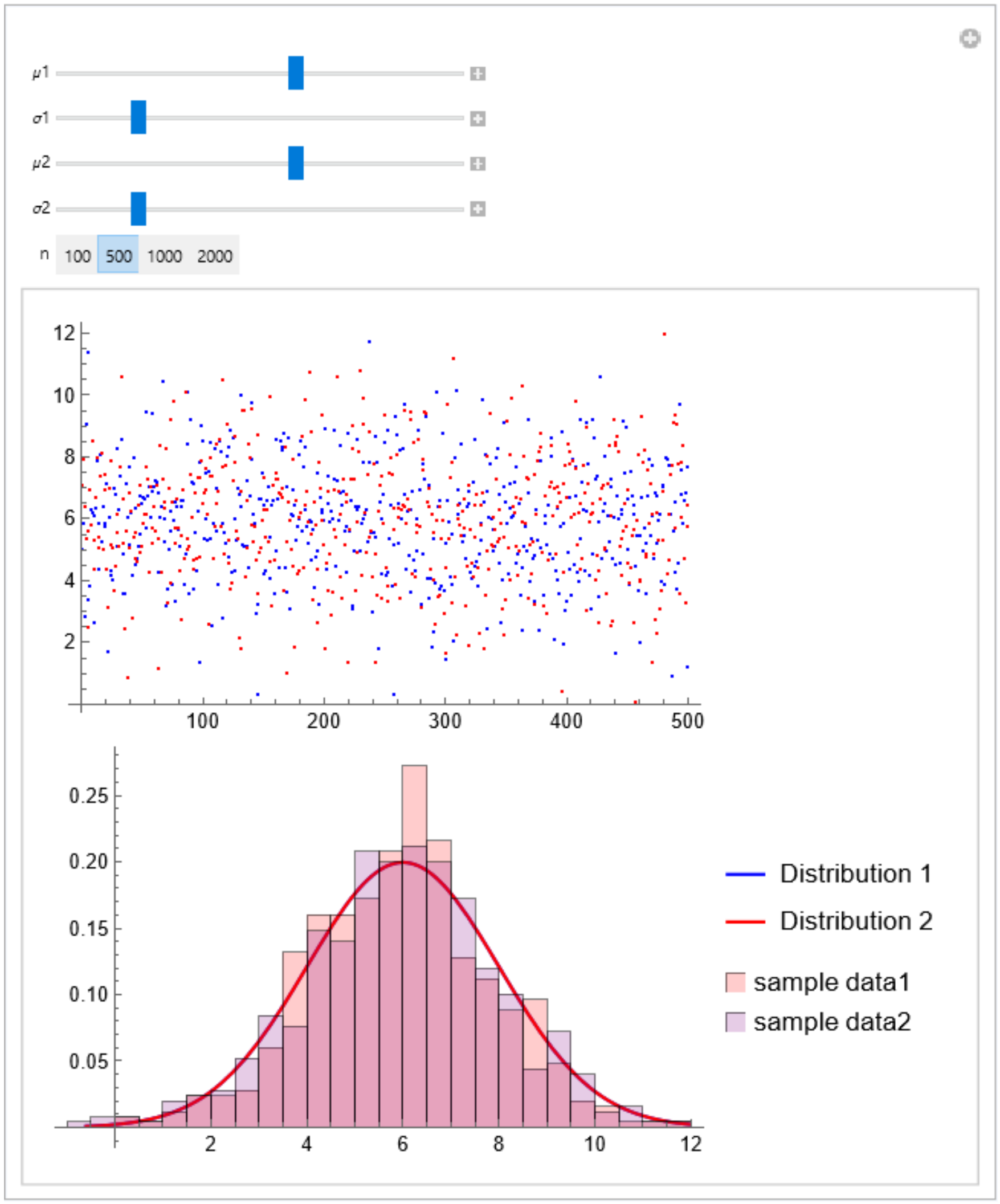

*Mathematica Examples 13.95*

Input

```
(* Student T Distribution goes to a normal distribution as v goes to ∞: *)

Limit[
 PDF[
  StudentTDistribution[v],
  x
  ],
 v->∞
 ]

PDF[
 NormalDistribution[0,1],
 x
 ]
```

Output

$$\frac{e^{-\frac{x^2}{2}}}{\sqrt{2\pi}}$$

Output

$$\frac{e^{-\frac{x^2}{2}}}{\sqrt{2\pi}}$$

***Mathematica Examples 13.96***

Input
```
(* A battery's lifespan has a mean of 1000 hours and a standard deviation of 50 hours, which is roughly normal distributed. What percentage has a lifetime between 800 and 1000 hours: *)

p=N[
  Probability[
   Quantity[800,"Hours"]<x<Quantity[1000,"Hours"],
   Distributed[x,NormalDistribution[Quantity[1000,"Hours"],Quantity[50,"Hours"]]]
   ]
  ]
```

Output
```
0.499968
```

***Mathematica Examples 13.97***

Input
```
(* The normal approximation to the binomial distribution is a method of approximating the binomial distribution with a normal distribution, which is a continuous probability distribution that can be characterized by its mean and standard deviation. This approximation is possible when the number of trials is large and the probability of success in each trial is not too close to 0 or 1. To apply the normal approximation, we first calculate the mean and standard deviation of the binomial distribution. The mean is equal to the product of the number of trials and the probability of success in each trial, while the standard deviation is the square root of the product of the number of trials, the probability of success, and the probability of failure. Once we have the mean and standard deviation, we can use the normal distribution to approximate the binomial distribution.
The code generates two sets of random data points, one from a binomial distribution with n and p and another from a normal distribution with μ=n*p and σ=Sqrt[n*p*(1-p)]. It then plots the data points as a scatter plot for the normal distribution and as a histogram with a probability density function (PDF) plot overlay for the binomial distribution. The plot for the normal distribution shows a bell-shaped curve, which is a characteristic of the normal distribution. The code also includes a slider for adjusting the value of n and p and another slider for adjusting the sample size n. This allows the user to interactively explore how the shape of the distributions change as the value of n and p and the sample size n are varied: *)

Manipulate[
 Module[
  {dataN,dataB},
  μ=n*p;
  σ=Sqrt[n*p*(1-p)];
  dataN=RandomVariate[NormalDistribution[μ,σ],no];
  dataB=RandomVariate[BinomialDistribution[n,p],no];
  Grid[
   {
    {
     ListPlot[
      dataN,
      PlotStyle->Blue,
      ImageSize->300,
      PlotRange->All,
      PlotLabel->Row[{"Normal Distribution μ = ",μ,", σ = ",σ}]
      ],
     ListPlot[
      dataB,
      PlotStyle->Red,
      ImageSize->300,
      PlotRange->All,
      PlotLabel->Row[{"Binomial Distribution n = ",n,", p = ",p}]
```

```
          ]
         },
        {
         Show[
          {
           Plot[
            PDF[NormalDistribution[μ,σ],x],
            {x,Min[dataN],Max[dataN]},
            PlotRange->All,
            PlotStyle->Blue,
            ImageSize->300
            ],
           Histogram[
            dataB,
            Automatic,
            "PDF",
            PlotRange->All,
            ChartStyle->Directive[Opacity[0.5],Purple],
            ImageSize->300
            ]
           }
          ]
         }
        }
       ]
      ],
     {{n,30,"n"},5,60,1,Appearance->"Labeled"},
     {{p,0.5,"p"},0.1,0.9,0.05,Appearance->"Labeled"},
     {{no,3000,"Sample Size"},400,5000,10,Appearance->"Labeled"}
     ]
```

Output

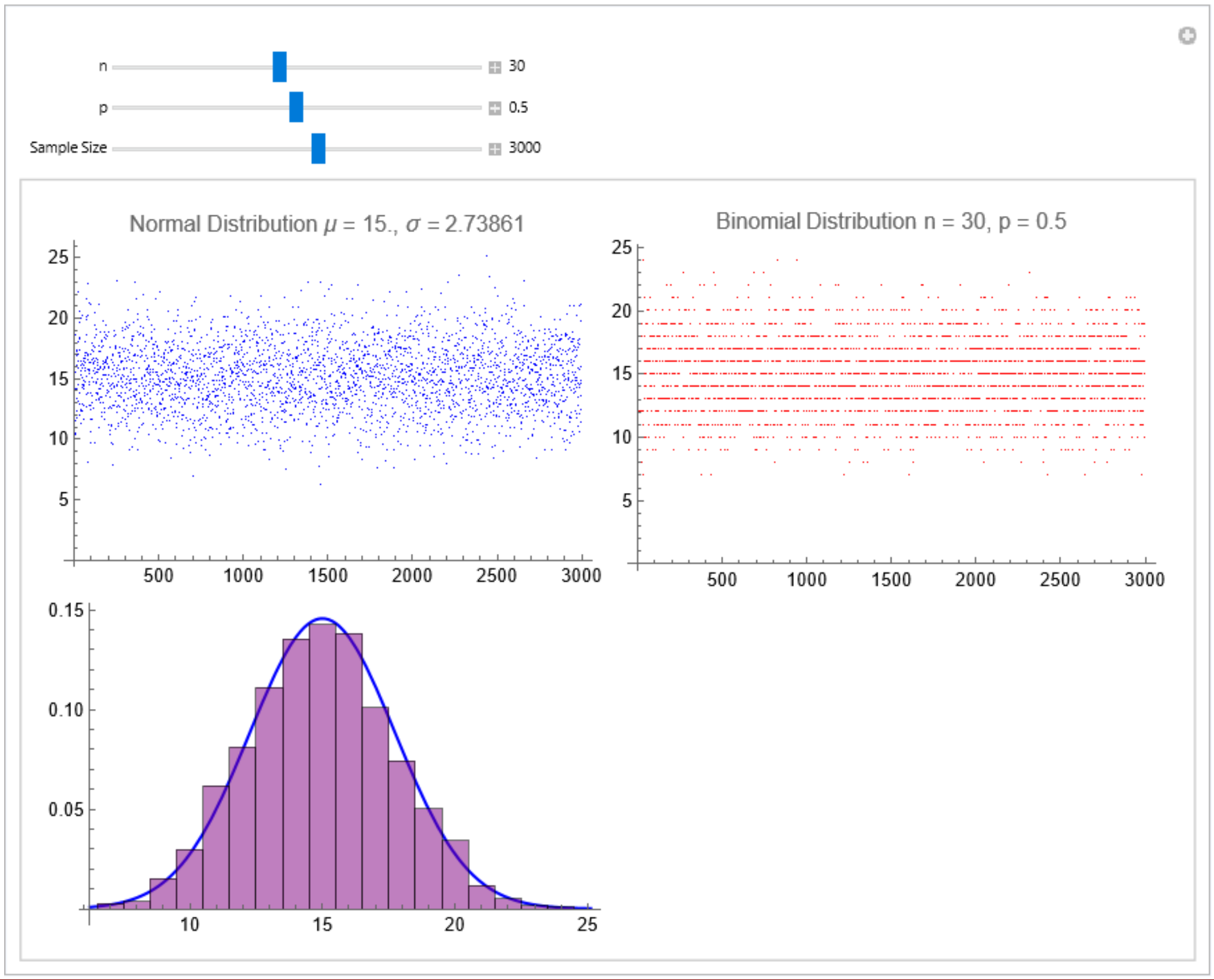

***Mathematica Examples 13.98***

Input

```
(* When the average rate of occurrence (λ) in a Poisson distribution is large, the distribution becomes increasingly bell-shaped and starts to resemble a normal distribution. Specifically, when λ is large (typically λ>20),the Poisson distribution can be approximated by a normal distribution with the same mean and standard deviation. The code generates two sets of random data points, one from a normal distribution and another from a Poisson distribution with the same mean value λ. It then plots the data points as a scatter plot for the normal distribution and as a histogram with a probability density function (PDF) plot overlay for the Poisson distribution. The plot for the normal distribution shows a bell-shaped curve, which is a characteristic of the normal distribution. The plot for the Poisson distribution shows a skewed curve with a peak at λ, which is a characteristic of the Poisson distribution. The code also includes a slider for adjusting the value of λ and another slider for adjusting the sample size n. This allows the user to interactively explore how the shape of the distributions change as the value of λ and the sample size n are varied: *)

Manipulate[
 Module[
  {dataN,dataP},
  dataN=RandomVariate[NormalDistribution[λ,Sqrt[λ]],n];
  dataP=RandomVariate[PoissonDistribution[λ],n];
  Grid[
   {
    {
     ListPlot[
      dataN,
      PlotStyle->Blue,
      ImageSize->300,
      PlotRange->All,
      PlotLabel->Row[{"Normal Distribution μ = ",λ,", σ = ",Sqrt[λ]}]
      ],
     ListPlot[
      dataP,
      PlotStyle->Purple,
      ImageSize->300,
      PlotRange->All,
      PlotLabel->Row[{"Poisson Distribution λ = ",λ}]
      ]
     },
    {
     Show[
      Plot[
       PDF[NormalDistribution[λ,Sqrt[λ]],x],
       {x,Min[dataN],Max[dataN]},
       PlotRange->All,
       PlotStyle->Blue,
       ImageSize->300
       ],
      Histogram[
       dataP,
       Automatic,
       "PDF",
       PlotRange->All,
       ChartStyle->Directive[Opacity[0.5],Purple],
       ImageSize->300
       ],
      Epilog->{Red,PointSize[Large],Point[{λ,0}]}
      ]
     }
```

```
      }
     ]
    ],
   {{λ,20,"λ"},4,40,0.1,Appearance->"Labeled"},
   {{n,3000,"n"},100,5000,100,Appearance->"Labeled"}
   ]
```

Output

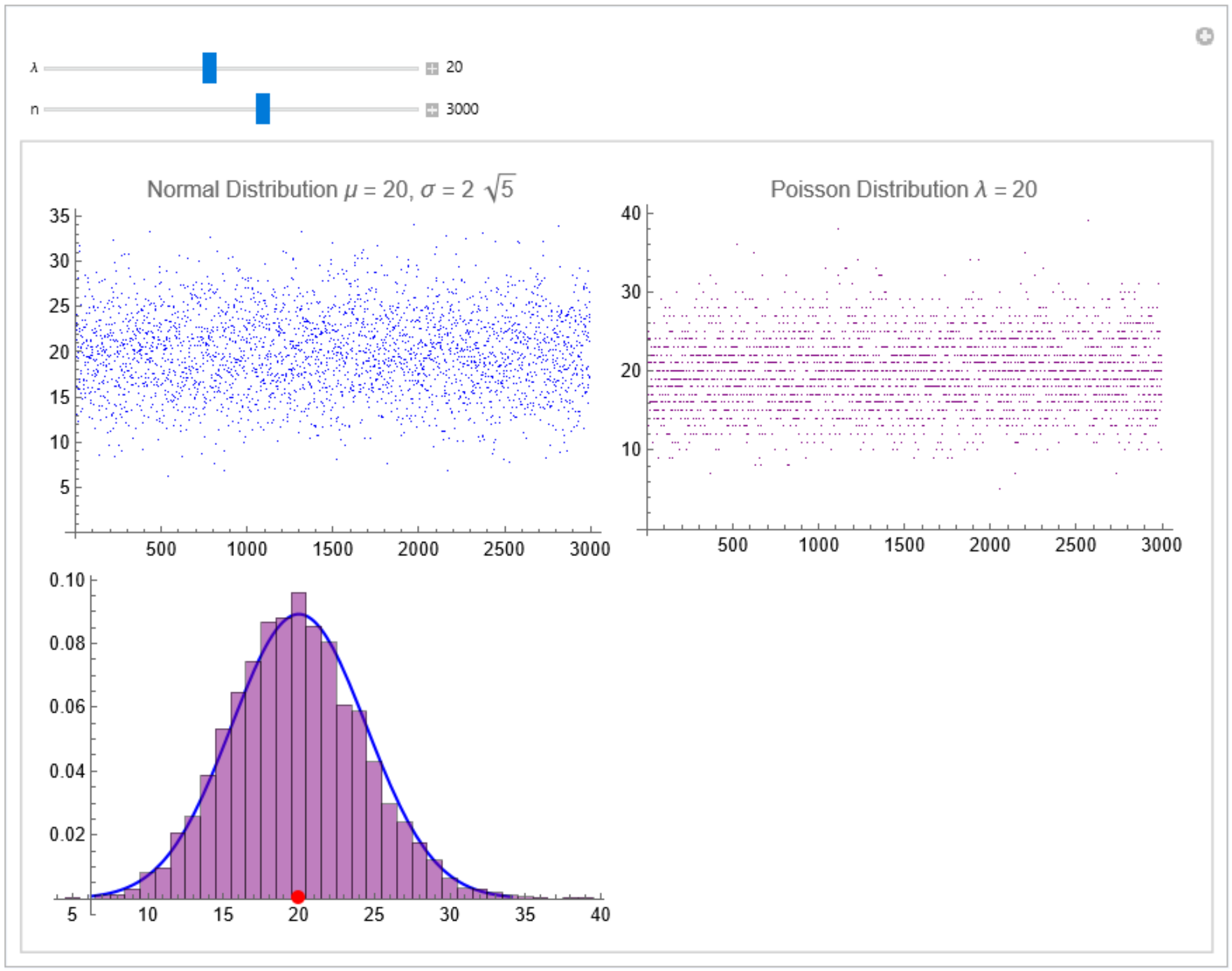

***Mathematica Examples 13.99***

Input

```
(* The code generates a Manipulate that displays the normal probability density function (PDF) of a normal distribution with mean μ=0 and σ=1 and computes the area under the curve for a given value of z in terms of the cumulative distribution function (CDF). The PDF is shown in two parts, where the shaded area represents the area under the curve from-∞ to z. The value of z can be changed using a slider, and the area under the curve from-∞ to z is computed and displayed on the top of the plot using the CDF function: *)

Manipulate[
 Module[
  {dist},
  μ=0;
  σ=1;
  dist=NormalDistribution[μ,σ];
  Show[
   {
    Plot[
     PDF[
      dist,
      x
      ],
     {x,μ+z*σ,μ+5σ},
     Filling->Axis,
     FillingStyle->Directive[Opacity[0.8],Purple],
```

```
          PlotLabel->Row[{"μ = ",μ,", σ = ",σ, ", z=",z,"  Area: ",N[CDF[dist,z]]}]
          ],
         Plot[
          PDF
           [
           dist,
           x
           ],
          {x,μ-5σ,μ+z*σ},
          Filling->Axis,
          FillingStyle->Directive[Opacity[0.8],RGBColor[0.12, 0.61, .78]]
          ]
         },
        PlotRange->{{-5,5},{0,.4}},Axes->{Automatic,False}
        ]
       ],
      {{z,-3.9},-3.9,4.9,0.1,Appearance->"Labeled"}
      ]
```

Output

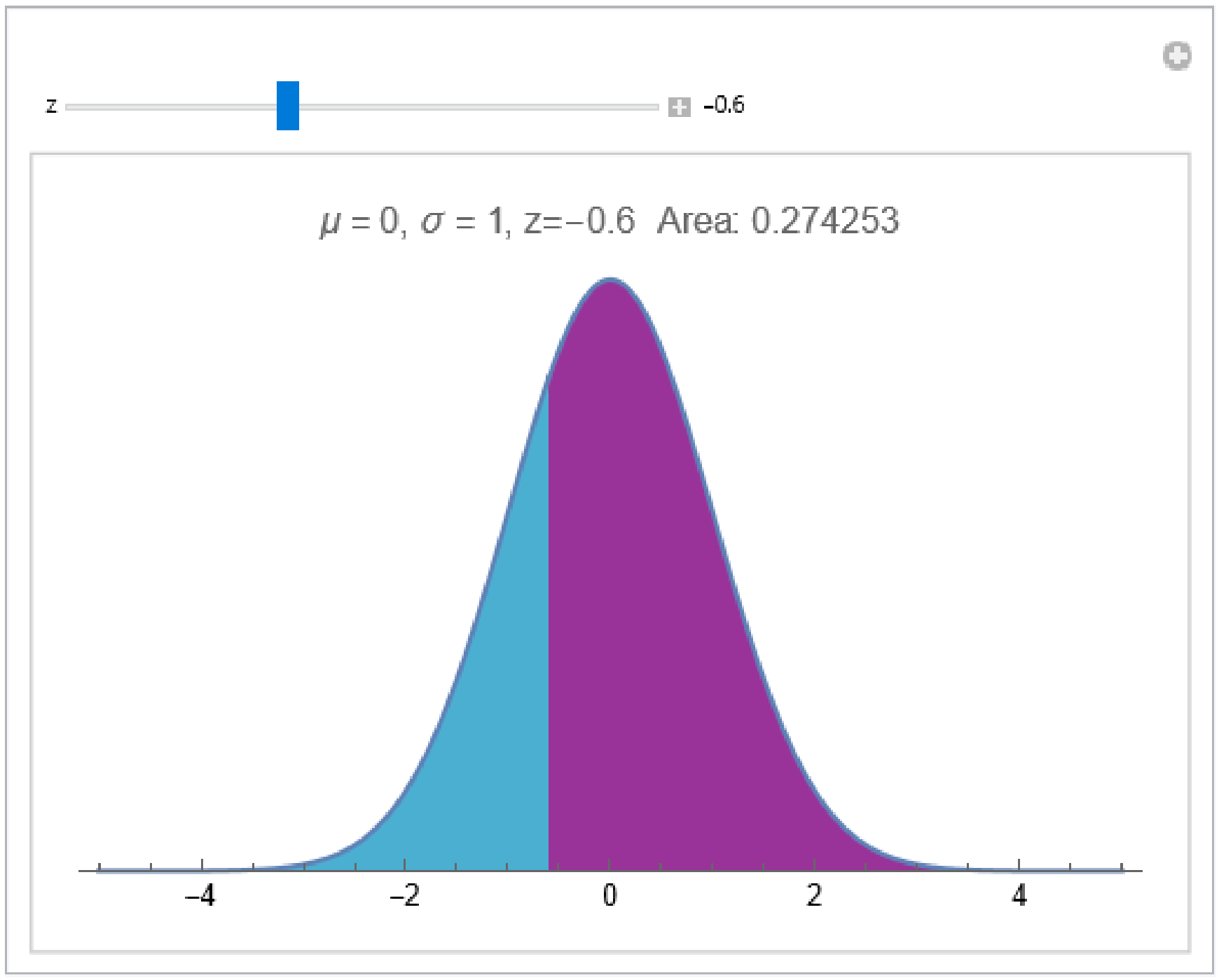

# UNIT 13.6

# PROBABILITY PLOTS

| | |
|---|---|
| `ProbabilityPlot[list]` | generates a plot of the CDF of list against the CDF of a normal distribution. |
| `ProbabilityPlot[dist]` | generates a plot of the CDF of the distribution dist against the CDF of a normal distribution. |
| `ProbabilityPlot[data,rdata]` | generates a plot of the CDF of data against the CDF of rdata. |
| `ProbabilityPlot[data,rdist]` | generates a plot of the CDF of data against the CDF of symbolic distribution rdist. |
| `ProbabilityPlot[{data1,data2,…}, ref]` | generates a plot of the CDF of datai against the CDF of a reference distribution ref. |

***Mathematica Examples 13.100***

Input
```
(* Generate 100 random variates from a uniform distribution on [0,2] and store them
in the variable 'originaldata': *)
originaldata=RandomVariate[UniformDistribution[{0,2}],100];

(* Estimate a normal distribution based on the generated data. The estimated
distribution is assumed to be a normal distribution with mean 'μ' and standard
deviation 'σ'. Store the estimated distribution in the variable 'ref': *)
ref=EstimatedDistribution[originaldata,NormalDistribution[μ,σ]];

(* Create two Probability plots , the first plot shows the CDFs of the original
data 'originaldata'. The default reference distribution used in this plot is the
closest estimated normal distribution. The second plot compares the CDFs of the
original data 'originaldata' with the CDFs of the estimated distribution 'ref'.
From the results, you can see that the default reference distribution is the
closest estimated NormalDistribution: *)
{
 ProbabilityPlot[
  originaldata,
  PlotStyle->Purple,
  ImageSize->250
  ],
 ProbabilityPlot[
  originaldata,
  ref,
  PlotStyle->Purple,
  ImageSize->250
  ]
 }
```

Output

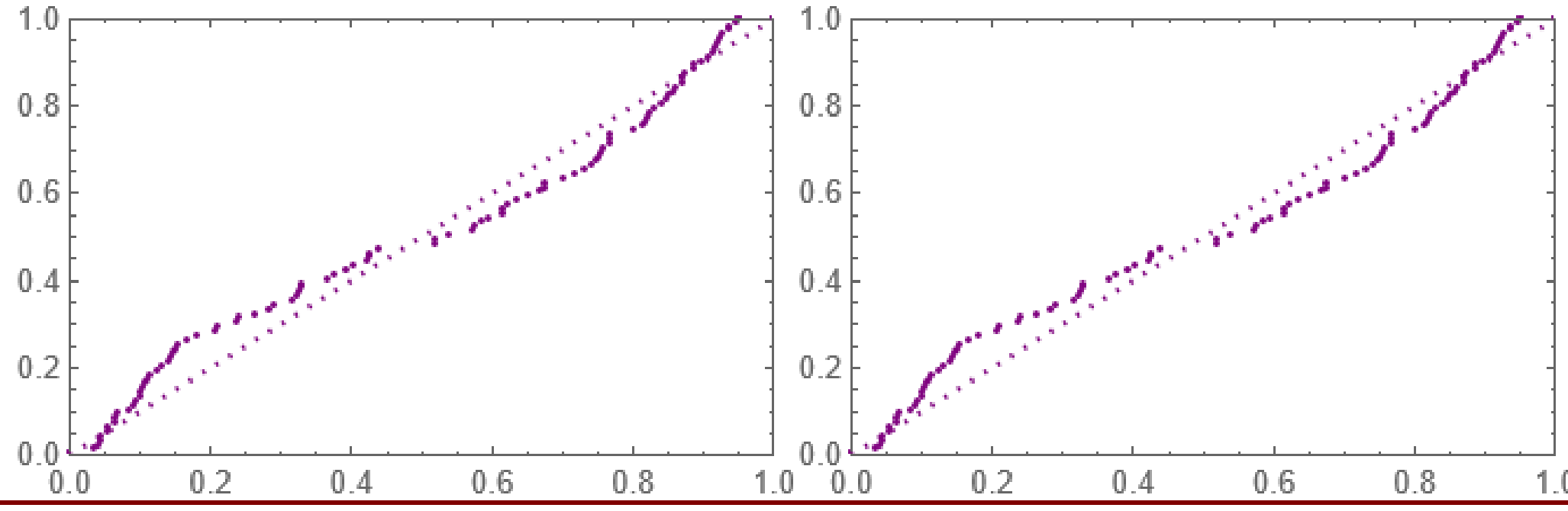

***Mathematica Examples 13.101***

Input
```
(* Compare data to a specific distribution: *)

(* Generate 100 random variates from a uniform distribution on [0,2] and store them
in the variable 'originaldata': *)
originaldata=RandomVariate[UniformDistribution[{0,2}],100];

(* Create a Probability plot to compare the CDFs of the data to the CDFs of the
specific distribution. In this case, the specific distribution is a uniform
distribution on [0,2]:*)
ProbabilityPlot[
 originaldata,
 UniformDistribution[{0,2}],
 PlotStyle->Purple,
 ImageSize->250
 ]
```

Output

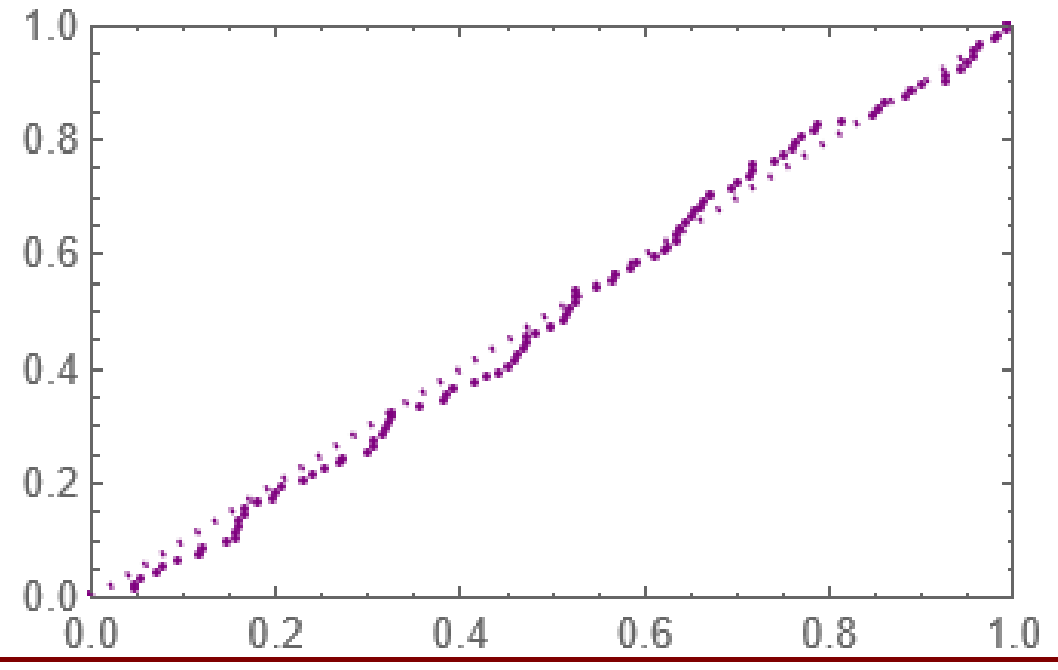

***Mathematica Examples 13.102***

Input
```
(* Compare data to an estimated distribution:*)

(* Generate 200 random variates from a Student's t-distribution with parameters
(5,3,2) and store them in the variable'data'*)
data=RandomVariate[StudentTDistribution[5,3,2],200];


(* ProbabilityPlot[data,dist[θ1,…]] with symbolic parameters θi is equivalent to
ProbabilityPlot[data,EstimatedDistribution[data,dist[θ1,…]]]. Using a symbolic
parameterized distribution in `ProbabilityPlot` is equivalent to using
`EstimatedDistribution` with the same distribution and the dataset'data'.*)

(* Create a Probability plot to compare the CDFs of the data to the CDFs of an
estimated Student's t-distribution.*)
ProbabilityPlot[
 data,
 StudentTDistribution[μ,σ,ν],
 PlotStyle->Purple,
 ImageSize->250
 ]

(*Estimate the parameters (μ,σ,ν) of the Student's t-distribution that best fit the
data.*)
EstimatedDistribution[data,StudentTDistribution[μ,σ,ν]]
```

Output 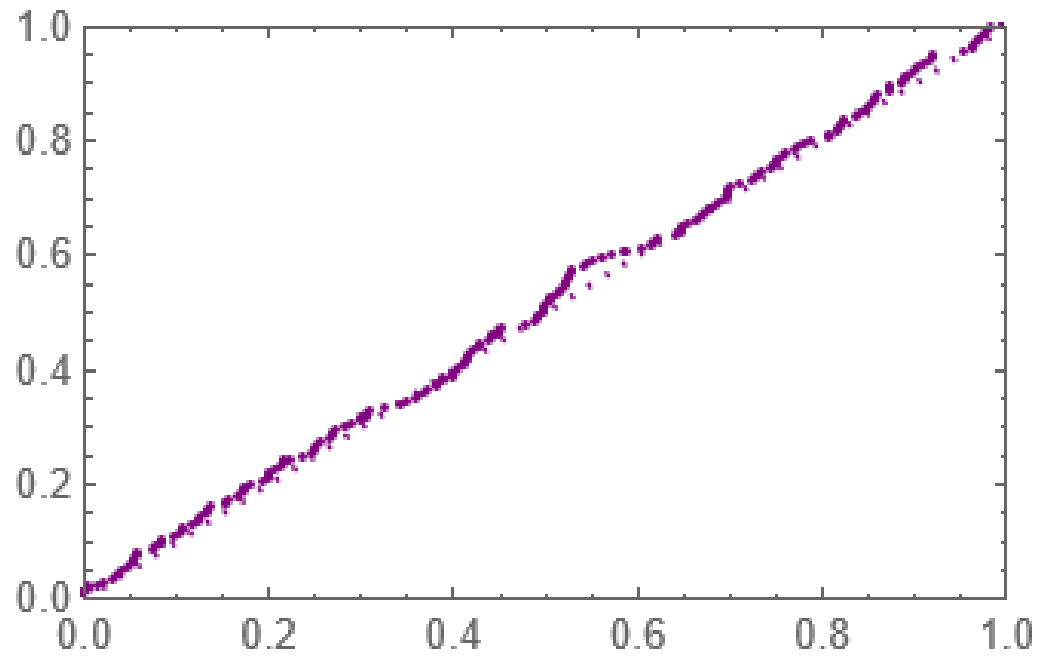

Output
```
StudentTDistribution[4.60278,3.04088,2.19351]
```

### *Mathematica Examples 13.103*

Input
```
(* Compare two datasets:*)
data1=RandomVariate[
    NormalDistribution[3,2],
    100
    ];

data2=RandomVariate[
    StudentTDistribution[5,3,2],
    200
    ];

ProbabilityPlot[
 data1,
 data2,
 PlotStyle->Purple,
 ImageSize->250
 ]
```

Output 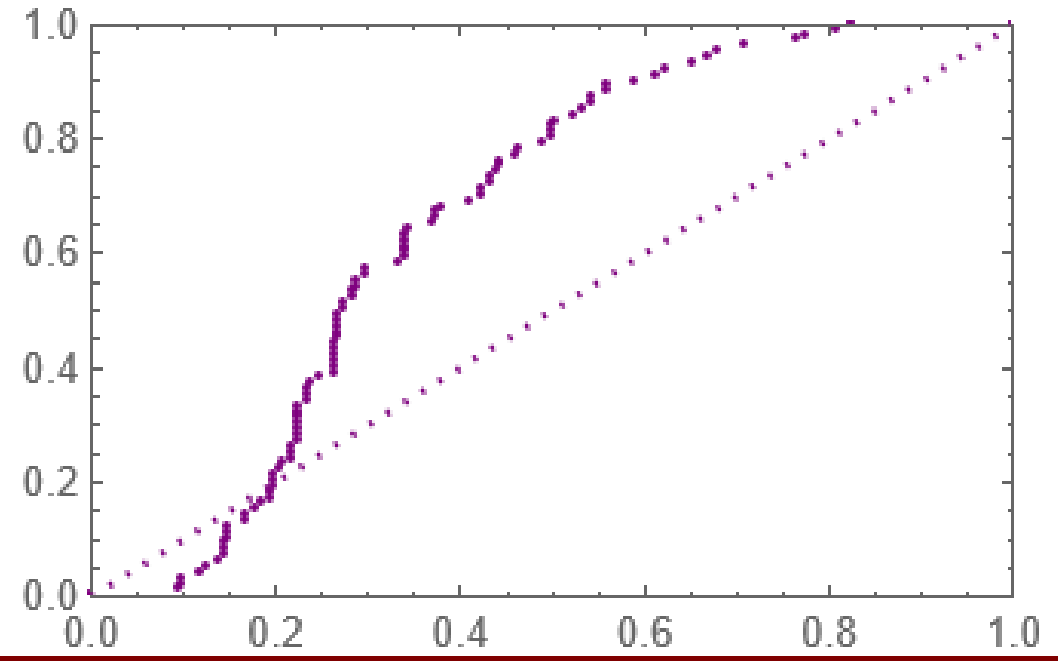

### *Mathematica Examples 13.104*

Input
```
(* The code demonstrates the flexibility of the ProbabilityPlot function by
showcasing how both numeric data and symbolic distributions can be used: *)

(* ProbabilityPlot works with numeric data:*)
ProbabilityPlot[
 RandomVariate[
  ExponentialDistribution[2],
  500
  ],
 PlotStyle->Purple,
 ImageSize->250
 ]
```

```
(* ProbabilityPlot works with symbolic distributions:*)
ProbabilityPlot[
 ExponentialDistribution[2],
 PlotStyle->Purple,
 ImageSize->250
 ]
```

Output

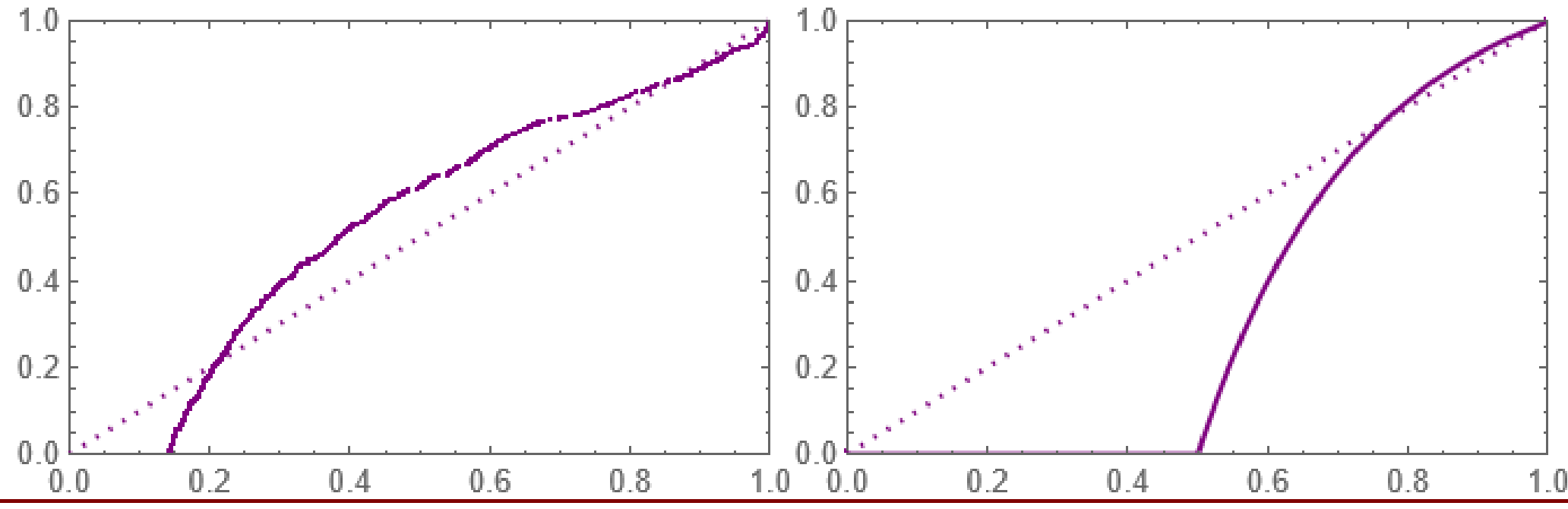

***Mathematica Examples 13.105***

Input

```
(* The code demonstrates the flexibility of the ProbabilityPlot function by
showcasing how both data and distributions can be used as references:*)
data=RandomVariate[
   UniformDistribution[{0,2}],
   200
   ];
{
 ProbabilityPlot[
  data,
  RandomVariate[NormalDistribution[0,1],200],
  PlotStyle->Purple,
  ImageSize->250
  ],
 ProbabilityPlot[
  data,
  NormalDistribution[0,1],
  PlotStyle->Purple,
  ImageSize->250
  ]
```

Output

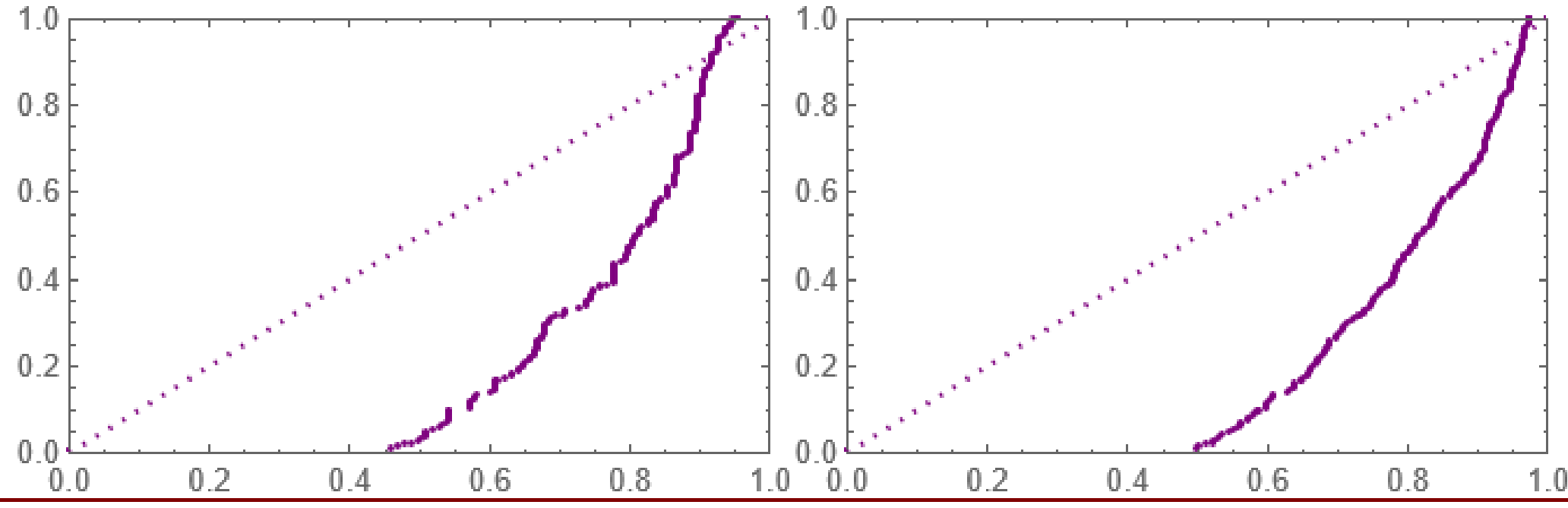

| | |
|---|---|
| `QuantilePlot[list]` | generates a plot of quantiles of list against the quantiles of a normal distribution. |
| `QuantilePlot[dist]` | generates a plot of quantiles of the distribution dist against the quantiles of a normal distribution. |
| `QuantilePlot[data,rdata]` | generates a plot of the quantiles of data against the quantiles of rdata. |
| `QuantilePlot[data,rdist]` | generates a plot of the quantiles of data against the quantiles of a symbolic distribution rdist. |
| `QuantilePlot[{data1,data2,…}, ref]` | generates a plot of quantiles of datai against the quantiles of a reference distribution ref. |

***Mathematica Examples 13.106***

Input
```
(* Generate 100 random variates from a uniform distribution on [0,2] and store them in the variable 'originaldata': *)
originaldata=RandomVariate[UniformDistribution[{0,2}],100];

(* Estimate a normal distribution based on the generated data. The estimated distribution is assumed to be a normal distribution with mean 'μ' and standard deviation 'σ'. Store the estimated distribution in the variable 'ref': *)
ref=EstimatedDistribution[originaldata,NormalDistribution[μ,σ]];

(* Create two quantile plots, the first plot shows the quantiles of the original data 'originaldata'. The default reference distribution used in this plot is the closest estimated normal distribution. The second plot compares the quantiles of the original data 'originaldata' with the quantiles of the estimated distribution 'ref'. From the results, you can see that the default reference distribution is the closest estimated NormalDistribution: *)
{
 QuantilePlot[
  originaldata,
  PlotStyle->Purple,
  ImageSize->250
  ],
 QuantilePlot[
  originaldata,
  ref,
  PlotStyle->Purple,
  ImageSize->250
  ]
 }
```

Output

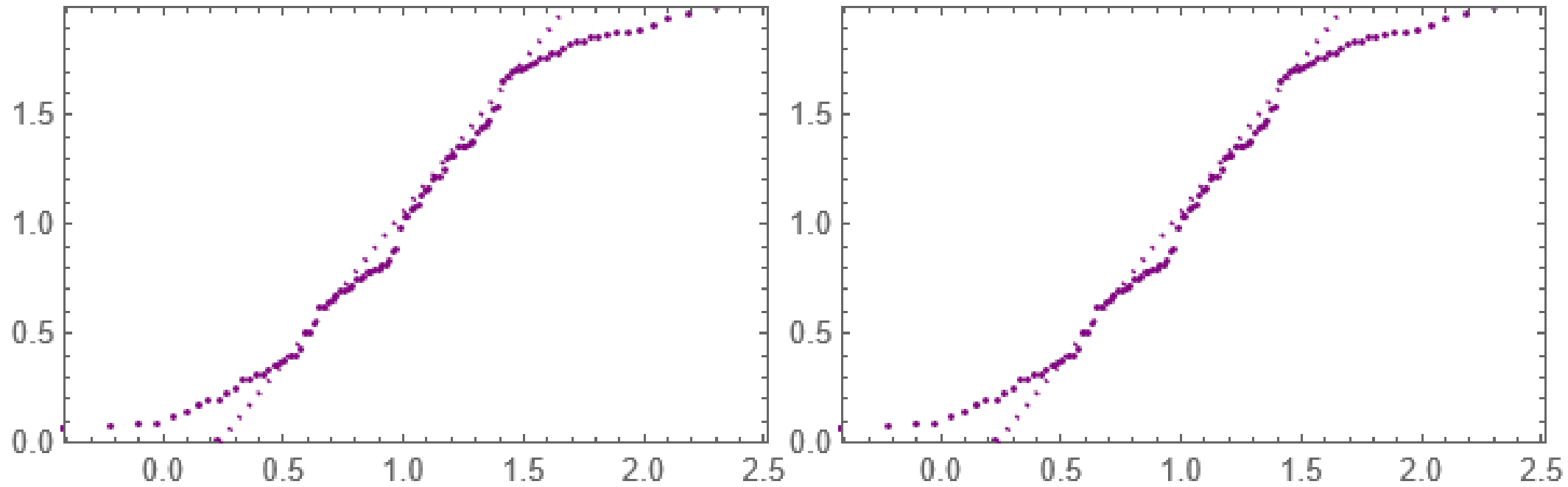

***Mathematica Examples 13.107***

Input
```
(* Compare data to a specific distribution: *)

(* Generate 100 random variates from a uniform distribution on [0,2] and store them in the variable 'originaldata': *)
originaldata=RandomVariate[UniformDistribution[{0,2}],100];

(* Create a quantile plot to compare the quantiles of the data to the quantiles of the specific distribution. In this case,the specific distribution is a uniform distribution on [0,2]:*)
QuantilePlot[
 originaldata,
 UniformDistribution[{0,2}],
 PlotStyle->Purple,
 ImageSize->250
 ]
```

Output

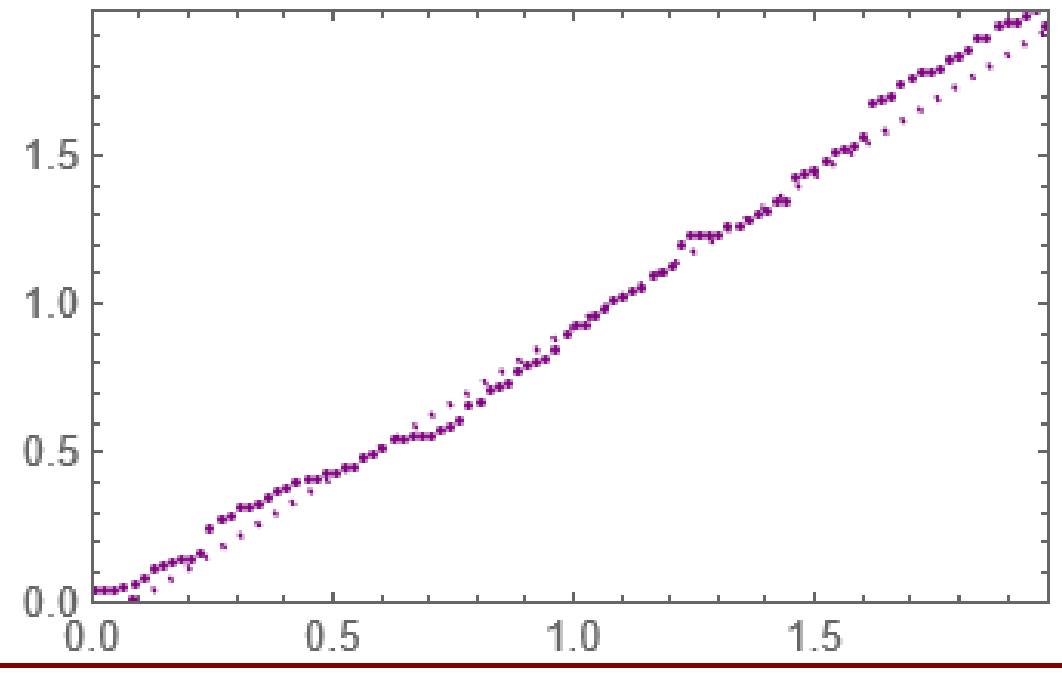

### *Mathematica Examples 13.108*

Input
```
(* Compare data to an estimated distribution:*)

(* Generate 200 random variates from a Student's t-distribution with parameters
(5,3,2) and store them in the variable'data'*)
data=RandomVariate[StudentTDistribution[5,3,2],200];


(* QuantilePlot[data,dist[θ1,…]] with symbolic parameters θi is equivalent to
QuantilePlot[data,EstimatedDistribution[data,dist[θ1,…]]]. Using a symbolic
parameterized distribution in `QuantilePlot` is equivalent to using
`EstimatedDistribution` with the same distribution and the dataset'data'.*)

(* Create a quantile plot to compare the quantiles of the data to the quantiles of
an estimated Student's t-distribution.*)
QuantilePlot[
 data,
 StudentTDistribution[μ,σ,ν],
 PlotStyle->Purple,
 ImageSize->250
 ]

(*Estimate the parameters (μ,σ,ν) of the Student's t-distribution that best fit the
data.*)
EstimatedDistribution[data,StudentTDistribution[μ,σ,ν]]
```

Output

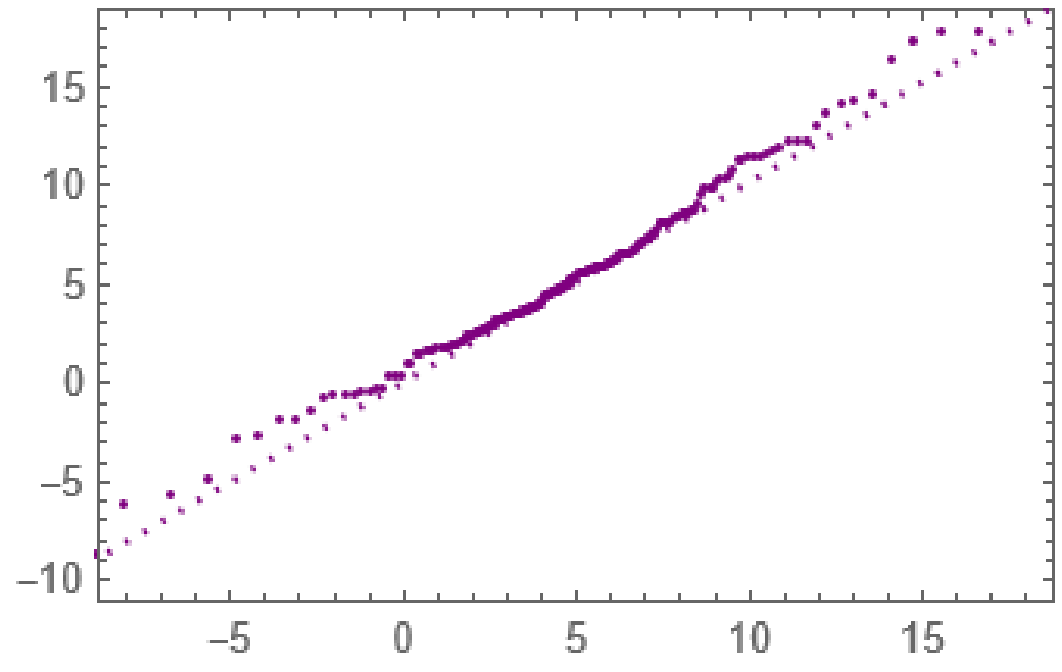

Output
```
StudentTDistribution[5.02885,2.965,2.29611]
```

### *Mathematica Examples 13.109*

Input
```
(* Compare two datasets:*)
data1=RandomVariate[
    NormalDistribution[3,2],
    100
    ];
```

```
data2=RandomVariate[
   StudentTDistribution[5,3,2],
   200
   ];

QuantilePlot[
 data1,
 data2,
 PlotStyle->Purple,
 ImageSize->250
 ]
```

Output

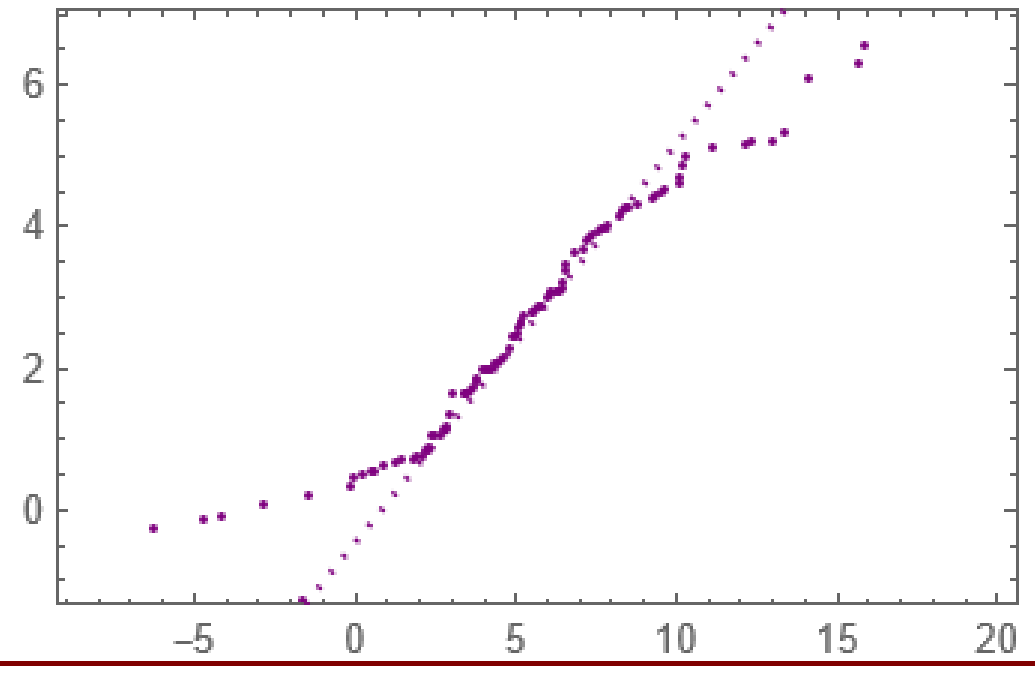

### Mathematica Examples 13.110

Input

```
(* The code demonstrates the flexibility of the QuantilePlot function by showcasing
how both numeric data and symbolic distributions can be used: *)

(*QuantilePlot works with numeric data:*)
QuantilePlot[
 RandomVariate[
  ExponentialDistribution[2],
  100
  ],
 PlotStyle->Purple,
 ImageSize->250
 ]

(* QuantilePlot works with symbolic distributions:*)
QuantilePlot[
 ExponentialDistribution[2],
 PlotStyle->Purple,
 ImageSize->250
 ]
```

Output

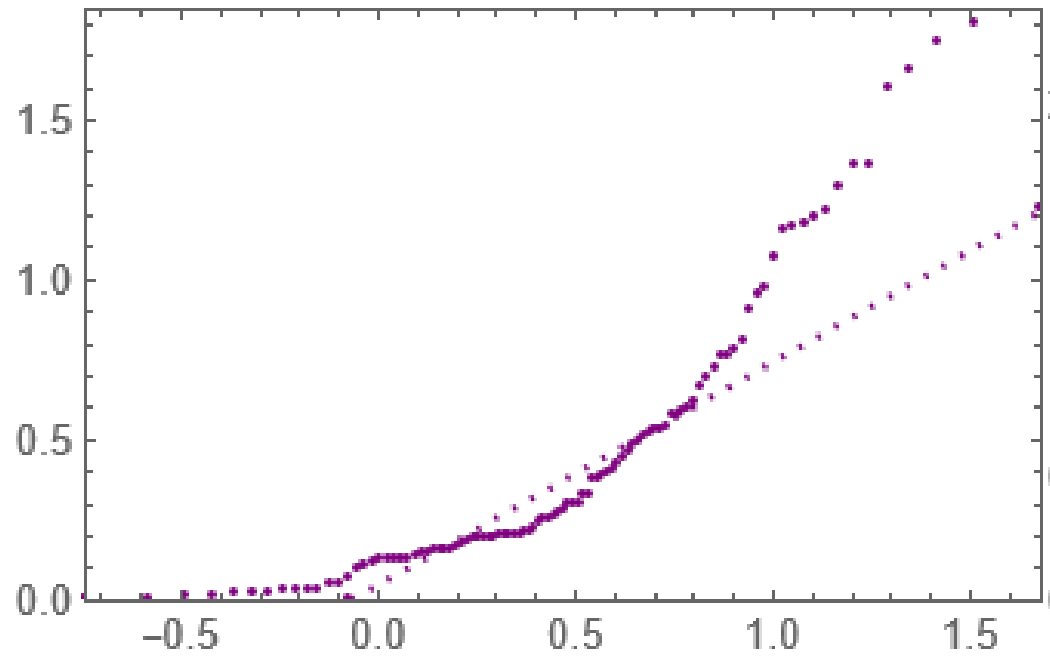

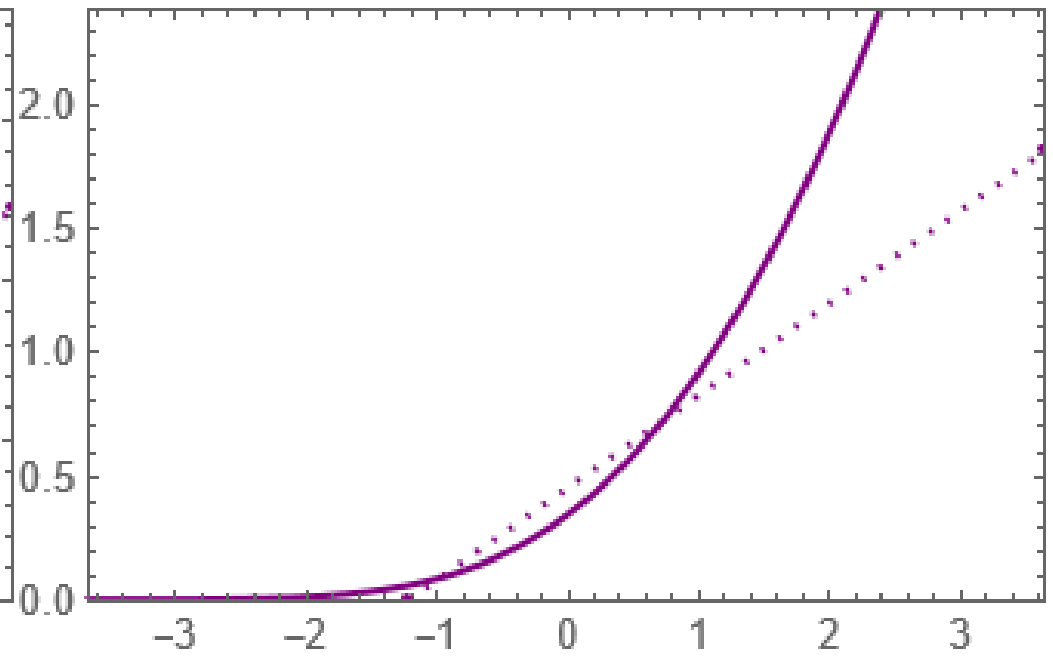

***Mathematica Examples 13.111***

Input
```
(* The code demonstrates the flexibility of the QuantilePlot function by showcasing
how both data and distributions can be used as references:*)
data=RandomVariate[
    UniformDistribution[{0,2}],
    200
    ];

{
 QuantilePlot[
  data,
  RandomVariate[NormalDistribution[0,1],200],
  PlotStyle->Purple,
  ImageSize->250
  ],
 QuantilePlot[
  data,
  NormalDistribution[0,1],
  PlotStyle->Purple,
  ImageSize->250
  ]
```

Output

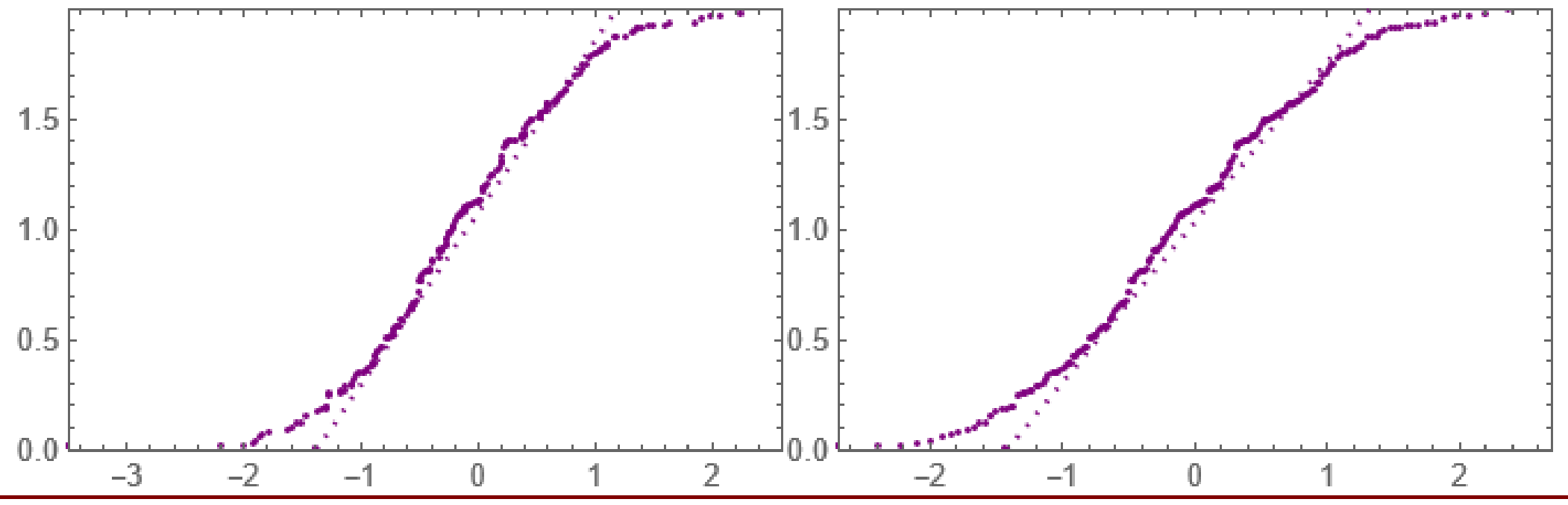

# UNIT 13.7

# DERIVED STATISTICAL DISTRIBUTIONS

- In the realm of probability theory and statistics, derived distributions play a fundamental role in understanding and analyzing random phenomena. Derived distributions are modifications or transformations of existing distributions, obtained through various techniques such as functions of random variables or weighted mixtures of distributions. These modifications allow us to explore new properties, uncover hidden patterns, and model complex systems with greater flexibility.
- By manipulating existing distributions, researchers and practitioners can create derived distributions that capture specific characteristics or behaviors of interest. These modified distributions serve as powerful tools in diverse fields, including physics, economics, engineering, and social sciences. They enable us to tackle complex problems and make informed decisions by providing a deeper understanding of the underlying probabilistic structures.
- One common approach to constructing derived distributions is through functions of random variables. By applying mathematical operations, such as addition, multiplication, exponentiation, or composition, to the values of random variables, we can generate new random variables with different distributions. These derived distributions often possess distinct properties or offer insights into the relationship between variables.
- Another method for obtaining derived distributions is through weighted mixtures of distributions. By combining multiple distributions with appropriate weights, we can create a composite distribution that represents a combination of different phenomena or subpopulations. This approach is particularly useful when dealing with heterogeneous data or complex systems with multiple interacting components.

### Transformed Distribution

| | |
|---|---|
| `TransformedDistribution[expr,x \[Distributed]dist]` | represents the transformed distribution of expr where the random variable x follows the distribution dist. |
| `TransformedDistribution[expr,{x1,x2,…} \[Distributed]dist]` | represents the transformed distribution of expr where {x1,x2,…} follows the multivariate distribution dist. |
| `TransformedDistribution[expr,x \[Distributed]proc]` | represents the transformed distribution where expr contains expressions of the form x[t], referring the value at time t from the random process proc. |
| `TransformedDistribution[expr,{x1 \[Distributed]dist1,x2 \[Distributed]dist2 ,…}]` | represents a transformed distribution where x1, x2, … are independent and follow the distributions dist1, dist2, …. |

***Mathematica Examples 13.112***

```
Input    (* Simple transformations of random variables: *)
         TransformedDistribution[2 u+1,u\[Distributed]NormalDistribution[μ,σ]]

         TransformedDistribution[3 x+2,x\[Distributed]UniformDistribution[{2,3}]]

         TransformedDistribution[a u+b,u\[Distributed]NormalDistribution[μ,σ]]

         TransformedDistribution[Exp[u],u\[Distributed]NormalDistribution[μ,σ]]

Output    NormalDistribution[1+2 μ,2 σ]
          UniformDistribution[{8,11}]
          NormalDistribution[b+a μ,σ Abs[a]]
          LogNormalDistribution[μ,σ]
```

*Mathematica Examples 13.113*

```
Input    (* Automatic Simplifications: *)
         TransformedDistribution[x+y,{x\[Distributed]NormalDistribution[μ1,σ1],y\[Distribute
         d]NormalDistribution[μ2,σ2]}]

         TransformedDistribution[u+v,{u\[Distributed]PoissonDistribution[μ1],v\[Distributed]
         PoissonDistribution[μ2]}]

         TransformedDistribution[x^2+y^2,{x\[Distributed]NormalDistribution[0,1],y\[Distribu
         ted]NormalDistribution[0,1]}]

         TransformedDistribution[x/y,{x\[Distributed]NormalDistribution[0,1],y\[Distributed]
         NormalDistribution[0,1]}]

         TransformedDistribution[Min[x,y],{x\[Distributed]GeometricDistribution[p1],y\[Distr
         ibuted]GeometricDistribution[p2]}]
```

Output NormalDistribution[μ1+μ2, $\sqrt{\sigma1^2+\sigma1^2}$ ]
PoissonDistribution[μ1+μ2]
ChiSquareDistribution[2]
CauchyDistribution[0,1]
GeometricDistribution[1-(1-p1) (1-p2)]

*Mathematica Examples 13.114*

```
Input    (* Applying the identity transformation to a distribution leaves it unchanged: *)
         TransformedDistribution[x,x\[Distributed]ExponentialDistribution[2]]

Output   ExponentialDistribution[2]
```

*Mathematica Examples 13.115*

```
Input    (* Transformed distributions can be used like any other distribution. PDF and CDF of
         a Transformed Distribution: *)
         dist=TransformedDistribution[u^2,u\[Distributed]BetaDistribution[3,2]];
         PDF[dist,x]
         CDF[dist,x]
```

Output $\begin{cases} 6\sqrt{x}-6x & 0<\sqrt{x}<1 \&\& 0 <= x <= 1 \\ 0 & \text{True} \end{cases}$

Output $\begin{cases} 1 & (\sqrt{x} >= 1 \&\& x >= 0) || x > 1 \\ 0 & x > 1 || x < 0 || \sqrt{x} <= 0 \\ \text{BetaRegularized}[\sqrt{x},3,2] & \text{True} \end{cases}$

*Mathematica Examples 13.116*

```
Input    (* Transformed distributions can be used like any other distribution. Statistical
         Analysis of a Transformed Distribution: *)
         dist=TransformedDistribution[x+y,{x\[Distributed]NormalDistribution[0,1],y\[Distrib
         uted]NormalDistribution[0,2]}];
         mean=Mean[dist];
         variance=Variance[dist];
         skewness=Skewness[dist];
         kurtosis=Kurtosis[dist];

         Grid[
          {{"    Mean:",mean},
           {"Variance:",variance},
           {"Skewness:",skewness},
           {"Kurtosis:",kurtosis}}
          ]
```

```
Output   {
          {     Mean:, 0},
          {Variance:, 5},
          {Skewness:, 0},
          {Kurtosis:, 3}
         }
```

***Mathematica Examples 13.117***

```
Input    (* Transformed distributions can be used like any other distribution. Compute the
         probability of an event for a transformed distribution: *)
         Probability[x^2+x<6,x\[Distributed]TransformedDistribution[y^2+1,y\[Distributed]Nor
         malDistribution[0,1]]]
         (* Substituting transformation into the event:*)
         Probability[(y^2+1)^2+(y^2+1)<6,y\[Distributed]NormalDistribution[0,1]]

Output   1/2 (Erfc[-(1/√2 )]-Erfc[1/√2 ])
Output   1/2 (Erfc[-(1/√2 )]-Erfc[1/√2 ])
```

***Mathematica Examples 13.118***

```
Input    (* Transformed distributions can be used like any other distribution. Compute the
         expectation of an expression for a transformed distribution: *)
         Expectation[x^2+x,x\[Distributed]TransformedDistribution[y^2+1,y\[Distributed]Norma
         lDistribution[0,1]]]

         (* Substituting transformation into the expression: *)
         Expectation[(y^2+1)^2+(y^2+1),y\[Distributed]NormalDistribution[0,1]]

Output   8
Output   8
```

***Mathematica Examples 13.119***

```
Input    (* Curve Plot of a Transformed Distribution: *)
         dist=TransformedDistribution[1/x,x\[Distributed]GammaDistribution[2,1/2]];

         Plot[
          PDF[dist,x],
          {x,0,5},
          Filling->Axis,
          PlotStyle->Purple,
          PlotRange->All,
          PlotLabel->"PDF of Transformed Distribution",
          ImageSize->250
          ]
Output
```

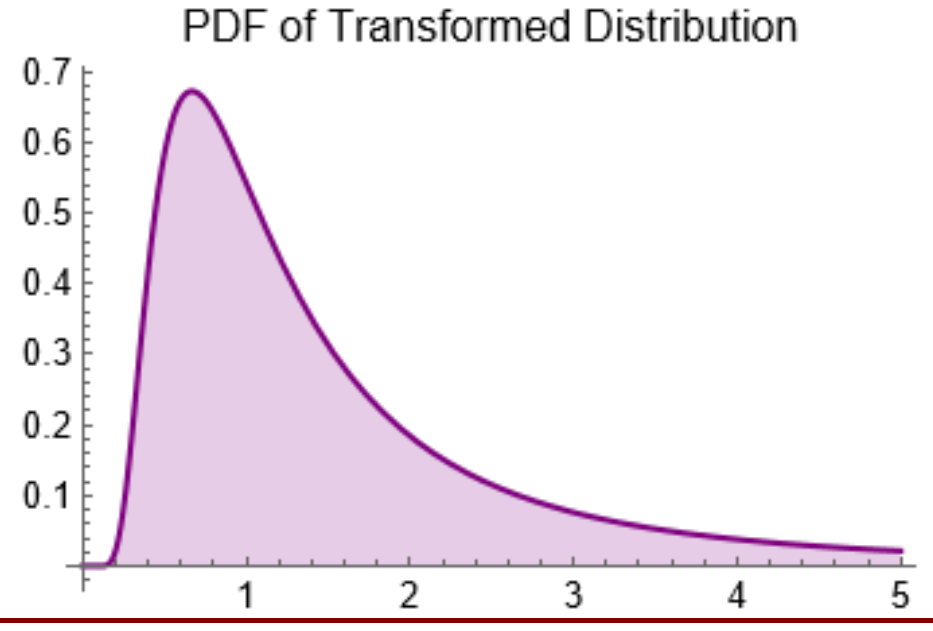

***Mathematica Examples 13.120***

```
Input    (* Simulation and Visualization of a Transformed Distribution: *)
         dist=TransformedDistribution[x^2,x\[Distributed]NormalDistribution[0,1]];
```

```
data=RandomVariate[dist,300];

ListPlot[
 data,
 Filling->Axis,
 PlotStyle->Directive[Purple,PointSize[0.007],Opacity[0.5]],
 Frame->True,
 FrameLabel->{"Index","Transformed Value"},
 PlotLabel->"Simulation of Transformed Distribution",
 ImageSize->250
 ]
```

Output

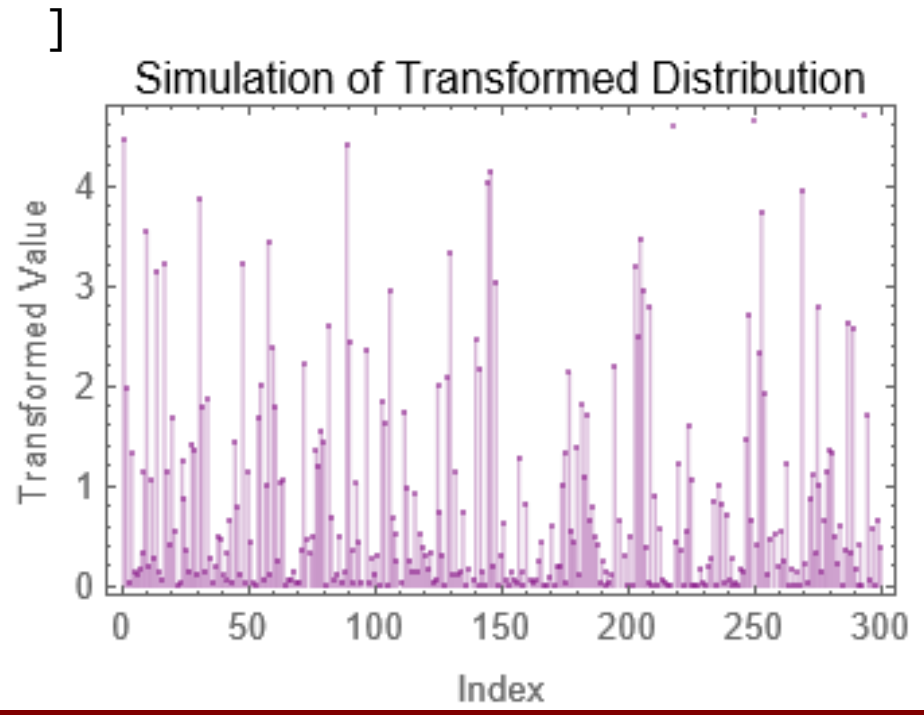

***Mathematica Examples 13.121***

Input

```
(* The code compares the original normal distribution [0,1] with two scaled distributions. It calculates the means and variances of the distributions. Additionally, it generates a plot showing the PDFs of the original distribution and the two scaled distributions: *)

(* Original distribution: *)
dist=NormalDistribution[0,1];

(* Scaled distributions: *)
dist1=TransformedDistribution[2 x,x\[Distributed]dist];
dist2=TransformedDistribution[x/2,x\[Distributed]dist];

(* Compare mean: *)
N[Map[Mean,{dist,dist1,dist2}]]
(* Compare variance: *)
N[Map[Variance,{dist,dist1,dist2}]]

(* Compare the PDFs with the probability density function of the original distribution: *)
Plot[
 {PDF[dist,x],PDF[dist1,x],PDF[dist2,x]},
 {x,-5,5},
 Filling->Axis,
 PlotStyle->{Purple,RGBColor[0.88,0.61,0.14],RGBColor[0.37,0.5,0.7]},
 PlotRange->All,
 PlotLegends->Placed[{"X","2 X","X / 2"},{0.8,0.75}],
 ImageSize->250
 ]
```

Output {0.,0.,0.}
Output {1.,4.,0.25}

Output

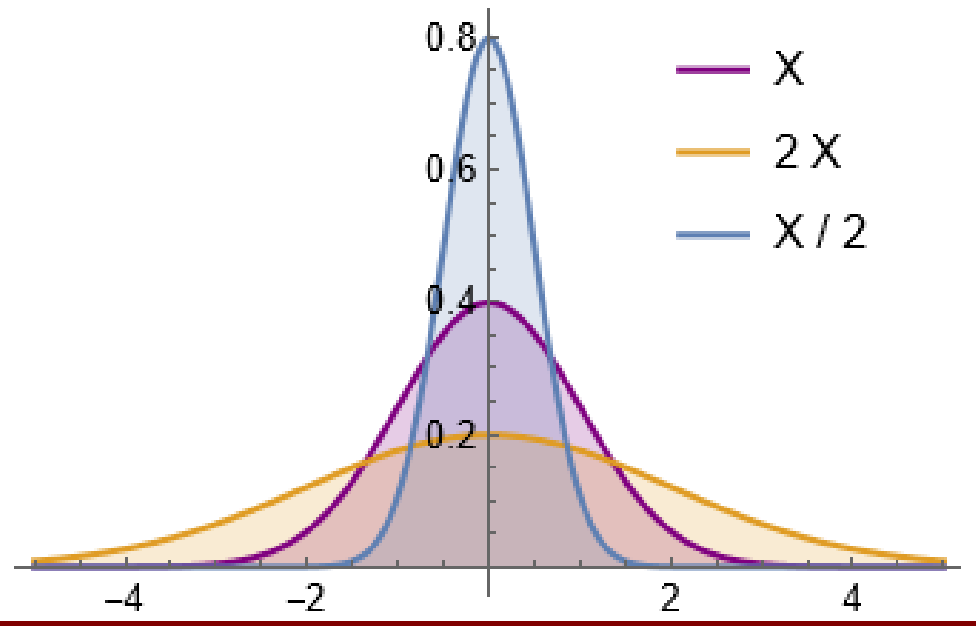

***Mathematica Examples 13.122***

Input

```
(* In this code, we first define the original distribution as a normal distribution with mean 2 and standard deviation 1. Then, we use the TransformedDistribution function to create a new distribution transformedDistribution. After that, we generate 1000 random samples from the transformed distribution using the RandomVariate function. Finally, we plot the PDF and histogram of the transformed distribution and display a histogram of the original distribution: *)

(* Define the original distribution: *)
originalDistribution=NormalDistribution[2,1];

(* Create the transformed distribution: *)
transformedDistribution=TransformedDistribution[3*x+1,x\[Distributed]originalDistri
bution];

(* Generate random samples from the original distribution: *)
randomSampleso=RandomVariate[originalDistribution,500];

(* Generate random samples from the transformed distribution: *)
randomSamplest=RandomVariate[transformedDistribution,500];

(* Display the transformed distribution and random samples: *)
Show[
 Histogram[
  randomSamplest,
  Automatic,
  "PDF",
  ChartStyle->Directive[Purple,Opacity[0.5]],
  PlotRange->All,
  PlotLabel->"Original and Transformed Distributions",
  ImageSize->250
  ],
 Histogram[
  randomSampleso,
  Automatic,
  "PDF",
  ChartStyle->RGBColor[0.88,0.61,0.14],
  PlotRange->All,
  ImageSize->250
  ],
 Plot[
  PDF[transformedDistribution,x],
  {x,-10,30},
  PlotStyle->Red,
  PlotRange->All,
  ImageSize->250
  ]
 ]
```

Output

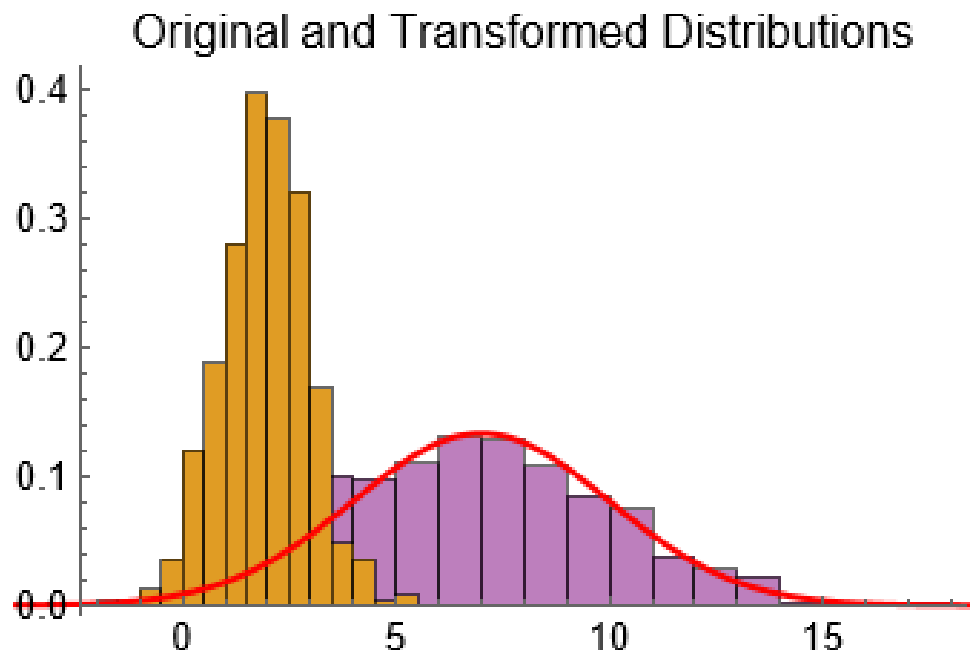

***Mathematica Examples 13.123***

Input
```
(* The code compares an original Poisson distribution with a shifted version of the distribution. It visualizes the PMFs of the two distributions and generates random numbers from each distribution for further comparison. Additionally, the code generates random numbers from both distributions and plots them as points on a graph. It also includes lines representing the means of each distribution. This allows for a visual comparison of the mean values between the original and shifted distributions: *)

(* Original and shifted distribution: *)
originaldistribution=PoissonDistribution[5];
shifteddistribution=TransformedDistribution[x+3,x\[Distributed]originaldistribution
];

(* Compare the PDFs: *)
DiscretePlot[
 {PDF[originaldistribution,x],PDF[shifteddistribution,x]},
 {x,0,16},
 PlotStyle->{Purple,RGBColor[0.88,0.61,0.14]},
 PlotLegends->Placed[{"X","X + 3"},{0.8,0.75}],
 ImageSize->250
 ]

(* Simulation of original and shifted distributions: *)
ListPlot[
 {
  RandomVariate[originaldistribution,100],
  RandomVariate[shifteddistribution,100],
  {{0,Mean[originaldistribution]},{100,Mean[originaldistribution]}},
  {{0,Mean[shifteddistribution]},{100,Mean[shifteddistribution]}}
  },
 PlotStyle->{Purple,RGBColor[0.88,0.61,0.14],RGBColor[0.37,0.5,0.7],Red},
 PlotLegends->{"Original Distribution","Shifted Distribution","Mean Original","Mean Shifted"},
 Joined->{False,False,True,True},
 Filling->{1->Axis,2->Axis},
 ImageSize->250
 ]
```

Output

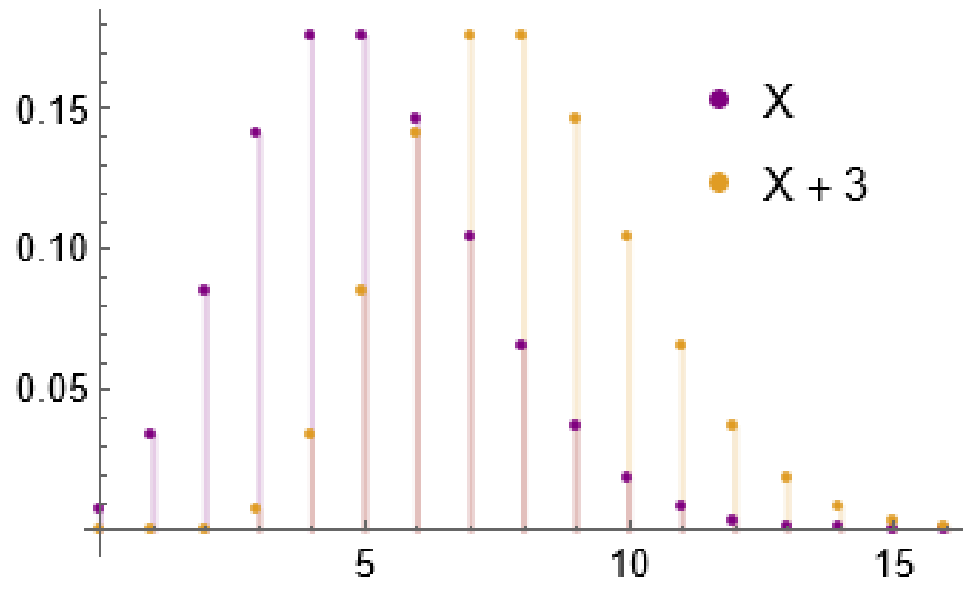

Output

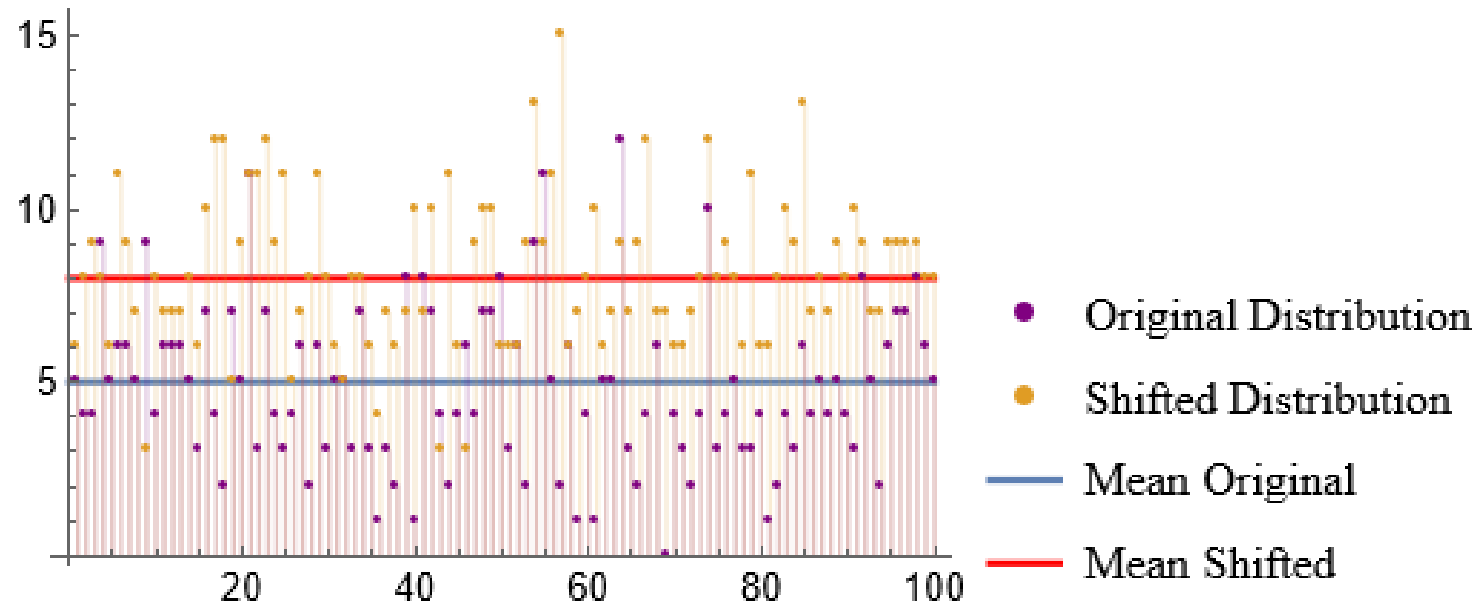

***Mathematica Examples 13.124***

Input

```
(* Find the distribution of the sum of two different variables: *)
originaldistribution1=GammaDistribution[α,β];
originaldistribution2=NormalDistribution[μ,σ];
transformedDistribution=TransformedDistribution[u+v,{u\[Distributed]originaldistrib
ution1,v\[Distributed]originaldistribution2}];

(* Probability density function: *)
pdf=Simplify[PDF[transformedDistribution,x]]

(* Compare the resulting distribution with the summands: *)
Block[
 {α=3,β=1,μ=0,σ=1},
 Plot[
  {pdf,PDF[originaldistribution1,x],PDF[originaldistribution2,x]},
  {x,-4,10},
  Filling->Axis,
  PlotStyle->{Purple,RGBColor[0.88,0.61,0.14],RGBColor[0.37,0.5,0.7]},
  PlotLegends->Placed[{"Transformed Dist.","Gamma Dist.","Normal
Dist."},{0.8,0.78}],
  ImageSize->300
  ]
 ]

(*The mean of dist is the sum of the means: *)
Mean[transformedDistribution]
Total[Map[Mean,{originaldistribution1,originaldistribution2}]]
```

Output

$$\frac{1}{\sqrt{\pi}\sigma^3\text{Gamma}[\alpha]} 2^{\frac{1}{2}(-3+\alpha)} e^{-\frac{(x-\mu)^2}{2\sigma^2}} \left(\frac{\sigma}{\beta}\right)^{1+\alpha} \times \left(\beta\,\sigma\,\text{Gamma}\left[\frac{\alpha}{2}\right]\text{Hypergeometric1F1}\left[\frac{\alpha}{2},\frac{1}{2},\frac{1}{2}\left(\frac{1}{\beta}+\frac{-x+\mu}{\sigma^2}\right)^2\sigma^2\right]+\sqrt{2}(x\,\beta-\beta\,\mu-\sigma^2)\text{Gamma}\left[\frac{1+\alpha}{2}\right]\text{Hypergeometric1F1}\left[\frac{1+\alpha}{2},\frac{3}{2},\frac{1}{2}\left(\frac{1}{\beta}+\frac{-x+\mu}{\sigma^2}\right)^2\sigma^2\right]\right)$$

Output 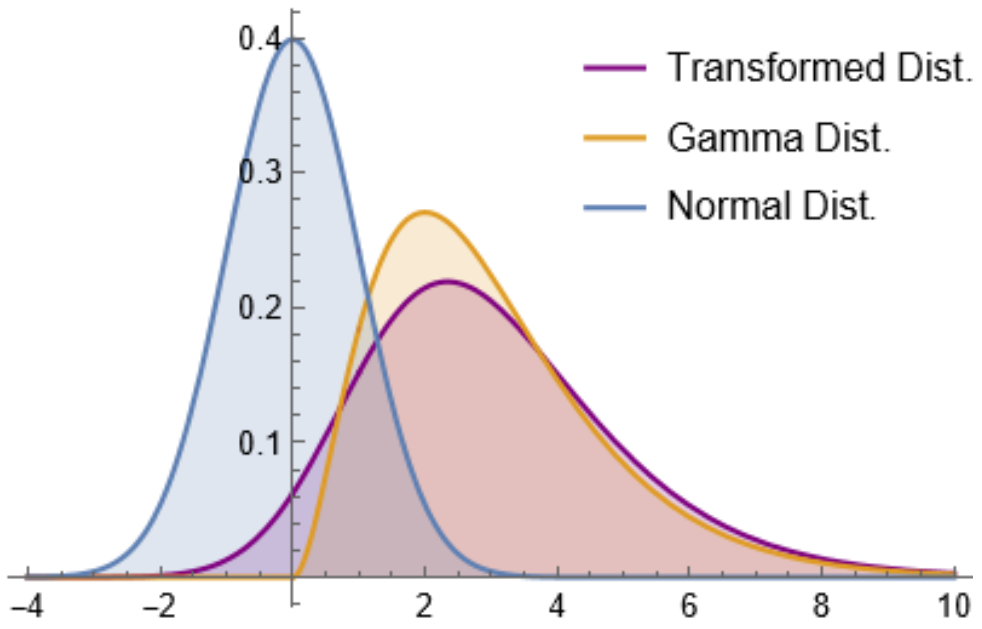

Output α β+μ
Output α β+μ

***Mathematica Examples 13.125***

Input

```
(* In this code, you can specify the original distribution by selecting one of the options provided in the drop-down menu for the dist0 parameter. You can also choose a transformation function by selecting one of the options provided in the drop-down menu for the expr parameter. The dist variable represents the transformed distribution, and samples stores the randomly generated samples from that distribution. The plot range is determined based on these samples to ensure the histogram visualization is properly scaled. The resulting histogram plot will update dynamically as you change the parameters, allowing you to explore the effects of different distributions and transformations: *)

Manipulate[
 Module[
  {dist,plotRange},

  (* Define the transformed distribution: *)
  dist=TransformedDistribution[expr,x\[Distributed]dist0];

  (* Generate random samples from the distribution: *)
  samples=RandomVariate[dist,2000];

  (* Determine the plot range based on the samples: *)
  plotRange={Min[samples],Max[samples]};

  (* Create the plot: *)
  Histogram[
   samples,
   Automatic,
   "PDF",
   PlotRange->Automatic,
   Frame->True,
   FrameLabel->{"x","PDF"},
   ColorFunction->Function[{height},Opacity[height]],
   ChartStyle->Purple,
   ImageSize->300
   ]
  ],
 (* Parameters: *)
 {
  {dist0,NormalDistribution[0,1],"Original Distribution"},
  {
   NormalDistribution[0,1],
   GammaDistribution[1,1],
   ExponentialDistribution[1],
   UniformDistribution[{-1,1}]
   }
```

```
  },
 {{expr,x,"Transformation"},{x,x^2,Sin[x],Exp[x]}}
 ]
```

Output

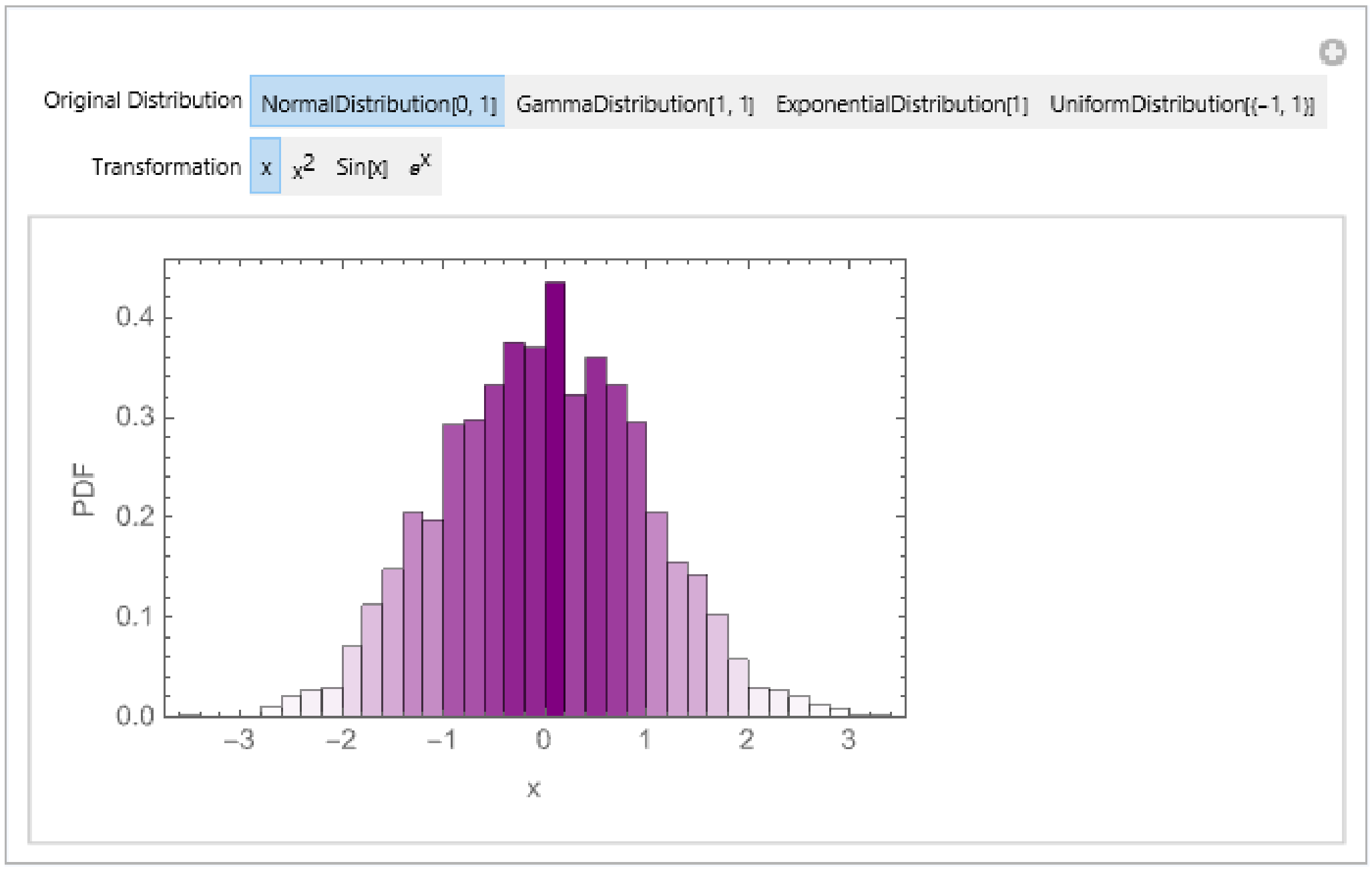

***Mathematica Examples 13.126***

Input

```
(* In this example, the TransformedDistribution takes an original distribution dist
(in this case, a normal distribution) and applies the transformation a+b*x to it.
The Manipulate function allows you to interactively change the parameters mu, sigma,
a, b, and numSamples. It generates random samples from the transformed distribution
and displays a histogram of these samples: *)

Manipulate[
 Module[
  {dist,randomSamples,plot},

  (* Define the original distribution: *)
  dist=NormalDistribution[mu,sigma];

  (* Generate random samples from the transformed distribution: *)
  randomSamples=RandomVariate[
    TransformedDistribution[a+b*x,x\[Distributed]dist],
    numSamples
    ];

  (* Create a histogram of the random samples: *)
  Histogram[
   randomSamples,
   Automatic,
   "PDF",
   ColorFunction->Function[{height},Opacity[height]],
   ChartStyle->Purple,
   ImageSize->300
   ]
  ],
 (* Manipulate parameters: *)
 {{mu,0,"Mean of Original Distribution"},-10,10,0.1},
```

```
 {{sigma,1,"Standard Deviation of Original Distribution"},0.1,5,0.1},
 {{a,0,"Shift Parameter"},-5,5,0.1},
 {{b,1,"Scaling Parameter"},0.1,5,0.1},
 {{numSamples,1000,"Number of Random Samples"},100,5000,100}
 ]
```

Output

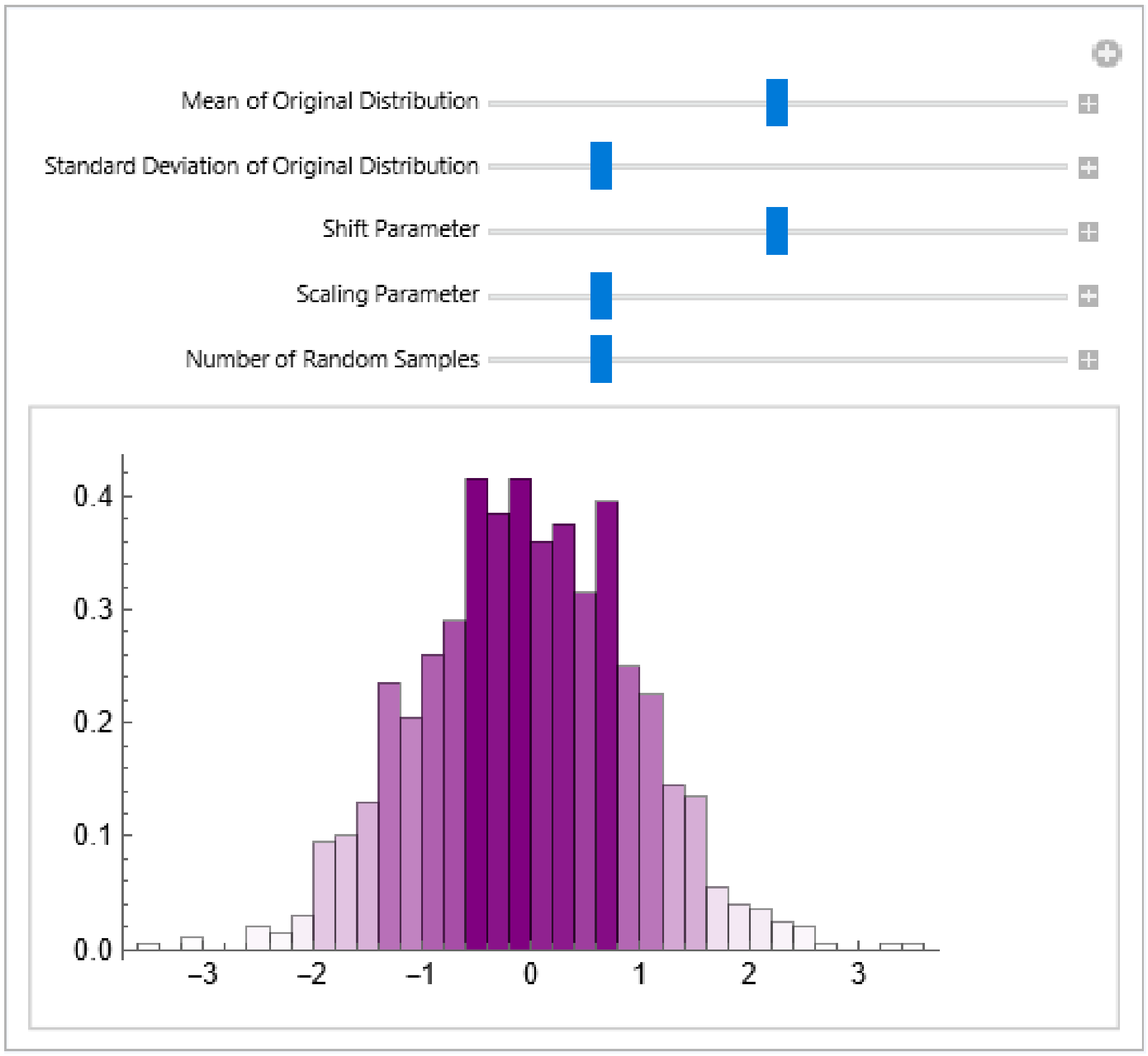

***Mathematica Examples 13.127***

Input

```
(* Scatter Plot and Histogram of Original and Transformed Distributions: *)
distn=NormalDistribution[0,1];
distu=UniformDistribution[{5,9}];
dist=TransformedDistribution[2 x+y,{x\[Distributed]distn,y\[Distributed]distu}];

n=500;(* Sample size *)

datan=RandomVariate[distn,n];
datau=RandomVariate[distu,n];
datat=RandomVariate[dist,n];

Histogram[
 {datat,datan,datau},
 {0.25},
 "PDF",
 ChartStyle->{Purple,RGBColor[0.88,0.61,0.14],RGBColor[0.37,0.5,0.7]},
 PlotLabel->"Histogram of Original and Transformed Distributions",
 ChartLegends->{"Transformed","Normal","Uniform"},
 ImageSize->300
 ]
```

```
ListPlot[
 {
  datat,
  datan,
  datau,
  {{0,Mean[dist]},{n,Mean[dist]}},
  {{0,Mean[distn]},{n,Mean[distn]}},
  {{0,Mean[distu]},{n,Mean[distu]}}
  },
 Joined->{False,False,False,True,True,True},

 PlotStyle->{
   Directive[Purple,PointSize[0.008]],
   Directive[RGBColor[0.88,0.61,0.14],PointSize[0.007]],
   Directive[RGBColor[0.37,0.5,0.7],PointSize[0.007]],
   Red,Red,Red
   },
 PlotLabel->"Scatter Plot of Original and Transformed Distributions",
 PlotLegends->{"Transformed","Normal","Uniform"},
 ImageSize->300
 ]
```

Output

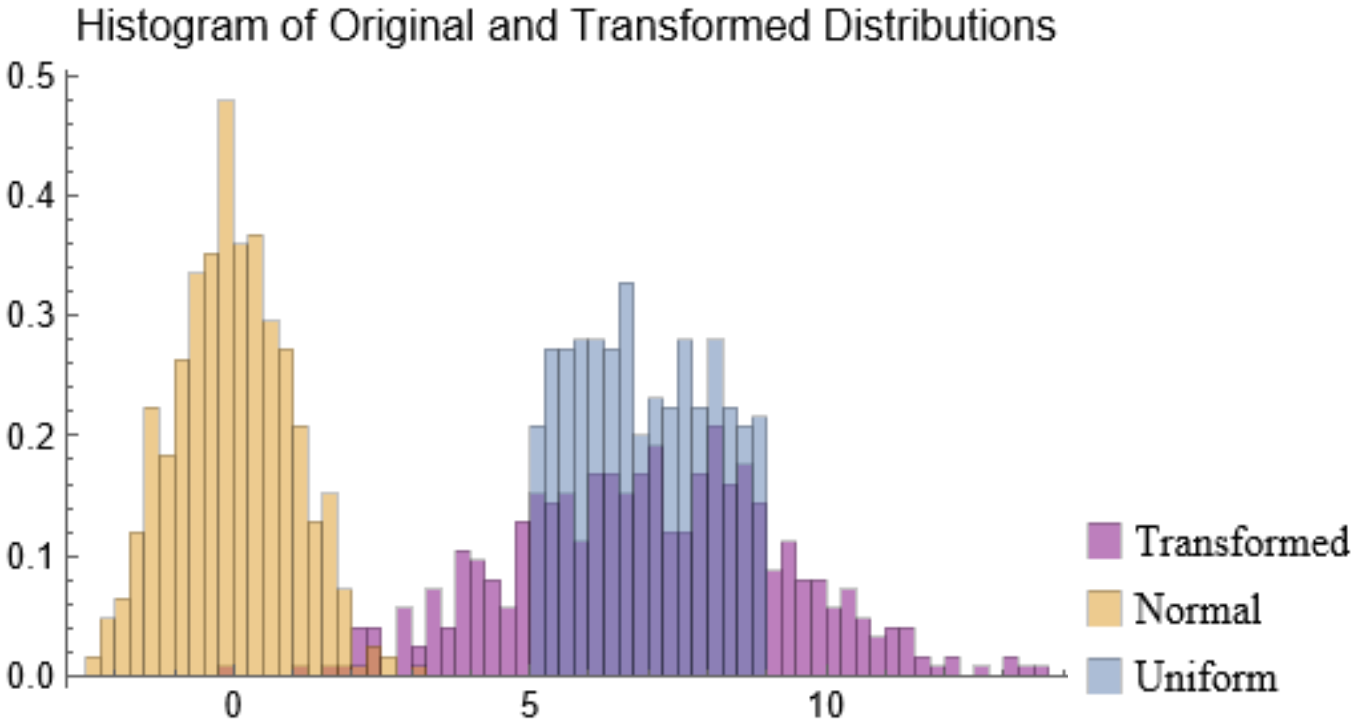

Output

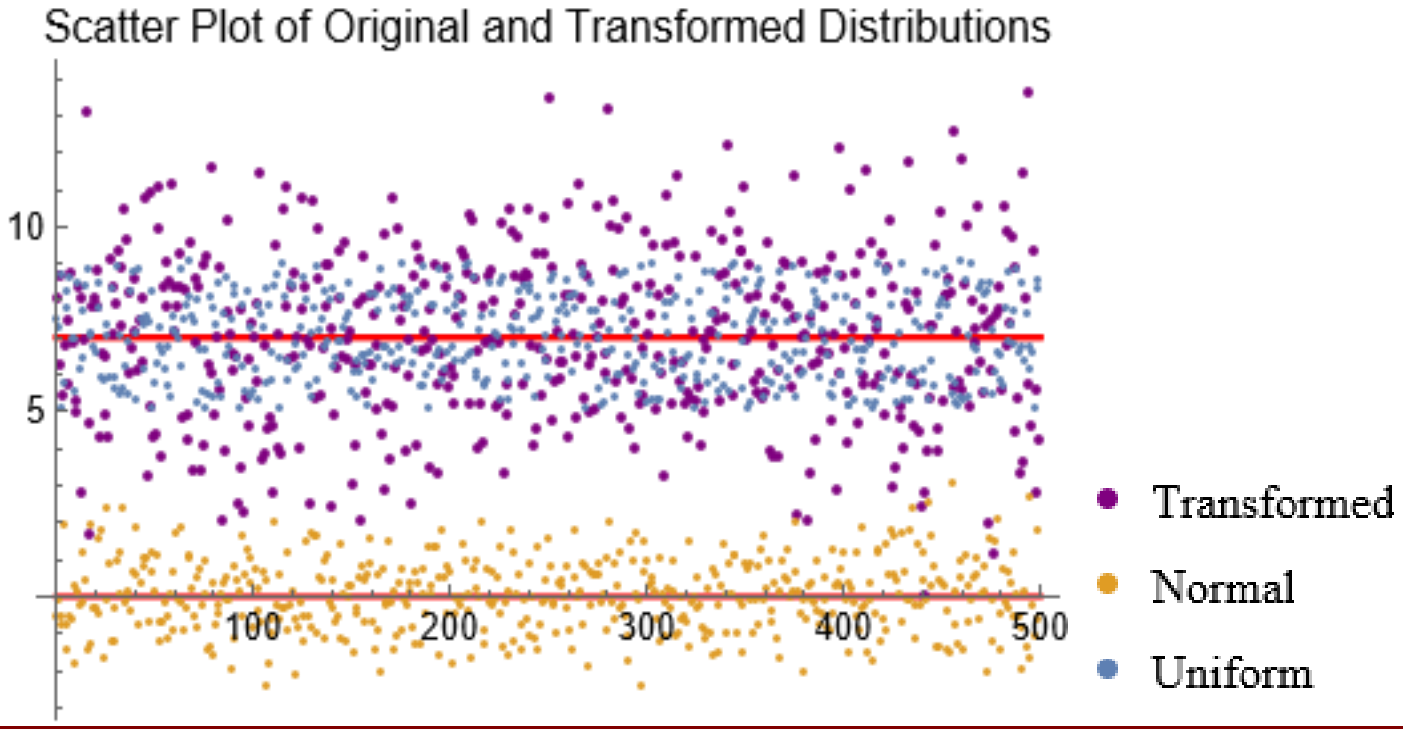

## Mixture Distribution

| | |
|---|---|
| `MixtureDistribution[{w1,…,wn},{dist1,…,distn}]` | represents a mixture distribution whose CDF is given as a sum of the CDFs of the component distributions disti, each with weight wi. |

### *Mathematica Examples 13.128*

Input

```
(* Define a mixture of two continuous distributions: *)
mixdist=MixtureDistribution[
    {3,1},
    {NormalDistribution[0,1],
```

```
     NormalDistribution[3,1/2]}
    ];

Plot[
 PDF[mixdist,x],
 {x,-4,6},
 Filling->Axis,
 ImageSize->250,
 PlotStyle->Purple
 ]
```

Output

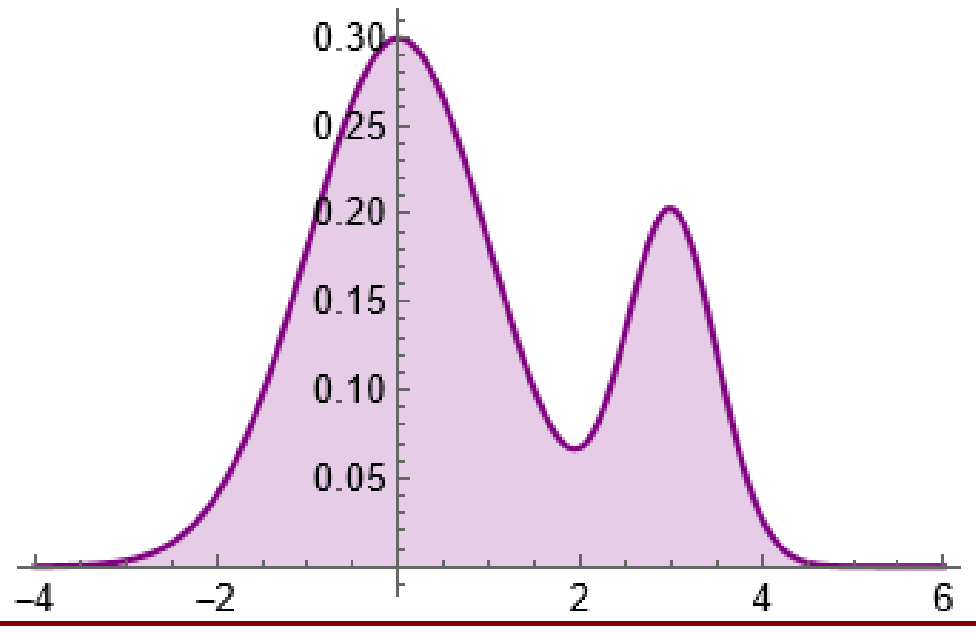

### *Mathematica Examples 13.129*

Input

```
(* Define a mixture of two discrete distributions: *)
mixdist=MixtureDistribution[
    {5,1},
    {PoissonDistribution[7],
     GeometricDistribution[0.5]}
    ];
DiscretePlot[
 PDF[mixdist,x],
 {x,0,20},
 ExtentSize->1/2,
 ImageSize->250,
 PlotStyle->Purple
 ]
```

Output

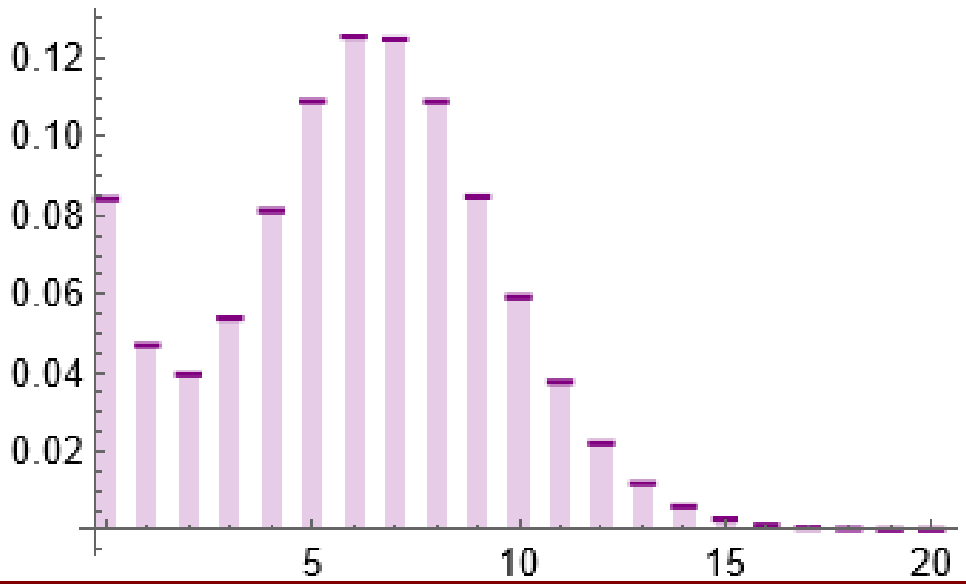

### *Mathematica Examples 13.130*

Input

```
(* A mixture with symbolic weights: *)
mixdist=MixtureDistribution[
    {a,b},
    {NormalDistribution[2,3],
     NormalDistribution[1,2]}
    ];
(* Probability density function: *)
PDF[mixdist,x]
```

Output
$$\frac{a e^{-\frac{1}{18}(-2+x)^2}}{3(a+b)\sqrt{2\pi}}+\frac{b e^{-\frac{1}{8}(-1+x)^2}}{2(a+b)\sqrt{2\pi}}$$

***Mathematica Examples 13.131***

Input

```
(* In this code, we added a plot that shows the densities of the component distributions (dist1 and dist2) individually. This plot helps visualize the individual distributions before they are combined into the mixture distribution. The subsequent plot shows the PDF of the mixture distribution, where the densities of the component distributions are combined according to their weights (weight1 and weight2). The resulting PDF represents the overall density of the mixture distribution, which is a weighted combination of the densities of the component distributions. By comparing the two plots, you can observe how the densities of the component distributions are combined based on their respective weights in the resulting mixture distribution: *)

(*Define the component distributions*)
dist1=NormalDistribution[0,1];
dist2=NormalDistribution[3,2];

(*Define the weights*)
weight1=0.7;
weight2=0.3;

(*Define the mixture distribution*)
mixtureDist=MixtureDistribution[{weight1,weight2},{dist1,dist2}];

(*Plot the densities of the component distributions*)
Plot[
 {PDF[dist1,x],PDF[dist2,x]},
 {x,-5,10},
 PlotRange->All,
 PlotLabel->"Component Densities",
 PlotLegends->{"Component 1","Component 2"},
 Filling->Axis,
 ImageSize->250
 ]

(*Plot the PDF of the mixture distribution*)
Plot[
 PDF[mixtureDist,x],
 {x,-5,10},
 PlotRange->All,
 Filling->Axis,
 PlotLabel->"Mixture Distribution PDF",
 PlotStyle->Purple,
 ImageSize->250
 ]
```

Output

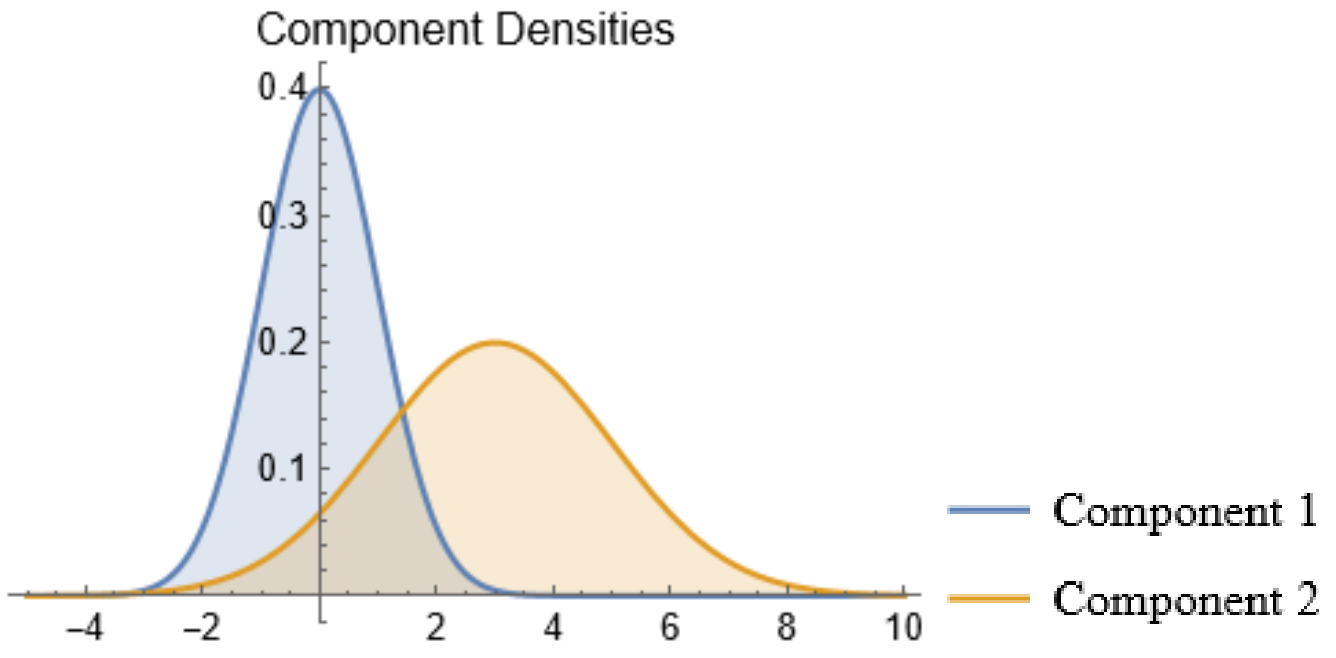

Output

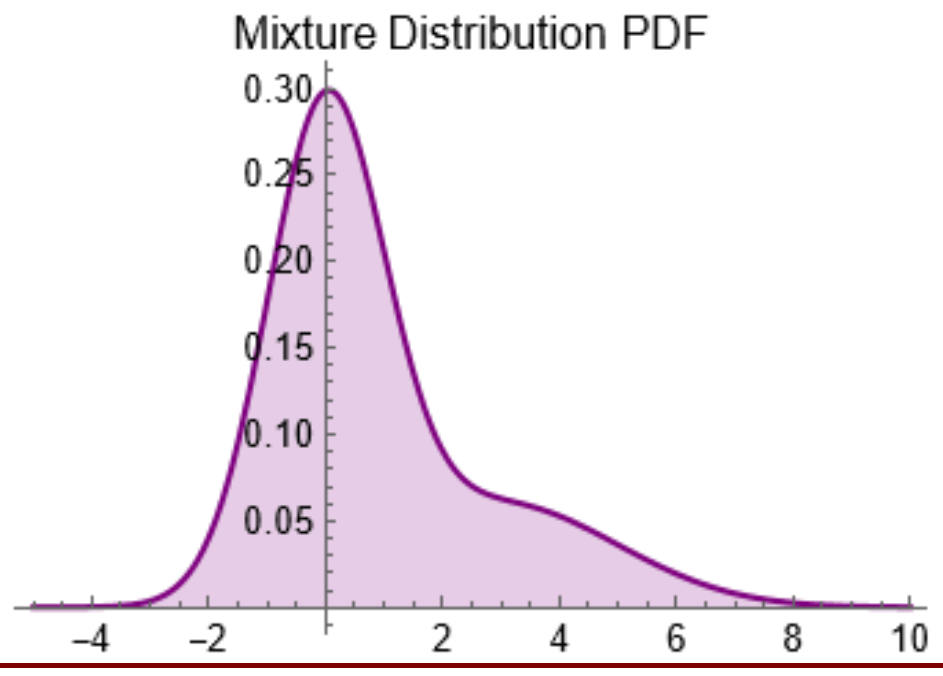

***Mathematica Examples 13.132***

Input

```
(* Two plots are created to visualize the individual component distributions (dist1 and dist2) separately. These plots are stored in dist1Plot and dist2Plot. The GraphicsGrid function is used to arrange all three plots (dist1Plot, dist2Plot, and mixturePlot) in a grid layout, allowing you to see the individual component distributions and the mixture distribution side by side. As you adjust the weights using the sliders, all three plots will update dynamically, providing a visual representation of how the weights control the contribution of each distribution to the overall mixture distribution: *)
(*Define the component distributions*)
dist1=NormalDistribution[-1,1/2];
dist2=NormalDistribution[1,1/2];
(*Define the initial weights*)
initialWeight1=0.7;
initialWeight2=0.3;
(*Define the Manipulate*)
Manipulate[
 (*Update the mixture distribution based on the current weights*)
 mixtureDist=MixtureDistribution[
   {weight1,weight2},
   {dist1,dist2}
   ];
 (*Plot the PDF of the component distributions*)
 dist1Plot=Plot[
   PDF[dist1,x],
   {x,-5,5},
   PlotRange->All,
   PlotLabel->"Component Distribution 1",
   Filling->Axis,
   ImageSize->300,
   PlotStyle->Purple
   ];
 dist2Plot=Plot[
   PDF[dist2,x],
   {x,-5,5},
   PlotRange->All,
   PlotLabel->"Component Distribution 2",
   Filling->Axis,
   ImageSize->300,
   PlotStyle->Purple
   ];
 (*Plot the PDF of the mixture distribution*)
 mixturePlot=Plot[
   PDF[mixtureDist,x],
   {x,-5,5},
   PlotRange->All,
   PlotLabel->"Mixture Distribution PDF",
   Filling->Axis,
```

```
      ImageSize->300,
      PlotStyle->Purple
      ];
    (*Display the plots*)
    GraphicsGrid[
     {{dist1Plot,dist2Plot},{mixturePlot}}
     ],
    (*Controls for the weights*)
    {{weight1,initialWeight1,"Weight 1"},0,1,0.1},
    {{weight2,initialWeight2,"Weight 2"},0,1,0.1}
    ]
```

Output

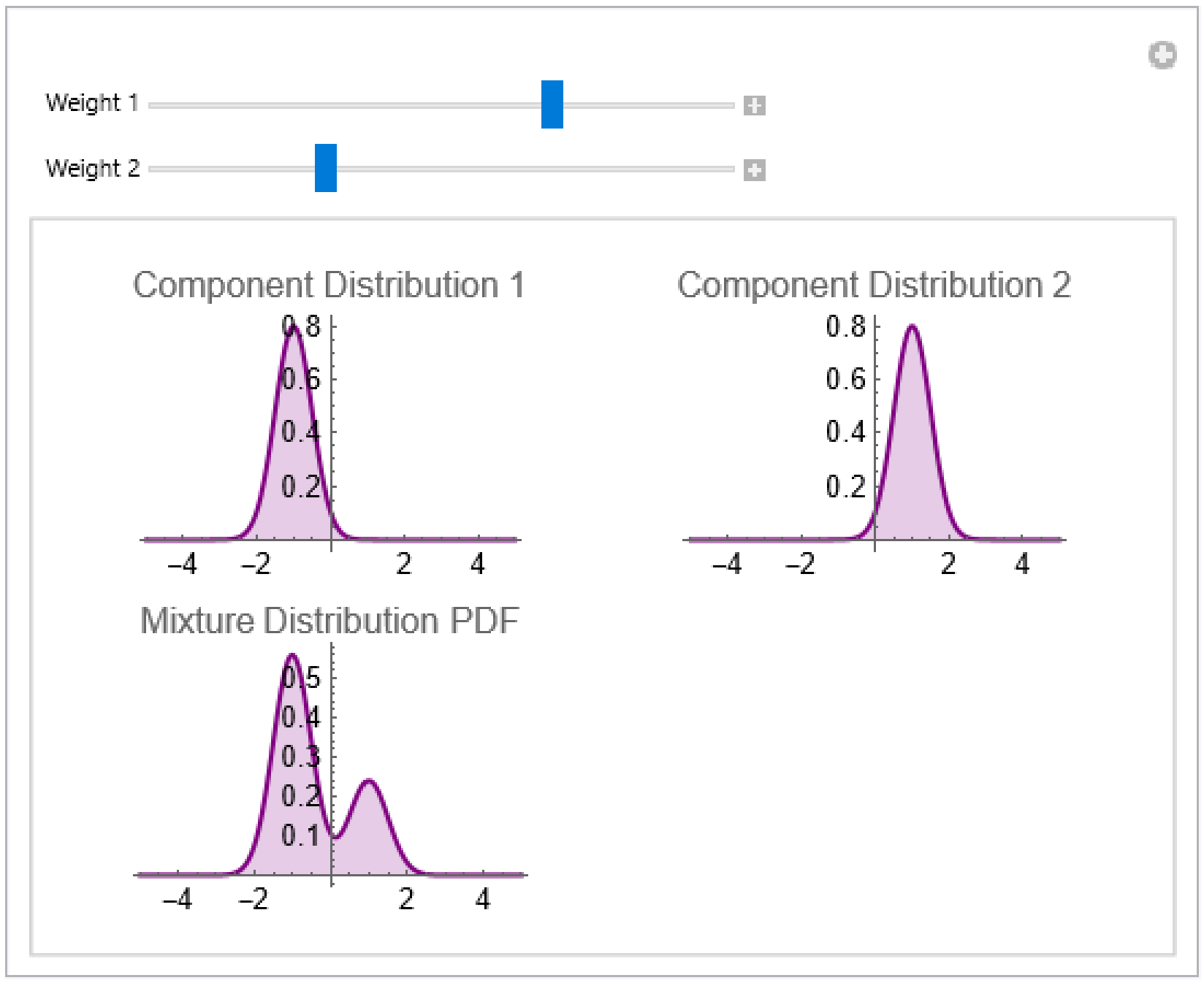

### *Mathematica Examples 13.133*

Input

```
(* Several univariate continuous distributions: *)
mixdist=MixtureDistribution[
    {1,5,1},
    {NormalDistribution[-4,3/4],
     NormalDistribution[0,1],
     NormalDistribution[4,3/4]}
    ];

(* Curve Plot of a mixture distribution: *)
Plot[
 PDF[mixdist,x],
 {x,-7,7},
 Filling->Axis,
 ImageSize->200,
 PlotStyle->Purple
 ]

(* Mixture distributions can be used like any other distribution. Statistical Analysis
of a mixture distribution: *)
mean=N[Mean[mixdist]];
variance=N[Variance[mixdist]];
skewness=N[Skewness[mixdist]];
kurtosis=N[Kurtosis[mixdist]];
```

```
        Grid[
         {{"    Mean:",mean},
          {"Variance:",variance},
          {"Skewness:",skewness},
          {"Kurtosis:",kurtosis}}
         ]

        (* Generating functions: *)
        MomentGeneratingFunction[mixdist,t]
Output
```

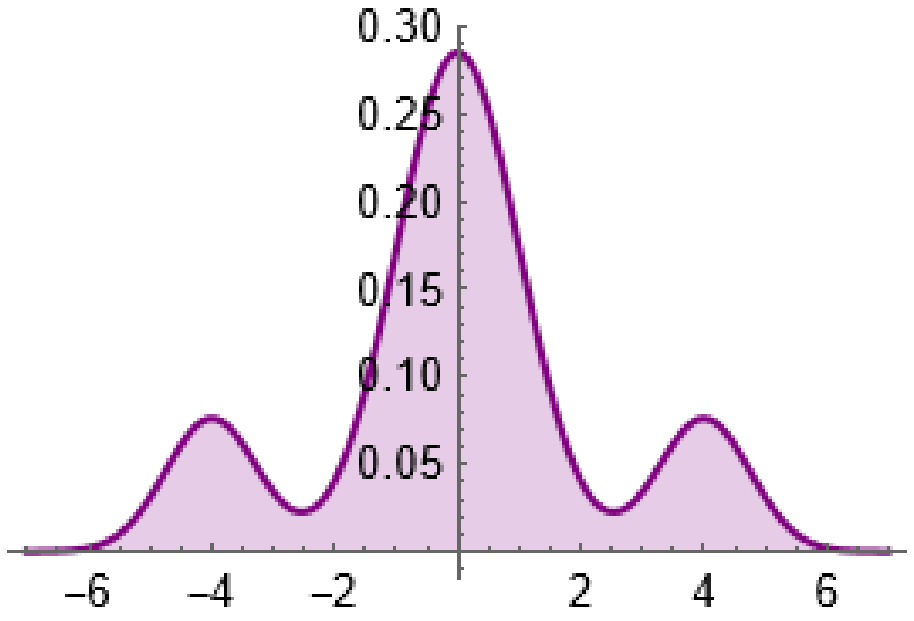

```
Output  {
         {     Mean:,  0.},
         {Variance:,  5.44643},
         {Skewness:,  0.},
         {Kurtosis:,  3.06725}
        }
Output  1/7 (5 E^(t^2/2) + E^(-4 t + (9 t^2)/32) + E^(4 t + (9 t^2)/32))
```

***Mathematica Examples 13.134***

```
Input   (* This code defines a Manipulate function that allows you to interactively change
        the weights (w1 and w2), means (μ1 and μ2),and standard deviations (σ1 and σ2) of a
        MixtureDistribution. The function displays a plot of the PDF and CDF of the
        distribution for the chosen parameter values: *)

        (*Define the Manipulate function*)
        Manipulate[
         Module[
          {dist},
          (*Create a MixtureDistribution with user-defined parameters*)
          dist=MixtureDistribution[
             {w1,w2},
             {
              NormalDistribution[μ1,σ1],
              NormalDistribution[μ2,σ2]
              }
             ];

          (*Generate random data from the MixtureDistribution*)
          data=RandomVariate[dist,500];

          (*Generate a plot of the PDF and CDF*)
          Column[
           {
            Plot[
              {PDF[dist,x],CDF[dist,x]},
              {x,-10,10},
              PlotRange->All,
              Filling->{1->Axis,2->Axis},
              PlotStyle->{Blue,Purple},
```

```
          Frame->True,
          FrameLabel->{"x","Probability"},
          PlotLegends->{"PDF","CDF"},
          ImageSize->300
          ],
         (*Generate a plot of the data*)
         ListPlot[
          data,
          PlotRange->All,
          PlotStyle->Directive[PointSize[0.01],Purple],
          ImageSize->300
          ]
         }
        ]
       ],
      (*Define the parameters and their ranges*)
      {{w1,0.5,"Weight 1"},0,1,0.1},
      {{w2,0.5,"Weight 2"},0,1,0.1},
      {{μ1,0,"Mean 1"},-5,5,0.1},
      {{μ2,0,"Mean 2"},-5,5,0.1},
      {{σ1,1,"Standard Deviation 1"},0.1,5,0.1},
      {{σ2,1,"Standard Deviation 2"},0.1,5,0.1}
      ]
```

Output

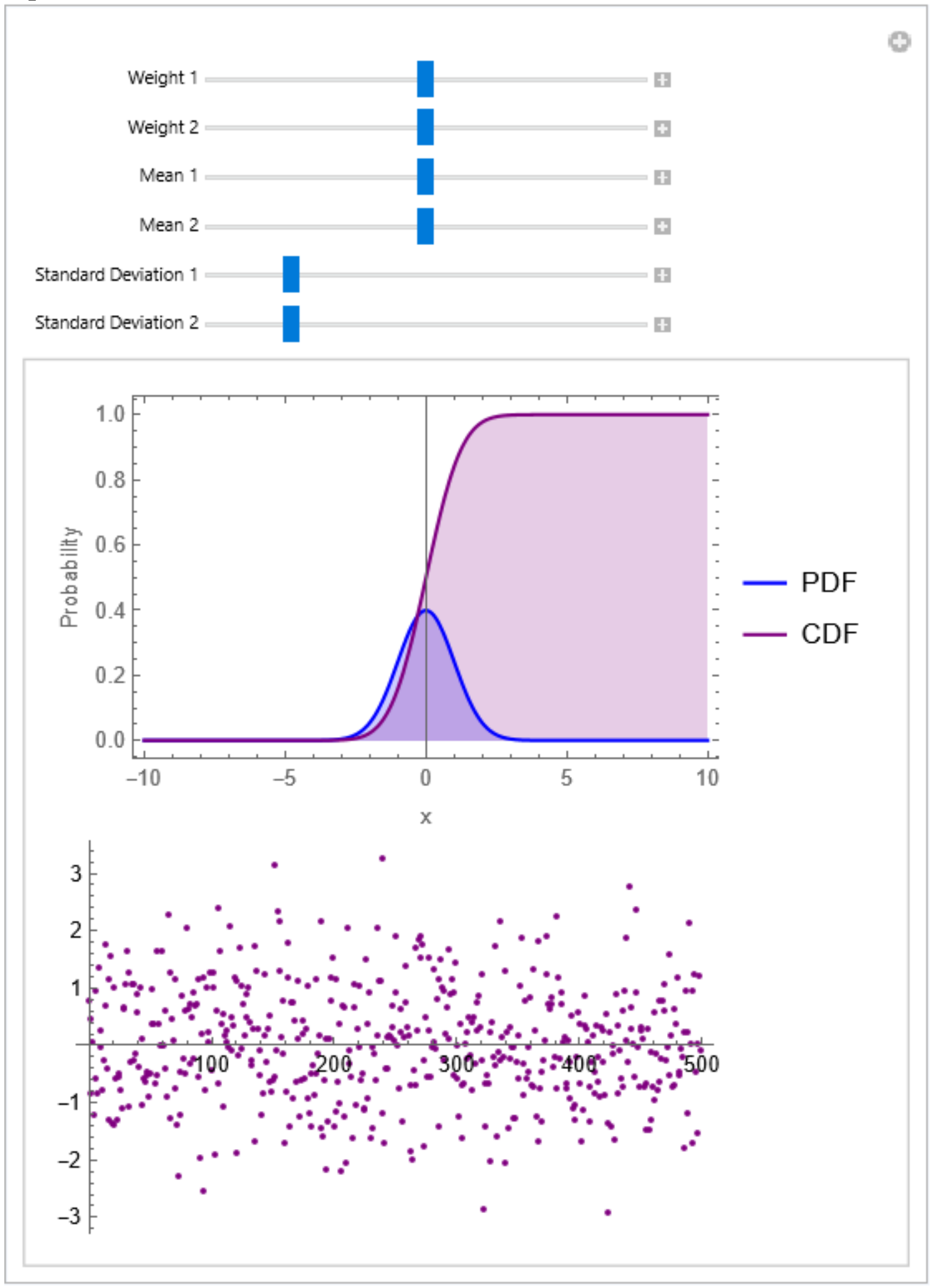

***Mathematica Examples 13.135***

Input

```
(*In this example, we create a MixtureDistribution with weights w1 and w2,and two
component distributions: a GammaDistribution and an ExponentialDistribution. The
Manipulate function allows you to change the weights and parameters of these
distributions. It also generates random data points from the MixtureDistribution
using the RandomVariate function and displays a histogram of the data along with the
PDF of the MixtureDistribution: *)

(* Define the Manipulate function: *)
Manipulate[
 Module[
  {dist,data,plot},
  (* Create a MixtureDistribution with user-defined parameters: *)
  dist=MixtureDistribution[
    {w1,w2},
    {
     GammaDistribution[a1,b1],
     ExponentialDistribution[λ2]
     }
    ];
  (* Generate random data from the MixtureDistribution: *)
  data=RandomVariate[dist,n];
  (*Generate a plot of the data and PDF*)
  plot=Show[
    Histogram[
     data,
     Automatic,
     "PDF",
     ChartStyle->Directive[Purple,Opacity[0.7]]
     ],
    Plot[
     PDF[dist,x],
     {x,0,200},
     PlotRange->All,
     PlotStyle->Red
     ]
    ];
  (* Display the plot: *)
  Show[
   plot,
   ImageSize->300
   ]
  ],

 (* Define the parameters and their ranges: *)
 {{w1,0.5,"Weight 1"},0,1,0.1},
 {{w2,0.5,"Weight 2"},0,1,0.1},
 {{a1,2,"Shape parameter 1"},0.1,10,0.1},
 {{b1,1,"Scale parameter 1"},0.1,10,0.1},
 {{λ2,1,"Rate parameter 2"},0.1,10,0.1},
 {{n,100,"Number of data points"},10,1000,10}
 ]
```

Output

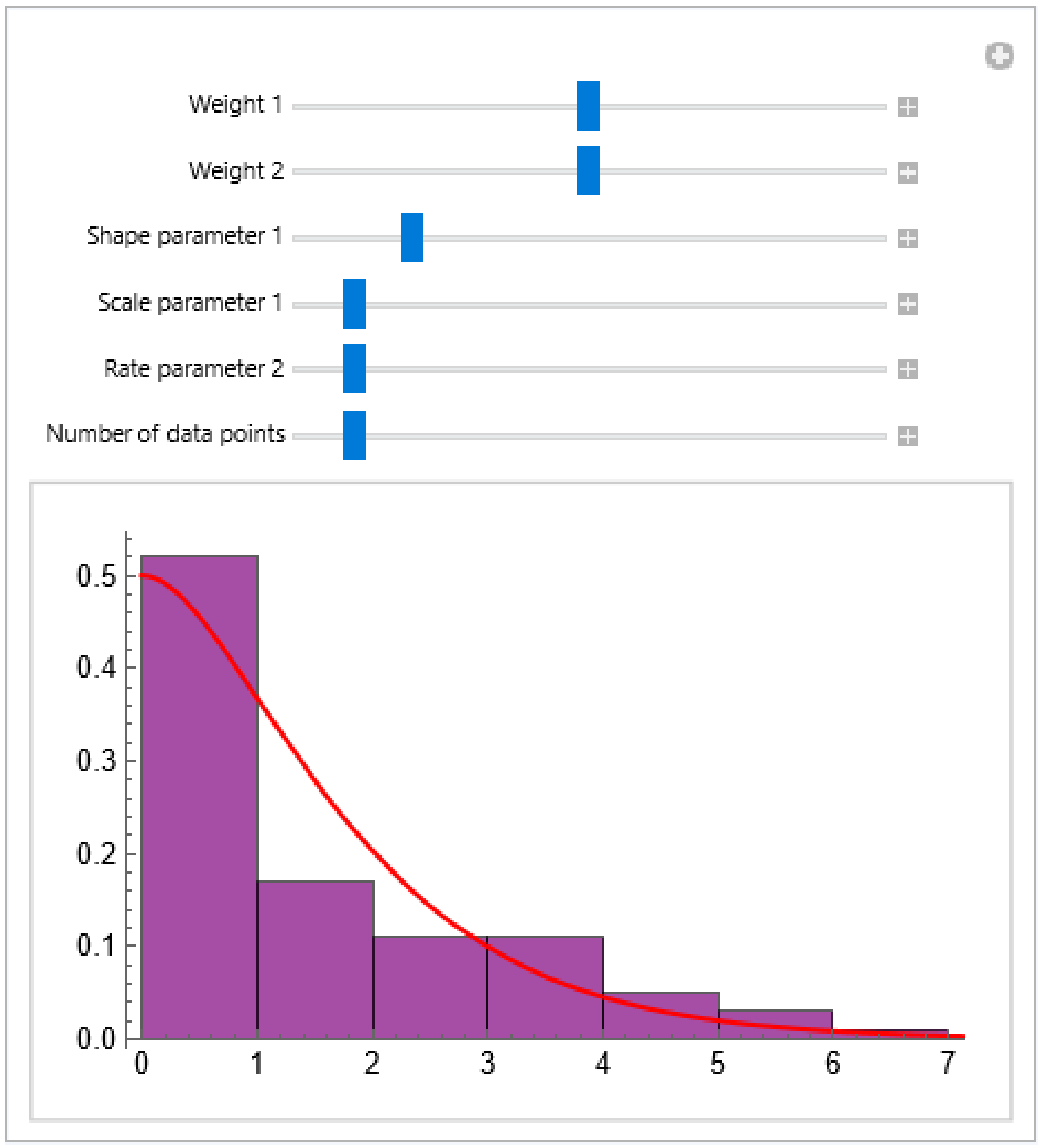

# CHAPTER 14

# BIVARIATE RANDOM VARIABLES AND DISTRIBUTIONS

In probability theory and statistics, the study of RVs is essential for understanding the uncertainty and variability in various phenomena. While the analysis of single RVs provides valuable insights, many real-world phenomena involve multiple variables that are interrelated. To explore such scenarios, this chapter focuses on bivariate RVs and their distributions. Bivariate RVs find applications in various fields, including finance, risk management, econometrics, environmental studies, quality control, biostatistics, and social sciences.

In this chapter, we will investigate the following topics.

- Joint distribution functions:
  The joint distribution function serves as a fundamental tool for analyzing bivariate RVs. It captures the probabilities associated with different combinations of values taken by two RVs.
- Discrete bivariate RVs:
  Discrete bivariate RVs deal with situations where both variables can only take on a countable set of values. We investigate the joint PMF of discrete bivariate RVs, examining concepts such as marginal PMFs, independence, and conditional probabilities.
- Continuous bivariate RVs:
  In contrast to discrete bivariate RVs, continuous bivariate RVs involve variables that can assume any value within a specified range. We study the joint PDF associated with continuous bivariate RVs, exploring concepts such as joint CDFs, marginal PDFs, and independence.
- Covariance and correlation:
  To measure the degree of association between two RVs, we need to focus on the concepts of covariance and correlation. Covariance quantifies the linear relationship between variables, while correlation provides a standardized measure of both linear and non-linear relationships. We explore the covariance matrix and its role in characterizing bivariate RVs.
- Independence:
  Independence between RVs is a crucial concept. When two RVs are independent, their joint behavior can be analyzed by examining the properties of each variable separately. It allows for simpler calculations and modeling assumptions.
- Joint moments:
  Moments provide valuable insights into the distribution of RVs. Joint moments offer additional information about the joint behavior of RVs. By studying moments such as means, variances, and higher-order moments, we gain a deeper understanding of the joint characteristics and properties of the variables.

Finally, we conclude the chapter by studying two joint distributions:

- Multinomial distribution:
  Multinomial distribution extends the notion of binomial distribution to scenarios where we have more than two outcomes. We investigate the properties and applications of multinomial distribution, highlighting its relevance in analyzing experiments involving multiple categorical variables.
- Bivariate normal distribution:
  Bivariate normal distribution is a fundamental distribution for understanding the behavior of correlated continuous RVs. It is a generalization of the univariate normal distribution to two dimensions.

## 14.1 Bivariate RV

The bivariate RV represents the outcomes of two random phenomena simultaneously. Let $X$ and $Y$ be two RVs defined on the same probability space. A bivariate RV $(X,Y)$ represents a pair of outcomes $(x,y)$, where $x$ is a possible outcome of $X$ and $y$ is a possible outcome of $Y$. They are particularly useful when analyzing data sets with two variables that may be related or dependent on each other. The behavior of a bivariate RV is typically described by its joint probability distribution, which provides information about the probability of observing specific values or ranges of values for both $X$ and $Y$ simultaneously.

If the RVs $X$ and $Y$ are each, by themselves, discrete RVs, then $(X,Y)$ is called a discrete bivariate RV. Similarly, if $X$ and $Y$ are each, by themselves, continuous RVs, then $(X,Y)$ is called a continuous bivariate RV. If one of $X$ and $Y$ is discrete while the other is continuous, then $(X,Y)$ is called a mixed bivariate RV.

Some examples of bivariate RVs:

- Height and weight:
  Suppose we have a dataset of individuals where we measure their height $X$ and weight $Y$. The pair $(X,Y)$ represents a bivariate RV and we can study their joint behavior.
- Temperature and humidity:
  Consider a weather station that measures the temperature $X$ and humidity $Y$ at a particular location. The bivariate RV $(X,Y)$ represents the temperature and humidity pairs observed at different times. We can analyze their relationship and study how changes in temperature relate to changes in humidity.
- Stock prices:
  Suppose we have two stocks, $A$ and $B$, and we are interested in their daily price movements. Let $X$ represent the daily price change of stock $A$, and $Y$ represent the daily price change of stock $B$. The bivariate RV $(X,Y)$ captures the joint behavior of the two stocks, and we can analyze their correlation and dependency.
- Age and income:
  In a survey of a population, we measure the age $X$ and income $Y$ of individuals. The bivariate RV $(X,Y)$ captures the joint distribution of age and income. This can be used to study various aspects, such as income patterns across different age groups.
- Exam scores:
  Consider a classroom where students take two exams, $X$ and $Y$. Let $X$ represent the score on the first exam and $Y$ represent the score on the second exam. The bivariate RV $(X,Y)$ represents the pair of scores for each student. We can analyze the relationship between the two exam scores and assess whether performance on one exam predicts performance on the other.
- Time spent studying and exam score:
  Suppose we have data on the time spent studying $X$ and the corresponding exam score $Y$ for a group of students. The bivariate RV $(X,Y)$ captures the study time and exam score pairs.
- Number of heads and tails in two-coin tosses:
  Let $X$ be the number of heads and $Y$ be the number of tails obtained when tossing two fair coins. The possible values of the bivariate RV $(X,Y)$ are $(0,2)$, $(1,1)$, and $(2,0)$.
- Sum and product of two dice rolls:
  Let $X$ be the sum and $Y$ be the product of two fair dice rolls. The possible values of the bivariate RV $(X,Y)$ are $(2,1)$, $(3,2)$, $\ldots$, $(12,36)$.
- Number of rainy and sunny days in a week:
  Let $X$ be the number of rainy days and $Y$ be the number of sunny days in a week. The possible values of $(X,Y)$ range from $(0,7)$ to $(7,0)$, depending on the weather patterns.

***Example 14.1***

Let us consider an example involving a bivariate RV related to flipping a coin. We will define two RVs: $X$, representing the number of heads obtained in two-coin flips, and $Y$, representing the number of tails.

***Solution***

The possible outcomes and their probabilities are as follows: The sample space $S$ of the experiment is

$$S = \{HH, HT, TH, TT\},$$
$$X = \{0, 1, 2\},$$
$$Y = \{0, 1, 2\}.$$

Since the coin is fair, we have

$$HH \equiv (X = 2, Y = 0): \quad P(X = 2, Y = 0) = P(HH) = 1/4 = 0.25.$$
$$TT \equiv (X = 0, Y = 2): \quad P(X = 0, Y = 2) = P(TT) = 1/4 = 0.25.$$
$$\{HT, TH\} \equiv (X = 1, Y = 1): \quad P(X = 1, Y = 1) = P(TH) + P(HT) = 2/4 = 0.5.$$

The range of a bivariate RV ($X$,$Y$) is {(2, 0), (1, 1), (0,2)}.

***Example 14.2***

When rolling a pair of fair six-sided dice, we can define a bivariate RV that represents the outcomes of the two dice. Let us denote the RV for the first die as $X$ and for the second die as $Y$. Each die can take on values from 1 to 6.

***Solution***

The sample space $S$ of the experiment can be written as the following.

| $Y =$ | 1 | 2 | 3 | 4 | 5 | 6 |
|---|---|---|---|---|---|---|
| $X = 1$ | (1, 1) | (1, 2) | (1, 3) | (1, 4) | (1, 5) | (1, 6) |
| $X = 2$ | (2, 1) | (2, 2) | (2, 3) | (2, 4) | (2, 5) | (2, 6) |
| $X = 3$ | (3, 1) | (3, 2) | (3, 3) | (3, 4) | (3, 5) | (3, 6) |
| $X = 4$ | (4, 1) | (4, 2) | (4, 3) | (4, 4) | (4, 5) | (4, 6) |
| $X = 5$ | (5, 1) | (5, 2) | (5, 3) | (5, 4) | (5, 5) | (5, 6) |
| $X = 6$ | (6, 1) | (6, 2) | (6, 3) | (6, 4) | (6, 5) | (6, 6) |

The probabilities associated with each pair of outcomes $(X, Y)$ can be written as the following.

| $Y =$ | 1 | 2 | 3 | 4 | 5 | 6 | $P(X)$ |
|---|---|---|---|---|---|---|---|
| $X = 1$ | 1/36 | 1/36 | 1/36 | 1/36 | 1/36 | 1/36 | 6/36 |
| $X = 2$ | 1/36 | 1/36 | 1/36 | 1/36 | 1/36 | 1/36 | 6/36 |
| $X = 3$ | 1/36 | 1/36 | 1/36 | 1/36 | 1/36 | 1/36 | 6/36 |
| $X = 4$ | 1/36 | 1/36 | 1/36 | 1/36 | 1/36 | 1/36 | 6/36 |
| $X = 5$ | 1/36 | 1/36 | 1/36 | 1/36 | 1/36 | 1/36 | 6/36 |
| $X = 6$ | 1/36 | 1/36 | 1/36 | 1/36 | 1/36 | 1/36 | 6/36 |
| $P(Y)$ | 6/36 | 6/36 | 6/36 | 6/36 | 6/36 | 6/36 | 1 |

Each cell in the table represents a specific outcome (combination) of $X$ and $Y$, and the corresponding probability is listed in that cell. For example, $P(X = 1, Y = 1)$ represents the probability of getting a 1 on the first die and a 1 on the second die. Since we assume that the dice are fair, we have equal probabilities for all possible outcomes.

***Example 14.3***

Let us consider an example involving a bivariate RV related to rolling a pair of fair six-sided dice. We will define two RVs: $X$, representing the sum of the two dice, and $Y$, representing the difference between the numbers on the first and second dice.

***Solution***

The possible values for $X$ range from 2 (when both dice show a 1) to 12 (when both dice show a 6). The probability distribution of $X$ is determined by calculating the probabilities of all possible sums.

$$P(X=2)=P\{(1,1)\}=1/36$$
$$P(X=3)=P\{(1,2),(2,1)\}=2/36$$
$$P(X=4)=P\{(1,3),(2,2),(3,1)\}=3/36$$
$$P(X=5)=P\{(1,4),(2,3),(3,2),(4,1)\}=4/36$$
$$P(X=6)=P\{(1,5),(2,4),(3,3),(4,2),(5,1)\}=5/36$$
$$P(X=7)=P\{(1,6),(2,5),(3,4),(4,3),(5,2),(6,1)\}=6/36$$
$$P(X=8)=P\{(2,6),(3,5),(4,4),(5,3),(6,2)\}=5/36$$
$$P(X=9)=P\{(3,6),(4,5),(5,4),(6,3)\}=4/36$$
$$P(X=10)=P\{(4,6),(5,5),(6,4)\}=3/36$$
$$P(X=11)=P\{(5,6),(6,5)\}=2/36$$
$$P(X=12)=P\{(6,6)\}=1/36$$

The possible values for $Y$ range from $-5$ to 5.

$$P(-5)=P\{(1,6)\}=1/36$$
$$P(-4)=P\{(1,5),(2,6)\}=2/36$$
$$P(-3)=P\{(1,4),(2,5),(3,6)\}=3/36$$
$$P(-2)=P\{(1,3),(2,4),(3,5),(4,6)\}=4/36$$
$$P(-1)=P\{(1,2),(2,3),(3,4),(4,5),(5,6)\}=5/36$$
$$P(0)=P\{(1,1),(2,2),(3,3),(4,4),(5,5),(6,6)\}=6/36$$
$$P(1)=P\{(2,1),(3,2),(4,3),(5,4),(6,5)\}=5/36$$
$$P(2)=P\{(3,1),(4,2),(5,3),(6,4)\}=4/36$$
$$P(3)=P\{(4,1),(5,2),(6,3)\}=3/36$$
$$P(4)=P\{(5,1),(6,2)\}=2/36$$
$$P(5)=P\{(6,1)\}=1/36$$

The sample space $S$ of the experiment can be written as the following.

| $Y=$ | $-5$ | $-4$ | $-3$ | $-2$ | $-1$ | $0$ | $1$ | $2$ | $3$ | $4$ | $5$ |
|---|---|---|---|---|---|---|---|---|---|---|---|
| $X=2$ | | | | | | $(1,1)$ | | | | | |
| $X=3$ | | | | | $(1,2)$ | | $(2,1)$ | | | | |
| $X=4$ | | | | $(1,3)$ | | $(2,2)$ | | $(3,1)$ | | | |
| $X=5$ | | | $(1,4)$ | | $(2,3)$ | | $(3,2)$ | | $(4,1)$ | | |
| $X=6$ | | $(1,5)$ | | $(2,4)$ | | $(3,3)$ | | $(4,2)$ | | $(5,1)$ | |
| $X=7$ | $(1,6)$ | | $(2,5)$ | | $(3,4)$ | | $(4,3)$ | | $(5,2)$ | | $(6,1)$ |
| $X=8$ | | $(2,6)$ | | $(3,5)$ | | $(4,4)$ | | $(5,3)$ | | $(6,2)$ | |
| $X=9$ | | | $(3,6)$ | | $(4,5)$ | | $(5,4)$ | | $(6,3)$ | | |
| $X=10$ | | | | $(4,6)$ | | $(5,5)$ | | $(6,4)$ | | | |
| $X=11$ | | | | | $(5,6)$ | | $(6,5)$ | | | | |
| $X=12$ | | | | | | $(6,6)$ | | | | | |

The probabilities associated with each pair of outcomes $(X,Y)$ can be written as the following.

| $Y=$ | $-5$ | $-4$ | $-3$ | $-2$ | $-1$ | $0$ | $1$ | $2$ | $3$ | $4$ | $5$ | $P(X)$ |
|---|---|---|---|---|---|---|---|---|---|---|---|---|
| $X=2$ | | | | | | $1/36$ | | | | | | $1/36$ |
| $X=3$ | | | | | $1/36$ | | $1/36$ | | | | | $2/36$ |
| $X=4$ | | | | $1/36$ | | $1/36$ | | $1/36$ | | | | $3/36$ |
| $X=5$ | | | $1/36$ | | $1/36$ | | $1/36$ | | $1/36$ | | | $4/36$ |
| $X=6$ | | $1/36$ | | $1/36$ | | $1/36$ | | $1/36$ | | $1/36$ | | $5/36$ |
| $X=7$ | $1/36$ | | $1/36$ | | $1/36$ | | $1/36$ | | $1/36$ | | $1/36$ | $6/36$ |
| $X=8$ | | $1/36$ | | $1/36$ | | $1/36$ | | $1/36$ | | $1/36$ | | $5/36$ |
| $X=9$ | | | $1/36$ | | $1/36$ | | $1/36$ | | $1/36$ | | | $4/36$ |
| $X=10$ | | | | $1/36$ | | $1/36$ | | $1/36$ | | | | $3/36$ |
| $X=11$ | | | | | $1/36$ | | $1/36$ | | | | | $2/36$ |
| $X=12$ | | | | | | $1/36$ | | | | | | $1/36$ |
| $P(Y)$ | $1/36$ | $2/36$ | $3/36$ | $4/36$ | $5/36$ | $6/36$ | $5/36$ | $4/36$ | $3/36$ | $2/36$ | $1/36$ | 1 |

***Example 14.4***

In Example 14.3, let $Y$ be the absolute value of the difference between the numbers on the two dice. In this case, the possible values for $X$ range from 2 to 12. However, the possible values for $Y$ range from 0 to 5.

***Solution***

The sample space $S$ of the experiment can be written as the following.

| $Y =$ | 0 | 1 | 2 | 3 | 4 | 5 |
|---|---|---|---|---|---|---|
| $X = 2$ | $(1,1)$ | | | | | |
| $X = 3$ | | $(2,1),(1,2)$ | | | | |
| $X = 4$ | $(2,2)$ | | $(3,1),(1,3)$ | | | |
| $X = 5$ | | $(3,2),(2,3)$ | | $(4,1),(1,4)$ | | |
| $X = 6$ | $(3,3)$ | | $(4,2),(2,4)$ | | $(5,1),(1,5)$ | |
| $X = 7$ | | $(4,3),(3,4)$ | | $(5,2),(2,5)$ | | $(6,1),(1,6)$ |
| $X = 8$ | $(4,4)$ | | $(5,3),(3,5)$ | | $(6,2),(2,6)$ | |
| $X = 9$ | | $(5,4),(4,5)$ | | $(6,3),(3,6)$ | | |
| $X = 10$ | $(5,5)$ | | $(6,4),(4,6)$ | | | |
| $X = 11$ | | $(6,5),(5,6)$ | | | | |
| $X = 12$ | $(6,6)$ | | | | | |

The probabilities associated with each pair of outcomes $(X,Y)$ can be written as the following.

| $Y =$ | 0 | 1 | 2 | 3 | 4 | 5 | $P(X)$ |
|---|---|---|---|---|---|---|---|
| $X = 2$ | 1/36 | | | | | | 1/36 |
| $X = 3$ | | 2/36 | | | | | 2/36 |
| $X = 4$ | 1/36 | | 2/36 | | | | 3/36 |
| $X = 5$ | | 2/36 | | 2/36 | | | 4/36 |
| $X = 6$ | 1/36 | | 2/36 | | 2/36 | | 5/36 |
| $X = 7$ | | 2/36 | | 2/36 | | 2/36 | 6/36 |
| $X = 8$ | 1/36 | | 2/36 | | 2/36 | | 5/36 |
| $X = 9$ | | 2/36 | | 2/36 | | | 4/36 |
| $X = 10$ | 1/36 | | 2/36 | | | | 3/36 |
| $X = 11$ | | 2/36 | | | | | 2/36 |
| $X = 12$ | 1/36 | | | | | | 1/36 |
| $P(Y)$ | 6/36 | 10/36 | 8/36 | 6/36 | 4/36 | 2/36 | 1 |

## 14.2 Joint Distribution Functions

**Definition (Joint CDF):** The joint CDF of $X$ and $Y$, denoted by $F_{XY}(x,y)$, is the function defined by

$$F_{XY}(x,y) = P(X \le x, Y \le y). \tag{14.1}$$

**Definition (Independent RVs):** Two RVs $X$ and $Y$ will be called independent if

$$F_{XY}(x,y) = F_X(x)F_Y(y), \tag{14.2}$$

for every value of $x$ and $y$.

**Definition (Properties of $F_{XY}(x,y)$):** The joint CDF of two RVs has many properties analogous to those of the CDF of a single RV.

$$0 \le F_{XY}(x,y) \le 1. \tag{14.3}$$

If $x_1 \le x_2$ and $y_1 \le y_2$, then

$$F_{XY}(x_1,y_1) \le F_{XY}(x_2,y_1) \le F_{XY}(x_2,y_2), \tag{14.4.1}$$
$$F_{XY}(x_1,y_1) \le F_{XY}(x_1,y_2) \le F_{XY}(x_2,y_2). \tag{14.4.2}$$

$$\lim_{\substack{x\to\infty \\ y\to\infty}} F_{XY}(x,y) = F_{XY}(\infty,\infty) = 1. \tag{14.5}$$

$$\lim_{x\to-\infty} F_{XY}(x,y) = F_{XY}(-\infty,y) = 0, \tag{14.6.1}$$
$$\lim_{y\to-\infty} F_{XY}(x,y) = F_{XY}(x,-\infty) = 0. \tag{14.6.2}$$

$$\lim_{x\to a^+} F_{XY}(x,y) = F_{XY}(a^+,y) = F_{XY}(a,y), \tag{14.7.1}$$
$$\lim_{y\to b^+} F_{XY}(x,y) = F_{XY}(x,b^+) = F_{XY}(x,b). \tag{14.7.2}$$

$$P(x_1 \le X \le x_2, Y \le y) = F_{XY}(x_2,y) - F_{XY}(x_1,y), \tag{14.8.1}$$
$$P(X \le x, y_1 \le Y \le y_2) = F_{XY}(x,y_2) - F_{XY}(x,y_1). \tag{14.8.2}$$

If $x_1 \le x_2$ and $y_1 \le y_2$, then

$$F_{XY}(x_2,y_2) - F_{XY}(x_1,y_2) - F_{XY}(x_2,y_1) + F_{XY}(x_1,y_1) \ge 0. \tag{14.9}$$

**Proof:** (14.9)

$$(x_1 \le X \le x_2, Y \le y_2) = (x_1 \le X \le x_2, Y \le y_1) \cup (x_1 \le X \le x_2, y_1 \le Y \le y_2).$$

The two events on the right-hand side are disjoint; hence

$$P(x_1 \le X \le x_2, Y \le y_2) = P(x_1 \le X \le x_2, Y \le y_1) + P(x_1 \le X \le x_2, y_1 \le Y \le y_2).$$

Then using (14.8), we obtain

$$\begin{aligned} P(x_1 \le X \le x_2, y_1 \le Y \le y_2) &= P(x_1 \le X \le x_2, Y \le y_2) - P(x_1 \le X \le x_2, Y \le y_1) \\ &= F_{XY}(x_2,y_2) - F_{XY}(x_1,y_2) - [F_{XY}(x_2,y_1) - F_{XY}(x_1,y_1)] \\ &= F_{XY}(x_2,y_2) - F_{XY}(x_1,y_2) - F_{XY}(x_2,y_1) + F_{XY}(x_1,y_1). \end{aligned}$$

Since the probability must be nonnegative, we conclude that

$$F_{XY}(x_2,y_2) - F_{XY}(x_1,y_2) - F_{XY}(x_2,y_1) + F_{XY}(x_1,y_1) \ge 0,$$

if $x_1 \le x_2$ and $y_1 \le y_2$.

■

**Theorem 14.1:** Consider a bivariate RV $(X,Y)$. If $X$ and $Y$ are independent, then every event of the form $(a < X < b)$ is independent of every event of the form $(c < Y < d)$.

**Proof:**

By definition, if $X$ and $Y$ are independent, we have

$$F_{XY}(x,y) = F_X(x)F_Y(y).$$

Using

$$P(x_1 \le X \le x_2, y_1 \le Y \le y_2) = F_{XY}(x_2,y_2) - F_{XY}(x_1,y_2) - F_{XY}(x_2,y_1) + F_{XY}(x_1,y_1),$$

such that $x_1 = a$, $x_2 = b$, $y_1 = c$, and $y_2 = d$, we get

$$\begin{aligned} P(a \le X \le b, c \le Y \le d) &= F_{XY}(b,d) - F_{XY}(a,d) - F_{XY}(b,c) + F_{XY}(a,c) \\ &= F_X(b)F_Y(d) - F_X(a)F_Y(d) - F_X(b)F_Y(c) + F_X(a)F_Y(c) \\ &= [F_X(b)F_Y(d) - F_X(a)F_Y(d)] - [F_X(b)F_Y(c) - F_X(a)F_Y(c)] \\ &= [F_X(b) - F_X(a)]F_Y(d) - [F_X(b) - F_X(a)]F_Y(c) \\ &= [F_X(b) - F_X(a)][F_Y(d) - F_Y(c)] \\ &= P(a \le X \le b)P(c \le Y \le d), \end{aligned}$$

which indicates that event $(a \le X \le b)$ and event $(c \le Y \le d)$ are independent.

■

If more than one RV is defined in a random experiment, it is important to distinguish between the joint probability distribution of $X$ and $Y$ (CDF that describes the behavior of these variables together) and the probability distribution of each variable individually (the behavior of each variable individually, regardless of the other). The individual CDF of a RV is referred to as its marginal CDF.

The probability of one event in the presence of all outcomes of the other RV is called the marginal probability. It considers the union of all events for the second variable. It is called the marginal probability because if all outcomes and probabilities for the two variables were laid out together in a table ($X$ as columns, $Y$ as rows), then the marginal probability of one variable $X$ would be the sum of probabilities for the other variable, $Y$ rows, on the margin of the table, see for instance, tables of Examples 14.2, 14.3, and 14.4. The marginal probability distribution of $X$ can be determined from the joint probability distribution.

**Definition (Marginal CDF):** Now

$$\lim_{y\to\infty}(X \le x, Y \le y) = (X \le x, Y \le \infty) = (X \le x),$$

since the condition $y \le \infty$ is always satisfied. Then

$$\lim_{y\to\infty} F_{XY}(x,y) = F_{XY}(x,\infty) = F_X(x), \quad (14.10.1)$$

$$\lim_{x\to\infty} F_{XY}(x,y) = F_{XY}(\infty,y) = F_Y(y). \quad (14.10.2)$$

The CDFs $F_X(x)$ and $F_Y(y)$ when obtained by (14.10.1) and (14.10.2), are referred to as the marginal CDFs of $X$ and $Y$, respectively.

## 14.3 Discrete Bivariate RVs

If $X$ and $Y$ are discrete RVs, the joint probability distribution of $X$ and $Y$ is a description of the set of points $(x_i, y_j)$ in the range of $(X, Y)$ along with the probability of each point. The joint probability distribution of two RVs is sometimes referred to as the bivariate probability distribution or bivariate distribution of the RVs. One way to describe the joint probability distribution of two discrete RVs is through a joint PMF $f(x_i, y_j) = f_{XY}(x_i, y_j) = P(X = x_i, Y = y_j)$.

**Definition (Joint PMF):** Let $(X, Y)$ be a discrete bivariate RV and let $(X, Y)$ take on the values $(x_i, y_j)$ for a certain allowable set of integers $i$ and $j$. Let

$$f_{XY}(x_i, y_j) = P(X = x_i, Y = y_j).$$

The function $f_{XY}(x_i, y_j)$ is called the joint PMF of $(X, Y)$ and satisfies

$$f_{XY}(x_i, y_j) \ge 0, \quad (14.11.1)$$

$$\sum_{x_i}\sum_{y_j} f_{XY}(x_i, y_j) = 1, \quad (14.11.2)$$

$$P[(X,Y) \in A] = \sum\sum_{(x_i,y_j)\in R_A} f_{XY}(x_i, y_j), \quad (14.11.3)$$

where the summation is over the points $(x_i, y_j)$ in the range space $R_A$ corresponding to the event $A$.

**Definition (Joint CDF):** The joint CDF of a discrete bivariate RV $(X, Y)$ is given by

$$F_{XY}(x,y) = \sum_{x_i \le x}\sum_{y_j \le y} f_{XY}(x_i, y_j). \quad (14.12)$$

**Definition (Marginal PMF):** Suppose that discrete RVs $X$ and $Y$ have joint PMF $f_{XY}(x_i, y_j)$. Let $x_1, x_2, \dots, x_i, \dots$ denote the possible values of $X$, and let $y_1, y_2, \dots, y_j, \dots$ denote the possible values of $Y$. The marginal PMFs of $X$ and $Y$ are respectively given by the following:

$$f_X(x_i) = \sum_{y_j} f_{XY}(X = x_i, y_j) \text{ (fix a value of } X \text{ and sum over possible values of } Y\text{)}, \quad (14.13.1)$$

$$f_Y(y_j) = \sum_{x_i} f_{XY}(x_i, Y = y_j) \text{ (fix a value of } Y \text{ and sum over possible values of } X\text{)}. \quad (14.13.2)$$

In some random experiments, knowledge of the values of $X$ does not change any of the probabilities associated with the values for $Y$.

**Definition (Independence):** If $X$ and $Y$ are independent discrete RVs,

$$f_{XY}(x_i, y_j) = f_X(x_i) f_Y(y_j). \quad (14.14)$$

***Example 14.5***

Consider an experiment of drawing randomly three balls from an urn containing two red, three white, and five blue balls. Let $(X,Y)$ be a bivariate where $X$ and $Y$ denote, respectively, the number of red and white balls chosen.

***Solution***

Since, we draw randomly three balls, the possible values for the RV $X$ is $\{0,1,2\}$. However, the possible values for $Y$ is $\{0,1,2,3\}$. Hence, the range of $(X,Y)$ is $\{(0,0),(0,1),(0,2),(0,3),(1,0),(1,1),(1,2),(2,0),(2,1)\}$.

Let us start by listing all possible outcomes of choosing three balls from the urn. Since the total number of balls is 2 red + 3 white + 5 blue = 10 balls, there are a total of $C(10,3) = 120$ possible outcomes. We have

$$P(X = x, Y = y) = \frac{\binom{2}{x}\binom{3}{y}\binom{5}{3-x-y}}{\binom{10}{3}},$$

where $x$ represents the number of red balls chosen (0, 1, or 2) and $y$ represents the number of white balls chosen (0, 1, 2, or 3).

$$P_{XY}(0,0) = \frac{\binom{2}{0}\binom{3}{0}\binom{5}{3}}{\binom{10}{3}} = \frac{10}{120}, \qquad P_{XY}(0,1) = \frac{\binom{2}{0}\binom{3}{1}\binom{5}{2}}{\binom{10}{3}} = \frac{30}{120},$$

$$P_{XY}(0,2) = \frac{\binom{2}{0}\binom{3}{2}\binom{5}{1}}{\binom{10}{3}} = \frac{15}{120}, \qquad P_{XY}(0,3) = \frac{\binom{2}{0}\binom{3}{3}\binom{5}{0}}{\binom{10}{3}} = \frac{1}{120},$$

$$P_{XY}(1,0) = \frac{\binom{2}{1}\binom{3}{0}\binom{5}{2}}{\binom{10}{3}} = \frac{20}{120}, \qquad P_{XY}(1,1) = \frac{\binom{2}{1}\binom{3}{1}\binom{5}{1}}{\binom{10}{3}} = \frac{30}{120}, \qquad P_{XY}(1,2) = \frac{\binom{2}{1}\binom{3}{2}\binom{5}{0}}{\binom{10}{3}} = \frac{6}{120},$$

$$P_{XY}(2,0) = \frac{\binom{2}{2}\binom{3}{0}\binom{5}{1}}{\binom{10}{3}} = \frac{5}{120}, \qquad P_{XY}(2,1) = \frac{\binom{2}{2}\binom{3}{1}\binom{5}{0}}{\binom{10}{3}} = \frac{3}{120}.$$

The following table represents the joint and marginal PMF of $(X,Y)$.

| | $y = 0$ | $y = 1$ | $y = 2$ | $y = 3$ | $P_X(x_i)$ |
|---|---|---|---|---|---|
| $x = 0$ | $\frac{10}{120}$ | $\frac{30}{120}$ | $\frac{15}{120}$ | $\frac{1}{120}$ | $\frac{56}{120}$ |
| $x = 1$ | $\frac{20}{120}$ | $\frac{30}{120}$ | $\frac{6}{120}$ | 0 | $\frac{56}{120}$ |
| $x = 2$ | $\frac{5}{120}$ | $\frac{3}{120}$ | 0 | 0 | $\frac{8}{120}$ |
| $P_Y(y_j)$ | $\frac{35}{120}$ | $\frac{63}{120}$ | $\frac{21}{120}$ | $\frac{1}{120}$ | 1 |

Since,

$$P_{XY}(0,0) = \frac{10}{120} \neq P_X(0)P_Y(0) = \left(\frac{56}{120}\right)\left(\frac{35}{120}\right),$$

$X$ and $Y$ are not independent.

***Example 14.6***

The joint PMF of a bivariate RV $(X, Y)$ is given by

$$f_{XY}(x_i, y_j) = \begin{cases} 2kx_iy_j, & x_i = 1,2,3;\, y_j = 1,2,3, \\ 0, & \text{otherwise}, \end{cases}$$

where $k$ is a constant.

(a) Find the value of $k$.

(b) Find the marginal PMF of $X$ and $Y$.

(c) Are $X$ and $Y$ independent?

***Solution***

(a)

The joint PMF of a bivariate RV $(X, Y)$ should be satisfy the condition $\sum_{x_i}\sum_{y_j} f_{XY}(x_i, y_j) = 1$. Hence,

$$\begin{aligned}\sum_{x_i}\sum_{y_i} f_{XY}(x_i, y_i) &= \sum_{x_i=1}^{3}\sum_{y_j=1}^{3} k(2x_iy_j) \\ &= 2k\big(1(1) + 1(2) + 1(3) + 2(1) + 2(2) + 2(3) + 3(1) + 3(2) + 3(3)\big) \\ &= 2k(1 + 2 + 3 + 2 + 4 + 6 + 3 + 6 + 9) = 72k = 1.\end{aligned}$$

Thus $k = \frac{1}{72}$.

(b)

The marginal PMF of $X$ is

$$\begin{aligned} f_X(x_i) &= \sum_{y_j} f_{XY}(x_i, y_i) \\ &= \sum_{y_i=1}^{3} \frac{1}{72}(2x_iy_j) \\ &= \frac{1}{36}\big(x_i(1) + x_i(2) + x_i(3)\big) \\ &= \frac{x_i}{6}. \end{aligned}$$

The marginal PMF of $Y$ is

$$\begin{aligned} f_Y(y_j) &= \sum_{x_i} f_{XY}(x_i, y_i) \\ &= \sum_{x_i=1}^{3} \frac{1}{72}(2x_iy_j) \\ &= \frac{1}{36}\big(1(y_j) + 2(y_j) + 3(y_j)\big) \\ &= \frac{y_j}{6}. \end{aligned}$$

(c)

Now

$$f_{XY}(x_i, y_i) = \frac{x_iy_j}{36} = f_X(x_i)f_Y(y_j),$$

hence $X$ and $Y$ are independent.

## 14.4 Continuous Bivariate RVs

The joint probability distribution of two continuous RVs $X$ and $Y$ can be specified by providing a method for calculating the probability that $X$ and $Y$ assume a value in any region $R$ of two-dimensional space. Analogous to the PDF of a single continuous RV, a joint PDF can be defined over two-dimensional space. The double integral of $f_{XY}(x,y)$ over a region $R$ provides the probability that $(X,Y)$ assumes a value in $R$. This integral can be interpreted as the volume under the surface $f_{XY}(x,y)$ over the region $R$. Typically, $f_{XY}(x,y)$ is defined over all of two-dimensional space by assuming that $f_{XY}(x,y) = 0$ for all points for which $f_{XY}(x,y)$ is not specified.

**Definition (Joint PDF):** A joint PDF of a continuous bivariate RV $(X,Y)$, denoted as $f_{XY}(x,y)$, satisfies the following properties:

$$f_{XY}(x,y) \geq 0, \tag{14.15.1}$$

$$\int_{-\infty}^{\infty}\int_{-\infty}^{\infty} f_{XY}(x,y)dxdy = 1, \tag{14.15.2}$$

$f_{XY}(x,y)$ is continuous for all values of $x$ or $y$ except possibly a finite set. (14.15.3)

$$P\big((X,Y) \in A\big) = \iint_{R_A} f_{XY}(x,y)dxdy, \tag{14.15.4}$$

$$\begin{aligned} P(a < X \leq b, c < Y \leq d) &= P(a \leq X \leq b, c \leq Y \leq d) \\ &= P(a \leq X < b, c \leq Y < d) \\ &= P(a < X < b, c < Y < d) \\ &= \int_c^d \int_a^b f_{XY}(x,y)dxdy. \end{aligned} \tag{14.15.5}$$

**Definition (Joint CDF):** The joint CDF of a continuous bivariate RV $(X,Y)$ is given by

$$F_{XY}(x,y) = \int_{-\infty}^{x}\int_{-\infty}^{y} f_{XY}(u,v)dvdu. \tag{14.16}$$

**Definition (Marginal PDF):** If the joint PDF of a continuous bivariate RV $(X,Y)$ is $f_{XY}(x,y)$, the marginal PDFs of $X$ and $Y$ are,

$$f_X(x) = \int_{-\infty}^{\infty} f_{XY}(x,y)dy, \tag{14.17.1}$$

$$f_Y(y) = \int_{-\infty}^{\infty} f_{XY}(x,y)dx. \tag{14.17.2}$$

**Definition (Independence):** If $X$ and $Y$ are independent continuous RVs

$$f_{XY}(x,y) = f_X(x)f_Y(y) \text{ for all } x \text{ and } y. \tag{14.18}$$

***Example 14.7***

The joint PDF of a bivariate RV $(X,Y)$ is given by

$$f_{XY}(x,y) = \begin{cases} kxy, & 0 < x < 1, 0 < y < 1, \\ 0, & \text{otherwise}, \end{cases}$$

where $k$ is a constant.

(a) Determine the value of $k$.

(b) Find the marginal PDFs of $X$ and $Y$.

(c) Find $P(X + Y < 1)$.

***Solution***

(a)

$$\int_{-\infty}^{\infty}\int_{-\infty}^{\infty} f_{XY}(x,y)dxdy = \int_0^1\int_0^1 kxydxdy = k\int_0^1\left(\frac{x^2}{2}\right)_0^1 ydy = k\int_0^1 \frac{y}{2}dy = k\left(\frac{y^2}{4}\right)_0^1 = \frac{k}{4} = 1.$$

Thus $k = 4$.

(b) The marginal PDF of $X$ is

$$f_X(x) = \int_0^1 4xydy = 4x\int_0^1 ydy = 4x\left(\frac{y^2}{2}\right)_0^1 = 2x.$$

Hence,

$$f_X(x) = \begin{cases} 2x, & 0 < x < 1, \\ 0, & \text{otherwise.} \end{cases}$$

The marginal PDF of $Y$ is

$$f_Y(y) = \int_0^1 4xydx = 4y\int_0^1 xdx = 4y\left(\frac{x^2}{2}\right)_0^1 = 2y.$$

Hence,

$$f_Y(y) = \begin{cases} 2y, & 0 < y < 1, \\ 0, & \text{otherwise.} \end{cases}$$

Since,

$$f_{XY}(x,y) = f_X(x)f_Y(y),$$

$X$ and $Y$ are independent.

(c) The region in the $xy$ plane corresponding to the event $(X + Y < 1)$ is shown in the following figure as a shaded area.

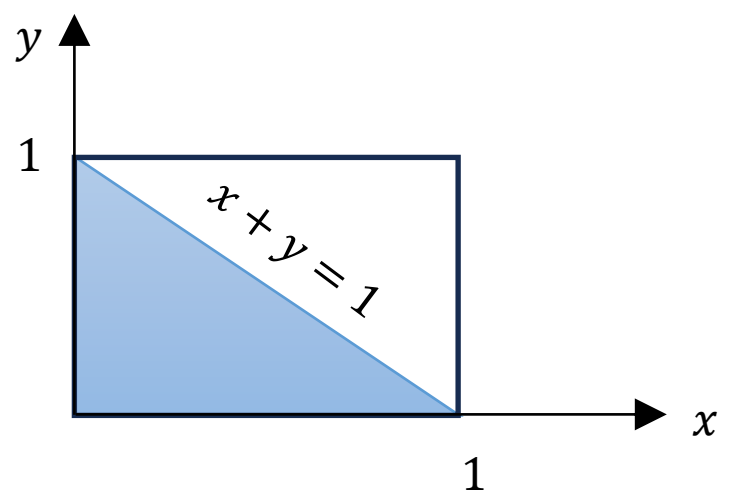

Then

$$\begin{aligned} P(X + Y < 1) &= \int_0^1\int_0^{1-y} 4xydxdy \\ &= \int_0^1 4y\left(\frac{x^2}{2}\right)_0^{1-y} dy \\ &= \int_0^1 4y\frac{(1-y)^2}{2}dy \\ &= \int_0^1 2y(1-y)^2dy \\ &= \int_0^1 2y(1-2y+y^2)dy \\ &= \int_0^1 (2y - 4y^2 + 2y^3)dy \\ &= 2\left(\frac{y^2}{2}\right)_0^1 - 4\left(\frac{y^3}{3}\right)_0^1 + 2\left(\frac{y^4}{4}\right)_0^1 \\ &= 1 - \frac{4}{3} + \frac{1}{2} \\ &= \frac{1}{6}. \end{aligned}$$

## 14.5 Covariance and Correlation

When two or more RVs are defined on a probability space, it is useful to describe how they vary together; that is, it is useful to measure the relationship between the variables. Joint moments of bivariate RV provide information about the relationship between two variables simultaneously. They describe the combined behavior of the variables and can be used to analyze the joint distribution.

**Definition (Joint Moment):** The $(k,n)$th joint moment of a bivariate RV $(X,Y)$ is defined by

$$m_{kn} = E[X^kY^n] = \begin{cases} \sum_{y_j}\sum_{x_i} x_i^k y_j^n f_{XY}(x_i,y_j), & \text{Discrete RV,} \\ \int_{-\infty}^{\infty}\int_{-\infty}^{\infty} x^k y^n f_{XY}(x,y)\,dxdy, & \text{Continuous RV.} \end{cases} \quad (14.19)$$

If $n = 0$, we obtain the $k$th moment of $X$, and if $k = 0$, we obtain the $n$th moment of $Y$. Thus,

$$m_{10} = E[X] = \mu_X \quad \text{and} \quad m_{01} = E[Y] = \mu_Y. \quad (14.20)$$

If $(X,Y)$ is a discrete bivariate RV, then using (14.19), and (14.13), we. obtain

$$\mu_X = E[X] = \sum_{y_j}\sum_{x_i} x_i f_{XY}(x_i,y_j) = \sum_{x_i} x_i\left[\sum_{y_j} f_{XY}(x_i,y_j)\right] = \sum_{x_i} x_i f_X(x_i), \quad (14.21.1)$$

$$\mu_Y = E[Y] = \sum_{x_i}\sum_{y_j} y_j f_{XY}(x_i,y_j) = \sum_{y_j} y_j\left[\sum_{x_i} f_{XY}(x_i,y_j)\right] = \sum_{y_j} y_j f_Y(y_j). \quad (14.21.2)$$

Similarly, we have

$$E[X^2] = \sum_{y_j}\sum_{x_i} x_i^2 f_{XY}(x_i,y_j) = \sum_{x_i} x_i^2 f_X(x_i), \quad (14.22.1)$$

$$E[Y^2] = \sum_{x_i}\sum_{y_j} y_j^2 f_{XY}(x_i,y_j) = \sum_{y_j} y_j^2 f_Y(y_j). \quad (14.22.2)$$

If $(X,Y)$ is a continuous bivariate RV, then using (14.19), and (14.17), we obtain

$$\mu_X = E[X] = \int_{-\infty}^{\infty}\int_{-\infty}^{\infty} x f_{XY}(x,y)dxdy = \int_{-\infty}^{\infty} x\left[\int_{-\infty}^{\infty} f_{XY}(x,y)dy\right]dx = \int_{-\infty}^{\infty} x f_X(x)dx, \quad (14.23.1)$$

$$\mu_Y = E[Y] = \int_{-\infty}^{\infty}\int_{-\infty}^{\infty} y f_{XY}(x,y)\,dxdy = \int_{-\infty}^{\infty} y\left[\int_{-\infty}^{\infty} f_{XY}(x,y)\,dx\right]dy = \int_{-\infty}^{\infty} y f_Y(y)\,dy. \quad (14.23.2)$$

Similarly, we have

$$E[X^2] = \int_{-\infty}^{\infty}\int_{-\infty}^{\infty} x^2 f_{XY}(x,y)dxdy = \int_{-\infty}^{\infty} x^2 f_X(x)dx, \quad (14.24.1)$$

$$E[Y^2] = \int_{-\infty}^{\infty}\int_{-\infty}^{\infty} y^2 f_{XY}(x,y)\,dxdy = \int_{-\infty}^{\infty} y^2 f_Y(y)dy. \quad (14.24.2)$$

The variances of $X$ and $Y$ are given by

$$\mathrm{Var}(X) = E[X^2] - (E[X])^2, \quad (14.25.1)$$
$$\mathrm{Var}(Y) = E[Y^2] - (E[Y])^2. \quad (14.25.2)$$

**Definition (Correlation):** The (1,1)th joint moment of $(X,Y)$,

$$m_{11} = E[X\,Y], \quad (14.26)$$

is called the correlation of $X$ and $Y$.

**Definition (Orthogonal RVs):** If $E[X\,Y] = 0$, then we say that $X$ and $Y$ are orthogonal.

A common measure of the relationship between two RVs is the covariance. Covariance is a measure of the joint variability of two RVs. Intuitively, the covariance between $X$ and $Y$ indicates how the values of $X$ and $Y$ move relative to each other. Let us provide the definition, then discuss the properties and applications of covariance.

**Definition (Covariance):** The covariance of $X$ and $Y$, denoted by $\mathrm{Cov}(X,Y)$ or $\sigma_{XY}$, is defined by

$$\begin{aligned}\mathrm{Cov}(X,Y) &= \sigma_{XY} \\ &= E[(X-E[X])(Y-E[Y])] \\ &= E[XY]-E[X]E[Y]. \end{aligned} \tag{14.27}$$

**Proof:**

$$\begin{aligned}\mathrm{Cov}(X,Y) &= E[(X-E[X])(Y-E[Y])] \\ &= E\big[XY-E[X]Y-XE[Y]+E[X]E[Y]\big] \\ &= E[XY]-E[X]E[Y]-E[X]E[Y]+E[X]E[Y] \\ &= E[XY]-E[X]E[Y]. \end{aligned}$$

■

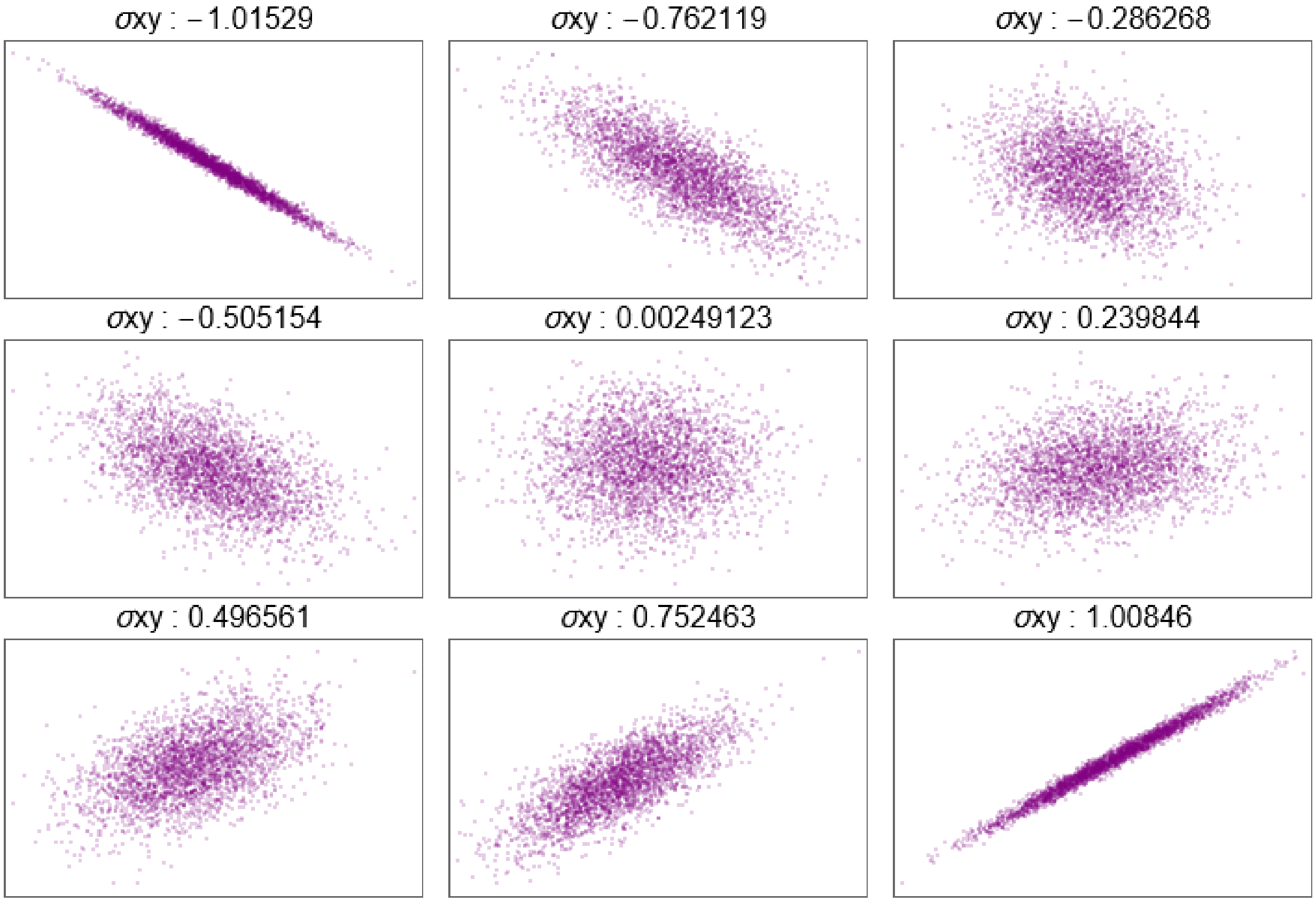

**Figure 14.1.** The sign of the covariance of two RVs $X$ and $Y$.

The resulting covariance value can provide insight into the relationship between the variables:

- Positive covariance:
  A positive covariance ($\mathrm{Cov} > 0$) indicates that when $X$ tends to be above its mean ($X-\mu_X > 0$), $Y$ tends to be above its mean ($Y-\mu_Y > 0$), and vice versa. It suggests a positive relationship between the variables, meaning that they tend to move together in the same direction. (i.e., if the greater values of one variable mainly correspond with the greater values of the other variable, and the same holds for the lesser values, the covariance is positive). For example, if $X$ represents hours of study and $Y$ represents exam scores, a positive covariance would imply that as study time increases, exam scores also tend to increase. See Figure 14.1.

- Negative covariance:
  A negative covariance ($\text{Cov} < 0$) implies that when $X$ tends to be above its mean, $Y$ tends to be below its mean (and vice versa). It indicates a negative relationship between the variables, suggesting that they tend to move in opposite directions. (i.e., the greater values of one variable mainly correspond to the lesser values of the other, the covariance is negative.) For instance, in the context of temperature and sales of winter clothing, a negative covariance would indicate that as temperatures rise ($X$ increases), sales of winter clothing decrease ($Y$ decreases).
- Zero covariance:
  A covariance of zero ($\text{Cov} = 0$) suggests that changes in one variable are unrelated to changes in the other.

**Definition (Uncorrelated RVs):** If $\text{Cov}(X, Y) = 0$, then we say that $X$ and $Y$ are uncorrelated. From (14.27), we see that $X$ and $Y$ are uncorrelated if

$$E[X\,Y] = E[X]E[Y]. \tag{14.28}$$

Note that if $X$ and $Y$ are independent, then it can be shown that they are uncorrelated, but the converse is not true in general; that is, the fact that $X$ and $Y$ are uncorrelated does not, in general, imply that they are independent.

**Theorem 14.2:** Let $(X, Y)$ be a bivariate RV. If $X$ and $Y$ are independent, then $X$ and $Y$ are uncorrelated.

**Proof:**

If $(X, Y)$ is a discrete bivariate RV, then by (14.19), and (14.13)

$$\begin{aligned}
E[XY] &= \sum_{y_j}\sum_{x_i} x_i y_j f_{XY}(x_i, y_j) \\
&= \sum_{y_j}\sum_{x_i} x_i y_j f_X(x_i) f_Y(y_j) \\
&= \left(\sum_{x_i} x_i f_X(x_i)\right)\left(\sum_{y_j} y_j f_Y(y_j)\right) \\
&= E[X]E[Y].
\end{aligned}$$

If $(X, Y)$ is a continuous bivariate RV, then by (14.19), and (14.17)

$$\begin{aligned}
E[XY] &= \int_{-\infty}^{\infty}\int_{-\infty}^{\infty} xy f_{XY}(x, y)\, dxdy \\
&= \int_{-\infty}^{\infty}\int_{-\infty}^{\infty} xy f_X(x) f_Y(y)\, dxdy \\
&= \left(\int_{-\infty}^{\infty} x f_X(x) dx\right)\left(\int_{-\infty}^{\infty} y f_Y(y) dy\right) \\
&= E[X]E[Y].
\end{aligned}$$

Thus, $X$ and $Y$ are uncorrelated by (14.28).

■

**Remarks:**

- The variance is a special case of the covariance in which the two variables are identical (that is, in which one variable always takes the same value as the other):
$$\begin{aligned}\text{Cov}(X,X) &= E[XX]-E[X]E[X]\\ &= E[X^2]-(E[X])^2\\ &= \sigma_X^2\\ &= \text{Var}(X).\end{aligned} \tag{14.29}$$
- If $X$, $Y$, $W$, and $V$ are real-valued RVs and $a,b,c,d$ are real-valued constants, then the following facts are a consequence of the definition of covariance:
$$\text{Cov}(X,a)=0, \tag{14.30}$$
$$\text{Cov}(X,X)=\text{Var}(X), \tag{14.31}$$
$$\text{Cov}(X,Y)=\text{Cov}(Y,X), \tag{14.32}$$
$$\text{Cov}(aX,bY)=ab\,\text{Cov}(X,Y), \tag{14.33}$$
$$\text{Cov}(X+a,Y+b)=\text{Cov}(X,Y), \tag{14.34}$$
$$\text{Cov}(aX+bY,cW+dV)=ac\,\text{Cov}(X,W)+ad\,\text{Cov}(X,V)+bc\,\text{Cov}(Y,W)+bd\,\text{Cov}(Y,V). \tag{14.35}$$
- The variance of the sum $X+Y$ is given by
$$\begin{aligned}\text{Var}(X+Y) &= E[(X+Y-E[X\ +\ Y])^2]\\ &= E[(X+Y-E[X]-E[Y])^2]\\ &= E[(X-E[X]+Y-E[Y])^2]\\ &= E[(X-E[X])^2+2(X-E[X])(Y-E[Y])+(Y-E[Y])^2]\\ &= E[(X-E[X])^2]+E[(Y-E[Y])^2]+2E[(X-E[X])(Y-E[Y])]\\ &= \text{Var}(X)+\text{Var}(Y)+2\text{Cov}(X,Y).\end{aligned} \tag{14.36}$$
- For a sequence $X_1, X_2,\ldots, X_n$ of RVs in real-valued, and constants $a_1, a_2,\ldots, a_n$, we have
$$\begin{aligned}\text{Var}\left(\sum_{i=1}^{n} a_i X_i\right) &= \sum_{i=1}^{n} a_i^2\sigma_{X_i}^2 + 2\sum_{i,j;i<j}^{n} a_i a_j \text{Cov}(X_i,X_j)\\ &= \sum_{i,j}^{n} a_i a_j \text{Cov}(X_i,X_j).\end{aligned} \tag{14.37}$$

***Example 14.8***

Let $X$ and $Y$ be two independent $N(0,1)$ RVs and $Z=1+X+XY^2$, $W=1+X$. Find $\text{Cov}(Z,W)$.

***Solution***

$$\begin{aligned}\text{Cov}(Z,W) &= \text{Cov}(1+X+XY^2,1+X)\\ &= \text{Cov}(X+XY^2,X)\\ &= \text{Cov}(X,X)+\text{Cov}(XY^2,X)\\ &= \text{Var}(X)+E[X^2Y^2]-E[XY^2]E[X]\\ &= 1+E[X^2]E[Y^2]-(E[X])^2E[Y^2]\\ &= 1+1-0\\ &= 2.\end{aligned}$$

**Definition (Correlation Coefficient):** The correlation coefficient, denoted by $\rho(X,Y)$ or $\rho_{XY}$, or $\text{Corr}(X,Y)$ is defined by
$$\rho(X,Y)=\rho_{XY}=\text{Corr}(X,Y)=\frac{\text{Cov}(X,Y)}{\sigma_X\sigma_Y}=\frac{\sigma_{XY}}{\sigma_X\sigma_Y}. \tag{14.38}$$
For any two RVs $X$ and $Y$,
$$|\rho_{XY}|\le 1 \quad \text{or} \quad -1\le\rho_{XY}\le 1. \tag{14.39}$$

***Example 14.9***

Suppose the joint PMF of a bivariate RV $(X, Y)$ is given by

$$f_{XY}(x_i, y_j) = \begin{cases} \frac{1}{3}, & (0,1), (1,0), (2,1), \\ 0, & \text{otherwise.} \end{cases}$$

Find $\text{Cov}(X, Y)$.

***Solution***

| | $x = 0$ | $x = 1$ | $x = 2$ | $f_Y(y)$ |
|---|---|---|---|---|
| $y = 0$ | 0 | $\frac{1}{3}$ | 0 | $\frac{1}{3}$ |
| $y = 1$ | $\frac{1}{3}$ | 0 | $\frac{1}{3}$ | $\frac{2}{3}$ |
| $f_X(x)$ | $\frac{1}{3}$ | $\frac{1}{3}$ | $\frac{1}{3}$ | 1 |

Since,

$$f_{XY}(0,1) = \frac{1}{3} \neq f_X(0) f_Y(1) = \frac{2}{9},$$

$X$ and $Y$ are not independent. We have also

$$E[X] = \sum_{x_i} x_i f_X(x_i) = (0)\left(\frac{1}{3}\right) + (1)\left(\frac{1}{3}\right) + (2)\left(\frac{1}{3}\right) = 1,$$

$$E[Y] = \sum_{y_j} y_j f_Y(y_j) = (0)\left(\frac{1}{3}\right) + (1)\left(\frac{2}{3}\right) = \frac{2}{3},$$

$$E[XY] = \sum_{y_j} \sum_{x_i} x_i y_j f_{XY}(x_i, y_j)$$
$$= (0)(1)\left(\frac{1}{3}\right) + (1)(0)\left(\frac{1}{3}\right) + (2)(1)\left(\frac{1}{3}\right) = \frac{2}{3}.$$

So,

$$\text{Cov}(X, Y) = E(XY) - E(X)E(Y)$$
$$= \frac{2}{3} - (1)\left(\frac{2}{3}\right)$$
$$= 0.$$

Thus, $X$ and $Y$ are uncorrelated.

**Definition (Covariance Matrix):** A covariance matrix, also known as a variance-covariance matrix, is a square matrix that summarizes the variances and covariances of multiple RVs. If we have $n$ RVs, $X_1$, $X_2$, ..., $X_n$, the covariance matrix is an $n \times n$ matrix denoted by $\Sigma$ where the element in the $i$th row and $j$th column represents the covariance between $X_i$ and $X_j$. Hence,

$$\Sigma_{ij} = \text{Cov}(X_i, X_j) = E\big[(X_i - E[X_i])(X_j - E[X_j])\big]. \tag{14.40}$$

It is important to note that the covariance between $X$ and $Y$ is the same as the covariance between $Y$ and $X$, which means that the covariance matrix is symmetric. Therefore,

$$\Sigma_{ij} = \Sigma_{ji}. \tag{14.41}$$

A covariance matrix provides several important pieces of information:

- Variances: The diagonal elements of the covariance matrix give the variances of the individual variables, providing a measure of their dispersion or variability. $\Sigma_{ii} = \text{Cov}(X_i, X_i) = \text{Var}(X_i)$.
- Covariances: The off-diagonal elements of the covariance matrix represent the covariances between pairs of variables.

## 14.6 Conditional Probability Distributions

When two RVs are defined in a random experiment, knowledge of one can change the probabilities that we associate with the values of the other. The RVs $X$ and $Y$ are expected to be dependent. Knowledge of the value obtained for $X$ changes the probabilities associated with the values of $Y$. The conditional probability distribution describes the probability distribution of a RV given that another RV has taken on a specific value or falls within a certain range. It provides information about the probability distribution of one variable, taking into account the values of another variable. Recall that the definition of conditional probability for events $A$ and $B$ is $P(B|A) = P(A \cap B) / P(A)$. This definition can be applied with the event $A$ defined to be $X = x$ and event $B$ define $Y = y$.

**Definition (Conditional PMF):** Given discrete RVs $X$ and $Y$ with joint PMF $f_{XY}(x, y)$, the conditional PMF of $Y$ given $X = x_i$ is

$$f_{Y|X}(y_j|x_i) = \frac{f_{XY}(x_i, y_j)}{f_X(x_i)}, \quad f_X(x_i) > 0. \tag{14.42}$$

Properties of $f_{Y|X}(y_j|x_i)$:

$$0 \le f_{Y|X}(y_j|x_i) \le 1, \tag{14.43.1}$$

$$\sum_{y_j} f_{Y|X}(y_j|x_i) = 1. \tag{14.43.2}$$

Notice that if $X$ and $Y$ are independent, then

$$f_{Y|X}(y_j|x_i) = f_Y(y_j), \quad f_{X|Y}(x_i|y_j) = f_X(x_i). \tag{14.44}$$

**Definition (Conditional Mean and Variance):** If $(X, Y)$ is a discrete bivariate RV with joint PMF $f_{XY}(x_i, y_j)$, then the conditional mean (or conditional expectation) of $Y$, given that $X = x_i$, is defined by

$$\mu_{Y|x_i} = E(Y|x_i) = \sum_{y_j} y_j\, f_{Y|X}(y_j|x_i). \tag{14.45}$$

The conditional variance of $Y$, given that $X = x_i$, is defined by

$$\sigma^2_{Y|x_i} = \mathrm{Var}(Y|x_i) = E\left(\left(Y - \mu_{Y|x_i}\right)^2 \middle| x_i\right) = \sum_{y_j} \left(y_j - \mu_{Y|x_i}\right)^2 f_{Y|X}(y_j|x_i) = E(Y^2|x_i) - [E(Y|x_i)]^2 \tag{14.46}$$

**Definition (Conditional PDF):** If $(X, Y)$ is a continuous bivariate RV with joint PDF $f_{XY}(x, y)$, then the conditional PDF of $Y$, given that $X = x$, $f_{Y|x}(y|x)$, is defined by

$$f_{Y|x}(y|x) = \frac{f_{XY}(x, y)}{f_X(x)}, \quad f_X(x) > 0. \tag{14.47}$$

Properties of $f_{Y|X}(y|x)$:

$$f_{Y|X}(y|x) \ge 0, \tag{14.48.1}$$

$$\int_{-\infty}^{\infty} f_{Y|X}(y|x)dy = 1. \tag{14.48.2}$$

Notice that if $X$ and $Y$ are independent, then

$$f_{Y|X}(y|x) = f_Y(y), \quad f_{X|Y}(x|y) = f_X(x). \tag{14.49}$$

**Definition (Conditional Mean and Variance):** If $(X, Y)$ is a continuous bivariate RV with joint PDF $f_{XY}(x, y)$, the conditional mean of $Y$, given that $X = x$, is defined by

$$\mu_{Y|x} = E(Y|x) = \int_{-\infty}^{\infty} y f_{Y|x}(y|x)\, dy. \tag{14.50}$$

The conditional variance of $Y$ given $X = x$, is defined by

$$\sigma_{Y|x}^2 = \mathrm{Var}(Y|x) = E\left(\left(Y - \mu_{Y|x}\right)^2 \middle| x_i\right) = \int_{-\infty}^{\infty} \left(y - \mu_{Y|x}\right)^2 f_{Y|x}(y|x)\, dy = E(Y^2|x) - [E(Y|x)]^2. \tag{14.51}$$

***Example 14.10***

Consider the bivariate RV $(X, Y)$ Example 14.2, find the conditional PMFs $f_{Y|X}(y_j|x_i)$ and $f_{X|Y}(x_i|y_j)$.

***Solution***

The conditional PMF of $Y$ given $X = x_i$, $f_{Y|X}(y_j|x_i)$, can be written as the following.

| $Y =$ | 1 | 2 | 3 | 4 | 5 | 6 | $\sum_{y_j} f_{Y\|X}(y_j\|x_i)$ |
|---|---|---|---|---|---|---|---|
| $X = 1$ | 1/6 | 1/6 | 1/6 | 1/6 | 1/6 | 1/6 | 1 |
| $X = 2$ | 1/6 | 1/6 | 1/6 | 1/6 | 1/6 | 1/6 | 1 |
| $X = 3$ | 1/6 | 1/6 | 1/6 | 1/6 | 1/6 | 1/6 | 1 |
| $X = 4$ | 1/6 | 1/6 | 1/6 | 1/6 | 1/6 | 1/6 | 1 |
| $X = 5$ | 1/6 | 1/6 | 1/6 | 1/6 | 1/6 | 1/6 | 1 |
| $X = 6$ | 1/6 | 1/6 | 1/6 | 1/6 | 1/6 | 1/6 | 1 |

Each cell in the table represents a specific outcome (combination) of $X$ and $Y$, and the corresponding conditional probability, $f_{Y|X}$, is listed in that cell. For example, $f_{Y|X}(Y = 1|X = 2)$ represents the conditional probability of $Y = 1$ given $X = 2$.

Note that in this case, since the dice are fair and independent, the $f_{Y|X}(y_j|x_i)$ is the same as the marginal PMF $f_Y(y_j)$. This means that the conditional PMF does not depend on the value of $X$, and each outcome of $Y$ has an equal probability of $1/6$ regardless of the value of $X$. Similarly, the conditional PMF of $X$ given $Y = y_j$, $f_{X|Y}(x_i|y_j)$, can be written as the following.

| $Y =$ | 1 | 2 | 3 | 4 | 5 | 6 |
|---|---|---|---|---|---|---|
| $X = 1$ | 1/6 | 1/6 | 1/6 | 1/6 | 1/6 | 1/6 |
| $X = 2$ | 1/6 | 1/6 | 1/6 | 1/6 | 1/6 | 1/6 |
| $X = 3$ | 1/6 | 1/6 | 1/6 | 1/6 | 1/6 | 1/6 |
| $X = 4$ | 1/6 | 1/6 | 1/6 | 1/6 | 1/6 | 1/6 |
| $X = 5$ | 1/6 | 1/6 | 1/6 | 1/6 | 1/6 | 1/6 |
| $X = 6$ | 1/6 | 1/6 | 1/6 | 1/6 | 1/6 | 1/6 |
| $\sum_{x_i} f_{X\|Y}(x_i\|y_j)$ | 1 | 1 | 1 | 1 | 1 | 1 |

***Example 14.11***

Consider the bivariate RV $(X, Y)$ Example 14.3, find the conditional PMFs $f_{Y|X}(y_j|x_i)$ and $f_{X|Y}(x_i|y_j)$.

***Solution***

The conditional PMF of $Y$ given $X = x_i$, $f_{Y|X}(y_j|x_i)$, can be written as the following.

| $Y =$ | −5 | −4 | −3 | −2 | −1 | 0 | 1 | 2 | 3 | 4 | 5 | $\sum_{y_j} f_{Y\|X}(y_j\|x_i)$ |
|---|---|---|---|---|---|---|---|---|---|---|---|---|
| $X = 2$ | | | | | | 1 | | | | | | 1 |
| $X = 3$ | | | | | 1/2 | | 1/2 | | | | | 1 |
| $X = 4$ | | | | 1/3 | | 1/3 | | 1/3 | | | | 1 |
| $X = 5$ | | | 1/4 | | 1/4 | | 1/4 | | 1/4 | | | 1 |
| $X = 6$ | | 1/5 | | 1/5 | | 1/5 | | 1/5 | | 1/5 | | 1 |
| $X = 7$ | 1/6 | | 1/6 | | 1/6 | | 1/6 | | 1/6 | | 1/6 | 1 |
| $X = 8$ | | 1/5 | | 1/5 | | 1/5 | | 1/5 | | 1/5 | | 1 |
| $X = 9$ | | | 1/4 | | 1/4 | | 1/4 | | 1/4 | | | 1 |
| $X = 10$ | | | | 1/3 | | 1/3 | | 1/3 | | | | 1 |
| $X = 11$ | | | | | 1/2 | | 1/2 | | | | | 1 |
| $X = 12$ | | | | | | 1 | | | | | | 1 |

Similarly, the conditional PMF of $X$ given $Y = y_j$, $f_{X|Y}(x_i|y_j)$, can be written as the following.

| $Y =$ | $-5$ | $-4$ | $-3$ | $-2$ | $-1$ | $0$ | $1$ | $2$ | $3$ | $4$ | $5$ |
|---|---|---|---|---|---|---|---|---|---|---|---|
| $X = 2$ | | | | | | $1/6$ | | | | | |
| $X = 3$ | | | | | $1/5$ | | $1/5$ | | | | |
| $X = 4$ | | | | $1/4$ | | $1/6$ | | $1/4$ | | | |
| $X = 5$ | | | $1/3$ | | $1/5$ | | $1/5$ | | $1/3$ | | |
| $X = 6$ | | $1/2$ | | $1/4$ | | $1/6$ | | $1/4$ | | $1/2$ | |
| $X = 7$ | $1$ | | $1/3$ | | $1/5$ | | $1/5$ | | $1/3$ | | $1$ |
| $X = 8$ | | $1/2$ | | $1/4$ | | $1/6$ | | $1/4$ | | $1/2$ | |
| $X = 9$ | | | $1/3$ | | $1/5$ | | $1/5$ | | $1/3$ | | |
| $X = 10$ | | | | $1/4$ | | $1/6$ | | $1/4$ | | | |
| $X = 11$ | | | | | $1/5$ | | $1/5$ | | | | |
| $X = 12$ | | | | | | $1/6$ | | | | | |
| $\sum_{x_i} f_{X|Y}(x_i|y_j)$ | $1$ | $1$ | $1$ | $1$ | $1$ | $1$ | $1$ | $1$ | $1$ | $1$ | $1$ |

***Example 14.12***

Consider the bivariate RV $(X, Y)$ Example 14.4, find the conditional PMFs $f_{Y|X}(y_j|x_i)$ and $f_{X|Y}(x_i|y_j)$.

***Solution***

The conditional PMF of $Y$ given $X = x_i$, $f_{Y|X}(y_j|x_i)$, can be written as the following.

| $Y =$ | $0$ | $1$ | $2$ | $3$ | $4$ | $5$ | $\sum_{y_j} f_{Y|X}(y_j|x_i)$ |
|---|---|---|---|---|---|---|---|
| $X = 2$ | $1$ | | | | | | $1$ |
| $X = 3$ | | $1$ | | | | | $1$ |
| $X = 4$ | $1/3$ | | $2/3$ | | | | $1$ |
| $X = 5$ | | $2/4$ | | $2/4$ | | | $1$ |
| $X = 6$ | $1/5$ | | $2/5$ | | $2/5$ | | $1$ |
| $X = 7$ | | $2/6$ | | $2/6$ | | $2/6$ | $1$ |
| $X = 8$ | $1/5$ | | $2/5$ | | $2/5$ | | $1$ |
| $X = 9$ | | $2/4$ | | $2/4$ | | | $1$ |
| $X = 10$ | $1/3$ | | $2/3$ | | | | $1$ |
| $X = 11$ | | $1$ | | | | | $1$ |
| $X = 12$ | $1$ | | | | | | $1$ |

Similarly, the conditional PMF of $X$ given $Y = y_j$, $f_{X|Y}(x_i|y_j)$, can be written as the following.

| $Y =$ | $0$ | $1$ | $2$ | $3$ | $4$ | $5$ |
|---|---|---|---|---|---|---|
| $X = 2$ | $1/6$ | | | | | |
| $X = 3$ | | $2/10$ | | | | |
| $X = 4$ | $1/6$ | | $2/8$ | | | |
| $X = 5$ | | $2/10$ | | $2/6$ | | |
| $X = 6$ | $1/6$ | | $2/8$ | | $2/4$ | |
| $X = 7$ | | $2/10$ | | $2/6$ | | $1$ |
| $X = 8$ | $1/6$ | | $2/8$ | | $2/4$ | |
| $X = 9$ | | $2/10$ | | $2/6$ | | |
| $X = 10$ | $1/6$ | | $2/8$ | | | |
| $X = 11$ | | $2/10$ | | | | |
| $X = 12$ | $1/6$ | | | | | |
| $\sum_{x_i} f_{X|Y}(x_i|y_j)$ | $1$ | $1$ | $1$ | $1$ | $1$ | $1$ |

***Example 14.13***

The joint PMF of a bivariate RV $(X, Y)$ is given by

$$f_{XY}(x,y) = \begin{cases} k(2x_i + y_j), & x_i = 1,2; y_j = 1,2, \\ 0, & \text{otherwise,} \end{cases}$$

where $k$ is a constant.

(a) Find the value of $k$.

(b) Find the marginal PMF of $X$ and $Y$.

(c) Find the conditional PMFs.

***Solution***

$$\sum_{x_i}\sum_{y_j} f_{XY}(x_i, y_j) = \sum_{x_i=1}^{2}\sum_{y_j=1}^{2} k(2x_i + y_j)$$
$$= k[(2+1) + (2+2) + (4+1) + (4+2)] = k(18) = 1.$$

Thus, $k = \frac{1}{18}$.

The marginal PMF of $X$ are

$$f_X(x_i) = \sum_{y_j} f_{XY}(x_i, y_j)$$
$$= \sum_{y_j=1}^{2} \frac{1}{18}(2x_i + y_j)$$
$$= \frac{1}{18}(2x_i + 1) + \frac{1}{18}(2x_i + 2)$$
$$= \frac{1}{18}(4x_i + 3), \quad x_i = 1,2.$$

The marginal PMF of $Y$ are

$$f_Y(y_j) = \sum_{x_i} f_{XY}(x_i, y_j)$$
$$= \sum_{x_i=1}^{2} \frac{1}{18}(2x_i + y_j)$$
$$= \frac{1}{18}(2 + y_j) + \frac{1}{18}(4 + y_j)$$
$$= \frac{1}{18}(2y_j + 6), \quad y_j = 1,2.$$

The conditional PMFs

$$f_{Y|X}(y_j|x_i) = \frac{2x_i + y_j}{4x_i + 3}, \quad x_i = 1,2; y_j = 1,2,$$
$$f_{X|Y}(x_i|y_j) = \frac{2x_i + y_j}{2y_j + 6}, \quad x_i = 1,2; y_j = 1,2,$$

$$P(Y = 2|X = 2) = P_{Y|X}(2|2) = \frac{2(2) + 2}{4(2) + 3} = \frac{6}{11},$$
$$P(X = 2|Y = 2) = P_{X|Y}(2|2) = \frac{2(2) + 2}{2(2) + 6} = \frac{3}{5}.$$

***Example 14.14***

Consider the bivariate RV $(X, Y)$ Example 14.7, find the conditional PDFs $f_{Y|X}(y|x)$ and $f_{X|Y}(x|y)$.

***Solution***

We have,

$$f_{XY}(x,y) = \begin{cases} 4xy, & 0 < x < 1, 0 < y < 1, \\ 0, & \text{otherwise,} \end{cases}$$

$$f_X(x) = \begin{cases} 2x, & 0 < x < 1, \\ 0, & \text{otherwise,} \end{cases}$$

$$f_Y(y) = \begin{cases} 2y, & 0 < y < 1, \\ 0, & \text{otherwise.} \end{cases}$$

Hence,

$$f_{Y|X}(y|x) = 2y = f_Y(y), \qquad 0 < x < 1, 0 < y < 1,$$
$$f_{X|Y}(x|y) = 2x = f_X(x), \qquad 0 < x < 1, 0 < y < 1,$$

Note that in this case, the $f_{Y|X}(y|x)$ is the same as the marginal PDF $f_Y(y)$. This means that the conditional PDF $f_{Y|X}(y|x)$ does not depend on the value of $X$. Similarly, the $f_{X|Y}(x|y)$ is the same as the marginal PDF $f_X(x)$. This means that the conditional PDF $f_{X|Y}(x|y)$ does not depend on the value of $Y$.

## 14.7 Joint Distributions

### 14.7.1 Bivariate Normal Distribution

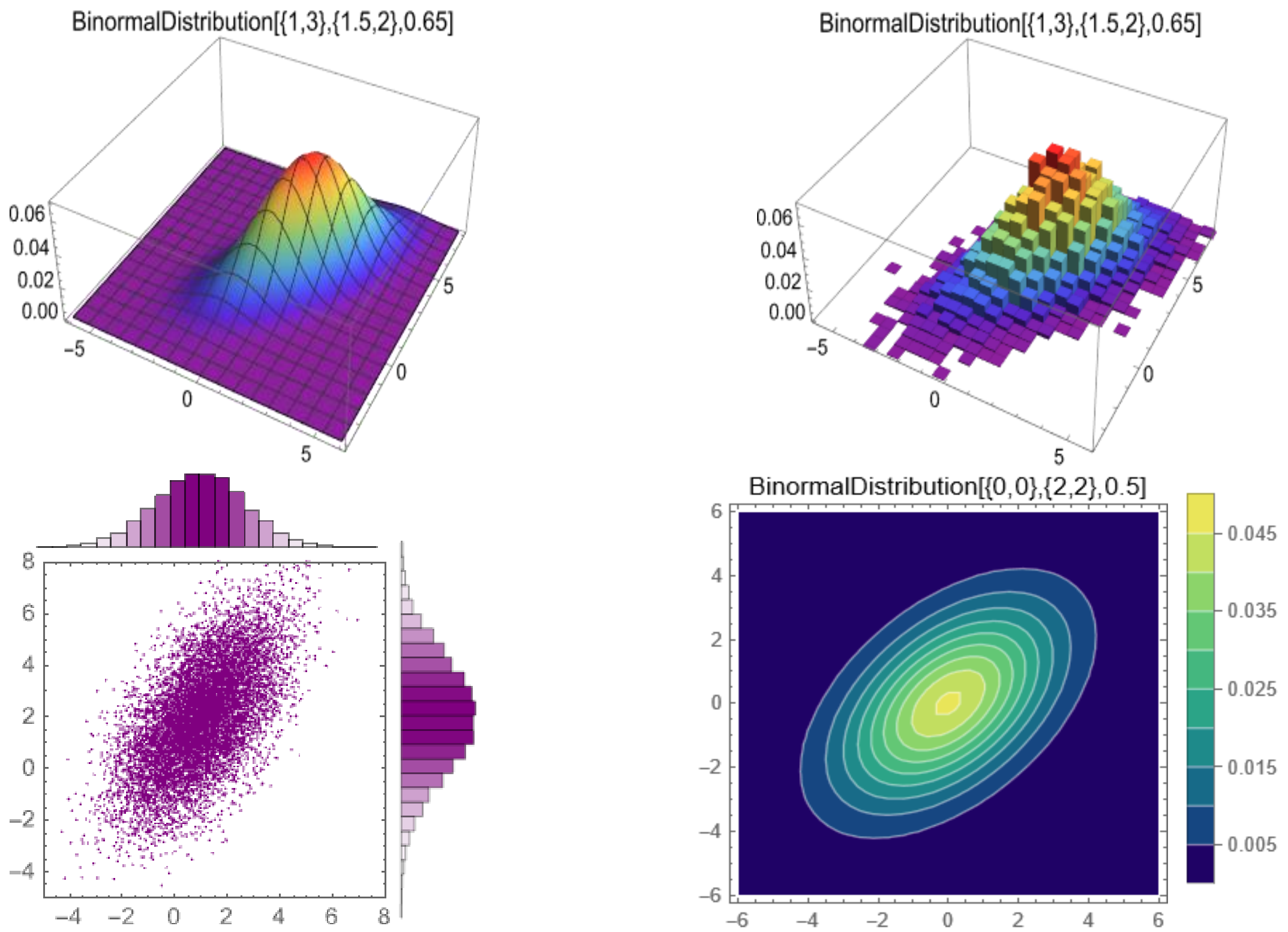

**Figure 14.2.** 3d plot (upper left panel), 3d histogram (upper right panel), scatter plot (lower left panel) and 2d contour plot (lower right panel) of bivariate normal distributions. The density function of bivariate normal distribution is a generalization of the familiar bell curve and graphs in three dimensions as a sort of bell-shaped hump. A bivariate normal density has elliptical contours. For each height $c > 0$ the set $\{(x, y): f_{XY}(x, y) = c\}$ is an ellipse. A scatter plot of $(x, y)$ pairs generated from a bivariate normal distribution will have a rough linear association and the cloud of points will resemble an ellipse. If $X$ and $Y$ have a bivariate normal distribution, then the marginal distributions are also normal: $X$ has a normal $(\mu_X, \sigma_X)$ distribution and $Y$ has a Normal $(\mu_Y, \sigma_Y)$.

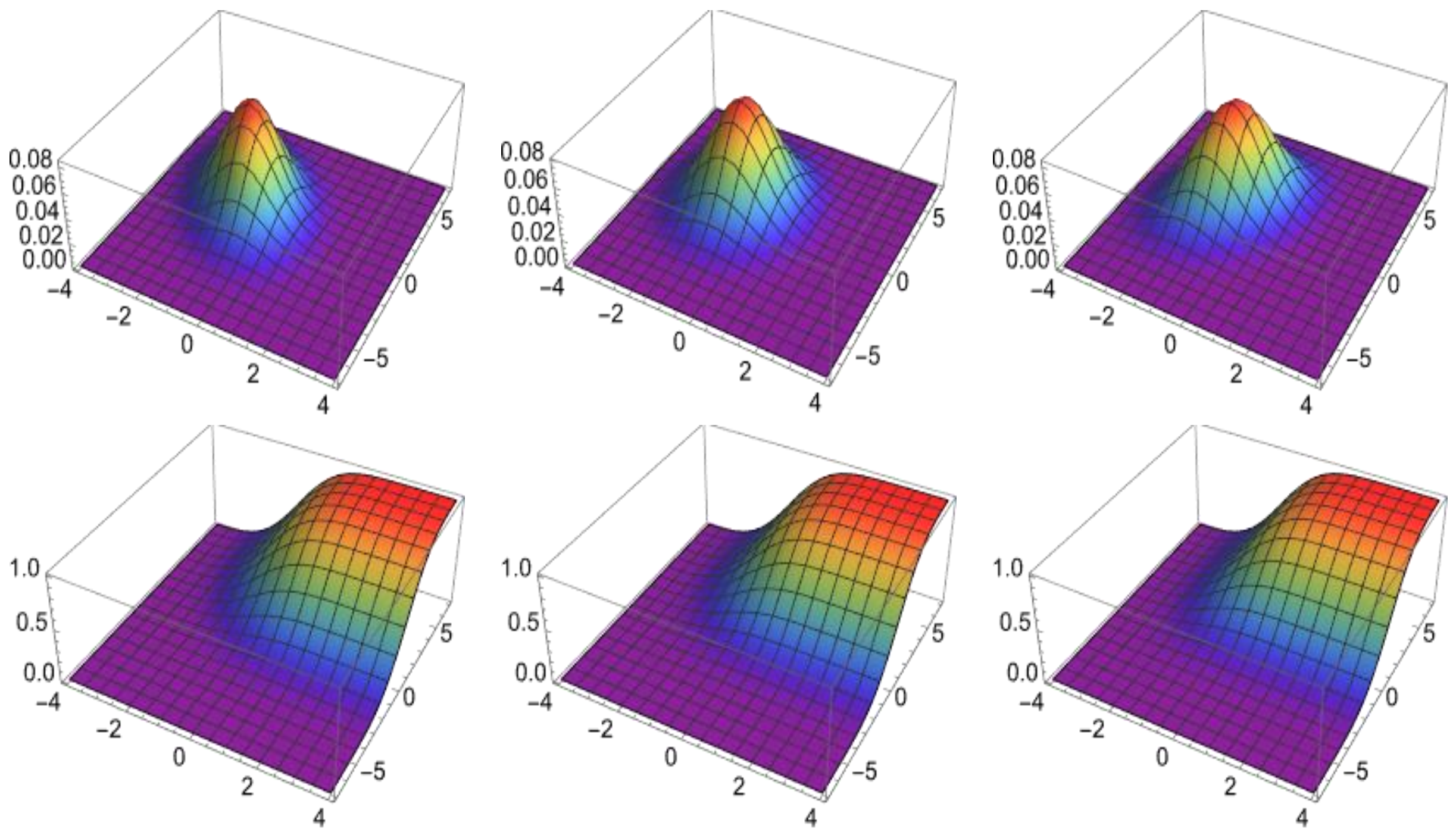

**Figure 14.3.** 3d plot of PDFs of the bivariate normal distributions with $\{\mu_1, \mu_2\} = \{-1,1\}$, $\{\sigma_1^2, \sigma_2^2\} = \{1,2\}$, and $\rho = \{-0.3,0,0.3\}$ (upper panel). 3d plot of CDFs of the same bivariate normal distributions (lower panel).

The bivariate normal distribution is an extension of the univariate normal distribution. It allows us to analyze and model relationships between two continuous RVs, taking into account both their individual characteristics and their dependence on each other. It is widely used in various fields, including finance, economics, engineering, and social sciences, to study phenomena where two variables are interrelated.

The bivariate normal distribution is characterized by its shape, mean vector, variance-covariance matrix, and correlation coefficient. Understanding these characteristics is crucial for comprehending the properties and applications of the distribution see Figures 14.2 and 14.3.

- Shape:
  The bivariate normal distribution forms an elliptical shape in two dimensions. The contours of the distribution are typically centered around the mean vector and symmetrically spread along the axes. The elliptical shape can be stretched or compressed, depending on the correlation between the variables.
- Mean vector:
  The mean vector of the bivariate normal distribution represents the average values of the two variables. It is denoted as a column vector,
  $$\begin{pmatrix} \mu_1 \\ \mu_2 \end{pmatrix},$$
  where $\mu_1$ is the mean of the first variable and $\mu_2$ is the mean of the second variable. The mean vector determines the location of the center of the distribution.
- Variance-covariance matrix:
  The variance-covariance matrix of the bivariate normal distribution characterizes the variability and relationship between the two variables. It is a $2 \times 2$ symmetric matrix that contains variances, and covariances. The matrix has the form:
  $$\begin{pmatrix} \sigma_1^2 & \sigma_{12} \\ \sigma_{12} & \sigma_2^2 \end{pmatrix},$$
  where $\sigma_1^2$ and $\sigma_2^2$ represent the variances of the first and second variables, respectively, and $\sigma_{12}$ is the covariance between the two variables.

- Correlation coefficient:
  The correlation coefficient measures the linear relationship between the two variables in the bivariate normal distribution. It is denoted as $\rho$ and ranges between $-1$ and $1$.

**Definition (Bivariate Normal PDF):** The PDF of a bivariate normal distribution, BVN $(\mu_1, \mu_2, \sigma_1^2, \sigma_2^2, \rho)$, is given by

$$f_{XY}(x, y; \sigma_1, \sigma_2; \mu_1, \mu_2, \rho) = \frac{1}{2\pi\sigma_1\sigma_2\sqrt{1-\rho^2}} e^{-\frac{1}{2(1-\rho^2)}\left[\frac{(x-\mu_1)^2}{\sigma_1^2} - \frac{2\rho(x-\mu_1)(y-\mu_2)}{\sigma_1\sigma_2} + \frac{(y-\mu_2)^2}{\sigma_2^2}\right]}, \quad (14.52)$$

for $-\infty < x < \infty$ and $-\infty < y < \infty$, with parameters $\sigma_1 > 0$, $\sigma_2 > 0$, $-\infty < \mu_1 < \infty$, $-\infty < \mu_2 < \infty$, and $-1 < \rho < 1$.

**Theorem 14.3 (MGF of Bivariate Normal Distribution):**
Let $(X, Y) \sim$ BVN $(\mu_1, \mu_2, \sigma_1^2, \sigma_2^2, \rho)$. Then

$$M_{XY}(t_1, t_2) = e^{t_1\mu_1 + t_2\mu_2 + \frac{1}{2}(t_1^2\sigma_1^2 + t_2^2\sigma_2^2 + 2\rho t_1 t_2 \sigma_1\sigma_2)}. \quad (14.53)$$

**Proof:**

By definition,

$$\begin{aligned} M_{XY}(t_1, t_2) &= E[e^{t_1X + t_2Y}] \\ &= \int_{-\infty}^{\infty}\int_{-\infty}^{\infty} e^{t_1x + t_2y} f_{XY}(x, y)dxdy \\ &= \int_{-\infty}^{\infty}\int_{-\infty}^{\infty} e^{t_1x+t_2y} \frac{1}{2\pi\sigma_1\sigma_2\sqrt{1-\rho^2}} e^{-\frac{1}{2(1-\rho^2)}\left[\frac{(x-\mu_1)^2}{\sigma_1^2} - \frac{2\rho(x-\mu_1)(y-\mu_2)}{\sigma_1\sigma_2} + \frac{(y-\mu_2)^2}{\sigma_2^2}\right]} dxdy. \end{aligned}$$

Put $\frac{x-\mu_1}{\sigma_1} = u, \frac{y-\mu_2}{\sigma_2} = v, -\infty < (u, v) < \infty$, then

$$x = u\sigma_1 + \mu_1 \quad \Rightarrow dx = \sigma_1 du,$$

$$y = v\sigma_2 + \mu_2 \quad \Rightarrow dy = \sigma_2 dv,$$

$$\begin{aligned} M_{XY}(t_1, t_2) &= \int_u \int_v e^{t_1(u\sigma_1+\mu_1)+t_2(v\sigma_2+\mu_2)} \frac{1}{2\pi\sqrt{1-\rho^2}} e^{\left(-\frac{1}{2(1-\rho^2)}[u^2 - 2\rho uv + v^2]\right)} dudv \\ &= \int_u \int_v e^{(ut_1\sigma_1 + t_1\mu_1)+(vt_2\sigma_2+t_2\mu_2)} \frac{1}{2\pi\sqrt{1-\rho^2}} e^{\left(-\frac{1}{2(1-\rho^2)}[u^2-2\rho uv+v^2]\right)} dudv \\ &= \frac{e^{t_1\mu_1+t_2\mu_2}}{2\pi\sqrt{1-\rho^2}} \int_u \int_v e^{\left(ut_1\sigma_1 + vt_2\sigma_2 - \frac{1}{2(1-\rho^2)}[u^2-2\rho uv+v^2]\right)} dudv \\ &= \frac{e^{t_1\mu_1+t_2\mu_2}}{2\pi\sqrt{1-\rho^2}} \int_u \int_v e^{-\frac{1}{2(1-\rho^2)}\left([u^2-2\rho uv+v^2] - 2(1-\rho^2)(ut_1\sigma_1+vt_2\sigma_2)\right)} dudv. \end{aligned}$$

We have

$$\begin{aligned} &[u^2 - 2\rho uv + v^2] - 2(1-\rho^2)(ut_1\sigma_1 + vt_2\sigma_2) \\ &\quad = (u - \rho v - (1-\rho^2)t_1\sigma_1)^2 + (1-\rho^2)[(v - \rho t_1\sigma_1 - t_2\sigma_2)^2 - t_1^2\sigma_1^2 - t_2^2\sigma_2^2 - 2\rho t_1 t_2\sigma_1\sigma_2]. \end{aligned}$$

By taking

$$u - \rho v - (1-\rho^2)t_1\sigma_1 = w\sqrt{1-\rho^2},$$

$$v - \rho t_1\sigma_1 - t_2\sigma_2 = z.$$

We have

$$dudv = \sqrt{1-\rho^2}dwdz.$$

$$\begin{aligned}
M_{XY}(t_1,t_2) &= \frac{e^{t_1\mu_1+t_2\mu_2}}{2\pi\sqrt{1-\rho^2}}\int_u\int_v e^{-\frac{1}{2(1-\rho^2)}\left([u^2-2\rho uv+v^2]-2(1-\rho^2)(ut_1\sigma_1+vt_2\sigma_2)\right)}dudv \\
&= \frac{e^{t_1\mu_1+t_2\mu_2}}{2\pi\sqrt{1-\rho^2}}\int_{-\infty}^{\infty}\int_{-\infty}^{\infty} e^{-\frac{1}{2(1-\rho^2)}\left(w^2(1-\rho^2)+(1-\rho^2)[z^2-t_1^2\sigma_1^2-t_2^2\sigma_2^2-2\rho t_1t_2\sigma_1\sigma_2]\right)}\sqrt{1-\rho^2}dwdz \\
&= \frac{e^{t_1\mu_1+t_2\mu_2}}{2\pi}\int_{-\infty}^{\infty}\int_{-\infty}^{\infty} e^{-\frac{1}{2}\left(w^2+z^2-t_1^2\sigma_1^2-t_2^2\sigma_2^2-2\rho t_1t_2\sigma_1\sigma_2\right)}dwdz \\
&= \frac{e^{t_1\mu_1+t_2\mu_2}}{2\pi}e^{-\frac{1}{2}\left(-t_1^2\sigma_1^2-t_2^2\sigma_2^2-2\rho t_1t_2\sigma_1\sigma_2\right)}\int_{-\infty}^{\infty}\int_{-\infty}^{\infty} e^{-\frac{1}{2}(w^2+z^2)}dwdz \\
&= \frac{e^{t_1\mu_1+t_2\mu_2}}{2\pi}e^{\frac{1}{2}\left(t_1^2\sigma_1^2+t_2^2\sigma_2^2+2\rho t_1t_2\sigma_1\sigma_2\right)}\int_{-\infty}^{\infty} e^{-\frac{1}{2}w^2}dw\int_{-\infty}^{\infty} e^{-\frac{1}{2}z^2}dz \\
&= e^{t_1\mu_1+t_2\mu_2+\frac{1}{2}\left(t_1^2\sigma_1^2+t_2^2\sigma_2^2+2\rho t_1t_2\sigma_1\sigma_2\right)}\left(\frac{1}{\sqrt{2\pi}}\int_{-\infty}^{\infty} e^{-\frac{1}{2}w^2}dw\right)\left(\frac{1}{\sqrt{2\pi}}\int_{-\infty}^{\infty} e^{-\frac{1}{2}z^2}dz\right) \\
&= e^{t_1\mu_1+t_2\mu_2+\frac{1}{2}\left(t_1^2\sigma_1^2+t_2^2\sigma_2^2+2\rho t_1t_2\sigma_1\sigma_2\right)}.
\end{aligned}$$

In particular if $(X,Y)\sim$BVN $(0,0,1,1,\rho)$, then

$$M_{XY}(t_1,t_2) = e^{\frac{1}{2}(t_1^2+t_2^2+2\rho t_1t_2)}.$$

■

**Theorem 14.4:** Let $(X,Y)\sim$BVN $(\mu_1,\mu_2,\sigma_1^2,\sigma_2^2,\rho)$. Then $X$ and $Y$ are independent if and only if $\rho = 0$.

**Proof:**

Let $(X,Y)\sim$BVN $(\mu_1,\mu_2,\sigma_1^2,\sigma_2^2,\rho)$. Then

$$M_{XY}(t_1,t_2) = e^{t_1\mu_1+t_2\mu_2+\frac{1}{2}\left(t_1^2\sigma_1^2+t_2^2\sigma_2^2+2\rho t_1t_2\sigma_1\sigma_2\right)}.$$

If $\rho = 0$, then

$$\begin{aligned}
M_{XY}(t_1,t_2) &= e^{t_1\mu_1+t_2\mu_2+\frac{1}{2}t_1^2\sigma_1^2+\frac{1}{2}t_2^2\sigma_2^2} \\
&= e^{t_1\mu_1+\frac{1}{2}t_1^2\sigma_1^2}e^{t_2\mu_2+\frac{1}{2}t_2^2\sigma_2^2} \\
&= M_X(t_1)M_Y(t_2).
\end{aligned}$$

Moreover, if $\rho = 0$, then

$$\begin{aligned}
f_{XY}(x,y;\sigma_1,\sigma_2;\mu_1,\mu_2,0) &= \frac{1}{2\pi\sigma_1\sigma_2}e^{-\frac{1}{2}\left[\frac{(x-\mu_1)^2}{\sigma_1^2}+\frac{(y-\mu_2)^2}{\sigma_2^2}\right]} \\
&= \frac{1}{\sqrt{2\pi}\sigma_1}e^{\left[-\frac{1}{2}\frac{(x-\mu_1)^2}{\sigma_1^2}\right]}\frac{1}{\sqrt{2\pi}\sigma_2}e^{\left[-\frac{1}{2}\frac{(y-\mu_2)^2}{\sigma_2^2}\right]} \\
&= f_X(x)f_Y(y).
\end{aligned}$$

From $M_{XY}(t_1,t_2) = M_X(t_1)M_Y(t_2)$ and $f_{XY}(x,y) = f_X(x)f_Y(y)$, $X$ and $Y$ are independent.

Conversely if $X$ and $Y$ are independent, then $\rho = 0$.

■

**Theorem 14.5 (Marginal Distribution of Bivariate Normal Distribution):**
Let $(X, Y) \sim \text{BVN}\ (\mu_1, \mu_2, \sigma_1^2, \sigma_2^2, \rho)$, then the marginal PDFs of $X$ and $Y$ are also normal. The marginal PDFs of RVs $X$ and $Y$ are given by

$$f_X(x) = \frac{1}{\sqrt{2\pi}\sigma_1} e^{-\frac{1}{2}\left(\frac{x-\mu_1}{\sigma_1}\right)^2}, \quad (14.54.1)$$

$$f_Y(y) = \frac{1}{\sqrt{2\pi}\sigma_2} e^{-\frac{1}{2}\left(\frac{y-\mu_2}{\sigma_2}\right)^2}. \quad (14.54.2)$$

**Proof:**

The marginal distribution of RV $X$ is given by

$$f_X(x) = \int_{-\infty}^{\infty} f_{XY}(x, y)dy$$
$$= \frac{1}{2\pi\sigma_1\sigma_2\sqrt{1-\rho^2}} \int_{-\infty}^{\infty} e^{-\frac{1}{2(1-\rho^2)}\left[\frac{(x-\mu_1)^2}{\sigma_1^2} - \frac{2\rho(x-\mu_1)(y-\mu_2)}{\sigma_1\sigma_2} + \frac{(y-\mu_2)^2}{\sigma_2^2}\right]} dy.$$

Put $\frac{y-\mu_2}{\sigma_2} = u$, then $dy = \sigma_2 du$. Therefore,

$$f_X(x) = \frac{1}{2\pi\sigma_1\sigma_2\sqrt{1-\rho^2}} \int_{-\infty}^{\infty} e^{-\frac{1}{2(1-\rho^2)}\left[\frac{(x-\mu_1)^2}{\sigma_1^2} - \frac{2\rho(x-\mu_1)u}{\sigma_1} + u^2\right]} \sigma_2 du$$
$$= \frac{1}{2\pi\sigma_1\sqrt{1-\rho^2}} e^{-\frac{1}{2(1-\rho^2)}\left(\frac{x-\mu_1}{\sigma_1}\right)^2} \int_{-\infty}^{\infty} e^{-\frac{1}{2(1-\rho^2)}\left[u^2 - 2\rho\left(\frac{x-\mu_1}{\sigma_1}\right)u\right]} du$$
$$= \frac{1}{2\pi\sigma_1\sqrt{1-\rho^2}} e^{-\frac{1}{2(1-\rho^2)}\left(\frac{x-\mu_1}{\sigma_1}\right)^2} \int_{-\infty}^{\infty} e^{-\frac{1}{2(1-\rho^2)}\left[u^2 - 2\rho\left(\frac{x-\mu_1}{\sigma_1}\right)u + \rho^2\left(\frac{x-\mu_1}{\sigma_1}\right)^2 - \rho^2\left(\frac{x-\mu_1}{\sigma_1}\right)^2\right]} du$$
$$= \frac{1}{2\pi\sigma_1\sqrt{1-\rho^2}} e^{-\frac{1}{2(1-\rho^2)}\left(\frac{x-\mu_1}{\sigma_1}\right)^2} \int_{-\infty}^{\infty} e^{-\frac{1}{2(1-\rho^2)}\left[u^2 - 2\rho\left(\frac{x-\mu_1}{\sigma_1}\right)u + \rho^2\left(\frac{x-\mu_1}{\sigma_1}\right)^2\right]} e^{-\frac{1}{2(1-\rho^2)}\left[-\rho^2\left(\frac{x-\mu_1}{\sigma_1}\right)^2\right]} du$$
$$= \frac{1}{2\pi\sigma_1\sqrt{1-\rho^2}} e^{-\frac{1}{2(1-\rho^2)}\left(\frac{x-\mu_1}{\sigma_1}\right)^2} e^{-\frac{1}{2(1-\rho^2)}\left[-\rho^2\left(\frac{x-\mu_1}{\sigma_1}\right)^2\right]} \int_{-\infty}^{\infty} e^{-\frac{1}{2(1-\rho^2)}\left[u - \rho\left(\frac{x-\mu_1}{\sigma_1}\right)\right]^2} du$$
$$= \frac{1}{2\pi\sigma_1\sqrt{1-\rho^2}} e^{-\frac{1}{2(1-\rho^2)}\left(\frac{x-\mu_1}{\sigma_1}\right)^2} e^{\frac{\rho^2}{2(1-\rho^2)}\left(\frac{x-\mu_1}{\sigma_1}\right)^2} \int_{-\infty}^{\infty} e^{-\frac{1}{2(1-\rho^2)}\left[u - \rho\left(\frac{x-\mu_1}{\sigma_1}\right)\right]^2} du$$
$$= \frac{1}{2\pi\sigma_1\sqrt{1-\rho^2}} e^{-\frac{1}{2(1-\rho^2)}\left(\frac{x-\mu_1}{\sigma_1}\right)^2 + \frac{\rho^2}{2(1-\rho^2)}\left(\frac{x-\mu_1}{\sigma_1}\right)^2} \int_{-\infty}^{\infty} e^{-\frac{1}{2(1-\rho^2)}\left[u - \rho\left(\frac{x-\mu_1}{\sigma_1}\right)\right]^2} du$$
$$= \frac{1}{2\pi\sigma_1\sqrt{1-\rho^2}} e^{-\frac{1-\rho^2}{2(1-\rho^2)}\left(\frac{x-\mu_1}{\sigma_1}\right)^2} \int_{-\infty}^{\infty} e^{-\frac{1}{2(1-\rho^2)}\left[u - \rho\left(\frac{x-\mu_1}{\sigma_1}\right)\right]^2} du$$
$$= \frac{1}{2\pi\sigma_1\sqrt{1-\rho^2}} e^{-\frac{1}{2}\left(\frac{x-\mu_1}{\sigma_1}\right)^2} \int_{-\infty}^{\infty} e^{-\frac{1}{2(1-\rho^2)}\left[u - \rho\left(\frac{x-\mu_1}{\sigma_1}\right)\right]^2} du.$$

Put $\frac{1}{\sqrt{(1-\rho^2)}}\left[u - \rho\left(\frac{x-\mu_1}{\sigma_1}\right)\right] = t$, then $du = \sqrt{(1-\rho^2)}dt$. We get,

$$f_X(x) = \frac{1}{2\pi\sigma_1\sqrt{1-\rho^2}} e^{-\frac{1}{2}\left(\frac{x-\mu_1}{\sigma_1}\right)^2} \int_{-\infty}^{\infty} e^{-\frac{1}{2(1-\rho^2)}\left[u - \rho\left(\frac{x-\mu_1}{\sigma_1}\right)\right]^2} du$$
$$= \frac{1}{2\pi\sigma_1\sqrt{1-\rho^2}} e^{-\frac{1}{2}\left(\frac{x-\mu_1}{\sigma_1}\right)^2} \int_{-\infty}^{\infty} e^{-\frac{t^2}{2}} \sqrt{(1-\rho^2)}dt$$

$$= \frac{1}{2\pi\sigma_1} e^{-\frac{1}{2}\left(\frac{x-\mu_1}{\sigma_1}\right)^2} \int_{-\infty}^{\infty} e^{-\frac{t^2}{2}} dt$$
$$= \frac{1}{2\pi\sigma_1} e^{-\frac{1}{2}\left(\frac{x-\mu_1}{\sigma_1}\right)^2} \sqrt{2\pi}$$
$$= \frac{1}{\sqrt{2\pi}\sigma_1} e^{-\frac{1}{2}\left(\frac{x-\mu_1}{\sigma_1}\right)^2},$$

where we used $\int_{-\infty}^{\infty} e^{-\frac{t^2}{2}} dt = \sqrt{2\pi}$.

Similarly, we shall get

$$f_Y(y) = \int_{-\infty}^{\infty} f_{XY}(x,y)dx = \frac{1}{\sqrt{2\pi}\sigma_2} e^{-\frac{1}{2}\left(\frac{y-\mu_2}{\sigma_2}\right)^2}.$$

Hence, $X \sim N(\mu_1, \sigma_1^2)$ and $Y \sim N(\mu_2, \sigma_2^2)$.

■

**Remark:**

We have proved that if $(X,Y) \sim \text{BVN}\ (\mu_1, \mu_2, \sigma_1^2, \sigma_2^2, \rho)$, then the marginal PDFs of $X$ and $Y$ are also normal. However, the converse is not true, i.e., if the marginal distributions of $X$ and $Y$ are normal, it does not necessarily imply that the joint distribution of $(X,Y)$ is bivariate normal.

**Theorem 14.6 (Conditional Distributions):**

Let $(X,Y) \sim \text{BVN}\ (\mu_1, \mu_2, \sigma_1^2, \sigma_2^2, \rho)$, then the conditional distribution of $X$ for a fixed $Y$, is given by

$$f_{X|Y}(x|y) = \frac{1}{\sqrt{2\pi}\sigma_1\sqrt{1-\rho^2}} e^{-\frac{1}{2(1-\rho^2)\sigma_1^2}\left[(x-\mu_1)-\rho\frac{\sigma_1}{\sigma_2}(y-\mu_2)\right]^2}, \quad (14.55.1)$$

and the conditional distribution of $Y$ for a fixed $X$, is given by

$$f_{Y|X}(y|x) = \frac{1}{\sqrt{2\pi}\sigma_2\sqrt{1-\rho^2}} e^{-\frac{1}{2(1-\rho^2)\sigma_2^2}\left[(y-\mu_2)-\rho\frac{\sigma_2}{\sigma_1}(x-\mu_1)\right]^2}. \quad (14.55.2)$$

**Proof:**

Conditional distribution of $X$ for a fixed $Y$, is given by

$$f_{X|Y}(x|y) = \frac{f_{XY}(x,y)}{f_Y(y)}$$
$$= \frac{\sqrt{2\pi}\sigma_2}{2\pi\sigma_1\sigma_2\sqrt{1-\rho^2}} e^{-\frac{1}{2(1-\rho^2)}\left[\frac{(x-\mu_1)^2}{\sigma_1^2}-\frac{2\rho(x-\mu_1)(y-\mu_2)}{\sigma_1\sigma_2}+\frac{(y-\mu_2)^2}{\sigma_2^2}\right]+\frac{1}{2}\left(\frac{y-\mu_2}{\sigma_2}\right)^2}$$
$$= \frac{1}{\sqrt{2\pi}\sigma_1\sqrt{1-\rho^2}} e^{-\frac{1}{2(1-\rho^2)}\left[\frac{(x-\mu_1)^2}{\sigma_1^2}-\frac{2\rho(x-\mu_1)(y-\mu_2)}{\sigma_1\sigma_2}+\frac{(y-\mu_2)^2}{\sigma_2^2}\right]+\frac{(1-\rho^2)}{2(1-\rho^2)}\left(\frac{y-\mu_2}{\sigma_2}\right)^2}$$
$$= \frac{1}{\sqrt{2\pi}\sigma_1\sqrt{1-\rho^2}} e^{-\frac{1}{2(1-\rho^2)}\left[\frac{(x-\mu_1)^2}{\sigma_1^2}-\frac{2\rho(x-\mu_1)(y-\mu_2)}{\sigma_1\sigma_2}+\frac{(y-\mu_2)^2}{\sigma_2^2}-(1-\rho^2)\left(\frac{y-\mu_2}{\sigma_2}\right)^2\right]}$$
$$= \frac{1}{\sqrt{2\pi}\sigma_1\sqrt{1-\rho^2}} e^{-\frac{1}{2(1-\rho^2)}\left[\frac{(x-\mu_1)^2}{\sigma_1^2}-\frac{2\rho(x-\mu_1)(y-\mu_2)}{\sigma_1\sigma_2}+\frac{(y-\mu_2)^2}{\sigma_2^2}\{1-(1-\rho^2)\}\right]}$$
$$= \frac{1}{\sqrt{2\pi}\sigma_1\sqrt{1-\rho^2}} e^{-\frac{1}{2(1-\rho^2)}\left[\frac{(x-\mu_1)^2}{\sigma_1^2}-\frac{2\rho(x-\mu_1)(y-\mu_2)}{\sigma_1\sigma_2}+\frac{(y-\mu_2)^2}{\sigma_2^2}\rho^2\right]}$$
$$= \frac{1}{\sqrt{2\pi}\sigma_1\sqrt{1-\rho^2}} e^{-\frac{1}{2(1-\rho^2)}\left[\left(\frac{x-\mu_1}{\sigma_1}\right)-\rho\left(\frac{y-\mu_2}{\sigma_2}\right)\right]^2}$$

$$= \frac{1}{\sqrt{2\pi}\sigma_1\sqrt{1-\rho^2}} e^{-\frac{1}{2(1-\rho^2)\sigma_1^2}\left[(x-\mu_1)-\rho\frac{\sigma_1}{\sigma_2}(y-\mu_2)\right]^2}$$

$$= \frac{1}{\sqrt{2\pi}\sigma_1\sqrt{1-\rho^2}} e^{-\frac{1}{2(1-\rho^2)\sigma_1^2}\left[x-\left(\mu_1+\rho\frac{\sigma_1}{\sigma_2}(y-\mu_2)\right)\right]^2},$$

which is the probability function of a univariate normal distribution with mean and variance given by

$$E(X|Y=y) = \mu_1 + \rho\frac{\sigma_1}{\sigma_2}(y-\mu_2),$$

$$V(X|Y=y) = \sigma_1^2(1-\rho^2).$$

Hence, the conditional distribution of $X$ for fixed $Y$ is given by

$$(X|Y=y) \sim N\left[\mu_1 + \rho\frac{\sigma_1}{\sigma_2}(y-\mu_2), \sigma_1^2(1-\rho^2)\right].$$

Similarly, the conditional distribution of RVs $Y$ for, a fixed $X$ is

$$f_{Y|X}(y|x) = \frac{f_{XY}(x,y)}{f_X(x)} = \frac{1}{\sqrt{2\pi}\sigma_2\sqrt{1-\rho^2}} e^{-\frac{1}{2(1-\rho^2)\sigma_2^2}\left[y-\left(\mu_2+\rho\frac{\sigma_2}{\sigma_1}(x-\mu_1)\right)\right]^2}.$$

Thus, the conditional distribution of $Y$ for fixed $X$ is given by

$$(Y|X=x) \sim N\left[\mu_2 + \rho\frac{\sigma_2}{\sigma_1}(x-\mu_1), \sigma_2^2(1-\rho^2)\right].$$

■

Note that when $X$ and $Y$ are independent, then $\rho = 0$ and $E(Y|X=x) = \mu_2 = E(Y)$.

### 14.7.2 Multinomial Distribution

The Multinomial experiment

- The experiment consists of $n$ identical trials.
- The outcome of each trial falls into one of $k$ categories.
- The probability that the outcome of a single trial falls into a particular category—say, category $i$—is $p_i$ and remains constant from trial to trial. This probability must be between 0 and 1, for each of the $k$ categories, and the sum of all $k$ probabilities is $\sum p_i = 1$.
- The trials are independent.
- The experimenter counts the observed number of outcomes in each category, written as $O_1$, $O_2$, ..., $O_k$, with $O_1 + O_2 + \cdots + O_k = n$.

You have probably noticed the similarity between the multinomial experiment and the binomial experiment. In fact, when there are $k = 2$ categories, the two experiments are identical, except for notation. Instead of $p$ and $q$, we write $p_1$ and $p_2$ to represent the probabilities for the two categories, "success" and "failure." Instead of $x$ and $(n - x)$, we write $O_1$ and $O_2$ to represent the observed number of "successes" and "failures."

Hence, the multinomial distribution is a probability distribution that generalizes the concept of the binomial distribution to more than two categories or outcomes. It is used to model experiments or situations where there are multiple possible outcomes, and each outcome has a certain probability of occurring.

It is defined by the following parameters:

- Number of categories or outcomes $k$:
  The multinomial distribution deals with $k$ mutually exclusive and exhaustive categories or outcomes.
- Number of trials $n$:
  The fixed number of independent trials or experiments.
- Probability vector $p$:
  A vector of probabilities, denoted as $\mathbf{p} = (p_1, p_2, \dots, p_k)$, where each $p_i$ represents the probability of occurrence for outcome $i$. The probabilities must satisfy the following conditions: $p_i \geq 0$ for all $i$ and $p_1 + p_2 + \cdots + p_k = 1$.

**Definition (Multinomial Distribution):**
The RVs $X_1, X_2, \dots, X_k$ that denote the number of trials that result in class 1, class 2, … , class $k$, respectively, have a multinomial distribution and the joint PMF is

$$P(X_1 = x_1, X_2 = x_2, \dots, X_p = x_p) = \frac{n!}{x_1!\,x_2!\dots x_k!} p_1^{x_1} p_2^{x_2} \dots p_k^{x_k}, \qquad (14.56)$$

for $x_1 + x_2 + \cdots + x_k = n$ and $p_1 + p_2 + \cdots + p_k = 1$. See Figure 14.4.

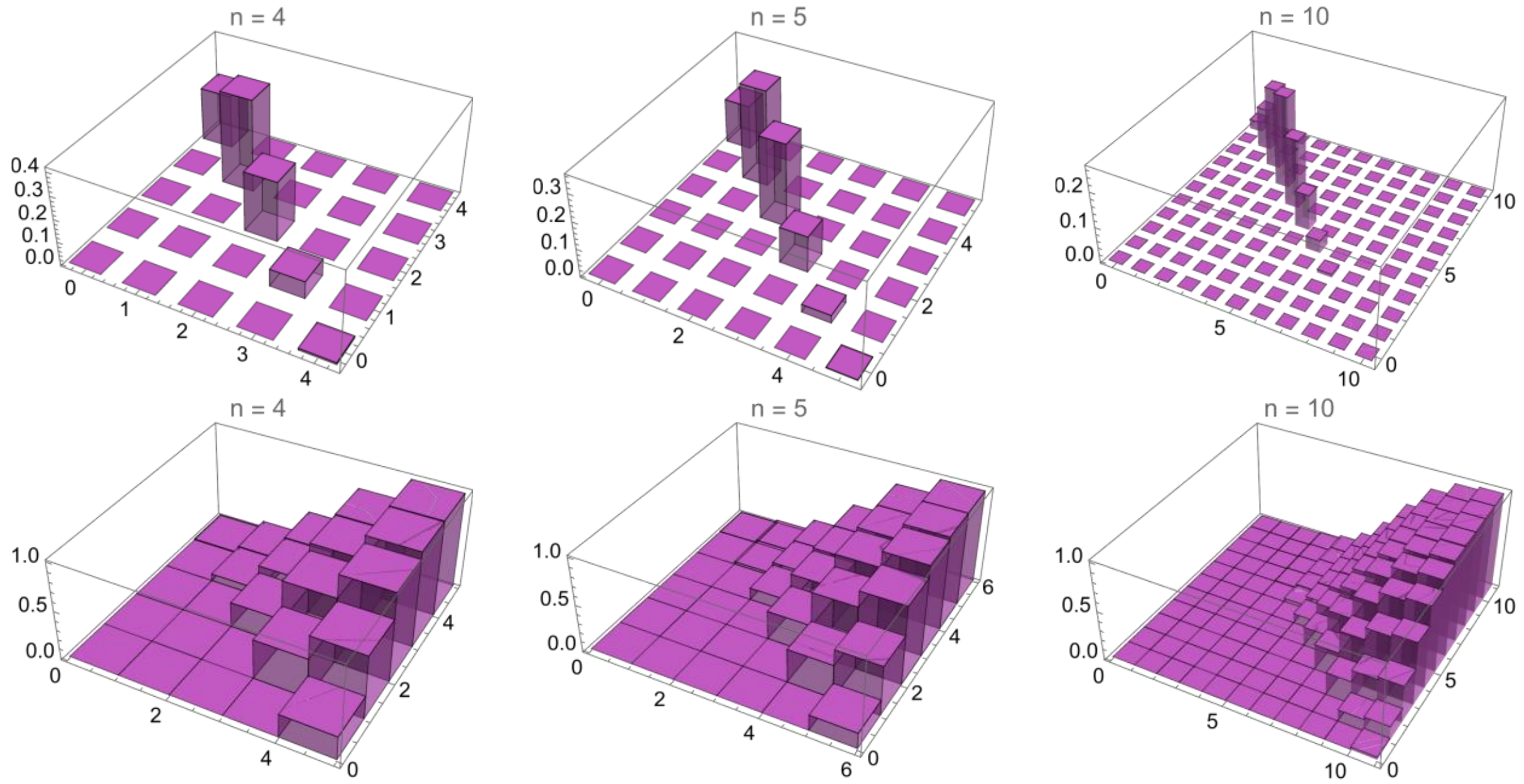

**Figure 14.4.** 3d plot of PDFs of the multinomial distributions with $n = 4{,}5{,}10$, $p_1 = 0.7$ and $p_2 = 0.3$ (upper panel). 3d plot of CDFs of the same multinomial distributions (lower panel).

***Example 14.15***

Suppose that a fair die is rolled seven times. Find the probability that 1 and 2 dots appear twice each; 3,4, and 5 dots once each; and 6 dots not at all.

***Solution***

Let $(X_1, X_2, \dots, X_6)$ be a six-dimensional random vector, where $X_i$ denotes the number of times $i$ dots appear in seven rolls of a fair die. Then $(X_1, X_2, \dots, X_6)$ is a multinomial RV with parameters $n = 7$ and $(p_1, p_2, \dots, p_6)$ where $p_i = \frac{1}{6}$ $(i = 1,2,\dots,6)$. Hence,

$$P(X_1 = 2, X_2 = 2, X_3 = 1, X_4 = 1, X_5 = 1, X_6 = 0) = \frac{7!}{2!\,2!\,1!\,1!\,1!\,0!}\left(\frac{1}{6}\right)^2\left(\frac{1}{6}\right)^2\left(\frac{1}{6}\right)^1\left(\frac{1}{6}\right)^1\left(\frac{1}{6}\right)^1\left(\frac{1}{6}\right)^0$$

$$= \frac{7!}{2!\,2!}\left(\frac{1}{6}\right)^7 = \frac{35}{6^5} \approx 0.0045.$$

**Theorem 14.7:** If $X_1, X_2, \ldots, X_k$ have a multinomial distribution, the marginal probability distribution of $X_i$ is binomial with

$$E[X_i] = np_i \text{ and } V(X_i) = np_i(1-p_i). \quad (14.57)$$

**Proof:**

The moment generating function is given by

$$\begin{aligned} M_X(t) &= M_{X_1,X_2,\ldots,X_k}(t_1,t_2,\ldots,t_k) \\ &= E\left[e^{\sum_{i=1}^k t_i X_i}\right] \\ &= \sum_x \frac{n!}{x_1!\,x_2!\ldots x_k!} p_1^{x_1} p_2^{x_2} \ldots p_k^{x_k} \left[e^{\sum_{i=1}^k t_i X_i}\right] \\ &= \sum_x \frac{n!}{x_1!\,x_2!\ldots x_k!} (p_1 e^{t_1})^{x_1} (p_2 e^{t_2})^{x_2} \ldots (p_k e^{t_k})^{x_k} \\ &= (p_1 e^{t_1} + p_2 e^{t_2} + \cdots + p_k e^{t_k})^n, \end{aligned}$$

where $x = (x_1, x_2, \ldots, x_k)$. Now,

$$\begin{aligned} M_{X_1}(t) &= M_X(t_1, 0, \ldots, 0) \\ &= (p_1 e^{t_1} + p_2 + \cdots + p_k)^n \\ &= \left((1-p_1) + p_1 e^{t_1}\right)^n. \end{aligned}$$

Since, $p_1 + p_2 + \cdots + p_k = 1 \quad \Rightarrow p_2 + \cdots + p_k = 1 - p_1$. Hence, $X_1 \sim B(n, p_1)$.

Similarly, we shall get:

$$X_i \sim B(n, p_i); \quad i = 1,2,\ldots,k,$$

$$E[X_i] = np_i,$$

$$V(X_i) = np_i(1-p_i).$$

■

**Theorem 14.8:** If $X_1, X_2, \ldots, X_k$ have a multinomial distribution, we have

$$\mathrm{Cov}(X_i, X_j) = -np_i p_j, \quad (14.58)$$

$$\rho(X_i, X_j) = -\sqrt{\frac{p_i p_j}{(1-p_i)(1-p_j)}}. \quad (14.59)$$

**Proof:**

$$\begin{aligned} E[X_i X_j] &= \left[\frac{\partial^2 M}{\partial t_i \partial t_j}\right]_{t=0} \\ &= \left[\frac{\partial}{\partial t_i}\{np_j e^{t_j}(p_1 e^{t_1} + p_2 e^{t_2} + \cdots + p_k e^{t_k})^{n-1}\}\right]_{t=0} \\ &= \left[(n-1)p_j e^{t_j} np_i e^{t_i}(p_1 e^{t_1} + p_2 e^{t_2} + \cdots + p_k e^{t_k})^{n-2}\right]_{t=0} \\ &= n(n-1)p_i p_j. \end{aligned}$$

Covariance,

$$\begin{aligned} \mathrm{Cov}(X_i, X_j) &= E[X_i X_j] - E[X_i]E[X_j] \\ &= n(n-1)p_i p_j - np_i np_j \\ &= (n^2 - n - n^2)p_i p_j \end{aligned}$$

$$= -np_i p_j.$$

Correlation coefficient,

$$\begin{aligned}
\rho(X_i, X_j) &= \frac{\mathrm{Cov}(X_i, X_j)}{\sigma_{X_i}\sigma_{X_j}} \\
&= \frac{-np_i p_j}{\sqrt{np_i(1-p_i)np_j(1-p_j)}} \\
&= \frac{-p_i p_j}{\sqrt{p_i p_j}\sqrt{(1-p_i)(1-p_j)}} \\
&= -\frac{\sqrt{p_i p_j}}{\sqrt{(1-p_i)(1-p_j)}} \\
&= -\sqrt{\frac{p_i p_j}{(1-p_i)(1-p_j)}}.
\end{aligned}$$

■

# CHAPTER 15

# MATHEMATICA LAB: BIVARIATE RANDOM VARIABLES AND DISTRIBUTIONS

Always remember that one of the notable advantages of Mathematica's functions is their ability to handle both discrete and continuous random variables seamlessly not only for univariate random variables but also for bivariate and multivariate random variables. This means that you can apply the `Probability`, `NProbability`, `PDF`, `CDF`, `Expectation`, `MomentGeneratingFunction`, and `CentralMomentGeneratingFunction` functions to a wide range of probability distributions, regardless of whether they are discrete, continuous, univariate, bivariate or multivariate random variables.

- We start by examining the marginal distribution, which allows us to focus on a single variable of interest while disregarding the other variables in a multivariate distribution. Mathematica offers efficient functions to compute the marginal distribution of a multivariate distribution, for example `MarginalDistribution`, which can be extremely useful for data analysis and modeling.
- Next, we focus on the concepts of covariance and correlation, two measures that quantify the linear relationship between two random variables. Mathematica provides built-in functions (`Covariance and Correlation`) to calculate these measures, making it effortless to explore the strength and direction of the relationship between variables in a dataset.
- Moving forward, we explore the binormal distribution, which is a bivariate probability distribution commonly used in applications where two variables are correlated. We demonstrate how to compute probabilities, generate random samples, and visualize the binormal distribution using Mathematica's functions.
- Lastly, we discuss multinomial distribution, which models scenarios where multiple outcomes occur simultaneously with specific probabilities. Mathematica offers versatile tools to work with multinomial distributions, allowing us to compute probabilities, generate random samples, and perform various analyses.
- Throughout the chapter, we will provide step-by-step explanations and practical examples to demonstrate the usage of Mathematica functions for these concepts. By the end of this chapter, you will have a solid understanding of how to leverage Mathematica's capabilities to analyze and work with probability distributions effectively.

In the following table, we list the built-in functions that are used in this chapter.

| | |
|---|---|
| `MarginalDistribution` | `BinormalDistribution` |
| `Covariance` | `MultinomialDistribution` |
| `Correlation` | |

Therefore, we divided this chapter into three units to cover the above topics.

**Chapter 15 Outline**

Unit 15.1. Marginal Distribution, Covariance, and Correlation
Unit 15.2. Binormal Distribution
Unit 15.3. Multinomial Distribution

# UNIT 15.1

# MARGINAL DISTRIBUTION, COVARIANCE AND CORRELATION

| | |
|---|---|
| `MarginalDistribution[dist,k]` | represents a univariate marginal distribution of the k^(th) coordinate from the multivariate distribution dist. |
| `MarginalDistribution[dist, {k1,k2,…}]` | represents a multivariate marginal distribution of the {k1,k2,…} coordinates. |

***Mathematica Examples 15.1***

```
Input    (* One-dimensional marginal distributions: *)
         MarginalDistribution[BinormalDistribution[{0,0},{1,1},ρ],1]
         MarginalDistribution[MultinomialDistribution[n,{p1,p2,p3}],1]

         (* Two-dimensional marginal distributions: *)
         MarginalDistribution[
          MultinormalDistribution[
            {0,0,0},
            {
             {1,0.5,0.3},
             {0.5,2,-0.2},
             {0.3,-0.2,3}
             }
            ],
          {1,2}
          ]

Output   NormalDistribution[0,1]
Output   BinomialDistribution[n,p1]
Output   MultinormalDistribution[{0,0},{{1,0.5},{0.5,2}}]
```

***Mathematica Examples 15.2***

```
Input    (* Univariate marginals behave as univariate distributions: *)
         dist=BinormalDistribution[{μ1,μ2},{σ1,σ2},ρ];
         m1=MarginalDistribution[dist,1]

         (* Find distribution functions and moments of marginal distribution: *)
         {PDF[m1,x],CDF[m1,x],Mean[m1],Variance[m1],Skewness[m1],Kurtosis[m1]}

Output   NormalDistribution[μ1,σ1]
Output
```

$\left\{\frac{e^{-\frac{(x-\mu1)^2}{2\,\sigma1^2}}}{\sqrt{2\pi}\,\sigma1},\ 1/2\ \mathrm{Erfc}[(-x+\mu1)/(\sqrt{2}\ \sigma1)],\ \mu1,\ \sigma1\text{^}2,\ 0,\ 3\right\}$

***Mathematica Examples 15.3***

```
Input    (* Multivariate marginals behave like a multivariate distribution: *)
         dist=MultinormalDistribution[
            {μ1,μ2,μ3},
```

```
             {
              {σ11,σ12,σ13},
              {σ21,σ22,σ21},
              {σ31,σ32,σ33}
              }
             ];
        m12=MarginalDistribution[dist,{1,2}]

        (* Find moments of marginal distribution: *)
        {Mean[m12],Variance[m12],Skewness[m12],Kurtosis[m12]}

Output  MultinormalDistribution[{μ1,μ2},{{σ11,σ12},{σ21,σ22}}]
Output  {{μ1,μ2},{σ11,σ22},{0,0},{3,3}}
```

***Mathematica Examples 15.4***

```
Input   (* In some cases, marginal distributions will not automatically simplify: *)
        dist=MultinomialDistribution[100,{.3,.5,.2}];

        (* Univariate marginals simplify to a BinomialDistribution: *)
        MarginalDistribution[dist,1]

        (* The multivariate marginals do not simplify: *)
        m12=MarginalDistribution[dist,{1,2}]

        (* The resulting marginal can still be used like any other distribution: *)
        DiscretePlot3D[
         Evaluate[
          PDF[m12,{x,y}]
          ],
         {x,10,50},
         {y,35,60},
         ExtentSize->1,
         PlotStyle->Directive[Purple,Opacity[.5]],
         ImageSize->250
         ]

Output  BinomialDistribution[100,0.3]
Output  MarginalDistribution[MultinomialDistribution[100,{0.3,0.5,0.2}],{1,2}]
Output
```

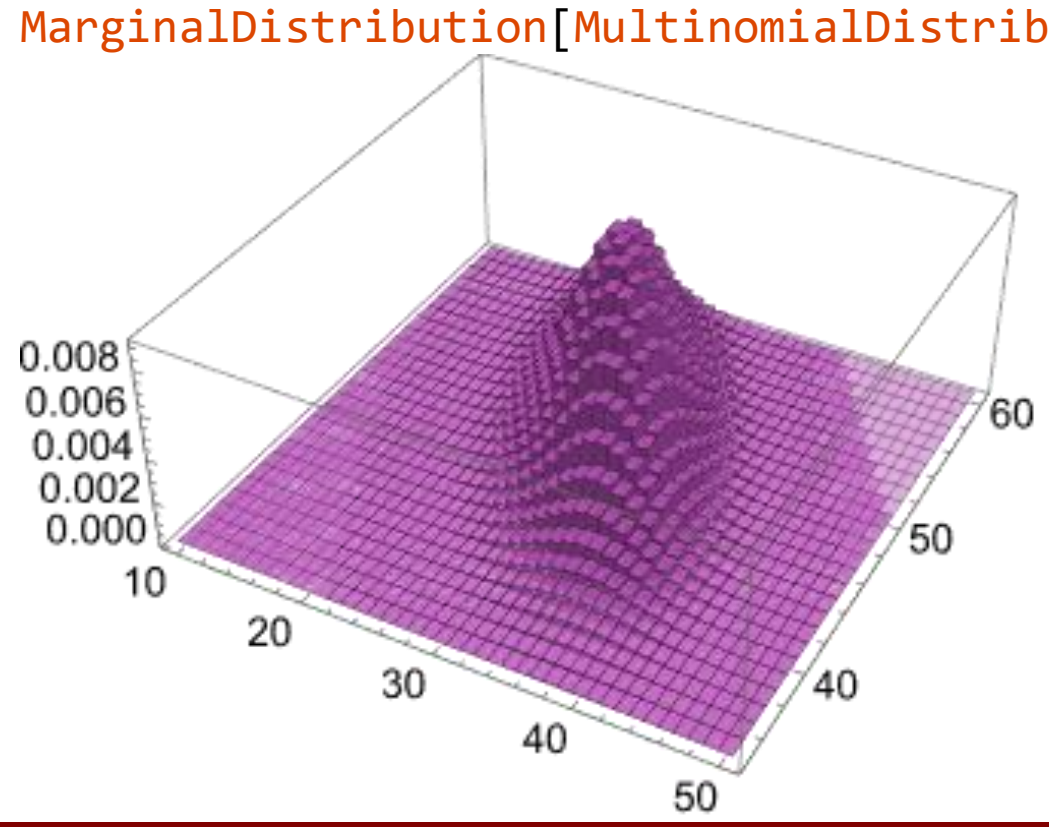

***Mathematica Examples 15.5***

```
Input   (* In this example, we first define a joint distribution called jointDist using
        MultinormalDistribution. We then use MarginalDistribution to compute the marginal
        distribution of the first variable (marginalDist1) and the second variable
        (marginalDist2). Next,we generate 1000 random samples from the joint distribution
        using RandomVariate and store them in the samples variable. We also generate 1000
        random samples from the marginal distributions of the first and second variables,
```

```
respectively, and store them in marginalSamples1 and marginalSamples2. Finally, we
create plots to visualize the joint distribution and the marginal distributions using
Histogram and Histogram3D: *)

(* Define a joint distribution: *)
jointDist=MultinormalDistribution[{0,0},{{1,0.5},{0.5,2}}];

(* Compute the marginal distribution of the first variable: *)
marginalDist1=MarginalDistribution[jointDist,1];

(* Compute the marginal distribution of the second variable: *)
marginalDist2=MarginalDistribution[jointDist,2];

(*Generate random samples from the joint distribution: *)
samples=RandomVariate[jointDist,1000];

(* Generate random samples from the marginal distribution of the first variable: *)
marginalSamples1=RandomVariate[marginalDist1,1000];

(* Generate random samples from the marginal distribution of the second variable: *)
marginalSamples2=RandomVariate[marginalDist2,1000];

(* Plot the joint distribution: *)
Histogram3D[
 samples,
 Automatic,
 "PDF",
 PlotRange->All,
 ColorFunction->Function[{height},Opacity[height]],
 ChartStyle->Purple,
 ImageSize->270
 ]

(* Plot the marginal distribution of the first variable: *)
Histogram[
 marginalSamples1,
 Automatic,
 "PDF",
 ColorFunction->Function[{height},Opacity[height]],
 ChartStyle->Purple,
 ImageSize->220,
 PlotRange->All
 ]

(* Plot the marginal distribution of the second variable: *)
Histogram[
 marginalSamples2,
 Automatic,
 "PDF",
 PlotRange->All,
 ColorFunction->Function[{height},Opacity[height]],
 ChartStyle->Purple,
 ImageSize->220
 ]
```

Output

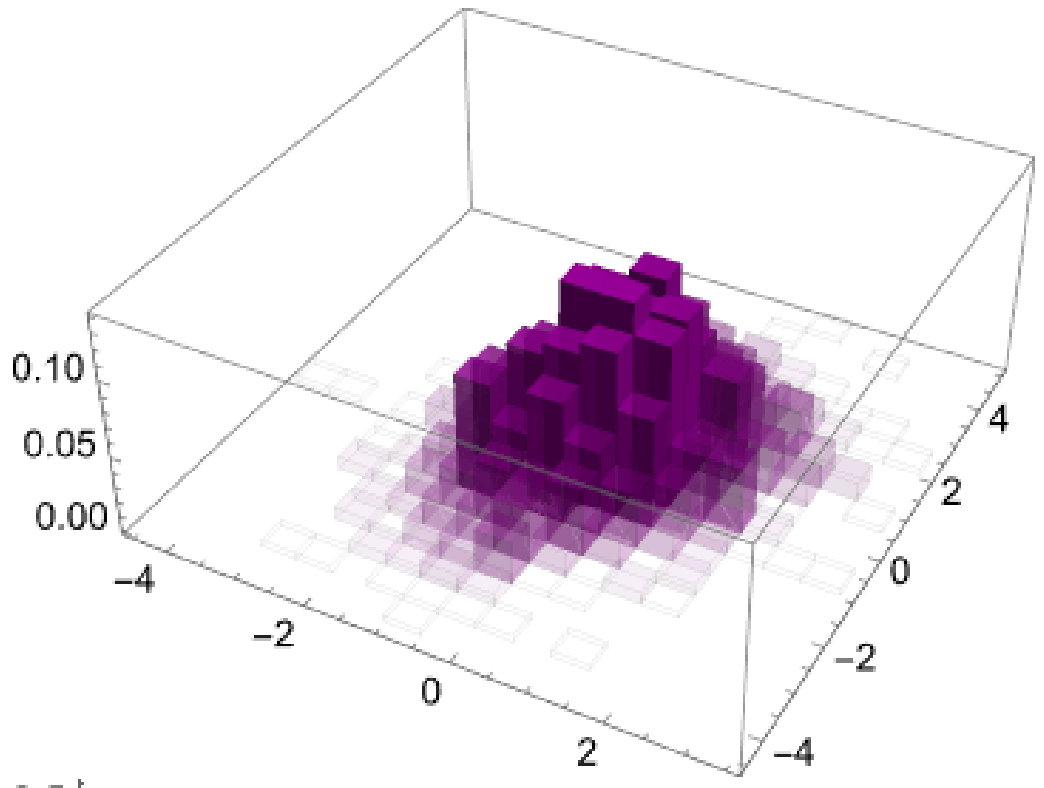

Output

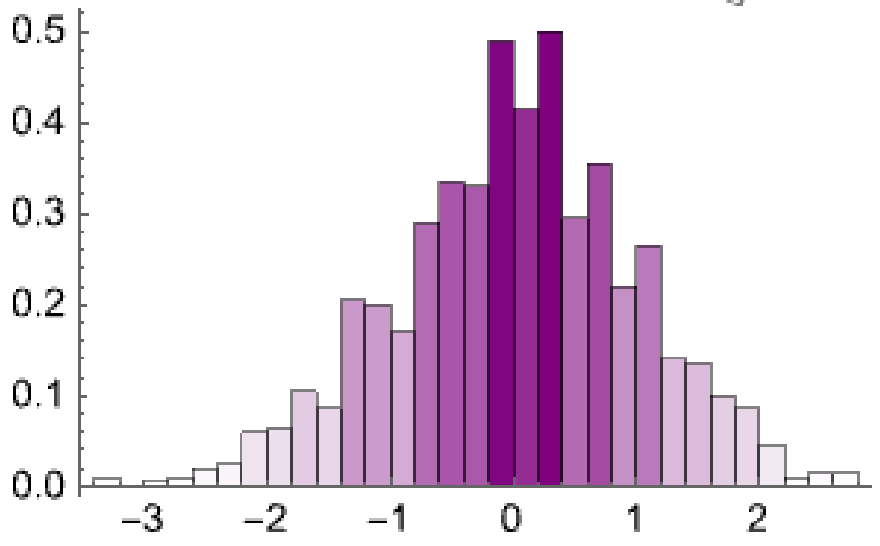

Output

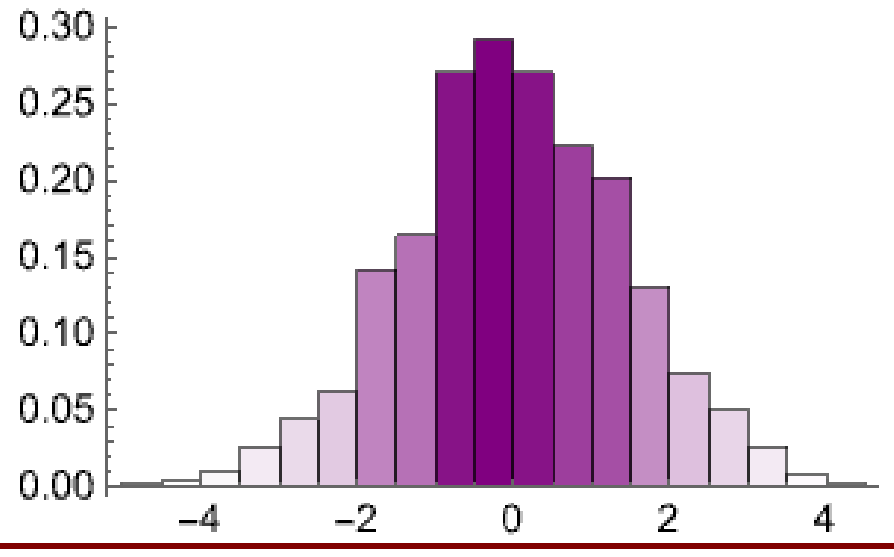

***Mathematica Examples 15.6***

Input

```
(* Step 1: Define the joint distribution*)
jointDistribution=MultinormalDistribution[{0,0,0},{{2,1,0.5},{1,3,-0.2},{0.5,-
0.2,1.5}}];

(* Step 2: Calculate the marginal distributions*)
marginalX=MarginalDistribution[jointDistribution,1];
marginalY=MarginalDistribution[jointDistribution,2];
marginalZ=MarginalDistribution[jointDistribution,3];

(* Step 3: Plot the marginal distributions*)
Plot[
 {
  PDF[marginalX,x],
  PDF[marginalY,x],
  PDF[marginalZ,x]
  },
 {x,-6,6},
 PlotLegends->{"Marginal X","Marginal Y","Marginal Z"},
 PlotRange->All,
 Frame->True,
 FrameLabel->{"X, Y, or Z","Density"},
 ImageSize->250
 ]
```

Output

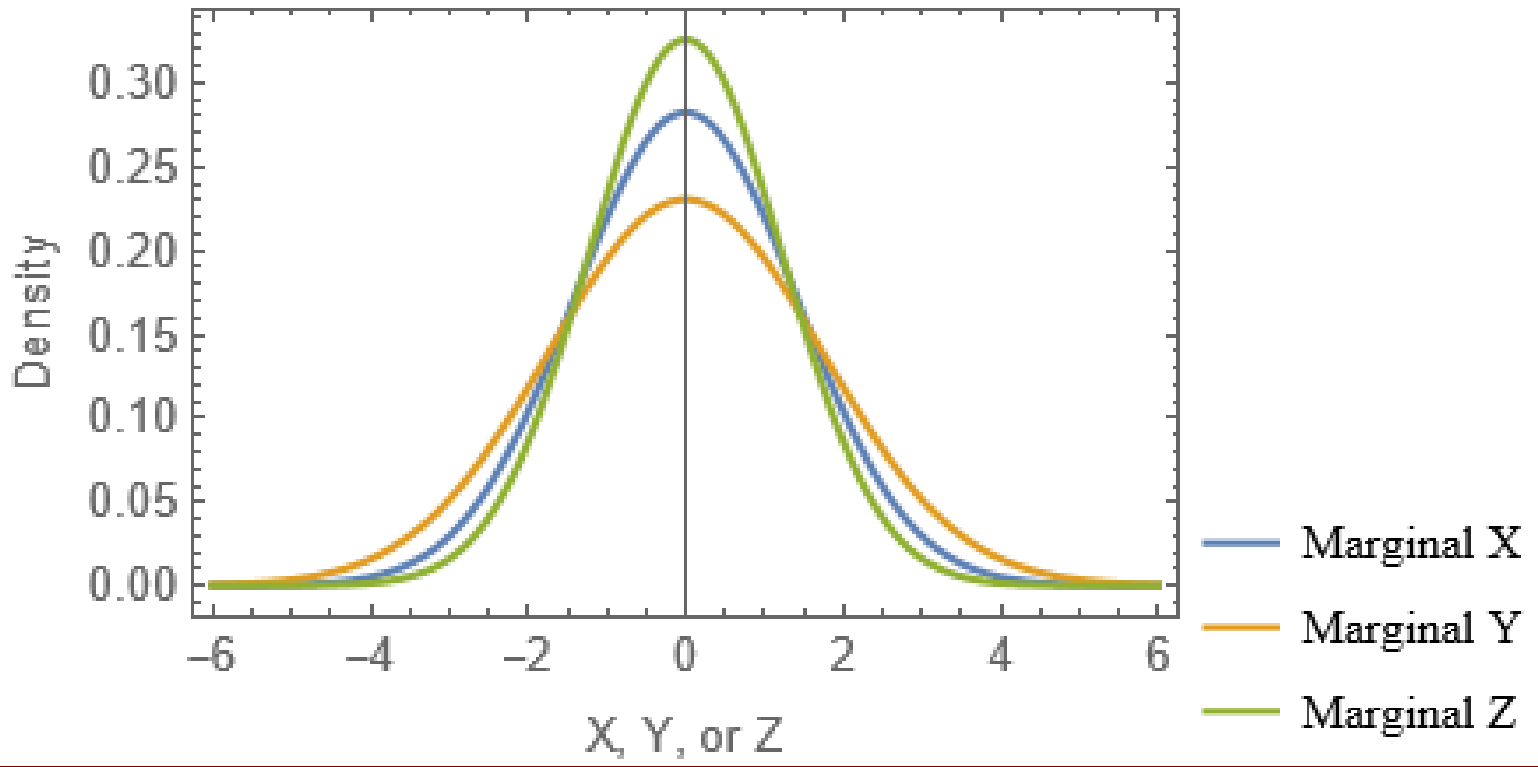

***Mathematica Examples 15.7***

Input

```
(* In this example, we define a multivariate distribution using MultinormalDistribution, which represents a multivariate normal distribution with three variables. We generate random samples from the multivariate distribution using RandomVariate and then extract the marginal distribution of the first two variables using MarginalDistribution. We plot a scatter plot of the marginal distribution using ListPlot. Finally, we calculate the means of the marginal distribution using Mean and display them: *)

(* Define a multivariate distribution: *)
multivariateDist=MultinormalDistribution[
    {0,0,0},
    {
     {1,0.5,0.3},
     {0.5,2,-0.2},
     {0.3,-0.2,3}
     }
    ];

(* Generate random samples from the multivariate distribution: *)
samples=RandomVariate[multivariateDist,1000];

(* Extract the marginal distribution of the first two variables: *)
marginalDist=MarginalDistribution[multivariateDist,{1,2}]

(* Calculate the means of the marginal distribution: *)
mean=Mean[marginalDist];

(* Display the means: *)
Row[{"Mean of Marginal Distribution: ",mean}]

(* Plot the scatter plot of the marginal distribution: *)
ListPlot[
 samples[[All,{1,2}]],
 PlotStyle->Directive[PointSize[0.02],Purple,Opacity[0.3]],
 AspectRatio->1,
 PlotLabel->"Marginal Distribution",
 ImageSize->220
 ]
```

Output `MultinormalDistribution[{0,0},{{1,0.5},{0.5,2}}]`

Output `Mean of Marginal Distribution:  {0,0}`

Output

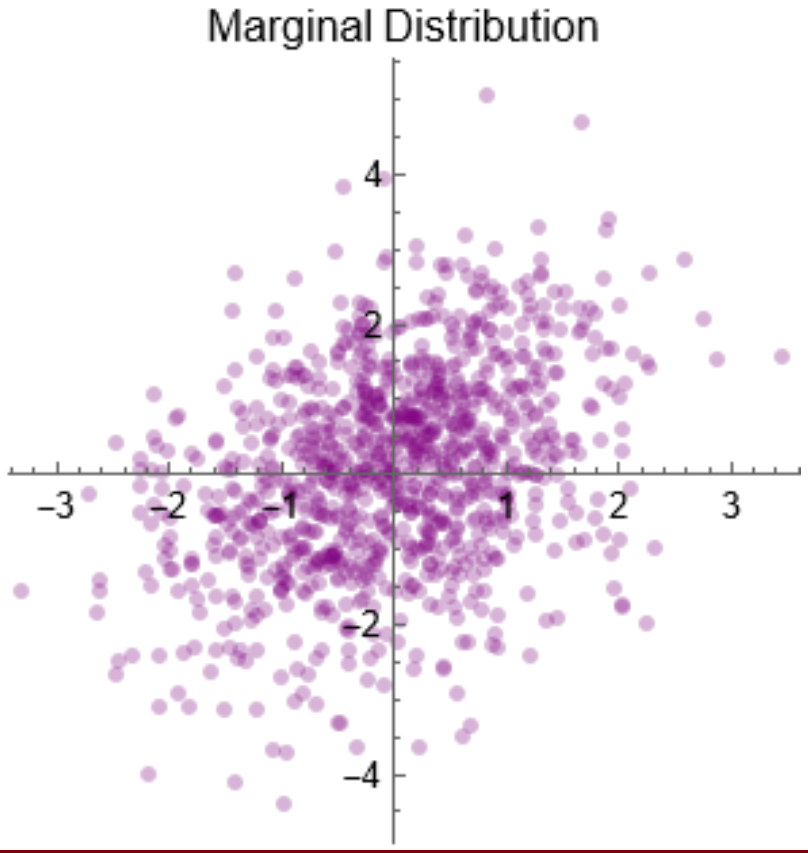

| | |
|---|---|
| Covariance[v1,v2] | gives the covariance between the vectors v1 and v2. |
| Covariance[m] | gives the sample covariance matrix for observations in matrix m. |
| Covariance[m1,m2] | gives the covariance matrix for the matrices m1 and m2. |
| Covariance[dist] | gives the covariance matrix for the multivariate symbolic distribution dist. |
| Covariance[dist,i,j] | gives the (i,j)^(th) covariance for the multivariate symbolic distribution dist. |
| Correlation[v1,v2] | gives the correlation between the vectors v1 and v2. |
| Correlation[m] | gives the sample correlation matrix for observations in matrix m. |
| Correlation[m1,m2] | gives the correlation matrix for the matrices m1 and m2. |
| Correlation[dist] | gives the correlation matrix for the multivariate symbolic distribution dist. |
| Correlation[dist,i,j] | gives the (i,j)^(th) correlation for the multivariate symbolic distribution dist. |

***Mathematica Examples 15.8***

```
Input    (* Covariance and Correlation Work with a large arrays: *)
         array1=RandomVariate[ExponentialDistribution[1/2],10^3];
         array2=RandomVariate[NormalDistribution[2,5],10^3];
         Covariance[array1,array2]
         Correlation[RandomReal[1,10^7],RandomReal[1,10^7]]

Output   -0.0500528
Output   -0.000519733
```

***Mathematica Examples 15.9***

```
Input    (* Covariance and Correlation Work with a continuous multivariate distribution: *)
         dist=BinormalDistribution[ρ];
         MatrixForm[
          Covariance[dist]
          ]
         MatrixForm[
          Correlation[dist]
          ]

Output   ({
           {1, ρ},
           {ρ, 1}
          })
Output   ({
           {1, ρ},
           {ρ, 1}
          })
```

***Mathematica Examples 15.10***

Input
```
(* Covariance and Correlation Work with a discrete multivariate distribution: *)
MatrixForm[
 Covariance[
  MultivariatePoissonDistribution[μ,{2,4,5}]
  ]
 ]

MatrixForm[
 Correlation[
  MultivariatePoissonDistribution[μ,{2,4}]
  ]
 ]
```

Output
$$\begin{pmatrix} 2+\mu & \mu & \mu \\ \mu & 4+\mu & \mu \\ \mu & \mu & 5+\mu \end{pmatrix}$$

Output
$$\begin{pmatrix} 1 & \frac{\mu}{\sqrt{2+\mu}\sqrt{4+\mu}} \\ \frac{\mu}{\sqrt{2+\mu}\sqrt{4+\mu}} & 1 \end{pmatrix}$$

***Mathematica Examples 15.11***

Input
```
(* Covariance and Correlation Work with a derived distributions: *)
MatrixForm[
 Covariance[
  ProductDistribution[
   ExponentialDistribution[1/2],
   NormalDistribution[2,5]]
  ]
 ]

MatrixForm[
 Correlation[
  ProductDistribution[
   ExponentialDistribution[1/2],
   NormalDistribution[2,5]]
  ]
 ]
```

Output
```
({
  {4, 0},
  {0, 25}
 })
```

Output
```
({
  {1, 0},
  {0, 1}
 })
```

***Mathematica Examples 15.12***

Input
```
(* The diagonal elements of a correlation matrix are equal to 1:*)
Diagonal[
 Correlation[
  RandomReal[20,{50,4}]
  ]
 ]
```

Output
```
{1.,1.,1.,1.}
```

***Mathematica Examples 15.13***

```
Input   (* In this code, data1 and data2 represent the two sets of data for which you want
        to calculate the covariance. The Mean function is used to compute the mean of each
        dataset. Then, the covariance is calculated by subtracting the mean of each dataset
        from their respective data points, multiplying the differences together, taking the
        average of these products and scale the average by (length/(length-1) for the
        covariance of sample data: *)

        (*Define the function*)
        covariance[data1_,data2_]:=Module[
          {mean1,mean2,covariance},
          mean1=Mean[data1];
          mean2=Mean[data2];
          length=Length[data1];
          covariance=Mean[(data1-mean1) (data2-mean2)]*(length/(length-1));
          covariance]
        (*Test the function*)
        data1=RandomReal[1,10^7];
        data2=RandomReal[1,10^7];

        covariance[data1,data2]
        Covariance[data1,data2]

Output  6.28732*10^-6
Output  6.28732*10^-6
```

***Mathematica Examples 15.14***

```
Input   (* The covariance of a list with itself is the variance: *)

        (* Define a sample list: *)
        list={a,b,c,d,e};

        (* Compute the covariance matrix: *)
        covariance=Covariance[list,list];

        (* Compute the variance: *)
        variance=Variance[list];

        (* Check if the covariance and variance are equal: *)
        covariance==variance

Output  True
```

***Mathematica Examples 15.15***

```
Input   (* The covariance matrix is symmetric and positive semidefinite: *)
        (* Define a sample dataset: *)
        data=RandomVariate[BinormalDistribution[1/3],10^3];

        (* Compute the covariance matrix: *)
        covMatrix=Covariance[data];

        (* Display the covariance matrix: *)
        MatrixForm[covMatrix]

        (* Check if the covariance matrix is symmetric: *)
        symmetric=SymmetricMatrixQ[covMatrix]

        (* Check if the covariance matrix is positive semidefinite: *)
        positiveSemidefinite=PositiveSemidefiniteMatrixQ[covMatrix]
```

```
          (* Matrix plot of the covariance matrix: *)
          MatrixPlot[
           covMatrix,
           ColorFunction->"Temperature",
           ImageSize->170
           ]

Output    ({
            {0.987536, 0.339421},
            {0.339421, 1.03269}
           })
Output    True
Output    True
Output
```

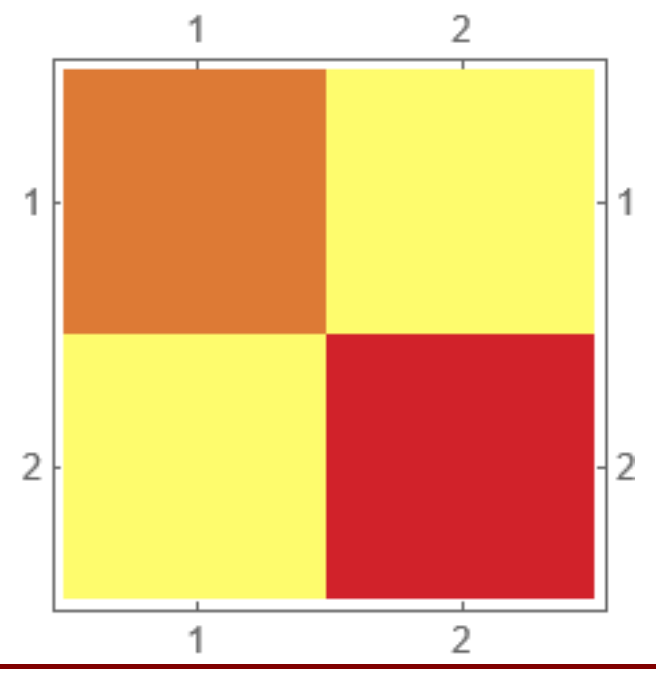

*Mathematica Examples 15.16*

```
Input     (* The covariance tends to be large only on the diagonal of a random matrix: *)
          ArrayPlot[
           Covariance[
            RandomReal[{-3,3},{20,20}]
            ],
           ColorFunction->"Rainbow",
           ImageSize->170,
           PlotLegends->Automatic
           ]

Output
```

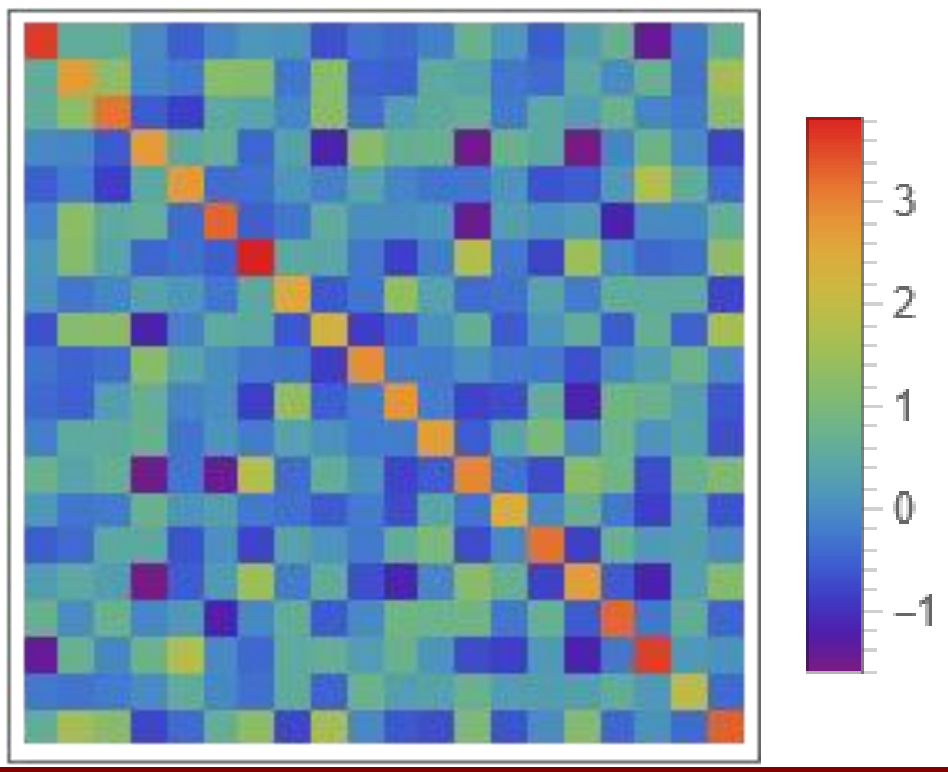

*Mathematica Examples 15.17*

```
Input     (* A covariance matrix scaled by standard deviations is a correlation matrix: *)
          data=RandomReal[5,{20,5}];
          s=DiagonalMatrix[1/StandardDeviation[data]];
          Correlation[data]==s.Covariance[data].s
Output    True
```

**Mathematica Examples 15.18**

Input
```
(*Covariance can be used to measure linear association:*)
data=Table[
   RandomVariate[BinormalDistribution[i],3000],
   {i,{-.99,-.75,-.25,-.5,0.,.25,.5,.75,.99}}
   ];

Grid[
 Partition[
  Table[
   ListPlot[
    i,
    PlotStyle->Directive[Purple,PointSize[0.01],Opacity[0.2]],
    FrameTicks->None,
    Frame->True,
    Axes->None,
    ImageSize->170,
    PlotLabel->Row[{"σxy : ",Covariance[i][[1,2]]}]
    ],
   {i,data}
   ],
  3
  ]
 ]
```

Output

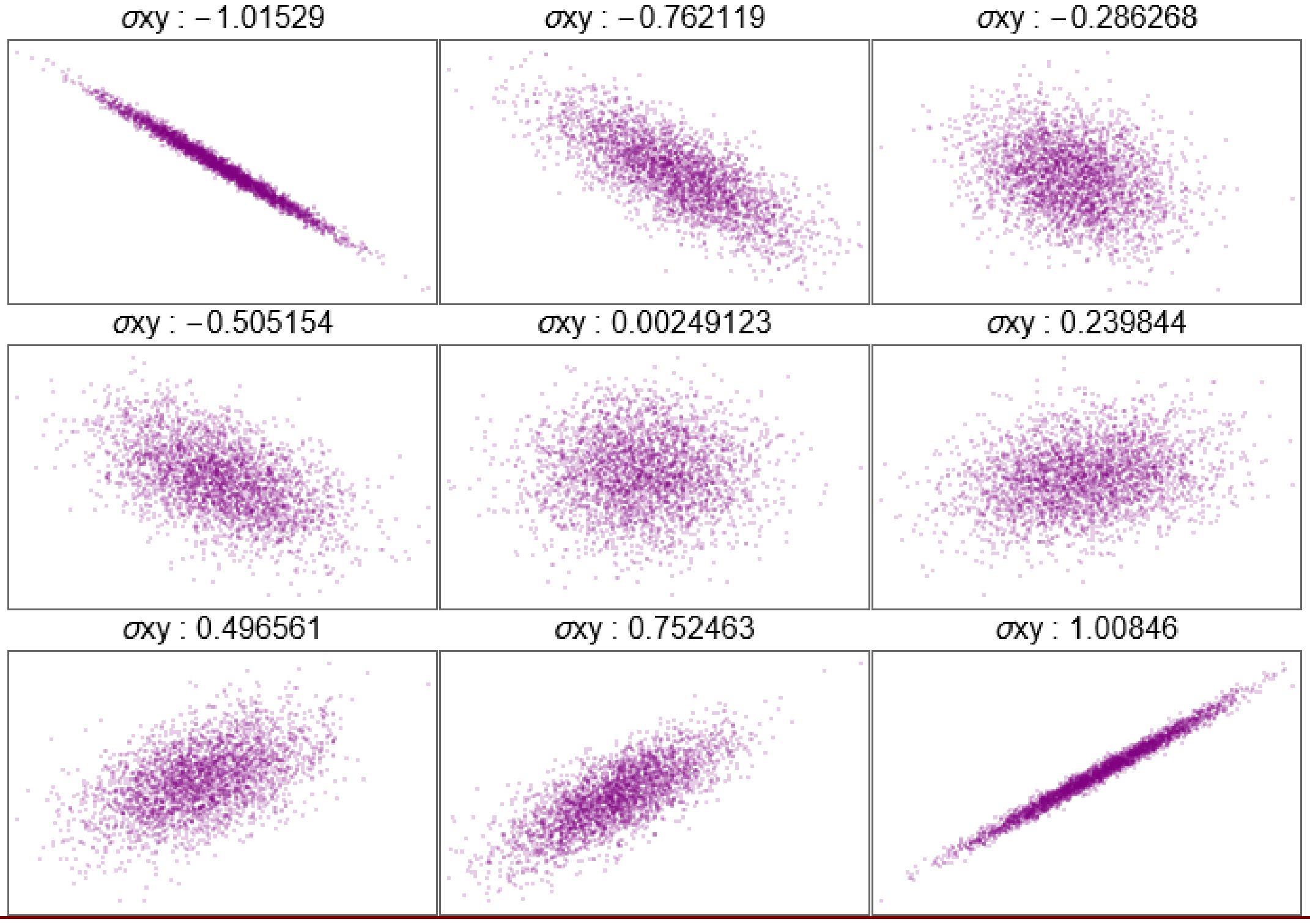

**Mathematica Examples 15.19**

Input
```
Manipulate[
 Module[
  {data,cov,plot},

  (*Generate random data from a BinormalDistribution*)
  data=RandomVariate[BinormalDistribution[{0,0},{1,1},ρ],1000];
```

```
(*Calculate the covariance matrix*)
cov=Covariance[data];

(*Create the ListPlot*)
plot=ListPlot[
  data,
  PlotStyle->Directive[Purple,PointSize[0.015],Opacity[0.5]],
  Frame->True,
  Axes->False,
  PlotLabel->StringForm["ρ = ``\nCovariance = ``",ρ,cov[[1,2]]]
  ];

(*Return the plot*)
plot
],
{{ρ,-0.9},-0.9,0.9,0.01,Appearance->"Labeled"}
]
```

Output

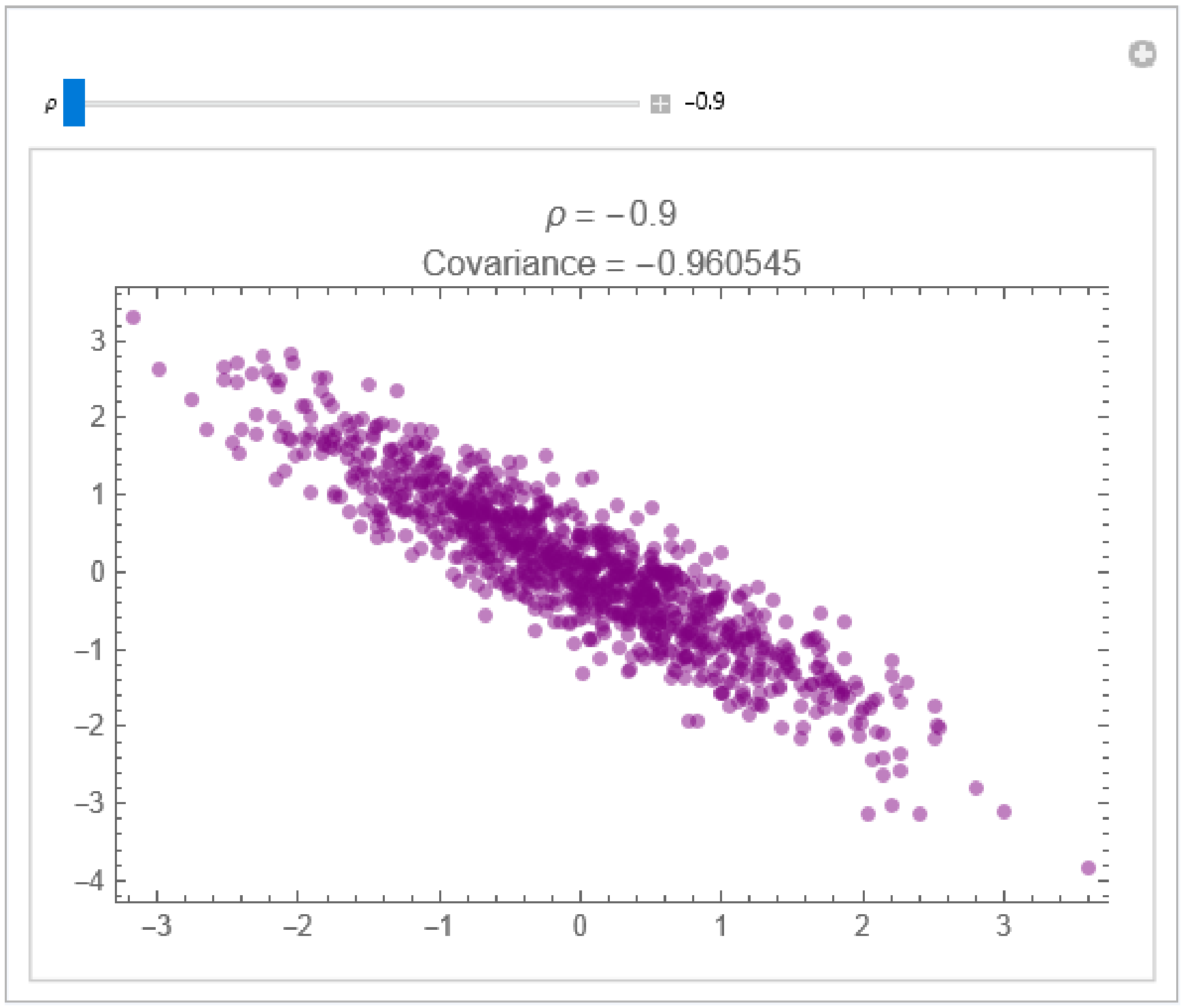

***Mathematica Examples 15.20***

Input

```
(* In this code, the distPlot function generates a sample of 1000 random variates
from the specified distribution and calculates the covariance matrix. It then displays
a 3D histogram of the data and a matrix plot of the covariance matrix. The Manipulate
function allows the user to interactively change the parameters μ1,μ2,σ1,σ2,and ρ,
which define the covariance matrix of a Binormal distribution. The resulting changes
are immediately reflected in the plots: *)

distPlot[dist_,covRange_]:=Module[
  {data,cov},
  SeedRandom[123];
  (*for reproducibility*)
  data=RandomVariate[dist,1000];
  cov=Covariance[data];
```

```
    GraphicsRow[
     {
      Histogram3D[
       data,
       Automatic,
       "PDF",
       ColorFunction->"Rainbow",
       ImageSize->300
       ],
      MatrixPlot[
       cov,
       ColorFunction->"TemperatureMap",
       FrameTicks->Automatic,
       Mesh->True,
       ImageSize->300
       ]
      }
     ]
    ]

  Manipulate[
   dist=BinormalDistribution[{μ1,μ2},{σ1,σ2},ρ];
   covRange=Range[1,1000,1];
   distPlot[
    dist,
    covRange
    ],
   {{μ1,1},1,2,Appearance->"Labeled"},
   {{μ2,1},1,2,Appearance->"Labeled"},
   {{σ1,1},1,2,Appearance->"Labeled"},
   {{σ2,1},1,2,Appearance->"Labeled"},
   {{ρ,-0.9},-0.9,0.9,Appearance->"Labeled"}
   ]
```

Output

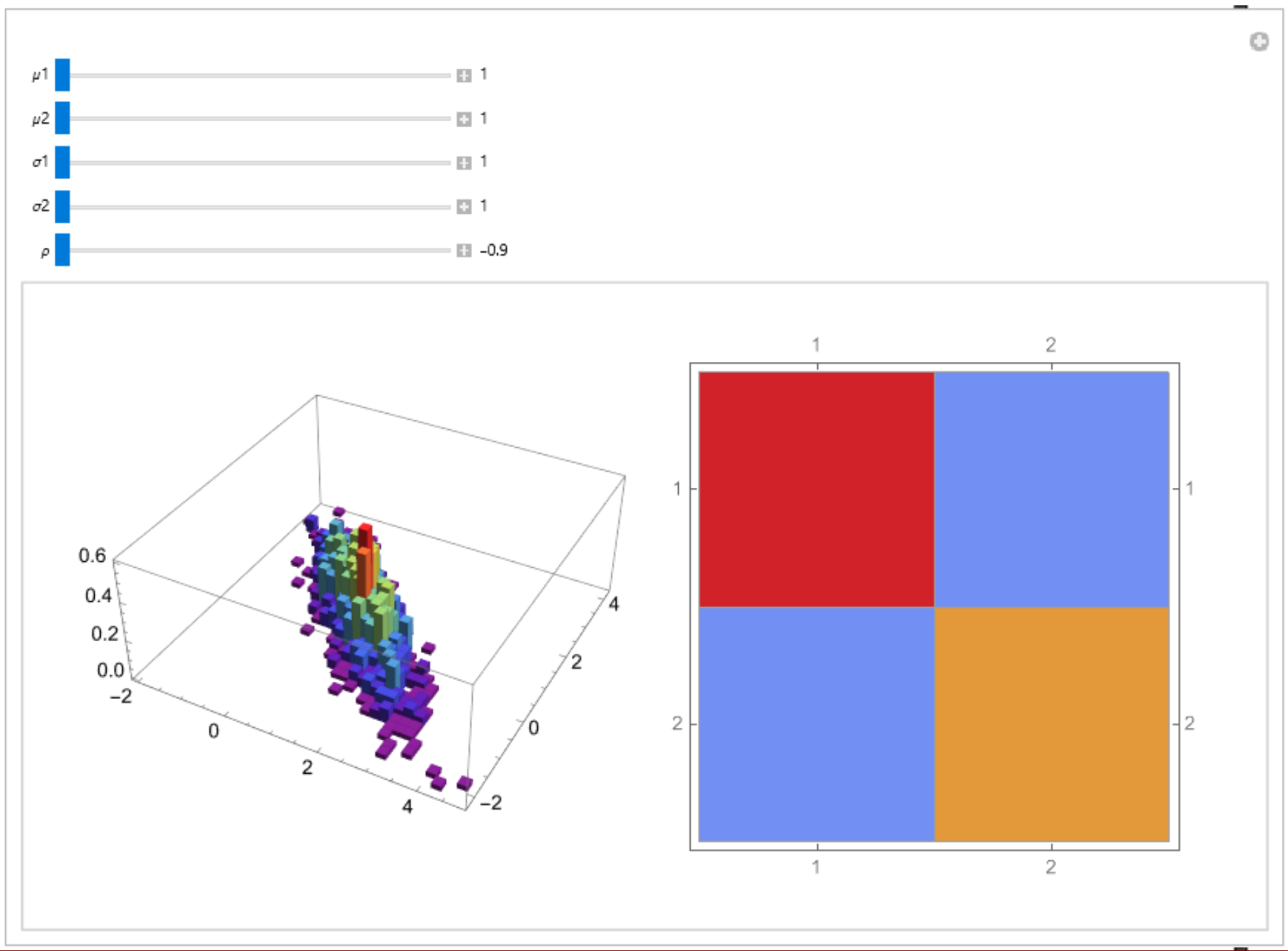

**Mathematica Examples 15.21**

```
Input    (* Covariance and Correlation are the same for standardized vectors: *)
         sample=RandomVariate[DiscreteUniformDistribution[{{2,3},{4,5}}],200];
         Covariance[sample]===Correlation[sample]
         Covariance[Standardize[sample]]===Correlation[sample]

Output   False
Output   True
```

**Mathematica Examples 15.22**

```
Input    (*The diagonal of a covariance matrix is the variance:*)
         data=RandomReal[5,{20,5}];
         Diagonal[Covariance[data]]
         Variance[data]

Output   {2.11181,2.41371,2.21731,2.19794,2.85778}
Output   {2.11181,2.41371,2.21731,2.19794,2.85778}
```

# UNIT 15.2

# BINORMAL DISTRIBUTION

| | |
|---|---|
| `BinormalDistribution[{μ1,μ2}, {σ1,σ2},ρ]` | represents a bivariate normal distribution with mean {μ1,μ2} and covariance matrix {{σ12,ρ σ1 σ2},{ρ σ1 σ2,σ22}}. |
| `BinormalDistribution[{σ1,σ2},ρ]` | represents a bivariate normal distribution with zero mean. |
| `BinormalDistribution[ρ]` | represents a bivariate normal distribution with zero mean and covariance matrix {{1,ρ },{ρ,1}}. |
| `Covariance[dist]` | gives the covariance matrix for the multivariate symbolic distribution dist. |
| `Covariance[dist,i,j]` | gives the (i,j)^(th) covariance for the multivariate symbolic distribution dist. |

***Mathematica Examples 15.23***

Input
```
(* BinormalDistribution[{μ1,μ2},{σ1,σ2},ρ] represents a bivariate normal distribution defined over pairs of real numbers with the property that each of the first and second marginal distributions is normal distribution, i.e. the variables x and y satisfy x\[Distributed]NormalDistribution[μ1,σ1] and y\[Distributed]NormalDistribution[μ2,σ2], respectively: *)

PDF[
 BinormalDistribution[{μ1,μ2},{σ1,σ2},ρ],
 {x,y}
 ]
```

Output

$$\frac{e^{\frac{-\frac{(x-\mu 1)^2}{\sigma 1^2}-\frac{(y-\mu 2)^2}{\sigma 2^2}+\frac{2(x-\mu 1)(y-\mu 2)\rho}{\sigma 1\,\sigma 2}}{2(1-\rho^2)}}}{2\,\pi\sqrt{1-\rho^2}\,\sigma 1\,\sigma 2}$$

***Mathematica Examples 15.24***

Input
```
(* The code generates a series of 3D plots, each representing the PDF of a binormal distribution with {μ1,μ2}={0,0}, {σ1,σ2}={2,2}, and different correlation coefficients (ρ values): *)

Table[
 Plot3D[
  PDF[
   BinormalDistribution[{0,0},{2,2},ρ],
   {x,y}
   ],
  {x,-7,7},
  {y,-7,7},
  PlotPoints->25,
  PlotRange->All,
  ColorFunction->"Rainbow",
  ImageSize->200
  ],
 {ρ,{-0.7,0,0.7}}
 ]
```

Output

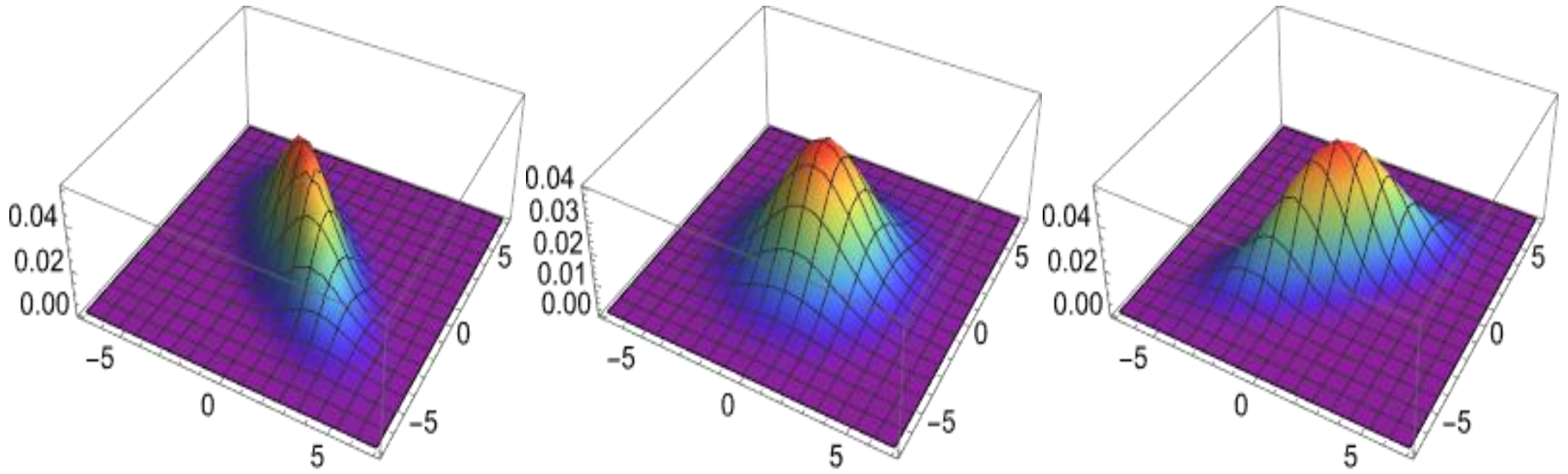

### *Mathematica Examples 15.25*

Input

```
(* The code generates a series of 3D plots, each representing the CDF of a binormal distribution with {μ1,μ2}={0,0}, {σ1,σ2}={2,2}, and different correlation coefficients (ρ values): *)

ParallelTable[
 Plot3D[
  CDF[
   BinormalDistribution[{0,0},{2,2},ρ],
   {x,y}
   ],
  {x,-7,7},
  {y,-7,7},
  PlotPoints->25,
  ColorFunction->"Rainbow",
  ImageSize->200
  ],
 {ρ,{-0.7,0,0.7}}
 ]
```

Output

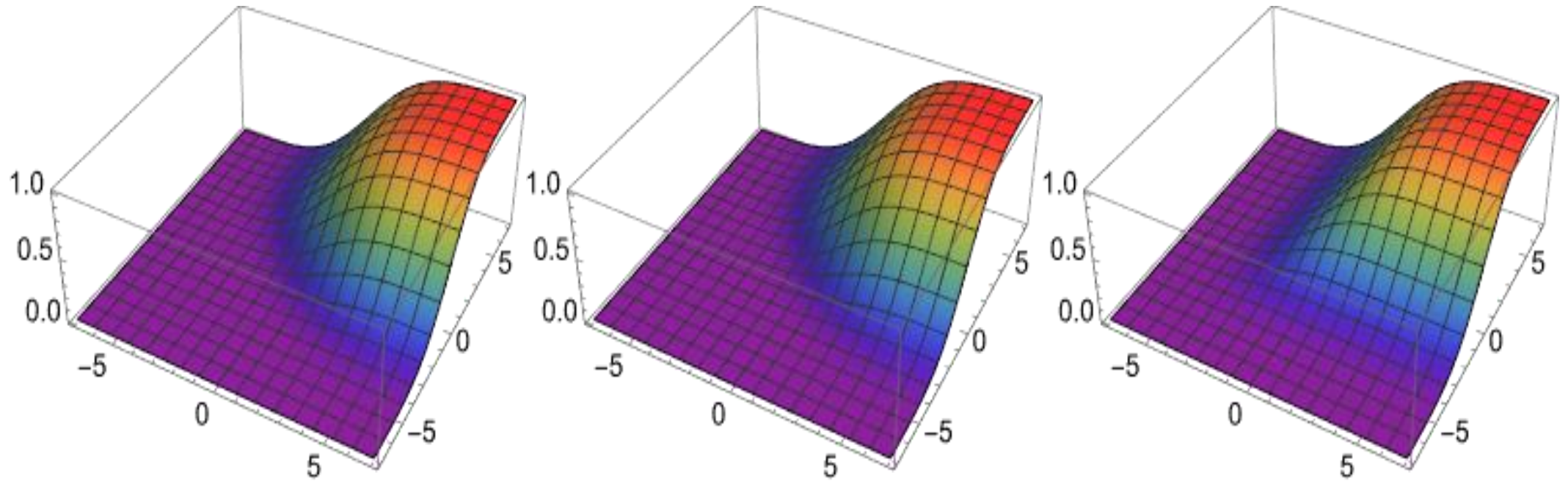

### *Mathematica Examples 15.26*

Input

```
(* The code generates a 3D-histogram and a 3D-plot of the PDF for a binormal distribution with parameters {μ1,μ2}={0,0}, {σ1,σ2}={2,2}, ρ=0.7 and sample size 10000: *)

sample=RandomVariate[
   d=BinormalDistribution[{0,0},{2,2},0.7],
   10^4
   ];
{
 Histogram3D[
  sample,
  30,
  "PDF",
  ColorFunction->"Rainbow",
```

```
   PlotRange->{{-6,6},{-6,6},All},
   ImageSize->250,
   PlotLabel->"BinormalDistribution[{0,0},{2,2},0.7]"
   ],

  Plot3D[
   PDF[
    d,
    {x,y}
    ],
   {x,-6,6},
   {y,-6,6},
   ColorFunction->"Rainbow",
   PlotPoints->35,
   PlotRange->All,
   ImageSize->250,
   PlotLabel->"BinormalDistribution[{0,0},{2,2},0.7]"
   ]
```

Output

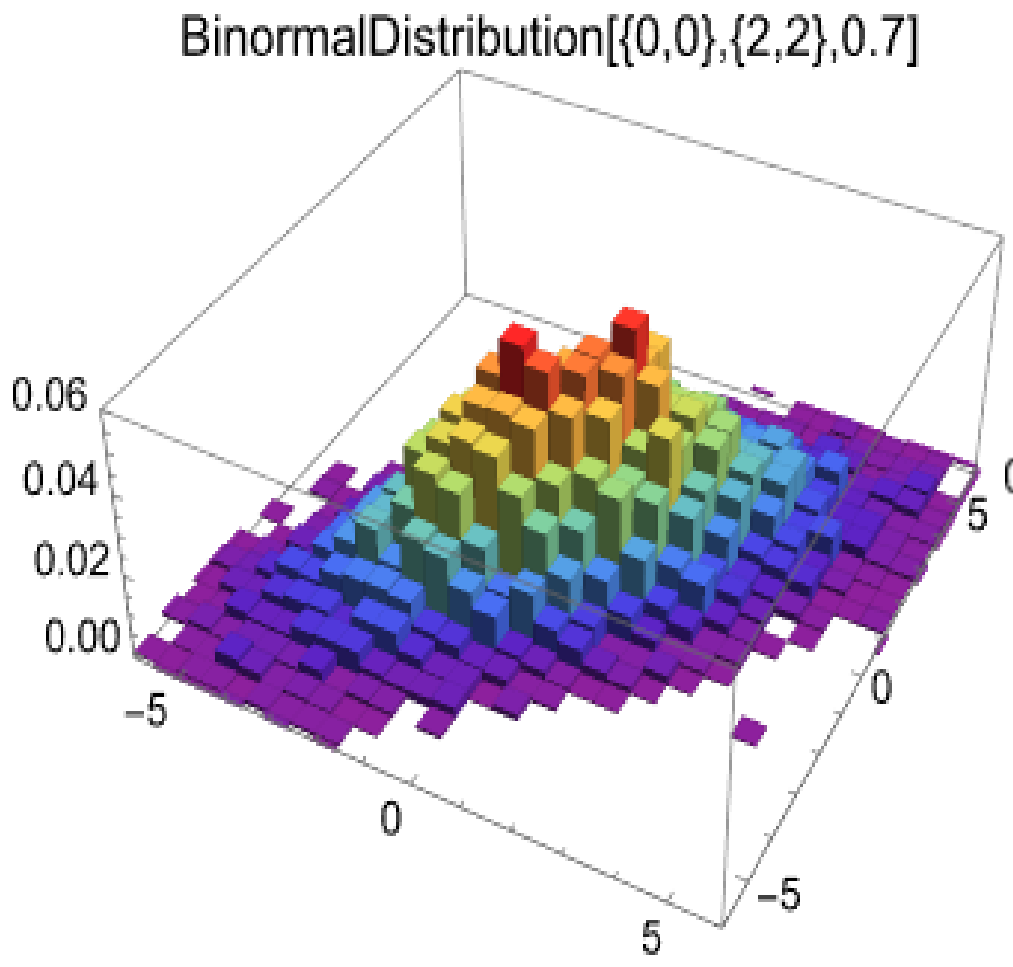

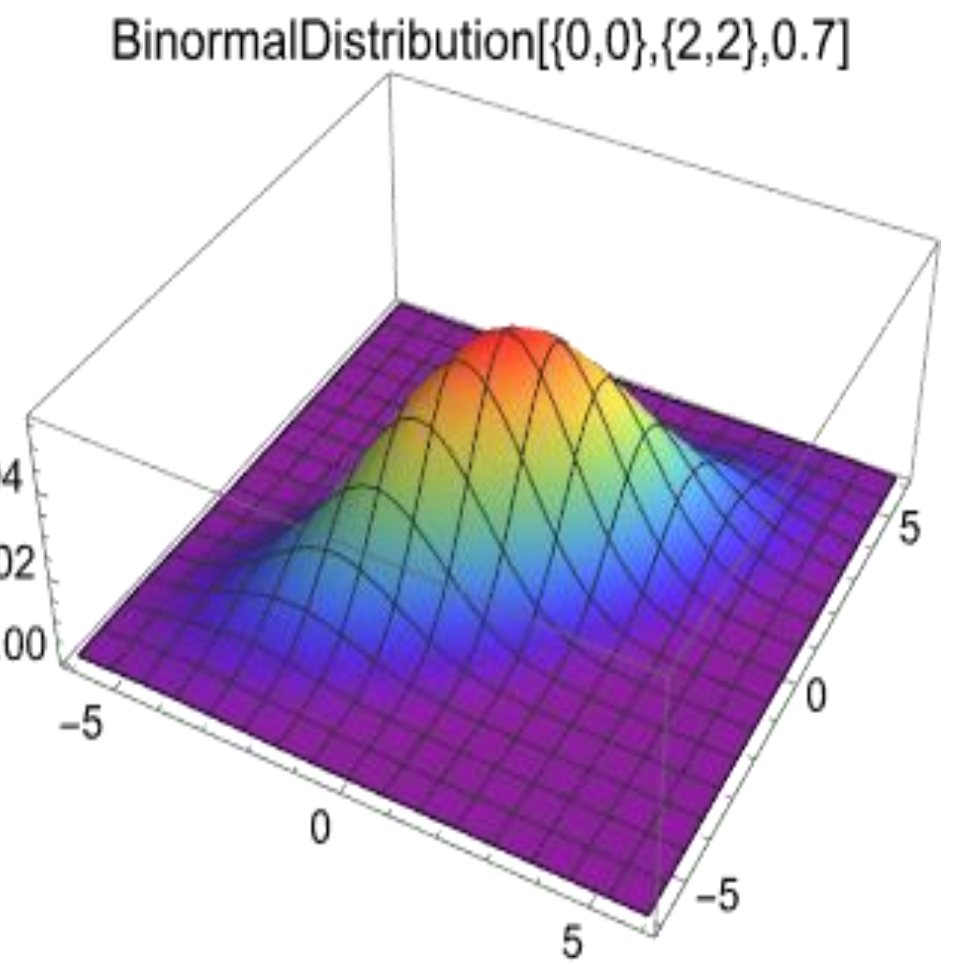

### Mathematica Examples 15.27

Input

```
(* The code calculates and displays some descriptive statistics (mean, variance, standard deviation, kurtosis and skewness) for a binormal distribution with parameters {μ1,μ2}, {σ1,σ2}, ρ: *)
Grid[
 Table[
  {
   statistics,
   FullSimplify[statistics[BinormalDistribution[{μ1,μ2},{σ1,σ2},ρ]]]
   },
  {statistics,{Mean,Variance,StandardDeviation,Kurtosis,Skewness}}
  ],
 ItemStyle->12,
 Alignment->{{Right,Left}},
 Frame->All,
 Spacings->{Automatic,0.8}
 ]
```

Output

| | |
|---|---|
| Mean | {μ1,μ2} |
| Variance | {σ1^2,σ2^2} |
| StandardDeviation | {σ1,σ2} |
| Kurtosis | {3,3} |
| Skewness | {0,0} |

**Mathematica Examples 15.28**

Input
```
(* The code calculates and displays some additional descriptive statistics (moments, central moments, and factorial moments) for a binormal distribution with parameters{μ1,μ2}, {σ1,σ2}, and ρ: *)

Grid[
 Table[
  {
   statistics,
   FullSimplify[statistics[BinormalDistribution[{μ1,μ2},{σ1,σ2},ρ],{1,1}]],
   FullSimplify[statistics[BinormalDistribution[{μ1,μ2},{σ1,σ2},ρ],{1,2}]]
   },
  {statistics,{Moment,CentralMoment,FactorialMoment}}
  ],
 ItemStyle->12,
 Alignment->{{Right,Left}},
 Frame->All,
 Spacings->{Automatic,0.8}
 ]
```

Output

| | | |
|---|---|---|
| Moment | μ1 μ2+ρ σ1 σ2 | 2 μ2 ρ σ1 σ2+μ1 (μ2^2+σ2^2) |
| CentralMoment | ρ σ1 σ2 | 0 |
| FactorialMoment | μ1 μ2+ρ σ1 σ2 | (-1+2 μ2) ρ σ1 σ2+μ1 ((-1+μ2) μ2+σ2^2) |

**Mathematica Examples 15.29**

Input
```
(* Covariance matrix of a binormal distribution: *)
MatrixForm[
 Covariance[BinormalDistribution[{μ1,μ2},{σ1,σ2},ρ]]
 ]
```

Output
```
σ1^2        ρ σ1 σ2
ρ σ1 σ2     σ2^2
```

**Mathematica Examples 15.30**

Input
```
(* Correlation of a standard binormal distribution: *)
MatrixForm[
 Correlation[
  BinormalDistribution[ρ]
  ]
 ]
```

Output
```
1  ρ
ρ  1
```

**Mathematica Examples 15.31**

Input
```
(* Marginal distributions of binormal distribution are normal: *)
dist=BinormalDistribution[{μ1,μ2},{σ1,σ2},ρ];
MarginalDistribution[dist,1]
MarginalDistribution[dist,2]
```

Output
```
NormalDistribution[μ1,σ1]
```
Output
```
NormalDistribution[μ2,σ2]
```

**Mathematica Examples 15.32**

Input
```
(* The code generates a dataset of 1000 observations from a binormal distribution with parameters {μ1,μ2}={1,1}, {σ1,σ2}={2,2}, ρ=0.6. Then, it computes the sample mean and quartiles of the data and plots a histogram of the data. The code adds vertical lines to the plot corresponding to the sample mean and quartiles: *)
```

```
data=RandomVariate[
    BinormalDistribution[{1,1},{2,2},0.6],
    1000
    ];

sampleMean=Mean[data]
quartiles=Quartiles[data]

histogram=Histogram3D[
    data,
    ColorFunction->"Rainbow",
    PlotRange->All,
    ImageSize->400,
    PlotLabel->"3D Histogram of Data"
    ];

meanLine=Graphics3D[
    {
     Red,
     Thickness[0.005],
     Line[{{sampleMean[[1]],-10,0},{sampleMean[[1]],10,0}}],
     Line[{{-10,sampleMean[[2]],0},{10,sampleMean[[2]],0}}]
     }
    ];

quartileLines=Graphics3D[
    {
     Blue,
     Thickness[0.003],
     Line[{{quartiles[[1]][[1]],-10,0},{quartiles[[1]][[1]],10,0}}],
     Line[{{quartiles[[1]][[2]],-10,0},{quartiles[[1]][[2]],10,0}}],
     Line[{{quartiles[[1]][[3]],-10,0},{quartiles[[1]][[3]],10,0}}],
     Green,
     Thickness[0.003],
     Line[{{-10,quartiles[[2]][[1]],0},{10,quartiles[[2]][[1]],0}}],
     Line[{{-10,quartiles[[2]][[2]],0},{10,quartiles[[2]][[2]],0}}],
     Line[{{-10,quartiles[[2]][[3]],0},{10,quartiles[[2]][[3]],0}}]
     }
    ];
Show[histogram,meanLine,quartileLines]
Output  {0.84332,0.8155}
Output  {{-0.440248,0.795172,2.32133},{-0.519394,0.857103,2.15124}}
Output
```

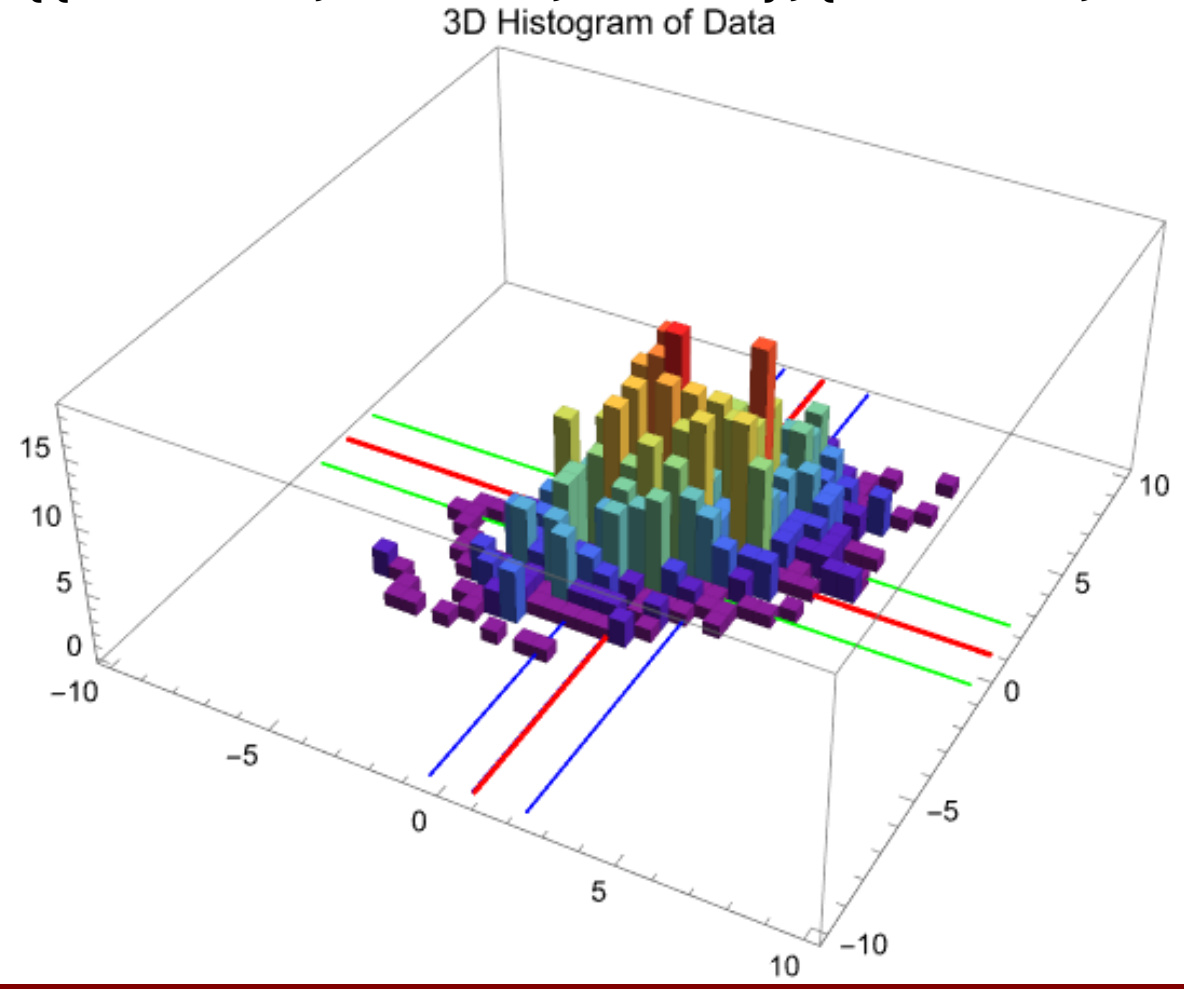

***Mathematica Examples 15.33***

Input
```
(* The code generates a contour plot of the PDF of a binormal distribution with {μ1,μ2}={0,0}, {σ1,σ2}={2,2}, ρ=0.5: *)

ContourPlot[
 PDF[
  BinormalDistribution[{0,0},{2,2},0.5],
  {x,y}
  ],
 {x,-6,6},
 {y,-6,6},
 ContourStyle->{White},
 ClippingStyle->Automatic,
 ColorFunction->"BlueGreenYellow",
 PlotLegends->Automatic,
 PlotLabel->"BinormalDistribution[{0,0},{2,2},0.5]",
 ImageSize->220
 ]
```

Output

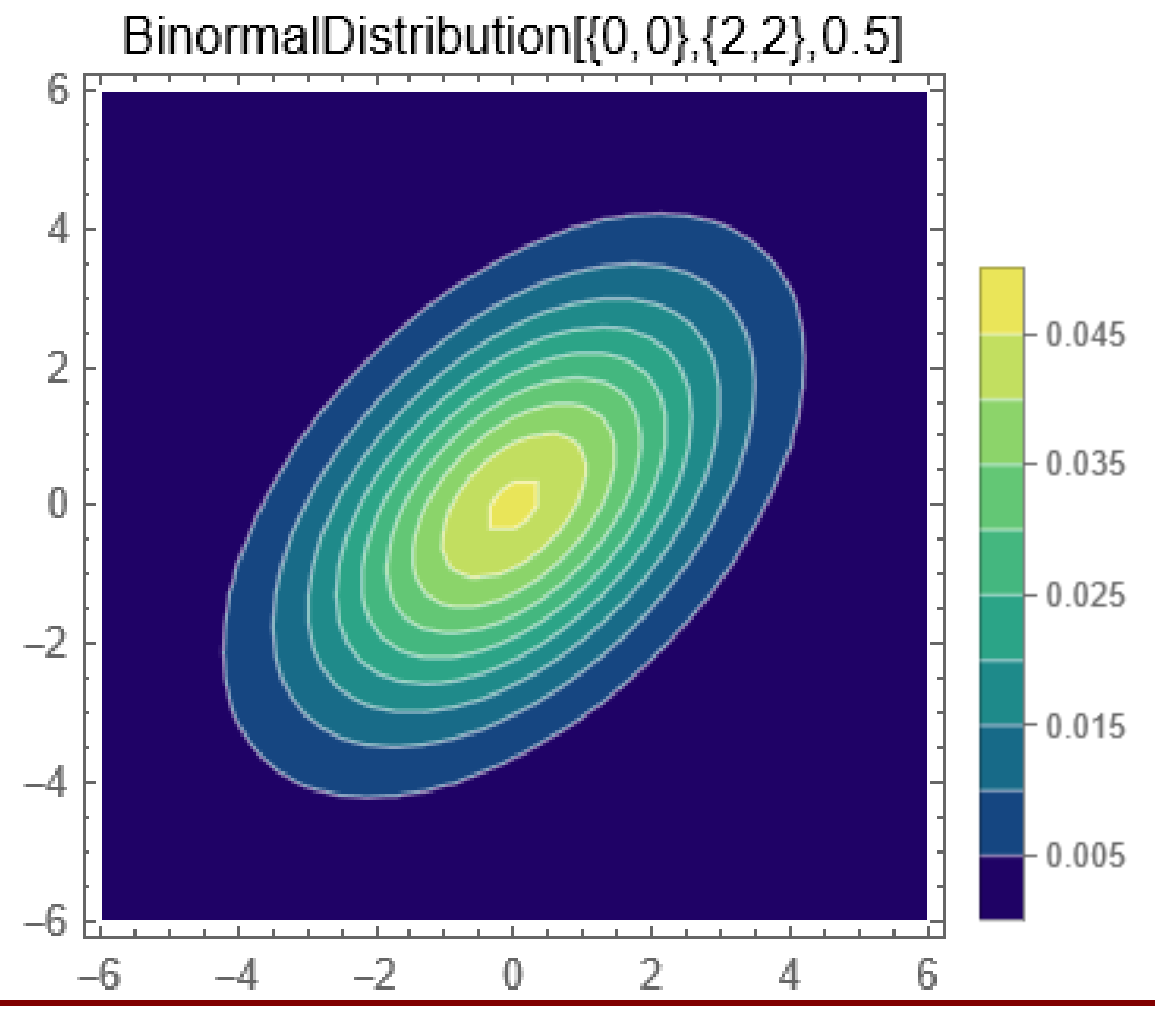

***Mathematica Examples 15.34***

Input
```
(* The code generates a visualization (3D Plot + Contour Plot) of the PDF of a binormal distribution with {μ1,μ2}={0,0}, {σ1,σ2}={2,2}, ρ=0: *)

(* Define the PDF using a binormal distribution with mean (0,0), standard deviations (2,2), and correlation coefficient 0: *)
dist=PDF[
    BinormalDistribution[{0,0},{2,2},0],
    {x,y}
    ];

(* Create a 3D plot of the distribution with specified ranges for x and y axes: *)
plot1=Plot3D[
    dist,
    {x,-5,5},
    {y,-5,5},
    PlotRange->Full,
    ClippingStyle->None,
    MeshFunctions->{#3&},
    Mesh->15,
    MeshStyle->Opacity[.5],
    MeshShading->{{Opacity[.3],Blue},{Opacity[.5],Yellow}},
```

```
    Lighting->"Neutral"
    ];

(* Create a 3D contour plot of the distribution at z=0: *)
slice=SliceContourPlot3D[
    dist,
    z==0,
    {x,-5,5},
    {y,-5,5},
    {z,-1,1},
    Contours->15,
    Axes->False,
    PlotPoints->50,
    PlotRangePadding->0,
    ContourStyle->{White},
    ClippingStyle->Automatic,
    ColorFunction->"BlueGreenYellow"
    ];

(* Combine the 3D plot and the contour plot into a single visualization: *)
Show[
 plot1,
 slice,
 PlotRange->All,
 BoxRatios->{1,1,0.7},
 FaceGrids->{Back,Left},
 ImageSize->300
 ]
```

Output

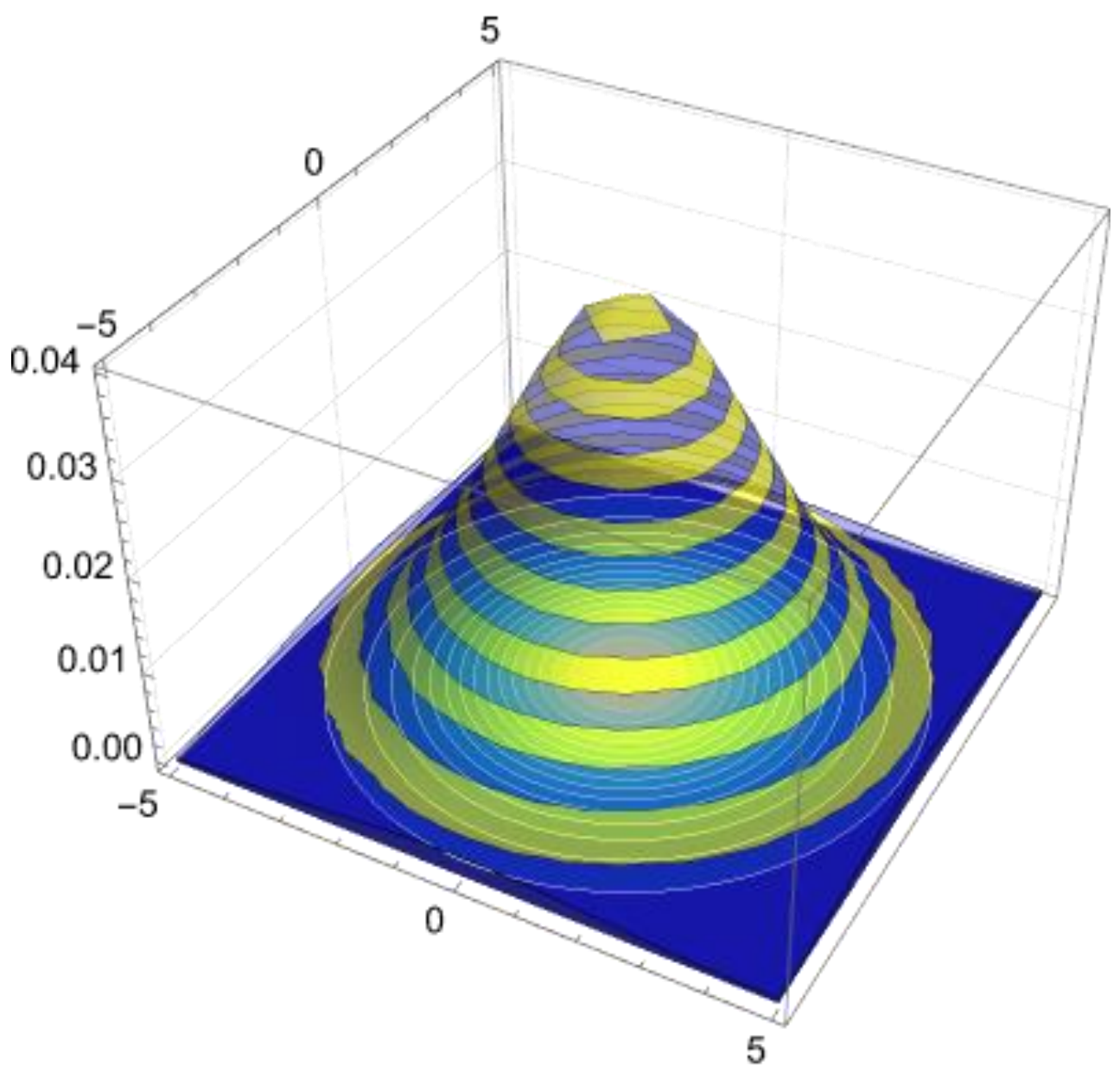

***Mathematica Examples 15.35***

Input

```
(* The code generates a random sample of size 10,000 from a binormal distribution with parameters {μ1,μ2}={1,1}, {σ1,σ2}={2,2}, ρ=0.6, estimates the distribution parameters using the EstimatedDistribution function, and then compares the histogram of the sample with the estimated PDF of the binormal distribution using a 3D histogram and a 3D plot of the PDF: *)

sampledata=RandomVariate[
```

```
    BinormalDistribution[{1,1},{2,2},0.6],
    10^4
    ];

(* Estimate the distribution parameters from sample data: *)
ed=EstimatedDistribution[
  sampledata,
  BinormalDistribution[{μ1,μ2},{σ1,σ2},ρ]
  ]

(* Compare the histogram of the sample with the estimated PDF of the binormal
distribution using a 3D histogram and a 3D plot of the PDF: *)
Show[
 Histogram3D[
  sampledata,
  {1},
  "PDF",
  ColorFunction->"Rainbow",
  ImageSize->300
  ],
 Plot3D[
  PDF[ed,{x,y}],
  {x,-10,10},
  {y,-10,10},
  ImageSize->300,
  ColorFunction->"Rainbow",
  PlotRange->All
  ]
 ]
 BinormalDistribution[{1.01831,1.02068},{2.01061,1.99799},0.602]
```

Output

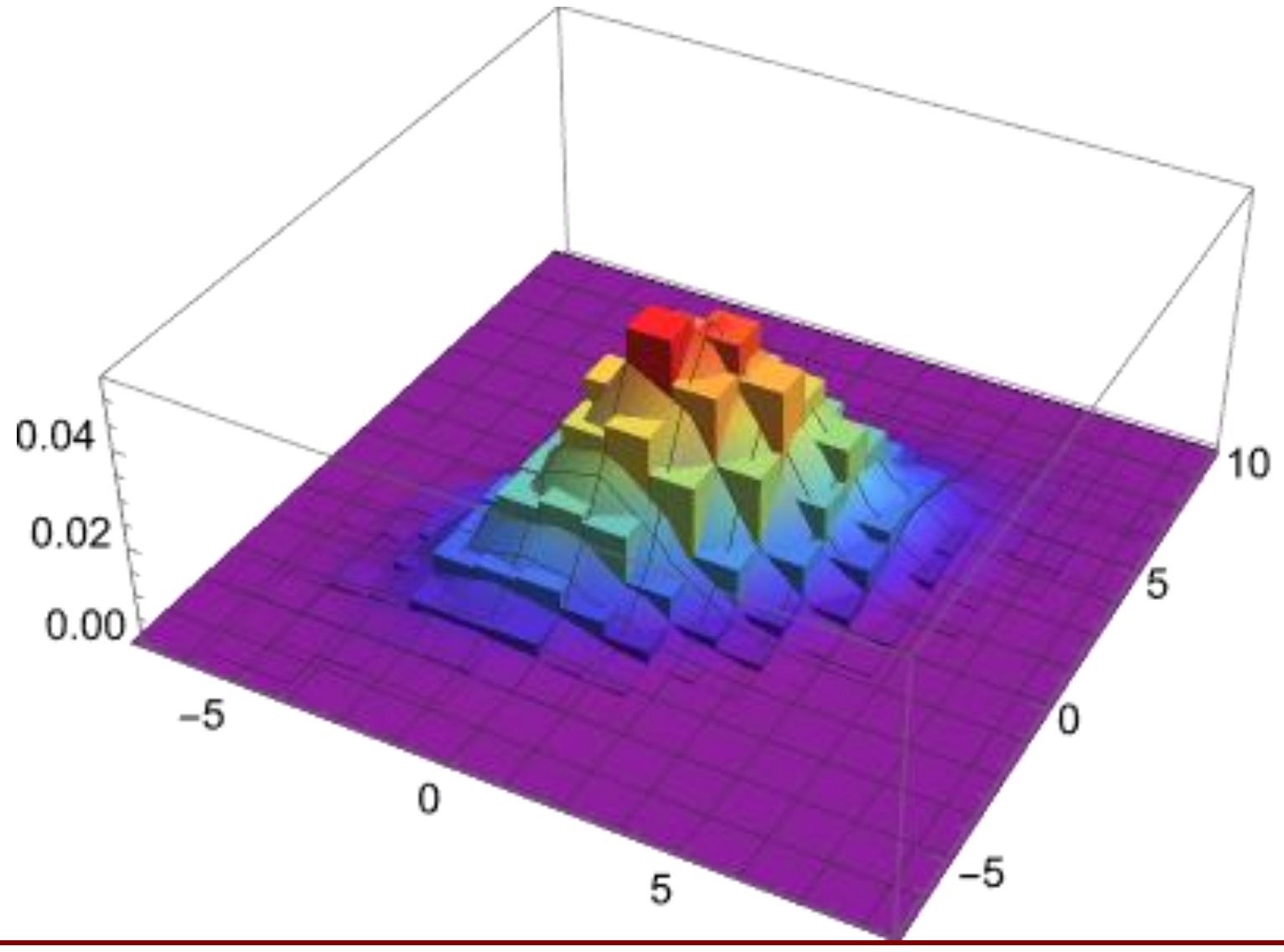

***Mathematica Examples 15.36***

Input

```
(* The code generates a 2D dataset with 2000 random points that follow a
BinormalDistribution[{1,0},{2,1},0.6]. The dataset is then used to create a row of
three plots. The first plot is a histogram of the X-axis values of the dataset and
PDF of marginal distribution in x-axis. The second plot is a histogram of the Y-axis
values of the dataset and PDF of marginal distribution in Y-axis. It is similar to
the first plot, but shows the distribution of the Y-axis values instead. The third
plot is a scatter plot of the dataset, with the X-axis values on the horizontal axis
```

```
and the Y-axis values on the vertical axis. Each point in the plot represents a pair
of X and Y values from the dataset: *)

d=BinormalDistribution[{1,0},{2,1},0.6];
data=RandomVariate[
   d,
   2000
   ];
GraphicsRow[
 {
  (* First plot, Histogram of X-axis values and PDF of marginal distribution: *)
  Show[
   Histogram[
    data[[All,1]],
    {0.2},
    PDF,
    PlotLabel->"X-axis",
    ColorFunction->Function[{height},Opacity[height]],
    ChartStyle->Purple,
    ImageSize->320
    ],
   Plot[
    PDF[MarginalDistribution[d,1],x],
    {x,-10,10},
    ImageSize->320,
    ColorFunction->"Rainbow",
    PlotRange->All
    ]
   ],

  (* Second plot, Histogram of Y-axis values and PDF of marginal distribution: *)
  Show[
   Histogram[
    data[[All,2]],
    {0.2},
    PDF,
    PlotLabel->"Y-axis",
    ColorFunction->Function[{height},Opacity[height]],
    ChartStyle->Purple,
    ImageSize->320
    ],
   Plot[
    PDF[MarginalDistribution[d,2],y],
    {y,-10,10},
    ImageSize->320,
    ColorFunction->"Rainbow",
    PlotRange->All
    ]
   ],
  (* Third plot, Scatter plot of the dataset: *)
  ListPlot[
   data,
   PlotStyle->{Purple,PointSize[0.015]},
   AspectRatio->1,
   Frame->True,
   Axes->False
   ]
  }
 ]
```

Output

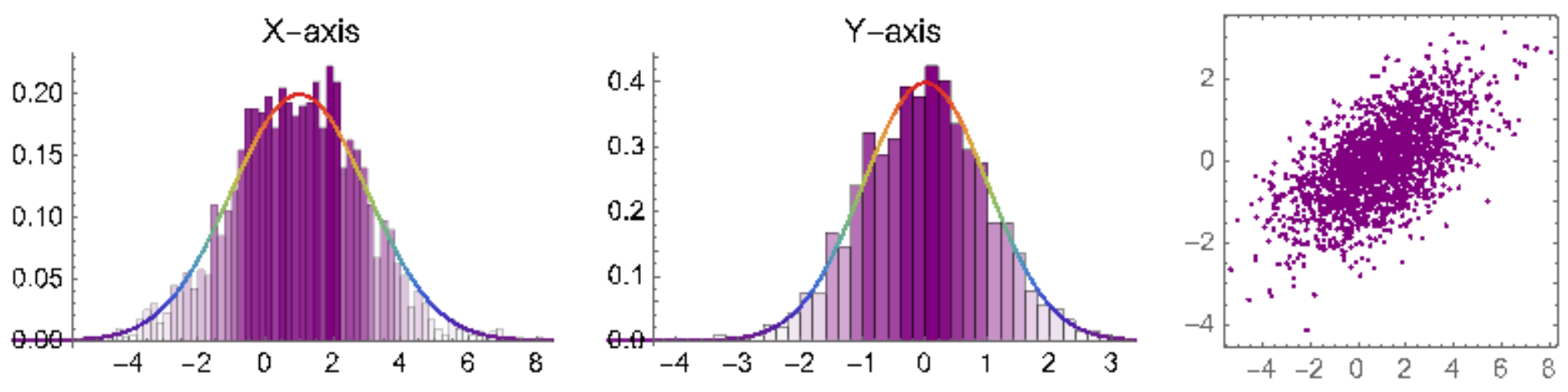

***Mathematica Examples 15.37***

Input

```
(* The code generates a bivariate dataset consisting of 10,000 random samples drawn
from a binormal distribution with mean (1,2), standard deviations (1.5,2), and
correlation coefficient 0.6. The code generates a scatter plot of the bivariate
dataset (each point in the plot represents a pair of X and Y values from the dataset)
with additional histograms as frame labels to provide insights into the individual
variable distributions (marginal distribution in x-axis and marginal distribution in
y-axis). The three plots are integrated into one plot: *)

data=RandomVariate[
   BinormalDistribution[{1,2},{1.5,2},0.6],
   10000
   ];

ListPlot[
 data,
 PlotRange->{{-5,8},{-5,8}},
 Axes->None,
 Frame->True,
 AspectRatio->1,
 ImageSize->200,
 PlotStyle->Purple,
 FrameLabel->{
   {
    Automatic,
    Histogram[
     data[[All,2]],
     {-5,8,0.5},
     AspectRatio->0.2,
     Axes->None,
     BarOrigin->Top,
     ImageSize->170,
     ColorFunction->Function[{height},Opacity[height]],
     ChartStyle->Purple
     ]
    },
   {
    Automatic,
    Histogram[
     data[[All,1]],
     {-5,8,0.5},
     AspectRatio->0.2,
     Axes->None,
     BarOrigin->Bottom,
     ImageSize->170,
     ColorFunction->Function[{height},Opacity[height]],
     ChartStyle->Purple
     ]
    }
   }
 ]
```

Output

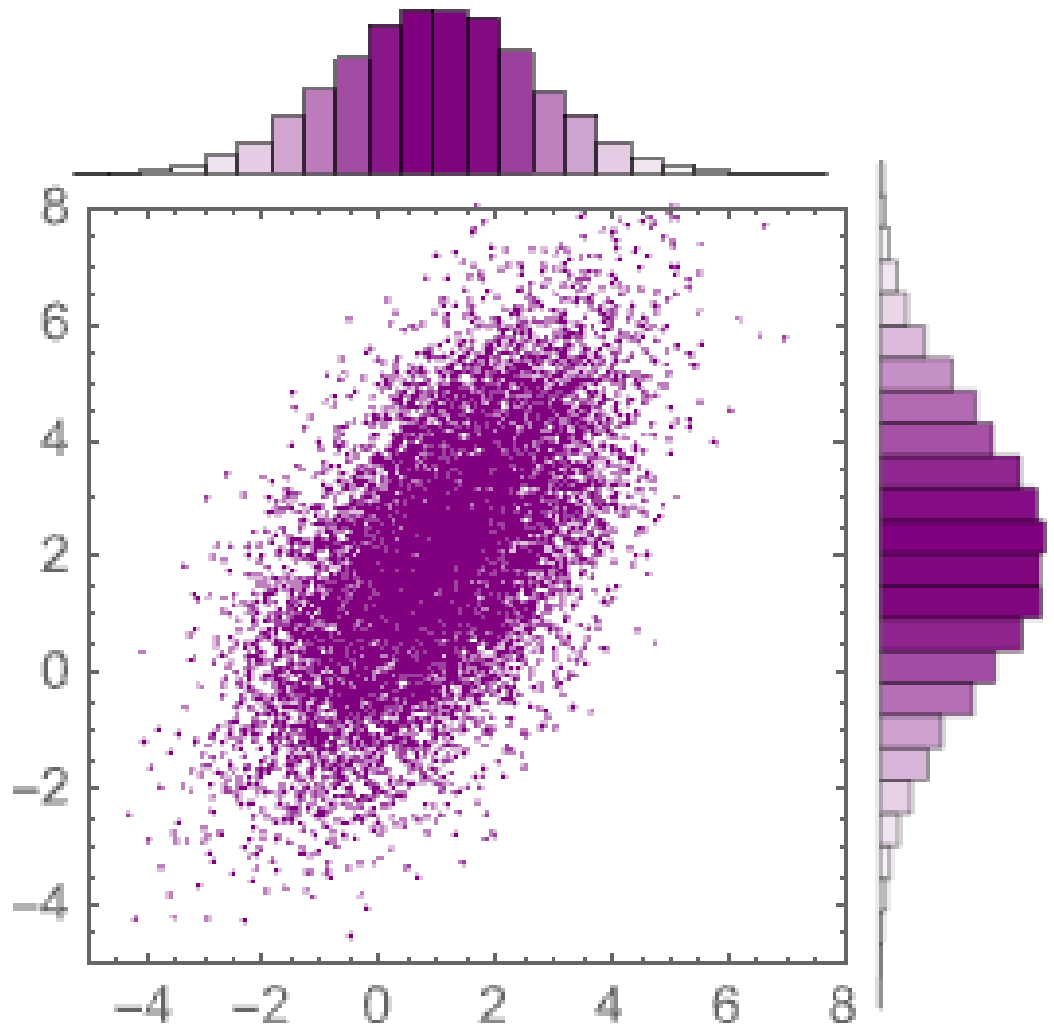

***Mathematica Examples 15.38***

Input

```
(* The code generates 3D histograms and scatter plot for a bivariate dataset. The dataset data is generated using the RandomVariate function, creating 1,000 random samples from a binormal distribution. The binormal distribution is characterized by a mean of (0,0), standard deviations of (2,2), and a correlation coefficient of 0.7. The histogram2dplot function takes two arguments, xrange and yrange, which determine the range of the plot along the x-axis and y-axis, respectively. Inside the function, three sets of data points are created for plotting. dataforListPointPlot3D contains the original data points in 3D space, setting the z-coordinate to 0. dataforxHistogram3D and dataforyHistogram3D are used to create histograms along the x-axis and y-axis, respectively. The Show function is then used to combine the plots of the data points and the histograms. The ListPointPlot3D function creates a scatter plot of the data points. The Histogram3D function is used to generate the 3D histograms for the x-axis and y-axis data separately: *)

data=RandomVariate[
    BinormalDistribution[{0,0},{2,2},0.7],
    10^3
    ];

histogram2dplot[xrange_,yrange_]:=Module[
   {dataforListPointPlot3D,dataforxHistogram3D,dataforyHistogram3D},

   dataforListPointPlot3D=Table[
     {data[[i]][[1]],data[[i]][[2]],0},
     {i,1,Length[data]}
     ];

   dataforxHistogram3D=Table[
     {data[[i]][[1]],yrange},
     {i,1,Length[data]}
     ];

   dataforyHistogram3D=Table[
     {-xrange,data[[i]][[2]]},
     {i,1,Length[data]}
     ];

   Show[
```

```
        {
         ListPointPlot3D[
          dataforListPointPlot3D,
          PlotStyle->Directive[Purple,PointSize[0.01],Opacity[.5]]
          ],
         Histogram3D[
          dataforxHistogram3D,
          {0.3},
          ColorFunction->Function[{height},Opacity[height]],
          ChartStyle->Red
          ],
         Histogram3D[
          dataforyHistogram3D,
          {.2},
          ColorFunction->Function[{height},Opacity[height]],
          ChartStyle->Blue
          ]
         },
        PlotRange->{{-xrange,xrange},{-yrange,yrange},All},
        AxesEdge->{{-1,-1},{1,-1},{-1,1}},
        FaceGrids->{{-1,0,0},{0,0,-1},{0,1,0}},
        Boxed->False
        ]
       ]

histogram2dplot[10,10]
```

Output

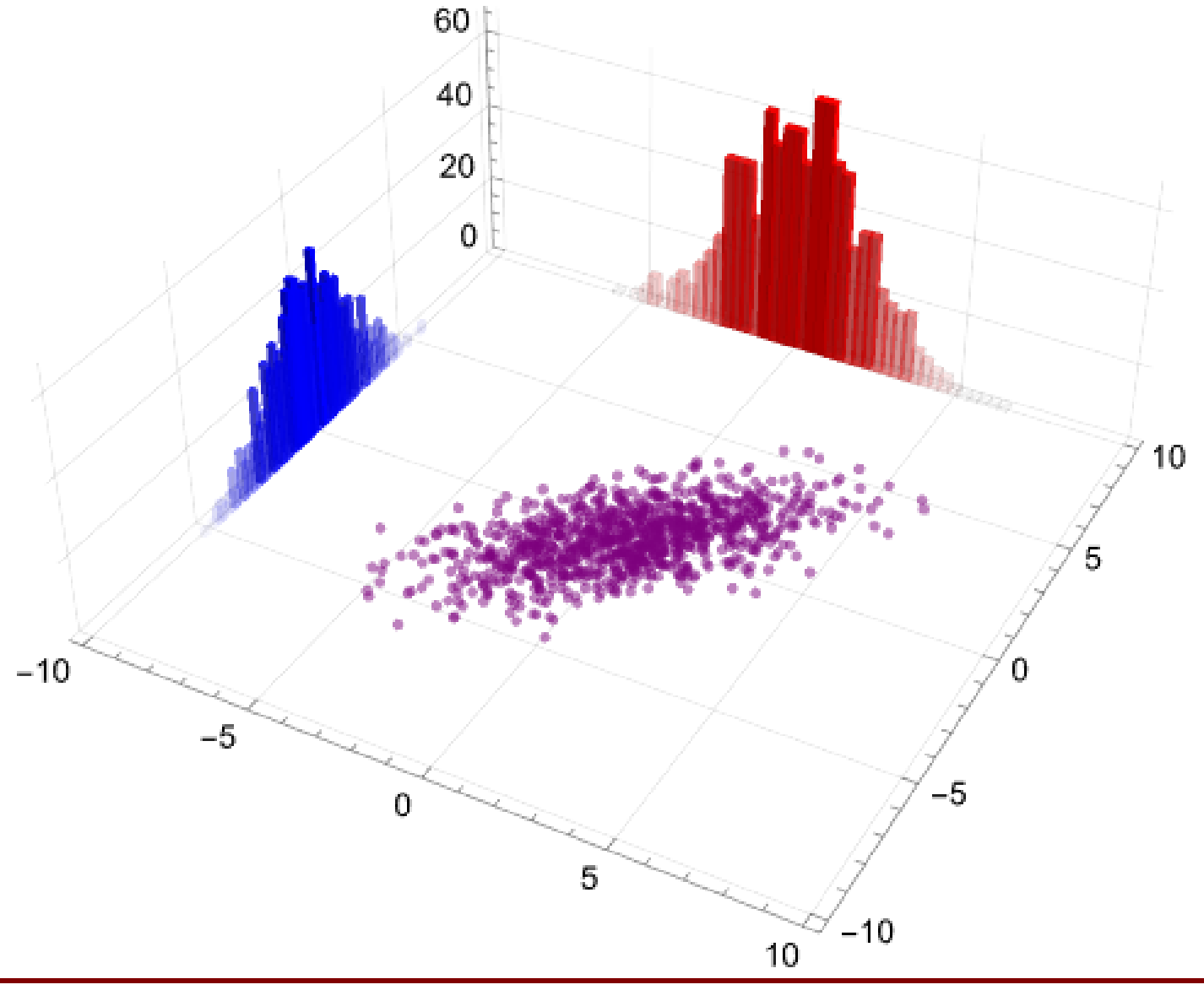

***Mathematica Examples 15.39***

Input

```
(* The code demonstrates a common technique in statistics and data analysis, which is the use of random sampling to estimate population parameters. The code generates random samples from a BinormalDistribution[{1,0},{0.5,1},0.6],and then using these samples to estimate the parameters of another binormal distribution with unknown {μ1,μ2},{σ1,σ2}, and ρ. This process is repeated 20 times, resulting in 20 different estimated distributions. The code plots the PDFs of the marginal distribution of estimated distributions using the PDF function and the estimated parameters: *)

dist=BinormalDistribution[{1,0},{0.5,1},0.6];

estim0distributions=Table[
```

```
      sampledata=RandomVariate[
        dist,
        1000
        ];

      ed=EstimatedDistribution[
        sampledata,
        BinormalDistribution[{μ1,μ2},{σ1,σ2}, ρ]
        ],
      {i,1,20}
      ];

   pdf0ed1=Table[
      PDF[MarginalDistribution[estim0distributions[[i]],1],x],
      {i,1,20}
      ];
   pdf0ed2=Table[
      PDF[MarginalDistribution[estim0distributions[[i]],2],x],
      {i,1,20}
      ];

   (* Visualizes the resulting PDF marginals distributions of estimated distributions
   *)
   Plot[
    {pdf0ed1,pdf0ed2},
    {x,-3,4},
    PlotRange->Full,
    ImageSize->320,
    PlotStyle-
   >{Directive[Purple,Opacity[0.3],Thickness[0.001]],Directive[Red,Opacity[0.3],Thickn
   ess[0.001]]}
    ]
```

Output

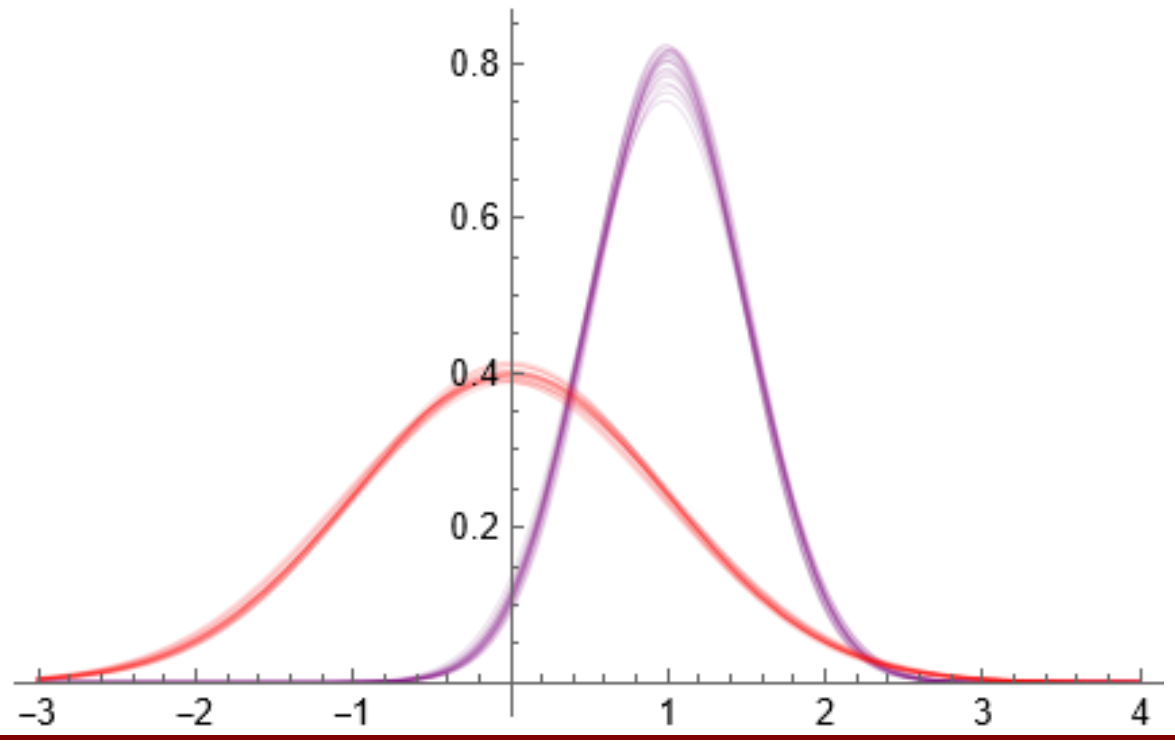

***Mathematica Examples 15.40***

Input

```
(* The code creates a dynamic histogram of data and a plot of the PDF generated from
a binormal distribution using the Manipulate function. The Manipulate function
creates interactive controls for the user to adjust the values of μ1,μ2,σ1,σ2, ρ and
n, which are the parameters of the normal distribution and the sample size: *)

Manipulate[
 Module[
  {
   data=RandomVariate[
     d=BinormalDistribution[{μ1,μ2},{σ1,σ2},ρ],
     n
     ]
```

```
          },
         {
          Histogram3D[
           data,
           30,
           "PDF",
           PlotRange->{{-5,15},{-5,15},All},
           ColorFunction->"Rainbow",
           ImageSize->320
           ],
          Plot3D[
           PDF[
            d,
            {x,y}
            ],
           {x,-5,15},
           {y,-5,15},
           ColorFunction->"Rainbow",
           PlotPoints->35,
           PlotRange->All,
           ImageSize->320
           ]
          }
         ],
        {{μ1,1,"μ1"},0,1.5,0.1},
        {{μ2,3,"μ2"},2,4,0.1},
        {{σ1,1.5,"σ1"},1,2,0.1},
        {{σ2,2.5,"σ2"},2,3,0.1},
        {{ρ,0.5,"ρ"},0.1,0.9,0.1},
        {{n,2000,"n"},500,4000,100}
        ]
```

Output

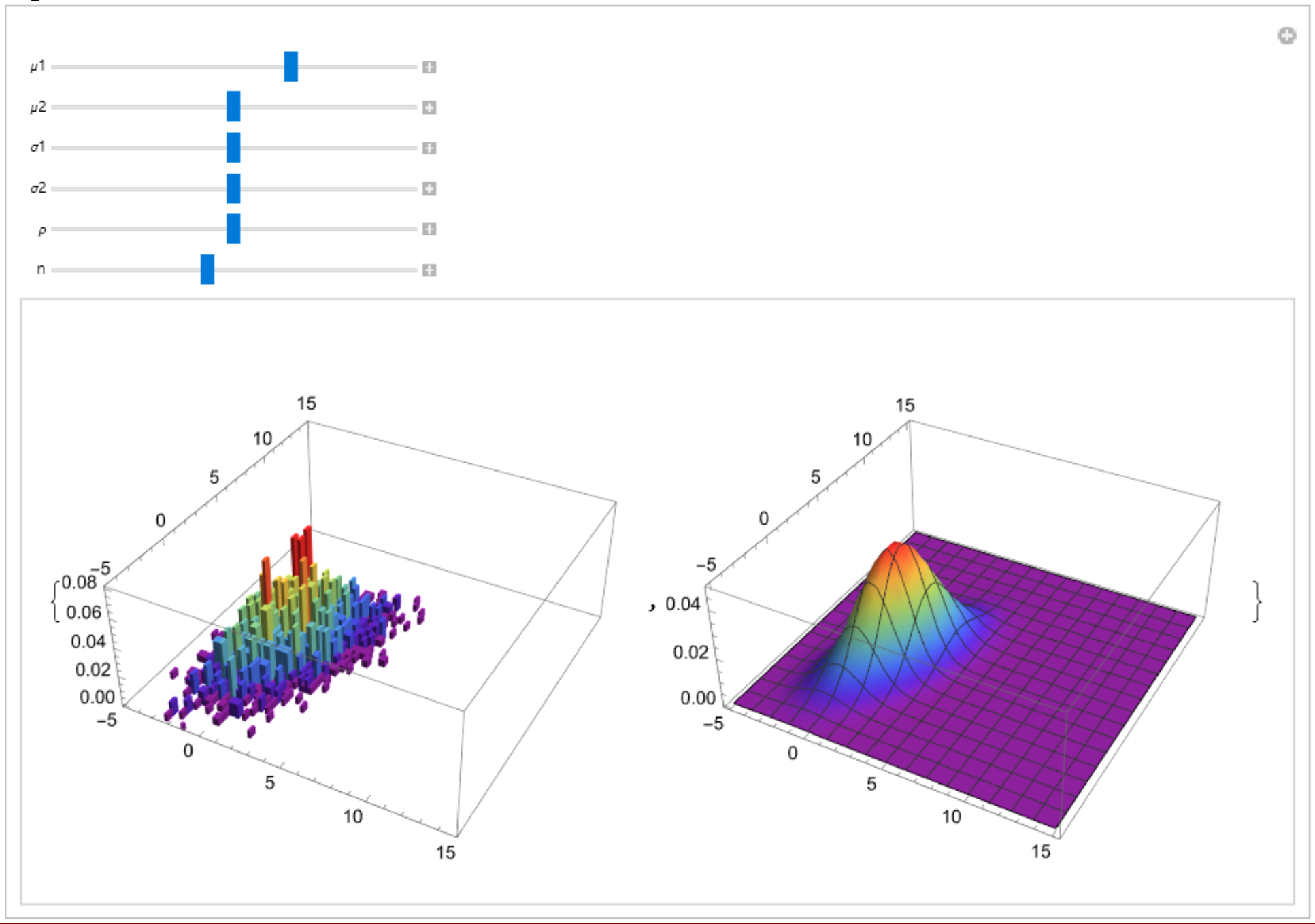

***Mathematica Examples 15.41***

Input (* The code generates a Manipulate interface that allows interactive exploration of a bivariate normal distribution. The interface consists of two plots displayed side by side. The left plot is a 3D plot, and the right plot is a 2D density plot. The

```
left plot, created using Plot3D, shows the PDF of the bivariate normal distribution.
The right plot, created using DensityPlot, shows a 2D density plot of the PDF. It
represents the same PDF as the left plot but in a 2D format: *)

Manipulate[
 GraphicsRow[
  {
   Plot3D[
    PDF[BinormalDistribution[{μ1,μ2},{σ1,σ2},ρ],{x,y}],
    {x,-4,4},
    {y,-4,4},
    ColorFunction->"Rainbow",
    PlotRange->Full,
    AxesLabel->{x,y,None},
    Boxed->False,
    AxesEdge->{{-1,-1},{1,-1},{-1,-1}},
    ImageSize->{275,275}],
   DensityPlot[
    PDF[BinormalDistribution[{μ1,μ2},{σ1,σ2},ρ],{x,y}],
     {x,-4,4},
    {y,-4,4},
    ColorFunction->"Rainbow",
    PlotRange->Full,
     FrameLabel->{x,y},
    ImageSize->{275,275}
    ]
   }
  ],
 {{μ1, 0, "μx"}, -2,2, Appearance->"Labeled"},
 {{μ2, 0, "μy"}, -2,2, Appearance->"Labeled"},
 {{σ1, 1, "σx"}, .5,2, Appearance->"Labeled"},
 {{σ2, 1, "σy"}, .5,2, Appearance->"Labeled"},
 {{ρ, 0, "ρ"}, -0.9,0.9, Appearance->"Labeled"}
 ]
```

Output

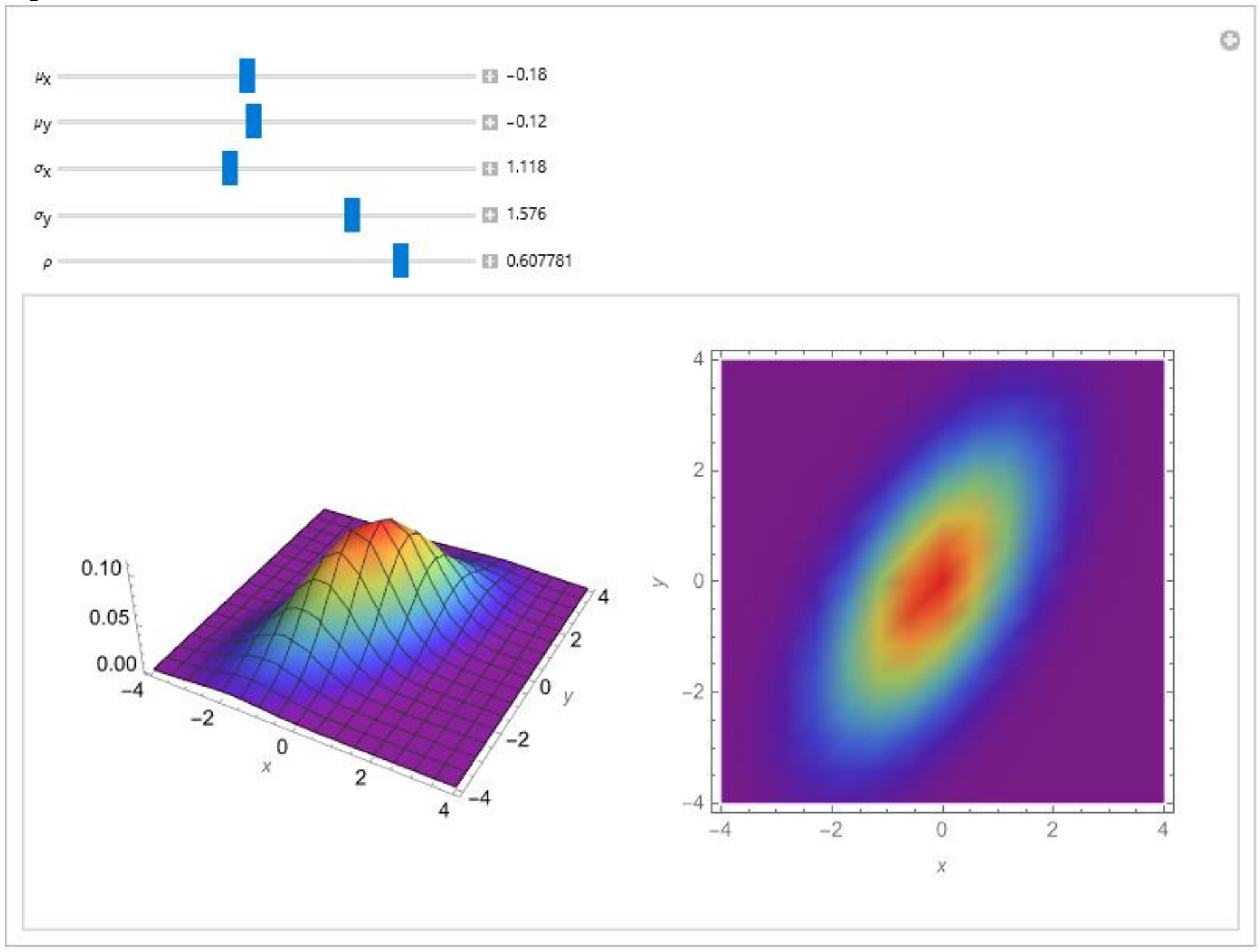

***Mathematica Examples 15.42***

```
Input    (* Continuous multivariate distribution:*)
         Probability[(x<1)&&(y<1),{x,y}\[Distributed]BinormalDistribution[{1,1},{2,2},0.5]]
         Probability[x+y<1,{x,y}\[Distributed]BinormalDistribution[{1,1},{2,2},0.5]]
         Probability[0<x+y,{x,y}\[Distributed]BinormalDistribution[{1,1},{2,2},0.5]]
         Probability[x+y<7/10\[Conditioned]y>1/3,{x,y}\[Distributed]BinormalDistribution[{1,
         1},{2,2},0.5]]

Output   0.333333
Output   0.386415
Output   0.718149
Output   0.111865
```

# UNIT 15.3

# MULTINOMIAL DISTRIBUTION

| | |
|---|---|
| `MultinomialDistribution[n,{p1,p2,…,pm}]` | represents a multinomial distribution with n trials and probabilities pi. |

***Mathematica Examples 15.43***

Input

```
(* MultinomialDistribution[n,{p1,p2,…,pm}] represents a discrete multivariate statistical distribution each of the variables x1,x2,...xm satisfies xj\[Distributed]BinomialDistribution[n,pj] for j=1,2,...,m: *)

PDF[
 MultinomialDistribution[n,{p1,p2,p3}],
 {x1,x2,x3}
 ]
```

Output

$$\begin{cases} p1^{x1}p2^{x2}p3^{x3}\text{Binomial}[n,x3]\text{Binomial}[x1+x2,x2] & x1+x2+x3==n\ \&\&x1>=0\&\&x2>=0\&\&x3>=0 \\ 0 & \text{True} \end{cases}$$

***Mathematica Examples 15.44***

Input

```
(* The code generates a series of Discrete 3D plots, each representing the PMF of a multinomial distribution with {p1,p2}={0.7,0.3}, and different values of n= (4, 5, 10): *)

Table[
 DiscretePlot3D[
  PDF[
   MultinomialDistribution[n,{0.7,0.3}],
   {x,y}
   ],
  {y,0,n},
  {x,0,n},
  PlotLabel->Row[{"n = ",n}],
  ExtentSize->0.6,
  PlotStyle->Purple,
  ImageSize->220
  ],
 {n,{4,5,10}}
 ]
```

Output

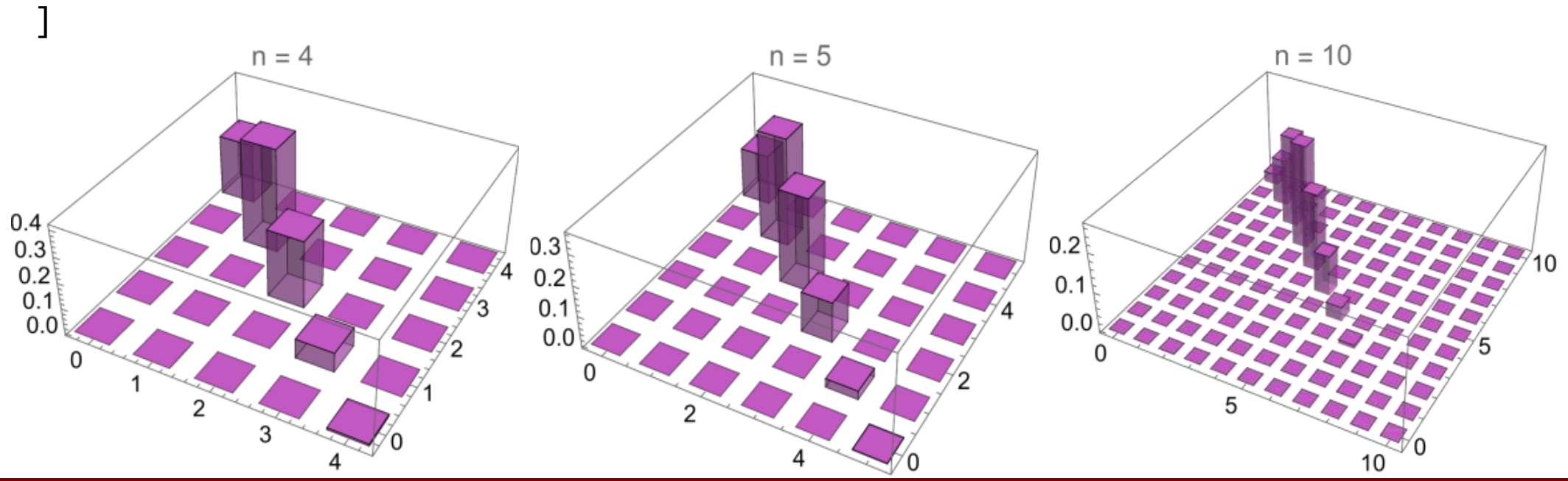

**Mathematica Examples 15.45**

Input

```
(* The code generates a series of Discrete 3D plots, each representing the CDF of a multinomial distribution with {p1, p2}={0.7, 0.3}, and different values of n= (4, 5, 10): *)

Table[
 DiscretePlot3D[
  CDF[
   MultinomialDistribution[n,{0.7,0.3}],
   {x,y}
   ],
  {x,0,n},
  {y,0,n},
  PlotLabel->Row[{"n = ",n}],
  ExtentSize->Right,
  PlotStyle->Purple,
  ImageSize->220
  ],
 {n,{4,5,10}}
 ]
```

Output

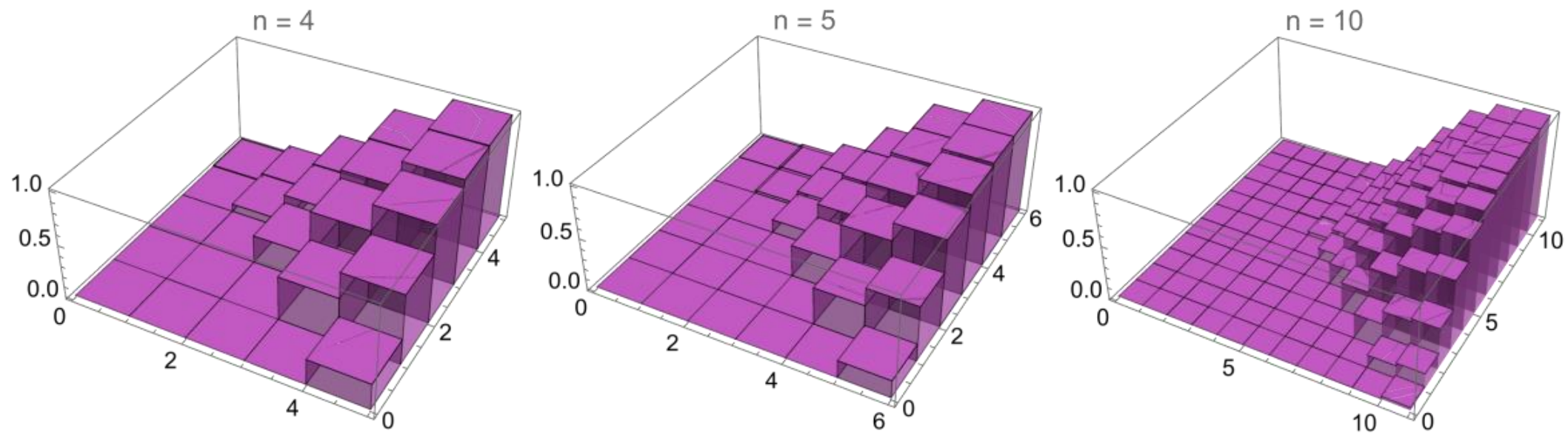

**Mathematica Examples 15.46**

Input

```
(* The code generates a 3D-histogram and a Discrete 3D Plot of the PMF for a multinomial distribution with parameters p1=0.6, p2=0.4, n=12 and sample size 10000: *)

sample=RandomVariate[
   d=MultinomialDistribution[12,{0.6,0.4}],
   10^4
   ];
{
 Histogram3D[
  sample,
  30,
  "PDF",
  ColorFunction->"Rainbow",
  ImageSize->250,
  PlotLabel->"MultinomialDistribution[12,{0.6,0.4}]"
  ],

 DiscretePlot3D[
  PDF[
   d,
   {x,y}
   ],
  {y,0,12},
  {x,0,12},
  ExtentSize->1,
  ImageSize->250,
```

```
  ColorFunction->"Rainbow",
  PlotRange->All,
  PlotLabel->"MultinomialDistribution[12,{0.6,0.4}]"
  ]
 }
```

Output

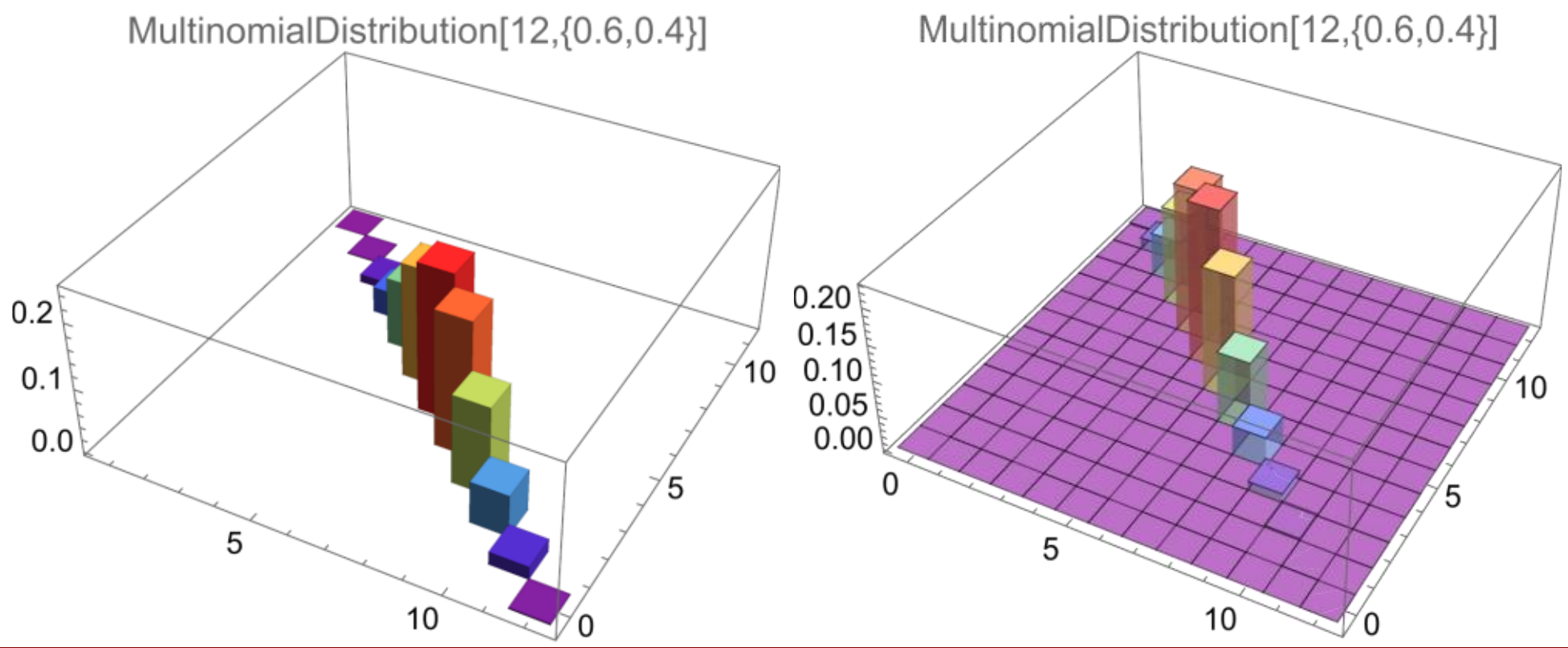

***Mathematica Examples 15.47***

Input

```
(* The code calculates and displays some descriptive statistics (mean, variance,
standard deviation, kurtosis and skewness) for a multinomial distribution with
parameters p1, p2 and n: *)

Grid[
 Table[
  {
   statistics,
   FullSimplify[statistics[MultinomialDistribution[n,{p1,p2}]]]
   },
  {statistics,{Mean,Variance,StandardDeviation,Kurtosis,Skewness}}
  ],
 ItemStyle->12,
 Alignment->{{Right,Left}},
 Frame->All,
 Spacings->{Automatic,0.8}
 ]
```

Output

| | |
|---|---|
| Mean | {n p1, n p2} |
| Variance | {-n (-1 + p1) p1, -n (-1 + p2) p2} |
| StandardDeviation | $\{\sqrt{-n(-1+p1)p1}, \sqrt{-n(-1+p2)p2}\}$ |
| Kurtosis | {3- 6/n + 1/(n p1-n p1^2),3- 6/n + 1/(n p2-n p2^2)} |
| Skewness | $\{(1-2\ p1)/\sqrt{-n(-1+p1)p1}\ ,(1-2\ p2)/\sqrt{-n(-1+p2)p2}\ \}$ |

***Mathematica Examples 15.48***

Input

```
(* Covariance matrix of a multinomial distribution: *)
MatrixForm[
 Covariance[MultinomialDistribution[n,{p1,p2}]]
 ]
```

Output

$$\begin{pmatrix} n\,(p1-p1\text{^}2) & -n\,p1\,p2 \\ -n\,p1\,p2 & n\,(p2-p2\text{^}2) \end{pmatrix}$$

***Mathematica Examples 15.49***

Input

```
(* Correlation of a multinomial distribution: *)
MatrixForm[
```

```
            Correlation[
             MultinomialDistribution[n,{p1,p2}]
             ]
            ]
```

Output

$$\begin{pmatrix} 1 & -\frac{p1\,p2}{\sqrt{p1-p1^2}\sqrt{p2-p2^2}} \\ -\frac{p1\,p2}{\sqrt{p1-p1^2}\sqrt{p2-p2^2}} & 1 \end{pmatrix}$$

***Mathematica Examples 15.50***

```
Input   (* Marginal distributions of multinomial distribution: *)
        d=MultinomialDistribution[n,{p1,p2}];
        MarginalDistribution[d,1]
        MarginalDistribution[d,2]

Output  BinomialDistribution[n,p1]
Output  BinomialDistribution[n,p2]
```

***Mathematica Examples 15.51***

```
Input   (* The code generates a dataset of 1000 observations from a multinomial distribution
        with parameters p1=0.6, p2=0.4 and n=10. Then, it computes the sample mean and
        quartiles of the data and plots a histogram of the data. The code adds vertical lines
        to the plot corresponding to the sample mean and quartiles: *)

        n=10;
        data=RandomVariate[
            MultinomialDistribution[n,{0.6,0.4}],
            1000
            ];

        sampleMean=N[Mean[data]]
        quartiles=N[Quartiles[data]]

        histogram=Histogram3D[
            data,
            ColorFunction->"Rainbow",
            PlotRange->All,
            ImageSize->400,
            PlotLabel->"3D Histogram of Data"
            ];
        meanLine=Graphics3D[
            {
             Red,
             Thickness[0.005],
             Line[{{sampleMean[[1]],0,0},{sampleMean[[1]],n,0}}],
             Line[{{0,sampleMean[[2]],0},{n,sampleMean[[2]],0}}]
             }
            ];
        quartileLines=Graphics3D[
            {
             Blue,
             Thickness[0.003],
             Line[{{quartiles[[1]][[1]],0,0},{quartiles[[1]][[1]],n,0}}],
             Line[{{quartiles[[1]][[2]],0,0},{quartiles[[1]][[2]],n,0}}],
             Line[{{quartiles[[1]][[3]],0,0},{quartiles[[1]][[3]],n,0}}],
             Green,
             Thickness[0.003],
```

```
         Line[{{0,quartiles[[2]][[1]],0},{n,quartiles[[2]][[1]],0}}],
         Line[{{0,quartiles[[2]][[2]],0},{n,quartiles[[2]][[2]],0}}],
         Line[{{0,quartiles[[2]][[3]],0},{n,quartiles[[2]][[3]],0}}]
         }
        ];

        Show[histogram,meanLine,quartileLines]
```

Output {5.931,4.069}
Output {{5.,6.,7.},{3.,4.,5.}}
Output

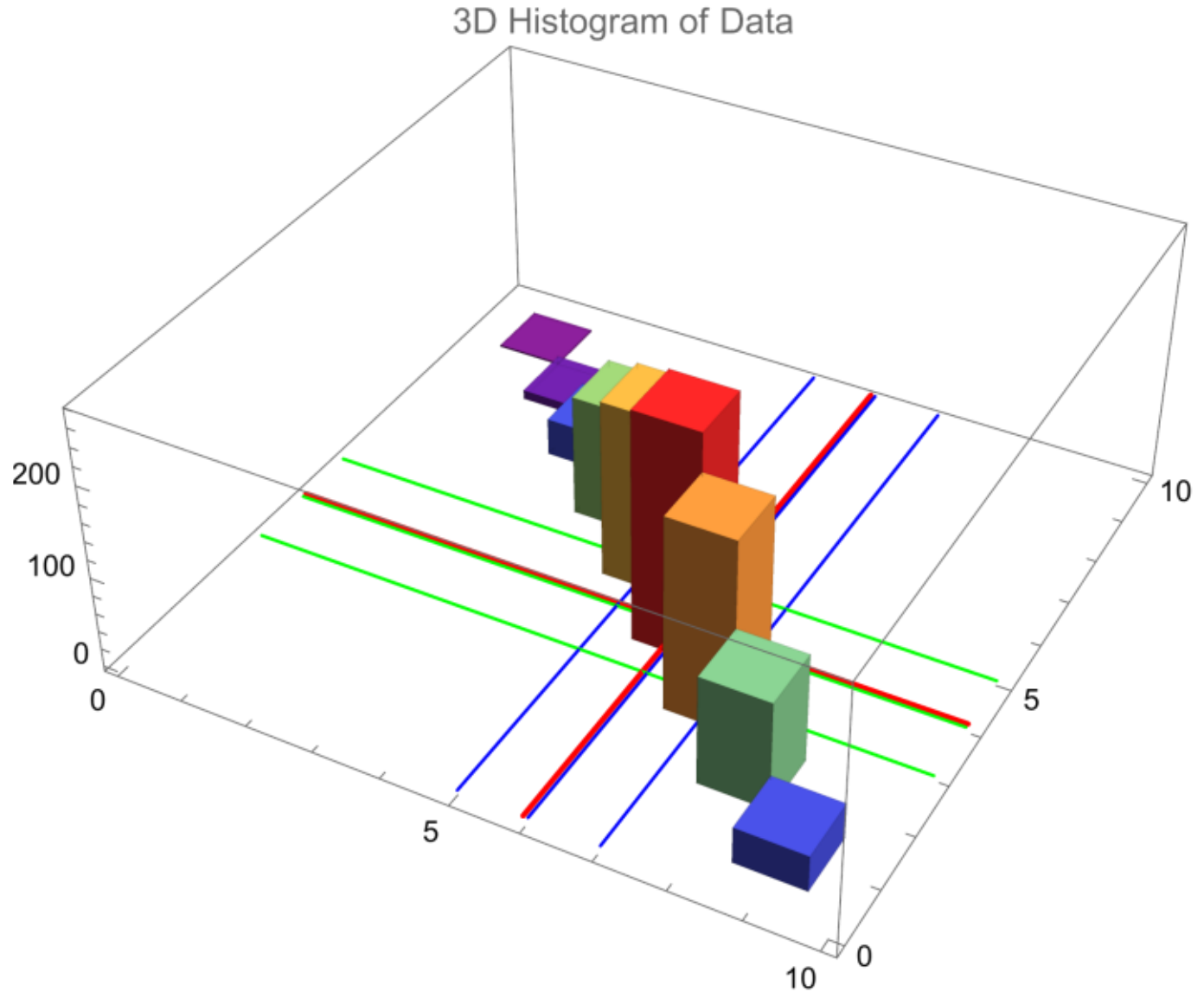

***Mathematica Examples 15.52***

Input

```
(* The code generates a random sample of size 10,000 from a multinomial distribution
with parameters n=10, p1=0.6 and p2=0.4, estimates the distribution parameters using
the EstimatedDistribution function, and then compares the histogram of the sample
with the estimated PDF of the multinomial distribution using a 3D histogram and a
DiscretePlot3D of the PDF: *)

sampledata=RandomVariate[
   MultinomialDistribution[10,{0.6,0.4}],
   10^4
   ];

(* Estimate the distribution parameters from sample data: *)
ed=EstimatedDistribution[
  sampledata,
  MultinomialDistribution[n,{p1,p2}]
  ]

(* Compare the histogram of the sample with the estimated PDF of the multinomial
distribution using a 3D histogram and a DiscretePlot3D of the PDF: *)
{
 Histogram3D[
  sampledata,
  {1},
  "PDF",
  ColorFunction->"Rainbow",
  ImageSize->300
  ],
 DiscretePlot3D[
```

```
             PDF[ed,{x,y}],
             {x,0,10},
             {y,0,10},
             ExtentSize->1,
             ImageSize->300,
             ColorFunction->"Rainbow",
             PlotRange->All
             ]
            }
```

Output MultinomialDistribution[10,{0.60163,0.39837}]

Output

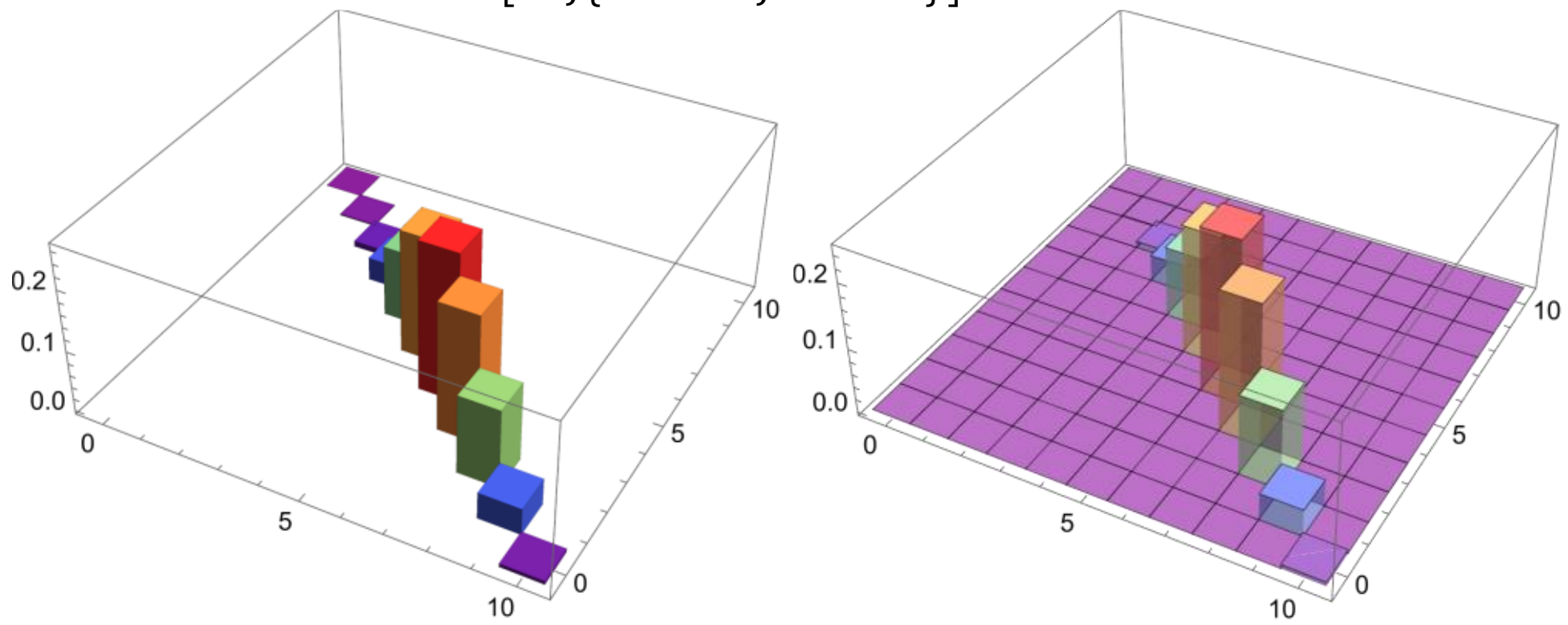

***Mathematica Examples 15.53***

Input

```
(* The code generates a 2D dataset with 2000 random points that follow a
MultinomialDistribution[10,{0.6,0.4}]. The dataset is then used to create a row of
three plots. The first plot is a histogram of the X-axis values of the dataset and
PMF of marginal distribution in x-axis. The second plot is a histogram of the Y-axis
values of the dataset and PMF of marginal distribution in Y-axis. It is similar to
the first plot but shows the distribution of the Y-axis values instead. The third
plot is a scatter plot of the dataset, with the X-axis values on the horizontal axis
and the Y-axis values on the vertical axis. Each point in the plot represents a pair
of X and Y values from the dataset: *)

d=MultinomialDistribution[10,{0.6,0.4}];
data=RandomVariate[
    d,
    2000
    ];
GraphicsRow[
 {
  (* First plot, Histogram of X-axis values and PDF of marginal distribution: *)
  Show[
   Histogram[
    data[[All,1]],
    Automatic,
    "Probability",
    PlotLabel->"X-axis",
    ColorFunction->Function[{height},Opacity[height]],
    ChartStyle->Purple,
    ImageSize->320
    ],
   DiscretePlot[
    PDF[MarginalDistribution[d,1],x],
    {x,0,10},
    ImageSize->320,
    ColorFunction->"Rainbow",
    PlotRange->All,
```

```
            PlotStyle->PointSize[Medium]
            ]
           ],
          (* Second plot, Histogram of Y-axis values and PDF of marginal distribution: *)
          Show[
           Histogram[
            data[[All,2]],
            Automatic,
            "Probability",
            PlotLabel->"Y-axis",
            ColorFunction->Function[{height},Opacity[height]],
            ChartStyle->Purple,
            ImageSize->320
            ],
           DiscretePlot[
            PDF[MarginalDistribution[d,2],y],
            {y,0,10},
            ImageSize->320,
            ColorFunction->"Rainbow",
            PlotRange->All,
            PlotStyle->PointSize[Medium]
            ]
           ],
          (* Third plot, Scatter plot of the dataset: *)
          ListPlot[
           data,
           PlotStyle->Directive[Purple,PointSize[0.05],Opacity[0.002]],
           AspectRatio->1,
           Frame->True,
           Axes->False
           ]
          }
         ]
```

Output

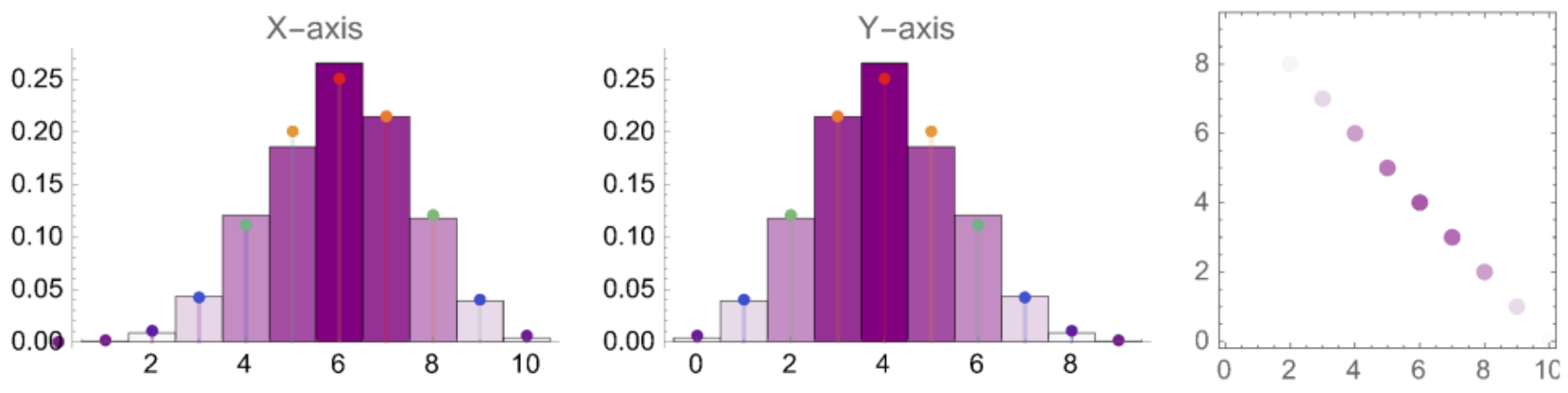

***Mathematica Examples 15.54***

Input

```
(* The code generates a bivariate dataset consisting of 10,000 random samples drawn
from a multinomial distribution with mean n=10, p1=0.6 and p2=0.4. The code generates
a scatter plot of the bivariate dataset (each point in the plot represents a pair of
X and Y values from the dataset) with additional histograms as frame labels to provide
insights into the individual variable distributions (marginal distribution in x-axis
and marginal distribution in y-axis). The three plots are integrated into one plot:
*)

data=RandomVariate[
   MultinomialDistribution[10,{0.6,0.4}],
   10000
   ];

ListPlot[
 data,
```

```
          PlotRange->{{0,10},{0,10}},
          Axes->None,
          Frame->True,
          AspectRatio->1,
          ImageSize->200,
          PlotStyle->Directive[Purple,PointSize[0.05],Opacity[0.002]],
          FrameLabel->{
            {
             Automatic,
             Histogram[
              data[[All,2]],
              {0,10,1},
              AspectRatio->0.2,
              Axes->None,
              BarOrigin->Top,
              ImageSize->170,
              ColorFunction->Function[{height},Opacity[height]],
              ChartStyle->Purple
              ]
             },
            {
             Automatic,
             Histogram[
              data[[All,1]],
              {0,10,1},
              AspectRatio->0.2,
              Axes->None,
              BarOrigin->Bottom,
              ImageSize->170,
              ColorFunction->Function[{height},Opacity[height]],
              ChartStyle->Purple
              ]
             }
            }
          ]
```

Output

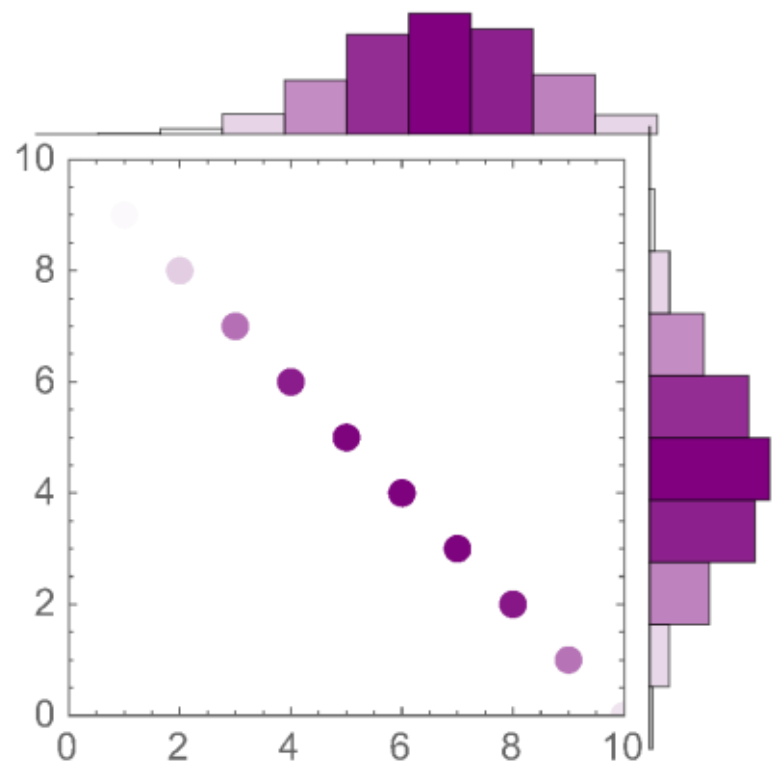

***Mathematica Examples 15.55***

Input

```
(* The code creates a dynamic histogram of data and a plot of the PDF generated from a multinomial distribution using the Manipulate function. The Manipulate function creates interactive controls for the user to adjust the values of n and p1, which are the parameters of the multinomial distribution and the sample size: *)

Manipulate[
 Module[
  {
   data=RandomVariate[
```

```
          d=MultinomialDistribution[n,{p1,1-p1}],
          Size
          ]
        },
      Histogram3D[
       data,
       Automatic,
       "PDF",
       ColorFunction->"Rainbow",
       ImageSize->320
       ]
      ],
    {{n,10,"n"},1,100,1},
    {{p1,0.5,"Probability 1"},0.1,0.9,0.1},
    {{Size,1000,"sample size"},500,5000,100}
    ]
```

Output

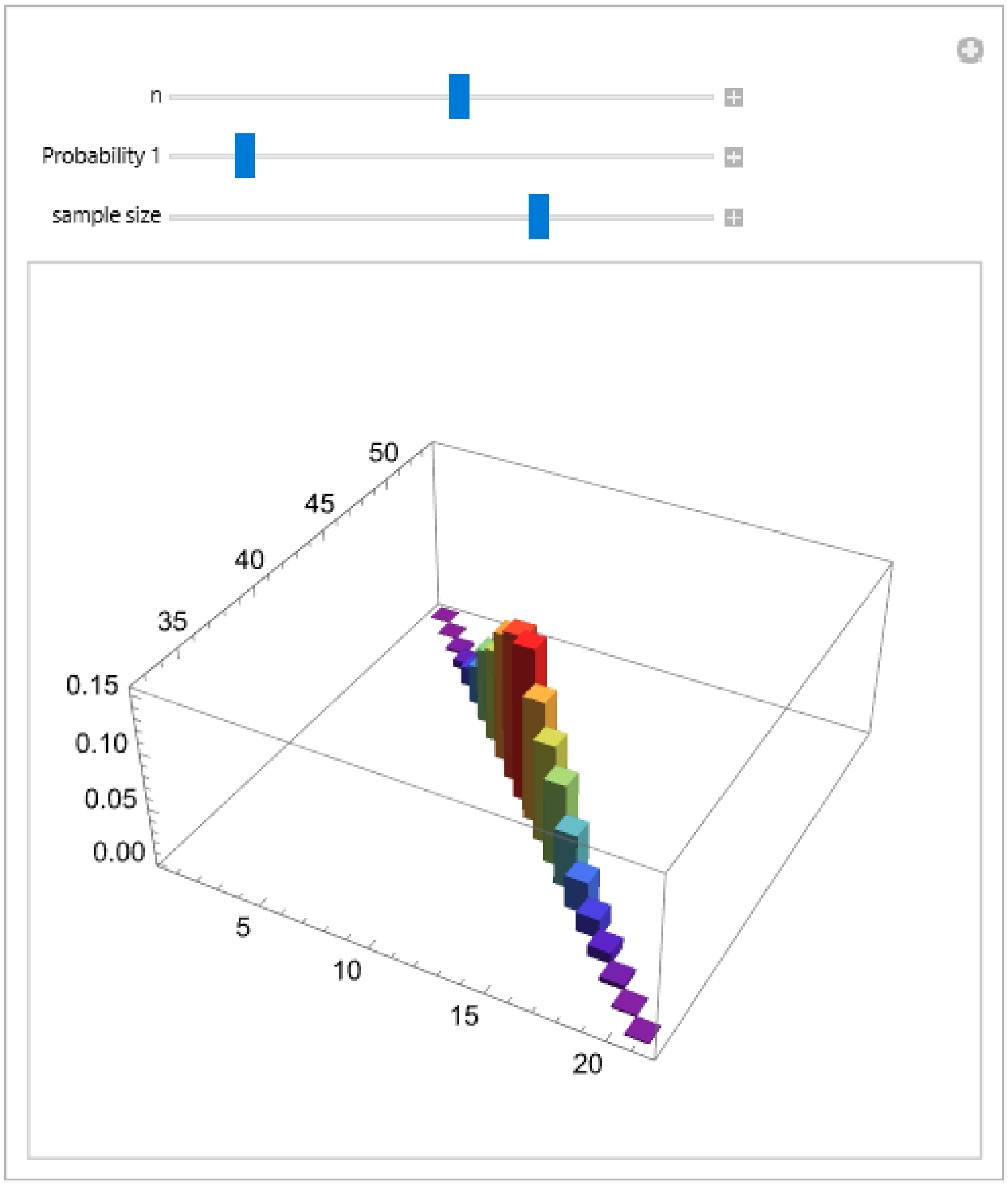

# CHAPTER 16

# SAMPLING THEORY

In this chapter, we will study the fundamental concepts of sampling theory and explore various distributions that play a crucial role in statistical inference. Understanding these concepts is essential for drawing accurate conclusions from a sample and making inferences about the population. We will cover the following topics and concepts.

- Sampling theory:
  Statistical inference refers to the process of drawing conclusions or making predictions about a population based on sample data. Sampling theory focuses on the process of selecting a subset, or sample, from a larger population. By studying samples, we can make inferences about the population as a whole, which can save time and resources.
- Sampling distribution of the sample mean:
  The sampling distribution of the sample mean is a key concept in statistics. It describes the distribution of all possible sample means that could be obtained from repeated sampling from a population. Understanding this distribution is crucial for estimating population parameters, such as the population mean, and making inferences about the population based on the sample mean.
- Central limit theorem (CLT):
  The CLT is a fundamental result in statistics that states that, under certain conditions, the distribution of sample means will approximate a normal distribution, regardless of the shape of the population distribution. This theorem is widely used in statistical inference and plays a pivotal role in hypothesis testing and confidence interval estimation.
- Chi-square distribution:
  The chi-square distribution is a probability distribution that arises in various statistical applications. It is commonly used for hypothesis testing and constructing confidence intervals for the population variance and assessing goodness-of-fit. We will explore the properties and applications of the chi-square distribution and its relationship with other statistical distributions.
- Student t-distribution:
  The student t-distribution is employed when the population standard deviation is unknown and needs to be estimated from the sample. We will explore the characteristics of the t-distribution and understand how it differs from the standard normal distribution.
- Fisher F-distribution:
  The Fisher F-distribution is employed in statistical inference to compare the variances of two or more populations. It is commonly used in the analysis of variance. We will examine the properties and applications of the F-distribution and how it relates to the chi-square distribution.
- Sampling distribution of sample variance:
  The sampling distribution of the sample variance provides insights into the variability of sample variances obtained from repeated sampling. It follows a chi-square distribution and is crucial for making inferences about the population variance.
- Sampling distribution of sample proportion:
  The sampling distribution of the sample proportion is relevant when studying categorical data and making inferences about population proportions. We will investigate the properties and applications of this distribution.

By understanding the concepts covered in this chapter, you will gain a solid foundation in sampling theory and the distributions associated with sampling. These concepts are vital in statistical analysis and will enable you to draw meaningful conclusions and make accurate inferences about populations based on sample data.

## 16.1 Sampling Theory

**Definition (Sampling Theory):** Sampling theory studies relationships between a population and samples drawn from the population.

Sampling theory is of great value in many connections:

- For example, it is useful in estimating unknown population quantities (such as population mean and variance), often called population parameters or briefly parameters, from knowledge of corresponding sample quantities (such as sample mean and variance), often called sample statistics or briefly statistics.
- Sampling theory is also helpful in determining whether the observed differences between two samples are due to chance variation or whether they are really significant. Such questions arise, for example, in deciding whether one production process is better than another.

**Definition (Statistical Inference):** Statistical inference, in general, is the study of the inferences made about a population using samples taken from it, together with indications of the accuracy of such inferences by using probability theory.

For the conclusions of sampling theory to be valid, samples must be chosen to represent a population. A study of sampling methods and related problems is called the design of the experiment.

A simple random sample is a subset of a statistical population in which each member of the subset has an equal probability of being chosen. The main attribute of this sampling plan is that every sample of size $n$ has the same chance of being chosen. For example, suppose you want to select a sample of size $n = 2$ from a population containing $N = 4$ objects. If the four objects are identified by the symbols $x_1$, $x_2$, $x_3$, and $x_4$, there are six distinct pairs that could be selected, as listed in Table 16.1. If the sample of $n = 2$ observations is chosen so that each of these six samples has the same chance—one out of six or $1/6$—of selection, then the resulting sample is called a simple random sample or just a random sample.

**Table 16.1**

| Sample | Observations in Sample |
|---|---|
| 1 | $x_1x_2$ |
| 2 | $x_1x_3$ |
| 3 | $x_1x_4$ |
| 4 | $x_2x_3$ |
| 5 | $x_2x_4$ |
| 6 | $x_3x_4$ |

**Definition (Random Sample 1):** If a sample of $n$ elements is selected from a population of $N$ elements using a sampling plan in which each member of a population has an equal chance of being included in the sample, then the sampling is said to be random, and the resulting sample is a simple random sample.

To use sample data to make inferences about an entire population, it is necessary to make some assumptions about the relationship between the two. One such assumption, which is often quite reasonable, is that there is an underlying (population) probability distribution such that the measurable values of the items in the population can be thought of as being independent RVs having this distribution. If the sample data are then chosen in a random fashion, then it is reasonable to suppose that they, too, are independent values from the distribution. The RVs are usually assumed to be independent and identically distributed.

So, in selecting a random sample of size $n$ from a population $F(x)$, let us define the RV $X_i$, $i = 1,2,\ldots,n$, to represent the $i$th measurement or sample value that we observe. The RVs $X_1, \ldots, X_n$ will then constitute a random sample from the population $F(x)$ with numerical values $x_1, \ldots, x_n$ if the measurements are obtained by repeating the experiment $n$ independent times under essentially the same conditions. Because of the identical conditions under which the elements of the sample are selected, it is reasonable to assume that the $n$ RVs $X_1,\ldots,X_n$ are independent and that each has the same probability distribution $F(x)$. That is, the probability distributions of $X_1,\ldots,X_n$ are, respectively, $F(x_1),\ldots,F(x_n)$, and their joint probability distribution is $F(x_1,\ldots,x_n) = F(x_1)\cdots F(x_n)$. Hence, the concept of a random sample is described formally by the following definition.

**Definition (Random Sample 2):** If $X_1, \ldots, X_n$ are independent RVs having a common distribution $F$, then we say that they constitute a random sample from the distribution $F$ and write its joint probability distribution as

$$F(x_1,\ldots,x_n) = F(x_1)\cdots F(x_n). \tag{16.1}$$

Such a collection of RVs is also referred to as being independent and identically distributed (IID).

In other words, the terms random sample and IID are basically one and the same. In statistics, "random sample" is the typical terminology, but in probability it is more common to say "IID".

By assuming independence, we can treat each observation as a unique source of information, unaffected by the other observations. The assumption of identical distribution ensures that the sample accurately represents the population, allowing us to make generalizations and draw conclusions about the population based on the sample.

In most applications, the population distribution $F$ will not be completely specified, and one will attempt to use the data to make inferences about $F$. Sometimes it will be supposed that $F$ is specified up to some unknown parameters (for instance, one might suppose that $F$ was a normal distribution function having an unknown mean and variance), and at other times it might be assumed that almost nothing is known about $F$ (except maybe for assuming that it is a continuous, or a discrete, distribution). Problems in which the form of the underlying distribution is specified up to a set of unknown parameters are called parametric inference problems, whereas those in which nothing is assumed about the form of $F$ are called nonparametric inference problems.

**Remarks:**

- If we draw a number from an urn, we have the choice of replacing or not replacing the number into the urn before a second drawing. In the first case, the number can come up again and again, whereas in the second it can only come up once.

  **Definition (Sampling with and without Replacement):** Sampling where each member of the population may be chosen more than once is called sampling with replacement, while if each member cannot be chosen more than once, it is called sampling without replacement.
- Populations are either finite or infinite. For example, if we draw 10 balls successively without replacement from an urn containing 100 balls, we are sampling from a finite population; while if we toss a coin 50 times and count the number of heads, we are sampling from an infinite population.
- A finite population in which sampling is with replacement can theoretically be considered infinite since any number of samples can be drawn without exhausting the population.
- For many practical purposes, sampling from a finite population that is very large can be considered to be sampling from an infinite population.

## Sampling Distribution

If a number of samples, each of the same size $n$, is drawn from a given population (either with or without replacement) and for each sample, the value of the statistic (such as the mean and the standard deviation) is calculated, a series of values of a statistic will be obtained. If the number of samples is large, this may be arranged into a frequency table. The frequency distribution of the statistic that will be obtained if the number of samples, each of the same size, were large is called the sampling distribution of a statistic. If, for example, the particular statistic used is the sample mean, then the distribution is called the sampling distribution of means, or the sampling distribution of the mean. Similarly, we could have sampling distributions of standard deviations, variances, medians, proportions, etc. For each sampling distribution, we can compute the mean, standard deviation, etc. Thus, we can speak of the mean and standard deviation of the sampling distribution of means, etc.

**Definition (Sampling Distribution):** The probability distribution of a statistic is called a sampling distribution.

**Remarks:**

- The standard deviation of a sampling distribution of a statistic is often called its standard error (SE).
- The standard deviation measures the dispersion or amount of variability of individual data values from its mean. While standard error measures how far the value of the statistic is likely to be from the true parameter value. For example, the standard error of the sample mean measures how far the sample mean of the data is likely to be from the true population mean.
- There is an important way to find the sampling distribution of a statistic: use a simulation to approximate the distribution. That is, draw a large number of samples of size $n$, calculating the value of the statistic for each sample, and tabulate the results in a relative frequency histogram. When the number of samples is large, the histogram will be very close to the theoretical sampling distribution.

***Example 16.1***

A population consists of $N = 5$ numbers: 3, 6, 9, 12, 15. If a random sample of size $n = 3$ is selected without replacement, find the sampling distributions for the sample mean $\bar{X}$ and the sample median $m$.

***Solution***

The population contains five distinct numbers, and each is equally likely, with probability $p(X) = 1/5$. We can easily find the population mean and median as

$$\mu = \frac{3+6+9+12+15}{5}, \qquad M = 9.$$

To find the sampling distribution, we need to know what values of $\bar{X}$ and $m$ can occur when the sample is taken. There are 10 possible random samples of size $n = 3$ and each is equally likely, with a probability $1/10$. These samples, along with the calculated values of $\bar{X}$ and $m$ for each, are listed in Table 16.2.

**Table 16.2**

| Sample | Sample elements | $\bar{X}$ | $m$ |
|---|---|---|---|
| 1 | 3,6,9 | 6 | 6 |
| 2 | 3,6,12 | 7 | 6 |
| 3 | 3,6,15 | 8 | 6 |
| 4 | 3,9,12 | 8 | 9 |
| 5 | 3,9,15 | 9 | 9 |
| 6 | 3,12,15 | 10 | 12 |
| 7 | 6,9,12 | 9 | 9 |
| 8 | 6,9,15 | 10 | 9 |
| 9 | 6,12,15 | 11 | 12 |
| 10 | 9,12,15 | 12 | 12 |

Using the values in Table 16.2, we can find the sampling distribution of $\bar{X}$ and $m$, shown in Table 16.3 and graphed in Figure 16.1.

**Table 16.3**

| $\bar{X}$ | 6 | 7 | 8 | 9 | 10 | 11 | 12 | $m$ | 6 | 9 | 12 |
|---|---|---|---|---|---|---|---|---|---|---|---|
| $p(\bar{X})$ | 0.1 | 0.1 | 0.2 | 0.2 | 0.2 | 0.1 | 0.1 | $p(m)$ | 0.3 | 0.4 | 0.3 |

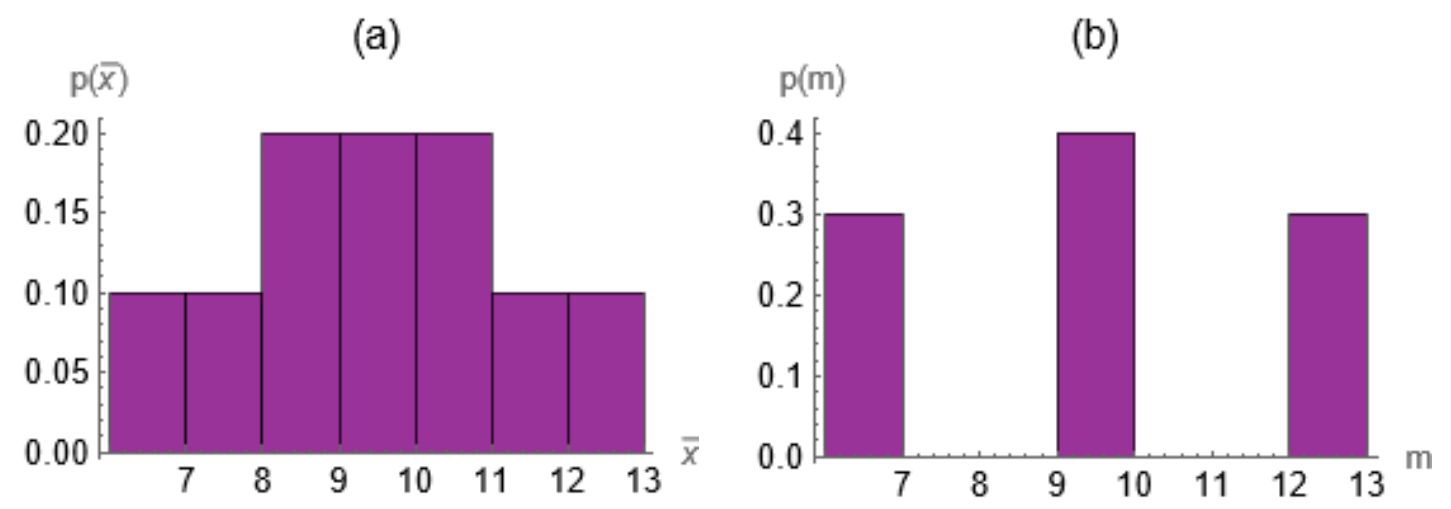

**Figure 16.1.** The sampling distribution of $\bar{X}$ and $m$.

## 16.2 Sampling Distribution of Means

**Theorem 16.1 (The Sampling Distribution of the Sample Mean, $\bar{X}$):** If a random sample of $n$ measurements is selected from a population with mean $\mu$ and standard deviation $\sigma$, the sampling distribution of the sample mean $\bar{X}$ will have

$$\mu_{\bar{X}} = \mu, \tag{16.2.1}$$

$$\sigma_{\bar{X}} = \frac{\sigma}{\sqrt{n}}. \tag{16.2.2}$$

As a result, we can conclude that $\bar{X}$ is also centered on the population mean $\mu$, but its spread becomes more and more reduced as the sample size increases.

**Proof:**

Let $X_1, X_2,..., X_n$ be a sample of values from the population. The sample mean is defined by

$$\bar{X} = \frac{X_1 + X_2 + \ldots + X_n}{n}.$$

Since the value of the sample mean $\bar{X}$ is determined by the values of the RVs in the sample, it follows that $\bar{X}$ is also a RV. Its expected value and variance are obtained as follows:

$$\begin{aligned} \mu_{\bar{X}} &= E[\bar{X}] \\ &= E\left(\frac{X_1 + X_2 + \cdots + X_n}{n}\right) \\ &= E\left(\frac{X_1}{n}\right) + E\left(\frac{X_2}{n}\right) + \cdots + E\left(\frac{X_n}{n}\right) \\ &= \frac{1}{n}[E(X_1) + E(X_2) + \cdots + E(X_n)] \\ &= \frac{1}{n}[n\mu] \\ &= \mu, \end{aligned}$$

where $E(X_1) = E(X_2) = \cdots = E(X_n) = \mu$ and

$$\begin{aligned} \sigma_{\bar{X}}^2 &= \mathrm{Var}(\bar{X}) \\ &= \mathrm{Var}\left(\frac{X_1 + X_2 + \cdots + X_n}{n}\right) \\ &= \frac{1}{n^2}[\mathrm{Var}(X_1) + \cdots + \mathrm{Var}(X_n)] \\ &= \frac{1}{n^2}[n\sigma^2] \\ &= \frac{\sigma^2}{n}, \end{aligned}$$

where $\mathrm{Var}(X_1) = \mathrm{Var}(X_2) = \cdots = \mathrm{Var}(X_n) = \sigma^2$ and $\mu$ and $\sigma^2$ are the population mean and variance, respectively. Hence, the expected value of the sample mean is the population mean $\mu$ whereas its variance is $1/n$ times the population variance.

■

**Remarks:**

- According to property (16.2.1), the distribution of $\bar{X}$ is centered precisely at the mean of the population from which the sample has been selected. Figure 16.2 plots the PDF of the sample mean from a normal population $N(0,2)$ for a variety of sample sizes.
- Property (16.2.2) shows that the $\bar{X}$ distribution becomes more concentrated about $\mu$ as the sample size $n$ increases, because its standard deviation decreases.
- The expression $\frac{\sigma}{\sqrt{n}}$ for the standard deviation of $\bar{X}$ is called the standard error of the mean, and it indicates the typical amount by which a value of $\bar{X}$ will deviate from the true mean, $\mu$ (in contrast, $\sigma$ itself represents the typical difference between an individual $X_i$ and $\mu$).

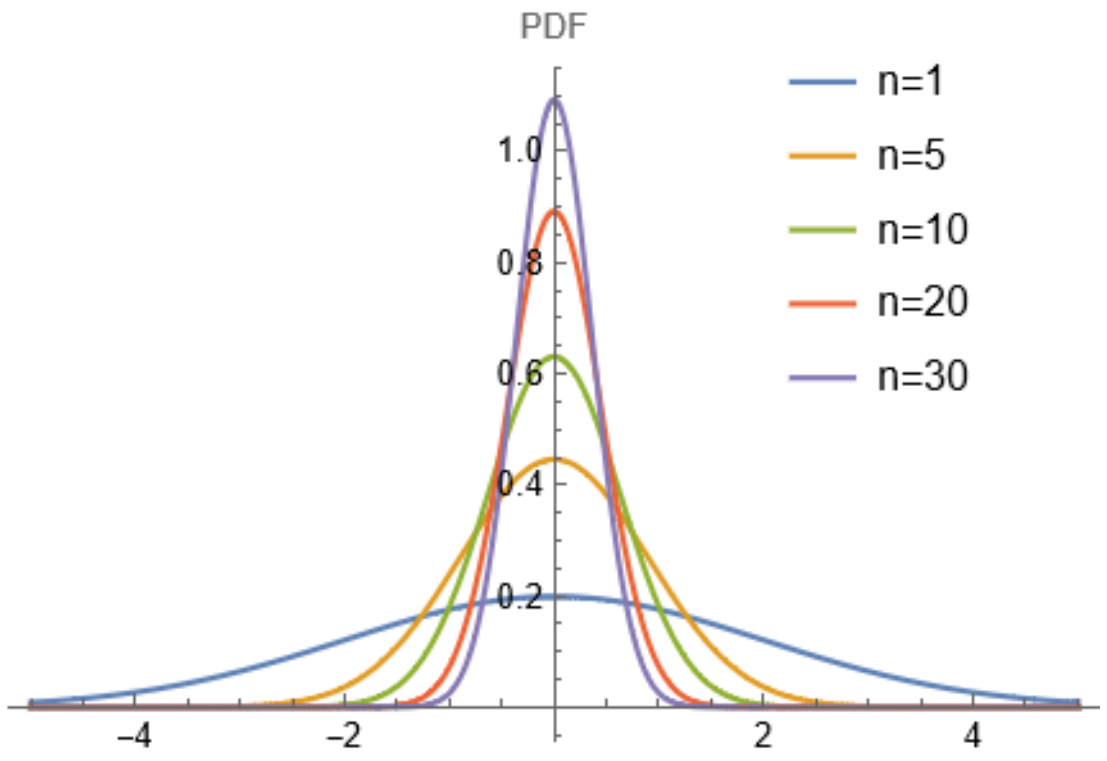

**Figure 16.2.** Densities of sample means from a normal population $N(0,2)$.

Before going into the details, let us first consider the following theorem which explains the MGF of a linear combination of independent RVs.

**Theorem 16.2:** The MGF of the linear combination of independent RVs, $X_1, X_2, \ldots, X_n$, is given by the product of the individual MGFs evaluated at the respective constants multiplied by $t$.

$$M_{\sum_{i=1}^{n} a_i X_i}(t) = \prod_{i=1}^{n} M_{X_i}(a_i t). \tag{16.3}$$

**Proof:**

Consider a linear combination of independent RVs $X_1, X_2, \ldots, X_n$ with respective MGFs $M_1(t), M_2(t), \ldots, M_n(t)$, and constants $a_1, a_2, \ldots, a_n$:

$$\begin{aligned} Y &= a_1 X_1 + a_2 X_2 + \cdots + a_n X_n \\ &= \sum_{i=1}^{n} a_i X_i. \end{aligned}$$

Now, we want to find the MGF of $Y$, denoted as $M_Y(t) = M_{\sum_{i=1}^{n} a_i X_i}(t)$. Using the definition of MGF, we have:

$$M_Y(t) = E[e^{tY}].$$

Substitute the expression for $Y$:

$$M_Y(t) = E\left[e^{t(a_1X_1+a_2X_2+\cdots+a_nX_n)}\right]$$
$$= E\left[e^{ta_1X_1} \cdot e^{ta_2X_2} \ldots e^{ta_nX_n}\right].$$

Since $X_1$, $X_2$, ..., $X_n$ are independent RVs, the joint MGF can be expressed as the product of the individual MGFs:

$$M_Y(t) = E[e^{ta_1X_1}] \cdot E[e^{ta_2X_2}] \ldots E[e^{ta_nX_n}].$$

Using the definition of the MGF, the expectation of $E[e^{taX}]$ for a RV $X$ is simply the MGF evaluated at $ta$:

$$M_X(ta) = E[e^{taX}].$$

Applying this result to each term in the previous equation, we obtain:

$$M_Y(t) = M_{X_1}(a_1t) \cdot M_{X_2}(a_2t) \ldots M_{X_n}(a_nt)$$
$$= \prod_{i=1}^{n} M_{X_i}(a_it).$$

■

**Theorem 16.3:** Let $X_1$, $X_2$,..., $X_n$ be a random sample of size $n$ from $N(\mu, \sigma^2)$. The sampling distribution of $\bar{X} = \frac{1}{n}\sum_{i=1}^{n} X_i$ will be exactly normally distributed, regardless of the sample size, $n$. Then,

$$\bar{X} \sim N\left(\mu, \frac{\sigma^2}{n}\right), \quad (16.4)$$

i.e., when the population is $N(\mu, \sigma^2)$, $\bar{X} \sim N(\mu, \frac{\sigma^2}{n})$ for any sample of size $n$.

**Proof:**

Since $X_1$, $X_2$,..., $X_n$ is a random sample of size $n$ from $N(\mu, \sigma^2)$, we can consider them as IID RVs having the same distribution $N(\mu, \sigma^2)$. Therefore,

$$M_{X_i}(t) = e^{\mu t + \frac{1}{2}t^2\sigma^2} \quad ; i = 1, 2, \ldots, n,$$

and

$$M_{\bar{X}}(t) = M_{\frac{1}{n}\sum_{i=1}^{n} X_i}(t)$$
$$= M_{\sum_{i=1}^{n} X_i}\left(\frac{t}{n}\right)$$
$$= \prod_{i=1}^{n} M_{X_i}\left(\frac{t}{n}\right)$$
$$= \prod_{i=1}^{n} e^{\mu\frac{t}{n} + \frac{1}{2}\left(\frac{t}{n}\right)^2\sigma^2}$$
$$= \left(e^{\mu\frac{t}{n} + \frac{1}{2}\left(\frac{t}{n}\right)^2\sigma^2}\right)^n$$
$$= e^{\mu t + \frac{1}{2}\frac{\sigma^2}{n}t^2},$$

which is the MGF of $N(\mu, \frac{\sigma^2}{n})$. Therefore,

$$\bar{X} \sim N\left(\mu, \frac{\sigma^2}{n}\right),$$

and its PDF is given by

$$f_{\bar{X}}(x) = \frac{\sqrt{n}}{\sigma\sqrt{2\pi}} e^{-\frac{n(x-\mu)^2}{2\sigma^2}} \quad ; -\infty < x < \infty.$$

■

**Theorem 16.4 (CLT):** If $X_1, X_2, \ldots, X_n$ is a random sample of size $n$ taken from a population (either finite or infinite) with mean $\mu$ and finite variance $\sigma^2$ and if $\bar{X}$ is the sample mean, the limiting form of the distribution of

$$Z = \frac{\bar{X} - \mu}{\frac{\sigma}{\sqrt{n}}}, \tag{16.5}$$

as $n \to \infty$, is the standard normal distribution.

**Proof:**

Recall that the MGF of a standard normal distribution is given by $e^{\frac{1}{2}t^2}$. Let $M(t) = e^{\frac{1}{2}t^2}$. Let $M_Z(t)$ denote the MGF of $Z$. It is our purpose to show that $M_Z(t)$ must approach $M(t)$ when $n$, the sample size, becomes large. Now

$$\begin{aligned}
M_Z(t) &= E(e^{tZ}) \\
&= E\left[\exp\left(t\frac{\bar{X}-\mu}{\frac{\sigma}{\sqrt{n}}}\right)\right] \\
&= E\left[\exp\left(t\frac{\sum_{i=1}^{n}\frac{X_i}{n}-\mu}{\frac{\sigma}{\sqrt{n}}}\right)\right] \\
&= E\left[\exp\left(\frac{t}{n}\sum_{i=1}^{n}\frac{X_i-\mu}{\frac{\sigma}{\sqrt{n}}}\right)\right] \\
&= E\left[\prod_{i=1}^{n}\exp\left(\frac{t}{n}\frac{X_i-\mu}{\frac{\sigma}{\sqrt{n}}}\right)\right] \\
&= \prod_{i=1}^{n} E\left[\exp\left(\frac{t}{\sqrt{n}}\frac{X_i-\mu}{\sigma}\right)\right],
\end{aligned}$$

using the independence of $X_1 \ldots, X_n$. Since $X_i$'s are identically distributed, we have

$$\begin{aligned}
M_Z(t) &= \left[E\left[\exp\left(\frac{t}{\sqrt{n}}\frac{X}{\sigma}\right)\exp\left(\frac{t}{\sqrt{n}}\frac{-\mu}{\sigma}\right)\right]\right]^n \\
&= \left[E\left[\exp\left(\frac{t}{\sqrt{n}}\frac{X}{\sigma}\right)\right]\exp\left(\frac{t}{\sqrt{n}}\frac{-\mu}{\sigma}\right)\right]^n \\
&= e^{n\frac{-\mu t}{\sigma\sqrt{n}}}\left[M_X\left(\frac{t}{\sigma\sqrt{n}}\right)\right]^n \\
&= e^{\frac{-\sqrt{n}\mu t}{\sigma}}\left[M_X\left(\frac{t}{\sigma\sqrt{n}}\right)\right]^n.
\end{aligned}$$

Therefore,

$$M_Z(t) = e^{\frac{-\sqrt{n}\mu t}{\sigma}}\left[1 + \mu_1\frac{t}{\sigma\sqrt{n}} + \frac{\mu_2}{2!}\left(\frac{t}{\sigma\sqrt{n}}\right)^2 + O\left(\frac{1}{n^{3/2}}\right)\right]^n.$$

Since, $M_X\left(\frac{t}{\sigma\sqrt{n}}\right) = 1 + \mu_1\frac{t}{\sigma\sqrt{n}} + \frac{\mu_2}{2!}\left(\frac{t}{\sigma\sqrt{n}}\right)^2 + O\left(\frac{1}{n^{3/2}}\right)$, where $O\left(\frac{1}{n^{3/2}}\right)$ denotes terms with $n^{3/2}$ and its higher powers.

$$
\begin{aligned}
\ln M_Z(t) &= \ln\left(e^{\frac{-\sqrt{n}\mu t}{\sigma}}\left[1+\mu_1\frac{t}{\sigma\sqrt{n}}+\frac{\mu_2}{2!}\left(\frac{t}{\sigma\sqrt{n}}\right)^2+O\left(\frac{1}{n^{3/2}}\right)\right]^n\right)\\
&= \ln e^{\frac{-\sqrt{n}\mu t}{\sigma}}+\ln\left[1+\mu_1\frac{t}{\sigma\sqrt{n}}+\frac{\mu_2}{2!}\left(\frac{t}{\sigma\sqrt{n}}\right)^2+O\left(\frac{1}{n^{3/2}}\right)\right]^n\\
&= \frac{-\sqrt{n}\mu t}{\sigma}+n\ln\left[1+\mu_1\frac{t}{\sigma\sqrt{n}}+\frac{\mu_2}{2!}\left(\frac{t}{\sigma\sqrt{n}}\right)^2+O\left(\frac{1}{n^{3/2}}\right)\right]\\
&= \frac{-\sqrt{n}\mu t}{\sigma}+n\ln\left[1+\left\{\mu_1\frac{t}{\sigma\sqrt{n}}+\frac{\mu_2}{2!}\left(\frac{t}{\sigma\sqrt{n}}\right)^2+O\left(\frac{1}{n^{3/2}}\right)\right\}\right]\\
&= \frac{-\sqrt{n}\mu t}{\sigma}+n\left[\left\{\mu_1\frac{t}{\sigma\sqrt{n}}+\frac{\mu_2}{2!}\left(\frac{t}{\sigma\sqrt{n}}\right)^2+O\left(\frac{1}{n^{3/2}}\right)\right\}-\frac{1}{2}\left\{\mu_1\frac{t}{\sigma\sqrt{n}}+\frac{\mu_2}{2!}\left(\frac{t}{\sigma\sqrt{n}}\right)^2+O\left(\frac{1}{n^{3/2}}\right)\right\}^2+\cdots\right]\\
&= \frac{-\sqrt{n}\mu t}{\sigma}+\left[n\left\{\mu_1\frac{t}{\sigma\sqrt{n}}+\frac{\mu_2}{2!}\left(\frac{t}{\sigma\sqrt{n}}\right)^2+O\left(\frac{1}{n^{3/2}}\right)\right\}-\frac{n}{2}\left\{\mu_1\frac{t}{\sigma\sqrt{n}}+\frac{\mu_2}{2!}\left(\frac{t}{\sigma\sqrt{n}}\right)^2+O\left(\frac{1}{n^{3/2}}\right)\right\}^2+\cdots\right]\\
&= \frac{-\sqrt{n}\mu t}{\sigma}+\left[\frac{\sqrt{n}\mu_1 t}{\sigma}+\frac{1}{2}\frac{\mu_2 t^2}{\sigma^2}+O\left(\frac{1}{n^{1/2}}\right)-\frac{n}{2}\left\{\mu_1\frac{t}{\sigma\sqrt{n}}+\frac{\mu_2}{2!}\left(\frac{t}{\sigma\sqrt{n}}\right)^2+O\left(\frac{1}{n^{3/2}}\right)\right\}^2+\cdots\right]\\
&= \frac{-\sqrt{n}\mu t}{\sigma}+\frac{\sqrt{n}\mu_1 t}{\sigma}+\frac{1}{2}\frac{\mu_2 t^2}{\sigma^2}+O\left(\frac{1}{n^{1/2}}\right)+\left[-\frac{n}{2}\left\{\mu_1\frac{t}{\sigma\sqrt{n}}+\frac{\mu_2}{2!}\left(\frac{t}{\sigma\sqrt{n}}\right)^2+O\left(\frac{1}{n^{3/2}}\right)\right\}^2+\cdots\right]\\
&= \frac{-\sqrt{n}\mu t}{\sigma}+\frac{\sqrt{n}\mu_1 t}{\sigma}+\frac{1}{2}\frac{t^2}{\sigma^2}(\mu_2-\mu_1^2)+O\left(\frac{1}{n^{1/2}}\right)\\
&= \frac{t^2}{2}+O\left(\frac{1}{n^{1/2}}\right),
\end{aligned}
$$

where $\ln(1+x)=x-\frac{x^2}{2}+\frac{x^3}{3}-\frac{x^4}{4}+\ldots$, $\mu_1=\mu$ and $\mu_2-\mu_1^2=\sigma^2$. Now, we have $\lim_{n\to\infty} M_Z(t)=e^{t^2/2}$, so that in the limit, $Z$ has the same MGF as a standard normal.

■

A practical difficulty in applying the CLT is in knowing when $n$ is "sufficiently large." The problem is that the accuracy of the approximation for a particular $n$ depends on the shape of the original underlying distribution being sampled. If the underlying distribution is symmetric and there is not much probability far out in the tails, then the approximation will be good even for a small $n$, whereas if it is highly skewed or has "heavy" tails, then a large $n$ will be required. For example, if the distribution is uniform on an interval, then it is symmetric with no probability in the tails, and the normal approximation is very good for $n$ as small as 10. However, at the other extreme, a distribution can have such fat tails that its mean fails to exist, and the CLT does not apply, so no $n$ is big enough. A popular, although frequently somewhat conservative, convention is that the CLT may be safely applied when $n > 30$. Of course, there are exceptions, but this rule applies to most distributions of real data. So, we have the following fact,

If the population distribution is nonnormal, the sampling distribution of $\bar{X}$ will be approximately normally distributed for large samples $n \geq 30$.

It is easy to demonstrate the CLT with a computer simulation experiment. We used computer software to draw 100,000 samples at random from the uniform distribution with parameters $a = 1$ and $b = 5$, each of size $n = 1, 2, 3, 10$. Figure 16.3 is a clear demonstration of the CLT, which states that as the sample size increases, the distribution of sample means approaches a normal distribution with a mean equal to the population mean and standard deviation equal to the population standard deviation divided by the square root of the sample size. In this case, the population distribution is uniform, which is not a normal distribution, but the CLT still applies. As the sample size increases, the histograms become more bell-shaped and symmetric, indicating that the sample means are approaching a normal distribution. Additionally, the histograms become narrower, indicating that the standard deviation of the sample means is decreasing as the sample size increases. This demonstrates the practical usefulness of the CLT in allowing us to make inferences about the population mean based on sample means, even when the population distribution is not normal. This type of sampling experiment can be used to investigate the sampling distribution of any statistic. Figure 16.4 provides a good illustration of how the CLT can be used to approximate the distribution of sample means for different distributions.

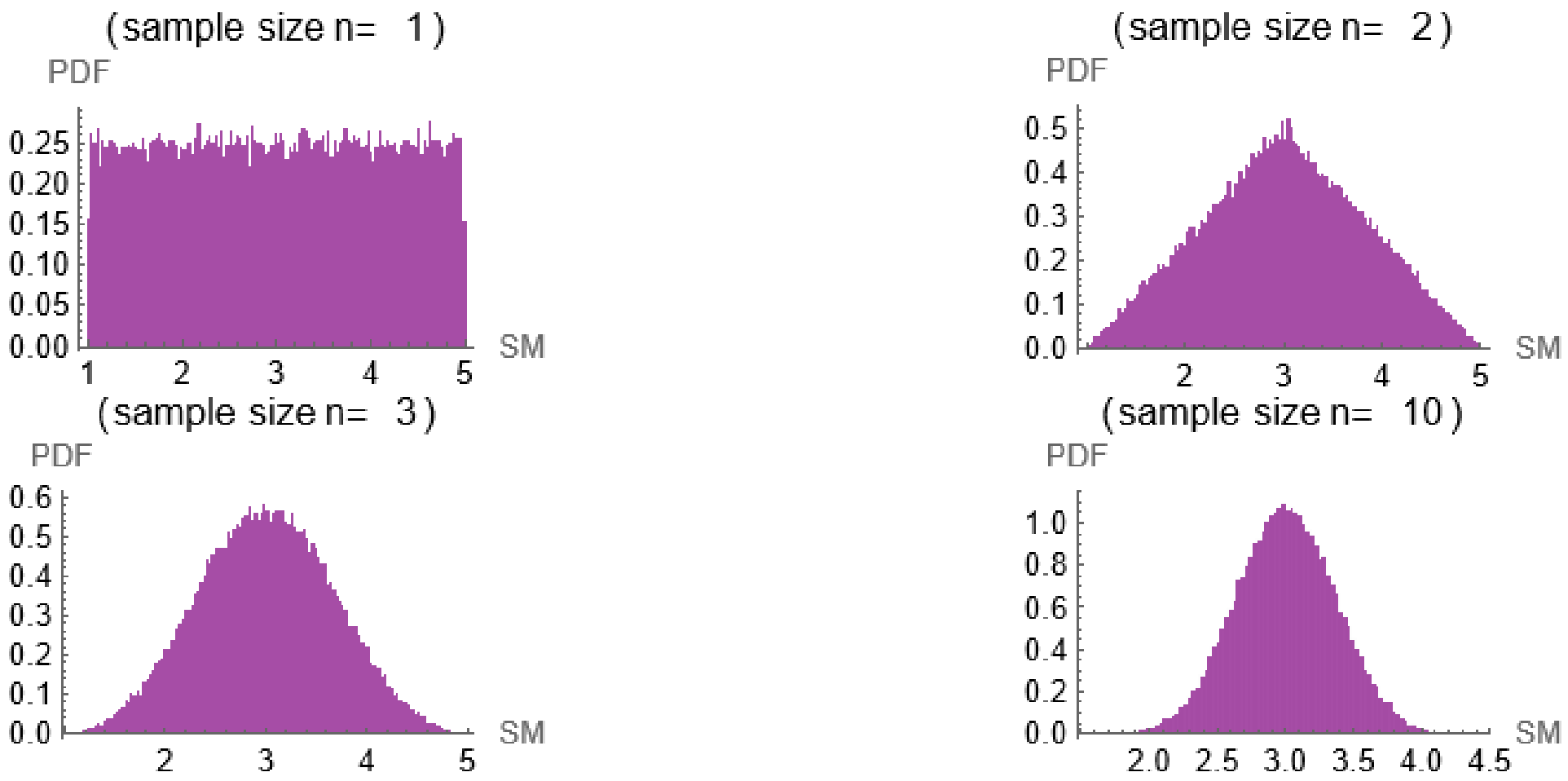

**Figure 16.3.** The figure represents histograms of sample means drawn from a uniform distribution on the interval [1,5]. The sample sizes $n = 1, 2, 3, 10$ and the number of samples is set to $100{,}000$. The histograms display the PDF of the sample means. The plot labels indicate the sample size for each histogram.

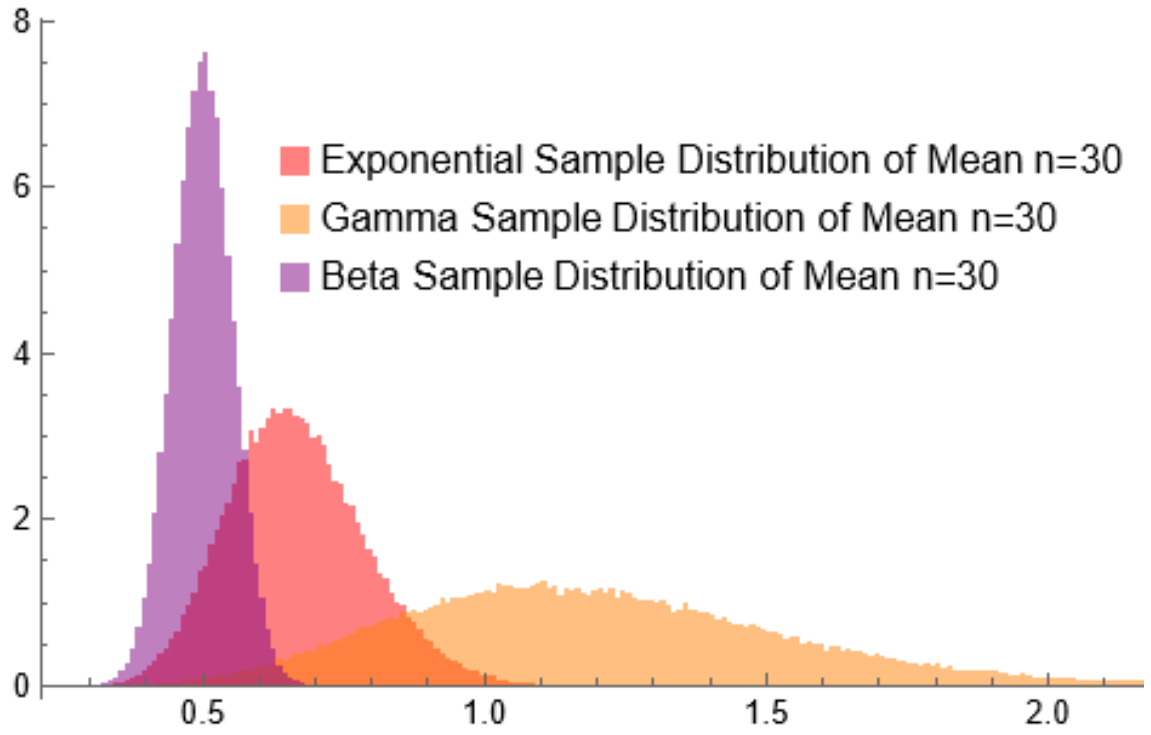

**Figure 16.4.** Histograms of a sample mean from three different distributions: exponential, gamma, and beta. The sample size is fixed at $n = 30$, and the number of samples is set to $100{,}000$.

**Theorem 16.5:** Let $\bar{X}$ be the sample mean of a sample of size $n$ from any population. Then, if $\sigma^2$ is unknown, when $n$ is large ($n \geq 30$), we can replace $\sigma^2$ by

$$S^2 = \frac{1}{n-1}\sum_{i}^{n}(X_i - \bar{X})^2, \tag{16.6.1}$$

the sample variance, and

$$\bar{X} \sim N\left(\mu, \frac{S^2}{n}\right) \quad \text{as } n \to \infty. \tag{16.6.2}$$

See Figure 16.5.

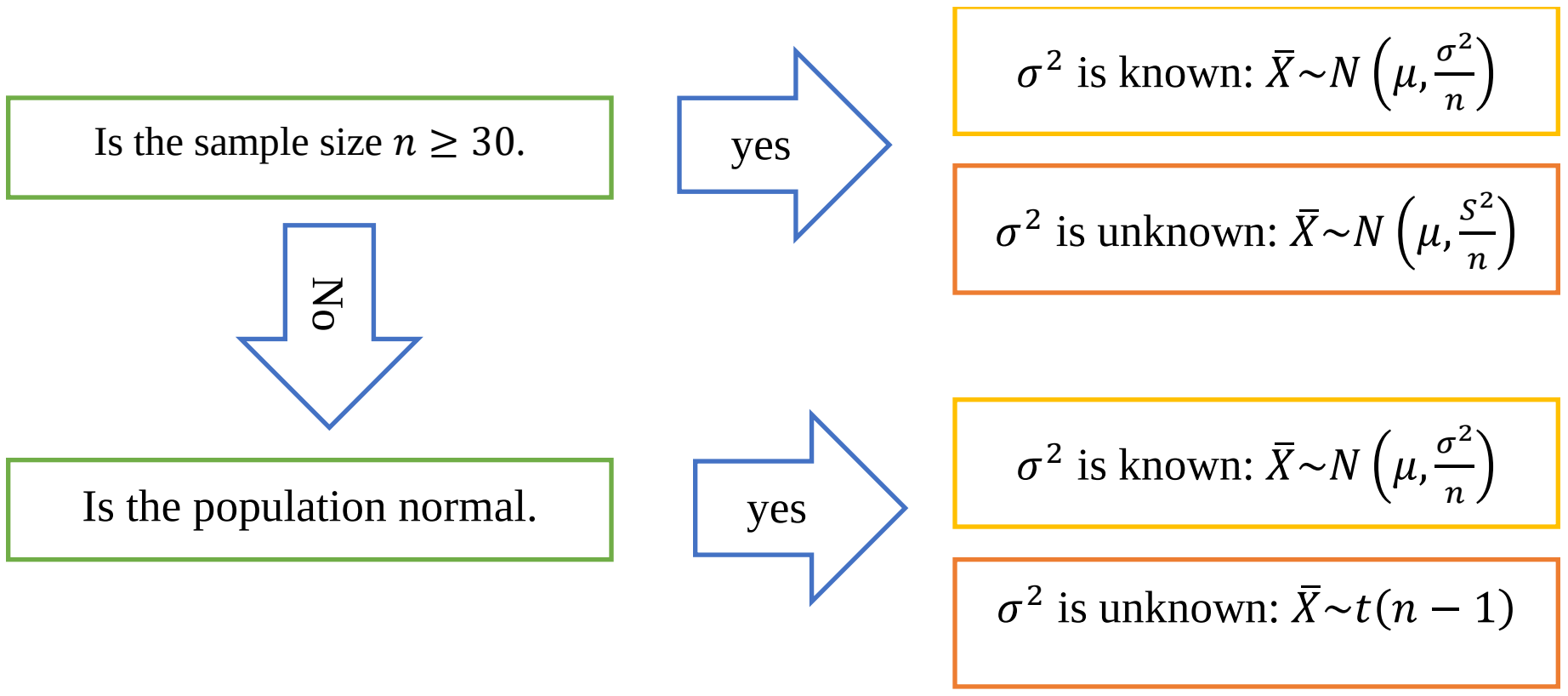

**Figure 16.5.** Flow chart of sample distribution of mean, $\bar{X}$. It shows when you should use normal and student distributions.

### *Example 16.2*

A company that makes electrical products produces light bulbs with a roughly normally distributed lifespan, with a mean of 780 hours and a standard deviation of 89 hours. Calculate the probability that 31 random bulbs will have an average lifespan of more than 814 hours.

***Solution***

The sampling distribution of $\bar{X}$ will be approximately normal, with $\mu_{\bar{X}} = 780$ and $\sigma_{\bar{X}} = 89/\sqrt{31}$. Corresponding to $\bar{x} = 814$, we find that

$$z = \frac{814 - 780}{89/\sqrt{31}} = 2.13,$$

and therefore

$$P(\bar{X} > 814) = P(Z > 2.13) = 1 - P(Z < 2.13) = 1 - 0.9834 = 0.0166.$$

```
NProbability[2.13<z,z\[Distributed]NormalDistribution[0,1]]
1-CDF[NormalDistribution[0,1],2.13]
 0.0165858
 0.0165858
```

### *Example 16.3*

Resistors with a mean resistance of 150 ohms and a standard variation of 15 ohms are produced by an electronics manufacturer. Resistance is distributed normally. Calculate the probability that a random sample of $n = 30$ resistors will have an average resistance of less than 145 ohms.

***Solution***

The sampling distribution of $\bar{X}$ will be normal, with $\mu_{\bar{X}} = 150$ ohms and a standard deviation of

$$\sigma_{\bar{X}} = \frac{\sigma}{\sqrt{n}} = \frac{15}{\sqrt{30}} = 2.73861.$$

Standardizing point $\bar{x} = 145$, we find that

$$z = \frac{145 - 150}{2.73861} = -1.82574,$$

and therefore,

$$P(\bar{X} < 145) = P(Z < -1.82574) = 0.03394.$$

```
NProbability[z<-1.82574,z\[Distributed]NormalDistribution[0,1]]
CDF[NormalDistribution[0,1],-1.82574]
 0.0339447
 0.0339447
```

***Example 16.4***

Consider the case where a RV, $X$, has a continuous uniform distribution

$$f(x) = \begin{cases} 2, & 1 \le x \le 5, \\ 0, & \text{otherwise.} \end{cases}$$

For a random sample of size $n = 50$, determine the distribution of the sample mean.

***Solution***

The mean and variance of $X$ are

$$\mu = \frac{a+b}{2} = \frac{1+5}{2} = 3,$$

and

$$\sigma^2 = \frac{(b-a)^2}{12} = \frac{(5-1)^2}{12} = \frac{4}{3}.$$

The CLT indicates that the distribution of the sample mean $\bar{X}$ is approximately normal with mean $\mu_{\bar{X}} = 3$ and variance,

$$\sigma_{\bar{X}}^2 = \frac{\sigma^2}{n} = \frac{\frac{4}{3}}{50} = \frac{2}{75}.$$

***Example 16.5***

An average of 15 ounces per bottle is used to package a particular brand of drink. There will be slight variances in the amount of liquid that each bottle really holds due to chance. The liquid content of these bottles is normally distributed with $\sigma = 0.8$ oz. The amount of liquid, in ounces, is measured in each of the 20 randomly chosen bottles of this particular type of soda. Find the probability that the sample mean will be within 0.3 oz of 15 oz.

***Solution***

We know that $X$ is normally distributed with mean $\mu = 15$ and variance $\sigma^2 = 0.64$. From CLT, $\bar{X}$ possesses a normal distribution with a mean 15 and a variance $\frac{\sigma^2}{n} = \frac{0.64}{20} = 0.032$. We find that:

$$\begin{aligned} P(|\bar{X} - 15| \le 0.3) &= P(-0.3 \le (\bar{X} - 15) \le 0.3) \\ &= P\left(-\frac{0.3}{\frac{\sigma}{\sqrt{n}}} \le \frac{(\bar{X}-15)}{\frac{\sigma}{\sqrt{n}}} \le \frac{0.3}{\frac{\sigma}{\sqrt{n}}}\right) \\ &= P\left(-\frac{0.3}{0.1789} \le Z \le \frac{0.3}{0.1789}\right) \\ &= P(-1.6769 \le Z \le 1.6769) \\ &= 0.906438. \end{aligned}$$

Therefore, there is a 0.90 percent probability that the average drink volume in any 20 randomly selected bottles will be between 14.7 and 15.3 oz.

```
NProbability[-1.6769<z<1.6769,z\[Distributed]NormalDistribution[0,1]]
CDF[NormalDistribution[0,1],1.6769]-CDF[NormalDistribution[0,1],-1.6769]
 0.906438
 0.906438
```

***Example 16.6***

A car manufacturer claims that the new line of its hybrid cars has an average gas mileage of 70 miles per gallon with a standard deviation of 5 miles per gallon. 68 miles per gallon was the average found in a randomly selected sample of 25 cars. What is the probability that the sample mean is less than or equal to 68 miles per gallon, assuming the company's claim is true?

***Solution***

If the company's claim is true, then from CLT, $\bar{X}$ is normally distributed with mean $\mu = 70$ and variance $\frac{\sigma^2}{n} = \frac{25}{25} = 1$. Hence,

$$\begin{aligned} P(\bar{X} \le 68) &= P(\bar{X} - 70 \le 68 - 70) \\ &= P(Z \le -2) \\ &= 0.022. \end{aligned}$$

It is, therefore, extremely unlikely that the mean value of the random sample of 25 cars will equal 68 miles per gallon if the company's claim is true. We come to the conclusion that the company's claim is most likely incorrect because the mean is, in fact, 68 miles per gallon.

```
NProbability[z<-2,z\[Distributed]NormalDistribution[0,1]]
N[CDF[NormalDistribution[0,1],-2]]
 0.0227501
 0.0227501
```

**Theorem 16.6:** Suppose that all possible samples of size $n$ are drawn without replacement from a finite population of size $N > n$. If we denote the mean and standard deviation of the sampling distribution of means by $\mu_{\bar{X}}$ and $\sigma_{\bar{X}}$ and the population mean and standard deviation by $\mu$ and $\sigma$, respectively, then

$$\mu_{\bar{X}} = \mu, \tag{16.7.1}$$

$$\sigma_{\bar{X}} = \frac{\sigma}{\sqrt{n}}\sqrt{\frac{N-n}{N-1}}. \tag{16.7.2}$$

The term $\sqrt{\frac{N-n}{N-1}}$, when we have a finite population and sampling is without replacement, is called the finite population correction factor for the standard error of the sample mean.

### Sampling Distribution of Differences

Suppose that we are given two populations. For each sample of size $n_1$ drawn from the first population, let us compute a statistic $\text{sta}_1$; this yields a sampling distribution for the statistic $\text{sta}_1$, whose mean and standard deviation we denote by $\mu_{\text{sta}_1}$ and $\sigma_{\text{sta}_1}$, respectively. Similarly, for each sample of size $n_2$ drawn from the second population, let us compute a statistic $\text{sta}_2$; this yields a sampling distribution for the statistic $\text{sta}_2$, whose mean and standard deviation are denoted by $\mu_{\text{sta}_2}$ and $\sigma_{\text{sta}_2}$. From all possible combinations of these samples from the two populations, we can obtain a distribution of the differences, $\text{sta}_1 - \text{sta}_2$, which is called the sampling distribution of differences of the statistics. The mean and standard deviation of this sampling distribution, denoted respectively by $\mu_{\text{sta}_1-\text{sta}_2}$ and $\sigma_{\text{sta}_1-\text{sta}_2}$, are given by

$$\mu_{\text{sta}_1-\text{sta}_2} = \mu_{\text{sta}_1} - \mu_{\text{sta}_2}, \qquad \sigma_{\text{sta}_1-\text{sta}_2} = \sqrt{\sigma^2_{\text{sta}_1} + \sigma^2_{\text{sta}_2}}, \tag{16.8}$$

provided that the samples chosen do not in any way depend on each other (i.e., the samples are independent).

If $\text{sta}_1$, and $\text{sta}_2$ are the sample means from the two populations—which means we denote by $\bar{X}_1$ and $\bar{X}_2$, respectively—then the sampling distribution of the differences of means is given for infinite populations with means and standard deviations $(\mu_1, \sigma_1)$ and $(\mu_2, \sigma_2)$, respectively, by

$$\mu_{\bar{X}_1-\bar{X}_2} = \mu_{\bar{X}_1} - \mu_{\bar{X}_2} = \mu_1 - \mu_2, \qquad \sigma_{\bar{X}_1-\bar{X}_2} = \sqrt{\sigma^2_{\bar{X}_1} + \sigma^2_{\bar{X}_2}} = \sqrt{\frac{\sigma_1^2}{n_1} + \frac{\sigma_2^2}{n_2}}, \tag{16.9}$$

using equations (16.2).

**Remarks:**

- The result also holds for finite populations if sampling is with replacement.

- Similar results can be obtained for finite populations in which sampling is without replacement by using equations (16.7).

***Example 16.7***

The average lifespan of plasma TV made by company $A$ is 7.5 years, with a standard deviation of 0.9 years, while that of company $B$ is 6.6 years, with a standard deviation of 0.7 years. What is the probability that a randomly selected sample of 50 plasma TV from company $A$ will have a mean lifespan that is at least one year longer than a similarly selected sample 40 plasma TV from company $B$?

***Solution***

We are given the following information:

| Population A | Population B |
|---|---|
| $\mu_1 = 7.5$ | $\mu_2 = 6.6$ |
| $\sigma_1 = 0.9$ | $\sigma_2 = 0.7$ |
| $n_1 = 50$ | $n_2 = 40$ |

The sampling distribution of $\bar{X}_1 - \bar{X}_2$ will be approximately normal and will have a mean and standard deviation

$$\mu_{\bar{X}_1-\bar{X}_2} = 7.5 - 6.6 = 0.9,$$

$$\sigma_{\bar{X}_1-\bar{X}_2} = \sqrt{\frac{0.81}{50} + \frac{0.49}{40}} = 0.02845.$$

Corresponding to the value $\bar{x}_1 - \bar{x}_2 = 1.0$, we find that

$$z = \frac{1.0 - 0.9}{0.02845} = 3.51494,$$

and hence

$$\begin{aligned} P(\bar{X}_1 - \bar{X}_2 \geq 1.0) &= P(Z > 3.51494) \\ &= 1 - P(Z < 3.51494) \\ &= 1 - 0.99978 \\ &= 0.0002199. \end{aligned}$$

```
NProbability[3.51494<z,z\[Distributed]NormalDistribution[0,1]]
1-N[CDF[NormalDistribution[0,1],3.51494]]
 0.000219927
 0.000219927
```

## 16.3 Chi-square Distribution

Karl Pearson in about 1900 described a well-known probability distribution called "chi-square distribution" or "distribution of chi-square". The square of a standard normal variate is known as the chi-square ($\chi^2$) variate with 1 degree of freedom. Thus, if $X \sim N(\mu, \sigma^2)$, then

$$Z = \frac{X - \mu}{\sigma} \sim N(0,1), \tag{16.10}$$

and

$$Z^2 = \left(\frac{X - \mu}{\sigma}\right)^2 \sim \chi^2(1). \tag{16.11}$$

Let $X_1, X_2, \ldots, X_n$ be a random sample of size $n$ from $N(\mu, \sigma^2)$. Then,

$$X_i \sim N(\mu, \sigma^2) \Rightarrow Z_i = \frac{X_i - \mu}{\sigma} \sim N(0,1). \tag{16.12}$$

**Definition ($\chi^2$ Distribution):** We define $\chi^2$ with $n$ degrees of freedom as the sum of the squares of $n$ independent standard normal variates. That is,

$$\chi^2 = \sum_{i=1}^{n} Z_i^2 = \sum_{i=1}^{n} \left(\frac{X_i - \mu}{\sigma}\right)^2 \sim \chi^2(n). \tag{16.13}$$

Now, we shall use MGF to obtain the distribution of $\chi^2$.

$$\begin{aligned} M_{\chi^2}(t) &= M_{\sum_{i=1}^{n} Z_i^2}(t) \\ &= \prod_{i=1}^{n} M_{Z_i^2}(t) \\ &= [M_{Z^2}(t)]^n. \end{aligned}$$

Since $M_{g(X)}(t) = \int_x \; e^{tg(x)} f_X(x)dx$,

$$\begin{aligned} M_{Z^2}(t) &= E\left(e^{tZ^2}\right) \\ &= \int_{-\infty}^{\infty} e^{tz^2} f_Z(z)dz \\ &= \int_{-\infty}^{\infty} e^{tz^2} \frac{1}{\sqrt{2\pi}} e^{-\frac{z^2}{2}} dz \\ &= \int_{-\infty}^{\infty} \frac{1}{\sqrt{2\pi}} e^{-\frac{z^2}{2}(1-2t)} dz \\ &= \int_{-\infty}^{\infty} \frac{1}{\sqrt{2\pi}} e^{-\frac{u^2}{2}} \frac{du}{\sqrt{1-2t}} \\ &= \frac{1}{\sqrt{1-2t}} \\ &= (1-2t)^{-\frac{1}{2}}, \end{aligned}$$

where $z\sqrt{1-2t} = u$, $\frac{du}{dz} = \sqrt{1-2t}$ and $\int_{-\infty}^{\infty} \frac{1}{\sqrt{2\pi}} e^{-\frac{u^2}{2}} du = 1$. Therefore, we have,

$$M_{\chi^2}(t) = (1-2t)^{-\frac{n}{2}}, \tag{16.14}$$

which is the MGF of a gamma distribution with $\alpha = n/2$ (shape parameter) and $\beta = 2$ (scale parameter) (shape-scale parameters). Hence, the PDF of $\chi^2$ is

$$f_{\chi^2}(x) = \frac{(2)^{-\frac{n}{2}}}{\Gamma(n/2)} e^{-\frac{x}{2}} x^{\left(\frac{n}{2}-1\right)}; \quad 0 < x < \infty. \tag{16.15}$$

**Definition (PDF of Chi-Square Distribution):** A RV $X$ is said to follow chi-square distribution with $n$ degrees of freedom if its PDF is given by

$$f_X(x) = \frac{(1/2)^{\frac{n}{2}}}{\Gamma(n/2)} e^{-\frac{x}{2}} x^{\left(\frac{n}{2}-1\right)} \quad ; 0 < x < \infty, \tag{16.16}$$

and we write $X \sim \chi^2(n)$.

### Properties of Chi-square Curve

- Chi-square distribution is a particular case of Gamma distribution with parameters $\alpha = n/2$ and $\beta = 2$.
- The chi-square distribution has only one parameter, the degrees of freedom $n$. It is usually denoted by $\nu$. The shape of a specific chi-square distribution depends on the number of degrees of freedom.
- The RV $\chi^2$ assumes nonnegative values only. Hence, a chi-square distribution curve starts at the origin (zero point) and lies entirely to the right of the vertical axis, see Figure 16.6.
- As we can see from Figure 16.6, the shape of a chi-square distribution curve is skewed for very small degrees of freedom, and it changes drastically as the degrees of freedom increase. Eventually, for large degrees of freedom, the chi-square distribution curve looks like a normal distribution curve. The peak of a chi-square distribution curve with 1 or 2 degrees of freedom occurs at zero and for a curve with 3 or more degrees of freedom occurs at $\nu - 2$. For instance, the peak of the chi-square distribution curve with $\nu = 3$ in Figure 16.6 occurs at $3 - 2 = 1$. The peak for the curve with $\nu = 5$ occurs at $5 - 2 = 3$. Finally, the peak for the curve with $\nu = 10$ occurs at $10 - 2 = 8$.

- The total area under a $\chi^2$-curve equals 1.

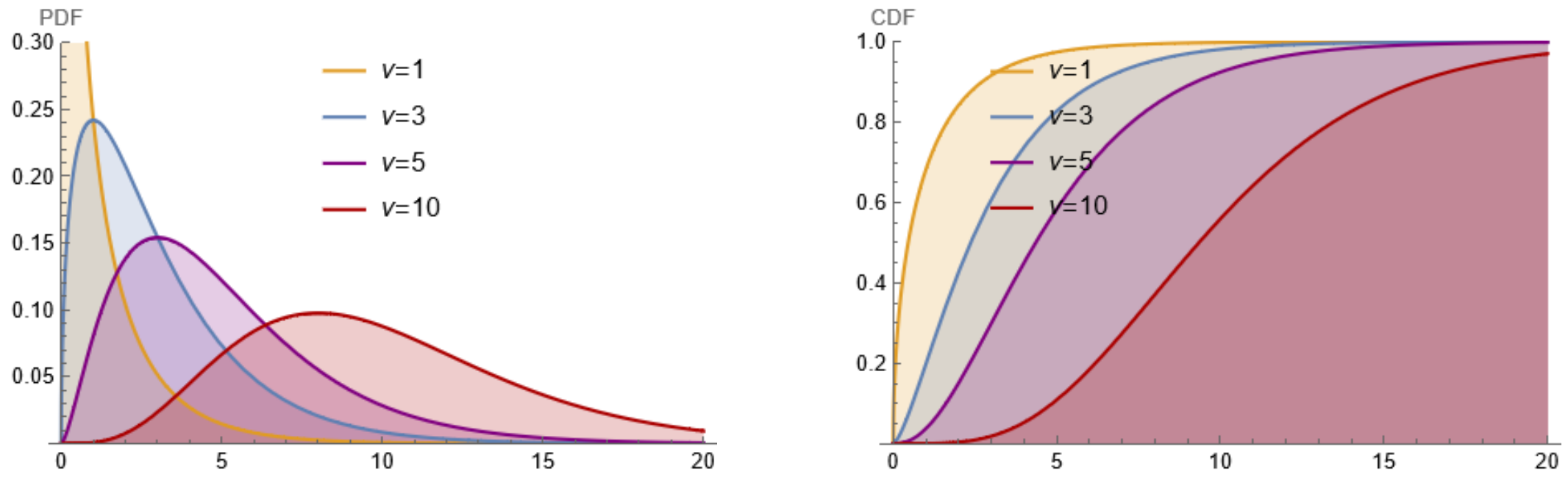

**Figure 16.6.** The shape of the Chi-Square distribution depends on the degrees of freedom parameter. When the degrees of freedom parameter is small, the Chi-Square distribution is skewed and has a long tail. As the degrees of freedom parameter increases, the distribution becomes more symmetric and bell-shaped. Specifically, as the degrees of freedom become large, the Chi-Square distribution approaches a normal distribution. Also, the shape of the CDF depends on the number of degrees of freedom $\nu$ of the Chi-Square distribution.

**Theorem 16.7:** The mean and variance of the RV $X$ having a chi-squared distribution with $n$ degrees of freedom are

$$E[X] = n, \tag{16.17.1}$$

$$\mathrm{Var}(X) = 2n. \tag{16.17.2}$$

**Proof:**

The expected value of a chi-square RV $X$ is

$$\begin{aligned} E[X] &= \int_0^\infty x f_X(x)dx \\ &= \int_0^\infty x \frac{\left(\frac{1}{2}\right)^{\frac{n}{2}}}{\Gamma\left(\frac{n}{2}\right)} e^{-\frac{x}{2}} x^{\left(\frac{n}{2}-1\right)} dx \\ &= \frac{\left(\frac{1}{2}\right)^{\frac{n}{2}}}{\Gamma\left(\frac{n}{2}\right)} \int_0^\infty e^{-\frac{x}{2}} x^{\left(\frac{n}{2}+1-1\right)} dx \\ &= \frac{\left(\frac{1}{2}\right)^{\frac{n}{2}}}{\Gamma\left(\frac{n}{2}\right)} \frac{\Gamma\left(\frac{n}{2}+1\right)}{\left(\frac{1}{2}\right)^{\frac{n}{2}+1}} \\ &= \frac{n/2}{1/2} \\ &= n. \end{aligned}$$

The variance of a Chi-square RV $X$ is

$$\mathrm{Var}(X) = E[X^2] - (E[X])^2,$$

$$E[X^2] = \int_0^\infty x^2 \frac{\left(\frac{1}{2}\right)^{\frac{n}{2}}}{\Gamma\left(\frac{n}{2}\right)} e^{-\frac{x}{2}} x^{\left(\frac{n}{2}-1\right)} dx$$
$$= \frac{\left(\frac{1}{2}\right)^{\frac{n}{2}}}{\Gamma\left(\frac{n}{2}\right)} \int_0^\infty e^{-\frac{x}{2}} x^{\left(\frac{n}{2}+2-1\right)} dx$$
$$= \frac{\left(\frac{1}{2}\right)^{\frac{n}{2}}}{\Gamma\left(\frac{n}{2}\right)} \frac{\Gamma\left(\frac{n}{2}+2\right)}{\left(\frac{1}{2}\right)^{\frac{n}{2}+2}}$$
$$= \frac{\left(\frac{n}{2}+1\right)\frac{n}{2}}{\left(\frac{1}{2}\right)^2}$$
$$= n(n+2).$$

Therefore,

$$\begin{aligned}\mathrm{Var}(X) &= E[X^2] - (E[X])^2 \\ &= n(n+2) - n^2 \\ &= 2n.\end{aligned}$$

■

**Theorem 16.8:** The MGF of the RV $X$ having a chi-squared distribution with $n$ degrees of freedom is

$$M_X(t) = (1-2t)^{-\frac{n}{2}}. \qquad (16.18)$$

**Proof:**

The MGF is

$$M_X(t) = E(e^{tX})$$
$$= \int_0^\infty e^{tx} \frac{\left(\frac{1}{2}\right)^{\frac{n}{2}}}{\Gamma\left(\frac{n}{2}\right)} e^{-\frac{x}{2}} x^{\left(\frac{n}{2}-1\right)} dx$$
$$= \frac{\left(\frac{1}{2}\right)^{\frac{n}{2}}}{\Gamma\left(\frac{n}{2}\right)} \int_0^\infty e^{-(1-2t)\frac{x}{2}} x^{\left(\frac{n}{2}-1\right)} dx$$
$$= \frac{\left(\frac{1}{2}\right)^{\frac{n}{2}}}{\Gamma\left(\frac{n}{2}\right)} \frac{\Gamma\left(\frac{n}{2}\right)}{\left(\frac{1-2t}{2}\right)^{\frac{n}{2}}}$$
$$= (1-2t)^{-\frac{n}{2}}.$$

■

**Theorem 16.9 (Additive Property):** The chi-square distribution has the additive property that if $X_1$ and $X_2$ are independent chi-square RVs with $n_1$ and $n_2$ degrees of freedom, respectively, then $X_1 + X_2$ is chi-square with $n_1 + n_2$ degrees of freedom. i.e., if $X_1 \sim \chi^2(n_1)$, $X_2 \sim \chi^2(n_2)$ and $X_1$ is independent of $X_2$, then

$$X_1 + X_2 \sim \chi^2(n_1 + n_2). \qquad (16.19)$$

**Proof:**

$X_1 \sim \chi^2(n_1)$, $X_2 \sim \chi^2(n_2)$ implies $M_{X_1}(t) = (1-2t)^{-n_1/2}$ and $M_{X_2}(t) = (1-2t)^{-n_2/2}$, respectively. Since $X_1$ and $X_2$ are independent,

$$\begin{aligned} M_{X_1+X_2}(t) &= M_{X_1}(t) M_{X_2}(t) \\ &= (1-2t)^{-\frac{n_1}{2}}(1-2t)^{-\frac{n_2}{2}} \\ &= (1-2t)^{-\frac{n_1+n_2}{2}}, \end{aligned}$$

which is the MGF of $\chi^2(n_1+n_2)$. Therefore, $X_1 + X_2 \sim \chi^2(n_1+n_2)$.

■

**Remarks:**

(1) Let $X_1, X_2, \dots, X_n$ be a random sample of size $n$ from $N(\mu, \sigma^2)$. Then,

$$\bar{X} \sim N\left(\mu, \frac{\sigma^2}{n}\right) \Rightarrow \frac{\bar{X}-\mu}{\frac{\sigma}{\sqrt{n}}} \sim N(0,1), \tag{16.20}$$

and

$$\left(\frac{\bar{X}-\mu}{\sigma/\sqrt{n}}\right)^2 \sim \chi^2(1). \tag{16.21}$$

(2) If $X \sim \chi^2(n)$, then $E[X] = n$ and $V(X) = 2n$ (a mean equal to its degrees of freedom and a variance equal to twice its degrees of freedom.)

**Theorem 16.10:** If $X \sim \chi^2(n)$, then as $n \to \infty$, we have

$$\frac{X-n}{\sqrt{2n}} \to N(0,1). \tag{16.22}$$

**Proof:**

The MGF of a chi-squared distribution with $n$ degrees of freedom is given by:

$$M_X(t) = (1-2t)^{-\frac{n}{2}}.$$

To prove that $(X-n)/\sqrt{2n}$ converges in distribution to a standard normal distribution as $n \to \infty$, we need to show that the MGF of $(X-n)/\sqrt{2n}$ converges pointwise to the MGF of a standard normal distribution as $n \to \infty$. Let $Y = (X-n)/\sqrt{2n}$. Then the MGF of $Y$ is given by:

$$M_Y(t) = E[e^{tY}].$$

Using the definition of $Y$, we can rewrite this as:

$$\begin{aligned} M_Y(t) &= E\left[e^{\frac{t(X-n)}{\sqrt{2n}}}\right] \\ &= E\left[e^{\frac{tX}{\sqrt{2n}}} e^{-\frac{tn}{\sqrt{2n}}}\right] \\ &= E\left[e^{\frac{tX}{\sqrt{2n}}}\right] e^{-\frac{tn}{\sqrt{2n}}} \\ &= M_X\left(\frac{t}{\sqrt{2n}}\right) e^{-\frac{tn}{\sqrt{2n}}}. \end{aligned}$$

Substituting the MGF of $X$, we have:

$$M_Y(t) = \left(1 - 2\left(\frac{t}{\sqrt{2n}}\right)\right)^{-\frac{n}{2}} e^{-\frac{tn}{\sqrt{2n}}}.$$

Now, taking the limit as $n \to \infty$, we have:

$$\lim_{n\to\infty} M_Y(t) = \lim_{n\to\infty} \left(1 - 2\left(\frac{t}{\sqrt{2n}}\right)\right)^{-\frac{n}{2}} e^{\frac{-tn}{\sqrt{2n}}}$$
$$= e^{-\frac{t^2}{2}},$$

which is the MGF of a standard normal distribution. Therefore, we have shown that the MGF of $(X-n)/\sqrt{2n}$ converges pointwise to the MGF of a standard normal distribution as $n \to \infty$, which implies that $(X-n)/\sqrt{2n}$ converges in distribution to a standard normal distribution as $n \to \infty$.

■

## 16.4 Sampling Distribution of Sample Variance

Let $X_1, X_2, \dots, X_n$ be a random sample of size $n$ from $N(\mu, \sigma^2)$. Let $\bar{X}$ be the sample mean. Then,

$$S^2 = \frac{1}{n-1}\sum_{i=1}^{n}(X_i - \bar{X})^2, \tag{16.23}$$

is called the sample variance. If a random sample of size $n$ is drawn from a normal population with mean $\mu$ and variance $\sigma^2$, and the sample variance is computed, we obtain a value of the statistic $S^2$. We shall proceed to consider the distribution of the statistic $\frac{(n-1)S^2}{\sigma^2}$.

**Theorem 16.11:** Let $X_1, X_2, \dots, X_n$ be IID $N(\mu, \sigma^2)$ RVs. Then $\bar{X}$ and $((X_1 - \bar{X}), (X_2 - \bar{X}), \dots (X_n - \bar{X}))$ are independent.

**Proof:**

Let $M(t, t_1, t_2, \dots, t_n)$ be the MGF of $(\bar{X}, (X_1 - \bar{X}), (X_2 - \bar{X}), \dots (X_n - \bar{X}))$.

$$\begin{aligned}
M(t, t_1, t_2, \dots, t_n) &= E[\exp\{t\bar{X} + t_1(X_1 - \bar{X}) + t_2(X_2 - \bar{X}) + \cdots + t_n(X_n - \bar{X})\}] \\
&= E\left[\exp\left\{\sum_{i=1}^{n} t_i X_i - \left(\sum_{i=1}^{n} t_i - t\right)\bar{X}\right\}\right] \\
&= E\left[\exp\left\{\sum_{i=1}^{n} t_i X_i - \left(\sum_{i=1}^{n} t_i - t\right)\frac{\sum_{i=1}^{n} X_i}{n}\right\}\right] \\
&= E\left[\exp\left\{\sum_{i=1}^{n} t_i X_i - \left(\frac{n\frac{\sum_{i=1}^{n} t_i}{n} - t}{n}\sum_{i=1}^{n} X_i\right)\right\}\right] \\
&= E\left[\exp\left\{\sum_{i=1}^{n} t_i X_i - \left(\frac{n\bar{t} - t}{n}\sum_{i=1}^{n} X_i\right)\right\}\right] \\
&= E\left[\exp\left\{\sum_{i=1}^{n} X_i\left(t_i - \frac{n\bar{t} - t}{n}\right)\right\}\right] \\
&= E\left[\exp\left\{\sum_{i=1}^{n} X_i\left(\frac{nt_i - n\bar{t} + t}{n}\right)\right\}\right]
\end{aligned}$$

$$= \prod_{i=1}^{n} E\left[\exp\left\{X_i\left(\frac{nt_i - n\bar{t} + t}{n}\right)\right\}\right]$$
$$= \prod_{i=1}^{n} \exp\left\{\mu\left(\frac{nt_i - n\bar{t} + t}{n}\right) + \left(\frac{nt_i - n\bar{t} + t}{n}\right)^2 \frac{\sigma^2}{2}\right\}$$
$$= \prod_{i=1}^{n} \exp\left\{\frac{\mu}{n}(n(t_i - \bar{t}) + t) + \frac{\sigma^2}{2n^2}(n(t_i - \bar{t}) + t)^2\right\}$$
$$= \exp\left\{\frac{\mu}{n}\left(n\sum_{i=1}^{n}(t_i - \bar{t}) + nt\right) + \frac{\sigma^2}{2n^2}\sum_{i=1}^{n}(n(t_i - \bar{t}) + t)^2\right\}$$
$$= \exp\left\{\frac{\mu}{n}\left(n\sum_{i=1}^{n} t_i - n^2\bar{t} + nt\right) + \frac{\sigma^2}{2n^2}\sum_{i=1}^{n}(n(t_i - \bar{t}) + t)^2\right\}$$
$$= \exp\left\{\frac{\mu}{n}(n^2\bar{t} - n^2\bar{t} + nt) + \frac{\sigma^2}{2n^2}\sum_{i=1}^{n}(n(t_i - \bar{t}) + t)^2\right\}$$
$$= \exp\left\{\mu t + \frac{\sigma^2}{2n^2}\sum_{i=1}^{n}\{t^2 + 2nt(t_i - \bar{t}) + n^2(t_i - \bar{t})^2\}\right\}$$
$$= \exp\left\{\mu t + \frac{\sigma^2}{2n^2}\left\{\sum_{i=1}^{n} t^2 + \sum_{i=1}^{n} 2nt(t_i - \bar{t}) + \sum_{i=1}^{n} n^2(t_i - \bar{t})^2\right\}\right\}$$
$$= \exp\left\{\mu t + \frac{\sigma^2}{2n^2}\left\{nt^2 + 2nt\sum_{i=1}^{n}(t_i - \bar{t}) + n^2\sum_{i=1}^{n}(t_i - \bar{t})^2\right\}\right\}$$
$$= \exp\left\{\mu t + \frac{\sigma^2}{2n^2}\left\{nt^2 + 2nt\left(\sum_{i=1}^{n} t_i - \sum_{i=1}^{n}\bar{t}\right) + n^2\sum_{i=1}^{n}(t_i - \bar{t})^2\right\}\right\}$$
$$= \exp\left\{\mu t + \frac{\sigma^2}{2n^2}\left\{nt^2 + 2nt(n\bar{t} - n\bar{t}) + n^2\sum_{i=1}^{n}(t_i - \bar{t})^2\right\}\right\}$$
$$= \exp \mu t \exp\left[\frac{\sigma^2}{2n^2}\left\{nt^2 + n^2\sum_{i=1}^{n}(t_i - \bar{t})^2\right\}\right]$$
$$= \exp\left(\mu t + \frac{t^2\sigma^2}{2n}\right)\exp\left[\frac{\sigma^2}{2}\sum_{i=1}^{n}(t_i - \bar{t})^2\right]$$
$$= M_{\bar{X}}(t)M_{(X_1-\bar{X}),(X_2-\bar{X}),\dots (X_n-\bar{X})}(t_1, t_2, \dots, t_n)$$
$$= M(t, 0, 0, \dots, 0)M(0, t_1, t_2, \dots, t_n).$$

Therefore, $\bar{X}$, $(X_1 - \bar{X})$, $(X_2 - \bar{X})$,… $(X_n - \bar{X})$ are independent.

■

**Theorem 16.12:** Let $X_1, X_2, \dots, X_n$ be IID $N(\mu, \sigma^2)$ RVs. Then $\bar{X}$ and $S^2$ are independent.

**Theorem 16.13:** If $S^2$ is the variance of a random sample of size $n$ taken from a normal population having the variance $\sigma^2$, then the statistic

$$\frac{(n-1)S^2}{\sigma^2} = \sum_{i=1}^{n}\frac{(X_i - \bar{X})^2}{\sigma^2}, \tag{16.24}$$

has a chi-squared distribution with $n - 1$ degrees of freedom. i.e.,

$$\frac{(n-1)S^2}{\sigma^2} \sim \chi^2(n-1). \qquad (16.25)$$

Hence, the PDF of $\frac{(n-1)S^2}{\sigma^2}$ is given by

$$f_{\frac{(n-1)S^2}{\sigma^2}}(x) = \frac{(1/2)^{\frac{n-1}{2}}}{\Gamma\big((n-1)/2\big)} e^{-\frac{x}{2}} x^{\left(\frac{n-1}{2}-1\right)} \quad ; 0 < x < \infty. \qquad (16.26)$$

**Proof:**

We have, $X_i \sim N(\mu, \sigma^2)$, $i = 1,2, \dots, n$. Then

$$\sum_{i=1}^{n} \left(\frac{X_i - \mu}{\sigma}\right)^2 \sim \chi^2(n).$$

Also, as $\bar{X} \sim N(\mu, \frac{\sigma^2}{n})$,

$$\left(\frac{\bar{X} - \mu}{\sigma/\sqrt{n}}\right)^2 \sim \chi^2(1).$$

Consider,

$$\begin{aligned}
\sum_{i=1}^{n} (X_i - \mu)^2 &= \sum_{i=1}^{n} (X_i - \bar{X} + \bar{X} - \mu)^2 \\
&= \sum_{i=1}^{n} (X_i - \bar{X})^2 + \sum_{i=1}^{n} (\bar{X} - \mu)^2 + \sum_{i=1}^{n} 2(\bar{X} - \mu)(X_i - \bar{X}) \\
&= (n-1)S^2 + n(\bar{X} - \mu)^2 + 2(\bar{X} - \mu)\left(\sum_{i=1}^{n} X_i - \sum_{i=1}^{n} \bar{X}\right) \\
&= (n-1)S^2 + n(\bar{X} - \mu)^2 + 2(\bar{X} - \mu)(n\bar{X} - n\bar{X}) \\
&= (n-1)S^2 + n(\bar{X} - \mu)^2.
\end{aligned}$$

Dividing each term by $\sigma^2$, we have

$$\sum_{i=1}^{n} \left(\frac{X_i - \mu}{\sigma}\right)^2 = \frac{(n-1)S^2}{\sigma^2} + \left(\frac{\bar{X} - \mu}{\sigma/\sqrt{n}}\right)^2.$$

Since $\bar{X}$ and $S^2$ are independent, we have

$$M_{\sum_{i=1}^{n}\left(\frac{X_i - \mu}{\sigma}\right)^2}(t) = M_{\frac{(n-1)S^2}{\sigma^2}} \times M_{\left(\frac{\bar{X} - \mu}{\sigma/\sqrt{n}}\right)^2},$$

$$(1-2t)^{-n/2} = M_{\frac{(n-1)S^2}{\sigma^2}} \times (1-2t)^{-1/2}.$$

Therefore,

$$M_{\frac{(n-1)S^2}{\sigma^2}} = (1-2t)^{-(n-1)/2}.$$

Hence,

$$\frac{(n-1)S^2}{\sigma^2} \sim \chi^2(n-1).$$

and

$$f_{\frac{(n-1)S^2}{\sigma^2}}(x) = \frac{(1/2)^{\frac{n-1}{2}}}{\Gamma\big((n-1)/2\big)} e^{-\frac{x}{2}} x^{\left(\frac{n-1}{2}-1\right)} \quad ; 0 < x < \infty.$$

■

**Theorem 16.14:** If $S^2$ is the variance of a random sample of size $n$ taken from a normal population having the variance $\sigma^2$, then

$$S^2 \sim \text{Gamma}\left(\alpha = \frac{n-1}{2}, \beta = \frac{2\sigma^2}{n-1}\right), \tag{16.27}$$

(shape–scale parametrization) and the PDF of $S^2$ is given by

$$f_{S^2}(x) = \frac{\big((n-1)/2\sigma^2\big)^{\frac{n-1}{2}}}{\Gamma\big((n-1)/2\big)} e^{-\frac{(n-1)x}{2\sigma^2}} x^{\left(\frac{n-1}{2}-1\right)}; 0 < x < \infty. \tag{16.28}$$

**Proof:**

**Method1:** let $X = \frac{(n-1)S^2}{\sigma^2} \sim \chi^2(n-1)$. Then, $S^2 = \frac{\sigma^2 X}{n-1}$ and

$$\begin{aligned} M_{S^2}(t) &= M_{\frac{\sigma^2 X}{n-1}}(t) \\ &= M_X\left(\frac{\sigma^2 t}{n-1}\right) \\ &= \left(1 - \frac{2\sigma^2}{n-1} t\right)^{-\frac{(n-1)}{2}}, \end{aligned}$$

which is the MGF of Gamma $\left(\alpha = \frac{n-1}{2}, \beta = \frac{2\sigma^2}{n-1}\right)$. Therefore, $S^2 \sim \text{Gamma}\left(\alpha = \frac{n-1}{2}, \beta = \frac{2\sigma^2}{n-1}\right)$.

**Method2:** If $X \sim \text{Gamma}(\nu/2, 2)$ (in shape–scale parametrization), then $X$ is identical to $\chi^2(\nu)$, the chi-squared distribution with $\nu$ degrees of freedom. Conversely, if $Q \sim \chi^2(\nu)$ and $c$ is a positive constant, then $cQ \sim \text{Gamma}(\nu/2, 2c)$. In our case $Q = \frac{(n-1)S^2}{\sigma^2} \sim \chi^2(n-1)$, therefore,

$$S^2 = \frac{\sigma^2}{n-1}\left(\frac{n-1}{\sigma^2} S^2\right) \sim \text{Gamma}\left(\frac{n-1}{2}, \frac{2\sigma^2}{n-1}\right),$$

with PDF

$$f_{S^2}(x) = \frac{\big((n-1)/2\sigma^2\big)^{\frac{n-1}{2}}}{\Gamma\big((n-1)/2\big)} e^{-\frac{(n-1)x}{2\sigma^2}} x^{\left(\frac{n-1}{2}-1\right)}; 0 < x < \infty.$$

■

**Remarks:**

In shape–scale parametrization of Gamma distribution, $\alpha = \frac{n-1}{2}$ (shape parameter) and $\beta = \frac{2\sigma^2}{n-1}$ (scale parameter), mean= $\alpha\beta$ and variance= $\alpha\beta^2$, so we have

1. $E[S^2] = \frac{n-1}{2}\frac{2\sigma^2}{n-1} = \sigma^2$,
2. $V(S^2) = \frac{n-1}{2}\left(\frac{2\sigma^2}{n-1}\right)^2 = \frac{2\sigma^4}{n-1}$.

As a summary of the above mathematical analysis; for describing the sampling distribution of the sample variance, we consider all possible sample of same size, say, $n$ taken from the population having variance $\sigma^2$ and for each sample we calculate sample variance $S^2$. The values of $S^2$ may vary from sample to sample so we construct the probability distribution of sample variances. The probability distribution thus obtained is known as sampling distribution of the sample variance.

**Definition (Sampling Distribution of Sample Variance):** The probability distribution of all values of the sample variance would be obtained by drawing all possible sample of same size from the parent population is called the sampling distribution of the sample variance.

We can summarize this section as the following, let $X_1, X_2, \dots, X_n$ be a random sample of size $n$ taken from normal population with mean $\mu$ and variance $\sigma^2$. The distribution of sample variance can not be obtained directly therefore in this case, some transformation is made by multiplying $S^2$ by $(n-1)$ and then dividing the product by $\sigma^2$. The obtained new variable follows the chi-square distribution with $(n-1)$ degrees of freedom, $\frac{(n-1)S^2}{\sigma^2} \sim \chi^2(n-1)$, where, $S^2$ is sample variance $S^2 = \frac{1}{n-1}\sum_{i=1}^{n}(X_i - \bar{X})^2$, for all $i = 1,2, \dots, n$. See Figure 16.7.

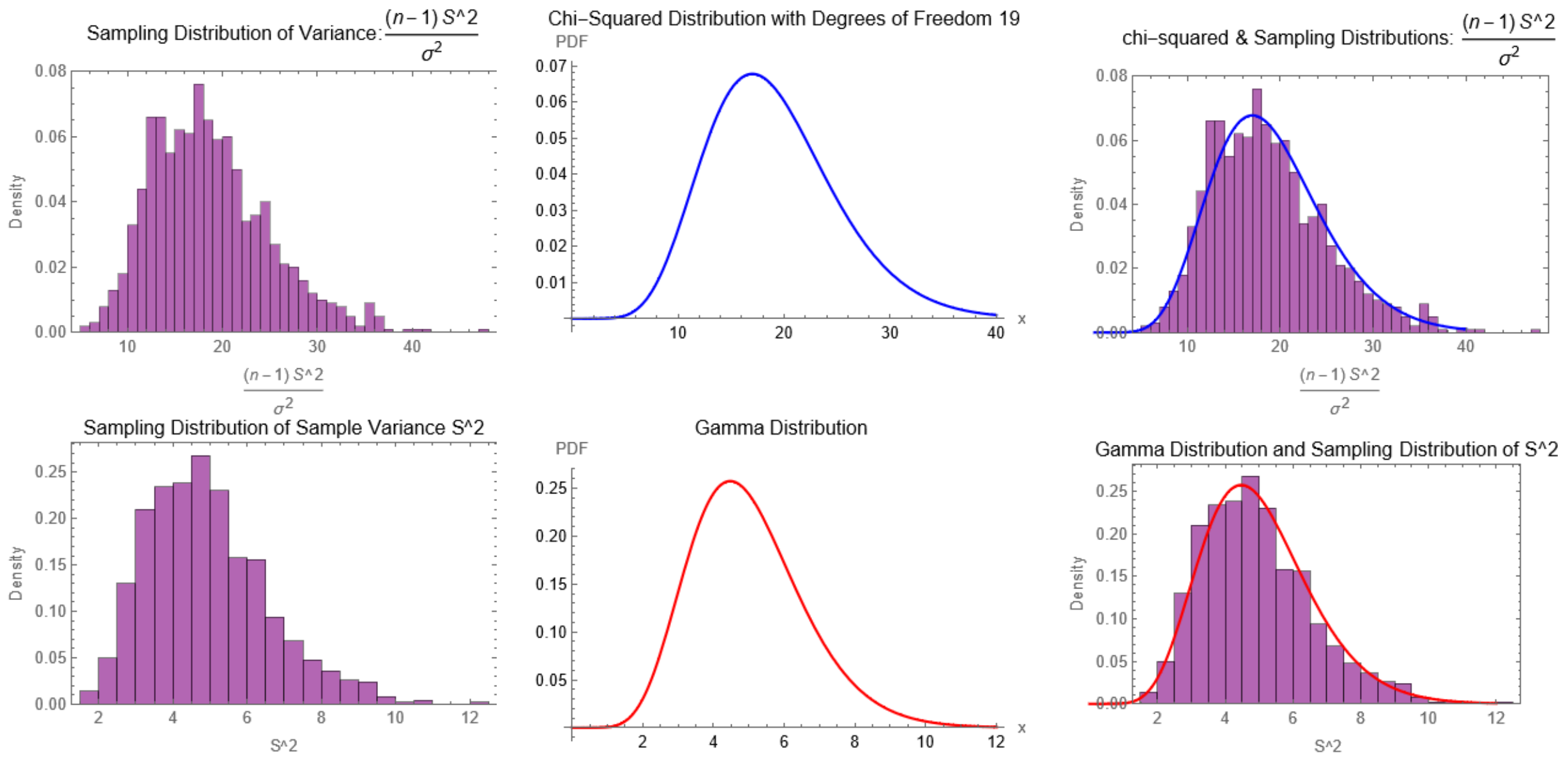

**Figure 16.7.** The relation between the distribution of the sample variance, denoted as $S^2$, and the transformed variable $((n-1)/\sigma^2)S^2$. The figure demonstrates how the sampling distribution of the sample variance follows the chi-squared distribution and how the gamma distribution can also be used to model it.

**Definition ($\chi^2_\alpha$):** The symbol $\chi^2_\alpha$ indicates that the $\chi^2$-value has an area $\alpha$ to its right.

***Example 16.8***

Find the $\chi^2_\alpha$, $\alpha = \{0.025, 0.05, 0.10, 0.5, 0.90, 0.95, 0.975\}$.

***Solution***

**Part 1**

```
(*The symbol αleft is used to represent the significance levels, which denote the areas under
the Chi Square Distribution curve to the left of the corresponding Chi Square-score: *)
αleft={0.025,0.05,0.10,0.5,0.90,0.95,0.975}

(*The variable αright represents the complement of the significance levels, which denotes the
areas under the Chi Square Distribution curve to the right of the corresponding Chi Square-
score: *)
αright=1-αleft
```

```
(*The Quantile function is used to calculate the Chi Square-scores corresponding to the specified significance levels (αleft) under the Chi Square curve. This provides the critical values for confidence intervals or hypothesis testing: *)
v=15;
N[Quantile[ChiSquareDistribution[v],αleft]]

(*The InverseCDF function is an alternative method to compute the Chi Square-scores corresponding to the significance levels (αleft) under the Chi Square Distributioncurve. It yields the same critical values as the Quantile function: *)
InverseCDF[ChiSquareDistribution[v],αleft]
 {0.025,0.05,0.1,0.5,0.9,0.95,0.975}
 {0.975,0.95,0.9,0.5,0.1,0.05,0.025}
 {6.26214,7.26094,8.54676,14.3389,22.3071,24.9958,27.4884}
 {6.26214,7.26094,8.54676,14.3389,22.3071,24.9958,27.4884}
```

**Part 2**

```
v=10;
dist=ChiSquareDistribution[v]; (*Define a Chi Square distribution with v=10: *)
αleft={0.025,0.05,0.10,0.5,0.90,0.95,0.975}; (*List of significance levels. *)
αright=1-αleft; (* Complement of significance levels. *)
q=N[Quantile[dist,αleft]]; (* Calculate quantiles based on the significance levels. *)

Plot[
 PDF[dist,x],(* Plot the PDF of the Chi Square distribution with v=10: *)
 {x,0,30},
 (* The 'Epilog' option adds white vertical lines at the quantiles corresponding to the significance levels, and blue text labels indicating the complement significance levels: *)
 Epilog->{
   White,
   Thickness[0.008],
   Apply[
    Sequence,
    Table[
     Line[{{q[[i]],0},{q[[i]],PDF[dist,q[[i]]]}}],
     {i,1,7}
     ]
    ],
   Blue,
   Table[
    Text[αright[[i]],{q[[i]],0.03+PDF[dist,q[[i]]]}],
    {i,1,7}
    ]
   },
 PlotRange->{0,0.15},(*Set the y-axis plot range*)
 Filling->Axis,
 PlotStyle->Purple,
 ImageSize->250
 ]
```

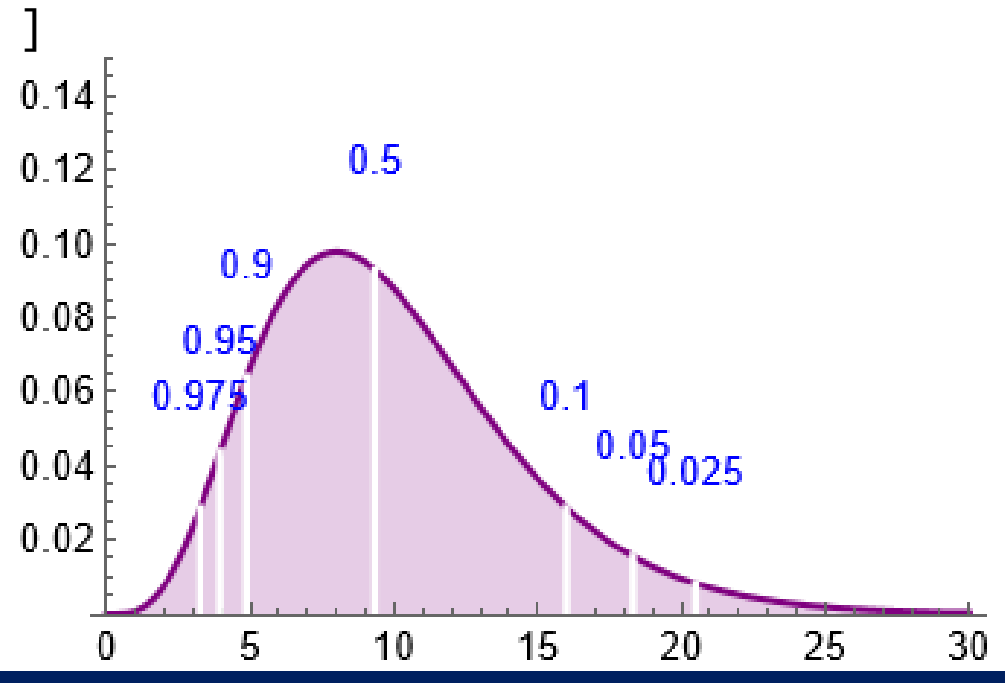

**Example 16.9**

Batteries for motorcycles are warranted to last, on average, for three years with a standard deviation of one year by the manufacturer. Should the manufacturer still be assured that the batteries have a standard deviation of 1 year if seven of these batteries had lives of 1.9, 2.4, 3.0, 3.5, 4.2, 4, and 2 years? Assume that the battery's lifespan is distributed normally.

***Solution***

We have $\mu = 3, \sigma^2 = 1$. Then, we find the sample mean $\bar{X} = 3$, sample size $n = 7$, and

$$S^2 = \frac{(1.9-3)^2 + (2.4-3)^2 + (3.-3)^2 + (3.5-3)^2 + (4.2-3)^2 + (4-3)^2 + (2-3)^2}{6} = 0.877.$$

Then

$$\chi^2 = \frac{(n-1)S^2}{\sigma^2} = \frac{(6)(0.877)}{1} = 5.26,$$

is a value from a chi-squared distribution with 6 degrees of freedom. The desired values can be found by setting the upper tail area and lower tail area each equal to 0.025. Since 95% of the $\chi^2$ values with 6 degrees of freedom fall between $1.2374 - 14.4494$, the computed value with $\sigma^2 = 1$ is reasonable, and therefore the manufacturer has no reason to suspect that the standard deviation is other than 1 year.

```
αleft={0.025,0.05,0.10,0.5,0.90,0.95,0.975}
αright=1-αleft
v=6;
N[Quantile[ChiSquareDistribution[v],αleft]]
InverseCDF[ChiSquareDistribution[v],αleft]
 {0.025,0.05,0.1,0.5,0.9,0.95,0.975}
 {0.975,0.95,0.9,0.5,0.1,0.05,0.025}
 {1.23734,1.63538,2.20413,5.34812,10.6446,12.5916,14.4494}
 {1.23734,1.63538,2.20413,5.34812,10.6446,12.5916,14.4494}
```

**Example 16.10**

Let $X_1, X_2, ., X_{13}$ be a random sample from a normal distribution with $\sigma^2 = 0.7$. Find two positive numbers $a$ and $b$ such that the sample variance $S^2$ satisfies

$$P(a \le S^2 \le b) = 0.90.$$

***Solution***

1. We know that $\frac{(n-1)S^2}{\sigma^2}$ follows a chi-squared distribution with $(n-1)$ degrees of freedom, where $n$ is the sample size.
2. We can transform the given probability into a probability involving the chi-squared distribution as follows:
$$P(a \le S^2 \le b) = P\left(\frac{(n-1)a}{\sigma^2} \le \frac{(n-1)S^2}{\sigma^2} \le \frac{(n-1)b}{\sigma^2}\right).$$
3. Since the chi-squared distribution is a continuous distribution, we can find the values of $a$ and $b$ that satisfy the given probability by finding the critical values (x1 and x2) of the chi-squared distribution for a given probability.
4. Since $P(a \le S^2 \le b) = 0.90$, we need to find the critical values that leave a probability of 0.10 in the tails of the distribution. We can split the probability between the two tails, giving us a probability of $(1 - 0.90)/2 = 0.05$ for each tail.
5. We can use the InverseCDF function in Mathematica to find the critical values of the chi-squared distribution with $(n-1)$ degrees of freedom corresponding to a probability 0.05. Specifically, we can find the critical values x1 and x2 such that $P(X \le \text{x1}) = 0.05$ and $P(X \ge \text{x2}) = 0.05$, where $X$ is the chi-squared distribution with $(n-1)$ degrees of freedom.
6. Once we have the critical values x1 and x2, we can calculate the values of $a$ and $b$ by multiplying them by $\frac{\sigma^2}{(n-1)}$.
7. Therefore, the values of $a$ and $b$ that satisfy the given probability are $a = \text{x1}\frac{\sigma^2}{(n-1)}$ and $b = \text{x2}\frac{\sigma^2}{(n-1)}$, where x1 and x2 are the critical values of the chi-squared distribution with 12 degrees of freedom corresponding to a probability 0.05.

8. Hence, we have:

$$\frac{(n-1)b}{\sigma^2} = \frac{12b}{0.7} = \text{InverseCDF[ChiSquareDistribution[v],0.95]} = 21.0261,$$

which implies $b = 1.22652$. Similarly,

$$\frac{(n-1)a}{\sigma^2} = \frac{12a}{0.7} = \text{InverseCDF[ChiSquareDistribution[v],0.05]} = 5.22603,$$

So, we have $a = 0.304852$. Hence,

$$P(0.305 \le S^2 \le 1.227) = 0.90.$$

It is important to note that this is not the only interval that would satisfy: $P(a \le S^2 \le b) = 0.90$. but it is a convenient one.

```
αleft={0.025,0.05,0.10,0.5,0.90,0.95,0.975}
αright=1-αleft
v=12;
N[Quantile[ChiSquareDistribution[v],αleft]]
InverseCDF[ChiSquareDistribution[v],αleft]
 {0.025,0.05,0.1,0.5,0.9,0.95,0.975}
 {0.975,0.95,0.9,0.5,0.1,0.05,0.025}
 {4.40379,5.22603,6.3038,11.3403,18.5493,21.0261,23.3367}
 {4.40379,5.22603,6.3038,11.3403,18.5493,21.0261,23.3367}
```

## 16.5 Student $t$-Distribution

According to CLT, if a simple random sample of size $n$ is taken from a population whose mean and variance are $\mu$ and $\sigma^2$ respectively, then the sample mean $\bar{X}$ will be distributed normally with mean $\mu$ and variance $\frac{\sigma^2}{n}$, for large $n$. In other words, for a population which is not normal

$$\frac{\bar{X}-\mu}{\sigma/\sqrt{n}} \to N(0,1) \quad \text{as } n \to \infty. \tag{16.29}$$

When the population standard deviation $\sigma$ is not known and $S$ is the sample standard deviation, then also

$$\frac{\bar{X}-\mu}{S/\sqrt{n}} \to N(0,1), \tag{16.30}$$

provided $n$, the sample size, is sufficiently large (i.e., $n \ge 30$). But, in the case when the population is normal, $\sigma$ is unknown and sample size is small (i.e., $n < 30$) the distribution of $\frac{\bar{X}-\mu}{S/\sqrt{n}}$ will not be normal. The distribution of the statistic in such cases is known as student $t$-distribution. The probability of student $t$-distribution was first published in 1908 in a paper written by W. S. Gosset. At the time, Gosset was employed by an Irish brewery that prohibited publication of research by members of its staff. To circumvent this restriction, he published his work secretly under the name "Student." Consequently, the $t$-distribution is usually called the student $t$-distribution.

**Definition ($t$-Distribution):** Let $Z \sim N(0,1)$ and $Y \sim \chi^2(n)$, and let $Z$ and $Y$ be independent. Then, the statistic

$$T = \frac{Z}{\sqrt{Y/n}}, \tag{16.31}$$

is said to have a $t$-distribution with $n$ degrees of freedom and we write $T \sim t(n)$.

Therefore, we can say that the standard normal distribution is the parent distribution of the chi-squared distribution and the $t$-distribution. The chi-squared distribution is obtained by summing up the squares of independent standard normal RVs, while the $t$-distribution is obtained by dividing a standard normal RV by the square root of a chi-squared RV divided by its degrees of freedom.

**Theorem 16.15:** The PDF of the RV $T$ with $n$ degrees of freedom is given by

$$f_T(t) = \frac{\Gamma\left(\frac{n+1}{2}\right)}{\sqrt{n\pi}\,\Gamma\left(\frac{n}{2}\right)}\left(1+\frac{t^2}{n}\right)^{-\frac{(n+1)}{2}} ; 0 < x < \infty. \tag{16.32}$$

**Proof:**

A $T$ variable is defined in terms of a standard normal $Z$ and a $\chi^2(n)$ variable $Y$. They are independent, so their joint PDF $f(y,z)$ is the product of their individual PDFs. We first find the CDF of $T$ and then differentiate to obtain the PDF:

$$\begin{aligned} F(t) &= P(T \le t) \\ &= P\left(\frac{Z}{\sqrt{Y/n}} \le t\right) \\ &= P\left(Z \le t\sqrt{Y/n}\right) \\ &= \int_0^\infty \int_{-\infty}^{t\sqrt{y/n}} f(y,z)dzdy. \end{aligned}$$

Differentiating with respect to $t$ using the Fundamental Theorem of Calculus,

$$\begin{aligned} f(t) &= \frac{d}{dt}F(t) \\ &= \int_0^\infty \frac{\partial}{\partial t} \int_{-\infty}^{t\sqrt{y/n}} f(y,z)dzdy \\ &= \int_0^\infty f(y, t\sqrt{\frac{y}{n}})\sqrt{\frac{y}{n}}dy. \end{aligned}$$

Now substitute the joint PDF—that is, the product of the marginal PDFs of $Y$ and $Z$—and integrate:

$$\begin{aligned} f(t) &= \int_0^\infty \frac{y^{\frac{n}{2}-1}}{2^{\frac{n}{2}}\Gamma\left(\frac{n}{2}\right)} e^{-\frac{y}{2}} \frac{1}{\sqrt{2\pi}} e^{-\frac{\left(t\sqrt{\frac{y}{n}}\right)^2}{2}} \sqrt{\frac{y}{n}}dy \\ &= \int_0^\infty \frac{y^{\frac{n}{2}-1}\sqrt{\frac{y}{n}}}{2^{\frac{n}{2}}\Gamma(n/2)} \frac{1}{\sqrt{2\pi}} e^{-\frac{y}{2}} e^{-\frac{t^2\frac{y}{n}}{2}} dy \\ &= \frac{1}{\sqrt{2\pi n}2^{\frac{n}{2}}\Gamma(n/2)} \int_0^\infty y^{\frac{n+1}{2}-1} e^{-\left(\frac{1}{2}+\frac{t^2}{2n}\right)y} dy. \end{aligned}$$

The integral can be evaluated using the gamma function:

$$\begin{aligned} f(t) &= \frac{1}{\sqrt{2\pi n}2^{\frac{n}{2}}\Gamma\left(\frac{n}{2}\right)} \frac{\Gamma\left(\frac{n+1}{2}\right)}{\left(\frac{1}{2}+\frac{t^2}{2n}\right)^{\frac{n}{2}+\frac{1}{2}}} \\ &= \frac{1}{2^{\frac{n}{2}+\frac{1}{2}}\sqrt{\pi n}\Gamma(n/2)} \frac{\Gamma\left(\frac{n+1}{2}\right)}{\left(\frac{1}{2}\right)^{\frac{n}{2}+\frac{1}{2}}\left(1+\frac{t^2}{n}\right)^{\frac{n}{2}+\frac{1}{2}}} \\ &= \frac{\Gamma\left(\frac{n+1}{2}\right)}{\sqrt{\pi n}\Gamma(n/2)}\left(1+\frac{t^2}{n}\right)^{-\frac{n+1}{2}} \quad ; -\infty < t < \infty. \end{aligned}$$

■

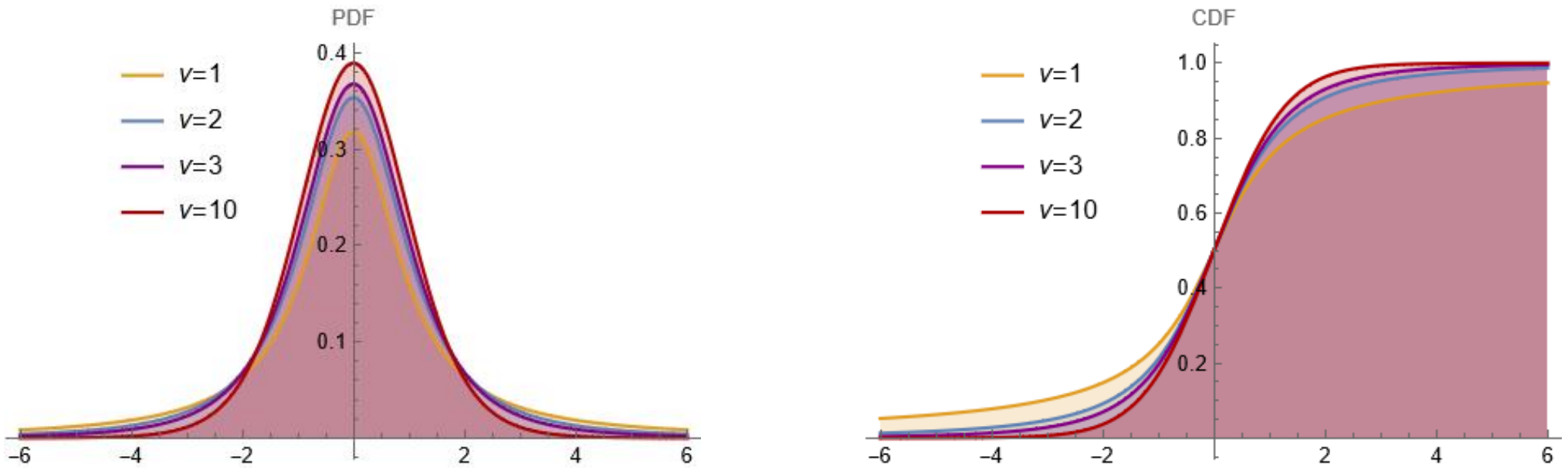

**Figure 16.8.** The shape of the Student $t$-distribution depends on the degrees of freedom ($n = \nu$). When the $\nu$ is large (i.e., greater than 30), the $t$-distribution approaches a normal distribution, with a bell-shaped curve and symmetrical about the mean. However, when the $\nu$ is small, the $t$-distribution is more spread out and has heavier tails than the normal distribution. Also, the shape of the CDF depends on the number of degrees of freedom $\nu$ of the Student t-distribution.

### Properties of $t$-distribution Curve

- The number of degrees of freedom is the only parameter of the $t$ distribution. There is a different $t$ distribution for each number of degrees of freedom.
- The $t$ distribution is similar to the normal distribution in some respects. Like the normal distribution curve, the $t$ distribution curve is symmetric (bell shaped) about the mean and never meets the horizontal axis, Figure 16.8.
- The total area under a $t$ distribution curve is 1.0.
- The $t$ distribution curve is flatter and wider than the standard normal distribution curve. In other words, the $t$-distribution curve has a lower height and a greater spread (or, we can say, a larger standard deviation) than the standard normal distribution. Figure 16.9 shows a graph of the $t$-density function with 5 degrees of freedom compared with the standard normal density. Notice that the $t$-density has thicker "tails," indicating greater variability, than does the normal density.
- As the sample size increases, the $t$ distribution approaches the standard normal distribution. To understand why, recall that $Y \sim \chi^2(n)$ can be expressed as the sum of the squares of $n$ standard normals, and so

$$\frac{Y}{n} = \frac{Z_1^2 + Z_2^2 + \cdots + Z_n^2}{n}, \tag{16.33}$$

where $Z_1, \ldots, Z_n$ are independent standard normal RVs. For large $n$, $Y/n$ will, with probability close to 1, be approximately equal to $E[Z_i^2] = 1$. Hence, for $n$ large, $T = Z/\sqrt{Y/n}$ will have approximately the same distribution as $Z$.

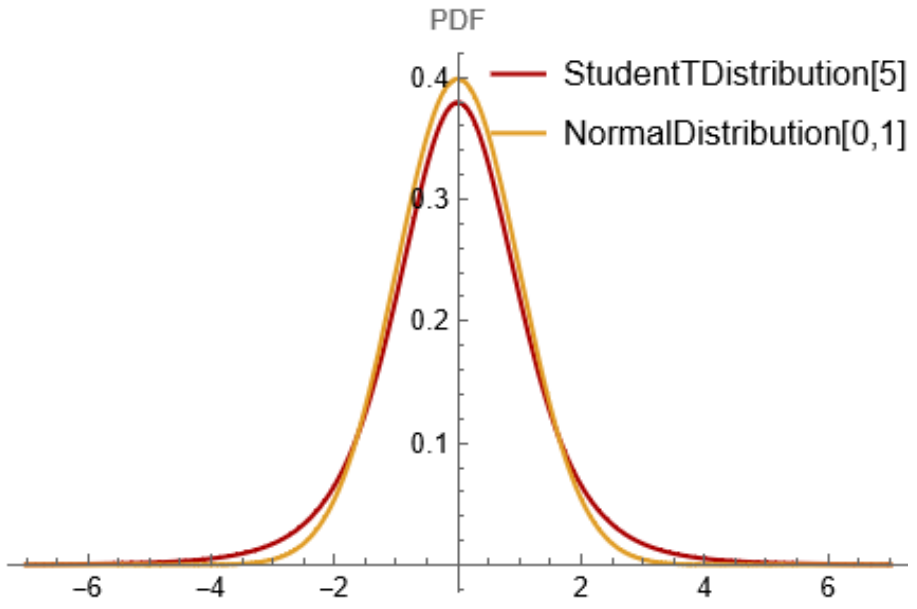

**Figure 16.9.** As the degrees of freedom increase, the Student $t$-distribution approaches the normal distribution, with the heavier tails becoming less pronounced.

**Theorem 16.16:** If $X$ is a RV having a student $t$-distribution with $n$ degrees of freedom, then

$$E[X] = 0; \;\; n > 1, \tag{16.34.1}$$

$$\text{Var}(X) = \frac{n}{n-2}; \;\; n > 2. \tag{16.34.2}$$

**Proof:**

The mean and variance of a $t$ variable can be obtained directly from the PDF, but it is instructive to derive them through the definition in terms of independent standard normal and chi-squared variables, $T = \frac{Z}{\sqrt{Y/n}}$. Recall that $E[UV] = E[U]E[V]$ if $U$ and $V$ are independent and the expectations of $U$ and $V$ both exist. Thus,

$$E[T] = E[Z]E[\frac{1}{\sqrt{Y/n}}] = E[Z]n^{\frac{1}{2}}E[Y^{-\frac{1}{2}}].$$

Of course, $E[Z] = 0$, so $E[T] = 0$ if $E[Y^{-1/2}]$ exists. Let's compute $E[Y^{-k}]$ for any $k$ if $Y$ is chi-squared:

$$\begin{aligned}
E[Y^k] &= \int_0^\infty y^k \frac{y^{\frac{n}{2}-1}}{2^{\frac{n}{2}}\Gamma\left(\frac{n}{2}\right)} e^{-\frac{y}{2}} dy \\
&= \frac{1}{2^{\frac{n}{2}}\Gamma\left(\frac{n}{2}\right)} \int_0^\infty y^{\left(k+\frac{n}{2}\right)-1} e^{-\frac{y}{2}} dy \\
&= \frac{1}{2^{\frac{n}{2}}\Gamma\left(\frac{n}{2}\right)} 2^{k+\frac{n}{2}}\Gamma\left(k+\frac{n}{2}\right) \\
&= \frac{2^k\Gamma(k+n/2)}{\Gamma(n/2)}; k+\frac{n}{2} > 0.
\end{aligned}$$

If $k + \frac{n}{2} \leq 0$, the integral does not converge and $E[Y^k]$ does not exist. When $k = -1/2$, we require that $n > 1$ for the integral to converge. Thus, the mean of a $t$ variable fails to exist if $n = 1$ and the mean is indeed 0 otherwise.

For the variance of $T$ we need $E[T^2] = E[Z^2]\,E[1/(Y/n)] = nE[Y^{-1}]$. We obtain, with the help of the property $\Gamma(a+1) = a\Gamma(a)$,

$$\begin{aligned}
E[Y^{-1}] &= \frac{2^{-1}\Gamma\left(-1+\frac{n}{2}\right)}{\Gamma\left(\frac{n}{2}\right)} \\
&= \frac{2^{-1}\Gamma\left(\frac{n}{2}-1\right)}{\Gamma\left(\frac{n}{2}-1+1\right)} \\
&= \frac{2^{-1}}{\frac{n}{2}-1} \\
&= \frac{1}{n-2}.
\end{aligned}$$

Hence,

$$\text{Var}(T) = n\frac{1}{n-2} = \frac{n}{n-2}.$$

provided that $-1 + n/2 > 0$, or $n > 2$. For $n = 1$ or $2$ the variance of $T$ does not exist. For $n > 2$, the variance always exceeds 1, and for large $n$ the variance is close to 1.

■

**Remarks:**

(1) When $n = 1$, the PDF (16.32) reduces to

$$f(t) = \frac{1}{\pi(1+t^2)}; 0 < x < \infty, \tag{16.35}$$

which is the standard Cauchy distribution $C(1,0)$.

(2) Let $X \sim t(n)$, $n > 1$. Then, $E(X^r)$ exists for $r < n$.

(a) If $r$ is odd, then

$$E(X^r) = 0. \tag{16.36}$$

(b) If $r$ is even, then

$$E(X^r) = n^{r/2} \frac{\Gamma\left(\frac{r+1}{2}\right)\Gamma\left(\frac{n-r}{2}\right)}{\Gamma\left(\frac{1}{2}\right)\Gamma\left(\frac{n}{2}\right)}. \tag{16.37}$$

(3) If $X \sim t(1)$, then $E(X)$ does not exist. (Cauchy)

(4) Mean = Median = Mode.

(5) For $t$ distribution MGF does not exist.

### Statistic following $t$-distribution

**Theorem 16.17:** Let $X_1, X_2, \dots, X_n$ be a random sample of size $n$ from $N(\mu, \sigma^2)$, where $\sigma^2$ is unknown. Let $\bar{X}$ be the sample mean and $S^2$ be the sample variance. Then,

$$T = \frac{\bar{X} - \mu}{S/\sqrt{n}} \sim t(n-1). \tag{16.38}$$

**Proof:**

Here, $\bar{X} \sim N(\mu, \frac{\sigma^2}{n})$ and $\frac{(n-1)S^2}{\sigma^2} \sim \chi^2(n-1)$. Also $\bar{X}$ and $S^2$ are independent. Then, by the definition of $T$- statistic.

$$\begin{aligned} T &= \frac{\bar{X} - \mu}{\frac{\sigma}{\sqrt{n}}} \Big/ \sqrt{\frac{(n-1)S^2}{\sigma^2} \frac{1}{(n-1)}} \\ &= \frac{\bar{X} - \mu}{\frac{\sigma}{\sqrt{n}}} \frac{\sigma}{S} \\ &= \frac{\bar{X} - \mu}{S/\sqrt{n}} \sim t(n-1). \end{aligned}$$

■

**Remarks:**

- The $T$-values depend on the fluctuations of two quantities, $\bar{X}$ and $S^2$, whereas the $Z$-values depend only on the changes in $\bar{X}$ from sample to sample.
- The $t$-distribution is used extensively in problems that deal with inference about the population mean or in problems that involve comparative samples (i.e., in cases where one is trying to determine if means from two samples are significantly different).

**Theorem 16.18:** Let $X_1, X_2, \dots, X_m$ be a random sample of size $m$ from $N(\mu_1, \sigma^2)$, and $Y_1, Y_2, \dots, Y_n$ be a random sample of size $n$ from $N(\mu_2, \sigma^2)$ where $\sigma^2$ is unknown. Then,

$$T = \frac{(\bar{X} - \bar{Y}) - (\mu_1 - \mu_2)}{\sqrt{\frac{(m-1)S_1^2 + (n-1)S_2^2}{m+n-2}\left(\frac{1}{m} + \frac{1}{n}\right)}} \sim t(m+n-2). \tag{16.39}$$

If $\mu_1 = \mu_2$, then

$$T = \frac{(\bar{X} - \bar{Y})}{\sqrt{\frac{(m-1)S_1^2 + (n-1)S_2^2}{m+n-2}\left(\frac{1}{m} + \frac{1}{n}\right)}} \sim t(m+n-2). \tag{16.40}$$

**Proof:**

Here, $\bar{X} \sim N(\mu_1, \frac{\sigma^2}{m})$ and $\bar{Y} \sim N(\mu_2, \frac{\sigma^2}{n})$. Therefore,

$$(\bar{X} - \bar{Y}) \sim N\left(\mu_1 - \mu_2, \frac{\sigma^2}{m} + \frac{\sigma^2}{n}\right) \qquad \Rightarrow \frac{(\bar{X} - \bar{Y}) - (\mu_1 - \mu_2)}{\sqrt{\frac{\sigma^2}{m} + \frac{\sigma^2}{n}}} \sim N(0,1),$$

$$\frac{(m-1)S_1^2}{\sigma^2} \sim \chi^2(m-1) \text{ and } \frac{(n-1)S_2^2}{\sigma^2} \sim \chi^2(n-1) \qquad \Rightarrow \frac{(m-1)S_1^2}{\sigma^2} + \frac{(n-1)S_2^2}{\sigma^2} \sim \chi^2(m+n-2).$$

Hence, by the definition of $T$- statistic

$$\begin{aligned} T &= \frac{(\bar{X} - \bar{Y}) - (\mu_1 - \mu_2)}{\sqrt{\frac{\sigma^2}{m} + \frac{\sigma^2}{n}}} \frac{1}{\sqrt{\frac{\frac{(m-1)S_1^2}{\sigma^2} + \frac{(n-1)S_2^2}{\sigma^2}}{m+n-2}}} \\ &= \frac{(\bar{X} - \bar{Y}) - (\mu_1 - \mu_2)}{\sigma\sqrt{\frac{1}{m} + \frac{1}{n}}} \frac{1}{\frac{1}{\sigma}\sqrt{\frac{(m-1)S_1^2 + (n-1)S_2^2}{m+n-2}}} \\ &= \frac{(\bar{X} - \bar{Y}) - (\mu_1 - \mu_2)}{\sqrt{\frac{(m-1)S_1^2 + (n-1)S_2^2}{m+n-2}\left(\frac{1}{m} + \frac{1}{n}\right)}} \sim t(m+n-2). \end{aligned}$$

If $\mu_1 = \mu_2$, then

$$T = \frac{(\bar{X} - \bar{Y})}{\sqrt{\frac{(m-1)S_1^2 + (n-1)S_2^2}{m+n-2}\left(\frac{1}{m} + \frac{1}{n}\right)}} \sim t(m+n-2).$$

■

**Definition ($t_\alpha$):** The symbol $t_\alpha$ indicates that the $t$-value has an area $\alpha$ to its right.

***Example 16.11***

Find the $t_\alpha$, $\alpha = \{0.025, 0.05, 0.10, 0.5, 0.90, 0.95, 0.975\}$.

***Solution***

**Part 1**

```
(* The symbol αleft is used to represent the significance levels, which denote the areas under the student T distribution curve to the left of the corresponding T-score: *)
αleft={0.025,0.05,0.10,0.5,0.90,0.95,0.975}

(* The variable αright represents the complement of the significance levels, which denotes the areas under the student T distribution curve to the right of the corresponding T-score: *)
αright=1-αleft
```

```
(* The Quantile function is used to calculate the student T-scores corresponding to the specified significance levels (αleft) under the student T distribution curve. This provides the critical values for confidence intervals or hypothesis testing: *)
n=15;
N[Quantile[StudentTDistribution[n],αleft]]

(*The InverseCDF function is an alternative method to compute the student T-scores corresponding to the significance levels (αleft) under the student T distribution curve. It yields the same critical values as the Quantile function: *)
InverseCDF[StudentTDistribution[n],αleft]
 {0.025,0.05,0.1,0.5,0.9,0.95,0.975}
 {0.975,0.95,0.9,0.5,0.1,0.05,0.025}
 {-2.13145,-1.75305,-1.34061,0.,1.34061,1.75305,2.13145}
 {-2.13145,-1.75305,-1.34061,0,1.34061,1.75305,2.13145}
```

**Part 2**

```
n=15;
dist=StudentTDistribution[n]; (* Define astudent T distribution with n=15: *)
αleft={0.025,0.05,0.10,0.5,0.90,0.95,0.975}; (* List of significance levels. *)
αright=1-αleft; (* Complement of significance levels. *)

q=N[Quantile[dist,αleft]]; (* Calculate quantiles based on the significance levels. *)
Plot[
 PDF[dist,x],(* Plot the PDF of the student T distribution with v=10: *)
 {x,-5,5},
 (* The'Epilog' option adds white vertical lines at the quantiles corresponding to the significance levels, and blue text labels indicating the complement significance levels: *)
 Epilog->{
   White,
   Thickness[0.008],
   Apply[
    Sequence,
    Table[
     Line[{{q[[i]],0},{q[[i]],PDF[dist,q[[i]]]}}],
     {i,1,7}
     ]
    ],
   Blue,
   Table[
    Text[αright[[i]],{q[[i]],0.05+PDF[dist,q[[i]]]}],
    {i,1,7}
    ]
   },
 PlotRange->{0,0.5},(*Set the y-axis plot range*)
 Filling->Axis,
 PlotStyle->Purple,
 ImageSize->250
 ]
```

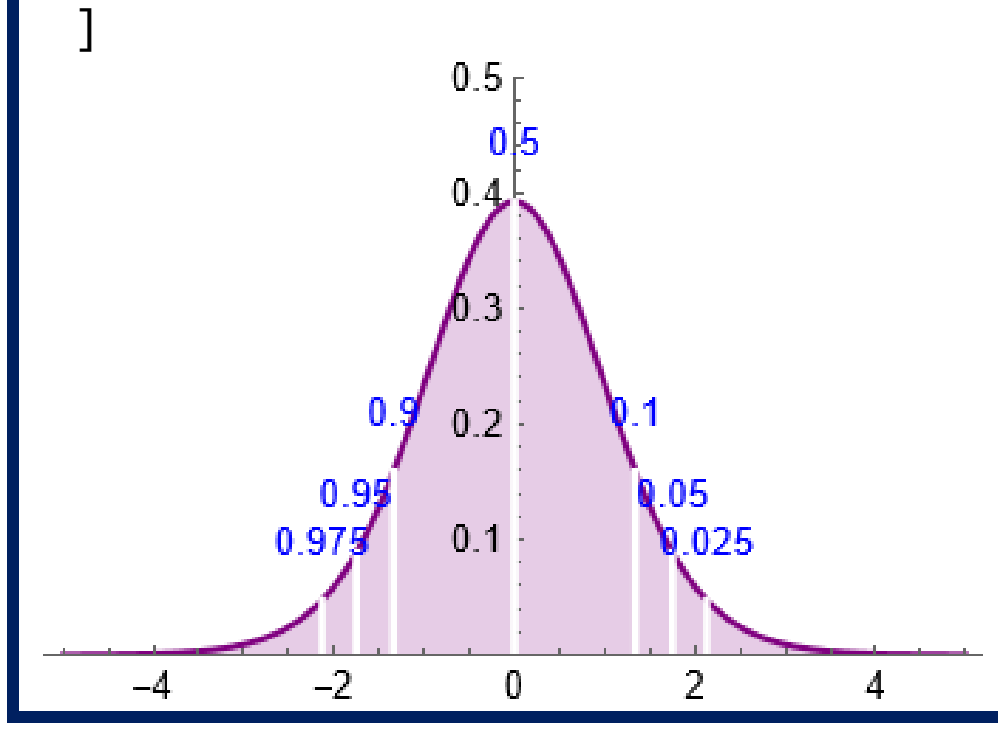

***Example 16.12***

Figure 16.10 displays the student $t$ distribution with 20 degrees of freedom. Find the values of $a$ using Mathematica code so that
(a) the area to the right of $a$ is 0.05,
(b) the total unshaded area is 0.05,
(c) the total shaded area is 0.99,
(d) the area on the right of $a$ is 0.01, and
(e) the area to the left of $a$ is 0.90.

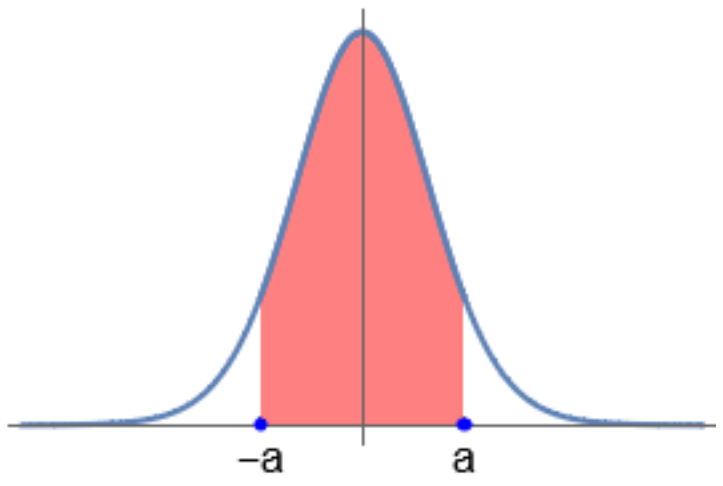

**Figure 16.10** Student $t$ distribution with 20 degrees

***Solution***

```
n=20 (* degrees of freedom *);
N[
 Quantile[
  StudentTDistribution[n],
  {0.05,0.95,0.025,0.975,0.005,0.995,0.01,0.99,0.90}
  ]
 ]
InverseCDF[
 StudentTDistribution[n],
 {0.05,0.95,0.025,0.975,0.005,0.995,0.01,0.99,0.90}
 ]
Output
 {-1.72472,1.72472,-2.08596,2.08596,-2.84534,2.84534,-2.52798,2.52798,1.32534}
 {-1.72472,1.72472,-2.08596,2.08596,-2.84534,2.84534,-2.52798,2.52798,1.32534}
```

(a) If the region on the right of $a$ is 0.05, the region to the left of $a$ is $1-0.05=0.95$ and $a$ stands for the 95th percentile. The result, $t=1.72472$.
(b) If the total unshaded region is 0.05, then the unshaded region on the right is 0.025 by symmetry. Thus, the region to the left of $a$ is $1-0.025=0.975$ and $a$ stands for the 97.5th percentile. The result, $t=2.08596$.
(c) If the total shaded region is 0.99, then the total unshaded area is $1-0.99=$ 0.01 and the region to the right is $0.01/2=0.005$, the region to the left of $a$ is $1-0.005=0.995$. The result, $t=2.84534$.
(d) If the region on the right of $a$ is 0.01. The result, $t=2.52798$.
(e) If the area to the left of $a$ is 0.90, the $a$ stands for the 90th percentile. The result, $t=1.32534$.

## 16.6 Fisher $F$-Distribution

The $F$-distribution was developed by Fisher to study the behavior of two variances from random samples taken from two independent normal populations. In applied problems we may be interested in knowing whether the population variances are equal, based on the response of the random samples. Knowing the answer to such a question is also important in selecting the appropriate statistical methods to study their true means. $F$-distribution is the ratio of two independent $\chi^2$ RVs divided by their respective degrees of freedom.

**Definition ($F$-Distribution) :** Let $U$ and $V$ be independent $\chi^2$ RVs with $\nu_1$ and $\nu_2$ degrees of freedom, respectively. Then the RV

$$F=\frac{U}{\nu_1}/\frac{V}{\nu_2}, \tag{16.41}$$

is said to have an $F$-distribution with $(\nu_1,\nu_2)$ degrees of freedom, and we write $F\sim F(\nu_1,\nu_2)$.

**Definition (PDF of $F$-distribution):** Let $U$ and $V$ be two independent RVs having chi-squared distributions with $\nu_1$ and $\nu_2$ degrees of freedom, respectively. Then the distribution of the RV $F = \frac{U}{\nu_1} / \frac{V}{\nu_2}$, is given by the PDF

$$f(x) = \begin{cases} \frac{\Gamma[(\nu_1+\nu_2)/2]}{\Gamma[\nu_1/2]\Gamma[\nu_2/2]} \left(\frac{\nu_1}{\nu_2}\right)^{\nu_1/2} \frac{x^{\frac{\nu_1}{2}-1}}{(1+\frac{\nu_1}{\nu_2}x)^{(\nu_1+\nu_2)/2}}, & x > 0, \\ 0, & \text{elsewhere.} \end{cases} \quad (16.42)$$

This is known as the F-distribution with $\nu_1$ and $\nu_2$ degrees of freedoms.

### Basic Properties of F-Curves

(1) The total area under an $F$-curve equals 1.

(2) An $F$ -curve starts at 0 on the horizontal axis and extends indefinitely to the right, approaching, but never touching, the horizontal axis as it does so.

(3) An $F$-curve is right skewed.

(4) Actually, there are infinitely many $F$-distributions, and we identify the $F$-distribution (and $F$-curve) in question by its number of degrees of freedom, just as we did for $t$-distributions and chi-square distributions. An $F$-distribution, however, has two numbers of degrees of freedom instead of one. The first number of degrees of freedom for an $F$-curve is called the degrees of freedom for the numerator and the second the degrees of freedom for the denominator.

(5) The curve of the $F$-distribution depends not only on the two parameters $\nu_1$ and $\nu_2$ but also on the order in which we state them since the density of the $F$ distribution is not symmetrical in $\nu_1$ and $\nu_2$. Once these two values are given, we can identify the curve. Typical $F$ -distributions are shown in Figure 16.11.

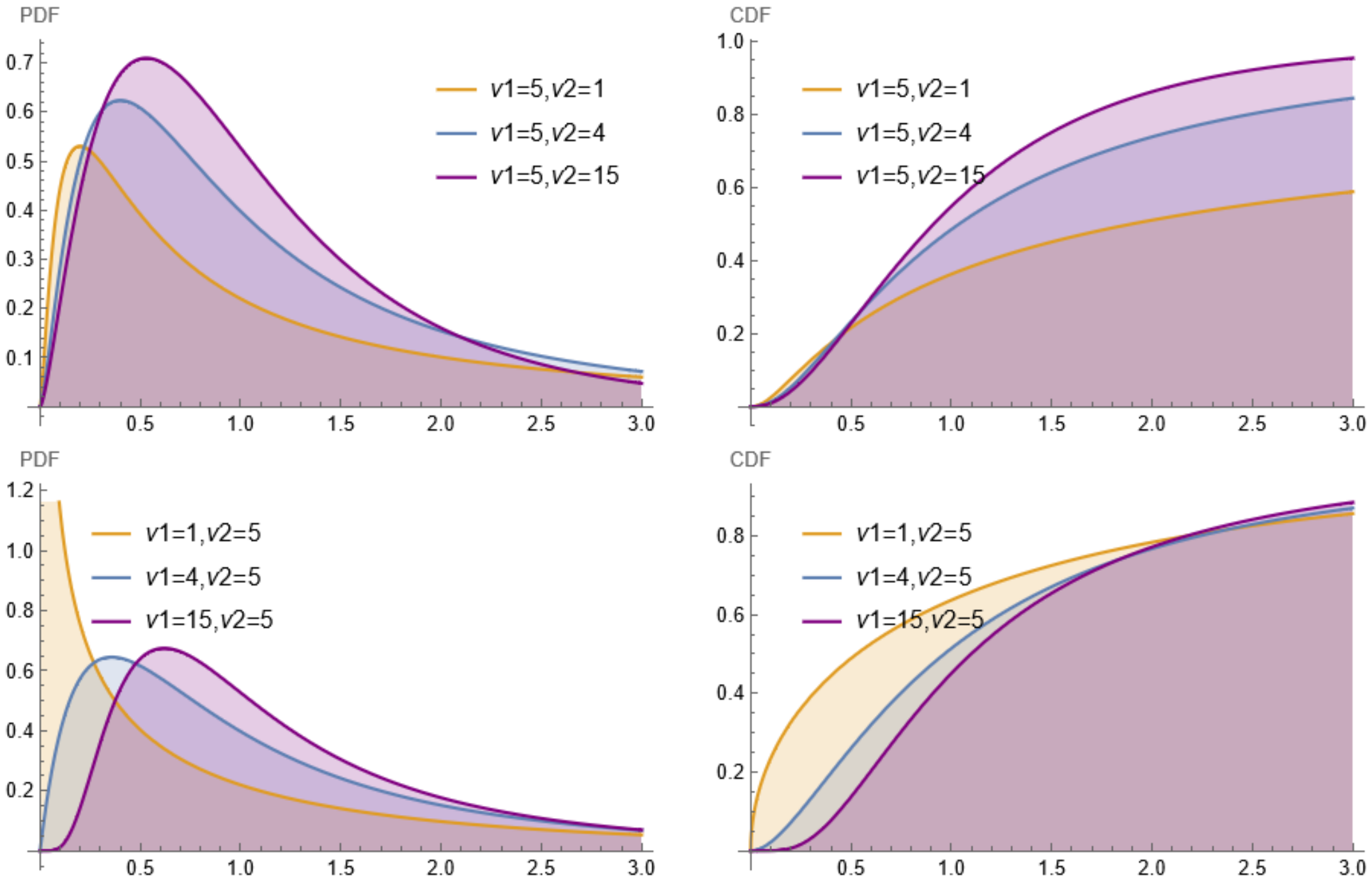

**Figure 16.11.** PDFs CDFs of the $F$-distribution. The curve of the $F$-distribution depends not only on the two parameters $\nu_1$ and $\nu_2$ but also on the order in which we state them since the density of the $F$ distribution is not symmetrical in $\nu_1$ and $\nu_2$.

(6) What happens to $F$ if the degrees of freedom are large? If $\nu_2$ is large, then the denominator of expression (16.41) will be close to 1, and approximately the $F$ will be just the numerator chi-squared over its degrees of freedom. Similarly, if both $\nu_1$ and $\nu_2$ are large, then both the numerator and denominator will be close to 1, and the $F$ ratio therefore will be close to 1.

**Theorem 16.19:** If $X$ is an $F$-distributed RV with $m$ and $n$ degrees of freedom, then

$$E[X] = \frac{n}{n-2} \quad ; \text{for } n > 2, \tag{16.43}$$

and

$$\text{Var}(X) = \frac{2n^2(m+n-2)}{m(n-2)^2(n-4)} \quad ; \text{for } n > 4. \tag{16.44}$$

**Proof:**

At first it might be surprising that the mean depends only on the degrees of freedom of the denominator. Write $X$ as

$$X = \left(\frac{U}{m}\right) / \left(\frac{V}{n}\right),$$

then

$$\begin{aligned} E[X] &= E\left[\left(\frac{U}{m}\right) / \left(\frac{V}{n}\right)\right] \\ &= \frac{n}{m}\, E[U]E[\frac{1}{V}]. \end{aligned}$$

But $E(U) = m$ and

$$\begin{aligned} E[\frac{1}{V}] &= \frac{1}{\Gamma\left(\frac{n}{2}\right)}\left(\frac{1}{2}\right)^{\frac{n}{2}} \int_0^{\infty} \frac{1}{x} x^{\frac{(n-2)}{2}} e^{-\frac{1}{2}x} dx \\ &= \frac{1}{\Gamma\left(\frac{n}{2}\right)}\left(\frac{1}{2}\right)^{\frac{n}{2}} \int_0^{\infty} x^{\frac{(n-4)}{2}} e^{-\frac{1}{2}x} dx \\ &= \frac{1}{\Gamma\left(\frac{n}{2}\right)}\left(\frac{1}{2}\right)^{\frac{n}{2}} \left[\Gamma\left(\frac{n-2}{2}\right)\left(\frac{1}{2}\right)^{-\frac{(n-2)}{2}}\right] \\ &= \frac{\Gamma\left(\frac{n-2}{2}\right)}{\Gamma\left(\frac{n}{2}\right)}\left(\frac{1}{2}\right)^{\frac{n}{2}}\left(\frac{1}{2}\right)^{-\frac{(n-2)}{2}} \\ &= \frac{2}{n-2}\left(\frac{1}{2}\right)^{\frac{n}{2}}\left(\frac{1}{2}\right)^{-\frac{n}{2}+1} \\ &= \frac{1}{n-2}, \end{aligned}$$

where $\int_0^{\infty} x^{\frac{(n-4)}{2}} e^{-\frac{1}{2}x} dx = \Gamma\left(\frac{n-2}{2}\right)\left(\frac{1}{2}\right)^{-\frac{(n-2)}{2}}$, and $\frac{\Gamma\left(\frac{n-2}{2}\right)}{\Gamma\left(\frac{n}{2}\right)} = \frac{2}{n-2}$. Hence,

$$E[X] = \frac{n}{m}\, E[U]E[1/V] = \left(\frac{n}{m}\right)(m)\left(\frac{1}{n-2}\right) = \frac{n}{n-2}.$$

The variance formula is similarly derived.

■

**Remarks:**

- If $X \sim F(m,n)$, then $\frac{1}{X} \sim F(n,m)$. If $X = \frac{U}{m}/\frac{V}{n} \sim F(m,n)$ then $\frac{1}{X} = \frac{V}{n}/\frac{U}{m} \sim F(n,m)$ i.e., the reciprocal of an $X$ variable also has an $F$ distribution, but with the degrees of freedom reversed.
- From the definition $T = Z/\sqrt{Y/n}$ of a $t$ RV, it follows that,

$$T^2 = \frac{Z^2}{\frac{Y}{n}} \sim \frac{\frac{\chi^2(1)}{1}}{\frac{[\chi^2(n)]}{n}} = F(1,n). \tag{16.45}$$

### Sampling Distribution of Ratio of Sample Variance

It is important in some applications to know the sampling distribution of the difference in means $\bar{X}_1 - \bar{X}_2$ of two samples. Similarly, we may need the sampling distribution of the difference in variances $S_1^2 - S_2^2$. It turns out, however, that this distribution is rather complicated. Because of this, we consider instead the statistic $S_1^2/S_2^2$, since a large or small ratio would indicate a large difference, while a ratio nearly equal to 1 would indicate a small difference. The sampling distribution in such a case can be found by using the $F$ distribution. Suppose there are two normal populations, say, population-I and population-II under study, the population-I has variance $\sigma_1^2$ and population-II has variance $\sigma_2^2$. For describing the sampling distribution for ratio of population variances, we consider all possible samples of same size $n_1$ from population-I and for each sample we calculate sample variance $S_1^2$. Similarly, calculate sample variance $S_2^2$ from each sample of same size $n_2$ drawn from population-II. Then we consider all possible values of the ratio of the variances $S_1^2$ and $S_2^2$, The values of $S_1^2/S_2^2$ may vary from sample to sample so we construct the probability distribution of the ratio of the sample variances. The probability distribution thus obtained is known as sampling distribution of the ratio of sample variances. Therefore, the sampling distribution of ratio of sample variances can be defined as:

**Definition (Sampling Distribution of Ratio of Sample Variances):** The probability distribution of all values of the ratio of two sample variances would be obtained by drawing all possible samples from both the populations is called sampling distribution of ratio of sample variances.

The ratio of two independent chi-square variates when divided by their respective degrees of freedom follows $F$-distribution as

$$F = \frac{[\chi^2(n_1-1)]/(n_1-1)}{[\chi^2(n_2-1)]/(n_2-1)}. \tag{16.46}$$

Since $\frac{(n-1)S^2}{\sigma^2} \sim \chi^2(n-1)$, we have

$$F = \frac{\frac{(n_1-1)S_1^2}{\sigma_1^2(n_1-1)}}{\frac{(n_2-1)S_2^2}{\sigma_2^2(n_2-1)}} = \frac{\frac{S_1^2}{\sigma_1^2}}{\frac{S_2^2}{\sigma_2^2}} \sim F(n_1-1, n_2-1). \tag{16.47}$$

If $\sigma_1^2 = \sigma_2^2$ then

$$F = \frac{S_1^2}{S_2^2} \sim F(n_1-1, n_2-1). \tag{16.48}$$

Therefore, the sampling distribution of ratio of sample variances follows $F$-distribution with $(n_1-1, n_2-1)$ degrees of freedom.

**Definition ($F_\alpha(n_1, n_2)$):** The symbol $F_\alpha(n_1, n_2)$ indicates that the $F(n_1, n_2)$-value has an area $\alpha$ to its right.

***Example 16.13***

Find the $F_\alpha(n_1, n_2)$, $\alpha = \{0.025, 0.05, 0.10, 0.5, 0.90, 0.95, 0.975\}$.

***Solution***

**Part 1**

```
(* The symbol αleft is used to represent the significance levels, which denote the areas under the FRatio distribution curve to the left of the corresponding F-score: *)
αleft={0.025,0.05,0.10,0.5,0.90,0.95,0.975}

(* The variable αright represents the complement of the significance levels, which denotes the areas under the FRatio distribution curve to the right of the corresponding F-score: *)
αright=1-αleft

(* The Quantile function is used to calculate the F-scores corresponding to the specified significance levels (αleft) under the FRatio distribution curve. This provides the critical values for confidence intervals or hypothesis testing.*)
n1=5;
n2=15;
N[Quantile[FRatioDistribution[n1,n2],αleft]]

(*The InverseCDF function is an alternative method to compute the F-scores corresponding to the significance levels (αleft) under the FRatio distribution curve. It yields the same critical values as the Quantile function: *)
InverseCDF[FRatioDistribution[n1,n2],αleft]
 {0.025,0.05,0.1,0.5,0.9,0.95,0.975}
 {0.975,0.95,0.9,0.5,0.1,0.05,0.025}
 {0.155576,0.216508,0.308832,0.910724,2.27302,2.90129,3.57642}
 {0.155576,0.216508,0.308832,0.910724,2.27302,2.90129,3.57642}
```

**Part 2**

```
n1=5;
n2=15;
dist=FRatioDistribution[n1,n2]; (*Define a FRatio distribution with n1=5, n2=15: *)

αleft={0.025,0.05,0.10,0.5,0.90,0.95,0.975}; (*List of significance levels*)
αright=1-αleft; (*Complement of significance levels*)

q=N[Quantile[dist,αleft]]; (*Calculate quantiles based on the significance levels*)
Plot[
 PDF[dist,x],(*Plot the PDF of the FRatio distribution with n1=5, n2=15: *)
 {x,0,6},
 (* The'Epilog' option adds white vertical lines at the quantiles corresponding to the significance levels, and blue text labels indicating the complement significance levels: *)
 Epilog->{
   White,
   Thickness[0.008],
   Apply[
    Sequence,
    Table[
     Line[{{q[[i]],0},{q[[i]],PDF[dist,q[[i]]]}}],
     {i,1,7}
     ]
    ],
   Blue,
   Table[
    Text[αright[[i]],{q[[i]],0.05+PDF[dist,q[[i]]]}],
    {i,1,7}
    ]
   },
 PlotRange->{0,0.9},(*Set the y-axis plot range*)
```

```
 Filling->Axis,
 PlotStyle->Purple,
 ImageSize->250
 ]
```

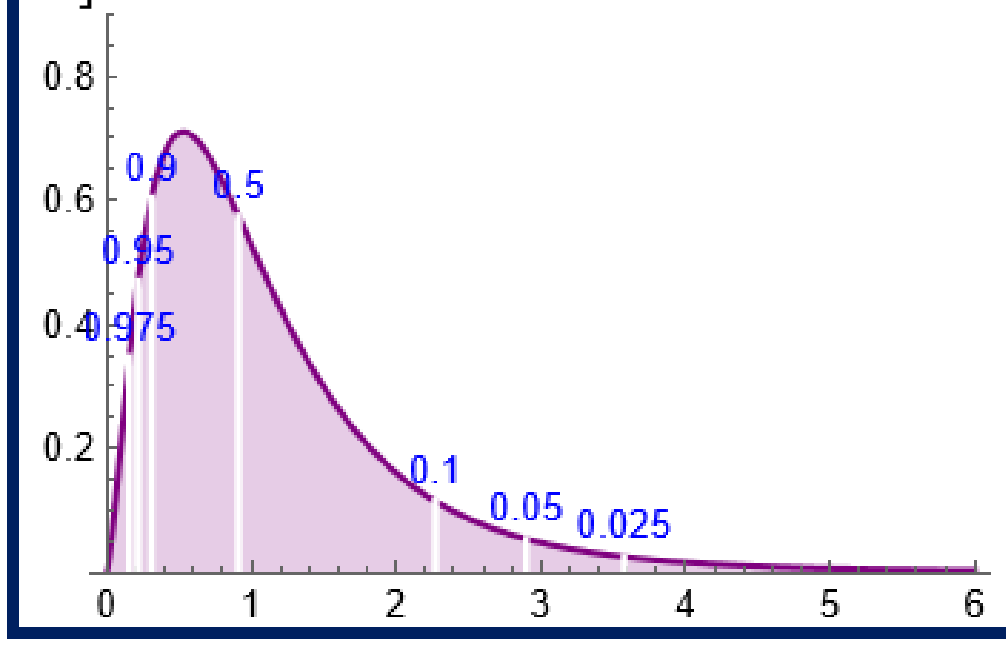

### *Example 16.14*

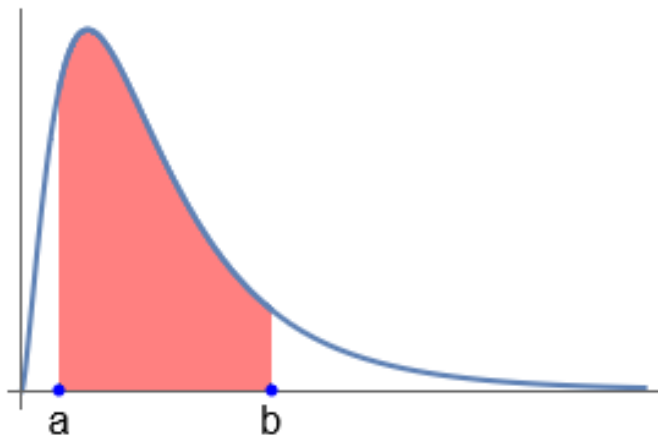

**Figure 16.12** Fisher $F$ distribution, $F(5,15)$.

Figure 16.12 displays the Fisher $F$ distribution, $F(5,15)$. Find the values of $a$ and/or b using Mathematica code so that

(a) the area to the right of $a$ is 0.95,
(b) the total shaded area is 0.95,
(c) the total unshaded area is 0.1,
(d) the area on the left of $a$ is 0.01, and
(e) the area to the right of $a$ is 0.1.

***Solution***

```
Input
N[
 Quantile[
  FRatioDistribution[5,15],
  {0.05,0.95,0.025,0.975,0.05,0.95,0.01,0.9}
  ]
 ]
InverseCDF[
 FRatioDistribution[5,15],
 {0.05,0.95,0.025,0.975,0.05,0.95,0.01,0.9}
 ]
Output

 {0.216508,2.90129,0.155576,3.57642,0.216508,2.90129,0.102857,2.27302}
 {0.216508,2.90129,0.155576,3.57642,0.216508,2.90129,0.102857,2.27302}
```

(a) If the area to the right of $a$ is 0.95, the region to the left of $a$ is $1 - 0.95 = 0.05$ and $a$ stands for the 5th percentile. The result, $F = 0.216508$.
(b) If the total shaded area is 0.95, then the unshaded region is $1 - 0.95 = 0.05$. Thus, the region to the left of $a$ is 0.025 and the region to the right to $b$ is 0.025. Hence, the region to the left to $b$ is $0.95 + 0.025 = 0.975$. The result, at $a$, $F = 1.55576$ and at $b$, $F = 3.57642$.

(c) If the total unshaded area is 0.1, then the region to the left of $a$ is 0.05 and the region to the left to $b$ is 0.95. The result, at $a$, $F = 0.216508$ and at $b$, $F = 2.90129$.
(d) If the area on the left of $a$ is 0.01. The result, at $a$, $F = 0.102857$.
(e) If the area to the right of $a$ is 0.1, then the region to the left of $a$ is $1 - 0.1 = 0.9$. The result, at $a$, $F = 2.27302$.

## 16.7 Sampling Distribution of Sample Proportion

In Section 16.2, we have discussed the sampling distribution of sample mean. But in many real situations, the data collected in form of counts or the collected data classified into two categories or groups according to an attribute. For example, the peoples living in a colony may be classified into two groups (male and female) with respect to the characteristic sex, the patients in a hospital may be classified into two groups as cancer and non-cancer patients, etc. Generally, such types of data are considered in terms of proportion of elements, individuals, items possess (success) or not possess (failure) a given characteristic or attribute. For example, the proportion of female in the population, proportion of cancer patents in a hospital, etc. In such situations, we deal with population proportion instead of population mean.

The population proportion, denoted by $p$, is obtained by taking the ratio of the number of elements in a population with a specific characteristic to the total number of elements in the population. The sample proportion, denoted by $\hat{p}$, gives a similar ratio for a sample.

**Definition (Population and Sample Proportions):** The population and sample proportions, denoted by $p$ and $\hat{p}$, respectively, are calculated as

$$p = \frac{Y}{N} \text{ and } \hat{p} = \frac{X}{n}, \tag{16.49}$$

where
$N$ = total number of elements in the population
$n$ = total number of elements in the sample
$Y$ = number of elements in the population that possess a specific characteristic
$X$ = number of elements in the sample that possess the same specific characteristic

The sample proportion $\hat{p}$ is a RV. In other words, the population proportion $p$ is a constant as it assumes one and only one value. However, the sample proportion $\hat{p}$ can assume one of a large number of possible values depending on which sample is selected. Hence, $\hat{p}$ is a RV and it possesses a probability distribution, which is called its sampling distribution of the sample proportion. For sampling distribution of sample proportion, we draw all possible samples from the population and for each sample we calculate the sample proportion $\hat{p}$.

**Definition (Sampling Distribution of Sample Proportion):** The probability distribution of all values of the sample proportion that obtained by drawing all possible samples of same size from the population is called the sampling distribution of the sample proportion.

For a better understanding of the process, we consider the following example.

***Example 16.15***

A company consists of five employees. The names and details about their knowledge of math are provided in Table 16.4.

**Table 16.4.** Information on the Five Employees.

| Sample | Knows Math |
|---|---|
| Ahmed | Yes |
| Mohamed | No |
| Eman | No |
| Ali | Yes |
| Ayman | Yes |

If we define the population proportion, $p$, as the proportion of employees who know math, then

$$p = 3/5 = 0.60.$$

Note that this population proportion, $p = 0.60$, is a constant.

Let's now assume that we randomly select three employees for each of the possible samples, and then we compute the proportion of employees in each sample that are mathematicians. From the population of five employees, a maximum of 10 samples of size three can be obtained. Table 16.5 displays these 10 possible samples along with the proportion of workers that are mathematicians for each sample.

Using Table 16.5, we prepare the frequency distribution of $\hat{p}$ as recorded in Table 16.6, along with the relative frequencies of classes, which are obtained by dividing the frequencies of classes by the total number of possible samples.

**Table 16.5.** All possible samples of size 3.

| Sample | Proportion who know math $\hat{p}$ |
|---|---|
| Ahmed, Mohamed, Eman | $1/3 = 0.33$ |
| Ahmed, Mohamed, Ali | $2/3 = 0.67$ |
| Ahmed, Mohamed, Ayman | $2/3 = 0.67$ |
| Ahmed, Eman, Ali | $2/3 = 0.67$ |
| Ahmed, Eman, Ayman | $2/3 = 0.67$ |
| Ahmed, Ali, Ayman | $3/3 = 1.00$ |
| Mohamed, Eman, Ali | $1/3 = 0.33$ |
| Mohamed, Eman, Ayman | $1/3 = 0.33$ |
| Mohamed, Ali, Ayman | $2/3 = 0.67$ |
| Eman, Ali, Ayman | $2/3 = 0.67$ |

**Table 16.6.** Frequency and relative frequency.

| $\hat{p}$ | $f$ | Relative frequency $P(\hat{p})$ |
|---|---|---|
| 0.33 | 3 | $3/10 = 0.30$ |
| 0.67 | 6 | $6/10 = 0.60$ |
| 1.00 | 1 | $1/10 = 0.10$ |

**Remark:**

Consider a population of $N$ elements and suppose that $p$ is the proportion of the population that has a certain characteristic of interest; that is, $Np$ elements have this characteristic, and $N(1-p)$ do not.

Suppose now that a random sample of size $n$ has been chosen from a population of size $N$. For $i = 1, \ldots, n$, let

$$X_i = \begin{cases} 1, & \text{if the ith member of the sample has the characteristic} \\ 0, & \text{otherwise} \end{cases}. \tag{16.50}$$

Consider now the sum of the $X_i$; that is, consider

$$X = X_1 + X_2 + \cdots + X_n = \sum_{i=1}^{n} X_i. \tag{16.51}$$

Because the term $X_i$ contributes 1 to the sum if the ith member of the sample has the characteristic and 0 otherwise, it follows that $X$ is equal to the number of members of the sample that possess the characteristic. In addition, the sample mean

$$\bar{X} = \sum_{i=1}^{n} \frac{X_i}{n} = \frac{X}{n}, \tag{16.52}$$

is equal to the proportion, $\hat{p}$, of the members of the sample that possess the characteristic $\bar{X} = \hat{p}$. Recall that a binomial variable $X$ is the number of successes in a binomial experiment consisting of $n$ independent success/failure trials with $p$ = P(success) for any particular trial. In other words, designating a failure in each binomial trial by the value 0 and a success by the value 1, the number of successes, $X$, can be interpreted as the sum of $n$ values consisting only of 0 and 1s, and $\hat{p}$ is just the sample mean of these $n$ values.

Let us now consider the probabilities associated with the statistics $X$. To begin, note that since each of the $N$ members of the population is equally likely to be the ith member of the sample, it follows that

$$P(X_i = 1) = \frac{Np}{N} = p, \tag{16.53}$$

$$P(X_i = 0) = 1 - P(X_i = 1) = 1 - p. \tag{16.54}$$

That is, each $X_i$ is equal to either 1 or 0 with respective probabilities $p$ and $1 - p$.

**Theorem 16.20:** The mean $\mu_{\hat{p}}$ and standard deviation $\sigma_{\hat{p}}$ of a sampling distribution of proportions, $\hat{p}$, are

$$\mu_{\hat{p}} = p, \qquad \sigma_{\hat{p}} = \sqrt{\frac{pq}{n}} = \sqrt{\frac{p(1-p)}{n}}. \tag{16.55}$$

**Proof:**

If a population whose elements are divided into two mutually exclusive groups− one containing the elements which possess a certain attribute (success) and other containing elements which do not possess the attribute (failure), then number of successes (elements possess a certain attribute), follows a binomial distribution with,

$$E(X) = np,$$

$$\text{Var}(X) = npq, \qquad q = 1 - p,$$

where, $p$ is the probability or proportion of success in the population. Now, we can easily find the mean and variance of the sampling distribution of sample proportion by using the above expression as

$$\begin{aligned} E(\hat{p}) &= E\left(\frac{X}{n}\right) \\ &= \frac{1}{n}E(X) \\ &= \frac{1}{n}np = p, \end{aligned}$$

and variance

$$\begin{aligned} \text{Var}(\hat{p}) &= \text{Var}\left(\frac{X}{n}\right) \\ &= \frac{1}{n^2}\text{Var}(X) \\ &= \frac{1}{n^2}npq = \frac{pq}{n}. \end{aligned}$$

Also, standard error of sample proportion can be obtained as

$$S = \sqrt{\text{Var}(\hat{p})} = \sqrt{\frac{pq}{n}}.$$

■

**Theorem 16.21 (CLT for Sample Proportion):** If $n$ is large and $X$ is a binomial RV with parameters $n$ and $p$,

$$Z = \frac{X - np}{\sqrt{np(1-p)}}, \tag{16.56}$$

is approximately standard normal. Hence, if sample size is sufficiently large, such that $np > 5$ and $nq > 5$ then by CLT, the sampling distribution of sample proportion $\hat{p}$ is approximately normally distributed with mean $p$ and variance $pq/n$ where, $q = 1 - p$, i.e.,

$$Z = \frac{X - np}{\sqrt{np(1-p)}} = \frac{n\left(\frac{X}{n} - p\right)}{\sqrt{np(1-p)}} = \frac{\frac{X}{n} - p}{\sqrt{\frac{p(1-p)}{n}}} = \frac{\hat{p} - p}{\sqrt{\frac{pq}{n}}}. \tag{16.57}$$

is approximately standard normal.

**Remark:**

- Note that the population is binomially distributed.
- If the sampling is done without replacement from a finite population then the mean and variance of sample proportion is given by

$$E[\hat{p}] = p, \tag{16.58}$$

and variance

$$\mathrm{Var}(\hat{p}) = \frac{N - n}{N - 1}\frac{pq}{n}, \tag{16.59}$$

where, $N$ is the population size and the factor $(N - n)/(N - 1)$ is called finite population correction.

### Sampling Distributions of Differences of Proportions

Suppose there are two populations, say, population-I and population-II under study and the population-I having population proportion $p_1$ and population-II having population proportion $p_2$ according to an attribute. For describing the sampling distribution of difference of two sample proportions, we consider all possible samples of same size $n_1$ taken from population-I and for each sample calculate the sample proportion $\hat{p}_1$ of success. Similarly, determine the sample proportion $\hat{p}_2$ of success by considering all possible sample of same size $n_2$ from population-II. Then we consider all possible differences of proportions $\hat{p}_1$ and $\hat{p}_2$. The difference of these proportions may or may not be differ so we construct the probability distribution of these differences. The probability distribution thus obtained is called the sampling distribution of the difference of sample proportions.

**Definition (Sampling Distribution of Difference of Two Sample Proportions):** The probability distribution of all values of the difference of two sample proportions that have been obtained by drawing all possible samples of same sizes from both the populations is called sampling distribution of difference between two sample proportions.

The results can be obtained for the sampling distributions of differences of proportions from two binomially distributed populations with parameters $(p_1, q_1)$ and $(p_2, q_2)$, respectively. We have

$$\mu_{\hat{p}_1 - \hat{p}_2} = \mu_{\hat{p}_1} - \mu_{\hat{p}_2} = p_1 - p_2, \qquad \sigma_{\hat{p}_1 - \hat{p}_2} = \sqrt{\sigma^2_{\hat{p}_1} + \sigma^2_{\hat{p}_2}} = \sqrt{\frac{p_1 q_1}{n_1} + \frac{p_2 q_2}{n_2}}. \tag{16.60}$$

As we have seen in case of single proportion, if sample size is sufficiently large, such that $np > 5$ and $nq > 5$ then by CLT, the sampling distribution of sample proportion $\hat{p}$ is approximately normally distributed with mean $p$ and variance $pq/n$ where, $q = 1 - p$. Therefore, if $n_1$ and $n_2$ are sufficiently large, such that $n_1 p_1 > 5$, $n_1 q_1 > 5$, $n_2 p_2 > 5$ and $n_2 q_2 > 5$, then

$$\hat{p}_1 \sim N\left(p_1, \frac{p_1 q_1}{n_1}\right) \text{ and } \hat{p}_2 \sim N\left(p_2, \frac{p_2 q_2}{n_2}\right), \tag{16.61}$$

where, $q_1 = 1 - p_1$ and $q_2 = 1 - p_2$. Hence, we have

$$\hat{p}_1 - \hat{p}_2 \sim N\left(p_1 - p_2, \frac{p_1 q_1}{n_1} + \frac{p_2 q_2}{n_2}\right). \tag{16.62}$$

***Example 16.16***

Determine the probability that less than 45% or more than 55% of 100 fair coin tosses will result in a head.

***Solution***

The 100 coin tosses are regarded as a sample from the infinite population of all possible coin tosses. In this population the probability of heads is $p = 1/2$ and $q = 1 - 1/2 = 1/2$.

$$\mu_{\hat{p}} = p = 0.5, \quad \sigma_{\hat{p}}^2 = \sqrt{\frac{pq}{n}} = \sqrt{\frac{0.5 * 0.5}{100}} = 0.05,$$

$$45\% \text{ in standard units} = \frac{0.45 - 0.5}{0.05} = -1,$$

$$55\% \text{ in standard units} = \frac{0.55 - 0.5}{0.05} = 1.$$

$$P(1 < |Z|) = 2(\text{area to the left of } -1) = 2(0.158655) = 0.317311.$$

```
NProbability[1<Abs[z],z\[Distributed]NormalDistribution[0,1]]
NProbability[z<-1||1<z,z\[Distributed]NormalDistribution[0,1]]
N[2*CDF[NormalDistribution[0,1],-1]]
 0.317311
 0.317311
 0.317311
```

***Example 16.17***

$A$ and $B$ play "heads and tails," tossing 40 coins each. If $A$ throws 4 or more heads than $B$, he wins the game; otherwise, $B$ wins. Determine the probabilities of $A$ winning any given game.

***Solution***

Let $\hat{p}_A$ and $\hat{p}_B$ denote the proportion of heads obtained by $A$ and $B$. Assuming that the coins are all fair, the probability, $p$, of heads equals $1/2$. Then

$$\mu_{\hat{p}_A - \hat{p}_B} = \mu_{\hat{p}_A} - \mu_{\hat{p}_B} = 0.5 - 0.5 = 0, \qquad \sigma_{\hat{p}_A - \hat{p}_B} = \sqrt{\frac{0.5 * 0.5}{40} + \frac{0.5 * 0.5}{40}} = 0.1118.$$

The standardized variable for the difference in proportions is

$$z = \frac{\hat{p}_A - \hat{p}_B - 0}{0.1118}.$$

On a continuous-variable basis, 4 or more heads means 3.5 or more heads, so that the difference in proportions should be $\frac{3.5}{40} = 0.0875$ or more; that is, $z$ is greater than or equal to $(0.0875 - 0)/0.1118 = 0.783$ (or $z \geq 0.783$). The probability of this is the area under the normal curve to the right of $z = 0.783$, which is 0.216814. Thus, the odds against $A$ winning are $(1 - 0.216814){:}\, 0.216814 = 0.783186{:}\, 0.216814$, or 3.61225 to 1.

```
NProbability[0.783<=z,z\[Distributed]NormalDistribution[0,1]]
 0.216814
```

# CHAPTER 17

# MATHEMATICA LAB: SAMPLING THEORY

This chapter explores the Mathematica functions related to several important probability distributions, namely the chi-square distribution, student t distribution, F ratio distribution, and the sampling distributions of mean and variance. Understanding these distributions and their corresponding functions is essential for performing statistical analyses and drawing meaningful conclusions from data.

- Sampling distributions of the mean and variance are essential concepts in inferential statistics. These distributions describe the behavior of sample means and variances when repeatedly drawn from a population. Mathematica provides functions to explore the properties of these sampling distributions, such as calculating probabilities, quantiles, and moments. By utilizing Mathematica built-in functions, we can generate random samples and analyze the behavior of these sampling distributions. Throughout the chapter, we provide an in-depth explanation of the key functions, such as `RandomVariate`, `PDF`, `CDF`, `Quantile`, `Histogram`, `ChiSquareDistribution`, `StudentTDistribution` and `FRatioDistribution` and demonstrate their usage in computing and visualizing sampling distributions of mean and variance.
- The Chi-square distribution plays a fundamental role in statistical inference, particularly in hypothesis testing and constructing confidence intervals for the population variance. The Student t distribution is widely used when the sample size is small, and the population standard deviation is unknown. It plays a central role in hypothesis testing and constructing confidence intervals for the population mean. The F ratio distribution is employed in statistical tests that involve comparing variances or testing the overall significance of a linear regression model. We will discuss the visualization capabilities of Mathematica and how they can enhance our understanding of these distributions. Mathematica offers various plotting functions that enable us to create histograms, probability density plots, and cumulative distribution plots, providing valuable insights into the distribution behavior and shape.
- Throughout this chapter, we will provide clear explanations of the concepts, practical examples, and demonstrate how to utilize Mathematica functions effectively. By mastering these functions, you will enhance your ability to analyze data, and draw valid statistical conclusions.

In the following table, we list the built-in functions that are used in this chapter.

| | | |
|---|---|---|
| `ChiSquareDistribution` | `StudentTDistribution` | `FRatioDistribution` |

Therefore, we divided this chapter into seven units to cover the above topics.

**Chapter 17 Outline**

Unit 17.1. Sampling Distributions of Mean for Large Sample with Normal Distribution
Unit 17.2. Chi-Square Distribution
Unit 17.3. Sampling Distributions of Sample Variance
Unit 17.4. Student t-Distribution
Unit 17.5. Sampling Distributions of Mean for Small Sample with Student Distribution
Unit 17.6. FRatio Distribution
Unit 17.7. Sampling Distributions of Ratio of Two Sample Variances with FRatio Distribution

# UNIT 17.1

# SAMPLING DISTRIBUTIONS OF MEANS FOR LARGE SAMPLE WITH NORMAL DISTRIBUTION

***Mathematica Examples 17.1***

Input

```
(* The code generates a plot of the theoretical sample distributions with different sample sizes n (1, 5, 10, 20 and 30) from a normal population with mean (μ=0 and σ=2), using the probability density function (PDF) of a normal distribution with mean 0 and variance 2/sqrt(n). The plot demonstrates how the variance of the sample distribution decreases as the sample size increases, where we can see that the PDFs become more concentrated around the mean as n increases, indicating a decrease in variance which is a fundamental principle of statistical inference: *)

Plot[
 Evaluate[
  Table[
   PDF[
    NormalDistribution[0,2/Sqrt[n]],
    x
    ],
   {n,{1,5,10,20,30}}
   ]
  ],
 {x,-5,5},
 PlotRange->All,
 PlotLegends->Placed[{"n=1","n=5","n=10","n=20","n=30"},{0.8,0.75}],
 ImageSize->320,
 AxesLabel->{None,"PDF"}
 ]
```

Output

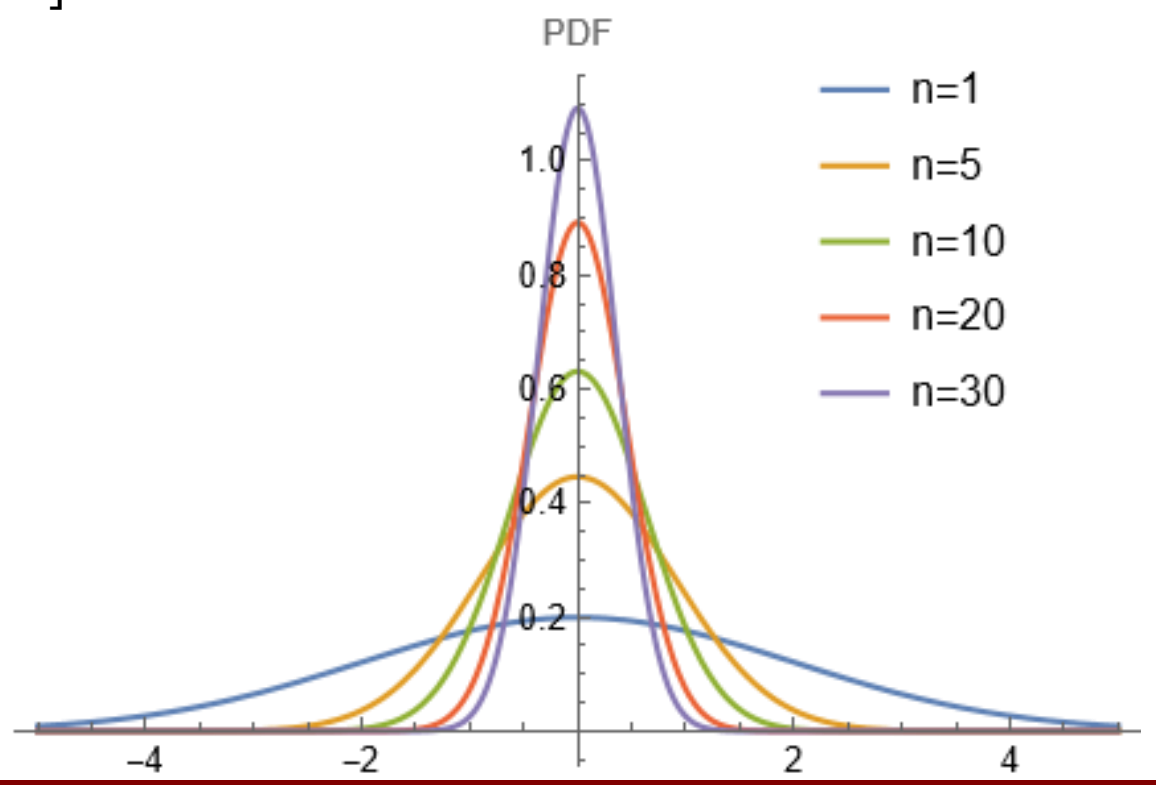

***Mathematica Examples 17.2***

Input

```
(* The code demonstrates the concept of the sampling distribution of the mean and how it can be used to make inferences about the population mean. The code generates a population of 1000 values from a normal distribution with mean 170 and standard deviation 5. It then takes a sample of size 100 and calculates the sample mean. The code repeats this sampling process 1000 times and calculates the sample means. It then calculates the mean of the sampling distribution of the mean and plots the sampling distribution of the mean using Histogram. The plot includes vertical lines
```

```
representing the population mean, the mean of the sampling distribution and the
dashed green line indicating the actual population mean: *)

(* Generate population data: *)
population=RandomVariate[
    NormalDistribution[170,5],
    1000
    ];

(* Calculate population mean: *)
populationMean=Mean[population]

(* Take a sample of size 100 and calculate the sample mean: *)
sample=RandomSample[population,100];
sampleMean=Mean[sample]

(* Repeat the sampling process 1000 times and calculate the sample means: *)
sampleMeans=Table[
    Mean[
     RandomSample[population,100]
     ],
    {i,1,1000}
    ];

(* The mean of sampling distribution of the mean: *)
m=Mean[sampleMeans]

(* Plot the sampling distribution of the mean: *)
Histogram[
 sampleMeans,
 Automatic,
 "ProbabilityDensity",
 Epilog->{
    Directive[Green,Dashed,Thickness[0.006]],
    Line[{{populationMean,0},{populationMean,0.8}}],
    Directive[Red,Dashed,Thickness[0.006]],
    Line[{{m,0},{m,0.8}}]
    },
 PlotLabel->"Sampling Distribution of the Mean",
 ColorFunction->Function[{height},Opacity[height]],
 ImageSize->320,
 ChartStyle->Purple,
 PlotRange->Automatic
 ]

Output   169.914
Output   169.579
Output   169.901
Output
```

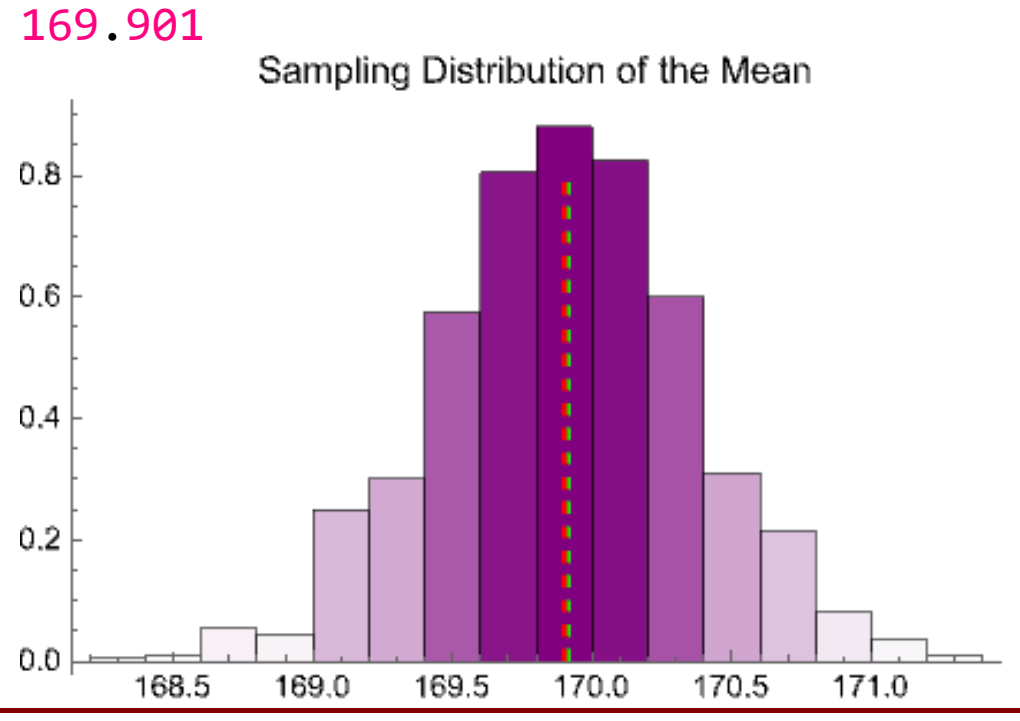

***Mathematica Examples 17.3***

Input

```
(* The code generates a series of sample means of different sizes from a normal distribution with mean 5 and standard deviation 2, and comparing each sample mean to the mean of the original distribution. The absolute difference between each sample mean and the original mean is calculated and stored in a list. Finally, the list is plotted using ListPlot, where the x-axis shows the sample size and the y-axis shows the absolute difference between the sample mean and the population mean. The plot generated by this code could provide useful insights into the behavior of the mean of sample distributions as the sample size increases. It seems that the plot shows a decreasing trend in the absolute difference between the sample means and the population mean as the sample size increases, which is consistent with the central limit theorem (CLT): *)

parent=NormalDistribution[5,2];

means=Table[
   n=i;
   sampleMeans=Table[Mean[RandomVariate[parent,n]],{1000}];
   meanOfSampleMeans=Mean[sampleMeans];
   meanOfParent=Mean[parent];
   Abs[N[meanOfSampleMeans-meanOfParent]],
   {i,1,100}
   ];

ListPlot[
 means,
 AxesLabel->{"Sample size","Absolute difference"},
 Joined->True,
 PlotRange->All,
 PlotStyle->Purple
 ]
```

Output

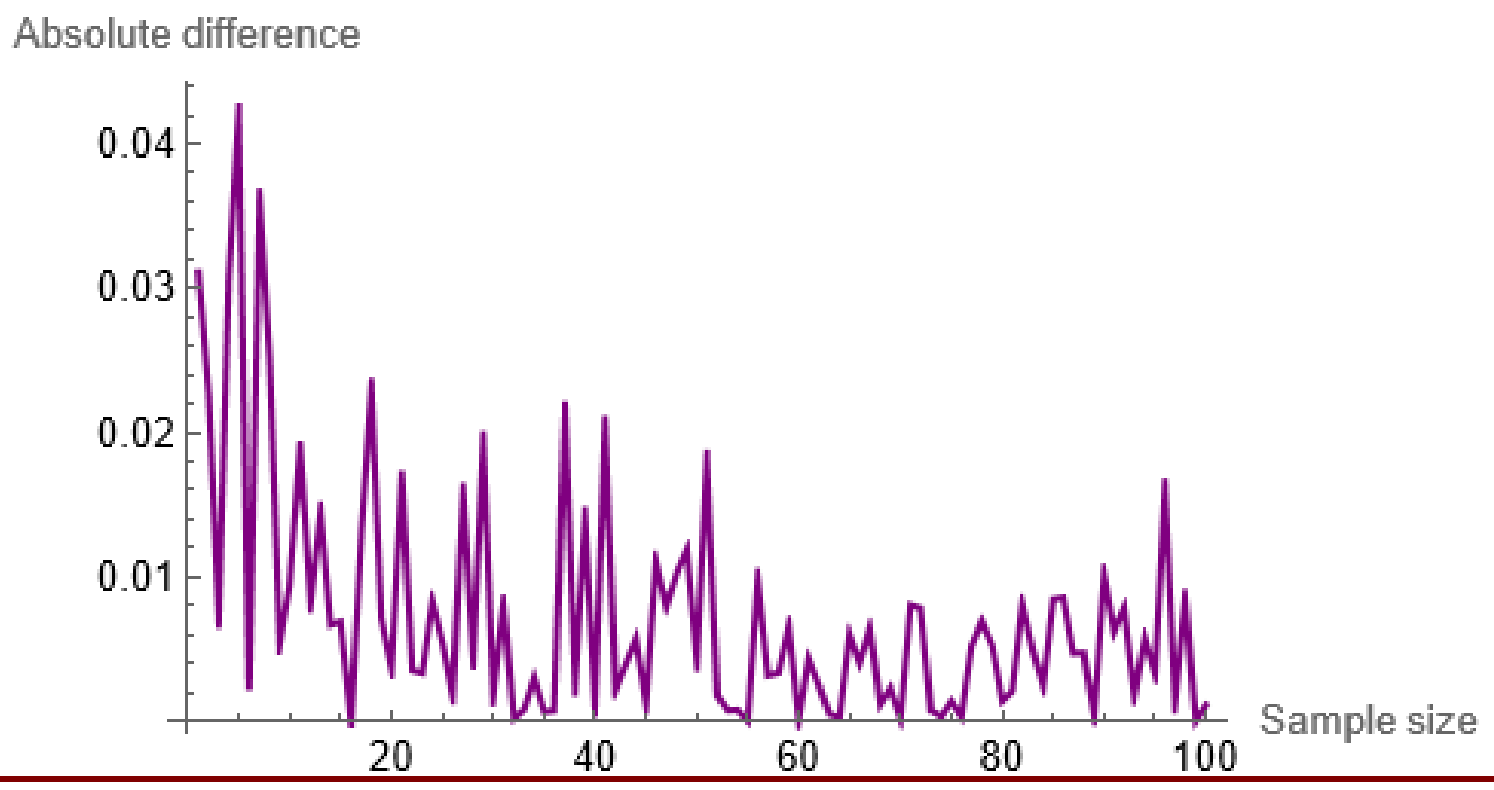

***Mathematica Examples 17.4***

Input

```
(* This code is a demonstration of the relationship between sample size and experimental standard error, and how it compares to the theoretical standard error. This code generates a normal distribution population and then proceeds to generate 200 samples for different sample sizes, ranging from 10 to 200 with step 10. For each sample size, the sample means and experimental standard errors are calculated and compared to the theoretical standard errors, which are also calculated. It is worth noting that the theoretical standard error calculation assumes a population standard deviation of 10, which is the same as the standard deviation used to generate the population. The results are plotted in a ListPlot graph, where the experimental and theoretical standard errors are shown as blue and red points, respectively: *)
```

```
(* Define a population with a normal distribution: *)
population=RandomVariate[
    NormalDistribution[100,10],
    100000
    ];

(* Define a range of sample sizes to test: *)
sampleSizes=Range[10,200,10];

(* Generate 1000 samples for each sample size and calculate the sample means for
each sample size in the range {10,1000,10}: *)
sampleMeans=Table[
    Table[
     Mean[
      RandomSample[population,n]
      ],
     {i,1,200}
     ],
    {n,sampleSizes}
    ];

(* Calculate the experimental standard error for each sample size in the range
{10,1000,10}: *)
standardErrors=Table[
    StandardDeviation[
     sampleMeans[[i]]
     ],
    {i,1,Length[sampleMeans]}
    ];

(* Calculate the theoretical standard error for each sample size in the range
{10,1000,10}: *)
thstandardErrors=Table[
    N[10/Sqrt[n]],
    {n,sampleSizes}
    ];
(* The points (sample size, experimental standard error): *)
exprdata=Transpose[
    {sampleSizes,standardErrors}
    ];

(* The points (sample size, theoretical standard error): *)
thdata=Transpose[
    {sampleSizes,thstandardErrors}
    ];

(* Plot the relationship between sample size and standard error: *)
ListPlot[
 {exprdata,thdata},
 PlotRange->All,
 AxesLabel->{"Sample Size","Standard Error"},
 ImageSize->320,
 PlotStyle->{Directive[Blue,PointSize[0.01]],Directive[Red,PointSize[0.01]]},
 PlotLegends->Placed[{"Experimental standard error","Theoretical standard
error"},{0.6,0.8}]
 ]
```

Output

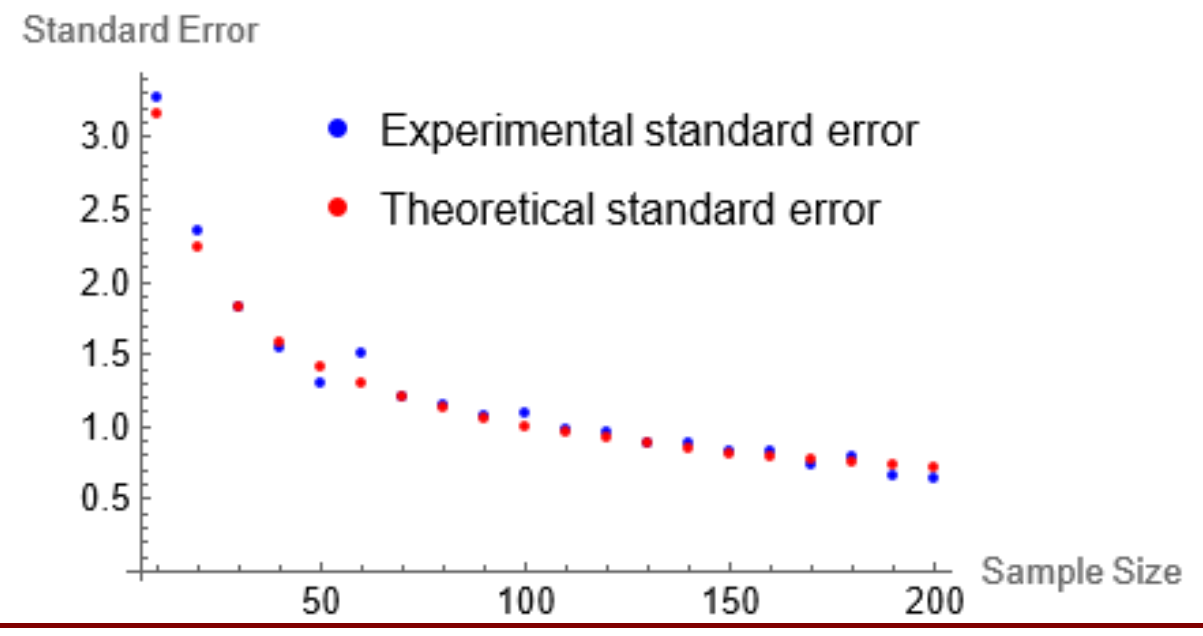

***Mathematica Examples 17.5***

Input

```
(* The code explores the CLT by comparing the standard deviation of sample means across three different population distributions: normal, exponential, and gamma. Specifically, the code generates 1000 samples of size n from three population distributions (normal, exponential, and gamma) using the RandomChoice function, and calculates the mean of each sample using the Mean function. This generates 1000 sample means for each value of n. Then, the standard deviation of the sample means is calculated for each value of n using the StandardDeviation function. The code plots the relationship between sample size (n) and the standard deviation of the sample means. The x-axis represents the sample size (n), and the y-axis represents the standard deviation of the sample means. The plot shows that as the sample size increases, the standard deviation of the sample means decreases. This relationship is in line with the CLT, which states that as the sample size increases, the distribution of the sample means approaches a normal distribution with a mean equal to the population mean and a standard deviation equal to the population standard deviation divided by the square root of the sample size: *)

(* Define a normal population distribution: *)
normalpopulation=RandomVariate[
    NormalDistribution[1,1],
    10000
    ];

(* Define an exponential population distribution: *)
exponentialpopulation=RandomVariate[
    ExponentialDistribution[1.5],
    10000
    ];

(* Define a gamma population distribution: *)
gammapopulation=RandomVariate[
    GammaDistribution[0.4,3],
    10000
    ];

population={normalpopulation,exponentialpopulation,gammapopulation};

(* Define a function to generate sample means: *)
sampleMeans[n_,j_]:=Module[
   {samples,means},
   samples=Table[
     RandomChoice[population[[j]],n],
     {i,1,1000}
     ];
   means=Map[Mean,samples];
   means
   ]
```

```
(* Use Manipulate to explore the CLT:*)
Manipulate[
 ListPlot[
  Table[
   Table[
    {n,StandardDeviation[sampleMeans[n,j]]},
    {n,2,max,1}
    ],
   {j,1,3}
   ],
  PlotLegends->Placed[{"Normal standard error","Exponential standard error","Gamma
standard error"},{0.6,0.8}],
  PlotRange->{{0,101},{0,1.5}},
  AxesLabel->{"Sample Size","Standard Deviation"},
  PlotLabel->"Standard Deviation of Sample Means",
  PlotStyle->{Darker[Red],Darker[Blue],Green},
  Joined->True
  ],
 {{max,20},2,100,1}
 ]
```

Output

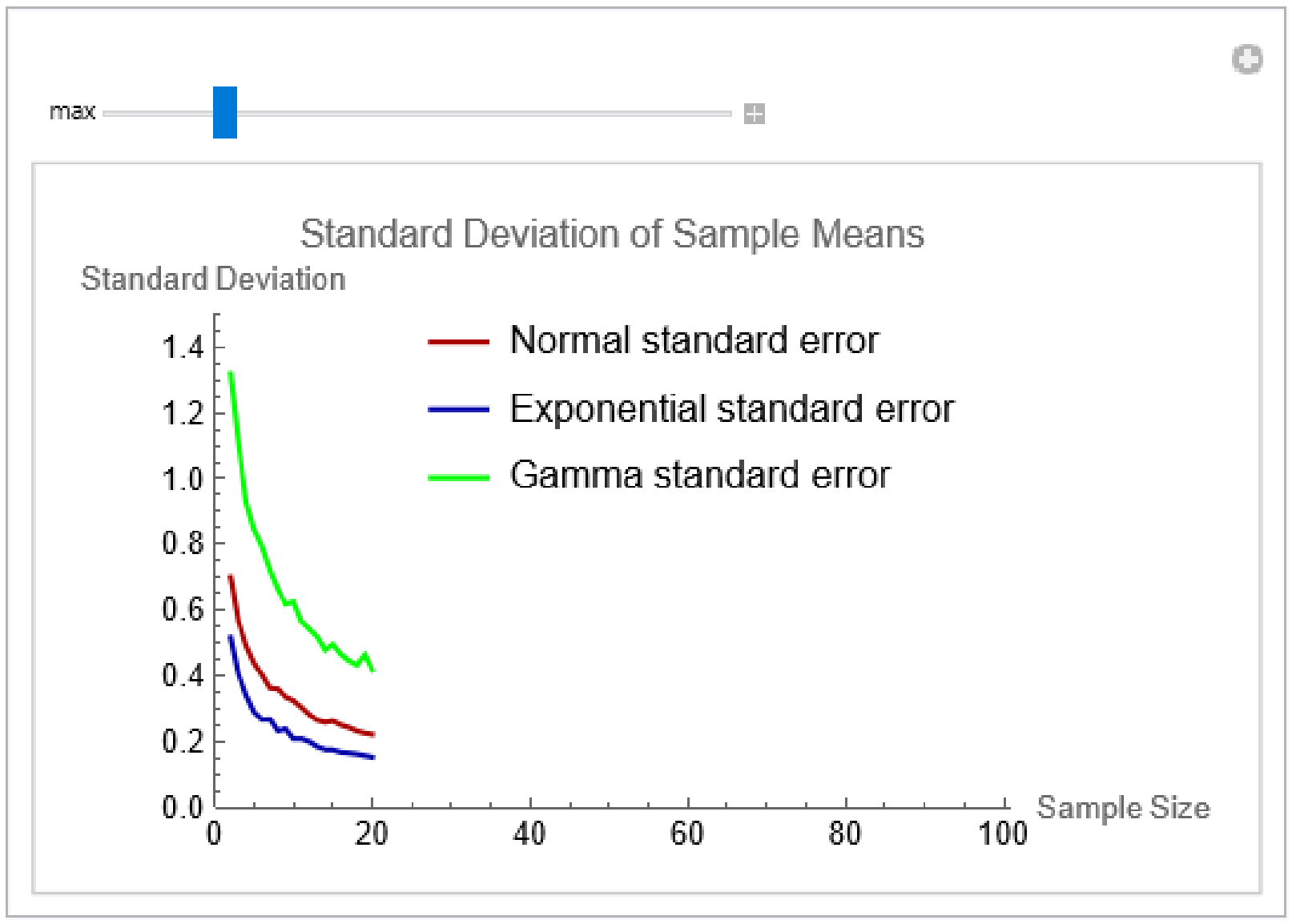

***Mathematica Examples 17.6***

Input

```
(* This code demonstrates the CLT, which states that as the sample size increases,
the distribution of the sample means approaches a normal distribution, regardless of
the underlying distribution of the population. The histogram of the sample means
approximates a bell curve, and the normal distribution curve overlaid on top shows
how well the sample means fit a normal distribution. The code generates a dataset of
100,000 random values sampled from a uniform distribution with parameters {1,5}.
Then, it creates a histogram of the data with bin widths of 0.02, displaying the PDF.
Next, the code generates 100,000 sample means, each taken from a sample of size 10
randomly drawn from the same uniform distribution as before. It then creates a
histogram of the sample means with bin widths of 0.01, displaying the PDF. Finally,
the code creates a plot of the normal distribution with mean equal to the mean of
the sample means and standard deviation equal to the standard deviation of the sample
means. The plot shows the PDF of the normal distribution on the same scale as the
histograms: *)

data=RandomVariate[
   UniformDistribution[{1,5}],
```

```
   10^5
   ];

Histogram[
 data,
 {0.02},
 "PDF",
 ColorFunction->Function[Opacity[0.7]],
 ChartStyle->Purple,
 ImageSize->170,
 AxesLabel->{None,"PDF"}
 ]

samplemeans=Table[
   Mean[
    RandomVariate[
     UniformDistribution[{1,5}],
     10
     ]
    ],
   100000
   ];

hist=Histogram[
  samplemeans,
  {0.01},
  "PDF",
  ColorFunction->Function[Opacity[0.7]],
  ChartStyle->Purple,
  ImageSize->170
  ]

plot=Plot[
  PDF[
   NormalDistribution[Mean[samplemeans],StandardDeviation[samplemeans]],
   x
   ],
  {x,1,5},
  ImageSize->170,
  PlotStyle->RGBColor[0.88,0.61,0.14]
  ]

Show[hist,plot]
```

Output

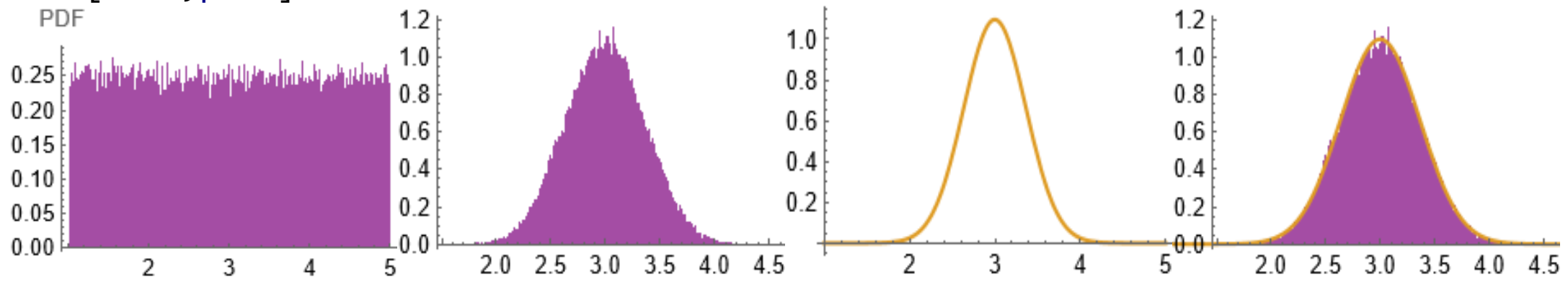

***Mathematica Examples 17.7***

Input

```
(* This code is a demonstration of the CLT, the population distribution is uniform,
which is not a normal distribution, but the CLT still applies. As the sample size
increases, the histograms become more bell-shaped and symmetric, indicating that the
sample means are approaching a normal distribution. Additionally, the histograms
become narrower, indicating that the standard deviation of the sample means is
decreasing as the sample size increases. This demonstrates the practical usefulness
of the CLT in allowing us to make inferences about the population mean based on
sample means, even when the population distribution is not normal. The code generates
```

```
histograms of sample means drawn from a uniform distribution on the interval [1,5].
The sample sizes n range from 1 to 10. The histograms display the PDF of the sample
means. The plot labels indicate the sample size for each histogram: *)

Table[
 Histogram[
  Table[
   Mean[
    RandomVariate[
     UniformDistribution[{1,5}],
     n
     ]
    ],
   100000
   ],
  {0.03},
  "PDF",
  ColorFunction->Function[Opacity[0.7]],
  ChartStyle->Purple,
  ImageSize->170,
  PlotLabel->{{"sample size n=",n}},
  AxesLabel->{"SM","PDF"}(* Sample Mean=SM. *)
  ],
 {n,{1,2,3,10}}
 ]
```

Output

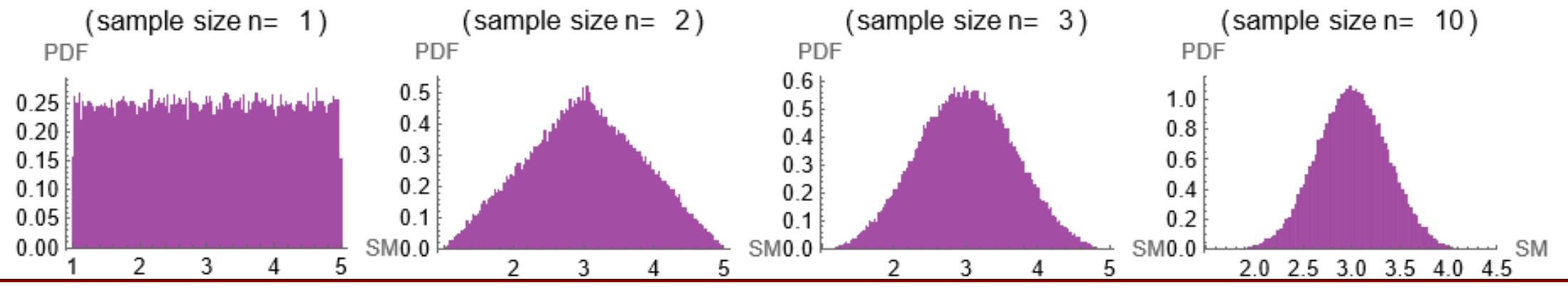

## *Mathematica Examples 17.8*

Input

```
(* The code generates histograms of sample means drawn from a gamma distribution with
shape parameter 4 and scale parameter 2. The sample sizes n range from 1 to 30. The
histograms display the PDF of the sample means. The plot labels indicate the sample
size for each histogram. Similar to the previous code, this code also demonstrates
the CLT, in this case, the population distribution is a gamma distribution, which is
not a normal distribution. However, the CLT still applies, and as the sample size
increases, the histograms become more bell-shaped and symmetric, indicating that the
sample means are approaching a normal distribution. Additionally, the histograms
become narrower, indicating that the standard deviation of the sample means is
decreasing as the sample size increases: *)

Table[
 Histogram[
  Table[
   Mean[
    RandomVariate[
     GammaDistribution[4,2],
     n
     ]
    ],
   100000
   ],
  {0.06},
  "PDF",
  ColorFunction->Function[Opacity[0.7]],
```

```
          ChartStyle->Purple,
          ImageSize->170,
          PlotLabel->{{"sample size n=",n}},
          AxesLabel->{"SM","PDF"}(*Sample Mean=SM*)
          ],
        {n,{1,2,3,10,20,30}}
        ]
```

Output

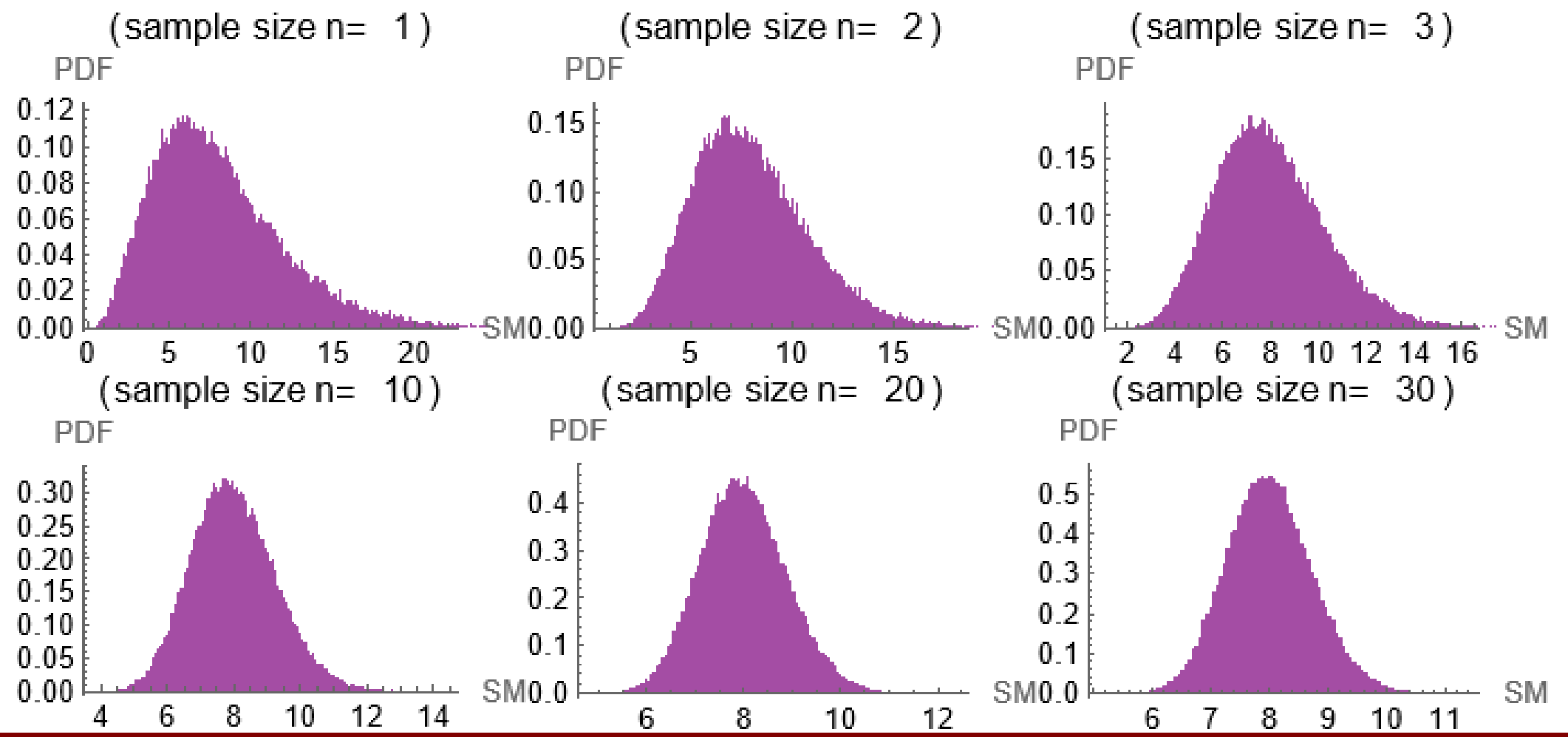

***Mathematica Examples 17.9***

Input

```
(* This code generates histograms of sample means from three different distributions: exponential, gamma, and beta. The sample size is fixed at n=30, and the number of samples is set to numSamples=100,000. For each distribution, the code generates a list of sample means using the Mean function and RandomVariate to generate random numbers from the specified distribution. The histograms display the PDF of the sample means for each distribution, with bars colored red, orange, and purple to represent the exponential, gamma, and beta distributions, respectively. This code provides an illustration of how the CLT can be used to approximate the distribution of sample means for different distributions. It also highlights the fact that the sample mean distribution becomes more normal as the sample size increases, regardless of the underlying distribution: *)


n=30;(* sample size. *)
numSamples=100000;(* number of samples. *)

exponentialMeans=Table[
   Mean[
    RandomVariate[
     ExponentialDistribution[1.5],
     n
     ]
    ],
   {i,1,numSamples}
   ];

gammaMeans=Table[
   Mean[
    RandomVariate[
     GammaDistribution[0.4,3],
     n
     ]
    ],
```

```
    {i,1,numSamples}
    ];

betaMeans=Table[
    Mean[
     RandomVariate[
      BetaDistribution[1,1],
      n
      ]
     ],
    {i,1,numSamples}
    ];

Histogram[
 {
  exponentialMeans,
  gammaMeans,
  betaMeans
  },
 {0.01},
 "PDF",
 ChartLegends->Placed[{"Exponential Sample Distribution of Mean n=30","Gamma Sample Distribution of Mean n=30","Beta Sample Distribution of Mean n=30"},{0.6,0.7}],
 ChartStyle->{Red,Orange,Purple},
 ImageSize->350
 ]
```

Output

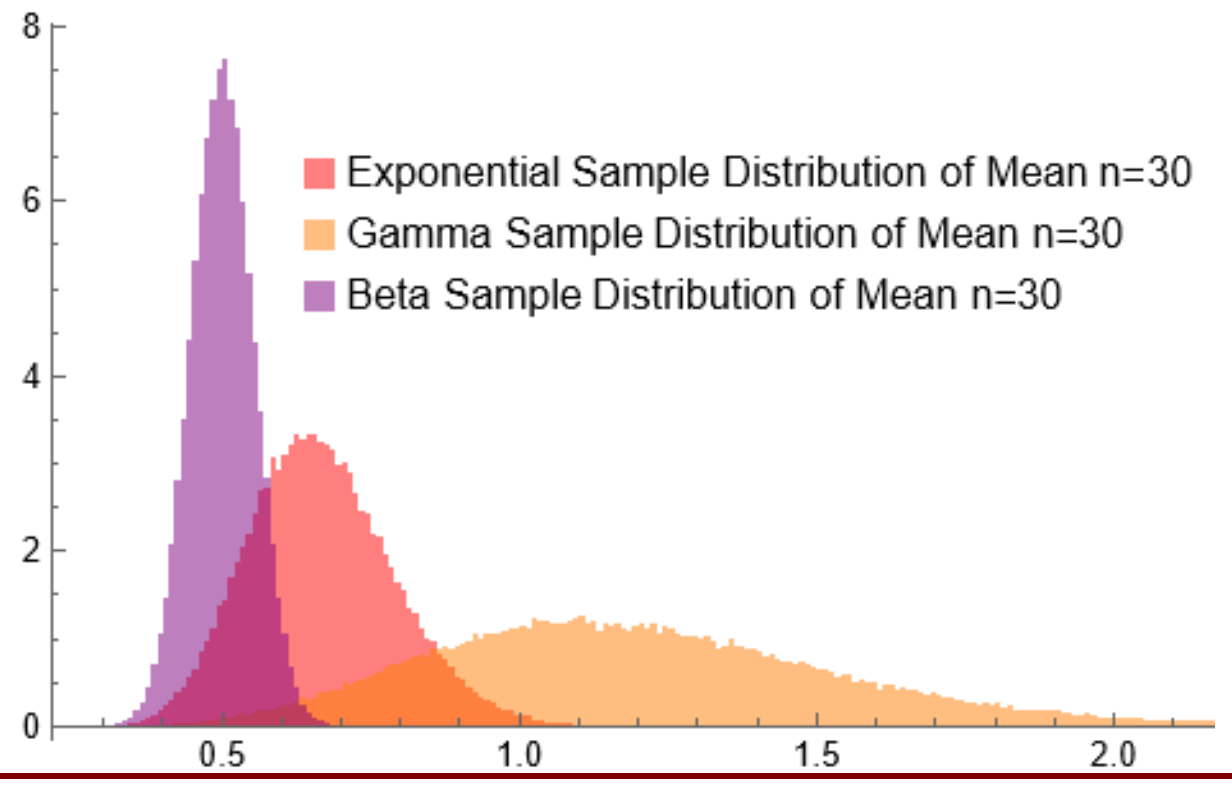

### *Mathematica Examples 17.10*

Input

```
(* This code defines a non-normal population distribution and then generates and plots the sampling distribution of the mean for different sample sizes using the Manipulate function. The population distribution is defined as an Exponential distribution with a mean of 5. The sampleMeans0 function generates 10,000 random samples of size n from this population, calculates the mean of each sample, and then plots the histogram of the means using Histogram function. The Manipulate function allows the user to vary the sample size from 5 to 100 in increments of 1 and observe how the sampling distribution of the mean changes as the sample size increases. The CLT states that the sampling distribution of the mean approaches a normal distribution as the sample size increases, regardless of the shape of the population distribution. However, the convergence to normality may be slower for non-normal population distributions. In particular, when the population distribution is highly skewed or has heavy tails, a larger sample size may be required for the CLT to apply. In our case, since the population distribution is non-normal, the sampling distribution of the mean is also non-normal for small sample size. However, as the sample size increases, the sampling distribution becomes more symmetric and less skewed, and its peak becomes more concentrated around the population mean: *)
```

```
(* Define a non-normal population distribution: *)
population0=RandomVariate[
    ExponentialDistribution[1/5],
    10000
    ];

(* Define a function to generate and plot sample means: *)
sampleMeans0[n_]:=Module[
  {samples,means},

  samples0=Table[
    RandomChoice[
     population0,n
     ],
    {i,1,10000}
    ];

  means0=Map[Mean,samples0];

  Histogram[
   means0,
   {0.05},
   "PDF",
   PlotRange->All,
   AxesLabel->{"Sample Mean","Probability Density"},
   ColorFunction->Function[Opacity[0.8]],
   ChartStyle->Purple,
   PlotLabel->Row[{"n = ",n}]
   ]
  ]

(* Use Manipulate to explore the sampling distributions of the mean: *)
Manipulate[
 sampleMeans0[n],
 {n,5,100,1}
 ]
```

Output

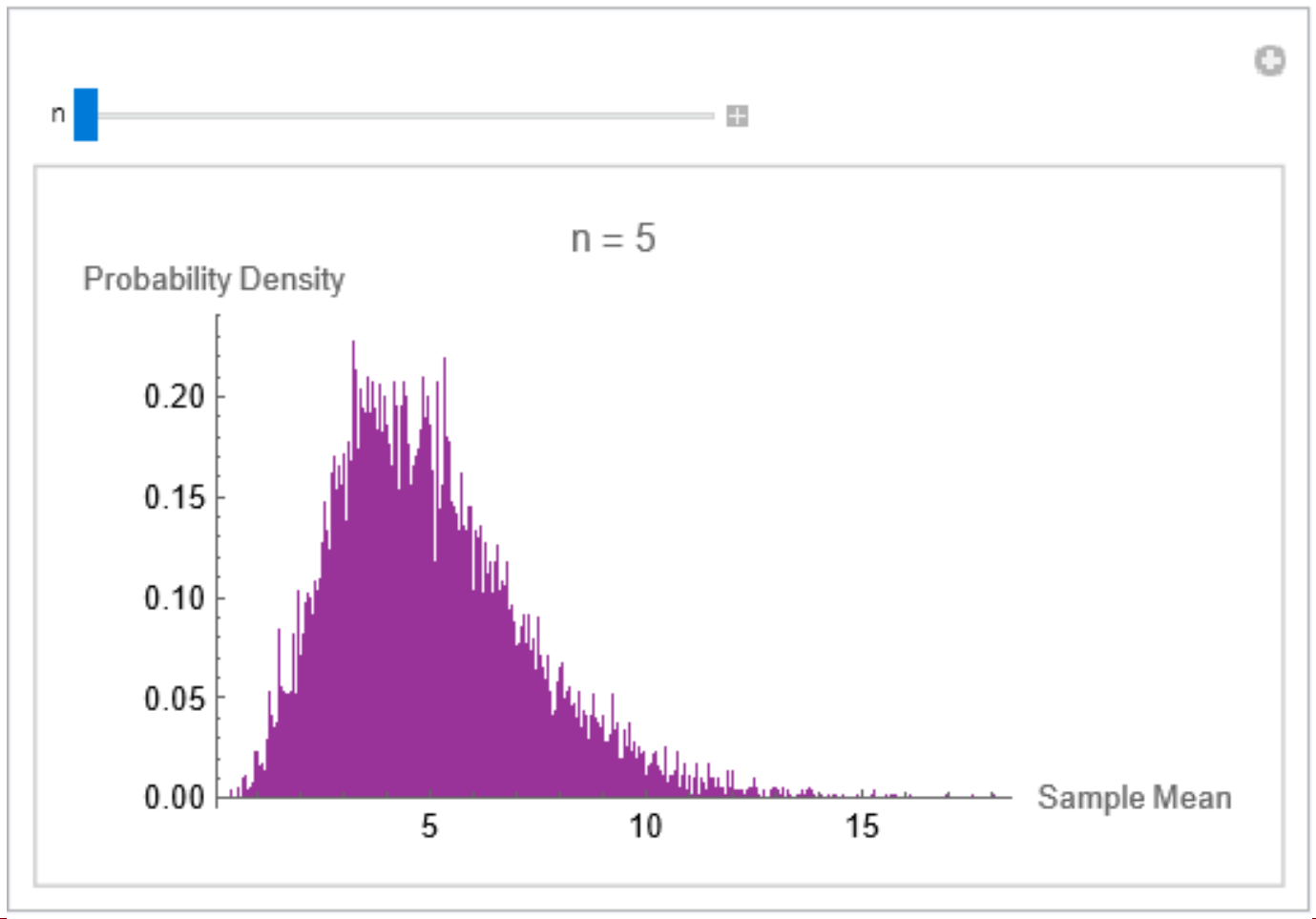

***Mathematica Examples 17.11***

Input (* This code defines a non-normal population distribution (an Exponential distribution with a mean of 5), generates random samples of different sizes from this distribution, and then plots the sampling distribution of the mean for each sample size. The Manipulate function allows the user to interactively explore how the

```
sampling distribution of the mean converges to a normal distribution as the sample
size increases. The sampleMeans function generates 1000 random samples of size n from
the population, calculates the mean of each sample, and then plots the histogram of
the means using Histogram function. The Manipulate function shows both the histogram
of the population distribution and the histogram of the sampling distribution of the
mean for each sample size. The green dashed line represents the population mean, and
its position is fixed across all sample sizes. The code demonstrates the CLT, as the
sample size increases, the sampling distribution becomes more symmetric and less
skewed, and its peak becomes more concentrated around the population mean: *)

(* Define a non-normal population distribution: *)
population=RandomVariate[
   ExponentialDistribution[1/5],
   10000
   ];
populationMean=Mean[population];

(* Define a function to generate and plot sample means: *)
sampleMeans[n_]:=Module[
  {samples,means},
  samples=Table[
    RandomChoice[population,n],
    {i,1,1000}
    ];

  means=Map[Mean,samples];

  Histogram[
   means,
   {0.08},
   "PDF",
   PlotRange->All,
   AxesLabel->{"Sample Mean","Probability Density"},
   ColorFunction->Function[Opacity[0.8]],
   ChartStyle->Purple,
   PlotLabel->Row[{"n = ",n}]
   ]
  ]

(* Use Manipulate to explore the CLT: *)
Manipulate[
 Show[
  Histogram[
   RandomVariate[
    ExponentialDistribution[1/5],
    10000
    ],
   {0.08},
   "PDF",
   PlotRange->{0,0.8},
   ColorFunction->Function[Opacity[0.3]],
   ChartStyle->Blue,
   Epilog->{
     Directive[Green,Dashed,Thickness[0.006]],
     Line[{{populationMean,0},{populationMean,0.8}}]
     }
   ],
  sampleMeans[n]
  ],
 {n,5,100,5}
 ]
```

Output

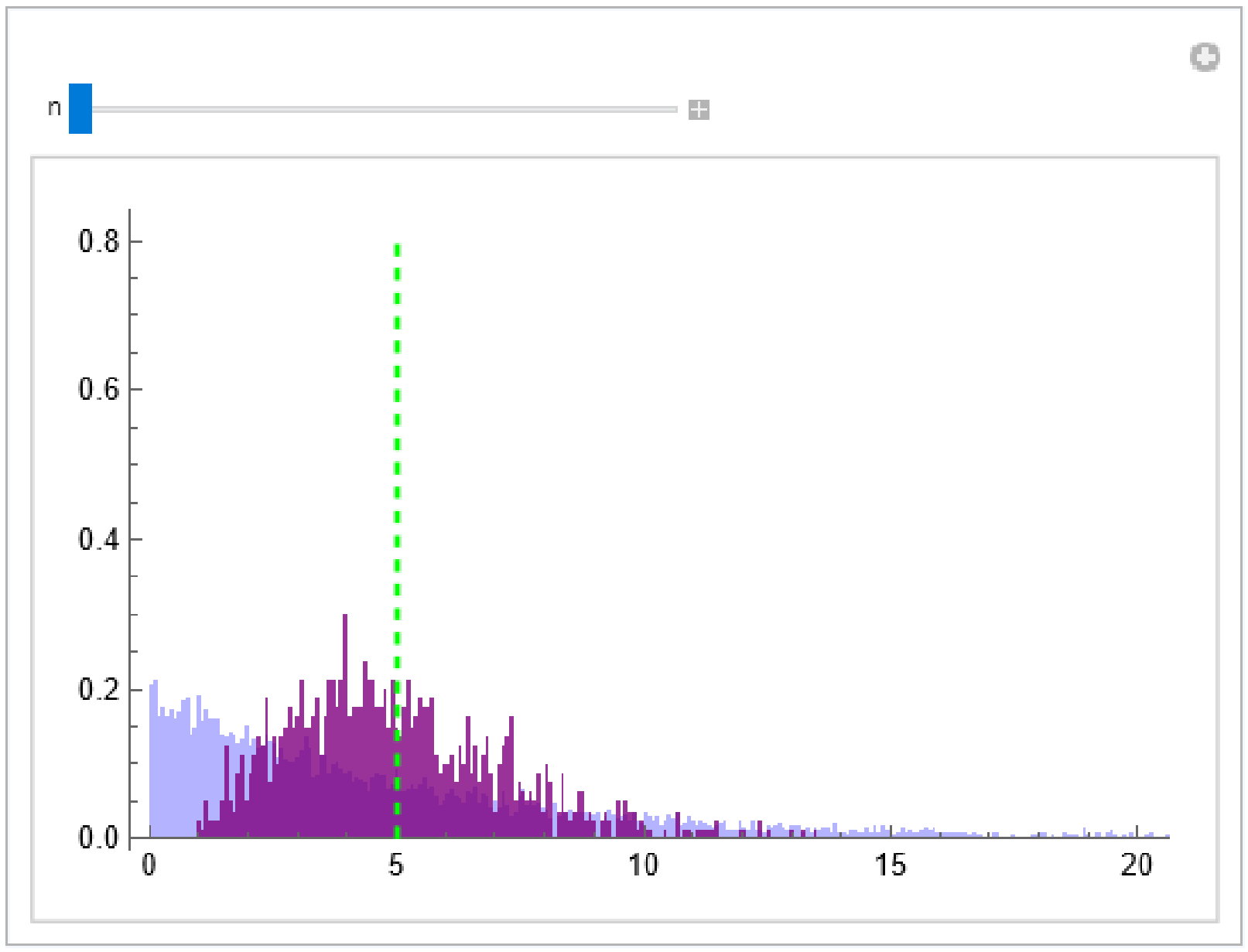

***Mathematica Examples 17.12***

Input

```
(* The code generates sample data from four different probability distributions (Normal, Beta, Gamma, and Exponential) and creating a graph that compares the actual observations with the sample mean and true mean of the distributions. The graph includes the following features:
a horizontal line indicating the true mean of the distribution; the area between the observations and the true mean line is filled with a purple color, while the sample means line is blue.

Accumulate is a built-in function in Mathematica that generates a list of the cumulative sums of a list of numbers. In this case, Accumulate[observations] generates a list of cumulative sums of the observations list. Range[n] generates a list of integers from 1 to n, which represents the sample sizes. Dividing the cumulative sums of the observations by the sample sizes (which is represented by Range[n]) gives the sample means for each corresponding sample size.

Note that the sampling is still random even when using the Accumulate[observations] and Range[n] method to calculate the sample means. The RandomVariate function is used to generate the initial observations from the given probability distribution, and this function generates random samples based on the specified distribution. After generating the observations, the Accumulate[observations] and Range[n] method is used to calculate the sample means. This method does not change the random nature of the original observations. Instead, it simply performs a mathematical operation on the observed data to calculate the sample means: *)

comparedata[distribution_,n_]:=Module[
  {title=distribution,observations,sampleMeans,μ},

  observations=RandomVariate[
    distribution,
    n
    ];

  sampleMeans=Accumulate[observations]/Range[n];
  μ=Mean[distribution];
```

```
        ListPlot[
         {
          Table[
           {i,observations[[i]]},
           {i,n}
           ],
          Table[
           {i,sampleMeans[[i]]},
           {i,n}
           ]
          },
         Joined->{False,True},
         Filling->{1->μ},
         PlotStyle-
>{Directive[Purple,Opacity[0.8]],Directive[Blue,Thickness[0.008],Opacity[0.5]]},
         Epilog->{Red,Line[{{0,μ},{n,μ}}]},
         Frame->True,
         FrameLabel->{"Sample Size","Observations & Sample Mean"},
         FrameStyle->Directive[Black],
         PlotLabel->title,
         ImageSize->230
         ]
        ]

distributions={
    NormalDistribution[1,2],
    BetaDistribution[2,2],
    GammaDistribution[5,2],
    ExponentialDistribution[1]
    };

Table[
 comparedata[distributions[[i]],100],
 {i,1,4}
 ]
```

Output

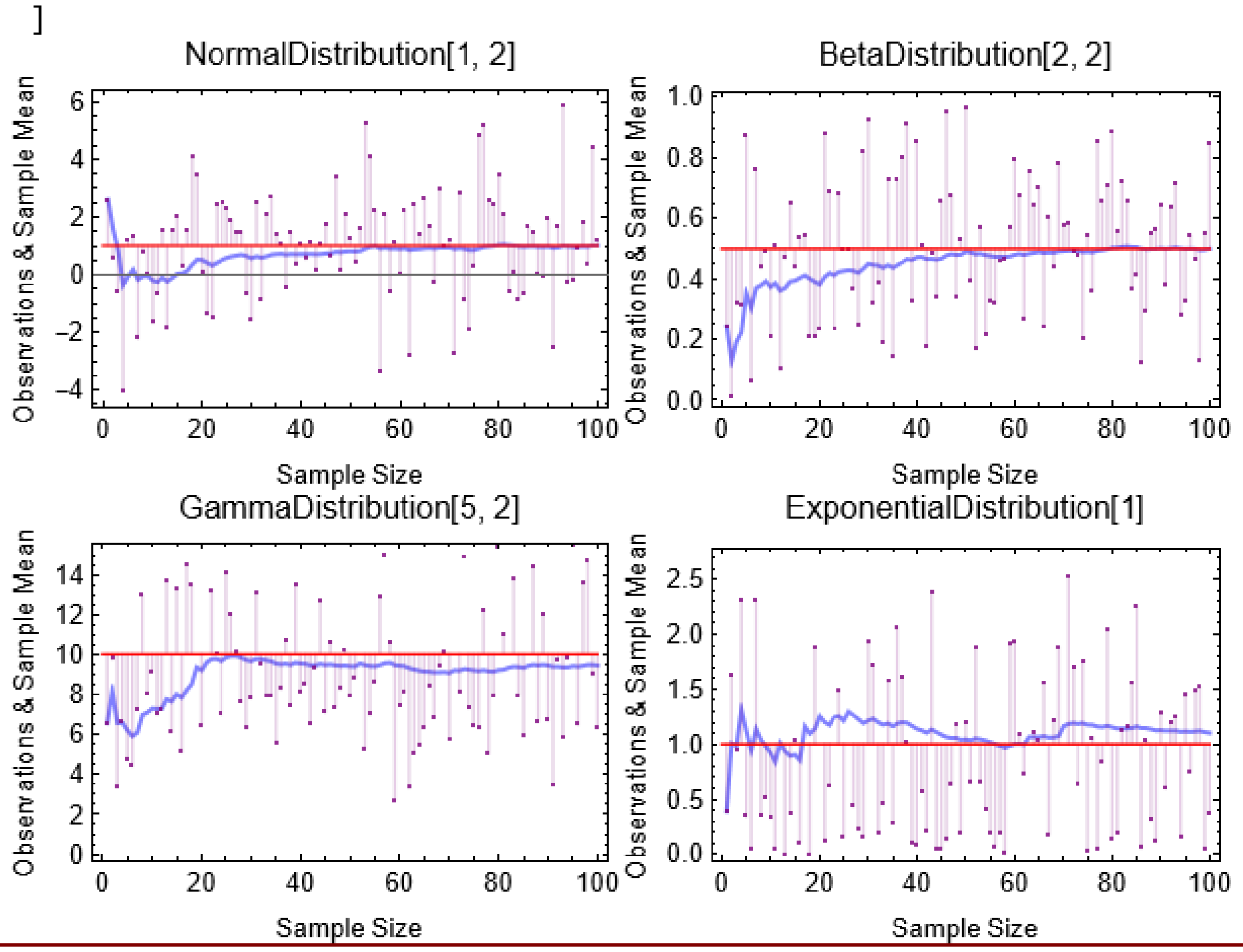

***Mathematica Examples 17.13***

Input

```
(* This code is a demonstration of the CLT in 2D, which states that as the sample
size increases, the distribution of sample means approaches a normal distribution
with mean equal to the population mean and standard deviation equal to the population
standard deviation divided by the square root of the sample size. In this case, the
population distribution is uniform, which is not a normal distribution, but the CLT
still applies. As the sample size increases, the 3D histograms become more bell-
shaped and symmetric, indicating that the sample means are approaching a normal
distribution. Additionally, the 3D histograms become narrower,indicating that the
standard deviation of the sample means is decreasing as the sample size increases.
The code generates 3D histograms of sample means drawn from a uniform distribution
on the interval [0,6]. The sample sizes n range from 1 to 10. The 3D histograms
display the PDF of the sample means. The plot labels indicate the sample size for
each 3D histogram: *)

Table[
 Histogram3D[
  Table[
   Mean[
    RandomVariate[
     UniformDistribution[{0,6}],
     {n,2}
     ]
    ],
   100000
   ],
  Automatic,
  "PDF",
  ColorFunction->"Rainbow",
  ImageSize->220,
  PlotLabel->{{"sample size n=",n}},
  AxesLabel->{"SM","PDF"}(* Sample Mean=SM. *)
  ],
 {n,{1,2,3,10}}
 ]
```

Output

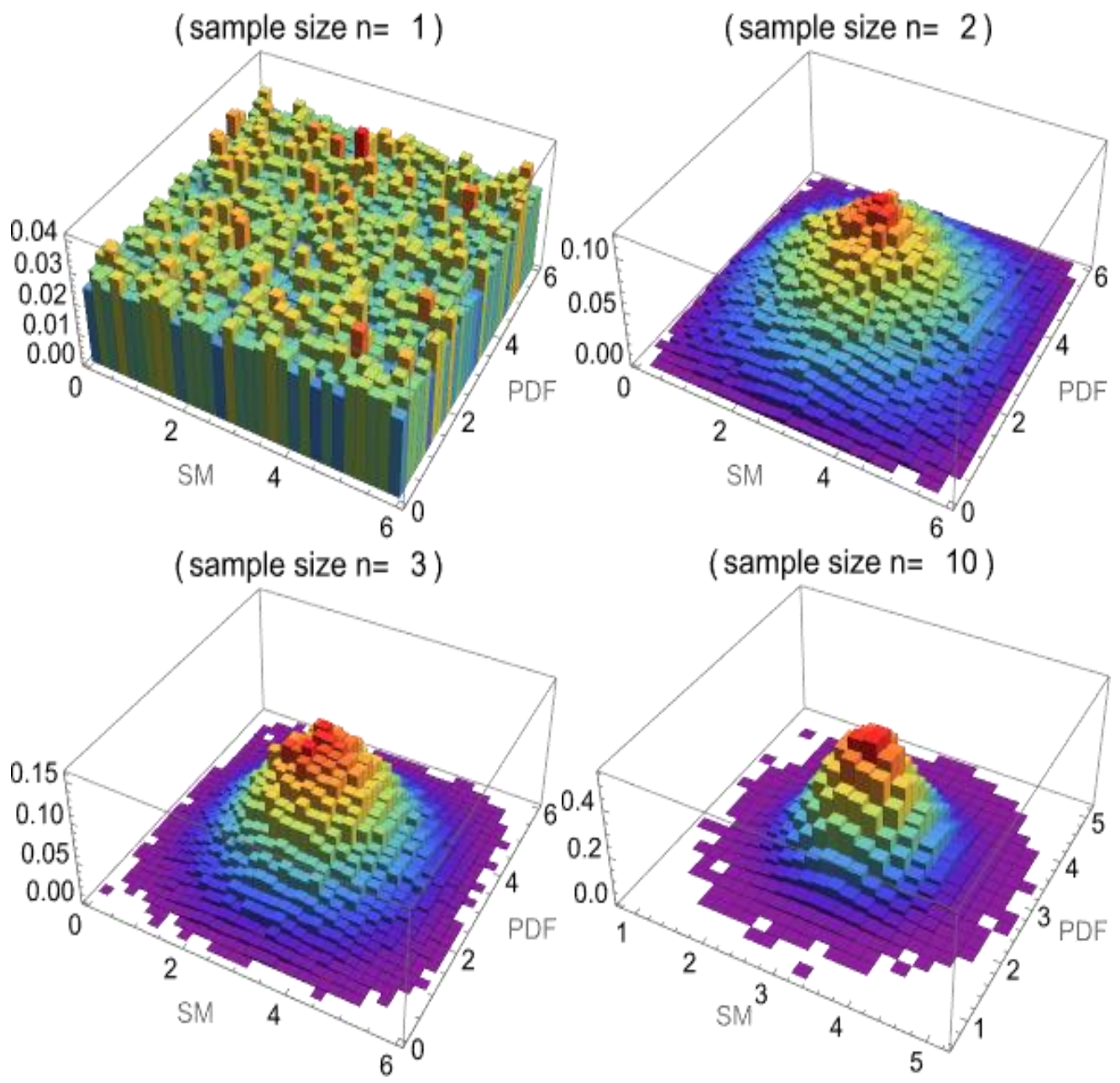

***Mathematica Examples 17.14***

Input

```
(* This code defines a 2D non-normal population and then generates and plots the
sampling distribution of the mean for different sample sizes using the Manipulate
function. The population distribution is defined as 2D uniform distribution on the
interval [0,6]. The sampleMeans3d function generates 10,000 random samples of size n
from this population, calculates the mean of each sample, and then plots the 3D
histogram of the means using Histogram3D function. The Manipulate function allows
the user to vary the sample size from 1 to 50 in increments of 1 and observe how the
sampling distribution of the mean changes as the sample size increases. In our case,
since the population distribution is non-normal, the sampling distribution of the
mean is also non-normal for small sample size. However, as the sample size increases,
the sampling distribution becomes more symmetric, and its peak becomes more
concentrated around the population mean: *)

population3d=RandomVariate[
    UniformDistribution[{0,6}],
    {10000,2}
    ];

(* Define a function to generate and plot sample means: *)
sampleMeans3d[n_]:=Module[
  {samples,means},

  samples3d=Table[
    RandomChoice[
     population3d,n
     ],
    {i,1,10000}
    ];

  means3d=Map[Mean,samples3d];

  Row[
   Histogram3D[
    means3d,
    Automatic,
    "PDF",
    PlotRange->All,
    AxesLabel->{"Sample Mean","Probability Density"},
    ColorFunction->"Rainbow",
    PlotLabel->Row[{"n = ",n}],
    ImageSize->320
    ],
   SmoothDensityHistogram[
    means3d,
    Automatic,
    "PDF",
    ColorFunction->"Rainbow",
    ImageSize->200,
    PlotLabel->{{"sample size n=",n}}
    ]
   ]
  ]
(* Use Manipulate to explore the sampling distributions of the mean: *)
Manipulate[
 sampleMeans3d[n],
 {n,5,50,1}
 ]
```

Output

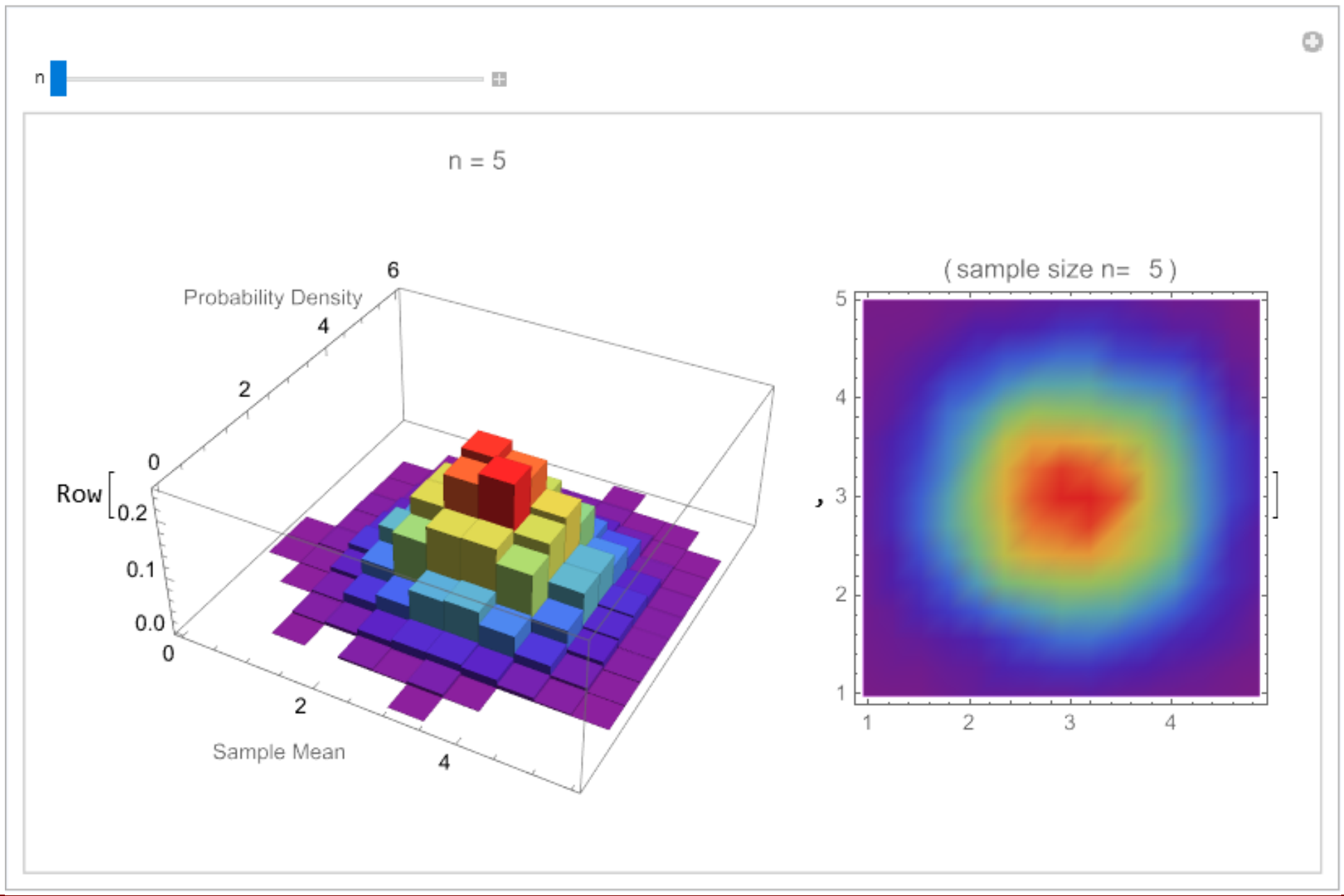

***Mathematica Examples 17.15***

Input

```
(* The code generates a 2D dataset with 5000 random points that follow a uniform
distribution on the interval [0,6]. The scatter plot of the dataset shows the X-axis
values on the horizontal axis and the Y-axis values on the vertical axis. Each point
in the plot represents a pair of X and Y values from the dataset. The code also
defines a function to generate and plot sample means. The function takes a sample
size 'n', draws 5000 samples from the population with replacement, calculates the
mean for each sample, and plots the means as purple points on a scatter plot. The
opacity of the points is set to 0.3 to indicate overlapping points. The Manipulate
function is then used to explore the sample means point. Manipulate allows the user
to interactively adjust the sample size 'n' and observe the changes in the sample
means point, as the sample size 'n' increases, the spread of the sample means points
around the population mean decreases: *)

population1=RandomVariate[
   UniformDistribution[{0,6}],
   {5000,2}
   ];

(* Define a function to generate and plot sample means: *)
sampleMeans1[n_]:=Module[
  {samples,means},

  samples1=Table[
    RandomChoice[
     population1,n
     ],
    {i,1,5000}
    ];

  means1=Map[Mean,samples1];

  Graphics[
   {Purple,PointSize[0.008],Opacity[0.3],Point[means1]},
   ImageSize->250
   ]
```

```
    ]

(* Use Manipulate to explore the sample means points: *)
Manipulate[
 sampleMeans1[n],
 {n,1,30,1}
 ]
```

Output

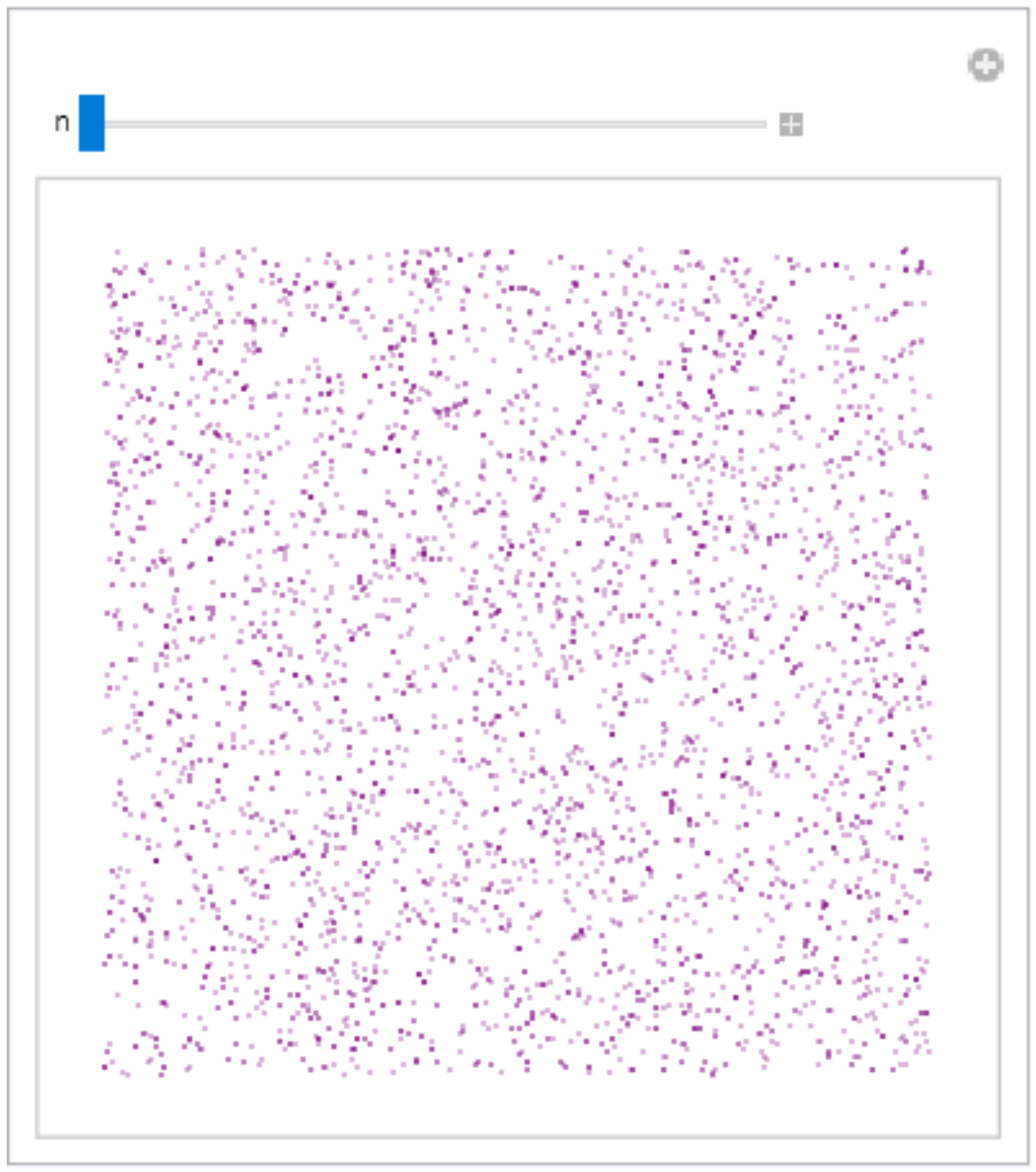

*Mathematica Examples 17.16*

Input

```
(* The CLT also applies to three-dimensional datasets, and the same general principles
of the theorem hold true: *)

populationx=RandomVariate[
    UniformDistribution[{0,6}],
    {5000,3}
    ];

(* Define a function to generate and plot sample means: *)
sampleMeansx[n_]:=Module[
   {samples,means},

   samplesx=Table[
     RandomChoice[
      populationx,n
      ],
     {i,1,5000}
     ];

   meansx=Map[Mean,samplesx];

   Graphics3D[
    {Purple,PointSize[0.008],Opacity[0.3],Point[meansx]},
    ImageSize->250
    ]
   ]
```

```
(* Use Manipulate to explore the sample means points: *)
Manipulate[
 sampleMeansx[n],
 {n,1,30,1}
 ]
```

Output

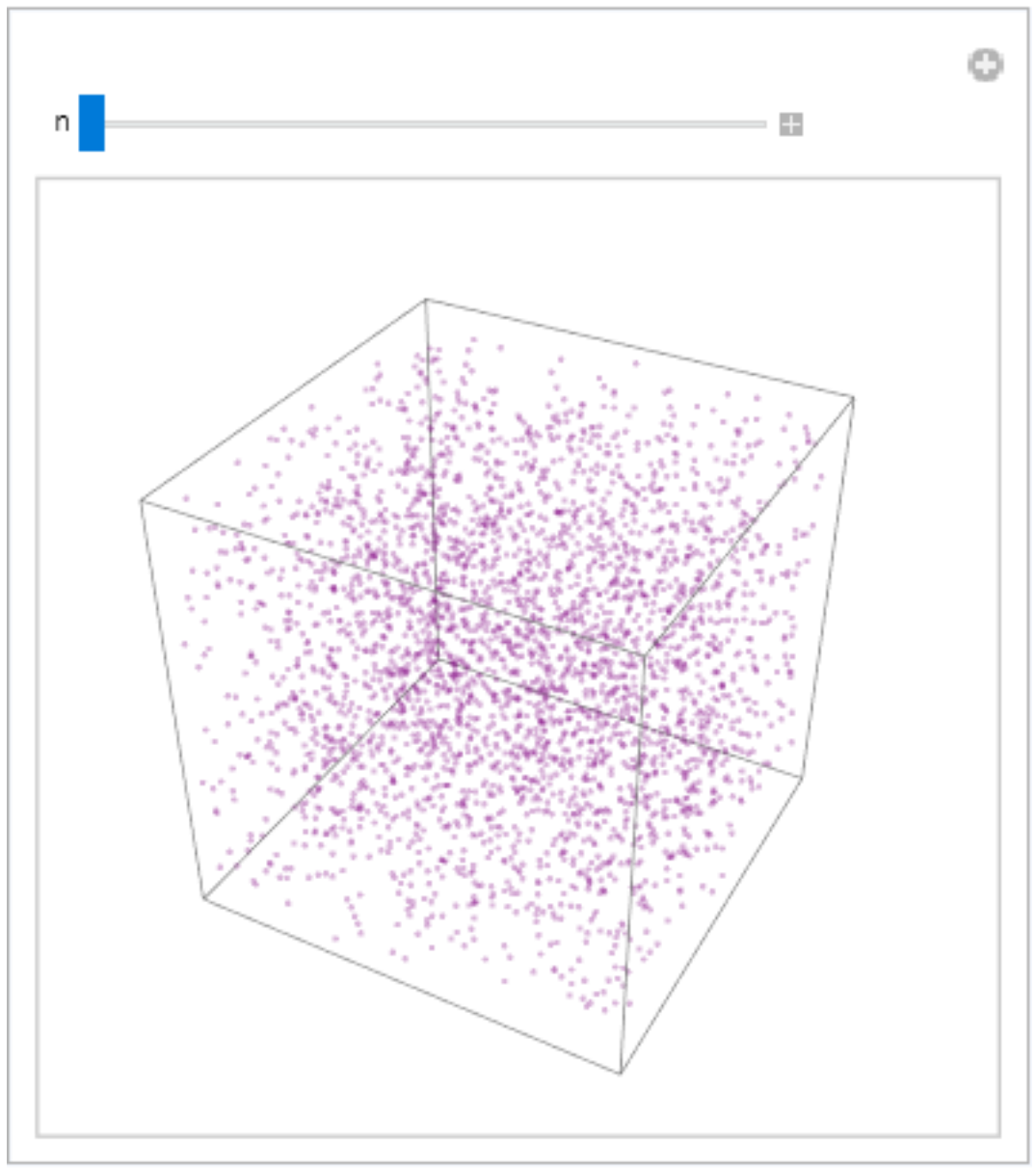

*Mathematica Examples 17.17*

Input

```
(* Assume we have two populations with means μ1=10 and μ2=12, and standard deviations
σ1=2 and σ2=3. We will take a sample of size n=30 from each population and calculate
the difference between the sample means: *)

μ1=10;
μ2=12;
σ1=2;
σ2=3;
n=30;

SeedRandom[1234];

sample1=RandomVariate[NormalDistribution[μ1,σ1],n];
sample2=RandomVariate[NormalDistribution[μ2,σ2],n];

difference=Mean[sample1]-Mean[sample2];

(* We can repeat this process multiple times to create a distribution of
differences between sample means: *)
numTrials=10000;
differences=Table[
    sample1=RandomVariate[NormalDistribution[μ1,σ1],n];
    sample2=RandomVariate[NormalDistribution[μ2,σ2],n];
    Mean[sample1]-Mean[sample2],
    {i,1,numTrials}
    ];

(* The mean of the Sampling Distribution of Differences of Means: *)
xbarofdifference=Mean[differences];
```

```
(* The mean of the distribution of differences can be calculated using the formula: *)
mean=μ1-μ2;

(* The standard deviation of the distribution of differences can be calculated using the formula: *)
standardError=Sqrt[(σ1^2+σ2^2)/n];

samplingDist=NormalDistribution[mean,standardError];

(* The code creates a histogram of the distribution of differences between sample means. The histogram is approximately normal, as expected based on the CLT. The mean of the distribution is close to the true difference between the population means (μ1-μ2): *)

Show[
 Histogram[
  differences,
  50,
  "PDF",
  PlotLabel->"Sampling Distribution of Differences of Means",
  Epilog->{
    Text["μ1 = 10, σ1 = 2",{0,0.6}],
    Text["μ2 = 12, σ2 = 3",{0,0.5}],
    Text["n = 30",{0,0.4}],
    Text["Difference = "<>ToString[xbarofdifference],{-0.2,0.3}]
    },
  ColorFunction->Function[Opacity[0.7]],
  ChartStyle->Purple,
  ImageSize->300
  ],
 Plot[
  PDF[samplingDist,x],
  {x,-6,1},
  PlotStyle->RGBColor[0.88,0.61,0.14]
  ]
 ]
```

Output

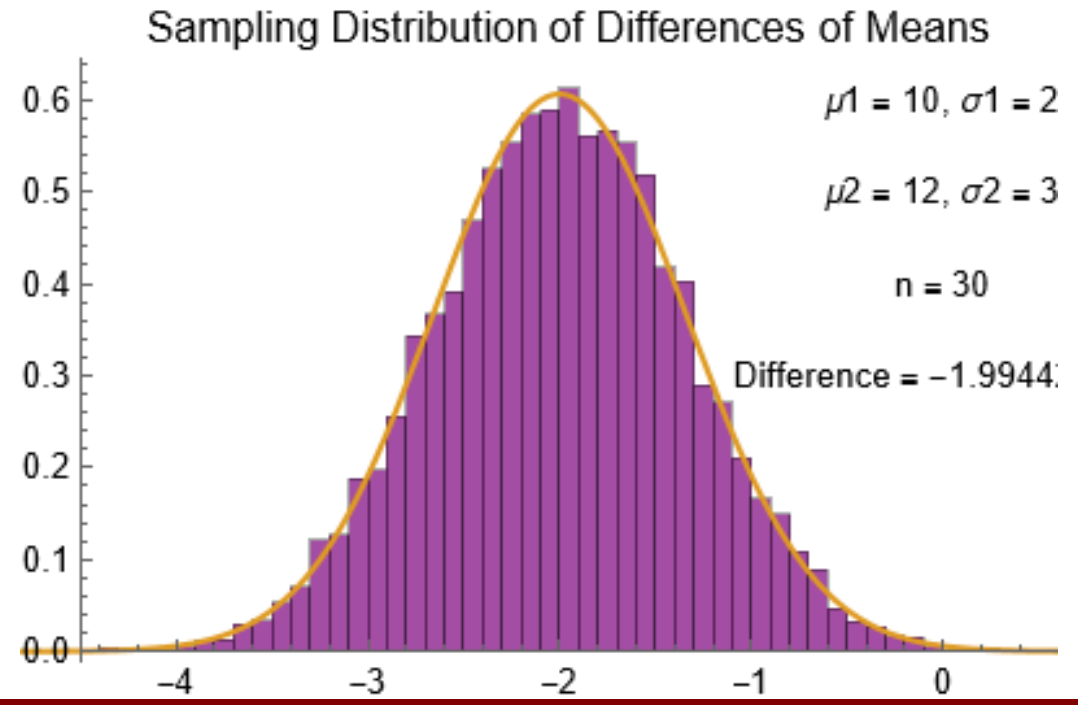

### *Mathematica Examples 17.18*

Input

```
(* This code calculates and visualizes the differences between sample means from two populations and provides histograms, plots, and sliders to manipulate the parameters. The code generates random samples from two normal distributions and computes the mean difference for each trial. It then calculates the sample mean of the differences and the standard error. Histograms and plots are displayed to illustrate the differences, sampling distribution, and sample means. The Manipulate function enables users to adjust parameters such as population means, standard deviations, and sample size,
```

```
dynamically updating the visualizations. This allows users to observe how changes in
the input values impact the resulting distributions and statistics: *)

Manipulate[
 SeedRandom[1234];
 numTrials=10000;
 differences=Table[
   sample1=RandomVariate[NormalDistribution[μ1,σ1],n];
   sample2=RandomVariate[NormalDistribution[μ2,σ2],n];
   Mean[sample1]-Mean[sample2],
   {i,1,numTrials}
   ];

 xbarofdifference=Mean[differences];
 standardError=Sqrt[(σ1^2+σ2^2)/n];

 samplingDist=NormalDistribution[μ1-μ2,standardError];
 sampledata1=Table[
   Mean[RandomVariate[NormalDistribution[μ1,σ1],n]],
   {i,1,numTrials}
   ];
 sampledata2=Table[
   Mean[RandomVariate[NormalDistribution[μ2,σ2],n]],
   {i,1,numTrials}
   ];

 Show[
  Histogram[
   differences,
   {0.01},
   "PDF",
   PlotRange->{{-10,10},{0,1}},
   PlotLabel->{
     Text["μ1 = "<>ToString[μ1]<>", σ1 = "<>ToString[σ1]],
     Text["μ2 = "<>ToString[μ2]<>", σ2 = "<>ToString[σ2]],
     Text["n = "<>ToString[n]],
     Text["Difference = "<>ToString[xbarofdifference]]
     },
   ColorFunction->Function[Opacity[0.7]],
   ChartStyle->Purple
   ],
  Plot[
   PDF[samplingDist,x],
   {x,-10,10},
   PlotRange->{{-10,10},{0,1}},
   PlotStyle->RGBColor[0.88,0.61,0.14]
   ],
  Histogram[
   sampledata1,
   {0.01},
   "PDF",
   PlotRange->{{-10,10},{0,1}},
   ColorFunction->Function[Opacity[0.5]],
   ChartStyle->Blue
   ],
  Histogram[
   sampledata2,
   {0.01},
   "PDF",
   PlotRange->{{-10,10},{0,1}},
   ColorFunction->Function[Opacity[0.5]],
```

```
        ChartStyle->Red
        ]
      ],
    {{μ1,2,"Mean 1"},2,7,0.1},
    {{μ2,5,"Mean 2"},3,7,0.1},
    {{σ1,3,"SD 1"},3,5,0.1},
    {{σ2,3,"SD 2"},3,5,0.1},
    {{n,30,"Sample Size"},10,50,10}
    ]
```

Output

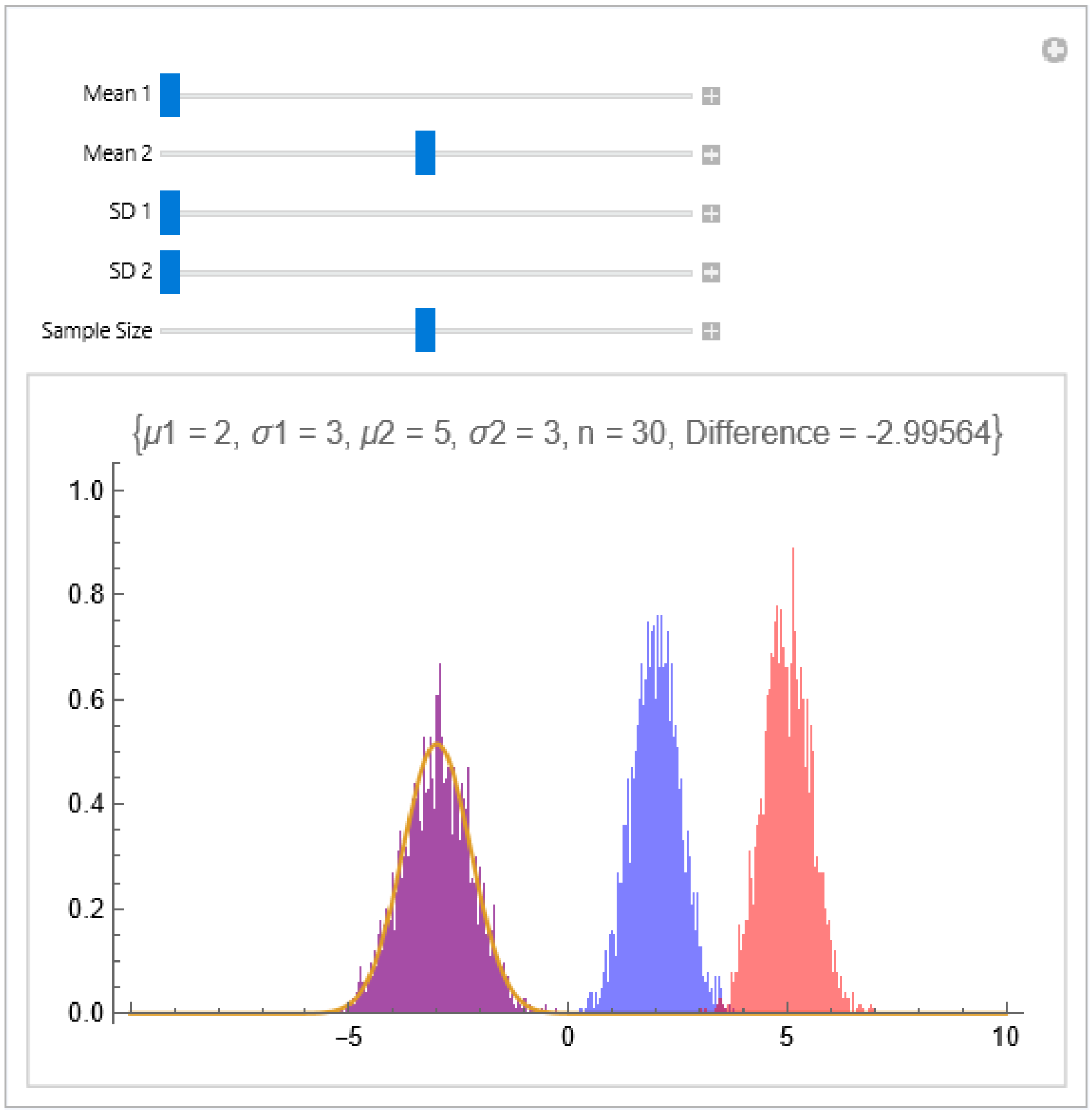

# UNIT 17.2

# CHI-SQUARE DISTRIBUTION

### Mathematica Examples 17.19

Input
```
(* This code generates a plot of the probability density function (PDF) for a chi
square distribution with different values of the parameter v=(1, 3, 5, 10). The PDF
is evaluated at various values of the random variable x between 0 and 20: *)

Plot[
 Evaluate[
  Table[
   PDF[
    ChiSquareDistribution[v],
    x
    ],
   {v,{1,3,5,10}}
   ]
  ],
 {x,0,20},
 PlotRange->{0,0.3},
 Filling->Axis,
 PlotLegends->Placed[{"v=1","v=3","v=5", "v=10"},{0.5,0.75}],
 PlotStyle->{RGBColor[0.88,0.61,0.14],RGBColor[0.37,0.5,0.7],Purple, Darker[Red]},
 ImageSize->320,
 AxesLabel->{None,"PDF"}
 ]
```

Output

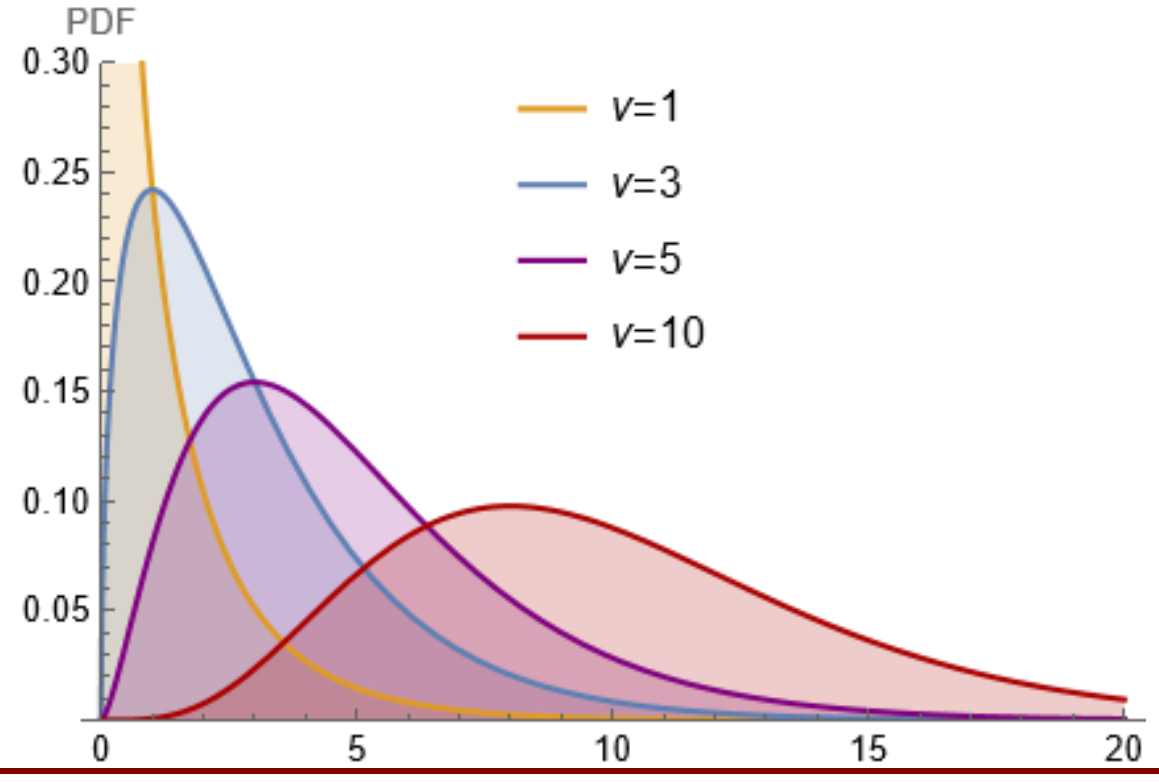

### Mathematica Examples 17.20

Input
```
(* The code generates a plot of the cumulative distribution function (CDF) of the
chi square distribution with different values of the parameter v= (1, 3, 5, 10). The
CDF is evaluated at various values of the random variable x between 0 and 20: *)

Plot[
 Evaluate[
  Table[
   CDF[
    ChiSquareDistribution[v],
    x
    ],
```

```
      {v,{1,3,5,10}}
      ]
     ],
    {x,0,20},
    PlotRange->{0,1},
    Filling->Axis,
    PlotLegends->Placed[{"v=1","v=3","v=5", "v=10"},{0.25,0.75}],
    PlotStyle->{RGBColor[0.88,0.61,0.14],RGBColor[0.37,0.5,0.7],Purple, Darker[Red]},
    ImageSize->320,
    AxesLabel->{None,"CDF"}
    ]
```

Output

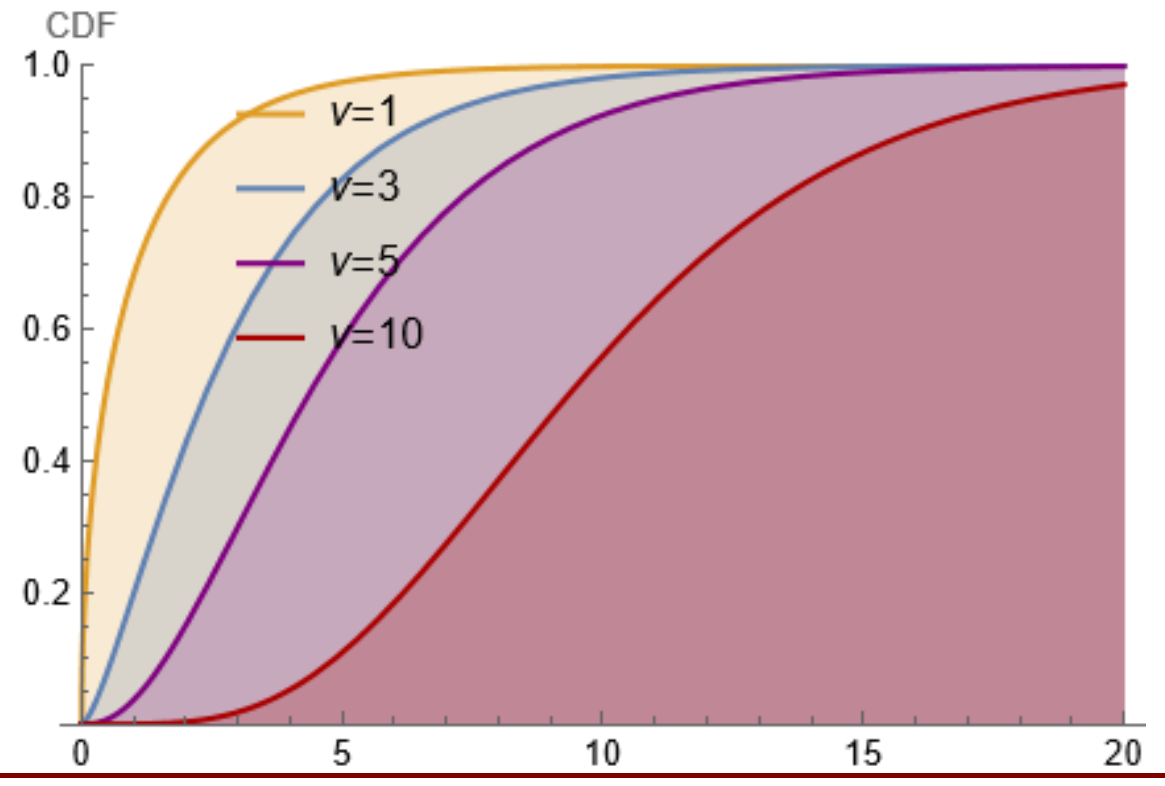

*Mathematica Examples 17.21*

Input

```
(* The code generates a histogram and a plot of the PDF for a chi square distribution
with parameters v=3 and sample size 10000: *)

data=RandomVariate[
    ChiSquareDistribution[3],
    10^4
    ];

Show[
 Histogram[
  data,
  {0,6,0.1},
  "PDF",
  ColorFunction->Function[{height},Opacity[height]],
  ChartStyle->Purple,
  ImageSize->320,
  AxesLabel->{None,"PDF"}
  ],

 Plot[
  PDF[
   ChiSquareDistribution[3],
   x
   ],
  {x,0,6},
  PlotStyle->RGBColor[0.88,0.61,0.14],
  PlotRange->{0,4}
  ]
 ]
```

Output

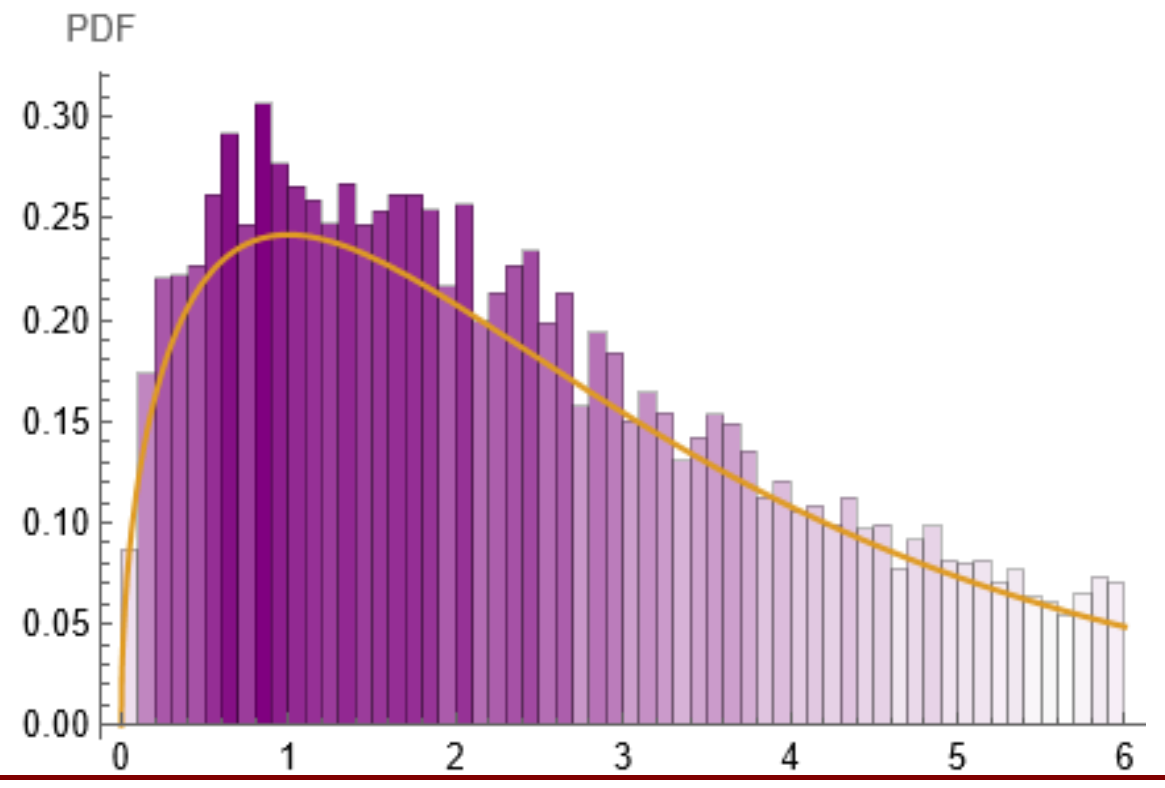

**Mathematica Examples 17.22**

Input

```
(* The code creates a dynamic histogram of data and a plot of the PDF generated from
a chi square distribution using the Manipulate function. The Manipulate function
creates interactive controls for the user to adjust the values of v and n, which are
the parameter of the chi square distribution and the sample size: *)

Manipulate[
 Module[
  {
   data=RandomVariate[
     ChiSquareDistribution[v],
     n
     ]
   },

  Show[
   Histogram[
    data,
    Automatic,
    "PDF",
    PlotRange->{{0,150},Automatic},
    ColorFunction->Function[{height},Opacity[height]],
    ImageSize->320,
    ChartStyle->Purple
    ],
   Plot[
    PDF[
     ChiSquareDistribution[v],
     x
     ],
    {x,0,150},
    PlotRange->Automatic,
    ColorFunction->"Rainbow"
    ]
   ]
  ],
 {{v,1,"v"},1,100,0.1},
 {{n,300,"n"},100,1000,10}
 ]
```

Output

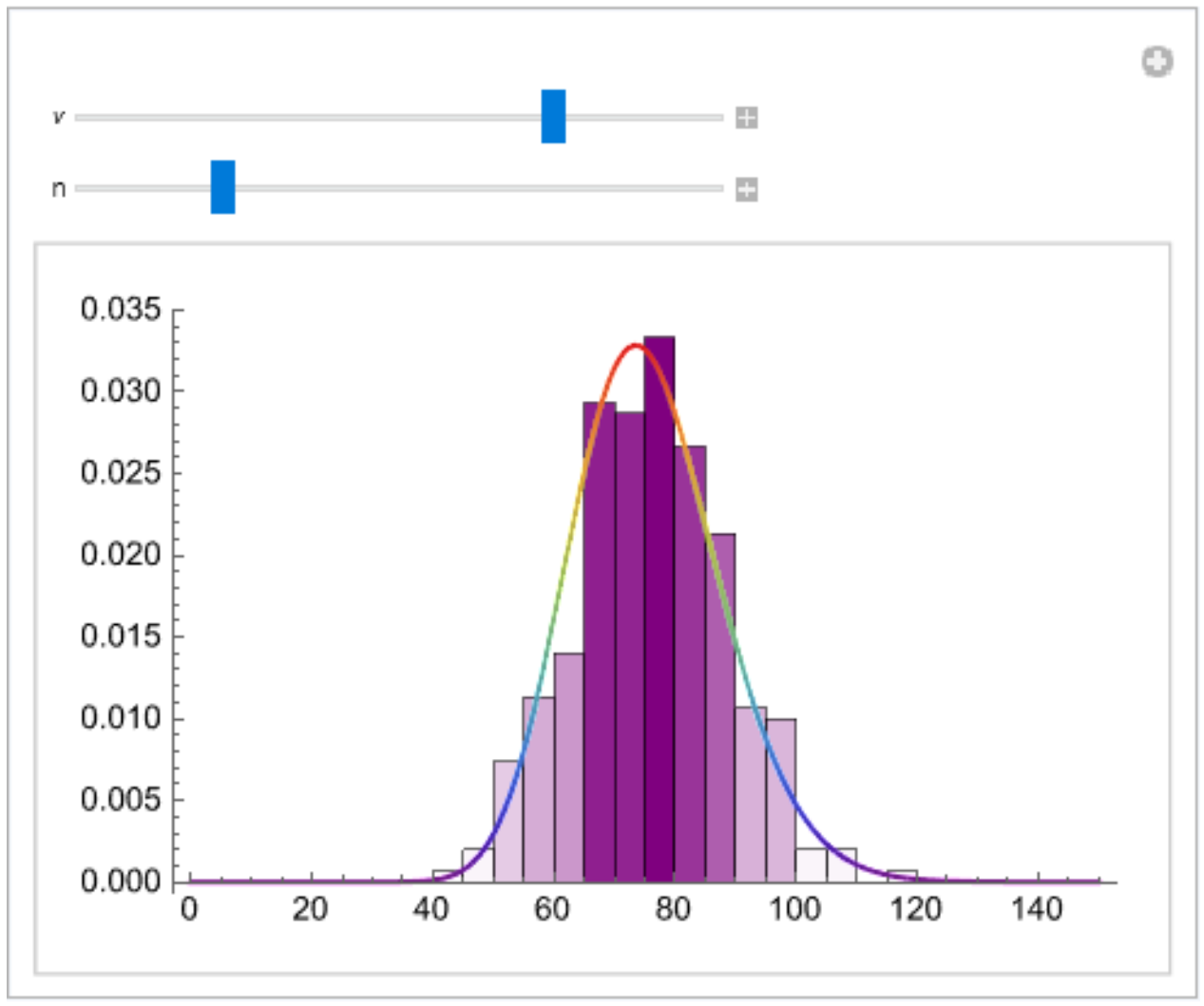

***Mathematica Examples 17.23***

Input

```
(* The code creates a plot of the CDF of a Chi Square distribution using the Manipulate function. The Manipulate function allows you to interactively change the values of the parameters v: *)
Manipulate[
 Plot[
  CDF[
   ChiSquareDistribution[v],
   x
   ],
  {x,0,7},
  Filling->Axis,
  FillingStyle->LightPurple,
  PlotRange->Automatic,
  AxesLabel->{"x","CDF"},
  ImageSize->320,
  PlotStyle->Purple,
  PlotLabel->Row[{"v = ",v}]
  ],
 {{v,1},1,6,0.1}
 ]
```

Output

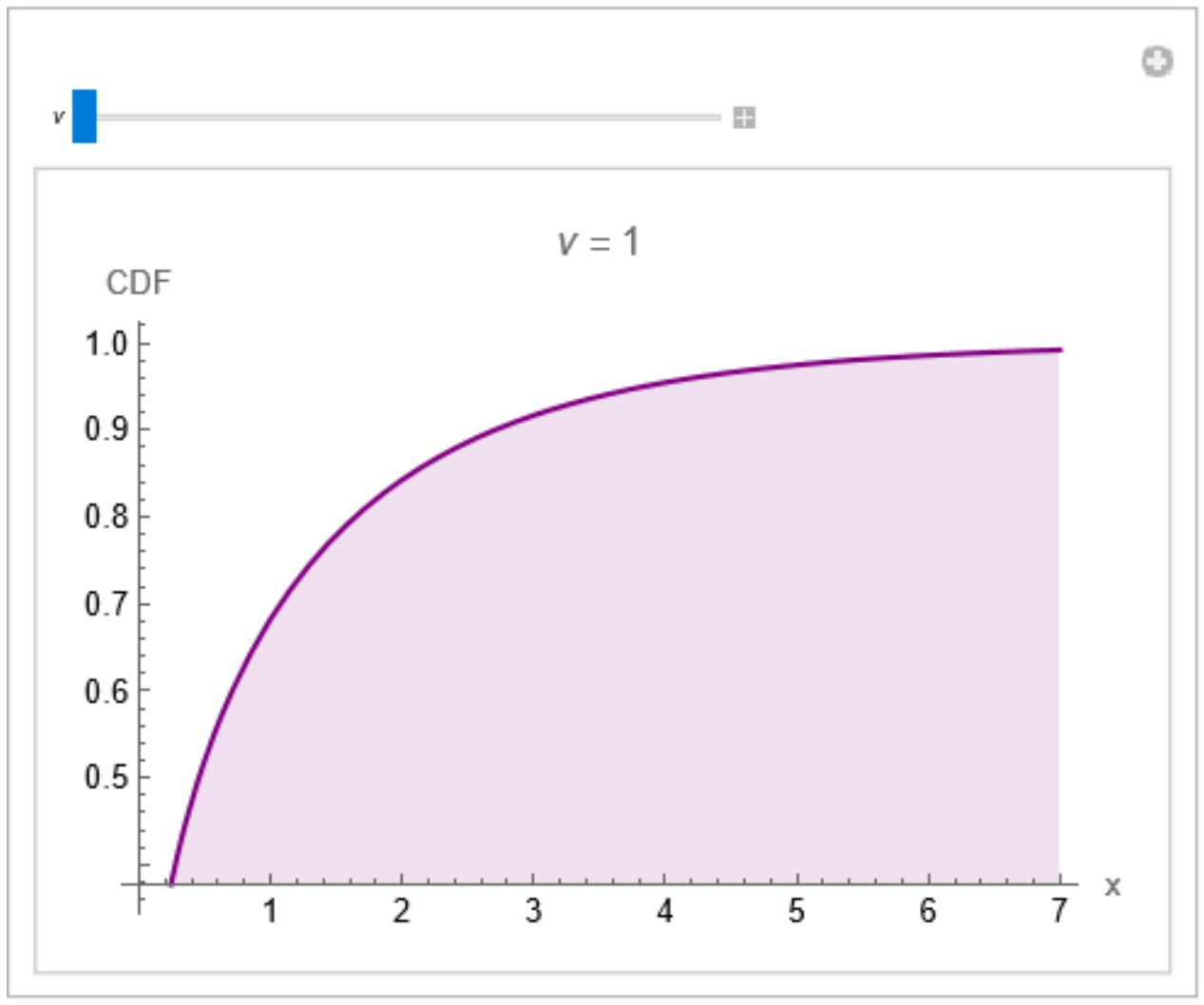

## *Mathematica Examples 17.24*

Input

```
(* The code uses the Grid function to create a grid of two plots, one for the PDF and one for the CDF of chi square distribution. The code uses slider controls to adjust the values of v: *)

Manipulate[
 Grid[
  {
   {Plot[
     PDF[
      ChiSquareDistribution[v],
      x
      ],
     {x,0,20},
     PlotRange->Automatic,
     PlotStyle->{Purple,PointSize[0.03]},
     PlotLabel->"PDF of Chi Square distribution",
     AxesLabel->{"x","PDF"}
     ],
    Plot[
     CDF[
      ChiSquareDistribution[v],
      x
      ],
     {x,0,20},
     PlotRange->Automatic,
     PlotStyle->{Purple,PointSize[0.03]},
     PlotLabel->"CDF of Chi Square distribution",
     AxesLabel->{"x","CDF"}
     ]
    }
   },
  Spacings->{5,5}
  ],
 {{v,1},1,6,0.1}
 ]
```

Output

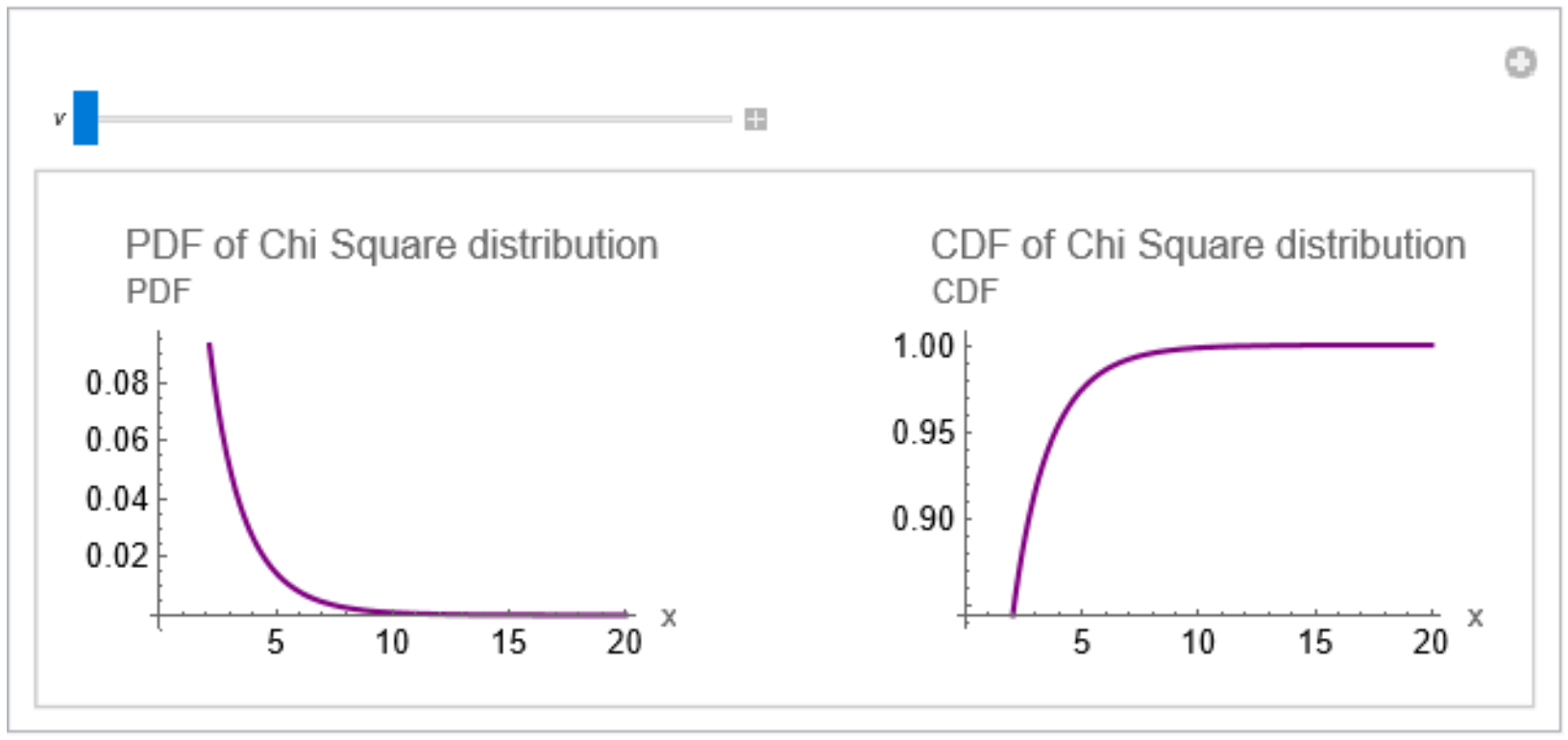

## *Mathematica Examples 17.25*

Input

```
(* The code calculates and displays some descriptive statistics (mean, variance, standard deviation, kurtosis and skewness) for a chi square distribution with parameter v: *)

Grid[
 Table[
  {
```

```
      statistics,
      FullSimplify[statistics[ChiSquareDistribution[v]]]
      },
     {statistics,{Mean,Variance,StandardDeviation,Kurtosis,Skewness}}
     ],
    ItemStyle->12,
    Alignment->{{Right,Left}},
    Frame->All,
    Spacings->{Automatic,0.8}
    ]
```

Output

| | |
|---|---|
| Mean | v |
| Variance | 2 v |
| StandardDeviation | Sqrt[2] Sqrt[v] |
| Kurtosis | (3 (4+v))/v |
| Skewness | 2 Sqrt[2] Sqrt[1/ v] |

**Mathematica Examples 17.26**

Input

```
(* The code calculates and displays some additional descriptive statistics (moments, central moments, and factorial moments) for a chi square distribution with parameter v: *)

Grid[
 Table[
  {
   statistics,
   FullSimplify[statistics[ChiSquareDistribution[v],1]],
   FullSimplify[statistics[ChiSquareDistribution[v],2]]
   },
  {statistics,{Moment,CentralMoment,FactorialMoment}}
  ],
 ItemStyle->12,
 Alignment->{{Right,Left}},
 Frame->All,
 Spacings->{Automatic,0.8}
 ]
```

Output

| | | |
|---|---|---|
| Moment | v | v (2+v) |
| CentralMoment | 0 | 2 v |
| FactorialMoment | v | v (1+v) |

**Mathematica Examples 17.27**

Input

```
(* The code generates a dataset of 10000 observations from a chi square distribution with parameter v=3. Then, it computes the sample mean and quartiles of the data, and plots a histogram of the data and plot of the PDF. Additionally, the code adds vertical lines to the plot corresponding to the sample mean and quartiles: *)

data=RandomVariate[
    ChiSquareDistribution[3],
    10000
    ];

mean=Mean[data];
quartiles=Quantile[
    data,
    {0.25,0.5,0.75}
    ];

Show[
```

```
 Histogram[
  data,
  Automatic,
  "PDF",
  Epilog->{
    Directive[Red,Thickness[0.006]],
    Line[{{mean,0},{mean,0.25}}],
    Directive[Green,Dashed],
    Line[{{quartiles[[1]],0},{quartiles[[1]],0.25}}],
    Line[{{quartiles[[2]],0},{quartiles[[2]],0.25}}],
    Line[{{quartiles[[3]],0},{quartiles[[3]],0.25}}]
    },
  ColorFunction->Function[{height},Opacity[height]],
  ImageSize->320,
  ChartStyle->Purple,
  PlotRange->Automatic
  ],
 Plot[
  PDF[
   ChiSquareDistribution[3],
   x
   ],
  {x,0,20},
  PlotRange->Automatic,
  ImageSize->320,
  ColorFunction->"Rainbow"
  ]
 ]
```

Output

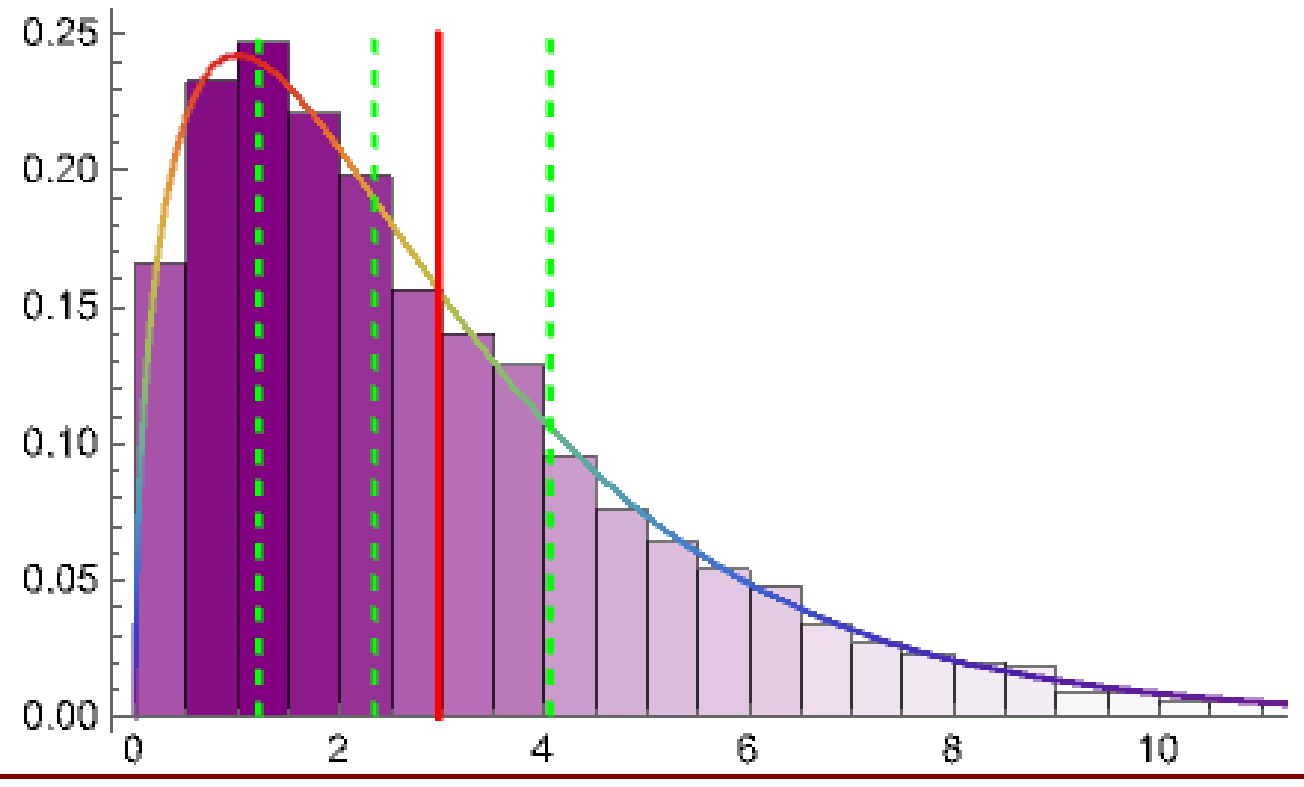

*Mathematica Examples 17.28*

Input

```
(* The code generates a random sample of size 10,000 from a chi square distribution
with parameter v=3, estimates the distribution parameters using the
EstimatedDistribution function, and then compares the histogram of the sample with
the estimated PDF of the chi Square distribution using a histogram and a plot of the
PDF: *)

sampledata=RandomVariate[
   ChiSquareDistribution[3],
   10^4
   ];
(* Estimate the distribution parameters from sample data: *)
ed=EstimatedDistribution[
  sampledata,
  ChiSquareDistribution[v]
  ]
```

```
(* Compare a density histogram of the sample with the PDF of the estimated
distribution: *)
Show[
 Histogram[
  sampledata,
  Automatic,
  "PDF",
  ColorFunction->Function[{height},Opacity[height]],
  ChartStyle->Purple,
  ImageSize->320
  ],
 Plot[
  PDF[ed,x],
  {x,0,20},
  ImageSize->320,
  ColorFunction->"Rainbow"
  ]
 ]
```

Output ChiSquareDistribution[2.99044]

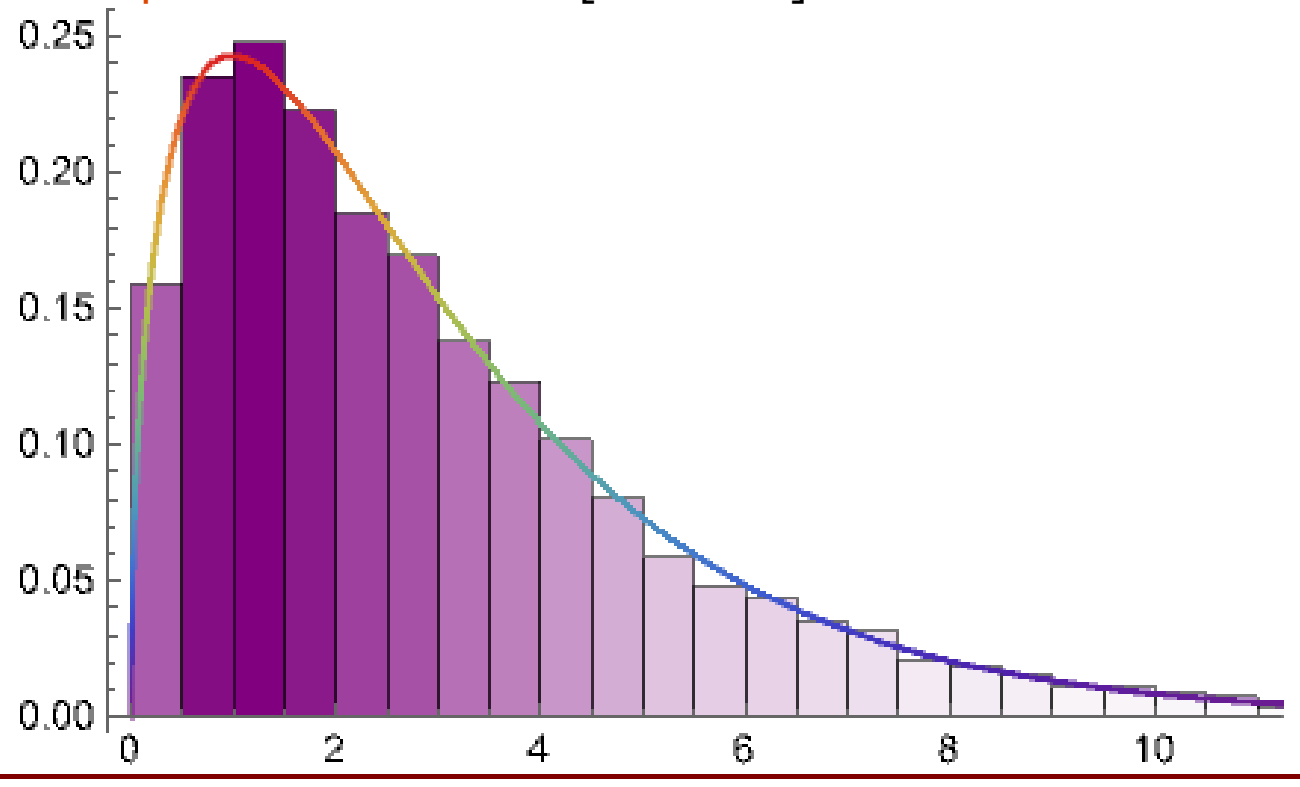

***Mathematica Examples 17.29***

Input

```
(* The code generates a 2D dataset with 1000 random points that follow a chi square
distribution with v=3. The dataset is then used to create a row of three plots. The
first plot is a histogram of the X-axis values of the dataset. The second plot is a
histogram of the Y-axis values of the dataset. It is similar to the first plot, but
shows the distribution of the Y-axis values instead. The third plot is a scatter plot
of the dataset, with the X-axis values on the horizontal axis and the Y-axis values
on the vertical axis. Each point in the plot represents a pair of X and Y values from
the dataset: *)

data=RandomVariate[
    ChiSquareDistribution[3],
    {1000,2}
    ];
GraphicsRow[
 {
  Histogram[
   data[[All,1]],
   {0.1},
   ImageSize->170,
   PlotLabel->"X-axis",
   ColorFunction->Function[{height},Opacity[height]],
   ChartStyle->Purple
   ],
  Histogram[
   data[[All,2]],
```

```
            {0.1},
            ImageSize->170,
            PlotLabel->"Y-axis",
            ColorFunction->Function[{height},Opacity[height]],
            ChartStyle->Purple
            ],
           ListPlot[
            data,
            ImageSize->170,
            PlotStyle->{Purple,PointSize[0.015]},
            AspectRatio->1,
            Frame->True,
            Axes->False
            ]
           }
          ]
```

Output

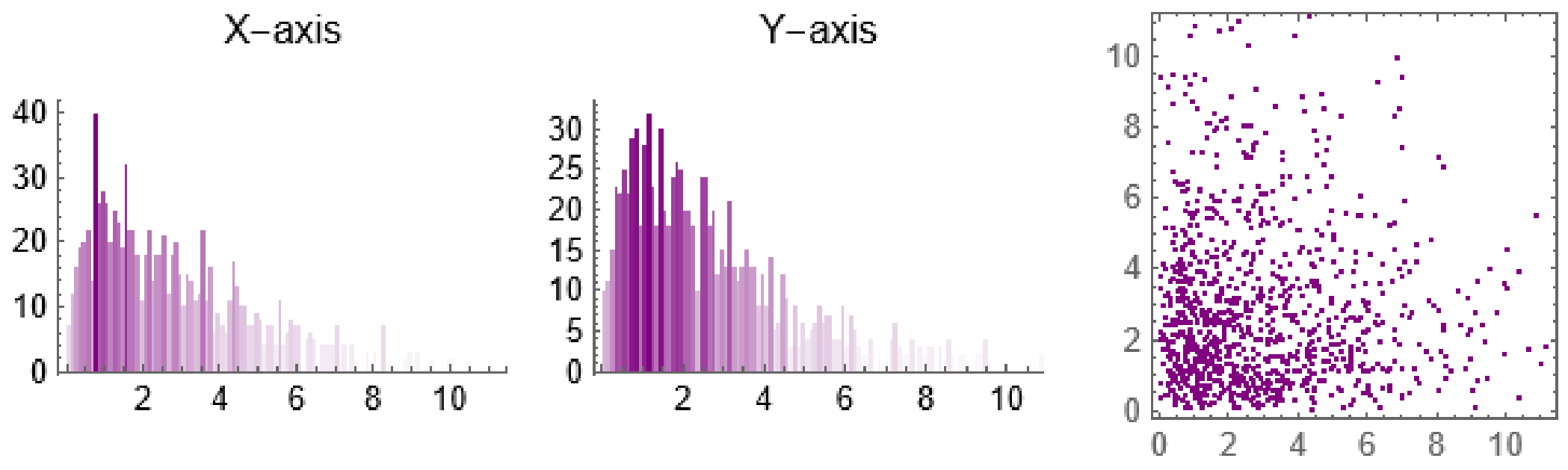

***Mathematica Examples 17.30***

Input

```
(* The code generates a set of random data points with a chi square distribution with
v=3 in three dimensions, and then creates three histograms, one for each dimension,
showing the distribution of the points along that axis. Additionally, it creates a
3D scatter plot of the data points: *)

data=RandomVariate[
    ChiSquareDistribution[3],
    {1000,3}
    ];

GraphicsGrid[
 {
  {
   Histogram[
    data[[All,1]],
    Automatic,
    "PDF",
    PlotLabel->"X-axis",
    ColorFunction->Function[{height},Opacity[height]],
    ChartStyle->Purple
    ],
   Histogram[
    data[[All,2]],
    Automatic,
    "PDF",
    PlotLabel->"Y-axis",
    ColorFunction->Function[{height},Opacity[height]],
    ChartStyle->Purple
```

```
          ],
         Histogram[
          data[[All,3]],
          Automatic,
          "PDF",
          PlotLabel->"Z-axis",
          ColorFunction->Function[{height},Opacity[height]],
          ChartStyle->Purple
          ],
         ListPointPlot3D[
          data,
          BoxRatios->{1,1,1},
          PlotStyle->{Purple,PointSize[0.015]}
          ]
         }
        }
       ]
```

Output

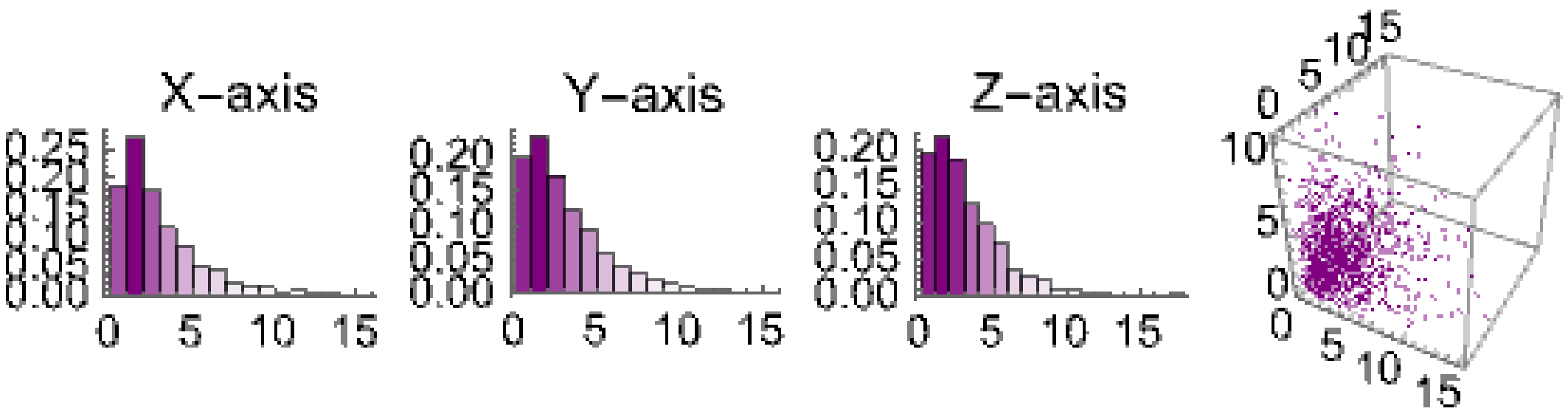

*Mathematica Examples 17.31*

Input

```
(* The code generates a 3D scatter plot of a chi square distribution points with v=3,
where the x-axis is red, y-axis is green, and z-axis is blue: *)

data=RandomVariate[
   ChiSquareDistribution[3],
   {2000,3}
   ];
Graphics3D[
 {
  {PointSize[0.006],Purple,Opacity[0.6],Point[data]},
  Thin,
  {Red,Opacity[0.4],Line[{{#,0,0},{#,0,-0.9}}]&/@data[[All,1]]},
  Thin,
  {Green,Opacity[0.4],Line[{{0,#,0},{0,#,-0.9}}]&/@data[[All,2]]},
  Thin,
  {Blue,Opacity[0.4],Line[{{0,0,#},{0,-0.9,#}}]&/@data[[All,3]]}
  },
 BoxRatios->{1,1,1},
 Axes->True,
 AxesLabel->{"X","Y","Z"},
 ImageSize->320
 ]
```

Output

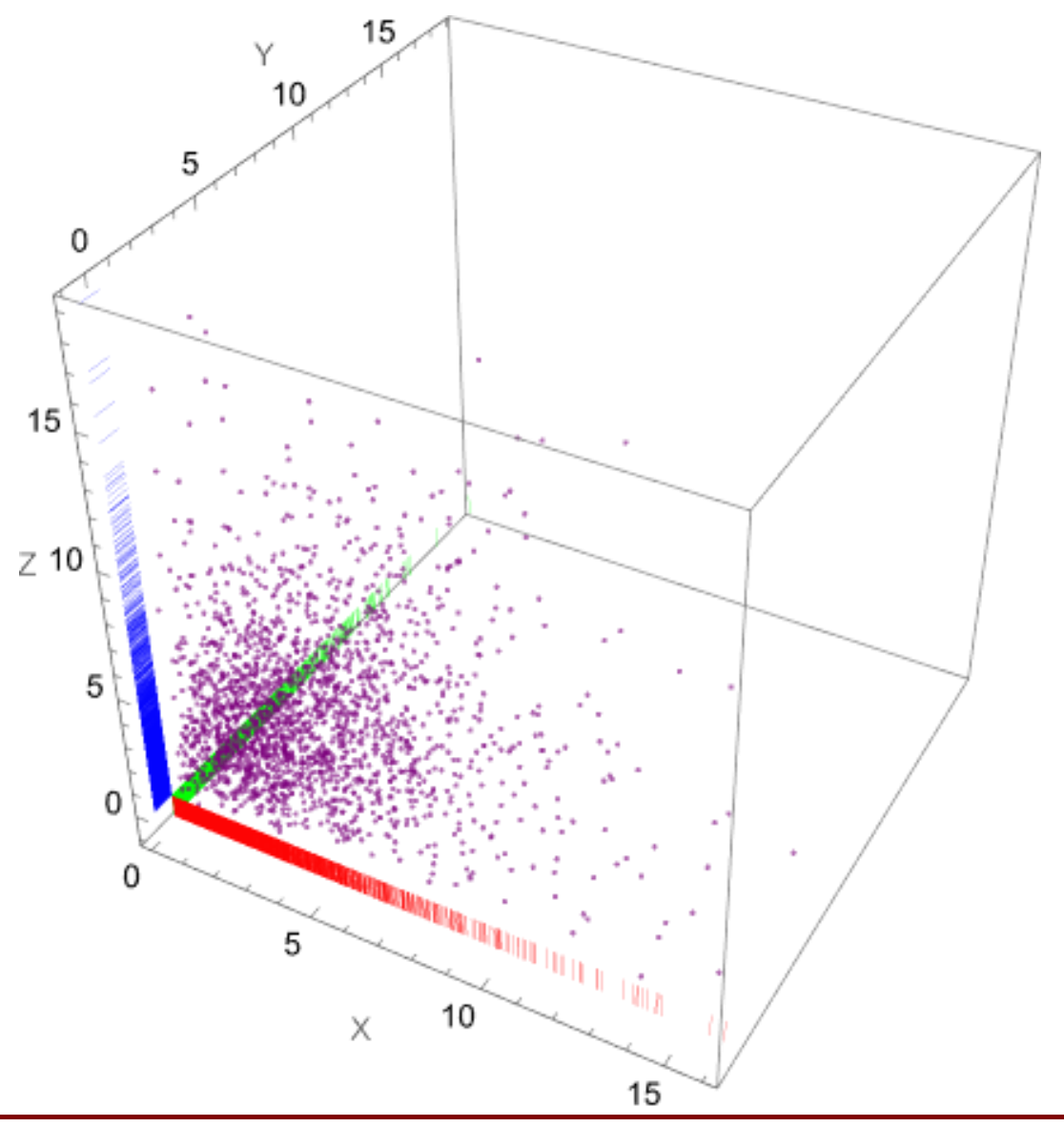

***Mathematica Examples 17.32***

Input

```
(* The code demonstrates a common technique in statistics and data analysis, which
is the use of random sampling to estimate population parameters. The code generates
random samples from chi square distribution with v=3, and then using these samples
to estimate the parameters of another chi square distribution with unknown v. This
process is repeated 20 times, resulting in 20 different estimated distributions. The
code also visualizes the resulting estimated distributions using the PDF function.
The code plots the PDFs of these estimated distributions using the PDF function and
the estimated parameters. The plot shows the PDFs in a range from-0 to 20. The code
also generates a list plot of 2 sets of random samples from the chi square distribution
with v=3. It shows the 100 random points generated from two random samples. The code
generates also a histogram of the PDF for Chi Square distribution of the two samples:
*)

estim0distributions=Table[
  dist=ChiSquareDistribution[3];

  sampledata=RandomVariate[
    dist,
    100
    ];

  ed=EstimatedDistribution[
    sampledata,
    ChiSquareDistribution[v]
    ],
  {i,1,20}
  ]

pdf0ed=Table[
   PDF[estim0distributions[[i]],x],
   {i,1,20}
   ];

(* Visualizes the resulting estimated distributions: *)
Plot[
 pdf0ed,
 {x,0,20},
 PlotRange->Full,
```

```
 ImageSize->400,
 PlotStyle->Directive[Purple,Opacity[0.3],Thickness[0.002]]
 ]

(* Visualizes 100 random points generated from two random samples: *)
table =Table[
   dist=ChiSquareDistribution[3];
   sampledata=RandomVariate[
     dist,
     100],
   {i,1,2}
   ];

ListPlot[
 table,
 ImageSize->320,
 Filling->Axis,
 PlotStyle->Directive[Opacity[0.5],Thickness[0.003]]
 ]

Histogram[
 table,
 30,
 LabelingFunction->Above,
 ChartLegends->{"Sample 1","Sample 2"},
 ChartStyle->{Directive[Opacity[0.2],Red],Directive[Opacity[0.2],Purple]},
 ImageSize->320
 ]
```

Output

```
{ChiSquareDistribution[3.24012],ChiSquareDistribution[3.22257],ChiSquareDistributio
n[3.06599],ChiSquareDistribution[3.24458],ChiSquareDistribution[2.81057],ChiSquareD
istribution[2.78842],ChiSquareDistribution[2.67399],ChiSquareDistribution[2.63232],
ChiSquareDistribution[2.95929],ChiSquareDistribution[3.04553],ChiSquareDistribution
[2.61087],ChiSquareDistribution[2.89354],ChiSquareDistribution[2.77373],ChiSquareDi
stribution[2.83582],ChiSquareDistribution[3.08538],ChiSquareDistribution[2.6934],Ch
iSquareDistribution[2.80758],ChiSquareDistribution[3.18608],ChiSquareDistribution[2
.93024],ChiSquareDistribution[3.12473]}
```

Output

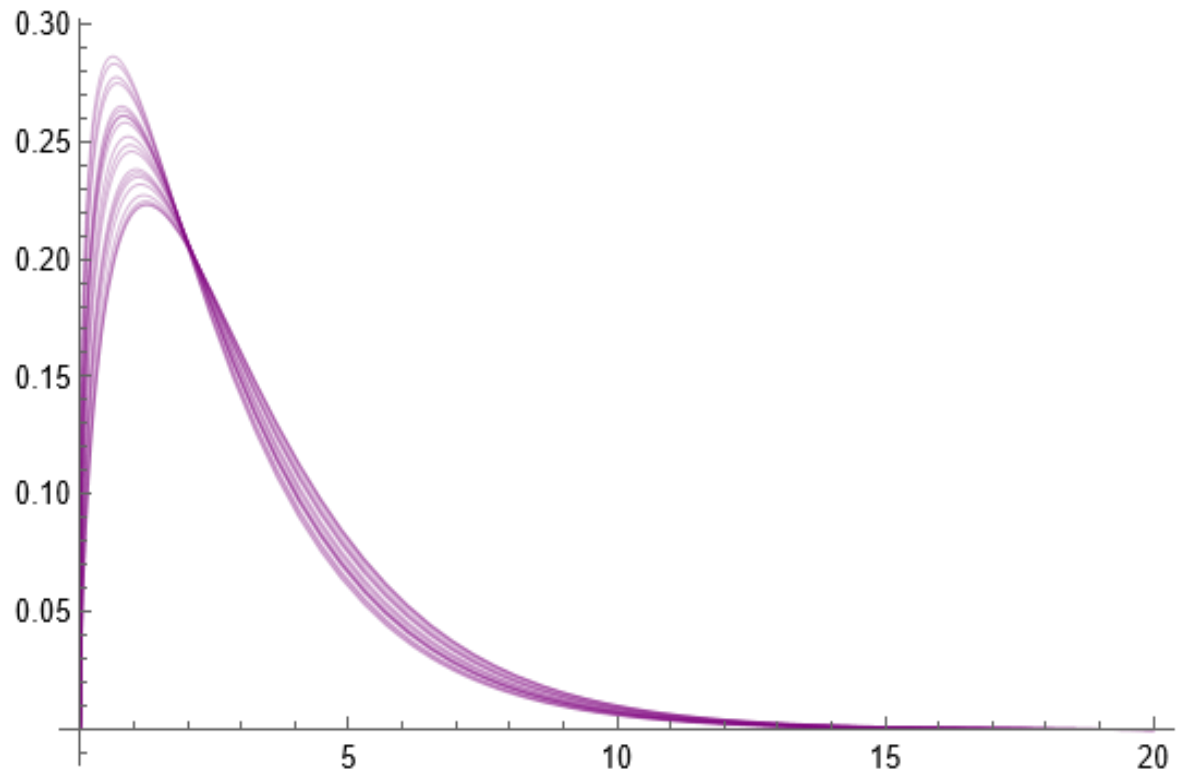

Output

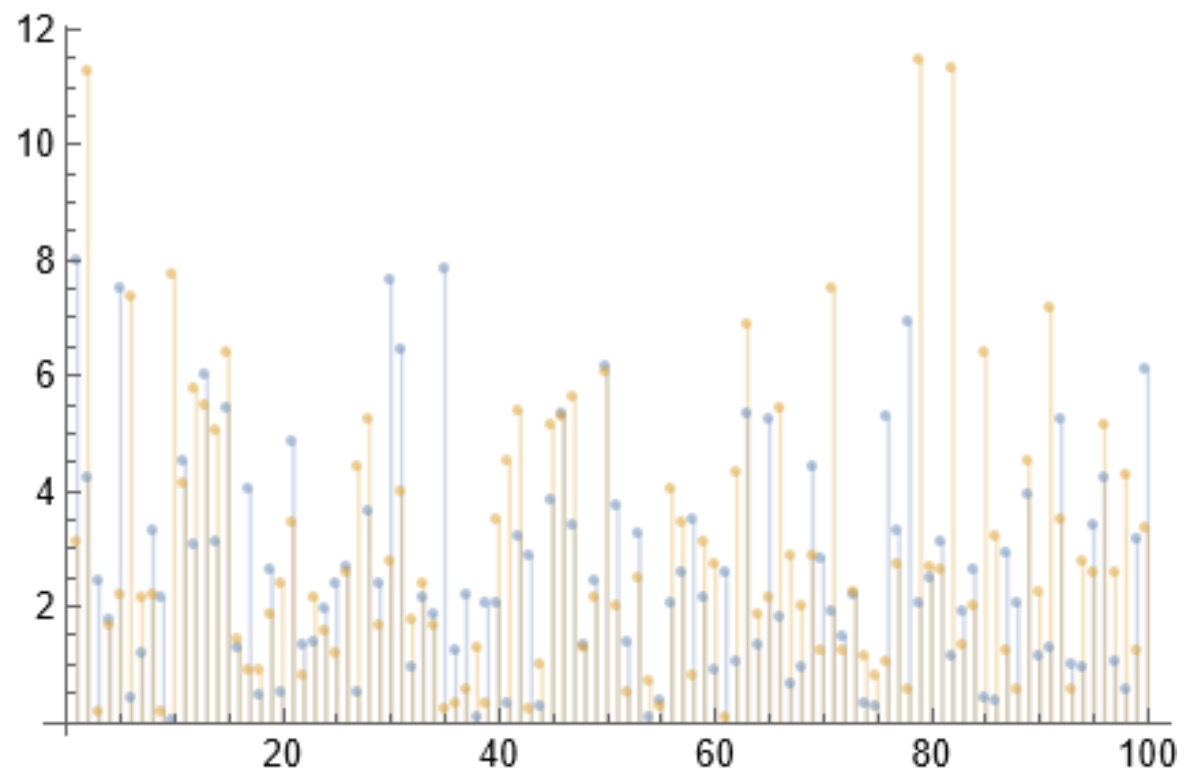

Output

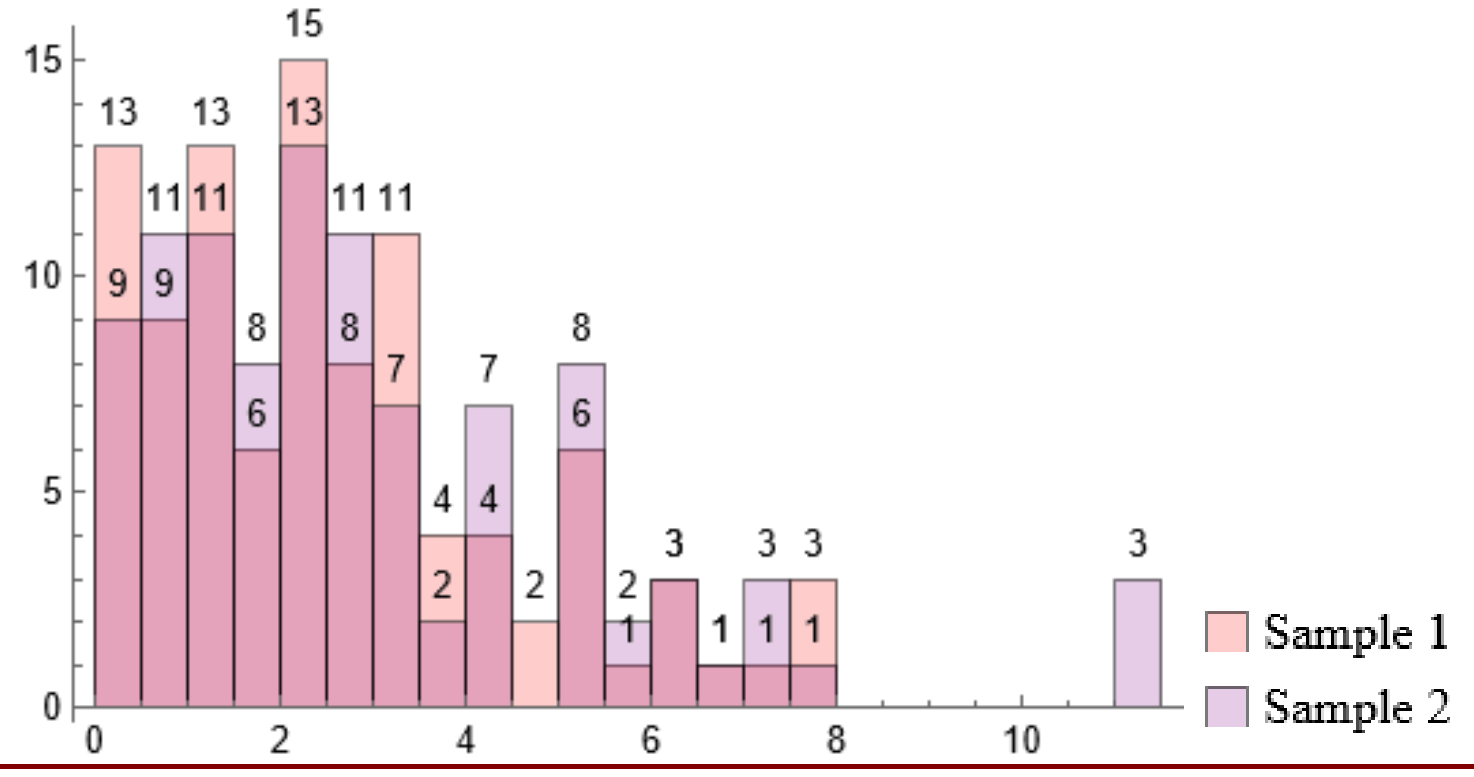

***Mathematica Examples 17.33***

Input
```
(* The code generates and compares the means of random samples drawn from a chi
square distribution with the given parameter v. The code uses the Manipulate function
to create a user interface with sliders to adjust the values of the parameter v,
number of samples, and sample size. By varying the values of "Number of Samples" and
"Sample Size" sliders, the code allows the user to explore how changing these
parameters affects the means of the random samples: *)

Manipulate[
 Module[
  {means,dist},

  dist=ChiSquareDistribution[v];

  means=Map[
    Mean,
    RandomVariate[dist,{n,samples}]
    ];
  m=N[Mean[means]];

  ListPlot[
   {means,{{0,m},{n,m}}},
   Joined->{False,True},
   Filling->Axis,
   PlotRange->{{1,50},{0,10}},
   PlotStyle->{Purple,Red},
   AxesLabel->{"Number of Samples","Sample Mean"},
   PlotLabel->Row[{"v = ",v}],
   ImageSize->320
   ]
  ],
 {{v,2,"v"},0.6,4,0.1},
```

```
{{n,50,"Number of Samples"},3,50,1},
{{samples,50,"Sample Size"},1,100,1},
TrackedSymbols:>{n,samples,v}
]
```

Output

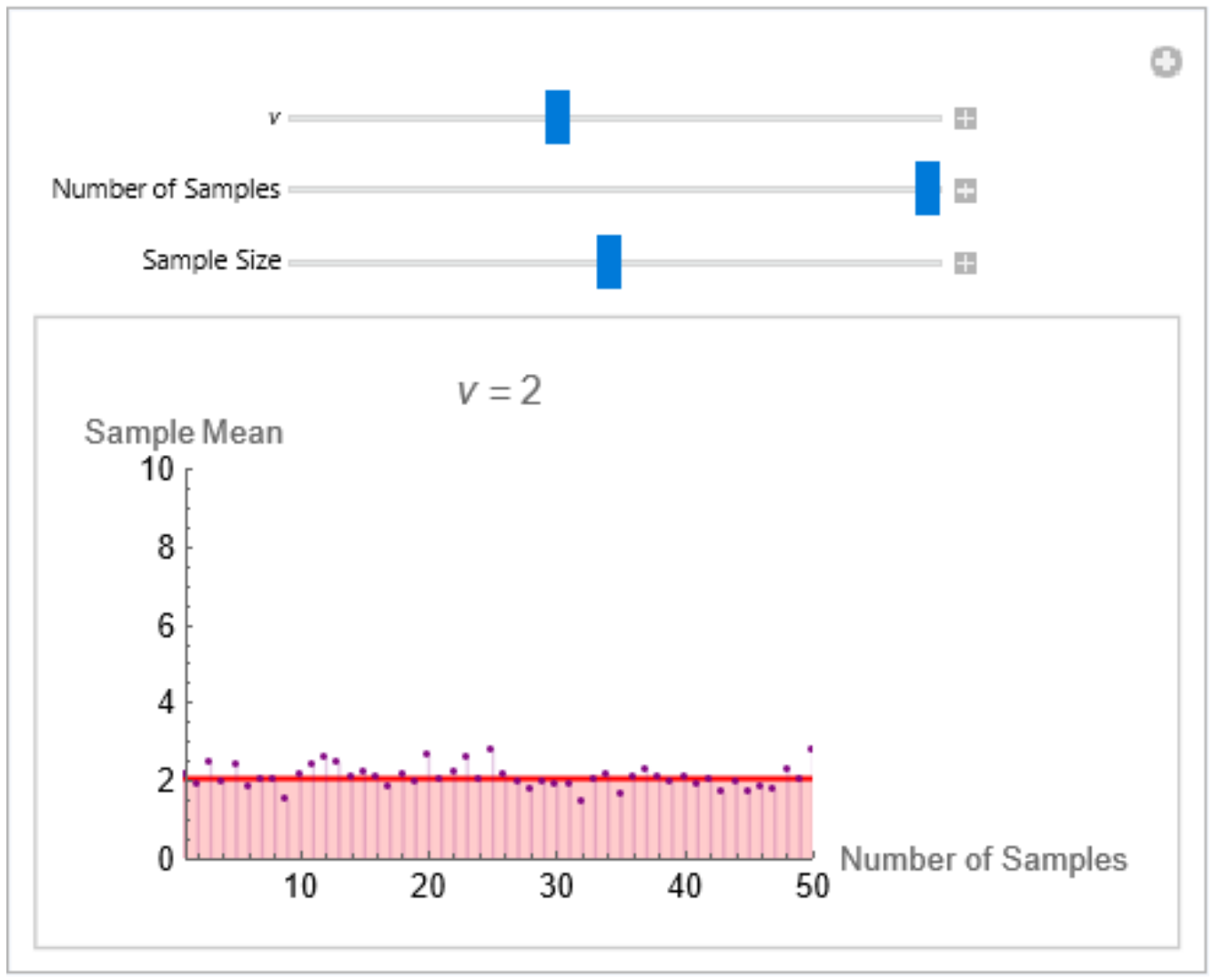

***Mathematica Examples 17.34***

Input

```
(* The code is designed to compare two chi Square distributions. It does this by
generating random samples from each distribution and displaying them in a histogram,
as well as plotting the probability density functions of the two distributions. The
code allows the user to manipulate the parameters v1 and v2 of both chi square
distributions through the sliders for v1 and v2. By changing these parameters, the
user can see how the distributions change and how they compare to each other. The
histograms display the sample data for each distribution, with the first histogram
showing the sample data for the first chi square distribution and the second histogram
showing the sample data for the second chi square distribution. The histograms are
overlaid on each other, with the opacity of each histogram set to 0.2 to make it
easier to see where the data overlap. The probability density functions of the two
distributions are also plotted on the same graph, with the first distribution shown
in blue and the second distribution shown in red. The legend indicates which color
corresponds to which distribution. By looking at the histograms and the probability
density functions, the user can compare the two chi square distributions and see how
they differ in terms of shape, scale, and overlap of their sample data: *)
Manipulate[
 Module[
  {dist1,dist2,data1,data2},
  SeedRandom[seed];
  dist1=ChiSquareDistribution[v1];
  dist2=ChiSquareDistribution[v2];
  data1=RandomVariate[dist1,n];
  data2=RandomVariate[dist2,n];
  Column[
   {
    Show[
     ListPlot[
      data1,
      ImageSize->320,
      PlotStyle->Blue,
      PlotRange->All
      ],
     ListPlot[
      data2,
```

```
        ImageSize->320,
        PlotStyle->Red,
        PlotRange->All
        ]
       ],
      Show[
       Plot[
        {PDF[dist1,x],PDF[dist2,x]},
        {x,Min[{data1,data2}],Max[{data1,data2}]},
        PlotLegends->{"Distribution 1","Distribution 2"},
        PlotRange->All,
        PlotStyle->{Blue,Red},
        ImageSize->320
        ],
       Histogram[
        {data1,data2},
        Automatic,
        "PDF",
        ChartLegends->{"sample data1","sample data2"},
        ChartStyle->{Directive[Opacity[0.2],Red],Directive[Opacity[0.2],Purple]},
        ImageSize->320
        ]
       ]
      }
     ]
    ],
  {{v1,6},0.1,10,0.1},
  {{v2,6},0.1,10,0.1},
  {{n,500},{100,500,1000,2000}},
  {{seed,1234},ControlType->None}
  ]
```

Output

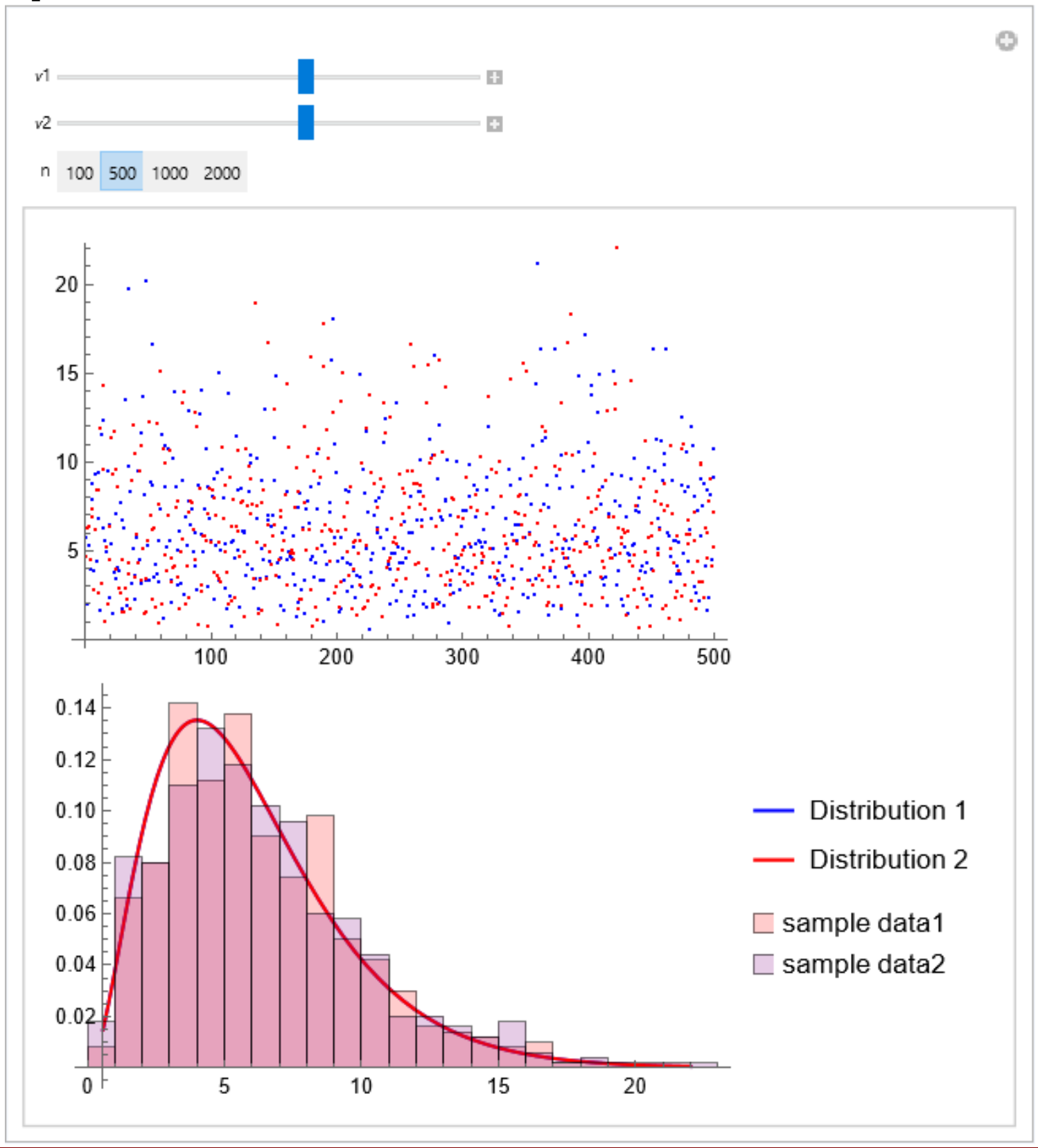

***Mathematica Examples 17.35***

Input
```
(*Sum of squares of n variables from normal distribution follows chi square distribution: *)
dist=TransformedDistribution[
  x^2+y^2+z^2,
  Distributed[{x,y,z},ProductDistribution[{NormalDistribution[],3}]]
  ]
```

Output
```
ChiSquareDistribution[3]
```

***Mathematica Examples 17.36***

Input
```
(*Chi square distribution is a special case of gamma distribution: *)
PDF[ChiSquareDistribution[v],x]
PDF[GammaDistribution[v/2,2],x]
```

Output

$$\begin{cases} \frac{2^{-\frac{\nu}{2}} e^{-\frac{x}{2}} x^{-1+\frac{\nu}{2}}}{\text{Gamma}[\nu/2]} & x > 0 \\ 0 & \text{True} \end{cases}$$

$$\begin{cases} \frac{2^{-\nu/2} e^{-x/2} x^{-1+\frac{\nu}{2}}}{\text{Gamma}[\nu/2]} & x > 0 \\ 0 & \text{True} \end{cases}$$

***Mathematica Examples 17.37***

Input
```
(* As you can see from the resulting plot, the chi-squared distributions with larger degrees of freedom (such as 20, 50, and 100) start to look more and more like the standard normal distribution. In the case of the chi-squared distribution, it is the sum of squares of independent, standard normal random variables that tends towards the chi-squared distribution with large degrees of freedom, which in turn tends towards the normal distribution: *)

n=10000;(* number of samples. *)
k={1,2,5,10,20,50,100};(* different degrees of freedom. *)
x=Table[
    RandomVariate[
     ChiSquareDistribution[k[[i]]],
     n
     ],
    {i,1,Length[k]}
    ];(* generate random samples from chi-squared distributions with different degrees of freedom. *)
Histogram[
 {x[[1]],x[[2]],x[[3]],x[[4]],x[[5]],x[[6]],x[[7]]},
 {0.5},
 "PDF",
 PlotLabel->"Chi-squared distributions with different degrees of freedom",
 ChartLegends->{"v = 1","v = 2","v = 5","v = 10","v = 20","v = 50","v = 100"},
 AxesLabel->{"x","PDF"},
 PlotRange->{{0,150},{0,0.15}}
 ]

(* compare with normal distribution. *)
y=RandomVariate[
    NormalDistribution[0,1],
    n
    ];(* generate random samples from the standard normal distribution. *)
Histogram[
 y,
```

```
{0.05},
"PDF",
ColorFunction->Function[Opacity[0.7]],
ChartStyle->Purple,
PlotRange->{{-5,5},{0,0.5}},
AxesLabel->{"x","PDF"},
Epilog->{Text["Normal",{3,0.4}]}
]
```

Output

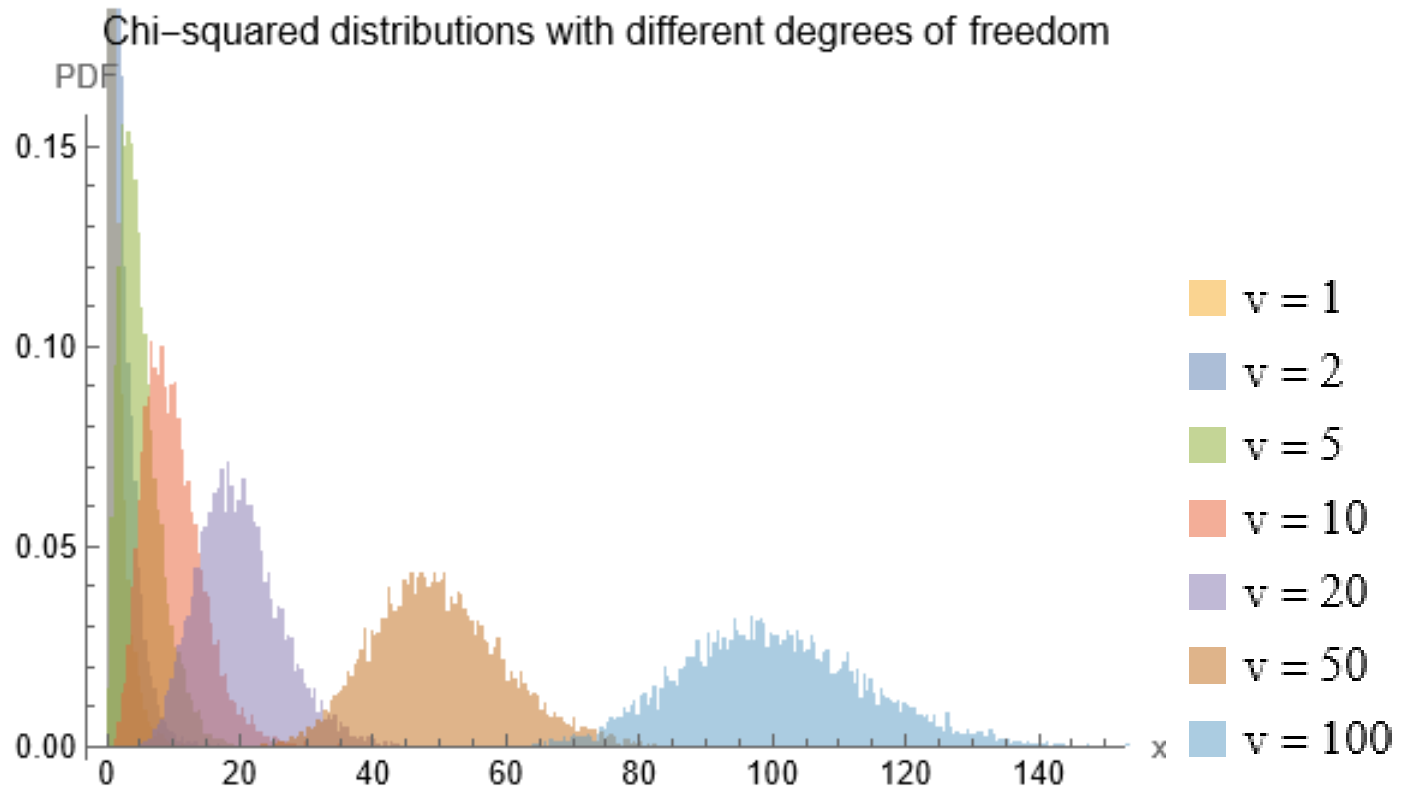

Output

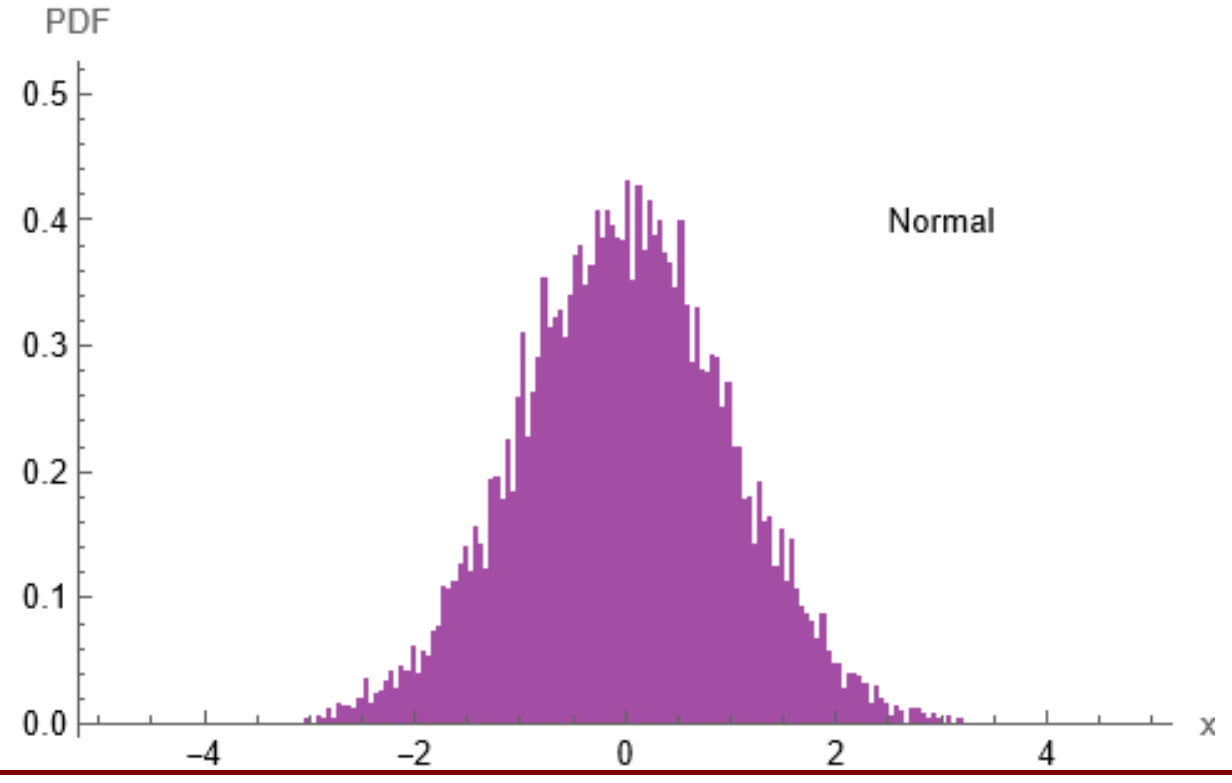

# UNIT 17.3

# SAMPLING DISTRIBUTIONS OF SAMPLE VARIANCE

**Mathematica Examples 17.38**

Input

```
(* Suppose X~N(0,1) (standard normal distribution), then X^2~chi-squared distribution with 1 degree of freedom. The code generates random samples from the chi-squared distribution with 1 degree of freedom using the relationship between the standard normal distribution and the chi-squared distribution. It does so by generating 10,000 independent random samples from the standard normal distribution using the RandomVariate function, and then squaring each sample to get a new sample set that follows a chi-squared distribution with 1 degree of freedom: *)

n=10000; (* Number of samples. *)
x=RandomVariate[
   NormalDistribution[],
   n
   ];(* Generate random samples from the standard normal distribution. *)
y=x^2;(* Calculate squared values. *)
Histogram[
 y,
 Automatic,
 "PDF",
 PlotLabel->"Chi-squared distribution with 1 degree of freedom",
 AxesLabel->{"x","PDF"},
 ColorFunction->Function[Opacity[0.6]],
 ChartStyle->Purple,
 ImageSize->300
 ]
```

Output

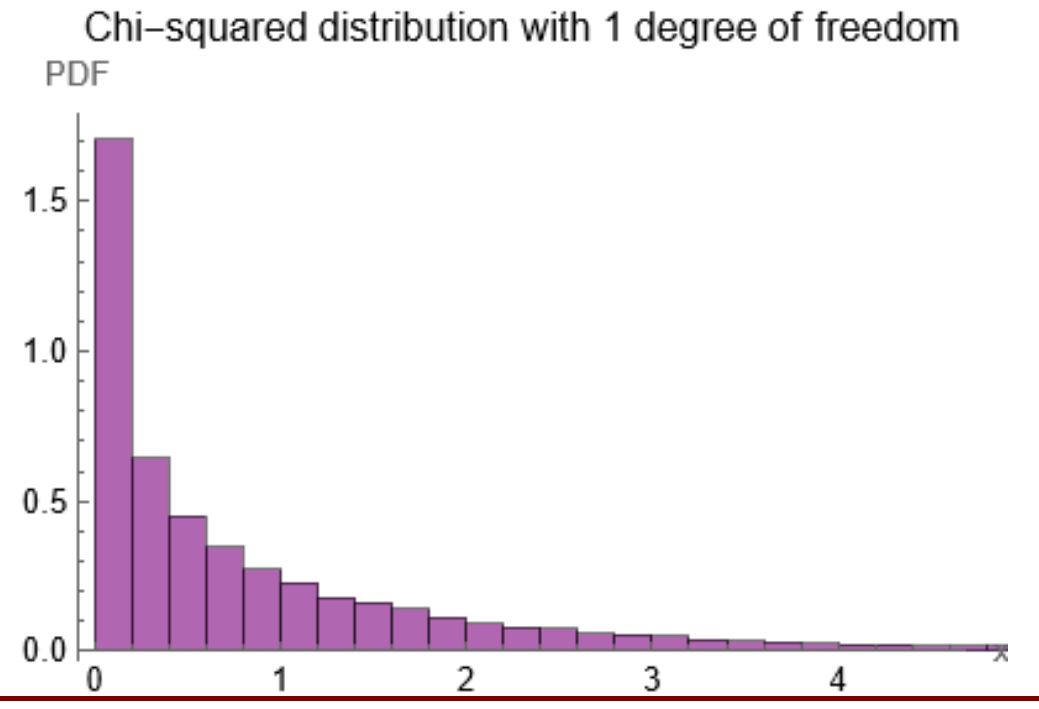

**Mathematica Examples 17.39**

Input

```
(* The code generates five independent standard normal random variables, squares each of them, adds up the squares, and repeats the process to generate 10,000 samples of the resulting chi-squared distribution with five degrees of freedom. Finally, it plots a histogram of the chi-squared samples along with the probability density function of the chi-squared distribution with five degrees of freedom: *)

(* Generate five independent standard normal random variables: *)
normalVars=RandomVariate[
   NormalDistribution[],
   5
```

```
    ];

(* Square each of the normal variables: *)
squaredVars=normalVars^2;

(* Add up the squares of the five normal variables: *)
chiSquared=Total[squaredVars];

(* Repeat the above steps to generate more samples: *)
nSamples=10000;
chiSquaredSamples=Table[
    normalVars=RandomVariate[NormalDistribution[],5];
    squaredVars=normalVars^2;
    Total[squaredVars],
    {i,nSamples}
    ];

(* Plot the resulting chi-squared distribution: *)
Show[
 Histogram[
   chiSquaredSamples,
   {0.3},
   "PDF",
   ColorFunction->Function[Opacity[0.6]],
   ChartStyle->Purple,
   ImageSize->250
   ],
 Plot[
   PDF[
    ChiSquareDistribution[5],
    x
    ],
   {x,0,20},
   PlotStyle->Darker[Red]
   ]
 ]
```

Output

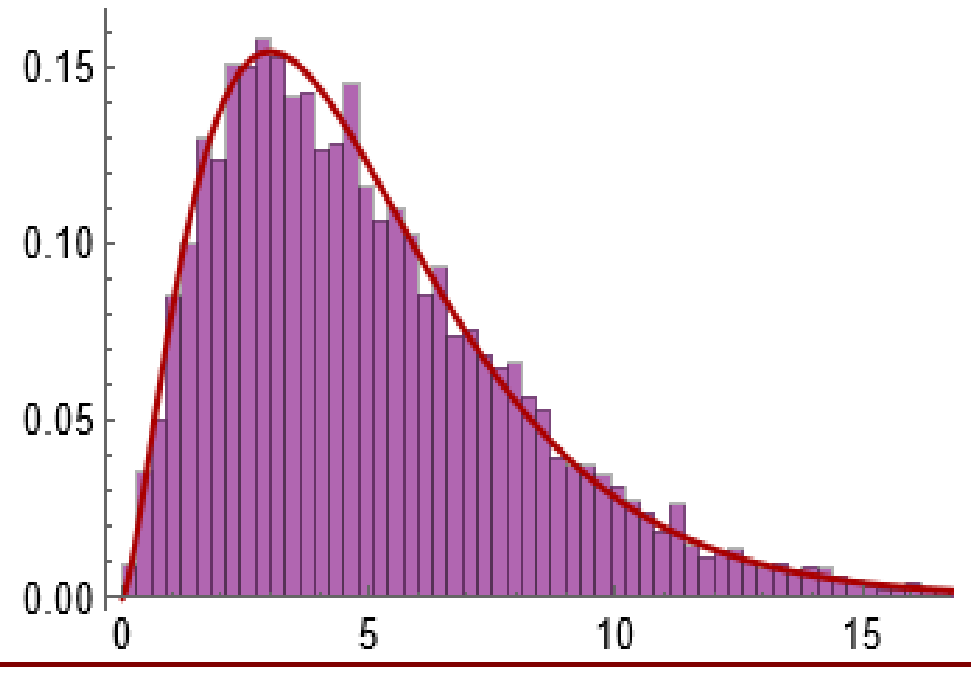

*Mathematica Examples 17.40*

Input

```
(* The code generates a Chi-squared distribution with a specified number of degrees
of freedom by squaring independent and identically distributed random variables drawn
from a standard normal distribution. This relationship is a consequence of the fact
that the sum of the squares of n independent standard normal random variables follows
a chi-squared distribution with n degrees of freedom. The generated samples are used
to create a histogram, and the Chi-squared PDF is plotted to visualize the
distribution. The Manipulate function allows the user to easily manipulate the number
of variables and the sample size to generate the desired distribution and histogram:
*)
```

```
Manipulate[
 samples=Table[
   Total[
    RandomVariate[NormalDistribution[0,1],n]^2
    ],
   {m}
   ];
 (* Plot the chi-squared PDF*)
 plot=Plot[
   PDF[
    ChiSquareDistribution[n],
    x
    ],
   {x,0,60},
   PlotRange->{{0,60},{0,0.2}},
   PlotLabel->StringJoin["Chi-Squared Distribution with ",ToString[n]," Degrees of
Freedom"],
   PlotStyle->Darker[Red]
   ];
 (* Plot a histogram of the generated samples: *)
 histogram=Histogram[
   samples,
   Automatic,
   "PDF",
   PlotRange->{{0,60},{0,0.2}},
   PlotLabel->"Histogram of Samples",
   ColorFunction->Function[Opacity[0.7]],
   ChartStyle->Purple,
   ImageSize->300
   ];
 (* Combine the plot and histogram into a single figure: *)
 Show[
  {histogram,plot}
  ],
 {{n,5,"Number of Variables"},1,30,1,Appearance->"Labeled"},
 {{m,500,"Number of trials"},500,10000,100,Appearance->"Labeled"}
 ]
```

Output

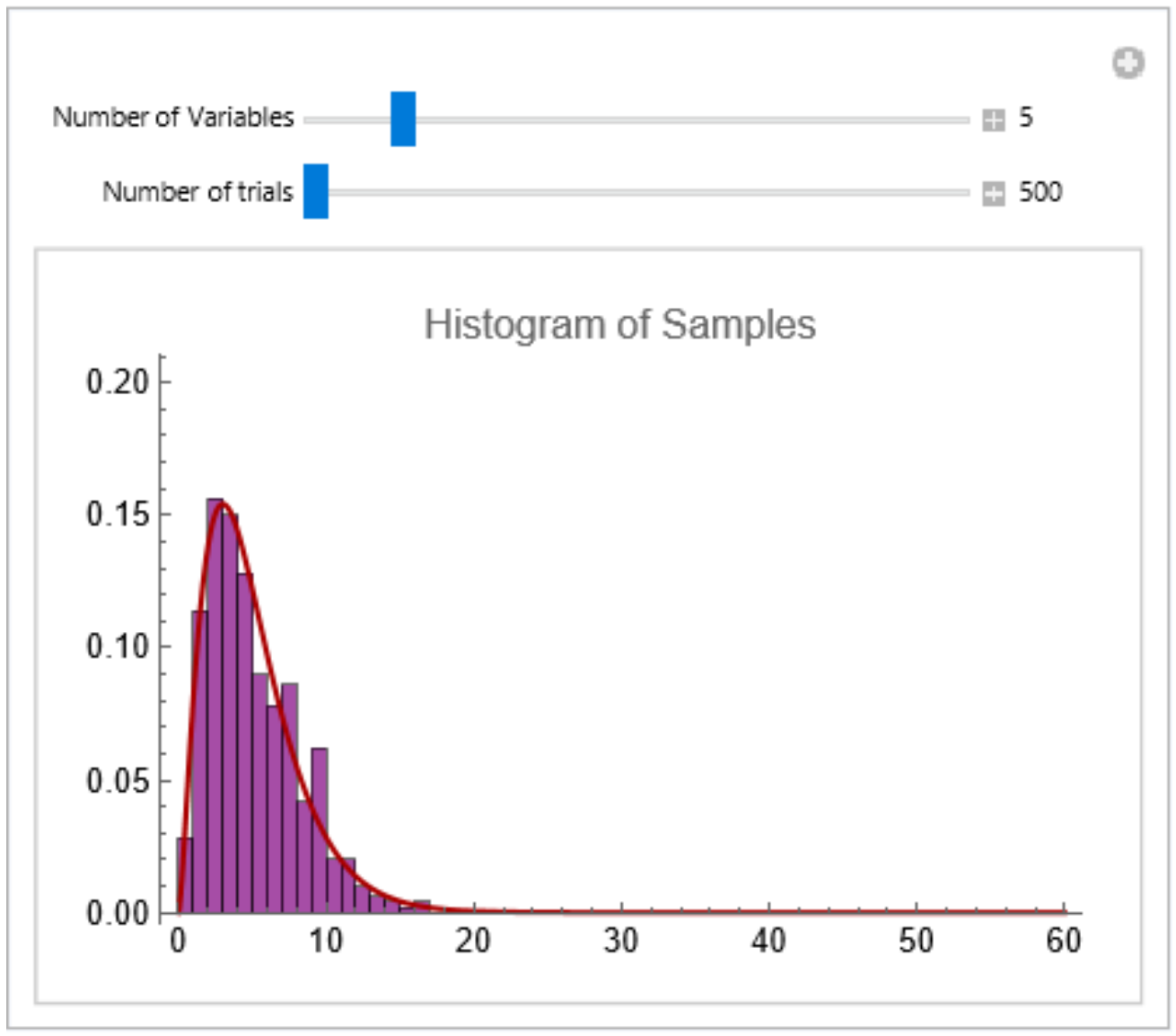

***Mathematica Examples 17.41***

Input

```
(* The code demonstrates the concept of the chi-squared distribution and how it arises from the standard normal distribution through the central limit theorem and sample distribution of mean. Specifically, the code generates 10,000 samples of size 5 from a normal distribution with mean 5 and standard deviation 3. Then, for each sample, the sample mean is calculated and standardized using the central limit theorem, resulting in a standard normal variable. The square of the standardized variable is then calculated, giving a chi-squared variable with 1 degree of freedom. Finally, the code plots histogram of the resulting chi-squared distribution along with the PDF of the chi-squared distribution with 1 degree of freedom: *)

μ=5;
σ=3;
n=5;
nSamples=10000;
chiSquaredSamples=Table[
    normalsample=RandomVariate[NormalDistribution[μ,σ],n];(* Normal sample of size n. *)
    samplemeans=Mean[normalsample];(* Mean of normal sample. *)
    snm=(samplemeans-μ)/(σ/Sqrt[n]);(* Using central limit theory (CLT), standardizing normal sample means. *)
    squaredVars=snm^2,(* Square of the standard normal sample means. (chi-squared distribution with 1 degrees of freedom).*)
    {i,nSamples}
    ];

(* Plot the resulting chi-squared distribution: *)
Show[
 Histogram[
  chiSquaredSamples,
  Automatic,
  "PDF",
  ColorFunction->Function[Opacity[0.6]],
  ChartStyle->Purple,
  ImageSize->300
  ],
 Plot[
  PDF[
   ChiSquareDistribution[1],
   x
   ],
  {x,0,5},
  PlotRange->{{0,5},{0,2}},
  PlotStyle->Darker[Red]
  ]
 ]
```

Output

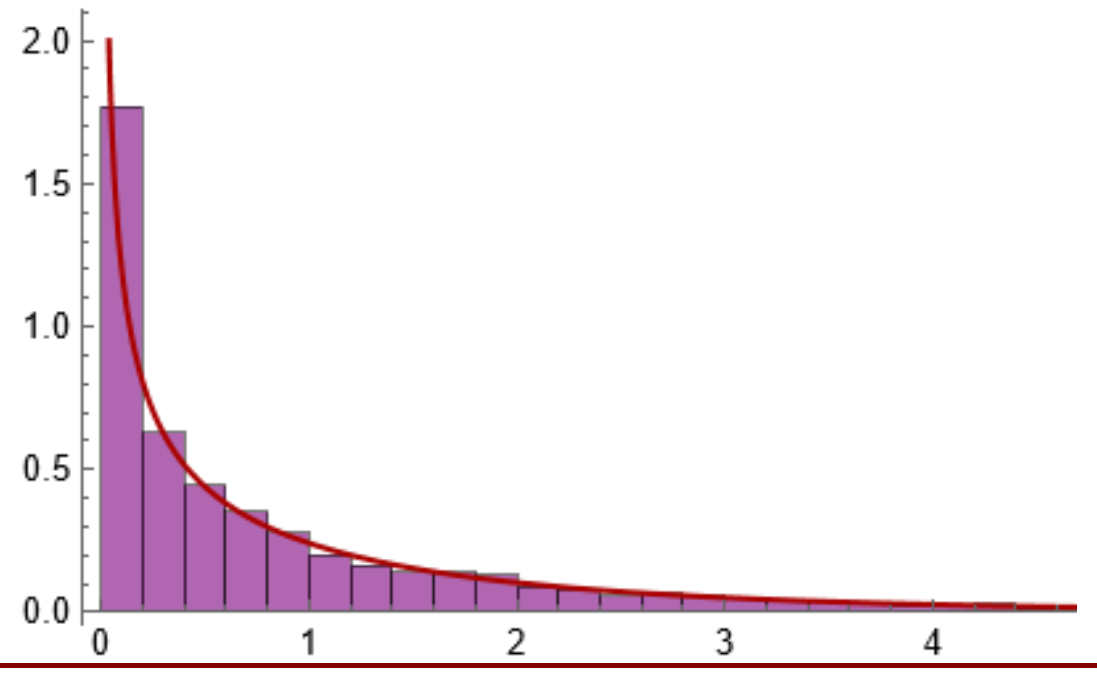

***Mathematica Examples 17.42***

Input

```
(* By generating m samples from the chi-squared distribution with n degrees of freedom
and computing their standardized values, the code is simulating the distribution of
(X-n)/\[Sqrt]2n. The resulting histogram of the standardized chi-squared distribution
is then compared to the standard normal distribution. As n and m increase, the
histogram of the standardized chi-squared distribution becomes increasingly similar
to the standard normal distribution: *)

(* Set the parameters: *)
n=1000;(* Number of degrees of freedom. *)
m=10000;(* Number of simulations. *)
(* Generate samples from the chi-squared distribution. *)
samples=Table[
    RandomVariate[ChiSquareDistribution[n]],
    {i,m}
    ];

standardizedChiSquare=Table[
    (samples[[i]]-n)/Sqrt[2*n],
    {i,m}
    ];

(* Compare the histogram of the standardized Chi Square to the standard normal
distribution: *)
Show[
 Histogram[
  standardizedChiSquare,
  {0.1},
  "PDF",
  PlotRange->{{-4,4},Automatic},
  ColorFunction->Function[Opacity[0.6]],
  ChartStyle->Purple,
  ImageSize->300
  ],
 Plot[
  PDF[NormalDistribution[0,1],x],
  {x,-4,4},
  PlotStyle->{Thick,Darker[Red]}
  ],
 PlotLabel->"Standardized Chi Square
vs Standard Normal Distribution",
 AxesLabel->{"x","Density"},
 PlotRange->{{-4,4},Automatic}
 ]
```

Output

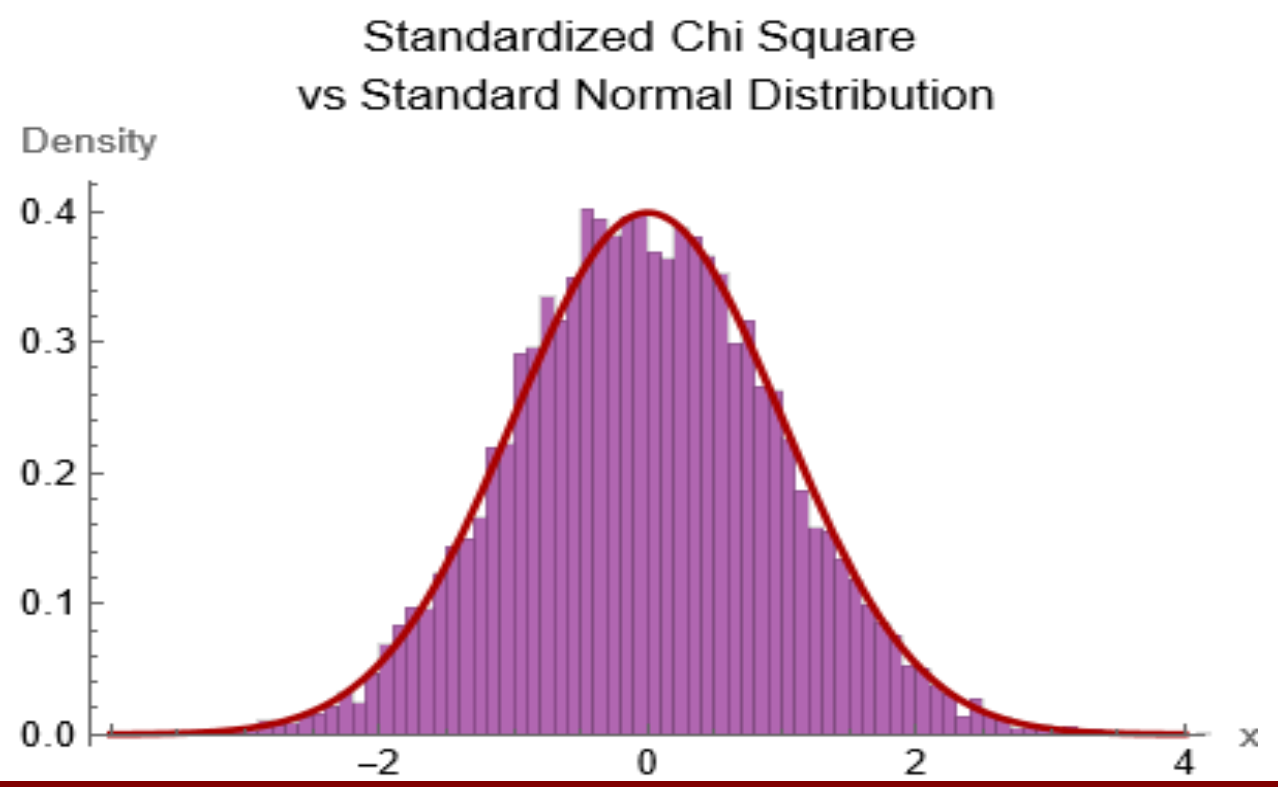

***Mathematica Examples 17.43***

```
Input    (* Scaled ChiSquareDistribution follows GammaDistribution: *)
         TransformedDistribution[c*X,X\[Distributed]ChiSquareDistribution[v],Assumptions-
         >c>0]

Output   GammaDistribution[v/2,2 c]
```

***Mathematica Examples 17.44***

```
Input    (* Scaled ChiSquareDistribution follows GammaDistribution: *)
         TransformedDistribution[(v/σ^2)*X,X\[Distributed]ChiSquareDistribution[v],Assumptio
         ns->(v/σ^2)>0]

Output   GammaDistribution[v/2,(2 v)/σ^2]
```

***Mathematica Examples 17.45***

```
Input    (* ChiSquareDistribution is a special case of gamma distribution: *)
         PDF[ChiSquareDistribution[v],x]
         PDF[GammaDistribution[v/2,2],x]
```

Output

$$\begin{cases} \frac{2^{-\nu/2} e^{-x/2} x^{-1+\frac{\nu}{2}}}{\text{Gamma}[\nu/2]} & x > 0 \\ 0 & \text{True} \end{cases}$$

Output

$$\begin{cases} \frac{2^{-\nu/2} e^{-x/2} x^{-1+\frac{\nu}{2}}}{\text{Gamma}[\nu/2]} & x > 0 \\ 0 & \text{True} \end{cases}$$

***Mathematica Examples 17.46***

```
Input    (* The code explains the relation between the distribution of the sample variance,
         denoted as S^2, and the transformed variable ((n-1)/σ^2)S^2. In other words, the code
         demonstrate how the sampling distribution of the sample variance follows the chi-
         squared distribution and how the gamma distribution can also be used to model it.
         The use of both distributions provides a visual comparison of how well each
         distribution fits the sampling distribution of the sample variance. The code first
         generates 1000 random samples from the normal distribution and then calculates the
         sample variance for each sample. In the code, the variable df is calculated as n-1,
         which represents the degrees of freedom of the chi-squared distribution used to model
         ((n-1)/σ^2)S^2. The transformed variable ((n-1)/σ^2)S^2 is calculated as
         (df/popvar)*samplevars in the code. The code then plots a histogram of the transformed
         variable ((n-1)/σ^2)S^2 and overlays the corresponding chi-squared distribution with
         df degrees of freedom. The comparison of the two plots illustrates how the chi-
         squared distribution provides a good fit for the transformed variable ((n-1)/σ^2)S^2.
         Finally, it plots the gamma distribution of S^2 and superimposes it on the histogram
         of the sample variances: *)

         (* Set the sample size and population variance: *)
         n=20;
         popvar=5;

         (* Calculate the degrees of freedom for the chi-squared distribution: *)
         df=n-1;

         (* Generate 1000 random samples from a normal distribution with mean 0 and variance
         popvar: *)
         samples=RandomVariate[
             NormalDistribution[0,Sqrt[popvar]],
             {1000,n}
             ];
```

```
(* Calculate the sample variance for each sample: *)
samplevars=Map[Variance,samples];

(* Calculate the mean and variance of the sample variances S^2: *)
mean=Mean[samplevars];
variance=Variance[samplevars];

(* Print the mean and variance of the sample variances S^2: *)
Print["Mean of sample variances: ",mean]
Print["Variance of sample variances: ",variance]

(* Plot a histogram of the sample variances ((n-1)S^2)/σ^2:(df/popvar)*samplevars:
*)
Histogram[
 (df/popvar)*samplevars,
 30,
 "PDF",
 Frame->True,
 ColorFunction->Function[Opacity[0.6]],
 ChartStyle->Purple,
 FrameLabel->{"((n-1)S^2)/σ2","Density"},
 PlotLabel->"Sampling Distribution of Variance:((n-1)S^2)/σ2",
 ImageSize->300
 ]

(* Plot the chi-squared distribution with the same degrees of freedom: *)
Plot[
 PDF[
  ChiSquareDistribution[df],
  x
  ],
 {x,0,40},
 PlotRange->All,
 PlotStyle->Blue,
 PlotLabel->"Chi-Squared Distribution with Degrees of Freedom "<>ToString[df],
 AxesLabel->{"x","PDF"},
 ImageSize->300
 ]

(* Superimpose the chi-squared distribution on the histogram of sample variances
((n-1)S^2)/σ^2: *)
Show[
 Histogram[
  (df/popvar)*samplevars,
  30,
  "PDF",
  Frame->True,
  ColorFunction->Function[Opacity[0.6]],
  ChartStyle->Purple,
  FrameLabel->{"((n-1)S^2)/σ2","Density"},
  PlotLabel->"chi-squared & Sampling Distributions: ((n-1)S^2)/σ2",
  ImageSize->300
  ],
 Plot[
  PDF[
   ChiSquareDistribution[df],
   x
   ],
  {x,0,40},
  PlotRange->All,
  PlotStyle->Blue
```

```
    ]
   ]

  (* Plot a histogram of the sample variances S^2: *)
  Histogram[
   samplevars,
   30,
   "PDF",
   Frame->True,
   ColorFunction->Function[Opacity[0.6]],
   ChartStyle->Purple,
   FrameLabel->{"S^2","Density"},
   PlotLabel->"Sampling Distribution of Sample Variance S^2",
   ImageSize->300
   ]

  (* Plot the Gamma distribution(df/2, 2 popvar/df): *)
  Plot[
   PDF[
    GammaDistribution[df/2,2popvar/df],
    x
    ],
   {x,0,12},
   PlotRange->All,
   PlotStyle->Red,
   PlotLabel->"Gamma Distribution ",
   AxesLabel->{"x","PDF"},
   ImageSize->300
   ]

  (* Superimpose the Gamma distribution on the histogram of sample variances S^2: *)
  Show[
   Histogram[
    samplevars,
    30,
    "PDF",
    Frame->True,
    ColorFunction->Function[Opacity[0.6]],
    ChartStyle->Purple,
    FrameLabel->{" S^2","Density"},
    PlotLabel->"Gamma Distribution and Sampling Distribution of S^2",
    ImageSize->300
    ],
   Plot[
    PDF[
     GammaDistribution[df/2,2popvar/df],
     x
     ],
    {x,0,12},
    PlotRange->All,
    PlotStyle->Red,
    AxesLabel->{"x","PDF"}
    ]
   ]


Output   Mean of sample variances:   4.91822
Output   Variance of sample variances:   2.66941
```

Output

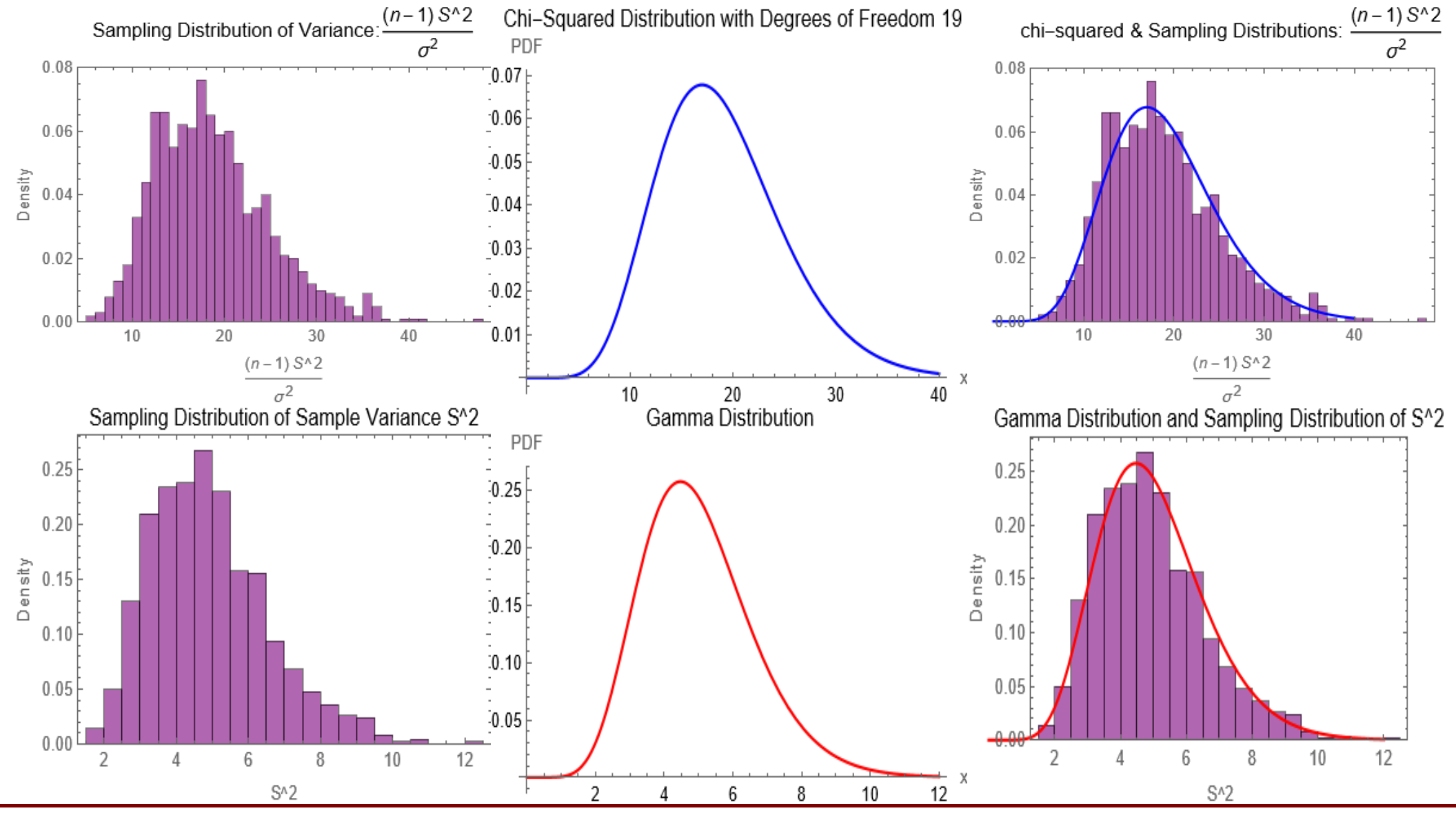

***Mathematica Examples 17.47***

Input

```
(* This code defines a population with a normal distribution and uses a Manipulate
function to create a histogram of sample variances. The code generates a random
sample from the population, computes the variance of the sample, and repeats this
process multiple times (10,000 times in this case) for different sample sizes. The
resulting histogram visualizes the distribution of sample variances. The Manipulate
function allows the user to control the sample size, which in turn affects the spread
of the distribution of sample variances. As the sample size increases, the spread of
the distribution decreases, reflecting the fact that larger samples provide more
precise estimates of the population variance. Note that, the shape of the histogram
will resemble the shape of the chi-squared distribution with the appropriate degrees
of freedom. As the sample size increases, the shape of the histogram transforms from
a skewed distribution to a more symmetrical distribution that resembles a normal
distribution: *)

(* Define a population with normal distribution: *)
pop=RandomVariate[
    NormalDistribution[0,1],
    10000
    ];

Manipulate[
  (* Create a histogram of the sample variances: *)
  Histogram[
   Table[
    Variance[RandomSample[pop,n]],
    {i,1,10000}
    ],
   {0,3,0.01},
   "PDF",
   PlotRange->{{0,3},{0,3}},
   ColorFunction->Function[Opacity[0.6]],
   ChartStyle->Purple
   ],
  (* Manipulate controls: *)
  {{n,10,"Sample size"},2,100,1,Appearance->"Labeled"}
  ]
```

Output

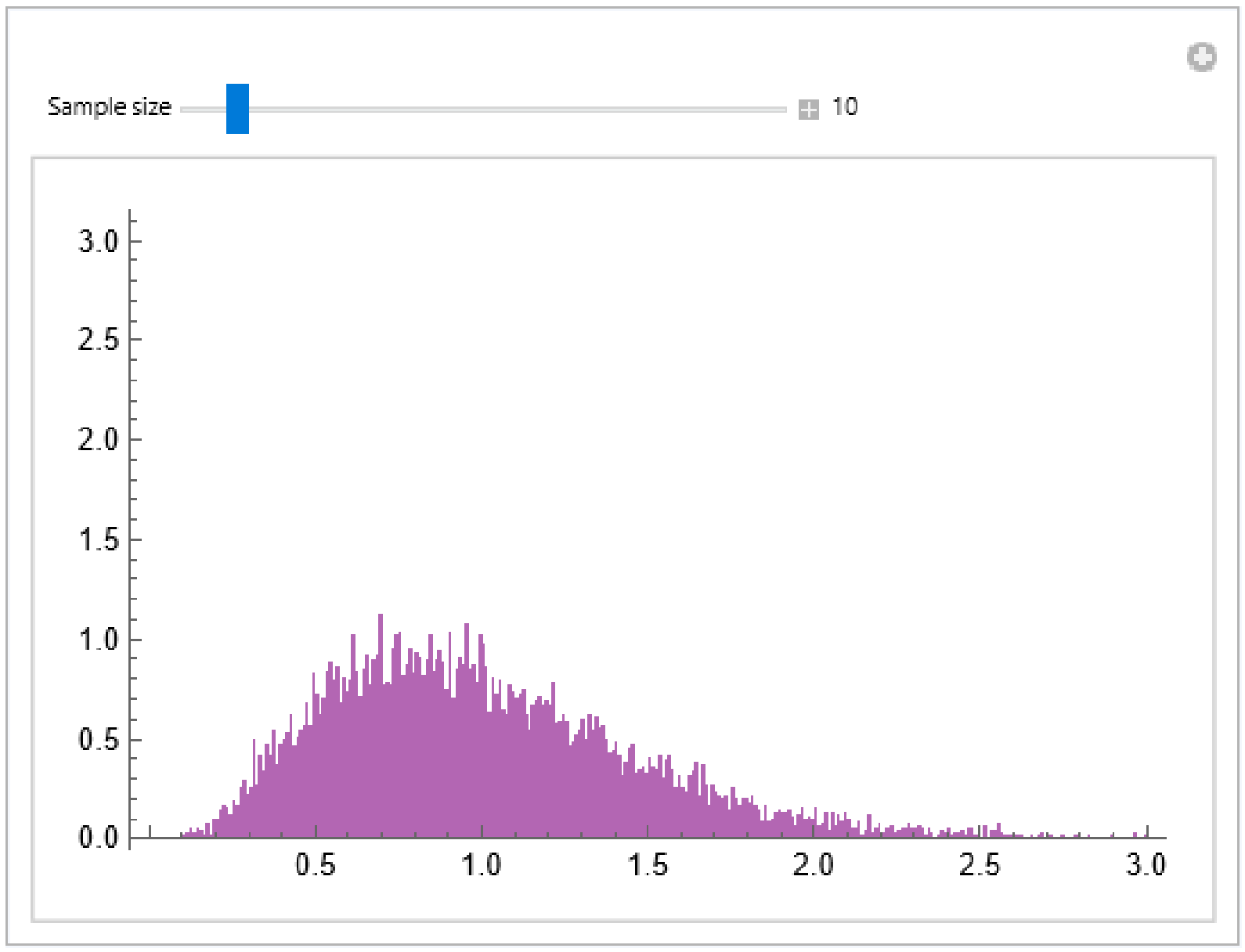

***Mathematica Examples 17.48***

Input

```
(* This code generates data from a normal distribution with mean 0 and standard
deviation 3, calculates the variance of the data,and then computes the sample
variances for m samples. It repeats this process numTrials times and calculates the
mean of the sample variances. Finally, it creates a scatter plot of the sample
variances, with the red line representing the population variance and the blue line
representing the mean of the sample variances. A histogram of the sample variances
is also created, showing the distribution of the sample variances: *)

Module[
 {data,variances,n=5000,m=10,numTrials=200},

 (* Generate data from normal distribution with mean 0 and standard deviation 1: *)
 data=RandomVariate[NormalDistribution[0,3],n];

 (* Calculate variances of data: *)
 var=Variance[data];

 (* Calculate sample variances for m samples: *)
 variances=Table[
   Variance[
    RandomSample[data,m]
    ],
   {i,1,numTrials}
   ];

 (* Calculate the mean of sample variances for m sample size: *)
 samplevar=Mean[variances];

 Row[
  (* Create a scatter plot of sample variances: *)
  ListPlot[
   variances,
   PlotRange->All,
   Epilog->{Red,Line[{{0,var},{numTrials,var}}],Blue,
Line[{{0,samplevar},{numTrials,samplevar}}]},
   Filling->Axis,
   PlotStyle->Directive[Purple,Opacity[0.5]],
```

```
     Frame->True,
     FrameLabel->{"Sample Number","Sample Variance"},
     PlotLabel->"Points of Sample Variances",
     FrameStyle->Directive[Black],
     ImageSize->250
     ],
    (* Create a histogram of sample variances: *)
    Histogram[
     variances,
     Automatic,
     "PDF",
     Frame->True,
     ColorFunction->Function[Opacity[0.6]],
     ChartStyle->Purple,
     FrameLabel->{" S^2","Density"},
     PlotLabel->"Sampling Distribution of S^2",
     ImageSize->250
     ]
    ]
   ]
```

Output

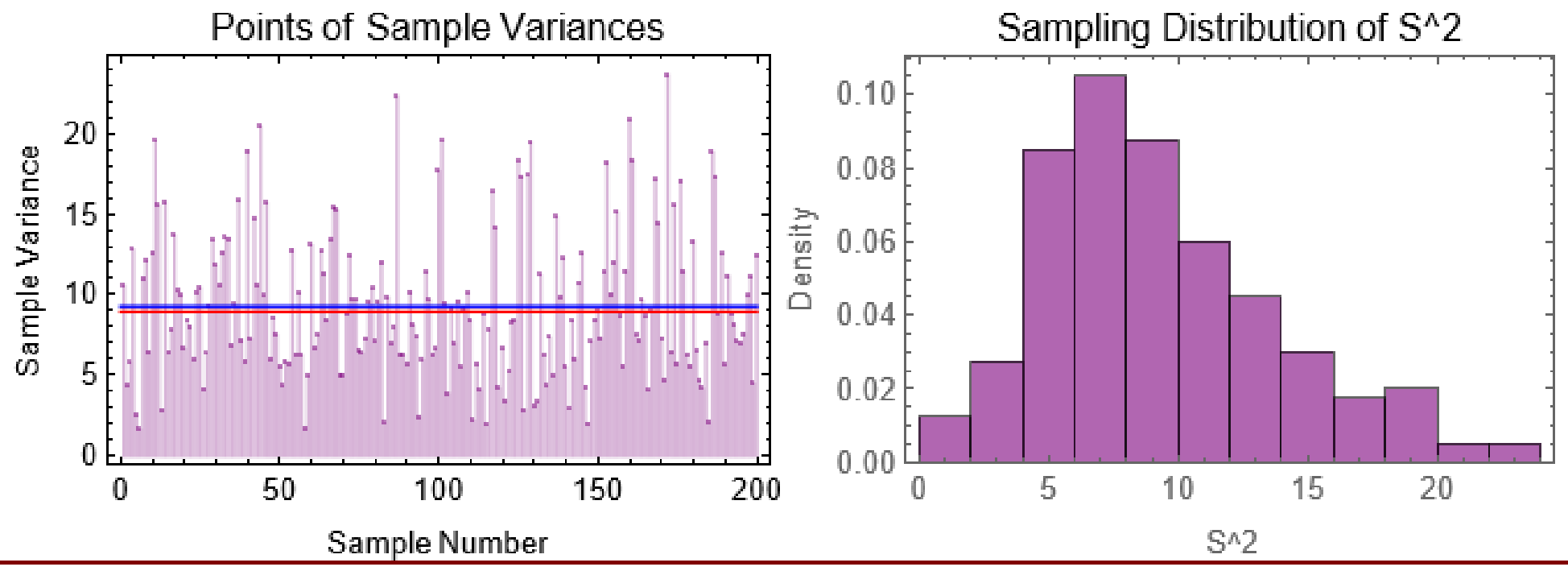

### *Mathematica Examples 17.49*

Input

```
(* The code is a statistical simulation that generates samples from a normal distribution with given mean and standard deviation, calculates their sample variances for a given sample size, and then creates a scatter plot and histogram of the resulting sample variances. The Manipulate function allows the user to adjust the parameters of the simulation, such as the mean, standard deviation, sample size, and number of samples. The scatter plot shows the sample variances for each trial as points, with a red line indicating the mean of the sample variances across all trials. The histogram shows the distribution of the sample variances across all trials: *)

Manipulate[
 Module[
  {data,variances},
  (* Generate data from normal distribution with mean mu and standard deviation sigma: *)
   data=RandomVariate[
     NormalDistribution[mu,sigma],
     10000
     ];
   (* Calculate sample variances for m Sample Size: *)
   variances=Table[
     Variance[
      RandomSample[data,m]],
     {i,1,numTrials}
     ];
   (* Create a scatter plot of sample variances: *)
```

```
      Row[
       ListPlot[
        variances,
        PlotRange->{{0,1000},{0,40}},
        Epilog->{Red,Line[{{0,Mean[variances]},{numTrials,Mean[variances]}}]},
        AxesOrigin->{0,0},
        FrameLabel->{"Sample Number","Sample Variance"},
        ColorFunction->Function[Opacity[0.2]],
        PlotStyle->Purple,
        ImageSize->250
        ],
       (* Create a histogram of sample variances: *)
       Histogram[
        variances,
        Automatic,
        "PDF",
        Frame->True,
        ColorFunction->Function[Opacity[0.6]],
        ChartStyle->Purple,
        FrameLabel->{" S^2","Density"},
        PlotLabel->"Sampling Distribution of S^2",
        ImageSize->250
        ]
       ]
      ],
     (* Manipulate controls*)
     {{mu,3,"Mean"},0,5,0.1,Appearance->"Labeled"},
     {{sigma,3,"Standard Deviation"},0.1,5,0.1,Appearance->"Labeled"},
     {{m,20,"Sample Size"},5,100,1,Appearance->"Labeled"},
     {{numTrials,700,"Number of Samples"},10,1000,10,Appearance->"Labeled"}
     ]
```

Output

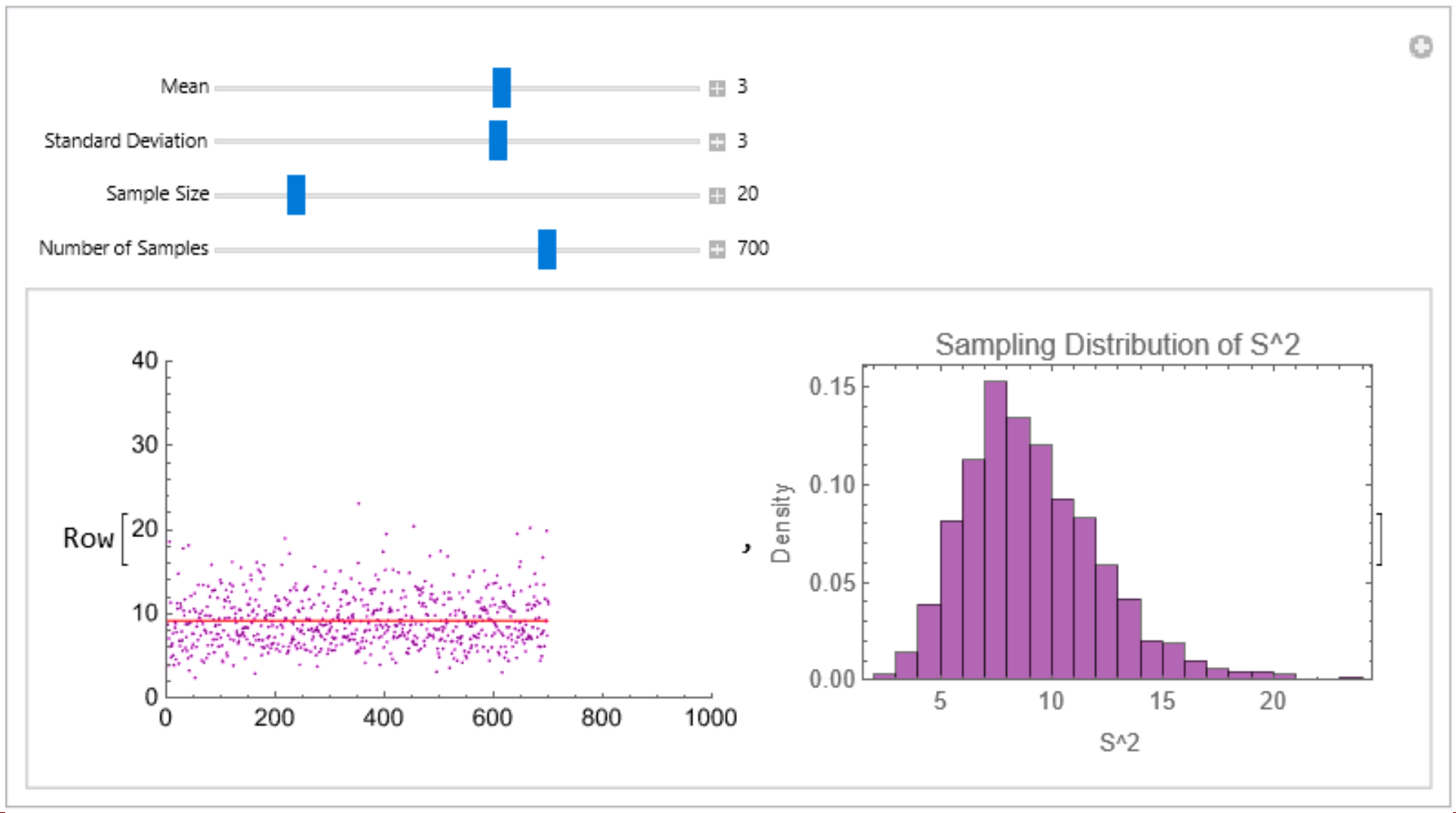

***Mathematica Examples 17.50***

Input

```
(* The code generates a series of sample variances of different sizes from a normal distribution with mean 1 and standard deviation 6, and comparing for each sample size the sample variance to the variance of the original distribution. For each sample size, the absolute difference between mean of sample variance and the original variance is calculated and stored in a list. Finally, the list is plotted using ListPlot, where the x-axis shows the sample size and the y-axis shows the absolute difference between the sample variance and the population variance. The plot generated
```

```
by this code could provide useful insights into the behavior of the variance of
sample distributions as the sample size increases. It seems that the plot shows a
decreasing trend in the absolute difference between the sample variances and the
population variance as the sample size increases: *)

parent=NormalDistribution[1,6];

variances=Table[
   n=i;
   samplevariances=Table[
     Variance[
      RandomVariate[parent,n]],
     {1000}
     ];
   meanOfSamplevariances=Mean[samplevariances];
   varianceOfParent=Variance[parent];
   Abs[N[meanOfSamplevariances-varianceOfParent]],
   {i,2,100}
   ];

ListPlot[
 variances,
 AxesLabel->{"Sample size","Absolute difference"},
 Joined->True,
 PlotRange->All,
 PlotStyle->Purple
 ]
```

Output

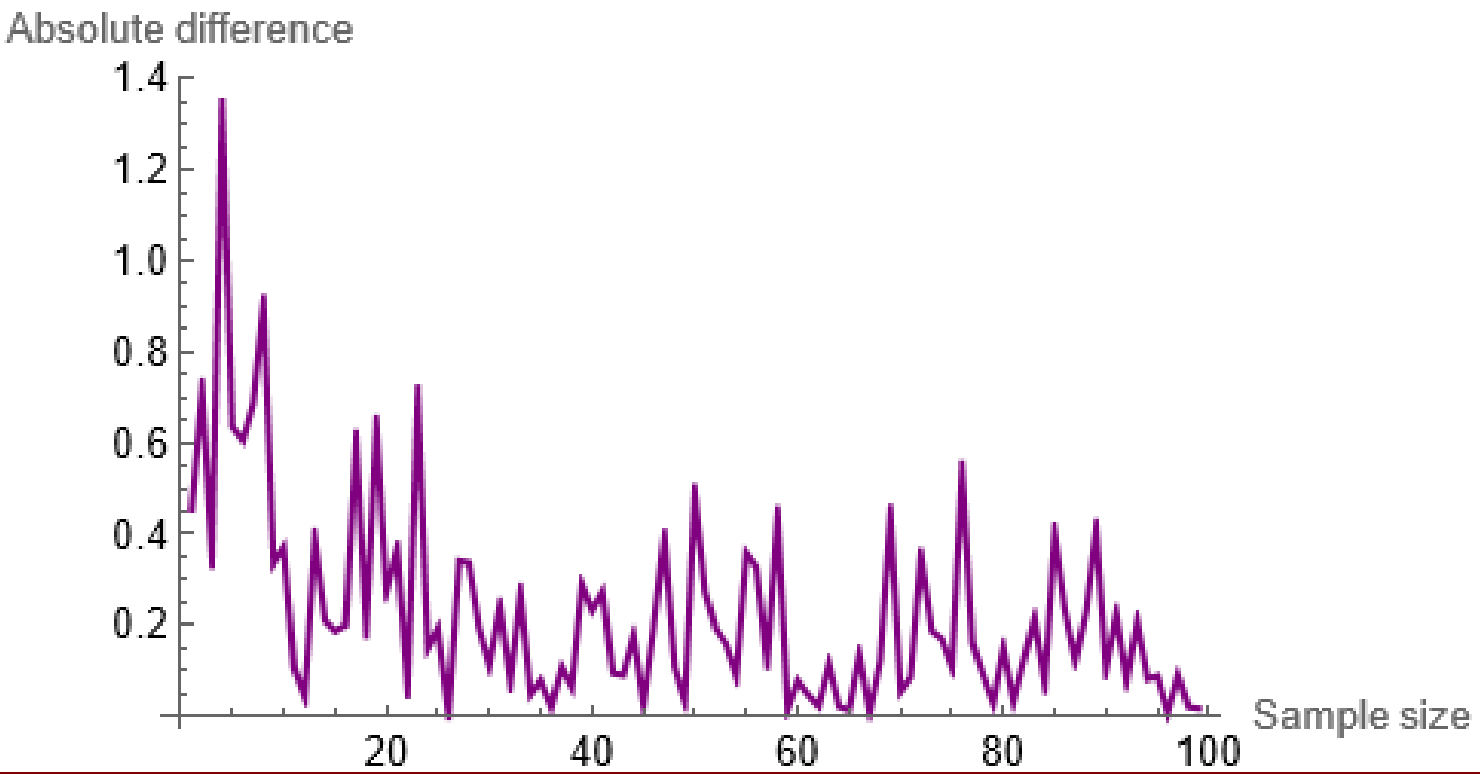

***Mathematica Examples 17.51***

Input

```
(* This code generates a grid of histograms and density plots to visualize the 2D
distribution of the sample variances for different sample sizes. Specifically, the
code computes the variance of 100,000 samples of size n drawn from a normal
distribution with mean 1 and standard deviation 6. The resulting variances are then
plotted in a 3D histogram using the "Rainbow" color scheme and a PDF plot type. The
code iterates over several sample sizes, namely n=3,10,20, and 50, and generates a
separate 3D histogram for each sample size. The plot label of each histogram indicates
the sample size being plotted, while the x-axis, y-axis and z-axis labels indicate
"SV X", "SV Y" and "PDF", respectively, where "SV" stands for sample variance. For
small sample size (n=3), the shape of the 3D histogram is highly skewed and elongated
along the z-axis. As the sample size increases (n=10, n=20, and n=50), the shape of
the 3D histograms becomes more symmetrical and bell-shaped. The gamma distribution
is closely related to the distribution of sample variances. As such, as the sample
size increases, the gamma distribution becomes more bell-shaped, indicating that the
distribution of sample variances is becoming more normally distributed: *)
```

```
Grid[Table[
   variances=Table[
     Variance[
      RandomVariate[
       NormalDistribution[1,6],
       {n,2}
       ]
      ],
     100000
     ];
   Column[
    {
     Histogram3D[
      variances,
      Automatic,
      "PDF",
      ColorFunction->"Rainbow",
      ImageSize->200,
      PlotLabel->{{"sample size n=",n}},
      AxesLabel->{"SV X","SV Y","PDF"}(* SV=Sample Variance. *)
      ],
     SmoothDensityHistogram[
      variances,
      Automatic,
      "PDF",
      ColorFunction->"Rainbow",
      ImageSize->170,
      PlotLabel->{{"sample size n=",n}}
      ]
     }
    ],
   {n,{3,10,20,50}}
   ]
  ]
```

Output

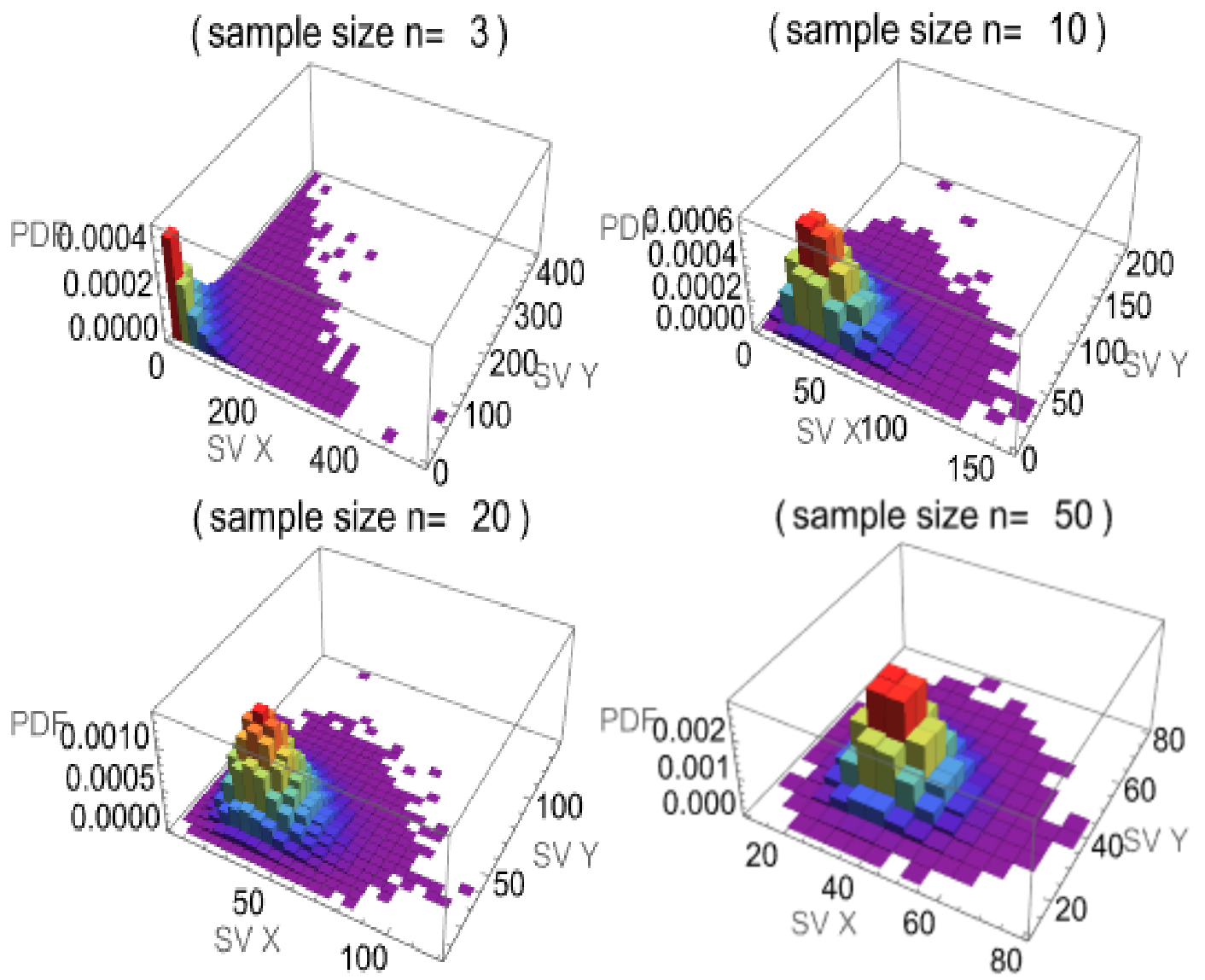

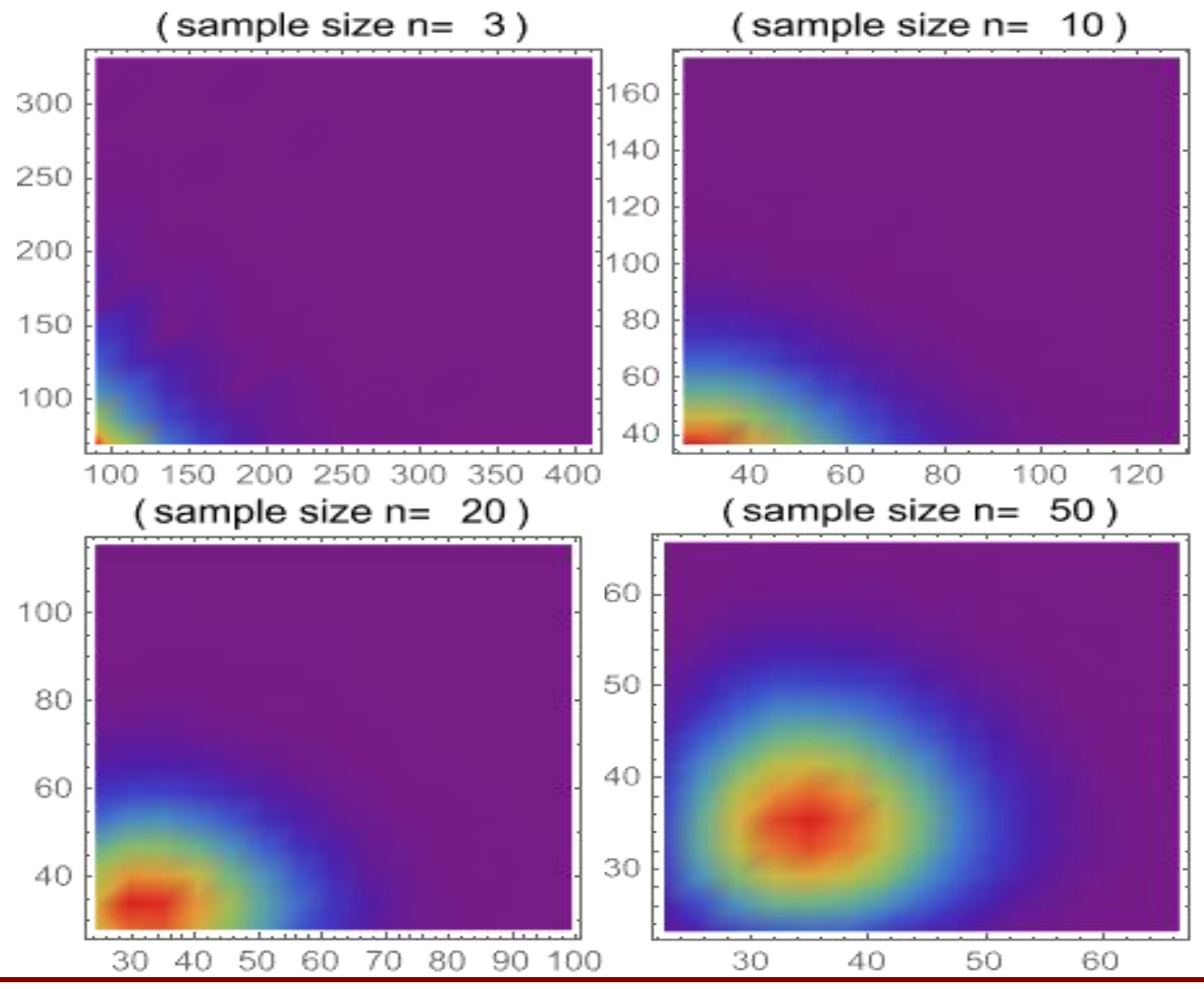

**Mathematica Examples 17.52**

Input
```
(* The code uses Histogram3D and SmoothDensityHistogram to visualize the 2D sample
variance distributions and Manipulate to explore the effect of sample size on the
distributions: *)

population3d=RandomVariate[
    NormalDistribution[1,6],
    {10000,2}
    ];

(* Define a function to generate and plot sample variance: *)
samplevariance3d[n_]:=Module[
  {samples,means},

  samples3d=Table[
    RandomChoice[
     population3d,n
     ],
    {i,1,10000}
    ];

  variance3d=Map[Variance,samples3d];

  Row[
   Histogram3D[
    variance3d,
    Automatic,
    "PDF",
    PlotRange->All,
    AxesLabel->{"SV X","SV Y","PDF"}(* SV=Sample variance. *),
    ColorFunction->"Rainbow",
    PlotLabel->Row[{"n = ",n}],
    ImageSize->250
    ],
   SmoothDensityHistogram[
    variance3d,
    Automatic,
    "PDF",
    ColorFunction->"Rainbow",
```

```
      ImageSize->170,
      PlotLabel->{{"sample size n=",n}}
      ]
     ]
    ]

(* Use Manipulate to explore the sampling distributions of the sample variance: *)
Manipulate[
 samplevariance3d[n],
 {n,2,50,1}
 ]
```

Output

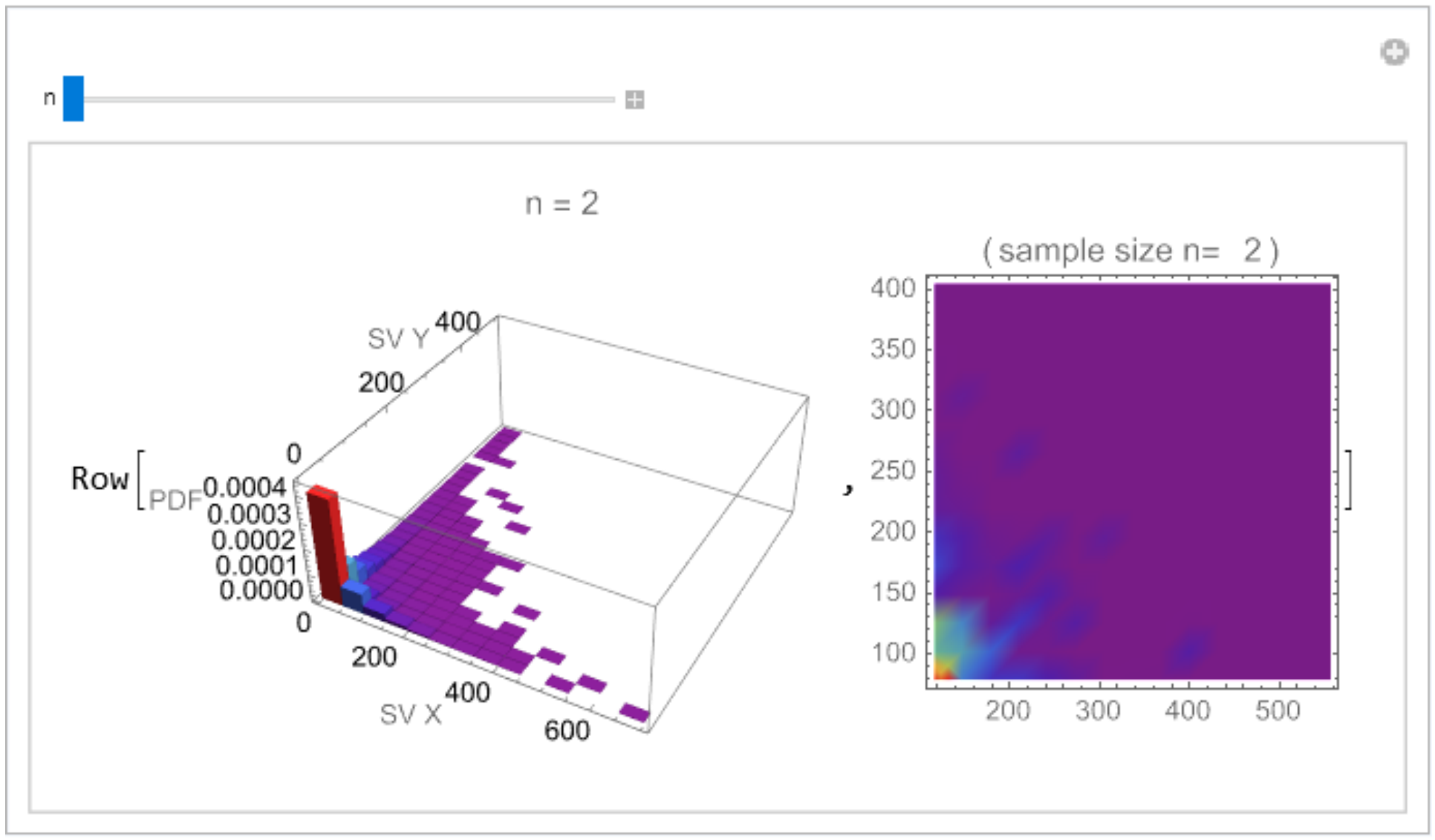

***Mathematica Examples 17.53***

Input

```
(* The code generates a 2D dataset with 5000 random points that follow a Normal
distribution with mean 1 and standard deviation 6. The code also defines a function
to generate and plot sample variances. The function takes a sample size 'n', draws
5000 samples from the population with replacement, calculates the variance for each
sample, and plots the variances as purple points on a scatter plot. The opacity of
the points is set to 0.3 to indicate overlapping points. The Manipulate function is
then used to explore the sample variance points. Manipulate allows the user to
interactively adjust the sample size 'n' and observe the changes in the sample
variance points, as the sample size 'n' increases: *)

population1=RandomVariate[
   NormalDistribution[1,6],
   {5000,2}
   ];

(* Define a function to generate and plot sample variance: *)
samplevariances[n_]:=Module[
  {samples,variances},

  samples=Table[
    RandomChoice[
     population1,n
     ],
    {i,1,5000}
    ];
```

```
    variances=Map[Variance,samples];

    Graphics[
     {Purple,PointSize[0.008],Opacity[0.3],Point[variances]},
     ImageSize->250
     ]
    ]

(* Use Manipulate to explore the sample variances points: *)
Manipulate[
 samplevariances[n],
 {n,2,30,1}
 ]
```

Output

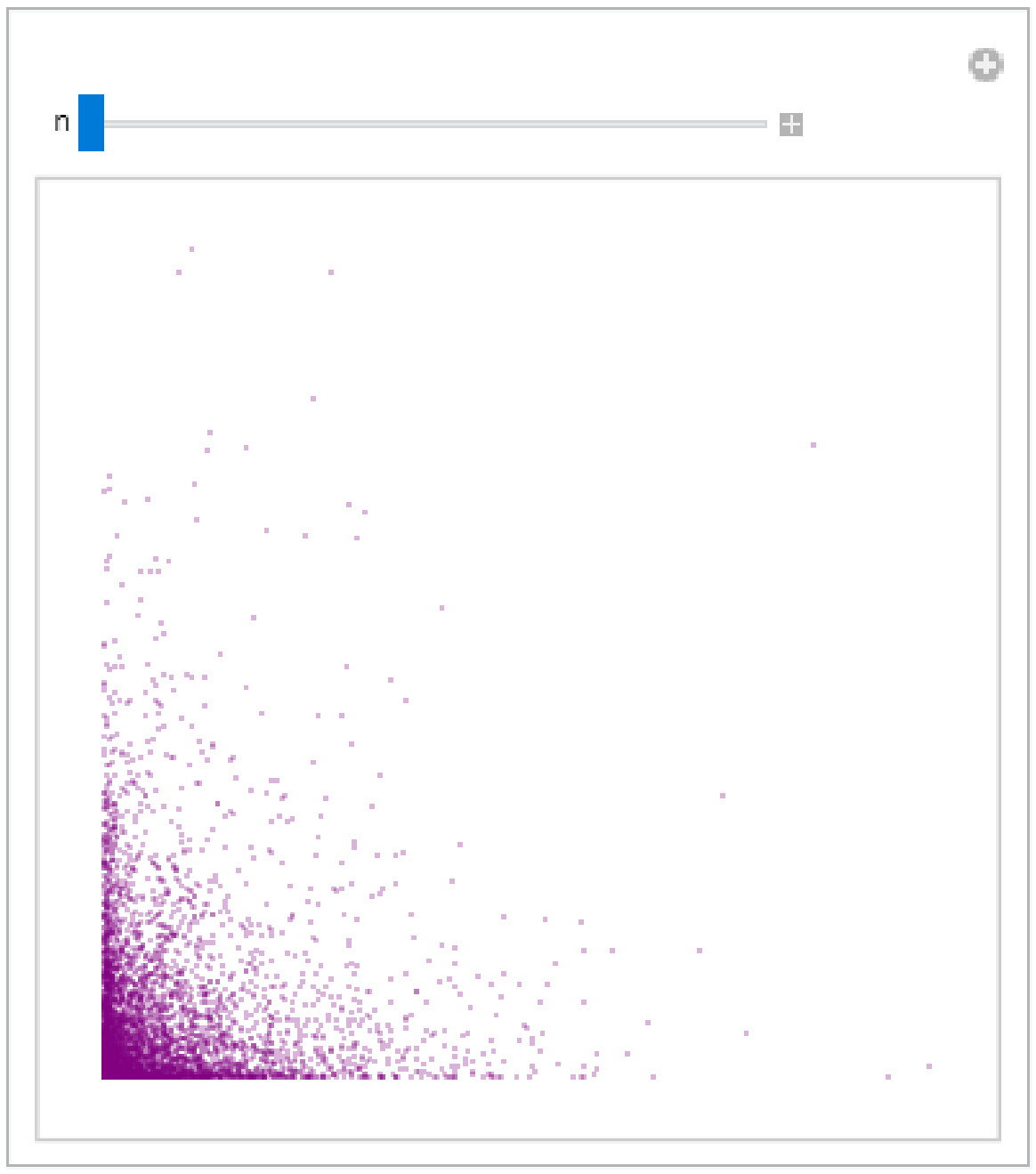

***Mathematica Examples 17.54***

Input

```
(* The code generates and plots sample variances for a 3D dataset with normal distribution using the Manipulate function to explore changes in the sample size: *)

populationx=RandomVariate[
    NormalDistribution[1,6],
    {5000,3}
    ];

(* Define a function to generate and plot sample variance: *)
samplevariance[n_]:=Module[
   {samples,variance},

   samples=Table[
     RandomChoice[
      populationx,n
      ],
     {i,1,5000}
     ];

   variance=Map[Variance,samples];
```

```
    Graphics3D[
     {Purple,PointSize[0.008],Opacity[0.3],Point[variance]},
     ImageSize->250
     ]
    ]

(* Use Manipulate to explore the sample variances points: *)
Manipulate[
 samplevariance[n],
 {n,2,50,1}
 ]
```

Output

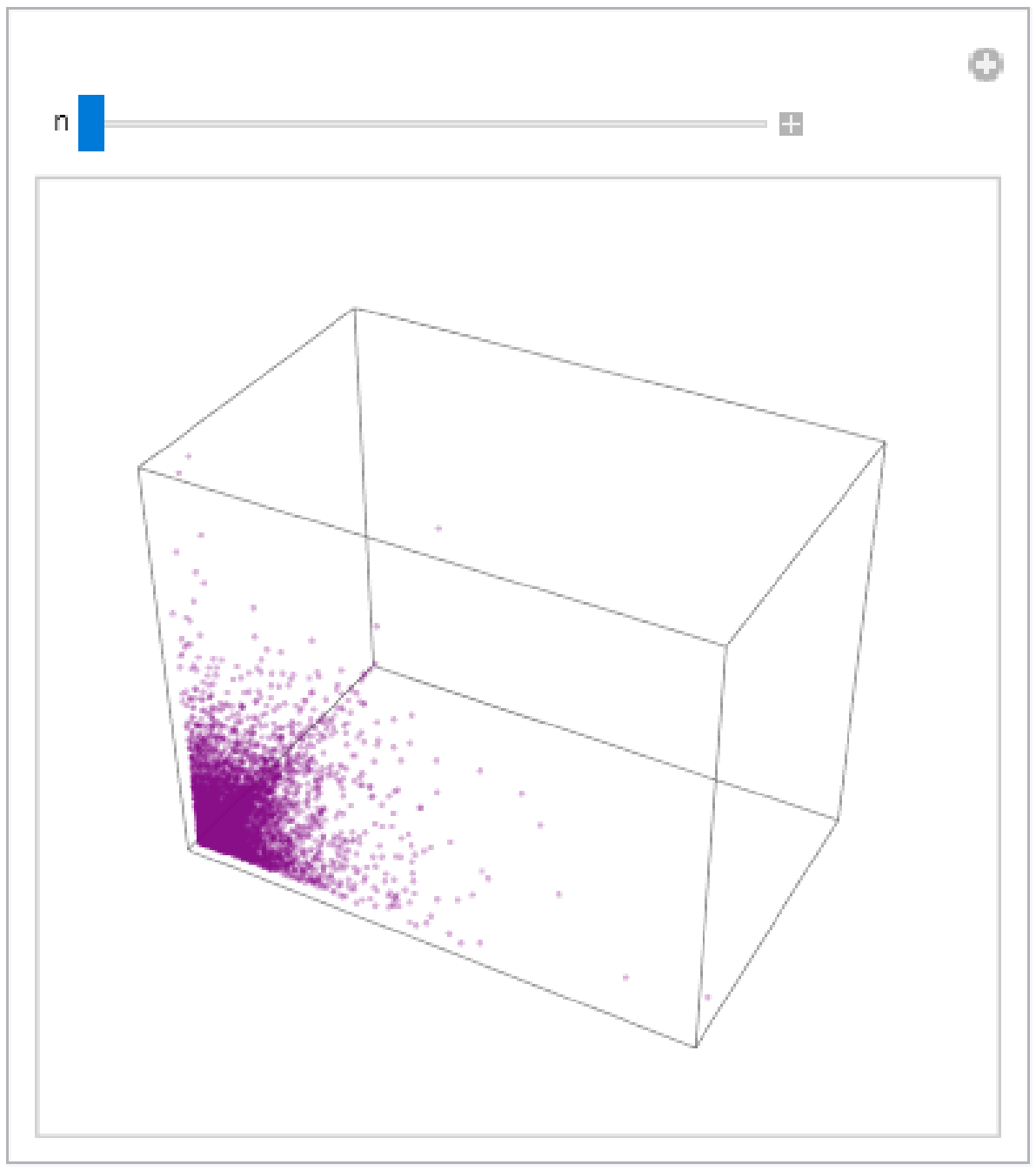

# UNIT 17.4

# STUDENT T DISTRIBUTION

*Mathematica Examples 17.55*

Input
```
(* This code generates a plot of the probability density function (PDF) for a student t distribution with different values of the parameter v= (1, 2, 3, 10). The PDF is evaluated at various values of the random variable x between -6 and 6: *)

Plot[
 Evaluate[
  Table[
   PDF[
    StudentTDistribution[v],
    x
    ],
   {v,{1,2,3,10}}
   ]
  ],
 {x,-6,6},
 PlotRange->Automatic,
 Filling->Axis,
 PlotLegends->Placed[{"v=1","v=2","v=3","v=10"},{0.25,0.75}],
 PlotStyle->{RGBColor[0.88,0.61,0.14],RGBColor[0.37,0.5,0.7],Purple, Darker[Red]},
 ImageSize->320,
 AxesLabel->{None,"PDF"}
 ]
```

Output

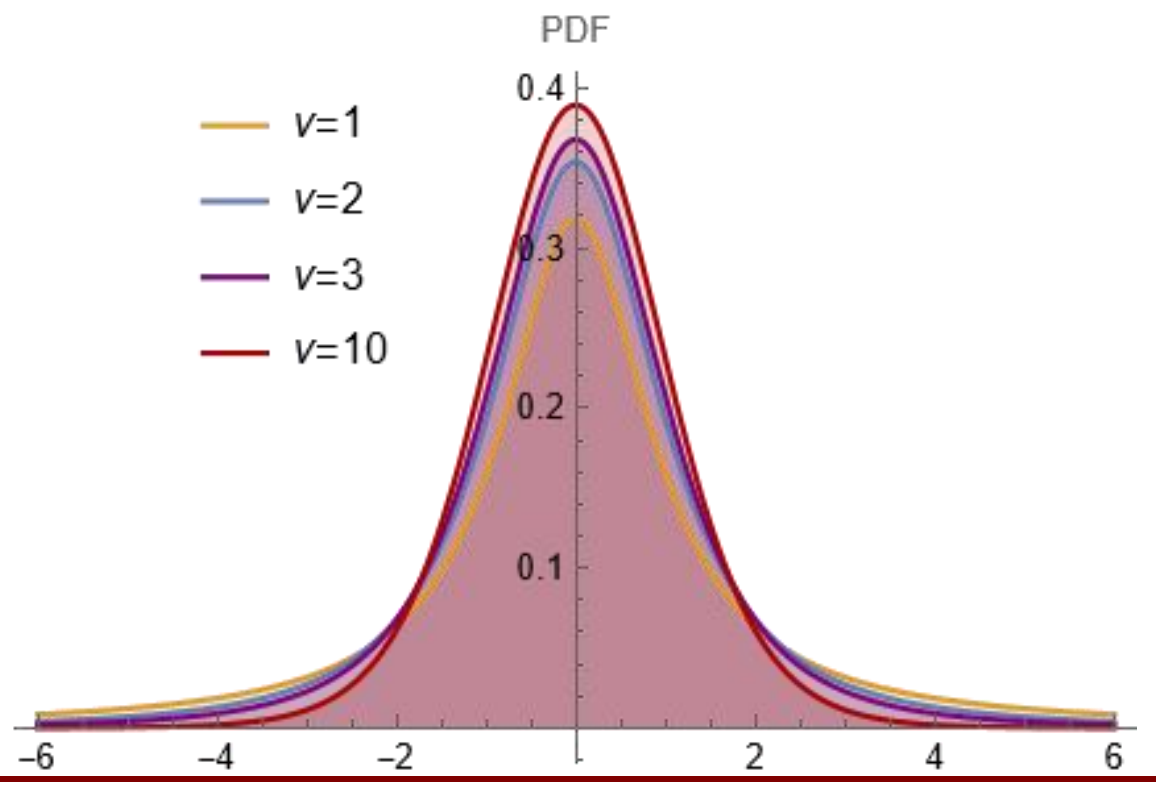

*Mathematica Examples 17.56*

Input
```
(* The code generates a plot of the cumulative distribution function (CDF) of the student t distribution with different values of the parameter v= (1, 2, 3, 10). The CDF is evaluated at various values of the random variable x between -6 and 6: *)

Plot[
 Evaluate[
  Table[
   CDF[
    StudentTDistribution[v],
    x
    ],
```

```
    {v,{1,2,3,10}}
    ]
   ],
  {x,-6,6},
  PlotRange->Automatic,
  Filling->Axis,
  PlotLegends->Placed[{"v=1","v=2","v=3","v=10"},{0.25,0.75}],
  PlotStyle->{RGBColor[0.88,0.61,0.14],RGBColor[0.37,0.5,0.7],Purple, Darker[Red]},
  ImageSize->320,
  AxesLabel->{None,"CDF"}
  ]
```

Output

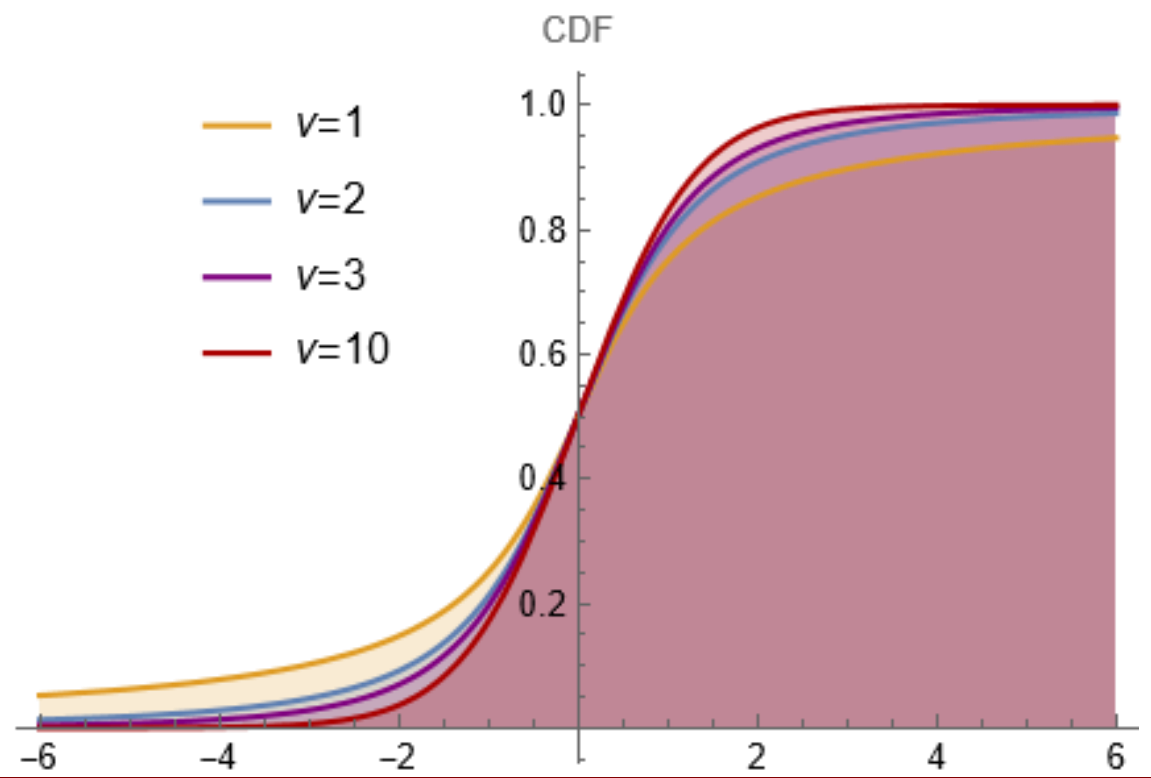

***Mathematica Examples 17.57***

Input

```
(* The code generates a histogram and a plot of the PDF for a student t distribution
with parameters v=3 and sample size 10000: *)

data=RandomVariate[
    StudentTDistribution[3],
    10^4
    ];

Show[
 Histogram[
  data,
  {-6,6,0.1},
  "PDF",
  ColorFunction->Function[{height},Opacity[height]],
  ChartStyle->Purple,
  ImageSize->320,
  AxesLabel->{None,"PDF"}
  ],

 Plot[
  PDF[
   StudentTDistribution[3],
   x
   ],
  {x,-6,6},
  PlotStyle->RGBColor[0.88,0.61,0.14],
  PlotRange->{-6,6}
  ]
 ]
```

Output

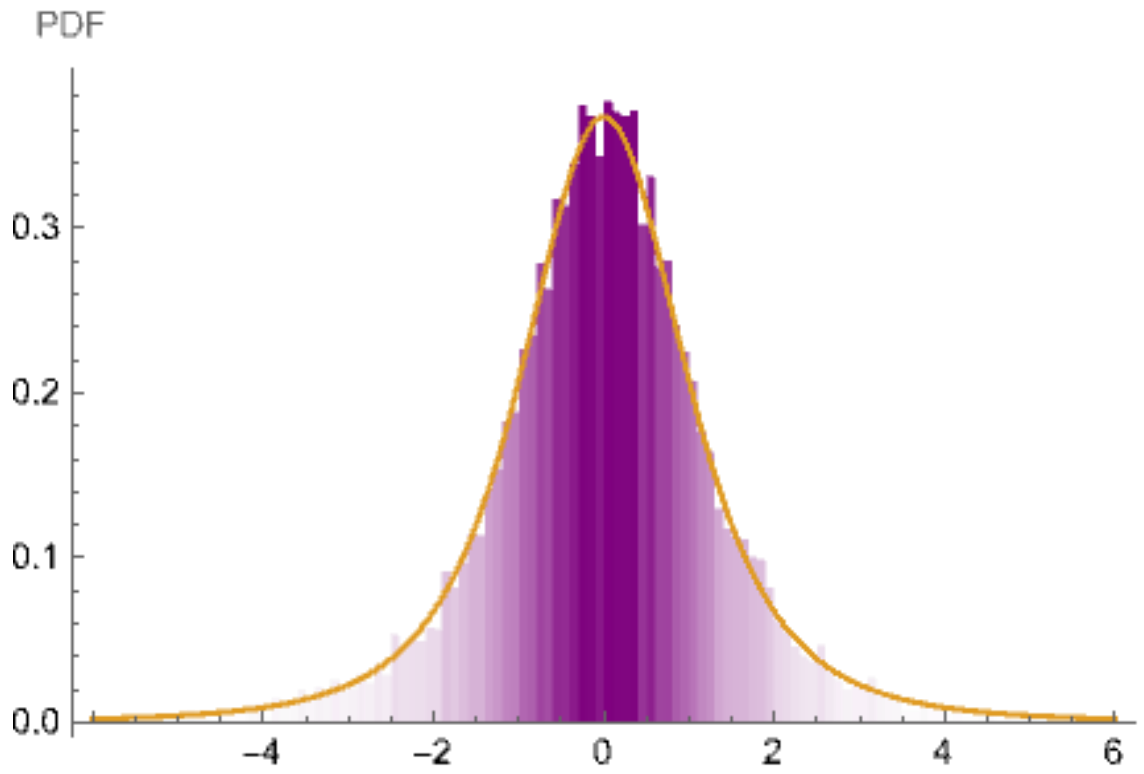

***Mathematica Examples 17.58***

Input (* The code creates a dynamic histogram of data and a plot of the PDF generated from a student t distribution using the Manipulate function. The Manipulate function creates interactive controls for the user to adjust the values of v and n, which are the parameter of the student t distribution and the sample size: *)

```
Manipulate[
 Module[
  {
   data=RandomVariate[
     StudentTDistribution[v],
     n
     ]
   },

  Show[
   Histogram[
    data,
    Automatic,
    "PDF",
    PlotRange->Automatic,
    ColorFunction->Function[{height},Opacity[height]],
    ImageSize->320,
    ChartStyle->Purple
    ],
   Plot[
    PDF[
     StudentTDistribution[v],
     x
     ],
    {x,-5,5},
    PlotRange->All,
    ColorFunction->"Rainbow"
    ]
   ]
  ],
 {{v,3,"v"},0.5,6,0.1},
 {{n,300,"n"},100,1000,10}
 ]
```

Output

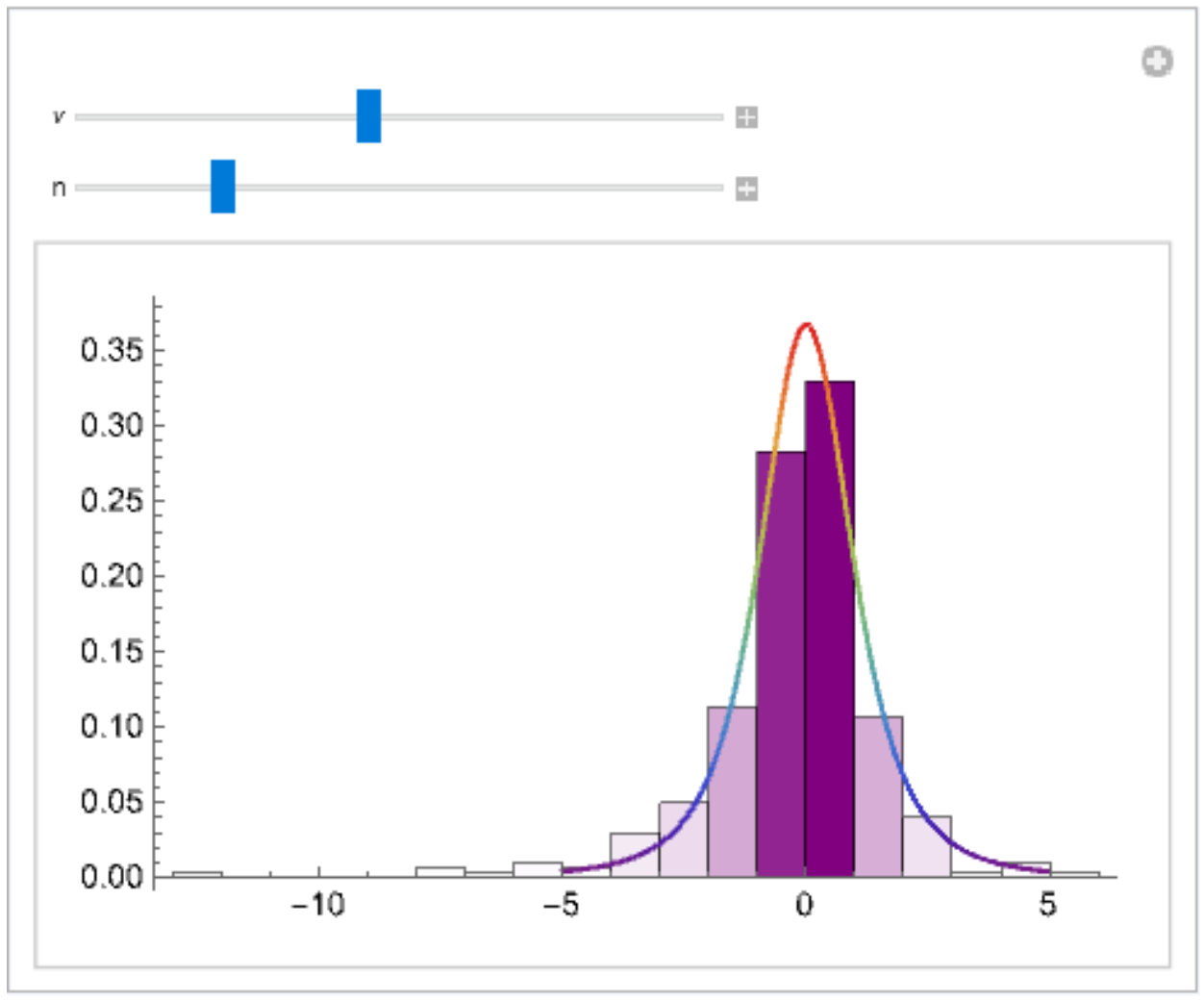

**Mathematica Examples 17.59**

Input

```
(* The code creates a plot of the CDF of a student t distribution using the Manipulate
function. The Manipulate function allows you to interactively change the values of
the parameters v: *)
Manipulate[
 Plot[
  CDF[
   StudentTDistribution[v],
   x
   ],
  {x,-5,5},
  Filling->Axis,
  FillingStyle->LightPurple,
  PlotRange->Automatic,
  AxesLabel->{"x","CDF"},
  ImageSize->320,
  PlotStyle->Purple,
  PlotLabel->Row[{"v = ",v}]
  ],
 {{v,1},1,6,0.1}
 ]
```

Output

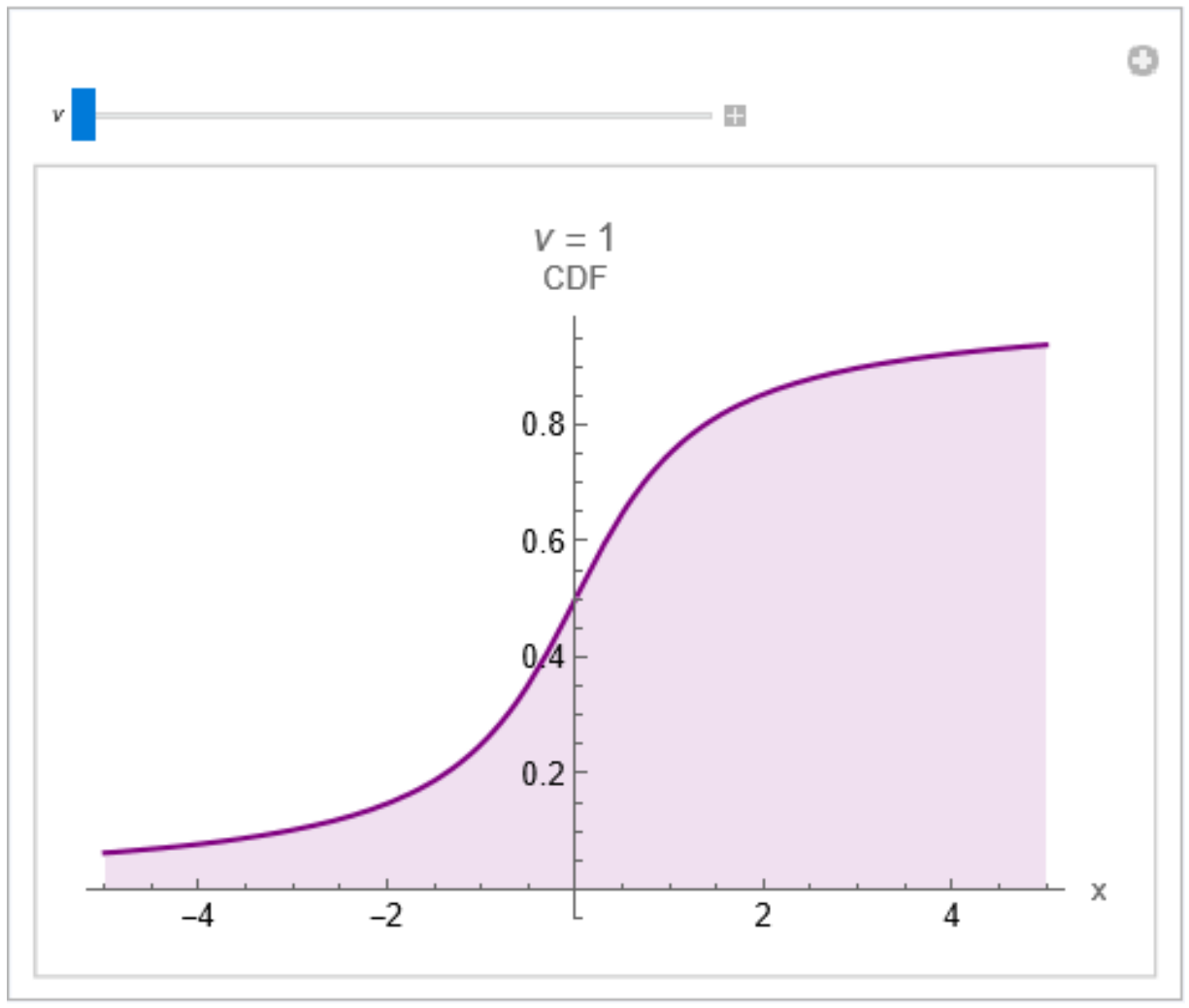

***Mathematica Examples 17.60***

Input

```
(* The code uses the Grid function to create a grid of two plots, one for the PDF and one for the CDF of student t distribution. The code uses slider controls to adjust the values of v: *)

Manipulate[
 Grid[
  {
   {Plot[
     PDF[
      StudentTDistribution[v],
      x
      ],
     {x,-7,7},
     PlotRange->Automatic,
     PlotStyle->{Purple,PointSize[0.03]},
     PlotLabel->"PDF of Student T distribution",
     AxesLabel->{"x","PDF"}
     ],
    Plot[
     CDF[
      StudentTDistribution[v],
      x
      ],
     {x,-7,7},
     PlotRange->Automatic,
     PlotStyle->{Purple,PointSize[0.03]},
     PlotLabel->"CDF of Student T distribution",
     AxesLabel->{"x","CDF"}
     ]
    }
   },
  Spacings->{5,5}
  ],
 {{v,1},1,6,0.1}
 ]
```

Output

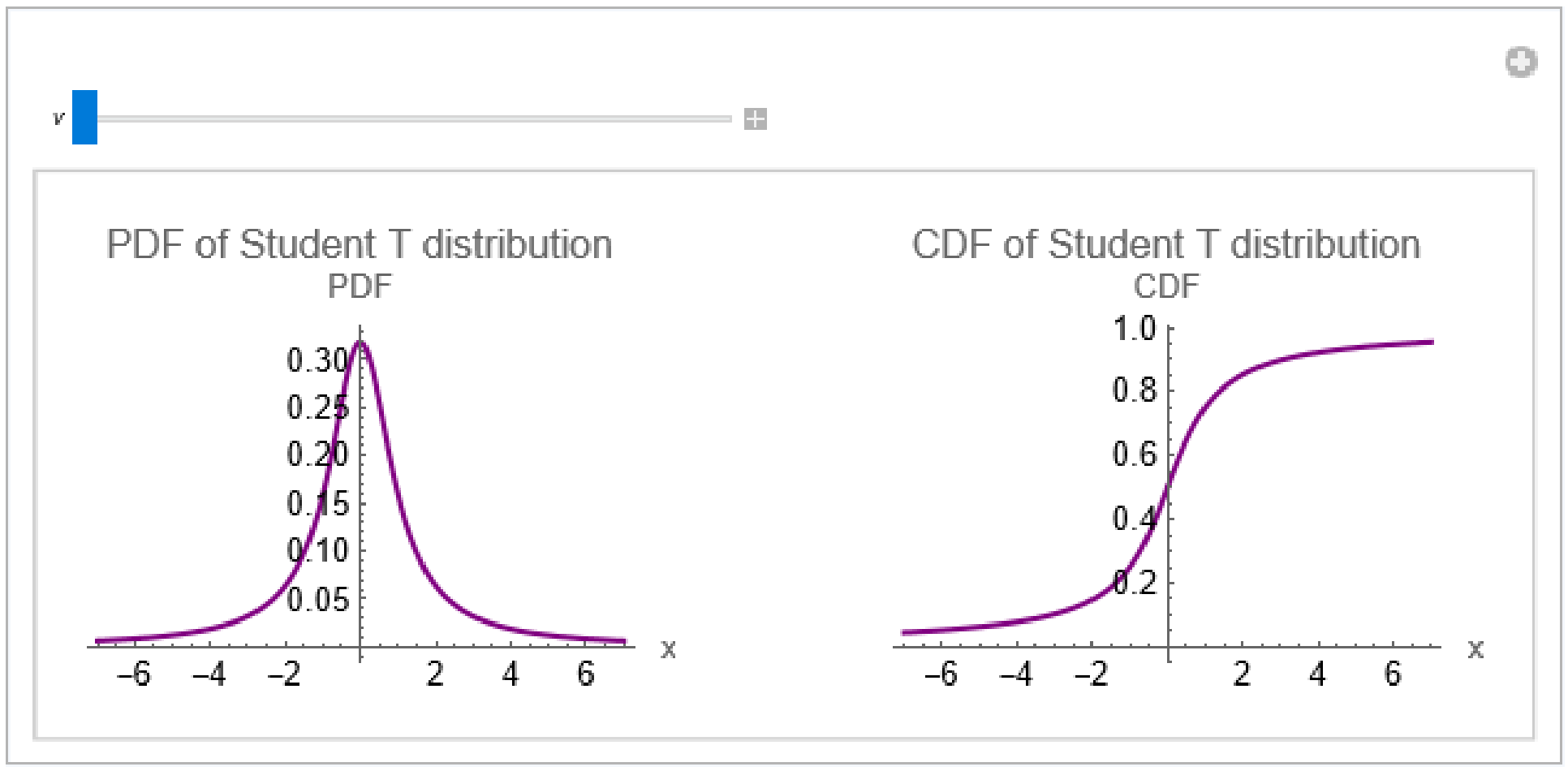

***Mathematica Examples 17.61***

Input

```
(* The code calculates and displays some descriptive statistics (mean, variance, standard deviation, kurtosis and skewness) for a student t distribution with parameter v: *)
```

```
Grid[
 Table[
  {
   statistics,
   FullSimplify[statistics[StudentTDistribution[v]]]
   },
  {statistics,{Mean,Variance,StandardDeviation,Kurtosis,Skewness}}
  ],
 ItemStyle->12,
 Alignment->{{Right,Left}},
 Frame->All,
 Spacings->{Automatic,0.8}
 ]
```

Output

| | |
|---|---|
| Mean | $\begin{cases} 0 & \nu > 1 \\ \text{Indeterminate} & \text{True} \end{cases}$ |
| Variance | $\begin{cases} \frac{\nu}{-2+\nu} & \nu > 2 \\ \text{Indeterminate} & \text{True} \end{cases}$ |
| StandardDeviation | $\begin{cases} \sqrt{\frac{\nu}{-2+\nu}} & \nu > 2 \\ \text{Indeterminate} & \text{True} \end{cases}$ |
| Kurtosis | $\begin{cases} 3+\frac{6}{-4+\nu} & \nu > 4 \\ \text{Indeterminate} & \text{True} \end{cases}$ |
| Skewness | $\begin{cases} 0 & \nu > 3 \\ \text{Indeterminate} & \text{True} \end{cases}$ |

***Mathematica Examples 17.62***

Input

```
(* The code calculates and displays some additional descriptive statistics (moments, central moments, and factorial moments) for a student t distribution with parameter v: *)

Grid[
 Table[
  {
   statistics,
   FullSimplify[statistics[StudentTDistribution[v],1]],
   FullSimplify[statistics[StudentTDistribution[v],2]]
   },
  {statistics,{Moment,CentralMoment,FactorialMoment}}
  ],
 ItemStyle->12,
 Alignment->{{Right,Left}},
 Frame->All,
 Spacings->{Automatic,0.8}
 ]
```

Output

| | | |
|---|---|---|
| Moment | $\begin{cases} 0 & 1 < \nu \\ \text{Indeterminate} & \text{True} \end{cases}$ | $\begin{cases} \nu/(-2+\nu) & \nu > 2 \\ \text{Indeterminate} & \text{True} \end{cases}$ |
| CentralMoment | $\begin{cases} 0 & 1 < \nu \\ \text{Indeterminate} & \text{True} \end{cases}$ | $\begin{cases} \nu/(-2+\nu) & \nu > 2 \\ \text{Indeterminate} & \text{True} \end{cases}$ |
| FactorialMoment | $\begin{cases} 0 & 1 < \nu \\ \text{Indeterminate} & \text{True} \end{cases}$ | $\begin{cases} \nu/(-2+\nu) & \nu > 2 \\ \text{Indeterminate} & \text{True} \end{cases}$ |

***Mathematica Examples 17.63***

Input

```
(* The code generates a dataset of 10000 observations from a student t distribution with parameter v=3. Then, it computes the sample mean and quartiles of the data, and plots a histogram of the data and plot of the PDF. Additionally, the code adds vertical lines to the plot corresponding to the sample mean and quartiles: *)
```

```
data=RandomVariate[
    StudentTDistribution[3],
    10000
    ];

mean=Mean[data];
quartiles=Quantile[
    data,
    {0.25,0.5,0.75}
    ];

Show[
 Histogram[
  data,
  Automatic,
  "PDF",
  Epilog->{
    Directive[Red,Thickness[0.006]],
    Line[{{mean,0},{mean,0.35}}],
    Directive[Green,Dashed],
    Line[{{quartiles[[1]],0},{quartiles[[1]],0.35}}],
    Line[{{quartiles[[2]],0},{quartiles[[2]],0.35}}],
    Line[{{quartiles[[3]],0},{quartiles[[3]],0.35}}]
    },
  ColorFunction->Function[{height},Opacity[height]],
  ImageSize->320,
  ChartStyle->Purple,
  PlotRange->Automatic
  ],
 Plot[
  PDF[
   StudentTDistribution[3],
   x
   ],
  {x,-7,7},
  PlotRange->Automatic,
  ImageSize->320,
  ColorFunction->"Rainbow"
  ]
 ]
```

Output

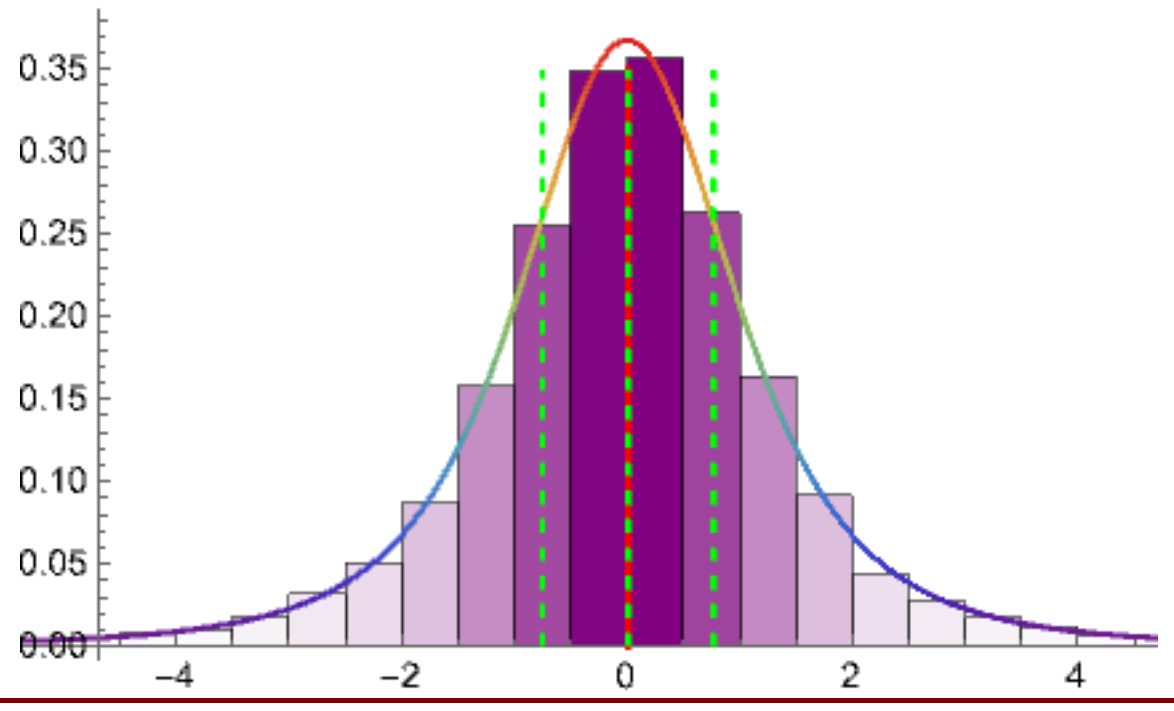

***Mathematica Examples 17.64***

Input

```
(* The code generates a random sample of size 10,000 from a student t distribution
with parameter v=3, estimates the distribution parameters using the
EstimatedDistribution function, and then compares the histogram of the sample with
```

```
the estimated PDF of the student t distribution using a histogram and a plot of the
PDF: *)

sampledata=RandomVariate[
    StudentTDistribution[3],
    10^4
    ];
(* Estimate the distribution parameters from sample data: *)
ed=EstimatedDistribution[
  sampledata,
  StudentTDistribution[v]
  ]
(* Compare a density histogram of the sample with the PDF of the estimated
distribution: *)
Show[
 Histogram[
  sampledata,
  Automatic,
  "PDF",
  ColorFunction->Function[{height},Opacity[height]],
  ChartStyle->Purple,
  ImageSize->320
  ],
 Plot[
  PDF[ed,x],
  {x,-7,7},
  ImageSize->320,
  ColorFunction->"Rainbow"
  ]
 ]
```

Output

```
StudentTDistribution[2.96133]
```

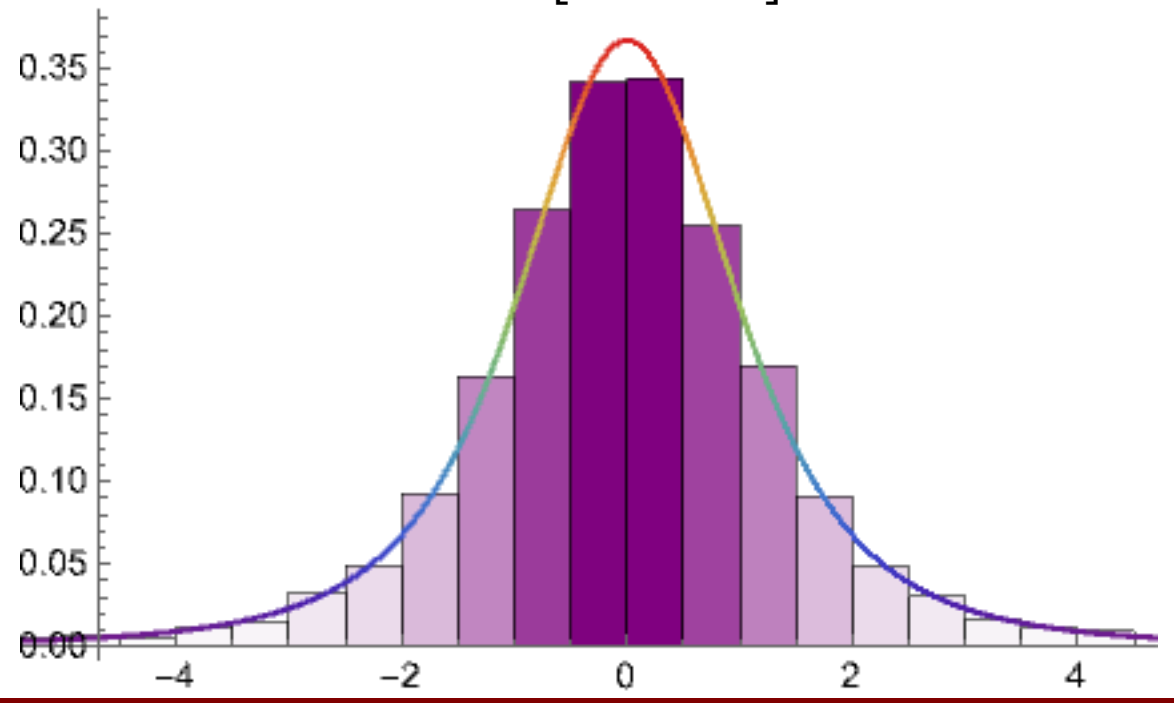

***Mathematica Examples 17.65***

Input

```
(* The code generates a 2D dataset with 1000 random points that follow a student t
distribution with v=3. The dataset is then used to create a row of three plots. The
first plot is a histogram of the X-axis values of the dataset. The second plot is a
histogram of the Y-axis values of the dataset. It is similar to the first plot, but
shows the distribution of the Y-axis values instead. The third plot is a scatter plot
of the dataset, with the X-axis values on the horizontal axis and the Y-axis values
on the vertical axis. Each point in the plot represents a pair of X and Y values from
the dataset: *)

data=RandomVariate[
    StudentTDistribution[3],
    {1000,2}
    ];
GraphicsRow[
```

```
    {
     Histogram[
      data[[All,1]],
      {0.1},
      ImageSize->170,
      PlotLabel->"X-axis",
      ColorFunction->Function[{height},Opacity[height]],
      ChartStyle->Purple
      ],
     Histogram[
      data[[All,2]],
      {0.1},
      ImageSize->170,
      PlotLabel->"Y-axis",
      ColorFunction->Function[{height},Opacity[height]],
      ChartStyle->Purple
      ],
     ListPlot[
      data,
      ImageSize->170,
      PlotStyle->{Purple,PointSize[0.015]},
      AspectRatio->1,
      Frame->True,
      Axes->False
      ]
     }
    ]
```

Output

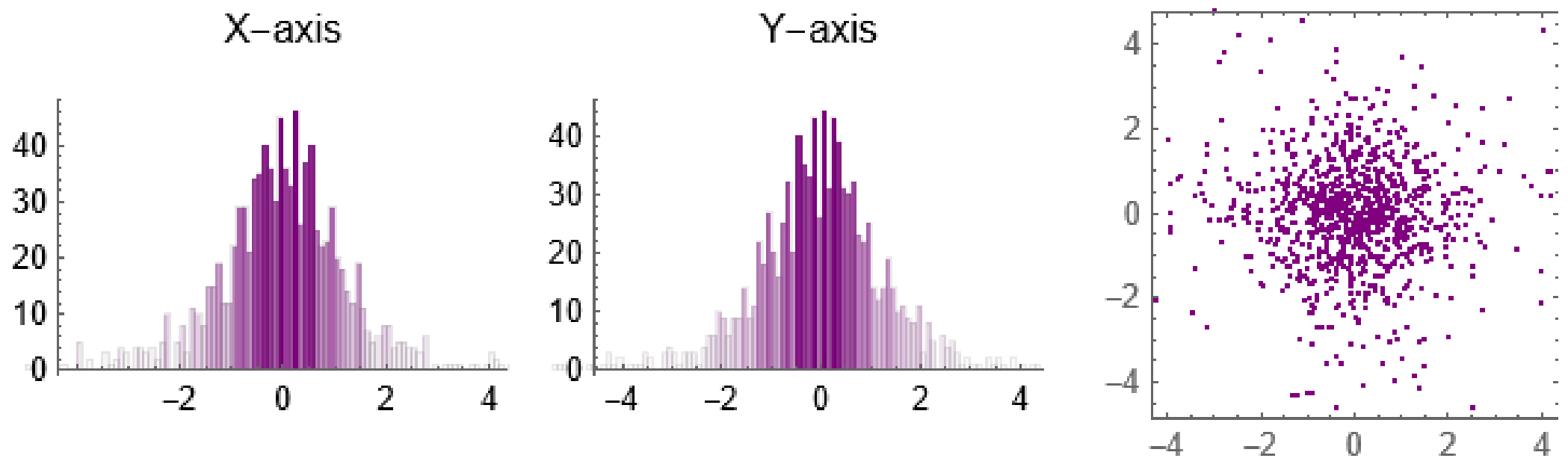

*Mathematica Examples 17.66*

Input

```
(* The code generates a set of random data points with a student t distribution with
v=3 in three dimensions, and then creates three histograms, one for each dimension,
showing the distribution of the points along that axis. Additionally, it creates a
3D scatter plot of the data points: *)

data=RandomVariate[
   StudentTDistribution[3],
   {1000,3}
   ];

GraphicsGrid[
 {
  {
   Histogram[
    data[[All,1]],
    Automatic,
    "PDF",
```

```
            PlotLabel->"X-axis",
            ColorFunction->Function[{height},Opacity[height]],
            ChartStyle->Purple
            ],
           Histogram[
            data[[All,2]],
            Automatic,
            "PDF",
            PlotLabel->"Y-axis",
            ColorFunction->Function[{height},Opacity[height]],
            ChartStyle->Purple
            ],
           Histogram[
            data[[All,3]],
            Automatic,
            "PDF",
            PlotLabel->"Z-axis",
            ColorFunction->Function[{height},Opacity[height]],
            ChartStyle->Purple
            ],
           ListPointPlot3D[
            data,
            BoxRatios->{1,1,1},
            PlotStyle->{Purple,PointSize[0.015]}
            ]
           }
          }
         ]
```

Output

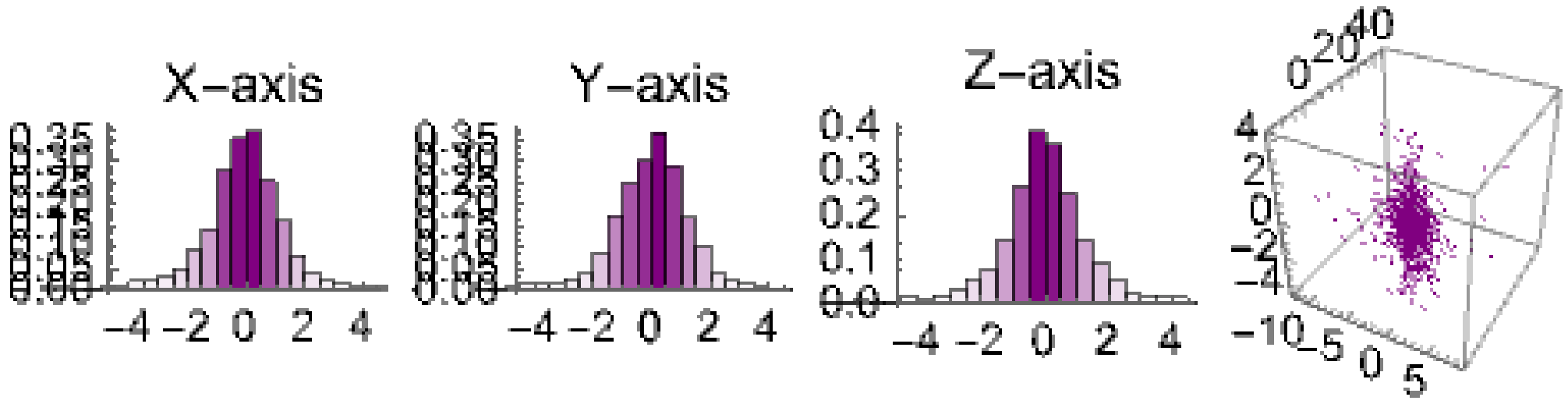

***Mathematica Examples 17.67***

Input

```
(* The code generates a 3D scatter plot of a student t distribution points with v=15,
where the x-axis is red, y-axis is green, and z-axis is blue: *)

data=RandomVariate[
   StudentTDistribution[15],
   {2000,3}
   ];
Graphics3D[
 {
  {PointSize[0.006],Purple,Opacity[0.6],Point[data]},
  Thin,
  {Red,Opacity[0.4],Line[{{#,0,0},{#,0,-0.5}}]&/@data[[All,1]]},
  Thin,
  {Green,Opacity[0.4],Line[{{0,#,0},{0,#,-0.5}}]&/@data[[All,2]]},
  Thin,
  {Blue,Opacity[0.4],Line[{{0,0,#},{0,-0.5,#}}]&/@data[[All,3]]}
  },
 BoxRatios->{1,1,1},
 Axes->True,
```

```
            AxesLabel->{"X","Y","Z"},
            ImageSize->320
            ]
```

Output

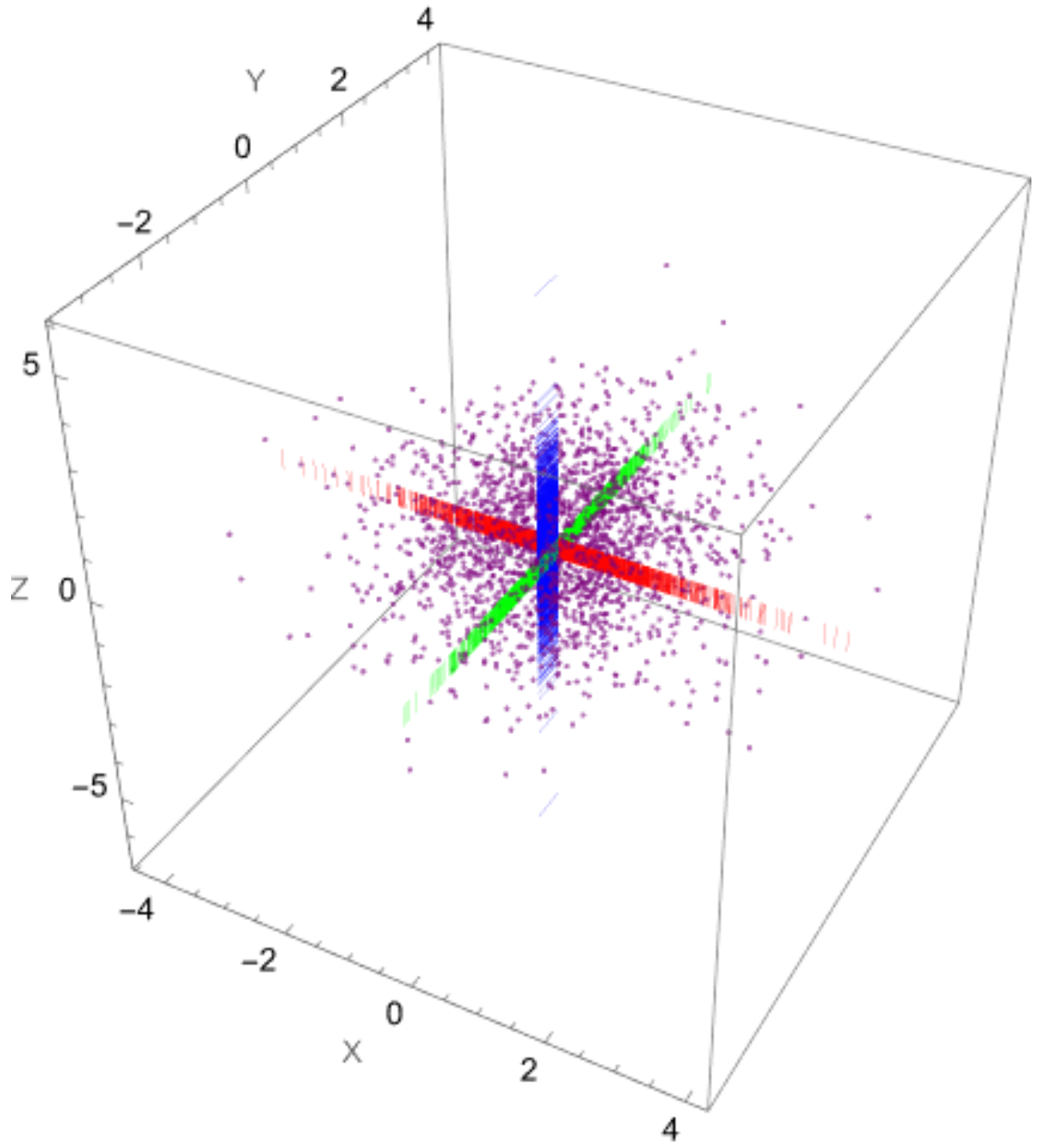

***Mathematica Examples 17.68***

Input

```
(* The code demonstrates a common technique in statistics and data analysis, which
is the use of random sampling to estimate population parameters. The code generates
random samples from student t distribution with v=3,and then using these samples to
estimate the parameters of another student t distribution with unknown v. This process
is repeated 20 times, resulting in 20 different estimated distributions. The code
also visualizes the resulting estimated distributions using the PDF function. The
code plots the PDFs of these estimated distributions using the PDF function and the
estimated parameters. The plot shows the PDFs in a range from -5 to 5. The code also
generates a list plot of 2 sets of random samples from the student t distribution
with v=3. The plot shows the 100 random points generated from two random samples.
The code generates also a histogram of the PDF for student t distribution of the two
samples: *)

estim0distributions=Table[
  dist=StudentTDistribution[3];

  sampledata=RandomVariate[
    dist,
    100
    ];

  ed=EstimatedDistribution[
    sampledata,
    StudentTDistribution[v]
    ],
  {i,1,20}
  ]

pdf0ed=Table[
   PDF[estim0distributions[[i]],x],
```

```
    {i,1,20}
    ];

(* Visualizes the resulting estimated distributions. *)
Plot[
 pdf0ed,
 {x,-5,5},
 PlotRange->Full,
 ImageSize->400,
 PlotStyle->Directive[Purple,Opacity[0.3],Thickness[0.002]]
 ]

(* Visualizes 100 random points generated from two random samples. *)
table =Table[
    dist=StudentTDistribution[3];
    sampledata=RandomVariate[
      dist,
      100],
    {i,1,2}
    ];

ListPlot[
 table,
 ImageSize->320,
 Filling->Axis,
 PlotStyle->Directive[Opacity[0.5],Thickness[0.003]]
 ]

Histogram[
 table,
 30,
 LabelingFunction->Above,
 ChartLegends->{"Sample 1","Sample 2"},
 ChartStyle->{Directive[Opacity[0.2],Red],Directive[Opacity[0.2],Purple]},
 ImageSize->320
 ]
```

Output {StudentTDistribution[3.40769],StudentTDistribution[3.81502],StudentTDistribution[3.23998],StudentTDistribution[3.01578],StudentTDistribution[3.3598],StudentTDistribution[3.64623],StudentTDistribution[3.47706],StudentTDistribution[2.32619],StudentTDistribution[4.31413],StudentTDistribution[2.94904],StudentTDistribution[3.825],StudentTDistribution[3.52121],StudentTDistribution[3.25712],StudentTDistribution[3.70916],StudentTDistribution[4.35236],StudentTDistribution[7.60725],StudentTDistribution[2.90244],StudentTDistribution[2.24026],StudentTDistribution[5.8475],StudentTDistribution[2.98665]}

Output

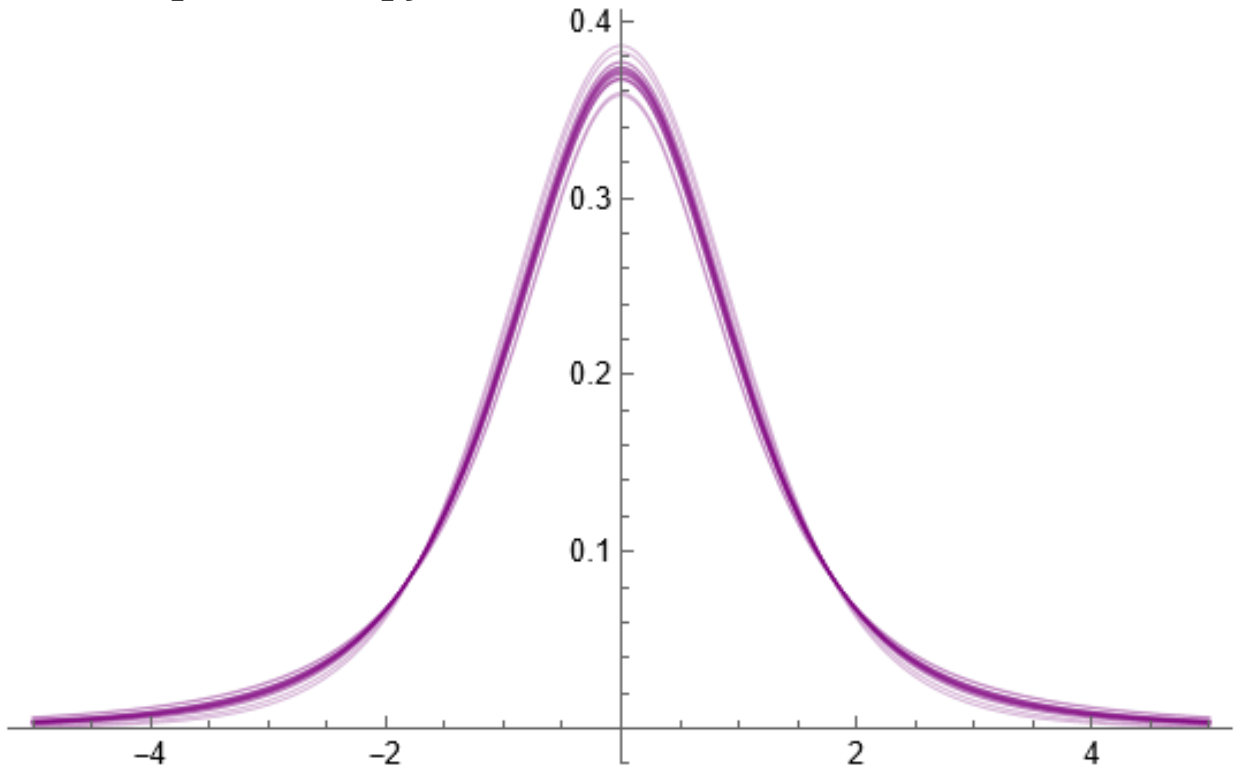

Output

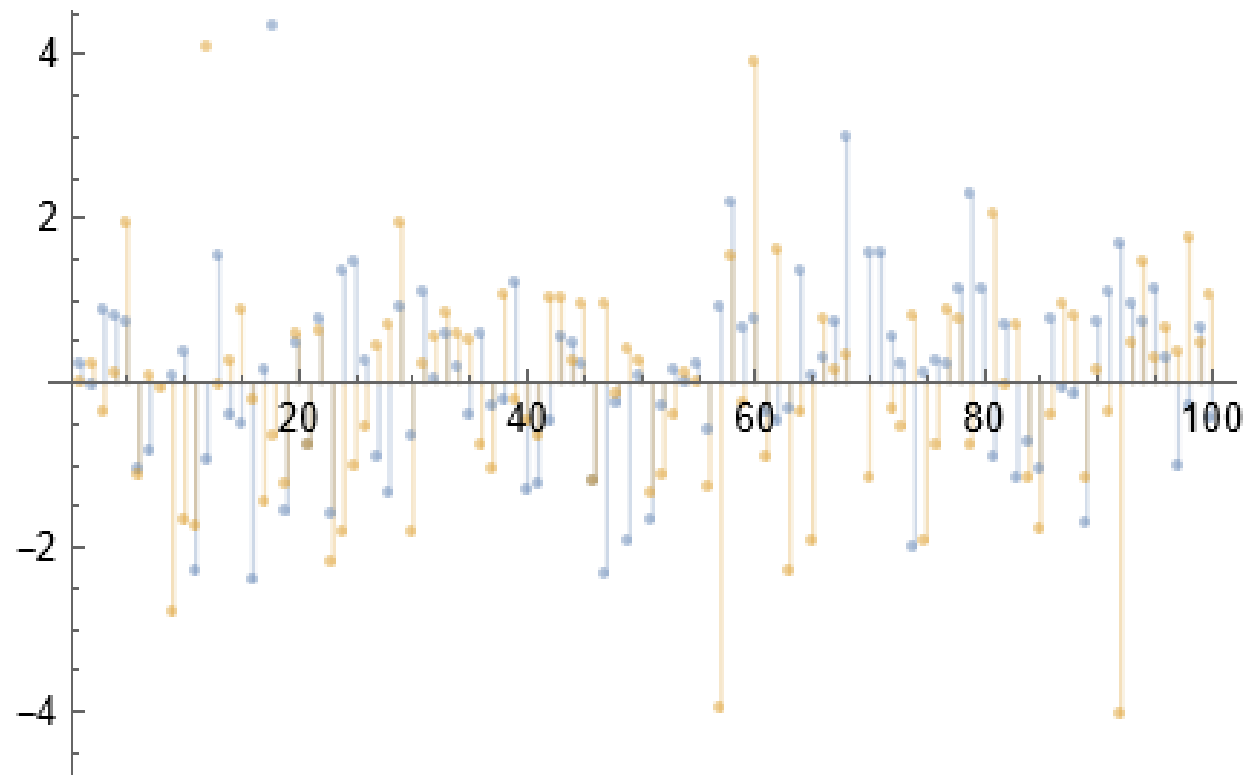

Output

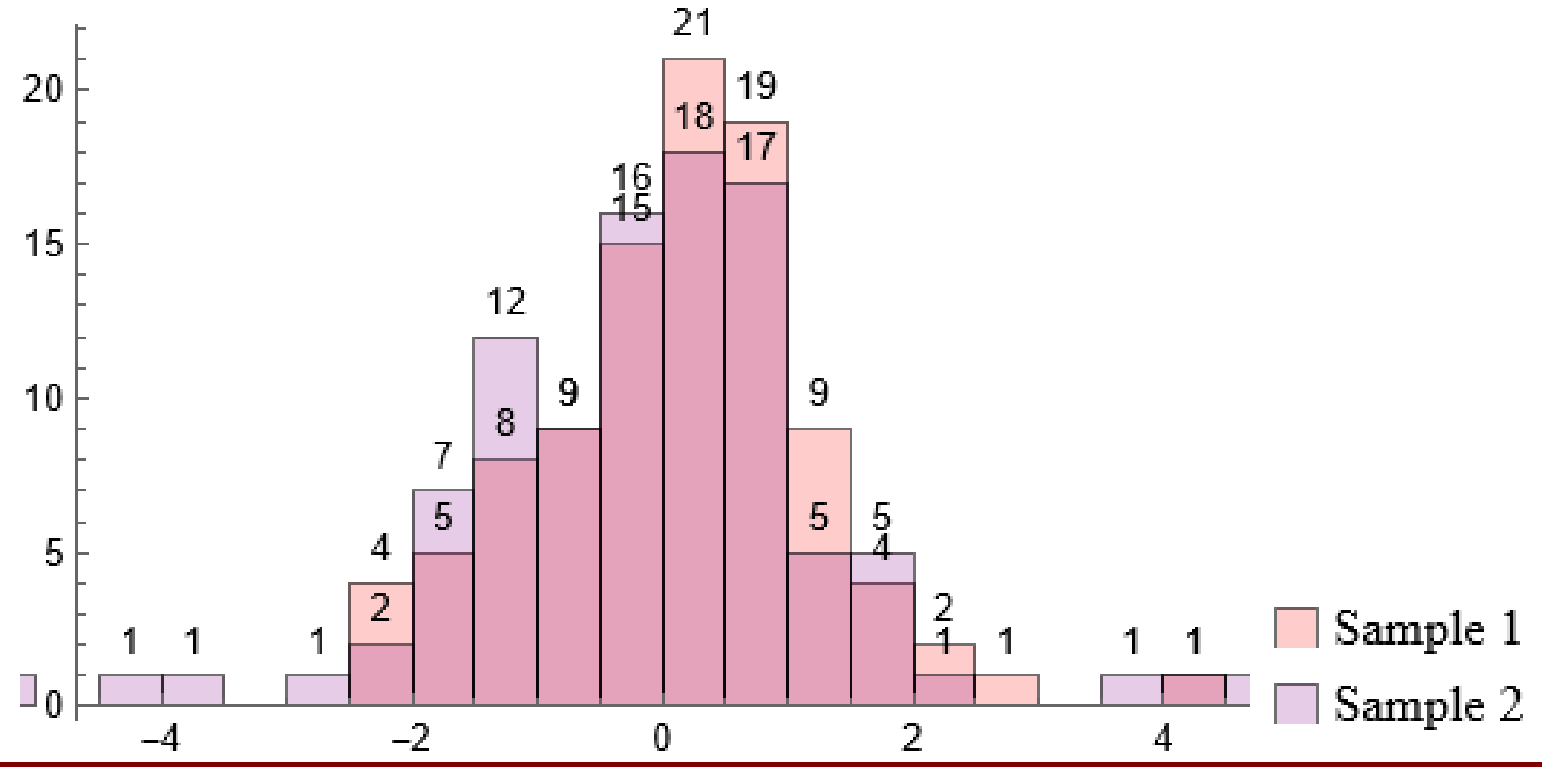

***Mathematica Examples 17.69***

Input

```
(* The code generates and compares the means of random samples drawn from a student
t distribution with the given parameter v. The code uses the Manipulate function to
create a user interface with sliders to adjust the values of the parameter v, number
of samples, and sample size. By varying the values of "Number of Samples" and "Sample
Size" sliders, the code allows the user to explore how changing these parameters
affects the means of the random samples: *)

Manipulate[
 Module[
  {means,dist},

  dist=StudentTDistribution[v];

  means=Map[
    Mean,
    RandomVariate[dist,{n,samples}]
    ];
  m=N[Mean[means]];

  ListPlot[
   {means,{{0,m},{n,m}}},
   Joined->{False,True},
   PlotRange->{-5,5},
   Filling->Axis,
   PlotStyle->{Purple,Red},
   AxesLabel->{"Number of Samples","Sample Mean"},
   PlotLabel->Row[{"v = ",v}],
   ImageSize->320
   ]
  ],
 {{v,2,"v"},1,6,0.1},
```

```
{{n,30,"Number of Samples"},3,50,1},
{{samples,10,"Sample Size"},1,50,1},
TrackedSymbols:>{n,samples,v}
]
```

Output

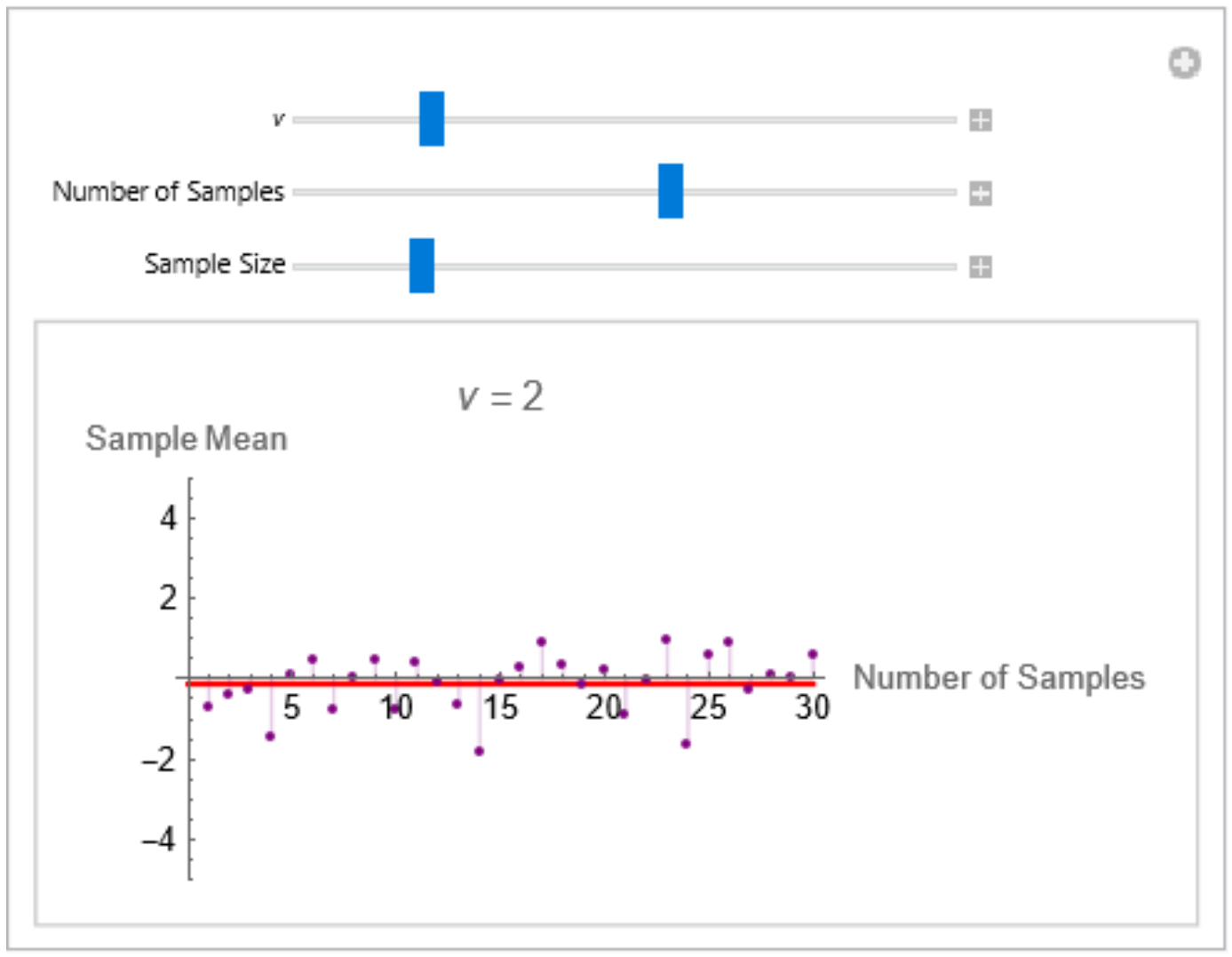

*Mathematica Examples 17.70*

Input

```
(* The code is designed to compare two student t distributions. It does this by generating random samples from each distribution and displaying them in a histogram, as well as plotting the PDFs of the two distributions. The code allows the user to manipulate the parameters v1 and v2 of both student t distributions through the sliders for v1 and v2. By changing these parameters, the user can see how the distributions change and how they compare to each other. The histograms display the sample data for each distribution, with the first histogram showing the sample data for the first student t distribution and the second histogram showing the sample data for the second student t distribution. The histograms are overlaid on each other, with the opacity of each histogram set to 0.2 to make it easier to see where the data overlap. The PDFs of the two distributions are also plotted on the same graph, with the first distribution shown in blue and the second distribution shown in red. The legend indicates which color corresponds to which distribution. By looking at the histograms and the PDFs, the user can compare the two student t distributions and see how they differ in terms of shape, scale, and overlap of their sample data: *)
Manipulate[
 Module[
  {dist1,dist2,data1,data2},
  SeedRandom[seed];
  dist1=StudentTDistribution[v1];
  dist2=StudentTDistribution[v2];
  data1=RandomVariate[dist1,n];
  data2=RandomVariate[dist2,n];
  Column[
   {
    Show[
     ListPlot[
      data1,
      ImageSize->320,
      PlotStyle->Blue,
      PlotRange->{-10,10}
      ],
     ListPlot[
      data2,
      ImageSize->320,
```

```
          PlotStyle->Red,
          PlotRange->{-10,10}
          ]
         ],
       Show[
        Plot[
         {PDF[dist1,x],PDF[dist2,x]},
         {x,Min[{data1,data2}],Max[{data1,data2}]},
         PlotLegends->{"Distribution 1","Distribution 2"},
         PlotRange->All,
         PlotStyle->{Blue,Red},
         ImageSize->320
         ],
        Histogram[
         {data1,data2},
         Automatic,
         "PDF",
         ChartLegends->{"sample data1","sample data2"},
         ChartStyle->{Directive[Opacity[0.2],Red],Directive[Opacity[0.2],Purple]},
         ImageSize->320
         ]
        ]
       }
     ]
    ],
   {{v1,6},0.5,6,0.1},
   {{v2,3},2,7,0.1},
   {{n,500},{100,500,1000,2000}},
   {{seed,1234},ControlType->None}
   ]
```

Output

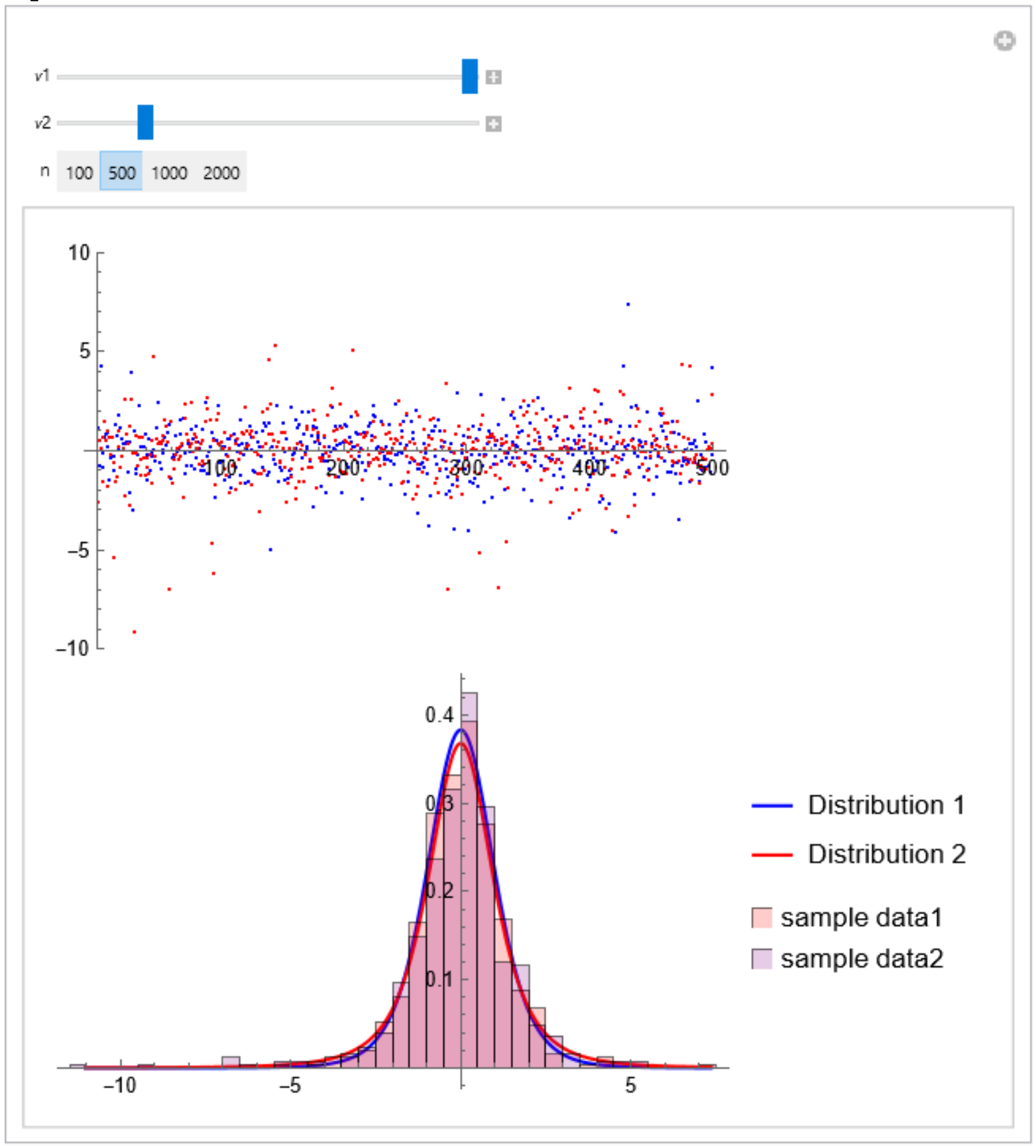

### Mathematica Examples 17.71

Input
```
(* The Student distribution converges to the standard normal distribution as v tends to infinity: *)
Limit[
 PDF[StudentTDistribution[v],x],
 v->∞
 ]

PDF[NormalDistribution[0,1],x]
```

Output $\frac{e^{-\frac{x^2}{2}}}{\sqrt{2\pi}}$

Output $\frac{e^{-\frac{x^2}{2}}}{\sqrt{2\pi}}$

### Mathematica Examples 17.72

Input
```
(* StudentTDistribution[1] is equivalent to CauchyDistribution[0,1]: *)
PDF[StudentTDistribution[1],x]
PDF[CauchyDistribution[0,1],x]
```

Output `1/(π (1+x^2))`

Output `1/(π (1+x^2))`

### Mathematica Examples 17.73

Input
```
(* Student distribution can be obtained from normal distribution and chi square distribution: *)

dist=TransformedDistribution[
    z Sqrt[v/w],
    {z\[Distributed]NormalDistribution[],w\[Distributed]ChiSquareDistribution[v]}
    ];

PDF[dist,x]//FullSimplify[#,v>0]&
FullSimplify[
 PDF[StudentTDistribution[v],x]//FunctionExpand//PowerExpand
 ]
```

Output $\begin{cases} \frac{\nu^{\nu/2}(x^2+\nu)^{-\frac{1}{2}-\frac{\nu}{2}}\text{Gamma}[(1+\nu)/2]}{\sqrt{\pi}\text{Gamma}[\nu/2]} & x! = 0 \\ 0 & \text{True} \end{cases}$

Output $\frac{\nu^{\nu/2}(x^2+\nu)^{-\frac{1}{2}-\frac{\nu}{2}}\text{Gamma}[(1+\nu)/2]}{\sqrt{\pi}\text{Gamma}[\nu/2]}$

### Mathematica Examples 17.74

Input
```
(* A square of a Student distributed variable has FRatioDistribution: *)

TransformedDistribution[x^2,x\[Distributed]StudentTDistribution[v]]
(* An inverse square of Student distributed variable has FRatioDistribution: *)
TransformedDistribution[1/z^2,z\[Distributed]StudentTDistribution[v]]
```

Output `FRatioDistribution[1,v]`

Output `FRatioDistribution[v,1]`

***Mathematica Examples 17.75***

```
Input    (* Compute the probability of t<=1.5 for a t distribution with 20 degrees of freedom:
         *)
         n=20;
         dist=StudentTDistribution[n];

         p1=N[CDF[dist,1.5]];
         (* or*)
         p2=NProbability[t<=1.5,t\[Distributed]dist];
         {p1,p2}

         (* Compute the probability of t>1.5 for the same distribution: *)
         p3=N[1-CDF[dist,1.5]];
         (* or*)
         p4=N[CDF[dist,-1.5]];
         (* or*)
         p5=NProbability[t>1.5,t\[Distributed]dist];
         {p3,p4,p5}

         (* Compute the probability of \[LeftBracketingBar]t\[RightBracketingBar]<1.5: *)
         p6=N[CDF[dist,1.5]-CDF[dist,-1.5]];
         (* or*)
         p7=N[2 CDF[dist,1.5]-1];
         (* or*)
         p8=N[1-2 CDF[dist,-1.5]];
         (* or*)
         p9=NProbability[-1.5<t<1.5,t\[Distributed]dist];
         {p6,p7,p8,p9}

         (* Compute the probability of \[LeftBracketingBar]t\[RightBracketingBar]>1.5: *)
         p10=N[CDF[dist,-1.5]+(1-CDF[dist,1.5])];
         (* or*)
         p11=N[1-(CDF[dist,1.5]-CDF[dist,-1.5])];
         (* or*)
         p12=N[2-2 CDF[dist,1.5]];
         (* or*)
         p13=N[2 CDF[dist,-1.5]];
         (* or*)
         p14=NProbability[Abs[t]>1.5,t\[Distributed]dist];
         {p10,p11,p12,p13,p14}

Output   {0.925382,0.925382}
Output   {0.0746179,0.0746179,0.0746179}
Output   {0.850764,0.850764,0.850764,0.850764}
Output   {0.149236,0.149236,0.149236,0.149236,0.149236}
```

***Mathematica Examples 17.76***

```
Input    (* The code calculates quantiles (also known as percentiles) and inverse CDF of a
         Student's t-distribution with degrees of freedom equal to 20. The Quantile function
         calculates the quantiles of the t-distribution at probability levels of 0.005,
         0.995,0.025, 0.975, 0.050, and 0.950, which are the extreme 0.5% tails, the central
         99% interval, the extreme 2.5% tails, the central 95% interval, the 5% left tail and
         the 5% right tail of the distribution, respectively. The InverseCDF function
         calculates the inverse CDF of the t-distribution at the same probability levels: *)

         n=20 (* degrees of freedom. *);

         N[
          Quantile[
           StudentTDistribution[n],
```

```
            {0.005,0.995,0.025,0.975,0.050,0.950}
            ]
           ]

          InverseCDF[
           StudentTDistribution[n],
           {0.005,0.995,0.025,0.975,0.050,0.950}
           ]

Output    {-2.84534,2.84534,-2.08596,2.08596,-1.72472,1.72472}
Output    {-2.84534,2.84534,-2.08596,2.08596,-1.72472,1.72472}
```

# UNIT 17.5

# SAMPLING DISTRIBUTIONS OF MEANS FOR SMALL SAMPLE WITH STUDENT DISTRIBUTION

***Mathematica Examples 17.77***

Input

```
(* The code demonstrates the generation and visualization of the t-distribution with a given number of degrees of freedom (n). It begins by generating 10,000 random variables from the standard normal and chi-squared distributions. These variables are then used to compute the t-variable by dividing the standard normal variables by the square root of the chi-squared variables divided by n. The code proceeds to create a histogram of the t-variable using the Histogram function. The histogram is configured to represent the probability density function (PDF) by setting the option "PDF". Additionally, the code plots the PDF of the t-distribution using the Plot function. It uses the built-in StudentTDistribution function to calculate the PDF. Finally, the code combines the histogram and the PDF plot using the Show function. This allows both visualizations to be displayed together, facilitating a comparison between the empirical distribution represented by the histogram and the theoretical distribution represented by the PDF plot: *)

(* Set the degrees of freedom: *)
n=5;

(* Generate random variables from the standard normal and chi-squared distributions: *)
z=RandomVariate[NormalDistribution[0,1],10000];
q=RandomVariate[ChiSquareDistribution[n],10000];

(* Compute the t-variable: *)
t=z/Sqrt[q/n];

(* Plot a histogram of the t-variable: *)
Histogram[
 t,
 {0.2},
 "PDF",
 Frame->True,
 FrameLabel->{"t","PDF"},
 PlotLabel->"t-Distribution with n Degrees of Freedom",
 ColorFunction->Function[Opacity[0.7]],
 ImageSize->250,
 ChartStyle->Purple
 ]

(* Plot the PDF of the t-distribution: *)
Plot[
 PDF[StudentTDistribution[n],x],
 {x,-4,4},
 PlotRange->All,
 Frame->True,
 FrameLabel->{"t","PDF"},
 PlotLabel->"t-Distribution with n Degrees of Freedom",
 PlotStyle->Darker[Red],
```

```
  ImageSize->250
  ]
 Show[
  Histogram[
   t,
   {0.2},
   "PDF",
   Frame->True,
   FrameLabel->{"t","PDF"},
   PlotLabel->"t-Distribution with n Degrees of Freedom",
   ColorFunction->Function[Opacity[0.7]],
   ImageSize->250,
   ChartStyle->Purple
   ],
  Plot[
   PDF[StudentTDistribution[n],x],
   {x,-4,4},
   PlotRange->All,
   Frame->True,
   FrameLabel->{"t","PDF"},
   PlotLabel->"t-Distribution with n Degrees of Freedom",
   PlotStyle->Darker[Red]
   ]
  ]
```

Output

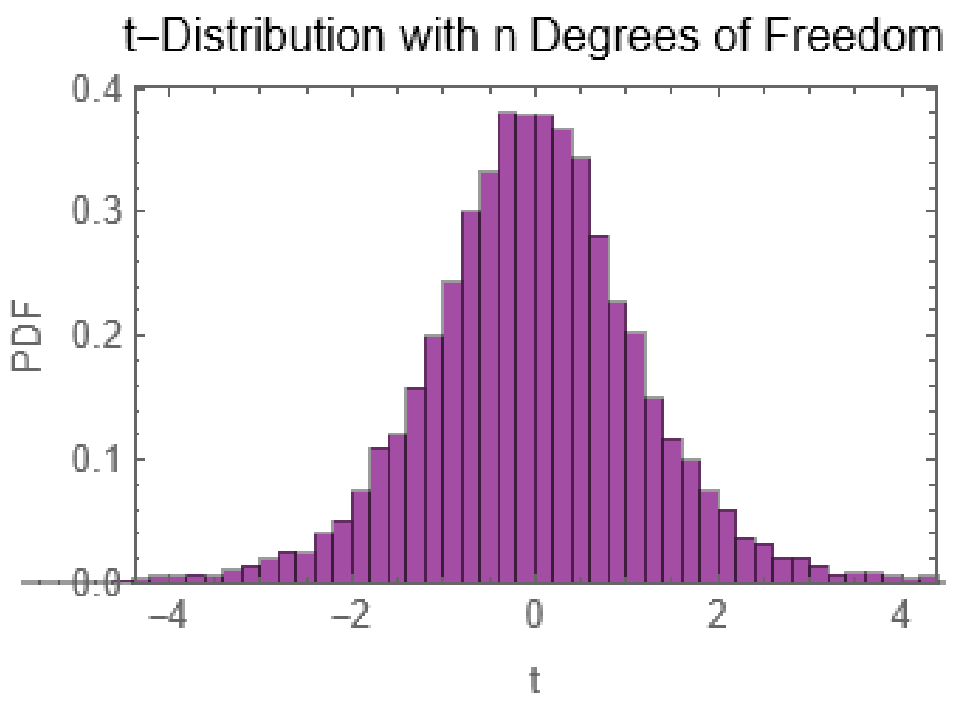

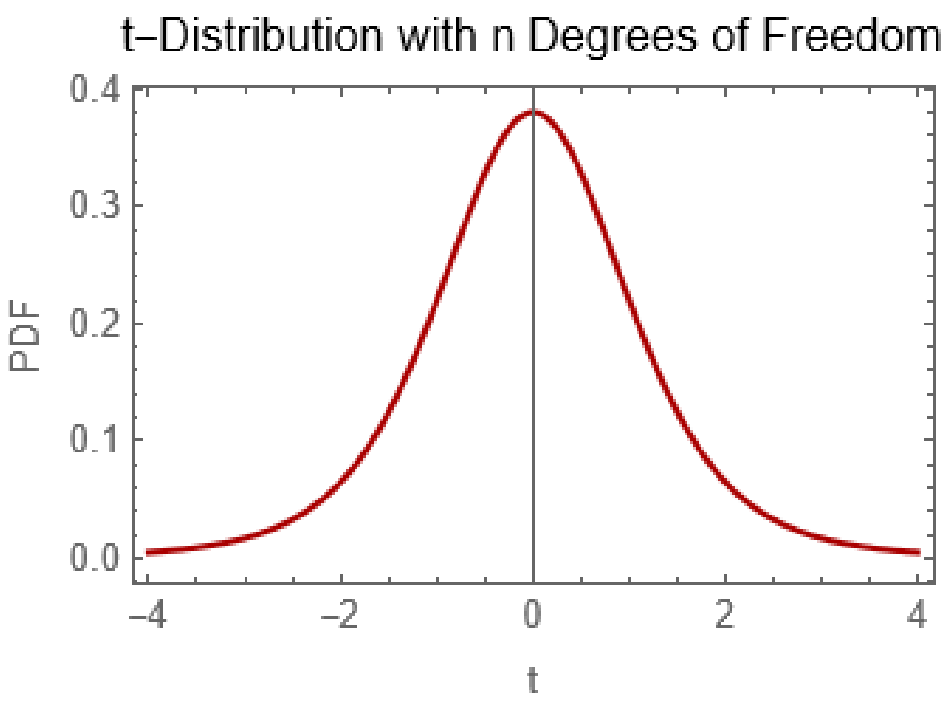

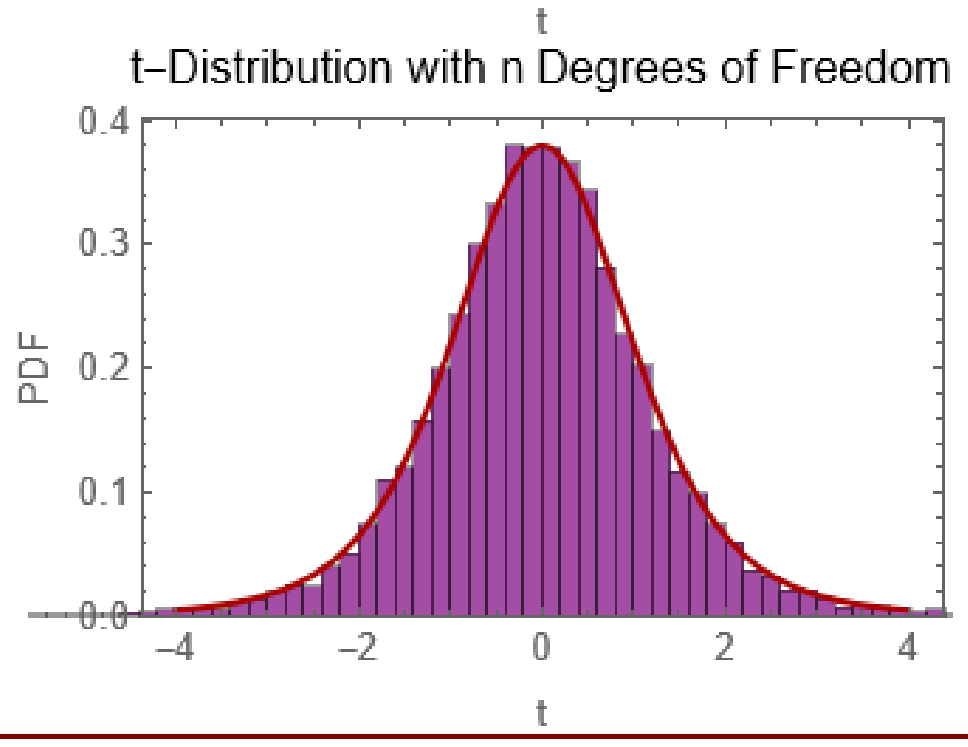

***Mathematica Examples 17.78***

Input

```
(* The code generates a population dataset of 10,000 random variables from a standard
normal distribution. It then repeats the sampling process 1000 times, each time
taking a random sample of size n (specified as 10) from the population. The sample
means and sample standard deviations are computed for each sample using the Map
function, which applies the Mean and StandardDeviation functions to each sample. The
t-variables are then calculated by subtracting the population mean (assumed to be 0)
from the sample means, multiplying by the square root of n, and dividing by the
sample standard deviations. The code proceeds to display a visualization that combines
a histogram of the t-variables and the PDF plot of the t-distribution. The PDF plot
```

```
is generated using the Plot function, utilizing the StudentTDistribution function
with n-1 degrees of freedom. Finally, the Show function is used to combine the
histogram and the PDF plot into a single visualization: *)

(* Generate population data: *)
population=RandomVariate[
    NormalDistribution[0,1],
    10000
    ];
populationMean=Mean[population];
(* Repeat the sampling process 1000 times: *)
n=10;
m=1000;
samples=Table[
    RandomSample[population,n],
    {i,1,m}
    ];

(* Compute the sample means: *)
xbar=Map[Mean,samples];
stanDev=Map[StandardDeviation,samples];

(* Compute the t-variables: *)
t=Sqrt[n]*(xbar-0)/stanDev;

Show[
 (* Plot a histogram of the t-variables: *)
 Histogram[
  t,
  Automatic,
  "PDF",
  Frame->True,
  FrameLabel->{"t","PDF"},
  PlotLabel->"t-Distribution with n Degrees of Freedom",
  ColorFunction->Function[Opacity[0.7]],
  ImageSize->250,
  ChartStyle->Purple
  ],
 (* Plot the PDF of the t-distribution: *)
 Plot[
  PDF[StudentTDistribution[n-1],x],
  {x,-5,5},
  PlotRange->All,
  Frame->True,
  PlotStyle->Darker[Red]
  ]
 ]
```

Output

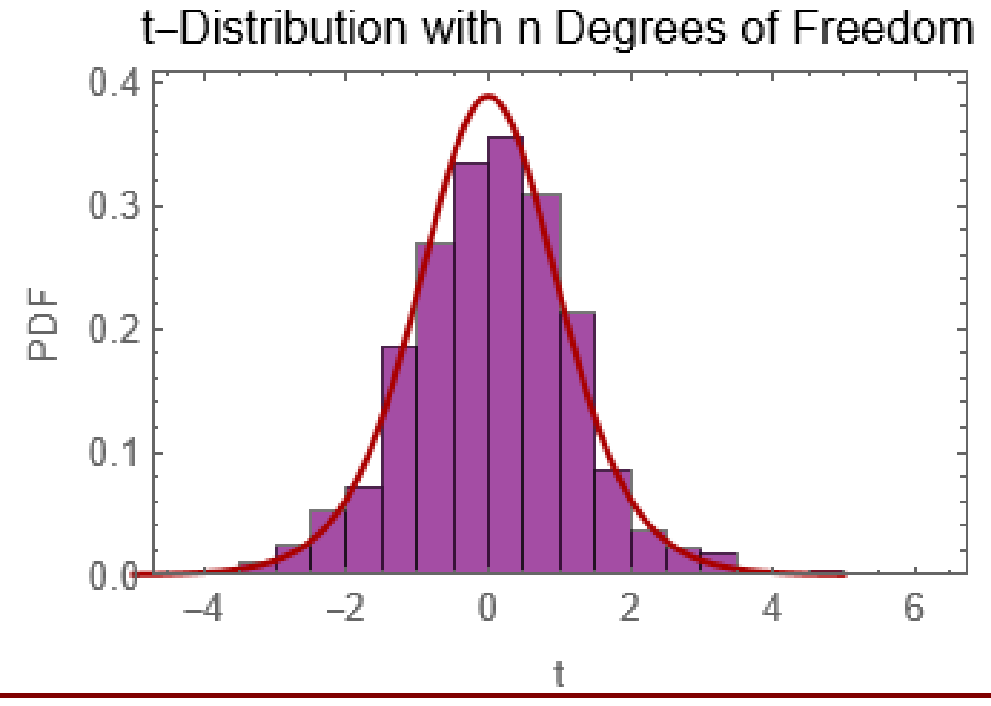

***Mathematica Examples 17.79***

Input

```
(* The code utilizes the Manipulate function to create an interactive visualization. It allows the user to adjust the number of variables (n) and the number of trials (m) and observe the resulting t-distribution. The code begins by generating m samples of n random variables from a standard normal distribution using the RandomVariate function. The sample means and standard deviations are then computed using the Map function and stored in the variables xbar and stanDev, respectively. Next, the t-variables are calculated by subtracting 0 (assumed population mean) from the sample means, multiplying by the square root of n, and dividing by the sample standard deviations. The code proceeds to create a plot that consists of two PDFs: one for the t-distribution with n-1 degrees of freedom and another for the standard normal distribution. A histogram is then generated using the Histogram function with the t-variables as input. The histogram represents the PDF of the t-distribution. Finally, the Show function is used to combine the histogram and the PDF plot into a single figure for display: *)

Manipulate[
 samples=Table[
   RandomVariate[NormalDistribution[0,1],n],
   {m}
   ];
 (* Compute the sample means: *)
 xbar=Map[Mean,samples];
 stanDev=Map[StandardDeviation,samples];

 (* Compute the t-variables: *)
 t=Sqrt[n]*(xbar-0)/stanDev;

 (* Plot the chi-squared PDF: *)
 plot=Plot[
   {PDF[StudentTDistribution[n-1],x],PDF[NormalDistribution[0,1],x]},
   {x,-10,10},
   PlotRange->{{-10,10},{0,0.4}}
   ];
 (* Plot a histogram of the generated samples: *)
 histogram=Histogram[
   t,
   Automatic,
   "PDF",
   PlotRange->{{-10,10},{0,0.4}},
   PlotLabel->"t-Distribution with n Degrees of Freedom and NormalDistribution[0,1]",
   ColorFunction->Function[Opacity[0.7]],
   ChartStyle->Purple,
   ImageSize->300
   ];
 (* Combine the plot and histogram into a single figure: *)
 Show[
  {histogram,plot}
  ],
 {{n,2,"Number of Variables"},2,50,1,Appearance->"Labeled"},
 {{m,500,"Number of trials"},500,10000,100,Appearance->"Labeled"}
 ]
```

Output

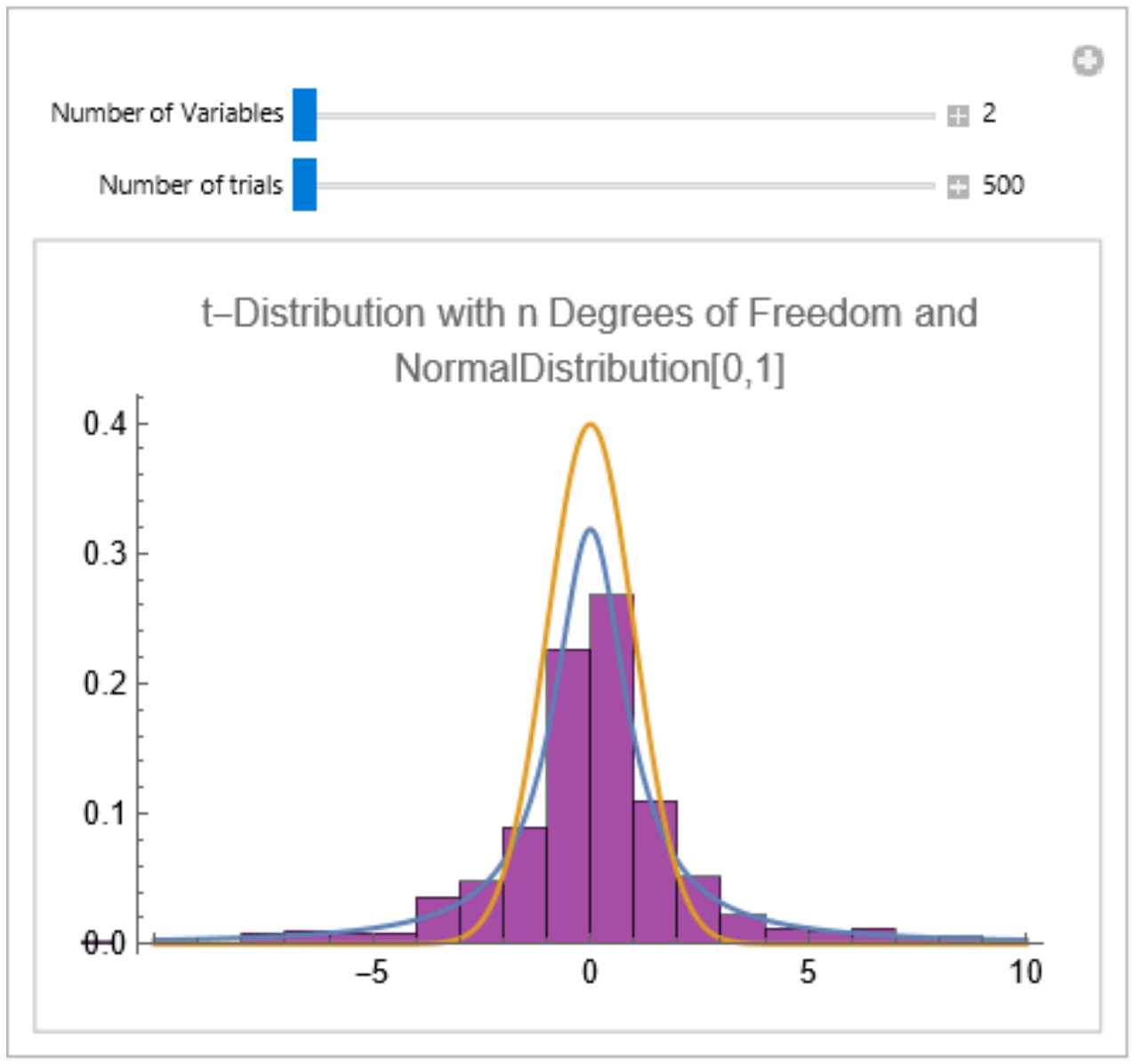

***Mathematica Examples 17.80***

Input

```
(* This code generates a grid of histograms and density plots to visualize the 2D distribution of the sample means for different sample sizes. Specifically, the code computes the mean and variance of 100,000 samples of size n drawn from a normal distribution with mean 0 and standard deviation 1. The resulting means are then plotted in a 3D histogram using the "Rainbow" color scheme and a PDF plot type. The code iterates over several sample sizes, namely n=3,10,20, and 50, and generates a separate 3D histogram for each sample size. The plot label of each histogram indicates the sample size being plotted, while the x-axis, y-axis and z-axis labels indicate "T X", "T Y" and "PDF", respectively, where "T" stands for 2D t-variable: *)

Grid[
 Table[

  samples=Table[
    RandomVariate[NormalDistribution[0,1],{n,2}],
    {100000}
    ];

  (* Compute the sample means: *)
  xbar=Map[Mean,samples];
  stanDev=Map[StandardDeviation,samples];

  (* Compute the t-variables: *)
  t=Sqrt[n]*(xbar-0)/stanDev;

  Column[
   {
    Histogram3D[
     t,
     Automatic,
     "PDF",
     ColorFunction->"Rainbow",
     ImageSize->200,
     PlotLabel->{{"sample size n=",n}},
     AxesLabel->{"T X","T Y","PDF"}(* T=t-variable*)
     ],
    SmoothDensityHistogram[
```

```
                t,
                Automatic,
                "PDF",
                ColorFunction->"Rainbow",
                ImageSize->170,
                PlotLabel->{{"sample size n=",n}}
                ]
              }
            ],
          {n,{3,5,10,50}}
          ]
        ]
```

Output

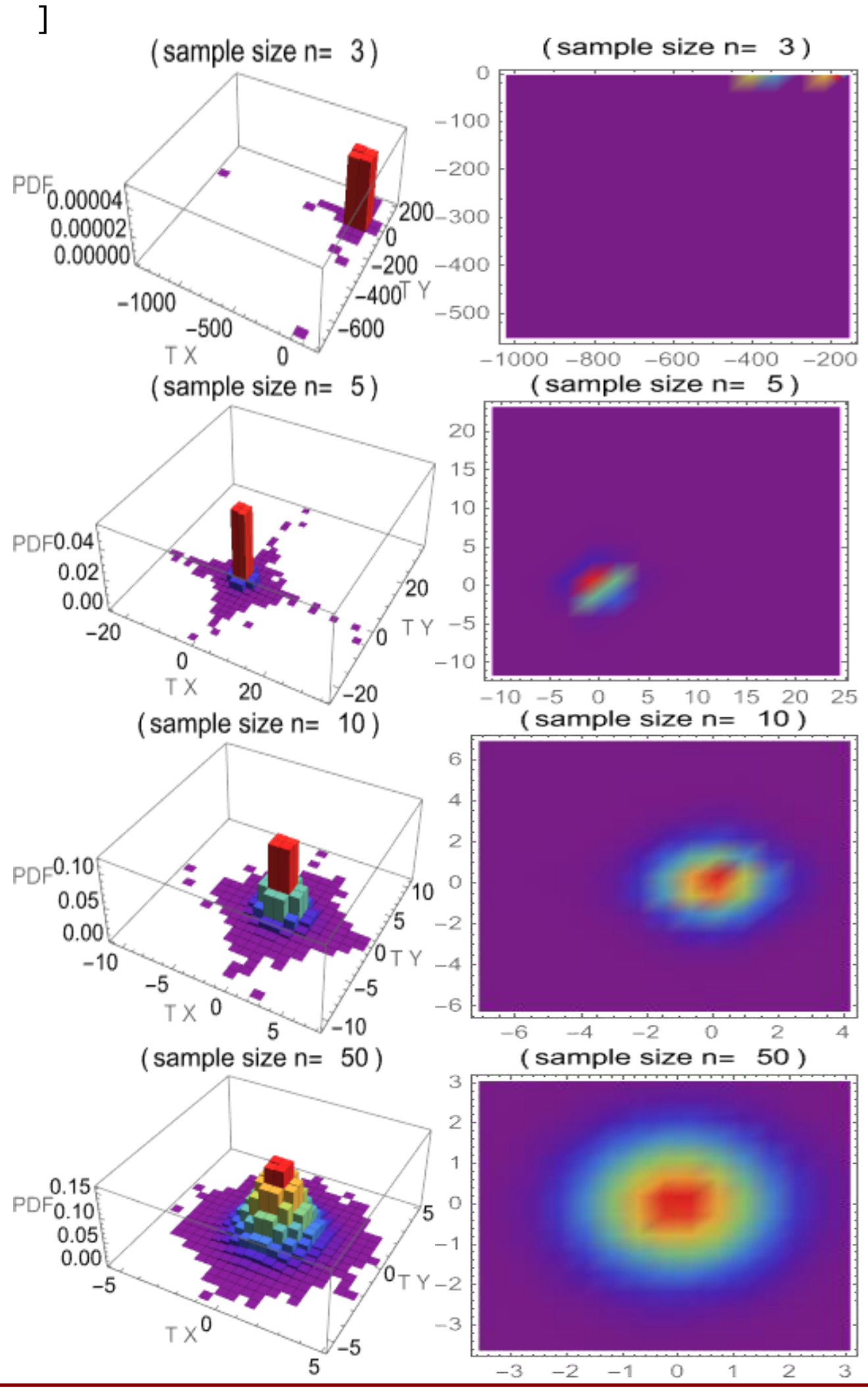

***Mathematica Examples 17.81***

Input

```
(* The code uses Histogram3D and SmoothDensityHistogram to visualize the 2D sample
mean student t-distributions and Manipulate to explore the effect of sample size on
the distributions: *)
population3d=RandomVariate[
    NormalDistribution[0,1],
    {100000,2}
    ];

(* Define a function to generate and plot t-variable: *)
samplevariance3d[n_]:=Module[
```

```
           {samples3d,mean3d,stanDev,t},

           samples3d=Table[
             RandomChoice[
              population3d,n
              ],
             {i,1,10000}
             ];
           (* Compute the sample means: *)
           mean3d=Map[Mean,samples3d];
           stanDev=Map[StandardDeviation,samples3d];

           (* Compute the t-variables: *)
           t=Sqrt[n]*(mean3d-0)/stanDev;

           Row[
            Histogram3D[
             t,
             Automatic,
             "PDF",
             PlotRange->All,
             AxesLabel->{"T X","T Y","PDF"}(* T=t-variable. *),
             ColorFunction->"Rainbow",
             PlotLabel->Row[{"n = ",n}],
             ImageSize->250
             ],
            SmoothDensityHistogram[
             t,
             Automatic,
             "PDF",
             ColorFunction->"Rainbow",
             ImageSize->170,
             PlotLabel->{{"sample size n=",n}}
             ]
            ]
           ]
         (* Use Manipulate to explore the t-variable 2D points: *)
         Manipulate[
          samplevariance3d[n],
          {n,2,50,1}
          ]
```

Output

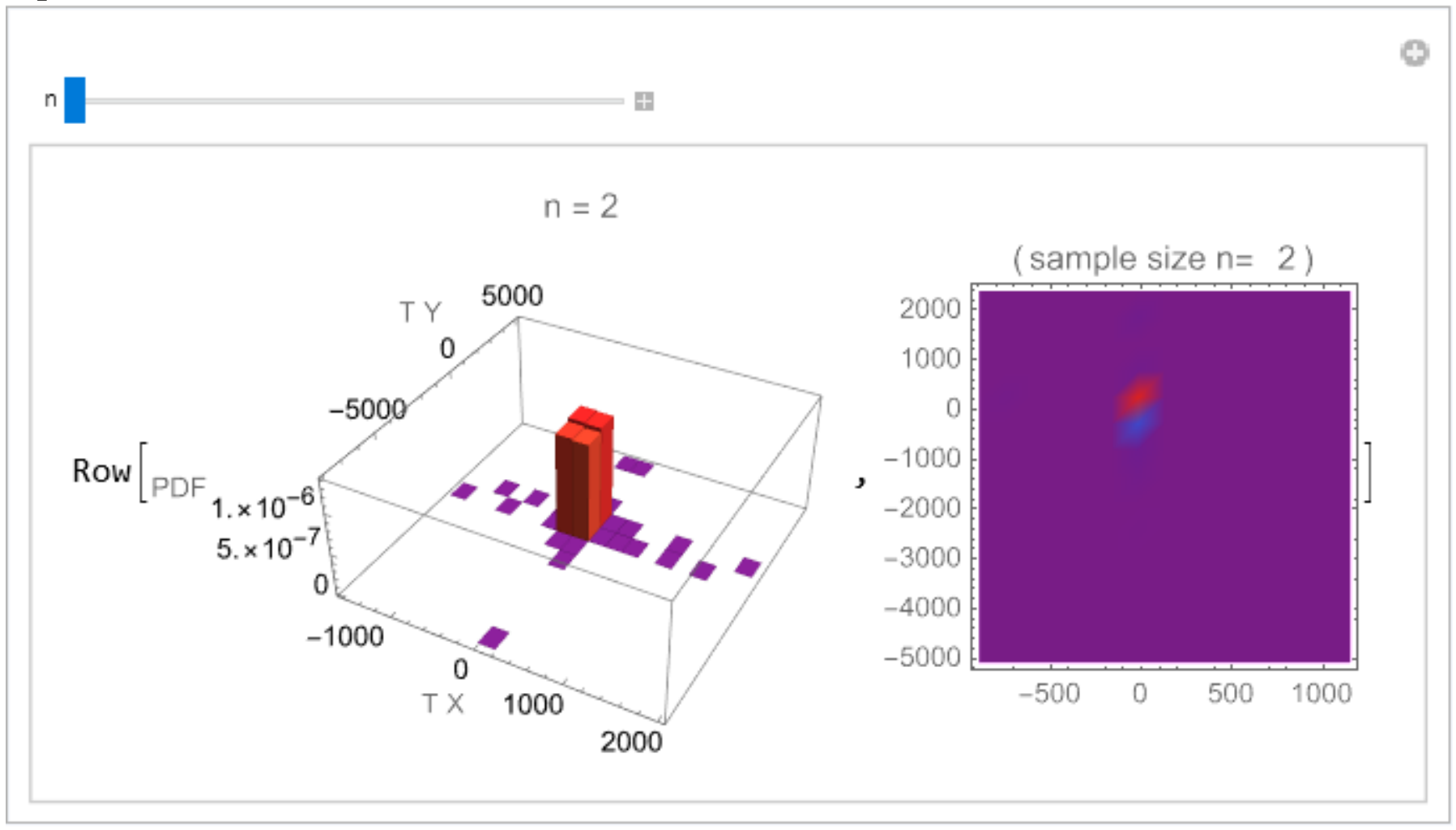

*Mathematica Examples 17.82*

Input
```
(* The code generates a 2D dataset with 5000 random points that follow a Normal distribution with mean 0 and standard deviation 1. The code also defines a function to generate and plot sample means. The function takes a sample size 'n', draws 5000 samples from the population with replacement, calculates the mean for each sample, and plots the t-variable as purple points on a scatter plot. The opacity of the points is set to 0.3 to indicate overlapping points. The Manipulate function is then used to explore the 2D t-variable points. Manipulate allows the user to interactively adjust the sample size 'n' and observe the changes in the 2D t-variable points, as the sample size 'n' increases: *)
SeedRandom[1234];
population1=RandomVariate[
    NormalDistribution[0,1],
    {5000,2}
    ];
(* Define a function to generate and plot t-variable: *)
samplemeans[n_]:=Module[
   {samples,means},
   samples=Table[
      RandomChoice[
       population1,n
       ],
      {i,1,5000}
      ];
   (* Compute the sample means: *)
   means=Map[Mean,samples];
   stanDev=Map[StandardDeviation,samples];
   (* Compute the t-variables: *)
   t=Sqrt[n]*(means-0)/stanDev;

   Graphics[
    {Purple,PointSize[0.008],Opacity[0.3],Point[t]},
    ImageSize->250
    ]
   ]
(* Use Manipulate to explore the t-variable 2D points: *)
Manipulate[
 samplemeans[n],
 {n,3,30,1}
 ]
```

Output

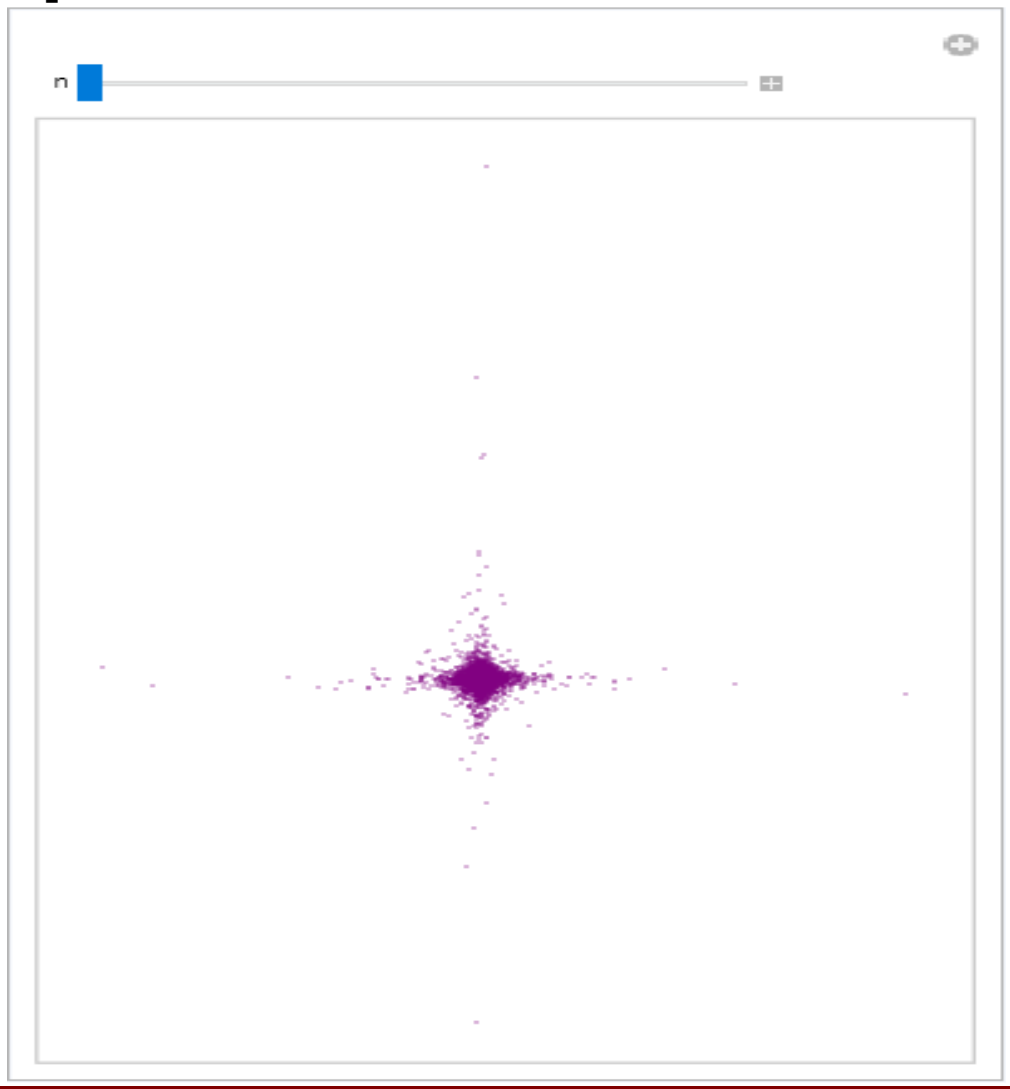

*Mathematica Examples 17.83*

Input

```
(* The code generates and plots sample means for a 3D dataset with normal distribution using the Manipulate function to explore changes in the sample size: *)

SeedRandom[1236];
populationx=RandomVariate[
    NormalDistribution[0,1],
    {5000,3}
    ];

(* Define a function to generate and plot t-variable: *)
samplemeansx[n_]:=Module[
   {samplesx,meansx},

   samplesx=Table[
      RandomChoice[
        populationx,n
        ],
      {i,1,5000}
      ];
   (* Compute the sample means: *)
   meansx=Map[Mean,samplesx];
   stanDevx=Map[StandardDeviation,samplesx];

   (* Compute the t-variables*)
   tx=Sqrt[n]*(meansx-0)/stanDevx;

   Graphics3D[
    {Purple,PointSize[0.008],Opacity[0.3],Point[tx]},
    ImageSize->250
    ]
   ]

(* Use Manipulate to explore the t-variable 3D points: *)
Manipulate[
 samplemeansx[n],
 {n,3,50,1}
 ]
```

Output

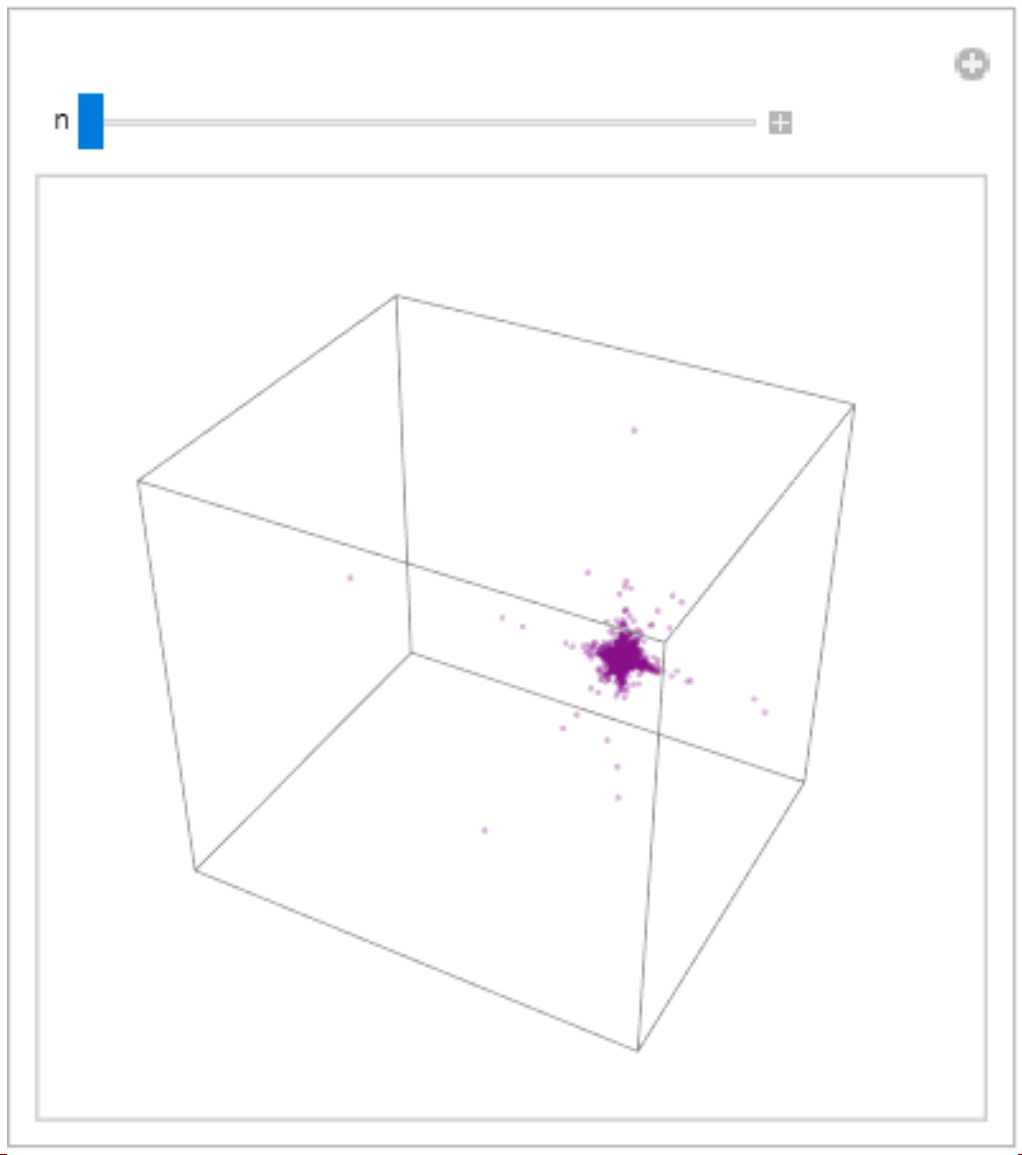

# UNIT 17.6

# FRATIO DISTRIBUTION

***Mathematica Examples 17.84***

Input
```
(* The code generates a plot of the probability density function (PDF) for a FRatio
distribution with different values of m= (1, 4 and 15) and a fixed n=5. The plot
shows the values of the PDF for all possible values of x between 0 and 3: *)

Plot[
 Evaluate[
  Table[
   PDF[
    FRatioDistribution[5,m],
    x
    ],
   {m,{1,4,15}}
   ]
  ],
 {x,0,3},
 PlotRange->Automatic,
 Filling->Axis,
 PlotLegends->Placed[{"v1=5,v2=1","v1=5,v2=4","v1=5,v2=15"},{0.8,0.75}],
 PlotStyle->{RGBColor[0.88,0.61,0.14],RGBColor[0.37,0.5,0.7],Purple},
 ImageSize->320,
 AxesLabel->{None,"PDF"}
 ]
```

Output

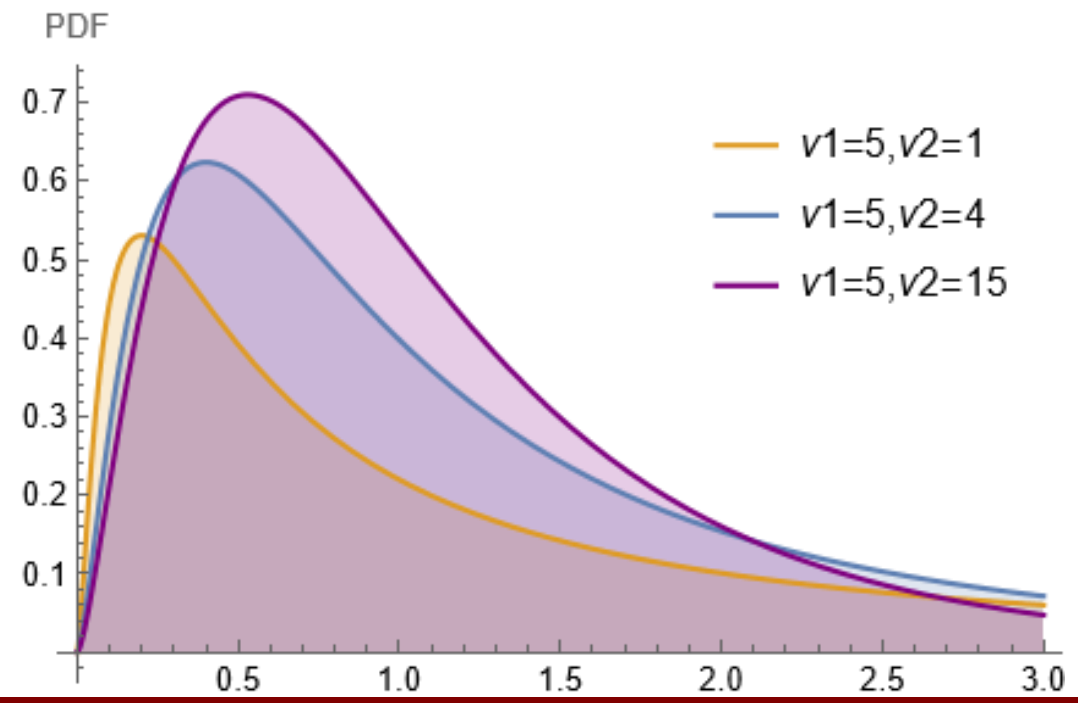

***Mathematica Examples 17.85***

Input
```
(* The given code generates a plot of the cumulative distribution function (CDF) of
the FRatio distribution with different values of m=(1, 4 and 15) and a fixed n=5: *)

Plot[
 Evaluate[
  Table[
   CDF[
    FRatioDistribution[5,m],
    x
    ],
   {m,{1,4,15}}
   ]
  ],
 {x,0,3},
```

```
PlotRange->Automatic,
Filling->Axis,
PlotLegends->Placed[{"v1=5,v2=1","v1=5,v2=4","v1=5,v2=15"},{0.25,0.75}],
PlotStyle->{RGBColor[0.88,0.61,0.14],RGBColor[0.37,0.5,0.7],Purple},
ImageSize->320,
AxesLabel->{None,"CDF"}
]
```

Output

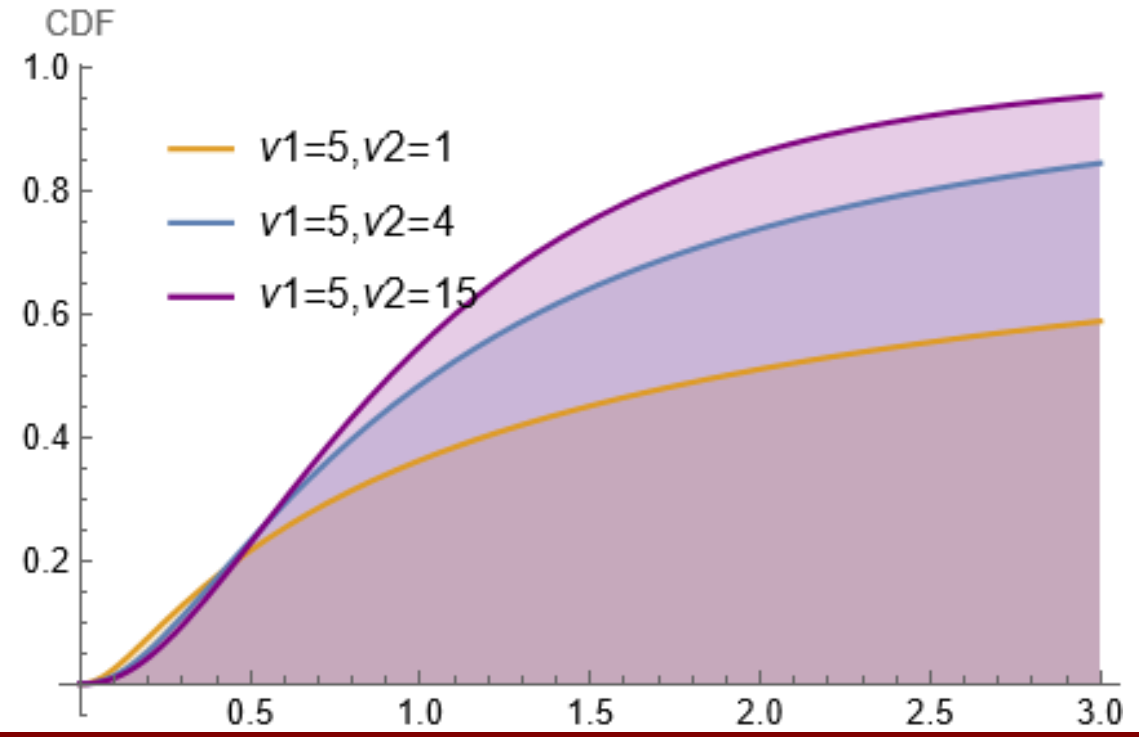

***Mathematica Examples 17.86***

Input

```
(* The code generates a plot of the PDF for a FRatio distribution for a fixed value of m=5 and different values of n= (1, 4 and 15). The plot shows the values of the PDF for all possible values of x between 0 and 3: *)

Plot[
 Evaluate[
  Table[
   PDF[
    FRatioDistribution[n,5],
    x
    ],
   {n,{1,4,15}}
   ]
  ],
 {x,0,3},
 PlotRange->Automatic,
 Filling->Axis,
 PlotLegends->Placed[{"v1=1,v2=5","v1=4,v2=5","v1=15,v2=5"},{0.25,0.75}],
 PlotStyle->{RGBColor[0.88,0.61,0.14],RGBColor[0.37,0.5,0.7],Purple},
 ImageSize->320,
 AxesLabel->{None,"PDF"}
 ]
```

Output

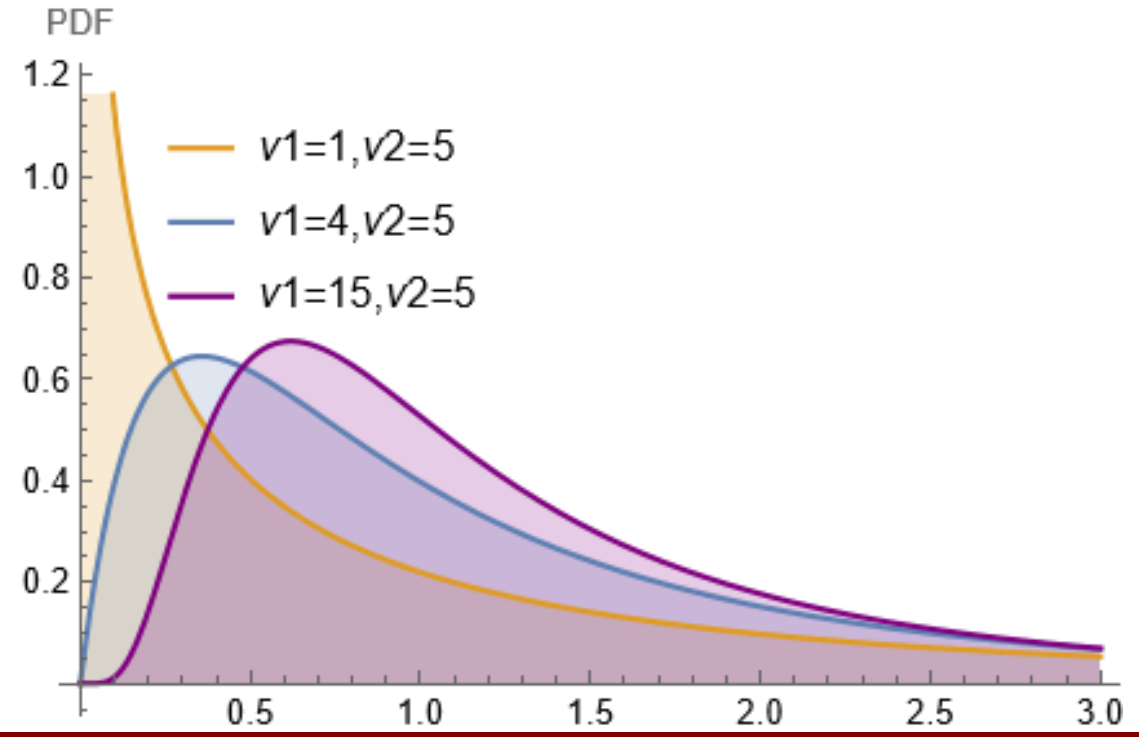

**Mathematica Examples 17.87**

Input

```
(* The code generates a plot of the CDF of the FRatio distribution for a fixed value
of m=5 and different values of n= (1, 4 and 15): *)

Plot[
 Evaluate[
  Table[
   CDF[
    FRatioDistribution[n,5],
    x
    ],
   {n,{1,4,15}}
   ]
  ],
 {x,0,3},
 PlotRange->Automatic,
 Filling->Axis,
 PlotLegends->Placed[{"v1=1,v2=5","v1=4,v2=5","v1=15,v2=5"},{0.25,0.75}],
 PlotStyle->{RGBColor[0.88,0.61,0.14],RGBColor[0.37,0.5,0.7],Purple},
 ImageSize->320,
 AxesLabel->{None,"CDF"}
 ]
```

Output

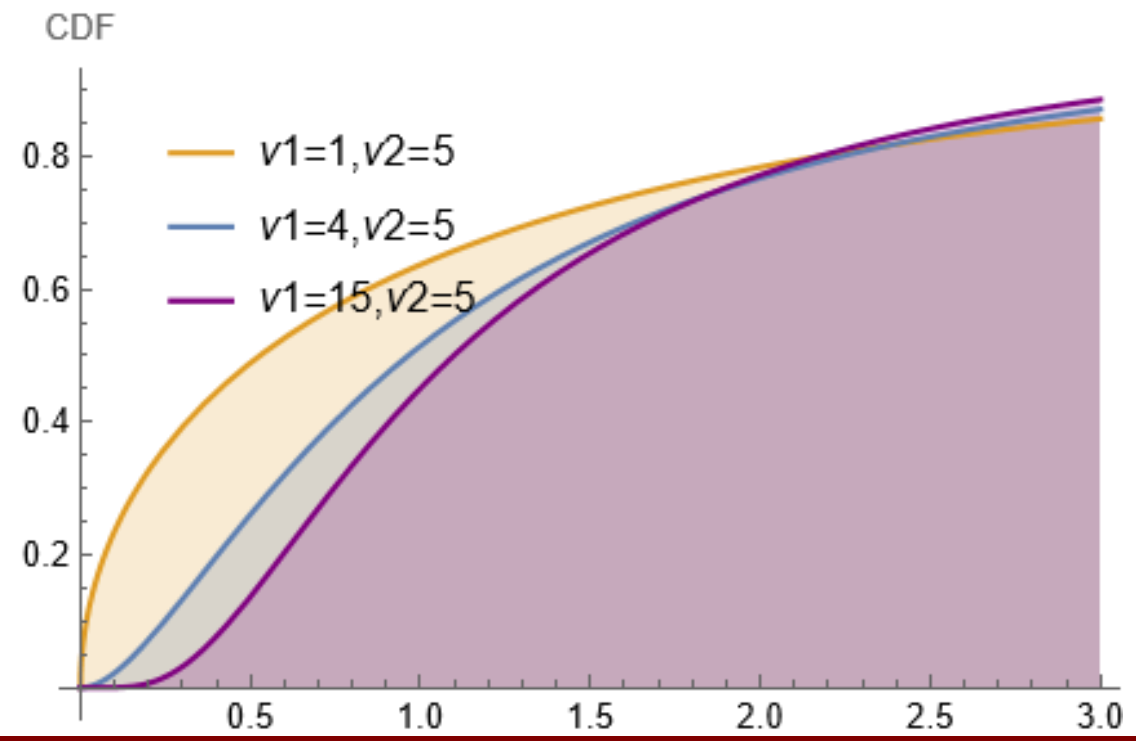

**Mathematica Examples 17.88**

Input

```
(* The code generates a histogram and a plot of the PDF for a FRatio Distribution
with parameters n=5 and m=10 and sample size 10000: *)

data=RandomVariate[
    FRatioDistribution[5,10],
    10^4
    ];
Show[
 Histogram[
  data,
  100,
  "PDF",
  ColorFunction->Function[{height},Opacity[height]],
  ChartStyle->Purple,
  ImageSize->320,
  AxesLabel->{None,"PDF"}
  ],
 Plot[
  PDF[
   FRatioDistribution[5,10],
   x
   ],
```

```
      {x,0,9},
      PlotStyle->RGBColor[0.88,0.61,0.14],
      PlotRange->{0,4}
      ]
     ]
```

Output

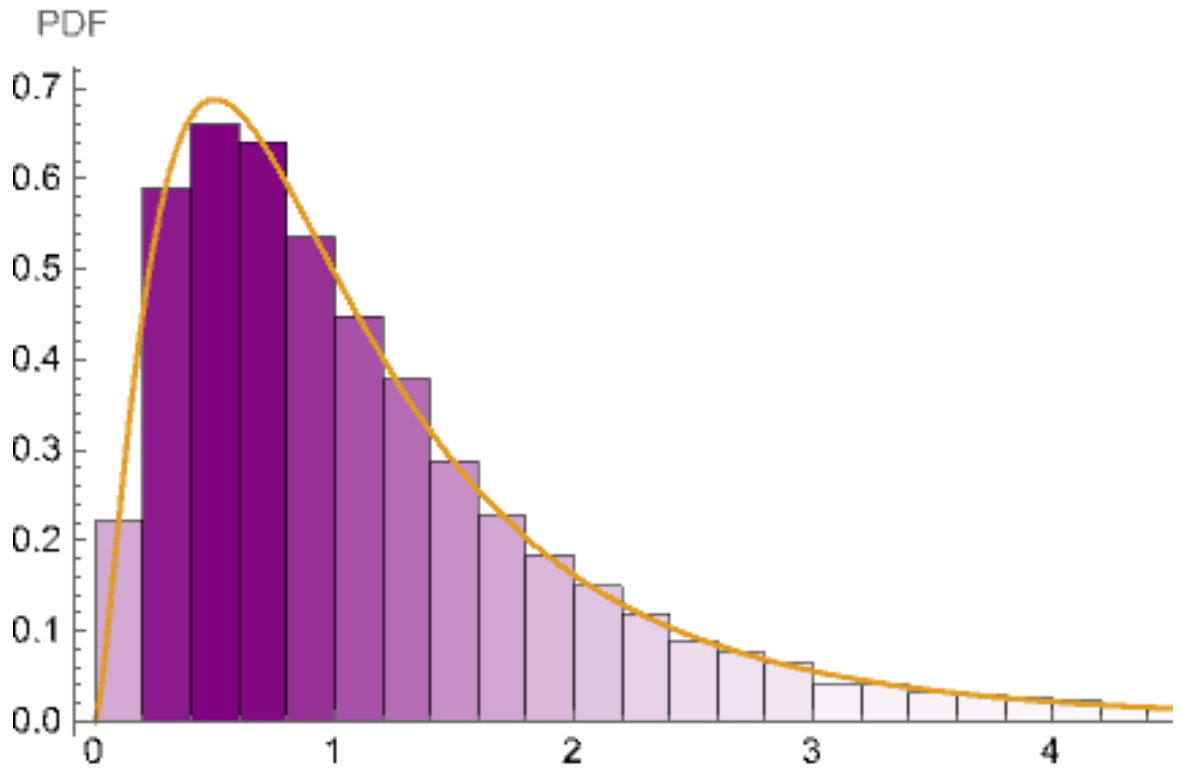

***Mathematica Examples 17.89***

Input

```
(* The code creates a dynamic histogram of data and a plot of the PDF generated from
a FRatio distribution using the Manipulate function. The Manipulate function creates
interactive controls for the user to adjust the values of n, m, and no, which are
the parameters of the FRatio distribution and the sample size: *)

Manipulate[
 Module[
  {
   data=RandomVariate[
     FRatioDistribution[n,m],
     no
     ]
   },

  Show[
   Histogram[
    data,
    Automatic,
    "PDF",
    PlotRange->{{0,7},All},
    ColorFunction->Function[{height},Opacity[height]],
    ImageSize->320,
    ChartStyle->Purple
    ],
   Plot[
    PDF[
     FRatioDistribution[n,m],
     x
     ],
    {x,0,7},
    PlotRange->All,
    ColorFunction->"Rainbow"
    ]
   ]
  ],
 {{n,3,"n"},2,10,0.1},
 {{m,5,"m"},3,20,0.1},
 {{no,100,"no"},100,1000,10}
 ]
```

Output

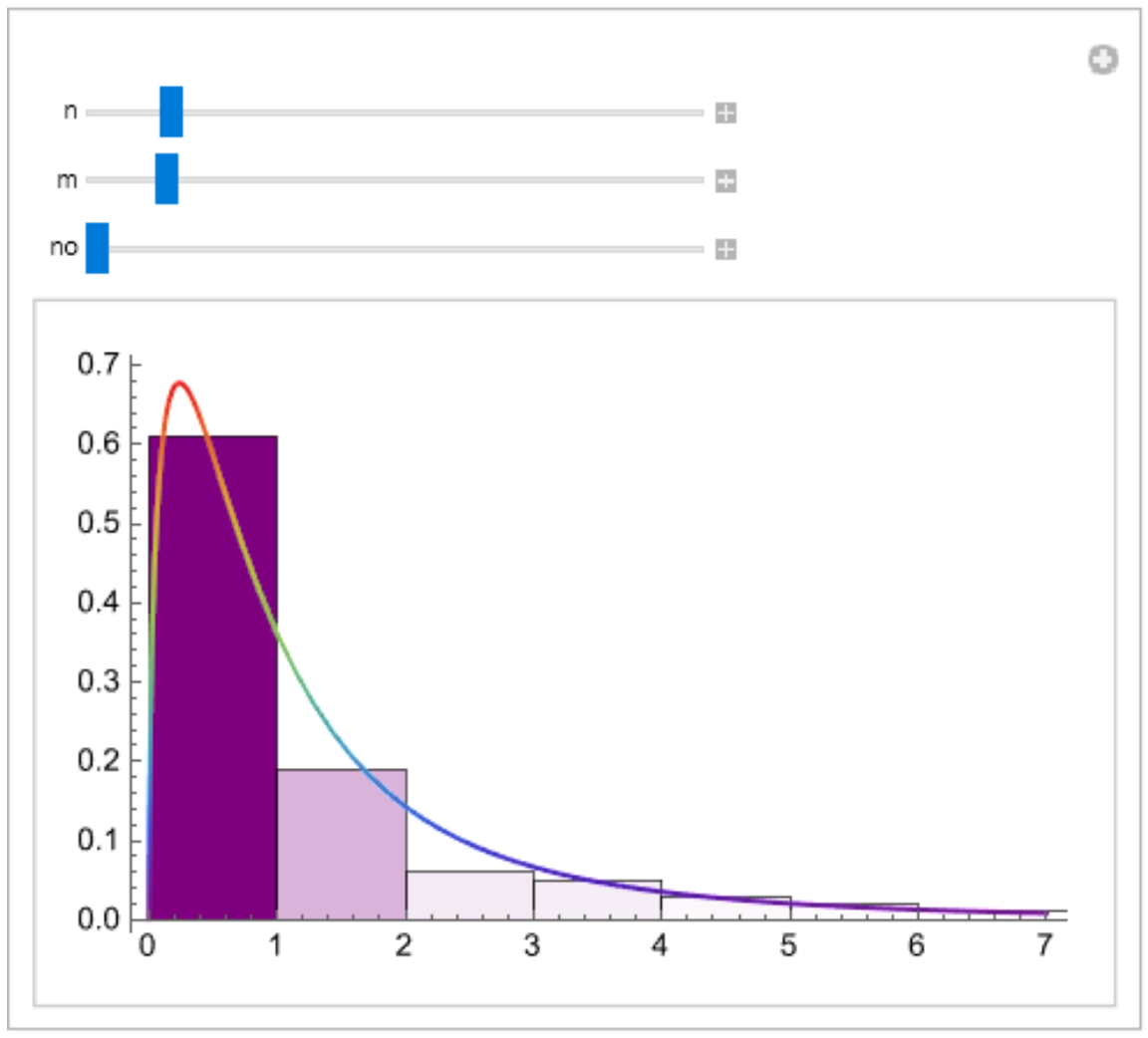

*Mathematica Examples 17.90*

Input

```
(* The code creates a plot of the CDF of FRatio distribution using the Manipulate
function. The Manipulate function allows you to interactively change the values of
the parameters n and m, respectively: *)
Manipulate[
 Plot[
  CDF[
   FRatioDistribution[n,m],
   x
   ],
  {x,0,1},
  Filling->Axis,
  FillingStyle->LightPurple,
  PlotRange->All,
  Epilog->{Text[StringForm["n = `` & m = ``",n,m],{n/2,0.9}]},
  AxesLabel->{"x","CDF"},
  ImageSize->320,
  PlotStyle->Purple
  ],
 {{n,4},1,5,0.1},
 {{m,3},1,5,0.1}
 ]
```

Output

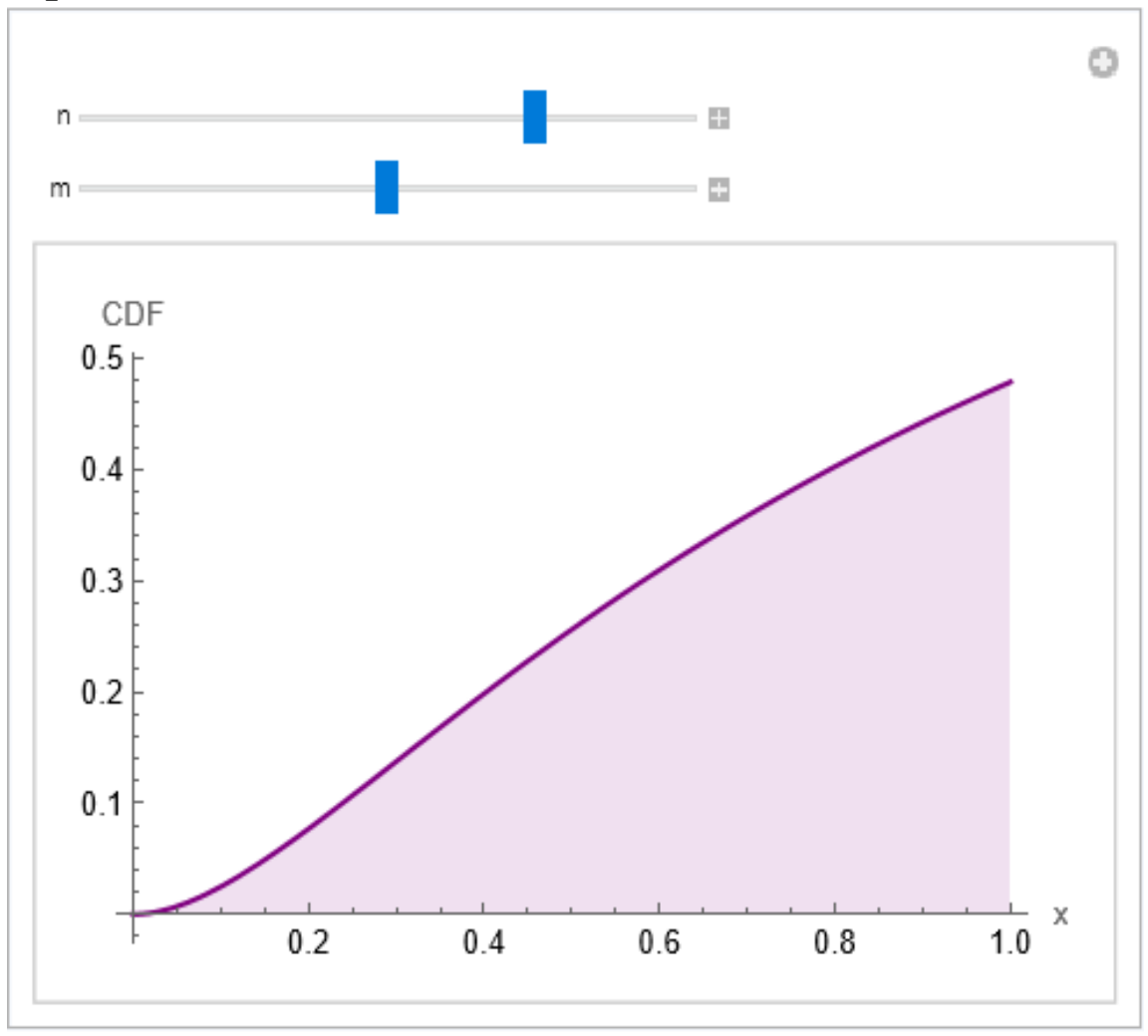

***Mathematica Examples 17.91***

Input

```
(* The code uses the Grid function to create a grid of two plots, one for the PDF and one for the CDF of FRatio distribution. The code uses slider controls to adjust the values of n and m: *)

Manipulate[
 Grid[
  {
   {Plot[
     PDF[
      FRatioDistribution[n,m],
      x
      ],
     {x,0,1},
     PlotRange->All,
     PlotStyle->{Purple,PointSize[0.03]},
     PlotLabel->"PDF of FRatio Distribution",
     AxesLabel->{"x","PDF"}
     ],
    Plot[
     CDF[
      FRatioDistribution[n,m],
      x
      ],
     {x,0,1},
     PlotRange->All,
     PlotStyle->{Purple,PointSize[0.03]},
     PlotLabel->"CDF of FRatio Distribution",
     AxesLabel->{"x","CDF"}
     ]
    }
   },
  Spacings->{5,5}
  ],
 {{n,2.5},1,5,0.1},
 {{m,2},2,5,0.1}
 ]
```

Output

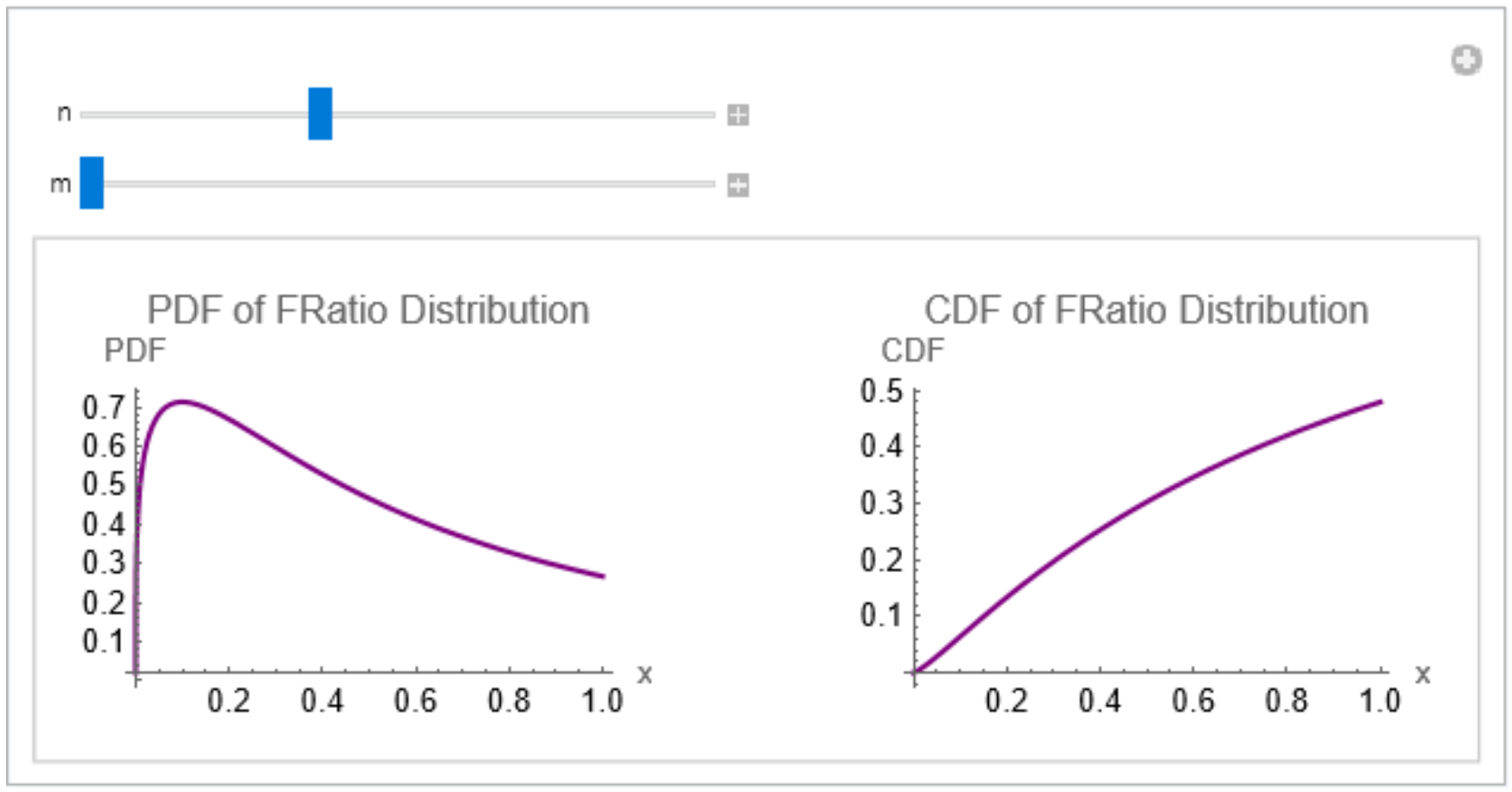

***Mathematica Examples 17.92***

Input

```
(* The code calculates and displays some descriptive statistics (mean, variance, standard deviation, kurtosis and skewness) for the FRatio distribution with parameters n and m: *)
```

```
Grid[
 Table[
  {
   statistics,
   FullSimplify[statistics[FRatioDistribution[n,m]]]
   },
  {statistics,{Mean,Variance,StandardDeviation,Kurtosis,Skewness}}
  ],
 ItemStyle->12,
 Alignment->{{Right,Left}},
 Frame->All,
 Spacings->{Automatic,0.8}
 ]
```

Output

| | |
|---|---|
| Mean | $\begin{cases} \frac{m}{-2+m} & m>2 \\ \text{Indeterminate} & \text{True} \end{cases}$ |
| Variance | $\begin{cases} \frac{2\,m^2\,(-2+m+n)}{(-4+m)(-2+m)^2\,n} & m>4 \\ \text{Indeterminate} & \text{True} \end{cases}$ |
| StandardDeviation | $\begin{cases} \frac{\sqrt{2}\,m\sqrt{-2+m+n}}{(-2+m)\sqrt{(-4+m)\,n}} & m>4 \\ \text{Indeterminate} & \text{True} \end{cases}$ |
| Kurtosis | $\begin{cases} 3+\frac{12\left((-4+m)(-2+m)^2+(-22+5\,m)\,n\,(-2+m+n)\right)}{(-8+m)(-6+m)\,n\,(-2+m+n)} & m>8 \\ \text{Indeterminate} & \text{True} \end{cases}$ |
| Skewness | $\begin{cases} \frac{2\sqrt{2}\sqrt{-4+m}\,(-2+m+2\,n)}{(-6+m)\sqrt{n}\sqrt{-2+m+n}} & m>6 \\ \text{Indeterminate} & \text{True} \end{cases}$ |

***Mathematica Examples 17.93***

Input

```
(* The code calculates and displays some additional descriptive statistics (moments, central moments, and factorial moments) for the FRatio Distribution with parameters n and m: *)
Grid[
 Table[
  {
   statistics,
   FullSimplify[statistics[FRatioDistribution[n,m],1]],
   FullSimplify[statistics[FRatioDistribution[n,m],2]]
   },
  {statistics,{Moment,CentralMoment,FactorialMoment}}
  ],
 ItemStyle->12,
 Alignment->{{Right,Left}},
 Frame->All,
 Spacings->{Automatic,0.8}
 ]
```

Output

| | | |
|---|---|---|
| Moment | $\begin{cases} \frac{m}{-2+m} & m>2 \\ \infty & \text{True} \end{cases}$ | $\begin{cases} \frac{m^2\,(2+n)}{(-4+m)(-2+m)\,n} & m>4 \\ \infty & \text{True} \end{cases}$ |
| CentralMoment | $\begin{cases} 0 & m>2 \\ \text{Indeterminate} & \text{True} \end{cases}$ | $\begin{cases} \frac{2\,m^2\,(-2+m+n)}{(-4+m)(-2+m)^2\,n} & m>4 \\ \text{Indeterminate} & \text{True} \end{cases}$ |
| FactorialMoment | $\begin{cases} \frac{m}{-2+m} & m>2 \\ \text{Indeterminate} & \text{True} \end{cases}$ | $\begin{cases} \frac{2\,m\,(m+2\,n)}{(-4+m)(-2+m)\,n} & m>4 \\ \text{Indeterminate} & \text{True} \end{cases}$ |

### Mathematica Examples 17.94

Input
```
(* The code generates a dataset of 1000 observations from the FRatio distribution with parameters n=5 and m=10. Then, it computes the sample mean and quartiles of the data, and plots a histogram of the data and plot of the PDF. Additionally, the code adds vertical lines to the plot corresponding to the sample mean and quartiles: *)

data=RandomVariate[
    FRatioDistribution[5,10],
    1000
    ];

mean=Mean[data];
quartiles=Quantile[
    data,
    {0.25,0.5,0.75}
    ];

Show[
 Histogram[
  data,
  Automatic,
  "PDF",
  Epilog->{
    Directive[Red,Thickness[0.006]],
    Line[{{mean,0},{mean,0.75}}],
    Directive[Green,Dashed],
    Line[{{quartiles[[1]],0},{quartiles[[1]],0.75}}],
    Line[{{quartiles[[2]],0},{quartiles[[2]],0.75}}],
    Line[{{quartiles[[3]],0},{quartiles[[3]],0.75}}]
    },
  ColorFunction->Function[{height},Opacity[height]],
  ImageSize->320,
  ChartStyle->Purple,
  PlotRange->{{0,6},{0,0.8}}
  ],
 Plot[
  PDF[FRatioDistribution[5,7],x],
  {x,0,6},
  ImageSize->320,
  ColorFunction->"Rainbow"
  ]
 ]
```
Output

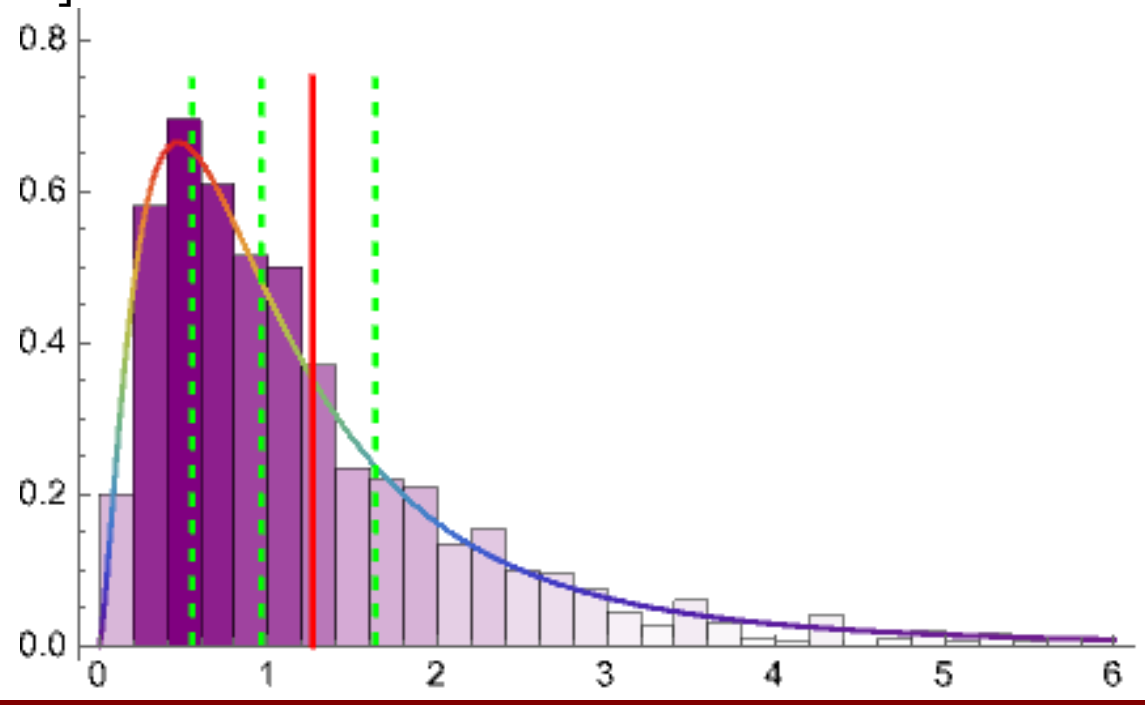

### Mathematica Examples 17.95

Input
```
(* The code generates a random sample of size 10,000 from the FRatio distribution with parameters n=5 and m=10, estimates the distribution parameters using the EstimatedDistribution function, and then compares the histogram of the sample with
```

```
the estimated PDF of the FRatio distribution using a histogram and a plot of the PDF:
*)

sampledata=RandomVariate[
    FRatioDistribution[5,10],
    10^4
    ];
(* Estimate the distribution parameters from sample data: *)
ed=EstimatedDistribution[
   sampledata,
   FRatioDistribution[n,m]
   ]
(* Compare a density histogram of the sample with the PDF of the estimated
distribution: *)
Show[
 Histogram[
   sampledata,
   Automatic,
   "PDF",
   ColorFunction->Function[{height},Opacity[height]],
   ChartStyle->Purple,
   ImageSize->320
   ],
 Plot[
   PDF[ed,x],
   {x,0,5},
   ImageSize->320,
   ColorFunction->"Rainbow"
   ]
 ]
```

Output FRatioDistribution[5.0238,9.77424]

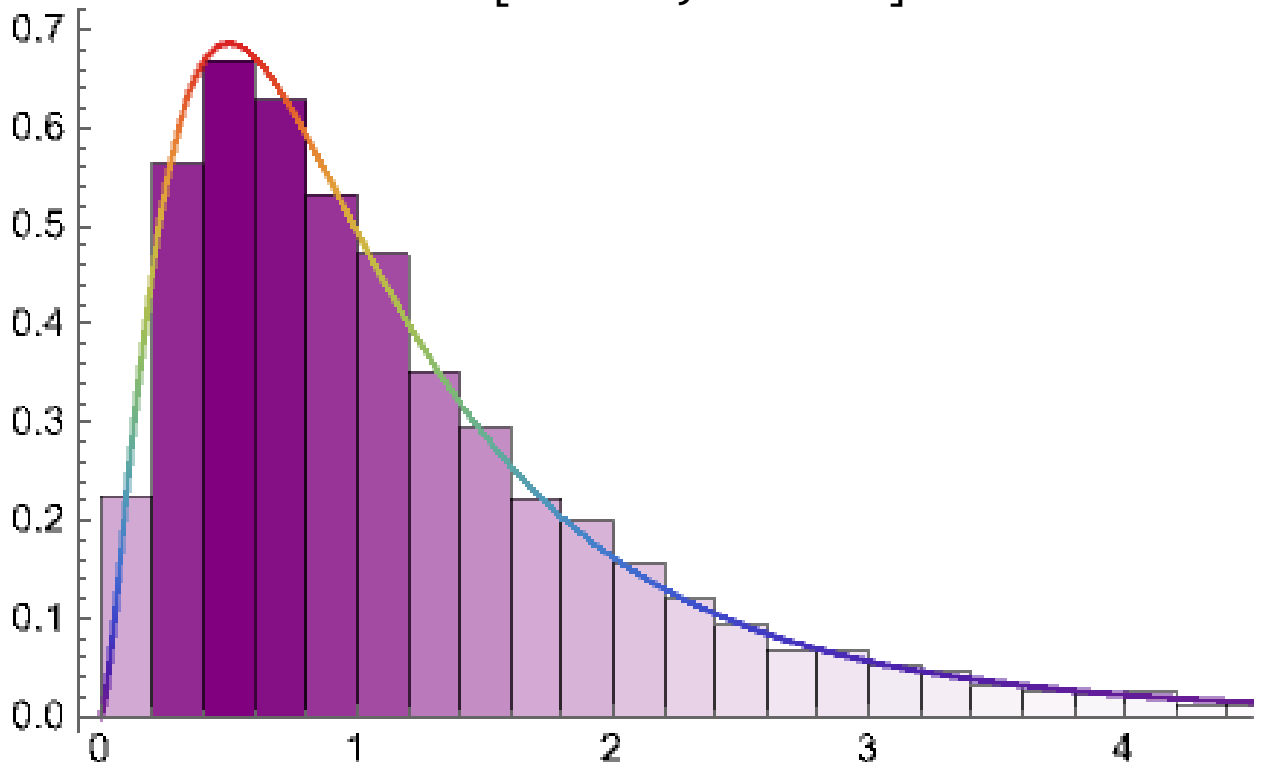

***Mathematica Examples 17.96***

Input

```
(* The code generates a 2D dataset with 1000 random points that follow the FRatio
distribution with n=5 and m=10. The dataset is then used to create a row of three
plots. The first plot is a histogram of the X-axis values of the dataset. The second
plot is a histogram of the Y-axis values of the dataset. It is similar to the first
plot, but shows the distribution of the Y-axis values instead. The third plot is a
scatter plot of the dataset, with the X-axis values on the horizontal axis and the
Y-axis values on the vertical axis. Each point in the plot represents a pair of X
and Y values from the dataset: *)

data=RandomVariate[
    FRatioDistribution[5,10],
    {1000,2}
    ];
```

```
GraphicsRow[
 {
  Histogram[
   data[[All,1]],
   {0.1},
   PlotLabel->"X-axis",
   ColorFunction->Function[{height},Opacity[height]],
   ChartStyle->Purple
   ],
  Histogram[
   data[[All,2]],
   {0.1},
   PlotLabel->"Y-axis",
   ColorFunction->Function[{height},Opacity[height]],
   ChartStyle->Purple
   ],
  ListPlot[
   data,
   PlotStyle->{Purple,PointSize[0.015]},
   AspectRatio->1,
   Frame->True,
   Axes->False
   ]
  }
 ]
```

Output

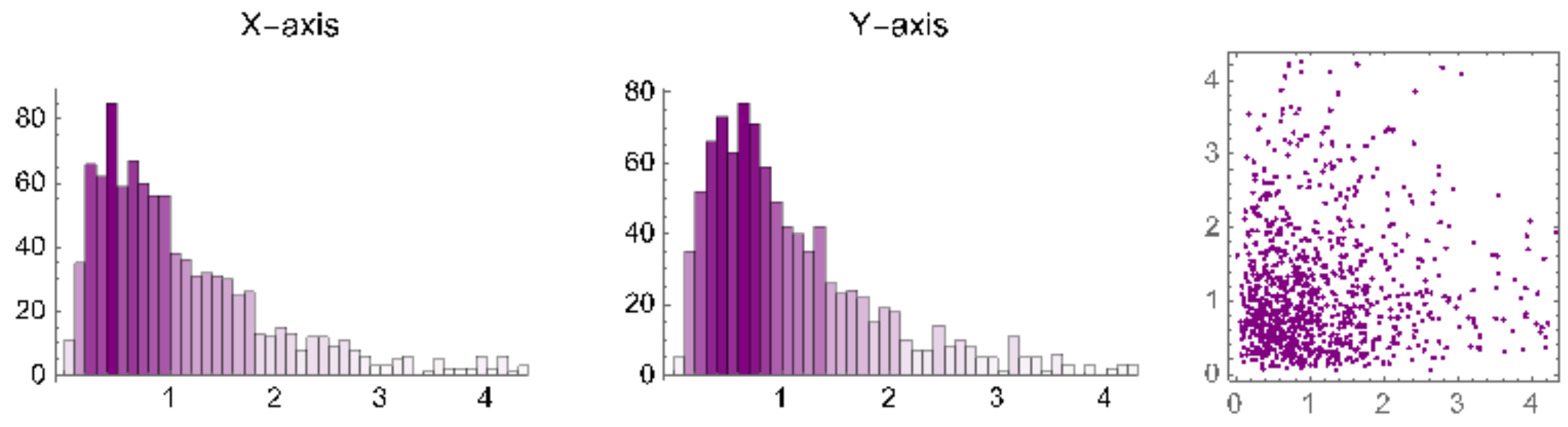

***Mathematica Examples 17.97***

Input

```
(* The code generates a set of random data points with the FRatio distribution in three dimensions, and then creates three histograms, one for each dimension, showing the distribution of the points along that axis. Additionally, it creates a 3D scatter plot of the data points: *)

data=RandomVariate[
   FRatioDistribution[5,10],
   {1000,3}
   ];

GraphicsGrid[
 {
  {
   Histogram[
    data[[All,1]],
    Automatic,
    "PDF",
    PlotLabel->"X-axis",
    ColorFunction->Function[{height},Opacity[height]],
    ChartStyle->Purple
    ],
```

```
          Histogram[
           data[[All,2]],
           Automatic,
           "PDF",
           PlotLabel->"Y-axis",
           ColorFunction->Function[{height},Opacity[height]],
           ChartStyle->Purple
           ],
          Histogram[
           data[[All,3]],
           Automatic,
           "PDF",
           PlotLabel->"Z-axis",
           ColorFunction->Function[{height},Opacity[height]],
           ChartStyle->Purple
           ],
          ListPointPlot3D[
           data,
           BoxRatios->{1,1,1},
           PlotStyle->{Purple,PointSize[0.015]}
           ]
          }
         }
        ]
```

Output

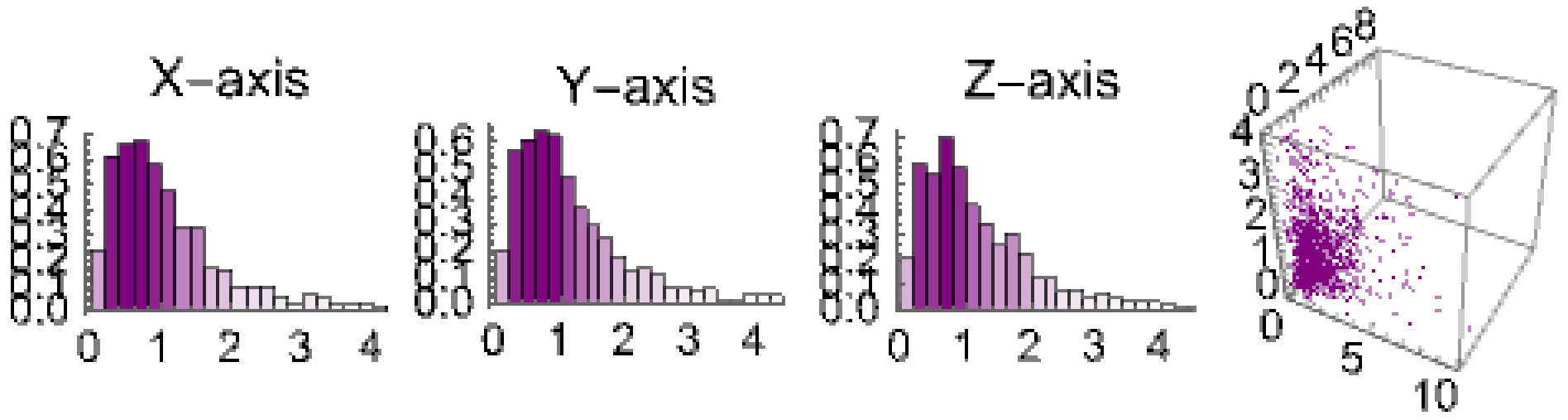

*Mathematica Examples 17.98*

Input

```
(* The code generates a 3D scatter plot of FRatio distributed points, where the x-
axis is red, y-axis is green, and z-axis is blue: *)

data=RandomVariate[
    FRatioDistribution[5,10],
    {2000,3}
    ];
Graphics3D[
 {
  {PointSize[0.006],Purple,Point[data]},
  Thin,
  {Red,Opacity[0.4],Line[{{#,0,0},{#,0,-0.6}}]&/@data[[All,1]]},
  Thin,
  {Green,Opacity[0.4],Line[{{0,#,0},{0,#,-0.6}}]&/@data[[All,2]]},
  Thin,
  {Blue,Opacity[0.4],Line[{{0,0,#},{0,-0.6,#}}]&/@data[[All,3]]}
  },
 BoxRatios->{0.4,0.4,0.4},
 Axes->True,
 AxesLabel->{"X","Y","Z"},
 ImageSize->320
 ]
```

Output

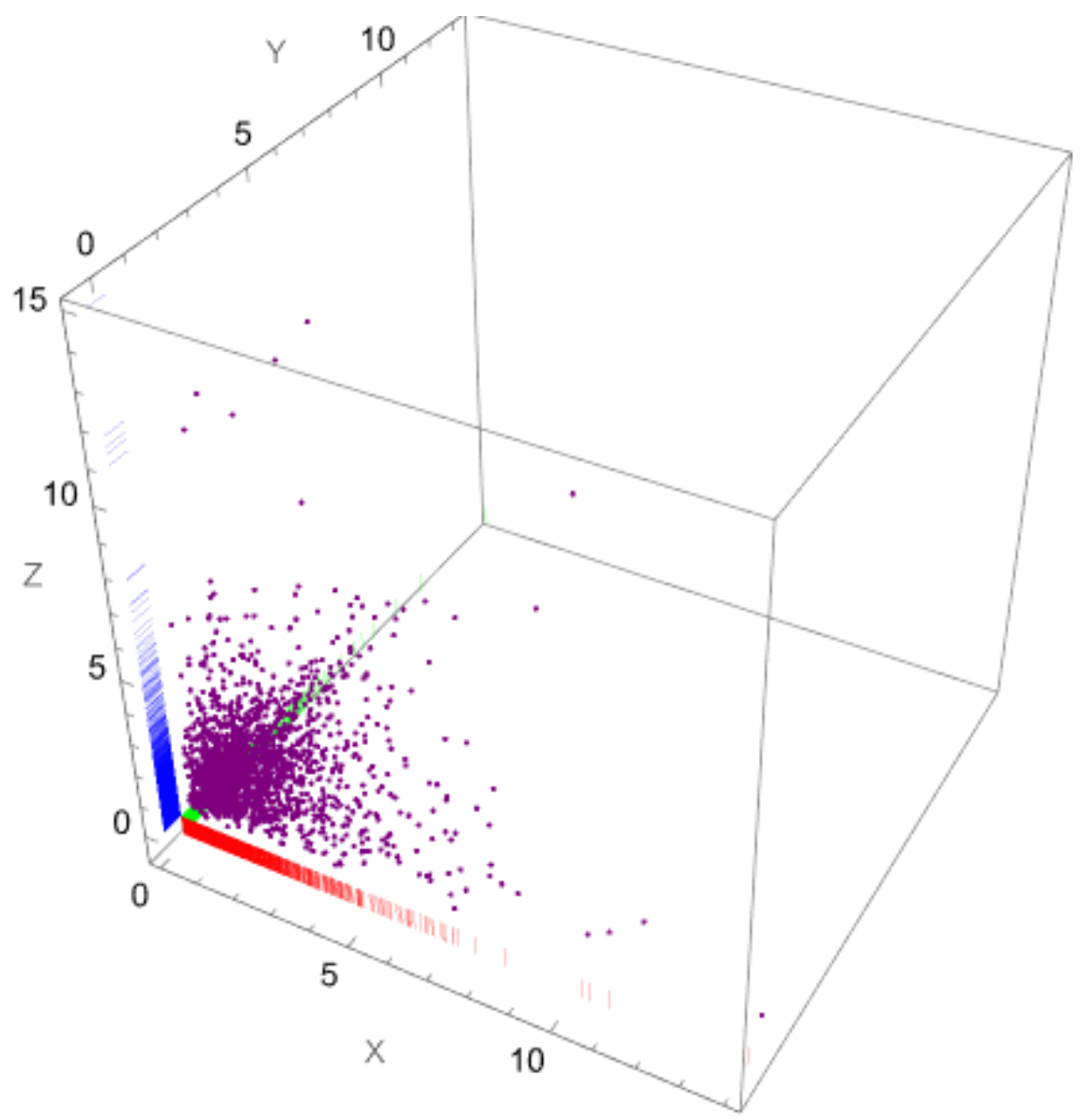

***Mathematica Examples 17.99***

Input

```
(* The code demonstrates a common technique in statistics and data analysis, which
is the use of random sampling to estimate population parameters. The code generates
random samples from FRatio distribution with n=5 and m=10,and then using these samples
to estimate the parameters of another FRatio distribution with unknown n and m. This
process is repeated 20 times, resulting in 20 different estimated distributions. The
code also visualizes the resulting estimated distributions using the PDF function.
The code plots the PDFs of these estimated distributions using the PDF function and
the estimated parameters. The plot shows the PDFs in a range from-0 to 3. The code
also generates a list plot of 2 sets of random samples from the normal distribution
with n=5 and m=10. The plot shows the 100 random points generated from two random
samples. The code generates also a histogram of the PDF for FRatio distribution of
the two samples: *)

estim0distributions=Table[
  dist=FRatioDistribution[5,10];

  sampledata=RandomVariate[
    dist,
    100
    ];

  ed=EstimatedDistribution[
    sampledata,
    FRatioDistribution[n,m]
    ],
  {i,1,20}
  ]

pdf0ed=Table[
   PDF[estim0distributions[[i]],x],
   {i,1,20}
   ];

(* Visualizes the resulting estimated distributions *)
Plot[
 pdf0ed,
 {x,0,3},
```

```
 PlotRange->Full,
 ImageSize->400,
 PlotStyle->Directive[Purple,Opacity[0.3],Thickness[0.002]]
 ]

(* Visualizes 100 random points generated from two random samples *)
table=Table[
   dist=FRatioDistribution[5,10];
   sampledata=RandomVariate[
     dist,
     100],
   {i,1,2}
   ];

ListPlot[
 table,
 ImageSize->320,
 Filling->Axis,
 PlotStyle->Directive[Opacity[0.5],Thickness[0.003]]
 ]

Histogram[
 table,
 Automatic,
 LabelingFunction->Above,
 ChartLegends->{"Sample 1","Sample 2"},
 ChartStyle->{Directive[Opacity[0.2],Red],Directive[Opacity[0.2],Purple]},
 ImageSize->320
 ]
```

Output

```
{FRatioDistribution[4.53532,23.311],FRatioDistribution[1674.59,2.45633],FRatioDistribution[4.80903,8.12755],FRatioDistribution[7.93864,6.22059],FRatioDistribution[468.925,2.83437],FRatioDistribution[5.19685,11.3133],FRatioDistribution[4.67207,84.9669],FRatioDistribution[4.73129,9.44774],FRatioDistribution[4.85831,15.1224],FRatioDistribution[4.2187,25.2685],FRatioDistribution[1750.13,2.58544],FRatioDistribution[6.51629,9.85254],FRatioDistribution[4.84708,11.4022],FRatioDistribution[6.18244,8.14405],FRatioDistribution[1835.29,1.78427],FRatioDistribution[3.79382,7.76494],FRatioDistribution[15627.4,2.50619],FRatioDistribution[6.56693,12.1631],FRatioDistribution[5.24017,11.4662],FRatioDistribution[2472.12,2.8793]}
```

Output

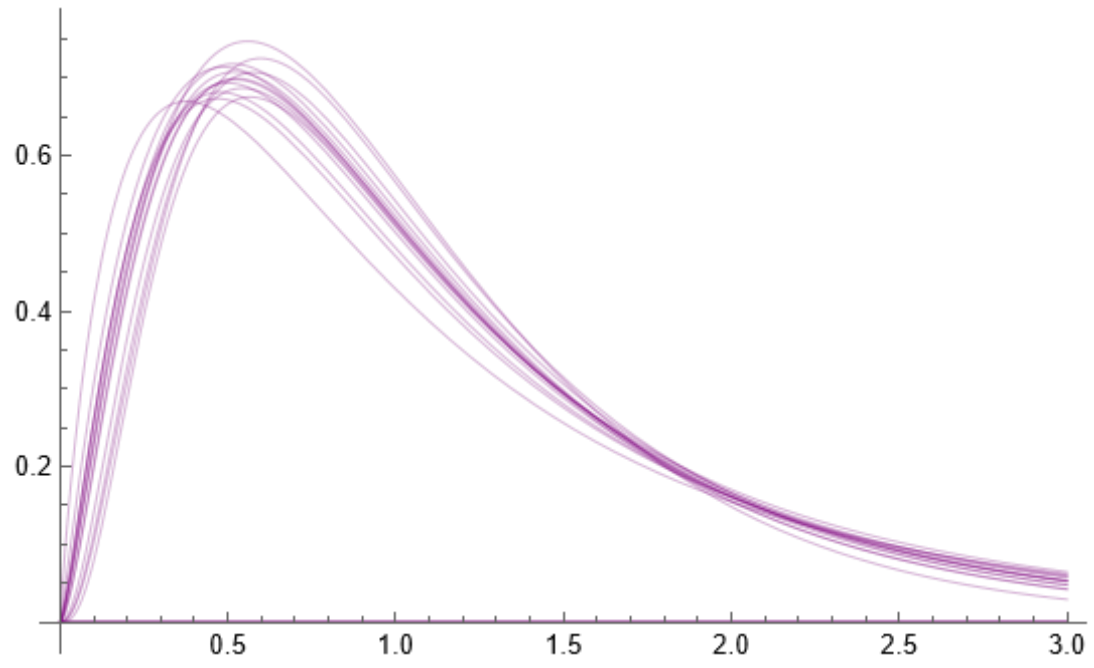

Output

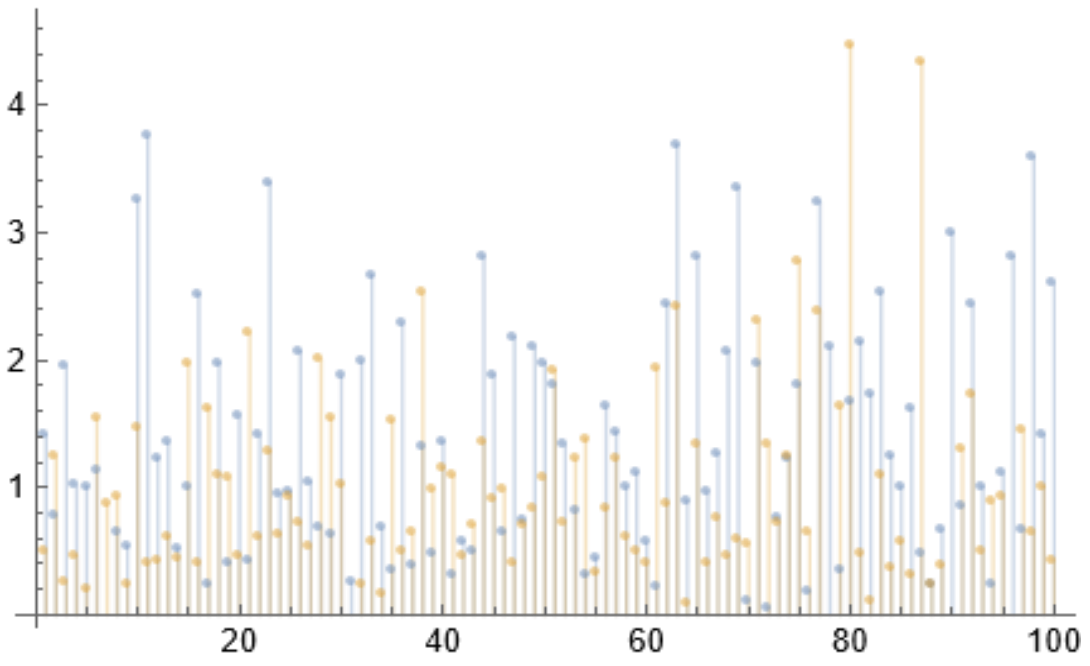

Output

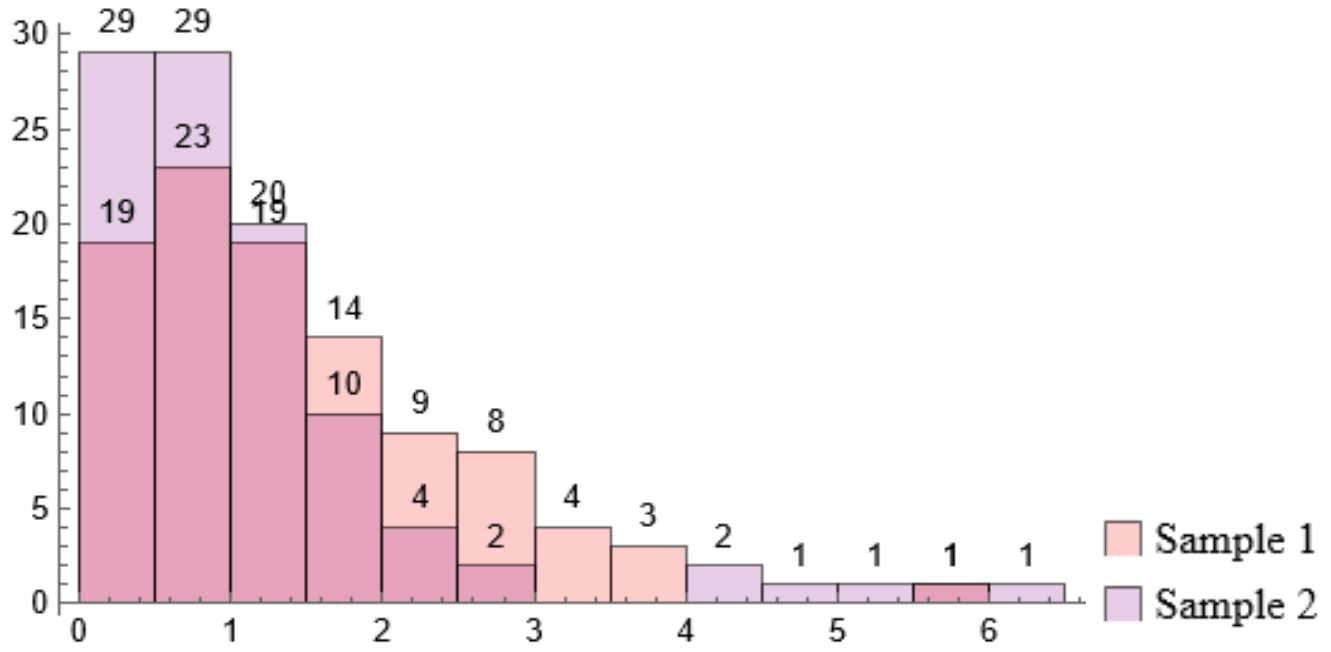

*Mathematica Examples 17.100*

Input

```
(* The code is designed to compare two FRatio distributions. It does this by generating random samples from each distribution and displaying them in a histogram, as well as plotting the PDFs of the two distributions. The code allows the user to manipulate the parameters n and m  of both FRatio distributions through the sliders for n1, m1, n2 and m2. By changing these parameters, the user can see how the distributions change and how they compare to each other. The histograms display the sample data for each distribution, with the first histogram showing the sample data for the first FRatio distribution and the second histogram showing the sample data for the second FRatio distribution. The histograms are overlaid on each other, with the opacity of each histogram set to 0.2 to make it easier to see where the data overlap. The PDFs of the two distributions are also plotted on the same graph, with the first distribution shown in blue and the second distribution shown in red. The legend indicates which color corresponds to which distribution. By looking at the histograms and the PDFs, the user can compare the two FRatio distributions and see how they differ in terms of n, m, and overlap of their sample data: *)
Manipulate[
 Module[
  {dist1,dist2,data1,data2},
  SeedRandom[seed];
  dist1=FRatioDistribution[n1,m1];
  dist2=FRatioDistribution[n2,m2];
  data1=RandomVariate[dist1,n];
  data2=RandomVariate[dist2,n];
  Column[
   {
    Show[
     ListPlot[
      data1,
      ImageSize->320,
      PlotStyle->Blue,
      PlotRange->{0,8}
      ],
     ListPlot[
      data2,
      ImageSize->320,
```

```
           PlotStyle->Red
           ]
          ],
         Show[
          Plot[
           {PDF[dist1,x],PDF[dist2,x]},
           {x,Min[{data1,data2}],Max[{data1,data2}]},
           PlotLegends->{"Distribution 1","Distribution 2"},
           PlotRange->{{0,20},{0,1}},
           PlotStyle->{Blue,Red},
           ImageSize->320
           ],
          Histogram[
           {data1,data2},
           Automatic,
           "PDF",
           ChartLegends->{"sample data1","sample data2"},
           ChartStyle->{Directive[Opacity[0.2],Red],Directive[Opacity[0.2],Purple]},
           ImageSize->320
           ]
          ]
         }
        ]
       ],
      {{n1,6},3,10,0.1},
      {{m1,5},4,10,0.1},
      {{n2,6},3,10,0.1},
      {{m2,5},4,10,0.1},
      {{n,500},{100,500,1000,2000}},
      {{seed,1234},ControlType->None}
      ]
```

Output

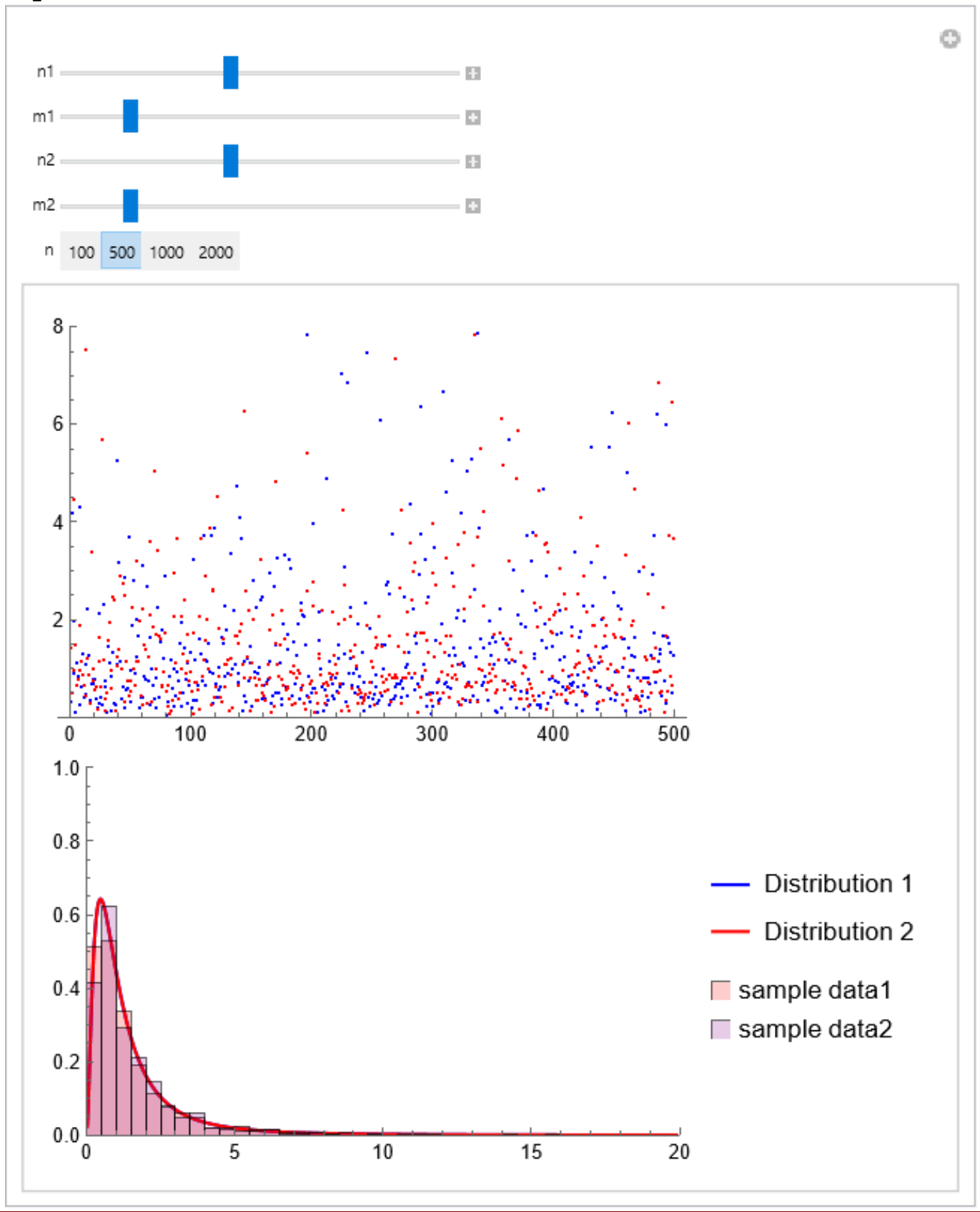

***Mathematica Examples 17.101***

Input
```
(* Chi square distribution is a limiting case of F-ratio distribution: *)
Limit[
 PDF[
  FRatioDistribution[1,m],
  x
  ],
 m->Infinity
 ]
PDF[ChiSquareDistribution[1],x]
```

Output

$$\begin{cases} \frac{e^{-x/2}}{\sqrt{2\pi}\sqrt{x}} & x>0 \\ 0 & \text{True} \end{cases}$$

Output

$$\begin{cases} \frac{e^{-x/2}}{\sqrt{2\pi}\sqrt{x}} & x>0 \\ 0 & \text{True} \end{cases}$$

***Mathematica Examples 17.102***

Input
```
(* F-ratio is the ratio of two Chi square distribution variables: *)
TransformedDistribution[
 (u/n)/(v/m),
 {Distributed[u,ChiSquareDistribution[n]],Distributed[v,ChiSquareDistribution[m]]}
 ]
```

Output
```
FRatioDistribution[n,m]
```

***Mathematica Examples 17.103***

Input
```
(* A square of student T distribution has F-ratio distribution: *)
TransformedDistribution[
 u^2,
 Distributed[u,StudentTDistribution[m]]
 ]
```

Output
```
FRatioDistribution[1,m]
```

# UNIT 17.7

# SAMPLING DISTRIBUTIONS OF RATIO OF TWO SAMPLE VARIANCES WITH FRATIO DISTRIBUTION

*Mathematica Examples 17.104*

Input

```
(* The code generates samples from two chi-squared distributions, calculates the F ratio for each pair of samples, and visualizes the F ratio distribution using a PDF plot and a histogram. The code begins by specifying the degrees of freedom for the chi-squared distributions. It then generates samples from each distribution using the RandomVariate function. The F ratio is calculated by dividing the samples from the first distribution by the scaled samples from the second distribution. The code proceeds to create a PDF plot of the F ratio distribution using the Plot function. The PDF is computed using the FRatioDistribution function with the specified degrees of freedom. Additionally, a histogram of the F ratios is generated using the Histogram function: *)

(* Define the degrees of freedom for the chi-squared distributions: *)
dof1=5;
dof2=10;

(* Generate samples from two chi-squared distributions: *)
samples1=RandomVariate[ChiSquareDistribution[dof1],1000];
samples2=RandomVariate[ChiSquareDistribution[dof2],1000];

(* Calculate the F ratio for each pair of samples: *)
fratios=samples1/(dof1*samples2/dof2);

(* Plot the probability density function of the F ratio distribution: *)
Plot[
 PDF[FRatioDistribution[dof1,dof2],x],
 {x,0,5},
 PlotRange->All,
 ImageSize->250,
 AxesLabel->{"F ratio","Probability density"},
 Filling->Axis,
 FillingStyle->LightPurple,
 PlotRange->All,
 PlotStyle->Purple
 ]

(* Create a histogram of the F ratios: *)
Histogram[
 fratios,
 "FreedmanDiaconis",
 AxesLabel->{"F ratio","Frequency"},
 ImageSize->250,
 ColorFunction->Function[{height},Opacity[height]],
 ChartStyle->Purple
 ]
```

Output

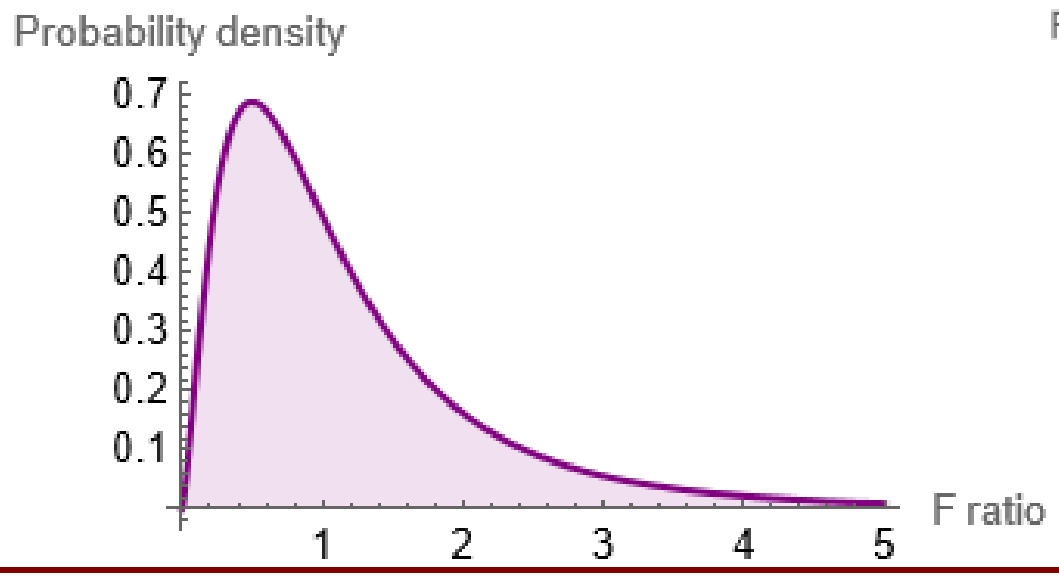

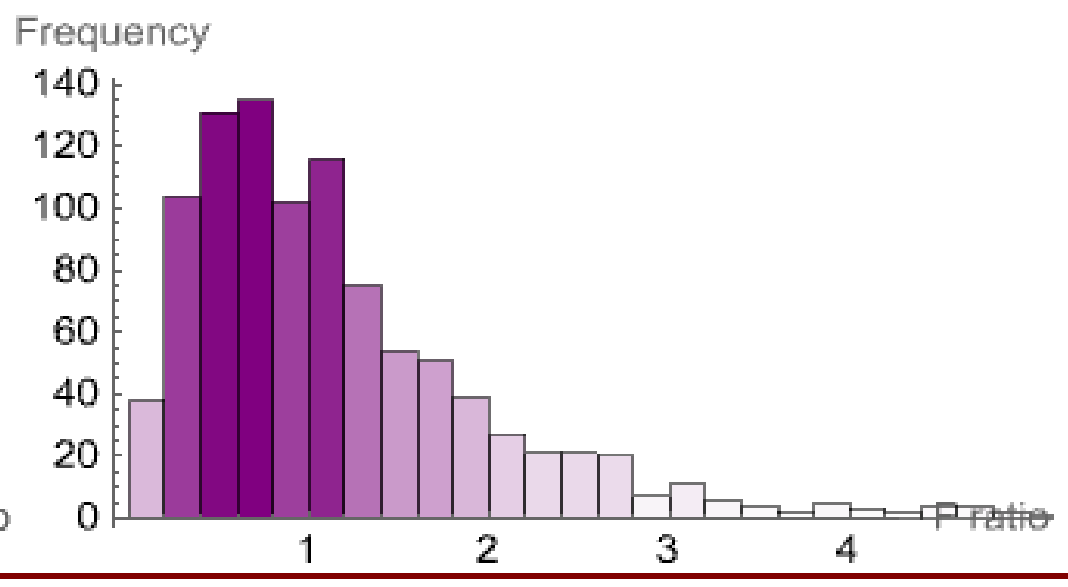

### *Mathematica Examples 17.105*

Input

```
(* The code generates a large number of sample ratios by sampling from two populations
with different sample sizes and standard deviations. It calculates the ratio of the
sample variances for each pair of samples and plots the resulting sampling
distribution using both a histogram and a PDF plot. The code starts by defining the
sample sizes and standard deviations for the two populations. It then generates
sample data by sampling from normal distributions with mean 0 and the specified
standard deviations. The sample variances are calculated, and the ratio of the
variances is normalized by the squared population standard deviations. Next, the code
creates a histogram of the sample ratios using the Histogram function. The binning
is automatically determined, and the histogram represents the probability density
function (PDF) of the sample ratios. Additionally, the code generates a PDF plot of
the F ratio distribution using the Plot function. The Show function combines the
histogram and the PDF plot into a single visualization, ensuring both elements are
displayed: *)

n1=20;   (* Sample size for population 1. *)
n2=25;   (* Sample size for population 2. *)
σ1=1;    (* Standard deviation of population 1. *)
σ2=2;    (* Standard deviation of population 2. *)

(* Generate a large number of sample ratios: *)
sampleRatios=Table[
    Module[
     {sample1,sample2,var1,var2},
     sample1=RandomVariate[NormalDistribution[0,σ1],n1];
     sample2=RandomVariate[NormalDistribution[0,σ2],n2];
     var1=Variance[sample1];
     var2=Variance[sample2];
     (var1/σ1^2)/(var2/σ2^2)
     ],
    10000
    ];

(* Plot the sampling distribution: *)
Show[
 Histogram[
  sampleRatios,
  Automatic,
  "PDF",
  ColorFunction->Function[{height},Opacity[height]],
  AxesLabel->{"Ratio of Sample Variances","Probability density"},
  PlotLabel->"Sampling Distribution of\n Ratio of Sample Variances",
  ChartStyle->Purple,
  PlotRange->All,
  ImageSize->350
  ],
 Plot[
  PDF[FRatioDistribution[n1-1,n2-1],x],
```

```
             {x,0,5},
             PlotRange->All,
             ImageSize->350,
             AxesLabel->{"F ratio","Probability density"},
             PlotRange->All,
             PlotStyle->Purple
             ]
            ]
```

Output

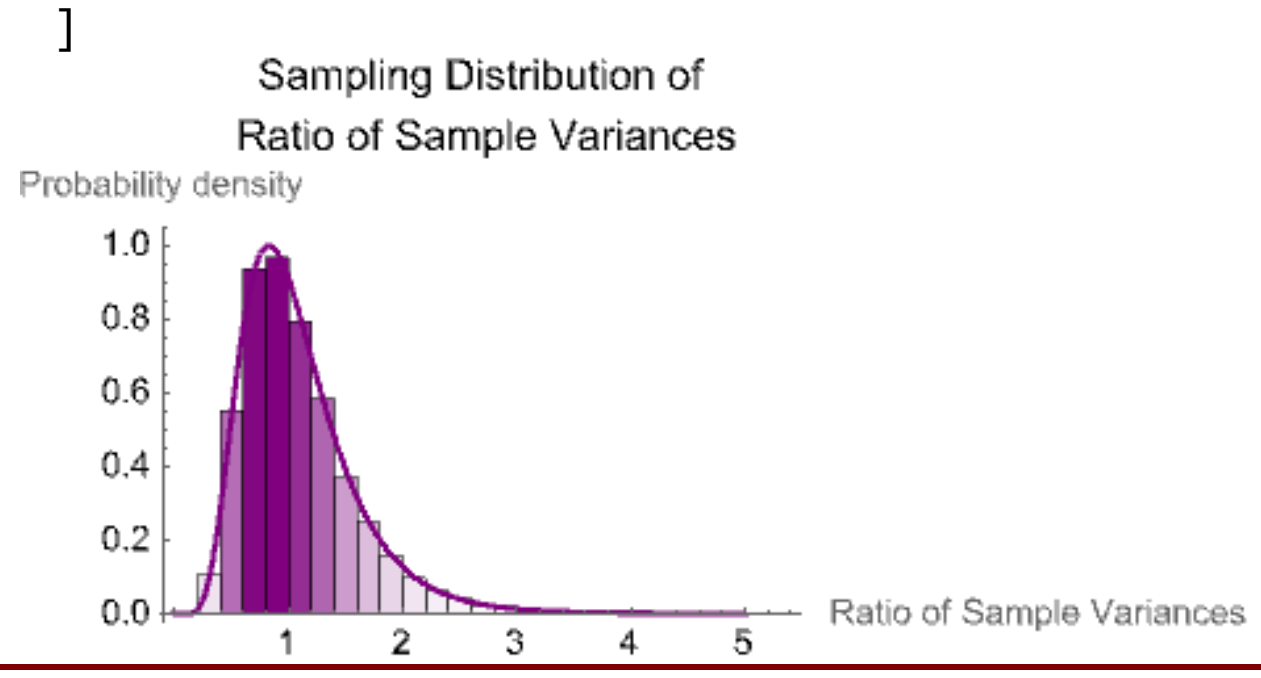

***Mathematica Examples 17.106***

Input

```
(* The code demonstrates a dynamic Manipulate interface that allows users to explore the effects of changing the parameters on the sample distribution of the ratio of sample variances using the F distribution. The code sets up sliders for adjusting the sample sizes and standard deviations of two populations. Inside the Manipulate, a large number of sample ratios are generated by sampling from the populations, calculating the sample variances, and computing the ratio of the variances. The code creates a visualization that consists of a histogram and a PDF plot. The histogram represents the probability density function (PDF) of the sample ratios, while the PDF plot displays the PDF of the F ratio distribution. Users can observe the impact of parameter changes on the resulting distribution and gain insights into the relationship between sample variances: *)

Manipulate[
 Module[
  {n1,n2,σ1,σ2,sampleRatios},
  n1=sampleSize1;
  n2=sampleSize2;
  σ1=stdDev1;
  σ2=stdDev2;
  (* Generate a large number of sample ratios: *)
  sampleRatios=Table[
    Module[
     {sample1,sample2,var1,var2},
     sample1=RandomVariate[NormalDistribution[0,σ1],n1];
     sample2=RandomVariate[NormalDistribution[0,σ2],n2];
     var1=Variance[sample1];
     var2=Variance[sample2];
     (var1/σ1^2)/(var2/σ2^2)
     ],
    10000
    ];
  (* Plot the sampling distribution: *)
  Show[
   Histogram[
    sampleRatios,
    Automatic,
    "PDF",
    PlotRange->All,
    ColorFunction->Function[{height},Opacity[height]],
    AxesLabel->{"Ratio of Sample Variances","Probability density"},
```

```
        PlotLabel->"Sampling Distribution of Ratio of Sample Variances",
        ChartStyle->Purple
        ],
       Plot[
        PDF[FRatioDistribution[n1-1,n2-1],x],
        {x,0,2.5},
        PlotRange->All,
        ImageSize->350,
        AxesLabel->{"F ratio","Probability density"},
        PlotRange->All,
        PlotStyle->Blue
        ]
       ]
      ],
     {{sampleSize1,30,"Sample Size 1"},5,50,1,Appearance->"Labeled"},
     {{sampleSize2,30,"Sample Size 2"},5,50,1,Appearance->"Labeled"},
     {{stdDev1,1,"Standard Deviation 1"},0.1,5,0.1,Appearance->"Labeled"},
     {{stdDev2,2,"Standard Deviation 2"},0.1,5,0.1,Appearance->"Labeled"}
     ]
```

Output

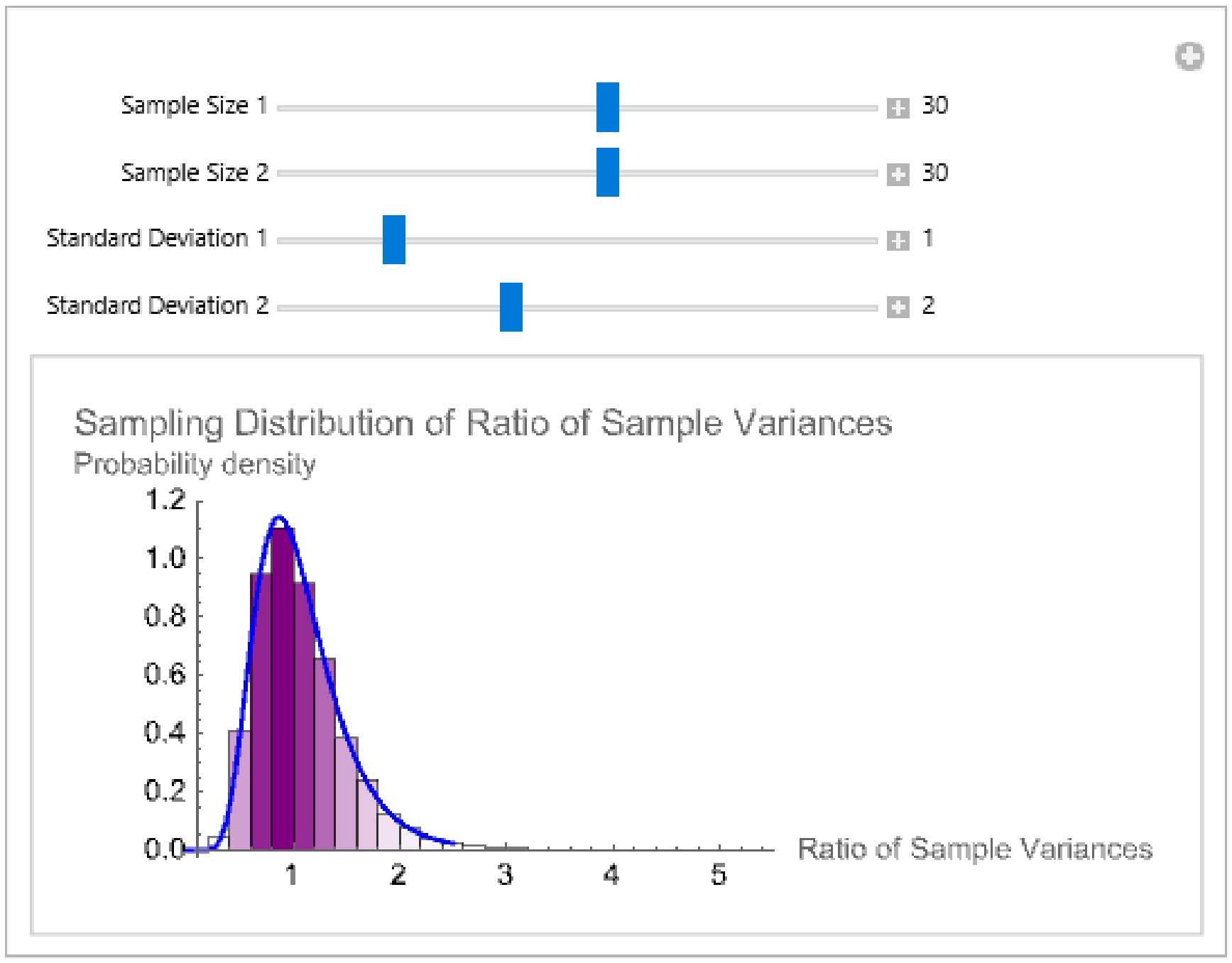

# CHAPTER 18

# ESTIMATION THEORY

Estimation theory is a branch of statistics that deals with the problem of estimating unknown parameters from observed data. It provides a framework for making informed decisions based on limited information and quantifying the uncertainty associated with those decisions.

Estimation theory plays a crucial role in various fields, including engineering, economics, physics, and social sciences, where accurate parameter estimation is essential for understanding and predicting phenomena. Throughout this chapter, we will explore the theoretical foundations and practical applications of estimation theory, enabling us to make reliable inferences about unknown population parameters based on sample data.

At its core, estimation theory addresses the following fundamental questions: How can we use the available data to estimate unknown parameters? How confident can we be in the accuracy of our estimates? What are the best methods for estimation given certain assumptions and constraints?

- The process of estimation involves formulating mathematical models that describe the relationship between the observed data and the unknown parameters of interest. It involves two main methods: point estimation and confidence intervals (CIs).
- A point estimator is a statistic that provides an estimate or guess for an unknown population parameter based on sample data. It summarizes the information from the sample and provides a single value as the estimate for the parameter of interest. This chapter explores various point estimation techniques, including the method of moments (MOM), and maximum likelihood estimation (MLE), along with their properties and applications.
- However, a point estimator alone does not convey any information about the accuracy of the estimate. To address the uncertainty associated with point estimators, CIs are used. A CI is an interval estimate that provides a range of plausible values for the unknown population parameter, along with a corresponding level of confidence. It quantifies the uncertainty in the estimation process and provides a measure of the precision of the estimate. These intervals incorporate both the point estimate and the variability associated with it, offering a more comprehensive understanding of the true value of the parameter. Throughout this chapter, we delve into the construction, interpretation, and calculation of CIs, exploring different approaches such as the t-distribution, F-ratio distribution, and z-distribution.
- In addition, we will explore various methods and techniques for constructing CIs in the following cases:
    - Large-sample CI for mean and $\sigma$ known.
    - Large-sample CI for mean and $\sigma$ unknown.
    - CI for mean in the case of the normal population and $\sigma$ unknown.
    - CI for $\mu_1$-$\mu_2$, $\sigma_1^2$ and $\sigma_2^2$ known.
    - CI for $\mu_1$-$\mu_2$, $\sigma_1^2 = \sigma_2^2$ but both unknown.
    - CI for $\mu_1$-$\mu_2$, $\sigma_1^2 \neq \sigma_2^2$ and both unknown.
    - CI for $\sigma^2$.
    - CI for $\sigma_1^2/\sigma_2^2$.
    - Large-sample CIs for a population proportion $p$.
    - Large-sample CI for $p_1 - p_2$.

## 18.1 Point Estimate

**Definition (Point Estimate):** An estimate of a population parameter given by a single number is called a point estimate of the parameter.

**Definition (Interval Estimate):** An estimate of a population parameter given by two numbers between which the parameter may be considered to lie is called an interval estimate of the parameter.

In a practical situation, there may be several statistics that could be used as point estimators for a population parameter. To decide which of several choices is best, you need to know how the estimator behaves in repeated sampling, described by its sampling distribution.

Sampling distributions provide information that can be used to select the best estimator. What characteristics would be valuable? First, the sampling distribution of the point estimator should be centered over the true value of the parameter to be estimated. That is, the estimator should not constantly underestimate or overestimate the parameter of interest. Such an estimator is said to be unbiased.

**Definition (Unbiased and Biased Estimator):** If the mean of the sampling distribution of a statistic equals the corresponding population parameter, the statistic is called an unbiased estimator of the parameter; otherwise, it is called a biased estimator. The corresponding values of such statistics are called unbiased or biased estimates, respectively.

The sampling distributions for an unbiased estimator and a biased estimator are shown in Figure 18.1. The sampling distribution for the biased estimator is shifted to the right of the true value of the parameter.

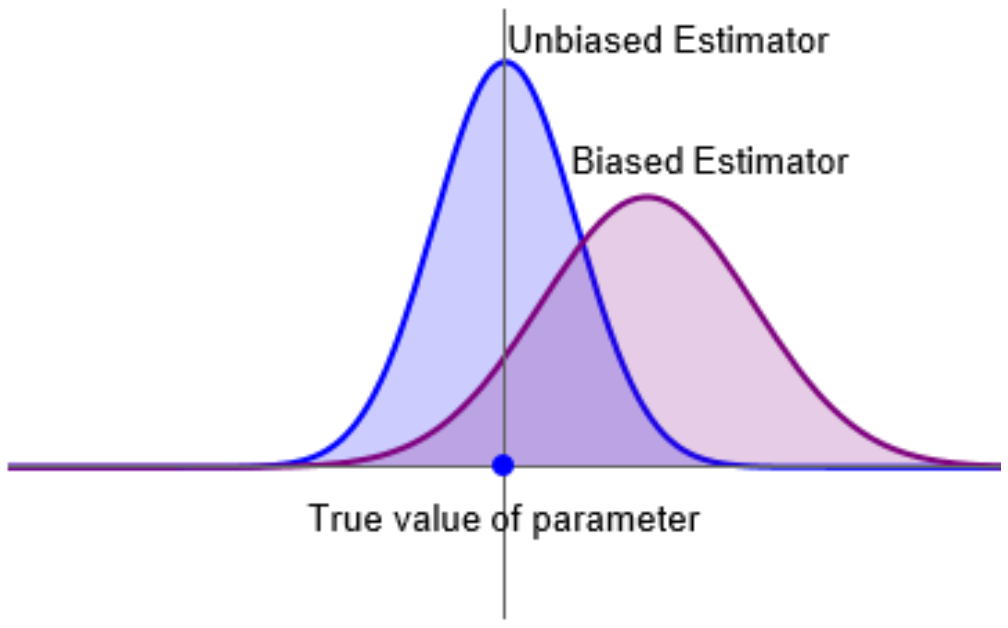

**Figure 18.1.** Distributions for biased and unbiased estimators.

***Example 18.1***

The mean of the sampling distribution of means $\mu_{\bar{X}}$ is $\mu$, the population mean. Hence the sample mean $\bar{X}$ is an unbiased estimate of the population mean $\mu$.

***Example 18.2***

The mean of the sampling distribution of variances is

$$\mu_{S^2} = \frac{n-1}{n}\sigma^2,$$

where $\sigma^2$ is the population variance and $n$ is the sample size. Thus, the sample variance $S^2$ is a biased estimate of the population variance $\sigma^2$. By using the modified variance

$$\hat{S}^2 = \frac{n}{n-1}S^2,$$

we find $\mu_{\hat{S}^2} = \sigma^2$, so that $\hat{S}^2$ is an unbiased estimate of $\sigma^2$. However, $\hat{S}$ is a biased estimate of $\sigma$.

In the language of expectation, we could say that a statistic is unbiased if its expectation equals the corresponding population parameter. Thus $\bar{X}$ and $\hat{S}^2$ are unbiased since $E[\bar{X}] = \mu$ and $E[\hat{S}^2] = \sigma^2$.

A second important characteristic is that the spread (as measured by the variance) of the estimator sampling distribution should be as small as possible. This ensures that, with a high probability, an individual estimate will fall close to the true value of the parameter. The sampling distributions for two unbiased estimators, one with a small variance and the other with a larger variance, are shown in Figure 18.2. Naturally, you would prefer the estimator with the smaller variance because the estimates tend to lie closer to the true value of the parameter than in the distribution with the larger variance.

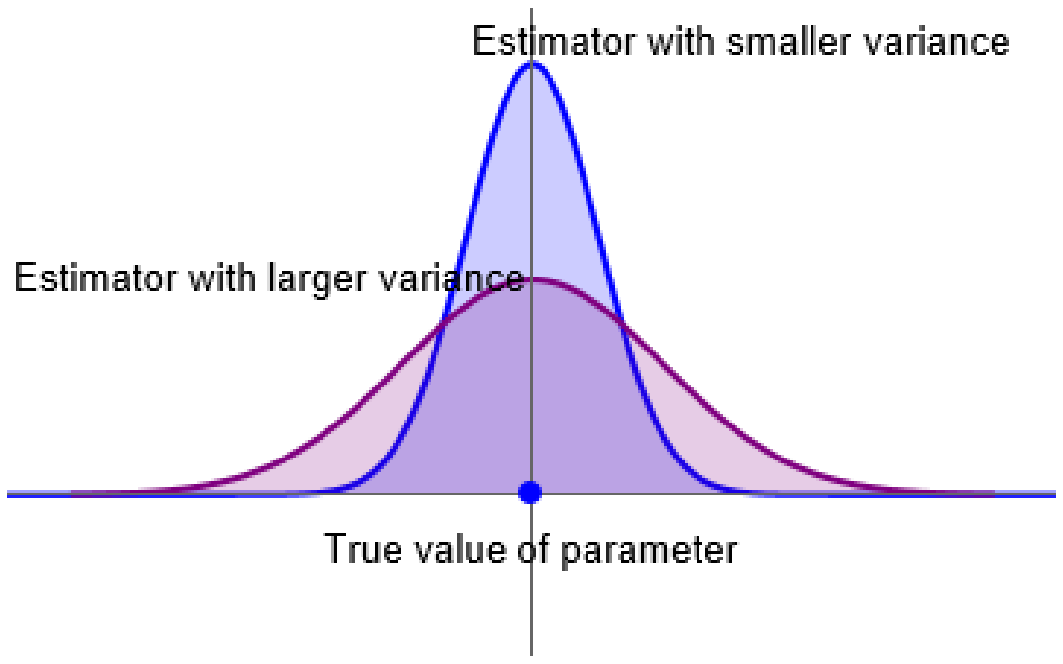

**Figure 18.2.** Comparison of estimator variability.

**Definition (Efficient and Inefficient Estimator):** If the sampling distributions of two statistics have the same mean (or expectation), then the statistic with the smaller variance is called an efficient estimator of the mean, while the other statistic is called an inefficient estimator. The corresponding values of the statistics are called efficient and inefficient estimates.

**Definition (Most Efficient, or Best, Estimator):** If we consider all possible statistics whose sampling distributions have the same mean, the one with the smallest variance is sometimes called the most efficient, or best, estimator of this mean.

**Definition (Error of Estimation):** The distance between an estimate and the true value of the parameter is called the error of estimation.

***Example 18.3***

The sampling distributions of the mean and median both have the same mean, namely, the population mean. However, the variance of the sampling distribution of means is smaller than the variance of the sampling distribution of medians. Hence the sample mean gives an efficient estimate of the population mean, while the sample median gives an inefficient estimate of it. Of all statistics estimating the population mean, the sample mean provides the best (or most efficient) estimate.

There are many methods available for estimating the true value(s) of the parameter(s) of interest, for example, method of MLE, MOM, method of least square, minimum-variance mean-unbiased estimator, median unbiased estimator, and best linear unbiased estimator. In this chapter, two of the more popular methods for obtaining point estimators will be considered, namely, the MLE and MOM.

### 18.1.1 MLEs

Any statistic used to estimate the value of an unknown parameter $\theta$ is called an estimator of $\theta$. The observed value of the estimator is called the estimate. For instance, as we shall see, the usual estimator of the mean of a normal population, based on a sample $X_1, \ldots, X_n$ from that population, is the sample mean $\bar{X} = \sum_i X_i/n$. If a sample of size 3 yields the data $x_1 = 2$, $x_2 = 3$, $x_3 = 4$, then the estimate of the population mean, resulting from the estimator $\bar{X}$, is the value 3.

MLE is a method of estimating the parameters of an assumed probability distribution, given some observed data. The MLE method is based on the principle that the values of the parameters should be chosen to make the observed data most probable. This is achieved by maximizing a likelihood function so that, under the assumed statistical model, the observed data is most probable.

Let us suppose we have a statistical model with a set of parameters $\boldsymbol{\theta}$ ($\boldsymbol{\theta}$ is a vector of parameters) and a sample of observed data $X = \{x_1, x_2, \ldots, x_n\}$. Let $f(x_1, x_2, \ldots, x_n; \boldsymbol{\theta})$ denote the joint PMF of the RVs $X_1, \ldots, X_n$ when they are discrete, and let it be their joint PDF when they are jointly continuous RVs. Because $\boldsymbol{\theta}$ is assumed unknown, we also write $f$ as a function of $\boldsymbol{\theta}$. The likelihood function, denoted by $\mathcal{L}(X|\boldsymbol{\theta})$, is a measure of how likely the observed data is under the given parameter values. Now since $\mathcal{L}(X|\boldsymbol{\theta}) = f(x_1, x_2, \ldots, x_n; \boldsymbol{\theta})$ represents the likelihood that the values $x_1, \ldots, x_n$ will be observed when $\boldsymbol{\theta}$ is the true value of the parameter, it would seem that a reasonable estimate of $\boldsymbol{\theta}$ would be that value yielding the largest likelihood of the observed values. The goal of MLE is to find the values of $\boldsymbol{\theta}$ that maximize the likelihood function. The likelihood function is typically defined as:

$$\mathcal{L}(X|\boldsymbol{\theta}) = f(x_1; \boldsymbol{\theta}) f(x_2; \boldsymbol{\theta}) \ldots f(x_n; \boldsymbol{\theta}), \tag{18.1.1}$$

where $f(x_i; \boldsymbol{\theta})$ is the PDF or PMF of the model. To find the MLEs, we seek the value of $\boldsymbol{\theta}$ that maximize the likelihood function. In determining the value of $\boldsymbol{\theta}$, it is often useful to use the fact that $f(x_1, x_2, \ldots, x_n; \boldsymbol{\theta})$ and $\log[f(x_1, x_2, \ldots, x_n; \boldsymbol{\theta})]$ have their maximum at the same value of $\boldsymbol{\theta}$. The log-likelihood function, denoted by $\ell(X|\boldsymbol{\theta})$, is:

$$\ell(X|\boldsymbol{\theta}) = \log[\mathcal{L}(X|\boldsymbol{\theta})] = \log[f(x_1; \boldsymbol{\theta})] + \log[f(x_2; \boldsymbol{\theta})] + \ldots + \log[f(x_n; \boldsymbol{\theta})]. \tag{18.1.2}$$

Once we have the log-likelihood function, we differentiate it with respect to $\boldsymbol{\theta}$, set the derivative to zero, and solve for $\boldsymbol{\theta}$.

A few more important aspects and properties of MLEs are:

- Consistency:
  Under certain regularity conditions, MLEs are consistent, meaning that as the sample size increases, the estimated parameter values converge to the true values. This property ensures that the estimates become more accurate with larger amounts of data.
- Efficiency:
  MLEs are often asymptotically efficient, which means that they achieve the smallest possible asymptotic variance among all consistent estimators. In simpler terms, MLEs tend to have smaller SEs compared to other estimators, making them more precise.
- Computational methods:
  In some cases, finding the MLE analytically may be challenging or impossible. In such situations, numerical optimization algorithms, such as the Newton-Raphson method or the expectation-maximization algorithm, are commonly used to find the MLEs.
- Applications:
  MLE is a versatile method used in various fields, including statistics, econometrics, machine learning, and many other areas of research. It is employed to estimate parameters in a wide range of models, including linear regression, logistic regression, survival analysis, mixed-effects models, and more.

***Example 18.4***

For the normal distribution $N(\mu, \sigma^2)$ which has PDF

$$f(x; \mu, \sigma^2) = \frac{1}{\sqrt{2\pi\sigma^2}} e^{-\frac{(x-\mu)^2}{2\sigma^2}},$$

the corresponding PDF for a sample of $n$ IID normal RVs (the likelihood) is

$$f(x_1, x_2, \dots, x_n; \mu, \sigma^2) = \prod_{i=1}^{n} f(x_i; \mu, \sigma^2) = \left(\frac{1}{2\pi\sigma^2}\right)^{n/2} e^{-\frac{\sum_{i=1}^{n}(x_i-\mu)^2}{2\sigma^2}}.$$

This family of distributions has two parameters: $\boldsymbol{\theta} = (\mu, \sigma)$; so, we maximize the likelihood, $\mathcal{L}(\mu, \sigma^2) = f(x_1, x_2, \dots, x_n; \mu, \sigma^2)$, over both parameters simultaneously, or if possible, individually. The log-likelihood can be written as follows:

$$\begin{aligned}\log[\mathcal{L}(\mu, \sigma^2)] &= \log\left[\left(\frac{1}{2\pi\sigma^2}\right)^{\frac{n}{2}} e^{-\frac{\sum_{i=1}^{n}(x_i-\mu)^2}{2\sigma^2}}\right] \\ &= \log\left[\left(\frac{1}{2\pi\sigma^2}\right)^{\frac{n}{2}}\right] + \log\left[e^{-\frac{\sum_{i=1}^{n}(x_i-\mu)^2}{2\sigma^2}}\right] \\ &= \log\left[(2\pi\sigma^2)^{-\frac{n}{2}}\right] - \frac{\sum_{i=1}^{n}(x_i-\mu)^2}{2\sigma^2} \\ &= -\frac{n}{2}\log[2\pi\sigma^2] - \frac{1}{2\sigma^2}\sum_{i=1}^{n}(x_i-\mu)^2.\end{aligned}$$

We now compute the derivatives of this log-likelihood with respect to $\mu$ and equate to zero:

$$\begin{aligned}\frac{\partial}{\partial\mu}\log[\mathcal{L}(\mu, \sigma^2)] &= 0 - \frac{2}{2\sigma^2}\sum_{i=1}^{n}(x_i-\mu)(-1) \\ &= \frac{2}{2\sigma^2}\left[\left(\sum_{i=1}^{n} x_i\right) - n\mu\right] \\ &= \frac{2}{2\sigma^2}\left[n\left(\frac{\sum_{i=1}^{n} x_i}{n}\right) - n\mu\right] \\ &= \frac{n}{\sigma^2}[\bar{x} - \mu] \\ &= 0,\end{aligned}$$

where $\bar{x}$ is the sample mean. This is solved by

$$\frac{n}{\sigma^2}[\bar{x} - \mu] = 0 \quad \Rightarrow \mu = \bar{x} = \sum_{i=1}^{n}\frac{x_i}{n} \quad \Rightarrow \hat{\mu} = \bar{X}.$$

This is indeed the maximum of the function, since it is the only turning point in $\mu$ and the second derivative is strictly less than zero. Its expected value is equal to the parameter $\mu$ of the given distribution,

$$E[\hat{\mu}] = \mu,$$

which means that the MLE $\hat{\mu}$ is unbiased.

Similarly, we differentiate the log-likelihood with respect to σ and equate to zero:

$$\begin{aligned}\frac{\partial}{\partial\sigma}\log[\mathcal{L}(\mu, \sigma^2)] &= \frac{\partial}{\partial\sigma}\left[-\frac{n}{2}\log[2\pi\sigma^2] - \frac{1}{2\sigma^2}\sum_{i=1}^{n}(x_i-\mu)^2\right] \\ &= -\frac{n}{\sigma} + \frac{1}{\sigma^3}\sum_{i=1}^{n}(x_i-\mu)^2 \\ &= 0,\end{aligned}$$

which is solved by,

$$\frac{n}{\sigma} = \frac{1}{\sigma^3}\sum_{i=1}^{n}(x_i-\mu)^2 \qquad \Rightarrow n = \frac{1}{\sigma^2}\sum_{i=1}^{n}(x_i-\mu)^2$$

$$\Rightarrow \sigma^2 = \frac{1}{n}\sum_{i=1}^{n}(x_i-\mu)^2.$$

Hence, we have

$$\hat{\sigma}^2 = \frac{1}{n}\sum_{i=1}^{n}(X_i-\mu)^2.$$

Inserting the estimate $\mu = \hat{\mu}$ we obtain

$$
\begin{aligned}
n\hat{\sigma}^2 &= \sum_{i=1}^{n}(X_i - \bar{X})^2 \\
&= \sum_{i=1}^{n}(X_i - \mu + \mu - \bar{X})^2 \\
&= \sum_{i=1}^{n}\big((X_i - \mu) - (\bar{X} - \mu)\big)^2 \\
&= \sum_{i=1}^{n}[(X_i - \mu)^2 - 2(X_i - \mu)(\bar{X} - \mu) + (\bar{X} - \mu)^2] \\
&= \sum_{i=1}^{n}(X_i - \mu)^2 - 2(\bar{X} - \mu)\sum_{i=1}^{n}(X_i - \mu) + (\bar{X} - \mu)^2\sum_{i=1}^{n}1 \\
&= \sum_{i=1}^{n}(X_i - \mu)^2 - 2(\bar{X} - \mu)\left(\left(\sum_{i=1}^{n}X_i\right) - n\mu\right) + n(\bar{X} - \mu)^2 \\
&= \sum_{i=1}^{n}(X_i - \mu)^2 - 2(\bar{X} - \mu)\left(n\left(\sum_{i=1}^{n}\frac{X_i}{n}\right) - n\mu\right) + n(\bar{X} - \mu)^2 \\
&= \sum_{i=1}^{n}(X_i - \mu)^2 - 2n(\bar{X} - \mu)(\bar{X} - \mu) + n(\bar{X} - \mu)^2 \\
&= \sum_{i=1}^{n}(X_i - \mu)^2 - 2n(\bar{X} - \mu)^2 + n(\bar{X} - \mu)^2 \\
&= \sum_{i=1}^{n}(X_i - \mu)^2 - n(\bar{X} - \mu)^2,
\end{aligned}
$$

now

$$
\begin{aligned}
E[\hat{\sigma}^2] &= E\left[\frac{1}{n}\sum_{i=1}^{n}(X_i - \bar{X})^2\right] \\
&= \frac{1}{n}E\left[\sum_{i=1}^{n}(X_i - \mu)^2 - n(\bar{X} - \mu)^2\right] \\
&= \frac{1}{n}\left(\sum_{i=1}^{n}E[(X_i - \mu)^2] - nE[(\bar{X} - \mu)^2]\right) \\
&= \frac{1}{n}\left(\sum_{i=1}^{n}\sigma_{X_i}^2 - n\sigma_{\bar{X}}^2\right).
\end{aligned}
$$

However

$$\sigma_{X_i}^2 = \sigma^2, \qquad \text{for } i = 1,2,\ldots,n,$$

and

$$\sigma_{\bar{X}}^2 = \frac{\sigma^2}{n}.$$

Therefore,

$$
\begin{aligned}
E[\hat{\sigma}^2] &= \frac{1}{n}\left(n\sigma^2 - n\frac{\sigma^2}{n}\right) \\
&= \frac{n-1}{n}\sigma^2.
\end{aligned}
$$

This means that the estimator $\hat{\sigma}^2$ is biased for $\sigma^2$.

**Procedure 18.1.**

Steps to apply MLE:

1. Define the probability distribution:
   Start by defining the probability distribution that you believe represents the data you are working with. This distribution could be Gaussian (normal), binomial, Poisson, etc., depending on the nature of your data.
2. Set up the likelihood function:
   The likelihood function represents the probability of observing your data given a set of parameters. It is derived from the probability distribution you defined in step 1. The likelihood function is typically denoted as $\mathcal{L}(X|\boldsymbol{\theta}) = f(x_1;\boldsymbol{\theta})f(x_2;\boldsymbol{\theta})\ldots f(x_n;\boldsymbol{\theta})$, where $f(x_i;\boldsymbol{\theta})$ is the PDF or PMF of the model and $\boldsymbol{\theta}$ represents the parameters of the distribution.
3. Take the natural logarithm:
   To simplify the calculations, it is common to take the natural logarithm of the likelihood function. This step does not change the location of the maximum, as the logarithm is a monotonic function.
4. Differentiate the log-likelihood function:
   Differentiate the logarithm of the likelihood function with respect to the parameters $\boldsymbol{\theta}$. This step helps find the maximum point in the parameter space.
5. Set the derivative to zero:
   Set the derivative obtained in step 4 to zero and solve for the parameters. This identifies the values of $\boldsymbol{\theta}$ that maximize the likelihood function.
6. Check the second derivative and positive definiteness:
   Calculate the second derivative of the log-likelihood function with respect to the parameters. This is known as the Hessian matrix. Evaluate the second derivative at the values of $\boldsymbol{\theta}$ obtained in step 5. Verify that the Hessian matrix is negative definite or negative semi-definite. This condition ensures that the maximum point found in step 5 is indeed a maximum and not a minimum or saddle point. You can use also Mathematica `FindMaximum` function to find values of $\boldsymbol{\theta}$ that maximize the likelihood function.
7. Solve for MLEs:
   Solve the equations obtained from step 5 to obtain the MLEs for the parameters of the distribution.

***Example 18.5***

```
(* Define the Probability Distribution: *)
dist=NormalDistribution[μ,σ];

(* Set up the Likelihood Function: *)
likelihood[data_,μ_,σ_]:=Product[PDF[dist,x],{x,data}];

(* Take the Natural Logarithm: *)
logLikelihood[data_,μ_,σ_]:=Log[likelihood[data,μ,σ]];

(* Differentiate the Log-Likelihood Function: *)
logLikelihoodDerivative[data_,μ_,σ_]:=D[logLikelihood[data,μ,σ],{{μ,σ}}];

(* Set the Derivative to Zero: *)
solutions=Solve[logLikelihoodDerivative[data,μ,σ]=={0,0},{μ,σ}];
(* Solve for Maximum Likelihood Estimates: *)
maximumLikelihoodEstimates={μ,σ}/. solutions[[1]];
(* Example usage: *)
data=RandomVariate[NormalDistribution[10,1],10]; (* Sample data.*)
result=FindMaximum[{logLikelihood[data,μ,σ],σ>0},{μ,σ}] ;(* Alternative approach using
FindMaximum. *)

Print["Maximum Likelihood Estimates: {μ,σ}= ",maximumLikelihoodEstimates];
Print["Alternative approach using FindMaximum: ",result];

 Maximum Likelihood Estimates: {μ,σ}=  {9.49631,-0.802853}
 Alternative approach using FindMaximum:  {-11.7935,{μ->9.34077,σ->0.786954}}
```

### 18.1.2. MOM

The MOM is a very simple procedure for finding an estimator for one or more population parameters. It starts by expressing the population moments (i.e., the expected values of powers of the RV under consideration) as functions of the parameters of interest. Those expressions are then set equal to the sample moments. The number of such equations is the same as the number of parameters to be estimated. Those equations are then solved for the parameters of interest. The solutions are estimates of those parameters. Let $\mu_k' = E[X^k]$ be the $k$th moment about the origin of a RV $X$, whenever it exists. Let $m_k' = \frac{1}{n}\sum_{i=1}^{n} X_i^k$ be the corresponding $k$th sample moment. Then, the estimator of $\mu_k'$ by the MOMs is $m_k'$.

**Example 18.6**

Let $X_1, \ldots, X_n$ be a random sample from a gamma probability distribution with parameters $\alpha$ and $\beta$. Find moment estimators for the unknown parameters $\alpha$ and $\beta$.

***Solution***

For the gamma distribution,

$$E[X] = \alpha\beta, \quad \text{and} \quad E[X^2] = \alpha\beta^2 + \alpha^2\beta^2.$$

Because there are two parameters, we need to find the first two moment estimators. Equating sample moments to distribution (theoretical) moments, we have:

$$\frac{1}{n}\sum_{i=1}^{n} x_i = \bar{x} = \alpha\beta,$$

and

$$\frac{1}{n}\sum_{i=1}^{n} x_i^2 = \alpha\beta^2 + \alpha^2\beta^2.$$

Solving for $\alpha$ and $\beta$, we obtain the estimates as $\alpha = \bar{x}/\beta$ and $\beta = \left(\frac{1}{n}\sum_{i=1}^{n} x_i^2 - \bar{x}^2\right)/\bar{x}$.

Therefore, the MOMs estimators for $\alpha$ and $\beta$ are:

$$\hat{\alpha} = \frac{\bar{X}}{\hat{\beta}}, \qquad \text{and} \qquad \hat{\beta} = \frac{\frac{1}{n}\sum_{i=1}^{n} X_i^2 - \bar{X}^2}{\bar{X}}.$$

**Procedure 18.2.**

Step-by-step explanation of the MOMs for finding point estimators:

1. Determine the number of parameters:
   First, determine the number of parameters needed to fully specify the probability distribution you are working with. For example, the normal distribution has two parameters (mean and standard deviation), while the exponential distribution has only one parameter (rate).
2. Calculate the sample moments:
   Calculate the sample moments from the observed data. The $k$-th sample moment is calculated by taking the average of the $k$-th powers of the data values. For example, the first sample moment (mean) is calculated as the average of the data, while the second sample moment is calculated as the average of the squared data values.
3. Set up equations:
   Set up a system of equations equating the theoretical moments to the sample moments. For example, if you are working with a distribution that has two parameters (mean and variance), you would set up two equations: one equating the theoretical mean to the sample mean, and another equating the theoretical variance to the sample variance.
4. Solve the equations:
   Solve the system of equations to find the values of the parameters that satisfy the equations. This can be done analytically or numerically depending on the complexity of the equations and the distribution.
5. Obtain the point estimators:
   Once the equations are solved, the values of the parameters obtained are the point estimators based on the MOMs.

***Example 18.7***

```
(* Identify the unknown parameters: *)
parameters={α,β};
dist=GammaDistribution[α,β];

(* Sample data: *)
data=RandomVariate[GammaDistribution[2,5],10]

(* Determine the sample moments: *)
sampleMoments={Moment[data,1],Moment[data,2]}

(* Equate sample moments to population moments: *)
populationMoments={Moment[dist,1],Moment[dist,2]}
equations=Thread[sampleMoments==populationMoments]

(* Solve the equations: *)
estimatedParameters=Solve[equations[[1]]&&equations[[2]],parameters,PositiveReals]

(* Obtain point estimates: *)
pointEstimates=parameters/. estimatedParameters

 {11.3119,24.1617,20.4791,13.041,7.3545,22.9465,5.03575,0.751959,3.31305,6.35594}
 {11.4751,195.913}
 {α β, α (1+α) β²}
 {11.4751==α β,195.913==α (1+α) β²}
 Solve::ratnz: Solve was unable to solve the system with inexact coefficients. The answer
was obtained by solving a corresponding exact system and numericizing the result.
 {{α->2.04996,β->5.59773}}
 {{2.04996,5.59773}}
```

***Example 18.8***

```
(* Identify the unknown parameters: *)
parameters={μ,σ};
dist=NormalDistribution[μ,σ];

(* Sample data: *)
data=RandomVariate[NormalDistribution[10,2],10]

(* Determine the sample moments: *)
sampleMoments={Moment[data,1],Moment[data,2]}

(* Equate sample moments to population moments: *)
populationMoments={Moment[dist,1],Moment[dist,2]}
equations=Thread[sampleMoments==populationMoments]

(* Solve the equations: *)
estimatedParameters=Solve[equations[[1]]&&equations[[2]]&&σ>0,parameters][[1]]

(* Obtain point estimates: *)
pointEstimates=parameters/. estimatedParameters

 {10.4598,8.95604,10.8539,8.566,13.1516,12.8169,8.08352,11.1618,8.20832,12.3804}
 {10.4638,112.862}
 {μ,μ²+σ²}
 {10.4638==μ,112.862==μ²+σ²}
```

```
Solve::ratnz: Solve was unable to solve the system with inexact coefficients. The answer was obtained by solving a corresponding exact system and numericizing the result.
{μ->10.4638,σ->1.8358}
{10.4638,1.8358}
```

## 18.2 Interval Estimate

Even the most efficient unbiased estimator is unlikely to estimate the population parameter exactly. It is true that estimation accuracy increases with large samples, but there is still no reason we should expect a point estimate from a given sample to be exactly equal to the population parameter it is supposed to estimate. There are many situations in which it is preferable to determine an interval within which we would expect to find the value of the parameter. Such an interval is called an interval estimate.

### CI Estimates of Population Parameters

Let $\mu_{\text{sta}}$ and $\sigma_{\text{sta}}$ be the mean and standard deviation (SE), respectively, of the sampling distribution of a statistic sta. Then if the sampling distribution of sta is approximately normal (which as we have seen is true for many statistics if the sample size $n \geq 30$), we can expect to find an actual sample statistic sta lying in the intervals $\mu_{\text{sta}} - \sigma_{\text{sta}}$ to $\mu_{\text{sta}} + \sigma_{\text{sta}}$, $\mu_{\text{sta}} - 2\sigma_{\text{sta}}$ to $\mu_{\text{sta}} + 2\sigma_{\text{sta}}$, or $\mu_{\text{sta}} - 3\sigma_{\text{sta}}$ to $\mu_{\text{sta}} + 3\sigma_{\text{sta}}$ about 68.27%, 95.45%, and 99.73% of the time, respectively.

Equivalently, we can expect to find (or we can be confident of finding) $\mu_{\text{sta}}$ in the intervals $\text{sta} - \sigma_{\text{sta}}$ to $\text{sta} + \sigma_{\text{sta}}$, $\text{sta} - 2\sigma_{\text{sta}}$ to $\text{sta} + 2\sigma_{\text{sta}}$, or $\text{sta} - 3\sigma_{\text{sta}}$ to $\text{sta} + 3\sigma_{\text{sta}}$ about 68.27%, 95.45%, and 99.73% of the time, respectively. Because of this, we call these respective intervals the 68.27%, 95.45%, and 99.73% CIs for estimating $\mu_{\text{sta}}$. The end numbers of these intervals ($\text{sta} \pm \sigma_{\text{sta}}$, $\text{sta} \pm 2\sigma_{\text{sta}}$, and $\text{sta} \pm 3\sigma_{\text{sta}}$) are then called the 68.27%, 95.45%, and 99.73% confidence limits.

Similarly, $\text{sta} \pm 1.96\sigma_{\text{sta}}$, and $\text{sta} \pm 2.58\sigma_{\text{sta}}$ are the 95% and 99% (or 0.95 and 0.99) confidence limits for sta. The percentage confidence is often called the confidence level. The numbers 1.96, 2.58, etc., in the confidence limits are called confidence coefficients, or critical values, and are denoted by $z_c$. From confidence levels we can find confidence coefficients, and vice versa.

**Definition (Confidence Coefficient):** The probability that a CI will contain the estimated parameter is called the confidence coefficient.

For example, experimenters often construct 95% CIs. This means that the confidence coefficient, or the probability that the interval will contain the estimated parameter, is 0.95, see Figure 18.3. You can increase or decrease your amount of certainty by changing the confidence coefficient. Some values typically used by experimenters are 0.90, 0.95, 0.98, and 0.99. Table 18.1 shows the values of $z_c$ corresponding to various confidence levels used in practice.

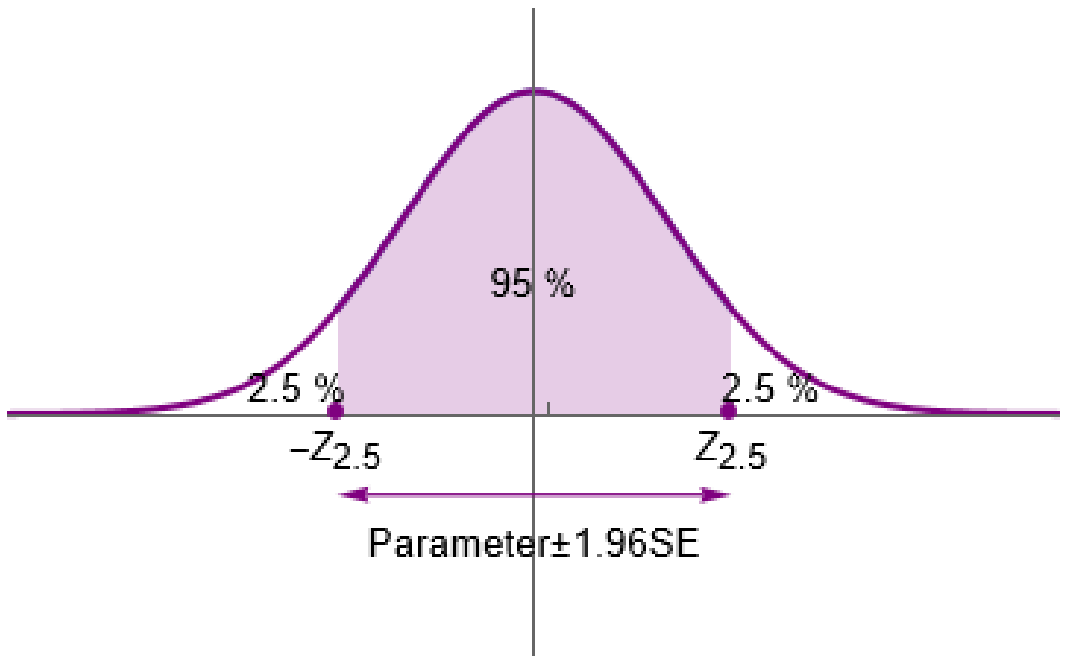

**Figure 18.3.** Parameter±1.96SE.

**Table 18.1.**

| confidence level | 99.73% | 99% | 98% | 96% | 95.45% | 95% | 90% | 80% | 68.27% | 50% |
|---|---|---|---|---|---|---|---|---|---|---|
| $z_c$ | 3.00 | 2.58 | 2.33 | 2.05 | 2.00 | 1.96 | 1.645 | 1.28 | 1.00 | 0.6745 |

How often will this interval work properly and enclose the parameter of interest? Refer to Figure 18.4.

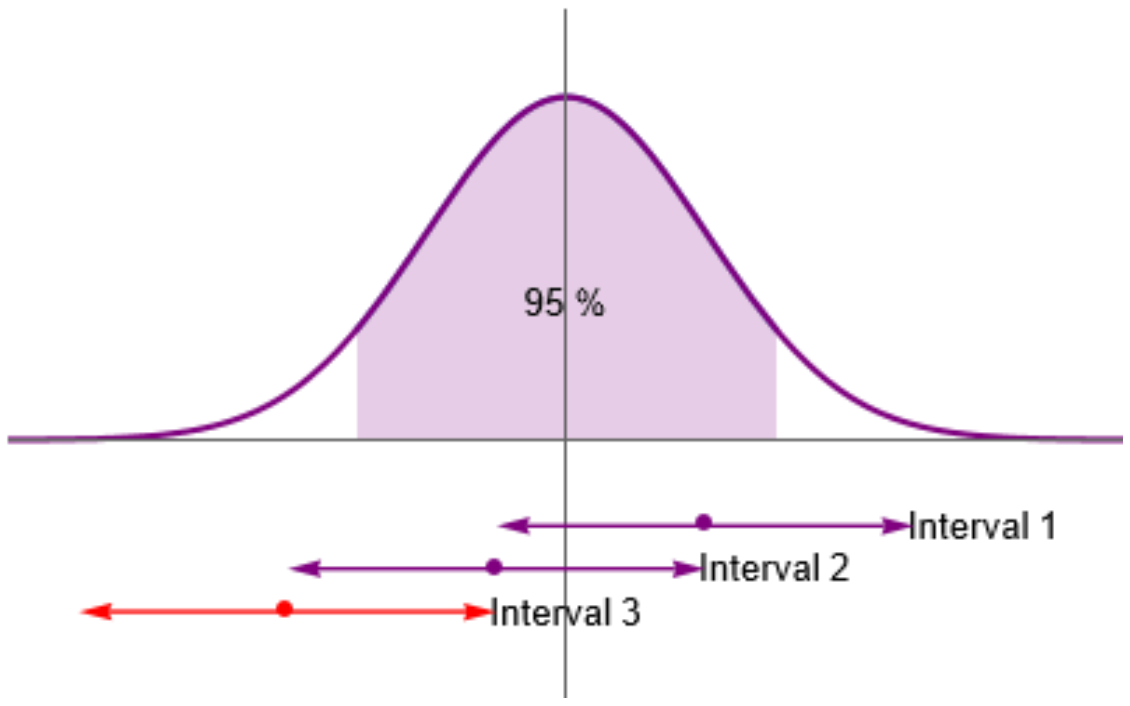

**Figure 18.4.** Some 95% CIs.

Intervals 1 and 2 work properly—the parameter (the center line) is contained within both intervals. Interval 3 does not work, because it fails to enclose the parameter. This happened because the value of the point estimator at the center of the interval was too far away from the parameter. Fortunately, values of the point estimator only fall this far away 5% of the time—our procedure will work properly 95% of the time!

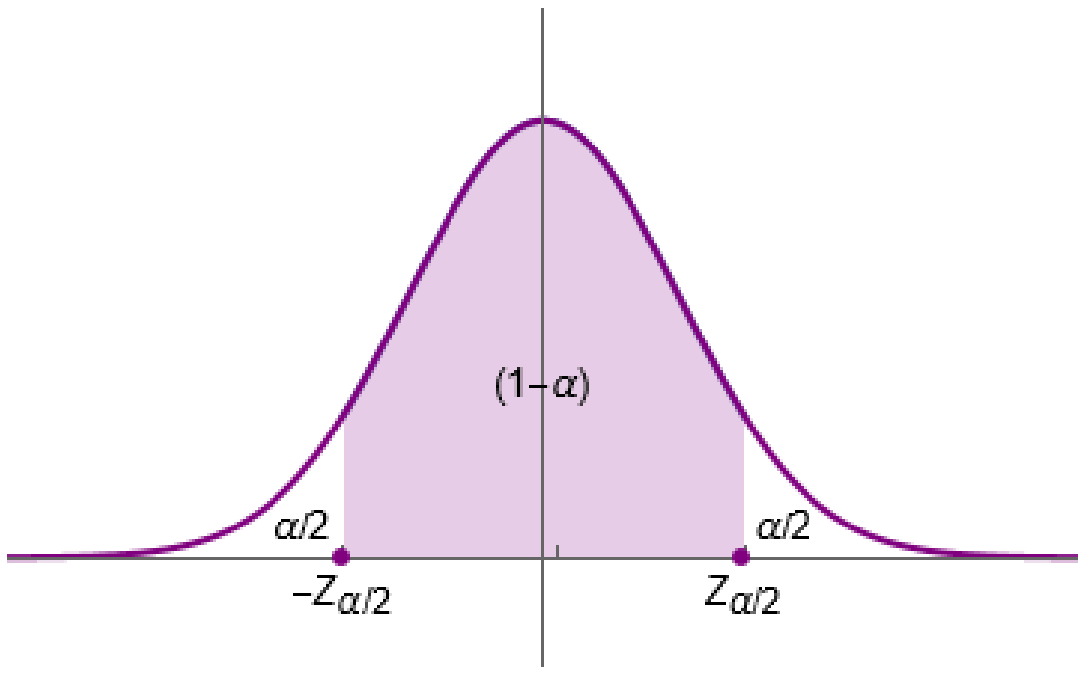

**Figure 18.5.** Location of $Z_{\alpha/2}$.

Since the total area under the curve is 1, the remaining area in the two tails is $\alpha$, and each tail contains area $\alpha/2$. The value of $z$ that has "tail area" $\alpha/2$ to its right is called $z_{\alpha/2}$, and the area between $-z_{\alpha/2}$ and $z_{\alpha/2}$ is the confidence coefficient $(1-\alpha)$, see Figure 18.5.

**A $(1-\alpha)$ 100% Large-sample CI**

(Point estimator) $\pm z_{\alpha/2}$ (SE of the estimator)

where $z_{\alpha/2}$ is the z-value with an area $\alpha/2$ in the right tail of a standard normal distribution. This formula generates two values: the lower confidence limit and the upper confidence limit.

The reader should notice that while point and interval estimation represent different approaches to gaining information regarding a parameter, they are related in the sense that CI estimators are based on point estimators. In the following section, for example, we will see that $\bar{X}$ is a very reasonable point estimator of $\mu$. As a result, the important CI estimator of $\mu$ depends on knowledge of the sampling distribution of $\bar{X}$.

### Interpreting the CI

What does it mean to say you are "90% confident" that the true value of the population mean $\mu$ is within a given interval? If you were to construct 20 such intervals, each using different sample information, your intervals might look like those shown in Figure 18.6. Of the 20 intervals, you might expect that 90% of them, or 18 out of 20, will perform as planned and contain $\mu$ within their upper and lower bounds. Remember that you cannot be absolutely sure that any one particular interval contains the mean $\mu$. You will never know whether your particular interval is one of the 18 that "worked," or whether it is the one interval that "missed." Your confidence in the estimated interval follows from the fact that when repeated intervals are calculated, 90% of these intervals will contain $\mu$.

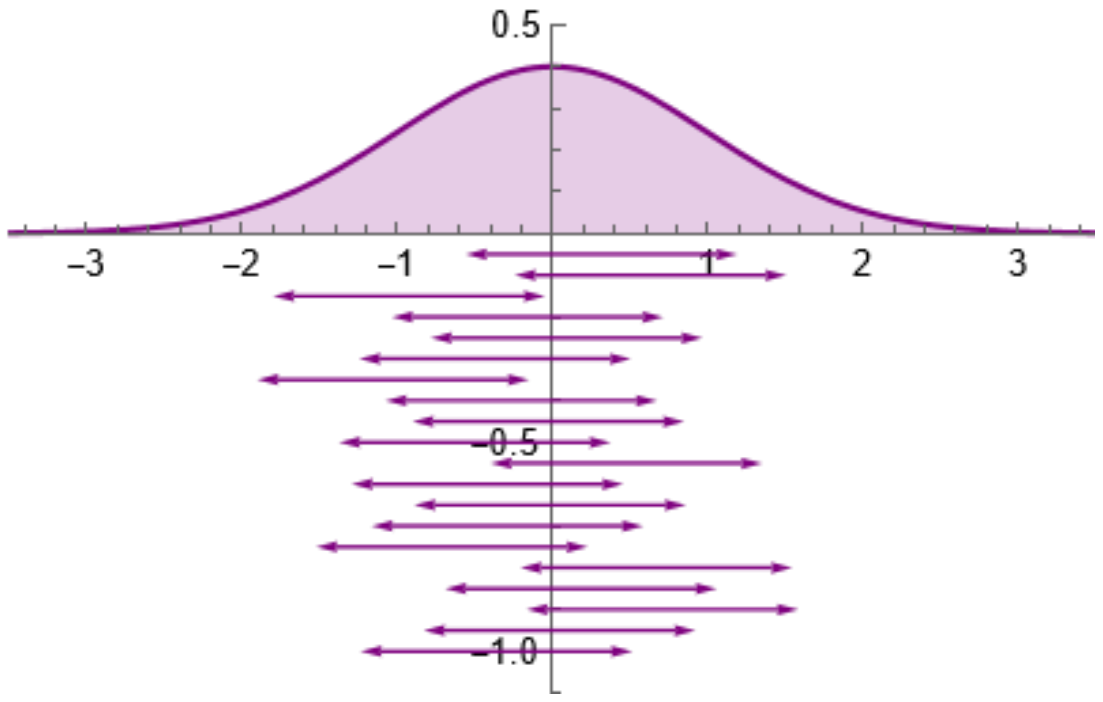

**Figure 18.6.** Interpreting CIs.

Statistical inference is concerned with making decisions or predictions about parameters. Three parameters that you have seen so far are the population mean $\mu$, the population standard deviation $S$ , and the binomial proportion $p$. In statistical inference, we state the practical problem in terms of one of these parameters.

**Procedure 18.3.**

The process of constructing a CI typically involves the following steps:

1. Determine the confidence level:
   Decide on the desired level of confidence for the interval, typically expressed as a percentage (e.g., 95% confidence level).
2. Collect sample data:
   Obtain a representative sample from the population of interest. The sample should be randomly selected and sufficiently large to meet the assumptions of the chosen inference method.
3. Calculate sample statistics:
   Compute the relevant sample statistics that will be used to estimate the population parameter. The specific statistic will depend on the type of data and the parameter of interest (e.g., mean, proportion, standard deviation).
4. Identify the sampling distribution:
   Determine the appropriate sampling distribution that corresponds to the sample statistic. The choice of distribution depends on the sampling method used and the characteristics of the data (e.g., normal distribution for large samples or t-distribution for small samples).
5. Determine the margin of error (ME):
   Calculate the ME based on the sampling distribution and the desired confidence level. The ME represents the maximum amount of error expected in the estimate.
6. Calculate the CI:
   Using the sample statistic, the ME, and the desired confidence level, compute the lower and upper bounds of the CI. The interval is typically expressed as an estimate of the parameter, plus or minus the ME.
7. Interpret the CI:
   Interpret the CI in the context of the problem. It represents a range of plausible values for the population parameter, with the chosen confidence level indicating the likelihood of capturing the true parameter.

## 18.3 Single Sample: CIs for Means

Before going into details of this section, it is important to remember when and how to use normal and student distributions with sample distribution of mean, see Figure 18.7.

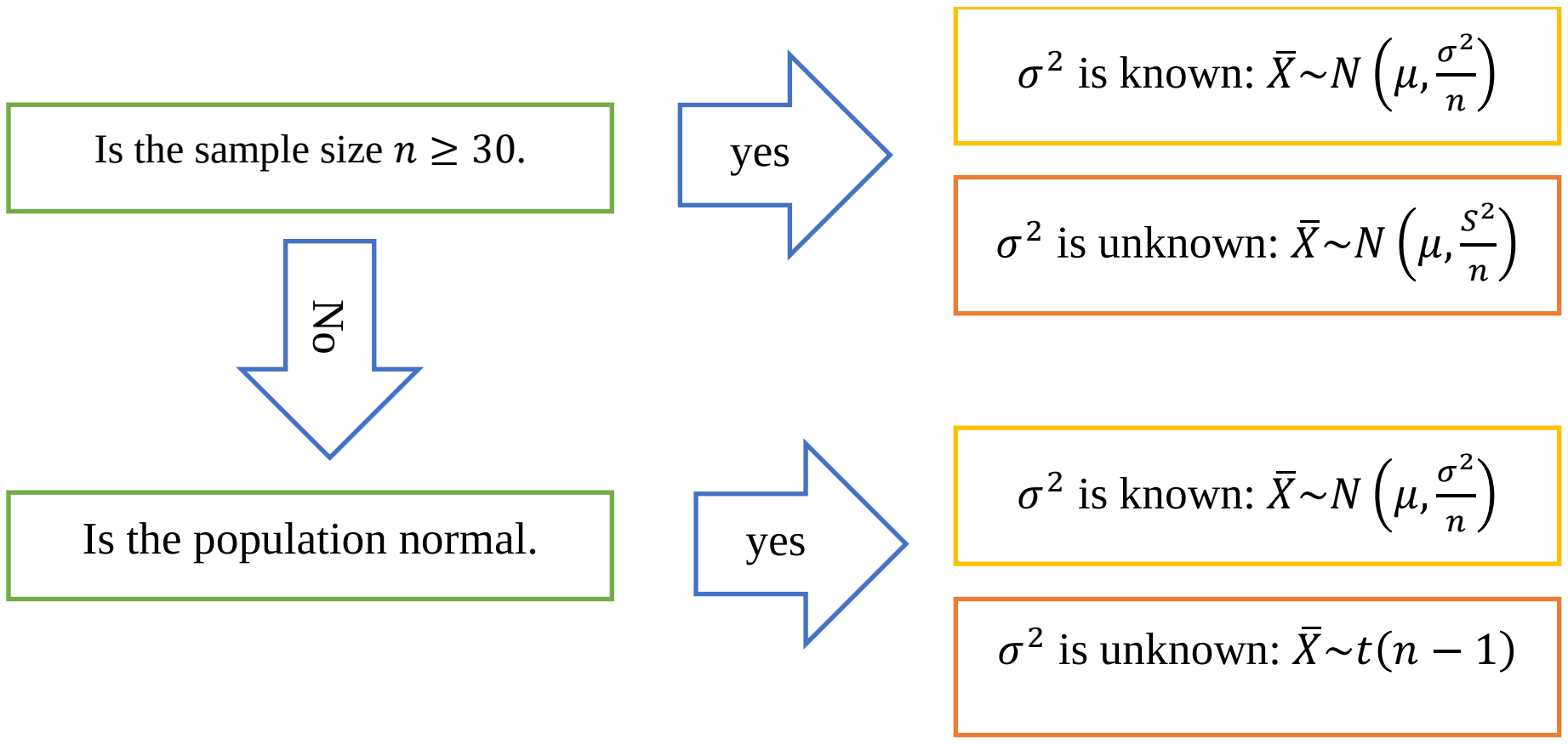

**Figure 18.7.** Flow chart of the sample distribution of mean, $\bar{X}$. It shows when you should use normal and student distributions.

We begin with the simplest case of a CI. The scenario is simple and yet unrealistic. We are interested in estimating a population mean $\mu$ and yet $\sigma$ is known. Clearly, if $\mu$ is unknown, it is quite unlikely that $\sigma$ is known. Any historical results that produced enough information to allow the assumption that $\sigma$ is known would likely have produced similar information about $\mu$. In practice, it is rare to know the true value of $\sigma$ for a population, as it typically needs to be estimated from sample data. However, by considering this simpler scenario, we can focus on the fundamental concepts and mechanics of constructing a CI, which remain the same in more realistic situations where both $\mu$ and $\sigma$ are unknown.

### Large-sample CI and $\sigma$ Known

The sampling distribution of $\bar{X}$ is centered at $\mu$, and in most applications the variance is smaller than that of any other estimators of $\mu$. Thus, the sample mean $\bar{x}$ will be used as a point estimate for the population mean $\mu$. Recall that $\sigma_{\bar{X}}^2 = \sigma^2/n$, so a large sample will yield a value of $\bar{X}$ that comes from a sampling distribution with a small variance. Hence, $\bar{x}$ is likely to be a very accurate estimate of $\mu$ when $n$ is large.

If the statistic is the sample mean $\bar{X}$, then the 95% and 99% confidence limits for estimating the population mean $\mu$, are given by $\bar{x} \pm 1.96\sigma_{\bar{X}}$ and $\bar{x} \pm 2.58\sigma_{\bar{X}}$, respectively. More generally, the confidence limits are given by

$$\bar{x} \pm z_c \sigma_{\bar{X}}, \tag{18.2.1}$$

where $z_c$ (which depends on the particular level of confidence desired) can be read from Table 18.1. Using the values of $\sigma_{\bar{X}} = \frac{\sigma}{\sqrt{n}}$, we see that the confidence limits for the population mean are given by

$$\bar{x} \pm z_c \frac{\sigma}{\sqrt{n}}, \tag{18.2.2}$$

if the sampling is either from an infinite population or with replacement from a finite population, and are given by

$$\bar{x} \pm z_c \frac{\sigma}{\sqrt{n}} \sqrt{\frac{N-n}{N-1}}, \tag{18.3}$$

if the sampling is without replacement from a population of finite size $N$.

## Large-sample CI and $\sigma$ Is Unknown

Often statisticians recommend that even when normality cannot be assumed, $\sigma$ is unknown, and $n \geq 30$, $S$ can replace $\sigma$ and the CI

$$\bar{x} \pm z_{\alpha/2} \frac{s}{\sqrt{n}}, \tag{18.4}$$

may be used. This is often referred to as a large sample CI. The justification lies only in the presumption that with a sample as large as 30 and the population distribution not too skewed, $S$ will be very close to the true $\sigma$ and thus the CLT prevails. It should be emphasized that this is only an approximation, and the quality of the result becomes better as the sample size grows larger.

**Theorem 18.1: A $(1 - \alpha)$ 100% Large-sample CI for a Population Mean $\mu$**

**If $\sigma$ is known**

$$\bar{x} \pm z_{\alpha/2} \frac{\sigma}{\sqrt{n}}, \tag{18.5}$$

where $z_{\alpha/2}$ is the $z$-value corresponding to an area $\alpha/2$ in the upper tail of a standard normal $z$ distribution, $n$ is sample size, and $\sigma$ is standard deviation of the population.

**If $\sigma$ is unknown**

It can be approximated by the sample standard deviation $s$ when the sample size is large $(n \geq 30)$ and the approximate CI is

$$\bar{x} \pm z_{\alpha/2} \frac{s}{\sqrt{n}}. \tag{18.6}$$

**Proof:**

To find the large-sample CI for a population mean $\mu$, we begin with the statistic

$$Z = \frac{\bar{X} - \mu}{\sigma/\sqrt{n}},$$

which has a standard normal distribution. If you write $z_{\alpha/2}$ as the value of $z$ with area $\alpha/2$ to its right, then you can write

$$P\left(-z_{\alpha/2} < \frac{\bar{X} - \mu}{\sigma/\sqrt{n}} < z_{\alpha/2}\right) = 1 - \alpha.$$

You can rewrite this inequality as

$$P\left(-z_{\frac{\alpha}{2}} \frac{\sigma}{\sqrt{n}} < \bar{X} - \mu < z_{\frac{\alpha}{2}} \frac{\sigma}{\sqrt{n}}\right) = 1 - \alpha,$$

$$P\left(-\bar{X} - z_{\frac{\alpha}{2}} \frac{\sigma}{\sqrt{n}} < -\mu < -\bar{X} + z_{\frac{\alpha}{2}} \frac{\sigma}{\sqrt{n}}\right) = 1 - \alpha,$$

$$P\left(-\left(\bar{X} + z_{\frac{\alpha}{2}} \frac{\sigma}{\sqrt{n}}\right) < -\mu < -\left(\bar{X} - z_{\frac{\alpha}{2}} \frac{\sigma}{\sqrt{n}}\right)\right) = 1 - \alpha,$$

$$P\left(\bar{X} - z_{\frac{\alpha}{2}} \frac{\sigma}{\sqrt{n}} < \mu < \bar{X} + z_{\frac{\alpha}{2}} \frac{\sigma}{\sqrt{n}}\right) = 1 - \alpha,$$

so that

$$P\left(\bar{X} - z_{\frac{\alpha}{2}} \frac{\sigma}{\sqrt{n}} < \mu < \bar{X} + z_{\frac{\alpha}{2}} \frac{\sigma}{\sqrt{n}}\right) = 1 - \alpha.$$

■

**Procedure 18.4.**

The steps for calculating the $100(1-\alpha)\%$ CI for population mean, $n \geq 30$:

1- Collect a random sample of data from the population of interest. Let the sample size be denoted by $n$.

2- Calculate the sample mean, denoted by $\bar{x}$, using the following formula:

$$\bar{x} = \sum \frac{x_i}{n},$$

where $x_i$ represents the ith observation in the sample.

3- Determine the level of significance $\alpha$.

4- Use a normal distribution Mathematica function to find the critical value, $z_{\alpha/2}$, for the given level of significance.

5- **If σ is known**, calculate the CI for the population mean using the following formula:

$$\left(\bar{x} - z_{\alpha/2}\frac{\sigma}{\sqrt{n}}, \bar{x} + z_{\alpha/2}\frac{\sigma}{\sqrt{n}}\right),$$

where $\sigma$ is the known population standard deviation, and $\sqrt{n}$ is the square root of the sample size.

**If σ is unknown**, it can be approximated by the sample standard deviation $s$ when the sample size is large ($n \geq 30$) and the approximate CI is

$$\left(\bar{x} - z_{\alpha/2}\frac{s}{\sqrt{n}}, \bar{x} + z_{\alpha/2}\frac{s}{\sqrt{n}}\right).$$

***Example 18.9***

A professor of math wants to estimate average scores for a math course. A random sample of 40 scores on the math exam produced the following results:

$$\bar{x} = 79, \qquad s = 7.5.$$

Calculate 95% CI for the average.

***Solution***

```
(* Sample size: *)
n=40;
(* Level of significance 95%: *)
alpha=0.05;
(* Sample mean: *)
sampleMean=79;
(* Sample standard deviation: *)
sampleStdDev=7.5;
(* Standard error: *)
standardError=sampleStdDev/Sqrt[n];
(* Calculate the critical value*)
criticalValue=Quantile[NormalDistribution[0,1],1-alpha/2]
(* Print the confidence interval: *)
confidenceInterval={sampleMean-
criticalValue*standardError,sampleMean+criticalValue*standardError}
 1.95996
 {76.6758,81.3242}
```

***Example 18.10***

The following is a random data set from a Gamma distribution with $\alpha = 4$, $\beta = 2$ population:

{3.04925, 7.01648, 6.03157, 10.061, 5.41684, 11.7751, 8.00373,
10.0007, 19.9241, 4.48058, 15.9479, 2.93503, 10.0726, 6.68935,
6.1183, 4.76766, 7.49241, 6.49076, 8.8584, 2.98887, 10.4784, 12.9221,
9.87839, 3.95537, 9.31186, 9.81395, 11.6124, 6.78275, 5.79358,
6.06747, 10.4627, 7.58096, 8.85288, 6.05002, 8.17595, 8.85313,
6.91628, 9.72203, 4.46266, 14.9402, 6.23094, 9.63511, 14.5473,
10.9694, 8.93981, 5.30347, 5.99925, 1.0471, 6.55006, 2.86364}.

Construct a 95% CI for the population mean $\mu$,

(a) assuming that the variance of population is 16.

(b) assuming that the variance of population is unknown.

***Solution***

```
SeedRandom[12345];
normaldata=RandomVariate[GammaDistribution[4,2],50];
populationMean=Mean[GammaDistribution[4,2]]
populationStdDev=StandardDeviation[GammaDistribution[4,2]]

(* Sample size: *)
n=Length[normaldata];
(* Level of significance in the case of 95% CI: *)
alpha=0.05;
(* Sample mean: *)
sampleMean=N[Mean[normaldata]]
(* Sample standard deviation: *)
sampleStdDev=N[StandardDeviation[normaldata]]
(* Standard error unknown population variance: *)
standardError1=sampleStdDev/Sqrt[n];
(* Calculate the critical value*)
criticalValue1=Quantile[NormalDistribution[0,1],1-alpha/2]
(* Print the confidence interval: *)
confidenceInterval1={sampleMean-
criticalValue1*standardError1,sampleMean+criticalValue1*standardError1}

(* Standard error known population variance: *)
standardError2=populationStdDev/Sqrt[n];
(* Calculate the critical value*)
criticalValue2=Quantile[NormalDistribution[0,1],1-alpha/2];
(* Print the confidence interval: *)
confidenceInterval2={sampleMean-
criticalValue2*standardError2,sampleMean+criticalValue2*standardError2}

 8
 4
 8.0568
 3.64598
 1.95996
 {7.0462,9.06739}
 {6.94807,9.16552}
```

### The Case of Normal Population and σ Unknown

**Theorem 18.2: CI of $\mu$, $\sigma^2$ Unknown**

If $\bar{x}$ and $s$ are the mean and standard deviation of a random sample from a normal population with unknown variance $\sigma^2$ and sample size is small, a $100(1-\alpha)\%$ CI for $\mu$ is

$$\bar{x} - t_{\alpha/2}\frac{s}{\sqrt{n}} < \mu < \bar{x} + t_{\alpha/2}\frac{s}{\sqrt{n}}, \tag{18.7}$$

where $t_{\alpha/2}$ is the $t$-value with $\nu = n-1$ degrees of freedom, leaving an area of $\alpha/2$ to the right.

**Proof:**

If we have a random sample from a normal distribution with unknown variance $\sigma^2$ and sample size is small, then the RV

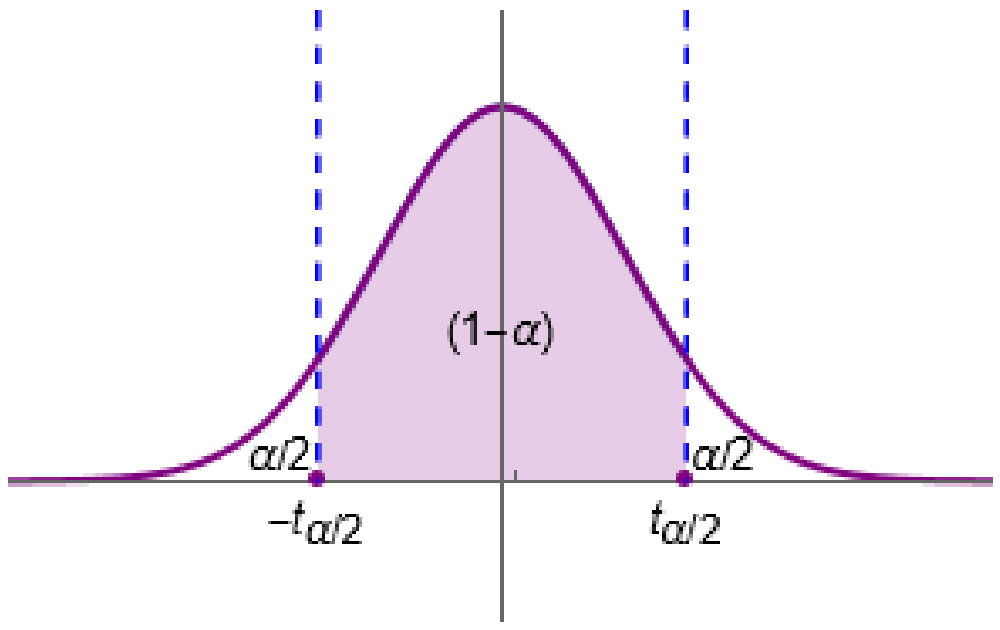

**Figure 18.8.** $P(-t_{\alpha/2} < T < t_{\alpha/2}) = 1 - \alpha$.

$$T = \frac{\bar{X} - \mu}{S/\sqrt{n}},$$

has a student $t$-distribution with $n-1$ degrees of freedom. Here $S$ is the sample standard deviation. In this situation, with $\sigma$ unknown, $T$ can be used to construct a CI on $\mu$. The procedure is the same as that with $\sigma$ known except that $\sigma$ is replaced by $S$ and the standard normal distribution is replaced by the $t$-distribution, see Figure 18.8.

$$P(-t_{\alpha/2} < T < t_{\alpha/2}) = 1 - \alpha,$$

where $t_{\alpha/2}$ is the $t$-value with $n-1$ degrees of freedom. Substituting for $T$, we write

$$P(-t_{\alpha/2} < \frac{\bar{X} - \mu}{S/\sqrt{n}} < t_{\alpha/2}) = 1 - \alpha,$$

and

$$P\left(\bar{X} - t_{\frac{\alpha}{2}}\frac{S}{\sqrt{n}} < \mu < \bar{X} + t_{\frac{\alpha}{2}}\frac{S}{\sqrt{n}}\right) = 1 - \alpha.$$

■

It is important to note that the use of the $t$-distribution assumes that the underlying population follows a normal distribution. This assumption is necessary for the validity of using the $t$-distribution in constructing CIs. If the population distribution is not approximately normal, the use of the $t$-distribution may not be appropriate. When the normality assumption is violated, other methods or statistical techniques may be more suitable for constructing CIs.

**Procedure 18.5.**

The steps for calculating the $100(1 - \alpha)\%$ CIs for normal population with $n \leq 30$:

1- Collect a random sample of data from the population of interest. Let the sample size be denoted by $n$.
2- Calculate the sample mean, denoted by $\bar{x}$, and the sample standard deviation, denoted by $s$, using the following formulas:

$$\bar{x} = \sum \frac{x_i}{n},$$

$$s = \frac{1}{n-1}\sum (x_i - \bar{x})^2,$$

where $x_i$ represents the $i$th observation in the sample.
3- Determine the level of significance, denoted by $\alpha$ and degree of freedom, denoted by $df = n - 1$.
4- Use a $t$-distribution Mathematica function to find the critical values, $-t_{\alpha/2}$ and $t_{\alpha/2}$, for the given level of significance and degree of freedom.
5- **If σ is known**, calculate the CIs for the population mean using the following formula:

$$\left(\bar{x} - z_{\alpha/2}\frac{\sigma}{\sqrt{n}}, \bar{x} + z_{\alpha/2}\frac{\sigma}{\sqrt{n}}\right).$$

where $\sigma$ is the known population standard deviation, and $\sqrt{n}$ is the square root of the sample size.

**If σ is unknown**, calculate the CI for the population mean using the following formula:

$$\left(\bar{x} - t_{\frac{\alpha}{2}}\frac{s}{\sqrt{n}}, \bar{x} + t_{\frac{\alpha}{2}}\frac{s}{\sqrt{n}}\right).$$

### *Example 18.11*

On a highway with a posted speed limit of 90 mph, fifteen random cars were measured for their speeds (in mph), and it was discovered that their average speed was 92.5 mph. Assume that we may infer from prior experience that the car speeds are normally distributed with $\sigma = 4.1$. Create a 90% CI for the actual mean speed of the traffic on this highway.

***Solution***

```
(* Sample size: *)
n=15;
(* Level of significance in the case of 90% CI: *)
alpha=0.10;
(* Sample mean: *)
sampleMean=92.5;
(* Population standard deviation: *)
populationStdDev=4.1;
(* Standard error: *)
standardError=populationStdDev/Sqrt[n];
(* Calculate the critical value*)
criticalValue=Quantile[NormalDistribution[0,1],1-alpha/2]
(* Print the confidence interval: *)
confidenceInterval={sampleMean-
criticalValue*standardError,sampleMean+criticalValue*standardError}
 1.64485
 {90.7587,94.2413}
```

### *Example 18.12*

The following is a random data set from a normal population:

{4.86232,2.73884,3.49852,2.87937,0.730487,3.48178,0.334349,5.33097,3.35013,1.08246}.

Construct a 95% CI for the population mean $\mu$.

***Solution***

```
SeedRandom[123];
normaldata=RandomVariate[NormalDistribution[3,1.5],10]
(* Sample size: *)
n=Length[normaldata];
(* Level of significance in the case of 95% CI: *)
alpha=0.05;
(* Sample mean: *)
sampleMean=N[Mean[normaldata]];
(* Sample standard deviation: *)
sampleStdDev=N[StandardDeviation[normaldata]];
(* Standard error: *)
standardError=sampleStdDev/Sqrt[n];
(* degrees of freedom *)
df=n-1;
(* Calculate the critical value*)
criticalValue=Quantile[StudentTDistribution[df],1-alpha/2]
(* Print the confidence interval: *)
confidenceInterval={sampleMean-
criticalValue*standardError,sampleMean+criticalValue*standardError}
 {4.86232,2.73884,3.49852,2.87937,0.730487,3.48178,0.334349,5.33097,3.35013,1.08246}
 2.26216
 {1.63123,4.02662}
```

***Example 18.13***

The results of a random sample of 14 TOEFL test takers are as follows:
{509.667, 527.176, 466.244, 591.464, 637.121, 477.327, 486.217, 535.718, 534.167, 463.076, 531.43, 536.807, 567.387, 496.925}.
Construct a 95% CI for the population mean $\mu$ of the TOEFL score, assuming that the scores are normally distributed.

***Solution***

```
SeedRandom[1234];
normaldata=RandomVariate[NormalDistribution[530,40],14]

(* Sample size: *)
n=Length[normaldata];
(* Level of significance in the case of 95% CI: *)
alpha=0.05;
(* Sample mean: *)
sampleMean=N[Mean[normaldata]];
(* Sample standard deviation: *)
sampleStdDev=N[StandardDeviation[normaldata]];
(* Standard error: *)
standardError=sampleStdDev/Sqrt[n];
(* degrees of freedom *)
df=n-1;
(* Calculate the critical value*)
criticalValue=Quantile[StudentTDistribution[df],1-alpha/2]
(* Print the confidence interval: *)
confidenceInterval={sampleMean-
criticalValue*standardError,sampleMean+criticalValue*standardError}

{509.667,527.176,466.244,591.464,637.121,477.327,486.217,535.718,534.167,463.076,531.43,536.
807,567.387,496.925}

 2.16037
 {497.458,554.074}
```

### One-Sided Confidence Bounds

In the case of one-sided confidence bounds, the focus is only on one side of the parameter value, either the lower bound or the upper bound. This is typically done when researchers are interested in determining the minimum or maximum value that a parameter can take, rather than estimating the full range. One-sided confidence bounds are useful in situations where researchers have specific hypotheses or expectations about the direction of the effect.

To understand one-sided confidence bounds, let us consider an example. Suppose we want to estimate the average height of a certain population of adults. We collect a sample of heights and calculate the sample mean. Using this sample mean, we can construct a one-sided confidence bound to estimate the true population mean height either from above or below. For instance, if we are interested in estimating the population mean height from above, we can construct an upper one-sided confidence bound. This interval will provide an upper limit within which the true population mean height is likely to lie with a certain level of confidence. Similarly, if we are interested in estimating the population mean height from below, we can construct a lower one-sided confidence bound, which provides a lower limit for the parameter.

**Theorem 18.3: One-Sided Confidence Bounds for $\mu$, $\sigma^2$ Known**
If $\bar{x}$ is the mean of a random sample of size $n$ from a population with variance $\sigma^2$, the one-sided $100(1-\alpha)\%$ confidence bounds for $\mu$ are given by

$$\bar{x} + z_\alpha \frac{\sigma}{\sqrt{n}}, \tag{18.8.1}$$

for upper one-sided bound and

$$\bar{x} - z_\alpha \frac{\sigma}{\sqrt{n}}, \tag{18.8.2}$$

in the case of lower one-sided bound.

**Proof:**

One-sided confidence bounds are developed in the same fashion as two-sided intervals. However, the source is a one-sided probability statement that makes use of the CLT:

$$P\left(\frac{\bar{X} - \mu}{\sigma/\sqrt{n}} < z_\alpha\right) = 1 - \alpha,$$

gives

$$P\left(\mu > \bar{X} - z_\alpha \frac{\sigma}{\sqrt{n}}\right) = 1 - \alpha.$$

In the same manner,

$$P\left(\frac{\bar{X} - \mu}{\sigma/\sqrt{n}} > -z_\alpha\right) = 1 - \alpha,$$

gives

$$P\left(\mu < \bar{X} + z_\alpha \frac{\sigma}{\sqrt{n}}\right) = 1 - \alpha.$$

■

**Procedure 18.6.**

To calculate one-sided confidence bounds for $\mu$ when the population variance $\sigma^2$ is known, you can follow these steps:

1. Obtain a sample from the population of interest, with size $n$.
2. Calculate the sample mean, denoted as $\bar{x}$.
3. Specify the desired confidence level for the interval. For example, if you want a 95% confidence level, you would have $\alpha = 0.05$ for a one-sided confidence bound.
4. Determine the critical value corresponding to the chosen confidence level and the chosen tail. For a one-sided upper confidence bound, use the z-value from the standard normal distribution that corresponds to the desired significance level $\alpha$.
5. Multiply the critical value by the SE to obtain the ME, where the SE is given by $\frac{\sigma}{\sqrt{n}}$.
$$\text{ME} = z_\alpha \times \frac{\sigma}{\sqrt{n}}.$$
6. Add the ME to the sample mean to calculate the upper bound of the CI.
$$\text{Upper bound} = \bar{x} + \text{ME}.$$

***Example 18.14***

Random sample from normal population with variance known produced the following data:
{9.8325,13.0659,9.18854,7.92993,9.26685,11.8435,11.7678,10.8883,10.808,8.27212,9.07788,11.9172,10.7207,8.8947,13.734,11.6947,10.0054,10.9551,11.28,7.99447,8.26085,12.5101,10.9733,9.1944,6.4452,10.8001,10.4735,8.49494,9.44695,9.66986}.
Obtain a 95% one-sided CI for $\mu$.

***Solution***

```
SeedRandom[2123];
(* Sample data: *)
sample=RandomVariate[NormalDistribution[10,2],30]
```

```
(* Compute the sample mean: *)
xbar=Mean[sample]

(* Determine confidence level: *)
confidenceLevel=0.95 ;
alpha=1-confidenceLevel;

(* Find critical value using standard normal distribution: *)
criticalValue=Quantile[NormalDistribution[0,1],1-alpha];

(* Calculate standard error: *)
s=StandardDeviation[sample]
n=Length[sample];
SE=s/Sqrt[n];

(* Calculate margin of error and upper bound: *)
ME=criticalValue*SE

(* Display the upper confidence bound: *)
upperBound=xbar+ME

{9.8325,13.0659,9.18854,7.92993,9.26685,11.8435,11.7678,10.8883,10.808,8.27212,9.07788,11.91
72,10.7207,8.8947,13.734,11.6947,10.0054,10.9551,11.28,7.99447,8.26085,12.5101,10.9733,9.194
4,6.4452,10.8001,10.4735,8.49494,9.44695,9.66986}
 10.1802
 1.67788
 0.50388
 10.6841
```

**One-Sided Confidence Bounds on $\mu$, $\sigma^2$ Unknown**

As long as the distribution is approximately bell shaped, CIs can be computed when $\sigma^2$ is unknown by using the $t$-distribution and we may expect very good results. Computed one-sided confidence bounds for $\mu$ with $\sigma$ unknown are

$$\bar{x} + t_\alpha \frac{s}{\sqrt{n}}, \tag{18.9.1}$$

$$\bar{x} - t_\alpha \frac{s}{\sqrt{n}}. \tag{18.9.2}$$

They are the upper and lower $100(1-\alpha)\%$ bounds, respectively. Here $t_\alpha$ is the $t$-value having an area of $\alpha$ to the right.

**Procedure 18.7.**

To calculate one-sided confidence bounds on $\mu$ when the population variance $\sigma^2$ is unknown, you can follow these steps:

1. Obtain a sample from the population of interest, with size $n$.
2. Calculate the sample mean, denoted as $\bar{x}$.
3. Calculate the sample standard deviation, denoted as $s$.
4. Specify the desired confidence level for the interval. For example, if you want a 95% confidence level, you would have $\alpha = 0.05$ for a one-sided confidence bound.
5. Determine the critical value corresponding to the chosen confidence level and the chosen tail. For a one-sided upper confidence bound, use the $t$-value from the $t$-distribution with $(n-1)$ degrees of freedom that corresponds to the desired significance level $\alpha$.
6. Compute the SE, which is given by $\frac{s}{\sqrt{n}}$.
7. Multiply the critical value by the SE to obtain the ME.

$$\text{ME} = t_\alpha \times \frac{s}{\sqrt{n}}.$$

8. Add the ME to the sample mean to calculate the upper bound of the CI.

$$\text{Upper bound} = \bar{x} + \text{ME}.$$

***Example 18.15***

Random sample from normal population with variance unknown produced the following data:

{9.47841,12.3643,10.0894,8.89491,7.82545,14.8155,8.52305,8.07484,10.3272,8.40925,9.14009,9.78196,8.92245,9.19725,10.4759}.

Obtain a 95% one-sided CI for $\mu$.

***Solution***

```
SeedRandom[21234];
(* Sample data: *)
sample=RandomVariate[NormalDistribution[10,2],15]

(* Compute the sample mean: *)
xbar=Mean[sample]

(* Compute the sample standard deviation: *)
s=StandardDeviation[sample]

(* Determine confidence level: *)
confidenceLevel=0.95 ;
alpha=1-confidenceLevel;

(* Find critical value using t-distribution: *)
df=Length[sample]-1;
criticalValue=Quantile[StudentTDistribution[df],1-alpha];

(* Calculate standard error: *)
n=Length[sample];
SE=s/Sqrt[n]

(* Calculate margin of error and upper bound: *)
ME=criticalValue*SE

(* Display the upper confidence bound: *)
upperBound=xbar+ME

{9.47841,12.3643,10.0894,8.89491,7.82545,14.8155,8.52305,8.07484,10.3272,8.40925,9.14009,9.78196,8.92245,9.19725,10.4759}

 9.75466
 1.80179
 0.465221
 0.819399
 10.5741
```

## 18.4 Two Samples: CIs for Differences between Two Means

### CI for $\mu_1 - \mu_2$, $\sigma_1^2$ and $\sigma_2^2$ Known

The following notations represent the parameters for two populations and their corresponding samples statistics.

| | Population 1 | Population 2 |
|---|---|---|
| Mean | $\mu_1$ | $\mu_2$ |
| Variance | $\sigma_1^2$ | $\sigma_2^2$ |

| | Sample 1 | Sample 2 |
|---|---|---|
| Mean | $\bar{X}_1$ | $\bar{X}_2$ |
| Variance | $S_1^2$ | $S_2^2$ |
| Sample Size | $n_1$ | $n_2$ |

If $\text{sta}_1$ and $\text{sta}_2$ are two sample statistics with approximately normal sampling distributions, confidence limits for the difference of the population parameters corresponding to $\text{sta}_1$ and $\text{sta}_2$ are given by

$$(\text{sta}_1 - \text{sta}_2) \pm z_c\sigma_{\text{sta}_1 - \text{sta}_2} = (\text{sta}_1 - \text{sta}_2) \pm z_c\sqrt{\sigma_{\text{sta}_1}^2 + \sigma_{\text{sta}_2}^2}, \tag{18.10}$$

while confidence limits for the sum of the population parameters are given by

$$(\text{sta}_1 + \text{sta}_2) \pm z_c\sigma_{\text{sta}_1 + \text{sta}_2} = (\text{sta}_1 - \text{sta}_2) \pm z_c\sqrt{\sigma_{\text{sta}_1}^2 + \sigma_{\text{sta}_2}^2}, \tag{18.11}$$

provided that the samples are independent.

For example, confidence limits for the difference of two population means, in the case where the populations are infinite, are given by

$$(\bar{x}_1 - \bar{x}_2) \pm z_c\sigma_{\bar{X}_1 - \bar{X}_2} = \bar{x}_1 - \bar{x}_2 \pm z_c\sqrt{\frac{\sigma_1^2}{n_1} + \frac{\sigma_2^2}{n_2}}. \tag{18.12}$$

**Theorem 18.4: Large-sample CI for $\mu_1 - \mu_2$, $\sigma_1^2$ and $\sigma_2^2$ Known**
If $\bar{x}_1$ and $\bar{x}_2$ are the means of independent random samples of sizes $n_1$ and $n_2$ from populations with known variances $\sigma_1^2$ and $\sigma_2^2$, respectively, a $100(1-\alpha)\%$ CI for $\mu_1 - \mu_2$ is given by

$$(\bar{x}_1 - \bar{x}_2) - z_{\frac{\alpha}{2}}\sqrt{\frac{\sigma_1^2}{n_1} + \frac{\sigma_2^2}{n_2}} < \mu_1 - \mu_2 < (\bar{x}_1 - \bar{x}_2) + z_{\frac{\alpha}{2}}\sqrt{\frac{\sigma_1^2}{n_1} + \frac{\sigma_2^2}{n_2}}, \tag{18.13}$$

where $z_{\alpha/2}$ is the $z$-value leaving an area of $\alpha/2$ to the right.

**Proof:**

According to (16.9), we can expect the sampling distribution of $\bar{X}_1 - \bar{X}_2$ to be approximately normally distributed with mean $\mu_{\bar{X}_1 - \bar{X}_2} = \mu_1 - \mu_2$ and standard deviation $\sigma_{\bar{X}_1 - \bar{X}_2} = \sqrt{\sigma_1^2/n_1 + \sigma_2^2/n_2}$. Therefore, we can assert with a probability of $1-\alpha$ that the standard normal variable

$$Z = \frac{(\bar{X}_1 - \bar{X}_2) - (\mu_1 - \mu_2)}{\sqrt{\frac{\sigma_1^2}{n_1} + \frac{\sigma_2^2}{n_2}}},$$

will fall between $-z_{\alpha/2}$ and $z_{\alpha/2}$. Hence, we have

$$P\left(-z_{\alpha/2} < Z < z_{\alpha/2}\right) = 1 - \alpha.$$

Substituting for $Z$, we state equivalently that

$$P\left(-z_{\frac{\alpha}{2}} < \frac{(\bar{X}_1-\bar{X}_2)-(\mu_1-\mu_2)}{\sqrt{\frac{\sigma_1^2}{n_1}+\frac{\sigma_2^2}{n_2}}} < z_{\frac{\alpha}{2}}\right) = 1-\alpha,$$

$$P\left(-z_{\frac{\alpha}{2}}\sqrt{\frac{\sigma_1^2}{n_1}+\frac{\sigma_2^2}{n_2}} < (\bar{X}_1-\bar{X}_2)-(\mu_1-\mu_2) < z_{\frac{\alpha}{2}}\sqrt{\frac{\sigma_1^2}{n_1}+\frac{\sigma_2^2}{n_2}}\right) = 1-\alpha,$$

$$P\left(-\left[(\bar{X}_1-\bar{X}_2)+z_{\frac{\alpha}{2}}\sqrt{\frac{\sigma_1^2}{n_1}+\frac{\sigma_2^2}{n_2}}\right] < -(\mu_1-\mu_2) < -\left[(\bar{X}_1-\bar{X}_2)-z_{\frac{\alpha}{2}}\sqrt{\frac{\sigma_1^2}{n_1}+\frac{\sigma_2^2}{n_2}}\right]\right) = 1-\alpha,$$

$$P\left(\left[(\bar{X}_1-\bar{X}_2)-z_{\alpha/2}\sqrt{\frac{\sigma_1^2}{n_1}+\frac{\sigma_2^2}{n_2}}\right] < \mu_1-\mu_2 < \left[(\bar{X}_1-\bar{X}_2)+z_{\frac{\alpha}{2}}\sqrt{\frac{\sigma_1^2}{n_1}+\frac{\sigma_2^2}{n_2}}\right]\right) = 1-\alpha,$$

which leads to the $100(1-\alpha)\%$ CI for $\mu_1-\mu_2$.

■

**Procedure 18.8.**

The steps for calculating the $100(1-\alpha)\%$ large-sample CI for estimating the difference between two means of two samples when the population standard deviations ($\sigma_1$ and $\sigma_2$) are known (unknown):

1. Obtain two independent samples from the populations of interest. Let us call them Sample 1 and Sample 2, with sizes $n_1$ and $n_2$, respectively.
2. Calculate the sample mean for Sample 1, denoted as $\bar{x}_1$, and the sample mean for Sample 2, denoted as $\bar{x}_2$.
3. Specify the desired confidence level for the interval. Common choices include 90%, 95%, or 99%.
4. Determine the critical value (z) corresponding to the chosen confidence level.
5. The SE of the difference between means is given by:

$$\sqrt{\frac{\sigma_1^2}{n_1}+\frac{\sigma_2^2}{n_2}},$$

   where $\sigma_1$ and $\sigma_2$ are known or

$$\text{SE} = \sqrt{\frac{s_1^2}{n_1}+\frac{s_2^2}{n_2}},$$

   where $\sigma_1$ and $\sigma_2$ are unknown.
6. Multiply the SE by the critical value $z_{\frac{\alpha}{2}}$ to obtain the ME:

$$\text{ME} = z_{\frac{\alpha}{2}} \times \text{SE}.$$

7. Subtract the ME from the sample mean difference $(\bar{x}_1-\bar{x}_2)$ to calculate the lower bound of the CI. Similarly, add the ME to the sample mean difference to obtain the upper bound:

$$\text{Lower bound} = (\bar{x}_1-\bar{x}_2) - \text{ME},$$
$$\text{Upper bound} = (\bar{x}_1-\bar{x}_2) + \text{ME}.$$

***Example 18.16***

Independent random samples from two normal populations with equal variances but unknown produced the following data:

Sample 1:
{119.977,81.7225,72.2266,101.779,80.7805,82.0576,79.3684,70.6839,85.0673,59.4836,58.0297,57.1463,57.9815,82.0756,81.3336,80.1593,84.2602,63.0735,131.321,74.0186,80.4339,58.8487,102.679,86.4697,86.468,87.6855,115.773,110.453,88.8101,56.0426}.

Sample 2:
{32.3799,67.2285,59.0925,61.5934,72.9079,64.962,43.2411,64.5692,71.6843,92.4318,74.7615,54.0313,63.882,71.5596,64.7811,59.8014,78.0505,63.4725,72.5987,84.9059,65.2094,84.7679,71.1616,116.283,52.7115,43.0102,59.1317,53.9341,109.216,40.3552,46.5045,70.2688,75.1751,83.4396,74.9343,75.8223,72.776,57.6744,55.0488,45.7263}.
Obtain a 95% CI for $\mu_1 - \mu_2$.

***Solution***

```
SeedRandom[12345678];
(* Data of two samples: *)
sample1=RandomVariate[NormalDistribution[80,18],30]
sample2=RandomVariate[NormalDistribution[70,19],40]

(* Compute sample means: *)
xbar1=Mean[sample1]
xbar2=Mean[sample2]

(* Determine confidence level: *)
confidenceLevel=0.95
alpha=1-confidenceLevel;

(* Find critical value using standard normal distribution: *)
criticalValue=Quantile[NormalDistribution[0,1],1-alpha/2]

(* Calculate standard error: *)
s1=StandardDeviation[sample1]
s2=StandardDeviation[sample2]
n1=Length[sample1]
n2=Length[sample2]
SE=Sqrt[(s1^2/n1)+(s2^2/n2)]

(* Calculate margin of error: *)
ME=criticalValue*SE

(* Compute lower and upper bounds: *)
lowerBound=(xbar1-xbar2)-ME
upperBound=(xbar1-xbar2)+ME

(* Display the confidence interval: *)
{lowerBound,upperBound}

{119.977,81.7225,72.2266,101.779,80.7805,82.0576,79.3684,70.6839,85.0673,59.4836,58.0297,57.1463,57.9815,82.0756,81.3336,80.1593,84.2602,63.0735,131.321,74.0186,80.4339,58.8487,102.679,86.4697,86.468,87.6855,115.773,110.453,88.8101,56.0426}

{32.3799,67.2285,59.0925,61.5934,72.9079,64.962,43.2411,64.5692,71.6843,92.4318,74.7615,54.0313,63.882,71.5596,64.7811,59.8014,78.0505,63.4725,72.5987,84.9059,65.2094,84.7679,71.1616,116.283,52.7115,43.0102,59.1317,53.9341,109.216,40.3552,46.5045,70.2688,75.1751,83.4396,74.9343,75.8223,72.776,57.6744,55.0488,45.7263}

 82.5403
 66.7771
 0.95
 1.95996
 19.4
 16.9937
 30
 40
 4.44578
```

```
8.71357
7.04962
24.4768
{7.04962,24.4768}
```

### Variances Unknown but Equal

**Theorem 18.5: CI for $\mu_1 - \mu_2$, $\sigma_1^2 = \sigma_2^2$ but Both Unknown**

If $\bar{x}_1$ and $\bar{x}_2$ are the means of independent random samples of sizes $n_1$ and $n_2$, respectively, from approximately normal populations with unknown but equal variances, a $100(1-\alpha)\%$ CI for $\mu_1 - \mu_2$ is given by

$$(\bar{x}_1 - \bar{x}_2) - t_{\frac{\alpha}{2}} s_p \sqrt{\frac{1}{n_1} + \frac{1}{n_2}} < \mu_1 - \mu_2 < (\bar{x}_1 - \bar{x}_2) + t_{\frac{\alpha}{2}} s_p \sqrt{\frac{1}{n_1} + \frac{1}{n_2}}, \tag{18.14.1}$$

where $s_p$ is the pooled estimate of the population standard deviation,

$$s_p^2 = \frac{(n_1 - 1)s_1^2 + (n_2 - 1)s_2^2}{(n_1 + n_2 - 2)}, \tag{18.14.2}$$

and $t_{\alpha/2}$ is the $t$-value with $\nu = n_1 + n_2 - 2$ degrees of freedom, leaving an area of $\alpha/2$ to the right.

**Proof:**

If $\sigma_1^2 = \sigma_2^2 = \sigma^2$, we obtain a standard normal variable of the form

$$Z = \frac{(\bar{X}_1 - \bar{X}_2) - (\mu_1 - \mu_2)}{\sqrt{\sigma^2 \left(\frac{1}{n_1} + \frac{1}{n_2}\right)}}.$$

According to Theorem 16.3, the two RVs

$$\frac{(n_1 - 1)S_1^2}{\sigma^2} \text{ and } \frac{(n_2 - 1)S_2^2}{\sigma^2},$$

have chi-squared distributions with $n_1 - 1$ and $n_2 - 1$ degrees of freedom, respectively. Furthermore, they are independent chi-squared variables, since the random samples were selected independently. According to Theorem 16.9,

$$V = \frac{(n_1 - 1)S_1^2}{\sigma^2} + \frac{(n_2 - 1)S_2^2}{\sigma^2} = \frac{(n_1 - 1)S_1^2 + (n_2 - 1)S_2^2}{\sigma^2},$$

has a chi-squared distribution with $\nu = n_1 + n_2 - 2$ degrees of freedom. Since the preceding expressions for $Z$ and $V$ can be shown to be independent, it follows from (16.31) that the statistic

$$T = \frac{\dfrac{(\bar{X}_1 - \bar{X}_2) - (\mu_1 - \mu_2)}{\sqrt{\sigma^2 \left(\frac{1}{n_1} + \frac{1}{n_2}\right)}}}{\sqrt{\dfrac{(n_1 - 1)S_1^2 + (n_2 - 1)S_2^2}{\sigma^2 (n_1 + n_2 - 2)}}},$$

has the $t$-distribution with $\nu = n_1 + n_2 - 2$ degrees of freedom.

A point estimate of the unknown common variance $\sigma^2$ can be obtained by pooling the sample variances. Denoting the pooled estimator by $S_p^2$, we have

$$S_p^2 = \frac{(n_1 - 1)S_1^2 + (n_2 - 1)S_2^2}{(n_1 + n_2 - 2)}.$$

Substituting $S_p^2$ in the $T$ statistic, we obtain

$$T = \frac{(\bar{X}_1 - \bar{X}_2) - (\mu_1 - \mu_2)}{S_p\sqrt{\left(\frac{1}{n_1} + \frac{1}{n_2}\right)}}.$$

Using the $T$ statistic, we have

$$P(-t_{\alpha/2} < T < t_{\alpha/2}) = 1 - \alpha,$$

where $t_{\alpha/2}$ is the $t$-value with $n_1 + n_2 - 2$ degrees of freedom, above which we find an area of $\alpha/2$. Substituting for $T$ in the inequality, we write

$$P\left(-t_{\frac{\alpha}{2}} < \frac{(\bar{X}_1 - \bar{X}_2) - (\mu_1 - \mu_2)}{S_p\sqrt{\left(\frac{1}{n_1} + \frac{1}{n_2}\right)}} < t_{\frac{\alpha}{2}}\right) = 1 - \alpha,$$

$$P\left(-t_{\frac{\alpha}{2}}S_p\sqrt{\left(\frac{1}{n_1} + \frac{1}{n_2}\right)} < (\bar{X}_1 - \bar{X}_2) - (\mu_1 - \mu_2) < t_{\frac{\alpha}{2}}S_p\sqrt{\left(\frac{1}{n_1} + \frac{1}{n_2}\right)}\right) = 1 - \alpha,$$

$$P\left(-(\bar{X}_1 - \bar{X}_2) - t_{\frac{\alpha}{2}}S_p\sqrt{\left(\frac{1}{n_1} + \frac{1}{n_2}\right)} < -(\mu_1 - \mu_2) < -(\bar{X}_1 - \bar{X}_2) + t_{\frac{\alpha}{2}}S_p\sqrt{\left(\frac{1}{n_1} + \frac{1}{n_2}\right)}\right) = 1 - \alpha,$$

$$P\left(-\left[(\bar{X}_1 - \bar{X}_2) + t_{\frac{\alpha}{2}}S_p\sqrt{\left(\frac{1}{n_1} + \frac{1}{n_2}\right)}\right] < -(\mu_1 - \mu_2) < -\left[(\bar{X}_1 - \bar{X}_2) - t_{\frac{\alpha}{2}}S_p\sqrt{\left(\frac{1}{n_1} + \frac{1}{n_2}\right)}\right]\right) = 1 - \alpha,$$

$$P\left(\left[(\bar{X}_1 - \bar{X}_2) - t_{\alpha/2}S_p\sqrt{\left(\frac{1}{n_1} + \frac{1}{n_2}\right)}\right] < \mu_1 - \mu_2 < \left[(\bar{X}_1 - \bar{X}_2) + t_{\frac{\alpha}{2}}S_p\sqrt{\left(\frac{1}{n_1} + \frac{1}{n_2}\right)}\right]\right) = 1 - \alpha.$$

■

The procedure for constructing CIs for $\mu_1 - \mu_2$ with $\sigma_1 = \sigma_2 = \sigma^2$ unknown requires the assumption that the populations are normal. Slight departures from either the equal variance or the normality assumption do not seriously alter the degree of confidence for our interval. If the population variances are considerably different, we still obtain reasonable results when the populations are normal, provided that $n_1 = n_2$. Therefore, in planning an experiment, one should make every effort to equalize the size of the samples.

**Procedure 18.9.**

To calculate the CI for the difference between two means, $\mu_1$ and $\mu_2$, when the variances, $\sigma_1^2$ and $\sigma_2^2$, are unknown but assumed to be equal, you can use the following steps:

1. Obtain two independent samples from the populations of interest. Let us call them Sample 1 and Sample 2, with sizes $n_1$ and $n_2$, respectively.
2. Calculate the sample mean for Sample 1, denoted as $\bar{x}_1$, and the sample mean for Sample 2, denoted as $\bar{x}_2$. Also, calculate the sample standard deviation for Sample 1, denoted as $s_1$, and the sample standard deviation for Sample 2, denoted as $s_2$.
3. Since the variances are assumed to be equal, you can pool the sample variances to estimate the common variance. The pooled standard deviation, $s_p$, is calculated using the formula:

$$s_p = \sqrt{\frac{(n_1 - 1)s_1^2 + (n_2 - 1)s_2^2}{(n_1 + n_2 - 2)}}.$$

4. The SE of the difference between means is given by:

$$\text{SE} = s_p\sqrt{\left(\frac{1}{n_1}+\frac{1}{n_2}\right)}.$$

5. Specify the desired confidence level for the interval. Common choices include 90%, 95%, or 99%.
6. Determine the critical value ($t$) corresponding to the chosen confidence level and the degrees of freedom ($df$). The degrees of freedom can be calculated as:
$$df = n_1 + n_2 - 2.$$
7. Look up the critical value by using the t-distribution Mathematica function to find the appropriate t-value. Determine the critical value $t_{\frac{\alpha}{2}}$ corresponding to the chosen confidence level.
8. Multiply the SE by the critical value $t_{\frac{\alpha}{2}}$ to obtain the ME:
$$\text{ME} = t_{\frac{\alpha}{2}} \times \text{SE}.$$
9. Subtract the ME from the sample mean difference $(\bar{x}_1 - \bar{x}_2)$ to calculate the lower bound of the CI. Similarly, add the ME to the sample mean difference to obtain the upper bound:
$$\text{Lower bound} = (\bar{x}_1 - \bar{x}_2) - \text{ME},$$
$$\text{Upper bound} = (\bar{x}_1 - \bar{x}_2) + \text{ME}.$$

***Example 18.17***

Independent random samples from two normal populations with equal variances but unknown produced the following data:

Sample 1: {4.19081, 2.94181, 3.63375, 4.21619, 3.05675, 1.70492, 4.87008, 4.60378, 1.9967, 2.32584}.
Sample 2: {2.49942, 1.31607, 3.30208, 1.75315, 0.833795, 2.42829, 2.41348}.

(a) Calculate the pooled estimate $S_p^2$.
(b) Obtain a 90% CI for $\mu_1 - \mu_2$.

***Solution***

```
SeedRandom[123456];
(* Data of two samples: *)
sample1=RandomVariate[NormalDistribution[3,1],10]
sample2=RandomVariate[NormalDistribution[2,1],7]

(* Compute sample means and sample standard deviations*)
n1=Length[sample1];
n2=Length[sample2];
xbar1=Mean[sample1];
xbar2=Mean[sample2];
s1=StandardDeviation[sample1];
s2=StandardDeviation[sample2];

(*Compute pooled standard deviation*)
sPooled=Sqrt[((n1-1)*s1^2+(n2-1)*s2^2)/(n1+n2-2)]

(* Calculate standard error*)
SE=sPooled*Sqrt[1/n1+1/n2]

(* Determine confidence level*)
confidenceLevel=0.90 ;
alpha=1-confidenceLevel;
(*Step 6:Find critical value using t-distribution*)
df=n1+n2-2
criticalValue=Quantile[StudentTDistribution[df],1-alpha/2]

(*Step 7:Calculate margin of error*)
ME=criticalValue*SE
```

```
(*Step 8:Compute lower and upper bounds*)
lowerBound=(xbar1-xbar2)-ME
upperBound=(xbar1-xbar2)+ME

(*Display the confidence interval*)
{lowerBound,upperBound}

 {4.19081, 2.94181, 3.63375, 4.21619, 3.05675, 1.70492, 4.87008, 4.60378, 1.9967, 2.32584}
 {2.49942, 1.31607, 3.30208, 1.75315, 0.833795, 2.42829, 2.41348}
 1.0134
 0.49941
 15
 1.75305
 0.875491
 0.400532
 2.15151
 {0.400532,2.15151}
```

### Unknown and Unequal Variances

**Theorem 18.6: CI for $\mu_1 - \mu_2$, $\sigma_1^2 \neq \sigma_2^2$ and Both Unknown**
If $\bar{x}_1$, $s_1^2$, $\bar{x}_2$, and $s_2^2$ are the means and variances of independent random samples of sizes $n_1$ and $n_2$, respectively, for normal populations with unknown and unequal variances, an approximate $100(1-\alpha)\%$ CI for $\mu_1 - \mu_2$ is given by

$$(\bar{x}_1 - \bar{x}_2) - t_{\frac{\alpha}{2}}\sqrt{\frac{s_1^2}{n_1} + \frac{s_2^2}{n_2}} < \mu_1 - \mu_2 < (\bar{x}_1 - \bar{x}_2) + t_{\frac{\alpha}{2}}\sqrt{\frac{s_1^2}{n_1} + \frac{s_2^2}{n_2}}, \tag{18.15.1}$$

where $t_{\alpha/2}$ is the $t$-value with

$$v = \frac{\left(\frac{s_1^2}{n_1} + \frac{s_2^2}{n_2}\right)^2}{\sqrt{\frac{1}{(n_1-1)}\left(\frac{s_1^2}{n_1}\right)^2 + \frac{1}{(n_2-1)}\left(\frac{s_2^2}{n_2}\right)^2}}, \tag{18.15.2}$$

degrees of freedom, leaving an area of $\alpha/2$ to the right.

**Proof:**

The statistic most often used in this case is

$$T' = \frac{(\bar{X}_1 - \bar{X}_2) - (\mu_1 - \mu_2)}{\sqrt{\left(\frac{S_1^2}{n_1} + \frac{S_2^2}{n_2}\right)}},$$

which has approximately a $t$-distribution with $v$ degrees of freedom, where

$$v = \frac{\left(\frac{S_1^2}{n_1} + \frac{S_2^2}{n_2}\right)^2}{\frac{1}{(n_1-1)}\left(\frac{S_1^2}{n_1}\right)^2 + \frac{1}{(n_2-1)}\left(\frac{S_2^2}{n_2}\right)^2}.$$

Since $v$ is seldom an integer, we round it down to the nearest whole number. Using the statistic $T'$, we write

$$P\left(-t_{\alpha/2} < T' < t_{\alpha/2}\right) \approx 1 - \alpha,$$

where $t_{\alpha/2}$ is the value of the $t$-distribution with $v$ degrees of freedom, above which we find an area of $\alpha/2$.

$$P\left(-t_{\frac{\alpha}{2}} < [(\bar{X}_1 - \bar{X}_2) - (\mu_1 - \mu_2)]/\sqrt{\left(\frac{S_1^2}{n_1} + \frac{S_2^2}{n_2}\right)} < t_{\frac{\alpha}{2}}\right) \approx 1 - \alpha,$$

$$P\left(-t_{\frac{\alpha}{2}}\sqrt{\left(\frac{S_1^2}{n_1} + \frac{S_2^2}{n_2}\right)} < (\bar{X}_1 - \bar{X}_2) - (\mu_1 - \mu_2) < t_{\frac{\alpha}{2}}\sqrt{\left(\frac{S_1^2}{n_1} + \frac{S_2^2}{n_2}\right)}\right) \approx 1 - \alpha,$$

$$P\left(-(\bar{X}_1 - \bar{X}_2) - t_{\frac{\alpha}{2}}\sqrt{\left(\frac{S_1^2}{n_1} + \frac{S_2^2}{n_2}\right)} < -(\mu_1 - \mu_2) < -(\bar{X}_1 - \bar{X}_2) + t_{\frac{\alpha}{2}}\sqrt{\left(\frac{S_1^2}{n_1} + \frac{S_2^2}{n_2}\right)}\right) \approx 1 - \alpha,$$

$$P\left(-\left[(\bar{X}_1 - \bar{X}_2) + t_{\frac{\alpha}{2}}\sqrt{\left(\frac{S_1^2}{n_1} + \frac{S_2^2}{n_2}\right)}\right] < -(\mu_1 - \mu_2) < -\left[(\bar{X}_1 - \bar{X}_2) - t_{\frac{\alpha}{2}}\sqrt{\left(\frac{S_1^2}{n_1} + \frac{S_2^2}{n_2}\right)}\right]\right) \approx 1 - \alpha,$$

$$P\left(\left[(\bar{X}_1 - \bar{X}_2) - t_{\alpha/2}\sqrt{\left(\frac{S_1^2}{n_1} + \frac{S_2^2}{n_2}\right)}\right] < \mu_1 - \mu_2 < \left[(\bar{X}_1 - \bar{X}_2) + t_{\frac{\alpha}{2}}\sqrt{\left(\frac{S_1^2}{n_1} + \frac{S_2^2}{n_2}\right)}\right]\right) \approx 1 - \alpha.$$

■

**Procedure 18.10.**

When the variances, $\sigma_1^2$ and $\sigma_2^2$, are unknown and assumed to be unequal, you can calculate the CI for the difference between two means, $\mu_1 - \mu_2$, using the following steps:

1. Obtain two independent samples from the populations of interest. Let us call them Sample 1 and Sample 2, with sizes $n_1$ and $n_2$, respectively.
2. Calculate the sample mean for Sample 1, denoted as $\bar{x}_1$, and the sample mean for Sample 2, denoted as $\bar{x}_2$. Also, calculate the sample standard deviation for Sample 1, denoted as $s_1$, and the sample standard deviation for Sample 2, denoted as $s_2$.
3. The SE of the difference between means is given by:

$$\text{SE} = \sqrt{\left(\frac{s_1^2}{n_1} + \frac{s_2^2}{n_2}\right)}.$$

4. Specify the desired confidence level for the interval. Common choices include 90%, 95%, or 99%.
5. Determine the critical value (t) corresponding to the chosen confidence level and the degrees of freedom ($df$). The degrees of freedom can be approximated using the Welch-Satterthwaite equation:

$$df \approx \frac{\left(\frac{s_1^2}{n_1} + \frac{s_2^2}{n_2}\right)^2}{\frac{1}{(n_1 - 1)}\left(\frac{s_1^2}{n_1}\right)^2 + \frac{1}{(n_2 - 1)}\left(\frac{s_2^2}{n_2}\right)^2}.$$

6. Look up the critical value by using the t-distribution Mathematica function to find the appropriate t-value. Determine the critical value $t_{\frac{\alpha}{2}}$ corresponding to the chosen confidence level.
7. Multiply the SE by the critical value $t_{\frac{\alpha}{2}}$ to obtain the ME:

$$\text{ME} = t_{\frac{\alpha}{2}} \times \text{SE}.$$

8. Subtract the ME from the sample mean difference $(\bar{x}_1 - \bar{x}_2)$ to calculate the lower bound of the CI. Similarly, add the ME to the sample mean difference to obtain the upper bound:

$$\text{Lower bound} = (\bar{x}_1 - \bar{x}_2) - \text{ ME},$$
$$\text{Upper bound} = (\bar{x}_1 - \bar{x}_2) + \text{ ME}.$$

***Example 18.18***

Independent random samples from two normal populations with unknown and unequal variances produced the following data:

Sample 1: {3.38546,5.68173,4.73423,5.35505,5.68232,2.59485,4.8497,5.69644,6.47435,2.90192,6.00894}

Sample 2: {4.09719,2.49517,2.57922,2.90767,2.55753,2.30692,2.20041,2.90017}.
Obtain a 95% CI for $\mu_1 - \mu_2$.

***Solution***

```
SeedRandom[1234567];
(* Data of two samples: *)
sample1=RandomVariate[NormalDistribution[5,1.5],11]
sample2=RandomVariate[NormalDistribution[3,0.5],8]

(* Compute sample means and sample standard deviations: *)
n1=Length[sample1];
n2=Length[sample2];
xbar1=Mean[sample1];
xbar2=Mean[sample2];
s1=StandardDeviation[sample1];
s2=StandardDeviation[sample2];

(* Compute standard error: *)
SE=Sqrt[(s1^2/n1)+(s2^2/n2)]

(* Determine confidence level: *)
confidenceLevel=0.95 ;
alpha=1-confidenceLevel;

(* Find critical value using t-distribution: *)
df=Floor[(s1^2/n1+s2^2/n2)^2/((s1^2/n1)^2/(n1-1)+(s2^2/n2)^2/(n2-1))]
criticalValue=Quantile[StudentTDistribution[df],1-alpha/2]

(* Calculate margin of error: *)
ME=criticalValue*SE

(* Compute lower and upper bounds: *)
lowerBound=(xbar1-xbar2)-ME
upperBound=(xbar1-xbar2)+ME

(* Display the confidence interval: *)
{lowerBound,upperBound}

 {3.38546,5.68173,4.73423,5.35505,5.68232,2.59485,4.8497,5.69644,6.47435,2.90192,6.00894}
 {4.09719,2.49517,2.57922,2.90767,2.55753,2.30692,2.20041,2.90017}
 0.449959
 14
 2.14479
 0.965066
 1.13076
 3.06089
 {1.13076,3.06089}
```

## 18.5 Single Sample: CIs for Standard Deviations

**Theorem 18.7: CI for $\sigma^2$**

If $s^2$ is the variance of a random sample of size $n$ from a normal population, a $100(1-\alpha)\%$ CI for $\sigma^2$ is

$$\frac{(n-1)s^2}{\chi^2_{\frac{\alpha}{2}}} < \sigma^2 < \frac{(n-1)s^2}{\chi^2_{1-\frac{\alpha}{2}}}, \tag{18.16}$$

where $\chi^2_{\alpha/2}$ and $\chi^2_{1-\alpha/2}$ are $\chi^2$-values with $\nu = n-1$ degrees of freedom, leaving areas of $\alpha/2$ and $1-\alpha/2$, respectively, to the right.

**Proof:**

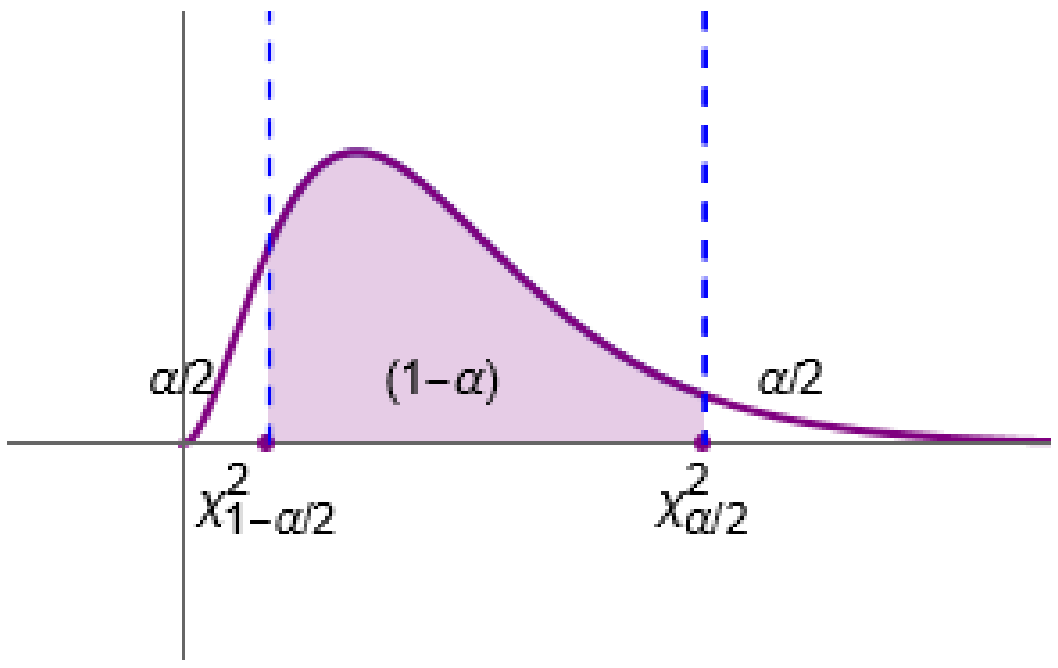

**Figure 18.9.** $P\left(\chi^2_{1-\alpha/2} < X^2 < \chi^2_{\alpha/2}\right) = 1 - \alpha$.

An interval estimate of $\sigma^2$ can be established by using the statistic

$$X^2 = \frac{(n-1)S^2}{\sigma^2}.$$

According to Theorem 16.13, the statistic $X^2$ has a chi-squared distribution with $n-1$ degrees of freedom when samples are chosen from a normal population, see Figure 18.9. We may write

$$P\left(\chi^2_{1-\frac{\alpha}{2}} < X^2 < \chi^2_{\frac{\alpha}{2}}\right) = 1 - \alpha,$$

where $\chi^2_{1-\alpha/2}$ and $\chi^2_{\alpha/2}$ are values of the chi-squared distribution with $n-1$ degrees of freedom, leaving areas of $1 - \alpha/2$ and $\alpha/2$, respectively, to the right. Substituting for $X^2$, we write

$$P\left(\chi^2_{1-\frac{\alpha}{2}} < \frac{(n-1)S^2}{\sigma^2} < \chi^2_{\frac{\alpha}{2}}\right) = 1 - \alpha,$$

$$P\left(\frac{\chi^2_{1-\frac{\alpha}{2}}}{(n-1)S^2} < \frac{1}{\sigma^2} < \frac{\chi^2_{\frac{\alpha}{2}}}{(n-1)S^2}\right) = 1 - \alpha,$$

$$P\left(\frac{(n-1)S^2}{\chi^2_{\frac{\alpha}{2}}} < \sigma^2 < \frac{(n-1)S^2}{\chi^2_{1-\frac{\alpha}{2}}}\right) = 1 - \alpha.$$

Finally,

$$\sqrt{\frac{(n-1)S^2}{\chi^2_{\frac{\alpha}{2}}}} < \sigma < \sqrt{\frac{(n-1)S^2}{\chi^2_{1-\frac{\alpha}{2}}}}.$$

■

**Procedure 18.11.**

To construct a CI for the standard deviation, follow the following steps:

1- Calculate the sample variance, denoted as $s^2$, from the data.
2- Determine the degrees of freedom $df$ for the chi-square distribution. It is given by $df = n - 1$, where $n$ is the sample size.
3- Choose the desired level of confidence.
4- Look up the critical values for the chi-square distribution with $df$ degrees of freedom and the desired $\alpha$ level. You can use a chi-square Mathematica function.
5- Calculate the lower and upper bounds for the CI using the formula:

$$\sqrt{\frac{(n-1)s^2}{\chi^2_{\frac{\alpha}{2}}}} < \sigma < \sqrt{\frac{(n-1)s^2}{\chi^2_{1-\frac{\alpha}{2}}}}.$$

***Example 18.19***

Random sample from normal population with standard deviation of 8 produced the following data:
{10.5779, 14.6391, 26.6495, 17.4986, 11.3655, 35.9454, 19.8406, 26.2044, 9.25341, 16.2295, 17.7125, 18.8675, 40.7876, 31.9293, 17.9794}.
Determine a 95% CI for $S^2$.

***Solution***

```
SeedRandom[51234];
(* Sample data: *)
sample=RandomVariate[NormalDistribution[20,8],15]

(* Compute the sample standard deviation: *)
s=StandardDeviation[sample]

(* Determine confidence level: *)
confidenceLevel=0.95 ;

(* Find critical values: *)
n=Length[sample];
df=n-1;
alpha=1-confidenceLevel;

chi2upper=Quantile[ChiSquareDistribution[df],1-alpha/2](* Subsuperscript[χ, α/2, 2] *)
chi2lower=Quantile[ChiSquareDistribution[df],alpha/2](* Subsuperscript[χ, 1-α/2, 2] *)

(* Calculate lower and upper bounds*)
lowerBound=Sqrt[(n-1)*s^2/chi2upper]
upperBound=Sqrt[(n-1)*s^2/chi2lower]

(* Display the confidence interval for the standard deviation: *)
{lowerBound,upperBound}
{lowerBound,upperBound}^2

{10.5779,14.6391,26.6495,17.4986,11.3655,35.9454,19.8406,26.2044,9.25341,16.2295,17.7125,18.8675,40.7876,31.9293,17.9794}
 9.38875
 26.1189
 5.62873
 6.87375
 14.807
 {6.87375,14.807}
 {47.2485,219.247}
```

## 18.6 Two Samples: CIs for Standard Deviations

**Theorem 18.8: CI for $\sigma_1^2/\sigma_2^2$**

If $s_1^2$ and $s_2^2$ are the variances of independent samples of sizes $n_1$ and $n_2$, respectively, from normal populations, then a $100(1-\alpha)\%$ CI for $\sigma_1^2/\sigma_2^2$ is

$$\frac{s_1^2}{s_2^2}\frac{1}{F_{\alpha/2}(\nu_1,\nu_2)} < \frac{\sigma_1^2}{\sigma_2^2} < \frac{s_1^2}{s_2^2}F_{\alpha/2}(\nu_2,\nu_1), \tag{18.17}$$

where $F_{\alpha/2}(\nu_1,\nu_2)$ is an $F$-value with $\nu_1 = n_1 - 1$ and $\nu_2 = n_2 - 1$ degrees of freedom, leaving an area of $\alpha/2$ to the right, and $F_{\alpha/2}(\nu_2,\nu_1)$ is a similar $F$-value with $\nu_2 = n_2 - 1$ and $\nu_1 = n_1 - 1$ degrees of freedom.

**Proof:**

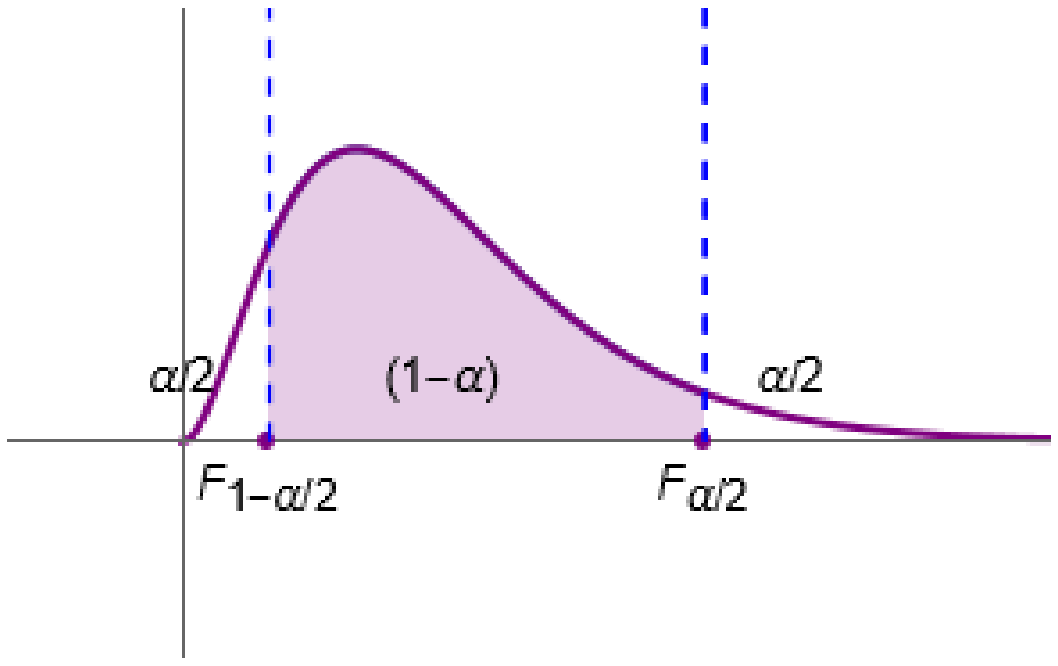

**Figure 18.10.** $P\left(F_{1-\alpha/2}(\nu_1,\nu_2) < F < F_{\alpha/2}(\nu_1,\nu_2)\right) = 1-\alpha.$

If $\sigma_1^2$ and $\sigma_2^2$ are the variances of normal populations, we can establish an interval estimate of $\sigma_1^2/\sigma_2^2$ by using the statistic

$$F = \frac{\sigma_2^2 S_1^2}{\sigma_1^2 S_2^2}.$$

According to 16.47, the RV $F$ has an $F$-distribution with $\nu_1 = n_1 - 1$ and $\nu_2 = n_2 - 1$ degrees of freedom, see Figure 18.10. Therefore, we may write

$$P\left(F_{1-\alpha/2}(\nu_1,\nu_2) < F < F_{\alpha/2}(\nu_1,\nu_2)\right) = 1-\alpha,$$

where $F_{1-\alpha/2}(\nu_1,\nu_2)$ and $F_{\alpha/2}(\nu_1,\nu_2)$ are the values of the $F$-distribution with $\nu_1$ and $\nu_2$ degrees of freedom, leaving areas of $1-\alpha/2$ and $\alpha/2$, respectively, to the right. Substituting for $F$, we write

$$P\left(F_{1-\frac{\alpha}{2}}(\nu_1,\nu_2) < \frac{\sigma_2^2 S_1^2}{\sigma_1^2 S_2^2} < F_{\frac{\alpha}{2}}(\nu_1,\nu_2)\right) = 1-\alpha,$$

$$P\left(\frac{S_2^2}{S_1^2}F_{1-\frac{\alpha}{2}}(\nu_1,\nu_2) < \frac{\sigma_2^2}{\sigma_1^2} < \frac{S_2^2}{S_1^2}F_{\frac{\alpha}{2}}(\nu_1,\nu_2)\right) = 1-\alpha,$$

$$P\left(\frac{S_1^2}{S_2^2}\frac{1}{F_{\frac{\alpha}{2}}(\nu_1,\nu_2)} < \frac{\sigma_1^2}{\sigma_2^2} < \frac{S_1^2}{S_2^2}\frac{1}{F_{1-\frac{\alpha}{2}}(\nu_1,\nu_2)}\right) = 1-\alpha,$$

$$P\left(\frac{S_1^2}{S_2^2}\frac{1}{F_{\alpha/2}(\nu_1,\nu_2)} < \frac{\sigma_1^2}{\sigma_2^2} < \frac{S_1^2}{S_2^2}F_{\alpha/2}(\nu_2,\nu_1)\right) = 1-\alpha,$$

where, we used $F_{1-\frac{\alpha}{2}}(\nu_1,\nu_2) = \frac{1}{F_{\alpha/2}(\nu_2,\nu_1)}$. Finally,

$$\sqrt{\frac{S_1^2}{S_2^2}\frac{1}{F_{\alpha/2}(\nu_1,\nu_2)}} < \frac{\sigma_1}{\sigma_2} < \sqrt{\frac{S_1^2}{S_2^2}F_{\alpha/2}(\nu_2,\nu_1)}.$$

■

**Procedure 18.12.**

To construct a CI for the ratio of two population variances $\frac{\sigma_1^2}{\sigma_2^2}$, follow the following steps:

1- Obtain the sample variances $s_1^2$ and $s_2^2$ from two independent samples taken from the two populations of interest. Also, record the sample sizes $n_1$ and $n_2$ for each group.
2- Calculate the degrees of freedom for each sample. The degrees of freedom for the numerator variance $S_1^2$ is equal to $n_1 - 1$, and the degrees of freedom for the denominator variance $S_2^2$ is equal to $n_2 - 1$.
3- Choose the desired level of confidence for your interval, such as 95%, 99%, etc. This will determine the critical values from the $F$-distribution.
4- Using the degrees of freedom for the numerator and denominator variances, find the critical values $F$-lower and $F$-upper from the $F$-distribution Mathematica function.
5- Finally, compute the CI using the formula:

$$\frac{s_1^2}{s_2^2}\frac{1}{F_{\frac{\alpha}{2}}(\nu_1,\nu_2)} < \frac{\sigma_1^2}{\sigma_2^2} < \frac{s_1^2}{s_2^2}\frac{1}{F_{1-\frac{\alpha}{2}}(\nu_1,\nu_2)}.$$

***Example 18.20***

Independent random samples from two normal populations produced the following data:

Sample 1:
{15.5651,28.1794,18.6429,15.8798,19.387,29.1141,26.2035,22.5669,26.5818,25.8155,27.5758,18.6239,27.4347,22.2549,22.5815,17.6626,18.3488,14.3448,9.50322,22.6582,15.4067,21.5116,27.2083,14.2074,16.7788}
Sample 2:
{19.3166,11.9135,12.8707,16.7202,21.5351,17.6614,14.0704,16.3901,14.6752,15.5305,18.8852,19.3216,13.5793,19.2083,14.7849,14.7475,16.0208,17.3395,15.458,19.7956}.

Obtain a 95% CI for $\mu_1 - \mu_2$.

***Solution***

```
SeedRandom[71567];
(* Data of two samples: *)
sample1=RandomVariate[NormalDistribution[19,5],25]
sample2=RandomVariate[NormalDistribution[16,3],20]

(* Compute the sample variances: *)
s1Squared=Variance[sample1]
s2Squared=Variance[sample2]

(* Determine confidence level: *)
confidenceLevel=0.95 ;

(* Find critical values using F-distribution: *)
n1=Length[sample1];
n2=Length[sample2];
df1=n1-1;
df2=n2-1;
alpha=1-confidenceLevel;
Fupper=Quantile[FRatioDistribution[df1,df2],1-alpha/2](* Subsuperscript[χ, α/2, 2] *)
Flower=Quantile[FRatioDistribution[df1,df2],alpha/2](* Subsuperscript[χ, 1-α/2, 2] *)

(* Calculate lower and upper bounds: *)
lowerBound=s1Squared/(s2Squared*Fupper)
upperBound=s1Squared/(s2Squared*Flower)

(* Display the confidence interval for the ratio of variances*)
{lowerBound,upperBound}

{15.5651,28.1794,18.6429,15.8798,19.387,29.1141,26.2035,22.5669,26.5818,25.8155,27.5758,18.6239,27.4347,22.2549,22.5815,17.6626,18.3488,14.3448,9.50322,22.6582,15.4067,21.5116,27.2083,14.2074,16.7788}
```

```
{19.3166,11.9135,12.8707,16.7202,21.5351,17.6614,14.0704,16.3901,14.6752,15.5305,18.8852,19.
3216,13.5793,19.2083,14.7849,14.7475,16.0208,17.3395,15.458,19.7956}

 28.7889
 6.70976
 2.45232
 0.426411
 1.74961
 10.0621
 {1.74961,10.0621}
```

## 18.7 Single Sample: CIs for Proportions

A point estimator of the proportion $p$ in a binomial experiment is given by the $\hat{P} = X/n$, where $X$ represents the number of successes in $n$ trials. Therefore, the sample proportion $\hat{p} = x/n$ will be used as the point estimate of the parameter $p$.

**Theorem 18.9: Large-Sample CIs for $p$**
If $\hat{p}$ is the proportion of successes in a random sample of size $n$ and $\hat{q} = 1 - \hat{p}$, an approximate $100(1-\alpha)\%$ CI, for the binomial parameter $p$ is given by

$$\hat{p} - z_{\frac{\alpha}{2}}\sqrt{\frac{\hat{p}\hat{q}}{n}} < p < \hat{p} + z_{\frac{\alpha}{2}}\sqrt{\frac{\hat{p}\hat{q}}{n}}, \tag{18.18.1}$$

if the sampling is either from an infinite population or with replacement from a finite population and are given by

$$\hat{p} \pm z_{\frac{\alpha}{2}}\sqrt{\frac{pq}{n}}\sqrt{\frac{N-n}{N-1}}, \tag{18.18.2}$$

if the sampling is without replacement from a population of finite size $N$, where $z_{\alpha/2}$ is the $z$-value leaving an area of $\alpha/2$ to the right.

**Proof:**

If the unknown proportion $p$ is not expected to be too close to 0 or 1, we can establish a CI for $p$ by considering the sampling distribution of $\hat{P}$. Designating a failure in each binomial trial by the value 0 and a success by the value 1, the number of successes, $x$, can be interpreted as the sum of $n$ values consisting only of 0 and 1s, and $\hat{p}$ is just the sample mean of these $n$ values. Hence, by the CLT, for $n$ sufficiently large, $\hat{P}$ is approximately normally distributed with mean

$$\mu_{\hat{P}} = E(\hat{P}) = E\left(\frac{X}{n}\right) = \frac{np}{n} = p,$$

and variance

$$\sigma^2_{\hat{P}} = \sigma^2_{X/n} = \frac{\sigma^2_X}{n^2} = \frac{npq}{n^2} = \frac{pq}{n}.$$

Therefore, we can assert that

$$P(-z_{\alpha/2} < Z < z_{\alpha/2}) = 1 - \alpha, \qquad \text{with } Z = \frac{(\hat{P} - p)}{\sqrt{\frac{pq}{n}}},$$

and $z_{\alpha/2}$ is the value above which we find an area of $\alpha/2$ under the standard normal curve. Substituting for $Z$, we write

$$P\left(-z_{\alpha/2} < \frac{(\hat{P}-p)}{\sqrt{\frac{pq}{n}}} < z_{\alpha/2}\right) = 1-\alpha.$$

When $n$ is large, very little error is introduced by substituting the point estimate $\hat{p} = x/n$ for the $p$ under the radical sign. Then we can write

$$P\left(\hat{P} - z_{\alpha/2}\sqrt{\frac{\hat{p}\hat{q}}{n}} < p < \hat{P} + z_{\alpha/2}\sqrt{\frac{\hat{p}\hat{q}}{n}}\right) \approx 1-\alpha.$$

■

It is important to note that this method assumes that the sample follows a large enough sample size and that the observations are independent (one should require both $n\hat{p}$ and $n\hat{q}$ to be greater than or equal to 5). If the sample size is small or the independence assumption is violated, other methods may be more appropriate.

**Procedure 18.13.**

To calculate a CI for proportions, you can follow these steps:

1. Start with a sample of data and calculate the proportion of successes ($\hat{p}$). This is the number of successes divided by the total number of observations.
2. Identify the total number of observations or sample size, $n$.
3. Choose the desired confidence level for the interval. Common choices include 90%, 95%, or 99% confidence levels.
4. Look up the critical value corresponding to the chosen confidence level by using standard normal distribution Mathematica function to find the appropriate z-value.
5. The SE represents the standard deviation of the sampling distribution of proportions and is calculated using the formula:
$$\text{SE} = \sqrt{\frac{\hat{p}(1-\hat{p})}{n}}.$$
6. The ME is the product of the SE and the critical value ($z_{\frac{\alpha}{2}}$):
$$\text{ME} = z_{\frac{\alpha}{2}} \times \text{SE}.$$
7. Subtract and add the ME from the sample proportion to get the lower and upper bounds of the CI:
$$\text{Lower bound} = \hat{p} - \text{ME},$$
$$\text{Upper bound} = \hat{p} + \text{ME}.$$

***Example 18.21***

For four years or 50,000 miles, an automaker offers a bumper-to-bumper warranty on all of its new cars. 15 of its vehicles in a random sample of 50 required five or more major warranty repairs during the warranty term. With a confidence coefficient of 0.95, determine the actual percentage of vehicles from this manufacturer that require five or more major repairs throughout the warranty period.

***Solution***

```
(* Determine the sample proportion (p-hat): *)
phat=15/50;

(* Determine the sample size (n): *)
n=50;

(* Determine the confidence level (C): *)
confidenceLevel=0.95;
```

```
alpha=1-confidenceLevel;

(* Determine the critical value (z): *)
criticalValue=Quantile[NormalDistribution[],1-alpha/2];

(* Calculate the standard error (SE): *)
SE=Sqrt[phat*(1-phat)/n];

(* Calculate the margin of error (ME): *)
ME=criticalValue*SE;

(* Calculate the lower and upper bounds: *)
lowerBound=phat-ME;
upperBound=phat+ME;

(* Display the confidence interval for the Proportion: *)
{lowerBound,upperBound}

 {0.17298,0.42702}
```

## 18.8 Two Samples: CIs the Difference between Two Proportions

A point estimator of the difference between the two proportions, $p_1 - p_2$, is given by the statistic $\hat{P}_1 - \hat{P}_2$. Therefore, the difference of the sample proportions, $\hat{p}_1 - \hat{p}_2$, will be used as the point estimate of $p_1 - p_2$.

**Theorem 18.10: Large-Sample CI for $p_1 - p_2$**
If $\hat{p}_1$ and $\hat{p}_2$ are the proportions of successes in random samples of sizes $n_1$ and $n_2$, respectively, $\hat{q}_1 = 1 - \hat{p}_1$, and $\hat{q}_2 = 1 - \hat{p}_2$, an approximate $100(1-\alpha)\%$ CI for the difference of two binomial parameters, $p_1 - p_2$, is given by

$$(\hat{p}_1 - \hat{p}_2) - z_{\frac{\alpha}{2}}\sqrt{\frac{\hat{p}_1\hat{q}_1}{n_1} + \frac{\hat{p}_2\hat{q}_2}{n_2}} < p_1 - p_2 < (\hat{p}_1 - \hat{p}_2) + z_{\frac{\alpha}{2}}\sqrt{\frac{\hat{p}_1\hat{q}_1}{n_1} + \frac{\hat{p}_2\hat{q}_2}{n_2}}, \tag{18.19}$$

where $z_{\alpha/2}$ is the $z$-value leaving an area of $\alpha/2$ to the right.

**Proof:**

A CI for $p_1 - p_2$ can be established by considering the sampling distribution of $\hat{P}_1 - \hat{P}_2$. We know that $\hat{P}_1$ and $\hat{P}_2$ are each approximately normally distributed, with means $p_1$ and $p_2$ and variances $p_1q_1/n_1$ and $p_2q_2/n_2$, respectively. Choosing independent samples from the two populations ensures that the variables $\hat{P}_1$ and $\hat{P}_2$ will be independent, and then by the reproductive property of the normal distribution, we conclude that $\hat{P}_1 - \hat{P}_2$ is approximately normally distributed with mean

$$\mu_{\hat{P}_1 - \hat{P}_2} = p_1 - p_2,$$

and variance

$$\sigma^2_{\hat{P}_1 - \hat{P}_2} = \frac{p_1q_1}{n_1} + \frac{p_2q_2}{n_2}.$$

Therefore, we can assert that

$$P\left(-z_{\alpha/2} < Z < z_{\alpha/2}\right) = 1 - \alpha,$$

where

$$Z = \frac{(\hat{P}_1 - \hat{P}_2) - (p_1 - p_2)}{\sqrt{\frac{p_1 q_1}{n_1} + \frac{p_2 q_2}{n_2}}},$$

and $z_{\alpha/2}$ is the value above which we find an area of $\alpha/2$ under the standard normal curve. Substituting for $Z$, we write

$$P\left(-z_{\frac{\alpha}{2}} < \frac{(\hat{P}_1 - \hat{P}_2) - (p_1 - p_2)}{\sqrt{\frac{p_1 q_1}{n_1} + \frac{p_2 q_2}{n_2}}} < z_{\frac{\alpha}{2}}\right) = 1 - \alpha,$$

$$P\left(-z_{\frac{\alpha}{2}}\sqrt{\frac{p_1 q_1}{n_1} + \frac{p_2 q_2}{n_2}} < (\hat{P}_1 - \hat{P}_2) - (p_1 - p_2) < z_{\frac{\alpha}{2}}\sqrt{\frac{p_1 q_1}{n_1} + \frac{p_2 q_2}{n_2}}\right) = 1 - \alpha,$$

$$P\left(-(\hat{P}_1 - \hat{P}_2) - z_{\frac{\alpha}{2}}\sqrt{\frac{p_1 q_1}{n_1} + \frac{p_2 q_2}{n_2}} < -(p_1 - p_2) < -(\hat{P}_1 - \hat{P}_2) + z_{\frac{\alpha}{2}}\sqrt{\frac{p_1 q_1}{n_1} + \frac{p_2 q_2}{n_2}}\right) = 1 - \alpha,$$

$$P\left(-\left[(\hat{P}_1 - \hat{P}_2) + z_{\frac{\alpha}{2}}\sqrt{\frac{p_1 q_1}{n_1} + \frac{p_2 q_2}{n_2}}\right] < -(p_1 - p_2) < -\left[(\hat{P}_1 - \hat{P}_2) - z_{\frac{\alpha}{2}}\sqrt{\frac{p_1 q_1}{n_1} + \frac{p_2 q_2}{n_2}}\right]\right) = 1 - \alpha,$$

$$P\left((\hat{P}_1 - \hat{P}_2) - z_{\alpha/2}\sqrt{\frac{p_1 q_1}{n_1} + \frac{p_2 q_2}{n_2}} < p_1 - p_2 < (\hat{P}_1 - \hat{P}_2) + z_{\frac{\alpha}{2}}\sqrt{\frac{p_1 q_1}{n_1} + \frac{p_2 q_2}{n_2}}\right) = 1 - \alpha.$$

Now, we replace $p_1$, $p_2$, $q_1$, and $q_2$ under the radical sign by their estimates $\hat{p}_1 = x_1/n_1$, $\hat{p}_2 = x_2/n_2$, $\hat{q}_1 = 1 - \hat{p}_1$, and $\hat{q}_2 = 1 - \hat{p}_2$, provided that $n_1\hat{p}_1$, $n_1\hat{q}_1$, $n_2\hat{p}_2$, and $n_2\hat{q}_2$ are all greater than or equal to 5, and the following approximate $100(1-\alpha)\%$ CI for $p_1 - p_2$ is obtained.

$$(\hat{p}_1 - \hat{p}_2) - z_{\frac{\alpha}{2}}\sqrt{\frac{\hat{p}_1\hat{q}_1}{n_1} + \frac{\hat{p}_2\hat{q}_2}{n_2}} < p_1 - p_2 < (\hat{p}_1 - \hat{p}_2) + z_{\frac{\alpha}{2}}\sqrt{\frac{\hat{p}_1\hat{q}_1}{n_1} + \frac{\hat{p}_2\hat{q}_2}{n_2}}.$$

■

It is important to note that these steps assume that the samples follow large enough sample sizes and that the observations are independent.

**Procedure 18.14.**

To calculate a CI for the difference between two proportions, you can follow these steps:

1. Start with two independent samples and calculate the proportions of successes ($\hat{p}_1$ and$\hat{p}_2$) for each sample. These are the numbers of successes divided by their respective total numbers of observations.
2. Identify the total number of observations for each sample ($n_1$ and $n_2$).
3. Choose the desired confidence level for the interval. Common choices include 90%, 95%, or 99% confidence levels.
4. Look up the critical value corresponding to the chosen confidence level by using standard normal distribution Mathematica function to find the appropriate z-value.
5. The SE represents the standard deviation of the sampling distribution of the difference between proportions and is calculated using the formula:

$$\text{SE} = \sqrt{\frac{\hat{p}_1(1-\hat{p}_1)}{n_1} + \frac{\hat{p}_2(1-\hat{p}_2)}{n_2}}.$$

6. The ME is the product of the SE and the critical value ($z_{\frac{\alpha}{2}}$):

$$\text{ME} = z_{\frac{\alpha}{2}} \times \text{SE}.$$

7. Subtract and add the ME from the difference in sample proportions to get the lower and upper bounds of the CI:

$$\text{Lower bound} = (\hat{p}_1 - \hat{p}_2) - \text{ME},$$
$$\text{Upper bound} = (\hat{p}_1 - \hat{p}_2) + \text{ME}.$$

***Example 18.22***

Males and females were questioned about what they would do if they accidentally received a $100 bill in the mail that was intended for their neighbors. Would they give their neighbors it back? 52 of the 69 sampled males and 120 of the 131 sampled females responded "yes," respectively.

Calculate a 95% confidence range for the proportional difference between men and women who answered "yes."

***Solution***

```
(* Determine the sample proportions: *)
phat1=70/79;
phat2=80/141;

(* Determine the sample sizes: *)
n1=79;
n2=141;

(* Determine the confidence level (C): *)
confidenceLevel=0.95;
alpha=1-confidenceLevel;

(* Determine the critical value (z): *)
criticalValue=Quantile[NormalDistribution[],1-alpha/2];

(* Calculate the standard error (SE): *)
SE=Sqrt[(phat1*(1-phat1)/n1)+(phat2*(1-phat2)/n2)];

(* Calculate the margin of error (ME): *)
ME=criticalValue*SE;

(* Calculate the lower and upper bounds: *)
diffProportions=phat1-phat2;
lowerBound=diffProportions-ME;
upperBound=diffProportions+ME;

(* Display the confidence interval for the difference between two proportions: *)
{lowerBound,upperBound}

 {0.211015,0.426385}
```

# CHAPTER 19

# MATHEMATICA LAB: ESTIMATION THEORY

In this chapter, we will explore two important topics in statistics: point estimate and interval estimate. These concepts play a crucial role in understanding and making inferences from data. We will specifically focus on the implementation of various mathematical functions in the Mathematica that can aid in performing point and interval estimation.

- **Point Estimate:**
  Point estimate is a statistical estimate of an unknown population parameter based on observed data. It provides a single value that represents the best estimate of the parameter. In this section, we will discuss the following Mathematica functions related to point estimation:
  - The `Likelihood` function in Mathematica calculates the likelihood of obtaining the observed data given a specific set of parameter values. It helps in assessing the plausibility of different parameter values and determining the maximum likelihood estimates.
  - The `LogLikelihood` function is used to compute the log-likelihood of the data. Taking the logarithm of the likelihood function simplifies calculations and aids in maximizing the likelihood using optimization techniques.
  - The `EstimatedDistribution` function is used to estimate the probability distribution of a dataset. It automatically selects the best-fitting distribution based on the data and provides parameter estimates for the chosen distribution.
  - The `FindDistributionParameters` function is used to estimate the parameters of a specific probability distribution given a dataset. It optimizes the parameters to maximize the likelihood of the observed data.
  - The `FindDistribution` function helps in finding the best-fitting distribution for a dataset by searching through a predefined list of distributions. It compares the goodness-of-fit for each distribution and returns the best match.
- **Interval Estimate:**
  Interval estimate involves determining a range of values that is likely to contain the unknown population parameter. It provides a measure of uncertainty and allows for the quantification of the estimation accuracy. Mathematica provides a comprehensive Hypothesis Testing Package that includes functions for constructing confidence intervals. These functions allow us to perform hypothesis tests and compute confidence intervals for various parameters based on different statistical distributions.

In the following table, we list the built-in functions that are used in this chapter.

| | | |
|---|---|---|
| `Likelihood` | `MeanCI` | `NormalCI` |
| `ogLikelihood` | `MeanDifferenceCI` | `StudentTCI` |
| `EstimatedDistribution` | `VarianceCI` | `ChiSquareCI` |
| `FindDistributionParameters` | `VarianceRatioCI` | `FRatioCI` |
| `FindDistribution` | | |

Therefore, we divided this chapter into three units to cover the above topics.

**Chapter 19 Outline**

# UNIT 19.1

# POINT ESTIMATE

| | |
|---|---|
| `Likelihood[dist,{x1,x2,…}]` | gives the likelihood function for observations x1, x2, … from the distribution dist. |
| `LogLikelihood[dist,{x1,x2,…}]` | gives the log-likelihood function for observations x1, x2, … from the distribution dist. |

***Mathematica Examples 19.1***

```
Input    (* The code calculates the likelihood for a given dataset using a normal distribution.
         It defines the dataset, creates a normal distribution object with unknown parameters,
         and defines a likelihood function that computes the product of the PDF values for
         each data point. The code then calculates the likelihood using the defined function
         and compares it with the built-in Likelihood function: *)

         (* Likelihood is a product of PDF values for the data: *)
         data={1.2,1.5,1.8,2.1,1.9};
         dist=NormalDistribution[μ,σ];

         (* Define the likelihood function: *)
         likelihood[data_,μ_,σ_]:=Apply[Times,PDF[dist,data]];

         (* Calculate the likelihood for the given data: *)
         likelihoodValue=Simplify[likelihood[data,μ,σ]]

         (* Using built-in function Likelihood: *)
         Likelihood[dist,data]

         (* Compute a likelihood for numeric data: *)
         μ=1;σ=1;
         likelihoodValue=Simplify[likelihood[data,μ,σ]]
         Likelihood[dist,data]

Output   E^(-(2.5 (2.99 -3.4 μ+μ^2))/σ^2 )/(4 √2  π^(5/2)  σ^5 )
Output   E^(-(2.5 (2.99 -3.4 μ+μ^2))/σ^2 )/(4 √2  π^(5/2)  σ^5 )
Output   0.00231188
Output   0.00231188
```

***Mathematica Examples 19.2***

```
Input    (* The code demonstrates the calculation of the log-likelihood for a dataset following
         a normal distribution. It defines the dataset and a normal distribution object with
         unknown parameters. The code then calculates the log-likelihood using different
         approaches, including directly applying the Log function to the built-in Likelihood
         function and using the LogLikelihood function. It also presents an alternative method
         using PowerExpand and Total to compute the sum of the logarithms of the PDF values:
         *)

         (* LogLikelihood is the log of Likelihood:*)
         data={1.2,1.5,1.8,2.1,1.9};
         dist=NormalDistribution[μ,σ];
```

```
        Log[Likelihood[dist,data]]
        LogLikelihood[dist,data]

        (* LogLikelihood is the sum of logs of PDF values for data: *)
        LogLikelihood[dist,data]

        PowerExpand[
         Total[
          Log[PDF[dist,data]]
          ]
         ]

Output  Log[E^(-(2.5 (2.99 -3.4 μ + μ^2))/σ^2 )/(4 √2  π^(5/2)  σ^5 )]
Output  -(2.5 (2.99 -3.4 μ+μ^2 ))/σ^2 -5 (1/2(Log[2]+Log[π])+Log[σ])
Output  -(2.5 (2.99 -3.4 μ+μ^2 ))/σ^2 -5 (1/2(Log[2]+Log[π])+Log[σ])
Output  -(1.2 -μ)^2/(2 σ^2 ) - (1.5 -μ)^2/(2 σ^2)-(1.8 -μ)^2/(2 σ^2)-(1.9 -μ)^2/(2 σ^2)-
        (2.1 -μ)^2/(2 σ^2)+5(1/2(-Log[2]-Log[π])-Log[σ])
```

### *Mathematica Examples 19.3*

```
Input   (* The code showcases the calculation of likelihood and log-likelihood for a normal
        distribution based on sample data. It defines the normal distribution with unknown
        parameters and the sample data as a list of variables. The code demonstrates the
        usage of the Likelihood and LogLikelihood functions to compute the likelihood and
        log-likelihood, respectively, for both symbolic and numeric data: *)

        (* Sample data: *)
        dist=NormalDistribution[μ,σ];
        data={x,y,z};

        (* Get the likelihood function for a normal distribution: *)
        Likelihood[dist,data]

        (* Get the log-likelihood function for a normal distribution: *)
        LogLikelihood[dist,data]

        (* Compute a likelihood for numeric values of μ and σ=1: *)
        μ=1;σ=1;
        Likelihood[dist,data]
        LogLikelihood[dist,data]

        (* Compute a likelihood for numeric data: *)
        x=1;
        y=1.2;
        z=2;
        Likelihood[dist,data]
        LogLikelihood[dist,data]

Output  E^(-((x-μ)^2-(y-μ)^2-(z-μ)^2)/(2 σ^2))/(2 √2 π^(3/2) σ^3)
Output  -((x-μ)^2-(y-μ)^2-(z-μ)^2)/(2 σ^2)-3 (1/2  (Log[2] + Log[π]) +Log[σ])
Output  E^(1/2(-(-1+x)^2-(-1+y)^2-(-1+z)^2))/(2 √2 π^(3/2) )
Output  1/2 (-(-1+x)^2-(-1+y)^2-(-1+z)^2)- 3/2 (Log[2] + Log[π])
Output  0.0377483
Output  -3.27682
```

### *Mathematica Examples 19.4*

```
Input   (* The code generates visualizations of log-likelihood contours and a 3D plot for a
        gamma distribution based on numeric data. It first creates a sample of random values
```

```
from a gamma distribution and then uses the ContourPlot and Plot3D functions to
display the log-likelihood surface over a range of shape and scale parameters: *)

data=RandomVariate[GammaDistribution[2,3],100];
ContourPlot[
 LogLikelihood[GammaDistribution[α,β],data],
 {α,1,4},
 {β,1,4},
 ContourStyle->{White},
 ClippingStyle->Automatic,
 ColorFunction->"BlueGreenYellow",
 PlotLegends->Automatic,
 ImageSize->220
 ]

Plot3D[
 LogLikelihood[GammaDistribution[α,β],data],
 {α,1,4},
 {β,1,4},
 ClippingStyle->Automatic,
 ColorFunction->"BlueGreenYellow",
 PlotLegends->Automatic,
 ImageSize->220
 ]
```

Output

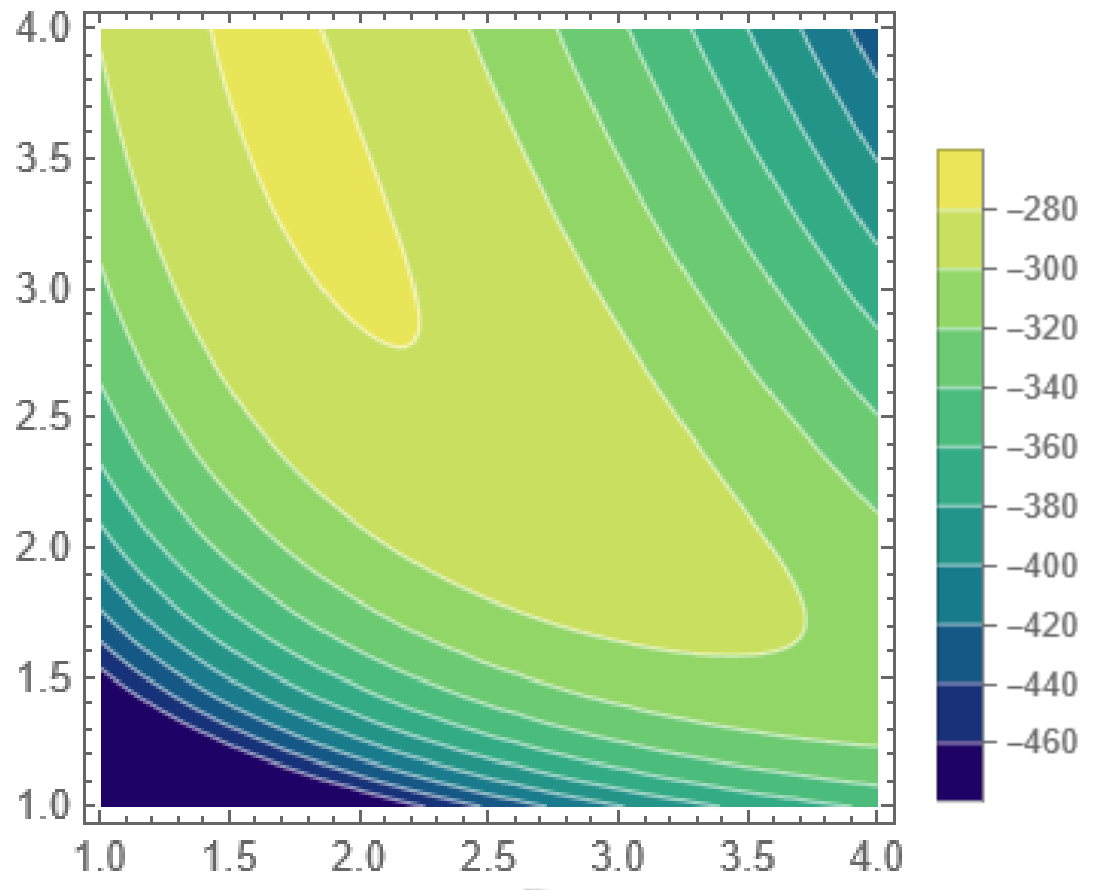

Output

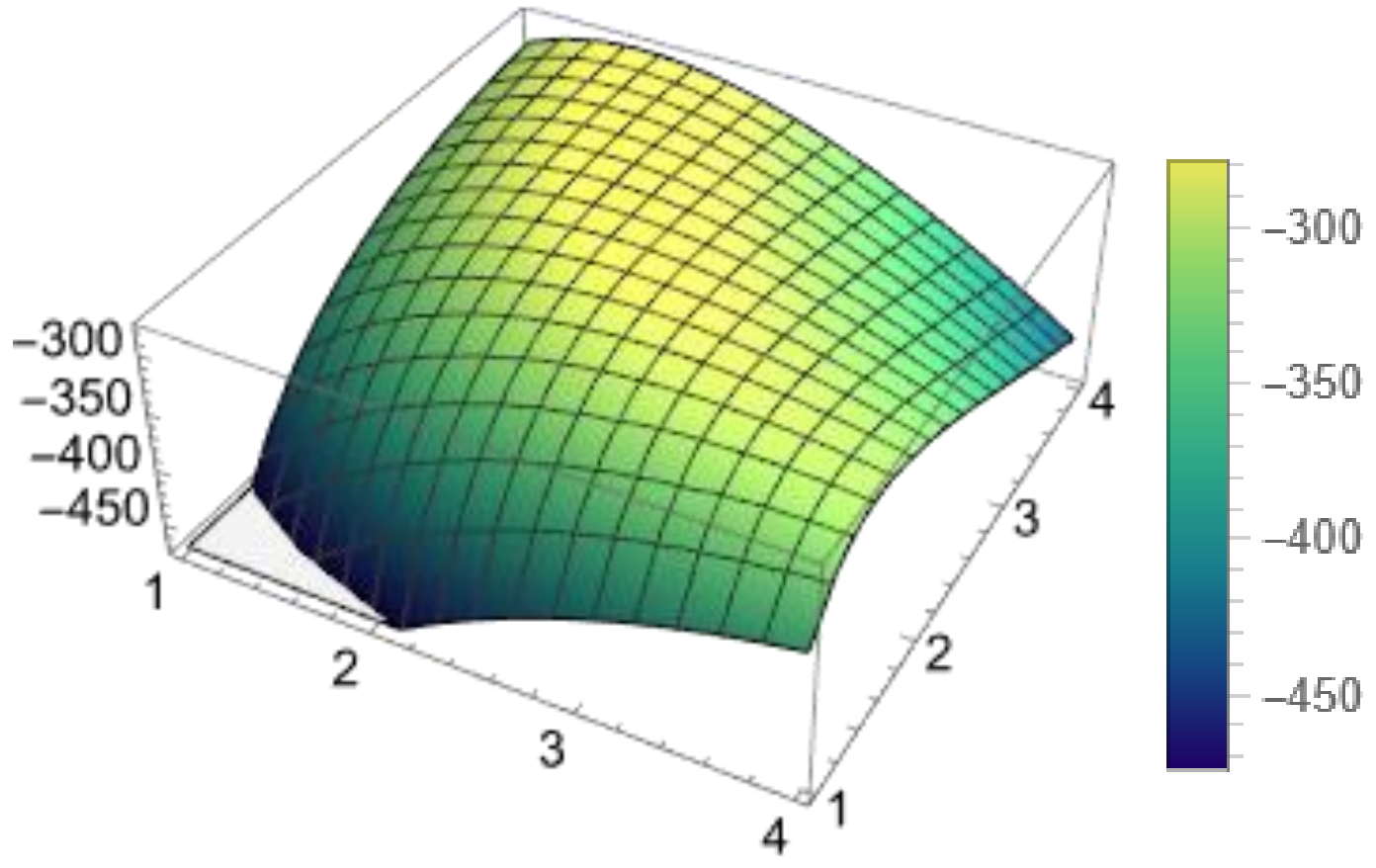

***Mathematica Examples 19.5***

Input

```
(* The code demonstrates the calculation of likelihood for multivariate data generated
from a binormal distribution. It generates a sample from the binormal distribution,
```

```
calculates the likelihood using the Likelihood function, and visualizes the log-
likelihood using contour plots and a 3D plot: *)

data=RandomVariate[
   BinormalDistribution[{1,1},{1,1},0.9],
   50
   ];

Likelihood[BinormalDistribution[{1,1},{σ1,σ2},0.9],data]

ContourPlot[
 Log[
  Likelihood[BinormalDistribution[{1,1},{σ1,σ2},0.9],data]
  ],
 {σ1,1,7},
 {σ2,1,7},
 ContourStyle->{White},
 ClippingStyle->Automatic,
 ColorFunction->"BlueGreenYellow",
 PlotLegends->Automatic,
 ImageSize->220
 ]

Plot3D[
 Log[
  Likelihood[BinormalDistribution[{1,1},{σ1,σ2},0.3],data]
  ],
 {σ1,1,7},
 {σ2,1,7},
 ClippingStyle->Automatic,
 ColorFunction->"BlueGreenYellow",
 PlotLegends->Automatic,
 ImageSize->220
 ]
```

Output (1.32485×10^(-22) E^(2.6(-45.3922/σ1^2 -56.8822/σ2^2 +81.7317/(σ1 σ2))))/(σ1^50 σ2^50)

Output

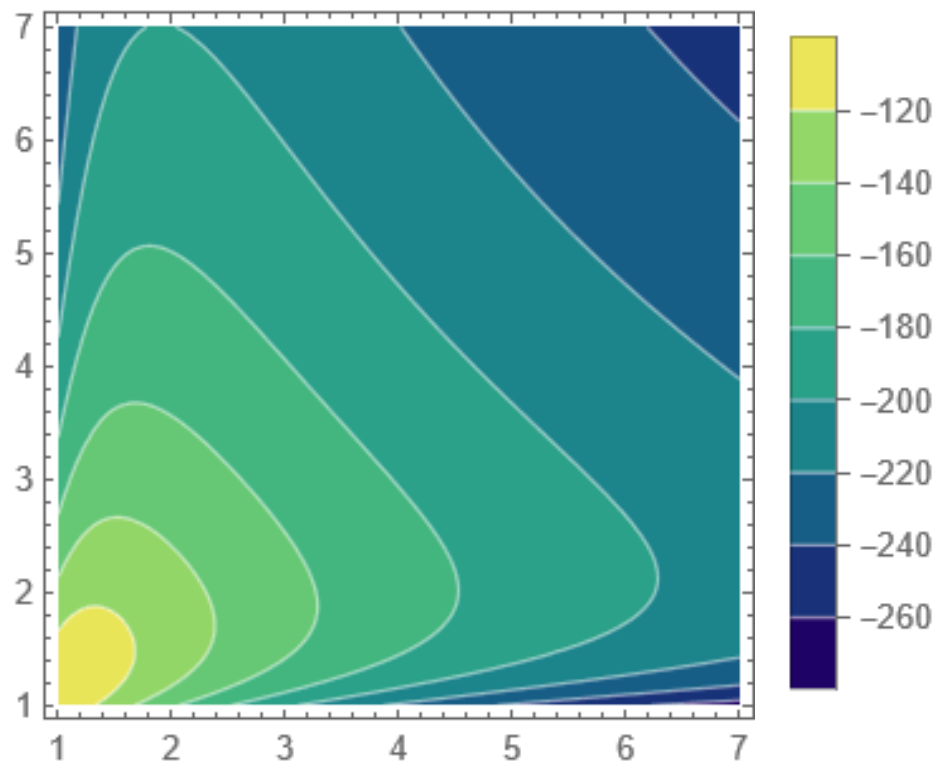

Output

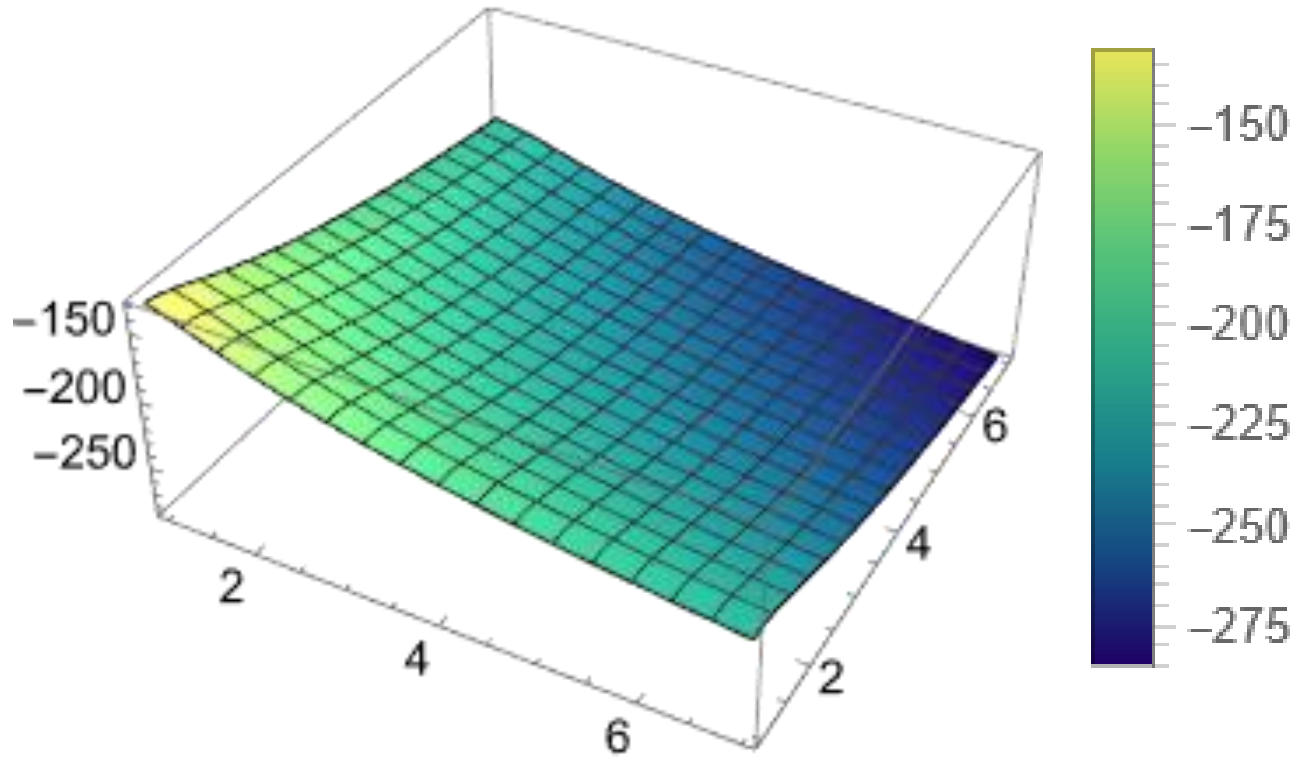

***Mathematica Examples 19.6***

Input

```
(* The code showcases the computation of the maximum likelihood estimate (MLE) and maximum log-likelihood estimate (MLLE) for the parameterμ in a Poisson distribution. The code defines a Poisson distribution and calculates the likelihood and log-likelihood based on the observed data. It then utilizes the Solve function to find the values of μ that satisfy the corresponding derivative equations. By solving the derivative equations for μ with the constraint that the observed data variables (x,y,z) are positive, the code obtains closed-form solutions for both the MLE and MLLE: *)

(* Solve for the Poisson maximum likelihood estimate in closed form: *)
Dist=PoissonDistribution[μ];
like=Likelihood[Dist,{x,y,z}];
loglike=LogLikelihood[Dist,{x,y,z}];

Solve[
 Refine[
  D[like,μ]==0,
  x>0&&y>0&&z>0
  ],
 μ
 ]

(* Solve for the Poisson maximum log-likelihood estimate in closed form: *)
Solve[
 Refine[
  D[loglike,μ]==0,
  x>0&&y>0&&z>0
  ],
 μ
 ]
```

Output

```
Solve::ifun: Inverse functions are being used by Solve, so some solutions may not be found; use Reduce for complete solution information.
```

Output `{{μ->1/3 (x+y+z)}}`
Output `{{μ->1/3 (x+y+z)}}`

***Mathematica Examples 19.7***

Input

```
(* The code demonstrates the direct computation of the maximum log-likelihood estimate (MLE) for a geometric distribution. The code generates a sample from a geometric distribution, finds the MLE using the FindMaximum function, and visualizes the likelihood and log-likelihood functions with the optimal point labeled. By optimizing the log-likelihood function with respect to the parameter p using FindMaximum, the
```

```
code obtains the MLE for the geometric distribution. The LogPlot and Plot functions
create plots of the likelihood and log-likelihood functions, respectively: *)
(* Compute a maximum log-likelihood estimate directly: *)
data=RandomVariate[
    GeometricDistribution[1/2],
    100
    ];

(* Maximize using the LogLikelihood for numeric value: *)
max=FindMaximum[
   {LogLikelihood[GeometricDistribution[p],data],0<p<1},
   p
   ]

(* Label the optimal point on a plot of the likelihood function: *)
LogPlot[
 Likelihood[GeometricDistribution[p],data],
 {p,.01,.99},
 Epilog->{Directive[Purple,PointSize[0.04]],Point[{p/. max[[2]],max[[1]]}]},
 ImageSize->220,
 PlotStyle->Purple
 ]
(* Label the optimal point on a plot of the LogLikelihood function: *)
Plot[
 LogLikelihood[GeometricDistribution[p],data],
 {p,0.01,0.99},
 Epilog->{Directive[Purple,PointSize[0.04]],Point[{p/. max[[2]],max[[1]]}]},
 ImageSize->220,
 PlotStyle->Purple
 ]
```

Output {-141.363,{p->0.490196}}

Output

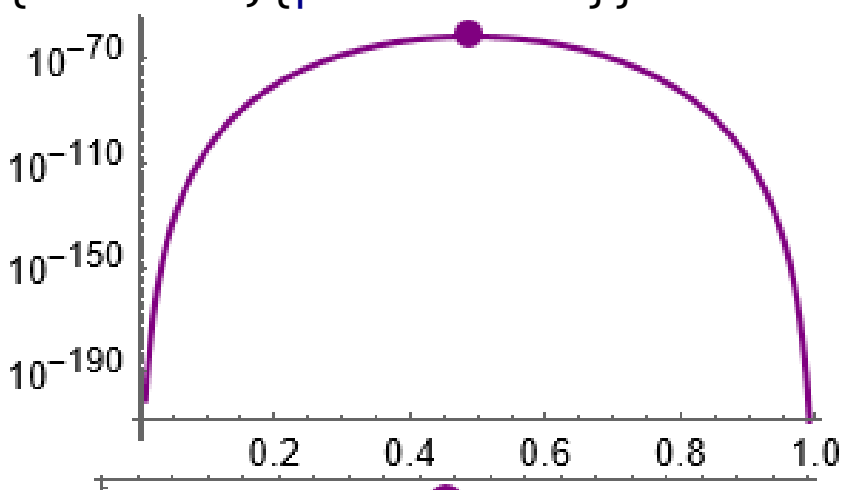

Output

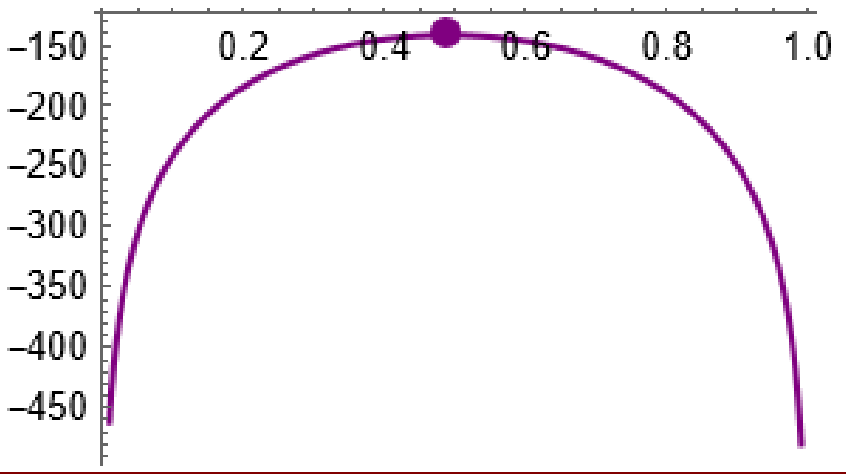

***Mathematica Examples 19.8***

Input

```
(* The code demonstrates the use of the Manipulate function to create an interactive
tool for exploring and analyzing a normal distribution. The code allows users to vary
the parameters of the distribution and observe the resulting changes in the likelihood
function, data visualization, and statistical summary. The Manipulate block includes
data simulation, likelihood function plotting, scatter plot and histogram generation
for data visualization, and the computation of statistical measures such as mean,
standard deviation, skewness, and kurtosis. The controls provided within the
Manipulate block allow users to interactively adjust the mean, standard deviation,
and number of data points: *)

(* Manipulate to vary parameters: *)
```

```
Manipulate[

 (* Simulating Data: *)
 data=RandomVariate[
   NormalDistribution[mu,sigma],
   n
   ];

 (* Plotting the Likelihood Function: *)
 contourPlot=ContourPlot[
   LogLikelihood[NormalDistribution[μ,σ],data],
   {μ,-2,4},
   {σ,1,3},
   ContourStyle->{White},
   ClippingStyle->Automatic,
   ColorFunction->"BlueGreenYellow",
   PlotLegends->Automatic,
   ImageSize->220
   ];

 plot3d=Plot3D[
   LogLikelihood[NormalDistribution[μ,σ],data],
   {μ,-2,4},
   {σ,1,3},
   ClippingStyle->Automatic,
   ColorFunction->"BlueGreenYellow",
   PlotLegends->Automatic,
   ImageSize->220
   ];

 (* Scatter Plot of Data: *)
 scatter=ListPlot[
   data,
   Filling->Axis,
   ColorFunction->"BlueGreenYellow",
   Frame->True,
   FrameStyle->Directive[Black],
   ImageSize->230
   ];
 (* Histogram of Data: *)
 histogram=Histogram[
   data,
   Automatic,
   "PDF",
   ColorFunction->"BlueGreenYellow",
   ImageSize->220,
   ChartStyle->Purple,
   PlotRange->Automatic
   ];

 (* Statistical Summary: *)
 stats=Grid[
   {
    {"Mean:",Mean[data]},
    {"Standard Deviation:",StandardDeviation[data]},
    {"Skewness:",Skewness[data]},
    {"Kurtosis:",Kurtosis[data]}
    }
   ];
 (* Displaying Plots and Statistics: *)
 Column[
```

```
   {{contourPlot,plot3d},{scatter,histogram},stats}
   ],
  (* Manipulate Controls: *)
  {{mu,0.5,"Mean"},0,1,Appearance->"Labeled"},
  {{sigma,1.5,"Standard Deviation"},1,2,Appearance->"Labeled"},
  {{n,70,"Number of data points"},10,100,10}
  ]
```

Output

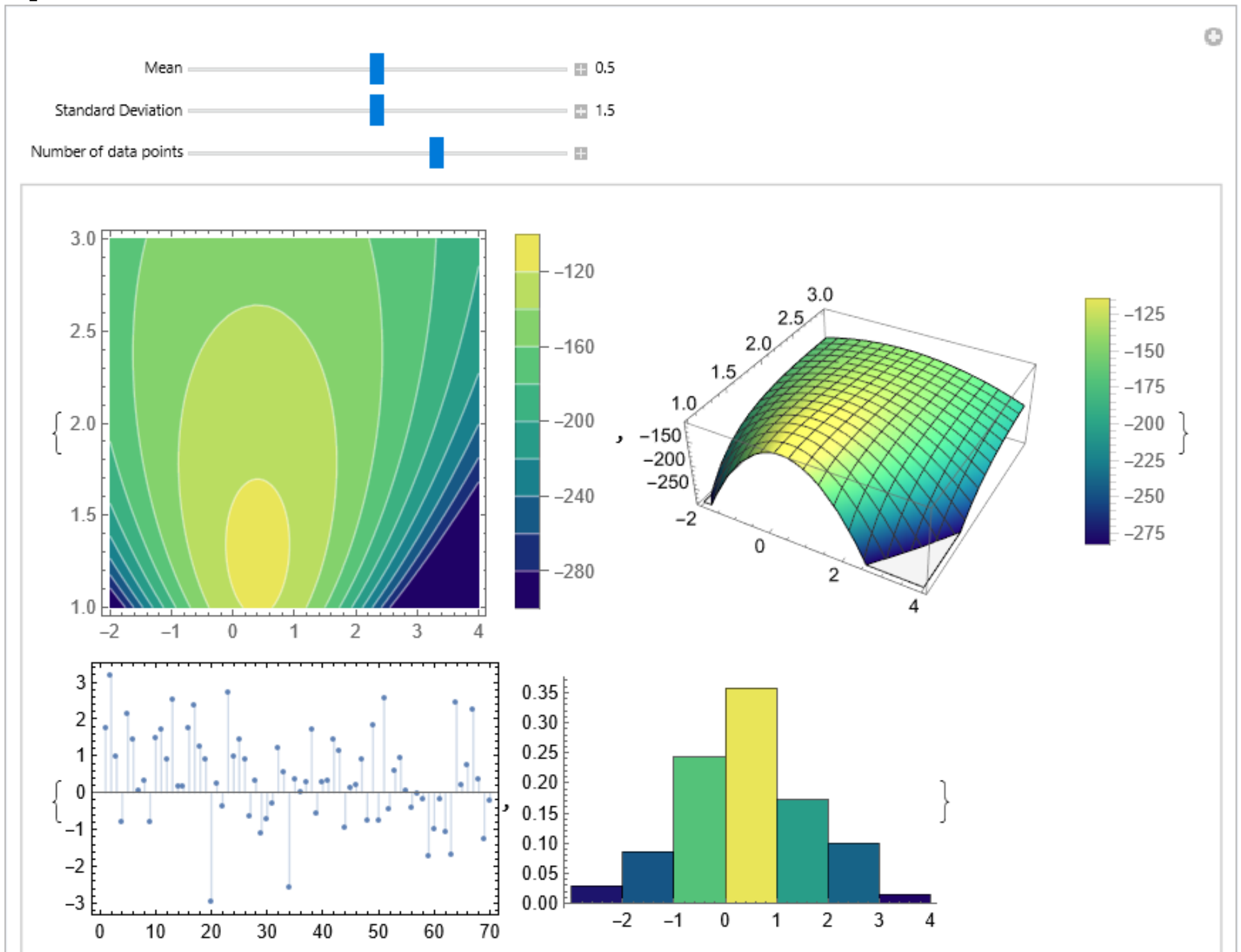

```
          Mean:          0.402053
Standard Deviation:     1.24788
       Skewness:       -0.111024
       Kurtosis:        2.97208
```

## *Mathematica Examples 19.9*

Input

```
(* The code showcases an interactive visualization of the likelihood function for a
geometric distribution using the Manipulate function. Inside the Manipulate block, a
local module generates random data from a geometric distribution with the parameter
p and a specified number of data points n. The LogPlot function is then used to
create a logarithmic plot of the likelihood function, representing the likelihood
over a range of values for the parameter μ. The Manipulate function provides
interactive controls for adjusting the parameters p and n, enabling users to
dynamically explore and visualize the impact of different parameter values and sample
sizes on the likelihood: *)

Manipulate[
 Module[
  {data,plot},
  data=RandomVariate[GeometricDistribution[p],n];(* Generate random data *)
  plot=LogPlot[
    Likelihood[GeometricDistribution[μ],data],
    {μ,0.01,0.99},
    ClippingStyle->Automatic,
    ColorFunction->"BlueGreenYellow",
    ImageSize->220
```

```
      ]
    ],
   {{p,0.5,"p"},0.01,0.99,0.01},
   {{n,10,"Number of data points"},10,100,10}
   ]
```

Output

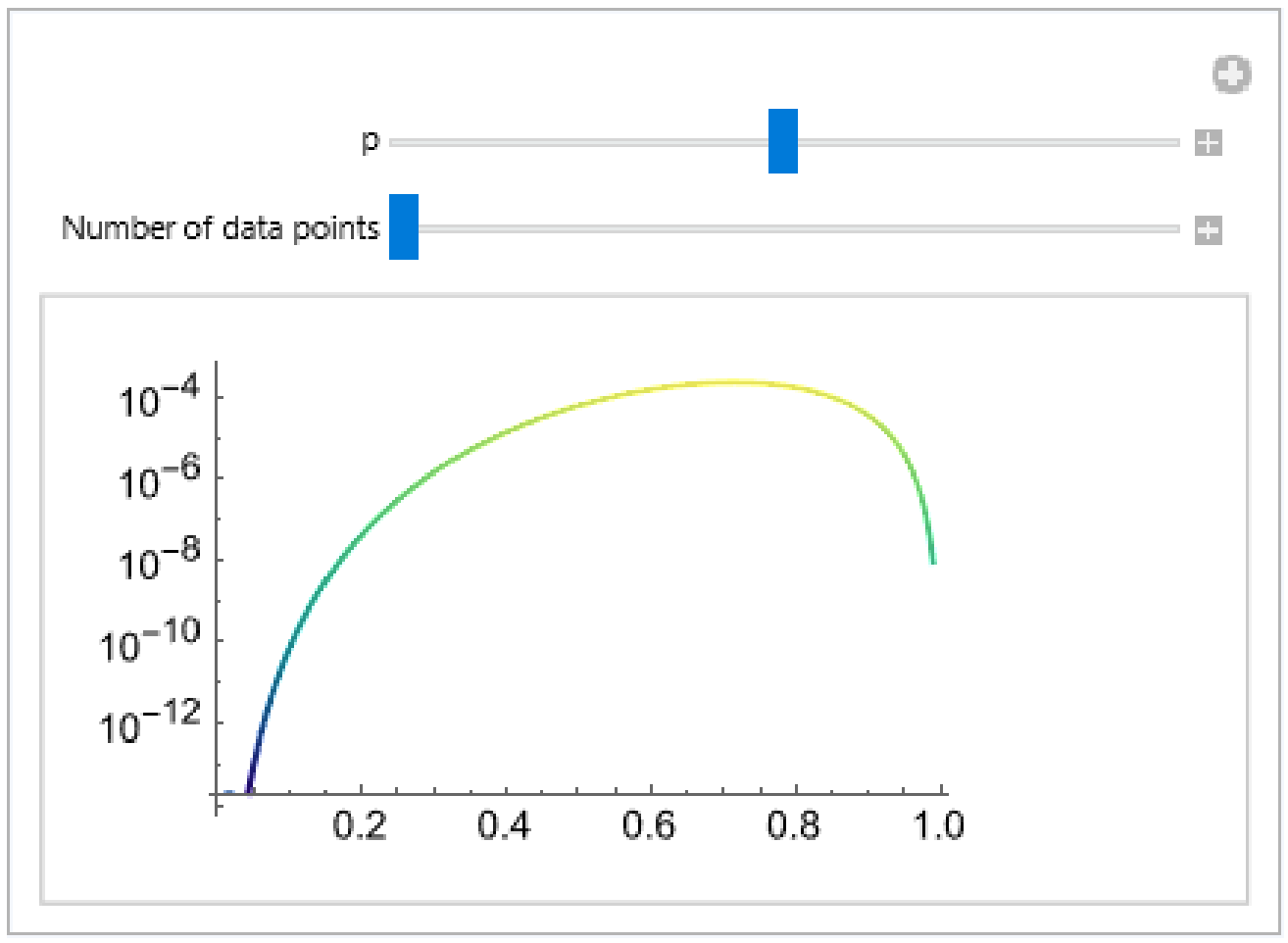

| | |
|---|---|
| `EstimatedDistribution[data,dist]` | estimates the parametric distribution dist from data. |
| `EstimatedDistribution[data,dist, {{p,p0}, {q,q0},…}]` | estimates the parameters p, q, … with starting values p0, q0, …. |
| `EstimatedDistribution[data,dist,idist]` | estimates distribution dist with starting values taken from the instantiated distribution idist. |
| `FindDistributionParameters[data,dist]` | finds the parameter estimates for the distribution dist from data. |
| `FindDistributionParameters[data,dist, {{p,p0}, {q,q0},…}]` | finds the parameters p, q, … with starting values p0, q0, …. |

The following basic settings can be used for `ParameterEstimator`:

| | |
|---|---|
| "MaximumLikelihood" | maximize the log-likelihood function |
| "MethodOfMoments" | match raw moments |
| "MethodOfCentralMoments" | match central moments |
| "MethodOfCumulants" | match cumulants |
| "MethodOfFactorialMoments" | match factorial moments |

***Mathematica Examples 19.10***

Input

```
(* The code performs parameter estimation for a normal distribution using maximum likelihood estimation and the method of moments. It generates random data from a specified normal distribution and estimates the parameters based on the generated data. The code then visually compares the probability density functions (PDFs) of the original and estimated distributions using a plot. The plot allows for a visual assessment of the accuracy of the estimation: *)

(* Obtain the maximum likelihood parameter estimates, assuming a normal distribution: *)
dist=NormalDistribution[0,1];
data=RandomVariate[dist,500];
estimatedist=EstimatedDistribution[
```

```
    data,
    NormalDistribution[μ,σ]
    ]

(* Estimate parameters for a normal distribution: *)
FindDistributionParameters[data,NormalDistribution[μ,σ]]

(* Visually compare the PDFs for the original and estimated distributions: *)
Plot[
 {PDF[dist,x],PDF[estimatedist,x]},
 {x,-5,5},
 PlotStyle->{Blue,Purple},
 PlotLegends->Placed[{"dist","estimatedist"},{0.25,0.75}]
 ]

(* Obtain the method of moments estimates: *)
EstimatedDistribution[
 data,
 NormalDistribution[μ,σ],
 ParameterEstimator->"MethodOfMoments"
 ]
```

Output NormalDistribution[-0.0323259,0.89941]
Output {μ->-0.0323259,σ->0.89941}
Output

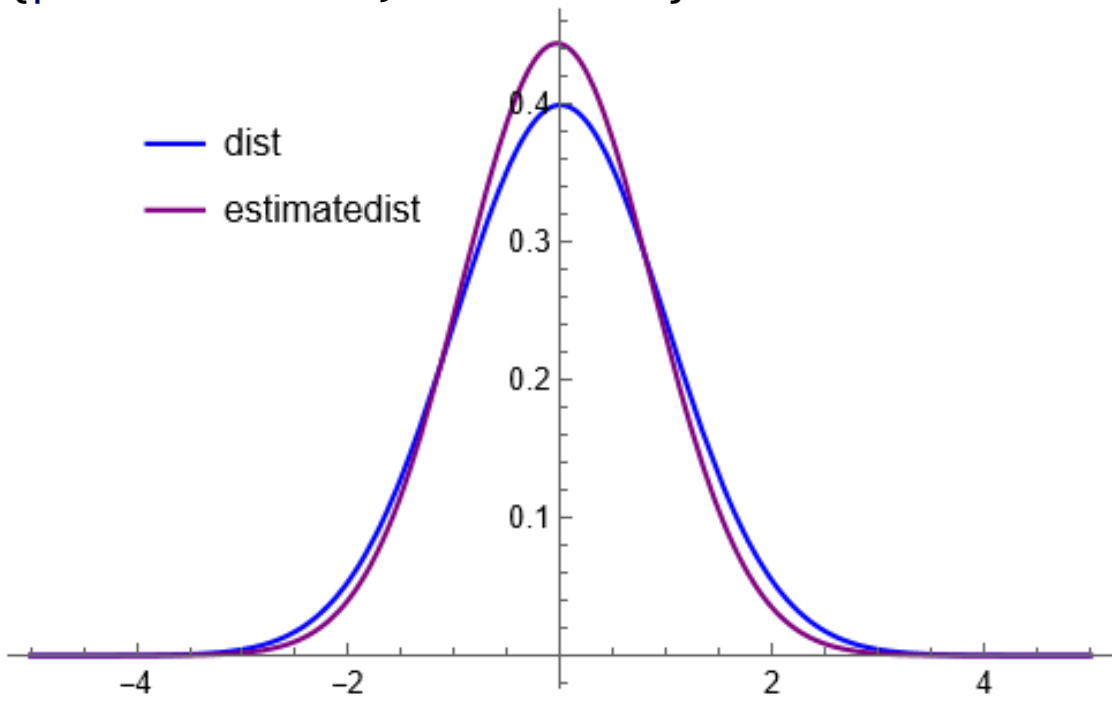

Output NormalDistribution[-0.0323259,0.89941]

***Mathematica Examples 19.11***

Input

```
(* The code generates a random sample of size 10,000 from a normal distribution with parameters μ=1 and σ=3, estimates the distribution parameters using the EstimatedDistribution and FindDistributionParameters functions, and then compares the histogram of the sample with the estimated PDF of the normal distribution using a histogram and a plot of the PDF: *)

sampledata=RandomVariate[
    NormalDistribution[1,3],
    10^4
    ];

(* Estimate the distribution parameters from sample data: *)
ed=EstimatedDistribution[
   sampledata,
   NormalDistribution[μ,σ]
   ]

(* Estimate parameters for a normal distribution: *)
FindDistributionParameters[sampledata,NormalDistribution[μ,σ]]
```

```
(* Compare a density histogram of the sample with the PDF of the estimated
distribution: *)
Show[
 Histogram[
  sampledata,
  {1},
  "PDF",
  ColorFunction->Function[{height},Opacity[height]],
  ChartStyle->Purple,
  ImageSize->320
  ],
 Plot[
  PDF[ed,x],
  {x,-10,10},
  ImageSize->320,
  ColorFunction->"Rainbow"
  ]
 ]
```

Output NormalDistribution[1.01916,2.99725]
Output {μ->1.01916,σ->2.99725}
Output

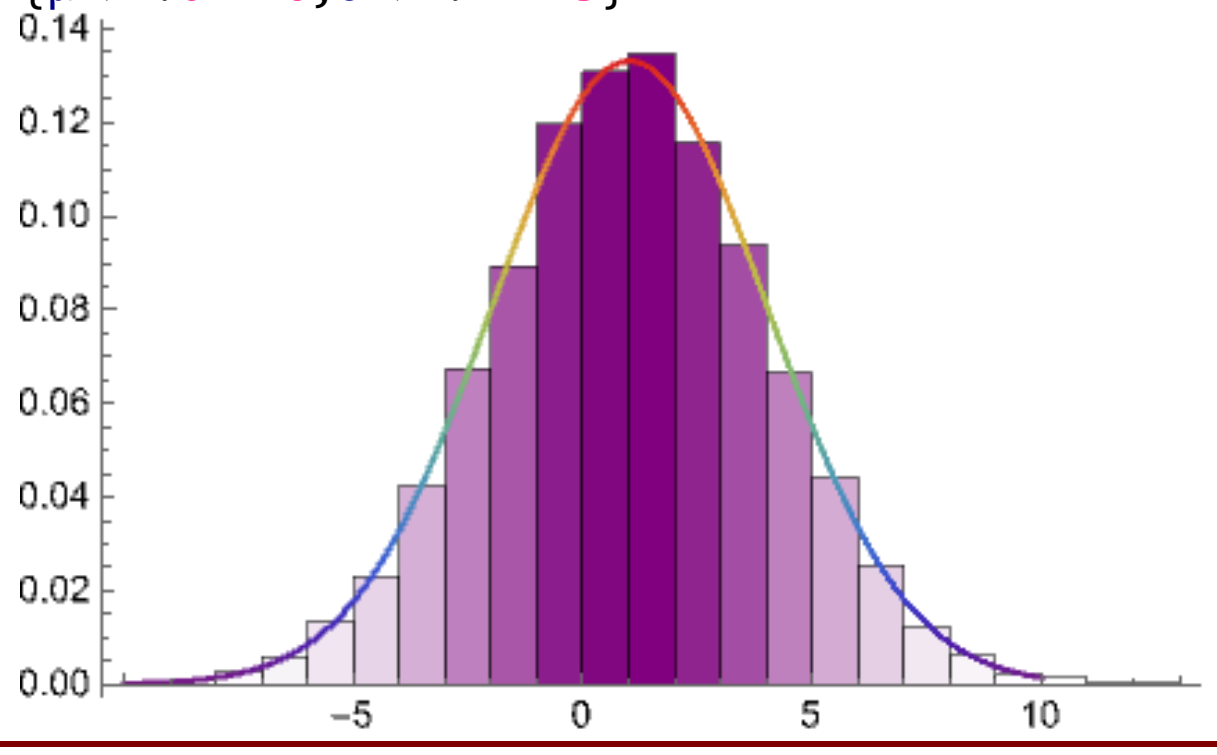

***Mathematica Examples 19.12***

Input

```
(* The code demonstrates a common technique in statistics and data analysis, which
is the use of random sampling to estimate population parameters. The code generates
random samples from a normal distribution with mean 0 and standard deviation 1,and
then using these samples to estimate the parameters of another normal distribution
with unknown mean and standard deviation. This process is repeated 20 times, resulting
in 20 different estimated distributions. The code also visualizes the resulting
estimated distributions using the PDF function. The code plots the PDFs of these
estimated distributions using the PDF function and the estimated parameters. The plot
shows the PDFs in a range from-3.5 to 3.5: *)

estim0distributions=Table[
  dist=NormalDistribution[0,1];

  sampledata=RandomVariate[
    dist,
    100
    ];

  ed=EstimatedDistribution[
    sampledata,
    NormalDistribution[α,β]
    ],
  {i,1,20}
  ]
```

```
        pdf0ed=Table[
            PDF[estim0distributions[[i]],x],
            {i,1,20}
            ];

        (* Visualizes the resulting estimated distributions: *)
        Plot[
         pdf0ed,
         {x,-3.5,3.5},
         PlotRange->Full,
         ImageSize->400,
         PlotStyle->Directive[Purple,Opacity[0.3],Thickness[0.002]]
         ]
Output  {NormalDistribution[0.0275173,1.01755],NormalDistribution[-0.0844031,0.879151],NormalDistribution[0.115425,0.884917],NormalDistribution[-0.0296073,1.02259],NormalDistribution[0.0487188,0.946338],NormalDistribution[0.0986574,1.09812],NormalDistribution[-0.0340808,1.08848],NormalDistribution[-0.0506773,1.0725],NormalDistribution[-0.138139,1.09712],NormalDistribution[0.0369571,1.05663],NormalDistribution[0.0432512,0.95691],NormalDistribution[-0.124249,0.90998],NormalDistribution[0.045595,0.933994],NormalDistribution[0.0184127,0.923313],NormalDistribution[-0.0303541,1.10383],NormalDistribution[0.0355708,1.03015],NormalDistribution[-0.0596503,0.992024],NormalDistribution[0.0828858,1.05965],NormalDistribution[-0.208048,0.9647],NormalDistribution[0.113367,0.931704]}
Output
```

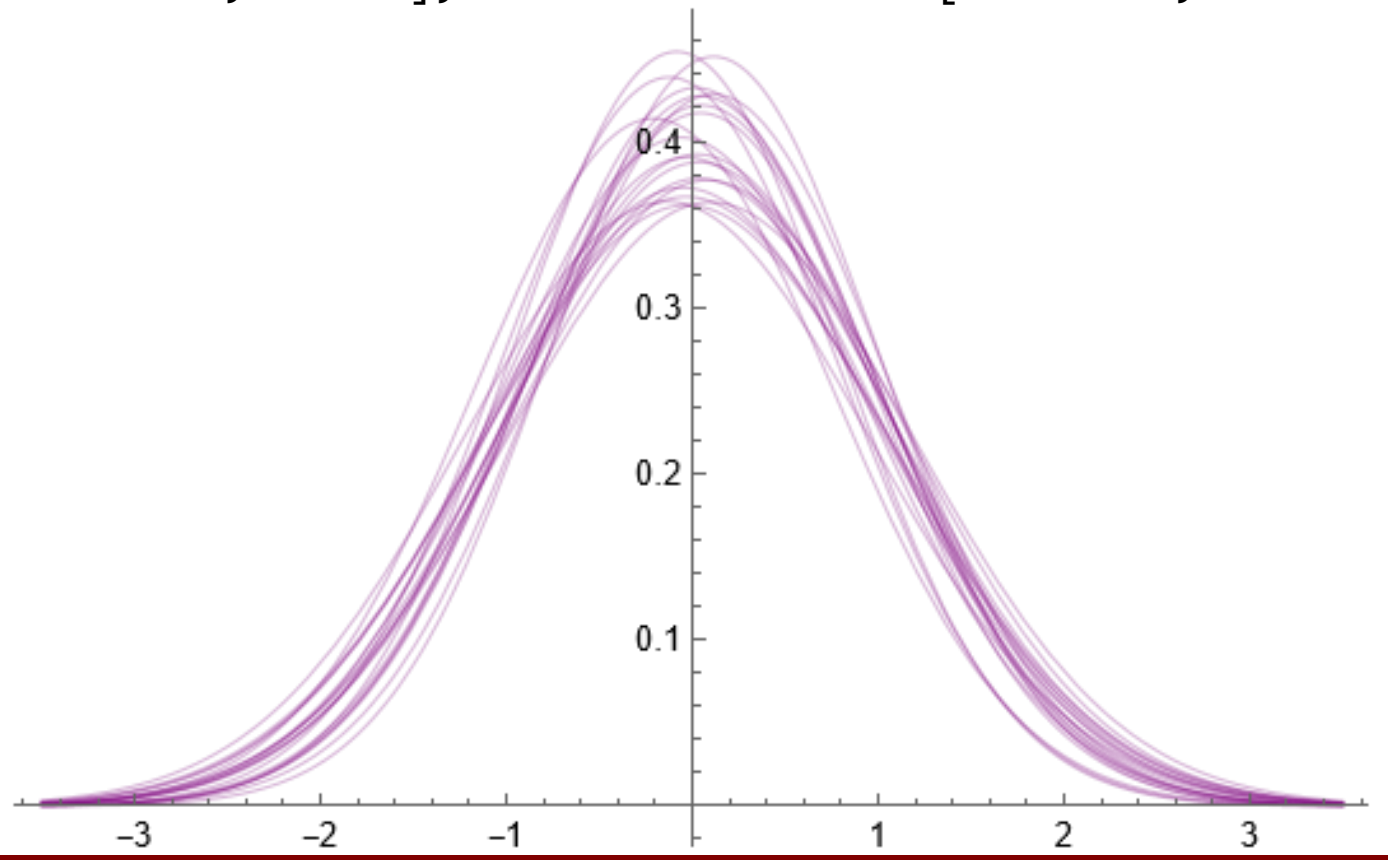

*Mathematica Examples 19.13*

```
Input   (* The code generates random data from a normal distribution, estimates the parameters of the distribution, and creates a plot of the estimated values. It iterates 100 times, generating 20 random data points each time. The FindDistributionParameters function is used to estimate the parameters of a normal distribution that best fit the generated data. The estimated values of the mean and standard deviation are stored in separate lists, and a line plot is created to visualize the changes in these estimated values over the iterations: *)

        params=Table[
            parent=NormalDistribution[2,1];
            data=RandomVariate[parent,20];
            FindDistributionParameters[
             data,
             NormalDistribution[μ,σ]
             ],
            {i,100}
            ];
```

```
eparams={μ,σ}/. params;
estimatedfirstparamμ=Table[eparams[[i]][[1]],{i,1,100}];
estimatedsecondparamσ=Table[eparams[[i]][[2]],{i,1,100}];

ListPlot[
 {estimatedfirstparamμ,estimatedsecondparamσ},
 Joined->True,
 PlotLegends->Placed[{"eμ","eσ"},{0.8,0.15}],
 ImageSize->300
 ]
```

Output

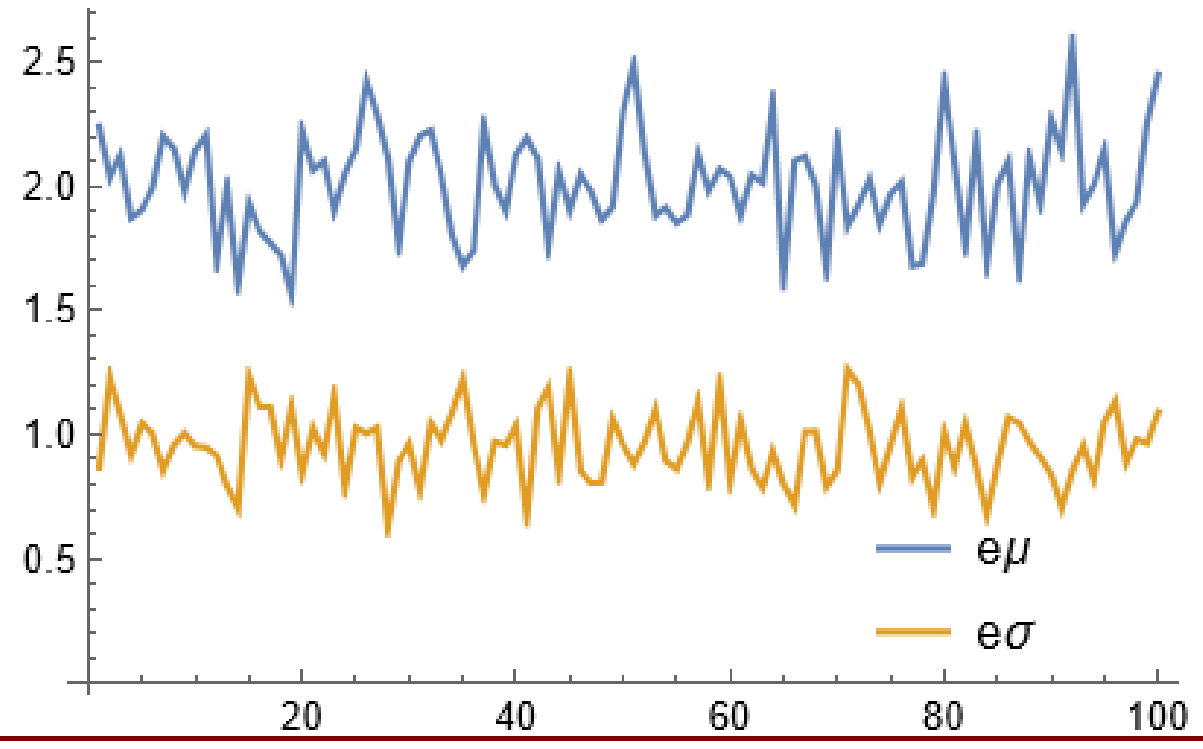

***Mathematica Examples 19.14***

Input

```
(* The code implements a Manipulate interface that enables interactive exploration
of different probability distributions and sample sizes. It generates random data
from the selected distribution and estimates the distribution based on the generated
data. The code then plots the probability density functions (PDFs) of both the true
distribution and the estimated distribution, allowing for a visual comparison. The
interface includes controls to adjust the distribution type and the sample size: *)

Manipulate[
 data=RandomVariate[dist[1,2],n];(* Generate random data from the distribution *)
 estDist=EstimatedDistribution[data,dist[u,v]];(* Estimate the distribution *)
 Plot[
  {PDF[dist[1,2],x],PDF[estDist,x]},
  {x,-10,10},
  PlotLegends->{"True Distribution","Estimated Distribution"},
  PlotRange->All,
  ImageSize->300
  ],

{{dist,NormalDistribution,"Distribution"},{NormalDistribution,GammaDistribution,Wei
bullDistribution}},
 {{n,10,"Sample Size"},5,100,1}
 ]
```

Output

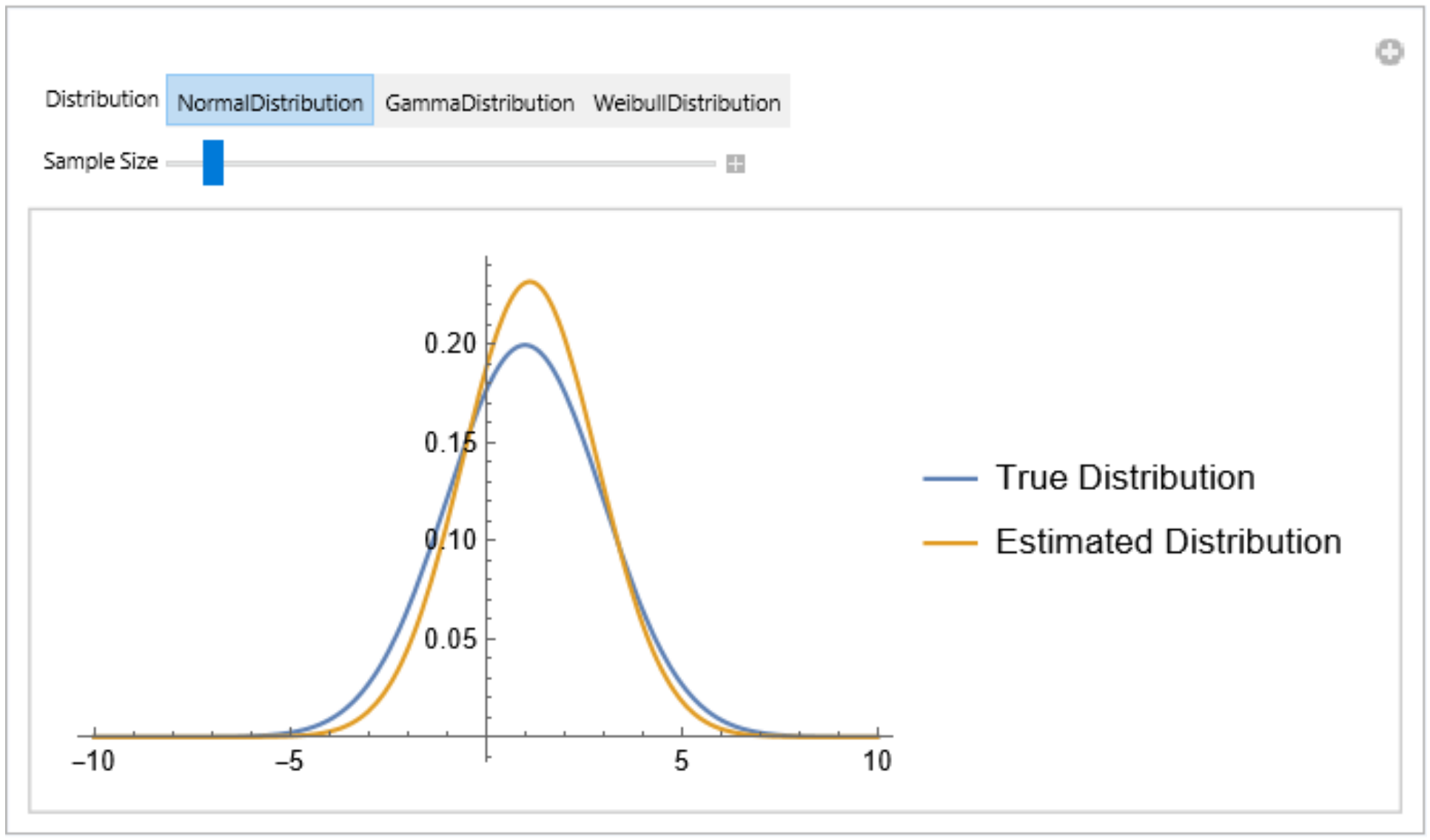

### *Mathematica Examples 19.15*

Input

```
(* The code creates a Manipulate interface that allows interactive exploration of different distributions and sample sizes. It generates random data based on the selected distribution and sample size and presents it as a histogram with overlaid PDF. The interface dynamically updates the visualization as you adjust the parameters. It provides a visual tool to understand how different distribution types and sample sizes affect the data distribution and estimated PDF: *)

Manipulate[
 Module[
  {data},
  data=RandomVariate[dist[1,2],n];
  Show[
   Histogram[
    data,
    Automatic,
    "PDF",
    PlotLabel->Row[FindDistributionParameters[data,dist[u,v]],"  "],
    ColorFunction->Function[{height},Opacity[height]],
    ImageSize->320,
    ChartStyle->Purple
    ],

   Plot[
    PDF[EstimatedDistribution[data,dist[u,v]],x],
    {x,Min[data],Max[data]},
    PlotStyle->Blue,
    ImageSize->300
    ]
   ]
  ],
 {{n,500,"Sample Size"},10,1000,10},
 {{dist,NormalDistribution,"Distribution"},{NormalDistribution,GammaDistribution}},
 Initialization:>(SeedRandom[123];)
 ]
```

Output

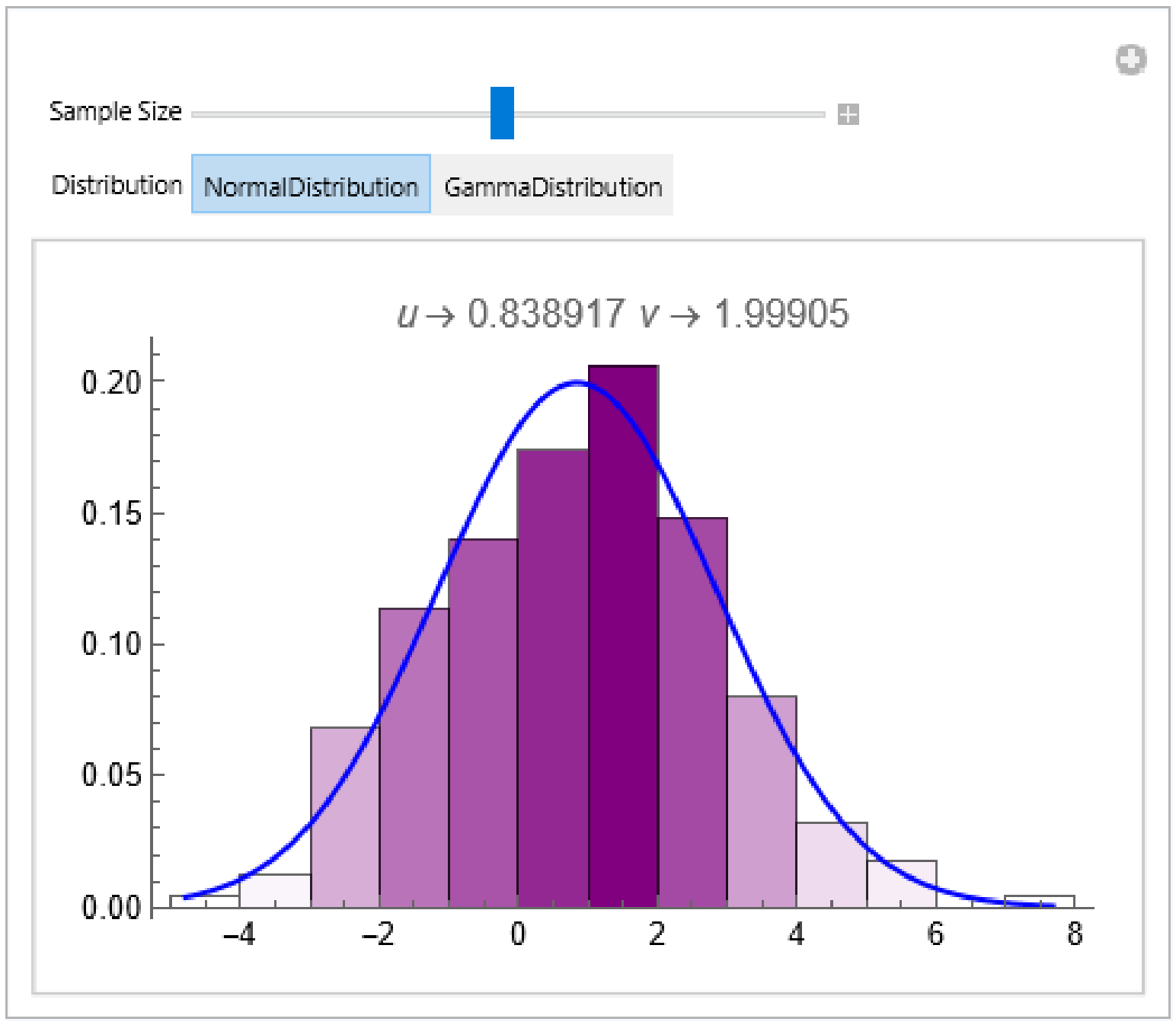

***Mathematica Examples 19.16***

Input

```
(* The code creates a Manipulate interface that allows interactive exploration of parameter estimation for a normal distribution. It generates random data from a normal distribution with adjustable parameters (mu and segma) and sample size (n). The code iterates multiple times to estimate the parameters of the normal distribution from the generated data using FindDistributionParameters. The estimated values of mu and segma are stored and plotted using ListPlot. The interface allows for dynamic updates of the plot as the parameter values and sample size are adjusted: *)

Manipulate[
 SeedRandom[123];
 params=Table[
    parent=NormalDistribution[mu,segma];
    data=RandomVariate[parent,n];
    FindDistributionParameters[
     data,
     NormalDistribution[μ,σ]
     ],
    {i,100}
    ];

 eparams={μ,σ}/. params;
 eμ=Table[eparams[[i]][[1]],{i,1,100}];
 eσ=Table[eparams[[i]][[2]],{i,1,100}];

 ListPlot[
  {eμ,eσ},
  Joined->True,
  PlotRange->All,
  PlotLegends->Placed[{"eμ","eσ"},{0.8,0.15}],
  ImageSize->300
  ],
 {{mu,1.5,"mu"},1,3,0.1},
 {{segma,0.7,"segma"},0.5,1,0.1},
 {{n,100,"Sample Size"},10,1000,10}
```

```
]
```

Output

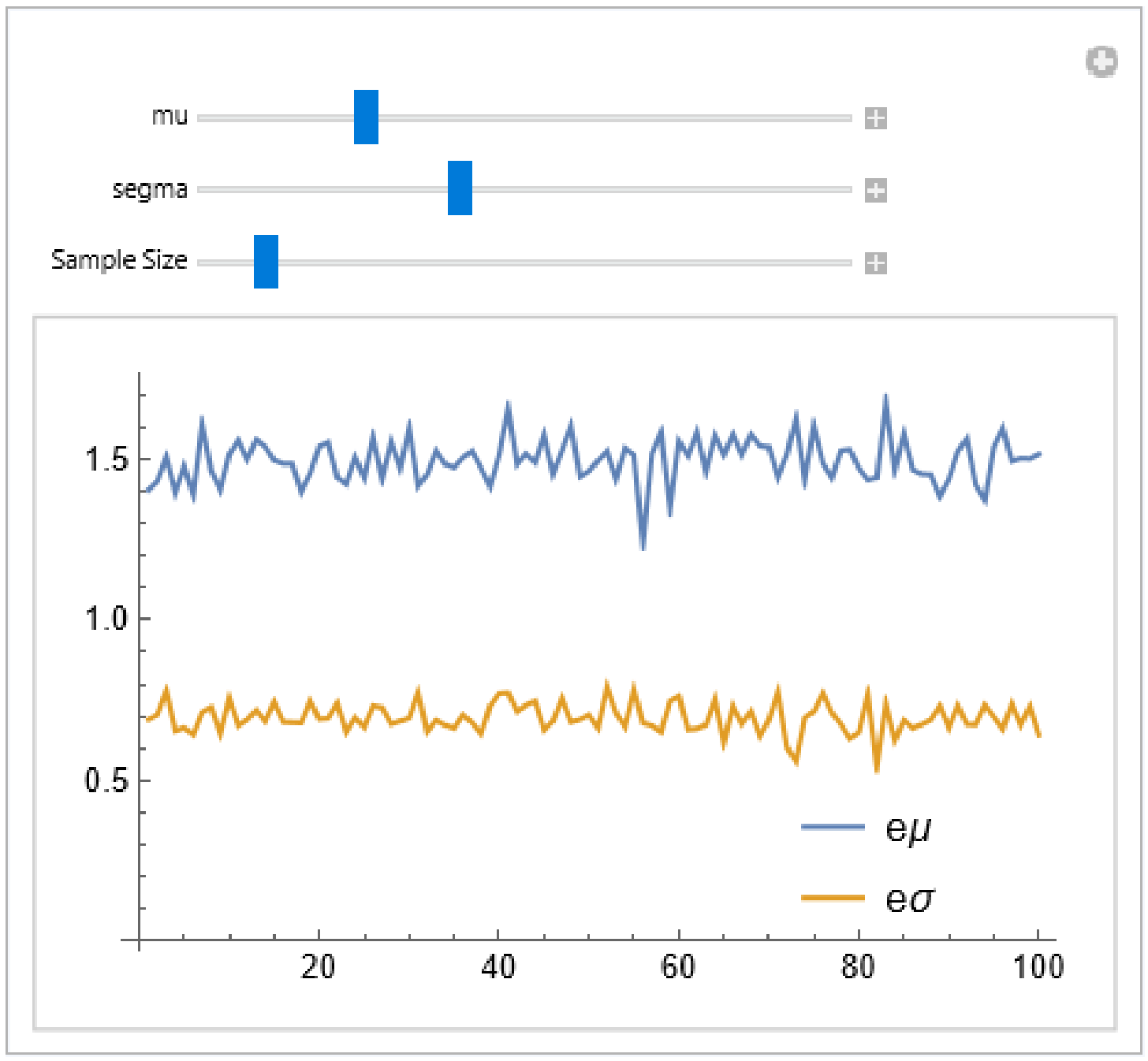

***Mathematica Examples 19.17***

Input

```
(* The code performs parameter estimation for a binormal distribution based on random data. It estimates the distribution parameters using both the EstimatedDistribution and FindDistributionParameters functions. The code then visualizes the difference between the original and estimated probability density functions (PDFs) using both a 3D plot (Plot3D) and a contour plot (ContourPlot). These visualizations allow for a comparison of the accuracy of the estimated distribution and provide insights into the differences between the original and estimated PDFs: *)

(* Estimate parameters for a multivariate distribution: *)
data=RandomVariate[BinormalDistribution[{4,5},{1,2},0.8],100];
edist=EstimatedDistribution[
   data,
   BinormalDistribution[{μ1,μ2},{σ1,σ2},ρ]
   ]

(* Estimate parameters for a Binormal distribution: *)
FindDistributionParameters[data,BinormalDistribution[{μ1,μ2},{σ1,σ2},ρ]]

(* Compare the difference between the original and estimated PDFs (Plot3D): *)
Plot3D[
 PDF[BinormalDistribution[{4,5},{1,2},0.8],{x,y}]-PDF[edist,{x,y}],
 {x,0,12},
 {y,0,12},
 PlotRange->All,
 ColorFunction->"Rainbow",
 ImageSize->270
 ]

(* Compare the difference between the original and estimated PDFs (ContourPlot): *)
ContourPlot[
 PDF[BinormalDistribution[{4,5},{1,2},0.8],{x,y}]-PDF[edist,{x,y}],
 {x,0,12},
```

```
          {y,0,12},
          ContourStyle->{White},
          ClippingStyle->Automatic,
          ColorFunction->"BlueGreenYellow",
          PlotLegends->Automatic,
          ImageSize->220
          ]

Output   BinormalDistribution[{4.04619,5.20397},{1.12139,1.96933},0.837671]
Output   {μ1->4.04619,μ2->5.20397,σ1->1.12139,σ2->1.96933,ρ->0.837671}
Output
```

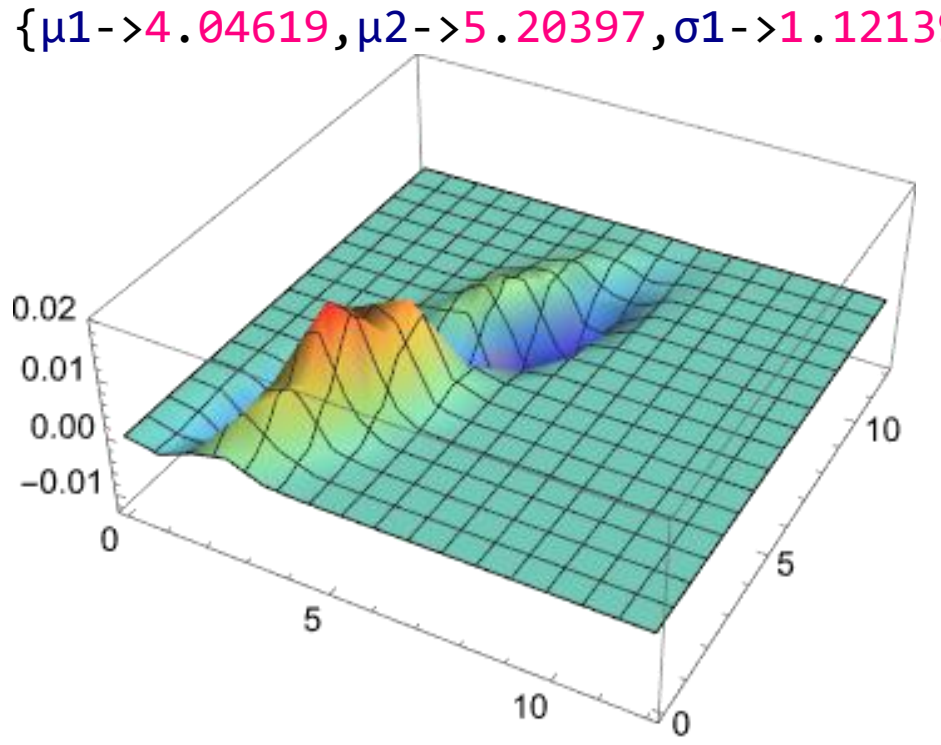

Output

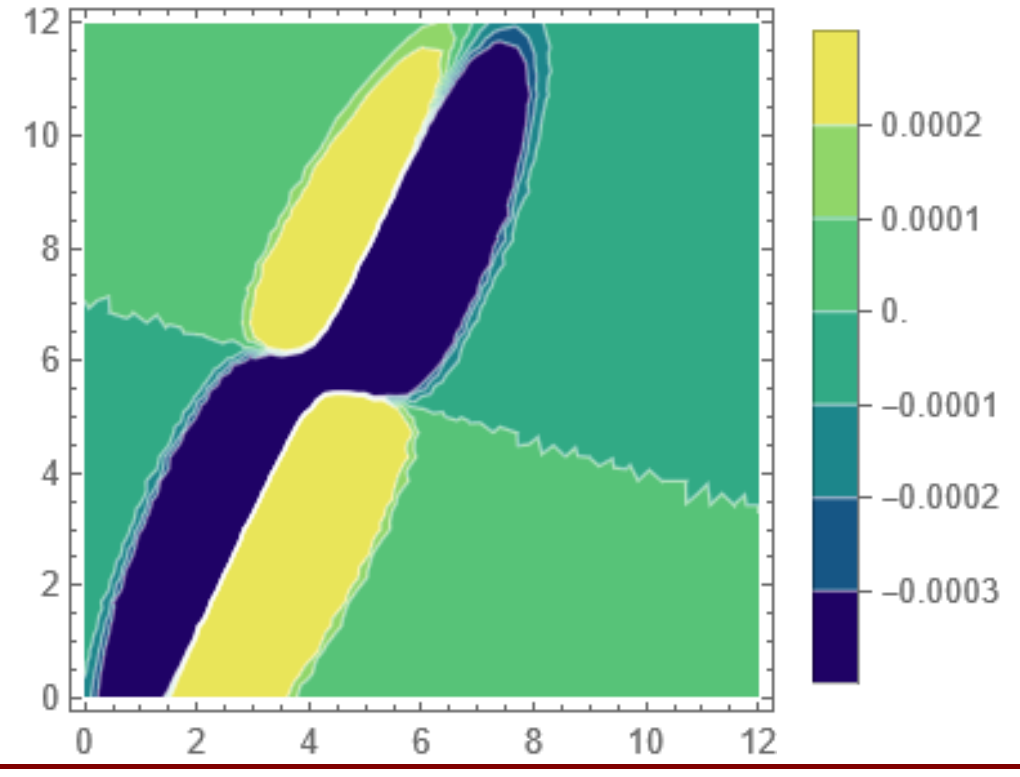

***Mathematica Examples 19.18***

```
Input    (* The code demonstrates the process of estimating parameters for a gamma distribution
         using a combination of visualizing the log-likelihood surface and conducting
         parameter estimation. It generates random data from a gamma distribution and uses
         rough parameter values obtained from the visualized log-likelihood surface as
         starting points for the estimation. The code visualizes the log-likelihood surface
         using a contour plot and overlays the estimated parameter values as a red point. This
         approach provides a useful way to explore the parameter space and refine the
         estimation process based on the observed log-likelihood values: *)

         (* Visualize a log-likelihood surface to find rough values for the parameters: *)
         data=RandomVariate[GammaDistribution[4,5],100];

         (* Supply those rough values as starting values for the estimation: *)
         EstimatedDistribution[
          data,
          GammaDistribution[α,β],
          GammaDistribution[4.2,4.2]
          ]

         (* Estimate parameters for a normal distribution: *)
         params=FindDistributionParameters[
           data,
```

```
           GammaDistribution[α,β],
           GammaDistribution[4.2,4.2]
           ]

         cp=ContourPlot[
            LogLikelihood[
             GammaDistribution[α,β],
             data
             ],
            {α,1,7},
            {β,1,10},
            ContourStyle->{White},
            ClippingStyle->Automatic,
            ColorFunction->"BlueGreenYellow",
            PlotLegends->Automatic,
            ImageSize->220
            ];

         Show[
          cp,
          Graphics[{PointSize[Large],Red,Point[{α,β}/. params]}]
          ]

Output   GammaDistribution[3.93712,5.3958]
Output   {α->3.93712,β->5.3958}
Output
```

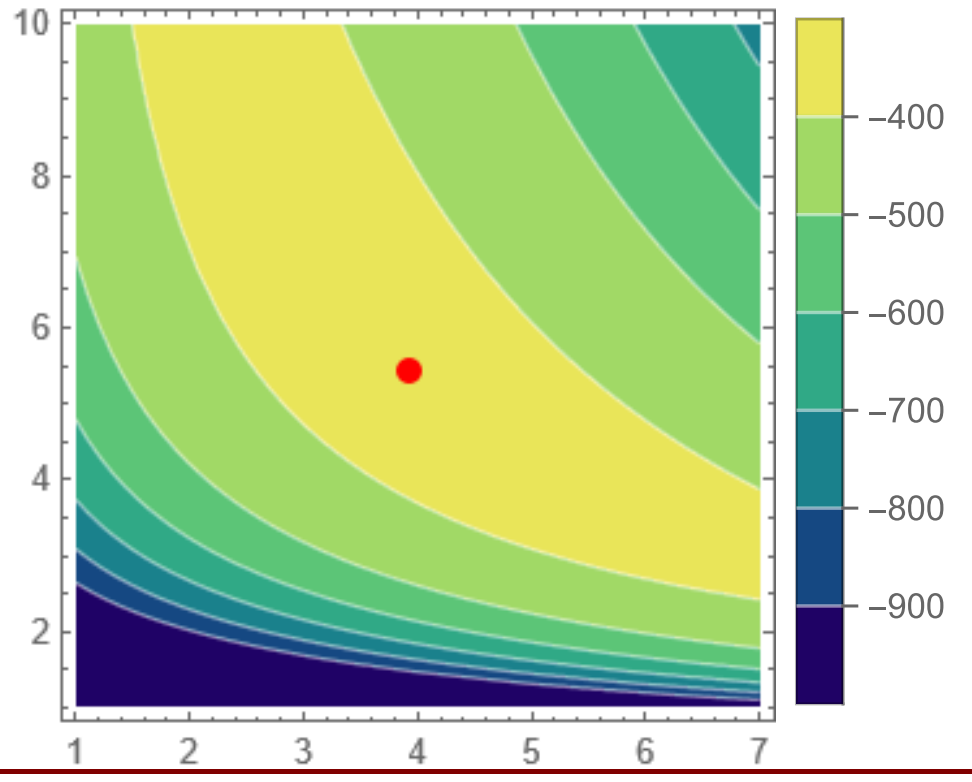

***Mathematica Examples 19.19***

```
Input    (* The code focuses on estimating the parameters for a normal distribution that
         approximates Poisson-distributed data. It generates random data from a Poisson
         distribution and then estimates the parameters of a normal distribution using both
         the EstimatedDistribution and FindDistributionParameters functions: *)

         (* Estimate the normal approximation of Poisson data: *)
         data=RandomVariate[
            PoissonDistribution[20],
            100
            ];
         EstimatedDistribution[
          data,
          NormalDistribution[μ,σ]
          ]

         (* Estimate parameters for a normal distribution: *)
         FindDistributionParameters[
          data,
          NormalDistribution[μ,σ]
```

```
          ]

Output   NormalDistribution[20.14,4.49893]
Output   {μ->20.14,σ->4.49893}
```

| | |
|---|---|
| `FindDistribution[data]` | finds a simple functional form to fit the distribution of data. |
| `FindDistribution[data,n]` | finds up to n best distributions. |
| `FindDistribution[data,n,prop]` | returns up to n best distributions associated with property prop. |
| `FindDistribution[data,n,{prop1, prop2,…}]` | returns up to n best distributions associated with properties prop1, prop2, etc. |

Possible settings for `TargetFunctions` include:

| | |
|---|---|
| `Automatic` | automatically chosen distributions |
| `All` | all built-in distributions |
| `"Continuous"` | all continuous distributions |
| `"Discrete"` | all discrete distributions |
| `{dist1,dist_(2),...}` | distributions disti |
| `{{w_(1),w_(2),...}→{dist1,dist_(2),...}}` | distributions disti using weights wi |

- Possible continuous distributions for `TargetFunctions` are: `BetaDistribution, CauchyDistribution, ChiDistribution, ChiSquareDistribution, ExponentialDistribution, ExtremeValueDistribution, FrechetDistribution, GammaDistribution, GumbelDistribution, HalfNormalDistribution, InverseGaussianDistribution, LaplaceDistribution, LevyDistribution, LogisticDistribution, LogNormalDistribution, MaxwellDistribution, NormalDistribution, ParetoDistribution, RayleighDistribution, StudentTDistribution, UniformDistribution, WeibullDistribution, HistogramDistribution.`

- Possible discrete distributions for `TargetFunctions` are: `BenfordDistribution, BinomialDistribution, BorelTannerDistribution, DiscreteUniformDistribution, GeometricDistribution, LogSeriesDistribution, NegativeBinomialDistribution, PascalDistribution, PoissonDistribution, WaringYuleDistribution, ZipfDistribution, HistogramDistribution, EmpiricalDistribution.`

***Mathematica Examples 19.20***

```
Input    (* The code generates a list of uniformly distributed random integers, finds the
         best-fit distribution for the data, and visualizes the data through a list
         plot,histogram, and PDF plot:*)

         (* Create a list of uniformly distributed random integers: *)
         data=RandomInteger[10,100]

         (* Find the underlying distribution from the data: *)
         bestdist=FindDistribution[data]

         (* Plot the data as a list plot: *)
         ListPlot[
          data,
          ImageSize->250,
          PlotStyle->Directive[Purple,Opacity[0.5],PointSize[0.01]]
          ]

         (* Display the data as a histogram with PDF representation: *)
         Show[
          Histogram[
```

```
data,
Automatic,
"PDF",
ColorFunction->Function[{height},Opacity[height]],
ChartStyle->Purple,
ImageSize->250
],
Plot[
PDF[bestdist[[1]],x],
{x,0,10},
PlotRange->All,
PlotStyle->{Directive[Thickness[0.005],Opacity[0.9]]},
ImageSize->250,
ColorFunction->"Rainbow"
]
]
```

Output {4,2,5,3,3,7,10,8,1,5,0,8,10,1,0,7,2,10,3,7,8,6,2,10,0,4,0,0,5,6,9,5,0,4,3,7,10,5,9,3,4,0,10,5,6,10,3,7,5,7,5,3,2,4,10,8,0,6,5,3,5,1,6,0,9,1,5,4,8,7,4,1,8,3,0,8,5,5,9,4,10,3,6,4,6,2,0,2,10,8,7,2,1,10,9,5,0,1,3,5}

Output DiscreteUniformDistribution[{0,10}]

Output

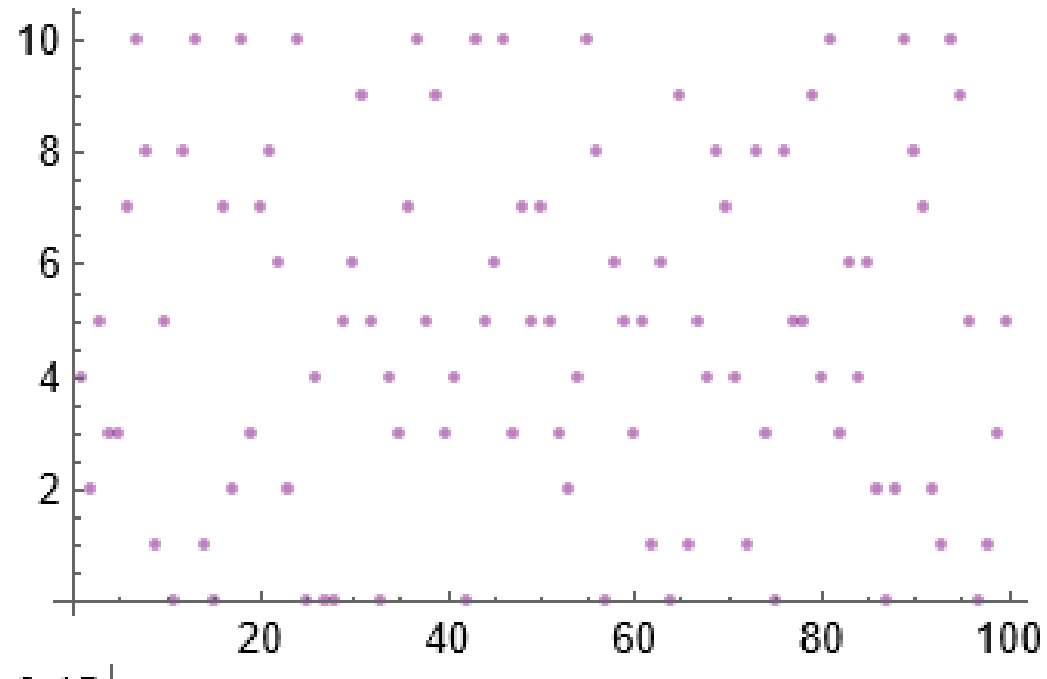

Output

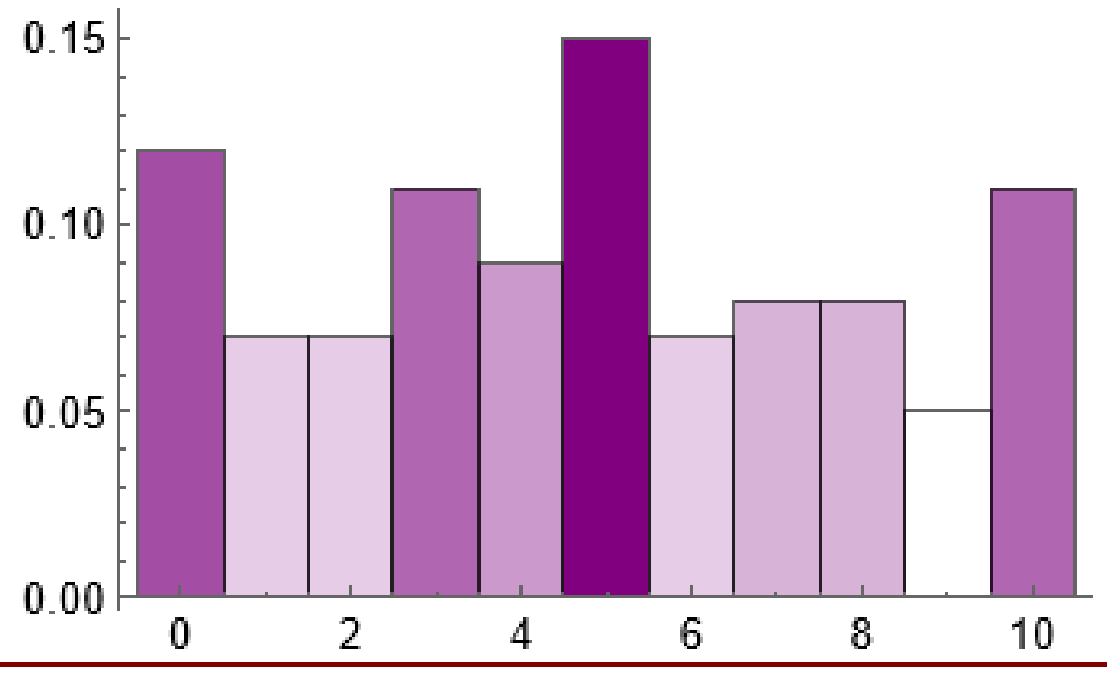

***Mathematica Examples 19.21***

Input

```
(* The code generates a random sample of size 1000 from an Exponential distribution. The FindDistribution function is used to identify the best three distributions that fit the data. The data is visualized through a list plot, displaying the individual data points. Additionally, a histogram is created, representing the data in PDF format. The PDFs of the best three distributions are plotted on the same graph, each with a different color and opacity. The resulting visualization combines the histogram and the PDF plots: *)

(* Set the seed for reproducible random numbers: *)
SeedRandom[8123];

(* Create a random sample from an Exponential distribution: *)
```

```
dist=ExponentialDistribution[1];
data=RandomVariate[dist,1000];

(* Find the best three distributions that fit the data: *)
bestdist=FindDistribution[data,3]

(* Plot the data as a list plot: *)
ListPlot[
 data,
 ImageSize->250,
 PlotStyle->Directive[Purple,Opacity[0.5],PointSize[0.01]]
 ]

(* Display the data as a histogram with PDF representation: *)
Show[
 Histogram[
  data,
  {0.1},
  "PDF",
  ColorFunction->Function[{height},Opacity[height]],
  ChartStyle->Purple,
  ImageSize->250
  ],
 Plot[
  {PDF[bestdist[[1]],x],PDF[bestdist[[2]],x],PDF[bestdist[[3]],x]},
  {x,0,10},
  PlotRange->All,
  PlotStyle->{Directive[Thickness[0.005],Opacity[0.9]]},
  ImageSize->250,
  ColorFunction->"Rainbow"
  ]
 ]
```

Output {ExponentialDistribution[1.04196],GammaDistribution[1.05726,0.891243],WeibullDistribution[1.0445,0.958565]}

Output

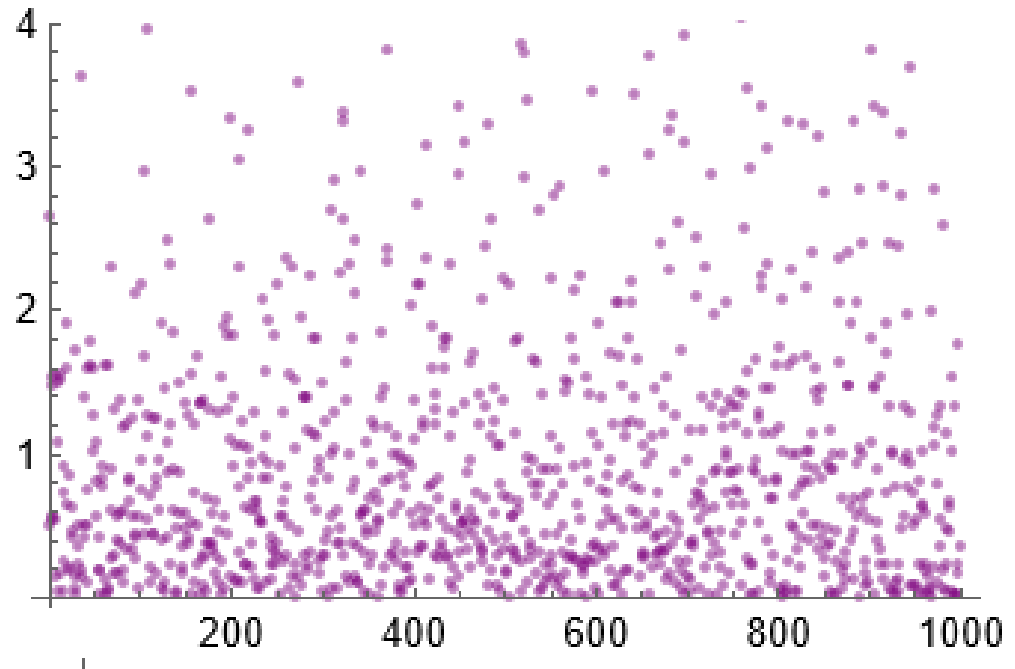

Output

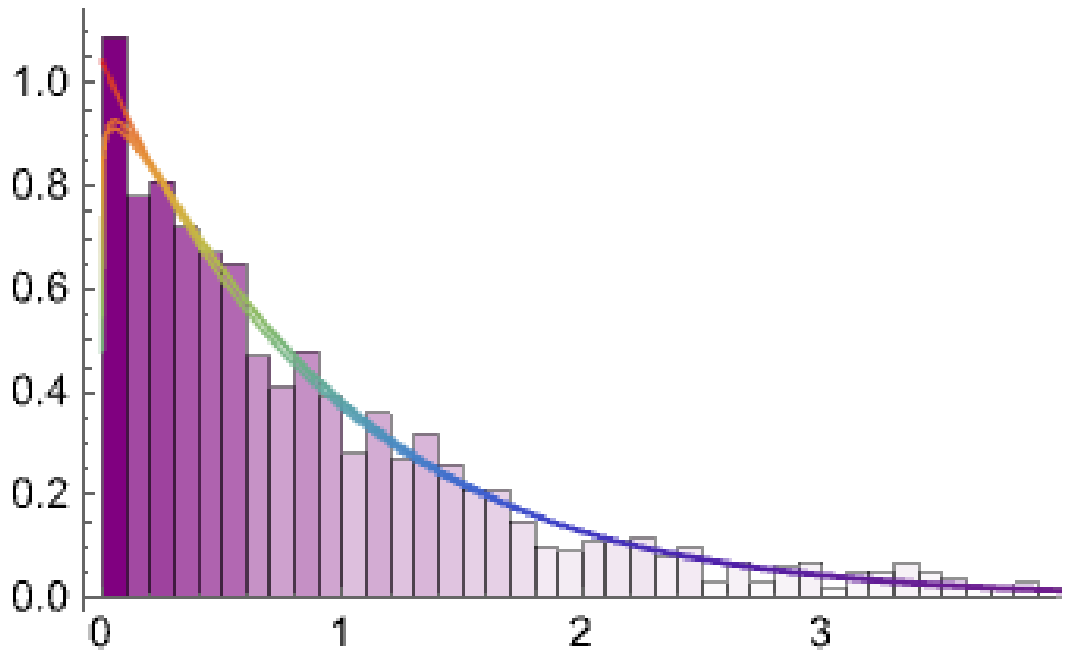

***Mathematica Examples 19.22***

Input
```
(* The code generates a random sample from a mixture distribution consisting of a
Gamma and Normal distribution. It then estimates the best distribution that fits the
generated data. The data is visualized through a list plot, and a comparison is made
between the PDFs of the original and estimated distributions using a histogram and
line plots: *)

(* Set the seed for reproducible random numbers: *)
SeedRandom[7123];
(* Define a mixture distribution with a Gamma and Normal distribution: *)
dist=MixtureDistribution[
   {1,0.8},
   {
    GammaDistribution[3,1],
    NormalDistribution[5,0.8]
    }
   ];

(* Generate a random sample from the mixture distribution: *)
data=RandomVariate[dist,1000];

(* Estimate the best distribution from the generated data: *)
ed=FindDistribution[data]

(* Plot the data as a list plot: *)
ListPlot[
 data,
 ImageSize->250,
 PlotStyle->Directive[Purple,Opacity[0.5],PointSize[0.01]]
 ]

(* Compare the PDFs of the original and estimated distributions: *)
Show[
 Histogram[
  data,
  Automatic,
  "PDF",
  ColorFunction->Function[{height},Opacity[height]],
  ChartStyle->Purple,
  ImageSize->250
  ],
 Plot[
  {PDF[d,x],PDF[ed,x]},
  {x,0,10},
  PlotRange->All,
  PlotStyle->{Directive[Thickness[0.005],Opacity[0.9]]},
  ImageSize->250,
  PlotLegends->Placed[{"d","ed"},{0.8,0.7}]
  ]
 ]
```

Output
```
MixtureDistribution[{0.285018,0.714982},{NormalDistribution[1.71568,0.745499],Norma
lDistribution[4.76374,1.19483]}]
```

Output

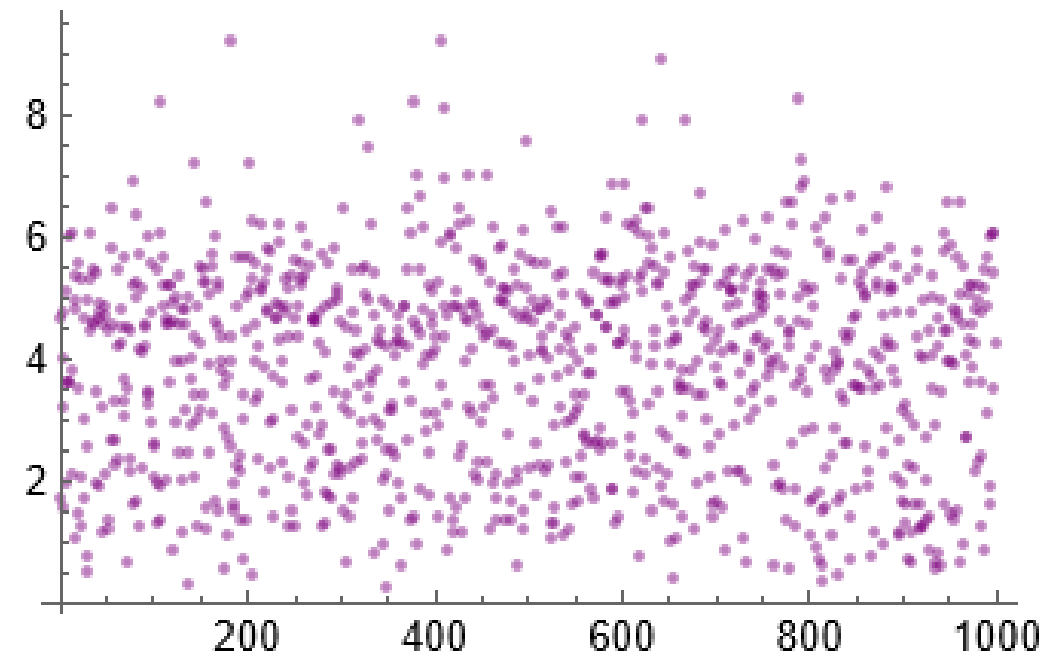

Output

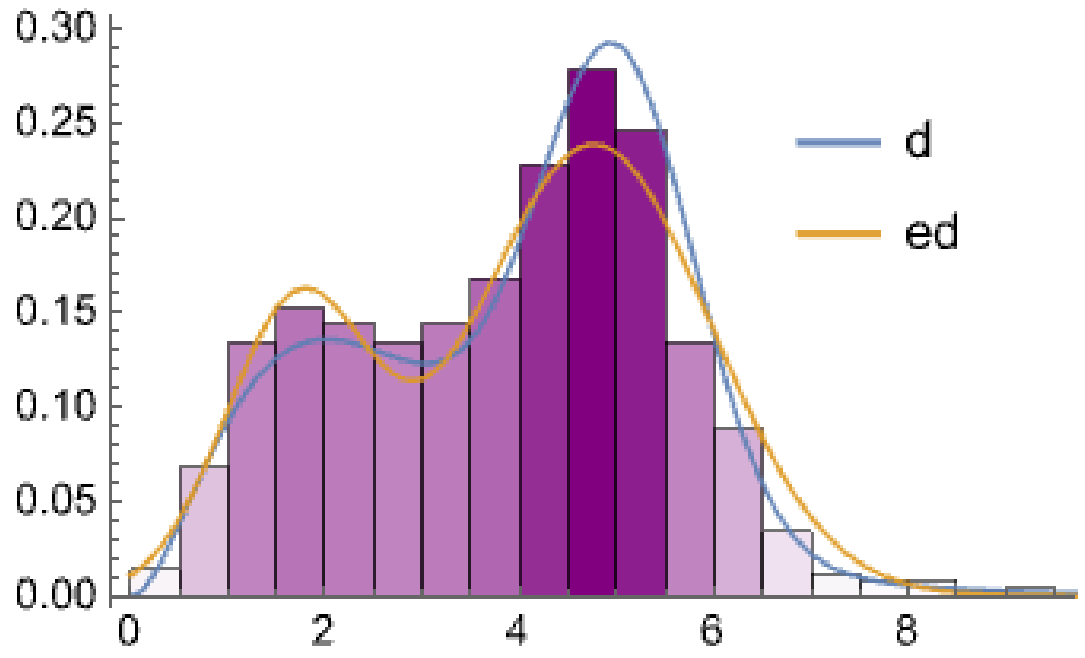

# UNIT 19.2

# INTERVAL ESTIMATE

### Hypothesis Testing Package

| | |
|---|---|
| `MeanCI[list]` | gives a confidence interval for the population mean estimated from list. |
| `MeanDifferenceCI[list1,list2]` | gives a confidence interval for the difference between the population means estimated from list1 and list2. |
| `VarianceCI[list]` | gives a confidence interval for the population variance estimated from list. |
| `VarianceRatioCI[list1,list2]` | gives a confidence interval for the ratio of the population variances estimated from list1 and from list2. |
| `NormalCI[μ,σ]` | gives a confidence interval based on a normal distribution. |
| `StudentTCI[μ,σ,df]` | gives a confidence interval based on Student's t distribution with df degrees of freedom. |
| `ChiSquareCI[var,df]` | gives a confidence interval based on a $\chi^2$ distribution with df degrees of freedom. |
| `FRatioCI[ratio,n,m]` | gives a confidence interval based on an F-ratio distribution with n and m degrees of freedom. |

### Remarks:

- Assumptions about variances of the populations from which the data were sampled will affect the distribution of the parameter estimate. The `KnownVariance` and `EqualVariances` options can be used to specify assumptions about population variances.
- Confidence intervals for the mean and for the difference between means are based on a normal distribution if the population variances are assumed known.
- Intervals for the mean are based on Student's t distribution with n-1 degrees of freedom when the population variance must be estimated from a list of n elements.
- Confidence intervals for the difference between means are also based on Student's t distribution if the variances are not known. If the variances are assumed equal, `MeanDifferenceCI` is based on Student's t distribution with `Length[list1]+Length[list2]-2` degrees of freedom. If the population variances are not assumed equal, Welch's approximation for the degrees of freedom is used.
- The variance confidence interval is based on a $\chi^2$ distribution and the variance ratio confidence interval is based on an F-ratio distribution.
- Confidence intervals can also be obtained for normal, chi-square, Student's t, or F-ratio distributions.

***Mathematica Examples 19.23***

Input
```
(* In this example, we first load the HypothesisTesting package using
Needs["HypothesisTesting"]. Then, we generate a sample dataset data` consisting of
100 random values drawn from a normal distribution with a mean of 10 and a standard
deviation of 2. Next, we specify the desired confidence level as confidenceLevel=0.95,
which corresponds to a 95% confidence interval. We pass the dataset data and the
```

```
        confidence level to the MeanCI function, which returns a list representing the
        confidence interval. Finally, we extract the lower and upper bounds of the confidence
        interval from the list and display them using Print: *)

        Needs["HypothesisTesting`"]

        (* Generate a sample dataset: *)
        data=RandomVariate[NormalDistribution[10,2],100];

        (* Compute a 95% confidence interval for the mean: *)
        ci=MeanCI[
            data,
            ConfidenceLevel->.95
            ];

        (* Display the confidence interval: *)
        lowerBound=ci[[1]];
        upperBound=ci[[2]];
        Print["Confidence Interval: ",{lowerBound,upperBound}]

Output  Confidence Interval:  {9.46578,10.2623}
```

***Mathematica Examples 19.24***

```
Input   (* A 95% confidence interval assuming the population variance is 1: *)
        MeanCI[{1,2,4,6,3},KnownVariance->1]

Output  {2.32348,4.07652}
```

***Mathematica Examples 19.25***

```
Input   (* The code generates a sample dataset, compute the 95% confidence interval for the
        difference in means between two populations, and then display the confidence interval.
        The code starts by generating two sample datasets, data1 and data2, from normal
        distributions with specific mean and standard deviation parameters. Then, it
        calculates the 95% confidence interval for the difference in means between the two
        populations using the MeanDifferenceCI function from the HypothesisTesting package.
        The ConfidenceLevel option is set to 0.95 to specify the desired confidence level.
        Finally, the code prints the confidence interval by displaying the lower and upper
        bounds: *)

        Needs["HypothesisTesting`"]

        (* Generate a sample dataset: *)
        data1=RandomVariate[NormalDistribution[5,1],100]; (* Sample data for group 1 *)
        data2=RandomVariate[NormalDistribution[3,2],50];(* Sample data for group 2 *)

        (* The 95% confidence interval for the difference in two population means: *)
        ci=MeanDifferenceCI[
            data1,
            data2,
            ConfidenceLevel->.95
            ];

        (* Display the confidence interval: *)
        lowerBound=ci[[1]];
        upperBound=ci[[2]];
        Print["Confidence Interval: ",{lowerBound,upperBound}]

Output  Confidence Interval:  {0.506878,1.7266}
```

**Mathematica Examples 19.26**

Input (* In this example, we first load the HypothesisTesting package using Needs["HypothesisTesting"]. Then we generate two sets of sample data stored in data1anddata2`. These data represent samples from two populations with known variances of 4 and 9,respectively. Next, we use the MeanDifferenceCI function to calculate the confidence interval for the difference between the means of the two populations. The KnownVariance option is specified to indicate that the variances of both populations are known. We set the confidenceLevel variable to 0.95 to calculate a 95% confidence interval: *)

```
(* Load the HypothesisTesting package: *)
Needs["HypothesisTesting`"]

(* Generate sample data: *)
data1=RandomVariate[NormalDistribution[10,2],100];
data2=RandomVariate[NormalDistribution[12,3],120];

(* Calculate the confidence interval: *)
ci=MeanDifferenceCI[
    data1,
    data2,
    ConfidenceLevel->.95,
    KnownVariance->{4,9}
    ];

(* Display the confidence interval: *)
lowerBound=ci[[1]];
upperBound=ci[[2]];
Print["Confidence Interval: ",{lowerBound,upperBound}]
```

Output Confidence Interval: {-2.41583,-1.08652}

**Mathematica Examples 19.27**

Input

```
(* A confidence interval assuming equal but unknown variances: *)
MeanDifferenceCI[{1,2,4,6,3},{4,10,6,8,5,8},EqualVariances->True]
```

Output {-6.50787,-0.758799}

**Mathematica Examples 19.28**

Input (* In this example, we generate a sample dataset with 100 data points from a standard normal distribution. We then specify a confidence level of 0.99 (corresponding to a 99% confidence interval). The VarianceCI function computes the confidence interval for the variance of the dataset, and the result is stored in the variable ci: *)

```
(* Load the HypothesisTesting` package: *)
Needs["HypothesisTesting`"]

(* Generate a sample dataset*)
data=RandomVariate[NormalDistribution[0,1],100];

(* Compute the confidence interval for the variance: *)
ci=VarianceCI[data,ConfidenceLevel->.99];

(* Display the confidence interval: *)
lowerBound=ci[[1]];
upperBound=ci[[2]];
Print["Confidence Interval: ",{lowerBound,upperBound}]
```

Output Confidence Interval: {0.687791,1.43728}

***Mathematica Examples 19.29***

Input

```
(* The code will compute the 90% confidence interval for the ratio of two population variances using the VarianceRatioCI function in Mathematica's HypothesisTesting` package. The code begins by loading the HypothesisTesting package using the Needs function. Then, two sample datasets, data1 and data2, are generated. The VarianceRatioCI function is used to compute the confidence interval for the ratio of variances, with a specified confidence level of 90% (ConfidenceLevel->0.90). Next, the lower and upper bounds of the confidence interval are extracted from the result, stored in ci. Finally, the Print statement displays the confidence interval to the user, showing both the lower and upper bounds: *)

(* Load the HypothesisTesting` package: *)
Needs["HypothesisTesting`"]

(* Generate two sample datasets: *)
data1=RandomVariate[NormalDistribution[0,1],50];
data2=RandomVariate[NormalDistribution[0,2],30];

(* The 90% confidence interval for the ratio of two population variances: *)
ci=VarianceRatioCI[data1,data2, ConfidenceLevel->0.90];

(* Display the confidence interval: *)
lowerBound=ci[[1]];
upperBound=ci[[2]];
Print["Confidence Interval: ",{lowerBound,upperBound}]
```

Output

```
Confidence Interval: {0.26373,0.796351}
```

***Mathematica Examples 19.30***

Input

```
(* A 90% confidence interval based on the normal distribution: *)
NormalCI[3,1,ConfidenceLevel->.90]
```

Output

```
{1.35515,4.64485}
```

***Mathematica Examples 19.31***

Input

```
(* A 90% confidence interval based on Student's distribution with 10 degrees of freedom: *)
StudentTCI[3,1,10,ConfidenceLevel->.90]
```

Output

```
{1.18754,4.81246}
```

***Mathematica Examples 19.32***

Input

```
(* A 90% confidence interval based on a ChiSquare distribution: *)
ChiSquareCI[4,6,ConfidenceLevel->.90]
```

Output

```
{1.90603,14.6755}
```

***Mathematica Examples 19.33***

Input

```
(* A 90% confidence interval based on an F-ratio distribution: *)
FRatioCI[2,5,20,ConfidenceLevel->.90]
```

Output

```
{0.737765,9.11626}
```

# UNIT 19.3

# INTERVAL ESTIMATE SIMULATION

*Mathematica Examples 19.34*

Input

```
(* The code calculates a 95% confidence interval for the population mean based on a random sample of 20 observations with a known population variance =9. It follows a logical sequence of steps, including data collection, calculation of sample mean and standard deviation, determination of the level of significance, and the computation of critical values and the confidence interval. The code also includes additional features such as using a built-in function for calculating the confidence interval and visualizing the interval with a probability density function plot: *)

(* Step 1. Collect your data: *)
data={7,5,11,9,10,7,6,8,9,10,11,6,8,9,10,11,12,8,9,10};

(* Step 2. Calculate the sample mean: *)
sampleMean=N[Mean[data]]
sampleStdDev=N[StandardDeviation[data]]

(* Step 3. Determine the level of significance: *)
alpha=0.05;

(* Step 4. Calculate the critical value: *)
criticalValue=Quantile[NormalDistribution[0,1],1-alpha/2]

(* Step 5. Calculate the confidence interval for the mean: *)
populationVariance=9;
n=Length[data];
standardError=Sqrt[populationVariance/n];
confidenceInterval={sampleMean-
criticalValue*standardError,sampleMean+criticalValue*standardError};

(* Step 6. Print the confidence interval: *)
confidenceInterval

(* Step 7. Using built-in function: *)
<<HypothesisTesting`;
MeanCI[data,ConfidenceLevel->0.95,KnownVariance->9]

(* Step 8. Visualize the confidence interval: *)
dist=NormalDistribution[sampleMean,sampleStdDev/Sqrt[n]];

Plot[
 PDF[dist,x],
 {x,sampleMean-4 sampleStdDev/Sqrt[n],sampleMean+4 sampleStdDev/Sqrt[n]},
 PlotRange->All,
 Filling->Axis,
 PlotStyle->Purple,
 AxesLabel->{"x","PDF"},
 ImageSize->250,
 Epilog->{
   Darker[Blue],
   PointSize[Large],
```

```
            Point[{sampleMean,PDF[dist,confidenceInterval[[1]]]}],
            {Arrowheads[{-0.035,0.035}],
             Arrow[
              {
               {confidenceInterval[[1]],PDF[dist,confidenceInterval[[1]]]},
               {confidenceInterval[[2]],PDF[dist,confidenceInterval[[2]]]}
               }
              ]
             }
            }
          ]

Output   8.8
Output   1.90843
Output   1.95996
Output   {7.48522,10.1148}
Output   {7.48522,10.1148}
Output
```

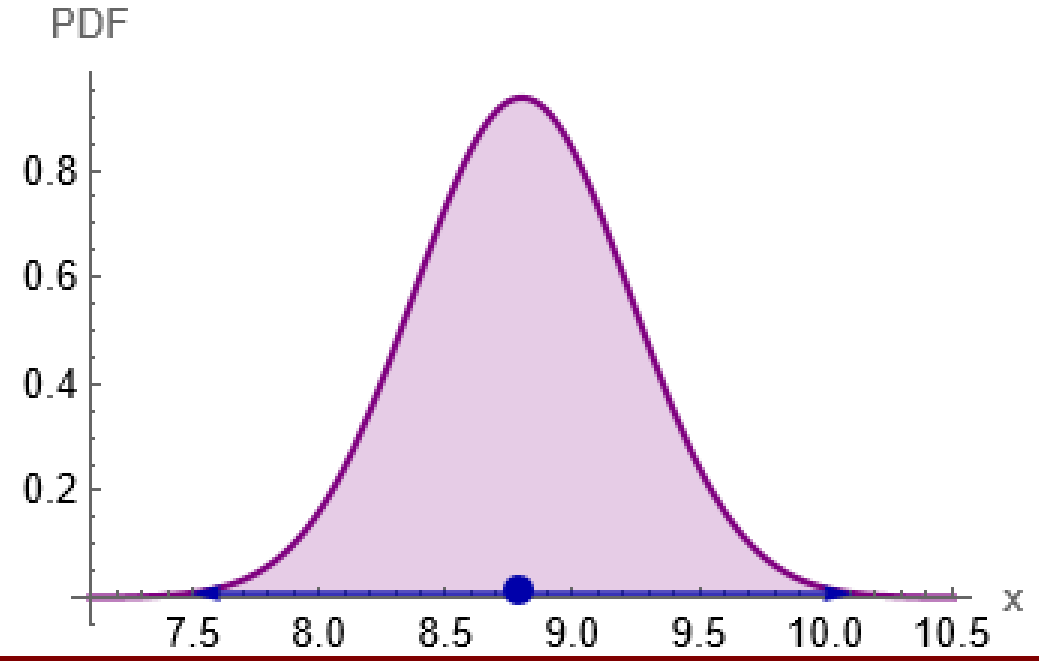

***Mathematica Examples 19.35***

```
Input    (* Similar to the above code but in the case of unknown Variance. The code calculates
         a 95% confidence interval for the population mean based on a random sample of 20
         observations with an unknown population variance. It follows a logical sequence of
         steps, including data collection, calculation of sample mean and standard deviation,
         determination of the level of significance, and the computation of critical values
         and the confidence interval. The code also includes additional features such as using
         a built-in function for calculating the confidence interval and visualizing the
         interval with a probability density function plot: *)

         (* Step 1. Collect your data: *)
         data={7,5,11,9,10,7,6,8,9,10,11,6,8,9,10,11,12,8,9,10};

         (* Step 2. Calculate the sample mean and standard deviation: *)
         sampleMean=N[Mean[data]]
         sampleStdDev=N[StandardDeviation[data]]

         (* Step 3. Determine the level of significance: *)
         alpha=0.05;

         (* Step 4. Calculate the critical value: *)
         degreesOfFreedom=Length[data]-1;
         criticalValue=Quantile[StudentTDistribution[degreesOfFreedom],1-alpha/2]

         (* Step 5. Calculate the confidence interval for the mean: *)
         standardError=sampleStdDev/Sqrt[Length[data]];
         confidenceInterval={sampleMean-
         criticalValue*standardError,sampleMean+criticalValue*standardError};

         (* Step 6. Print the confidence interval: *)
```

```
        confidenceInterval

        (* Step 7. Using built-in function: *)
        <<HypothesisTesting`;
        MeanCI[data,ConfidenceLevel->0.95]

        (* Step 8. Visualize the confidence interval: *)
        dist=NormalDistribution[sampleMean,sampleStdDev/Sqrt[Length[data]]];
        Plot[
         PDF[dist,x],
         {x,sampleMean-4 sampleStdDev/Sqrt[Length[data]],sampleMean+4
        sampleStdDev/Sqrt[Length[data]]},
         PlotRange->All,
         PlotStyle->Purple,
         Filling->Axis,
         AxesLabel->{"x","PDF"},
         ImageSize->250,
         Epilog->{
           Darker[Blue],
           PointSize[Large],
           Point[{sampleMean,PDF[dist,confidenceInterval[[1]]]}],
           {
            Arrowheads[{-0.035,0.035}],
            Arrow[
             {
              {confidenceInterval[[1]],PDF[dist,confidenceInterval[[1]]]},
              {confidenceInterval[[2]],PDF[dist,confidenceInterval[[2]]]}
              }
             ]
            }
           }
         ]

Output  8.8
Output  1.90843
Output  2.09302
Output  {7.90683,9.69317}
Output  {7.90683,9.69317}
Output
```

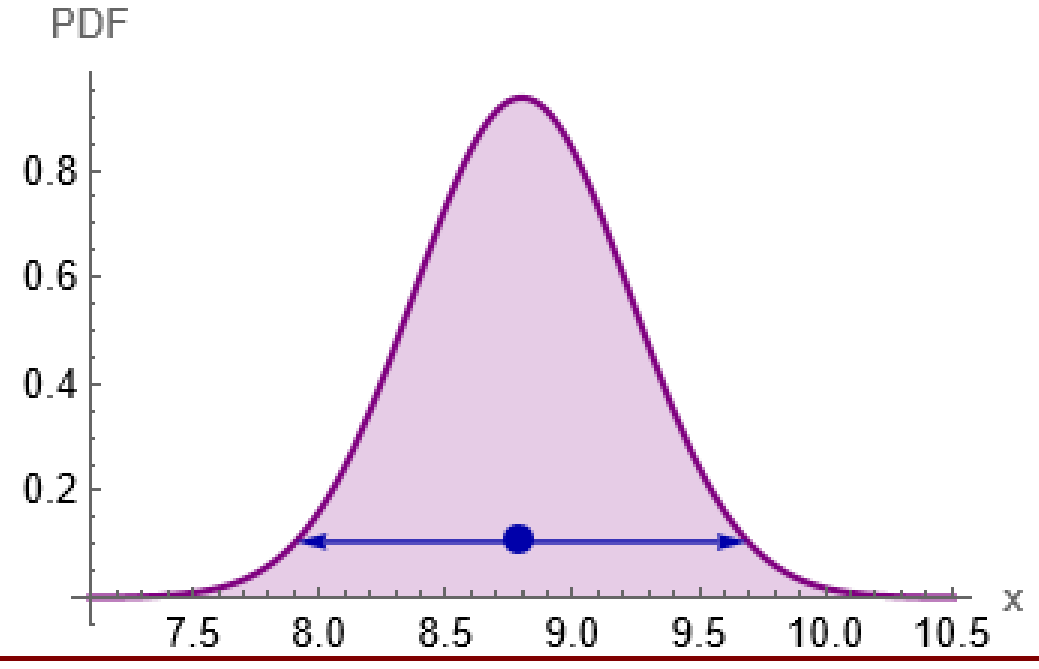

***Mathematica Examples 19.36***

```
Input   (* The code is an effective implementation for calculating and visualizing a
        confidence interval for the mean of a dataset. The code begins by collecting the
        data, assuming it is provided as a string with comma-separated values. It then
        calculates the sample mean and standard deviation, which are essential for estimating
        the population mean. Next, the code determines the level of significance (alpha)
        based on user input. This allows flexibility in selecting the desired confidence
        level for the interval. Using the level of significance, the code calculates the
        critical value from the Student's t-distribution. This value is crucial for
        constructing the confidence interval. The confidence interval is then computed using
```

```
the sample mean, sample standard deviation, and critical value. To visualize the
confidence interval, the code generates a probability density plot (PDF) of a normal
distribution centered around the sample mean. The confidence interval is displayed
with vertical lines indicating the lower and upper bounds. The sample mean is marked
with a blue dot for reference. The code also includes an interactive interface created
using the Manipulate function. This interface allows users to input their own data
and adjust the level of significance through a slider, providing a dynamic and
interactive experience for exploring different datasets and confidence levels: *)

Manipulate[
 (* Step 1. Collect your data: *)
 data=ToExpression@StringSplit[dataString,","];

 (* Step 2. Calculate the sample mean and standard deviation: *)
 sampleMean=Mean[data];
 sampleStdDev=StandardDeviation[data];

 (* Step 3. Determine the level of significance: *)
 alpha=alphaInput;

 (* Step 4. Calculate the critical value: *)
 degreesOfFreedom=Length[data]-1;
 criticalValue=Quantile[StudentTDistribution[degreesOfFreedom],1-alpha/2];

 (* Step 5. Calculate the confidence interval for the mean: *)
 standardError=sampleStdDev/Sqrt[Length[data]];
 confidenceInterval={sampleMean-
criticalValue*standardError,sampleMean+criticalValue*standardError};

 (* Step 6. Visualize the confidence interval: *)
 dist=NormalDistribution[sampleMean,sampleStdDev/Sqrt[Length[data]]];
 plot=Plot[
    PDF[dist,x],
    {x,sampleMean-4 sampleStdDev/Sqrt[Length[data]],sampleMean+4
sampleStdDev/Sqrt[Length[data]]},
    PlotRange->All,
    PlotStyle->Purple,
    Filling->Axis,
    AxesLabel->{"x","PDF"},
    ImageSize->250,
    Epilog->{
      Darker[Blue],
      PointSize[Large],
      Point[{sampleMean,0}],
      {Arrowheads[{-0.05,0.05}],
       Arrow[
        {{confidenceInterval[[1]],PDF[dist,confidenceInterval[[1]]]},
         {confidenceInterval[[2]],PDF[dist,confidenceInterval[[2]]]}}
        ]}}
    ];
 lines={
    Darker[Blue],
    Line[
     {{confidenceInterval[[1]],0},
      {confidenceInterval[[1]],PDF[dist,confidenceInterval[[1]]]}}
     ],
    Line[
     {{confidenceInterval[[2]],0},
      {confidenceInterval[[2]],PDF[dist,confidenceInterval[[2]]]}}
     ]
```

```
  };
point={
  Red,
  PointSize[Large],
  Point[{sampleMean,PDF[dist,confidenceInterval[[1]]]}]
  };

Show[
 plot,
 Graphics[{lines,point}]
 ],
{{dataString,"7,5,11,9,10,7,6,8,9,10,11,6,8,9,10,11,12,8,9,10"},"Data:"},
{{alphaInput,0.05},0.01,0.99,Appearance->"Labeled",ImageSize->Small},
ControlPlacement->Left,
FrameMargins->10
]
```

Output

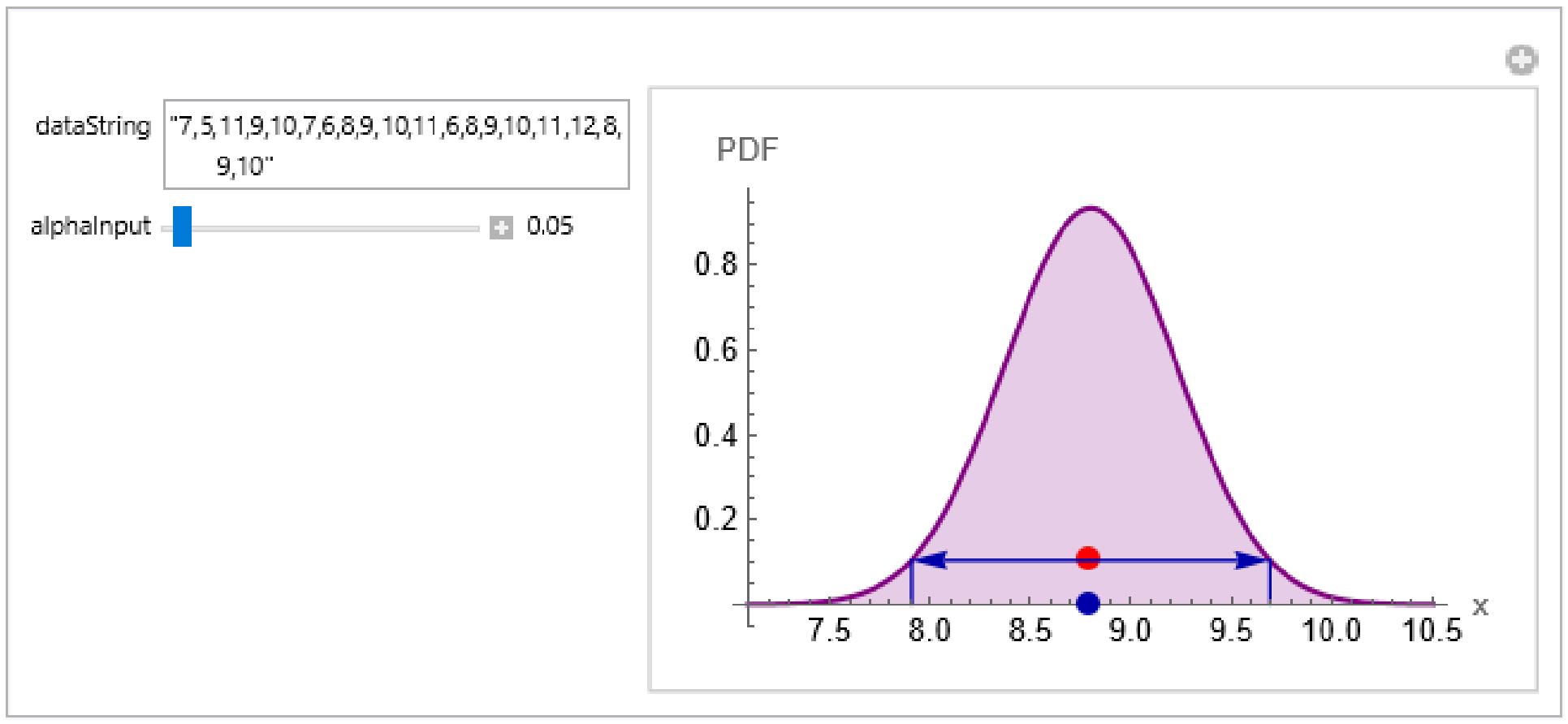

***Mathematica Examples 19.37***

Input

```
(* The code generates random data from a normal distribution and calculates and visualizes confidence intervals for the means of each row in the data. The code generates a Normal distribution with mean 0 and standard deviation 1. It then generates a random sample of size 20 by 5 from this distribution. Next, it loads the HypothesisTesting package, which contains functions for hypothesis testing and confidence intervals. The code calculates confidence intervals for the means of each row in the data using the MeanCI function from the package. The confidence level is set to 0.95, and it assumes a known variance of 1.
After calculating the confidence intervals, the code plots the probability density function (PDF) of the normal distribution using the PDF function. It also adds arrows to represent the confidence intervals on the plot using the Arrow function and the confIntervals data calculated earlier: *)

dist=NormalDistribution[0,1];
data=RandomVariate[dist,{20,5}];

<<HypothesisTesting`
(* Calculate confidence intervals: *)
confIntervals=Table[MeanCI[data[[i]],ConfidenceLevel->0.95,KnownVariance->1],{i,1,20}]

Plot[
 PDF[dist,x],
 {x,-3.5,3.5},
 PlotRange->{{-3.5,3.5},{-1.1,0.5}},
```

```
  Filling->Axis,
  PlotStyle->Purple,
  FillingStyle->Directive[Opacity[0.2],Purple],
  AxesOrigin->{0,0},
  Epilog->{
    Directive[PointSize[0.015],Purple],
    Flatten[Table[{Arrowheads[{-.02,.02}],Arrow[{{confIntervals[[i]][[1]],-
i/20},{confIntervals[[i]][[2]],-i/20}}]},{i,1,20}]]
    },
  AxesOrigin->{0,0},
  ImageSize->300
  ]
```

Output {{-0.559556,1.19349},{-0.243681,1.50936},{-1.79944,-0.0463922},{-1.03608,0.716965},{-0.784552,0.968493},{-1.24552,0.507524},{-1.90203,-0.148982},{-1.0763,0.676748},{-0.902017,0.851028},{-1.37529,0.377758},{-0.398764,1.35428},{-1.29461,0.45844},{-0.890759,0.862286},{-1.16586,0.587188},{-1.52349,0.229552},{-0.206374,1.54667},{-0.69151,1.06153},{-0.166193,1.58685},{-0.829442,0.923603},{-1.23619,0.516857}}

Output

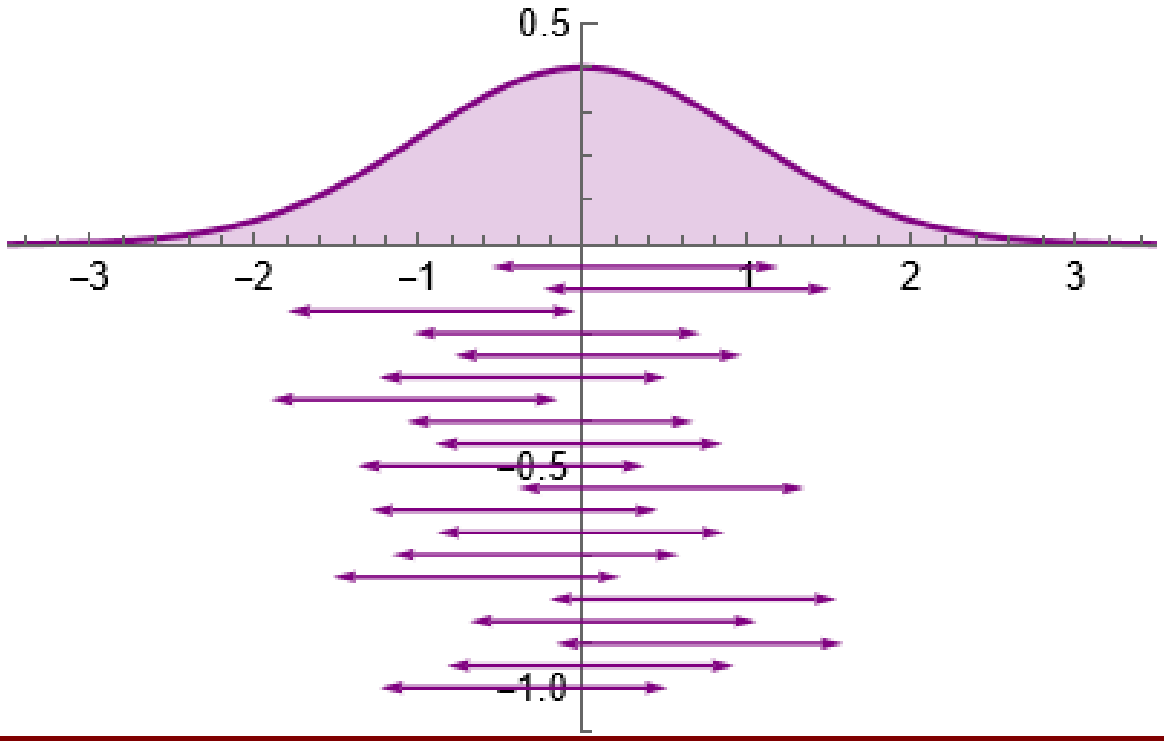

***Mathematica Examples 19.38***

Input (* This code sets up a Manipulate environment to visualize a normal distribution. The Manipulate controls allow you to adjust the parameters of the distribution, such as the mean (mu), standard deviation (sigma), sample size (n), confidence level, and random seed. The generated data is displayed as a plot with the probability density function (PDF) of the distribution and confidence intervals. The code uses the HypothesisTesting package to calculate the confidence intervals with Known Variance, and it plots the PDF with filled areas and arrows representing the confidence intervals. By interacting with the controls, you can observe how changing the parameters affects the distribution and confidence intervals: *)

```
Manipulate[
 SeedRandom[seedno];
 dist=NormalDistribution[mu,sigma];
 data=RandomVariate[dist,{20,n}];

 <<HypothesisTesting`;
 (* Calculate confidence intervals: *)
 confIntervals=Table[MeanCI[data[[i]],ConfidenceLevel-
>confidenceLevel,KnownVariance->1],{i,1,20}];

 Plot[
  PDF[dist,x],
  {x,-4.5,4.5},
  PlotRange->{{-4.5,4.5},{-1.1,0.5}},
  Axes->True,
  Filling->Axis,
```

```
    PlotStyle->Purple,
    FillingStyle->Directive[Opacity[0.2],Purple],
    AxesOrigin->{Mean[dist],0},
    Epilog->{
      Directive[PointSize[0.015],Purple],
      Flatten[Table[{Arrowheads[{-.02,.02}],Arrow[{{confIntervals[[i]][[1]],-i/20},{confIntervals[[i]][[2]],-i/20}}]},{i,1,20}]]
      },
    ImageSize->300
    ],
  {{mu,0},-2,2},
  {{sigma,1},1,2},
  {{n,5},5,100,1},
  {{confidenceLevel,0.90},0.80,0.99},
  {{seedno,20},16,25,1},
  ControlPlacement->Left,
  FrameMargins->10
  ]
```

Output

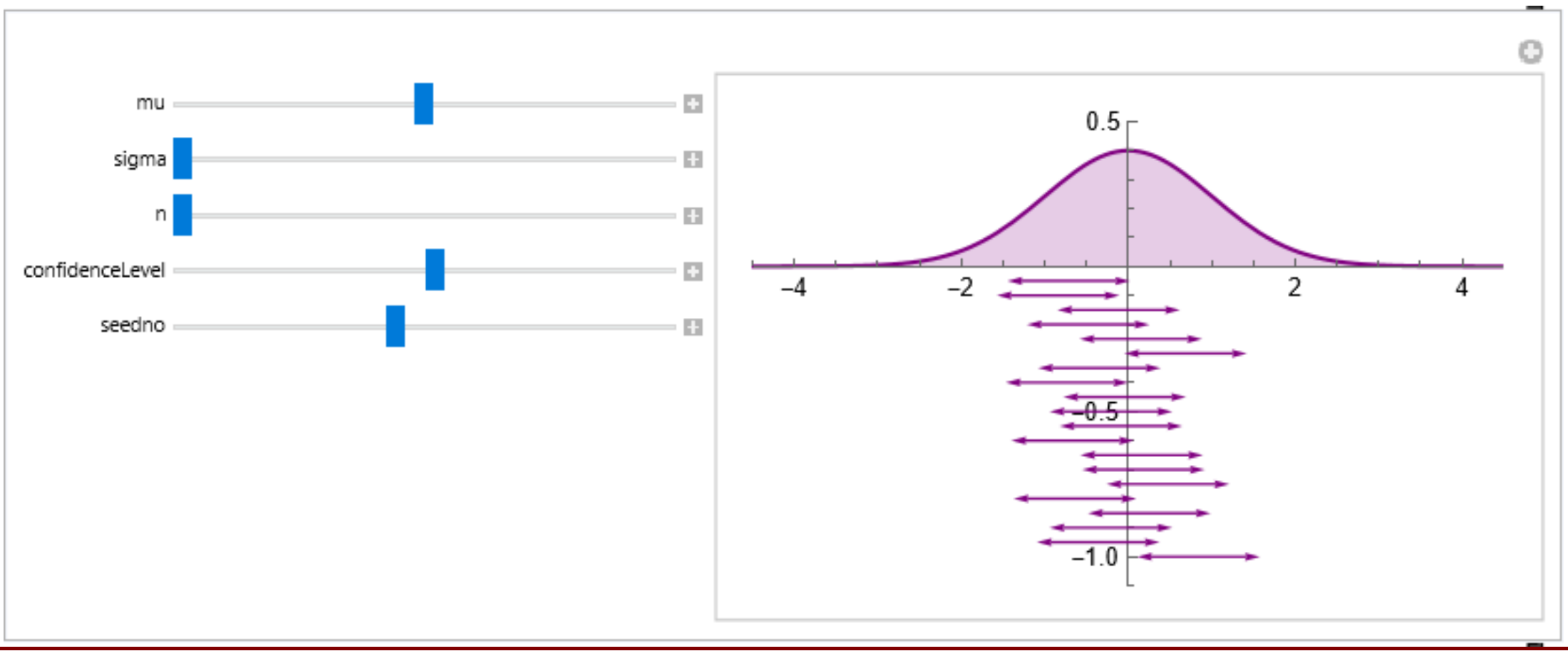

***Mathematica Examples 19.39***

Input

```
(* Similar to the above code but in the case of unknown Variance: *)

Manipulate[
  SeedRandom[seedno];

  dist=NormalDistribution[mu,sigma];
  data=RandomVariate[dist,{20,n}];

  <<HypothesisTesting`;
  (* Calculate confidence intervals:*)
  confIntervals=Table[MeanCI[data[[i]],ConfidenceLevel->confidenceLevel],{i,1,20}];

  Plot[
    PDF[dist,x],
    {x,-4.5,4.5},
    PlotRange->{{-4.5,4.5},{-1.1,0.5}},
    Axes->True,
    Filling->Axis,
    PlotStyle->Purple,
    FillingStyle->Directive[Opacity[0.2],Purple],
    AxesOrigin->{Mean[dist],0},
    Epilog->{
      Directive[PointSize[0.015],Purple],
```

```
        Flatten[Table[{Arrowheads[{-.02,.02}],Arrow[{{confIntervals[[i]][[1]],-
i/20},{confIntervals[[i]][[2]],-i/20}}]},{i,1,20}]]
        },
      ImageSize->300
      ],
     {{mu,0},-2,2},
     {{sigma,1},1,2},
     {{n,5},5,100,1},
     {{confidenceLevel,0.90},0.80,0.99},
     {{seedno,10},5,15,1},
     ControlPlacement->Left,
     FrameMargins->10
     ]
```

Output

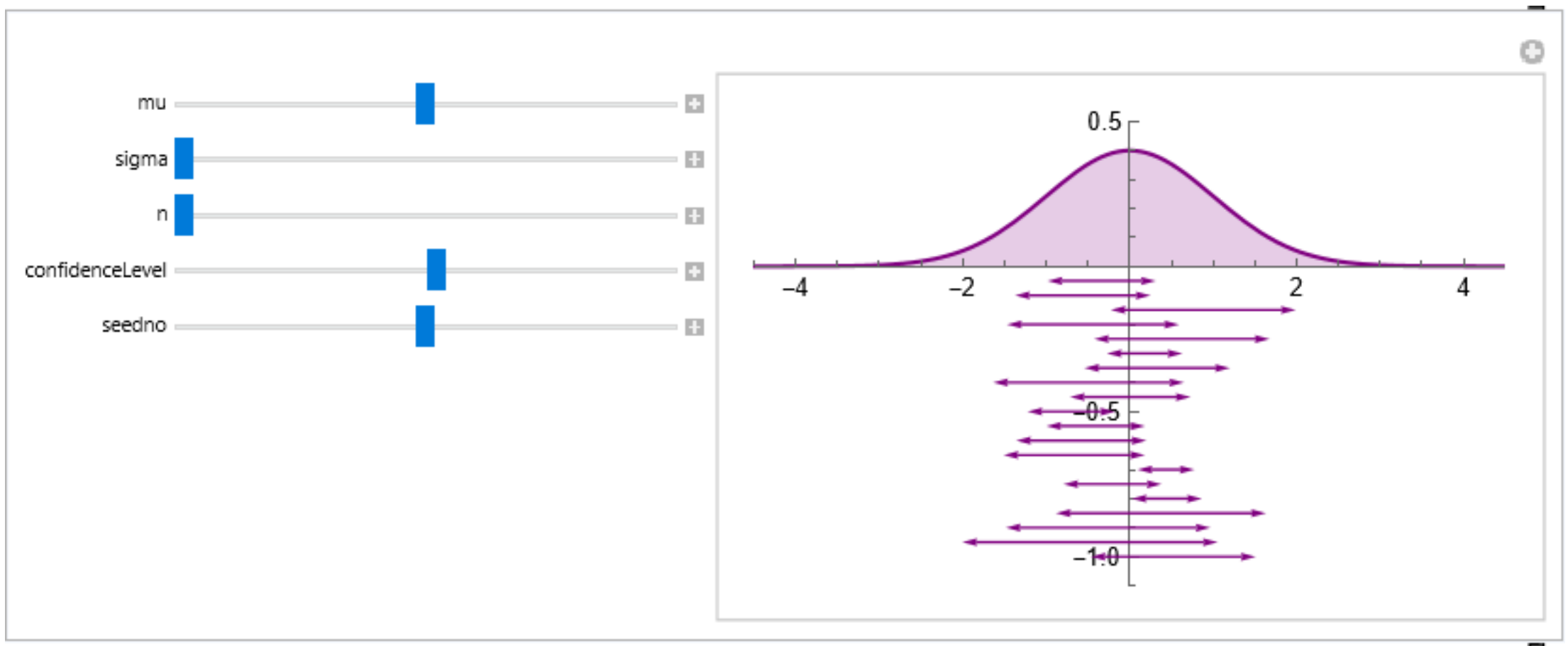

*Mathematica Examples 19.40*

Input

```
(* The code is a Mathematica implementation of a confidence interval for the mean based on a random sample from a normal distribution. It performs the following steps: Generates a random sample from a normal distribution with user-defined parameters for the mean and standard deviation. Calculates the sample mean and standard deviation of the generated data. Determines the level of significance (alpha) for the confidence interval. Calculates the critical value based on the student's t-distribution and the alpha level. Computes the confidence interval for the mean using the sample mean, critical value, and standard error. Visualizes the original normal distribution, the distribution based on the sample mean, and the confidence interval using plots and graphics. The code also includes sliders and controls for adjusting the parameters and allows for interactive exploration of the confidence interval: *)

Manipulate[
 (* Step 1, Generate a random sample from a normal distribution: *)
 data=RandomVariate[NormalDistribution[mu,sigma],n];

 (* Step 2, Calculate the sample mean and standard deviation: *)
 sampleMean=Mean[data];
 sampleStdDev=StandardDeviation[data];

 (* Step 3, Determine the level of significance: *)
 alpha=alphaInput;

 (* Step 4, Calculate the critical value: *)
  degreesOfFreedom=Length[data]-1;
 criticalValue=Quantile[StudentTDistribution[degreesOfFreedom],1-alpha/2];

 (* Step 5, Calculate the confidence interval for the mean: *)
```

```
 standardError=sampleStdDev/Sqrt[Length[data]];
 confidenceInterval={sampleMean-
criticalValue*standardError,sampleMean+criticalValue*standardError};

 (* Step 6, Visualize the sample and confidence interval: *)
 dist=NormalDistribution[sampleMean,sampleStdDev/Sqrt[Length[data]]];
 plot0=Plot[
   PDF[NormalDistribution[mu,sigma],x],
   {x,mu-4 sigma,mu+4 sigma},
   Axes->True,
   PlotRange->All,
   PlotStyle->Directive[Thick,Purple],
   Filling->Axis,
   AxesLabel->{"x","PDF"},
   ImageSize->300
   ];

 plot2=Plot[
   PDF[dist,x],
   {x,sampleMean-4 sampleStdDev/Sqrt[Length[data]],sampleMean+4
sampleStdDev/Sqrt[Length[data]]},
   Axes->True,
   PlotRange->All,
   PlotStyle->Directive[Thick,Orange],
   Filling->Axis,
   AxesLabel->{"x","PDF"},
   ImageSize->300
   ];

 lines={
   Red,

Line[{{confidenceInterval[[1]],0},{confidenceInterval[[1]],PDF[dist,confidenceInter
val[[1]]]}}],

Line[{{confidenceInterval[[2]],0},{confidenceInterval[[2]],PDF[dist,confidenceInter
val[[2]]]}}]
   };

 point={
   Red,
   PointSize[Large],
   Point[{sampleMean,0}],
   Blue,
   Point[{Mean[NormalDistribution[mu,sigma]],0}]
   };

 Show[
  plot0,
  plot2,
  Graphics[{lines,point}]
  ],
 {{mu,0},-10,10},
 {{sigma,1},0.1,5},
 {{n,5},5,100,1},
 {{alphaInput,0.05},0.01,0.99},
 ControlPlacement->Left,
 FrameMargins->10
 ]
```

Output

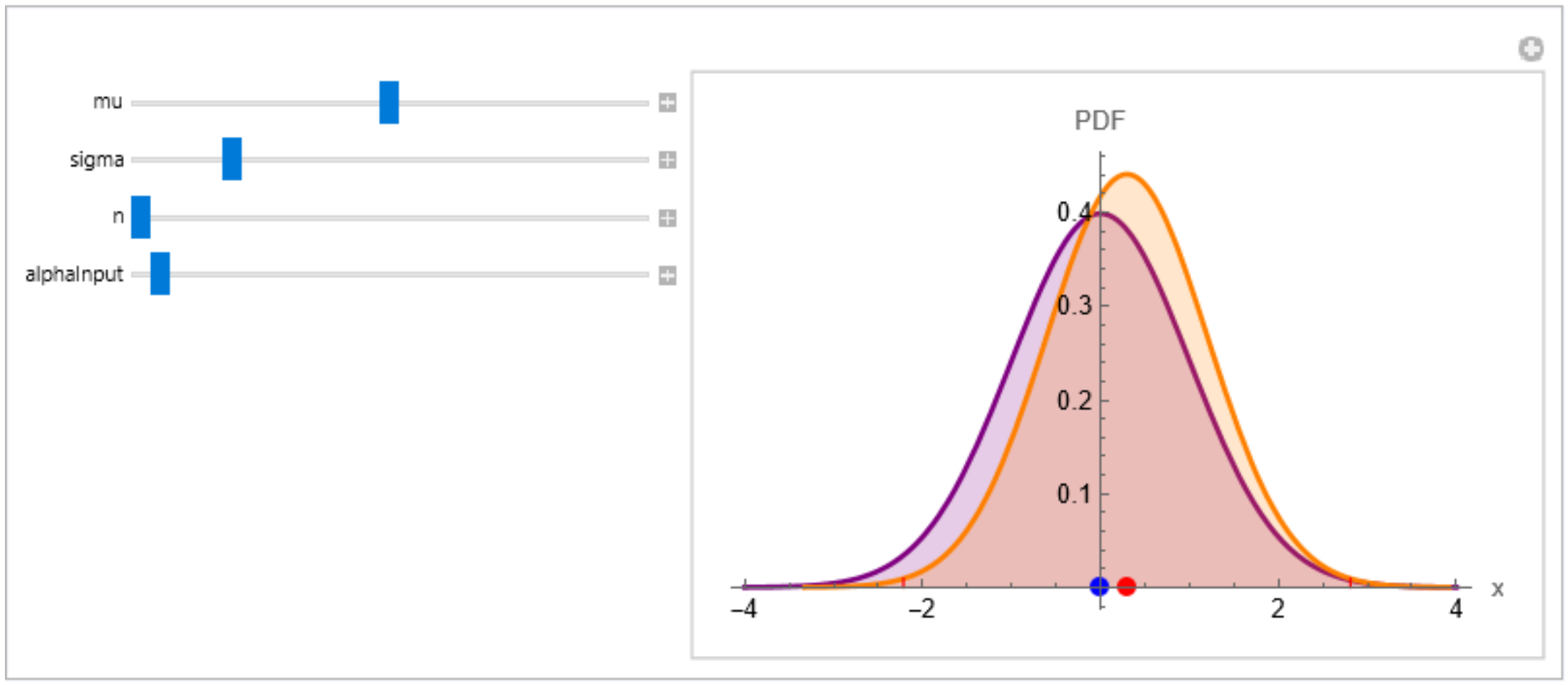

# CHAPTER 20

# DECISION THEORY AND HYPOTHESIS TESTING

In statistics and data analysis, the process of drawing conclusions and making inferences about a population based on a sample is of utmost importance. Hypothesis testing is a widely used technique that allows researchers, scientists, and data analysts to make well-founded decisions and validate their assumptions using empirical evidence.

- The fundamental idea behind hypothesis testing is to evaluate the plausibility of a claim or statement about a population parameter, such as the mean, variance, proportion, or correlation. These claims, known as hypotheses, are formulated based on prior knowledge, observations, or theories. The hypothesis testing process enables us to determine if the evidence from the sample data supports or contradicts these hypotheses.
- The two primary types of hypotheses involved in hypothesis testing are the null hypothesis ($H_0$) and the alternative hypothesis ($H_a \equiv H_1$). The null hypothesis typically represents a default assumption. On the other hand, the alternative hypothesis proposes a specific change or effect that the researcher is interested in detecting.
- Hypothesis testing involves collecting sample data and using statistical methods to assess the probability of obtaining such data if the null hypothesis were true. The process aims to determine whether the observed data provide enough evidence to reject the null hypothesis in favor of the alternative hypothesis.

Throughout this chapter, we will explore the key concepts and steps involved in conducting hypothesis tests. Some of the main topics covered will include:

- Formulating hypotheses: Understanding the significance of null and alternative hypotheses, and how to properly structure them for various scenarios.
- Test statistics: Introducing different test statistics, such as t-tests, z-tests, chi-square tests, and others, which serve as the basis for hypothesis testing.
- $P$-values: Explaining the concept of $P$-values and their role in hypothesis testing. $P$-values provide a measure of the strength of evidence against the null hypothesis.
- Type I and Type II errors: Understanding the risks associated with hypothesis testing, including the possibility of making incorrect decisions.
- We will discuss various types of hypothesis tests (one-sample, and two-sample), including tests for means, proportions, and variances. We will consider the following cases:
    - Large-sample hypothesis testing for mean and $\sigma$ known.
    - Large-sample hypothesis testing for mean and $\sigma$ unknown.
    - Hypothesis testing for mean in the case of the normal population and $\sigma$ unknown.
    - Hypothesis testing for $\mu_1$-$\mu_2$, $\sigma_1^2$ and $\sigma_2^2$ known.
    - Hypothesis testing for $\mu_1$-$\mu_2$, $\sigma_1^2 = \sigma_2^2$ but both are unknown.
    - Hypothesis testing for $\mu_1$-$\mu_2$, $\sigma_1^2 \neq \sigma_2^2$ and both unknown.
    - Hypothesis testing for $\sigma^2$.
    - Hypothesis testing for $\sigma_1^2/\sigma_2^2$.
    - Large-sample hypothesis testing for a population proportion $p$.
    - Large-sample hypothesis testing for $p_1 - p_2$.
    - Goodness-of-Fit Test

## 20.1 Hypothesis Testing

### 20.1.1 Null and alternative hypotheses

Very often in practice, we are called upon to make decisions about populations based on sample information. Such decisions are called statistical decisions. For example, we may wish to decide based on sample data whether a new serum is really effective in curing a disease, whether one educational procedure is better than another, or whether a given coin is loaded.

In attempting to reach decisions, it is useful to make assumptions (or guesses) about the populations involved. Such assumptions, which may or may not be true, are called statistical hypotheses. They are general statements about the probability distributions of the populations.

Procedures that help us decide whether to accept or reject hypotheses are called tests of hypotheses.

The structure of hypothesis testing will be formulated with the use of the term null hypothesis, which refers to any hypothesis we wish to test and is denoted by $H_0$. The null hypothesis is the default assumption made in hypothesis testing. It is the hypothesis that researchers assume to be true initially, and they seek evidence to either reject or fail to reject it. It is often written in terms of population parameters, such as population means, proportions, standard deviations, correlation coefficients, etc., depending on the type of data and the research question. For example, if you are comparing the mean scores of two groups (group A and group B) on a certain variable (e.g., test scores), the null hypothesis might be: "The population mean of group A is equal to the population mean of group B."

The value of the population parameter specified in the null hypothesis is usually determined in one of three ways.

- First, it may result from past experience or knowledge of the process or even from previous tests or experiments. The objective of hypothesis testing, then, is usually to determine whether the parameter value has changed.
- Second, this value may be determined from some theory or model regarding the process under study. Here the objective of hypothesis testing is to verify the theory or model.
- A third situation arises when the value of the population parameter results from external considerations, such as design or engineering specifications, or from contractual obligations. In this situation, the usual objective of hypothesis testing is conformance testing.

The alternative hypothesis, also known as the research hypothesis, is the statement that contradicts the null hypothesis. It represents what the researchers want to find evidence for. The alternative hypothesis $H_a$ usually represents the question to be answered or the theory to be tested, and thus its specification is crucial.

We will always state the null hypothesis as an equality claim. However, when the alternative hypothesis is stated with the $<$ sign, the implicit claim in the null hypothesis can be taken as $\geq$ and when the alternative hypothesis is stated with the $>$ sign, the implicit claim in the null hypothesis can be taken as $\leq$.

| Sign in $H_0$ | $=$ | $=$ or $\geq$ | $=$ or $\leq$ |
|---|---|---|---|
| Sign in $H_a$ | $\neq$ | $<$ | $>$ |

***Example 20.1***

Some examples of null hypotheses $H_0$ and alternative hypotheses $H_a$ in different scenarios:

Research question:
Is the average weight of a certain product different from a specified value of 10 kg?
$H_0$: The average weight of the product is equal to 10 kg. $H_0: \mu = 10$.
$H_a$: The average weight of the product is not equal to 10 kg. $H_a: \mu \neq 10$.

Research question:
Is there a significant difference in the exam scores of two study groups (group A and group B)?
$H_0$: There is no significant difference in the exam scores of the two groups. $H_0: \mu_A = \mu_B$.
$H_a$: There is a significant difference in the exam scores of the two groups. $H_a: \mu_A \neq \mu_B$.

Research question:
Is there a relationship between gender and voting preference?
$H_0$: There is no association between gender and voting preference. $H_0$: Gender and voting preference are independent.
$H_a$: There is a relationship between gender and voting preference. $H_a$: Gender and voting preference are dependent.

Research question:
Is there a significant difference in blood pressure before and after a new treatment?
$H_0$: There is no significant difference in blood pressure before and after the treatment. $H_0: \mu(\text{before}) = \mu(\text{after})$.
$H_a$: There is a significant difference in blood pressure before and after the treatment. $H_a: \mu(\text{before}) \neq \mu(\text{after})$.

Research question:
Is there a correlation between the number of hours spent studying and the exam scores?
$H_0$: There is no correlation between the number of hours spent studying and exam scores. $H_0: \rho = 0$ (where $\rho$ is the population correlation coefficient)
$H_a$: There is a correlation between the number of hours spent studying and exam scores. $H_a: \rho \neq 0$.

Research question:
Is the proportion of people who prefer brand $X$ over brand $Y$ significantly different from 0.5?
$H_0$: The proportion of people who prefer brand $X$ over brand $Y$ is equal to 0.5. $H_0: p = 0.5$ (where $p$ is the population proportion)
$H_a$: The proportion of people who prefer brand $X$ over brand $Y$ is not equal to 0.5. $H_a: p \neq 0.5$.

Research Question:
Is the average time it takes to complete a task using Method A less than the average time using Method B?
$H_0$: The average time to complete the task using Method A is greater than or equal to the average time using Method B.
$H_a$: The average time to complete the task using Method A is less than the average time using Method B.

Research Question:
Is the average time it takes to complete a task using Method A greater than the average time using Method B?
$H_0$: The average time to complete the task using Method A is less than or equal to the average time using Method B.
$H_a$: The average time to complete the task using Method A is greater than the average time using Method B.

Consider a population having a distribution $F_\theta$, where $\theta$ is an unknown parameter, and suppose we want to test a specific hypothesis about $\theta$. For example, if $F_\theta$ is a normal distribution function with mean $\theta$ and variance equal to 1, then, for example, two possible null hypotheses about $\theta$ are

$$\text{(a) } H_0: \theta = 1, \qquad \text{(b) } H_0: \theta \leq 1.$$

Thus, the first of these hypotheses states that the population is normal with mean 1 and variance 1, whereas the second states that it is normal with variance 1 and a mean less than or equal to 1. Note that the null hypothesis in (a), when true, completely specifies the population distribution, whereas the null hypothesis in (b) does not. A hypothesis that, when true, completely specifies the population distribution is called a simple hypothesis; one that does not is called a composite hypothesis.

To test a specific null hypothesis $H_0$, a population sample of size $n$ — say $X_1, \ldots, X_n$ — is to be observed. Hypothesis-testing procedures rely on using the information in a random sample from the population of interest. If this information is consistent with the null hypothesis, we will not reject it; however, if this information is inconsistent with the null hypothesis, we will conclude that the null hypothesis is false and reject it in favor of the alternative. We emphasize that the truth or falsity of a particular hypothesis can never be known with certainty unless we can examine the entire population. This is usually impossible in most practical situations. Therefore, a hypothesis-testing procedure should be developed with the probability of reaching a wrong conclusion in mind.

Testing the hypothesis involves taking a random sample, computing a test statistic from the sample data, and then using the test statistic to make a decision about the null hypothesis.

### 20.1.2 One-tailed and two-tailed tests

One-tailed and two-tailed tests are two different types of hypothesis tests used in statistical analysis. The choice between these two types of tests depends on the research question and the directionality of the effect or difference being investigated.

One-tailed test:

Also known as a one-sided test, a one-tailed test is used when the research question specifically focuses on whether the population parameter is greater than or less than a certain value. The alternative hypothesis $H_a$ in a one-tailed test is directional, and it is formulated to detect an effect in one specific direction.

For example, in this book, the null hypothesis for a hypothesis test concerning a population mean, $\mu$, always specifies a single value for that parameter. Hence, we can express the null hypothesis as

$H_0: \mu = \mu_0$,

where $\mu_0$ is some number. If the primary concern is deciding whether a population mean, $\mu$, is less than a specific value $\mu_0$, we express the alternative hypothesis as

$H_a: \mu < \mu_0$.

A hypothesis test whose alternative hypothesis has this form is called a left-tailed test.

On the other hand, if the primary concern is deciding whether a population mean, $\mu$, is greater than a specified value $\mu_0$, we express the alternative hypothesis as

$H_a: \mu > \mu_0$.

A hypothesis test whose alternative hypothesis has this form is called a right-tailed test. A hypothesis test is called a one-tailed test if it is either left-tailed or right-tailed.

***Example 20.2***

Consider the following research question:
"Does a new drug significantly increase the average test scores of students?"

In a one-tailed test, the hypotheses would be as follows:
$H_0$: The new drug has no effect on the average test scores of students.
$H_a$ : The new drug significantly increases the average test scores of students.

Two-tailed test:

A two-tailed test, also known as a two-sided test, is used when the research question is concerned with whether the population parameter is different from a specific value, without specifying the direction of the difference. The alternative hypothesis $H_a$ in a two-tailed test is non-directional, and it is formulated to detect an effect in either direction.

For example, if the primary concern is deciding whether a population mean, $\mu$, is different from a specified value $\mu_0$, we express the alternative hypothesis as

$H_a: \mu \neq \mu_0.$

A hypothesis test whose alternative hypothesis has this form is called a two-tailed test.

***Example 20.3***

Consider the following research question:
"Is there a significant difference in the average test scores between two groups of students?"

In a two-tailed test, the hypotheses would be as follows:
$H_0$: There is no significant difference in the average test scores between the two groups of students.
$H_a$: There is a significant difference in the average test scores between the two groups of students.

| Sign in $H_0$ | $=$ | $=$ or $\geq$ | $=$ or $\leq$ |
|---|---|---|---|
| Sign in $H_a$ | $\neq$ | $<$ | $>$ |
| Tests | Two-tailed test | Left-tailed test | Right-tailed test |

### 20.1.3 Type I and type II errors

Any decision we make based on a hypothesis test may be incorrect because we have used partial information obtained from a sample to draw conclusions about the entire population. There are two types of incorrect decisions—Type I error and Type II error, as indicated in the following table. These errors are associated with the acceptance or rejection of a null hypothesis.

| | $H_0$ is True | $H_0$ is False |
|---|---|---|
| Do not reject $H_0$ | Correct decision | Type II error |
| Reject $H_0$ | Type I error | Correct decision |

**Definition (Type I Error):** Rejecting the null hypothesis when it is in fact true.

**Definition (Type II error):** Not rejecting the null hypothesis when it is in fact false.

Type I error (False Positive):

- A Type I error occurs when the null hypothesis $H_0$ is incorrectly rejected when it is true. In other words, it is a false positive error where we conclude that there is a significant effect or difference when there is none in reality.
- The probability of making a Type I error is denoted by the symbol "$\alpha$" and is known as the significance level of the test. It represents the maximum allowable probability of incorrectly rejecting the null hypothesis when it is true. Commonly used significance levels include 0.05 (5%) and 0.01 (1%).
- If, for example, the 0.05 (or 5%) significance level is chosen in designing a decision rule, then there are about 5 chances in 100 that we would reject the hypothesis when it should be accepted; that is, we are about 95% confident that we have made the right decision. In such a case, we say that the hypothesis has been rejected at the 0.05 significance level, which means that the hypothesis has a 0.05 probability of being wrong.
- By setting a lower significance level, you reduce the chance of committing a Type I error but increase the risk of committing a Type II error.

Type II error (False Negative):

- A Type II error occurs when the null hypothesis $H_0$ is incorrectly failed to be rejected when it is false. In other words, it is a false negative error where we fail to detect a significant effect or difference that exists in reality. The probability of making a Type II error is denoted by the symbol "$\beta$". It represents the probability of failing to reject the null hypothesis when it is false, i.e., the probability of not detecting a true effect or difference.
- The power of a statistical test is equal to $(1-\beta)$ and represents the probability of correctly rejecting the null hypothesis when it is false. In other words, it is the probability of correctly detecting a real effect or difference.
- Researchers can control the Type II error rate by increasing the sample size, which generally improves the power of the test. A larger sample size allows for more precise estimates and increases the likelihood of detecting true effects. Additionally, researchers can choose more sensitive statistical tests or modify experimental designs to enhance the chances of detecting effects if they exist.

***Example 20.4***

Scenario: A new medical test is developed to diagnose a particular disease. The null hypothesis $H_0$ states that the patient does not have the disease.

Type I error:
Situation:
The test results show that a healthy patient has the disease (reject $H_0$), but in reality, the patient is disease-free.
Consequence:
The patient may undergo unnecessary and potentially harmful treatments or surgeries.

Type II error:
Situation:
The test results show that a patient does not have the disease (fail to reject $H_0$), but in reality, the patient has the disease.
Consequence:
The disease goes undetected, and the patient misses the opportunity for early treatment and intervention.

***Example 20.5***

Scenario: A manufacturing process produces a batch of products and the null hypothesis $H_0$ states that the batch meets the required quality standard.

Type I error:
Situation:
The quality control test rejects the batch (reject $H_0$), but in reality, the batch meets the required standard.
Consequence:
The entire batch is discarded or reworked, leading to unnecessary costs and waste.

Type II error:
Situation:
The quality control test accepts the batch (fail to reject $H_0$), but in reality, the batch does not meet the required standard.
Consequence:
Substandard products are released into the market, potentially leading to customer dissatisfaction and recalls.

***Example 20.6***

Scenario: A researcher investigates the effects of a new drug on a certain condition and the null hypothesis $H_0$ states that the drug has no effect.

Type I error:
Situation:
The study shows that the drug has a significant effect (reject $H_0$), but in reality, the drug has no effect on the condition.
Consequence:
False claims of the drug's effectiveness may lead to unwarranted prescriptions and wasted resources.

Type II error:
Situation:
The study fails to find a significant effect of the drug (fail to reject $H_0$), but in reality, the drug does have a beneficial effect.
Consequence:
The potential benefits of the drug remain undiscovered, and patients miss out on a useful treatment.

**Remarks:**

- The probabilities of Type I and Type II errors are related, and decreasing one type of error typically increases the other type. Statisticians and researchers need to balance these error probabilities depending on the context of the study and the potential consequences of each error. This balance is usually achieved by adjusting the sample size or the significance level of the hypothesis test.
- In practice, one type of error may be more serious than the other, and so a compromise should be reached in favor of limiting the more serious error. The only way to reduce both types of error is to increase the sample size, which may or may not be possible.
- Type I errors are generally considered more serious than type II errors. For example, it is mostly agreed that finding an innocent person guilty is a more serious error than finding a guilty person innocent. Here, the null hypothesis is that the person is innocent, and the alternative hypothesis is that the person is guilty. “Not rejecting the null hypothesis” is equivalent to acquitting a defendant. It does not prove that the null hypothesis is true, or that the defendant is innocent.
- The significance level, $\alpha$, is the probability of making a Type I error, that is, of rejecting a true null hypothesis. Therefore, if the hypothesis test is conducted at a small significance level (e.g., $\alpha = 0.05$), the chance of rejecting a true null hypothesis will be small. In this book, we generally specify a small significance level. Thus, if we do reject the null hypothesis, we can be reasonably confident that the null hypothesis is false. In other words, if we do reject the null hypothesis, we conclude that the data provide sufficient evidence to support the alternative hypothesis.
- However, we usually do not know the probability, $\beta$, of making a Type II error, that is, of not rejecting a false null hypothesis. Consequently, if we do not reject the null hypothesis, we simply reserve judgment about which hypothesis is true. In other words, if we do not reject the null hypothesis, we conclude only that the data do not provide sufficient evidence to support the alternative hypothesis; we do not conclude that the data provide sufficient evidence to support the null hypothesis.

### 20.1.4 Rejection and non-rejection regions: Tests involving normal distribution

To illustrate the ideas presented above, suppose that under a given hypothesis the sampling distribution of a statistic Sta is a normal distribution with mean $\mu_{\text{Sta}}$ and standard deviation $\sigma_{\text{Sta}}$. Thus, the distribution of the standardized variable (or $z$ score), given by $z = (\text{Sta} - \mu_{\text{Sta}})/\sigma_{\text{Sta}}$, is the standardized normal distribution (mean 0, variance 1), as shown in Figure 20.1.

As indicated in Figure 20.1, we can be 95% confident that if the hypothesis is true, then the $z$ score of an actual sample statistic Sta will lie between $-1.96$ and $1.96$ (since the area under the normal curve between these values is 0.95). However, if on choosing a single sample at random we find that the $z$ score of its statistic lies outside the range $-1.96$ to $1.96$, we would conclude that such an event could happen with a probability of only 0.05 (the total non-shaded area in the figure) if the given hypothesis were true. We would then say that this $z$ score differed significantly from what would be expected under the hypothesis, and we would then be inclined to reject the hypothesis.

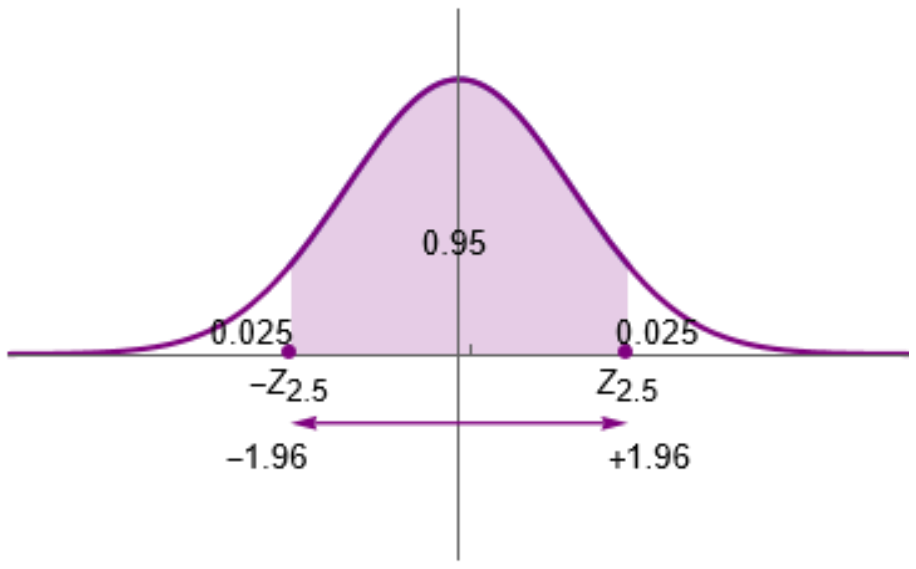

**Figure 20.1.** Standard normal curve with the critical region (0.05) and acceptance region (0.95).

The total non-shaded area, 0.05, is the significance level of the test. It represents the probability of our being wrong in rejecting the hypothesis (i.e., the probability of making a Type I error). Thus, we say that the hypothesis is rejected at the 0.05 significance level or that the $z$ score of the given sample statistic is significant at the 0.05 level. The set of $z$ scores outside the range $-1.96$ to 1.96 constitutes what is called the critical region of the hypothesis, the region of rejection of the hypothesis, or the region of significance.

**Definition (Rejection Region):** The rejection region is the set of values of the test statistic for which the null hypothesis $H_0$ is rejected in favor of the alternative hypothesis $H_a$. In other words, it represents the range of test statistic values that are unlikely to occur if the null hypothesis is true, assuming a specific significance level $\alpha$.

For a one-tailed test, the rejection region is located in one tail of the distribution of the test statistic. For a two-tailed test, the rejection region is divided between both tails of the distribution. The size and location of the rejection region depend on the chosen significance level $\alpha$ of the test.

The set of $z$ scores inside the range $-1.96$ to 1.96 is thus called the region of acceptance of the hypothesis, or the region of nonsignificance.

**Definition (Non-Rejection Region):** The non-rejection region, also called the acceptance region, is the complementary set of the rejection region. It includes the values of the test statistic for which the null hypothesis $H_0$ is not rejected, and the test does not provide sufficient evidence to support the alternative hypothesis $H_a$.

The non-rejection region includes the values of the test statistic that are considered likely to occur if the null hypothesis is true, given the chosen significance level $\alpha$. If the test statistic falls within this region, the decision is to fail to reject the null hypothesis.

**Definition (Critical Value(s)):** Critical values are specific values or boundaries on a distribution that define the boundary between the rejection region and the non-rejection region. These values are determined based on the chosen significance level $\alpha$ and the distribution of the test statistic under the null hypothesis.

| | | | |
|---|---|---|---|
| Sign in $H_0$ | $=$ | $=$ or $\geq$ | $=$ or $\leq$ |
| Sign in $H_a$ | $\neq$ | $<$ | $>$ |
| Rejection region | In both tails | In the left tail | In the right tail |
| Figure | 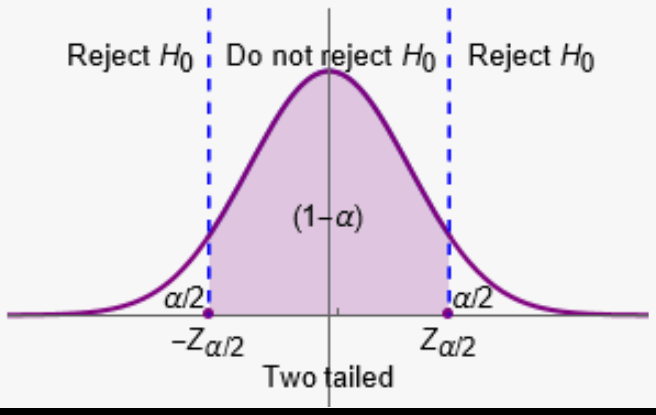 | 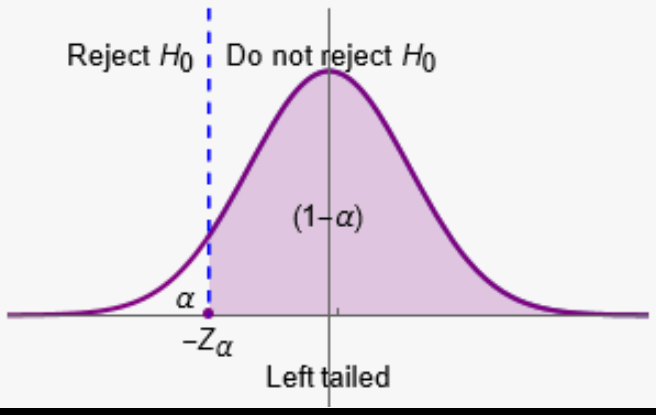 | 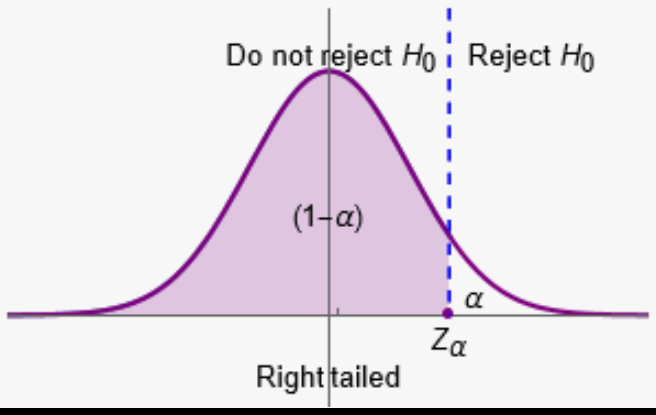 |

For example, in a one-tailed test, the critical value might be the value of the test statistic corresponding to the $\alpha$th percentile of the null distribution. In a two-tailed test, there are two critical values, one for the lower tail and one for the upper tail, each corresponding to $\alpha/2$ percentiles of the null distribution.

To make a decision in hypothesis testing, researchers compare the observed value of the test statistic (calculated from the sample data) to the critical values. If the observed value falls within the rejection region, the null hypothesis is rejected in favor of the alternative hypothesis. If it falls within the non-rejection region, the null hypothesis is not rejected.

Based on the above remarks, we can formulate the following decision rule (or test of hypothesis or significance):

- Reject the hypothesis at the $0.05$ significance level if the $z$ score of the statistic Sta lies outside the range $-1.96$ to $1.96$ (i.e., either $z > 1.96$ or $z < -1.96$). This is equivalent to saying that the observed sample statistic is significant at the $0.05$ level.
- Accept the hypothesis otherwise (or, if desired, make no decision at all).

### 20.1.5 *P*-value for hypotheses tests

**Definition (P-Value):** The $P$-value is the probability of obtaining a test statistic as extreme as, or more extreme than, the one calculated from the sample data, assuming that the null hypothesis is true. In other words, it quantifies the strength of evidence against the null hypothesis.

***Example 20.7***

For testing means, using large samples ($n > 30$), calculate the $P$-value as follows:

1. For $H_0: \mu = \mu_0$ and $H_1: \mu < \mu_0$, $P$-value $= P(Z < \text{computed test statistic})$,
2. For $H_0: \mu = \mu_0$ and $H_1: \mu > \mu_0$, $P$-value $= P(Z > \text{computed test statistic})$, and
3. For $H_0: \mu = \mu_0$ and $H_1: \mu \neq \mu_0$,
   $P$-value $= P(Z < -|\text{computed test statistic}|) + P(Z > |\text{computed test statistic}|)$.

The computed test statistic is $(\bar{x} - \mu_0)/(s/\sqrt{n})$, where $\bar{x}$ is the mean of the sample, $s$ is the standard deviation of the sample, and $\mu_0$ is the value specified for $\mu$ in the null hypothesis. Note that if $\sigma$ is unknown, it is estimated from the sample by using $s$. This method of testing the hypothesis is equivalent to the method of finding a critical value or values and if the computed test statistic falls in the rejection region, reject the null hypothesis. The same decision will be reached using either method.

**Definition (Decision Criterion for a Hypothesis Test Using the *P*-Value):** Typically, a significance level $\alpha$ is chosen before conducting the test. The $P$-value is then compared to this significance level. If the $P$-value is less than or equal to the specified significance level, reject the null hypothesis; otherwise, do not reject the null hypothesis. In other words, if $P \leq \alpha$, reject $H_0$; otherwise, do not reject $H_0$.

It is essential to understand that a small $P$-value does not prove that the null hypothesis is false or that the effect is practically significant. It only indicates that the observed data is unlikely to occur under the assumption of the null hypothesis, leading to the rejection of the null hypothesis in favor of the alternative.

| | | | |
|---|---|---|---|
| Sign in $H_0$ | $=$ | $=$ or $\geq$ | $=$ or $\leq$ |
| Sign in $H_a$ | $\neq$ | $<$ | $>$ |
| Rejection region | In both tails | In the left tail | In the right tail |

Figure

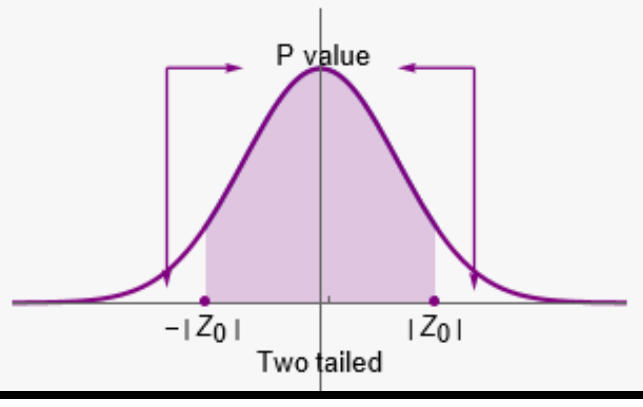

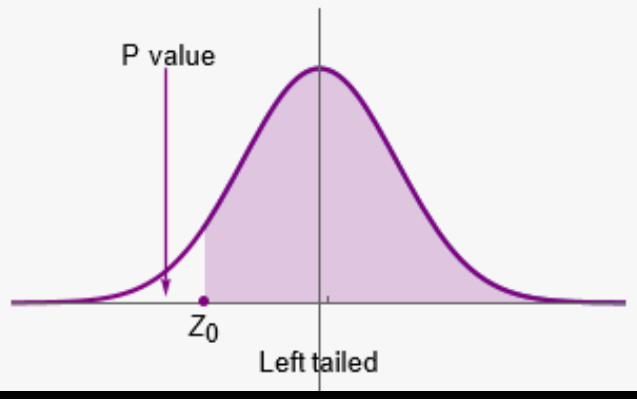

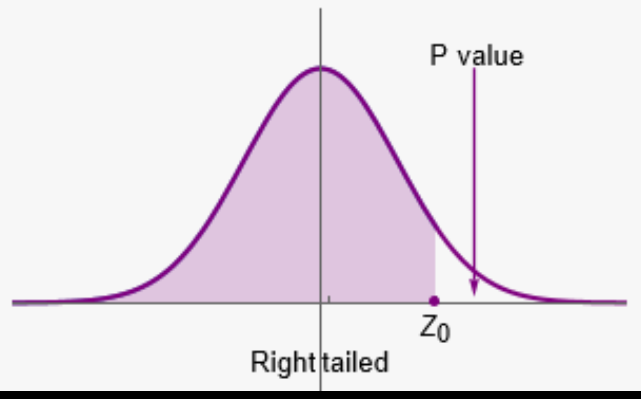

The $P$-value of a hypothesis test is also referred to as the observed significance level. To understand why, suppose that the $P$-value of a hypothesis test is $P = 0.07$. Then, for instance, we can reject the null hypothesis at the 10% significance level (because $P \leq 0.10$), but we cannot reject the null hypothesis at the 5% significance level (because $P > 0.05$). Here, the null hypothesis can be rejected at any significance level of at least $0.07$ and cannot be rejected at any significance level less than $0.07$. More generally, we have the following fact.

**Definition ($P$-Value as the Observed Significance Level):** The $P$-value of a hypothesis test equals the smallest significance level at which the null hypothesis can be rejected, that is, the smallest significance level for which the observed sample data results in rejection of $H_0$.

**Procedure 20.1.**

The general steps involved in hypothesis testing are as follows:

1. State the null hypothesis $H_0$ and the alternative hypothesis $H_a$:
   The null hypothesis represents the default assumption, while the alternative hypothesis represents the claim or the hypothesis to be tested.
2. Choose the significance level ($\alpha$):
   The significance level, often denoted by $\alpha$, determines the probability of rejecting the null hypothesis when it is actually true. Commonly used significance levels are 0.05 (5%) and 0.01 (1%).
3. Select an appropriate test statistic:
   The choice of test statistic depends on the nature of the problem, the type of data, and the hypothesis being tested. Common test statistics include t-tests, z-tests, chi-square tests, and F-tests.
4. Determine the critical region:
   The critical region is the set of values of the test statistic that leads to the rejection of the null hypothesis. It is determined based on the significance level and the distribution of the test statistic under the null hypothesis.
5. Collect and analyze the sample data:
   Obtain a sample of data from the population of interest. Calculate the test statistic using the sample data.
6. Make a decision 1: (Critical value approach)
   Compare the calculated test statistic with the critical values from the distribution.
   - If the test statistic falls within the critical region, reject the null hypothesis.
   - Otherwise, fail to reject the null hypothesis.
7. Calculate the $P$-value associated with the test statistic. This is the probability of obtaining a test statistic as extreme as the observed one, assuming the null hypothesis is true.
   - For a two-tailed test, calculate the area in both tails of the distribution.
   - For a one-tailed test, calculate the area in the appropriate tail.
8. Make a decision 2: ($P$ value approach)
   Compare the $P$-value to the significance level ($\alpha$).
   - If the $P$-value is less than $\alpha$, reject the null hypothesis.
   - Otherwise, fail to reject the null hypothesis.
9. Draw conclusions:
   Based on the decision in steps 6 and 8, draw conclusions about the population.
   - If the null hypothesis is rejected, it suggests evidence in favor of the alternative hypothesis.
   - If the null hypothesis is not rejected, there is insufficient evidence to support the alternative hypothesis.
10. Report the results:
    Report the test statistic, the critical values, the decision, and the conclusions in a clear and concise manner.

## 20.2 Large-Sample Hypothesis Tests for One Population Mean When $\sigma$ Is Known (Unknown)

| | | | |
|---|---|---|---|
| Assumptions | 1. Simple random sample<br>2. Normal population or large sample<br>3. $\sigma$ known (unknown) | | |
| Null hypothesis | $H_0: \mu = \mu_0$ | | |
| | Two-tailed | Left tailed | Right tailed |
| Alternative hypothesis | $H_a: \mu \neq \mu_0$ | $H_a: \mu < \mu_0$ | $H_a: \mu > \mu_0$ |
| Significance level | $\alpha$ | | |
| Test statistic:<br>For details see Chapter 18. | $z_0 = \dfrac{\bar{x} - \mu_0}{\sigma/\sqrt{n}}$<br>($\sigma$ is known) | $z_0 = \dfrac{\bar{x} - \mu_0}{s/\sqrt{n}}$<br>($\sigma$ is unknown) | |
| **Critical value approach** | $\pm z_{\alpha/2}$ | $-z_\alpha$ | $z_\alpha$ |

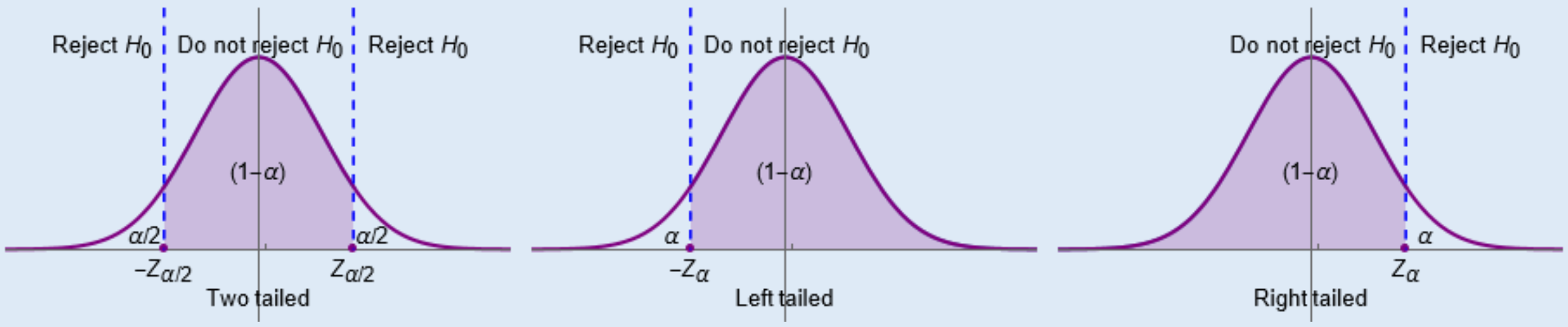

$$\begin{cases} \text{reject } H_0, & \text{If the value of the test statistic falls in the rejection region} \\ \text{do not reject } H_0, & \text{otherwise} \end{cases}$$

***P* value approach**

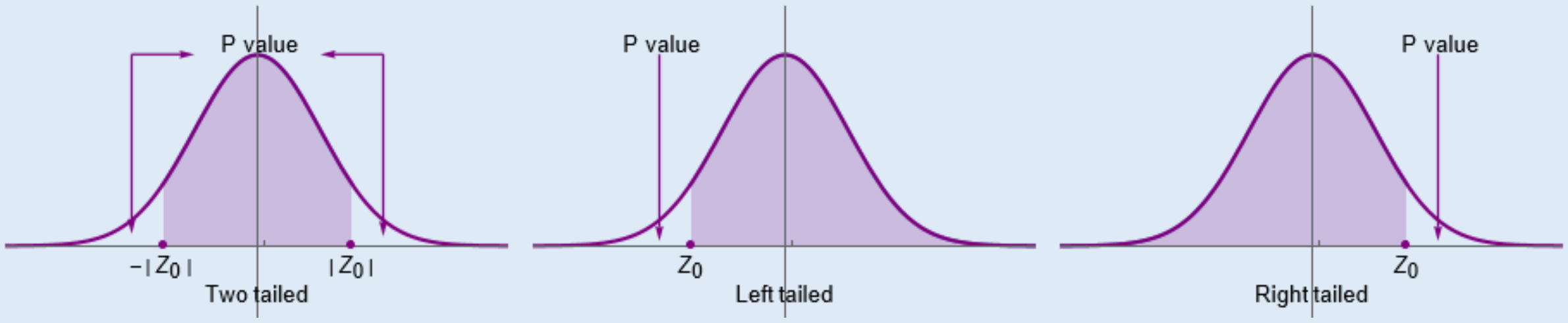

$$\begin{cases} \text{reject } H_0, & \text{If } P \leq \alpha \\ \text{do not reject } H_0, & \text{otherwise} \end{cases}$$

**Procedure 20.2.**
The steps for conducting a hypothesis test for one population mean when the population standard deviation ($\sigma$) is known:

1. State the null and alternative hypotheses:
   - $H_0$: The population mean is equal to a specified value.
   - $H_a$: The population mean is not equal to the specified value (two-tailed test), or it is greater than/less than the specified value (one-tailed test).
2. Set the significance level, $\alpha$, for the test.
3. Collect a random sample from the population and calculate the sample mean $\bar{x}$.
4. Determine the critical value(s) for the test statistic.

- For a two-tailed test, divide the significance level by 2 and find the corresponding $z$-score(s) using the standard Mathematica normal distribution.
- For a one-tailed test, find the $z$-score corresponding to the desired tail area.

5. Calculate the test statistic using the formula:
   For a $z$-test:
   $$z = \frac{\bar{x} - \mu_0}{\sigma/\sqrt{n}},$$
   where $\bar{x}$ is the sample mean, $\mu_0$ is the hypothesized population mean, $\sigma$ is the known population standard deviation, and $n$ is the sample size.
6. Compare the test statistic to the critical value(s).
   - If the test statistic falls within the critical region, reject the null hypothesis.
   - Otherwise, fail to reject the null hypothesis.
7. Calculate the $P$-value associated with the test statistic.
   - For a two-tailed test, calculate the area in both tails of the distribution.
   - For a one-tailed test, calculate the area in the appropriate tail.
8. Compare the $P$-value to $\alpha$.
   - If the $P$-value is less than $\alpha$, reject the null hypothesis.
   - Otherwise, fail to reject the null hypothesis.
9. Draw a conclusion based on the results:
   - If the null hypothesis is rejected, conclude that there is sufficient evidence to support the alternative hypothesis.
   - If the null hypothesis is not rejected, conclude that there is not enough evidence to support the alternative hypothesis.

***Example 20.8***

```
(* Two tailed case: *)

(* The code demonstrates the steps for conducting a hypothesis test for one population mean
when the population standard deviation (σ) is known, n>30: *)

(* data: *)
μ0=10;
σ0=5;

(* Step 1. State the null and alternative hypotheses: *)
"H0:μ==μ0"; (* Null hypothesis:Population mean is equal to μ0. *)
"H1:μ!=μ0"; (* Alternative hypothesis:Population mean is not equal to μ0. *)

(* Step 2. Set the significance level (α): *)
α=0.05; (* Significance level of 0.05. *)

(* Step 3. Collect a random sample and calculate the sample mean: *)
SeedRandom[123];
sample=RandomVariate[NormalDistribution[10,5],30]

n=Length[sample]; (* Sample size.*)
xbar=Mean[sample]; (* Sample mean. *)

(* Critical value approach: *)
(* Step 4. Determine the critical value(s): *)
zCritical=Quantile[NormalDistribution[0,1],1-α/2]; (* For two-tailed test. *)

(* Step 5. Calculate the test statistic: *)
σ=σ0; (* Known population standard deviation. *)
testStatistic=(xbar-μ0)/(σ/Sqrt[n]);
```

```
(* Step 6. Compare the test statistic to the critical value(s): *)
criticalvalueapproach=If[
    Abs[testStatistic]>zCritical,
    (* Reject the null hypothesis: *)
    conclusion="Reject the null hypothesis, there is sufficient evidence to support the
alternative hypothesis, at the significance level of "<>ToString[α],
    (* Fail to reject the null hypothesis: *)
    conclusion="Fail to reject the null hypothesis, there is not enough evidence to support
the alternative hypothesis, at the significance level of "<>ToString[α]
    ];

(* P value approach: *)
(* Step 7. Calculate the P-value: *)
pValue=2*(1-CDF[NormalDistribution[0,1],Abs[testStatistic]]);

(* Step 8. Compare the P-value to the significance level: *)
pvalueapproach=If[
    pValue<α,
    (* Reject the null hypothesis: *)
    conclusion="Reject the null hypothesis, there is sufficient evidence to support the
alternative hypothesis, at the significance level of "<>ToString[α],
    (* Fail to reject the null hypothesis: *)
    conclusion="Fail to reject the null hypothesis, there is not enough evidence to support
the alternative hypothesis, at the significance level of "<>ToString[α]
    ];

(* Step 9. Draw a conclusion based on the results: *)
Print["Sample Mean: ",xbar];
Print["Test Statistic: ",testStatistic];
Print["Critical Value: ",zCritical];
Print["p-value: ",pValue];
Print["Conclusion using critical value approach: ",criticalvalueapproach];
Print["Conclusion using P value approach: ",pvalueapproach];

{16.2077,9.12947,11.6617,9.59791,2.43496,11.6059,1.1145,17.7699,11.1671,3.60819,1.48501,1.12
807,16.7492,14.2961,11.9865,11.6247,17.0985,6.01639,17.6307,15.5132,4.95817,13.5086,9.5237,1
1.102,8.58062,20.082,7.63687,5.58636,1.59344,10.4334}
 Sample Mean:  10.0277
 Test Statistic:  0.0303455
 Critical Value:  1.95996
 p-value:  0.975791
 Conclusion using critical value approach:  Fail to reject the null hypothesis, there is
not enough evidence to support the alternative hypothesis, at the significance level of
0.05
 Conclusion using P value approach:  Fail to reject the null hypothesis, there is not
enough evidence to support the alternative hypothesis, at the significance level of 0.05
```

***Example 20.9***

```
(* Left tailed case: *)

(* The code demonstrates the steps for conducting a hypothesis test for one population mean
when the population standard deviation (σ) is known, n>30: *)

(* data: *)
μ0=10;
σ0=5;

(* Step 1. State the null and alternative hypotheses: *)
```

```
"H0:μ==μ0"; (* Null hypothesis:Population mean is greater than or equal to μ0. *)
"H1:μ<μ0"; (* Alternative hypothesis:Population mean is less than μ0. *)

(* Step 2. Set the significance level (α): *)
α=0.05; (* Significance level of 0.05. *)

(* Step 3. Collect a random sample and calculate the sample mean: *)
SeedRandom[123];
sample=RandomVariate[NormalDistribution[10,5],30]
n=Length[sample]; (* Sample size. *)
xbar=N[Mean[sample]]; (*S ample mean. *)

(* Critical value approach: *)
(* Step 4. Determine the critical value(s): *)
zCritical=Quantile[NormalDistribution[0,1],α]; (* For a left-tailed test. *)

(* Step 5. Calculate the test statistic: *)
σ=σ0; (* Known population standard deviation. *)
testStatistic=N[(xbar-μ0)/(σ/Sqrt[n])];

(* Step 6. Compare the test statistic to the critical value(s): *)
criticalvalueapproach=If[
    testStatistic<zCritical,
    (* Reject the null hypothesis: *)
    conclusion="Reject the null hypothesis, there is sufficient evidence to support the
alternative hypothesis, at the significance level of "<>ToString[α],
    (* Fail to reject the null hypothesis: *)
    conclusion="Fail to reject the null hypothesis, there is not enough evidence to support
the alternative hypothesis, at the significance level of "<>ToString[α]
    ];

(* Step 7. Calculate the p-value: *)
pValue=N[CDF[NormalDistribution[0,1],testStatistic]];
(* Step 8. Compare the p-value to the significance level: *)
pvalueapproach=If[
    pValue<α,
    (* Reject the null hypothesis: *)
    conclusion="Reject the null hypothesis, there is sufficient evidence to support the
alternative hypothesis, at the significance level of "<>ToString[α],
    (* Fail to reject the null hypothesis: *)
    conclusion="Fail to reject the null hypothesis, there is not enough evidence to support
the alternative hypothesis, at the significance level of "<>ToString[α]
    ];
(* Step 9. Draw a conclusion based on the results: *)
Print["Sample Mean: ",xbar];
Print["Test Statistic: ",testStatistic];
Print["Critical Value: ",zCritical];
Print["p-value: ",pValue];
Print["Conclusion using critical value approach: ",criticalvalueapproach];
Print["Conclusion using P value approach: ",pvalueapproach];

{16.2077,9.12947,11.6617,9.59791,2.43496,11.6059,1.1145,17.7699,11.1671,3.60819,1.48501,1.12
807,16.7492,14.2961,11.9865,11.6247,17.0985,6.01639,17.6307,15.5132,4.95817,13.5086,9.5237,1
1.102,8.58062,20.082,7.63687,5.58636,1.59344,10.4334}
 Sample Mean:  10.0277
 Test Statistic:  0.0303455
 Critical Value:  -1.64485
 p-value:  0.512104
```

```
 Conclusion using critical value approach:  Fail to reject the null hypothesis, there is not enough evidence to support the alternative hypothesis, at the significance level of 0.05
 Conclusion using P value approach:  Fail to reject the null hypothesis, there is not enough evidence to support the alternative hypothesis, at the significance level of 0.05
```

***Example 20.10***

```
(* Right tailed case: *)

(* The code demonstrates the steps for conducting a hypothesis test for one population mean when the population standard deviation (σ) is known, n>30: *)

(* data: *)
μ0=10;
σ0=5;

(* Step 1. State the null and alternative hypotheses: *)
"H0:μ<=μ0"; (* Null hypothesis:Population mean is less than or equal to μ0. *)
"H1:μ>μ0"; (* Alternative hypothesis:Population mean is greater than μ0. *)

(* Step 2. Set the significance level (α): *)
α=0.05; (* Significance level of 0.05.*)

(* Step 3. Collect a random sample and calculate the sample mean: *)
SeedRandom[123];
sample=RandomVariate[NormalDistribution[10,5],30]
n=Length[sample]; (* Sample size. *)
xbar=Mean[sample]; (* Sample mean. *)

(* Critical value approach: *)
(* Step 4. Determine the critical value(s): *)
zCritical=Quantile[NormalDistribution[0,1],1-α]; (* For a right-tailed test. *)

(* Step 5. Calculate the test statistic: *)
σ=σ0; (* Known population standard deviation. *)
testStatistic=(xbar-μ0)/(σ/Sqrt[n]);

(* Step 6. Compare the test statistic to the critical value(s): *)
criticalvalueapproach=If[
    testStatistic>zCritical,
    (* Reject the null hypothesis: *)
    conclusion="Reject the null hypothesis, there is sufficient evidence to support the alternative hypothesis, at the significance level of "<>ToString[α],
    (* Fail to reject the null hypothesis: *)
    conclusion="Fail to reject the null hypothesis, there is not enough evidence to support the alternative hypothesis, at the significance level of "<>ToString[α]
    ];

(* Step 7. Calculate the p-value: *)
pValue=1-CDF[NormalDistribution[0,1],testStatistic];
(* Step 8. Compare the p-value to the significance level: *)
pvalueapproach=If[
    pValue<α,
    (* Reject the null hypothesis: *)
    conclusion="Reject the null hypothesis, there is sufficient evidence to support the alternative hypothesis, at the significance level of "<>ToString[α],
    (* Fail to reject the null hypothesis: *)
```

```
    conclusion="Fail to reject the null hypothesis, there is not enough evidence to support the alternative hypothesis, at the significance level of "<>ToString[α]
    ];
(* Step 9. Draw a conclusion based on the results: *)
Print["Sample Mean: ",xbar];
Print["Test Statistic: ",testStatistic];
Print["Critical Value: ",zCritical];
Print["p-value: ",pValue];
Print["Conclusion using critical value approach: ",criticalvalueapproach];
Print["Conclusion using P value approach: ",pvalueapproach];

{16.2077,9.12947,11.6617,9.59791,2.43496,11.6059,1.1145,17.7699,11.1671,3.60819,1.48501,1.12807,16.7492,14.2961,11.9865,11.6247,17.0985,6.01639,17.6307,15.5132,4.95817,13.5086,9.5237,11.102,8.58062,20.082,7.63687,5.58636,1.59344,10.4334}
 Sample Mean:  10.0277
 Test Statistic:  0.0303455
 Critical Value:  1.64485
 p-value:  0.487896
 Conclusion using critical value approach:  Fail to reject the null hypothesis, there is not enough evidence to support the alternative hypothesis, at the significance level of 0.05
 Conclusion using P value approach:  Fail to reject the null hypothesis, there is not enough evidence to support the alternative hypothesis, at the significance level of 0.05
```

***Example 20.11***

```
(* General case: *)

(* The code demonstrates the steps for conducting a hypothesis test for one population mean when the population standard deviation s is unknown, n>30. The code will output the sample mean, test statistic, critical value, p-value, and the conclusion based on the user-specified type of hypothesis test: *)

(* Specify the type of hypothesis test, choose one of the following,  "two-tailed","left-tailed", or "right-tailed": *)
hypothesisType="two-tailed";
μ0=10;

(* Step 1. State the null and alternative hypotheses: *)
"H0:μ==μ0"; (* Null hypothesis:Population mean is equal to μ0. *)
alternativeHypothesis=Switch[
    hypothesisType,
    "two-tailed",
    "H1:μ!=μ0",
    "left-tailed",
    "H1:μ<μ0",
    "right-tailed",
    "H1:μ>μ0"
    ];

(* Step 2. Set the significance level (α): *)
α=0.05; (* Significance level of 0.05. *)

(* Step 3. Collect random samples from the populations and calculate sample mean, sample standard deviation, and sample size: *)
SeedRandom[123];
sample=RandomVariate[NormalDistribution[10,5],30]
n=Length[sample]; (* Sample size. *)
xbar=Mean[sample]; (* Sample mean. *)
```

```
s=StandardDeviation[sample]; (* Sample standard deviation. *)

(* Critical value approach: *)
(* Step 4. Determine the critical value(s): *)
criticalValue=Switch[
    hypothesisType,
    "two-tailed",
    Quantile[NormalDistribution[0,1],1-α/2],
    "left-tailed",
    Quantile[NormalDistribution[0,1],α],
    "right-tailed",
    Quantile[NormalDistribution[0,1],1-α]
    ];

(* Step 5. Calculate the test statistic: *)
testStatistic=(xbar-μ0)/(s/Sqrt[n]);

(* Step 6. Compare the test statistic to the critical value(s): *)
criticalvalueapproach=Switch[
    hypothesisType,
    "two-tailed",
    If[
     Abs[testStatistic]>criticalValue,
     (* Reject the null hypothesis: *)
     conclusion="Reject the null hypothesis, there is sufficient evidence to support the
alternative hypothesis, at the significance level of "<>ToString[α],
     (* Fail to reject the null hypothesis: *)
     conclusion="Fail to reject the null hypothesis, there is not enough evidence to support
the alternative hypothesis, at the significance level of "<>ToString[α]
     ],
    "left-tailed",
    If[
     testStatistic<criticalValue,
     (* Reject the null hypothesis: *)
     conclusion="Reject the null hypothesis, there is sufficient evidence to support the
alternative hypothesis, at the significance level of "<>ToString[α],
     (* Fail to reject the null hypothesis: *)
     conclusion="Fail to reject the null hypothesis, there is not enough evidence to support
the alternative hypothesis, at the significance level of "<>ToString[α]
     ],
    "right-tailed",
    If[
     testStatistic>criticalValue,
     (* Reject the null hypothesis: *)
     conclusion="Reject the null hypothesis, there is sufficient evidence to support the
alternative hypothesis, at the significance level of "<>ToString[α],
     (* Fail to reject the null hypothesis: *)
     conclusion="Fail to reject the null hypothesis, there is not enough evidence to support
the alternative hypothesis, at the significance level of "<>ToString[α]
     ]
    ];
(* P value approach: *)
(* Step 7. Calculate the p-value: *)
pValue=
   Switch[
    hypothesisType,
    "two-tailed",
    2*(1-CDF[NormalDistribution[0,1],Abs[testStatistic]]),
    "left-tailed",
    CDF[NormalDistribution[0,1],testStatistic],
```

```
    "right-tailed",
    1-CDF[NormalDistribution[0,1],testStatistic]
    ];

(* Step 8. Compare the P-value to the significance level: *)
pvalueapproach=If[
    pValue<α,
    (* Reject the null hypothesis: *)
    conclusion="Reject the null hypothesis, there is sufficient evidence to support the alternative hypothesis, at the significance level of "<>ToString[α],
    (* Fail to reject the null hypothesis: *)
    conclusion="Fail to reject the null hypothesis, there is not enough evidence to support the alternative hypothesis, at the significance level of "<>ToString[α]
    ];

(* Step 9. Draw a conclusion based on the results: *)
Print["Sample Mean: ",xbar];
Print["Test Statistic: ",testStatistic];
Print["Critical Value: ",criticalValue];
Print["p-value: ",pValue];
Print["Conclusion using critical value approach: ",criticalvalueapproach];
Print["Conclusion using P value approach: ",pvalueapproach];

(* Built-in function in Mathematica: *)
LocationTest[
  sample,
  10,
  {"TestDataTable",All},
  Switch[
   hypothesisType,
   "two-tailed",
   AlternativeHypothesis->"Unequal",
   "left-tailed",
   AlternativeHypothesis->"Less",
   "right-tailed",
   AlternativeHypothesis->"Greater"
   ]
  ]

{16.2077,9.12947,11.6617,9.59791,2.43496,11.6059,1.1145,17.7699,11.1671,3.60819,1.48501,1.12807,16.7492,14.2961,11.9865,11.6247,17.0985,6.01639,17.6307,15.5132,4.95817,13.5086,9.5237,11.102,8.58062,20.082,7.63687,5.58636,1.59344,10.4334}
 Sample Mean:  10.0277
 Test Statistic:  0.0272512
 Critical Value:  1.95996
 p-value:  0.978259
 Conclusion using critical value approach:  Fail to reject the null hypothesis, there is not enough evidence to support the alternative hypothesis, at the significance level of 0.05
 Conclusion using P value approach:  Fail to reject the null hypothesis, there is not enough evidence to support the alternative hypothesis, at the significance level of 0.05
```

| | Statistic | P-Value |
|---|---|---|
| Paired T | 0.0272512 | 0.978446 |
| Paired Z | 0.0272512 | 0.978259 |
| Sign | 16 | 0.855536 |
| Signed-Rank | 238. | 0.918089 |
| T | 0.0272512 | 0.978446 |
| Z | 0.0272512 | 0.978259 |

## 20.3 Small Sample Hypothesis Tests for One Population Mean When $\sigma$ Is Unknown

| | | | |
|---|---|---|---|
| Assumptions | 1. Simple random sample<br>2. Normal population and $n < 30$<br>3. $\sigma$ unknown | | |
| Null hypothesis | $H_0: \mu = \mu_0$ | | |
| | Two-tailed | Left tailed | Right tailed |
| Alternative hypothesis | $H_a: \mu \neq \mu_0$ | $H_a: \mu < \mu_0$ | $H_a: \mu > \mu_0$ |
| Significance level | $\alpha$ | | |
| Test statistic:<br>For details see Chapter 18. | $t_0 = \dfrac{\bar{x} - \mu_0}{s/\sqrt{n}}$ | | |
| **Critical value approach** | $\pm t_{\alpha/2}$ | $-t_\alpha$ | $t_\alpha$ |

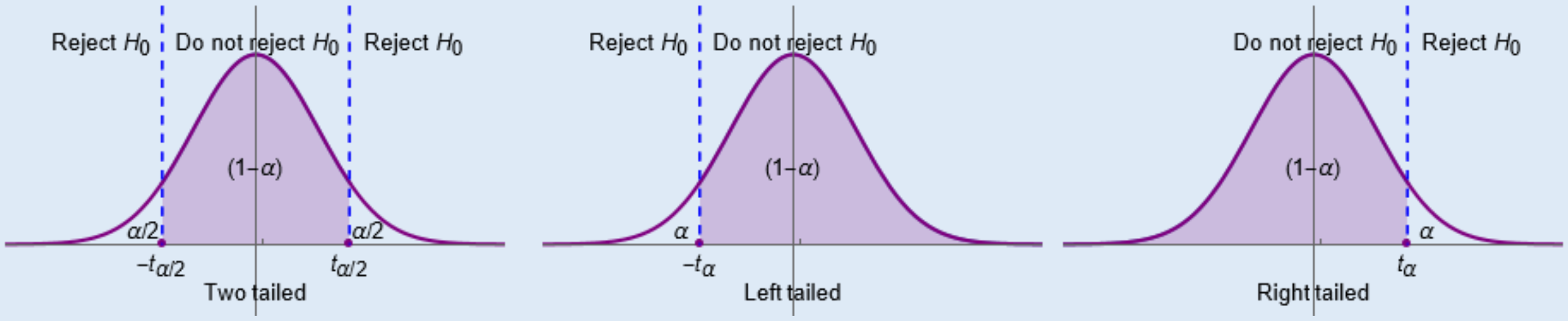

$$\begin{cases} \text{reject } H_0, & \text{If the value of the test statistic falls in the rejection region} \\ \text{do not reject } H_0, & \text{otherwise} \end{cases}$$

***P* value approach**

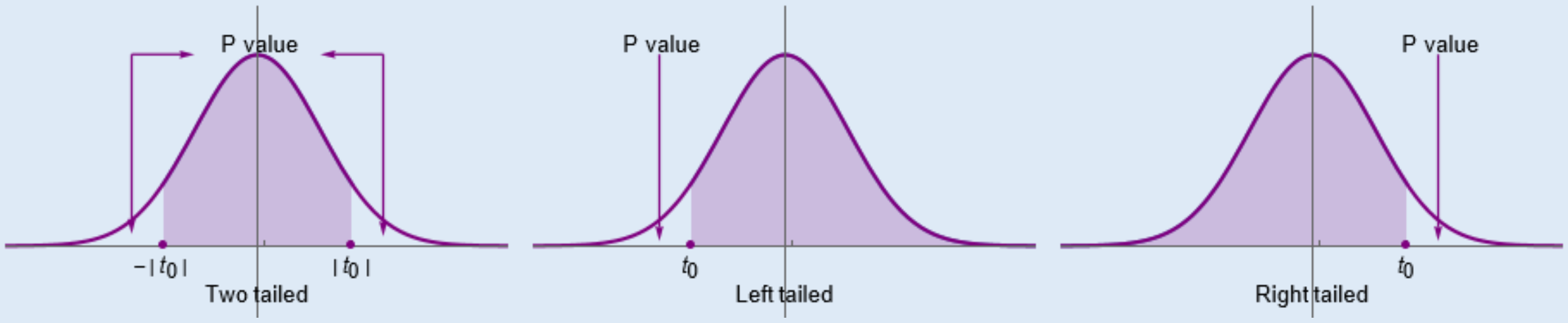

$$\begin{cases} \text{reject } H_0, & \text{If } P \le \alpha \\ \text{do not reject } H_0, & \text{otherwise} \end{cases}$$

**Procedure 20.3.**
The steps for conducting a hypothesis test for one population mean when the population standard deviation ($\sigma$) is unknown (small sample):

1. State the null and alternative hypotheses:
   - $H_0$: The population mean is equal to a specified value.
   - $H_a$: The population mean is not equal to the specified value (two-tailed test), or it is greater than/less than the specified value (one-tailed test).
2. Set the significance level, $\alpha$, for the test.

3. Collect a random sample from the population and calculate the sample mean ($\bar{x}$) and the sample standard deviation ($s$).
4. Determine the critical value(s) for the test statistic.
   - For a two-tailed test, divide the significance level by 2 and find the corresponding $t$-score(s) using Mathematica for t-distribution with $n-1$ degrees of freedom.
   - For a one-tailed test, find the $t$-score corresponding to the desired tail area.
5. Calculate the test statistic using the formula:
   For a $t$-test:
   $$t = \frac{\bar{x} - \mu_0}{s/\sqrt{n}},$$
   where $\bar{x}$ is the sample mean, $\mu_0$ is the hypothesized population mean, $s$ is the sample standard deviation, and $n$ is the sample size.
6. Compare the test statistic to the critical value(s).
   - If the test statistic falls within the critical region, reject the null hypothesis.
   - Otherwise, fail to reject the null hypothesis.
7. Calculate the $P$-value associated with the test statistic.
   - For a two-tailed test, calculate the area in both tails of the t-distribution with $n-1$ degrees of freedom.
   - For a one-tailed test, calculate the area in the appropriate tail.
8. Compare the $P$-value to $\alpha$.
   - If the $P$-value is less than $\alpha$, reject the null hypothesis.
   - Otherwise, fail to reject the null hypothesis.
9. Draw a conclusion based on the results:
   - If the null hypothesis is rejected, conclude that there is sufficient evidence to support the alternative hypothesis.
   - If the null hypothesis is not rejected, conclude that there is not enough evidence to support the alternative hypothesis.

***Example 20.12***

```
(* The code demonstrates the steps for conducting a hypothesis test for one population mean
when the population standard deviation (σ) is unknown n<30. The code will output the sample
mean, sample standard deviation, test statistic, critical value, p-value,and the conclusion
based on the user-specified type of hypothesis test: *)

(* Specify the type of hypothesis test, choose one of the following, "two-tailed","left-
tailed", or "right-tailed": *)
hypothesisType="two-tailed";
μ0=10;

(* Step 1. State the null and alternative hypotheses: *)

"H0:μ==μ0"; (* Null hypothesis:Population mean is equal to μ0. *)
alternativeHypothesis=Switch[
    hypothesisType,
    "two-tailed",
    "H1:μ!=μ0",
    "left-tailed",
    "H1:μ<μ0",
    "right-tailed",
    "H1:μ>μ0"
    ];

(* Step 2. Set the significance level (α): *)
α=0.05; (* Significance level of 0.05.*)
```

```
(* Step 3. Collect random samples from the populations and calculate sample mean, sample
standard deviation, and sample size: *)
SeedRandom[2123];
sample=RandomVariate[NormalDistribution[10,5],20]
n=Length[sample]; (* Sample size. *)
xbar=Mean[sample]; (* Sample mean. *)
s=StandardDeviation[sample]; (* Sample standard deviation. *)

(* Critical value approach: *)
(* Step 4. Determine the critical value(s): *)
criticalValue=Switch[
    hypothesisType,
    "two-tailed",
    Quantile[StudentTDistribution[n-1],1-α/2],
    "left-tailed",
    Quantile[StudentTDistribution[n-1],α],
    "right-tailed",
    Quantile[StudentTDistribution[n-1],1-α]
    ];

(* Step 5. Calculate the test statistic: *)
testStatistic=(xbar-μ0)/(s/Sqrt[n]);

(* Step 6. Compare the test statistic to the critical value(s): *)
criticalvalueapproach=Switch[
    hypothesisType,
    "two-tailed",
    If[
     Abs[testStatistic]>criticalValue,
     (* Reject the null hypothesis: *)
     conclusion="Reject the null hypothesis, there is sufficient evidence to support the
alternative hypothesis, at the significance level of "<>ToString[α],
     (* Fail to reject the null hypothesis: *)
     conclusion="Fail to reject the null hypothesis, there is not enough evidence to support
the alternative hypothesis, at the significance level of "<>ToString[α]
     ],
    "left-tailed",
    If[
     testStatistic<criticalValue,
     (* Reject the null hypothesis: *)
     conclusion="Reject the null hypothesis, there is sufficient evidence to support the
alternative hypothesis, at the significance level of "<>ToString[α],
     (* Fail to reject the null hypothesis: *)
     conclusion="Fail to reject the null hypothesis, there is not enough evidence to support
the alternative hypothesis, at the significance level of "<>ToString[α]
     ],
    "right-tailed",
    If[
     testStatistic>criticalValue,
     (* Reject the null hypothesis: *)
     conclusion="Reject the null hypothesis, there is sufficient evidence to support the
alternative hypothesis, at the significance level of "<>ToString[α],
     (* Fail to reject the null hypothesis: *)
     conclusion="Fail to reject the null hypothesis, there is not enough evidence to support
the alternative hypothesis, at the significance level of "<>ToString[α]
     ]
    ];

(* P value approach: *)
(* Step 7. Calculate the p-value: *)
```

```
pValue=
  Switch[
   hypothesisType,
   "two-tailed",
   2*(1-CDF[StudentTDistribution[n-1],Abs[testStatistic]]),
   "left-tailed",
   CDF[StudentTDistribution[n-1],testStatistic],
   "right-tailed",
   1-CDF[StudentTDistribution[n-1],testStatistic]
   ];

(* Step 8. Compare the P-value to the significance level: *)
pvalueapproach=If[
   pValue<α,
   (* Reject the null hypothesis: *)
   conclusion="Reject the null hypothesis, there is sufficient evidence to support the
alternative hypothesis, at the significance level of "<>ToString[α],
   (* Fail to reject the null hypothesis: *)
   conclusion="Fail to reject the null hypothesis, there is not enough evidence to support
the alternative hypothesis, at the significance level of "<>ToString[α]
   ];

(* Step 9. Draw a conclusion based on the results: *)
Print["Sample Mean: ",xbar];
Print["Sample Standard Deviation: ",s];
Print["Test Statistic: ",testStatistic];
Print["Critical Value: ",criticalValue];
Print["p-value: ",pValue];
Print["Conclusion using critical value approach: ",criticalvalueapproach];
Print["Conclusion using P value approach: ",pvalueapproach];

(* Built-in function in Mathematica: *)
LocationTest[
 sample,
 10,
 {"TestDataTable",All},
 Switch[
  hypothesisType,
  "two-tailed",
  AlternativeHypothesis->"Unequal",
  "left-tailed",
  AlternativeHypothesis->"Less",
  "right-tailed",
  AlternativeHypothesis->"Greater"
  ]
 ]

{9.58126,17.6647,7.97135,4.82482,8.16712,14.6088,14.4195,12.2207,12.0199,5.68029,7.69469,14.
7929,11.8017,7.23674,19.3351,14.2367,10.0136,12.3878,13.2001,4.98618}
 Sample Mean:  11.1422
 Sample Standard Deviation:  4.10952
 Test Statistic:  1.24299
 Critical Value:  2.09302
 p-value:  0.228994
 Conclusion using critical value approach:  Fail to reject the null hypothesis, there is
not enough evidence to support the alternative hypothesis, at the significance level of
0.05
 Conclusion using P value approach:  Fail to reject the null hypothesis, there is not
enough evidence to support the alternative hypothesis, at the significance level of 0.05
```

| | Statistic | P-Value |
|---|---|---|
| Paired T | 1.24299 | 0.228994 |
| Paired Z | 1.24299 | 0.213873 |
| Sign | 12 | 0.503445 |
| Signed-Rank | 132. | 0.322509 |
| T | 1.24299 | 0.228994 |
| Z | 1.24299 | 0.213873 |

## 20.4 Large-Sample Hypothesis Tests for Differences between Two Means, $\sigma_1^2$ and $\sigma_2^2$ Known (Unknown)

| | | | |
|---|---|---|---|
| Assumptions | 1. The samples are independent<br>2. Simple random samples<br>3. Normal population or large sample<br>4. $\sigma$ known (unknown) | | |
| Null hypothesis | $H_0: \mu_1 - \mu_2 = \delta_0$ | | |
| | Two-tailed | Left tailed | Right tailed |
| Alternative hypothesis | $H_a: \mu_1 - \mu_2 \neq \delta_0$ | $H_a: \mu_1 - \mu_2 < \delta_0$ | $H_a: \mu_1 - \mu_2 > \delta_0$ |
| Significance level | $\alpha$ | | |
| Test statistic:<br>For details see Chapter 18. | $z = \dfrac{(\bar{x}_1 - \bar{x}_2) - \delta_0}{\sqrt{\dfrac{\sigma_1^2}{n_1} + \dfrac{\sigma_2^2}{n_2}}}$,<br>($\sigma$ known) | $z = \dfrac{(\bar{x}_1 - \bar{x}_2) - \delta_0}{\sqrt{\dfrac{s_1^2}{n_1} + \dfrac{s_2^2}{n_2}}}$<br>($\sigma$ unknown) | |
| **Critical value approach** | $\pm z_{\alpha/2}$ | $-z_{\alpha}$ | $z_{\alpha}$ |

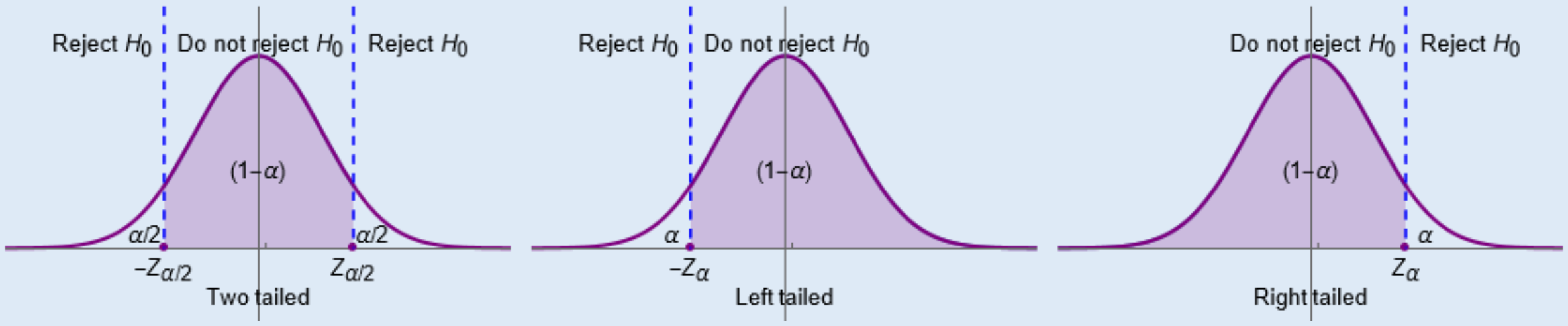

$$\begin{cases} \text{reject } H_0, & \text{If the value of the test statistic falls in the rejection region} \\ \text{do not reject } H_0, & \text{otherwise} \end{cases}$$

***P* value approach**

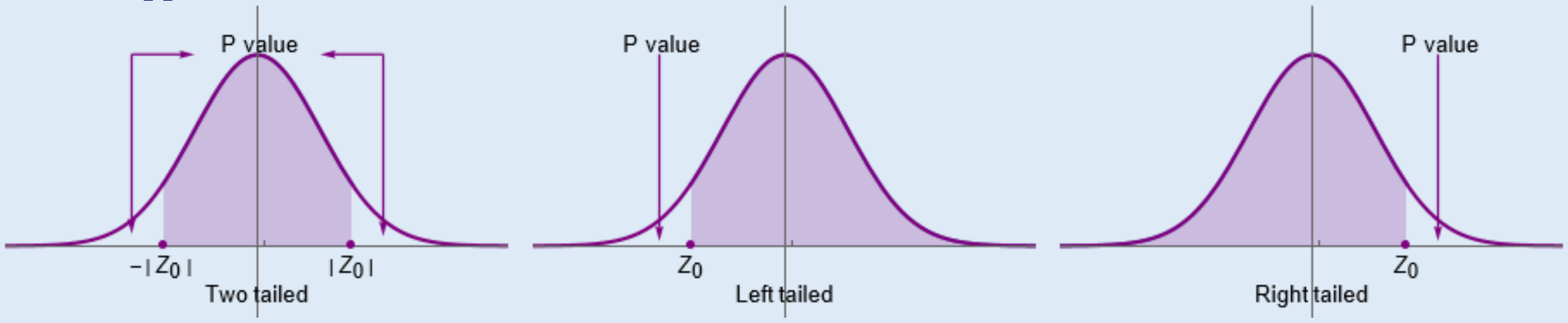

$$\begin{cases} \text{reject } H_0, & \text{If } P \leq \alpha \\ \text{do not reject } H_0, & \text{otherwise} \end{cases}$$

**Procedure 20.4.**

The steps for conducting a large sample hypothesis test for the differences between two means when the population standard deviations ($\sigma_1^2$ and $\sigma_2^2$) are known (unknown):

1. State the null and alternative hypotheses:
   - $H_0$: The difference between the population means is equal to a specified value ($\mu_1 - \mu_2 = \delta_0$).
   - $H_1$: The difference between the population means is not equal to the specified value ($\mu_1 - \mu_2 \neq \delta_0$, two-tailed test), or it is greater than/less than the specified value ($\mu_1 - \mu_2 > \delta_0$ or $\mu_1 - \mu_2 < \delta_0$, one-tailed test).
2. Set the significance level, $\alpha$, for the test.
3. Collect random samples from the two populations and calculate the sample means ($\bar{x}_1$ and $\bar{x}_2$), the sample standard deviations ($s_1$ and $s_2$), and the sample sizes ($n_1$ and $n_2$).
4. Determine the critical value(s) for the test statistic.
   - For a two-tailed test, divide the significance level by 2 and find the corresponding $z$-score(s) using the Mathematica standard normal distribution.
   - For a one-tailed test, find the z-score corresponding to the desired tail area.
5. Calculate the test statistic using the formula:
   For a $z$-test:
   $$z = \frac{(\bar{x}_1 - \bar{x}_2) - \delta_0}{\sqrt{\frac{\sigma_1^2}{n_1} + \frac{\sigma_2^2}{n_2}}}, (\sigma_1^2 \text{ and } \sigma_2^2 \text{ known}), \qquad z = \frac{(\bar{x}_1 - \bar{x}_2) - \delta_0}{\sqrt{\frac{s_1^2}{n_1} + \frac{s_2^2}{n_2}}}, (\sigma_1^2 \text{ and } \sigma_2^2 \text{ unknown})$$
   where $\bar{x}_1$ and $\bar{x}_2$ are the sample means, $\delta_0$ is the specified difference, $\sigma_1^2$ and $\sigma_2^2$ are the population variances, $n_1$ and $n_2$ are the sample sizes, and the sample standard deviations ($s_1$ and $s_2$),.
6. Compare the test statistic to the critical value(s).
   - If the test statistic falls within the critical region, reject the null hypothesis.
   - Otherwise, fail to reject the null hypothesis.
7. Calculate the $P$-value associated with the test statistic.
   - For a two-tailed test, calculate the area in both tails of the distribution.
   - For a one-tailed test, calculate the area in the appropriate tail.
8. Compare the $P$-value to $\alpha$.
   - If the $P$-value is less than $\alpha$, reject the null hypothesis.
   - Otherwise, fail to reject the null hypothesis.
9. Draw a conclusion based on the results.
   - If the null hypothesis is rejected, conclude that there is sufficient evidence to support the alternative hypothesis.
   - If the null hypothesis is not rejected, conclude that there is not enough evidence to support the alternative hypothesis.

***Example 20.13***

```
(* The code demonstrates the steps for conducting a hypothesis test for the differences
between two means. The code will output the sample means, sample standard deviations, test
statistic, critical value, p-value,and the conclusion based on the user-specified type of
hypothesis test: *)

(* Specify the type of hypothesis test, choose one of the following, "two-tailed","left-
tailed", or "right-tailed": *)
hypothesisType="two-tailed";
δ0=10;

(* Step 1. State the null and alternative hypotheses: *)

"H0:μ1-μ2==δ0"; (* Null hypothesis:Difference between population means is equal to δ0: *)
```

```
alternativeHypothesis=Switch[
    hypothesisType,
    "two-tailed",
    "H1:μ1-μ2!=δ0",
    "left-tailed",
    "H1:μ1-μ2<δ0",
    "right-tailed",
    "H1:μ1-μ2>δ0"
    ];

(* Step 2. Set the significance level (α): *)
α=0.05; (* Significance level of 0.05: *)

(* Step 3. Collect random samples from the populations and calculate sample means,sample
standard deviations, and sample sizes:*)
SeedRandom[3123];
sample1=RandomVariate[NormalDistribution[80,20],40]
sample2=RandomVariate[NormalDistribution[70,15],30]
n1=Length[sample1]; (* Sample size for population 1. *)
n2=Length[sample2]; (* Sample size for population 2. *)
xbar1=Mean[sample1]; (* Sample mean for population 1. *)
xbar2=Mean[sample2]; (* Sample mean for population 2. *)
s1=StandardDeviation[sample1]; (* Sample standard deviation for population 1. *)
s2=StandardDeviation[sample2]; (* Sample standard deviation for population 2. *)

(* Critical value approach: *)
(* Step 4. Determine the critical value(s) and alternative hypotheses: *)
criticalValue=Switch[
    hypothesisType,
    "two-tailed",
    Quantile[NormalDistribution[0,1],1-α/2],
    "left-tailed",
    Quantile[NormalDistribution[0,1],α],
    "right-tailed",
    Quantile[NormalDistribution[0,1],1-α]
    ];

(* Step 5. Calculate the test statistic: *)
testStatistic=((xbar1-xbar2)-δ0)/Sqrt[(s1^2/n1)+(s2^2/n2)];

(* Step 6. Compare the test statistic to the critical value(s): *)
criticalvalueapproach=Switch[
    hypothesisType,
    "two-tailed",
    If[
     Abs[testStatistic]>criticalValue,
     (* Reject the null hypothesis: *)
     conclusion="Reject the null hypothesis, there is sufficient evidence to support the
alternative hypothesis, at the significance level of "<>ToString[α],
     (* Fail to reject the null hypothesis: *)
     conclusion="Fail to reject the null hypothesis, there is not enough evidence to support
the alternative hypothesis, at the significance level of "<>ToString[α]
     ],
    "left-tailed",
    If[
     testStatistic<criticalValue,
     (* Reject the null hypothesis: *)
     conclusion="Reject the null hypothesis, there is sufficient evidence to support the
alternative hypothesis, at the significance level of "<>ToString[α],
     (* Fail to reject the null hypothesis: *)
```

```
    conclusion="Fail to reject the null hypothesis, there is not enough evidence to support
the alternative hypothesis, at the significance level of "<>ToString[α]
    ],
   "right-tailed",
   If[
    testStatistic>criticalValue,
    (* Reject the null hypothesis: *)
    conclusion="Reject the null hypothesis, there is sufficient evidence to support the
alternative hypothesis, at the significance level of "<>ToString[α],
    (* Fail to reject the null hypothesis: *)
    conclusion="Fail to reject the null hypothesis, there is not enough evidence to support
the alternative hypothesis, at the significance level of "<>ToString[α]
    ]
   ];

(* P value approach: *)
(* Step 7. Calculate the p-value: *)
pValue=
  Switch[
   hypothesisType,
   "two-tailed",
   2*(1-CDF[NormalDistribution[0,1],Abs[testStatistic]]),
   "left-tailed",
   CDF[NormalDistribution[0,1],testStatistic],
   "right-tailed",
   1-CDF[NormalDistribution[0,1],testStatistic]
   ];

(* Step 8. Compare the P-value to the significance level: *)
pvalueapproach=If[
   pValue<α,
   (* Reject the null hypothesis: *)
   conclusion="Reject the null hypothesis, there is sufficient evidence to support the
alternative hypothesis, at the significance level of "<>ToString[α],
   (* Fail to reject the null hypothesis: *)
   conclusion="Fail to reject the null hypothesis, there is not enough evidence to support
the alternative hypothesis, at the significance level of "<>ToString[α]
   ];

(* Step 9. Draw a conclusion based on the results: *)
Print["Sample 1 Mean: ",xbar1];
Print["Sample 2 Mean: ",xbar2];
Print["Sample 1 Standard Deviation: ",s1];
Print["Sample 2 Standard Deviation: ",s2];
Print["Test Statistic: ",testStatistic];
Print["Critical Value: ",criticalValue];
Print["p-value: ",pValue];
Print["Conclusion using critical value approach: ",criticalvalueapproach];
Print["Conclusion using P value approach: ",pvalueapproach];

(* Built-in function in Mathematica: *)
LocationTest[
 {sample1,sample2},
 10,
 {"TestDataTable",All},
 Switch[
  hypothesisType,
  "two-tailed",
  AlternativeHypothesis->"Unequal",
  "left-tailed",
```

```
   AlternativeHypothesis->"Less",
   "right-tailed",
   AlternativeHypothesis->"Greater"
   ]
  ]

{79.3279,80.3181,91.3605,90.4478,53.352,97.242,76.9837,89.6453,61.4194,66.3222,55.7355,104.0
84,85.063,71.3139,77.1027,87.6114,74.5493,68.3002,61.5715,119.571,100.425,53.3078,68.3738,84
.3623,117.101,92.1009,114.006,67.6695,73.5109,106.598,81.7854,82.0673,67.8093,98.4223,98.212
6,79.5442,91.1059,95.786,107.332,88.4231}

{77.157,65.7694,83.4907,76.5516,54.1441,86.9657,66.2608,48.7735,58.7336,67.2011,54.4628,72.0
166,61.0868,61.0107,56.9808,47.9187,60.5188,55.1392,90.028,74.4308,75.348,49.4431,60.8854,60
.5018,60.7718,78.6525,71.6264,72.2868,61.4825,59.3553}

 Sample 1 Mean:  83.9816
 Sample 2 Mean:  65.6331
 Sample 1 Standard Deviation:  17.1998
 Sample 2 Standard Deviation:  11.1927
 Test Statistic:  2.45419
 Critical Value:  1.95996
 p-value:  0.01412
 Conclusion using critical value approach:  Reject the null hypothesis, there is
sufficient evidence to support the alternative hypothesis, at the significance level of
0.05
 Conclusion using P value approach:  Reject the null hypothesis, there is sufficient
evidence to support the alternative hypothesis, at the significance level of 0.05
```

| | Statistic | P-Value |
|---|---|---|
| Mann-Whitney | 778. | 0.0351577 |
| T | 2.45419 | 0.0167323 |
| Z | 2.45419 | 0.01412 |

## 20.5 Hypothesis Tests for the Means of Two Populations with Equal Standard Deviations

| | | | |
|---|---|---|---|
| Assumptions | 1. Simple random samples<br>2. Independent samples<br>3. Normal populations<br>4. Equal population standard deviations | | |
| Null hypothesis | $H_0: \mu_1 = \mu_2$ | | |
| Alternative hypothesis | Two-tailed<br>$H_a: \mu_1 \neq \mu_2$ | Left tailed<br>$H_a: \mu_1 < \mu_2$ | Right tailed<br>$H_a: \mu_1 > \mu_2$ |
| Significance level | $\alpha$ | | |
| Test statistic:<br>For details see Chapter 18. | $t_0 = \dfrac{\bar{x}_1 - \bar{x}_2}{s_p\sqrt{\dfrac{1}{n_1} + \dfrac{1}{n_2}}}$,<br>$s_p = \sqrt{\dfrac{(n_1 - 1)s_1^2 + (n_2 - 1)s_2^2}{(n_1 + n_2 - 2)}}$ | | |
| **Critical value approach** | $\pm t_{\alpha/2}$ | $-t_{\alpha}$ | $t_{\alpha}$ |

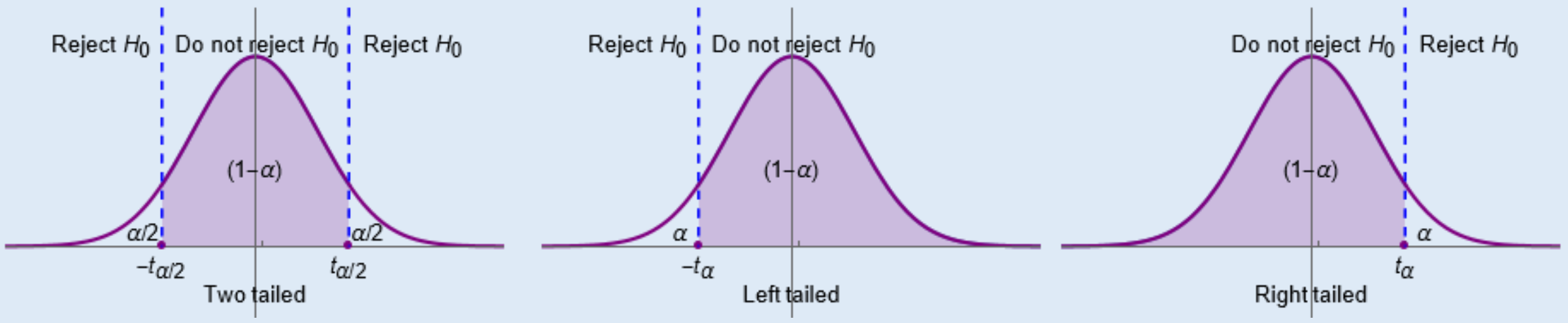

$$\begin{cases} \text{reject } H_0, & \text{If the value of the test statistic falls in the rejection region} \\ \text{do not reject } H_0, & \text{otherwise} \end{cases}$$

***P* value approach**

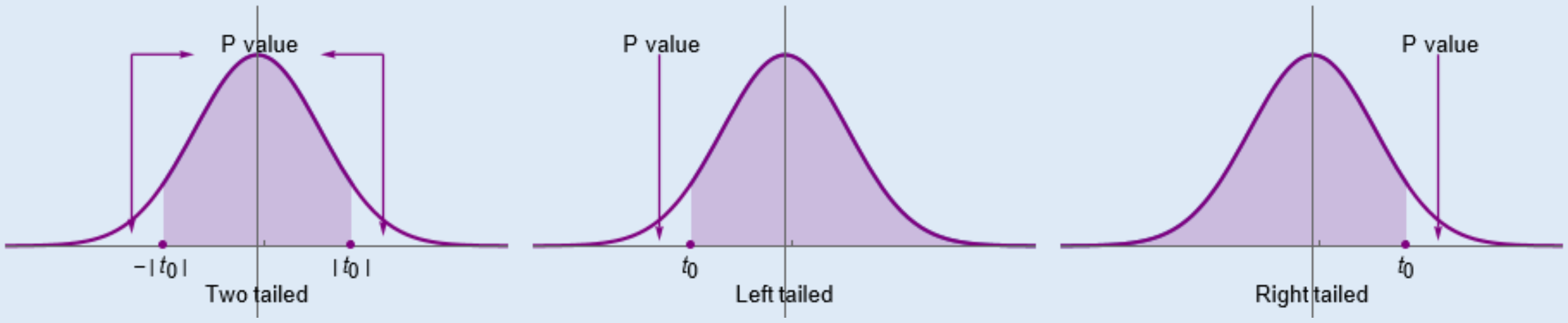

$$\begin{cases} \text{reject } H_0, & \text{If } P \leq \alpha \\ \text{do not reject } H_0, & \text{otherwise} \end{cases}$$

**Procedure 20.5.**
The steps for conducting a hypothesis test for the means of two populations with equal standard deviations:

1. State the null and alternative hypotheses:
   - $H_0$: The means of the two populations are equal ($\mu_1 = \mu_2$).
   - $H_1$: The means of the two populations are not equal ($\mu_1 \neq \mu_2$, two-tailed test), or one mean is greater than/less than the other ($\mu_1 > \mu_2$ or $\mu_1 < \mu_2$, one-tailed test).
2. Set the significance level, $\alpha$, for the test.
3. Collect random samples from both populations and calculate the sample means ($\bar{x}_1$ and $\bar{x}_2$), the sample standard deviations ($s_1$ and $s_2$), and the sample sizes ($n_1$ and $n_2$).
4. Calculate the pooled standard deviation (s) using the formula:
$$s_p = \sqrt{\frac{(n_1 - 1)s_1^2 + (n_2 - 1)s_2^2}{(n_1 + n_2 - 2)}}.$$
5. Calculate the test statistic using the formula:
For a t-test:
$$t_0 = \frac{\bar{x}_1 - \bar{x}_2}{s_p\sqrt{\frac{1}{n_1} + \frac{1}{n_2}}}.$$
6. Determine the degrees of freedom ($df$) for the t-distribution using the formula:
$$df = n_1 + n_2 - 2$$
7. Determine the critical value(s) for the test statistic.
   - For a two-tailed test, divide the significance level by 2 and find the corresponding t-score(s) from the Mathematica t-distribution.
   - For a one-tailed test, find the t-score corresponding to the desired tail area.
8. Compare the test statistic to the critical value(s).
   - If the test statistic falls within the critical region, reject the null hypothesis.
   - Otherwise, fail to reject the null hypothesis.

9. Calculate the $P$-value associated with the test statistic.
    - For a two-tailed test, calculate the area in both tails of the t-distribution with the degrees of freedom.
    - For a one-tailed test, calculate the area in the appropriate tail.
10. Compare the $P$-value to $\alpha$.
    - If the $P$-value is less than $\alpha$, reject the null hypothesis.
    - Otherwise, fail to reject the null hypothesis.
11. Draw a conclusion based on the results.
    - If the null hypothesis is rejected, conclude that there is sufficient evidence to support the alternative hypothesis.
    - If the null hypothesis is not rejected, conclude that there is not enough evidence to support the alternative hypothesis.

***Example 20.14***

```
(* The code demonstrates the steps for conducting a hypothesis test for the means of two
populations with equal standard deviations. The code will output the sample means, sample
standard deviations, test statistic, critical value, p-value, and the conclusion based on
the user-specified type of hypothesis test: *)

(* Specify the type of hypothesis test,choose one of the following, "two-tailed", "left-
tailed", or "right-tailed": *)
hypothesisType="two-tailed";

(* Step 1. State the null and alternative hypotheses: *)
(* Null hypothesis:Means of the two populations are equal: *)
"H0:μ1==μ2";

(* Alternative hypothesis: *)
alternativeHypothesis=Switch[
    hypothesisType,
    "two-tailed",
    "H1:μ1!=μ2",
    "left-tailed",
    "H1:μ1<μ2",
    "right-tailed",
    "H1:μ1>μ2"
    ];

(* Step 2. Set the significance level (α): *)
α=0.05; (* Significance level of 0.05. *)

(* Step 3. Collect random samples from both populations and calculate sample means, sample
standard deviations, and sample sizes: *)
SeedRandom[4123];
sample1=RandomVariate[NormalDistribution[80,20],30] ;
sample2=RandomVariate[NormalDistribution[80,20],20] ;
n1=Length[sample1]; (* Sample size for population 1. *)
n2=Length[sample2]; (* Sample size for population 2. *)
xbar1=Mean[sample1]; (* Sample mean for population 1. *)
xbar2=Mean[sample2]; (* Sample mean for population 2. *)
s1=StandardDeviation[sample1]; (* Sample standard deviation for population 1. *)
s2=StandardDeviation[sample2]; (* Sample standard deviation for population 2. *)

(* Step 4. Calculate the pooled standard deviation: *)
s=Sqrt[((n1-1)*s1^2+(n2-1)*s2^2)/(n1+n2-2)];

(* Step 5. Calculate the test statistic: *)
```

```
testStatistic=(xbar1-xbar2)/(s*Sqrt[1/n1+1/n2]);

(* Step 6. Determine the degrees of freedom: *)
df=n1+n2-2;

(* Critical value approach: *)
(* Step 7. Determine the critical value(s) and alternative hypotheses: *)
criticalValue=Switch[
    hypothesisType,
    "two-tailed",
    Quantile[StudentTDistribution[df],1-α/2],
    "left-tailed",
    Quantile[StudentTDistribution[df],α],
    "right-tailed",
    Quantile[StudentTDistribution[df],1-α]
    ];

(* Step 8. Compare the test statistic to the critical value(s): *)
criticalvalueapproach=Switch[
    hypothesisType,
    "two-tailed",
    If[
     Abs[testStatistic]>criticalValue,
     (* Reject the null hypothesis: *)
     conclusion="Reject the null hypothesis, there is sufficient evidence to support the alternative hypothesis, at the significance level of "<>ToString[α],
     (* Fail to reject the null hypothesis: *)
     conclusion="Fail to reject the null hypothesis, there is not enough evidence to support the alternative hypothesis, at the significance level of "<>ToString[α]
     ],
    "left-tailed",
    If[
     testStatistic<criticalValue,
     (* Reject the null hypothesis: *)
     conclusion="Reject the null hypothesis, there is sufficient evidence to support the alternative hypothesis, at the significance level of "<>ToString[α],
     (* Fail to reject the null hypothesis: *)
     conclusion="Fail to reject the null hypothesis, there is not enough evidence to support the alternative hypothesis, at the significance level of "<>ToString[α]
     ],
    "right-tailed",
    If[
     testStatistic>criticalValue,
     (* Reject the null hypothesis: *)
     conclusion="Reject the null hypothesis, there is sufficient evidence to support the alternative hypothesis, at the significance level of "<>ToString[α],
     (* Fail to reject the null hypothesis: *)
     conclusion="Fail to reject the null hypothesis, there is not enough evidence to support the alternative hypothesis, at the significance level of "<>ToString[α]
     ]
    ];

(* P value approach: *)
(* Step 9. Calculate the p-value: *)
pValue=Switch[
    hypothesisType,
    "two-tailed",
    2*(1-CDF[StudentTDistribution[df],Abs[testStatistic]]),
    "left-tailed",
    CDF[StudentTDistribution[df],testStatistic],
```

```
    "right-tailed",
    1-CDF[StudentTDistribution[df],testStatistic]
    ];

(* Step 10. Compare the p-value to the significance level: *)
pvalueapproach=If[
    pValue<α,
    (* Reject the null hypothesis: *)
    conclusion="Reject the null hypothesis, there is sufficient evidence to support the alternative hypothesis, at the significance level of "<>ToString[α],
    (* Fail to reject the null hypothesis: *)
    conclusion="Fail to reject the null hypothesis, there is not enough evidence to support the alternative hypothesis, at the significance level of "<>ToString[α]
    ];

(* Step 11. Draw a conclusion based on the results: *)
Print["Sample 1 Mean: ",xbar1];
Print["Sample 2 Mean: ",xbar2];
Print["Sample 1 Standard Deviation: ",s1];
Print["Sample 2 Standard Deviation: ",s2];
Print["Pooled Standard Deviation: ",s];
Print["Test Statistic: ",testStatistic];
Print["Degrees of Freedom: ",df];
Print["Critical Value: ",criticalValue];
Print["p-value: ",pValue];
Print["Conclusion using critical value approach: ",criticalvalueapproach];
Print["Conclusion using P value approach: ",pvalueapproach];

(* Built-in function in Mathematica: *)
LocationTest[
 {sample1,sample2},
 0,
 {"TestDataTable",All},
 VerifyTestAssumptions->"EqualVariance",
 Switch[
  hypothesisType,
  "two-tailed",
  AlternativeHypothesis->"Unequal",
  "left-tailed",
  AlternativeHypothesis->"Less",
  "right-tailed",
  AlternativeHypothesis->"Greater"
  ]
 ]
 Sample 1 Mean:  77.3221
 Sample 2 Mean:  78.9341
 Sample 1 Standard Deviation:  18.0174
 Sample 2 Standard Deviation:  16.1391
 Pooled Standard Deviation:  17.2983
 Test Statistic:  -0.322823
 Degrees of Freedom:  48
 Critical Value:  2.01063
 p-value:  0.748232
 Conclusion using critical value approach:  Fail to reject the null hypothesis, there is not enough evidence to support the alternative hypothesis, at the significance level of 0.05
 Conclusion using P value approach:  Fail to reject the null hypothesis, there is not enough evidence to support the alternative hypothesis, at the significance level of 0.05

              | Statistic  P-Value
```

| | | |
|---|---|---|
| Mann-Whitney | 286. | 0.774003 |
| T | -0.322823 | 0.748232 |
| Z | -0.330129 | 0.741302 |

## 20.6 Hypothesis Tests for the Means of Two Populations with Unknown and Unequal Variances

| | | | |
|---|---|---|---|
| Assumptions | 1. Simple random samples<br>2. Independent samples<br>3. Normal populations | | |
| Null hypothesis | $H_0: \mu_1 = \mu_2$ | | |
| Alternative hypothesis | Two-tailed<br>$H_a: \mu_1 \neq \mu_2$ | Left tailed<br>$H_a: \mu_1 < \mu_2$ | Right tailed<br>$H_a: \mu_1 > \mu_2$ |
| Significance level | $\alpha$ | | |

Test statistic:
For details see Chapter 18.

$$t_0 = \frac{\bar{x}_1 - \bar{x}_2}{\sqrt{\frac{s_1^2}{n_1} + \frac{s_2^2}{n_2}}}.$$

$$df = \frac{\left(\frac{s_1^2}{n_1} + \frac{s_2^2}{n_2}\right)^2}{\left(\frac{1}{n_1 - 1}\left(\frac{s_1^2}{n_1}\right) + \frac{1}{n_2 - 1}\left(\frac{s_2^2}{n_2}\right)\right)}.$$

**Critical value approach** $\pm t_{\alpha/2}$ $-t_\alpha$ $t_\alpha$

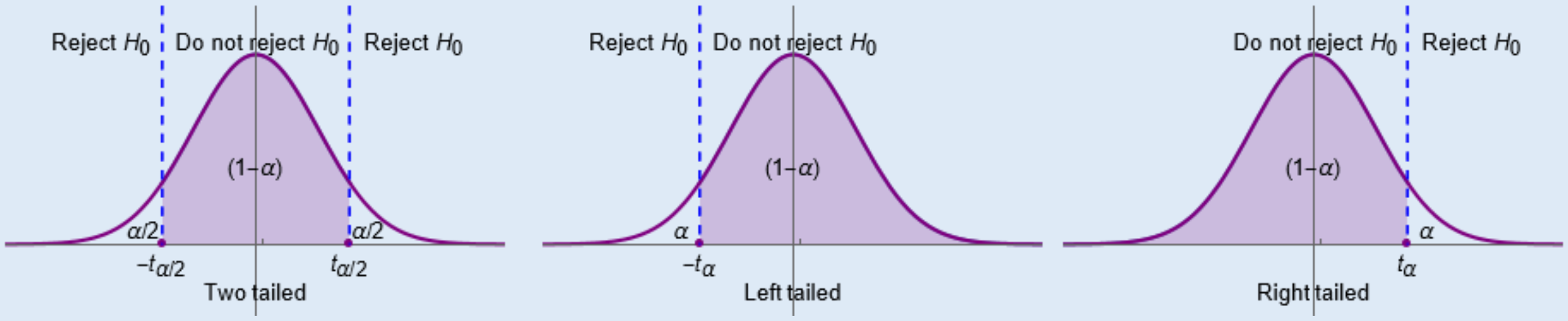

$\begin{cases} \text{reject } H_0, & \text{If the value of the test statistic falls in the rejection region} \\ \text{do not reject } H_0, & \text{otherwise} \end{cases}$

***P* value approach**

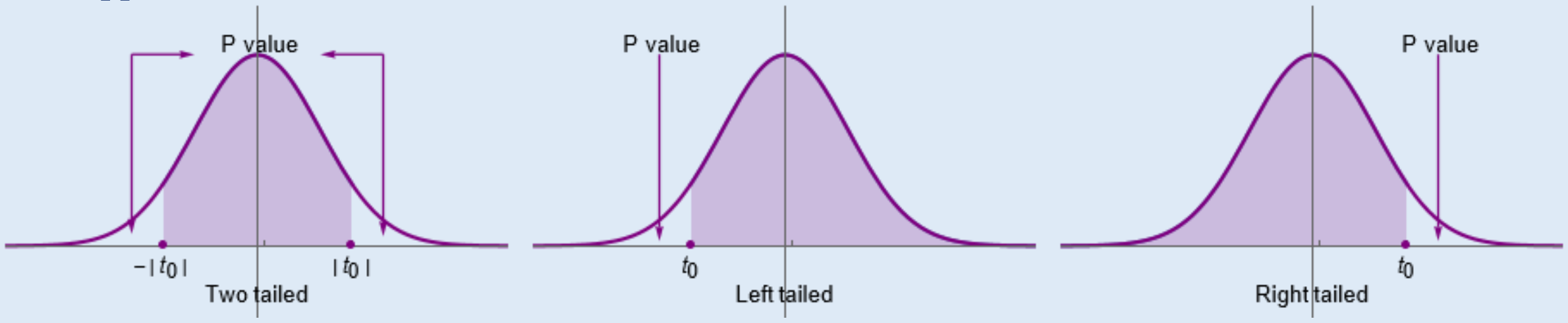

$\begin{cases} \text{reject } H_0, & \text{If } P \leq \alpha \\ \text{do not reject } H_0, & \text{otherwise} \end{cases}$

**Procedure 20.6.**
The steps for conducting a hypothesis test for the means of two populations using independent samples, specifically when the variances are unknown and unequal:

1. State the null and alternative hypotheses:
   - $H_0$: The means of the two populations are equal ($\mu_1 = \mu_2$).
   - $H_1$: The means of the two populations are not equal ($\mu_1 \neq \mu_2$, two-tailed test), or one mean is greater than/less than the other ($\mu_1 > \mu_2$ or $\mu_1 < \mu_2$, one-tailed test).
2. Set the significance level, $\alpha$, for the test.
3. Collect independent random samples from both populations and calculate the sample means ($\bar{x}_1$ and $\bar{x}_2$), the sample standard deviations ($s_1$ and $s_2$), and the sample sizes ($n_1$ and $n_2$).
4. Calculate the degrees of freedom ($df$) for the test statistic using the formula:
$$df = \frac{\left(\frac{s_1^2}{n_1} + \frac{s_2^2}{n_2}\right)^2}{\left(\frac{1}{n_1 - 1}\left(\frac{s_1^2}{n_1}\right) + \frac{1}{n_2 - 1}\left(\frac{s_2^2}{n_2}\right)\right)}.$$
5. Calculate the test statistic using the formula:
   For a t-test:
$$t_0 = \frac{\bar{x}_1 - \bar{x}_2}{\sqrt{\frac{s_1^2}{n_1} + \frac{s_2^2}{n_2}}}.$$
6. Determine the critical value(s) for the test statistic.
   - For a two-tailed test, divide the significance level by 2 and find the corresponding t-score(s) from the Mathematica t-distribution.
   - For a one-tailed test, find the t-score corresponding to the desired tail area.
7. Compare the absolute value of the test statistic to the critical value(s).
   - If the test statistic falls within the critical region, reject the null hypothesis.
   - Otherwise, fail to reject the null hypothesis.
8. Calculate the $P$-value associated with the test statistic.
   - For a two-tailed test, calculate the area in both tails of the t-distribution with the degrees of freedom.
   - For a one-tailed test, calculate the area in the appropriate tail.
9. Compare the $P$-value to the $\alpha$.
   - If the $P$-value is less than $\alpha$, reject the null hypothesis.
   - Otherwise, fail to reject the null hypothesis.
10. Draw a conclusion based on the results.
    - If the null hypothesis is rejected, conclude that there is sufficient evidence to support the alternative hypothesis.
    - If the null hypothesis is not rejected, conclude that there is not enough evidence to support the alternative hypothesis.

***Example 20.15***

```
(* The code demonstrates the steps for conducting a hypothesis test for the means of two
populations using independent samples with unknown and unequal variances.
The code will output the sample means, sample standard deviations, test statistic, critical
value, p-value, and the conclusion based on the user-specified type of hypothesis test: *)

(* Specify the type of hypothesis test, choose one of the following, "two-tailed","left-
tailed", or "right-tailed": *)
hypothesisType="two-tailed";

(* Step 1. State the null and alternative hypotheses: *)
(* Null hypothesis:Means of the two populations are equal: *)
```

```
"H0:μ1==μ2";

(* Alternative hypothesis: *)
alternativeHypothesis=Switch[
    hypothesisType,
    "two-tailed",
    "H1:μ1!=μ2",
    "left-tailed",
    "H1:μ1<μ2",
    "right-tailed",
    "H1:μ1>μ2"
    ];

(* Step 2. Set the significance level (α): *)
α=0.05; (* Significance level of 0.05. *)

(* Step 3. Collect independent random samples from both populations and calculate sample
means, sample standard deviations, and sample sizes: *)
SeedRandom[5123];
sample1=RandomVariate[NormalDistribution[50,20],30]
sample2=RandomVariate[NormalDistribution[40,10],35]
n1=Length[sample1]; (* Sample size for population 1. *)
n2=Length[sample2]; (* Sample size for population 2. *)
xbar1=Mean[sample1]; (* Sample mean for population 1. *)
xbar2=Mean[sample2]; (* Sample mean for population 2. *)
s1=StandardDeviation[sample1]; (* Sample standard deviation for population 1. *)
s2=StandardDeviation[sample2]; (* Sample standard deviation for population 2. *)

(* Step 4. Calculate the degrees of freedom: *)
df=((s1^2/n1+s2^2/n2)^2)/(((s1^2/n1)^2)/(n1-1)+((s2^2/n2)^2)/(n2-1));

(* Step 5. Calculate the test statistic: *)
testStatistic=(xbar1-xbar2)/Sqrt[(s1^2/n1)+(s2^2/n2)];

(* Critical value approach: *)
(* Step 6. Determine the critical value(s): *)
criticalValue=Switch[
    hypothesisType,
    "two-tailed",
    Quantile[StudentTDistribution[df],1-α/2],
    "left-tailed",
    Quantile[StudentTDistribution[df],α],
    "right-tailed",
    Quantile[StudentTDistribution[df],1-α]
    ];

(* Step 7. Compare the test statistic to the critical value(s): *)
criticalvalueapproach=Switch[
    hypothesisType,
    "two-tailed",
    If[
     Abs[testStatistic]>criticalValue,
     (* Reject the null hypothesis: *)
     conclusion="Reject the null hypothesis, there is sufficient evidence to support the
alternative hypothesis, at the significance level of "<>ToString[α],
     (* Fail to reject the null hypothesis: *)
     conclusion="Fail to reject the null hypothesis, there is not enough evidence to support
the alternative hypothesis, at the significance level of "<>ToString[α]
     ],
    "left-tailed",
```

```
    If[
     testStatistic<criticalValue,
     (* Reject the null hypothesis: *)
     conclusion="Reject the null hypothesis, there is sufficient evidence to support the alternative hypothesis, at the significance level of "<>ToString[α],
     (* Fail to reject the null hypothesis: *)
     conclusion="Fail to reject the null hypothesis, there is not enough evidence to support the alternative hypothesis, at the significance level of "<>ToString[α]
     ],
    "right-tailed",
    If[
     testStatistic>criticalValue,
     (* Reject the null hypothesis: *)
     conclusion="Reject the null hypothesis, there is sufficient evidence to support the alternative hypothesis, at the significance level of "<>ToString[α],
     (* Fail to reject the null hypothesis: *)
     conclusion="Fail to reject the null hypothesis, there is not enough evidence to support the alternative hypothesis, at the significance level of "<>ToString[α]
     ]
    ];

(* P value approach: *)
(* Step 8. Calculate the p-value: *)
pValue=Switch[
    hypothesisType,
    "two-tailed",
    2*(1-CDF[StudentTDistribution[df],Abs[testStatistic]]),
    "left-tailed",
    CDF[StudentTDistribution[df],testStatistic],
    "right-tailed",
    1-CDF[StudentTDistribution[df],testStatistic]
    ];

(* Step 9. Compare the p-value to the significance level: *)
pvalueapproach=If[
    pValue<α,
    (* Reject the null hypothesis: *)
    conclusion="Reject the null hypothesis, there is sufficient evidence to support the alternative hypothesis, at the significance level of "<>ToString[α],
    (* Fail to reject the null hypothesis: *)
    conclusion="Fail to reject the null hypothesis, there is not enough evidence to support the alternative hypothesis, at the significance level of "<>ToString[α]
    ];

(* Step 10. Draw a conclusion based on the results: *)
Print["Sample 1 Mean: ",xbar1];
Print["Sample 2 Mean: ",xbar2];
Print["Sample 1 Standard Deviation: ",s1];
Print["Sample 2 Standard Deviation: ",s2];
Print["Degrees of Freedom: ",df];
Print["Test Statistic: ",testStatistic];
Print["Critical Value: ",criticalValue];
Print["p-value: ",pValue];
Print["Conclusion using critical value approach: ",criticalvalueapproach];
Print["Conclusion using P value approach: ",pvalueapproach];

(* Built-in function in Mathematica: *)
LocationTest[
 {sample1,sample2},
 0,
```

```
 {"TestDataTable",All},
 Switch[
  hypothesisType,
  "two-tailed",
  AlternativeHypothesis->"Unequal",
  "left-tailed",
  AlternativeHypothesis->"Less",
  "right-tailed",
  AlternativeHypothesis->"Greater"
  ]
 ]

{65.8179,74.8467,32.1745,40.8303,53.3451,81.0876,21.4019,8.69666,47.3945,61.2536,51.7821,74.2816,52.0914,27.2412,53.6193,59.3016,88.9166,75.079,46.2268,86.4708,50.116,66.8337,35.0818,38.8135,53.7453,39.4238,43.8772,24.5794,52.4848,63.1908}

{26.6424,42.8022,52.3312,45.1386,43.5929,46.5988,38.338,37.284,65.5342,27.5743,52.3768,36.8759,55.52,39.4923,29.6158,66.981,35.9229,26.6001,35.4696,46.5576,49.3788,42.5244,22.6636,34.577,38.5991,34.3755,44.8854,34.0422,20.6511,50.5005,45.7123,17.0159,36.5962,33.1143,50.293}

 Sample 1 Mean:  52.3335
 Sample 2 Mean:  40.1765
 Sample 1 Standard Deviation:  19.5327
 Sample 2 Standard Deviation:  11.4583
 Degrees of Freedom:  45.2715
 Test Statistic:  2.99568
 Critical Value:  2.01377
 p-value:  0.0044301
 Conclusion using critical value approach:  Reject the null hypothesis, there is sufficient evidence to support the alternative hypothesis, at the significance level of 0.05
 Conclusion using P value approach:  Reject the null hypothesis, there is sufficient evidence to support the alternative hypothesis, at the significance level of 0.05
```

| | Statistic | P-Value |
|---|---|---|
| Mann-Whitney | 750. | 0.00300367 |
| T | 2.99568 | 0.0044301 |
| Z | 2.99568 | 0.00273833 |

## 20.7 Hypothesis Tests for a Population Standard Deviation

| | | | |
|---|---|---|---|
| Assumptions | 1. Simple random sample<br>2. Normal populations | | |
| Null hypothesis | $H_0: \sigma = \sigma_0$ | | |
| Alternative hypothesis | Two-tailed<br>$H_a: \sigma \neq \sigma_0$ | Left tailed<br>$H_a: \sigma < \sigma_0$ | Right tailed<br>$H_a: \sigma > \sigma_0$ |
| Significance level | $\alpha$ | | |
| Test statistic:<br>For details see Chapter 18. | $\chi_0^2 = (n-1)\dfrac{s^2}{\sigma_0^2}$.<br>$df = n - 1$. | | |

**Critical value approach** $\chi^2_{1-\alpha/2}$ and $\chi^2_{\alpha/2}$ $\chi^2_{1-\alpha}$ $\chi^2_{\alpha}$

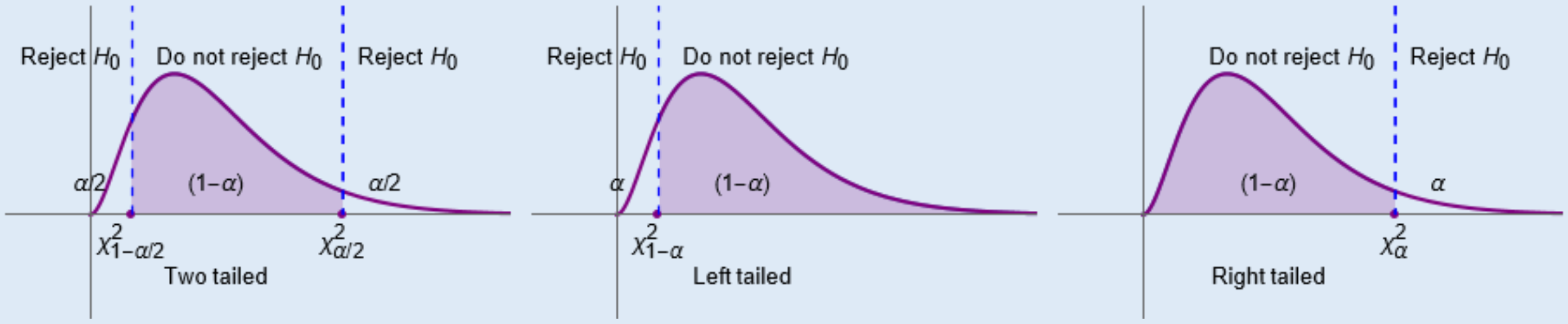

$$\begin{cases} \text{reject } H_0, & \text{If the value of the test statistic falls in the rejection region} \\ \text{do not reject } H_0, & \text{otherwise} \end{cases}$$

***P* value approach**

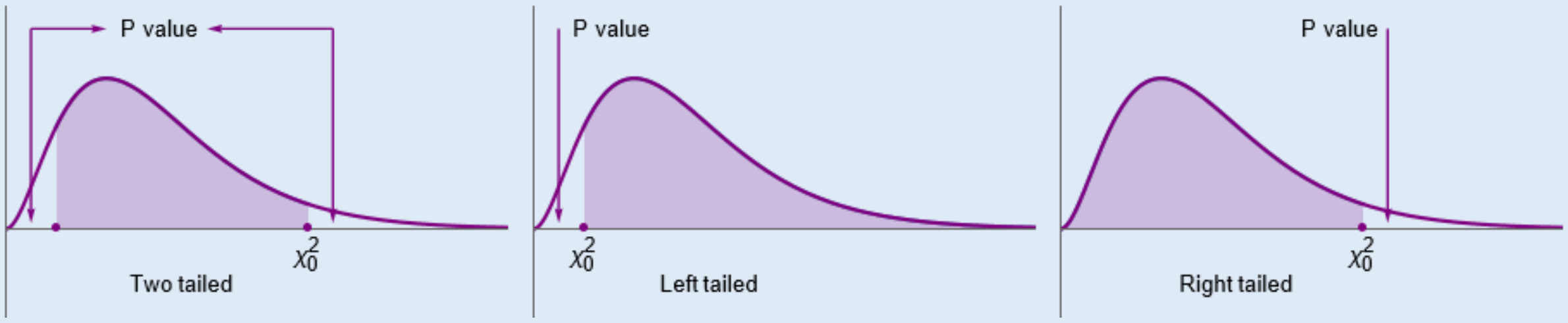

$$\begin{cases} \text{reject } H_0, & \text{If } P \leq \alpha \\ \text{do not reject } H_0, & \text{otherwise} \end{cases}$$

**Procedure 20.7.**

The steps for conducting a hypothesis test for a population standard deviation:

1. State the null and alternative hypotheses:
   - $H_0$: The population standard deviation is equal to a specific value ($\sigma = \sigma_0$).
   - $H_1$: The population standard deviation is not equal to the specific value ($\sigma \neq \sigma_0$, two-tailed test), or it is greater than/less than the specific value ($\sigma > \sigma_0$ or $\sigma < \sigma_0$, one-tailed test).
2. Set the significance level, $\alpha$, for the test.
3. Collect a random sample from the population of interest.
4. Calculate the sample standard deviation (s) of the sample.
5. Determine the degrees of freedom ($df$) for the chi-square distribution. For a sample size of $n$,
$$df = n - 1$$
6. Calculate the test statistic using the formula:
For a chi-square test:
$$\chi^2 = (n - 1) * \frac{s^2}{\sigma_0^2}$$
7. Determine the critical value(s) for the test statistic.
   - For a two-tailed test, divide the significance level by 2 and find the corresponding critical value(s) from the Mathematica chi-square distribution.
   - For a one-tailed test, find the critical value(s) corresponding to the desired tail area.
8. Compare the test statistic to the critical value(s).
   - If the test statistic falls within the critical region, reject the null hypothesis.
   - Otherwise, fail to reject the null hypothesis.
9. Calculate the p-value associated with the test statistic.

- For a two-tailed test, calculate the probability of the test statistic being greater than the observed value or less than the negative of the observed value from the chi-square distribution with the degrees of freedom.
- For a one-tailed test, calculate the probability of the test statistic being greater than the observed value or less than the observed value, depending on the alternative hypothesis.

10. Compare the p-value to $\alpha$.
    - If the p-value is less than α, reject the null hypothesis.
    - Otherwise, fail to reject the null hypothesis.
11. Draw a conclusion based on the results.
    - If the null hypothesis is rejected, conclude that there is sufficient evidence to support the alternative hypothesis.
    - If the null hypothesis is not rejected, conclude that there is not enough evidence to support the alternative hypothesis.

***Example 20.16***

```
(* The code demonstrates the steps for conducting a hypothesis test for a population
standard deviation. The code will output the sample standard deviations, test statistic,
critical value, p-value, and the conclusion based on the user-specified type of hypothesis
test: *)

(* Specify the type of hypothesis test, choose one of the following, "two-tailed","left-
tailed", or "right-tailed": *)
hypothesisType="two-tailed";
σ0=20;

(* Step 1. State the null and alternative hypotheses: *)
(* Null hypothesis:Population standard deviation is equal to σ0: *)
"H0:σ==σ0";

(* Alternative hypothesis: *)
alternativeHypothesis=Switch[
    hypothesisType,
    "two-tailed",
    "H1:σ!=σ0",
    "left-tailed",
    "H1:σ<σ0",
    "right-tailed",
    "H1:σ>σ0"
    ];

(* Step 2. Set the significance level (α): *)
α=0.05; (* Significance level of 0.05. *)

(* Step 3. Collect a random sample from the population: *)
SeedRandom[6123];
sample=RandomVariate[NormalDistribution[50,20],30]

(* Step 4. Calculate the sample standard deviation: *)
s=StandardDeviation[sample];

(* Step 5. Determine the degrees of freedom: *)
n=Length[sample];
df=n-1;

(* Step 6. Calculate the test statistic: *)
testStatistic=(n-1)*s^2/σ0^2;
```

```
(* Critical value approach: *)
(* Step 7. Determine the critical value(s) and alternative hypotheses: *)
criticalValue=Switch[
    hypothesisType,
    "two-tailed",
    Quantile[ChiSquareDistribution[df],1-α/2],
    "left-tailed",
    Quantile[ChiSquareDistribution[df],α],
    "right-tailed",
    Quantile[ChiSquareDistribution[df],1-α]
    ];

(* Step 8. Compare the test statistic to the critical value(s): *)
criticalvalueapproach=Switch[
    hypothesisType,
    "two-tailed",
    If[
     testStatistic>criticalValue,
     (* Reject the null hypothesis: *)
     conclusion="Reject the null hypothesis, there is sufficient evidence to support the
alternative hypothesis, at the significance level of "<>ToString[α],
     (* Fail to reject the null hypothesis: *)
     conclusion="Fail to reject the null hypothesis, there is not enough evidence to support
the alternative hypothesis, at the significance level of "<>ToString[α]
     ],
    "left-tailed",
    If[
     testStatistic<criticalValue,
     (* Reject the null hypothesis: *)
     conclusion="Reject the null hypothesis, there is sufficient evidence to support the
alternative hypothesis, at the significance level of "<>ToString[α],
     (* Fail to reject the null hypothesis: *)
     conclusion="Fail to reject the null hypothesis, there is not enough evidence to support
the alternative hypothesis, at the significance level of "<>ToString[α]
     ],
    "right-tailed",
    If[
     testStatistic>criticalValue,
     (* Reject the null hypothesis: *)
     conclusion="Reject the null hypothesis, there is sufficient evidence to support the
alternative hypothesis, at the significance level of "<>ToString[α],
     (* Fail to reject the null hypothesis: *)
     conclusion="Fail to reject the null hypothesis, there is not enough evidence to support
the alternative hypothesis, at the significance level of "<>ToString[α]
     ]
    ];

(* P value approach: *)
(* Step 9. Calculate the p-value: *)
pValue=Switch[
    hypothesisType,
    "two-tailed",
    2*(1-CDF[ChiSquareDistribution[df],testStatistic]),
    "left-tailed",
    CDF[ChiSquareDistribution[df],testStatistic],
    "right-tailed",
    1-CDF[ChiSquareDistribution[df],testStatistic]
    ];
```

```
(* Step 10. Compare the p-value to the significance level: *)
pvalueapproach=If[
    pValue<α,
    (* Reject the null hypothesis: *)
    conclusion="Reject the null hypothesis, there is sufficient evidence to support the alternative hypothesis, at the significance level of "<>ToString[α],
    (* Fail to reject the null hypothesis: *)
    conclusion="Fail to reject the null hypothesis, there is not enough evidence to support the alternative hypothesis, at the significance level of "<>ToString[α]
    ];

(* Step 11. Draw a conclusion based on the results: *)
Print["Sample Standard Deviation: ",s];
Print["Degrees of Freedom: ",df];
Print["Test Statistic: ",testStatistic];
Print["Critical Value: ",criticalValue];
Print["p-value: ",pValue];
Print["Conclusion using critical value approach: ",criticalvalueapproach];
Print["Conclusion using P value approach: ",pvalueapproach];

(* Built-in function in Mathematica: *)
VarianceTest[
 sample,
 20^2,
 {"TestDataTable",All},
 Switch[
  hypothesisType,
  "two-tailed",
  AlternativeHypothesis->"Unequal",
  "left-tailed",
  AlternativeHypothesis->"Less",
  "right-tailed",
  AlternativeHypothesis->"Greater"
  ]
 ]

{26.1734,81.8904,65.2959,70.916,54.9925,37.243,82.2368,49.7475,14.7791,42.6197,46.3899,42.4151,53.4866,64.0125,53.2706,34.4091,42.7251,26.9687,28.5789,63.2913,102.423,59.9912,60.2673,55.2909,-3.35907,56.5588,47.9882,65.3735,32.1185,17.093}

 Sample Standard Deviation:  21.9906
 Degrees of Freedom:  29
 Test Statistic:  35.0599
 Critical Value:  45.7223
 p-value:  0.405161
 Conclusion using critical value approach:  Fail to reject the null hypothesis, there is not enough evidence to support the alternative hypothesis, at the significance level of 0.05
 Conclusion using P value approach:  Fail to reject the null hypothesis, there is not enough evidence to support the alternative hypothesis, at the significance level of 0.05
```

| | Statistic | P-Value |
|---|---|---|
| Brown-Forsythe | 35.0599 | 0.405161 |
| Fisher Ratio | 35.0599 | 0.405161 |
| Levene | 35.0599 | 0.405161 |

## 20.8 Hypothesis Tests for Two Population Standard Deviations

| | | | |
|---|---|---|---|
| Assumptions | 1. Simple random samples<br>2. Independent samples<br>3. Normal populations | | |
| Null hypothesis | $H_0: \sigma_1 = \sigma_2$ | | |
| | Two-tailed | Left tailed | Right tailed |
| Alternative hypothesis | $H_a: \sigma_1 \neq \sigma_2$ | $H_a: \sigma_1 < \sigma_2$ | $H_a: \sigma_1 > \sigma_2$ |
| Significance level | $\alpha$ | | |
| Test statistic:<br>For details see Chapter 18. | $F_0 = \frac{s_1^2}{s_2^2}$.<br>$df = (n_1 - 1, n_2 - 1)$. | | |
| **Critical value approach** | $F_{1-\alpha/2}$ and $F_{\alpha/2}$ | $F_{1-\alpha}$ | $F_{\alpha}$ |

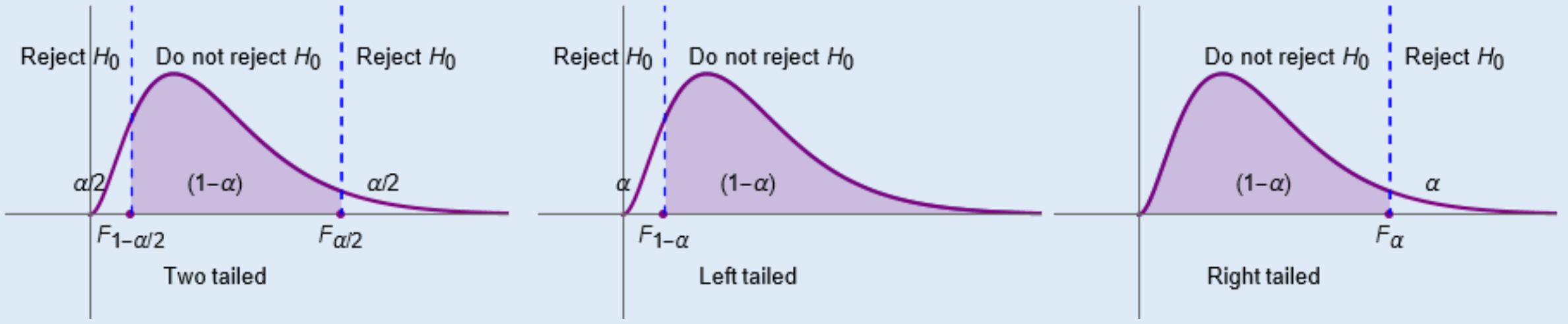

$$\begin{cases} \text{reject } H_0, & \text{If the value of the test statistic falls in the rejection region} \\ \text{do not reject } H_0, & \text{otherwise} \end{cases}$$

***P* value approach**

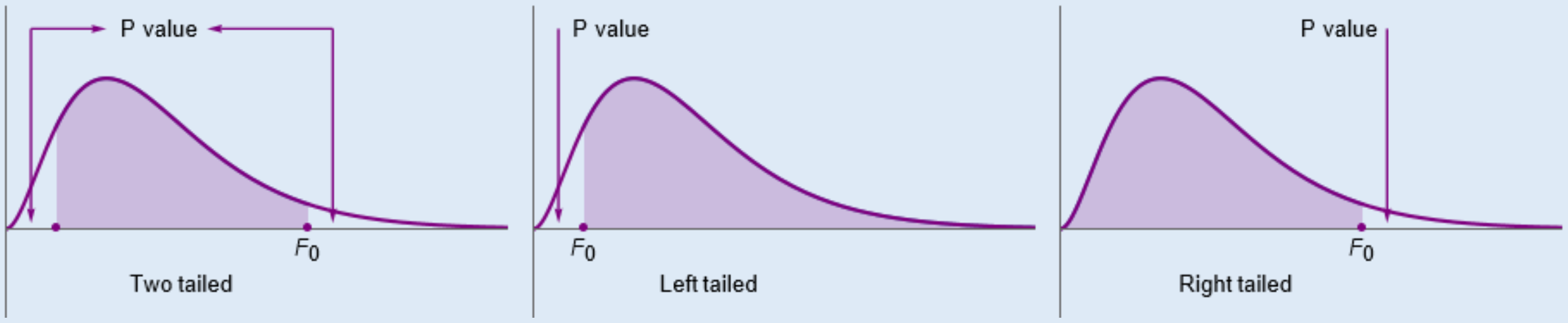

$$\begin{cases} \text{reject } H_0, & \text{If } P \leq \alpha \\ \text{do not reject } H_0, & \text{otherwise} \end{cases}$$

**Procedure 20.8.**
The steps for conducting a hypothesis test for two population standard deviations:

1. State the null and alternative hypotheses:
   - $H_0$: The two population standard deviations are equal ($\sigma_1 = \sigma_2$).
   - $H_1$: The two population standard deviations are not equal ($\sigma_1 \neq \sigma_2$, two-tailed test), or one standard deviation is greater than/less than the other ($\sigma_1 > \sigma_2$ or $\sigma_1 < \sigma_2$, one-tailed test).
2. Set the significance level, $\alpha$, for the test.

3. Collect independent random samples from both populations and calculate the sample standard deviations ($s_1$ and $s_2$), and the sample sizes ($n_1$ and $n_2$).
4. Determine the degrees of freedom ($df_1$ and $df_2$) for the test statistic. For sample sizes $n_1$ and $n_2$, $df_1 = n_1 - 1$ and $df_2 = n_2 - 1$.
5. Calculate the test statistic using the formula:
   For an F-test:
$$F = \frac{s_1^2}{s_2^2}.$$
6. Determine the critical value(s) for the test statistic.
   - For a two-tailed test, find the critical values for the upper and lower tails of the F-distribution with $df_1$ and $df_2$ degrees of freedom.
   - For a one-tailed test, find the critical value corresponding to the desired tail area.
7. Compare the test statistic to the critical value(s).
   - If the test statistic is greater than the upper critical value or less than the lower critical value, reject the null hypothesis.
   - Otherwise, fail to reject the null hypothesis.
8. Calculate the $P$-value associated with the test statistic.
   - For a two-tailed test, calculate the probability of the test statistic being greater than the observed value or less than the reciprocal of the observed value from the F-distribution with $df_1$ and $df_2$ degrees of freedom.
   - For a one-tailed test, calculate the probability of the test statistic being greater than the observed value or less than the observed value, depending on the alternative hypothesis.
9. Compare the $P$-value to $\alpha$.
   - If the $P$-value is less than $\alpha$, reject the null hypothesis.
   - Otherwise, fail to reject the null hypothesis.
10. Draw a conclusion based on the results.
   - If the null hypothesis is rejected, conclude that there is sufficient evidence to support the alternative hypothesis.
   - If the null hypothesis is not rejected, conclude that there is not enough evidence to support the alternative hypothesis.

***Example 20.17***

```
(* The code demonstrates the steps for conducting a hypothesis test for two population
standard deviations. The code will output the sample standard deviations, test statistic,
critical value, p-value, and the conclusion based on the user-specified type of hypothesis
test: *)

(* Specify the type of hypothesis test, choose one of the following, "two-tailed","left-
tailed", or "right-tailed": *)
hypothesisType="two-tailed";

(* Step 1. State the null and alternative hypotheses: *)
(* Null hypothesis:Population standard deviations are equal: *)
"H0:σ1==σ2";

(* Alternative hypothesis: *)
alternativeHypothesis=Switch[
    hypothesisType,
    "two-tailed",
    "H1:σ1!=σ2",
    "left-tailed",
    "H1:σ1<σ2",
    "right-tailed",
    "H1:σ1>σ2"
    ];
```

```
(* Step 2. Set the significance level (α): *)
α=0.05; (* Significance level of 0.05. *)

(* Step 3. Collect independent random samples from both populations and calculate sample
standard deviations and sample sizes: *)
SeedRandom[7123];
sample1=RandomVariate[NormalDistribution[50,20],30]
sample2=RandomVariate[NormalDistribution[50,10],40]
n1=Length[sample1]; (* Sample size for population 1. *)
n2=Length[sample2]; (* Sample size for population 2. *)
s1=StandardDeviation[sample1]; (* Sample standard deviation for population 1. *)
s2=StandardDeviation[sample2]; (* Sample standard deviation for population 2. *)

(* Step 4. Determine the degrees of freedom: *)
df1=n1-1; (* Degrees of freedom for population 1. *)
df2=n2-1; (* Degrees of freedom for population 2. *)

(* Step 5. Calculate the test statistic: *)
testStatistic=(s1^2/s2^2);

(* Critical value approach: *)
(* Step 6. Determine the critical value(s) and alternative hypotheses: *)
criticalValue=Switch[
    hypothesisType,
    "two-tailed",
    Quantile[FRatioDistribution[df1,df2],1-α/2],
    "left-tailed",
    Quantile[FRatioDistribution[df1,df2],α],
    "right-tailed",
    Quantile[FRatioDistribution[df1,df2],1-α]
    ];

(* Step 7. Compare the test statistic to the critical value(s): *)
criticalvalueapproach=Switch[
    hypothesisType,
    "two-tailed",
    If[
     testStatistic>criticalValue,
     (* Reject the null hypothesis: *)
     conclusion="Reject the null hypothesis, there is sufficient evidence to support the
alternative hypothesis, at the significance level of "<>ToString[α],
     (* Fail to reject the null hypothesis: *)
     conclusion="Fail to reject the null hypothesis, there is not enough evidence to support
the alternative hypothesis, at the significance level of "<>ToString[α]
     ],
    "left-tailed",
    If[
     testStatistic<criticalValue,
     (* Reject the null hypothesis: *)
     conclusion="Reject the null hypothesis, there is sufficient evidence to support the
alternative hypothesis, at the significance level of "<>ToString[α],
     (* Fail to reject the null hypothesis: *)
     conclusion="Fail to reject the null hypothesis, there is not enough evidence to support
the alternative hypothesis, at the significance level of "<>ToString[α]
     ],
    "right-tailed",
    If[
     testStatistic>criticalValue,
     (* Reject the null hypothesis: *)
```

```
     conclusion="Reject the null hypothesis, there is sufficient evidence to support the
alternative hypothesis, at the significance level of "<>ToString[α],
     (* Fail to reject the null hypothesis: *)
     conclusion="Fail to reject the null hypothesis, there is not enough evidence to support
the alternative hypothesis, at the significance level of "<>ToString[α]
     ]
    ];

(* P value approach: *)
(* Step 8. Calculate the p-value: *)
pValue=Switch[
    hypothesisType,
    "two-tailed",
    2*(1-CDF[FRatioDistribution[df1,df2],testStatistic]),
    "left-tailed",
    CDF[FRatioDistribution[df1,df2],testStatistic],
    "right-tailed",
    1-CDF[FRatioDistribution[df1,df2],testStatistic]
    ];

(* Step 9. Compare the p-value to the significance level: *)
pvalueapproach=If[
    pValue<α,
    (* Reject the null hypothesis: *)
    conclusion="Reject the null hypothesis, there is sufficient evidence to support the
alternative hypothesis, at the significance level of "<>ToString[α],
    (* Fail to reject the null hypothesis: *)
    conclusion="Fail to reject the null hypothesis, there is not enough evidence to support
the alternative hypothesis, at the significance level of "<>ToString[α]
    ];

(* Step 10. Draw a conclusion based on the results: *)
Print["Sample 1 Standard Deviation: ",s1];
Print["Sample 2 Standard Deviation: ",s2];
Print["Degrees of Freedom 1: ",df1];
Print["Degrees of Freedom 2: ",df2];
Print["Test Statistic: ",testStatistic];
Print["Critical Value: ",criticalValue];
Print["p-value: ",pValue];
Print["Conclusion using critical value approach: ",criticalvalueapproach];
Print["Conclusion using P value approach: ",pvalueapproach];

(* Built-in function in Mathematica: *)
VarianceTest[
 {sample1,sample2},
 1,
 {"TestDataTable",All},
 Switch[
  hypothesisType,
  "two-tailed",
  AlternativeHypothesis->"Unequal",
  "left-tailed",
  AlternativeHypothesis->"Less",
  "right-tailed",
  AlternativeHypothesis->"Greater"
  ]
 ]

{26.1987,50.7936,52.252,67.5175,74.4236,30.6144,48.0437,65.3422,51.9694,50.8177,68.1662,49.2
```

```
962,40.4375,38.0291,73.9759,50.8878,42.3107,35.2646,54.1942,50.2761,54.9061,62.7875,66.7046,
54.9125,73.4031,54.814,58.1381,16.465,23.2794,31.4406}

{45.2656,62.2532,59.3059,49.5688,57.2062,48.0694,48.7842,52.7968,65.186,31.3938,62.6442,40.0
455,69.6964,36.2182,42.7134,55.2215,66.6812,47.978,51.494,47.2708,45.4286,36.8856,36.2155,57
.889,59.4626,23.5343,46.124,41.5626,45.3514,63.135,44.9269,36.922,65.4477,49.6322,55.6351,49
.4908,40.0645,49.9312,55.7323,52.0636}

 Sample 1 Standard Deviation:  15.3471
 Sample 2 Standard Deviation:  10.4128
 Degrees of Freedom 1:  29
 Degrees of Freedom 2:  39
 Test Statistic:  2.17229
 Critical Value:  1.96187
 p-value:  0.0242904
 Conclusion using critical value approach:  Reject the null hypothesis, there is
sufficient evidence to support the alternative hypothesis, at the significance level of
0.05
 Conclusion using P value approach:  Reject the null hypothesis, there is sufficient
evidence to support the alternative hypothesis, at the significance level of 0.05

               | Statistic  P-Value
Brown-Forsythe | 3.24052    0.0762731
Conover        | 1.8013     0.0716555
Fisher Ratio   | 2.17229    0.0242904
Levene         | 3.3399     0.0720063
Siegel-Tukey   | 1.05624    0.290861
```

## 20.9 Hypothesis Tests for One Population Proportion

| | | | |
|---|---|---|---|
| Assumptions | 1. Simple random samples<br>2. both $np_0$ and $n(1-p_0)$ are 5 or greater | | |
| Null hypothesis | $H_0: p = p_0$ | | |
| Alternative hypothesis | Two-tailed<br>$H_a: p \neq p_0$ | Left tailed<br>$H_a: p < p_0$ | Right tailed<br>$H_a: p > p_0$ |
| Significance level | $\alpha$ | | |
| Test statistic:<br>For details see Chapter 18. | $z_0 = \dfrac{\hat{p} - p_0}{\sqrt{p_0(1-p_0)/n}}$, | | |
| **Critical value approach** | $\pm z_{\alpha/2}$ | $-z_\alpha$ | $z_\alpha$ |

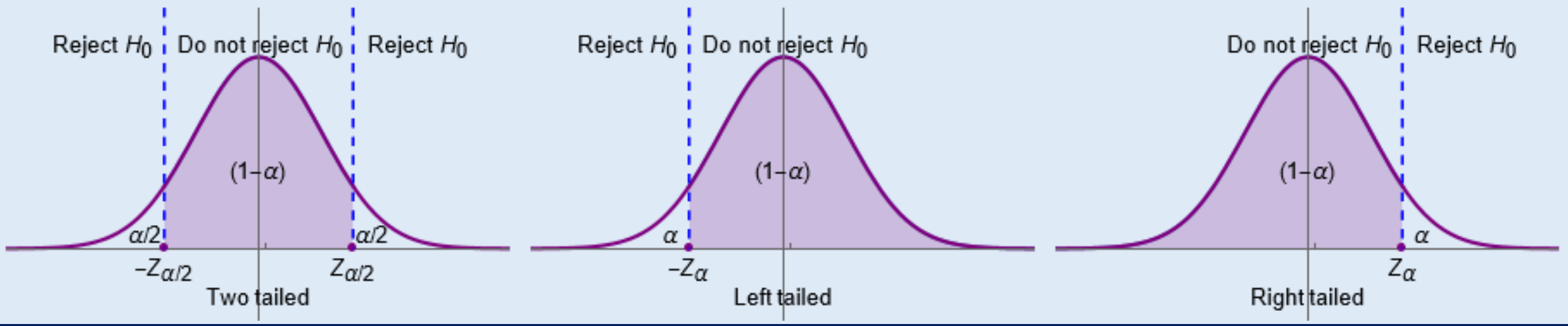

$$\begin{cases} \text{reject } H_0, & \text{If the value of the test statistic falls in the rejection region} \\ \text{do not reject } H_0, & \text{otherwise} \end{cases}$$

***P* value approach**

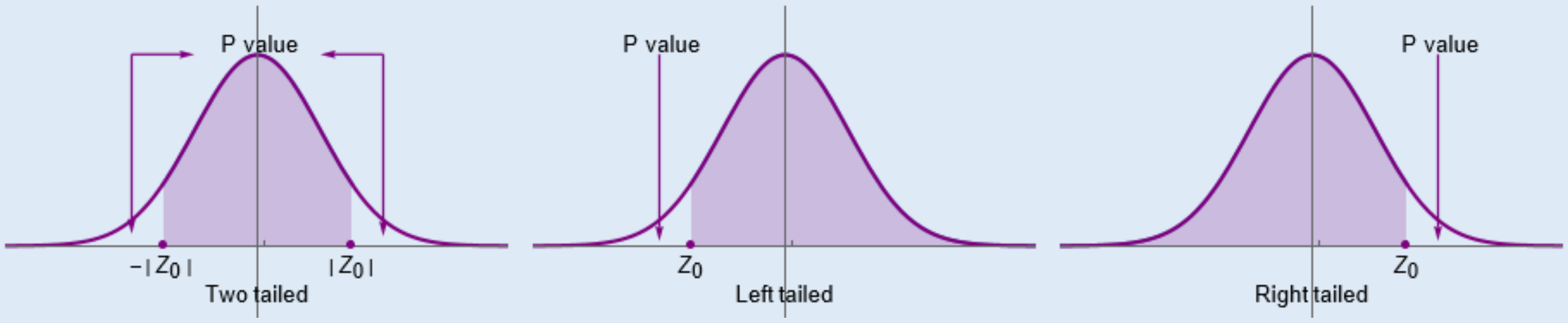

$$\begin{cases} \text{reject } H_0, & \text{If } P \leq \alpha \\ \text{do not reject } H_0, & \text{otherwise} \end{cases}$$

**Procedure 20.9.**
The steps for conducting a hypothesis test for one population proportion:

1. State the null and alternative hypotheses:
    - $H_0$: The population proportion is equal to a specific value ($p = p_0$).
    - $H_1$: The population proportion is not equal to the specific value ($p \neq p_0$, two-tailed test), or it is greater than/less than the specific value ($p > p_0$ or $p < p_0$, one-tailed test).
2. Set the significance level, $\alpha$, for the test.
3. Collect a random sample from the population of interest.
4. Calculate the sample proportion ($\hat{p}$) of the sample.
5. Calculate the SE of the sample proportion using the formula:

$$\text{SE} = \sqrt{\frac{p_0(1 - p_0)}{n}}$$

6. Calculate the test statistic using the formula:
   For a z-test:

$$z = \frac{\hat{p} - p_0}{\text{SE}}$$

7. Determine the critical value(s) for the test statistic.
    - For a two-tailed test, divide the significance level by 2 and find the corresponding critical value(s) from the standard normal distribution table.
    - For a one-tailed test, find the critical value(s) corresponding to the desired tail area.
8. Compare the test statistic to the critical value(s).
    - If the test statistic falls within the critical region, reject the null hypothesis.
    - Otherwise, fail to reject the null hypothesis.
9. Calculate the $P$-value associated with the test statistic.
    - For a two-tailed test, calculate the probability of the test statistic being greater than the observed value or less than the negative of the observed value from the standard normal distribution.
    - For a one-tailed test, calculate the probability of the test statistic being greater than the observed value or less than the observed value, depending on the alternative hypothesis.
10. Compare the $P$-value to $\alpha$.
    - If the $P$-value is less than α, reject the null hypothesis.
    - Otherwise, fail to reject the null hypothesis.
11. Draw a conclusion based on the results.
    - If the null hypothesis is rejected, conclude that there is sufficient evidence to support the alternative hypothesis.

- If the null hypothesis is not rejected, conclude that there is not enough evidence to support the alternative hypothesis.

***Example 20.18***

```
(* The code demonstrates the steps for conducting a hypothesis test for one population
proportion. The code will output the sample proportion, standard error,test statistic,
critical value, p-value, and the conclusion based on the results of the hypothesis test for
the population proportion: *)

(* Specify the type of hypothesis test, choose one of the following, "two-tailed", "left-
tailed", or "right-tailed": *)
hypothesisType="two-tailed";
p0=0.5;

(* Step 1. State the null and alternative hypotheses: *)
(* Null hypothesis:Population proportion is equal to p0: *)
"H0:p==p0";

(* Alternative hypothesis: *)
alternativeHypothesis=Switch[
    hypothesisType,
    "two-tailed",
    "H1:p!=p0",
    "left-tailed",
    "H1:p<p0",
    "right-tailed",
    "H1:p>p0"
    ];

(* Step 2. Set the significance level (α): *)
α=0.05; (* Significance level of 0.05. *)

(* Step 3. Collect a random sample from the population: *)
SeedRandom[8123];
sample=RandomInteger[1,50]
n=Length[sample]; (* Sample size. *)

(* Step 4. Calculate the sample proportion: *)
phat=N[Count[sample,1]/n]; (* Count the number of successes in the sample and divide by the
sample size. *)

(* Step 5. Calculate the standard error: *)
SE=Sqrt[(p0*(1-p0))/n];

(* Step 6. Calculate the test statistic: *)
testStatistic=(phat-p0)/SE;

(* Critical value approach: *)
(* Step 7. Determine the critical value(s) and alternative hypotheses: *)
criticalValue=Switch[
    hypothesisType,
    "two-tailed",
    Quantile[NormalDistribution[0,1],1-α/2],
    "left-tailed",
    Quantile[NormalDistribution[0,1],α],
    "right-tailed",
    Quantile[NormalDistribution[0,1],1-α]
    ];
```

```
(* Step 8. Compare the test statistic to the critical value(s): *)
criticalvalueapproach=Switch[
    hypothesisType,
    "two-tailed",
    If[
     Abs[testStatistic]>criticalValue,
     (* Reject the null hypothesis: *)
     conclusion="Reject the null hypothesis, there is sufficient evidence to support the
alternative hypothesis, at the significance level of "<>ToString[α],
     (* Fail to reject the null hypothesis: *)
     conclusion="Fail to reject the null hypothesis, there is not enough evidence to support
the alternative hypothesis, at the significance level of "<>ToString[α]
     ],
    "left-tailed",
    If[
     testStatistic<criticalValue,
     (* Reject the null hypothesis: *)
     conclusion="Reject the null hypothesis, there is sufficient evidence to support the
alternative hypothesis, at the significance level of "<>ToString[α],
     (* Fail to reject the null hypothesis: *)
     conclusion="Fail to reject the null hypothesis, there is not enough evidence to support
the alternative hypothesis, at the significance level of "<>ToString[α]
     ],
    "right-tailed",
    If[
     testStatistic>criticalValue,
     (* Reject the null hypothesis: *)
     conclusion="Reject the null hypothesis, there is sufficient evidence to support the
alternative hypothesis, at the significance level of "<>ToString[α],
     (* Fail to reject the null hypothesis: *)
     conclusion="Fail to reject the null hypothesis, there is not enough evidence to support
the alternative hypothesis, at the significance level of "<>ToString[α]
     ]
    ];

(* P value approach: *)
(* Step 9. Calculate the p-value: *)
pValue=Switch[
    hypothesisType,
    "two-tailed",
    2*(1-CDF[NormalDistribution[0,1],Abs[testStatistic]]),
    "left-tailed",
    CDF[NormalDistribution[0,1],testStatistic],
    "right-tailed",
    1-CDF[NormalDistribution[0,1],testStatistic]
    ];

(* Step 10. Compare the p-value to the significance level: *)
pvalueapproach=If[
    pValue<α,
    (* Reject the null hypothesis: *)
    conclusion="Reject the null hypothesis, there is sufficient evidence to support the
alternative hypothesis, at the significance level of "<>ToString[α],
    (* Fail to reject the null hypothesis: *)
    conclusion="Fail to reject the null hypothesis, there is not enough evidence to support
the alternative hypothesis, at the significance level of "<>ToString[α]
    ];

(* Step 11. Draw a conclusion based on the results: *)
```

```
Print["Sample Proportion: ",phat];
Print["Standard Error: ",SE];
Print["Test Statistic: ",testStatistic];
Print["Critical Value: ",criticalValue];
Print["p-value: ",pValue];
Print["Conclusion using critical value approach: ",criticalvalueapproach];
Print["Conclusion using P value approach: ",pvalueapproach];

{1,0,0,1,1,1,0,0,1,1,0,0,0,0,0,0,0,1,0,0,1,1,1,1,1,0,1,0,0,0,1,1,1,0,0,1,0,0,1,0,1,1,1,1,0,0
,0,1,0,0}

 Sample Proportion:  0.46
 Standard Error:  0.0707107
 Test Statistic:  -0.565685
 Critical Value:  1.95996
 p-value:  0.571608
 Conclusion using critical value approach:  Fail to reject the null hypothesis, there is
not enough evidence to support the alternative hypothesis, at the significance level of
0.05
 Conclusion using P value approach:  Fail to reject the null hypothesis, there is not
enough evidence to support the alternative hypothesis, at the significance level of 0.05
```

## 20.10 Hypothesis Tests for Two Population Proportions

| | | | |
|---|---|---|---|
| Assumptions | 1. Simple random samples<br>2. Independent samples<br>3. $x_1$, $n_1 - x_1$, $x_2$, and $n_2 - x_2$ are all 5 or greater | | |
| Null hypothesis | $H_0: p_1 = p_2$ | | |
| Alternative hypothesis | Two-tailed<br>$H_a: p_1 \neq p_2$ | Left tailed<br>$H_a: p_1 < p_2$ | Right tailed<br>$H_a: p_1 > p_2$ |
| Significance level | $\alpha$ | | |
| Test statistic:<br>For details see Chapter 18. | $z_0 = \dfrac{\hat{p}_1 - \hat{p}_2}{\sqrt{\hat{p}_p(1-\hat{p}_p)}\sqrt{\dfrac{1}{n_1} + \dfrac{1}{n_2}}}$,<br>$\hat{p}_p = \dfrac{x_1 + x_2}{n_1 + n_2}$ | | |
| **Critical value approach** | $\pm z_{\alpha/2}$ | $-z_\alpha$ | $z_\alpha$ |

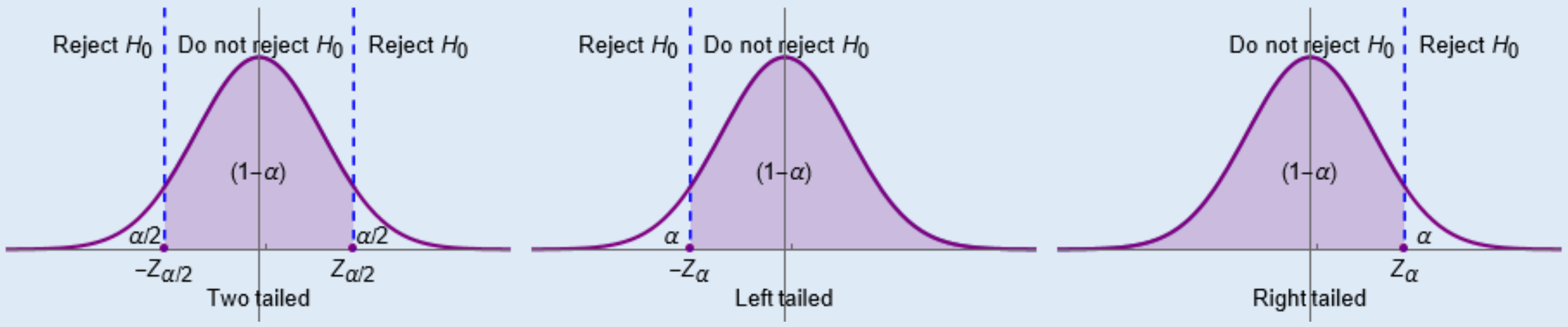

$$\begin{cases} \text{reject } H_0, & \text{If the value of the test statistic falls in the rejection region} \\ \text{do not reject } H_0, & \text{otherwise} \end{cases}$$

***P* value approach**

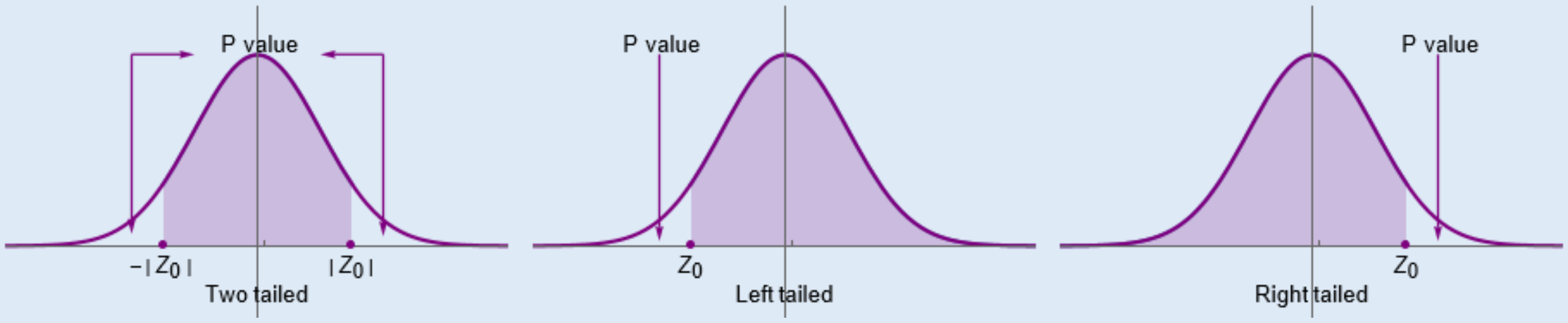

$$\begin{cases} \text{reject } H_0, & \text{If } P \leq \alpha \\ \text{do not reject } H_0, & \text{otherwise} \end{cases}$$

**Procedure 20.10.**

The steps for conducting a hypothesis test for two population proportions using the pooled sample proportion:

1. State the null and alternative hypotheses:
   - $H_0$: The two population proportions are equal ($p_1 = p_2$).
   - $H_1$: The two population proportions are not equal ($p_1 \neq p_2$, two-tailed test), or one is greater than/less than the other ($p_1 > p_2$ or $p_1 < p_2$, one-tailed test).
2. Set the significance level, $\alpha$, for the test.
3. Collect independent random samples from both populations.
4. Calculate the sample proportions ($\hat{p}_1$ and $\hat{p}_2$) for each sample.
5. Calculate the pooled sample proportion ($\hat{p}_p$), which is the weighted average of the two sample proportions, using the formula:

$$\hat{p}_p = \frac{x_1 + x_2}{n_1 + n_2}$$

   where $x_1$ and $x_2$ are the number of successes in each sample, and $n_1$ and $n_2$ are the sample sizes.
6. Calculate the SE of the pooled sample proportion using the formula:

$$\text{SE} = \sqrt{\hat{p}_p(1 - \hat{p}_p)}\sqrt{\frac{1}{n_1} + \frac{1}{n_2}}$$

7. Calculate the test statistic using the formula:
   For a z-test:

$$z = \frac{(\hat{p}_1 - \hat{p}_2) - 0}{\text{SE}}$$

8. Determine the critical value(s) for the test statistic.
   - For a two-tailed test, divide the significance level by 2 and find the corresponding critical value(s) from the standard normal distribution table.
   - For a one-tailed test, find the critical value(s) corresponding to the desired tail area.
9. Compare the test statistic to the critical value(s).
   - If the test statistic falls within the critical region, reject the null hypothesis.
   - Otherwise, fail to reject the null hypothesis.
10. Calculate the $P$-value associated with the test statistic.
    - For a two-tailed test, calculate the probability of the test statistic being greater than the observed value or less than the negative of the observed value from the standard normal distribution.
    - For a one-tailed test, calculate the probability of the test statistic being greater than the observed value or less than the observed value, depending on the alternative hypothesis.
11. Compare the $P$-value to $\alpha$.
    - If the $P$-value is less than α, reject the null hypothesis.
    - Otherwise, fail to reject the null hypothesis.
12. Draw a conclusion based on the results.
    - If the null hypothesis is rejected, conclude that there is sufficient evidence to support the alternative hypothesis.

- If the null hypothesis is not rejected, conclude that there is not enough evidence to support the alternative hypothesis.

***Example 20.19***

```
(* The code demonstrates the steps for conducting a hypothesis test for two population
proportions using the pooled sample proportion. The code will output the sample
proportions,pooled sample proportion, standard error, test statistic, critical value, p-
value,and the conclusion based on the results of the hypothesis test for the two population
proportions using the pooled sample proportion: *)

(* Specify the type of hypothesis test, choose one of the following, "two-tailed","left-
tailed", or "right-tailed": *)
hypothesisType="two-tailed";

(* Step 1. State the null and alternative hypotheses: *)
(* Null hypothesis:Population proportions are equal: *)
"H0:p1==p2";

(* Alternative hypothesis: *)
alternativeHypothesis=Switch[
    hypothesisType,
    "two-tailed",
    "H1:p1!=p2",
    "left-tailed",
    "H1:p1<p2",
    "right-tailed",
    "H1:p1>p2"
    ];

(* Step 2. Set the significance level (α): *)
α=0.05; (* Significance level of 0.05. *)

(* Step 3. Collect independent random samples from both populations: *)
SeedRandom[9123];
sample1=RandomInteger[1,50]
sample2=RandomInteger[1,70]
n1=Length[sample1]; (* Sample size of population 1. *)
n2=Length[sample2]; (* Sample size of population 2. *)

(* Step 4. Calculate the sample proportions: *)
phat1=N[Count[sample1,1]/n1]; (* Sample proportion of population 1. *)
phat2=N[Count[sample2,1]/n2]; (* Sample proportion of population 2. *)

(* Step 5. Calculate the pooled sample proportion: *)
phat=N[(phat1*n1+phat2*n2)/(n1+n2)];

(* Step 6. Calculate the standard error: *)
SE=N[Sqrt[phat*(1-phat)*(1/n1+1/n2)]];

(* Step 7. Calculate the test statistic: *)
testStatistic=N[((phat1-phat2)-0)/SE];

(* Critical value approach: *)
(* Step 8. Determine the critical value(s) and alternative hypotheses: *)
criticalValue=Switch[
    hypothesisType,
    "two-tailed",
    Quantile[NormalDistribution[0,1],1-α/2],
```

```
    "left-tailed",
    Quantile[NormalDistribution[0,1],α],
    "right-tailed",
    Quantile[NormalDistribution[0,1],1-α]
    ];

(* Step 9. Compare the test statistic to the critical value(s): *)
criticalvalueapproach=Switch[
    hypothesisType,
    "two-tailed",
    If[
     Abs[testStatistic]>criticalValue,
     (* Reject the null hypothesis: *)
     conclusion="Reject the null hypothesis, there is sufficient evidence to support the alternative hypothesis, at the significance level of "<>ToString[α],
     (* Fail to reject the null hypothesis: *)
     conclusion="Fail to reject the null hypothesis, there is not enough evidence to support the alternative hypothesis, at the significance level of "<>ToString[α]
     ],
    "left-tailed",
    If[
     testStatistic<criticalValue,
     (* Reject the null hypothesis: *)
     conclusion="Reject the null hypothesis, there is sufficient evidence to support the alternative hypothesis, at the significance level of "<>ToString[α],
     (* Fail to reject the null hypothesis: *)
     conclusion="Fail to reject the null hypothesis, there is not enough evidence to support the alternative hypothesis, at the significance level of "<>ToString[α]
     ],
    "right-tailed",
    If[
     testStatistic>criticalValue,
     (* Reject the null hypothesis: *)
     conclusion="Reject the null hypothesis, there is sufficient evidence to support the alternative hypothesis, at the significance level of "<>ToString[α],
     (* Fail to reject the null hypothesis: *)
     conclusion="Fail to reject the null hypothesis, there is not enough evidence to support the alternative hypothesis, at the significance level of "<>ToString[α]
     ]
    ];

(* P value approach: *)
(* Step 10. Calculate the p-value: *)
pValue=Switch[
    hypothesisType,
    "two-tailed",
    2*(1-CDF[NormalDistribution[0,1],Abs[testStatistic]]),
    "left-tailed",
    CDF[NormalDistribution[0,1],testStatistic],
    "right-tailed",
    1-CDF[NormalDistribution[0,1],testStatistic]
    ];

(* Step 11. Compare the p-value to the significance level: *)
pvalueapproach=If[
    pValue<α,
    (* Reject the null hypothesis: *)
    conclusion="Reject the null hypothesis, there is sufficient evidence to support the alternative hypothesis, at the significance level of "<>ToString[α],
    (* Fail to reject the null hypothesis: *)
```

```
    conclusion="Fail to reject the null hypothesis, there is not enough evidence to support the alternative hypothesis, at the significance level of "<>ToString[α]
    ];

(* Step 12. Draw a conclusion based on the results: *)
Print["Sample Proportion 1: ",phat1];
Print["Sample Proportion 2: ",phat2];
Print["Pooled Sample Proportion: ",phat];
Print["Standard Error: ",SE];
Print["Test Statistic: ",testStatistic];
Print["Critical Value: ",criticalValue];
Print["p-value: ",pValue];
Print["Conclusion using critical value approach: ",criticalvalueapproach];
Print["Conclusion using P value approach: ",pvalueapproach];

{0,0,1,1,0,1,1,1,0,0,0,1,0,0,0,1,0,1,0,1,1,0,1,0,1,1,0,1,1,0,0,0,1,1,1,0,0,0,0,0,0,1,1,1,1,1,1,0,0,1}

{1,0,1,0,1,1,0,0,0,1,1,1,0,0,0,1,1,1,0,0,0,1,1,1,1,1,1,0,1,1,1,0,0,1,0,1,1,0,0,1,1,0,0,0,1,1,1,0,0,1,1,1,0,0,1,1,0,0,1,1,0,1,1,1,0,0,0,1,0,0}

 Sample Proportion 1:  0.5
 Sample Proportion 2:  0.542857
 Pooled Sample Proportion:  0.525
 Standard Error:  0.0924662
 Test Statistic:  -0.46349
 Critical Value:  1.95996
 p-value:  0.643013
 Conclusion using critical value approach:  Fail to reject the null hypothesis, there is not enough evidence to support the alternative hypothesis, at the significance level of 0.05
 Conclusion using P value approach:  Fail to reject the null hypothesis, there is not enough evidence to support the alternative hypothesis, at the significance level of 0.05
```

## 20.11 Chi-Square Goodness-of-Fit Test

| | |
|---|---|
| Assumptions | 1. All expected frequencies are 1 or greater<br>2. At most 20% of the expected frequencies are less than 5<br>3. Simple random sample |
| Null hypothesis | H0: The variable has the specified distribution |
| Alternative hypothesis | Ha: The variable does not have the specified distribution. |
| Significance level | $\alpha$ |
| Test statistic:<br>For details see Chapter 18. | $$\chi_0^2 = \sum \frac{(\text{Observed} - \text{Expected})^2}{\text{Expected}}$$ where Observed and Expected represent observed and expected frequencies, respectively. |
| **Critical value approach** | |

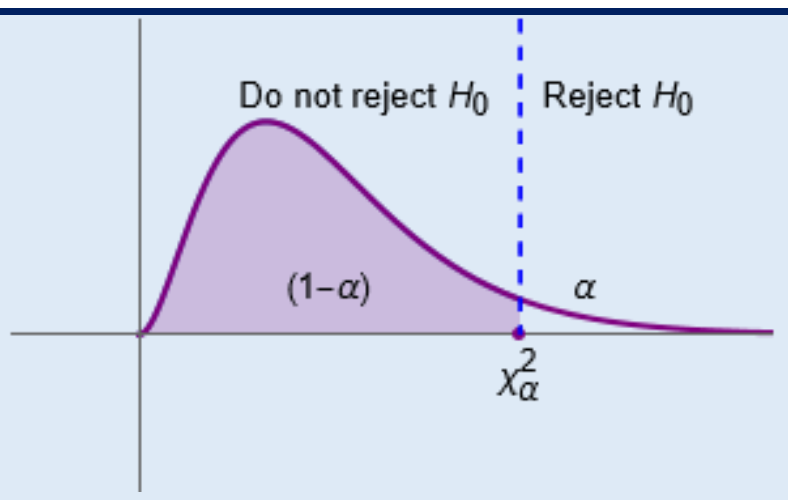

$$\begin{cases} \text{reject } H_0, & \text{If the value of the test statistic falls in the rejection region} \\ \text{do not reject } H_0, & \text{otherwise} \end{cases}$$

***P* value approach**

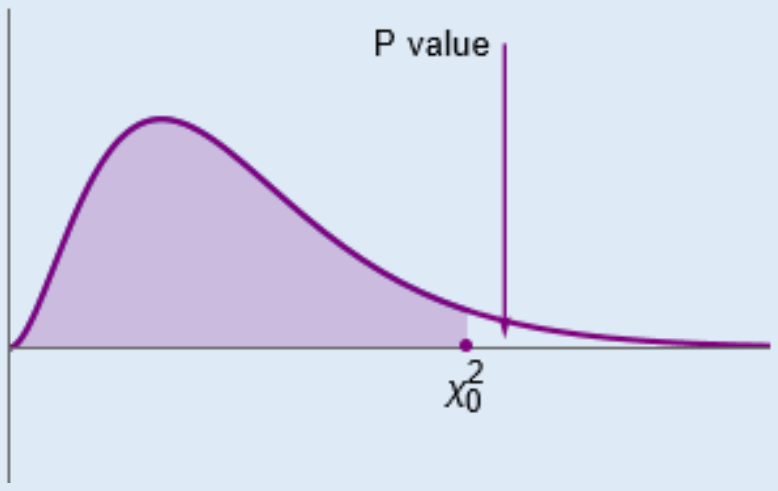

$$\begin{cases} \text{reject } H_0, & \text{If } P \leq \alpha \\ \text{do not reject } H_0, & \text{otherwise} \end{cases}$$

**Procedure 20.11.**

The steps for conducting a Chi-Square goodness-of-fit test:

1. State the null and alternative hypotheses:
    - $H_0$: The observed frequencies in each category follow the expected frequencies.
    - $H_1$: The observed frequencies in at least one category do not follow the expected frequencies.
2. Set the significance level, $\alpha$, for the test.
3. Define the observed frequencies in each category. These are the frequencies or counts you have obtained from your data.
4. Determine the expected frequencies in each category. These are the frequencies you would expect to observe if the null hypothesis is true. The expected frequencies can be calculated based on a specified distribution or a hypothesis about the proportions in each category.
5. Calculate the test statistic using the Chi-Square formula:
$$\chi^2 = \sum \frac{(\text{Observed} - \text{Expected})^2}{\text{Expected}}$$
where $\Sigma$ represents the sum of all categories.
6. Determine the degrees of freedom ($df$) for the test. This is equal to the number of categories minus 1 ($df = k - 1$), where $k$ is the number of categories.
7. Determine the critical value of the Chi-Square distribution for the specified significance level and degrees of freedom. This critical value separates the rejection region from the non-rejection region.
8. Compare the test statistic to the critical value.
    - If the test statistic is greater than the critical value, reject the null hypothesis.
    - Otherwise, fail to reject the null hypothesis.
9. Calculate the p-value associated with the test statistic. The p-value can be calculated using the Chi-Square distribution and the cumulative distribution function (CDF).
10. Compare the p-value to $\alpha$.
    - If the p-value is less than α, reject the null hypothesis.
    - Otherwise, fail to reject the null hypothesis.
11. Draw a conclusion based on the results.

- If the null hypothesis is rejected, conclude that there is sufficient evidence to support the alternative hypothesis.
- If the null hypothesis is not rejected, conclude that there is not enough evidence to support the alternative hypothesis.

***Example 20.20***

```
(* Step 1. State the null and alternative hypotheses: *)
"H0:The observed frequencies follow the expected frequencies";
"H1:The observed frequencies do not follow the expected frequencies";

(* Step 2. Set the significance level (α): *)
α=0.05; (* Significance level of 0.05. *)

(* Step 3. Define the observed frequencies: *)
observed={18,22,20,25}; (* Observed frequencies in each category. *)
n=Total[observed]; (* Total number of observations.*)

(* Step 4. Determine the expected frequencies: *)
expected={20,20,20,20}; (* Expected frequencies in each category. *)

(* Step 5. Calculate the test statistic: *)
chiSquare=N[Total[((observed-expected)^2)/expected]];

(* Step 6. Determine the degrees of freedom: *)
df=Length[observed]-1;

(* Step 7. Determine the critical value: *)
criticalValue=N[Quantile[ChiSquareDistribution[df],1-α]];

(* Step 8. Compare the test statistic to the critical value: *)
criticalvalueapproach=If[
    chiSquare>criticalValue,
    (* Reject the null hypothesis: *)
    conclusion="Reject the null hypothesis, there is sufficient evidence to support the
alternative hypothesis, at the significance level of "<>ToString[α],
    (* Fail to reject the null hypothesis: *)
    conclusion="Fail to reject the null hypothesis, there is not enough evidence to support
the alternative hypothesis, at the significance level of "<>ToString[α]
    ];

(* Step 9. Calculate the p-value: *)
pValue=N[1-CDF[ChiSquareDistribution[df],chiSquare]];

(* Step 10. Compare the p-value to the significance level: *)
pvalueapproach=If[
    pValue<α,
    (* Reject the null hypothesis: *)
    conclusion="Reject the null hypothesis, there is sufficient evidence to support the
alternative hypothesis, at the significance level of "<>ToString[α],
    (* Fail to reject the null hypothesis: *)
    conclusion="Fail to reject the null hypothesis, there is not enough evidence to support
the alternative hypothesis, at the significance level of "<>ToString[α]
    ];

(* Step 11. Draw a conclusion based on the results: *)
Print["Observed Frequencies: ",observed];
Print["Expected Frequencies: ",expected];
Print["Test Statistic: ",chiSquare];
```

```
Print["Degrees of Freedom: ",df];
Print["Critical Value: ",criticalValue];
Print["p-value: ",pValue];
Print["Conclusion using critical value approach: ",criticalvalueapproach];
Print["Conclusion using P value approach: ",pvalueapproach];

 Observed Frequencies:  {18,22,20,25}
 Expected Frequencies:  {20,20,20,20}
 Test Statistic:  1.65
 Degrees of Freedom:  3
 Critical Value:  7.81473
 p-value:  0.648107
 Conclusion using critical value approach:  Fail to reject the null hypothesis, there is
not enough evidence to support the alternative hypothesis, at the significance level of
0.05
 Conclusion using P value approach:  Fail to reject the null hypothesis, there is not
enough evidence to support the alternative hypothesis, at the significance level of 0.05
```

# CHAPTER 21

# MATHEMATICA LAB: DECISION THEORY AND HYPOTHESIS TESTING

Mathematica offers a wide range of built-in functions and tools to perform various statistical analyses and decision-making tasks. In this chapter, we will explore several essential Mathematica functions that are commonly used in decision theory and statistical tests. These functions enable researchers and data analysts to draw reliable inferences from data and make well-informed choices based on statistical evidence.

The following Mathematica functions will be covered in this chapter:

- The `LocationTest` function is used to test hypotheses related to the location or central tendency of a dataset. It helps determine whether there are statistically significant differences between the means of two or more groups of data. The function employs various statistical tests, such as Mann Whitney test, paired t test, paired z test, sign test, signed rank test, t test, and z test, depending on the nature of the data and the assumptions made.
- `VarianceTest` is another important function in Mathematica, which focuses on testing hypotheses related to the variability or spread of data. The function employs various statistical tests, such as Conover test, Brown-Forsythe test, Fisher ratio test, Levene test, and Siegel-Tukey test, depending on the nature of the data and the assumptions made.
- When dealing with real-world data, it is essential to ascertain whether the data follows a specific theoretical distribution. The `DistributionFitTest` function allows us to test whether a dataset fits a given probability distribution. This functionality is invaluable in decision-making processes where assumptions about the underlying distribution are crucial.
- The `IndependenceTest` function is specifically designed to assess the independence between two or more variables. By applying various statistical tests, such as chi-square tests, it aids in determining whether there is a significant relationship between the variables or if they are independent.
- `CorrelationTest` is used to evaluate the correlation between two numerical variables. Correlation measures the strength and direction of the relationship between variables, and this function helps in making decisions about the strength and significance of the correlation coefficient.

Throughout this chapter, we will provide detailed explanations of each function's usage, syntax, and interpretation of results. In the following table, we list the built-in functions that are used in this chapter.

| | |
|---|---|
| `LocationTest` | `DistributionFitTest` |
| `LocationEquivalenceTest` | `IndependenceTest` |
| `VarianceTest` | `CorrelationTest` |
| `VarianceEquivalenceTest` | |

Therefore, we divided this chapter into four units to cover the above topics.

**Chapter 21 Outline**

# UNIT 21.1

# LOCATION TESTS

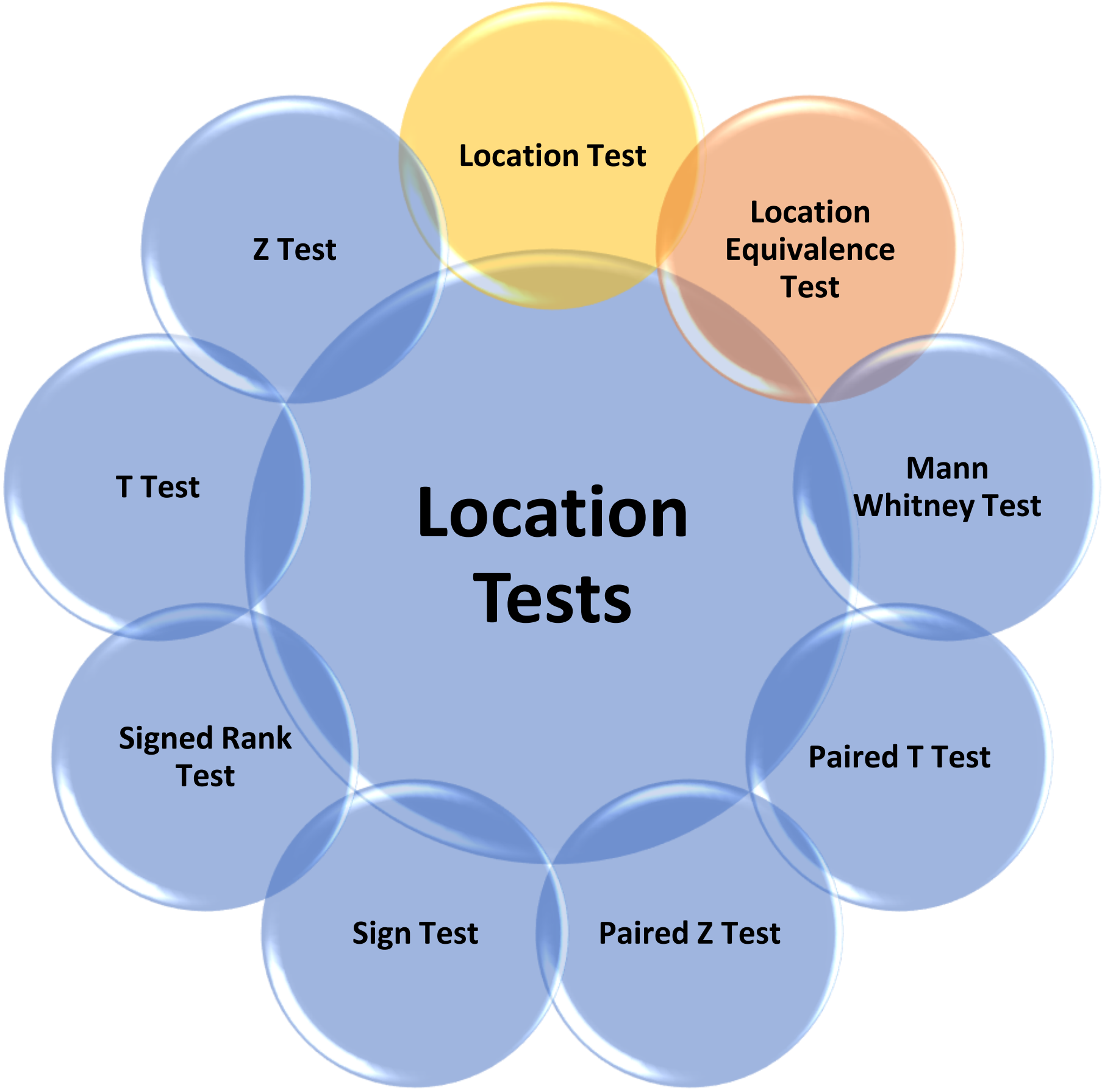

## LocationTest

| | |
|---|---|
| `LocationTest[data]` | tests whether the mean or median of the data is zero. |
| `LocationTest[{data1,data2}]` | tests whether the means or medians of data1 and data2 are equal. |
| `LocationTest[dspec,μ0]` | tests a location measure against μ0. |
| `LocationTest[dspec,μ0,"property"]` | returns the value of "property". |

The following tests can be used:

| | | |
|---|---|---|
| `"PairedT"` | normality | paired sample test with unknown variance |
| `"PairedZ"` | Normality | paired sample test with known variance |
| `"Sign"` | Robust | median test for one sample or matched pairs |
| `"SignedRank"` | Symmetry | median test for one sample or matched pairs |
| `"T"` | Normality | mean test for one or two samples |
| `"MannWhitney"` | Symmetry | median test for two independent samples |
| `"Z"` | normality | mean test with known variance |

## LocationEquivalenceTest

| | |
|---|---|
| `LocationEquivalenceTest[{data1,data2,…}]` | tests whether the means or medians of the datai are equal. |
| `LocationEquivalenceTest[{data1,…},"property"]` | returns the value of "property". |

The following tests can be used:

| | | |
|---|---|---|
| "CompleteBlockF" | normality, blocked | mean test for complete block design |
| "FriedmanRank" | blocked | median test for complete block design |
| "KruskalWallis" | symmetry | median test for two or more samples |
| "KSampleT" | normality | mean test for two or more samples |

## Reporting of test results

Properties related to the reporting of test results include:

| | |
|---|---|
| "AllTests" | list of all applicable tests |
| "AutomaticTest" | test chosen if Automatic is used |
| "DegreesOfFreedom" | the degrees of freedom used in a test |
| "PValue" | list of -values |
| "PValueTable" | formatted table of -values |
| "ShortTestConclusion" | a short description of the conclusion of a test |
| "TestConclusion" | a description of the conclusion of a test |
| "TestData" | list of pairs of test statistics and -values |
| "TestDataTable" | formatted table of -values and test statistics |
| "TestStatistic" | list of test statistics |
| "TestStatisticTable" | formatted table of test statistics |

## Options

The following options can be used:

| | | |
|---|---|---|
| AlternativeHypothesis | "Unequal" | the inequality for the alternative hypothesis |
| MaxIterations | Automatic | max iterations for multivariate median tests |
| Method | Automatic | the method to use for computing -values |
| SignificanceLevel | 0.05 | cutoff for diagnostics and reporting |
| VerifyTestAssumptions | Automatic | what assumptions to verify |

## Hypothesis testing package

| | |
|---|---|
| `NormalPValue[x]` | gives the cumulative density beyond x for a normal distribution with zero mean and unit variance. |
| `StudentTPValue[x,df]` | gives the cumulative probability beyond x for Student's t distribution with df degrees of freedom. |
| `ChiSquarePValue[x,df]` | gives the cumulative probability beyond x for the chi distribution with df degrees of freedom. |
| `FRatioPValue[x,n,m]` | gives the cumulative probability beyond x for the F-ratio distribution with n and m degrees of freedom. |

***Mathematica Examples 21.1***

```
Input    (* The P-value for -1.96 in a normal distribution: *)
         Needs["HypothesisTesting`"]
         NormalPValue[-1.96]
         NormalPValue[-1.96,TwoSided->True]

Output   OneSidedPValue->0.0249979
Output   TwoSidedPValue->0.0499958
```

***Mathematica Examples 21.2***

```
Input    (* The P-value for -3 in a t distribution with 6 degrees of freedom: *)
```

```
         Needs["HypothesisTesting`"]
         StudentTPValue[-3.,6]
         StudentTPValue[-3.,6,TwoSided->True]

Output   OneSidedPValue->0.0120041
Output   TwoSidedPValue->0.0240082
```

***Mathematica Examples 21.3***

```
Input    (* The P-value for 12 in a Chi square distribution with 6 degrees of freedom: *)
         Needs["HypothesisTesting`"]
         ChiSquarePValue[12,6]
         ChiSquarePValue[12,6,TwoSided->True]

Output   OneSidedPValue->0.0619688
Output   TwoSidedPValue->0.123938
```

***Mathematica Examples 21.4***

```
Input    (* The-value for 12 in an F-ratio distribution with 3 and 6 degrees of freedom: *)
         Needs["HypothesisTesting`"]
         FRatioPValue[12,3,6]
         FRatioPValue[12,3,6,TwoSided->True]

Output   OneSidedPValue->0.00602551
Output   TwoSidedPValue->0.012051
```

***Mathematica Examples 21.5***

```
Input    (* The code demonstrates how the p-values in one-sample location tests behave
         differently depending on the proximity of the dataset's mean to the hypothesized
         mean. For data1, where the mean is close to 0, the p-values are typically large.
         Conversely, for data2, where the mean is far from 0, the p-values are typically
         small: *)

         (* Test H0: µ=0 versus Ha: µ!=0: *)
         data1=RandomVariate[
             NormalDistribution[0,1],
             500
             ];

         data2=RandomVariate[
             NormalDistribution[3,1],
             500
             ];

         (* The P-values are typically large when the mean is close to 0: *)
         LocationTest[data1]

         (* The P-values are typically small when the mean is far from 0: *)
         LocationTest[data2]

Output   0.9475
Output   1.22001*10^-250
```

***Mathematica Examples 21.6***

```
Input    (* The code demonstrates how the p-values in one-sample location tests behave
         differently depending on the proximity of the dataset's mean to the hypothesized
         mean. In this case, for data1, where the mean is close to the hypothesized value 3,
         the p-values tend to be large. Conversely, for data2, where the mean is far from the
         value 3, the p-values tend to be small: *)
```

```
        (* Test Subscript[H, 0]:μ=3 versus Subscript[H, a]:μ!=3: *)
        data1=RandomVariate[
           NormalDistribution[3,1],
           500
           ];

        data2=RandomVariate[
           NormalDistribution[0,1],
           500
           ];

        (* The P-values are typically large when the mean is close to μ0: *)
        LocationTest[data1,3]

        (* The P-values are typically small when the mean is far from μ0: *)
        LocationTest[data2,3]

Output  0.207801
Output  5.93575*10^-263
```

***Mathematica Examples 21.7***

```
Input   (* The code generates a sample dataset from a normal distribution, calculates its
        mean, and performs a one-sample test to compare the mean with a hypothesized value
        of 0. The code then presents the test statistic, p-value, test conclusion and a test
        data table for further analysis. Note that, the P-values are typically large when
        the mean is close to 0: *)

        (* Generate a sample data set: *)
        data=RandomVariate[
           NormalDistribution[0,1],
           100
           ];

        (* Generate the mean of the data set: *)
        Mean[data]

        (* Perform one-sample test: *)
        result=LocationTest[
          data,
          0,
          {"TestStatistic","PValue","TestConclusion"}
          ]

        (* Print the test statistic, p-value and test conclusion: *)
        Print["Test Statistic: ",result[[1]]];
        Print["P-value: ",result[[2]]];
        Print["Test Conclusion: ",result[[3]]];

        (* Test data table: *)
        LocationTest[
         data,
         0,
         {"TestDataTable",All}
         ]

Output  -0.027955
Output  {-0.277438,0.782022, The null hypothesis that the mean of the population is equal
        to 0   is not rejected at the 5 percent level based on the T test.}
Output  Test Statistic:  -0.277438
Output  P-value:  0.782022
```

```
Output  Test Conclusion:  The null hypothesis that the mean of the population is equal
        to 0 is not rejected at the 5 percent level based on the T test.

Output               | Statistic  P-Value
         Paired T    | -0.277438  0.782022
         Paired Z    | -0.277438  0.781444
         Sign        | 51         0.920411
         Signed-Rank | 2502.      0.938335
         T           | -0.277438  0.782022
         Z           | -0.277438  0.781444
```

***Mathematica Examples 21.8***

```
Input   (* The code generates a sample dataset from a normal distribution, calculates its
        mean, and performs a one-sample test to compare the mean with a hypothesized value
        of 1. The code then presents the test statistic, p-value, test conclusion and a test
        data table for further analysis. Note that, the P-values are typically small when
        the mean is far from 1: *)

        (* Generate a sample data set: *)
        data=RandomVariate[
            NormalDistribution[0,1],
            100
            ];

        (* Generate the mean of the data set: *)
        Mean[data]

        (* Perform one-sample test: *)
        result=LocationTest[
          data,
          1,
          {"TestStatistic","PValue","TestConclusion"}
          ]

        (* Print the test statistic, p-value and test conclusion: *)
        Print["Test Statistic: ",result[[1]]];
        Print["P-value: ",result[[2]]];
        Print["Test Conclusion: ",result[[3]]];

        (* Test data table: *)
        LocationTest[
         data,
         1,
         {"TestDataTable",All}
         ]
Output  0.135152
Output  {-8.79096,4.71549*10^-14,The null hypothesis that  the  mean  of the population
        is equal to 1 is rejected at the 5 percent level  based on the T test.}
Output  Test Statistic:  -8.79096
Output  P-value:  4.71549*10^-14
Output  Test Conclusion:  The null hypothesis that the mean of the population is equal
        to 1 is rejected at the 5 percent level based on the T test.

Output               | Statistic  P-Value
         Paired T    | -8.79096   4.71549*10^-14
         Paired Z    | -8.79096   1.48293*10^-18
         Sign        | 16         2.60594*10^-12
         Signed-Rank | 552.       1.18445*10^-11
         T           | -8.79096   4.71549*10^-14
         Z           | -8.79096   1.48293*10^-18
```

***Mathematica Examples 21.9***

Input

```
(* The code generates a sample dataset from a normal distribution, calculates its mean, and performs both a one-sample z-test and a one-sample t-test to compare the mean with a hypothesized value of 0. The code then presents the test statistics, p-values, and a test data table for further analysis: *)

(* Generate a sample data set: *)
data=RandomVariate[
    NormalDistribution[0,1],
    100
    ];

(* Generate the mean of the data set: *)
Mean[data]

(* Perform one-sample test: *)
result=LocationTest[
   data,
   0,
   {{"Z","TestStatistic"},{"T","TestStatistic"},"TestConclusion"}
   ]

(* Print the test statistic and p-value: *)
Print["P-value: ",{result[[1]][[1]],result[[2]][[1]]}];
Print["Test Statistic: ",{result[[1]][[2]],result[[2]][[2]]}];

(* Test data table: *)
LocationTest[
 data,
 0,
 {"TestDataTable",All}
 ]
```

Output 0.0127526

Output {{0.894868,0.132148},{0.895136,0.132148}, The null hypothesis that the mean of the population is equal to 0 is not rejected at the 5 percent level based on the T test.}

Output P-value: {0.894868,0.895136}

Output Test Statistic: {0.132148,0.132148}

Output

| | Statistic | P-Value |
|---|---|---|
| Paired T | 0.132148 | 0.895136 |
| Paired Z | 0.132148 | 0.894868 |
| Sign | 57 | 0.193348 |
| Signed-Rank | 2585. | 0.8379 |
| T | 0.132148 | 0.895136 |
| Z | 0.132148 | 0.894868 |

***Mathematica Examples 21.10***

Input

```
(* The code demonstrates two approaches to perform the same test, one using the actual hypothesized value (0) and the other using the keyword "Automatic" as a shorthand for the same hypothesized mean. Both methods yield identical results in testing for a mean of zero: *)

data=RandomVariate[
    NormalDistribution[0,1],
    500
    ];
```

```
        LocationTest[data,0]
        LocationTest[data,Automatic]

Output  0.740653
Output  0.740653
```

***Mathematica Examples 21.11***

```
Input   (* The code performs a one-sample location test for the dataset data and then
        interprets the results based on the obtained p-values. The code iterates through a
        range of values from 1 to 3 with an increment of 0.25, representing potential
        locations to test. The function LocationTest is used to perform the test for each
        location, and it returns the test statistic and the p-value for each test. The results
        are stored in the testResult list. If the p-value is less than 0.05, it suggests
        rejecting the null hypothesis, indicating that there is a significant difference
        between the data and the tested location. Otherwise, it suggests failing to reject
        the null hypothesis, meaning there is not enough evidence to suggest a significant
        difference between the data and the tested location at the 0.05 significance level:
        *)

        (* Step 1, define your data: *)
        data={2.3,1.8,2.6,2.2,1.9,2.4,2.1,2.7,2.5,2.0};
        Mean[data]

        (* Step 2, perform the one-sample location test: *)
        testResult=Table [
          LocationTest[
            data,
            i,
            {"TestStatistic","PValue"}
            ],
          {i,1,3,0.25}
          ]

        (* Step 3, interpret the results: *)
        testStatistic=Table[
          testResult[[i]][[1]],
          {i,1,9}
          ]

        pValue=Table[
          testResult[[i]][[2]],
          {i,1,9}
          ]

        Table[
         If[
          pValue[[i]]<0.05,
          Print["Reject the null hypothesis."],
          Print["Fail to reject the null hypothesis."]
          ],
         {i,1,9}
         ]

Output  2.25
Output  {{13.0558,3.74269*10^-7}, {10.4447,2.4883*10^-6}, {7.83349,0.0000261747},
        {5.22233,0.000547495}, {2.61116,0.0282168}, {0.,1.}, {-2.61116,0.0282168}, {-
        5.22233,0.000547495}, {-7.83349,0.0000261747}}
Output  {13.0558,10.4447,7.83349,5.22233,2.61116,0.,-2.61116,-5.22233,-7.83349}
Output  {3.74269*10^-7, 2.4883*10^-6, 0.0000261747, 0.000547495, 0.0282168, 1., 0.0282168,
        0.000547495, 0.0000261747}
```

```
Output    Reject the null hypothesis.
          Reject the null hypothesis.
          Reject the null hypothesis.
          Reject the null hypothesis.
          Reject the null hypothesis.
          Fail to reject the null hypothesis.
          Reject the null hypothesis.
          Reject the null hypothesis.
          Reject the null hypothesis.
```

***Mathematica Examples 21.12***

```
Input     (* Perform a one-sided and two-sided test: *)
          data=RandomVariate[
              NormalDistribution[],
              100
              ];

          (* The code performs a two-sided location test, where the alternative hypothesis (Ha)
          is that the population mean (μ) is not equal to 0: *)
          LocationTest[data,0,AlternativeHypothesis->"Unequal"]
          LocationTest[data,0,AlternativeHypothesis->Automatic]

          (* The code performs a one-sided location test, where the alternative hypothesis (Ha)
          is that the population mean (μ) is less than 0: *)
          LocationTest[data,0,AlternativeHypothesis->"Less"]

          (* The code performs another one-sided location test, but with the alternative
          hypothesis (Ha) that the population mean (μ) is greater than 0: *)
          LocationTest[data,0,AlternativeHypothesis->"Greater"]

Output    0.673936
Output    0.673936
Output    0.663032
Output    0.336968
```

***Mathematica Examples 21.13***

```
Input     (* The code performs a simulation-based analysis to investigate the relationship
          between the sample mean and the test statistic (and p-value) resulting from one-
          sample location tests. The code generates random samples from normal distributions
          with known parameters (varying means and a constant standard deviation) and then
          conducts one-sample location tests for a fixed test location of 5. It collects the
          test statistics and p-values for each test and plots them against different sample
          means (μ) ranging from 1 to 10. The two list plots generated from the data illustrate
          how the test statistic and p-values change as the sample mean (μ) varies. These
          visualizations offer insights into the variability of test results across different
          sample means: *)

          (* Generate a random samples from a normal distribution with known parameters: *)
          SeedRandom[124];
          data=Table[
              RandomVariate[
               NormalDistribution[μ,2],
               100
               ],
              {μ,1,10,0.1}
              ];

          testResult=Table[
              LocationTest[
               data[[i]],
```

```
            5,
            {"TestStatistic","PValue"}
            ],
          {i,1,Length[data]}
          ];

    (* Extract the Test Statistic and P-value for plotting: *)
    testStatistic=Transpose[testResult][[1]];
    pValues=Transpose[testResult][[2]];
    meanRange=Range[1,10,0.1];

    testStatisticdata=Table[
        {meanRange[[i]],testStatistic[[i]]},
        {i,1,Length[data]}
        ];

    pValuesdata=Table[
        {meanRange[[i]],pValues[[i]]},
        {i,1,Length[data]}
        ];

    (* Plot the Test Statistic and P-value against Mean (μ): *)
    ListPlot[
     testStatisticdata,
     Frame->True,
     ImageSize->250,
     PlotStyle->Directive[Purple,Opacity[0.5]]
     ]

    ListPlot[
     pValuesdata,
     Frame->True,
     PlotRange->All,
     ImageSize->250,
     PlotStyle->Directive[Purple,Opacity[0.5]]
     ]
```

Output

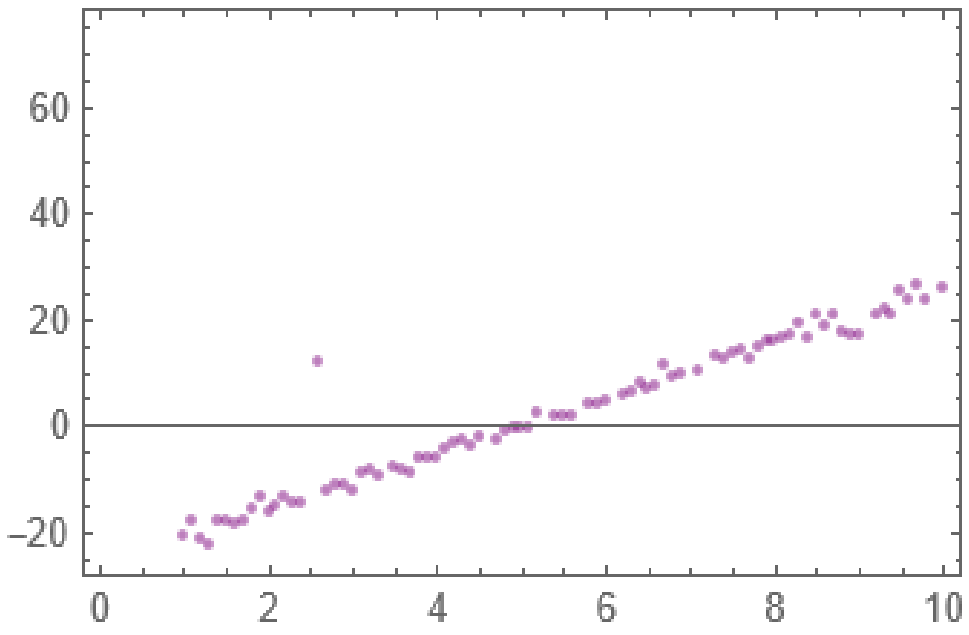

Output

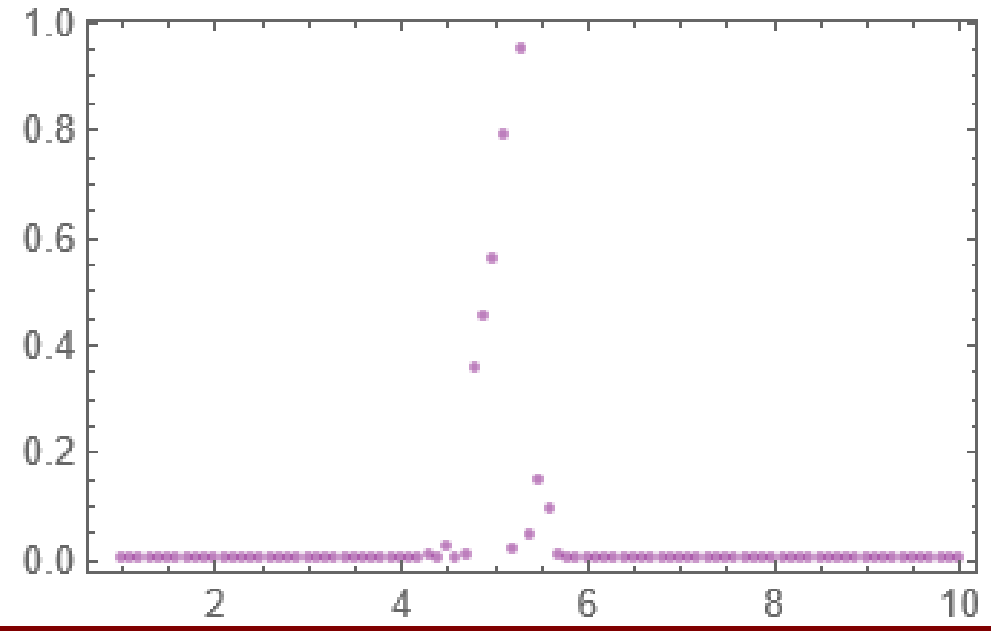

### *Mathematica Examples 21.14*

Input

```
(* The code creates a Manipulate function that allows interactive exploration of a
one-sample location test. The Manipulate function dynamically generates random
samples from a normal distribution with user-defined mean and standard deviation
parameters. It then performs a one-sample location test with a fixed test location
of 5 and displays the test statistic, p-value, and the result of the test (whether
the null hypothesis is rejected or not): *)

(* Define the Manipulate function: *)
Manipulate[
 Module[
  {testResult},

  (* Generate a random sample from a normal distribution with known parameters: *)
  data=RandomVariate[
    NormalDistribution[μ,σ],
    100
    ];

  (* Perform the LocationTest with user-defined parameters: *)
  testResult=LocationTest[
    data,
    5,
    {"TestStatistic","PValue"}
    ];

  (* Display the results*)
  Column[
   {
    Row[{"Test Statistic: ",testResult[[1]]}],
    Row[{"P-value: ",testResult[[2]]}],
    If[
     testResult[[2]]<0.05,
     Row[{"Reject the null hypothesis at 5% significance level"}],
     Row[{"Fail to reject the null hypothesis at 5% significance level"}]]}]
  ],

 (* Controls for the parameters: *)
 {{μ,5,"Mean"},0,10,0.1},
 {{σ,2,"Standard Deviation"},0.1,5,0.1}
 ]
```

Output

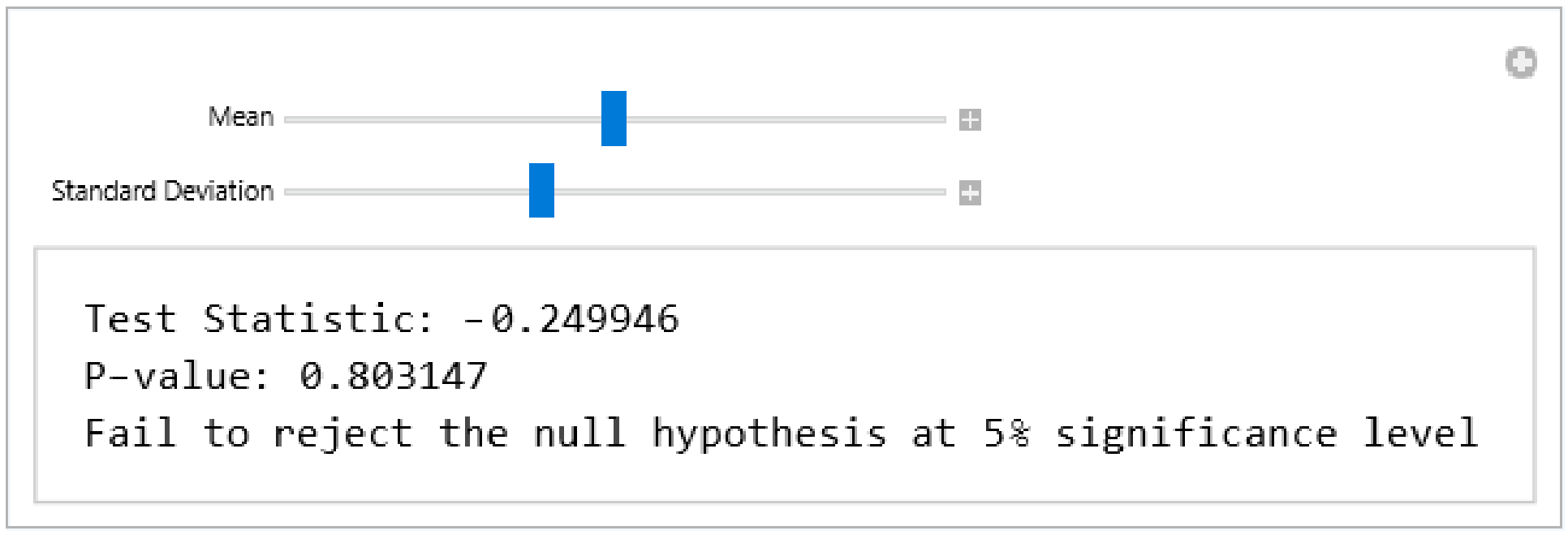

### *Mathematica Examples 21.15*

Input

```
(* The code provides an interactive tool for exploring the behavior of a one-sample
location test under different settings. Users can observe how the test conclusion
changes as they modify the distribution parameters, sample size, null hypothesis
mean, and significance level: *)
```

```
(* Create a Manipulate function: *)
Manipulate[
(* Generate a random sample from a normal distribution with specified parameters: *)
 data=RandomVariate[
   NormalDistribution[μ,σ],
   n
   ];

 (* Perform the LocationTest with the chosen parameters: *)
 testResult=LocationTest[
   data,
   μ0,
   {"TestStatistic","PValue","ShortTestConclusion"},
   SignificanceLevel->alpha
   ];

 (* Display the results*)
 Column[
  {
   Row[{"Test Statistic: ",testResult[[1]]}],
   Row[{"PValue: ",testResult[[2]]}],
   If[
    testResult[[3]]==="Do not reject",
    Row[{"Conclusion: The null hypothesis is not rejected at ",alpha*100,"%
significance level."}],
    Row[{"Conclusion: The null hypothesis is rejected at "  ,alpha*100,"%
significance level."}]
    ]
   }
  ],

 (* Manipulate Parameters*)
 {{μ,0,"Mean of Distribution"},-5,5,0.1},
 {{σ,1,"Standard Deviation"},0.1,5,0.1},
 {{n,20,"Sample Size"},10,100,10},
 {{μ0,0,"Null Hypothesis Mean"},-5,5,0.1},
 {{alpha,0.05,"Significance Level"},0.01,0.10,0.01}
 ]
```

Output

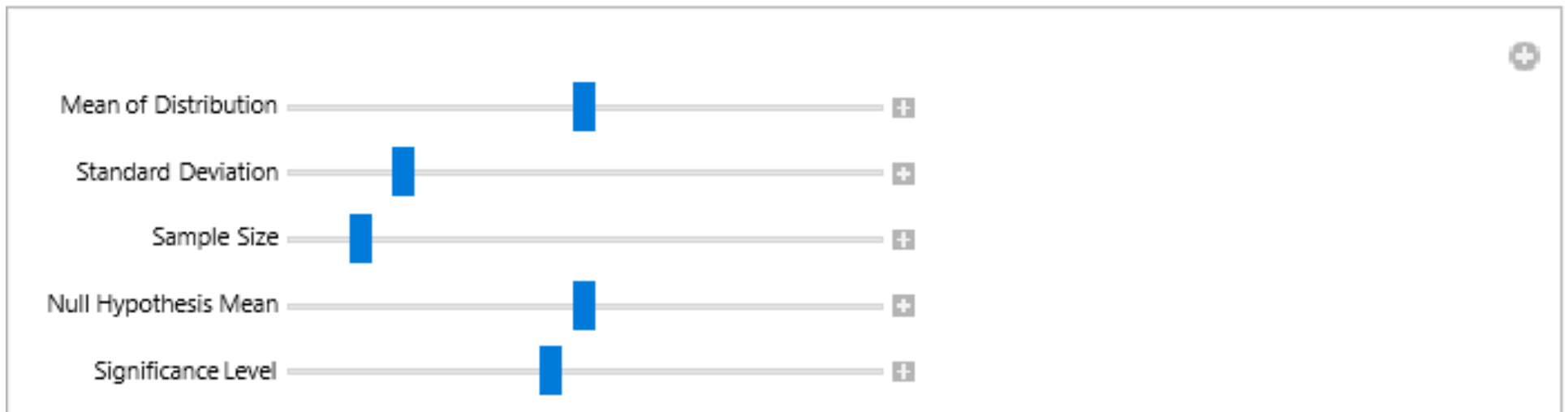

```
Test Statistic: 0.61252
PValue: 0.547455
Conclusion: The null hypothesis is not rejected at
 5.% significance level.
```

***Mathematica Examples 21.16***

Input

```
(* The code creates an interactive visualization using Manipulate function to conduct
and visualize a one-sample location test. Users can dynamically adjust the mean,
```

```
standard deviation, sample size, and significance level parameters and observe the
corresponding test results: *)

Manipulate[
 Module[
  {data,pValue},
  (* Generate sample data from a normal distribution: *)
  data=RandomVariate[
    NormalDistribution[mean,stdDev],
    sampleSize
    ];

  (* Perform a one-sample test against the null hypothesis mean=0: *)
  pValue=LocationTest[
    data,
    0,
    {"PValue"},
    SignificanceLevel->significanceLevel
    ];

  (* Display the histogram of the data along with the test p-value: *)
  Histogram[
   data,
   Automatic,
   "Probability",
   PlotLabel->Row[{"p-value: ",NumberForm[pValue,{4,3}]}],
   ColorFunction->Function[{height},Opacity[height]],
   ImageSize->300,
   ChartStyle->Purple
   ]
  ],
 {{mean,0,"Mean"},-6,6,0.1,Appearance->"Labeled"},
 {{stdDev,1,"Standard Deviation"},0.1,5,0.1,Appearance->"Labeled"},
 {{sampleSize,200,"Sample Size"},10,1000,10,Appearance->"Labeled"},
 {{significanceLevel,0.05,"Significance Level"},0.01,0.5,0.01,Appearance-
>"Labeled"}
 ]
```

Output

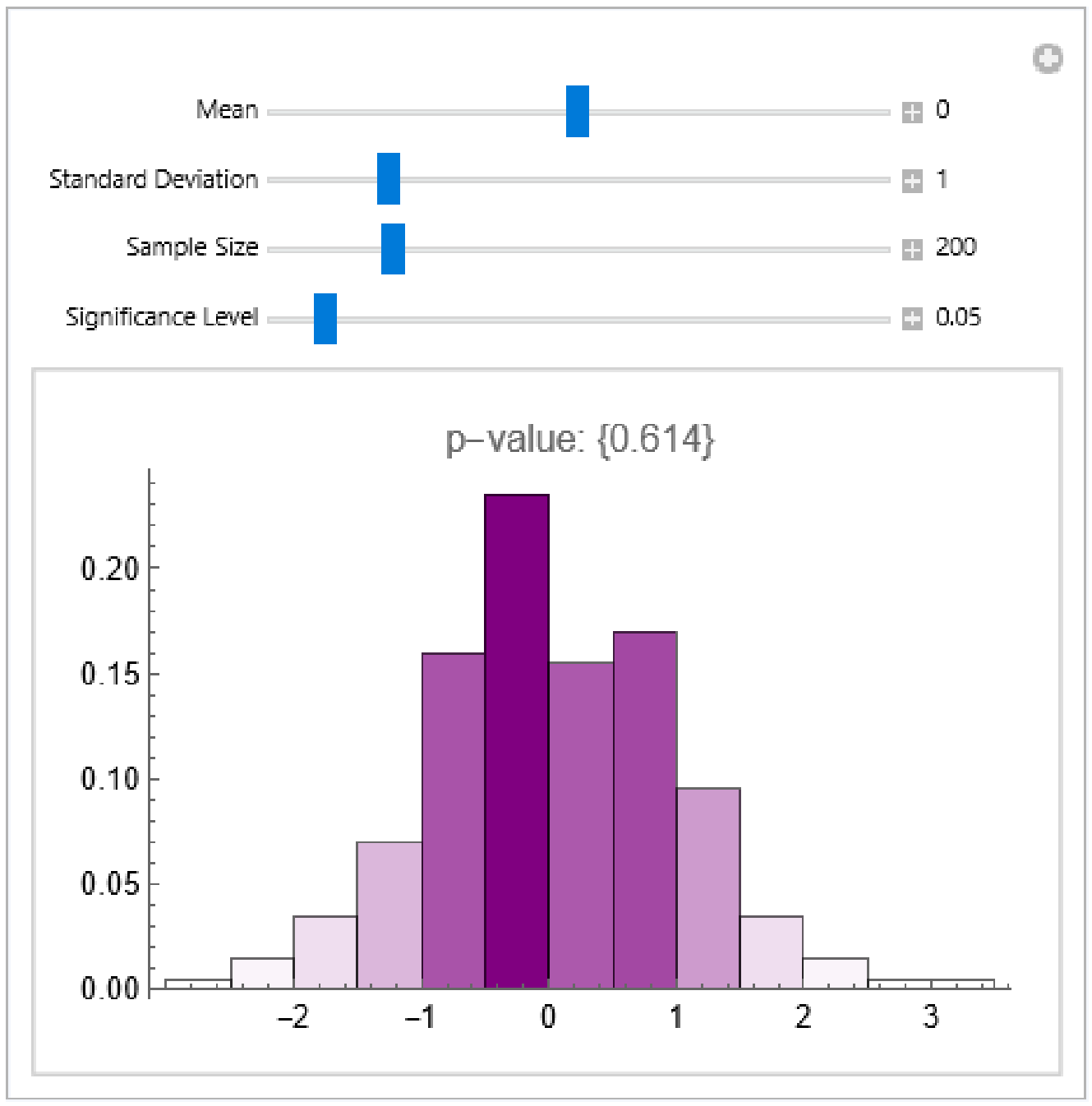

***Mathematica Examples 21.17***

Input

```
(* The code provides a tool to visualize the p-value resulting from the one-sample
test and understand how the test outcome changes as you vary the null hypothesis
mean. The p-value represents the probability of observing the data, assuming that
the null hypothesis is true (i.e., the data is drawn from a normal distribution with
mean 0). If the p-value is small (e.g., less than 0.05),it suggests that the null
hypothesis is unlikely, and you may reject it in favor of an alternative hypothesis.
If the p-value is large (e.g., greater than 0.05),it suggests that the null hypothesis
is plausible, and you may fail to reject it: *)

(* Generate sample data from a normal distribution: *)
data=RandomVariate[
    NormalDistribution[0,1],
    200
    ];

Manipulate[
 Module[
  {pValue},
  (* Perform a one-sample test against the null hypothesis mean=μ0: *)
  pValue=LocationTest[
     data,
     μ0,
     {"PValue"}
     ];

  (* Display the histogram of the data along with p-value: *)
  Histogram[
    data,
    Automatic,
    "Probability",
    PlotLabel->Row[{"p-value: ",NumberForm[pValue,{4,3}]}],
    ColorFunction->Function[{height},Opacity[height]],
    ImageSize->300,
    ChartStyle->Purple
    ]
  ],
 {{μ0,0,"μ0"},-6,6,0.1,Appearance->"Labeled"}
 ]
```

Output

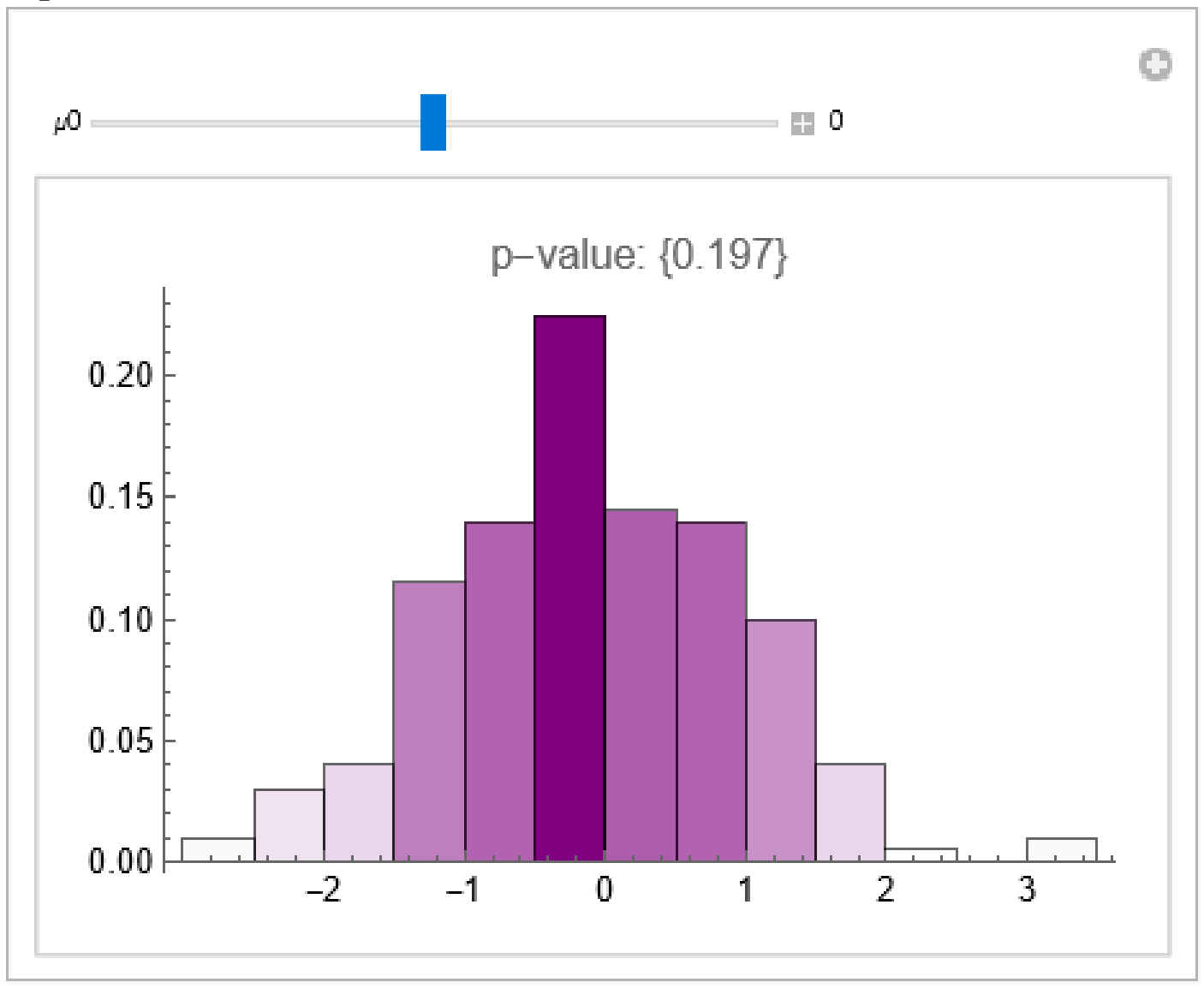

***Mathematica Examples 21.18***

Input

```
(* The code performs a hypothesis test to compare the locations of two datasets. Specifically, it uses the LocationTest function to compare the locations of data1 and data2 with a null hypothesis μ1-μ2=0, and data1 and data3 with a null hypothesis μ1-μ3=0. The LocationTest function calculates a P-value for each test. A P-value is a measure of the evidence against the null hypothesis. A small P-value (typically less than a chosen significance level, e.g., 0.05) suggests strong evidence to reject the null hypothesis in favor of the alternative hypothesis. Conversely, a large P-value indicates weak evidence against the null hypothesis, and it does not provide sufficient reason to reject it. In this case, when the datasets data1 and data2 are tested for equality of means, the P-value is generally small, indicating that their means are significantly different from each other. On the other hand, when data1 and data3 are tested for equality of means, the P-value is generally large, suggesting that their means are not significantly different, and there is no strong evidence to reject the null hypothesis of equal means: *)

(* Test Subscript[H, 0]: μ1-μ2=0 versus Subscript[H, a]: μ1-μ2!=0: *)
data1=RandomVariate[
    NormalDistribution[1,1],
    200
    ];

data2=RandomVariate[
    NormalDistribution[2,1],
    200
    ];

data3=RandomVariate[
    NormalDistribution[1,2],
    200
    ];

(* The P-values are generally small when the locations are not equal: *)
LocationTest[{data1,data2},0]

(* The P-values are generally large when the locations are equal: *)
LocationTest[{data1,data3},0]
```

Output `1.78649*10^-26`

Output `0.0969896`

***Mathematica Examples 21.19***

Input

```
(* The code performs a comparison of the locations (means) of three populations represented by datasets data1, data2,and data3. It then visualizes the data using a BoxWhiskerChart and conducts hypothesis tests using the LocationTest function to compare the locations of the populations. The first two populations (data1 and data2) have similar locations, as indicated by the LocationTest result, which does not provide strong evidence to reject the null hypothesis that their means are equal. The third population (data3) differs in location from the first, as indicated by the LocationTest result, which likely provides a small P-value,suggesting that there is strong evidence to reject the null hypothesis of equal means between data1 and data3. The box-and-whisker plot visually reinforces this information, showing that the boxes representing data1 and data2 are similar in location (median and spread), while the box representing data3 is shifted to a higher mean: *)

(* Test whether the locations of some populations are equivalent: *)
data1=RandomVariate[
    NormalDistribution[1,1],
    200
```

```
    ];

data2=RandomVariate[
    NormalDistribution[1,1],
    200
    ];

data3=RandomVariate[
    NormalDistribution[4,1],
    200
    ];

BoxWhiskerChart[
 {data1,data2,data3},
 "Notched",
 ChartStyle->"SolarColors",
 ImageSize->250
 ]

(* The first two populations have similar locations: *)
LocationTest[
 {data1,data2},
 Automatic
 ]

(* The third population differs in location from the first: *)
LocationTest[
 {data1,data3},
 Automatic
 ]
```

Output

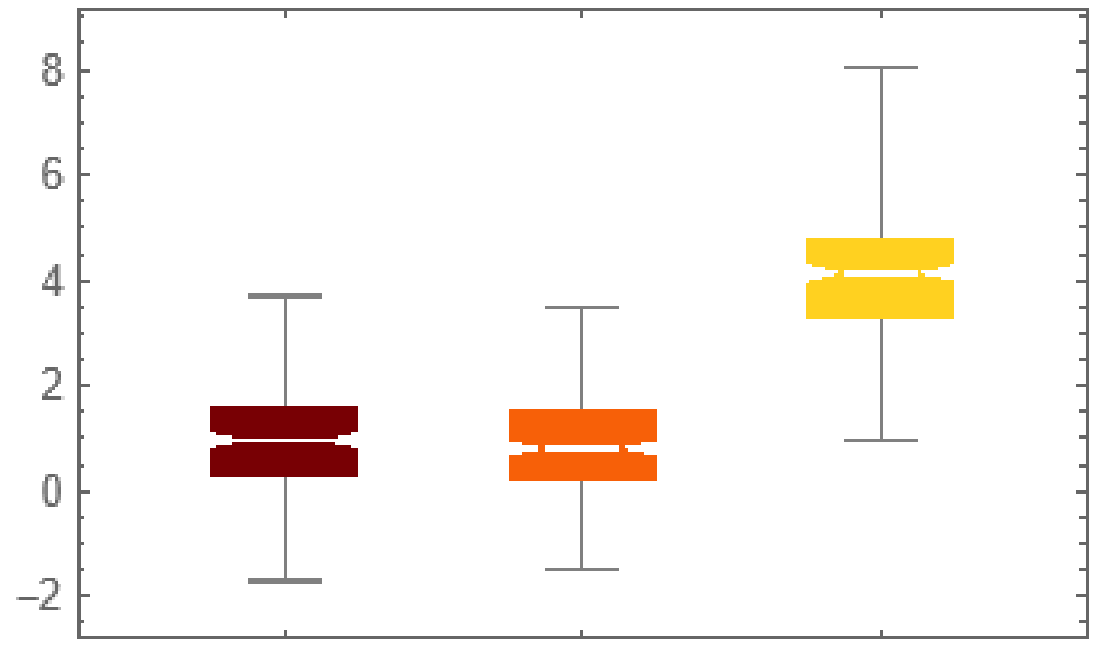

Output 0.277583

Output 8.77911*10^-100

*Mathematica Examples 21.20*

Input

```
(* The code performs a hypothesis test to compare the means of two populations represented by datasets data1 and data2. The code also calculates the mean difference between data1 and data2, plots the two populations' histograms, and then plots the histograms after adding a constant value of 2 to data2. The mean difference between data1 and data2 is calculated to be 2.5. The LocationTest result suggests that at the 0.05 significance level, the mean difference between data1 and data2 is significantly different from 2. The two sets of histograms are plotted to visually compare the distributions of data1 and data2. The second set of histograms, where 2 is added to data2, shows a shift to the right, indicating the effect of the constant difference of 2: *)

(* Test whether the means of two populations differ by 2: *)
data1=RandomVariate[
```

```
    NormalDistribution[5.5,2],
    700
    ];

data2=RandomVariate[
    NormalDistribution[3,2],
    500
    ];

(* The mean difference μ1-μ2: *)
Mean[data1]-Mean[data2]

(* At the 0.05 level, μ1-μ2 is significantly different from 2: *)
LocationTest[{data1,data2},2]

(* Plot the two populations *)
SmoothHistogram[
 {data1,data2},
 ImageSize->250,
 PlotStyle->{Purple,Blue},
 PlotLegends->{"Data 1","Data 2"}
 ]
SmoothHistogram[
 {data1,data2+2},
 ImageSize->250,
 PlotStyle->{Purple,Blue},
 PlotLegends->{"Data 1","Data 2"}
 ]
```

Output 2.55641
Output 2.55428*10^-6
Output

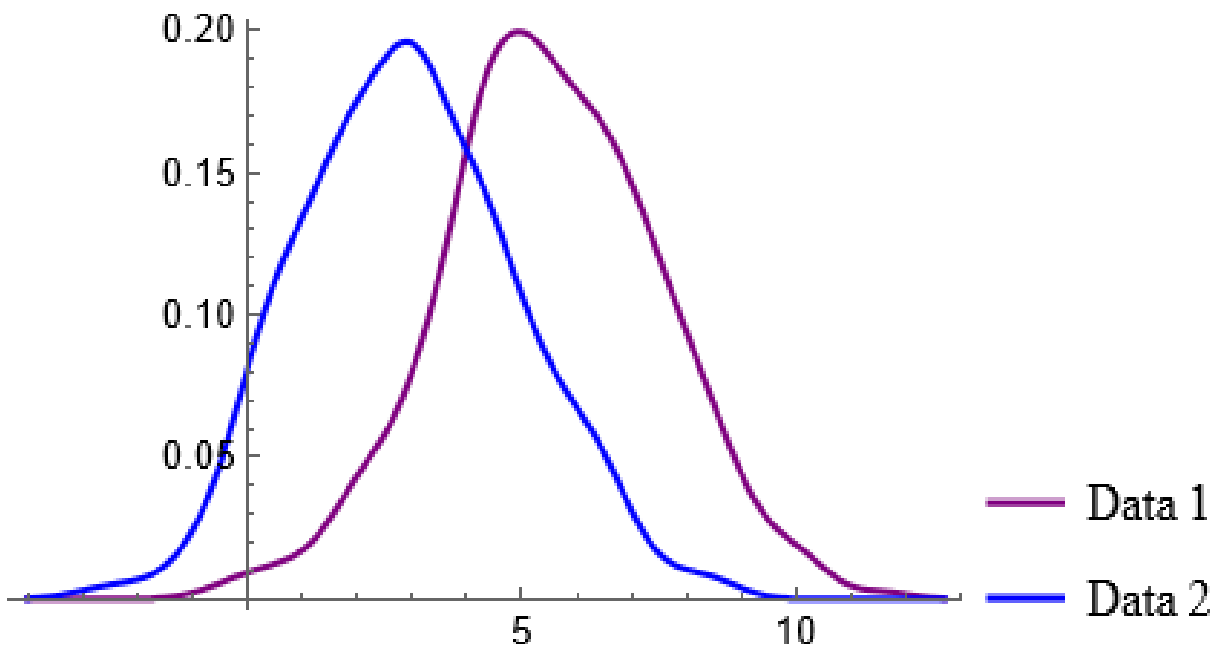

Output

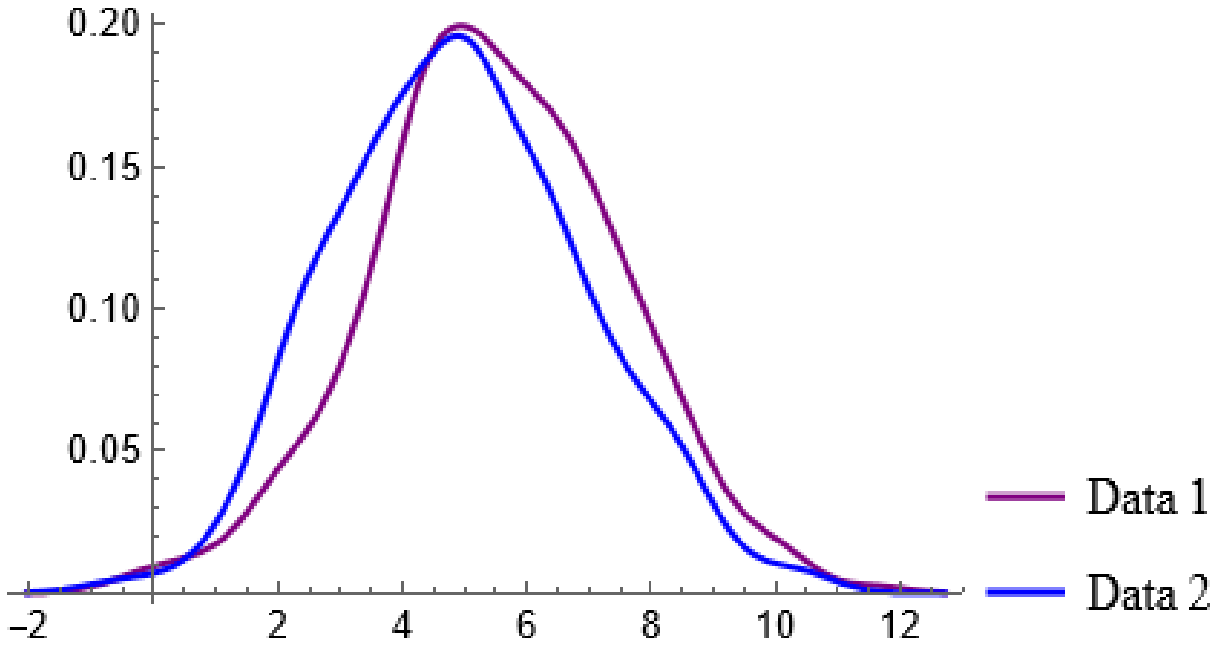

***Mathematica Examples 21.21***

Input (* The code provides a visual representation of the two datasets, the calculated mean difference, and the result of the hypothesis test. The code generates random data from two normal distributions, data1 and data2, with different means and standard deviations. The code plots histograms of the two datasets and visually shows the mean

```
difference along with the test result. The green dashed lines and arrows help to
understand the means of the datasets, and the test result demonstrates whether the
observed mean difference is significant at the chosen significance level: *)

(* Generate random data for demonstration: *)
data1=RandomVariate[
   NormalDistribution[5,1],
   50
   ];

data2=RandomVariate[
   NormalDistribution[7,1.5],
   60
   ];

(* The mean difference µ1-µ2: *)
Mean[data1]-Mean[data2]

(* Perform the test of mean difference: *)
testResult=LocationTest[{data1,data2},-2]

(* Plot the two populations and the test results: *)
SmoothHistogram[
 {data1,data2},
 Automatic,
 "PDF",
 PlotStyle->{Directive[Red,Opacity[0.5]],Directive[Blue,Opacity[0.5]]},
 PlotLegends->{"Data 1","Data 2"},
 ImageSize->250,
 PlotRange->All,
 PlotLabel->"Histogram of Data 1 and Data 2",
 Epilog->{
   Text["µ1-µ2",{Mean[data1]+0.3,0.2}],
   Arrowheads[{-0.035,0.035}],
   Arrow[{{Mean[data1],0.15},{Mean[data2],0.15}}],
   Text[Round[Mean[data1]-Mean[data2],0.01],{Mean[data2]+0.3,0.2}],
   Directive[Green,Dashed,Thickness[0.006]],
   Line[{{Mean[data1],0},{Mean[data1],0.3}}],
   Line[{{Mean[data2],0},{Mean[data2],0.3}}]
   }
 ]
SmoothHistogram[
 {data1,data2-2},
 Automatic,
 "PDF",
 PlotStyle->{Directive[Red,Opacity[0.5]],Directive[Blue,Opacity[0.5]]},
 PlotLegends->{"Data 1","Data 2"},
 ImageSize->250,
 PlotRange->All,
 PlotLabel->"Histogram of Data 1 and shifted Data 2",
 Epilog->{
   Text["PValue",{Mean[data1],0.2}],
   Arrowheads[{-0.035,0.035}],
   Arrow[{{Mean[data1],0.15},{Mean[data2-2],0.15}}],
   Text[Round[testResult,0.01],{Mean[data2],0.2}],
   Directive[Green,Dashed,Thickness[0.006]],
   Line[{{Mean[data1],0},{Mean[data1],0.3}}],
   Line[{{Mean[data2-2],0},{Mean[data2-2],0.3}}]
   }
 ]
```

```
Output   -1.89616
Output   0.658189
Output
```

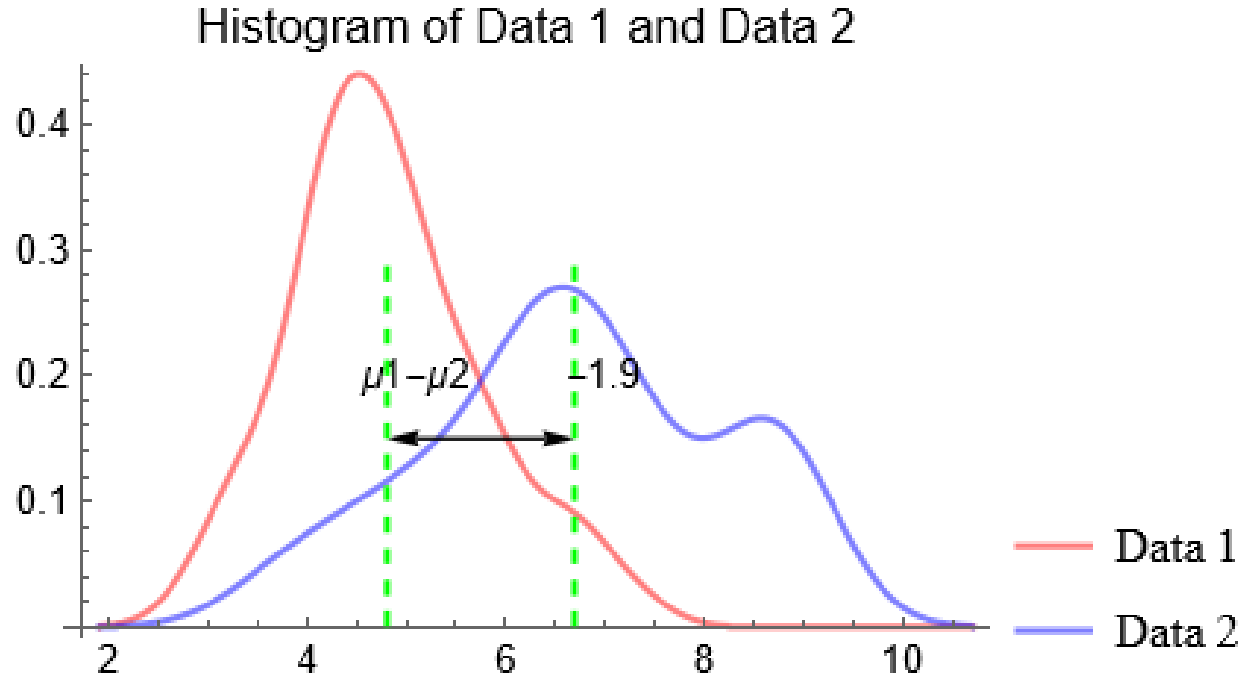

```
Output
```

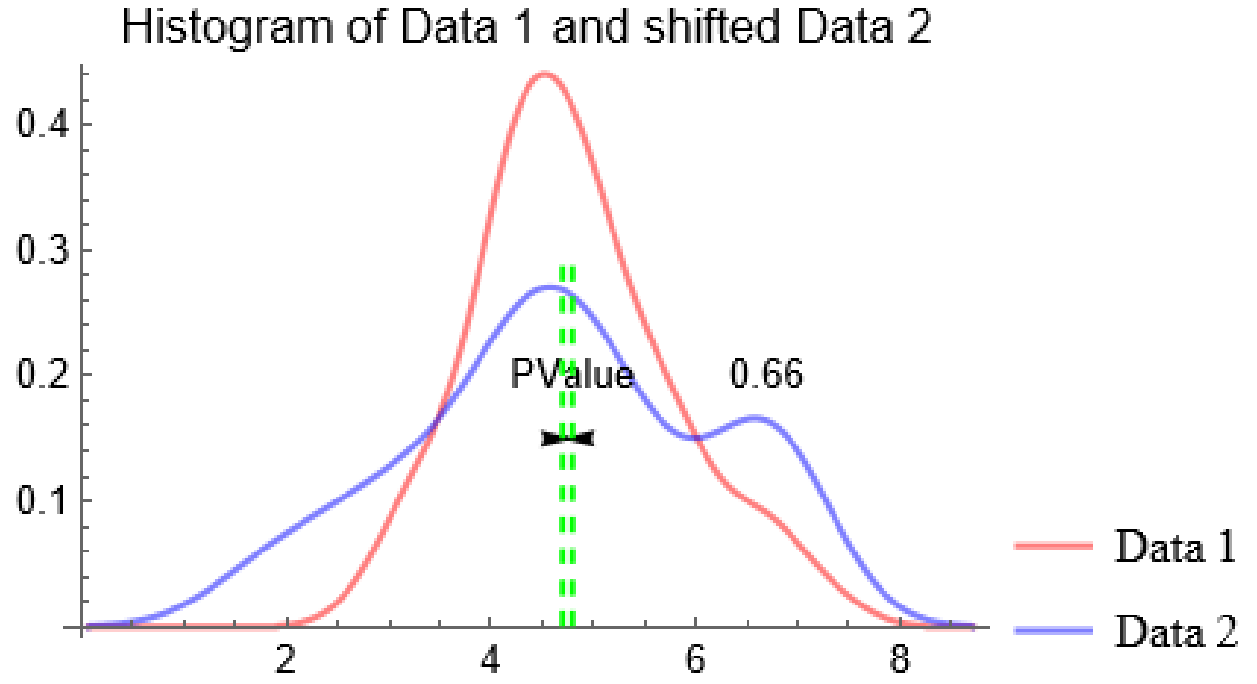

## Mathematica Examples 21.22

```
Input    (* The code demonstrates an interactive Manipulate that allows users to explore the
         comparison of means between two populations. Users can modify parameters such as the
         means, standard deviations, and sample sizes of the two datasets. The code generates
         random data based on the specified parameters, performs a hypothesis test for mean
         difference, and visualizes the datasets and test results using histograms. This
         interactive visualization provides an insightful way for users to understand how
         changes in the parameters affect the distributions of the two populations and how
         the mean difference test outcome varies accordingly: *)

         Manipulate[
          (* Generate data for two populations based on specified parameters: *)
          data1=RandomVariate[
             NormalDistribution[mean1,stdDev1],
             sampleSize1
             ];

          data2=RandomVariate[
             NormalDistribution[mean2,stdDev2],
             sampleSize2
             ];

          (* Perform the test of mean difference: *)
          testResult=LocationTest[{data1,data2},-2];

          (* Plot the two populations and the test results: *)
          Show[
           Histogram[
             data1,
             Automatic,
             "PDF",
             ImageSize->300,
```

```
    ChartStyle->Directive[Blue,Opacity[0.3]],
    ChartLegends->{"Data 1"}
    ],
   Histogram[
    data2,
    Automatic,
    "PDF",
    ImageSize->300,
    ChartStyle->Directive[Red,Opacity[0.3]],
    ChartLegends->{"Data 2"}
    ],
   PlotRange->All,
   PlotLabel->Row[{"Histogram of Data 1 and Data 2\n p-value: ",testResult}],
   Epilog->{
     Arrowheads[{-0.035,0.035}],
     Arrow[{{mean1,0.15},{mean2,0.15}}],
     Directive[Green,Dashed,Thickness[0.006]],
     Line[{{mean1,0},{mean1,0.25}}],
     Line[{{mean2,0},{mean2,0.25}}]
     }
   ],
  (* Controls for user to modify parameters: *)
  {{mean1,5,"Mean of Data 1"},0,10,0.1},
  {{stdDev1,1,"Standard Deviation of Data 1"},0.1,5,0.1},
  {{sampleSize1,50,"Sample Size of Data 1"},10,100,10},
  {{mean2,7,"Mean of Data 2"},0,10,0.1},
  {{stdDev2,1.5,"Standard Deviation of Data 2"},0.1,5,0.1},
  {{sampleSize2,100,"Sample Size of Data 2"},10,500,10}
  ]
```

Output

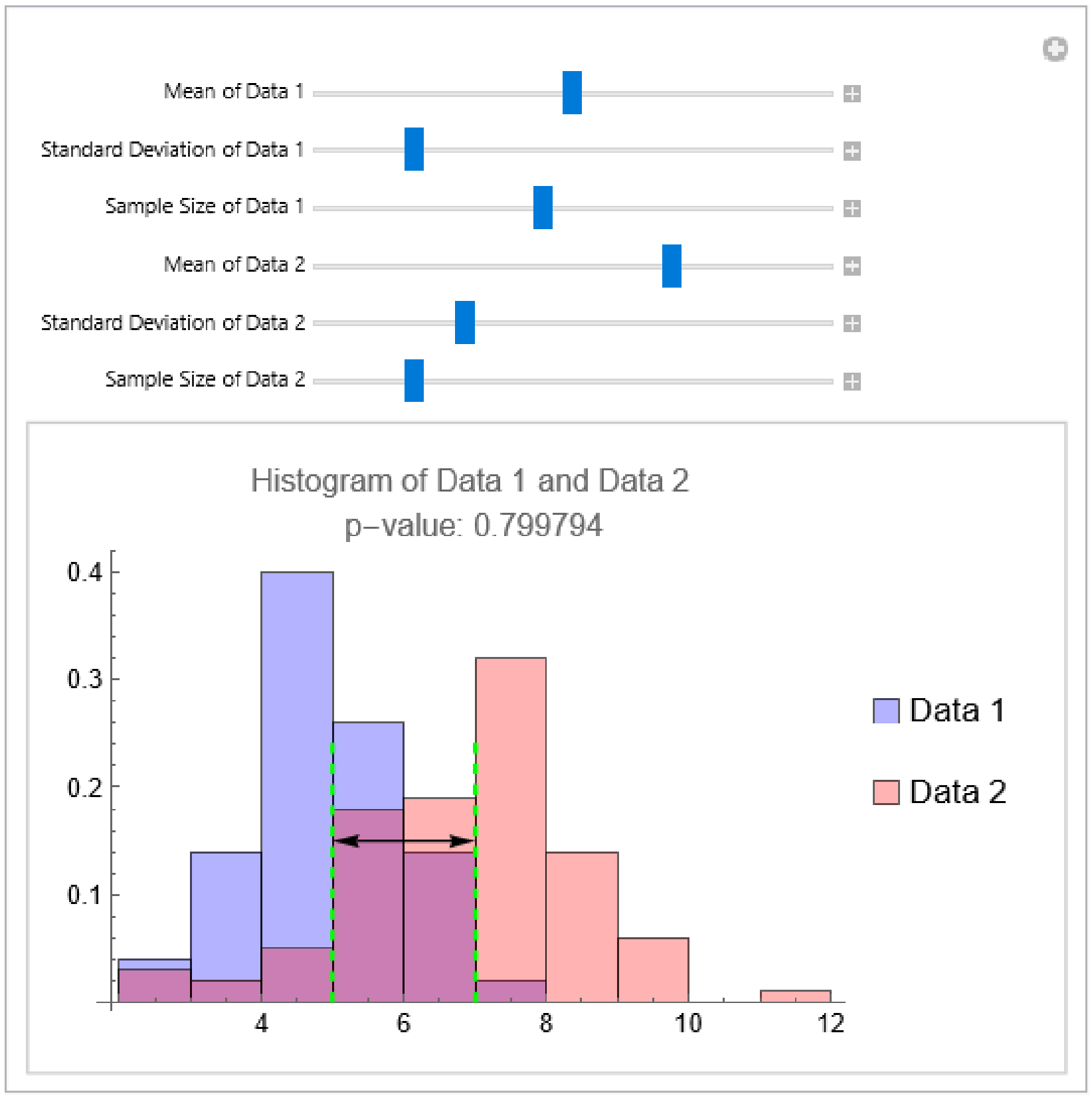

***Mathematica Examples 21.23***

Input

```
(* The code utilizes the LocationEquivalenceTest function to test whether the means
of three populations represented by datasets data1, data2, and data3 are statistically
equivalent. These datasets are randomly generated from normal distributions with
different means but the same standard deviation. By performing the equivalence test,
the code aims to determine if the means of the three populations are similar enough
to be considered equivalent or if there are significant differences between them: *)

data1=RandomVariate[
    NormalDistribution[1,1],
    100
    ];

data2=RandomVariate[
    NormalDistribution[0,1],
    150
    ];

data3=RandomVariate[
    NormalDistribution[0,1],
    135
    ];

LocationEquivalenceTest[
 {data1,data2,data3},
 {"TestDataTable",All}
 ]
```

Output

| | Statistic | P-Value |
|---|---|---|
| Kruskal-Wallis | 41.3764 | 3.49008*10^-10 |
| K-Sample T | 22.7025 | 4.82771*10^-10 |

***Mathematica Examples 21.24***

Input

```
(* The code generates and analyzes four datasets (data) from normal distributions
with different means but the same standard deviation. The datasets are visualized as
overlapping smooth histograms, allowing for a visual comparison of their
distributions. Next, the mean of each dataset is calculated and displayed. This
provides an overview of the central tendencies of the four populations. Finally, the
code performs a test of equivalence for the means of all four datasets using the
LocationEquivalenceTest function: *)

data={
    RandomVariate[
     NormalDistribution[3,1],
     500
     ],
    RandomVariate[
     NormalDistribution[5,1],
     650
     ],
    RandomVariate[
     NormalDistribution[5.2,1],
     400
     ],
    RandomVariate[
     NormalDistribution[5.5,1],
     500
     ]
    };
```

```
SmoothHistogram[
 data,
 ImageSize->250
 ]

Map[Mean,data]

LocationEquivalenceTest[data,{"TestDataTable",All}]
```

Output

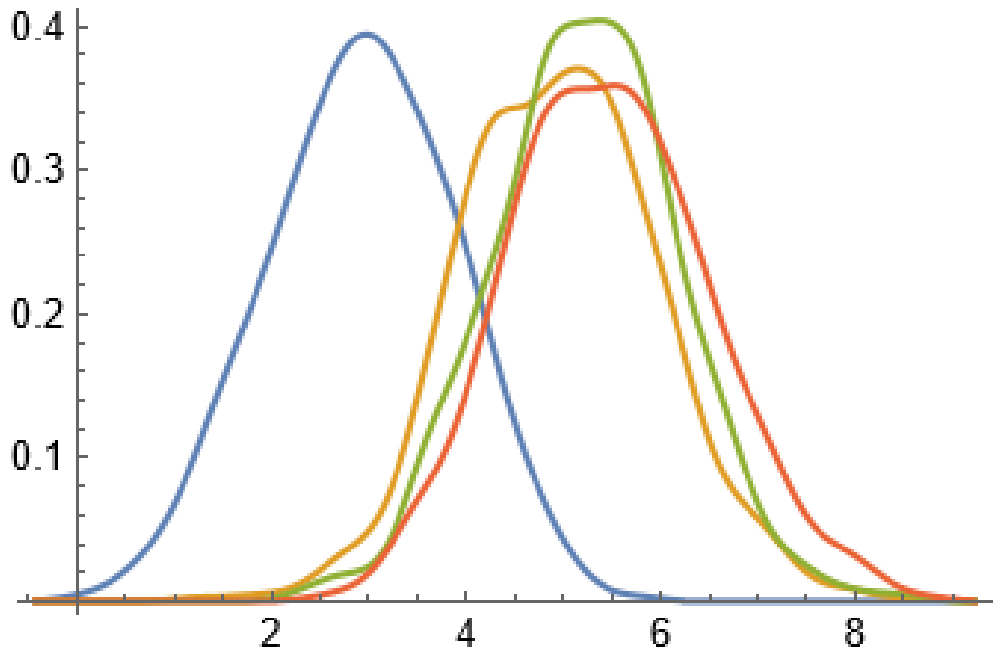

Output {2.92949,4.95899,5.22184,5.46358}

Output

| | Statistic | P-Value |
|---|---|---|
| Kruskal-Wallis | 955.471 | 2.55834*10^-278 |
| K-Sample T | 674.123 | 1.06067*10^-304 |

***Mathematica Examples 21.25***

Input

```
(* The code is a Manipulate function that allows users to explore the comparison of means among three populations. Users can interactively modify parameters to generate random data for these populations with different means. The code then performs the LocationEquivalenceTest to determine whether the means are statistically equivalent or not based on a specified significance level. By manipulating the sliders for the mean parameters and the significance level, users can dynamically observe the changes in the test results. This interactive visualization allows users to explore the impact of different mean values and significance levels on the comparison of means among the three populations: *)

(* Create the Manipulate function: *)
Manipulate[
 (*Generate data for multiple populations with different parameters: *)
 data1=RandomVariate[
   NormalDistribution[μ1,1],
   100
   ];
 data2=RandomVariate[
   NormalDistribution[μ2,1],
   150
   ];
 data3=RandomVariate[
   NormalDistribution[μ3,1],
   200
   ];

 (* Perform the LocationEquivalenceTest for the selected populations: *)
 testResult=LocationEquivalenceTest[
   {data1,data2,data3},
   {"PValue","ShortTestConclusion"},
   SignificanceLevel->alpha
   ];
 (* Display the test result and other information: *)
```

```
Column[
 {
  Row[{"P-Value: ",testResult[[1]]}],
  Row[{"LocationEquivalenceTest Conclusion: ",testResult[[2]]}]
  }
 ],
(* Controls for parameters*)
{{μ1,0,"Mean 1:"},-5,5,0.1},
{{μ2,0,"Mean 2:"},-5,5,0.1},
{{μ3,0,"Mean 3:"},-5,5,0.1},
(* Control for the significance level*)
{{alpha,0.05,"SignificanceLevel:"},0.01,0.1,0.01}
]
```

Output

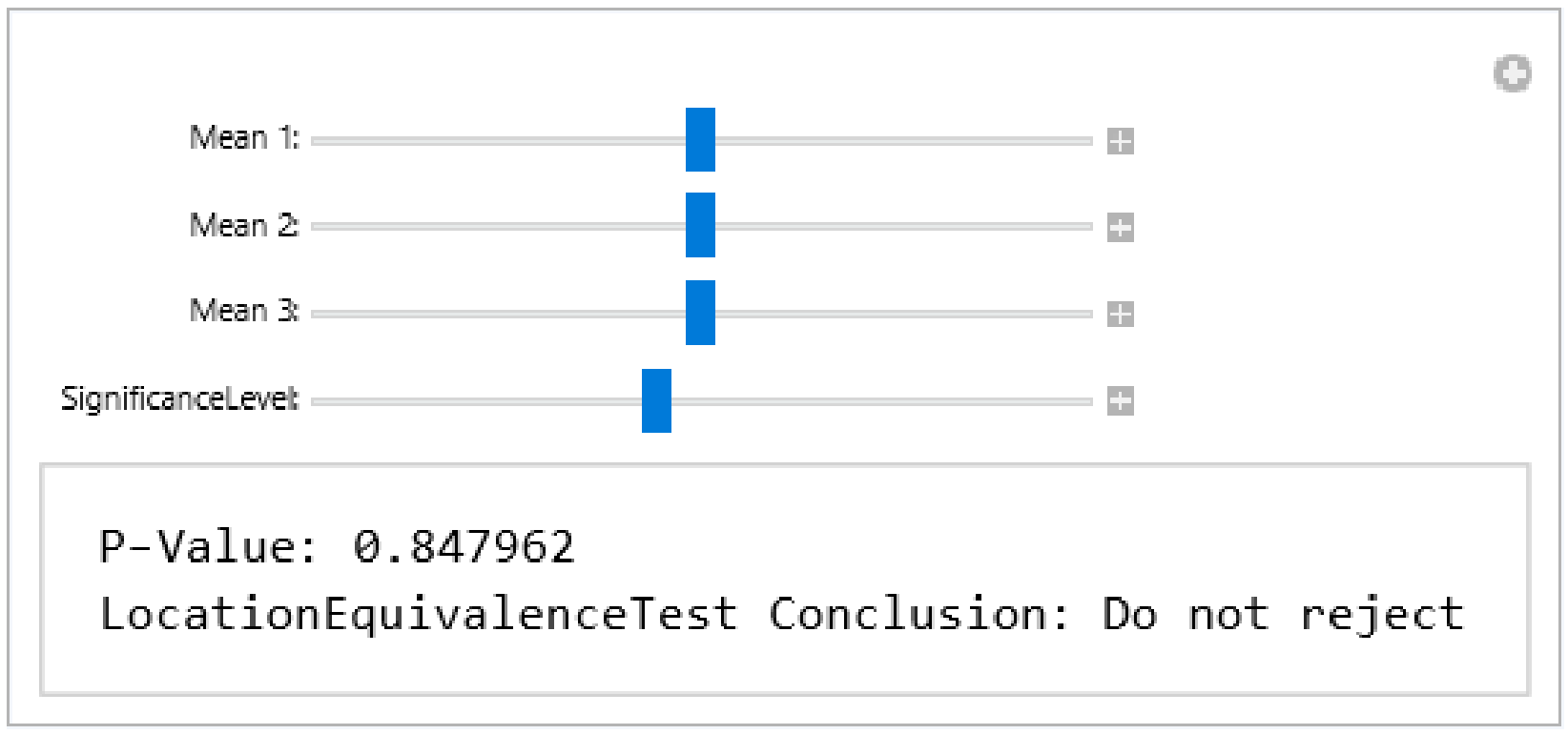

# UNIT 21.2

# VARIANCE TESTS

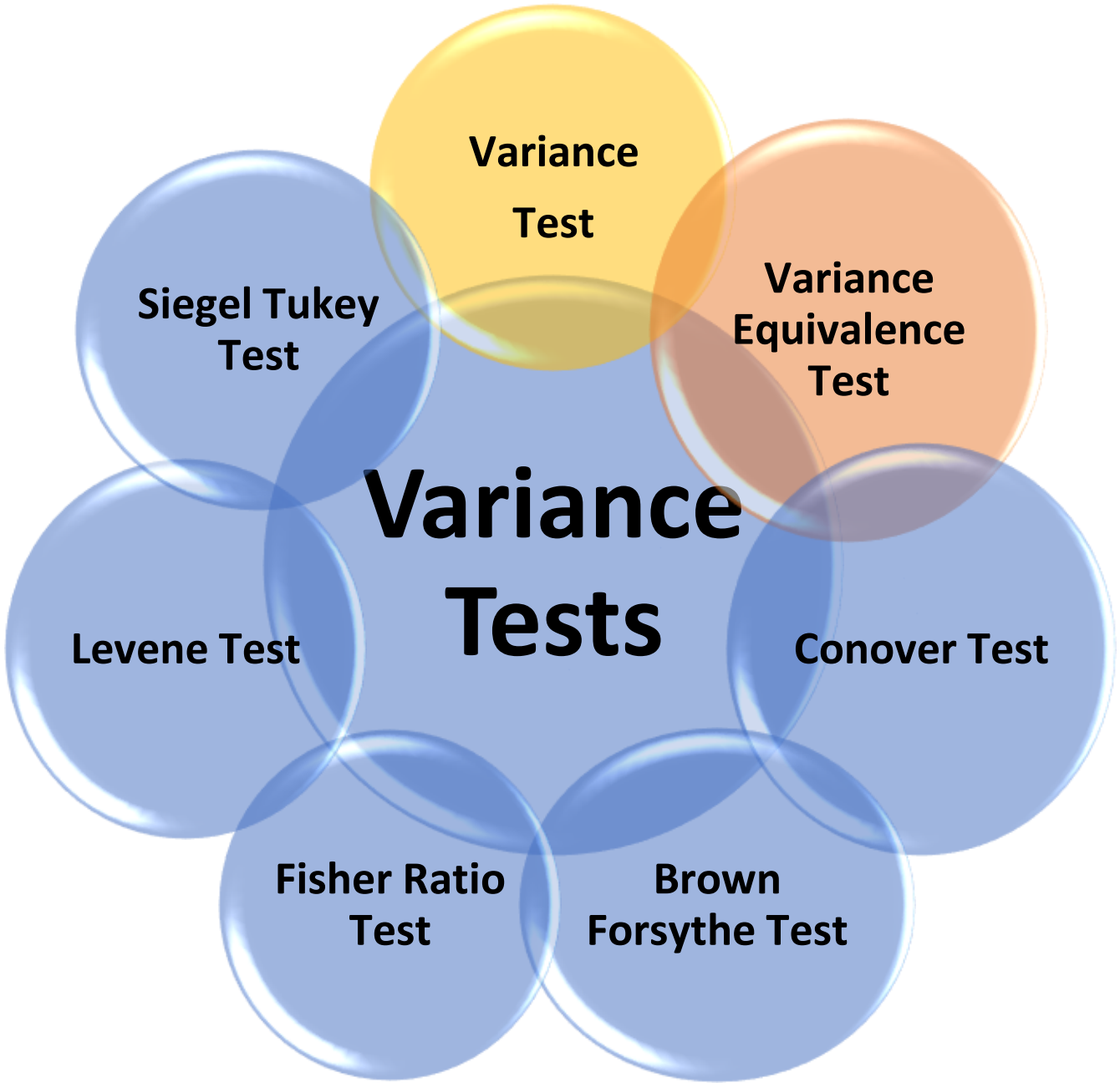

## VarianceTest

| | |
|---|---|
| `VarianceTest[data]` | tests whether the variance of the data is one. |
| `VarianceTest[{data1,data2}]` | tests whether the variances of data1 and data2 are equal. |
| `VarianceTest[dspec,σ02]` | tests a dispersion measure against σ02. |
| `VarianceTest[dspec,σ02,"property"]` | returns the value of "property". |

The following tests can be used:

| | | |
|---|---|---|
| `"BrownForsythe"` | Robust | robust Levene test |
| `"Conover"` | Symmetry | based on squared ranks of data |
| `"FisherRatio"` | Normality | based on $\sigma_1^2/\sigma_2^2$ |
| `"Levene"` | Robust, Symmetry | compare individual and group variances |
| `"SiegelTukey"` | Symmetry | based on ranks of pooled data |

## VarianceEquivalenceTest

| | |
|---|---|
| `VarianceEquivalenceTest[{data1,data2,…}]` | tests whether the variances of the datai are equal. |
| `VarianceEquivalenceTest[{data1,…},"property"]` | returns the value of "property". |

The following tests can be used:

| | | |
|---|---|---|
| `"Bartlett"` | normality | modified likelihood ratio test |
| `"BrownForsythe"` | robust | robust Levene test |
| `"Conover"` | symmetry | Conover's squared ranks test |
| `"FisherRatio"` | normality | based on $\sigma_1^2/\sigma_2^2$ |
| `"Levene"` | robust,symmetry | compares individual and group variances |

***Mathematica Examples 21.26***

Input
```
(* The code performs variance tests on two datasets, data1 and data2, to compare their variances with the null hypothesis variance of one. The results indicate that the variance of data1 is consistent with one, while the variance of data2 is significantly different from one. For data1: The P-value is typically large, indicating that the observed variance is consistent with the null hypothesis variance of one. There is no significant evidence to reject the null hypothesis for data1. For data2: The P-value is typically small, suggesting that the observed variance is significantly different from the null hypothesis variance of one. There is evidence to reject the null hypothesis for data2, indicating that its variance differs significantly from one: *)

(* Test whether the variance of a population is one: *)
data1=RandomVariate[
    NormalDistribution[4,1],
    500
    ];

data2=RandomVariate[
    NormalDistribution[4,2],
    500
    ];

(* The P-values are typically large under H0: *)
VarianceTest[data1,Automatic]

(* The P-values are typically small when H0 is false: *)
VarianceTest[data2,Automatic]
```

Output `0.614953`
Output `2.07289*10^-186`

***Mathematica Examples 21.27***

Input
```
(* The code compares the variances of two populations, data1 and data2, to a specified value of 9. Each dataset contains 500 random samples generated from normal distributions with different variances (4 for data1 and 9 for data2). The code uses the VarianceTest function to perform hypothesis tests and obtain p-values indicating whether the population variances significantly differ from 9: *)

data1=RandomVariate[
    NormalDistribution[3,2],
    500
    ];

data2=RandomVariate[
    NormalDistribution[3,3],
    500
    ];
(* The P-values are typically small when H0 is false: *)
VarianceTest[data1,3^2]
(* The P-values are typically large under H0: *)
VarianceTest[data2,3^2]
```

Output `3.69672*10^-33`
Output `0.570897`

***Mathematica Examples 21.28***

```
Input	(* The code demonstrates the use of the VarianceTest function to compare the variance
	of a dataset (data) with a specified value of 4. The code includes three examples,
	all of which use the default test selection (Automatic) for comparing variances. The
	dataset contains 500 random samples from a normal distribution with a mean of 0 and
	a variance of 4. The VarianceTest function calculates a p-value indicating the
	probability of obtaining the observed variance (or more extreme) assuming a population
	variance of 4. Using Automatic or AutomaticTest as options allows the function to
	automatically select the most suitable and powerful test for the comparison, enhancing
	flexibility and ease of use: *)

	data=RandomVariate[
	    NormalDistribution[0,2],
	    500
	    ];

	VarianceTest[data,4]
	VarianceTest[data,4,Automatic]
	VarianceTest[data,4,"AutomaticTest"]

Output	0.561231
Output	0.561231
Output	FisherRatio
```

***Mathematica Examples 21.29***

```
Input	(* The code performs three different variance tests on a sample dataset data, which
	is generated from a standard normal distribution. Each test compares the sample
	variance to a specific value or range of values, evaluating whether there is enough
	evidence to support the alternative hypothesis over the null hypothesis: *)

	data=RandomVariate[
	    NormalDistribution[0,1],
	    100];

	(* Two-Sided Test (H0:σ^2=1 versus Ha:σ^2!=1):This test checks if the population
	variance is significantly different from 1. It aims to detect any departure from the
	null hypothesis and determine whether the variance is either larger or smaller than
	1: *)
	VarianceTest[data,Automatic,AlternativeHypothesis->"Unequal"]
	VarianceTest[data,Automatic,AlternativeHypothesis->Automatic]

	(* One-Sided Test for Smaller Variance (H0: σ^2>=1 versus Ha:σ^2<1): *)
	VarianceTest[data,Automatic,AlternativeHypothesis->"Less"]

	(* One-Sided Test for Larger Variance (H0: σ^2<=1 versus Ha:σ^2>1): *)
	VarianceTest[data,Automatic,AlternativeHypothesis->"Greater"]

Output	0.0437699
Output	0.0437699
Output	0.978115
Output	0.0218849
```

***Mathematica Examples 21.30***

```
Input	(* The code performs a variance test to compare the variances of three populations
	generated from normal distributions. The populations are represented by data1, data2,
	and data3. The code does the following: Generates three datasets (data1, data2, and
	data3) with 200 random samples each from normal distributions with specific mean and
	standard deviation values. Creates a BoxWhiskerChart to visualize the distributions
	of the three datasets, with notches around the medians for better comparison. Uses
	the VarianceTest function to compare the variances between the populations. It tests
```

```
whether the variances of data1 and data2 are equivalent and whether the variance of
data1 is different from that of data3: *)

data1=RandomVariate[
   NormalDistribution[0,1],
   200
   ];
data2=RandomVariate[
   NormalDistribution[0,1],
   200
   ];
data3=RandomVariate[
   NormalDistribution[0,3],
   200
   ];

BoxWhiskerChart[
 {data1,data2,data3},
 "Notched",
 ChartStyle->"SolarColors",
 ImageSize->250
 ]

(* The first two populations have similar variances: *)
VarianceTest[
 {data1,data2},
 Automatic
 ]

(* The third population differs in variance from the first: *)
VarianceTest[
 {data1,data3},
 Automatic
 ]
```

Output

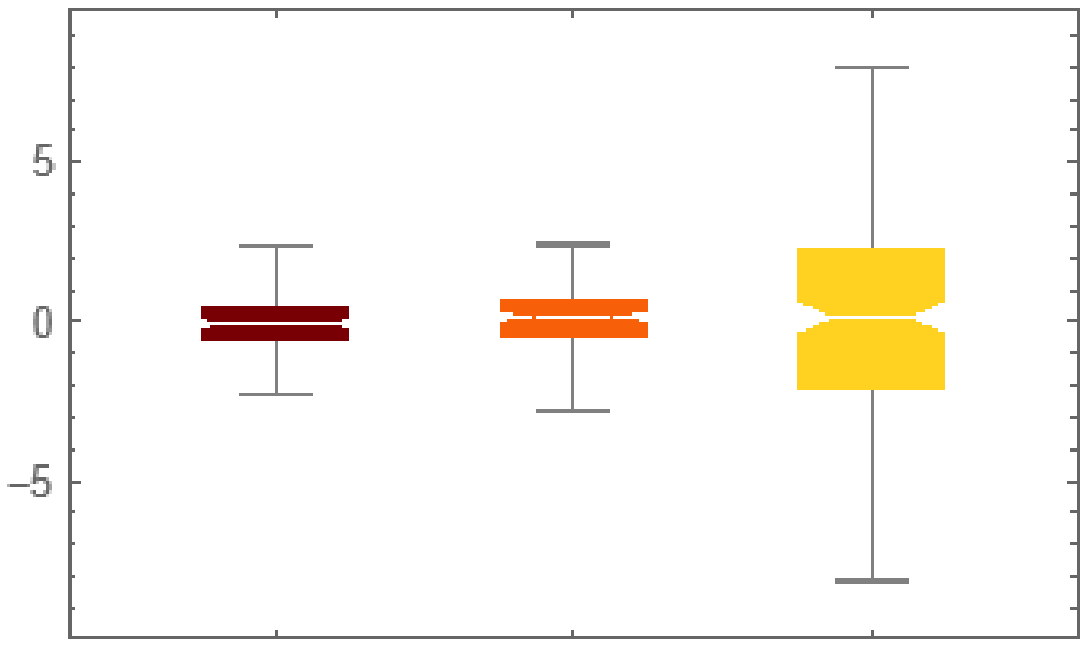

Output 0.423764

Output 3.04957*10^-58

***Mathematica Examples 21.31***

Input

```
(* The code performs a variance test to compare the ratio of variances between two
populations (data1 and data2) against a specified value (σ0). The two datasets are
generated with different standard deviations: data1 with a standard deviation of 2,
and data2 with a standard deviation of 1. The variance test is performed using the
VarianceTest function with two equivalent forms: VarianceTest[{data2, data1},1/σ0]:
Tests whether the ratio of the variance of data2 to the variance of data1 is equal
to 1/σ0, i.e.,1/4 or 0.25. VarianceTest[{data1, data2},σ0]: Tests whether the ratio
of the variance of data1 to the variance of data2 is equal to σ0,i.e.,4. The output
of both tests is a p-value, representing the probability of obtaining the observed
```

```
        results (or more extreme results) under the assumption that the null hypothesis is
        true: *)

        SeedRandom[1];
        data1=RandomVariate[
            NormalDistribution[0,2],
            500
            ];
        data2=RandomVariate[
            NormalDistribution[0,1],
            500
            ];

        σ0=4;
        (* The following forms are equivalent: *)
        VarianceTest[{data2,data1},1/σ0]
        VarianceTest[{data1,data2},σ0]

        (* The order of the datasets should be considered when determining σ0: *)
        VarianceTest[{data2,data1},σ0]

Output  0.703526
Output  0.703526
Output  9.69572*10^-169
```

***Mathematica Examples 21.32***

```
Input   (* The code performs a variance test to compare the variances of two populations
        represented by the data matrix. Each row of the matrix contains data points from a
        separate population, and the test is done to check if the variances of these
        populations are equal. The code first generates two datasets, each with 100 data
        points, sampled from a standard normal distribution (mean=0, standard deviation=1).
        These datasets are stored in a 2x100 matrix called data. Next, the variance test is
        conducted using the VarianceTest function on the data matrix. By default, the function
        tests the null hypothesis that the variances of the two populations are equal: *)

        (* Test variances from two populations for equality: *)
        data=RandomVariate[
            NormalDistribution[],
            {2,100}
            ];

        VarianceTest[data]

        (* A HypothesisTestData object h is created to store detailed information about the
        variance test. The TestDataTable property of the HypothesisTestData object h is then
        extracted, which contains valuable information about the test, including the test
        statistic, p-value, and the method used: *)
        h=VarianceTest[
            data,
            Automatic,
            "HypothesisTestData"
            ];

        h["TestDataTable",All]

Output  0.236337
Output
```

| | Statistic | P-Value |
|---|---|---|
| Brown-Forsythe | 0.386022 | 0.535112 |
| Conover | -0.272482 | 0.785251 |
| Fisher Ratio | 0.787481 | 0.236337 |
| Levene | 0.382049 | 0.537219 |
| Siegel-Tukey | -0.207688 | 0.835473 |

***Mathematica Examples 21.33***

Input
```
(* The code generates a sample data set from a standard normal distribution, calculates the variance of the data set, performs a one-sample variance test against the expected variance of 1, and then prints the test statistic, p-value, and test conclusion. Additionally, it generates a test data table for the variance test: *)

(* Generate a sample data set: *)
data=RandomVariate[
    NormalDistribution[0,1],
    500
    ];

(* Generate the variance of the data set: *)
Variance[data]

(* Perform one-sample test: *)
result=VarianceTest[
   data,
   1,
   {"TestStatistic","PValue","TestConclusion"}
   ]

(* Print the test statistic and p-value: *)
Print["Test Statistic: ",result[[1]]];
Print["P-value: ",result[[2]]];
Print["Test Conclusion: ",result[[3]]];

(* Test data table: *)
VarianceTest[
 data,
 1,
 {"TestDataTable",All}
 ]
```

Output 0.944181

Output {471.146, 0.380243, The null hypothesis that the variance of the population is equal to 1 is not rejected at the 5 percent level based on the Fisher Ratio test.}

Output Test Statistic: 471.146

Output P-value: 0.380243

Output Test Conclusion: The null hypothesis that the variance of the population is equal to 1 is not rejected at the 5 percent level based on the Fisher Ratio test.

Output

| | Statistic | P-Value |
|---|---|---|
| Brown-Forsythe | 471.146 | 0.380243 |
| Fisher Ratio | 471.146 | 0.380243 |
| Levene | 471.146 | 0.380243 |

***Mathematica Examples 21.34***

Input
```
(* The code generates a sample data set from a normal distribution, calculates its variance, performs a one-sample variance test, and then prints the test statistic, p-value, and test conclusion. Finally, it displays the test data table: *)

(* Generate a sample data set: *)
data=RandomVariate[
    NormalDistribution[0,2],
    100
    ];
```

```
(* Generate the variance of the data set: *)
Variance[data]

(* Perform one-sample test: *)
result=VarianceTest[
  data,
  1,
  {"TestStatistic","PValue","TestConclusion"}
  ]

(* Print the test statistic and p-value: *)
Print["Test Statistic: ",result[[1]]];
Print["P-value: ",result[[2]]];
Print["Test Conclusion: ",result[[3]]];

(* Test data table: *)
VarianceTest[
 data,
 1,
 {"TestDataTable",All}
 ]
```

Output 4.07316

Output {403.243, 4.88289*10^-38, The null hypothesis that the variance of the population is equal to 1 is rejected at the 5 percent level based on the Fisher Ratio test.}

Output Test Statistic: 403.243

Output P-value: 4.88289*10^-38

Output Test Conclusion: The null hypothesis that the variance of the population is equal to 1 is rejected at the 5 percent level based on the Fisher Ratio test.

Output

| | Statistic | P-Value |
|---|---|---|
| Brown-Forsythe | 403.243 | 4.88289*10^-38 |
| Fisher Ratio | 403.243 | 4.88289*10^-38 |
| Levene | 403.243 | 4.88289*10^-38 |

***Mathematica Examples 21.35***

Input

```
(* The code performs a one-sample variance test on a dataset using different null hypotheses. The data consists of a list of numerical values. The code calculates the variance test results for various null hypotheses with values ranging from 0.1 to 1 in increments of 0.1. The test results include test statistics and p-values, which are stored in separate lists. The code then iterates through the p-values and interprets the results by comparing them to a significance level of 0.05. If the p-value is less than 0.05, it concludes that there is strong evidence against the null hypothesis and prints "Reject the null hypothesis." Otherwise, it prints "Fail to reject the null hypothesis.": *)

(* Step 1, define your data: *)
data={2.3,1.8,2.6,2.2,1.9,2.4,2.1,2.7,2.5,2.0};
Variance[data]

(* Step 2, perform the one-sample variance test: *)
testResult=Table [
  VarianceTest[
   data,
   i^2,
   {"TestStatistic","PValue"}
   ],
  {i,0.1,1,0.1}
  ]
```

```
        (* Step 3, interpret the results: *)
        testStatistic=Table[testResult[[i]][[1]],{i,1,10}]
        pValue=Table[testResult[[i]][[2]],{i,1,10}]

        Table[
         If[
          pValue[[i]]<0.05,
          Print["Reject the null hypothesis."],
          Print["Fail to reject the null hypothesis."]
          ],
         {i,1,10}
         ]

Output  0.0916667
Output  {{82.5, 1.02854*10^-13}, {20.625,0.0288473}, {9.16667,0.844068},
        {5.15625,0.359039}, {3.3,0.0975906}, {2.29167,0.0280564}, {1.68367,0.00891787},
        {1.28906,0.00313971}, {1.01852,0.00121274}, {0.825,0.00050801}}

Output  {82.5,20.625,9.16667,5.15625,3.3,2.29167,1.68367,1.28906,1.01852,0.825}

Output  {1.02854*10-^13, 0.0288473, 0.844068, 0.359039, 0.0975906, 0.0280564, 0.00891787,
        0.00313971, 0.00121274, 0.00050801}

Output   Reject the null hypothesis.
         Reject the null hypothesis.
         Fail to reject the null hypothesis.
         Fail to reject the null hypothesis.
         Fail to reject the null hypothesis.
         Reject the null hypothesis.
         Reject the null hypothesis.
         Reject the null hypothesis.
         Reject the null hypothesis.
         Reject the null hypothesis.
```

***Mathematica Examples 21.36***

```
Input   (* The code implements an interactive Manipulate function that allows users to explore
        the results of a one-sample variance test on a random sample generated from a normal
        distribution with user-defined mean and standard deviation. The interactive interface
        enables users to adjust the parameters (μ and σ) of the normal distribution and
        observe how the variance test results change in real-time. The code generates a
        random data sample from the specified normal distribution and then performs the
        variance test against a user-defined null hypothesis variance (5^2 or 25). The output
        of the test, including the test statistic and p-value, is displayed in a column.
        Additionally, the code uses an If statement to interpret the results by comparing
        the p-value to a significance level of 0.05. It informs the user whether to "Reject
        the null hypothesis " if the p-value is less than 0. 05, or "Fail to reject the null
        hypothesis" if the p-value is greater than or equal to 0.05: *)

        (* Define the Manipulate function: *)
        SeedRandom[123];
        Manipulate[
         Module[
          {testResult},

          (* Generate a random sample from a normal distribution with known parameters: *)
          data=RandomVariate[
            NormalDistribution[μ,σ],
            100
            ];
```

```
          (* Perform the VarianceTest: *)
          testResult=VarianceTest[
            data,
            5^2,
            {"TestStatistic","PValue"}
            ];

          (* Display the results: *)
          Column[
           {
            Row[{"Test Statistic: ",testResult[[1]]}],
            Row[{"P-value: ",testResult[[2]]}],
            If[
             testResult[[2]]<0.05,
             Row[{"Reject the null hypothesis at 5% significance level"}],
             Row[{"Fail to reject the null hypothesis at 5% significance level"}]
             ]
            }
           ]
          ],

         (* Controls for the parameters: *)
         {{μ,5,"Mean"},0,10,0.1},
         {{σ,3,"Standard Deviation"},2,7,0.1}
         ]
```

Output

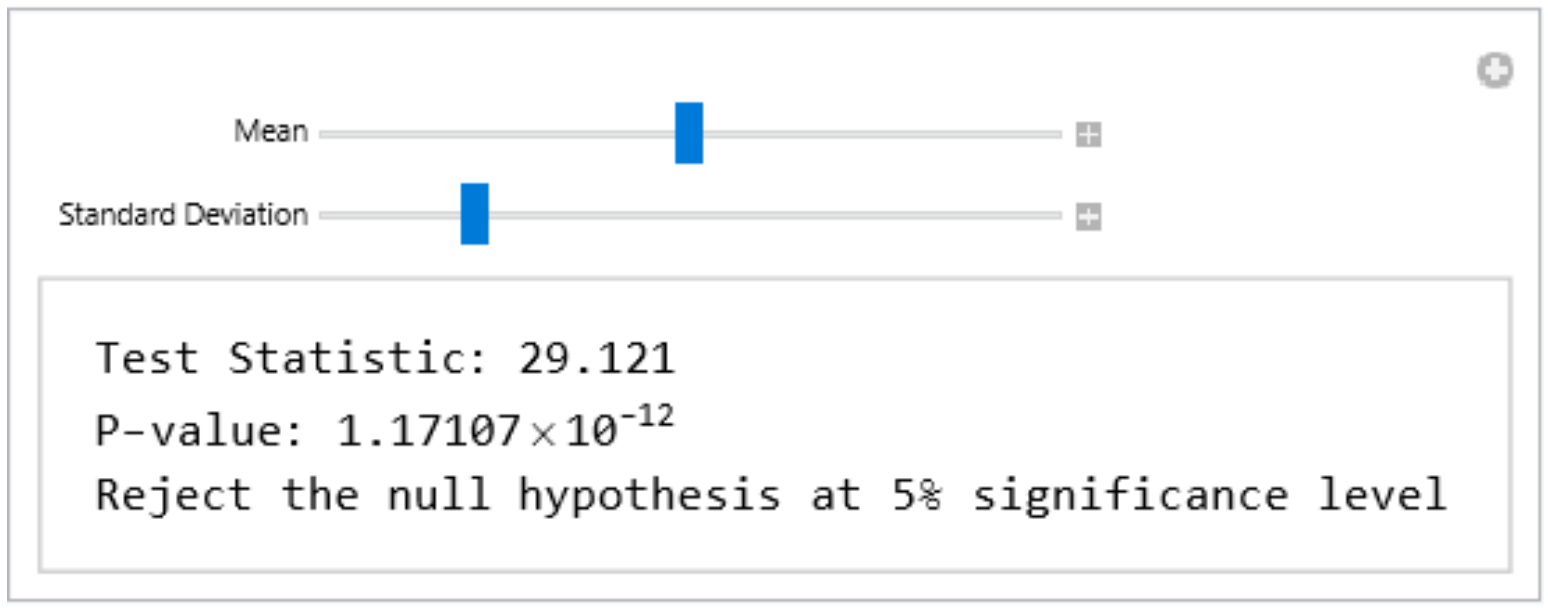

***Mathematica Examples 21.37***

Input

```
(* The code creates an interactive Manipulate function for conducting a one-sample
variance test on a random sample generated from a normal distribution with user-
defined parameters. Users can dynamically adjust various parameters, such as the
mean, standard deviation, sample size, null hypothesis variance, and significance
level. The interactive interface allows users to explore how changes in these
parameters influence the variance test results and the conclusion regarding the null
hypothesis. The code generates a random sample based on the specified normal
distribution parameters and performs a variance test against the user-defined null
hypothesis variance. The output displays the test statistic, p-value, and a conclusion
indicating whether the null hypothesis is rejected or not at the chosen significance
level. The "Conclusion" statement is based on the comparison between the p-value and
the specified significance level (alpha): *)

(* Create a Manipulate function: *)
Manipulate[
 (* Generate a random sample from a normal distribution with specified parameters:
*)
 data=RandomVariate[
   NormalDistribution[μ,σ],
   n
```

```
    ];

  (* Perform the VarianceTest with the chosen parameters: *)
  testResult=VarianceTest[
    data,
    σ0^2,
    {"TestStatistic","PValue","ShortTestConclusion"},
    SignificanceLevel->alpha
    ];

  (* Display the results: *)
  Column[
   {
    Row[{"Test Statistic: ",testResult[[1]]}],
    Row[{"PValue: ",testResult[[2]]}],
    If[
     testResult[[3]]==="Do not reject",
     Row[{"Conclusion: The null hypothesis is not rejected at ",alpha*100,"%
significance level."}],
     Row[{"Conclusion: The null hypothesis is rejected at "  ,alpha*100,"%
significance level."}]
     ]
    }
   ],

  (* Manipulate Parameters: *)
  {{μ,0,"Mean of Distribution"},-5,5,0.1},
  {{σ,1,"Standard Deviation"},0.1,5,0.1},
  {{n,20,"Sample Size"},10,100,10},
  {{σ0,1,"Null Hypothesis Mean"},-5,5,0.1},
  {{alpha,0.05,"Significance Level"},0.01,0.10,0.01}
  ]
```

Output

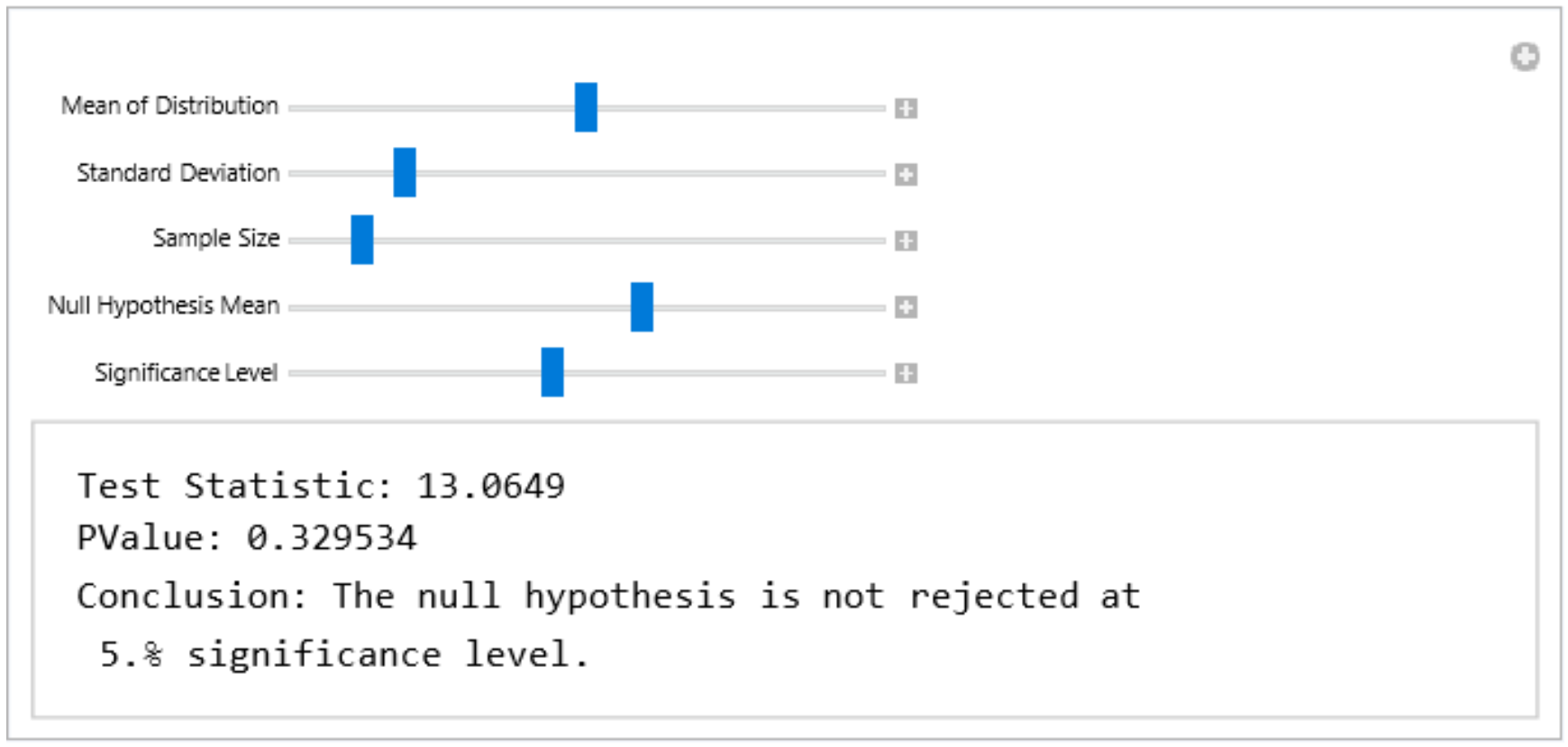

***Mathematica Examples 21.38***

Input

```
(* The code creates an interactive Manipulate function that allows users to explore a one-sample variance test on a randomly generated sample from a normal distribution. Users can interactively adjust the mean, standard deviation, sample size, and significance level to observe how these parameters influence the results of the variance test. The Manipulate function generates random sample data based on user-defined mean and standard deviation parameters. It then performs a one-sample variance test against a null hypothesis variance of 1, using the chosen significance level. The output is displayed as a histogram of the generated data, with probabilities on the y-axis. The plot's title includes the p-value rounded to three decimal places, indicating the statistical significance of the variance test's result: *)
```

```
Manipulate[
 Module[
  {data,pValue},
  (* Generate sample data from a normal distribution: *)
  data=RandomVariate[
    NormalDistribution[mean,stdDev],
    sampleSize
    ];

  (* Perform a one-sample test against the null hypothesis: *)
  pValue=VarianceTest[
    data,
    1,
    {"PValue"},
    SignificanceLevel->significanceLevel
    ];

  (* Display the histogram of the data along with the test p-value: *)
  Histogram[
   data,
   Automatic,
   "Probability",
   PlotLabel->Row[{"p-value: ",NumberForm[pValue,{4,3}]}],
   ColorFunction->Function[{height},Opacity[height]],
   ImageSize->300,
   ChartStyle->Purple
   ]
  ],
 {{mean,0,"Mean"},-6,6,0.1,Appearance->"Labeled"},
 {{stdDev,1,"Standard Deviation"},0.1,5,0.1,Appearance->"Labeled"},
 {{sampleSize,200,"Sample Size"},10,1000,10,Appearance->"Labeled"},
 {{significanceLevel,0.05,"Significance Level"},0.01,0.5,0.01,Appearance-
>"Labeled"}
 ]
```

Output

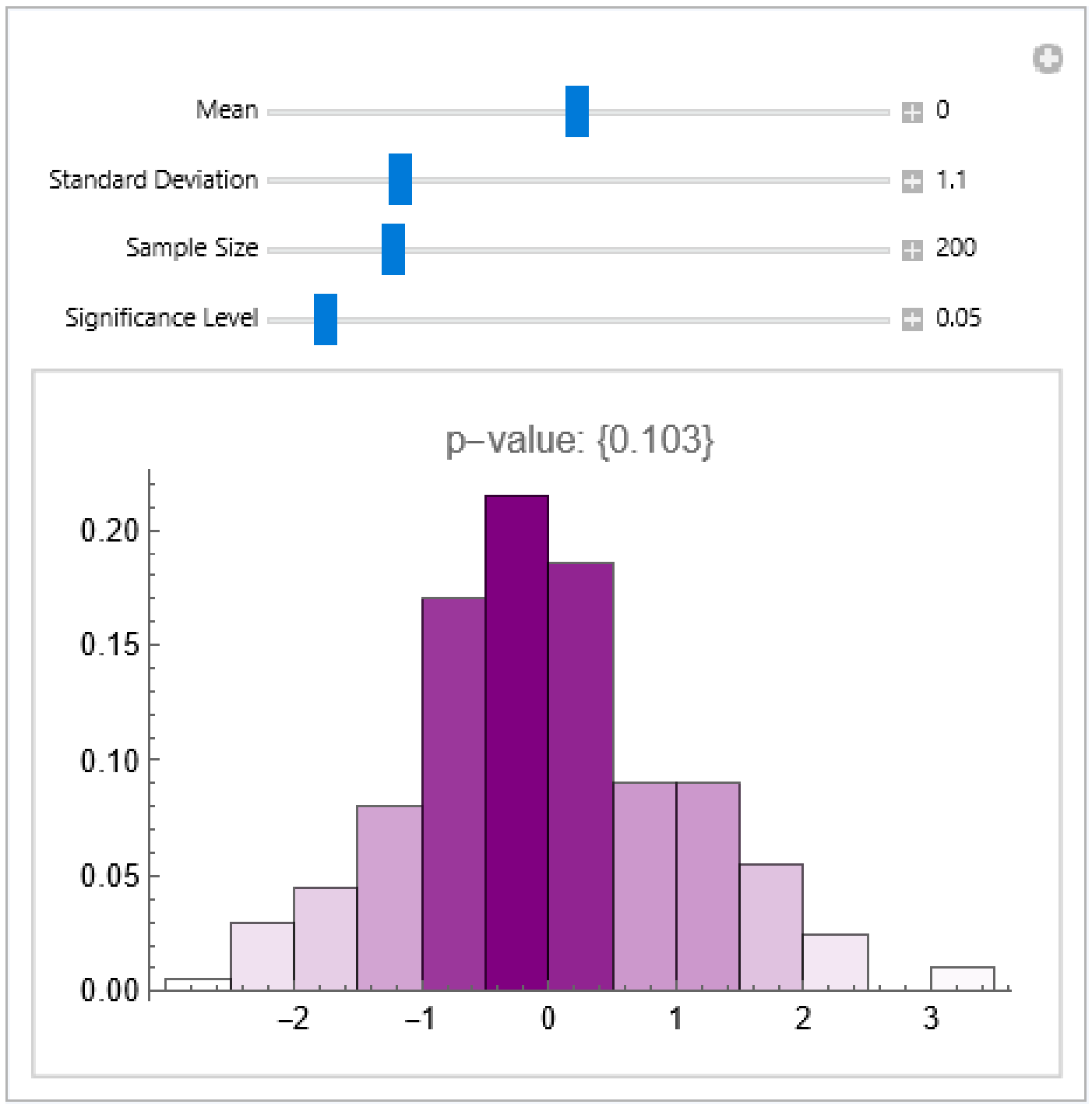

***Mathematica Examples 21.39***

Input
```
(* The code generates sample data from a normal distribution and creates an interactive Manipulate function for performing a one-sample variance test. Users can dynamically adjust the null hypothesis variance (σ0) and observe its effect on the variance test results. The interactive interface allows users to explore how changes in the null hypothesis variance impact the p-value, indicating the statistical significance of the variance test's outcome. The output displays a histogram of the generated data, providing a visual representation of the data's distribution. The plot's title includes the p-value, rounded to three decimal places, providing information about the statistical significance of the variance test's result: *)

(* Generate sample data from a normal distribution: *)
data=RandomVariate[
    NormalDistribution[0,1],
    200
    ];

Manipulate[
 Module[
  {pValue},
  (* Perform a one-sample test against the null hypothesis: *)
  pValue=VarianceTest[
    data,
    σ0,
    {"PValue"}
    ];

  (* Display the histogram of the data along with p-value: *)
  Histogram[
   data,
   Automatic,
   "Probability",
   PlotLabel->Row[{"p-value: ",NumberForm[pValue,{4,3}]}],
   ColorFunction->Function[{height},Opacity[height]],
   ImageSize->300,
   ChartStyle->Purple
   ]
  ],
 {{σ0,0.5,"σ0"},0.1,6,0.1,Appearance->"Labeled"}
 ]
```
Output

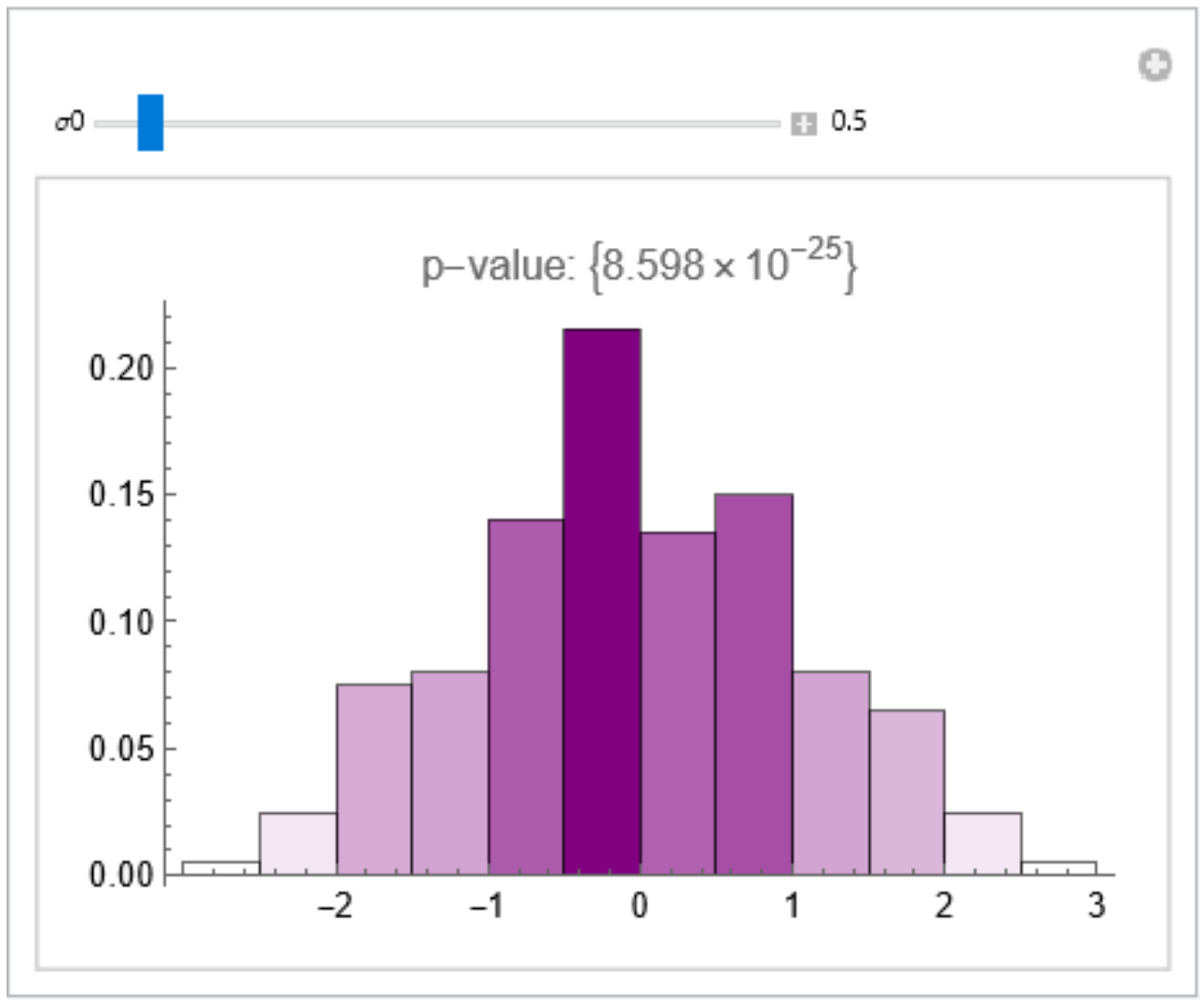

***Mathematica Examples 21.40***

Input

```
(* The code performs a hypothesis test to determine whether the ratio of variances σ1^2/σ2^2 is equal to 0.5. It does so by generating two sets of random data samples, data1 and data2, from normal distributions with different standard deviations. It then calculates the variance ratio σ1^2/σ2^2 and performs a variance test to assess if this ratio significantly differs from 0.5. The output of the code includes: The calculated ratio of variances σ1^2/σ2^2. The results of the variance test, including the test statistic and p-value. This helps determine whether there is sufficient evidence to either reject or fail to reject the null hypothesis that σ1^2/σ2^2=0.5. A plot with two smooth histograms representing the two data populations, data1 and data2. Each population is visualized with a distinct color (purple and blue) and is accompanied by corresponding plot legends. By analyzing the p-value obtained from the variance test, one can draw conclusions about the statistical significance of the difference in variances between the two datasets. The smooth histogram plot helps visualize the distributions of the two populations and provides insights into their variability: *)

(* Test whether the ratio σ1^2/σ2^2=0.5: *)
data1=RandomVariate[
    NormalDistribution[5,2],
    700
    ];
data2=RandomVariate[
    NormalDistribution[5,4],
    500
    ];

(* The ratio σ1^2/σ2^2: *)
Variance[data1]/Variance[data2]

(* At the 0.05 level, σ1^2/σ2^2 is significantly different from 0.5: *)
VarianceTest[{data1,data2},0.5]

(* Plot the two populations: *)
SmoothHistogram[
 {data1,data2},
 ImageSize->250,
 PlotStyle->{Purple,Blue},
 PlotLegends->{"Data 1","Data 2"}
 ]
```

Output 0.248251

Output 1.4786*10^-17

Output

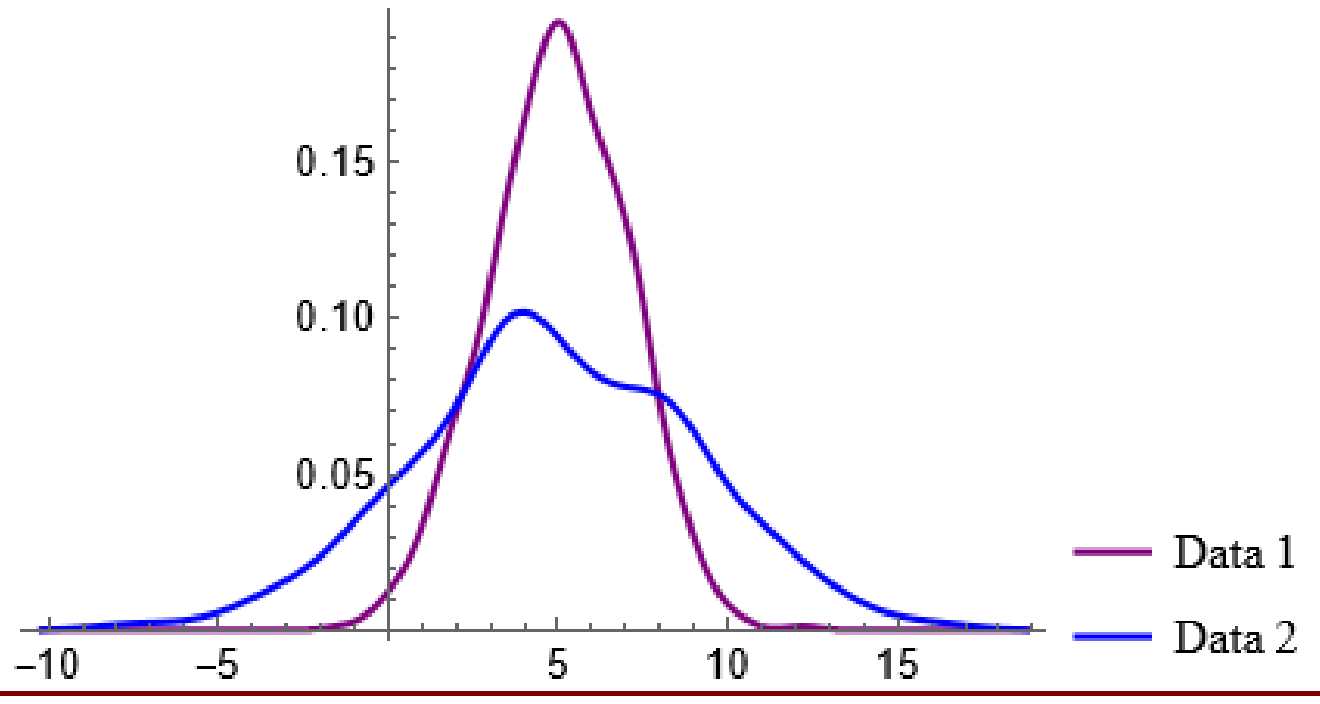

***Mathematica Examples 21.41***

Input

```
(* The code creates an interactive Manipulate function for comparing two data populations using a hypothesis test for the ratio of their variances (σ1^2/σ2^2).
```

```
The code generates random data samples from normal distributions based on user-
defined parameters for both populations. The output displays two histograms, each
representing one population (Data 1 and Data 2), with different colors (blue and red)
and transparency for visual differentiation. The plot's title includes the p-value
obtained from the variance test, providing information about the statistical
significance of the difference in variances between the two datasets. This interactive
visualization and hypothesis testing allow users to gain insights into the variability
comparison between two data populations and understand how different combinations of
mean, standard deviation, and sample size impact the variance test's outcome. The
histogram visualization enhances the understanding of data distribution
characteristics for each population: *)

Manipulate[
 (* Generate data for two populations based on specified parameters: *)
 data1=RandomVariate[
   NormalDistribution[mean1,stdDev1],
   sampleSize1
   ];

 data2=RandomVariate[
   NormalDistribution[mean2,stdDev2],
   sampleSize2
   ];

 (* Perform the test of σ1^2/σ2^2: *)
 testResult=VarianceTest[{data1,data2},1];

 (* Plot the two populations and the test results: *)
 Show[
  Histogram[
   data1,
   Automatic,
   "PDF",
   ImageSize->300,
   ChartStyle->Directive[Blue,Opacity[0.3]],
   ChartLegends->{"Data 1"}
   ],
  Histogram[
   data2,
   Automatic,
   "PDF",
   ImageSize->300,
   ChartStyle->Directive[Red,Opacity[0.3]],
   ChartLegends->{"Data 2"}
   ],
  PlotRange->All,
  PlotLabel->Row[{"Histogram of Data 1 and Data 2\n p-value: ",testResult}]
  ],
 (* Controls for user to modify parameters: *)
 {{mean1,5,"Mean of Data 1"},0,10,0.1},
 {{stdDev1,1,"Standard Deviation of Data 1"},0.1,5,0.1},
 {{sampleSize1,50,"Sample Size of Data 1"},10,100,10},
 {{mean2,7,"Mean of Data 2"},0,10,0.1},
 {{stdDev2,1.5,"Standard Deviation of Data 2"},0.1,5,0.1},
 {{sampleSize2,100,"Sample Size of Data 2"},10,500,10}
 ]
```

Output

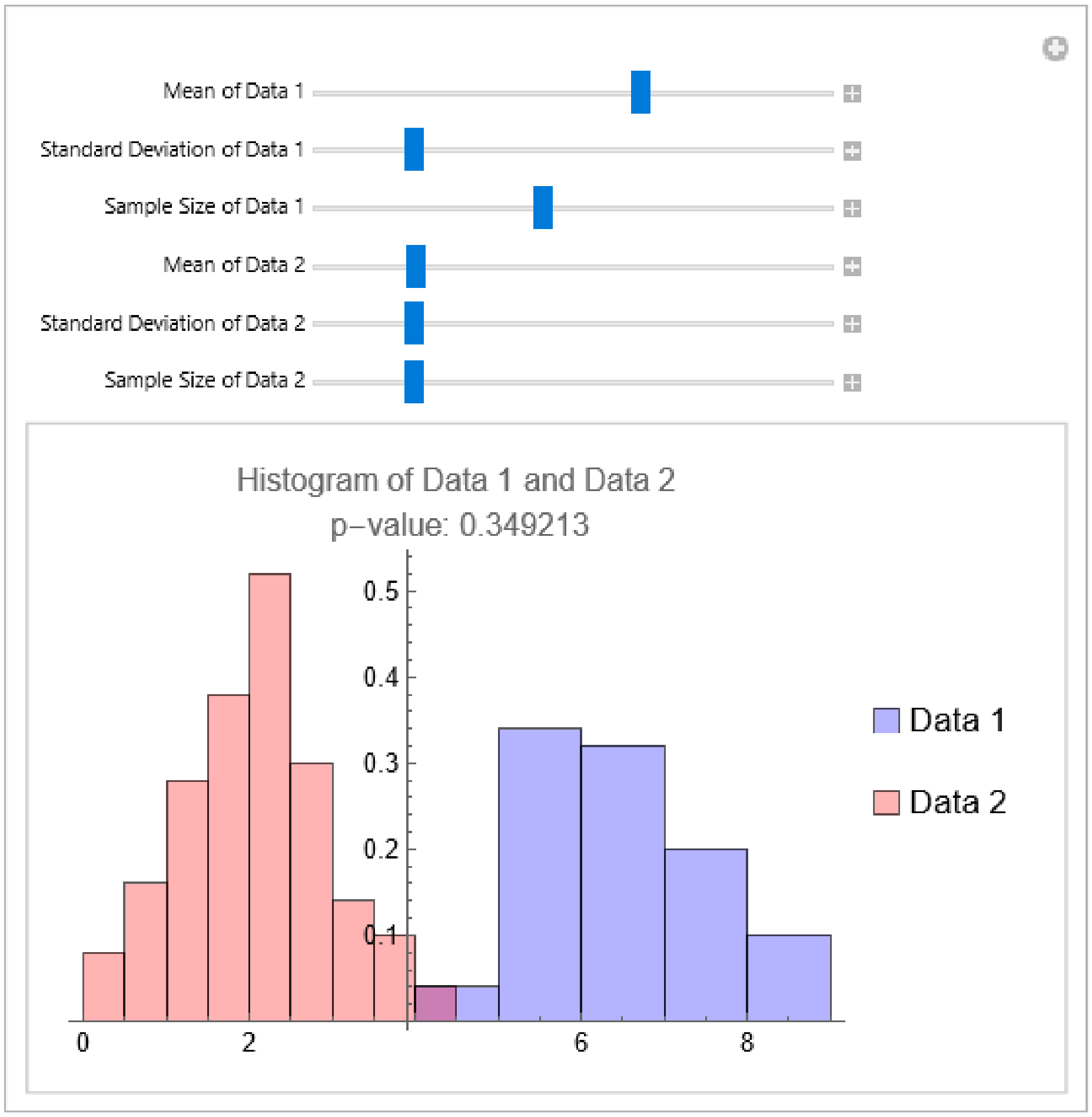

*Mathematica Examples 21.42*

Input

```
(* The code conducts an analysis of four datasets (data) generated from normal
distributions with different standard deviations but the same mean (mean of 3). It
includes a histogram visualization of the four datasets, calculates the variances of
each dataset, and performs a variance equivalence test to assess whether the variances
are equivalent: *)

data={
    RandomVariate[
     NormalDistribution[3,1],
     500
     ],
    RandomVariate[
     NormalDistribution[3,1.2],
     650
     ],
    RandomVariate[
     NormalDistribution[3,1.5],
     400
     ],
    RandomVariate[
     NormalDistribution[3,2],
     500
     ]
    };
```

```
        SmoothHistogram[
         data,
         ImageSize->250
         ]

        Map[Variance,data]

        VarianceEquivalenceTest[data,{"TestDataTable",All}]
```

Output

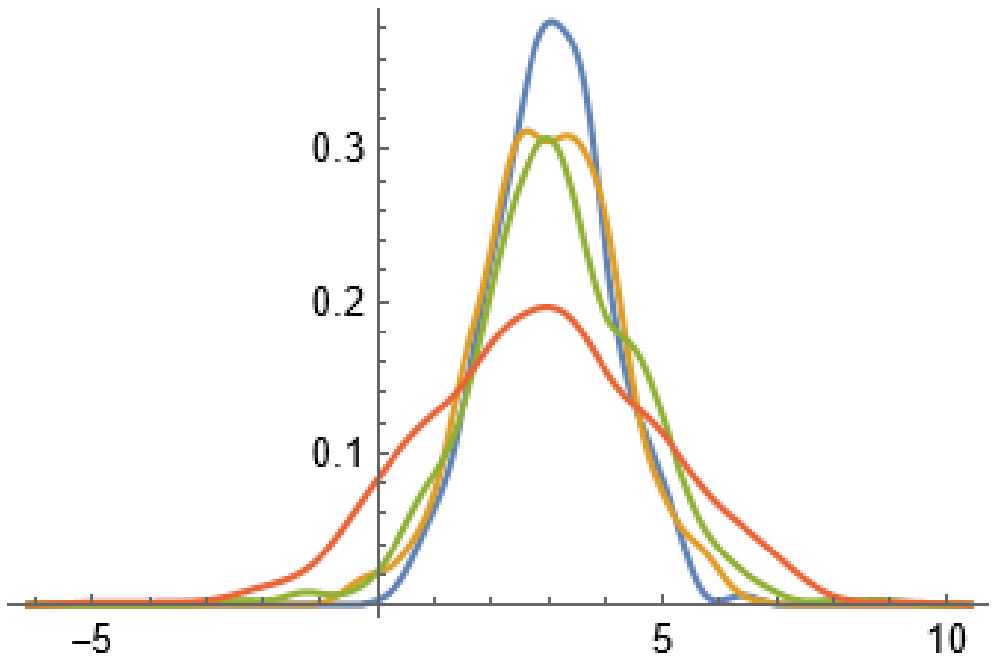

Output {1.05544,1.42735,2.13569,4.15982}

Output

| | Statistic | P-Value |
|---|---|---|
| Bartlett | 285.601, | 1.29988*10^-61 |
| Brown-Forsythe | 79.0742, | 2.13438*10^-48 |
| Conover | 181.43, | 4.33216*10^-39 |
| Levene | 79.3068, | 1.5635*10^-48 |

### *Mathematica Examples 21.43*

Input

```
(* The code analyzes four datasets (data) generated from normal distributions with
varying means and the same standard deviation. It includes a histogram visualization
of the datasets, calculates the variance of each dataset, and performs a variance
equivalence test to determine whether the variances are equivalent: *)

data={
    RandomVariate[
     NormalDistribution[1,1],
     500
     ],
    RandomVariate[
     NormalDistribution[2,1],
     650
     ],
    RandomVariate[
     NormalDistribution[3,1],
     400
     ],
    RandomVariate[
     NormalDistribution[4,1],
     500
     ]
    };

SmoothHistogram[
 data,
 ImageSize->250
 ]

Map[Variance,data]
```

```
VarianceEquivalenceTest[data,{"TestDataTable",All}]
```

Output

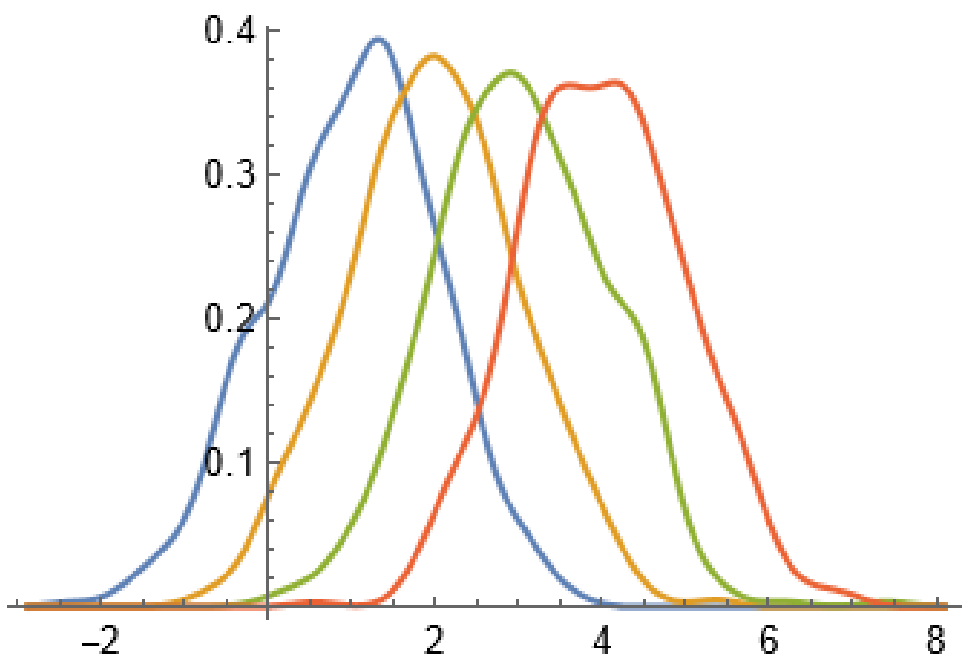

Output {1.04011,1.05668,1.07323,0.989129}

Output

| | Statistic | P-Value |
|---|---|---|
| Bartlett | 0.898073 | 0.825893 |
| Brown-Forsythe | 0.125963 | 0.94476 |
| Conover | 0.38328 | 0.943674 |
| Levene | 0.131019 | 0.941661 |

***Mathematica Examples 21.44***

Input

```
(* The code creates an interactive Manipulate function for conducting a variance equivalence test on multiple populations. Users can dynamically adjust the parameters (standard deviations) of three populations and the significance level to observe how these changes impact the test results: *)

(* Create the Manipulate function: *)
Manipulate[
 (* Generate data for multiple populations with different parameters: *)
 data1=RandomVariate[
   NormalDistribution[10,σ1],
   100
   ];
 data2=RandomVariate[
   NormalDistribution[10,σ2],
   150
   ];
 data3=RandomVariate[
   NormalDistribution[10,σ3],
   200
   ];

 (* Perform the LocationEquivalenceTest for the selected populations: *)
 testResult=VarianceEquivalenceTest[
   {data1,data2,data3},
   {"PValue","ShortTestConclusion"},
   SignificanceLevel->alpha
   ];

 (* Display the test result and other information: *)
 Column[
  {
   Row[{"P-Value: ",testResult[[1]]}],
   Row[{"VarianceEquivalenceTest Conclusion: ",testResult[[2]]}]
   }
  ],

 (* Controls for parameters*)
```

```
{{σ1,2,"σ1:"},1,5,0.1},
{{σ2,2,"σ2:"},1,5,0.1},
{{σ3,2,"σ3:"},1,5,0.1},

(* Control for the significance level*)
{{alpha,0.05,"SignificanceLevel:"},0.01,0.1,0.01}
]
```

Output

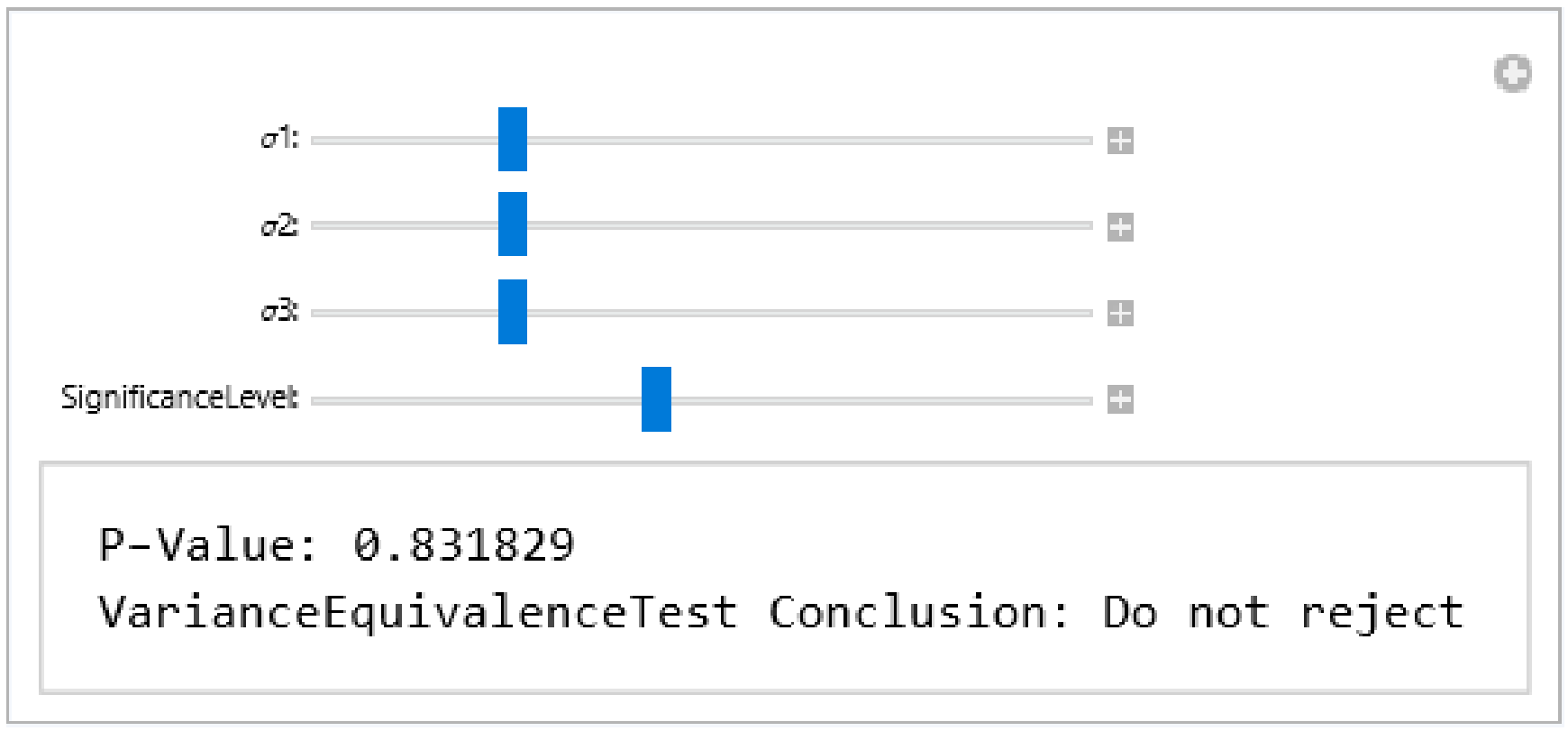

# UNIT 21.3

# GOODNESS-OF-FIT TESTS

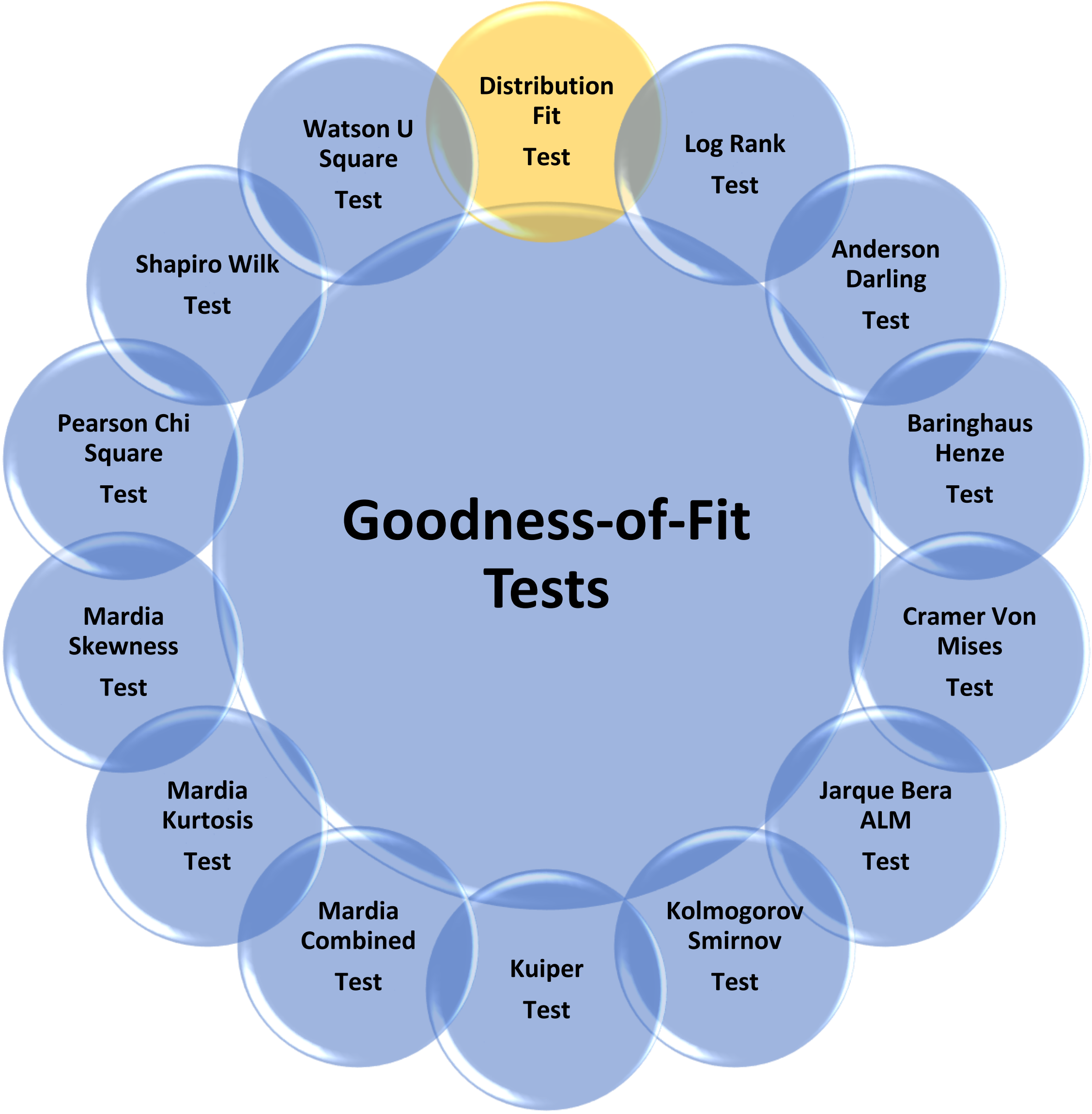

## DistributionFitTest

| | |
|---|---|
| `DistributionFitTest[data]` | tests whether data is normally distributed. |
| `DistributionFitTest[data,dist]` | tests whether data is distributed according to dist. |
| `DistributionFitTest[data, dist,"property"]` | returns the value of "property". |

***Mathematica Examples 21.45***

```
Input   (* The code generates two random datasets, dataA and dataB, and performs a goodness-
        of-fit test to assess if they follow a Normal distribution. dataA contains 100 random
        numbers drawn from a Normal distribution with mean 0 and standard deviation 1. dataB
        contains 100 random numbers drawn from a Gamma distribution with shape parameter 3
        and scale parameter 2. The goodness-of-fit test calculates p-values for both datasets:
        testResultA: The p-value for dataA is typically large (close to 1), indicating that
        the data is consistent with a Normal distribution. testResultB: The p-value for dataB
        is typically small (close to 0), suggesting that the data significantly deviates from
        a Normal distribution: *)

        (* Generate a random dataset following a Normal distribution: *)
        dataA=RandomVariate[
            NormalDistribution[0,1],
            100
            ];

        (* Perform the goodness-of-fit test for Normal distribution: *)
        testResultA=DistributionFitTest[
            dataA
            ];

        (* The P-values for the normally distributed data are typically large: *)
        Print["p-value: ",testResultA]

        (* Generate a random dataset following a Gamma distribution: *)
        dataB=RandomVariate[
            GammaDistribution[3,2],
            100
            ];

        (* Perform the goodness-of-fit test for Normal distribution: *)
        testResultB=DistributionFitTest[
            dataB
            ];

        (* The P-values for data that is not normally distributed are typically small: *)
        Print["p-value: ",testResultB]

Output  p-value:  0.994018
Output  p-value:  0.000341889
```

***Mathematica Examples 21.46***

```
Input   (* The code generates a random dataset of 1000 data points following a normal
        distribution with mean 0 and standard deviation 1. Perform a goodness-of-fit test
        using DistributionFitTest to test whether the data fits a normal distribution with
        unknown parameters mu (mean) and sigma (standard deviation). Print the test conclusion
        and the p-value, providing insights into whether the data can be considered normally
        distributed based on the significance level. Output the full test table, which
        contains additional details about the goodness-of-fit test. Create a plot that shows
        the histogram of the data as well as the fitted PDF of the test distribution. This
        visualization allows for a visual comparison of how well the fitted distribution
        matches the data: *)

        (* Generate a random dataset following a Normal distribution: *)
        data=RandomVariate[
            NormalDistribution[0,1],
            1000
            ];
```

```
(* Perform the goodness-of-fit test for Normal distribution: *)
testResult=DistributionFitTest[
    data,
    NormalDistribution[mu,sigma],
    "HypothesisTestData"
    ];

(* Print the test conclusion and p-value: *)
Print[testResult["TestConclusion"]]
Print["p-value: ",testResult["PValue"]]

(* The fitted Distribution: *)
fDist=testResult["FittedDistribution"]

(* The full test table: *)
testResult["TestDataTable",All]

(* Compare the histogram of the data to the PDF of the test distribution: *)
Show[
 Histogram[
  data,
  Automatic,
  "ProbabilityDensity",
  ImageSize->250,
  ColorFunction->Function[{height},Opacity[height]],
  ChartStyle->Purple
  ],
 Plot[
  PDF[fDist,x],
  {x,-5,5},
  PlotStyle->Directive[Thick,RGBColor[0.88,0.61,0.14]],
  ImageSize->250
  ]
 ]
```

Output The null hypothesis that the data is distributed according to the NormalDistribution[mu,sigma] is not rejected at the 5 percent level based on the Cramér-von Mises test.

Output p-value: 0.454032

Output NormalDistribution[0.00128788,0.958332]

Output

| | Statistic | P-Value |
|---|---|---|
| Anderson-Darling | 0.369162 | 0.434603 |
| Baringhaus-Henze | 0.454337 | 0.502564 |
| Cramér-von Mises | 0.0545171 | 0.454032 |
| Jarque-Bera ALM | 4.70093 | 0.0936709 |
| Kolmogorov-Smirnov | 0.0202813 | 0.42076 |
| Kuiper | 0.0326863 | 0.514794 |
| Mardia Combined | 4.70093 | 0.0936709 |
| Mardia Kurtosis | 2.10359 | 0.0354141 |
| Mardia Skewness | 0.0419931 | 0.837633 |
| Pearson $\chi^2$ | 26.048 | 0.622925 |
| Shapiro-Wilk | 0.997525 | 0.134977 |
| Watson $U^2$ | 0.0517416 | 0.438954 |

Output

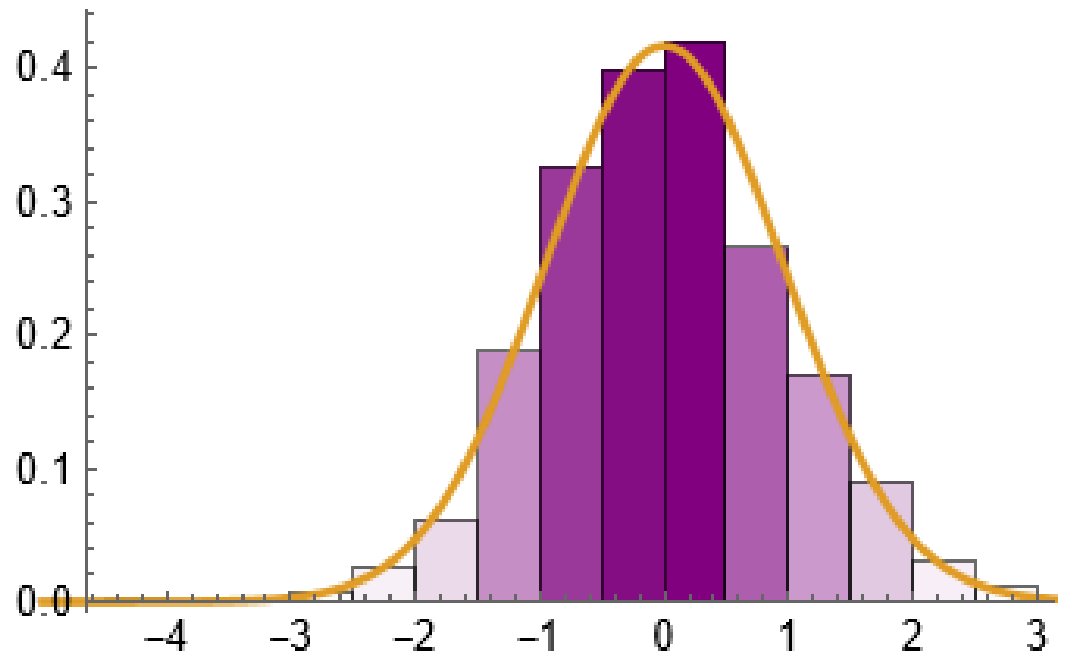

***Mathematica Examples 21.47***

Input

```
(* The code generates a random dataset following a gamma distribution with shape parameter (alpha) 2 and scale parameter (beta) 1. It then performs a goodness-of-fit test to determine if the data follows a gamma distribution with the same parameters (alpha=2, beta=1). Finally, it prints the test conclusion and the p-value and outputs the full test table: *)

(* Generate a random dataset following a gamma distribution: *)
data=RandomVariate[
    GammaDistribution[2,1],
    500
    ];

(* Perform the goodness-of-fit test for gamma distribution: *)
testResult=DistributionFitTest[
    data,
    GammaDistribution[2,1],
    "HypothesisTestData"
    ];

(* Print the test result and p-value: *)
Print[testResult["TestConclusion"]]
Print["p-value: ",testResult["PValue"]]

(* The full test table: *)
testResult["TestDataTable",All]
```

Output

The null hypothesis that the data is distributed according to the GammaDistribution[2,1] is not rejected at the 5 percent level based on the Cramér-von Mises test.

Output

p-value: 0.70995

Output

| | Statistic | P-Value |
|---|---|---|
| Anderson-Darling | 0.443821 | 0.804343 |
| Cramér-von Mises | 0.0769044 | 0.70995 |
| Kolmogorov-Smirnov | 0.0283782 | 0.804567 |
| Kuiper | 0.0491871 | 0.596376 |
| Pearson $\chi^2$ | 18.7 | 0.767794 |
| Watson $U^2$ | 0.0550538 | 0.649466 |

***Mathematica Examples 21.48***

Input

```
(* The code generates a random dataset following a gamma distribution with shape parameter (alpha) 2 and scale parameter (beta) 1. It then performs a goodness-of-fit test to check if the data fits a gamma distribution with unknown parameters (alpha and beta). The code then prints the test conclusion, the p-value, and outputs the fitted distribution based on the data. Additionally, it provides the full test table
```

```
and creates a plot to compare the histogram of the data with the fitted PDF of the
test distribution: *)

(* Generate a random dataset following a gamma Distribution: *)
data=RandomVariate[
    GammaDistribution[2,1],
    500
    ];

(* Perform the goodness-of-fit test for gamma distribution with unknown parameters:
*)
testResult=DistributionFitTest[
    data,
    GammaDistribution[alpha,beta],
    "HypothesisTestData"
    ];

(* Print the test conclusion and p-value: *)
Print[testResult["TestConclusion"]]
Print["p-value: ",testResult["PValue"]]

(* The fitted Distribution: *)
fDist=testResult["FittedDistribution"]

(* The full test table: *)
testResult["TestDataTable",All]

(* Compare the histogram of the data to the PDF of the test distribution: *)
Show[
 Histogram[
  data,
  Automatic,
  "ProbabilityDensity",
  ImageSize->250,
  ColorFunction->Function[{height},Opacity[height]],
  ChartStyle->Purple
  ],
 Plot[
  PDF[fDist,x],
  {x,0,10},
  PlotStyle->Directive[Thick,RGBColor[0.88,0.61,0.14]],
  ImageSize->250
  ]
 ]
```

Output The null hypothesis that the data is distributed according to the GammaDistribution[alpha,beta] is not rejected at the 5 percent level based on the Cramér-von Mises test.

Output p-value: 0.332813

Output GammaDistribution[2.09843,0.955069]

Output

| | Statistic | P-Value |
|---|---|---|
| Anderson-Darling | 0.356504 | 0.468162 |
| Cramér-von Mises | 0.0649709 | 0.332813 |
| Kolmogorov-Smirnov | 0.0305299 | 0.315062 |
| Kuiper | 0.0526315 | 0.261953 |
| Pearson $\chi^2$ | 33.8 | 0.0514721 |
| Watson $U^2$ | 0.0648888 | 0.289252 |

Output

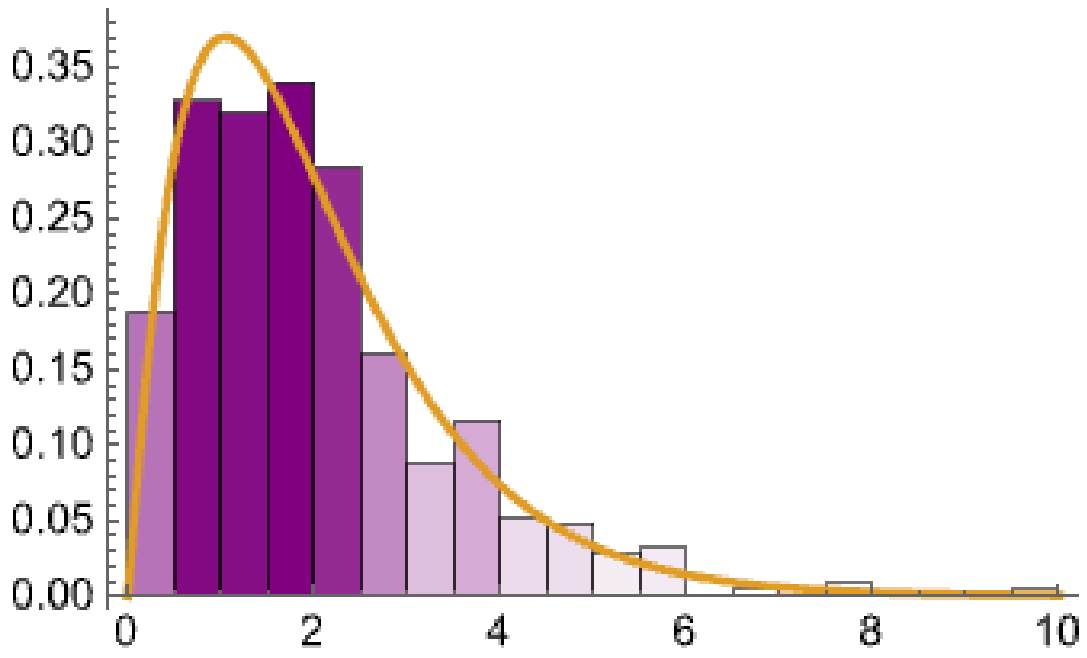

***Mathematica Examples 21.49***

Input

```
(* The code generates a random dataset following a multinormal distribution with specific means, variances, and correlation coefficients. It then performs a goodness-of-fit test to check if the data fits a multinormal distribution with unknown parameters. The code prints the test conclusion, the p-value, and outputs the fitted distribution based on the data. It also provides the full test table and creates several plots to compare the histogram of the data with the fitted PDF of the test distribution and to visualize the marginal PDFs of the test distribution: *)

(* Generate a random dataset following a multinormal distribution: *)
data=RandomVariate[
   MultinormalDistribution[{-1,1},{{1,0.5},{0.5,2}}],
   1000
   ];

(* Perform the goodness-of-fit test for a multinormal distribution: *)
testResult=DistributionFitTest[
   data,
   MultinormalDistribution[{mu1,mu2},{{sigma1,rho},{rho,sigma2}}],
   "HypothesisTestData"
   ];

(* Print the test result and p-value: *)
Print[testResult["TestConclusion"]]
Print["p-value: ",testResult["PValue"]]

(* The fitted Distribution: *)
fDist=testResult["FittedDistribution"]

(* The full test table: *)
testResult["TestDataTable",All]

(* Compare the histogram of the data to the PDF of the test distribution: *)
Histogram3D[
 data,
 Automatic,
 "ProbabilityDensity",
 ColorFunction->"Rainbow",
 PlotRange->All,
 ImageSize->300
 ]
ContourPlot[
 PDF[fDist,{x,y}],
 {x,-4,2},
 {y,-2,4},
 PlotRange->All,
 ContourStyle->{White},
```

```
    ClippingStyle->Automatic,
    ColorFunction->"BlueGreenYellow",
    PlotLegends->Automatic,
    ImageSize->250
    ]
   (* Plot the marginal PDFs of the test distribution against the data to confirm the
   test results: *)
   Table[
    Show[
     SmoothHistogram[
      data[[All,i]],
      PlotStyle->{Orange,Dashed,Thick},
      PlotRange->{0,.4},
      ImageSize->220
      ],
     Plot[
      PDF[MarginalDistribution[fDist,i],
       x
       ],
      {x,-5,5},
      ImageSize->220,
      PlotStyle->Directive[Thick,Purple]
      ]
     ],
    {i,2}
    ]
```

Output The null hypothesis that the data is distributed according to the MultinormalDistribution[{mu1,mu2},{{sigma1,rho},{rho,sigma2}}] is not rejected at the 5 percent level based on the Mardia Combined test.

Output p-value: 0.366222

Output MultinormalDistribution[{-1.01349,0.981946}, {{1.02115,0.497712}, {0.497712,2.02919}}]

Output

| | Statistic | P-Value |
|---|---|---|
| Anderson-Darling | 1.78502 | 0.976892 |
| Baringhaus-Henze | 0.533692 | 0.889212 |
| Cramér-von Mises | 1.79174 | 0.978314 |
| Jarque-Bera ALM | 0.523148 | 0.136842 |
| Kolmogorov-Smirnov | 1.86149 | 0.990408 |
| Kuiper | 1.67349 | 0.946696 |
| Mardia Combined | 5.42516 | 0.366222 |
| Mardia Kurtosis | 1.37424 | 0.142919 |
| Mardia Skewness | 3.51901 | 0.936024 |
| Pearson $\chi^2$ | 1.36847 | 0.800587 |
| Shapiro-Wilk | 0.998461 | 0.529355 |
| Watson $U^2$ | 1.59739 | 0.918954 |

Output

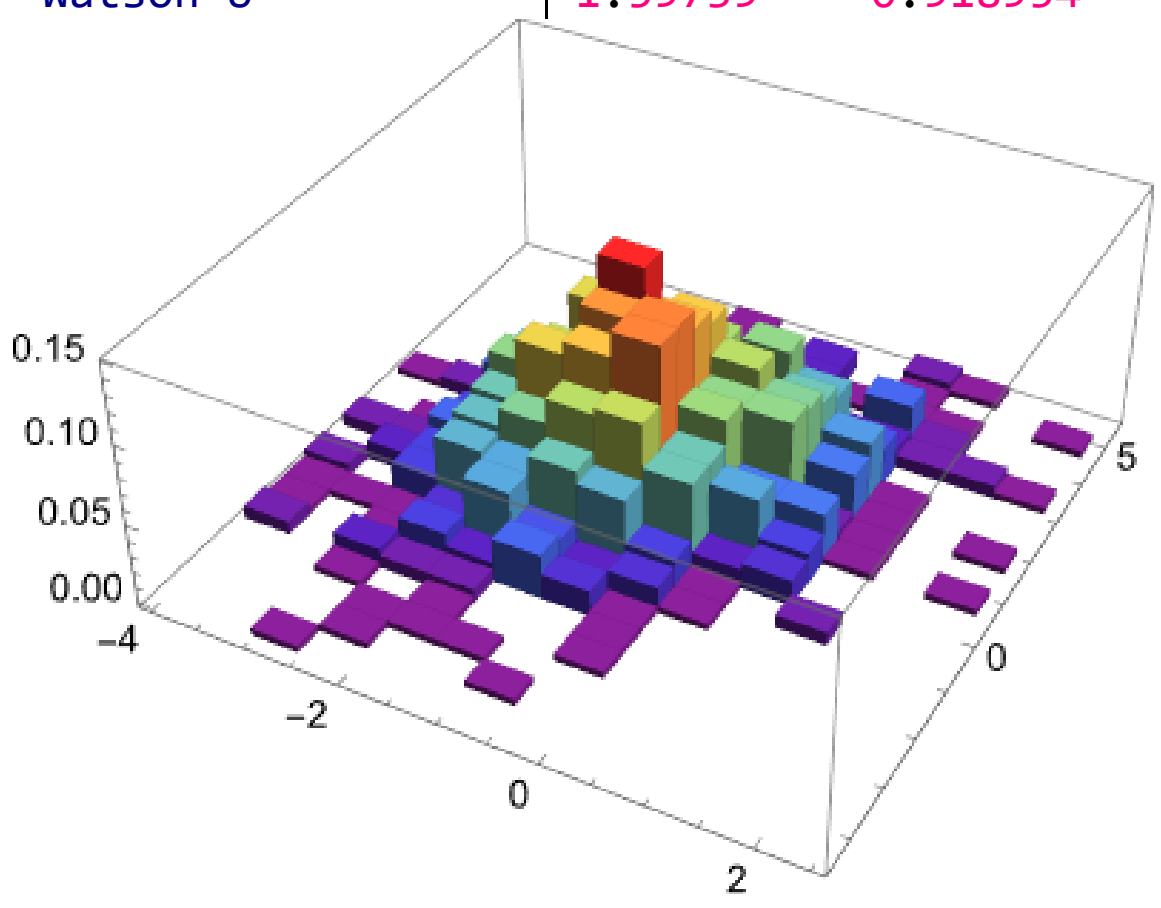

Output

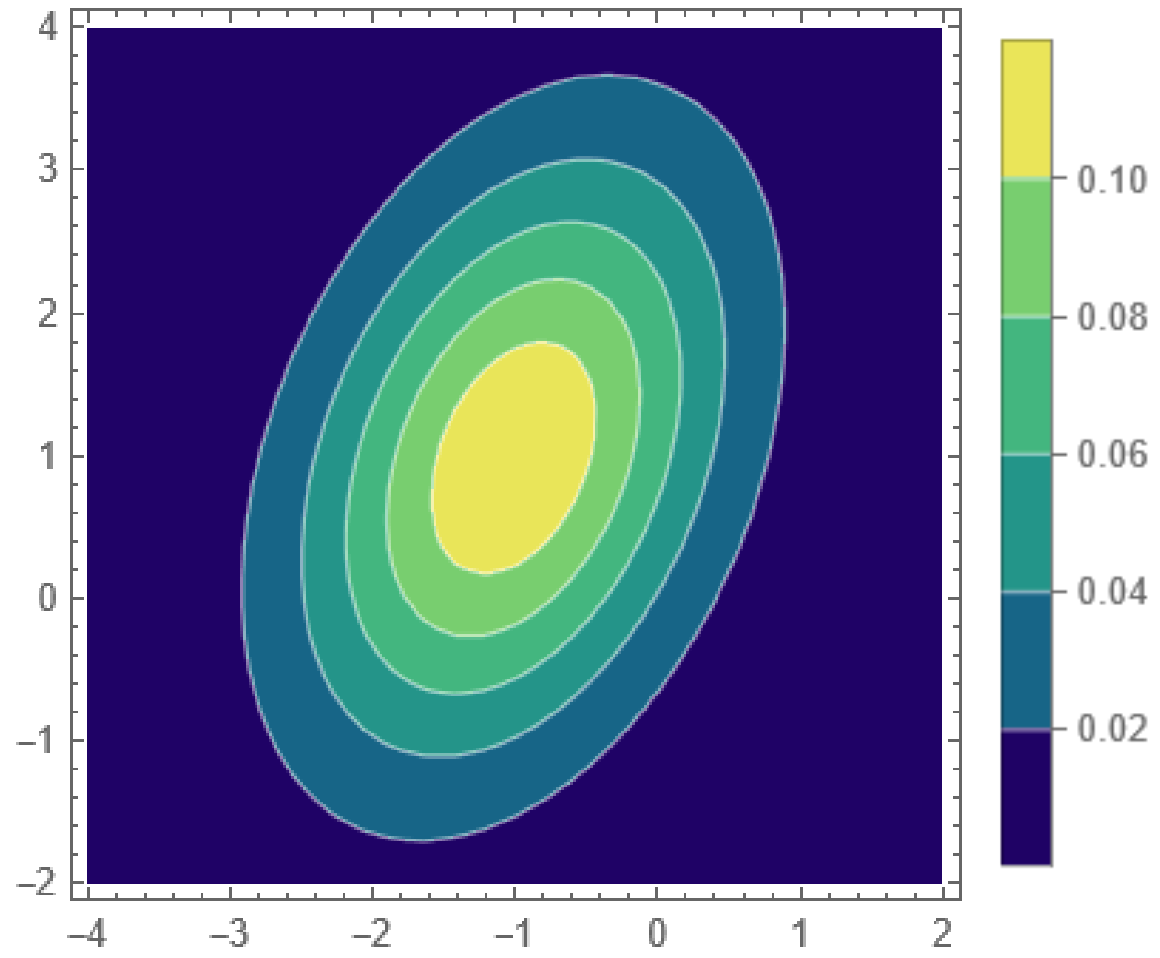

Output

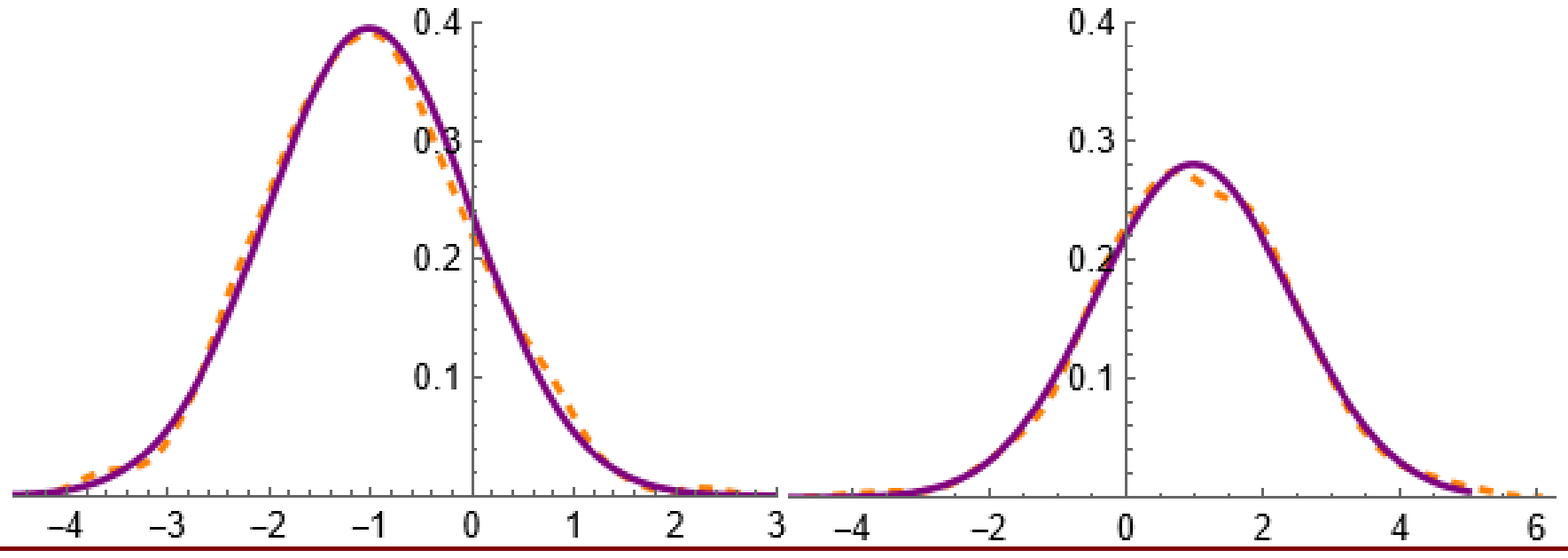

***Mathematica Examples 21.50***

Input
```
(* A random dataset of 1000 data points is generated following a standard normal distribution. The DistributionFitTest function is then used to perform a goodness-of-fit test on the data. By default, when the third argument of DistributionFitTest is left blank, it is set to "Automatic," which allows Mathematica to apply a generally powerful and appropriate test automatically based on the characteristics of the data: *)

data=RandomVariate[
    NormalDistribution[0,1],
    1000
    ];
DistributionFitTest[data]

(* The property "AutomaticTest" can be used to determine which test was chosen: *)
DistributionFitTest[
 data,
 Automatic,
 "AutomaticTest"
 ]
```

Output `0.74719`
Output `CramerVonMises`

***Mathematica Examples 21.51***

Input
```
(* Two random datasets, data1 and data2, each containing 1000 data points, are generated following a standard normal distribution. The DistributionFitTest function is then used to compare the distributions of the two datasets. If the p-value of the test is low (typically below 0.05), it suggests that the distributions of data1 and data2 are significantly different. On the other hand, a higher p-value indicates that
```

```
          there is not enough evidence to reject the null hypothesis, meaning the distributions
          are similar enough: *)

          (* Compare the distributions of two datasets: *)
          data1=RandomVariate[
              NormalDistribution[],
              10^3
              ];

          data2=RandomVariate[
              NormalDistribution[],
              10^3
              ];

          DistributionFitTest[data1,data2]

Output    0.846622
```

***Mathematica Examples 21.52***

```
Input     (* Two random datasets, data1 and data2, each containing 1000 data points, are
          generated following different normal distributions. The DistributionFitTest function
          is then used to compare the distributions of the two datasets. The code prints the
          test conclusion, and the p-value. If the p-value of the test is low (typically below
          0.05), it suggests that the distributions of data1 and data2 are significantly
          different. On the other hand, a higher p-value indicates that there is not enough
          evidence to reject the null hypothesis, meaning the distributions are similar enough:
          *)

          (* Generate some example data from two different distributions: *)
          data1=RandomVariate[
              NormalDistribution[0,1],
              100
              ];
          data2=RandomVariate[
              NormalDistribution[1,1],
              100
              ];

          (* Perform the distribution fit test: *)
          result=DistributionFitTest[data1,data2,"HypothesisTestData"];

          (* Print the test result and p-value: *)
          Print[result["TestConclusion"]]
          Print["p-value: ",result["PValue"]]

          (* The full test table: *)
          result["TestDataTable",All]
```

Output The null hypothesis that the datasets have the same distribution is rejected at the 5 percent level based on the Watson $U^2$ test.

Output p-value: 3.50148*10^-6

Output

| | Statistic | P-Value |
|---|---|---|
| Anderson-Darling | 20.0213 | 0. |
| Cramér-von Mises | 3.8878 | 8.35102*10^-10 |
| Kolmogorov-Smirnov | 0.41 | 6.61736*10^-8 |
| Kuiper | 0.41 | 1.77777*10^-6 |
| Pearson $\chi^2$ | 48.7336 | 2.32834*10^-6 |
| Watson $U^2$ | 0.656918 | 3.50148*10^-6 |

***Mathematica Examples 21.53***

Input
```
(* Similar to above code bout in the case multivariate data: *)
(* Generate some example multivariate data: *)
data1=RandomVariate[
    MultinormalDistribution[{0,0},{{1,0.5},{0.5,1}}],
    200
    ];
data2=RandomVariate[
    MultinormalDistribution[{1,1},{{1,0.5},{0.5,1}}],
    200
    ];

(* Perform the distribution fit test: *)
result=DistributionFitTest[data1,data2,"HypothesisTestData"];

(* Print the test result and p-value: *)
Print[result["TestConclusion"]]
Print["p-value: ",result["PValue"]]

(* The full test table: *)
result["TestDataTable",All]
```

Output The null hypothesis that the datasets have the same distribution is rejected at the 5 percent level based on the Szekely Energy test.

Output p-value: 0.

Output

| | Statistic | P-Value |
|---|---|---|
| Anderson-Darling | 0. | 0. |
| Cramér-von Mises | 6.66134*10^-16 | 2.21867*10^-31 |
| Kolmogorov-Smirnov | 8.75722*10^-18 | 3.83445*10^-35 |
| Kuiper | 3.66374*10^-15 | 6.71148*10^-30 |
| Pearson χ2 | 1.46184*10^-9 | 1.06848*10^-18 |
| Szekely Energy | 99.851 | 0. |
| Watson U2 | 0. | 0. |

***Mathematica Examples 21.54***

Input
```
(* The significance level is also used for "ShortTestConclusion": *)
data=RandomVariate[
    NormalDistribution[0,1],
    100
    ];

h1=DistributionFitTest[data,Automatic,"HypothesisTestData",SignificanceLevel->.05];
h2=DistributionFitTest[data,Automatic,"HypothesisTestData",SignificanceLevel-
>.001];
h1["ShortTestConclusion"]
h2["ShortTestConclusion"]
```

Output Reject

Output Do not reject

***Mathematica Examples 21.55***

Input
```
(* The random dataset data of 1000 data points is generated following a normal
distribution with a chosen mean of 1 and standard deviation of 2. The code then
creates a Manipulate function that allows you to interactively explore the
distribution fit test based on different mean (mu) and standard deviation (sigma)
values. Then the code displays the following results in a column format: Sample mean
of data, sample standard deviation of data, distribution fit test statistic, and
distribution fit test p-value: *)
```

```
(* Generate data following a normal distribution: *)
data=RandomVariate[
   NormalDistribution[1,2],
   1000
   ];

(* Create the Manipulate function: *)
Manipulate[
 (* Perform the distribution fit test: *)
 testResult=DistributionFitTest[
   data,
   NormalDistribution[mu,sigma],
   "HypothesisTestData"
   ];
 (* Extract the test statistic and p-value: *)
 {testStat,pValue}=testResult["TestStatistic","PValue"];

 (* Display the results: *)
 Column[
  {
   Row[{"Sample Mean: ",Mean[data]}],
   Row[{"Sample Standard Deviation: ",StandardDeviation[data]}],
   Row[{"Distribution Fit Test Statistic: ",testStat}],
   Row[{"Distribution Fit Test P-Value: ",pValue}]
   }
  ],
 (* Manipulate Parameters: *)
 {{mu,0,"Mean"},-3,3,0.1,Appearance->"Labeled"},
 {{sigma,1,"Standard Deviation"},0.1,3,Appearance->"Labeled"}
 ]
```

Output

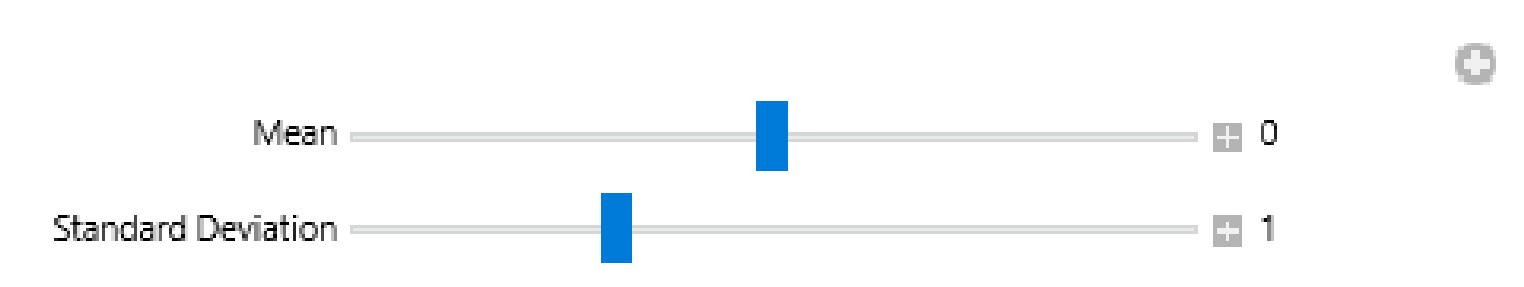

```
Sample Mean: 1.04129
Sample Standard Deviation: 2.04202
Distribution Fit Test Statistic: 50.3011
Distribution Fit Test P-Value: 1.11022×10^-16
```

***Mathematica Examples 21.56***

Input

```
(* The code contains a Manipulate function that allows interactive exploration of a goodness-of-fit test for a normal distribution. The Manipulate generates a random dataset (data) from a normal distribution with a given mean and standard deviation based on user-defined sliders. It performs a goodness-of-fit test using DistributionFitTest to compare the generated data with a normal distribution with the chosen parameters (mean and stdDev). The significance level of the test can be adjusted through the significanceLevel slider, enabling users to control the level of confidence in accepting or rejecting the null hypothesis. The Histogram of the generated data is displayed, and the plot title includes the p-value from the goodness-of-fit test with three decimal places precision. The interactive sliders allow users to observe how changes in the mean, standard deviation, sample size, and significance level impact the histogram and the p-value: *)
```

```
Manipulate[
 Module[
  {data,pValue},
  (* Generate sample data from a normal distribution: *)
  data=RandomVariate[
    NormalDistribution[1,3],
    sampleSize
    ];

  (* Perform the distribution fit test: *)
  testResult=DistributionFitTest[
    data,
    NormalDistribution[mean,stdDev],
    "HypothesisTestData",
    SignificanceLevel->significanceLevel
    ];

  (* Display the histogram of the data along with the test p-value: *)
  Histogram[
   data,
   Automatic,
   "Probability",
   PlotLabel->Row[{"p-value: ",NumberForm[testResult["PValue"],{4,3}]}],
   ColorFunction->Function[{height},Opacity[height]],
   ImageSize->300,
   ChartStyle->Purple
   ]
  ],
 {{mean,0,"Mean"},-6,6,0.1,Appearance->"Labeled"},
 {{stdDev,1,"Standard Deviation"},0.1,5,0.1,Appearance->"Labeled"},
 {{sampleSize,200,"Sample Size"},10,1000,10,Appearance->"Labeled"},
 {{significanceLevel,0.05,"Significance Level"},0.01,0.5,0.01,Appearance-
>"Labeled"}
 ]
```

Output

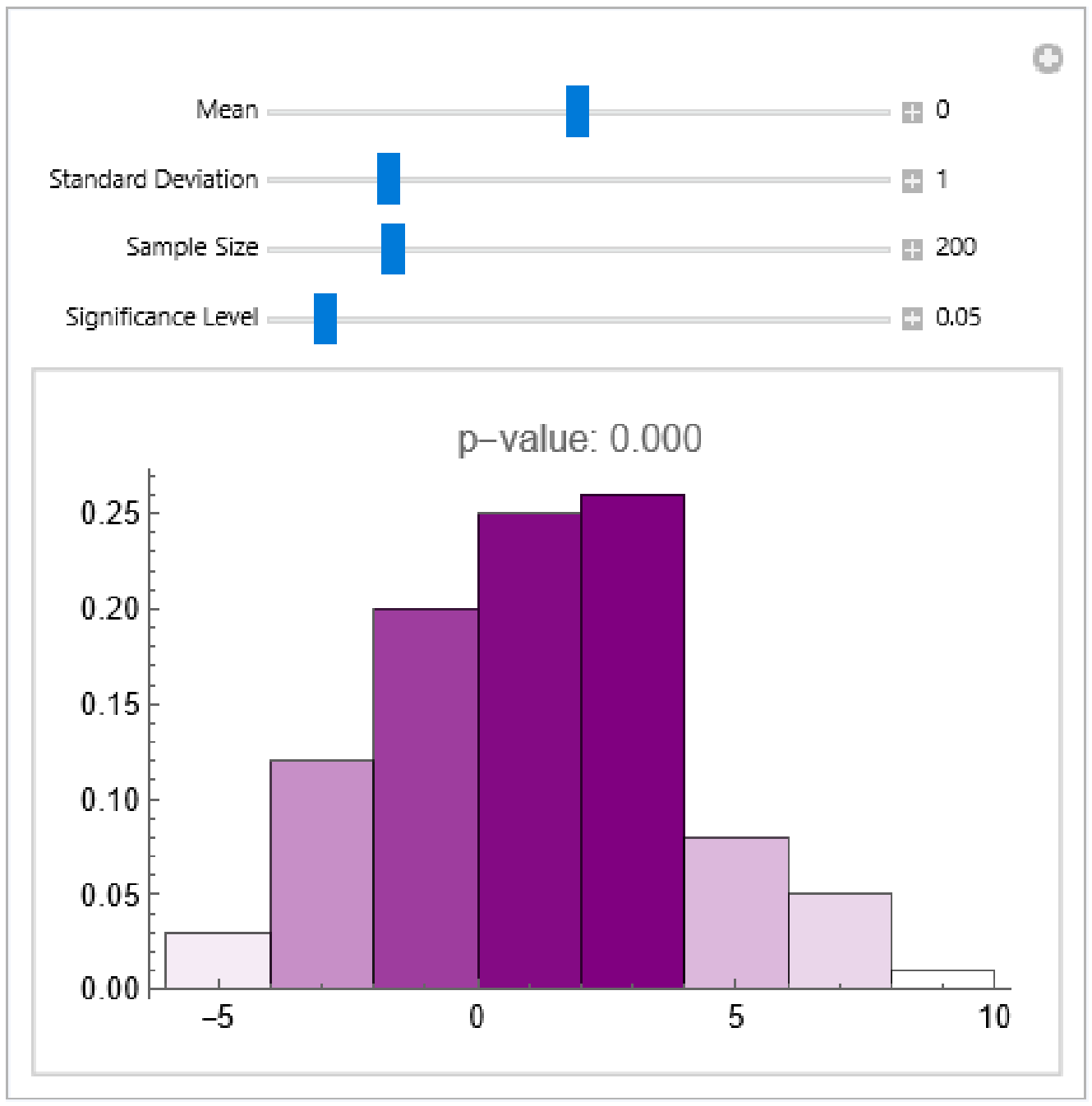

***Mathematica Examples 21.57***

Input
```
(* Three datasets (data1, data2, and data3) are generated, each containing 200 data
points from normal distributions with different mean and standard deviation values.
The BoxWhiskerChart is used to create a box-and-whisker plot that visually compares
the distributions of the three datasets. A goodness-of-fit test (DistributionFitTest)
is performed to compare the distributions of data1 and data2, confirming that the
first two populations have similar distributions. Another goodness-of-fit test is
performed to compare the distributions of data1 and data3, which reveals that the
third population differs significantly in distribution from the first: *)

(* Compare the distributions of three datasets of some populations: *)
data1=RandomVariate[
    NormalDistribution[1,1],
    200
    ];
data2=RandomVariate[
    NormalDistribution[1,1],
    200
    ];
data3=RandomVariate[
    NormalDistribution[4,1],
    200
    ];

BoxWhiskerChart[
 {data1,data2,data3},
 "Notched",
 ChartStyle->"SolarColors",
 ImageSize->250
 ]

(* The first two populations have similar distributions: *)
DistributionFitTest[
 data1,data2
 ]

(* The third population differs in distribution from the first: *)
DistributionFitTest[
 data1,data3
 ]
```

Output

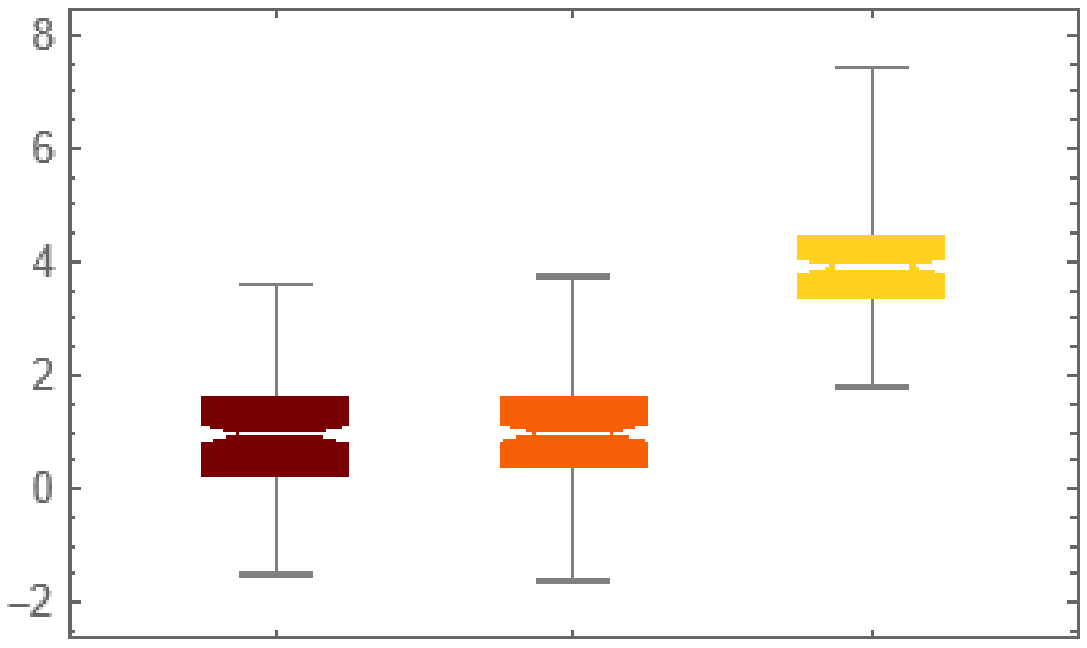

Output 0.41494

Output 0.

# UNIT 21.4

# DEPENDENCY TESTS

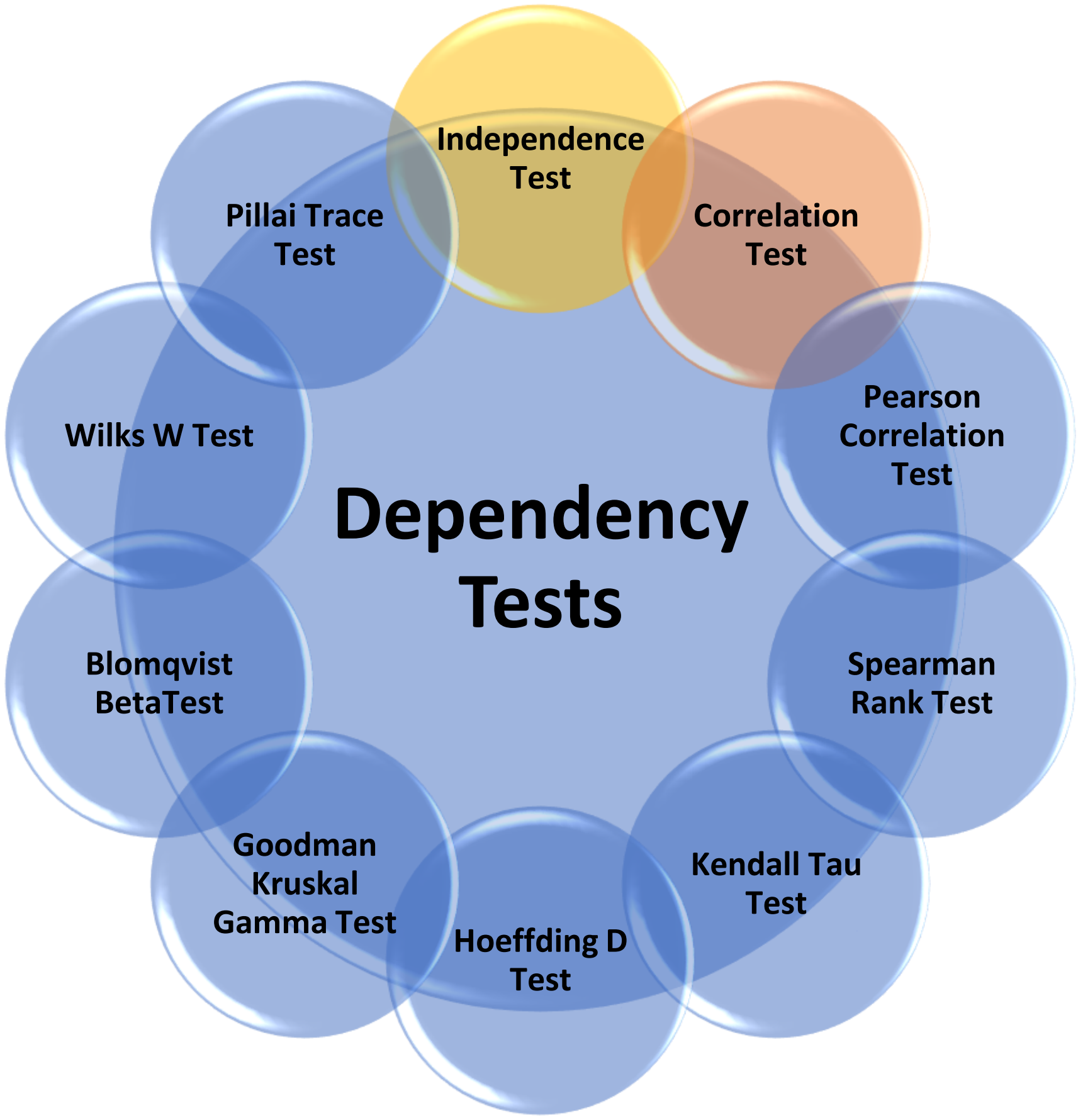

## IndependenceTest

| | |
|---|---|
| `IndependenceTest[v1,v2]` | tests whether the vectors v1 and v2 are independent. |
| `IndependenceTest[m1,m2]` | tests whether the matrices m1 and m2 are independent. |
| `IndependenceTest[…,"property"]` | returns the value of "property". |

## CorrelationTest

| | |
|---|---|
| `CorrelationTest[{{x1,y1},{x2,y2},…}]` | tests whether the correlation coefficient for a bivariate population is zero. |
| `CorrelationTest[{{x1,y1},{x2,y2},…}, ρ0]` | tests whether the correlation coefficient is ρ0. |
| `CorrelationTest[{{x1,y1},{x2,y2},…},{{u1,v1},{u2,v2},…}]` | tests whether the correlation coefficients for two populations are equal. |
| `CorrelationTest[…,"property"]` | returns the value of "property". |

*Mathematica Examples 21.58*

Input

```
(* The code is designed to test the independence between two sets of vectors (vector1
and vector2, and vector3 and vector4) using the "IndependenceTest" function. It
generates the vectors using the BinormalDistribution with different correlation
coefficients. The first set of vectors (vector1 and vector2) is generated from the
standard bivariate normal distribution with a correlation coefficient of 0,
indicating that the vectors should be independent. Thus, the P-value obtained from
the "IndependenceTest" should typically be large, as expected. The second set of
vectors (vector3 and vector4) is generated from the bivariate normal distribution
with a correlation coefficient of 0.6, indicating that there should be a dependency
between the vectors. Consequently, the P-value obtained from the "IndependenceTest"
should typically be small, indicating a rejection of the null hypothesis of
independence: *)

{vector1,vector2}=Transpose[
    RandomVariate[
     BinormalDistribution[0],
     100
     ]
    ];

{vector3,vector4}=Transpose[
    RandomVariate[
     BinormalDistribution[0.6],
     100
     ]
    ];

(* The P-values are typically large when the vectors are independent: *)
IndependenceTest[vector1,vector2]

(* The P-values are typically small when there are dependencies: *)
IndependenceTest[vector3,vector4]
```

Output 0.316656

Output 5.56862*10^-13

*Mathematica Examples 21.59*

Input

```
(* The code aims to perform an independence test on two datasets, data1 and data2,
and visualize the datasets using a ListPlot. The first part generates two datasets,
data1 and data2, each containing 100 random real numbers in the range[1,5]. These
datasets are created using the RandomReal function. The independence test is conducted
using the "IndependenceTest" function, which is applied to data1 and data2, and the
resulting P-value is stored in the variable pValue. The code then proceeds to display
the results. The Print function is used to show the calculated P-value from the
independence test. Additionally, the code performs the independence test again, but
this time with the option "TestDataTable" set to "All". This likely outputs a table
with more detailed information about the test results. Finally, a ListPlot is created
to visualize the relationship between data1 and data2. The two datasets are plotted
as points with a purple color and some opacity, making it easier to identify any
patterns or correlations between the two datasets: *)

(* Generate two datasets: *)
data1=RandomReal[{1,5},100];
data2=RandomReal[{1,5},100];

(* Perform the independence test: *)
pValue=IndependenceTest[data1,data2];
```

```
(* Display the results: *)
Print["Independence Test, P-Value: ",pValue];

IndependenceTest[data1,data2,{"TestDataTable",All}]

ListPlot[
 Transpose[{data1,data2}],
 PlotStyle->Directive[Purple,Opacity[0.9]],
 ImageSize->200
 ]
```

Output Independence Test, P-Value: 0.559367

Output

| | Statistic | P-Value |
|---|---|---|
| Blomqvist β | 0.04 | 0.548744 |
| Goodman-Kruskal γ | 0.0367677 | 0.58986 |
| Hoeffding | -0.00459182 | 0.910597 |
| Kendall τ | 0.0367677 | 0.587807 |
| Spearman Rank | 0.0590699 | 0.559367 |

Output

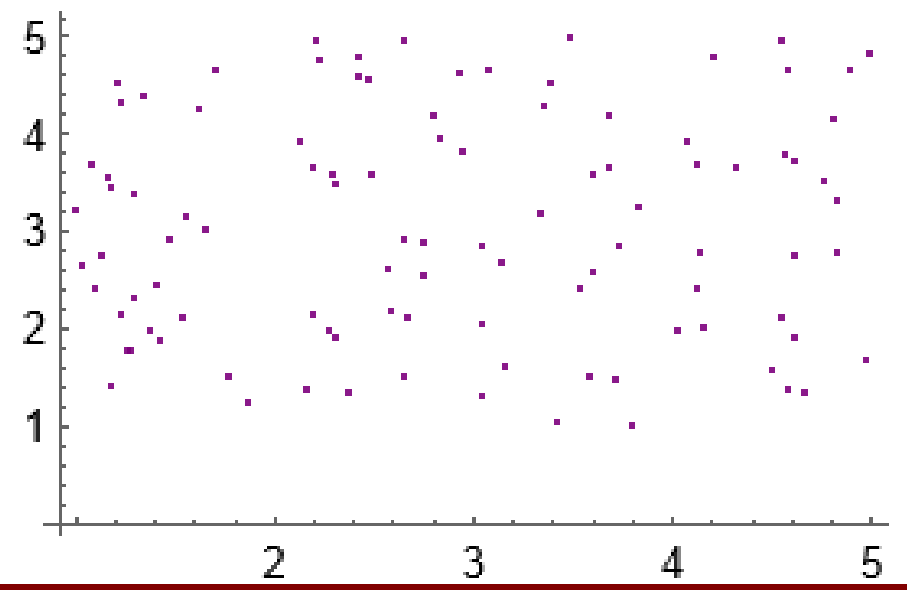

*Mathematica Examples 21.60*

Input

```
(* The code demonstrates a test for independence between two vectors, vector1 and
vector2, which are generated using RandomVariate with 500 samples each from the
standard normal distribution (NormalDistribution[]). The ListPlot function is then
used to visualize the relationship between vector1 and vector2 as a scatter plot.
The data points are displayed with a purple color and some opacity, which can be
helpful in identifying any patterns or correlations between the two vectors. Finally,
the code performs the independence test using the "IndependenceTest" function with
the options "TestDataTable" and "All" specified. The results of the test are stored
in the variable h: *)

vector1=RandomVariate[
    NormalDistribution[],
    500
    ];

vector2=RandomVariate[
    NormalDistribution[],
    500
    ];

datapoints=Transpose[{vector1,vector2}];

ListPlot[
 datapoints,
 PlotStyle->Directive[Purple,Opacity[0.9]],
 ImageSize->200
 ]
```

```
h=IndependenceTest[
  vector1,
  vector2,
  {"TestDataTable",All}
  ]
```

Output

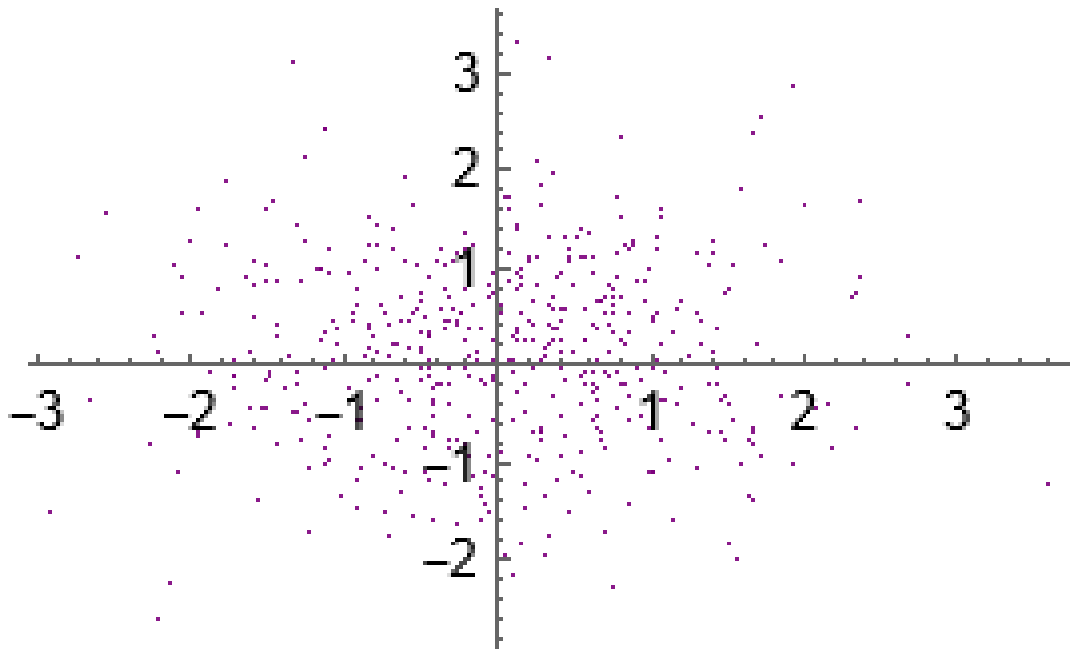

Output

| | Statistic | P-Value |
|---|---|---|
| Blomqvist β | 0.04 | 0.325183 |
| Goodman-Kruskal γ | -0.000529058 | 0.986105 |
| Hoeffding | -0.000849554 | 0.903562 |
| Kendall τ | -0.000529058 | 0.985892 |
| Pearson Correlation | 0.00287311 | 0.948903 |
| Pillai Trace | 8.25475*10^-6 | 0.948775 |
| Spearman Rank | 0.000491522 | 0.991253 |
| Wilks | 8.25475*10^-6 | 0.948775 |

### *Mathematica Examples 21.61*

Input

```
(* The code tests for independence between two matrices, datamatrix1 and datamatrix2, which are generated using RandomVariate and BinormalDistribution with a correlation coefficient of 0.7. The "IndependenceTest" function is then used to test the independence between datamatrix1 and datamatrix2. The test is performed at the 0.05 significance level, which means that if the p-value from the test is less than 0.05, there is evidence to reject the null hypothesis of independence. However, the code comment states that at this significance level, there is insufficient evidence to reject the null hypothesis of independence, implying that the two matrices are not significantly correlated: *)

datamatrix1=RandomVariate[
   BinormalDistribution[.7],
   100];
datamatrix2=RandomVariate[
   BinormalDistribution[.7],
   100
   ];

(* At the 0.05 level, there is insufficient evidence to reject independence: *)
IndependenceTest[
 datamatrix1,
 datamatrix2,
 {"TestDataTable",All}
 ]
```

Output

| | Statistic | P-Value |
|---|---|---|
| Blomqvist β | 0.60454 | 0.962558 |
| Kendall τ | 0.813984 | 0.936564 |

```
Pillai Trace    0.0145802   0.834051
Spearman Rank   0.832729    0.934007
Wilks           0.0145608   0.832509
```

***Mathematica Examples 21.62***

```
Input   (* The code aims to test whether the population correlation coefficient between two
        variables is zero or not, using the BinormalDistribution to generate two sets of data
        (data1 and data2) with different correlation coefficients. In the first part, data1
        is generated with 500 samples from the standard bivariate normal distribution,
        BinormalDistribution[0],where the correlation coefficient is 0. This ensures that
        the two variables in data1 are independent, and the true correlation between them is
        zero. Next, the code calculates the true population correlation coefficient for
        BinormalDistribution[0] using the Correlation function. It then performs a
        correlation test (CorrelationTest) on data1 to statistically test whether the sample
        correlation coefficient is significantly different from zero. The p-value obtained
        from the test will generally be large because the true correlation is indeed close
        to zero. In the second part, data2 is generated with 200 samples from the bivariate
        normal distribution, BinormalDistribution[0.6],where the correlation coefficient is
        0.6. This ensures that the two variables in data2 are correlated, and the true
        correlation between them is not zero. The code calculates the true population
        correlation coefficient for BinormalDistribution[0.6] using the Correlation function.
        It then performs a correlation test (CorrelationTest) on data2 to statistically test
        whether the sample correlation coefficient is significantly different from zero. The
        p-value obtained from the test will generally be small because the true correlation
        is not zero,indicating a rejection of the null hypothesis that the correlation is
        zero: *)

        data1=RandomVariate[
            BinormalDistribution[0],
            500
            ];
        data2=RandomVariate[
            BinormalDistribution[0.6],
            200
            ];

        Correlation[BinormalDistribution[0]][[1,2]]
        CorrelationTest[data1]

        Correlation[BinormalDistribution[0.6]][[1,2]]
        CorrelationTest[data2]

Output  0
Output  0.378423
Output  0.7
Output  3.39471*10^-179
```

***Mathematica Examples 21.63***

```
Input   (* The code compares the correlation coefficients from two samples (data1 and data2)
        and (data1 and data3) generated using the BinormalDistribution. The goal is to assess
        whether the correlations in these samples are similar or dissimilar. data1 is
        generated with 500 samples from the bivariate normal
        distribution,BinormalDistribution[0.4], which has a correlation coefficient of 0.4.
        This ensures that the variables in data1 are correlated. data2 is generated with 150
        samples from the same BinormalDistribution[0.4]. It also has a correlation
        coefficient of 0.4,similar to data1. data3 is generated with 250 samples from the
        standard bivariate normal distribution, BinormalDistribution[0],where the
        correlation coefficient is 0. This ensures that the variables in data3 are independent
        and uncorrelated. The code then performs correlation tests (CorrelationTest) between
```

```
        data1 and data2, and between data1 and data3, to compare their correlation
        coefficients: *)

        data1=RandomVariate[
            BinormalDistribution[0.4],
            500
            ];

        data2=RandomVariate[
            BinormalDistribution[0.4],
            150
            ];

        data3=RandomVariate[
            BinormalDistribution[0],
            250
            ];

        (* The p-values are typically large when the correlations are similar: *)
        CorrelationTest[data1,data2]

        (* The p-values are typically small when the correlations are dissimilar: *)
        CorrelationTest[data1,data3]

Output  0.875714
Output  1.19095*10^-9
```

## Further Reading (Some Statistical Built-in Functions in Mathematica)

RandomInteger
RandomReal
RandomPoint
RandomChoice
RandomSample
RandomVariate
SeedRandom
Gather
GatherBy
Sort
SortBy
Ordering
TakeLargest
TakeSmallest
DeleteDuplicates
Tally
Count
Counts
Accumulate
BinLists
BinCounts
HistogramList
Histogram
SmoothHistogram
Histogram3D
SmoothHistogram3D
DensityHistogram
SmoothDensityHistogram
Mean
Median
Commonest
TrimmedMean
WinsorizedMean
SpatialMedian
HarmonicMean
GeometricMean
RootMeanSquare
Quartiles
Quantile
BoxWhiskerChart
InterquartileRange
QuartileDeviation
MeanDeviation
StandardDeviation
Variance
TrimmedVariance
WinsorizedVariance
Moment
CentralMoment
FactorialMoment
Skewness
QuartileSkewness
Kurtosis
Tuples
Subsets
Subsequences
Permutations
Union
Intersection
Complement
IntersectingQ
DisjointQ
SubsetQ
ContainsAll
ContainsNone
ContainsAny
ContainsOnly
ContainsExactly
Factorial
Binomial
Multinomial
Distributed
Conditioned
Probability
NProbability
PDF
CDF
Expectation
NExpectation
MomentGeneratingFunction
CentralMomentGeneratingFunction
EstimatedDistribution
BinomialDistribution
PoissonDistribution
NegativeBinomialDistribution
GeometricDistribution
HypergeometricDistribution
DiscreteUniformDistribution
ProbabilityPlot
QuantilePlot
UniformDistribution
ExponentialDistribution
GammaDistribution
Normal Distribution
TransformedDistribution
MixtureDistribution
MarginalDistribution
Covariance
Correlation
BinormalDistribution
MultinomialDistribution
ChiSquareDistribution
StudentTDistribution
FRatioDistribution
Likelihood
LogLikelihood
EstimatedDistribution
FindDistributionParameters
FindDistribution
MeanCI
MeanDifferenceCI
VarianceCI
VarianceRatioCI
NormalCI
StudentTCI
ChiSquareCI
FRatioCI
LocationTest
LocationEquivalenceTest
VarianceTest
VarianceEquivalenceTest
DistributionFitTest
IndependenceTest
CorrelationTest